\documentclass[a4paper,12pt]{book}

\usepackage[utf8]{inputenc}
\usepackage[T1]{fontenc}
\usepackage[french]{babel}
\usepackage[babel=true]{csquotes}
\usepackage[top=3cm,bottom=3cm,left=3cm,right=3cm]{geometry}

\usepackage[french]{minitoc}
\mtcsetformat{minitoc}{tocrightmargin}{2.55em plus 1fil}

\usepackage{hyperref}
\usepackage{url}

\usepackage{fancyhdr}

\renewcommand{\sectionmark}[1]{}
\renewcommand{\chaptermark}[1]{}

\usepackage{titlesec, blindtext, color}
\definecolor{gray75}{gray}{0.75}
\newcommand{\hsp}{\hspace{20pt}}
\titleformat{\chapter}[hang]{\Huge\bfseries}{\thechapter\hsp\textcolor{gray75}{|}\hsp}{0pt}{\Huge\bfseries}

\usepackage{tabularx}
\usepackage{supertabular}
\usepackage{array}
\renewcommand{\arraystretch}{1.5}

\usepackage{amsmath, amsthm, amssymb}
\usepackage{xcolor}
\usepackage{tikz}
\usepackage{tabularx}
\usepackage{algorithmic}
\usepackage{algorithm}
\usepackage{array}
\usepackage{placeins}
\usepackage{eso-pic}
\usepackage{lscape}
\usepackage{lmodern}
\usepackage{graphicx}
\usepackage{titlesec}
\usepackage[round]{natbib}

\usepackage[font=footnotesize]{caption}
\usepackage[font=footnotesize]{subcaption}

\usepackage{enumitem}
\setitemize[1]{itemsep=10pt}
\setitemize[2]{itemsep=4pt}
\setitemize[3]{itemsep=4pt}
\setitemize[4]{itemsep=4pt}
\setlist{nosep,after=\vspace{\baselineskip}}

\begin{document}

\renewcommand{\labelitemi}{$\bullet$}

\dominitoc

\sloppy

\widowpenalty=10000
\clubpenalty=10000
\raggedbottom

\setcounter{mtc}{3}


\pagestyle{empty}

\begin{titlepage}

	\newgeometry{left=0cm,right=0cm,top=0cm}

	\AddToShipoutPicture*{
		\put(0,0){%
		\parbox[b][\paperheight]{\paperwidth}{%
		\vfill
		\centering
		\includegraphics[width=\paperwidth,height=\paperheight]{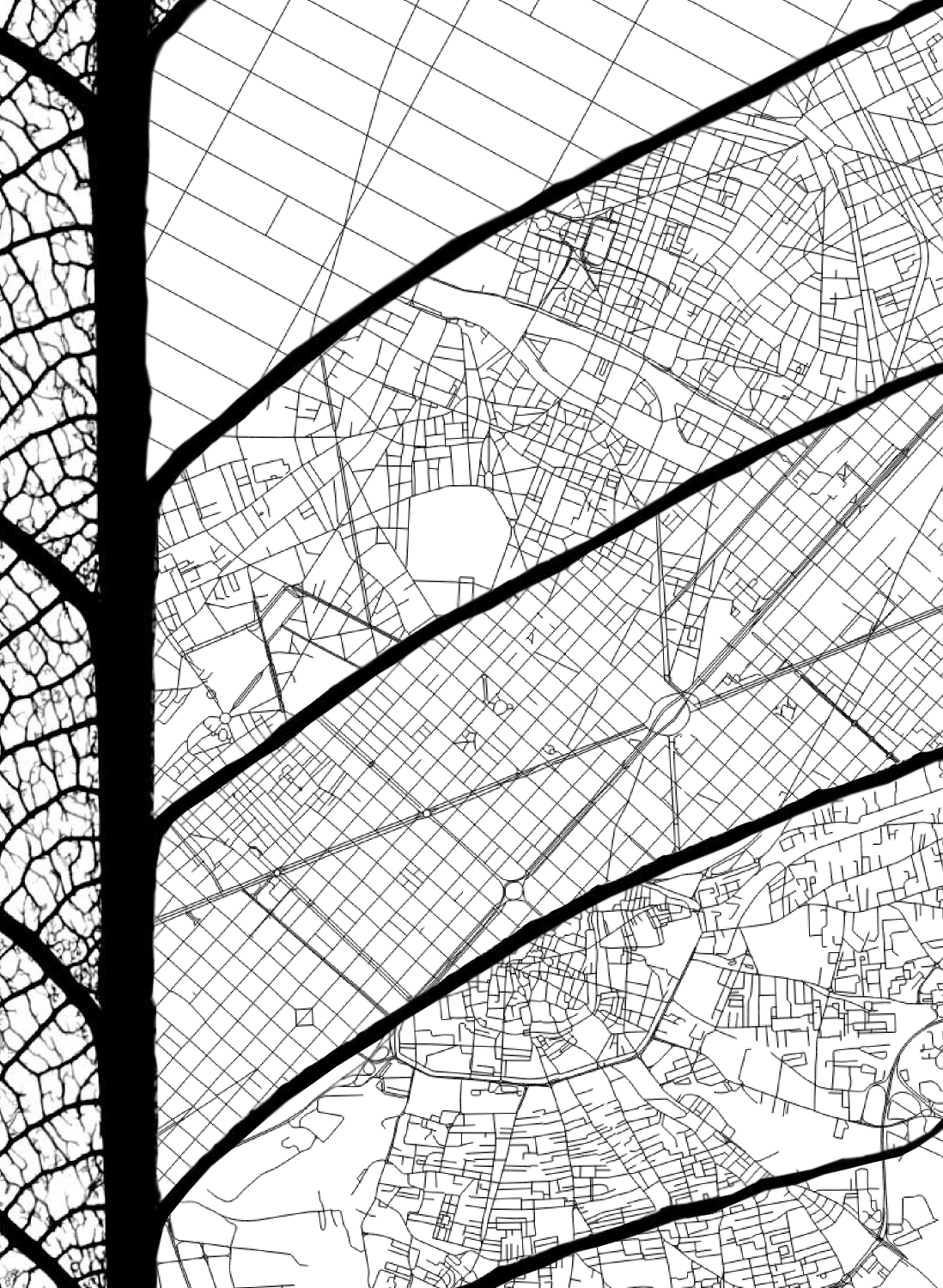}%
		\vfill
		}}	
	} 
	
	\flushleft
	
	\vspace*{10.8cm}
	
	\fcolorbox{white}{white}{

	\begin{minipage}{4cm}
		\centering
		\includegraphics[width=75pt]{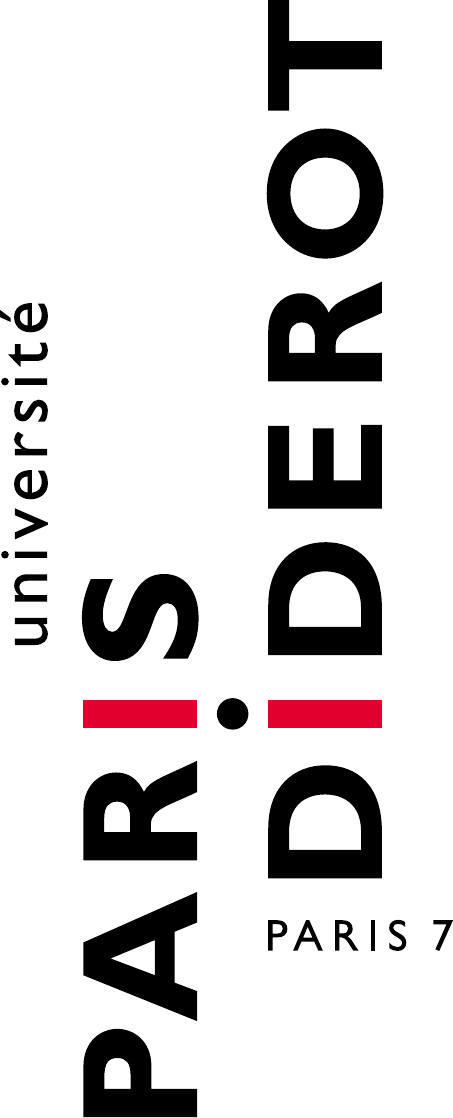}
	\end{minipage} \hfill
	\begin{minipage}{14cm}
		\sffamily
		\centering
		\vspace{1cm}
		
		{\Large Thèse de doctorat - spécialité Physique}
		
		\vspace{1cm}	
			
		{\Huge \textbf{Lire les Lignes de la Ville}}
		
		\vspace{0.5cm}	
			
		{\LARGE \textbf{Méthodologie de caractérisation des graphes spatiaux}}
		
		\vspace{1cm}
		
		{\huge Claire Lagesse}
		
		\vspace{1cm}
	
	\end{minipage} \hfill
	\begin{minipage}{5cm}
		\hspace{1cm}
	\end{minipage}	

	}
	
	\vfill 
	
	\restoregeometry

\end{titlepage}


\begin{titlepage}

	\sffamily

	\newgeometry{left=1cm,right=1cm,top=1cm, bottom=0cm}

	\center
	
	\begin{minipage}{0.45\textwidth}
		\flushleft
		UFR de physique
	\end{minipage}	\hfill
	\begin{minipage}{0.45\textwidth}
		\flushright
		École Doctorale 564 Physique en Île-de-France
	\end{minipage}
	\hrule
	\begin{minipage}{0.15\textwidth}
		\centering
		\includegraphics[width=0.9\textwidth]{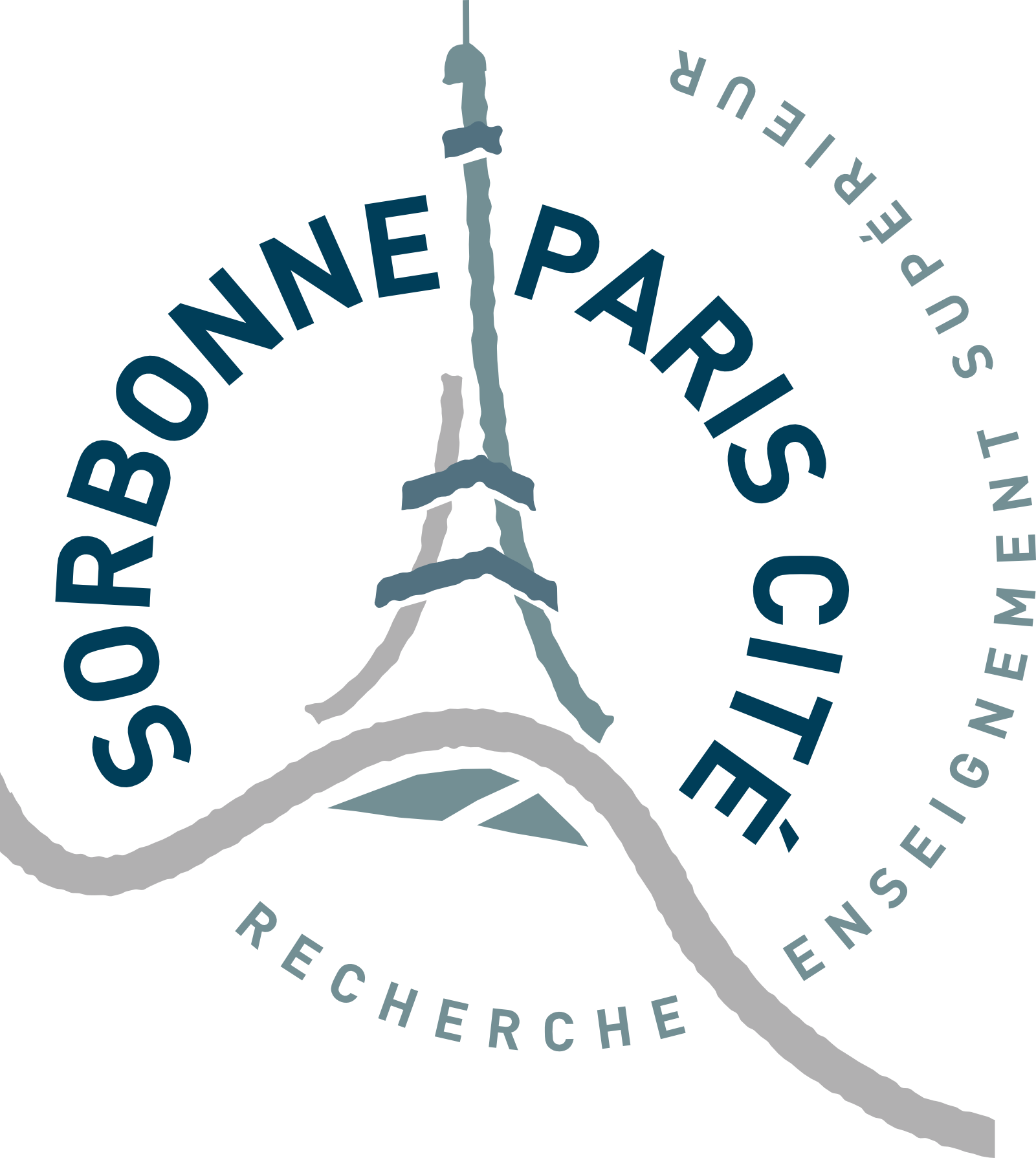}
	\end{minipage}
	\begin{minipage}{0.65\textwidth}
		\centering
		{\large 
		Université Paris-Diderot (Paris 7)
		
		Sorbonne Paris Cité
		}
	\end{minipage}
	\begin{minipage}{0.15\textwidth}
		\centering
		\includegraphics[width=0.5\textwidth]{images/couverture/Logo_P7.pdf}
	\end{minipage}
		
	\vspace*{1cm}

	\begin{minipage}{12cm}
		\centering
		
		{\huge Claire LAGESSE}		
		
		\vspace{0.5cm}
		\vspace{1cm}			
		{\large a présenté publiquement le 25 septembre 2015 ses travaux pour l'obtention du}
		
		\vspace{0.5cm}	
		
		{\LARGE titre de docteur - spécialité Physique}
	\end{minipage}
	
	\vspace*{2cm}
	
	\begin{minipage}{12cm}
		\centering
			
		{\LARGE \textbf{Lire les Lignes de la Ville}}
		
		\vspace{0.5cm}	
			
		{\Large \textbf{Méthodologie de caractérisation des graphes spatiaux}}
	
	\end{minipage}

	\vspace*{2cm}

	\begin{minipage}{8cm}
		\begin{center}

\renewcommand{\arraystretch}{0.5}		
		{\Large
		
		\begin{tabular}{l l}
			Directeur & Stéphane DOUADY \\
			Co-directrice & Patricia BORDIN \\
			Rapporteurs & Arnaud BANOS \\
			 & Alain BARRAT \\
			Examinateurs & Claude GRASLAND, président \\
			 & Jean-Loup GUILLAUME \\
			 & Renaud LAMBIOTTE \\
			 & Jacynthe POULIOT \\
		\end{tabular}				
		
		}
		
		\end{center}
	\end{minipage}
		
	\vspace*{2cm}
	
	\hrule
	\begin{minipage}[t]{3cm}
		\centering
		\includegraphics[width=0.5\textwidth]{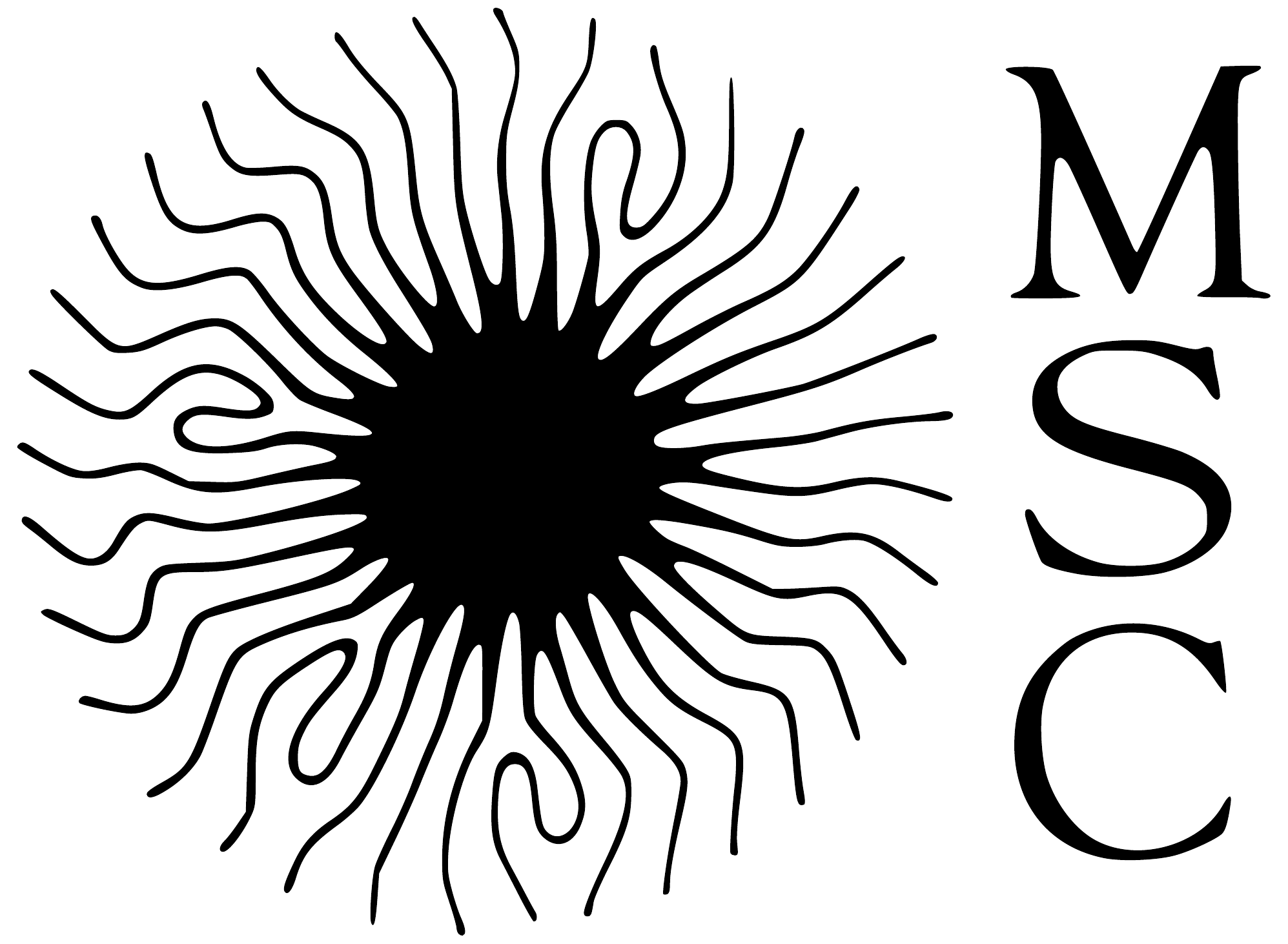}
		
		Laboratoire Matière et Systèmes Complexes
	\end{minipage} \hspace*{0.5cm} \hfill
	\begin{minipage}[t]{3cm}
		\centering
		\includegraphics[width=0.5\textwidth]{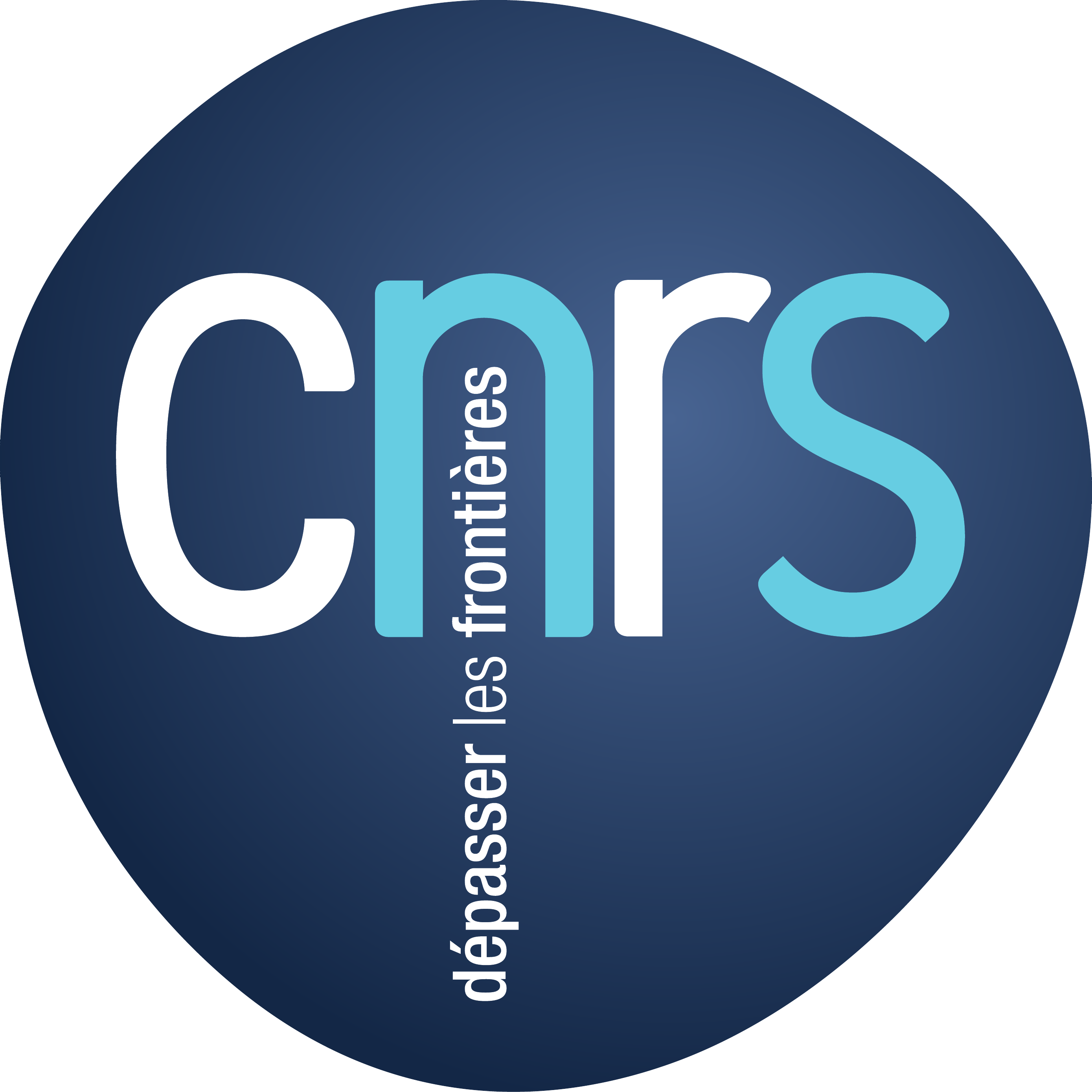}
		
		Centre Nationale de le Recherche Scientifique		
	\end{minipage} \hspace*{0.5cm} \hfill
	\begin{minipage}[t]{3cm}
		\centering
		\includegraphics[width=0.5\textwidth]{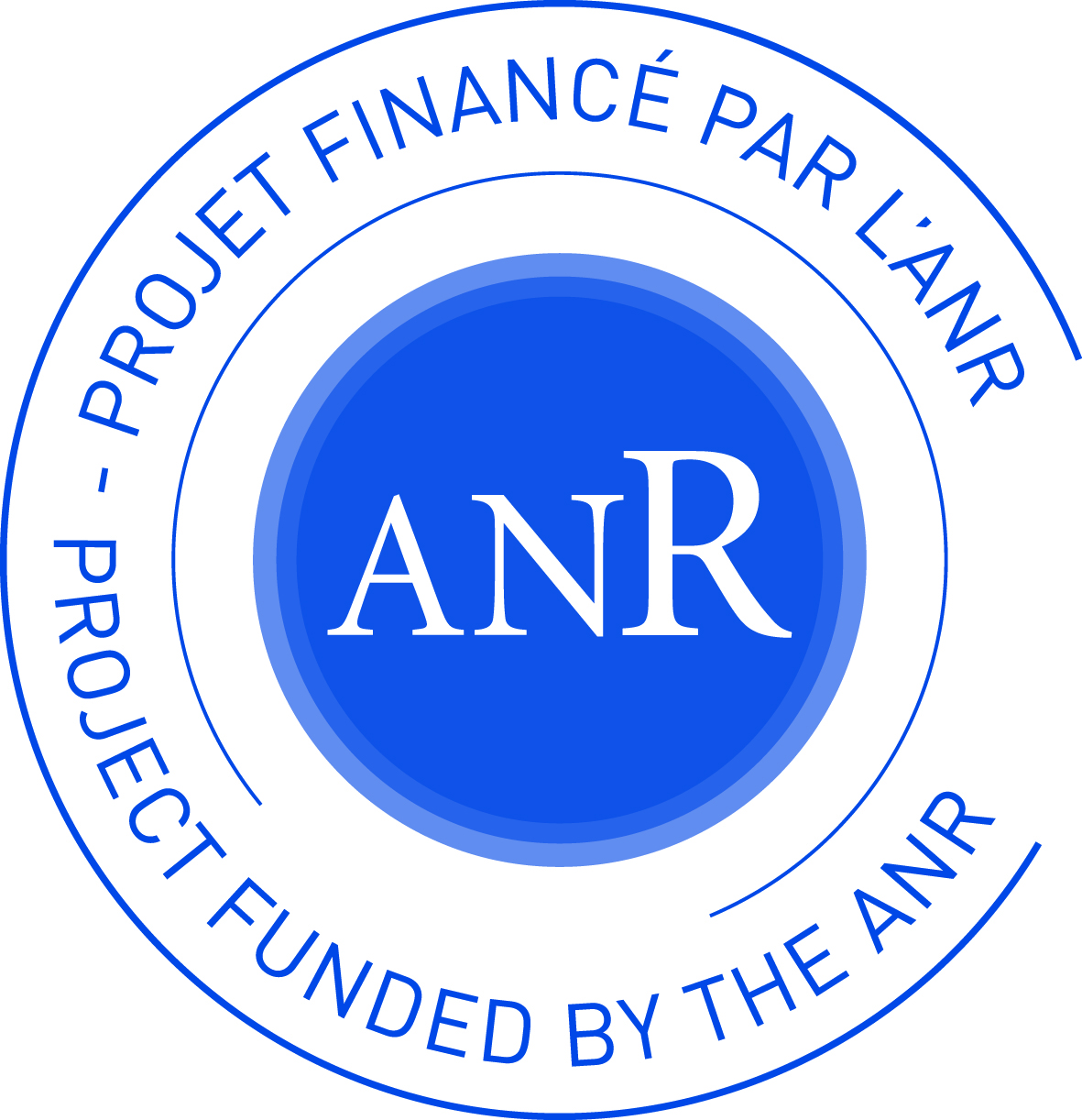}
		
		Agence Nationale de la Recherche
	\end{minipage} \hspace*{0.5cm} \hfill
	\begin{minipage}[t]{3cm}
		\centering
		\includegraphics[width=0.5\textwidth]{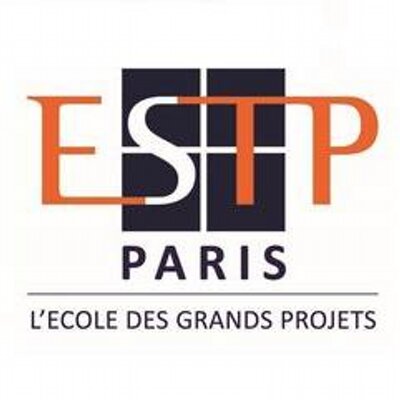}
		
		École Spéciale des Travaux Publics
	\end{minipage} \hspace*{0.5cm} \hfill
	\begin{minipage}[t]{3cm}
		\centering
		\includegraphics[width=0.5\textwidth]{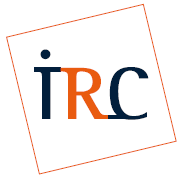}
		
		Institut de Recherche en Constructibilité
	\end{minipage}

	
	\restoregeometry

\end{titlepage}

\pagestyle{fancy}

\renewcommand{\arraystretch}{1.5}

\clearpage{\pagestyle{empty}\cleardoublepage}
~\vspace{2cm}
\section*{Résumé}
\addcontentsline{toc}{chapter}{Résumé \& Abstract}

La ville est un parfait exemple de système complexe. Elle regroupe une telle diversité de composants et d’interactions qu'il est impossible d'en faire une description exhaustive. Parmi sa pluralité, nous choisissons un élément qui structure son développement et son usage : le réseau de ses rues.

Suivant la piste initiée par les travaux en syntaxe spatiale, nous traduisons ce réseau sous forme de graphe. À partir de cette représentation, nous construisons un objet, la \textit{voie}, par des règles locales, indépendantes du sens de lecture du réseau, nous démarquant ainsi des travaux de \citep{porta2006network}. Cet objet se révèle être multi-échelle, rendant son analyse robuste au découpage du réseau. Nous étudions plusieurs indicateurs, certains existants et d'autres que nous proposons. Nous identifions plus particulièrement ceux qui apportent une information pertinente en établissant une grammaire de caractérisations non-redondantes. La \textit{voie} montre ainsi des propriétés spatiales particulières, en rendant équivalentes certaines analyses globales à d'autres locales. Cet objet géographique, construit à l'aide d'une paramétrisation appropriée, permet une étude approfondie des graphes spatiaux, indépendamment de l'emprise choisie.

L'application de cette méthodologie nous permet de mettre en évidence les propriétés particulières partagées par des graphes viaires de différents continents, et celles qui se retrouvent également dans d'autres réseaux spatiaux (biologiques, hydrographiques, etc). En nous concentrant sur les villes, nous montrons que les distributions de certains indicateurs suivent des comportements proches, malgré l'éloignement géographique et culturel.

Dans une approche diachronique, nous construisons une méthodologie de différentiation temporelle, permettant de quantifier les changements de proximité topologique entre les éléments du graphe. Cela nous permet d'avoir une première appréhension de la \textit{cinématique} de croissance des réseaux étudiés.

Cette recherche se termine par l'intégration de l'objet \textit{voie} et de ses indicateurs dans une approche qualitative. Nous montrons ainsi comment l'analyse de villes, à travers les propriétés topologiques et topographiques de leurs réseaux viaires, permet de retrouver une partie des contextes historiques et géographiques de leur construction. La mise en perspective de ces travaux, par une synthèse des échanges pluridisciplinaires qui les ont entourés, révèle le potentiel de leurs applications et ouvre sur de nouvelles pistes de recherches.

\newpage

\clearpage{\pagestyle{empty}\cleardoublepage}
~\vspace{2cm}
\section*{Abstract}

Cities can be seen as the epitome of complex systems. They arise from a set of  interactions and components so diverse that is almost impossible to describe them exhaustively. Amid this diversity, we chose an object which orchestrates the development and use of an urban area: the road network.

Following the established work on space syntax, we represent road networks as graphs. From this symbolic representation we can build a geographical object called the \textit{way}. Contrarily to previous approaches \citep{porta2006network}, the \textit{way} is defined by local rules independently from the direction in which the network is read. The resulting object is multi-scale, making its analysis robust against zoning. We evaluate several indicators, including some of our own, and identify those that give the most relevant and non-redundant information. The \textit{way}, appears to have unique spatial properties, revealing parallels between global and local analyses. This complex object, built upon appropriate parametrization, allows us to carry out deep analysis of spatial networks, independent from their borders.

With this methodology, we demonstrate how different road graphs, from various places in the world, show similar properties, and how some of those properties are also present in other networks (biological, hydrographical, \textit{etc.}). Focusing on cities, we point out that the distributions of some indicators have very similar behaviors despite geographic and cultural distances.

After considering the static properties of networks, we analyze how global characterization evolves through time. We establish a \textit{panchronic} database, allowing us to identify geometric changes over time. We define a model of temporal differentiation, where the change in accessibility of each object is highlighted. It is thus possible to have a first estimation of the growth \textit{kinematic} of the road networks studied.

This work culminates with the integration of the \textit{way} and its associated indicators into a qualitative approach. We show how such analysis, based on the topological and topographical properties of their road networks, allows us to trace back some aspects of the  historical and geographical contexts of city formation. Multidisciplinary discussions are synthesized to reveal the panel of research applications and future work.



\clearpage{\pagestyle{empty}\cleardoublepage}

\thispagestyle{empty}
~\vfill
\begin{flushright}
{\itshape À mes grand-mères,}
\end{flushright}
~\vfill

\clearpage{\pagestyle{empty}\cleardoublepage}
\chapter*{Un immense merci...}
\addcontentsline{toc}{chapter}{Remerciements}

Lorsque j'ai commencé ma thèse, je me suis sentie à la lisière d'une immense forêt vierge. Mon objectif était de la traverser, mais je n'avais en main qu'une minuscule machette et une torche de faible lueur. Autant dire que les débuts étaient difficiles : je n'avançais qu'à petits pas et je ne voyais pas grand chose. Heureusement, dans ce vaste environnement dense et périlleux, m'attendaient de belles rencontres. Ces personnes précieuses m'ont aidée à mieux regarder pour trouver mes chemins dans ce lieu inconnu. Elles m'ont appris à forger les outils dont j'avais besoin et à comprendre comment les utiliser. Sans elles, la forêt aurait eu raison de moi.

Mon premier grand merci va à mon directeur de thèse, Stéphane Douady. Pour ses mille et une idées, pour son humanité et son humilité. Ce sont, pour moi, les qualités d'un grand homme. Merci de m'avoir poussée à donner le meilleur de moi même (jusqu'à me demander de nouveaux calculs à une semaine du rendu final de mon manuscrit !). Merci d'avoir su trouver des mots apaisants dans les moments où la vie nous rappelle notre peu d'emprise sur son cours. Merci de m'avoir accompagnée en dehors des sentiers battus de la physique pour construire la jeune scientifique que je suis, ce fut un parcours d'une grande richesse.

Je tiens également à remercier ma co-directrice, Patricia Bordin. Lors de ma deuxième année d'école d'ingénieur, j'ai été initiée à la recherche par un projet qu'elle encadrait. Elle a donc été présente aux balbutiements de mes pas de chercheur qui firent naître ce goût si particulier qui ne m'a plus quittée depuis. Merci de m'avoir permis de prendre le recul nécessaire à la mise en perspective de mon travail de thèse, afin de mieux en comprendre les articulations. Merci d'avoir pris le temps de faire une relecture attentive de mon manuscrit, afin que chaque mot donne la bonne coloration au paragraphe qui l'entoure. Un grand merci également pour le partage de quelques anecdotes de vie...

J'ai eu la grande chance de construire mon travail au sein d'une équipe de recherche riche de la diversité des parcours et des connaissances des personnes qui la composent. Cette équipe, MorphoCity, m'a nourrie d'une multitude de regards et m'a permis d'apprendre des notions de champs disciplinaires qui m'étaient jusqu'alors inconnus : l’anthropologie, la sociologie, l'architecture, l'histoire, l'urbanisme, l'archéo-géographie. J'ai découvert ainsi tout l'enjeu de la pluridisciplinarité. Aider à une meilleure communication entre personnes parlant la même langue avec des mots différents a été un défi que j'ai adoré relever. Les divergences provoquent des débats passionnants dont l'aboutissement commun n'en est que plus précieux.

J'aimerais adresser ici un merci particulier à Philippe Bonnin, qui porte de toute sa passion cette équipe de recherche. Merci pour sa bienveillance, son énergie et sa sincérité. Merci pour sa présence et ses conseils qui m'ont permis de rendre possible une des choses qui me tenait le plus à cœur : parvenir à allier mes connaissances à celles des sciences humaines pour mettre en évidence la fertilité de leur association.

Un grand merci également à tous les autres membres de l'équipe, Clément-Noël Douady pour son regard poétique, Jean-Pierre Frey pour sa voix polémique, Pierre Vincent pour son approche pionnière en architecture. Merci à Romain Pousse et Clément Bresch, dont les stages m'ont permis de creuser plus loin en m'appuyant sur leurs recherches et questions. Merci à Wang Xi, de m'avoir épaulée dans les dernières semaines, les plus difficiles, de mon écriture. Merci à Babak Atashinbar et à Maryam Mansouri, ainsi qu'à leur famille, de m'avoir permis de faire un fabuleux voyage en Iran. Merci aux trois \enquote{Anes} pour l'atmosphère chaleureuse des moments partagés. Sans oublier tous les autres qui ont traversé mon chemin au sein de cette équipe, ils se reconnaîtront.

Un mot particulier pour Estelle Degouys, dont je ne sais plus bien si je dois la remercier ici ou dans le paragraphe dédié à mes amis... Il nous arrive parfois dans la vie de rencontrer des personnes qui nous comprennent d'une manière toute particulière. Ces rencontres sont rares et précieuses. Merci d'avoir été l'une d'entre elles.

À cette famille scientifique se sont ajoutées de nombreuses belles rencontres, au cours d'écoles thématiques ou de conférences. J'ai eu l'immense chance de rencontrer de grands chercheurs, avec lesquels j'ai pu échanger sur mes travaux. Ces discussions ont pour moi une valeur inestimable. Je tiens à leur exprimer ici toute ma gratitude. J'ai ainsi rencontré à ma première école d'été, au tout début de ma vie de chercheur, Arnaud Banos, que j'ai le grand plaisir d'avoir pour rapporteur de ce travail. Nous nous sommes recroisés ensuite, avec à chaque occasion quelques mots échangés, des conseils remplis de sagesse, des questions édifiantes de justesse. À la fin de ma deuxième année de thèse, en participant à l'école d'été organisée par l'institut de Santa Fe, j'ai eu la chance de rencontrer Sanders Bais. Échanger avec ce monsieur, d'une accessibilité déconcertante, fut un grand bonheur. J'ai pu également, durant ce voyage, discuter avec Geoffrey West. La vingtaine de minutes qu'il prit pour me donner son avis sur mon travail furent d'une richesse inouïe. Un peu plus tard, à Florence, j'ai pu discuter avec Tim Kohler, Chris Brunsdon et Martin Charlton qui m'ont guidée, chacun à leur manière, sur certains aspects de mes recherches. De manière plus approfondie, j'ai pu échanger avec Itzhak Benenson, dont le regard sur mon travail m'a donné un éclairage nouveau. Dans les mois qui suivirent, à Gand, j'ai rencontré Claire Lermercier et Ray Rivers dont les avis entrecroisés ont été très enrichissants. À Paris, Denis Eckert et Lena Sanders ont pris le temps de me recevoir pour discuter de mes recherches et répondre à mes questions. Merci à tous pour leur accessibilité, leurs conseils et leur passion.

D'autres belles rencontres se sont placées sur ma route, aux détours de séminaires, dont celles de Jean-Loup Guillaume et de Renaud Lambiotte, que j'ai le grand plaisir de réunir dans mon jury. Ce sont des échanges parfois anodins qui nous construisent. Merci à eux de m'avoir aidée à avancer, peut être même sans le savoir. Merci également à David Chavalarias et César Ducruet pour leur curiosité scientifique intarissable et les échanges que l'on a pu avoir.

Il y a beaucoup d'autres personnes, rencontrées au fil de ces trois années, que j'aimerais remercier. Certaines avec qui j'ai partagé une mezzanine, d'autres des moments clés ou de simples discussions, d'autres encore des projets aux colorations multiples. Il m'est impossible de tous les citer.

Une pensée à tous ceux de passage, plus ou moins long, au 9\textsuperscript{ème} étage du bâtiment Condorcet... Des plus anciens : Mathieu Génois, Sébastien Kosgodagan-Acharige, Kevin Sin Ronia, Juliette Pierre, Marie-Lys Beoutis... Aux plus récents : Olivier Lombard, Mathieu Rivière, Caroline Cohen, Tanguy Fardet... En passant par ceux avec qui nous nous sommes serrés les coudes tout au long de ces trois années : Amandine Garcia et Renaud Renault ! Merci pour leur aide et leur amitié ! La 911A n'aurait pas eu la même saveur sans eux...

Merci à la \enquote{\textit{Santa Fe team}} pour tous les moments scientifiques - et moins scientifiques - partagés, et ceux qui se profilent à l'horizon. \textit{Special thanks to} Diana LaScala-Gruenewald, Morgan Edwards, Fahad Khalid, Alireza Goudarzi, Alberto Antonioni, Leto Peel \textit{and} Massimo Stella.

Merci à tous les chercheurs que j'ai pu écouter, et qui m'ont beaucoup appris. J'ai une pensée particulière pour Melanie Mitchell, Aaron Clauset et Mark Newman.

Merci aux maîtres de conférence de l'IUT d'informatique de Villetaneuse de m'avoir accompagnée dans ma première expérience d'enseignement à l'université. Je pense notamment à Pierre Gérard, Franck Butelle et Jean-Christophe Dubacq, qui ont été présents tout au long de mon monitorat.

Je tiens aussi à remercier Alain Barrat, Claude Grasland et Jacynthe Pouliot, qui ont accepté de faire partie de mon jury de thèse. Leurs trois regards, issus de disciplines très différentes, posés sur mon travail, seront, j'en suis sûre, sources de discussions passionnantes.

Je garde pour la fin les piliers de tout cet édifice : ma famille et mes amis. Ils m'ont soutenue quelles que soient mes décisions, ont accepté avec bienveillance la mise entre parenthèse de certains moments et ont compris avec indulgence ma volonté de faire une thèse et de m'y consacrer pleinement. Merci à eux de m'avoir écoutée et encouragée. Merci d'avoir été là, tout simplement.

Un grand merci tout particulier à Cécile, pour la relecture orthographique complète de mon manuscrit (nous sommes venues à bout de nos interrogations sur les compléments du nom !). Et un autre grand merci à Timothée, qui m'a permis de faire de ma thèse un bel objet. Merci également à Véronique et à la fameuse Ricoh qui ont rendu l'entreprise possible. Se sentir entourée dans les moments critiques est d'un grand réconfort.

Un immense merci à mon frère et à mes parents. Pour avoir toujours été présents lorsque j'avais besoin d'eux, et pour m'avoir donné la volonté d'aller toujours plus loin. Merci à ma maman d'avoir relu chaque mot de ma thèse pour m'apporter toute l'aide possible. Merci pour les paroles réconfortantes, pour les gestes apaisants, pour l'écoute attentive et la présence chaleureuse.

 \textit{Merci à Kuzco pour la ronron-thérapie.}

Enfin, mon dernier merci, incommensurable, revient à celui qui partage ma vie... Pour avoir été là chaque jour, les bons comme les moins bons. Pour m'avoir offert son soutien infaillible et inconditionnel. Pour son aide, inestimable. Merci d'avoir absolument tout partagé avec moi, j'y ai puisé ma force. Ce moment de vie n'aura pas été de tout repos mais dessine un futur aux belles couleurs.

\vfill

\textit{Une pensée pour Bornia, qui était persuadée que je serais un jour docteur, sans se douter que ce ne serait pas en médecine... Merci d'avoir cru en moi.}

\newpage
\null
\thispagestyle{empty}
\newpage



\clearpage{\pagestyle{empty}\cleardoublepage}
\addcontentsline{toc}{chapter}{Sommaire}
\markboth{Sommaire}{Sommaire}
\tableofcontents

\clearpage{\pagestyle{empty}\cleardoublepage}
\chapter*{Glossaire}
\markboth{Glossaire}{Glossaire}
\addcontentsline{toc}{part}{Glossaire}
\section*{Sigles}

\textbf{ANR} : Agence Nationale de la Recherche

\textbf{CNRS} : Centre National de la Recherche Scientifique

\textbf{ESTP} : École Spéciale des Travaux Publics

\textbf{IGN} : Institut National de l'Information Géographique et Forestière 

\textbf{IRC} : Institut de Recherche en Constructibilité

\textbf{MSC} : laboratoire Matière et Systèmes Complexes

\textbf{OSM} : OpenStreetMap

\textbf{SIG} : Système d'Information Géographique

\textbf{STIF} : Syndicat des Transports d'Île-de-France

\section*{Notations}

\subsection*{Pour décrire un réseau}

$L_{tot}$ : longueur totale

$L_{voie}$ : longueur d'une voie

$L_{arc}$ : longueur d'un arc



$N_{arcs}$ : nombre d'arcs

$N_{arcs}(voie)$ : nombre d'arcs dans une voie

$N_{sommets}$ : nombre de sommets

$N_{voies}$ : nombre de voies

\subsection*{Statistiques}

$\overline{X}$ ou $\mu(X)$ : moyenne

$\sigma(X)$ : écart-type

$max \vert X \vert$ : maximum en valeur absolue

\section*{Représentation sous forme de graphe}

 \textbf{Graphe} $G = (S,A)$\\
  Abstraction mathématique ayant pour but de modéliser un ensemble d'objets et leurs interactions par une formalisation simple composée de \emph{sommets} et d'\emph{arcs}. Nous pourrons parler de graphes comme de \emph{réseaux}.
 	
 \textbf{Sommet} $S$\\
 Appelé aussi \emph{Nœud} ou \emph{Acteur} (sociologie). Il est l'unité fondamentale du graphe. Il symbolise un élément unitaire, pouvant avoir de nombreux attributs.
  		
 \textbf{Arc} $A$\\
  Appelé aussi \emph{Lien} ou \emph{Arête}. Il établit une connexion entre deux sommets. Il peut être orienté (d'un sommet vers un autre) ou non-orienté. Il est également possible de lui attribuer une valuation.
  
  \textbf{Face} $A$\\
  Appelée aussi \emph{Région} ou \emph{Domaine}. Surface pour un graphe représenté sous forme planaire délimitée par ses arcs.
   
 \textbf{Degré} $d_S >= 0$\\
  Attribut du sommet. Il équivaut au nombre d'arcs connectés au sommet. Un sommet peut avoir un \emph{degré entrant} (\textit{in-degree}) correspondant au nombre d'arcs connectés à ce sommet et orientés vers lui. De même, nous pouvons établir le \emph{degré sortant} (\textit{out-degree}) avec les arcs connectés au sommet et orientés vers l'extérieur.
  
  \textbf{Graphe Orienté} \\
  Graphe ne contenant que des arcs orientés.
  
  \textbf{Graphe Planaire} \\
  Graphe pour lequel il existe une configuration en deux dimensions telle qu'aucun arc ne croise un autre arc.  
  
  \textbf{Graphe dual} \\
  Graphe B issu d'un graphe A où les sommets de B correspondent aux faces de A et où il existe un arc entre deux sommets de B si les deux faces de A sont séparées par un même arc.
  
  \textbf{\textit{Line Graph}} \\
  Appelé aussi Graphe adjoint. Graphe B issu d'un graphe A où les sommets de B correspondent aux arcs de A et où il existe un arc entre deux sommets de B si les deux arcs de A aboutissent à un même sommet. Dans cette configuration, le graphe A est appelé \emph{graphe primal}.
  
  \textbf{Hyper-Arc} $HA$\\
Appelé aussi Hyper-Lien. Arc pouvant joindre plus de deux sommets.

\textbf{Hyper-Graphe} $HG$\\
Graphe constitué d'hyper-arcs.
 
 \textbf{Chemin dans un graphe} $C_{(o,d)}$\\
 Ensemble d'arcs contigus traversés pour aller d'un sommet origine ($o$) à un sommet destination ($d$).
    
 \textbf{Graphe Connexe}\\
  Graphe n'admettant que des sommets entre lesquels il est possible d'établir un chemin (tous les sommets sont liés entre eux).
   
 \textbf{Chemin Eulérien} $C_{E(o,d)}$\\
 Sous-partie du graphe n'admettant que des sommets entre lesquels il est possible d'établir un chemin où chaque arc n'est traversé qu'une unique fois. Si le sommet initial du chemin et le même que celui final, nous parlerons de \textit{Cycle Eulérien}.
   
 \textbf{Chemin Hamiltonien} $C_{H(o,d)}$\\
  Sous-partie du graphe n'admettant que des sommets entre lesquels il est possible d'établir un chemin où chaque sommet n'est traversé qu'une unique fois. Si le sommet initial du chemin est le même que celui final, nous parlerons de \textit{Cycle Hamiltonien}.
   
%
  
 \textbf{Matrice d'adjacence} \\
  Matrice symétrique où chaque ligne (et donc chaque colonne) représente un sommet du graphe. L'élément de la matrice $M(i,j)$ vaut 1 s'il existe un arc entre les sommets $i$ et $j$, sinon l'élément vaut 0. La représentation sous forme matricielle nous permet de lister exhaustivement tous les sommets du graphe et leurs connexions.

\clearpage{\pagestyle{empty}\cleardoublepage}
\chapter*{Guide de Lecture}
\markboth{Guide}{Guide}
\addcontentsline{toc}{part}{Guide de Lecture}

\setlength{\parskip}{5pt}
\newgeometry{left=2cm,right=2cm,top=2cm,bottom=2cm}

\section*{Notions utilisées sur les graphes}

\paragraph{Graphe primal}

Graphe où chaque arc relie deux sommets.
 
\paragraph{Graphe spatial}

Graphe où les sommets et les arcs ont des coordonnées.

Dans un graphe primal spatial :
\begin{itemize}
\item sommet : intersection
\item arc : tronçon entre deux intersections
\end{itemize}

\paragraph{Hypergraphe primal}

Graphe où les arcs peuvent relier plus de deux sommets.

Dans un hypergraphe primal spatial :
\begin{itemize}
\item sommet : intersection
\item voie : alignement d'arcs continus respectant les paramètres fixés
\end{itemize}

\begin{figure}[h]
	\begin{minipage}[c]{0.6\linewidth}
	\centering
		\includegraphics[width=\linewidth]{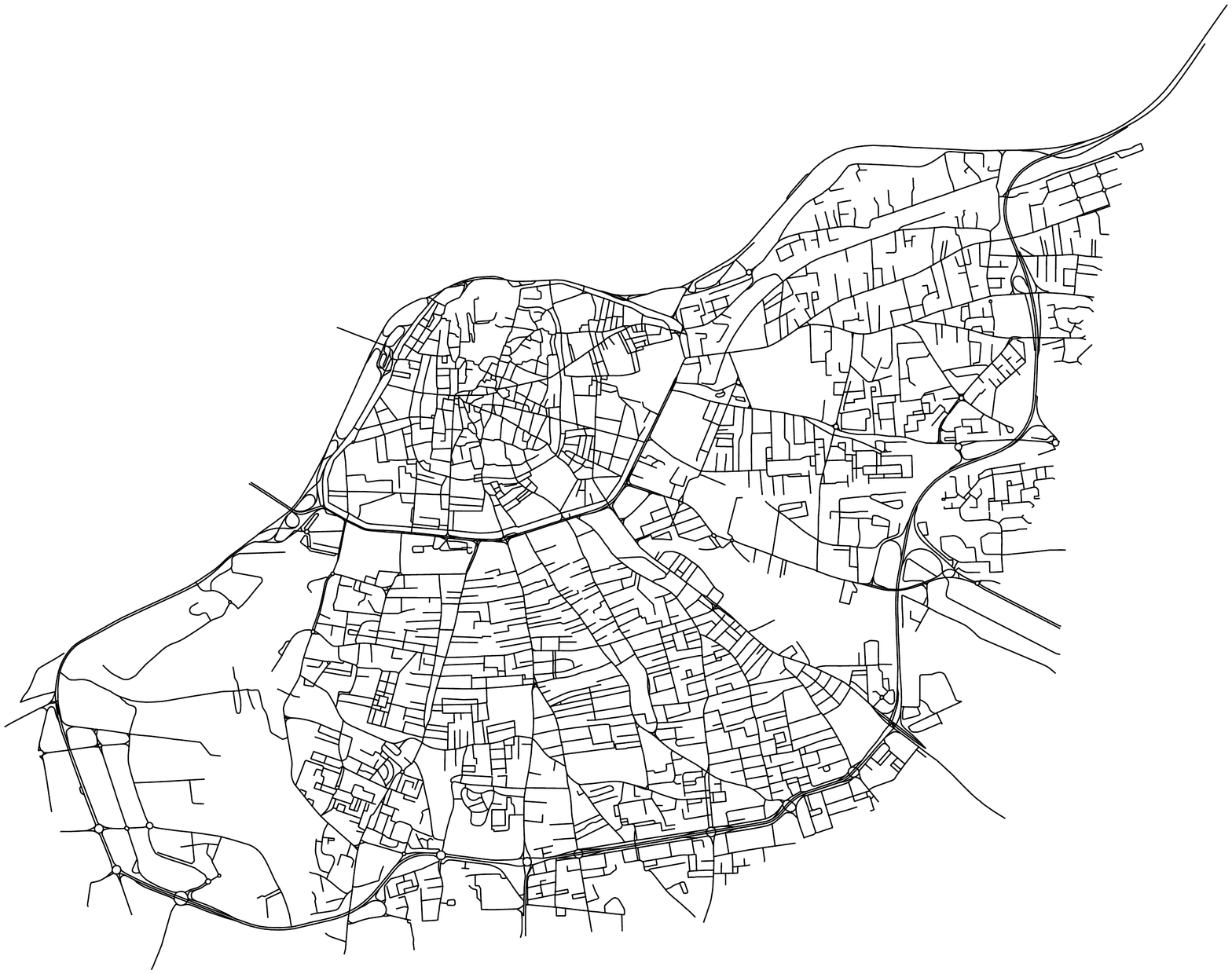}
	\end{minipage}
	\hfill
	\begin{minipage}[c]{0.3\linewidth}
		\centering
		\begin{itemize}
		\item 3127 sommets
		\item 4852 arcs
		\item 1493 voies
		\end{itemize}
	\end{minipage}
	\caption{Graphe viaire d'{\large \textbf{Avignon}}, utilisé dans l'ensemble du guide}
\end{figure}

\begin{figure}[h]
    \centering
    \begin{subfigure}[t]{0.31\textwidth}
    \centering
        \includegraphics[width=\linewidth]{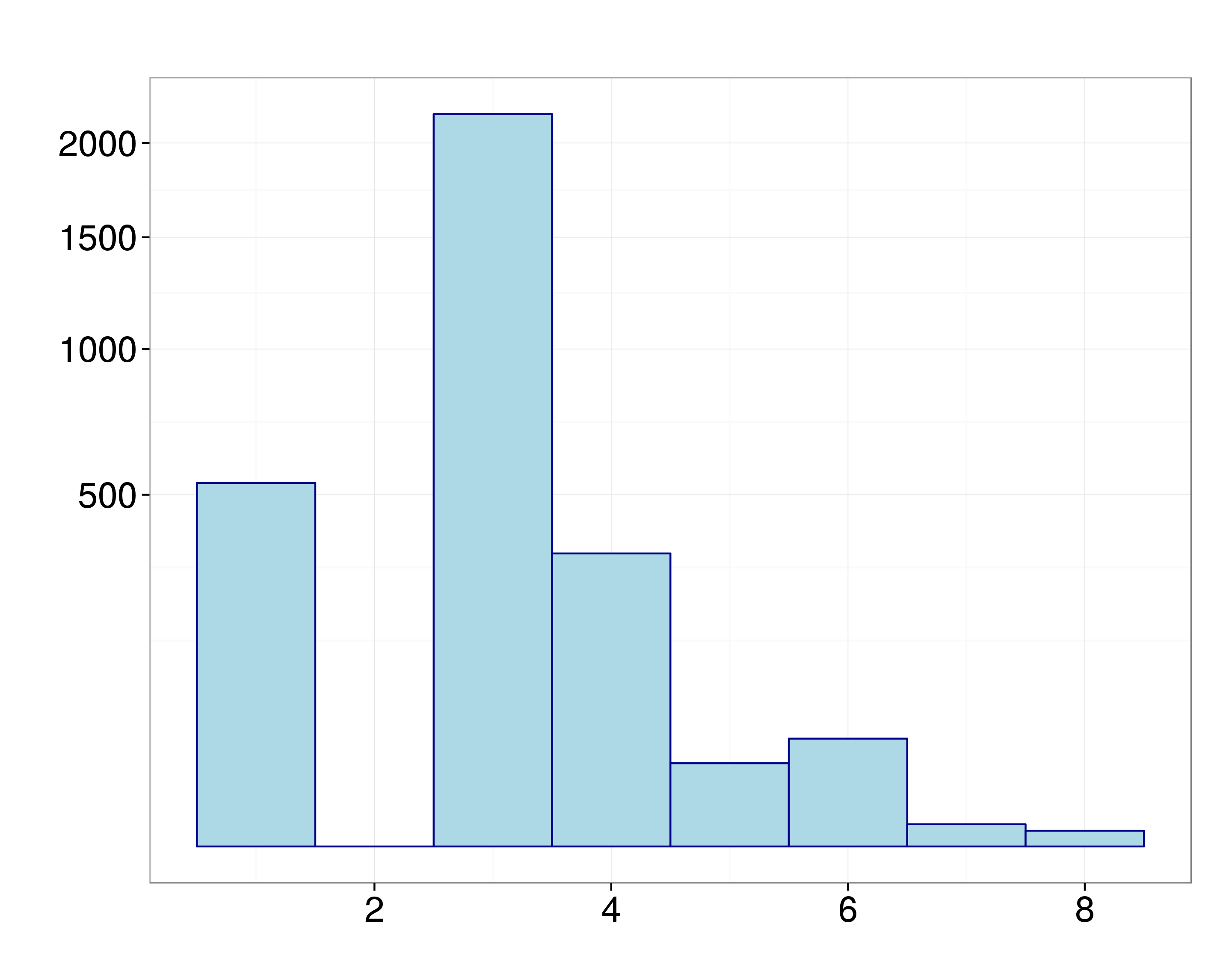}
        \caption{Degré des sommets}
    \end{subfigure}
	~
    \begin{subfigure}[t]{0.31\textwidth}
    \centering
        \includegraphics[width=\linewidth]{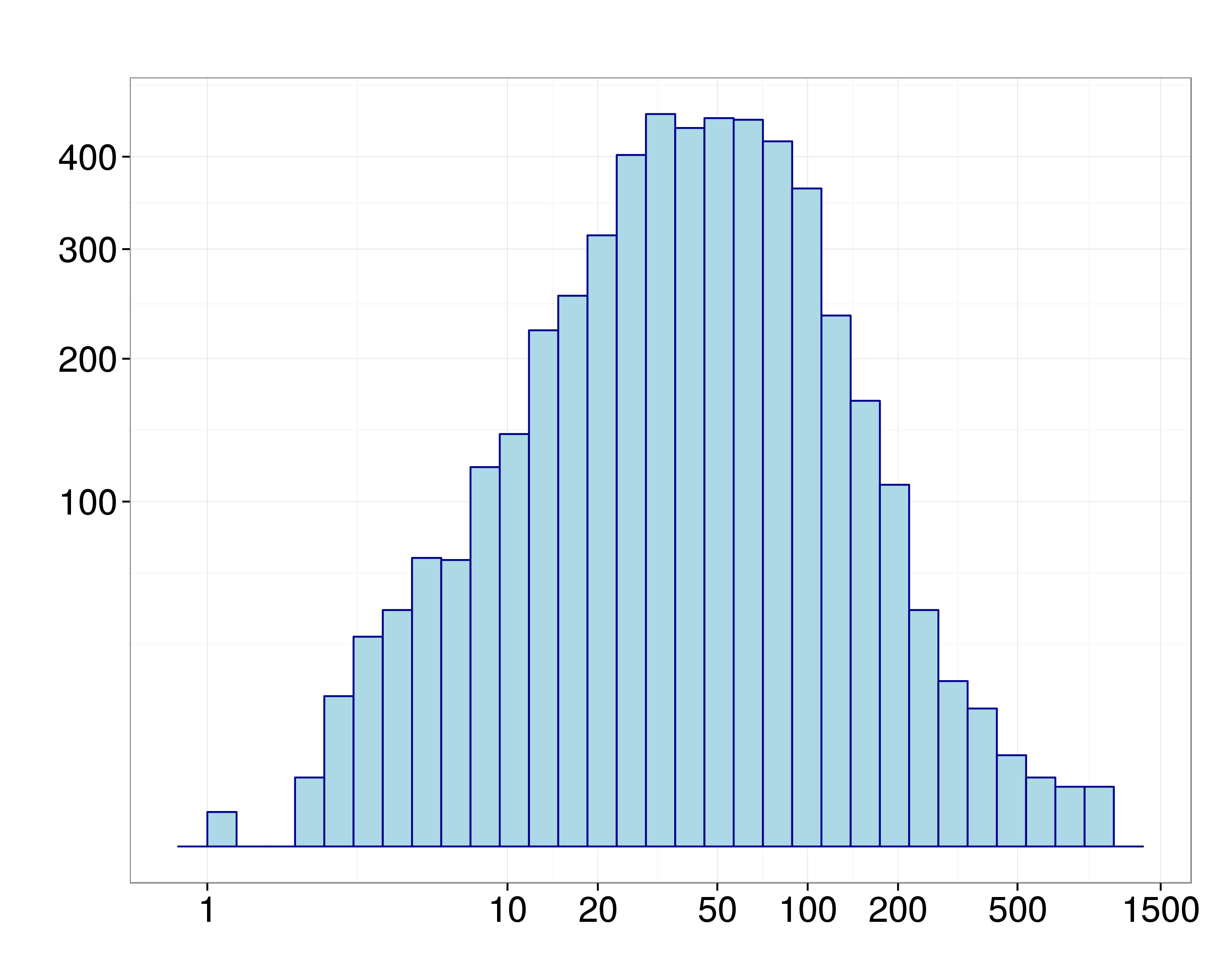}
        \caption{Longueur des arcs}
    \end{subfigure}
	~
    \begin{subfigure}[t]{0.31\textwidth}
    \centering
        \includegraphics[width=\linewidth]{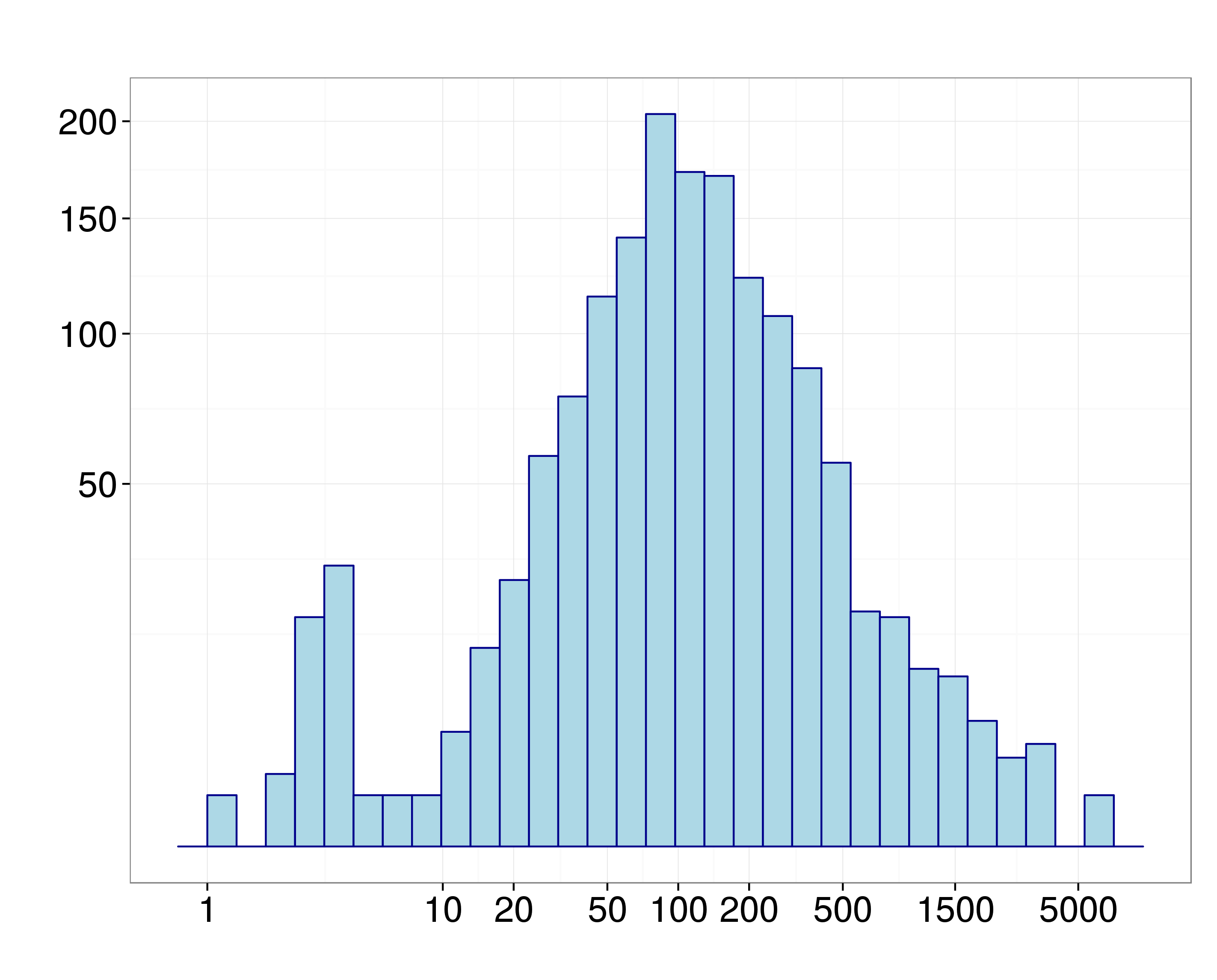}
        \caption{Longueur des voies}
    \end{subfigure}
	\caption{Distribution des éléments du graphe.}
\end{figure}

\clearpage
\paragraph{Line graph}

Graphe $LG(S',A')$ construit à partir d'un graphe ou d'un hypergraphe primal $G(S,A)$, où :
\begin{itemize}
\item sommet de $LG$ ($S'$) : arc ou voie de $G$ ($A$)
\item arc de $LG$ ($A'$) : relie deux sommets de $LG$ dont les arcs ou les voies correspondantes s'intersectent dans $G$
\end{itemize}

\begin{figure}[h]
    \centering
    \begin{subfigure}[t]{0.45\textwidth}
    \centering
        \includegraphics[scale=0.8]{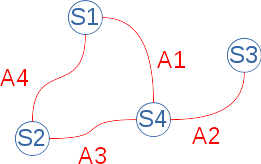}
        \caption{Graphe primal}
    \end{subfigure}
	~
    \begin{subfigure}[t]{0.45\textwidth}
    \centering
        \includegraphics[scale=0.8]{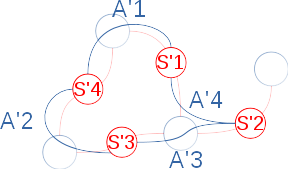}
        \caption{Line graphe associé}
    \end{subfigure}
	\caption{Construction du line graph}
\end{figure}

Un \textit{tournant} dans nos travaux équivaut à un changement de sommet sur le \textit{line graph}.

\paragraph{Distances}

\begin{enumerate}

\item \textit{Distance géodésique} : Nombre d'arcs contenu dans le plus court chemin entre deux sommets.

\item \textit{Distance euclidienne} : Distance \enquote{à vol d'oiseau} entre deux points du réseau, sans considérer la géométrie des arcs. Elle correspond à la ligne droite entre deux points.

\item \textit{Distance géographique} : Distance \textit{géodésique} appliquée au graphe primal dont les arcs sont pondérés par leurs longueurs.
Est mesurée ici la distance métrique parcourue. Nous appelons le chemin associé à cette distance le \emph{chemin le plus court}.

\item \textit{Distance topologique} (notée $d_{simple}$) : Distance \textit{géodésique} appliquée au \textit{line graph} dont les arcs ne sont pas pondérés. Est mesurée ici la distance en nombre d'éléments traversés. Chaque passage par un sommet du \textit{line graph} est équivalent à un changement d'arc ou de voie. Nous appelons le chemin associé à cette distance le \emph{chemin le plus simple}.

\end{enumerate}

\section*{Indicateurs sur les graphes}

\subsection*{Indicateurs locaux}

\FloatBarrier
\paragraph{Connectivité d'une voie} Nombre d'arcs du graphe qu'elle intersecte

\begin{equation}
 connectivite(v_{ref}) = \sum_{s \in v_{ref}} Card(a / [(s \in a) \wedge (a \notin v_{ref})])
\end{equation}

\clearpage

\paragraph{Degré d'une voie}

Nombre de voies de l'hypergraphe qu'elle intersecte.

\begin{equation}
 degre(v_{ref}) = Card(v \in G / v \cap v_{ref})
\end{equation}

\begin{figure}[h]
    \centering
    \begin{subfigure}[t]{0.50\textwidth}
    \centering
        \includegraphics[width=\linewidth]{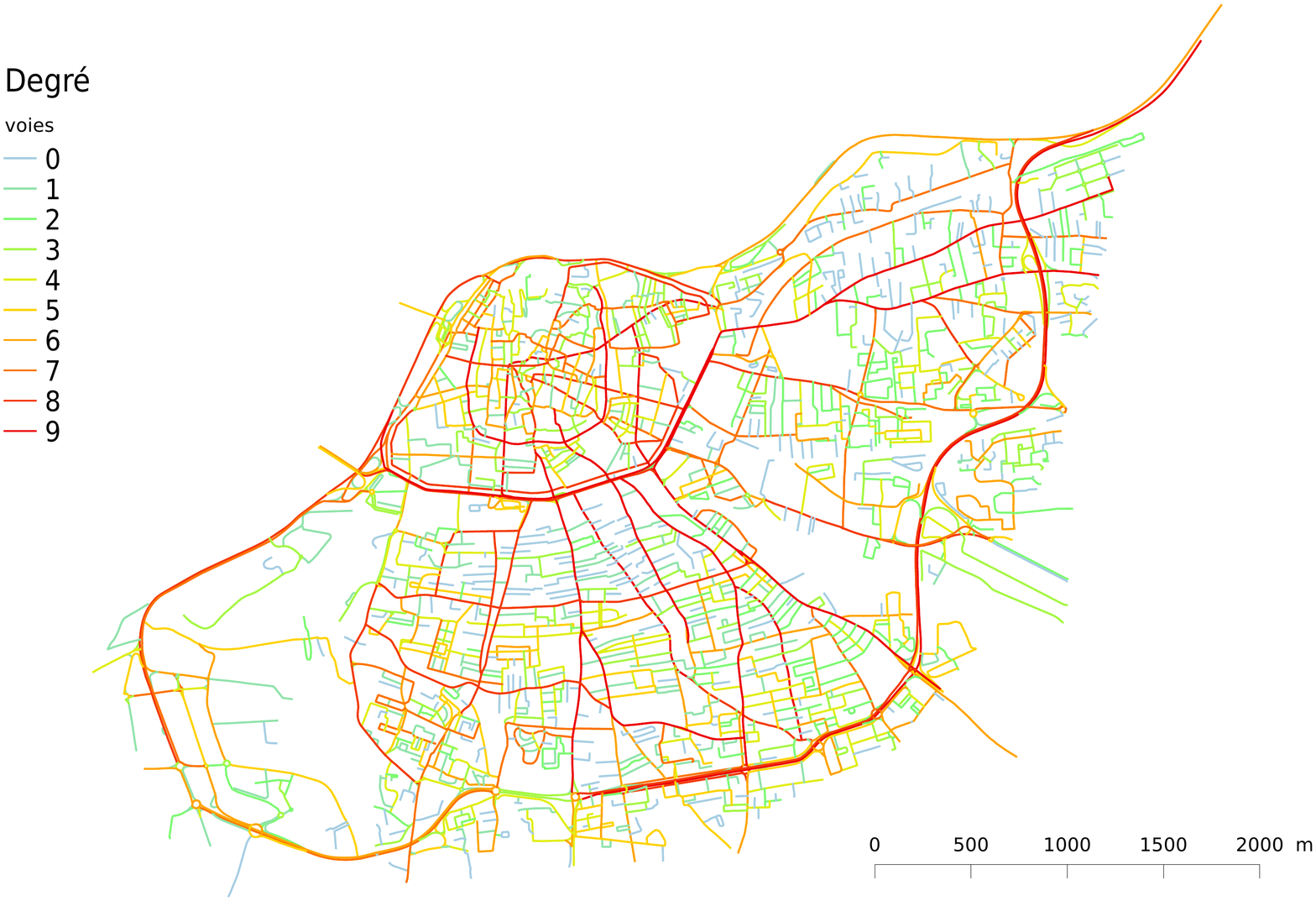}
        \caption{Représentation cartographique en 10 classes de longueur équivalente.}
    \end{subfigure}
	~
    \begin{subfigure}[t]{0.45\textwidth}
    \centering
        \includegraphics[width=\linewidth]{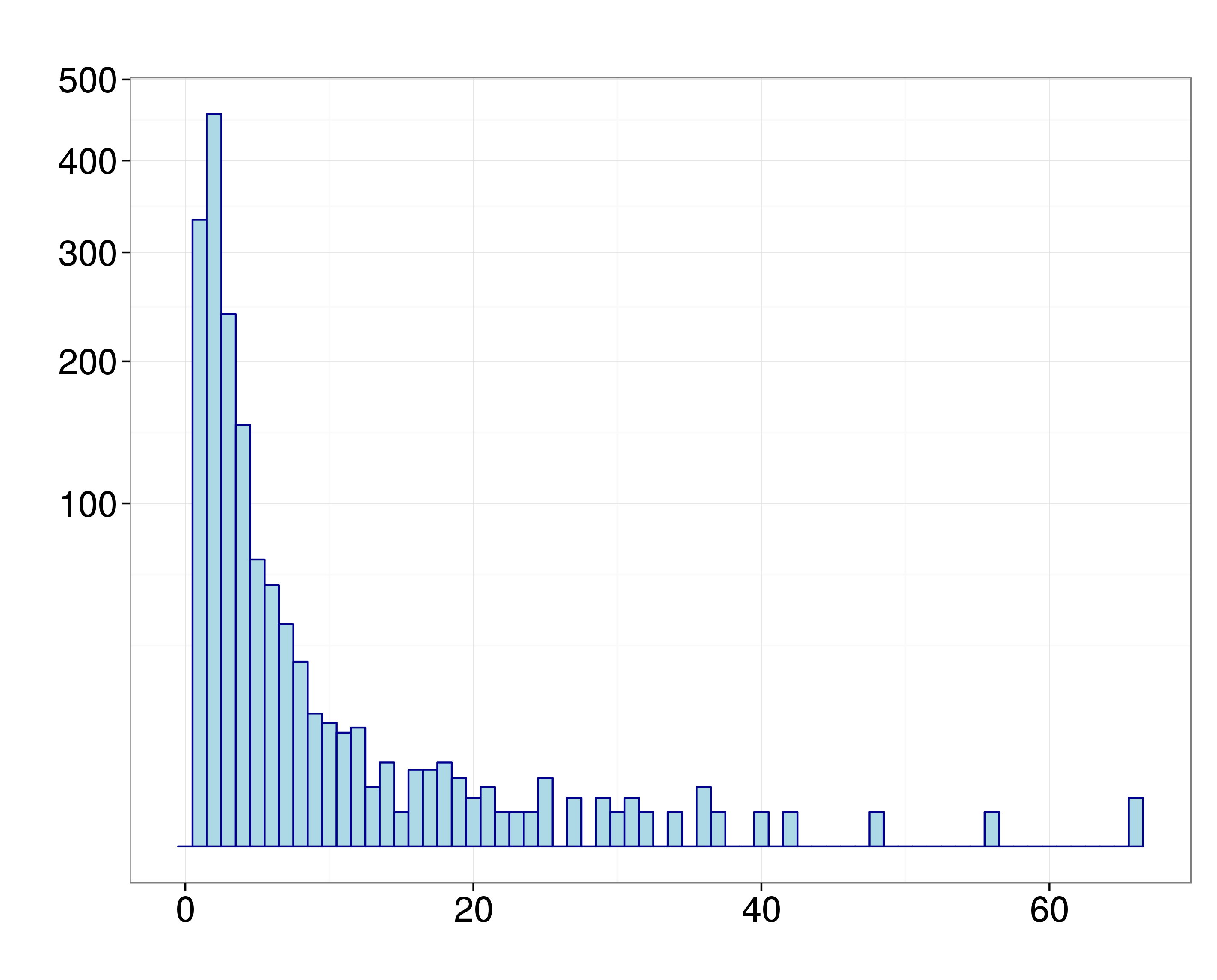}
        \caption{Distribution du degré des voies.}
    \end{subfigure}
	\caption{Degré des voies}
\end{figure}

\FloatBarrier
\paragraph{Degré de desserte d'une voie.}

Nombre de voies qu'elle n'intersecte pas à une extrémité

\begin{equation}
 degreDesserte(v_{ref}) = connectivite(v_{ref}) - degre(v_{ref})
\end{equation}

\begin{figure}[h]
    \centering
    \begin{subfigure}[t]{0.50\textwidth}
    \centering
        \includegraphics[width=\linewidth]{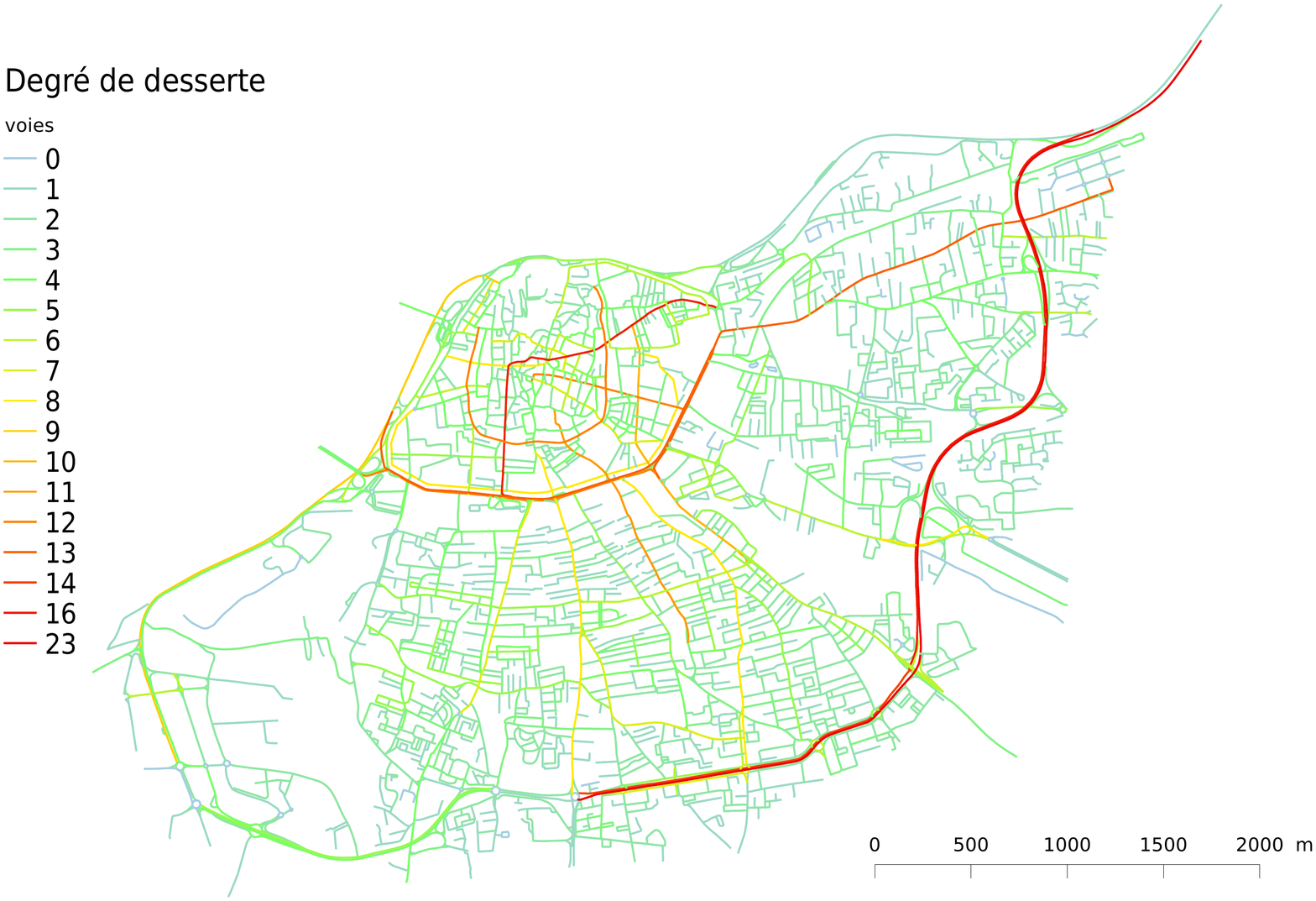}
        \caption{Représentation cartographique.}
    \end{subfigure}
	~
    \begin{subfigure}[t]{0.45\textwidth}
    \centering
        \includegraphics[width=\linewidth]{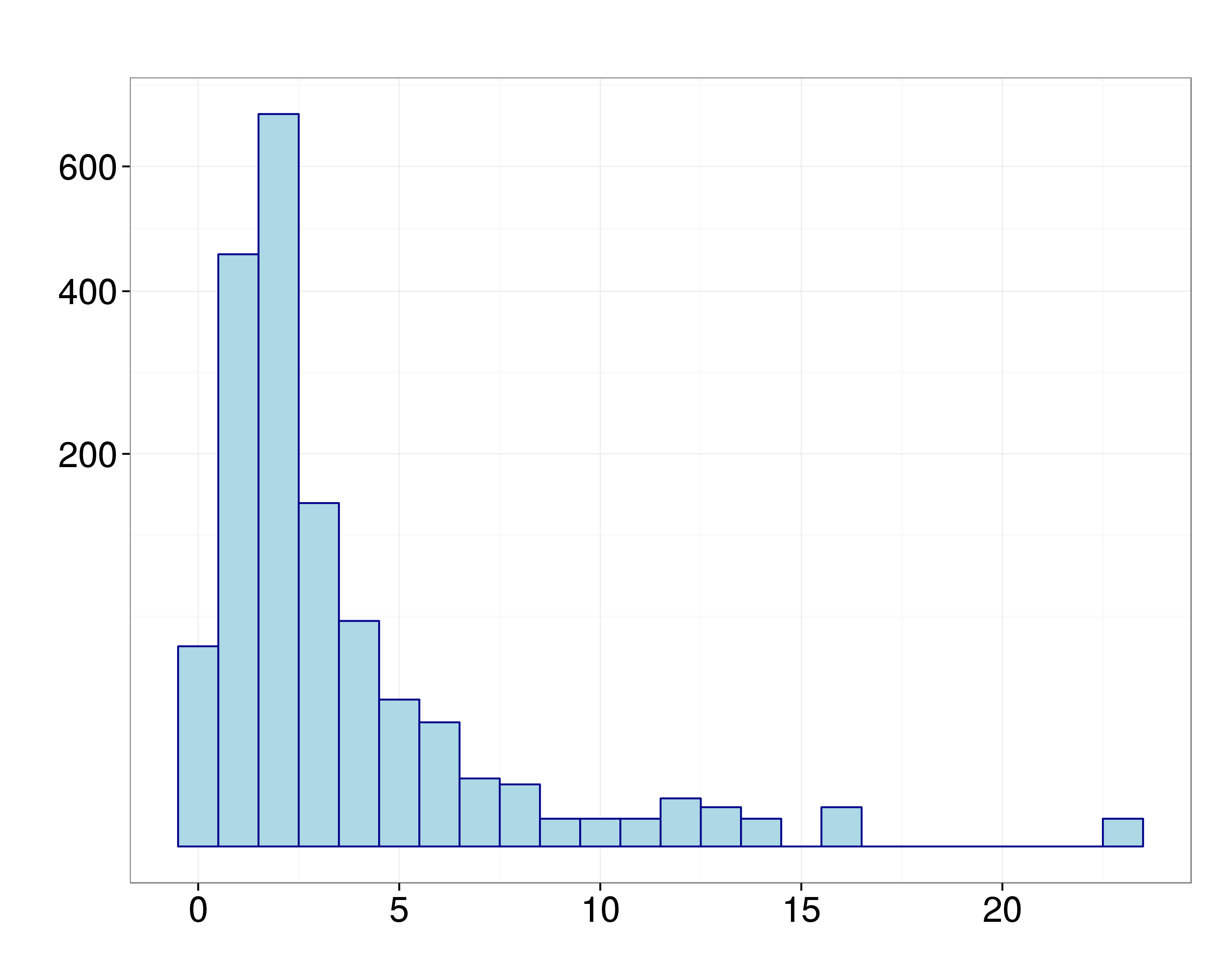}
        \caption{Distribution du degré de desserte des voies.}
    \end{subfigure}
	\caption{Degré de desserte des voies.}
\end{figure}

\clearpage

\FloatBarrier
\paragraph{Espacement d'une voie}

Espace moyen entre deux connexions (inverse d'une densité linéaire)

\begin{equation}
espacement(v_{ref}) = \frac {longueur(v_{ref})}{connectivite(v_{ref})}
\end{equation}

\begin{figure}[h]
    \centering
    \begin{subfigure}[t]{0.50\textwidth}
    \centering
        \includegraphics[width=\linewidth]{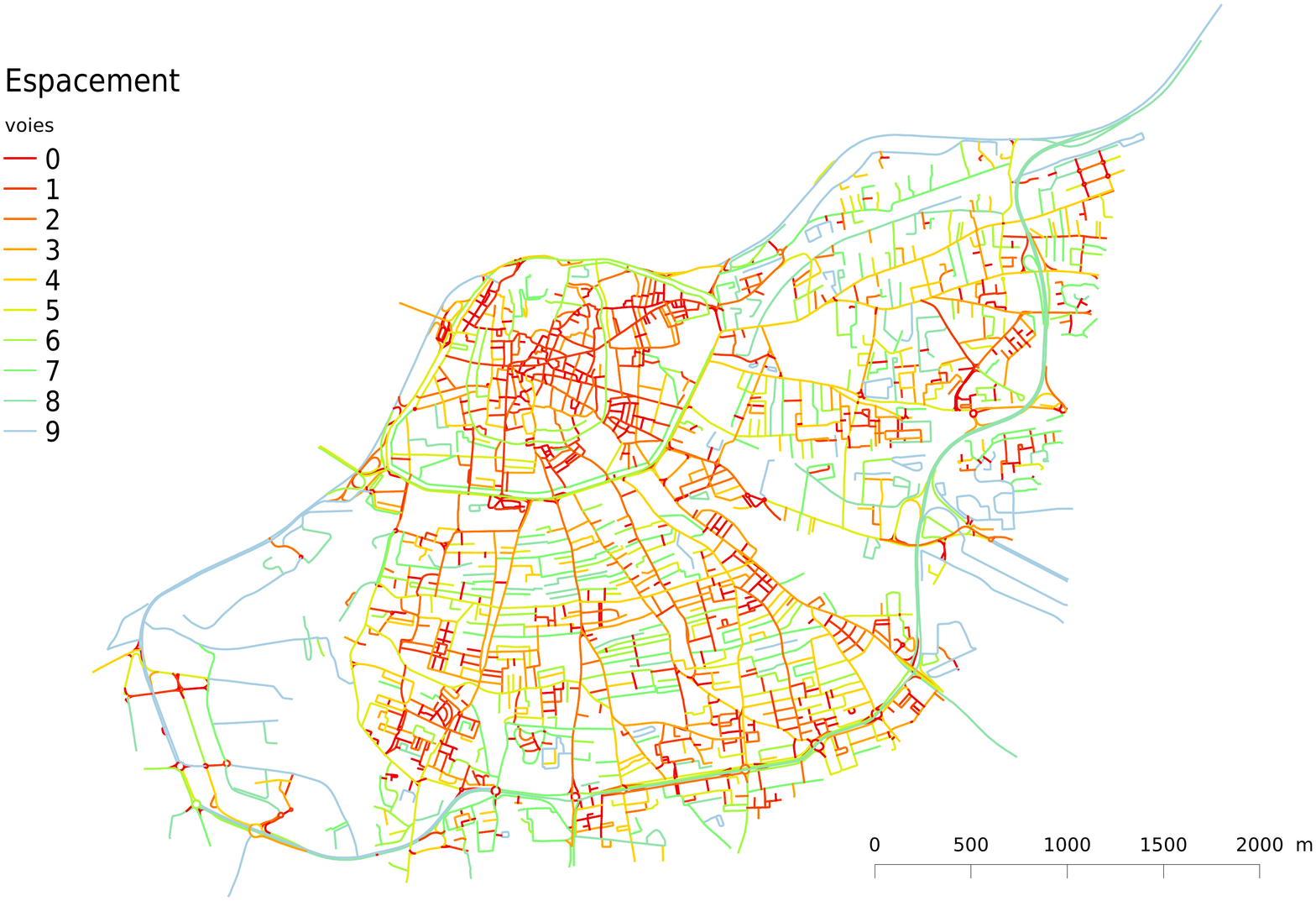}
        \caption{Représentation cartographique en 10 classes de longueur équivalente.}
    \end{subfigure}
	~
    \begin{subfigure}[t]{0.45\textwidth}
    \centering
        \includegraphics[width=\linewidth]{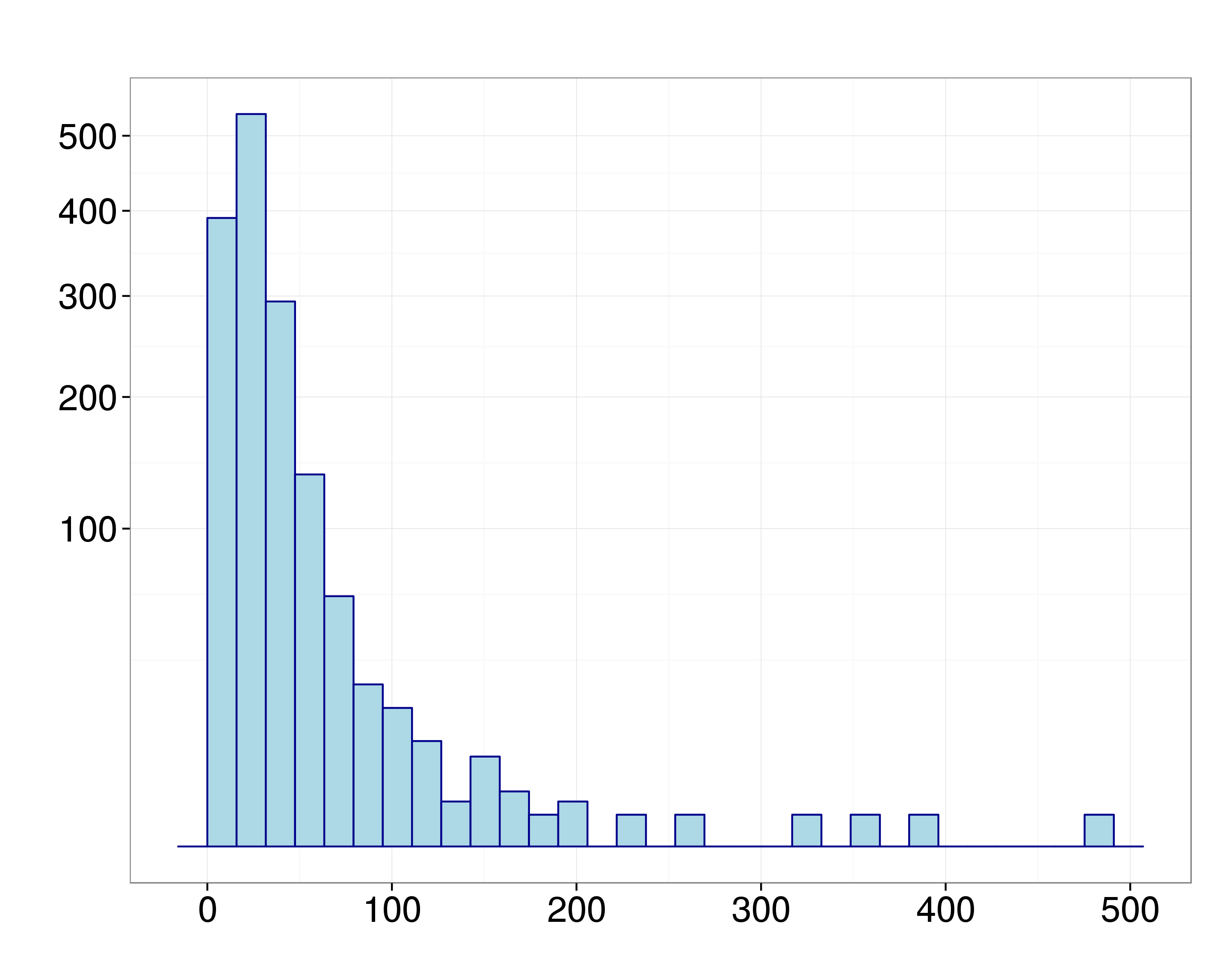}
        \caption{Distribution de l'espacement des voies.}
    \end{subfigure}
	\caption{Espacement des voies.}
\end{figure}

\FloatBarrier
\paragraph{Orthogonalité d'une voie}

Moyenne du sinus des angles de connexions avec les arcs qu'elle intersecte. Valeur entre 0 et 1 : plus elle sera proche de 1, plus les connexions sont faites orthogonalement

\begin{equation}
orthogonalite(v_{ref}) = \frac {\sum\limits^{}_{s \in v_{ref}} \  \sum\limits^{}_{arc_i \cap s \wedge arc_i \notin v_{ref}} \  \min(\sin(\theta_{arc_i arc_j})) / (arc_j \cap s \wedge arc_j \in v_{ref})}{connectivite(v_{ref}
)}
\end{equation}

\begin{figure}[h]
    \centering
    \begin{subfigure}[t]{0.50\textwidth}
    \centering
        \includegraphics[width=\linewidth]{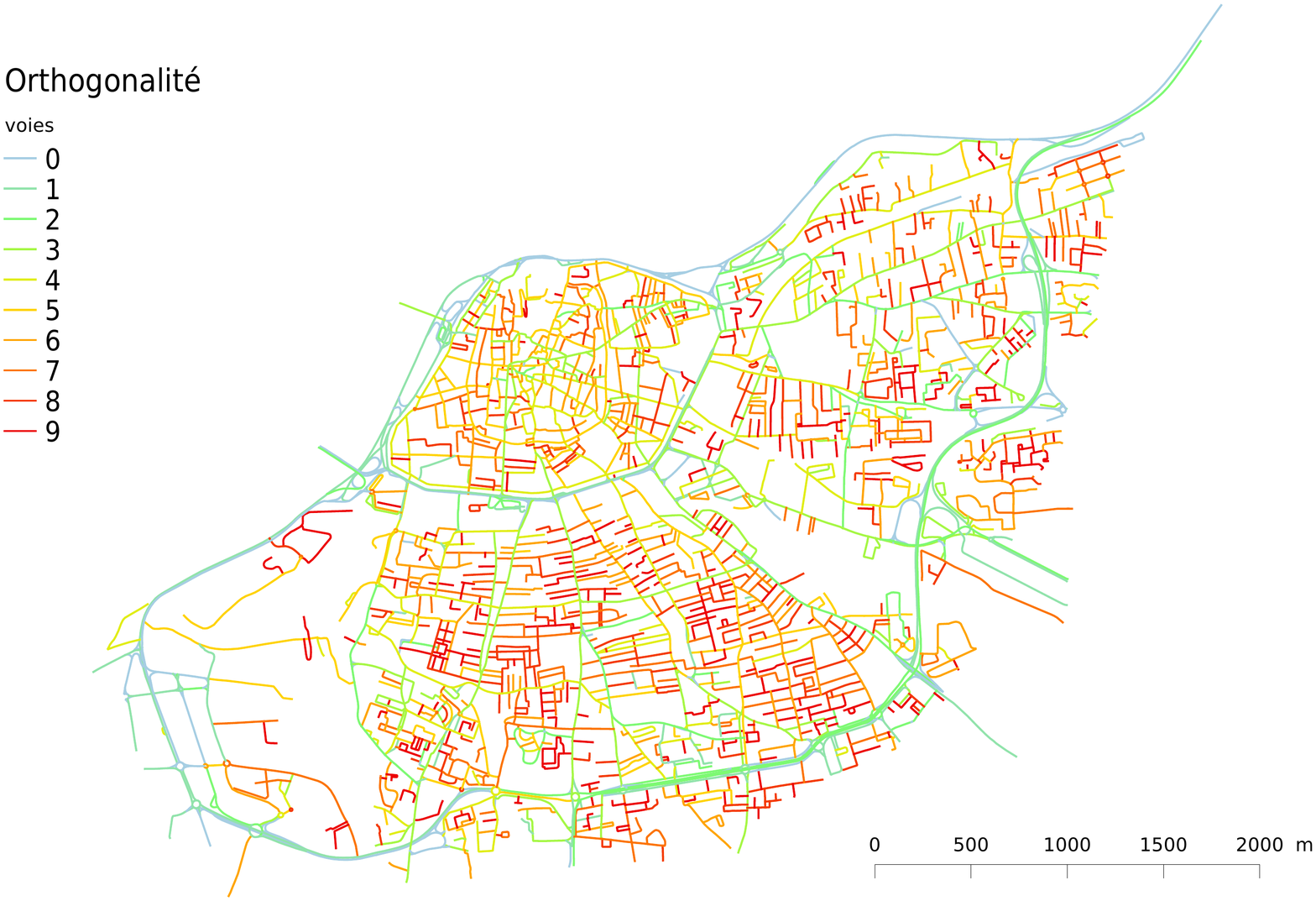}
        \caption{Représentation cartographique en 10 classes de longueur équivalente.}
    \end{subfigure}
	~
    \begin{subfigure}[t]{0.45\textwidth}
    \centering
        \includegraphics[width=\linewidth]{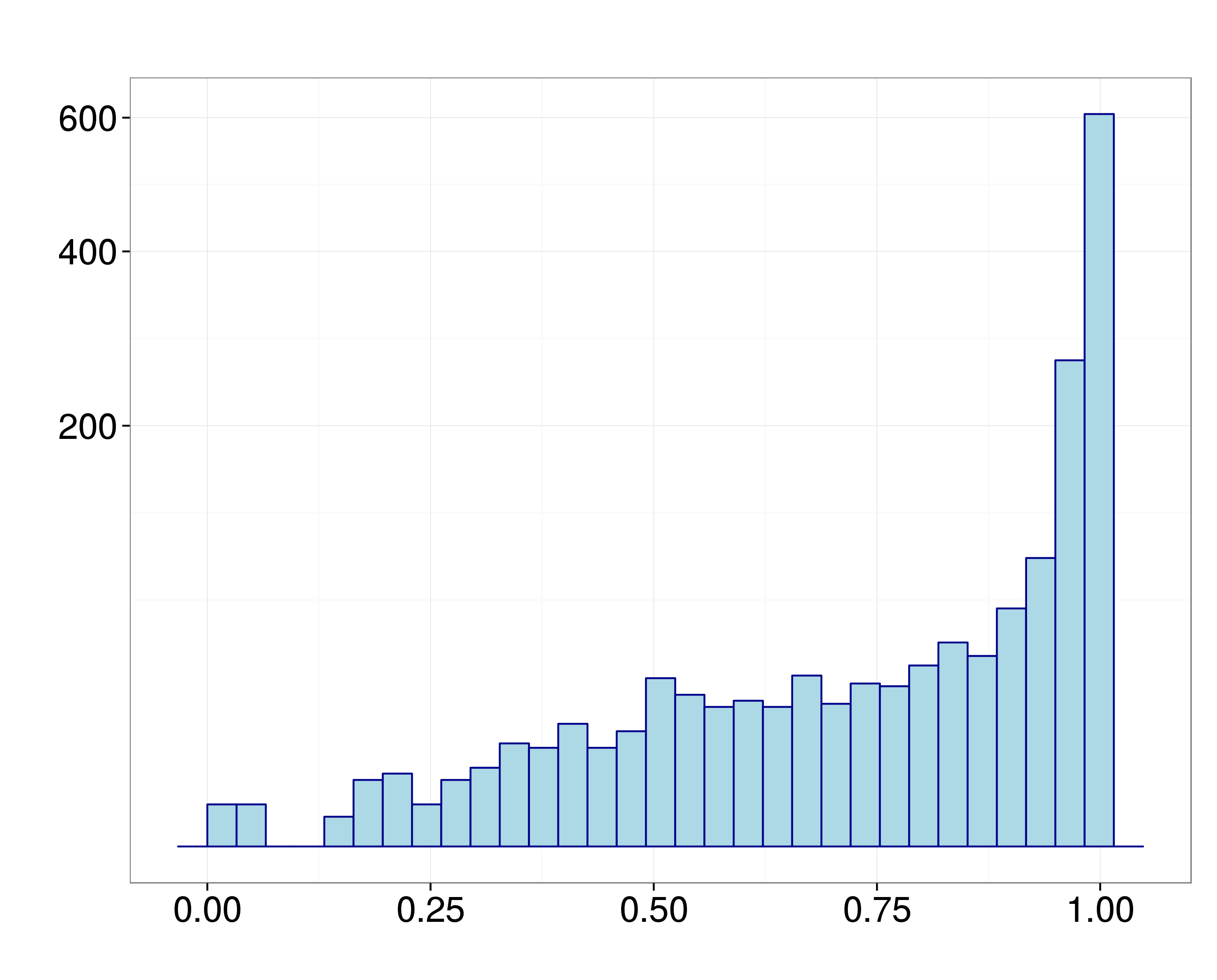}
        \caption{Distribution de l'orthogonalité des voies.}
    \end{subfigure}
	\caption{Orthogonalité des voies.}
\end{figure}

\clearpage
\subsection*{Indicateurs globaux}

\paragraph{Closeness d'une voie}

Proximité topologique de la voie avec l'ensemble du réseau. Plus la closeness aura une valeur forte, plus la voie permettra d'accéder à l'ensemble du réseau en un minimum de \textit{tournants}. Nous définition les notions d'\textit{efficacité} et de \textit{centralité} à partir de la valeur de cet indicateur.

\begin{equation}
closeness(v_{ref})=\frac{1}{\sum_{v \in G} d_{simple}(v,v_{ref})}
\end{equation}

\begin{figure}[h]
    \centering
    \begin{subfigure}[t]{0.50\textwidth}
    \centering
        \includegraphics[width=\linewidth]{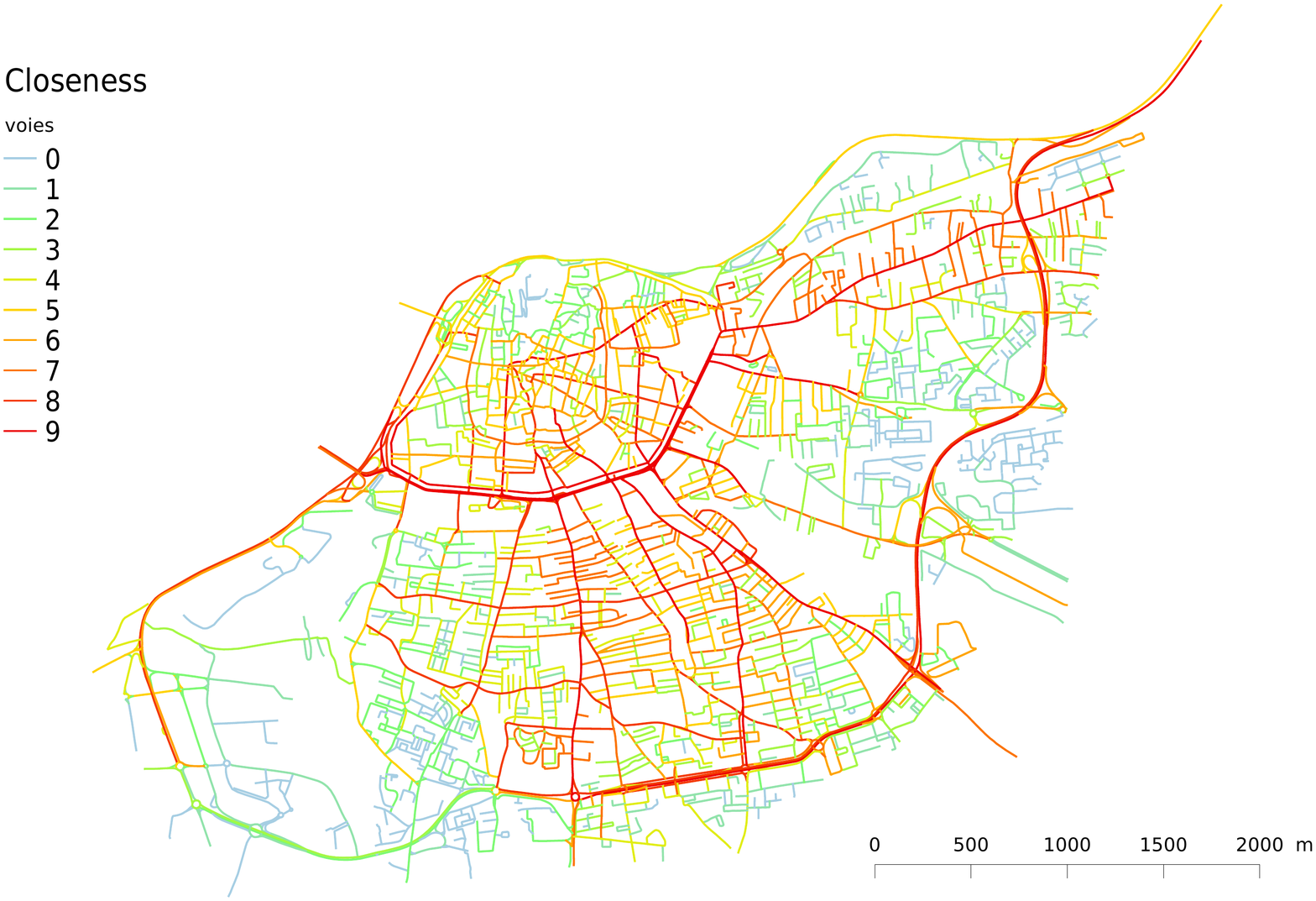}
        \caption{\scriptsize{Représentation cartographique en 10 classes de longueur équivalente.}}
    \end{subfigure}
	~
    \begin{subfigure}[t]{0.45\textwidth}
    \centering
        \includegraphics[width=\linewidth]{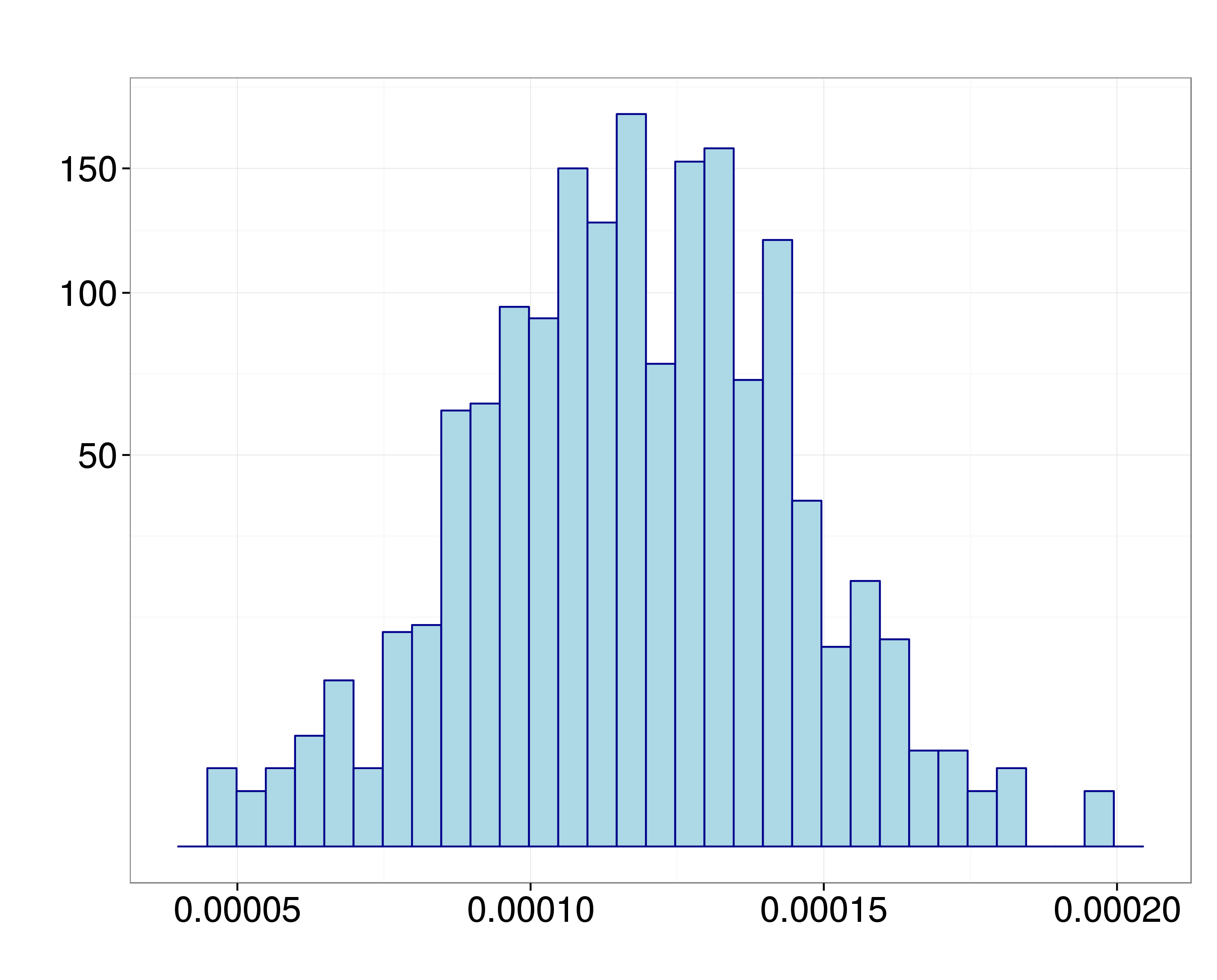}
        \caption{Distribution de la closeness des voies.}
    \end{subfigure}
	\caption{Closeness des voies.}
\end{figure}

\FloatBarrier
\paragraph{Accessibilité Maillée d'une voie}

Proximité topologique d'une voie associée à son orthogonalité : plus une voie sera \textit{centrale} et fera des angles proches de la perpendiculaire avec son voisinage, plus elle aura une accessibilité maillée forte.

\begin{equation}
accessibiliteMaillee(v_{ref})= closeness(v_{ref}) \times orthogonalite(v_{ref})
\end{equation}

\begin{figure}[h]
    \centering
    \begin{subfigure}[t]{0.50\textwidth}
    \centering
        \includegraphics[width=\linewidth]{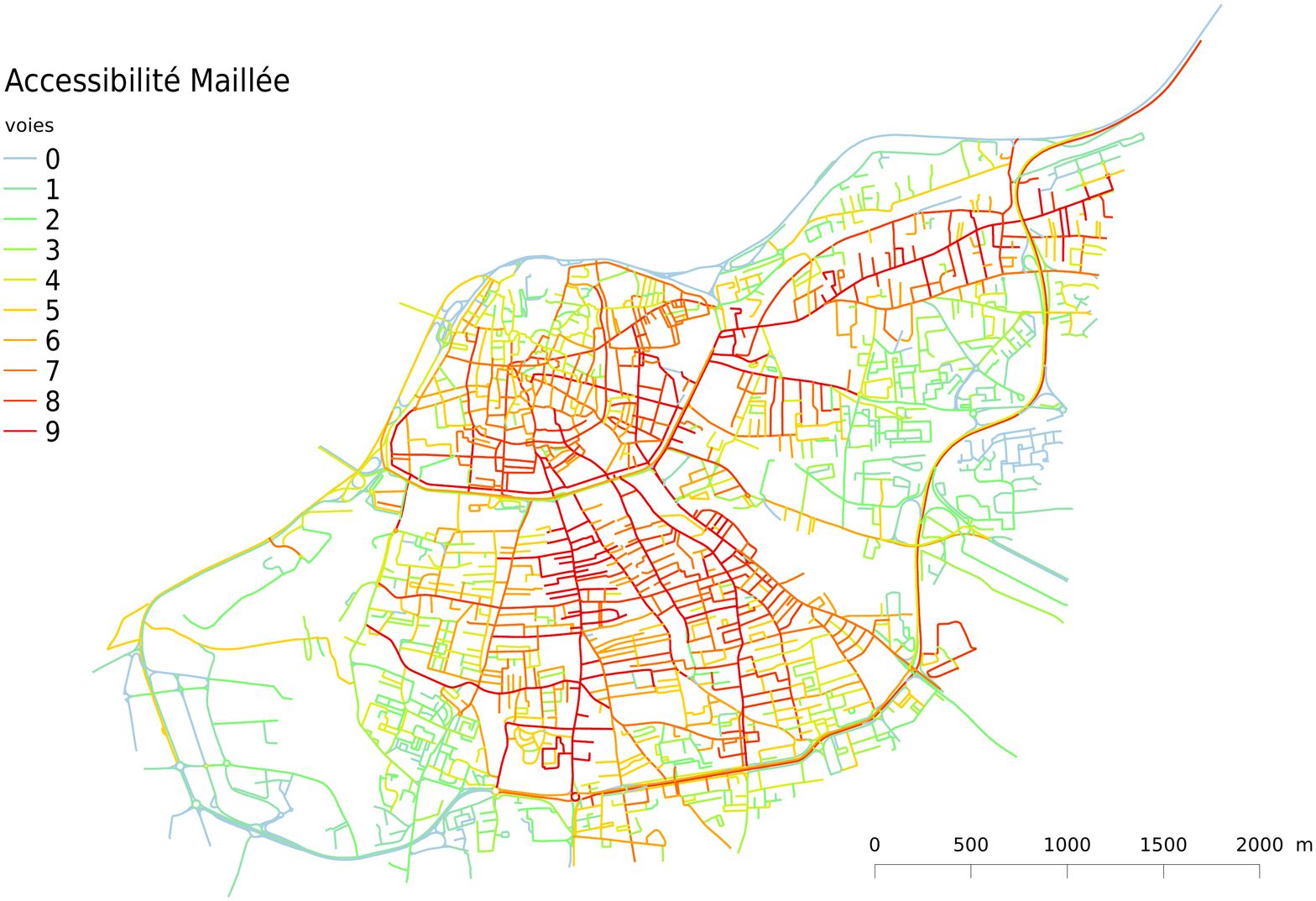}
        \caption{\scriptsize{Représentation cartographique en 10 classes de longueur équivalente.}}
    \end{subfigure}
	~
    \begin{subfigure}[t]{0.45\textwidth}
    \centering
        \includegraphics[width=\linewidth]{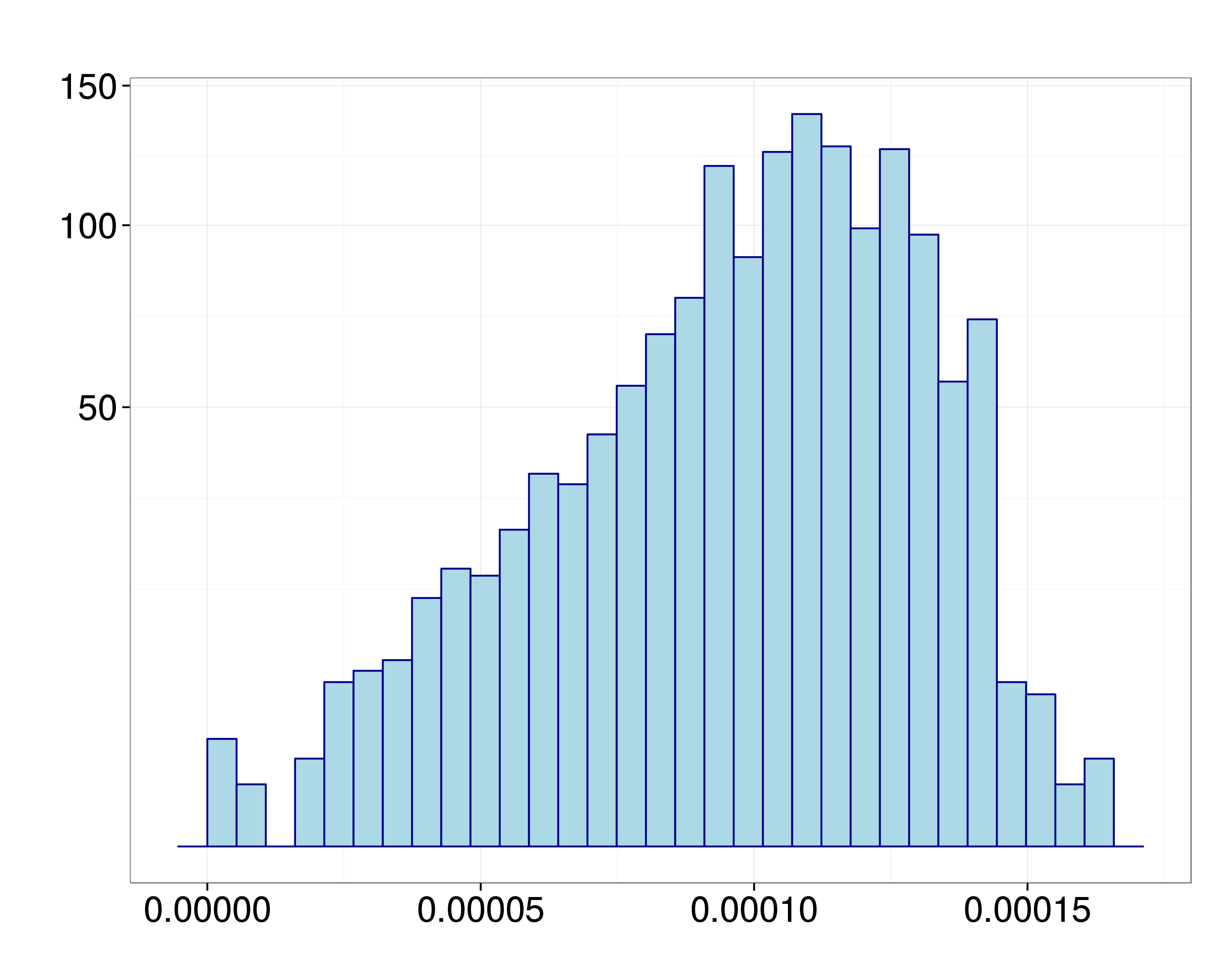}
        \caption{Distribution de l'accessibilité maillée des voies.}
    \end{subfigure}
	\caption{Accessibilité maillée des voies.}
\end{figure}

\restoregeometry
\setlength{\parskip}{10pt}

\clearpage{\pagestyle{empty}\cleardoublepage}

\thispagestyle{empty}
~\vfill
{\itshape Truth is ever to be found in simplicity, and not in the multiplicity and confusion of things.}
\begin{flushright}
  - Isaac \textsc{Newton}\label{cit:newton}
\end{flushright}
~\vfill
\clearpage{\pagestyle{empty}\cleardoublepage}
\chapter*{Introduction générale}
\markboth{Introduction générale}{Introduction générale}
\addcontentsline{toc}{part}{Introduction générale}

Ce sont les structures qui nous guident dans cette recherche, les trames porteuses de vie et d’histoire, nées d’éléments en interaction les uns avec les autres. Nous nous immergeons dans la science des systèmes complexes. Pour cela, nous nous appuyons sur plusieurs disciplines, qu’il est nécessaire de faire cohabiter pour tenter de comprendre. Comprendre comment un graphe inscrit dans l'espace porte en lui des informations structurelles essentielles. Comprendre comment il est possible de distinguer ces informations et avec quels outils. Comprendre comment, dans le cas d’une ville, nous pouvons avoir l’intuition de l'usage de ses rues grâce à leurs géométries. Comment des constructions viaires de différents continents peuvent avoir des propriétés structurelles identiques. Comment décrire les transformations de ce squelette urbain dans le temps. Comment les hommes, qui ont construit la ville, l’utilisent et se la représentent. Nous proposons ainsi la découverte de territoires en lisant à travers leurs lignes.

\FloatBarrier

\section*{La naissance de la théorie des graphes}

\begin{figure}
    \centering
    \begin{subfigure}[t]{0.45\textwidth}
        \includegraphics[width=\textwidth]{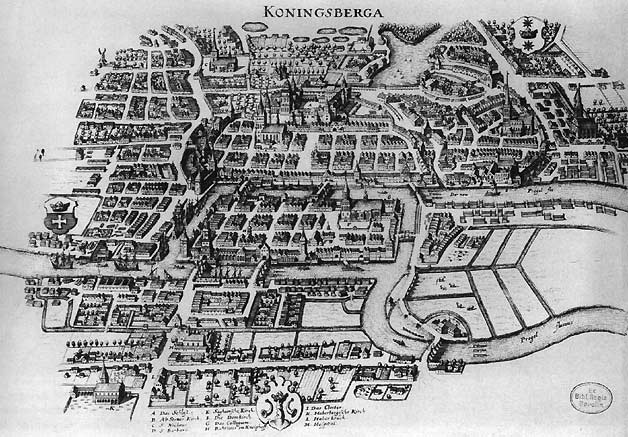}
    \end{subfigure}%
    ~ 
    \begin{subfigure}[t]{0.45\textwidth}
        \includegraphics[width=\textwidth]{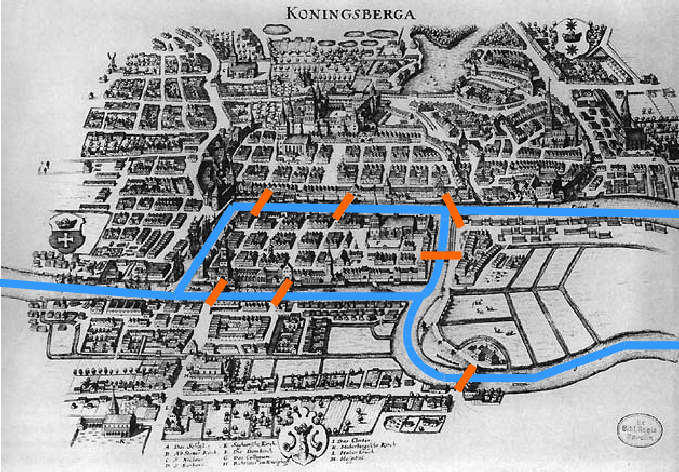}
    \end{subfigure}

    \begin{subfigure}[t]{0.45\textwidth}
        \includegraphics[width=\textwidth]{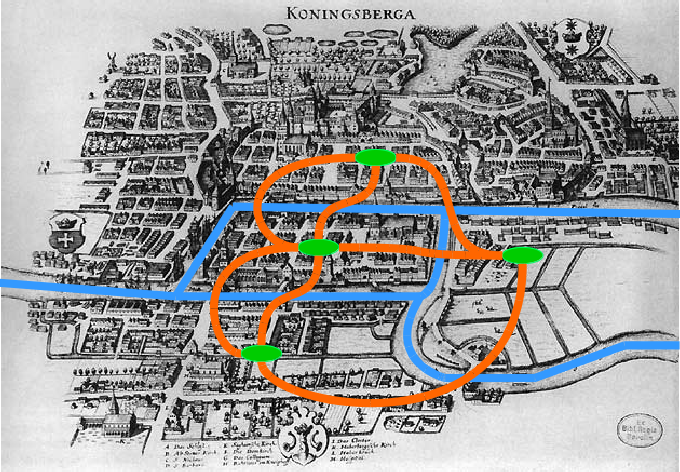}
    \end{subfigure}
    ~
    \begin{subfigure}[t]{0.45\textwidth}
        \includegraphics[width=\textwidth]{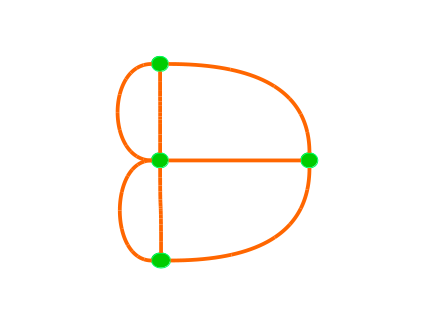}
    \end{subfigure}
    \caption{Abstraction topologique du problème des ponts de Königsberg.}
    \label{fig:konig_1}
\end{figure}

L'histoire commence en 1736, avec le grand mathématicien Léonard Euler, une ville fluviale et ses ponts. La ville de Königsberg, située dans l'empire Prusse de l'époque (en Russie actuellement, renommée depuis Kaliningrad), est construite sur les bords du fleuve Pregel, de son confluent, et sur une île située au milieu du fleuve. Pour accéder aux différentes berges, sept ponts ont été construits (figure \ref{fig:konig_1}). Une énigme populaire de l'époque était de savoir s'il était possible de traverser tous les ponts en passant une seule fois par chacun d'entre eux. Pour y répondre Euler dessina le problème sous la forme d'un graphe. Il représenta les terres avec des \emph{nœuds} et établit un lien entre deux nœuds si les berges étaient reliées par un pont. Il démontra ainsi que si on ne voulait traverser qu'une seule fois chacun des ponts, on devait pouvoir accéder et quitter chaque nœud avec un lien différent. Le graphe ne pouvait donc admettre que des nœuds avec un nombre pair de liens connectés, exception faite de deux d'entre eux : le premier et le dernier du chemin. Il était donc impossible à Königsberg de traverser une seule fois tous les ponts en un même chemin.

Depuis, le chemin dans un graphe ne passant qu'une unique fois par chaque lien est appelé \emph{chemin Eulérien}. Et si le graphe complet permet de traverser tous les arcs une unique fois en un chemin il se nomme \emph{graphe Eulérien}.

Thomas Kirkman, mathématicien, puis William Rowan Hamilton, astronome, sont venus compléter cette approche au XIXe siècle en transposant le problème aux nœuds. Le \emph{chemin Hamiltonien} sera ainsi celui ne passant qu'une unique fois par chaque nœud. De même, le graphe sera dit \emph{Hamiltonien} s'il permet que l'on traverse tous ses nœuds ainsi.

Une discipline est ainsi née, traduisant des environnements complexes en deux éléments descriptifs simples : des nœuds (entités) et des liens (connexions entre les entités). De nombreux scientifiques sont venus l'étayer par la suite, mais Euler est considéré par beaucoup comme le premier ayant établi une théorie en utilisant cette formalisation. Depuis, elle est considérée comme un outil essentiel pour décrire les propriétés d'un milieu réunissant des éléments en interaction les uns avec les autres \citep{freeman1977set, hage1995eccentricity}.

\FloatBarrier

\section*{Les graphes : représentation formelle du monde qui nous entoure}

La représentation sous forme de graphe permet le développement de nombreux d'indicateurs qui ont pour but de décrire et d'analyser les propriétés de l'objet qu'elle formalise. Ainsi, depuis Euler, de nombreux scientifiques ont exploité les possibilités offertes par cette théorisation.

Les premiers étaient issus des sciences sociales, au début du XX\textsuperscript{ème} siècle. Comprenant l'enjeu d'une telle modélisation, ils ont fait circuler des questionnaires pour récolter des données sur les interactions sociales. Ils ont ainsi évaluer la centralité de tels réseaux (par quels individus passent le plus de chemins les plus courts), leur connectivité (qui sont les agents ayant le plus de liens), et diverses autres propriétés. En sociologie, un graphe peut ainsi formaliser les interactions au sein d'un groupe \citep{wasserman1994social, john2000social}, des liens professionnels \citep{mariolis1975interlocking, mizruchi1982american, newman2001scientific}, d'amitiés \citep{moreno1934shall, rapoport1961study}, ou familiaux \citep{padgett1993robust}. Ces interactions peuvent être fondées sur un type d'échanges particulier : par exemple, certains travaux se sont penchés sur le graphe représentant les citations entre publications scientifiques \citep{yu1965networks, white2004does}.

À la fin du XX\textsuperscript{ème} siècle, le développement de l'informatique a permis l'étude de plus en plus vaste de données, donnant ainsi naissance à l'étude statistique des \textit{grands graphes}. Nous retrouvons alors des graphes réunissant plus d'un million de sommets. Les scientifiques ont analysé, entre autres, la robustesse de tels graphes : à quel point restent-ils connectés lorsqu'on leur enlève des sommets ou des arcs ; et leurs dynamiques \citep{albert2002statistical}. Ils se sont également intéressés à l'évolution de leurs propriétés \citep{dorogovtsev2002evolution}. La quantité de données à analyser a demandé aux chercheurs de développer des algorithmes performants pour pouvoir en extraire des structures pertinentes \citep{blondel2008fast}.

Les champs d'application de ces travaux sont multiples. Les réseaux technologiques (réseaux d'approvisionnement en énergie - \textit{power-grids}- ou réseaux de communication) ont pu ainsi être étudiés en détail, notamment en ce qui concerne leurs propriétés de \textit{petit monde} \citep{watts1998collective}. C'est dans ces réseaux technologiques que nous comptons également les réseaux de transports, ferrés \citep{latora2002boston, sen2003small, wang2009spatiotemporal}, aériens \citep{bryan1999hub, guimera2005worldwide}, maritimes \citep{hu2009empirical}, ou routiers \citep{amaral2000classes, kalapala2003structure}, sur lesquels nous reviendrons dans ce travail pour étudier leurs propriétés particulières.

Au début du XXI\textsuperscript{ème} siècle, la biologie s'est également emparée du sujet. Ainsi, nous retrouvons des études statistiques sur des réseaux métaboliques \citep{jeong2000large, podani2001comparable, stelling2002metabolic}. Mais également de nombreuses études s'appuyant sur des graphes modélisant des interactions mécaniques entre protéines \citep{uetz2000comprehensive, ito2001comprehensive, jeong2001lethality, maslov2002specificity, sole20067}. La théorie des réseaux a permis également d'étudier statistiquement les interactions entre espèces via des graphes de prédation (\textit{food webs}) \citep{sole2001complexity, dunne2002food, dunne2002network, camacho2002robust, montoya2002small, williams2002two}. Parmi les applications biologiques, celles réalisées sur les réseaux de neurones ouvrent un large champ d'étude. En effet, l'abondance d'objets dans de tels réseaux rend leur analyse très complexe. Une des premières modélisations connues est celle d'un cerveau de ver, avec ses 282 neurones \citep{white1986structure}. Les travaux en neurosciences se concentrent pour la plupart à une échelle plus grande que celle du neurone  \citep{sporns2000theoretical, sporns2002network}.

Ces différents champs disciplinaires ont en commun de travailler sur un graphe. C'est une formalisation très riche puisqu'elle peut s'appliquer à de nombreux domaines différents. Des méta-études ont été opérées sur cette théorie regroupant ses différentes avancées \citep{ben2004complex, newman2006structure, newman2010networks, boccaletti2006complex}. Bien que la topologie permette une première caractérisation, certains graphes sont également porteurs d'une information morphologique. L'inscription de leurs arcs dans l'espace leur donne des caractéristiques spécifiques \citep{barthelemy2011spatial}. Ce sont sur ces graphes que les travaux présentés ici se concentrent.

\FloatBarrier

\section*{La particularité des graphes spatiaux}

Pour formaliser le problème des sept ponts de Königsberg, Euler a placé des sommets sur chaque berge et a créé autant de connexions entre les sommets qu'il y avait de ponts entre les berges (figure \ref{fig:konig_1}). Faisant cela, il s'est détaché de l'information géographique qui était contenue dans le réseau des routes de Königsberg. La résolution de son problème ne nécessitait pas qu'il en tienne compte. Cependant, dans d'autres cas, c'est une information dont il aurait été dommage de se priver.

\begin{figure}
    \centering
    \includegraphics[width=0.6\textwidth]{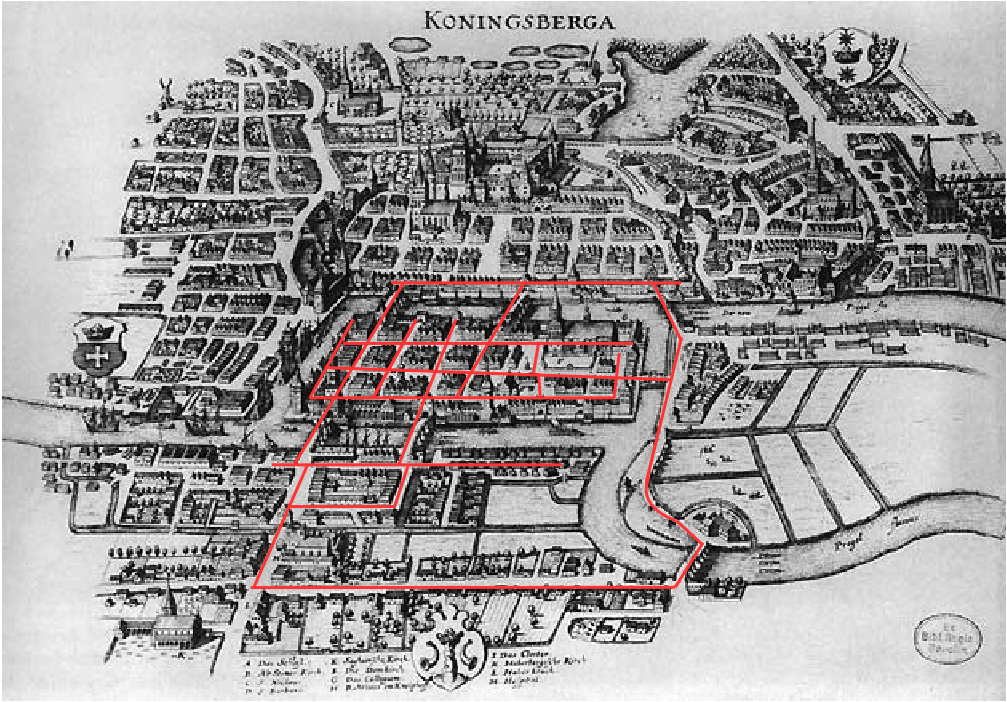}
    \caption{Tracé d'une partie du graphe viaire sur la carte de Königsberg.}
    \label{fig:konig_5}
\end{figure}

\begin{figure}
    \centering
    \includegraphics[width=0.8\textwidth]{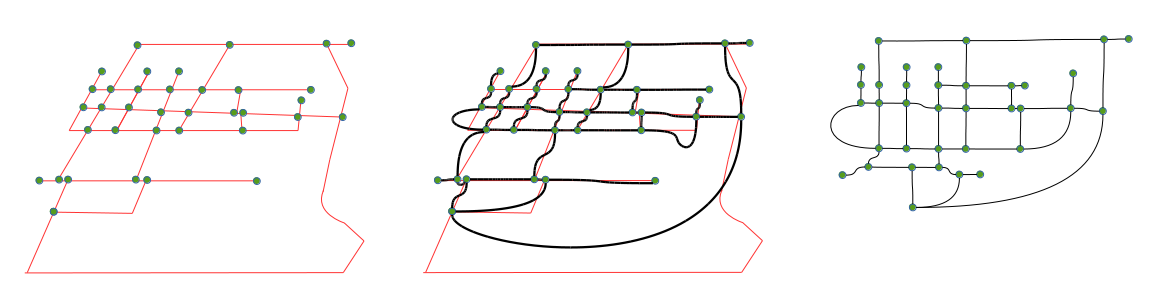}
    \caption{Illustration d'un graphe spatial et d'un graphe topologique représentant les mêmes données. À gauche : le graphe viaire de Königsberg tel qu'il a été tracé sur la figure \ref{fig:konig_5}. Des sommets sont placés aux intersections, les arcs matérialisent les tronçons de rue en respectant leur emprise au sol. Au centre : transition vers le graphe topologique : l'information de liaison est conservée mais pas celle d'inscription spatiale. À droite : le graphe topologique, abstraction du graphe viaire. Ici la position des sommets et la géométrie des arcs n'a pas d'importance. Seule compte l'information de connexion.}
    \label{fig:konig_6}
\end{figure}

En effet, il est possible de tracer sur l'île de Königsberg, et sur les rives opposées proches, le réseau viaire de l'époque (figure \ref{fig:konig_5} : le réseau possède la \textit{projection} de la carte sur laquelle il a été vectorisé). En plus de l'information de connexion entre les berges, nous pouvons alors décrire toute la circulation possible à l'intérieur de l'île et autour. Ainsi, nous positionnons des sommets aux intersections et nous créons un lien (arc) entre deux sommets s'il existe un tronçon de rue entre les deux intersections. L'information qui résulte de ce graphe peut être décrite de deux manières. La première résulte directement du tracé des rues : on conserve l'inscription spatiale. La seconde ne s'intéresse qu'aux connexions effectives entre deux intersections, elle ne conserve que la topologie. La figure \ref{fig:konig_6} illustre ces deux méthodes de représentation et la transition de l'une à l'autre. La représentation topologique (à droite) répond à la question\textit{ \enquote{Est-ce que ces deux intersections sont connectées ?}}. La représentation topographique (à gauche) ajoute à la réponse obtenue une deuxième information : \textit{\enquote{Comment ces deux intersections sont-elles connectées ?}}. Il est dès lors possible de s'intéresser à de nouvelles informations apportées par l'inscription spatiale, comme la distance entre deux intersections ou les différentes orientations du lien entre celles-ci. Ces informations pourraient être portées par le graphe topologique sous forme de poids attribué aux arcs. Il serait cependant difficile de les inclure toutes dans une même représentation.

Les graphes représentant des données spatialisées peuvent être qualifiés de \emph{graphes spatiaux} si, en tous points, leurs arcs et sommets peuvent être identifiés selon des coordonnées précises. Ces graphes, dont les réseaux de transports ne sont qu'un exemple, ont des propriétés géométriques qui viennent s'ajouter à celles topologiques. Cette spécificité a intéressé de nombreux scientifiques à travers le monde \citep{mohring1961land, gauthier1966highway, marshall2004streets}. Il est possible de représenter de nombreuses formes de données spatialisées. Ainsi S. Bohn et A. Perna ont travaillé sur les graphes de veinures de feuilles \citep{bohn2002constitutive, perna2011characterization} explorant les propriétés structurelles d'un réseau biologique. S. Bohn se pencha également avec S. Douady sur les réseaux de craquelures dans de l'argile \citep{bohn2005four, bohn2005hierarchical} essayant de comprendre la dynamique liée à leur évolution. D'autres se sont penchés sur les similitudes pouvant exister entre un réseau géographique de transport et un réseau biologique de croissance d'un champignon : le physarum \citep{tero2007mathematical}. Plus récemment, M. P. Viana \textit{et al.} étudièrent les similitudes et disparités entre différents types de réseaux spatiaux, considérant les propriétés liées à leur planarité \citep{viana2013simplicity}.

\begin{figure}
    \centering
    \includegraphics[width=0.6\textwidth]{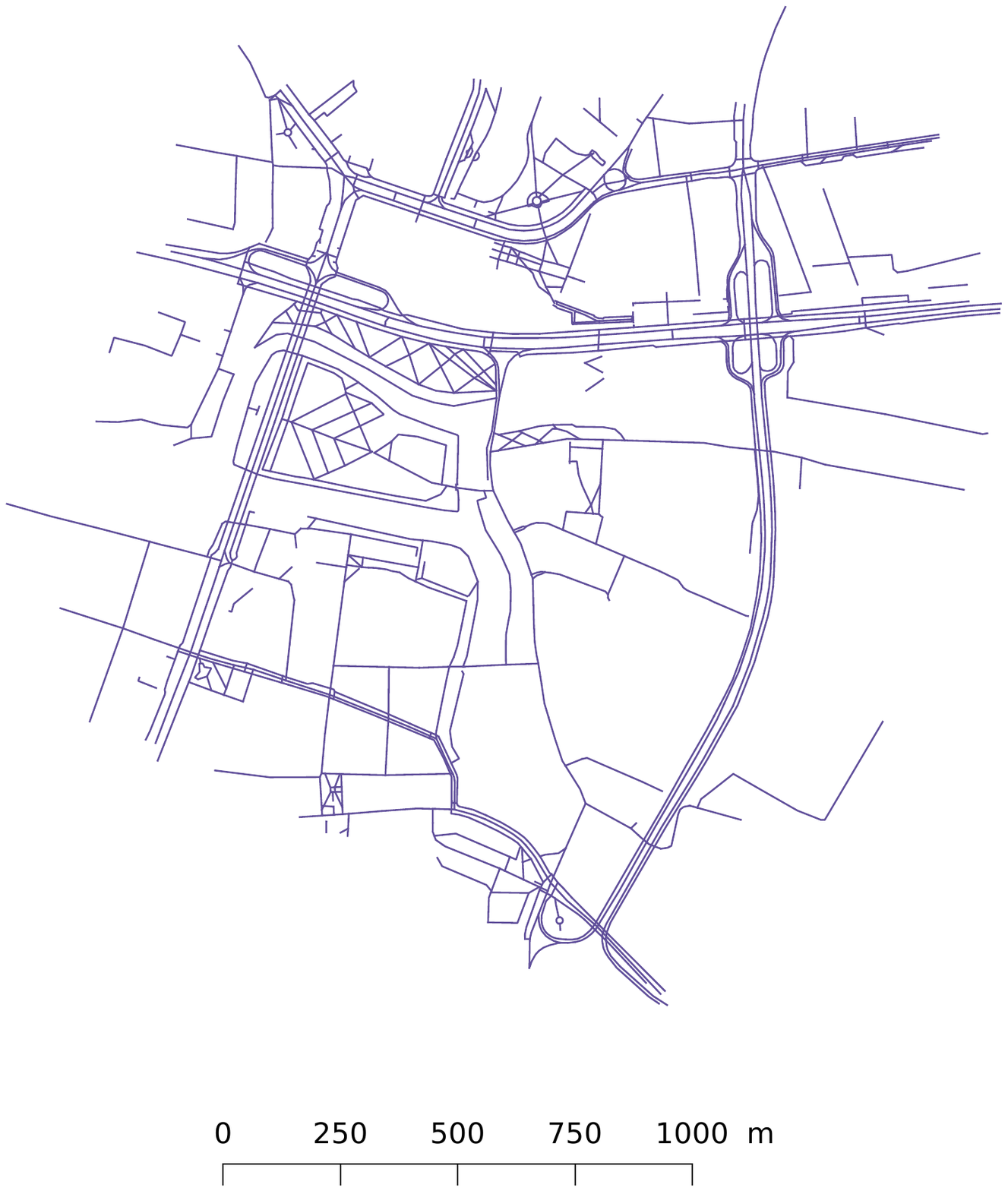}
    \caption{Extraction filaire du réseau viaire de Kaliningrad. Centre ville (autour de l'île). Données OSM (mai 2015).}
    \label{fig:kaliningrad_small}
\end{figure}

\begin{figure}
    \centering
    \includegraphics[width=0.8\textwidth]{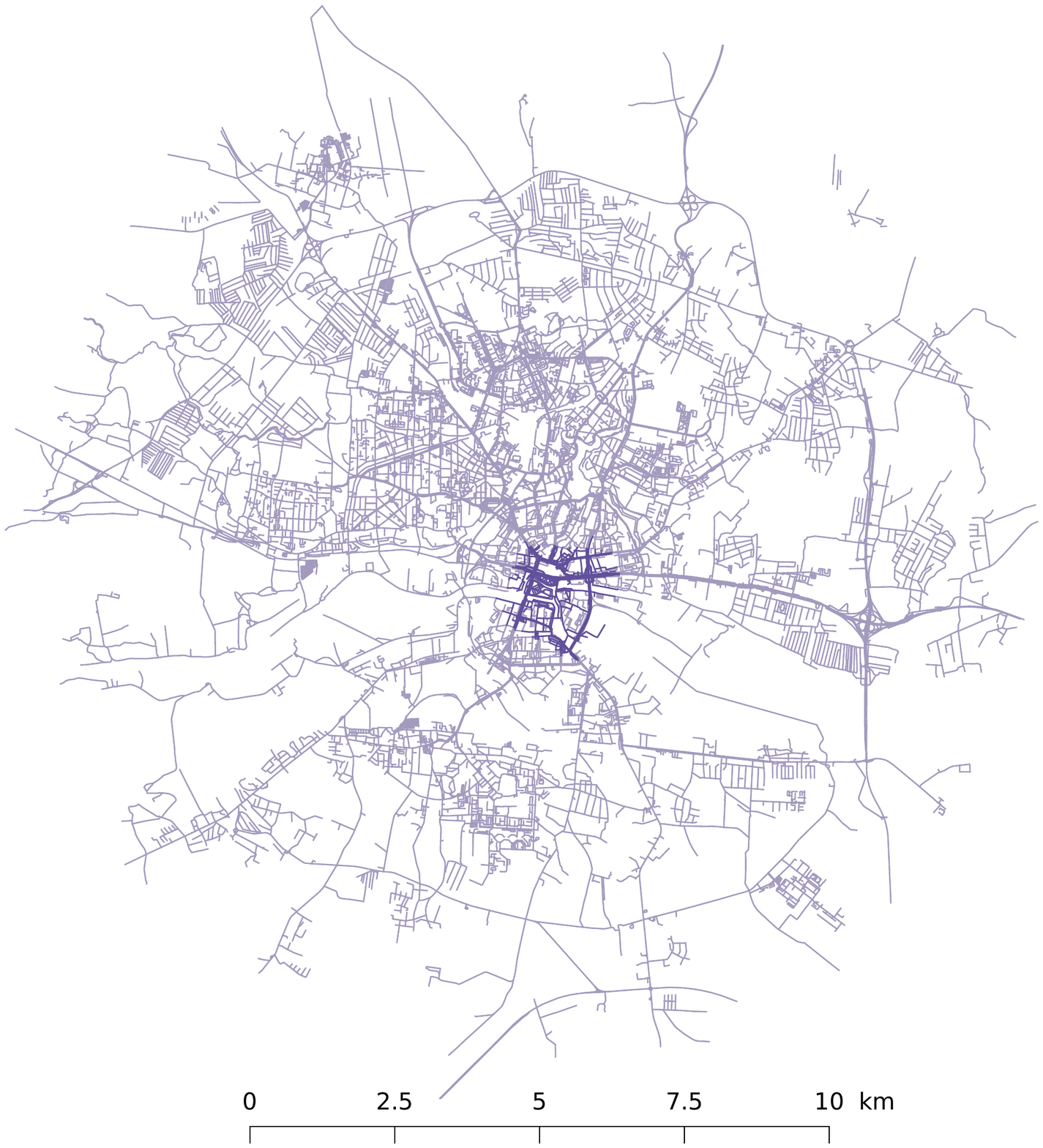}
    \caption{Extraction filaire du réseau viaire de Kaliningrad. Découpage circulaire autour de la ville. Données OSM (mai 2015).}
    \label{fig:kaliningrad_medium}
\end{figure}

Si l'on observe le réseau viaire d'aujourd'hui, à Kaliningrad, nous trouvons un réseau complexe et difficilement lisible au premier coup d’œil (figure \ref{fig:kaliningrad_small}). Celui-ci se complexifie d'autant plus si on l'élargit à la ville (figure \ref{fig:kaliningrad_medium}) ou même au pays (la Russie !). Il devient alors compliqué de comprendre les structures qui le composent. C'est la problématique qui nous intéresse dans cette thèse : comment faire apparaître de l'information structurelle sur un réseau spatial complexe. Nous nous intéressons à la lecture des graphes en hiérarchisant leurs liens. Puis, après avoir établie une méthodologie, nous comparons les caractéristiques de différents graphes afin de distinguer les propriétés spatiales structurelles universelles de celles plus spécifiques. Enfin, à travers la lecture des structures, nous cherchons à expliciter une \textit{cinématique} de développement à travers différents états d'un même graphe au cours du temps.

\FloatBarrier

\section*{La formalisation géographique}

Représenter une information sous forme de graphe spatial, nous offre la possibilité de compléter son étude grâce à des systèmes d'informations spécialisés : les SIG (Systèmes d'Information Géographique). En effet, ceux-ci incluent trois modes de représentation de l'espace physique : ponctuel (0D), linéaire (1D) ou surfacique (2D) \citep{bordin2002sig}. Des développements permettent désormais d'y intégrer de l'information volumique (3D) artificiellement codée par une emprise (2D) et une hauteur. La transcription d'une réalité physique en une information géométrique s'appelle la \textit{vectorisation}. L'information devient \textit{vectorielle} : les sommets ont une position particulière et les arcs une géométrie précise.

L'abstraction de l'espace physique en information vectorielle (ou vecteur) peut se faire à différentes échelles. Selon l'échelle choisie, la représentation ne se fera pas dans la même dimension. Par exemple, une ville peut être représentée par un élément ponctuel : sommet d'un graphe autoroutier ; linéaire : frontière qui la sépare de la commune voisine ; ou surfacique : emprise au sol sur un territoire. Les réseaux spatiaux, pour être étudiés en tant que graphes, seront représentés par une vectorisation linéaire. Chaque objet vectoriel possède un attribut géométrique : une liste de coordonnées qui correspond à son inscription spatiale. Cet attribut peut être complété par d'autres qui constitueront la table attributaire de notre objet géographique (par exemple, sa toponymie, sa largeur, etc). Dans ces travaux, nous ne conservons que deux informations de la table attributaire des données acquises : leur identifiant et leur géométrie. C'est l'information minimale dont nous avons besoin pour tracer les graphes correspondants.

\begin{figure}
    \centering
    \includegraphics[width=0.3\textwidth]{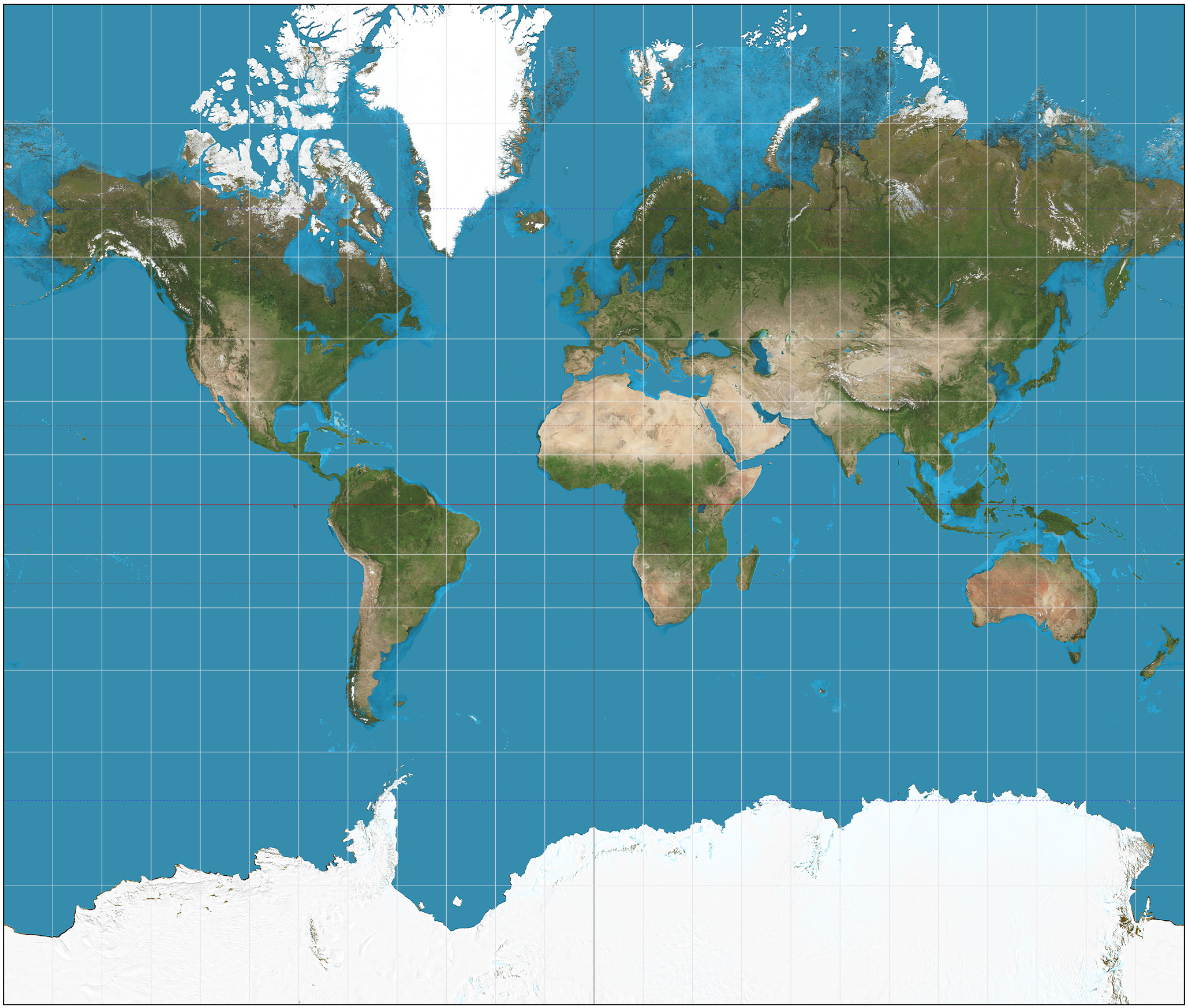}
    \caption{ Planisphère représenté avec la projection Mercator. \\source : Strebe - Travail personnel}
    \label{fig:Mercator_projection}
\end{figure}

Lorsque l'on observe de larges espaces, il est nécessaire de porter une attention particulière à la \emph{projection} des données. En effet, la terre n'étant pas plate, la représentation planisphérique comporte toujours des ajustements géométriques. On distinguera dans les projections les plus célèbres celle de Mercator qui appose les géométries sur un cylindre, conservant ainsi les angles mais pas les surfaces (Flemish, géographe, et Gerardus Mercator, cartographe, 1569). Nous remarquons ainsi sur la figure \ref{fig:Mercator_projection} que les carrés proches de l'équateur ne subissent aucune déformation alors que ceux proches des pôles se transforment en rectangles allongés. Cette projection est dite \emph{conforme}.

Nous distinguons trois types de projections, qui sont exclusives l'une de l'autre :
\begin{enumerate}
    \item projection équivalente : conserve localement les surfaces ;
    \item projection conforme : conserve localement les angles, donc les formes ;
    \item projection aphylactique : elle n'est ni conforme ni équivalente, mais peut être équidistante, c'est-à-dire conserver les distances sur les méridiens.
\end{enumerate}

\begin{figure}
    \centering
    \begin{subfigure}[t]{0.3\textwidth}
    	\centering
        \includegraphics[width=0.6\textwidth]{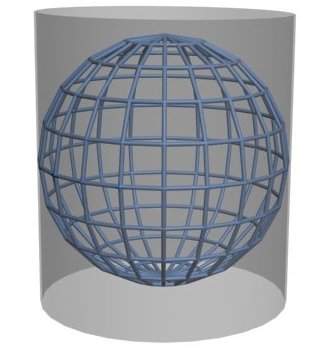}
    \end{subfigure}%
    ~ 
    \begin{subfigure}[t]{0.3\textwidth}
    	\centering
        \includegraphics[width=0.6\textwidth]{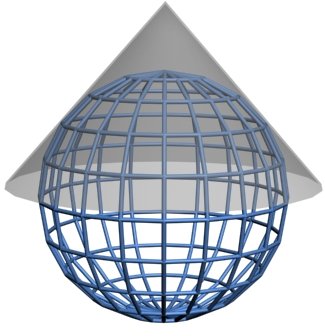}
    \end{subfigure}
    ~ 
    \begin{subfigure}[t]{0.3\textwidth}
    	\centering
        \includegraphics[width=0.9\textwidth]{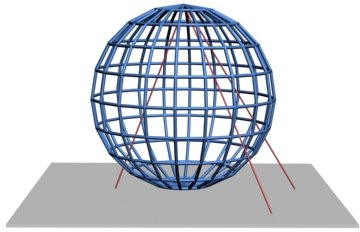}
    \end{subfigure}

    \caption{Différents types de surfaces développables, sur lesquelles s'appuient les différentes projections. \\source : Wikimedia Commons}
    \label{fig:surfaces_developpables}
\end{figure}

\begin{figure}
    \centering
    \begin{subfigure}[t]{0.3\textwidth}
    	\centering
        \includegraphics[width=0.7\textwidth]{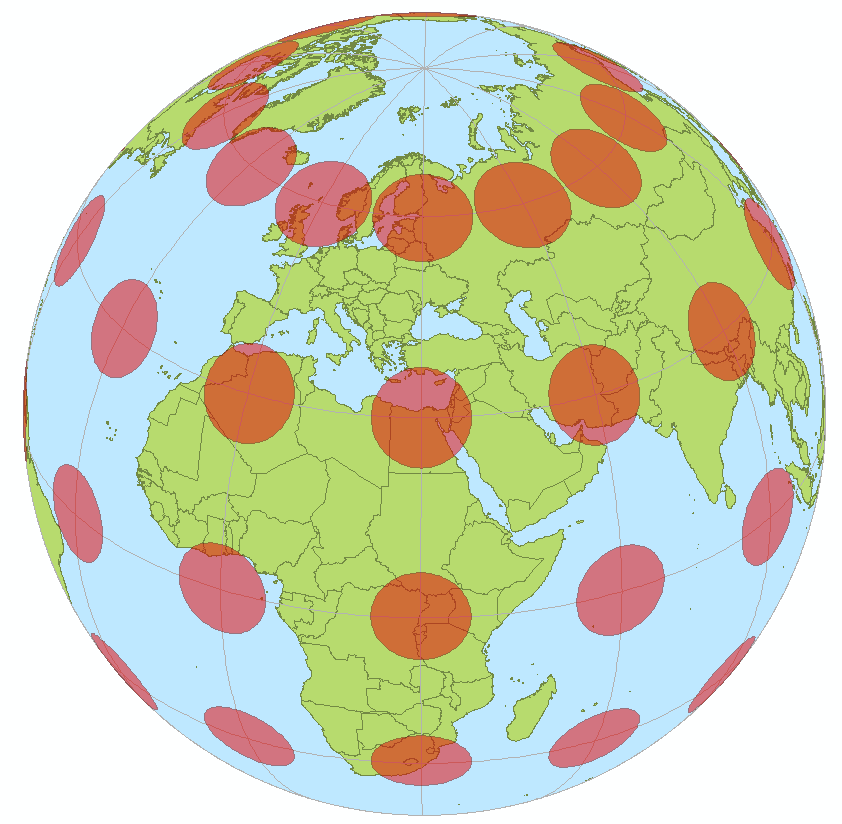}
        \caption{Représentation sans projection. \\source : Stefan Kühn - Travail personnel}
    \end{subfigure}
    ~
    \begin{subfigure}[t]{0.3\textwidth}
    	\centering
        \includegraphics[width=\textwidth]{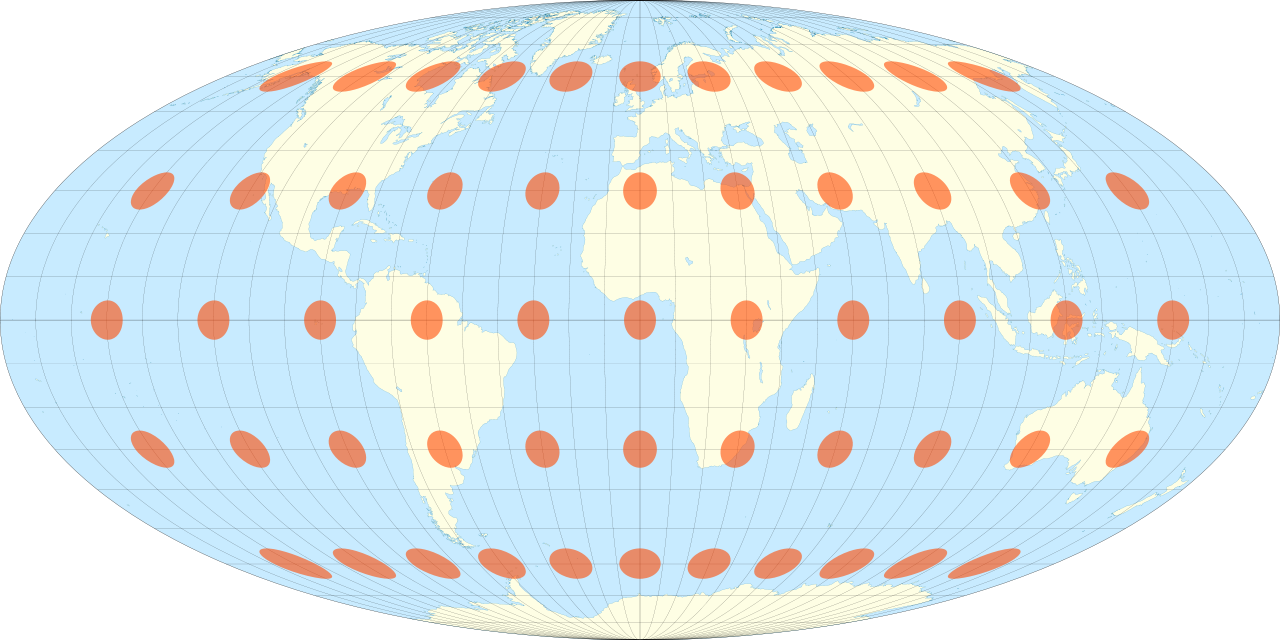}
        \caption{Projection de Mollweide. \\source : Eric Gaba - Travail personnel}
    \end{subfigure}
    ~
    \begin{subfigure}[t]{0.3\textwidth}
    	\centering
        \includegraphics[width=0.8\textwidth]{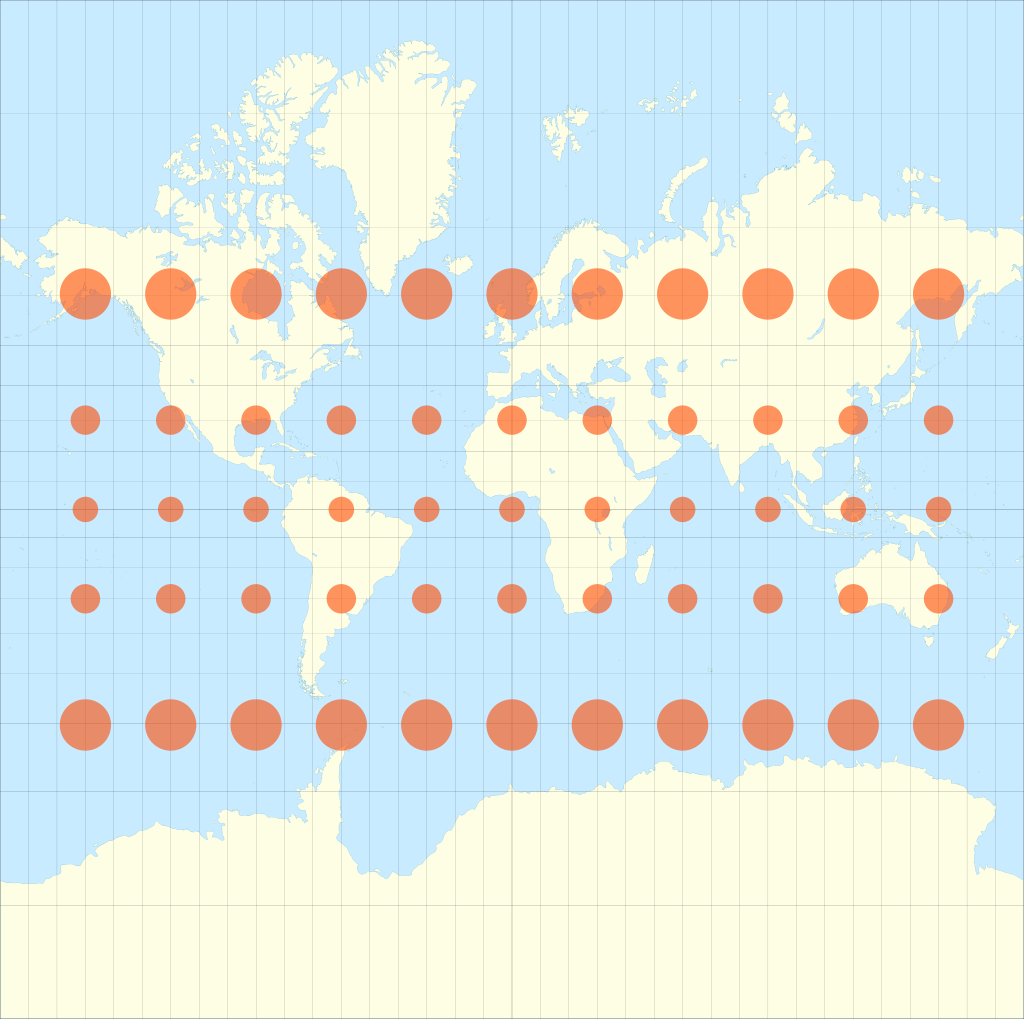}
        \caption{Projection de Mercator (arrêtée à 85\degres de latitude). \\source : Eric Gaba - Travail personnel}
    \end{subfigure}
    \caption{Indicatrices de Tissot, permettant de visualiser les déformations engendrées par les projections. Chaque cercle/ellipse rouge a un rayon de 500 km. 
    }
    \label{fig:tissot}
\end{figure}

Pour représenter des données géographiques sur un plan, il est nécessaire de définir à la fois un ellipsoïde de référence, auquel sera assimilée la surface terrestre, mais également une surface développable qui pourra être étalée sans déformation sur un plan. Elles peuvent être de trois types : cylindriques, coniques ou azimutales (figure \ref{fig:surfaces_developpables}). Une projection qui ne peut être classée dans un de ces types est appelée individuelle ou unique. L'indicatrice de Tissot est une forme géométrique (cercle ou ellipse) permettant d'apprécier le degré de conservation ou de déformation des formes ou des surfaces (figure \ref{fig:tissot} (Nicolas Auguste Tissot, \citep{ecole1856journal}).

Dans le présent travail, pour les réseaux géographiques, nous utiliserons l’ellipsoïde de référence WGS84 et une projection conique conforme (sur la France : le Lambert 93).

\FloatBarrier

\section*{Notre problématique : la caractérisation des structures spatiales}

Notre problématique porte sur la lecture des graphes spatiaux à travers la caractérisation de leur information spatiale. En effet, les structures des graphes ayant une inscription physique dans l'espace sont, pour la plupart, caractéristiques de leur croissance et de leur utilisation. Ainsi, le réseau des veinures d'une feuille est optimisé pour desservir l'ensemble du tissu végétal ; les craquelures dans de l'argile résultent d'un assèchement progressif ; le physarum optimise son parcours entre différents points d'alimentation ; et les réseaux de transports sont construits pour assurer la traversée et la desserte d'un espace. Autant d'exemples où le réseau se construit progressivement, de manière planifiée (avec un schéma général connu dès le début), ou organique (par développements successifs et réactions itératives par rapport à ces développements). Certains voient dans le schéma final une synthèse émergeant des différentes dynamiques de construction, fruit d'interactions non supervisées entre les différents composants du réseau.


C'est en adoptant ce point de vue, propre aux systèmes complexes, à l'intersection des sciences quantitatives et thématiques, que nous abordons notre problématique. Nous travaillons sur un graphe, avec des propriétés spatiales, qui s'est développé au cours du temps en hiérarchisant ses différents arcs en fonction de leur place dans le réseau et de leur utilisation. Pour comprendre la complexité des structures créées, nous construisons un hypergraphe que nous superposons au réseau. Celui-ci sera garant de la stabilité dans l'espace de nos résultats.

Dans la première partie de ce travail, nous expliquerons la modélisation des réseaux spatiaux que nous avons adoptée. Nous exposerons comment nous construisons un objet complexe multi-échelle pour créer un hypergraphe. Nous détaillerons les indicateurs que nous avons développés sur cet objet, qu'ils soient repris d'anciens travaux ou élaborés dans cette recherche. Nous comparerons ainsi les résultats obtenus, sur le graphe et l'hypergraphe, pour montrer l'apport de l'objet multi-échelle créé à la caractérisation spatiale. Nous montrerons également que s'il est possible de créer de très nombreux indicateurs, certains sont redondants, et qu'on peut ainsi se limiter à l'utilisation d'une \textit{grammaire de caractérisation de la spatialité}.

Dans la seconde partie de ce document, nous chercherons dans un premier temps à évaluer la sensibilité de notre modèle aux données utilisées. Puis, nous appliquerons notre travail à plusieurs réseaux spatiaux, de nature diverse. Nous nous appuierons ainsi sur un panel de quarante graphes, dont plus de la moitié sont des réseaux viaires, pour déceler leurs propriétés communes et identifier celles qui leur sont plus spécifiques. Enfin, nous ouvrirons l'étude de la caractérisation spatiale vers celle de la dynamique de développement des espaces urbains. Nous définirons pour cela une méthodologie de quantification diachronique des modifications structurelles. Nous concentrerons nos travaux sur deux villes dont nous disposons des données viaires historiques vectorisées (cartes panchroniques dont la spécification et la commande ont été faites pour cette étude) : Avignon (intra-muros) et Rotterdam (partie Nord). Nous examinerons le graphe de leurs réseaux de rues à travers une dizaines de dates (du XIV\textsuperscript{ème} siècle - pour Rotterdam - ou du XVIII\textsuperscript{ème} siècle - pour Avignon - à nos jours) et nous expliciterons ce que notre méthode de caractérisation peut apporter à la compréhension de leur développement.

La troisième et dernière partie de cette étude propose une mise en perspective pluridisciplinaire du travail développé. Nous nous concentrerons ainsi sur les graphes viaires pour proposer une lecture qualitative de la structure des villes à travers leur \emph{squelette}. Nous nous pencherons sur les enjeux portés par la forme et par sa transmission dans le temps. Nous croiserons notre analyse mathématique avec les descriptions d'historiens, d'urbanistes et d'archéo-géographes pour évaluer à quel point nous pouvons lire la ville entre ses lignes.

Dans chacune de ces parties, au travers des questions que nous nous posons, nous reviendrons sur les apports des différents chercheurs. Nous avons voulu permettre une lecture plurielle de ce document. Chacune des deux premières parties est donc terminée par une synthèse, regroupant l'essentiel de ce qui y a été développé. Pour faciliter la lecture, nous regroupons dans un glossaire en début de document les notations et le vocabulaire de théorie des graphes que nous utilisons. Nous avons également regroupé dans un \textit{guide de lecture} les notions principales que nous retrouverons au fil du texte.




\clearpage{\pagestyle{empty}\cleardoublepage}
\part{Modéliser les réseaux spatiaux : Construction d'une méthodologie de lecture}
\markboth{Modéliser les réseaux spatiaux}{Modéliser les réseaux spatiaux}


\clearpage{\pagestyle{empty}\cleardoublepage}
\chapter{Introduction}
\minitoc
\markright{Introduction de la première partie}

Les graphes spatialisés soulèvent différentes questions et s'appuient sur des disciplines diverses ; depuis la biologie (veinures de feuilles, réseaux neuronaux) ou la géomorphologie (réseaux fluviaux) jusqu'à l'urbanisme (réseaux de transports), sans oublier l'étude des réseaux d'énergie (réseaux électriques, de gaz ou de pétrole) ou de télécommunication. Nous nous concentrerons ici sur les exemples apportés par les réseaux viaires. C'est à travers ceux-ci que nous définirons dans cette partie un objet d'étude multi-échelle et des indicateurs liés, pour aboutir à une grammaire de lecture de la spatialité.

La \textit{spatialité}, la \textit{continuité} et la \textit{centralité} sont trois concepts au fondement de ce travail. La spatialité est établie par le caractère physique d'inscription dans l'espace des réseaux sur lesquels nous travaillons. C'est un état d'un système qui s'étend dans un espace pouvant être de une à trois dimensions. Le territoire correspond à une spatialité parmi d'autres. Dans notre cas, elle peut également s'entendre comme une feuille ou une plaque d'argile. La spatialité est le support qui impose des limites à la structure qu'elle contient.

Le concept de continuité, correspond empiriquement à un sentiment de perspective. Il peut être déterminé de plusieurs manières. Il pourrait se traduire par la notion de \textit{voie principale} sur laquelle viennent se connecter des \textit{voies secondaires} pour un réseau viaire, ou de \textit{fleuve} différenciable de ses \textit{affluents} pour un réseau hydrographique. Dans notre travail, nous ne considérons pas uniquement la continuité des éléments principaux du réseau mais celles de tous ses objets, comparés les uns aux autres, selon des critères quantifiés. Nous rejoindrons en cela les travaux faits en \textit{syntaxe spatiale}.

La centralité est un concept phare de la théorie des graphes \citep{freeman1977set}. Elle permet d'établir une hiérarchisation des différents objets d'un réseau afin d'en déterminer l'importance relative dans la structure de celui-ci. Calculer la centralité des objets dans un réseau peut se faire de plusieurs manières et ne nécessite pas que celui-ci soit spatialisé. Le caractère spatial de notre étude nous ouvre cependant à une approche pouvant être différente de celle réservée aux réseaux purement topologiques.

Les structures spatiales sur lesquelles nous travaillons ont une emprise et une forme. Leur emprise correspond au lieu où elles se situent, leur forme à leur géométrie propre. Cette géométrie, établie par le réseau physique sur lequel nous travaillons, peut se lire à plusieurs échelles. Nous appellerons échelle globale celle où ressortent les structures traversantes, dont la \textit{continuité} peut être dessinée d'un bout à l'autre du réseau. Les structures locales sont celles qui se limitent à une partie du réseau, pour desservir plus finement l'espace. Caractériser et quantifier ces différentes structures est un des enjeux de ce travail.

\FloatBarrier
\section{La spatialité}

Les études sur les graphes spatiaux, et notamment ceux représentant des réseaux de transports, ont débuté dans les années 1960. Dans un premier temps, elles se concentrèrent uniquement sur leurs propriétés topologiques, essayant de qualifier leurs structures en étudiant la centralité des nœuds. Les chercheurs disposaient à l'époque d'une quantité limitée de données, et étaient contraints par les techniques de modélisations et la puissance informatique \citep{garrison1962structure, haggett1969network}. Les réseaux de transports se développant, les recherches se sont rassemblées autour des flux dans des graphes aux géométries très différentes \citep{newell1980traffic, vaughan1987urban}. Les schémas de connexions des réseaux routiers ont intéressé également les chercheurs. W.L. Garrison s'est penché sur le réseau autoroutier des USA \citep{garrison1960connectivity} ; K.J. Kansky a affiné la caractérisation topologique des réseaux de transports \citep{kansky1963structure} ; Taylor a étudié les liaisons des voies rapides \citep{taylor1995hub}. Au début du XXI\textsuperscript{ème} siècle, des études statistiques de \textit{grands graphes} aux propriétés spatiales plus ou moins conservées ont également été réalisées \citep{albert1999internet, barabasi2002linked, barabasi2003scale, newman2003structure}. En 2011, M. Barthelemy fait une synthèse de l'avancée des recherches menées sur les réseaux spatiaux \citep{barthelemy2011spatial}. Il montre ainsi la pluralité des domaines d'application de cette modélisation et les caractéristiques propres aux contraintes spatiales.

En effet, la spatialisation des graphes ouvre un nouveau champ d'études. En s'intéressant plus spécifiquement à l'analyse des réseaux viaires, Rietveld suggère : \enquote{the quality of transport networks does not only depend on the features of the links, but also on the way the links are connected}\citep{rietveld1995some}. Il a étudié l’inter-connectivité entre différents modes de transports (plates-formes multimodales) mais également l'intra-connectivité d'un même réseau de transport \citep{rietveld1995some, rietveld1997policy}. Lee and Lee ont renforcé cette idée en mettant en avant la contrainte temporelle associée au transfert entre différents niveaux de routes \citep{lee1998new}. La réticence de l'utilisateur à la discontinuité a également été étudiée par L. Zhang qui décrit la sélection empirique des routes suivies comme fortement dépendante de l'alignement au carrefour \citep{zhang2006search}.

Le critère de continuité se dessine donc comme un facteur essentiel dans la caractérisation des réseaux spatiaux. Plus particulièrement, cela représente pour les réseaux viaires une possibilité d'étude de déplacements selon les tronçons \emph{continûment alignés}. Xie et Levinson ont étudié trois types différents d'itinéraires \citep{xie2007measuring}. Le premier se raccorde au plus vite aux voies les plus continûment alignées (appelées \emph{artères} par les deux chercheurs), pour ensuite quitter cette artère et rejoindre un réseau plus local. Cet itinéraire se rapproche de celui défini par Pailhous dans son étude des itinéraires des chauffeurs de taxis \citep{pailhous1970representation}. Le second itinéraire emprunte des tronçons de rues aléatoires. Le troisième se concentre uniquement sur les rues locales. Xie et Levinson montrèrent ainsi que le premier itinéraire, utilisant les \textit{artères}, est le plus efficace en terme de distance parcourue pour un temps donné. Ils comparent par la suite différentes abstractions géométriques de réseaux de rues, apposant sur certains d'entre eux des artères. Ils montrent ainsi l'impact de la géométrie sur des indicateurs, définis pour l'étude. Ils évaluent de cette manière la dépendance des caractéristiques du réseau à son inscription spatiale.

\FloatBarrier
\section{La continuité}

L'analyse de la continuité afin d'établir des relations entre les tronçons de graphes spatiaux est une problématique récurrente, dont plusieurs groupes de recherche se sont emparés. 

Nous retrouvons ainsi les travaux de \textit{syntaxe spatiale}, conçus et développés par B. Hillier, qui portent sur les réseaux viaires. À leurs débuts, ils ont cherché à reconstruire des lignes de perspectives pour étudier la structure du réseau \citep{hillier1976space, hillier1993natural, hillier1999centrality}. L'idée des chercheurs était alors une analyse géométrique du réseau viaire corrélée à une analyse psychologique du piéton évoluant dans celui-ci \citep{penn2003space, hillier2005network, hillier2006studying}. Ces études ont pour objet de comprendre comment l'usager perçoit son environnement et à quel niveau de \textit{profondeur} se situent les rues autour de lui, selon le nombre de tournants qui les séparent de leur destination. Les analyses mathématiques ont ainsi pour ambition d'aider à l'analyse des flux piétonniers. Nous reviendrons sur ces aspects qualitatifs dans la troisième partie de cette thèse.

La méthodologie utilisée pour définir la syntaxe spatiale d'un lieu fonctionne en quatre étapes. Tout d'abord, le graphe des rues est transformé en une carte d'axes correspondants aux différentes lignes de perspective. Ensuite est construit le \textit{line graph} de cette carte, où chaque ligne de perspective est représentée par un sommet et l'intersection entre deux lignes correspond à un arc entre les deux sommets. L'\textit{information géographique} disparaît ainsi derrière l'\textit{information topologique}. 

À partir de ce \textit{line graph} est calculé un indice d'intégration pour chaque sommet qui permet de hiérarchiser les lignes de perspectives. Cet indice, fondamental en syntaxe spatiale, est fondé sur la distance topologique la plus courte entre deux nœuds, appelée aussi distance géodésique \citep{hillier1996space, jiang2002integration}. Par la suite, il est représenté sur le graphe primal (où les nœuds correspondent aux intersections) en classant les axes selon une échelle de couleur. Cela permet aux chercheurs de cartographier la \textit{profondeur} du réseau évoquée plus haut. Cette méthodologie est ensuite élargie aux rues, reconstituées à partir de leur toponymie, par Jiang et Claramunt \citep{jiang2004topological}. Ils créent ainsi un nouveau \enquote{sur-réseau} (hypergraphe) pour caractériser le graphe primal spatialisé.


Avec le temps, les travaux en syntaxe spatiale se sont affinés autour de la notion d'angle à chaque carrefour qui impacte la perception du réseau de l'usager \citep{turner2001angular}. L'approche qu'ils développent, fondée sur les arcs du réseau, caractérise ces derniers selon une distance qualifiée d'\emph{angulaire}. Elle évalue ainsi la sensibilité des itinéraires aux angles entre segments \citep{hillier2007city}. Plus une combinaison de segments est continue et alignée, plus les tronçons routiers apparaîtront comme \enquote{centraux}. Les chercheurs utilisent leur théorie pour proposer des projets urbains qui permettraient théoriquement d'améliorer l'accès au centre ville et ainsi de décongestionner certains espaces problématiques \citep{hillier2009spatial}.

Ces travaux caractérisent chaque segment (tronçon routier) en fonction des autres segments auxquels il est relié en quantifiant les angles de connexion. C'est donc sur des unités élémentaires que porte le travail faisant ressortir des continuités transcendant la petite échelle lorsque les segments sont \textit{continûment alignés}.

Les travaux de Porta, Crucitti et Latora découlent de la théorie proposée par la syntaxe spatiale. Ils ont développé une méthode baptisée \emph{ICN} (\textit{Intersection Continuity Negociation}) afin de créer des \emph{lignes droites} dans le graphe \citep{porta2006network}. Pour chaque arc, l'algorithme recherche l'arc connecté qui lui est le plus aligné. Si l'angle entre les deux arcs est égal ou inférieur à 30\degres , ils sont identifiés comme appartenant à la même \emph{ligne droite}. Puis la procédure est répétée en partant de l'arc adjacent, jusqu'à ce que tout le réseau soit parcouru. Un fois qu'un arc est associé et ajouté à une ligne droite, il est retiré de l'ensemble des arcs restants pour le calcul. Cet algorithme est fondé sur la lecture des arcs et est donc dépendant du sens de lecture du réseau. Pour pallier ce problème (suivant le sens de lecture, différentes associations peuvent être faites), l'algorithme commence toujours par un arc appartenant à la plus longue ligne droite présumée du réseau (déterminée par une analyse succincte). Puis il prend la seconde ligne la plus longue, et ainsi de suite, jusqu'à avoir parcouru l'ensemble du réseau.

En se concentrant sur chaque arc, ces travaux permettent d'aboutir à un travail sur un \textit{line graph} du réseau viaire, où chaque sommet représente une \textit{ligne droite}. Les arcs entre sommets correspondent alors aux intersections entre ces lignes. Les changements de ligne droite se font dès que l'angle de déviation seuil, fixé à 30\degres , est dépassé entre deux arcs. Nous distinguerons également dans nos travaux le graphe spatial primal (abstraction du réseau physique observé) de sa représentation sous forme de \textit{line graph} où les arcs deviennent des sommets et inversement. 

L'importance de l'alignement a été soulevée dans de nombreux autres travaux, notamment ceux de Xie et Levinson qui créent des \emph{artères} sur le réseau, correspondant à un \textit{sur-réseau} (hypergraphe) \citep{xie2007measuring}. Ils ont pu ainsi établir une hiérarchie entre différentes parties de réseaux viaires et montrer que les graphes possédant de grands axes permettent le parcours de plus longues distances en un temps donné.

Au sein de l'IGN, fournisseur national de données géographiques, les laboratoires de recherche sont confrontés à cette notion d'alignement pour des questions de généralisation (simplification des géométries pour une meilleure visualisation à petite échelle). G. Touya construit ainsi des \textit{traits} (\textit{strockes}) réunissant les arcs par \textit{courbature continue} \citep{touya2010road}.

Qu'il s'agisse de \textit{lignes de perspectives}, de \textit{lignes droites}, d'\textit{artères} ou de \textit{traits}, le concept de continuité est fondamental dans toutes ces études portant sur des réseaux spatialisés. Nous poursuivrons ces recherches en construisant l'objet \textit{voie} selon différentes méthodes afin de déterminer quelle est celle répondant le mieux à la recherche de lignes \textit{continûment alignées}.

\FloatBarrier
\section{La centralité}

Le développement d'indicateurs, notamment pour mesurer la centralité des éléments dans un graphe a été une des premières préoccupations des scientifiques. Il s'agit en cela de déterminer quels sont les éléments les plus \enquote{influents} dans un réseau : quels sont ceux qui le perturberaient le plus en disparaissant ou bien participeraient à une contamination plus rapide dans un mécanisme de diffusion. La première conception de centralité a été proposée par Bavelas qui a étudié les réseaux de communication \citep{bavelas1948mathematical}. Il considère alors qu'un sommet est central dans le réseau si un grand nombre de chemins les plus courts passent par celui-ci. Il pose ainsi les fondements d'un indicateur structurel essentiel dans la théorie des graphes : la \emph{betweenness}. Il identifie dans son travail le potentiel \enquote{bloquant} des sommets à l'indicateur de betweenness élevé.

Cette idée fut reprise par Shimbel qui estime que les \enquote{stations intermédiaires} traversées sur un chemin ont un impact sur celui-ci \citep{shimbel1953structural}. Selon lui, elles imposent un \enquote{stress} au chemin. Il mesure ainsi le \enquote{stress} de chaque sommet en fonction du nombre de chemins les plus courts auxquels ils participent. Cohn et Marriott ont approfondi ce travail en considérant l'impact de la centralité différemment. Ils ne parlent plus en terme de potentiel \enquote{bloquant} ou \enquote{stressant} mais observent la capacité des sommets centraux à densifier le réseau autour d'eux et à coordonner l'activité des autres sommets \citep{cohn1958networks}.

Dans la plupart des premiers travaux sur la centralité les chercheurs considèrent la somme de toutes les distances minimales entre un point et les autres du graphe pour déterminer s'il est central ou non \citep{bavelas1950communication,beauchamp1965improved,sabidussi1966centrality}. Ceci pose le problème des calculs de centralité dans les graphes non connexes : la distance entre deux points situés dans deux parties non connectées étant infinie. Partant de ce constat, Freeman propose un formalisation mathématique de la betweenness \citep{freeman1977set} la rendant compatible avec des graphes non connexes.

Aux côtés de la betweenness, plusieurs autres mesures de centralités furent également explorées. Nous dénombrons ainsi parmi les plus connues : la \textit{degree centrality} (liée aux degrés des sommets), la \textit{closeness centrality} (déterminant la proximité à partir d'un élément du réseau vers tous les autres) et la \textit{Eigenvector centrality} (mesurant l'influence d'un nœud au sein du réseau, à partir des connexions des nœuds auxquels il est connecté). Nous développerons également la \textit{stress centrality} dont la construction est proche de celle de la betweenness. Toutes ces mesures ont été faites dans un premier temps sur des graphes formalisant des liens sociologiques, où seule la topologie est significative \citep{bavelas1951reseaux, leavitt1951some, parlebas1972centralite}. Certains scientifiques en ont déjà exploré l'application sur des graphes spatialisés \citep{crucitti2006centralityin, crucitti2006centralitymeasures, barthelemy2013self}. Nous verrons dans ce travail les informations particulières que peuvent apporter certains de ces indicateurs selon leur objet d'application dans un graphe spatialisé.

\FloatBarrier
\section{Méthodologie de caractérisation de la spatialité à travers la continuité}

Dans cette partie, nous commencerons par expliciter le découpage mis en œuvre pour extraire de réseaux spatiaux physiques une unité élémentaire de lecture : le segment. Cet objet mathématique à la géométrie très simple servira de \enquote{brique} de construction des arcs des graphes étudiés. À partir de ces définitions d'objets simples, aux géométries fixées par des coordonnées, nous montrerons comment construire un objet complexe multi-échelle, la \emph{voie}, selon une paramétrisation étudiée.

Une fois les paramètres de construction de la voie établis, nous définirons plusieurs indicateurs, certains déjà utilisés en théorie des graphes (essentiellement topologiques), d'autres que nous construirons à partir des propriétés topologiques et géométriques des graphes. Nous les différencierons en deux types : les indicateurs locaux, qui dépendent de l'objet d'étude et de son voisinage direct ; et les indicateurs globaux, qui sont calculés en tenant compte de l'ensemble du réseau. Nous comparerons leurs résultats sur le graphe des arcs et sur celui des voies. Nous effectuerons également des combinaisons entre indicateurs locaux et globaux afin de voir si cela aboutit à une caractérisation différente du réseau.

Pour assurer la non redondance d'informations apportées par les indicateurs développés, nous comparerons leurs résultats sur différents graphes viaires. Nous définirons grâce à cela une grammaire d'indicateurs primaires (de première génération) que nous composerons entre eux par division afin de comprendre si ce type de composition permet de créer de nouvelles caractérisations du réseau.

Ces recherches aboutissent à la définition d'une méthodologie de caractérisation des réseaux spatiaux à travers un objet géographique, la voie, et des calculs d'indicateurs qui lui sont appliqués. Le caractère multi-échelle de cet objet rend la caractérisation et la lecture du réseau à travers ses propriétés topologiques, particulièrement efficaces. Nous verrons comment il peut rendre une caractérisation globale équivalente à une opérée localement, et ainsi permettre de lire les structures de grands graphes en réduisant les temps de calcul.

L'analyse que nous réalisons ici est quantitative. Nous en présenterons les limites, les multiples applications possibles (ainsi que les potentialités de comparaisons que cela offre) et les perspectives diachroniques en deuxième partie. Enfin, nous réserverons l'étude qualitative des résultats pour la troisième partie de ce travail. Nous n'évoquerons donc ces aspects que très brièvement dans cette partie.

\clearpage{\pagestyle{empty}\cleardoublepage}
\chapter{Construction d'un hypergraphe appliqué aux réseaux spatiaux à travers la \textit{voie}}
\minitoc
\markright{Construction d'un hypergraphe appliqué aux réseaux spatiaux à travers la \textit{voie}}


Nous voulons ici compléter les travaux cités en introduction \citep{hillier2007city, porta2006network} en créant un objet complexe, dont la paramétrisation est optimisée par une étude statistique complète. Pour construire cet objet, nous nous appuyons sur des règles locales, que nous appliquons à chaque sommet. Notre objet sera ainsi robuste au sens de lecture du réseau. Nous explicitons dans ce chapitre comment, à travers trois méthodes de construction, nous aboutissons à la création de l'objet \emph{voie}, qui est au fondement de ce travail de recherche.

\FloatBarrier
\section{Abstraction d'un réseau spatial}

\begin{figure}
    \centering
    \begin{subfigure}[t]{0.48\textwidth}
       \includegraphics[width=\textwidth]{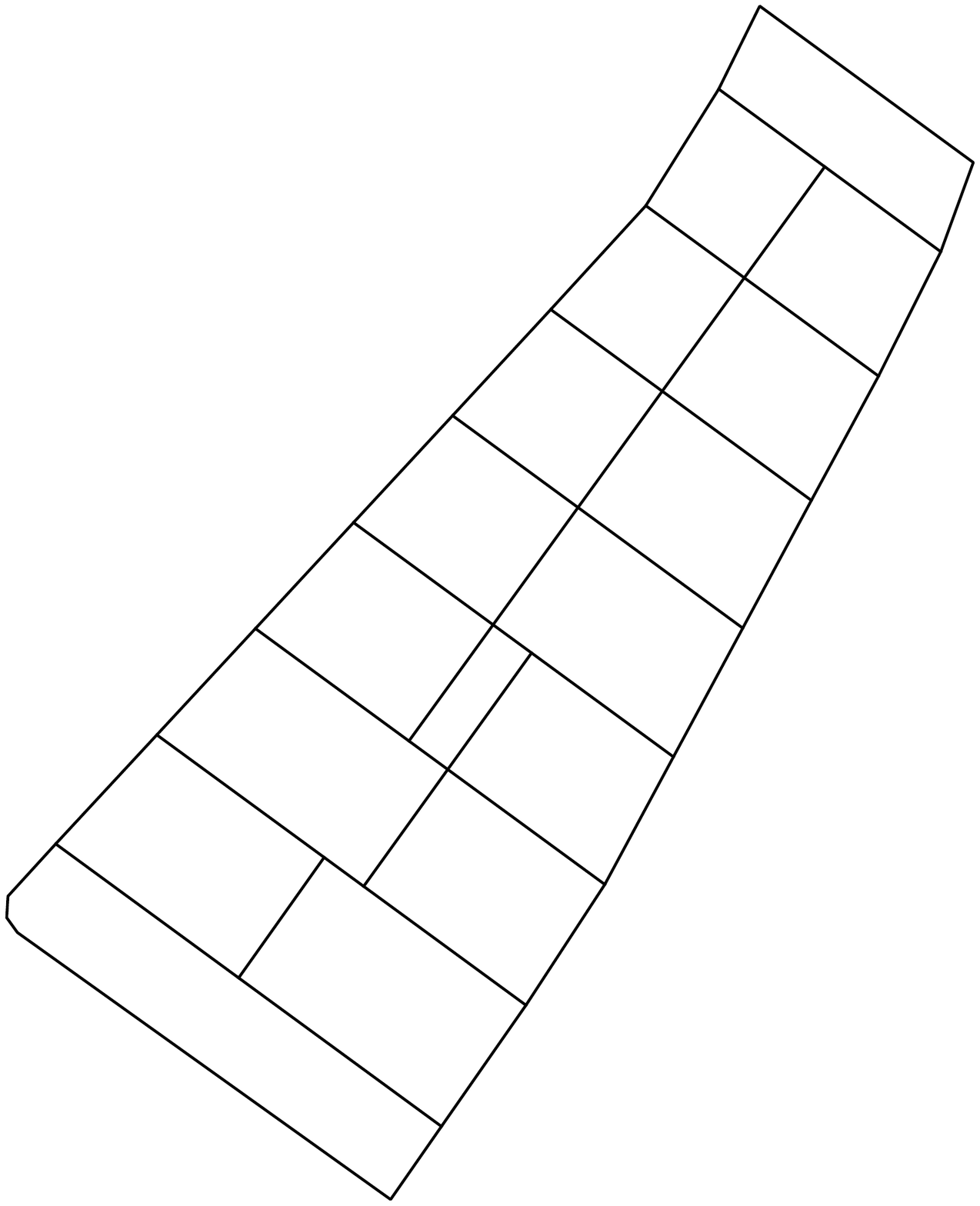}
    \caption{ Graphe spatial brut.}
    \label{fig:1_reseau_brut}
    \end{subfigure}
	~
    \begin{subfigure}[t]{0.48\textwidth}
        \includegraphics[width=\textwidth]{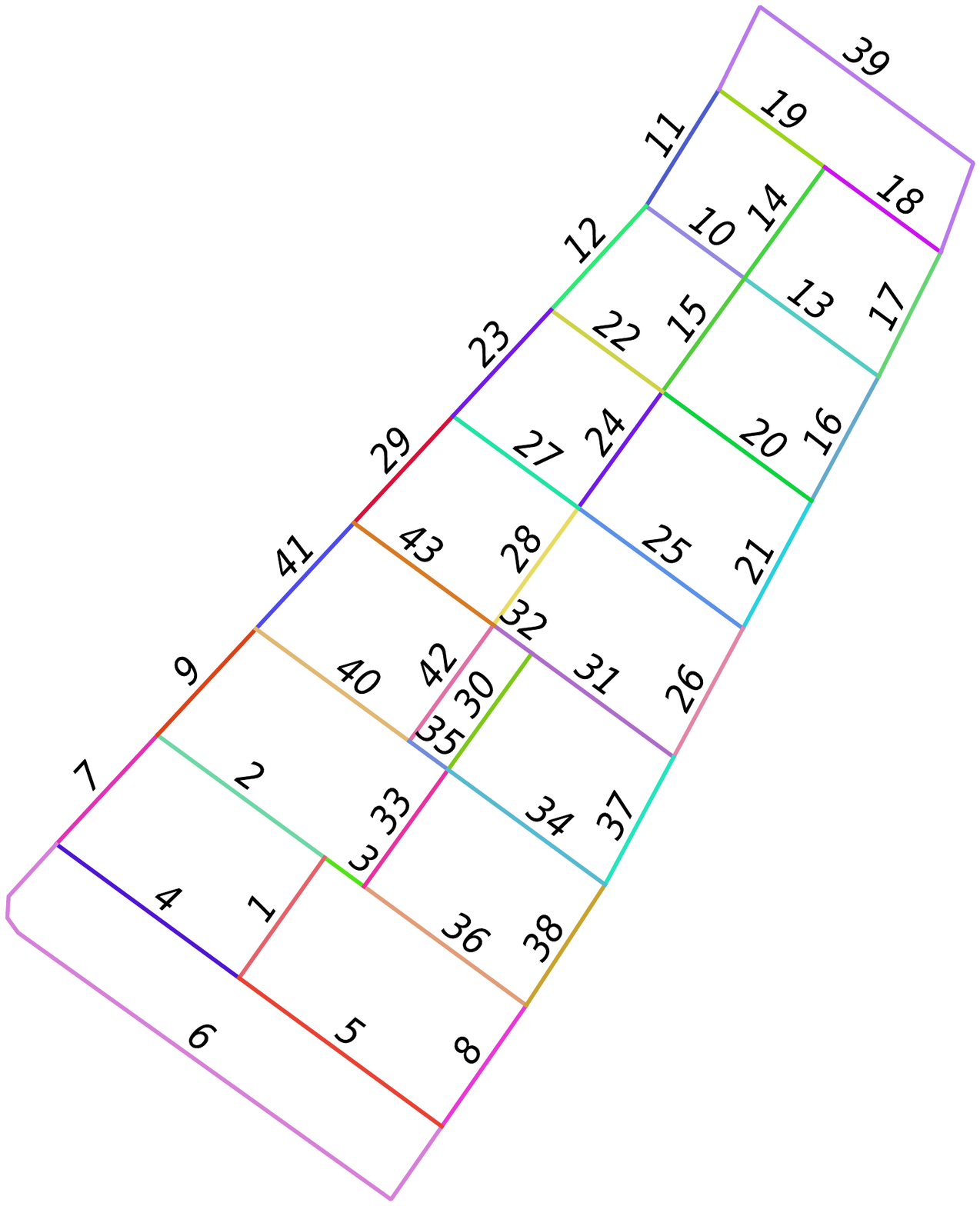}
        \caption{Création des arcs, par association des segments entre deux sommets.}
        \label{fig:4_reseau_arcs}
    \end{subfigure}
    \caption{Graphe spatial schématique extrait du réseau viaire de la ville de Téhéran.}
\end{figure}

\begin{figure}
    \centering
    \begin{subfigure}[t]{0.45\textwidth}
        \includegraphics[width=\textwidth]{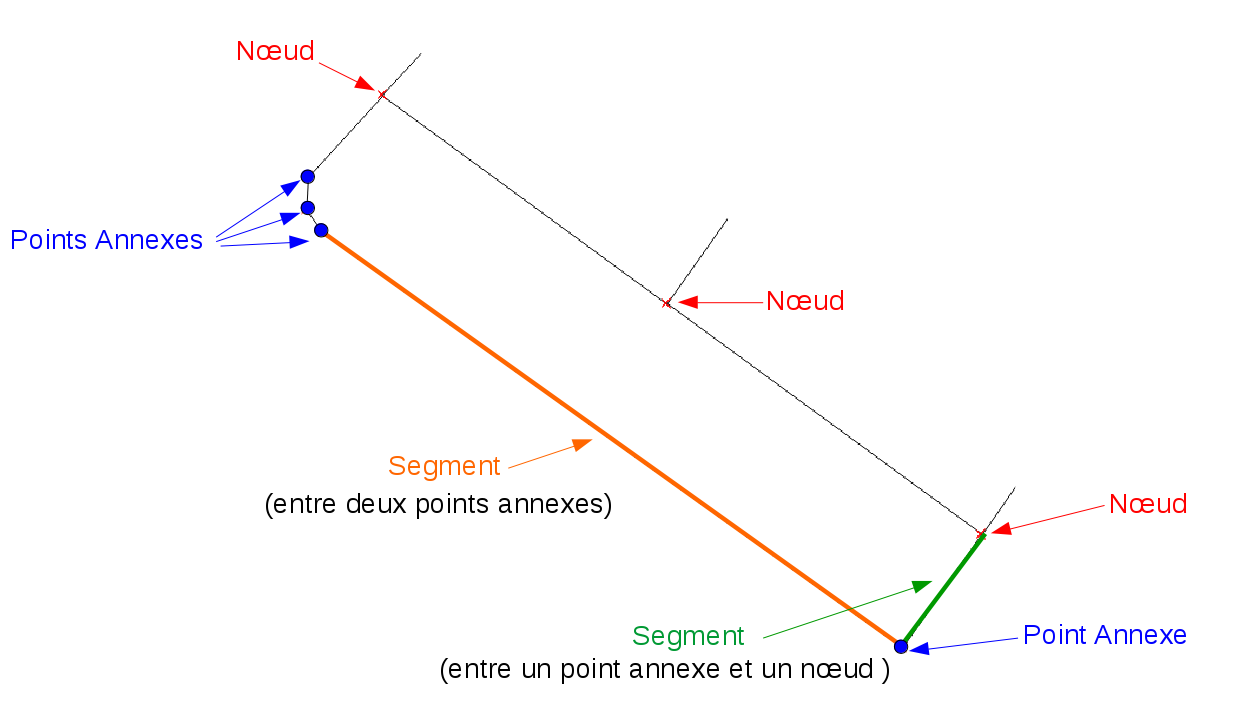}
        \caption{ Positionnement des sommets, points annexes et segments.}
        \label{fig:2_reseau_elem1}
    \end{subfigure}
	~
    \begin{subfigure}[t]{0.45\textwidth}
        \includegraphics[width=\textwidth]{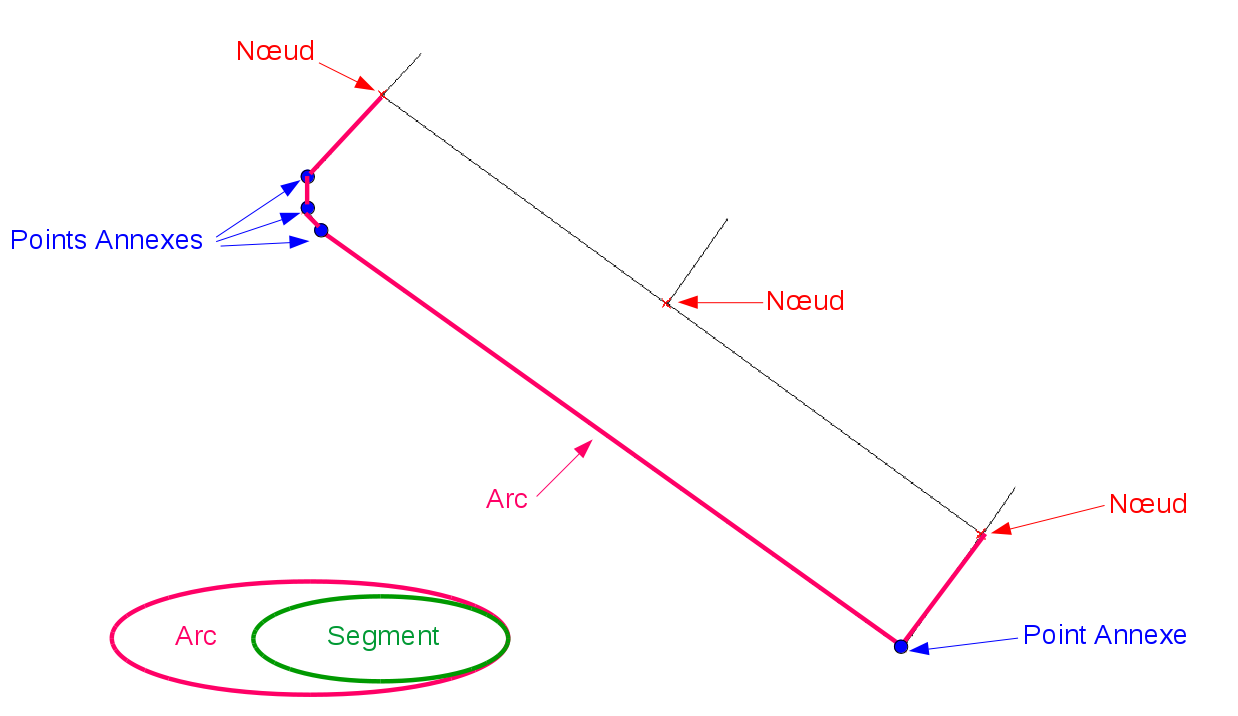}
        \caption{ Création des arcs, par association des segments entre deux sommets.}
        \label{fig:3_reseau_elem2}
    \end{subfigure}
    \caption{Détail de construction des objets élémentaires sur le graphe.}
\end{figure}

Pour construire un graphe spatial à partir de réseaux physiques (routes, veinures ou craquelures), nous commençons par les déconstruire afin de réduire l'information géométrique qu'ils portent à une unité élémentaire : le segment. Ainsi, dans de tels réseaux, nous considérons uniquement le squelette de la structure (figure \ref{fig:1_reseau_brut}). À partir de ce filaire, nous positionnons des \emph{sommets} aux intersections et des \emph{points annexes} aux changements de direction. Entre un sommet et un point annexe ou entre deux points annexes, nous avons donc des segments, au sens mathématique du terme : ligne droite entre deux points. Ce segment est l'unité spatiale linéaire la plus petite de notre graphe (figure \ref{fig:2_reseau_elem1}). Entre deux sommets les segments sont regroupés pour former des arcs (figures \ref{fig:3_reseau_elem2} et \ref{fig:4_reseau_arcs}).

Nous obtenons ainsi un graphe, constitué de sommets et d'arcs $G(S,A)$ ($S$ étant l'ensemble des sommets {$s$} et $A$ l'ensemble des arcs {$a$} du graphe $G$). Ce graphe a la particularité d'avoir des arcs avec une géométrie caractéristique propre : les points annexes sont garants de cette inscription spatiale en donnant des coordonnées à chaque changement de direction d'un arc. La construction de ce graphe spatial permettra dans la suite de l'étude d'associer l'information géométrique des arcs à celle topologique, contenue dans tout graphe.

Nous notons que, dans le cas des graphes routiers, le graphe créé n'est pas forcément planaire. En effet, un pont urbain (deux rues qui se croisent à des altitudes différentes : un échangeur autoroutier par exemple) ne donnera pas lieu à la création d'une intersection, et donc ne sera pas formalisé par un sommet. Si nous considérons d'autres types de graphes spatiaux comme celui des veinures de feuille ou de craquelures dans de l'argile, en revanche, le graphe considéré sera bien planaire.

Nous nous limitons à l'information minimale fournie par le filaire : la géométrie et la topologie du réseau. Quel que soit le cas d'application, nous ne considérons dans le réseau spatial que l'axe central de l'objet physique, dénué de toute autre information attributaire (pas de largeur ou de type d'utilisation...). Par exemple, dans un graphe routier, le sens de circulation n'est pas une donnée prise en compte : les arcs sont supposés mener d'une intersection à une autre indifféremment. Sur chaque échantillon spatial considéré, nous ne conservons que la plus grande partie connexe (nous l’identifions comme étant celle qui comprend le plus d'arcs). Les graphes sur lesquels nous travaillons sont donc non orientés et connexes. En revanche, ils ne sont pas nécessairement planaires.

\FloatBarrier
\section{Construction locale d'un objet multi-échelle : la voie}

\begin{figure}{H}
    \centering
    \begin{subfigure}[t]{0.4\textwidth}
        \includegraphics[width=\textwidth]{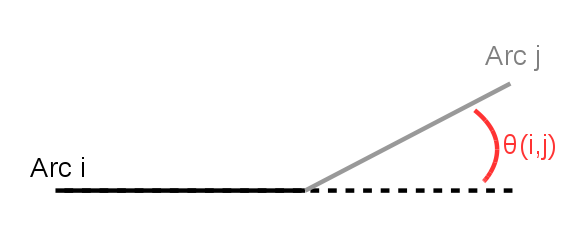}
        \caption{Illustration de l'angle de déviation $\theta(i,j)$ considéré entre les arcs $i$ et $j$.}
        \label{fig:6_deviation}
    \end{subfigure}
    ~
    \begin{subfigure}[t]{0.4\textwidth}
        \includegraphics[width=\textwidth]{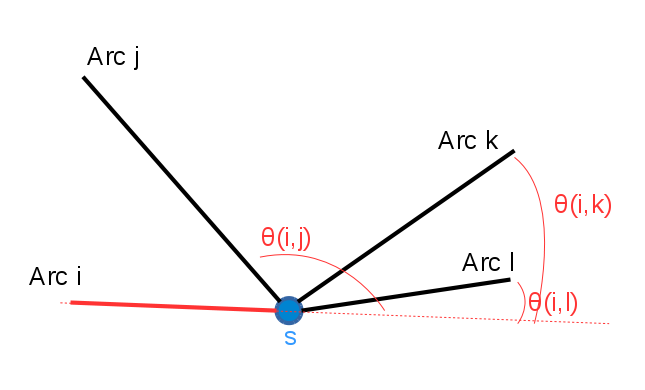}
        \caption{Exemple d'un sommet $s$ sur lequel se joignent 4 arcs $i$, $j$, $k$ et $l$. L'arc $i$ est pris comme arc de référence pour l'itération illustrée et le schéma fait apparaître les angles que l'on considère à partir de $i$ pour les autres arcs.}
        \label{fig:7_sommet_angles}
    \end{subfigure}
    \caption{Détail de l'analyse d'angles de déviation à chaque sommet.}
\end{figure}

La notion de voie est très polysémique. Elle peut avoir pour signification \enquote{route}, \enquote{rue}, \enquote{chemin}, \enquote{passage} mais également \enquote{direction prise}. La voie est un objet que nous construisons localement à chaque sommet $s$ d'un graphe $G(S,A)$. Pour chaque arc lié à un sommet $a_{ref} / s \in a_{ref}$ nous allons examiner tous les angles de déviation formés avec les autres arcs {$a_i / (s \in a_i) \wedge (a_i \neq a_{ref})$}. La déviation alors considérée est l'angle complémentaire à 180\degres  de celui fait par les deux arcs (figure \ref{fig:6_deviation}). Pour calculer l'angle entre deux arcs, nous retenons uniquement le segment de cet arc lié au sommet considéré (segment situé entre le sommet et le premier point annexe de l'arc). Pour chaque arc lié au sommet $s$, il y a donc autant d'angles de déviation calculés que d'autres arcs connectés (figure \ref{fig:7_sommet_angles}). Si $N$ représente le degré du sommet, l'analyse considère donc $\frac{N(N-1)}{2}$ angles. 

Nous développons trois méthodes pour la construction de la voie, chacune ayant des critères d'association différents :

\begin{itemize}
    \item [$M0$] : les arcs sont associés par ordre de déviation minimale, déviation restant inférieure à un angle seuil.
    \item [$M1$] : les arcs sont associés de façon à minimiser la somme de l'ensemble des déviations, tout en respectant un angle seuil.
    \item [$M2$] : les arcs sont associés aléatoirement à chaque sommet (méthode \enquote{blanche}), tout en respectant un angle seuil.
\end{itemize}

Pour chacune de ces trois méthodes nous testons l'impact du choix de l'angle seuil $\theta_{seuil}$ à partir duquel les deux arcs ne seront plus considérés comme alignés et ne pourront donc plus appartenir à la même voie. Si $\theta_{seuil} = 0\degres$, les arcs ne pourront être couplés que s'ils sont exactement alignés (aucune déviation possible). Au contraire, si $\theta_{seuil} = 180\degres$ alors il n'y a aucune contrainte d'alignement.

S. Porta utilisait un angle de déviation maximale de 30\degres  \citep{porta2006network} alors que T. Courtat le fixait à 90\degres  \citep{courtat2011mathematics}. Tous deux utilisaient une méthode de construction privilégiant l'alignement maximal par paires. Notre objectif ici est de déterminer si la construction de la voie est pertinente, à travers la comparaison avec la méthode blanche ($M2$). Le cas échéant, d'identifier l'angle de déviation le plus adapté et la méthode qui donne l'objet le plus robuste : celle qui privilégie l'alignement local ($M0$) ou global ($M1$).

La méthode $M0$ procède par itérations successives sur chaque sommet $s$. Cette méthode consiste à sélectionner itérativement le couple d'arcs liés au sommet $(s_{a_{i}}, s_{a_{j}})$ ayant l'angle de déviation minimum $\theta_{(i,j)}$. Si celui-ci est inférieur à l'angle seuil ($\theta_{(i,j)} < \theta_{seuil}$), les arcs sont appariés et retirés de l'ensemble des candidats restants pour le calcul. L'itération suivante est amorcée et ce jusqu'à ce que l'angle de déviation minimum dépasse l'angle seuil fixé. Quand cette situation est atteinte, les arcs restants sont identifiés comme étant \emph{célibataires}. Ils constitueront une extrémité de voie (algorithme \ref{alg:m0}).

La méthode $M1$ consiste à calculer pour chaque sommet $s$ toutes les combinaisons possibles de couples d'arcs qui respectent l'angle de déviation seuil. Pour chaque combinaison, l'ensemble des angles de déviation est sommé. La combinaison ayant la somme de déviation minimale est retenue. De même, les arcs non appariés restent \emph{célibataires} (algorithme \ref{alg:m1}).

La méthode $M2$ couple à chaque sommet les arcs aléatoirement tout en respectant l'angle seuil fixé. Cette méthode pourra donc également voir des arcs non appariés.

\begin{algorithm}
    \caption{Méthode M0}
    \label{alg:m0}
    \begin{algorithmic}

        \FORALL{$sommet \in graphe$}
        \STATE $tab_{arcs} \gets arcsAuSommet(sommet)$
        \WHILE{$taille(tab_{arcs}) > 1$}
        \STATE $\theta_{min} \gets min({\theta_{(arc_i, arc_j)}}/\{arc_i, arc_j\} \in tab_{arcs})$

        \IF{$\theta_{min} \leq \theta_{seuil}$}
        \STATE $ajouteCouple(sommet, arc_i, arc_j)$
        \STATE $tab_{arcs} \gets retire(tab_{arcs}, arc_i)$
        \STATE $tab_{arcs} \gets retire(tab_{arcs}, arc_j)$
        \ENDIF

        \IF{$\theta_{min} > \theta_{seuil}$}
        \FORALL{$arc \in tab_{arcs}$}
        \STATE $ajouteCouple(sommet, arc, 0)$
        \STATE sommet suivant
        \ENDFOR
        \ENDIF

        \ENDWHILE
        \IF{$tab_{arcs} \neq \emptyset$}
        \STATE $arc \gets tab_{arcs}[0]$
        \STATE $ajouteCouple(sommet, arc, 0)$
        \ENDIF
        \ENDFOR

    \end{algorithmic}
\end{algorithm}

\begin{algorithm}
    \caption{Méthode M1}
    \label{alg:m1}
    \begin{algorithmic}

        \FORALL{$sommet \in graphe$}
        \STATE $tab_{arcs} \gets arcsAuSommet(sommet)$
        \WHILE{$taille(tab_{arcs}) > 1$}

        \STATE $ {\Sigma\theta}_{min} \gets min({ \sum (\theta_{(arc_i, arc_j)}/\{arc_i, arc_j\} \in tab_{arcs} \wedge (\theta_{(arc_i, arc_j)} \leq \theta_{seuil}))})$
        \STATE $ EnsemblePaires_{{\Sigma\theta}_{min}} =  (arc_i, arc_j) / \sum (\theta_{(arc_i, arc_j)}/\{arc_i, arc_j\} \in tab_{arcs}) = {\Sigma\theta}_{min}$

        \IF{$(arc_i, arc_j) \in EnsemblePaires_{{\Sigma\theta}_{min}}) $}
        \STATE $ajouteCouple(sommet, arc_i, arc_j)$
        \STATE $tab_{arcs} \gets retire(tab_{arcs}, arc_i)$
        \STATE $tab_{arcs} \gets retire(tab_{arcs}, arc_j)$
        \ENDIF

        \ENDWHILE

        \IF{$tab_{arcs} \neq \emptyset$}
        \FORALL{$arc \in tab_{arcs}$}
        \STATE $ajouteCouple(sommet, arc, 0)$
        \STATE sommet suivant
        \ENDFOR
        \ENDIF

        \ENDFOR

    \end{algorithmic}
\end{algorithm}

Une fois la table d'appariement entre arcs faite, selon une des trois méthodes, les voies peuvent être construites à partir de celle-ci. À chaque sommet, chaque paire d'arcs est identifiée comme appartenant à la même voie. Il est ainsi possible d'élaborer, de proche en proche, un objet multi-échelle à partir des ces couples construits localement. Nous obtenons ainsi un hypergraphe $G(S,V)$ à partir du graphe $G(S,A)$, où l'ensemble des voies {$V$} est construit à partir d'associations d'arcs $\in \{A\}$. La dénomination d'hypergraphe a été mise en place par C. Berge, pour désigner un graphe où les arêtes ne relient plus deux sommets mais un nombre de sommets compris entre 1 et le nombre de sommets du graphe \citep{berge1973graphes}. Plus le nombre de voies créées est faible, plus elles réunissent d'arcs et peuvent potentiellement traverser le réseau. Ce point est au fondement de la pertinence de notre analyse (cf chapitre 2 de la deuxième partie).

\begin{figure}
    \centering
    \begin{subfigure}[t]{0.48\textwidth}
        \includegraphics[width=\textwidth]{images/schemas/teh_arcs.pdf}
        \caption{Création des arcs, par association des segments entre deux sommets.}
        \label{fig:4_reseau_arcs2}
    \end{subfigure}
    ~
    \begin{subfigure}[t]{0.48\textwidth}
       \includegraphics[width=\textwidth]{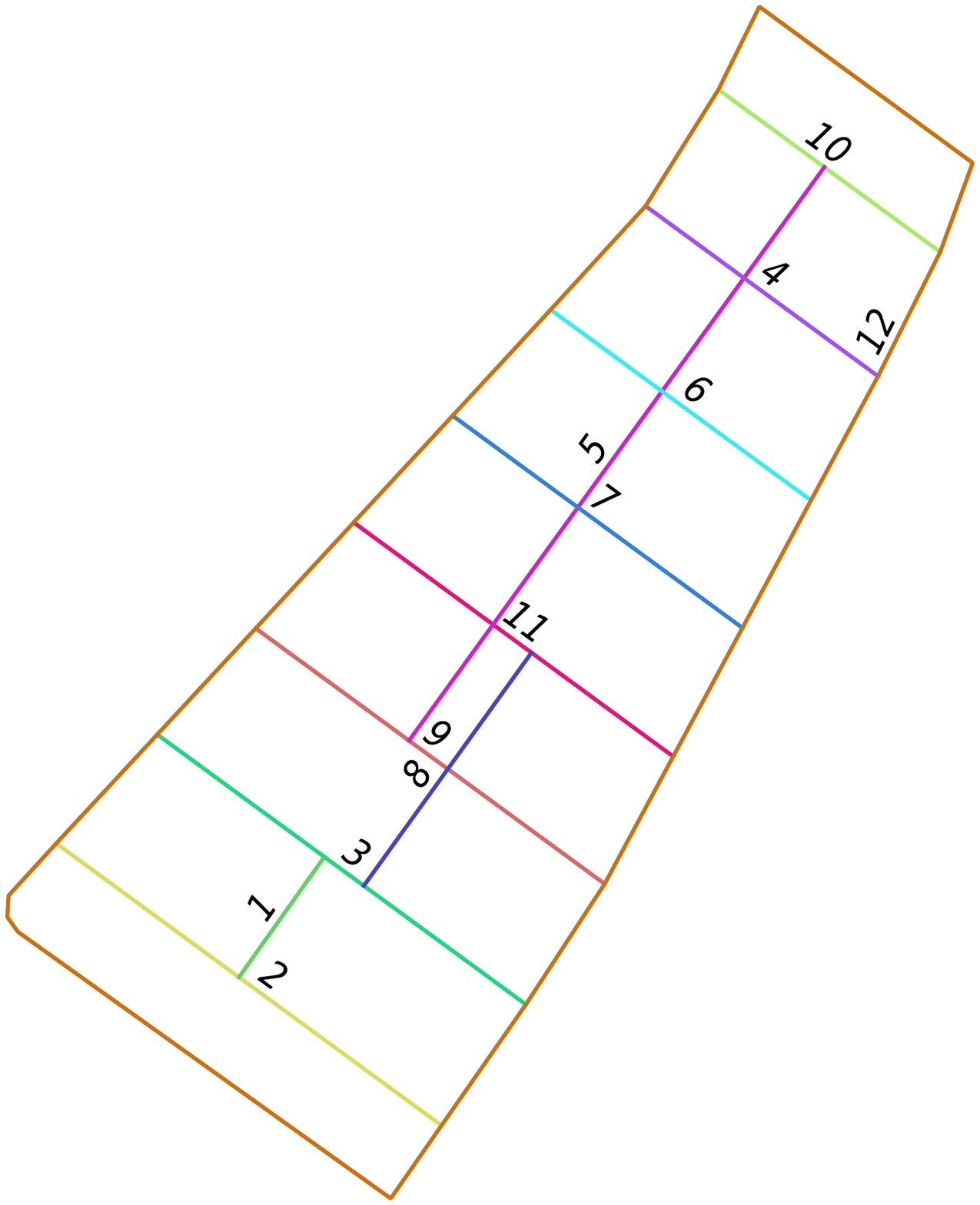}
    \caption{Création des voies, par association d'arcs à chaque sommet.}
    \label{fig:5_reseau_voie}
    \end{subfigure}
    \caption{Construction d'objets géographiques. Mise en parallèle de la création d'arcs et de voies.}
\end{figure}

\begin{figure}[h]
    \centering
    \includegraphics[width=0.6\textwidth]{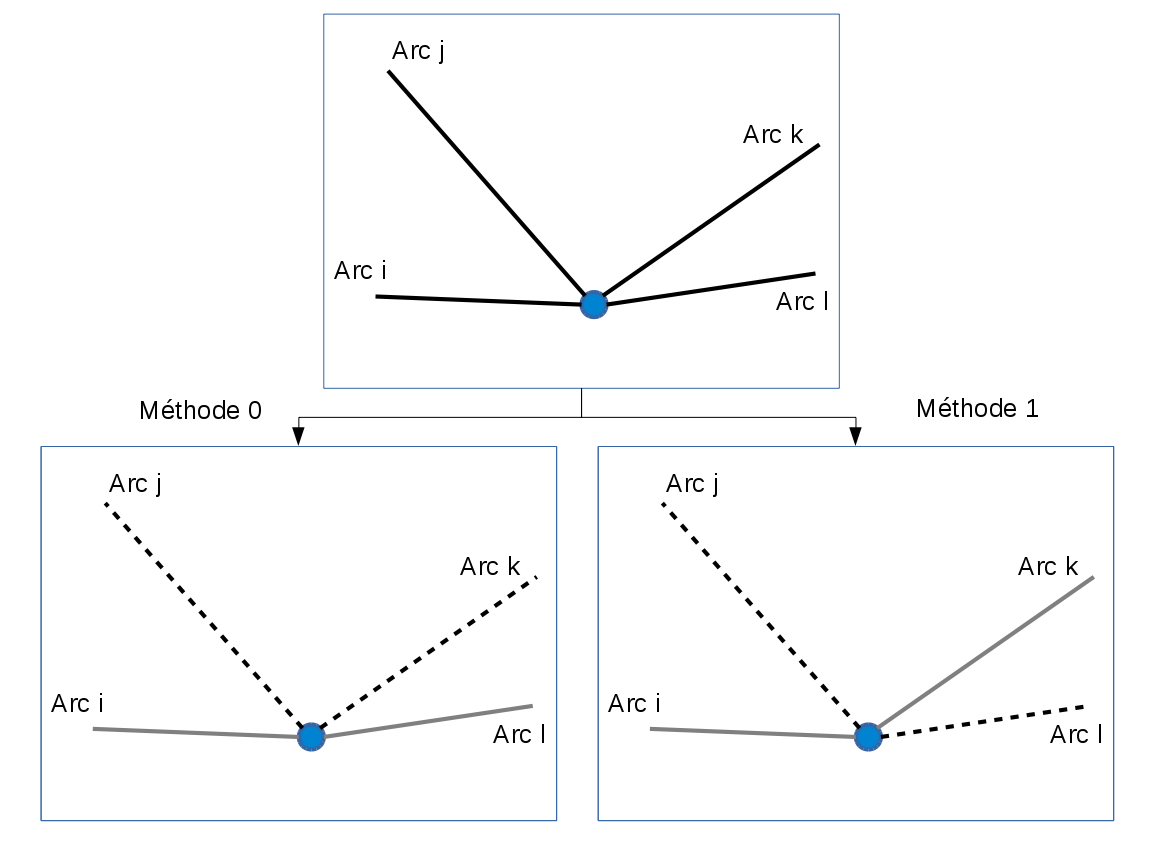}
    \caption{Résultat des associations faites avec chacune des méthodes $M0$ et $M1$. Avec $M0$ : les couples formés sont $(Arc_i, Arc_l)$ et $(Arc_j, Arc_k)$ ; Avec $M1$ : les couples formés sont $(Arc_i, Arc_k)$ et $(Arc_j, Arc_l)$.}
    \label{fig:8_comparaison_methodes}
\end{figure}

Si l'on crée les voies sur l'échantillon schématique que nous avons utilisé précédemment, la méthode choisie n'aura pas d'impact sur les géométries des voies construites (si l'angle seuil est choisi entre 40\degres  et 80\degres ). En effet, la déviation aux intersections entre arcs est soit proche de 0\degres , soit proche de 90\degres . Ceci équivaut donc à un choix binaire évident : les deux arcs peuvent clairement être identifiés comme appartenant à la même voie ou à deux voies différentes (figure \ref{fig:5_reseau_voie}). Cependant, si les angles aux intersections sont moins contrastés, les deux méthodes peuvent aboutir à des associations différentes (figure \ref{fig:8_comparaison_methodes}). Nous nommons cette configuration entre quatre arcs le \textit{problème du K}, où les déviations minimales ne sont pas celles qui aboutissent à une déviation globale minimale. En effet, cette configuration particulière résulte de deux arcs se connectant avec des angles faibles sur deux autres entre lesquels la déviation est quasi-nulle (ce qui est illustré par la forme de la lettre \enquote{K}).

\FloatBarrier
\section{Impact de la paramétrisation de la construction}

\begin{figure}[H]
    \centering
    \begin{subfigure}[t]{0.45\textwidth}
        \includegraphics[width=\textwidth]{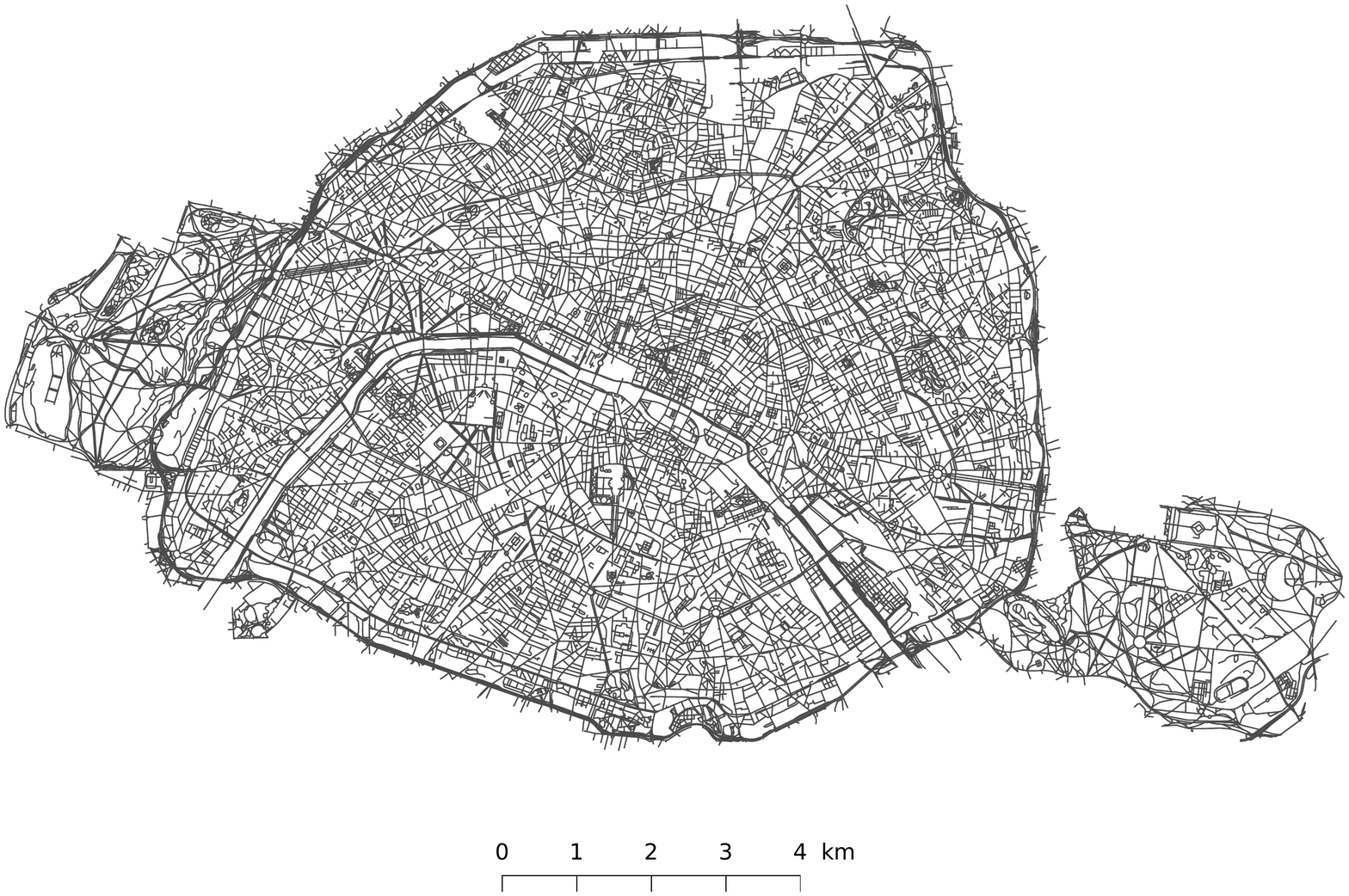}
        \caption{Commune de Paris en 2010. Données issues de la ©BDTOPO 2010 de l'IGN. 
        }
        \label{fig:9_paris_brut}
    \end{subfigure}
    ~
    \begin{subfigure}[t]{0.45\textwidth}
        \includegraphics[width=\textwidth]{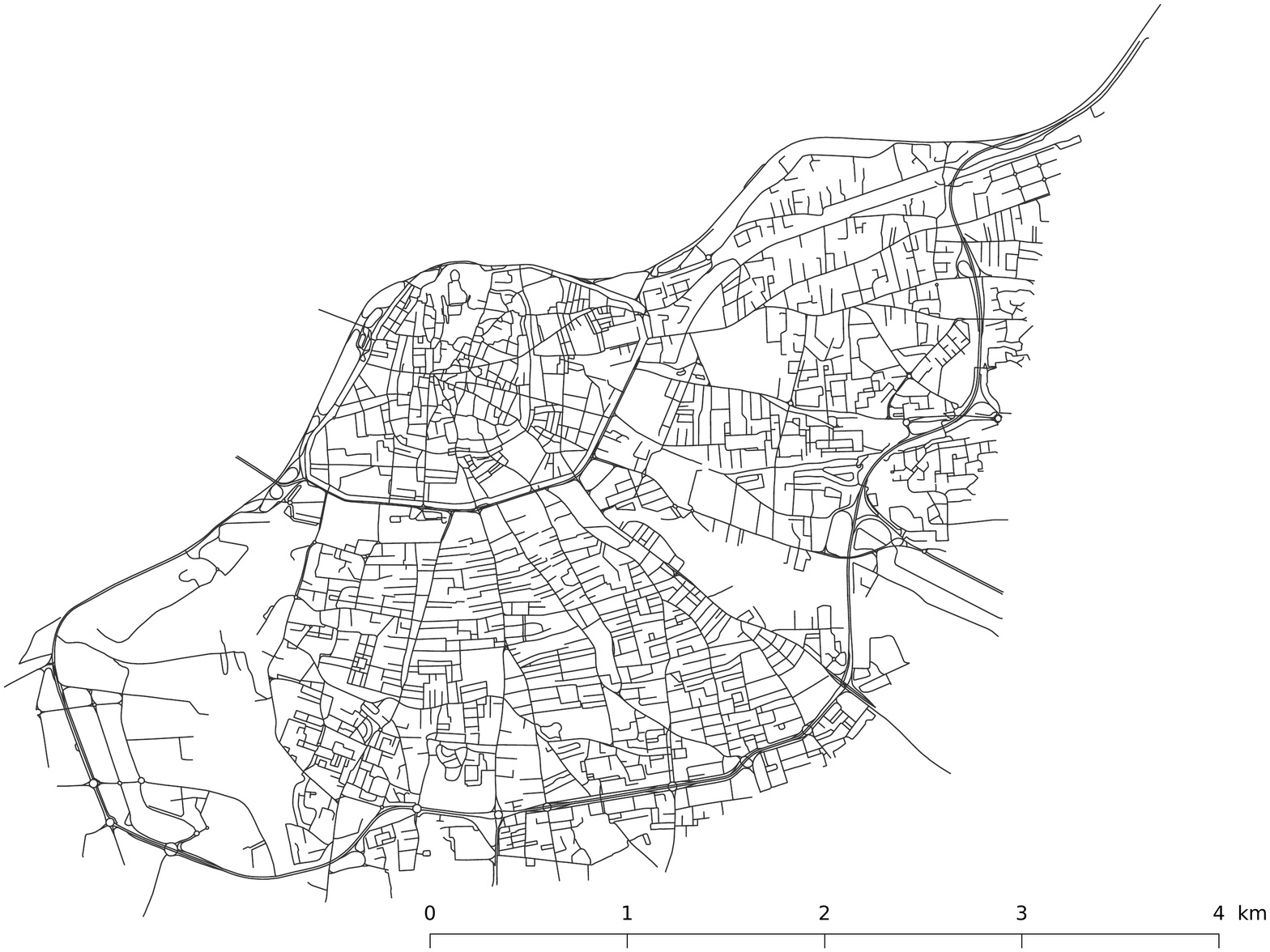}
        \caption{Majeure partie de la commune d'Avignon en 2014. Données issues de la ©BDTOPO 2014 de l'IGN. 
        }
        \label{fig:9_avignon_brut}     
    \end{subfigure}
    \caption{Réseaux viaires numérisés.} 
\end{figure}

Pour étudier l'impact des différentes méthodes de construction, nous travaillerons sur deux graphes extraits de réseaux routiers. Les graphes spatiaux utilisés regroupent indifféremment rues, chemins et passages. Nous avons donc tout le filaire qui peut être emprunté par un piéton ou une voiture. Nous choisissons d'étudier d'une part le réseau des rues de Paris en 2010, découpé selon les limites communales (figure \ref{fig:9_paris_brut}). D'autre part, nous travaillons sur le réseau des rues d'Avignon en 2014, découpé selon une cohérence globale (intra et extra murros) raisonnée (figure \ref{fig:9_avignon_brut}). Ces deux réseaux ont été choisis pour leurs différences et complémentarités (tableau \ref{tab:pres_ville0}). Paris, capitale, possède un réseau urbain dense, d'une envergure d'environ 12km. Avignon, ville provinciale, possède un graphe viaire aux structures mixtes d'une envergure d'environ 4km. Les deux réseaux sont extraits de villes non planifiées, dont la structure a crû au fil des siècles. Cependant Paris a subi, entre autres, une intervention globale au XIXème siècle avec les travaux dirigés par le préfet Haussmann quand Avignon n'a été modifiée que par des projets urbains locaux. Ces aspects qualitatifs et leurs impacts quantitatifs seront développés dans les parties suivantes.

\begin{table}[h]

\begin{tabular}{| c | r | r | r |}
\hline
& $L_{tot}$ (en mètres) & $N_{sommets}$ & $N_{arcs}$ \\
\hline
Avignon & 304 283 & 3 745 & 5 048 \\
\hline
Paris & 2 421 570 & 21 685 & 32 173 \\
\hline
\end{tabular}
\caption{Tableau descriptif des caractéristiques et classifications utilisées pour les deux villes étudiées}
\label{tab:pres_ville0}
\end{table}

\begin{figure}[H]

    \centering
    \begin{subfigure}[t]{0.48\textwidth}
        \includegraphics[width=\textwidth]{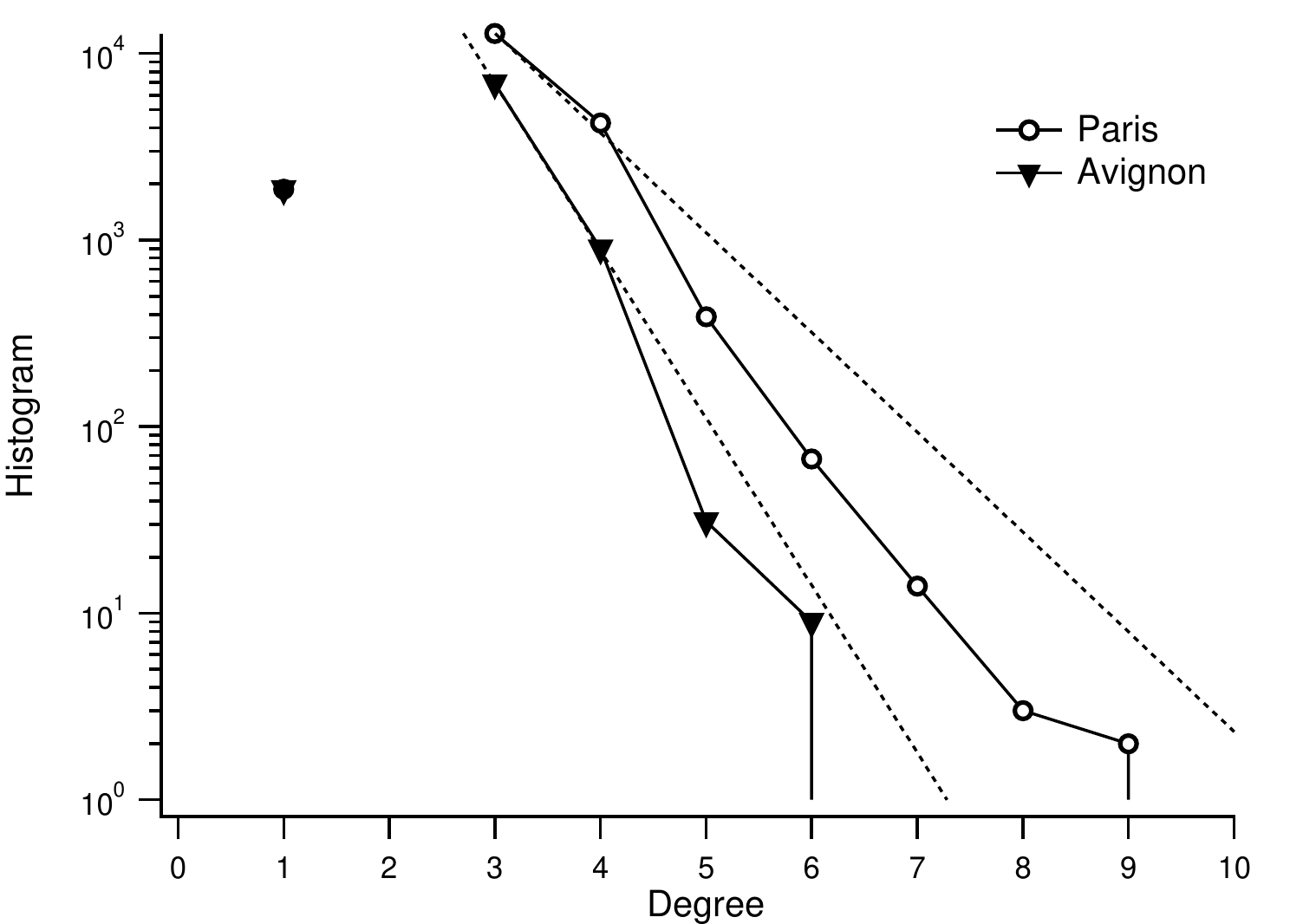}
        \caption{ Histogramme du degré des nœuds.}
        \label{fig:HistDegree}
    \end{subfigure}
    ~
    \begin{subfigure}[t]{0.48\textwidth}
        \centering
        \includegraphics[width=\textwidth]{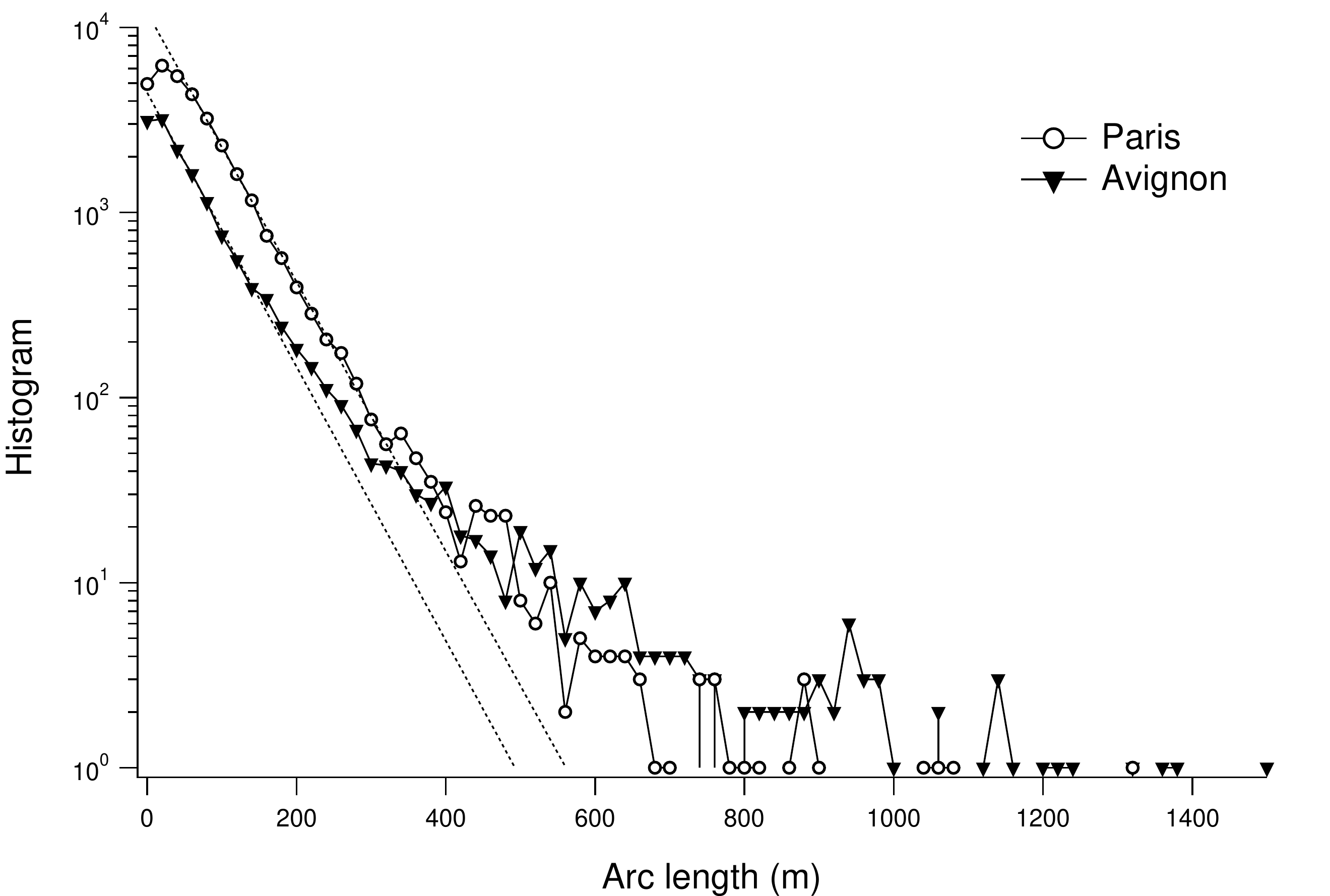}
        \caption{ Histogramme de la longueur des arcs.}
        \label{fig:HistLength}
    \end{subfigure}
    \caption{Étude des caractéristiques des objets élémentaires des graphes viaires de Paris et d'Avignon. \\ source : \citep{lagesse2015spatial}} 
\end{figure}

Une étude rapide des propriétés structurelles fondamentales de ces deux graphes révèle des similitudes aussi bien dans l'étude du degré des sommets, intersections de ces graphes spatiaux (figure \ref{fig:HistDegree}) que dans celle de la longueur de leurs arcs (figure \ref{fig:HistLength}).
L’histogramme représentant le degré des nœuds des deux échantillons montre qu'ils ont tous deux approximativement le même nombre d'impasses (nœuds de degré 1). Le degré le plus répandu dans une ville comme dans l'autre est 3, puis il décroît un peu plus rapidement qu'une fonction exponentielle avec un nombre caractéristique de 0,8 pour Paris et 0,5 pour Avignon. Cela signifie que les intersections de plus de trois tronçons sont plus rares pour Avignon que pour Paris. La distribution de la longueur des arcs peut être approximée par une exponentielle dans les deux échantillons avec une longueur caractéristique proche, autour de 60 mètres. Une comparaison plus poussée des propriétés structurelles de différents types de graphes spatiaux sera faite dans la deuxième partie de ce travail.

\begin{figure}[h]
    \centering
    \begin{subfigure}[t]{0.48\textwidth}
        \includegraphics[width=\textwidth]{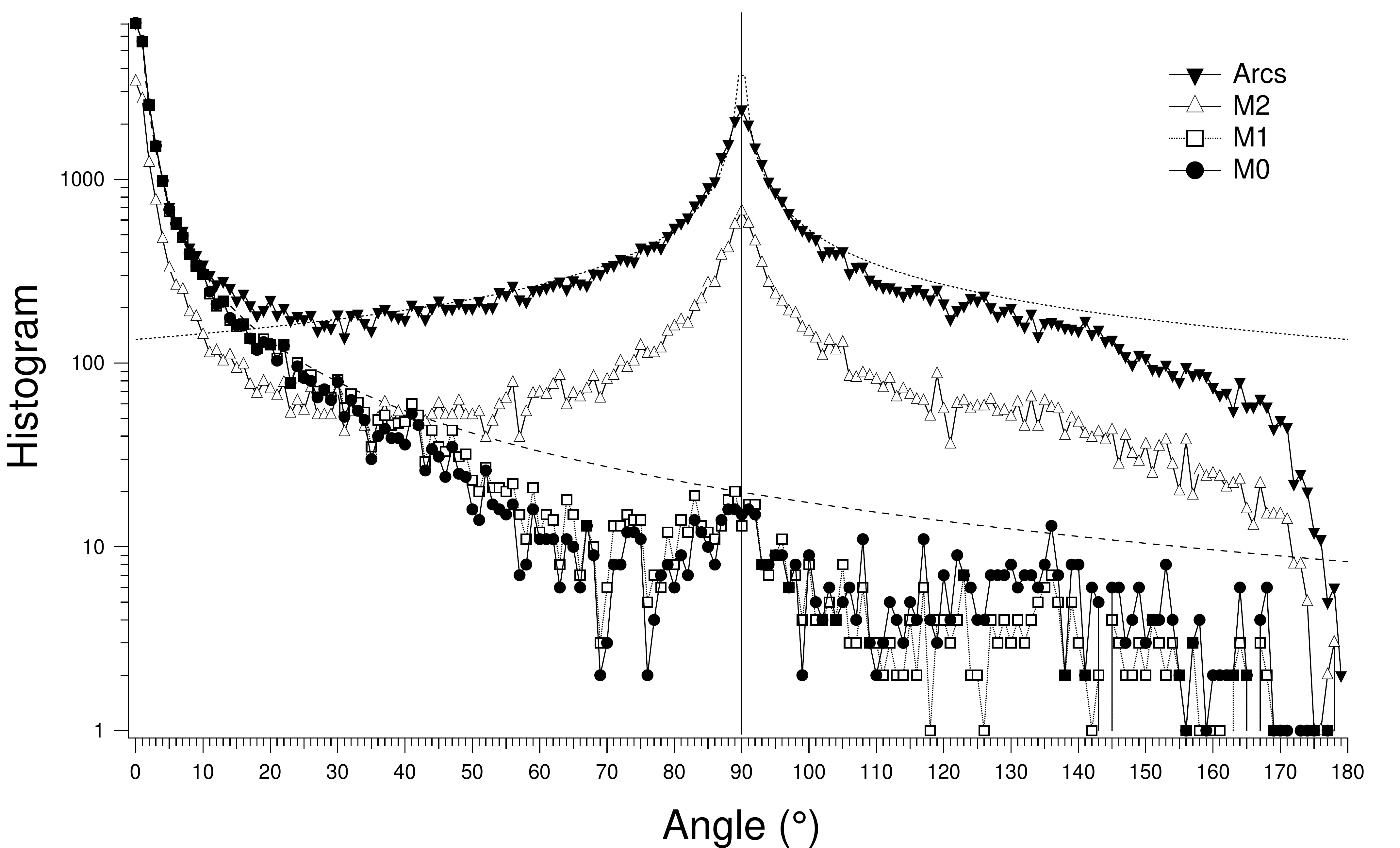}
        \caption{Histogramme des angles de déviation aux intersections du graphe. Les angles considérés sont ceux entre arcs bruts ; et ceux retenus pour les différentes méthodes de construction des voies ($\theta_{seuil} = 180\degres$).}
        \label{fig:10_angles}
    \end{subfigure}
    ~
    \begin{subfigure}[t]{0.48\textwidth}
        \includegraphics[width=\textwidth]{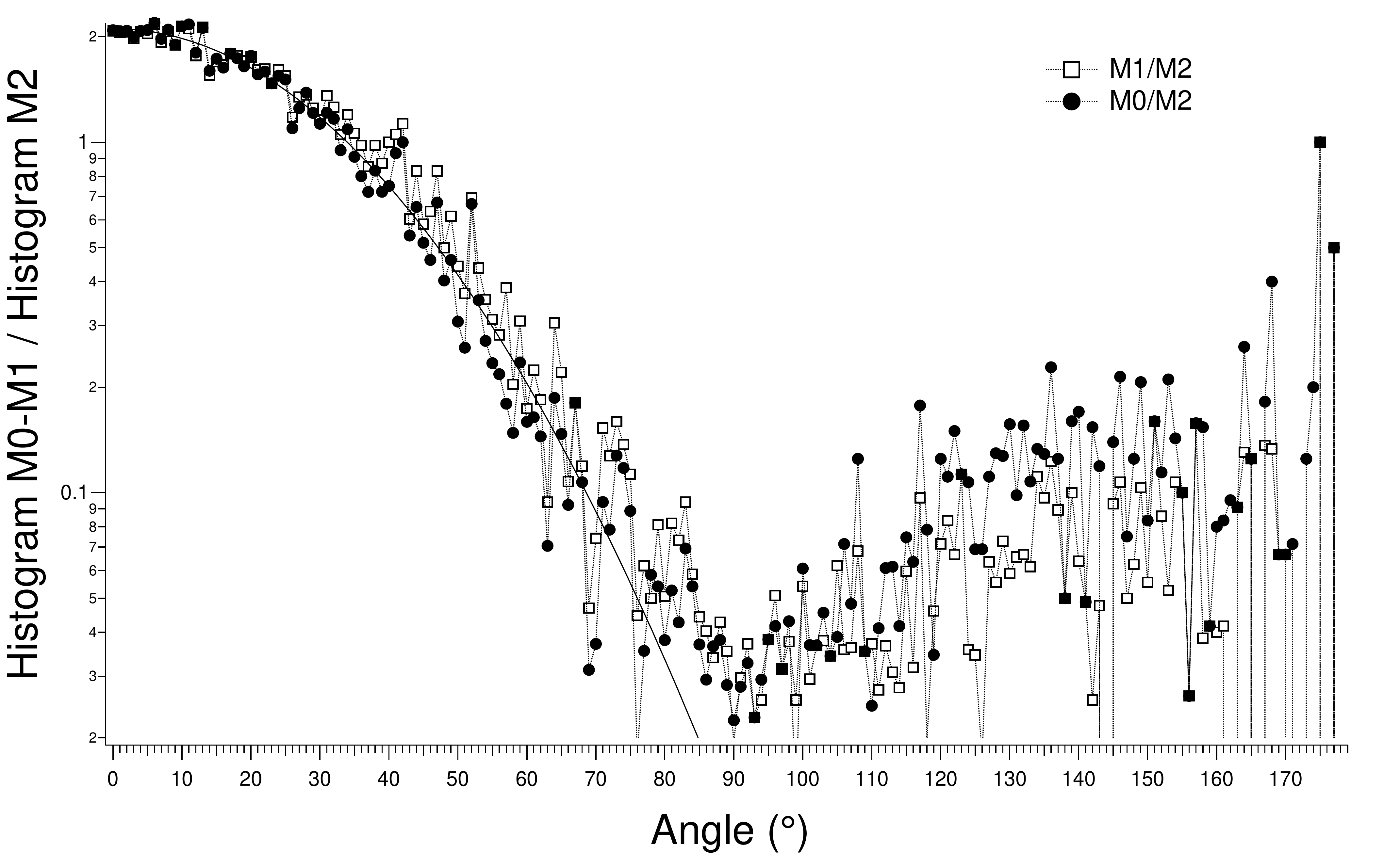}
        \caption{Histogramme des angles retenus à l'intérieur des voies construites avec les méthodes $M0$ et $M1$ normalisés par les valeurs obtenues avec la méthode blanche $M2$.}
        \label{fig:11_angles_norm}
    \end{subfigure}
    
    \caption{Étude des angles de déviation sur le graphe viaire de Paris. \\ source : \citep{lagesse2015spatial}}
\end{figure}

Nous commençons par observer les angles de déviation faits entre l'ensemble des arcs du graphe ainsi que dans les voies successivement créées par chacune des trois méthodes sur le réseau (figure \ref{fig:10_angles}). Les courbes présentées ici correspondent au réseau de Paris. En effet, ce graphe comprend plus d'éléments et donne donc une analyse statistique plus fine. Si l'on trace l'histogramme des angles entre les arcs bruts (non encore regroupés pour former les voies), on remarque deux pics significatifs (cf courbe représentée en triangles pleins). Le premier pic correspond aux angles de déviation de 0\degres , le second à ceux de 90\degres . Ces deux valeurs sont en effet prégnantes dans les réseaux viaires, les tronçons de rues se connectant entre eux la plupart du temps soit dans leur alignement soit perpendiculairement. La distribution asymétrique autour de ces valeurs nous montre que les continuités rectilignes entre deux arcs sont plus présentes dans le réseau urbain que celles perpendiculaires. Nous pouvons également noter un faible nombre de déviations de plus de 170\degres , ce qui est également compréhensible car les tronçons de route se rejoignent très rarement avec un angle très faible. La courbe peut être approximée par une loi de puissance d'exposant -1,25 autour de 0\degres  et une loi de puissance d'exposant -0,625 autour de 90\degres , représentées ici en traits pointillés.

Nous traçons également les histogrammes des angles de déviation à l'intérieur des voies construites successivement par les méthodes $M0$ (courbe avec ronds pleins), $M1$ (carrés vides) et $M2$ (triangles vides), sans angle seuil imposé ($\theta_{seuil} = 180\degres$). La méthode blanche ($M2$), regroupant les arcs aléatoirement à chaque sommet, construit des voies dont les angles de déviation internes suivent l'impact reconnu entre arcs bruts (alignement et perpendicularité). Cependant, la sélection de couples à chaque intersection enlève des angles potentiels pour les arcs restants ce qui rend cette distribution non exactement proportionnelle à la première. Les deux méthodes de construction $M0$ et $M1$ regroupent quant à elles des arcs dont les déviations sont majoritairement minimales.

Si l'on normalise les valeurs obtenues pour $M0$ et $M1$ par celles obtenues pour $M2$ nous obtenons, pour les angles inférieurs à 90\degres , une distribution gaussienne de maximum 0\degres  et d'écart type d'environ 40\degres  (figure \ref{fig:11_angles_norm}). Sur le réseau choisi, nous considérons donc que les arcs sont alignés jusqu'à 60\degres  de déviation (une fois et demi l'écart-type). L'angle de 90\degres  marque le passage a un mode bruité.

\begin{figure}[h]
    \centering
    \begin{subfigure}[t]{0.48\textwidth}
        \includegraphics[width=\textwidth]{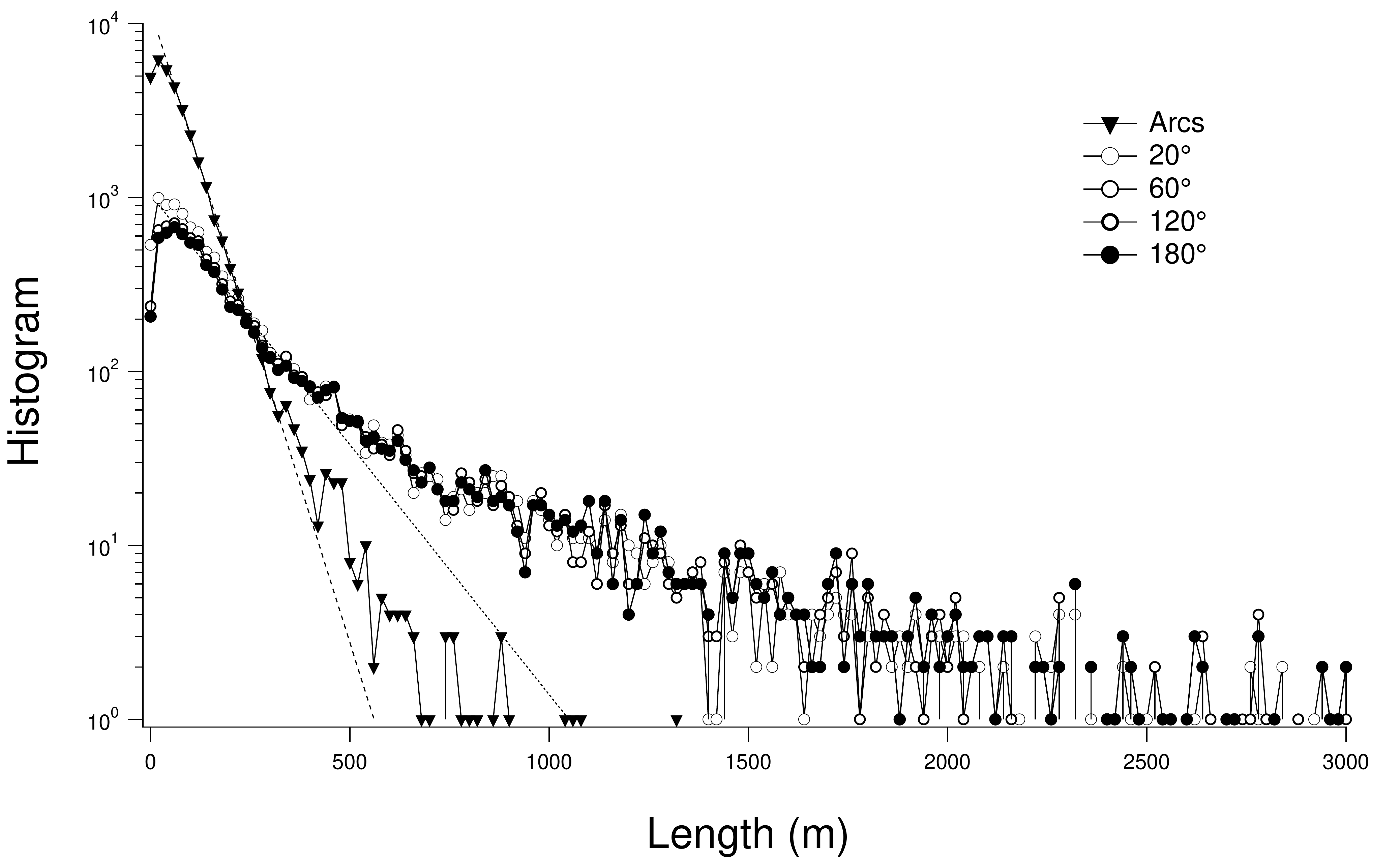}
        \caption{ Histogramme des longueurs.}
        \label{fig:12_lengthM0}
    \end{subfigure}
    ~
    \begin{subfigure}[t]{0.48\textwidth}
        \includegraphics[width=\textwidth]{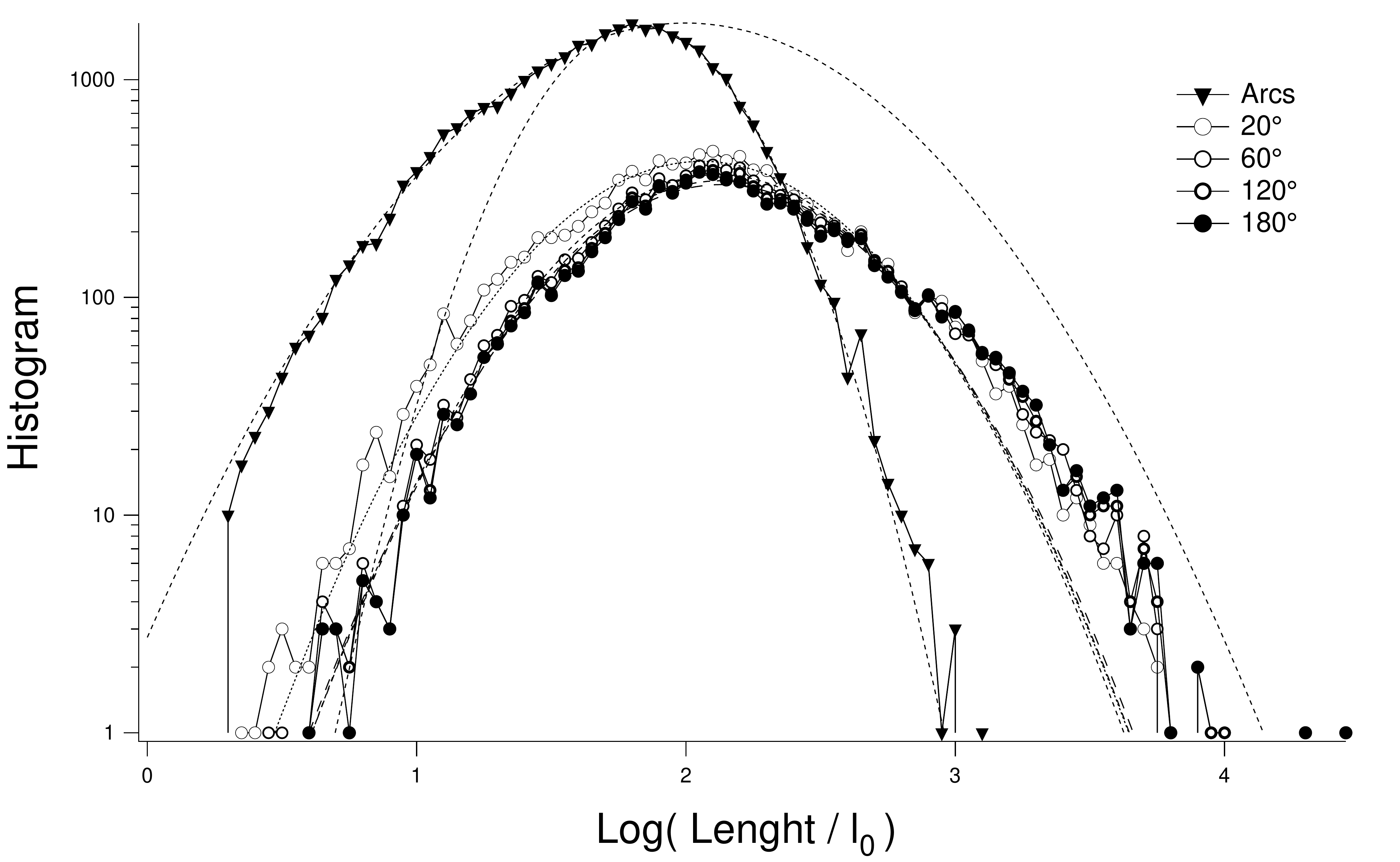}
        \caption{ Histogramme du logarithme des longueurs.}
        \label{fig:13_loglengthM0}
    \end{subfigure}
    \caption{Étude des longueurs des arcs et de celle des voies crées avec $M0$ selon quatre angles seuils différents : 20\degres , 60\degres , 120\degres et 180\degres . Graphe viaire de Paris. \\ source : \citep{lagesse2015spatial}}
\end{figure}

Nous poursuivons en comparant les différentes longueurs de voies données par les trois méthodes de construction, pour différents angles seuil. Nous fixons $\theta_{seuil}$ à quatre différentes valeurs : 20\degres , 60\degres , 120\degres  et 180\degres . Pour chacun de ces seuils nous comparons l'histogramme des longueurs des voies créées et nous ajoutons à la comparaison les longueurs d'arcs brut. Nous observons ainsi que l'histogramme des longueurs des voies a une distribution beaucoup plus large que celui de longueurs d'arcs (figure \ref{fig:12_lengthM0}). Les voies, comme nous pouvions l'attendre, sont de longueur nettement plus importante : la longueur moyenne passe ici de 60m à 150m. C'est là le principe même de l’agrégation : obtenir par association d'arcs un hypergraphe sur notre réseau.

\begin{figure}[h]
    \centering
    \begin{subfigure}[t]{0.48\textwidth}
        \includegraphics[width=\textwidth]{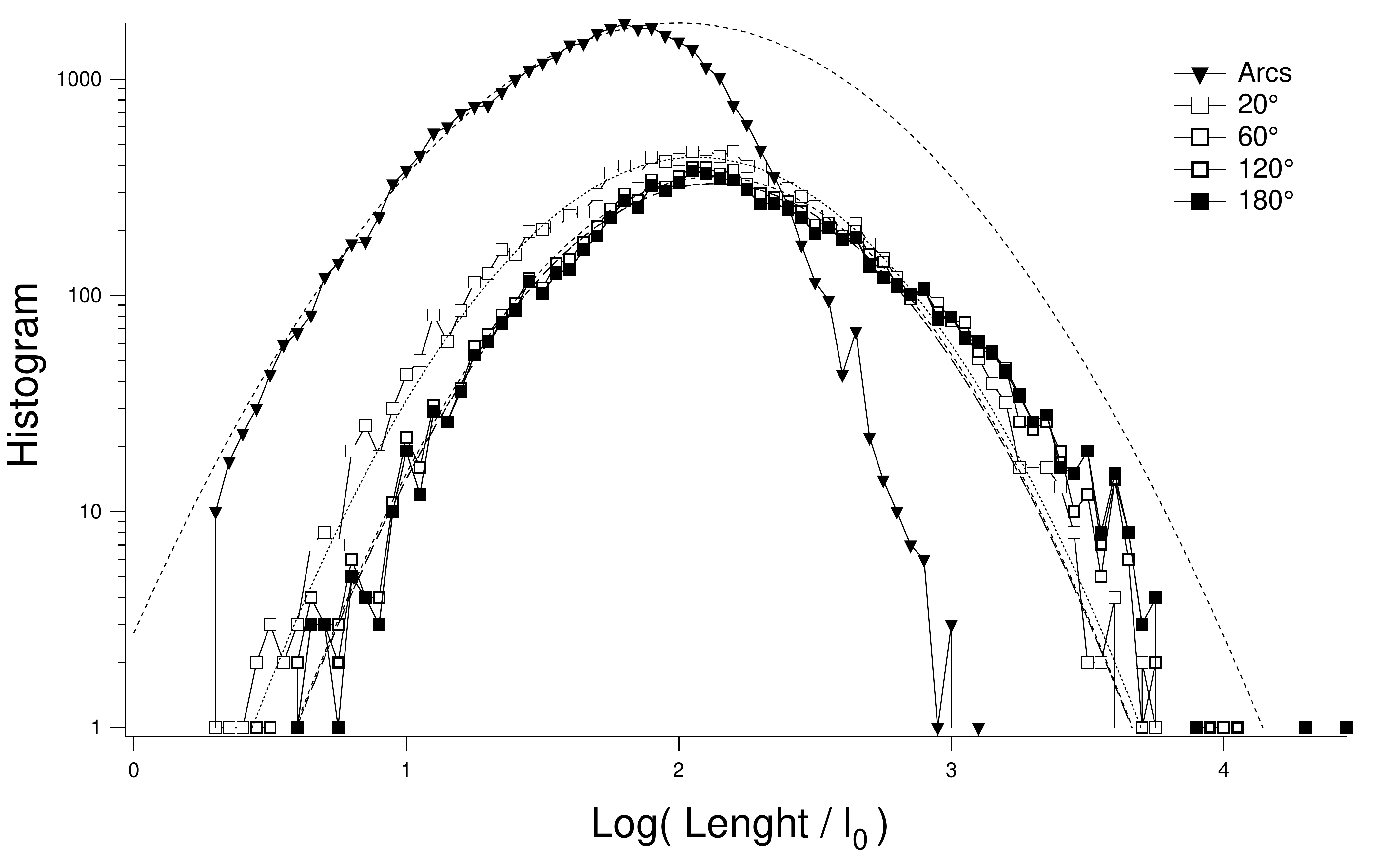}
        \caption{Histogramme du logarithme des longueurs. Voies créées avec la méthode $M1$.}
        \label{fig:14_loglengthM1}
    \end{subfigure}
    ~
    \begin{subfigure}[t]{0.48\textwidth}
        \includegraphics[width=\textwidth]{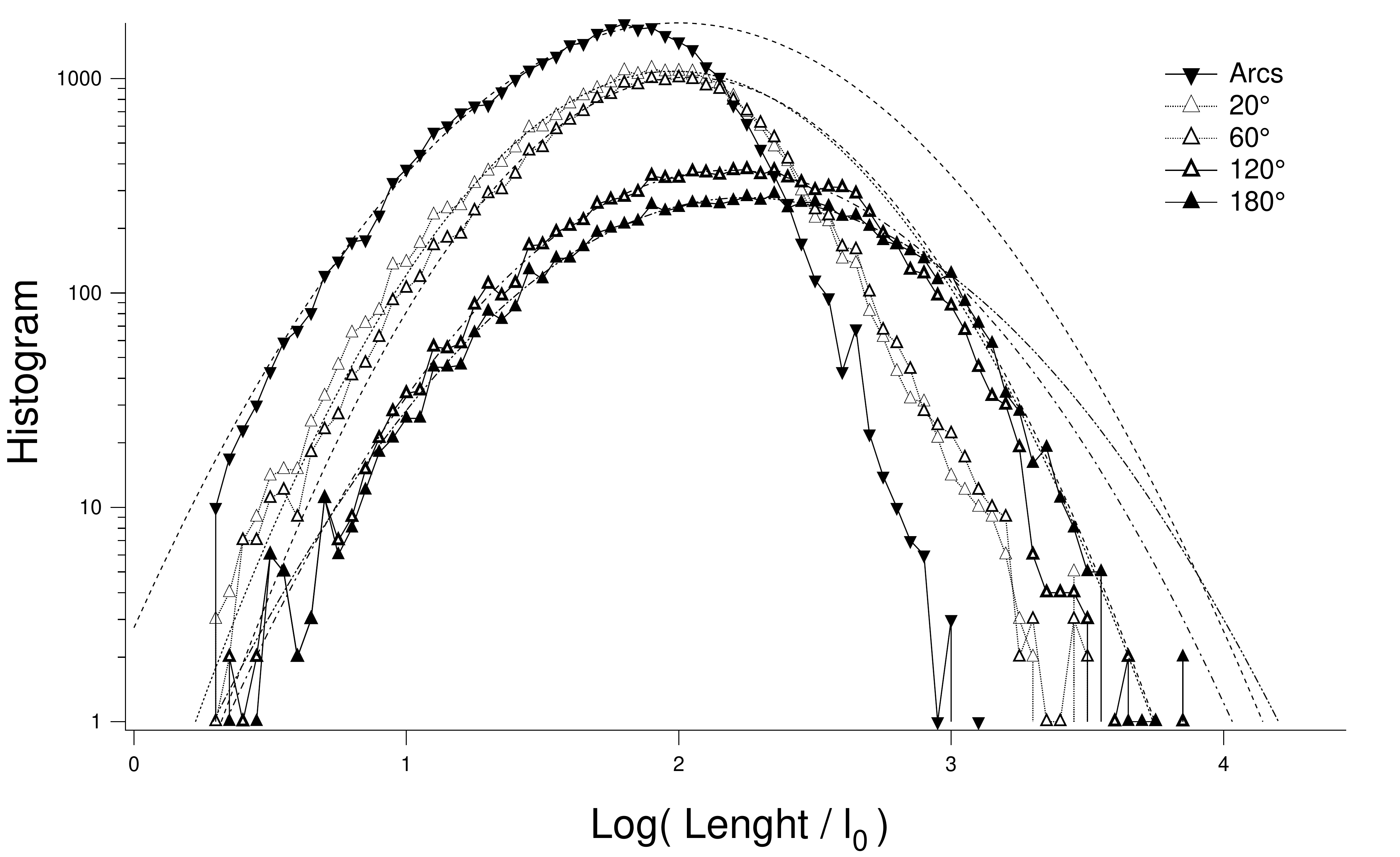}
        \caption{Histogramme du logarithme des longueurs. Voies créées avec la méthode $M2$.}
        \label{fig:15_loglengthM2}
    \end{subfigure}
    \caption{Étude des longueurs des arcs et de celle des voies crées avec $M1$ (gauche) et $M2$ (droite) selon quatre angles seuils différents : 20\degres , 60\degres , 120\degres et 180\degres . Graphe viaire de Paris. \\ source : \citep{lagesse2015spatial}}
\end{figure}

Les histogrammes des logarithmes des longueurs permettent d'approximer la courbe représentant le logarithme des longueurs des arcs par deux gaussiennes d'étendues très différentes. Ceci est dû à un nombre plus important d'arcs courts. Le logarithme de la longueur des voies peut être, lui, approximé par une gaussienne de laquelle se détachent les voies les plus longues de manière très marquée pour $M0$ (figure \ref{fig:13_loglengthM0}). La courbe correspondant aux valeurs calculées pour $M1$ fait apparaître une convergence plus lente et moins de longues voies créées (figure \ref{fig:14_loglengthM1}).

Enfin les valeurs calculées par $M2$ ne convergent que pour $\theta_{seuil} = 180\degres$. Contrairement aux deux méthodes précédentes, il y a moins de voies longues que ce que l'analyse de la gaussienne aurait fait prédire (figure \ref{fig:15_loglengthM2}). Cette distribution de longueurs de voies correspond à un processus aléatoire de divisions successives. Une voie en coupe une autre en deux éléments dont les longueurs sont une fraction de la première. Si cette fraction est purement aléatoire, alors les longueurs finales sont un produit de nombres aléatoires. Leur logarithme est donc la somme du logarithme de nombres aléatoires, ce qui donne une gaussienne. Pour ce type de processus des îlots quadrangulaires sont obtenus après un certain nombre de divisions, comme observé dans les craquelures d'argile \citep{bohn2005four}.

\begin{figure}[h]
    \centering
    \begin{subfigure}[t]{.48\linewidth}
        \includegraphics[width=\textwidth]{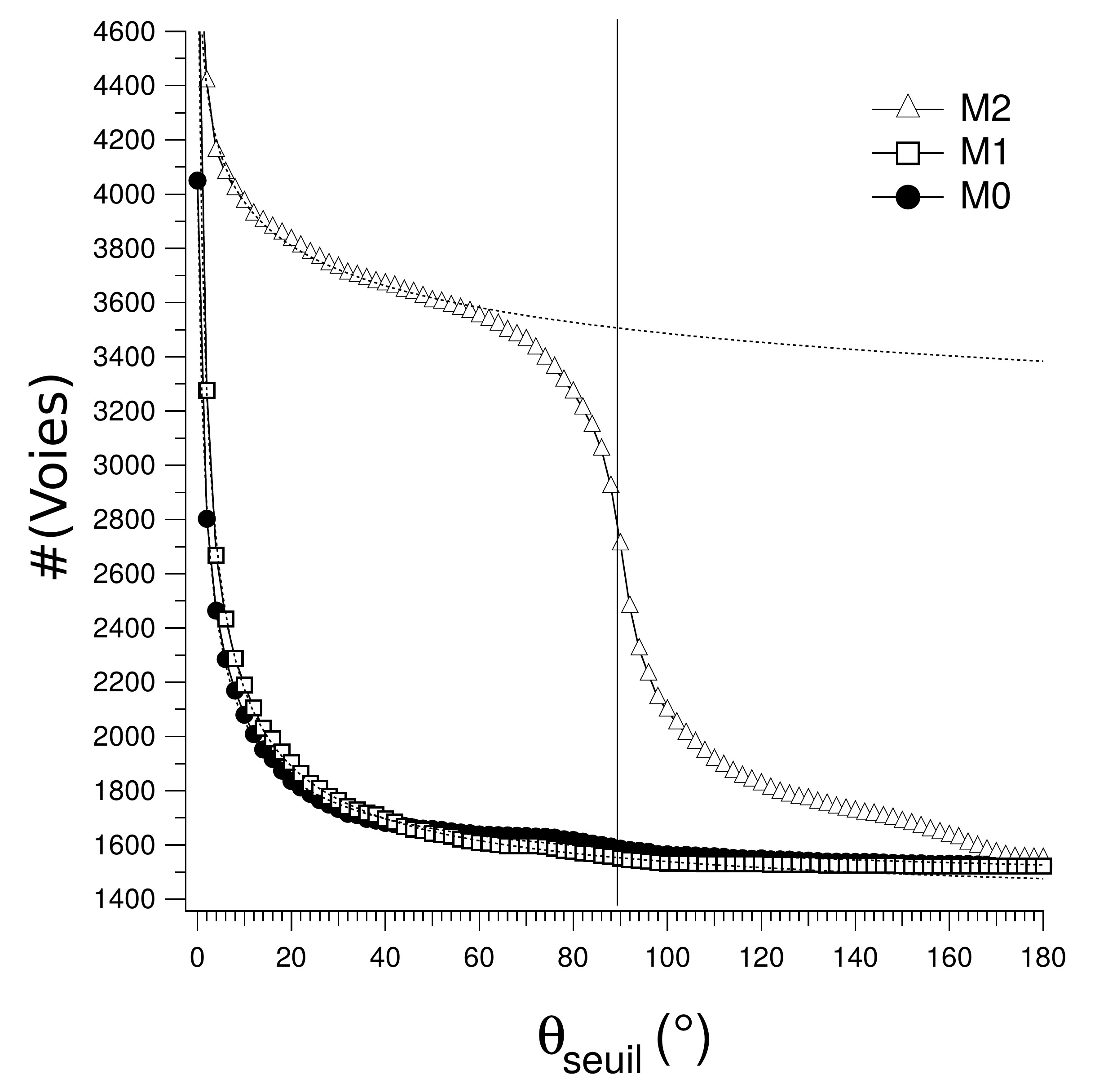}
        \caption{ Nombre de voies créées tracé en fonction de l'angle seuil choisi $\theta_{seuil}$.}
        \label{fig:16_nbways}
    \end{subfigure}
    ~
    \begin{subfigure}[t]{.48\linewidth}
        \includegraphics[width=\textwidth]{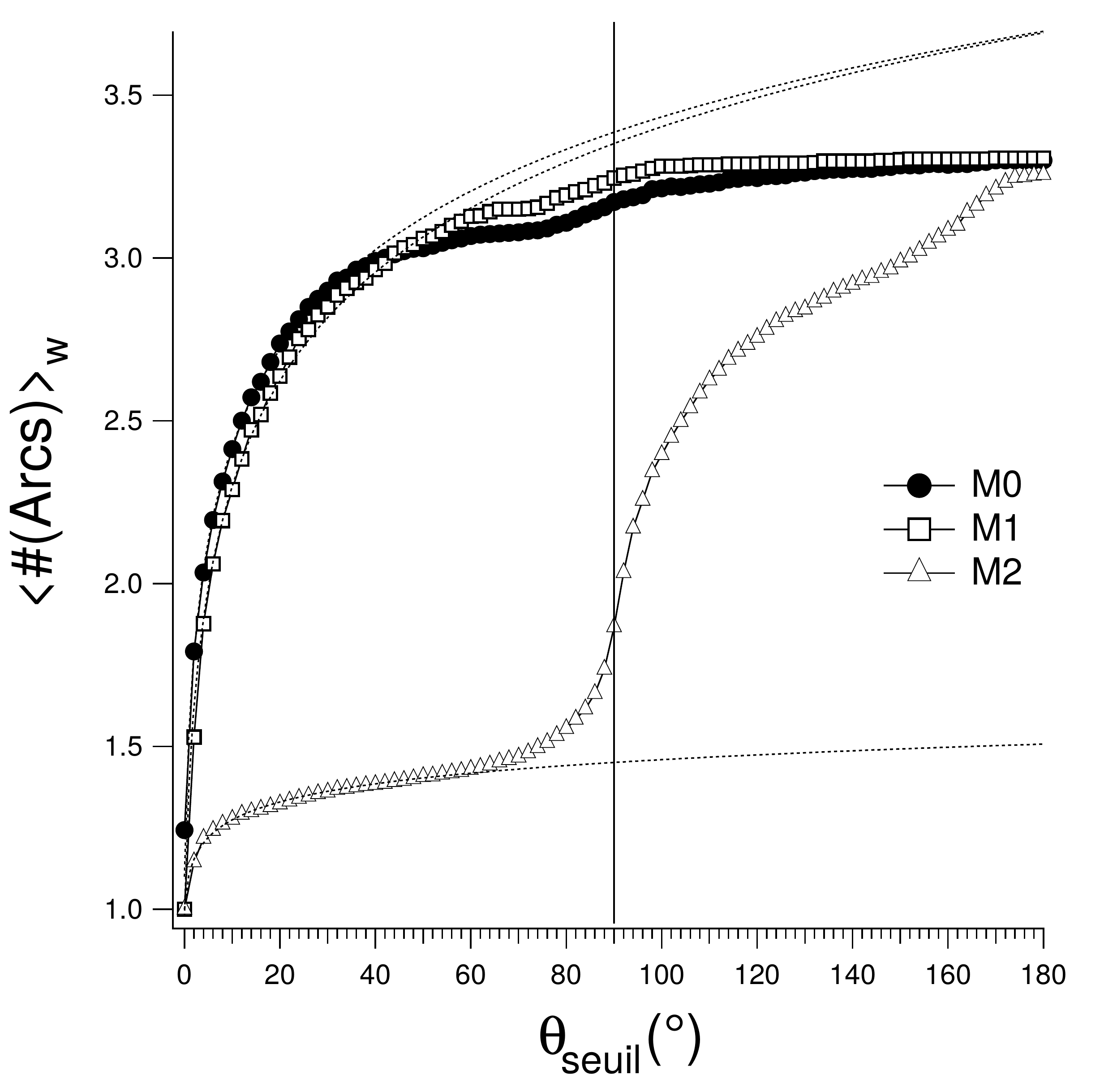}
        \caption{ Nombre moyen d'arcs par voies créées tracé en fonction de l'angle seuil choisi $\theta_{seuil}$.}
        \label{fig:17_nbarcs}
    \end{subfigure}
    
    \caption{Étude des changements de caractéristiques des voies construites selon l'angle seuil. Graphe viaire d'Avignon. \\ source : \citep{lagesse2015spatial}}
    \label{fig:carct_seuil}
\end{figure}

Nous étudions la variation du nombre et des caractéristiques des voies construites en fonction de l'angle seuil fixé. Nous comparons ainsi le nombre de voies créées, le nombre moyen d'arcs par voies et l'angle de déviation moyen de ceux sélectionnés à l'intérieur de chaque voie. Nous observons la vitesse de transition des valeurs de chaque caractéristique selon les angles seuils fixés pour chacune des trois méthodes (figure \ref{fig:carct_seuil}). Elle est beaucoup plus rapide pour les méthodes $M0$ et $M1$ que pour la méthode aléatoire $M2$. Pour illustrer cela nous nous appuierons ici sur l'exemple d'Avignon. En effet, le graphe étant plus restreint, il nous est possible de faire une analyse dont la granularité est réduite au degré, pour l'angle seuil choisi.

Sur la figure \ref{fig:16_nbways} nous observons qu'un angle seuil de 0\degres  ou proche correspond à un très grand nombre de voies : les arcs se retrouvent groupés en très petite quantité car la déviation autorisée est faible. D'un autre côté, l'angle seuil maximal ($\theta_{seuil} = 180\degres$) aboutit à un faible nombre de voies : un nombre important d'arcs sont appariés pour former un nombre minimal d'éléments. Le nombre de voies créées par les méthodes $M0$ et $M1$ converge rapidement entre 1500 et 1600. Le nombre de celles créées avec $M2$ suit deux phases décroissantes, dont la première peut être approximée par une loi de puissance. Elles sont entrecoupées d'une phase de transition brutale liée au grand nombre de connexions entre arcs perpendiculaires. L'impact de l'angle seuil fixé à 90\degres  se ressent également, à moindre échelle, sur les deux autres courbes. On observe une légère baisse du nombre de voies pour les angles seuils fixés entre 70\degres  et 110\degres .

Si l'on observe le nombre moyen d'arcs par voies selon la méthode de construction choisie (figure \ref{fig:17_nbarcs}), nous remarquons le même type de comportement. $M0$ et $M1$ aboutissent à une convergence rapide autour de 3.2 arcs, alors que $M2$ donne une courbe à la variation croissante impactée par l'alignement ($\theta_{seuil} = 0\degres$) ou la perpendicularité ($\theta_{seuil} = 90\degres$). De même entre 70\degres  et 110\degres  les courbes correspondant à $M0$ et $M1$ ont une variation à la régularité légèrement perturbées. Entre 0 et 40 \degres , les courbes peuvent être approximées par une fonction logarithmique, dont la croissance à l'origine est proche de la verticale. La transition est donc très sensible : le nombre d'arcs moyen par voie croît rapidement selon l'angle seuil choisi.

\begin{figure}[h]
    \centering
    \begin{subfigure}[c]{.48\linewidth}
        \includegraphics[width=\textwidth]{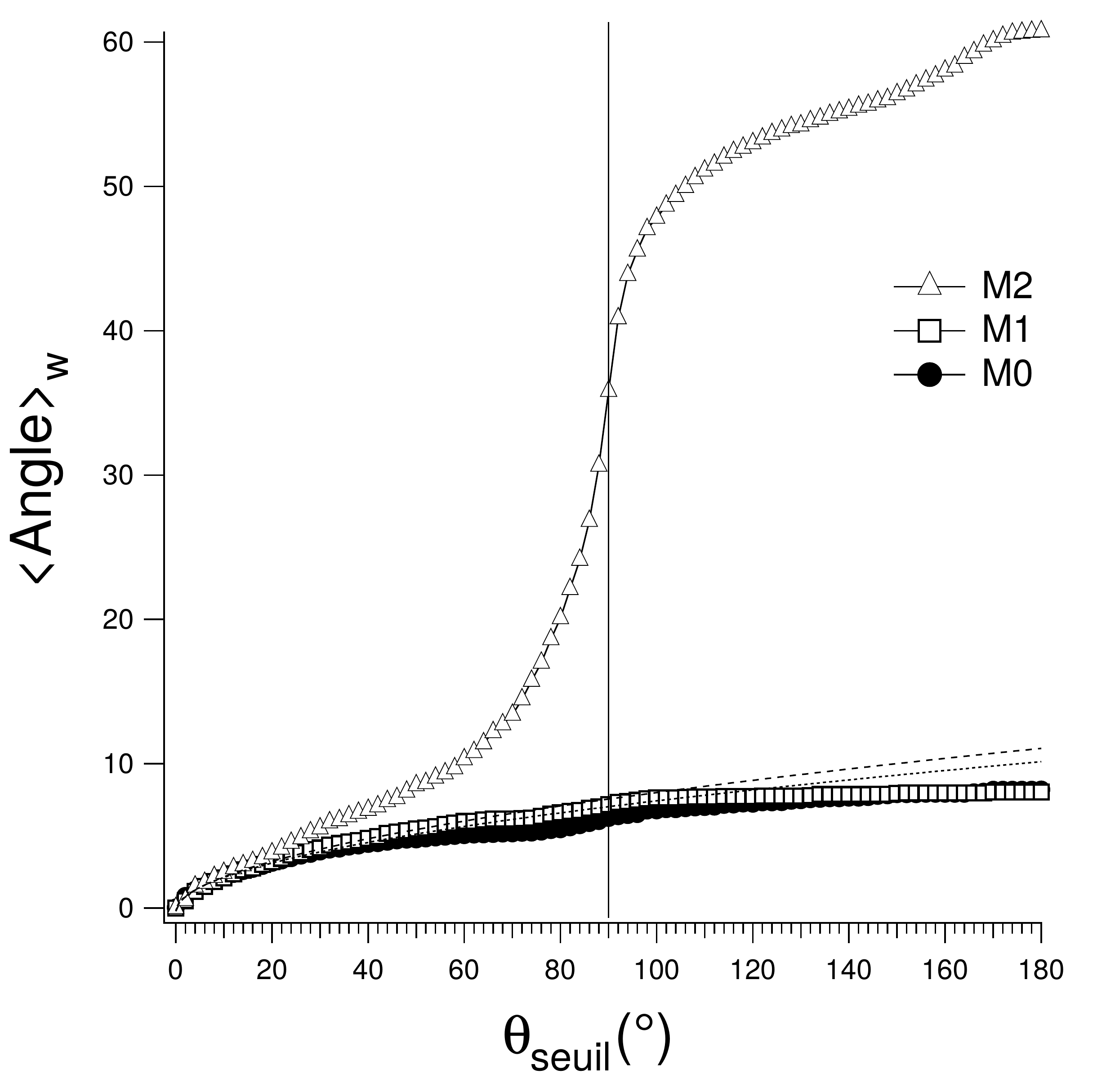}
    \end{subfigure}
    ~
    \begin{subfigure}[c]{.48\linewidth}
        \includegraphics[width=\textwidth]{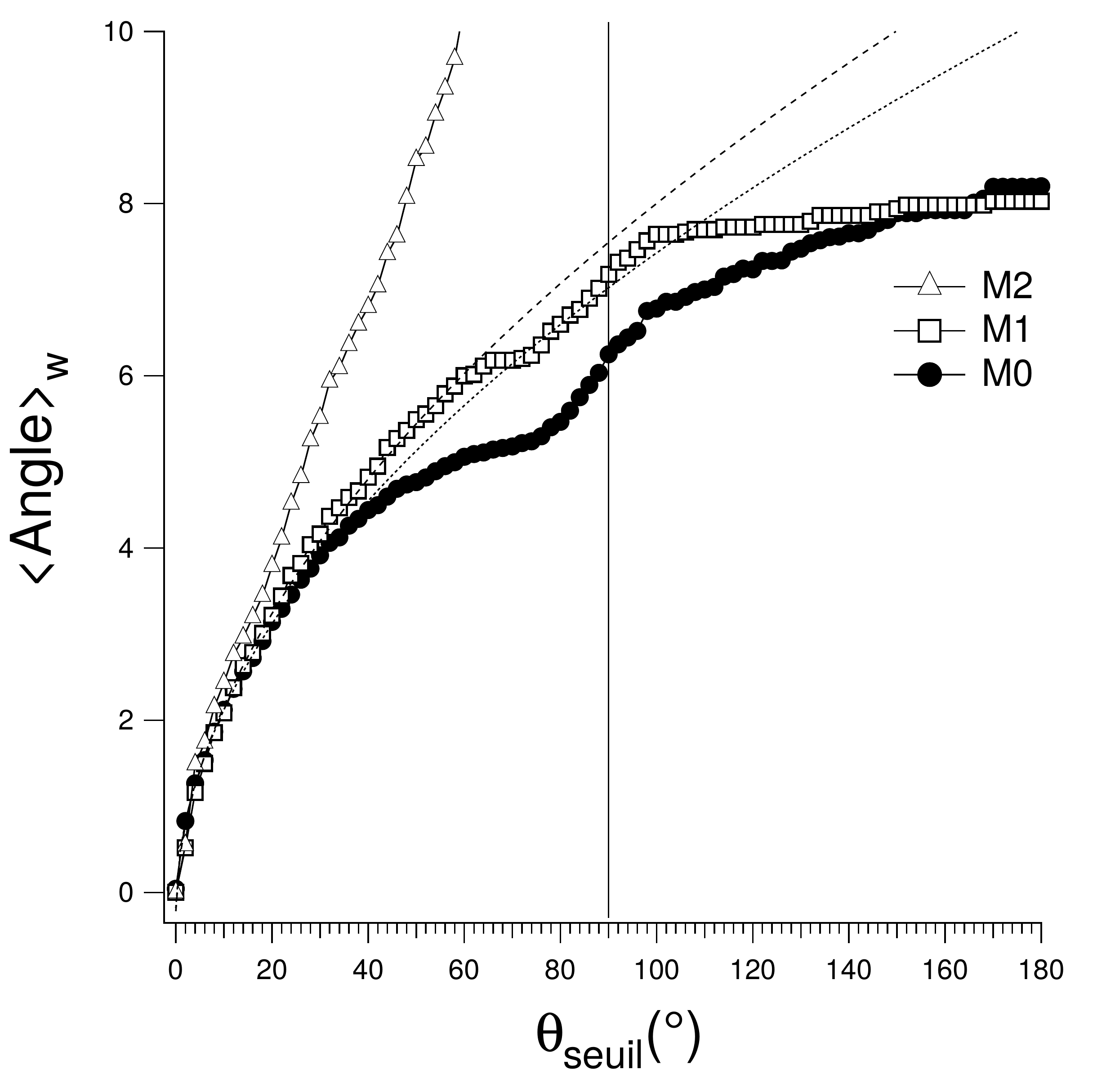}
    \end{subfigure}
    
    \caption{ Angle moyen de déviation entre arcs utilisés dans les voies créées, tracé en fonction de l'angle seuil choisi $\theta_{seuil}$. À gauche : courbe complète. À droite : Zoom entre 0\degres et 10\degres. Graphe viaire d'Avignon. \\ source : \citep{lagesse2015spatial}}
    \label{fig:18_anglemoy}
\end{figure}

Enfin, nous traçons la courbe de l'angle moyen de déviation entre arcs choisis pour la construction d'une même voie en respectant l'angle seuil fixé (figure \ref{fig:18_anglemoy}). La courbe de $M2$ croît rapidement et se stabilise autour de 60\degres , angle moyen de déviation sur une intersection de degré 3, qui est le degré moyen des intersections dans les graphes étudiés. Les courbes de $M0$ et $M1$ convergent dans un premier temps vers (respectivement) 5\degres  et 6\degres . $M0$ converge plus rapidement vers une valeur plus petite. Puis, les deux courbes subissent la perturbation liée au grand nombre de connexions entre arcs à 90\degres , pour enfin converger de nouveau autour de 8\degres   (figure \ref{fig:18_anglemoy}). Les deux courbes suivent au début de leur progression une croissance assimilable à celle de la fonction racine carrée (représentée en ligne pointillée).

\FloatBarrier

\section{Choix d'une méthode de construction}

De la comparaison des différentes méthodes et des angles seuils associés nous retenons trois points.\footnote{Cette analyse a été publiée dans \citep{lagesse2015spatial}.}

\begin{itemize}

    \item La construction de l'objet voie est pertinente puisque les propriétés statistiques de l'objet créé par $M0$ ou $M1$ suivent un comportement rationnel bien différent de celui de l'objet créé avec la méthode blanche $M2$. La voie, privilégiant une forme de continuité locale, crée donc une structure cohérente à travers les échelles. L'ensemble constitue un hypergraphe apposé sur le réseau.

    \item Les réseaux viaires admettent une prédominance d'angles de déviation de 0\degres  et 90\degres . Cette surabondance impacte la convergence des courbes de l'étude statistique autour de ces valeurs d'angles seuils (figures \ref{fig:16_nbways}, \ref{fig:17_nbarcs}, \ref{fig:18_anglemoy}). Nous voulons privilégier la création d'un nombre minimal de voies (et donc ainsi associer un nombre maximal d'arcs) sans pour autant avoir un résultat impacté par la perturbation créée par les angles seuils proches de 90\degres . Celle-ci commence à environ 70\degres . De plus, lorsque nous normalisons les angles présents dans les voies construites avec les méthodes $M0$ et $M1$ par ceux dans les voies construites avec $M2$, nous obtenons une courbe gaussienne dans laquelle l'angle de 60\degres  se détache comme étant celui au delà duquel les arcs ne sont plus alignés (figure \ref{fig:11_angles_norm}). Ces résultats nous conduisent à choisir un angle seuil de 60\degres . Celui-ci correspond également à l'angle moyen de déviation entre arcs pour un sommet de degré 3. Le degré moyen des sommets sur les graphes étudiés étant de 3, cela vient renforcer notre choix de seuil.

    \item Les deux méthodes observées, $M0$ et $M1$, donnent des résultats proches mais une analyse fine montre des différences dans le nombre de voies les plus longues créées (figures \ref{fig:13_loglengthM0} et \ref{fig:14_loglengthM1}) et dans la convergence autour des angles moyens dans les voies créées (figure \ref{fig:18_anglemoy}). Nous sélectionnons la méthode qui converge le plus rapidement et nous donne le plus d'éléments de longueur importante : $M0$. En effet, on lit sur \ref{fig:13_loglengthM0} un nombre important de voies longues, au delà de ce que fait prédire la gaussienne et de manière plus marquée qu'avec les voies créées selon $M1$ (figure \ref{fig:14_loglengthM1}). De plus, sur la figure \ref{fig:18_anglemoy} nous démontrons que les voies créées avec $M0$ et un seuil $\theta_{seuil} < 70\degres$ ont un angle moyen de déviation interne plus faible (donc une plus grande linéarité à l'intérieur même des voies) et une convergence plus rapide.

\end{itemize}

En appliquant ces paramètres $(M0, 60\degres)$ aux deux réseaux présentés ici, nous obtenons sur Paris un total de 8132 voies à partir de 32173 arcs ; et sur Avignon 1552 voies à partir de 5048 arcs. Le ratio $\frac{N_{voies}}{N_{arcs}}$ est donc de 0.25 pour Paris et de 0,30 pour Avignon. Le nombre d'éléments du graphe a donc été réduit d'un quart sur le réseau de Paris et d'un peu moins (un tiers) sur le réseau d'Avignon dans la construction des  voies. Cela signifie que le réseau des voies de Paris admet plus de structures traversantes continues. Nous étudierons l'évolution de ce facteur en fonction de la taille et du type du réseau étudié dans la deuxième partie de ce document.

\begin{figure}[h]
    \centering
    \begin{subfigure}[c]{.6\linewidth}
        \includegraphics[width=\textwidth]{images/schemas/teh_voies.pdf}
        \caption{Graphe primal.}
    \end{subfigure}
    ~
    \begin{subfigure}[c]{.6\linewidth}
        \includegraphics[width=\textwidth]{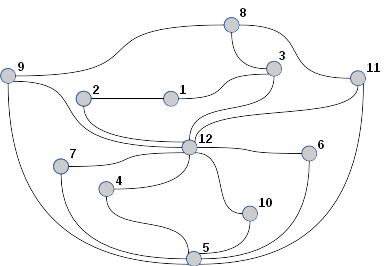}
        \caption{\textit{Line Graph}.}
    \end{subfigure}

    \caption{ Construction du \textit{line graph} des voies à partir du graphe spatial primal.}
    \label{fig:6_dual_voies}
\end{figure}

Nous construirons pour la suite de cette étude systématiquement les voies avec la méthode $M0$ et un angle seuil de 60\degres. Les voies ainsi créées représentent un hypergraphe du réseau spatial. Pour étudier certaines propriétés de cet hypergraphe, il est nécessaire de construire son \textit{line graph} : chaque voie sera représentée par un nœud et les connexions entre voies constitueront les arcs du \textit{line graph} (figure \ref{fig:6_dual_voies}). Le nom \textit{line graph} vient des travaux de Harary \citep{harary1960some}. Le concept avait déjà été utilisé précédemment par d'autres chercheurs \citep{whitney1932congruent, krausz1943demonstration}.

Cette méthode de représentation d'un graphe spatial a été utilisée sur des graphes spatialisés par B. Hillier puis S. Porta \citep{hillier1976space, porta2006network}, sous le nom de \textit{graphe dual}. Elle permet d'expliciter le calcul de certains indicateurs sur l'hypergraphe des voies, notamment ceux faisant intervenir les distances topologiques entre voies. Le développement de ces notions fait l'objet du chapitre suivant.

{\pagestyle{empty}\cleardoublepage}
\chapter{Indicateurs appliqués à la spatialité : connus ou construits}
\minitoc
\markright{Indicateurs appliqués à la spatialité}

\begin{figure}[h]
    \centering
    \begin{subfigure}[t]{.45\linewidth}
        \includegraphics[width=\textwidth]{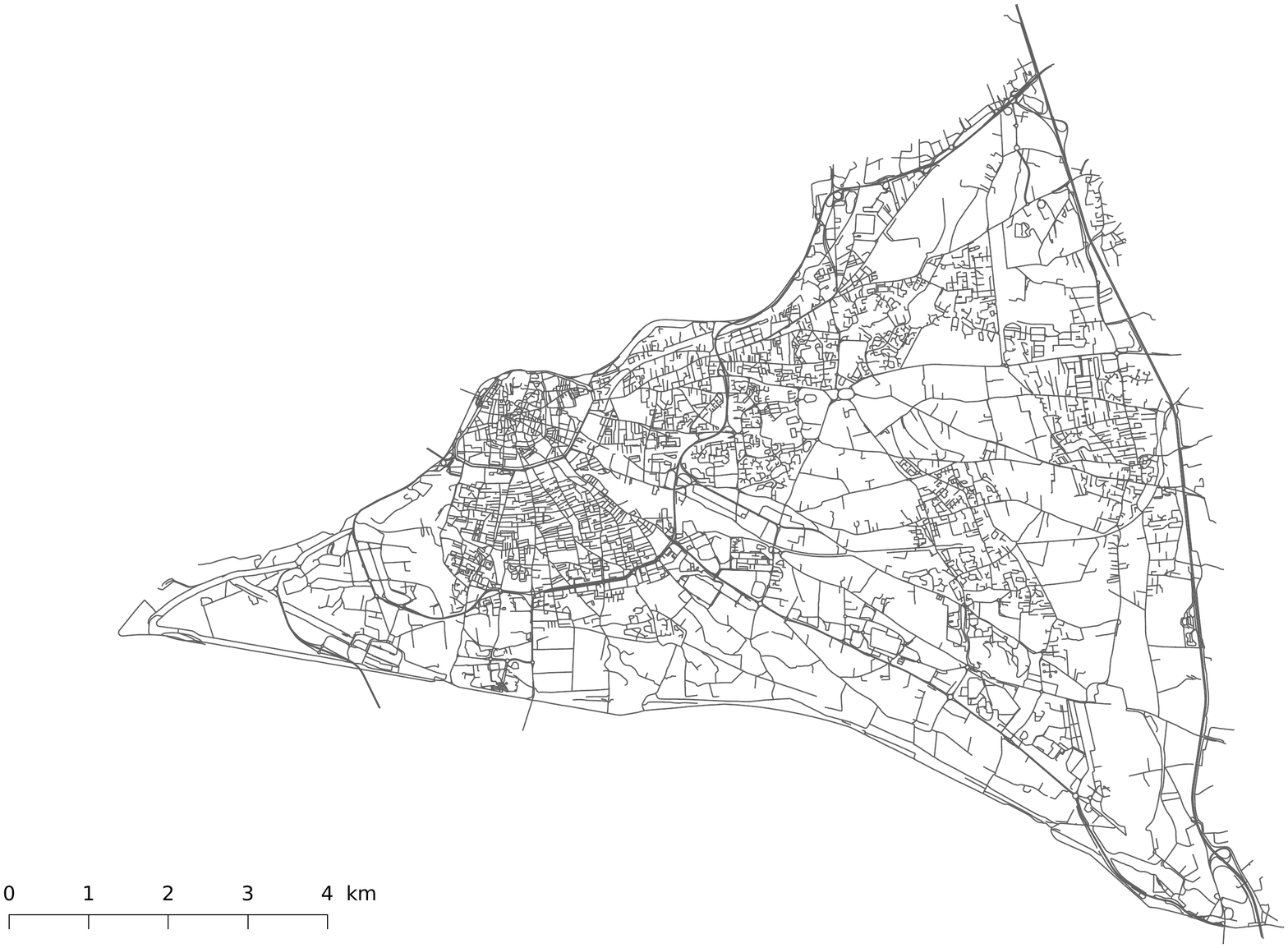}
        \caption{Réseau viaire numérisé de la ville d'Avignon et ses alentours. Données issues de la \copyright BDTOPO 2014 de l'IGN.}
        \label{fig:brut_avignon}
    \end{subfigure}
    ~
    \begin{subfigure}[t]{.45\linewidth}
        \includegraphics[width=\textwidth]{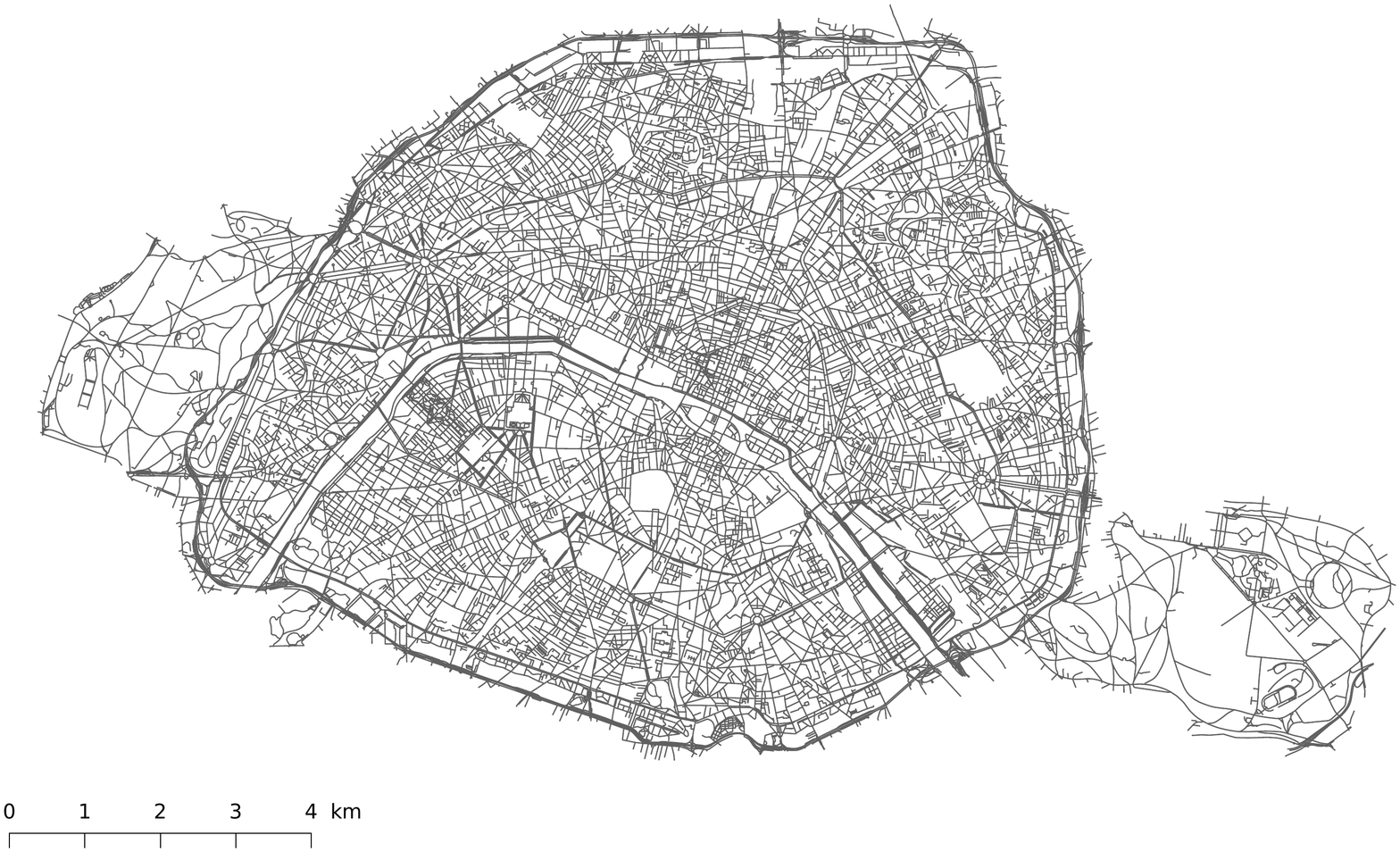}
        \caption{Réseau viaire numérisé de la commune de Paris. Données issues de la \copyright BDTOPO 2015 de l'IGN.}
        \label{fig:brut_paris}
    \end{subfigure}
\end{figure}

Pour la suite de cette étude nous travaillerons sur deux graphes spatiaux extraits de réseaux viaires. Pour cela nous utilisons des données issues de la \copyright BDTOPO de l'IGN. Nous extrayons le graphe viaire de la commune d'Avignon et de ses alentours (plus étendu que celui utilisé dans le chapitre précédent) ainsi que celui de la commune de Paris. Le premier extrait a la particularité d'avoir le centre de la ville dans sa périphérie et de regrouper des structures de types très différents (centre ville, banlieue, campagne) (figure \ref{fig:brut_avignon}). Le second échantillon spatial a une structure plus dense et regroupe essentiellement un tissu urbain (figure \ref{fig:brut_paris}). Nous développons dans ce chapitre le calcul d'indicateurs que nous appliquons au réseau d'arcs comme à celui de voies. Nous définissons ainsi des attributs de ces deux objets qui ont pour but de caractériser les structures des graphes considérés. Nous montrons les informations résultant des calculs sur les réseaux viaires, qui constituent le cœur de notre réflexion, des deux échantillons cités. Pour permettre une meilleure visibilité des résultats sans trop encombrer ce chapitre, nous présentons les cartes de Paris au fil du texte et réservons celles d'Avignon en annexe \ref{ann:chap_indicateurs}. Nous regroupons le nombre des différents objets contenus dans ces réseaux dans le tableau \ref{tab:pres_ville1}.

Nous traiterons tout d'abord les indicateurs locaux, c'est-à-dire qui sont calculés en fonction de la géométrie et de la topologie de l'objet considéré et de son entourage direct. Ils sont robustes aux découpages de l'échantillon spatial à condition que la géométrie et le voisinage de l'objet restent inchangés. L'effet de bord ne pourra donc avoir d'impact que sur les arcs (ou les voies) en limite de la zone de découpage. Puis nous expliciterons les indicateurs globaux, calculés pour chaque objet en tenant compte de l'ensemble du réseau. Cette distinction de mesures peut se faire sous différentes dénominations. Ainsi, C. Ducruet évoque dans ses travaux ces deux types de caractérisation sous le nom de \textit{mesures locales de voisinage} ou \textit{mesures locales d'ensemble} \citep{ducruet2010mesures}.

Nous nous concentrerons dans ce chapitre sur la construction des indicateurs, nous montrerons l'impact du découpage de l'échantillon sur leur calcul dans la deuxième partie.

\begin{table}[h]
\begin{tabular}{| c | r | r | r | r |}
\hline
& $L_{tot}$ (en mètres) & $N_{sommets}$ & $N_{arcs}$ & $N_{voies}$ \\
\hline
Avignon & 949 413 & 8 428 & 13 221 & 4 045 \\
\hline
Paris & 2 112 715 & 17 222 & 30 957 & 6 893 \\
\hline
\end{tabular}
\caption{Tableau regroupant les caractéristiques des deux réseaux étudiés dans ce chapitre}
\label{tab:pres_ville1}
\end{table}

Afin de représenter les indicateurs calculés, nous créons dix classes de longueur équivalente. Dans chacune des classes est regroupé approximativement un dixième du linéaire total du réseau. L'approximation réside dans le fait que chaque classe est remplie jusqu'à ce qu'elle contienne au moins un dixième de la longueur totale du réseau. Les voies sont classées selon l'ordre croissant de l'indicateur qui est considéré dans la représentation. Nous utilisons ces classes de longueurs équivalentes afin de faciliter la lecture sur la carte. Ainsi, chaque classe (et la couleur qui lui est associée) est représentée sur le filaire de manière égale. Nous conserverons la même échelle de couleur pour la représentation des différents indicateurs (exception faite de l'indicateur d'orthogonalité, pour une raison détaillée plus bas). L’œil arrive de cette manière à mieux hiérarchiser et comparer les différents éléments. Nous reviendrons sur les problématiques de représentation de l'information donnée par les indicateurs dans la troisième partie.

\FloatBarrier
\section{Distances sur un graphe}

Pour aborder le rôle d’un objet dans un réseau spatial, il est nécessaire de travailler sur la notion de distance. En théorie des graphes, lorsque l'on évoque la \textit{distance} entre deux sommets, il s'agit du nombre d'arcs contenu dans le plus court chemin entre ces deux sommets \citep{chartrand1998graph}. Cette distance est également appelée \textit{distance géodésique} \citep{bouttier2003geodesic}.

\begin{figure}[h]
    \centering
    \includegraphics[width=\textwidth]{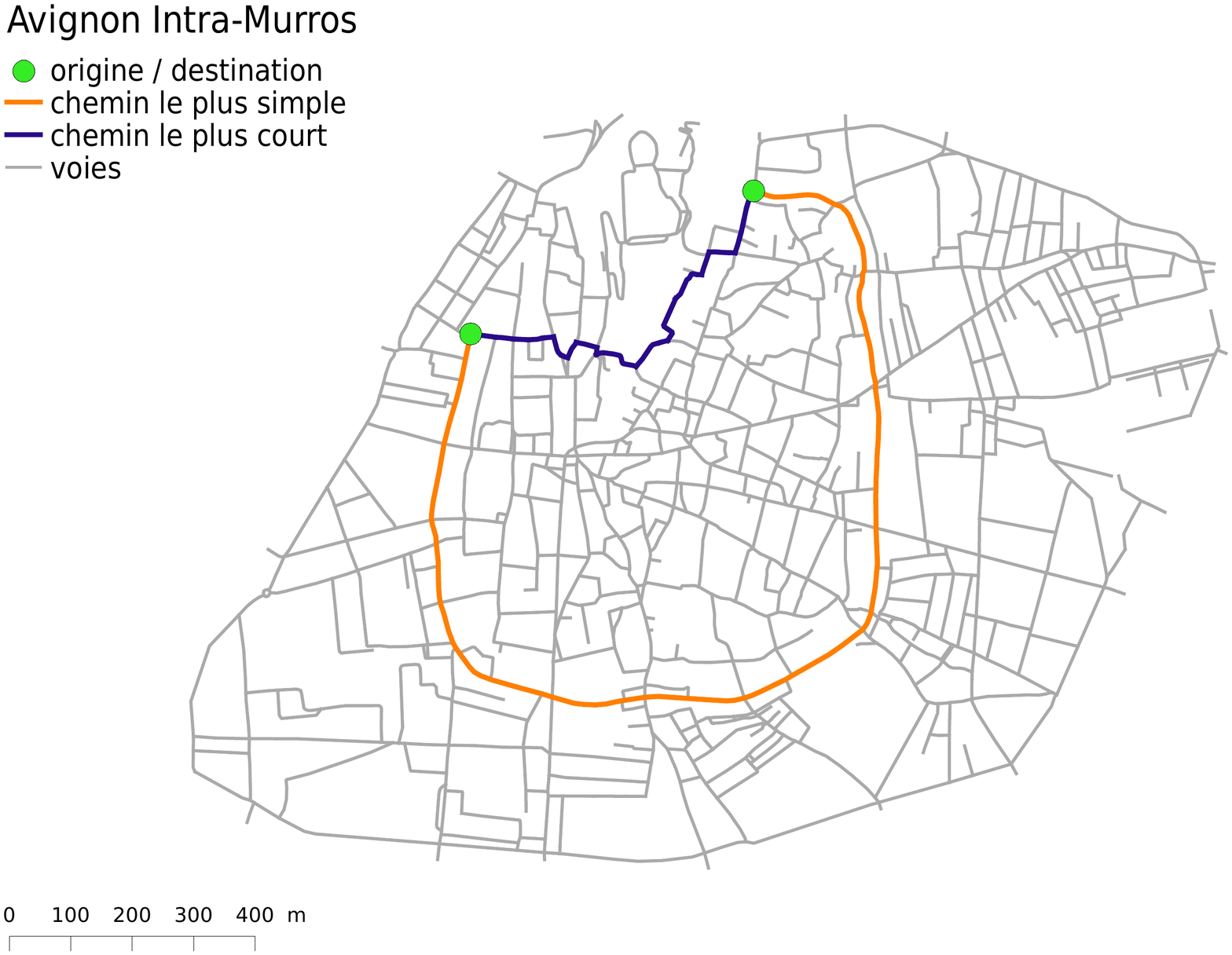}
    \caption{Illustration du chemin le plus court et du chemin le plus simple entre deux points de l'intra-murros de la ville d'Avignon.}
    \label{fig:distances}
\end{figure}

La notion de distance peut également se diversifier. P. Bonnin et S. Douady en explorent différentes définitions et interprétations dans \citep{bonnin2013distance}). Nous distinguerons ici trois types de distance :

\begin{enumerate}

\item La \textit{distance euclidienne} : Distance \enquote{à vol d'oiseau} entre deux points du réseau, sans considérer la géométrie des arcs. Elle correspond à la ligne droite entre deux points.
\item La \textit{distance géographique} : Distance \textit{géodésique} appliquée au graphe primal dont les arcs sont pondérés par leurs longueurs.
Est mesurée ici la distance métrique parcourue. Nous appellerons dans la suite le chemin associé à cette distance le \emph{chemin le plus court} (figure \ref{fig:distances}).
\item La \textit{distance topologique} (notée $d_{simple}$) : Distance \textit{géodésique} appliquée au \textit{line graph} dont les arcs ne sont pas pondérés. Est mesurée ici la distance en nombre d'éléments traversés. Chaque passage par un sommet du \textit{line graph} est équivalent à un changement d'arc ou de voie. Nous appellerons dans la suite le chemin associé à cette distance le \emph{chemin le plus simple} (figure \ref{fig:distances}).

\end{enumerate}

Ces notions de distances nous permettent de définir des parcours caractéristiques entre deux points. Ainsi, le chemin le plus court minimise la distance géographique entre deux objets du réseau. Il a cependant été montré dans plusieurs études que le facteur métrique n’est pas suffisant pour expliciter la structure des villes et leur usage \citep{marchand1974pedestrian, foltete2002structures}. Le chemin le plus court n’est donc pas le seul élément caractéristique des réseaux urbains. Les changements de direction sont également un facteur important \citep{turner2001angular, dalton2001secret}.
Nous définissons le chemin le plus simple comme étant celui qui optimise la distance topologique entre deux voies. Il s’agit du chemin mis en évidence par Pailhous en adoptant le point de vue d’un chauffeur de taxi qui cherche en priorité à se raccorder aux grands axes (réseau principal) avant de s’en détacher pour rentrer dans une desserte locale (sous-réseau) afin d'atteindre sa destination \citep{pailhous1970representation}. Le chemin topologique le plus simple, sur le graphe des voies, est ainsi celui qui minimise le nombre de voies distinctes nécessaires pour se rendre d'une voie de référence à une autre voie. T. Courtat suggère que le chemin le plus simple est une bonne approximation du chemin le plus court \citep{courtat2012walk}. M. P. Vianna \textit{et al.} se penchent sur leur association dans \citep{viana2013simplicity}. Ils les comparent et décrivent l’évolution de leurs combinaisons sur différents types de réseaux planaires (veinures de feuilles, réseau routier, physarum). Les différentes méthodes pour analyser un réseau spatial conduisent au calcul d'indicateurs que nous détaillerons dans la suite de ce chapitre.

\FloatBarrier
\section{Indicateurs locaux}

\FloatBarrier
\subsection{Indicateurs classiques}

\subsubsection{Longueur}

\begin{figure}[h]
    \centering
    \includegraphics[width=\textwidth]{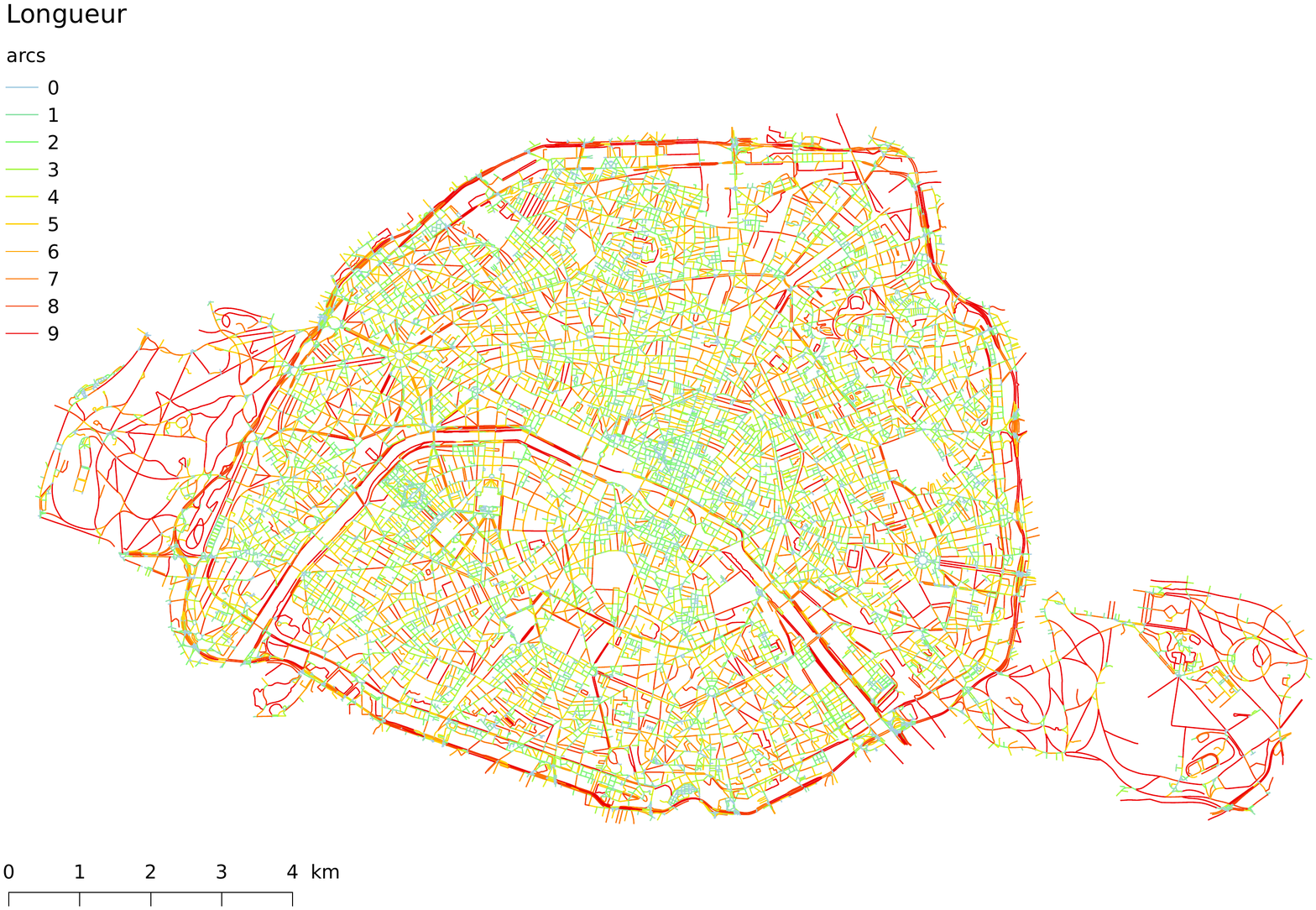}
    \caption{Indicateur de longueur calculé sur les arcs du graphe viaire de Paris.}
    \label{fig:arcs_longueur}
\end{figure}

La longueur des éléments du réseau est la première caractéristique dont on peut étudier les variations. Elle ne nécessite aucun calcul, utilisant directement un attribut fondamental des graphes spatialisés : leur géométrie. La longueur permet de calculer le plus court chemin, mais également de distinguer les grandes structures maillées de l'espace lorsqu'on la cartographie sur les voies. Nous avons déjà montré que la distribution du logarithme des longueurs des voies suit une gaussienne (Partie I, Chapitre 2). Nous pouvons à présent visualiser sur les graphes des deux terrains étudiés la longueur des arcs et celle des voies (figures \ref{fig:arcs_longueur}).

\begin{figure}[h]
    \centering
    \includegraphics[width=\textwidth]{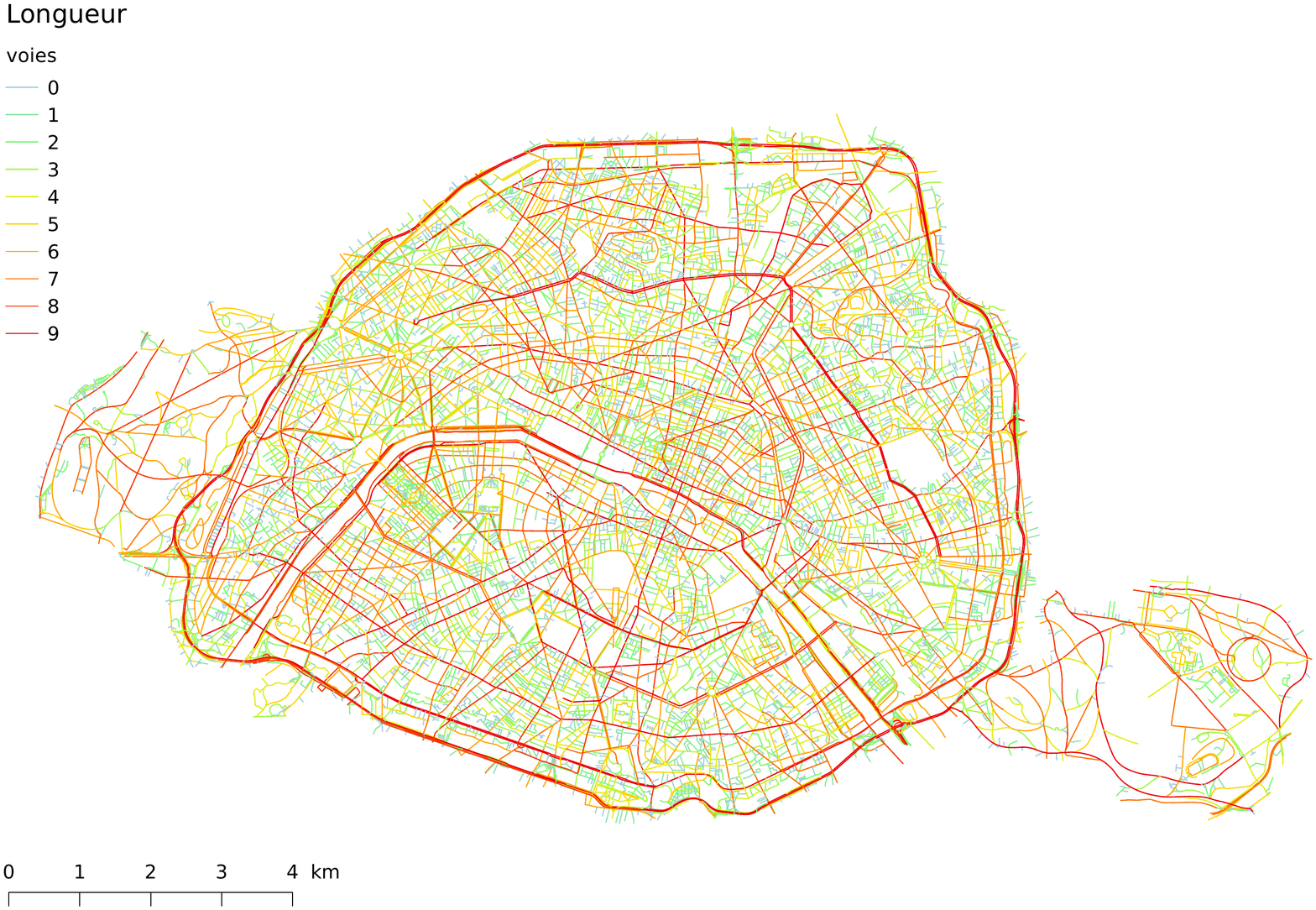}
    \caption{Indicateur de longueur calculé sur les voies du graphe viaire de Paris.}
    \label{fig:voies_longueur}
\end{figure}

Nous observons que les deux couples de cartes font ressortir des informations très différentes. Les cartes représentant les longueurs d'arcs (figure \ref{fig:arcs_longueur}) portent notre attention sur des espaces de concentration d'arcs très courts (en bleu). À l'inverse, celles représentant la longueur des voies (figure \ref{fig:voies_longueur}) font ressortir les voies les plus longues (en rouge) en dessinant ainsi un sur-réseau de distribution de l'espace. Nous qualifierons ces voies de \emph{maillantes}. L'étude de la longueur des arcs nous donne donc une information très localisée alors que celle des voies fait ressortir des structures qui traversent l'espace. C'est la fonction même de cet objet composé qui transcende les échelles et donne une compréhension globale du graphe à partir d'une construction locale. Nous approfondirons cette propriété dans l'étude qualitative de ces graphes, réservée en troisième partie.

\FloatBarrier
\subsubsection{Degré}

Pour un sommet, le degré correspond au nombre d'arcs qui lui sont reliés. Pour calculer le degré des arcs ou des voies dans un graphe spatial, nous traçons son \textit{line graph}. Un arc (ou une voie) devient donc un sommet et son degré correspond au nombre d'autres arcs (resp. voies) auxquels il est connecté (équation \ref{eq:degre}). Nous illustrons la construction du \textit{line graph} pour les voies avec la figure \ref{fig:6_dual_voies}. Il est utile d'observer le degré des éléments d'un réseau afin de définir leur importance dans la \emph{connectivité} de celui-ci. C'est-à-dire, à quel point il reste connexe si des sommets lui sont retirés (les sommets correspondant, dans notre cas, à des arcs ou des voies).

\begin{equation}
 degre(v_{ref}) = Card(v \in G / v \cap v_{ref})
 \label{eq:degre}
\end{equation}

La quantification du degré est utilisée pour explorer la topologie des réseaux. Elle a été étudiée sous différents aspects par différents chercheurs \citep{haggett1969network, taaffe1996geography, rodrigue2004transport}. Depuis les années 1960, elle est appliquée aux réseaux spatiaux, et notamment de transports. Un des premiers à se pencher sur la question a été W. Garrison qui a mesuré la connectivité du réseau autoroutier entre les états des USA \citep{garrison1960connectivity}. Puis K.J. Kansky a proposé quatorze indicateurs pour mesurer les caractéristiques topologiques des réseaux de transport \citep{kansky1963structure}. J. Dill s'est penché sur le point de vue des piétons et cyclistes et a étudié la connectivité du réseau viaire de Portland (Oregon) \citep{dill2004measuring}. F. Xie et D. Levinson, quant à eux, ont défini des schémas de connexions élémentaires entre quatre à six sommets (qu'ils nomment \textit{ring}, \textit{web}, \textit{star} et \textit{hub and spoke}) \citep{xie2007measuring}. Ils peuvent de cette manière comparer différents réseaux en sommant le nombre de ces schémas qui se retrouvent dans chacun d'entre eux. En 2006, S. Latora et son équipe se sont penchés également sur l'étude du degré des sommets du \textit{line graph} de leurs \enquote{lignes droites} \citep{crucitti2006centralitymeasures, cardillo2006structural, scellato2006backbone}.

\begin{figure}[h]
    \centering
    \begin{subfigure}[t]{.45\linewidth}
        \includegraphics[width=\textwidth]{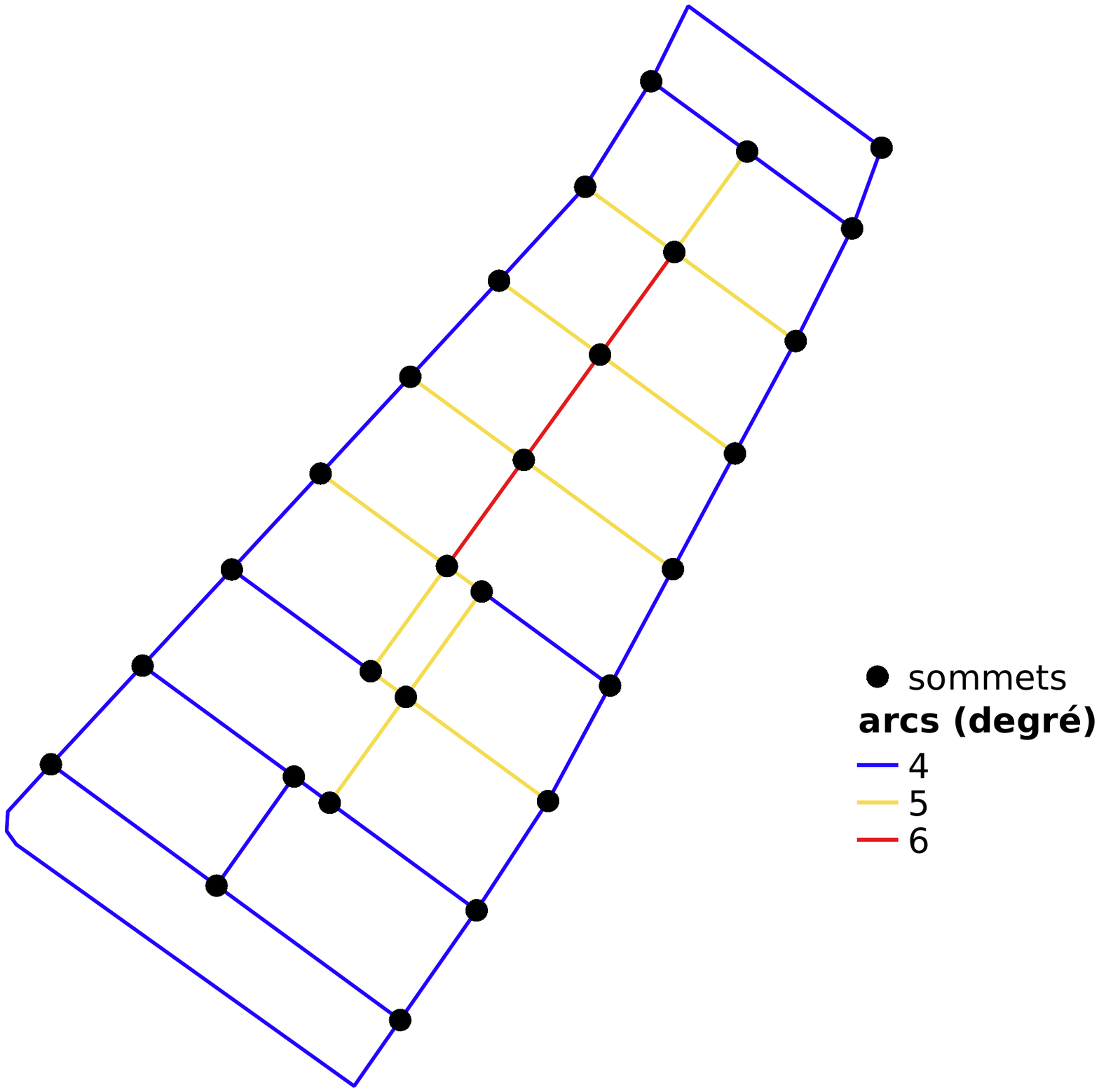}
        \caption{ Degrés des \textbf{arcs}.}
        \label{fig:teh_degre_arcs}
    \end{subfigure}
    ~
    \begin{subfigure}[t]{.45\linewidth}
        \includegraphics[width=\textwidth]{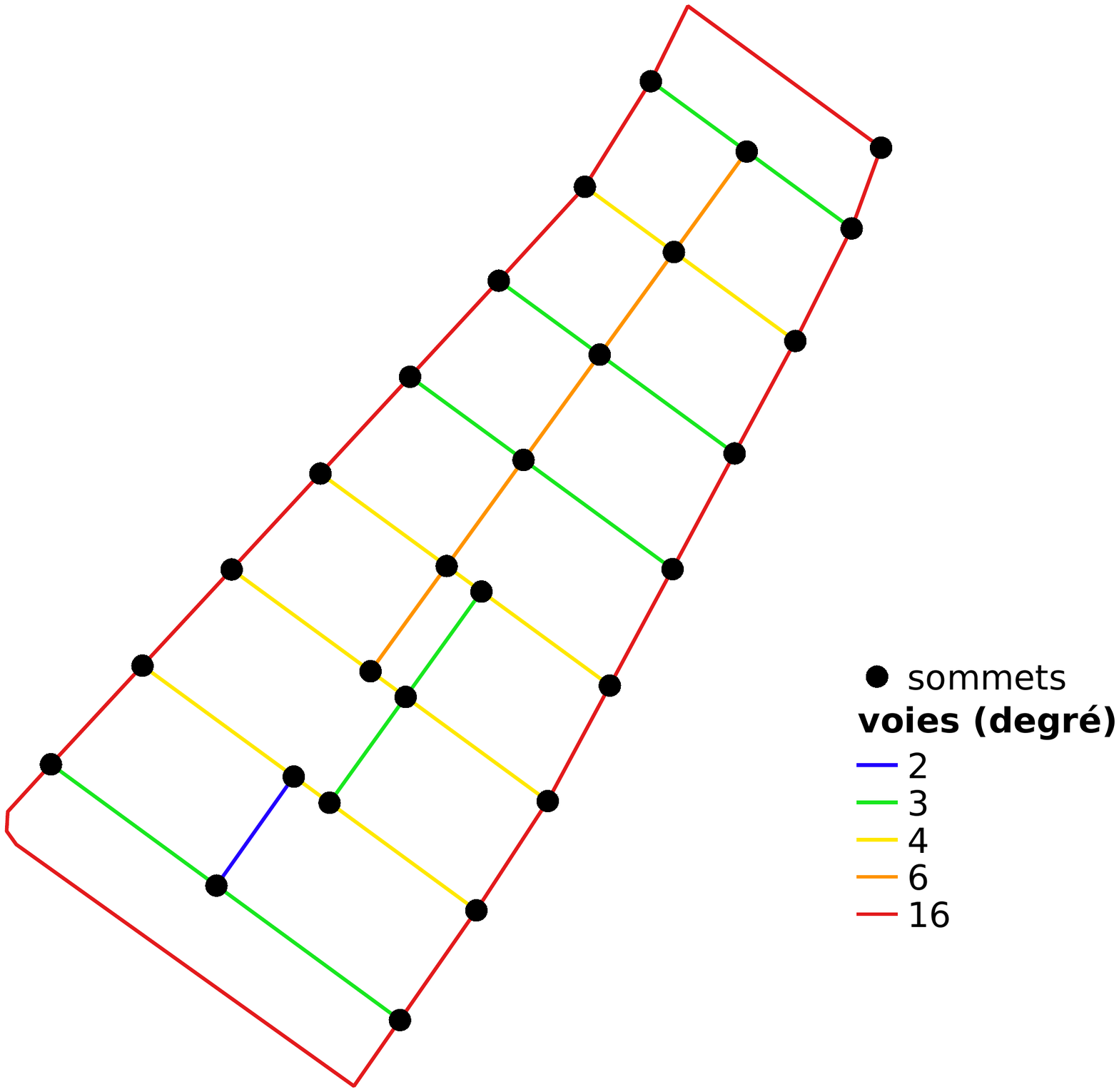}
        \caption{ Degrés des \textbf{voies}.}
        \label{fig:teh_degre_voies}
    \end{subfigure}
    \caption{ Indicateur de degré calculé sur un extrait schématique.}
     \label{fig:teh_degre_arcsvoies}
\end{figure}

\begin{figure}[h]
    \centering
    \includegraphics[width=\textwidth]{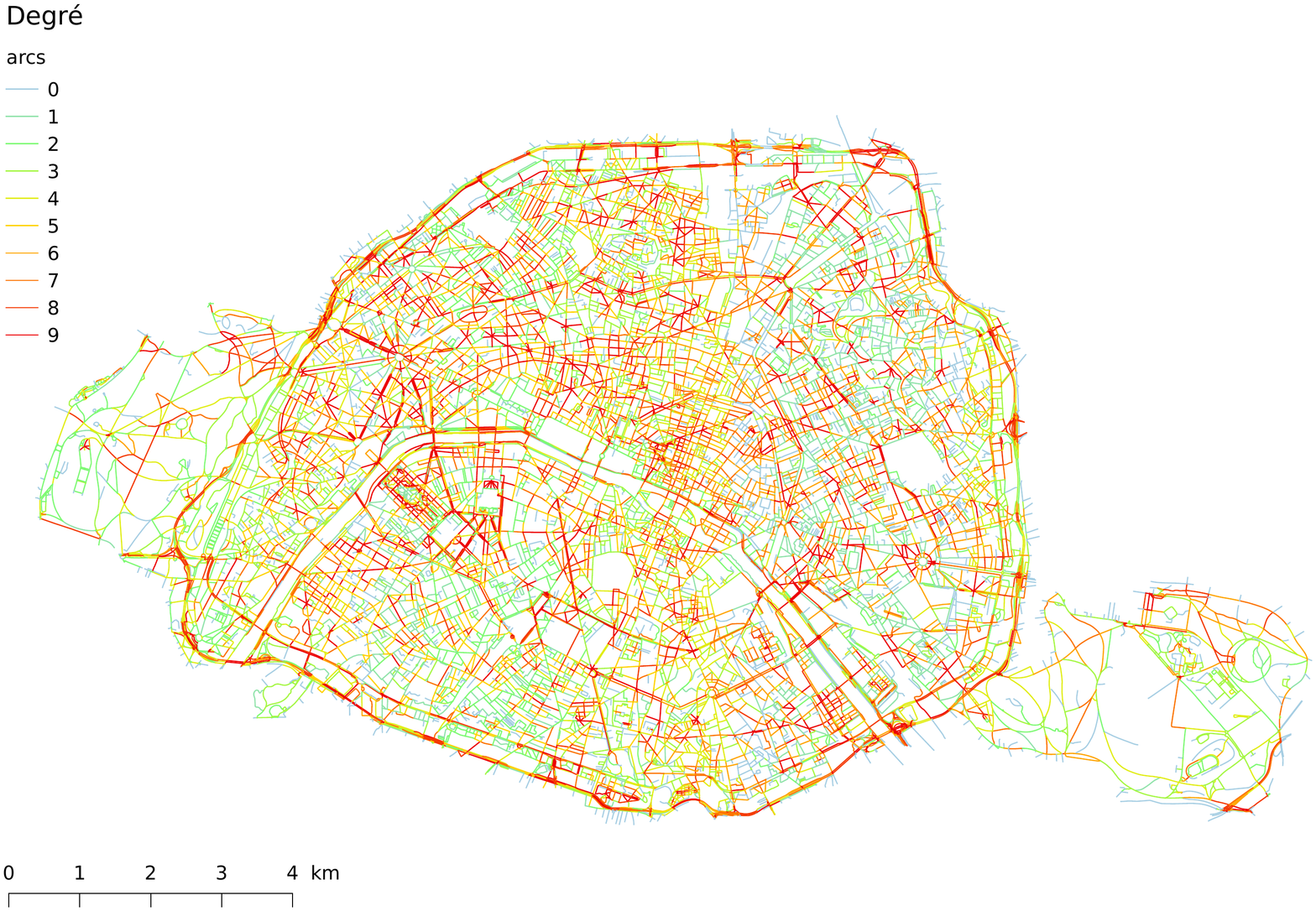}
    \caption{Indicateur de degré calculé sur les arcs du graphe viaire de Paris.}
    \label{fig:arcs_degre}
\end{figure}

\begin{figure}[h]
    \centering
    \includegraphics[width=\textwidth]{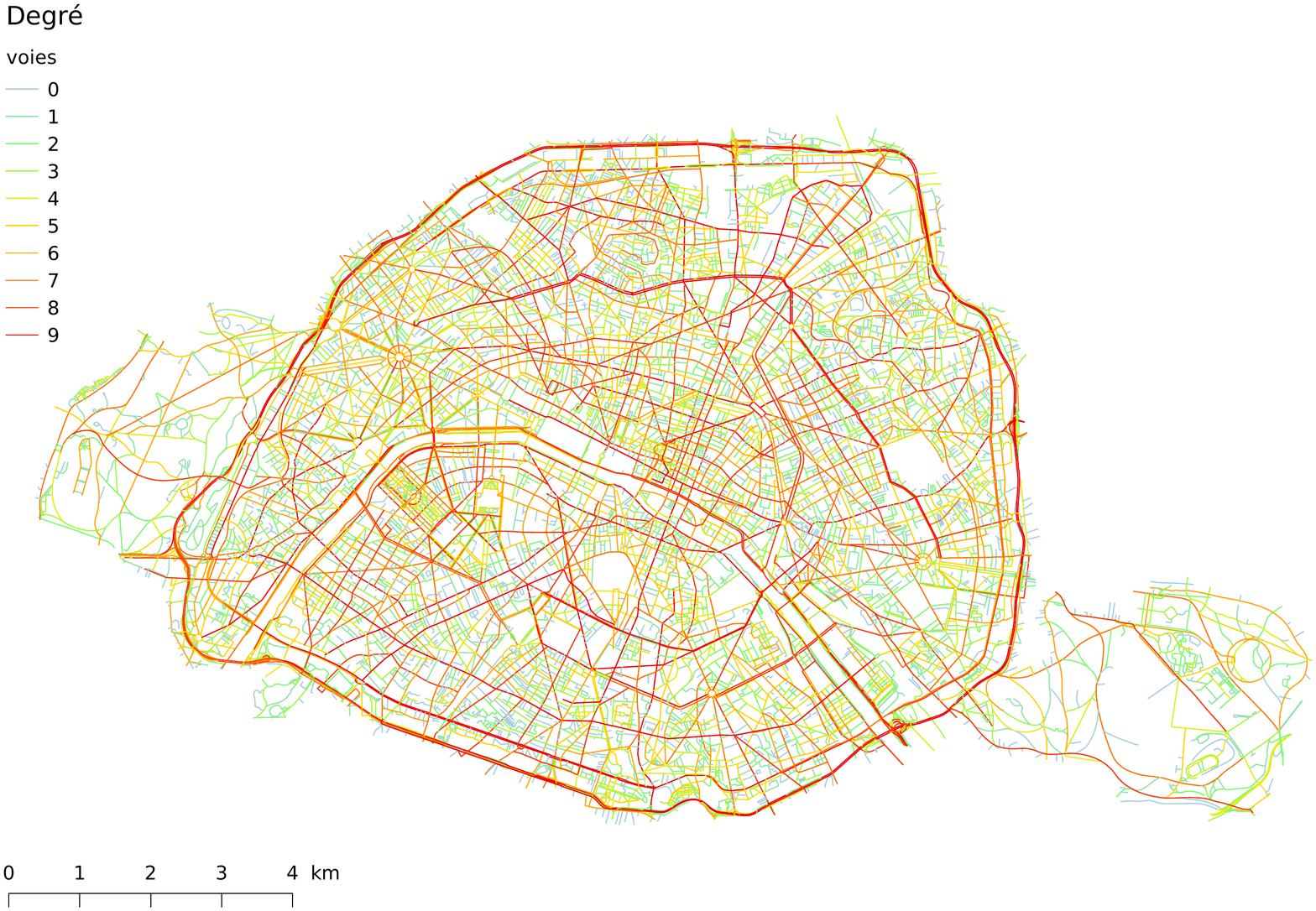}
    \caption{Indicateur de degré calculé sur les voies du graphe viaire de Paris.}
    \label{fig:voies_degre}
\end{figure}

La caractérisation du graphe par le degré des voies apporte une hiérarchisation nettement plus contrastée que celle faite sur les arcs. Nous illustrons cela à travers un exemple schématique pour expliciter la construction de l'indicateur (figure \ref{fig:teh_degre_arcsvoies}). L'application de cette caractérisation à nos réseaux d'études montre sa capacité, tout comme la longueur, à dissocier les structures traversantes sur le graphe des voies (figure \ref{fig:voies_degre}) alors que sa caractérisation est beaucoup moins lisible sur celui des arcs (figures \ref{fig:arcs_degre}).
La construction d'un objet complexe nous permet ainsi de hiérarchiser les structures. Les voies de degré important définissent un réseau principal, celles de degré plus faible, un sous-réseau. Nous verrons dans le chapitre suivant que le squelette mis en évidence est proche de celui caractérisé par la longueur pour les voies.

\FloatBarrier
\subsection{Indicateurs complémentaires}

\FloatBarrier
\subsubsection{Connectivité et Degré de desserte}

\begin{figure}[h]
    \centering
    \begin{subfigure}[t]{.45\linewidth}
        \includegraphics[width=\textwidth]{images/schemas/teh_degre_voies.pdf}
        \caption{\textbf{Degrés} des voies.}
        \label{fig:teh_deg_voies}
    \end{subfigure}
    ~
    \begin{subfigure}[t]{.45\linewidth}
        \includegraphics[width=\textwidth]{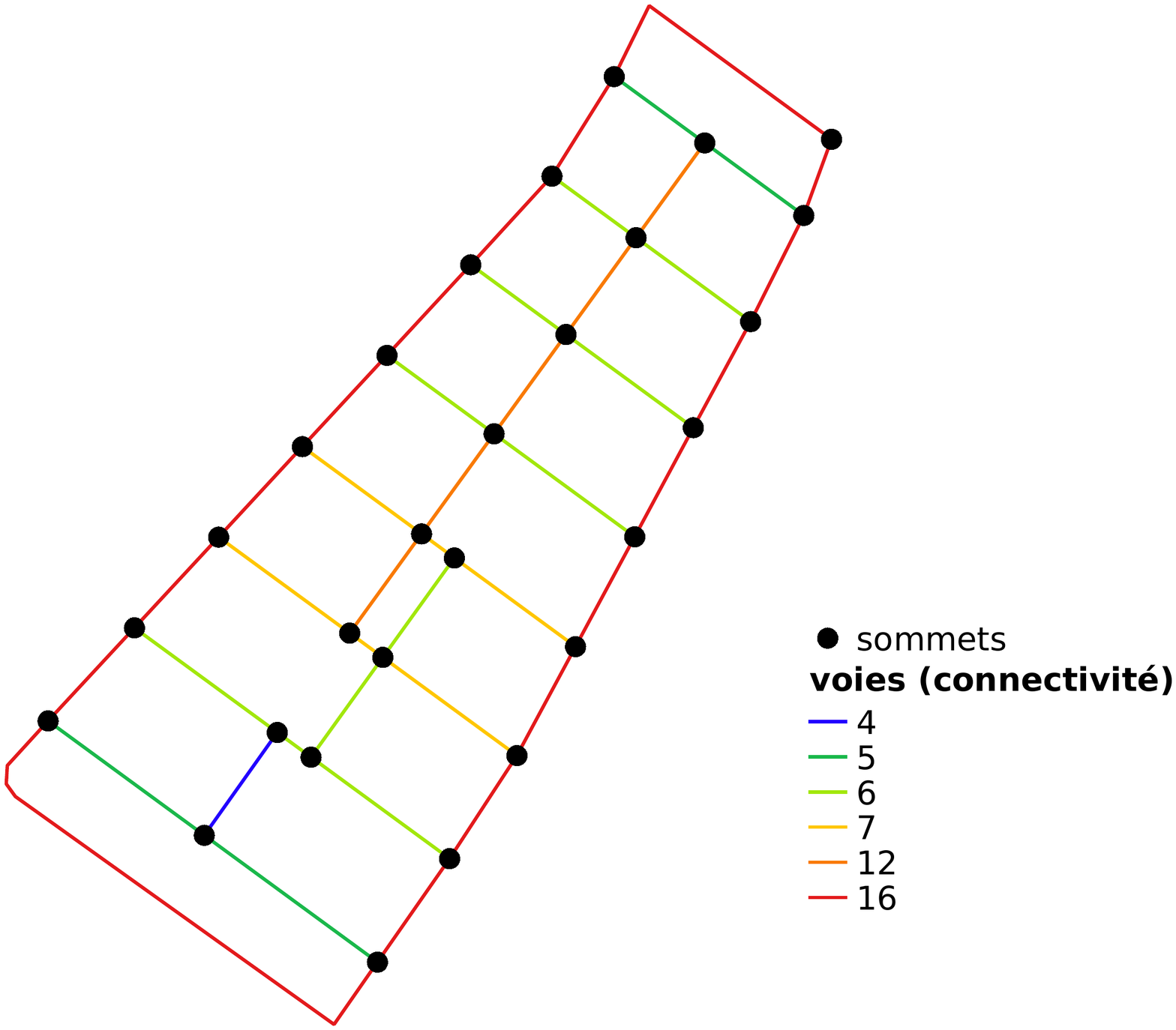}
        \caption{\textbf{Connectivités} des voies.}
        \label{fig:teh_nbc_voies}
    \end{subfigure}
    \caption{Comparaison du degré et de la connectivité des voies sur un extrait schématique.}
    \label{fig:teh_degnbc}
\end{figure}

En affinant le calcul du degré pour les voies, nous pouvons considérer une information supplémentaire : le nombre d'arcs auxquels une voie est connectée. Il diffère légèrement de son degré car une voie peut intersecter une autre voie en la coupant en son centre (et être donc reliée à deux arcs) ou à son extrémité (elle n'est dans ce cas relié qu'à un seul arc) (figure \ref{fig:teh_degnbc}). Nous appellerons cet indicateur \emph{connectivité} (ou \emph{connectivity}) car il fait référence aux travaux précédents effectués sur la topologie des réseaux spatiaux, où étaient considérées les connexions directes d'un objet aux autres. La nuance que nous apportons ici porte sur la différence de nature entre les objets dont on considère les connexions. En effet, la connectivité ne peut pas être calculée sur un \textit{line graph}, car dans celui-ci tous les sommets sont de même nature (dans le \textit{line graph} des voies, tous les sommets représentent des voies : l'information des arcs est perdue). Cet attribut, dans le cas des arcs, est toujours égal à leur degré.

Pour calculer l'indicateur de connectivité, nous nous plaçons dans le graphe spatial primal des voies $G(S, V)$. Nous considérons les sommets d'une voie $v_{ref}$. Pour chaque sommet $s \in v_{ref} $, nous dénombrons les arcs qui y sont liés et qui ne font pas partie de la voie pour laquelle nous faisons le calcul $a / [(s \in a) \wedge (a \notin v_{ref})] $ (équation \ref{eq:connectivite}).

\begin{equation}
 connectivite(v_{ref}) = \sum_{s \in v_{ref}} Card(a / [(s \in a) \wedge (a \notin v_{ref})])
 \label{eq:connectivite}
\end{equation}

\begin{figure}[h]
    \centering
    \includegraphics[width=\textwidth]{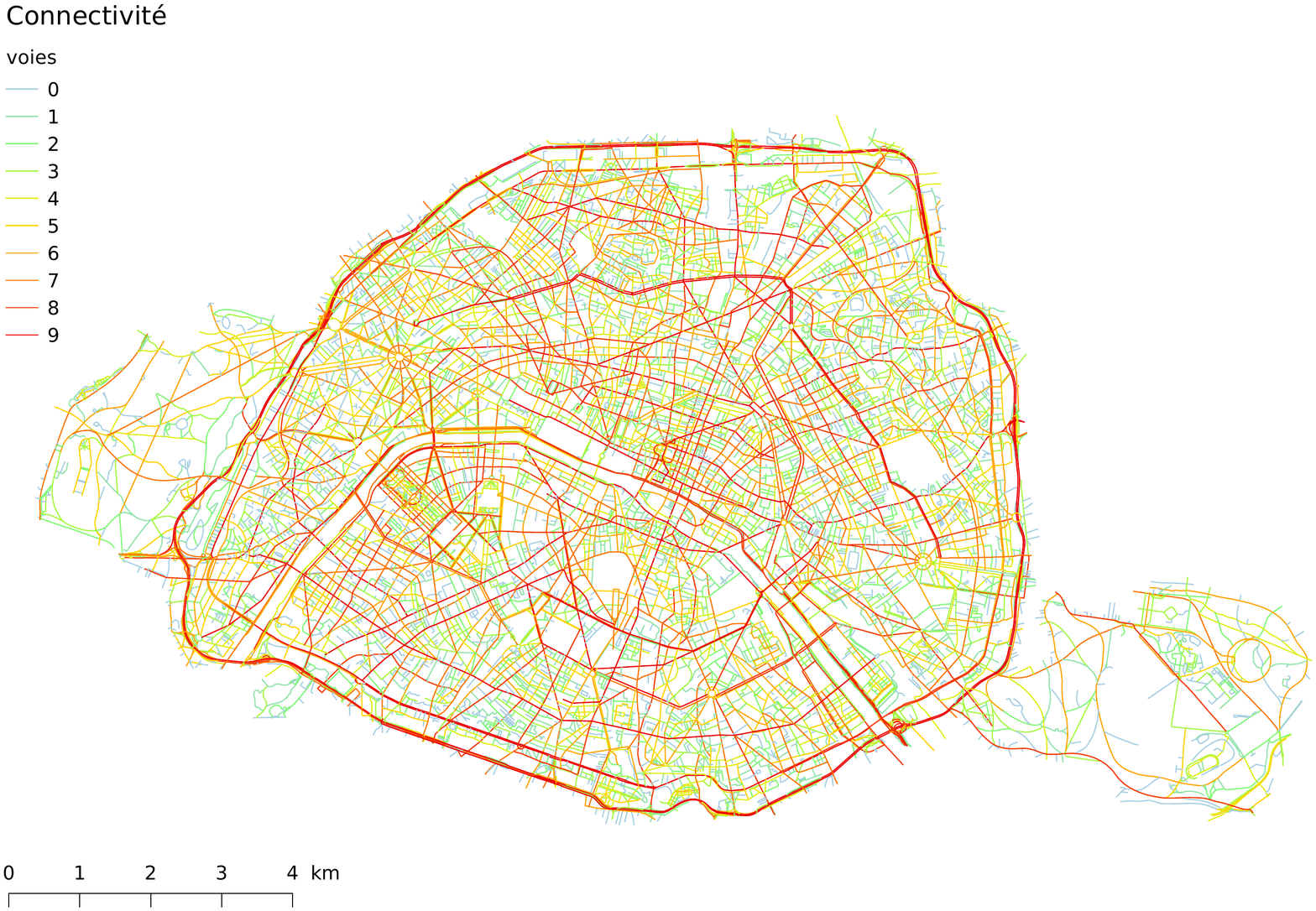}
    \caption{Indicateur de connectivité calculé sur les voies du graphe viaire de Paris.}
    \label{fig:voies_connectivite}
\end{figure}

\begin{figure}[h]
    \centering
    \includegraphics[width=\textwidth]{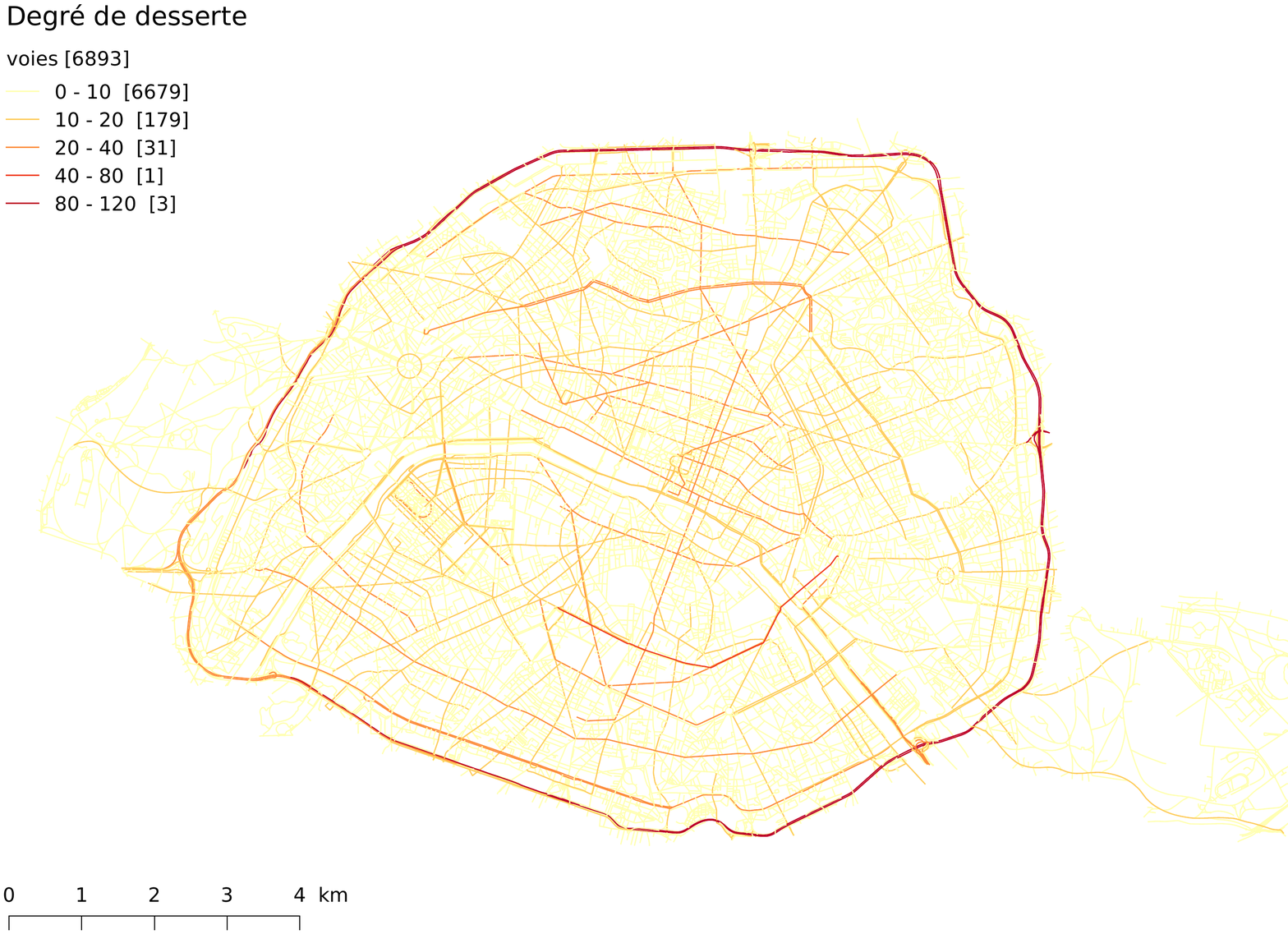}
    \caption{Indicateur de degré de desserte calculé sur les voies du graphe viaire de Paris.}
    \label{fig:voies_degdesserte}
\end{figure}

En pratique, ce calcul donne un résultat proche de celui obtenu pour les degrés (figure \ref{fig:voies_connectivite}). Il permet néanmoins de distinguer les voies dont l'\emph{inclusion} dans le réseau est la plus forte, en soustrayant le degré à la connectivité pour chaque voie. En effet, plus cette différence a une valeur élevée, plus cela signifie que la voie intersecte son voisinage périphérique direct \textit{centralement} (sans être à une de ses extrémités). En mettant en valeur les voies pour lesquelles le résultat de la soustraction est le plus important, nous retrouvons sur Avignon et sur Paris une hiérarchisation des principales voies d'accès au centre de ces villes qui met en avant les structures les plus importantes (figure \ref{fig:voies_degdesserte}). Nous qualifierons cet indicateur de \emph{degré de desserte} (ou \emph{access degree}) de l'espace (équation \ref{eq:degDesserte}).

\begin{equation}
 degreDesserte(v_{ref}) = connectivite(v_{ref}) - degre(v_{ref})
 \label{eq:degDesserte}
\end{equation}

\FloatBarrier
\subsubsection{Espacement}

La combinaison de la longueur et de la connectivité des voies permet de nous renseigner sur la densité linéaire du réseau. Afin d'explorer les informations fournies par cette combinaison, nous calculons un nouvel indicateur, appelé \emph{espacement} (ou \emph{spacing}) issu de la division de la longueur d'une voie par sa connectivité (équation \ref{eq:espacement}). Cette équation est facilement transposable pour les arcs en utilisant leur degré (égal à leur connectivité). L'indicateur d'espacement correspond à la distance moyenne entre deux intersections de l'objet auquel il est appliqué. 

\begin{equation}
espacement(v_{ref}) = \frac {longueur(v_{ref})}{connectivite(v_{ref})}
\label{eq:espacement}
\end{equation}

\begin{figure}[h]
    \centering
    \includegraphics[width=\textwidth]{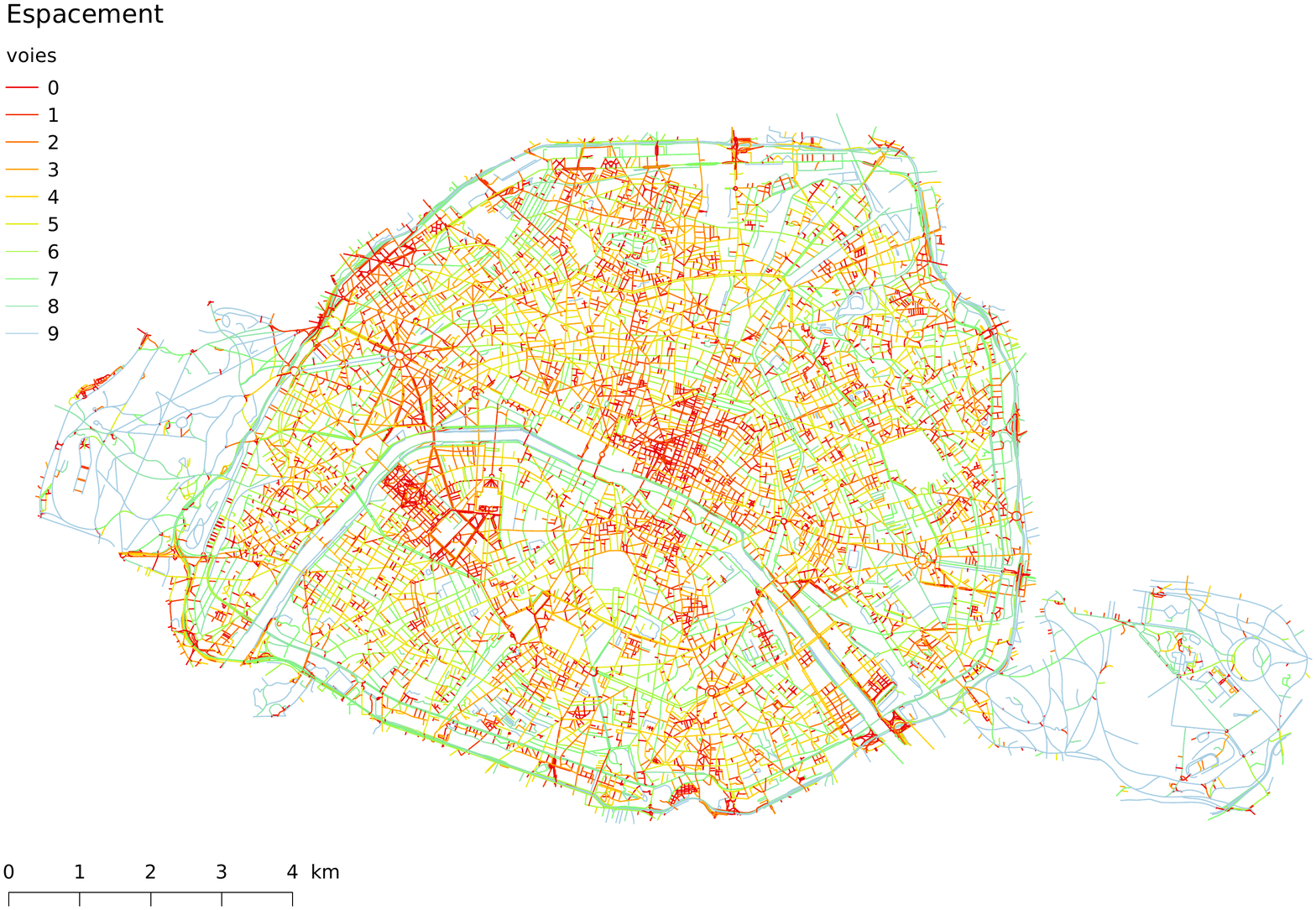}
    \caption{Indicateur d'espacement calculé sur les voies du graphe viaire de Paris.}
    \label{fig:voies_espacement}
\end{figure}

Quel que soit le réseau spatial considéré les voies courtes et très connectées servent à une desserte fine de l'espace alors que celles qui sont longues et peu connectées servent à des transitions rapides entre différents points de l'espace. En appliquant ce raisonnement aux réseaux viaires, nous retrouvons les voies courtes et très connectées comme révélatrices de centres villes denses ou de quartiers résidentiels et les voies longues ayant un degré faible de connexion comme étant des voies rapides. Sur les réseaux routiers de nos cas d'applications, nous mettons en valeur la densité des quartiers en inversant l'échelle de couleur. Les zones d'espacement faible ressortent, à Avignon, comme les quartiers résidentiels (soulignés en rouge) au même titre que le centre ville. À l'inverse, les structures de circulation rapide, d'espacement fort, sont mises au second plan. Il en est de même sur le réseau de Paris, où les zones de forte densité se retrouvent de manière plus diffuse puisque l'on se situe dans un tissu entièrement urbain. Il est néanmoins possible de remarquer le centre historique de la ville (quartier de l'hôtel de ville et des halles) et celui touristique du champ de Mars (dont le niveau de détail est particulièrement important). Les voies rapides, longues et peu connectées, ressortent avec un coefficient d'espacement fort, et une densité linéaire faible. À l'opposé, les voies qui permettent une desserte locale, et des déplacements piétons, sont souvent courtes et très connectées. Ces dernières ressortent avec un faible espacement et donc une densité linaire forte (figure \ref{fig:voies_espacement}).

\begin{figure}[h]
    \centering
    \includegraphics[width=\textwidth]{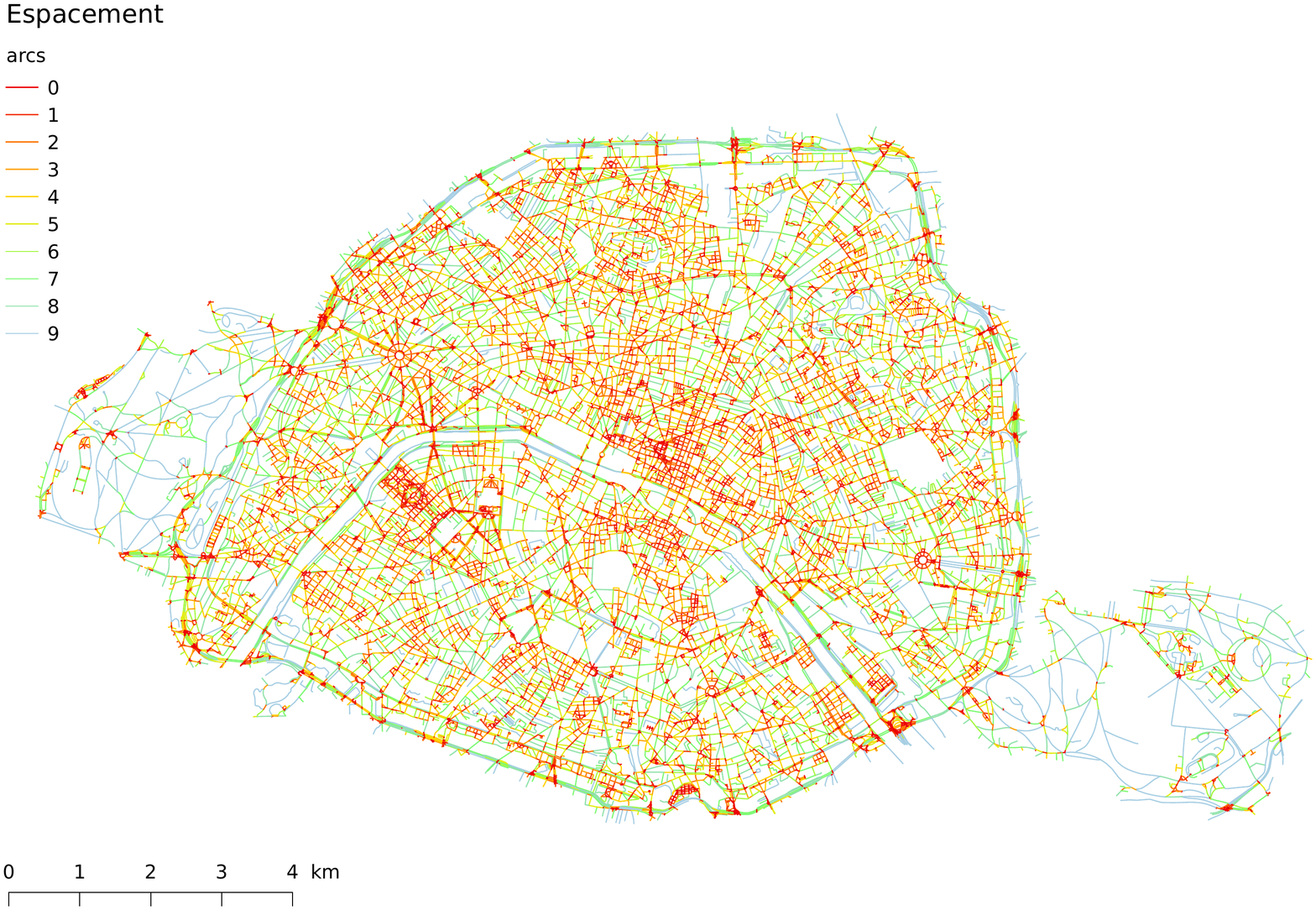}
    \caption{Indicateur d'espacement calculé sur les arcs du graphe viaire de Paris.}
    \label{fig:arcs_espacement}
\end{figure}

Si nous considérons l'indicateur d'espacement appliqué aux arcs, en combinant leur longueur et leur degré, nous observons un résultat très proche de celui des voies (figure \ref{fig:arcs_espacement}). En effet, cette mesure calculant un effet de densité locale, elle ne pâtit pas du fait d'être calculée à une échelle plus locale. Le caractère multi-échelle de la voie n'apporte donc pas pour l'espacement une caractérisation supplémentaire. Elle permet cependant d'accentuer légèrement la visualisation des zones denses, car les voies très courtes et très connectées ressortent dans leur intégralité.

\FloatBarrier
\subsubsection{Orthogonalité}

Le dernier indicateur local que nous présentons ici porte directement sur la géométrie des intersections entre voies. Pour construire l'indicateur d'\emph{orthogonalité} (ou \emph{orthogonality}) nous nous intéressons au sinus des angles de connexion entre voies (nous considérons toujours le plus petit angle fait entre deux voies à un sommet). Plus celui-ci s'approche de 1, plus la connexion sera proche de la perpendiculaire. Plus il tend vers 0, plus le changement de voie sera fait avec un angle faible. En sommant les sinus de tous les angles de connexions $\varphi_{arc_i,arc_j}$ et en normalisant par le nombre de connexions $connectivite(v_{ref})$, nous obtenons une valeur caractéristique de la manière dont la voie est incluse dans le réseau (équation \ref{eq:orthogonalite}).

\begin{equation}
orthogonalite(v_{ref}) = \frac {\sum\limits^{}_{s \in v_{ref}} \  \sum\limits^{}_{arc_i \cap s \wedge arc_i \notin v_{ref}} \  \min(\sin(\varphi_{arc_i arc_j})) / (arc_j \cap s \wedge arc_j \in v_{ref})}{connectivite(v_{ref}
)}
\label{eq:orthogonalite}
\end{equation}

\begin{figure}[h]
    \centering
    \includegraphics[width=0.6\textwidth]{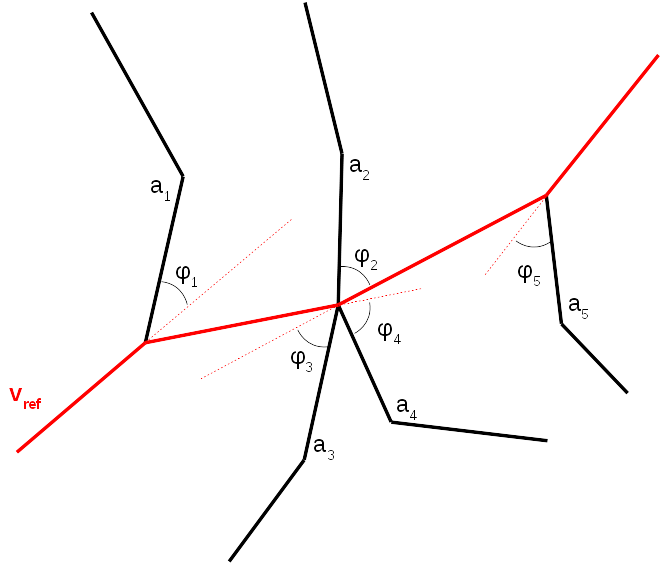}    
    \caption{Schéma illustrant le calcul de l'indicateur d'orthogonalité. \\ Dans cet exemple, $orthogonalite(v_{ref}) = \frac {\sum_{i=0}^5 \sin(\varphi_i)}{5}$}
    \label{fig:orthogonalite}
\end{figure}

\begin{figure}[h]
    \centering
    \includegraphics[width=\textwidth]{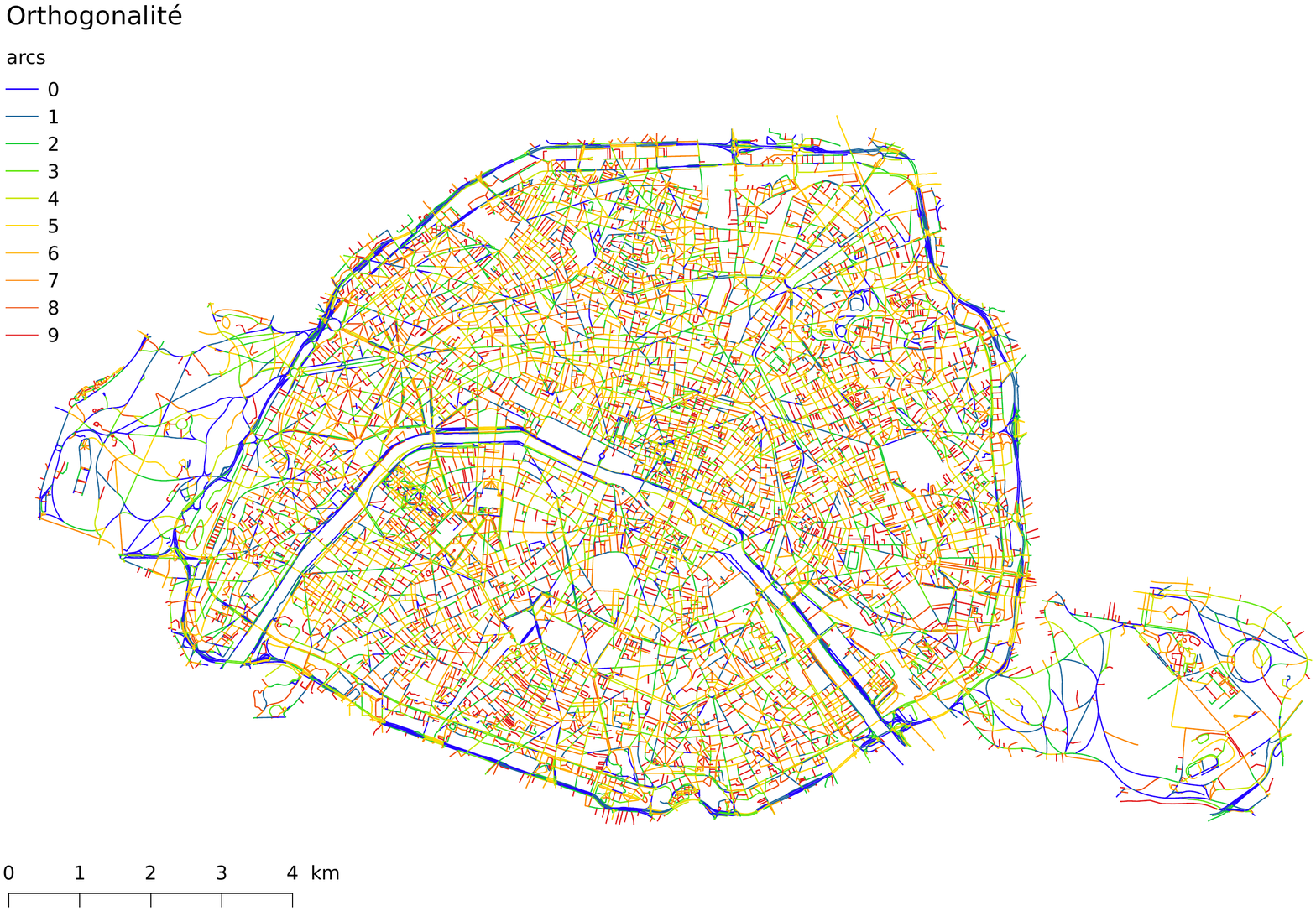}
    \caption{Indicateur d'orthogonalité calculé sur les arcs du graphe viaire de Paris.}
    \label{fig:arcs_ortho}
\end{figure}

\begin{figure}[h]
    \centering
    \includegraphics[width=\textwidth]{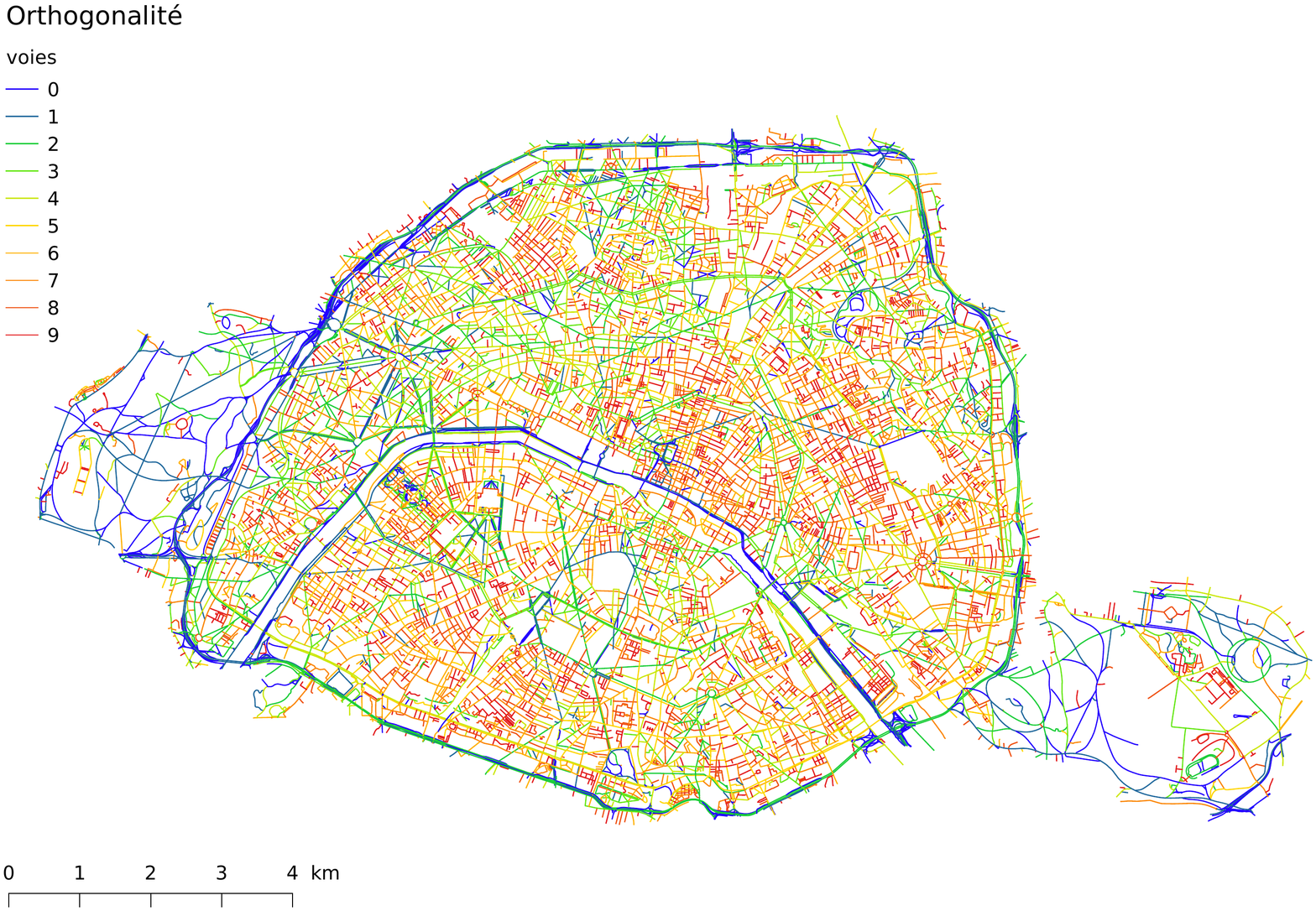}
    \caption{Indicateur d'orthogonalité calculé sur les voies du graphe viaire de Paris.}
    \label{fig:voies_ortho}
\end{figure}

Cet indicateur nous permet de mettre en évidence sur les graphes étudiés deux extrêmes : les objets inclus dans des structures très maillées et ceux aux connexions souples. Le résultat donné pour les arcs est très bruité (figure \ref{fig:arcs_ortho}) et ne fait pas apparaître de structures particulières. En effet, le calcul est fait à une échelle trop locale pour être pertinent. En revanche, appliqué aux voies, l'indicateur d'orthogonalité permet de faire ressortir les grands axes de circulation rapide (aux connexions souples) et les zones résidentielles ou anciennes qui ont un maillage plus marqué (figure \ref{fig:voies_ortho}). À Avignon, nous retrouvons ainsi les axes de contournement de la ville qui permettent la circulation vers son extérieur (en bleu) et les centres urbains (en rouge). Sur le graphe de Paris, les voies Haussmanniennes le découpant transversalement ressortent avec une orthogonalité moyenne. Nous approfondirons ces interprétations dans la troisième partie.

\FloatBarrier

\section{Indicateurs globaux}

\subsection{Betweenness et Utilisation (classiques)}

Un des indicateurs les plus célèbres en théorie des graphes est celui de \textit{betweenness}. Nous l'avons évoqué dans l'introduction de ce chapitre : le calcul de la \emph{centralité} des éléments d'un réseau était au cœur des préoccupations des premiers chercheurs dans ce domaine. L.C. Freeman \citep{freeman1977set} formalisa donc une théorie pour déterminer quels étaient les éléments les plus \textit{centraux}, dans tous types de graphes (connexes ou non). Ces éléments sont ceux les plus \textit{influents}, ils ont un potentiel bloquant ou de diffusion plus important puisqu'ils participent au plus de chemins sur le réseau.

La betweenness est usuellement calculée sur les sommets d'un graphe en considérant l'ensemble des chemins les plus courts entre chaque paire de sommets ($\sigma_{s_1s_2}$). Ainsi, la betweenness d'un sommet $s_{ref}$ représente la somme des chemins les plus courts, entre chaque couple de sommets du graphe $(s_1,s_2)$ passant par $s_{ref}$, divisée par le nombre total de chemins les plus courts existant entre $s_1$ et $s_2$ (équation \ref{eq:betw_sommets}). Il est possible d'attribuer un \textit{poids} à chaque arc du graphe et ainsi de calculer le chemin le plus court suivant ce poids. Les chemins les plus courts ne seront ainsi pas forcément les plus simples.

\begin{equation}
betweenness(s_{ref})= \sum_{s_1 \neq s_2 \neq s_{ref} \in G}\frac{\sigma_{s_1s_2}(s_{ref})}{\sigma_{s_1s_2}}
\label{eq:betw_sommets}
\end{equation}

E.W. Dijkstra propose une méthode pour trouver l'ensemble des chemins les plus courts sur un graphe \citep{dijkstra1959note}. Chaque chemin est trouvé par propagation dans le graphe, en partant d'un sommet initial, et en suivant les arcs de poids les plus faibles. Dans le cas le moins favorable, la complexité du calcul est en $O( \vert A \vert + \vert S \vert * ln(S)) $ où $A$ est le nombre d'arcs et $S$ le nombre de sommets du graphe.

Les algorithmes usuels proposés pour calculer le nombre de ces chemins les plus courts passant par un sommet ont une complexité en $O(A^3)$ (algorithme de Floyd–Warshall, \citep{weisstein2008floyd}). Si le graphe n'est pas trop important, la complexité peut être réduite à $O(A^2*ln(A) + \vert A \vert + \vert S \vert)$. Enfin, si le graphe n'est pas pondéré, Brandes propose un algorithme de complexité en $O(\vert A \vert * \vert S \vert)$ \citep{brandes2001faster}.

Si nous appliquons ce calcul sur le graphe primal des arcs, nous obtenons la betweenness des sommets. Nous calculons alors le chemin le plus court entre deux croisements, ou impasses, du graphe primal. Entre le point d'origine et le point de destination, nous avons une succession d'arcs de longueurs différentes qui peuvent constituer leur valuation. La caractérisation est ainsi faite sur les éléments ponctuels du graphe spatial, et non sur son linéaire.

Dans ces travaux, nous ne souhaitons pas étudier la caractérisation des sommets mais celle des arcs ou des voies. Nous nous positionnons donc dans le \textit{line graph} du graphe primal afin de transformer les arêtes en sommets.
Nous ne considérons donc plus chaque chemin selon leur distance métrique la plus courte (distance géographique), mais selon celle topologique minimale (distance topologique). Nous observons le nombre de changements d'objets entre chaque couple d'arcs ou de voies du graphe. Chacun de ces changements correspond à un \enquote{tournant} selon notre définition de l'alignement et incrémentera la distance topologique entre le couple d'objets.

La betweenness d'une voie de référence $v_{ref}$ est donc donnée par la somme des chemins les plus simples (chemins topologiques les plus courts), entre chaque couple de voies $(v_1,v_2)$ qui passent par $v_{ref}$, divisée par le nombre total de chemins les plus simples possibles entre ces deux voies ($\sigma_{v_1v_2}$) (équation \ref{eq:betw_voies}).

\begin{equation}
betweenness(v_{ref})= \sum_{v_1 \neq v_2 \neq v_{ref} \in G}\frac{\sigma_{v_1v_2}(v_{ref})}{\sigma_{v_1v_2}}
\label{eq:betw_voies}
\end{equation}

Le temps de calcul de la betweenness la rend contraignante à appliquer à des graphes importants. C'est pourquoi nous étudions un autre indicateur que nous appelons \textit{utilisation} (ou \textit{use}). Cet indicateur correspond à celui appelé \textit{stress centrality} en théorie des graphes \citep{brandes2008variants}. Si sa complexité est en $O(S^3)$ (avec $S$ le nombre de sommets du \textit{line graph}), son temps de calcul est nettement réduit. En effet, à chaque itération, le programme se concentre sur un échantillon restreint d'éléments. Le calcul de cet indicateur consiste à sommer pour chaque objet le nombre de chemins les plus simples auquel il appartient, en en considérant la totalité entre chaque couple d'objets sur le graphe. Cela revient donc à simplifier l'équation de la betweenness à la somme appliquée uniquement au numérateur (équation \ref{eq:use_voies}).

\begin{equation}
utilisation(v_{ref})= \sum_{v_1 \neq v_2 \neq v_{ref} \in G} \sigma_{v_1v_2}(v_{ref})
\label{eq:use_voies}
\end{equation}

La méthode que nous développons pour calculer cet indicateur s'appuie sur un processus incrémental qui utilise un raisonnement lié à la \textit{filiation} du chemin. Nous considérons chaque voie comme partie d'un \textit{arbre généalogique} (avec un ascendant et un descendant uniques) : une voie \textit{parente} et une voie \textit{fille}. Nous construisons ainsi une mémoire de la succession d'objets dans les chemins. Celle-ci nous permet pour chaque voie de savoir dans combien de généalogies elle apparaît, et donc combien de fois elle est utilisée dans les chemins les plus simples.

Afin de calculer l'utilisation de chaque objet $voie$ nous identifions à partir de chaque voie successivement placée en tant que voie de référence $v_{ref}$ l'ensemble des chemins les plus simples entre $(v_{ref})$ et $(v \in graphe / v \neq v_{ref})$. À chaque $voie$ rencontrée, nous incrémentons son \enquote{compteur d'utilisation}. La somme finale pour chaque objet $voie$ une fois toutes les voies placées en tant que $v_{ref}$ nous donne son coefficient d'utilisation (algorithme \ref{alg:use}). Nous procédons de même pour calculer l'indicateur sur les arcs, en remplaçant l'objet $voie$ par l'objet $arc$ et en considérant les connexions de celui-ci.

\begin{algorithm}
    \caption{Calcul de l'utilisation}
    \label{alg:use}
    \begin{algorithmic}
		\STATE $\forall voie, compteur_{use}[voie] = 0$

        \FOR{$v_{ref} \in graphe$}
        \STATE $dtopo = 0$
        \STATE $\forall v, tab_{dtopo}[v] = -1$
        
        \STATE $tab_{parent}[v_{ref}] = 0$
        \STATE $tab_{dtopo}[v_{ref}] = dtopo$
        
	        \FOR{$v \in graphe / tab_{dtopo}[v] = dtopo$}
		        
			        \FOR{$v_i \in graphe / (v_i \cap v \wedge (tab_{dtopo}[v_i] = -1 \vee tab_{dtopo}[v_i] = dtopo+1))$}
			        \STATE $tab_{parent}[v_i] = v$
			        \STATE $tab_{dtopo}[v_i] = dtopo + 1$
			        
			        \STATE $idv_{fille} = v_i $
			        \STATE $idv_{parent} = v $
			        
				        \WHILE{$tab_{parent}[idv_{fille}] \neq 0$}
				        \STATE $compteur_{use}[idv_{parent}] ++$
				        \STATE $idv_{fille} = idv_{parent} $
				        \STATE $idv_{parent} = tab_{parent}[idv_{fille}] $
				        \ENDWHILE
				        
				    \ENDFOR
					
					\IF {$\nexists v \in graphe / tab_{dtopo}[v] = dtopo$}
					\STATE $dtopo ++$
					\ENDIF
	        \ENDFOR  	      
	    \ENDFOR

    \end{algorithmic}
\end{algorithm}

Pour chaque objet, l'utilisation est donc le nombre de chemins les plus simples qui passent par celui-ci. Nous pouvons normaliser cette somme par le nombre total de chemins les plus simples sur l'ensemble du réseau afin d'obtenir une valeur comparable d'un réseau à l'autre (équation \ref{eq:usenorm_voies}).

\begin{equation}
utilisation_{normalisee}(v_{ref})= \frac{\sum_{v_1 \neq v_2 \neq v_{ref} \in G} \sigma_{v_1v_2}(v_{ref})}{\sum_{v_1 \neq v_2 \in G} \sigma_{v_1v_2}}
\label{eq:usenorm_voies}
\end{equation}

\begin{figure}[h]
    \centering
    \includegraphics[width=\textwidth]{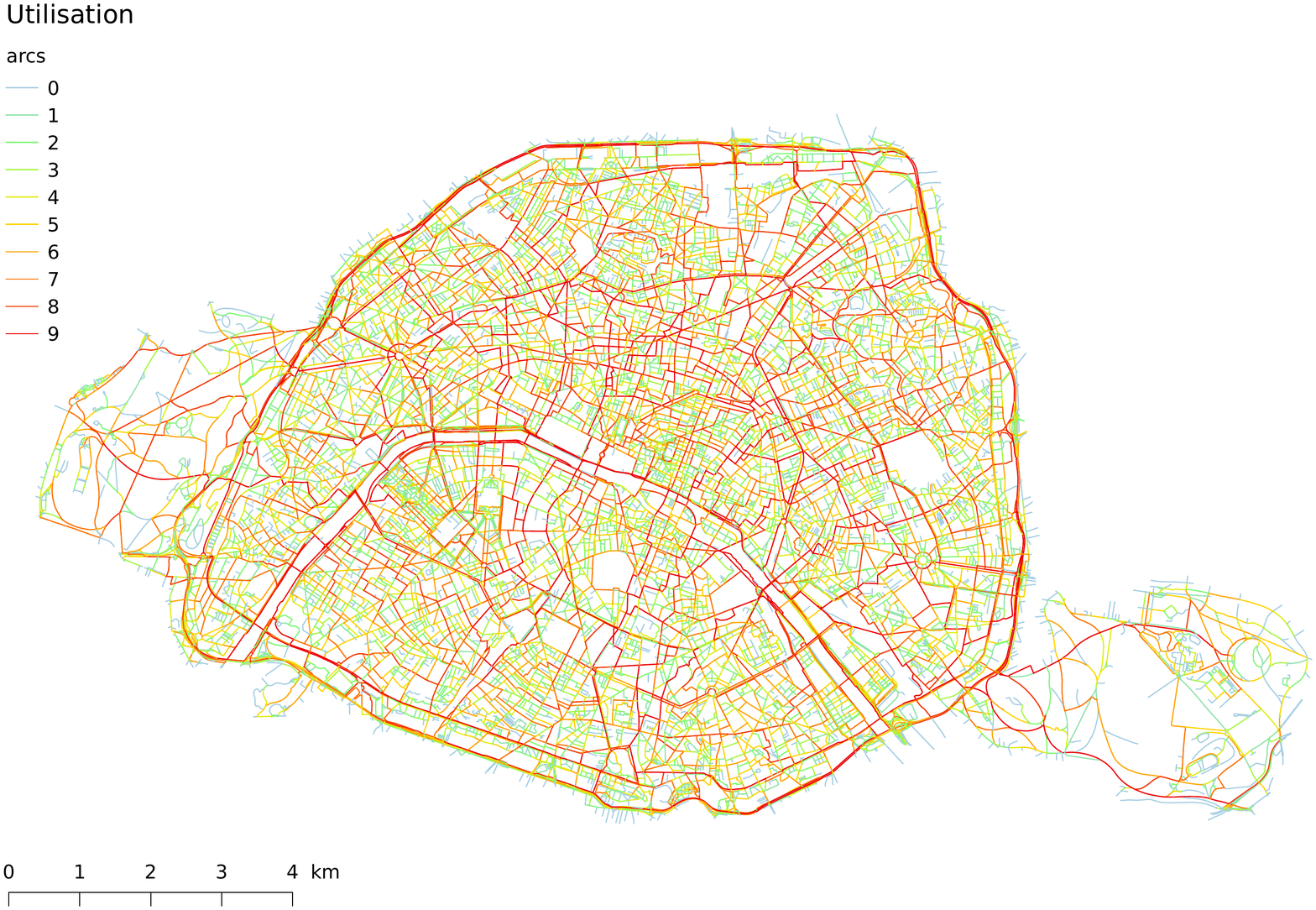}
    \caption{Indicateur d'utilisation calculé sur les arcs du graphe viaire de Paris.}
    \label{fig:arcs_use}
\end{figure}

\begin{figure}[h]
    \centering
    \includegraphics[width=\textwidth]{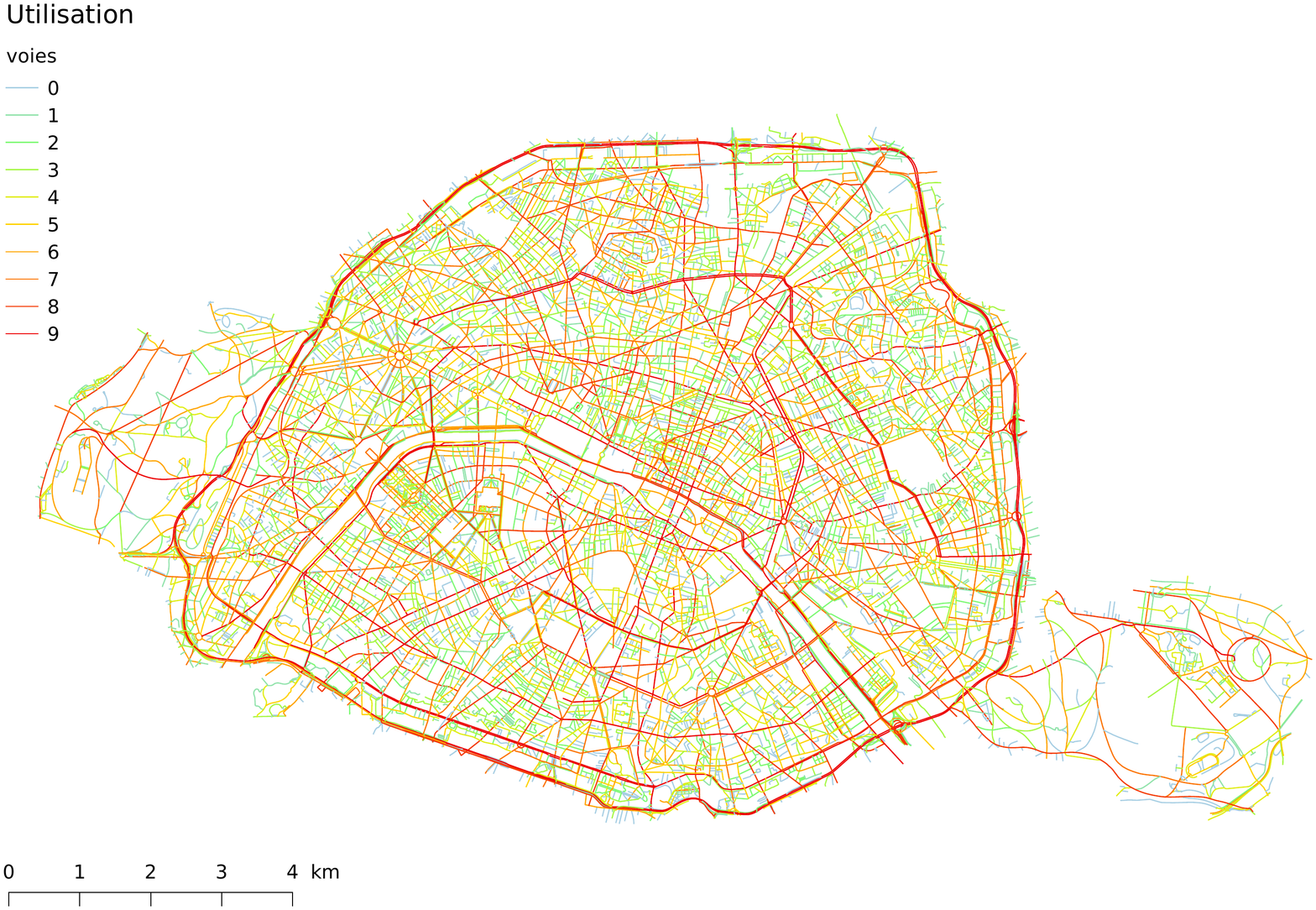}
    \caption{Indicateur d'utilisation calculé sur les voies du graphe viaire de Paris.}
    \label{fig:voies_use}
\end{figure}

Si l'on compare les résultats donnés par les indicateurs de betweenness et d'utilisation sur différents réseaux, nous observons une forte corrélation (que ce soit pour les voies ou les arcs, cf chapitre suivant). L'indicateur d'utilisation révèle des structures complexes sur le graphe des arcs comme sur celui des voies (figures \ref{fig:arcs_use}, \ref{fig:voies_use}). C'est un indicateur global capable de révéler une hiérarchie cohérente entre les différents objets. Pour le graphe des arcs, nous parvenons à lire des axes traversants à partir de cet objet local. C'est donc une caractérisation forte pour un graphe spatial. Sur le graphe des voies nous retrouvons les mêmes structures que celles mises en évidence par l'indicateur de degré et de longueur. Cette propriété de lecture de l'espace équivalente à partir d'une caractérisation locale à celle d'une caractérisation globale est due au caractère multi-échelle de la voie. Elle sera développée dans le chapitre suivant.

\FloatBarrier
\subsection{Rayon topologique, Closeness (classiques) et Accessibilité (complémentaire)}

\begin{figure}[h]
    \centering
    \includegraphics[width=\textwidth]{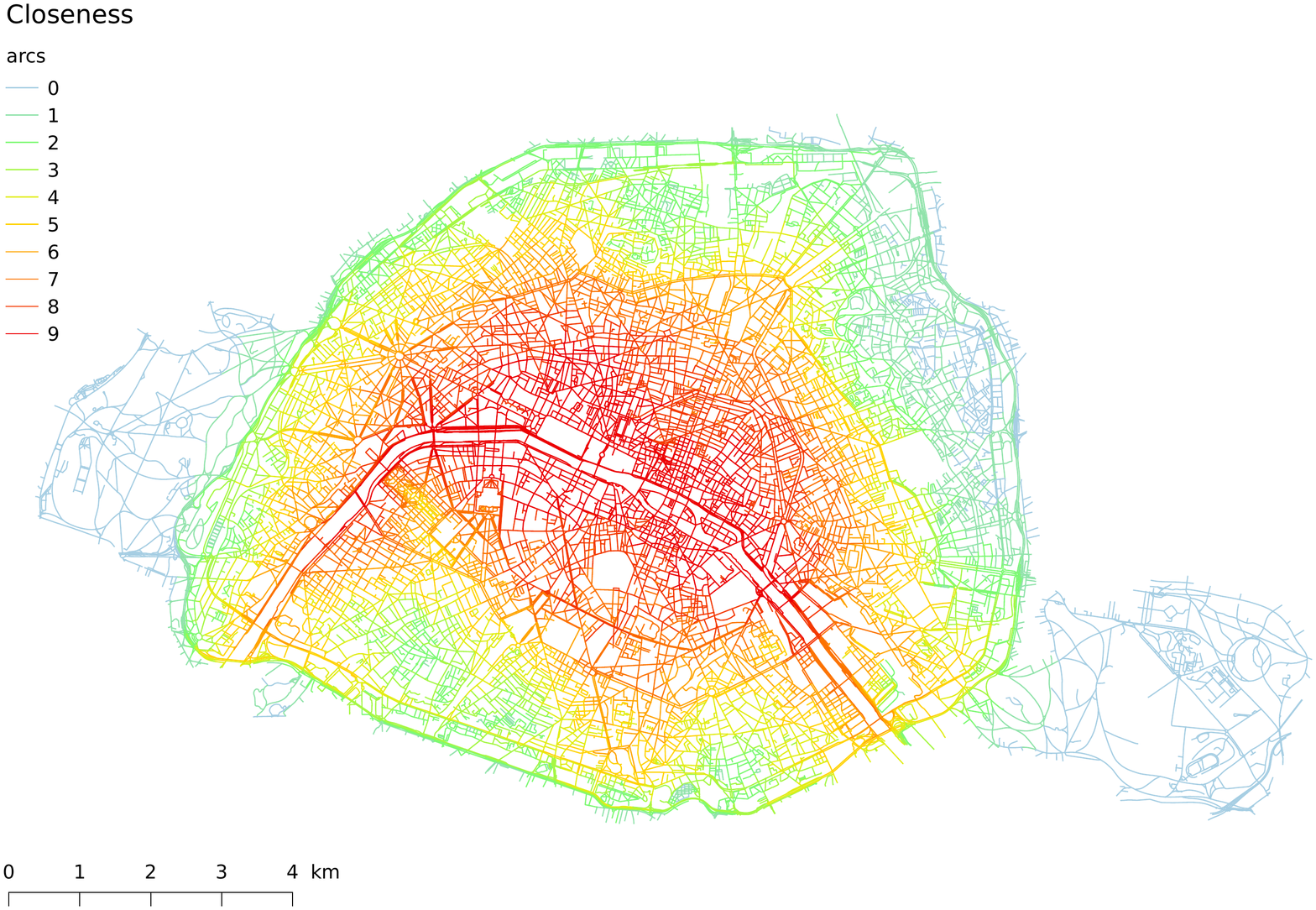}
    \caption{Indicateur de closeness calculé sur les arcs du graphe viaire de Paris.}
    \label{fig:arcs_closeness}
\end{figure}

Les indicateurs de cette sous-section caractérisent la proximité des éléments d'un graphe par rapport à tous les autres. Appliqués aux arcs, ils ont une forte dépendance au découpage du réseau et ne présentent donc aucun intérêt dans la lecture des structures de celui-ci. Ils nous renvoient dans la plupart des cas au centre de l'échantillon découpé (figure \ref{fig:arcs_closeness}). Nous nous penchons sur ces indicateurs essentiellement pour caractériser les voies.

La voie porte déjà dans sa construction l’importance de la continuité et de la linéarité. Pour appuyer le \textit{coût du tournant}, c'est-à-dire appliquer un poids à chaque changement de voie, nous calculons le rayon topologique d'une voie sur l'ensemble du réseau. Il est calculé dans le \textit{line graph} comme étant la somme des distances géodésiques à partir d'un sommet vers tous les autres. Dans le graphe primal cela correspond à la somme pour chaque voie, placée successivement en tant que voie de référence $v_{ref}$, du nombre de \textit{tournants} (distances topologiques les plus simples) depuis $v_{ref}$ vers chacune des autres voies du réseau $d_{simple}(v,v_{ref})$ (équation \ref{eq:rtopo}).

\begin{equation}
rtopo(v_{ref})=\sum_{v \in G} d_{simple}(v,v_{ref})
\label{eq:rtopo}
\end{equation}

En théorie des graphes, un indicateur célèbre de centralité est la \emph{closeness} qui est définie comme étant l'inverse du rayon topologique de l'objet \citep{bavelas1950communication, sabidussi1966centrality} (équation \ref{eq:closeness}). Selon les études, il est possible de le normaliser par le nombre de sommets $S$ du graphe sur lequel il est calculé (dans notre cas, le \textit{line graph} des voies). Dans notre étude, nous considérerons sa version non normalisée afin de rendre possible sa comparaison dans différents sous-échantillons d'un même graphe (cf partie 2). Les voies ayant l'indicateur de closeness le plus élevé seront les plus proches topologiquement de toutes les autres dans le graphe.

\begin{equation}
closeness(v_{ref})=\frac{1}{\sum_{v \in G} d_{simple}(v,v_{ref})}
\label{eq:closeness}
\end{equation}

Pour appuyer le caractère spatial de ce travail, nous combinons ce rayon topologique à la longueur des voies. Nous créons ainsi l’indicateur d'\emph{accessibilité} (\emph{accessibility}), qui représente la distance d’une voie par rapport à l’ensemble du réseau. Pour chaque voie, il est le fruit du calcul des chemins les plus simples à partir de celle-ci pour accéder à l'ensemble du réseau ($d_{simple}(v,v_{ref})$), pondérés par la longueur des voies. T. Courtat faisait référence à cet indicateur en le nommant \textit{centralité} dans \citep{courtat2011mathematics}, nous avons également travaillé dessus en le nommant \textit{structuralité} dans \citep{lagesse2015spatial}. Cependant, cet indicateur qualifie la vision de l’ensemble du réseau à partir de chaque voie. À partir de chaque point du réseau, il définit en combien de changements de voie et avec quelle distance géographique on peut accéder à toute autre partie du graphe. Nous lui préférons donc finalement la dénomination d'\textit{accessibilité}, puisqu'il en est la représentation pour chaque objet du graphe.

L'accessibilité est donnée pour chaque voie - dans le graphe $G(S,V)$ -, successivement placée en tant que voie de référence $v_{ref}$, par la somme du produit de la distance topologique $d_{simple}(v,v_{ref})$ de chaque autre voie par rapport à $v_{ref}$ et de sa longueur $longueur(v)$ (équation \ref{eq:access}).

\begin{equation}
accessibilite(v_{ref})=\sum_{v \in G} [d_{simple}(v,v_{ref}) \times longueur(v)]
\label{eq:access}
\end{equation}

\begin{figure}[h]
    \centering
    \includegraphics[width=\textwidth]{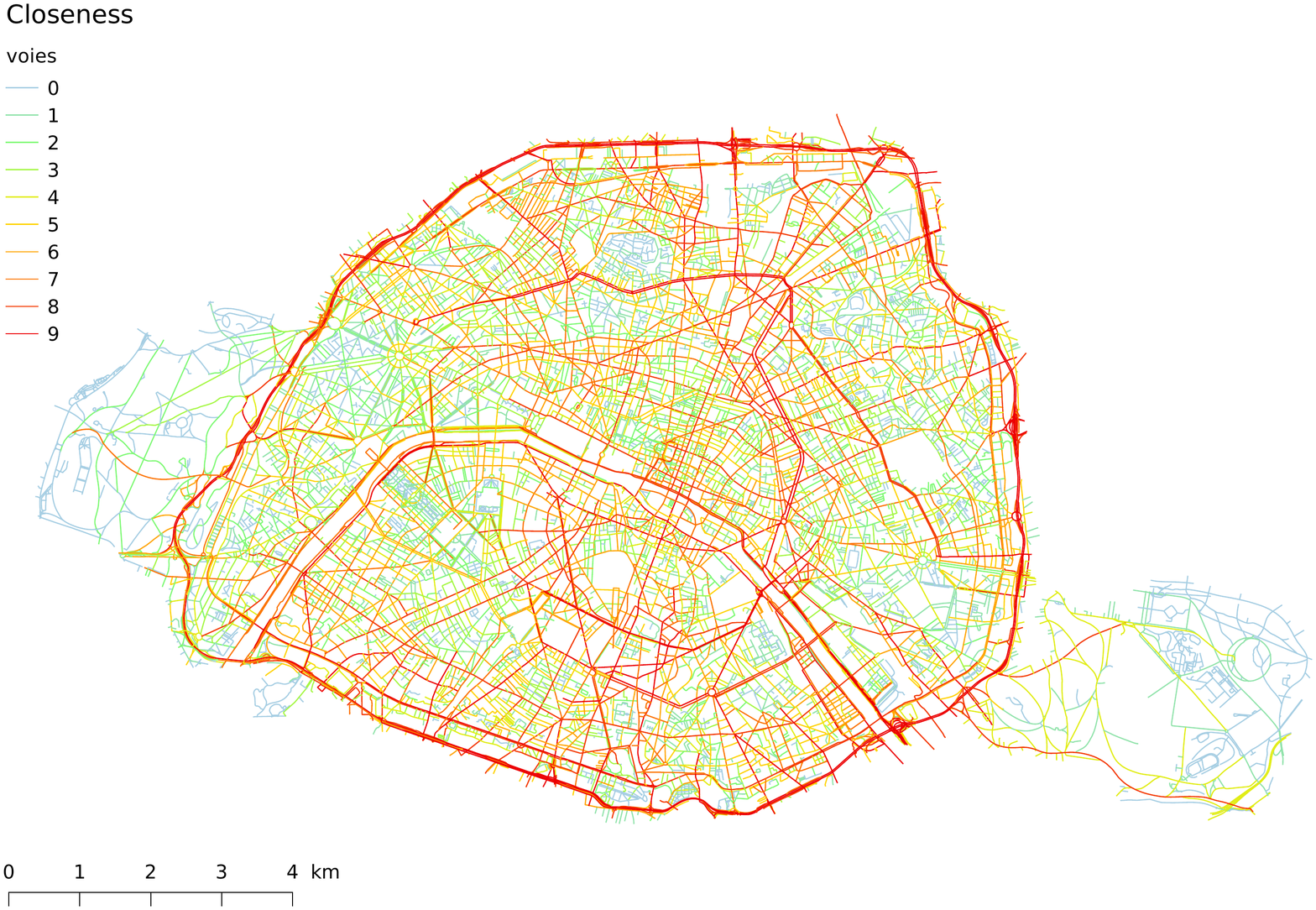}
    \caption{Indicateur de closeness calculé sur les voies du graphe viaire de Paris.}
    \label{fig:voies_closeness}
\end{figure}

En hiérarchisant les voies, cet indicateur fait apparaître celles depuis lesquelles il est le plus rapide (en terme de nombre de changements de voies) d'accéder à l'ensemble du réseau (figure \ref{fig:voies_closeness}). Et plus spécifiquement celles qui permettent d'accéder en un nombre minimum de tournants aux plus longues voies (celles dont le poids est le plus important). Les voies dont l’indicateur d'accessibilité est le plus faible minimisent le nombre de changements de voies, notamment vers les grandes structures, et sont ainsi les plus centrales au sens de cet indicateur. Dans la pratique, si nous comparons le rayon topologique des voies avec leur accessibilité, nous verrons que l'ajout de la longueur dans le calcul n'est pas significatif. Nous développerons le lien entre accessibilité et closeness dans le chapitre suivant.

En confrontant cette analyse mathématique au savoir d'urbanistes et anthropologues, nous observons que cet indicateur fait ressortir les objets au cœur des déplacements dans la ville \citep{degouysAPitineraire}, qui correspondent souvent au squelette de sa croissance. Nous développerons cet aspect dans la dernière partie de cette thèse.

\FloatBarrier
\subsection{Accessibilité maillée et Structuralité potentielle (complémentaires)}

Il est possible de combiner les indicateurs précédents, qu'ils soient calculés localement ou en tenant compte de tout le réseau. Le degré de desserte et l'espacement en sont déjà des exemples : ils combinent deux indicateurs locaux (degré et connectivité pour le degré de desserte ; longueur et connectivité pour l'espacement).

Nous nous penchons ici, dans un premier temps, sur la combinaison de la closeness avec l'orthogonalité. Le premier indicateur est calculé à partir d'une voie en prenant en compte tout le réseau. Le second est calculé pour chaque voie, à partir de son voisinage direct. Plus sa valeur de closeness est grande, plus la voie permet d'accéder rapidement à l'ensemble du réseau. Plus sa valeur d'orthogonalité est grande, plus la voie est incluse dans le réseau de manière \enquote{maillée} (coupant orthogonalement les autres voies).

En faisant le produit de ces indicateurs nous créons un nouvel indicateur qui met en valeur les voies \enquote{centrales} dans le réseau (au sens de la closeness, elles permettent un accès rapide à tout le réseau) et qui sont maillées avec le reste du réseau (orthogonalité proche de 1). Nous l’appellerons \emph{accessibilité maillée} (\emph{meshed accessibility}) (équation \ref{eq:accessmail}). Sa valeur est faible pour les voies dont la closeness est importante mais aux connexions \enquote{lisses} avec le réseau (voies rapides dans les réseaux viaires pour lesquelles l'orthogonalité est proche de 0). Les voies peu accessibles et dans des structures maillées auront une valeur intermédiaire forte dans la représentation. Celles peu accessibles et connectées avec des angles faibles seront les moins mises en avant dans la hiérarchisation (elles correspondent dans un réseau viaire le plus souvent à des voies de nouveaux quartiers résidentiels).

\begin{equation}
accessibiliteMaillee(v_{ref})= closeness(v_{ref}) \times orthogonalite(v_{ref})
\label{eq:accessmail}
\end{equation}

\begin{figure}[h]
    \centering
    \includegraphics[width=\textwidth]{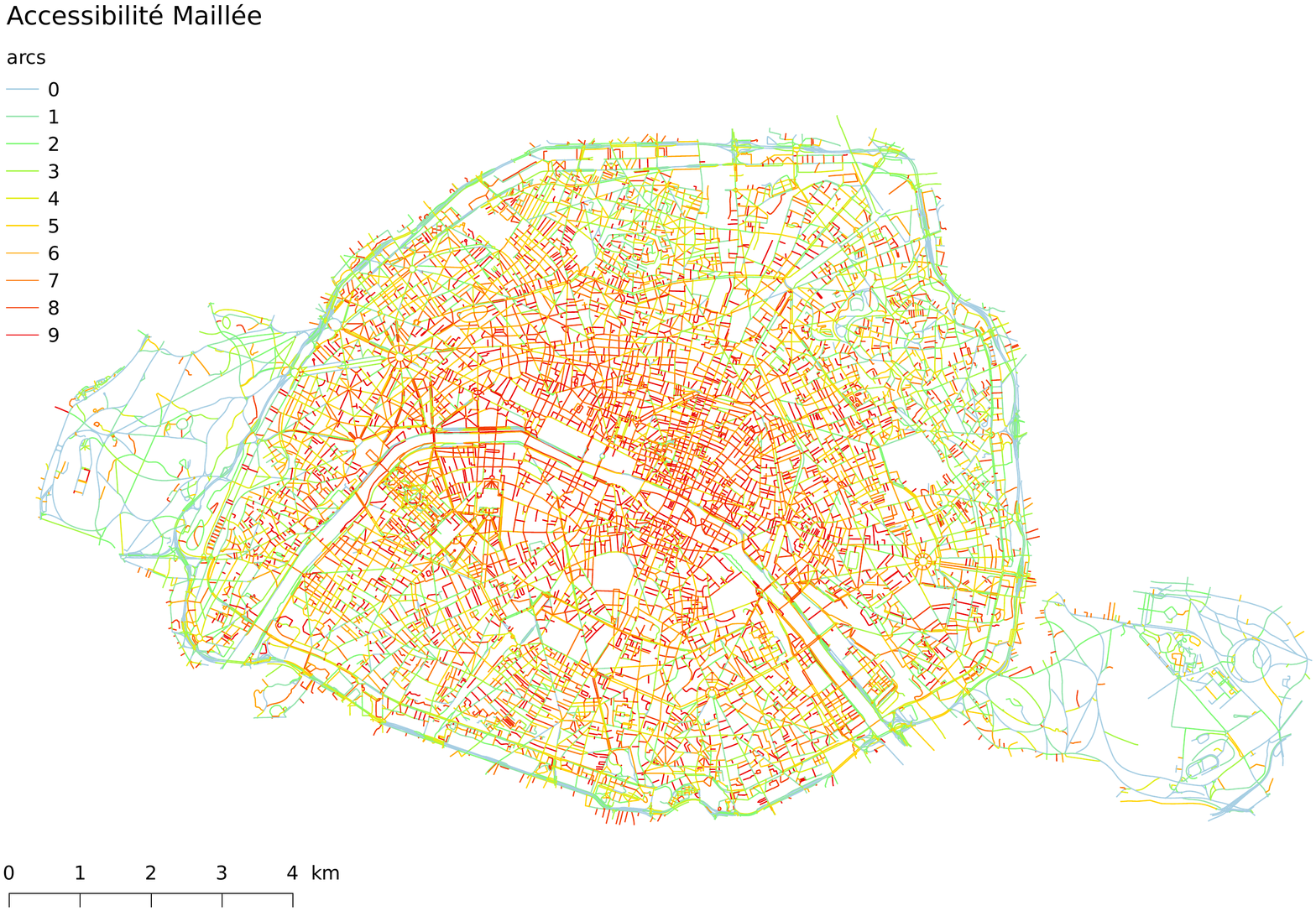}
    \caption{Indicateur d'accessibilité maillée calculé sur les arcs du graphe viaire de Paris.}
    \label{fig:arcs_roo}
\end{figure}

\begin{figure}[h]
    \centering
    \includegraphics[width=\textwidth]{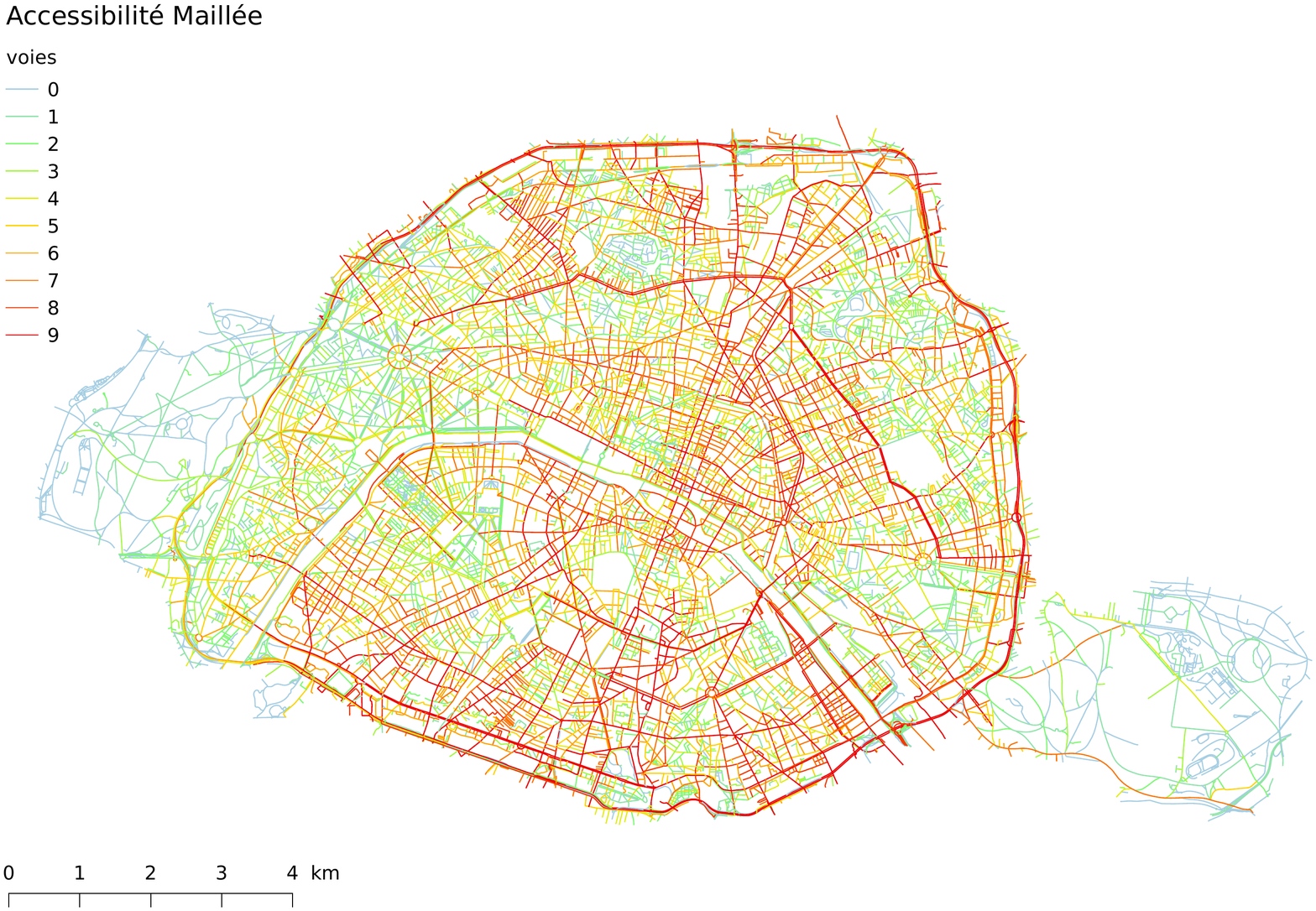}
    \caption{Indicateur d'accessibilité maillée calculé sur les voies du graphe viaire de Paris.}
    \label{fig:voies_roo}
\end{figure}

Nous appliquons cet indicateur sur les réseaux viaires qui nous servent de terrains dans cette partie (figure \ref{fig:voies_roo}). Pour Paris, les structures maillées très accessibles ressortent circulairement autour d'un centre ancien qui est également maillé. Les résultats sur les arcs ne révèlent aucune structure cohérente et ne seront donc pas retenus comme pertinents (figure \ref{fig:arcs_roo}). Nous reviendrons sur le cas d'Avignon, dont les cartes figurent en annexe (\ref{ann:chap_indicateurs}), ainsi que sur celui d'autres villes, lues à travers cet indicateur, en troisième partie.

\FloatBarrier

Nous pouvons penser à d'autres types de combinaisons ou de déclinaisons des indicateurs présentés ici. Pour caractériser la voie, l'indicateur d'accessibilité conjuguant topologie et distance métrique pourrait être renforcé. Appliqué à chaque objet, nous observons la façon dont il s'intègre dans le réseau à partir de l'accessibilité de son voisinage. Nous construisons ainsi ce que nous appelons la \emph{structuralité potentille}. Pour le calculer nous sommons pour un objet la valeur d'accessibilité de tous ceux qui lui sont connectés. Pour les voies, nous obtenons l'équation \ref{eq:potstruct}.

\begin{equation}
structuralitePotentielle(v_{ref})=\sum_{v \in G / v \cap v_{ref}} accessibilite(v)
\label{eq:potstruct}
\end{equation}

\begin{figure}[h]
    \centering
    \includegraphics[width=\textwidth]{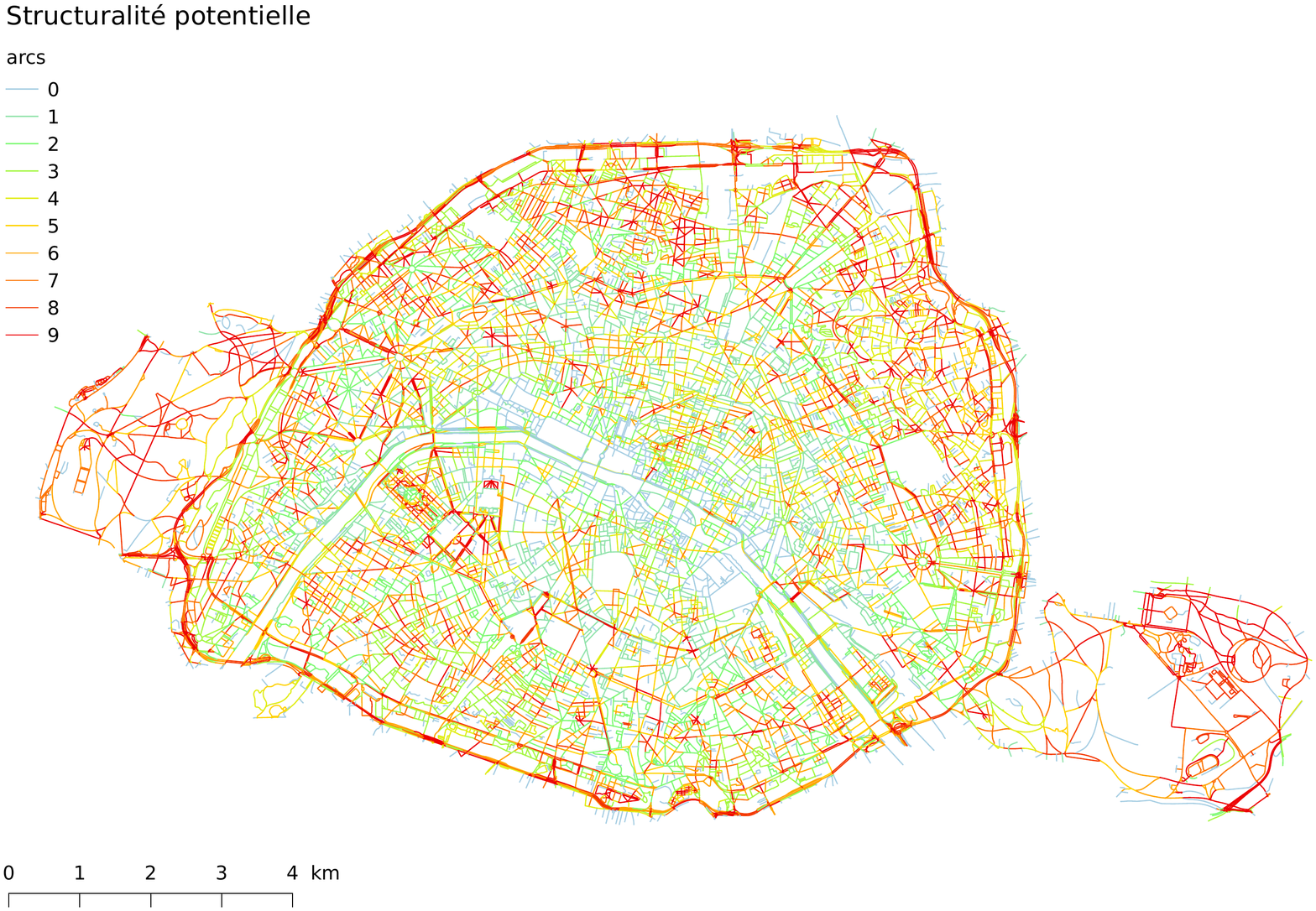}
    \caption{Indicateur de structuralité potentielle calculé sur les arcs du graphe viaire de Paris.}
    \label{fig:arcs_structpot}
\end{figure}

\begin{figure}[h]
    \centering
    \includegraphics[width=\textwidth]{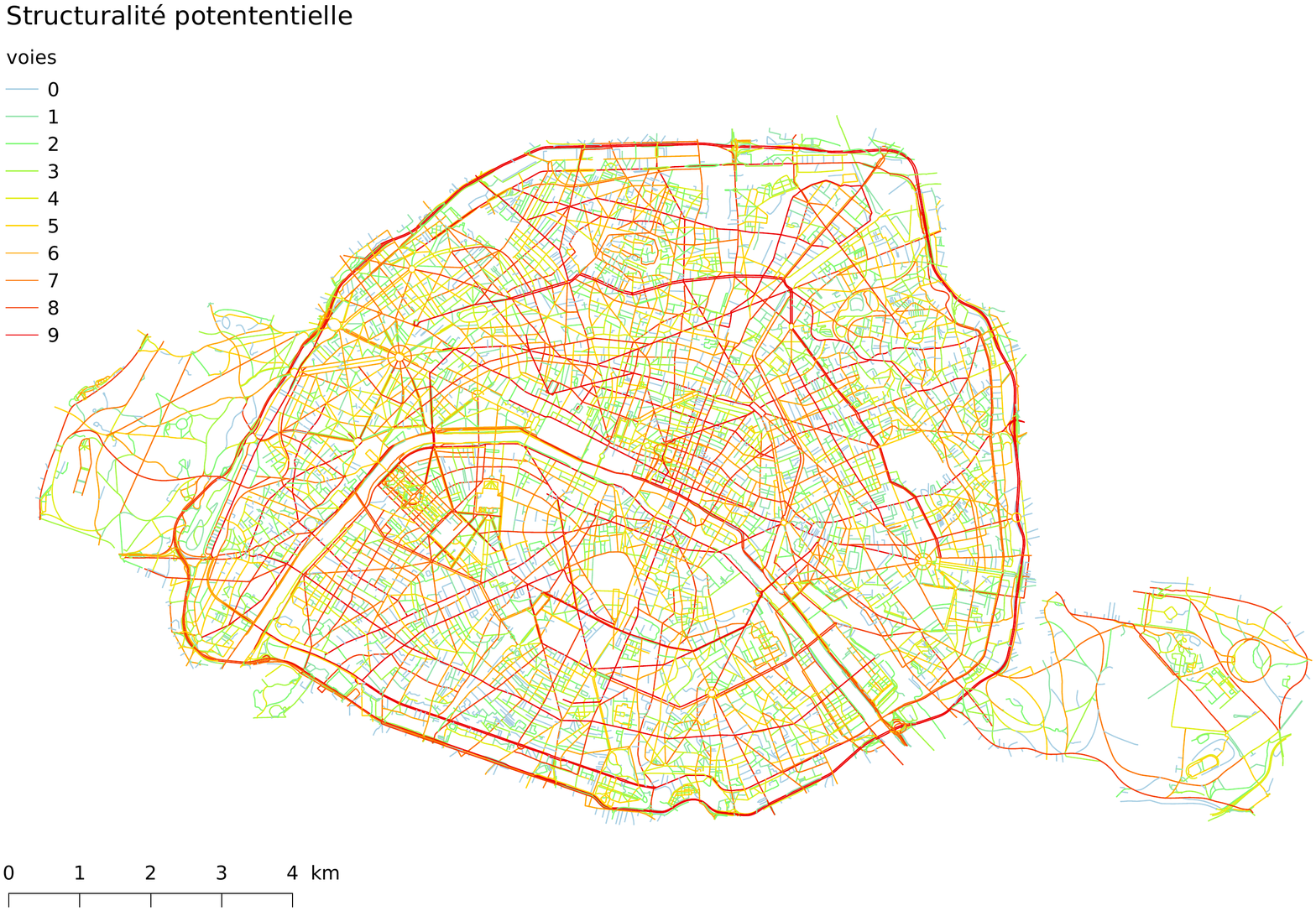}
    \caption{Indicateur de structuralité potentielle calculé sur les voies du graphe viaire de Paris.}
    \label{fig:voies_structpot}
\end{figure}

Cet indicateur nous donne une carte où la valeur d'accessibilité se diffuse moins sur les voies voisines de voies très accessibles. Elle est plus caractéristique de la voie elle-même. Dans la pratique, cet indicateur ne nous apporte que très peu d'informations supplémentaires par rapport au degré de la voie (figure \ref{fig:voies_structpot}). Sommer les accessibilités des voies connectées à une voie de référence revient, dans la hiérarchisation des objets, à considérer uniquement le nombre de connexions de la voie de référence. Cet indicateur nous offre un nouvel exemple d'une information calculée sur tout le réseau ramenée à une information locale par la voie. En effet, le phénomène est caractéristique de cet objet : sur le graphe des arcs, la caractérisation fait ressortir en grande partie les objets qui avaient un faible coefficient de closeness (et donc un fort indicateur d'accessibilité, puisque ces deux indicateurs sont anti-corrélés). La caractérisation est donc fortement dépendante de l'effet du global sur le local, et par conséquent aux effets de bord. Elle ne fait pas ressortir de structures pertinentes (figure \ref{fig:arcs_structpot}).

Il ne nous est pas possible de tester l'ensemble des combinaisons (linéraires ou non-linéraires) de l'ensemble des indicateurs présentés ici. Nous en testerons néanmoins une partie dans le chapitre suivant. Nous montrerons ainsi qu'il n'est pas possible, dans le cadre de l'étude, de créer des combinaisons à l'infini apportant de nouvelles informations pertinentes. En effet, notre travail montre que seuls certains indicateurs se distinguent comme étant révélateurs d'informations structurelles et que les autres leur sont directement corrélés. Nous montrerons également que leur nombre et leur nature varient suivant leurs objets d'application : les arcs et les voies.
\clearpage{\pagestyle{empty}\cleardoublepage}
\vspace{15cm}
\chapter{Grammaire de lecture de la spatialité}
\minitoc
\markright{Grammaire de lecture de la spatialité}

Nous avons exploré dans la partie précédente un certain nombre d'indicateurs qui nous permettent d'observer les structures spatiales. Il est possible de définir un grand nombre de nouveaux indicateurs en associant les informations topologiques et géométriques ou en combinant les indicateurs existants. L'originalité des résultats de ces nouvelles combinaisons pose alors question. Nous montrerons ici que les informations structurelles que l'on peut mettre en évidence ne sont pas infinies. Nous reviendrons sur les indicateurs décrits précédemment pour en étudier les corrélations afin de déceler les caractéristiques dominantes sur un graphe spatial. Nous déterminerons ainsi dans ce chapitre quels sont les critères forts de la caractérisation spatiale : la métrique, la topologie et/ou une combinaison particulière.

Des comparaisons entre indicateurs ont déjà été effectuées sur des graphes spatialisés. Ainsi, P. Crucitti, V. Latora et S. Porta ont comparé plusieurs indicateurs de centralité : la \textit{degree centrality}, la \textit{closeness centrality}, la \textit{betweenness centrality}, la \textit{straightness centrality} et l'\textit{information centrality} \citep{crucitti2006centralitymeasures}. Ils procèdent pour cela en quatre étapes. Ils fondent leur comparaison sur un réseau viaire, tels que ceux que nous utilisons ici. Ils construisent, comme nous le faisons, son graphe primal : les intersections sont des nœuds et les arcs sont les portions de rue entre chaque intersection. Le graphe, planaire, est non orienté et pondéré par la longueur métrique des arcs. Ils calculent leurs indicateurs sur ce graphe pour caractériser les intersections (nœuds du graphe). Nous pouvons également citer M. Barthelemy qui compare la distance la plus courte et la plus simple respectivement sur les arcs et les \textit{lignes droites} (construite avec la méthode \textit{ICN} évoquée précédemment) d'un graphe spatialisé \citep{barthelemy2013self}. Il s'intéresse également à la caractérisation de la centralité des intersections pour évaluer les changements opérés par Haussmann sur Paris \citep{barthelemy2013self} mais en se concentrant sur l'indicateur de betweenness.

Ici nous nous intéressons aux arcs et aux voies tels que nous les avons définis plus haut. Nous ne cherchons pas à caractériser les intersections, qui sont les points d'articulation du linéaire. Nous reviendrons sur les raisons de ce choix dans la troisième partie. Notre objectif est de définir une liste d'indicateurs pertinents pour les arcs d'une part et pour les voies d'autre part.

\FloatBarrier
\section{Méthodologie de comparaison}

Nous manipulons des indicateurs de différents types, combinant topologie et topographie. Il est donc impossible de comparer les variations de leurs valeurs brutes car elles ne sont pas toutes normalisées. Nous avons fait ce choix afin de permettre leur comparaison entre plusieurs échantillons de tailles différentes (voir partie 2).

Nous travaillons ici sur quatre échantillons viaires qui couvrent un large spectre d'organisations et d'histoires de construction. De la ville la plus \textit{organique} à la plus \textit{planifiée} nous prendrons en exemple un graphe viaire découpé autour d'Avignon (commune du Sud de la France), le réseau de la ville de Paris (qui a subi quelques rectifications par le Baron Haussmann au XIXème siècle), celui de la ville de Barcelone (dont une partie a été entièrement redessinée par Cerdà) et enfin, un des réseaux historiquement planifié selon un quadrillage précis : Manhattan. Nous illustrerons nos propos avec les résultats de calculs sur ces réseaux. Le graphe d'Avignon est celui dont la longueur totale est la plus faible, tandis que celui de Manhattan possède le plus petit nombre de voies (tableau \ref{tab:pres_ville2}). Pour le travail de corrélation sur les arcs nous utiliserons ces deux graphes aux caractéristiques de construction différentes. Sur les voies, nous y ajouterons ceux de Paris et de Barcelone, aux tailles comparables plus importantes que les deux premiers (tableau \ref{tab:pres_ville2}).

\begin{table}[h]

\begin{tabular}{| c | r | r | r | r | r |}
\hline
& $L_{tot}$ (en mètres) & $N_{arcs}$ & $N_{classes}(arcs)$ & $N_{voies}$ & $N_{classes}(voies)$ \\
\hline
Manhattan & 3 379 135 & 10 152 & 2 500 & 1 030 & 250 \\
\hline
Avignon & 949 413 & 13 221 & 2 500 & 4 045 & 1 000 \\
\hline
Paris & 2 112 715 & 30 957 & / & 6 893 & 1 500 \\
\hline
Barcelone & 2 066 335 & 30 003 & / & 6 028 & 1 500 \\
\hline
\end{tabular}
\caption{Tableau descriptif des caractéristiques et classifications utilisées pour les quatre villes étudiées. Paris et Barcelone ne participent qu'à l'étude de corrélation faite sur les voies.}
\label{tab:pres_ville2}
\end{table}

Afin de comparer les indicateurs calculés sur ces réseaux nous utiliserons une classification par longueurs. Cela nous permet de rendre homogènes les résultats (les valeurs brutes dans nos calculs n'étant pas toujours normalisées), et de nous affranchir des effets de non-linéarité des différentes valeurs considérées. Nous définissons le nombre de classes ($N_{classes}$) en fonction de la longueur totale ($L_{tot}$) du réseau étudié pour que le nombre moyen d'éléments par classe soit autour de quatre (cf tableau \ref{tab:pres_ville2}). Chaque classe est remplie selon l'ordre ascendant de l'indicateur classifié jusqu'à avoir atteint $L_{seuil} = \frac{L_{tot}}{N_{classes}}$. Le passage à la classe suivante est fait, si $L_{seuil}$ est dépassé, après l'ajout du dernier objet. La longueur cumulée de chaque classe sera donc légèrement supérieure à la valeur moyenne $L_{seuil}$ ce qui se traduira \textit{in fine} par un nombre de classes légèrement inférieur à celui défini avec $N_{classes}$. C'est pour cela que nous choisissons $N_{classes} \sim \frac{L_{tot}}{4}$ : cela nous permet de conserver un grand nombre de classes, en minimisant le nombre de classes laissées vides par l'effet du dépassement de seuil.

Cette classification nous permet d'effectuer par la suite un calcul de corrélation sur les valeurs de classe pour chaque objet avec la méthode de Pearson \citep{lee1988thirteen}. Celle-ci considère le produit des écarts à la moyenne de chaque valeur $X$ et $Y$, normalisé par le produit des écarts types. Ce calcul nous permet de déterminer si deux indicateurs apportent une information différente sur le graphe ($\mathrm{corr}(X,Y)$ proche de $0$) ou bien si leurs valeurs subissent les mêmes variations ($\mathrm{corr}(X,Y)$ proche de 1, dans le cas d'une corrélation, ou de -1 , dans celui d'une anti-corrélation).

\begin{equation}
\mathrm{corr}(X,Y)=\frac{1}{N} \sum (\frac{X-\bar{X}}{\sigma_X})(\frac{Y-\bar{Y}}{\sigma_Y})
\end{equation}

Nous considérons comme indicateurs \emph{primaires} les indicateurs suivants :
\begin{itemize}
\item la longueur
\item le nombre d'arcs (nba - uniquement pour les voies, égal à 1 pour les arcs)
\item le degré
\item la connectivité (nbc - uniquement pour les voies, égale au degré pour les arcs)
\item la structuralité potentielle
\item la betweenness
\item l'utilisation
\item la closeness
\item l'accessibilité
\item l'orthogonalité
\end{itemize}

Après avoir défini quels sont les indicateurs pertinents nous les combinerons entre eux afin de trouver quelles sont les combinaisons qui apportent une information supplémentaire. Nous retrouverons ainsi des \emph{indicateurs composés} tels que l'\emph{espacement} ou l'\emph{accessibilité maillée}.

\FloatBarrier

\section{Analyse et comparaison des indicateurs sur les arcs}

Nous nous positionnons tout d'abord dans le graphe primal des arcs. Unités élémentaires de référence, nous voulons observer l'originalité des résultats des indicateurs calculés sur ceux-ci. Nous chercherons également les combinaisons pertinentes utiles à ajouter aux premières caractérisations.

\subsection{Indicateurs primaires}

\begin{figure}[h]
    \centering

    \begin{subfigure}{.35\textwidth}
        \includegraphics[width=\textwidth]{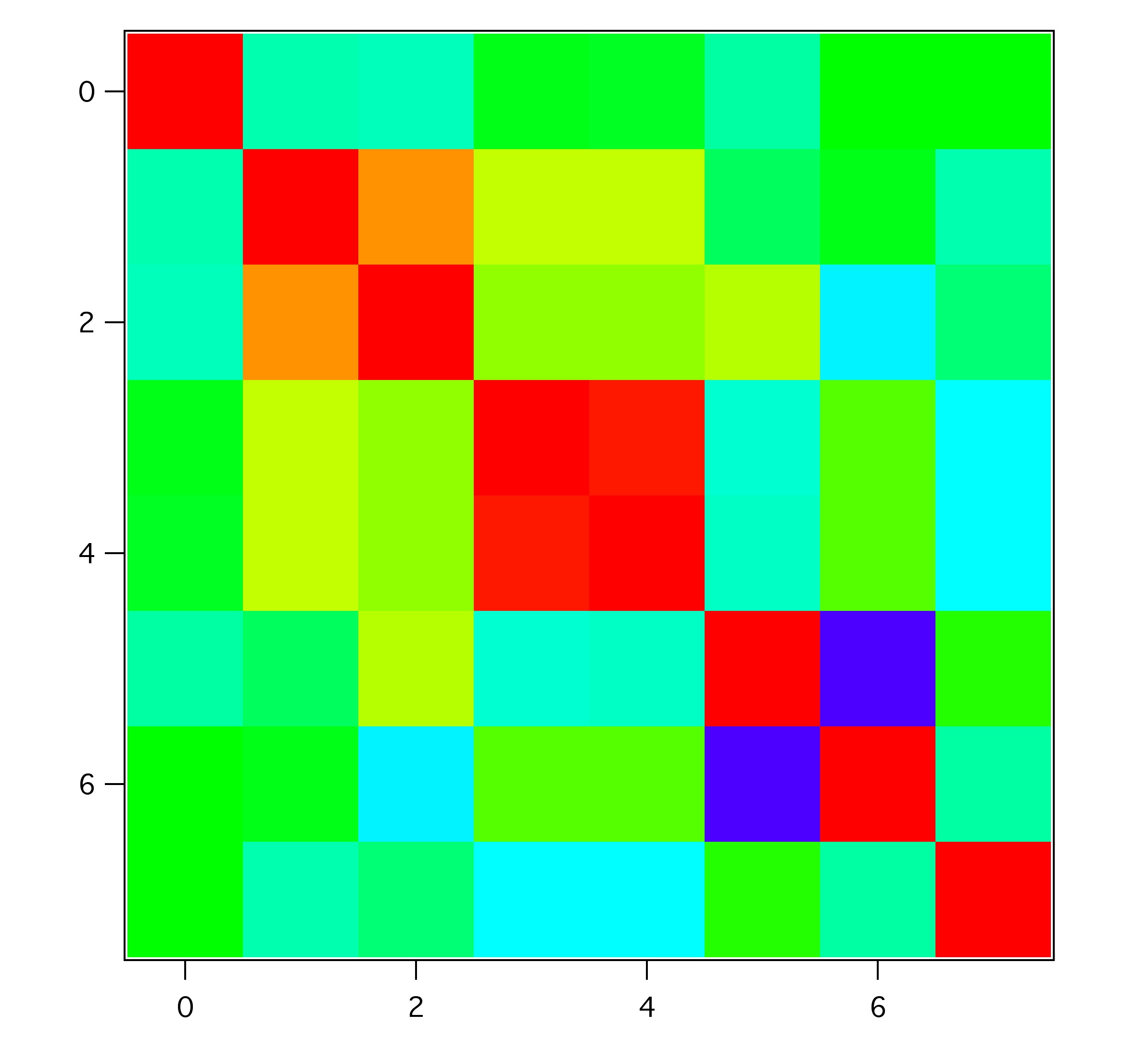}
        \caption{Représentation du tableau \ref{tab:corr_avignon_arc}, calculé sur les arcs d'Avignon.}
    \end{subfigure}
    ~
    \begin{subfigure}{.35\linewidth}
        \includegraphics[width=\textwidth]{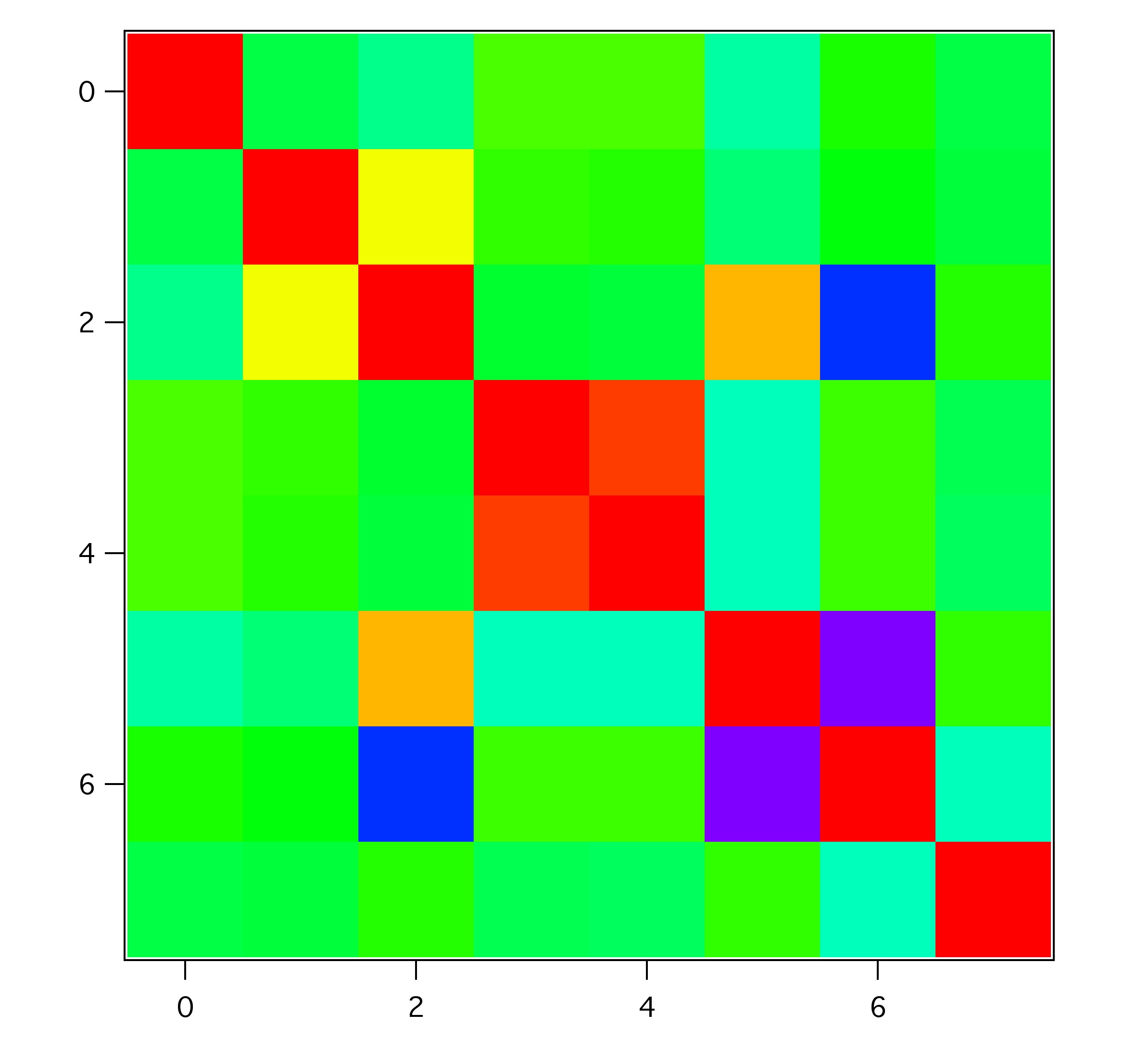}
        \caption{Représentation du tableau \ref{tab:corr_manhattan_arc}, calculé sur les arcs de Manhattan.}
    \end{subfigure}
   ~
    \begin{subfigure}{.2\linewidth}
        \includegraphics[width=\textwidth]{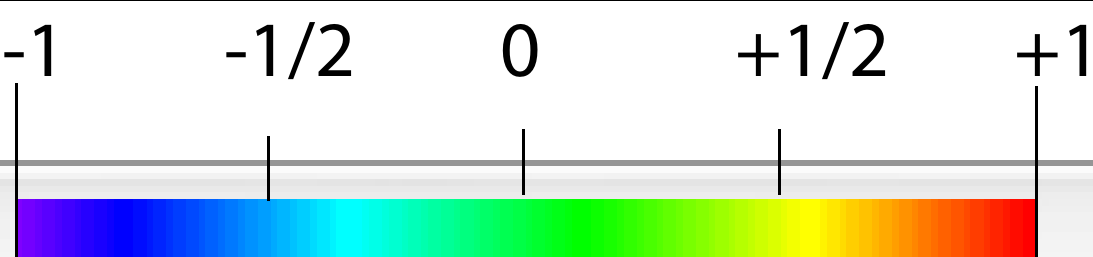}
        \caption{Échelle}
    \end{subfigure}

    \caption{Représentation de la corrélation entre indicateurs primaires calculés sur les arcs sous forme de matrice colorée. Rappel de l'ordre des indicateurs : 0 : longueur ; 1 : degré ; 2 : structuralité potentielle ; 3 : betweenness ; 4 : utilisation ; 5 : accessibilité ; 6 : closeness ; 7 : orthogonalité}
    \label{fig:mat_prim_arcs}

\end{figure}

Nous calculons sur les échantillons d'Avignon et de Manhattan l'ensemble des indicateurs primaires sur les arcs. Ces deux réseaux sont en effet ceux aux structures les plus éloignées et possèdent des nombres d'arcs proches. Ils nous permettront de déterminer un comportement caractéristique des indicateurs. Le calcul de corrélation donne le résultat suivant (tableaux en annexe \ref{tab:corr_avignon_arc}, \ref{tab:corr_manhattan_arc}), pour plus de lisibilité, nous le représentons également avec des matrices colorées permettant de distinguer les groupes d'indicateurs corrélés (figure \ref{fig:mat_prim_arcs}).

Nous pouvons distinguer sur la figure \ref{fig:mat_prim_arcs} et lire dans les tableaux \ref{tab:corr_avignon_arc} et \ref{tab:corr_manhattan_arc} (en annexe), six groupes d'un ou plusieurs indicateurs non corrélés entre eux, qui se retrouvent sur les deux réseaux :
\begin{enumerate}
\item la longueur
\item le degré
\item la structuralité potentielle
\item la betweenness et l'utilisation
\item la closeness et l'accessibilité
\item l'orthogonalité
\end{enumerate}

\begin{figure}[h]
    \centering

    \begin{subfigure}{.40\textwidth}
        \includegraphics[width=\textwidth]{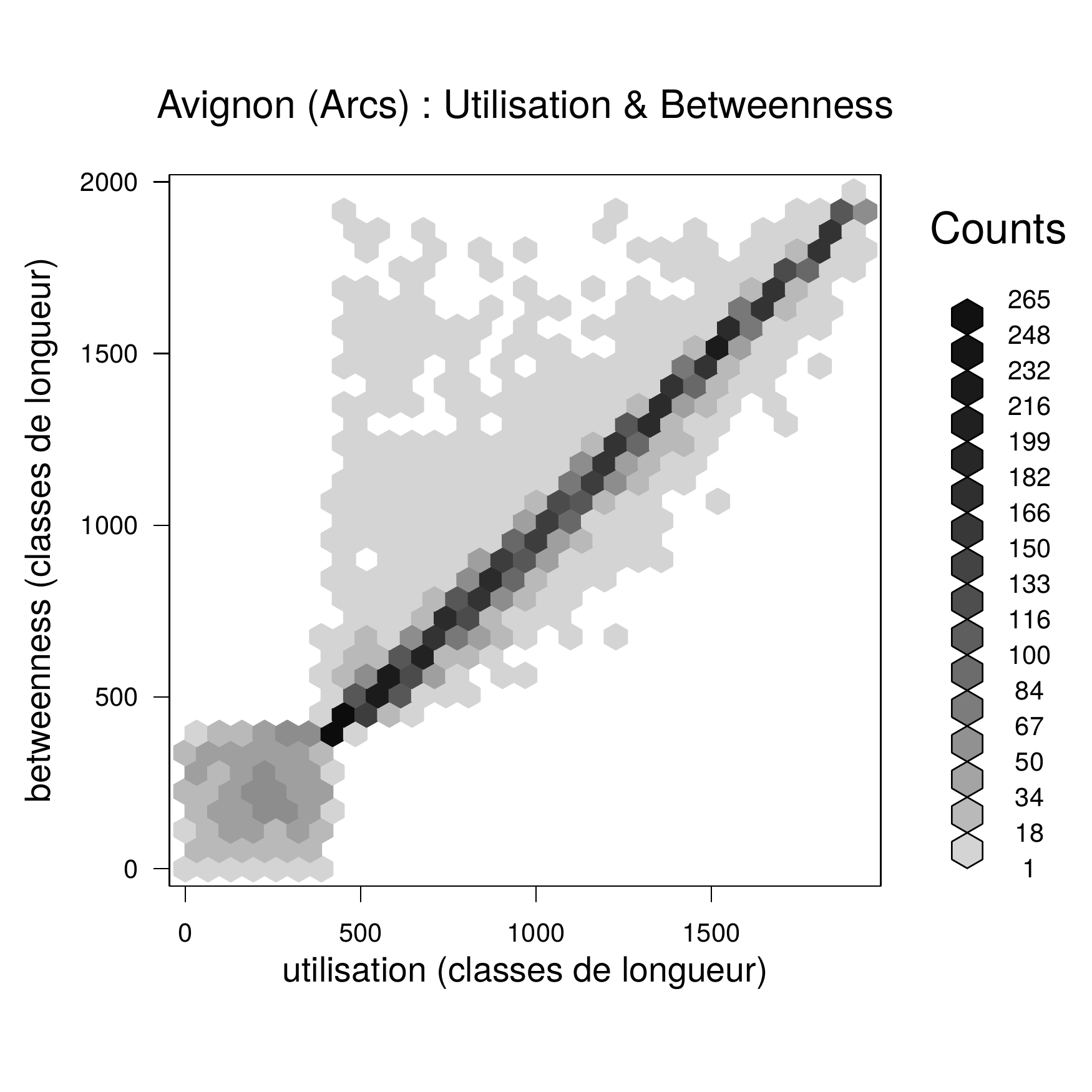}
        \caption{Sur le graphe viaire d'Avignon.}
    \end{subfigure}
    ~
    \begin{subfigure}{.40\textwidth}
        \includegraphics[width=\textwidth]{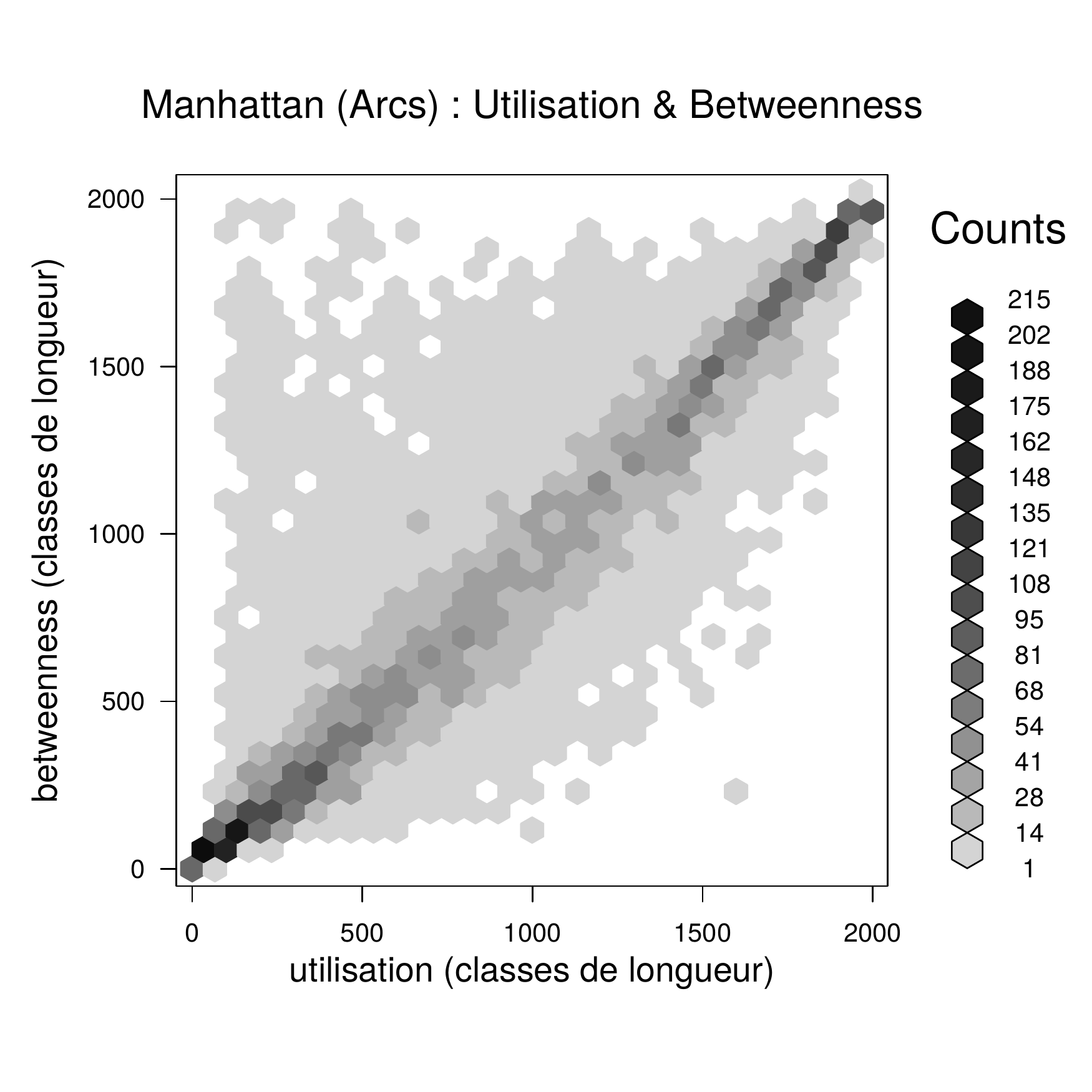}
        \caption{Sur le graphe viaire de Manhattan.}
    \end{subfigure}

    \caption{Cartes de corrélation croisée entre utilisation et betweenness calculés sur les arcs.}
    \label{fig:arcs_usebetw}

\end{figure}

Sur les arcs, tous les indicateurs dont les méthodes de calculs diffèrent donnent un résultat non redondant. Lorsque nous nous penchons sur la corrélation entre betweenness et utilisation, nous observons un pourcentage de plus de 90\%. Leur lien fort montre que considérer uniquement le nombre de chemins les plus simples passant par un arc sans le normaliser par leur nombre total entre chaque paire d'arcs du réseau est suffisante. Les cartes de corrélation ci dessous illustrent la similarité des informations fournies par les deux indicateurs (figure \ref{fig:arcs_usebetw}).

\begin{figure}[h]
    \centering

    \begin{subfigure}{.40\textwidth}
        \includegraphics[width=\textwidth]{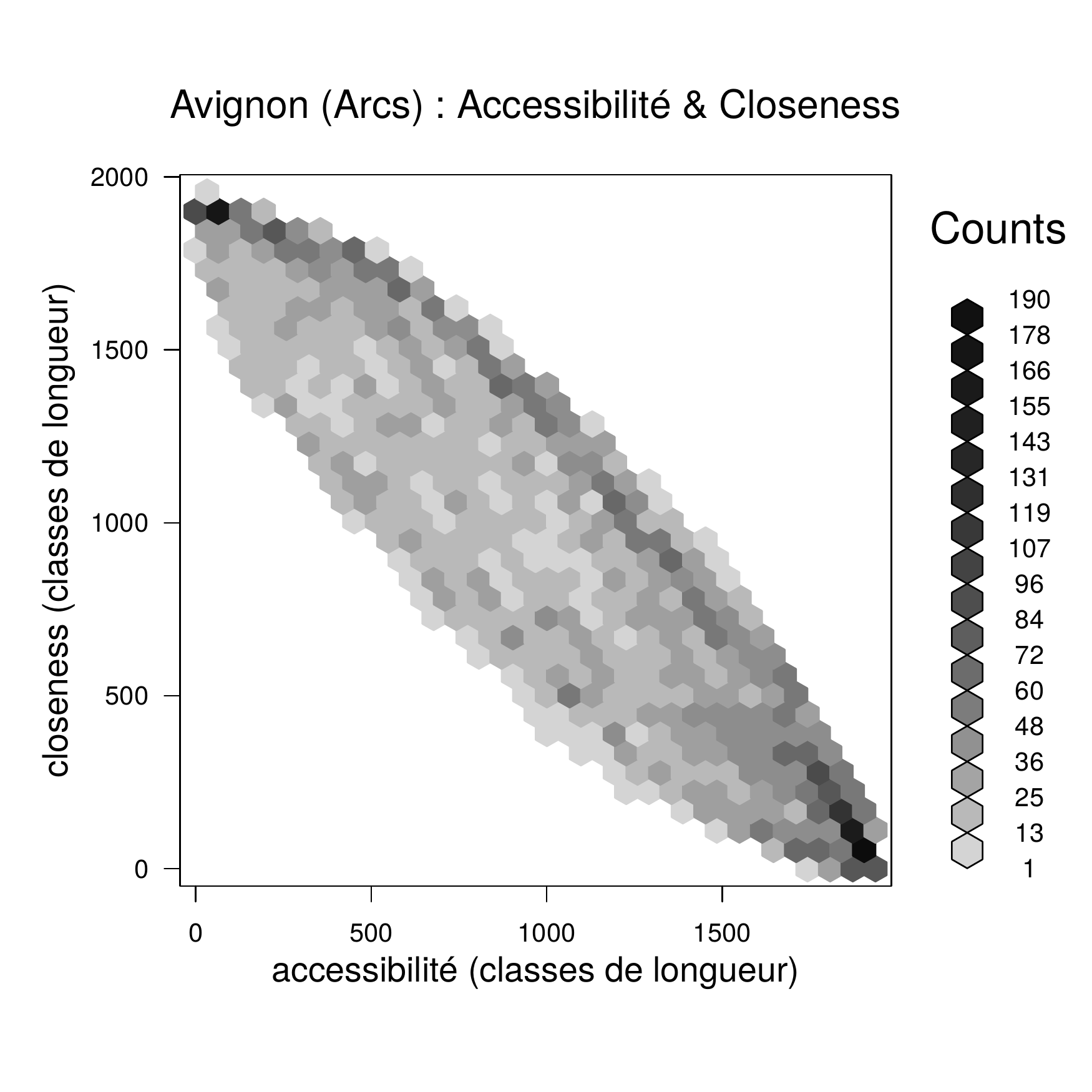}
        \caption{Sur le graphe viaire d'Avignon.}
    \end{subfigure}
    ~
    \begin{subfigure}{.40\textwidth}
        \includegraphics[width=\textwidth]{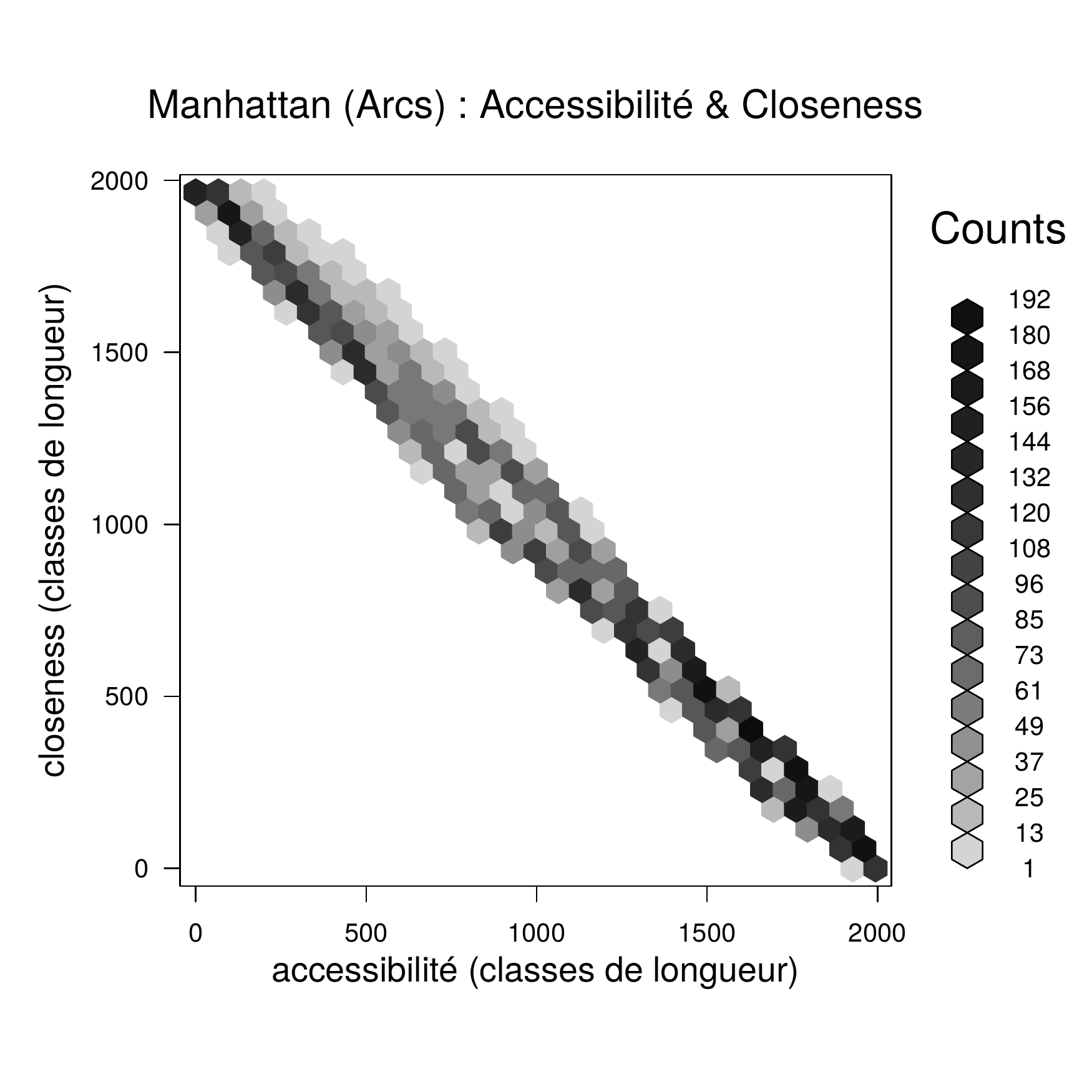}
        \caption{Sur le graphe viaire de Manhattan.}
    \end{subfigure}

    \caption{Cartes de corrélation croisée entre accessibilité et closeness calculés sur les arcs.}
    \label{fig:arcs_accessclo}

\end{figure}

De même entre closeness et accessibilité. Le calcul de l'accessibilité est fait en associant la longueur métrique à la distance topologique. La closeness ne considérant que la distance topologique, nous pouvons en déduire que celle-ci prime sur la métrique qui n'apporte pas un poids suffisant pour rendre les deux indicateurs pertinents indépendamment. On observe entre eux une anti-corrélation due au fait que la closeness est calculée selon l'inverse de la somme des distances topologiques (figure \ref{fig:arcs_accessclo}). Sur le graphe de Manhattan, la structure maillée rend ces deux indicateurs proches de la structuralité potentielle, ce qui n'est pas le cas sur le graphe d'Avignon. En effet, la régularité du graphe de Manhattan rend les distances topologiques à partir d'un arc comparables à celles de ses arcs voisins.

Nous ne retiendrons donc pour ces quatre indicateurs que l'utilisation (plus rapide à calculer que la betweenness) et la closeness (car l'introduction de la longueur des arcs n'apporte aucune information supplémentaire).

\begin{figure}[h]
    \centering
    \begin{subfigure}{\textwidth}
        \includegraphics[width=\textwidth]{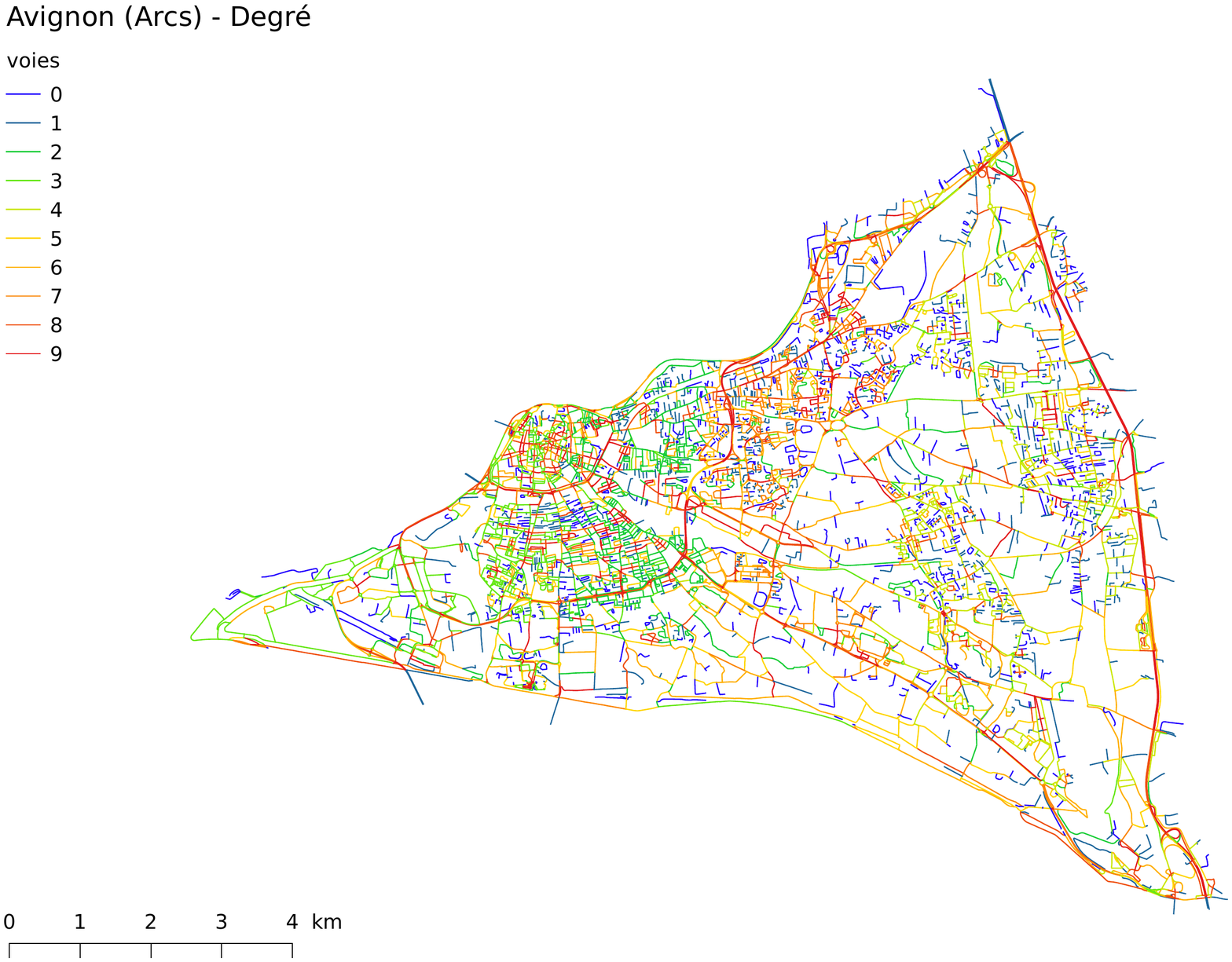}
    \end{subfigure}
    
    \begin{subfigure}{\textwidth}
        \includegraphics[width=\textwidth]{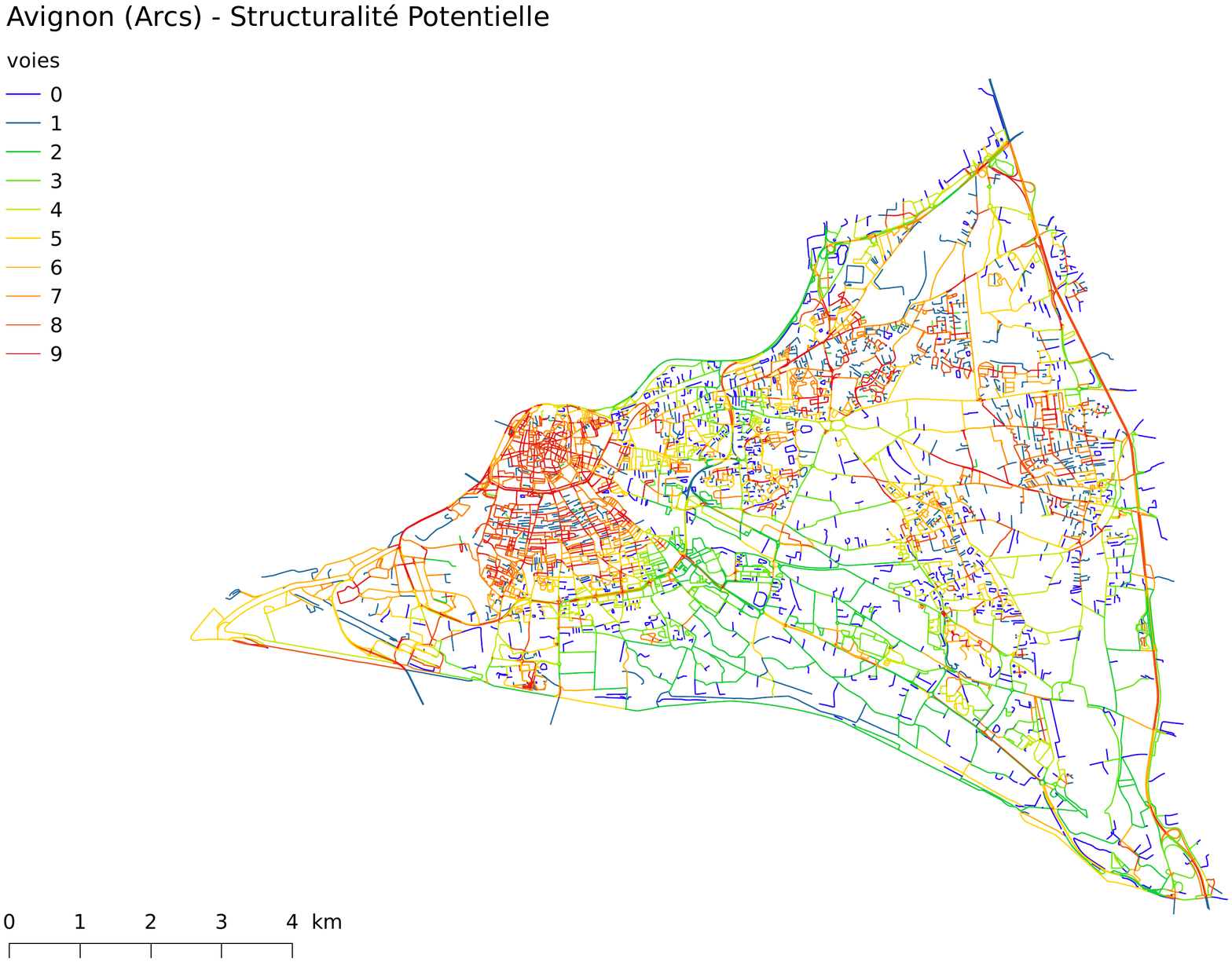}
    \end{subfigure}

    \caption{Comparaison entre degré et structuralité potentielle calculés sur les arcs du graphe viaire d'Avignon.}
    \label{fig:arcs_carte_degrestructpot_av}
\end{figure}

\begin{figure}[h]
    \centering
    
    \begin{subfigure}{\linewidth}
        \includegraphics[width=\textwidth]{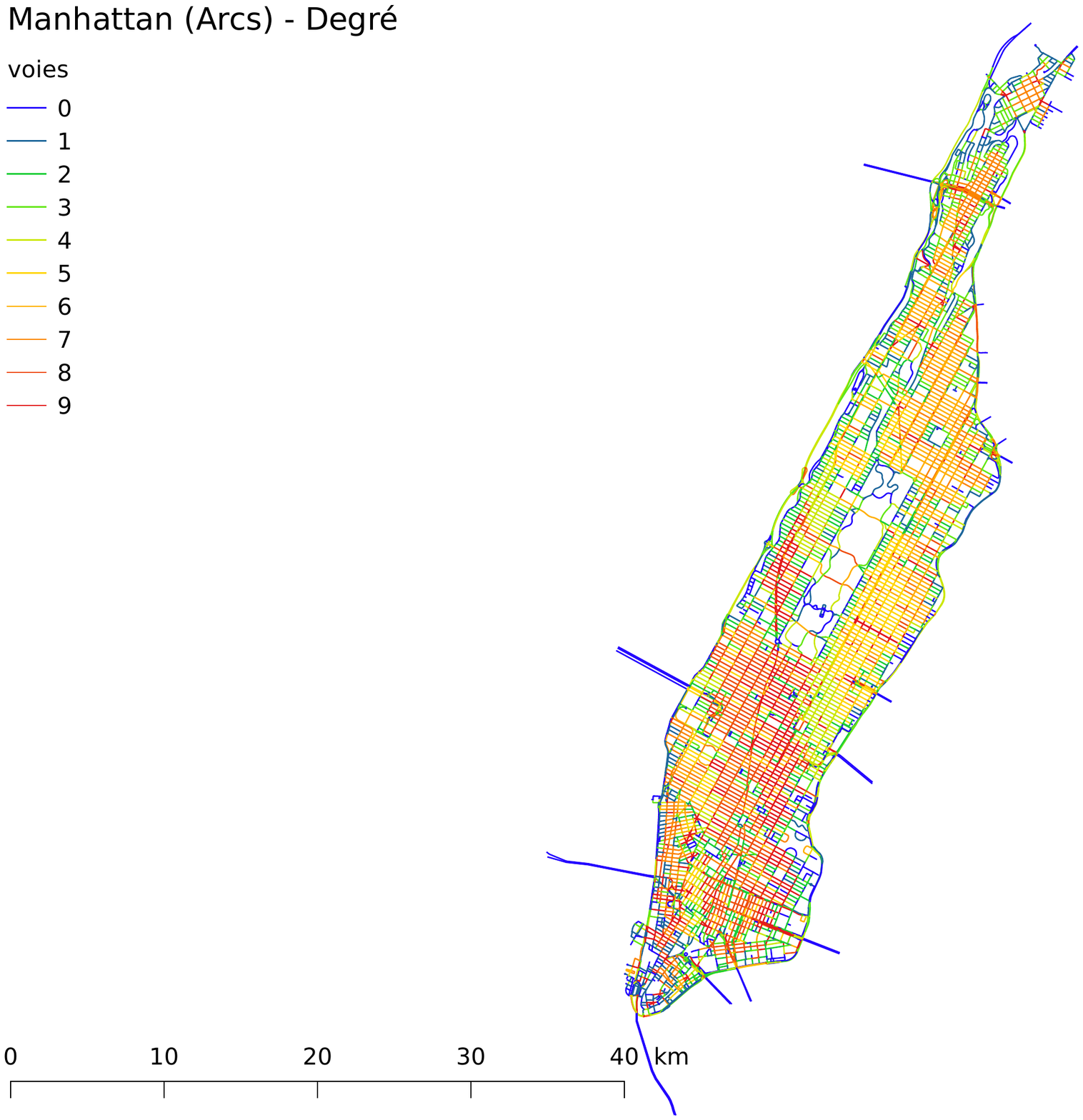}
    \end{subfigure}
    
    \begin{subfigure}{\linewidth}
        \includegraphics[width=\textwidth]{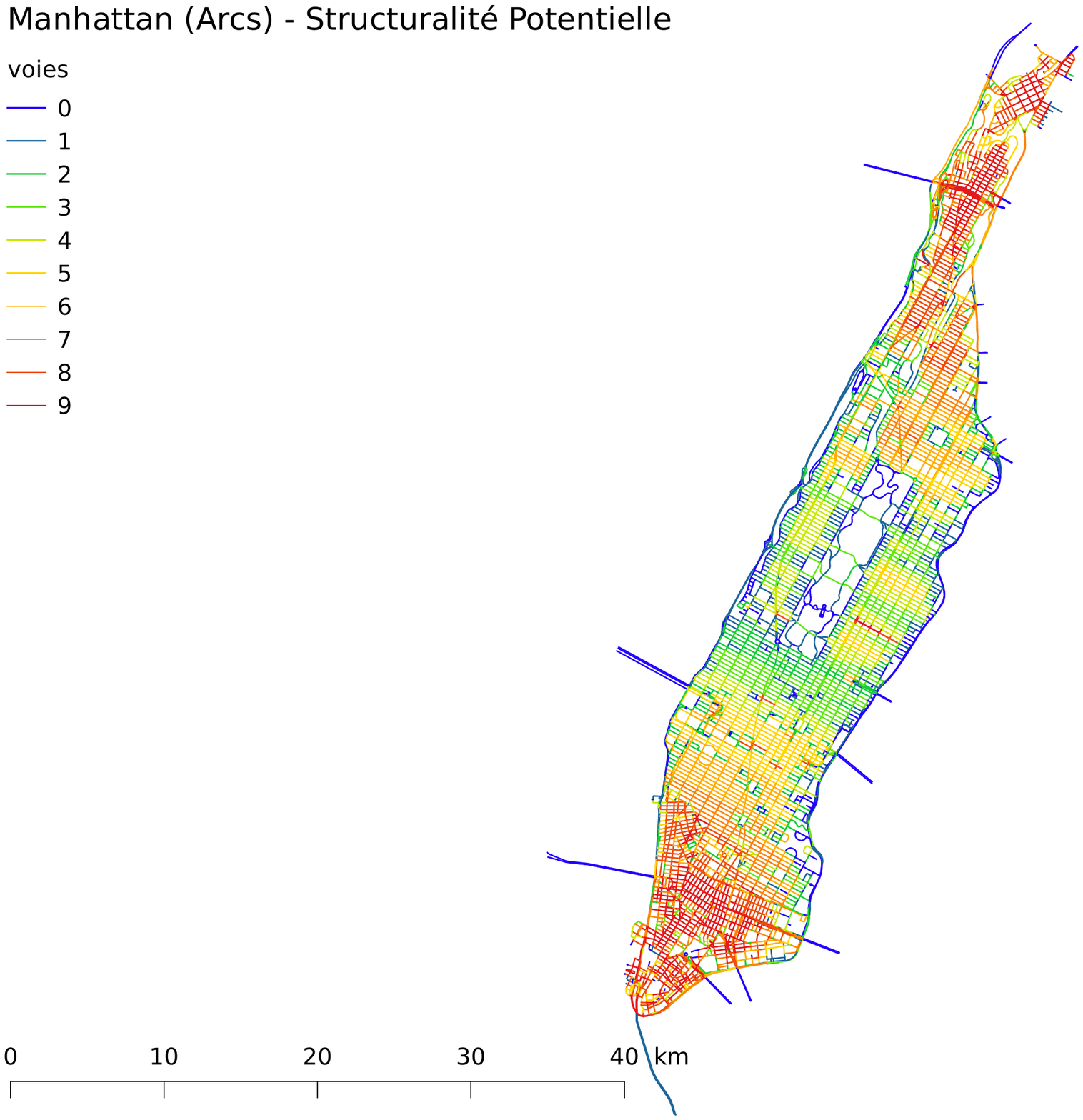}
    \end{subfigure}

    \caption{Comparaison entre degré et structuralité potentielle calculés sur les arcs du graphe viaire de Manhattan.}
    \label{fig:arcs_carte_degrestructpot_man}
\end{figure}

La corrélation entre degré et structuralité potentielle est plus forte sur le réseau d'Avignon que sur celui de Manhattan (tableaux en annexe  \ref{tab:corr_avignon_arc}, \ref{tab:corr_manhattan_arc}). En traçant les cartes de corrélation de ces deux indicateurs sur chacun des réseaux, nous observons un effet de seuil plus marqué sur Avignon que sur Manhattan (figures \ref{fig:arcs_carte_degrestructpot_av} et \ref{fig:arcs_carte_degrestructpot_man}). Celui-ci s'explique par le faible nombre d'arcs isolés à Manhattan. En effet, le calcul de structuralité potentielle consiste à sommer pour chaque arc les accessibilités des arcs qui lui sont connectés. Nous observons sur les cartes la différence des résultats donnés par les deux indicateurs sur les deux espaces considérés (figures \ref{fig:arcs_carte_degrestructpot_av} et \ref{fig:arcs_carte_degrestructpot_man}). Tous les arcs de faible degré (impasses) à Avignon sont également ceux dont la structuralité potentielle est faible. Les longueurs de ces arcs étant peu importantes, ils se retrouvent tous dans les premières classes, ce qui crée un effet de seuil dû à la discrétisation.

\begin{figure}[h]
    \centering

    \begin{subfigure}{.40\textwidth}
        \includegraphics[width=\textwidth]{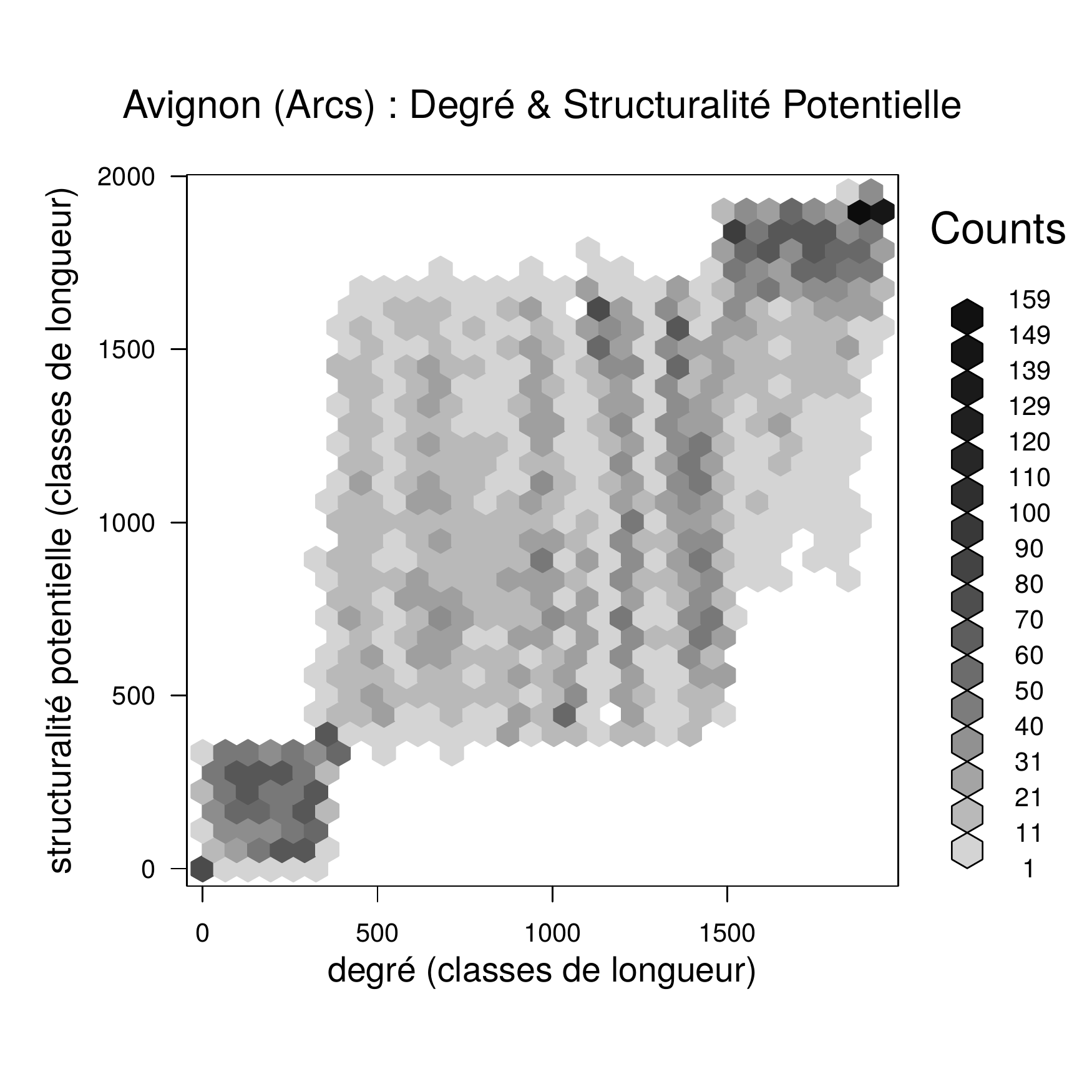}
        \caption{Sur le graphe viaire d'Avignon.}
    \end{subfigure}
    ~
    \begin{subfigure}{.40\textwidth}
        \includegraphics[width=\textwidth]{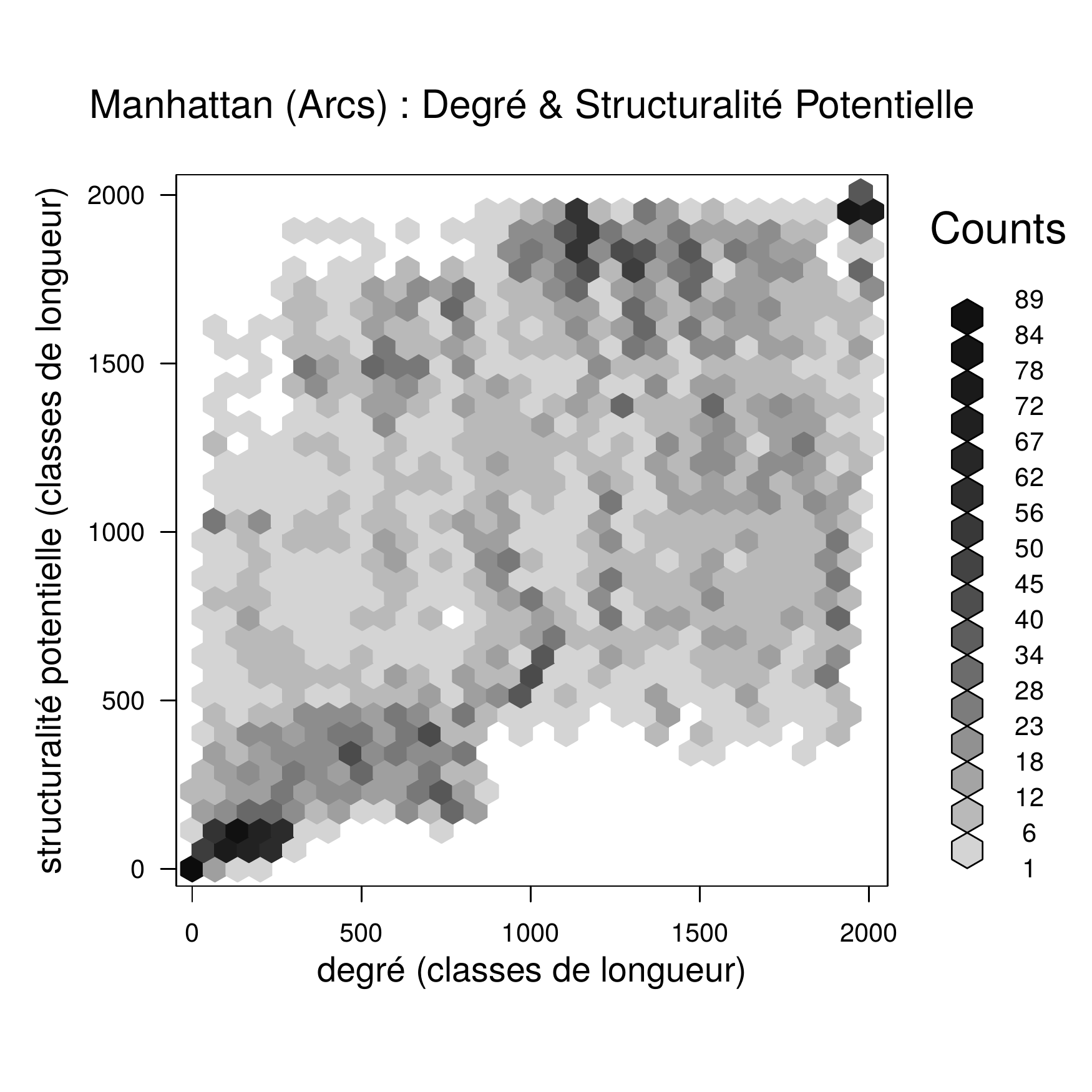}
        \caption{Sur le graphe viaire de Manhattan.}
    \end{subfigure}

    \caption{Cartes de corrélation croisée entre dégré et structuralité potentielle sur les arcs.}
    \label{fig:arcs_degrestructpot}

\end{figure}

\FloatBarrier

Les indicateurs de longueur et l'orthogonalité apportent quant à eux une information bien différente de tous les autres. Leurs valeurs ne sont corrélées à aucun autre résultat, ce qui appuie leur importance dans l'étude d'un graphe spatialisé. En effet, les deux indicateurs utilisent dans leur construction le caractère géométrique des objets sur lesquels ils sont appliqués. Pour les arcs, ils constituent une caractérisation spécifique liée à leur géométrie.

\begin{figure}[h]
    \centering

    \begin{subfigure}{.40\textwidth}
        \includegraphics[width=\textwidth]{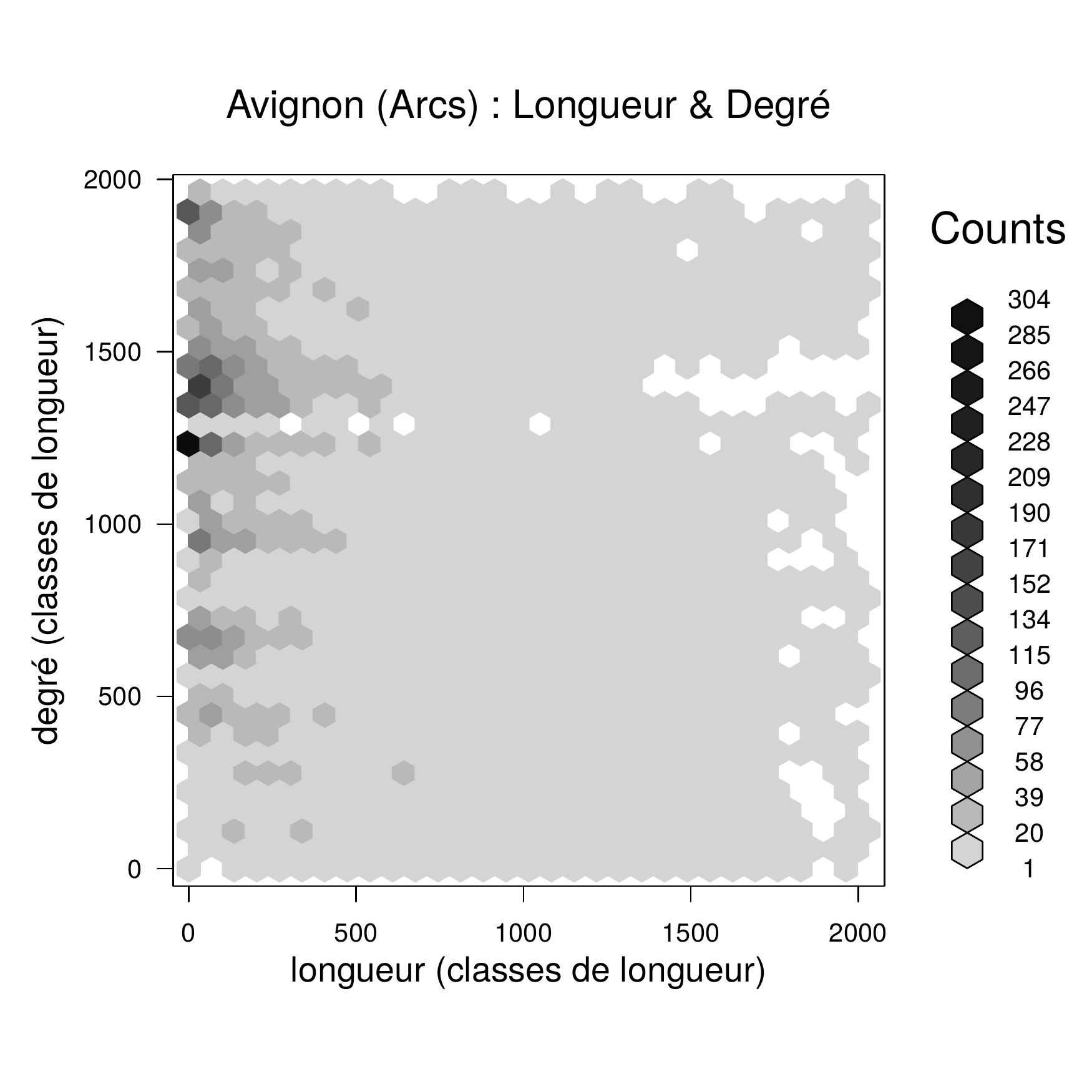}
    \end{subfigure}
    ~
    \begin{subfigure}{.40\textwidth}
        \includegraphics[width=\textwidth]{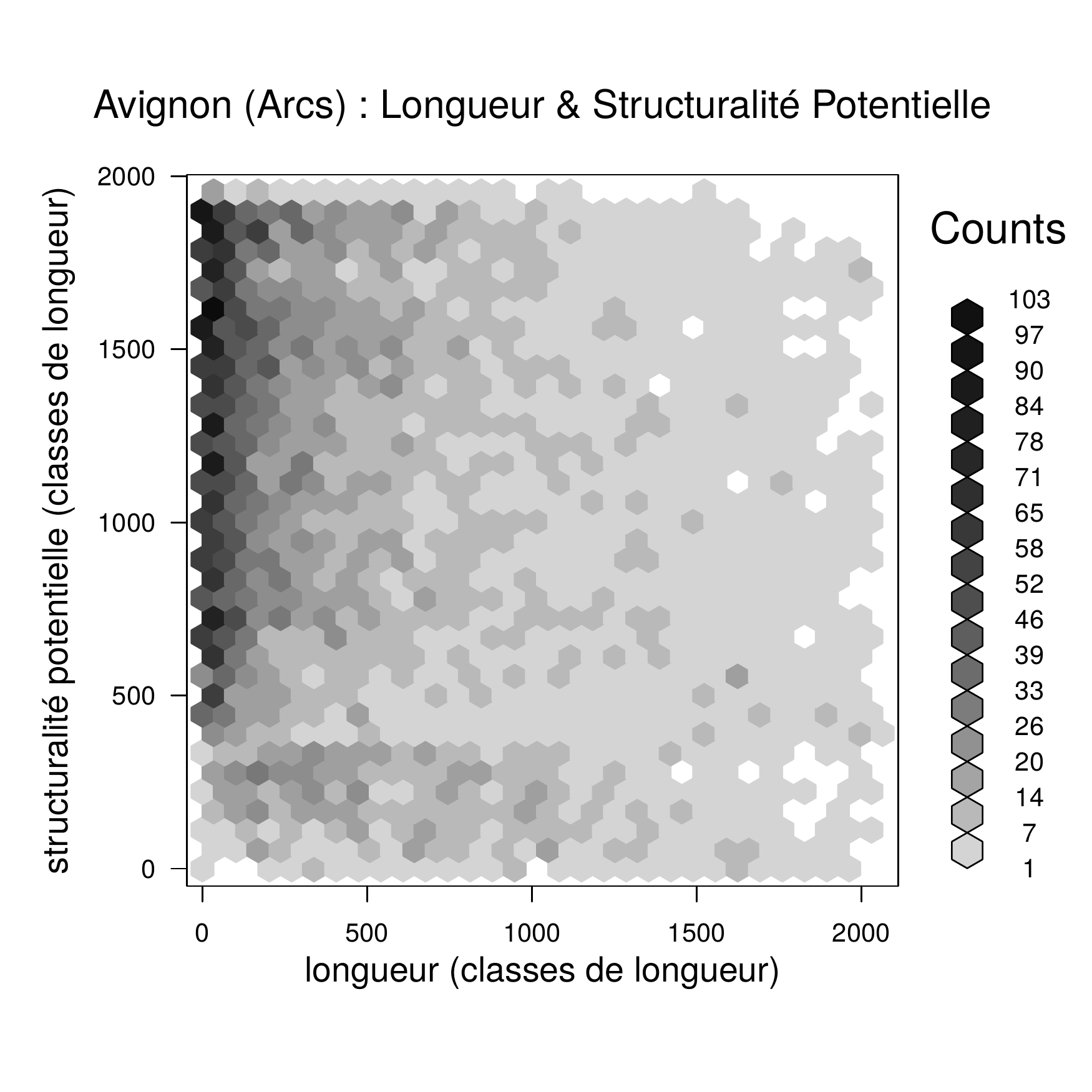}
    \end{subfigure}
    
    \begin{subfigure}{.40\textwidth}
        \includegraphics[width=\textwidth]{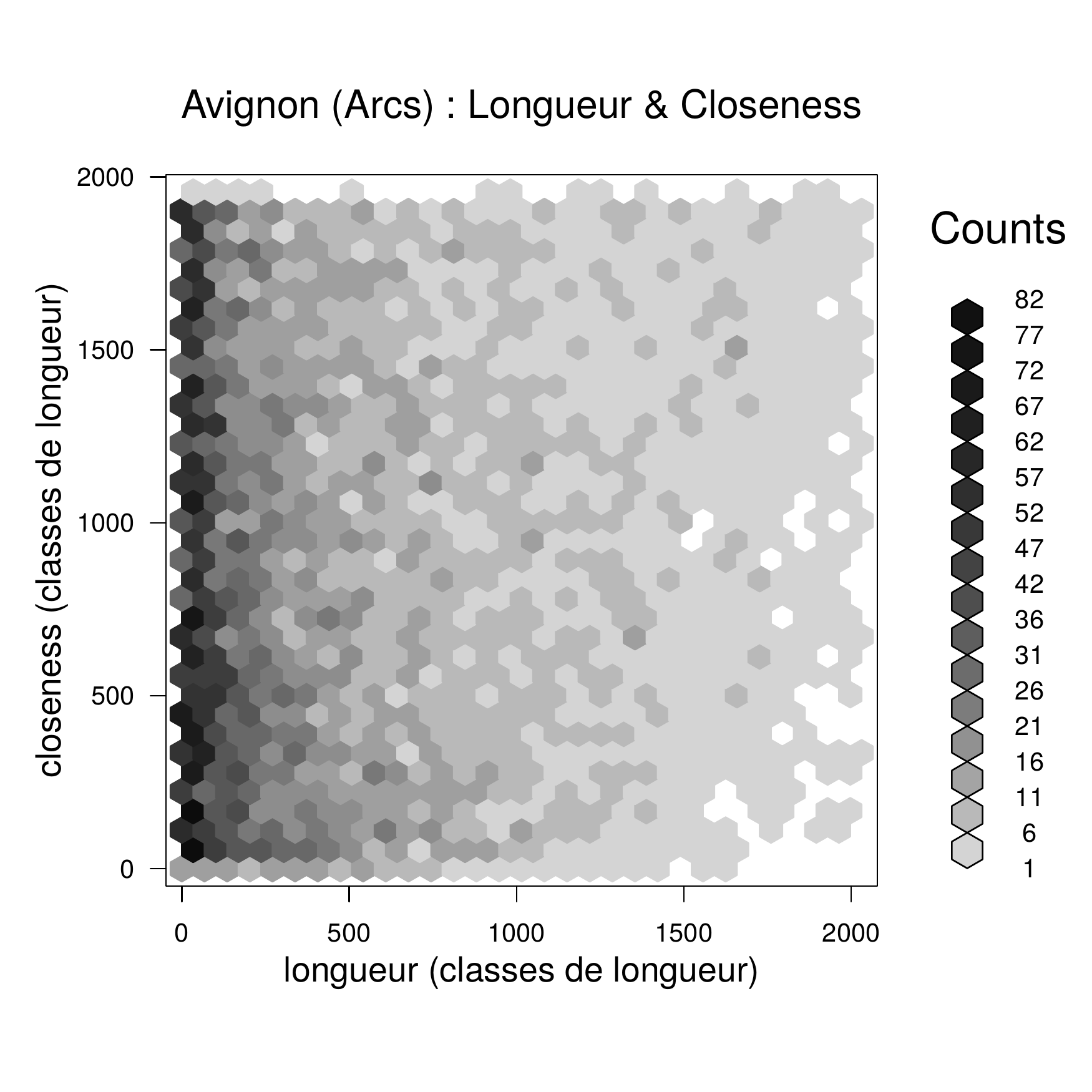}
    \end{subfigure}
     ~
    \begin{subfigure}{.40\textwidth}
        \includegraphics[width=\textwidth]{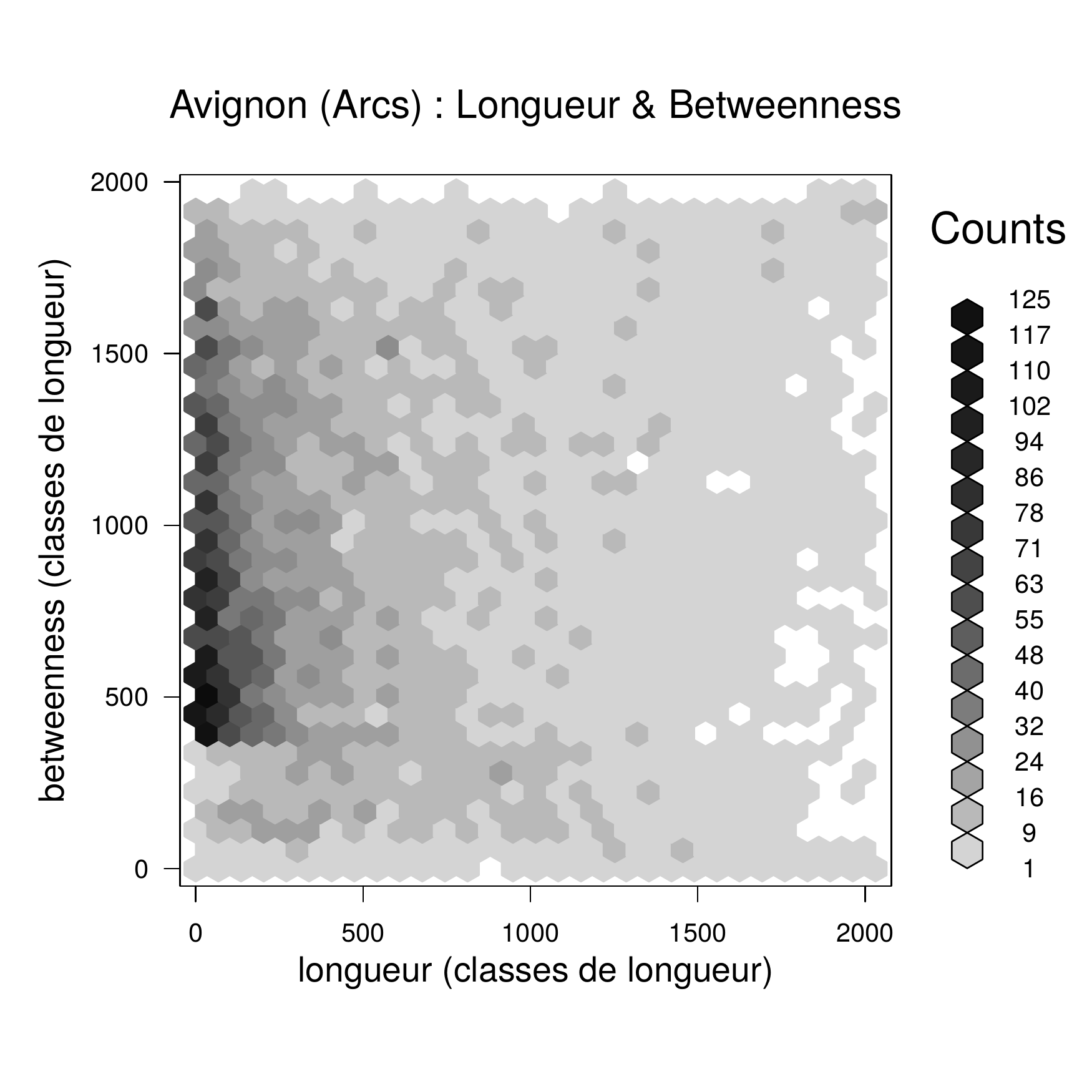}
    \end{subfigure}

    \caption{Cartes de corrélation croisée entre longueur et autres indicateurs sur le graphe viaire d'Avignon.}
    \label{fig:arcs_avi_lengthX}

\end{figure}

\begin{figure}[h]
    \centering

    \begin{subfigure}{.40\textwidth}
        \includegraphics[width=\textwidth]{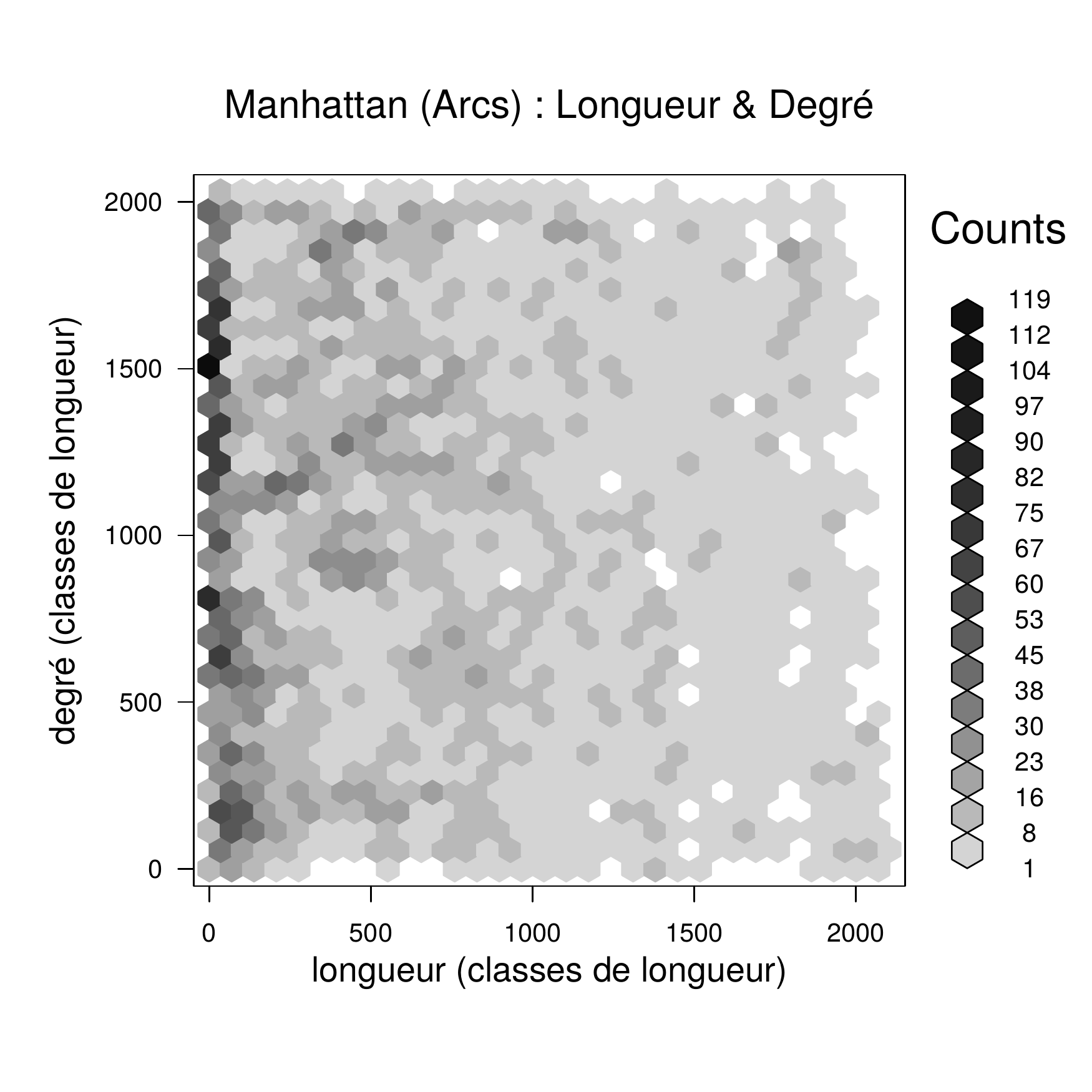}
    \end{subfigure}
    ~
    \begin{subfigure}{.40\textwidth}
        \includegraphics[width=\textwidth]{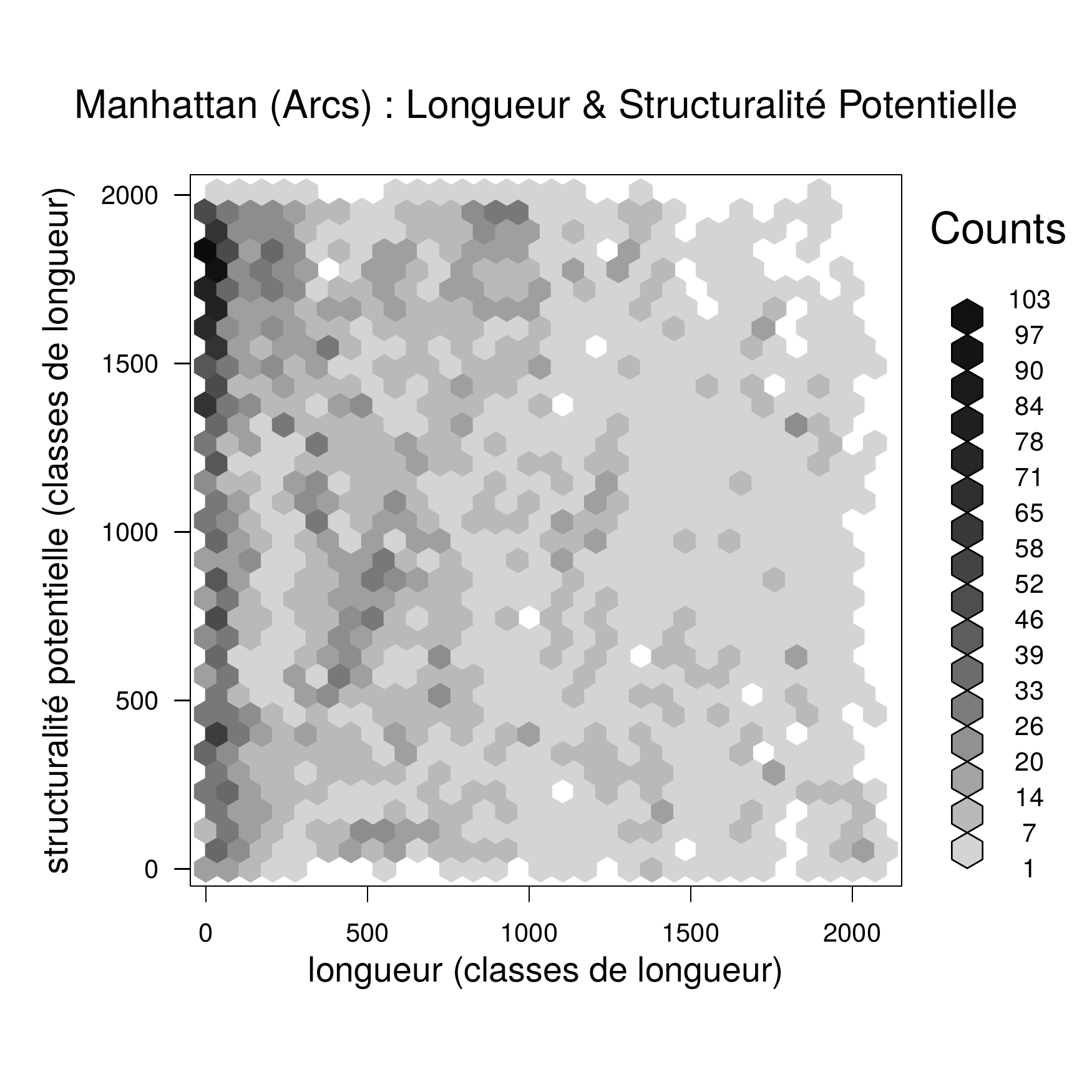}
    \end{subfigure}
    
    \begin{subfigure}{.40\textwidth}
        \includegraphics[width=\textwidth]{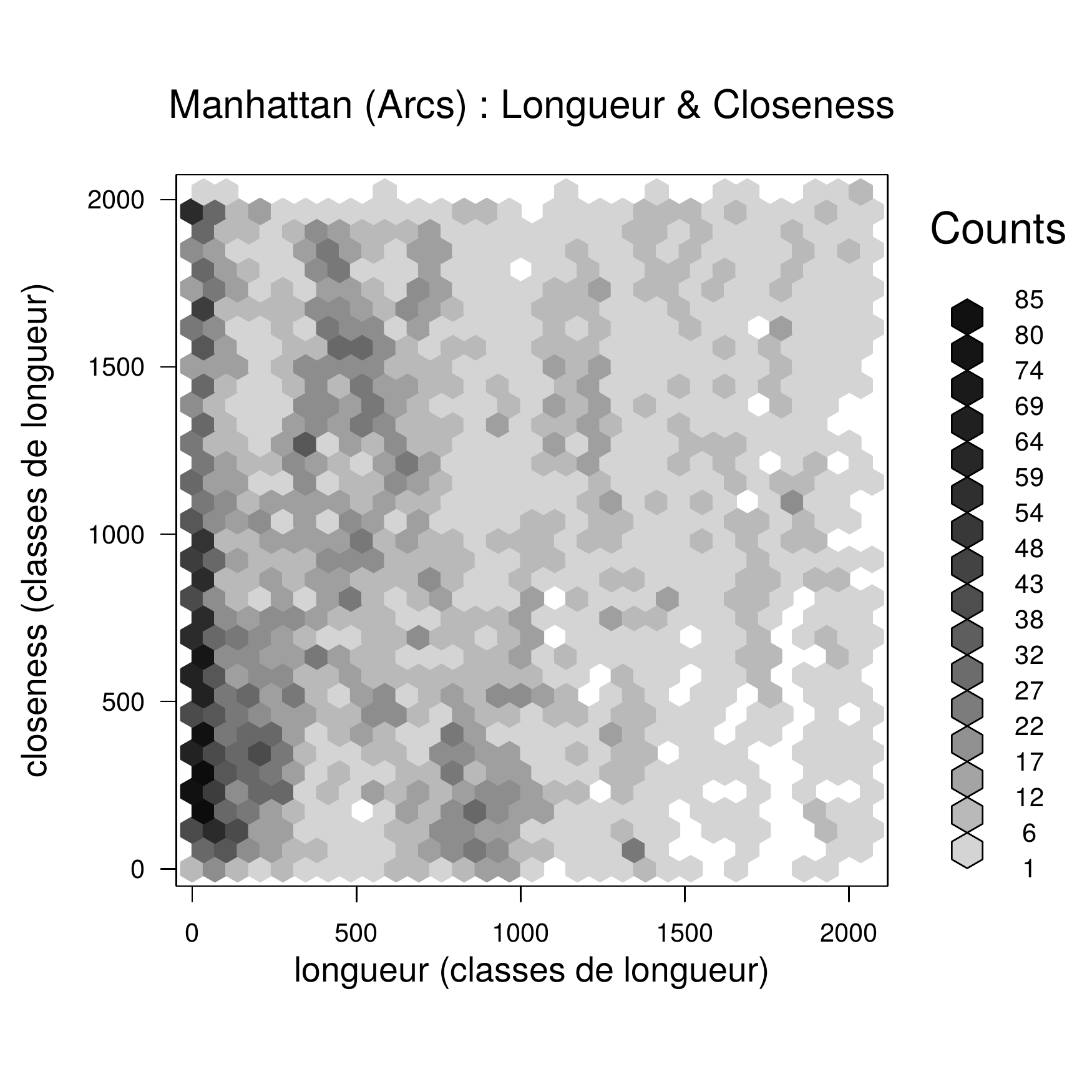}
    \end{subfigure}
     ~
    \begin{subfigure}{.40\textwidth}
        \includegraphics[width=\textwidth]{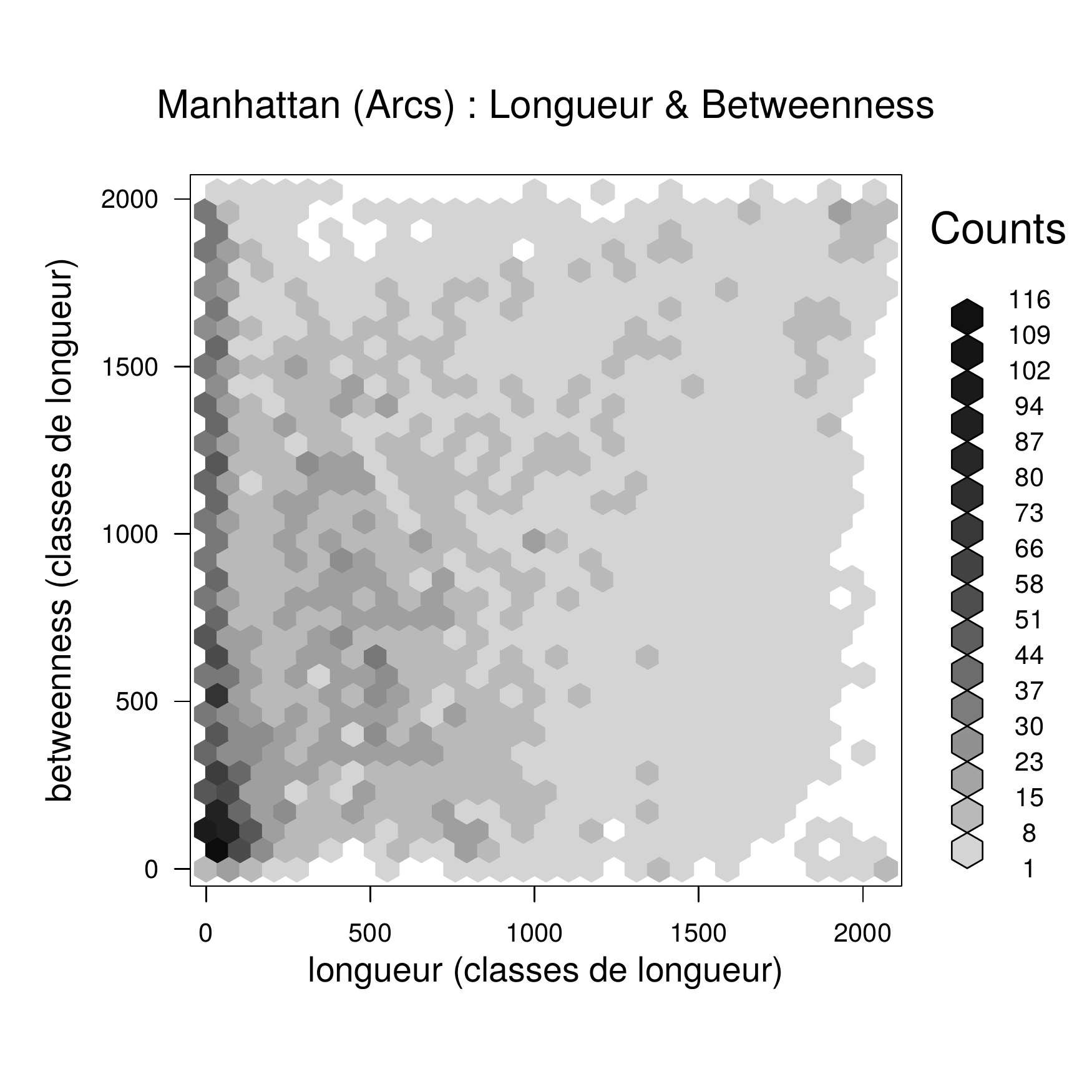}
    \end{subfigure}

    \caption{Cartes de corrélation croisée entre longueur et autres indicateurs sur le graphe viaire de Manhattan.}
    \label{fig:arcs_man_lengthX}

\end{figure}

La classification que nous utilisons dans cette étude regroupe les arcs de faible longueur dans les premières classes. Ils y sont donc en nombre beaucoup plus important, ce qui constitue un léger amas dans les premières classes qui se diffuse de manière homogène dans les suivantes (figures \ref{fig:arcs_avi_lengthX}, \ref{fig:arcs_man_lengthX}).

\begin{figure}[h]
    \centering

    \begin{subfigure}{.40\textwidth}
        \includegraphics[width=\textwidth]{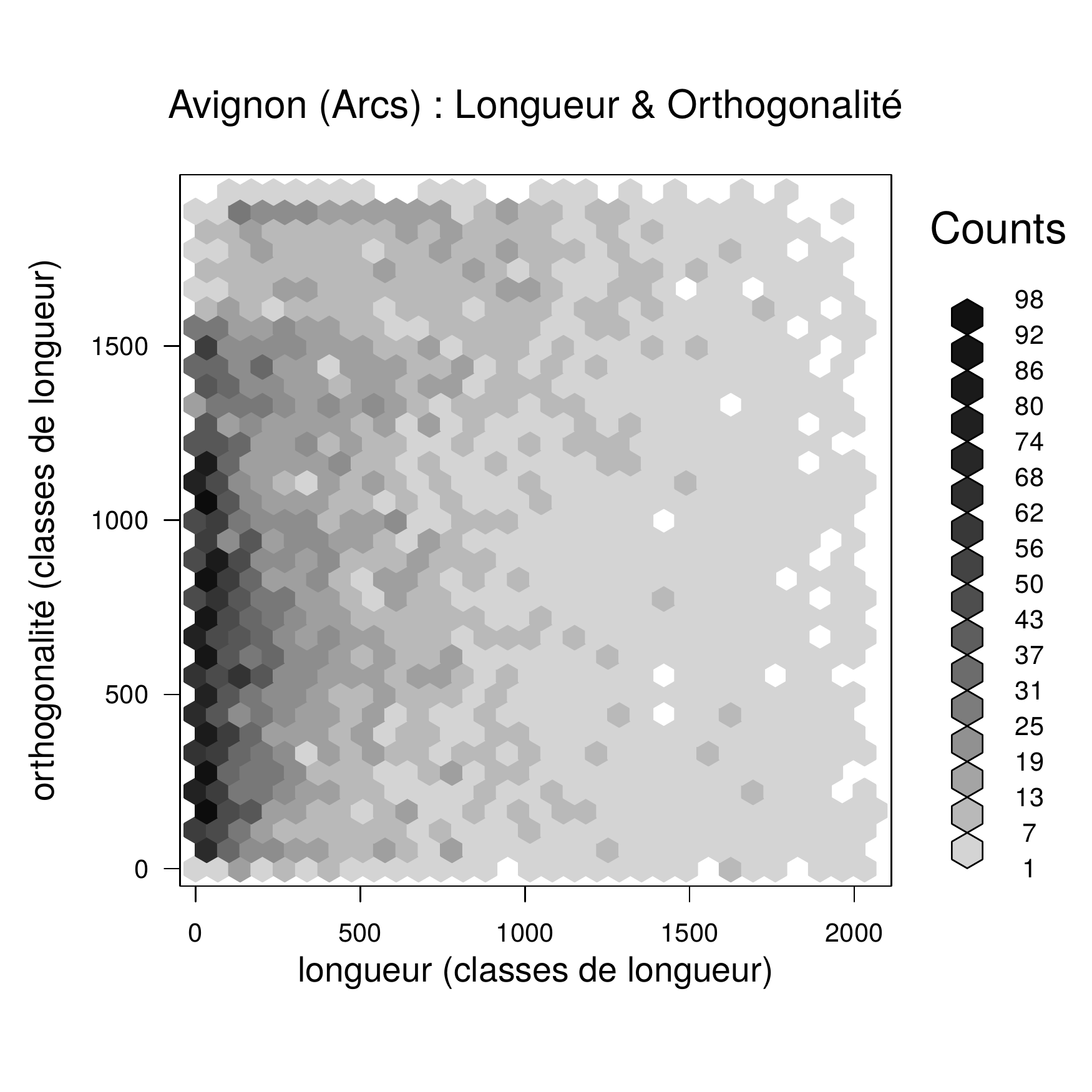}
    \end{subfigure}
    ~
    \begin{subfigure}{.40\textwidth}
        \includegraphics[width=\textwidth]{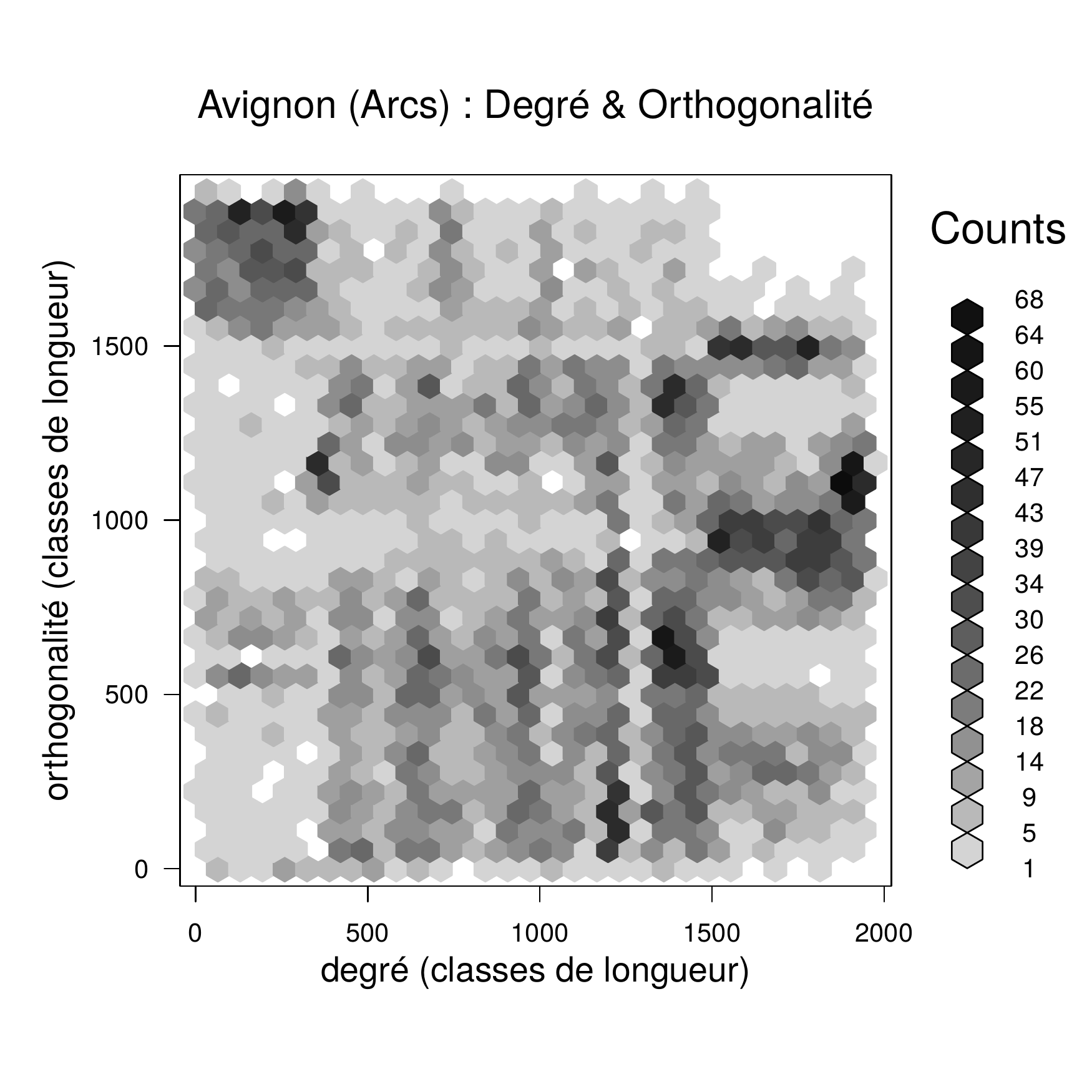}
    \end{subfigure}
    ~
    \begin{subfigure}{.40\textwidth}
        \includegraphics[width=\textwidth]{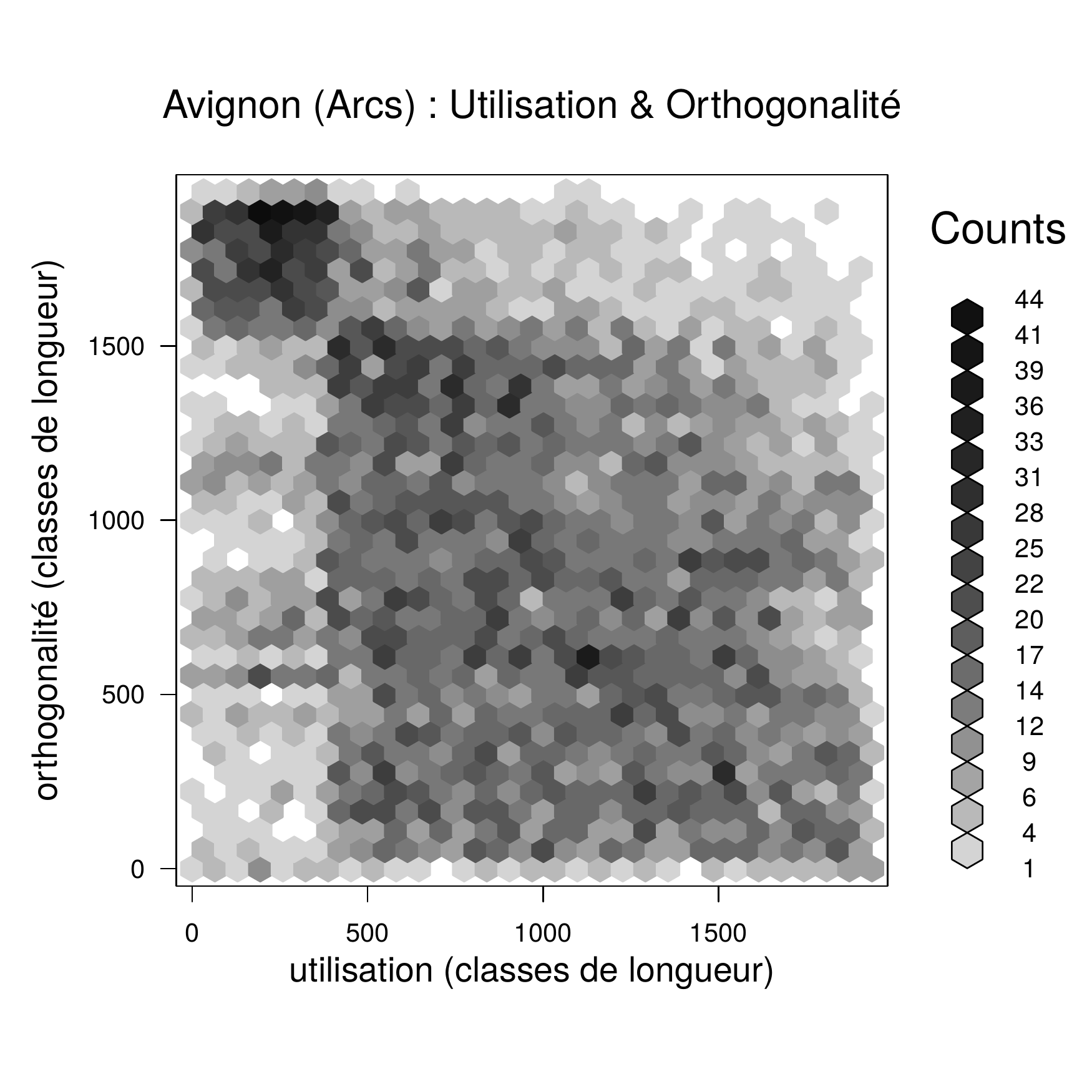}
    \end{subfigure}

    \caption{Cartes de corrélation croisée entre orthogonalité et autres indicateurs sur le graphe viaire d'Avignon.}
    \label{fig:arcs_avi_orthoX}

\end{figure}

\begin{figure}[h]
    \centering

     \begin{subfigure}{.40\textwidth}
        \includegraphics[width=\textwidth]{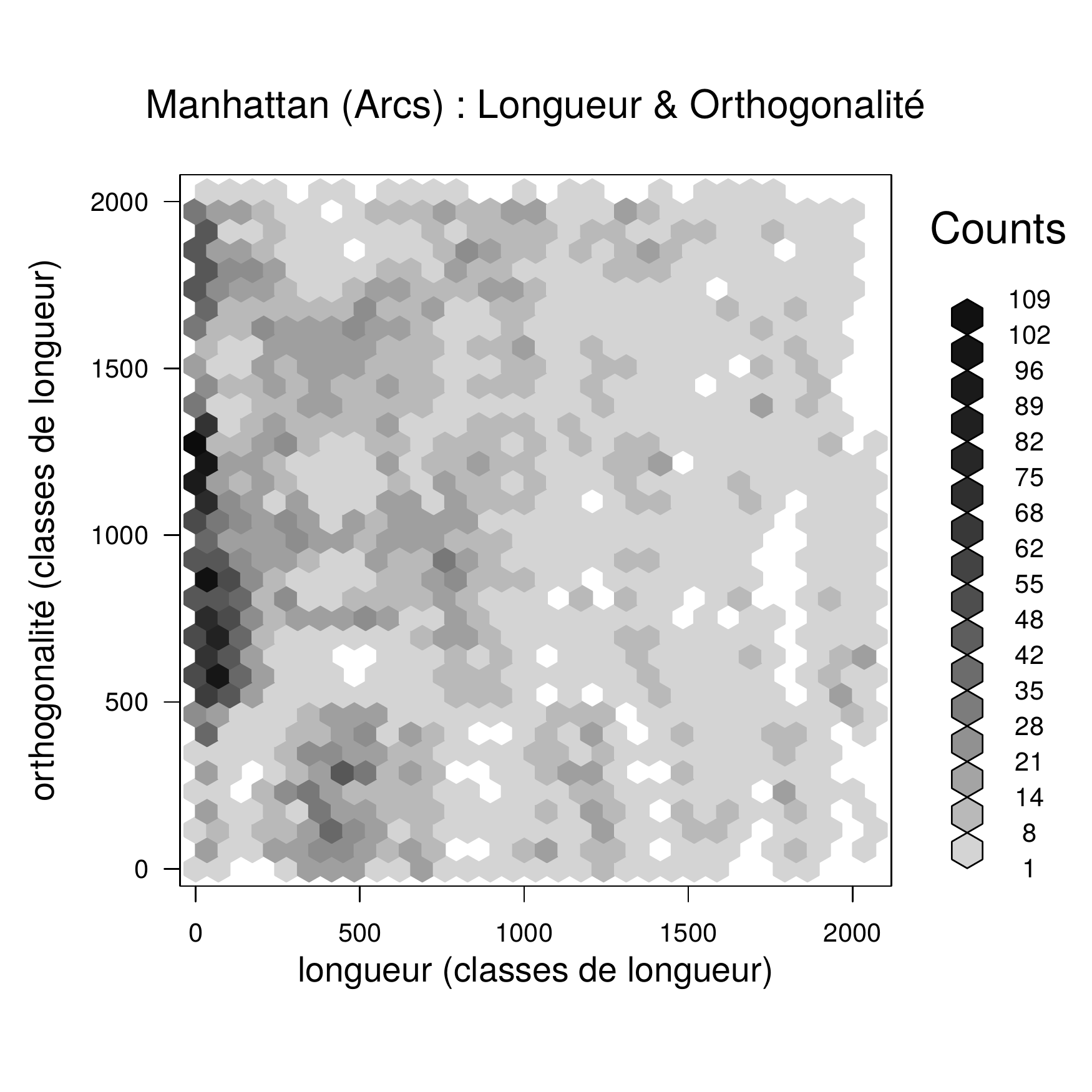}
    \end{subfigure}
    ~
    \begin{subfigure}{.40\textwidth}
        \includegraphics[width=\textwidth]{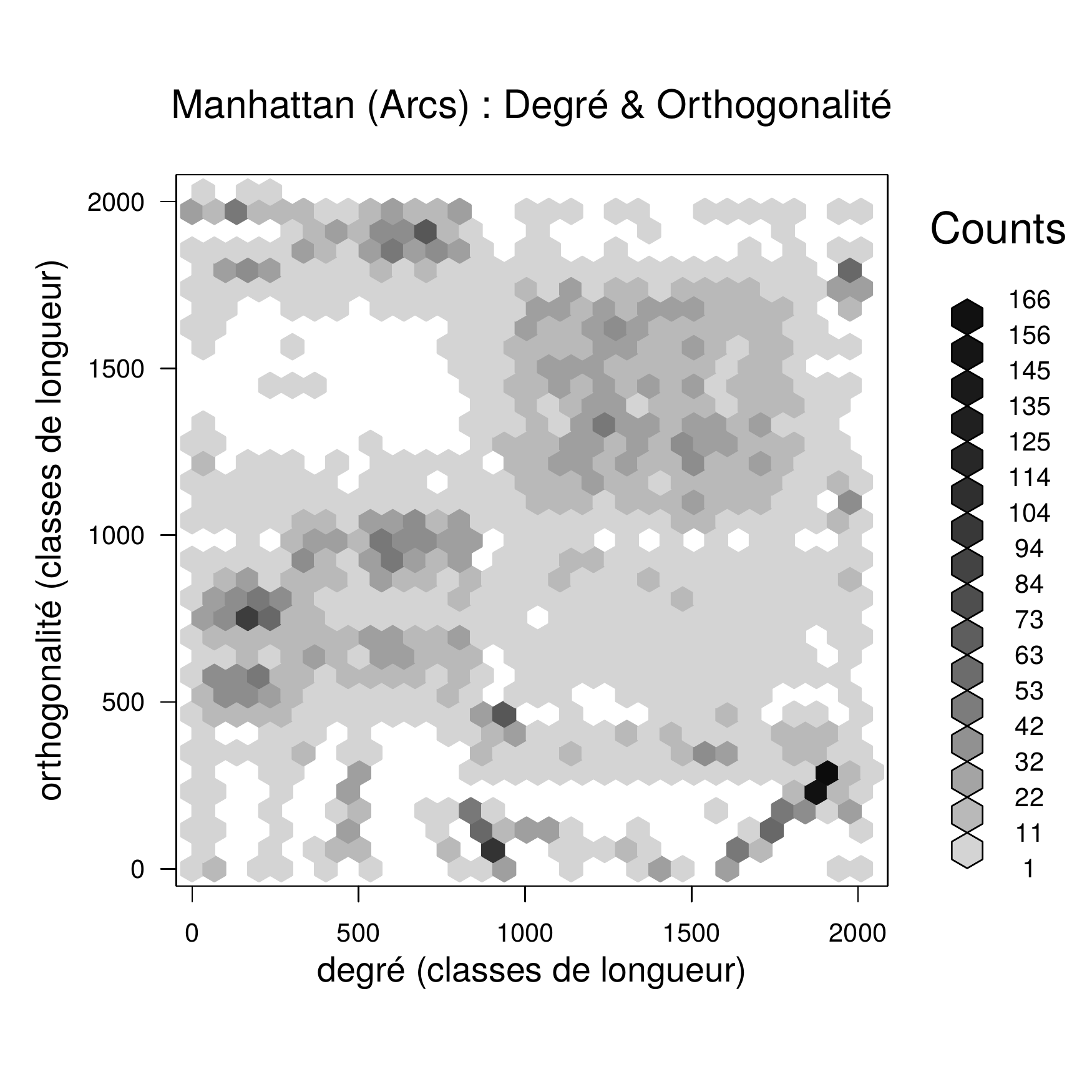}
    \end{subfigure}
    ~
    \begin{subfigure}{.40\textwidth}
        \includegraphics[width=\textwidth]{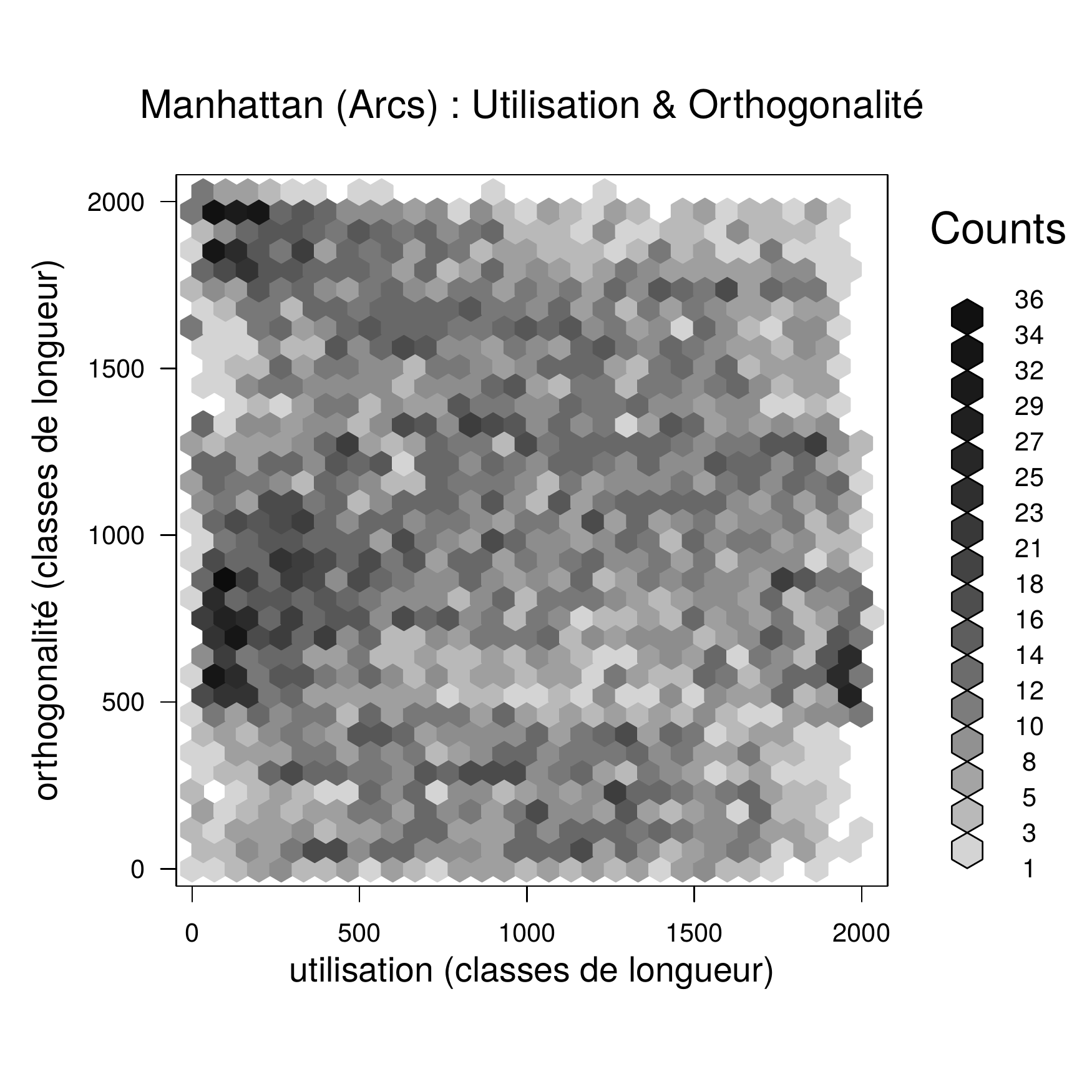}
    \end{subfigure}

    \caption{Cartes de corrélation croisée entre orthogonalité et autres indicateurs sur le graphe viaire de Manhattan.}
    \label{fig:arcs_man_orthoX}

\end{figure}

Pour l'orthogonalité, la répartition est complètement homogène quel que soit l'indicateur auquel elle est comparée et quel que soit l'échantillon observé (figures \ref{fig:arcs_avi_orthoX}, \ref{fig:arcs_man_orthoX}).

De cette étude, nous concluons que nous pouvons limiter notre observation des structures créées par les arcs à six indicateurs primaires : la \textbf{longueur}, le \textbf{degré}, la \textbf{structuralité potentielle}, l'\textbf{utilisation}, la \textbf{closeness} et l'\textbf{orthogonalité}. Les deux réseaux étudiés dans cette partie possèdent des caractéristiques de construction radicalement opposées ce qui nous permet de faire l'hypothèse que ces résultats restent valables quel que soit le réseau viaire considéré.

Chacun d'entre eux a une pertinence dans la représentation d'une information. Il est possible de les combiner de multiples manières afin de trouver une nouvelle méthode de caractérisation. Nous proposons des combinaisons par division dans le paragraphe suivant.

\FloatBarrier
\subsection{Indicateurs composés}

\begin{figure}[h]
    \centering

    \begin{subfigure}{.35\textwidth}
        \includegraphics[width=\textwidth]{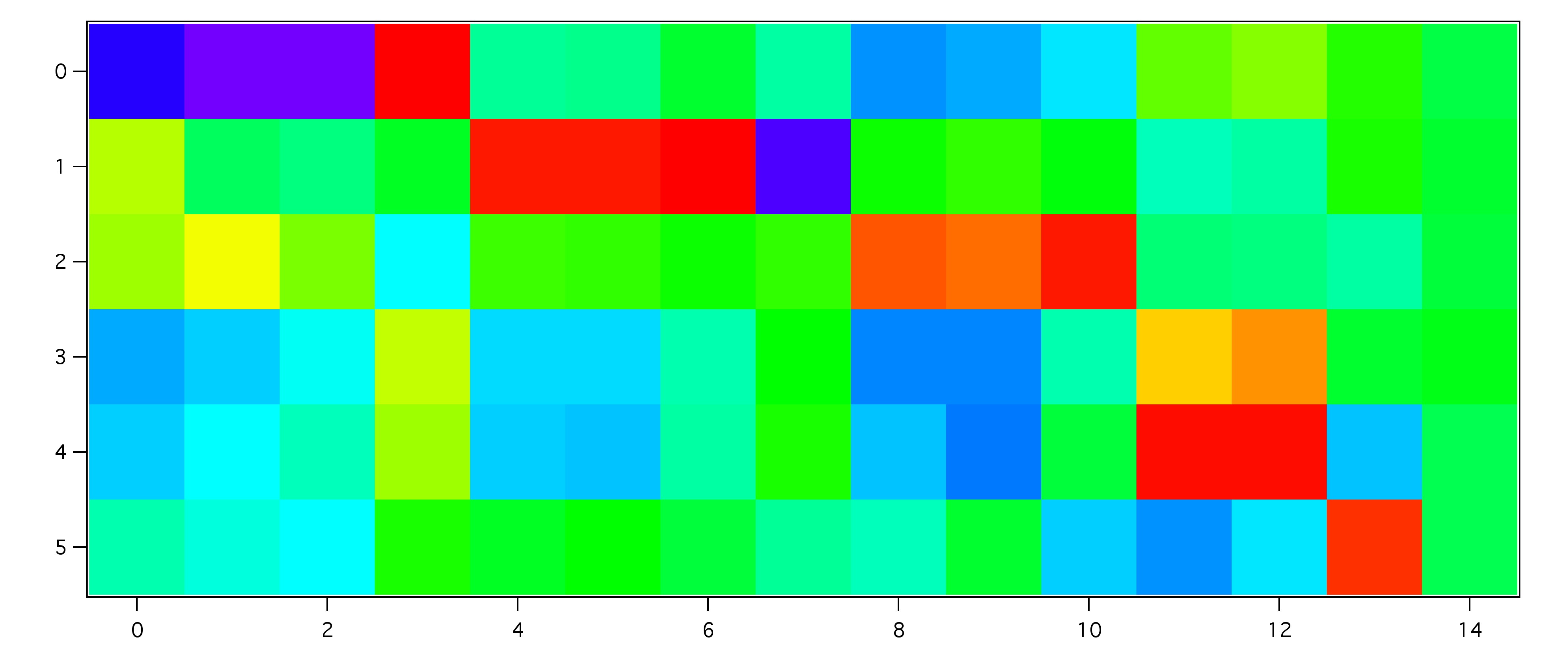}
        \caption{Représentation du tableau \ref{tab:corr_avignon_comb}, calculé sur les arcs d'Avignon.}
    \end{subfigure}
    ~
    \begin{subfigure}{.35\linewidth}
        \includegraphics[width=\textwidth]{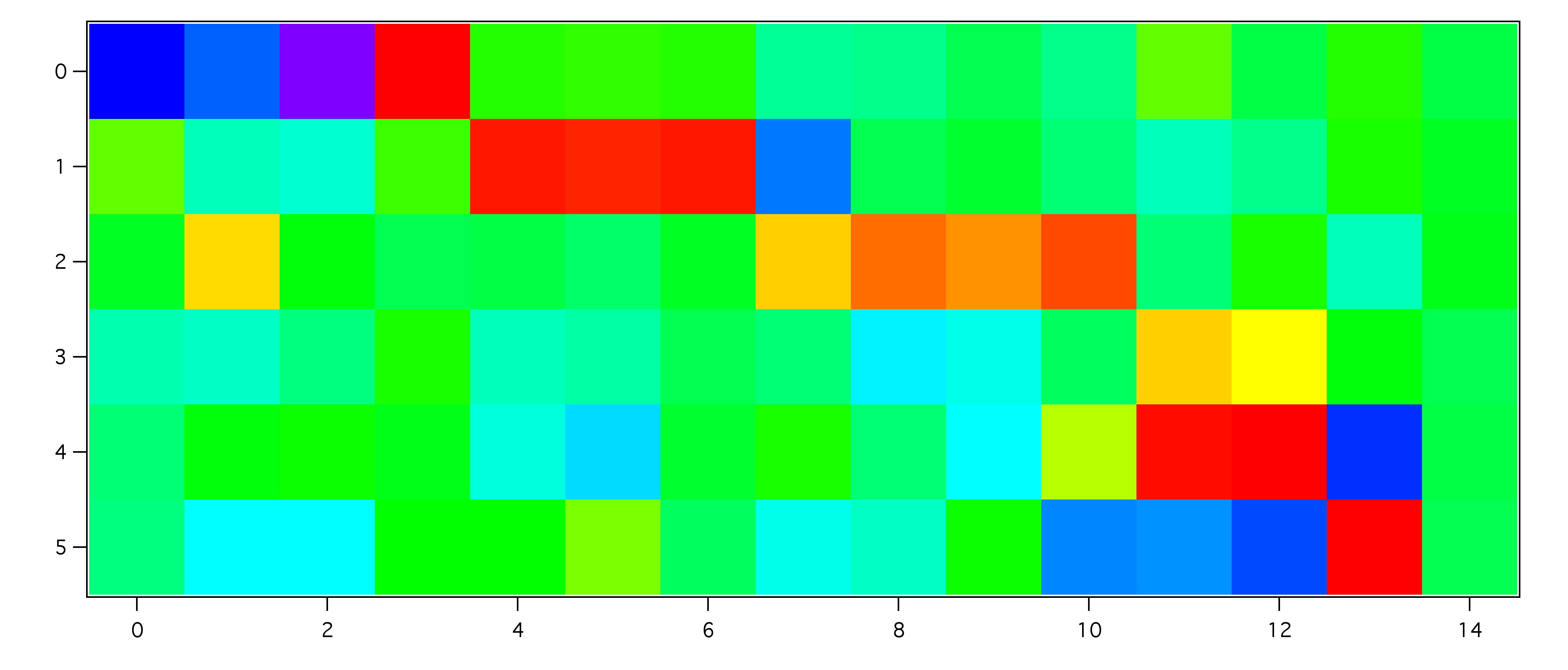}
        \caption{Représentation du tableau \ref{tab:corr_manhattan_comb}, calculé sur les arcs de Manhattan.}
    \end{subfigure}
   ~
    \begin{subfigure}{.2\linewidth}
        \includegraphics[width=\textwidth]{images/matrices_colorees/mat_echelle_dec.jpg}
        \caption{Échelle}
    \end{subfigure}

    \caption{Représentation de la corrélation entre indicateurs composés calculés sur les arcs sous forme de matrice colorée. Rappel de l'ordre des indicateurs : \\
    Vertical : 0 : utilisation ; 1 : longueur ; 2 : orthogonalité ; 3 : degré ; 4 : structuralité potentielle ; 5 : closeness \\
    Horizontal : 0 : $\frac{longueur}{utilisation}$ ; 1 : $\frac{orthogonalite}{utilisation}$ ; 2 : $\frac{structuralite}{utilisation}$ ; 3 : $\frac{utilisation}{closeness}$ ; 4 : $\frac{longueur}{degre}$ ; 5 : $\frac{longueur}{structuralite}$  ; 6 : $\frac{longueur}{closeness}$ ; 7 : $\frac{orthogonalite}{longueur}$ ; 8 : $\frac{orthogonalite}{degre}$ ; 9 : $\frac{orthogonalite}{structuralite}$ ; 10 : $\frac{orthogonalite}{closeness}$ ; 11 : $\frac{structuralite}{closeness}$ ; 12 : $\frac{degre}{closeness}$ ; 13 : $\frac{degre}{structuralite}$ ; 14 : $\frac{degre}{utilisation}$}
    \label{fig:mat_comb_arcs}

\end{figure}

Les six indicateurs primaires, définis dans le paragraphe précédent, permettent de caractériser finement les arcs du réseau. Nous les combinons ici par division afin de créer de nouveaux indicateurs, dont nous voulons étudier l'originalité. Pour cela, nous croisons les valeurs brutes des indicateurs en les multipliant ou en les divisant l'un par l'autre. Puis nous effectuons la même classification que précédemment, afin d'avoir une longueur de réseau sensiblement identique dans chaque classe. Enfin, nous effectuons des calculs de corrélation (avec la méthode de Pearson) que nous reportons dans les tableaux \ref{tab:corr_avignon_comb} et \ref{tab:corr_manhattan_comb} en annexe. Nous illustrons de la même manière que pour les indicateurs primaires ces corrélations à travers deux matrices colorées qui permettent de lire plus facilement les groupes d'indicateurs corrélés (figure \ref{fig:mat_comb_arcs}).

La caractérisation des arcs avec les indicateurs composés fait ressortir une hiérarchie de prévalence entre eux. En effet, les combinaisons faites avec l'indicateur d'utilisation, que ce soit avec la longueur, la structuralité potentielle, la closeness ou l'orthogonalité, sont corrélées à celui-ci (figure \ref{fig:mat_comb_arcs}). Seule la combinaison (utilisation, degré) se détache de cette tendance pour créer un indicateur qui n'est corrélé à aucun autre et qui apporte donc une nouvelle information sur le réseau. Les cartes montrent quelques artefacts pour les plus petites valeurs, notamment pour l'orthogonalité. Cela s'explique avec le grand nombre d'arcs qui ont un coefficient d'orthogonalité proche de 0. En effet, cela correspond aux arcs alignés avec leurs voisins ce qui est un cas courant. Cet artefact sera supprimé avec l'utilisation de la voie qui utilise cet alignement comme facteur de construction.

\begin{figure}[h]
    \centering

     \begin{subfigure}{.40\textwidth}
        \includegraphics[width=\textwidth]{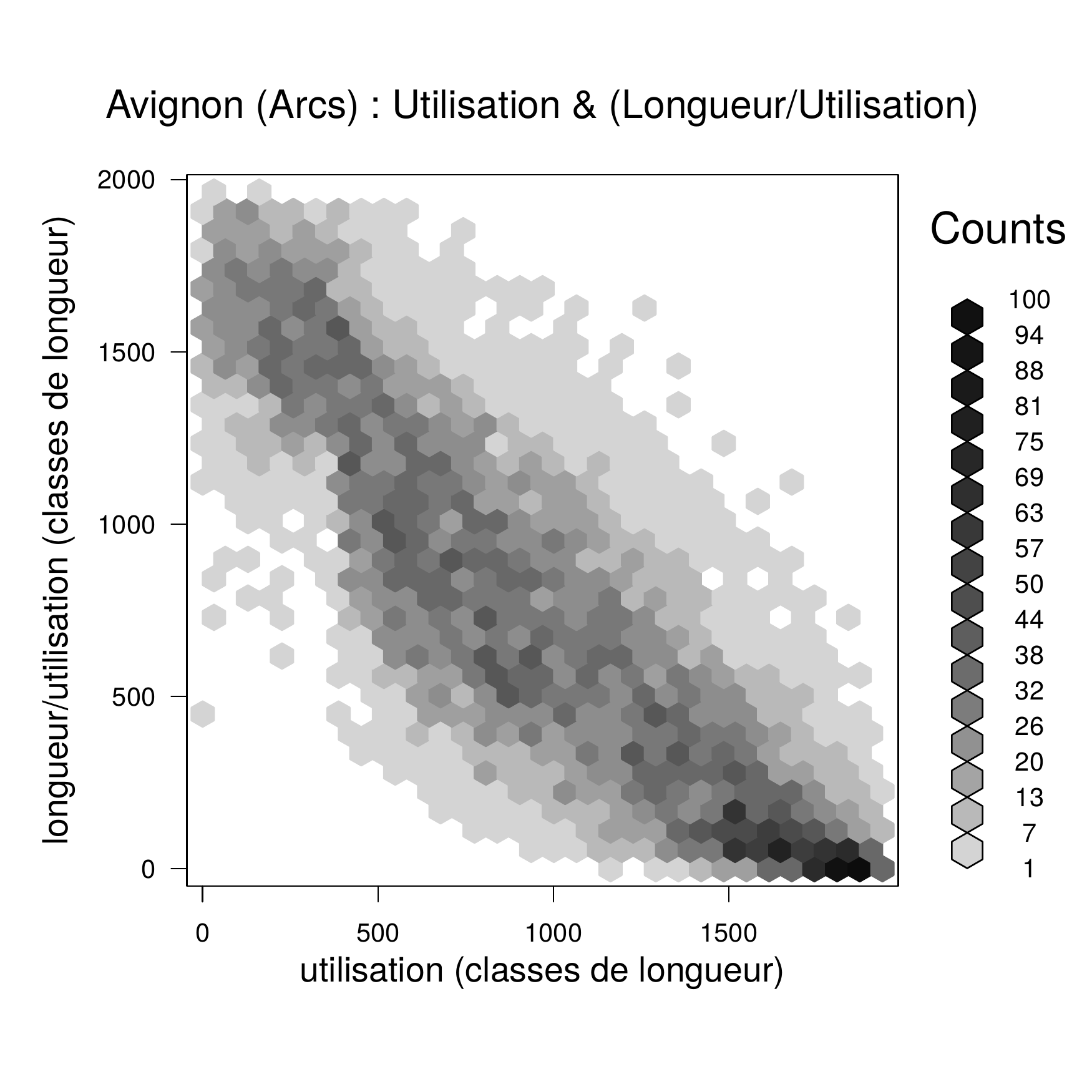}
    \end{subfigure}
    ~
    \begin{subfigure}{.40\textwidth}
        \includegraphics[width=\textwidth]{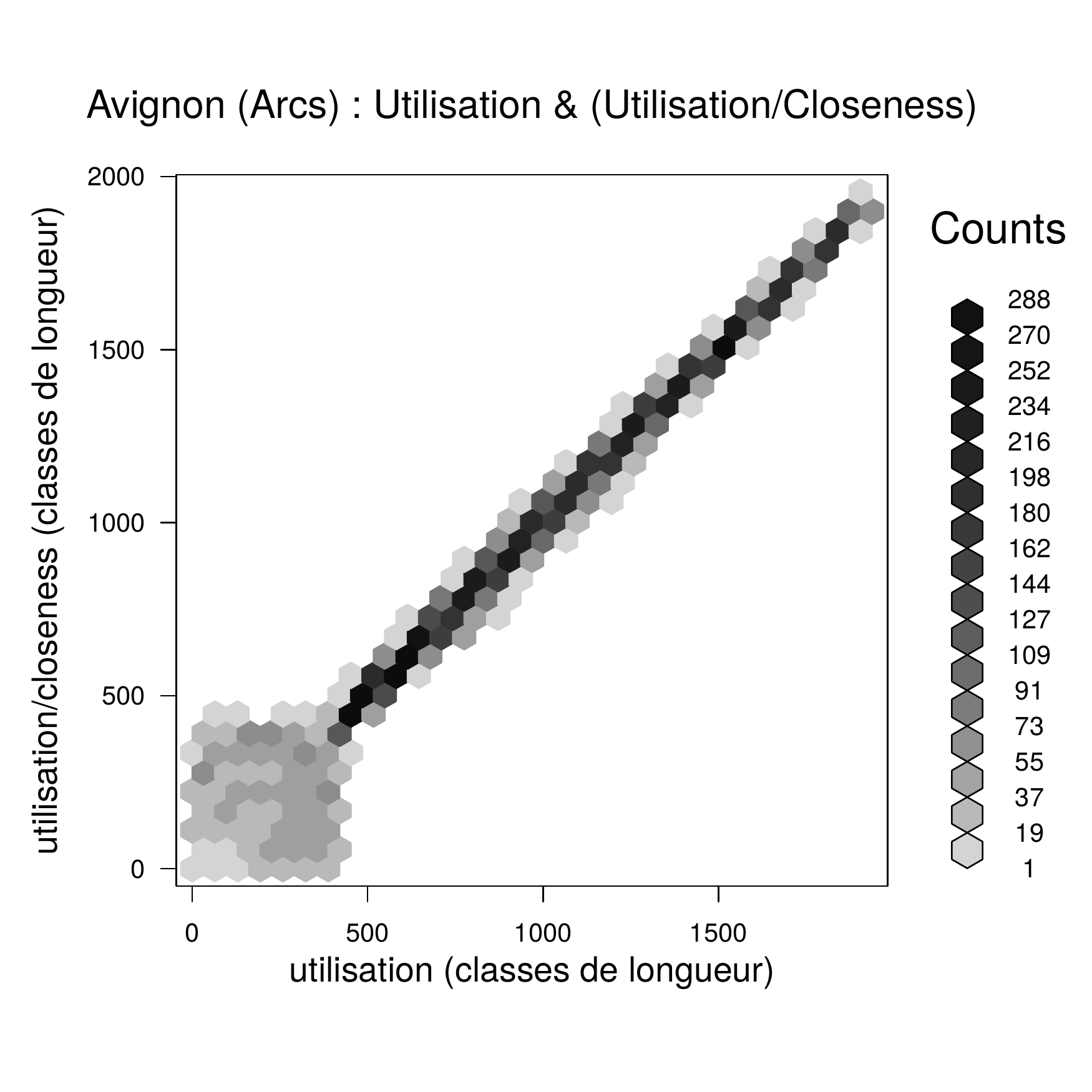}
    \end{subfigure}
   
   \begin{subfigure}{.40\textwidth}
        \includegraphics[width=\textwidth]{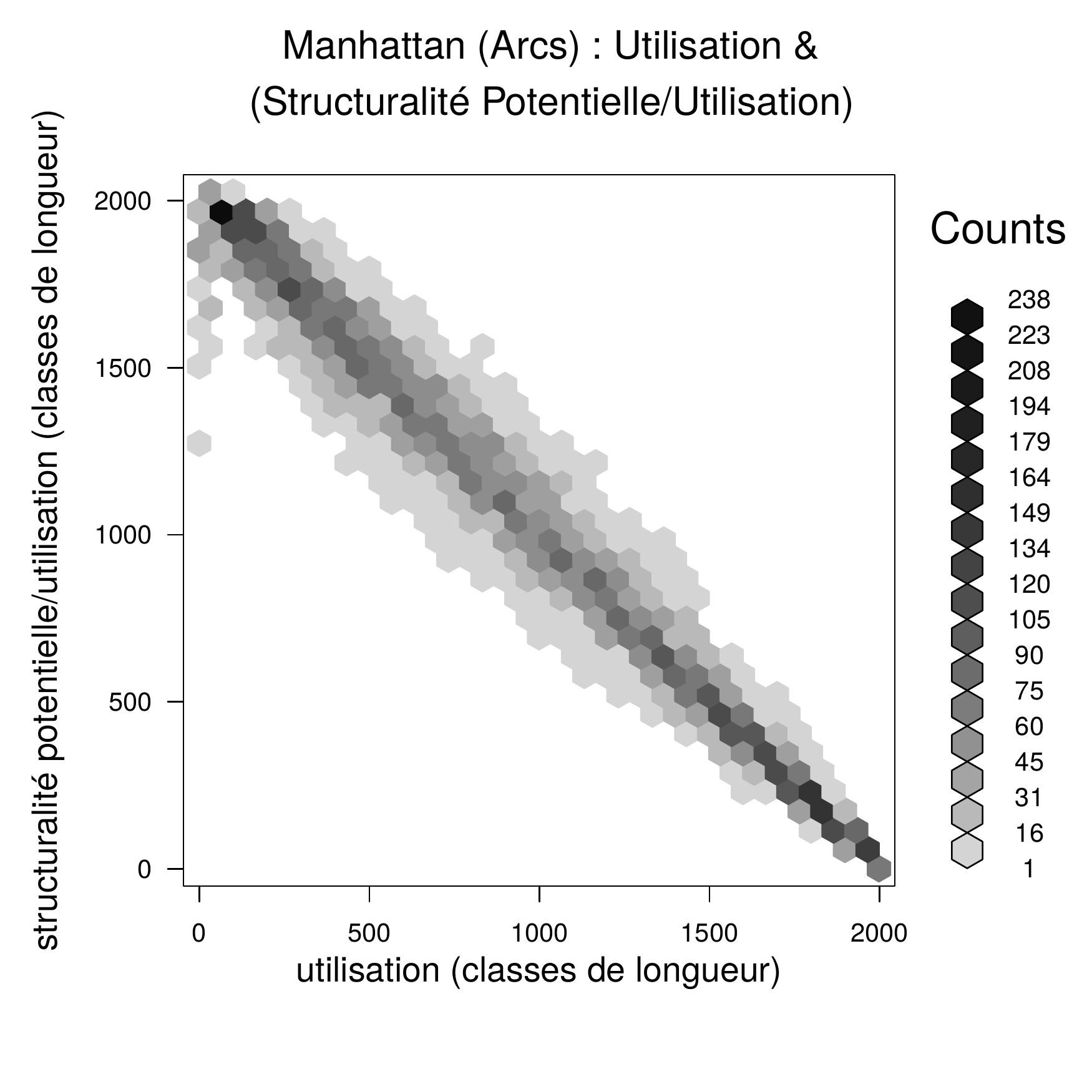}
    \end{subfigure}
    ~
    \begin{subfigure}{.40\textwidth}
        \includegraphics[width=\textwidth]{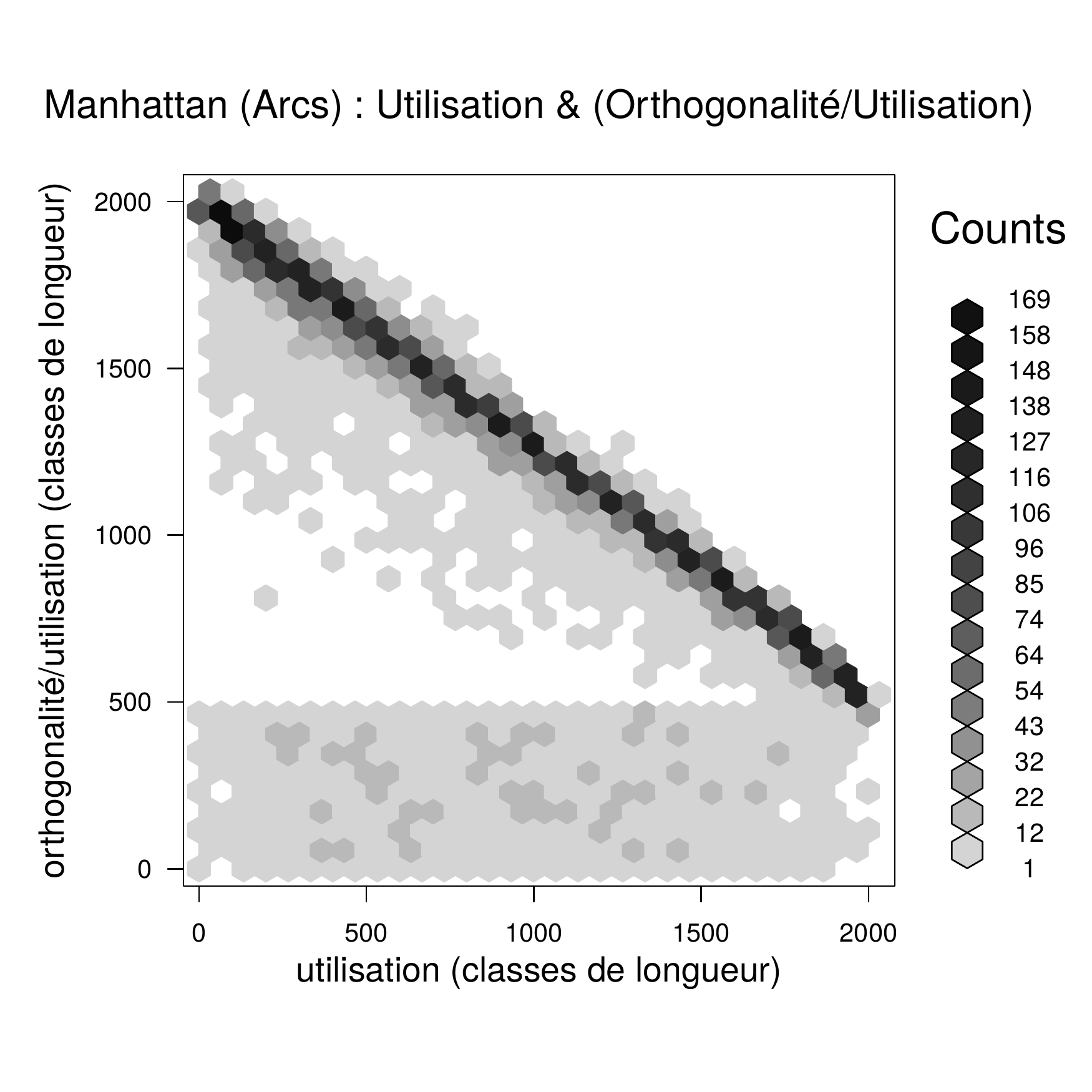}
    \end{subfigure}

    \caption{Cartes de corrélation croisée entre utilisation et utilisation combinée avec autres indicateurs par division sur les graphes viaires d'Avignon et de Manhattan. Les autres cartes de corrélations sont consultables en annexe \ref{ann:chap_cartes_corr}.}
    \label{fig:comp_arcs_useX}

\end{figure}

\begin{figure}[h]
    \centering

     \begin{subfigure}{.40\textwidth}
        \includegraphics[width=\textwidth]{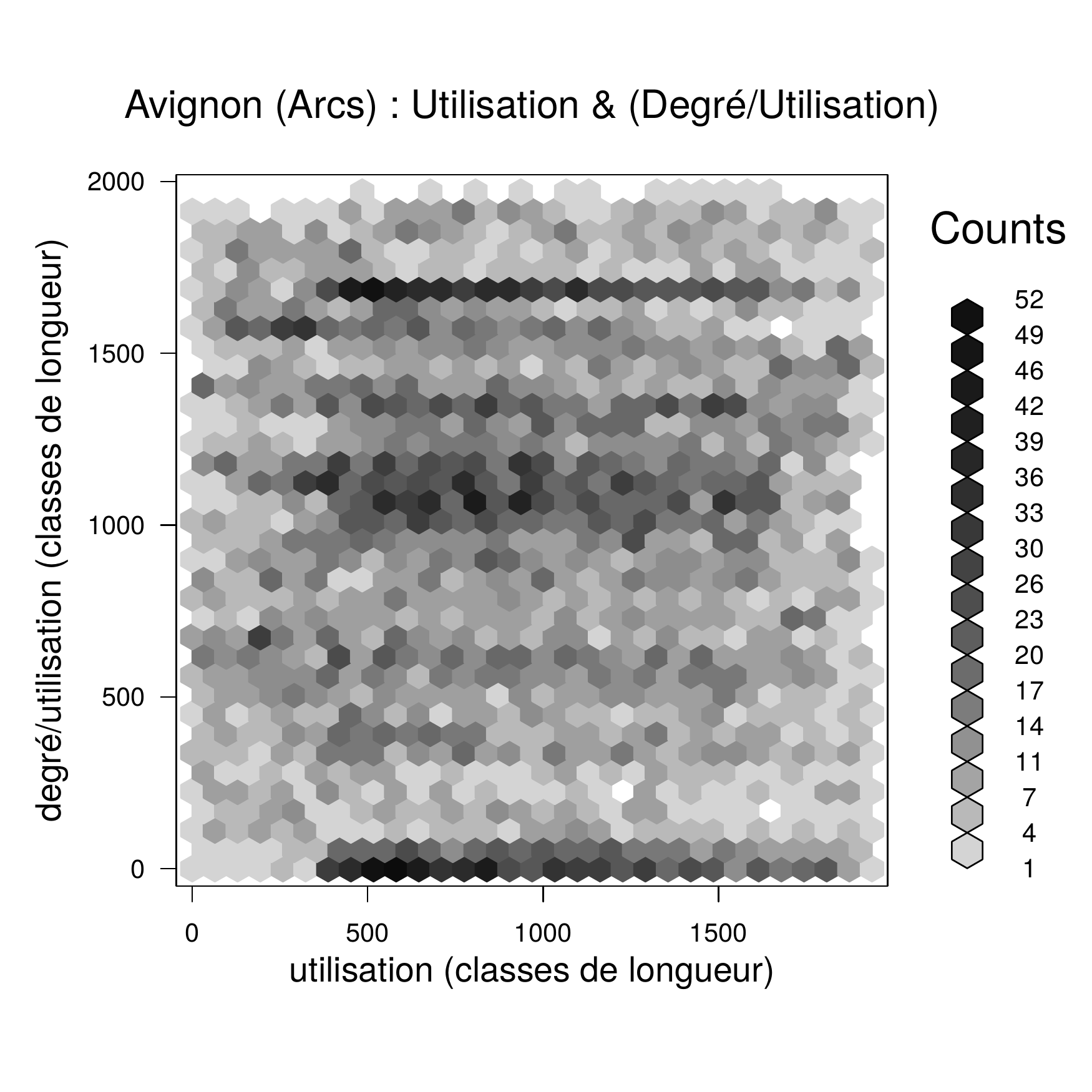}
    \end{subfigure}
    ~
    \begin{subfigure}{.40\textwidth}
        \includegraphics[width=\textwidth]{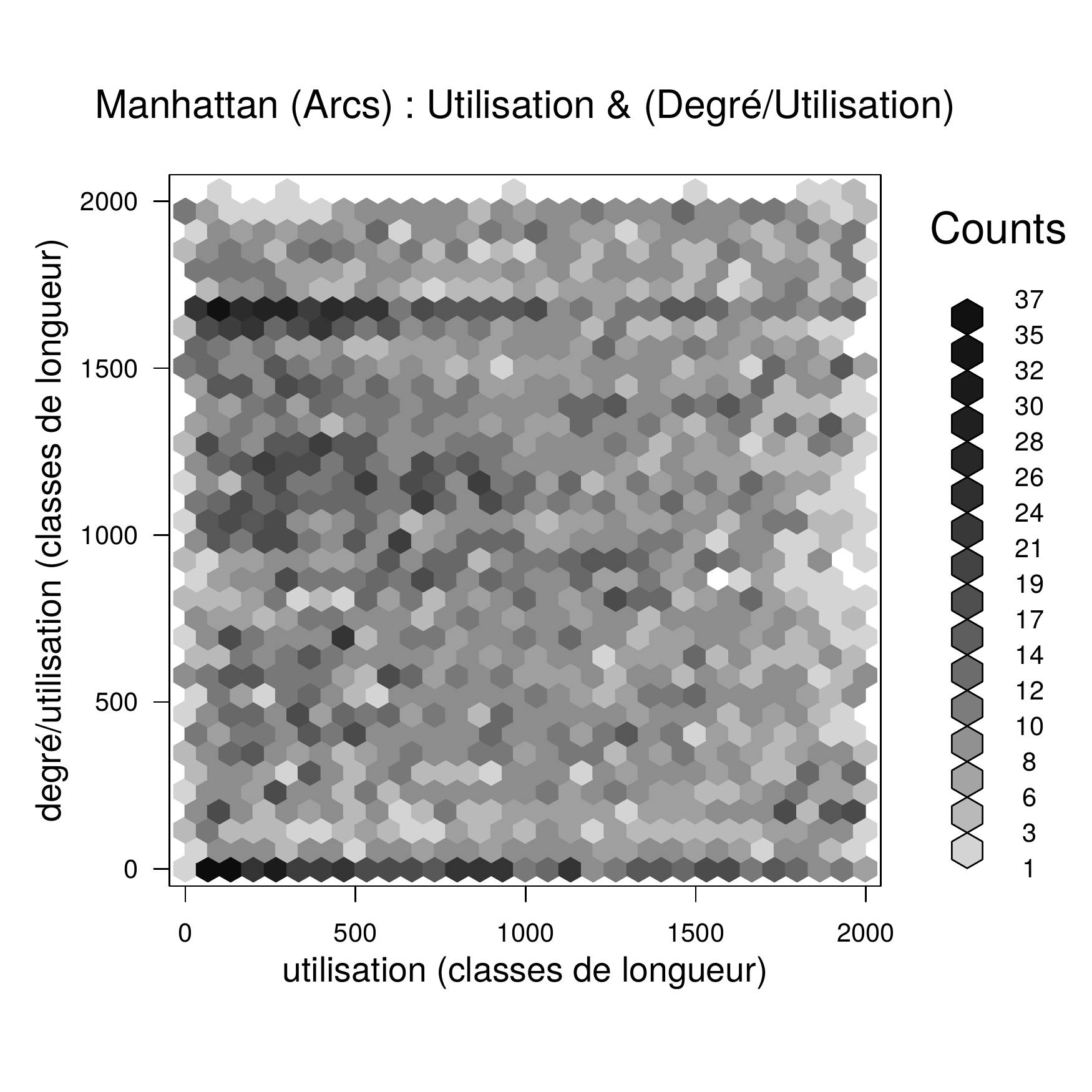}
    \end{subfigure}
    
    \caption{Cartes de corrélation croisée entre utilisation et utilisation combinée avec degré par division sur les graphes viaires d'Avignon et de Manhattan.}
    \label{fig:comp_arcs_usedou}

\end{figure}

\begin{figure}[h]
    \centering

     \begin{subfigure}{.40\textwidth}
        \includegraphics[width=\textwidth]{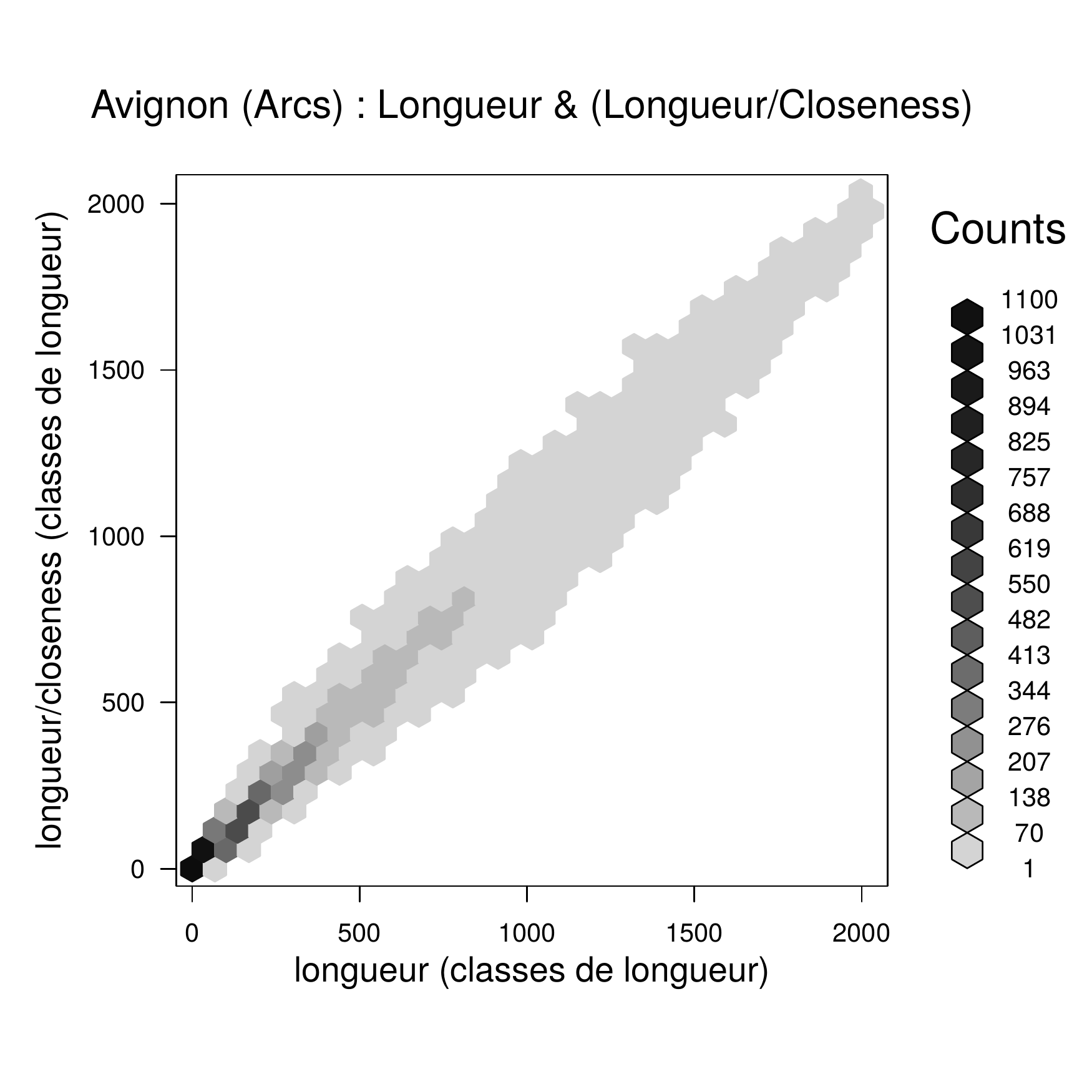}
    \end{subfigure}
    ~
    \begin{subfigure}{.40\textwidth}
        \includegraphics[width=\textwidth]{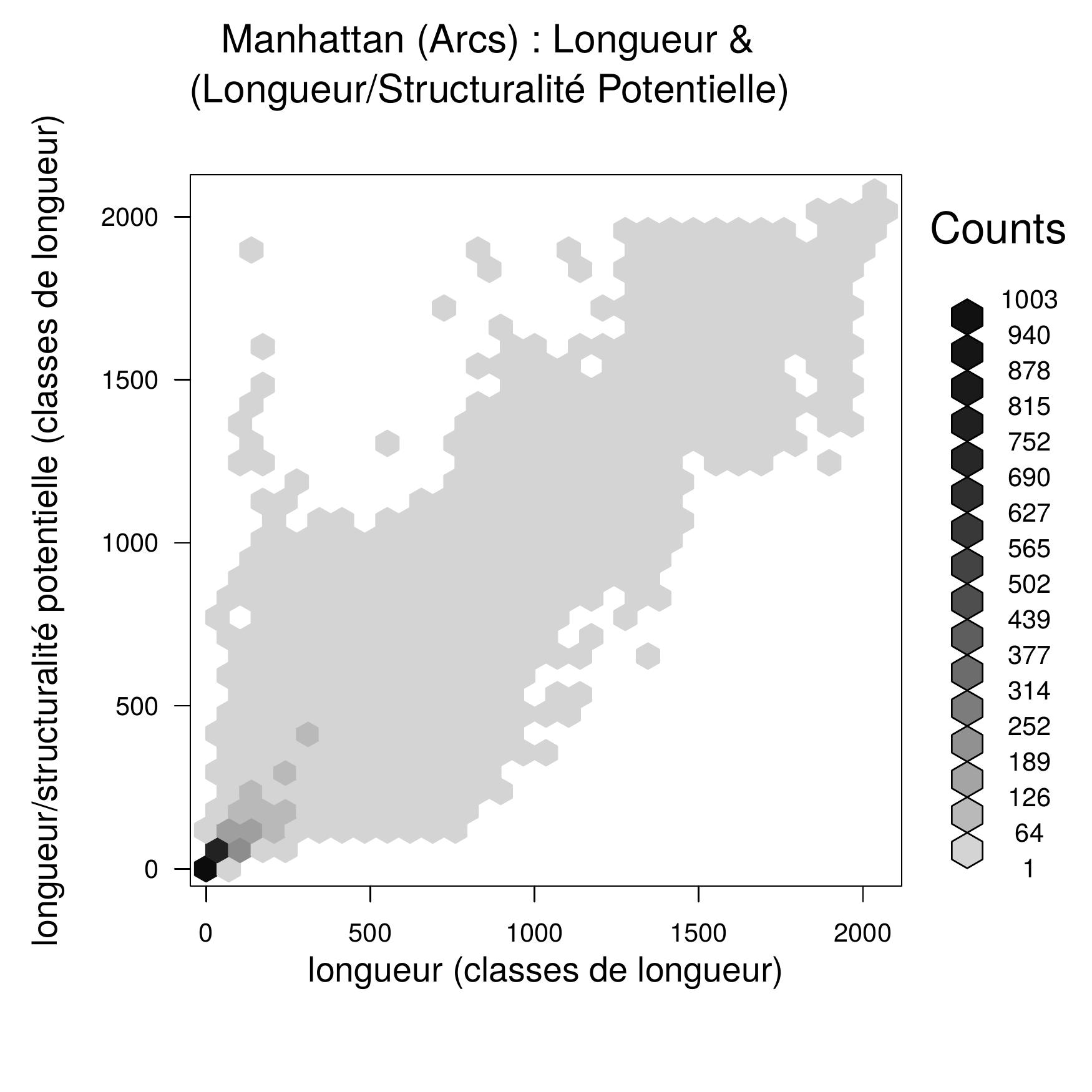}
    \end{subfigure}
    
    \caption{Cartes de corrélation croisée entre longueur et longueur combinée avec autres indicateurs par division sur les graphes viaires d'Avignon et de Manhattan. Les autres cartes de corrélations sont consultables en annexe \ref{ann:chap_cartes_corr}.}
    \label{fig:comp_arcs_lengthX}

\end{figure}

Les coefficients et cartes de corrélation montrent en deuxième lieu une prévalence de la longueur des arcs. En effet, celle-ci est corrélée à toutes les combinaisons dans lesquelles elle intervient (hors utilisation). Deux exemples sont reportés en figure \ref{fig:comp_arcs_lengthX}, illustrant la forte corrélation entre la longueur des arcs et les combinaisons faites avec celle-ci.

\begin{figure}[h]
    \centering

     \begin{subfigure}{.40\textwidth}
        \includegraphics[width=\textwidth]{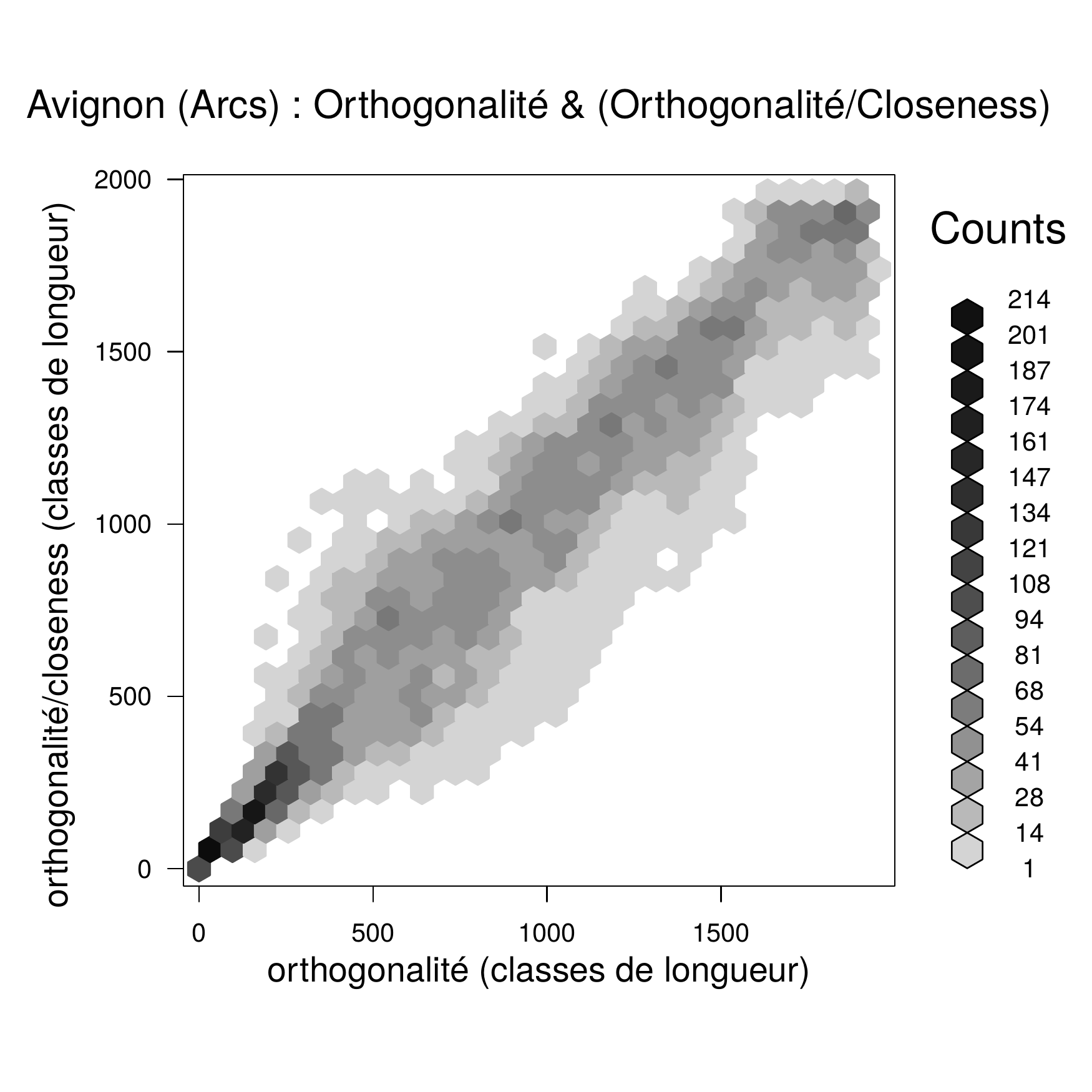}
    \end{subfigure}
    ~
    \begin{subfigure}{.40\textwidth}
        \includegraphics[width=\textwidth]{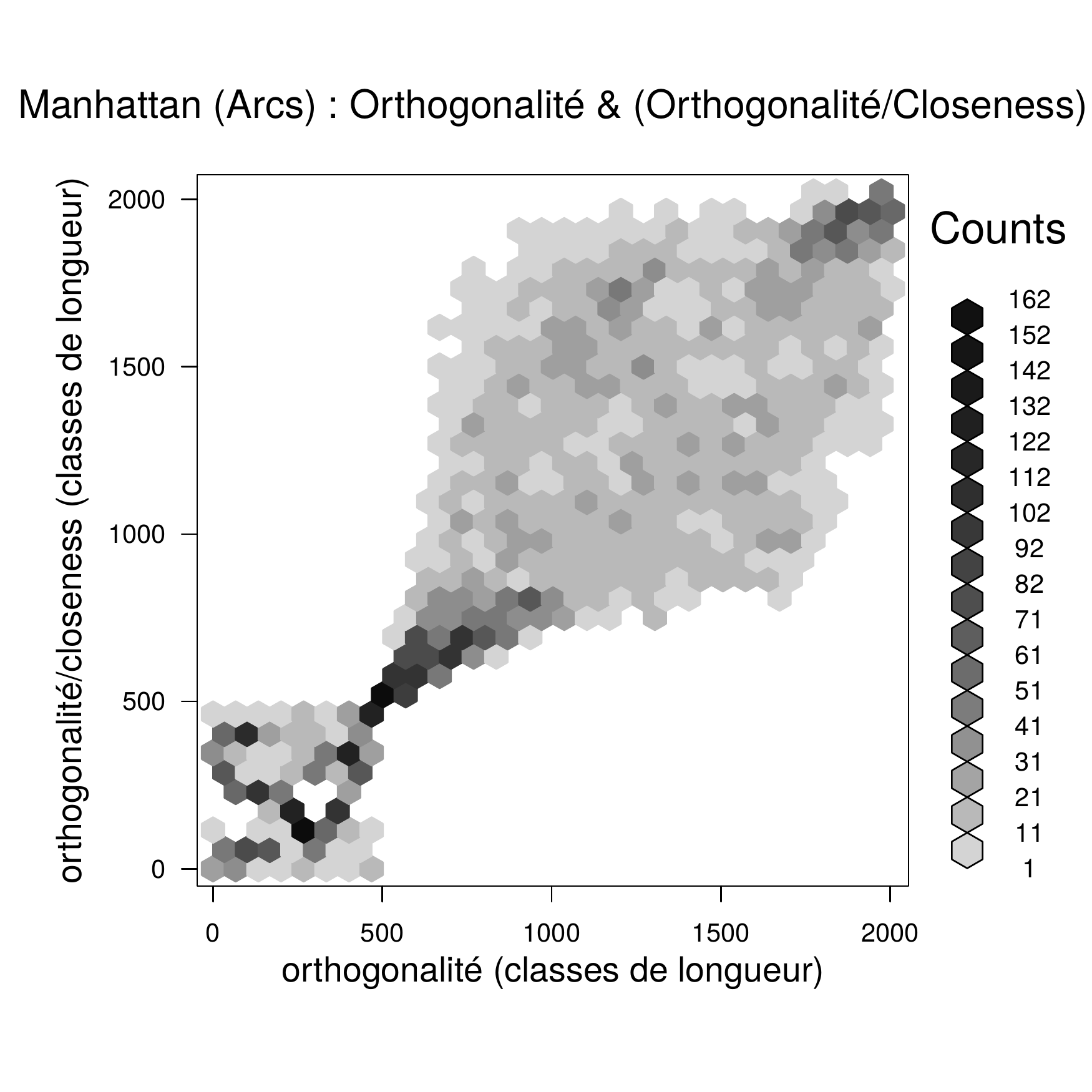}
    \end{subfigure}
    
    \caption{Cartes de corrélation croisée entre orthogonalité et orthogonalité combinée avec autres indicateurs par division sur les graphes viaires d'Avignon et de Manhattan. Les autres cartes de corrélations sont consultables en annexe \ref{ann:chap_cartes_corr}.}
    \label{fig:comp_arcs_orthoX}

\end{figure}

L'orthogonalité se positionne ensuite, car même si ses variations se plient à celles d'autres indicateurs, lorsqu'elle est combinée à l'utilisation ou la longueur, son quotient avec le degré, la structuralité potentielle ou la closeness lui est fortement corrélé (figure \ref{fig:comp_arcs_orthoX}).

\begin{figure}[h]
    \centering

    \begin{subfigure}{\textwidth}
        \includegraphics[width=\textwidth]{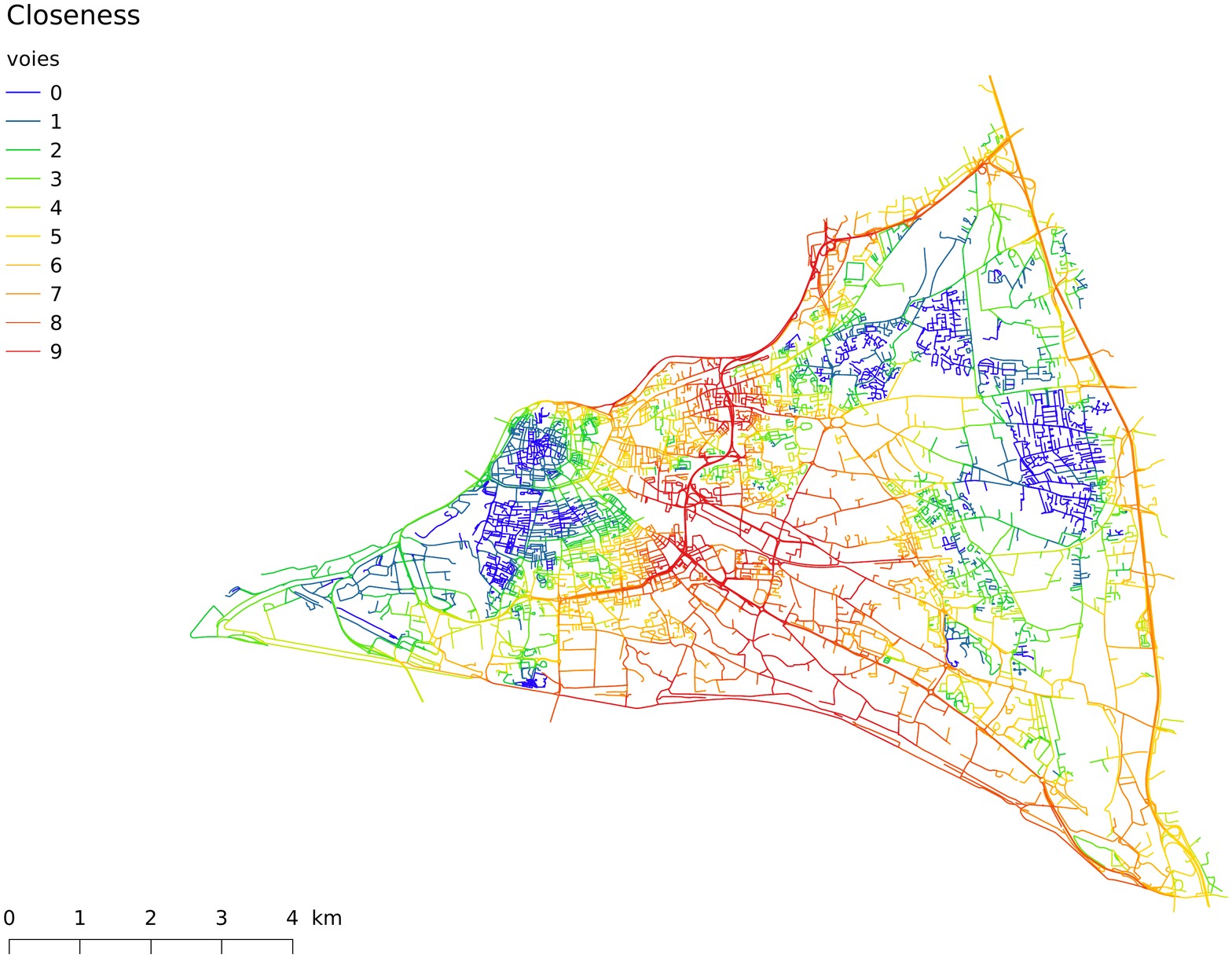}
        \caption{Avignon.}
    \end{subfigure}
    
    \begin{subfigure}{\linewidth}
        \includegraphics[width=\textwidth]{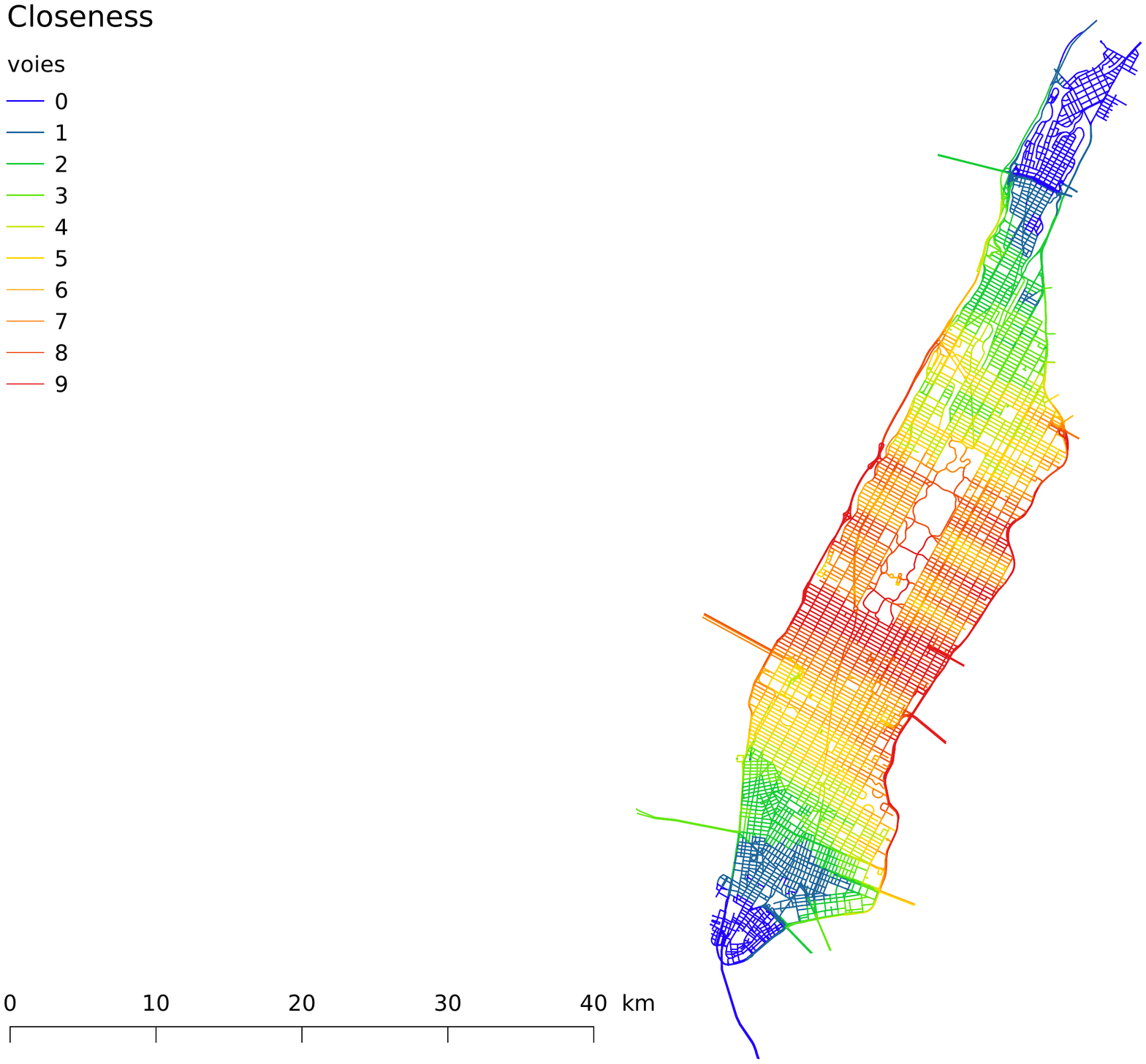}
        \caption{Manhattan.}
    \end{subfigure}

    \caption{Indicateur de closeness calculé sur les arcs des graphes viaires d'Avignon et de Manhattan.}
    \label{fig:arcs_carte_closeness}

\end{figure}

\begin{figure}[h]
    \centering

     \begin{subfigure}{.40\textwidth}
        \includegraphics[width=\textwidth]{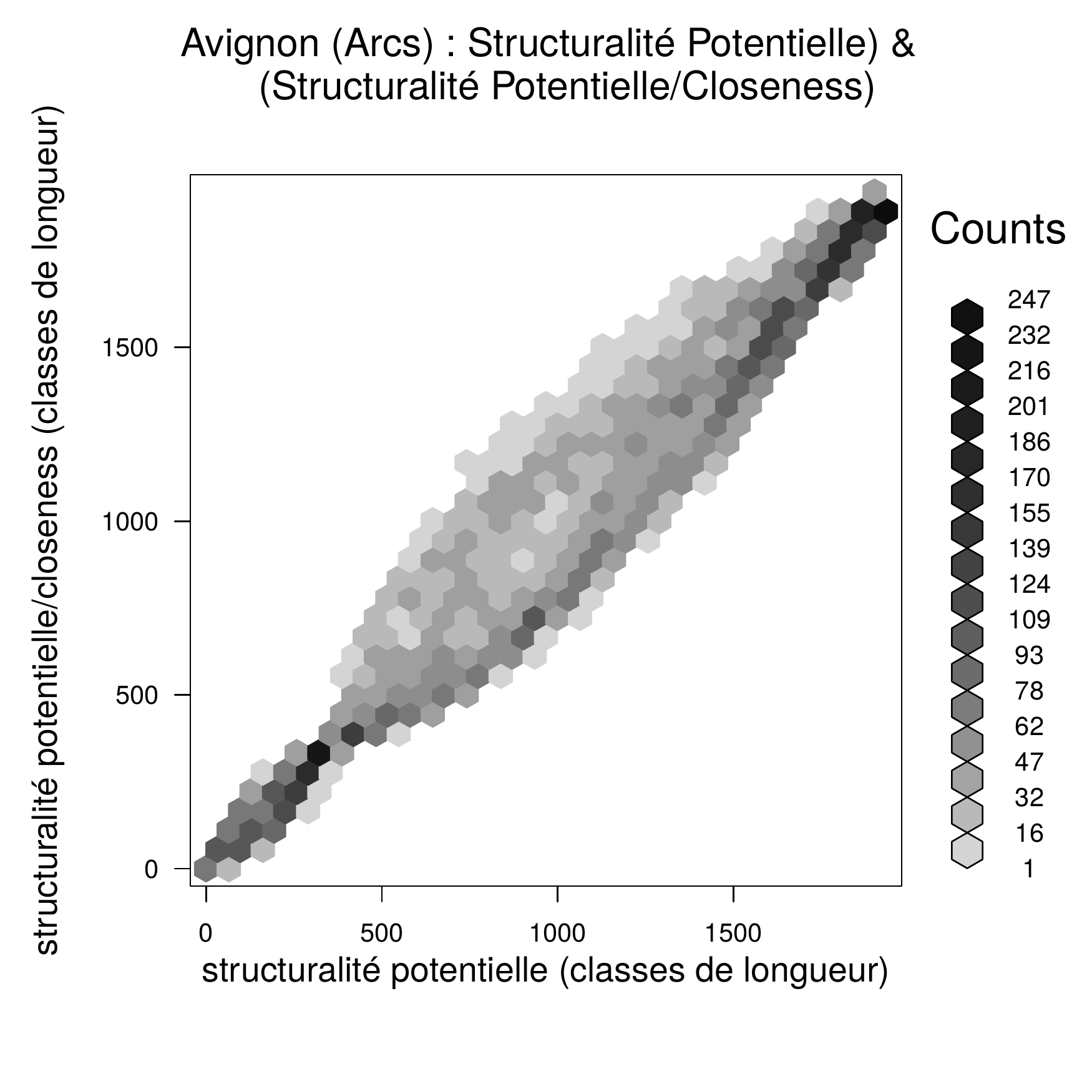}
    \end{subfigure}
    ~
    \begin{subfigure}{.40\textwidth}
        \includegraphics[width=\textwidth]{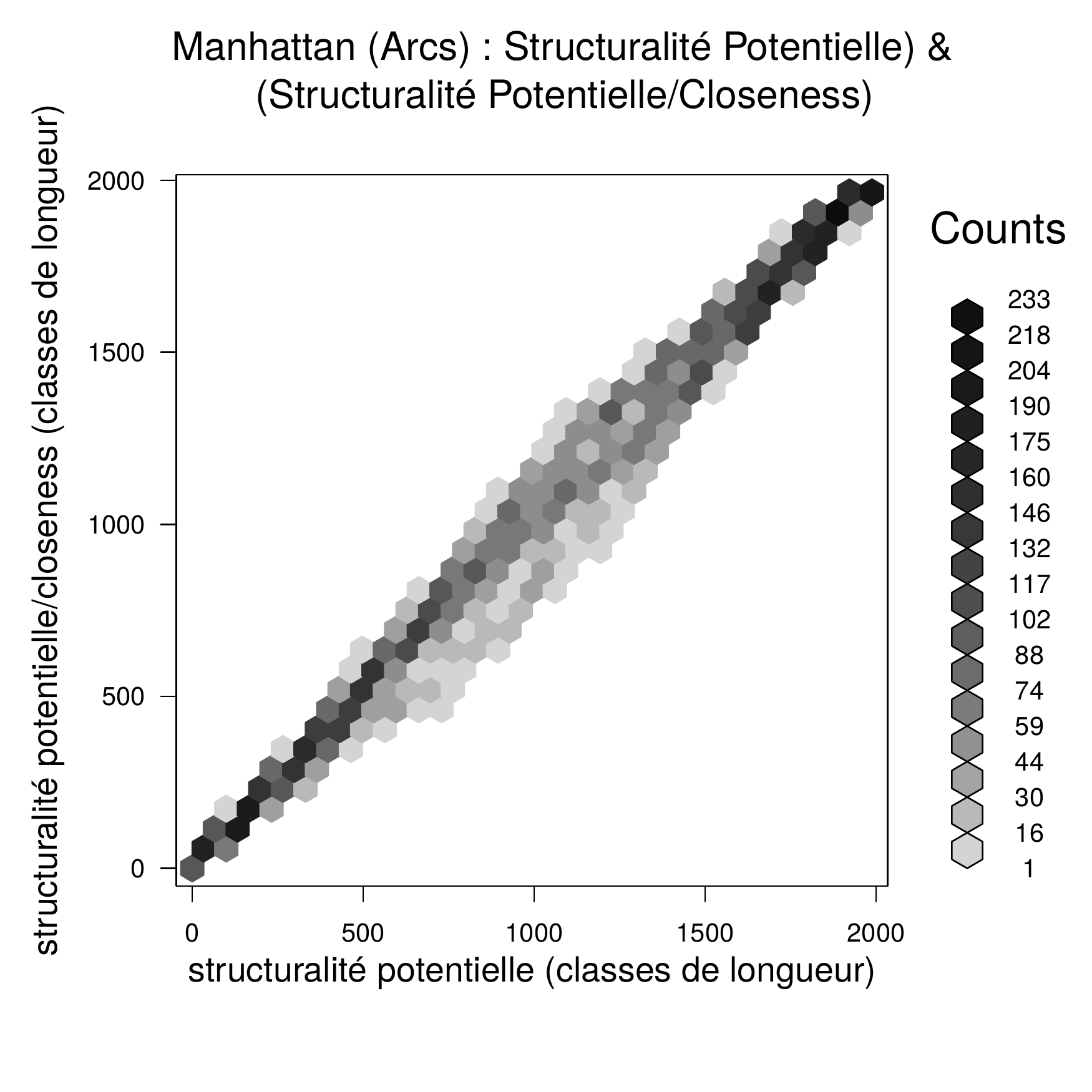}
    \end{subfigure}
    
    \caption{Cartes de corrélation croisée entre structuralité potentielle et structuralité potentielle combinée avec la closeness par division sur les graphes viaires d'Avignon et de Manhattan.}
    \label{fig:comp_arcs_structpotsoc}

\end{figure}

Les deux indicateurs dont les combinaisons se plient aux variations de l'indicateur auquel ils sont associés sont la structuralité potentielle et la closeness. En effet, leur calcul est fondé sur celui de distances topologiques. Lorsque celui-ci est effectué sur les arcs, l'importance des effets de bord est perceptible à l’œil nu (figures \ref{fig:arcs_carte_degrestructpot_av}, \ref{fig:arcs_carte_degrestructpot_man} \ref{fig:arcs_carte_closeness}). Ils sont influencés par la position du centre du graphe. Ce sont donc les indicateurs les moins stables, face à ceux calculés localement, comme la longueur ou l'orthogonalité. L'utilisation, en comptant le nombre de chemins les plus simples passant par chaque arc, prédomine dans sa combinaison avec toute autre caractérisation. Lorsque nous les combinons l'un à l'autre, c'est la structuralité potentielle qui se révèle être prégnante (figure \ref{fig:comp_arcs_structpotsoc}).

\begin{figure}[h]
    \centering

     \begin{subfigure}{.40\textwidth}
        \includegraphics[width=\textwidth]{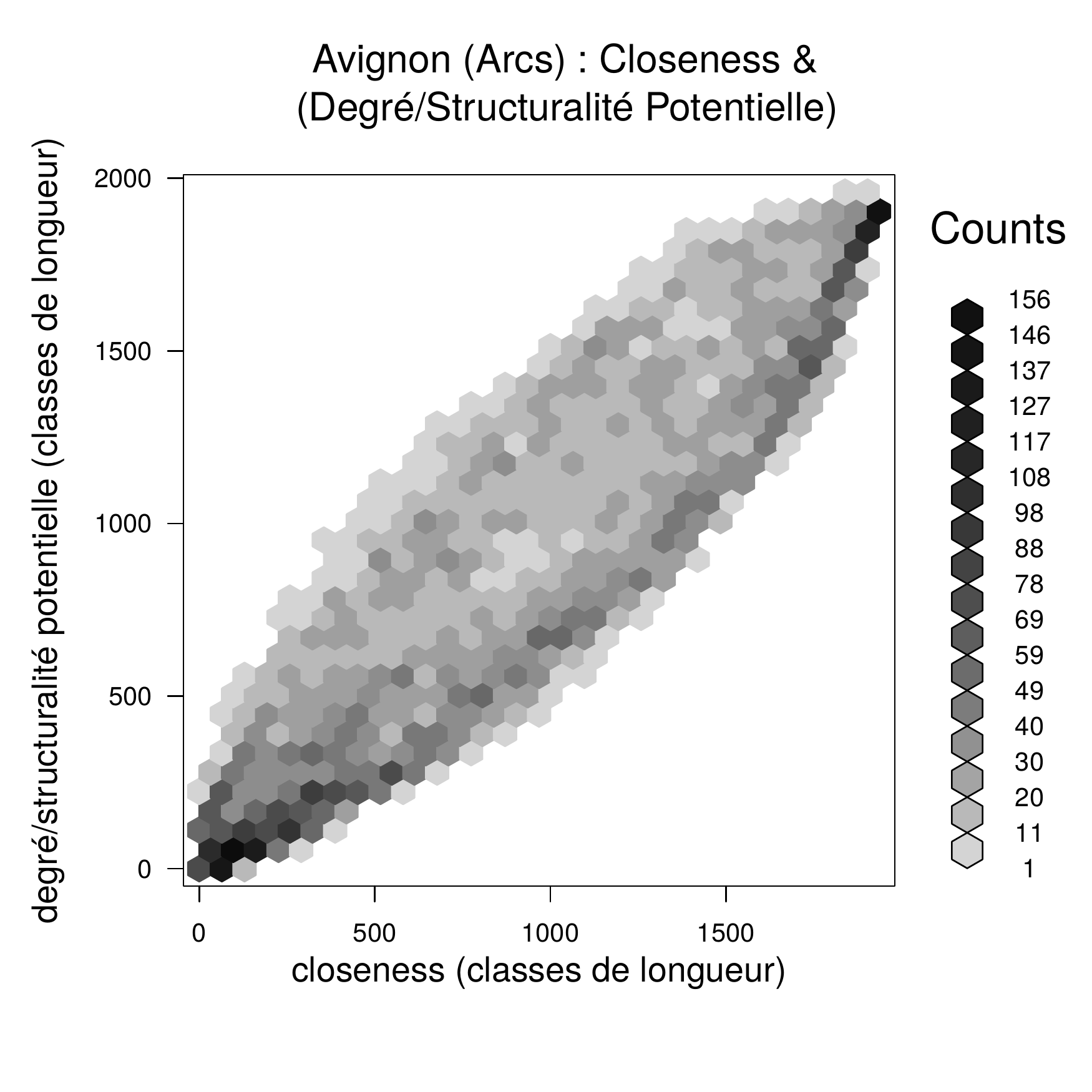}
    \end{subfigure}
    ~
    \begin{subfigure}{.40\textwidth}
        \includegraphics[width=\textwidth]{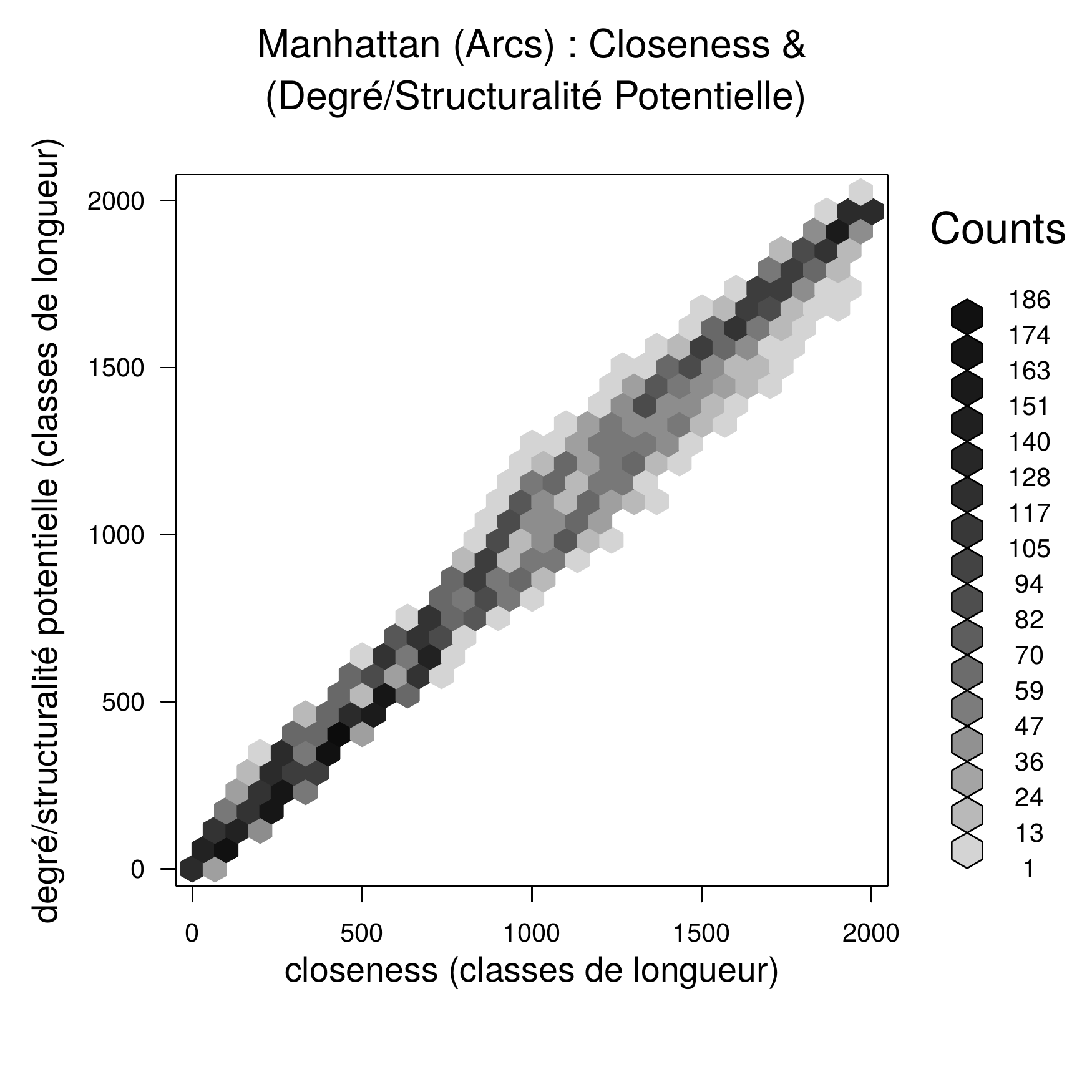}
    \end{subfigure}
    
    \caption{Cartes de corrélation croisée entre closeness et degré combiné avec la structuralité potentielle par division sur les graphes viaires d'Avignon et de Manhattan.}
    \label{fig:comp_arcs_degreestructpotclo}

\end{figure}

Lorsque l'on combine l'indicateur de degré avec l'un ou l'autre de ces deux derniers indicateurs, nous retrouvons le troisième. Ainsi le quotient $\frac{degre}{structuralitePotentielle}$ donne un résultat corrélé à la closeness et inversement (figure \ref{fig:comp_arcs_degreestructpotclo}). Nous pouvons en déduire que pour les arcs, l'accessibilité de son voisinage direct est équivalent à son propre rayon topologique. En effet, la closeness est l'inverse du rayon topologique de l'arc (cf \ref{eq:clo}) et la structuralité potentielle est la somme des accessibilités des arcs connectés à l'arc considéré (cf \ref{eq:structpot}). Cette somme a une valeur qui dépend du degré de l'arc car c'est celui-ci qui donne le nombre d'arcs connectés $a \wedge a_{ref}$. Ce qui nous donne l'équivalence \ref{eq:clostructpot}. Ce résultat est logique du fait du faible impact de l'ajout de la longueur dans le calcul des distances topologiques, démontré dans le paragraphe précédent. En effet, les arcs connectés à l'arc de référence n'ont une différence de distance topologique avec le reste du réseau que de 1 par rapport à $a_{ref}$.

\begin{equation}
closeness(a_{ref}) = \frac{1}{\sum_{\forall a \in G} dtopo(a_{ref},a)}
\label{eq:clo}
\end{equation}

\begin{equation}
structuralitePotentielle(a_{ref}) = \sum_{\forall a \cap a_{ref}}accessibilite(a)
\label{eq:structpot}
\end{equation}

\begin{equation}
degre \times \sum_{\forall a \in G} dtopo(a_{ref},a) \sim \sum_{\forall a_i \cap a_{ref}} [ \sum_{\forall a \in G} dtopo(a_i,a) \times longueur(a) ]
\label{eq:clostructpot}
\end{equation}

Ces calculs nous permettent de hiérarchiser les indicateurs primaires indépendants suivant leur caractère dominant dans les corrélations :
\begin{enumerate}
\item utilisation (équivalent à la betweenness)
\item longueur
\item orthogonalité
\item degré
\item structuralité potentielle
\item closeness (équivalent à l'accessibilité)
\end{enumerate}

Nous pouvons conclure qu'une seule combinaison par division de ces indicateurs apporte une information complémentaire à celles qu'ils nous fournissent : celle du degré et de l'utilisation. En divisant le nombre de connexions d'un arc avec le nombre de chemins les plus simples passant par celui-ci nous obtenons un indice permettant de différencier les arcs de fort degré qui sont peu utilisés de ceux d'un faible degré qui le sont beaucoup. Cela mettra en avant les arcs les plus vulnérables dans le réseau, ceux pour lesquels, si une connexion est coupée, un fort impact sur les distances géodésiques sera observé.


\FloatBarrier
\section{Analyse et comparaison des indicateurs sur les voies}

Un des axes principaux de la thèse que nous exposons ici est de construire une méthodologie pour s'émanciper de l'arc afin de construire un objet plus robuste aux découpages du réseau et plus significatif dans l'information qu'il porte. Nous avons donc décrit dans la partie précédente la construction de la voie, qui nous permet de construire à l'échelle locale un objet capable de traverser le réseau de part en part. Nous voulons donc élargir l'étude précédente faite sur les arcs à cet objet géographique afin de déterminer son potentiel d'extraction d'informations globales sur le réseau.

\subsection{Indicateurs primaires}

Nous utilisons la même méthodologie que celle décrite pour les arcs. Nous réalisons la classification explicitée en début de chapitre pour les voies, mais en réduisant le nombre de classes en conséquence (tableau \ref{tab:pres_ville2}). Nous étudions ici, en plus des indicateurs primaires évoqués pour les arcs, le nombre d'arcs par voies et la connectivité de chaque voie (telle que décrite dans le chapitre précédent).

La voie étant un objet complexe, nous ajoutons à notre étude deux réseaux supplémentaires : ceux de Paris et Barcelone. Comme évoqué en début de ce chapitre, ces réseaux ont des structures particulières qui permettent d'insérer des cas d'analyse entre les deux extrêmes que représentent Avignon et Manhattan. Cela nous permet également d'ajouter deux réseaux avec un nombre de voies plus important. Il nous sera ainsi possible de conclure plus justement de la corrélation ou non-corrélation de deux indicateurs sur les voies. Nous reportons le détail des coefficients de Pearson des calculs de corrélation dans les tableaux en annexe \ref{tab:corr_avignon_voie}, \ref{tab:corr_manhattan_voie}, \ref{tab:corr_paris_voie} et \ref{tab:corr_barcelone_voie}.

\begin{figure}[h]
    \centering
    
    \begin{subfigure}{.35\textwidth}
        \includegraphics[width=\textwidth]{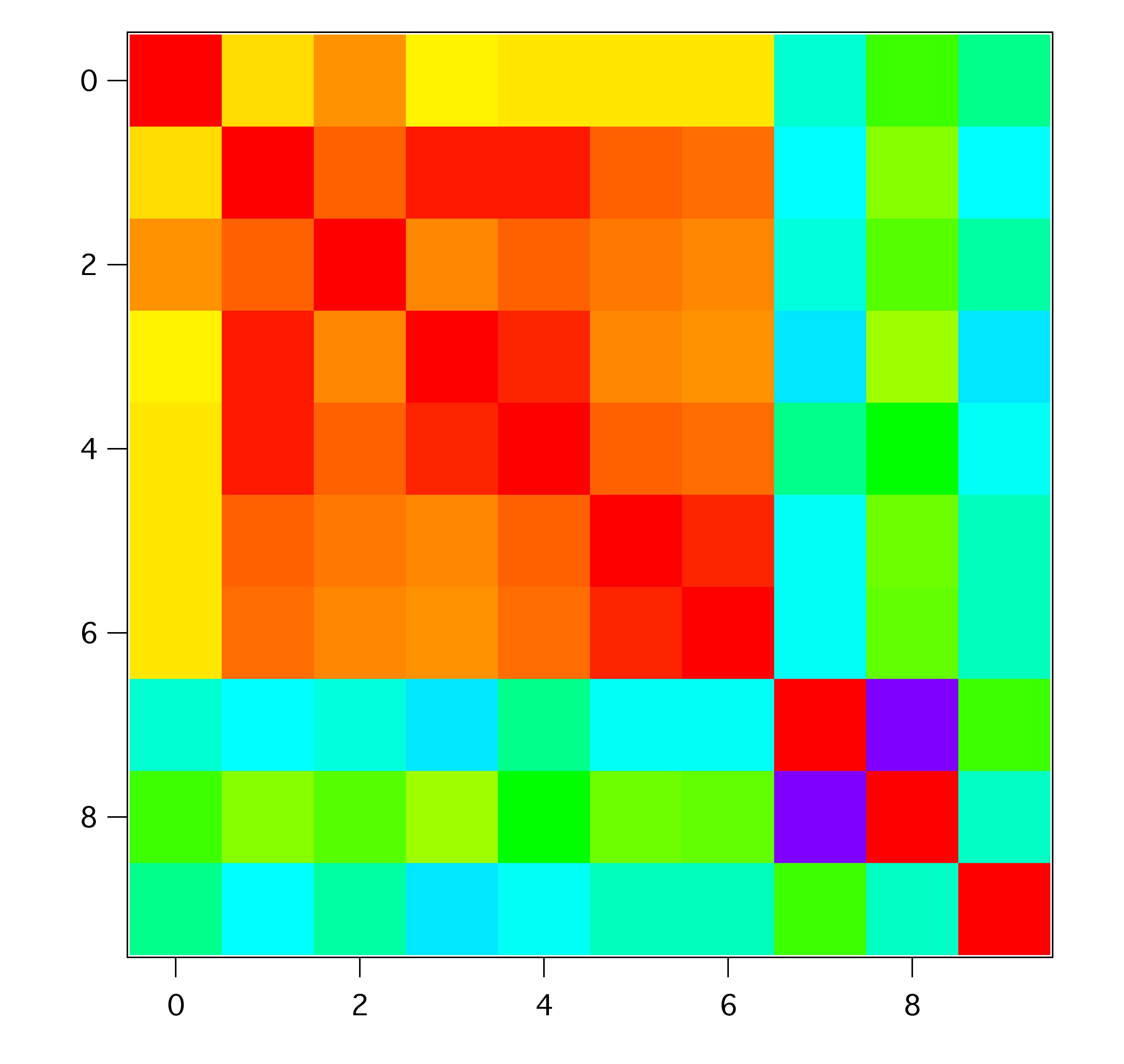}
        \caption{Représentation du tableau \ref{tab:corr_avignon_voie}, calculé sur les voies d'Avignon.}
    \end{subfigure}
    ~
    \begin{subfigure}{.35\linewidth}
        \includegraphics[width=\textwidth]{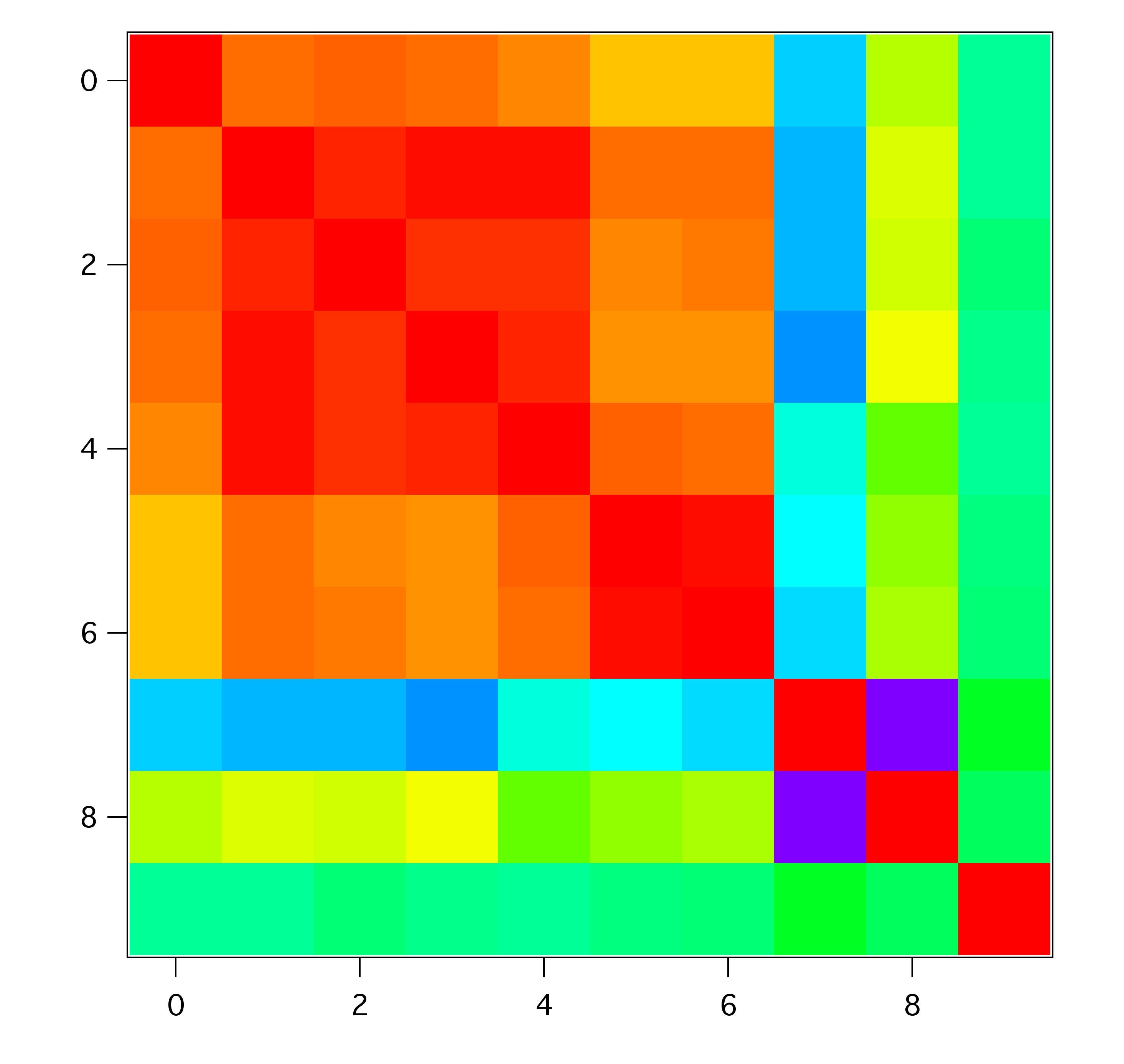}
        \caption{Représentation du tableau \ref{tab:corr_barcelone_voie}, calculé sur les voies de Barcelone.}
    \end{subfigure}
    
    \begin{subfigure}{.35\textwidth}
        \includegraphics[width=\textwidth]{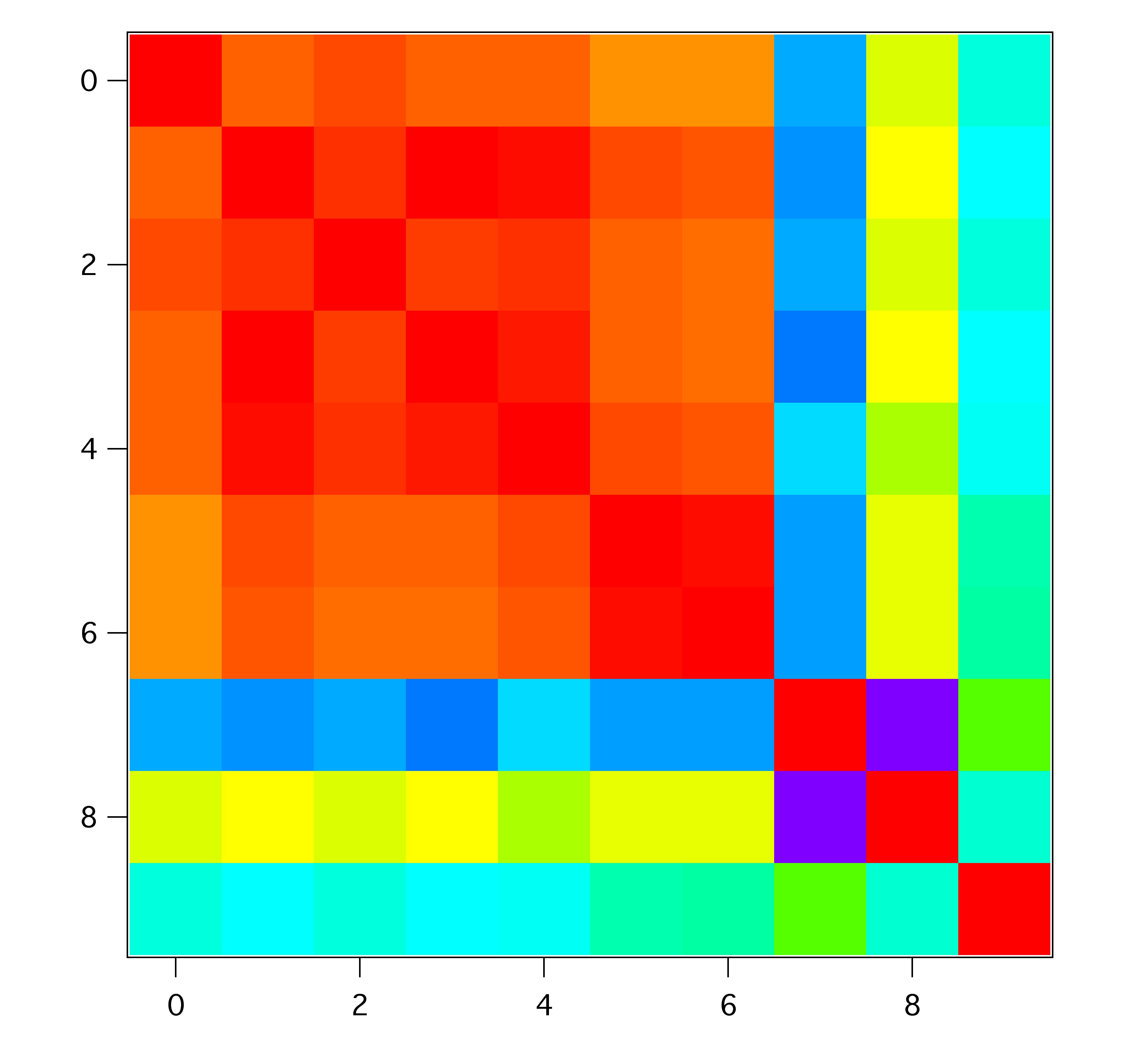}
        \caption{Représentation du tableau \ref{tab:corr_paris_voie}, calculé sur les voies de Paris.}
    \end{subfigure}
    ~
    \begin{subfigure}{.35\linewidth}
        \includegraphics[width=\textwidth]{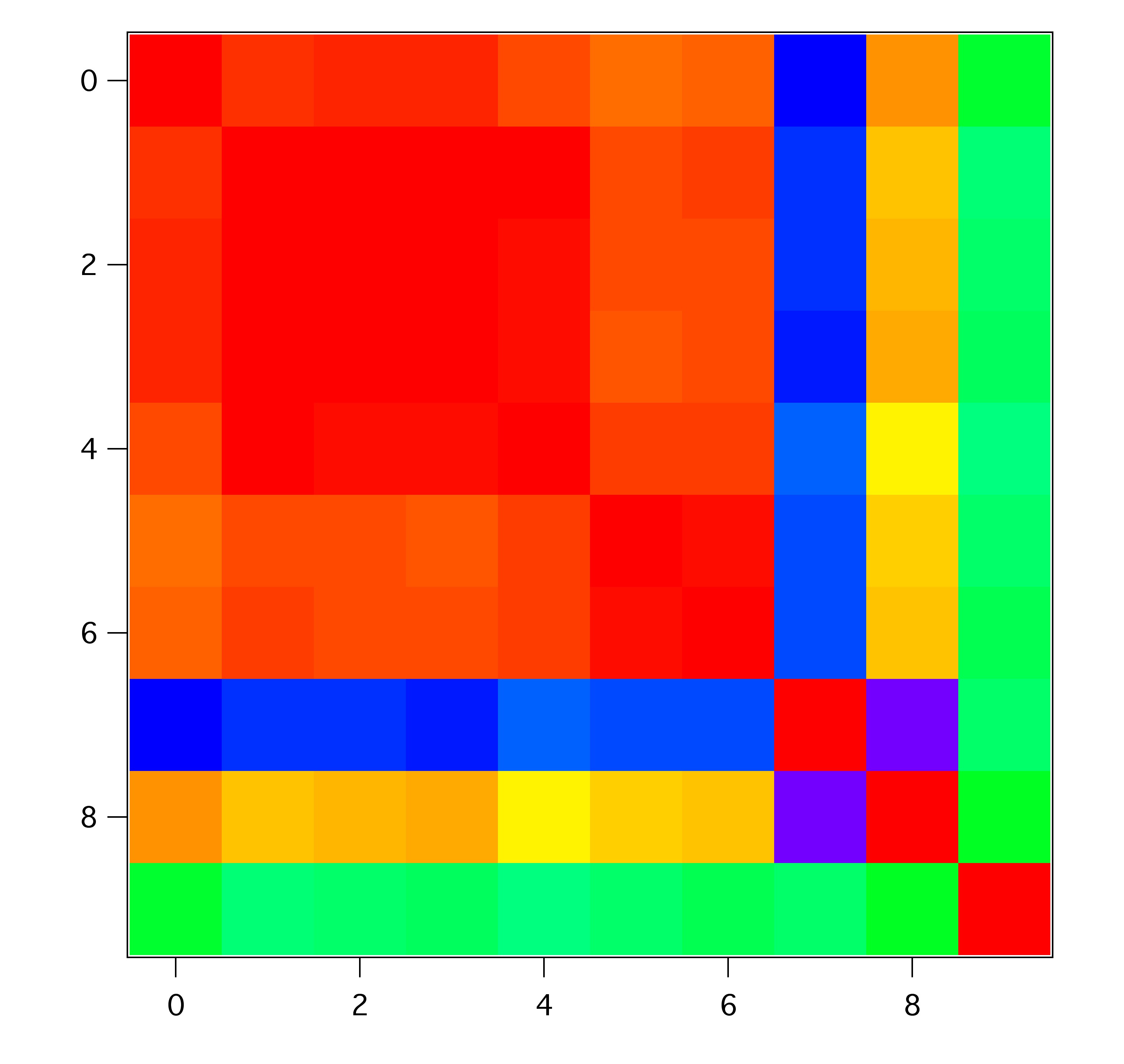}
        \caption{Représentation du tableau \ref{tab:corr_manhattan_voie}, calculé sur les voies de Manhattan.}
    \end{subfigure}
    ~
    \begin{subfigure}{.2\linewidth}
        \includegraphics[width=\textwidth]{images/matrices_colorees/mat_echelle_dec.jpg}
        \caption{Échelle}
    \end{subfigure}

    \caption{Représentation de la corrélation entre indicateurs primaires calculés sur les voies sous forme de matrice colorée. Rappel de l'ordre des indicateurs : 0 : longueur ; 1 : degré ; 2 : nombre d'arcs ; 3 : connectivité ; 4 : structuralité potentielle ; 5: betweenness ; 6 : utilisation ; 7 : accessibilité ; 8 : closeness ; 9 : orthogonalité}
    \label{fig:mat_prim_voies}

\end{figure}

La figure de corrélation que nous obtenons (figure \ref{fig:mat_prim_voies}) montre que la voie apporte une unification tangible dans l'information calculée. Elle permet de rendre équivalents des indicateurs qui apportaient une caractérisation différente sur les arcs. Son caractère multi-échelle allie des indicateurs locaux (comme le degré) à d'autres calculés globalement (comme l'utilisation).

Sur les quatre réseaux, nous observons un \enquote{bloc} d'indicateurs corrélés : longueur, degré, nombre d'arcs par voies, connectivité des voies, structuralité potentielle, betweenness et utilisation ne représentent pour la voie qu'une unique caractérisation, plus ou moins contrastée selon la méthode de calcul retenue. Se détachent de cet ensemble corrélé, la closeness (qui reste comme pour les arcs extrêmement proche du calcul d'accessibilité) et l'orthogonalité qui apparaît comme très spécifique. Si nous suivons les représentations des corrélations sous forme de matrices colorées (figure \ref{fig:mat_prim_voies}), un ordre de \textit{régularité} se dessine entre les quatre villes : Avignon, Barcelone, Paris et Manhattan (du moins au plus régulier). Dans les cas les plus réguliers, comme celui de Manhattan, les corrélations sont plus fortes, car la simplicité de la structure du réseau entraîne une redondance des informations.

\begin{figure}[h]
    \centering

    \begin{subfigure}{.40\textwidth}
        \includegraphics[width=\textwidth]{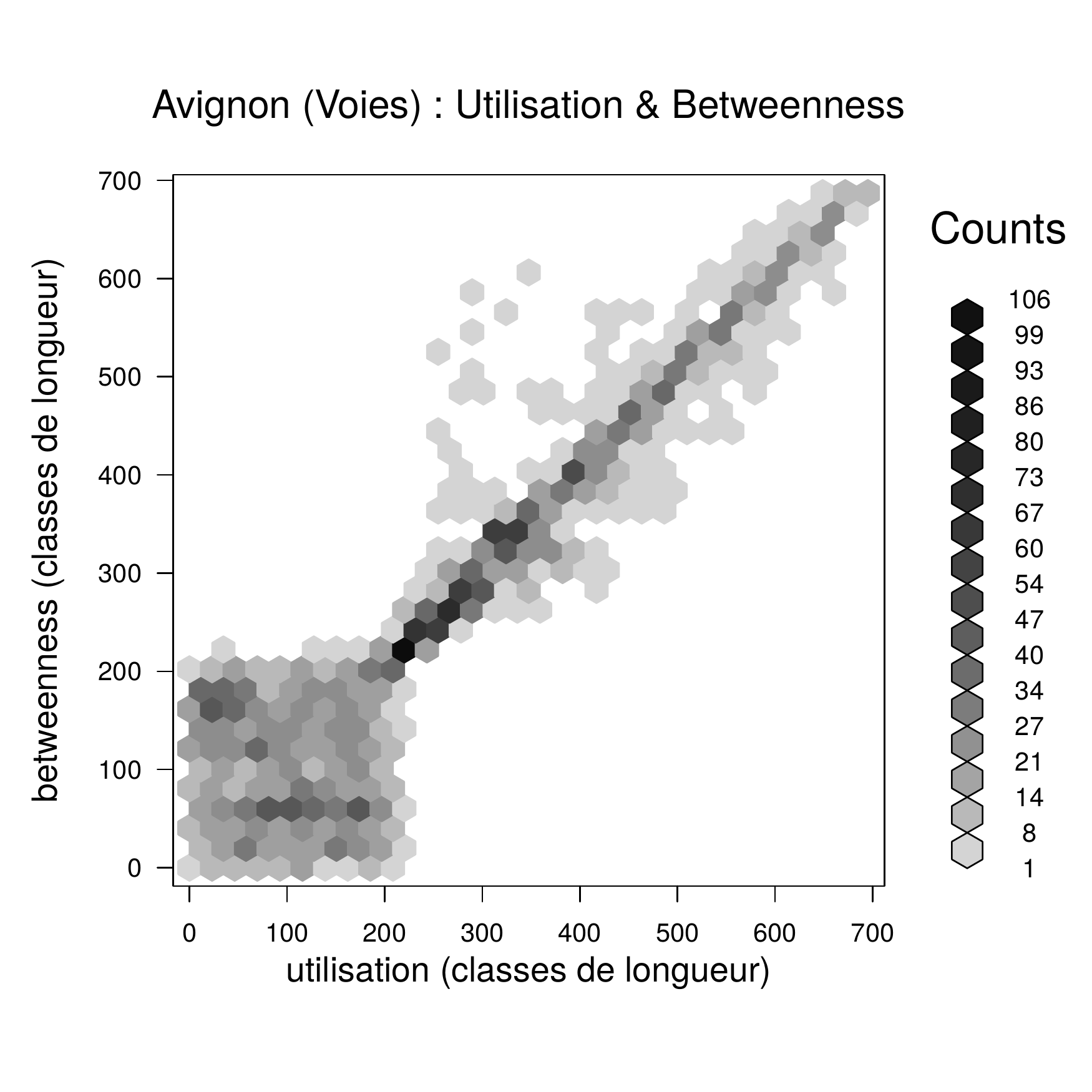}
        \caption{Sur le graphe viaire d'Avignon.}
    \end{subfigure}
    ~
     \begin{subfigure}{.40\textwidth}
        \includegraphics[width=\textwidth]{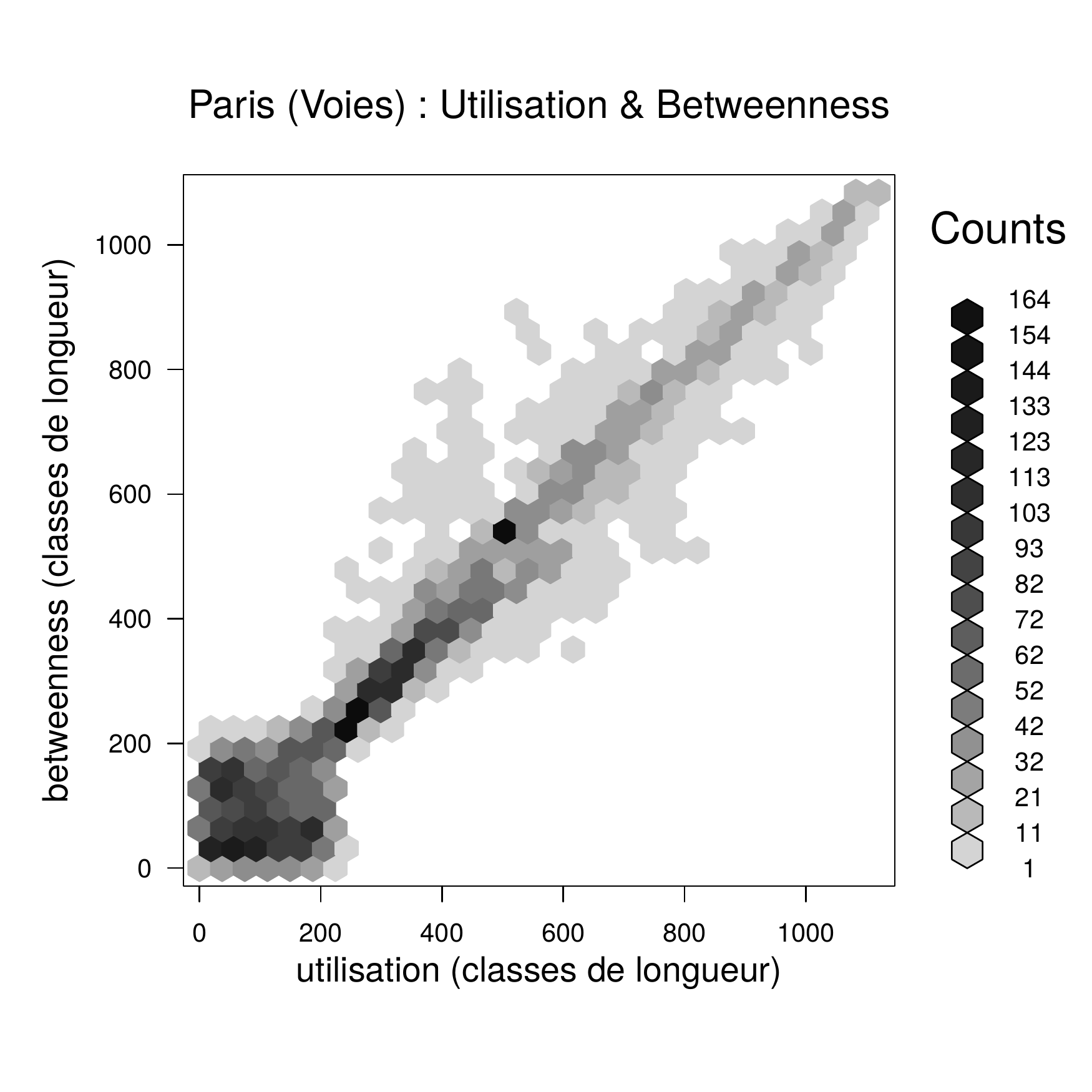}
        \caption{Sur le graphe viaire de Paris.}
    \end{subfigure}
    
    \begin{subfigure}{.40\textwidth}
        \includegraphics[width=\textwidth]{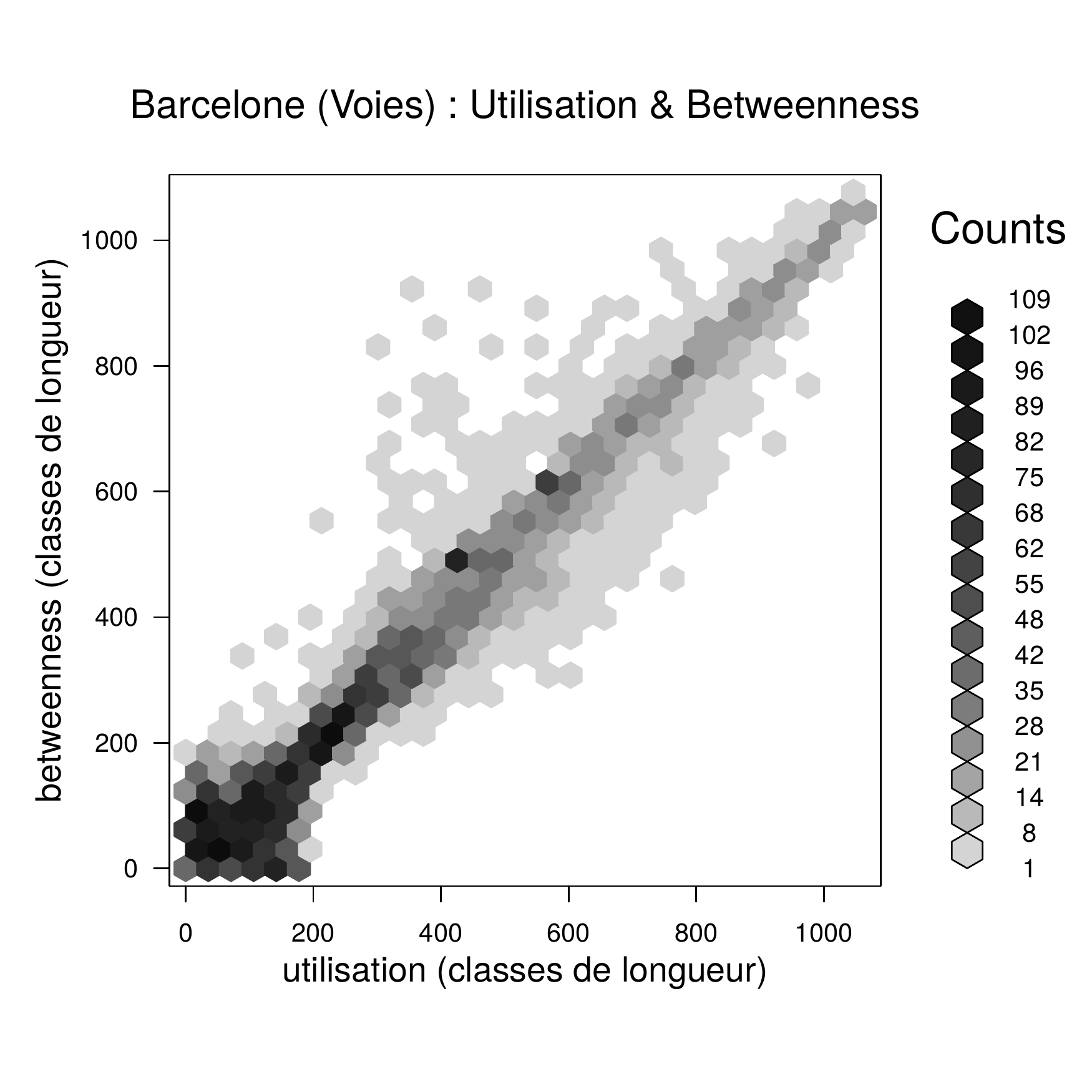}
        \caption{Sur le graphe viaire de Barcelone.}
    \end{subfigure}
     ~
    \begin{subfigure}{.40\textwidth}
        \includegraphics[width=\textwidth]{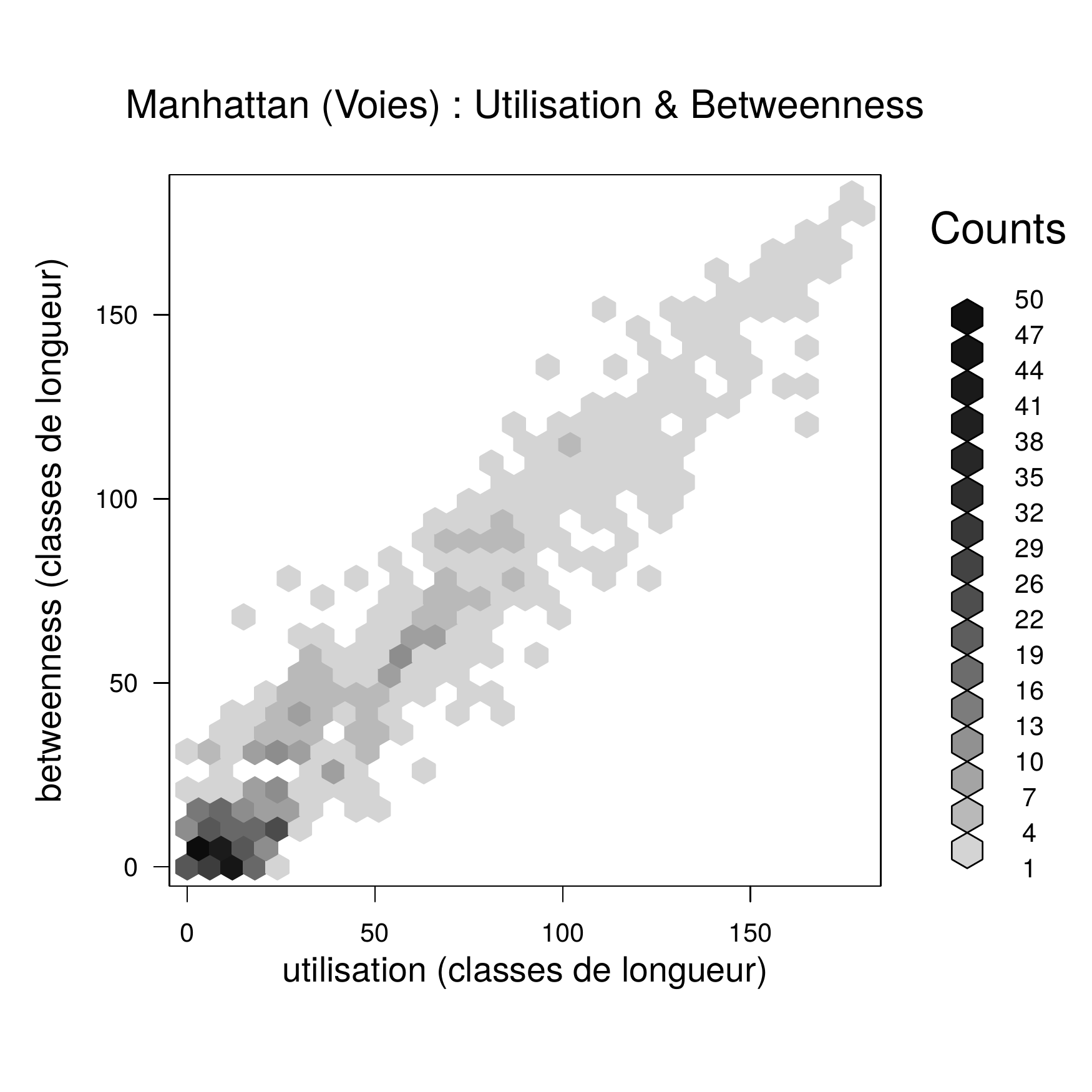}
        \caption{Sur le graphe viaire de Manhattan.}
    \end{subfigure}

    \caption{Cartes de corrélation croisée entre utilisation et betweenness calculés sur les voies.}
    \label{fig:voies_usebetw}

\end{figure}

Comme pour les arcs, la betweenness et l'utilisation renvoient une information similaire sur les voies. Ainsi, sur les graphes des voies de Manhattan, Barcelone et Paris, le pourcentage de corrélation est de 96\% avec le coefficient de Pearson. Sur celui des voies d'Avignon, il se situe autour de 92\% (le réseau étant moins régulier). Les cartes de corrélations illustrent le lien fort existant entre les deux indicateurs (figure \ref{fig:voies_usebetw}). Nous déduisons ainsi que l'utilisation est équivalente à la betweenness pour décrire les graphes. Comme son temps de calcul est moindre, nous la préférerons par la suite.

\begin{figure}[h]
    \centering

    \begin{subfigure}{.40\textwidth}
        \includegraphics[width=\textwidth]{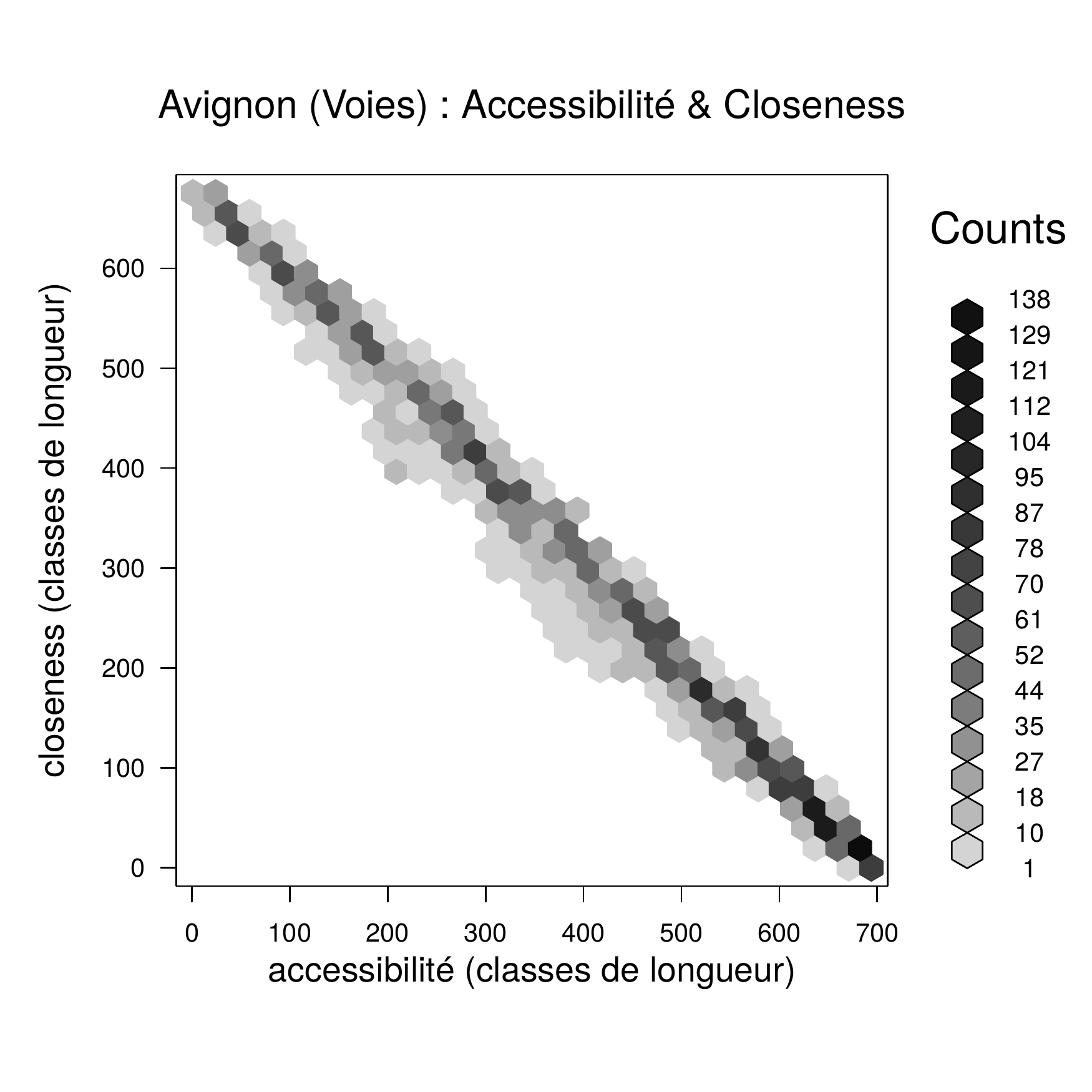}
        \caption{Sur le graphe viaire d'Avignon.}
    \end{subfigure}
    ~
     \begin{subfigure}{.40\textwidth}
        \includegraphics[width=\textwidth]{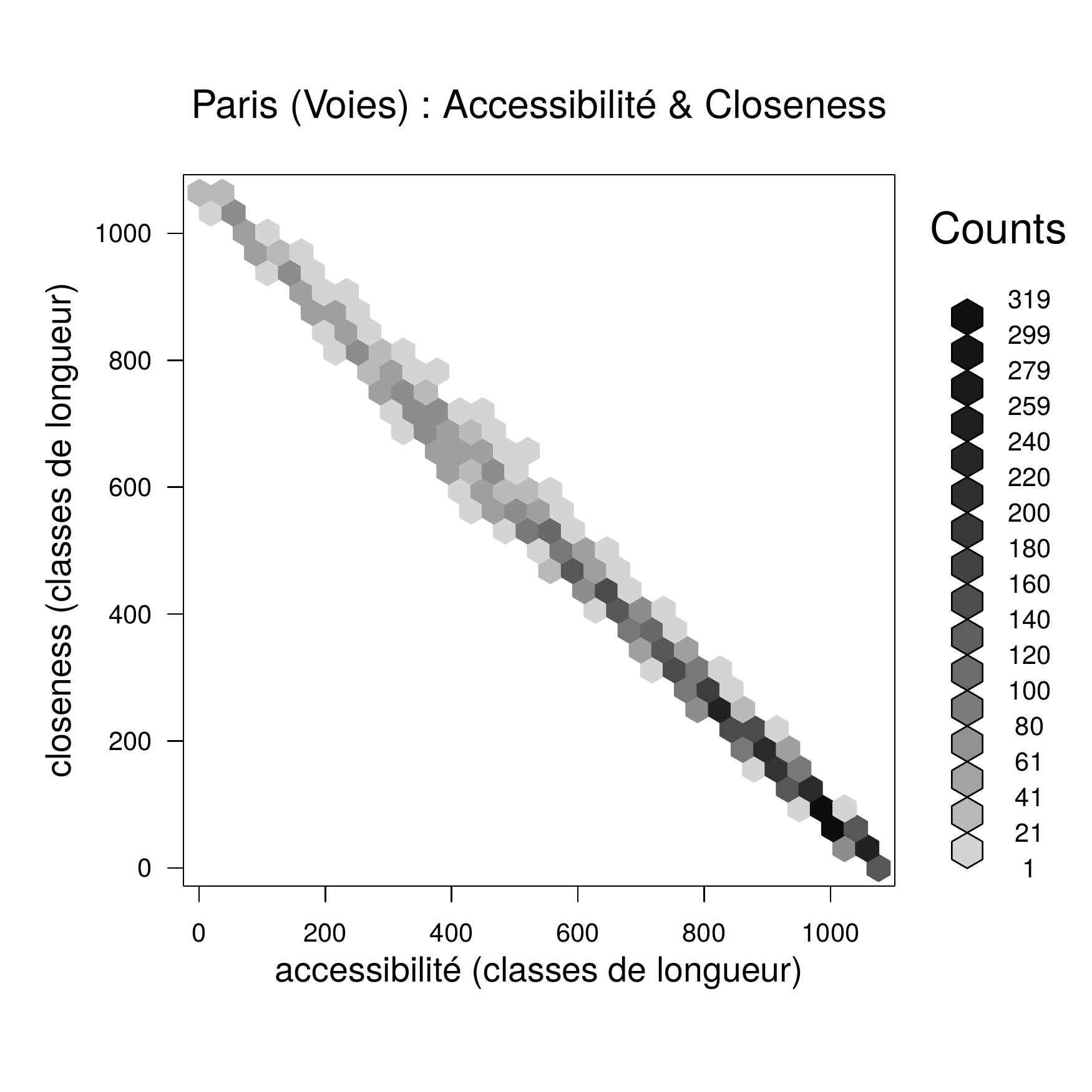}
        \caption{Sur le graphe viaire de Paris.}
    \end{subfigure}
    
    \begin{subfigure}{.40\textwidth}
        \includegraphics[width=\textwidth]{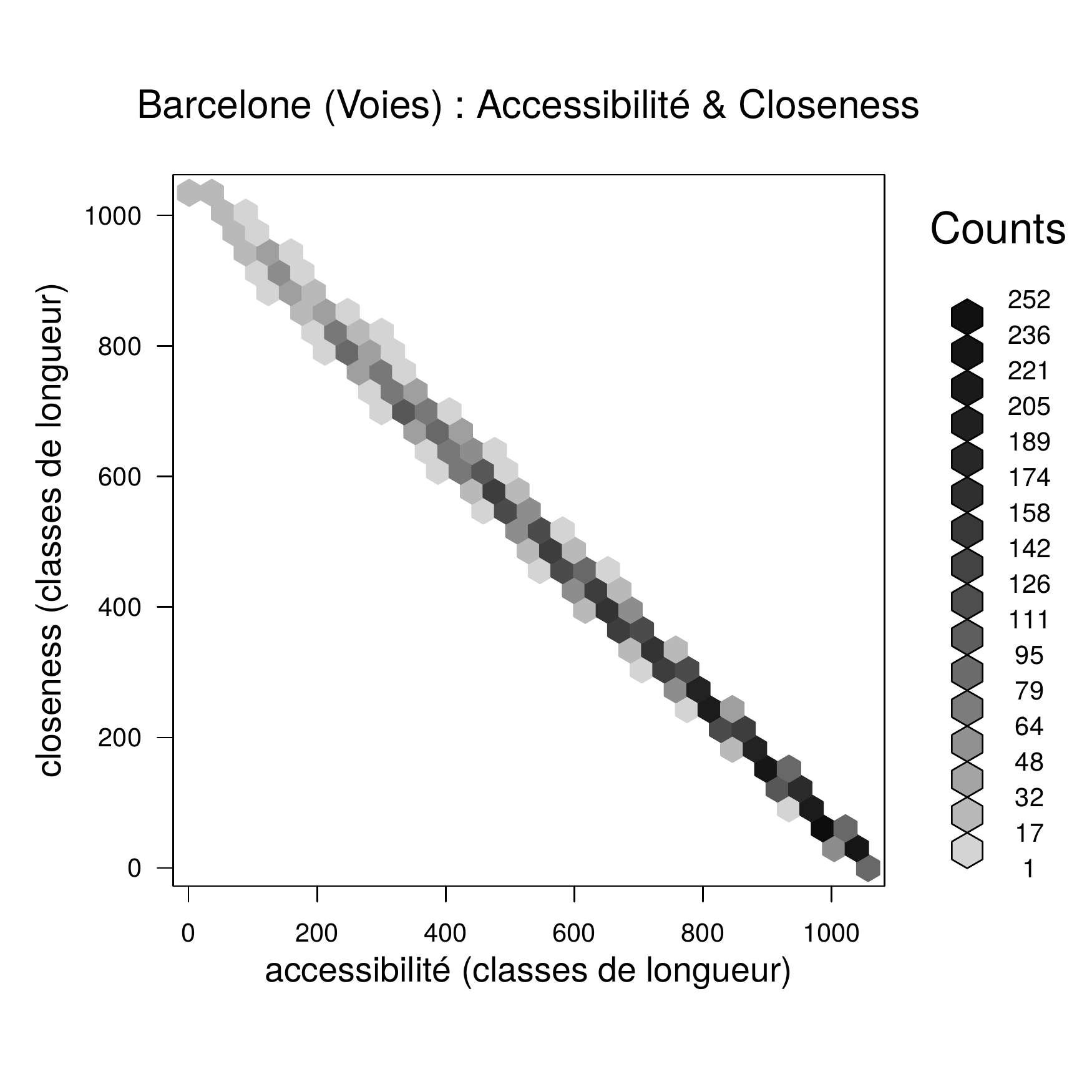}
        \caption{Sur le graphe viaire de Barcelone.}
    \end{subfigure}
     ~
    \begin{subfigure}{.40\textwidth}
        \includegraphics[width=\textwidth]{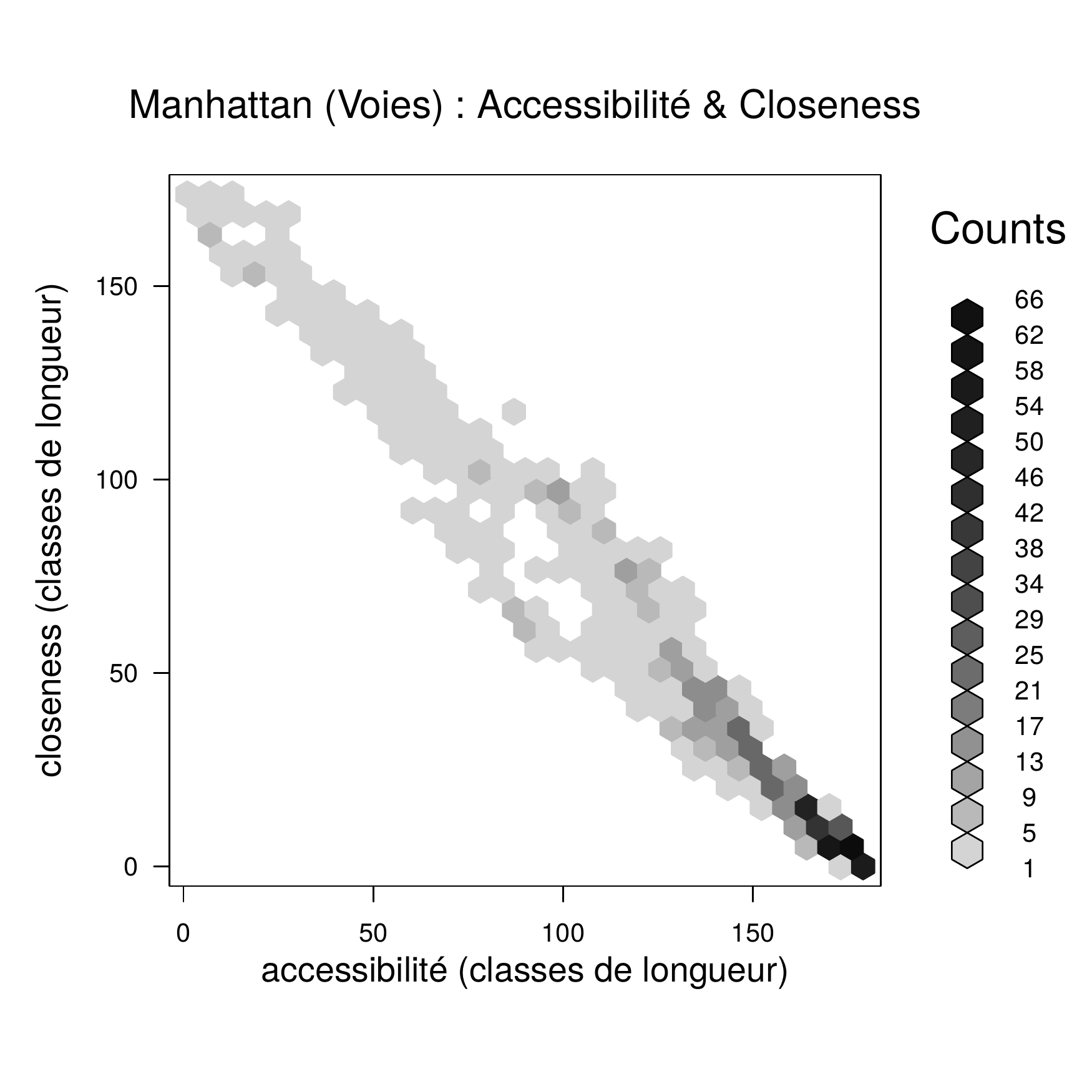}
        \caption{Sur le graphe viaire de Manhattan.}
    \end{subfigure}

    \caption{Cartes de corrélation croisée entre accessibilité et closeness calculés sur les voies.}
    \label{fig:voies_accessclo}

\end{figure}

De la même manière, nous constatons que pour les voies, l'ajout de la longueur à la distance topologique n'a pas un poids suffisant pour faire de l'accessibilité et de la closeness des indicateurs apportant une information différente. Les cartes de corrélation montrent la redondance des deux indicateurs sur tous les échantillons (figure \ref{fig:voies_accessclo}).

\begin{figure}[h]
    \centering

    \begin{subfigure}{.40\textwidth}
        \includegraphics[width=\textwidth]{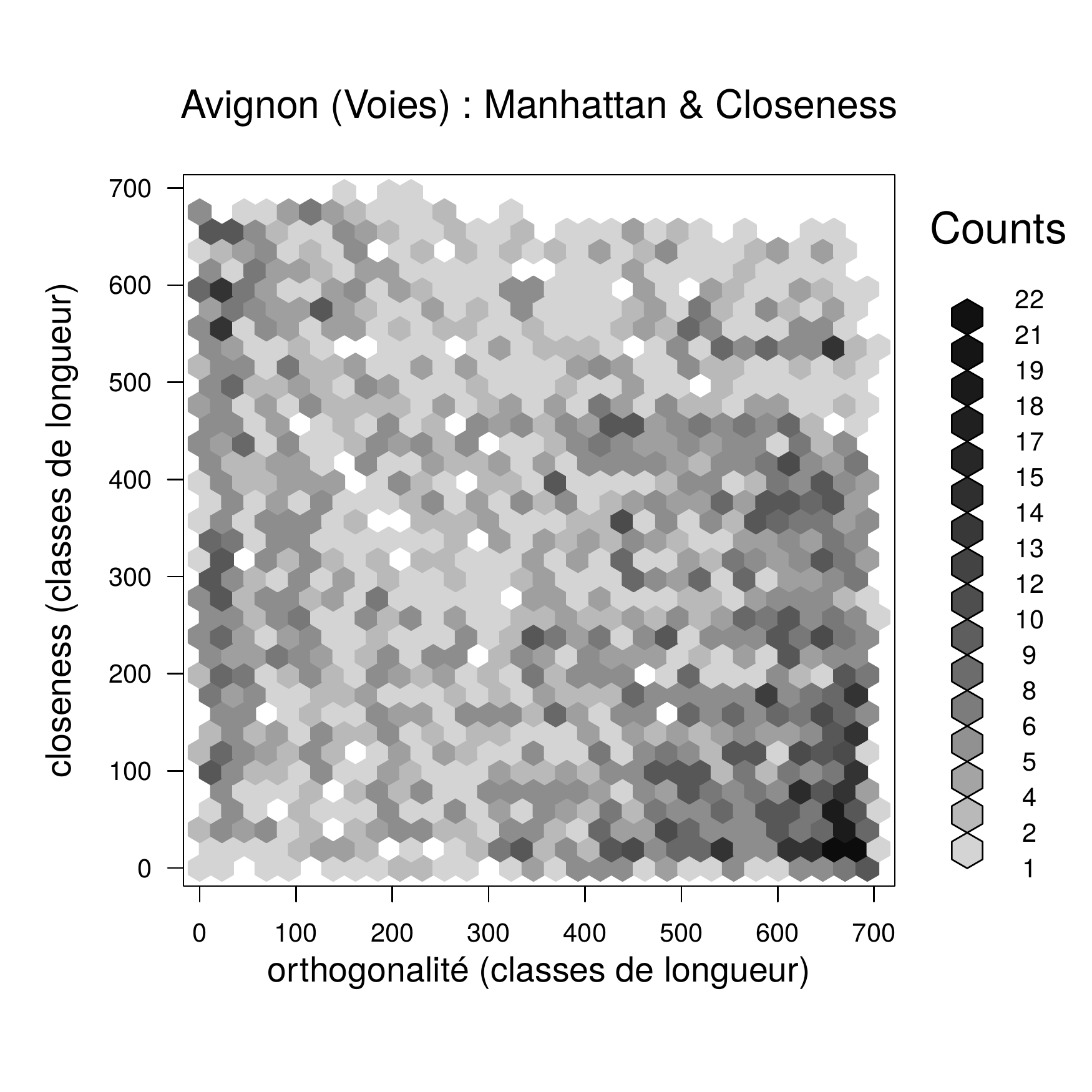}
        \caption{Sur le graphe viaire d'Avignon.}
    \end{subfigure}
    ~
    \begin{subfigure}{.40\textwidth}
        \includegraphics[width=\textwidth]{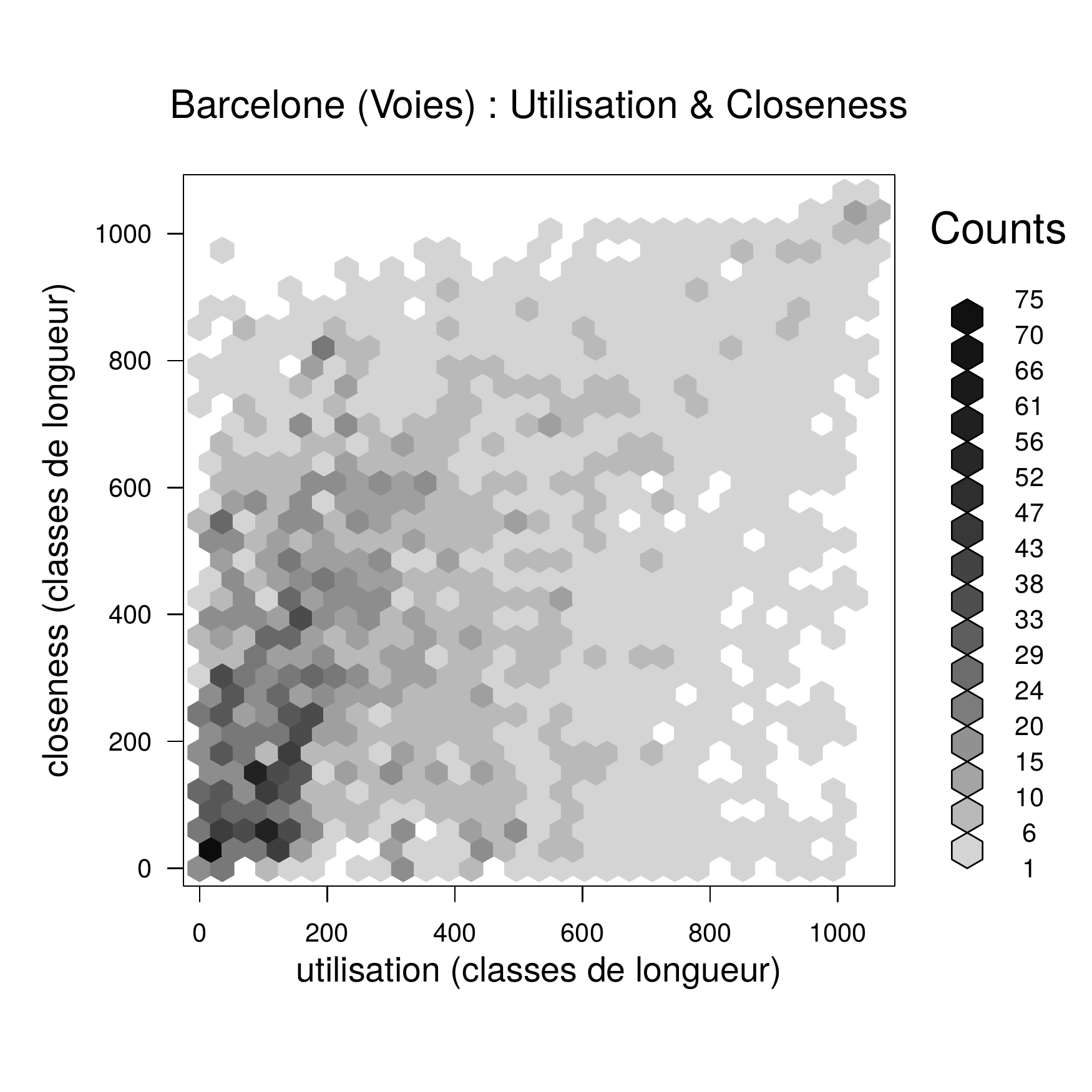}
        \caption{Sur le graphe viaire de Barcelone.}
    \end{subfigure}
    
     \begin{subfigure}{.40\textwidth}
        \includegraphics[width=\textwidth]{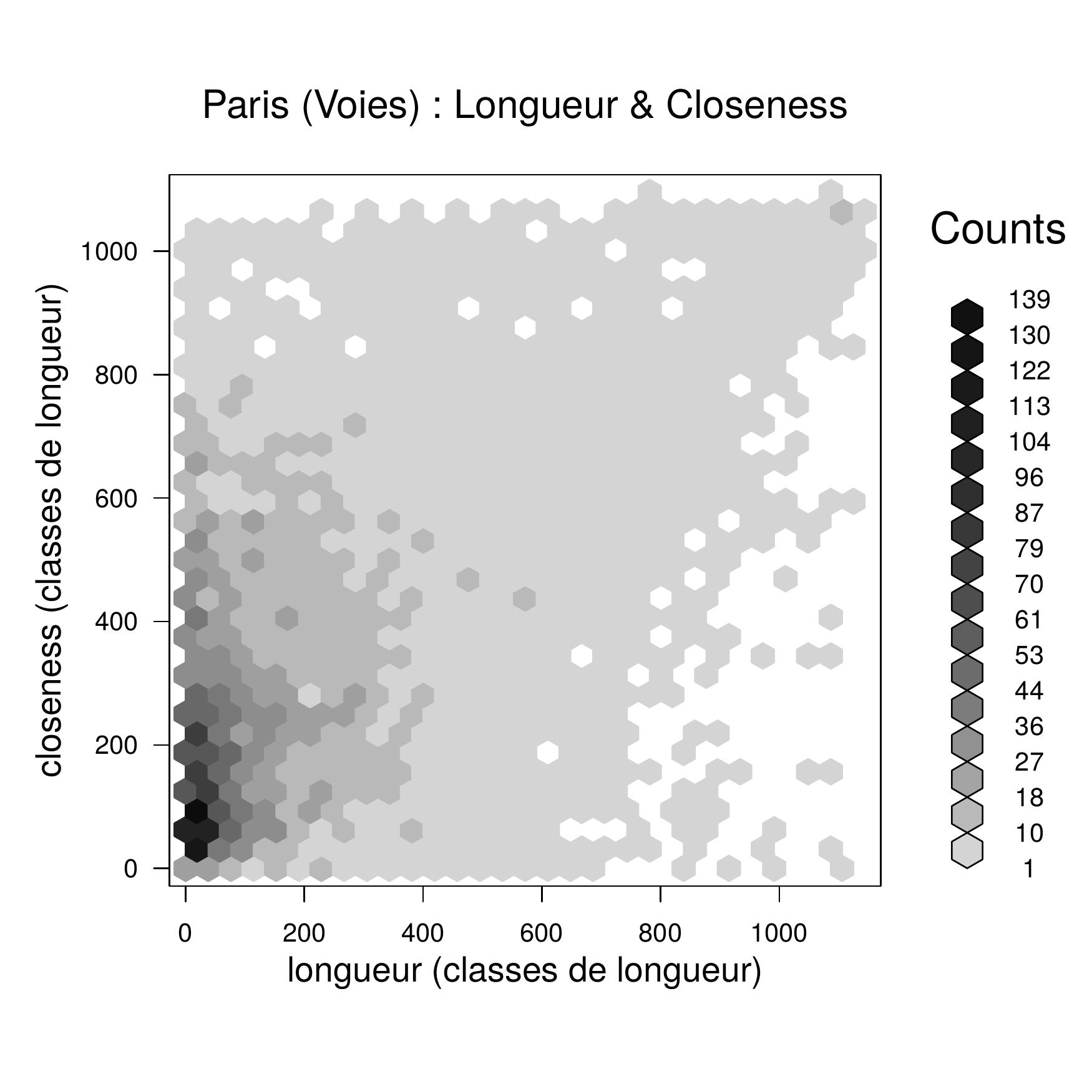}
        \caption{Sur le graphe viaire de Paris.}
    \end{subfigure}
    ~
    \begin{subfigure}{.40\textwidth}
        \includegraphics[width=\textwidth]{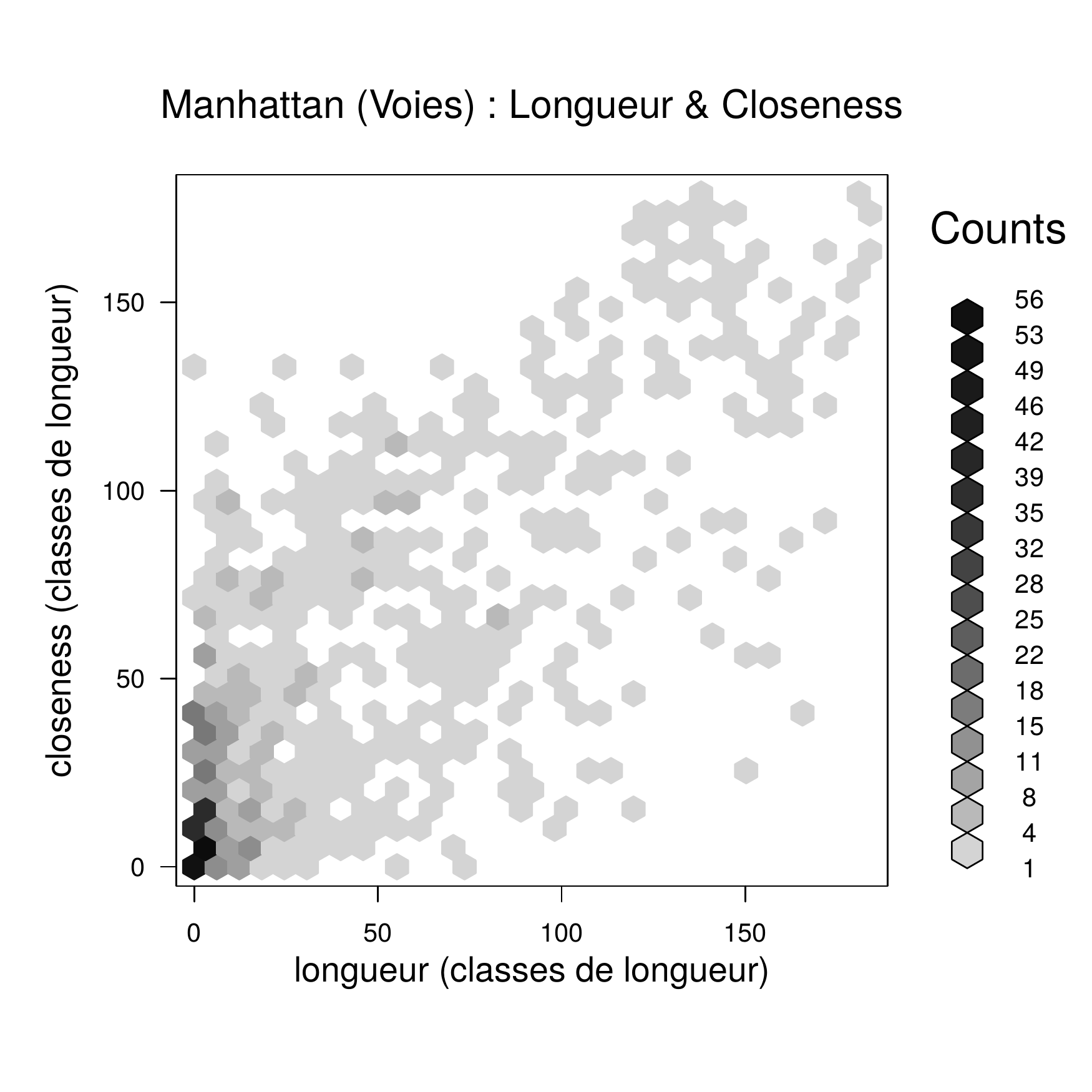}
        \caption{Sur le graphe viaire de Manhattan.}
        \label{fig:voies_cloX-man}
    \end{subfigure}

    \caption{Cartes de corrélation croisée entre closeness et orthogonalité, utilisation ou longueur calculés sur les voies. Les autres cartes de corrélations sont consultables en annexe \ref{ann:chap_cartes_corr}.}
    \label{fig:voies_cloX}

\end{figure}

Nous retiendrons la closeness comme étant un indicateur apportant une caractérisation particulière du réseau car non corrélée aux autres indicateurs (une fois l'accessibilité écartée). La figure \ref{fig:voies_cloX} montre la non corrélation entre closeness, orthogonalité, utilisation et longueur sur Avignon, Paris et Barcelone. Sur l'échantillon représentant Manhattan la closeness est corrélée à la longueur (avec un coefficient de 0,75 d'après le tableau en annexe \ref{tab:corr_manhattan_voie}). En effet, pour cette configuration de graphe précise, les voies ayant de faibles distances topologiques vers l'ensemble du réseau sont celles traversant l'île de part en part et sont donc également les plus longues.

L'orthogonalité apporte ici également une information bien singulière. Son calcul étant construit sur la géométrie de chaque intersection, c'est ce critère proprement géométrique qui fonde son caractère unique. En effet, tous les autres indicateurs font intervenir la topologie alors que l'orthogonalité s'en émancipe en normalisant la somme des sinus des angles par le degré de la voie. Elle ne se retrouve donc corrélée à aucun autre type d'indicateur (figure \ref{fig:comp_arcs_orthoX}).

\begin{figure}[h]
    \centering

    \begin{subfigure}{.40\textwidth}
        \includegraphics[width=\textwidth]{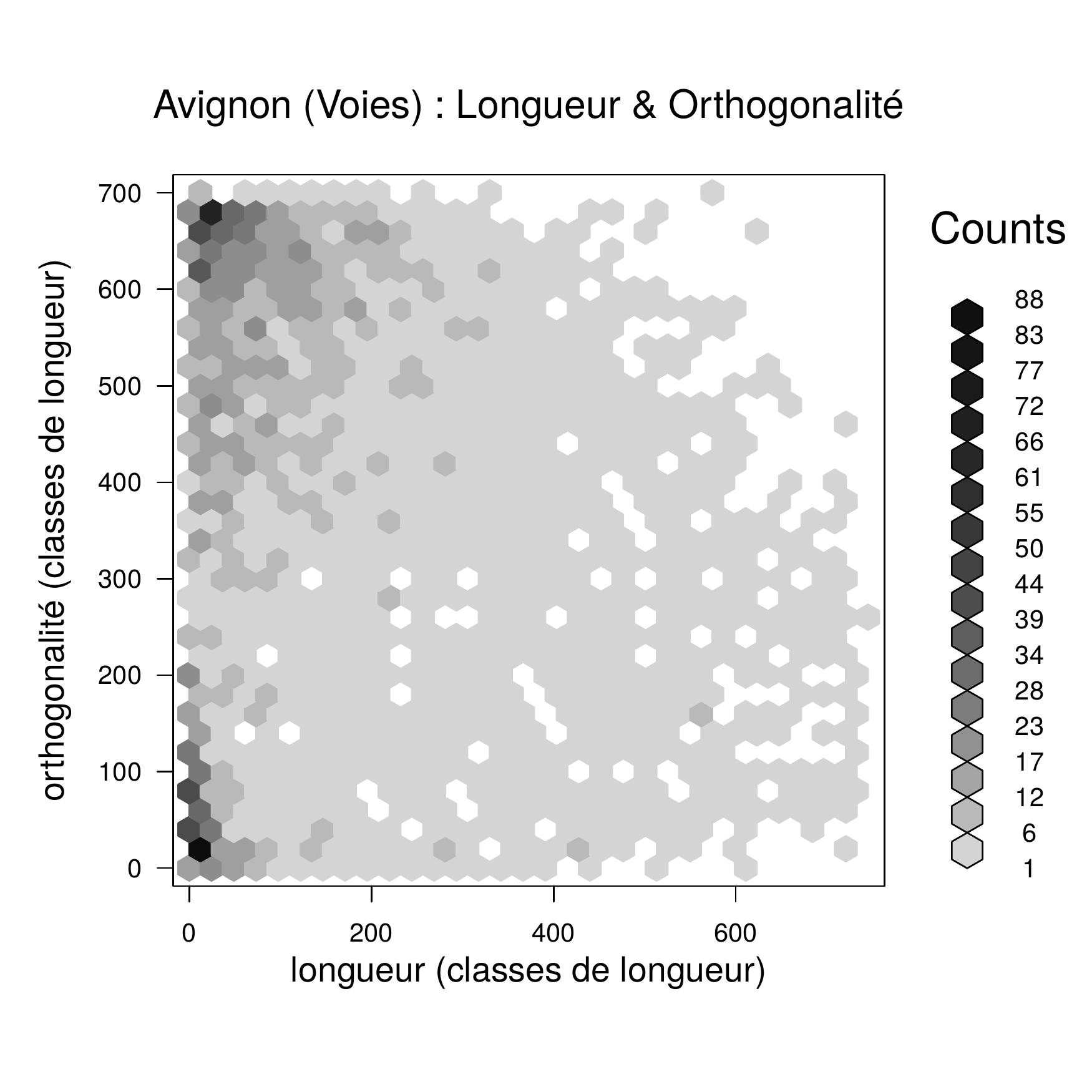}
        \caption{Sur le graphe viaire d'Avignon.}
    \end{subfigure}
    ~
    \begin{subfigure}{.40\textwidth}
        \includegraphics[width=\textwidth]{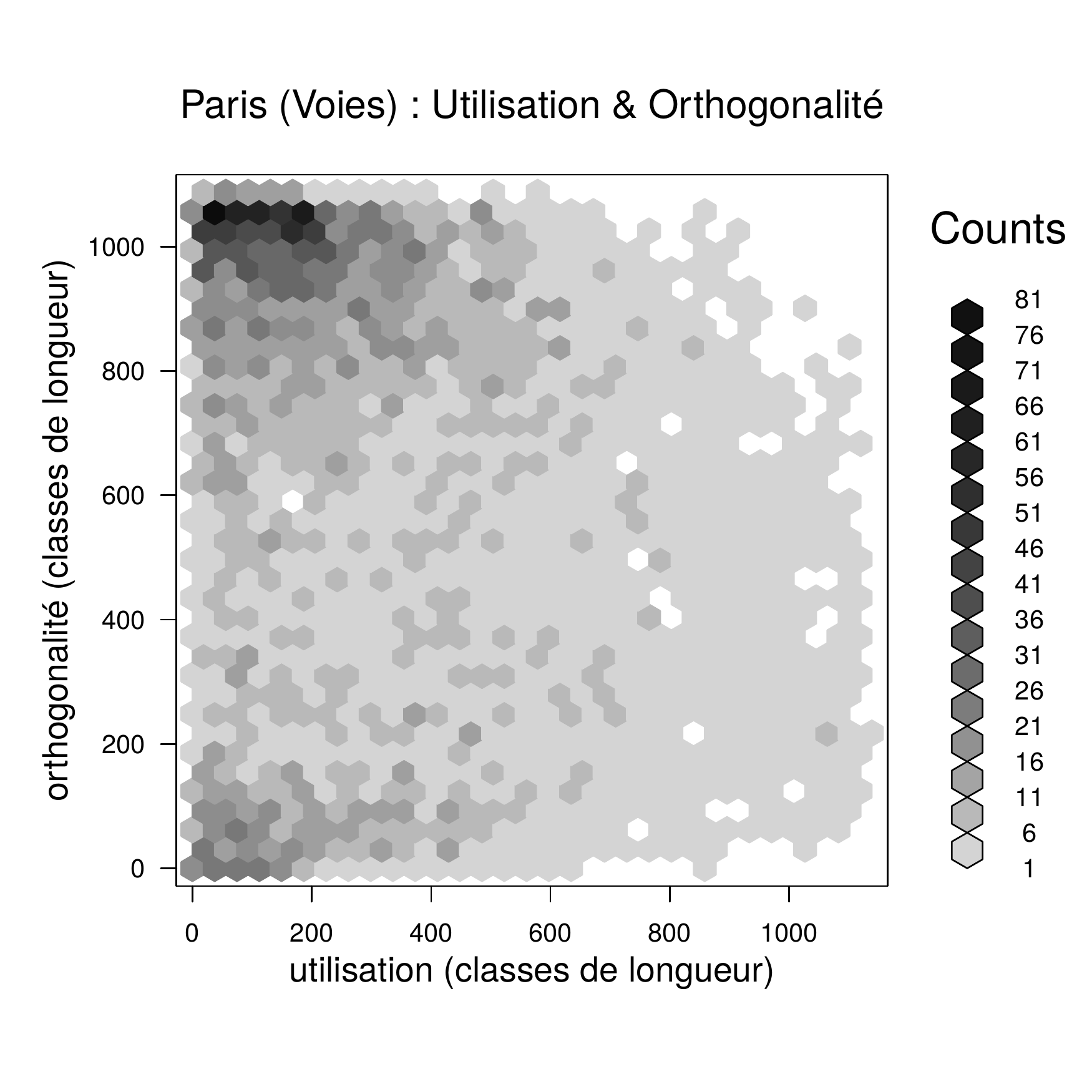}
        \caption{Sur le graphe viaire de Paris.}
    \end{subfigure}

    \caption{Cartes de corrélation croisée entre orthogonalité et utilisation ou longueur calculés sur les voies. Les autres cartes de corrélations sont consultables en annexe \ref{ann:chap_cartes_corr}.}
    \label{fig:voies_orthoX}

\end{figure}

Outre ces deux indicateurs bien différenciables des autres, le caractère multi-échelle de la voie groupe les informations apportées par les autres calculs en une unique caractérisation. Des indicateurs calculés localement (comme la longueur ou le degré) deviennent ainsi équivalents à ceux calculés en tenant compte de l'ensemble du réseau (comme la structuralité potentielle ou l'utilisation) en étant appliqués à l'hypergraphe. Cela rend l'objet \textit{voie} particulièrement intéressant pour la simplification de la complexité des calculs, pour la caractérisation des réseaux spatiaux. Ce caractère fort du global amené au local, apporté par l'analyse des réseaux spatiaux par la voie, amène également une hypothèse de stabilité aux effets de bord qui seront analysés plus précisément dans la seconde partie de ce travail.

\begin{figure}[h]
    \centering

    \begin{subfigure}{.40\textwidth}
        \includegraphics[width=\textwidth]{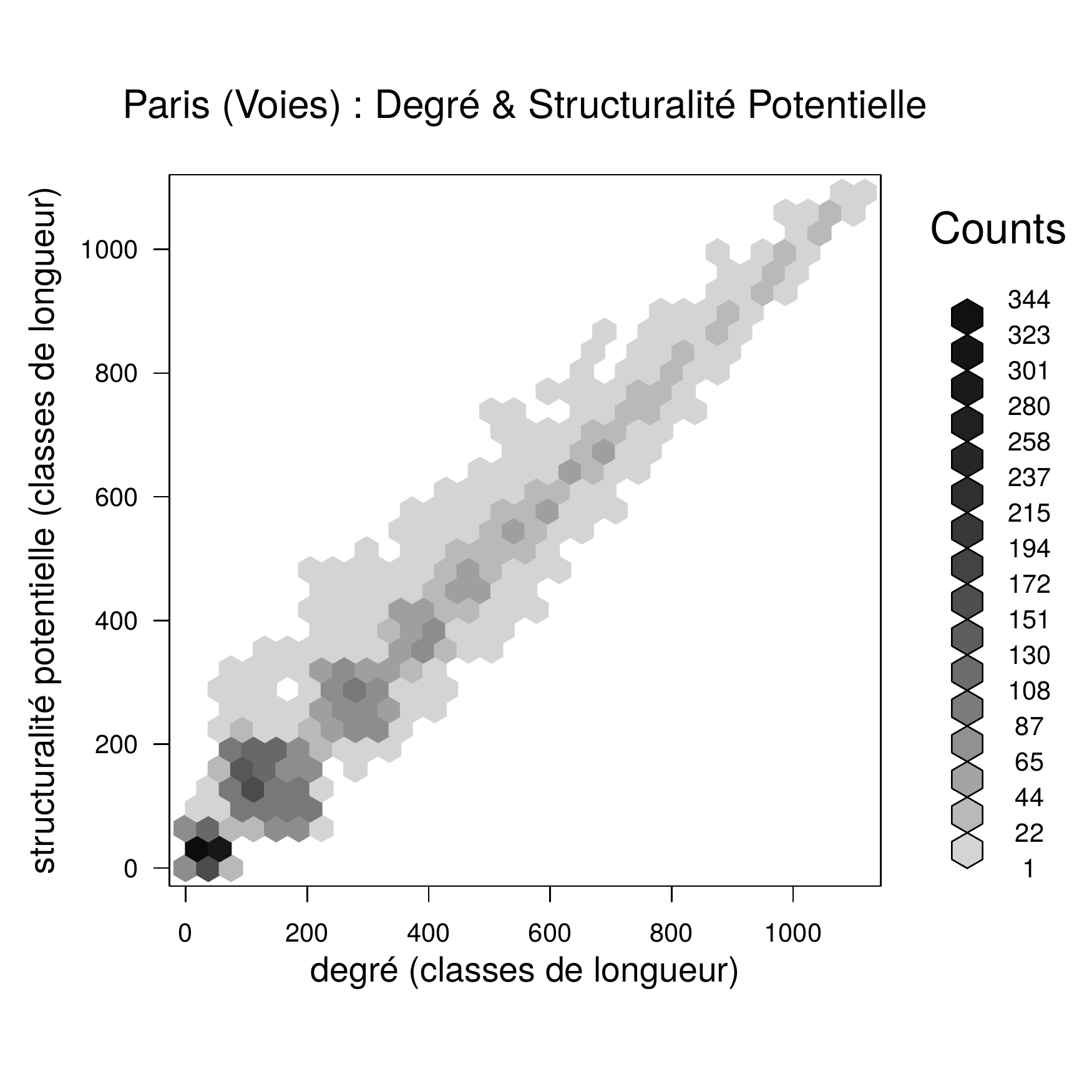}
    \end{subfigure}
    ~
    \begin{subfigure}{.40\textwidth}
        \includegraphics[width=\textwidth]{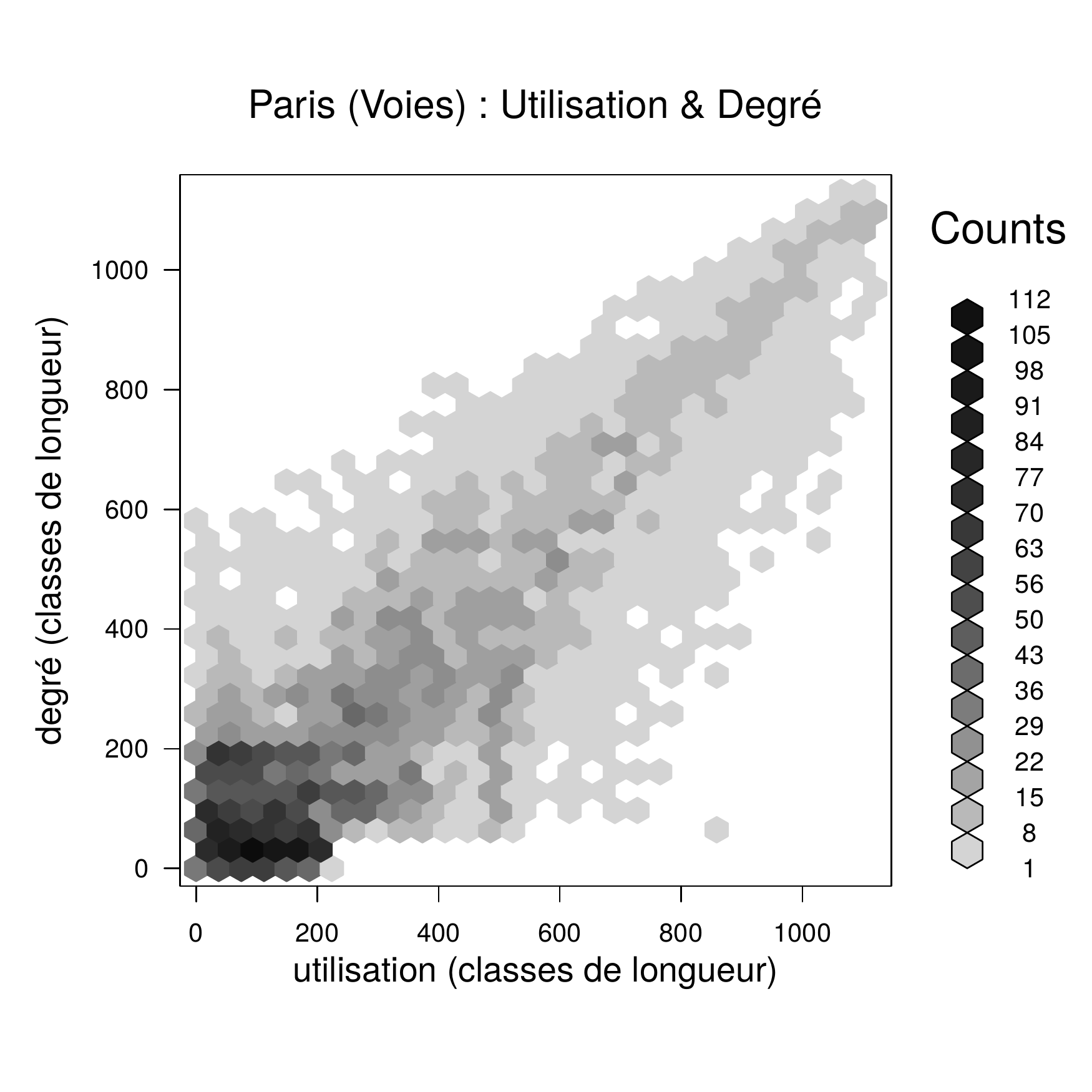}
    \end{subfigure}

    \caption{Cartes de corrélation croisée entre un indicateur local (le degré) et deux indicateurs globaux (structuralité potentielle et utilisation) calculés sur les voies du graphe de Paris. Les autres cartes de corrélations sont consultables en annexe \ref{ann:chap_cartes_corr}.}
    \label{fig:voies_degreeX}
\end{figure}

Nous distinguons donc trois groupes d'indicateurs :
\begin{enumerate}
\item la longueur, le nombre d'arcs, le degré, la structuralité potentielle, la betweenness et l'utilisation
\item la closeness et l'accessibilité
\item l'orthogonalité
\end{enumerate}

Pour chacun de ces groupes nous ne conservons qu'un indicateur. Pour le premier, nous identifions un \enquote{centre d'îlot}, c'est-à-dire l'indicateur qui est le plus corrélé à l'ensemble des autres de son groupe : il s'agit du degré. Pour le second groupe, nous conservons la closeness, car l'ajout de la longueur (comme pour les arcs) ne se révèle pas pertinente. Et enfin pour le dernier groupe nous avons l'orthogonalité.

\subsection{Indicateurs composés}

\begin{figure}[h]
    \centering

    \begin{subfigure}[t]{.23\textwidth}
        \includegraphics[width=\textwidth]{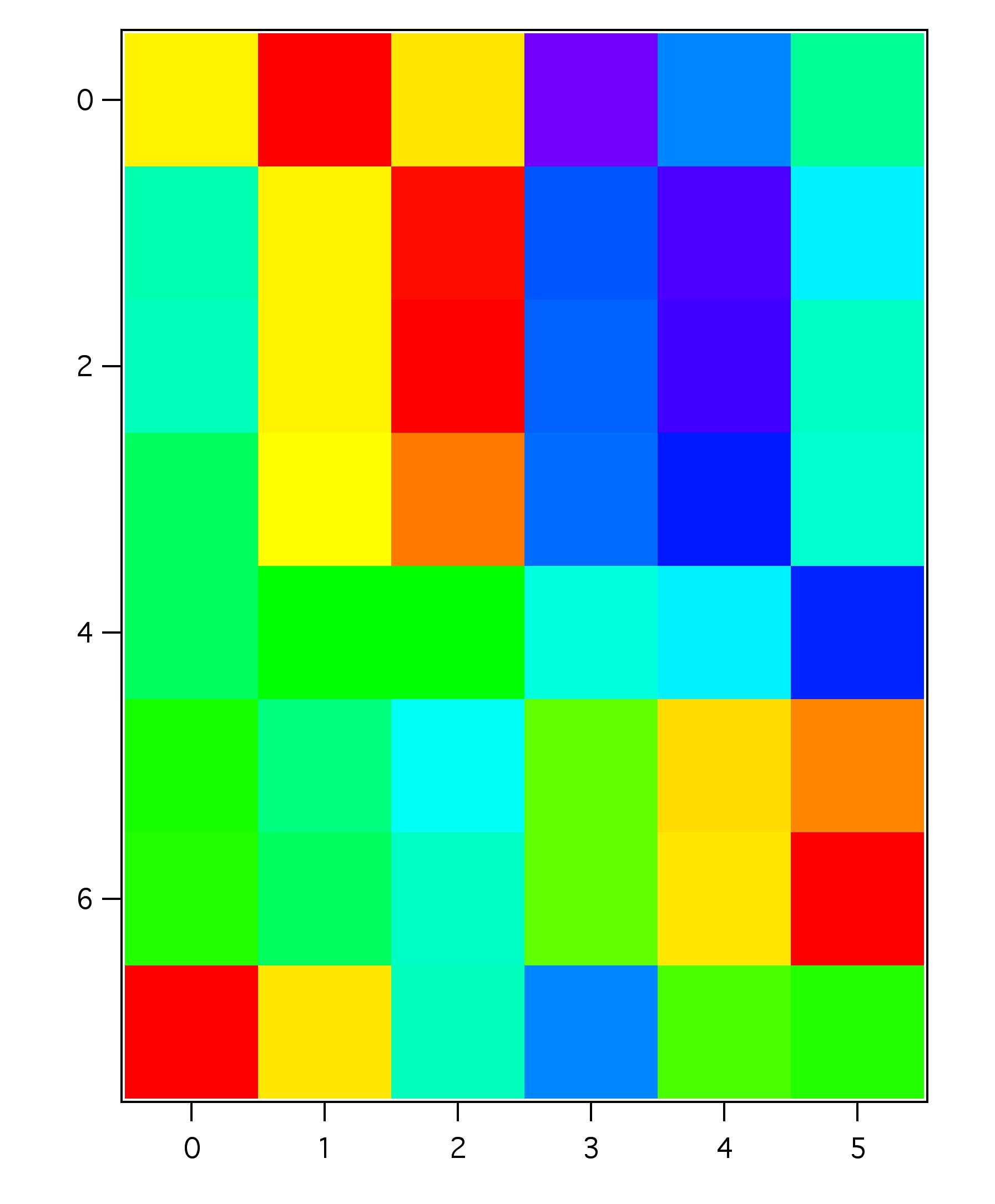}
        \caption{Représentation du tableau \ref{tab:corr_voies_avignon_comb}, calculé sur les voies d'Avignon.}
    \end{subfigure}
    ~
    \begin{subfigure}[t]{.23\linewidth}
        \includegraphics[width=\textwidth]{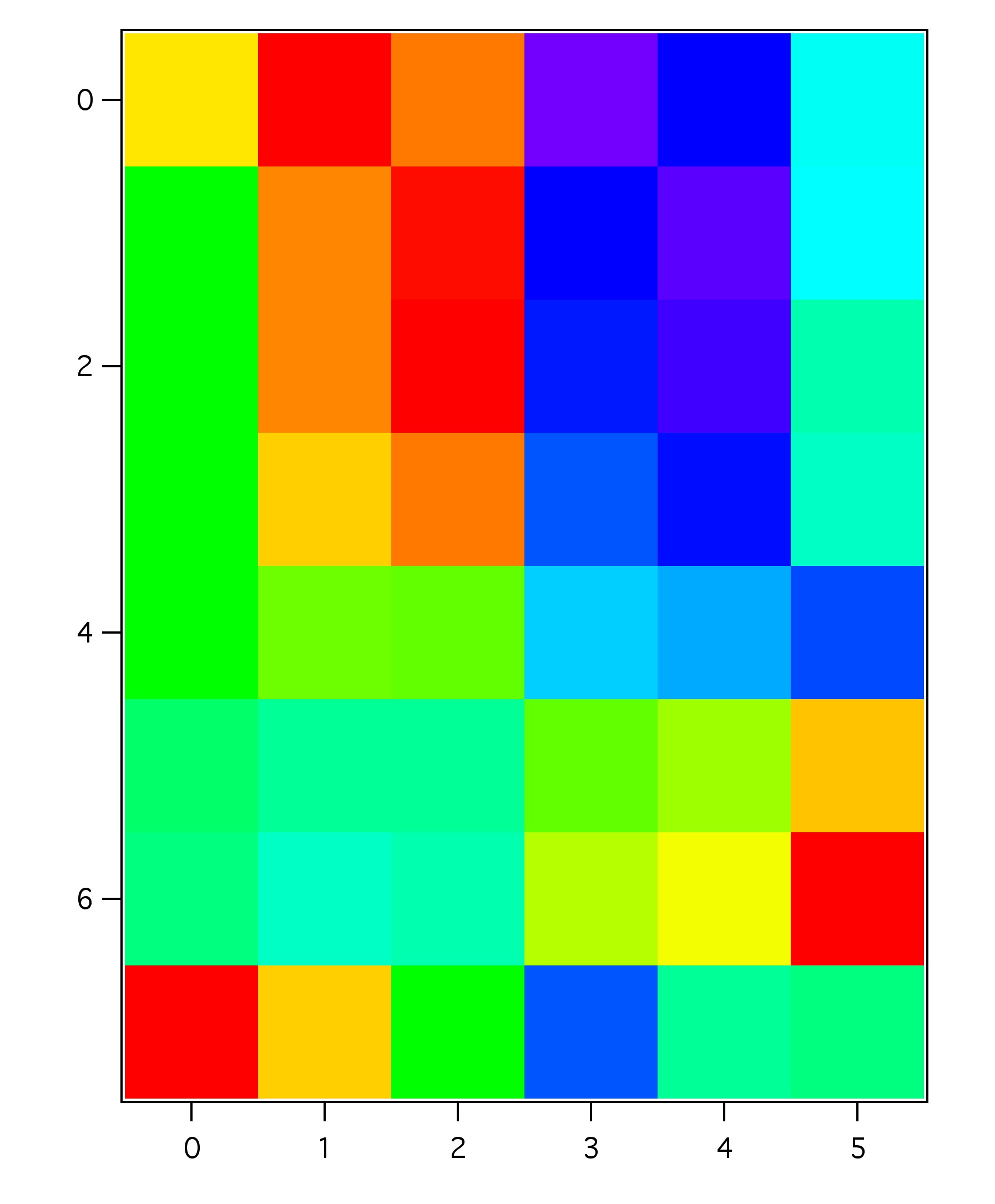}
        \caption{Représentation du tableau \ref{tab:corr_voies_barcelone_comb}, calculé sur les voies de Barcelone.}
    \end{subfigure}
    ~   
    \begin{subfigure}[t]{.23\textwidth}
        \includegraphics[width=\textwidth]{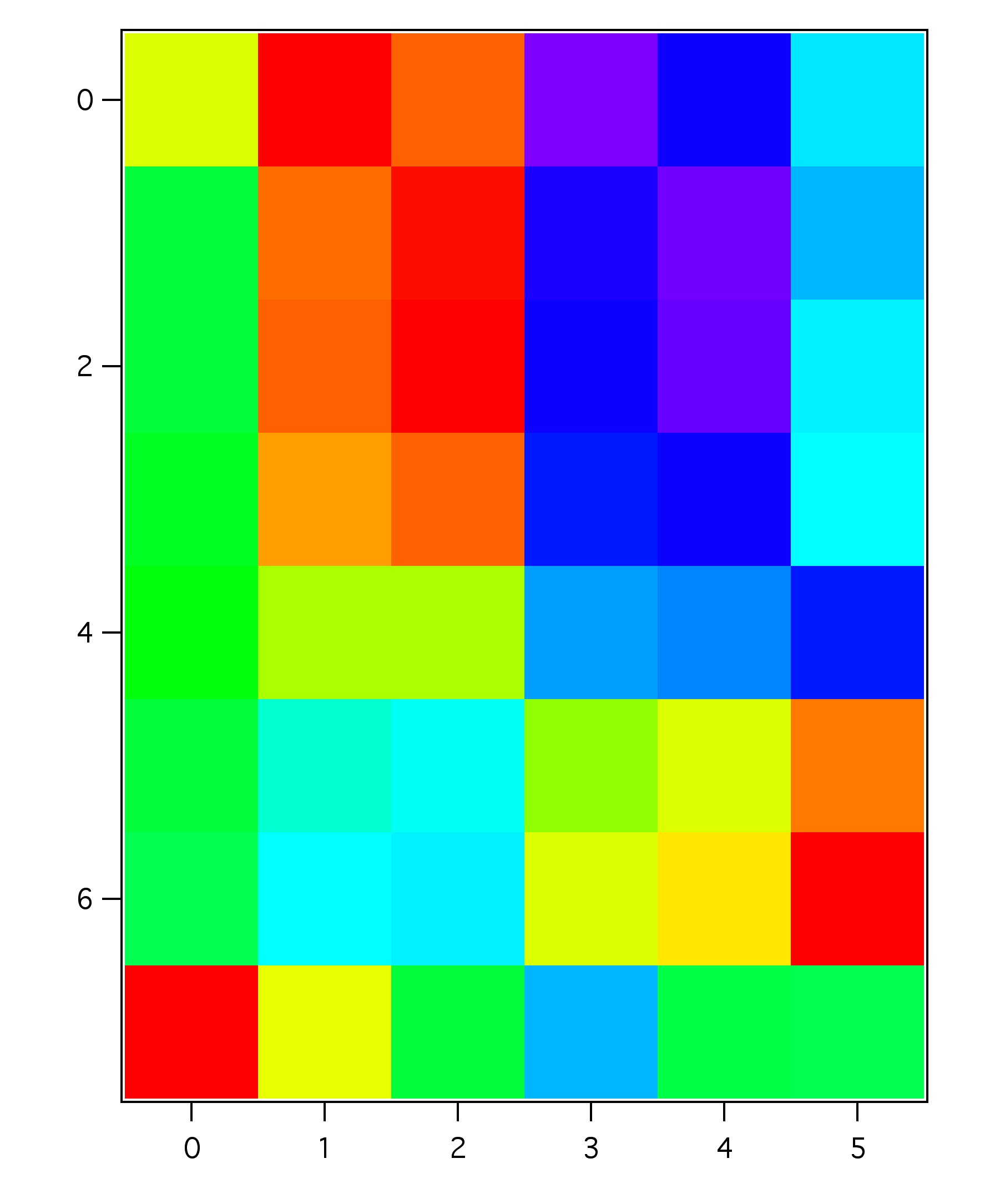}
        \caption{Représentation du tableau \ref{tab:corr_voies_paris_comb}, calculé sur les voies de Paris.}
    \end{subfigure}
    ~
    \begin{subfigure}[t]{.23\linewidth}
        \includegraphics[width=\textwidth]{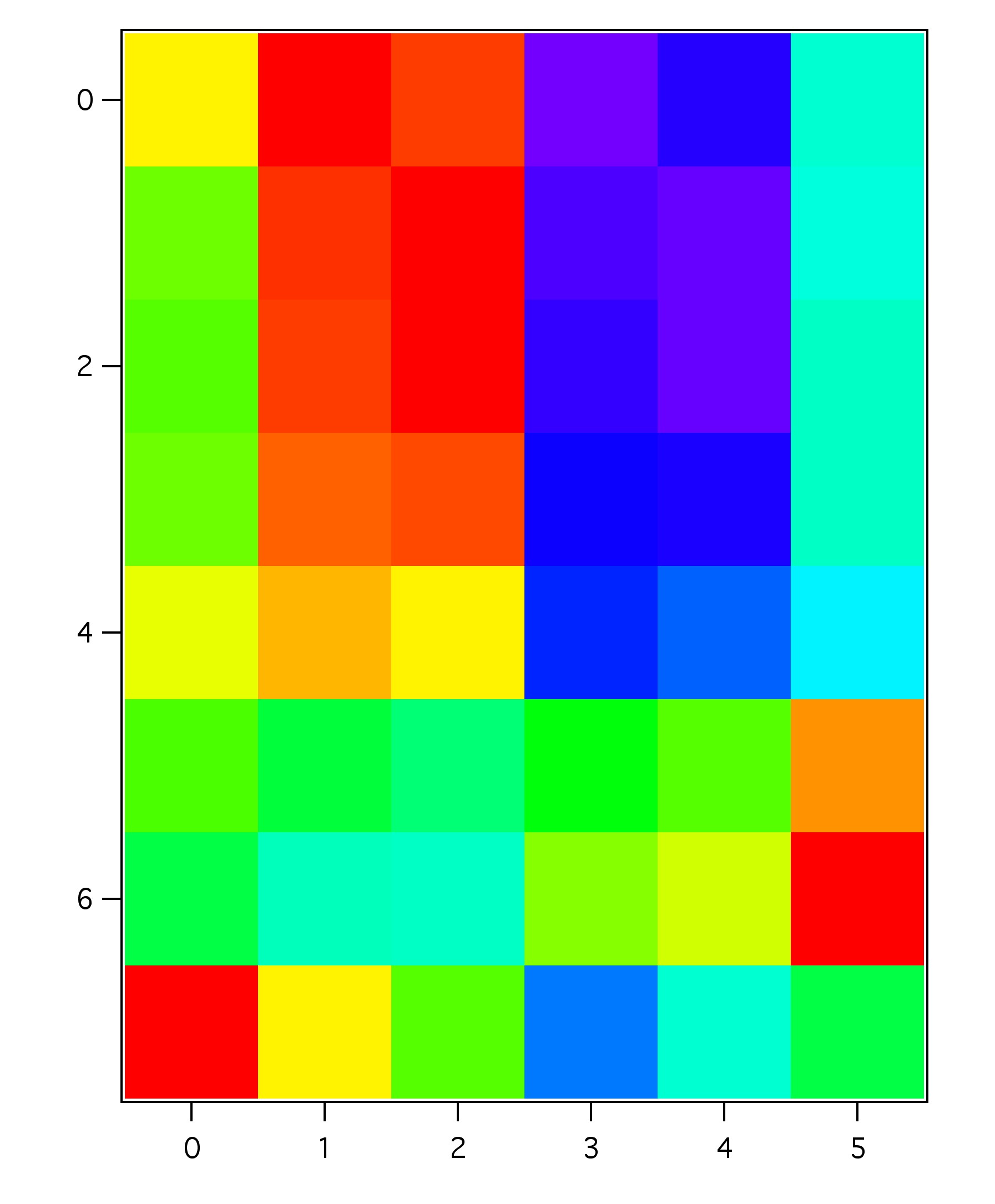}
        \caption{Représentation du tableau \ref{tab:corr_voies_manhattan_comb}, calculé sur les voies de Manhattan.}
    \end{subfigure}

    \begin{subfigure}{.2\linewidth}
        \includegraphics[width=\textwidth]{images/matrices_colorees/mat_echelle_dec.jpg}
        \caption{Échelle}
    \end{subfigure}
    
    \caption{Représentation de la corrélation entre indicateurs composés calculés sur les voies sous forme de matrice colorée. Rappel de l'ordre des indicateurs : \\
    Horizontal : 0 : $\frac{longueur}{degre}$ ; 1 : $\frac{longueur}{closeness}$ ; 2 : $\frac{degre}{closeness}$ ; 3 : $\frac{orthogonalite}{longueur}$ ; 4 : $\frac{orthogonalite}{degre}$ ; 5 : $\frac{orthogonalite}{closeness}$\\
     Vertical : 0 : longueur ; 1 : degré ; 2 : structuralité potentielle ; 3 : utilisation ; 4 : closeness ; 5 : orthogonalité ; 6 : $\frac{orthogonalite}{closeness}$ ; 7 : $\frac{longueur}{degre}$}
    \label{fig:mat_comb_voies}

\end{figure}

De la même manière que pour les arcs, nous composons par division les indicateurs primaires des voies (figure \ref{fig:mat_prim_voies}). Nous retenons pour les combinaisons quatre indicateurs primaires : la longueur, le degré, la closeness et l'orthogonalité. Nous procédons aux calculs de corrélation sur Avignon, Paris, Barcelone et Manhattan. Nous reportons les indicateurs de Pearson calculés dans les tableaux \ref{tab:corr_voies_avignon_comb}, \ref{tab:corr_voies_manhattan_comb}, \ref{tab:corr_voies_paris_comb} et \ref{tab:corr_voies_barcelone_comb} (en annexe). Nous représentons ces corrélations avec les matrices colorées de la figure \ref{fig:mat_comb_voies}.

Il apparaît dans ce tableau que les combinaisons de la longueur ou du degré des voies avec d'autres indicateurs sont corrélées avec les valeurs primaires de ces indicateurs. Sur les quatre échantillons spatiaux, combiner la longueur ou le degré de la voie avec sa closeness ou son orthogonalité n'apporte donc aucune information supplémentaire. Les cartes de corrélations montrent le lien fort qui existe entre ces indicateurs et leurs différentes combinaisons.

\begin{figure}[h]
    \centering

    \begin{subfigure}{.40\textwidth}
        \includegraphics[width=\textwidth]{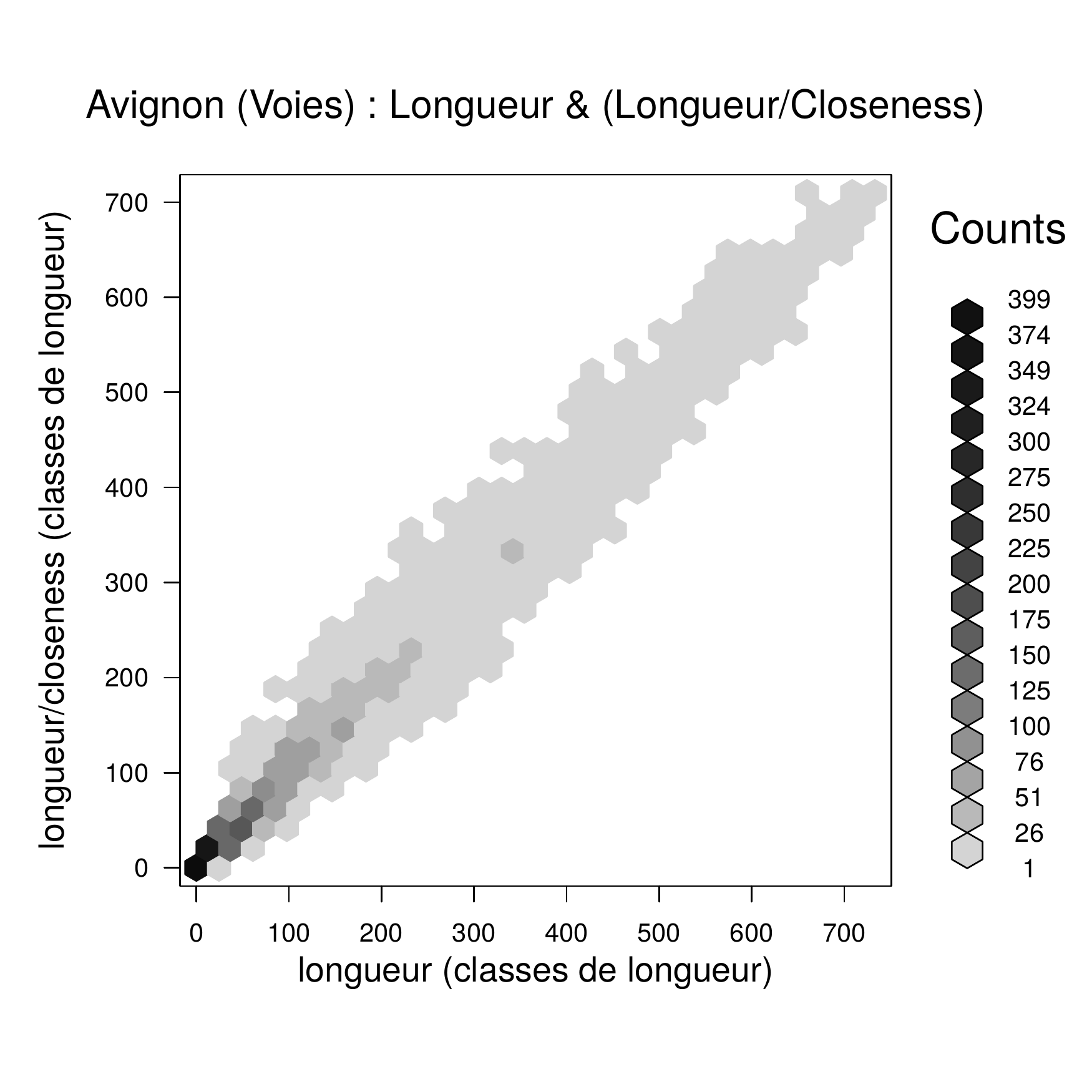}
    \end{subfigure}
    ~
    \begin{subfigure}{.40\textwidth}
        \includegraphics[width=\textwidth]{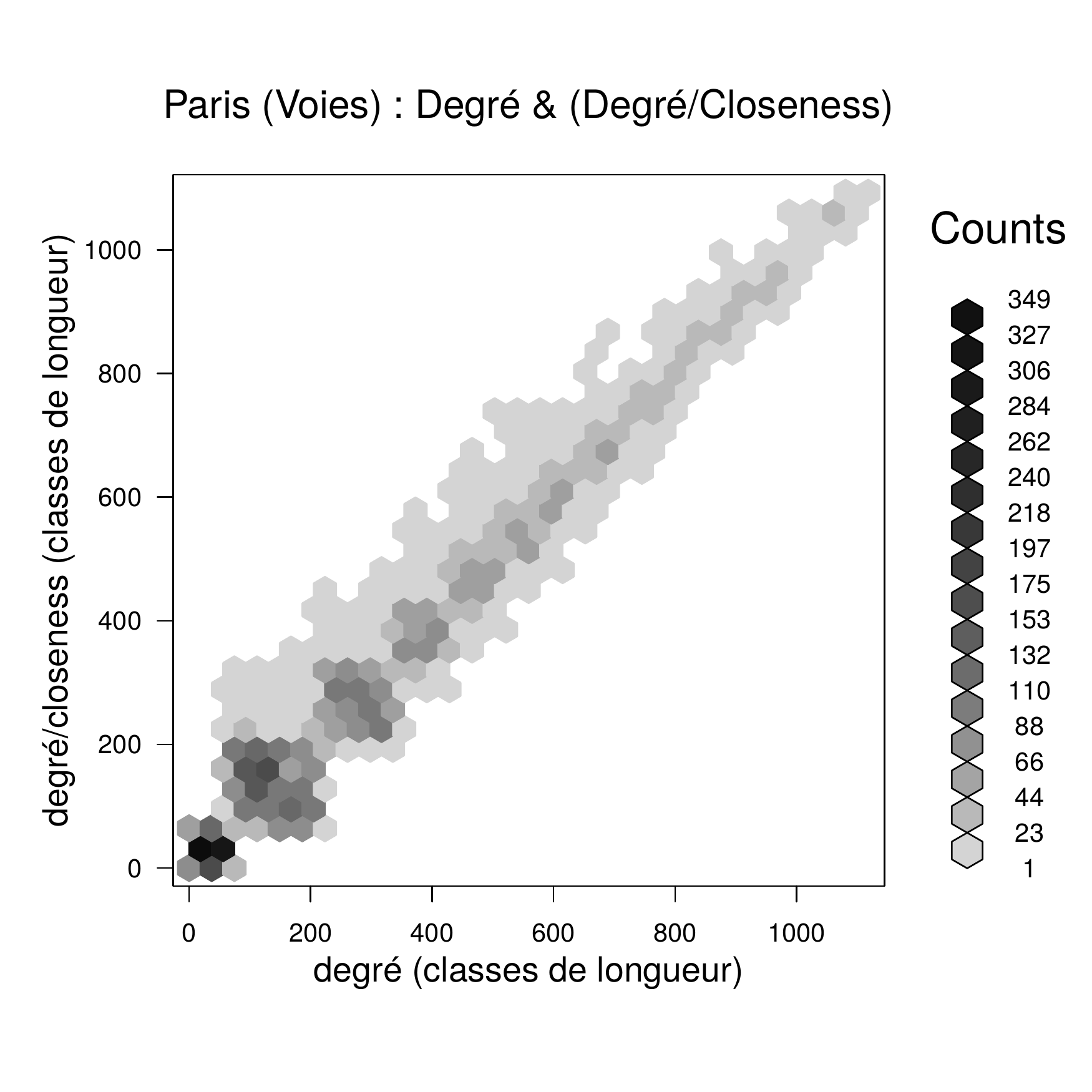}
    \end{subfigure}
    
    \begin{subfigure}{.40\textwidth}
        \includegraphics[width=\textwidth]{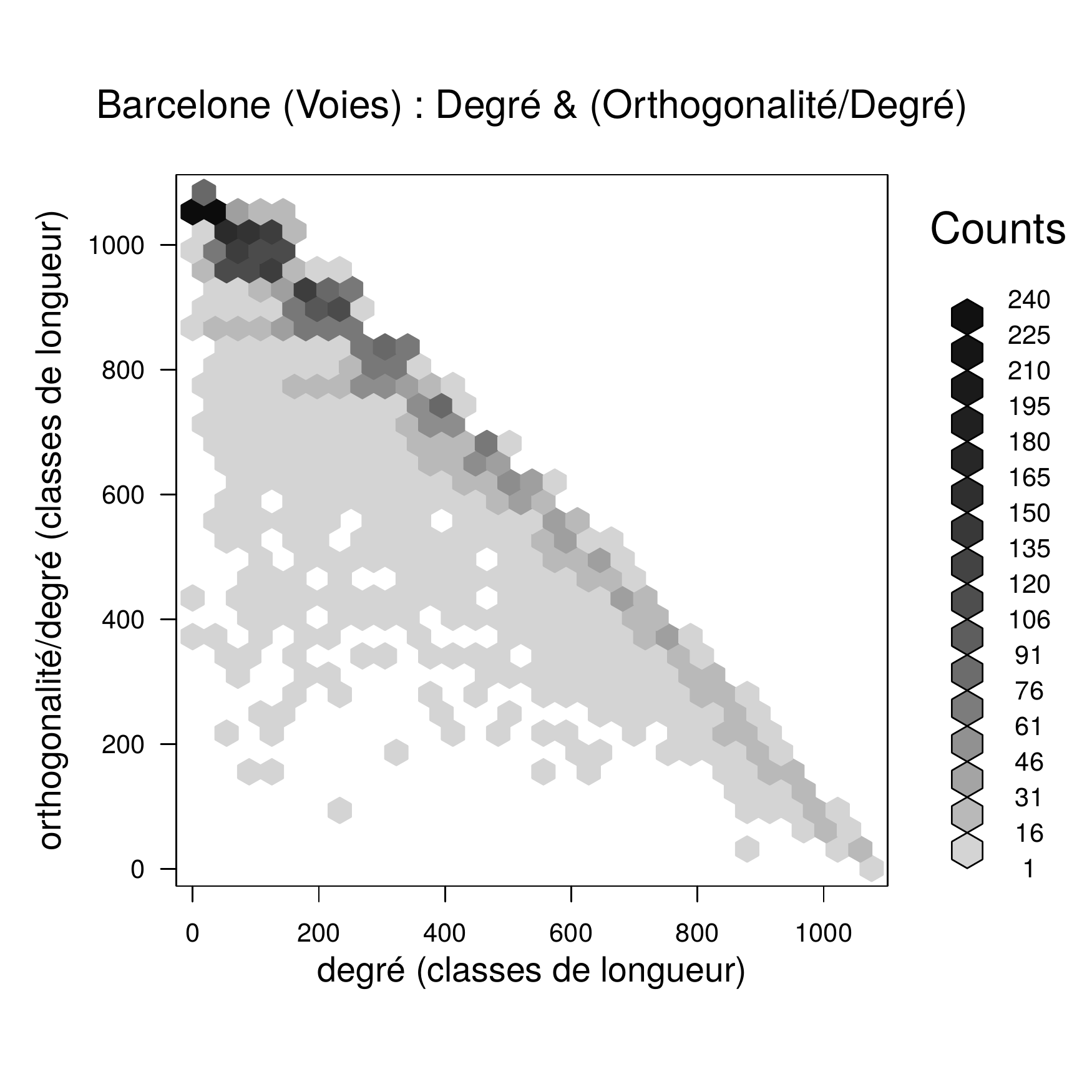}
    \end{subfigure}
    ~
    \begin{subfigure}{.40\textwidth}
        \includegraphics[width=\textwidth]{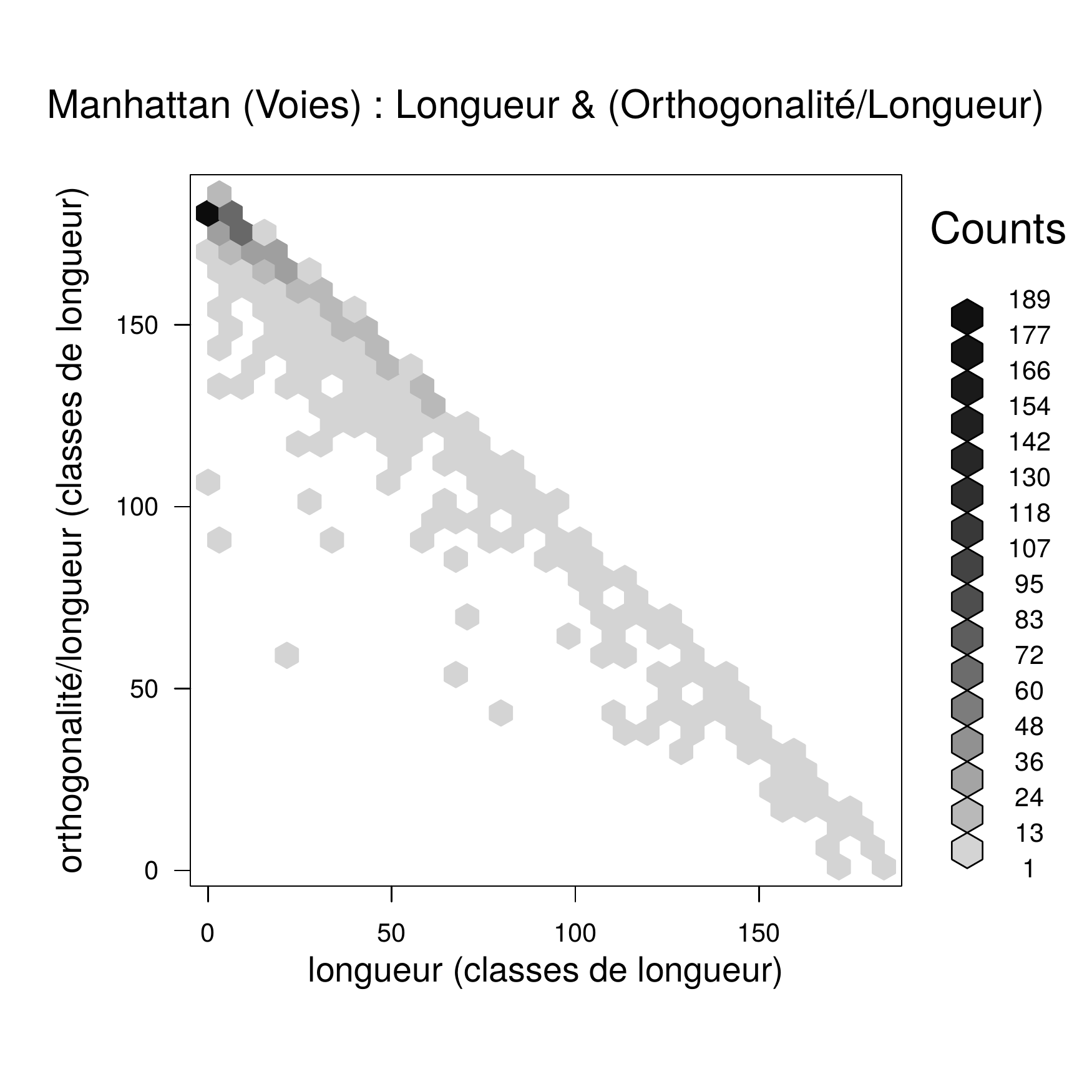}
    \end{subfigure}

    \caption{Cartes de corrélation croisée entre degré et longueur et leurs combinaisons par division avec closeness et orthogonalité sur les réseaux viaires d'Avignon, Paris, Barcelone et Manhattan. Les autres cartes de corrélations sont consultables en annexe \ref{ann:chap_cartes_corr}.}
    \label{fig:voies_deglen}
\end{figure}

\begin{figure}[h]
    \centering

    \begin{subfigure}{.40\textwidth}
        \includegraphics[width=\textwidth]{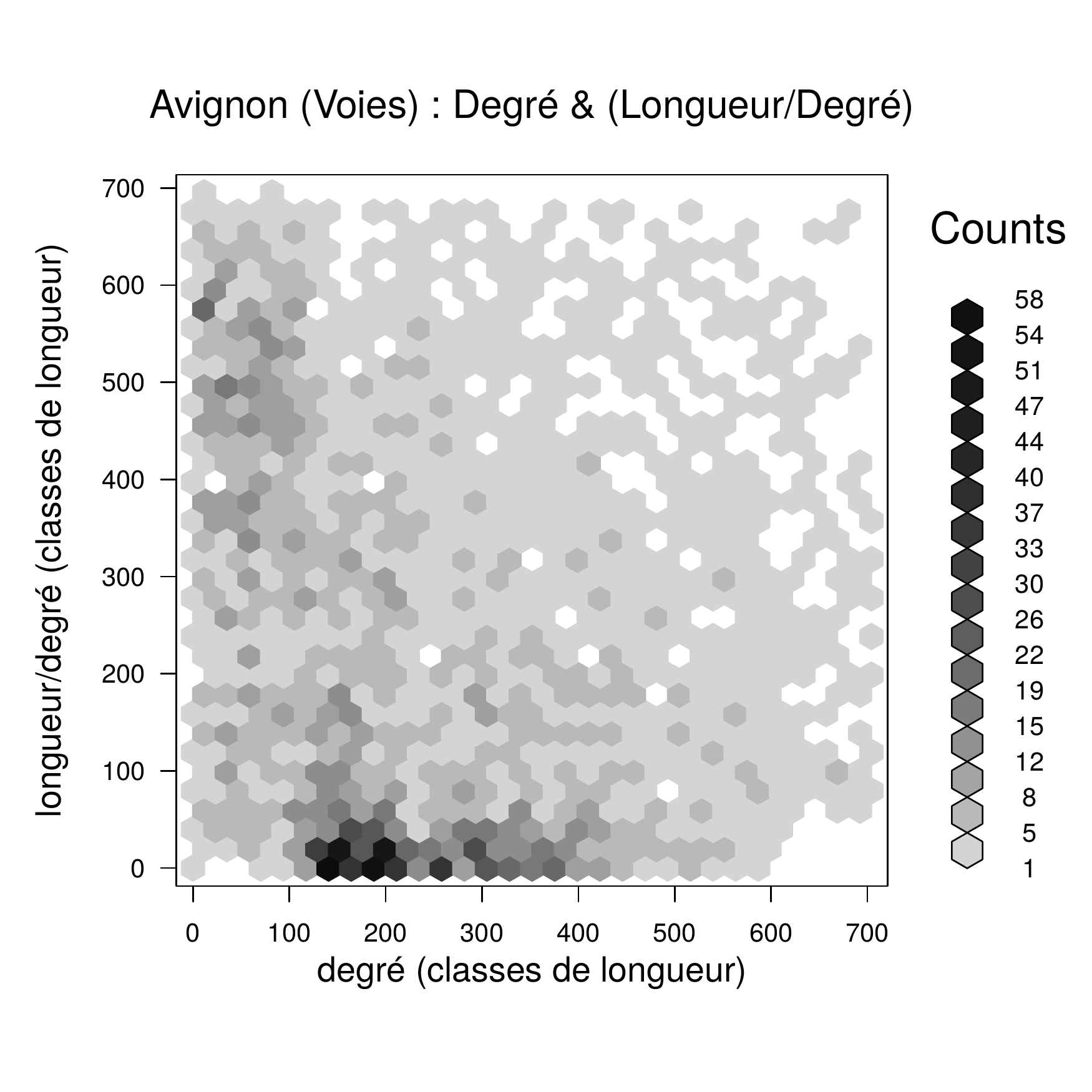}
    \end{subfigure}
    ~
    \begin{subfigure}{.40\textwidth}
        \includegraphics[width=\textwidth]{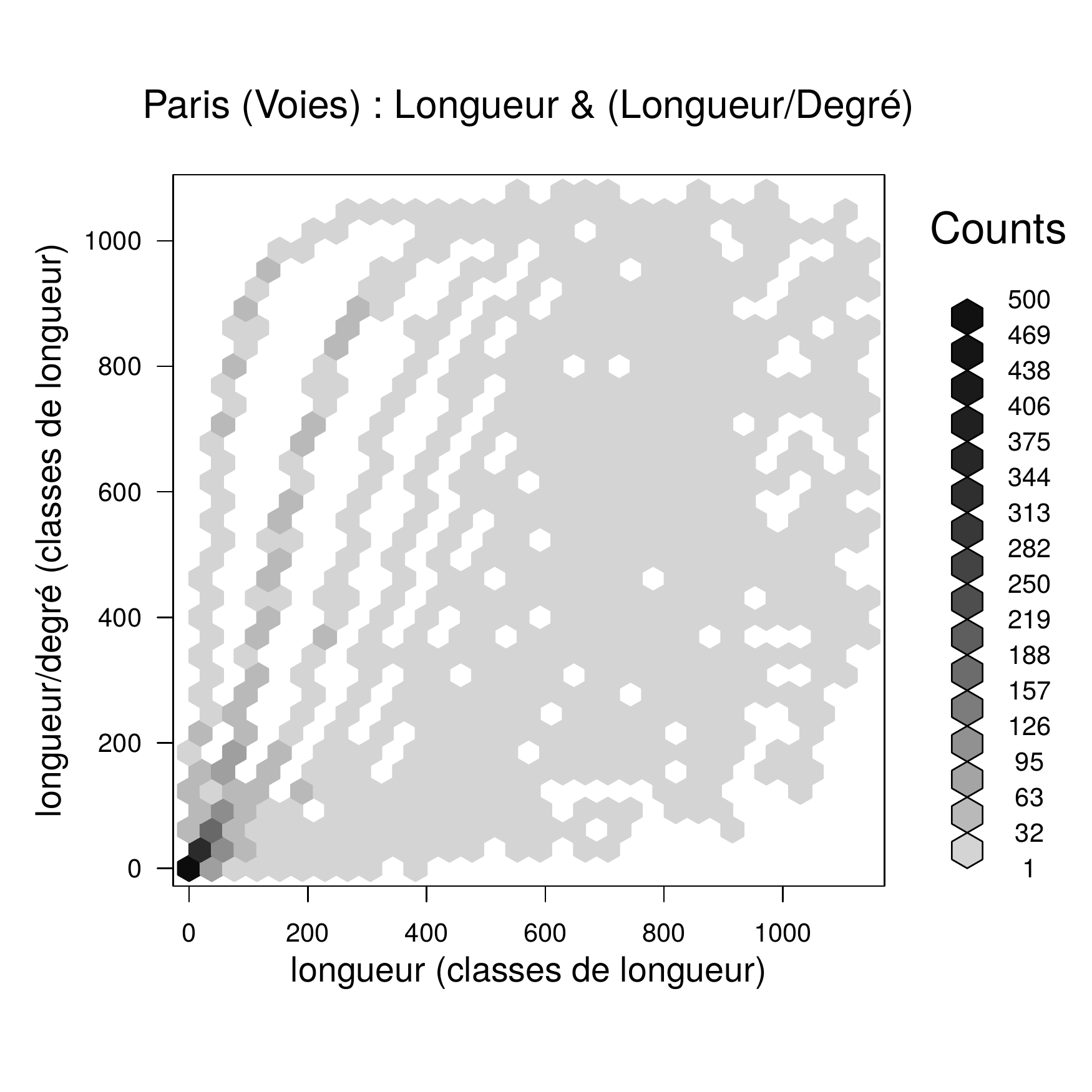}
    \end{subfigure}
    
    \begin{subfigure}{.40\textwidth}
        \includegraphics[width=\textwidth]{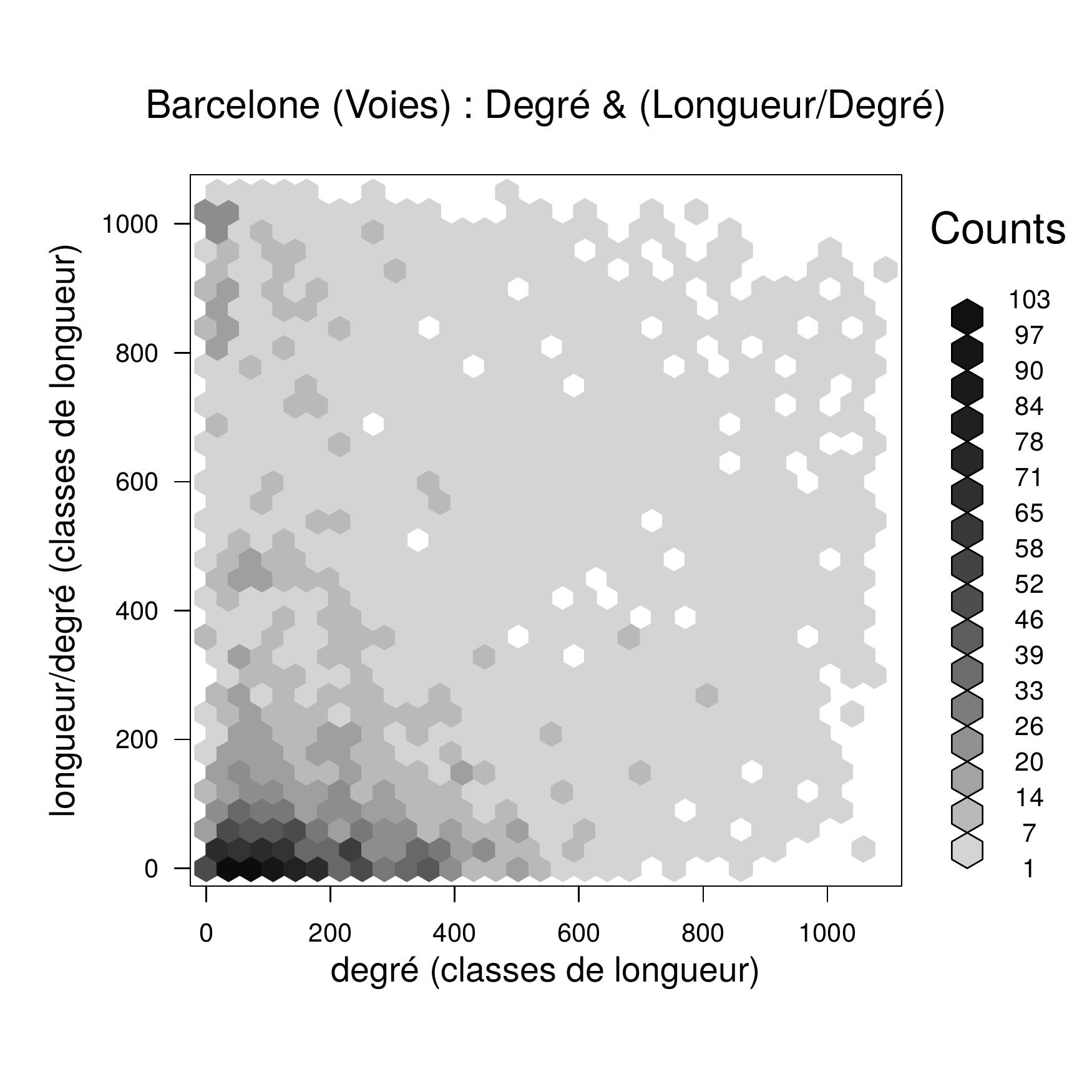}
    \end{subfigure}
    ~
    \begin{subfigure}{.40\textwidth}
        \includegraphics[width=\textwidth]{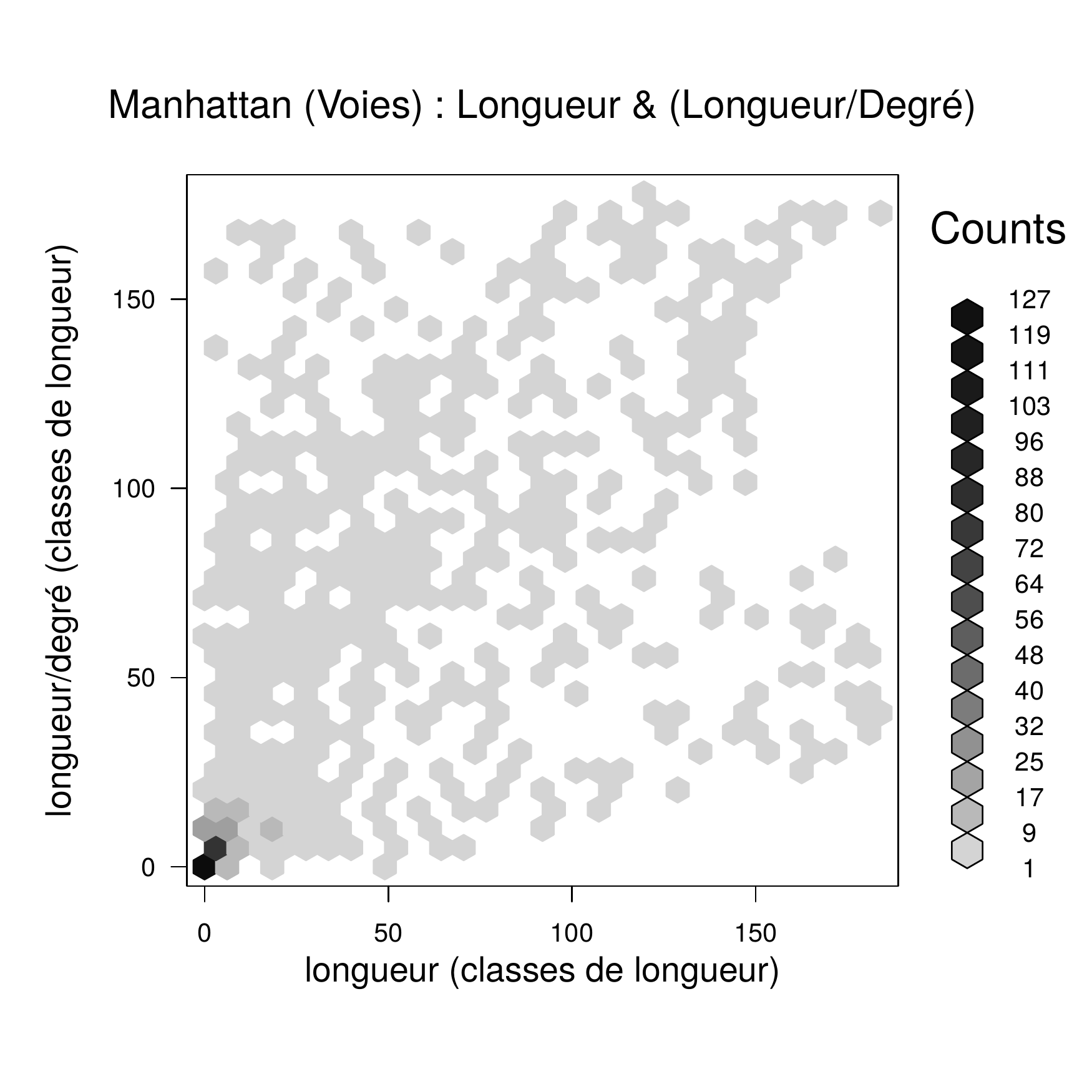}
    \end{subfigure}

    \caption{Cartes de corrélation croisée entre le degré et la longueur des voies calculés sur les réseaux viaires d'Avignon, Paris, Barcelone et Manhattan.}
    \label{fig:voies_lodX}
\end{figure}

Cependant, la combinaison de la longueur et du degré des voies entre eux n'est corrélée à aucune autre. En divisant la longueur des voies par leur degré nous obtenons un indicateur nouveau et pertinent. Nous le raffinons ensuite en remplaçant le degré par la connectivité pour obtenir la densité locale des voies. Nous avons appelé cet indicateur \emph{espacement} (cf Partie I, chapitre 3). Les cartes de corrélation appuient son unicité (figure \ref{fig:voies_lodX}).

\begin{figure}[h]
    \centering

    \begin{subfigure}{.40\textwidth}
        \includegraphics[width=\textwidth]{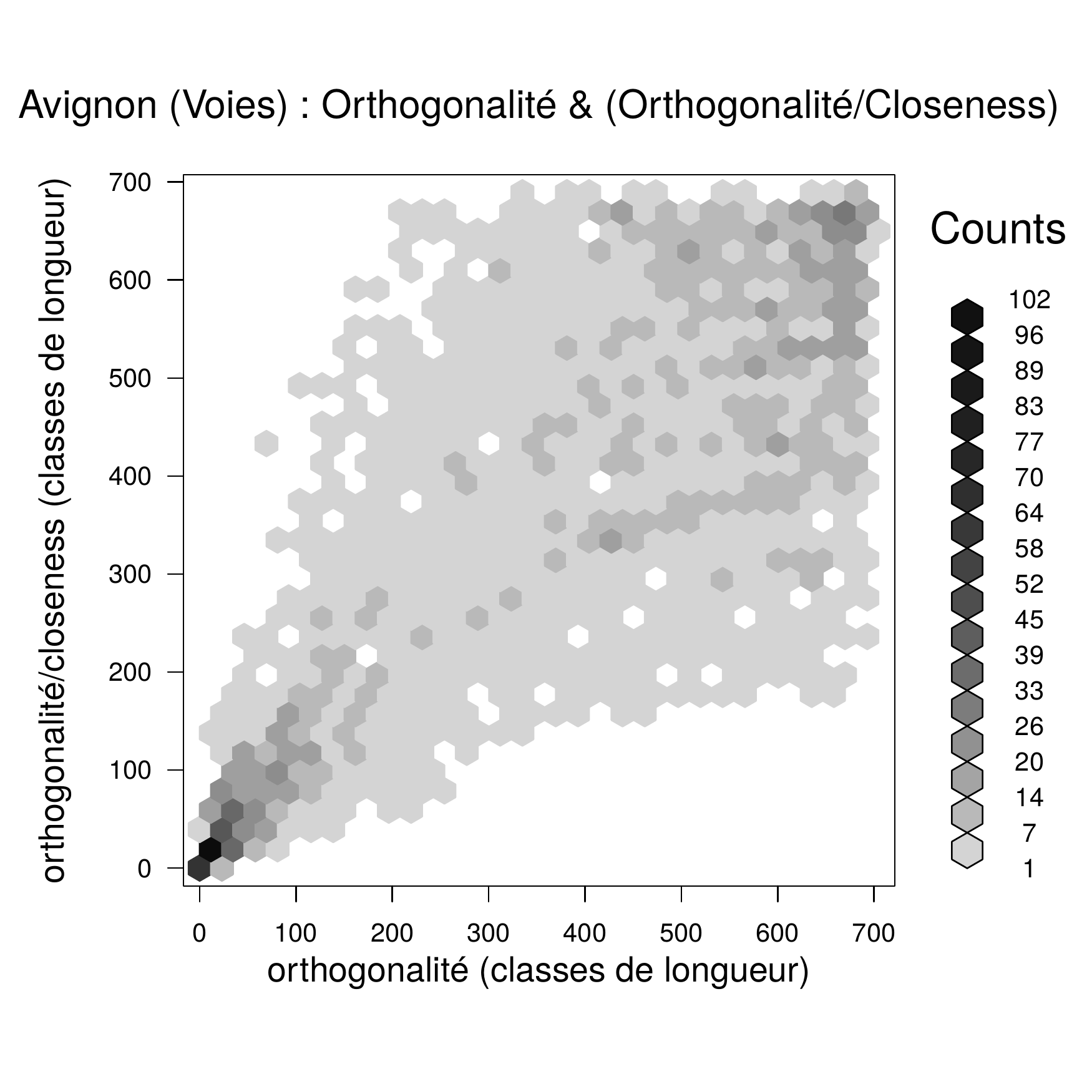}
    \end{subfigure}
    ~
    \begin{subfigure}{.40\textwidth}
        \includegraphics[width=\textwidth]{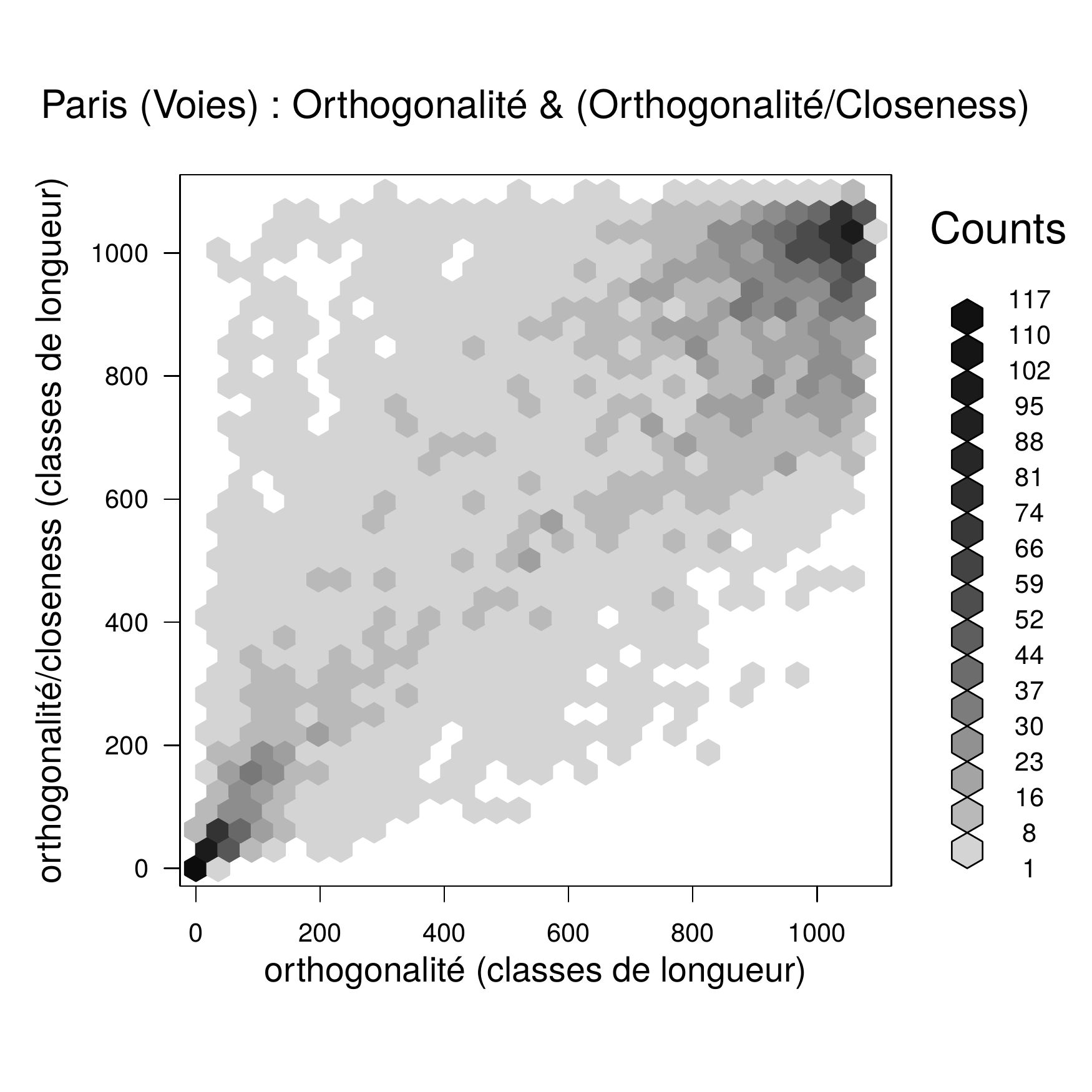}
    \end{subfigure}
    
    \begin{subfigure}{.40\textwidth}
        \includegraphics[width=\textwidth]{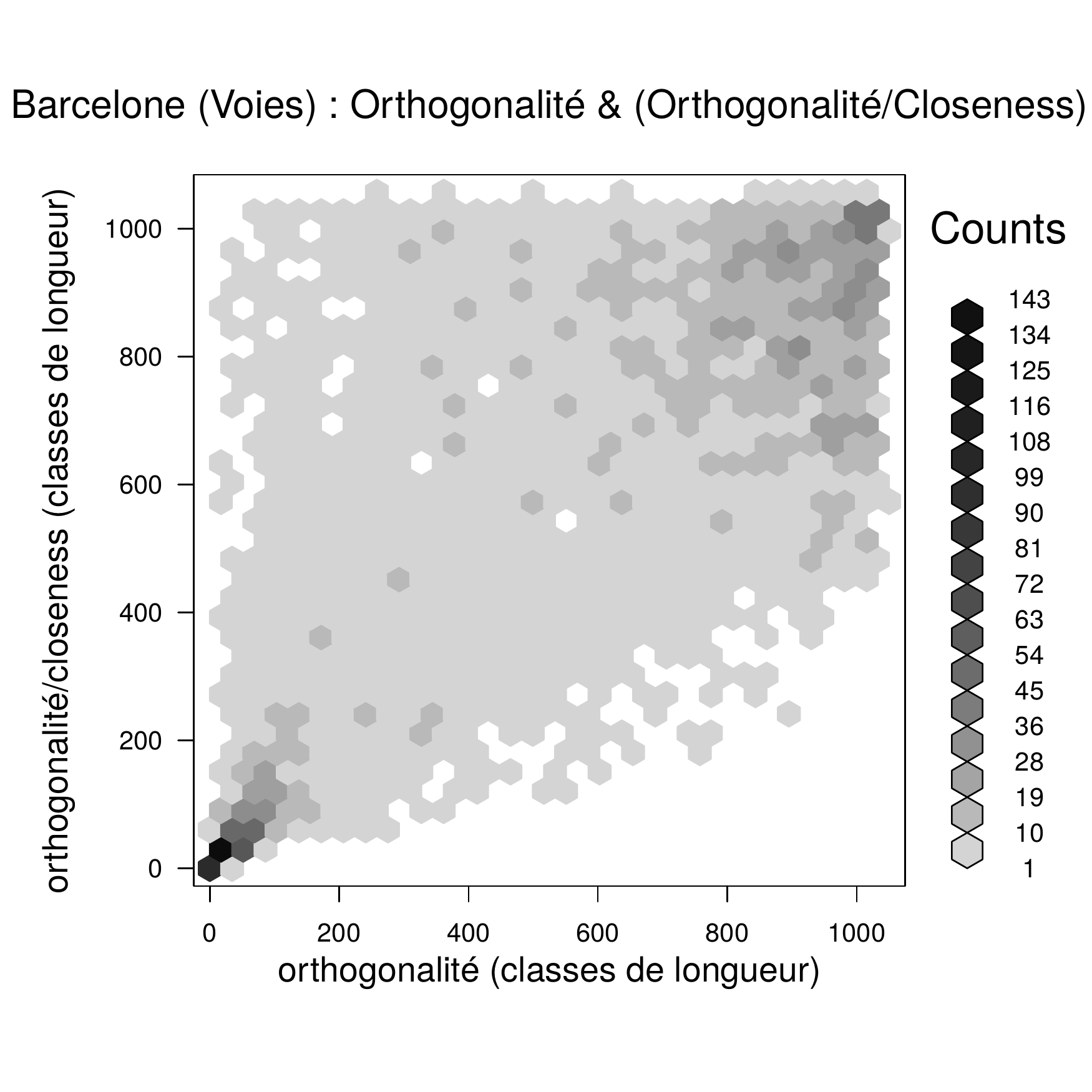}
    \end{subfigure}
    ~
    \begin{subfigure}{.40\textwidth}
        \includegraphics[width=\textwidth]{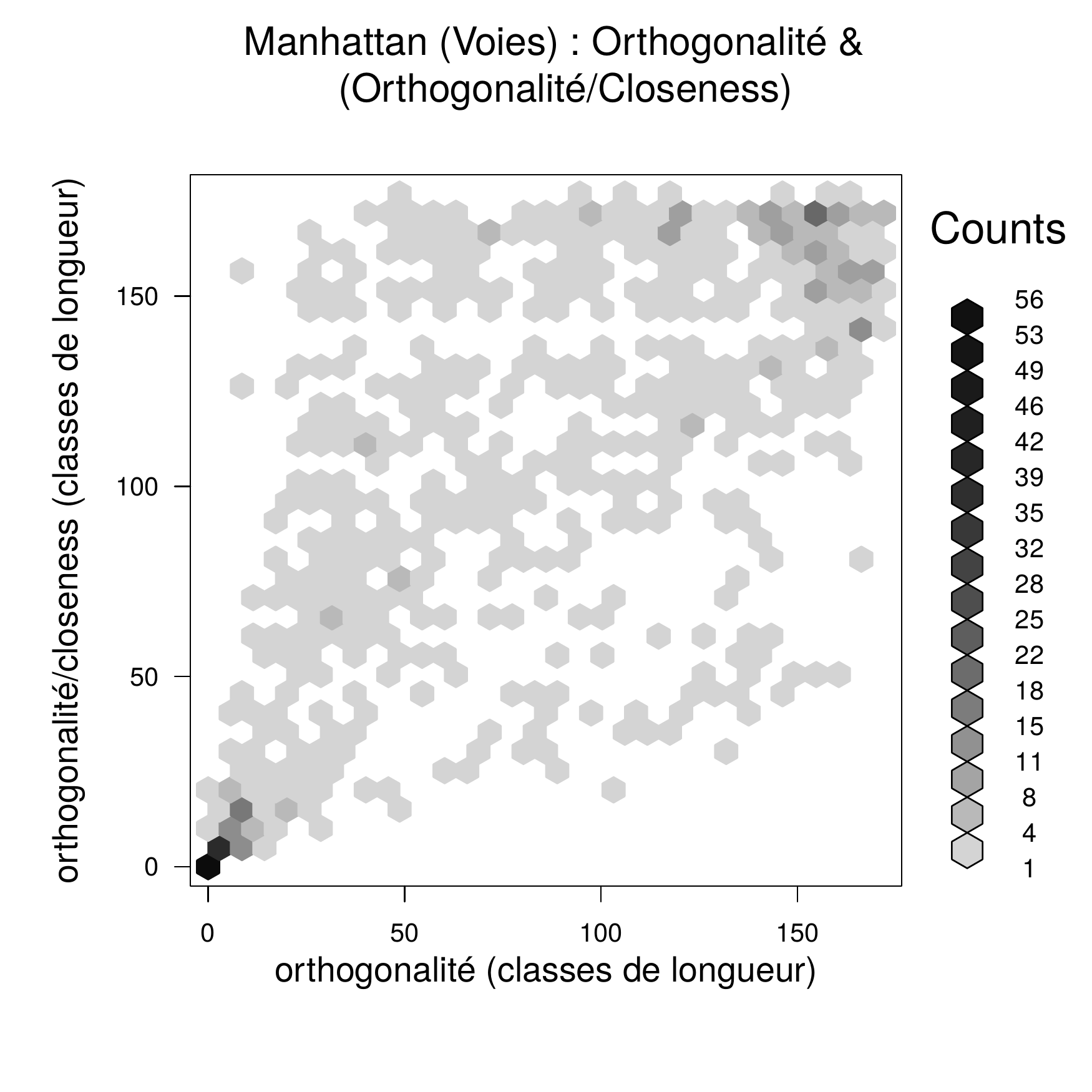}
    \end{subfigure}

    \caption{Cartes de corrélation croisée entre l'orthogonalité des voies et sa combinaison avec la closeness calculée sur les réseaux viaires d'Avignon, Paris, Barcelone et Manhattan.}
    \label{fig:voies_orthoooc}
\end{figure}

\begin{figure}[h]
    \centering

    \begin{subfigure}{.40\textwidth}
        \includegraphics[width=\textwidth]{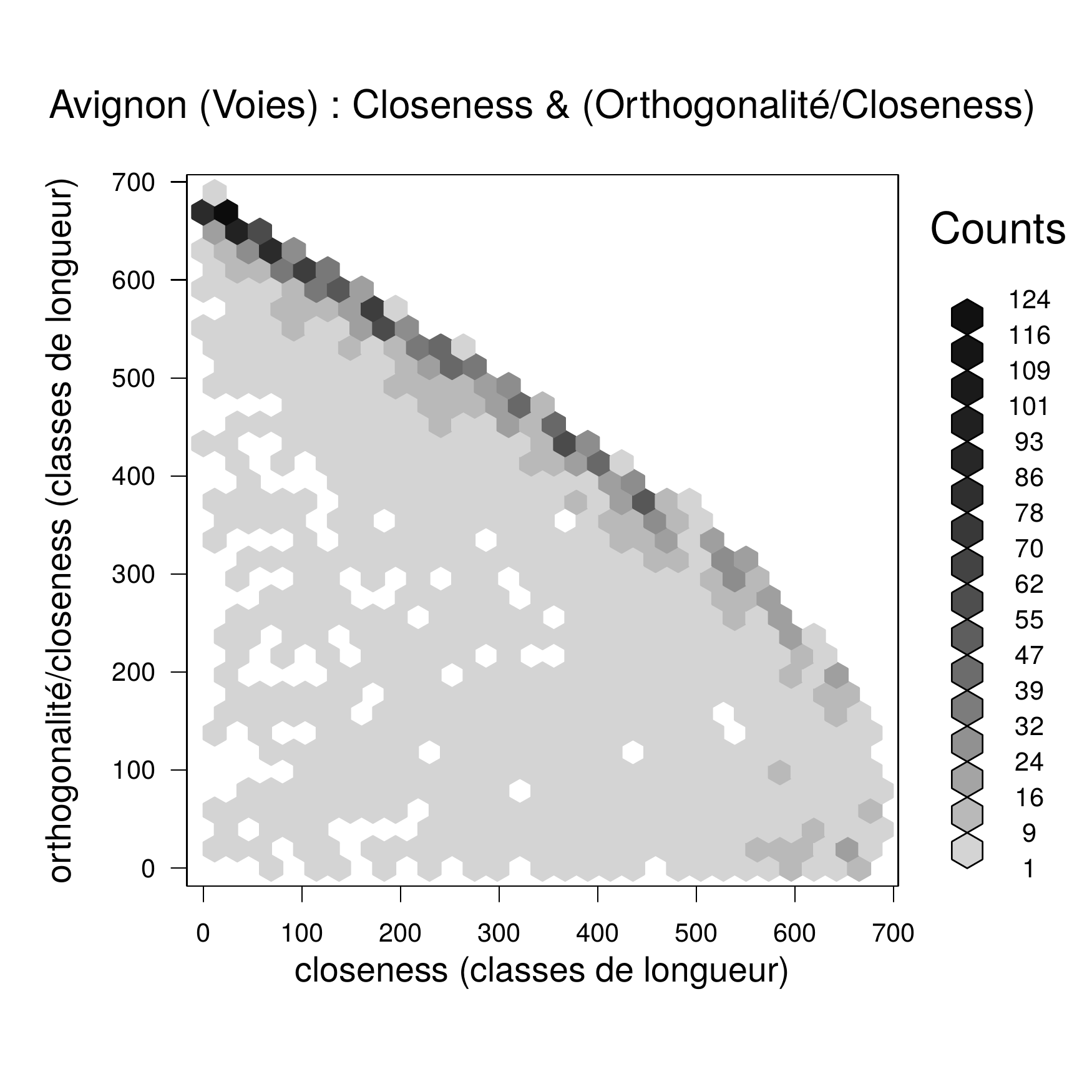}
    \end{subfigure}
    ~
    \begin{subfigure}{.40\textwidth}
        \includegraphics[width=\textwidth]{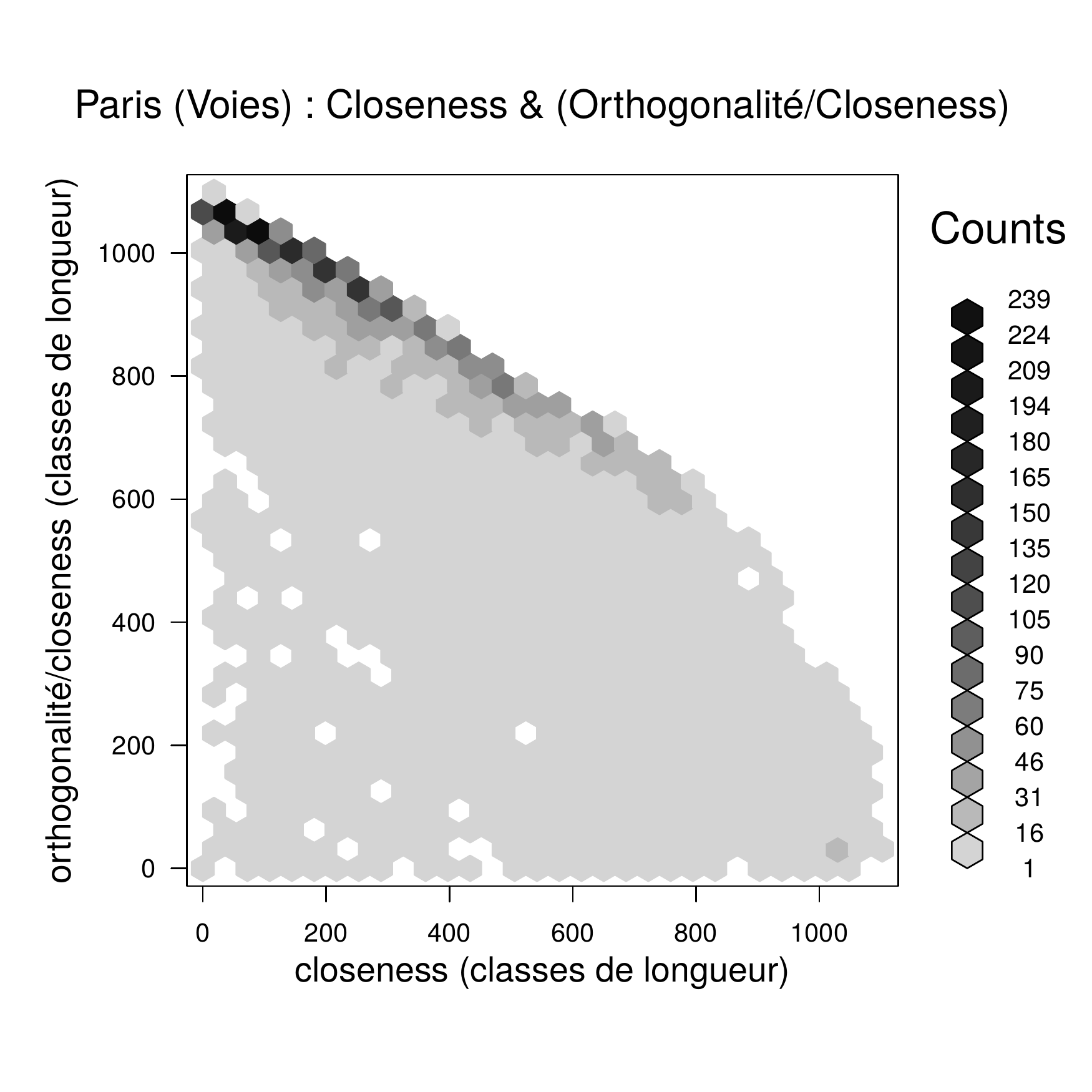}
    \end{subfigure}
    
    \begin{subfigure}{.40\textwidth}
        \includegraphics[width=\textwidth]{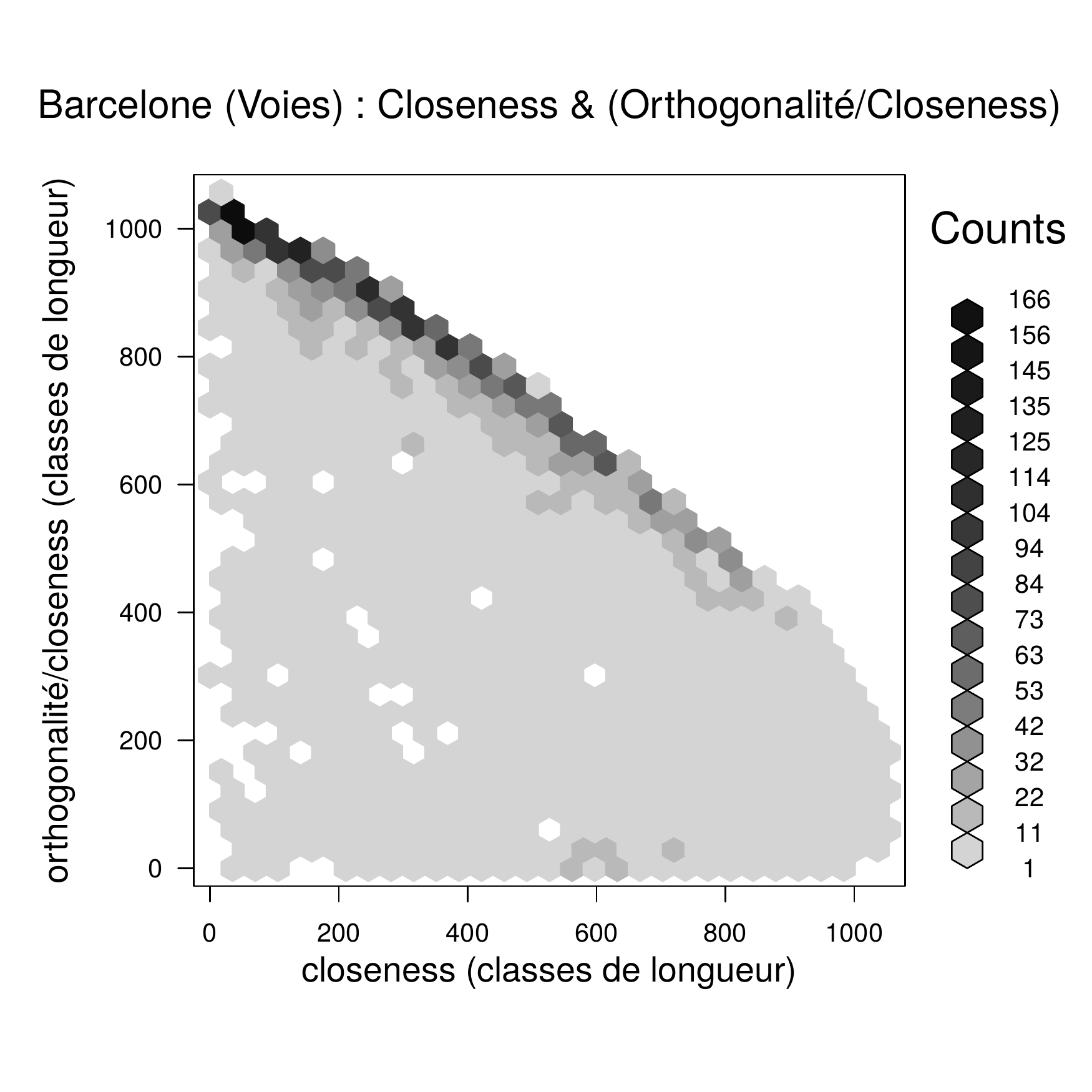}
    \end{subfigure}
    ~
    \begin{subfigure}{.40\textwidth}
        \includegraphics[width=\textwidth]{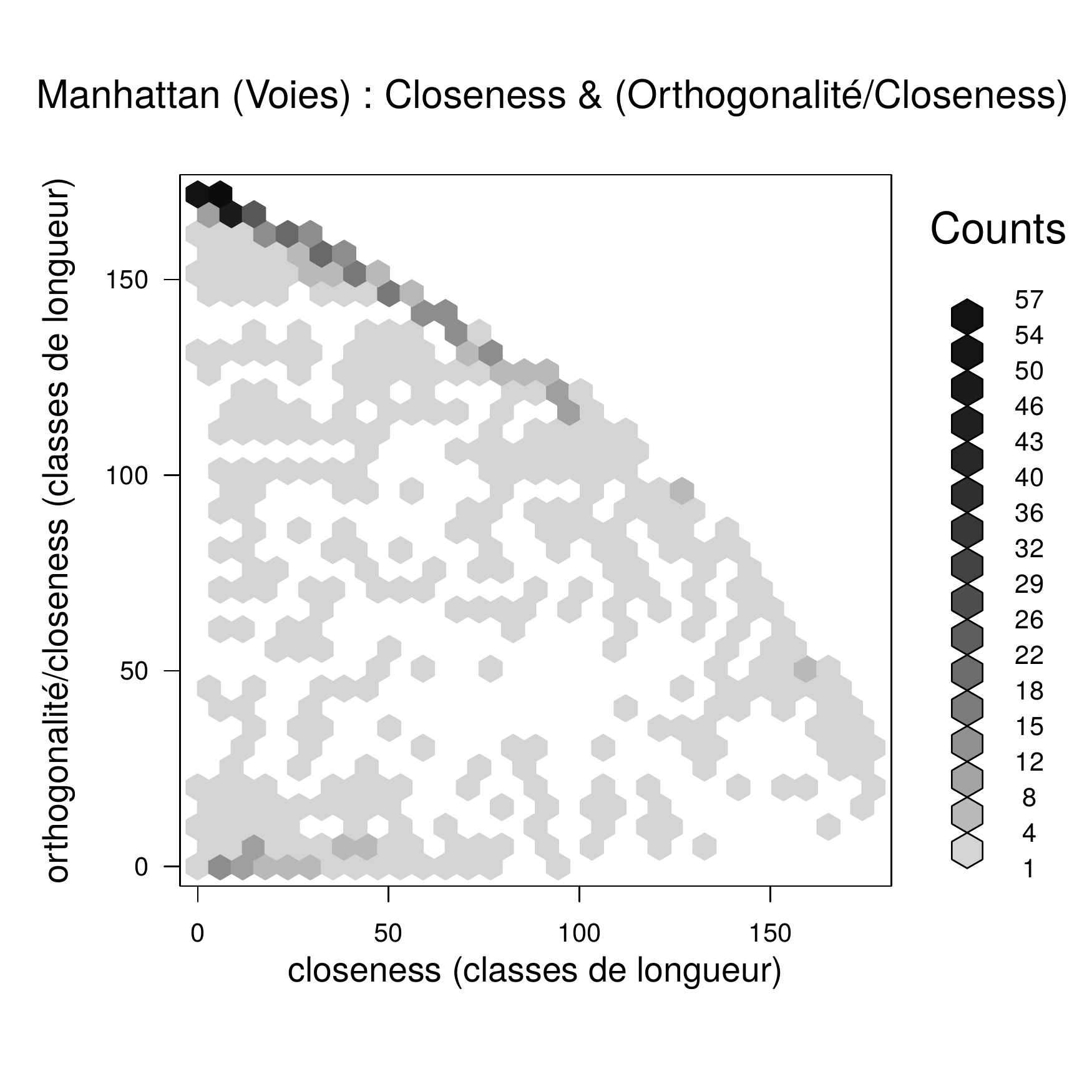}
    \end{subfigure}

    \caption{Cartes de corrélation croisée entre la closeness des voies et sa combinaison avec l'orthogonalité calculée sur les réseaux viaires d'Avignon, Paris, Barcelone et Manhattan.}
    \label{fig:voies_cloooc}
\end{figure}

Les deux indicateurs primaires restants, une fois la longueur et le degré écartés, sont la closeness et l'orthogonalité. Leur combinaison par division suit les mêmes variations que leurs valeurs respectives pour Avignon, Paris et Barcelone. Nous pouvons en déduire que l'information apportée est proche de celles que nous avions déjà. Si nous nous penchons sur les cartes de corrélations, nous observons que la corrélation évaluée autour de 0,75 par le coefficient de Pearson garde un caractère diffus (figures \ref{fig:voies_orthoooc}, \ref{fig:voies_cloooc}). Celui-ci s'observe d'autant plus sur les réseaux avec un faible nombre de voies, comme celui de Manhattan, où la structure quadrillée rend la combinaison de l'orthogonalité et de la closeness beaucoup plus corrélée à la première qu'à la seconde. Le fait que cette combinaison suive un comportement corrélé à deux indicateurs qui ne le sont pas entre eux le rend particulier. Nous pouvons en déduire que des indicateurs de closeness ou d'orthogonalité, aucun n'a une caractérisation dont la valeur domine l'autre. Nous avons observé le résultat de la multiplication de ces deux indicateurs. Le résultat obtenu donne un coefficient de Pearson, pour Avignon, de 0,67 lorsque la corrélation est faite avec l'orthogonalité, de 0,50 lorsqu'elle est faite avec la closeness. Elle est inférieure à 0,1 lorsqu'on compare cette multiplication à la longueur ou au degré.

La combinaison entre closeness et orthogonalité par multiplication nous permet de faire ressortir une nuance sur le réseau. La combinaison de ces deux indicateurs permet en effet de détecter les structures qui sont centrales (au sens de la distance topologique) et qui ont des connexions perpendiculaires. Il permet donc de distinguer dans les graphes viaires les voies rapides (qui se connectent avec des angles faibles, permettant l'insertion des véhicules) des voies d'accès anciennes (qui s'intègrent plus perpendiculairement au réseau). Nous consacrerons la troisième partie de ce travail à l'analyse qualitative de ces résultats sur les réseaux viaires.

Nous terminerons cette partie en évoquant l'indicateur de desserte locale, calculé en soustrayant le degré de chaque voie à sa connectivité. C'est une combinaison particulière, qui diffère par nature de celles considérées précédemment. Elle est construite sur deux indicateurs qui indiquent le même type d'information et qui sont donc du même ordre. Elle apporte une information quantifiée : le nombre de connexions entre une voie et ses voisines qui sont en milieu de voie et non à leur extrémité. Il n'est donc pas utile de comparer cette information quantifiée avec l'ensemble des indicateurs considérés ici. La corrélation entre degré de la voie et degré de desserte est liée au graphe sur lequel sont faits les calculs. En effet, un graphe où les voies ne s'intersectent qu'à leurs extrémités verra cette valeur s'approcher de 1 alors que celui qui est extrêmement maillé la verra diminuer.

Dans l'analyse de la combinaison d'indicateurs sur la voie, nous retenons donc l'espacement (combinant longueur et degré, ou, plus finement, longueur et connectivité), l'accessibilité maillée (combinant closeness et orthogonalité) et le degré de desserte qui est le fruit de la soustraction du degré et de la connectivité des voies.

\FloatBarrier

\section{Définition d'une grammaire pour la suite de l'étude}

Dans ce chapitre, nous avons comparé les indicateurs primaires que nous utilisons, pour définir ceux utiles à une caractérisation spatiale non redondante. Nous les avons ensuite combinés pour établir de nouvelles caractérisations pertinentes. 

Sur les arcs comme sur les voies, certains indicateurs apparaissent comme équivalents : l'utilisation et la betweenness, l'accessibilité et la closeness. Dans le calcul de l'utilisation, la normalisation par le nombre de chemins les plus simples entre deux sommets du \textit{line graph} n'a pas été prise en compte, ce qui simplifie le calcul mais qui n'en modifie pas l'information apportée par rapport à la betweenness. Dans le calcul de la closeness n'est pas prise en compte la distance métrique qui participe au calcul de l'accessibilité. Nous en déduisons donc que cette distance ne constitue pas un poids significatif. Nous écartons donc, pour les arcs comme pour les voies, les indicateurs de betweenness et d'accessibilité au profit de ceux d'utilisation et de closeness.

Pour la suite de l'étude les deux objets géographiques réagissent différemment. Les indicateurs primaires sur les arcs apparaissent comme étant décorrélés les uns des autres, et donc apportant tous une information significative. Pour les voies, le résultat est fondamentalement différent, ce qui appuie l'apport quantitatif de ce nouvel objet géographique. En effet, son caractère multi-échelle rend une caractérisation locale (par la longueur ou le degré de la voie) équivalente à une caractérisation globale (par la structuralité potentielle ou l'utilisation). Nous pouvons donc en déduire que la somme des accessibilités des voies connectées à une voie de référence est équivalente à son degré (caractériser une voie en sommant l'accessibilité des arcs qui l'intersectent revient tout simplement à sommer ce nombre de voies), de même que son utilisation. La caractérisation apportée par des indicateurs globaux (calculés sur l'ensemble du réseau) devient, grâce à la voie, équivalente à une information topologique locale.

Longueur, nombre d'arcs, degré, structuralité potentielle et utilisation apportent donc le même type d'informations sur le réseau de voies. Parmi ceux-ci nous choisirons le degré dans la suite de l'étude comme indicateur \enquote{primaire}. En effet, cet indicateur est celui dont la corrélation est la plus forte avec l'ensemble de ceux de son groupe. Nous pouvons ainsi lister les trois indicateurs primaires permettant de caractériser la voie : degré, closeness et orthogonalité, qui font ressortir des structures différentes sur le réseau.

Sur l'étude des combinaisons par division des indicateurs sur les arcs, une hiérarchie se dessine entre ceux-ci. La valeur de l'utilisation a un poids significatif dans ses combinaisons avec d'autres indicateurs. Viennent ensuite, dans l'ordre : la longueur, l'orthogonalité, le degré, la structuralité potentielle et la closeness. Une seule combinaison se détache et permet d'apporter une caractérisation supplémentaire au réseau d'arcs : celle entre degré et utilisation. Elle permet de pointer les éléments clés du réseau, qui font sa vulnérabilité : les arcs de faible degré et de forte utilisation.

Sur les voies, les résultats des combinaisons des indicateurs de longueur et de degré avec la closeness ou l'orthogonalité leur sont corrélés. Seule la combinaison de la longueur et du degré apporte une caractérisation pertinente : la densité linéaire de réseau. Nous l'affinerons dans la suite de ce travail en construisant l'indicateur d'espacement en divisant la longueur de chaque voie par sa connectivité (nombre d'arcs qui lui sont liés). La combinaison de la closeness et de l'orthogonalité nous permet d'affiner le calcul de centralité selon la distance topologique en mettant en valeur les structures centrales liées perpendiculairement au réseau. Ces structures caractéristiques seront analysées attentivement en troisième partie, portant sur l'étude qualitative des résultats. Le dernier indicateur que nous considérons pertinent pour l'étude du graphe des voies est la desserte locale, qui définit la manière dont la voie est incluse dans le réseau. Plus elle coupera les autres voies en leur centre et non en leur extrémité, plus cet indicateur aura une valeur importante.

Ces travaux nous permettent de définir plusieurs indicateurs non redondants, utiles pour la caractérisation spatiale.

Pour les arcs :

\begin{enumerate}
\item la longueur
\item le degré
\item la structuralité potentielle
\item l'utilisation
\item la closeness
\item l'orthogonalité
\item la composition entre degré et utilisation par division : $\frac{degre}{utilisation}$ 
\end{enumerate}

Pour les voies :

\begin{enumerate}
\item le degré
\item la closeness
\item l'orthogonalité
\item l'espacement (combinaison entre longueur et connectivité par division : $\frac{longueur}{connectivite}$)
\item l'accessibilité maillée (combinaison entre closeness et orthogonalité par multiplication : $closeness \times orthogonalite$)
\item le degré de desserte (combinaison entre connectivité et degré par soustraction : $connectivite - degre$)
\end{enumerate}

Dans la suite de ce travail, nous nous concentrerons sur l'objet \emph{voie}. Nous utiliserons les indicateurs définis ici pour le caractériser dans les graphes spatiaux. Ces indicateurs sont révélateurs de structures. Nous étudierons donc dans la deuxième partie trois aspects importants de leurs applications. D'une part, la dépendance entre le résultat obtenu et les données ayant servi à faire les calculs (différents types de graphes) ; d'autre part, celle avec la finesse de la numérisation (différents types de généralisation) ; et enfin celle avec le découpage de l'échantillon spatial. 
Nous ferons dans la troisième partie une analyse qualitative des résultats que ces indicateurs nous permettent de lire sur les graphes viaires.

\clearpage{\pagestyle{empty}\cleardoublepage}
\chapter{Synthèse}
\minitoc
\markright{Synthèse de la première partie}

\FloatBarrier

\section{Construction d'un objet complexe multi-échelle}

Nous avons retracé dans cette partie l'importance de l'analyse de la continuité et de l'alignement dans les graphes spatiaux, notamment pour la caractérisation des réseaux viaires. Nous approfondissons cette réflexion en construisant un objet complexe : la voie. L'objectif de notre démarche est d’offrir à un public pluridisciplinaire la possibilité de se saisir de cet objet géographique et d’en faire un élément de référence dans des domaines différents. Nous reconstruisons de manière objective une continuité qui, dans une ville, est fondée sur un sentiment de perspective. L'hypothèse sous-jacente est forte : elle suppose qu'un réseau spatialisé porte dans son inscription spatiale et dans sa topologie des informations liées à sa croissance et à son utilisation. La structure d'un réseau viaire est ainsi susceptible de participer aux stratégies de déplacement de ses utilisateurs \citep{pailhous1970representation, hillier1984social, xie2007measuring, degouysAPitineraire}. Nous tentons de comprendre dans nos travaux jusqu'où il est possible de suivre cette idée. Nous développerons le lien entre analyse quantitative et étude qualitative en troisième partie.

La voie servira de fondement à la suite de nos recherches. Construite localement, à chaque sommet, elle est indépendante du sens de lecture du graphe. C'est un objet aux multiples échelles qui permet une lecture globale approfondie du réseau à travers le développement d'indicateurs. Nous nous détachons des arcs pour montrer comment l'analyse de cet objet peut être révélatrice de structures stables. Son processus de construction est indépendant de la nature du graphe considéré et pourra être appliqué, de manière plus ou moins pertinente, sur tous les réseaux disposant d'une spatialisation.

\FloatBarrier

\section{Développement d'indicateurs}

Nous travaillons dans ce projet de recherche sur le squelette des réseaux spatiaux : nous construisons leur caractérisation à partir de leur géométrie, dénuée de toute autre information. Leur topologie est inhérente aux points de coordonnées partagées entre arcs, les sommets correspondant à leurs intersections. 

\begin{figure}[h]
    \centering
    \includegraphics[width=\textwidth]{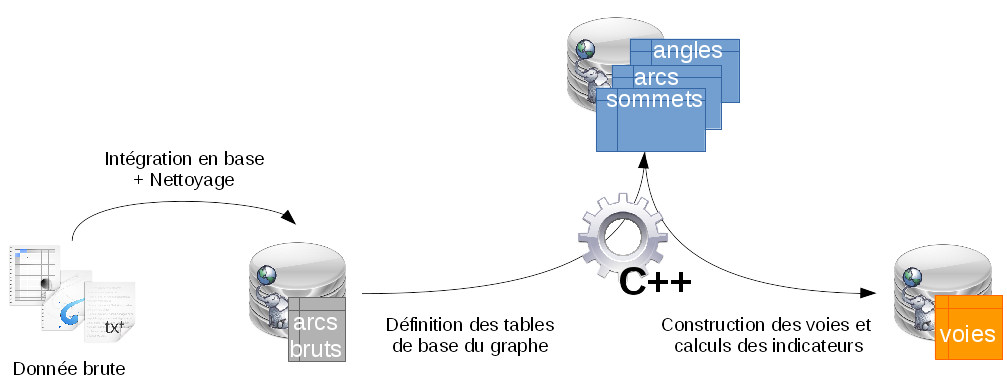}
    \caption{Processus de construction des voies à partir des données brutes.}
    \label{fig:structure_BDD}
\end{figure}

La voie est construite à partir de données brutes, issues de base de données vectorielle comme celle que propose l'IGN (la \copyright BDTOPO) sur la France ou OpenStreetMap sur l'ensemble de la planète. Après avoir été nettoyées (suppression de géométries parasites et découpage des arcs), les données sont intégrées dans un système de gestion de base de données adapté aux données spatialisées. Un programme développé en C++ (présenté en annexe \ref{ann:chap_tuto_1}) interagit avec la base de données pour y lire la table de données brutes et y écrire les nouvelles tables utiles à l'analyse de la géométrie du graphe. Ces nouvelles informations permettront de créer, via ce programme, une table de voies contenant leurs géométries (figure \ref{fig:structure_BDD}).

Nous pouvons paramétrer les différentes méthodes de construction de la voie étudiées dans le deuxième chapitre de cette partie. Nous choisirons pour la suite de ce travail de conserver une méthode de construction privilégiant les angles de déviation minimum entre arcs ($M0$) avec un angle seuil (définissant la déviation maximale autorisée) de 60\degres .

La table des voies ainsi construite permet, à travers le programme, de compléter ses attributs avec le développement de différents indicateurs. Ceux présentés dans cette recherche sont pour certains issus de travaux antérieurs en théorie des graphes. Nous retrouvons ainsi plusieurs calculs de centralités suivant différents critères : la \textbf{\emph{degree centrality}}, la \textbf{\emph{betweenness centrality}} et la \textbf{\emph{closeness centrality}}. Nous avons analysé leurs pertinences sur des réseaux spatialisés et nous les avons articulés avec de nouveaux indicateurs propres à nos travaux. Nous développons donc de nouvelles caractérisations, fondées sur le caractère spatial de nos recherches. L'indicateur le plus simple que nous puissions évoquer lié à cette propriété est celui de \emph{\textbf{longueur}}. 

Nous proposons ainsi l'indicateur d'\textbf{\emph{orthogonalité}}, qui caractérise un objet en fonction des angles de connexions que fait son voisinage direct avec lui. Calculé localement, pour chaque voie, il révèle une capacité à identifier des structures globales et des formes de tissus particulières. Appliqué aux arcs, il apporte une information complémentaire sur des structures plus locales. En combinant cet indicateur avec la closeness, nous créons l'indicateur d'\textbf{\emph{accessibilité maillée}}, qui, dans le contexte urbain, parvient à caractériser plus finement certains types de réseaux, en fonction de leurs géométries et de leurs centralités topologiques. Le lien avec l'histoire de ceux-ci sera développé par la suite.

Nous développons un indicateur combinant des objets à plusieurs niveaux de lecture d'un graphe. Ainsi, la \textbf{\emph{connectivité}} est donnée pour une voie par l'ensemble des arcs qu'elle intersecte. Ce double niveau de lecture nous permet d'affiner notre analyse de l'espace. Nous pouvons ainsi construire l'indicateur de \textbf{\emph{degré de desserte}} qui fait ressortir les voies dont la connectivité est plus importante que le degré. Ces voies intersectent leurs voisines en leurs centres, et non à leur extrémité. Elles sont révélatrices d'un certain mode de desserte du territoire.

Nous construisons également sur des critères géométriques, l'indicateur d'\textbf{\emph{espacement}}, en divisant la longueur des objets par leur connectivité. Cela correspond à la densité linéaire des voies. C'est-à-dire, combien d'arcs sont connectés à la voie par unité de longueur. Cette caractérisation permet de différencier les parties du graphe où le filaire se concentre en amas dense de celles où il s'étire.

Nous avons ajouté la distance métrique comme poids de la closeness pour observer l'impact de ce critère spatial. Nous avons ainsi développé l'indicateur d'\emph{\textbf{accessibilité}} en suivant le raisonnement de T. Courtat \citep{courtat2012walk}. Cet indicateur est révélateur de la proximité de l'ensemble du réseau par rapport à chaque objet, en tenant compte des longueurs de chacun d'eux. Appliquées aux arcs, la closeness et l'accessibilité donnent des résultats inexploitables, car trop influencés par le découpage de l'échantillon. Sur les voies, les deux indicateurs donnent des résultats très corrélés. La pondération par la longueur n'est donc pas pertinente pour créer une nouvelle caractérisation du graphe. 

Pour essayer de rendre l'information d'accessibilité innovante, nous la renforçons en sommant pour chaque objet l'accessibilité de ceux qui l'intersectent. Nous construisons ainsi l'indicateur de \emph{\textbf{structuralité potentielle}}. Sur les voies, celui-ci donne une information qui se retrouve très corrélée à celle apportée par leur degré. Nous pouvons en conclure que l'information topologique locale (de distance ou de voisinage) prime sur celle métrique globale pour les voies.

La \emph{\textbf{betweenness}} a également retenu notre attention. Le temps de calcul que nécessite cet indicateur sur de grands graphes nous a amené à lui préférer une autre méthode de construction, par \enquote{filiation}, qui ne tient compte que du nombre de chemins les plus simples passant par chaque sommet. L'indicateur obtenu, appelé ici \emph{\textbf{utilisation}} (plus connu sous le nom de \textit{stress centrality} en théorie des graphes), donne un résultat significativement corrélé à celui du célèbre indicateur. Nous pourrons donc l'utiliser de manière équivalente sur les arcs ou les voies.

\begin{figure}[h]
    \centering
    \includegraphics[width=\textwidth]{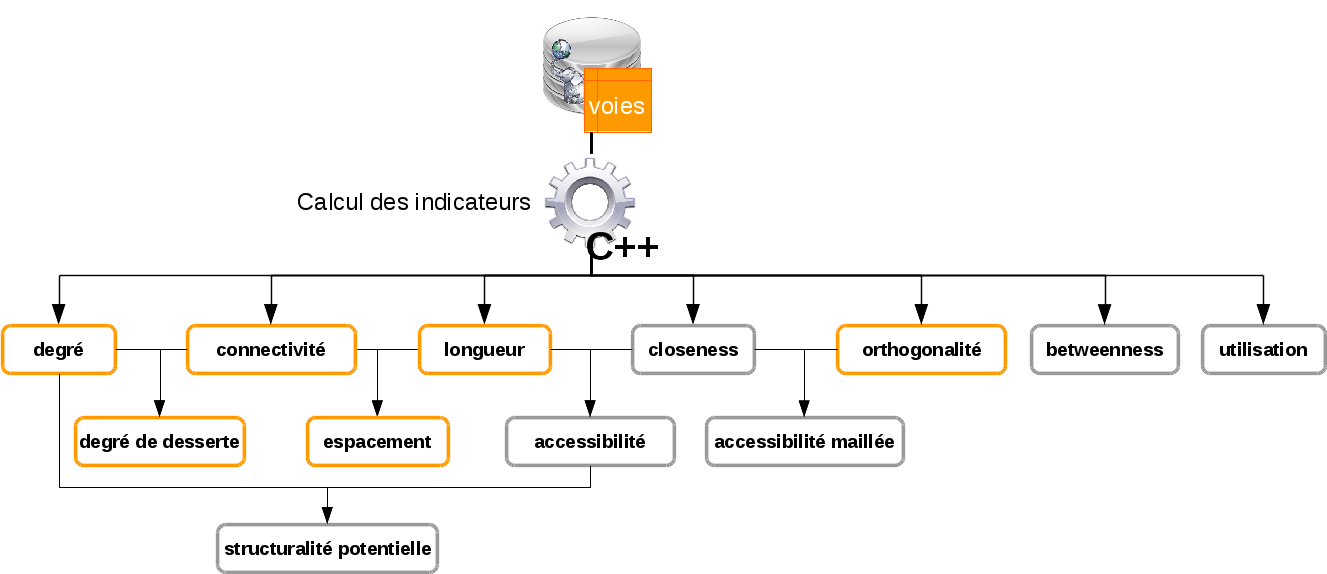}
    \caption{Arborescence d'indicateurs construits à partir de la voie. Les indicateurs calculés localement sont cerclés de orange, ceux dont la construction se fait à partir de l'ensemble du graphe le sont de gris.}
    \label{fig:BDD_indicateurs}
\end{figure}

Nous avons ainsi défini des indicateurs de différents \textit{niveaux} de constructions. Certains sont issus des propriétés propres aux objets, calculées localement (comme le degré ou la longueur) ou en tenant compte de l'ensemble du réseau (comme la closeness ou l'utilisation). D'autres utilisent les premiers dans leur construction afin d'apporter de nouvelles informations. Nous avons représenté l'arborescence complète des indicateurs construits sur la figure \ref{fig:BDD_indicateurs}.

Ce travail a ainsi pour objectif de pousser la caractérisation des réseaux spatiaux. Les nouveaux indicateurs que nous développons ont pour objectif d'apporter une nouvelle information pour caractériser le graphe, notamment à travers sa géométrie.

\FloatBarrier

\section{Grammaire de lecture de la spatialité}

L'ensemble des indicateurs développés et explorés permet de caractériser les arcs et les voies de réseaux spatiaux. Selon l'objet sur lequel ils sont appliqués, l'information apportée sera différente. Sur les arcs, hormis la closeness et l'accessibilité ainsi que la betweenness et l'utilisation qui ont été identifiées comme très corrélées, chaque indicateur apporte une caractérisation différente. Le caractère multi-échelle de la voie homogénéise ces différences lorsque les indicateurs lui sont appliqués. Plus le réseau sur lequel sont construites les voies est maillé (à l'image de celui de Manhattan), plus la corrélation entre les différents indicateurs sera forte.

\begin{figure}[h]
    \centering
    \includegraphics[width=0.8\textwidth]{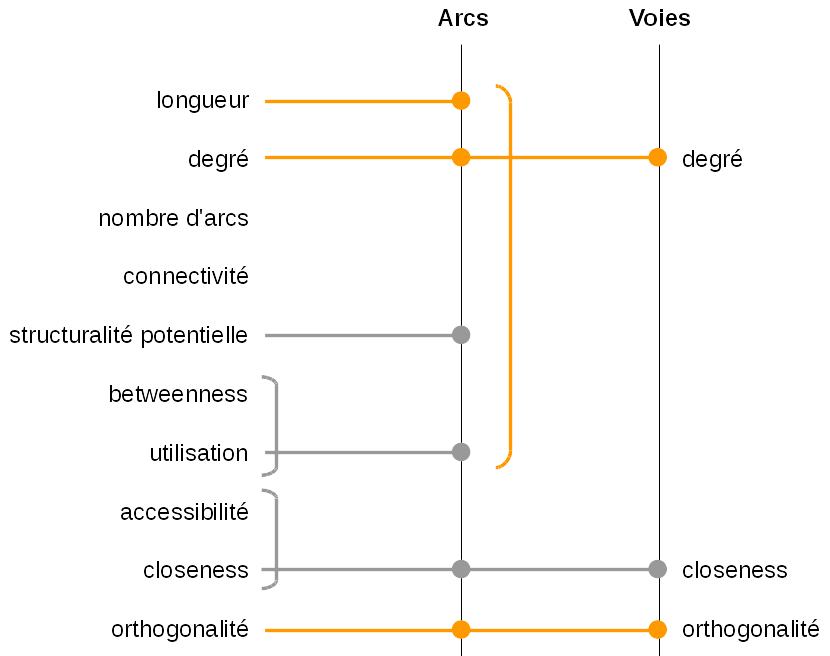}
    \caption{Indicateurs primaires retenus pour la caractérisation des arcs et des voies. Les lignes oranges font référence à un caractérisation locale, les lignes grises à une caractérisation globale.}
    \label{fig:indicateurs_primaires}
\end{figure}

Ainsi, si nous pouvons retenir six indicateurs \textit{de première génération} (que nous appelons également indicateurs \textit{primaires}) pour les arcs, leur nombre est réduit à trois pour la voie (figure \ref{fig:indicateurs_primaires}). Le calcul d'indicateurs sur cet objet géographique permet d'unifier des caractérisations qui lui sont propres (comme sa longueur, son nombre d'arcs ou son degré) à celles calculées sur l'ensemble du réseau (comme l'utilisation). La voie permet donc de rendre une information globale équivalente à une information locale.

Par ailleurs, le calcul de l'indicateur de closeness sur les voies permet de mettre en évidence des structures cohérentes, alors qu'elle n'était pas pertinente sur les arcs. Ceci permet de présager de la stabilité de la lecture proposée par l'objet.

Une liste de combinaisons exhaustive des indicateurs primaires est impossible à dresser. Nous pouvons cependant conjecturer l'étendue possible de nouvelles caractérisations à travers les corrélations plus ou moins importantes que nous observons. Ainsi, se révèle pertinente pour les arcs celle entre degré et utilisation. Pour les voies, nous retiendrons l'espacement, l'accessibilité maillée et le degré de desserte. Chacun des indicateurs retenus apporte une information différente sur la structure du réseau. Il est de cette manière possible de définir une grammaire de lecture de la spatialité que nous utiliserons dans la suite de ce travail. 

Les recherches que nous présentons dans les parties suivantes sont fondées sur les voies, hypergraphe auquel est apposée la grille de lecture définie par cette grammaire. Nous avons évoqué les différents aspects de construction, locale ou globale, des indicateurs. Il nous faudra donc porter une attention très particulière à l'impact que pourront avoir les données (nature, finesse de vectorisation ou découpage) sur les résultats obtenus.



\clearpage{\pagestyle{empty}\cleardoublepage}
\part{Analyser les structures : Caractérisation quantitative des graphes spatiaux}
\markboth{Analyser les structures}{Analyser les structures}


\clearpage{\pagestyle{empty}\cleardoublepage}
\chapter{Introduction}
\minitoc
\markright{Introduction de la deuxième partie}

\FloatBarrier
\section{La structure des réseaux viaires}

Au XVIII\textsuperscript{ème} siècle, un problème lié à un réseau de transport, a inspiré à Euler sa formalisation sous forme de graphe. W. Garrison et D. Marble ont introduit cette modélisation dans l'étude géographique de ces réseaux \citep{garrison1962structure}. Leur géométrie peut être considérée comme une structure d'accueil de flux, un canal de mouvements \citep{lowe1975geography} dont l'analyse peut se faire en conservant une information plus ou moins spatialisée. Ainsi dans les travaux de M.T. Gastner, les nœuds ont des coordonnées mais pas les arcs qui les connectent \citep{gastner2006spatial}. C'est la \textit{relation} entre différents points spatialisés qui est étudiée ici et non la géométrie avec laquelle cette relation est établie. L'attention peut alors être portée sur les temps de parcours \citep{janelle1969spatial} ou sur les prix pratiqués sur les routes privatisées \citep{zhang2005road}. Ces recherches impliquent un poids aux arcs (non nécessairement spatialisés). La forme de la connexion entre les sommets est ici une information secondaire.

Cependant, les informations portées par la géométrie peuvent se révéler riches de sens \citep{kansky1963structure}. 
La relation entre la géométrie des structures et le territoire dans lequel elles sont incluses est établie par l'histoire et la topographie du lieu. Parmi les réseaux spatiaux, les réseaux viaires se construisent selon deux logiques différentes : à travers le temps ou selon des programmes urbains spécifiques, liés à des périodes précises. Ces deux \textit{dynamiques} font intervenir des \textit{forces} différentes. Sur une longue durée, les réseaux se construisent par actions réciproques d'une partie du graphe sur une autre (influence des centres villes sur les quartiers avoisinants). La densification, fait naître des \textit{potentiels} autour des arcs existants permettant de déterminer, selon les politiques du lieu, comment chaque nouveau tronçon est ajouté au réseau pré-existant. Nous appelons \textit{villes organiques}, les villes qui se sont construites à travers l'Histoire, autour d'un noyau, par ajouts successifs. Celles-ci s'opposent aux \textit{villes planifiées} qui ont vu l'ensemble de leur plan urbain se construire selon un seul et même projet. Cela peut également ne concerner que certains quartiers.

Quelle que soit la manière dont elles sont créées, les structures de leurs réseaux viaires posent question. Tout d'abord, dans leur caractérisation, car ces réseaux peuvent être transcrits selon des graphes de natures très différentes. R. Hamaina \textit{et al.} ont ainsi approximé un réseau viaire selon trois types de graphes : un arbre de recouvrement minimal, une triangulation de Delaunay et le graphe correspondant au réseau physique tel que nous l'utilisons dans ce travail \citep{hamaina2012caracterisation}. En appliquant différents indicateurs de centralité à ces réseaux, ces travaux ont abouti à une analyse des tissus urbains indépendante des informations sémantiques qui leur sont liées. Cette approche est la même que celle que nous suivons pour cette recherche, mais son objet d'application est différent. Le caractère générique de la méthodologie utilisée permet de comparer différents réseaux pour en identifier les propriétés communes \citep{kansky1989measures, cardillo2006structural}.

Les structures des réseaux viaires introduisent également des problématiques de représentation. En effet, leur complexité peut nuire à leur lecture selon l'échelle de représentation et l'information conservée. G. Touya s'est ainsi penché sur les problématiques liées à leur généralisation \citep{touya2010road}. Il a développé un modèle de reconnaissance automatique de structures complexes et de réduction de celles-ci en des schémas simples (intersections en \enquote{T}, fourches, etc.).

Une fois la représentation du réseau établie, la hiérarchisation de structures à l'intérieur de celui-ci pose de nouvelles questions de recherche. S. Scellato \textit{et al.} définissent ainsi la structure principale d'un réseau avec un arbre couvrant minimum \citep{scellato2006backbone}. Puis P. Crucitti \textit{et al.} classent les intersections à l'aide d'indicateurs classiques de centralité, dont la \textit{closeness}, et la \textit{betweenness}, étudiées dans la première partie \citep{crucitti2006centralitymeasures}. E. Mermet, dans ses travaux de doctorat, propose une exploration des propriétés structurelles des réseaux de transports à travers des indicateurs relationnels, qu'il exploite à l'aide du développement d'un logiciel, GeoGraphLab \citep{mermet2011aide}. Ces travaux montrent que les motifs, qu'ils soient topologiques \citep{buhl2006topological} ou géométriques \citep{xie2007measuring}, sont révélateurs d'une hiérarchie entre les structures du réseau, non nécessairement planifiée \citep{levinson2006self}.

\FloatBarrier
\section{La philosophie des analogies}

La détection de structures dans les réseaux permet de disposer d'éléments de comparaison, et d'identifier d'éventuels schémas partagés \citep{milo2002network}. Cela ne concerne pas uniquement les réseaux spatialisés. A. Barrat \textit{et al.} se sont ainsi intéressés à la comparaison des propriétés structurelles de réseaux pondérés, à partir de données de transport aérien (aéroports connectés par des vols directs) et de données sur les collaborations scientifiques (définies à partir des co-auteurs d'articles en physique soumis sur un site d'archivage) \citep{barrat2004architecture}. Ces deux thématiques n'ont \textit{a priori} aucun lien entre elles, si ce n'est qu'elles peuvent être représentées et analysées sous forme de réseau, ce qui permet aux chercheurs de confronter leurs comportements.

Les réseaux spatialisés se retrouvent dans un nombre important de contextes : biologique, géo-morphologique, technique, etc. Il est ainsi également possible de comparer deux réseaux n'ayant \textit{a priori} aucune relation. Ainsi, en observant un motif de craquelures dans de l'argile, S. Bohn \textit{et al.} ont montré ses similitudes structurelles avec un extrait du réseau viaire de Paris \citep{bohn2005four}. En effet, le réseau de craquelures, formé sur le plat en céramique étudié, fait apparaître une hiérarchisation de ses arcs, et une division organisée de l'espace. Le réseau s'étend sur la surface qui lui est proposée par des divisions successives (que l'on peut représenter sous forme d'arbre), dont il est possible de retracer la progression \citep{bohn2005hierarchical}. Ce processus formant chaque nouvel arc à partir de ceux qui le précèdent peut être comparé à celui qui voit naître un tronçon de rue à partir d'un tronçon plus ancien.

Chez les plantes, au cours de leur croissance, les feuilles grandissent de 10 micromètres à 10 centimètres, créant le maillage de leur réseau de veinures. Les fougères, plantes au système de développement primitif, composent un réseau simple de veinures. Celui-ci apparaît dans d'autres plantes, de manière transformée, sous la forme d'un réseau réticulé, évolution du premier en un système plus complexe. Certaines plantes développent des réseaux avec des arbres de distribution, d'autres le complexifient. Par ailleurs, les gorgones, coraux aux réseaux de développement complexes, font apparaître une structure de réseau hiérarchisée localement mais pas globalement.

Les réseaux fractals, à l'image du chou romanesco qui au cours de sa croissance répète sa structure, sont également comparables sous un certain angle aux réseaux de rues \citep{batty2007cities}. On retrouve ainsi dans les réseaux viaires un caractère multi-échelle. La complémentarité entre le réseau d’autoroutes à l'échelle nationale et celui entre les communes importantes à l'échelle régionale en est une illustration. Cependant, cette analogie a ses limites, le réseau viaire étant, à l'échelle locale, contraint par son environnement géographique et l'aspect organique ou planifié de sa construction.

Qu'ils correspondent à des craquelures, des veinures de feuilles ou aux rues d'une ville, les mécanismes de croissance d'un réseau sont complexes. Et même si des analogies peuvent êtres faites entre différents réseaux spatiaux, celles-ci ne prennent en compte que leurs géométries. Quantifier les limites de cette comparaison est un des objets du travail que nous présenterons dans cette partie.

\FloatBarrier
\section{Modélisation diachronique}

Les dynamiques étudiées sur des réseaux peuvent être de types très différents : résilience à des attaques successives (par retrait d'arcs et/ou de nœuds), diffusion, émergence... Elles ont été principalement étudiées sur des réseaux non spatialisés \citep{barrat2008dynamical}, où le temps peut devenir un élément de la modélisation \citep{holme2012temporal}. La croissance des réseaux de transport a fait l'objet de travaux qui ouvrent sur de nombreuses questions de recherche \citep{xie2009modeling}  : qu'il s'agisse de réseaux routiers britanniques  \citep{fullerton1975development}, d'Amérique du Nord \citep{taaffe1996geography} ou de réseaux ferrés suédois \citep{morrill1965migration}. Les chercheurs se sont également questionnés sur les propriétés invariantes dans l'évolution de tels réseaux \citep{bon1979allometry}.

Liée à leur développement, leur utilisation à travers le temps a été retracée par W. Garrison \citep{garrison2005transportation}. Les géographes se sont concentrés d'abord sur leurs transformations topologiques (années 1960), avant de porter leur attention sur leurs flux (années 1970) lorsque les déplacements via le réseau routier s'intensifièrent, la prédiction du trafic devenant un enjeu économique. Les géographes se sont intéressés ensuite à la modélisation de leur croissance d'un point de vue topologique et structurel \citep{haggett1969network, lowe1975geography}. Le territoire était alors considéré comme le \textit{contenant} du développement économique \citep{lachene1965networks}.

Modéliser la croissance du réseau viaire implique de reconstituer un phénomène continu à partir de données discrètes. E. J. Taaffe a proposé un modèle en quatre étapes pour décrire l'expansion du réseau routier mis en place par les colonies dans des pays en voie de développement \citep{taaffe1963transport}. Ce modèle a été repris, tout d'abord par A. Pred sur le bord de mer Atlantique des USA \citep{pred1966spatial} ; puis par P. Rimmer sur une île au Sud de la Nouvelle-Zélande \citep{rimmer1967changing}. W. Black a proposé un modèle itératif appliqué au réseau ferré du Maine (USA) \citep{black1971iterative}. Il le décrit comme un arbre issu du réseau de Portland, s'étendant pour se connecter aux nœuds périphériques.

En 1990, la science des réseaux apporte à l'étude de leur croissance les notions d'attachement préférentiel, de centralité et d'émergence. Les réseaux de transports étaient considérés comme un système composé d'agents et dont les modélisations pouvaient être faites en ce sens. Les travaux menés se partagent en deux courants distincts : celui s'intéressant aux surfaces (parcelles, bâti) et celui s'intéressant aux lignes (axes des routes). Dans la majorité des études de croissance via des surfaces, les modélisations sont très descriptives (elles regroupent un grand nombre de paramètres). Nous retrouvons parmi celles-ci les travaux du laboratoire COGIT de l'IGN avec le projet GeOpenSim \citep{perret2010systeme, curie2010simulation}.

Les physiciens modélisent la croissance des réseaux de transport comme un processus d'optimisation, en s'intéressant au linéaire, dépourvu de toute autre information. Ils s'intéressent à un équilibre entre le coût et les bénéfices de tels réseaux \citep{gastner2006spatial}. Les chercheurs simulent une géométrie optimale, définie pour assurer la desserte efficace du territoire (connexions directes entre points stratégiques) tout en maintenant au plus bas le coût de l'exploitation (évalué selon la longueur métrique totale du réseau) \citep{schweitzer1997optimization, yamins2003growing}. La topologie des réseaux les plus optimisés révèle des structures particulières, comme par exemple celle d'un arbre couvrant minimal (\textit{minimum spanning tree}) mis en évidence par M. Barthelemy \textit{et al.} \citep{barthelemy2006optimal}. Les chercheurs ont montré qu'en faisant fluctuer les paramètres du modèle, différents types d'arbres peuvent être observés. Ils ont poursuivi ensuite leur idée en modélisant une croissance de réseau où le dernier nœud arrivé se raccorde au reste du réseau selon différents critères (au dernier nœud ou perpendiculairement à l'arc le plus proche selon les modèles) \citep{barthelemy2008modeling, barthelemy2009co}. Cette idée a été prolongée par T. Courtat qui a modélisé un potentiel autour de chaque arc afin de déterminer la direction de croissance d'un réseau \citep{courtat2011mathematics}. Ces travaux mettent en évidence deux processus corrélés dans la croissance d'un réseau viaire : l'exploration et la densification \citep{strano2012elementary}.

L'étude physique de la croissance des réseaux a pris un nouveau tournant avec les travaux de A.-L. Barabàsi \textit{et al.} \citep{barabasi2002linked}. Ils ont montré qu'un nouveau nœud introduit dans un réseau a plus de probabilité de se connecter à un nœud pré-existant de fort degré \citep{barabasi1999emergence}. Ce principe d'\textit{attachement préférentiel} aboutit à des réseaux dont la distribution des degrés suit une courbe significative : beaucoup de sommets ont un degré faible et peu sont de degré important (\textit{long tails}). Cette propriété donne une invariance d'échelle, quelle que soit la sous partie du réseau observée \citep{newman2003structure}. Ces réseaux sont qualifiés de \textit{scale-free}. Cependant, cette dynamique observée sur des graphes topologiques ne semble pas s'appliquer aux réseaux de transports, qui sont contraints par leur inscription spatiale. Dans ces derniers, l'importance d'un nœud n'est pas forcément liée au nombre d'arcs qui s'y raccordent (si l'on considère un nœud comme étant une intersection). Les caractéristiques de ces réseaux sont donc spécifiques, différents des graphes où seule la topologie compte \citep{csanyi2004fractal, jiang2004topological, lammer2006scaling}.

La complémentarité de l'étude des propriétés topologiques (et géométriques) des réseaux, en plus des flux qui y circulent, et de leurs impacts sociaux, a été mis en avant dans de nombreux travaux \citep{levinson2005evolution}. Ces études allient des points de vues pour aboutir à des modèles de croissance \citep{zhang2004model}. Les dynamiques qui régissent la création de structures et leur hiérarchisation ont fait l'objet d'un modèle créé par B. Yerra \citep{yerra2005emergence}. L'idée des chercheurs est que plus un lien du réseau est utilisé, plus il aura vocation à se développer et contribuer ainsi au développement du réseau autour de lui. Des travaux ont également étudié les mécaniques de croissance urbaine, et les effets de transition qu'elle peut faire apparaître \citep{louf2013modeling}.

Toutes ces analyses cherchent à établir un lien entre l'état présent du réseau et chacun de ses états passés. La dépendance d'un état à un autre établit un lien entre la géométrie du réseau et ce que l'on peut considérer comme les conditions initiales de son développement \citep{arthur1994increasing}. L'information dont nous disposons est imparfaite, et considérer un simple enchaînement d'états peut aboutir à des situations bloquées dont il n'est possible de se libérer qu'avec une rupture dans le modèle \citep{liebowitz1995path}. La croissance des réseaux de transports apparaît comme un processus séquentiel où les décisions prises localement peuvent être optimales si l'attention n'est portée qu'à une échelle locale (par manque d'informations sur celle globale) \citep{bertolini2007evolutionary, zhang2007economics}. La question de l'importance du lien entre les états passés et l'état présent reste ouverte.

La conceptualisation des suivis temporels de données géographiques a été étudiée en détail par P. Bordin \citep{bordin2006methode}. Elle identifie ainsi les différentes méthodes de classifications permettant de créer un lien entre des données appartenant à des dates différentes, et notamment une approche fondée sur un modèle par emprise constante \citep{bordin2010integration}. Elle définit particulièrement, avec une analogie venant de la physique, les différences conceptuelles entre des modèles d'étude \textit{statique}, \textit{cinématique} ou \textit{dynamique} de phénomènes.

Dans cette thèse nous observerons la croissance de réseaux viaires. Sans chercher à la modéliser, nous quantifierons les différences structurelles sur plusieurs périodes de temps. À travers les données de plusieurs années sur un même territoire, nous reconstituerons le fil d'une hiérarchisation des objets, dont l'ordre peut varier avec le temps. En juxtaposant des images \textit{statiques} prises dans un système en évolution constante, nous reconstituerons sa \textit{cinématique} à travers la quantification de ses différences structurelles. 

\FloatBarrier
\section{Application de la caractérisation de la spatialité à travers la continuité}

Dans la première partie, nous avons défini notre objet d'étude : la voie. Nous avons choisi les différents paramètres qui régissent sa construction selon des critères quantifiés. Nous avons également construit une grammaire non exhaustive d'indicateurs apportant des informations différentes selon l'objet sur lequel ils sont appliqués.

Nous établissons la suite de notre étude sur l'objet \textit{voie}. Nous l'utiliserons pour caractériser les réseaux spatiaux à travers les indicateurs primaires qui lui sont propres. Pour cela, nous commencerons par étudier la sensibilité de notre modèle à la qualité des données dont nous disposons, à leur granularité de vectorisation et au découpage de la zone d'étude. En effet, avant d'appliquer notre méthodologie à des réseaux spatialisés, nous voulons en assurer la pertinence par rapport à ses contraintes géométriques (imprécisions ou découpes).

À l'issu de ce travail, nous appliquerons le raisonnement construit sur un panel de recherche de quarante réseaux spatiaux, principalement des réseaux viaires, mais également des réseaux biologiques, hydrographiques, ferrés, etc. Tous partagent la même propriété : avoir une inscription spatiale. Cette comparaison quantitative nous permettra d'établir objectivement leurs différences et similitudes d'un point de vue géométrique et topologique.

Enfin, nous ferons progresser notre travail vers une analyse diachronique des changements sur les réseaux viaires. Nous prendrons pour cela en exemple deux réseaux viaires, dont nous avons établi une base de données géographique \textit{panchronique} : celle-ci regroupe des géométries correspondant aux différentes années de vectorisation de plans anciens. Nous étudierons ainsi les cartes de l'intra-muros de la ville d'Avignon et celles du nord de la ville de Rotterdam. Nous quantifierons à l'aide d'une méthode adaptée l'évolution de l'accessibilité des voies de ces réseaux.


\clearpage{\pagestyle{empty}\cleardoublepage}
\chapter{Étude de la robustesse du modèle développé}
\minitoc
\markright{Étude de la robustesse du modèle développé}

Avant d'appliquer notre méthodologie à plusieurs réseaux pour les étudier et les comparer, il est nécessaire de déterminer les limites de son utilisation et les influences auxquelles elle peut être soumise. En effet, les caractérisations dépendent des données sur lesquelles elles sont appliquées. Dans notre cas, nous travaillons sur des données géographiques. Elles retranscrivent une réalité physique : pour le réseau viaire, le graphe illustre un espace de circulation possédant de multiples attributs. Nous avons choisi de ne pas prendre en compte les informations autres que la géométrie du réseau. Cependant, cette dernière est sujette aux modulations de sa transcription sous forme vectorielle. Ainsi, l'opérateur ayant la mission de géoréférencer un plan (afin de lui attribuer des coordonnées) et de dessiner les axes de son réseau porte la responsabilité de la qualité de la donnée sur laquelle nous travaillons. Cette qualité étant variable, c'est donc à la méthode de s'y adapter pour être robuste.

La description d'une information spatiale à l'aide d'une carte implique un choix dans les données représentées. Ainsi, sur un graphe viaire, nous pouvons choisir de considérer uniquement les routes, ou d'y ajouter les chemins, etc. Nous observerons les différences dans les résultats de nos calculs selon la prise en compte ou pas de données attributaires minoritaires.

Les lignes vectorielles que nous utilisons représentent le centre axial du réseau sur lequel nous travaillons. Si nous considérons le réseau viaire, sur lequel nous nous appuyons majoritairement, ce centre peut varier en fonction de la perception de l'utilisateur. Ainsi, un piéton n'aura pas la même perception d'une intersection qu'un cycliste ou un automobiliste. Les ronds-points, décrochements et autres aménagements urbains ont été pensés pour briser la continuité du réseau mais il est parfois nécessaire pour l'analyse de \enquote{recoller les morceaux} et de reconstituer les alignements pour déceler des structures antérieures.

D'autre part, dans l'analyse de données continues dans l'espace, comme celles d'un réseau routier, nous sommes amenés à le découper pour en extraire une sous-partie qui constituera l'échantillon d'étude. Ce découpage peut avoir un impact non négligeable sur le calcul de certains indicateurs. Il est donc nécessaire de quantifier cet impact afin de savoir comment interpréter les résultats obtenus. Nous comparerons donc différents tissus selon des découpages variés, à plusieurs échelles, afin de définir la robustesse d'utilisation de notre méthodologie.

Enfin, en cartographie comme dans toute représentation de données, en fonction de ce qui veut être représenté et de l'information que l'on souhaite garder, il est parfois nécessaire d'adapter la finesse de représentation vectorielle. Cette opération se nomme la généralisation. Elle est utilisée, par exemple, pour réduire le nombre de tournants d'une rue de montagne afin de la rendre plus facilement lisible sur une carte à petite échelle. Nous testerons son influence sur le calcul de nos indicateurs.

\FloatBarrier

\section{La nature des données viaires utilisées}

Les données que nous utilisons pour l'étude des réseaux viaires proviennent de l'IGN (\copyright BDTOPO) et d'OpenStreetMap (site collaboratif donnant accès à des données élaborées par les utilisateurs eux-même, sur le même schéma de participation que Wikipédia). L'IGN propose des données sur la France entière (DOM/TOM compris) ; OpenStreetMap regroupe tous les pays où les utilisateurs ont réalisé une vectorisation. C'est donc une base qui couvre un territoire beaucoup plus large, mais qui est généralement moins structurée concernant les attributs liés aux géométries qu'elle propose. Nous avons donc choisi des données issues de la \copyright BDTOPO de l'IGN pour les villes françaises de notre corpus principal (Avignon et Paris), et des données issues d'OpenStreetMap pour les autres réseaux.

Nous extrayons de ces bases des données vectorielles. Celles-ci possèdent une table attributaire contenant, entre autres, pour chaque objet, leur identifiant, leur géométrie et leur \textit{type} (route, chemin, escalier...). Ce dernier attribut est renseigné de manière systématique et codifié dans les bases de données fournies par l'IGN ; il l'est moins lorsqu'on utilise les données collaboratives disponibles sur OpenStreetMap.

Nous utilisons néanmoins cet attribut de \textit{type} pour faire un premier traitement qualitatif de nos données. Ainsi, sur le réseau extrait de la \copyright BDTOPO, nous conservons les vecteurs qualifiés de route (à 1 ou 2 chaussées), de bretelle, quasi-autoroute ou autoroute. Nous supprimons en revanche les vecteurs identifiés comme des escaliers, sentiers, chemins, routes empierrées ou pistes cyclables. Nous faisons de même avec les données issues d'OpenStreetMap en utilisant les informations mises à disposition par les différents collaborateurs. Afin de quantifier l'influence de ce choix, nous évaluerons dans ce chapitre l'impact de l'ajout de données d'un type mineur, sur la calcul de l'accessibilité des voies du graphe.

\FloatBarrier

\section{Étude de la sensibilité à la géométrie des carrefours}

\begin{figure}[h]
    \centering
        \includegraphics[width=0.8\textwidth]{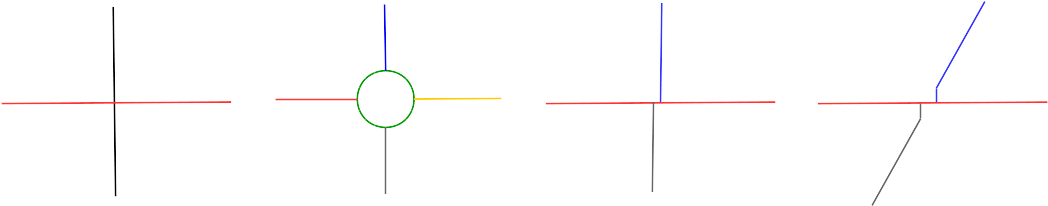}
    \caption{Schéma illustrant trois types de perturbations géométriques brisant la continuité d'une voie à un carrefour.}
    \label{fig:place_axes}
\end{figure}

La géométrie de la voie dépend des angles à chaque intersection. Ainsi, l'ajout d'un rond-point ou d'un décrochement brise sa continuité et la divise en plusieurs objets. Tout changement de la géométrie impacte donc le calcul des indicateurs, qu'il soit local ou global. La figure \ref{fig:place_axes} illustre trois types différents de perturbations qui peuvent intervenir à un carrefour entre deux voies. Dans le cas de l'ajout d'un rond-point, les deux voies sont dédoublées pour aboutir à un schéma de cinq voies, le rond-point en devenant une à part entière. Dans les deux autres perturbations (à droite) la voie horizontale garde son intégrité alors que la verticale est coupée en deux.

Ces perturbations peuvent être dues à une erreur de vectorisation comme à un aménagement urbain réfléchi. Dans le cas où nous voudrions les effacer, pour retrouver, par exemple, des continuités supprimées à tort ou antérieures à une construction viaire, nous avons mis au point une méthodologie faisant intervenir des \textit{zones tampons} que nous appellerons aussi \textit{places} dans le calcul de continuité. Cette méthode nous permet d'être robuste aux éventuels décrochements.

\begin{figure}[h]
    \centering
    \includegraphics[width=0.8\textwidth]{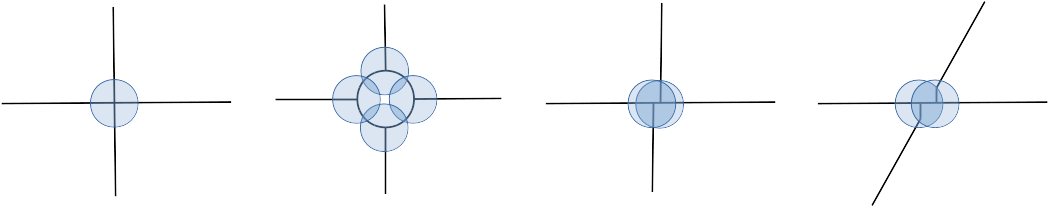}
    \caption{Étape 1 : Positionnement des zones tampons autour de chaque intersection.}
    \label{fig:place_buffers}
\end{figure}

\begin{figure}[h]
    \centering
    \includegraphics[width=0.8\textwidth]{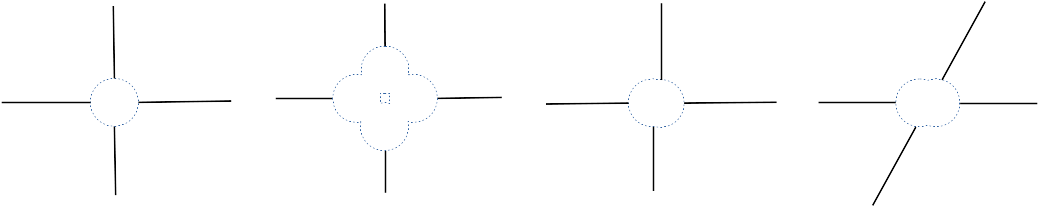}
    \caption{Étape 2 \& 3 : Fusion des zones tampons et suppression des segments recouverts.}
    \label{fig:place_merged}
\end{figure}

\begin{figure}[h]
    \centering
    \includegraphics[width=0.8\textwidth]{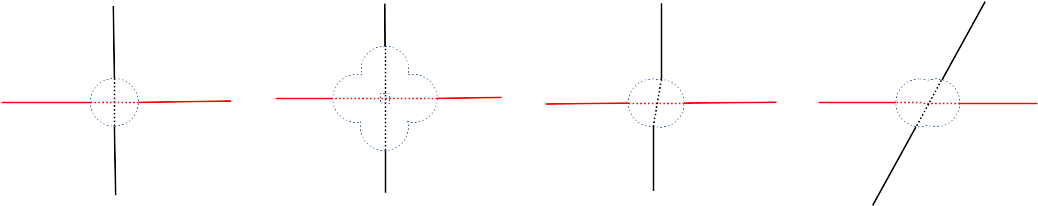}
    \caption{Étape 3 : Association des arcs pour construire les voies.}
    \label{fig:place_voies}
\end{figure}

\begin{figure}[h]
    \centering
    \includegraphics[width=0.8\textwidth]{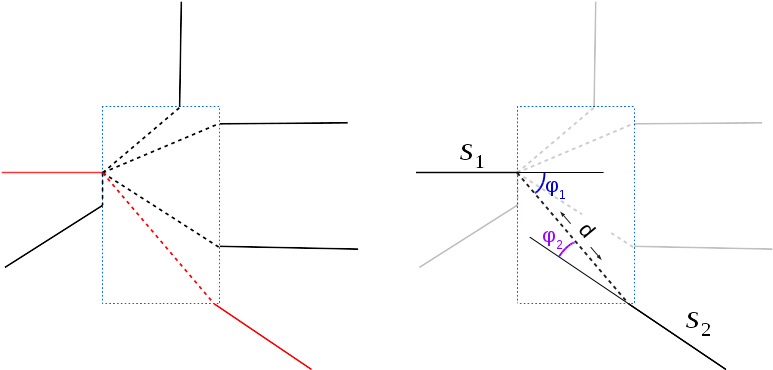}
    \caption{Afin d'associer les arcs $arc_1$ et $arc_2$, nous considérons les segments $s_1$ et $s_2$ leur appartenant qui intersectent la zone tampon. Nous calculons pour chaque couple d'arcs un facteur $\Gamma$ en fonction de la droite $s_{join}$ joignant les points d'intersection de $s_1$ et $s_2$ avec la zone tampon. Nous calculons les angles $\varphi_1$ et $\varphi_2$ respectivement entre $s_1$ et $s_{join}$ et $s_2$ et $s_{join}$. Nous prenons le sinus de ces angles que nous multiplions à la longueur de $s_{join}$ ($d$) afin de calculer la déviation entre $s_1$ et $s_2$ (équation \ref{eq:gamma}).
}
    \label{fig:place_associations}
\end{figure}

La méthodologie de construction des voies nécessite, dans ce cas, 4 étapes :

\begin{enumerate}
\item Construction d'une zone tampon, autour de chaque intersection, circulaire, avec un rayon paramétrable selon le tissu étudié (figure \ref{fig:place_buffers}).

\item Fusion de toutes les zones tampons qui s'intersectent.

\item Suppression de tous les segments, ou portions de segments, contenus à l'intérieur des zones tampons fusionnées (figure \ref{fig:place_merged}).

\item Calcul des associations d'arcs  (figure \ref{fig:place_associations}, voir explication ci dessous) pour construire les voies (figure \ref{fig:place_voies}) dans chaque zone tampon.
\end{enumerate}

\FloatBarrier

\begin{figure}[h]
    \centering
    \includegraphics[width=0.6\textwidth]{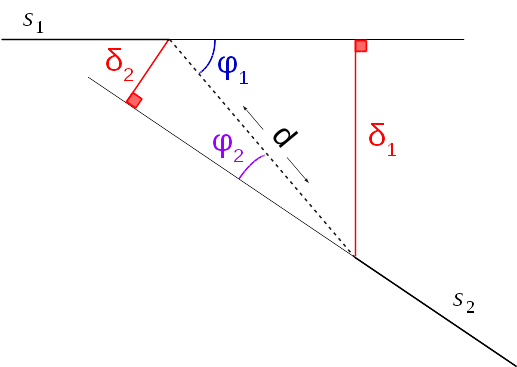}
    \caption{Détail du calcul de déviation sur les zones tampons. \\ $\delta_i = \sin(\varphi_i) \times d$ \\\ $\Gamma(arc_1, arc_2) = \sum_{i \in \{1,2\}}\delta_i$}
    \label{fig:place_associations2}
\end{figure}

Les associations entre arcs connectés à une même place sont faites en traçant la droite passant par les intersections des arcs avec la zone tampon (figure \ref{fig:place_associations}). Sont considérés les angles faits entre le dernier segment des arcs et cette droite ($\varphi_1$ et $\varphi_2$)(figure \ref{fig:place_associations}). Nous étudions la somme des valeurs absolues des sinus, multipliée à la distance entre les deux intersections, selon l'équation \ref{eq:gamma}. Le résultat est égal à la somme des distances de déviation (figure \ref{fig:place_associations2}). Ce calcul nous permettra d'identifier les déviations entre chaque couple d'arcs. Multiplier celle-ci par la distance permet d'être moins sensible à la variation de taille de la zone tampon. Les arcs proches avec un faible angle de déviation seront donc associés de manière privilégiée.

Selon la méthode choisie, les couples retenus seront ceux avec la déviation minimale deux à deux ou ceux participant à la configuration totale de déviation minimale (cf Partie I, chapitre 2, à propos de la construction de la voie). La figure \ref{fig:place_associations} illustre un choix complexe d'associations entre le segment de référence (représenté en rouge) et les autres segments intersectant une même zone tampon. La figure \ref{fig:place_voies} illustre le résultat donné dans les cas simples étudiés.

\begin{equation}
    \Gamma(arc_1, arc_2) = (\vert \sin(\varphi_1) \vert + \vert \sin(\varphi_2) \vert) \times d
    \label{eq:gamma}
\end{equation}

\FloatBarrier

Si l'on s'intéresse au réseau viaire, il est possible de compléter cette étude par l'ajout des places réelles, si nous en possédons la vectorisation sous forme de polygone. Nous considérons ainsi les places des villes comme des espaces ouverts où les arcs entrants peuvent être associés en tenant compte de l'ensemble de ceux connectés à la même place. Nous avons fait réaliser une vectorisation des places sur la partie intra-muros de la ville d'Avignon. Nous avons ainsi pu étudier les modifications apportées au réseau avec cette méthodologie de construction, moins dépendante de la vectorisation des intersections.

\begin{figure}[h]
    \centering
    \includegraphics[width=\textwidth]{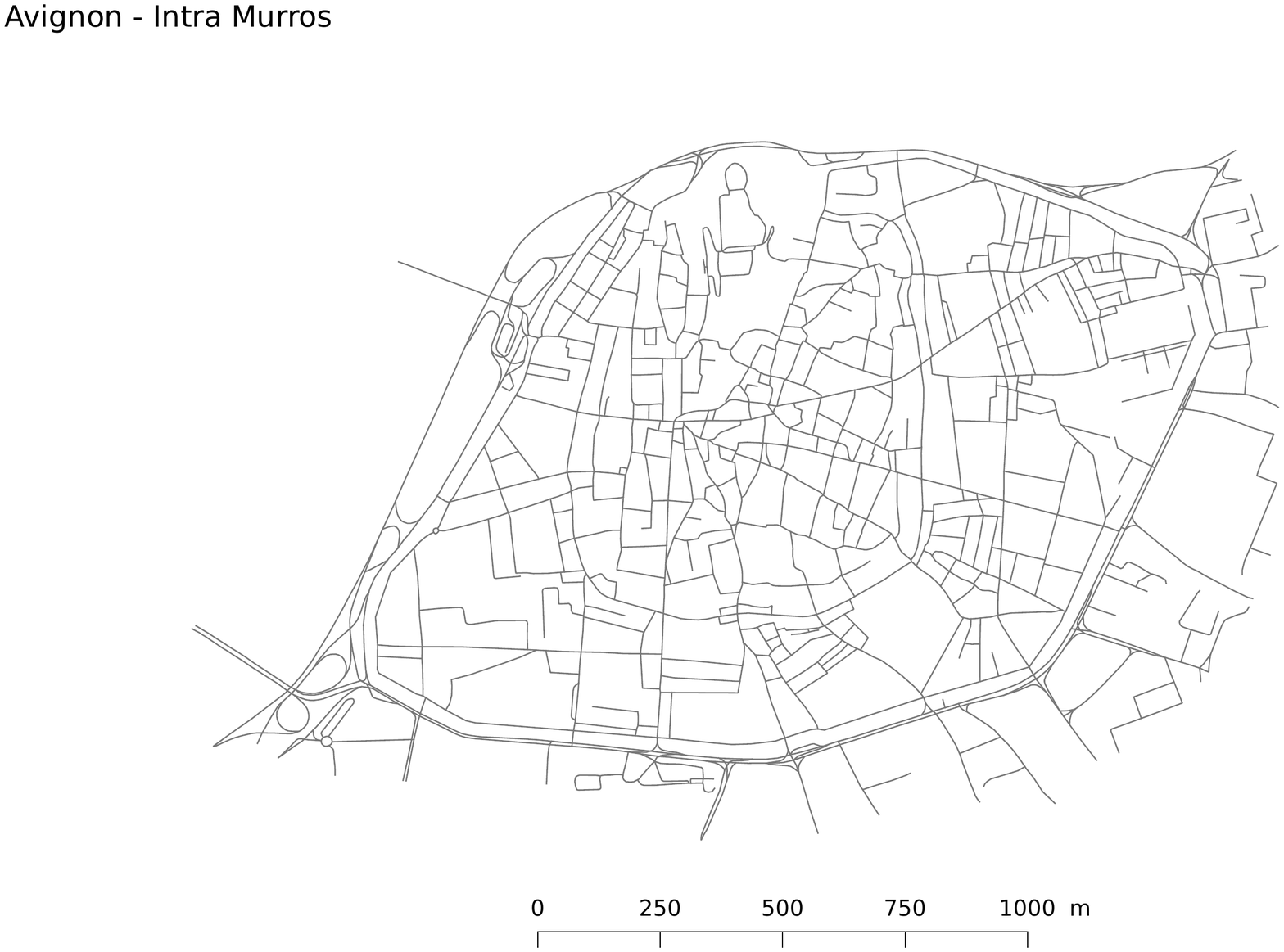}
    \caption{Réseau viaire brut d'Avignon intra-muros. Données fournies par l'IGN, \copyright BDTOPO 2014.}
    \label{fig:avignon_brut}
\end{figure}

\begin{figure}[h]
    \centering
    \includegraphics[width=\textwidth]{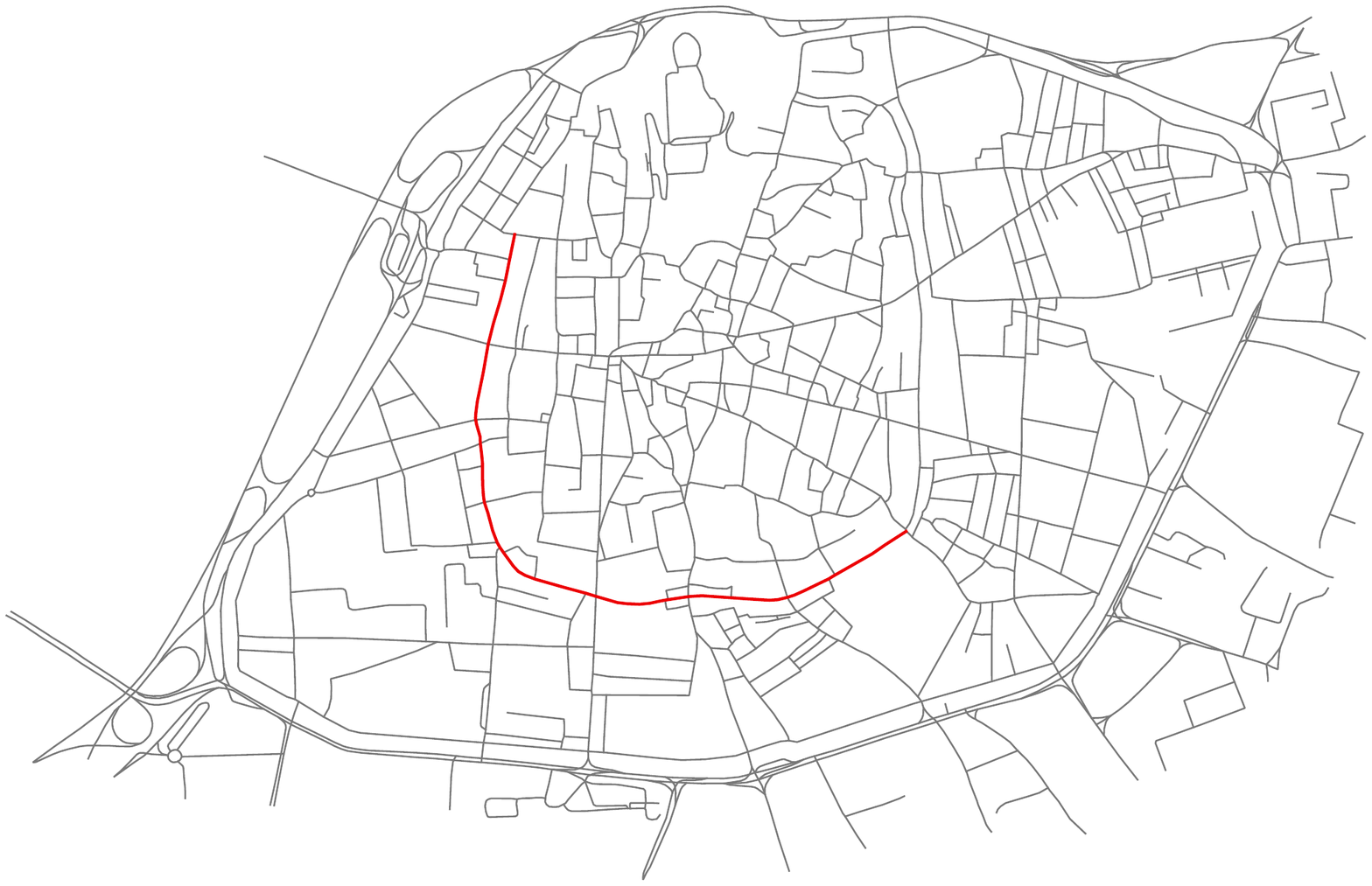}
    \caption{Exemple d'une voie qui perd sa continuité par un petit décrochement.}
    \label{fig:avignon_voieCoupee}
\end{figure}

\begin{figure}[h]
    \centering
    \includegraphics[width=0.6\textwidth]{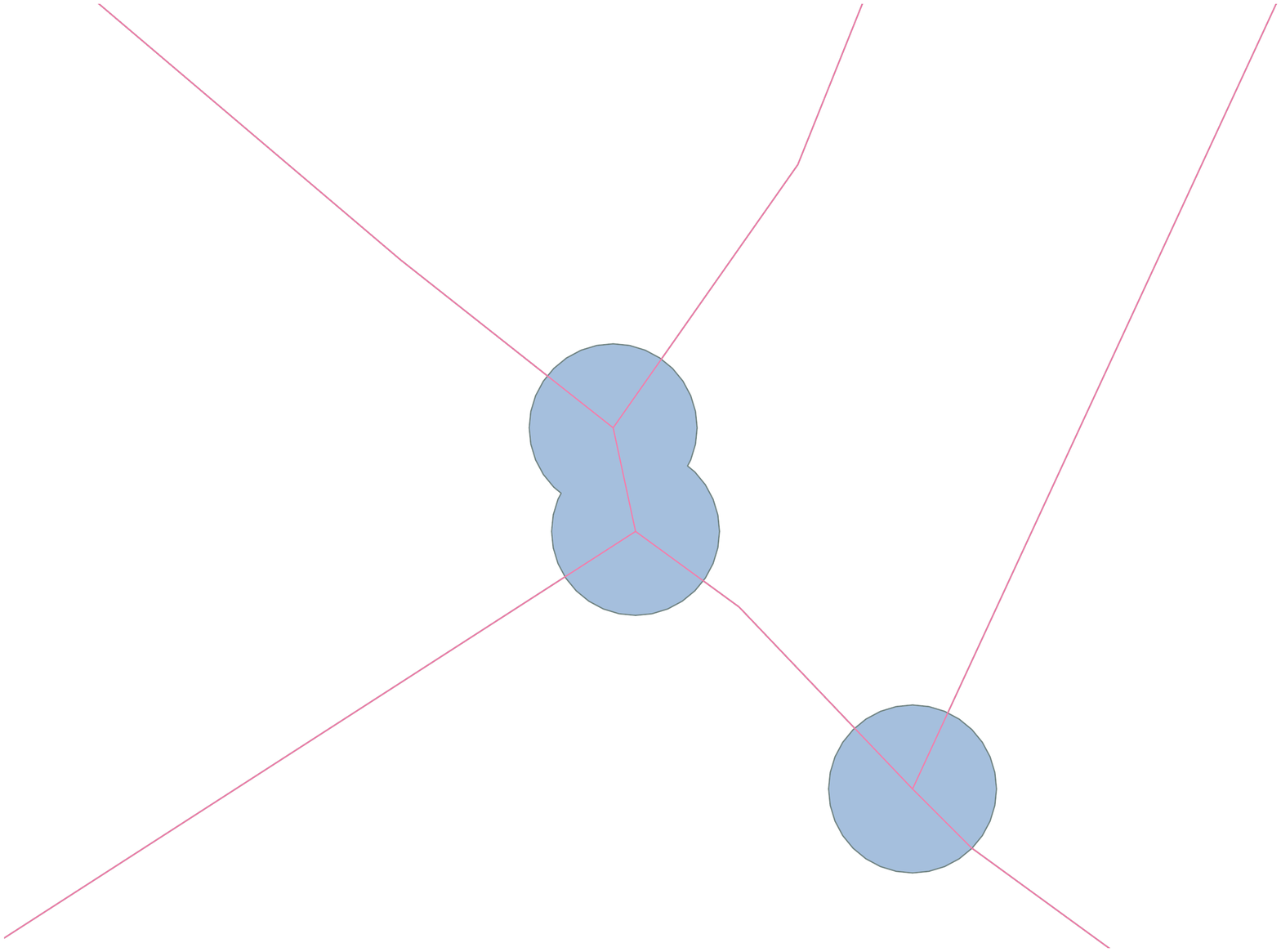}
    \caption{Intersection impliquée dans la rupture de la voie (figure \ref{fig:avignon_voieCoupee}). Application des zones tampons autour de chaque intersection.}
    \label{fig:avignon_places_zoom_max}
\end{figure}

\begin{figure}[h]
    \centering
    \includegraphics[width=0.8\textwidth]{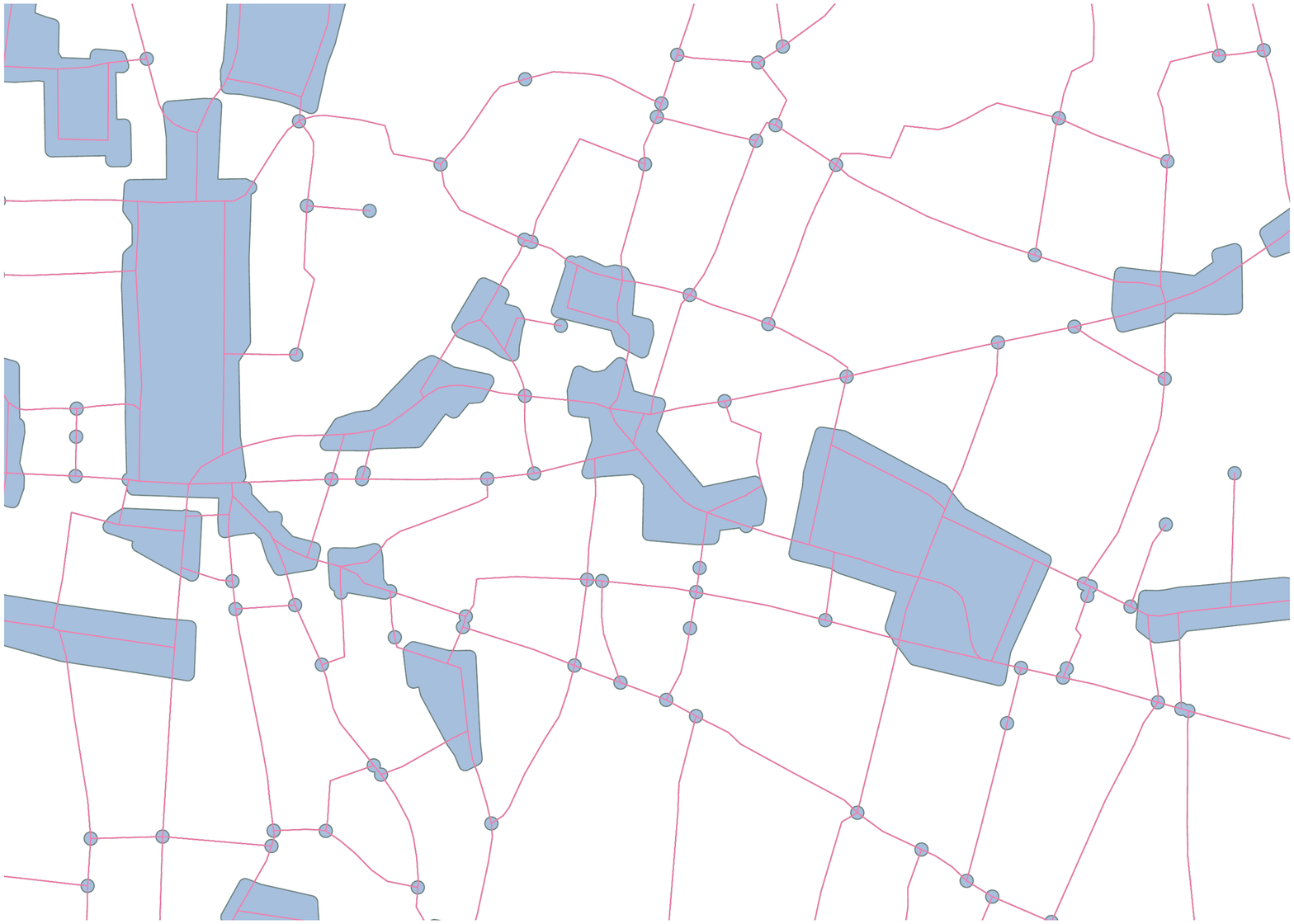}
    \caption{Ajout des places aux zones tampons.}
    \label{fig:avignon_places_zoom}
\end{figure}

\begin{figure}[h]
    \centering
    \includegraphics[width=\textwidth]{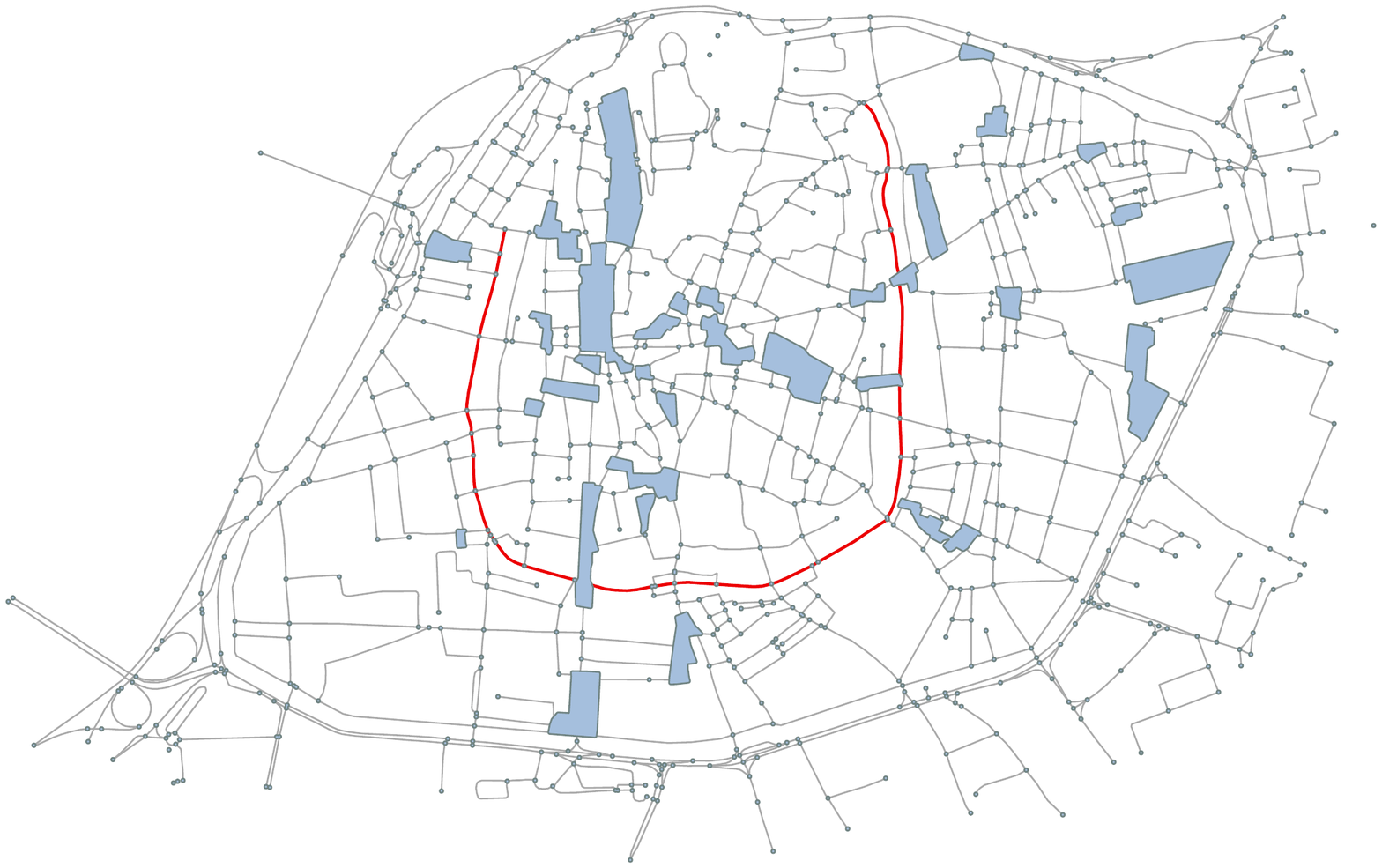}
    \caption{Vue générale des places (zones tampons créées ou ajoutées après vectorisation). La voie coupée de la figure \ref{fig:avignon_voieCoupee} est maintenant continue.}
    \label{fig:avignon_voieNonCoupee}
\end{figure}

Nous prenons cet exemple pour illustrer notre méthodologie d'ajustement de la continuité (figure \ref{fig:avignon_brut}). À Avignon, nous pouvons lire à l'intérieur des murailles une continuité d'arcs circulaires qui suivent une ancienne muraille (maintenant détruite) datant du XIème siècle. Cependant, dans la base de données, cette continuité est coupée par un léger décrochement (figures \ref{fig:avignon_voieCoupee} et \ref{fig:avignon_places_zoom_max}). En allant sur place, nous avons constaté qu'il s'agit dans ce cas d'un choix de vectorisation erroné. En effet, les deux tronçons sont bien physiquement alignés : la ligne des trottoirs correspond de part et d'autre de l'intersection à une même perspective. Pour rendre notre construction robuste à ce type d'erreur de saisie, nous plaçons une zone tampon autour de chacune de nos intersections (figure \ref{fig:avignon_places_zoom_max}) que nous complétons par l'ajout des places réelles vectorisées (figure \ref{fig:avignon_places_zoom}). En faisant ce travail sur l'ensemble du territoire étudié, nous reconstituons ainsi des continuités qui avaient été brisées (figure \ref{fig:avignon_voieNonCoupee}). À partir de ce nouvel hypergraphe, nous pouvons calculer nos indicateurs afin de le caractériser en utilisant les voies créées via les zones tampons.

\FloatBarrier

\begin{figure}[h]
    \centering
    \begin{subfigure}[t]{\linewidth}
        \includegraphics[width=\textwidth]{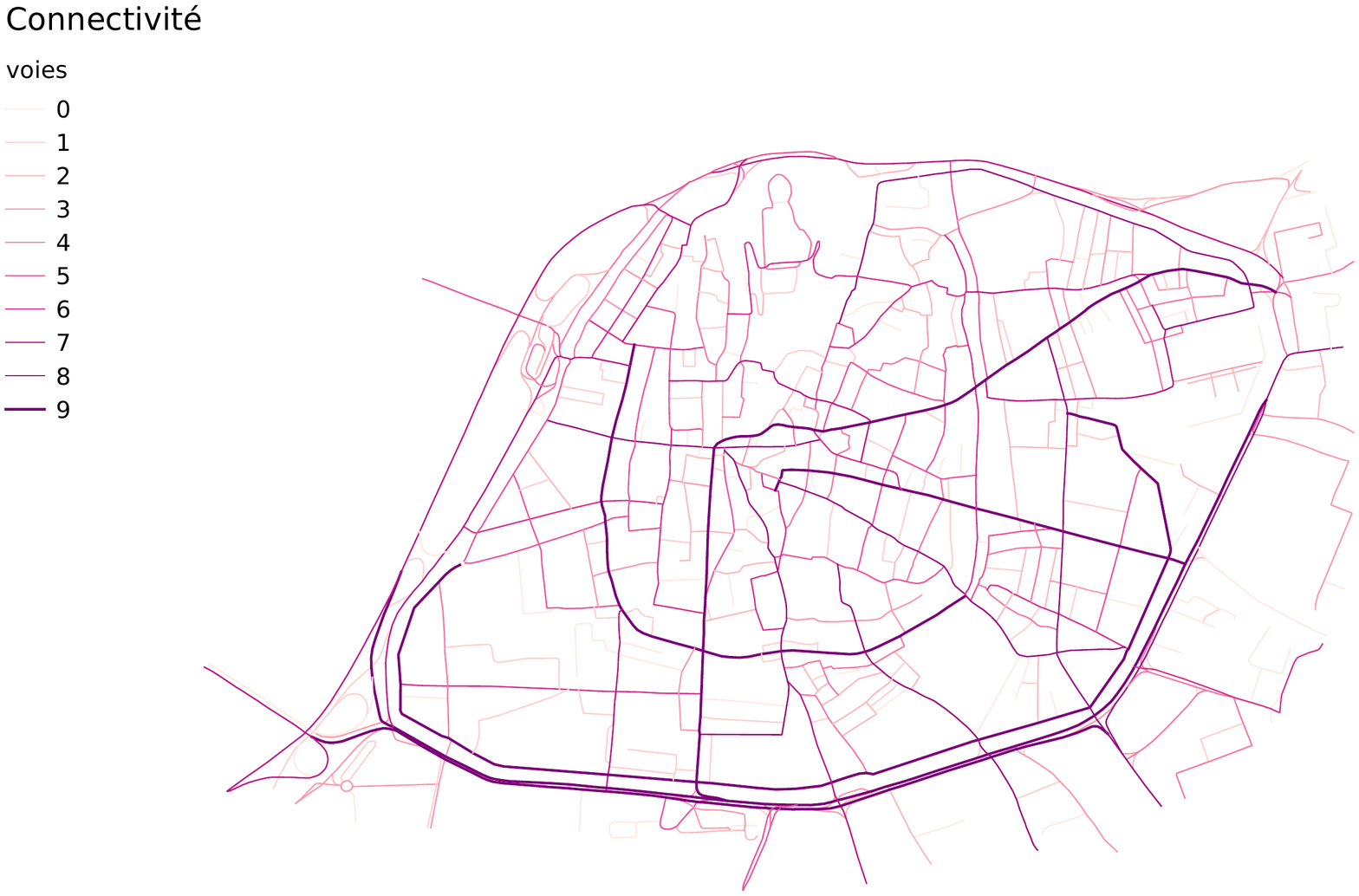}
        \caption{Sur le réseau brut, avec association d'arcs à chaque intersection selon leur déviation minimum d'après la méthode expliquée dans la partie 1 ($M0$).}
    \end{subfigure}

    \begin{subfigure}[t]{\linewidth}
        \includegraphics[width=\textwidth]{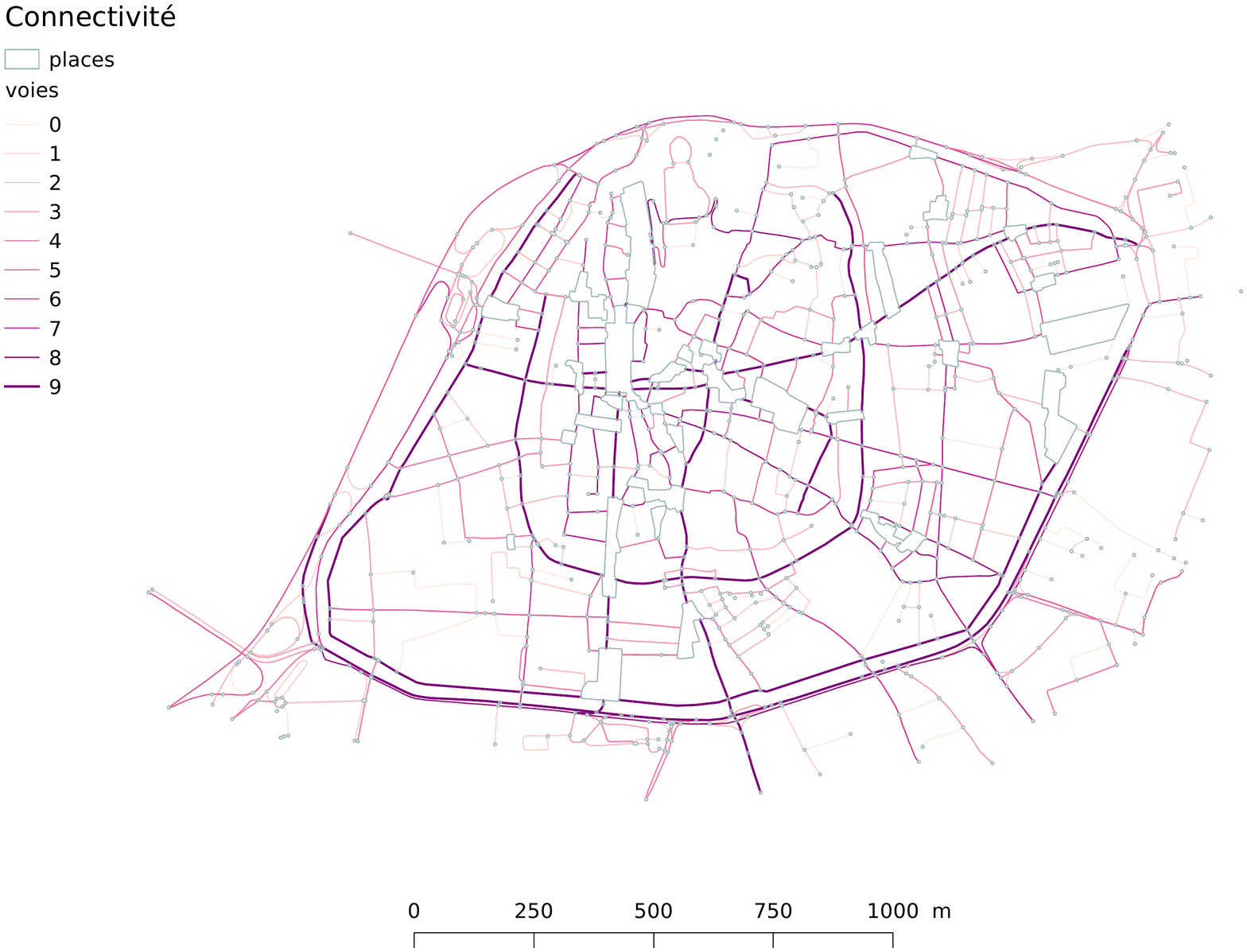}
        \caption{Sur le réseau \textit{amélioré} avec construction des voies avec les places.}
    \end{subfigure}
    \caption{Calcul de la connectivité}
    \label{fig:avignon_connectivite}
\end{figure}

\begin{figure}[h]
    \centering
    \begin{subfigure}[t]{\linewidth}
        \includegraphics[width=\textwidth]{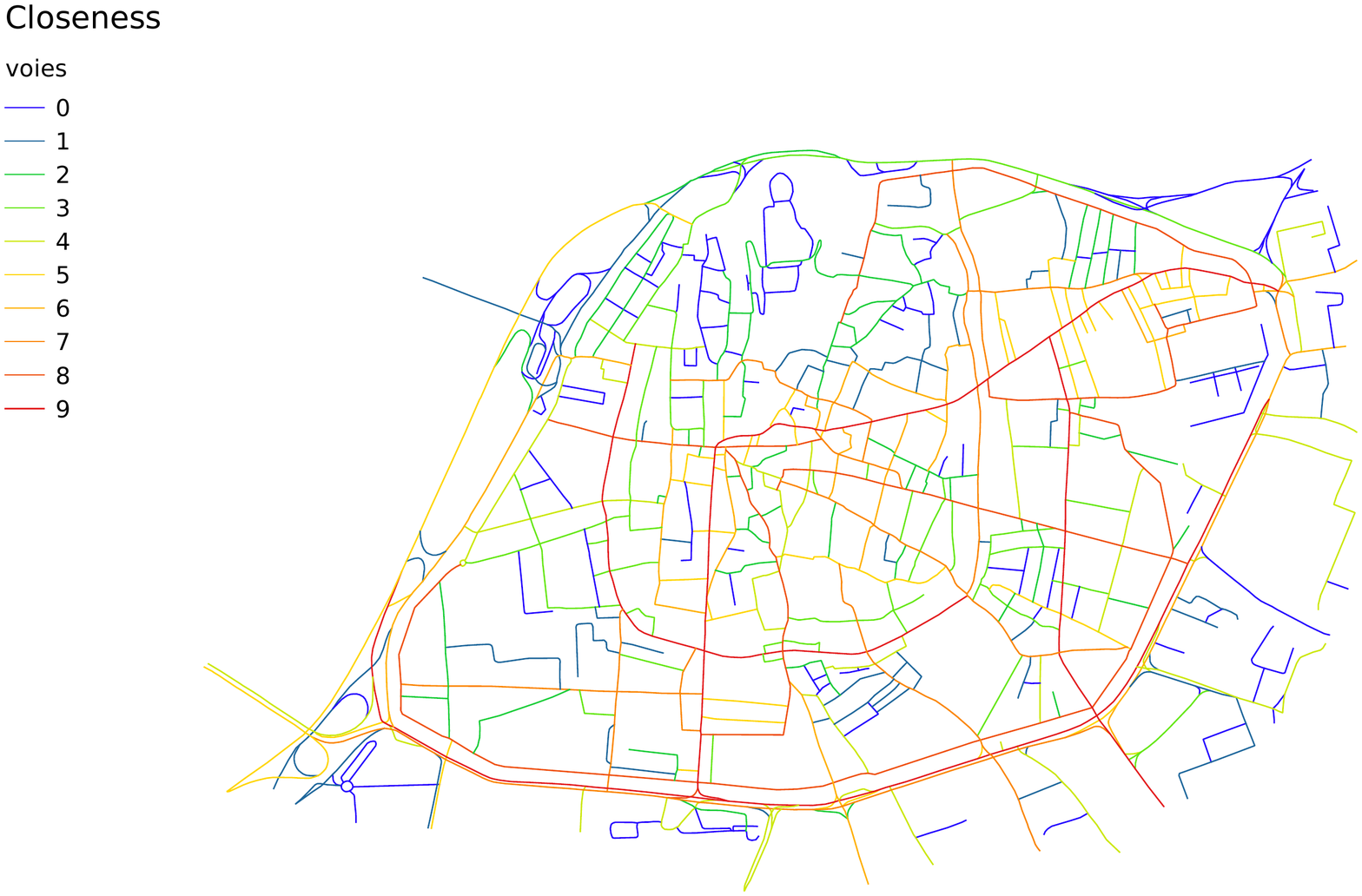}
        \caption{Sur le réseau brut, avec association d'arcs à chaque intersection.}
    \end{subfigure}

    \begin{subfigure}[t]{\linewidth}
        \includegraphics[width=\textwidth]{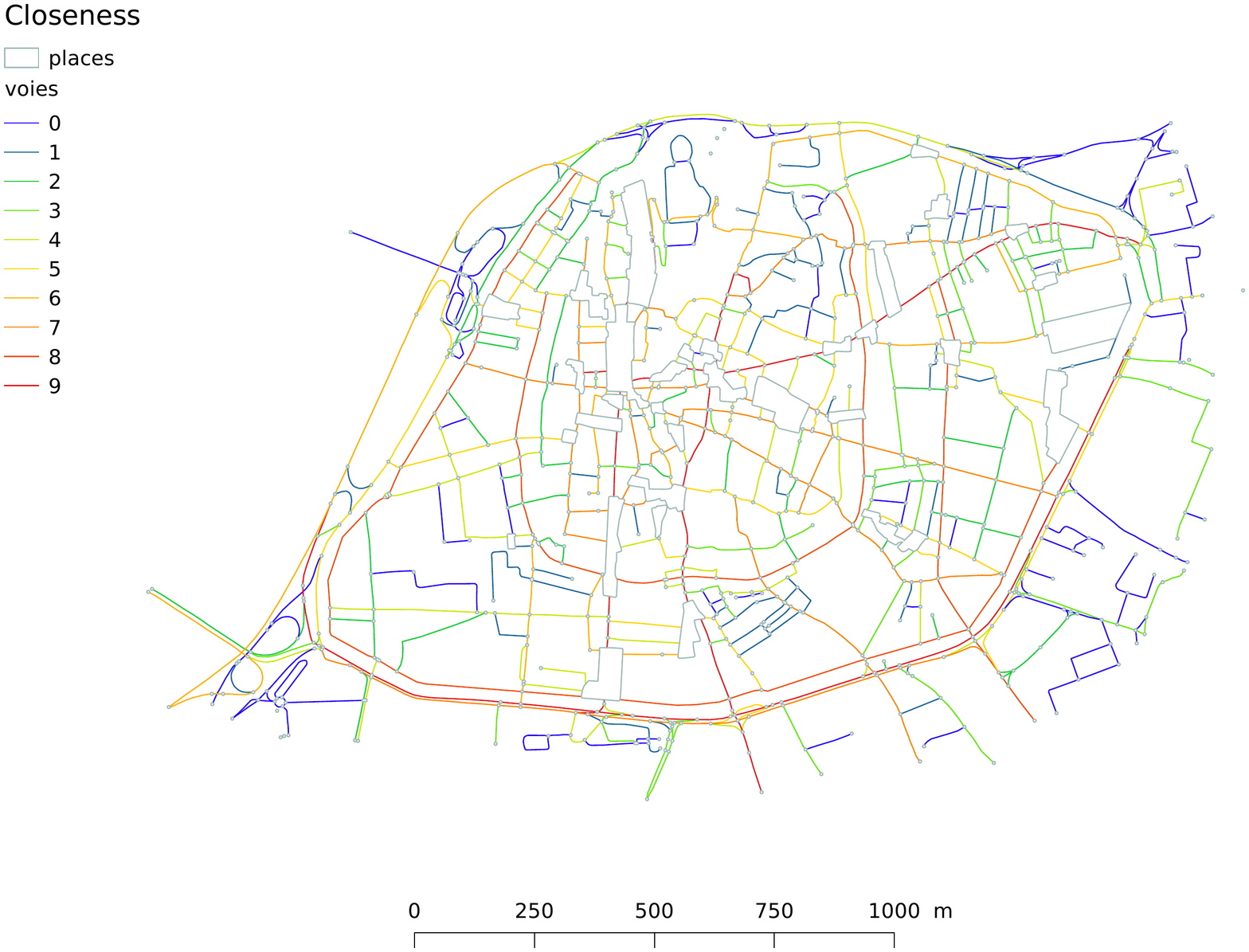}
        \caption{Sur le réseau \textit{amélioré} avec construction des voies avec les places.}
    \end{subfigure}
    \caption{Calcul de la closeness}
    \label{fig:avignon_rtopo}
\end{figure}

Nous illustrons les différences dans le calcul de deux indicateurs. Le premier tient compte d'une information locale  : la connectivité, illustrant finement les connexions de chaque voie (figure \ref{fig:avignon_connectivite}). Le second tient compte de la topologie de tout le réseau : la closeness (figure \ref{fig:avignon_rtopo}). Sur les deux exemples, nous observons une modification des axes principaux de la ville : Est-Ouest et Sud-Nord, ainsi que l'apparition de la petite ceinture de la ville comme un seul objet continu. L'étude quantitative des voies lorsqu'elles sont construites en tenant compte des places a donc une réelle incidence sur les résultats obtenus. L'information observée dans les deux cas possède un champ structurel commun (l'enceinte extérieure est toujours très proche topologiquement de l'ensemble du réseau, le palais des Papes en est toujours isolé) mais diffère pour certains grands axes (comme celui Nord-Sud formé par la rue de la République et le cours Jean Jaurès). Cependant, les données des places vectorisées sont rarement disponibles. Il n'est pas possible d'en disposer sur de larges espaces ni dans plusieurs pays. Nous conserverons cette méthode d'étude uniquement pour des analyses locales qui nécessitent la finesse de cette approche. Pour des analyses plus globales, les zones tampons créées automatiquement permettent de paramétrer l'analyse en fonction de la problématique étudiée.

En effet, l'utilisation de la méthode de construction des voies via les zones tampons dépend de la question de recherche à laquelle nous souhaitons répondre. Si le but est d'identifier les discontinuités créées par des aménagements urbains pensés pour cela, le diamètre des zones tampons paramétré pour le calcul devra être faible par rapport à la longueur moyenne des arcs. Au contraire, si nous voulons nous émanciper de ces transformations locales pour reconstruire des axes qui leur sont antérieurs, nous pouvons paramétrer un diamètre plus important, en restant attentifs à l'échelle du réseau. En effet, si nous considérons des zones tampons trop grandes autour des sommets, il est possible que des arcs soient supprimés, car complètement inclus dans cette zone, au détriment de l'information potentiellement pertinente qu'ils portent. La qualité des données étudiées s'en trouverait alors amoindrie. Le rayon de la zone tampon paramétré devra dépendre du tissu étudié. En effet, l'échelle locale de villes médiévales (dont le centre d'Avignon fait partie), n'est pas celle de villes d'Amérique du Nord qui profitent d'un large espace pour étendre leur réseau viaire.

\FloatBarrier

\section{Études de la sensibilité à la sélection des données}

Dans la définition d'une grammaire d'indicateurs utiles à la caractérisation spatiale (Partie I, chapitre 4), nous avons limité le nombre d'indicateurs primaires à 6 pour les arcs et 3 pour les voies. Parmi ces indicateurs, un seul pour les voies est calculé en tenant compte de l'ensemble du réseau : la proximité topologique ou \textit{closeness}. Les autres dépendent de la géométrie et des connexions propres à l’objet auquel ils sont appliqués. Ils ne seront modifiés que si l'arc ou la voie sur lesquels ils sont calculés sont coupés ou si son voisinage direct est transformé. Dans le cas de découpages d'emprises différentes, seuls les éléments en bord d'échantillons seront impactés. La voie pouvant être traversante, nous serons donc attentifs à ne pas couper brutalement des structures apparaissant \textit{a priori} comme traversantes sur l'espace étudié (grands boulevards, enceintes ou voies rapides).

Dans la première partie, nous avons montré que, lorsqu'elle est appliquée aux arcs, la closeness est sensible aux limites du réseau. Les cartes de Manhattan ou d'Avignon (figure \ref{fig:arcs_carte_closeness}) montrent qu'appliquée à un objet local, sa caractérisation fait essentiellement ressortir le centre du graphe. Nous voulons ici quantifier sa stabilité dans son application aux voies.

\subsection{Méthodologie}

Pour quantifier l'influence de la zone d'emprise sur le calcul des indicateurs, nous identifions les géométries d'arcs ou de voies qui ont été modifiées d'un échantillon à l'autre. Dans le cas d'étude des effets de bord, en élargissant le territoire d'étude, il s'agira d'arcs ajoutés qui viendront éventuellement compléter la géométrie des voies. Afin d'étudier les variations de l'indicateur de closeness nous reportons dans la table attributaire des arcs les identifiants des voies auxquelles ils appartiennent ainsi que la valeur du rayon topologique de celles-ci. Cette valeur, inverse de la closeness, somme pour chaque voie l'ensemble des distances topologiques qui la sépare du reste du réseau.

Nous ne normalisons pas ce résultat pour pouvoir apprécier les différences brutes entre deux réseaux de tailles différentes. En revanche, nous éliminons l'impact \textit{neutre} dû à l'ajout des arcs du nouveau découpage en retranchant à la différence brute des indicateurs, la somme des distances topologiques entre les arcs du graphe initial (couvrant une plus petite zone) et ceux ajoutés. Ce calcul, détaillé ci-dessous, nous permet d'observer les modifications d'accessibilité sur le graphe initial. Les variations de valeurs observées ne sont donc pas celles dues à la différence de taille des deux réseaux mais à la modification éventuelle de chemins topologiques les plus simples à l'intérieur de ceux-ci.

Nous considérons deux échantillons de tailles différentes, avec une partie commune. Nous appellerons le plus petit graphe $G_1$ et le plus grand $G_2$. Ces deux graphes ont respectivement $S_1$ et $S_2$ sommets et $A_1$ et $A_2$ arcs. Nous considérerons ici que tout arc de $G_1$ est également présent dans $G_2$ ($G_1 \subset G_2$). Les arcs appartenant à $G_2$ sans être dans $G_1$ seront notés $A_{2-1}$. Nous faisons de même pour les constructions de voies $V_1$ dans $G_1$ et $V_2$ dans $G_2$, $V_{2-1}$ sera l'ensemble des voies présentes dans $G_2$ et non dans $G_1$. Chaque arc $a_i \in v_i \in V_1$ possède donc en attribut l'identifiant $v_i$ de la voie à laquelle il appartient et une valeur $rtopo_{v_i}$. 

Afin de ne pas sommer pour chaque arc d'une même voie toutes les distances topologiques de celles-ci, nous ré-introduisons la longueur à travers le coefficient d'accessibilité. Chaque arc porte le rayon topologique de la voie à laquelle il appartient. En sommant le produit des distances topologiques et des longueurs de chaque arc nous reconstituons ainsi, à travers les arcs, les proximités entre voies. Le résultat donné par l'indicateur d'accessibilité étant très corrélé à celui de closeness, l'observation des variations de l'un sera équivalente à celle de l'autre.

Pour chaque arc $a_i \in A_1$, nous sommons alors le produit de sa distance topologique avec chaque arc de $A_{2-1}$ et de sa longueur. Nous soustrayons ensuite cette somme à la différence des accessibilités des voies auxquelles appartient $a_i$ respectivement calculées dans $G_2$ et $G_1$ (équation \ref{eq:delta}). Nous retirons ainsi l'effet différentiel dû à l'ajout de nouveaux arcs pour ne considérer que les modifications de distances topologiques à l'intérieur de $G_1$.

\begin{equation}
\Delta(a_{ref} \in v) = accessibilite_{G_2}(v) - accessibilite_{G_1}(v) - \sum_{a \in A_{2-1}} (d_{simple}(a_{ref}, a) \times longueur(a))
\label{eq:delta}
\end{equation}

Cette méthodologie sera développée dans le chapitre 4 de cette partie. Elle sera au fondement de la recherche diachronique que nous menons.

\FloatBarrier
\subsection{Étude des effets de bord}

Lorsque l'on étudie des réseaux spatiaux, et spécifiquement des réseaux viaires, il est impossible de considérer le réseau dans son intégralité. Les limites \enquote{naturelles} de tels réseaux sont les mers et les océans. Il est donc nécessaire de définir une emprise qui impose une limite au graphe. Si les résultats que nous obtenons dépendent de cette limite, il faudra donc les interpréter avec précaution.

Pour étudier l'influence du découpage du graphe sur l'accessibilité des voies calculées, nous utilisons quatre territoires différents, aux histoires distinctes et aux formes de réseaux différentes. Les quatre villes auxquelles nous appliquons notre calcul seront les mêmes que celles utilisées dans la définition de la grammaire de lecture de la spatialité : Avignon, Paris, Barcelone et New-York. Nous conservons ces quatre territoires car leurs disparités structurelles nous assurent un travail au delà de l'identité morphologique d'un espace. Nous découpons ces réseaux de manières très différentes afin d'avoir des résultats selon plusieurs stratégies de délimitation.

Pour Avignon, nous procédons par ajouts successifs d'échantillons en demi-cercles, en respectant les grandes structures viaires (voies rapides, autoroutes). Pour Paris, ce sont des découpes concentriques, autour du centre historique de la ville. Pour Barcelone, nous procédons par découpages quasi-rectangulaires qui s'élargissent depuis la vieille ville jusqu'à l'ensemble du réseau de la ville. Pour New-York, nous utilisons l'île de Manhattan et les territoires qui l'entourent, séparés par des baies ou rivières. Les communications de l'un à l'autre ne se font donc que par des ponts.

\begin{figure}[h]
    \centering
    \includegraphics[width=0.7\textwidth]{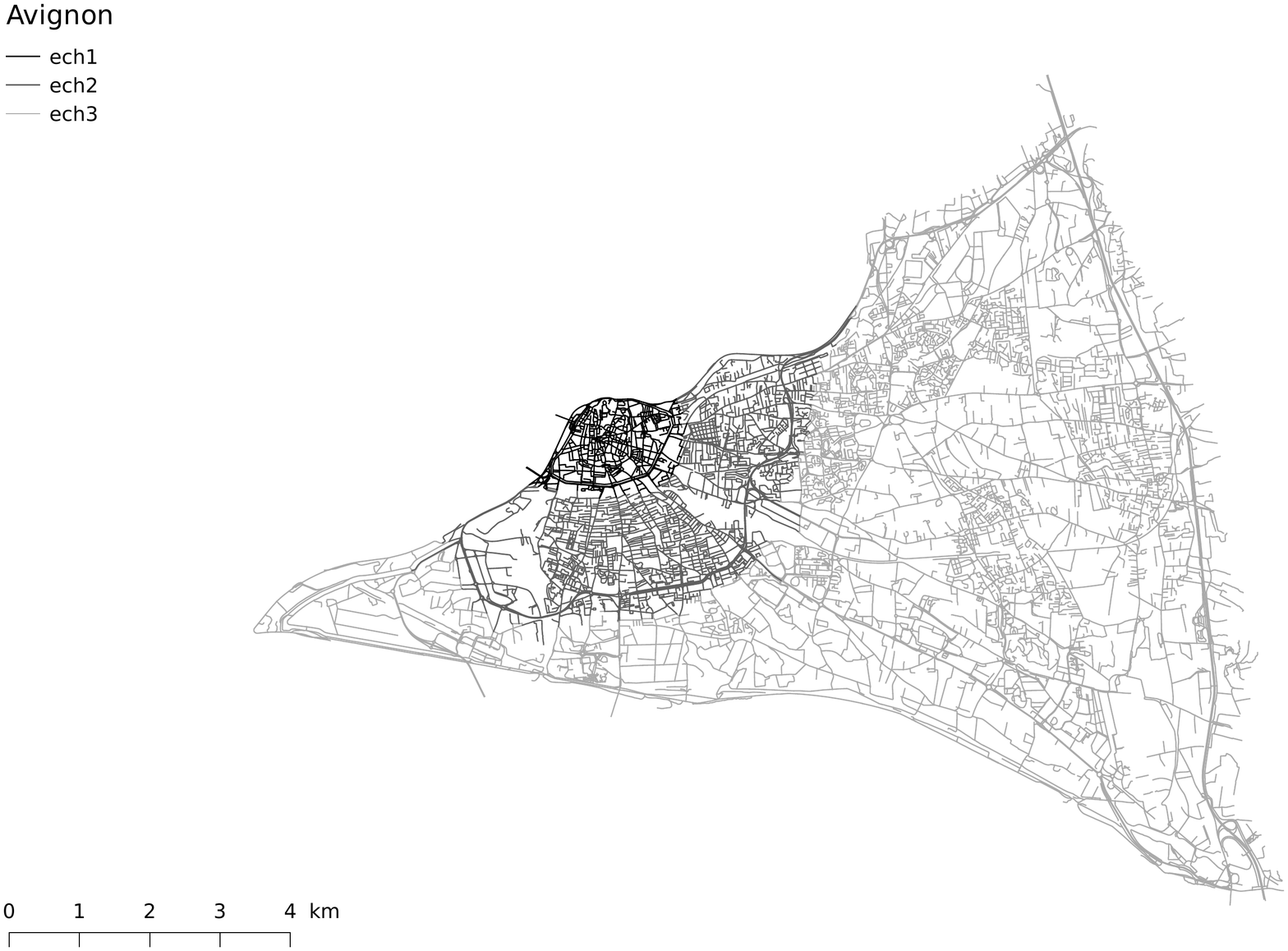}
    \caption{Carte des trois différents échantillons pris sur la ville d'Avignon et ses alentours. Tous les échantillons sont hiérarchiquement superposés.}
    \label{fig:brut_echs_avignon}
\end{figure}

\begin{figure}[h]
    \centering
    \includegraphics[width=0.7\textwidth]{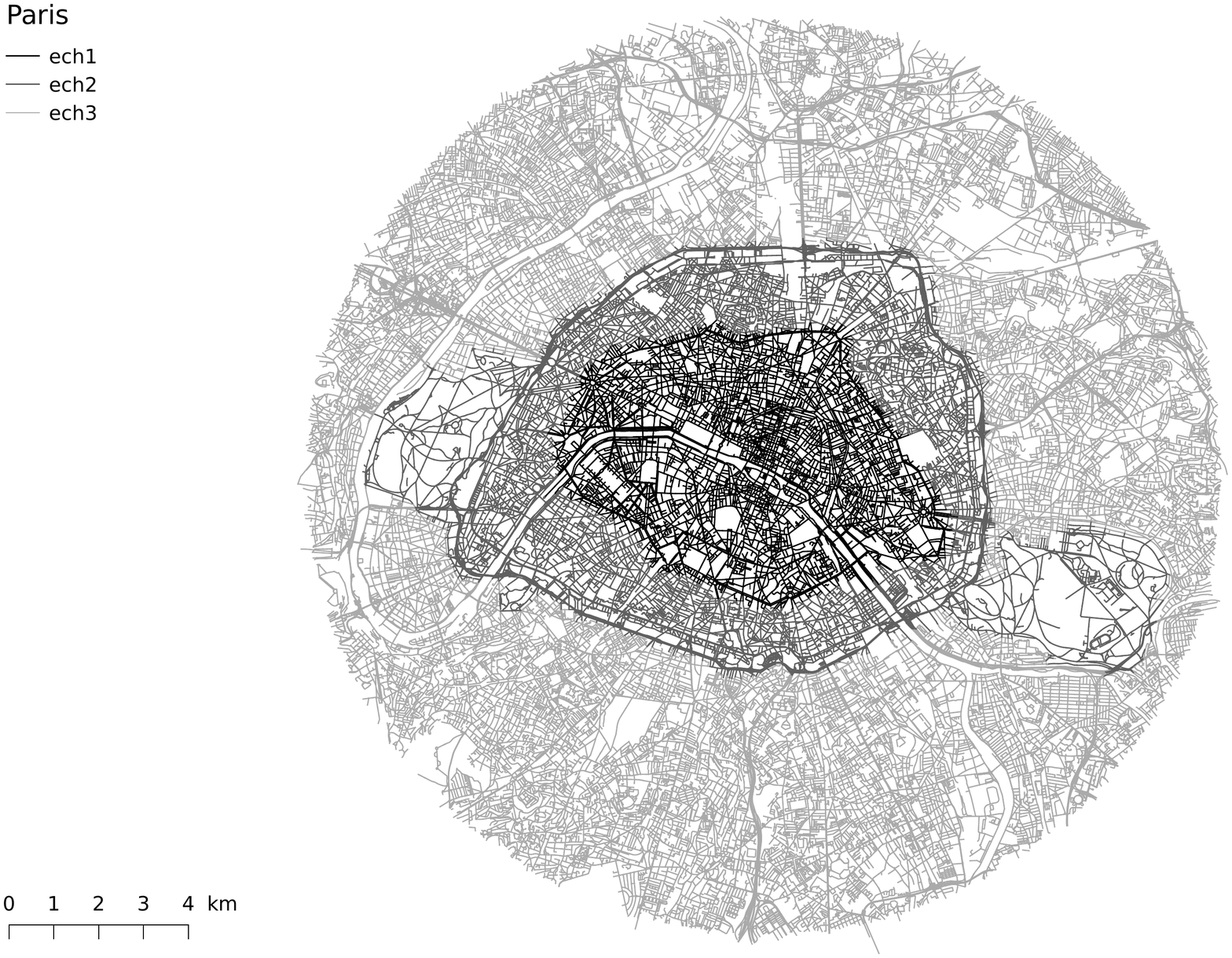}
    \caption{Carte des trois différents échantillons pris sur la ville de Paris et ses alentours. Tous les échantillons sont hiérarchiquement superposés.}
    \label{fig:brut_echs_paris}
\end{figure}

\begin{figure}[h]
    \centering
    \includegraphics[width=0.7\textwidth]{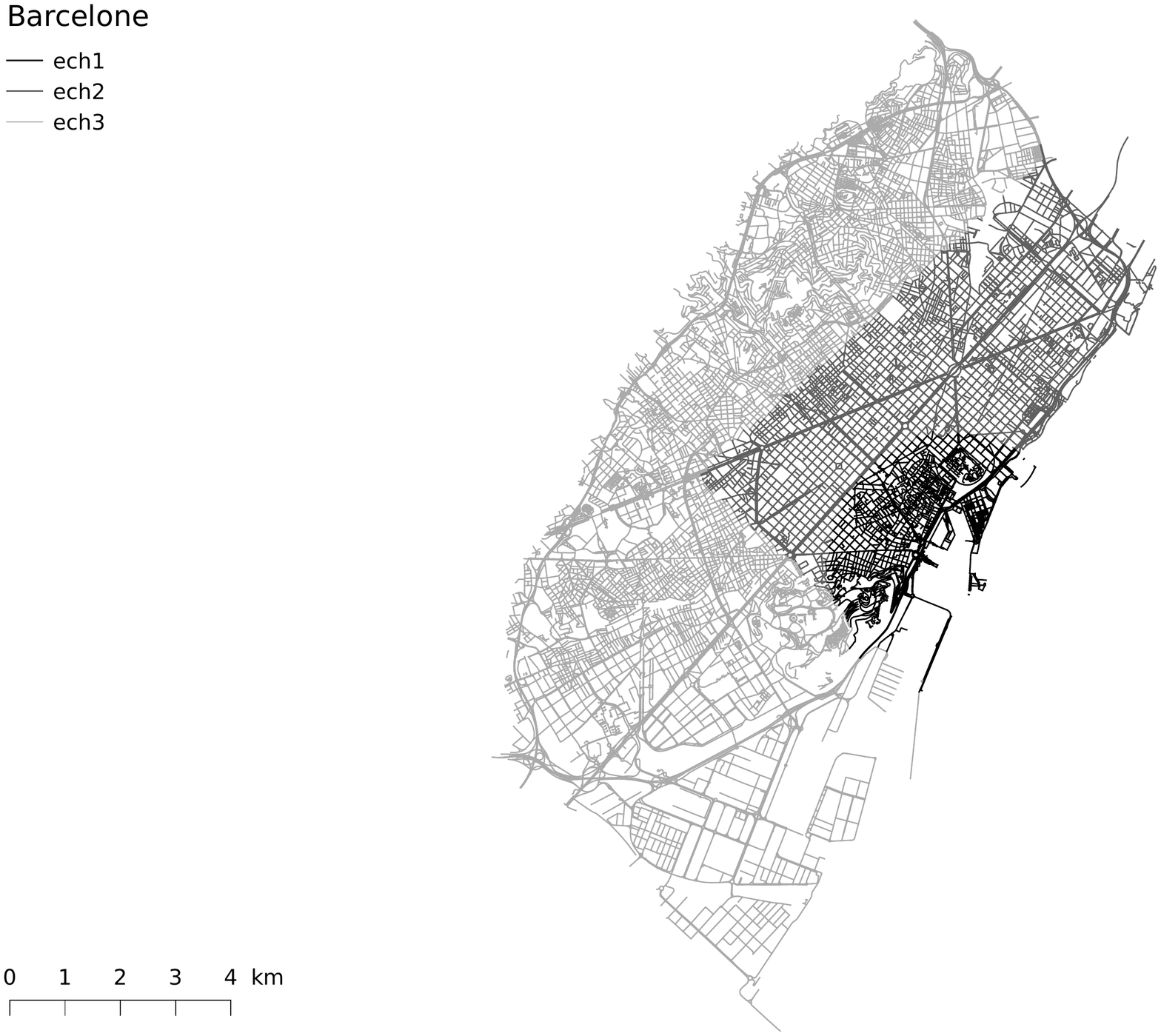}
    \caption{Carte des trois différents échantillons pris sur la ville de Barcelone (Espagne). Tous les échantillons sont hiérarchiquement superposés.}
    \label{fig:brut_echs_barcelone}
\end{figure}

\begin{figure}[h]
    \centering
    \includegraphics[width=0.7\textwidth]{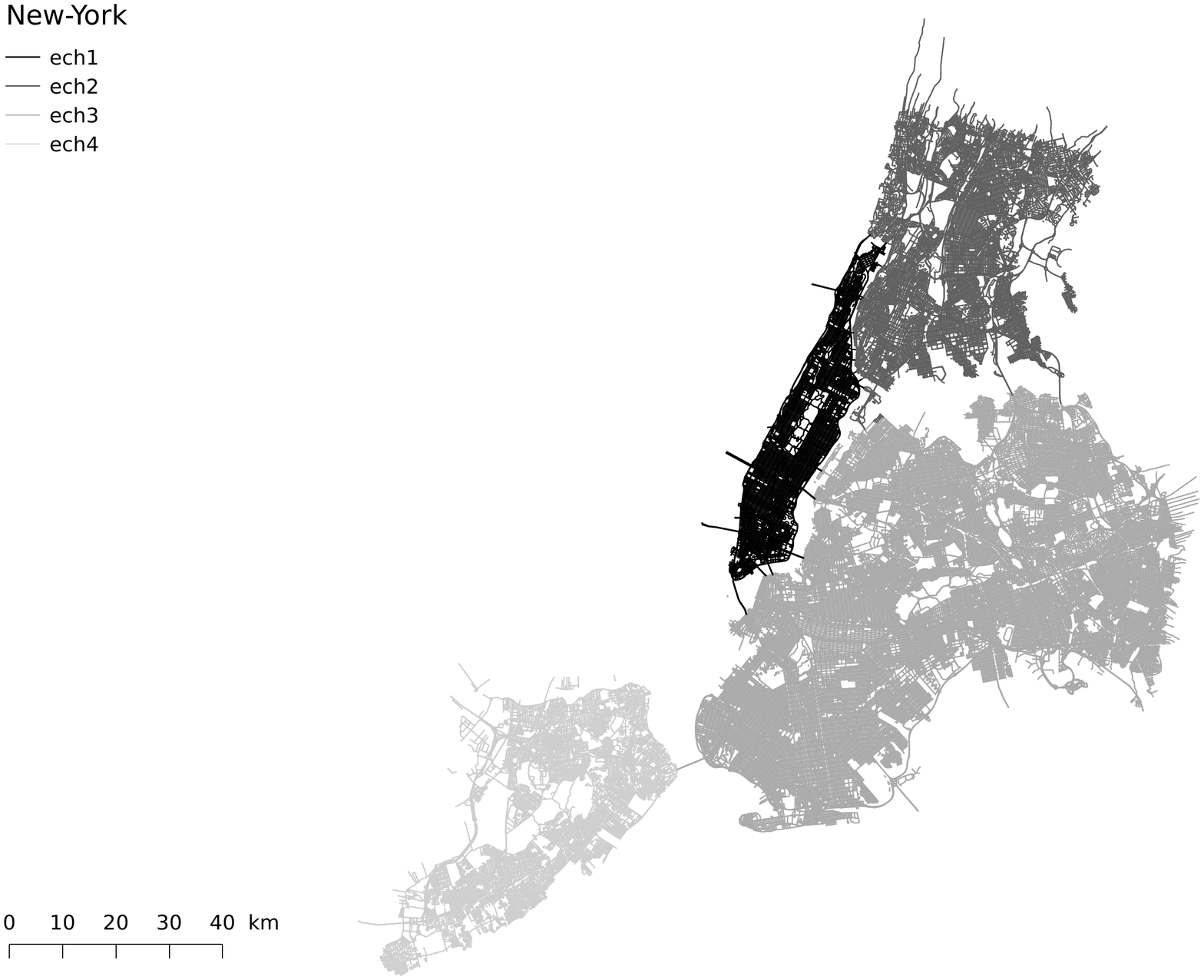}
    \caption{Carte des quatre différents échantillons pris sur la ville de New-York (USA). Tous les échantillons sont hiérarchiquement superposés. \\ \textit{NB} : pour l'étude des effets de bords sur les échantillons de New-York, quelques arcs dédoublés de manière particulière (robuste au nettoyage de données opéré) créent un artefact de variation. L'effet reste cependant très minoritaire.}
    \label{fig:brut_echs_manhattan}
\end{figure}

Nous appliquons notre méthode de quantification des effets de bord sur treize échantillons spatiaux, chacun d'entre eux appartenant à un des quatre espaces d'études choisis (figures \ref{fig:brut_echs_paris}, \ref{fig:brut_echs_avignon}, \ref{fig:brut_echs_barcelone} et \ref{fig:brut_echs_manhattan}).

Les échantillons spatiaux retenus sont donc :

\begin{itemize}
\item Avignon (France)
\begin{itemize}
\item échantillon 1 : découpage intra-muros
\item échantillon 2 : découpage jusqu'à la première ceinture de voies rapides
\item échantillon 3 : découpage jusqu'à l'autoroute à l'Est \footnotemark[1]
\end{itemize}

\item Paris (France)
\begin{itemize}
\item échantillon 1 : découpage intra-muros
\item échantillon 2 : découpage selon les limites communales \footnotemark[1]
\item échantillon 3 : découpage circulaire, coupant arbitrairement le Grand Paris
\end{itemize}

\item Barcelone (Espagne)
\begin{itemize}
\item échantillon 1 : vieille ville
\item échantillon 2 : échantillon 1 \& réseau planifié par Cerdà
\item échantillon 3 : découpage autour de la ville \footnotemark[1]
\end{itemize}

\item New-York (États-Unis)
\begin{itemize}
\item échantillon 1 : île de Manhattan \footnotemark[1]
\item échantillon 2 : échantillon 1 \& Bronx ( jusqu'au \textit{Cross County Parkway} au Nord)
\item échantillon 3 : échantillon 2 \& Brooklyn et le Queens 
\item échantillon 4 : échantillon 3 \& Staten Island
\end{itemize}

\end{itemize}

\footnotetext[1]{graphes utilisés dans la première partie de la thèse}

Les échantillons des trois espaces se recouvrent successivement. Les premiers évoqués dans la liste ci-dessus sont les plus restreints, les suivants étendent l'espace de manières différentes dans les quatre cas. Nous nommons successivement les échantillons ($ech1$, ..., $ech4$) pour chaque espace autour d'un lieu de référence (Avignon, Paris, Barcelone ou Manhattan).

Nous construisons les voies pour chacun de ces graphes et calculons leur indicateur d'accessibilité. Nous quantifions les différences entre ces valeurs, pour chaque couple de graphes d'un même espace, en cartographiant pour chaque arc la valeur de $\Delta_{relatif}$ calculée selon l'équation \ref{eq:delta_rel}. Nous obtenons ainsi la variation relative pour chaque arc du $\Delta$ d'accessibilité de la voie à laquelle il appartient et de sa valeur d'accessibilité calculée dans $G_2$ que nous multiplions par $(-1)$ pour qu'une amélioration d'accessibilité corresponde à une valeur positive. Nous obtenons des valeurs entre -1 et 1, interprétables comme des variations d'accessibilité de -100\% à 100\%  de la valeur calculée dans $G_2$.

\begin{figure}[h]
    \centering
    \includegraphics[width=0.8\textwidth]{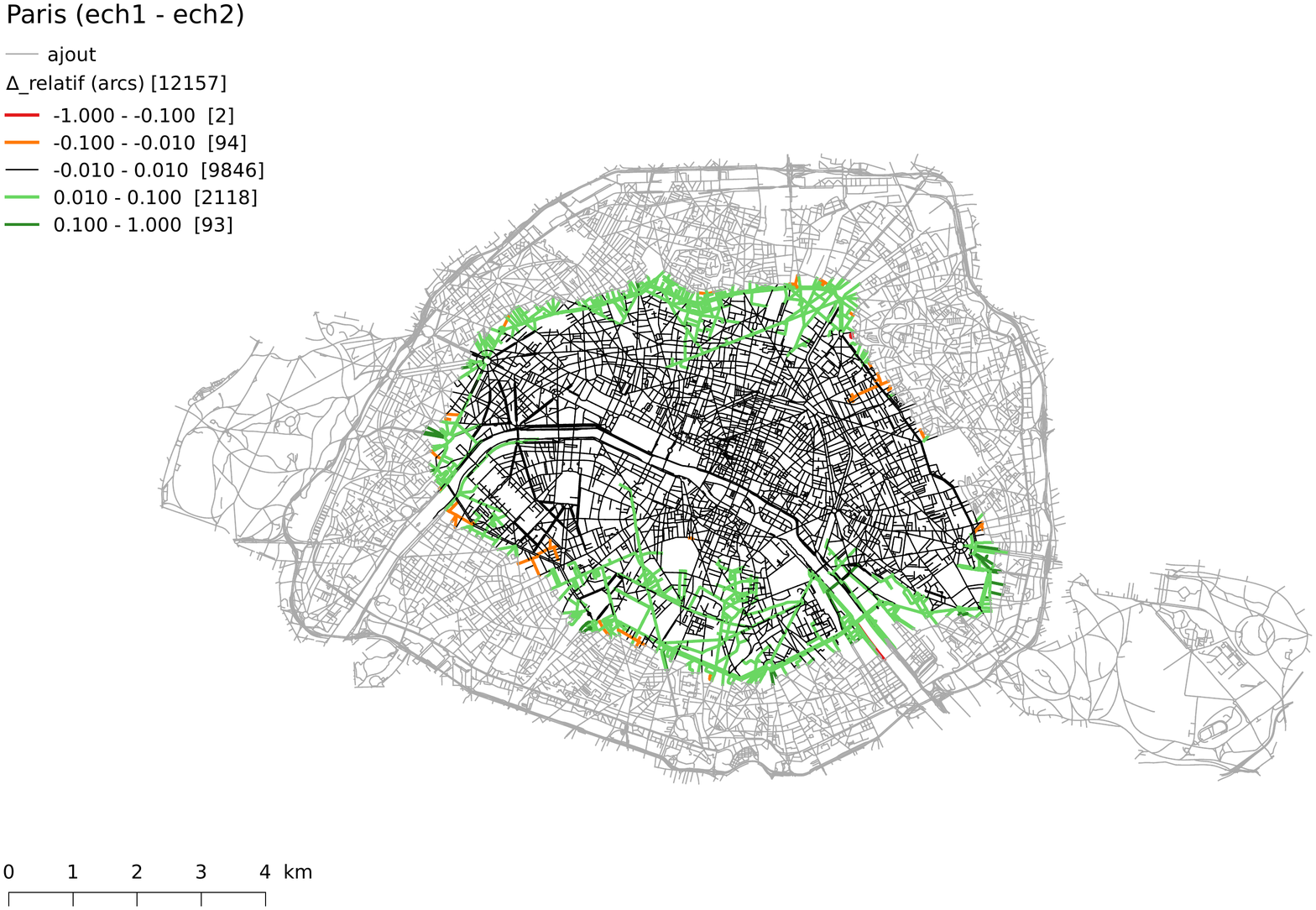}
    \caption{Carte de $\Delta_{relatif}$ calculé entre les échantillons 1 et 2 de Paris.}
    \label{fig:diff_paris_sm}
\end{figure}

\begin{figure}[h]
    \centering
    \includegraphics[width=0.8\textwidth]{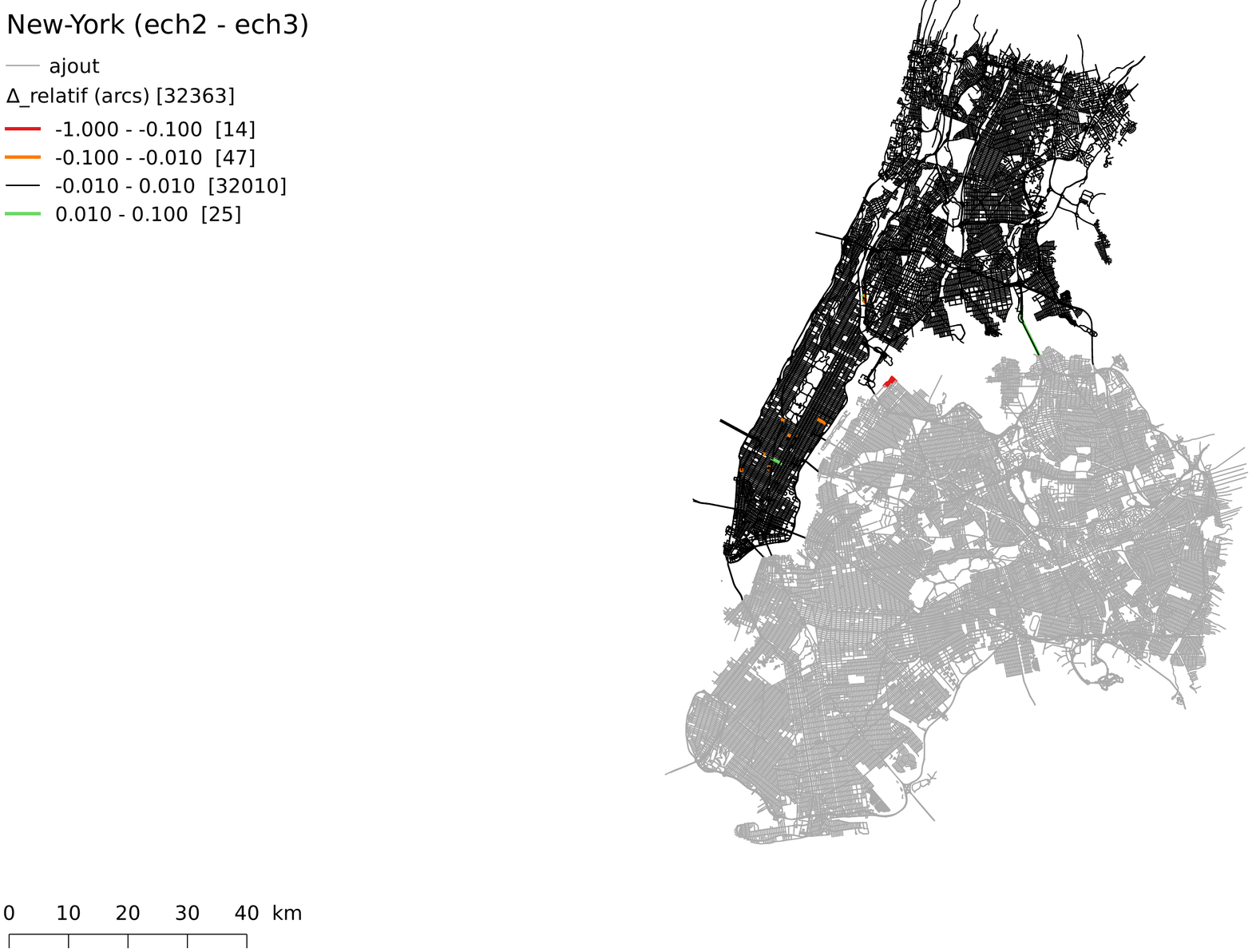}
    \caption{Carte de $\Delta_{relatif}$ calculé entre les échantillons 2 et 3 de New-York.}
    \label{fig:diff_man_sl}
\end{figure}

Nous représentons en rouge les diminutions d'accessibilité, et en vert ses augmentations. Les arcs représentés en noir sont ceux ayant subi moins de 1\% de variation : ils sont considérés comme stables. Pour chaque carte, nous indiquons le nombre d'arcs présents dans chaque intervalle de valeurs de $\Delta_{relatif}$ (figures \ref{fig:diff_paris_sm} à \ref{fig:diff_man_sl}). Nous plaçons quelques unes des cartes au fil du texte, l'ensemble étant réservé en annexe \ref{ann:chap_ebords}.

\begin{equation}
\Delta_{relatif}(a_{ref} \in v) = (-1) \times \frac{\Delta(a_{ref} \in v)}{accessibilite_{G_2}(v_{ref})}
\label{eq:delta_rel}
\end{equation}

Nous observons sur l'ensemble des espaces étudiés que les variations de l'indicateur sont très faibles (moins de 1\% de la valeur d'accessibilité calculée sur $G_2$). Nous regroupons dans le tableau \ref{tab:delta} les valeurs moyennes ($\overline{\Delta_{relatif}}$), les écarts types $\sigma(\Delta_{relatif})$ ainsi que les variations maximales en valeur absolue ($max \vert \Delta_{relatif}\vert$) pour chaque couple d'échantillons. Nous lisons dans ce tableau des variations moyennes de l'ordre de 1 à 3\% des valeurs d'accessibilité calculées sur $G_2$.

Les valeurs minimales sont obtenues pour Manhattan, où les échantillons ne sont reliés que pour des ponts. L’interaction est donc minimale dans ce cas. La plus grande variation moyenne (0.0075) est observée entre les échantillons 1 et 2 de Paris. Entre ceux-ci, des avenues ont été coupées, elles ressortent sur la carte \ref{fig:diff_paris_sm} avec des couleurs plus soutenues. En moyenne, les variations observées sont toujours positives (le réseau gagne en accessibilité en étendant le territoire observé). Les moyennes observées négatives, pour les échantillons 1 et 2 ainsi que 2 et 3 d'Avignon (figures en annexe \ref{ann:sec_eb_avignon}), les échantillons 2 et 3 de New-York (figure en annexe \ref{ann:sec_eb_newyork}) et les échantillons 2 et 4 de la même ville (figure \ref{fig:diff_man_sl}) le sont avec des valeurs très faibles et significativement plus petites que l'écart type lui-même. Le signe de la valeur n'est pas significatif : il s'agit de faibles variations autour d'un impact moyen proche de 0. Pour les échantillons aux variations moyennes plus élevées, l'impact observé sur la valeur de l'indicateur d'accessibilité est homogène. L'écart-type des variations est en effet très faible, ne dépassant pas 0.02 dans l'ensemble de notre étude (tableau \ref{tab:delta}).

Les variations les plus importantes sont observées lorsque le réseau ajouté vient compléter le réseau précédent. C'est notamment le cas à Barcelone, entre les échantillons 1 et 2 ainsi que 1 et 3 (figures en annexe \ref{ann:sec_eb_barcelone}). La ville reste très stable, alors que les portions de réseau quadrillé qui s'y insèrent ont un $\Delta_{relatif}$ positif plus marqué.

\begin{table}
\begin{center}
{ \small
\begin{tabular}{|c|c|r|r|r|r|r|}
\hline

ville & comparaison & $N_{arcs(ajoutés)}$ & $L_{ajoutée}$ & $\overline{\Delta_{relatif}}$ & $\sigma(\Delta_{relatif})$  & $max \vert \Delta_{relatif}\vert$  \\ \hline

Avignon & ech1 - ech2 & 3 710 & 243 774 m & -0.0007 & 0.0103 & 0.1941  \\ \hline
 & ech1 - ech3 & 13 193 & 1 058 804 m & 0.0016 & 0.0020 & 0.0357  \\ \hline
 & ech2 - ech3 & 9 483 & 815 030 m & -0.0012 & 0.0149 & 0.2424  \\ \hline

Paris & ech1 - ech2 & 18 098 & 1 240 912 m & 0.0075 & 0.0179 & 0.4126  \\ \hline
 & ech1 - ech3 & 75 971 & 5 276 349 m & 0.0034 & 0.0050 & 0.1208  \\ \hline
 & ech2 - ech3 & 58 163 & 4 050 920 m & 0.0060 & 0.0165 & 0.3100  \\ \hline

Barcelone & ech1 - ech2 & 7 398 & 550 264 m & 0.0026 & 0.0100 & 0.1597  \\ \hline
 & ech1 - ech3 & 26 451 & 1 906 460 m & 0.0032 & 0.0049 & 0.0959  \\ \hline
 & ech2 - ech3 & 19 053 & 1 356 196 m & 0.0030 & 0.0166 & 0.3520  \\ \hline

New-York & ech1 - ech2 & 21 832 & 7 862 045 m & 0.0007 & 0.0038 & 0.0667  \\ \hline
 & ech1 - ech3 & 81 349 & 29 829 101 m & 0.0002 & 0.0012 & 0.0193  \\ \hline
 & ech1 - ech4 & 95 709 & 35 289 761 m & 0.0003 & 0.0009 & 0.0148  \\ \hline
 & ech2 - ech3 & 59 517 & 21 967 056 m & -0.0002 & 0.0072 & 0.3548  \\ \hline
 & ech2 - ech4 & 73 877 & 27 427 716 m & -0.00009 & 0.0054 & 0.2625  \\ \hline
 & ech3 - ech4 & 14 360 & 5 460 659 m & 0.00006 & 0.0044 & 0.1467  \\ \hline

\end{tabular}
}
\end{center}
\caption{Détail statistique des variations relatives $\Delta_{relatif}$ pour chaque couple d'échantillons.}
\label{tab:delta}
\end{table}

Nous pouvons donc conclure que la voie apporte une grande stabilité dans la lecture des caractéristiques du réseau. Nous avions vu précédemment qu'elle rend équivalente une caractérisation du réseau avec un indicateur global (l'utilisation - ou betweenness) à celle d'un indicateur local (le degré). Nous pouvons ici ajouter à cette propriété celle de rendre la lecture globale du réseau (avec un indicateur calculé sur l'ensemble de celui-ci) stable au découpage spatial choisi. Ceci à condition qu'il soit fait \textit{en cohérence} avec l'espace étudié : sans couper de longues structures spatiales \textit{a priori} continûment alignées.


\begin{figure}[h]
    \centering
    \includegraphics[width=0.8\textwidth]{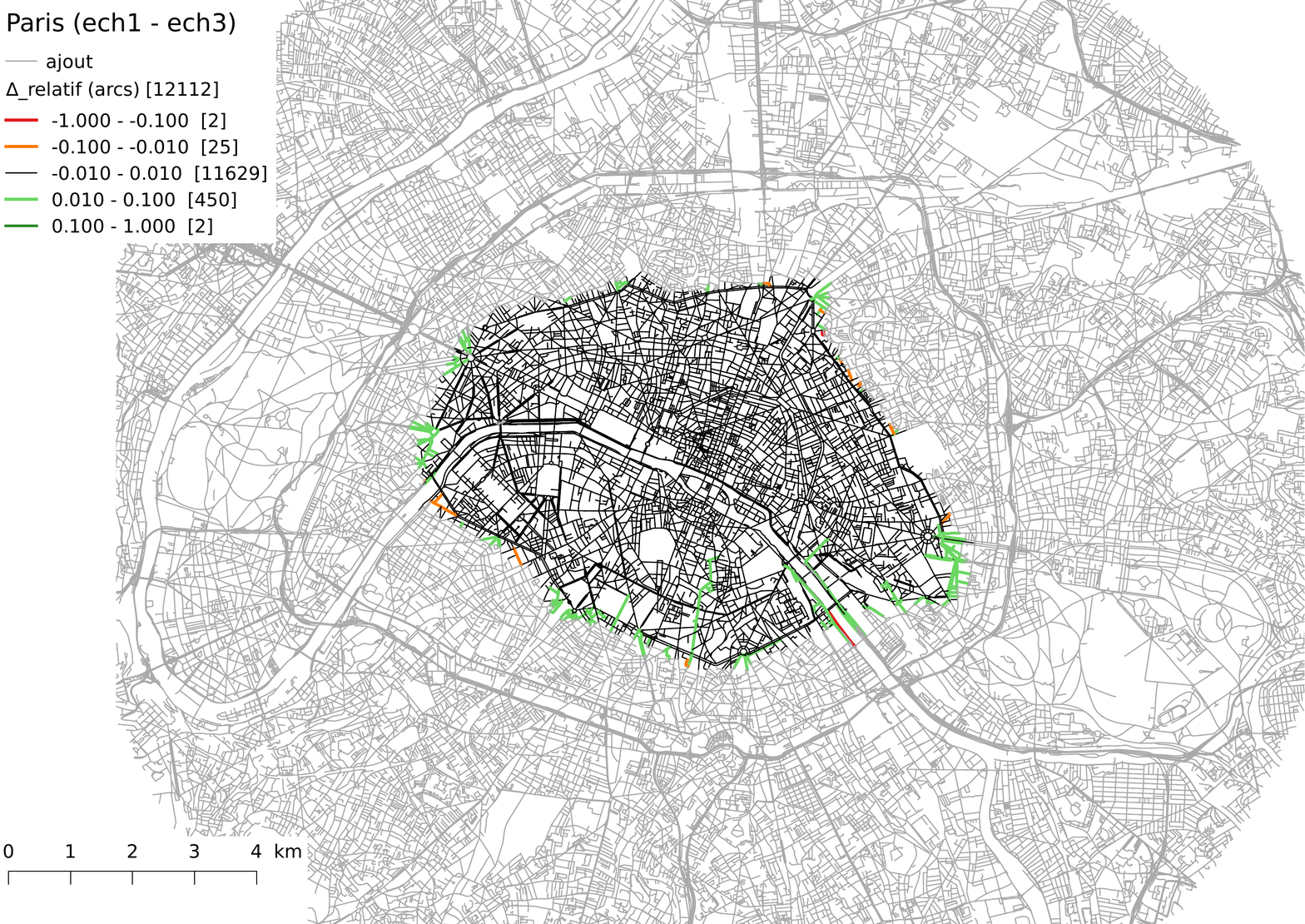}
    \caption{Carte de $\Delta_{relatif}$ calculé entre les échantillons 1 et 3 de Paris.}
    \label{fig:diff_paris_sl}
\end{figure}

\begin{figure}[h]
    \centering
    \includegraphics[width=0.8\textwidth]{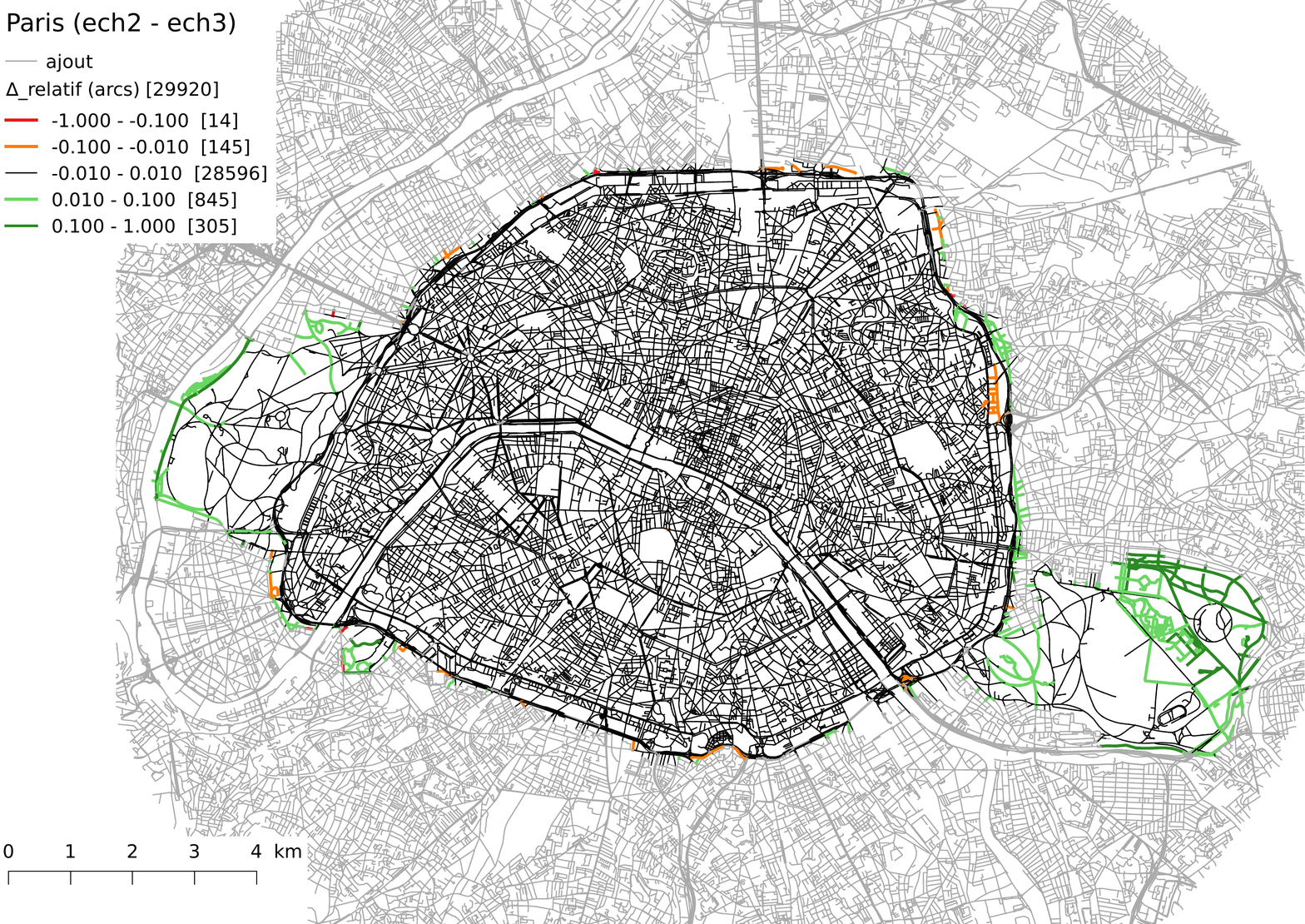}
    \caption{Carte de $\Delta_{relatif}$ calculé entre les échantillons 2 et 3 de Paris.}
    \label{fig:diff_paris_ml}
\end{figure}


\begin{figure}[h]
    \centering
    \includegraphics[width=0.8\textwidth]{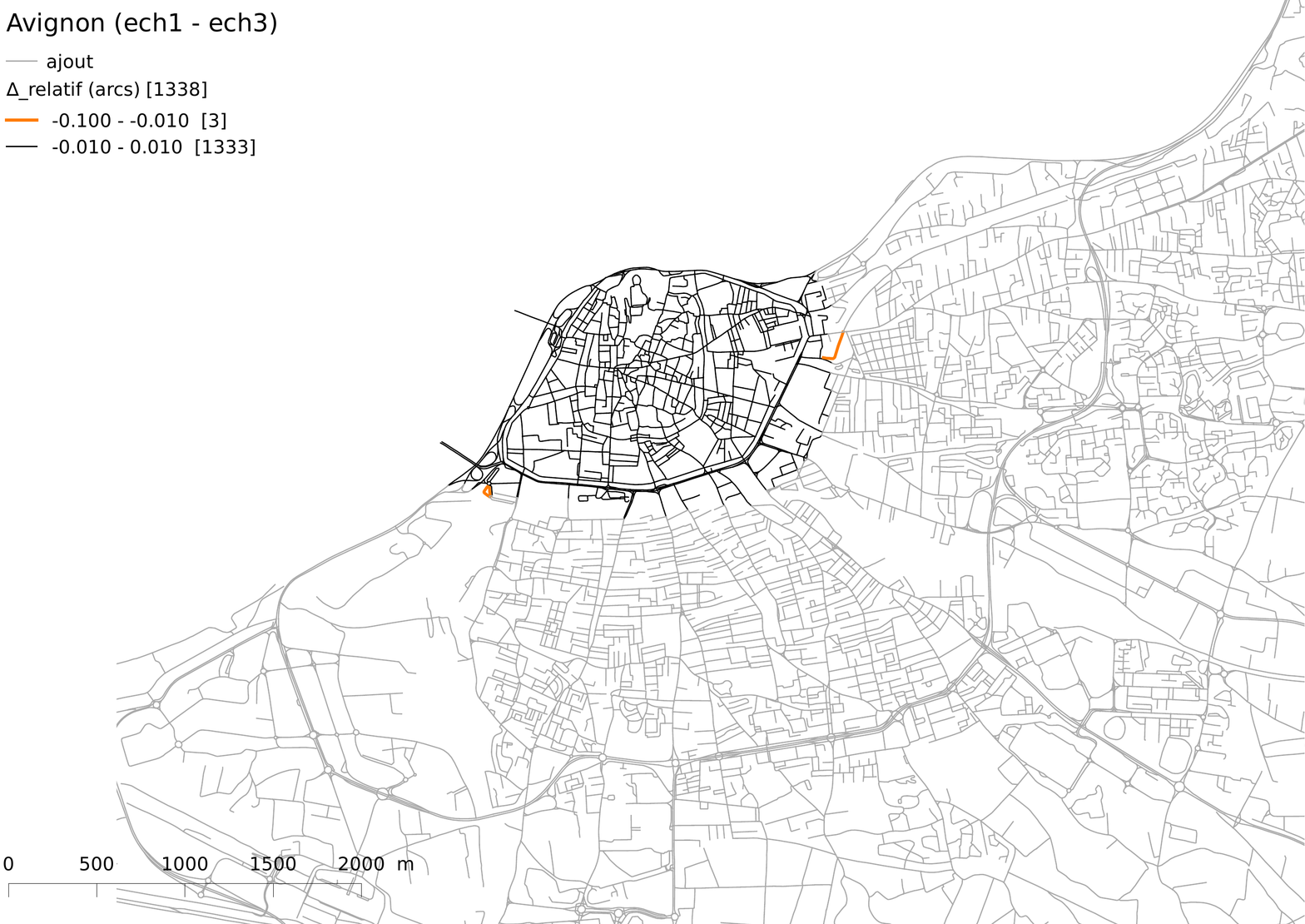}
    \caption{Carte de $\Delta_{relatif}$ calculé entre les échantillons 1 et 3 d'Avignon.}
    \label{fig:diff_avignon_sl}
\end{figure}


\begin{figure}[h]
    \centering
    \includegraphics[width=0.8\textwidth]{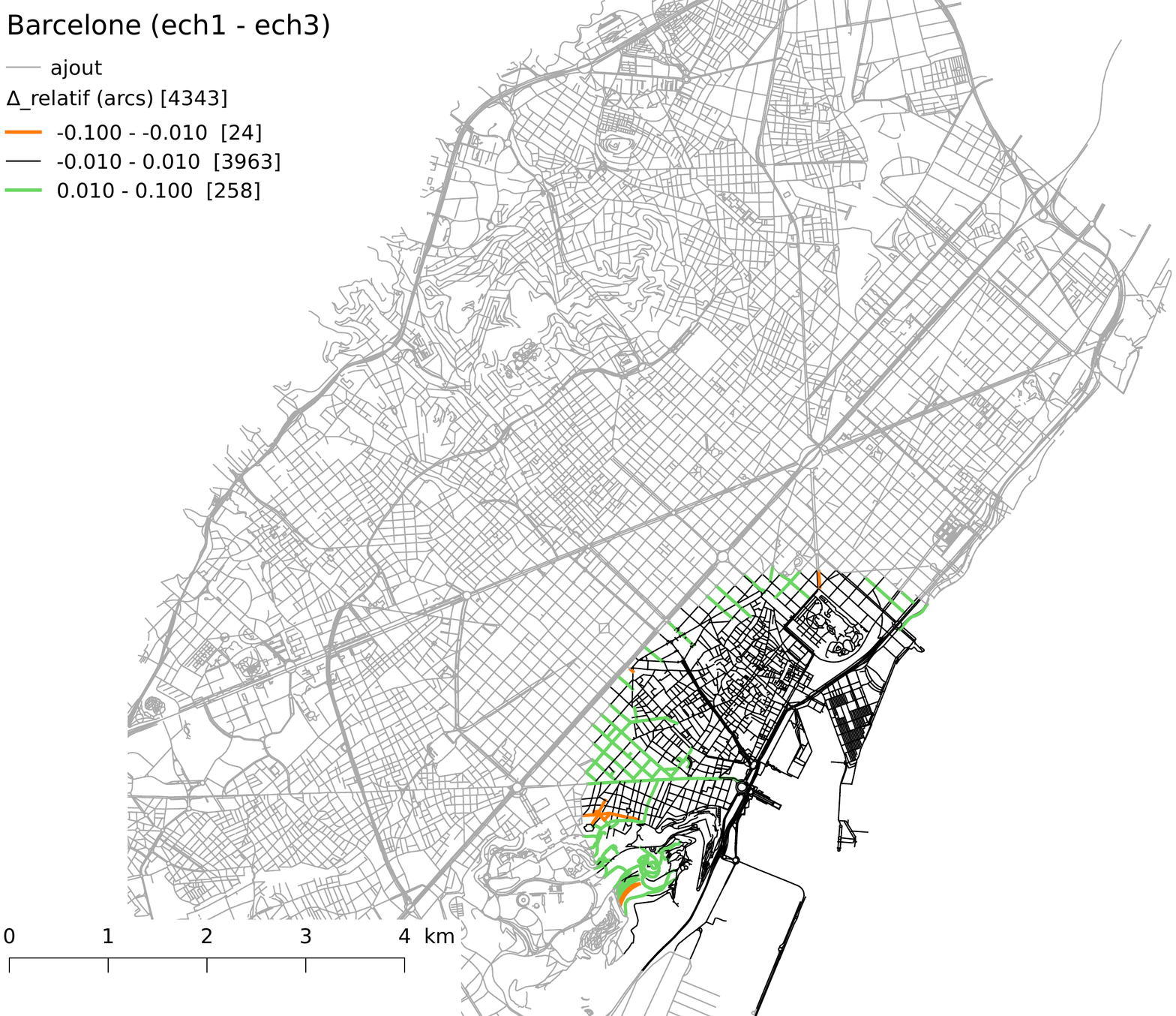}
    \caption{Carte de $\Delta_{relatif}$ calculé entre les échantillons 1 et 3 de Barcelone.}
    \label{fig:diff_barcelone_sl}
\end{figure}


\FloatBarrier
\subsection{Étude des effets de discrimination}

Nous avons vu que le découpage du réseau a très peu d'incidence sur le calcul de l'indicateur d'accessibilité. Cet indicateur, équivalent à la closeness, est le seul calculé globalement sur les voies dont le résultat diffère de ceux calculés localement. C'est donc celui dont nous voulons particulièrement étudier la stabilité.

Nous souhaitons maintenant déterminer l'influence de la mise à l'écart de certains types de données sur l'accessibilité du réseau. Nous étudions pour cela la commune de Paris, pour laquelle nous disposons des données de la \copyright BDTOPO. Dans notre étude, nous avons choisi de ne conserver que les données de type \enquote{route} (à 1 ou 2 chaussées), \enquote{bretelle}, \enquote{autoroute} ou \enquote{quasi-autoroute}. Nous appliquons notre méthode de quantification pour savoir si l'ajout des routes empierrées a un impact sur les accessibilités calculées sur les voies.

\begin{figure}[h]
    \centering
    \includegraphics[width=\textwidth]{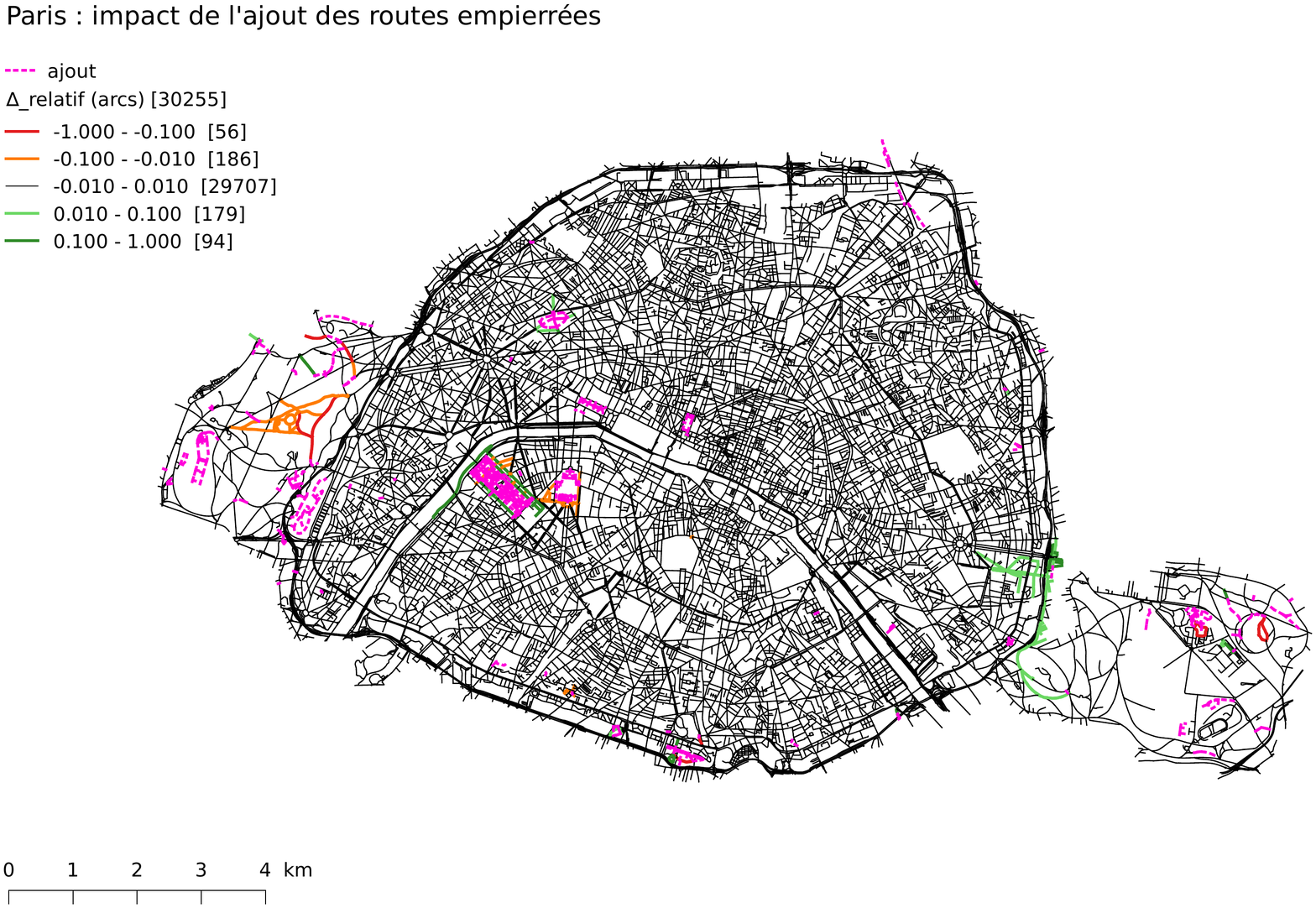}
    \caption{Carte de $\Delta_{relatif}$ calculé sur la commune de Paris avec et sans routes empierrées.}
    \label{fig:diff_paris_mm}
\end{figure}

La figure \ref{fig:diff_paris_mm} représente l'impact de l'ajout de ces données, de types différents. Nous voyons dans le détail de la classification des arcs que l'influence de ces données est quasiment nul. La valeur moyenne de la variation résultante est de -0.0002. En revanche, l'écart-type (de 0.0127) est parmi les plus forts que nous observons car la valeur maximale de variation (en valeur absolue) est de 0.4155. En effet, les arcs positionnés au contact direct de ceux ajoutés voient un changement important dans leur accessibilité.

Nous observons également sur ce réseau qu'un changement dans une structure principale, comme celui proche du boulevard périphérique au Sud-Est du réseau viaire, a un impact plus diffus que ceux isolés. Cette influence est à relativiser, car elle est en majorité inférieur à 10\%. Cependant, cela nous montre que, s'il est possible de ne pas se soucier de la prise en compte ou non de données secondaires dans le réseau, l'impasse ne peut bien évidemment pas être faite sur celles constituant sa forme structurelle. Cette morphologie, squelette de notre représentation des formes urbaines, se retrouve en mettant en relief sur le réseau les voies au degré le plus élevé (cf Partie I).

\FloatBarrier

\section{Étude de la sensibilité à la granularité de vectorisation}

Dans cette thèse nous travaillons sur la caractérisation des réseaux spatiaux. Nous élaborons celle-ci en travaillant sur des réseaux viaires, mais nous la concevons comme générique afin de l'appliquer à n'importe quel type de graphe spatialisé. Les données que nous utilisons sur les réseaux viaires sont de bonne qualité : les arcs ont une géométrie bien définie dans l'espace, ils ne se chevauchent pas et n'utilisent de points annexes que pour des changements de direction ayant une correspondance sur le réseau viaire physique.

Si l'on applique nos raisonnements à d'autres réseaux spatiaux (veinures de feuilles ou craquelures sur une plaque d'argile par exemple), le graphe d'application ne sera pas toujours d'une qualité équivalente. Pour ces réseaux, nous utilisons une méthode d'acquisition automatique développée par Annemiek Cornelissen, et poursuivie par Auguste Bonnin. Celle-ci vectorise à partir d'une photographie le réseau qu'elle représente en étudiant les contrastes radiométriques (zones de transition). L'idée (à l'origine d’Étienne Couturier) est de faire un pavage de Voronoï \citep{voronoi1908nouvelles} à partir des points du contour pour déterminer les points médians, et donc le squelette géométrique le plus précis possible. Grâce à ce travail nous disposons d'un réseau vectoriel pouvant être analysé avec les outils que nous avons développés.

Dans le cas du graphe obtenu à partir d'une plaque d'argile, la présence d'un grand nombre de points annexes, dont la position n'est pas significative, vient brouiller l'information représentée. Pour visualiser l'impact de la finesse de vectorisation, nous utiliserons cet exemple. Ce graphe spatial est planaire car, dans ce cas, chaque intersection d'arcs donne lieu à la création d'un sommet. Ce réseau a été créé par Philippe Bonnin à partir d'un mélange d'illite verte (2/3) et de maïzena (1/3), qu'il a réparti avec une épaisseur de 2mm environ sur une plaque et fait sécher à l'air libre.

\begin{figure}[h]
    \centering
    \includegraphics[width=\textwidth]{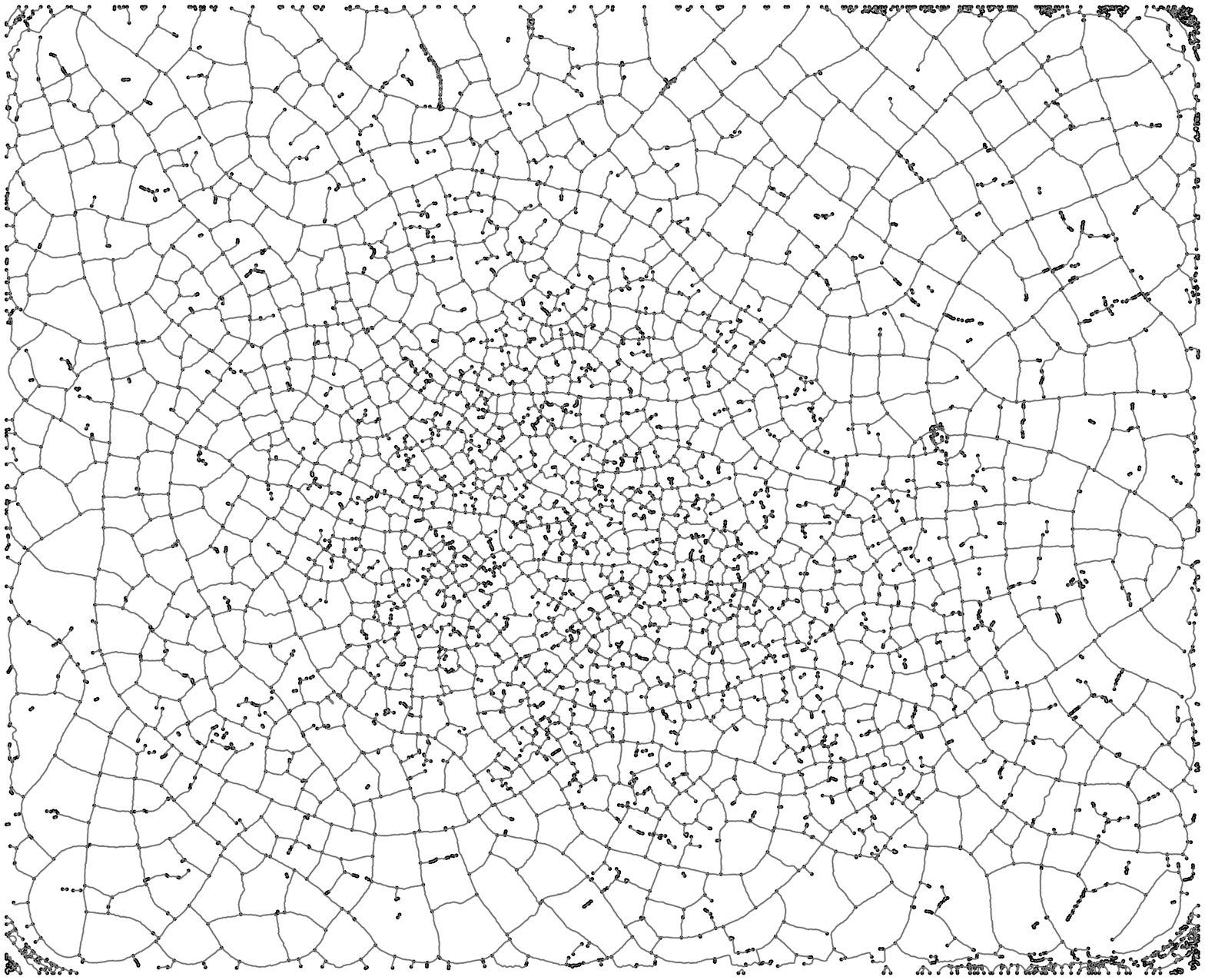}
    \caption{Réseau de craquelures sur une plaque d'argile. Numérisation à partir d'une photographie. Réalisation : A. \& Ph. Bonnin}
    \label{fig:craqs_brut}
\end{figure}

\begin{figure}[h]
    \centering
    \includegraphics[width=0.8\textwidth]{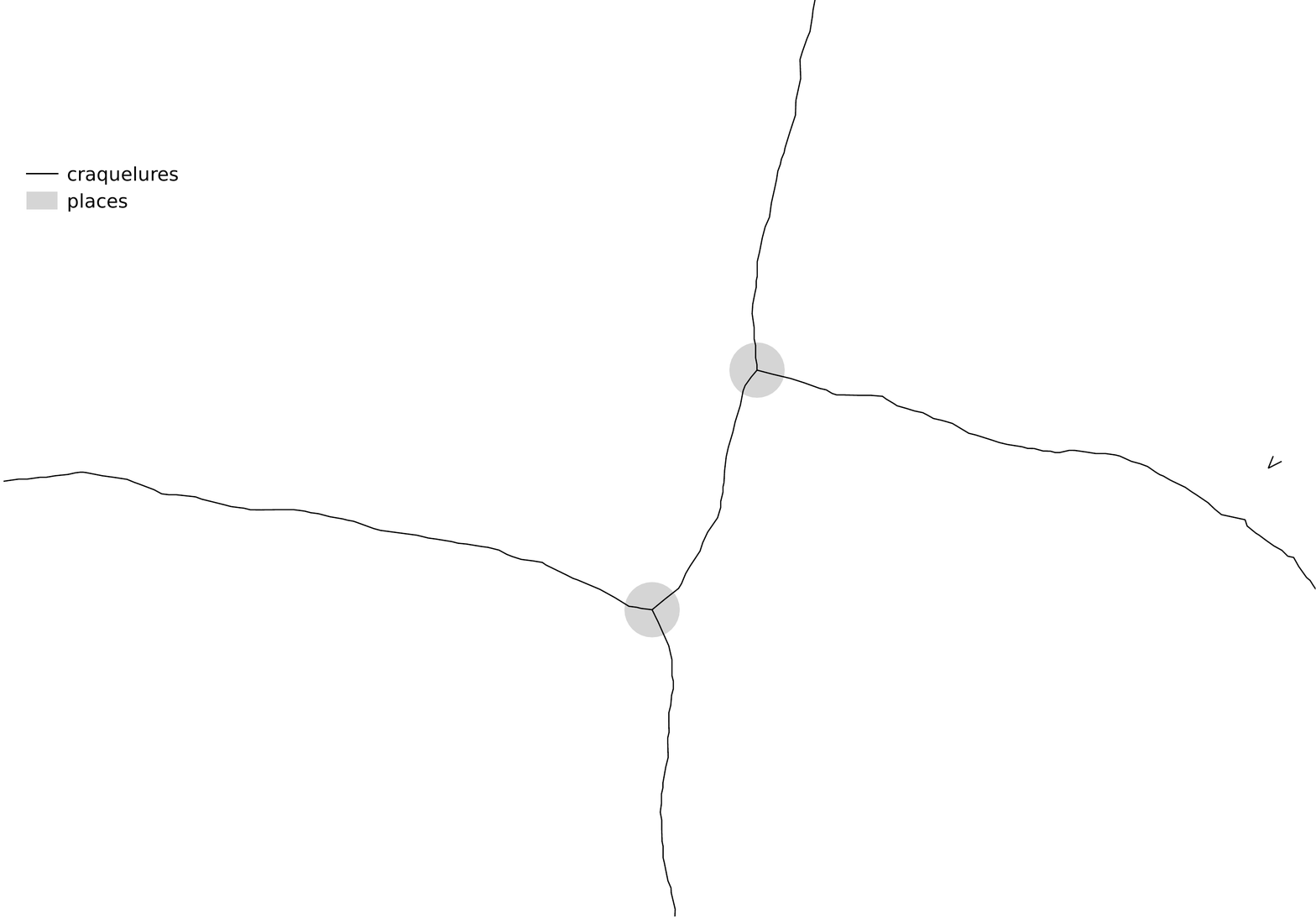}
    \caption{Effet \textit{bulle de savon} qui brisent les continuités aux intersections.}
    \label{fig:craqs_bulleDeSavon}
\end{figure}

\begin{figure}[h]
    \centering
    \includegraphics[width=0.8\textwidth]{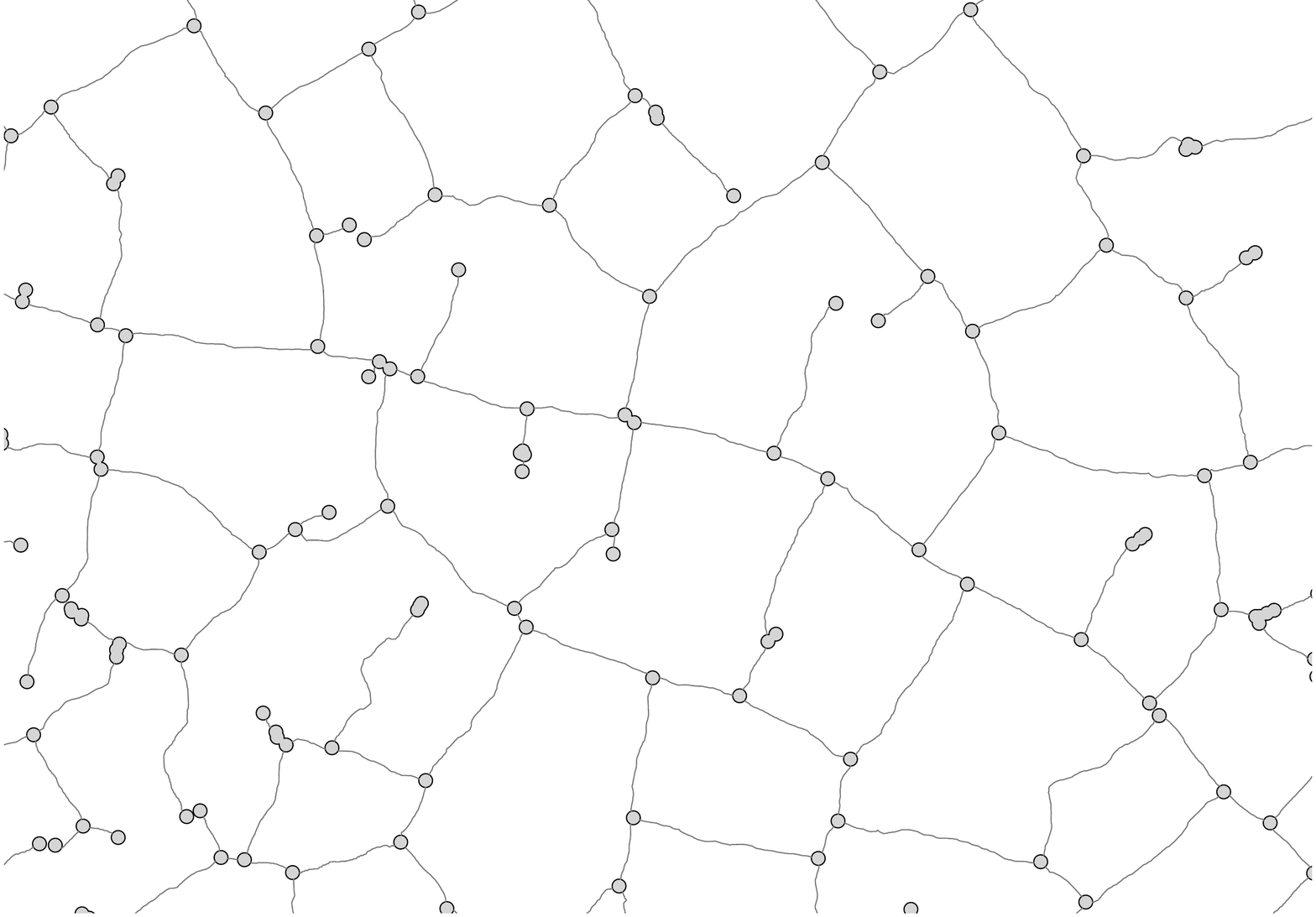}
    \caption{Création de zones tampons circulaires aux intersections.}
    \label{fig:craqs_brut_zoom}
\end{figure}

La vectorisation automatique du réseau de craquelures aboutit à un graphe à la géométrie bruitée (figure \ref{fig:craqs_brut}). À chaque intersection apparaît un effet \textit{bulle de savon}, typique de Voronoï, car en ces points (figure \ref{fig:craqs_bulleDeSavon}), les craquelures sont trop profondes pour être approximées par un simple trait. Cet effet a pour conséquence de briser la voie à chaque intersection car les angles, dans ce cas, dépassent le plus souvent  les 60\degres  de déviation seuil fixés. Pour pallier ce problème, nous utilisons la méthode expliquée au début de ce chapitre : nous créons des zones tampons autour de chaque sommet (figure \ref{fig:craqs_brut_zoom}). Cependant, dans le cas des craquelures numérisées, l'approximation vectorielle de la photo numérisée impose un bruit à la géométrie des arcs qui peut modifier la direction de leur dernier segment (à l'entrée de la zone tampon) et ainsi fausser les calculs de différences d'azimuts.

\begin{figure}[h]
    \centering
    \includegraphics[width=0.8\textwidth]{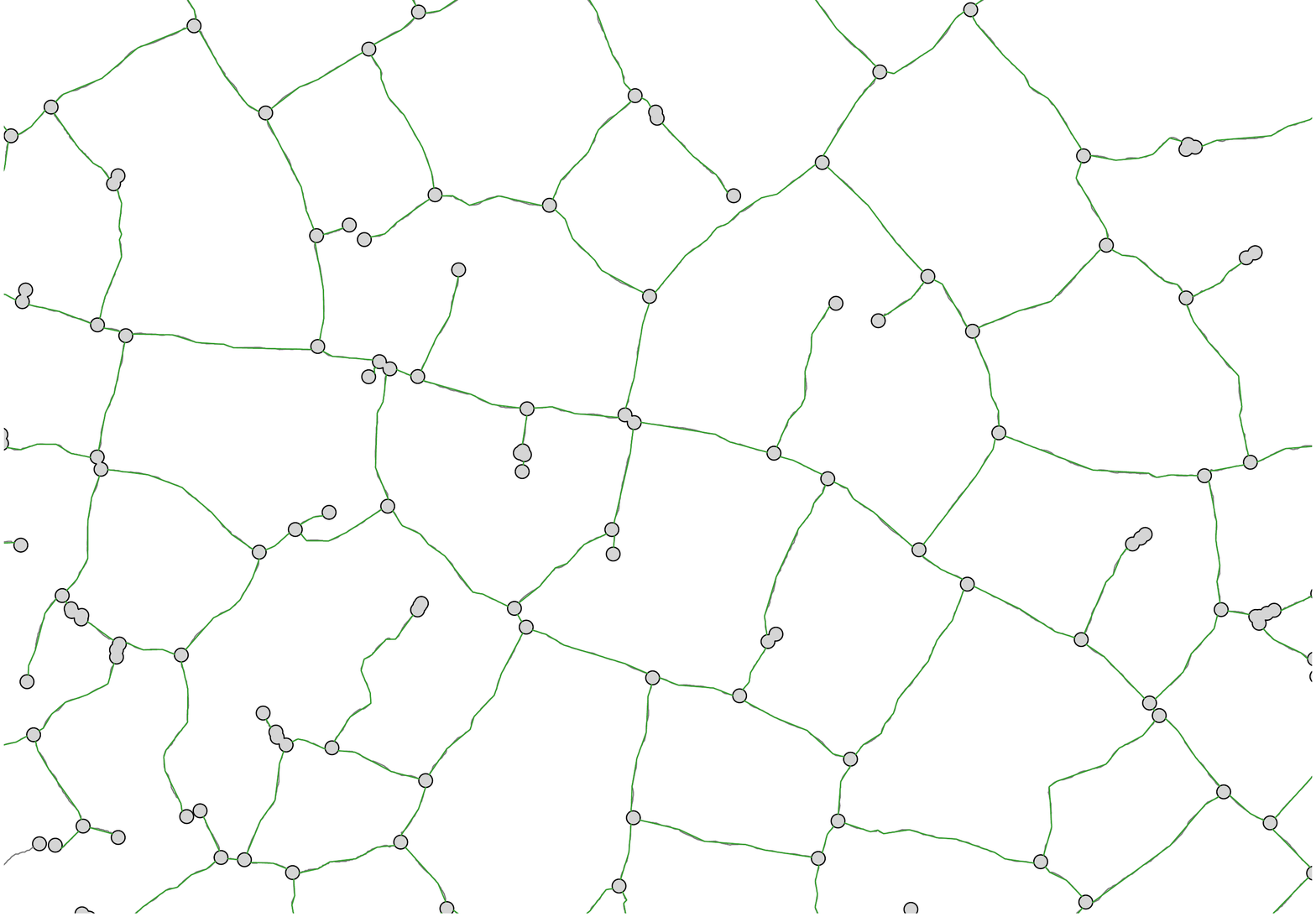}
    \caption{Première généralisation de la vectorisation : chaque arc est approximé par un arc de géométrie simplifiée. Dans ce cas, la nouvelle géométrie reste proche de celle numérisée (seuil d'écart maximal fixé à 0.5 unité de la carte) en tentant de réduire le nombre de points annexes au maximum.}
    \label{fig:craqs_brut_zoom_s1}
\end{figure}

\begin{figure}[h]
    \centering
    \includegraphics[width=0.8\textwidth]{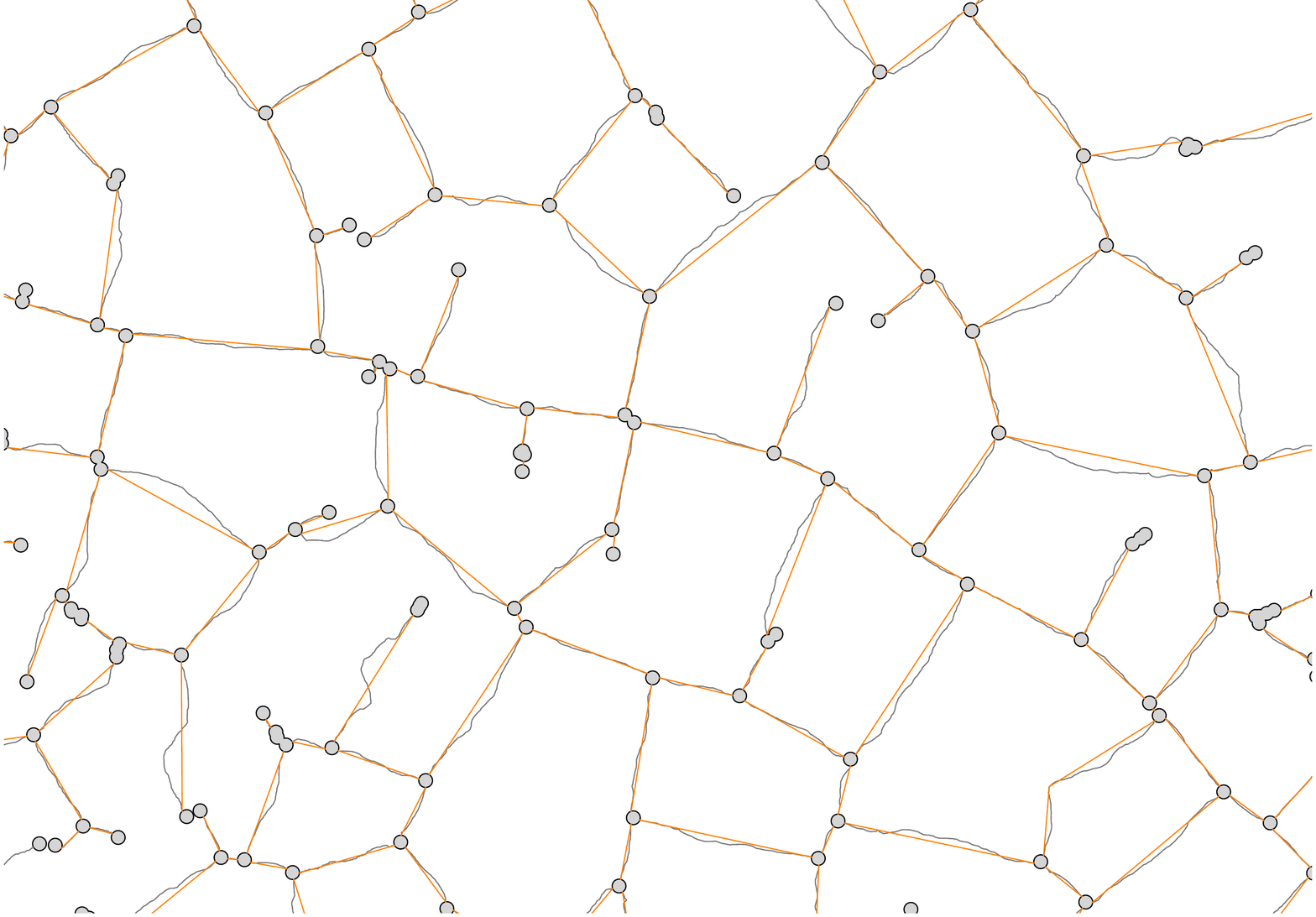}
    \caption{Deuxième généralisation de la vectorisation : plus tranchée que la précédente, seules les géométries des points initial et final de l'arc sont conservées. L'arc est approximé par un segment.}
    \label{fig:craqs_brut_zoom_s2}
\end{figure}

Afin d'obtenir une géométrie moins bruitée, nous effectuons un \textit{lissage} de celle-ci entre deux intersections, dans une opération de \textit{généralisation}. Ici, cela consiste à réduire le nombre de points annexes sur chaque arcs. Deux types de généralisation ont été mises en place. La première ($s1$) reste proche de la géométrie initiale avec un écart maximum fixé à 0.5 unités de la carte (figure \ref{fig:craqs_brut_zoom_s1}). La seconde ($s2$) ne conserve que les géométries des points d'entrée de l'arc dans les zones tampons et crée un segment entre ces deux points. Le réseau est donc approximé avec un ensemble de segments constituant les arcs (figure \ref{fig:craqs_brut_zoom_s2}).

Nous regroupons les principales caractéristiques des réseaux étudiés dans le tableau \ref{tab:craqs_reseaux} (les longueurs sont à considérer de manière relative car elles sont exprimées en unités de la carte, fixées arbitrairement lors de la définition de la projection du réseau).

\begin{figure}
    \centering
    { \small
    \begin{tabular}{|c|r|r|r|r|}
        \hline
        réseau & $L_{tot}$   & $N_{voies}$ & $\overline{N_{arcs}(voie)}$  & $\overline{L_{voie}}$ \\
        \hline
        brut & 109 309 & 1 667 & 2,75 & 65,6 \\
        \hline
        $s1$ & 107 831 & 1 673 & 2,74 & 64,5 \\
        \hline
        $s2$ & 105 404 & 1 684 & 2,72 & 62,6 \\
        \hline
    \end{tabular}
    }
    \caption{Statistiques sur les trois vectorisations de craquelures.}
    \label{tab:craqs_reseaux}
\end{figure}

Plus la géométrie du réseau est simplifiée, plus le linéaire total est réduit et plus le nombre de voies créées augmente. Les angles sont moins abrupts, les angles de déviation entre arcs respectent donc plus souvent l'angle seuil. Cependant, le nombre d'arcs moyen par voie reste approximativement constant. Cette moyenne est vérifiée en regardant dans le détail : sur les trois réseaux, le nombre de voies de plus de 10 arcs varie entre 53 et 57 et le nombre de voies de deux arcs ou moins varie de 1141 à 1187. Il y a donc plus de longues voies dans $s2$ mais également plus de voies réunissant très peu d'arcs. L'augmentation du nombre de voies se fait donc de manière homogène.

\begin{figure}[h]
    \centering
    \includegraphics[width=0.8\textwidth]{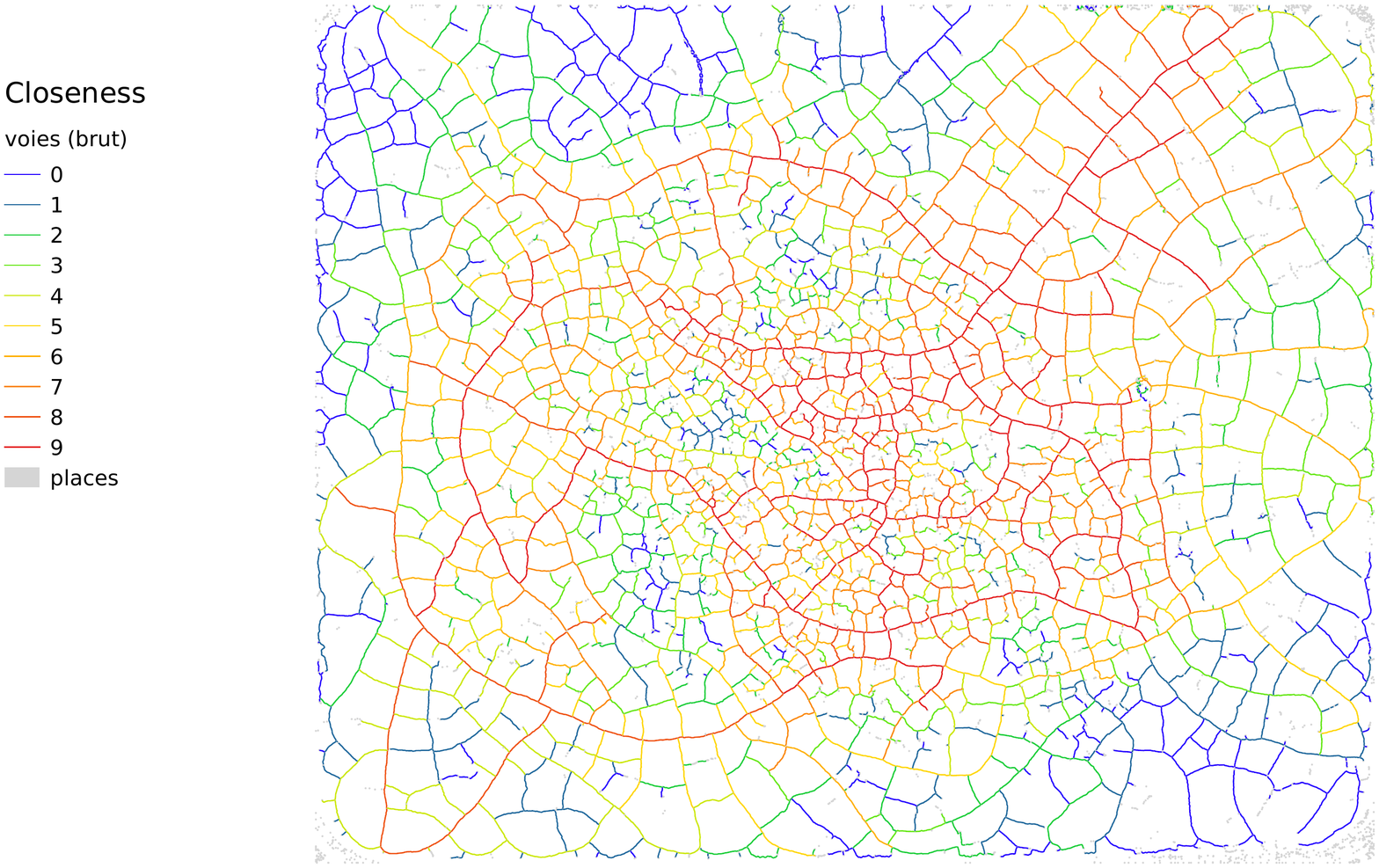}
    \caption{Calcul de la closeness sur le réseau brut vectorisé.}
    \label{fig:craqs_closeness_brut}
\end{figure}

\begin{figure}[h]
    \centering
    \includegraphics[width=0.8\textwidth]{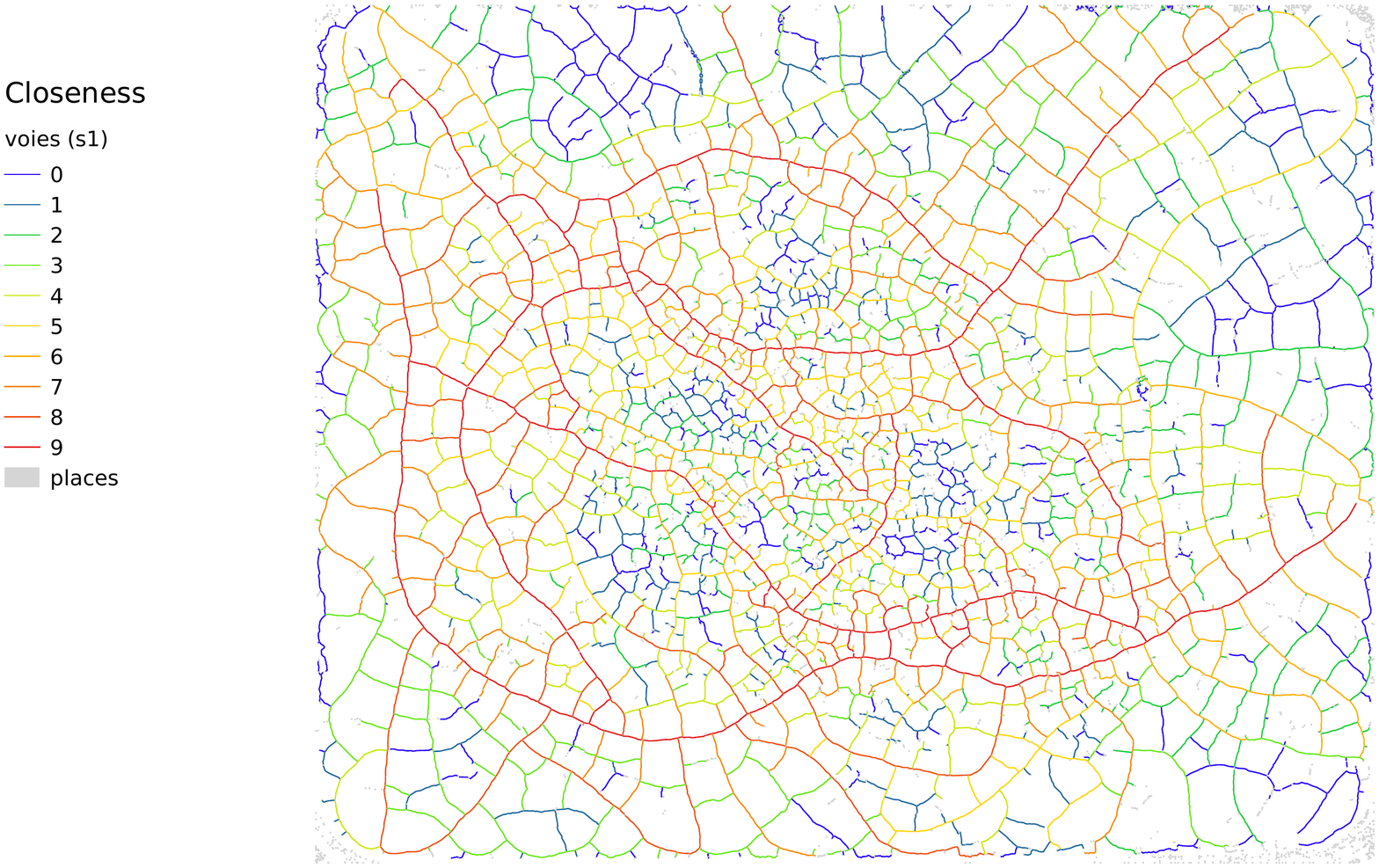}
    \caption{Calcul de la closeness sur le réseau transformé selon la première généralisation.}
    \label{fig:craqs_closeness_s1}
\end{figure}

\begin{figure}[h]
    \centering
    \includegraphics[width=0.8\textwidth]{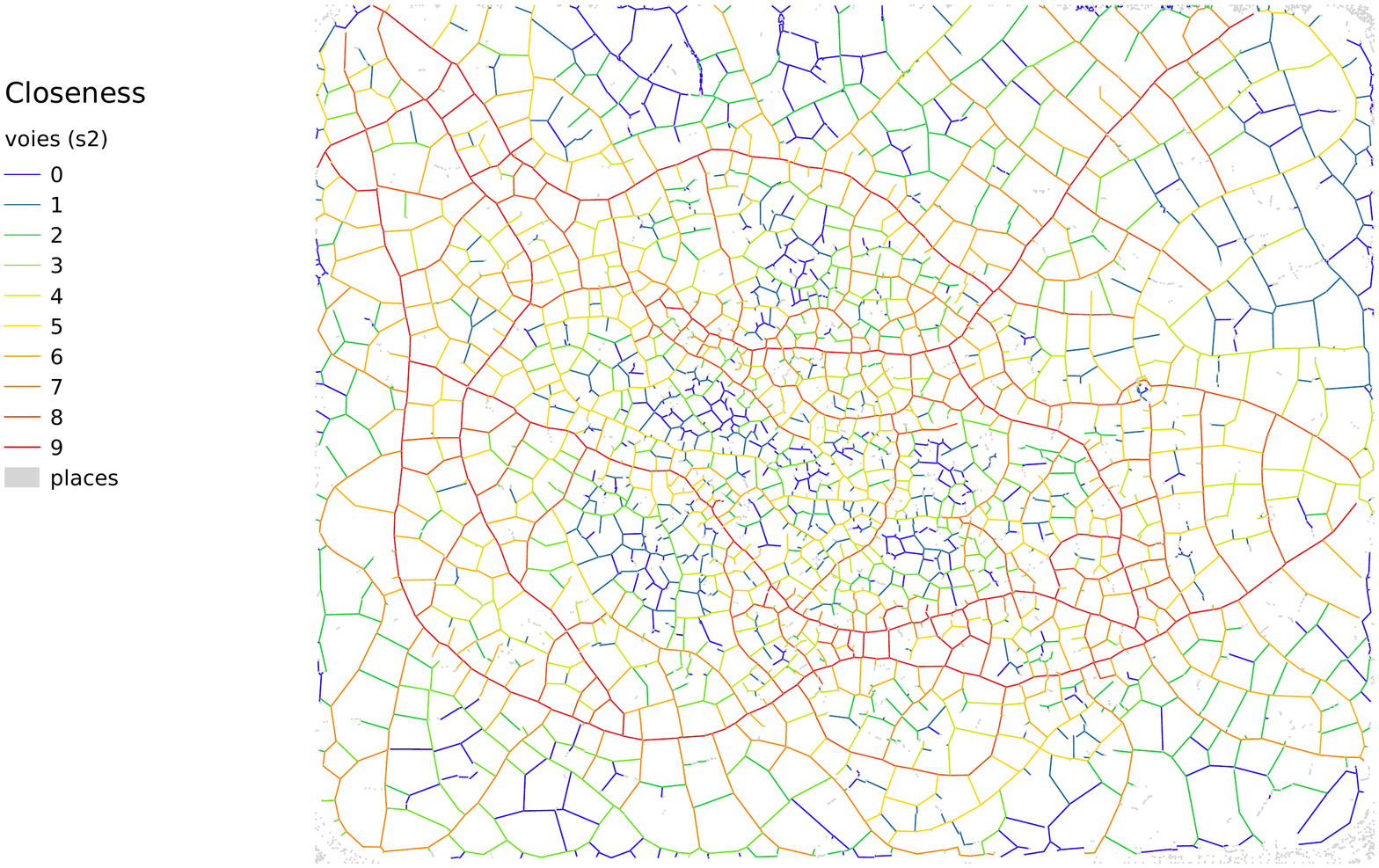}
    \caption{Calcul de la closeness sur le réseau transformé selon la deuxième généralisation.}
    \label{fig:craqs_closeness_s2}
\end{figure}

À partir de ces trois réseaux nous calculons la \textit{closeness} de leurs voies (figures \ref{fig:craqs_closeness_brut}, \ref{fig:craqs_closeness_s1} et \ref{fig:craqs_closeness_s2}). Nous observons sur ces cartes que plus le filaire est généralisé, plus l'effet de bord s'estompe : les éléments centraux au sens de la closeness sont de moins en moins concentrés au centre du graphe. En effet, plus le graphe est généralisé, plus les voies créées ont des géométries qui traversent l'ensemble de l'échantillon au lieu de se recourber dans une partie de celui-ci (effet dû aux géométries lissées à l'entrée des zones tampons). Nous appuyons donc ici l'importance d'un objet multi-échelle, continu et aligné pour permettre une analyse stable d'un graphe spatialisé. La stabilité réside donc plus dans la propriété \textit{traversante} de la voie que dans sa longueur ou son nombre d'arcs.

La granularité de numérisation du réseau n'impacte pas de manière notable ses valeurs statistiques (nombre de voies, nombre d'arcs par voie, longueur moyenne d'une voie...). Elle réduit la longueur totale du réseau (puisqu'il est moins détaillé), les autres paramètres s'ajustent proportionnellement à cette réduction. Il serait intéressant d'élargir l'analyse à partir de données d'autres graphes spatiaux de différents niveaux de généralisation.

En revanche, l'association des arcs à chaque intersection pour former les voies a une grande importance dans les résultats donnés par les indicateurs. Les nombreux points annexes présents sur un arc bruité peuvent fausser cette association par des changements de directions abrupts intervenant juste avant l'entrée dans une zone tampon. De ce fait, lorsque nous considérons des réseaux naturels (veinures d'une feuille, craquelures d'une plaque d'argile), numérisés automatiquement à partir de photos contrastées, nous avons recours à une généralisation optimale ($s2$) afin de pouvoir créer des objets couvrant la plus grande partie possible de l'échantillon.

Pour les réseaux viaires, les sources de données en assurent une meilleure qualité. De plus, sur les réseaux physiques, l'optimisation de la construction entre deux intersections (notamment pour réduire les coûts) lisse d'elle-même la géométrie. Seuls quelques projets urbains ponctuels (évoqués au début de ce chapitre) peuvent couper la continuité des voies. Ces discontinuités peuvent être effacées par la création des zones tampons, sans recourir à une généralisation de la géométrie.
\clearpage{\pagestyle{empty}\cleardoublepage}
\chapter{Comparaison quantitative}
\minitoc
\markright{Comparaison quantitative}

Dans le chapitre précédent, nous avons exploré la sensibilité de notre modèle aux données utilisées, selon leur qualité, leur choix ou leur découpage. Nous avons démontré l'apport de la voie à la stabilité de l'analyse. En effet, la  robustesse de la lecture que proposent les indicateurs développés sur l'hypergraphe est assurée par les géométries des voies, qui traversent le graphe à de multiples échelles. Nous développons notre analyse dans ce chapitre en comparant différents types de réseaux spatiaux. Pour ceux dont la vectorisation est faite automatiquement, nous veillons à faire une généralisation afin d'assurer le caractère \textit{traversant} des voies construites.

\FloatBarrier
\section{Définition d'un panel de recherche}

Dans ce chapitre, nous comparons différents réseaux spatiaux pour déterminer les propriétés communes et distinguer celles qui leur seraient spécifiques. Fidèles à notre champ d'application principal, nous considérons d'abord des réseaux viaires et nous introduisons quelques autres types de réseaux spatiaux pour tenter de saisir ce qui fait la spécificité des villes. Nous étudions ainsi des réseaux géographiques (hydrographiques, ferrés), biologiques (veinures de feuilles ou de coraux) et de craquelures dans de l'argile. Nous introduisons également des réseaux générés artificiellement.

\FloatBarrier
\subsection{Graphes viaires}

Nous reprenons ici les réseaux qui nous ont permis de définir une grammaire de lecture de la spatialité dans le chapitre 4 de la première partie. Nous retrouvons ainsi : le graphe viaire de la commune de Paris, à la structure dense et complexe, rectifiée par des projets urbains majeurs ; le graphe découpé largement autour d'Avignon, qui réunit plusieurs types de structures (centre-ville, banlieue, connexions entre différents villages, voies rapides et autoroutes) ; l'île de Manhattan, qui, par sa géométrie particulière, permet de confronter la structure d'un graphe planifié à celle des graphes organiques ; et enfin le réseau viaire de Barcelone, qui allie géométries régulières et organiques (réseau étendu par le plan Cerdà).

À ces quatre graphes nous ajoutons les réseaux de seize autres villes, réparties sur plusieurs continents. 

En France, nous joignons aux réseaux de Paris et d'Avignon, ceux de Bordeaux (grande ville construite sur le méandre d'un fleuve, la Garonne) ; Brive (ville moyenne provinciale à la topographie contrainte par le relief qui l'entoure) ; Cergy-Pontoise (ville nouvelle construite aux abords de la capitale) ; et Villers-sur-Mer (petite ville de bord de mer).

En Europe, nous ajoutons aux villes françaises et à Barcelone deux capitales : Londres et Bruxelles, à la croissance organique établie sur plusieurs siècles. Nous étudierons également Rotterdam, dont le réseau viaire a subi d'importants changements suite à une inondation majeure au XVIIIème siècle et au bombardement de son centre ville au début de la seconde guerre mondiale.

Nous complétons notre panel de dix villes européennes par dix autres choisies sur trois continents différents. En Amérique du Nord, à Manhattan nous ajoutons San-Francisco (qui appose sa régularité à la topographie sans s'y adapter) et Santa-Fe (une des plus anciennes villes établie par des hommes de culture européenne sur le nouveau continent). En Amérique du Sud, nous nous intéressons à Manaus (grande ville brésilienne située au cœur de la forêt amazonienne) et Cuzco (située au Sud du Pérou). Enfin, nous étudions trois villes d'Asie : Téhéran (en Iran), Varanasi (en Inde) et Kyoto (au Japon) ; et deux villes d'Afrique : Casablanca (au Maroc) et Nairobi (capitale du Kenya).

Le choix de ces villes a été fait sur différents critères. Leur pertinence géométrique en premier lieu : nous avons choisi des villes aux structures marquées, particulièrement planifiées ou au contraire, organiques voir arborescentes (Nairobi). Nous avons également tenu compte des contraintes géographiques : relief, fleuve ou mer. Nous avons choisi des villes avec des réseaux de tailles variées : le plus petit regroupe un linéaire d'environ 66 km (Villers-sur-Mer), le plus grand avoisine la dizaine de milliers de kilomètres (Téhéran). Nous avons également  tenu à regrouper des réseaux aux histoires de constructions différentes. Nous avons donc des graphes construits en un seul projet (Manhattan, San Francisco) qui sont comparés à des graphes millénaires (Londres, Casablanca). Entre ceux-ci, nous avons des réseaux qui ont été repensés en partie lors de grands projets urbains (Paris, Barcelone).

Les manières de qualifier les vingt graphes viaires choisis sont donc multiples. Leurs propriétés se croisent et s'opposent pour créer un panel de recherche pensé pour être aussi riche que possible. Lors de l'élaboration du panel, la connaissance des villes que nous pouvions avoir a également été un facteur décisif. Ainsi, nous avons établi des liens avec différentes municipalités, cabinets d'urbanismes ou chercheurs spécialistes d'un territoire. Ce qui nous a permis de mieux comprendre les données sur lesquelles nous travaillons. Nous reviendrons, dans un troisième temps, sur ces différents liens et leur importance. En effet ce sont les relations que l'on a pu établir avec les sciences thématiques et les professionnels de l'urbanisme qui nous ont permis de mieux comprendre les résultats obtenus avec nos indicateurs et ainsi de déterminer leurs pertinences sur l'objet urbain.

\begin{figure}[h]
    \centering
    \begin{subfigure}[t]{.45\linewidth}
        \includegraphics[width=\textwidth]{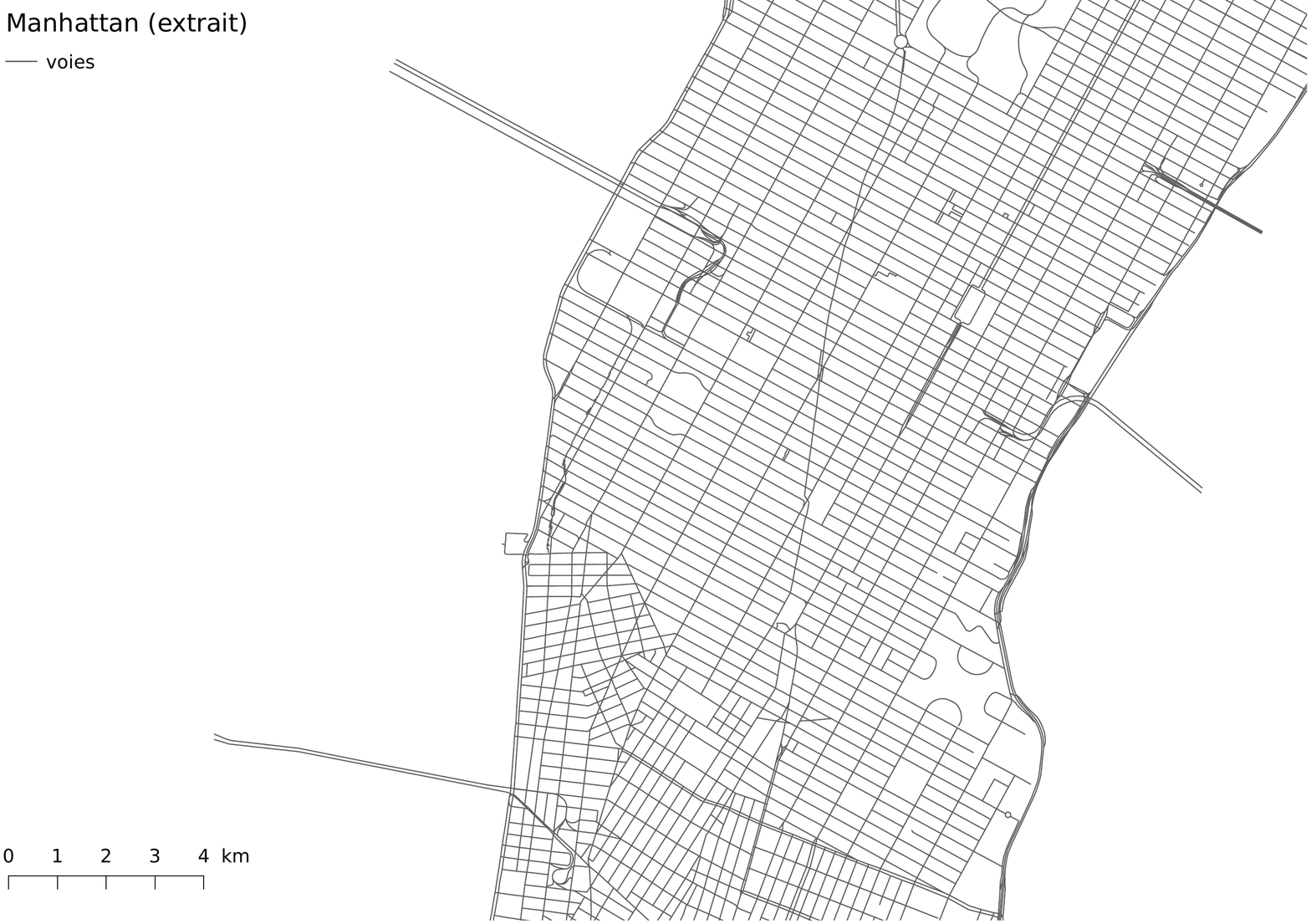}
        \caption{Manhattan.}
        \label{fig:brut_zoom_man}
    \end{subfigure}
    ~
    \begin{subfigure}[t]{.45\linewidth}
        \includegraphics[width=\textwidth]{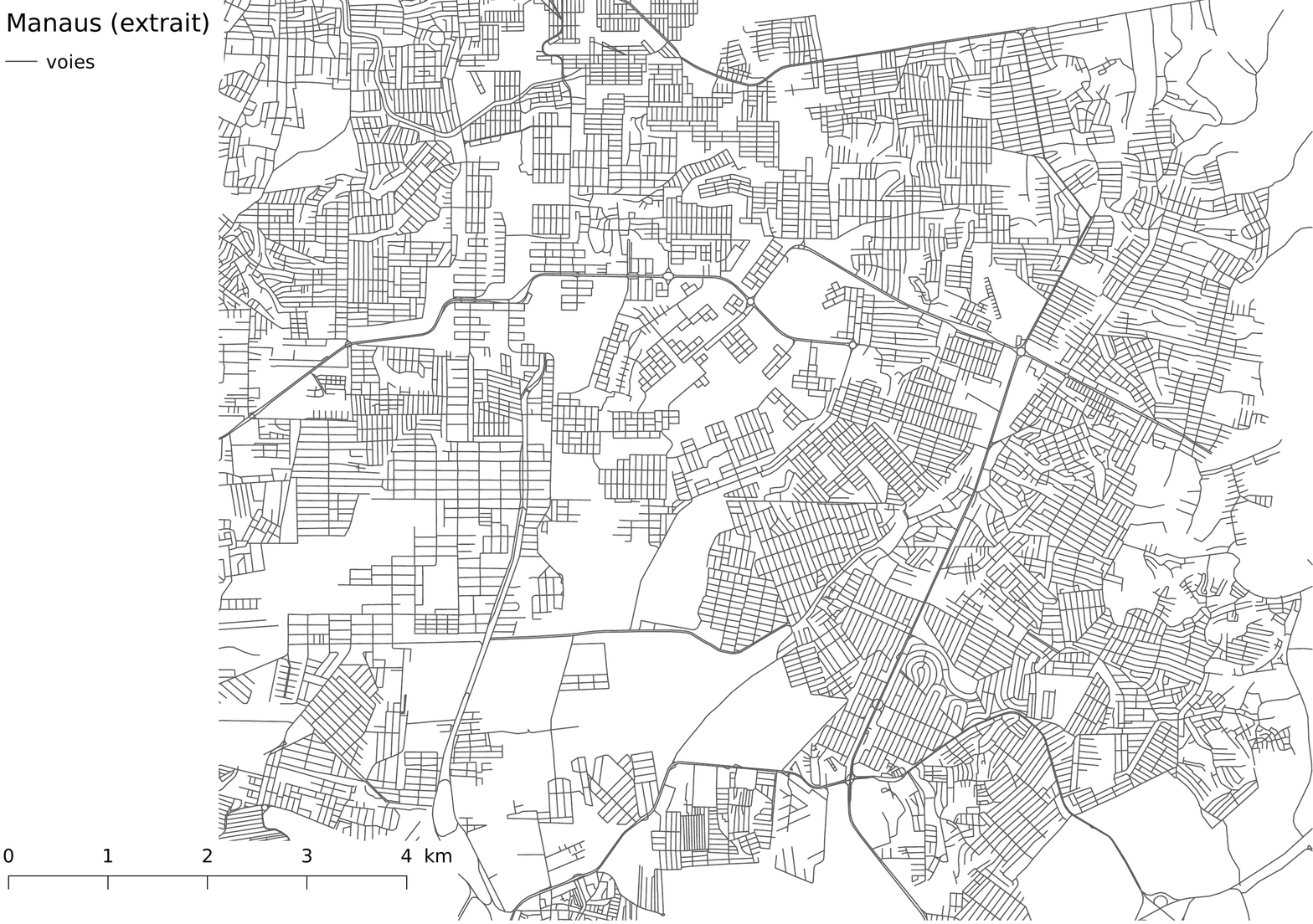}
        \caption{Manaus.}
        \label{fig:brut_zoom_manaus}
    \end{subfigure}
   
    \centering
    \begin{subfigure}[t]{.45\linewidth}
        \includegraphics[width=\textwidth]{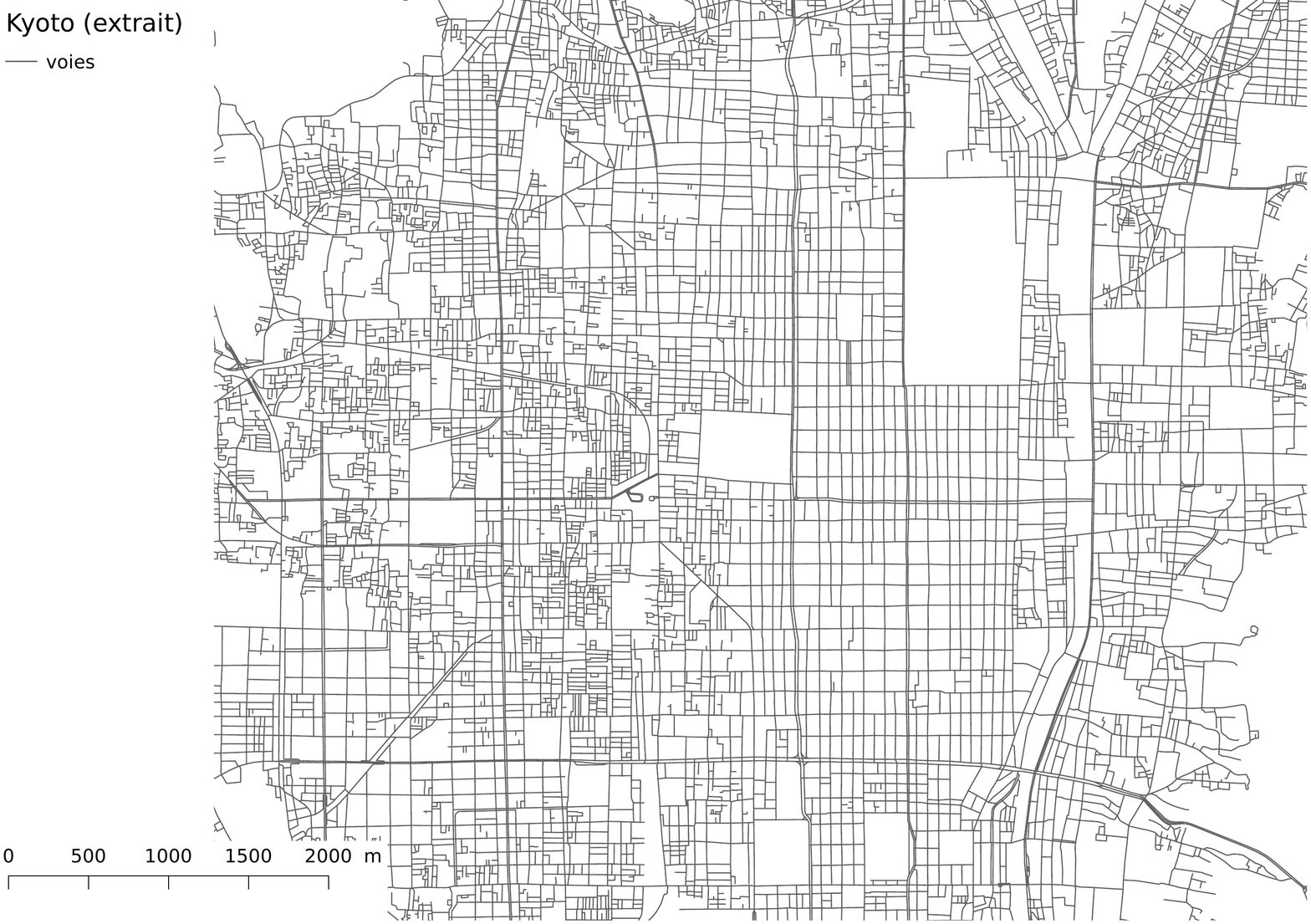}
        \caption{Kyoto.}
        \label{fig:brut_zoom_kyo}
    \end{subfigure}
    ~   
    \begin{subfigure}[t]{.45\linewidth}
        \includegraphics[width=\textwidth]{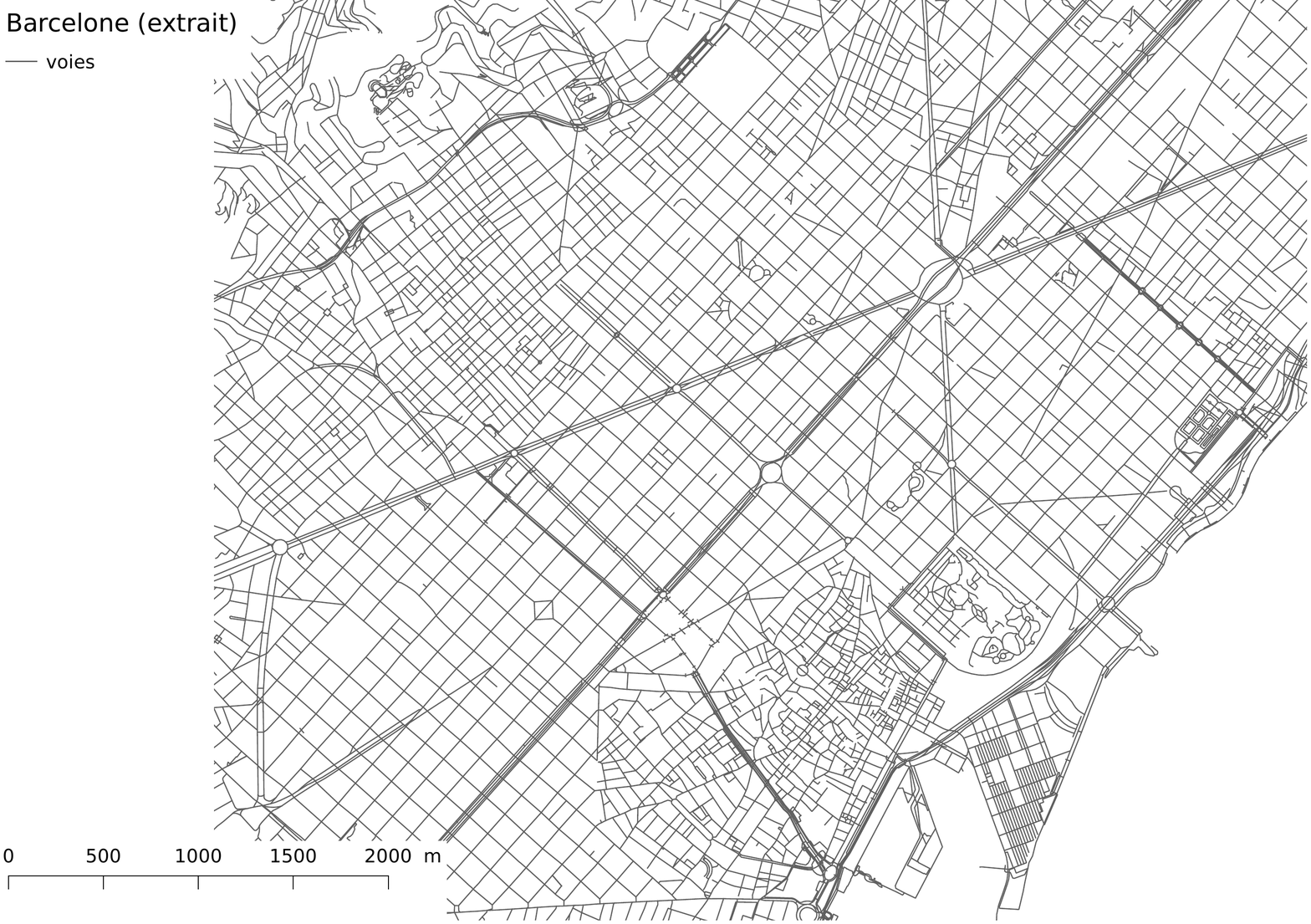}
        \caption{Barcelone.}
        \label{fig:brut_zoom_bar}
    \end{subfigure}
    
    \begin{subfigure}[t]{.45\linewidth}
        \includegraphics[width=\textwidth]{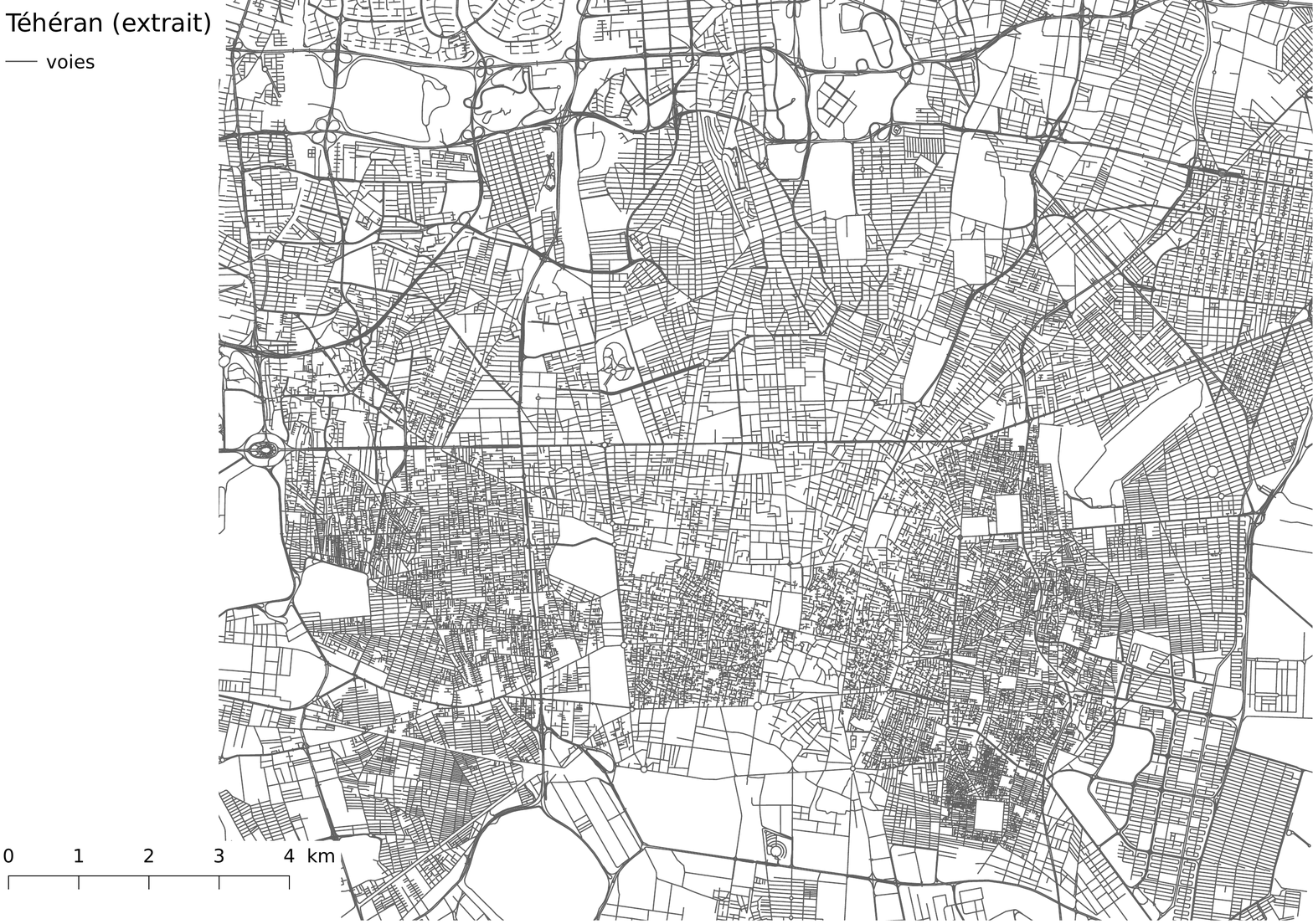}
        \caption{Téhéran.}
        \label{fig:brut_zoom_teh}
    \end{subfigure}
    ~
    \begin{subfigure}[t]{.45\linewidth}
        \includegraphics[width=\textwidth]{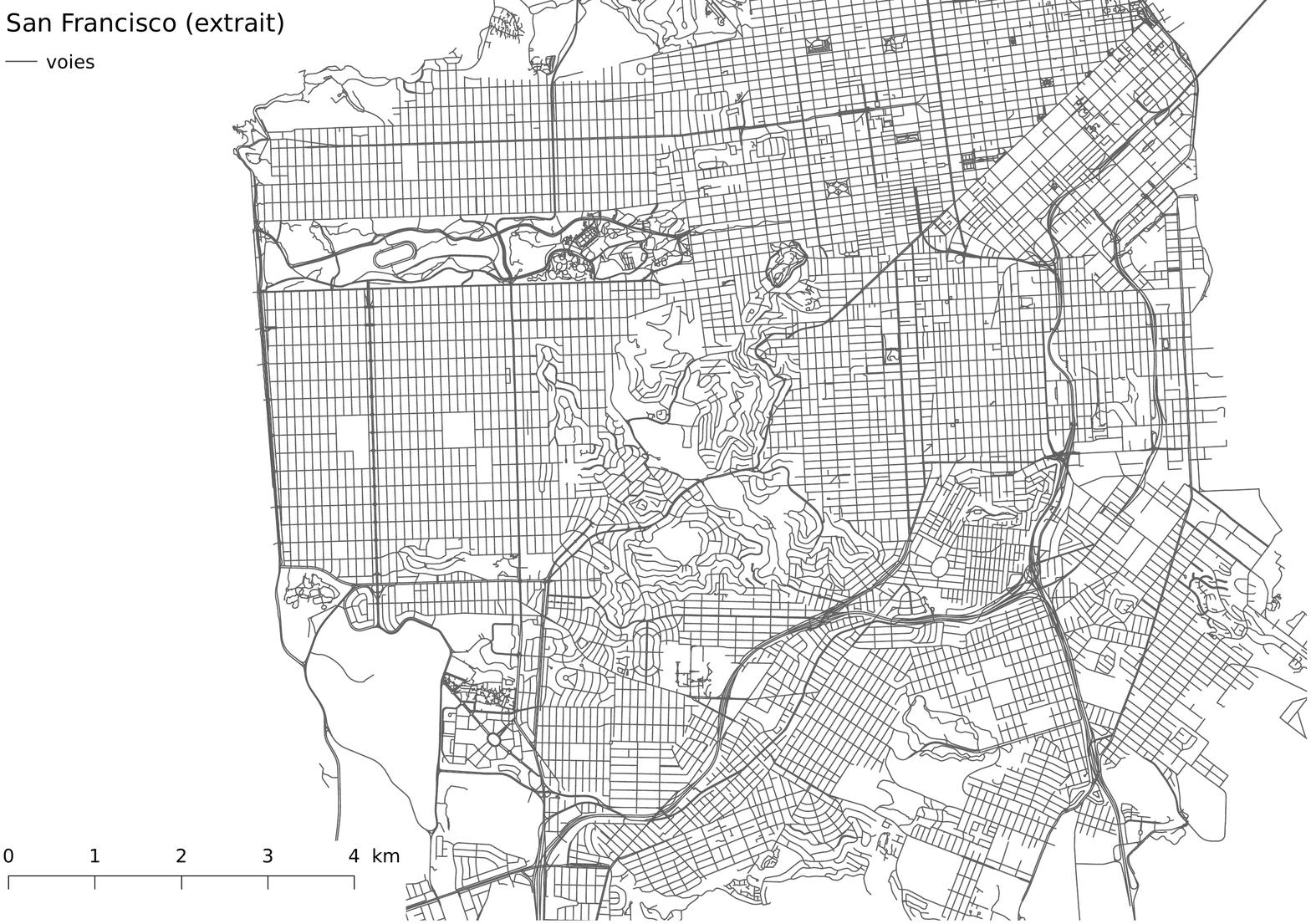}
        \caption{San-Francisco.}
        \label{fig:brut_zoom_san}
    \end{subfigure}
    
    \caption{Quadrillages. Graphes viaires extraits des villes du panel de recherche.}
    \label{fig:brut_zoom_reg}
\end{figure}

Entièrement ou par quartiers, nous retrouvons dans plusieurs villes des réseaux quadrillés, issus d'une construction planifiée (figures \ref{fig:brut_zoom_man}, \ref{fig:brut_zoom_manaus}, \ref{fig:brut_zoom_kyo}). Cependant, nous observons souvent ce type de quartiers juxtaposé à d'autres souvent plus anciens à la géométrie beaucoup moins régulière. C'est notamment le cas à Barcelone (figure \ref{fig:brut_zoom_bar}) où la vieille ville trace une délimitation dans le quadrillage régulier dessiné par Cerdà. Il en est de même à Téhéran (figure \ref{fig:brut_zoom_teh}) où les vieux quartiers abritant bazars et ruelles où se recroquevillent les habitations jouxtent les grandes avenues autoroutières et les géométries régulières de constructions plus récentes ; mais également à San Francisco (figure \ref{fig:brut_zoom_san}) où le relief perturbe la grille que l'on tente de lui imposer.

\begin{figure}[h]
    \centering
    \begin{subfigure}[t]{.45\linewidth}
        \includegraphics[width=\textwidth]{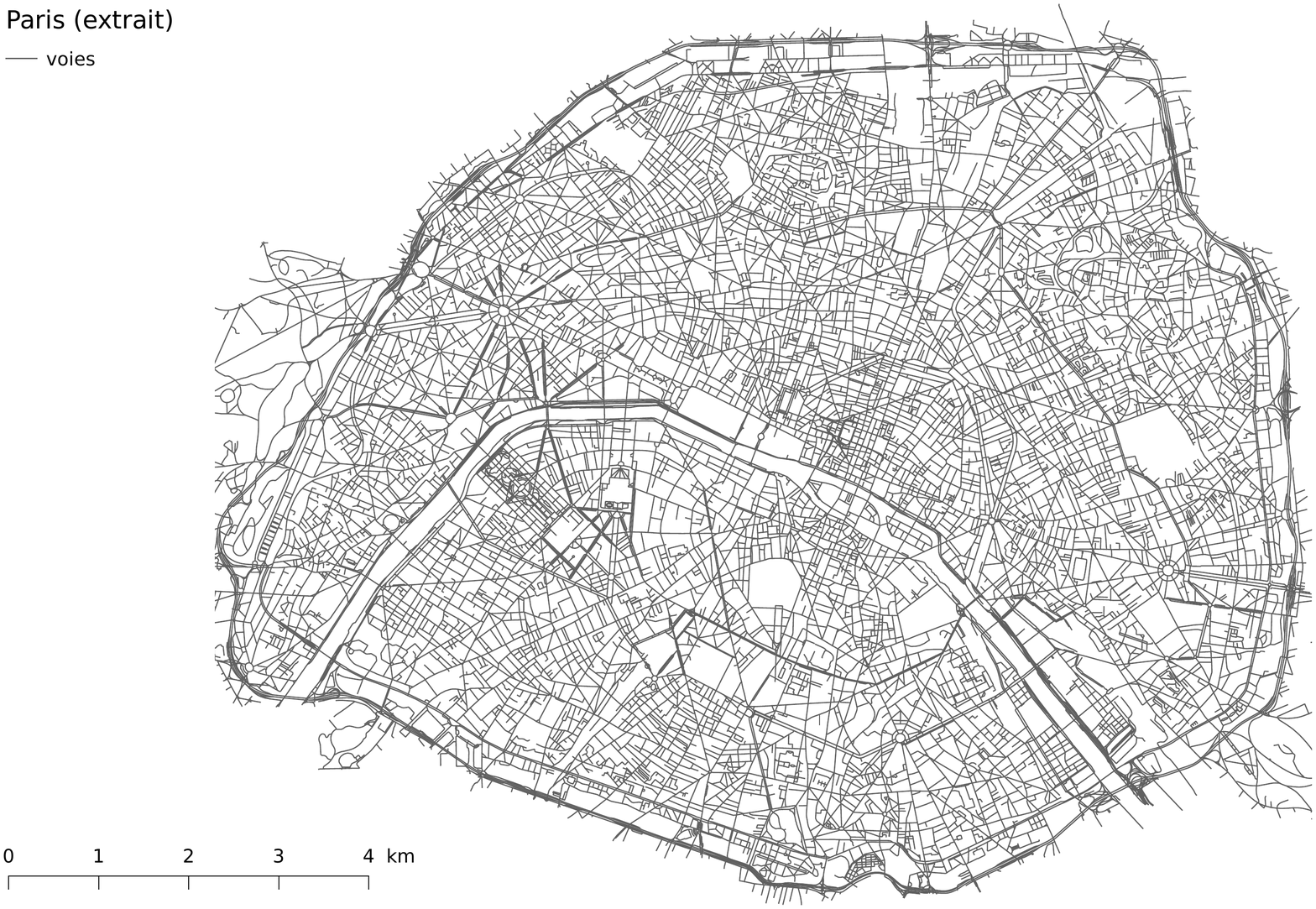}
        \caption{Paris.}
        \label{fig:brut_zoom_par}
    \end{subfigure}
    ~
    \begin{subfigure}[t]{.45\linewidth}
        \includegraphics[width=\textwidth]{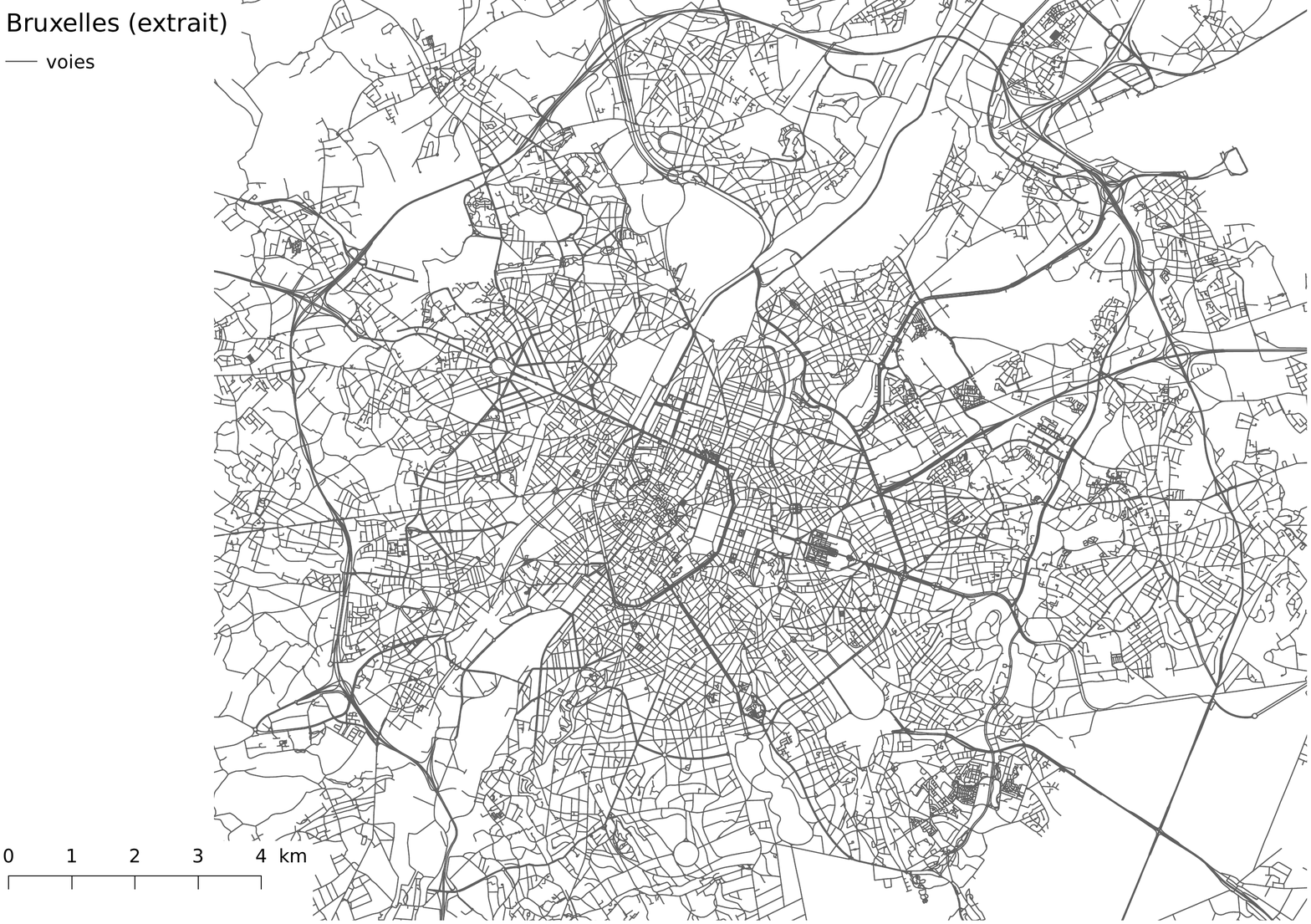}
        \caption{Bruxelles.}
        \label{fig:brut_zoom_bru}
    \end{subfigure}
    
    \caption{Formes circulaires. Graphes viaires extraits des villes du panel de recherche.}
    \label{fig:brut_zoom_circulaire}
\end{figure}

Autre que la grille, la forme circulaire se retrouve dans quelques uns des graphes choisis. Qu'elle soit issue d'une forme ancienne de protection (propos que nous développerons en troisième partie) ou simplement créée dans une stratégie de contournement, elle structure l'espace qu'elle traverse. Ainsi, le réseau viaire de Paris se développe à l'intérieur de son périphérique, selon des enceintes successives \citep{huard2013atlas} (figure \ref{fig:brut_zoom_par}) et la capitale belge se retrouve au centre d'un système concentrique (figure \ref{fig:brut_zoom_bru}).

\FloatBarrier

\begin{figure}[h]
    \centering
    \begin{subfigure}[t]{.45\linewidth}
        \includegraphics[width=\textwidth]{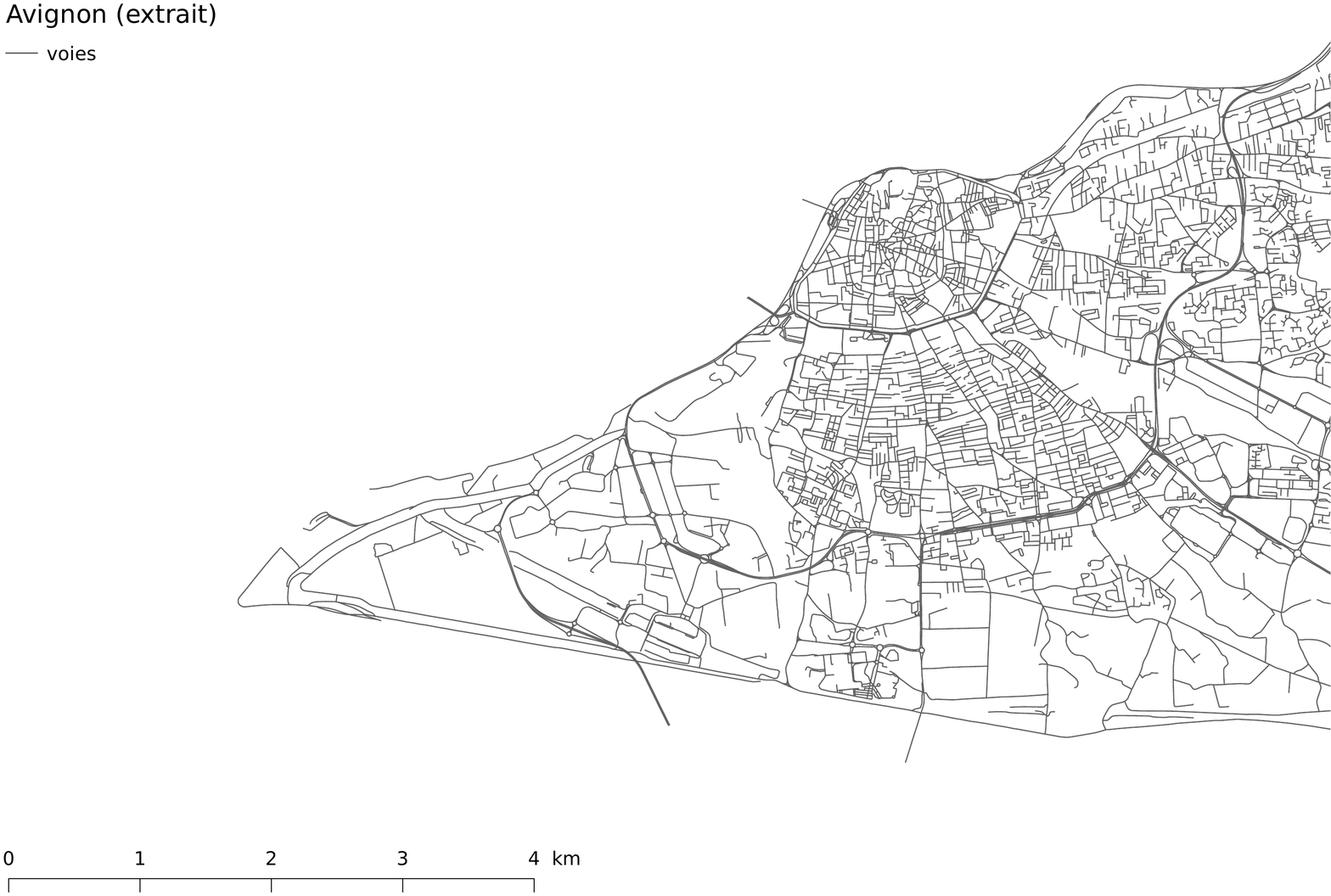}
        \caption{Avignon.}
        \label{fig:brut_zoom_av}
    \end{subfigure}
    ~
    \begin{subfigure}[t]{.45\linewidth}
        \includegraphics[width=\textwidth]{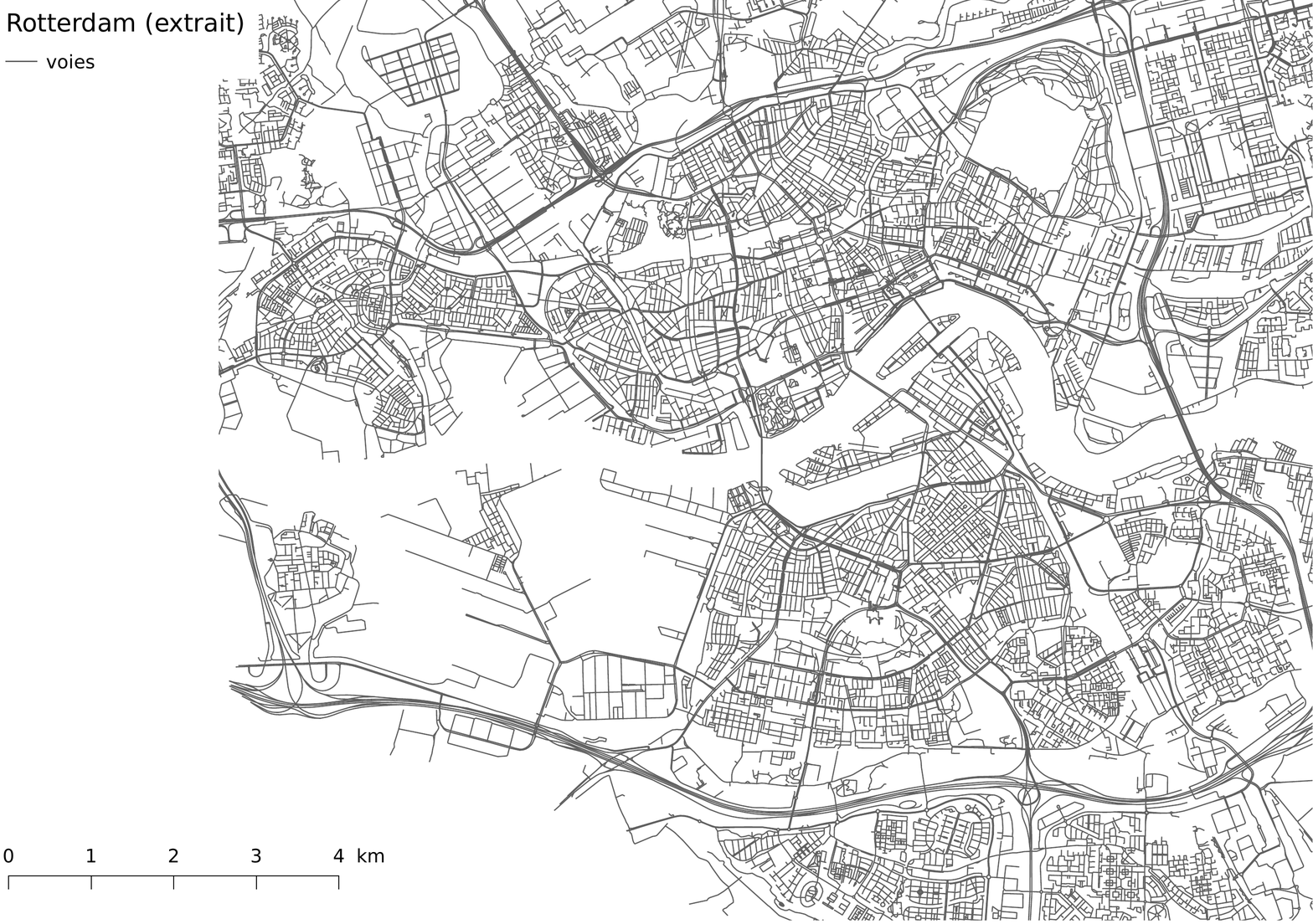}
        \caption{Rotterdam.}
        \label{fig:brut_zoom_rot}
    \end{subfigure}
    
    \begin{subfigure}[t]{.45\linewidth}
        \includegraphics[width=\textwidth]{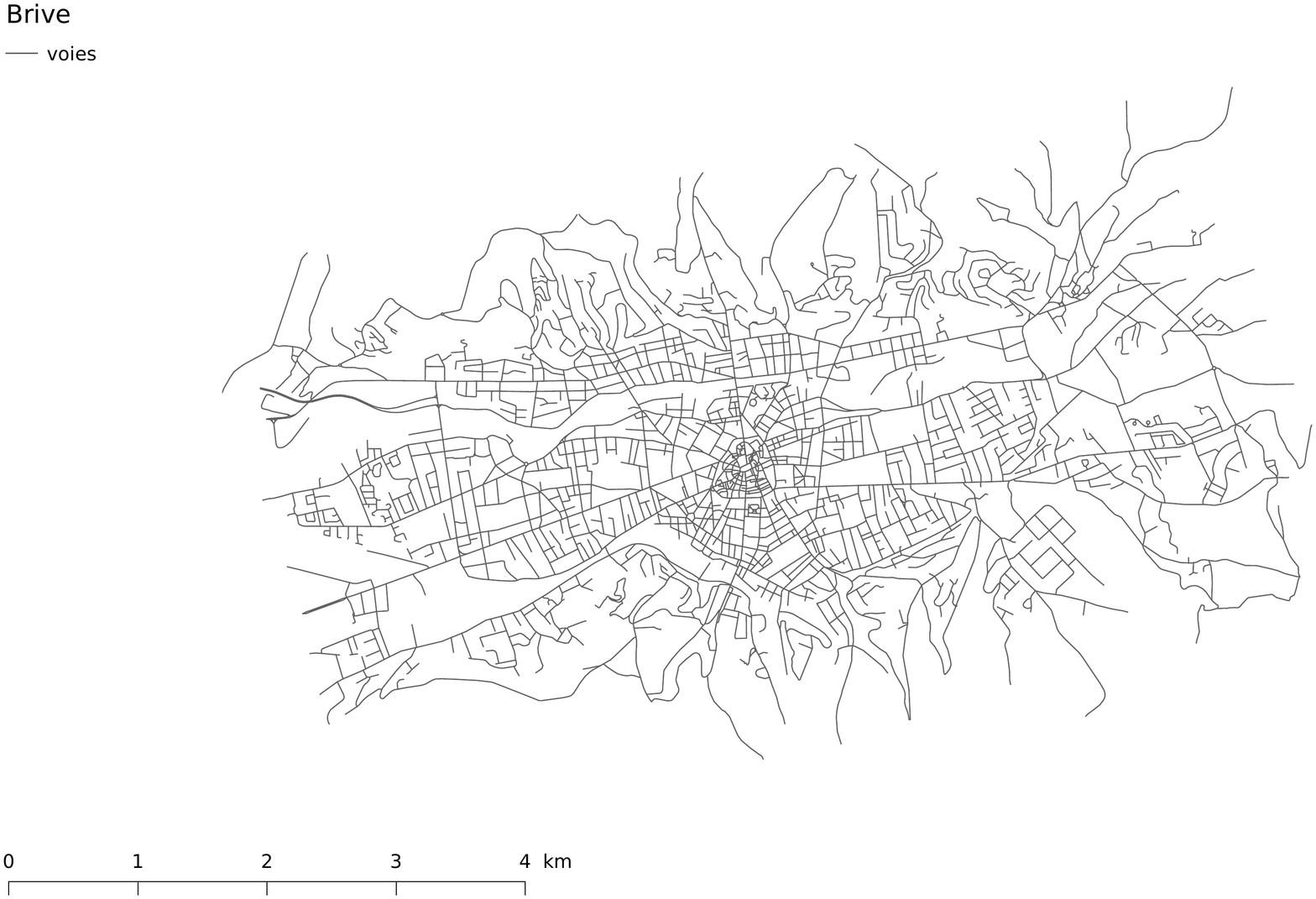}
        \caption{Brive-la-Gaillarde.}
        \label{fig:brut_zoom_bri}
    \end{subfigure}
    ~
    \begin{subfigure}[t]{.45\linewidth}
        \includegraphics[width=\textwidth]{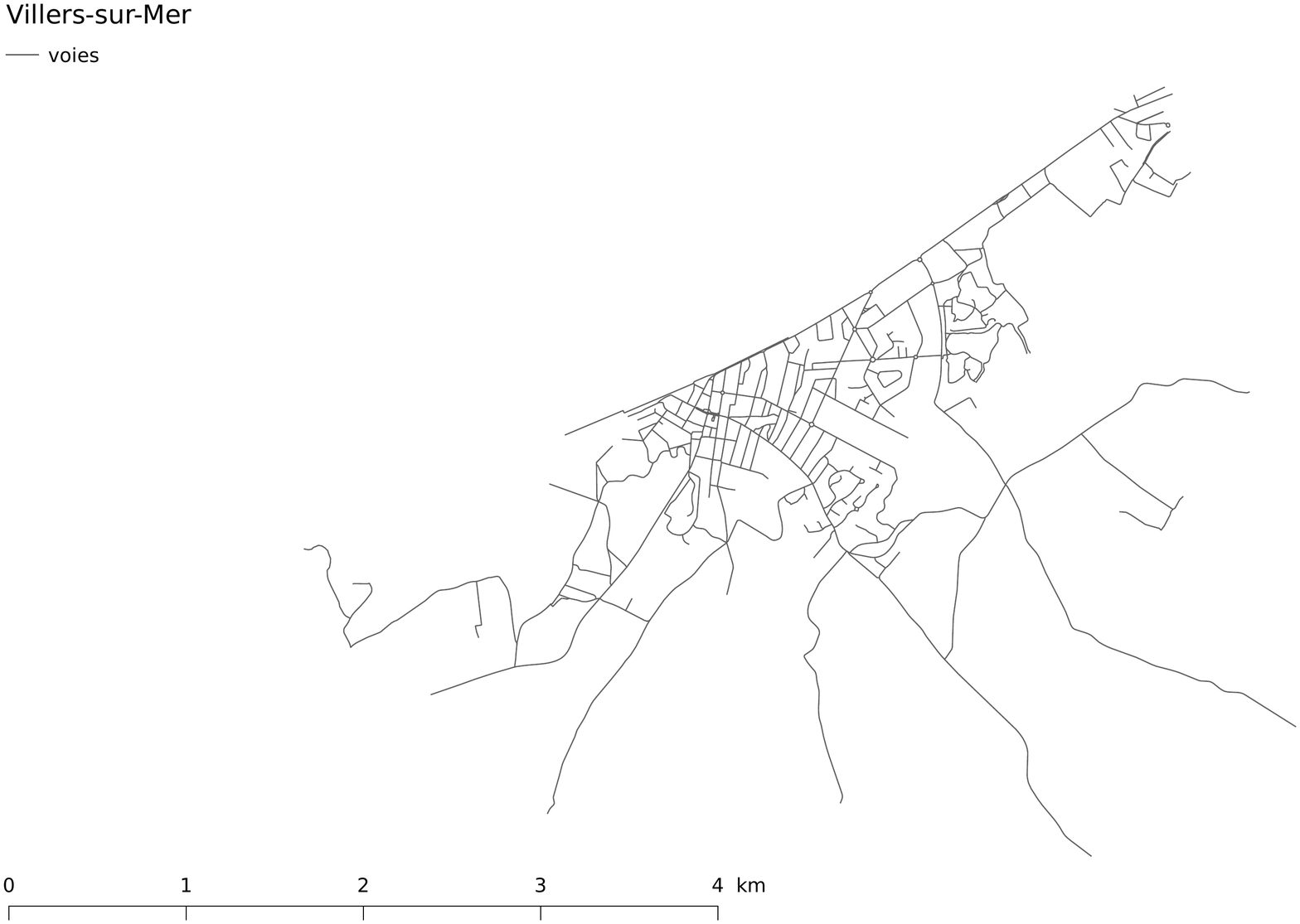}
        \caption{Villers-sur-Mer.}
        \label{fig:brut_zoom_vil}
    \end{subfigure}

    \caption{Formes imposées par la topographie. Graphes viaires extraits des villes du panel de recherche.}
    \label{fig:brut_zoom_topo}
\end{figure}

Parfois, nous observons dans les structures mêmes, les contraintes géographiques dans lesquelles elles se sont développées. Dans le cas des réseaux viaires, nous pouvons voir transparaître à travers certaines géométries le contexte topographique ou hydrographique dans lequel elles s'inscrivent. C'est particulièrement le cas à Avignon, construite entre le Rhône et la Durance. Les limites imposées par ces deux fleuves façonnent la géométrie du graphe viaire (figure \ref{fig:brut_zoom_av}). De même, à Rotterdam, où la Nouvelle Meuse sépare la ville en deux (figure \ref{fig:brut_zoom_rot}). Sur un territoire plus réduit, à Brive-la-Gaillarde, le relief contraint le développement de la ville à l'horizontale (figure \ref{fig:brut_zoom_bri}) ou à Villers-sur-Mer, la limite imposée par la Manche (figure \ref{fig:brut_zoom_vil}).

\begin{figure}[h]
    \centering
    \begin{subfigure}[t]{.45\linewidth}
        \includegraphics[width=\textwidth]{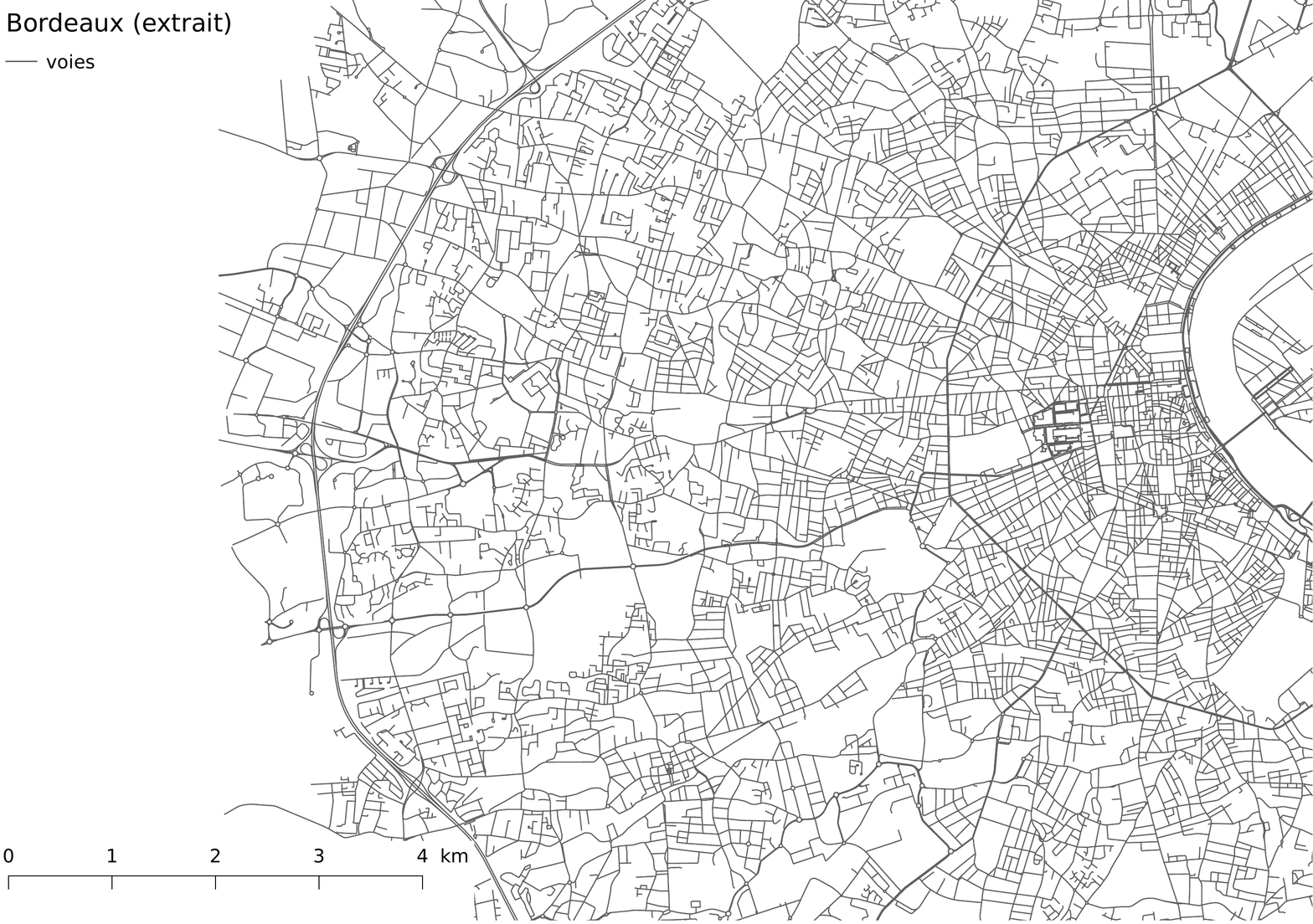}
        \caption{Bordeaux.}
        \label{fig:brut_zoom_bdx}
    \end{subfigure}
    ~
    \begin{subfigure}[t]{.45\linewidth}
        \includegraphics[width=\textwidth]{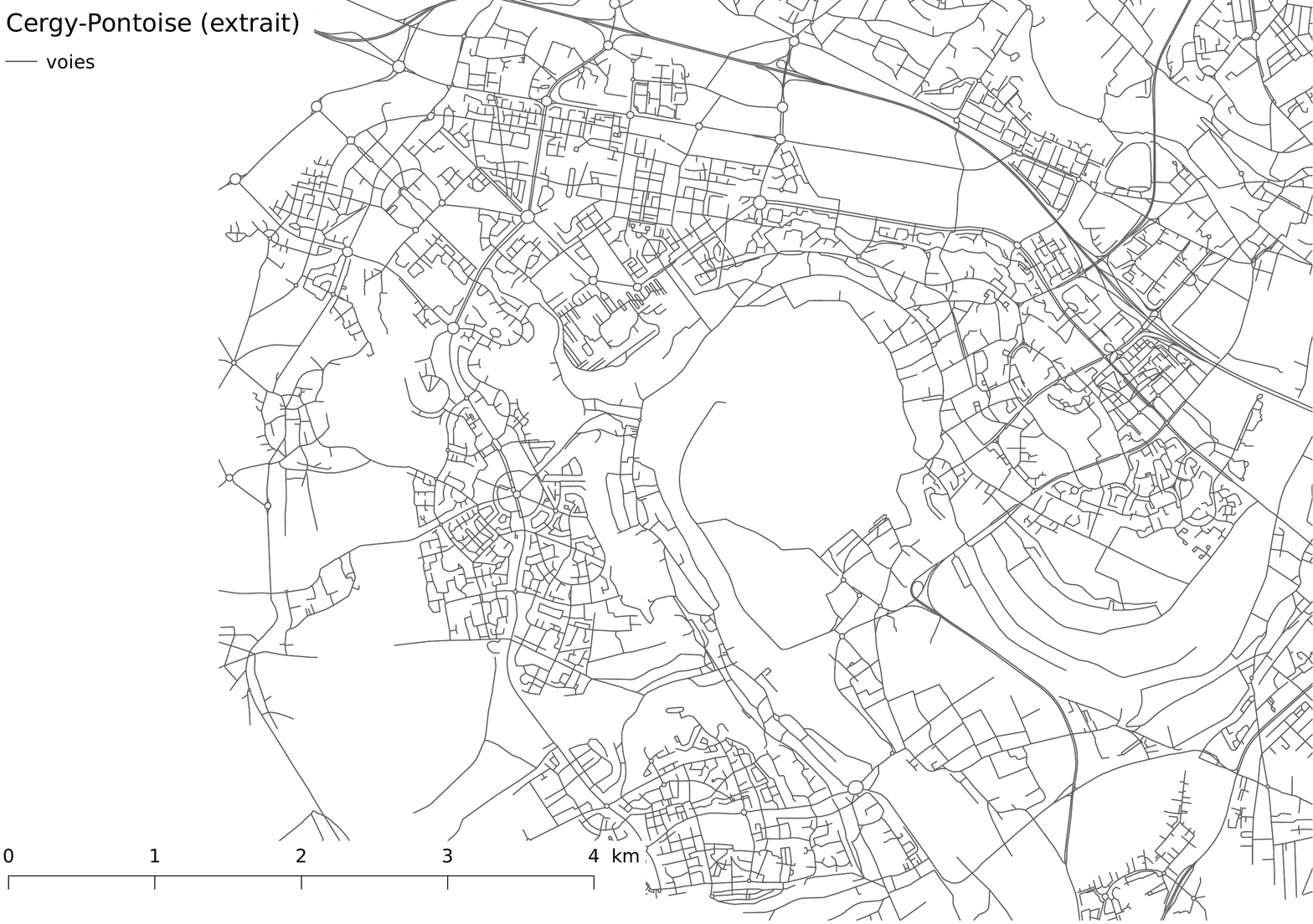}
        \caption{Cergy Pontoise.}
        \label{fig:brut_zoom_cer}
    \end{subfigure}    
    
    \caption{Formes imposées par la topographie. Graphes viaires extraits des villes du panel de recherche.}
    \label{fig:brut_zoom_topo2}
\end{figure}

Les exemples de formes particulières dues au contexte topographique sont multiples. Cependant, si parfois elles se répercutent sur le profil viaire, d'autres fois elles s'estompent sous celui-ci. Ainsi, à Bordeaux, la courbe de la Garonne impacte celle des boulevards principaux qui traversent ou contournent la ville (figure \ref{fig:brut_zoom_bdx}) ; alors qu'à Cergy-Pontoise, la boucle de l'Oise (affluent de la Seine) disparaît presque sous les nombreux aménagement dont elle fait l'objet (figure \ref{fig:brut_zoom_cer}).

\FloatBarrier

\begin{figure}[h]
    \centering
    \begin{subfigure}[t]{.45\linewidth}
        \includegraphics[width=\textwidth]{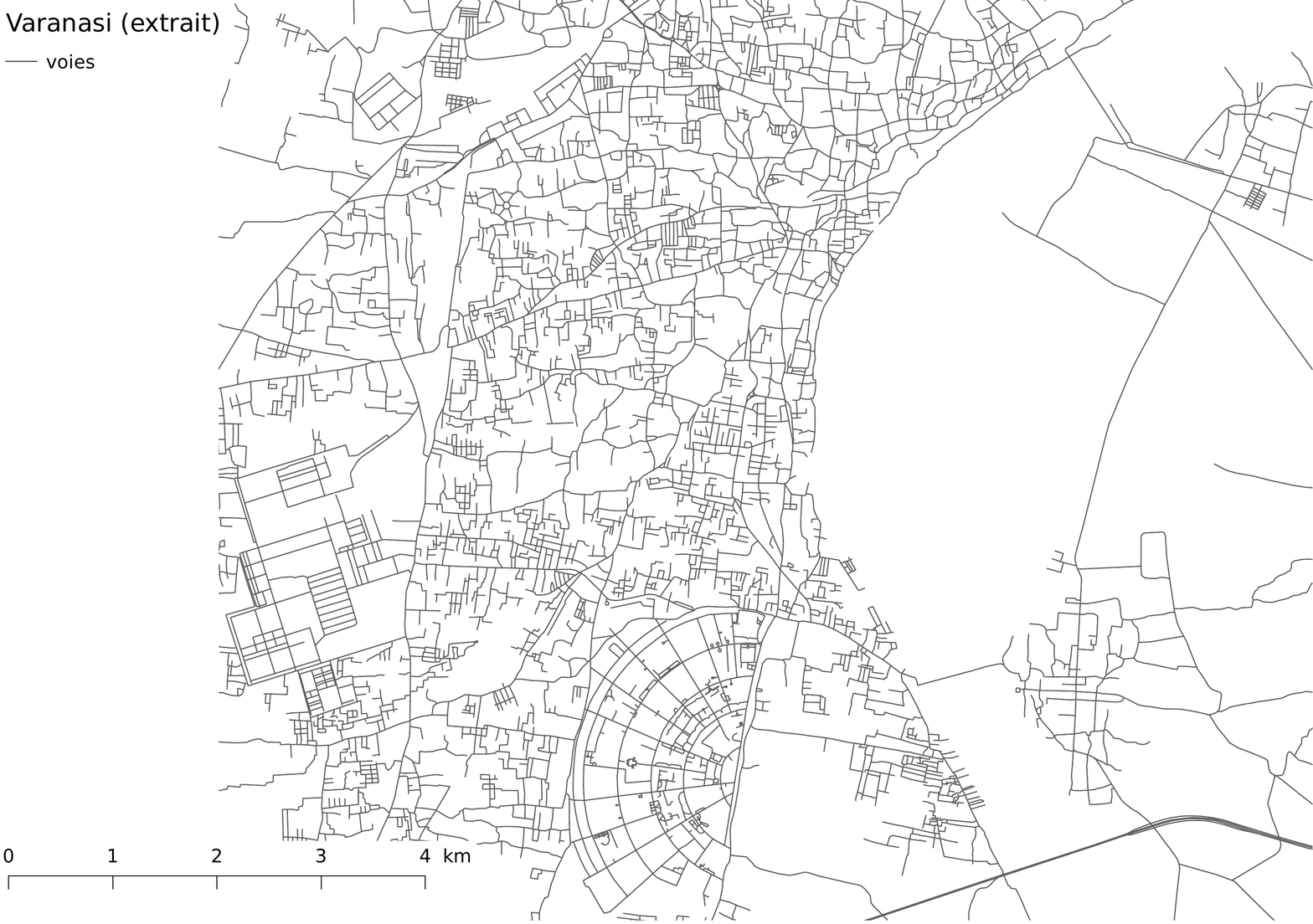}
        \caption{Varanasi.}
        \label{fig:brut_zoom_var}
    \end{subfigure}
    ~
    \begin{subfigure}[t]{.45\linewidth}
        \includegraphics[width=\textwidth]{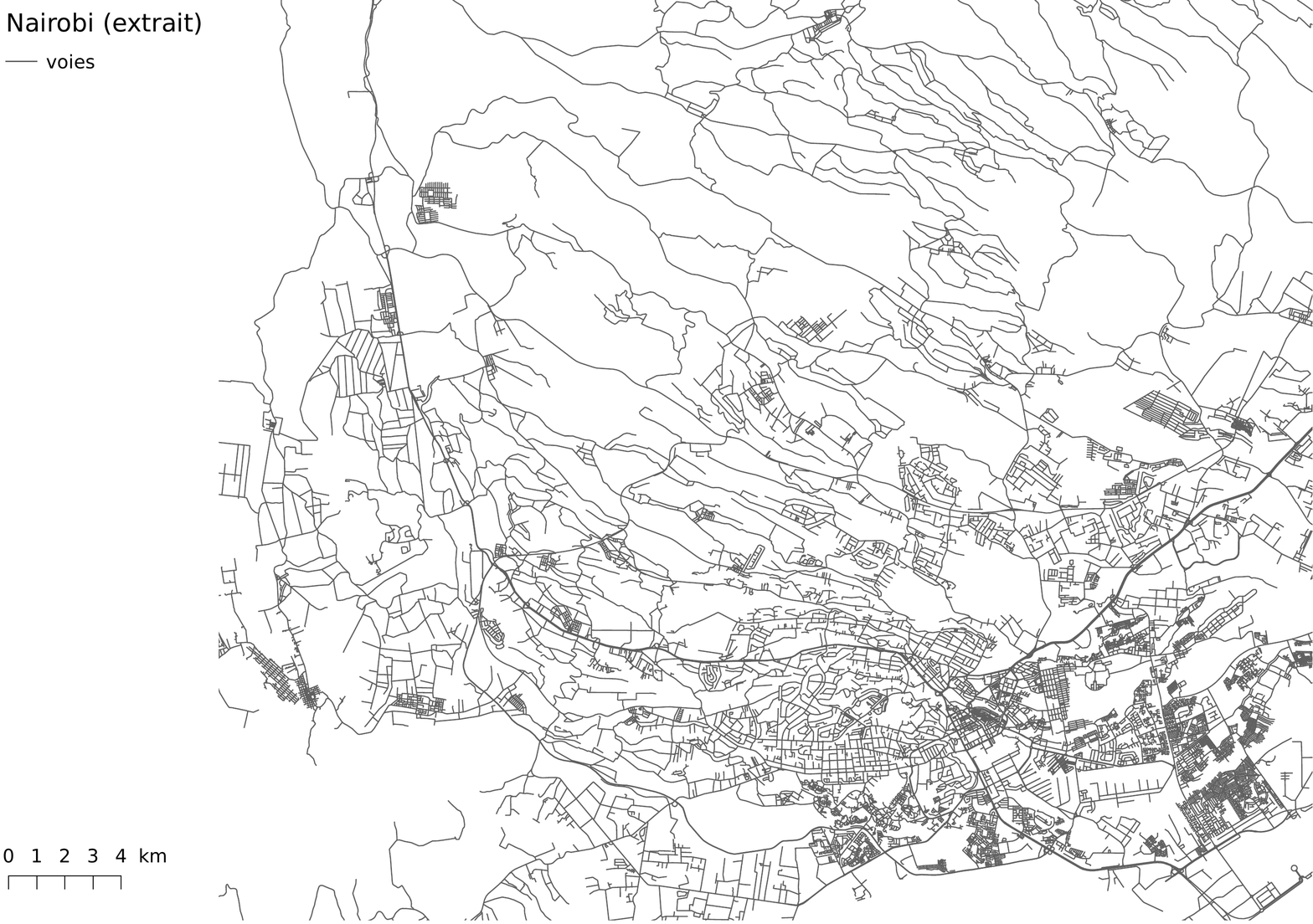}
        \caption{Nairobi.}
        \label{fig:brut_zoom_nai}
    \end{subfigure}

    \caption{Formes erratiques. Graphes viaires extraits des villes du panel de recherche.}
    \label{fig:brut_zoom_err}
\end{figure}

D'autres graphes viaires paraissent beaucoup plus erratiques. C'est notamment le cas en Inde, à Varanasi (figure \ref{fig:brut_zoom_var}) et au Kenya, à Nairobi (figure \ref{fig:brut_zoom_nai}) dont la topographie avoisinante donne des allures de réseau hydrographique. Les structures dominantes y sont plus difficilement décelables. À Varanasi, seul l'amphithéâtre se détache, empreinte coloniale en désaccord avec le reste de la géométrie du graphe.

\FloatBarrier
 
\begin{figure}[h]
    \centering
    \begin{subfigure}[t]{.45\linewidth}
        \includegraphics[width=\textwidth]{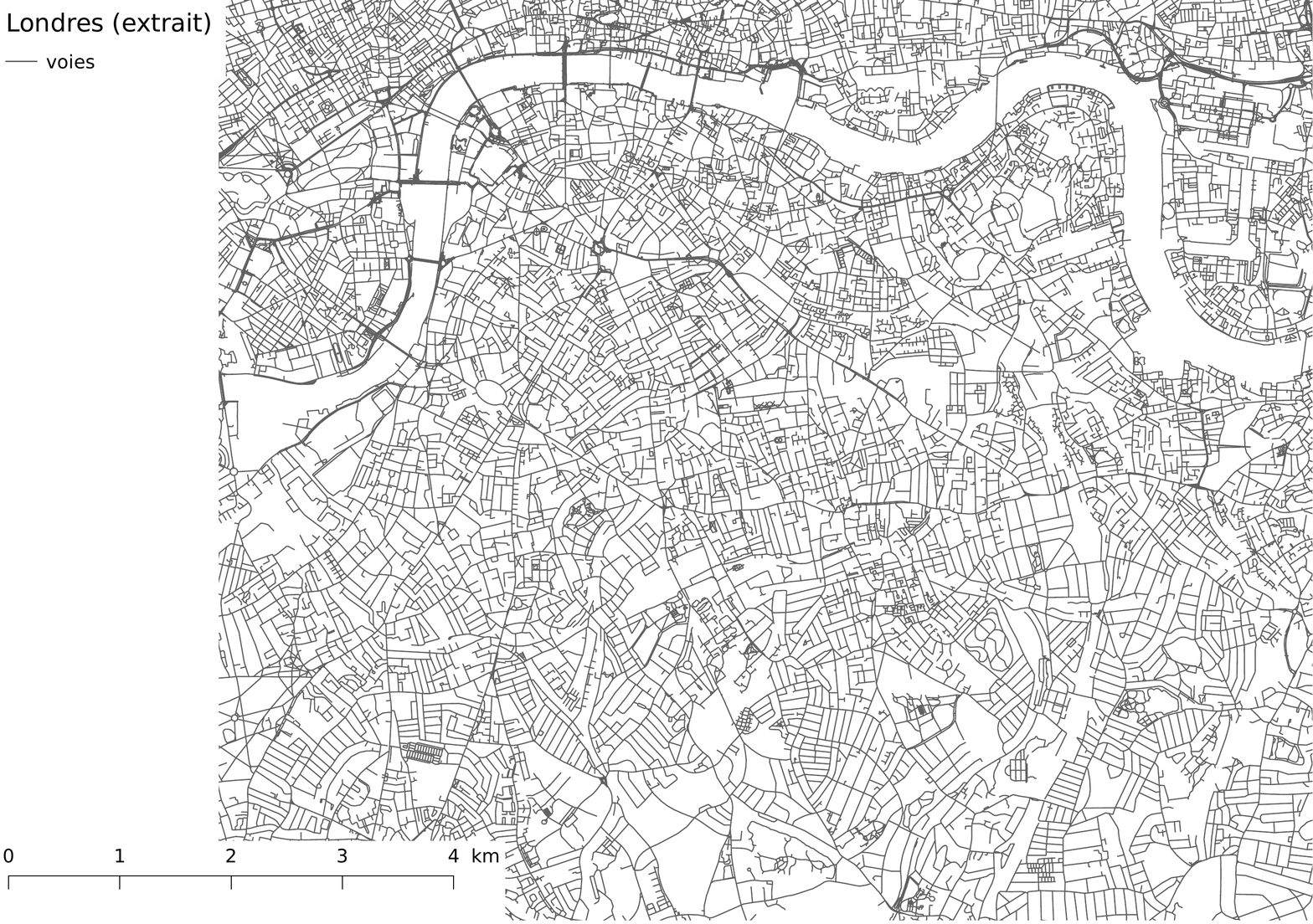}
        \caption{Londres.}
        \label{fig:brut_zoom_lon}
    \end{subfigure}
    ~
    \begin{subfigure}[t]{.45\linewidth}
        \includegraphics[width=\textwidth]{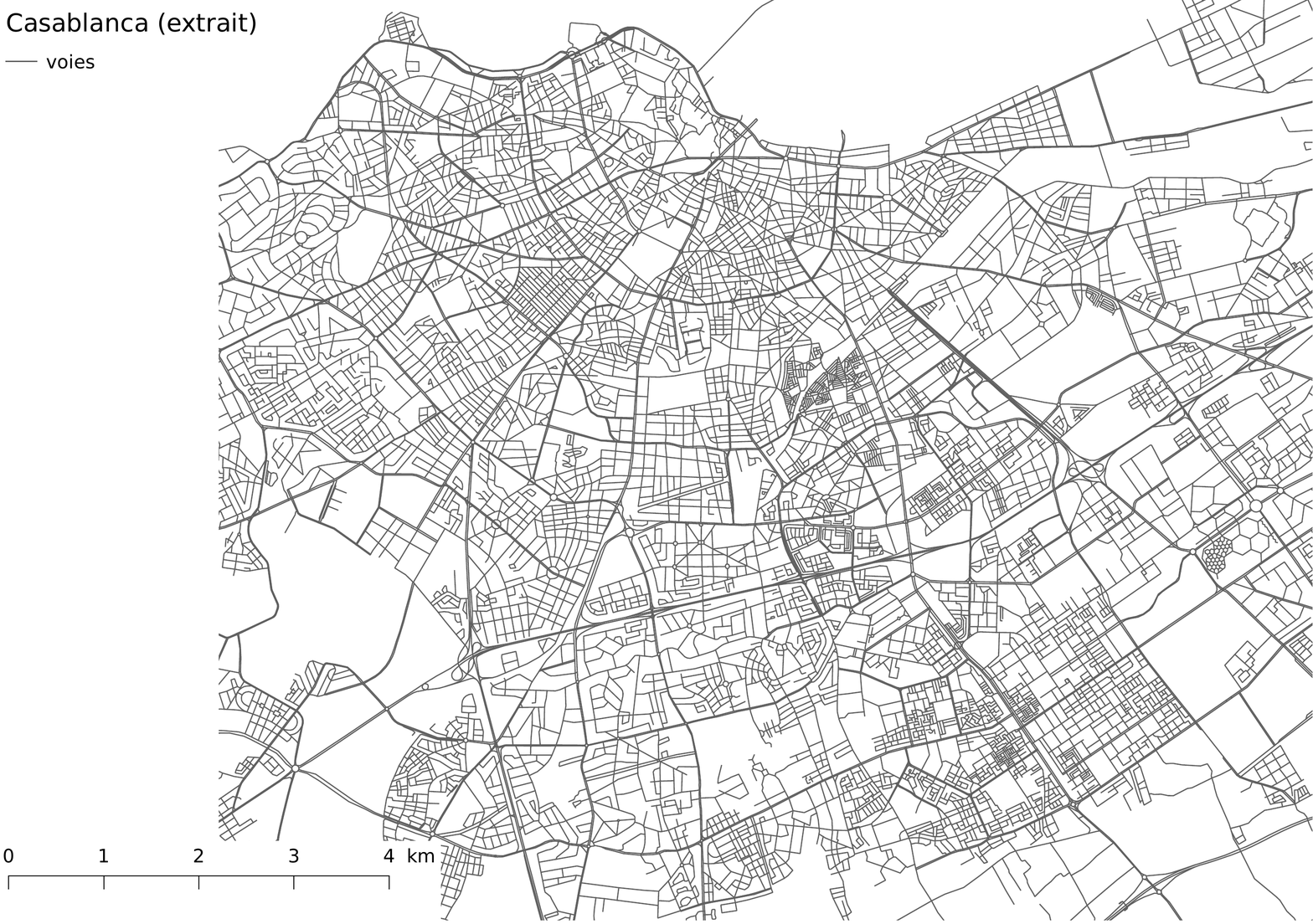}
        \caption{Casablanca.}
        \label{fig:brut_zoom_casa}
    \end{subfigure}

    \caption{Cohabitation de la courbe et de la grille. Graphes viaires extraits des villes du panel de recherche.}
    \label{fig:brut_zoom_coh}
\end{figure}

\begin{figure}[h]
    \centering
    \begin{subfigure}[t]{.45\linewidth}
        \includegraphics[width=\textwidth]{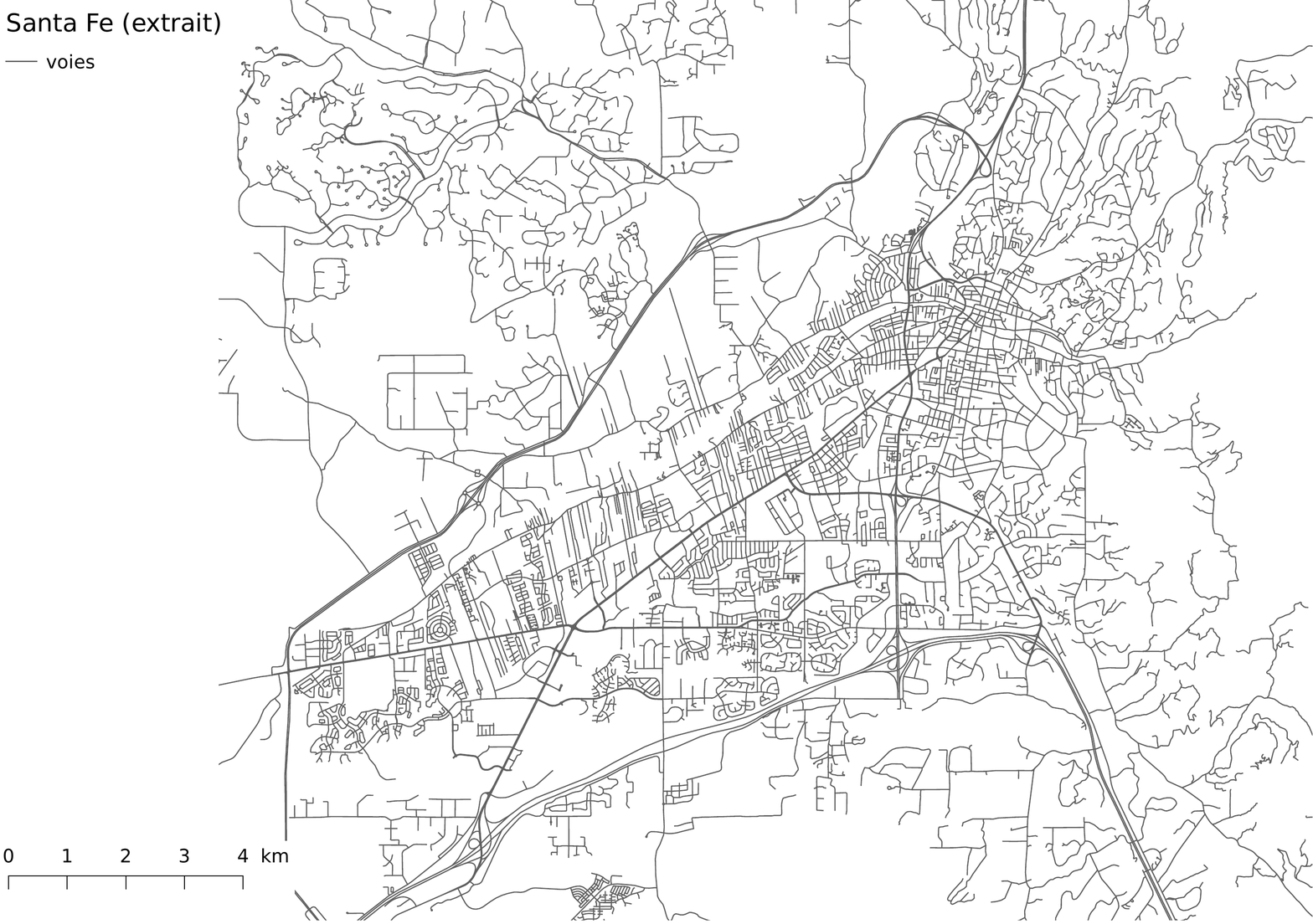}
        \caption{Santa Fe.}
        \label{fig:brut_zoom_santafe}
    \end{subfigure}
    ~
    \begin{subfigure}[t]{.45\linewidth}
        \includegraphics[width=\textwidth]{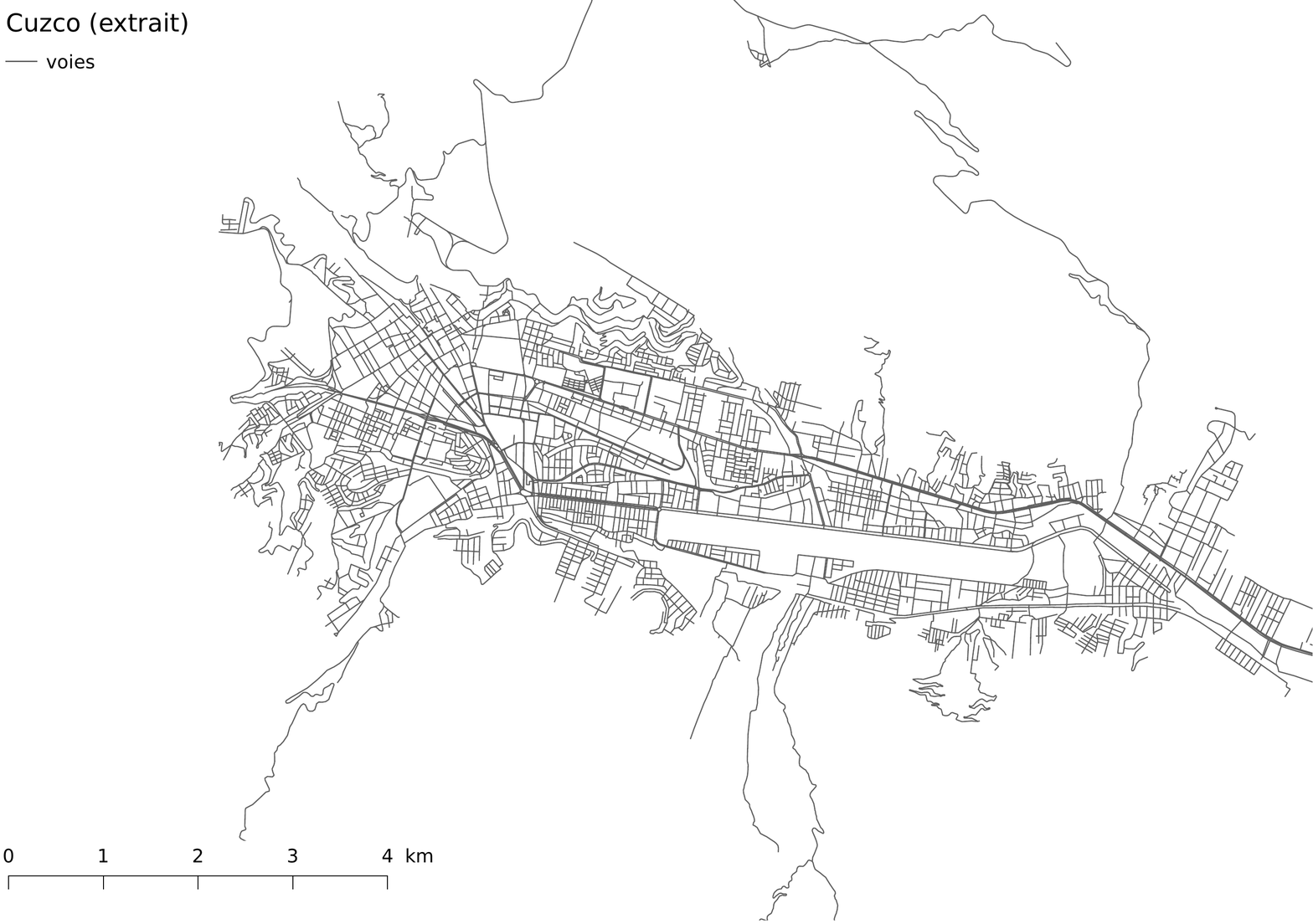}
        \caption{Cuzco.}
        \label{fig:brut_zoom_cuz}
    \end{subfigure}

    \caption{Formes imposées par la topographie. Graphes viaires extraits des villes du panel de recherche.}
    \label{fig:brut_zoom_topo3}
\end{figure}

À travers l'ensemble de ces graphes, les rapprochements de formes que l'on peut faire sont parfois étonnants. Ainsi, la similarité entre les graphes viaires de Londres et de Casablanca, tous deux avec des structures quadrillées entremêlées à des formes plus organiques (figure \ref{fig:brut_zoom_coh}), pose question. De même, au cœur des montagnes, les contraintes topographiques similaires entre Cuzco et Santa-Fe rapprochent leurs profils structurels (figure \ref{fig:brut_zoom_topo3}). Y aurait-il des formes aux tracés pré-définis ? L'esprit humain façonne-t-il des espaces selon des contraintes qui dépassent celles politiquement reconnues ?

\FloatBarrier

Les graphes viaires représentés dans leur intégralité sont disponibles en annexe \ref{ann:chap_panel_graphe}.

\begin{figure}[h]
    \centering
    \begin{subfigure}[t]{.45\linewidth}
        \includegraphics[width=\textwidth]{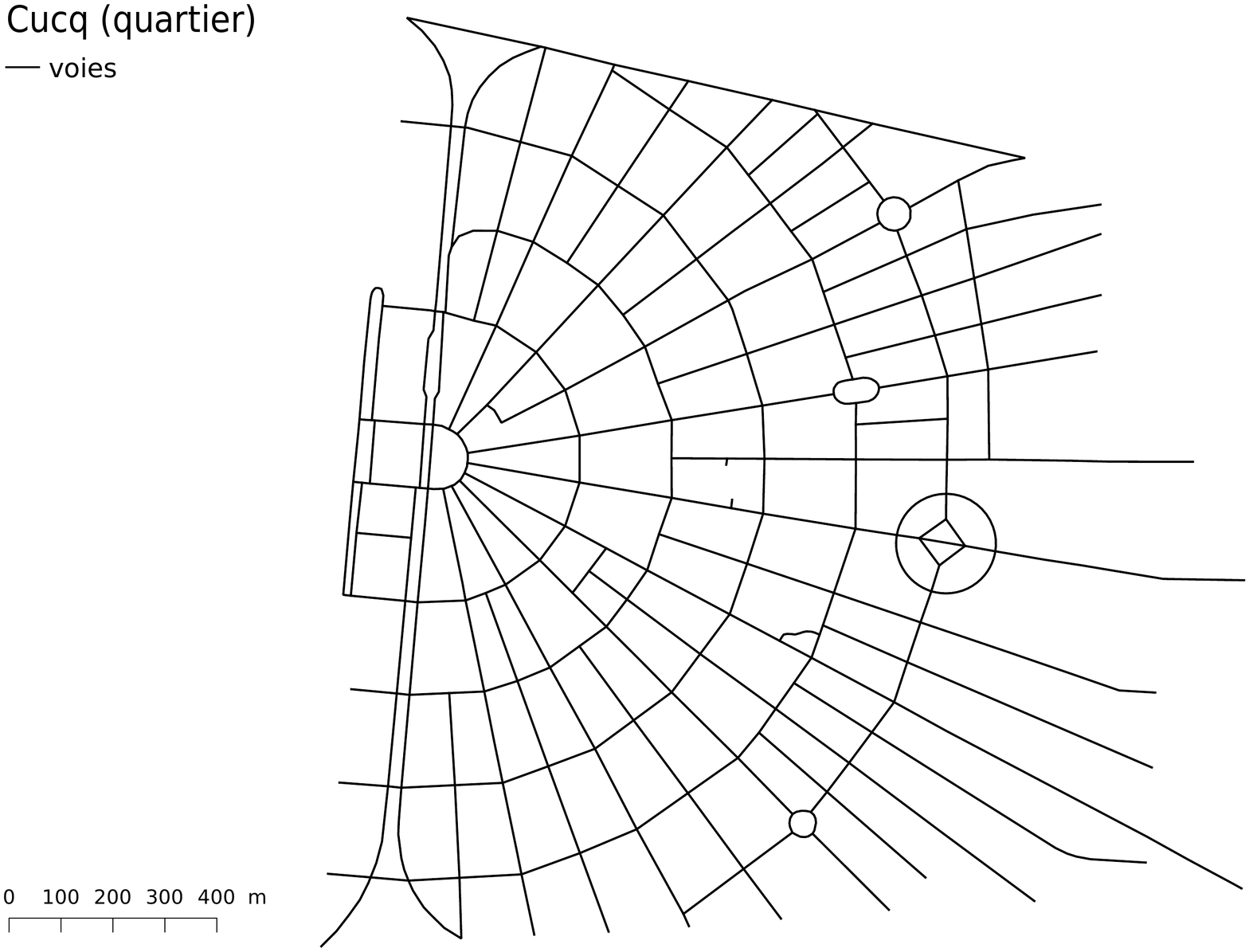}
        \caption{Graphe viaire issu de la ville de Cucq.}
        \label{fig:brut_cucq}
    \end{subfigure}
    ~
    \begin{subfigure}[t]{.45\linewidth}
        \includegraphics[width=\textwidth]{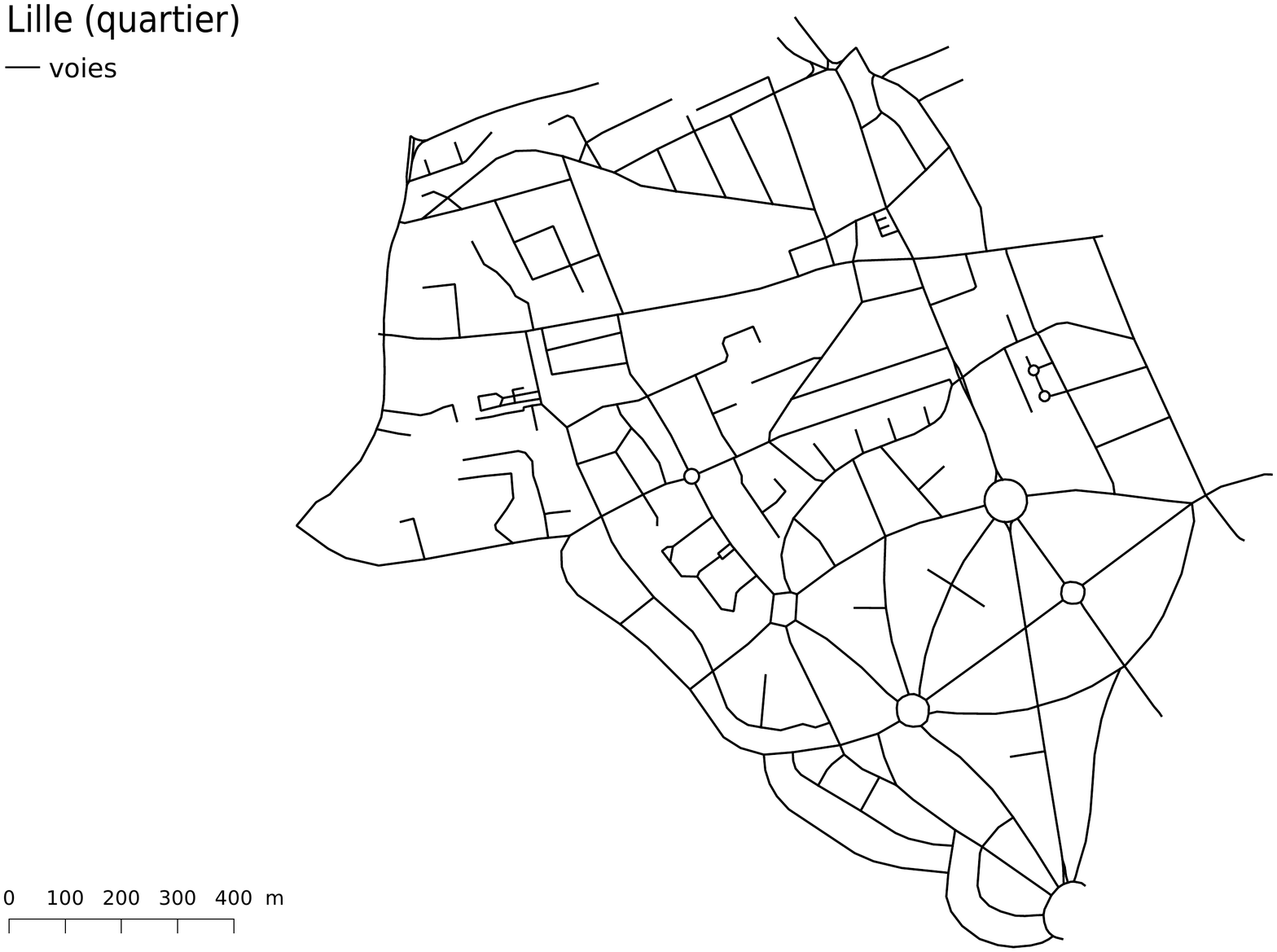}
        \caption{Graphe viaire issu de la ville de Lille.}
        \label{fig:brut_lille}
    \end{subfigure}
    
    \centering
    \begin{subfigure}[t]{.45\linewidth}
        \includegraphics[width=\textwidth]{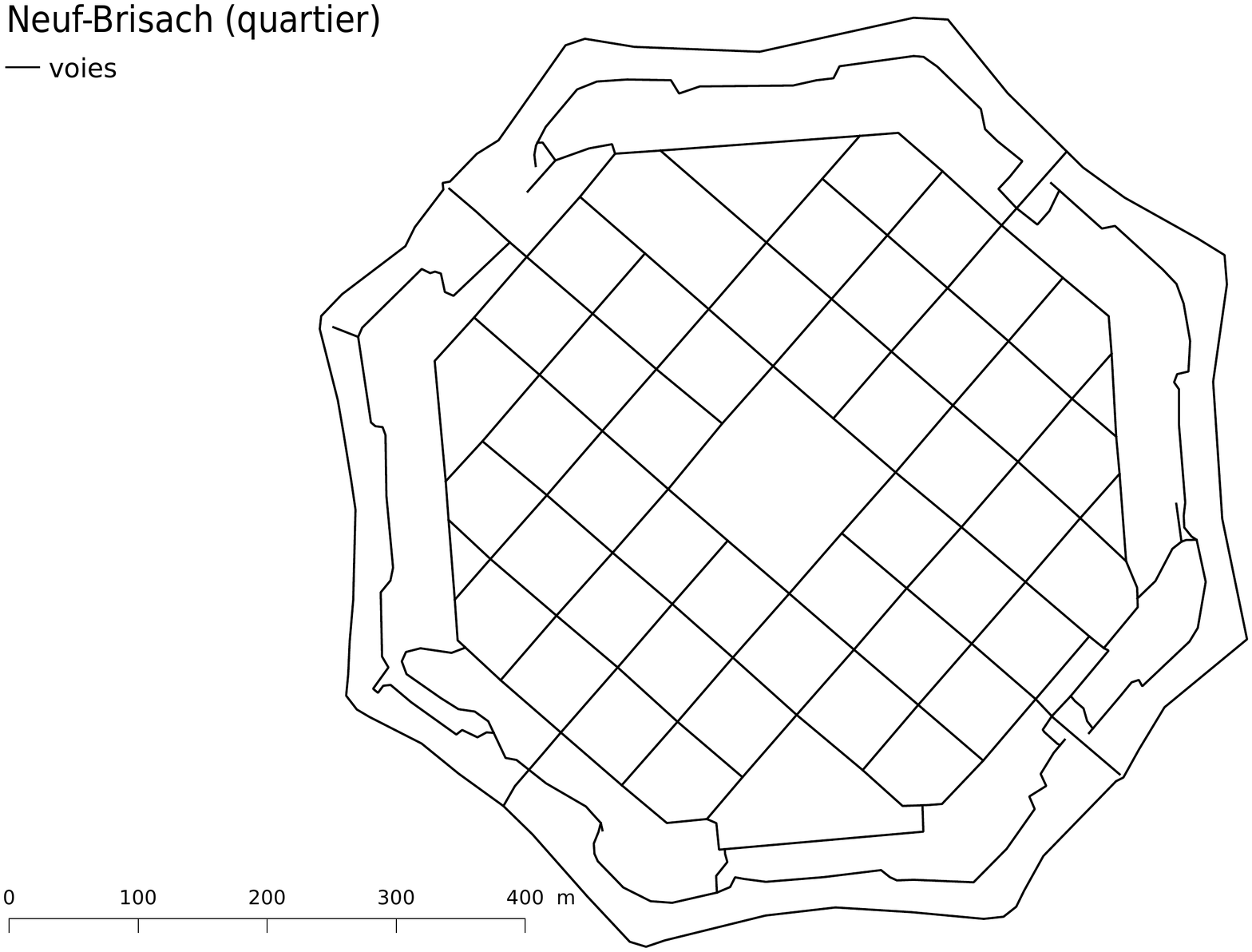}
        \caption{Graphe viaire issu de la ville de Neuf-Brisach.}
        \label{fig:brut_brisach}
    \end{subfigure}
    ~    
    \begin{subfigure}[t]{.45\linewidth}
        \includegraphics[width=\textwidth]{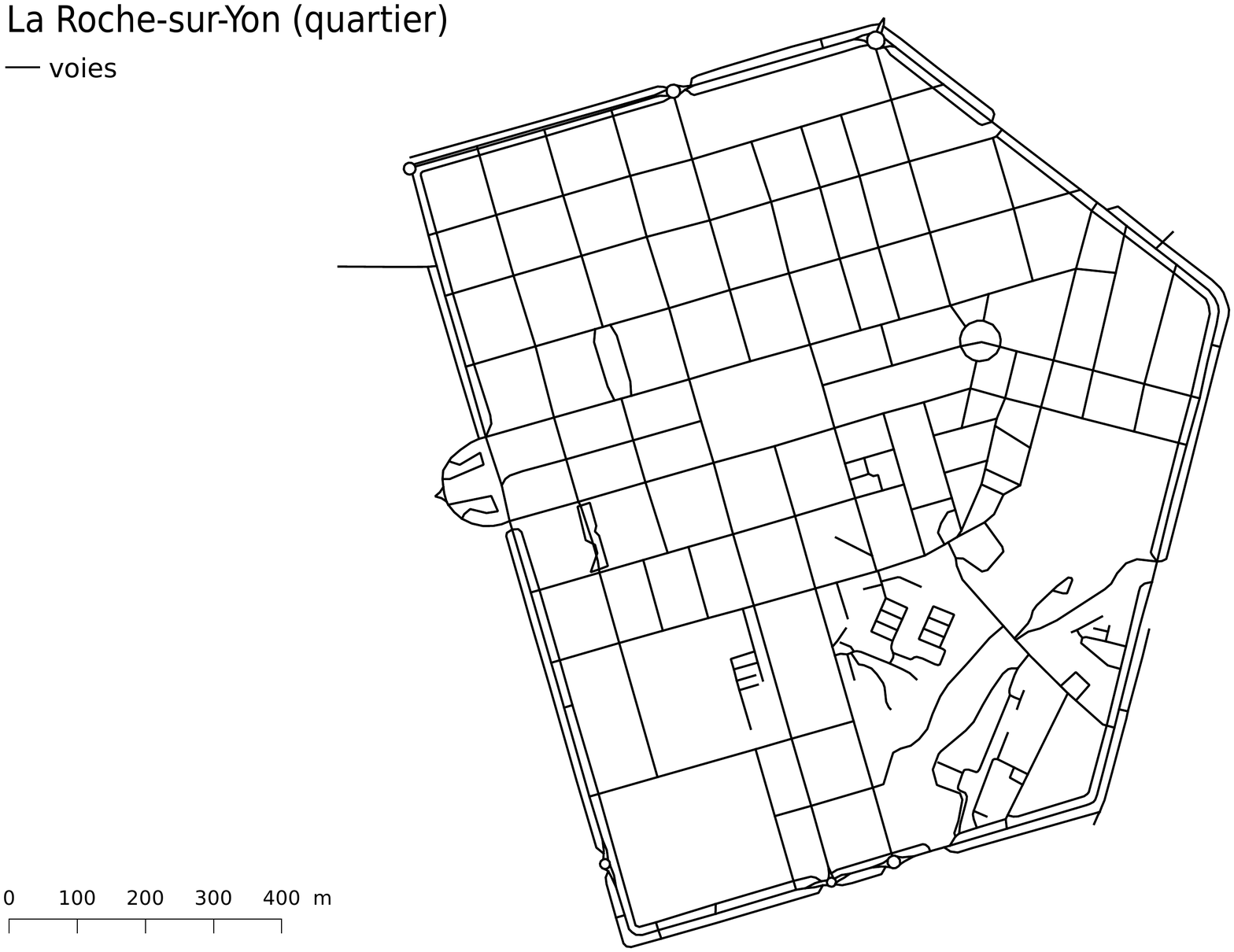}
        \caption{Graphe viaire issu de la ville de la Roche-sur-Yon.}
        \label{fig:brut_yon}
    \end{subfigure}
    
    \begin{subfigure}[t]{.45\linewidth}
        \includegraphics[width=\textwidth]{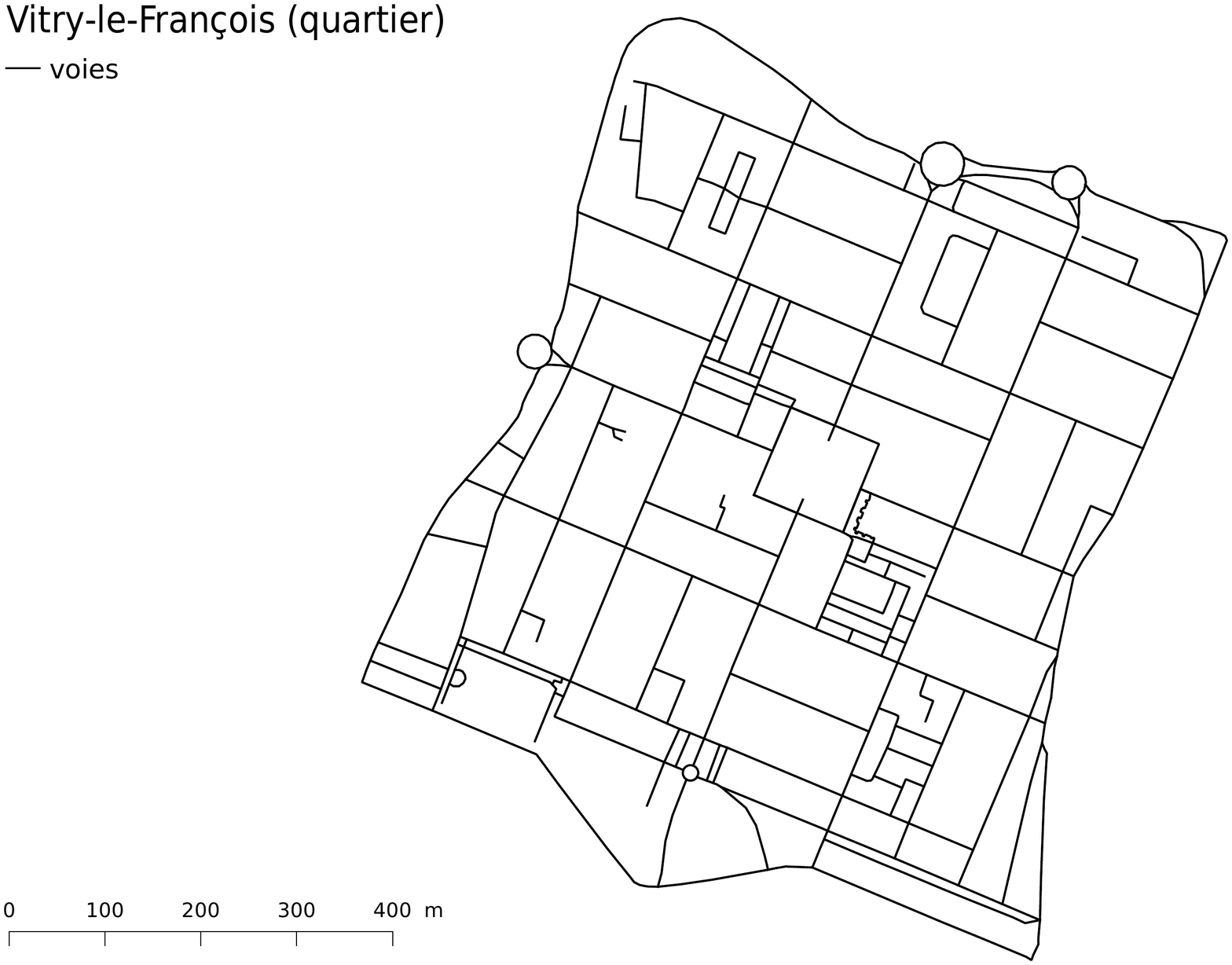}
        \caption{Graphe viaire issu de la ville de Vitry-le-François.}
        \label{fig:brut_vitry}
    \end{subfigure}
    
    \caption{Réseaux viaires des quartiers schématiques choisis. Données issues de la \copyright BDTOPO 2014 de l'IGN.}
    \label{fig:brut_quartiers}
\end{figure}

Nous complétons ce panel viaire, par l'ajout de cinq quartiers schématiques, choisis pour leur singularités géométriques. Ces quartiers ont été choisis à partir de l'Atlas élaboré par E. Degouys durant son stage de Master 2 \citep{estelledegouys2013}. Ils sont tous choisis en France : à Cucq (figure \ref{fig:brut_cucq}), Lille (figure \ref{fig:brut_lille}), Neuf-Brisach (figure \ref{fig:brut_brisach}), la Roche-sur-Yon (figure \ref{fig:brut_yon}) et Vitry-le-François (figure \ref{fig:brut_vitry}).

Nous avons donc un total de 25 graphes issus de réseaux viaires. Les graphes de Paris et d'Avignon sont issus de la \copyright BDTOPO. Les autres graphes viaires que nous utilisons sont extraits de la base de données d'OpenStreetMap

\FloatBarrier
\subsection{Graphes artificiels}

\begin{figure}[h]
	\centering
    \includegraphics[width=\textwidth]{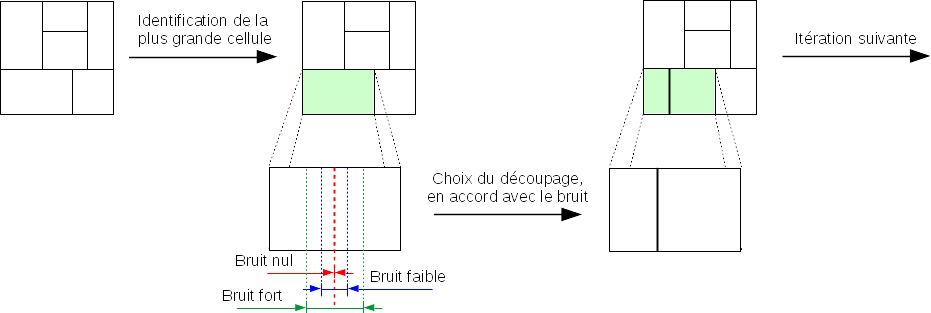}       
    \caption{Schéma de construction du réseau artificiel selon le modèle \emph{Grande longueur} sans angle}
    \label{fig:r_artif_GL}
\end{figure}

\begin{figure}[h]
	\centering
    \includegraphics[width=\textwidth]{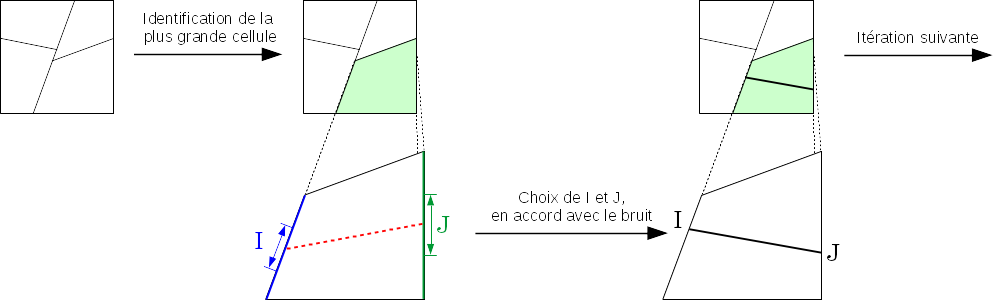}       
    \caption{Schéma de construction du réseau artificiel selon le modèle \emph{Grande longueur} avec angles}
    \label{fig:r_artif_GLA}
\end{figure}

\begin{figure}[h]
	\centering
    \includegraphics[width=\textwidth]{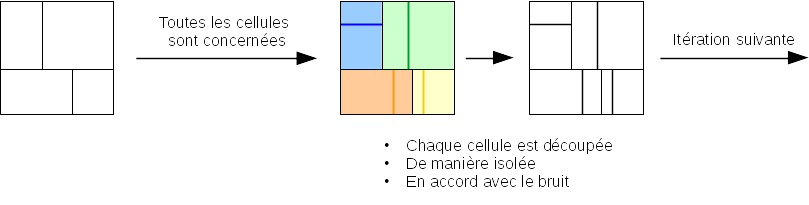}       
    \caption{Schéma de construction du réseau artificiel selon le modèle \emph{Génération} sans angle}
    \label{fig:r_artif_Gen}
\end{figure}

Nous considérons également cinq réseaux \textit{artificiels}, construits selon des règles simples. Ces réseaux ont été élaborés par R. Pousse dans le cadre de son stage de Master 2 \citep{romainpousse2015}. La construction de ces graphes consiste en un découpage successif de cellules, selon un modèle \enquote{Mondrian}. À chaque itération, tout ou une partie des cellules qui composent le graphe, sont coupées en deux par un nouveau segment (figures \ref{fig:r_artif_GL}, \ref{fig:r_artif_GLA} et \ref{fig:r_artif_Gen}). Ce découpage repose sur quatre paramètres :

\begin{enumerate}
\item le nombre de cellules final ($N_c$)
\item le bruit linéaire (autour du centre du bord de cellule) introduit à chaque nouveau découpage ($B_{lin}$)
\item l'indépendance des bruits pour chaque bord de cellule ($Indep$)
\item le choix des cellules à découper à chaque itération : 

\begin{enumerate}
\item seulement la plus grande cellule (modèle \emph{Grande longueur})
\item toutes les cellules (modèle \emph{Génération})
\end{enumerate}

\end{enumerate}

La position de la coupure dans la cellule suit une distribution gaussienne. Le paramètre $B_{lin}$ correspond à l'écart type de cette gaussienne, il détermine la position du point de découpage sur les bords verticaux de la cellule. S'il est nul, la courbe est réduite à un dirac : toutes les divisions de cellules se feront en leur centre. Plus $B_{lin}$ augmente, plus la position de la division pourra s'écarter du centre de la cellule, en conservant une forte probabilité de se rapprocher de celui-ci (selon la forme de la courbe). La valeur de ce bruit ne pourra néanmoins pas dépasser 1, au risque de retrouver une division de cellule à l'extérieur de celle-ci. La probabilité d'avoir une valeur de bruit supérieure à 1 étant non nulle (fixée par la gaussienne), les résultats seront filtrés pour ne conserver que des bruits qui y sont inférieurs. Le paramètre d'indépendance des bruits sur chaque bord de cellule ($Indep$) permet d'introduire des angles dans la construction des graphes. S'il n'est pas activé ($Indep = 0$) alors le décalage sera le même de chaque côté de la cellule (figure \ref{fig:r_artif_GL}). Dans le cas contraire ($Indep = 1$), la découpe de la cellule rejoindra deux points choisis indépendamment, selon le bruit $B_{lin}$ imposé, de chaque côté de la cellule (figure \ref{fig:r_artif_GLA}). 

\begin{figure}[h]
    \centering
    \begin{subfigure}[t]{.45\linewidth}
        \includegraphics[width=\textwidth]{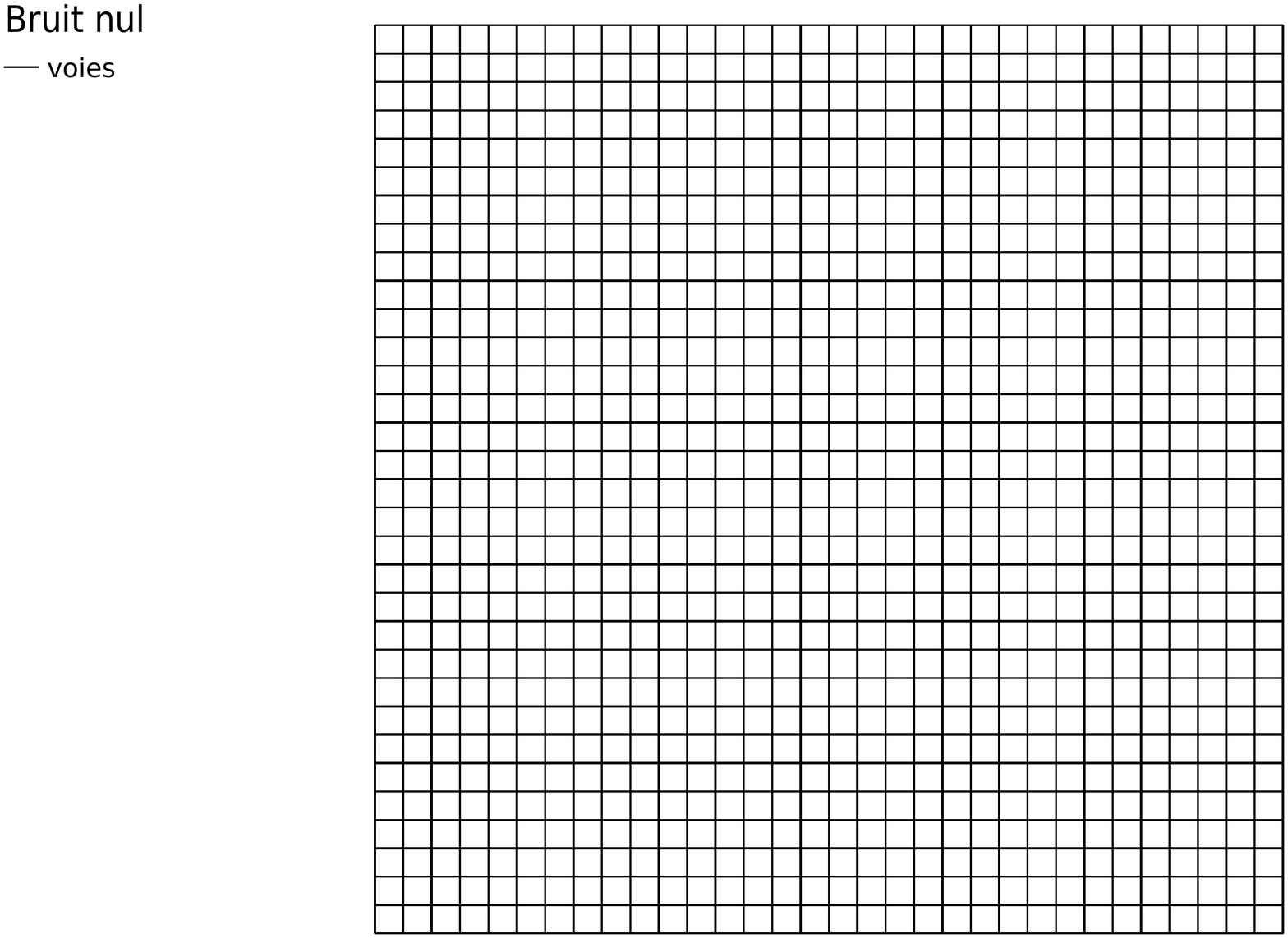}
        \caption{\enquote{Bruit nul}.}
        \label{fig:brut_bruitnul}
    \end{subfigure}
    ~
    \begin{subfigure}[t]{.45\linewidth}
        \includegraphics[width=\textwidth]{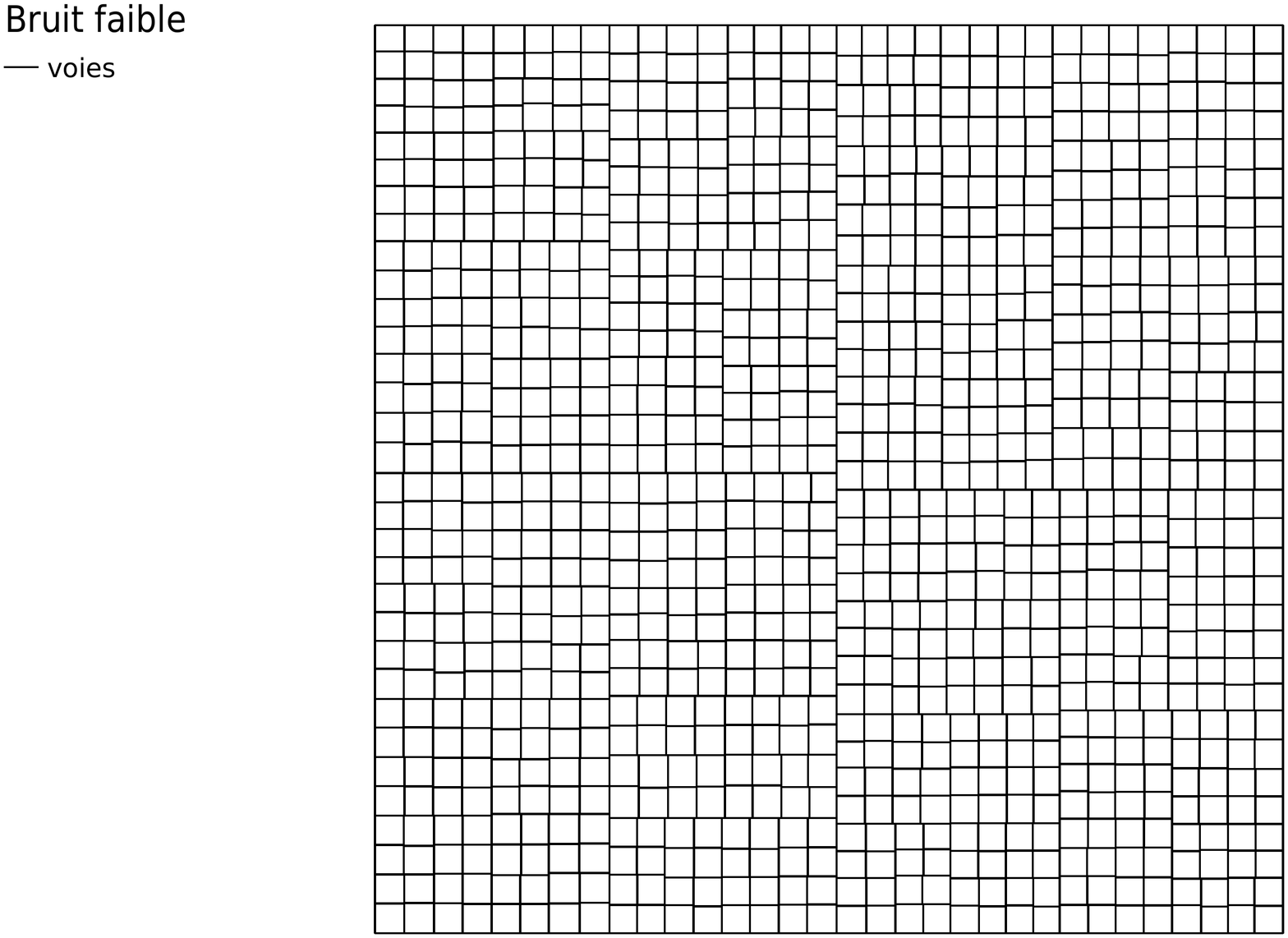}
        \caption{\enquote{Bruit faible}.}
        \label{fig:brut_bruitfaible}
    \end{subfigure}
    
    \centering
    \begin{subfigure}[t]{.45\linewidth}
        \includegraphics[width=\textwidth]{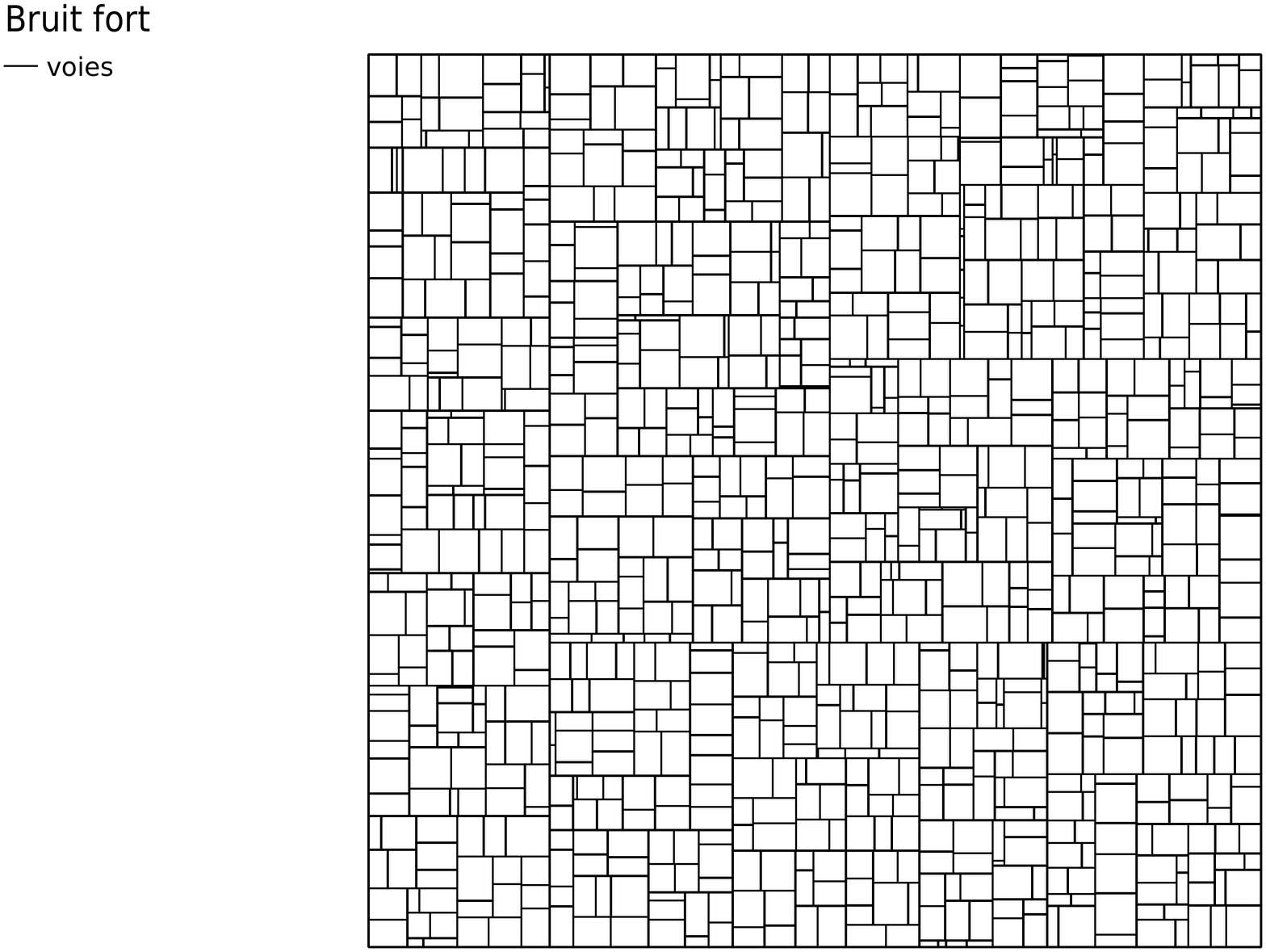}
        \caption{\enquote{Bruit fort}.}
        \label{fig:brut_bruitfort}
    \end{subfigure}
    ~    
    \begin{subfigure}[t]{.45\linewidth}
        \includegraphics[width=\textwidth]{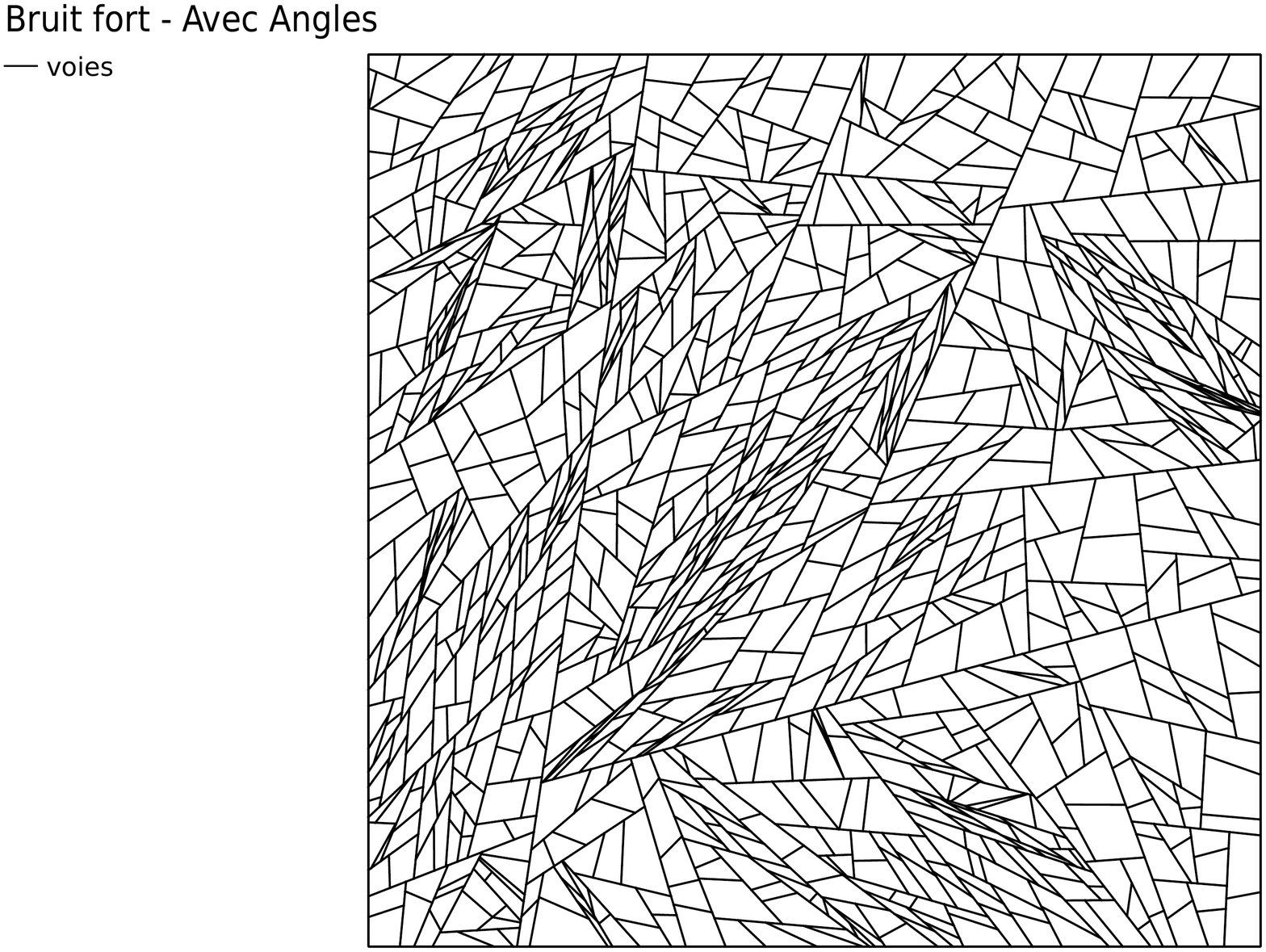}
        \caption{\enquote{Bruit fort, avec Angle}.}
        \label{fig:brut_angle}
    \end{subfigure}
    
    \begin{subfigure}[t]{.45\linewidth}
        \includegraphics[width=\textwidth]{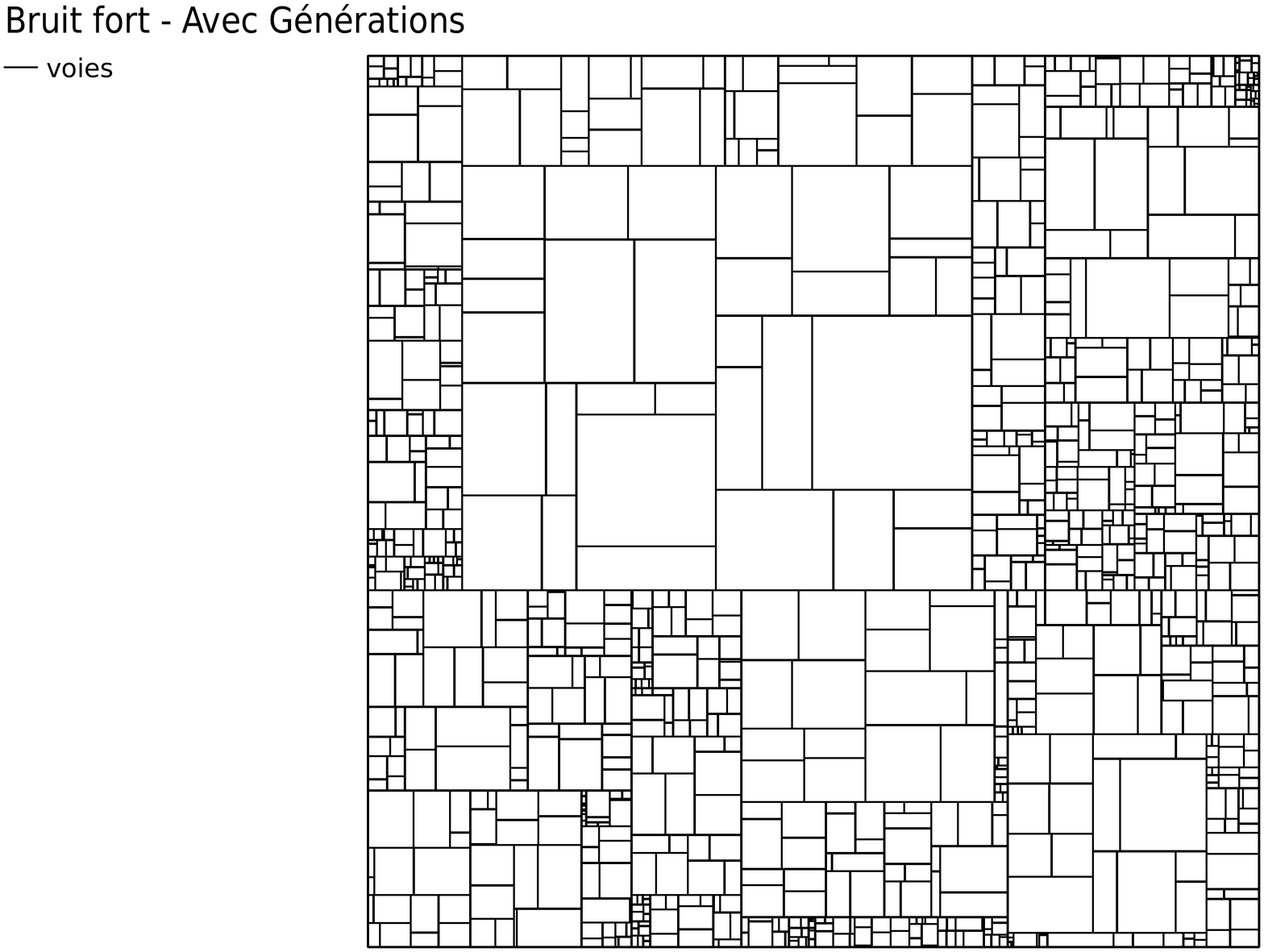}
        \caption{\enquote{Bruit fort, avec Générations}.}
        \label{fig:brut_generation}
    \end{subfigure}
    
    \caption{Réseaux artificiels. Données fournies par R. Pousse.}
    \label{fig:brut_reseauxartif}
\end{figure}

Les graphes artificiels sont générés par découpages successifs de cellules selon un bruit plus ou moins fort. Dans le modèle \emph{Grande longueur}, la plus grande cellule est découpée à chaque itération. La division aboutit (pour un nombre de cellules fixé) à une densité homogène (figures \ref{fig:brut_bruitnul}, \ref{fig:brut_bruitfaible}, \ref{fig:brut_bruitfort} et \ref{fig:brut_angle}). Dans le modèle \emph{Génération}, qui force le découpage de l'ensemble des cellules à chaque itération, le résultat sera d'autant plus hétérogène que le bruit paramétré sera fort (figure \ref{fig:brut_generation}).

\begin{figure}[h]
    \centering
    \begin{subfigure}[t]{.6\linewidth}
        \includegraphics[width=\textwidth]{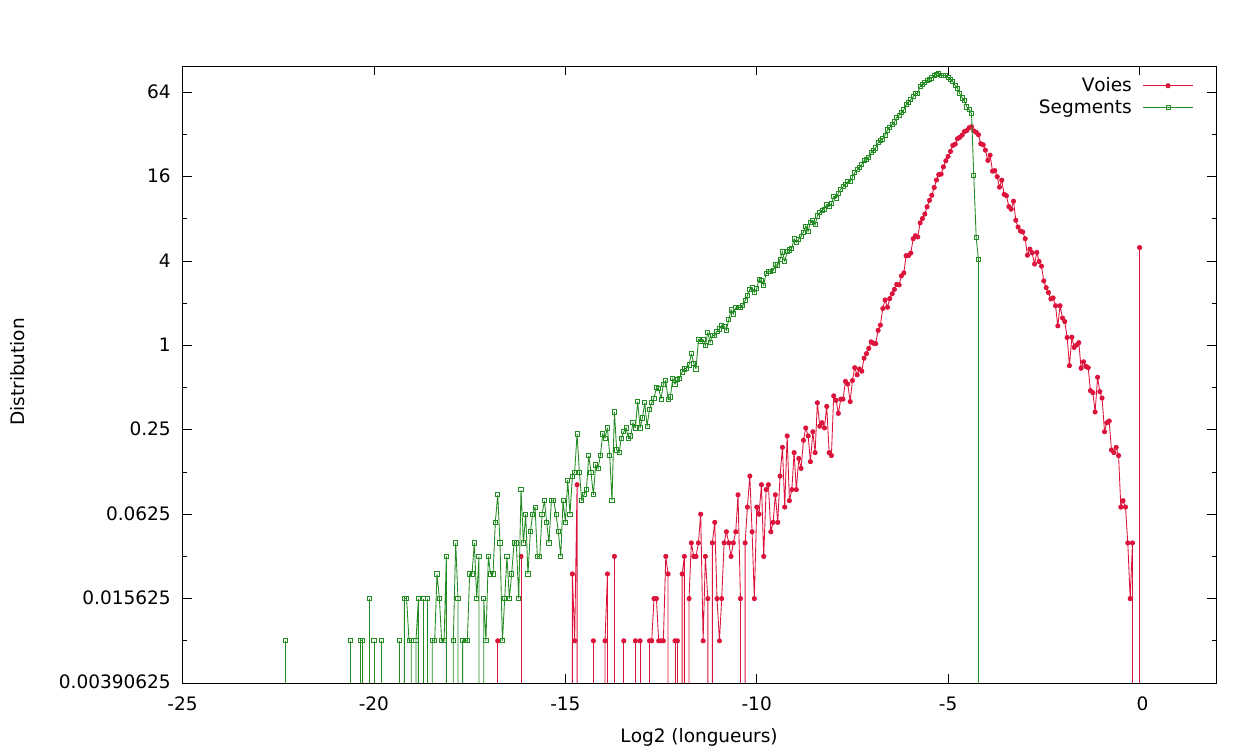}
        \caption{Modèle \emph{Grande longueur}, avec $N_c = 2^{10}$, $B_{lin} = \frac{1}{5}$ et $Indep = 0$.}
        \label{fig:courbe_bruitfort}
    \end{subfigure}
    ~
    \begin{subfigure}[t]{.6\linewidth}
        \includegraphics[width=\textwidth]{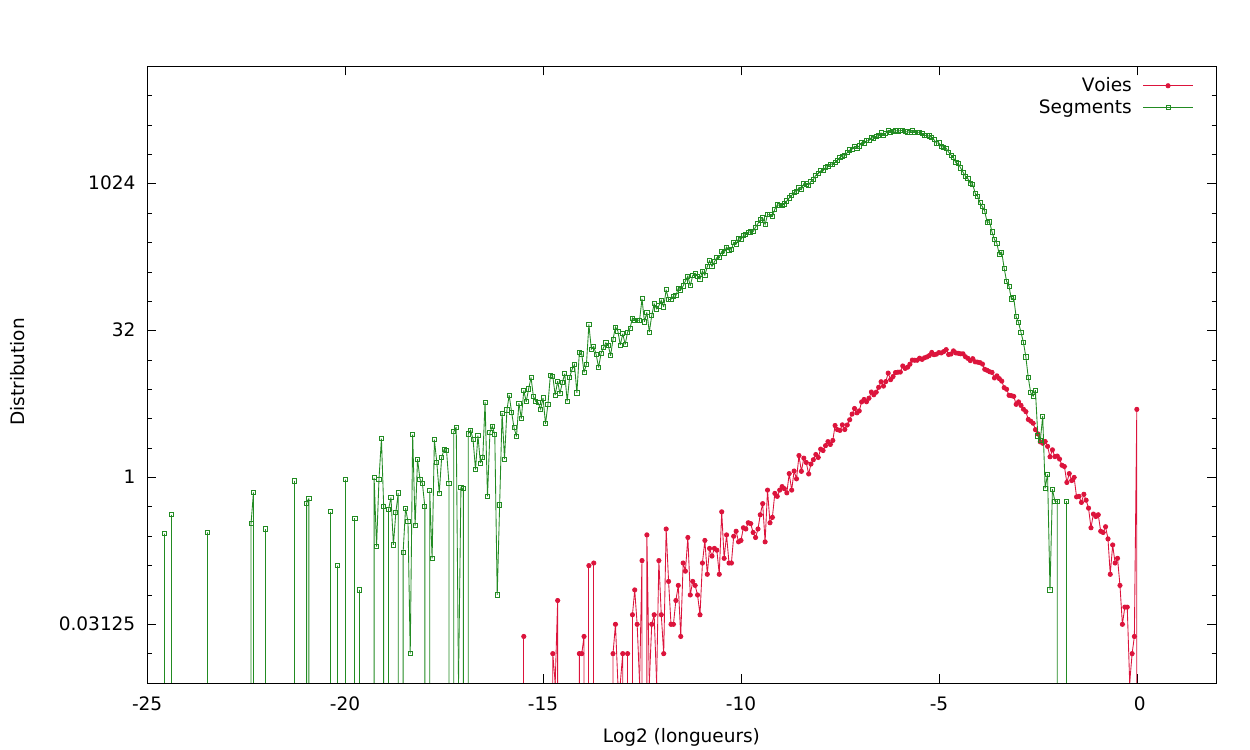}
        \caption{Modèle \emph{Génération} avec $N_c = 2^{10}$, $B_{lin} = \frac{1}{5} $ et $ Indep = 0$.}
        \label{fig:courbe_generation}
    \end{subfigure}
    
    \caption{Étude des distributions de longueurs de voies et d'arcs pour les réseaux artificiels. En ordonnée : moyenne d'apparition des valeurs dans un graphe sur un certain nombre de tirages (ici : 128).    
     \\ source : courbes réalisées par R. Pousse.}
    \label{fig:courbe_reseauxartif}
\end{figure}

La distribution du logarithme des longueurs de voies pour le modèle \emph{Grande longueur} est beaucoup plus piquée : l'homogénéité du réseau donne des voies de longueurs proches, dont les pentes de la distribution peuvent être approximées par une loi de puissance. L'hétérogénéité du réseau influence la distribution tracée pour le modèle \emph{Génération} en la rendant proche d'une courbe gaussienne (figure \ref{fig:courbe_reseauxartif}).

Les cinq réseaux que nous étudions ici ont un total de $N_c = 2^{10}$ cellules. Leurs différences résident dans les variations des trois autres paramètres. Nous introduisons dans notre étude :
\begin{itemize}
\item Un réseau \enquote{Bruit nul} : $B_{lin} = 0 \wedge Indep = 0$, modèle \emph{Grande longueur}
\item Un réseau \enquote{Bruit faible} : $B_{lin} = \frac{1}{100} \wedge Indep = 0$, modèle \emph{Grande longueur}
\item Un réseau \enquote{Bruit fort} : $B_{lin} = \frac{1}{5} \wedge Indep = 0$, modèle \emph{Grande longueur}
\item Un réseau \enquote{Bruit fort, avec Angles} : $B_{lin} = \frac{1}{5} \wedge Indep = 1$, modèle \emph{Grande longueur}
\item Un réseau \enquote{Bruit fort, avec Générations} : $B_{lin} = \frac{1}{5} \wedge Indep = 0$, modèle \emph{Génération}
\end{itemize}

\FloatBarrier
\subsection{Autres graphes}

Nous finalisons ce panel de recherche avec des graphes de natures différentes. Nous prenons deux graphes issus de la numérisation de craquelures dans de l'argile ; un graphe numérisé à partir d'une feuille ; un graphe numérisé à partir d'une gorgone (corail). Nous considérons également des graphes hydrographiques. Nous prenons un extrait de celui de la forêt amazonienne, un autre dans les montagnes du Nord de l'Italie, un troisième dans celles du sud de l'Inde. Enfin, nous étudions des graphes de réseaux ferrés : celui de Belgique, celui de Californie (USA) et celui de Russie.

\begin{figure}[h]
    \centering
    \begin{subfigure}[t]{.6\linewidth}
        \includegraphics[width=\textwidth]{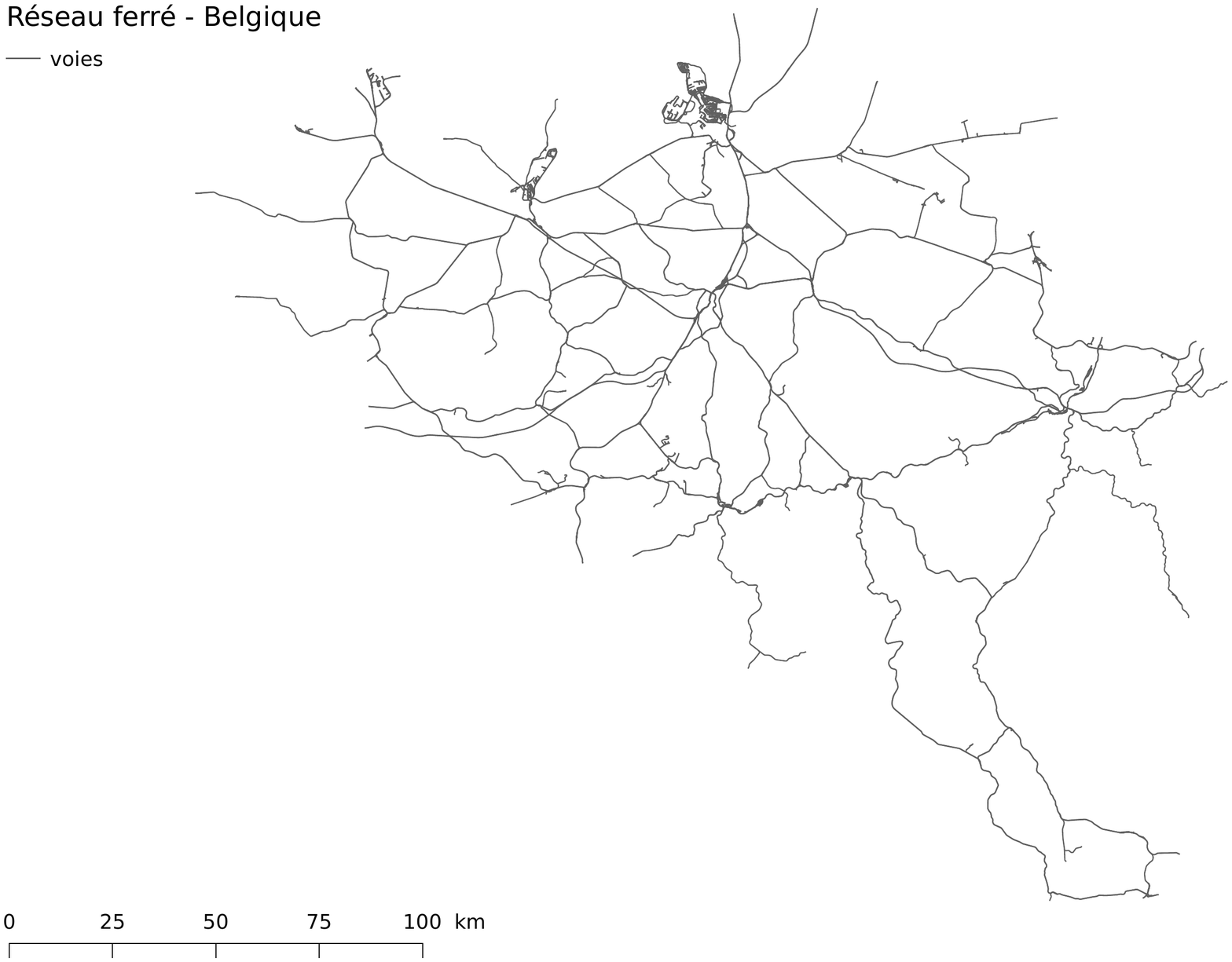}
        \caption{Belgique.}
    \end{subfigure}
   
    \begin{subfigure}[t]{.6\linewidth}
        \includegraphics[width=\textwidth]{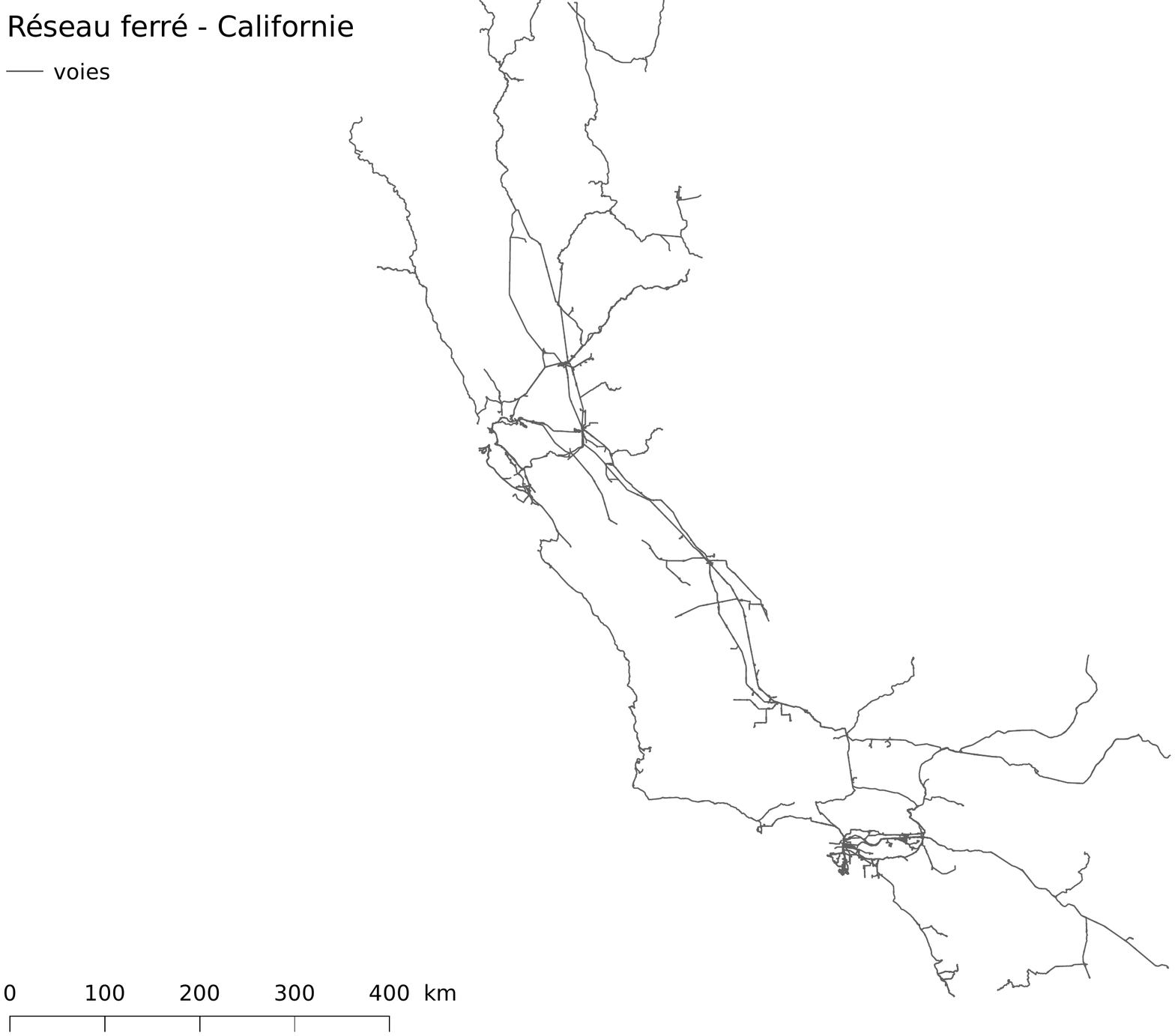}
        \caption{Californie.}
    \end{subfigure}
    
    \begin{subfigure}[t]{.6\linewidth}
        \includegraphics[width=\textwidth]{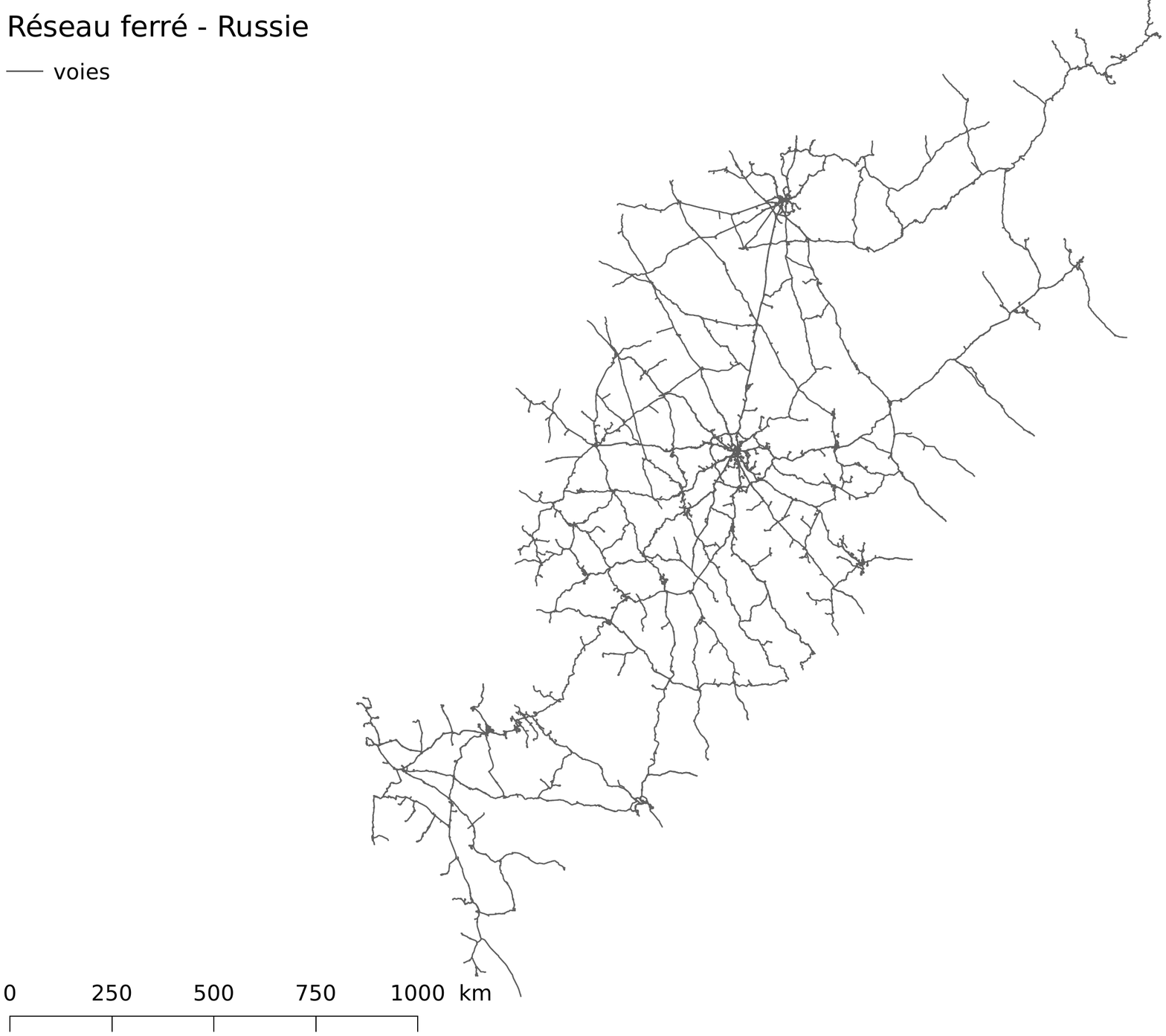}
        \caption{Russie.}
    \end{subfigure}

    \caption{Graphes issus de réseaux ferrés. Données : OpenStreetMap. \\ Nous retirons sur le linéaire ferré les vecteurs correspondant aux métros ou aux plate-formes de gare autant que possible (lorsque la donnée de type est renseignée).}
    \label{fig:brut_ferre}
\end{figure}

\begin{figure}[h]
    \centering
    \begin{subfigure}[t]{.6\linewidth}
        \includegraphics[width=\textwidth]{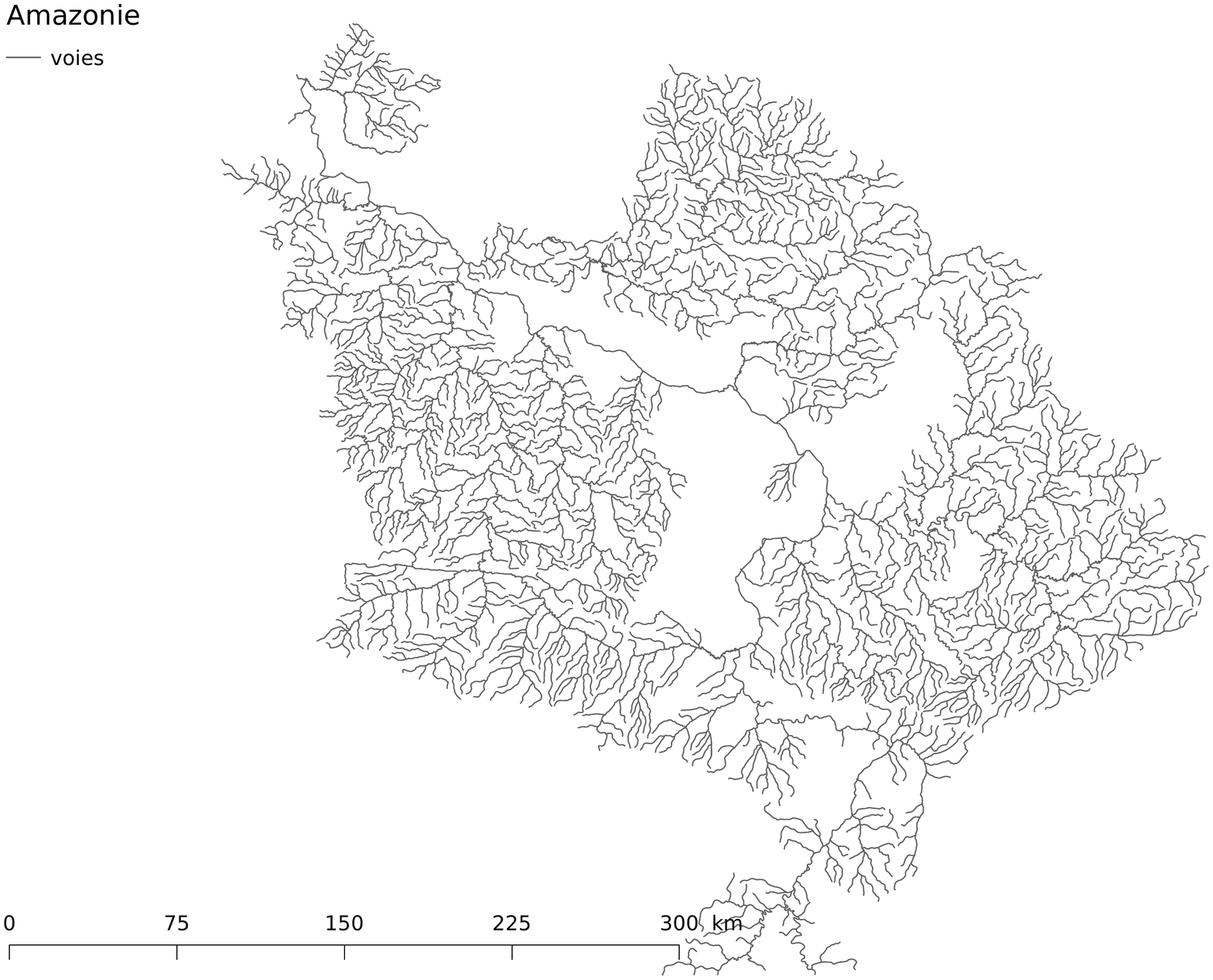}
        \caption{Amazonie.}
    \end{subfigure}
   
    \begin{subfigure}[t]{.6\linewidth}
        \includegraphics[width=\textwidth]{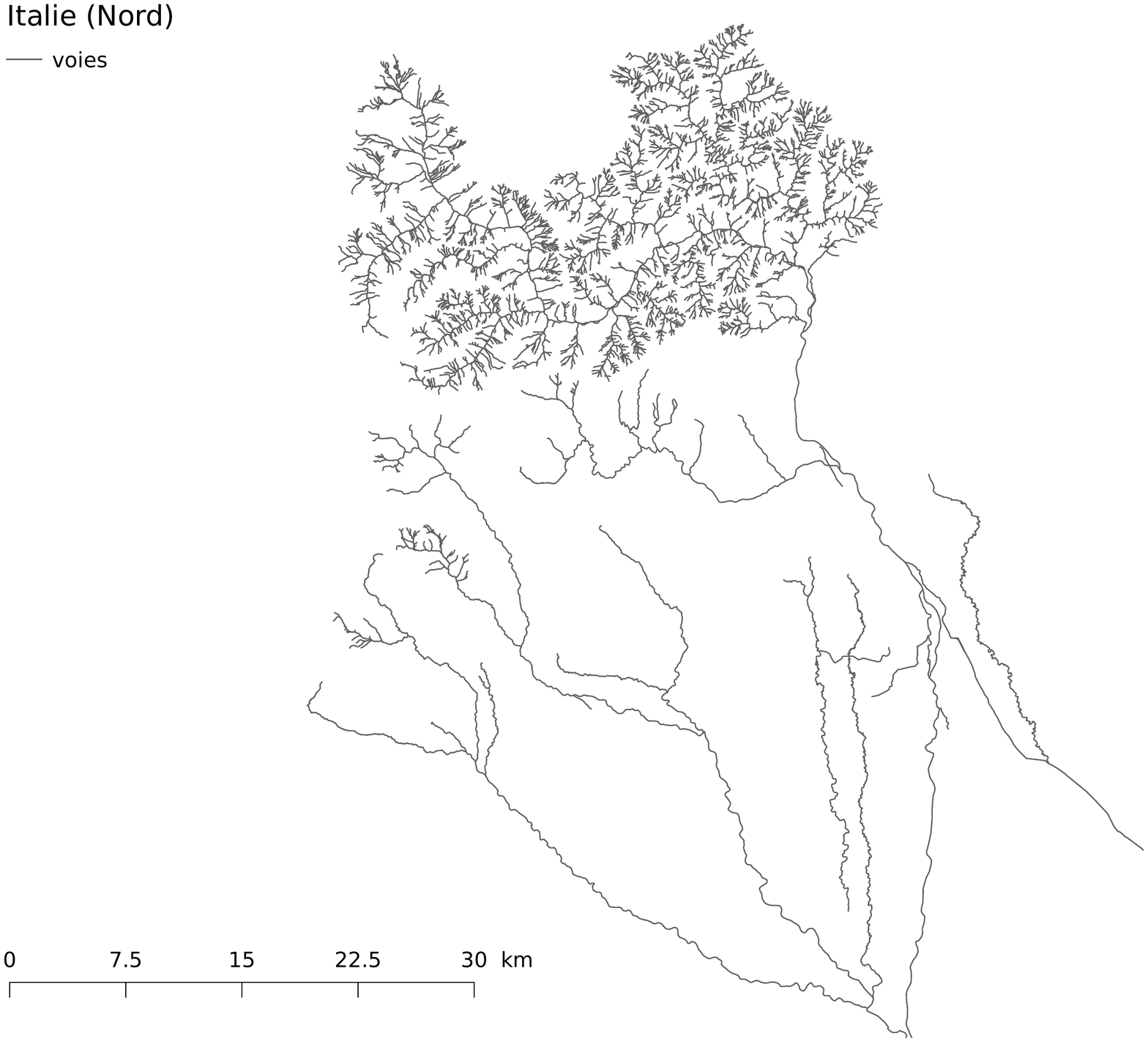}
        \caption{Italie (Nord).}
    \end{subfigure}
    
    \begin{subfigure}[t]{.6\linewidth}
        \includegraphics[width=\textwidth]{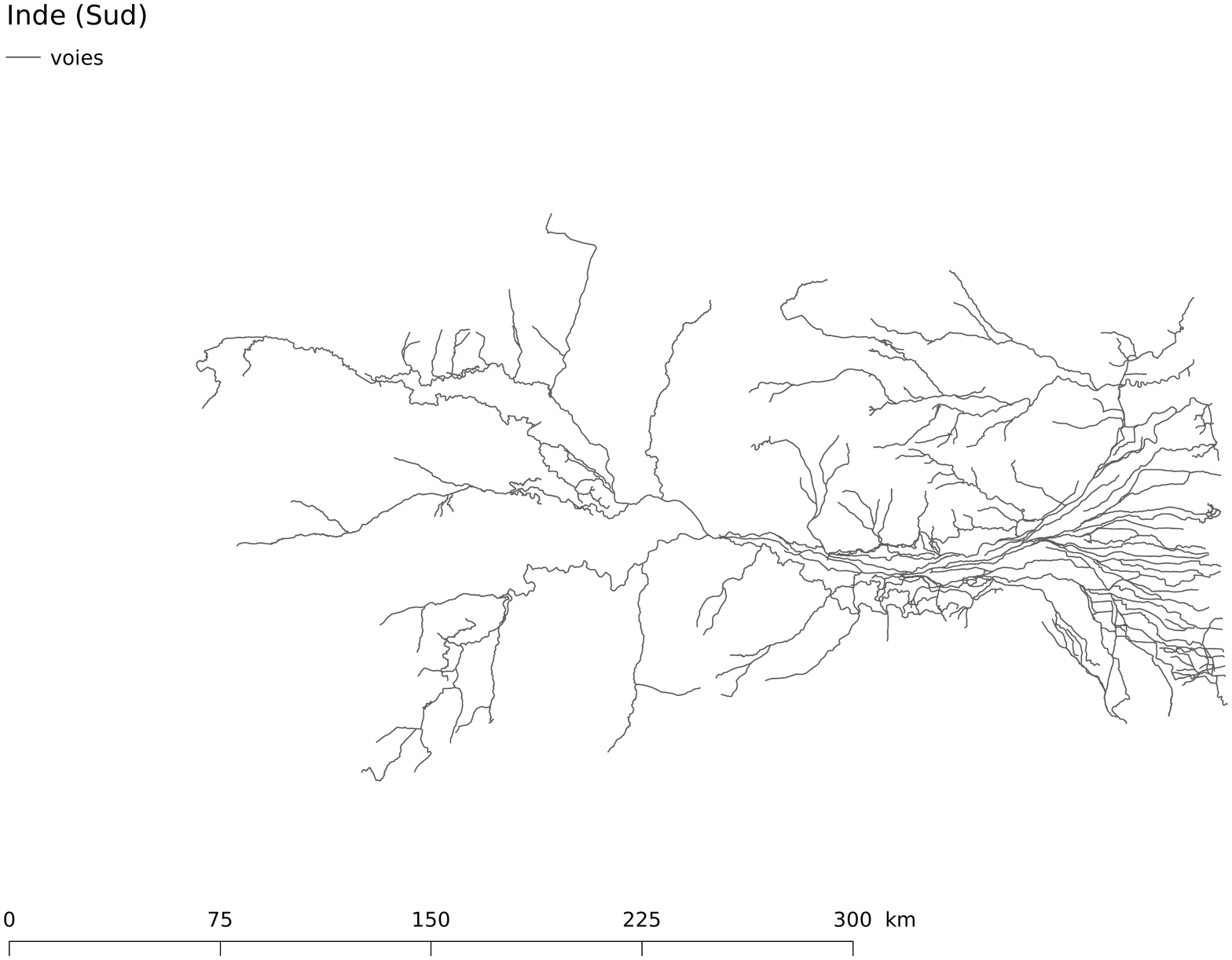}
        \caption{Inde (Sud).}
    \end{subfigure}

    \caption{Graphes issus de réseaux hydrographiques. Données : OpenStreetMap.}
    \label{fig:brut_water}
\end{figure}

\begin{figure}[h]
    \centering
        \includegraphics[width=\textwidth]{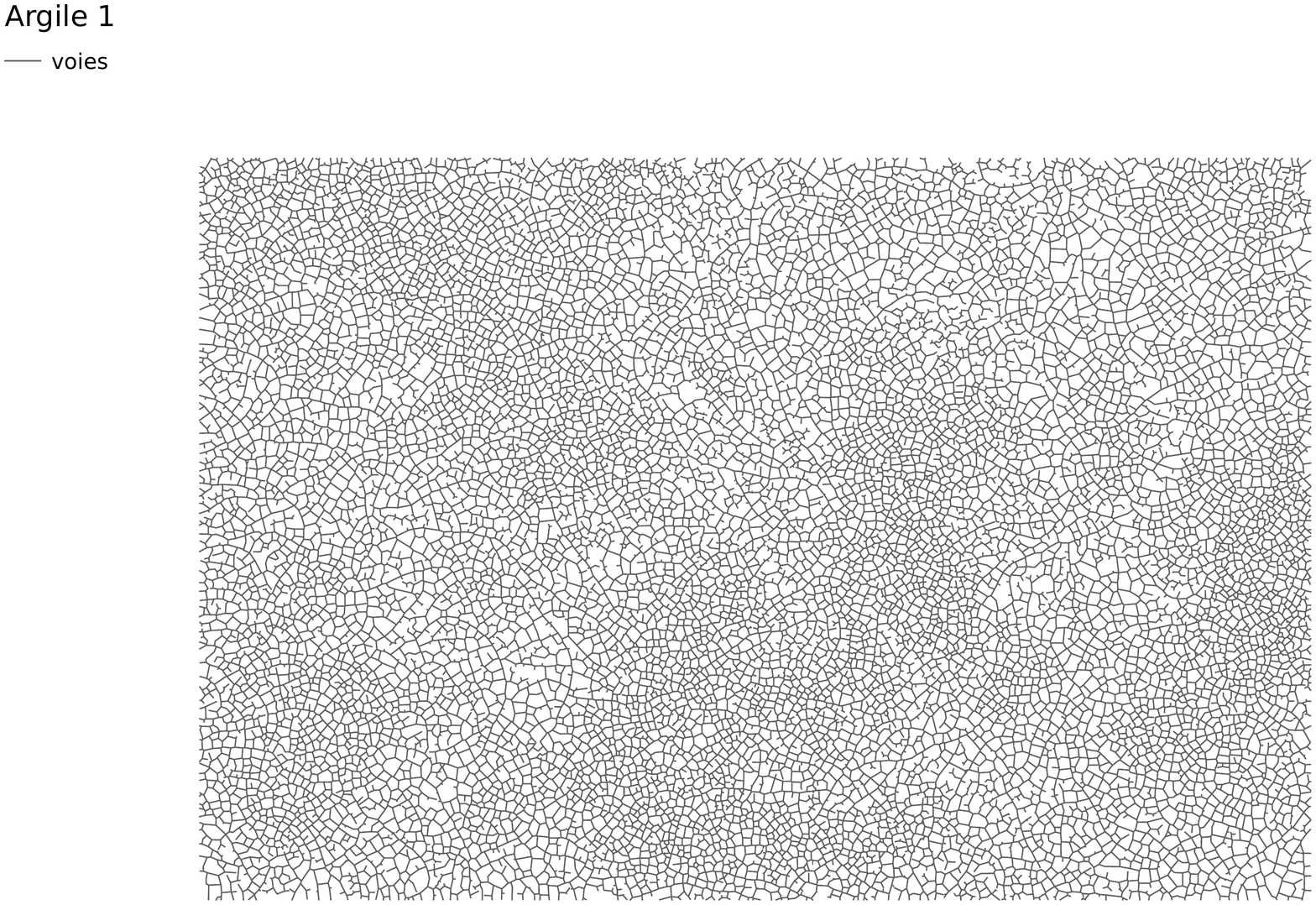}  
    
    \caption{Graphe issu de réseaux de craquelures sur des plaques d'argile (Échantillon 1).}
    \label{fig:brut_argile1}
\end{figure}

\begin{figure}[h]
    \centering
        \includegraphics[width=\textwidth]{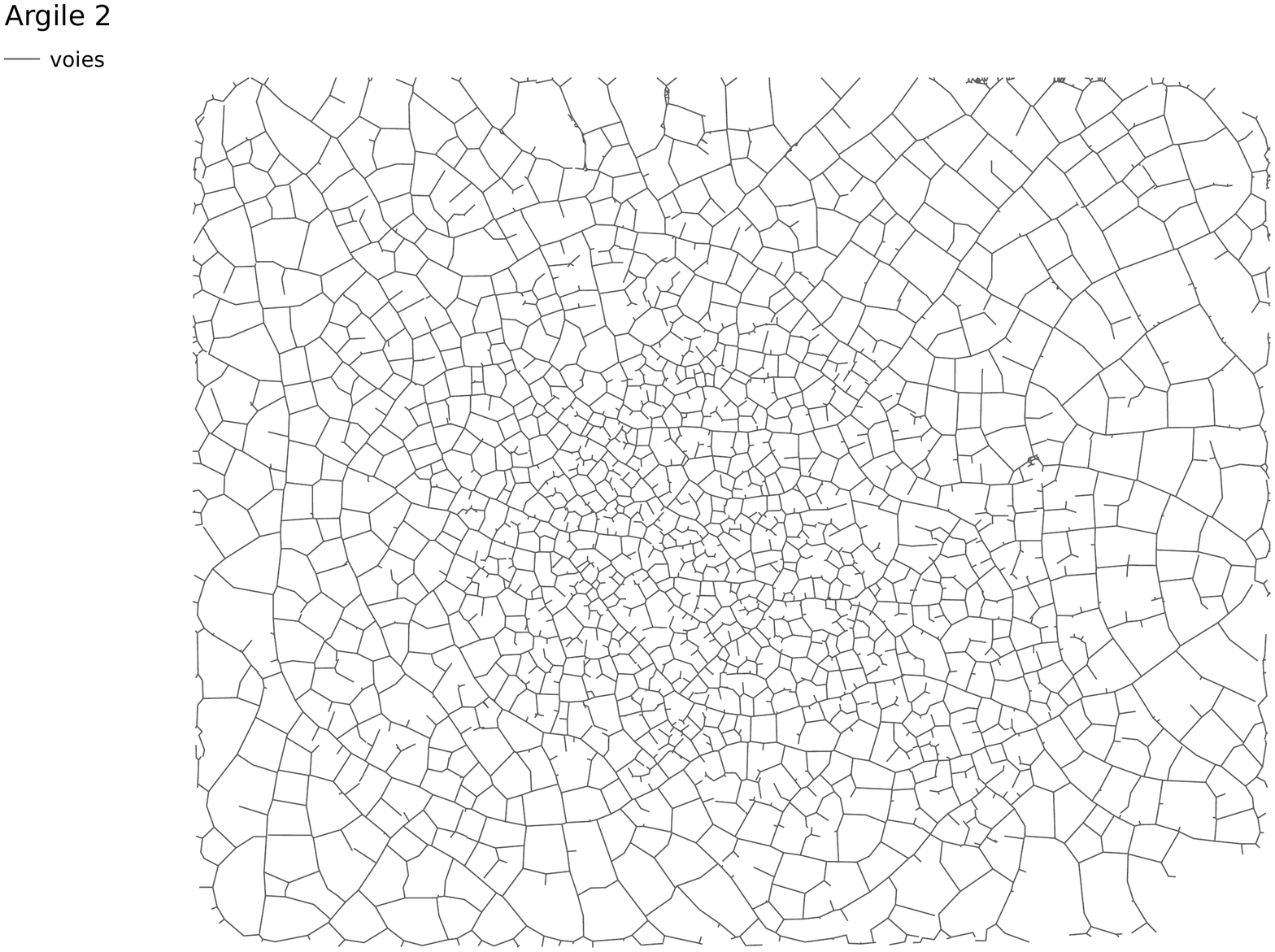}
    
    \caption{Graphe issu de réseaux de craquelures sur des plaques d'argile (Échantillon 2).}
    \label{fig:brut_argile2}
\end{figure}

\begin{figure}[h]
    \centering
        \includegraphics[width=\textwidth]{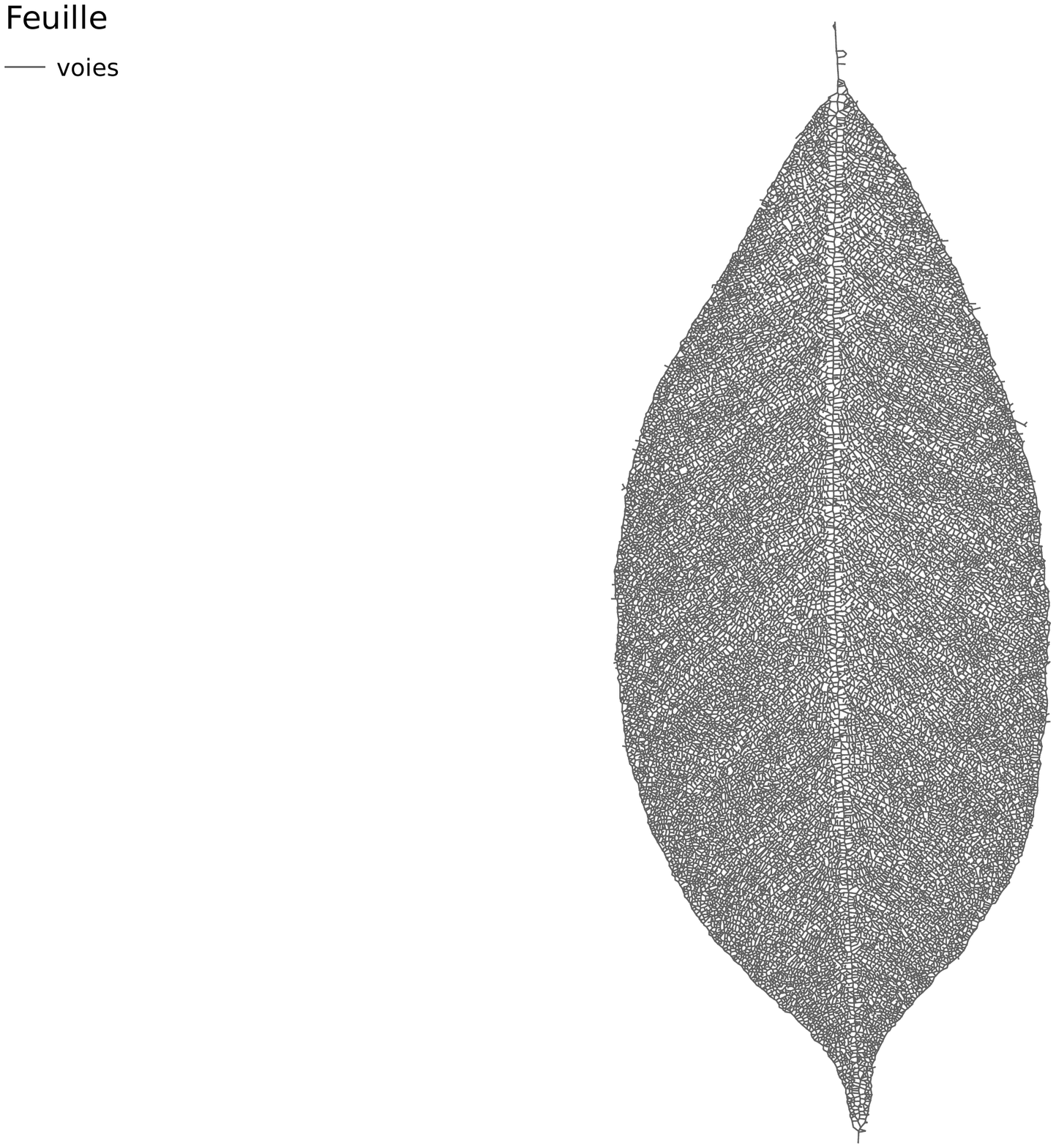}
        
    \caption{Graphe issu de réseaux de tissus biologiques : feuille.}
    \label{fig:brut_bio1}
\end{figure}

\begin{figure}[h]
    \centering
        \includegraphics[width=\textwidth]{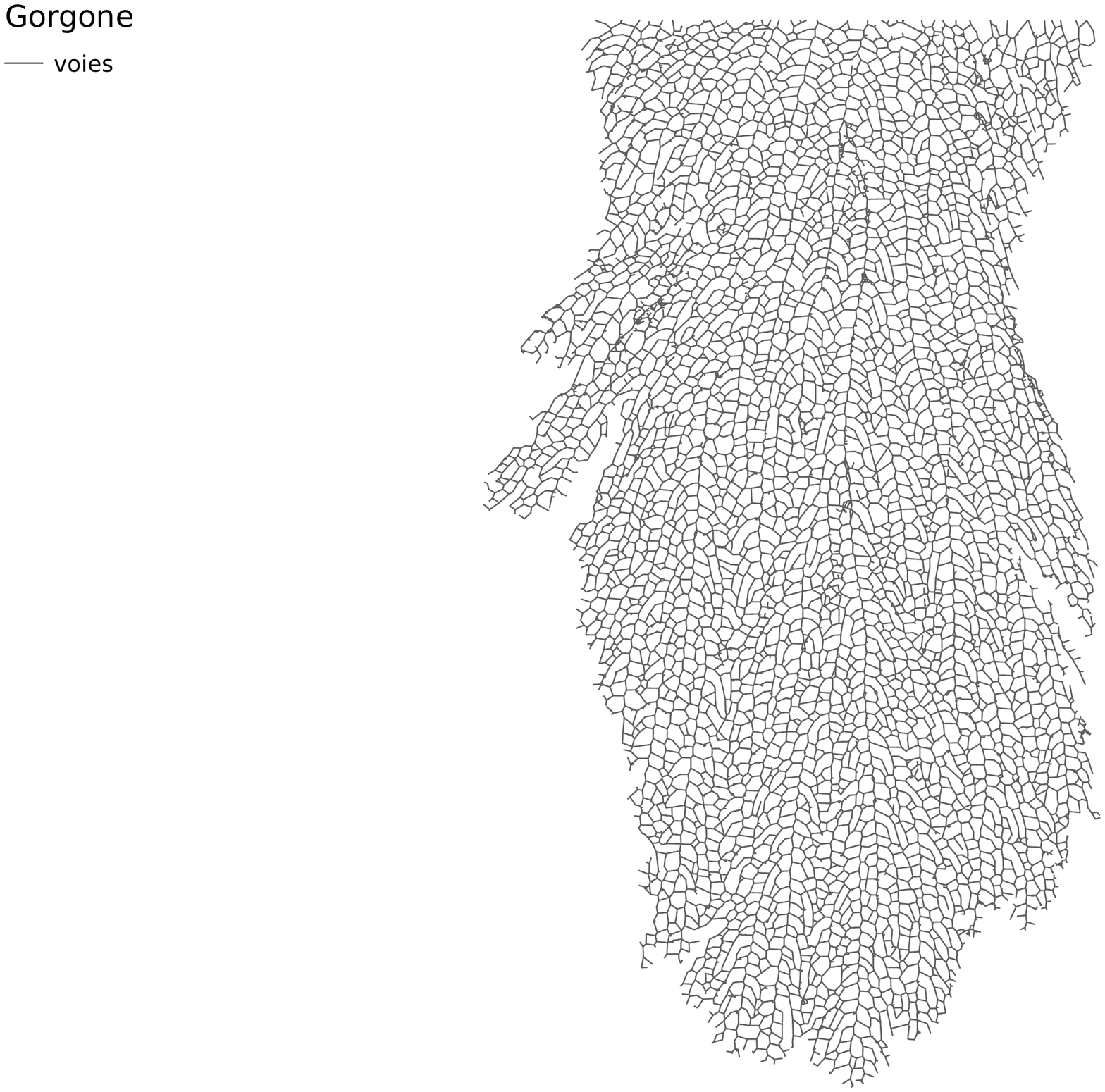}
        
    \caption{Graphe issu de réseaux de tissus biologiques : gorgone.}
    \label{fig:brut_bio2}
\end{figure}

Ces graphes ont pour but d'appuyer les différences ou similitudes topologiques ou géométriques pouvant exister entre réseaux spatiaux. Ils sont regroupés en quatre catégories :

\begin{enumerate}
\item Graphes issus de craquelures dans de l'argile (figures \ref{fig:brut_argile1}, \ref{fig:brut_argile2})
\item Graphes issus de veinures biologiques (figures \ref{fig:brut_bio1}, \ref{fig:brut_bio2})
\item Graphes issus de réseaux hydrographiques (figure \ref{fig:brut_water})
\item Graphes issus de réseaux ferrés (figure \ref{fig:brut_ferre})
\end{enumerate}

Les réseaux ferrés et hydrologiques rassemblent de très importantes longueurs métriques. Leur structures sont particulières, formées au cours des siècles par la circulation de l'eau (figure \ref{fig:brut_water}) ou pensées par l'homme pour desservir l'ensemble d'un État (figure \ref{fig:brut_ferre}). Tout comme pour le réseau viaire, ces réseaux sont une image d'un état à un moment fixé par la date d'acquisition des données. Celles que nous utilisons pour ces six réseaux sont issues d'OpenStreetMap. Elles sont donc le fruit d'une œuvre collaborative, avec ce que cela comporte d'approximations. Néanmoins, nous utilisons ces réseaux comme témoins d'une situation, leur précision de vectorisation a donc peu d'importance tant que la forme topologique et topographique reste caractéristique de l'objet étudié.

\FloatBarrier

Les réseaux de craquelures sur les plaques d'argile ont été créés par Philippe Bonnin selon le protocole expliqué dans le chapitre précédent. Le graphe de la figure \ref{fig:brut_argile2} est celui que nous avons utilisé pour étudier l'impact des généralisations. Pour ces réseaux ainsi que pour ceux issus de tissus biologiques (feuille et gorgone) nous avons procédé à une numérisation automatique, faite par Auguste Bonnin, à partir de photographies. Suite à cela, pour pouvoir traiter efficacement les réseaux via notre programme, nous les avons nettoyés en généralisant le filaire (voir chapitre précédent). Nous n'avons ainsi conservé qu'un segment entre chaque intersection et avons utilisé des places dont le rayon est égal à une unité de la carte. Ce traitement aboutit à des graphes dont la géométrie est moins bruitée (figures \ref{fig:brut_argile1}, \ref{fig:brut_argile2} et \ref{fig:brut_bio1}, \ref{fig:brut_bio2}).

\clearpage

\begin{figure}[h]
    \centering
        \includegraphics[width=\textwidth]{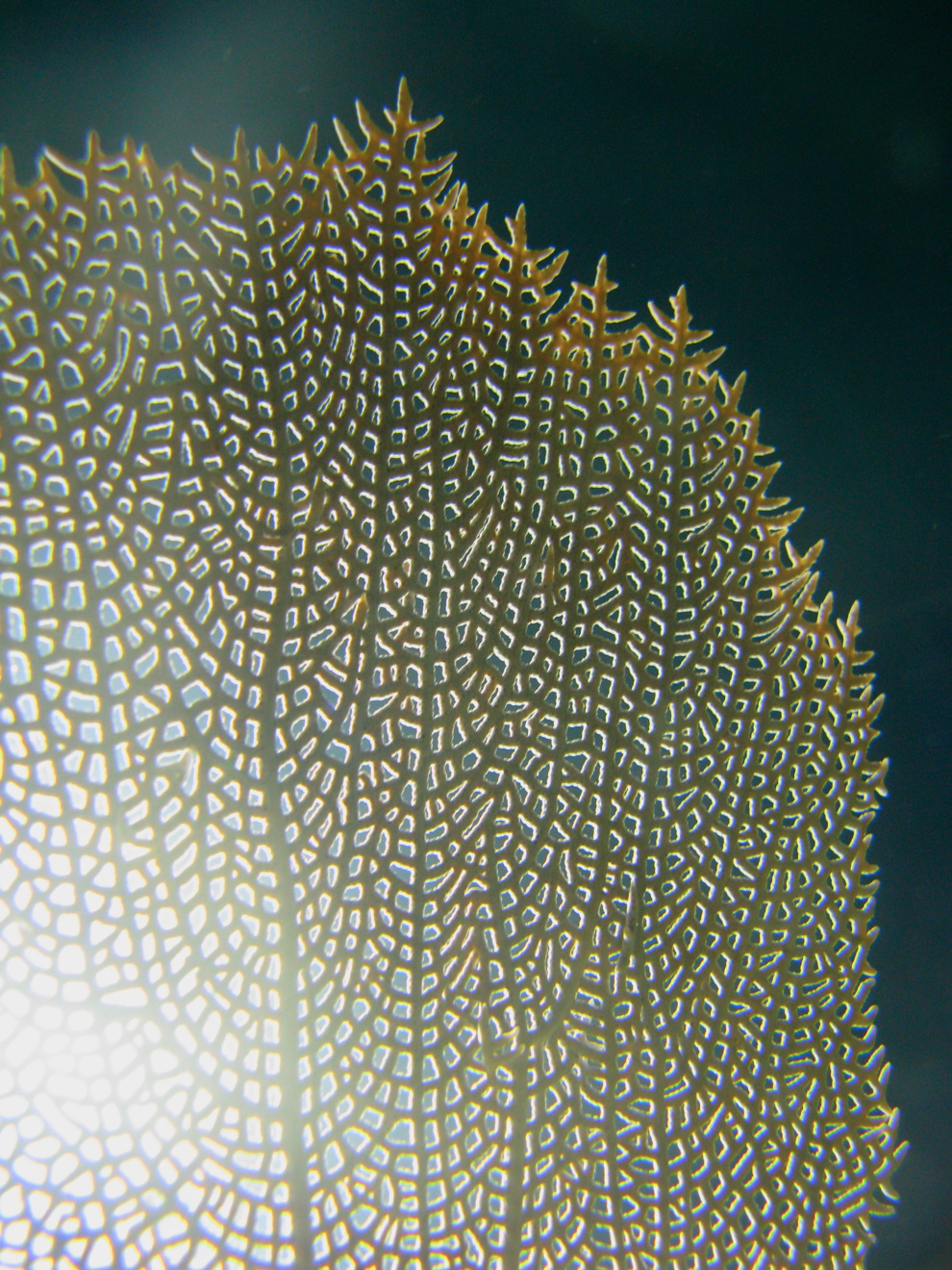}
        
    \caption{Photographie d'une gorgone. \\crédit : S. Douady}
    \label{fig:brut_bio_image}
\end{figure}

\clearpage

\FloatBarrier
\subsection{Vue d'ensemble}

Notre panel regroupe six grands types de graphes, dont, pour les graphes viaires, trois sous-types :

\begin{enumerate}
\item Graphes viaires (25)
\begin{enumerate}
\item Quartiers (Cucq, Lille, Neuf-Brisach, la Roche-sur-Yon et Vitry-sur-Seine)
\item Villes planifiées, entièrement ou en partie (Barcelone, Manhattan, San-Francisco, Kyoto)

\item Villes organiques 

\begin{enumerate}
\item Capitales (Bruxelles, Londres, Nairobi, Paris, Téhéran)
\item Grandes (Casablanca, Manaus, Rotterdam)
\item Moyennes (Bordeaux, Santa-Fe, Varanasi)
\item Petites (Avignon, Brive-la-Gaillarde, Cergy-Pontoise, Cuzco, Villers-sur-Mer)
\end{enumerate}

\end{enumerate}
\item Graphes artificiels (5)
\item Graphes issus de craquelures dans de l'argile (2)
\item Graphes issus de veinures biologiques (2)
\item Graphes issus de réseaux hydrographiques (3)
\item Graphes issus de réseaux ferrés (3)
\end{enumerate}

Ces 40 réseaux spatiaux étant de tailles et de découpages différents, il sera nécessaire de normaliser les indicateurs qui leur seront appliqués pour les rendre comparables. Les tableaux \ref{tab:caract_ville}, \ref{tab:caract_quartiers}, \ref{tab:caract_artif}, \ref{tab:caract_divers} regroupent l'ensemble des caractéristiques topologiques et métriques de ces graphes. Notre but ici est d'identifier les caractéristiques communes entre ces différents réseaux, et de mettre en évidence ce qui les oppose.

\begin{table}
\begin{center}
{ \small
\begin{tabular}{|c|r|r|r|r|}
\hline
nom & $L_{tot}$ & $N_{arcs}$ & $N_{sommets}$ & $N_{voies}$ \\ \hline 

\multicolumn{5}{|c|}{Réseaux viaires}   \\ \hline

 Paris & 2 112 km & 30 957 & 17 222 & 6 893		   \\ \hline
 Avignon & 949 km & 13 221 & 8 428 & 4 045   \\ \hline
 Bordeaux & 2 243 km & 23 614 & 15 285 & 6 839   \\ \hline
 Brive & 310 km & 3152 & 2 127 & 942   \\ \hline
 Cergy-Pontoise & 1 157 km & 14 714 & 9 189 & 4 223   \\ \hline
 Villers-sur-Mer & 66 km & 557 & 377 & 181   \\ \hline
 Bruxelles & 4 744 km & 57 624 & 37 706 & 15 307   \\ \hline
 Londres & 9 013 km & 142 288 & 98 105 & 42 384   \\ \hline
 Barcelone & 2 066 km & 30 003 & 17 491 & 6 029   \\ \hline
 Rotterdam & 3 326 km & 44 814 & 27 451 & 10 985   \\ \hline
 Manhattan & 3 379 km & 10 152 & 5 323 & 1 030   \\ \hline
 San-Francisco & 2 234 km & 24 784 & 14 886 & 4 493   \\ \hline
 Santa-Fe & 1 800 km & 9586 & 6 984 & 3 063   \\ \hline
 Manaus & 3 810 km & 36997 & 24 139 & 9 718   \\ \hline
 Cuzco & 644 km & 6225 & 4 059 & 1 552   \\ \hline
 Téhéran & 9 514 km & 101 057 & 72 112 & 29 908   \\ \hline
 Varanasi & 2 603 km & 9 952 & 7 531 & 3 310   \\ \hline
 Kyoto & 2 228 km & 36 091 & 23 170 & 9 222   \\ \hline
 Casablanca & 3 171 km & 29 193 & 18 006 & 7 139   \\ \hline
 Nairobi & 6 893 km & 29 596 & 21 485 & 9 582   \\ \hline
 
\end{tabular}
}
\end{center}
\caption{Détail des attributs topologiques et métriques de chaque réseau viaire issu de villes.}
\label{tab:caract_ville}
\end{table}
\begin{table}
\begin{center}
{ \small
\begin{tabular}{|c|r|r|r|r|}
\hline
nom & $L_{tot}$ & $N_{arcs}$ & $N_{sommets}$ & $N_{voies}$ \\ \hline 

\multicolumn{5}{|c|}{Réseaux viaires (quartiers)}   \\ \hline

Neuf-Brisach & 13 km & 189 & 99 & 30   \\ \hline
Cucq & 38 km & 318 & 193 & 56   \\ \hline
Lille & 29 km & 374 & 264 & 125   \\ \hline
Vitry-le-François & 17 km & 343 & 218 & 102   \\ \hline
La Roche-sur-Yon & 33 km & 540 & 359 & 133   \\ \hline
 
\end{tabular}
}
\end{center}
\caption{Détail des attributs topologiques et métriques de chaque réseau viaire issu de quartiers.}
\label{tab:caract_quartiers}
\end{table}
\begin{table}
\begin{center}
{ \small
\begin{tabular}{|c|r|r|r|r|}
\hline
nom & $L_{tot}$ & $N_{arcs}$ & $N_{sommets}$ & $N_{voies}$ \\ \hline 

\multicolumn{5}{|c|}{Réseaux artificiels}   \\ \hline

Bruit nul & 66 unités & 2 112 & 1 085 & 63   \\ \hline
Bruit faible & 66 unités & 3 072 & 2 045 & 1 023   \\ \hline
Bruit fort & 66 unités & 3 073 & 2 046 & 1 024   \\ \hline
Avec Générations & 53 unités & 3 073 & 2 046 & 1024    \\ \hline
Avec Angles & 80 unités & 3 073 & 2 046 & 1 024   \\ \hline

\end{tabular}
}
\end{center}
\caption{Détail des attributs topologiques et métriques de chaque réseau artificiel.}
\label{tab:caract_artif}
\end{table}
\begin{table}
\begin{center}
{ \small
\begin{tabular}{|c|r|r|r|r|}
\hline
nom & $L_{tot}$ & $N_{arcs}$ & $N_{sommets}$ & $N_{voies}$ \\ \hline 

\multicolumn{5}{|c|}{Réseaux de craquelures}   \\ \hline
argile 1 & 635 959 unités & 31 069 & 23 887 & 10 818   \\ \hline
argile 2 & 122 119 unités & 5 013 & 7 085 & 2 004   \\ \hline

\multicolumn{5}{|c|}{Réseaux biologiques}   \\ \hline
feuille & 496 380 unités & 71 349 & 48 156 & 22 823   \\ \hline
gorgone & 264 637 unités & 13 979 & 10 080 & 4 812   \\ \hline

\multicolumn{5}{|c|}{Réseaux hydrographiques}   \\ \hline
Amazonie & 21 437 km & 2 573 & 9 179 & 1 205   \\ \hline
Italie (Nord) & 1 923 km & 5 077 & 7 251 & 2 408   \\ \hline
Inde (Sud) & 6 539 km & 940 & 1 587 & 295   \\ \hline

\multicolumn{5}{|c|}{Réseaux ferrés}   \\ \hline
Belgique & 9 621 km & 28 873 & 15 585 & 6 346   \\ \hline
Californie & 16 914 km & 32 168 & 21 129 & 9 565   \\ \hline
Russie & 67 301 km & 90 441 & 69 385 & 31 890   \\ \hline

\end{tabular}
}
\end{center}
\caption{Détail des attributs topologiques et métriques des réseaux autres que viaires ou artificiels.}
\label{tab:caract_divers}
\end{table}

\FloatBarrier
\section{Indicateurs de comparaison quantitative}

Pour chacun des graphes du panel nous construisons les voies avec la paramétrisation retenue dans la première partie : méthode retenant les couples d'arcs dont l'angle de déviation est minimum à chaque sommet ($M0$) et angle seuil fixé à 60\degres . Sur cet hypergraphe de voies, nous calculons l'ensemble des indicateurs primaires définis dans la grammaire de lecture de la spatialité (Partie 1, chapitre 4). Nous effectuons ensuite une moyenne de la valeur de ces indicateurs sur l'ensemble du réseau. Pour la voie, le seul indicateur calculé en tenant compte de l'ensemble du réseau qui s'est révélé être pertinent est la closeness. Nous calculons donc sa valeur pour chaque voie en la multipliant par le nombre total de voies dans le réseau considéré afin d'obtenir des valeurs comparables d'un graphe à l'autre (équation \ref{eq:23_clo_n}).

\begin{equation}
closeness_{normalisee}(v_{ref})= \frac{N_{voies}}{\sum_{v \in G} d_{simple}(v,v_{ref})}
\label{eq:23_clo_n}
\end{equation}

Nous retenons pour chaque graphe, ses caractéristiques métriques et topologiques fondamentales :

\begin{itemize}
\item sa longueur totale : $L_{tot}$
\item son nombre d'arcs : $N_{arcs}$
\item son nombre de sommets : $N_{sommets}$
\end{itemize}

Pour caractériser les intersections des réseaux, nous calculons pour chaque graphe :

\begin{itemize}
\item son nombre d'impasses : $N_{deg_1}$ (pourcentage : $\%deg_1$)
\item son nombre de sommets de degré 3 : $N_{deg_3}$ (pourcentage : $\%deg_3$)
\item le degré moyen de l'ensemble de ses sommets : $\overline{deg(sommets)}$
\item le degré moyen de ses sommets sans prendre en compte les impasses : $\overline{deg_{\>3}(sommets)}$
\end{itemize}

Dans les graphes que nous étudions, il n'y a pas de sommets de degré 2 : les changements de directions sont symbolisés par des points annexes.

Pour caractériser les voies des réseaux, nous utilisons :

\begin{itemize}
\item leur nombre : $N_{voies}$
\item leur longueur moyenne : $\overline{L_{voies}}$
\item leur degré moyen :  $\overline{deg(voies)}$
\item la moyenne de leur indicateur de closeness (normalisé) : $\overline{clo(voies)}$
\item la moyenne de leur indicateur d'orthogonalité : $\overline{ortho(voies)}$
\end{itemize}

La valeur moyenne de l'indicateur de closeness sur le réseau décrit son efficacité en terme d'accessibilité topologique. Plus le réseau est homogène, plus elle sera révélatrice de l'efficacité d'un certain type de structure. Dans un graphe \enquote{idéal}, à l'accessibilité topologique optimale, toutes les voies s'intersectent et sont donc à une distance 1 les unes des autres. La somme des distances topologiques à partir d'une voie vers l'ensemble du graphe est donc égale au nombre de voies dans le graphe moins 1 (la voie pour laquelle nous faisons le calcul). La closeness, telle que nous l'utilisons ici, est normalisée avec le nombre d'objets du graphe (équation \ref{eq:23_clo_n}). Le nombre de voies important des graphes que nous considérons nous pousse à négliger le (-1) dans la normalisation pour utiliser la définition classique de l'indicateur. La closeness moyenne des voies du graphe \enquote{idéal} est donc de 1. Si nous considérons un graphe non connexe, où toutes les distances topologiques sont infinies, la closeness moyenne des voies est de 0. Les variations entre 0 et 1 de la moyenne de cet indicateur sur un graphe sont donc utilisées pour évaluer son efficacité.

Au delà de la voie, pour caractériser le graphe dans son ensemble, nous introduisons quatre coefficients :

\begin{itemize}
\item le coefficient d'organicité : $coef_{orga}$
\item le coefficient de maillance : $coef_{mail}$
\item le coefficient d'hétérogénéité : $coef_{hete}$
\item le coefficient de réduction : $coef_{red}$
\end{itemize}

Le coefficient d'organicité a été défini par T. Courtat dans \citep{courtat2011mathematics}. Il donne la prédominance d'impasses et de nœuds de degré 3 dans le réseau (équation \ref{eq:orga}). Il permet ainsi de les différencier des nœuds de degré 4 ou plus, qui, comme nous le verrons, sont souvent le signe d'une planification du réseau viaire, ou d'une administration urbaine centralisée.

\begin{equation}
coef_{orga} = \frac{N_{deg1} + N_{deg3}}{N_{sommets}}
\label{eq:orga}
\end{equation}

Le coefficient de maillance est construit en pondérant l'indicateur d'orthogonalité par la longueur des voies (équation \ref{eq:mail}). Ainsi, la valeur obtenue se rapprochera de la longueur totale $L_{tot}$ si les voies les plus longues sont connectées perpendiculairement au reste du réseau. Ce coefficient complète la moyenne de l'indicateur d'orthogonalité, qui, lui, ne tient compte que du nombre de voies orthogonales au réseau et non de leur longueur. 

\begin{figure}[h]
    \centering
        \includegraphics[width=0.8\textwidth]{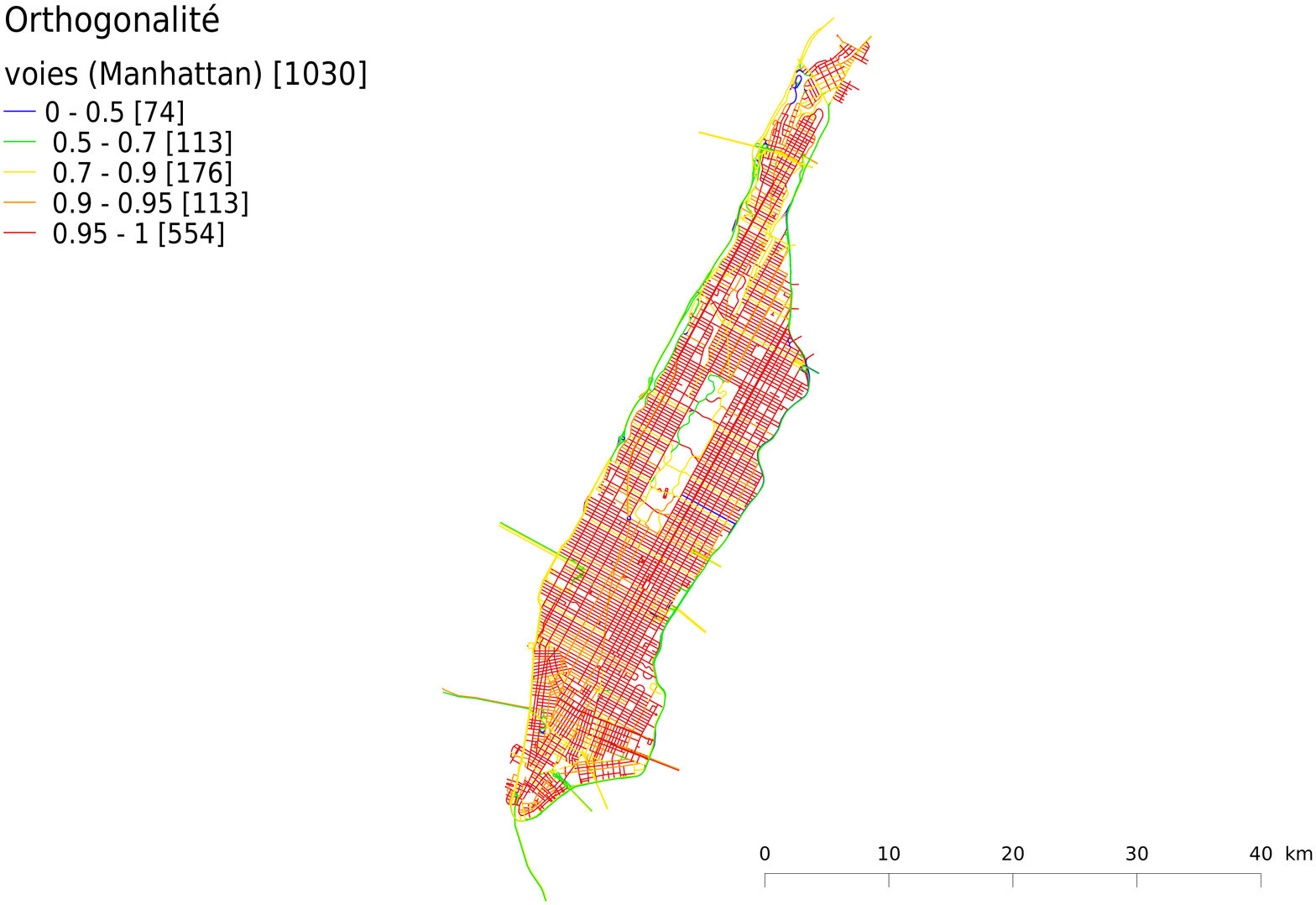}        
    \caption{Indicateur d'orthogonalité calculé sur le graphe viaire de Manhattan.}
    \label{fig:ortho_man}
\end{figure}

\begin{figure}[h]
    \centering
        \includegraphics[width=0.8\textwidth]{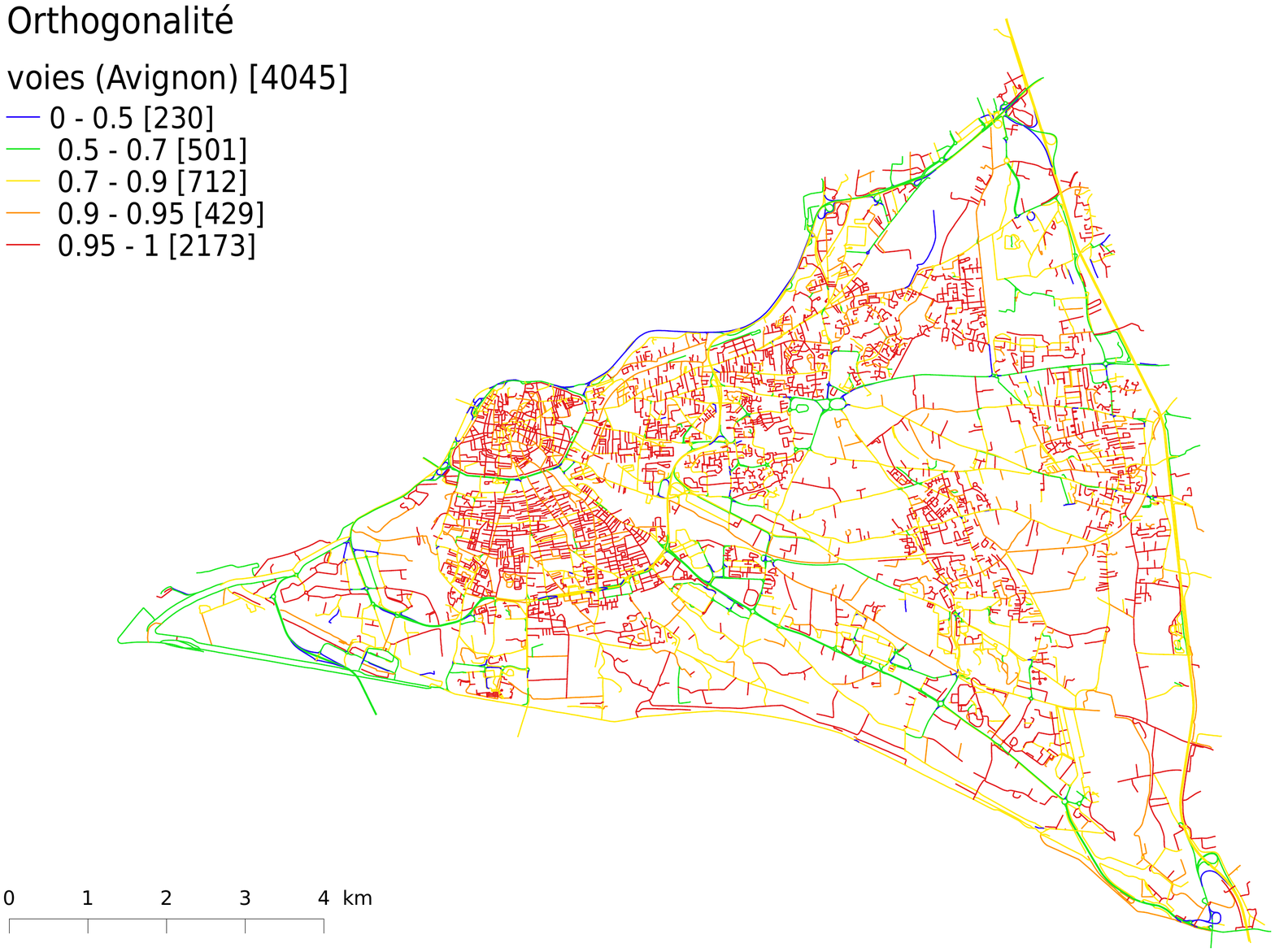}        
    \caption{Indicateur d'orthogonalité calculé sur le graphe viaire d'Avignon.}
    \label{fig:ortho_avi}
\end{figure}

\begin{figure}[h]
    \centering
        \includegraphics[width=0.8\textwidth]{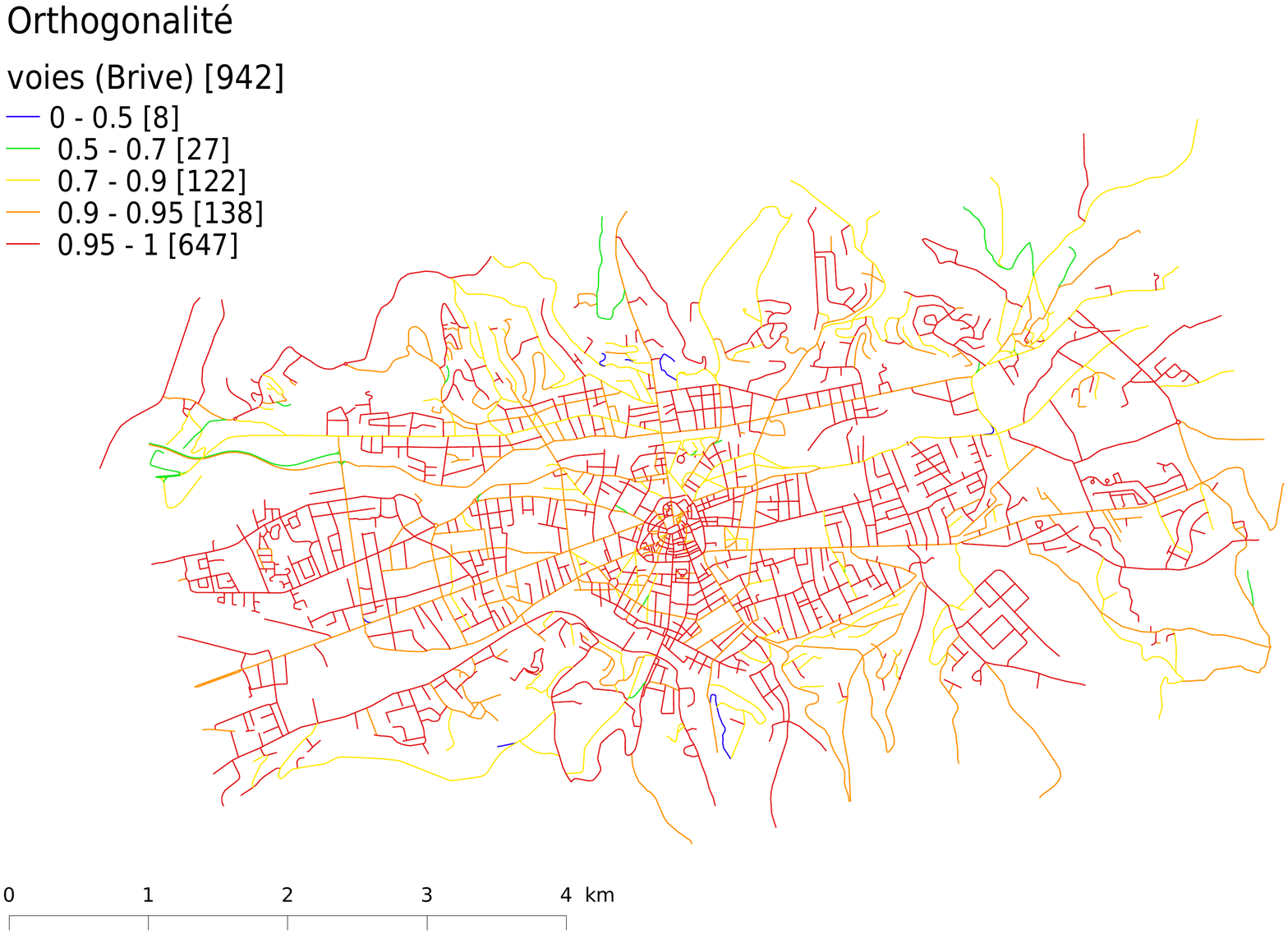}        
    \caption{Indicateur d'orthogonalité calculé sur le graphe viaire de Brive-la-Gaillarde.}
    \label{fig:ortho_bri}
\end{figure}

\begin{figure}[h]
    \centering
        \includegraphics[width=\textwidth]{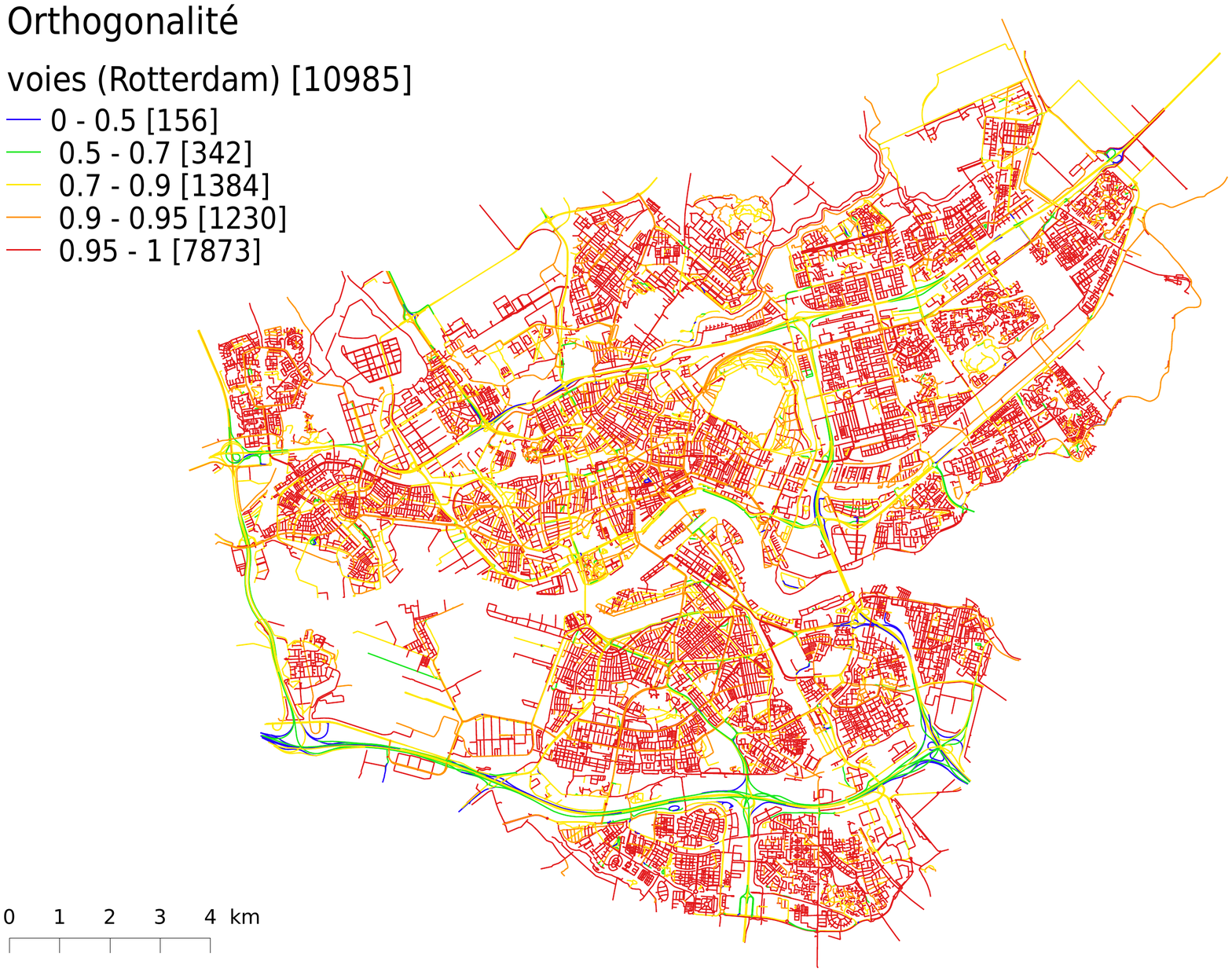}        
    \caption{Indicateur d'orthogonalité calculé sur le graphe viaire de Rotterdam.}
    \label{fig:ortho_rot}
\end{figure}

En effet, la taille du réseau et la longueur des voies peuvent jouer un rôle trompeur dans le calcul de la moyenne d'orthogonalité des voies. Ainsi, à Manhattan, nous ne considérons que l'île. Les voies qui dessinent son pourtour sont nombreuses, et forment des angles de connexions faibles avec le quadrillage régulier. Sur l'ensemble du graphe, ces voies de connexions, plus courtes, représentent environ la moitié du nombre total de voies. Lorsque nous calculons la moyenne de l'indicateur d'orthogonalité, où chaque voie a le même poids, Manhattan ressort avec un indice comparable à celui d'Avignon. En effet, si nous observons les cartes de ces deux villes et le dénombrement de leurs voies dans chaque classe d'orthogonalité, la répartition entre la classe des valeurs les plus fortes et les autres, est proche (figures \ref{fig:ortho_man} et \ref{fig:ortho_avi}). Alors que sur des villes comme Brive-la-Gaillarde ou Rotterdam, plus d'un tiers du réseau a une orthogonalité entre 0,95 et 1 (figures \ref{fig:ortho_bri} et \ref{fig:ortho_rot}). En calculant le coefficient de maillance, nous réintroduisons la longueur des voies. Nous appliquons ainsi un poids à chaque objet en fonction de sa part dans la longueur totale du réseau, ce qui estompe l'effet observé avec une moyenne non pondérée. 


\begin{equation}
coef_{mail} = \frac{\sum_{v \in G}(orthogonalite(v) \times longueur(v))}{L_{tot}}
\label{eq:mail}
\end{equation}

\FloatBarrier

La construction du coefficient d'hétérogénéité est particulière. Elle ne prend pas en compte les valeurs brutes des indicateurs mais celles des classes de longueurs équivalentes qui leur sont associées. Ces classes nous permettent de comparer des indicateurs qui, \textit{a priori}, n'ont pas le même ordre de grandeur. C'est le cas du degré et de la longueur de chaque voie. Ces deux indicateurs sont proches, nous les avons identifiés comme ayant un comportement corrélé dans la partie précédente. Leur différence est significative car elle témoigne des irrégularités du réseau. C'est notamment cela que nous avions mis en évidence avec le coefficient d'espacement. En effet, si les voies les plus longues ne sont pas les plus connectées cela signifie que la densité linéaire du réseau n'est pas homogène. Nous comparons les valeurs de dix classes de longueur équivalente dans lesquelles les voies ont été ordonnées selon la valeur de leur degré ($cl_{degre}(v)$) ou celle de leur longueur ($cl_{longueur}(v)$)(équation \ref{eq:hete}). Si la valeur absolue de la différence se rapproche de 0 ($ \vert cl_{longueur}(v) - cl_{degre}(v) \vert \rightarrow 0$), cela signifie que, pour le réseau étudié, la hiérarchisation des voies faite avec leur degré est proche de celle faite selon leur longueur. Le réseau est donc homogène. Si, au contraire, $ \vert cl_{longueur}(v) - cl_{degre}(v) \vert \rightarrow N_{classes} \times N_{voies}$ (où $N_{classes}$ représente le nombre de classes de longueur étudiées, ici 10), cela signifie que pour chaque voie, les classifications par degré et par longueur suivent un ordre différent. Cela sera donc un indice d'hétérogénéité linéaire du réseau. Nous voulons ainsi différencier les réseaux aux structures régulières de ceux dont la géométrie est plus erratique.

\begin{equation}
coef_{hete} = \frac{\sum_{v \in G} \vert cl_{longueur}(v) - cl_{degre}(v) \vert}{10 \times N_{voies}}
\label{eq:hete}
\end{equation}

Ces trois coefficients ont ainsi pour objectif de nous aider à déceler le comportement géométrique des réseaux. Que ce soit selon le degré de leurs sommets, avec le coefficient d'organicité ; selon les angles que font les voies en se connectant entre elles, avec le coefficient de maillance ; ou selon les variations de la densité linéaire du réseau, avec le coefficient d'hétérogénéité. Nous nous attendons \textit{a priori} à ce que les trois aient un comportement lié : les réseaux dont les sommets ont un degré supérieur à 3 devraient être également ceux aux connexions les plus perpendiculaires et dont l'hétérogénéité est la plus faible. Nous verrons dans la section suivante que ce n'est pas toujours le cas.

Nous considérons enfin un coefficient de réduction, évaluant la diminution du nombre d'objets du graphe lors du regroupement des arcs pour construire les voies. Nous le calculons, pour chaque graphe, en soustrayant à 1 la division de son nombre de voies par son nombre d'arcs (équation \ref{eq:red}). Plus la valeur du coefficient se rapproche de 0, moins il y aura d'arcs par voies. Cela signifiera donc que peu de continuités ont pu être reconstruites selon l'angle seuil choisi pour la construction des voies (ici de 60\degres ).

\begin{equation}
coef_{red} = 1 - (\frac{N_{voies}}{N_{arcs}})
\label{eq:red}
\end{equation}

\FloatBarrier
\section{Analyse Inter-Graphes}

\FloatBarrier
\subsection{Tableaux d'indicateurs}

\begin{table}
\begin{center}
{ \small
\begin{tabular}{|c|r|r|r|r|}
\hline
nom & $coef_{orga}$ & $coef_{mail}$ & $coef_{hete}$ & $coef_{red}$ \\ \hline 

	\multicolumn{5}{|c|}{Réseaux viaires}   \\ \hline

	Paris & 0,73 & 0,88 & 0,08 & 0,22   \\ \hline
	Avignon & 0,90 & 0,85 & 0,15 & 0,31   \\ \hline
	Bordeaux & 0,84 & 0,89 & 0,12 & 0,29   \\ \hline
	Brive & 0,88 & 0,94 & 0,14 & 0,30   \\ \hline
	Cergy-Pontoise & 0,88 & 0,90 & 0,12 & 0,29   \\ \hline
	Villers-sur-Mer & 0,90 & 0,91 & 0,16 & 0,32   \\ \hline
	Bruxelles & 0,81 & 0,89 & 0,12 & 0,27   \\ \hline
	Londres & 0,86 & 0,93 & 0,12 & 0,30   \\ \hline
	Barcelone & 0,66 & 0,87 & 0,09 & 0,20   \\ \hline
	Rotterdam & 0,81 & 0,92 & 0,10 & 0,25   \\ \hline
	Manhattan & 0,36 & 0,91 & 0,05 & 0,10   \\ \hline
	San-Francisco & 0,57 & 0,91 & 0,06 & 0,18   \\ \hline
	Santa-Fe & 0,86 & 0,90 & 0,12 & 0,32   \\ \hline
	Manaus & 0,79 & 0,93 & 0,11 & 0,26   \\ \hline
	Cuzco & 0,75 & 0,88 & 0,14 & 0,25   \\ \hline
	Téhéran & 0,81 & 0,86 & 0,10 & 0,30   \\ \hline
	Varanasi & 0,90 & 0,91 & 0,16 & 0,33   \\ \hline
	Kyoto & 0,79 & 0,96 & 0,07 & 0,26   \\ \hline
	Casablanca & 0,75 & 0,90 & 0,11 & 0,24   \\ \hline
	Nairobi & 0,92 & 0,89 & 0,15 & 0,32 \\ \hline
	
	\multicolumn{5}{|c|}{Réseaux viaires (quartiers)}   \\ \hline
	
	Neuf-Brisach & 0,59 & 0,95 & 0,18 & 0,16   \\ \hline
	Cucq & 0,52 & 0,98 & 0,11 & 0,18   \\ \hline
	Lille & 0,89 & 0,92 & 0,10 & 0,33   \\ \hline
	Vitry-le-François & 0,83 & 0,92 & 0,10 & 0,30   \\ \hline
	La Roche-sur-Yon & 0,77 & 0,94 & 0,08 & 0,25 \\ \hline

\end{tabular}
}
\end{center}
\caption{Détail des coefficients.}
\label{tab:caract_coef_viaire}
\end{table}

\begin{table}
\begin{center}
{ \small
\begin{tabular}{|c|r|r|r|r|}
\hline
nom & $coef_{orga}$ & $coef_{mail}$ & $coef_{hete}$ & $coef_{red}$ \\ \hline 
	
	\multicolumn{5}{|c|}{Réseaux artificiels}   \\ \hline
	
	Bruit nul & 0,11 & 1,00 & 0,00 & 0,03   \\ \hline
	Bruit faible & 1,00 & 1,00 & 0,09 & 0,33   \\ \hline
	Bruit fort & 1,00 & 1,00 & 0,07 & 0,33   \\ \hline
	Avec Générations & 1,00 & 1,00 & 0,16 & 0,33   \\ \hline
	Avec Angles & 1,00 & 0,68 & 0,10 & 0,33 \\ \hline
	
	\multicolumn{5}{|c|}{Réseaux de craquelures}   \\ \hline
	
	argile 1 & 0,98 & 0,95 & 0,06 & 0,35   \\ \hline
	argile 2 & 0,99 & 0,96 & 0,08 & 0,40 \\ \hline
	
	\multicolumn{5}{|c|}{Réseaux biologiques}   \\ \hline
	
	feuille & 0,93 & 0,92 & 0,06 & 0,32   \\ \hline
	gorgone & 0,96 & 0,92 & 0,08 & 0,34 \\ \hline
	
	\multicolumn{5}{|c|}{Réseaux hydrographiques}   \\ \hline
	
	Amazonie & 0,99 & 0,87 & 0,15 & 0,47   \\ \hline
	Italie (Nord) & 0,99 & 0,77 & 0,16 & 0,47   \\ \hline
	Inde (Sud) & 0,88 & 0,80 & 0,14 & 0,31 \\ \hline
	
	\multicolumn{5}{|c|}{Réseaux ferrés}   \\ \hline
	
	Belgique & 0,85 & 0,16 & 0,08 & 0,22   \\ \hline
	Californie & 0,96 & 0,16 & 0,07 & 0,30   \\ \hline
	Russie & 0,97 & 0,17 & 0,08 & 0,35 \\ \hline
 
\end{tabular}
}
\end{center}
\caption{Détail des coefficients.}
\label{tab:caract_coef_autre}
\end{table}

\begin{table}
\begin{center}
{ \small
\begin{tabular}{|c|r|r|r|r|r|r|r|}
\hline
nom & $N_{sommets}$ & $N_{deg_1}$ & $\%deg_1$ & $N_{deg_3}$ & $\%deg_3$ & $\overline{deg(sommets)}$ & $\overline{deg_{\>3}(sommets)}$ \\ \hline

	\multicolumn{8}{|c|}{Réseaux viaires}   \\ \hline
	Paris & 17 222 & 1 894 & 11,00 & 10 637 & 61,76 & 3,08 & 3,34   \\ \hline
	Avignon & 8 428 & 1 709 & 20,28 & 5 883 & 69,80 & 2,70 & 3,13   \\ \hline
	Bordeaux & 15 285 & 2 036 & 13,32 & 10 749 & 70,32 & 2,91 & 3,20   \\ \hline
	Brive & 2 127 & 436 & 20,50 & 1 434 & 67,42 & 2,72 & 3,16   \\ \hline
	Cergy-Pontoise & 9 189 & 2 554 & 27,79 & 5 491 & 59,76 & 2,57 & 3,18   \\ \hline
	Villers-sur-Mer & 377 & 82 & 21,75 & 258 & 68,44 & 2,67 & 3,14   \\ \hline
	Bruxelles & 37 706 & 5 691 & 15,09 & 24 992 & 66,28 & 2,89 & 3,23   \\ \hline
	Londres & 98 105 & 21 494 & 21,91 & 63 189 & 64,41 & 2,71 & 3,18   \\ \hline
	Barcelone & 17 491 & 1 739 & 9,94 & 9 741 & 55,69 & 3,16 & 3,40   \\ \hline
	Rotterdam & 27 451 & 5 078 & 18,50 & 17 023 & 62,01 & 2,83 & 3,24   \\ \hline
	Manhattan & 5 323 & 185 & 3,48 & 1 706 & 32,05 & 3,59 & 3,69   \\ \hline
	San-Francisco & 14 886 & 1 872 & 12,58 & 6 600 & 44,34 & 3,20 & 3,52   \\ \hline
	Santa-Fe & 6 984 & 1 868 & 26,75 & 4 118 & 58,96 & 2,61 & 3,20   \\ \hline
	Manaus & 24 139 & 3 448 & 14,28 & 15 618 & 64,70 & 2,93 & 3,25   \\ \hline
	Cuzco & 4 059 & 538 & 13,25 & 2 503 & 61,67 & 3,00 & 3,31   \\ \hline
	Téhéran & 72 112 & 16 698 & 23,16 & 41 910 & 58,12 & 2,73 & 3,25   \\ \hline
	Varanasi & 7 531 & 1 910 & 25,36 & 4 862 & 64,56 & 2,59 & 3,14   \\ \hline
	Kyoto & 23 170 & 2 402 & 10,37 & 15 808 & 68,23 & 3,01 & 3,24   \\ \hline
	Casablanca & 18 006 & 989 & 5,49 & 12 602 & 69,99 & 3,15 & 3,27   \\ \hline
	Nairobi & 21 485 & 5 921 & 27,56 & 13 775 & 64,11 & 2,53 & 3,12 \\ \hline
	
	\multicolumn{8}{|c|}{Réseaux viaires (quartiers)}   \\ \hline
	Neuf-Brisach & 99 & 10 & 10,10 & 48 & 48,48 & 3,21 & 3,46   \\ \hline
	Cucq & 193 & 24 & 12,44 & 76 & 39,38 & 3,23 & 3,55   \\ \hline
	Lille & 264 & 53 & 20,08 & 181 & 68,56 & 2,71 & 3,14   \\ \hline
	Vitry-le-François & 218 & 22 & 10,09 & 160 & 73,39 & 3,00 & 3,22   \\ \hline
	La Roche-sur-Yon & 359 & 39 & 10,86 & 239 & 66,57 & 3,01 & 3,26 \\ \hline
	
\end{tabular}
}
\end{center}
\caption{Détail des statistiques liées aux sommets.}
\label{tab:caract_som_viaire}
\end{table}

\begin{table}
\begin{center}
{ \small
\begin{tabular}{|c|r|r|r|r|r|r|r|}
\hline
nom & $N_{sommets}$ & $N_{deg_1}$ & $\%deg_1$ & $N_{deg_3}$ & $\%deg_3$ & $\overline{deg(sommets)}$ & $\overline{deg_{\>3}(sommets)}$ \\ \hline 
	
	\multicolumn{8}{|c|}{Réseaux artificiels}   \\ \hline
	Bruit nul & 1 085 & 0 & 0,00 & 124 & 11,43 & 3,89 & 3,89   \\ \hline
	Bruit faible & 2 045 & 0 & 0,00 & 2 044 & 99,95 & 3,00 & 3,00   \\ \hline
	Bruit fort & 2 046 & 0 & 0,00 & 2 046 & 100,00 & 3,00 & 3,00   \\ \hline
	Avec Générations & 2 046 & 0 & 0,00 & 2 046 & 100,00 & 3,00 & 3,00   \\ \hline
	Avec Angles & 2 046 & 0 & 0,00 & 2 046 & 100,00 & 3,00 & 3,00 \\ \hline
	
	\multicolumn{8}{|c|}{Réseaux de craquelures}   \\ \hline
	argile 1 & 23 887 & 6 241 & 26,13 & 17 263 & 72,27 & 2,49 & 3,02   \\ \hline
	argile 2 & 7 085 & 3 969 & 56,02 & 3 037 & 42,87 & 1,89 & 3,03 \\ \hline
	
	\multicolumn{8}{|c|}{Réseaux biologiques}   \\ \hline
	feuille & 48156 & 10 051 & 20,87 & 34 958 & 72,59 & 2,65 & 3,09   \\ \hline
	gorgone & 10080 & 12 02 & 11,92 & 8 520 & 84,52 & 2,80 & 3,04 \\ \hline
	
	\multicolumn{8}{|c|}{Réseaux hydrographiques}   \\ \hline
	Amazonie & 9 179 & 4 933 & 53,74 & 4 120 & 44,89 & 1,94 & 3,03   \\ \hline
	Italie (Nord) & 7 251 & 3734 & 51,50 & 3 471 & 47,87 & 1,98 & 3,01   \\ \hline
	Inde (Sud) & 1 587 & 930 & 58,60 & 467 & 29,43 & 1,95 & 3,29 \\ \hline
	
	\multicolumn{8}{|c|}{Réseaux ferrés}   \\ \hline
	Belgique & 15 585 & 3 641 & 23,36 & 9 643 & 61,87 & 2,69 & 3,20   \\ \hline
	Californie & 21 129 & 5 945 & 28,14 & 14 379 & 68,05 & 2,48 & 3,05   \\ \hline
	Russie & 69 385 & 23 657 & 34,10 & 43 325 & 62,44 & 2,36 & 3,06 \\ \hline

\end{tabular}
}
\end{center}
\caption{Détail des statistiques liées aux sommets.}
\label{tab:caract_som_autre}
\end{table}

\begin{table}
\begin{center}
{ \small
\begin{tabular}{|c|r|r|r|r|r|}
\hline
nom & $N_{voies}$ & $\overline{L_{voies}}$ & $\overline{deg(voies)}$ & $\overline{clo(voies)}$ & $\overline{ortho(voies)}$  \\ \hline
 
	\multicolumn{6}{|c|}{Réseaux viaires}   \\ \hline
	Paris & 6 893 & 306,50 & 4,92 & 0,17 & 0,90  \\ \hline
	Avignon & 4 045 & 234,71 & 3,47 & 0,13 & 0,87  \\ \hline
	Bordeaux & 6 839 & 328,03 & 3,98 & 0,13 & 0,91  \\ \hline
	Brive & 942 & 329,94 & 3,61 & 0,21 & 0,94  \\ \hline
	Cergy-Pontoise & 4 223 & 274,13 & 3,21 & 0,10 & 0,93  \\ \hline
	Villers-sur-Mer & 181 & 365,95 & 3,43 & 0,20 & 0,92  \\ \hline
	Bruxelles & 15 307 & 309,97 & 4,22 & 0,13 & 0,91  \\ \hline
	Londres & 42 384 & 212,67 & 3,71 & 0,10 & 0,93  \\ \hline
	Barcelone & 6 029 & 342,73 & 5,45 & 0,15 & 0,88  \\ \hline
	Rotterdam & 10 985 & 302,84 & 4,10 & 0,12 & 0,94  \\ \hline
	Manhattan & 1 030 & 3280,71 & 10,50 & 0,29 & 0,86  \\ \hline
	San-Francisco & 4 493 & 497,24 & 6,19 & 0,19 & 0,90  \\ \hline
	Santa-Fe & 3 063 & 587,75 & 3,29 & 0,17 & 0,92  \\ \hline
	Manaus & 9 718 & 392,12 & 4,31 & 0,12 & 0,95  \\ \hline
	Cuzco & 1 552 & 415,32 & 4,78 & 0,18 & 0,89  \\ \hline
	Téhéran & 29 908 & 318,14 & 3,79 & 0,12 & 0,92  \\ \hline
	Varanasi & 3 310 & 786,55 & 3,33 & 0,15 & 0,95  \\ \hline
	Kyoto & 9 222 & 241,64 & 4,56 & 0,19 & 0,97  \\ \hline
	Casablanca & 7 139 & 444,24 & 4,93 & 0,13 & 0,93  \\ \hline
	Nairobi & 9 582 & 719,46 & 3,22 & 0,11 & 0,93 \\ \hline
	
	\multicolumn{6}{|c|}{Réseaux viaires (quartiers)}   \\ \hline
	Neuf-Brisach & 30 & 439,93 & 5,93 & 0,47 & 0,94  \\ \hline
	Cucq & 56 & 694,29 & 6,21 & 0,41 & 0,98  \\ \hline
	Lille & 125 & 238,99 & 3,49 & 0,26 & 0,92  \\ \hline
	Vitry-le-François & 102 & 171,17 & 4,37 & 0,32 & 0,91  \\ \hline
	La Roche-sur-Yon & 133 & 253,09 & 4,71 & 0,29 & 0,93 \\ \hline

\end{tabular}
}
\end{center}
\caption{Détail des statistiques liées aux voies.}
\label{tab:caract_voies_viaire}
\end{table}

\begin{table}
\begin{center}
{ \small
\begin{tabular}{|c|r|r|r|r|r|}
\hline
nom & $N_{voies}$ & $\overline{L_{voies}}$ & $\overline{deg(voies)}$ & $\overline{clo(voies)}$ & $\overline{ortho(voies)}$  \\ \hline
	
	\multicolumn{6}{|c|}{Réseaux artificiels}   \\ \hline
	Bruit nul & 63 & 1 unité & 34,46 & 0,69 & 1,00  \\ \hline
	Bruit faible & 1 023 & 0,64 unité & 4,00 & 0,23 & 1,00  \\ \hline
	Bruit fort & 1 024 & 0,64 & 4,00 & 0,23 & 1,00  \\ \hline
	Avec Générations & 1 024 & 0,52 & 4,00 & 0,25 & 1,00  \\ \hline
	Avec Angles & 1 024 & 0,78 & 4,00 & 0,22 & 0,68 \\ \hline
	
	\multicolumn{6}{|c|}{Réseaux de craquelures}   \\ \hline
	argile 1 & 10 818 & 58,79 & 3,27 & 0,06 & 0,94  \\ \hline
	argile 2 & 2 004 & 60,94 & 2,81 & 0,16 & 0,93 \\ \hline
	
	\multicolumn{6}{|c|}{Réseaux biologiques}   \\ \hline
	feuille & 22 823 & 21,75 & 3,47 & 0,06 & 0,91  \\ \hline
	gorgone & 4 812 & 54 unité & 3,79 & 0,06 & 0,90 \\ \hline
	
	\multicolumn{6}{|c|}{Réseaux hydrographiques}   \\ \hline
	Amazonie & 1 205 & 17790,23 & 2,01 & 0,07 & 0,86  \\ \hline
	Italie (Nord) & 2 408 & 798,78 & 2,08 & 0,08 & 0,75  \\ \hline
	Inde (Sud) & 295 & 22169,47 & 2,74 & 0,13 & 0,83 \\ \hline
	
	\multicolumn{6}{|c|}{Réseaux ferrés}   \\ \hline
	Belgique & 6 346 & 1516,15 & 3,75 & 0,10 & 0,11  \\ \hline
	Californie & 9 565 & 1768,34 & 3,05 & 0,08 & 0,12  \\ \hline
	Russie & 31 890 & 2110,41 & 2,87 & 0,04 & 0,14 \\ \hline

\end{tabular}
}
\end{center}
\caption{Détail des statistiques liées aux voies.}
\label{tab:caract_voies_autre}
\end{table}

\FloatBarrier
\subsection{Graphes artificiels}

Les graphes que nous avons construits artificiellement nous permettent d'avoir un aperçu \enquote{neutre} sur les coefficients que nous utilisons afin de mieux apprécier les résultats qu'ils nous donnent.

Le graphe \enquote{Bruit nul} est une grille simple. Sur ce réseau, il n'y a aucune impasse. Le degré moyen des sommets est de 3.89 : tous les sommets à l'intérieur de la grille ayant un degré 4 et ceux sur le contour un degré 3. Le grand nombre de nœuds de degré 4 aboutit à un coefficient d'organicité faible (0,11). Le coefficient de réduction est de 0,97 : les voies créées regroupent un grand nombre d'arcs. Les degrés de toutes les voies sont égaux (de la valeur du nombre de mailles) exception faite de la voie \enquote{périphérique} faisant le tour de l'échantillon. De fait, cette voie oppose sa valeur unique de closeness (plus efficace) à celle de toutes les autres (pour lesquelles la valeur est la même). L'orthogonalité est ici de 1 pour toutes les voies, ce qui aboutit à un coefficient de maillance de 1 également. La longueur et le degré de toutes les voies, sauf de celles périphériques, étant égales, leur classification en classes de longueurs n'est pas significative, puisque complètement dépendante du sens de lecture du réseau. Le coefficient d'hétérogénéité sera donc mis à 0 en supposant que le classement des voies soit fait en lisant le graphe d'une unique manière pour les deux classifications.

Les caractéristiques que l'on retient pour la grille simple sont donc :
\begin{itemize}
\item $N_{deg1} = 0$
\item $\overline{deg(sommets)} \rightarrow 4$
\item $coef_{orga} \sim 0,1$
\item $coef_{mail} = 1$
\item $coef_{hete} = 0$
\item $coef_{red} \rightarrow 1$
\end{itemize}

Dans le graphe \enquote{Bruit faible} un léger décalage est introduit autour de chaque intersection.  Ceci a pour effet d'amener le degré de tous les sommets à 3, sans créer d'impasse. Le coefficient d'organicité est donc à sa valeur maximale : aucun nœud n'a un degré supérieur à 3. Le coefficient de réduction voit sa valeur diminuer à 0,67, car si le nombre d'arcs augmente, le nombre de voies créées également. Les voies créées sont dépendantes de la fluctuation du bruit à chaque intersection : dès qu'un décalage est introduit, l'angle formé de 90\degres  équivaudra à une fin de voie. Le degré moyen des voies sur le réseau trouvé est de 4. Ces valeurs sont équivalentes à celles obtenues avec les graphes \enquote{Bruit fort} avec ou sans introduction de générations (\enquote{Avec Générations}) ou d'indépendance de bruit (\enquote{Avec Angles}) dans le modèle. Les seules différences significatives observées entre les quatre modèles portent sur le coefficient de maillance qui, pour le modèle avec angles, voit sa valeur modifiée de 1 à 0,67 ; et sur le coefficient d'hétérogénéité dont la valeur augmente lorsque les générations sont introduites dans le modèle.

Les caractéristiques que l'on retient pour une structure régulière (sans être une grille parfaite) sont donc :
\begin{itemize}
\item $N_{deg1} = 0$
\item $\overline{deg(sommets)} = 3$
\item $coef_{orga} = 1$
\item $coef_{mail} = 1$ si aucun angle introduit, sinon $coef_{mail} = 0$
\item $coef_{hete} \le 0,1$ si le découpage est homogène, sinon $coef_{hete} \ge 0,1$
\item $coef_{red} \rightarrow 0,7$
\end{itemize}

Ces exemples simples nous permettent de percevoir les différences d'informations apportées par les coefficients que nous utilisons. Pour ce qui est de l'efficacité de chaque réseau au sens de la closeness, nous observons une valeur importante pour la grille régulière (0,69) qui diminue autour de 0,23 pour les quatre autres réseaux. Sur ceux-ci, le modèle avec génération propose la valeur moyenne la plus importante (0,25) et celui avec angles, la plus faible (0,22). Nous pouvons donc faire l'hypothèse, que si nous ne sommes pas sur un graphe parfaitement régulier (avec des voies entièrement traversantes), alors un découpage inhomogène de l'espace réduira les distances topologiques entre ses différents éléments.

\FloatBarrier
\subsection{Graphes viaires (quartiers)}

Dans l'étude des graphes choisis dans des quartiers aux motifs particuliers, nous remarquons que deux des cinq réseaux retenus ont des caractéristiques qui se détachent des trois autres : Cucq et Neuf-Brisach. Leur régularité leur donne un coefficient de réduction de plus de 0,8 contre une moyenne de 0,7 pour Lille, la Roche-sur-Yon et Vitry-le-François. De même, ces deux réseaux présentent un coefficient d'organicité plus faible (tableau \ref{tab:caract_coef_autre}).

Les cinq réseaux comportent un coefficient de maillance élevé (entre 0,92 et 0,98) dû à leurs régularités géométriques respectives. Les plus bas (0,92) correspondent aux réseaux de Lille et de Vitry-le-François qui ne comportent pas de schémas quadrillés. Le coefficient de maillance le plus élevé est celui calculé pour le graphe de Cucq. En effet, même si sa structure comporte des demi-cercles, ce réseau ne comprend aucune voie dont les connexions seraient plus souples.

\begin{figure}[h]
    \centering
    \begin{subfigure}[t]{.45\linewidth}
        \includegraphics[width=\textwidth]{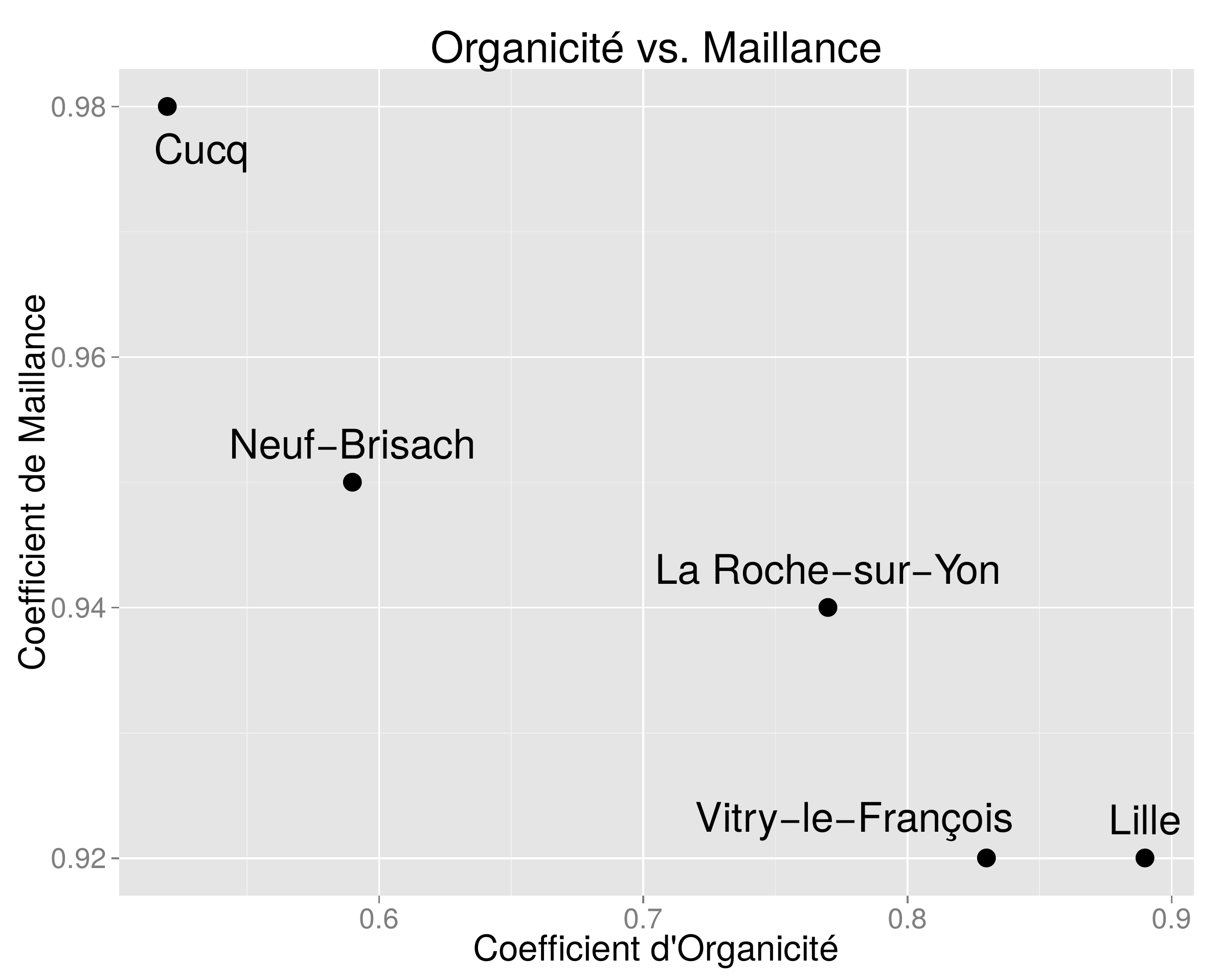}
        \caption{Coefficient de maillance tracé en fonction du coefficient d'organicité.}
        \label{fig:qu_orgamail}
    \end{subfigure}
    ~
    \begin{subfigure}[t]{.45\linewidth}
        \includegraphics[width=\textwidth]{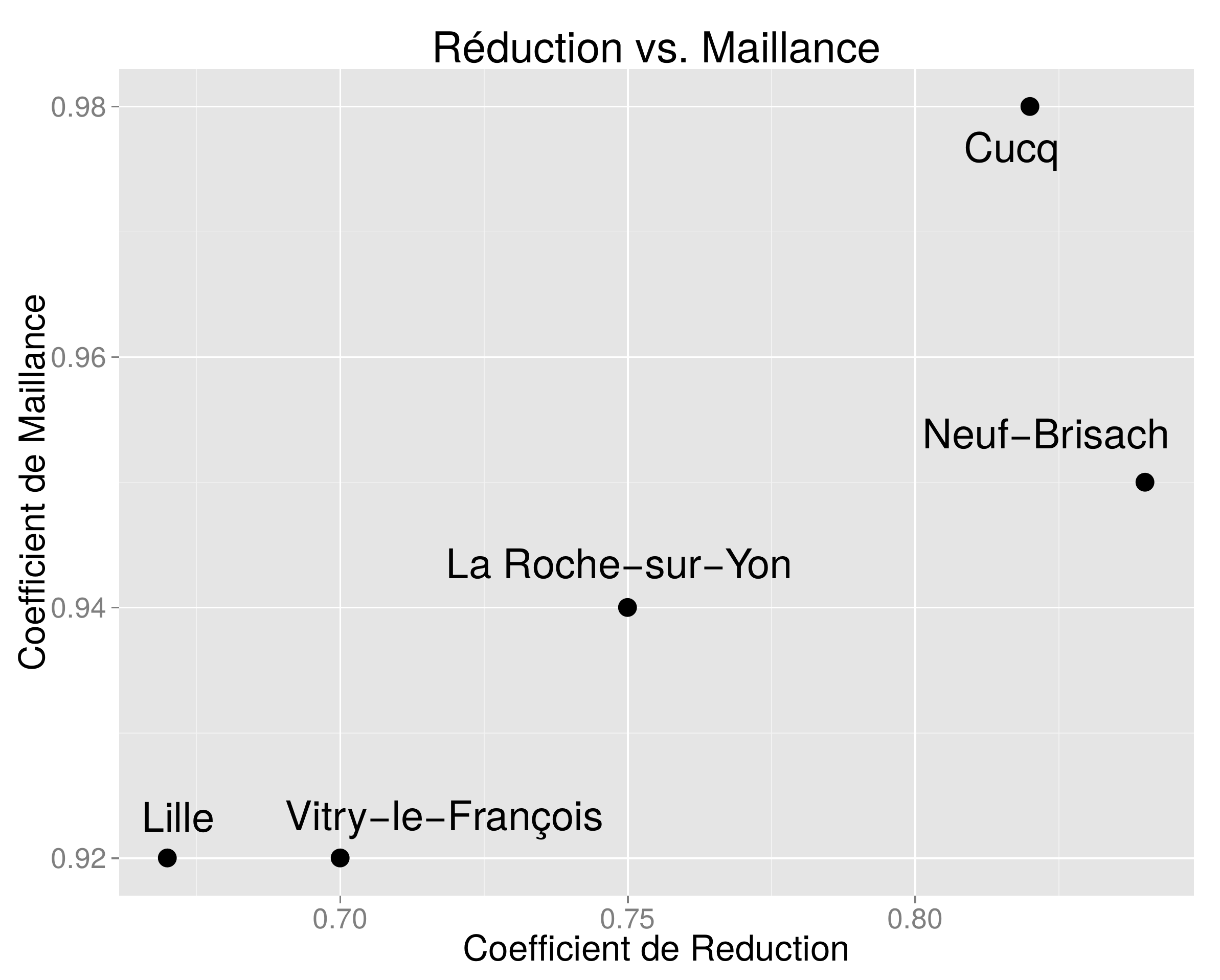}
        \caption{Coefficient de maillance tracé en fonction du coefficient de réduction.}
        \label{fig:qu_redmail}
    \end{subfigure}

    \caption{Étude sur les coefficients. Courbes tracées pour les graphes issus des motifs viaires (quartiers) du panel de recherche.}
    \label{fig:qu_coefs}
\end{figure}

Pour ces cinq échantillons, les coefficients d'organicité, de maillance et de réduction donnent des résultats corrélés (figure \ref{fig:qu_coefs}). Les coefficients d'hétérogénéité n'ont pas de variations significatives de valeurs entre les échantillons. Seul le graphe de Neuf-Brisach présente une hétérogénéité un peu plus importante due au contraste apporté par la structure circulaire à celle quadrillée.

Dans l'étude des propriétés liées aux sommets, le quartier issu de Lille se différencie des quatre autres : il comporte 20\% d'impasses (contre environ 10\% pour les autres graphes), ce qui lui donne une valeur de degré moyen plus faible (2,71 lorsque les autres réseaux sont entre 3 et 3,23).

\begin{figure}[h]
    \centering
    \begin{subfigure}[t]{.45\linewidth}
        \includegraphics[width=\textwidth]{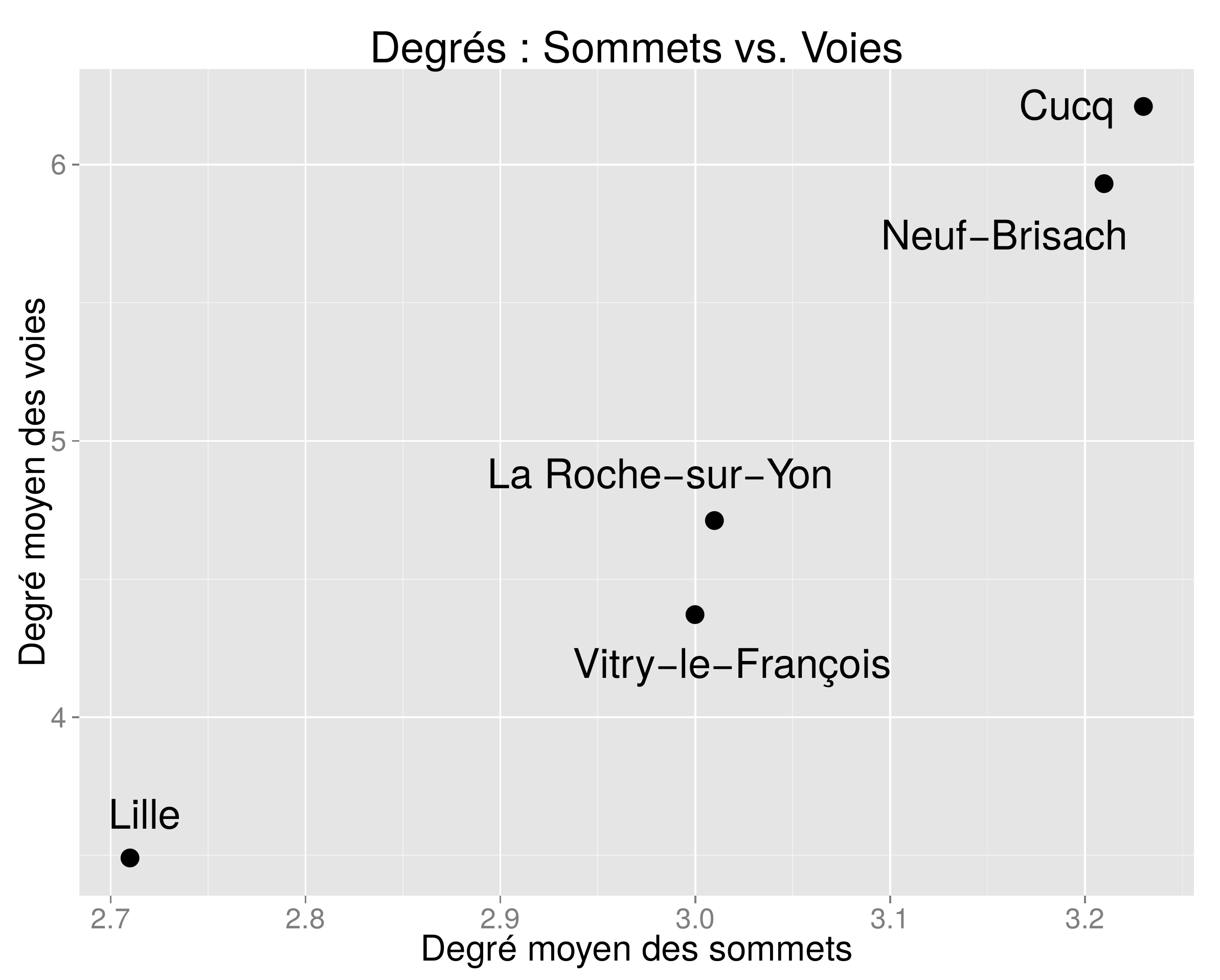}
        \caption{Degré moyen des voies tracé en fonction du degré moyen des sommets.}
        \label{fig:qu_degsomvoi}
    \end{subfigure}
    ~
    \begin{subfigure}[t]{.45\linewidth}
        \includegraphics[width=\textwidth]{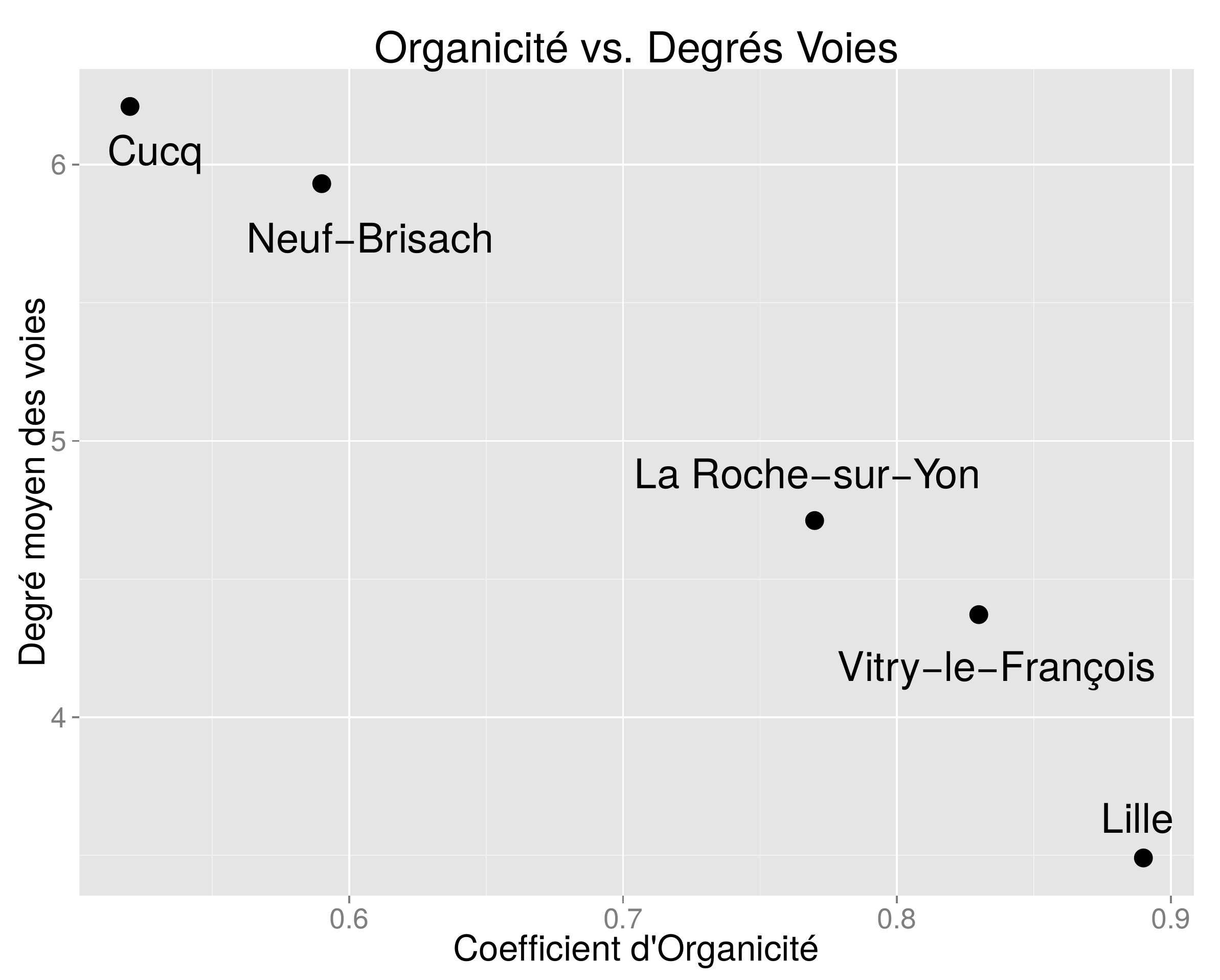}
        \caption{Degré moyen des voies tracé en fonction du coefficient d'organicité.}
        \label{fig:qu_orgadeg}
    \end{subfigure}

    \caption{Étude sur les degrés. Courbes tracées pour les graphes issus des motifs viaires (quartiers) du panel de recherche.}
    \label{fig:qu_deg}
\end{figure}

Si l'on trace la courbe du degré moyen des voies en fonction du degré moyen des sommets, les cinq graphes ont des valeurs qui se positionnent sur une droite. Nous pouvons donc présumer, à cette échelle, une proportionnalité entre le degré des intersections et celui des voies créées (figure \ref{fig:qu_degsomvoi}). Le lien est tangible car le degré des voies est fortement dépendant de celui des sommets qu'elles réunissent. Si l'on compare dans un second temps le coefficient d'organicité au degré moyen des voies (figure \ref{fig:qu_orgadeg}) nous observons également un comportement corrélé. Plus le degré des voies est faible, plus l'organicité est forte. En effet, les structures qui réunissent des voies de degré important sont souvent révélatrices d'organisations planifiées et de sommets de degré supérieur à 3. 

Les réseaux qui apparaissent comme les plus efficaces en terme d'accessibilité, à travers l’indicateur de closeness, sont ceux aux structures les plus régulières : Neuf-Brisach et Cucq. S'y oppose celui dont le graphe est le plus tourmenté : le quartier issu de Lille.

\FloatBarrier
\subsection{Graphes viaires (villes)}

Nous reportons dans le tableau \ref{tab:ville_num} les numéros attribués à chaque réseau viaire pour pouvoir les identifier sur les courbes présentées dans la suite.

\begin{table}
\begin{center}
{ \small
\begin{tabular}{|c|c|c|c|}
\hline
Continent & Pays & Ville & Numéro \\ \hline 
		
Europe & France & Paris & 1 \\
 & & Avignon & 2 \\
 & & Bordeaux & 3 \\
 & & Brive & 4 \\ 
 & & Cergy-Pontoise & 5  \\
 & & Villers-sur-Mer & 6 \\

 & Belgique & Bruxelles & 7 \\ 
 & Angleterre & Londres & 8 \\
 & Espagne & Barcelone & 9 \\ 
 & Pays-Bas & Rotterdam & 10 \\ \hline

Amérique & États-Unis & Manhattan & 11 \\ 
 & États-Unis & San-Francisco & 12 \\

 & États-Unis & Santa-Fe & 13 \\ 
 & Brésil & Manaus & 14 \\ 
 & Pérou & Cuzco & 15 \\ \hline 

Asie & Iran & Téhéran & 16 \\ 
 & Inde & Varanasi & 17 \\
 & Japon & Kyoto & 18 \\ \hline 

Afrique & Maroc & Casablanca & 19 \\
 & Kenya & Nairobi & 20 \\ \hline 
\end{tabular}
}
\end{center}
\caption{Détail des numéros attribués à chaque ville pour les identifier sur les courbes.}
\label{tab:ville_num}
\end{table}

\begin{figure}[h]
    \centering
    \begin{subfigure}[t]{.45\linewidth}
        \includegraphics[width=\textwidth]{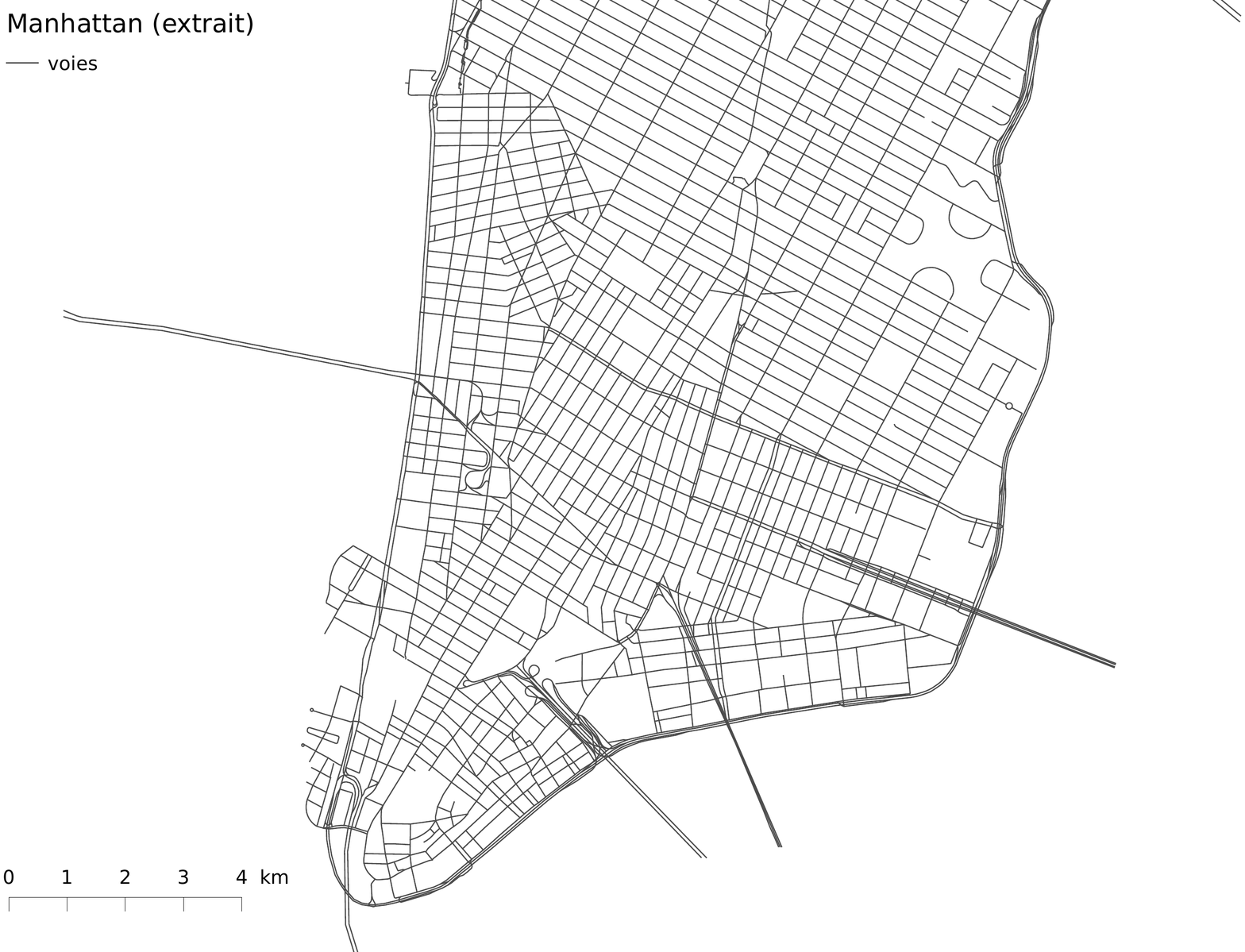}
        \caption{Manhattan.}
        \label{fig:brut_zoom_man2}
    \end{subfigure}
    ~
    \begin{subfigure}[t]{.45\linewidth}
        \includegraphics[width=\textwidth]{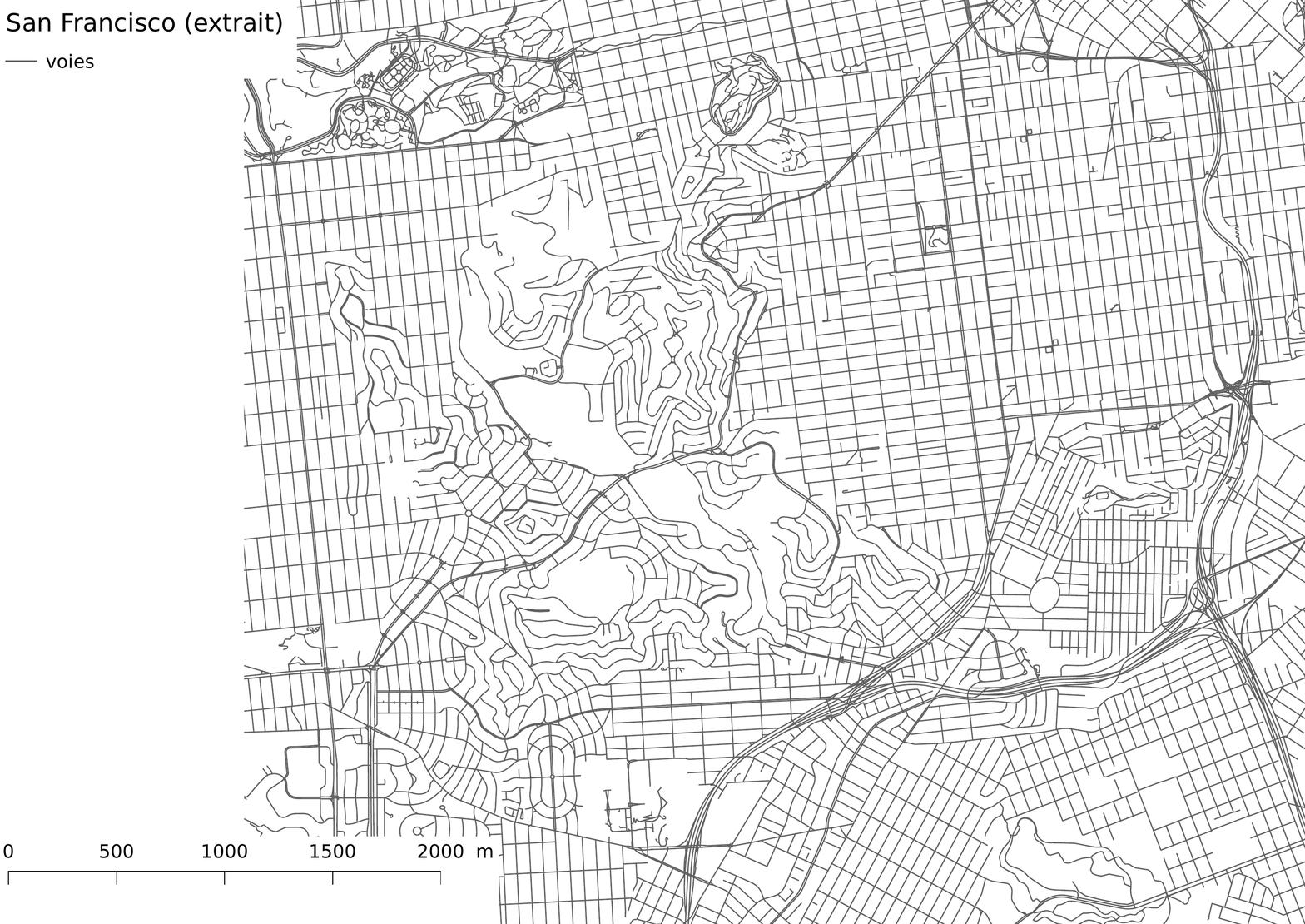}
        \caption{San Francisco.}
        \label{fig:brut_zoom_san2}
    \end{subfigure}
    
    \caption{Contraste entre connexions perpendiculaires et connexions souples.}
    \label{fig:brut_zoom_reg2}
\end{figure}

Si nous positionnons les vingt réseaux viaires étudiés les uns par rapport aux autres selon les quatre coefficients définis plus haut, nous remarquons en premier lieu le comportement particulier du réseau issu de Manhattan révélé par trois d'entre eux. En effet, les voies créées sur la célèbre île participent à une grille régulière. Celle-ci réunit un grand nombre d'arcs au sein des même éléments continus. Manhattan apparaît donc avec un coefficient de réduction très fort. La ville ne contient que 32\% de sommets de degré 3. C'est la valeur la plus basse, suivie de près par celle de la ville de San Francisco (44\%), quand les autre réseaux comptent entre 55 et 70\% de sommets de degré 3. Les valeurs de l'organicité de ces deux villes sont donc les plus basses. Elles ont également les valeurs d'hétérogénéité les plus faibles, qui sont suivies de près par celles des réseaux de Kyoto, Paris et Barcelone (réseaux dont les tailles sont comparables). En revanche, les résultats donnés par le coefficient de maillance sont complètements différents. Ce coefficient est révélateur de la trame du réseau. Plus il s'approche de 1, plus les voies se coupent perpendiculairement. Or à Manhattan comme à San Francisco, la présence de parties du graphe aux connexions beaucoup plus souples (contours ou parcs) diminue la valeur du coefficient (figure \ref{fig:brut_zoom_reg2}). Pour celui-ci, c'est la ville de Kyoto qui a la valeur la plus élevée. En effet, son graphe ne comporte aucune voie périphérique (figure \ref{fig:brut_zoom_kyo}). Sa régularité se rapproche de celle du modèle aléatoire avec générations. Le graphe issu de la ville d'Avignon est le moins maillé de tous ceux que nous étudions ici. À l'intersection entre deux fleuves, il s'est construit en suivant leurs formes arrondies (figures \ref{fig:brut_zoom_av}).

\begin{figure}[h]
    \centering    
    
    \begin{subfigure}[t]{.45\linewidth}
        \includegraphics[width=\textwidth]{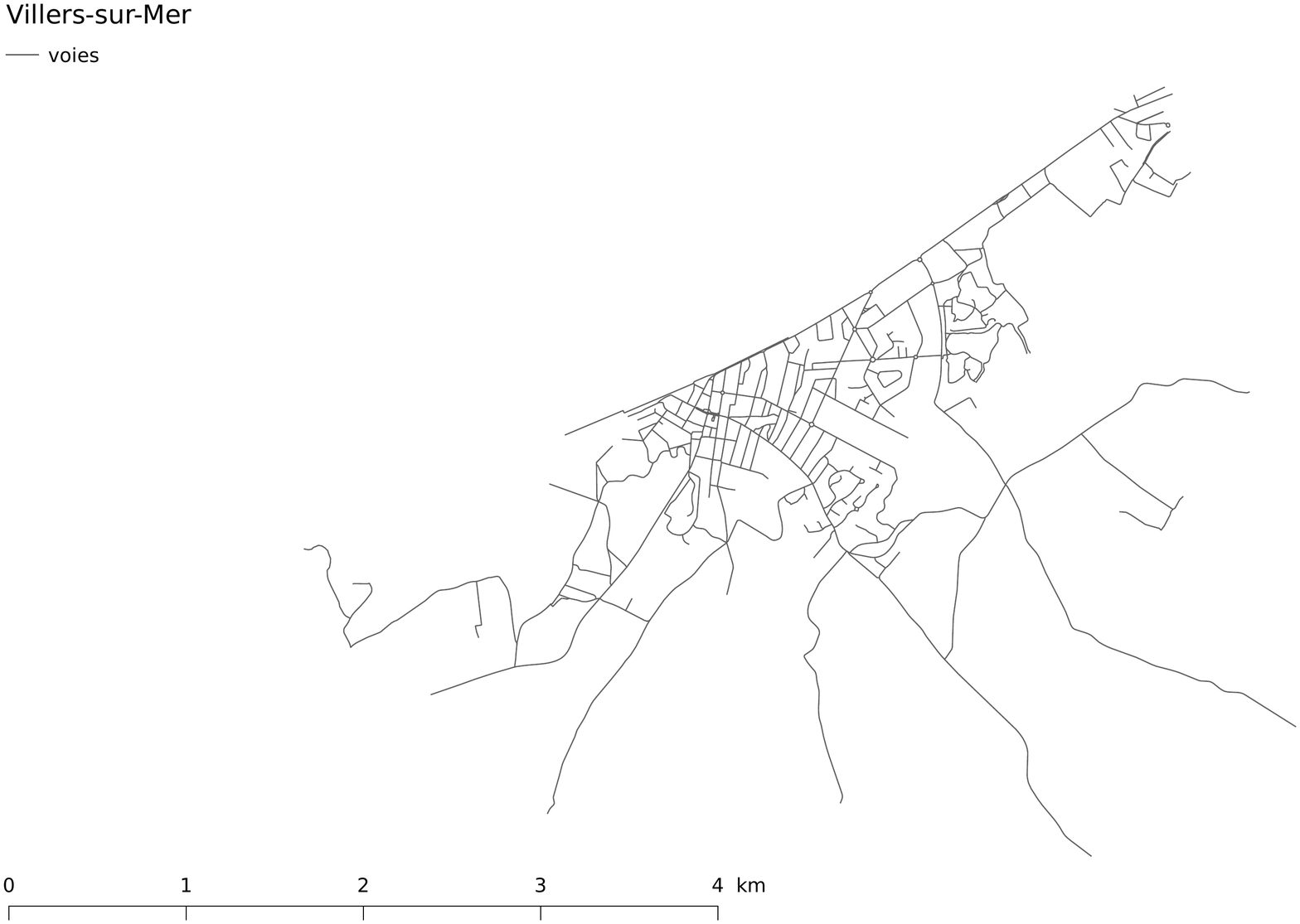}
        \caption{Villers-sur-Mer.}
        \label{fig:brut_zoom_vil2}
    \end{subfigure}
    ~
    \begin{subfigure}[t]{.45\linewidth}
        \includegraphics[width=\textwidth]{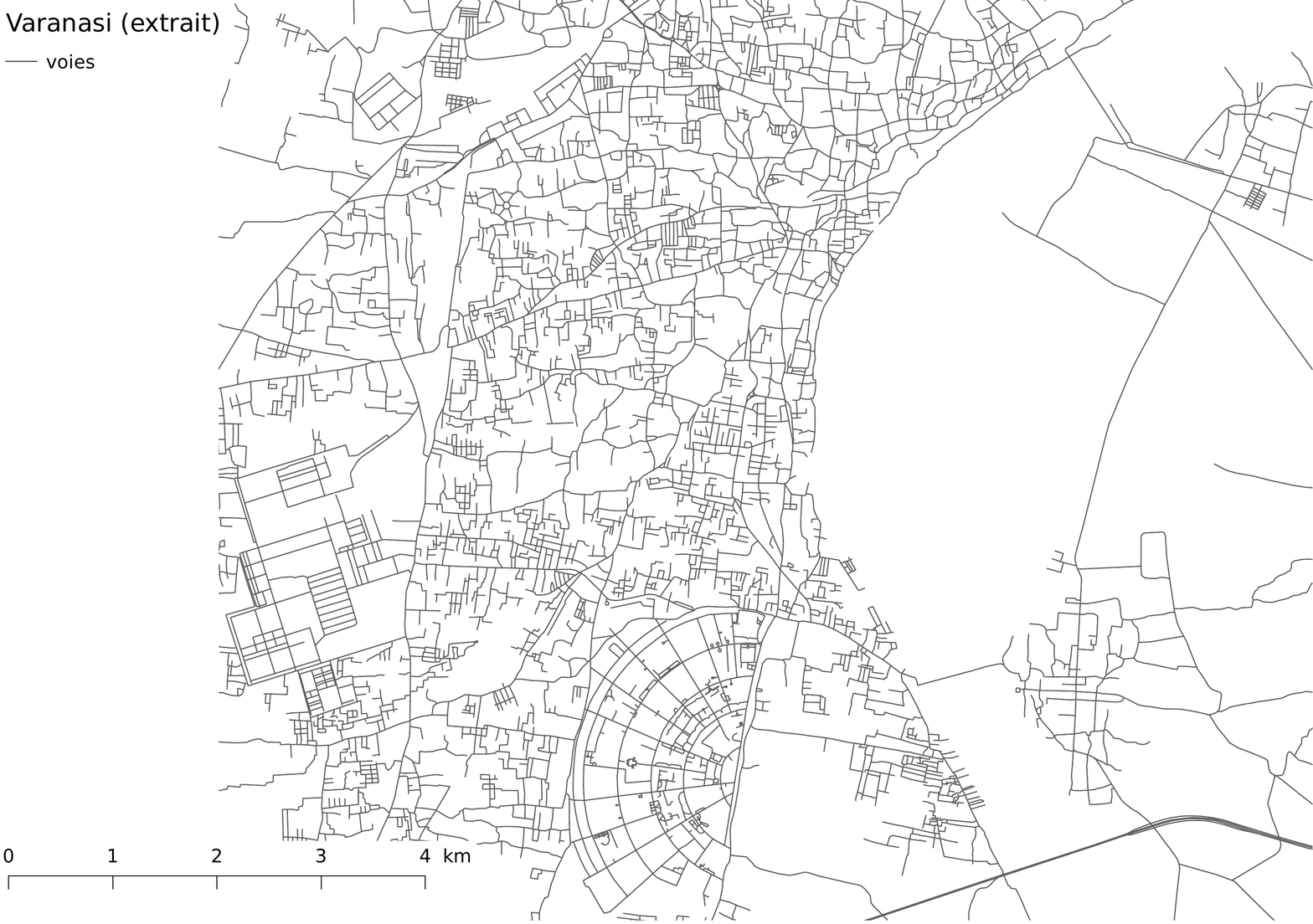}
        \caption{Varanasi.}
        \label{fig:brut_zoom_var2}
    \end{subfigure}

    \caption{Deux villes que l'échelle et le lieu opposent, avec une même logique géométrique.}
    \label{fig:comp_viller_vara}
\end{figure}

Nous pouvons également remarquer des parallèles étonnants. Ainsi, les villes de Villers-sur-Mer (au nord de la France) et de Varanasi (en Inde) ressortent avec des coefficients extrêmement proches. À deux échelles différentes (le réseau de Villers-sur-Mer étant d'une longueur environ 40 fois inférieure à celle de Varanasi), nous pouvons donc considérer la proximité de leurs structures et leurs logiques communes à travers les indicateurs étudiés (figure \ref{fig:comp_viller_vara}).

\begin{figure}[h]
    \centering
    \begin{subfigure}[t]{.45\linewidth}
        \includegraphics[width=\textwidth]{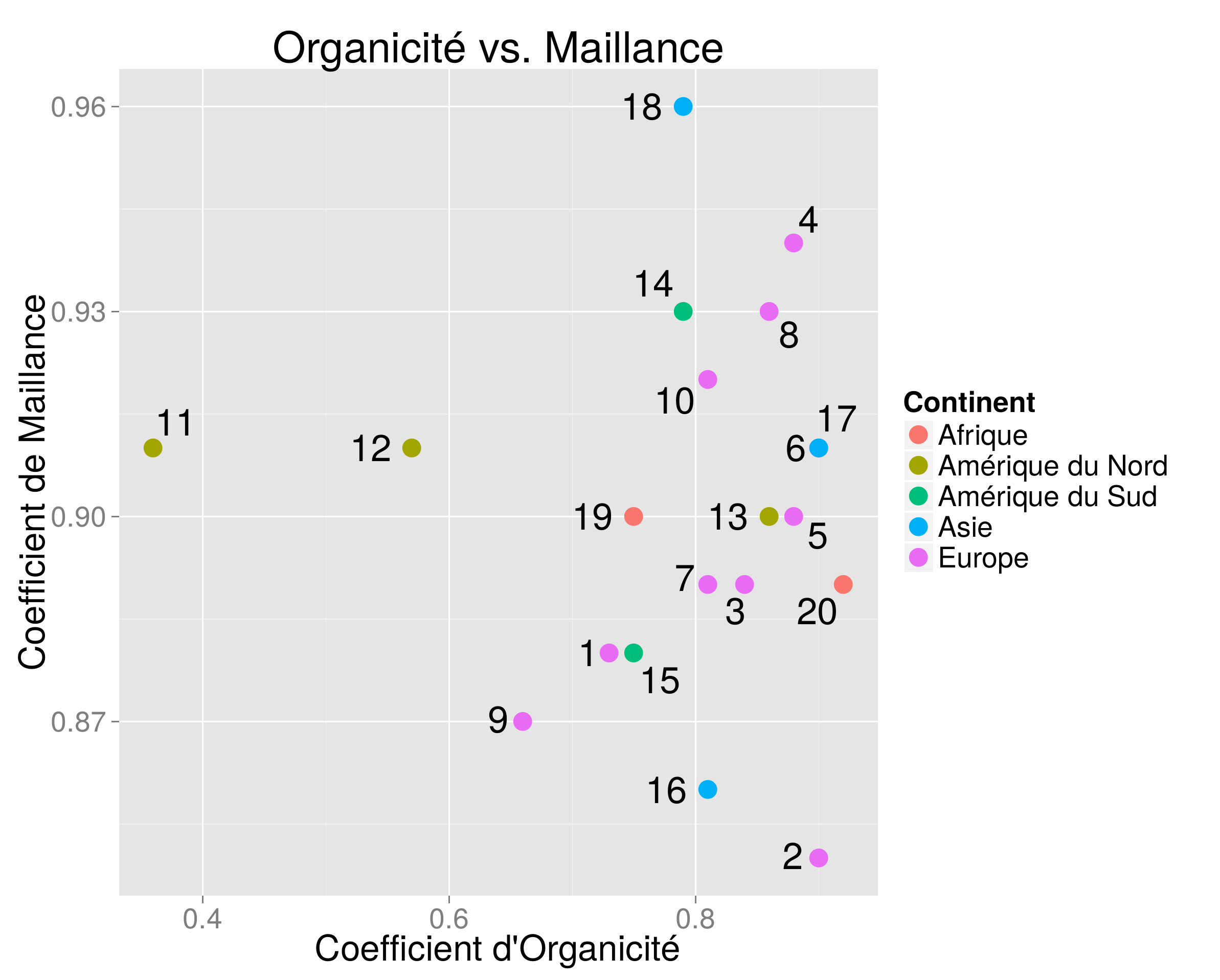}
        \caption{Coefficient de maillance tracé en fonction de celui d'organicité.}
        \label{fig:ci_orgamail}
    \end{subfigure}
    ~
    \begin{subfigure}[t]{.45\linewidth}
        \includegraphics[width=\textwidth]{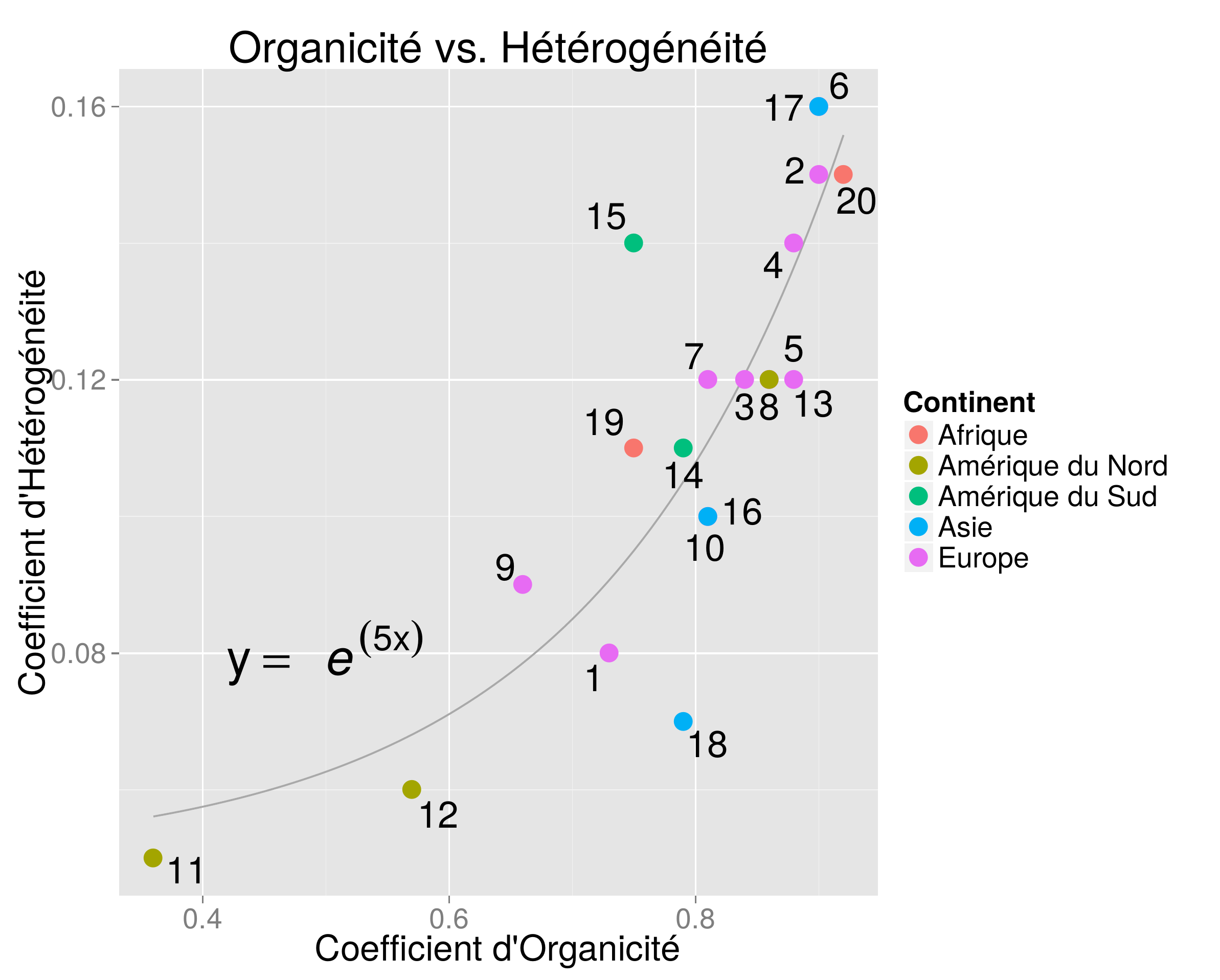}
        \caption{Coefficient d'hétérogénéité tracé en fonction de celui d'organicité.}
        \label{fig:ci_orgahete}
    \end{subfigure}    
    
    \caption{Étude sur les coefficients. Courbes tracées pour les graphes issus des 20 villes du panel de recherche.}
    \label{fig:ci_coef1}
\end{figure}

\begin{figure}[h]
    \centering
    \begin{subfigure}[t]{.45\linewidth}
        \includegraphics[width=\textwidth]{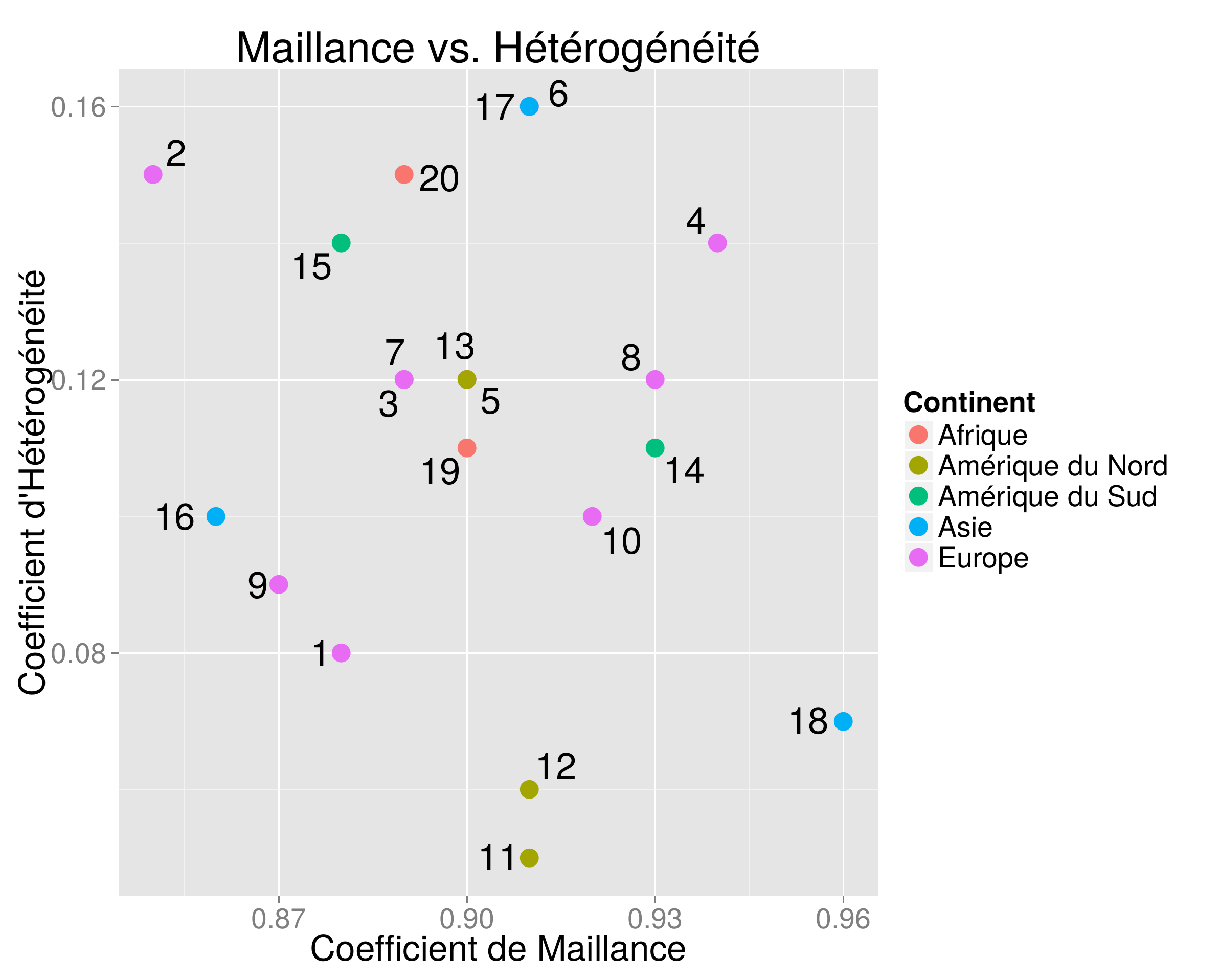}
        \caption{Coefficient d'hétérogénéité tracé en fonction de celui de maillance.}
        \label{fig:ci_mailhete}
    \end{subfigure}
    ~
    \begin{subfigure}[t]{.45\linewidth}
        \includegraphics[width=\textwidth]{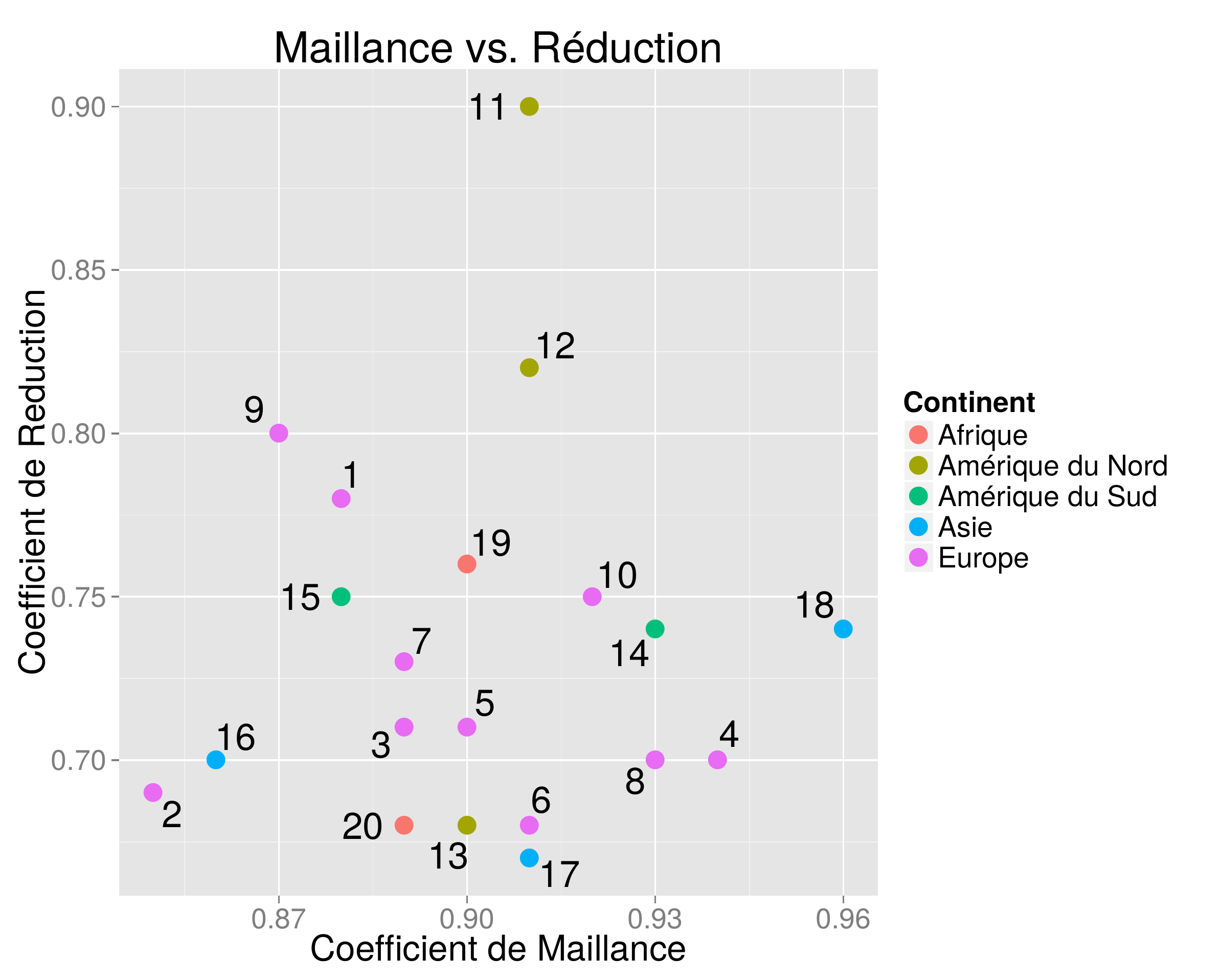}
        \caption{Coefficient de réduction tracé en fonction de celui de maillance.}
        \label{fig:ci_redmail}
    \end{subfigure}    
    
    \caption{Étude sur les coefficients. Courbes tracées pour les graphes issus des 20 villes du panel de recherche.}
    \label{fig:ci_coef2}
\end{figure}

Les coefficients de maillance, d'hétérogénéité et de réduction nous donnent donc des informations bien différentes selon les graphes étudiés. Les deux coefficients qui présentent un comportement lié sont ceux d'organicité et d'hétérogénéité. Le premier s'intéresse au degré des sommets et le second à la dispersion qu'il peut y avoir entre longueur et degré des voies. Nous pouvons donc en conclure que plus un graphe comporte de nœuds de degré 4, plus la longueur des voies créées sur celui ci sera corrélée à leur degré.

\begin{figure}[h]
    \centering
    \begin{subfigure}[t]{.45\linewidth}
        \includegraphics[width=\textwidth]{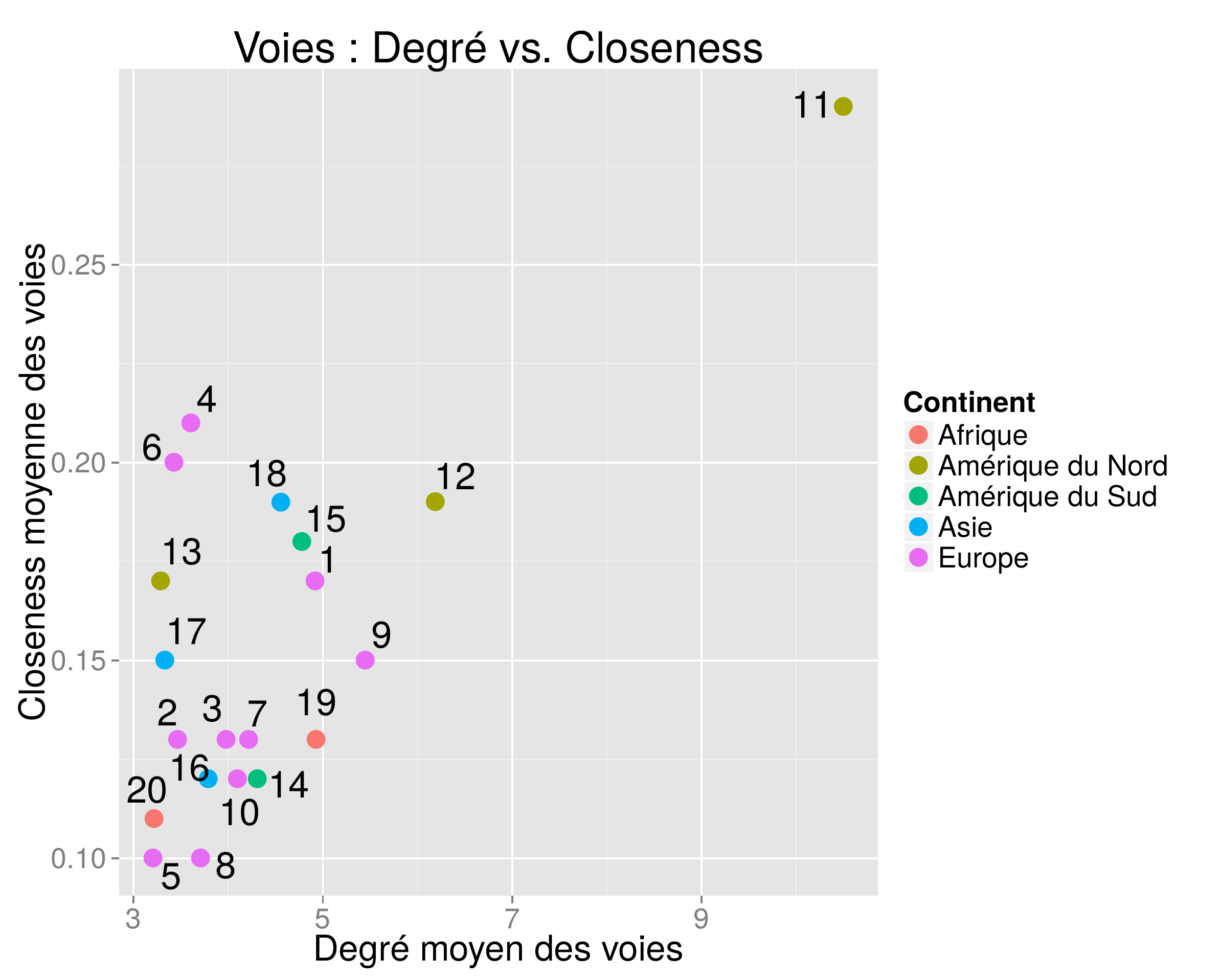}
        \caption{Moyenne de l'indicateur de closeness tracée en fonction de celle de l'indicateur de degré des voies.}
        \label{fig:ci_degclo}
    \end{subfigure}
    ~
    \begin{subfigure}[t]{.45\linewidth}
        \includegraphics[width=\textwidth]{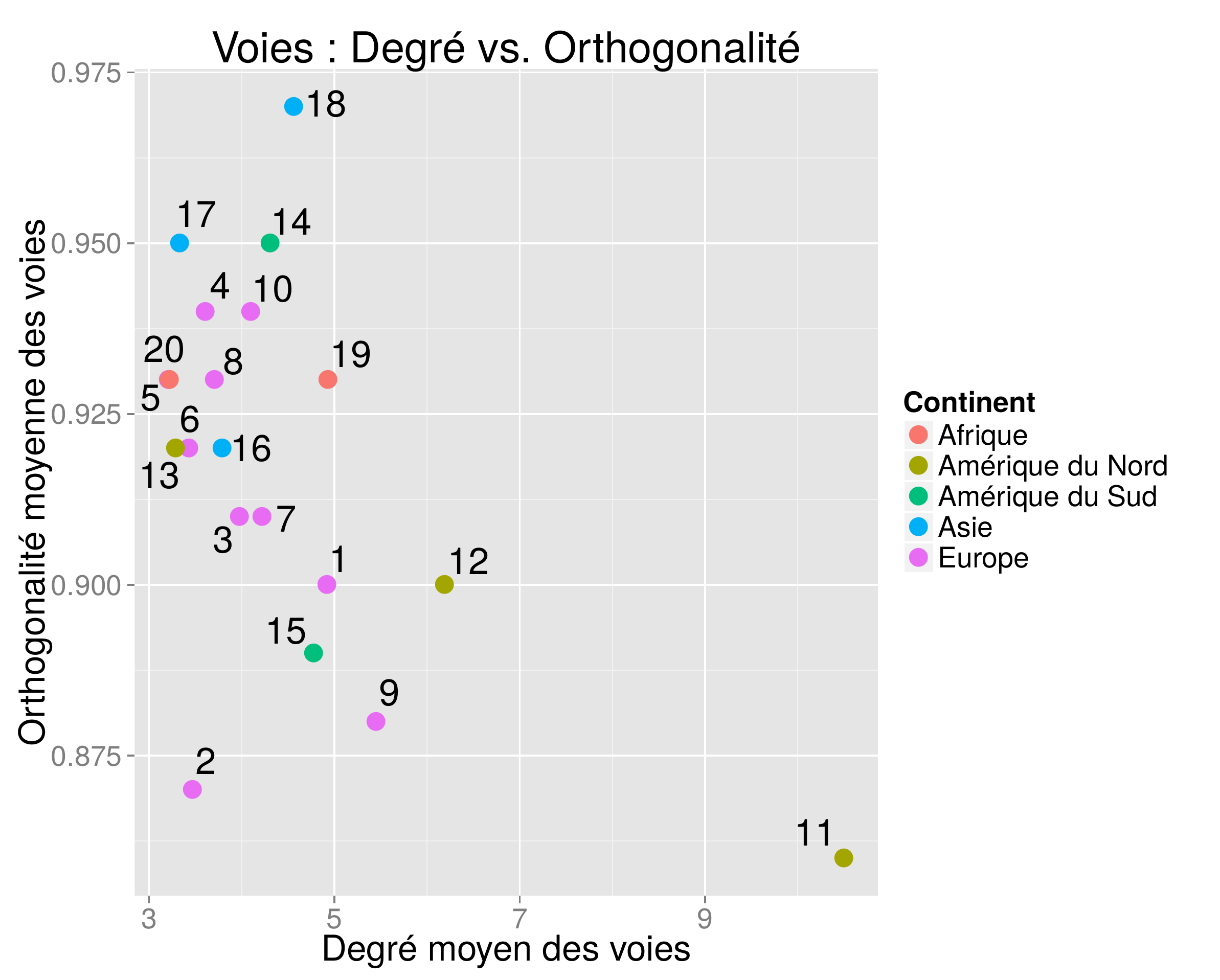}
        \caption{Moyenne de l'indicateur d'orthogonalité tracée en fonction de celle de l'indicateur de degré des voies.}
        \label{fig:ci_degortho}
    \end{subfigure}

    \caption{Étude sur les indicateurs calculés sur l'hypergraphe de voies. Courbes tracées pour les graphes issus des 20 villes du panel de recherche.}
    \label{fig:ci_ind1}
\end{figure}

\begin{figure}[h]
    \centering
    \begin{subfigure}[t]{.45\linewidth}
        \includegraphics[width=\textwidth]{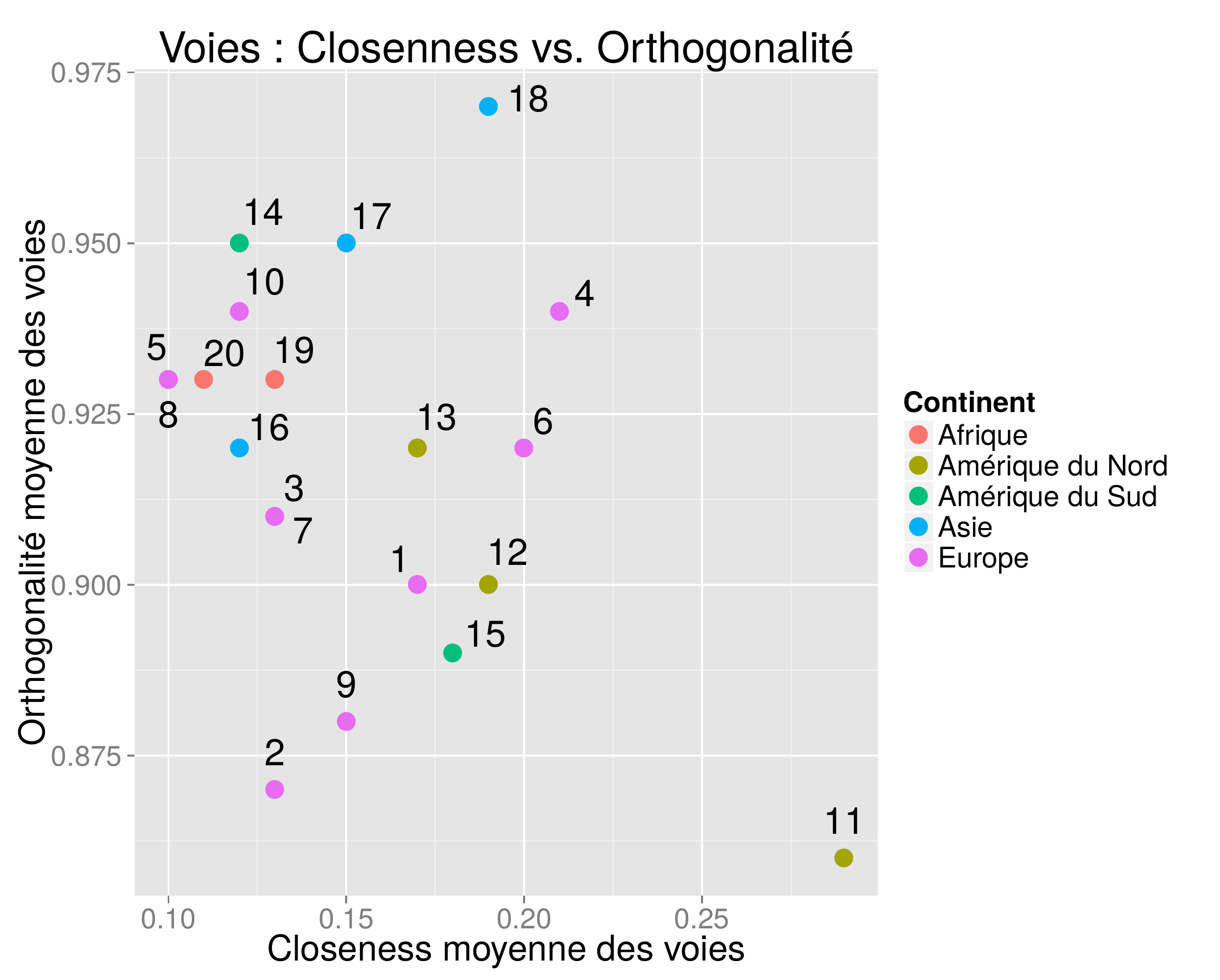}
        \caption{Moyenne de l'indicateur d'orthogonalité tracée en fonction de celle de l'indicateur de closeness.}
        \label{fig:ci_cloortho}
    \end{subfigure}
     ~
    \begin{subfigure}[t]{.45\linewidth}
        \includegraphics[width=\textwidth]{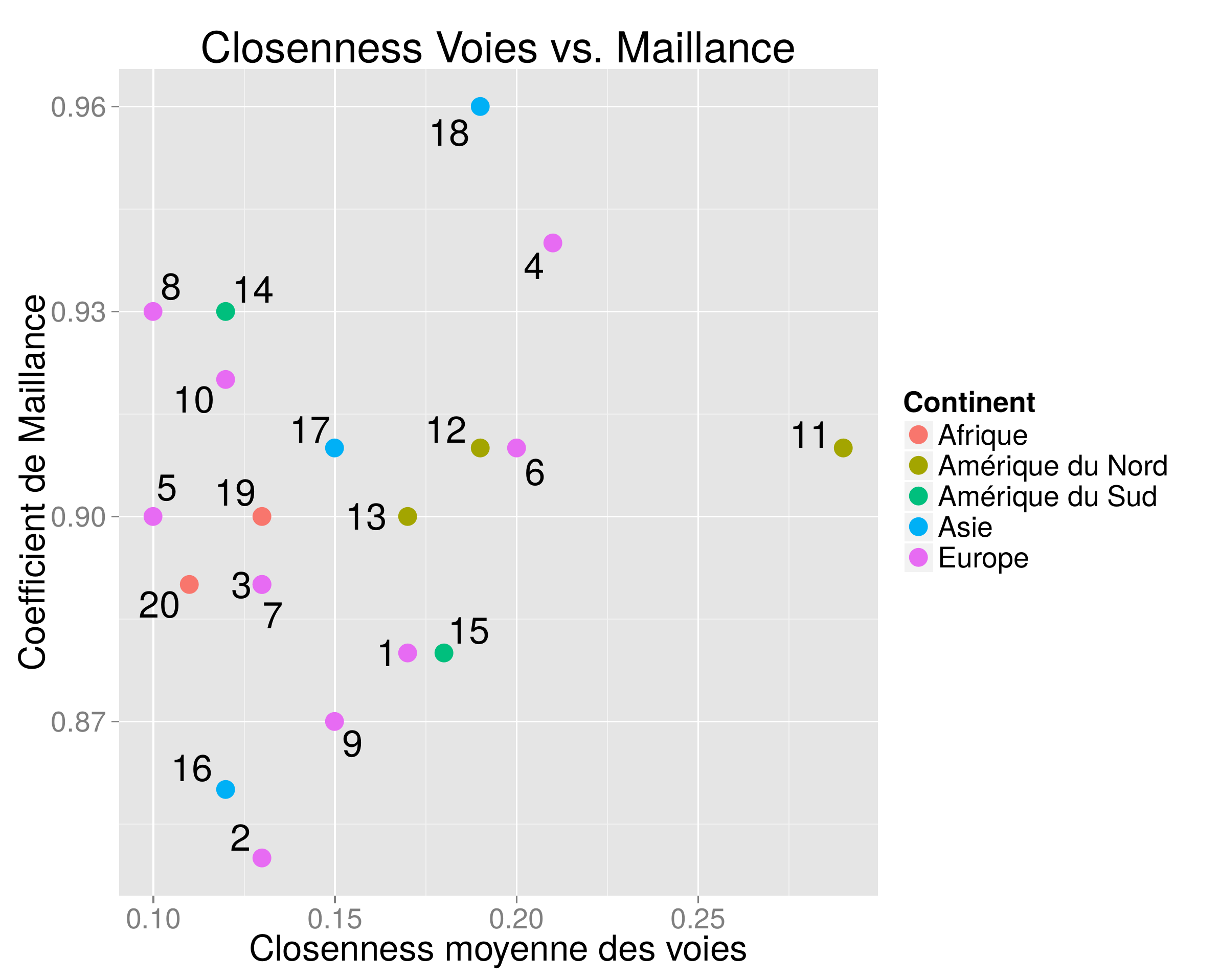}
        \caption{Coefficient de maillance tracé en fonction de la moyenne de l'indicateur de closeness.}
        \label{fig:ci_mailclo}
    \end{subfigure}

    \caption{Étude des corrélations avec l'indicateur de closeness. Courbes tracées pour les graphes issus des 20 villes du panel de recherche.}
    \label{fig:ci_ind2}
\end{figure}

Si nous nous penchons sur les trois indicateurs primaires que nous avons retenus pour qualifier les voies, nous pouvons tracer leurs comportements croisés pour les réseaux étudiés (figures \ref{fig:ci_ind1} et \ref{fig:ci_ind2}). Nous n'observons pas de groupement particulier, si ce n'est le détachement fort du réseau de Manhattan. Son motif particulier affecte aux voies des indicateurs de longueur, de degré et de closeness importants. Le réseau d'Avignon est celui dont le degré et l'orthogonalité des voies sont les plus faibles. En terme d'accessibilité de l'ensemble du réseau, l'indicateur de closeness évalue Manhattan comme étant la ville la plus efficace en terme de distances topologiques entre voies, suivie de Brive-la-Gaillarde. Ces réseaux apparaissant également avec un coefficient de maillance important, nous traçons l'indicateur de closeness en fonction de ce coefficient. Nous obtenons le résultat de la figure \ref{fig:ci_mailclo}). Ce graphique met en évidence qu'avoir un réseau de forte maillance est une condition nécessaire mais non suffisante pour que les voies aient un coefficient de closeness important. Ainsi, le réseau de Kyoto, dont le coefficient de maillance est plus fort que celui de Manhattan, apparaît avec une moyenne de l'indicateur de closeness sur ses voies, plus faible.


\FloatBarrier

\begin{figure}[h]
    \centering
    \begin{subfigure}[t]{.45\linewidth}
        \includegraphics[width=\textwidth]{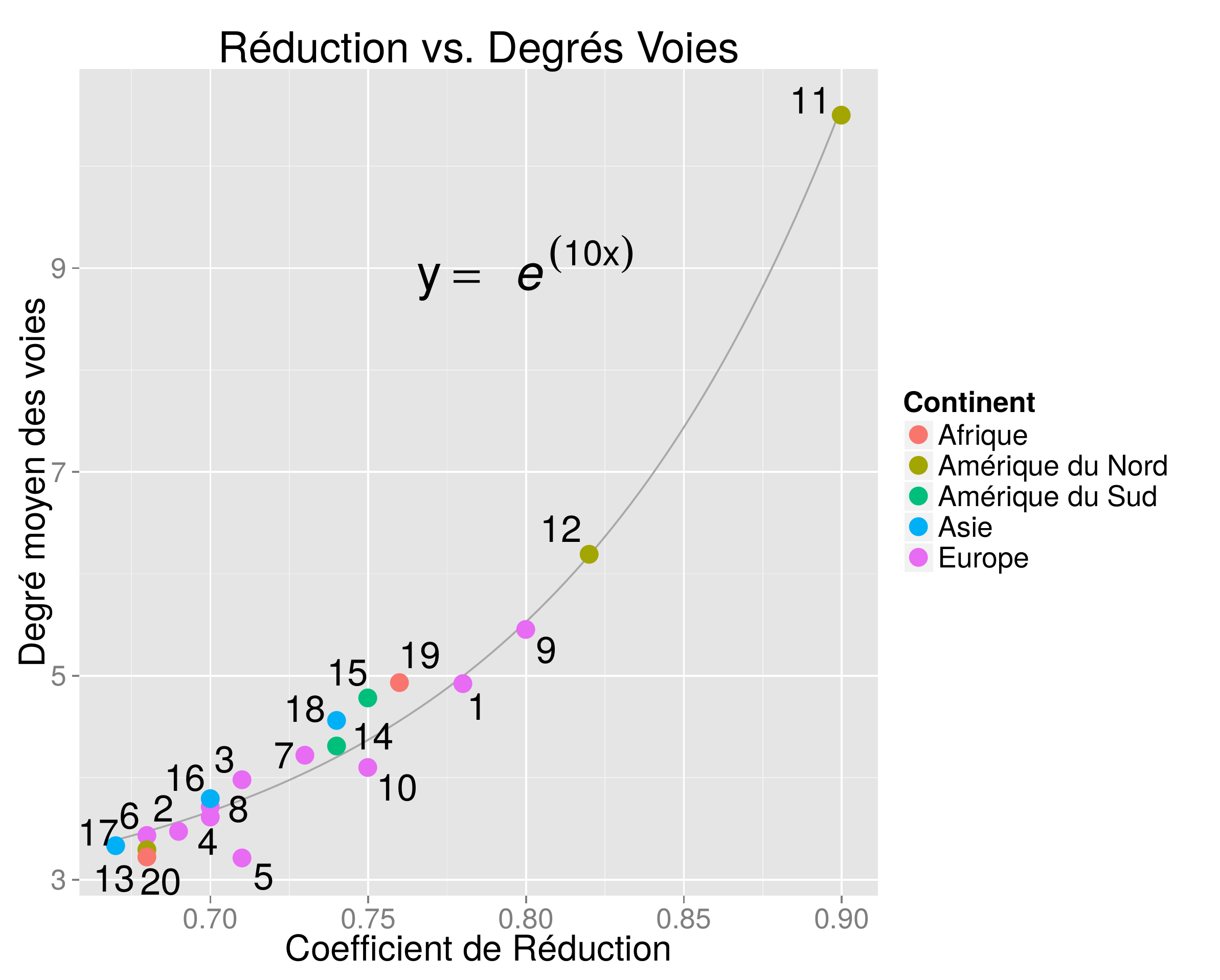}
        \caption{Moyenne de l'indicateur de degré des voies tracée en fonction du coefficient de réduction.}
        \label{fig:ci_reddeg}
    \end{subfigure}
   ~    
    \begin{subfigure}[t]{.45\linewidth}
        \includegraphics[width=\textwidth]{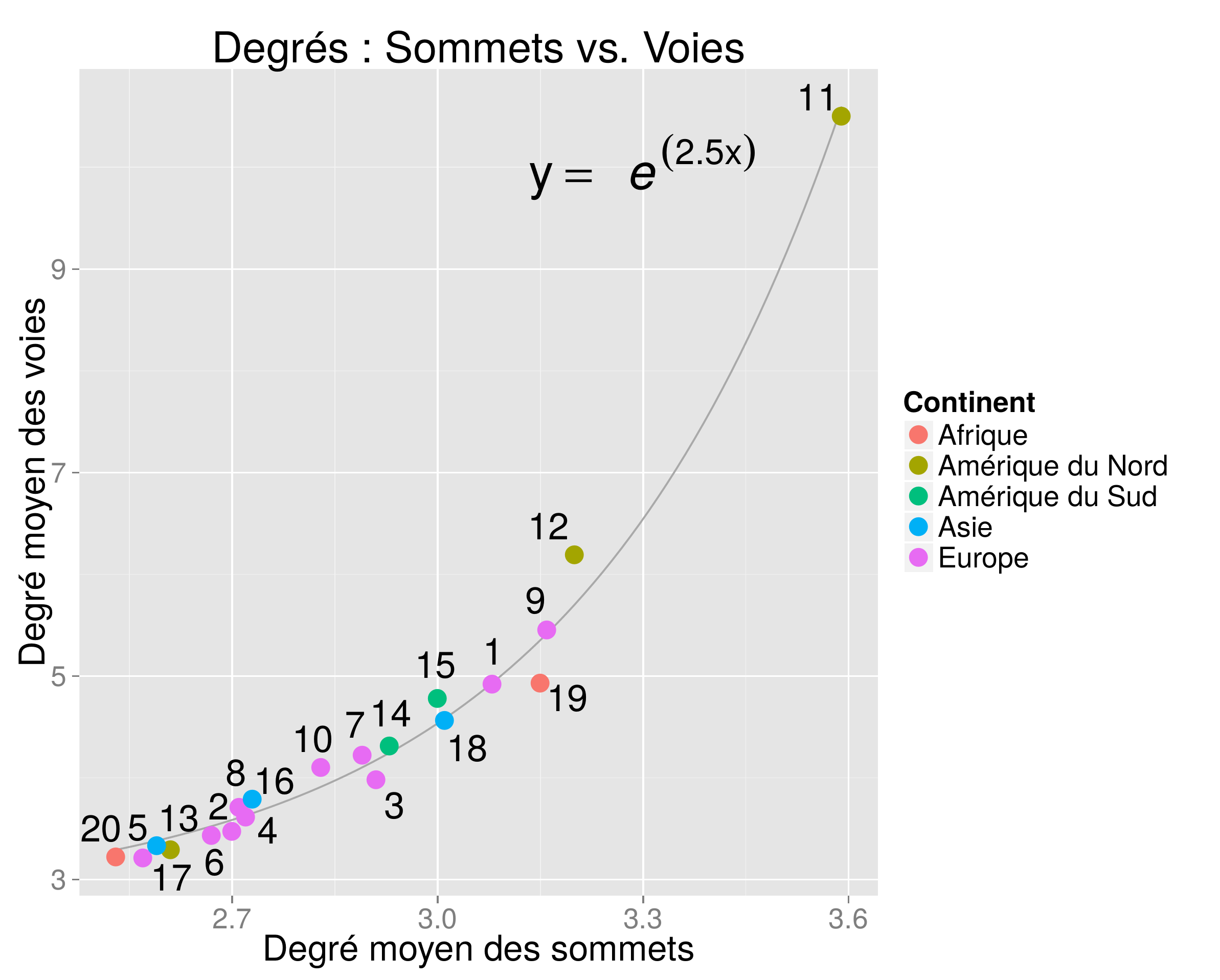}
        \caption{Moyenne de l'indicateur de degré des voies tracée en fonction du degré moyen des sommets.}
        \label{fig:ci_degsomvoie}
    \end{subfigure}
    
    \caption{Étude de corrélation avec le degré moyen des voies. Courbes tracées pour les graphes issus des 20 villes du panel de recherche.}
    \label{fig:ci_deg}
\end{figure}

Nous avons montré que le nombre d'arcs par voie était fortement corrélé à leur degré (Partie I, chapitre 4). Nous observons donc ici une forte corrélation entre degré moyen des voies et coefficient de réduction (figure \ref{fig:ci_reddeg}). Le degré moyen des sommets est lui aussi fortement corrélé au degré moyen des voies (figure \ref{fig:qu_degsomvoi}). En effet, plus une voie comporte des sommets de fort degré, plus elle aura d'occasions d'être connectée à d'autres voies. Ces deux résultats étaient donc prévisibles. 

\begin{figure}[h]
    \centering
    \begin{subfigure}[t]{.45\linewidth}
        \includegraphics[width=\textwidth]{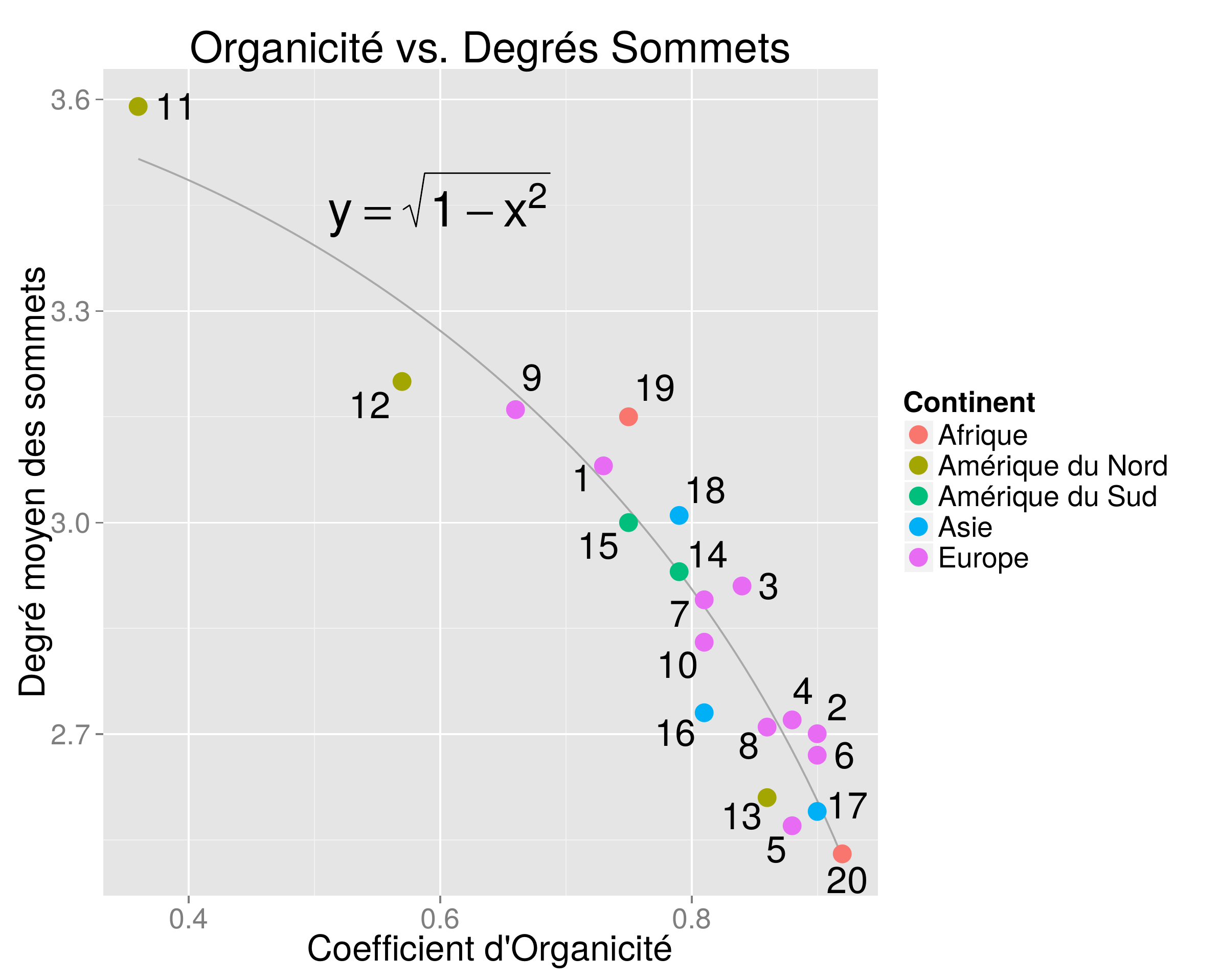}
        \caption{Degré moyen des sommets tracé en fonction du coefficient d'organicité.}
        \label{fig:ci_orgadegsom}
    \end{subfigure}
    ~    
    \begin{subfigure}[t]{.45\linewidth}
        \includegraphics[width=\textwidth]{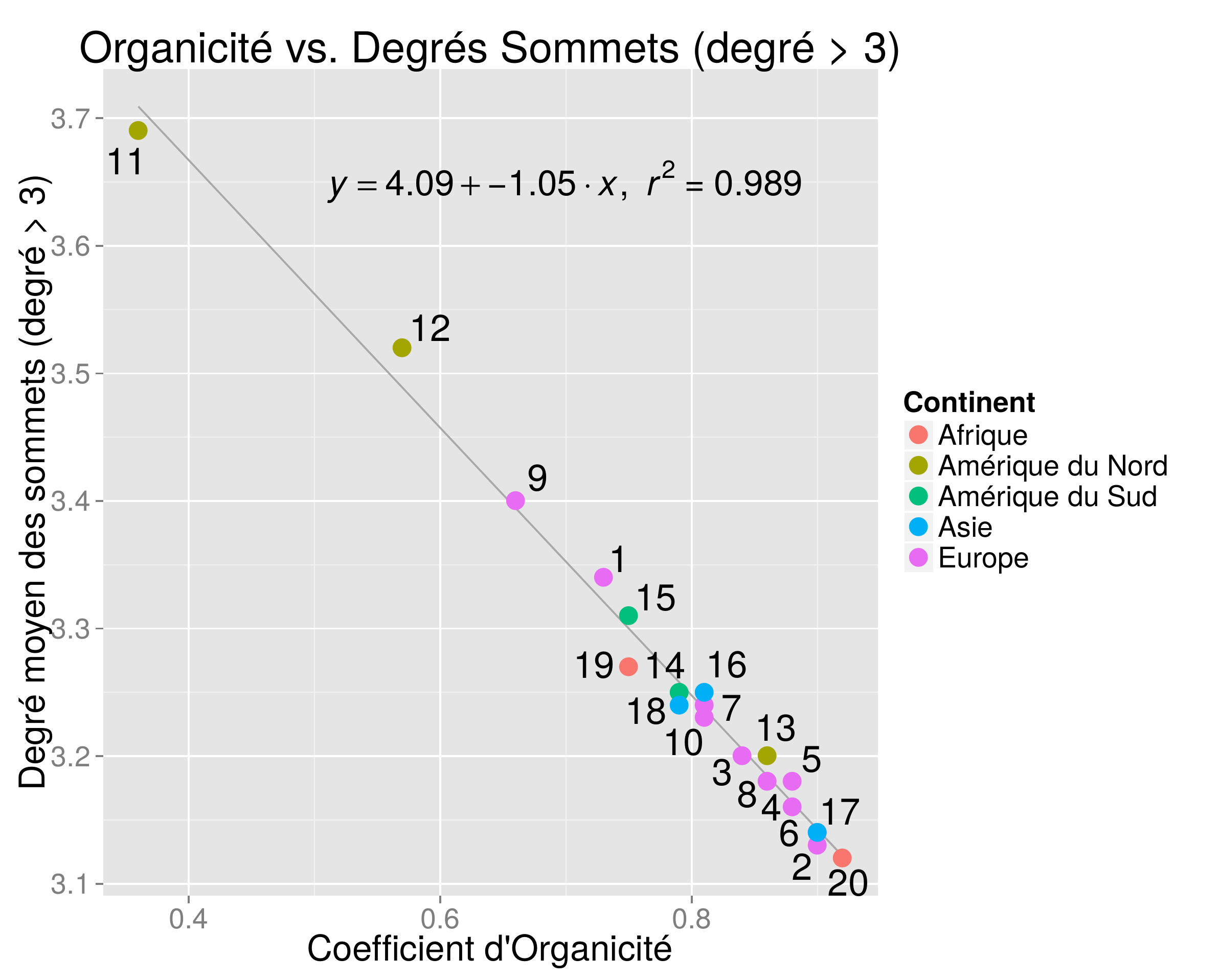}
        \caption{Degré moyen des sommets de degré supérieur ou égal à 3 tracé en fonction du coefficient d'organicité.}
        \label{fig:ci_orgadegsom3}
    \end{subfigure}
    
    \begin{subfigure}[t]{.45\linewidth}
        \includegraphics[width=\textwidth]{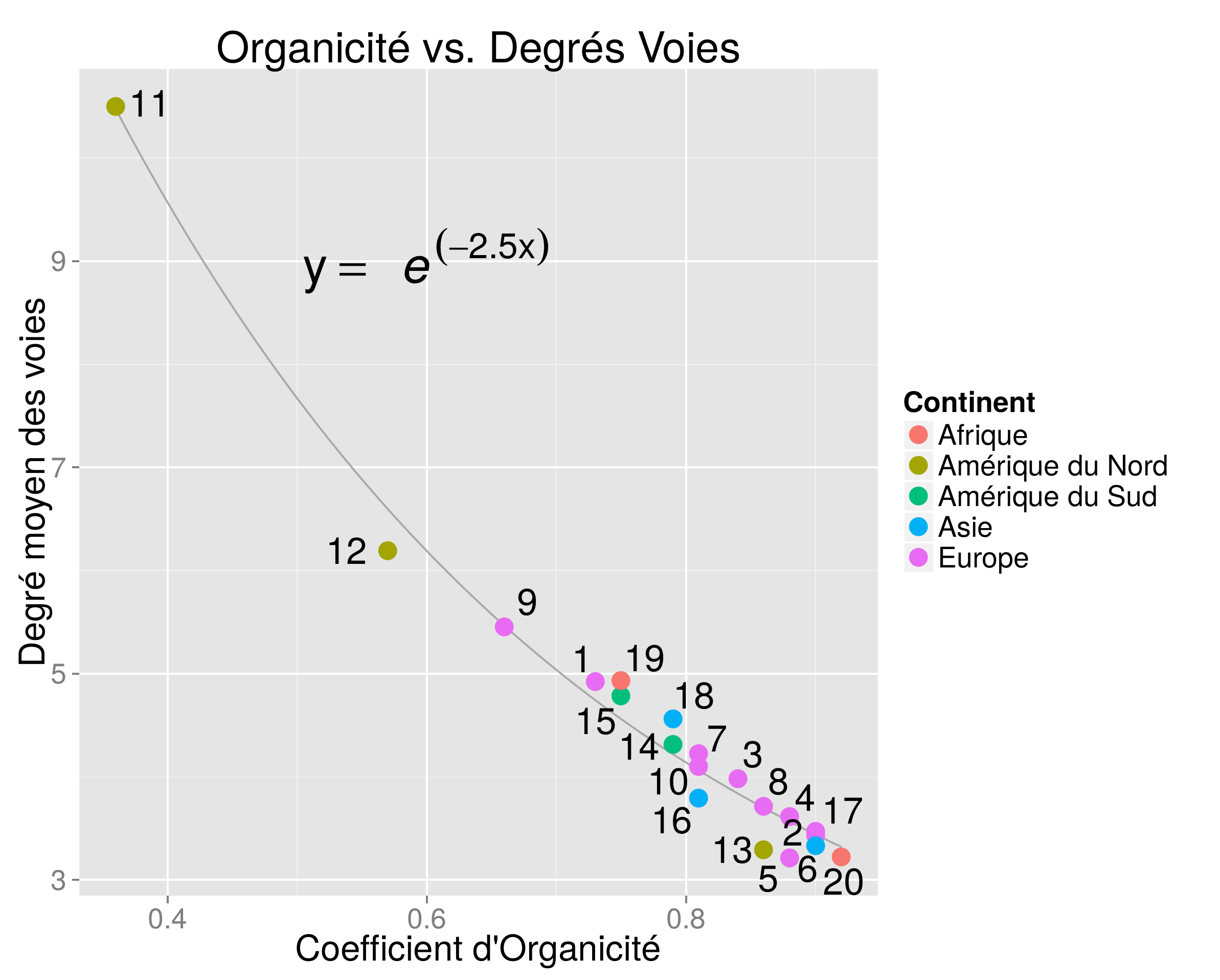}
        \caption{Degré moyen des voies tracé en fonction du coefficient d'organicité.}
        \label{fig:ci_orgadegvoie}
    \end{subfigure}

    \caption{Étude des corrélations avec le coefficient d'organicité. Courbes tracées pour les graphes issus des 20 villes du panel de recherche.}
    \label{fig:ci_orga}
\end{figure}

Nous traçons ensuite les courbes des valeurs du coefficient d'organicité en fonction du degré moyen des sommets (figure \ref{fig:ci_orgadegsom}). Leur comportement est lié : plus le graphe comporte de sommets de degrés importants, plus le coefficient d'organicité sera fort. En retirant toutes les impasses du calcul du degré moyen des graphes, nous obtenons une valeur proportionnelle au coefficient d'organicité (figure \ref{fig:ci_orgadegsom3}). Le degré moyen des nœuds une fois les impasses retirées est donc révélateur de la structure organique ou planifiée du réseau, de la même manière que ce coefficient.
Nous traçons enfin le coefficient d'organicité en fonction du degré moyen des voies (figure \ref{fig:ci_orgadegvoie}). Nous trouvons ainsi une anti-corrélation nette : plus le réseau comporte de sommets de degré 4 ou plus, plus les voies auront de possibilités d'avoir un degré important.

\FloatBarrier

\begin{figure}[h]
    \centering   
    \includegraphics[width=0.45\textwidth]{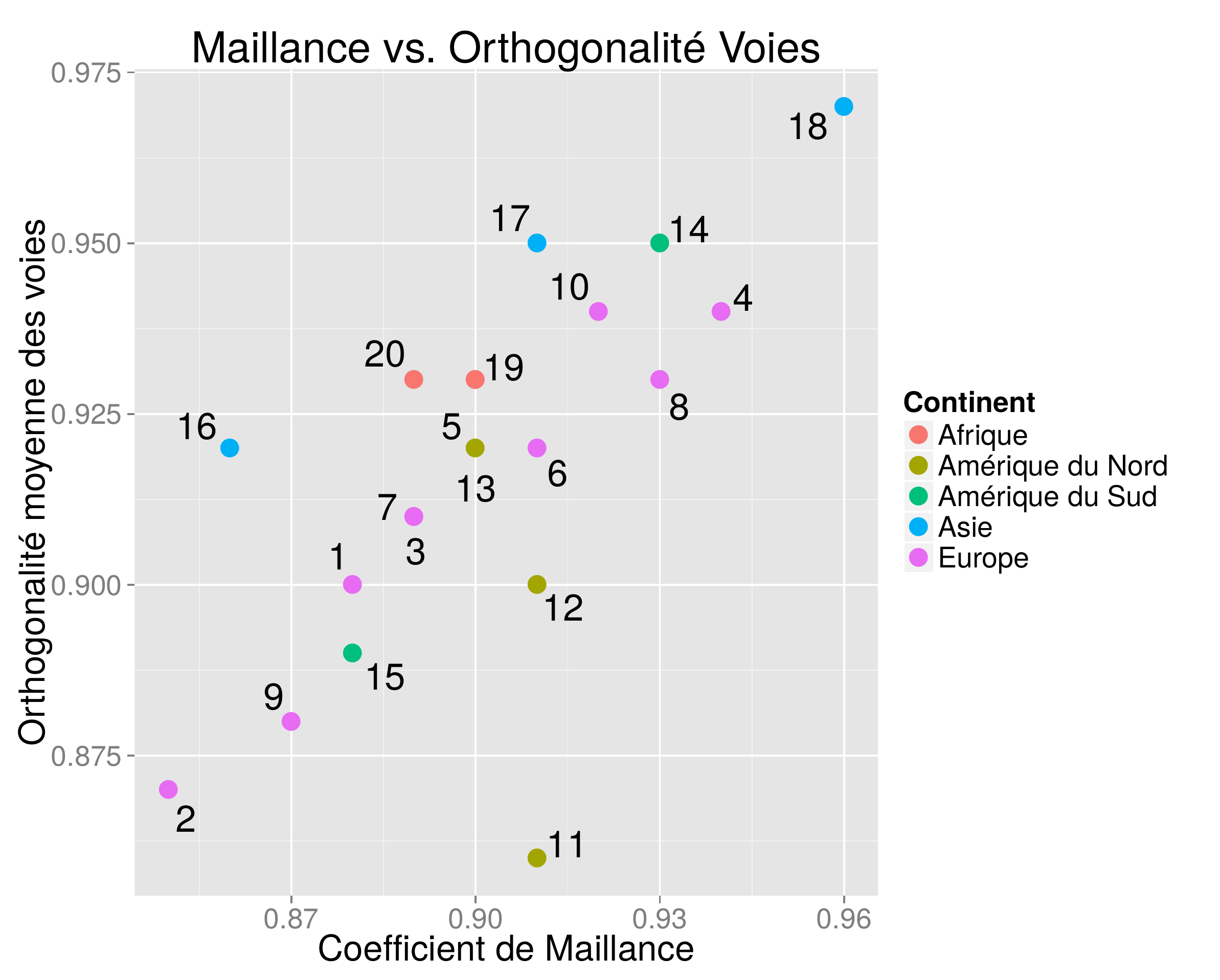}
      
    \caption{Moyenne de l'indicateur d'orthogonalité tracée en fonction du coefficient de maillance.}
    \label{fig:ci_mailortho}
\end{figure}

À la suite de l'étude entre villes, nous relevons des paramètres d'observations équivalents : le coefficient de réduction et le degré moyen des voies ; le coefficient d'organicité et la moyenne des degrés des sommets de degré 3 ou plus. Le coefficient de maillance et la moyenne des valeurs de l'indicateur d'orthogonalité calculés sur les voies apportent quant à eux des informations légèrement différentes (figure \ref{fig:ci_mailortho}). Le coefficient de maillance, en pondérant l'orthogonalité par la longueur des voies, rend l'étude de perpendicularité sur le réseau plus fine.

Nous retenons donc pour notre comparaison générale :

\begin{itemize}
\item le coefficient d'organicité : $coef_{orga}$
\item le coefficient de maillance : $coef_{mail}$
\item le coefficient d'hétérogénéité : $coef_{hete}$
\item le coefficient de réduction : $coef_{red}$
\item la moyenne de leur indicateur de closeness (normalisé) calculé sur les voies: $\overline{closeness_{voies}}$
\end{itemize}

\FloatBarrier
\subsection{Comparaison générale}

\begin{figure}[h]
    \centering
    \begin{subfigure}[t]{.45\linewidth}
        \includegraphics[width=\textwidth]{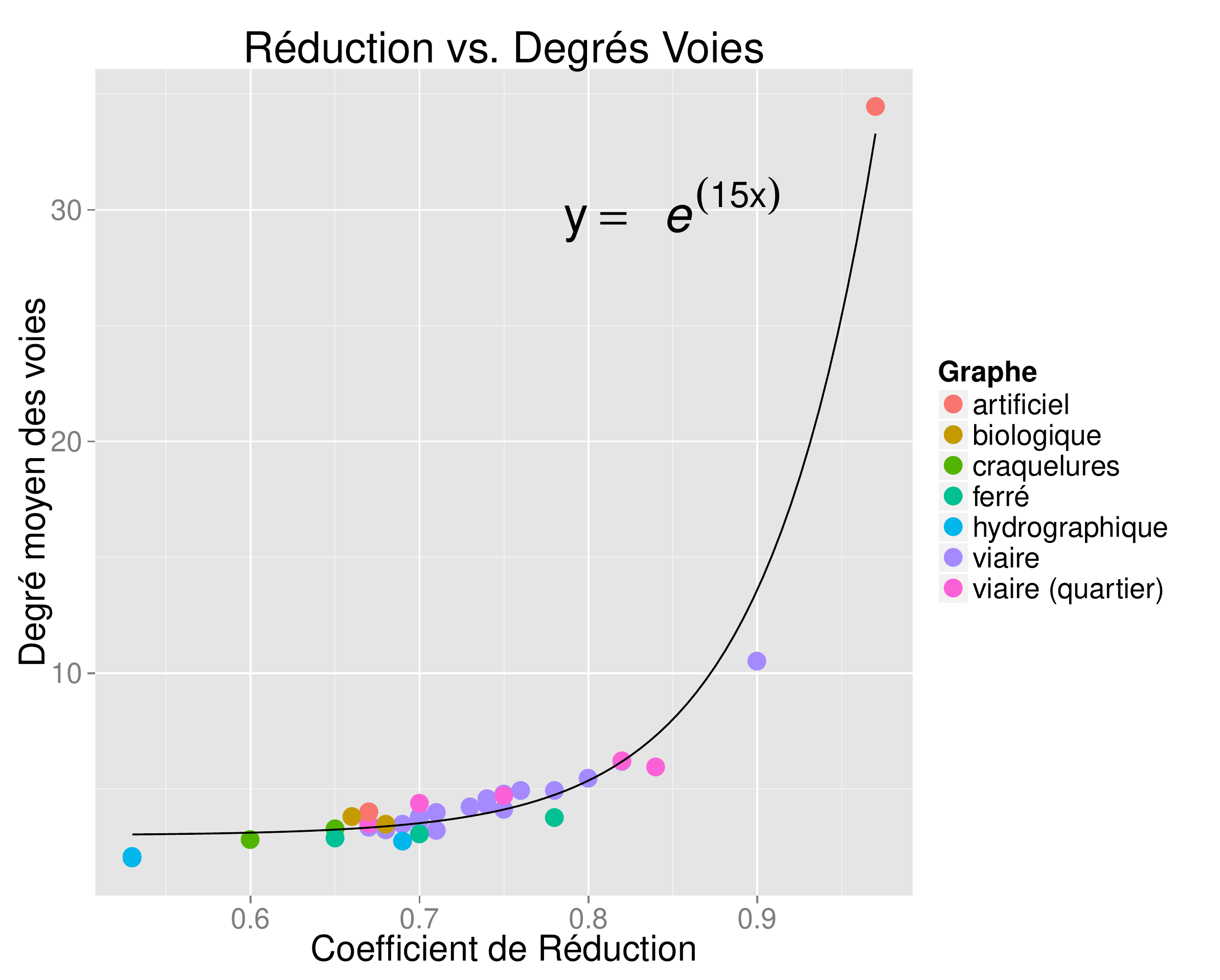}
        \caption{Moyenne de l'indicateur de degré des voies tracée en fonction du coefficient de réduction.}
        \label{fig:all_reddeg}
    \end{subfigure}
   ~    
    \begin{subfigure}[t]{.45\linewidth}
        \includegraphics[width=\textwidth]{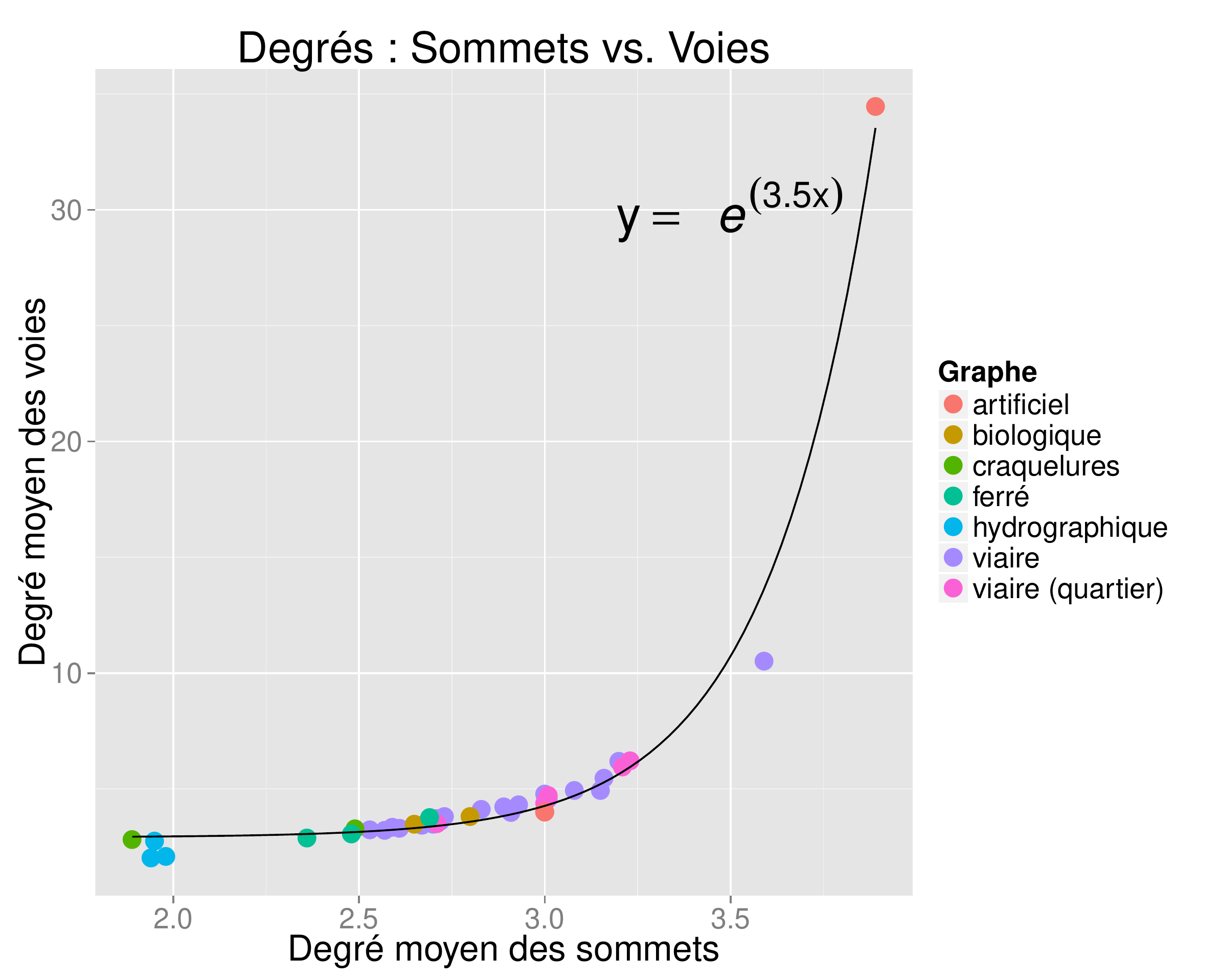}
        \caption{Moyenne de l'indicateur de degré des voies tracée en fonction du degré moyen des sommets.}
        \label{fig:all_degsomvoie}
    \end{subfigure}
      
    \begin{subfigure}[t]{.45\linewidth}
        \includegraphics[width=\textwidth]{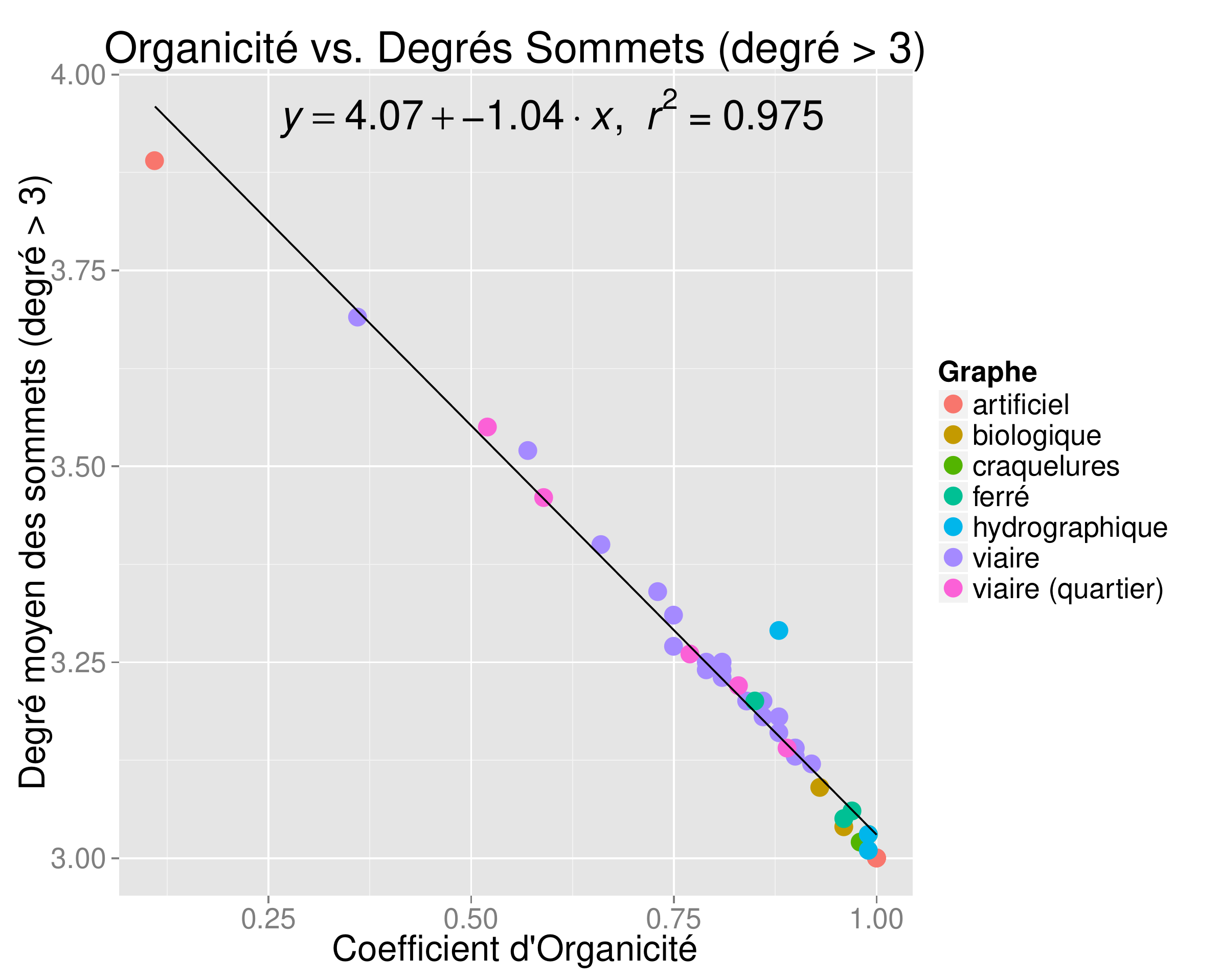}
        \caption{Degré moyen des sommets de degré supérieur ou égal à 3 tracé en fonction du coefficient d'organicité.}
        \label{fig:all_orgadegsom3}
    \end{subfigure}
    ~
    \begin{subfigure}[t]{.45\linewidth}
        \includegraphics[width=\textwidth]{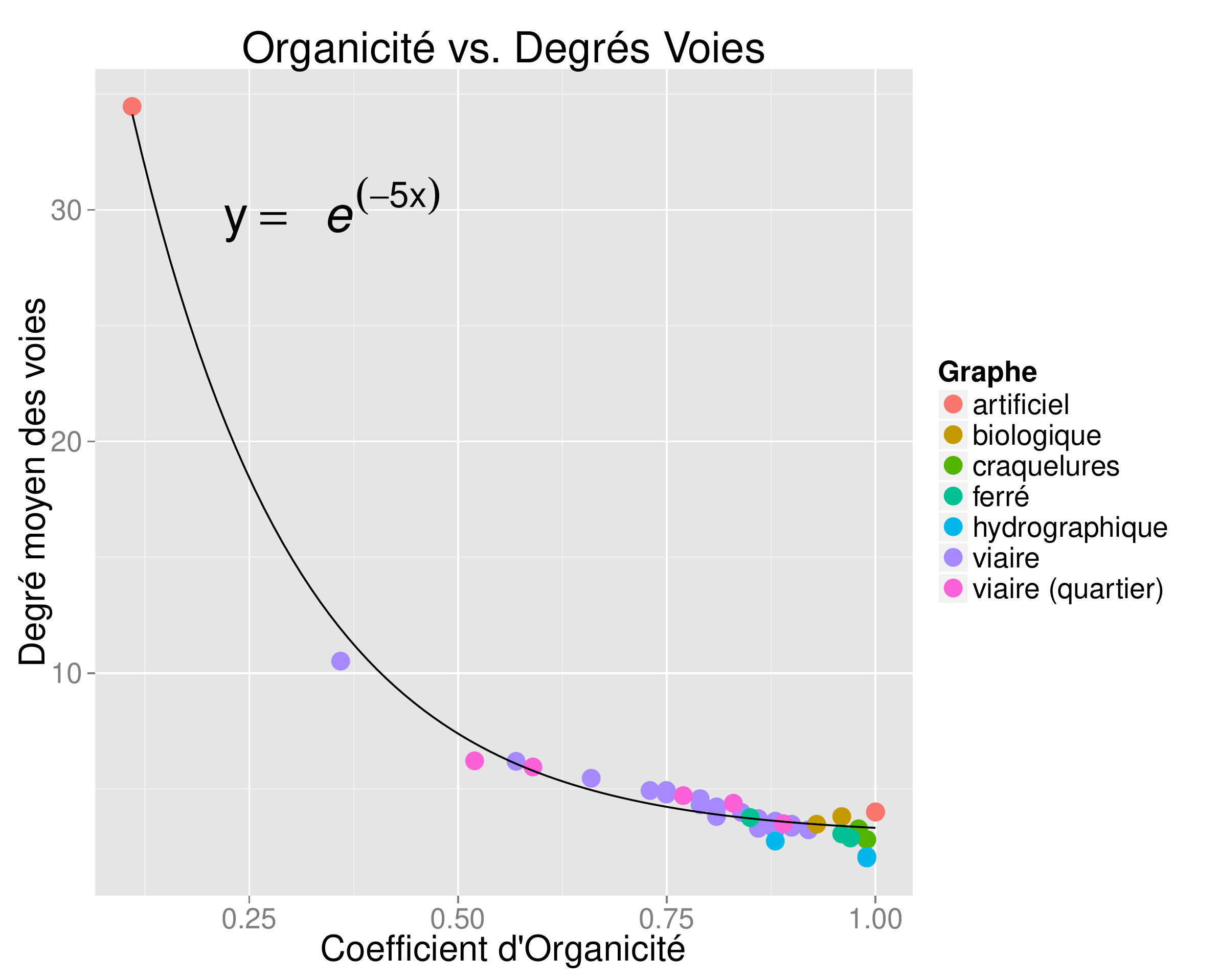}
        \caption{Degré moyen des voies tracé en fonction du coefficient d'organicité.}
        \label{fig:all_orgadegvoie}
    \end{subfigure}

    \caption{Étude de conservation des relations de corrélations. Courbes tracées pour l'ensemble du panel de recherche.}
    \label{fig:all_corr}
\end{figure}

Nous introduisons pour finir l'ensemble des graphes du panel de recherche dans la comparaison. Nous remarquons dans un premier temps que les relations de proportionnalité observées pour les graphes de villes seules se conservent une fois l'étude étendue à l'ensemble des graphes (figure \ref{fig:all_corr}). La grande majorité des graphes observés a des sommets de degré moyen entre 2,5 et 3,2. Seuls se distinguent, en dessous de cette moyenne, les graphes de réseaux hydrographiques et un des deux réseaux créés sur une plaque d'argile, avec des degrés moyens autour de 2. Au-dessus de cette moyenne, le graphe de Manhattan et (à l'extrême) celui créé artificiellement avec un bruit nul ont des degrés moyens supérieurs à 3,5. Cette moyenne de trois connexions par sommet s'explique par le caractère géographique de notre étude. En effet, dans un espace à deux dimensions, le nombre de connexions sur un sommet est limité par la contrainte planaire. L'échelle d'utilisation du réseau ne modifie pas le degré moyen car la taille de l'intersection est proportionnelle à celle du filaire s'y connectant. Ainsi, une jonction entre deux veinures de feuille aura le même rapport de taille que celle de deux rues se connectant dans un tissu urbain à échelle humaine. Nous verrons que cet aspect physique, spatialisé, des réseaux étudiés, est le vecteur du rapprochement de leurs logiques topologiques.

\FloatBarrier

\begin{figure}[h]
    \centering
    \begin{subfigure}[t]{.45\linewidth}
        \includegraphics[width=\textwidth]{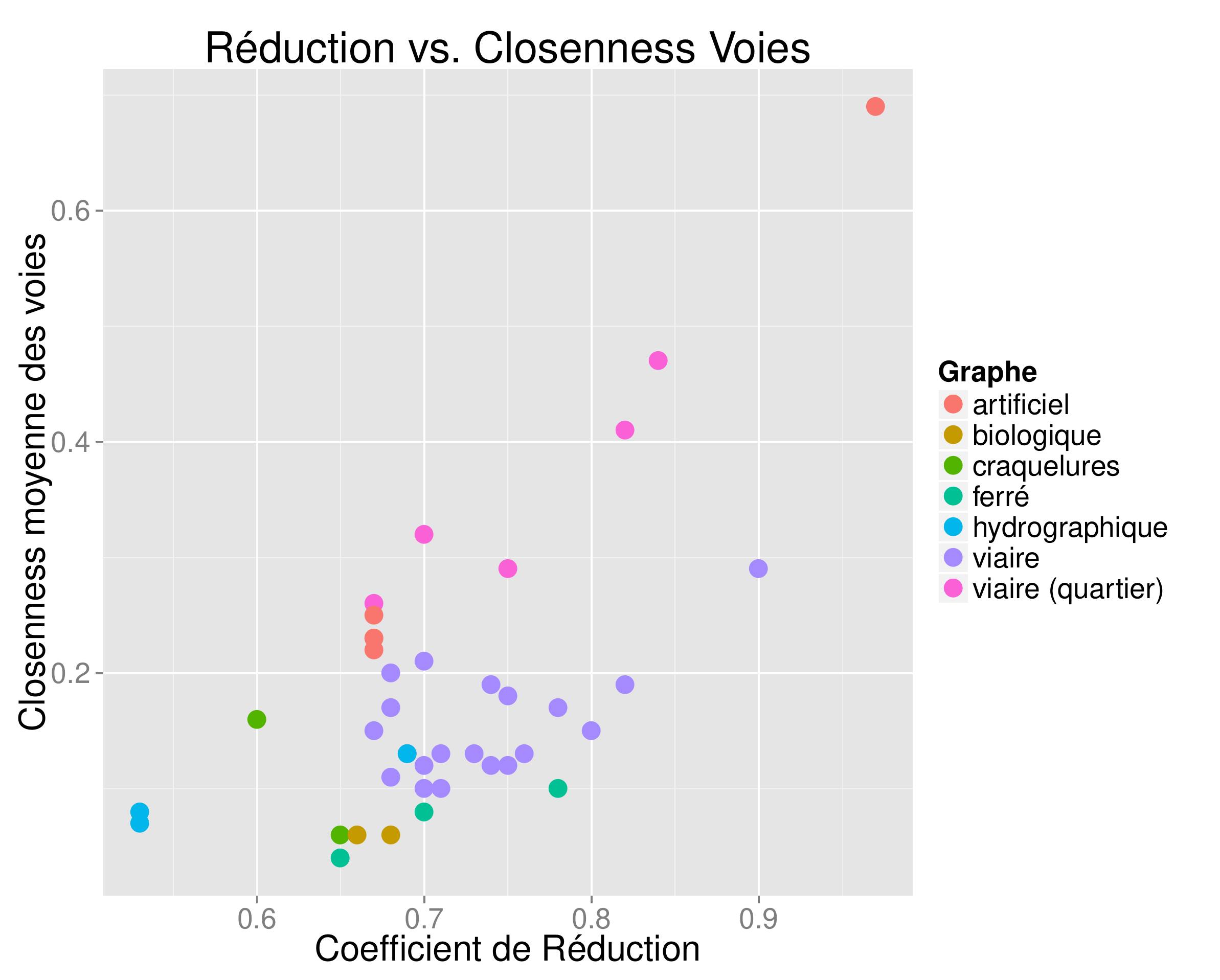}
        \caption{Moyenne de l'indicateur de closeness tracée en fonction du coefficient de réduction.}
        \label{fig:all_redclo}
    \end{subfigure}
    ~
    \begin{subfigure}[t]{.45\linewidth}
        \includegraphics[width=\textwidth]{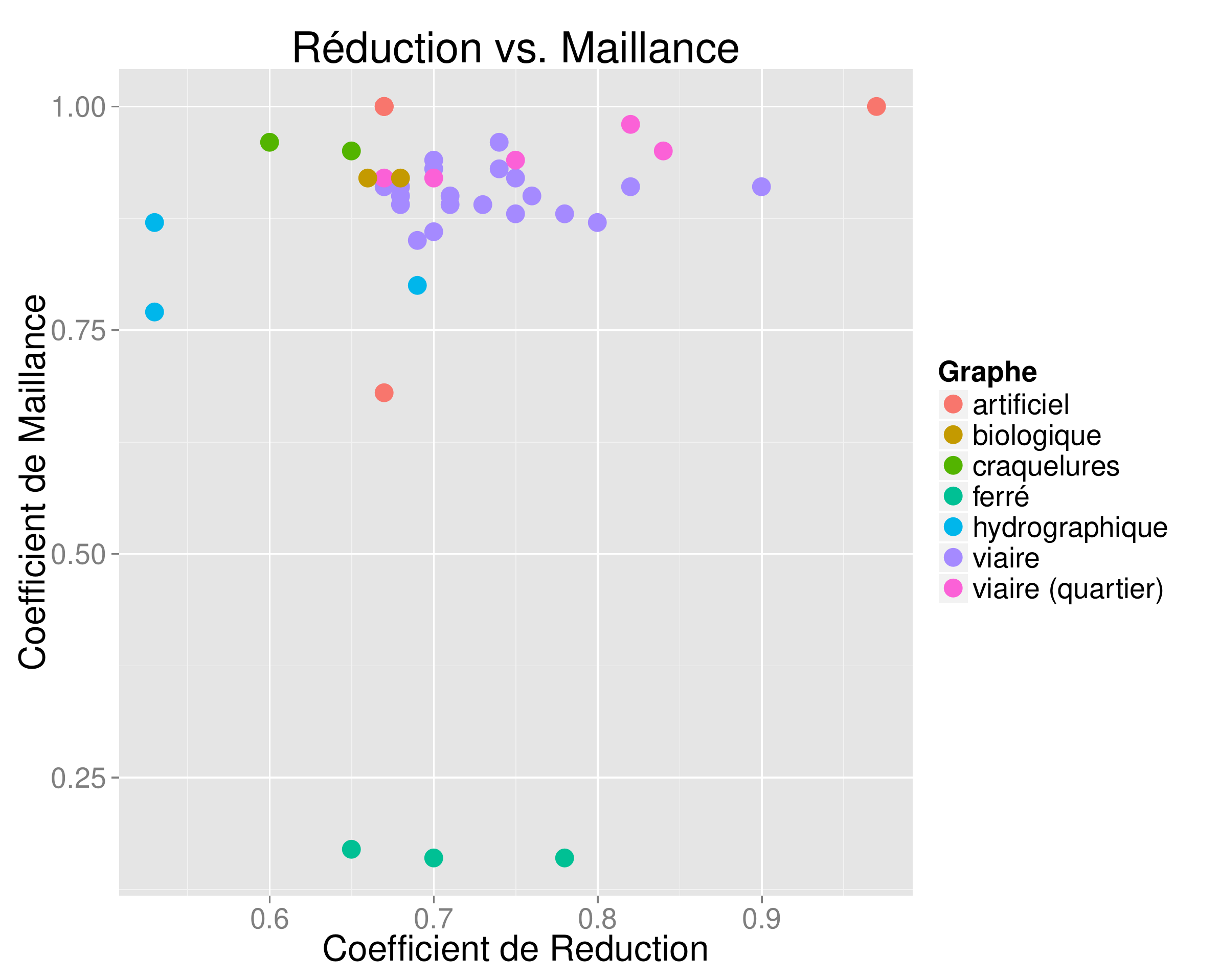}
        \caption{Coefficient de maillance tracé en fonction de celui de réduction.}
        \label{fig:all_redmail}
    \end{subfigure}

    \caption{Étude de variations des paramètres de comparaison en fonction du coefficient de réduction. Courbes tracées pour l'ensemble du panel de recherche. }
    \label{fig:all_compred}
\end{figure}

Comme pour la sélection de 20 villes, la closeness des voies et le coefficient de réduction sont liés (figure \ref{fig:all_redclo}). Le coefficient de Pearson donne une corrélation pour les deux valeurs de 0,69.  Les réseaux viaires ont une meilleure closeness comparativement à leur coefficient de réduction par rapport aux autres réseaux. Le réseau \enquote{Bruit nul} se détache de l'ensemble des autres avec un coefficient de closeness beaucoup plus important. À l'inverse, les réseaux hydrographiques se positionnent avec les coefficients de réduction et des moyennes pour les indicateurs de closeness les plus faibles. Sur ces réseaux, l'indicateur de proximité n'est pas révélateur de structures car les voies construites ne forment pas d'éléments multi-échelle. Sans objets traversants, le réseau de voie n'apporte pas plus de stabilité que celui d'arcs. La closeness ne révèle donc que le centre du réseau, dont la distance topologique avec les autres objets du réseau est d'autant plus grande que le linéaire étudié est important. Il en est de même pour les réseaux de voies ferrées qui, même s'ils ont un meilleur coefficient de réduction, dû à la conservation des continuités aux connexions, gardent un coefficient moyen de closeness très bas.

Les réseaux ferrés se distinguent des autres graphes par leur coefficient de maillance très bas (figure \ref{fig:all_redmail}). Les réseaux hydrographiques extraits d'Inde et d'Italie ont également cette particularité, moins exacerbée. Seul le réseau hydrographique extrait d'Amazonie se retrouve avec un coefficient de maillance comparable à ceux d'Avignon, de Téhéran ou même de Barcelone. Mais il se distingue de ceux-ci par un coefficient de réduction très faible. Le réseau artificiel construit avec le modèle avec angles a une maillance intermédiaire entre les réseaux ferrés et hydrographiques. Ceux construits sans y ajouter d'angles sont noyés dans la masse des réseaux viaires de la même manière que les réseaux issus de tissus biologiques. Ceux construits à partir de numérisations de craquelures dans de l'argile ont une maillance comparable à celle des villes, mais présentent un coefficient de réduction plus faible. Ceci est dû à l'effet observé, appelé \enquote{effet bulle de savon}, qui, malgré les redressements de géométrie, crée des \enquote{pétales} d'argiles et brise les voies.

\FloatBarrier

\begin{figure}[h]
    \centering
     \begin{subfigure}[t]{.45\linewidth}
        \includegraphics[width=\textwidth]{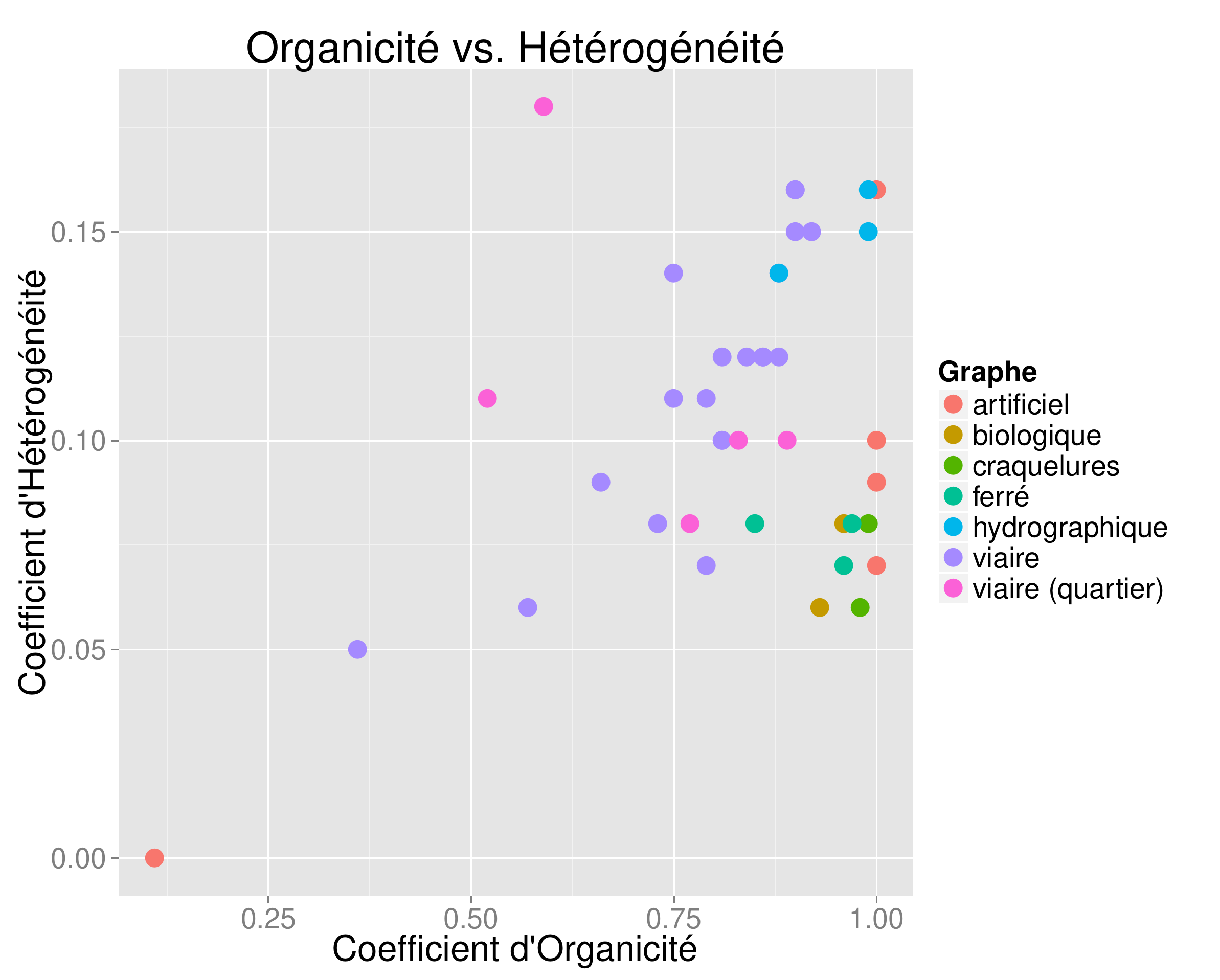}
        \caption{Coefficient d'hétérogénéité tracé en fonction de celui d'organicité.}
        \label{fig:all_orgahete}
    \end{subfigure}
    ~
    \begin{subfigure}[t]{.45\linewidth}
        \includegraphics[width=\textwidth]{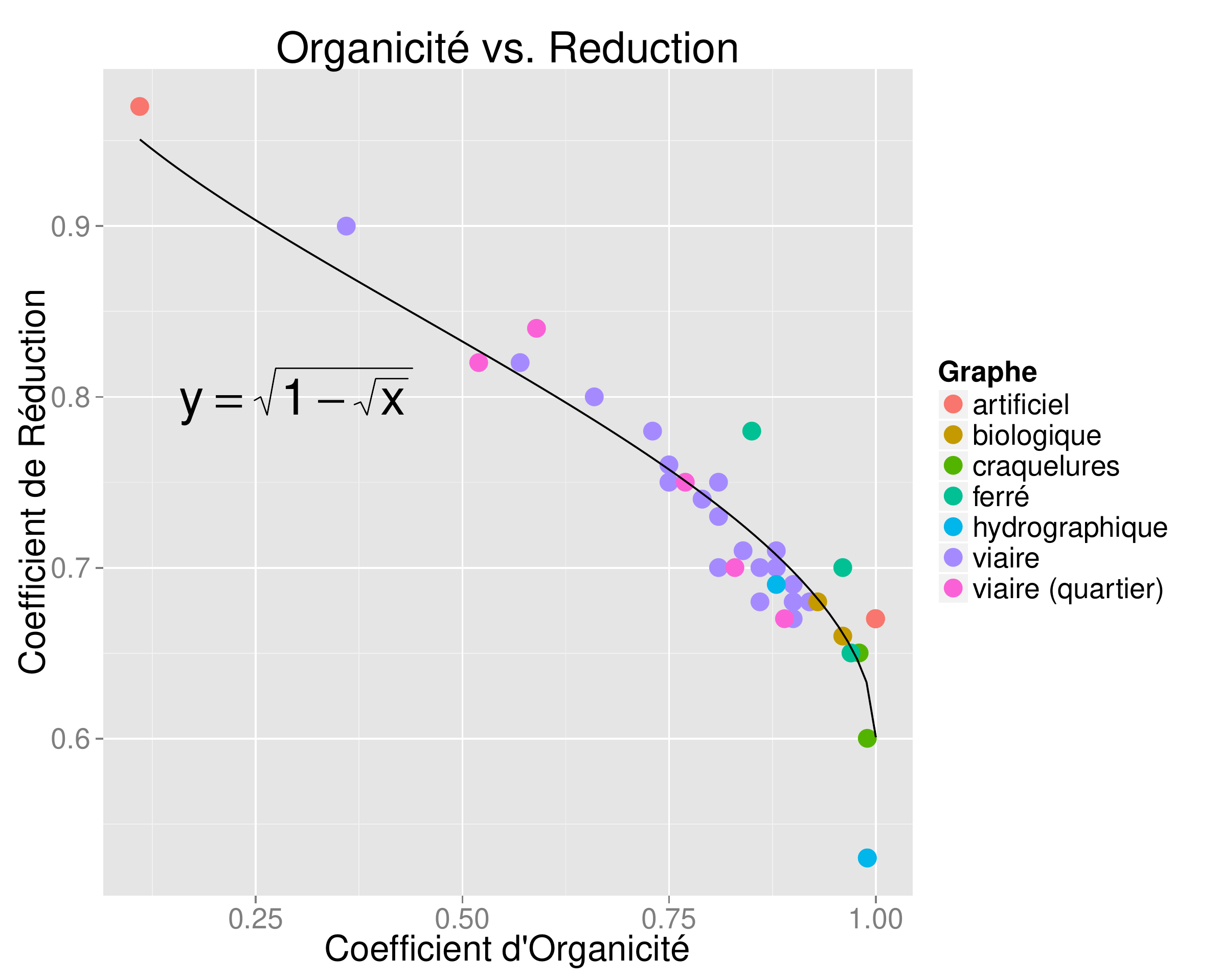}
        \caption{Coefficient de réduction tracé en fonction de celui de organicité.}
        \label{fig:all_orgared}
    \end{subfigure}
    
    \begin{subfigure}[t]{.45\linewidth}
        \includegraphics[width=\textwidth]{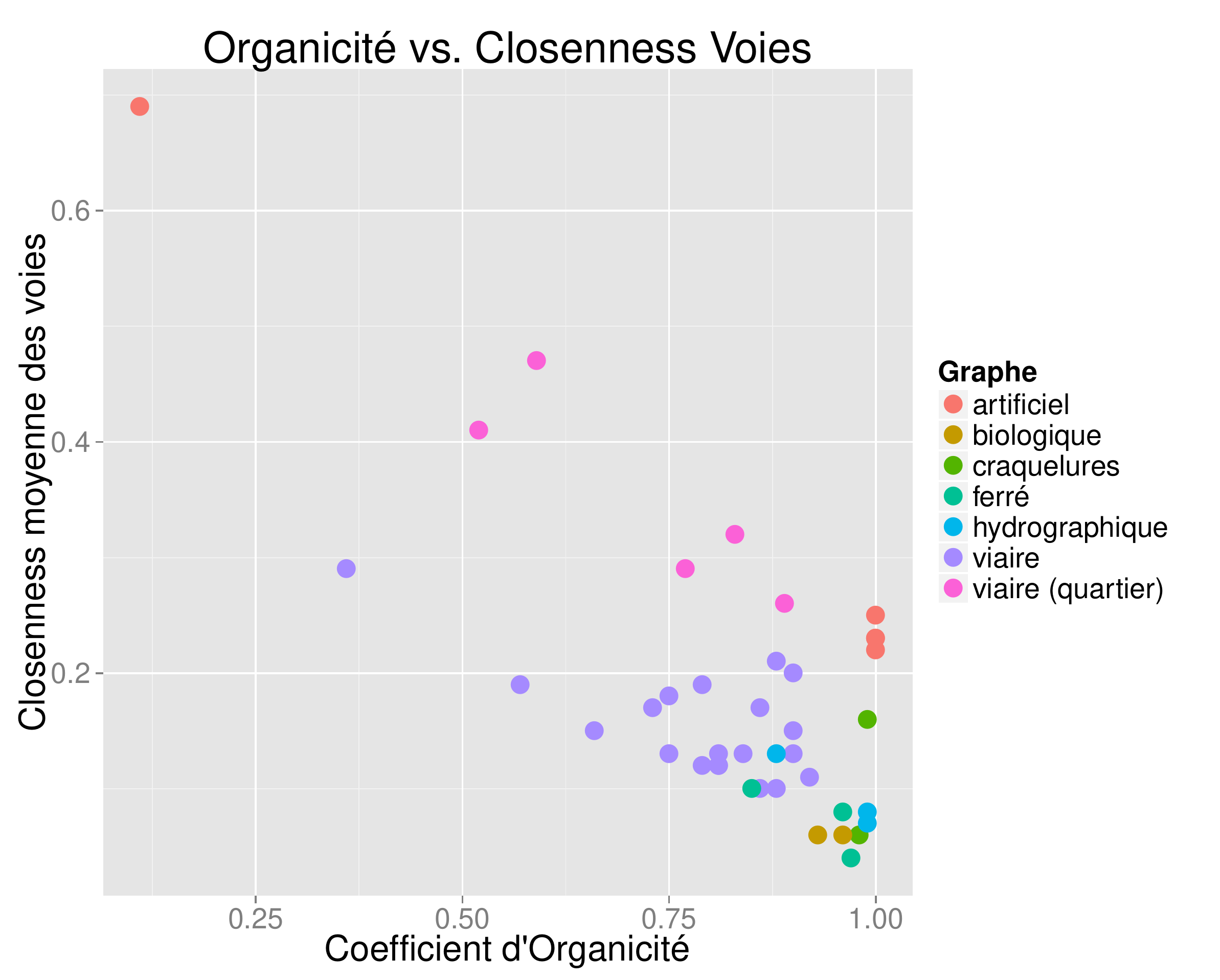}
        \caption{Moyenne de l'indicateur de closeness tracée en fonction du coefficient d'organicité.}
        \label{fig:all_orgaclo}
    \end{subfigure}

    \caption{Étude de variations des paramètres de comparaison en fonction du coefficient d'organicité. Courbes tracées pour l'ensemble du panel de recherche.}
    \label{fig:all_orga}
\end{figure}

Concernant le coefficient d'hétérogénéité, tous les graphes ont des valeurs regroupées entre 0,05 (pour les plus homogènes : Manhattan, San Francisco mais également les réseaux issus de l'argile et de la feuille) et 0,18 pour le plus hétérogène (le quartier de Neuf-Brisach qui fait coexister sur un nombre d'arcs très réduits deux structures opposées : une circulaire et l'autre quadrillée). Ceci en mettant à l'écart l'échantillon artificiel \enquote{Bruit nul} dont l'hétérogénéité est de 0 (figure \ref{fig:all_orgahete}).

L'organicité est ici également, anti-corrélée au coefficient de réduction du graphe (figure \ref{fig:all_orgared}). Nous pouvons en conclure que, quelle que soit la nature du graphe étudié, le nombre de sommets de degré 4 et plus du graphe est inversement proportionnel au nombre d'arcs par voie. L'organicité a donc un comportement inversé par rapport à celui du coefficient de réduction dans sa comparaison avec la closeness (figure \ref{fig:all_orgaclo}). De cette figure se détachent 3 réseaux viaires : ceux issus de Manhattan, Cucq et Neuf-Brisach. Ce sont les trois réseaux les plus réguliers de notre étude. Ils ont de fortes valeurs de closeness pour une basse organicité. 

Cette étude inter-graphes nous permet de comparer les propriétés géométriques de réseaux spatiaux aux histoires et objectifs de constructions très différents. Avec des paramètres définis uniquement à partir de la géométrie et de la topologie du réseau, nous parvenons à mettre en évidence des similitudes et des différences. Celles-ci montrent que l'orographie, comme les besoins de déplacement, contraignent la traversée d'un espace. Qu'il s'agisse d'eau, de trains, de voitures ou de nutriments tous ces réseaux veulent recouvrir un espace pour y faciliter la transition d'un point vers un autre. Leurs principales différences résident dans la géométrie de leurs intersections, qui isolent les réseaux ferrés et quelques réseaux hydrographiques des réseaux viaires et biologiques. Cependant, les logiques topologiques restent les mêmes : plus le réseau admet de nœuds au degré important, mieux les voies créées sur celui-ci pourront se connecter avec leur voisinage, et plus elles regrouperont d'arcs. En effet, la contrainte géographique dans un espace à deux dimensions implique que plus un sommet comporte d'arcs connectés, plus la probabilité de trouver un couple d'arcs ayant un angle de déviation inférieur à l'angle seuil fixé augmente. Cette propriété reste vraie pour tous les graphes s'inscrivant dans un espace à deux dimensions, quelle que soit leur nature (tous sont contraints par leur inscription sur le territoire et leur échelle d'utilisation).

\FloatBarrier
\section{Analyse Intra-Graphes}

Nous proposons de comparer les variations d'un même indicateur à l'intérieur d'un graphe. Pour cela, nous choisissons 3 graphes viaires du panel de recherche proposé. Nous fondons notre sélection sur les différences de propriétés structurelles entre ces graphes. Nous étudions donc ici le réseau viaire de Manhattan, celui de Paris et celui d'Avignon. Tous trois regroupent des propriétés différentes. Nous voulons étudier l'impact de ces différences sur les variations des indicateurs appliqués aux voies.

Nous étudions ainsi les variations des indicateurs définis dans la grammaire de lecture de la spatialité pour leurs voies :
\begin{itemize}
\item longueurs
\item degré
\item closeness
\item orthogonalité
\item accessibilité maillée
\item espacement
\end{itemize}

Nous regroupons les informations à propos des approximations des courbes dans les tableaux \ref{tab:recap_approx} et \ref{tab:valeurs_coef} à la fin du paragraphe.

\begin{figure}[h]
    \centering
     \begin{subfigure}[t]{.3\linewidth}
        \includegraphics[width=\textwidth]{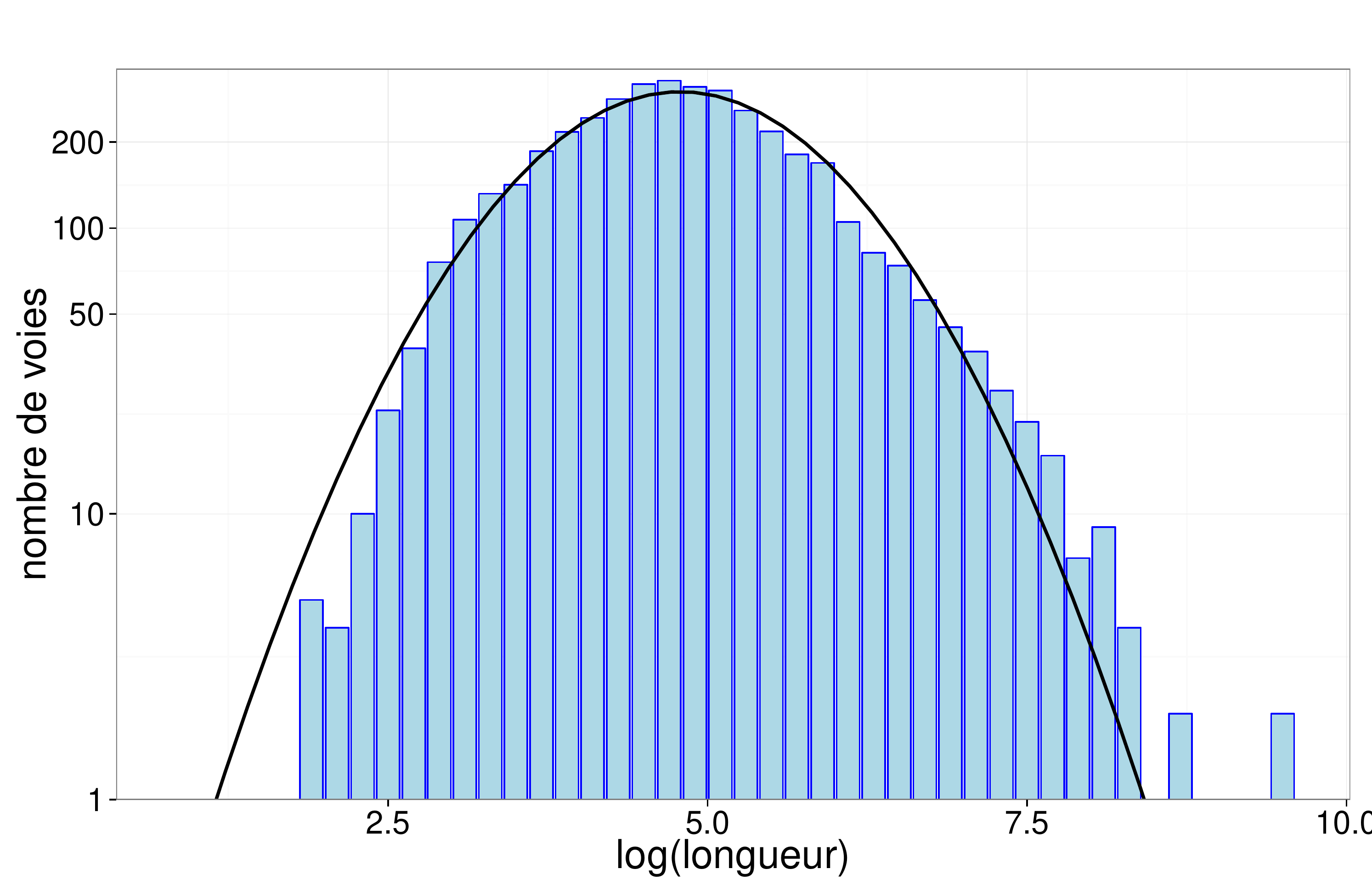}
        \caption{Avignon.}
        \label{fig:int_len_av}
    \end{subfigure}
    ~
    \begin{subfigure}[t]{.3\linewidth}
        \includegraphics[width=\textwidth]{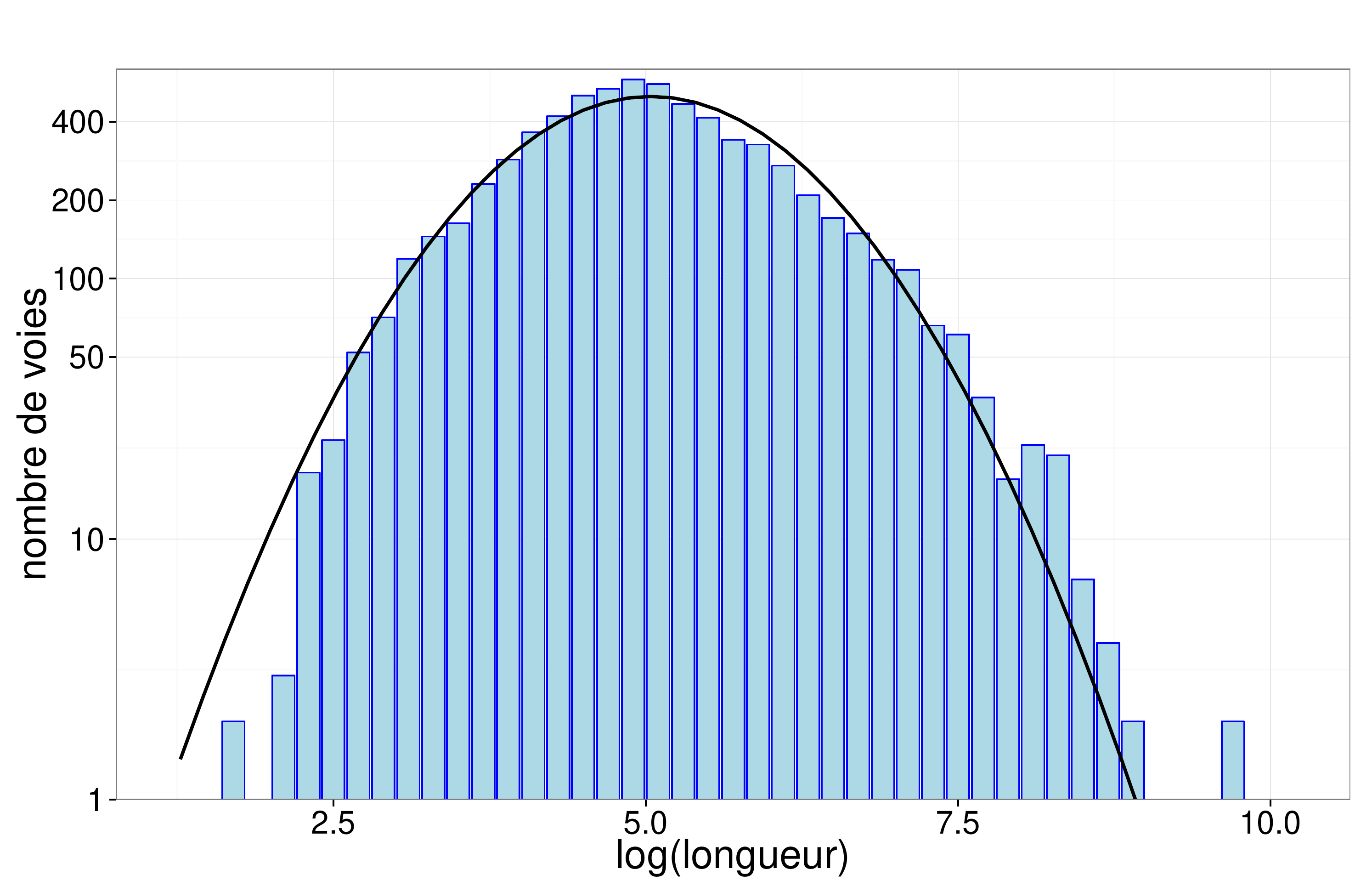}
        \caption{Paris.}
        \label{fig:int_len_par}
    \end{subfigure}
    ~
    \begin{subfigure}[t]{.3\linewidth}
        \includegraphics[width=\textwidth]{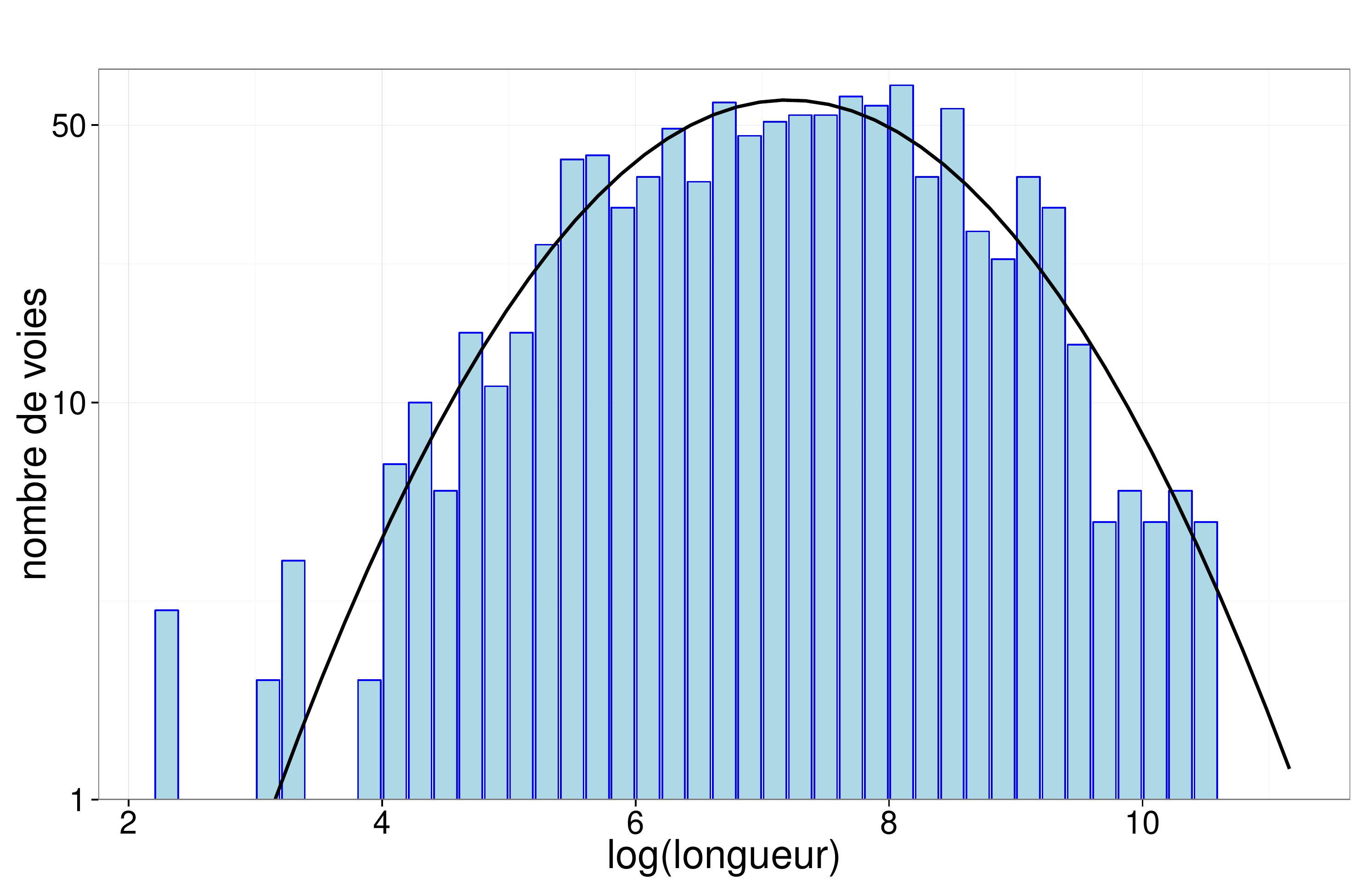}
        \caption{Manhattan.}
        \label{fig:int_len_man}
    \end{subfigure}

    \caption{Histogrammes représentant la distribution de la longueur des voies. Loi log normale tracée à partir des données.}
    \label{fig:int_len}
\end{figure}

Nous prenons le logarithme des longueurs de voies afin d'avoir une meilleure visibilité de leurs variations. En faisant cela, nous divisons les valeurs brutes par une longueur de référence, choisie ici à 1 mètre. Les tracés des histogrammes révèlent tous des distributions que l'on peut approximer avec une courbe log-normale (figure \ref{fig:int_len}). Celle de Manhattan est plus bruitée que les deux autres : la régularité des voies qui sont créées sur ce graphe augmente l'écart-type de la courbe. Sur Paris et Avignon, les deux courbes suivent le même tracé, malgré les différences géographiques entre ces deux villes : l'un est issu de la capitale, l'autre d'une ville de province et de ses alentours, ce qui leur donne des géométries très différentes. La distribution de ces courbes révèle un processus commun de découpage. 
Nous observons un comportement qui témoigne d'une même dynamique : une nouvelle voie créée dans un réseau viaire est souvent issue d'une voie précédente. Ce découpage successif est celui modélisé par les graphes artificiels que nous avons créés et dont les longueurs de voies suivent les mêmes distributions (figure \ref{fig:courbe_reseauxartif}). En découpant successivement et aléatoirement l'espace, chaque nouvelle longueur d'arc créée est une fraction de celle de l'arc dont elle est issue. En reconstituant les continuités nous sommons ces fractions successives aléatoires, ce qui nous donne la forme de distribution en log normale. Cependant, ce modèle ne correspond à ce qui est observé dans les villes que si la distribution des aires des faces du graphe est suffisamment hétérogène, avec un bruit sur la position de découpe assez grand (cf paragraphe \enquote{Graphes artificiels}).

\FloatBarrier

\begin{figure}[h]
    \centering
     \begin{subfigure}[t]{.3\linewidth}
        \includegraphics[width=\textwidth]{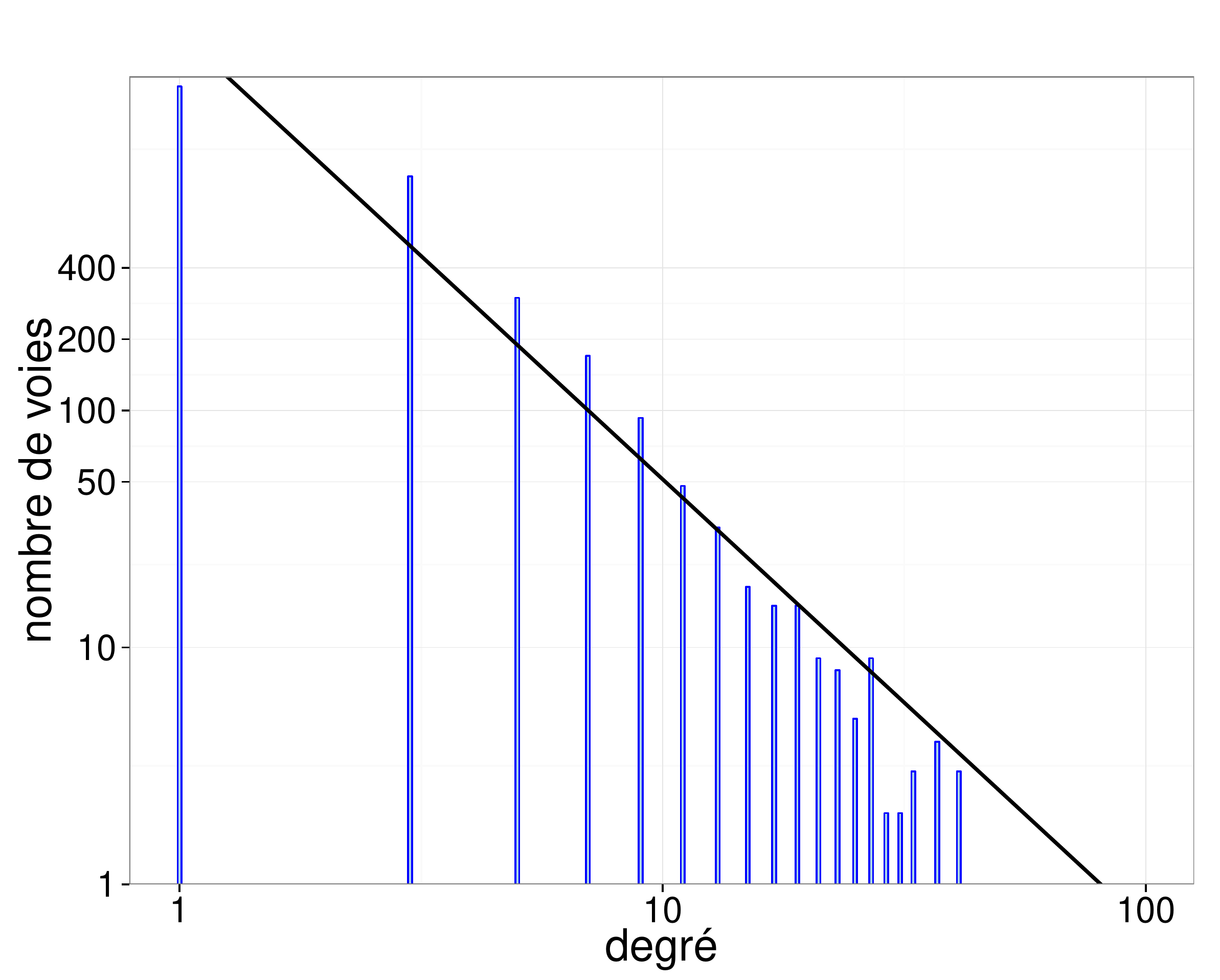}
        \caption{Avignon.}
        \label{fig:int_degree_av}
    \end{subfigure}
    ~
    \begin{subfigure}[t]{.3\linewidth}
        \includegraphics[width=\textwidth]{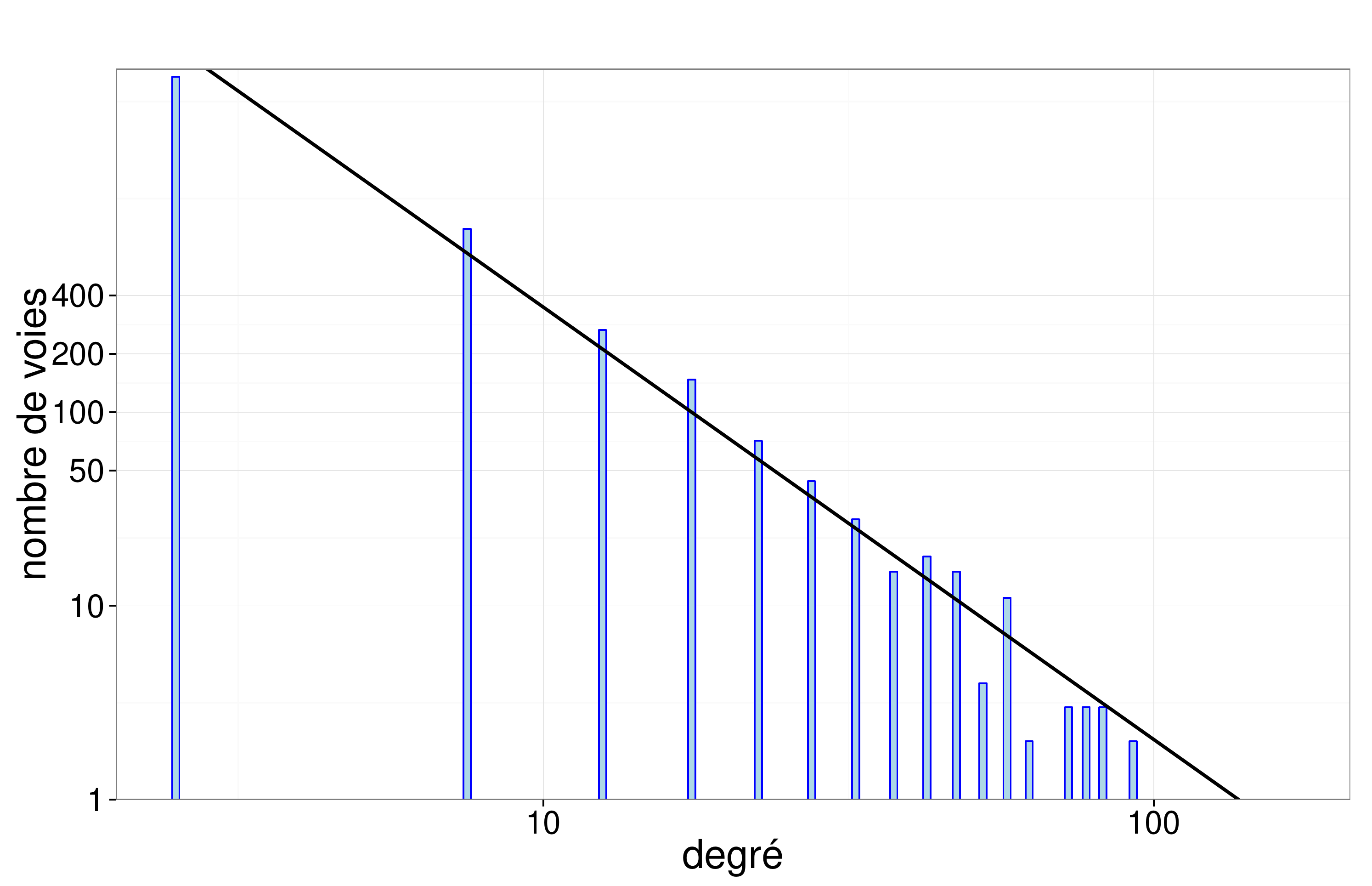}
        \caption{Paris.}
        \label{fig:int_degree_par}
    \end{subfigure}
    ~
    \begin{subfigure}[t]{.3\linewidth}
        \includegraphics[width=\textwidth]{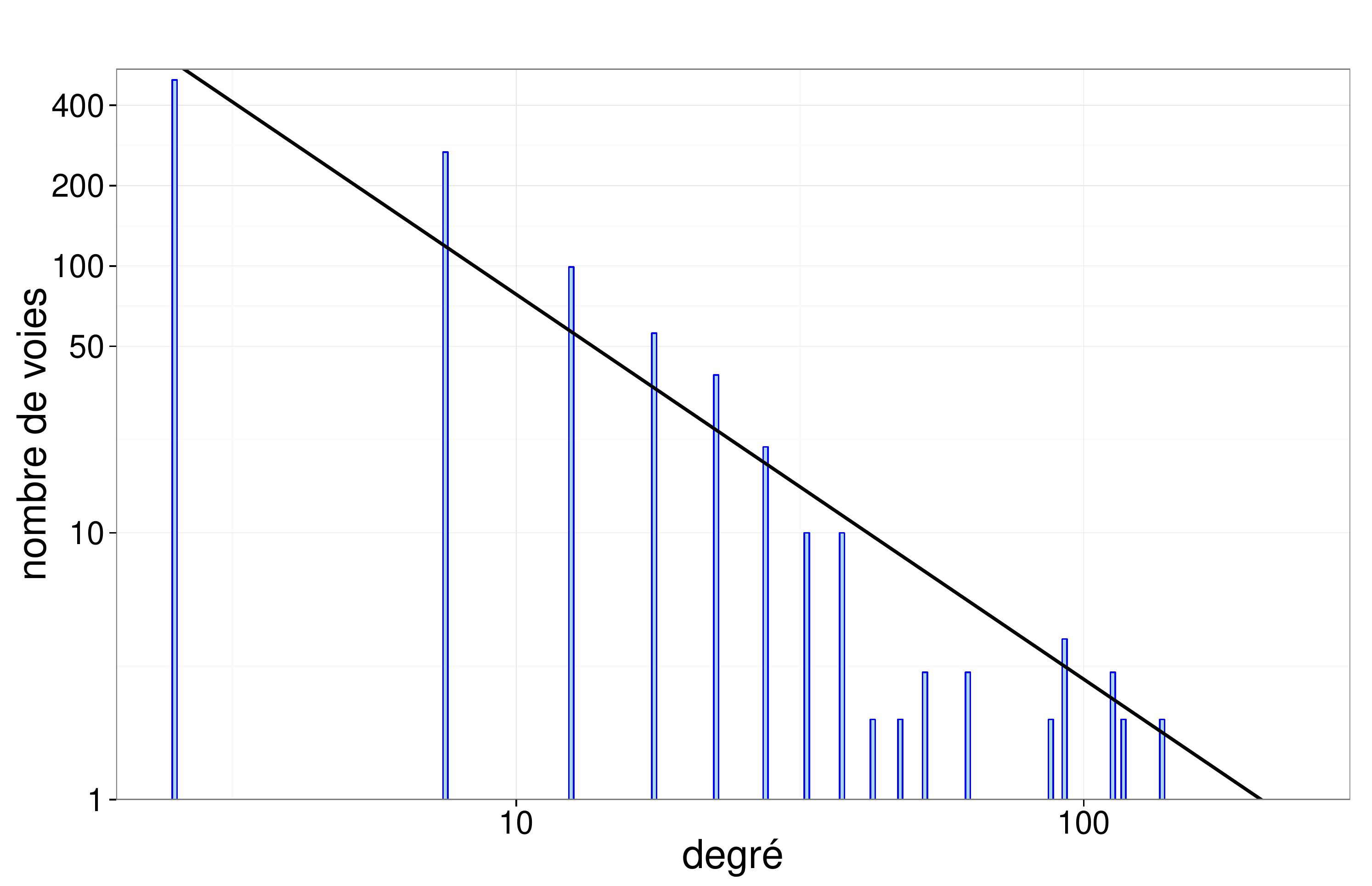}
        \caption{Manhattan.}
        \label{fig:int_degree_man}
    \end{subfigure}

    \caption{Histogrammes représentant la distribution du degré des voies. Approximation par une loi de puissance.}
    \label{fig:int_degree}
\end{figure}

Si nous traçons les distributions des autres indicateurs caractéristiques pour la description des voies, nous observons que même si leur valeur moyenne changent, les variations ont des comportements proches. Ainsi, dans tous les réseaux, peu de voies sont de degré important, et beaucoup de voies ont des degrés faibles (figure \ref{fig:int_degree}). C'est un comportement souvent observé dans les sciences des systèmes complexes. Il traduit un phénomène de \textit{clustering} dans le \textit{line graph} du réseau viaire : certaines voies, déjà très connectées à leur entourage, auront tendance à attirer de nouvelles connexions (\textit{the richest get richer}) et seront ainsi celles qui établiront des liens encore plus rapides (en terme de distance topologique) entre d'autres voies moins connectées.

\FloatBarrier

\begin{figure}[h]
    \centering
     \begin{subfigure}[t]{.3\linewidth}
        \includegraphics[width=\textwidth]{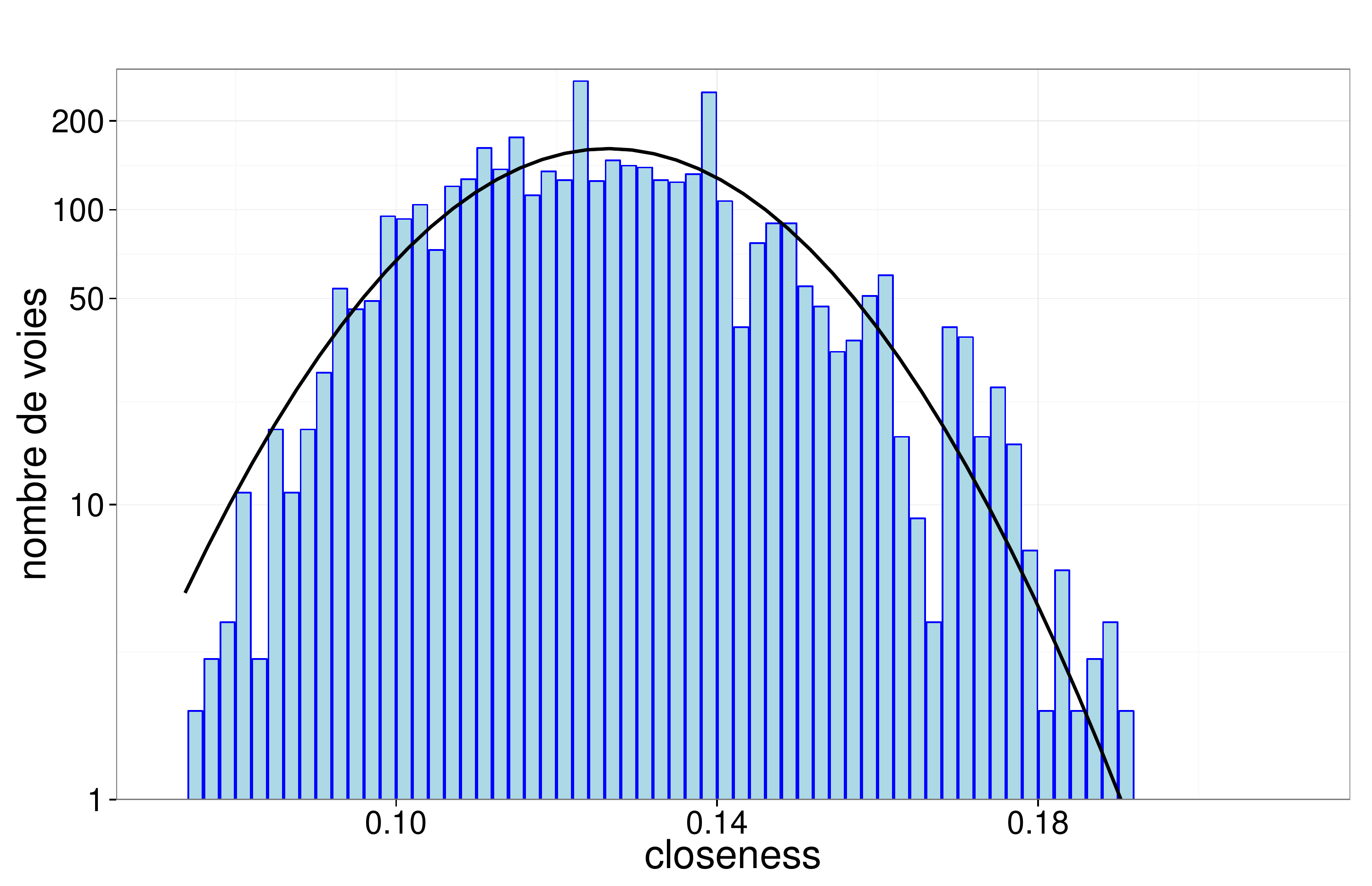}
        \caption{Avignon.}
        \label{fig:int_clo_av}
    \end{subfigure}
    ~
    \begin{subfigure}[t]{.3\linewidth}
        \includegraphics[width=\textwidth]{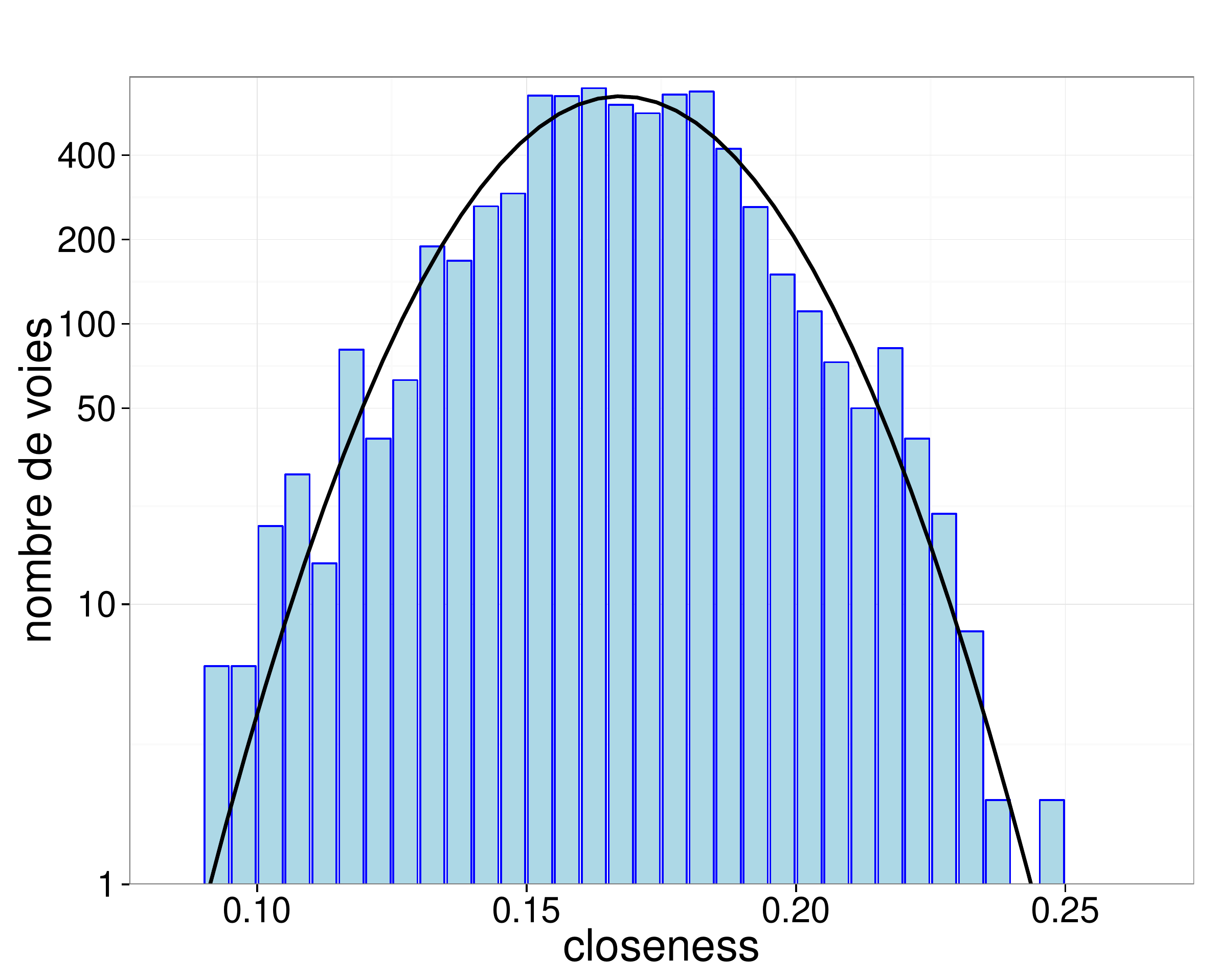}
        \caption{Paris.}
        \label{fig:int_clo_par}
    \end{subfigure}
    ~
    \begin{subfigure}[t]{.3\linewidth}
        \includegraphics[width=\textwidth]{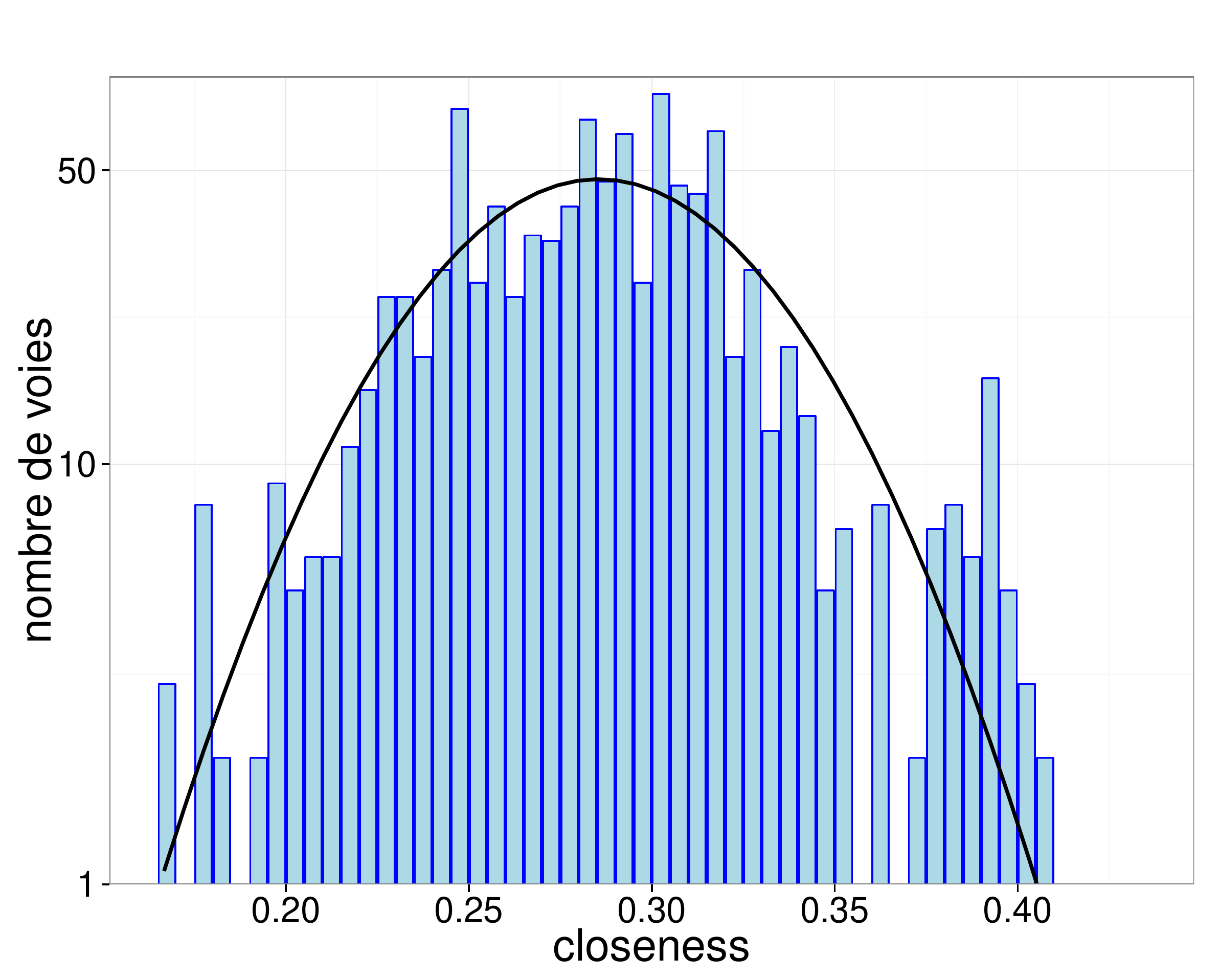}
        \caption{Manhattan.}
        \label{fig:int_clo_man}
    \end{subfigure}

    \caption{Histogrammes représentant la distribution de la closeness des voies. Loi normale tracée à partir des données.}
    \label{fig:int_clo}
\end{figure}

\begin{figure}[h]
    \centering
     \begin{subfigure}[t]{.3\linewidth}
        \includegraphics[width=\textwidth]{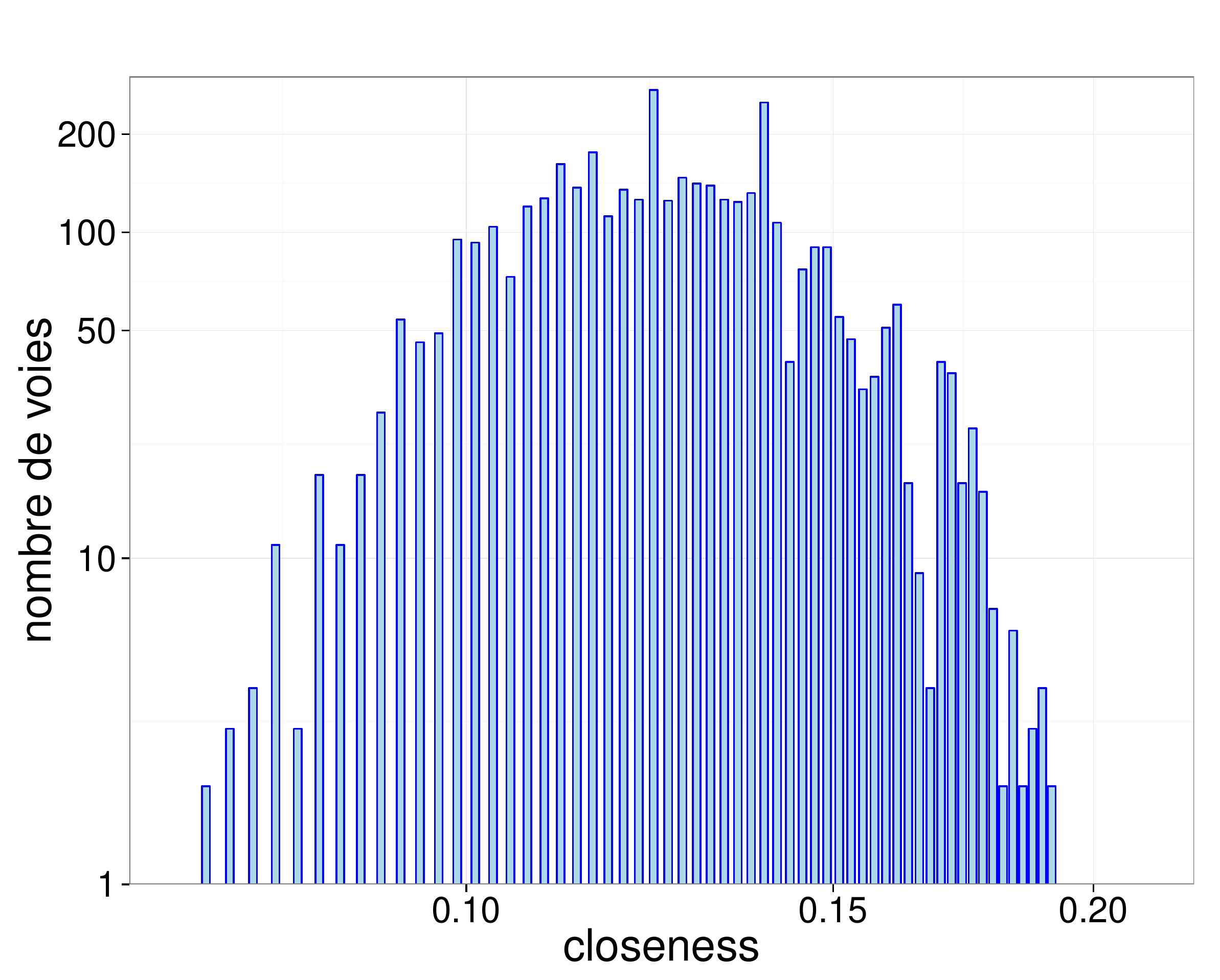}
        \caption{Avignon.}
        \label{fig:int_clo2_av}
    \end{subfigure}
    ~
    \begin{subfigure}[t]{.3\linewidth}
        \includegraphics[width=\textwidth]{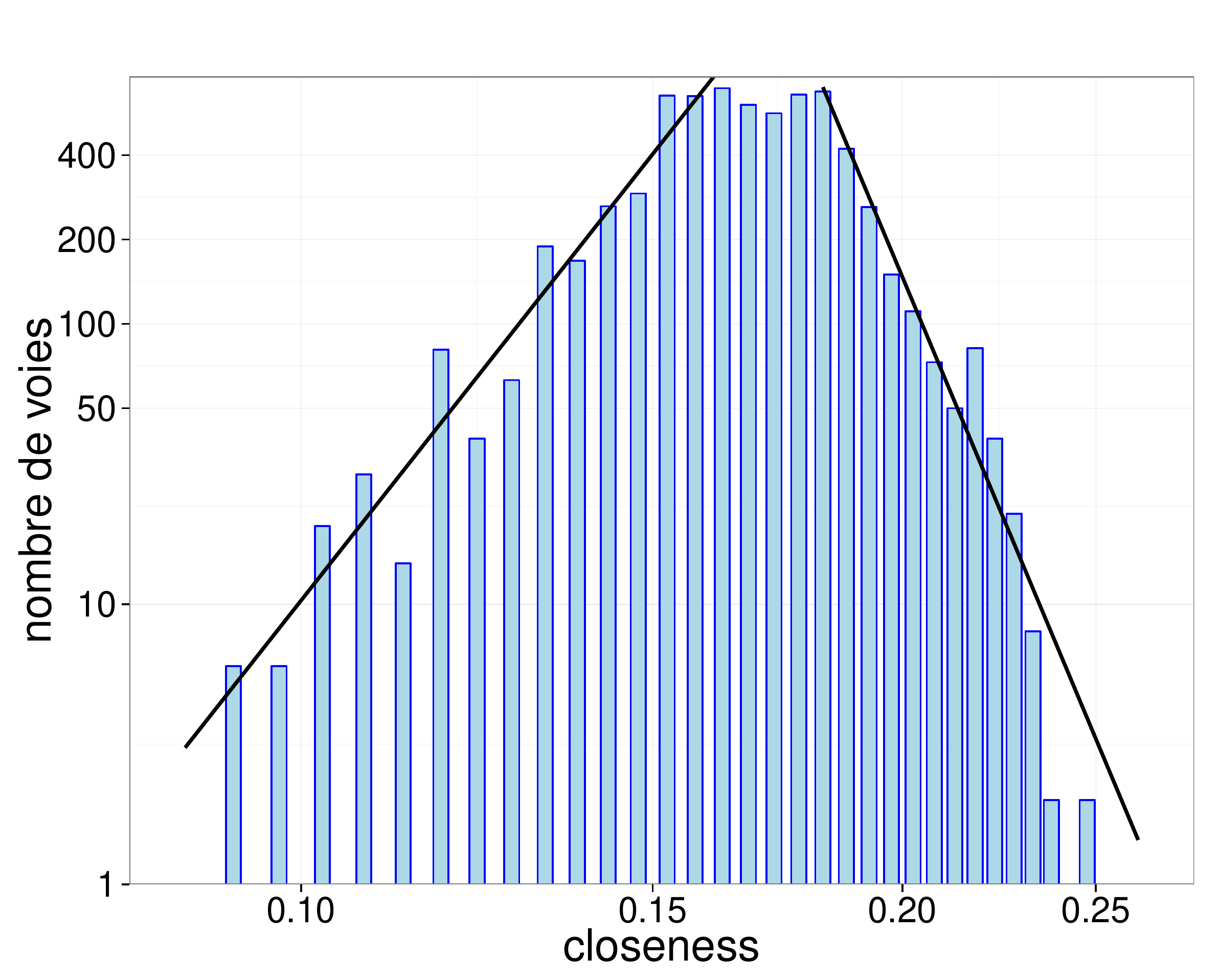}
        \caption{Paris.}
        \label{fig:int_clo2_par}
    \end{subfigure}
    ~
    \begin{subfigure}[t]{.3\linewidth}
        \includegraphics[width=\textwidth]{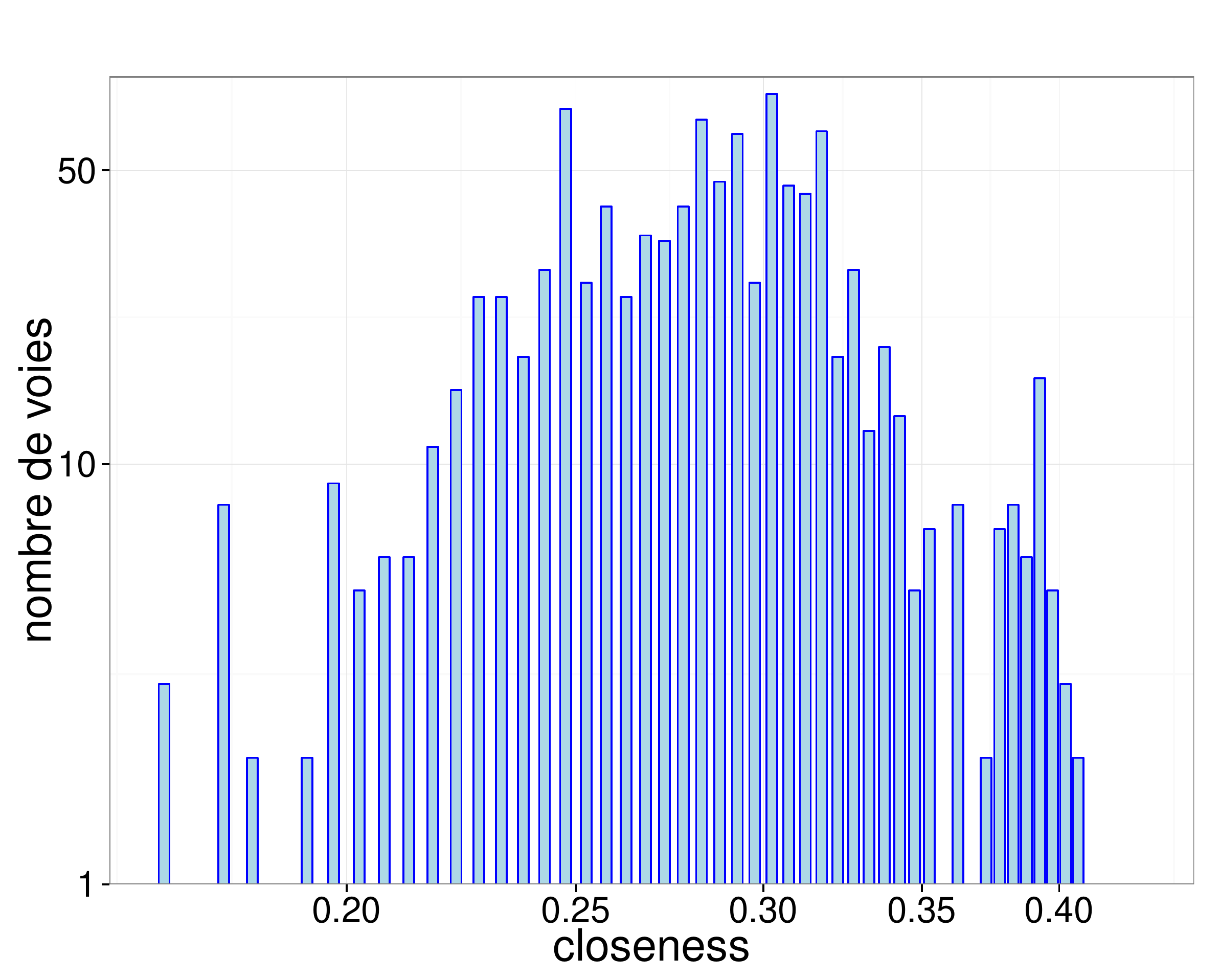}
        \caption{Manhattan.}
        \label{fig:int_clo2_man}
    \end{subfigure}

    \caption{Histogrammes représentant la distribution de la closeness des voies. Approximation par deux lois de puissance.}
    \label{fig:int_clo2}
\end{figure}

Les histogrammes représentant la distribution de la closeness des voies sur les trois réseaux peuvent être approximés de deux manières. Sur Avignon, comme sur Manhattan, la distribution gaussienne semble la plus appropriée (figure \ref{fig:int_clo}). Cependant, sur Paris, une approximation par deux lois de puissance paraît mieux convenir (figure \ref{fig:int_clo2}). Dans chaque cas, peu de voies ont une proximité très forte avec le reste du réseau, peu sont complètement mises à l'écart et nombreuses sont celles situées entre ces deux extrêmes. La distribution tracée pour Manhattan présente néanmoins un léger pic pour les voies de fortes closeness car toutes les voies traversant l'île selon l'axe Nord/Sud ont une proximité très forte avec l'ensemble du réseau.

\FloatBarrier

L'indicateur d'orthogonalité appliqué aux voies suit un comportement identique sur les trois villes : peu de voies ont une orthogonalité proche de 0 car la majorité des connexions entre voies se fait avec un angle de déviation proche de 90\degres . En effet, nous avons vu dans le chapitre 2 de la partie I que les connexions entre arcs, en considérant tout le réseau, se font avec une forte tendance pour l'alignement (angle de déviation proche de 0\degres) ou la perpendicularité (angle de déviation proche de 90\degres ). Les voies étant construites de manière à joindre les arcs de déviations minimales, les angles de déviation restant majoritairement entre voies sont ceux proches de 90\degres , l'orthogonalité des voies est donc proche de 1. Pour Paris, l'approximation de la variation des orthogonalités par une fonction exponentielle convient parfaitement. Nous remarquons sur Avignon, au contraire, plus d'angles de faible degré de connexion et moins d'angles de fort degré de connexion que ce que la loi aurait pu faire prédire. Nous retrouvons donc le caractère plus organique de ce graphe à travers sa distribution. Les valeurs caractéristiques des deux approximations (sur Paris et Avignon) sont identiques : les valeurs de l'indicateur appliqué aux voies ont des progressions similaires. Pour Manhattan, nous observons un très fort pic pour la dernière classe de l’histogramme : beaucoup de voies ont une orthogonalité proche de 1.

\begin{figure}[h]
    \centering
     \begin{subfigure}[t]{.3\linewidth}
        \includegraphics[width=\textwidth]{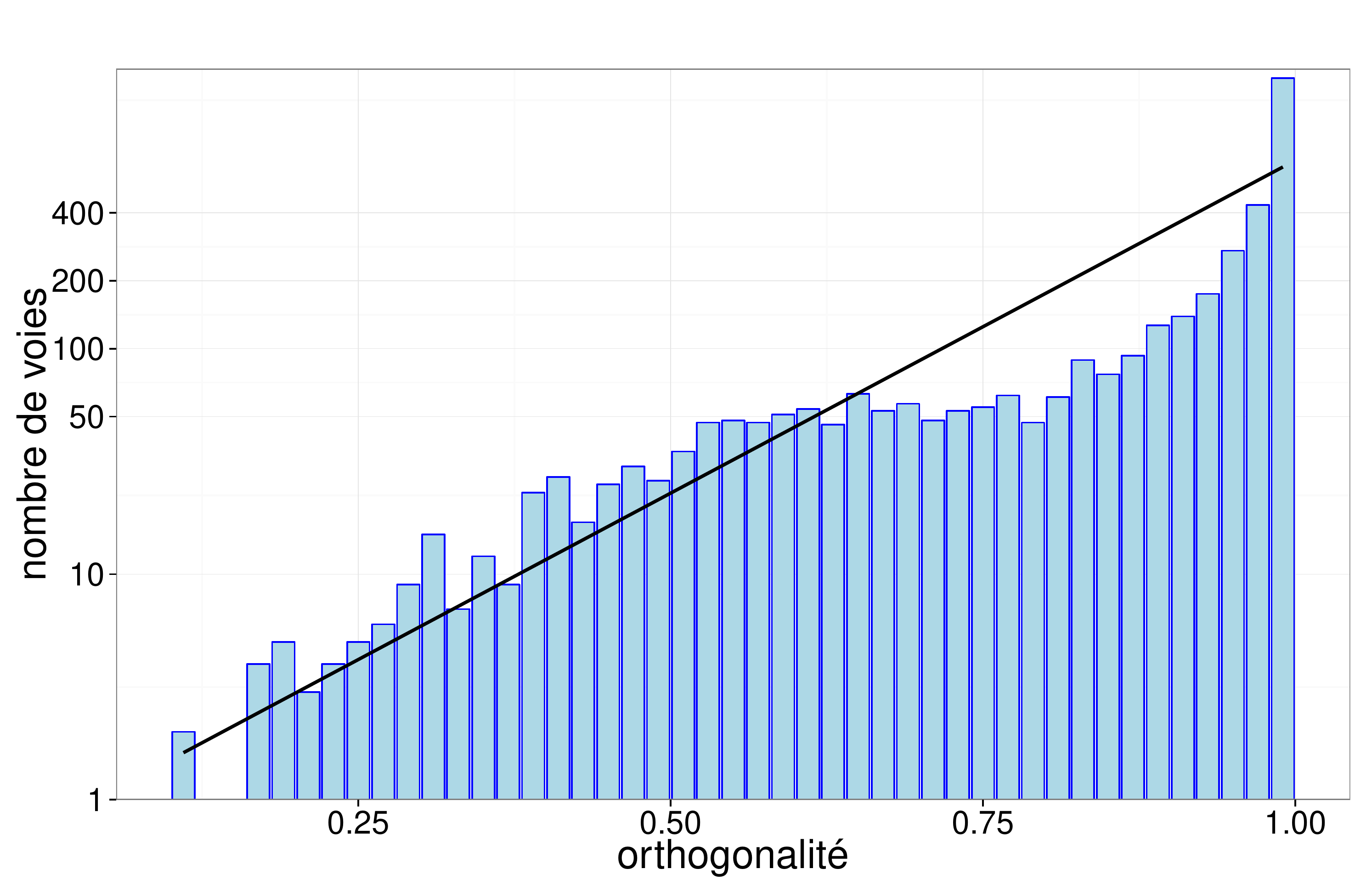}
        \caption{Avignon.}
        \label{fig:int_ortho_av}
    \end{subfigure}
    ~
    \begin{subfigure}[t]{.3\linewidth}
        \includegraphics[width=\textwidth]{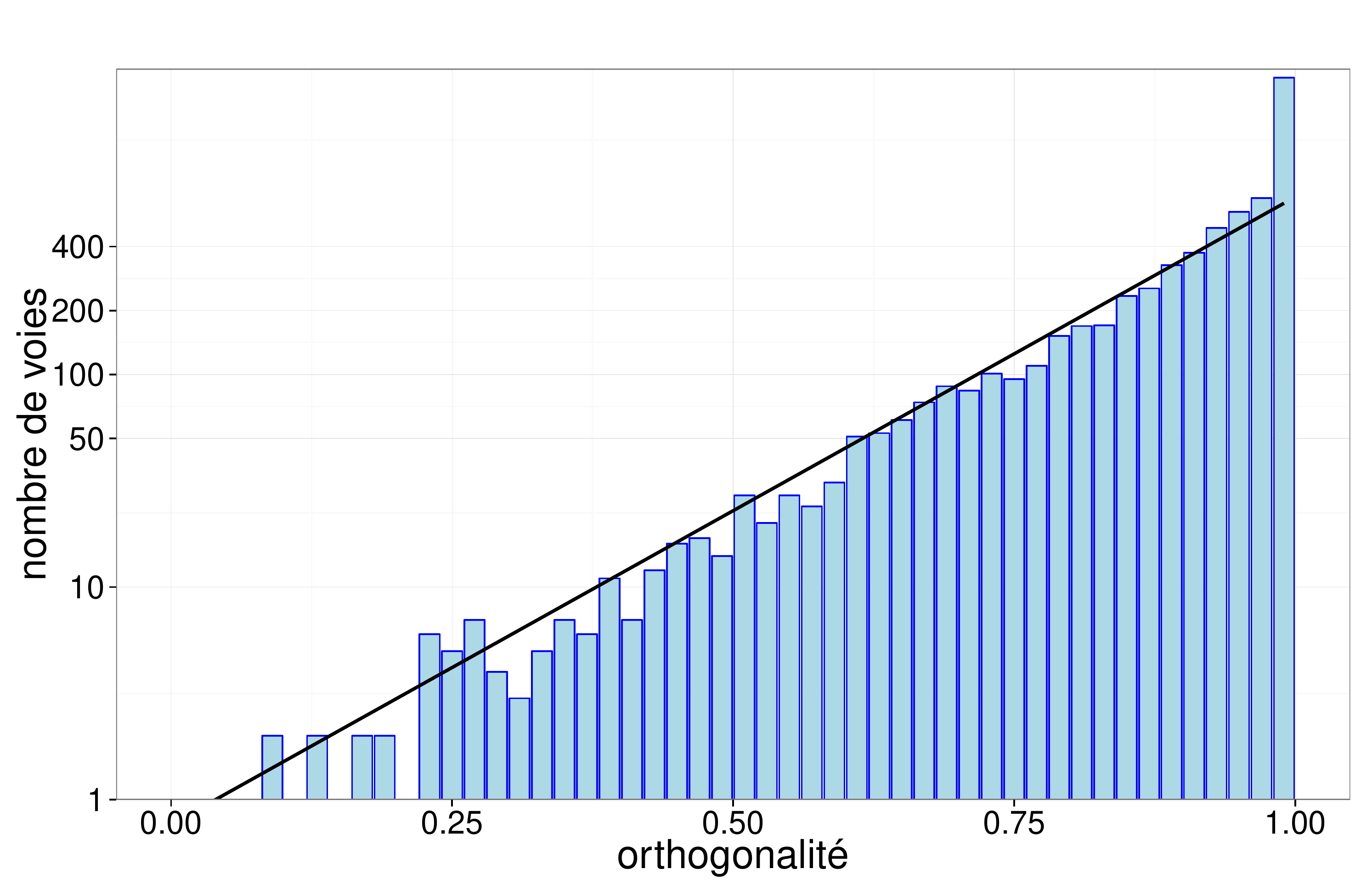}
        \caption{Paris.}
        \label{fig:int_ortho_par}
    \end{subfigure}
    ~
    \begin{subfigure}[t]{.3\linewidth}
        \includegraphics[width=\textwidth]{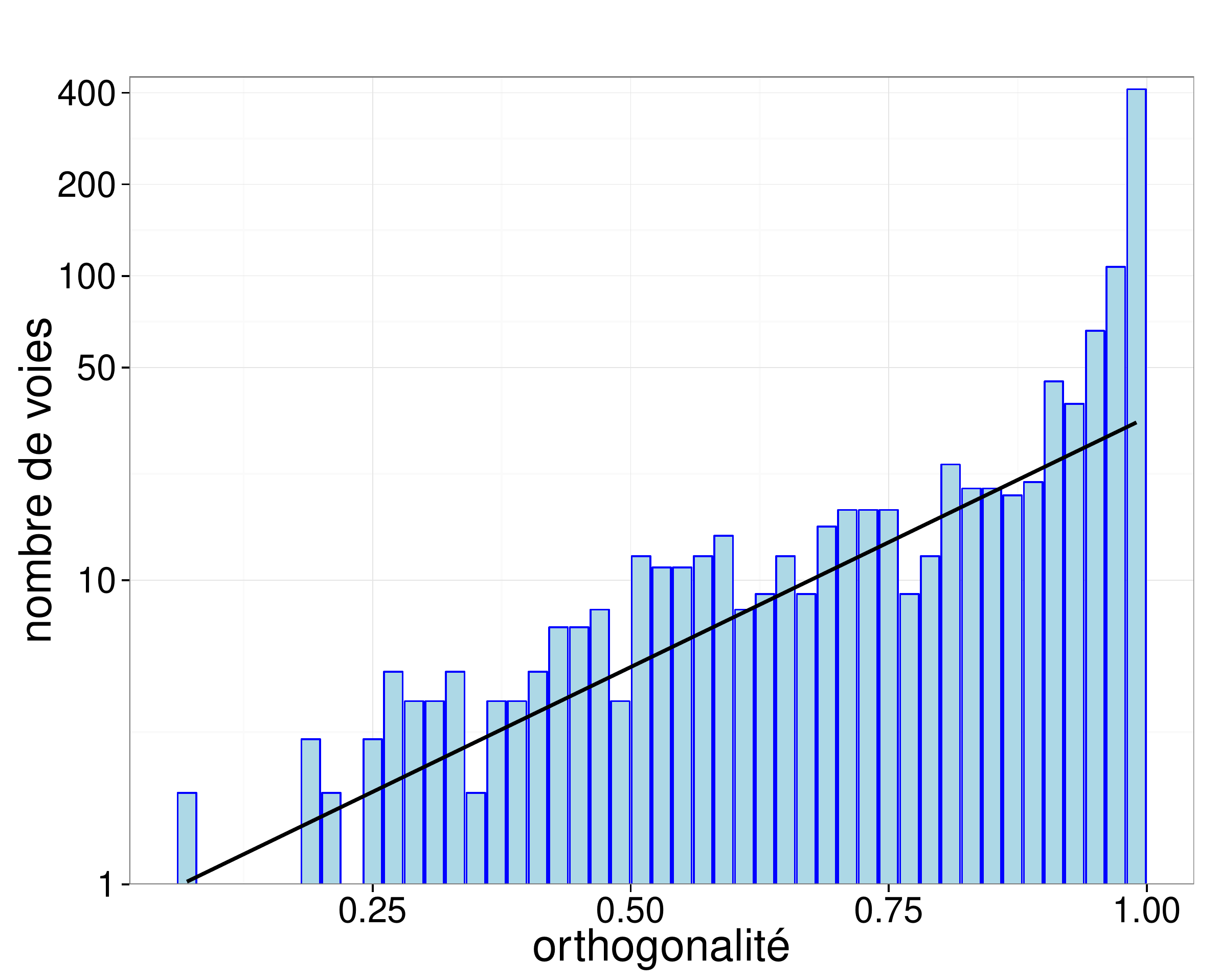}
        \caption{Manhattan.}
        \label{fig:int_ortho_man}
    \end{subfigure}

    \caption{Histogrammes représentant la distribution de l'orthogonalité des voies. Approximation par une fonction exponentielle.}
    \label{fig:int_ortho}
\end{figure}

\FloatBarrier

\begin{figure}[h]
    \centering
    \begin{subfigure}[t]{.45\linewidth}
        \includegraphics[width=\textwidth]{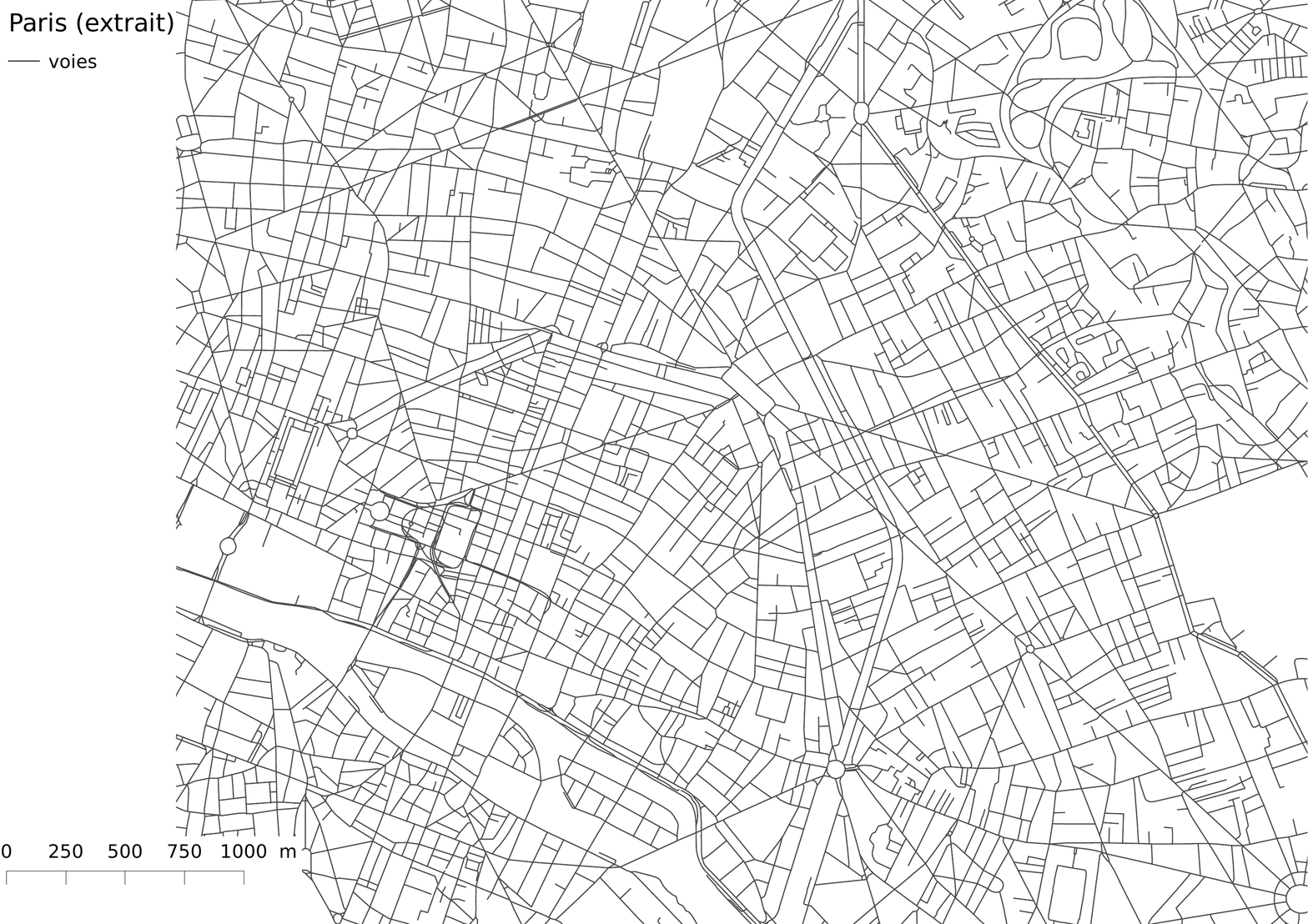}
        \caption{Paris : radiales autour de la place de la République.}
        \label{fig:brut_zoom_par3}
    \end{subfigure}
    ~
    \begin{subfigure}[t]{.45\linewidth}
        \includegraphics[width=\textwidth]{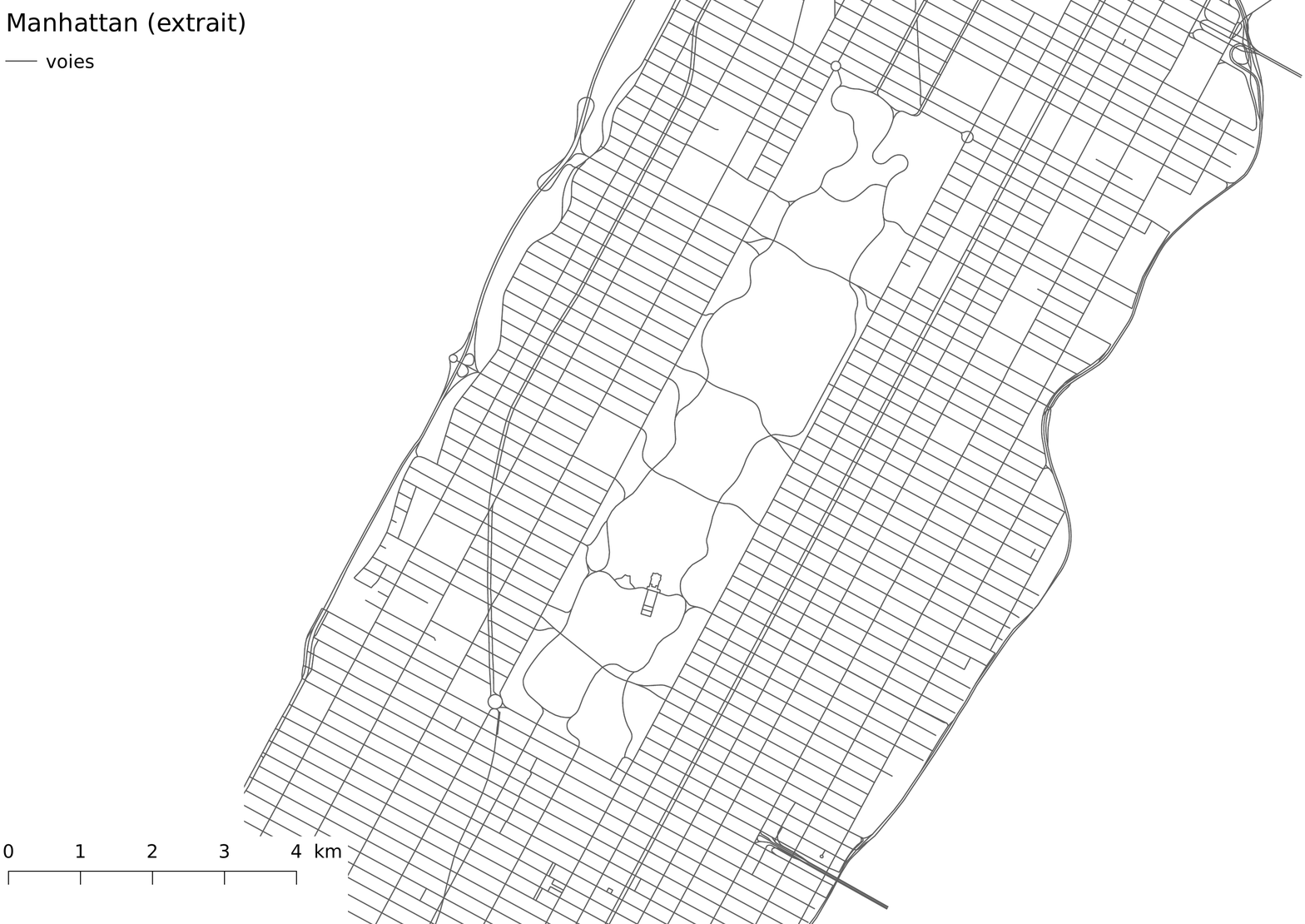}
        \caption{Manhattan : réseau maillé donnant au graphe des faces rectangulaires.}
        \label{fig:brut_zoom_man3}
    \end{subfigure}
    
    \caption{Détails extraits des graphes de Paris et de Manhattan.}
    \label{fig:brut_zoom_reg3}
\end{figure}

\begin{figure}[h]
    \centering
     \begin{subfigure}[t]{.3\linewidth}
        \includegraphics[width=\textwidth]{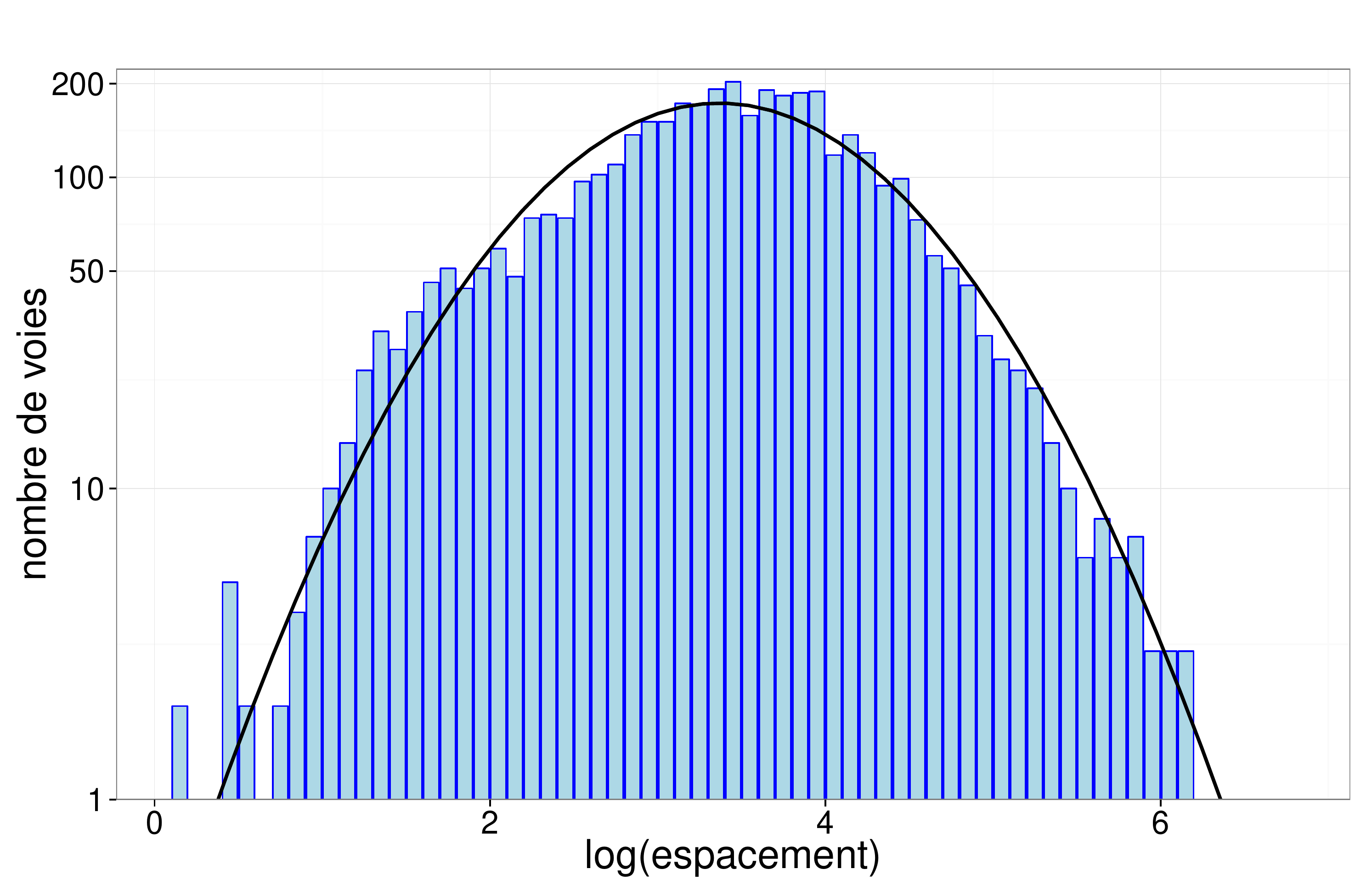}
        \caption{Avignon.}
        \label{fig:int_esp_av}
    \end{subfigure}
    ~
    \begin{subfigure}[t]{.3\linewidth}
        \includegraphics[width=\textwidth]{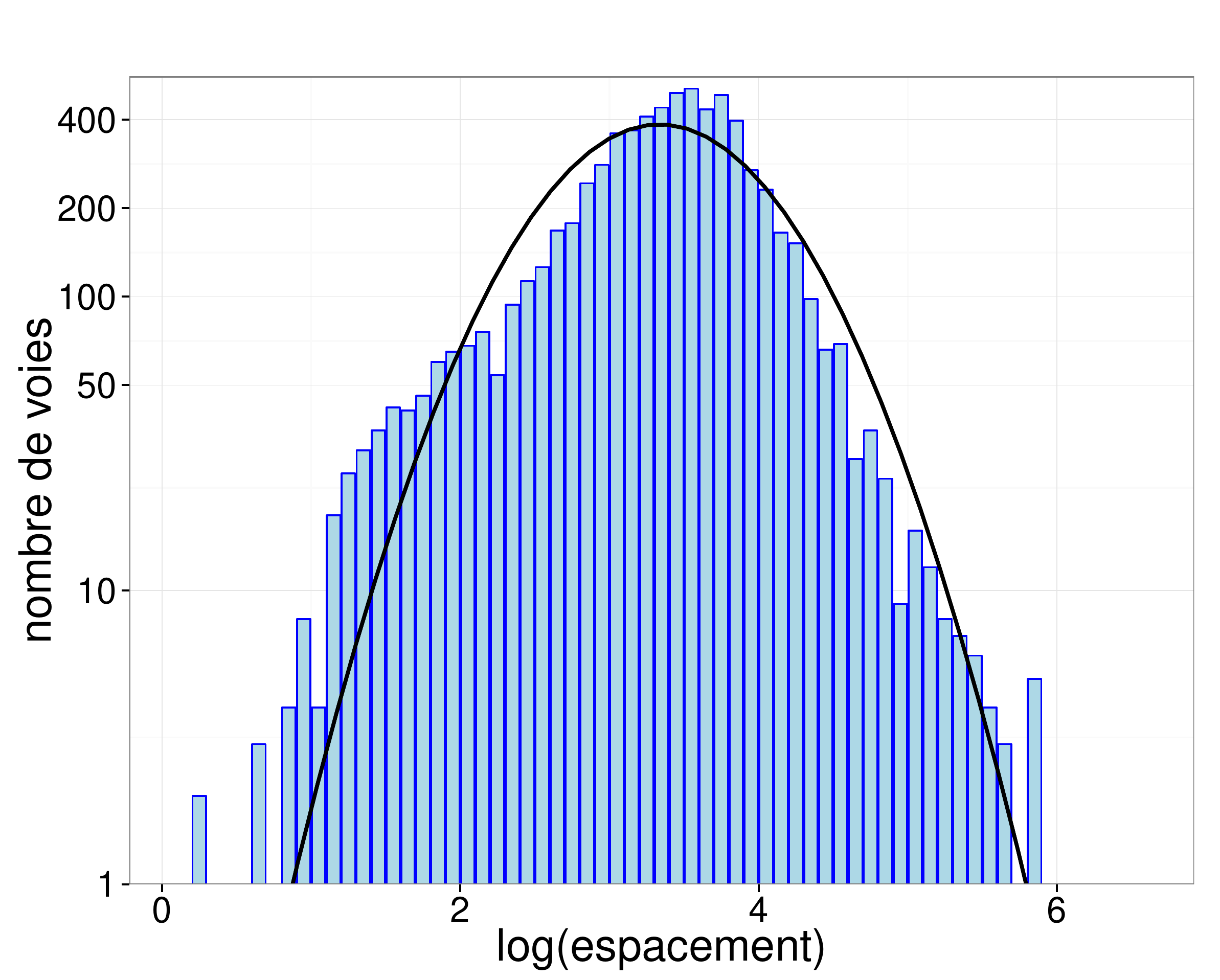}
        \caption{Paris.}
        \label{fig:int_esp_par}
    \end{subfigure}
    ~
    \begin{subfigure}[t]{.3\linewidth}
        \includegraphics[width=\textwidth]{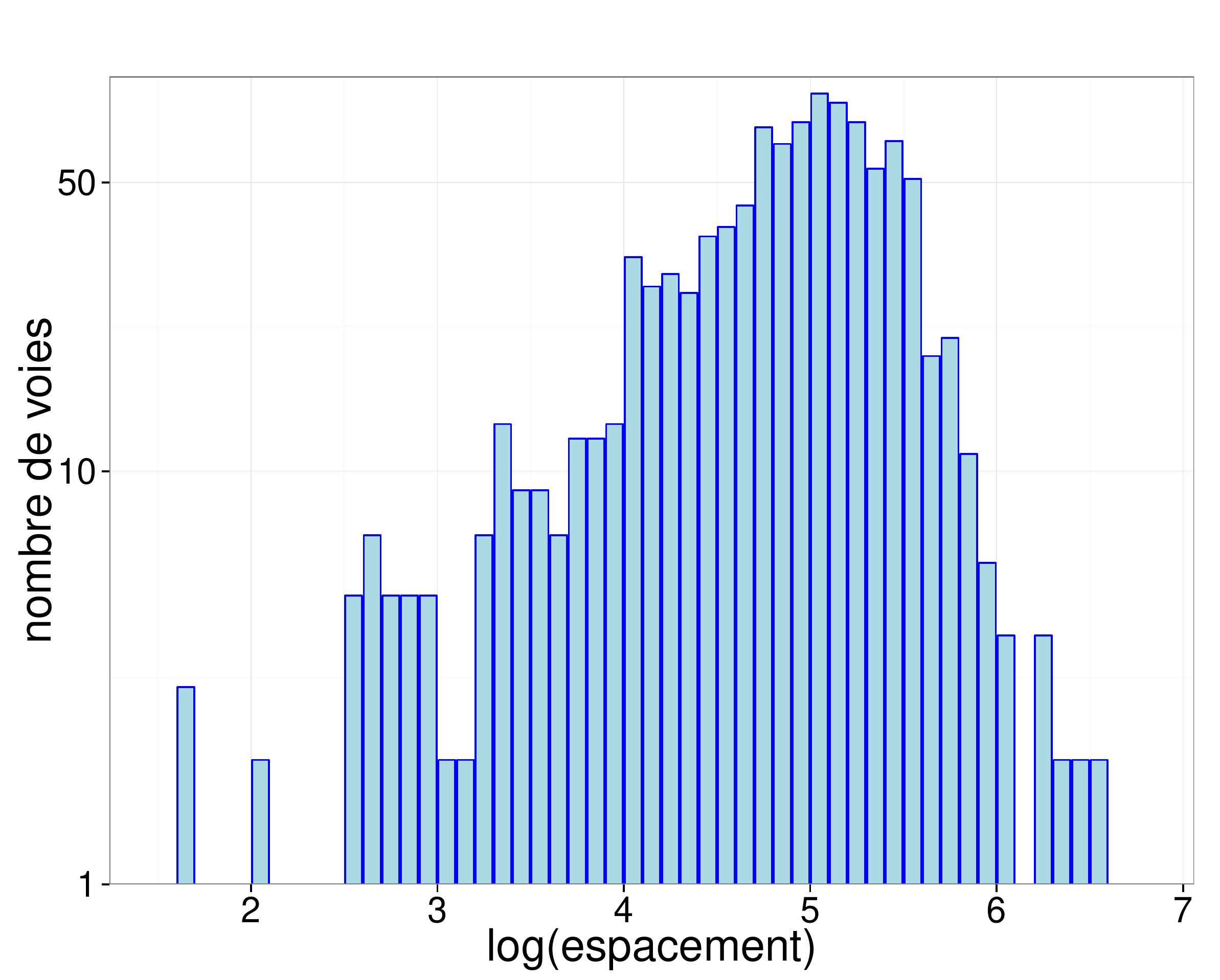}
        \caption{Manhattan.}
        \label{fig:int_esp_man}
    \end{subfigure}

    \caption{Histogrammes représentant la distribution de l'espacement des voies. Loi log normale tracée à partir des données.}
    \label{fig:int_esp}
\end{figure}

L'espacement des voies traduit la densité linéaire du réseau. Nous pouvons également faire deux approximations des courbes obtenues : une avec une loi log normale (figure \ref{fig:int_esp}), une avec deux lois exponentielles (figure \ref{fig:int_esp}). Pour Avignon, la loi log normale est celle qui décrit le mieux la distribution, surtout pour les voies de fort espacement. En effet, la ville regroupe des tissus d'espacement très différents : depuis les centres d'habitation denses, aux voies rapides longues et très peu connectées. Pour Paris, les deux lois exponentielles sont plus appropriées. Ainsi, sur le graphe de la capitale, les voies d'espacement intermédiaire sont les plus présentes. Elles correspondent, entre autres, aux percements qui traversent le tissu de part en part. Ceux-ci sont découpés en plusieurs voies par les places et les ronds-points qu'ils traversent (par exemple, autour de la place de la République, figure \ref{fig:brut_zoom_par3}). Pour Manhattan, les deux lois exponentielles se rapprochent le plus de la courbe, sans parvenir à la décrire exactement. Nous observons sur la courbe deux pics, qui correspondent aux deux directions du maillage. Le premier décrit les voies orientées selon l'axe Sud-Ouest / Nord-Est, moins nombreuses, plus longues, leurs connexions sont plus rapprochées. Le second correspond aux voies orientées en Sud-Est / Nord-Ouest, plus nombreuses, plus courtes, elles ont des connexions plus distantes (voir figure \ref{fig:brut_zoom_man3}).

\begin{figure}[h]
    \centering
     \begin{subfigure}[t]{.3\linewidth}
        \includegraphics[width=\textwidth]{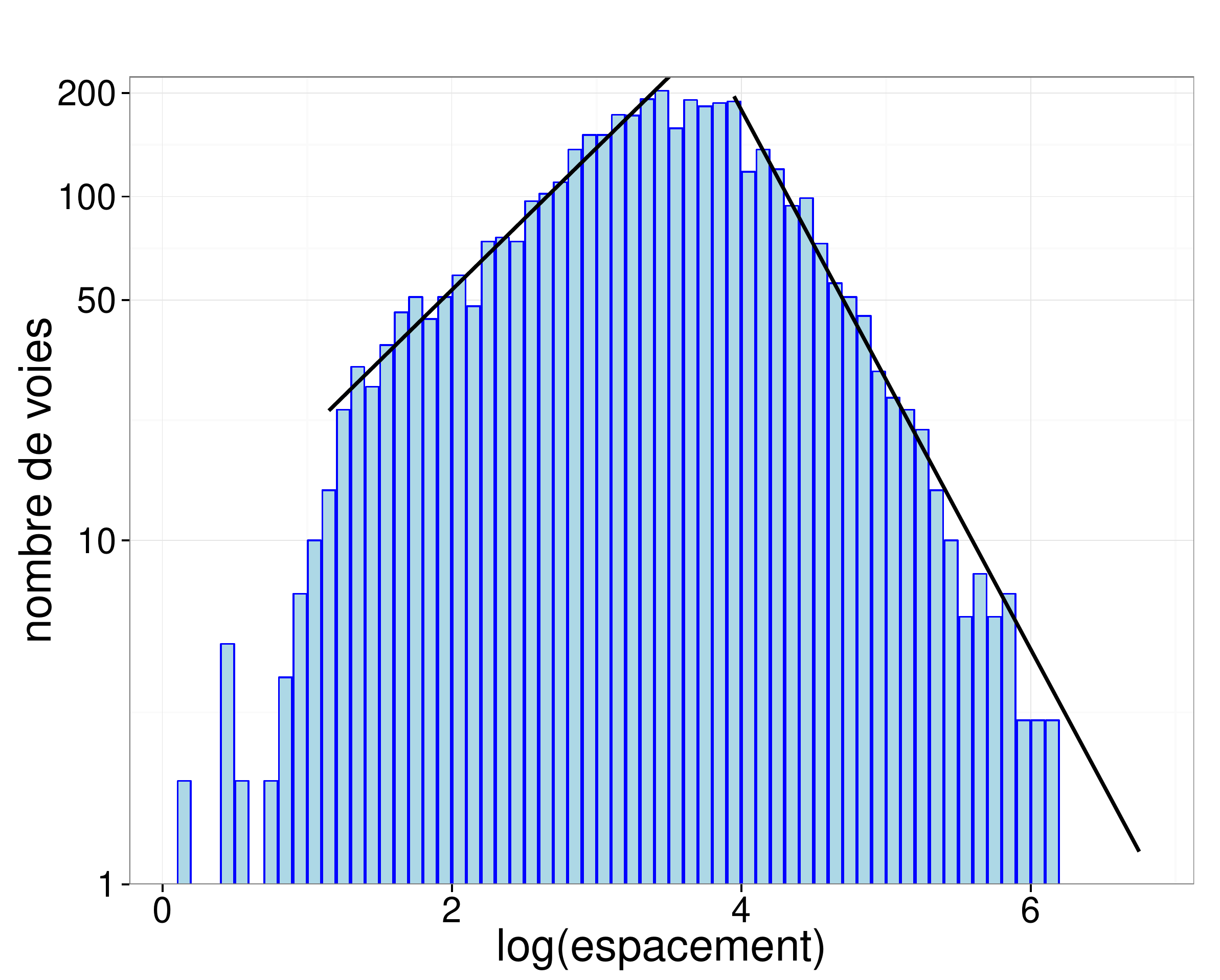}
        \caption{Avignon.}
        \label{fig:int_esp2_av}
    \end{subfigure}
    ~
    \begin{subfigure}[t]{.3\linewidth}
        \includegraphics[width=\textwidth]{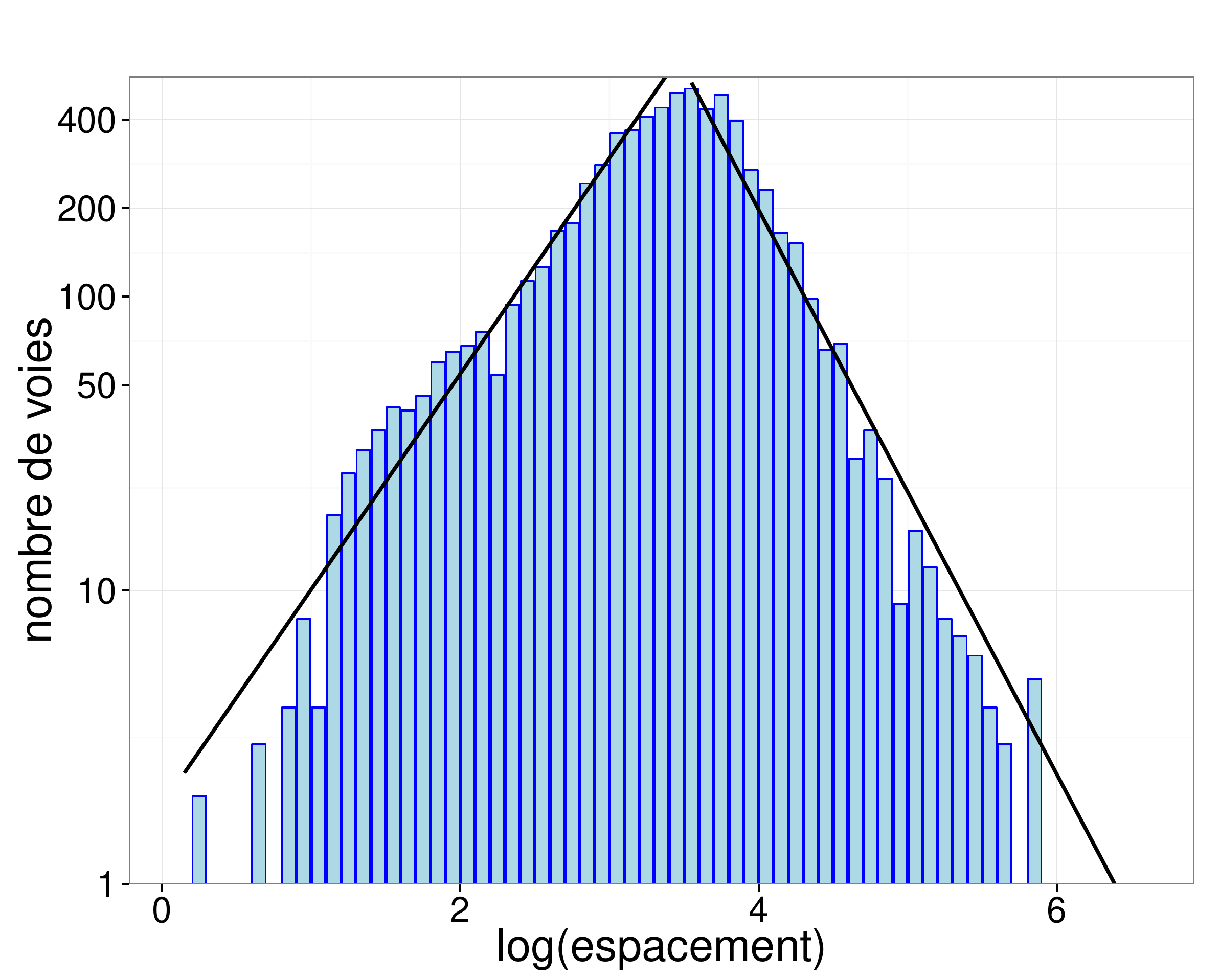}
        \caption{Paris.}
        \label{fig:int_esp2_par}
    \end{subfigure}
    ~
    \begin{subfigure}[t]{.3\linewidth}
        \includegraphics[width=\textwidth]{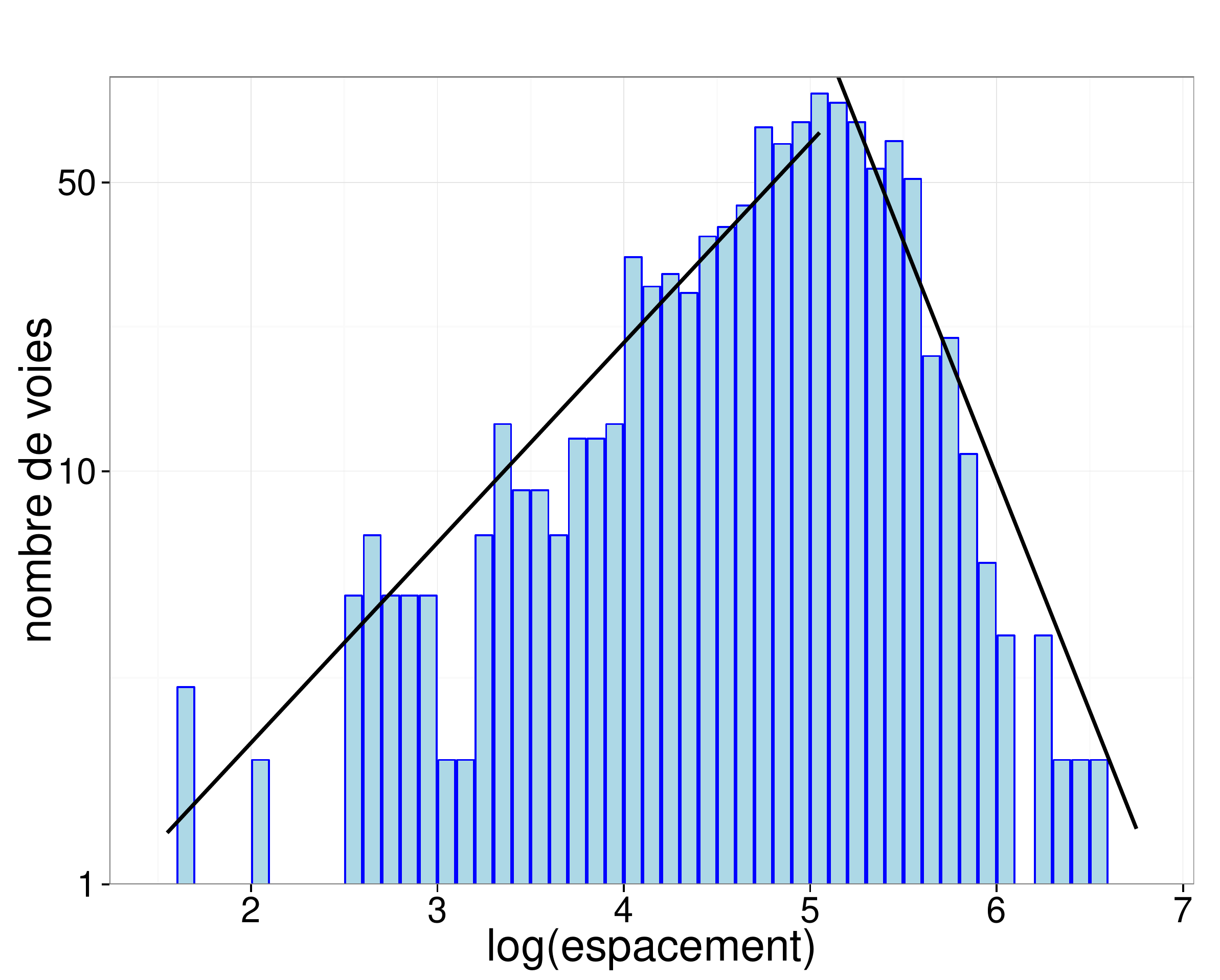}
        \caption{Manhattan.}
        \label{fig:int_esp2_man}
    \end{subfigure}

    \caption{Histogrammes représentant la distribution de l'espacement des voies. Approximation par deux lois exponentielles.}
    \label{fig:int_esp2}
\end{figure}

\FloatBarrier

\begin{figure}[h]
    \centering
     \begin{subfigure}[t]{.3\linewidth}
        \includegraphics[width=\textwidth]{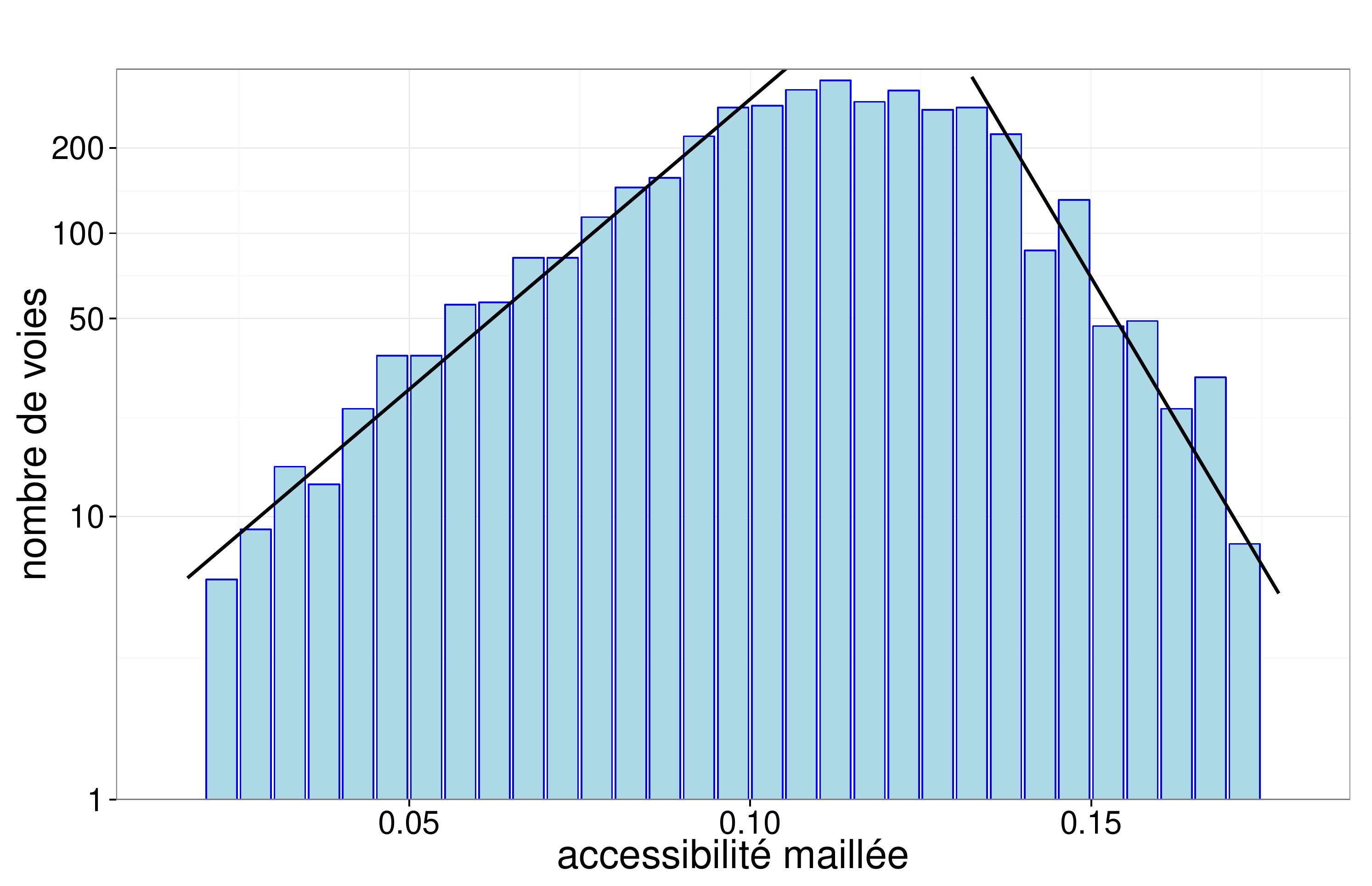}
        \caption{Avignon.}
        \label{fig:int_accmail_av}
    \end{subfigure}
    ~
    \begin{subfigure}[t]{.3\linewidth}
        \includegraphics[width=\textwidth]{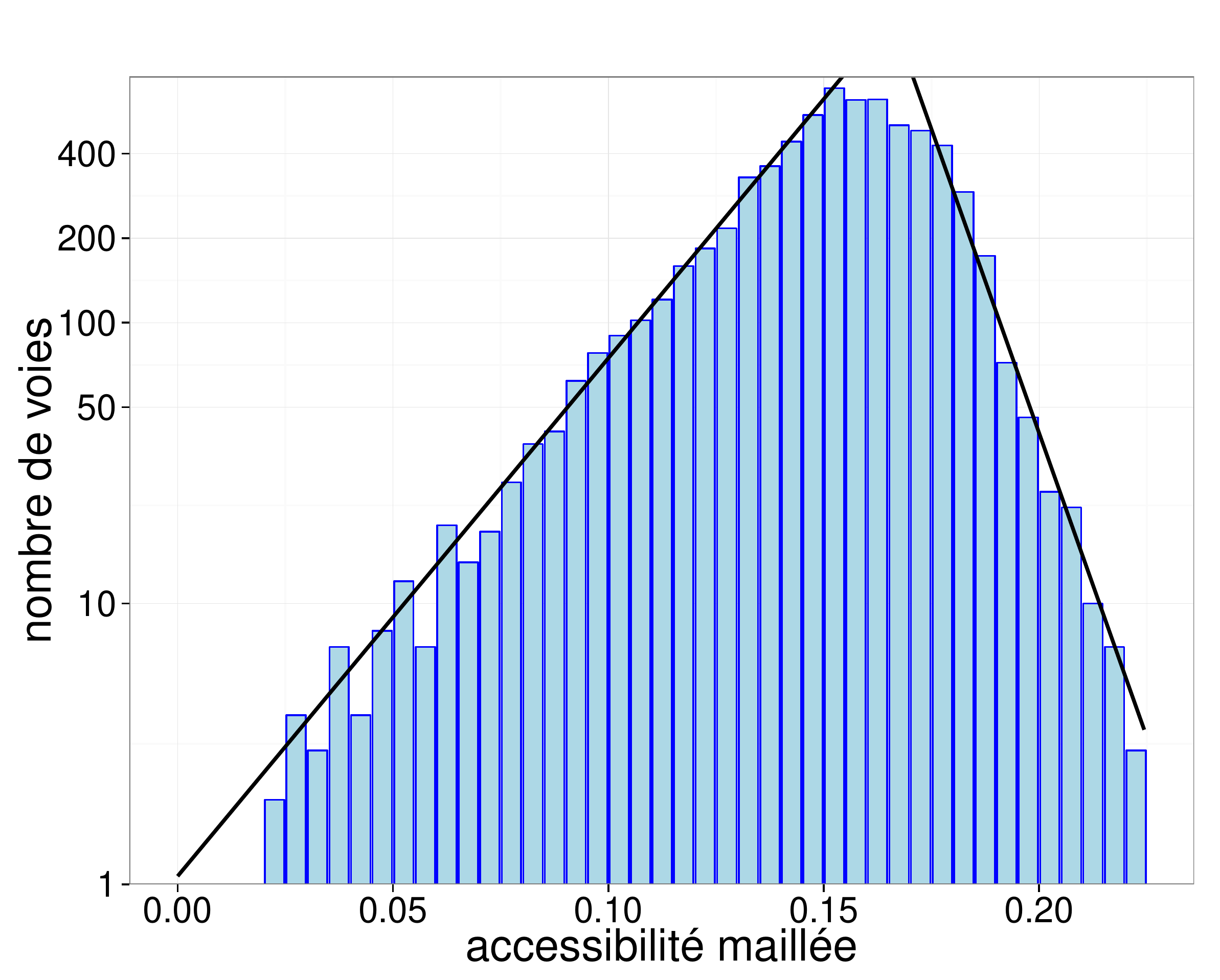}
        \caption{Paris.}
        \label{fig:int_accmail_par}
    \end{subfigure}
    ~
    \begin{subfigure}[t]{.3\linewidth}
        \includegraphics[width=\textwidth]{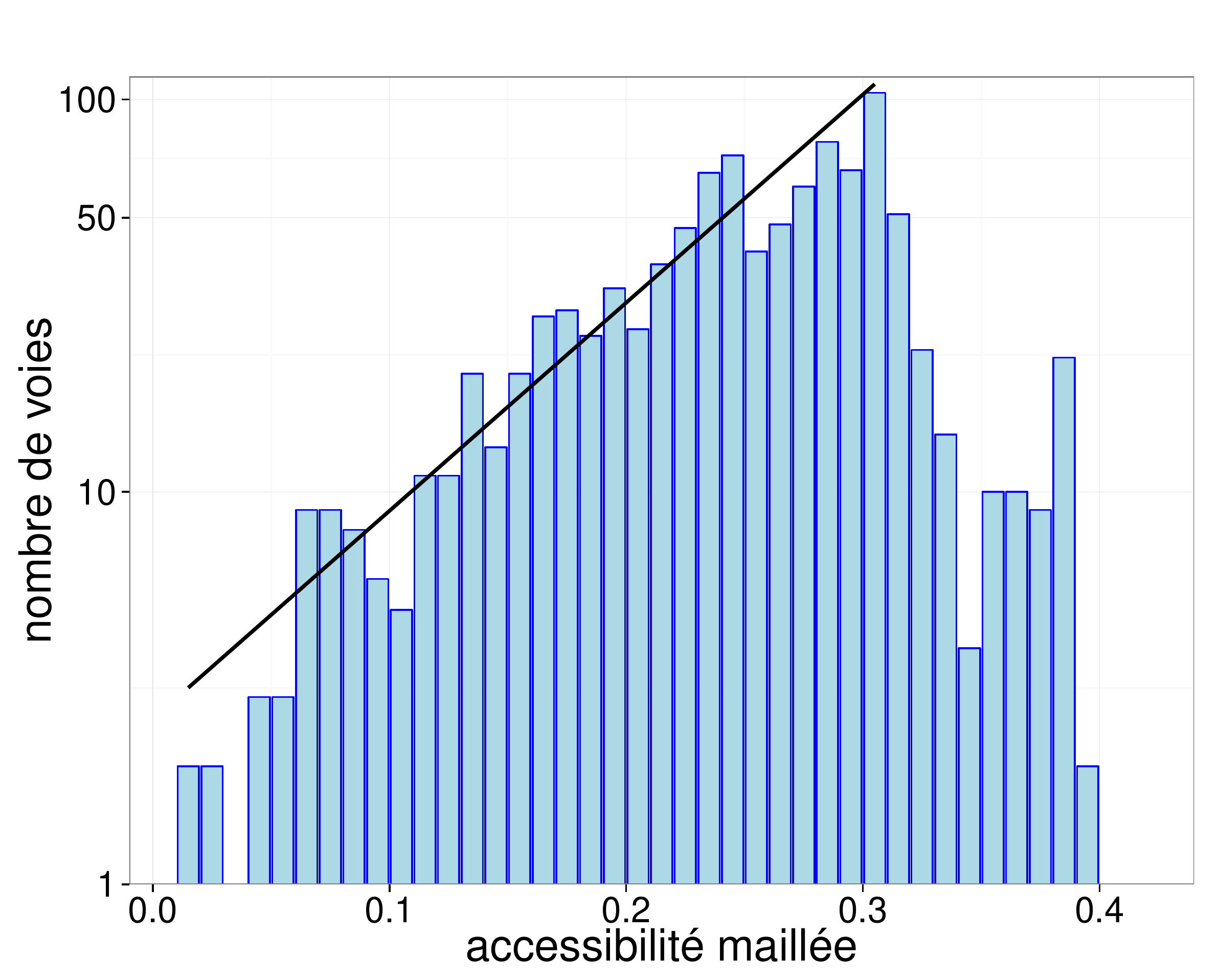}
        \caption{Manhattan.}
        \label{fig:int_accmail_man}
    \end{subfigure}

    \caption{Histogrammes représentant la distribution de l'accessibilité maillée des voies. Approximation par deux lois exponentielles.}
    \label{fig:int_accmail}
\end{figure}

L'accessibilité maillée est calculée pour chaque voie en multipliant la closeness et l'orthogonalité. Nous obtenons ainsi une distribution qu'il est possible d'approximer par deux lois exponentielles. Les répartitions sont légèrement accentuées pour les fortes valeurs (figure \ref{fig:int_accmail}).

\begin{figure}[h]
    \centering
     \begin{subfigure}[t]{.3\linewidth}
        \includegraphics[width=\textwidth]{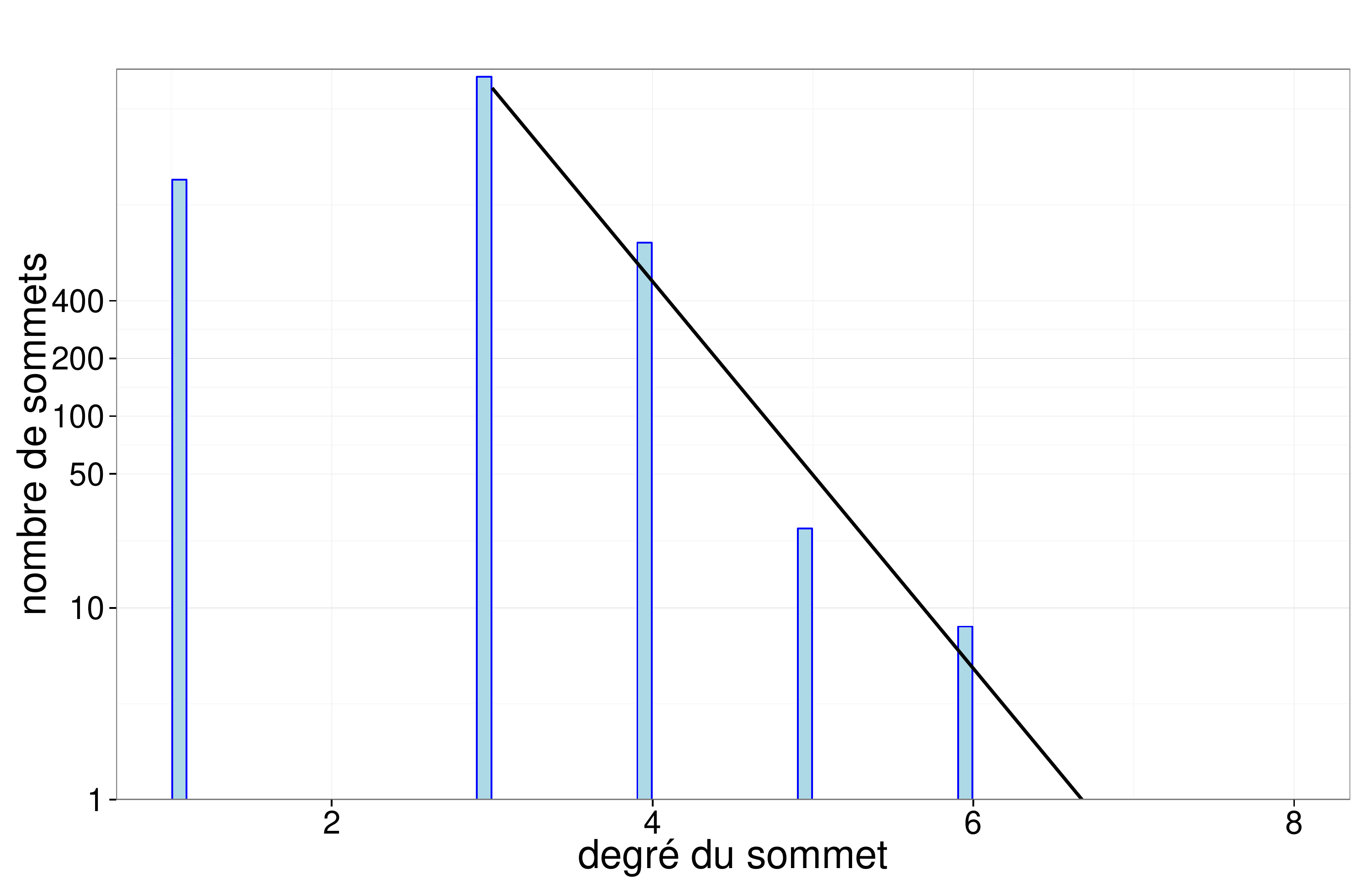}
        \caption{Avignon.}
        \label{fig:int_degsom_av}
    \end{subfigure}
    ~
    \begin{subfigure}[t]{.3\linewidth}
        \includegraphics[width=\textwidth]{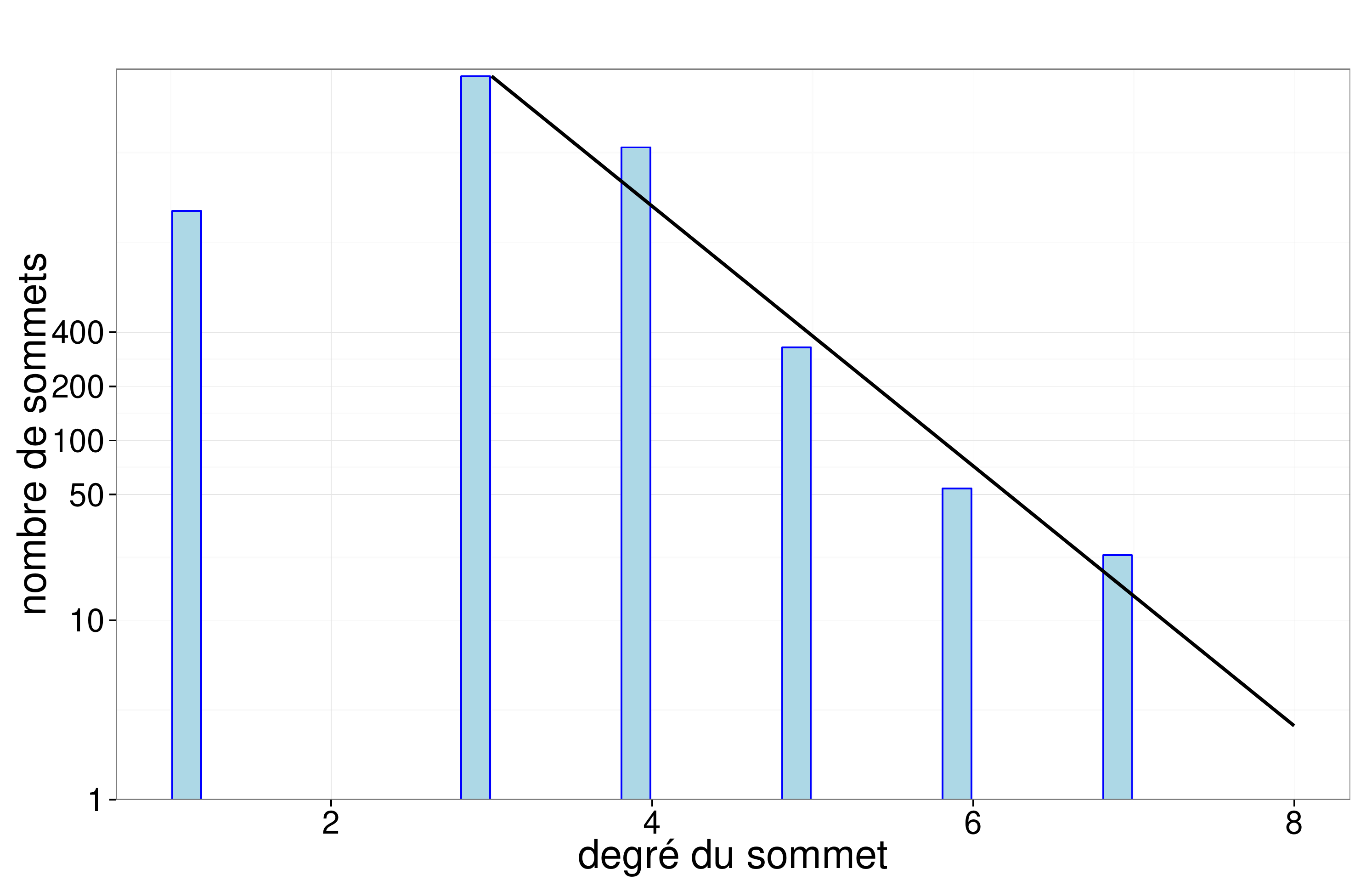}
        \caption{Paris.}
        \label{fig:int_degsom_par}
    \end{subfigure}
    ~
    \begin{subfigure}[t]{.3\linewidth}
        \includegraphics[width=\textwidth]{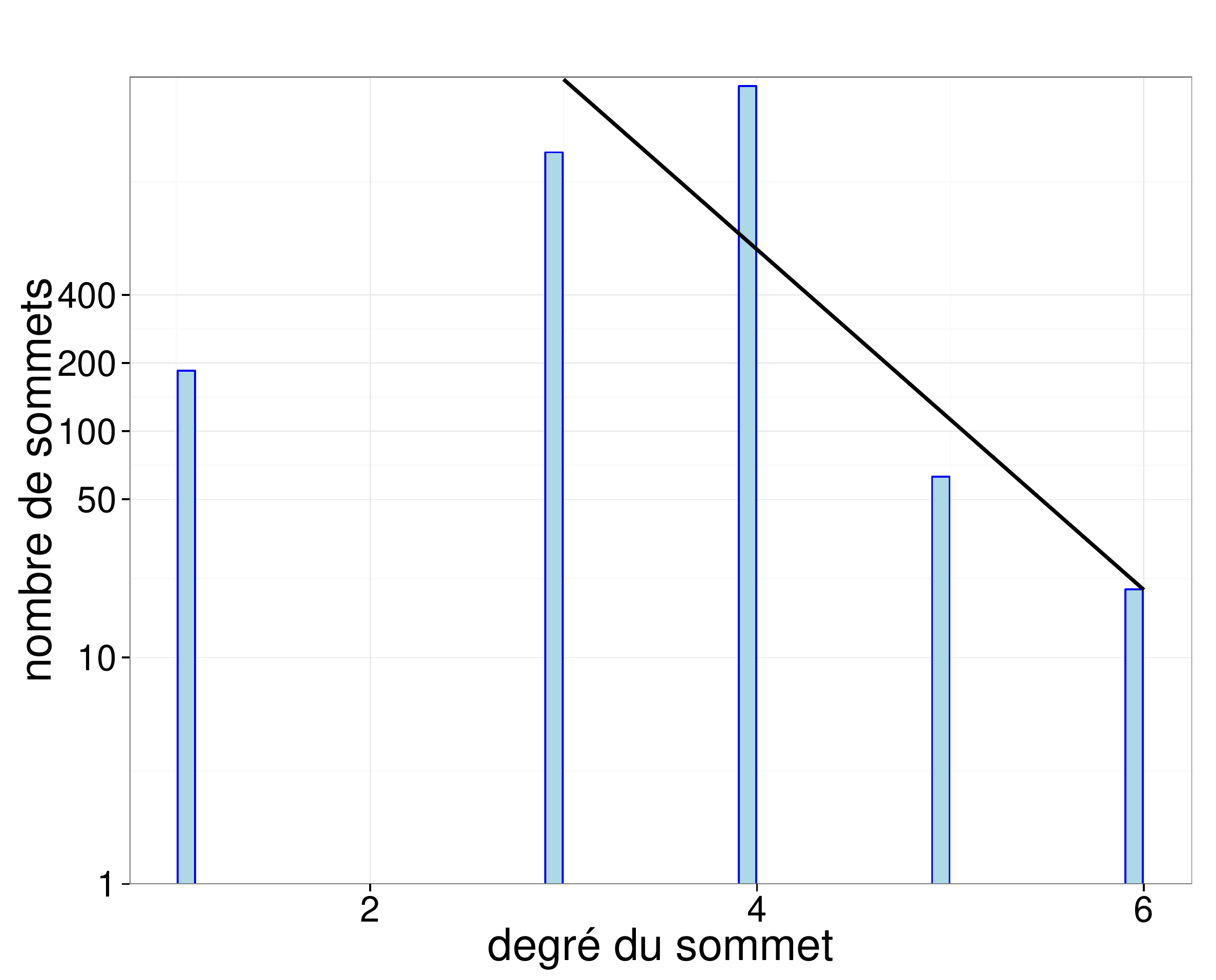}
        \caption{Manhattan.}
        \label{fig:int_degsom_man}
    \end{subfigure}

    \caption{Histogrammes représentant la distribution du degré des sommets. Approximation par une fonction exponentielle.}
    \label{fig:int_degsom}
\end{figure}

\begin{figure}[h]
    \centering
    \begin{subfigure}[t]{.45\linewidth}
        \includegraphics[width=\textwidth]{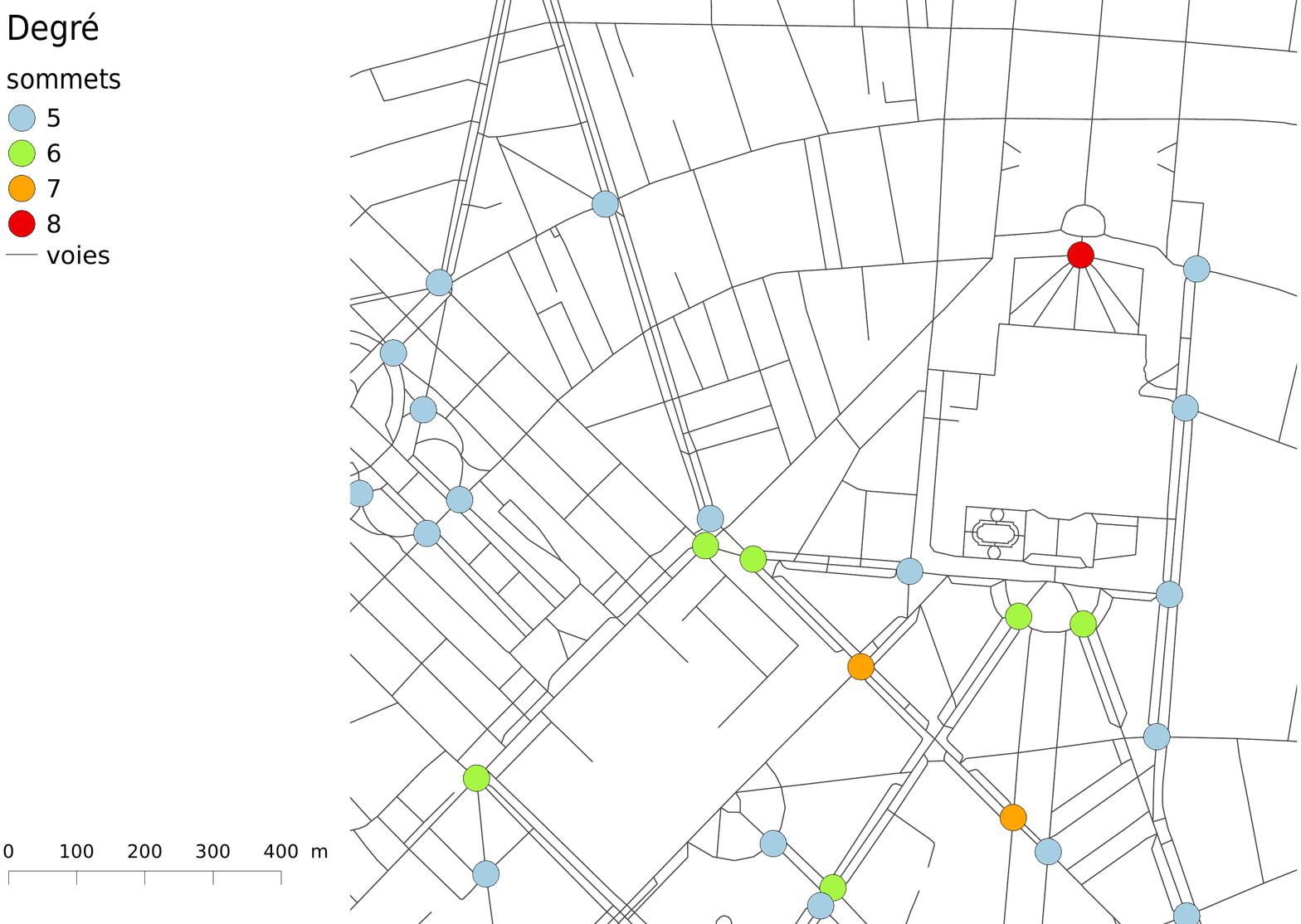}
        \caption{Paris.}
        \label{fig:int_degsomc_par}
    \end{subfigure}
    ~
    \begin{subfigure}[t]{.45\linewidth}
        \includegraphics[width=\textwidth]{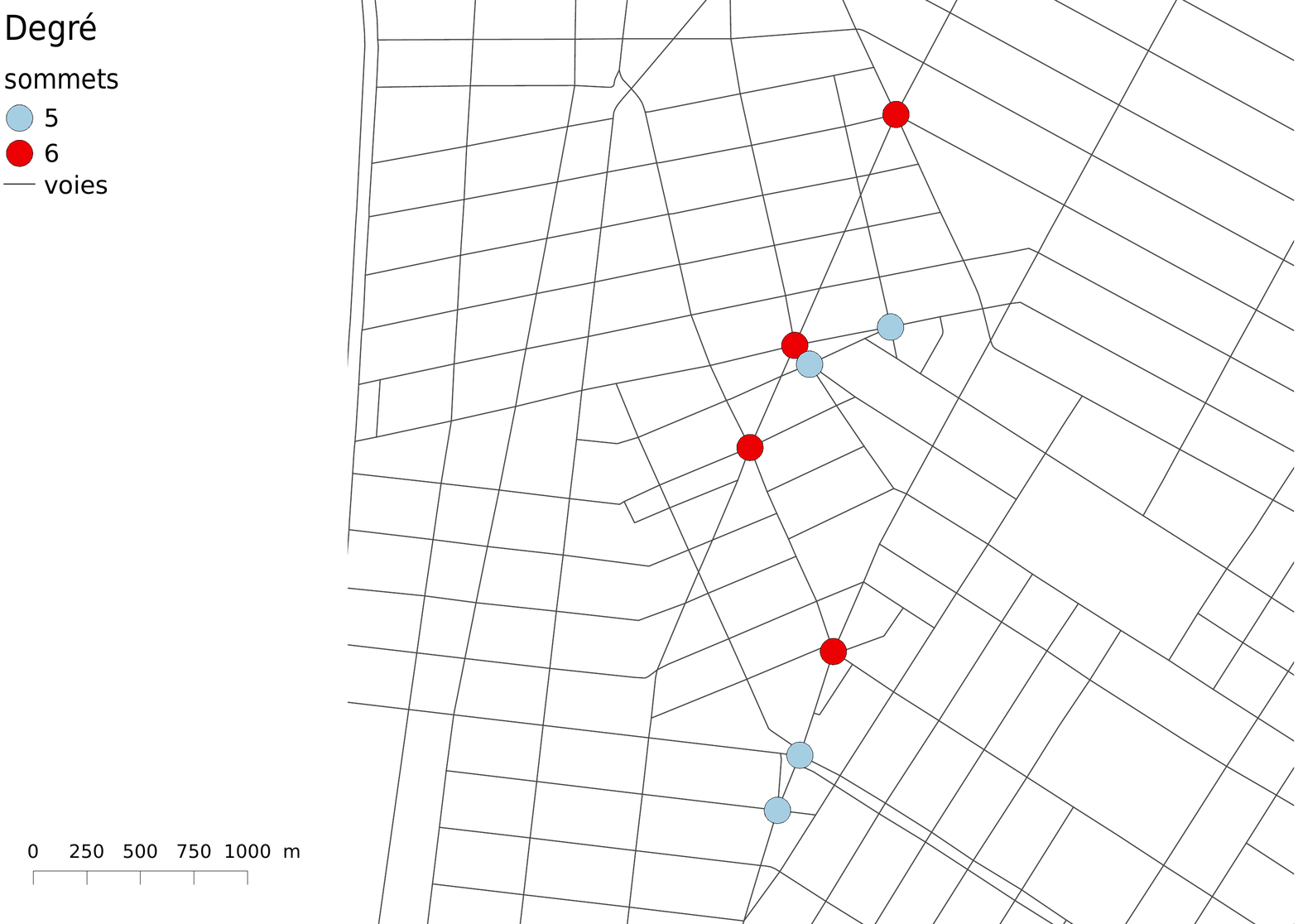}
        \caption{Manhattan.}
        \label{fig:int_degsomc_man}
    \end{subfigure}

    \caption{Cartes des nœuds de degré 5 ou plus.}
    \label{fig:int_degsom_carte}
\end{figure}

Nous étudions également les variations du degré des intersections des trois réseaux viaires. Nous avons vu précédemment que ces trois graphes ont des degrés moyens pour leurs sommets très différents : 2,70 pour Avignon ; 3,08 pour Paris ; 3,59 pour Manhattan. Nous observons sur la figure \ref{fig:int_degsom} que le degré majoritaire des nœuds du réseau issu de Manhattan est de 4 alors que celui des deux autres villes est de 3. Il y a également beaucoup plus d'impasses sur le graphe d'Avignon, ce qui appuie le caractère organique du réseau. Sur les trois réseaux, le degré maximum que l'on peut trouver n'excède pas 8 (il est de 6 à Avignon et Manhattan). Ce maximum est lié à la contrainte planaire imposée par l'aspect spatial de ces graphes. Telles que nous avons défini les intersections (croisement entre deux tronçons de rues), il n'est pas possible d'avoir un grand nombre d'arcs qui y sont liés (les ronds-points étant conservés dans l'analyse de ce panel de recherche). Les sommets dont le degré est important correspondent souvent à des aménagements planifiés (figure \ref{fig:int_degsom_carte}).

\begin{table}[h]
\centering

\begin{tabular}{|c|c|c|c|c|c|}
\hline
\multicolumn{ 1}{|c|}{Objet} & \multicolumn{ 1}{c|}{Indicateur} & \multicolumn{ 2}{c|}{Échelles} & \multicolumn{ 1}{c|}{Modèle} & \multicolumn{ 1}{c|}{Coefficients des approximations} \\ \cline{ 3- 4}
\multicolumn{ 1}{|c|}{} & \multicolumn{ 1}{c|}{} & x & y & \multicolumn{ 1}{c|}{} & \multicolumn{ 1}{c|}{} \\ \hline
\multicolumn{ 1}{|c|}{Voie} & log(longueur) & lin & log & Log normale & Moyenne $\mu$ et écart-type $\sigma$ \\ \cline{ 2- 6}
\multicolumn{ 1}{|c|}{} & degré & log & log & Puissance & a et b , tels que $y = b \times a^x$ \\ \cline{ 2- 6}
\multicolumn{ 1}{|c|}{} & closeness & lin & log & Normale & Moyenne $\mu$ et écart-type $\sigma$ \\ \cline{ 2- 6}
\multicolumn{ 1}{|c|}{} & closeness & log & log & Puissance & a et b , tels que $y = b \times a^x$ \\ \cline{ 2- 6}
\multicolumn{ 1}{|c|}{} & accessibilité& lin & log & Exponentielle & a et b , tels que $y = e^{ax+b}$ \\
\multicolumn{ 1}{|c|}{} & maillée & & & & \\ \cline{ 2- 6}
\multicolumn{ 1}{|c|}{} & orthogonalité & lin & log & Exponentielle & a et b , tels que $y = e^{ax+b}$ \\ \cline{ 2- 6}
\multicolumn{ 1}{|c|}{} & log(espacement) & lin & log & Log normale & Moyenne $\mu$ et écart-type $\sigma$ \\ \cline{ 2- 6}
\multicolumn{ 1}{|c|}{} & log(espacement) & lin & log & Exponentielle & a et b , tels que $y = e^{ax+b}$ \\ \hline
Sommet & degré & lin & log & Exponentielle & a et b , tels que $y = e^{ax+b}$ \\ \hline
\end{tabular}

\caption{Récapitulatif des indicateurs et de leurs approximations.}
\label{tab:recap_approx}
\end{table}

\begin{table}[h]
\centering
{ \small
\begin{tabular}{|c|c|c|c|c|}
\hline
Objet & Indicateur & Avignon & Paris & Manhattan \\ \hline
\multicolumn{ 1}{|c|}{Voie} & log(longueur) & $log : 4.79 \pm 1.08 $  & $log : 5.04 \pm 1.10$  & $log : 7.20 \pm 1.42$  \\
\multicolumn{ 1}{|c|}{}  &  & $reel : 119.74m {\times \atop \div} 2.93 $  & $reel : 154.72m {\times \atop \div} 3.01 $  & $reel : 1342.88m {\times \atop \div} 4.14 $  \\
 \cline{ 2- 5}

\multicolumn{ 1}{|c|}{} & degré & $e^{8.276} \times x^{-1.885}$ & $e^{10.990} \times x^{-2.231}$ & $e^{7.680} \times x^{-1.442}$ \\ \cline{ 2- 5}

\multicolumn{ 1}{|c|}{} & closeness & $ 0.126 \pm 0.020 $  & $ 0.167 \pm 0.021 $  & $ 0.285 \pm 0.043 $  \\
\cline{ 2- 5}

\multicolumn{ 1}{|c|}{} & closeness & $e^{32.15} \times x^{12.01}$ & $e^{23.17} \times x^{9.05}$ & $e^{10.883} \times x^{5.506}$ \\
\multicolumn{ 1}{|c|}{} &  & $e^{-17.48} \times x^{-11.39}$ & $e^{-22.40} \times x^{-17.02}$ & $e^{-6.89} \times x^{-8.93}$ \\ \cline{ 2- 5}

\multicolumn{ 1}{|c|}{} & accessibilité  & $e^{47.1801 \times x + 0.9772}$  & $e^{42.48051 \times x + 0.06681}$ & $e^{12.2133 \times x + 0.9703}$ \\ 
\multicolumn{ 1}{|c|}{} & maillée & $\frac{1}{a} = 0,0212$ & $\frac{1}{a} = 0,0235$ & $\frac{1}{a} =0,0819 $ \\ 
\multicolumn{ 1}{|c|}{} &  & $e^{-93.31 \times x + 18.24}$ & $e^{-99.70 \times x + 23.64}$ & \\
\multicolumn{ 1}{|c|}{} &  & $\frac{1}{a} = -0,0107$ & $\frac{1}{a} = -0,0100$ & \\ \cline{ 2- 5}

\multicolumn{ 1}{|c|}{} & orthogonalité & $e^{6.7950 \times x - 0.2673}$ & $e^{6.7950 \times x - 0.2673}$ & $e^{3.7781 \times x - 0.2443}$ \\
\multicolumn{ 1}{|c|}{} &  & $\frac{1}{a} = 0,1472  $ & $\frac{1}{a} = 0,1472 $ & $\frac{1}{a} = 0.2647 $  \\ \cline{ 2- 5}

\multicolumn{ 1}{|c|}{} & log(espacement) & $log : 3.37 \pm 0.93 $  & $log : 3.34 \pm 0.71$  & $log : 4.79 \pm 0.74$  \\
\multicolumn{ 1}{|c|}{}  &  & $reel : 29.01m {\times \atop \div} 2.54 $  & $reel : 28.07m {\times \atop \div} 2.04 $  & $reel : 120.51m {\times \atop \div} 2.09 $  \\
 \cline{ 2- 5}

\multicolumn{ 1}{|c|}{} & log(espacement) & $log : e^{1.4101 \times x + 0.9114}$ & $log : e^{-1.806 \times x + 12.410}$ & $log : e^{1.115 \times x  - 1.441}$ \\
\multicolumn{ 1}{|c|}{} &  & $reel : x^{0.4101} \times e^{0.9114}$ & $reel : x^{-2.806} \times e^{12.410}$ & $reel : x^{0.115} \times e^{- 1.441}$ \\ 

\multicolumn{ 1}{|c|}{} &  & $log : e^{-1.806 \times x + 12.410}$ & $log : e^{-2.213 \times x + 14.138}$ & $log : e^{-2.619 \times x + 17.988}$ \\
\multicolumn{ 1}{|c|}{} &  & $reel : x^{-2.806} \times e^{12.410}$ & $reel :  x^{-3.213} \times e^{14.138}$ & $reel :  x^{-3.619} \times e^{17.988}$ \\ \hline

\multicolumn{ 1}{|c|}{Sommet} & degré & $e^{-2.323 \times x + 15.514}$ & $e^{-1.665 \times x + 14.268}$ & $e^{-1.731 \times x + 13.378}$ \\
\multicolumn{ 1}{|c|}{} &  & $\frac{1}{a} = -0,4305$ & $\frac{1}{a} = -0,6006$ & $\frac{1}{a} = -0,5777$  \\ \hline

\end{tabular}
}
\caption{Valeurs des coefficients des approximations.}
\label{tab:valeurs_coef}
\end{table}

\clearpage{\pagestyle{empty}\cleardoublepage}
\chapter{Analyser le changement : \\ caractérisation de la cinématique des graphes}
\minitoc
\markright{Analyser le changement : caractérisation de la cinématique des graphes}

\FloatBarrier
\section{Méthodologie de quantification du changement}

\subsection{Explication de la méthodologie de quantification}

Arriver à exprimer la dimension temporelle dans l'analyse de la spatialité est un enjeu difficile. En effet, nous travaillons sur des données géographiques discrètes. Nous avons des coordonnées relevées à un instant précis. Comment, dès lors, comprendre l'évolution d'un paysage ? Comment décrire le mouvement d'un graphe ? Qui plus est d'un graphe spatialisé ?

Les SIG permettent une étude spatiale approfondie mais ne sont pas conçus pour intégrer une dimension temporelle. Patricia Bordin propose une classification des solutions permettant de répondre à cette problématique, afin de réaliser des observations temporelles \citep{bordin2006methode}. Elle distingue en particulier les solutions \textit{a priori} qui intègrent, dans la modélisation des données, les aspects propres à leurs évolutions (dates de création, de modification, lien vers l'objet successeur, etc) et les solution \textit{a posteriori} qui visent à identifier et à qualifier les changements à l'aide de méthodes d'appariement et de calculs différentiels. Elle met en évidence les difficultés de mise en œuvre de ces deux types de solutions. En effet, les premières sont complexes à manipuler et ne sont mobilisables que par les producteurs initiaux des bases de données, et les secondes, sophistiquées techniquement, ne sont de fait pas accessibles à la majorité des utilisateurs. P. Bordin propose une solution alternative, méthodologique, qui est fondée sur l'utilisation d'une \textit{emprise} (ou \textit{portion de territoire}) maintenue constante, servant de support aux calculs d'indicateurs au cours du temps. 

\begin{figure}[h]
    \centering
    
     \begin{subfigure}[t]{.45\linewidth}
        \includegraphics[width=\textwidth]{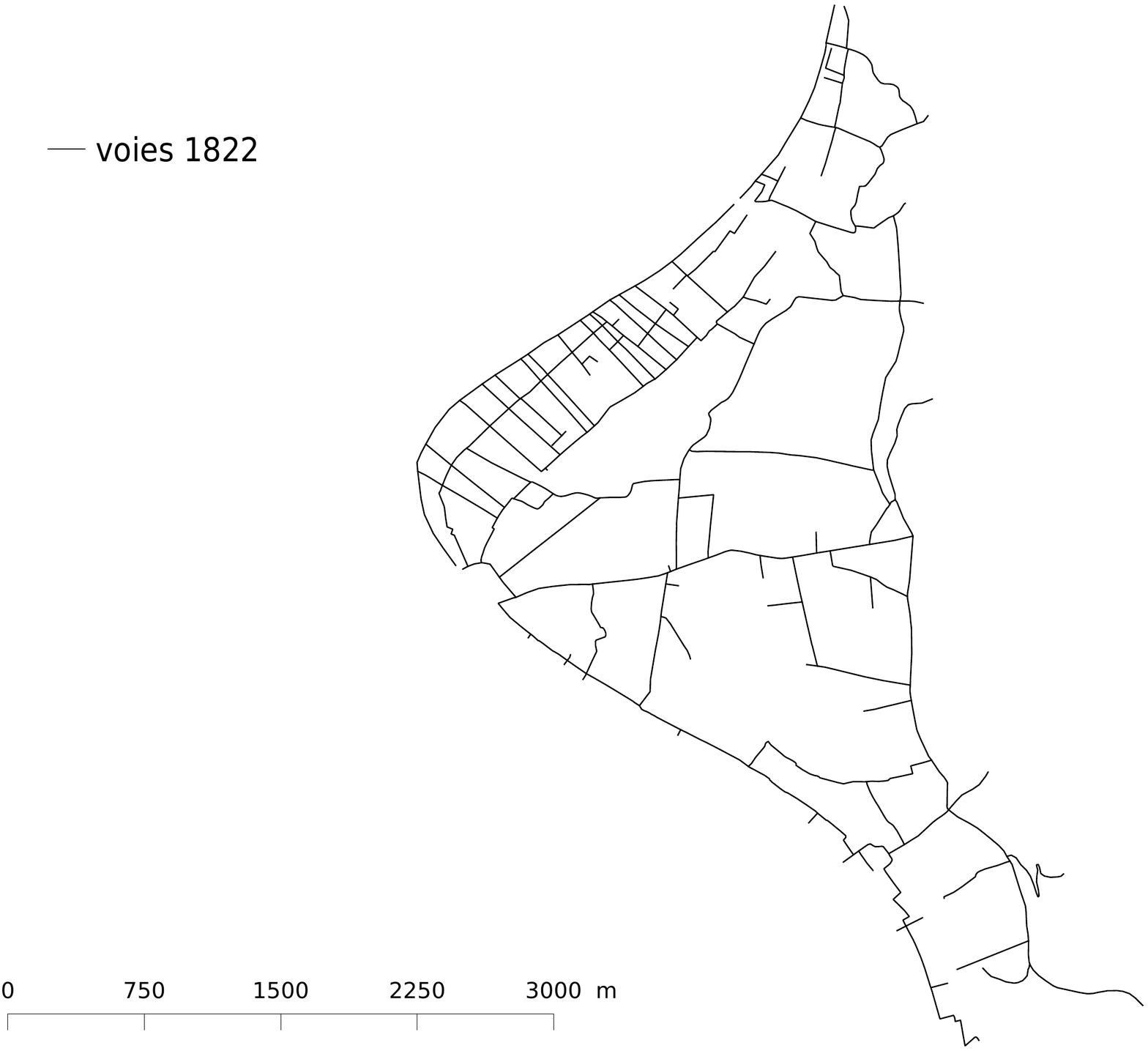}
        \caption{Données de 1822.}
        \label{fig:bdx_1822}
    \end{subfigure}
    ~    
    \begin{subfigure}[t]{.45\linewidth}
        \includegraphics[width=\textwidth]{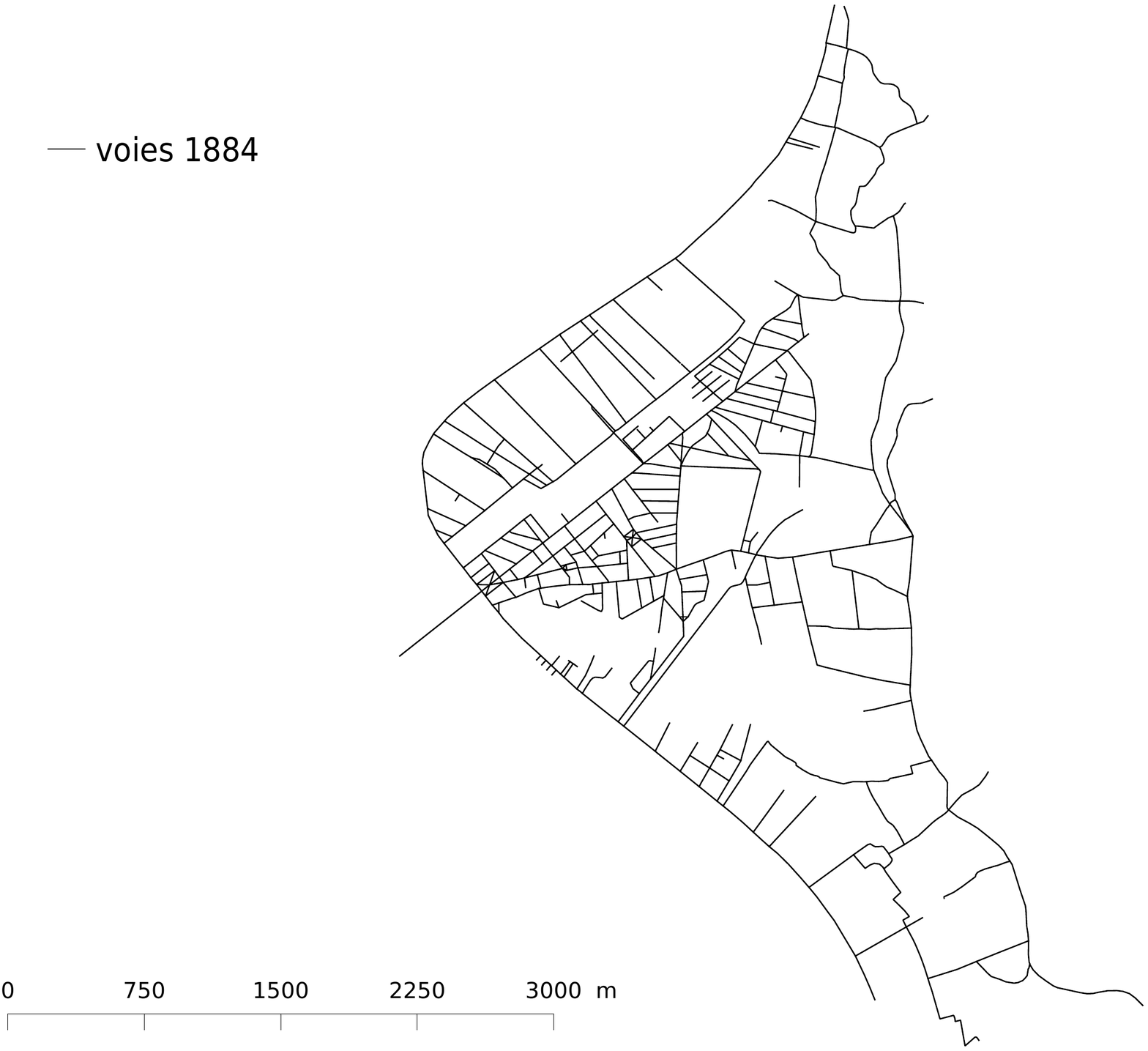}
        \caption{Données de 1884.}
        \label{fig:bdx_1884}
    \end{subfigure}
    
     \begin{subfigure}[t]{.45\linewidth}
        \includegraphics[width=\textwidth]{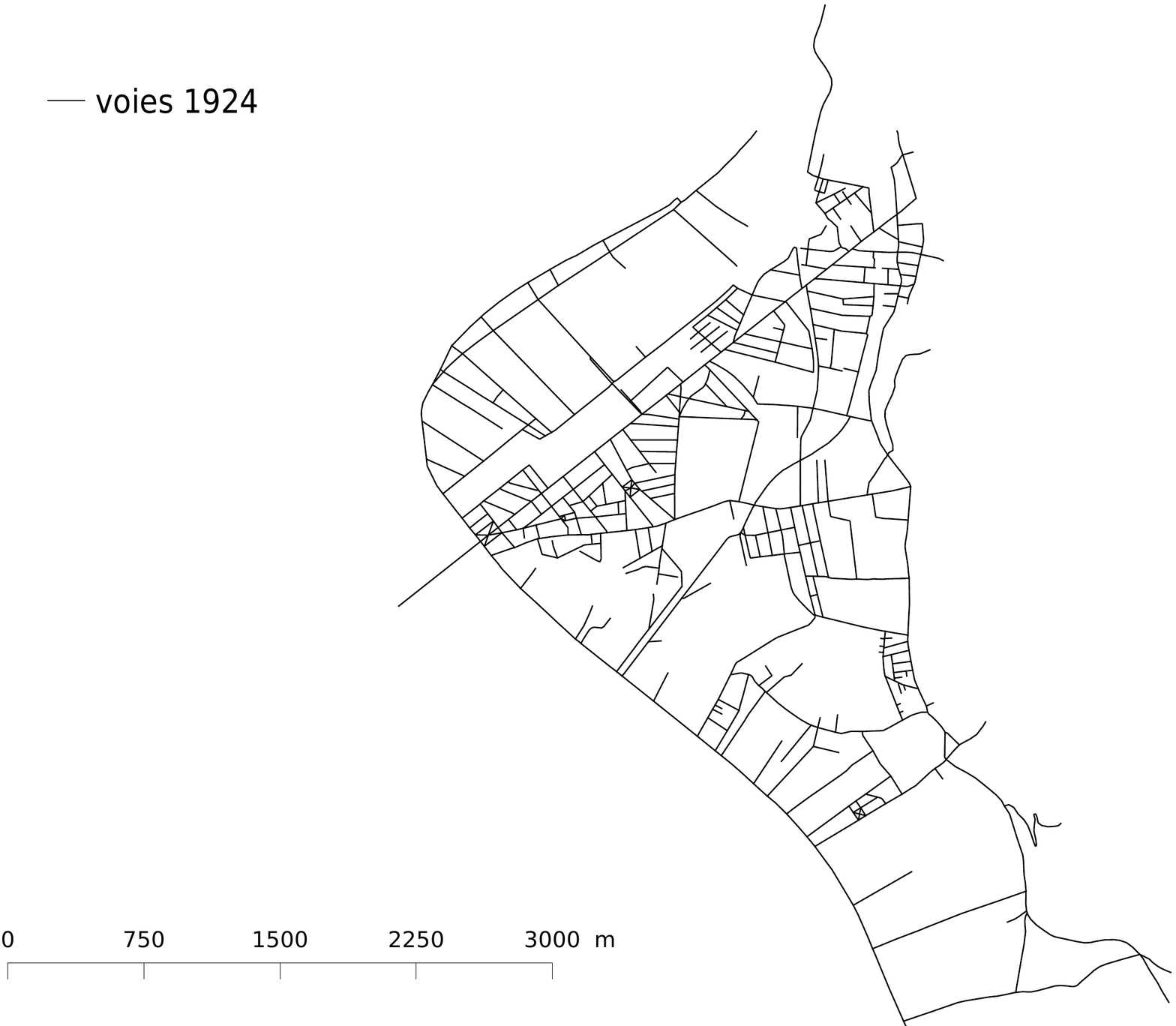}
        \caption{Données de 1924.}
        \label{fig:bdx_1924}
    \end{subfigure}
    ~    
    \begin{subfigure}[t]{.45\linewidth}
        \includegraphics[width=\textwidth]{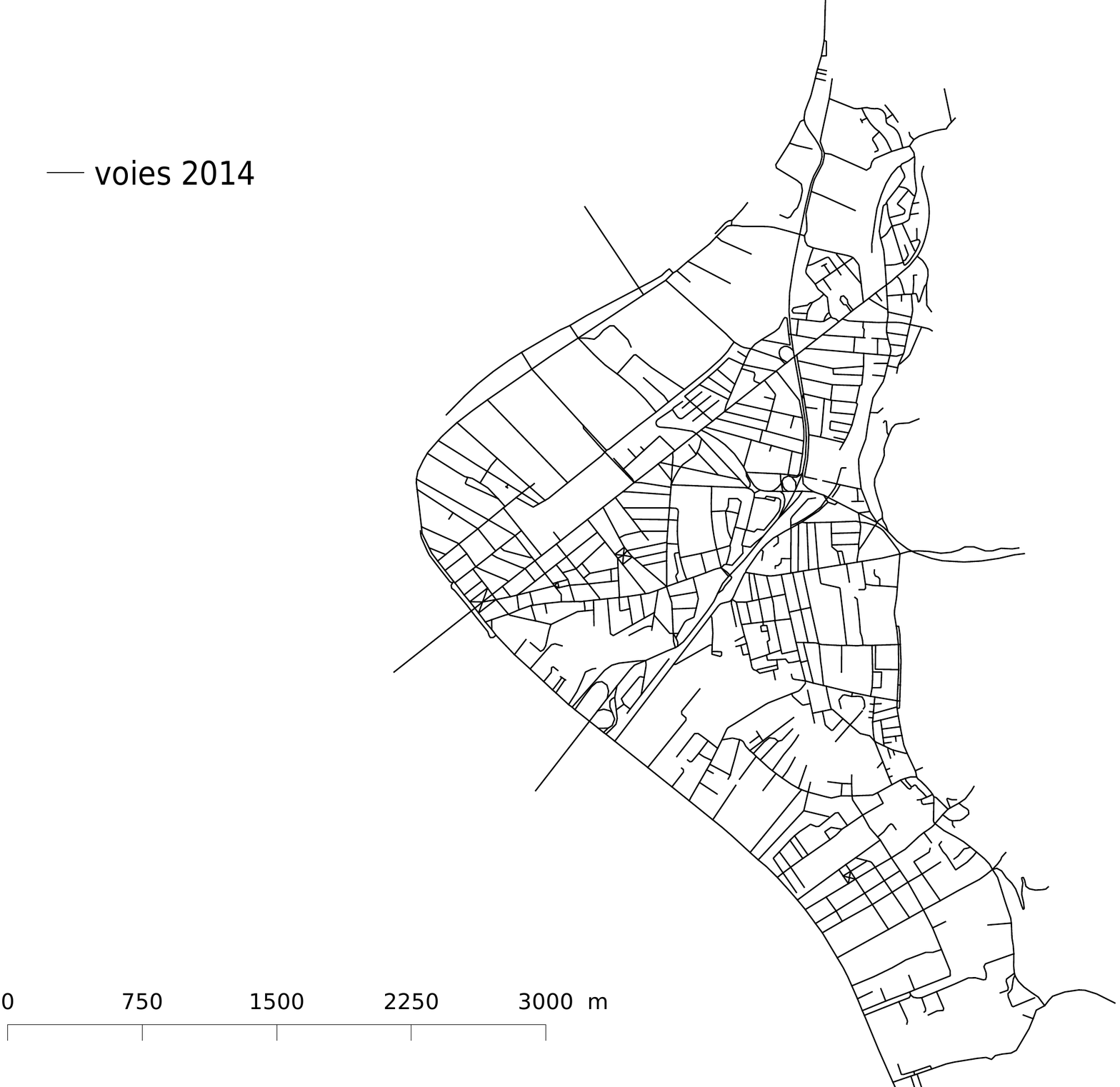}
        \caption{Données de 2014.}
        \label{fig:bdx_2014}
    \end{subfigure}

     \begin{subfigure}[t]{.45\linewidth}
        \includegraphics[width=\textwidth]{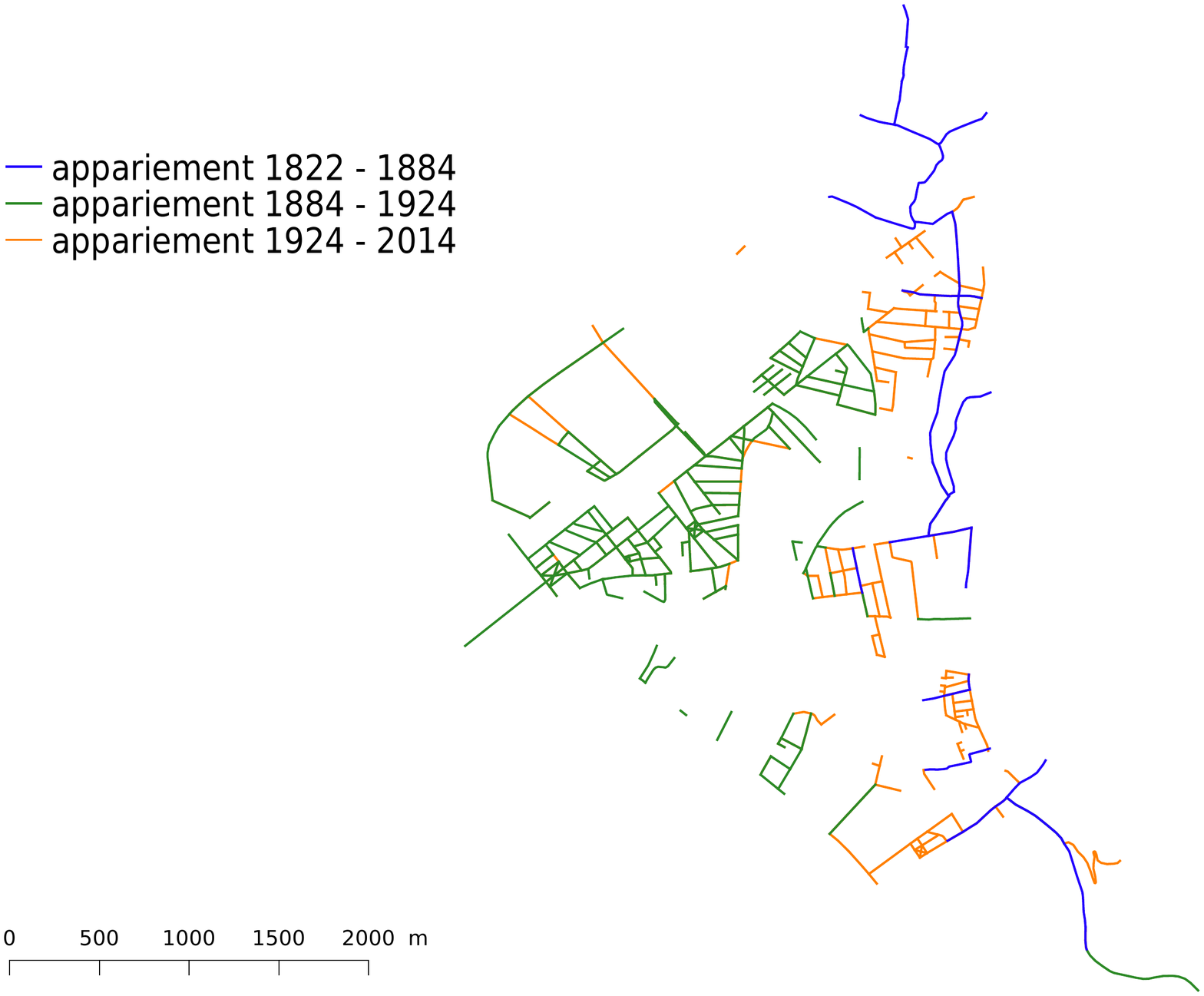}
        \caption{Données qui ont pu être appariées.}
        \label{fig:bdx_app}
    \end{subfigure}   
    
    \caption{Illustration du problème d’appariement sur Bordeaux (rive droite)}
    \label{fig:app_pb}
\end{figure}

Dans notre travail, nous proposons une analyse de l'évolution des graphes viaires extraits de deux villes : Avignon, ville du Sud de la France (intra-muros) et Rotterdam, ville des Pays Bas (partie Nord). Afin de pouvoir analyser automatiquement les changements de géométries d'une année à l'autre dans chacun de ces graphes, il est nécessaire d'avoir des données appariées : les géométries doivent avoir la même emprise, pour deux graphes d'un même territoire, représenté à deux instants $t$ différents. Ceci afin que nous puissions distinguer celles qui ont été conservées de celles qui ont été modifiées. Si les numérisations sont faites indépendamment les unes des autres, la conservation des géométries d'une année sur l'autre est compromise. Même si l’opérateur porte une attention particulière à la sélection des mêmes emprises, un léger décalage aboutira à une impossibilité de reconnaissance automatique de structures pourtant similaires (figure \ref{fig:app_pb}).

\FloatBarrier

Notre équipe de recherche, Morphocity, a commandé (via le financement de l'ANR MONUMOVI) la numérisation des réseaux viaires à différentes dates sur les deux territoires étudiés. Étant commanditaires des données, nous avons pu préciser leur forme pour l'adapter à l'étude que nous souhaitions mener. Nous avons recueilli des cartes anciennes auprès des archives d'Avignon et, pour Rotterdam, nous avons rencontré les créateur de la société Mapping History qui avait déjà dessiné vectoriellement des réseaux de rues anciens de la ville. Devant ces sources de données, nous avons fait attention à ne pas procéder à une simple numérisation par état (dite par \textit{snapshot}), indépendamment les uns des autres. Nous avons souhaité construire une base qui contient explicitement le lien temporel, évitant de devoir retrouver \textit{a posteriori} les arcs qui se correspondent.

Nous nous sommes appuyés sur la notion d'emprise constante, définie dans les travaux de P. Bordin, pour assurer un appariement des données à travers le temps. Le modèle que nous construisons propose un procédé de numérisation régressif : partant des données les plus récentes (temps $t_m$), considérées \textit{a priori} comme les plus justes géométriquement et les plus complètes, nous avons numérisé des représentations antérieures ($t_m-1, t_m-2... t_m-n$).

La méthodologie mise en place, à partir de $t_m$, consiste donc à supprimer les voies qui ont été créées entre $t_m-i-1$ et $t_m-i$ et à ajouter celles qui ont été supprimées. Lorsqu'une géométrie a existé sur une vectorisation, une emprise identique sera conservée sur les suivantes. Ainsi, nous pouvons retrouver la trace d'objets anciens disparus sur des cartes plus récentes avec conservation de leur géométrie.

\begin{figure}[h]
    \centering
        \includegraphics[width=\textwidth]{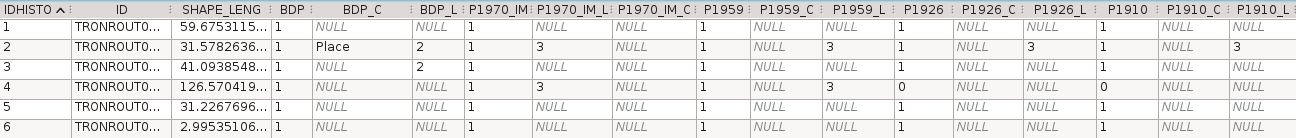}
    
    \caption{Aperçu de la base de données panchronique.}
    \label{fig:bdd_pan}
\end{figure}

Ce processus de vectorisation, réalisé par la société DIGITECH, aboutit à la construction d'une base de données que nous avons appelée \emph{panchronique} : regroupant des géométries numérisées sur plusieurs cartes d'époques différentes. Chaque géométrie y est référencée avec un identifiant unique $IDHISTO$. Un attribut est créé pour chaque année où une carte a été vectorisée. Cet attribut est rempli par un booléen : 1 si la géométrie est présente sur la carte, 0 si ce n'est pas le cas (figure \ref{fig:bdd_pan}). Cependant, la lecture des cartes anciennes, même après leur  géoréférencement, ne permet pas d'établir de manière certaine la présence ou l'absence de certains tronçons. Une codification a donc été introduite au modèle pour expliciter les cas de litige et ajouter des commentaires éventuels. La modification légère d'une géométrie (cas d'un redressement par exemple) peut également poser la question de \textit{l'objet suffisamment lui-même} \citep{bordin2006methode}. Nous reviendrons sur ces problématiques en dernière partie.

Pour Avignon, nous disposons ainsi des données vectorisées sur l'intra-muros pour dix dates entre 1760 et 2014. Pour Rotterdam, nous disposons de neuf dates entre 1374 et 1955. Nous reviendrons par la suite sur les caractéristiques de ces deux graphes.

\FloatBarrier

Comme nous l'avons vu précédemment, sur les trois attributs primaires liés aux voies (dégré, closeness et orthogonalité) un seul est calculé en tenant compte de l'ensemble du réseau : la closeness. Les deux autres traitent de caractéristiques locales topologiques (nombre de voies connectées) ou topographiques (avec quelles géométries elles sont connectées). Ces propriétés sont donc propres à chaque voie, et ne seront modifiées sur le graphe que si la géométrie de la voie ou de son voisinage direct est modifiée (ajout ou suppression d'un ou plusieurs arcs qui intersectent la voie, totalement ou partiellement). Dans ce cas, l'effet des changements sur le réseau est réduit à l'interaction directe des ajouts ou suppressions avec un ou plusieurs objets. Ce qui nous intéresse ici est de quantifier l'impact des transformations du réseau sur l’ensemble des proximités topologiques entre voies. Nous voulons comprendre les modifications liées à l'accessibilité topologique de chaque objet.

Pour cela, nous utilisons la même méthodologie que celle introduite pour la quantification des effets de bord. Pour l'adapter à notre problématique diachronique, nous lui ajoutons le poids des objets éventuellement retirés du réseau. Nous détaillons ici notre approche.

\begin{figure}[h]
    \centering
        \includegraphics[width=\textwidth]{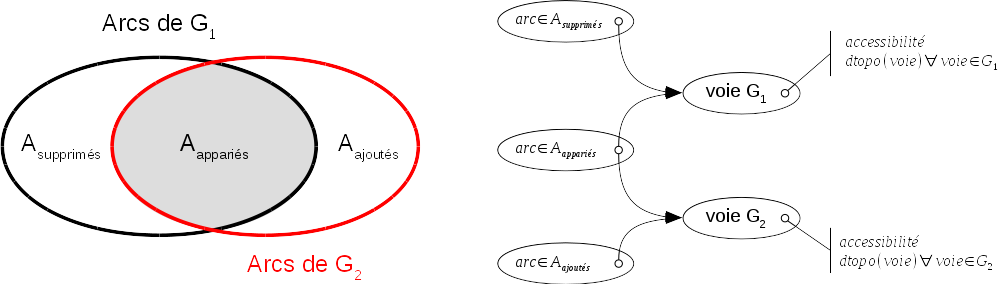}
    
    \caption{Méthodologie.}
    \label{fig:sch_diff}
\end{figure}

Pour deux graphes $G_1$ et $G_2$ d'une même ville à deux dates données $t_1$ et $t_2$, nous calculons l'ensemble des voies, les indicateurs d'accessibilité, et les distances topologiques entre elles. Pour étudier plus finement les changements d'accessibilité sur le graphe, nous reportons les valeurs calculées pour les voies sur les arcs qui les composent : chaque arc porte dans sa base attributaire la voie à laquelle il appartient ainsi que l'indicateur d'accessibilité calculé pour celle-ci. Cela permet, lors de l'identification des géométries inchangées entre $G_1$ et $G_2$ de considérer de plus petites portions de réseau en identifiant les changements d'arcs et non de voies (figure \ref{fig:sch_diff}).

Sur l'ensemble des arcs présents dans les deux graphes $\{a\} \in (G_1 \cup G_2)$ nous distinguons : 

\begin{itemize}
\item les arcs supprimés entre $t_1$ et $t_2$ ont une géométrie dans $G_1$ mais pas dans $G_2$ : $a_{sup} \in A_{supprimes}$
\item les arcs ajoutés entre $t_1$ et $t_2$ ont une géométrie dans $G_2$ mais pas dans $G_1$ : $a_{aj} \in A_{ajoutes}$
\item  les arcs inchangés entre $t_1$ et $t_2$ ont une géométrie dans $G_1$ et dans $G_2$ : $a_{ap} \in A_{apparies}$
\end{itemize}

\FloatBarrier

Pour déceler la variation de l'accessibilité de chaque objet nous voulons comparer les valeurs initiales et finales de cet indicateur. Nous appliquons cette différence sur le graphe apparié, sans tenir compte des distances topologiques et des longueurs des arcs qui ont été supprimés ou ajoutés.

Nous avons vu, dans le chapitre 3 de la première partie, que l’accessibilité se calcule pour chaque objet $o_{ref}$ sur un graphe $G$, en sommant le produit des distances topologiques ($d_{simple}$) et des longueurs selon l'équation \ref{eq:dtopo_a}. Nous utilisons dans cette étude l'indicateur d'accessibilité et non celui de closeness pour pouvoir travailler à partir des arcs. En effet, cela nous permet de sommer le produit des distances topologiques entre les voies, à partir des arcs qui les composent, en les multipliant pour chaque arc par la longueur de celui-ci. En faisant le calcul pour tous les arcs d'une voie nous retrouvons ainsi la longueur de la voie (somme des longueurs de ses arcs). Si nous ne travaillions qu'avec les rayons topologiques cela ne serait pas possible car pour chaque arc d'une voie nous sommerions les distances topologiques de cette voie vers les autres du réseau. En utilisant l'accessibilité, il est possible de calculer sur $G_1$ et $G_2$ une valeur où, les distances topologiques entre arcs qui ont été appariés, et les distances topologiques des arcs qui n'appartiennent qu'à un des deux graphes, sont dissociées (équations \ref{eq:dtopo_ad1} et \ref{eq:dtopo_ad2}).

\begin{equation}
accessibilite(o_{ref})=\sum_{o \in G} [d_{simple}(o,o_{ref}) * longueur(o)]
\label{eq:dtopo_a}
\end{equation}

\begin{eqnarray*}
accessibilite_{G_1}(a_{ref}) &=& \sum_{a_{ap} \in G_1} (d_{simple}(a_{ref}, a_{ap}) \times longueur(a_{ap})) \\
&+& \sum_{a_{sup} \in G_1} (d_{simple}(a_{ref}, a_{sup}) \times longueur(a_{sup}))
\label{eq:dtopo_ad1}
\end{eqnarray*}

\begin{eqnarray*}
accessibilite_{G_2}(a_{ref}) &=& \sum_{a_{ap} \in G_2} (d_{simple}(a_{ref}, a_{ap}) \times longueur(a_{ap})) \\
&+& \sum_{a_{sup} \in G_2} (d_{simple}(a_{ref}, a_{aj}) \times longueur(a_{aj}))
\label{eq:dtopo_ad2}
\end{eqnarray*}

La différence brute entre les deux valeurs d'accessibilité pour une voie (reportée sur un arc) telle qu'elle est donnée par l'équation \ref{eq:deltabrut} peut donc être affinée pour avoir une différence faite en ne tenant compte que des arcs appariés. Il faut pour cela retirer la somme des accessibilités des arcs qui ont été ajoutés (équation \ref{eq:deltaadd}) et ajouter celle des arcs qui ont été supprimés (équation \ref{eq:deltarem}). Nous obtenons ainsi une différence qui, si les arcs ajoutés ou supprimés n'ont aucun impact sur le réseau, sera nulle (équation \ref{eq:deltatot}).

\begin{equation}
\Delta _{brut}access(a_{ref}) = accessibilite_{G_2}(a_{ref}) - accessibilite_{G_1}(a_{ref})
\label{eq:deltabrut}
\end{equation}

\begin{equation}
\delta_{ajouts}(a_{ref}) = \sum_{a_{aj} \in G_2} (d_{simple}(a_{ref}, a_{aj}) \times longueur(a_{aj}))
\label{eq:deltaadd}
\end{equation}

\begin{equation}
\delta_{suppressions}(a_{ref}) = \sum_{a_{sup} \in G_1} (d_{simple}(a_{ref}, a_{sup}) \times longueur(a_{sup}))
\label{eq:deltarem}
\end{equation}

\begin{eqnarray*}
  \Delta access(a_{ref}) &=& (accessibilite{G_2}(a_{ref}) - \delta_{ajouts}(a_{ref})) \\
  & & - (accessibilite{G_1}(a_{ref}) - \delta_{suppressions}(a_{ref})) \\
  &=& \Delta _{brut}access(a_{ref}) -  \delta_{ajouts}(a_{ref}) + \delta_{suppressions}(a_{ref})
\label{eq:deltatot}
\end{eqnarray*}

Le $\Delta access$ que nous calculons avec l'équation \ref{eq:deltatot} est à comparer avec la valeur finale de l'accessibilité calculée pour les voies de $G_2$. Cela permet de quantifier l'ampleur de l'influence des modifications par rapport à la valeur de l'indicateur. Nous calculons donc le quotient de ces deux valeurs, dont nous inversons le signe afin qu'un impact positif sur l'accessibilité topologique d'un objet dans le réseau ait un $\Delta_{relatif}$ positif (équation \ref{eq:deltafin}).

\begin{equation}
  \Delta_{relatif}(a_{ref}) = (-1) \times \frac{\Delta access(a_{ref})}{accessibilite_{G_2}(a_{ref})}
\label{eq:deltafin}
\end{equation}

\FloatBarrier
\subsection{Illustration par deux exemples}

\paragraph{But de la comparaison}

L'objectif est de quantifier les différences d'accessibilité engendrées par des aménagements du réseau. Ces aménagements peuvent être :
\begin{itemize}
\item la création de nouvelles routes
\item la suppression de routes existantes
\item la modification de routes existantes, que nous traiterons comme une suppression / création
\end{itemize}

Ces aménagements, correspondent à des ajouts et suppressions d'arcs. Nous travaillons donc sur deux versions d'un même réseau, deux graphes $G_1$ et $G_2$. 

Trois ensembles apparaissent :
\begin{itemize}
\item les arcs dans $G_1$ uniquement, appelés les arcs supprimés
\item les arcs dans $G_2$ uniquement, appelés les arcs ajoutés
\item les arcs dans $G_1$ et $G_2$, appelés les arcs appariés. Le changement de l'accessibilité n'a de sens que pour ces arcs.
\end{itemize}

L'ajout ou la suppression d'un arc modifie l'accessibilité des voies d'un réseau dans l'absolu, par sa longueur et sa distance topologique aux voies. Dans nos travaux, nous cherchons à identifier les éventuels \textit{chemins les plus courts} ajoutés ou retirés du réseau, qui seront, eux, gage d'une meilleure ou moins bonne accessibilité. Nous étudions donc comment éliminer le \enquote{bruit} dû à la géométrie propre des arcs modifiés pour ne garder que les réelles évolutions d'accessibilité dans le réseau.

Nous utiliserons les notations suivantes :
\begin{itemize}
\item La distance topologique la plus simple entre les objets $o_1$ et $o_2$ ($d_{simple}(o_1,o_2))$ : $\widehat{o_1 o_2}$
\item La longueur d'un objet $o$ : $\|o\|$
\end{itemize}

\FloatBarrier
\paragraph{Exemple 1 : ajout et suppression d'arcs n'impactant pas l'accessibilité}

\begin{figure}[h]
    \centering
    \begin{subfigure}[b]{0.45\textwidth}
        \includegraphics[width=\textwidth]{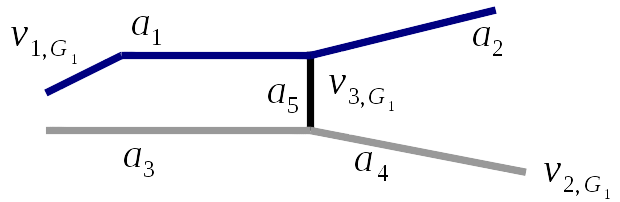}
        \caption{Version de référence du graphe, $G_1$}
    \end{subfigure}
	~
    \begin{subfigure}[b]{0.45\textwidth}
        \includegraphics[width=\textwidth]{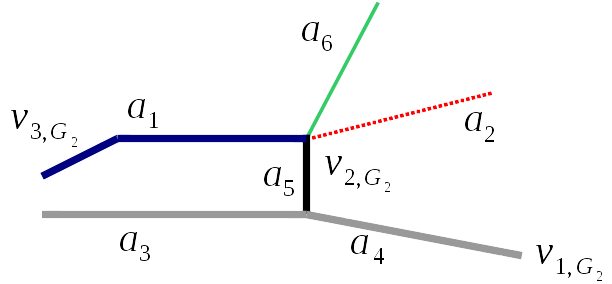}
        \caption{Évolution du graphe, $G_2$}
    \end{subfigure}
	\caption{Modification d'un réseau, sans impact sur l'accessibilité}
	\label{fig:access_diff_exemple1}
\end{figure}

Considérons le graphe suivant (figure \ref{fig:access_diff_exemple1}). Un arc est ajouté ($a_6$) et un arc est supprimé ($a_2$).

\FloatBarrier
\subparagraph{Analyse des deux graphes}

Les deux graphes sont analysés séparément, les voies sont construites et les indicateurs calculés.

\begin{table}[h]
\centering
{ \small
\begin{tabular}{|c|c|c|c|}
\hline
voie & arcs & accessibilité \\ \hline
$v_{1,G_1}$ & $a_1$, $a_2$ & $\widehat{v_{1,G_1}, v_{3,G_1}}*\|v_{3,G_1}\| + \widehat{v_{1,G_1}, v_{2,G_1}}*\|v_{2,G_1}\|$ \\
 & & $= 1*\|a_5\| + 2*(\|a_3\|+\|a_4\|)$  \\ \hline

$v_{2,G_1}$ & $a_3$, $a_4$ & $\widehat{v_{2,G_1}, v_{1,G_1}}*\|v_{1,G_1}\| + \widehat{v_{2,G_1}, v_{3,G_1}}*\|v_{3,G_1}\|$ \\
 & & $= 1*\|a_5\| + 2*(\|a_1\|+\|a_2\|)$  \\ \hline

$v_{3,G_1}$ & $a_5$ & $\widehat{v_{3,G_1}, v_{1,G_1}}*\|v_{1,G_1}\| + \widehat{v_{3,G_1}, v_{2,G_1}}*\|v_{2,G_1}\|$ \\
 & & $= 1*(\|a_1\|+\|a_2\|) + 1*(\|a_3\|+\|a_4\|)$  \\ \hline
\end{tabular}
}
\caption{Voies du graphe 1}
\end{table}

\begin{table}[h]
\centering
{ \small
\begin{tabular}{|c|c|c|c|}
\hline
voie & arcs & accessibilité \\ \hline
$v_{1,G_2}$ & $a_3$, $a_4$ & $\widehat{v_{1,G_2}, v_{3,G_2}}*\|v_{3,G_2}\| + \widehat{v_{1,G_2}, v_{2,G_2}}*\|v_{2,G_2}\|$ \\
 & & $= 2*\|a_1\| + 1*(\|a_5\|+\|a_6\|)$  \\ \hline

$v_{2,G_2}$ & $a_5$, $a_6$ & $\widehat{v_{2,G_2}, v_{1,G_2}}*\|v_{1,G_2}\| + \widehat{v_{2,G_2}, v_{3,G_2}}*\|v_{3,G_2}\|$ \\
 & & $= 1*(\|a_3\|+\|a_4\|) + 1*\|a_1\|$  \\ \hline

$v_{3,G_2}$ & $a_1$ & $\widehat{v_{3,G_2}, v_{1,G_2}}*\|v_{1,G_2}\| + \widehat{v_{3,G_2}, v_{2,G_2}}*\|v_{2,G_2}\|$ \\
 & & $= 2*(\|a_3\|+\|a_4\|) + 1*(\|a_5\|+\|a_6\|)$  \\ \hline

\end{tabular}
}
\caption{Voies du graphe 2}
\end{table}

\FloatBarrier
\subparagraph{Point de vue des arcs}

L'accessibilité est calculée sur les voies, puis reportée sur les arcs qui les composent. Tous les arcs constituant une même voie ont donc la même accessibilité. Cette opération peut être réalisée pour tous les arcs, qu'ils soient appariés ou non.

\begin{table}[h]
\centering
{ \small
\begin{tabular}{|c|c|c|c|c|}
\hline
arc & voie  & voie  & accessibilité dans $G_1$ & accessibilité dans $G_2$ \\
 & dans $G_1$ & dans $G_2$ & & \\ \hline
$a_1$ & $v_{1,G_1}$ & $v_{3,G_2}$ & $1*\|a_5\| + 2*(\|a_3\|+\|a_4\|)$ & $2*(\|a_3\|+\|a_4\|)$ \\
 & & & & $+ 1*(\|a_5\|+\|a_6\|)$  \\ \hline
$a_2$ & $v_{1,G_1}$ &  & $1*\|a_5\| + 2*(\|a_3\|+\|a_4\|)$ &  \\ \hline
$a_3$ & $v_{2,G_1}$ & $v_{1,G_2}$ & $1*\|a_5\| + 2*(\|a_1\|+\|a_2\|)$ & $2*(\|a_1\|) + 1*(\|a_5\|+\|a_6\|)$ \\ \hline
$a_4$ & $v_{2,G_1}$ & $v_{1,G_2}$ & $1*\|a_5\| + 2*(\|a_1\|+\|a_2\|)$ & $2*(\|a_1\|) + 1*(\|a_5\|+\|a_6\|)$ \\ \hline
$a_5$ & $v_{3,G_1}$ & $v_{2,G_2}$ & $1*(\|a_1\|+\|a_2\|) + 1*(\|a_3\|+\|a_4\|)$ & $1*(\|a_3\|+\|a_4\|) + 1*(\|a_1\|)$  \\ \hline
$a_6$ &  & $v_{2,G_2}$ & & $1*(\|a_3\|+\|a_4\|)$ \\
 & & & & $+ 1*(\|a_1\|+\|a_2\|)$  \\ \hline
\end{tabular}
}
\caption{Arcs des graphes $G_1$ et $G_2$}
\end{table}

\FloatBarrier
\subparagraph{Calcul du différentiel}

Qualitativement, dans cet exemple, les arcs ajoutés et modifiés ne changent pas l'accessibilité générale du réseau : l'arc ajouté ne permet pas de raccourcir les distances topologiques, et l'arc supprimé ne les allonge pas. Le différentiel attendu est donc nul. Nous le calculons selon la formule \ref{eq:calc_ex_diff}.

\begin{figure}
\begin{eqnarray*}
\forall a \in G_1 \cap G_2, \Delta_{access}(a)&=&\underbrace{access(a,G_2)-access(a,G_1)}_{\text{différentiel brut}} \\
&+&\underbrace{\sum_{a_{sup} \in G_1-G_2} \widehat{a, a_{sup}}*\|a_{sup}\|}_{\text{suppression du bruit dû aux arcs supprimés}}\\
&-& \underbrace{\sum_{a_{aj} \in G_2-G_1} \widehat{a, a_{aj}}*\|a_{aj}\|}_{\text{suppression du bruit dû aux arcs ajoutés}}
\end{eqnarray*}
\caption{Calcul du différentiel d'accessibilité}
\label{eq:calc_ex_diff}
\end{figure}

\FloatBarrier

Nous connaissons les distances topologiques entre les voies de $G_1$ et entre les voies de $G_2$. Pour les obtenir entre les arcs, il suffit de remonter à la voie à laquelle chacun appartient, dans leur graphe respectif. Par exemple, $\widehat{a_1, a_6} = \widehat{v_{3,G_2}, v_{2,G_2}} = 1$. Nous présentons le calcul de différentiel pour l'arc $a_1$ avec l'équation \ref{eq:calc_ex_delta}.

\begin{eqnarray*}
\Delta_{access}(a_1) &=& access(a_1, G_2) - access(a_1, G_1) + \widehat{a_1, a_2}*\|a_2\| - \widehat{a_1, a_6}*\|a_6\| \\
&=& 2*(\|a_3\|+\|a_4\|)+ 1*(\|a_5\|+\|a_6\|) - (1*\|a_5\| + 2*(\|a_3\|+\|a_4\|)) \\
&&+ 0*\|a_2\| - 1*\|a_6\| \\
&=& \|a_2\|*0 + \|a_3\|*(2 - 2) + \|a_4\|*(2 - 2) + \|a_5\|*(1 - 1) + \|a_6\|*(1 - 1) \\
&=& 0
\label{eq:calc_ex_delta}
\end{eqnarray*}

\FloatBarrier
\paragraph{Exemple 2 : ajout d'un arc impactant l'accessibilité}

\begin{figure}[h]
    \centering
    \begin{subfigure}[b]{0.45\textwidth}
        \includegraphics[width=\textwidth]{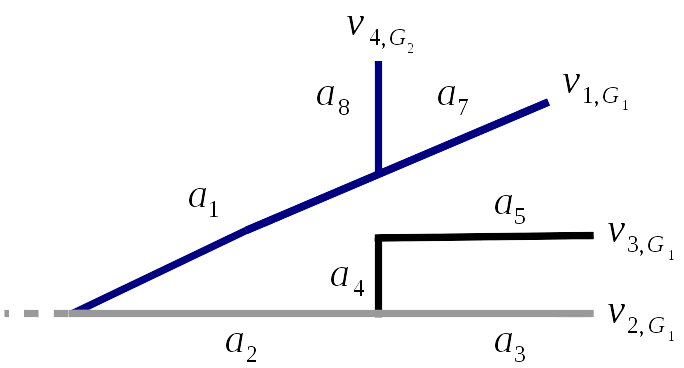}
        \caption{Version de référence du graphe, $G_1$}
    \end{subfigure}
	~
    \begin{subfigure}[b]{0.45\textwidth}
        \includegraphics[width=\textwidth]{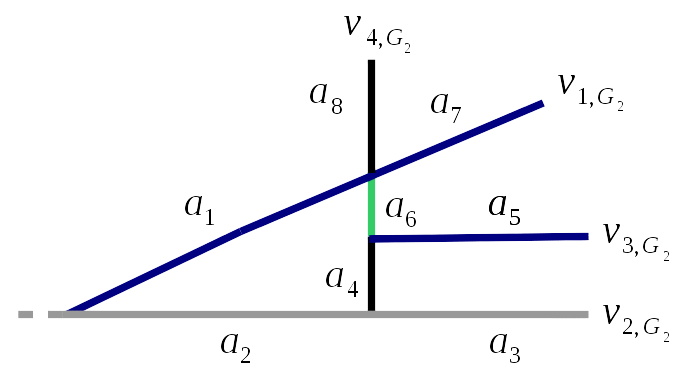}
        \caption{Évolution du graphe, $G_2$}
    \end{subfigure}
	\caption{Modification d'un réseaux améliorant l'accessibilité du réseau}
	\label{fig:access_diff_exemple2}
\end{figure}

Nous considérons dans un deuxième temps les graphes représentés par la figure \ref{fig:access_diff_exemple2}. Une seule modification est faite entre $G_1$ et $G_2$ : l'arc ($a_6$) a été ajouté.

\subparagraph{Analyse des deux graphes}

\begin{table}[h]
\centering
{ \small
\begin{tabular}{|c|c|c|c|}
\hline
voie & arcs & accessibilité \\ \hline
$v_{1,G_1}$ & $a_1$, $a_7$ & $\widehat{v_{1,G_1}, v_{2,G_1}}*\|v_{2,G_1}\| + \widehat{v_{1,G_1}, v_{3,G_1}}*\|v_{3,G_1}\| + \widehat{v_{1,G_1}, v_{4,G_1}}*\|v_{4,G_1}\|$ \\
 & & $= 1*(\|a_2\|+\|a_3\|) + 2*(\|a_4\| + \|a_5\|) + 1*\|a_8\|$  \\ \hline

$v_{2,G_1}$ & $a_2$, $a_3$ & $\widehat{v_{2,G_1}, v_{1,G_1}}*\|v_{1,G_1}\| + \widehat{v_{2,G_1}, v_{3,G_1}}*\|v_{3,G_1}\| + \widehat{v_{2,G_1}, v_{4,G_1}}*\|v_{4,G_1}\|$ \\
 & & $= 1*(\|a_1\|+\|a_7\|) + 1*(\|a_4\| + \|a_5\|) + 2*\|a_8\|$  \\ \hline

$v_{3,G_1}$ & $a_4$, $a_5$ & $\widehat{v_{3,G_1}, v_{1,G_1}}*\|v_{1,G_1}\| + \widehat{v_{3,G_1}, v_{2,G_1}}*\|v_{2,G_1}\| + \widehat{v_{3,G_1}, v_{4,G_1}}*\|v_{4,G_1}\|$ \\
 & & $= 2*(\|a_1\|+\|a_7\|) + 1*(\|a_2\|+\|a_3\|) + 3*\|a_8\|$  \\ \hline
 
$v_{4,G_1}$ & $a_8$ & $\widehat{v_{4,G_1}, v_{1,G_1}}*\|v_{1,G_1}\| + \widehat{v_{4,G_1}, v_{2,G_1}}*\|v_{2,G_1}\| + \widehat{v_{4,G_1}, v_{3,G_1}}*\|v_{3,G_1}\|$ \\
 & & $= 1*(\|a_1\|+\|a_7\|) + 2*(\|a_2\|+\|a_3\|) + 3*(\|a_4\| + \|a_5\|)$  \\ \hline
\end{tabular}
}
\caption{Voies du graphe 1}
\end{table}

\begin{table}[h]
\centering
{ \small
\begin{tabular}{|c|c|c|c|}
\hline
voie & arcs & accessibilité \\ \hline
$v_{1,G_2}$ & $a_1$, $a_7$ & $\widehat{v_{1,G_2}, v_{2,G_2}}*\|v_{2,G_2}\| + \widehat{v_{1,G_2}, v_{3,G_2}}*\|v_{3,G_2}\| + \widehat{v_{1,G_2}, v_{4,G_2}}*\|v_{4,G_2}\|$ \\
 & & $= 1*(\|a_2\|+\|a_3\|) + 2*\|a_5\| + 1*(\|a_4\| + \|a_6\| + \|a_8\|)$  \\ \hline

$v_{2,G_2}$ & $a_2$, $a_3$ & $\widehat{v_{2,G_2}, v_{1,G_2}}*\|v_{1,G_2}\| + \widehat{v_{2,G_2}, v_{3,G_2}}*\|v_{3,G_2}\| + \widehat{v_{2,G_2}, v_{4,G_2}}*\|v_{4,G_2}\|$ \\
 & & $= 1*(\|a_1\| + \|a_7\|) + 2*\|a_5\| + 1*(\|a_4\| + \|a_6\| + \|a_8\|)$  \\ \hline

$v_{3,G_2}$ & $a_5$ & $\widehat{v_{3,G_2}, v_{1,G_2}}*\|v_{1,G_2}\| + \widehat{v_{3,G_2}, v_{2,G_2}}*\|v_{2,G_2}\| + \widehat{v_{3,G_2}, v_{4,G_2}}*\|v_{4,G_2}\|$ \\
 & & $= 2*(\|a_1\| + \|a_7\|) + 2*(\|a_2\|+\|a_3\|) + 1*(\|a_4\|+\|a_6\| + \|a_8\|)$  \\ \hline
 
$v_{4,G_2}$ & $a_4$, $a_6$, $a_8$ & $\widehat{v_{4,G_2}, v_{1,G_2}}*\|v_{1,G_2}\| + \widehat{v_{4,G_2}, v_{2,G_2}}*\|v_{2,G_2}\| + \widehat{v_{4,G_2}, v_{3,G_2}}*\|v_{3,G_2}\|$ \\
 & & $= 1*(\|a_1\| + \|a_7\|) + 1*(\|a_2\|+\|a_3\|) + 1*\|a_5\|$  \\ \hline
\end{tabular}
}
\caption{Voies du graphe 2}
\end{table}

\subparagraph{Point de vue des arcs}

\begin{table}[h!]
\centering
{\footnotesize
\begin{tabular}{|c|c|c|c|c|}
\hline
arc & voie  & voie  & accessibilité dans $G_1$ & accessibilité dans $G_2$ \\
 & dans $G_1$ & dans $G_2$ & & \\ \hline
$a_1$ & $v_{1,G_1}$ & $v_{1,G_1}$  & $1*(\|a_2\|+\|a_3\|) + 2*(\|a_4\| $ & $1*(\|a_2\|+\|a_3\|) + 2*\|a_5\| $ \\
 & & & $+ \|a_5\|) + 1*\|a_8\|$ & $+ 1*(\|a_4\| + \|a_6\| + \|a_8\|)$ \\ \hline
$a_2$ & $v_{2,G_1}$ & $v_{2,G_2}$ &  $1*(\|a_1\|+\|a_7\|)  $ & $1*(\|a_1\| + \|a_7\|) + 2*\|a_5\| $ \\
 & & & $+ 1*(\|a_4\| + \|a_5\|) + 2*\|a_8\|$ & $+ 1*(\|a_4\| + \|a_6\| + \|a_8\|)$ \\ \hline
$a_3$ & $v_{2,G_1}$ & $v_{2,G_2}$ &  $1*(\|a_1\|+\|a_7\|) $ & $1*(\|a_1\| + \|a_7\|) + 2*\|a_5\| $ \\
 & & & $+ 1*(\|a_4\| + \|a_5\|) + 2*\|a_8\|$ & $+ 1*(\|a_4\| + \|a_6\| + \|a_8\|)$ \\ \hline 
$a_4$ & $v_{4,G_1}$ & $v_{4,G_2}$ &  $2*(\|a_1\|+\|a_7\|) $ & $1*(\|a_1\| + \|a_7\|) $ \\
 & & & $+ 1*(\|a_2\|+\|a_3\|) + 3*\|a_8\|$ & $+ 1*(\|a_2\|+\|a_3\|) + 1*\|a_5\|$ \\ \hline
$a_5$ & $v_{3,G_1}$ & $v_{3,G_2}$ &  $2*(\|a_1\|+\|a_7\|) $ & $2*(\|a_1\| + \|a_7\|) + 2*(\|a_2\|+\|a_3\|) $ \\
 & & & $ + 1*(\|a_2\|+\|a_3\|) + 3*\|a_8\|$ & $+ 1*(\|a_4\|+\|a_6\| + \|a_8\|)$ \\ \hline
$a_6$ &  & $v_{4,G_2}$ &  & $1*(\|a_1\| + \|a_7\|) $ \\
 & & & & $+ 1*(\|a_2\|+\|a_3\|) + 1*\|a_5\|$ \\ \hline
$a_7$ &  $v_{1,G_1}$ & $v_{1,G_2}$ &  $1*(\|a_2\|+\|a_3\|) $& $1*(\|a_2\|+\|a_3\|) + 2*\|a_5\| $ \\
 & & & $+ 2*(\|a_4\| + \|a_5\|) + 1*\|a_8\|$  & $+ 1*(\|a_4\| + \|a_6\| + \|a_8\|)$ \\ \hline
$a_8$ & $v_{4,G_1}$ & $v_{4,G_2}$ & $1*(\|a_1\|+\|a_7\|) + 2*(\|a_2\|+\|a_3\|) $ & $1*(\|a_1\| + \|a_7\|) $ \\
 & & & $+ 3*(\|a_4\| + \|a_5\|)$ & $+ 1*(\|a_2\|+\|a_3\|) + 1*\|a_5\|$ \\ \hline
\end{tabular}}
\caption{Arcs des graphes $G_1$ et $G_2$}
\end{table}

\subparagraph{Calcul du différentiel}

Qualitativement, dans cet exemple, nous voyons que la création du nouvel arc a amélioré l’accessibilité de tous les arcs : toutes les distances ont été réduites.

\begin{figure}[h]
\begin{eqnarray*}
\Delta_{access}(a_8) &=& access(a_8, G_2) - access(a_8, G_1) - \widehat{a_5, a_6}*\|a_6\| \\
&=& 1*(\|a_1\| + \|a_7\|) + 1*(\|a_2\|+\|a_3\|) + 1*\|a_5\| \\
&& - (1*(\|a_1\|+\|a_7\|) + 2*(\|a_2\|+\|a_3\|) + 3*(\|a_4\| + \|a_5\|)) - 1 * \|a_6\| \\
&=& \|a_1\|*(1-1) +\|a_7\|*(1-1) + \|a_2\|*(1-2) + \|a_3\|*(1-2) \\
&&+ \|a_4\|*(-3) + \|a_5\|*(1-3) + \|a_6\|*(-1) \\
&=& - (\|a_2\| + \|a_3\| + 3*\|a_4\| + 2*\|a_5\| + \|a_6\|) < 0
\end{eqnarray*}
\caption{Calcul du différentiel d'accessibilité pour l'arc $a_5$}
\end{figure}

D'après sa définition, plus la valeur de l'accessibilité est faible, plus la voie permet d'accéder rapidement à l'ensemble du réseau, et peut être considérée comme accessible. Dans notre cas, nous avons une baisse de la valeur de l'accessibilité, soit une \enquote{meilleure} accessibilité. C'était prévisible, l'ajout de l'arc ayant créé des \enquote{raccourcis}.

Nous avons calculé ici le différentiel d'accessibilité absolu. Pour finaliser notre quantification, nous divisons dans les graphes étudiés la valeur de $\Delta access$ pour chaque voie par sa valeur d'accessibilité dans $G_2$. En multipliant le résultat obtenu par (-1), pour qu'une amélioration d'accessibilité donne un résultat positif, nous obtenons ainsi un différentiel relatif $\Delta_{relatif}$ selon l'équation \ref{eq:deltafin}.

\FloatBarrier
\section{Analyses cinématiques de graphes}

\subsection{Avignon}

\begin{figure}[h]
    \centering
        \includegraphics[width=0.8\textwidth]{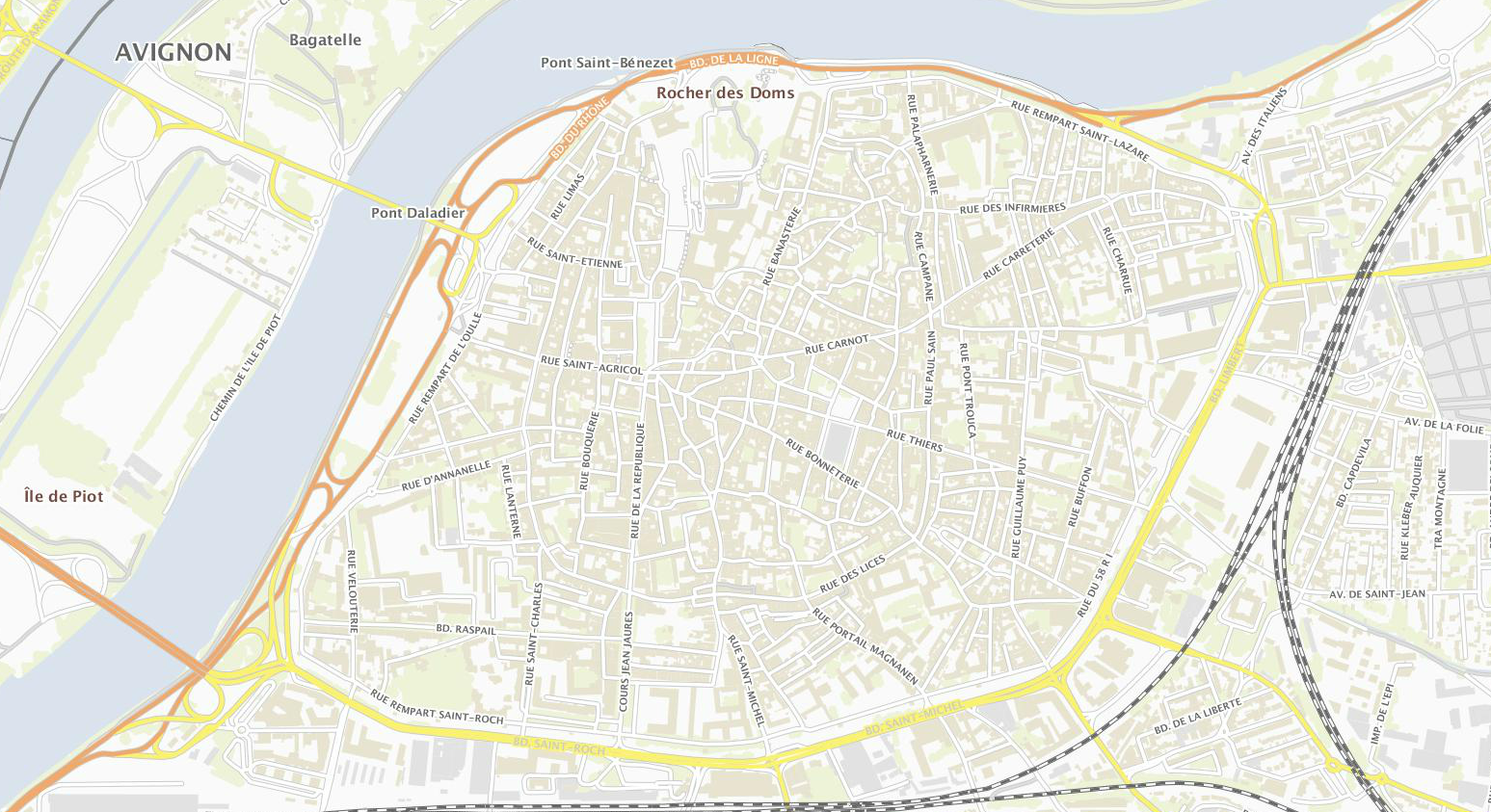}
    
    \caption{Carte de l'intra-muros d'Avignon (Plan IGN 2015).}
    \label{fig:av_plan}
\end{figure}

\begin{figure}[h]
    \centering
        \includegraphics[width=0.8\textwidth]{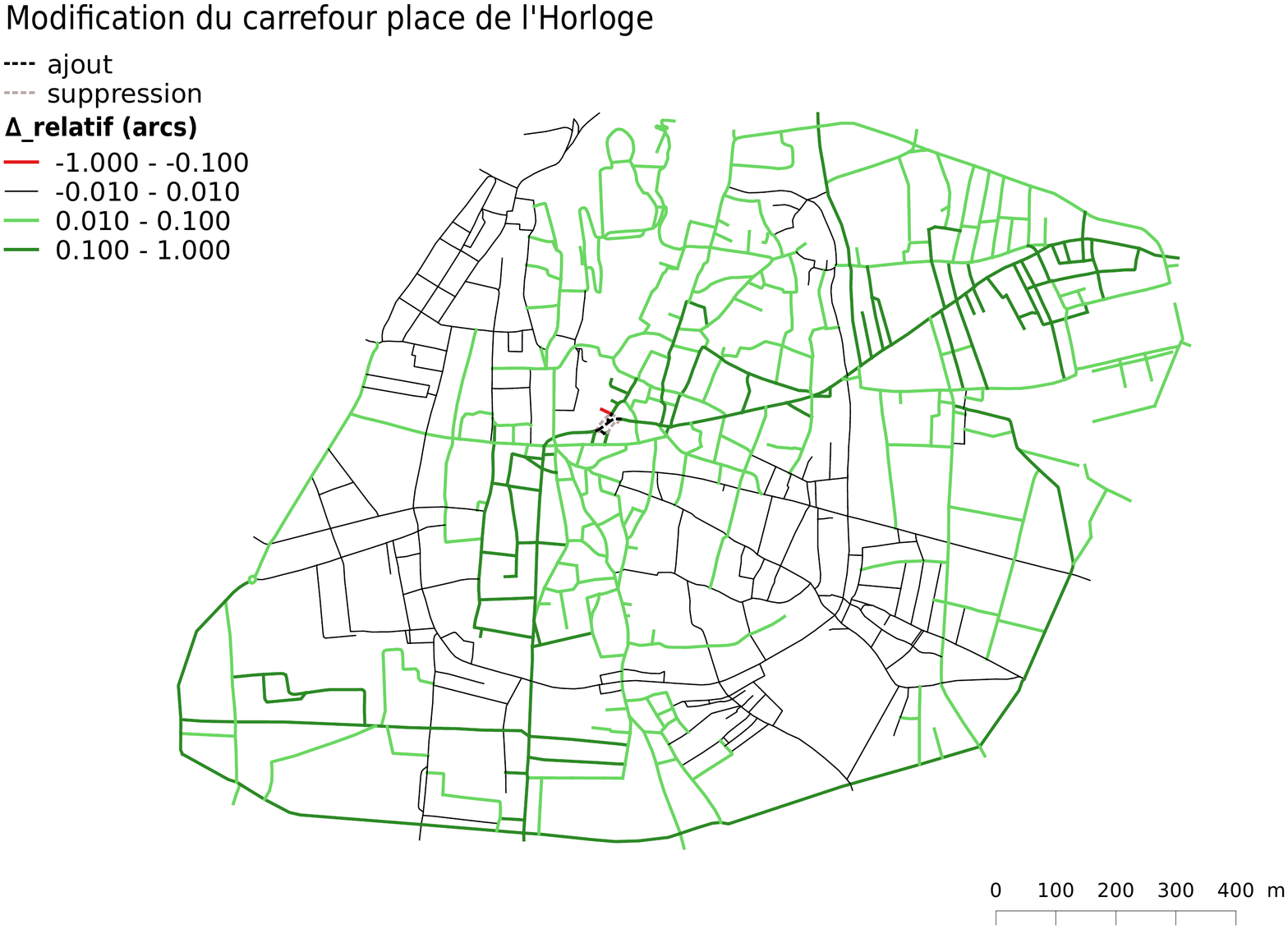}  
    
    \caption{Étude de $\Delta_{relatif}$ de la modification du centre du graphe d'Avignon intra-muros.}
    \label{fig:proj_avcentre}
\end{figure}

Nous commençons notre étude diachronique avec l'intra-muros d'Avignon. Nous donnons des repères toponymiques avec une vue du Géoportail de l'IGN sur la figure \ref{fig:av_plan}. Pour illustrer la quantification d'un changement sur le graphe via notre méthodologie nous commençons par étudier une modification précise : celle du carrefour central de la ville, à l'Est de la place de l'Horloge, entre la rue de la République et la rue Carnot. Figure \ref{fig:proj_avcentre} nous visualisation l'impact de cette modification ($\Delta_{relatif}$). Il se révèle totalement positif, sauf pour l'arc qui faisait partie de la voie construite sur la rue Carnot avant la modification.

Nous disposons pour Avignon d'une base de données \textit{panchroniques} de dix dates, numérisées dans l'intra-muros de la ville, exploitant la modélisation fondée sur le concept de \enquote{portion de territoire}. Les caractéristiques des graphes sont décrites dans le tableau \ref{tab:avignon_pres}. Nous lisons dans ce tableau l'évolution du graphe, dont la longueur totale ($L_{tot}$) et le nombre de sommets ($N_{sommets}$) augmentent avec les années. Le nombre d'arcs ($N_{arcs}$) et le nombre de voies ($N_{voies}$) ont eux une évolution complètement corrélée avec un nombre moyen d'arcs par voie constant sur toute la période étudiée.

\begin{table}
\begin{center}
{ \small
\begin{tabular}{|c|r|r|r|r|r|r|r|}
\hline
Année & $L_{tot}$ & $N_{sommets}$ & $N_{arcs}$ & $L_{moy}(arc)$ & $N_{voies}$ & $L_{moy}(voie)$ & $\overline{N_{arcs(voie)}}$ \\ \hline 		
1760 & 40 372 m & 574 & 1042 & 38.7 m & 260 & 155 m & 4.0 \\ \hline 
1819 & 40 705 m & 595 & 1060 & 38.4 m & 271 & 150 m & 3.9 \\ \hline 
1836 & 41 953 m & 623 & 1102 & 38.1 m & 284 & 148 m & 3.9 \\ \hline  
1854 & 43 559 m & 653 & 1136 & 38.3 m & 290 & 150 m & 3.9 \\ \hline 
1879 & 43 774 m & 660 & 1142 & 38.3 m & 294 & 149 m & 3.9 \\ \hline 

1910 & 44 592 m & 676 & 1161 & 38.4 m & 301 & 148 m & 3.9 \\ \hline 
1926 & 43 937 m & 662 & 1142 & 38.5 m & 293 & 150 m & 3.9 \\ \hline 
1959 & 44 793 m & 670 & 1153 & 38.8 m & 296 & 151 m & 3.9 \\ \hline 
1970 & 45 135 m & 674 & 1161 & 38.9 m & 298 & 151 m & 3.9 \\ \hline 

2014 & 46808 m & 693 & 1178 & 39.7 m & 307 & 152 m & 3.8 \\ \hline 

\end{tabular}
}
\end{center}
\caption{Détail des caractéristiques topologiques et métriques des graphes panchroniques du centre d'Avignon.}
\label{tab:avignon_pres}
\end{table}

\begin{figure}[c]
    \centering
        \includegraphics[width=0.8\textwidth]{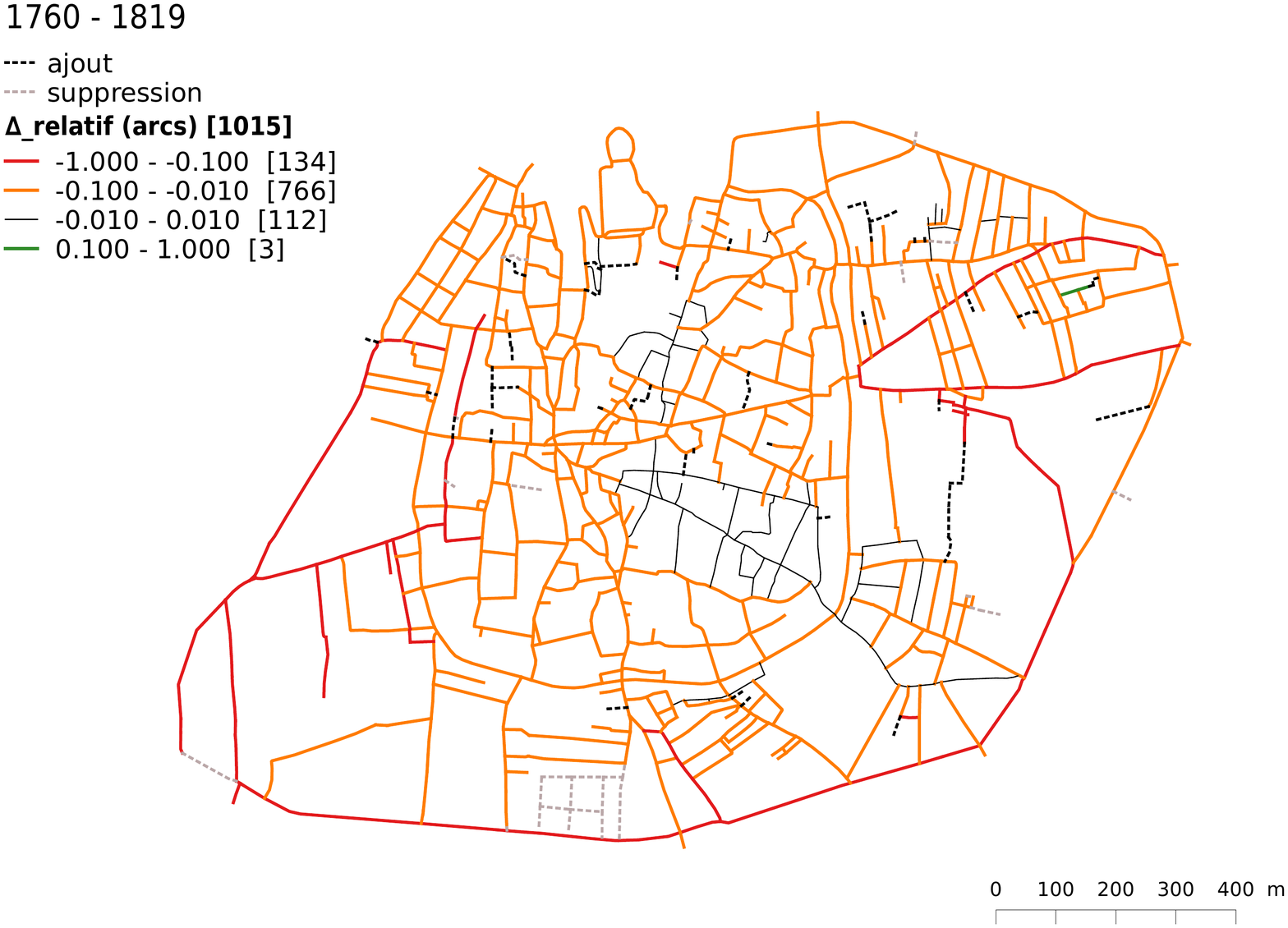}
        \caption{Étude cartographique de $\Delta_{relatif}$ sur la période 1760 - 1819.}
        \label{fig:diff_av_1760}
    \end{figure}

\begin{figure}[c]
    \centering
        \includegraphics[width=0.6\textwidth]{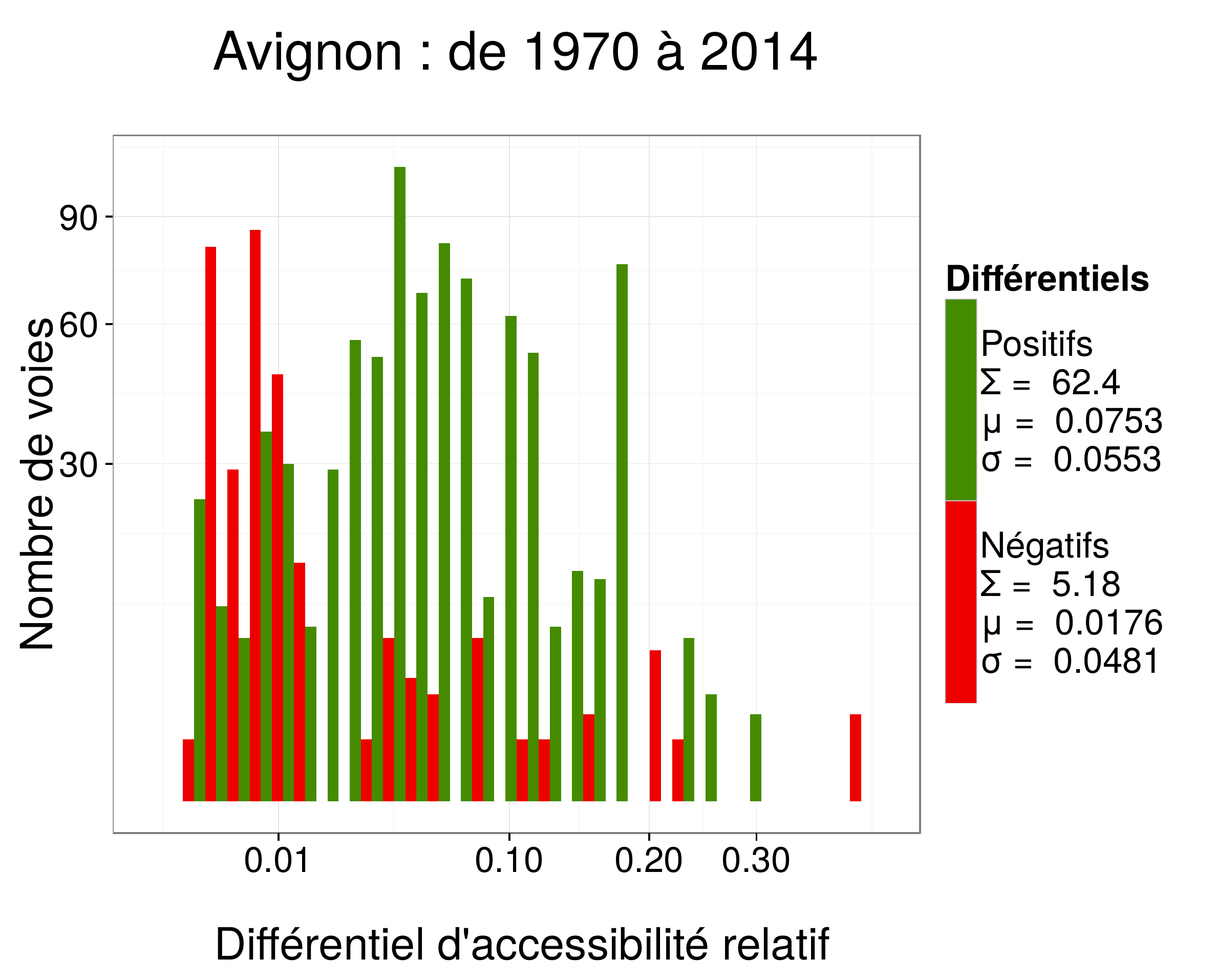}
        \caption{Étude statistique de $\Delta_{relatif}$ sur la période 1970 - 2014. \\ $\Sigma$ : somme ; $\mu$ : moyenne ; $\sigma$ : écart-type}
        \label{fig:diff_av_1970_stat}
    \end{figure}

Pour analyser les différences d'accessibilité sur le graphe, nous appliquons notre méthodologie de quantification du changement à ces réseaux (figures \ref{fig:diff_av_1760} à \ref{fig:diff_av_1970_stat}). Sur ces cartes, est représenté en rouge un impact négatif sur l'accessibilité de l'objet, en vert un impact positif. Nous avons également mis en valeur les arcs ajoutés (en pointillés noirs) et supprimés (en pointillés gris) entre les deux dates considérées.  Pour mieux comprendre comment est répartie l'influence des modifications sur les objets, nous avons noté sur cette carte le dénombrement des unités. Nous avons également tracé, pour chaque période, un histogramme représentant les variations positives et négatives d'accessibilité relative du graphe. Nous regroupons sur ceux-ci les valeurs statistiques de ces variations.

L'indicateur de closeness (indicateur primaire de la voie corrélé à son accessibilité) a été cartographié pour chacun de ces graphes et reporté en annexe \ref{ann:sec_cartedia_avignon} afin d'apprécier la hiérarchisation \enquote{brute} de la proximité topologique des voies sur chaque année.

\begin{figure}[c]
   \centering
        \includegraphics[width=0.8\textwidth]{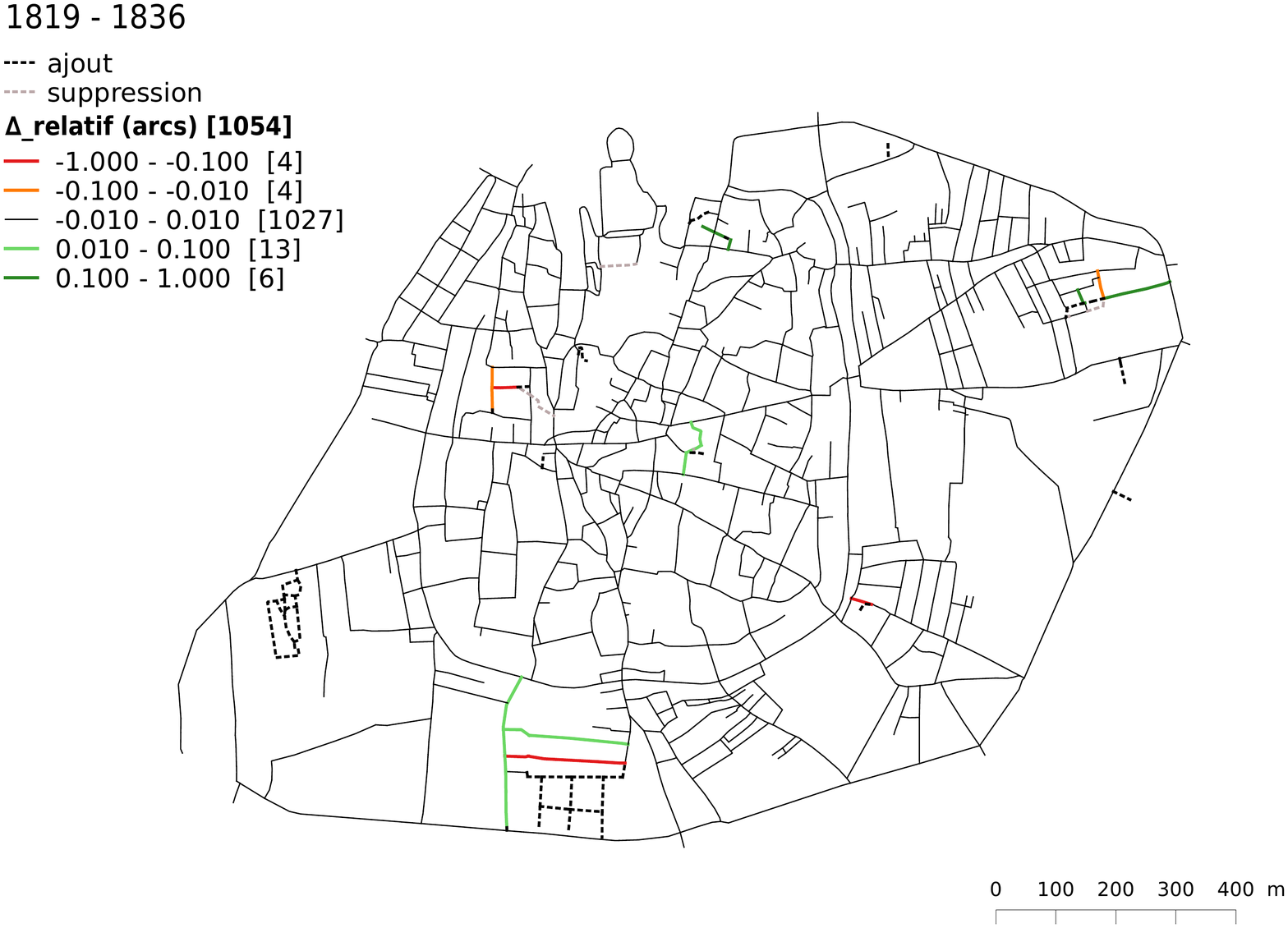}
        \caption{Étude cartographique de $\Delta_{relatif}$ sur la période 1819 - 1836.}
        \label{fig:diff_av_1819}
    \end{figure}

Nous voyons à travers les ajouts et retraits mis en valeur sur ces cartes que le filaire vectorisé dépend grandement des décisions de représentation de la carte créée à une année précise. L'exemple de la structure quadrillée au Sud de l'intra-muros, présente en 1760, non représentée en 1819, puis existant de nouveau en 1836, montre que les choix de représentation du réseau viaire sur ces cartes introduit un biais dans l'analyse du changement (figures \ref{fig:diff_av_1760} et \ref{fig:diff_av_1819}). Seule une confrontation avec le terrain et la consultation des archives de la ville permettront de pallier ce problème en faisant des corrections à la main. 

\begin{figure}[c]
     \centering
        \includegraphics[width=0.8\textwidth]{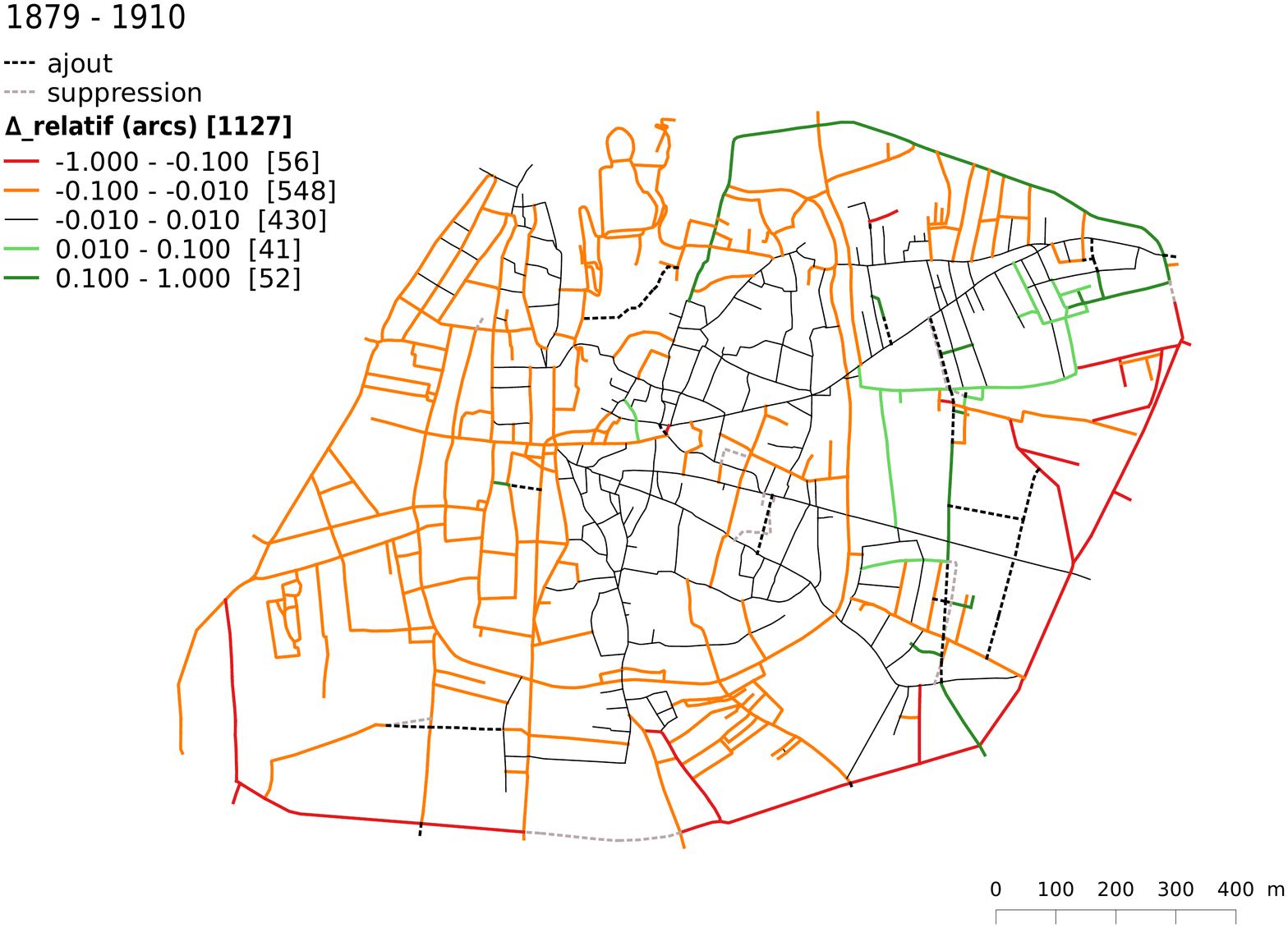}
        \caption{Étude cartographique de $\Delta_{relatif}$ sur la période 1879 - 1910.}
        \label{fig:diff_av_1879}
    \end{figure}

\begin{figure}[c]
    \centering
        \includegraphics[width=0.8\textwidth]{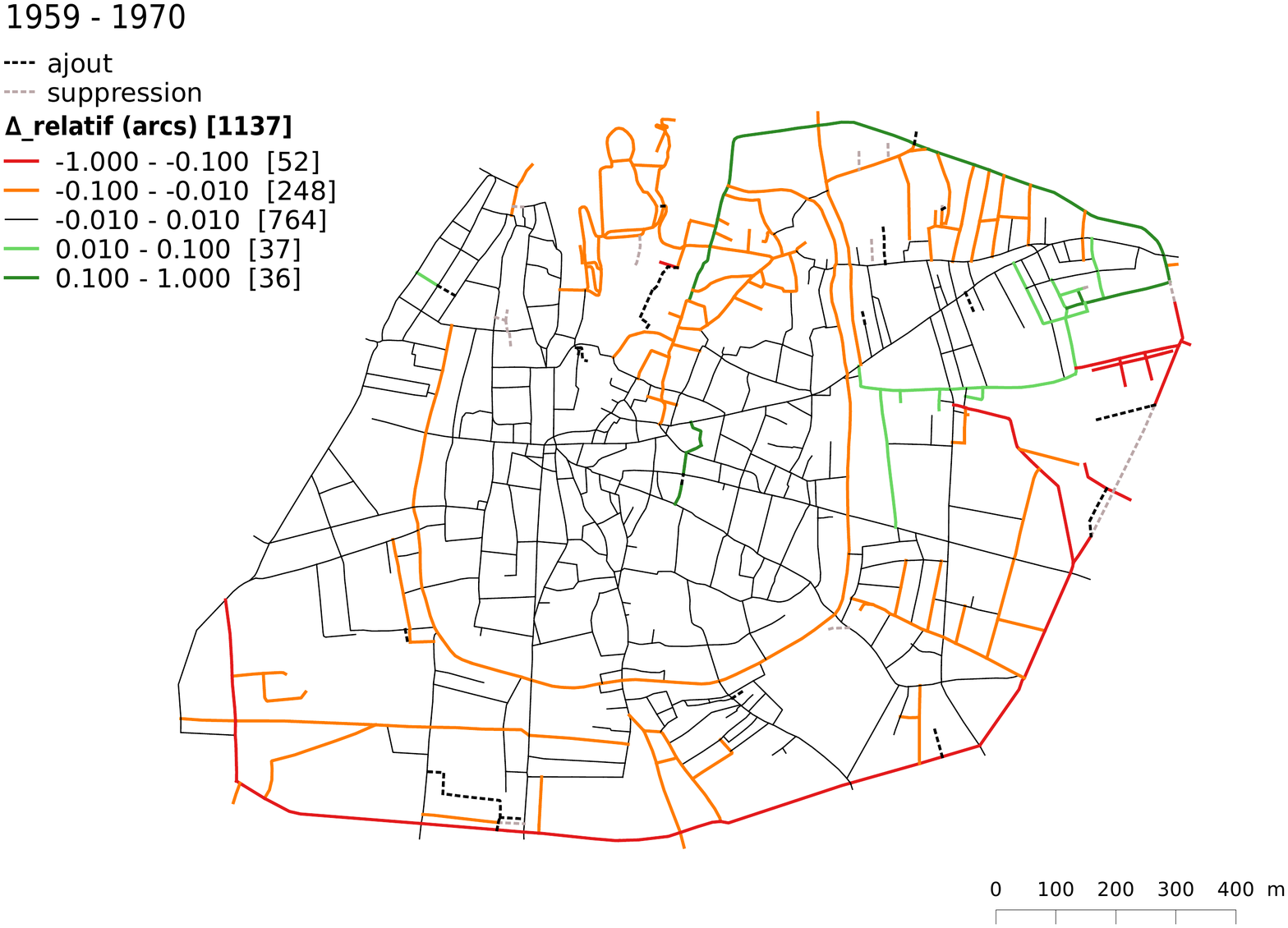}
        \caption{Étude cartographique de $\Delta_{relatif}$ sur la période 1959 - 1970.}
        \label{fig:diff_av_1959}
    \end{figure}

Cependant, le biais introduit par ces structures dont la représentation fluctue (souvent car elles passent du public au privé ou inversement) reste minime car elles ne participent pas à la connexion de structures importantes entre elles. Au contraire la petite jonction retirée au Sud-Ouest de la carte entre 1760 et 1819 coupe la structure de contournement de la ville et a donc un impact négatif sur l'accessibilité de la quasi-totalité du graphe (figure \ref{fig:diff_av_1760}). La partie de cette même structure extérieure, qui suit les remparts de la ville, a également été retirée entre 1879 et 1910 (figure \ref{fig:diff_av_1879}) et, à un endroit différent, entre 1959 et 1970 (figure \ref{fig:diff_av_1959}) avec le même impact. Une modification de la structure extérieure a un impact important car celle-ci permet de réduire efficacement les distances topologiques d'un bout à l'autre du graphe.

\begin{figure}[c]
   \centering
        \includegraphics[width=0.8\textwidth]{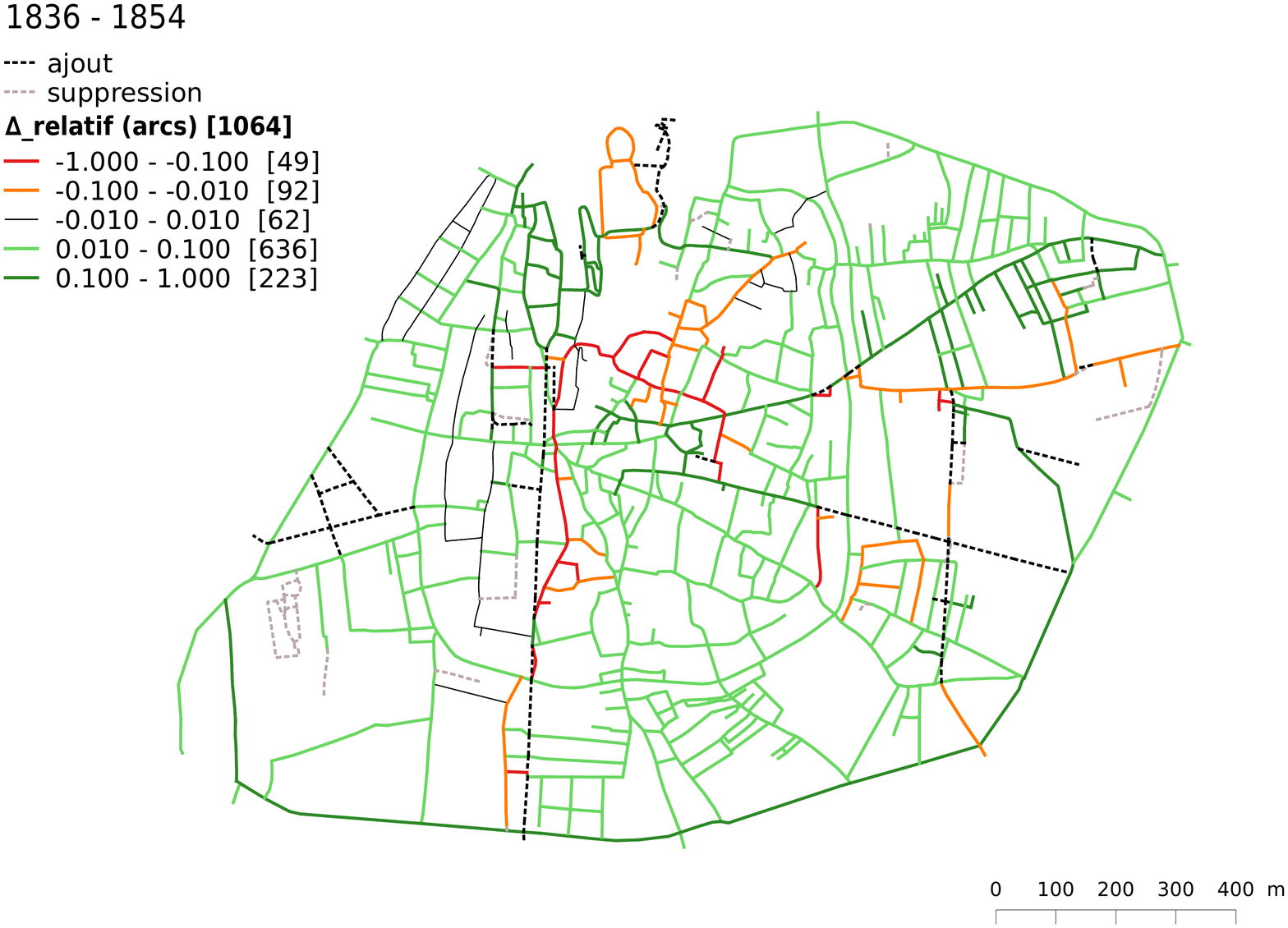}
        \caption{Étude cartographique de $\Delta_{relatif}$ sur la période 1836 - 1854.}
        \label{fig:diff_av_1836}
    \end{figure}

\begin{figure}[c]
    \centering
        \includegraphics[width=0.8\textwidth]{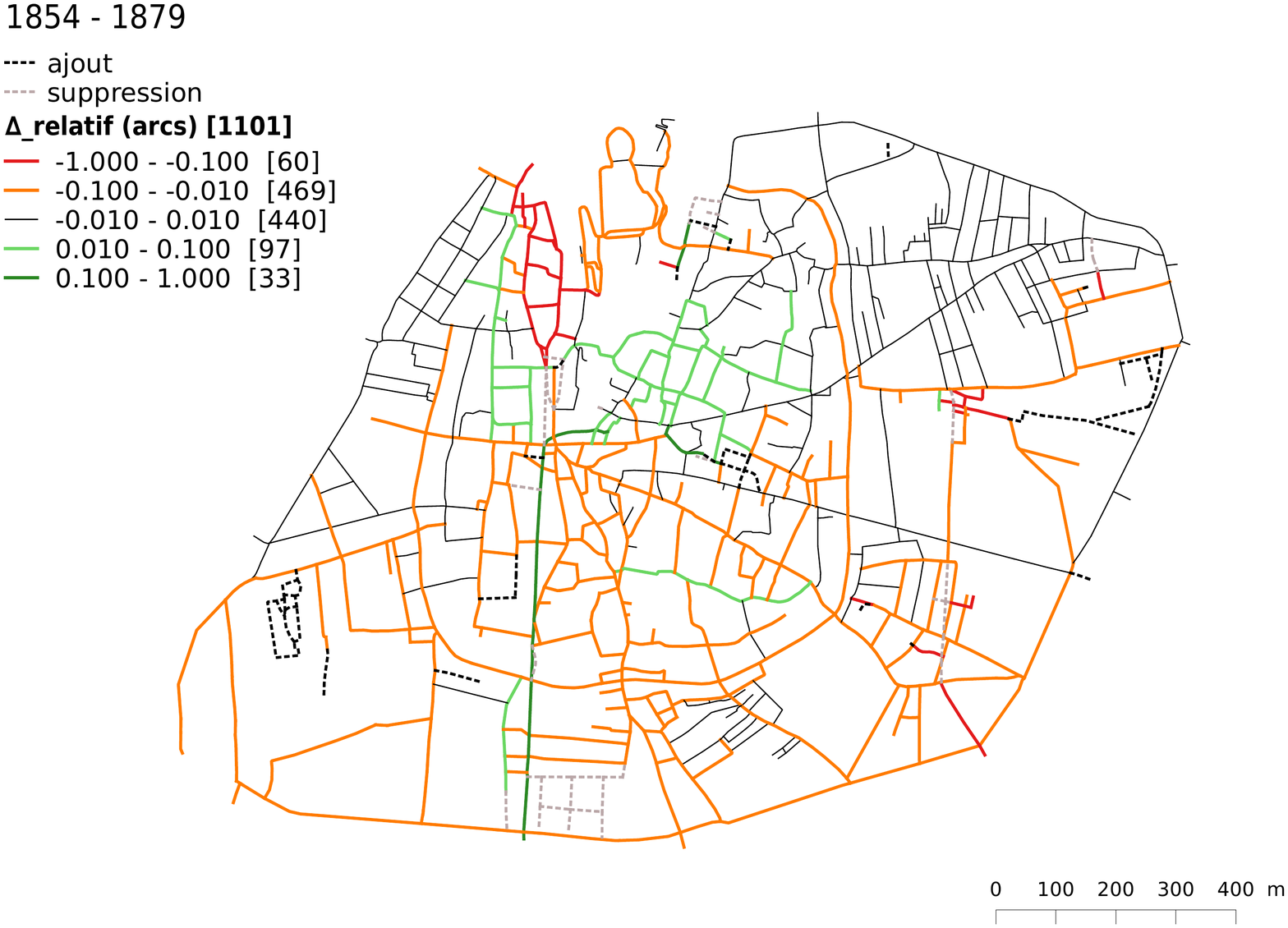}
        \caption{Étude cartographique de $\Delta_{relatif}$ sur la période 1854 - 1879.}
        \label{fig:diff_av_1854}
    \end{figure}

\begin{figure}[c]
    \centering
        \includegraphics[width=0.8\textwidth]{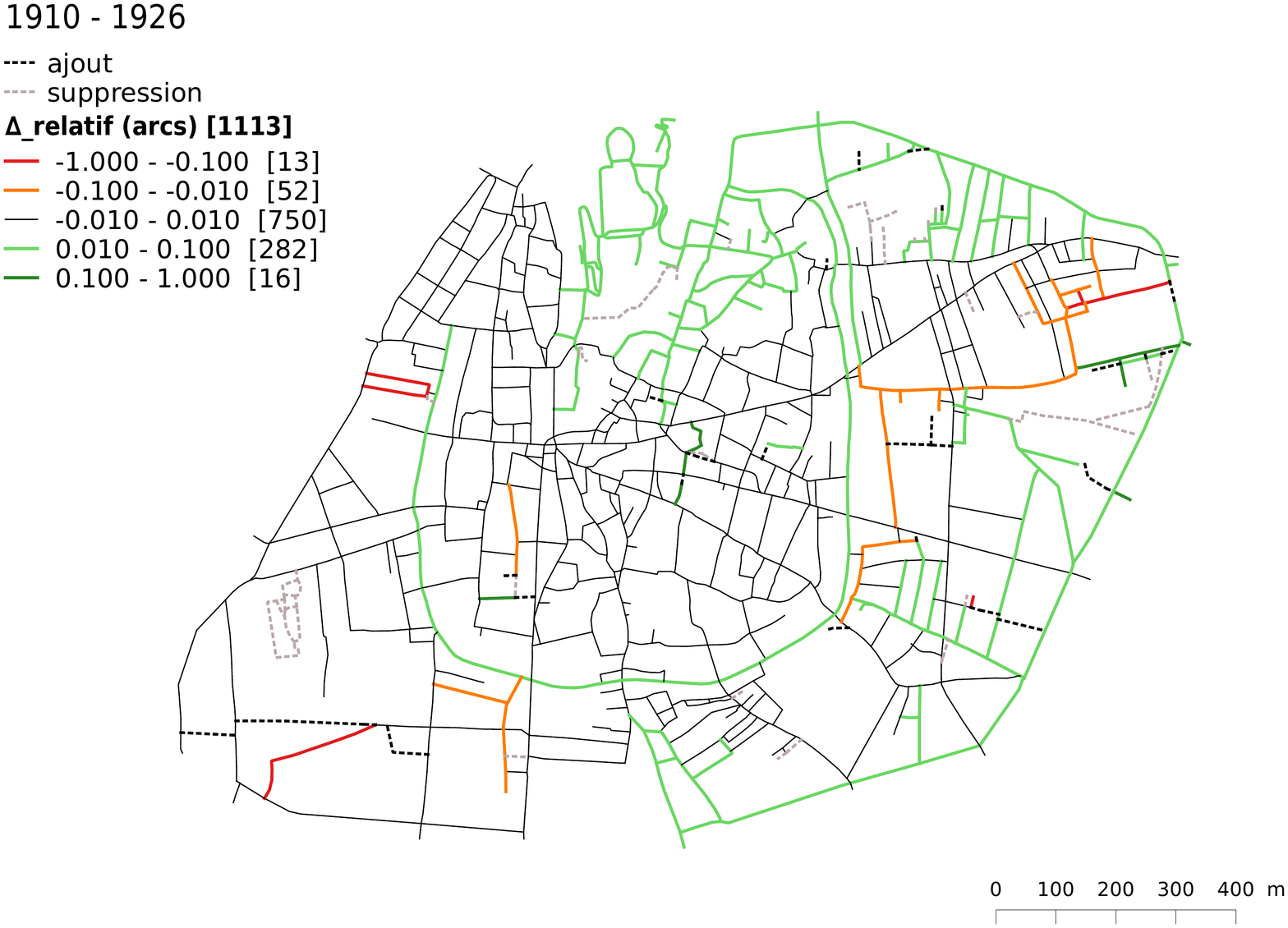}
        \caption{Étude cartographique de $\Delta_{relatif}$ sur la période 1910 - 1926.}
        \label{fig:diff_av_1910}
    \end{figure}

\begin{figure}[c]
     \centering
        \includegraphics[width=0.8\textwidth]{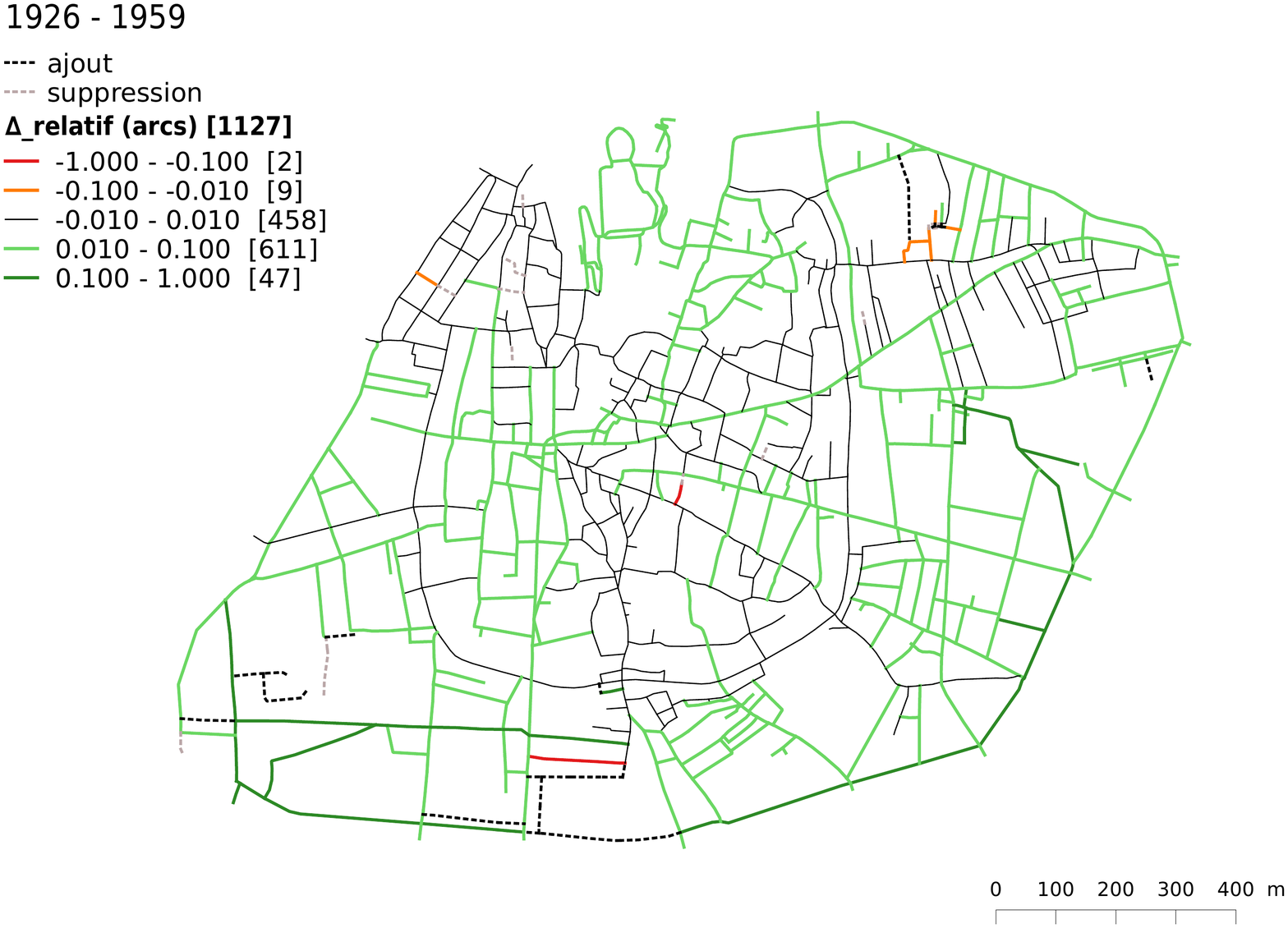}
        \caption{Étude cartographique de $\Delta_{relatif}$ sur la période 1926 - 1959.}
        \label{fig:diff_av_1926}
    \end{figure}

L'impact des modifications à l'intérieur des murs de la ville varie selon les périodes. Entre 1819 et 1836, il est quasi inexistant (figure \ref{fig:diff_av_1819}). Les structures retirées ou ajoutées n'ont pas d'impact sur les distances entre objets dans le graphe. De 1910 à 1959 (figures \ref{fig:diff_av_1910} et \ref{fig:diff_av_1926}) l'influence des modifications est globalement positive sur les distances entre voies : l'ajout de nouveaux arcs augmente la proximité entre les objets. Entre 1836 et 1854, l'avenue de la République est créée. Son impact est fortement positif, seules quelques voies au centre du graphe voient leur rayon topologique diminuer : elles étaient au cœur des chemins topologiques les plus simples avant l'arrivée de l'axe rectiligne Nord-Sud (figure \ref{fig:diff_av_1836}). La suppression d'éléments autour de cette nouvelle voie entre 1854 et 1879 a un impact négatif sur l'accessibilité des objets dans le graphe, même si celle de ces éléments \textit{centraux} est renforcée (\ref{fig:diff_av_1854}). En 2014, le carrefour central du graphe (à l'Est de la place de l'Horloge) est modifié, liant le Cours Jean Jaurès à la rue Carreterie, ce qui crée une voie longue qui dessert en un minimum de \textit{tournants} l'ensemble du réseau (un \textit{tournant} dans notre cas équivaut à un changement de voie).

Pour chacune des périodes étudiées nous notons dans le tableau \ref{tab:avignon_delta1} le détail des modifications des arcs du réseau et dans le tableau \ref{tab:avignon_delta2} la moyenne et l'écart type des $\Delta_{relatif}$ ainsi que le maximum de $\Delta_{relatif}$ observé en valeur absolue. En plus de cette description statistique, nous sommons l'ensemble des $\Delta_{relatif}$ en valeur absolue. Comme nous nous intéressons à un même espace pour toutes les périodes d'observation, cela nous permettra de quantifier l'impact relatif des transformations de chaque période.

\begin{table}
\begin{center}
{ \small
\begin{tabular}{|c|c|r|r|r|r|}
\hline

période & durée & $N_{arcs(ajoutés)}$ & $L_{ajoutée}$ & $N_{arcs(retirés)}$ & $L_{retirée}$ \\ \hline

1760 - 1819 & 59 ans & 45 & 1 413 m & 27 & 1 081 m  \\ \hline
1819 - 1836 & 17 ans & 48 & 1 448 m & 6 & 200 m  \\ \hline
1836 - 1854 & 18 ans & 74 & 3 195 m & 38 & 1 480 m \\ \hline
1854 - 1879 & 25 ans & 43 & 1 809 m & 37 & 1 594 m \\ \hline
1879 - 1910 & 31 ans & 34 & 1 664 m & 17 & 956 m \\ \hline
1910 - 1926 & 16 ans & 29 & 1 195 m & 48 & 1 850 m \\ \hline
1926 - 1959 & 33 ans & 26 & 1 285 m & 15 & 429 m \\ \hline
1959 - 1970 & 11 ans & 24 & 952 m & 16 & 610 m \\ \hline
1970 - 2014 & 44 ans & 54 & 2 555 m & 37 & 883 m \\ \hline

\end{tabular}
}
\end{center}
\caption{Détail du nombre de modifications du réseau pour chaque période d'étude du centre d'Avignon.}
\label{tab:avignon_delta1}
\end{table}

\begin{table}
\begin{center}
{ \small
\begin{tabular}{|c|r|r|r|r|r|}
\hline

période & $N_{arcs(modifiés)}$ & $\overline{\Delta_{relatif}}$ & $\sigma(\Delta_{relatif})$  & $max \vert \Delta_{relatif}\vert$ & $\sum \vert \Delta_{relatif} \vert$ \\ \hline

1760 - 1819 & 72 & -0.0502 & 0.0583 & 0.3644 & 51.87  \\ \hline
1819 - 1836 & 54 & 0.0007 & 0.0302 & 0.4569 & 5.18  \\ \hline
1836 - 1854 & 112 & 0.0552 & 0.1055 & 0.4999 & 80.96  \\ \hline
1854 - 1879 & 80 & -0.0145 & 0.0672 & 0.3473 & 35.99  \\ \hline
1879 - 1910 & 51 & -0.0130 & 0.0886 & 0.9611 & 49.05  \\ \hline
1910 - 1926 & 77 & 0.0105 & 0.0430 & 0.4514 & 21.76  \\ \hline
1926 - 1959 & 41 & 0.0250 & 0.0415 & 0.3301 & 30.19  \\ \hline
1959 - 1970 & 40 & -0.0101 & 0.0683 & 0.5273 & 33.73  \\ \hline
1970 - 2014 & 91 & 0.0509 & 0.0673 & 0.3980 & 67.59  \\ \hline

\end{tabular}
}
\end{center}
\caption{Détail statistique des variations relatives $\Delta_{relatif}$ pour chaque période d'étude du centre d'Avignon.}
\label{tab:avignon_delta2}
\end{table}

Nous pouvons lire dans le tableau \ref{tab:avignon_delta2} l'hétérogénéité des variations selon les périodes allant d'un $\sum \vert \Delta_{relatif} \vert$ de 5.18 entre 1819 et 1836 à une valeur de 80.96 entre 1836 et 1854. Les deux périodes concernent pourtant un nombre similaire d'années, mais les modifications du réseau viaire effectuées n'ont pas la même ampleur. La carte de 1854 voit notamment apparaître un axe majeur principal issu de l'association du cours Jean Jaurès et de la rue de la République.  $\overline{\Delta_{relatif}}$ et $\sigma(\Delta_{relatif})$ traduisent également cet important changement : leurs valeurs sont les plus élevées sur cette période temporelle. L'autre moyenne dont la variation est importante concerne la période 1760 - 1819 où une modification de la voie de contournement de l'intra-muros fait chuter les accessibilités. 

En sommant pour chaque période les $\Delta_{relatif}$ positifs d'une part, et négatifs d'autre part, nous pouvons observer l'asymétrie d'évolution de l'accessibilité dans le graphe selon les constructions sur une période donnée. Ces valeurs viennent appuyer la représentation cartographique détaillée ci-dessous.

    \clearpage 
    
    \begin{figure}[c]
    \centering
        \includegraphics[width=0.6\textwidth]{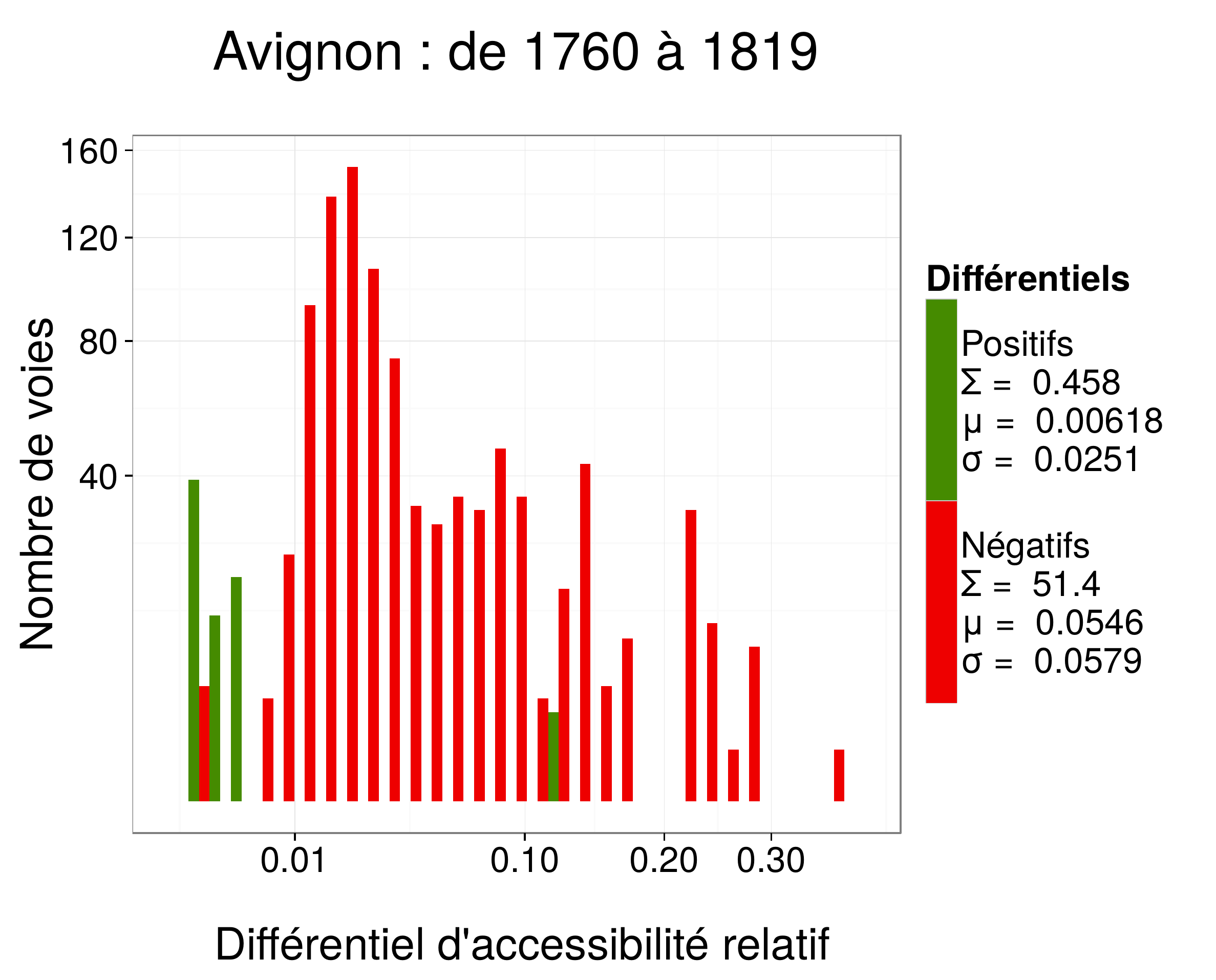}
        \caption{Étude statistique de $\Delta_{relatif}$ sur la période 1760 - 1819. \\ $\Sigma$ : somme ; $\mu$ : moyenne ; $\sigma$ : écart-type}
        \label{fig:diff_av_1760_stat}
    \end{figure}
   
   \clearpage  
    
    \begin{figure}[c]
    \centering
        \includegraphics[width=0.6\textwidth]{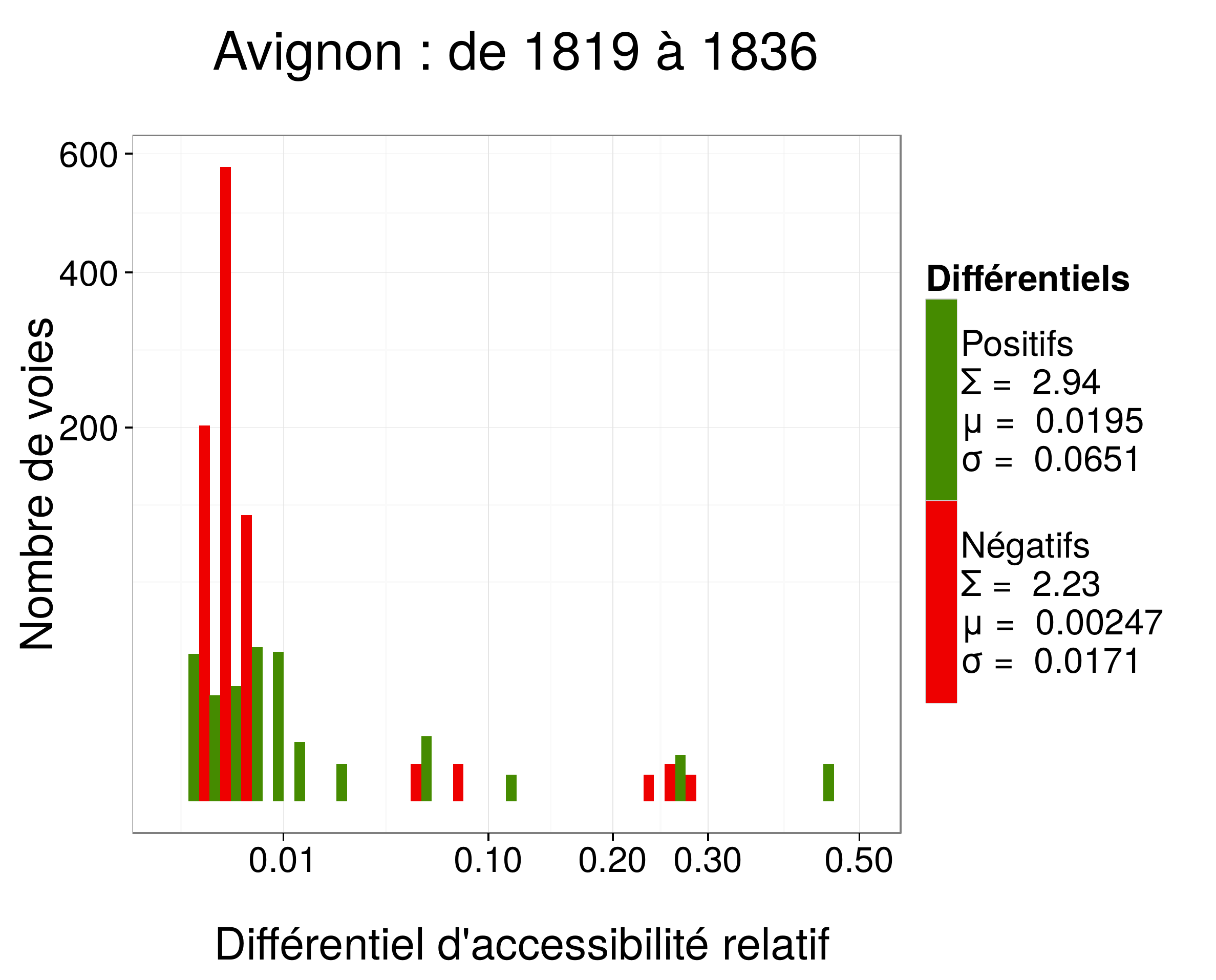}
        \caption{Étude statistique de $\Delta_{relatif}$ sur la période 1819 - 1836. \\ $\Sigma$ : somme ; $\mu$ : moyenne ; $\sigma$ : écart-type}
        \label{fig:diff_av_1819_stat}
    \end{figure}
    
   \clearpage 
    
    \begin{figure}[c]
    \centering
        \includegraphics[width=0.6\textwidth]{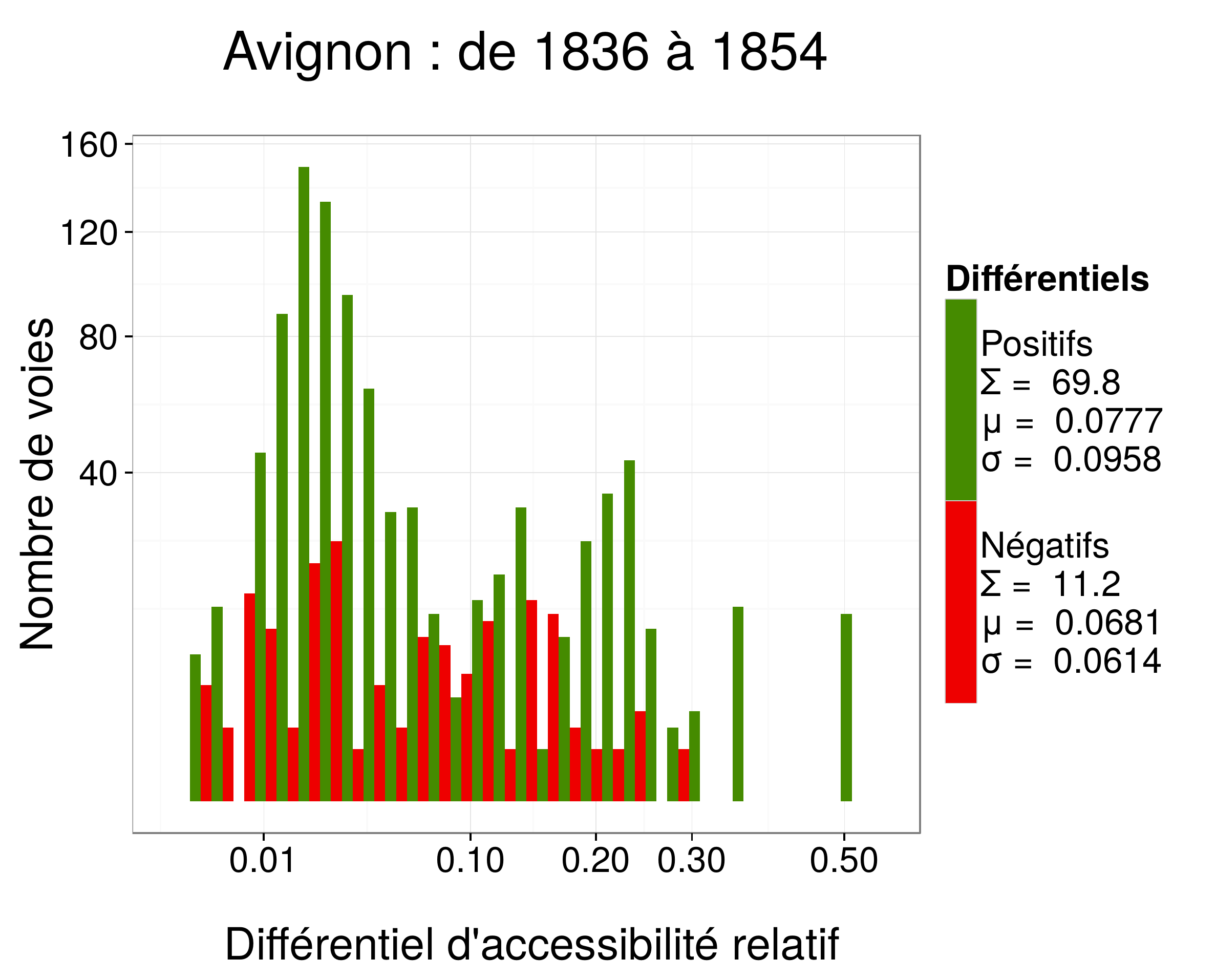}
        \caption{Étude statistique de $\Delta_{relatif}$ sur la période 1836 - 1854. \\ $\Sigma$ : somme ; $\mu$ : moyenne ; $\sigma$ : écart-type}
        \label{fig:diff_av_1836_stat}
    \end{figure}
        
    \clearpage  
    
    \begin{figure}[c]
    \centering
        \includegraphics[width=0.6\textwidth]{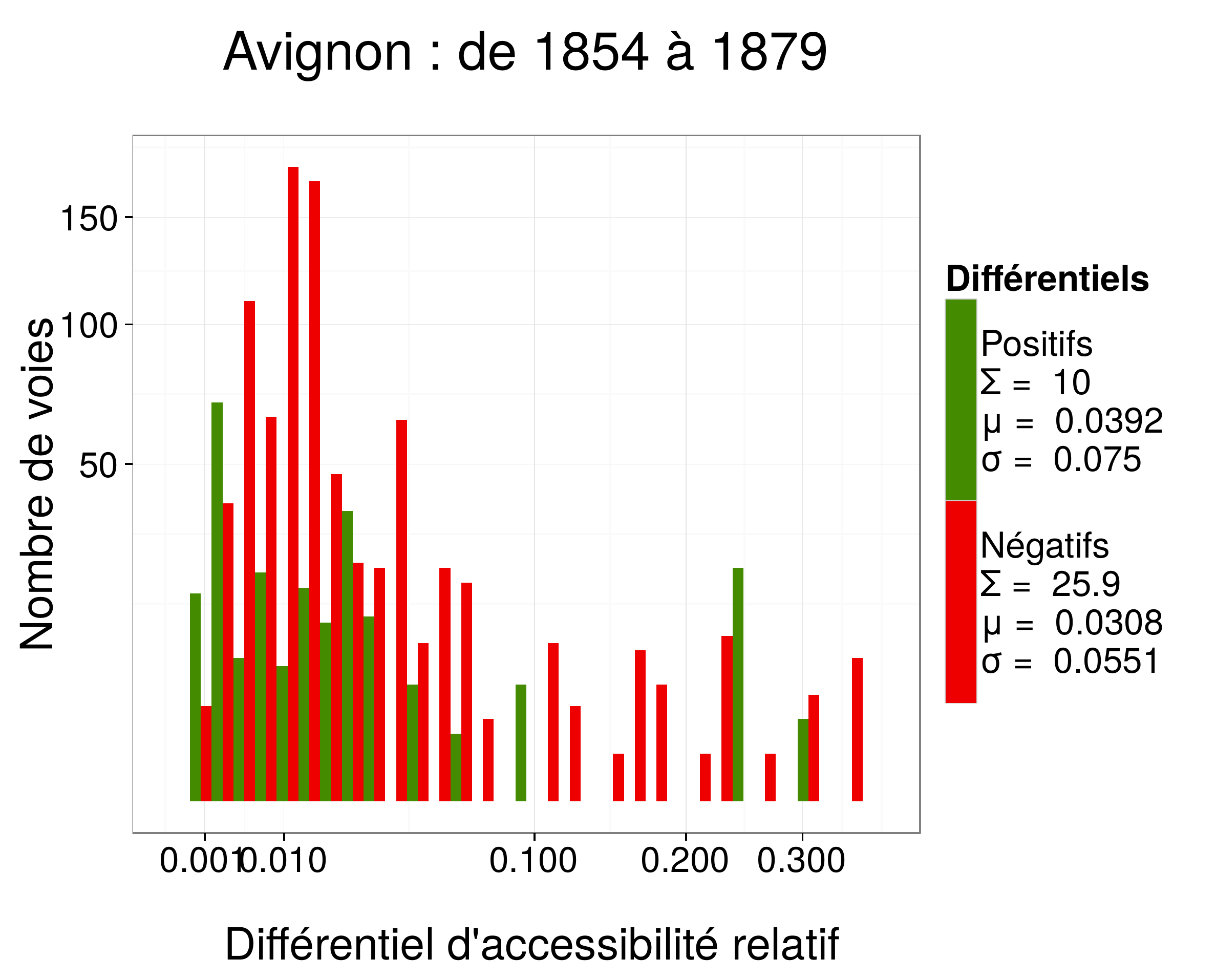}
        \caption{Étude statistique de $\Delta_{relatif}$ sur la période 1854 - 1879. \\ $\Sigma$ : somme ; $\mu$ : moyenne ; $\sigma$ : écart-type}
        \label{fig:diff_av_1854_stat}
    \end{figure}
    
    \clearpage 
    
    \begin{figure}[c]
    \centering
        \includegraphics[width=0.6\textwidth]{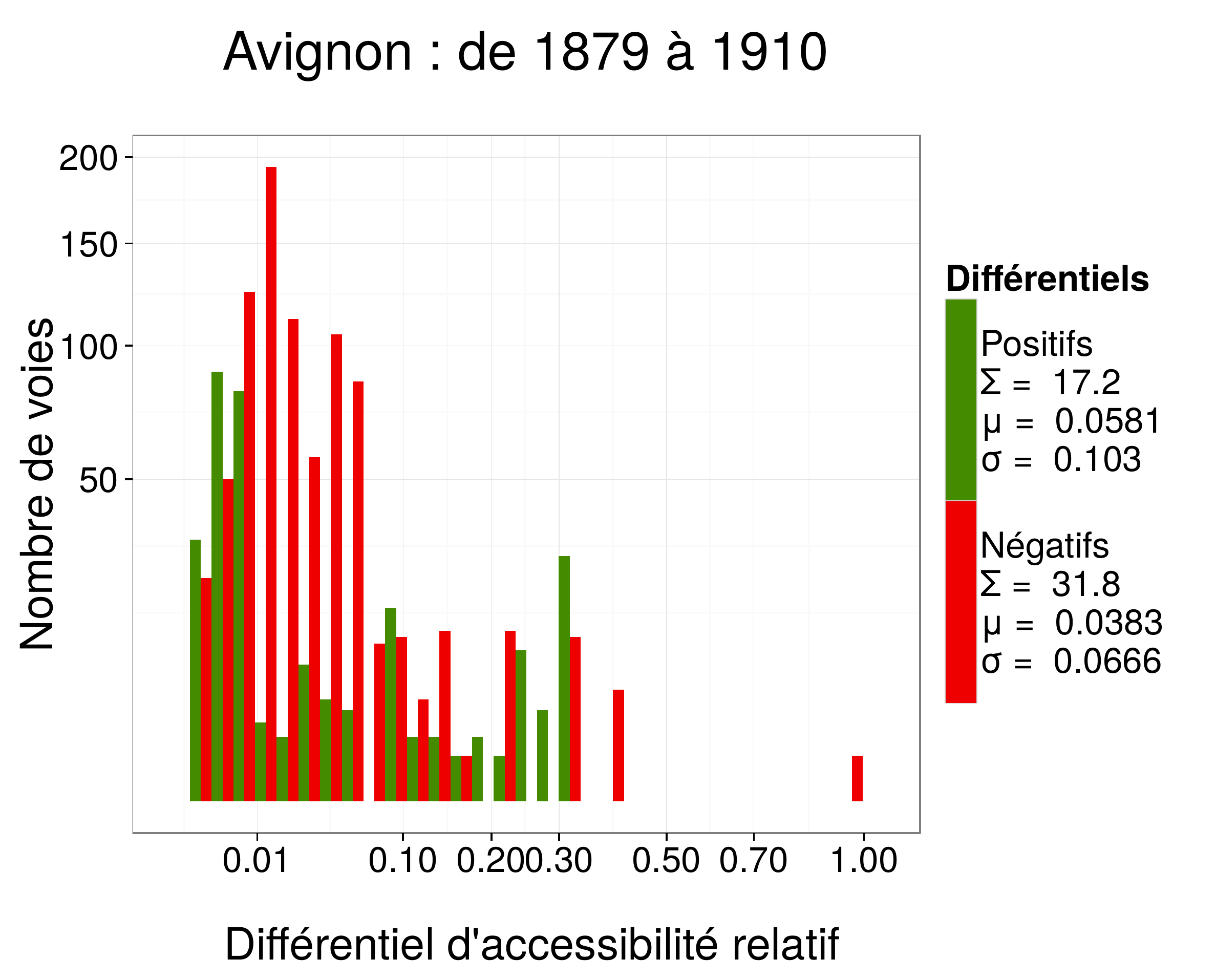}
        \caption{Étude statistique de $\Delta_{relatif}$ sur la période 1879 - 1910. \\ $\Sigma$ : somme ; $\mu$ : moyenne ; $\sigma$ : écart-type}
        \label{fig:diff_av_1879_stat}
    \end{figure}
    
    \clearpage    
    
    \begin{figure}[c]
    \centering
        \includegraphics[width=0.6\textwidth]{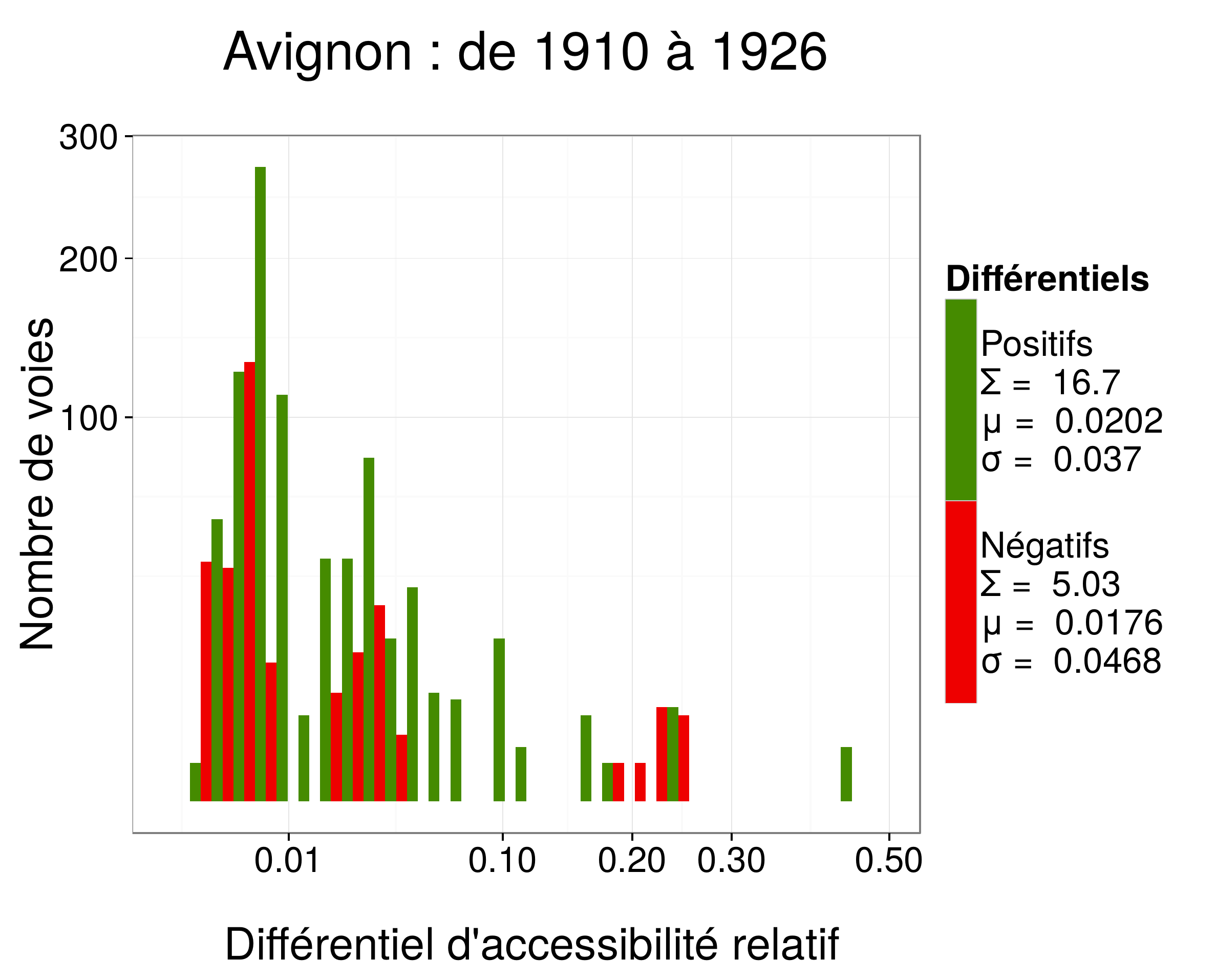}
        \caption{Étude statistique de $\Delta_{relatif}$ sur la période 1910 - 1926. \\ $\Sigma$ : somme ; $\mu$ : moyenne ; $\sigma$ : écart-type}
        \label{fig:diff_av_1910_stat}
    \end{figure}
    
    \clearpage 
    
    \begin{figure}[c]
    \centering
        \includegraphics[width=0.6\textwidth]{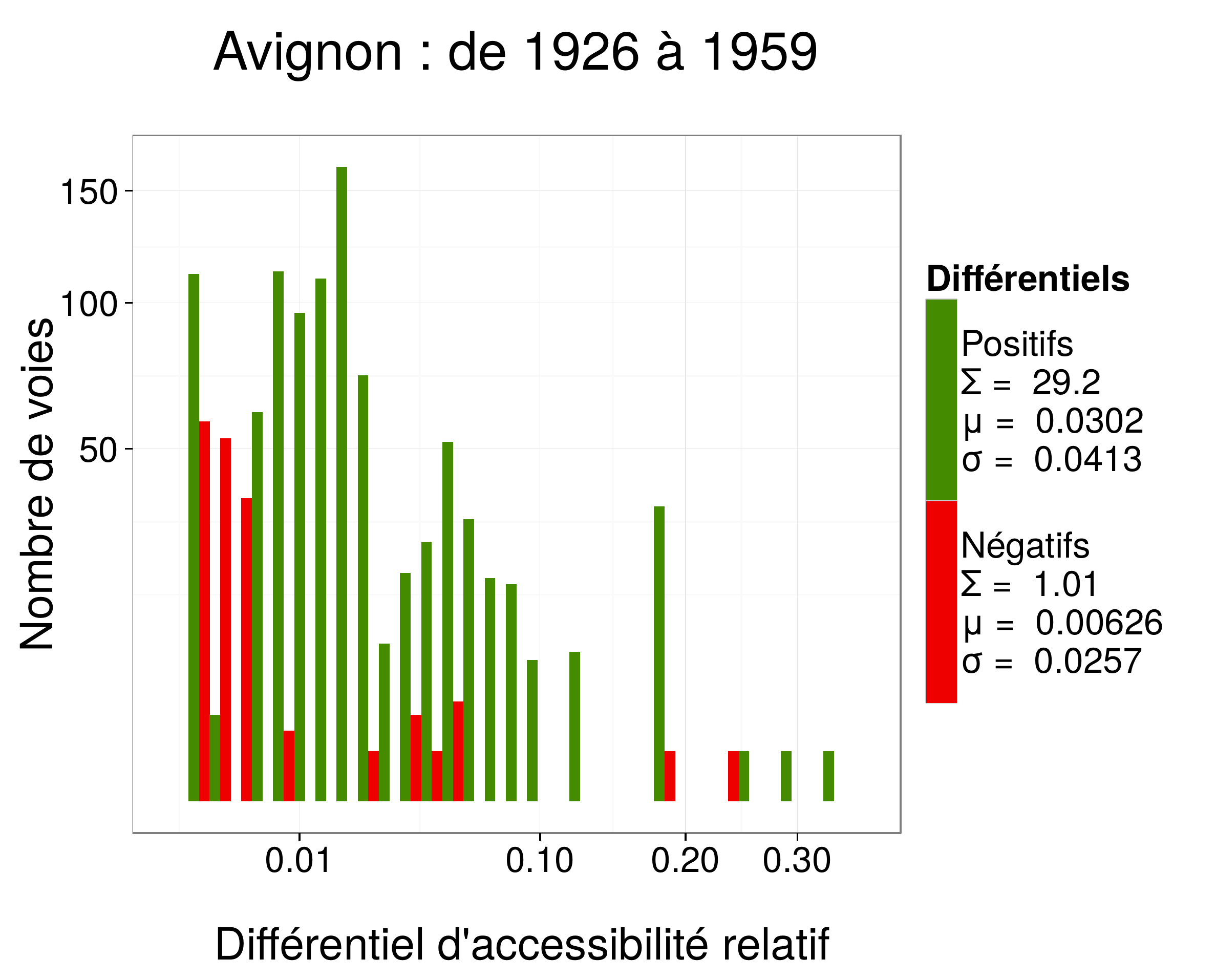}
        \caption{Étude statistique de $\Delta_{relatif}$ sur la période 1926 - 1959. \\ $\Sigma$ : somme ; $\mu$ : moyenne ; $\sigma$ : écart-type}
        \label{fig:diff_av_1926_stat}
    \end{figure}
   
   \clearpage    
    
    \begin{figure}[c]
    \centering
        \includegraphics[width=0.6\textwidth]{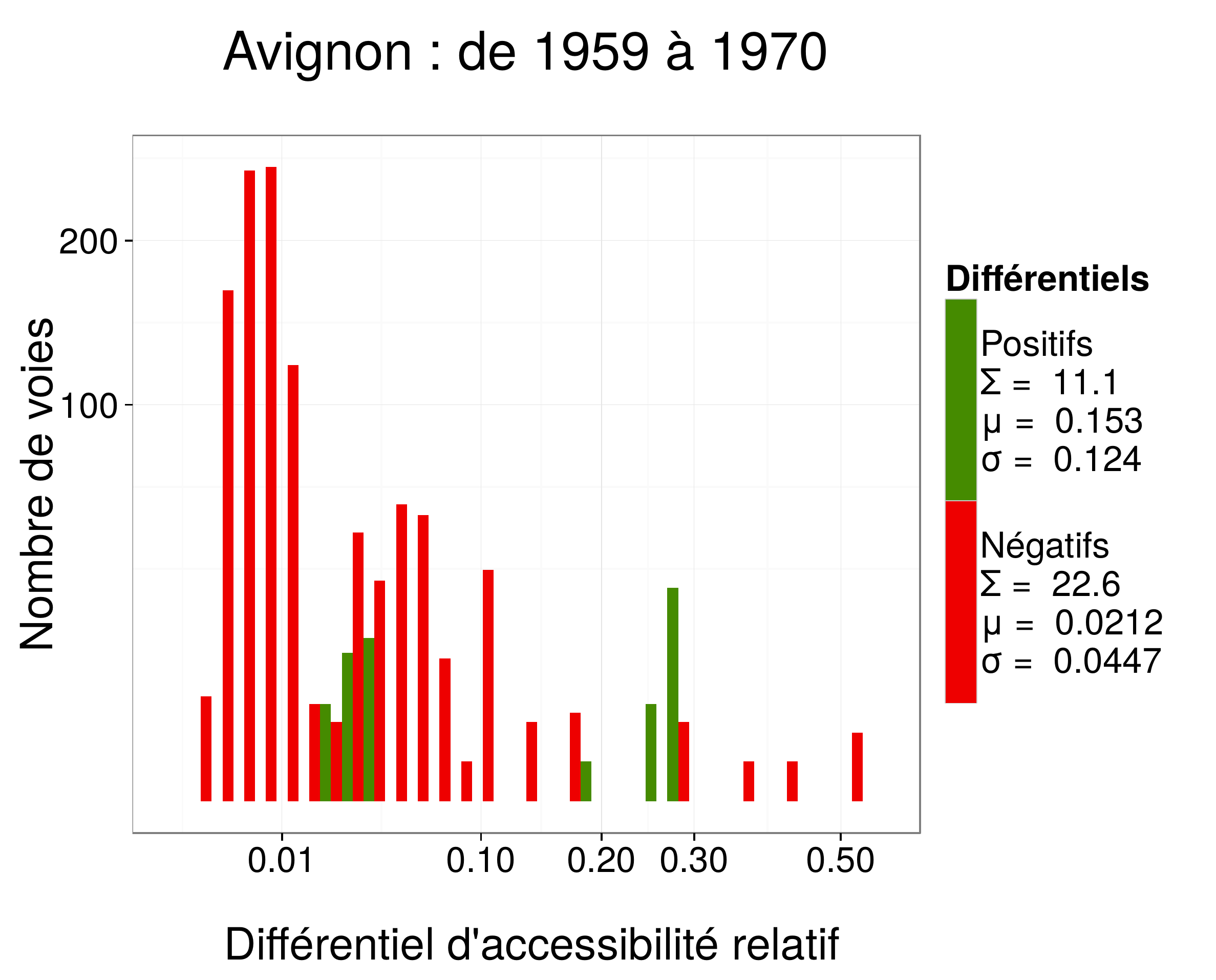}
        \caption{Étude statistique de $\Delta_{relatif}$ sur la période 1959 - 1970. \\ $\Sigma$ : somme ; $\mu$ : moyenne ; $\sigma$ : écart-type}
        \label{fig:diff_av_1959_stat}
    \end{figure}
    
    \clearpage 
     \begin{figure}[c]
     \centering
        \includegraphics[width=0.8\textwidth]{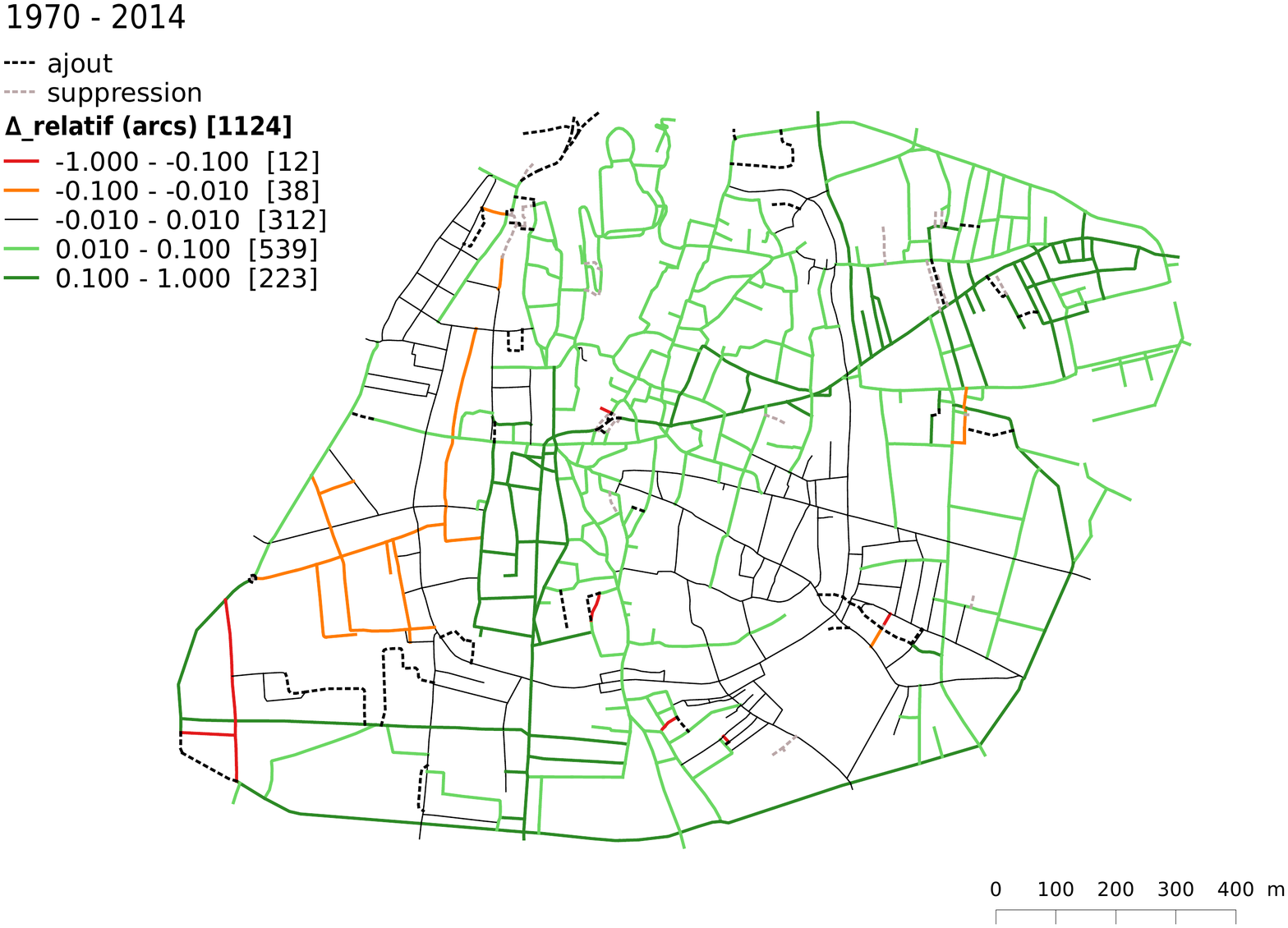}
        \caption{Étude cartographique de $\Delta_{relatif}$ sur la période 1970 - 2014.}
        \label{fig:diff_av_1970}
    \end{figure}
    
\FloatBarrier
\subsection{Rotterdam}

\begin{figure}[h]
    \centering
        \includegraphics[width=0.8\textwidth]{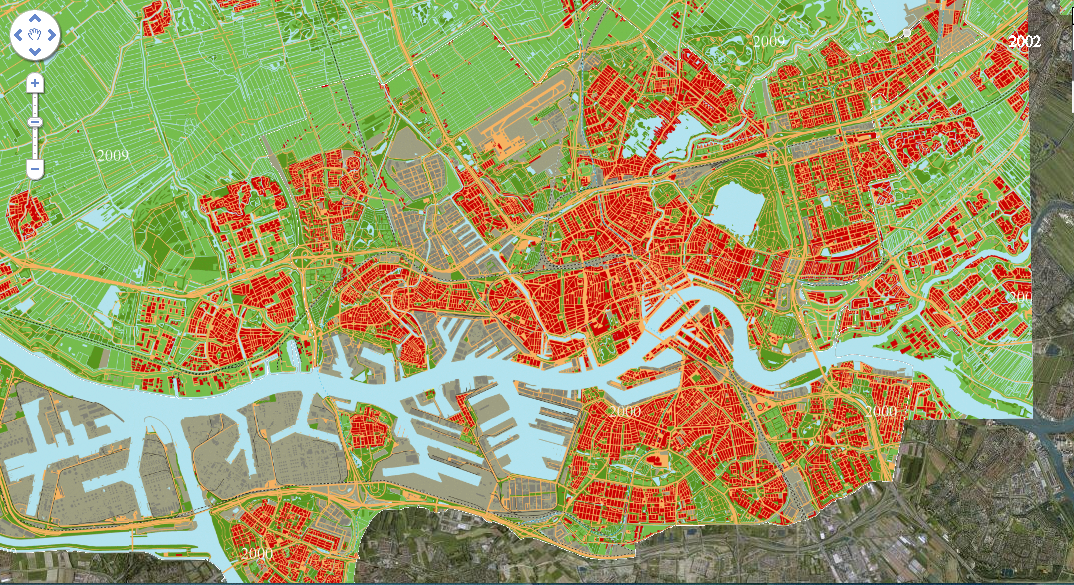}
    
    \caption{Carte du Nord de Rotterdam et Schiedam en 2000. \\ source : \citep{mappinghistory}}
    \label{fig:rot_plan2000}
\end{figure}

La seconde ville sur laquelle porte notre étude diachronique est Rotterdam (figure \ref{fig:rot_plan2000}). Grâce au travail des personnes de Mapping History (précédemment Mapsplusmotion), entreprise qui travaille pour les musées (ou autres collectivités) en numérisant des cartes anciennes \citep{mappinghistory}, nous avons pu recueillir des données vectorisées de plans à plusieurs dates (figure \ref{fig:rot_plan2000}). À partir de leur travail, réalisé sous un logiciel de dessin, nous avons commandé le géo-référencement et le nettoyage du filaire pour pouvoir l'exploiter avec un SIG, puis avec notre programme.

\begin{figure}[h]
    \centering
        \includegraphics[width=0.8\textwidth]{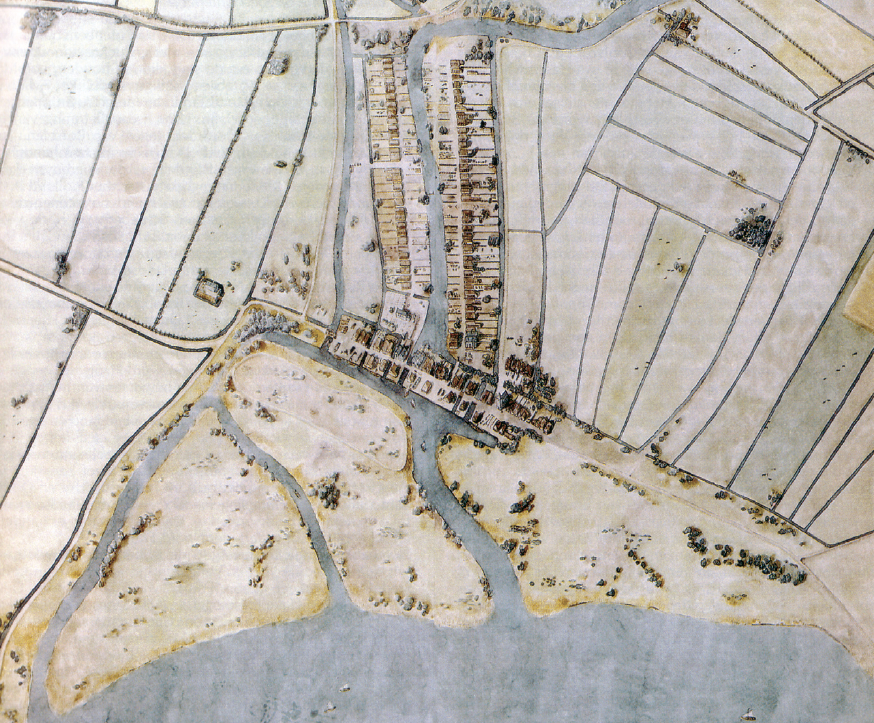}
    
    \caption{Carte du Nord de Rotterdam en 1340. \\ source : \citep{mappinghistory}}
    \label{fig:rot_plan1340}
\end{figure}

\begin{figure}[h]
    \centering
        \includegraphics[width=0.8\textwidth]{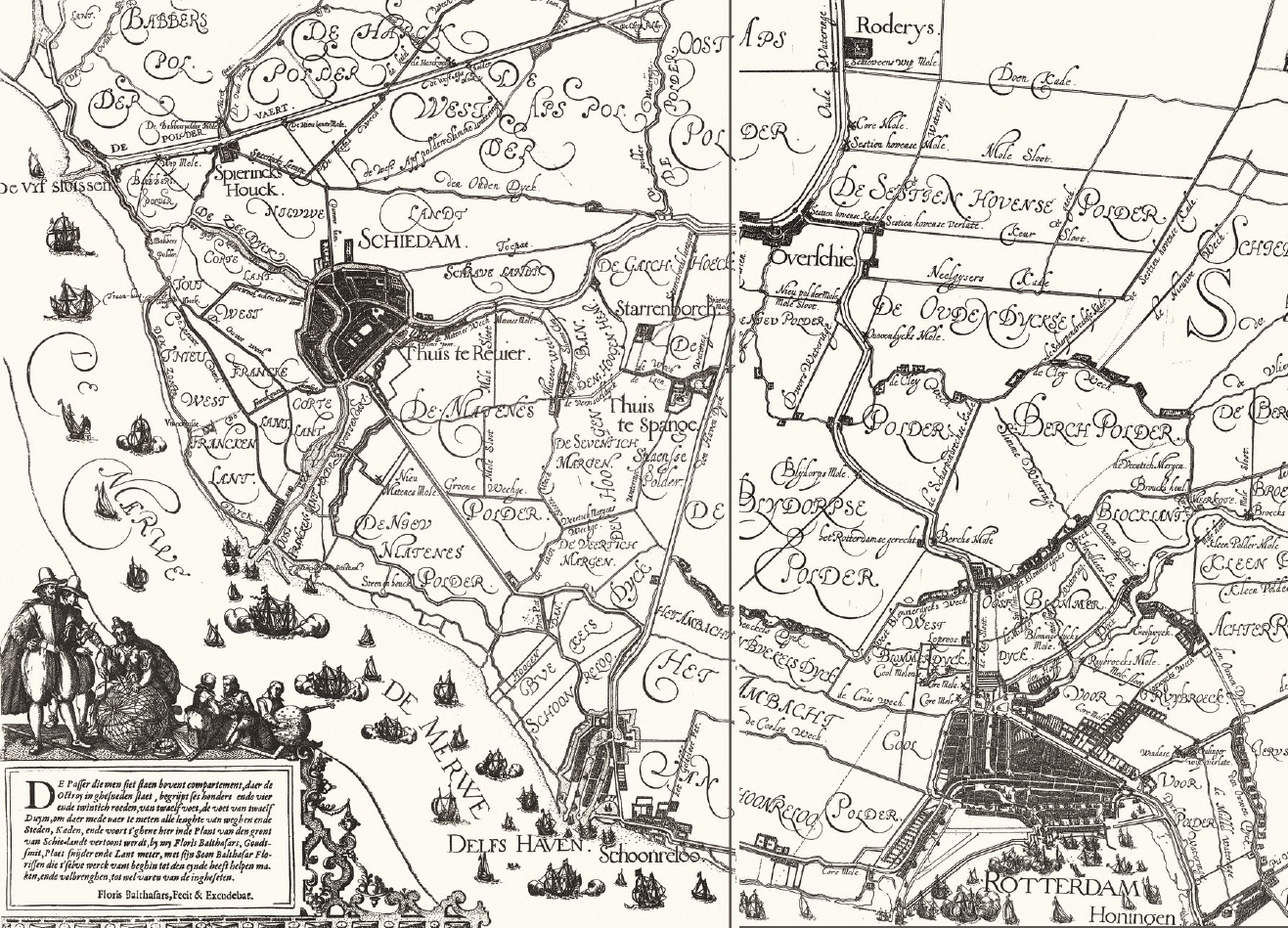}
    
    \caption{Carte du Nord de Rotterdam et alentours en 1600. \\ source : \citep{mappinghistory}}
    \label{fig:rot_plan1600}
\end{figure}

Nous disposons ainsi pour Rotterdam d'une base de données panchroniques de neuf dates, numérisées sur le Nord de la ville, dont les caractéristiques des graphes sont décrites dans le tableau \ref{tab:rott_entier_pres}. Nous lisons une densification forte du réseau entre 1374 et 1890 (le nombre d'arcs moyen par voie passe de 9.9 à 5.1). Celle-ci se ressent aussi sur la cartographie d'époque, où les représentations s'étoffent (figures \ref{fig:rot_plan1340}, \ref{fig:rot_plan1600}).

Les voies, dans un premier temps très longues et comprenant un grand nombre d'arcs eux même de longueur importante, partaient des centres-villes de Rotterdam et Schiedam vers la campagne. Au fil des années, les centres-villes se sont densifiés et la campagne s'est urbanisée pour donner un réseau beaucoup plus maillé, aux voies plus courtes (406 mètres en moyenne en 1955) mais également plus nombreuses. Le nombre d'arcs du graphe, en constante augmentation, passe de 1329 en 1374 à 8393 en 1955 : plus courts, ils assurent une desserte plus fine de l'espace (densification du réseau). Le territoire a subi deux événements marquants : une inondation majeure vers l'an 1740 et un bombardement du centre ville au début de la seconde guerre mondiale dont nous reparlerons plus loin.

\begin{table}
\begin{center}
{ \small
\begin{tabular}{|c|r|r|r|r|r|r|r|}
\hline
Année & $L_{tot}$ & $N_{sommets}$ & $N_{arcs}$ & $L_{moy}(arc)$ & $N_{voies}$ & $L_{moy}(voie)$ & $\overline{N_{arcs(voie)}}$ \\ \hline 		

1374 & 176 758 m & 306 & 1 329 & 133 m & 134 & 1 319 m & 9.9 \\ \hline
1570 & 234 701 m & 543 & 1 947 & 121 m & 228 & 1 029 m & 8.5 \\ \hline
1600 & 244 632 m & 651 & 2 156 & 113 m & 271 & 903 m & 8.0 \\ \hline
1625 & 257 674 m & 764 & 2 404 & 107 m & 314 & 821 m & 7.7 \\ \hline
1890 & 349 734 m & 1399 & 3 527 & 99 m & 578 & 605 m & 6.1 \\ \hline
1907 & 420 375 m & 2056 & 4 480 & 94 m & 840 & 500 m & 5.3 \\ \hline
1920 & 475 095 m & 2559 & 5 172 & 92 m & 1026 & 463 m & 5.0 \\ \hline
1940 & 640 179 m & 3797 & 7 193 & 89 m & 1506 & 425 m & 4.8 \\ \hline
1955 & 758 366 m & 4662 & 8 393 & 90 m & 1869 & 406 m & 4.5 \\ \hline

\end{tabular}
}
\end{center}
\caption{Détail des caractéristiques topologiques et métriques des graphes panchroniques du nord de Rotterdam.}
\label{tab:rott_entier_pres}
\end{table}

Nous procédons de la même manière que pour Avignon sur le réseau viaire du Nord de Rotterdam. Nous reportons en annexe \ref{ann:sec_cartedia_rotterdamsch} la cartographie de l'indicateur de closeness sur les graphes. Nous pouvons apprécier la densification du graphe au cours du temps et l'inclusion progressive de Schiedam (à gauche du graphe) au sein de Rotterdam (point de densification à droite). Pour chaque période, nous cartographions le  $\Delta_{relatif}$ calculé et représentons ses variations positives (gain d'accessibilité) et négatives (perte d'accessibilité) à l'aide d'un histogramme. La taille du réseau, dont le linéaire est beaucoup plus important que celui d'Avignon, entraîne des variations totales beaucoup plus importantes (tableau \ref{tab:rott_delta2}).

\begin{figure}[h]
    \centering
        \includegraphics[width=0.8\textwidth]{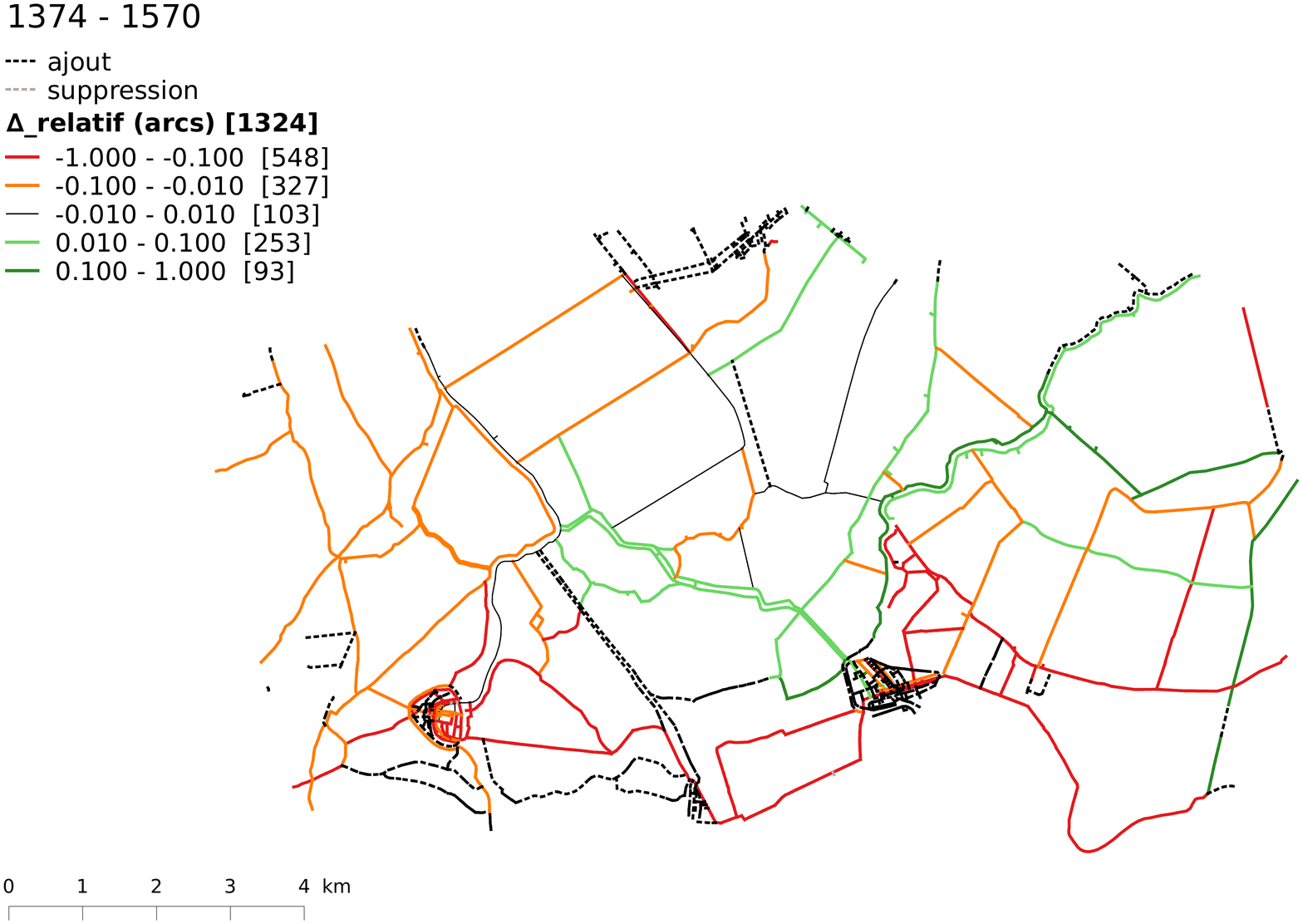}
        \caption{Étude cartographique de $\Delta_{relatif}$ sur la période 1374 - 1570.}
        \label{fig:diff_re2_1374}
    \end{figure}

La quantification des différences sur les huit périodes étudiées montre un résultat contrasté, où, dès la période 1374 - 1570, les axes joignant Rotterdam gagnent en accessibilité au détriment de ceux vers Schiedam qui en perdent (figure \ref{fig:diff_re2_1374}). Ainsi, les première cartes sont très révélatrices : les extensions liées à Rotterdam, vers le port sur le fleuve, ont un impact toujours positif pour la ville (même si certaines parties sont alors moins desservies), alors que les modifications internes à Schiedam peuvent au contraire diminuer l'accessibilité de ses voies et de celles qui y mène. Dès les premières quantifications, nous observons donc la distinction de caractère entre ces deux villages, initialement de tailles comparables. Nous pouvons percevoir leur différence d'évolution, et avoir l'intuition du développement de Rotterdam au détriment de celui de Schiedam.

\begin{figure}[h]
 		\centering
        \includegraphics[width=0.8\textwidth]{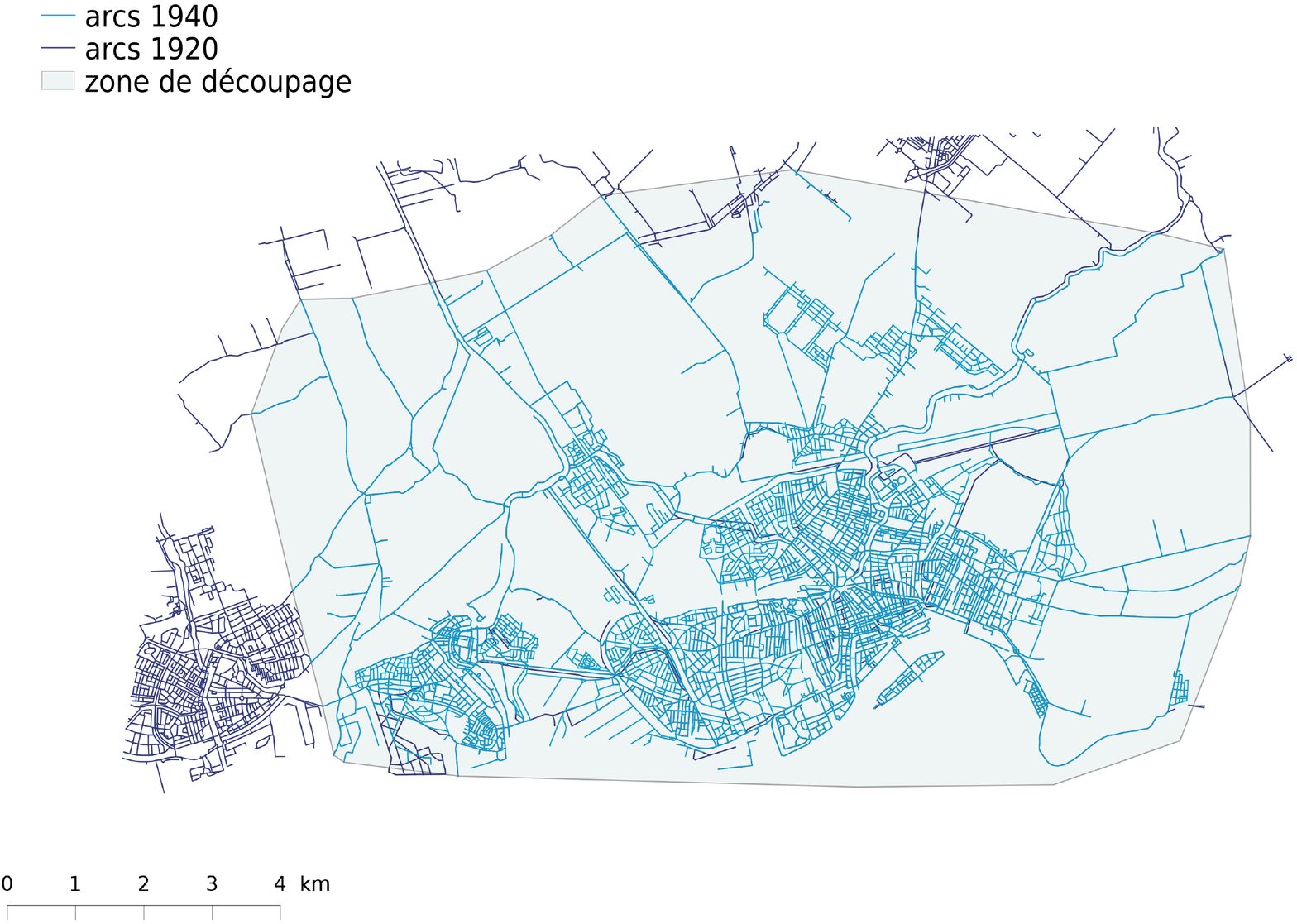}
        \caption{Comparaison des emprises des graphes entre 1920 et 1940. Visualisation de la zone de découpage choisie.}
        \label{fig:zone_emp}
\end{figure}

Nous portons une attention particulière dans notre étude à travailler sur un territoire dont nous avons les données pour chaque carte vectorisée. Nous avons dû ainsi découper les vectorisations selon les bords de la carte d'emprise minimale. Si nous tentons de comparer des cartes brutes, des arcs du graphe en bordure apparaissent comme ajoutés ou supprimés alors que cela ne correspond qu'à un découpage de carte différent. Ainsi, la partie Ouest de Schiedam n'apparaît sur les cartes vectorisées qu'entre 1890 et 1920. Nous l'avons donc supprimée sur l'ensemble des périodes (figure \ref{fig:zone_emp}).

Si nous comparons des cartes de dates trop éloignées (en passant sur une vectorisation par exemple) la somme des modifications est trop importante et nuit à la lisibilité des résultats. Nous avons de cette manière comparé les cartes de 1625 et 1940, mais la période de temps trop longue rend la carte illisible
. Les arcs ajoutés et supprimés sont trop nombreux et la valeur $\Delta_{relatif}$ n'est plus significative car le graphe comprend plus d'arcs modifiés que d'arcs appariés (1811 arcs appariés sur 7193 en 1940).

Pour pouvoir étudier à part l'évolution de Rotterdam (sans l'influence de Schiedam), nous réduisons notre étude autour de son centre de développement. Les propriétés topologiques et métriques de ce réseau sont réunies dans le tableau \ref{tab:rott_dec_pres}. Celui-ci fait ressortir le nombre plus important d'arcs par voie sur cette sous partie du réseau jusqu'à 1920, qui vient ensuite s'équilibrer avec celui du tableau précédent. En effet, nous réduisons ici notre étude à un seul centre entouré de campagne : les voies courtes et très connectées correspondent plutôt à des caractéristiques de centre-ville, elles sont donc moins présentes dans cet échantillon.

\begin{figure}[h]
    \centering
        \includegraphics[width=0.8\textwidth]{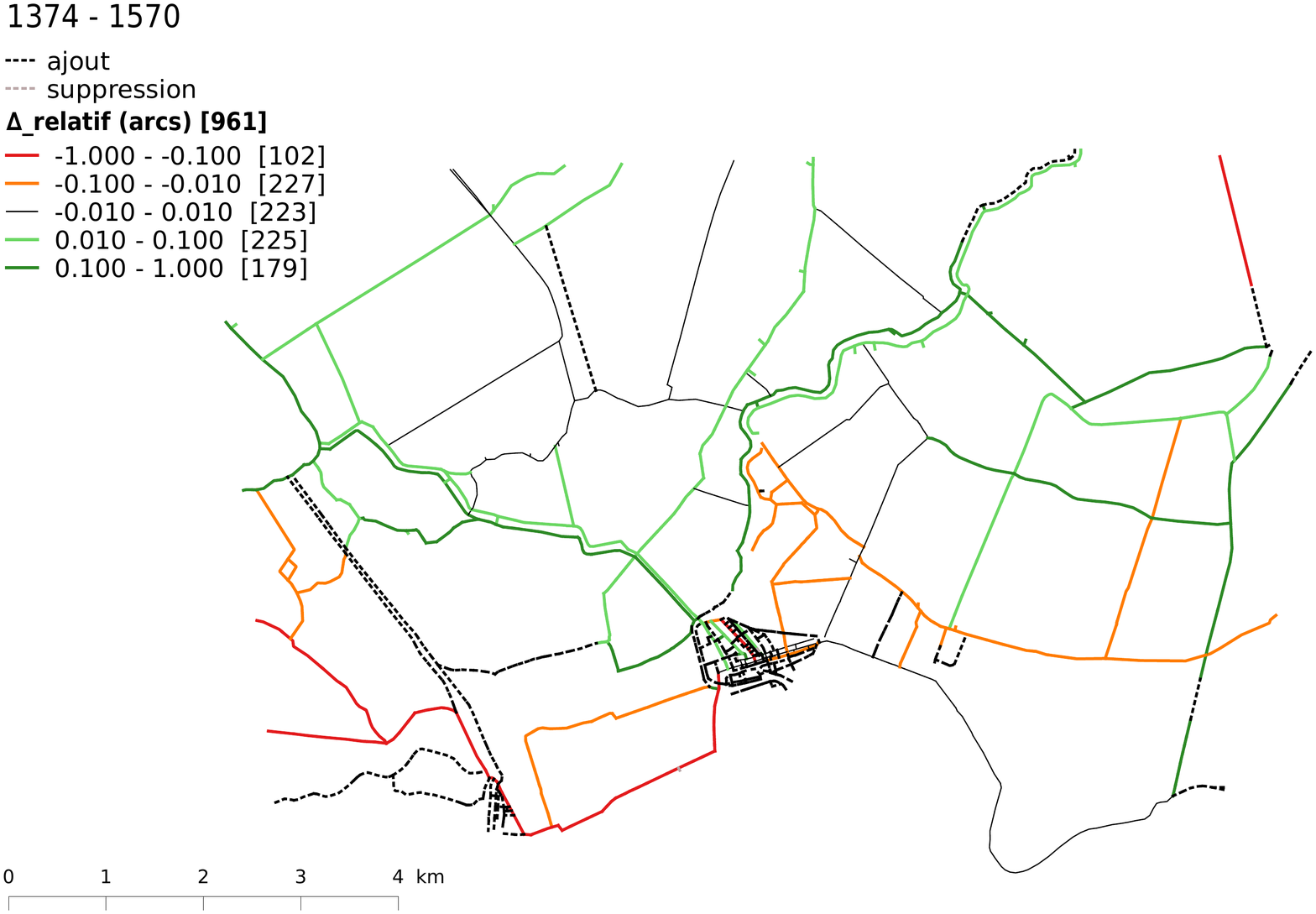}
        \caption{Étude cartographique de $\Delta_{relatif}$ sur la période 1374 - 1570.}
        \label{fig:diff_rd_1374}
    \end{figure}

\begin{figure}[c]
	 \centering
        \includegraphics[width=0.6\textwidth]{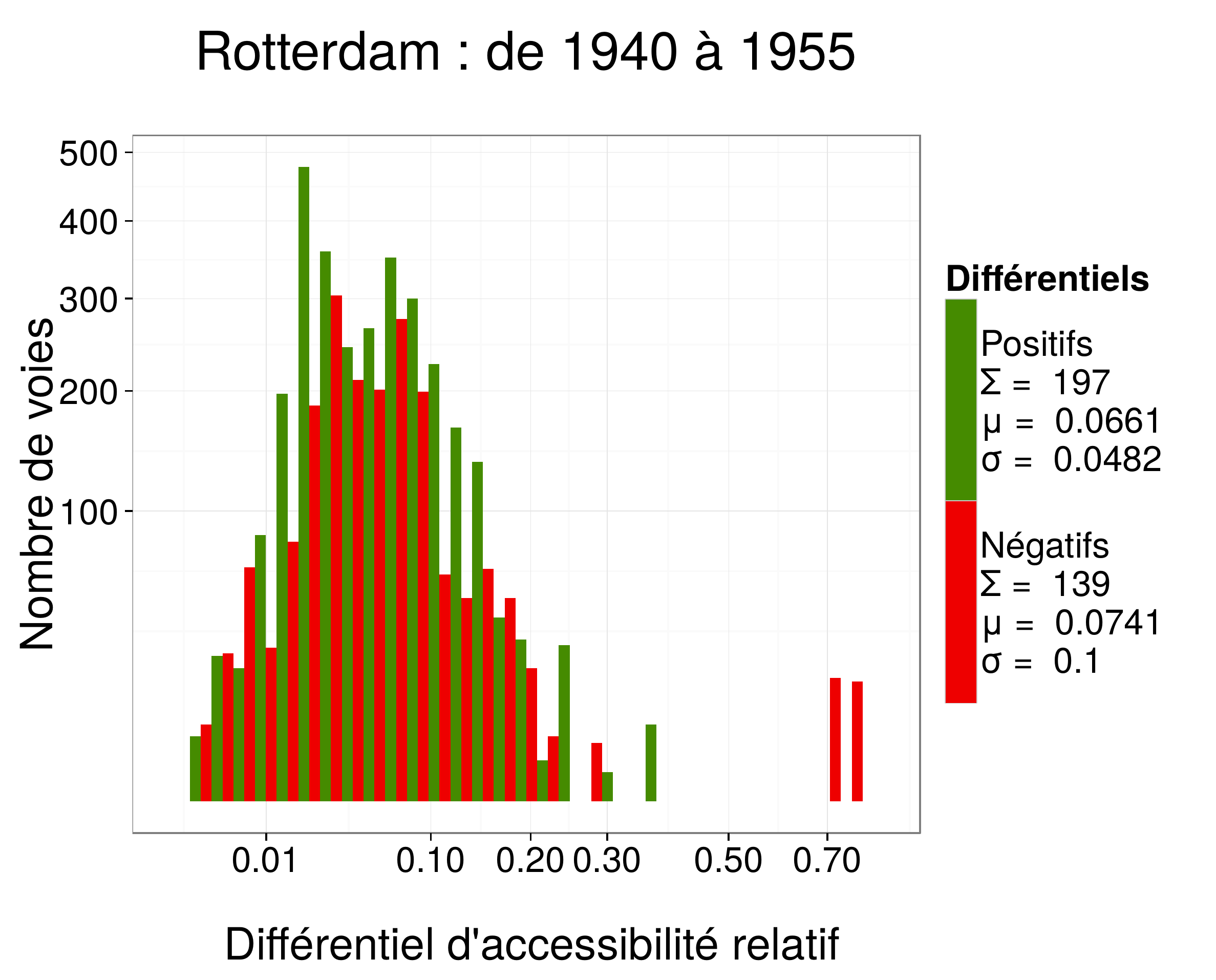}
        \caption{Étude statistique de $\Delta_{relatif}$ sur la période 1940 - 1955. \\ $\Sigma$ : somme ; $\mu$ : moyenne ; $\sigma$ : écart-type}
        \label{fig:diff_rd_1940_stat}
    \end{figure}

Le calcul de la closeness sur les voies de ce graphe réduit est également reporté en annexe \ref{ann:ssec_cartedia_rotterdam}. La quantification des changements d'accessibilité montre une évolution où l'ajout et la suppression d'arcs a une influence différente selon les quartiers (figures \ref{fig:diff_rd_1374} à \ref{fig:diff_rd_1940_stat}). Ce découpage du réseau nous permet de mieux apprécier la sectorisation des influences positives ou négatives sur le centre de Rotterdam.

\begin{table}
\begin{center}
{ \small
\begin{tabular}{|c|r|r|r|r|r|r|r|}
\hline
Année & $L_{tot}$ & $N_{sommets}$ & $N_{arcs}$ & $L_{moy}(arc)$ & $N_{voies}$ & $L_{moy}(voie)$ & $\overline{N_{arcs(voie)}}$ \\ \hline 		

1374 & 117 547 m & 187 & 959 & 123 m & 85 & 1 383 m & 11.3 \\ \hline
1570 & 151 551 m & 312 & 1350 & 112 m & 131 & 1 157 m & 10.3 \\ \hline
1600 & 159 702 m & 391 & 1521 & 105 m & 162 & 986 m & 9.4 \\ \hline
1625 & 172 745 m & 504 & 1769 & 98 m & 205 & 843 m & 8.6 \\ \hline
1890 & 255 946 m & 1058 & 2777 & 92 m & 437 & 586 m & 6.4 \\ \hline
1907 & 322 677 m & 1659 & 3666 & 88 m & 670 & 482 m & 5.5 \\ \hline
1920 & 363 048 m & 2034 & 4172 & 87 m & 808 & 449 m & 5.2 \\ \hline
1940 & 516 788 m & 3139 & 5905 & 88 m & 1236 & 418 m & 4.8 \\ \hline
1955 & 612 670 m & 3808 & 6854 & 89 m & 1525 & 402 m & 4.5 \\ \hline

\end{tabular}
}
\end{center}
\caption{Détail des caractéristiques topologiques et métriques des graphes panchroniques du nord de Rotterdam découpé.}
\label{tab:rott_dec_pres}
\end{table}

L'ensemble du détail statistique des résultats sur les huit périodes pour les deux territoires (entier et redécoupé) est reporté dans les tableaux \ref{tab:rott_delta1} et \ref{tab:rott_delta2}. Nous comparons le maximum de $\Delta_{relatif}$, observé en valeur absolue, sur les graphes de la partie Nord de Rotterdam (entier et redécoupé). Nous y reportons également, comme nous avons fait pour Avignon, la somme des $\Delta_{relatif}$ normalisée par le nombre d'arcs sur lesquels le calcul est appliqué (moyenne : $\overline{\Delta_{relatif}}$). Nous lisons dans ce tableau que les modifications apportées n'ont pas un impact équivalent d'une période à l'autre et que cela ne dépend pas de l'intervalle de temps observé. La somme pour chaque période des $\Delta_{relatif}$ positifs et négatifs appuie l'hétérogénéité du développement (cf histogrammes).

\begin{figure}[c]
    \centering
        \includegraphics[width=0.8\textwidth]{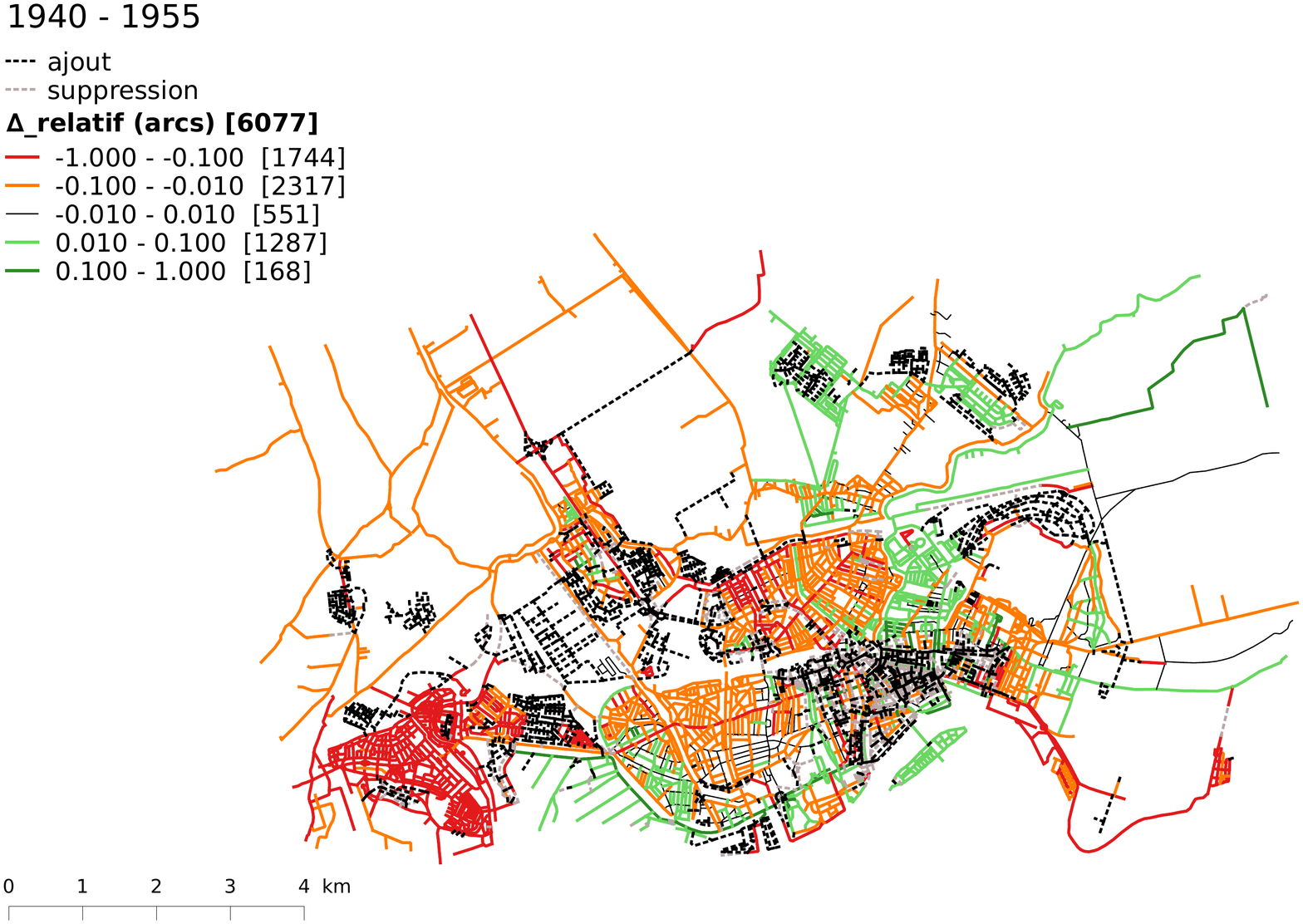}
        \caption{Étude cartographique de $\Delta_{relatif}$ sur la période 1940 - 1955.}
        \label{fig:diff_re2_1940}
    \end{figure}

\begin{figure}[c]
    \centering
        \includegraphics[width=0.8\textwidth]{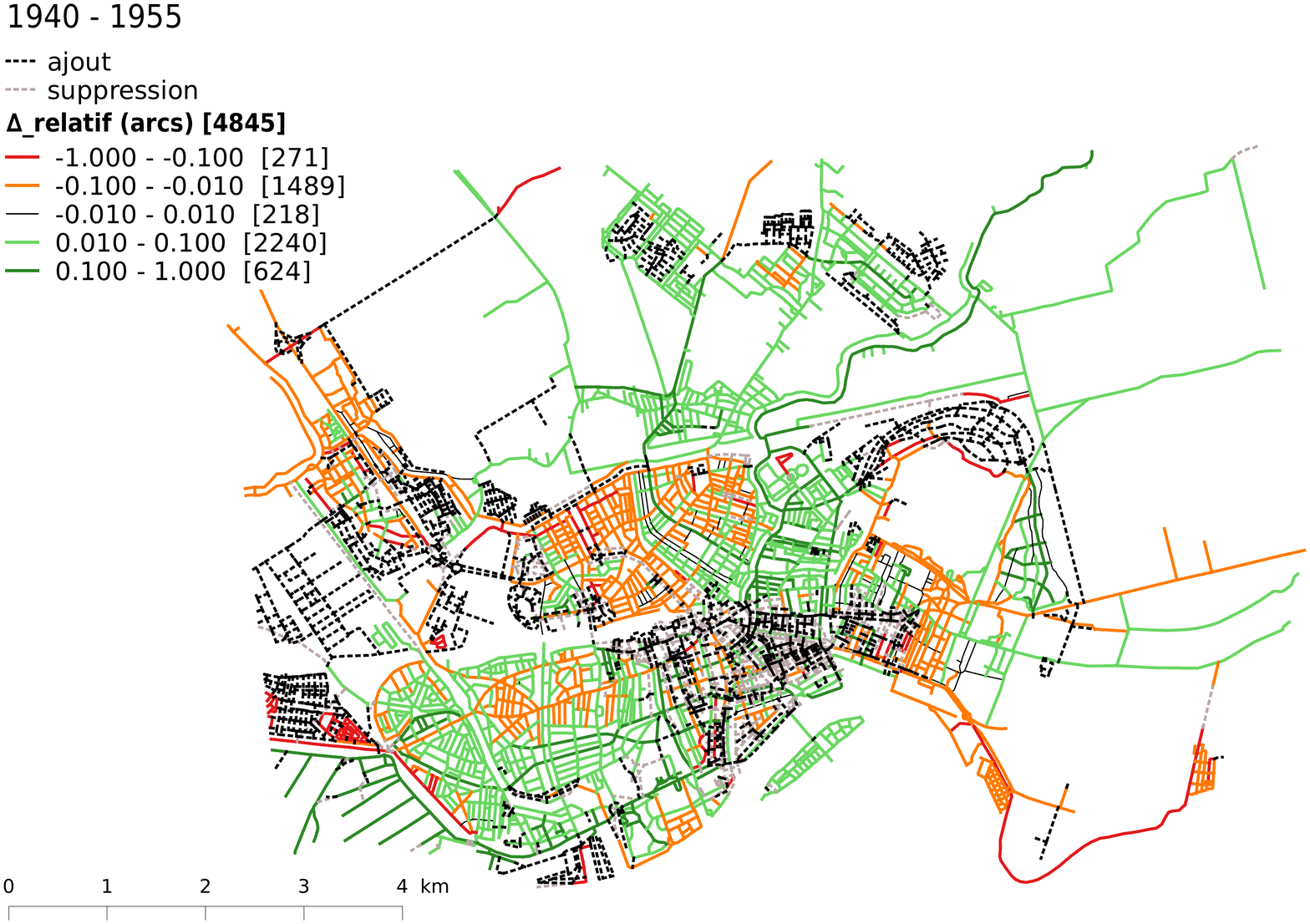}
        \caption{Étude cartographique de $\Delta_{relatif}$ sur la période 1940 - 1955.}
        \label{fig:diff_rd_1940}
    \end{figure}

Sur les deux réseaux, celui entier comme celui redécoupé, la période où les modifications ont eu l'impact le plus important est celle entre 1940 et 1955. Cela n'est pas justifié par un nombre plus important d'ajouts ou de retraits car ils sont proches de ceux des périodes précédentes. Les modifications interviennent en majorité sur le centre-ville, leur impact sur les accessibilités entre voies y sont les plus conséquentes (figures \ref{fig:diff_re2_1940} et \ref{fig:diff_rd_1940}). Le centre historique de Rotterdam a en effet été bombardé par les allemands au début de la seconde guerre mondiale, pour affaiblir la résistance à leur passage pour envahir la France. Il a été reconstruit par la suite sur un plan moderne très différent. On raconte que, dès la nuit du bombardement, les architectes ont commencé à refaire les plans. Les travaux de reconstruction du centre-ville furent finalisés en 1955.

\begin{figure}[c]
   \centering
        \includegraphics[width=0.8\textwidth]{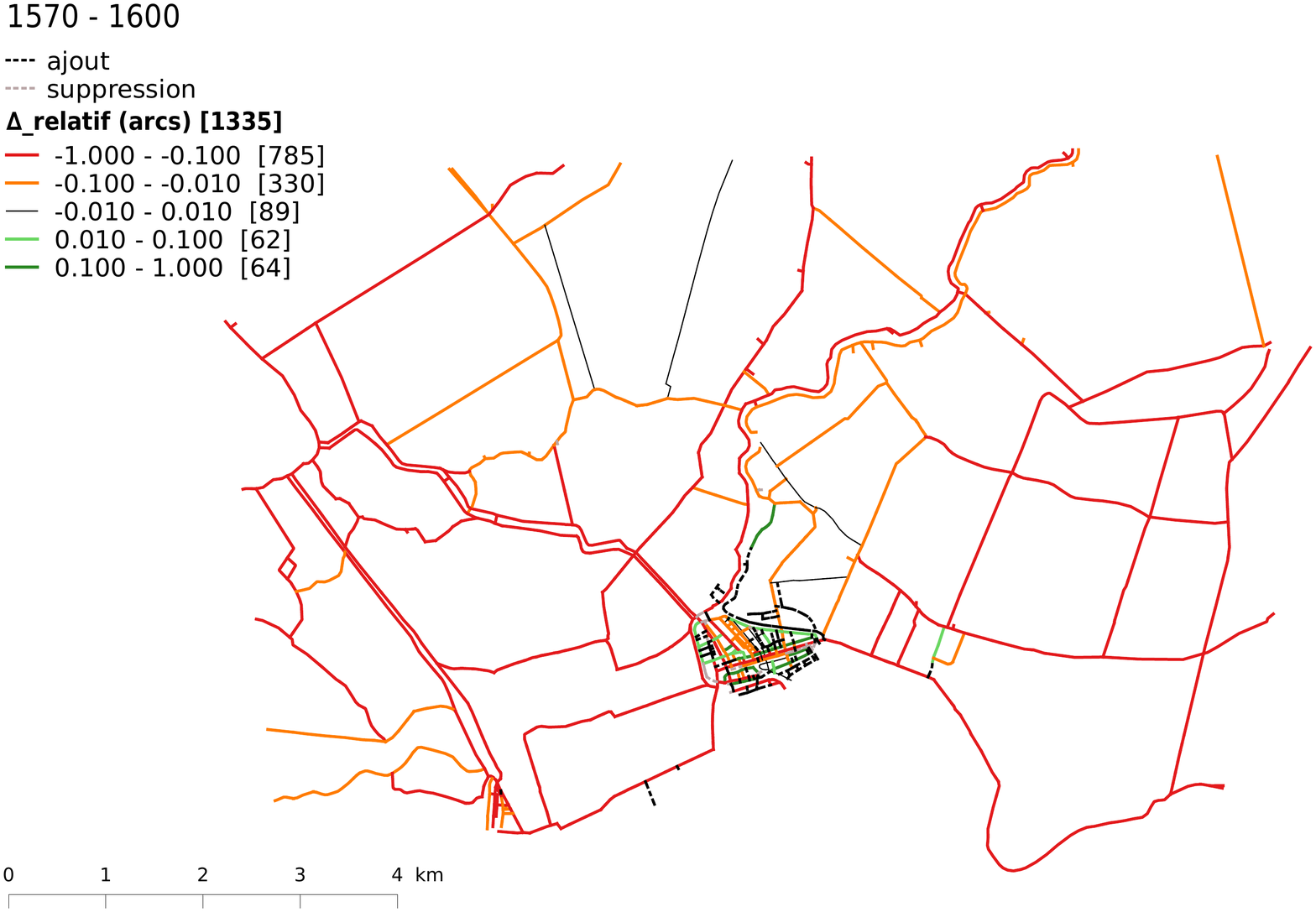}
        \caption{Étude cartographique de $\Delta_{relatif}$ sur la période 1570 - 1600.}
        \label{fig:diff_rd_1570}
    \end{figure}

\begin{figure}[c]
   \centering
        \includegraphics[width=0.8\textwidth]{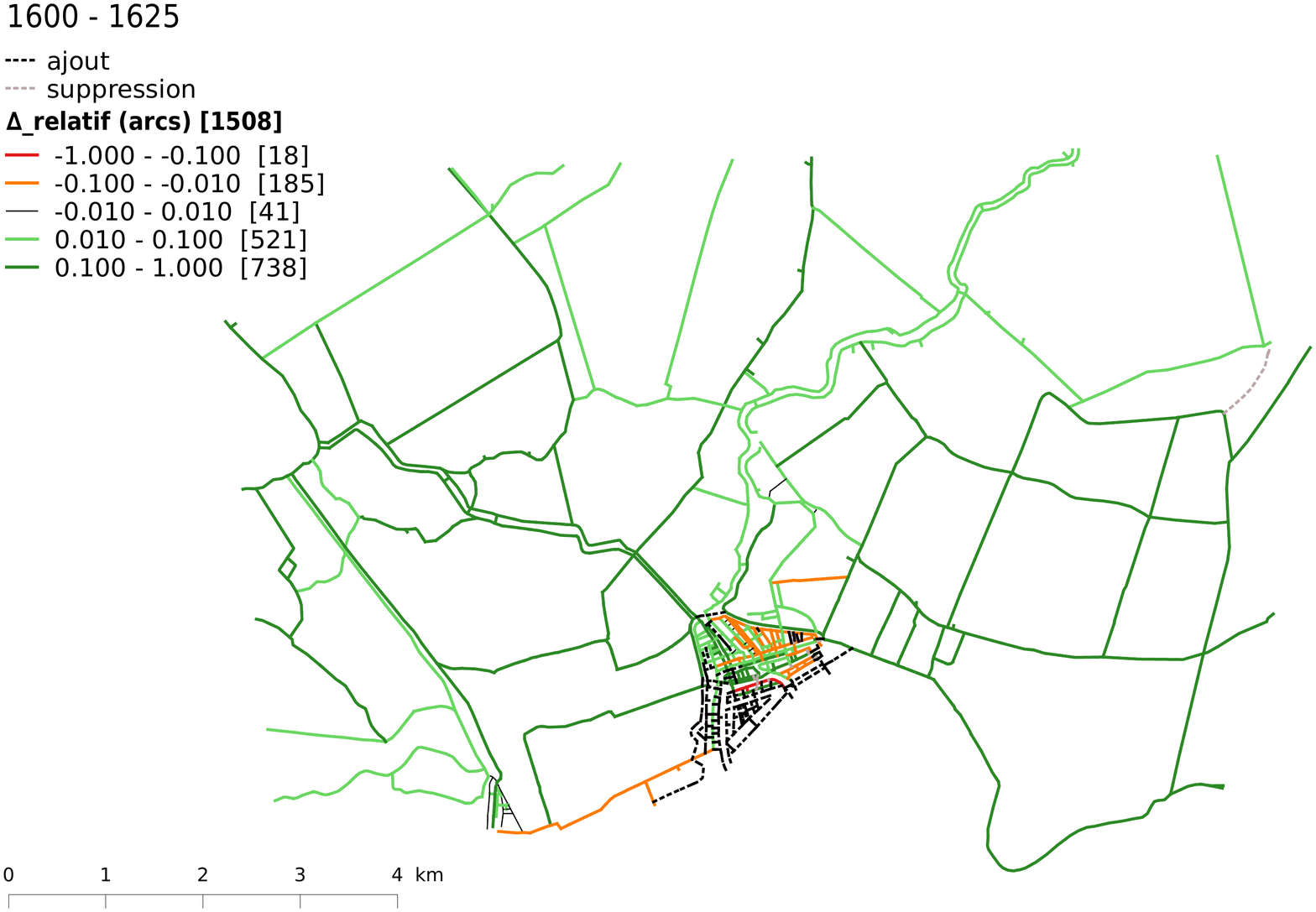}
        \caption{Étude cartographique de $\Delta_{relatif}$ sur la période 1600 - 1625.}
        \label{fig:diff_rd_1600}
    \end{figure}

\begin{figure}[c]
    \centering
        \includegraphics[width=0.8\textwidth]{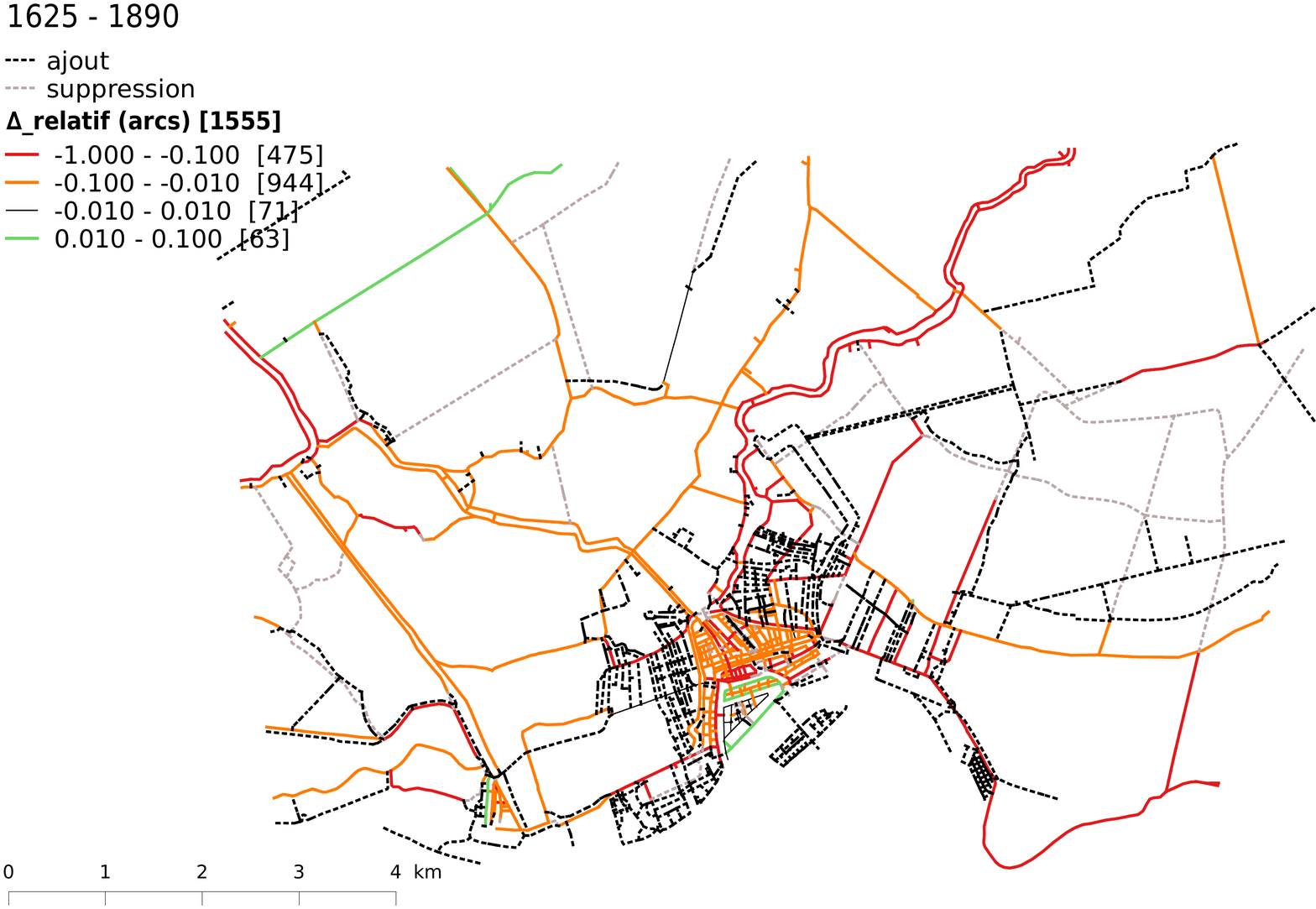}
        \caption{Étude cartographique de $\Delta_{relatif}$ sur la période 1625 - 1890.}
        \label{fig:diff_rd_1625}
    \end{figure}

\begin{figure}[c]
     \centering
        \includegraphics[width=0.8\textwidth]{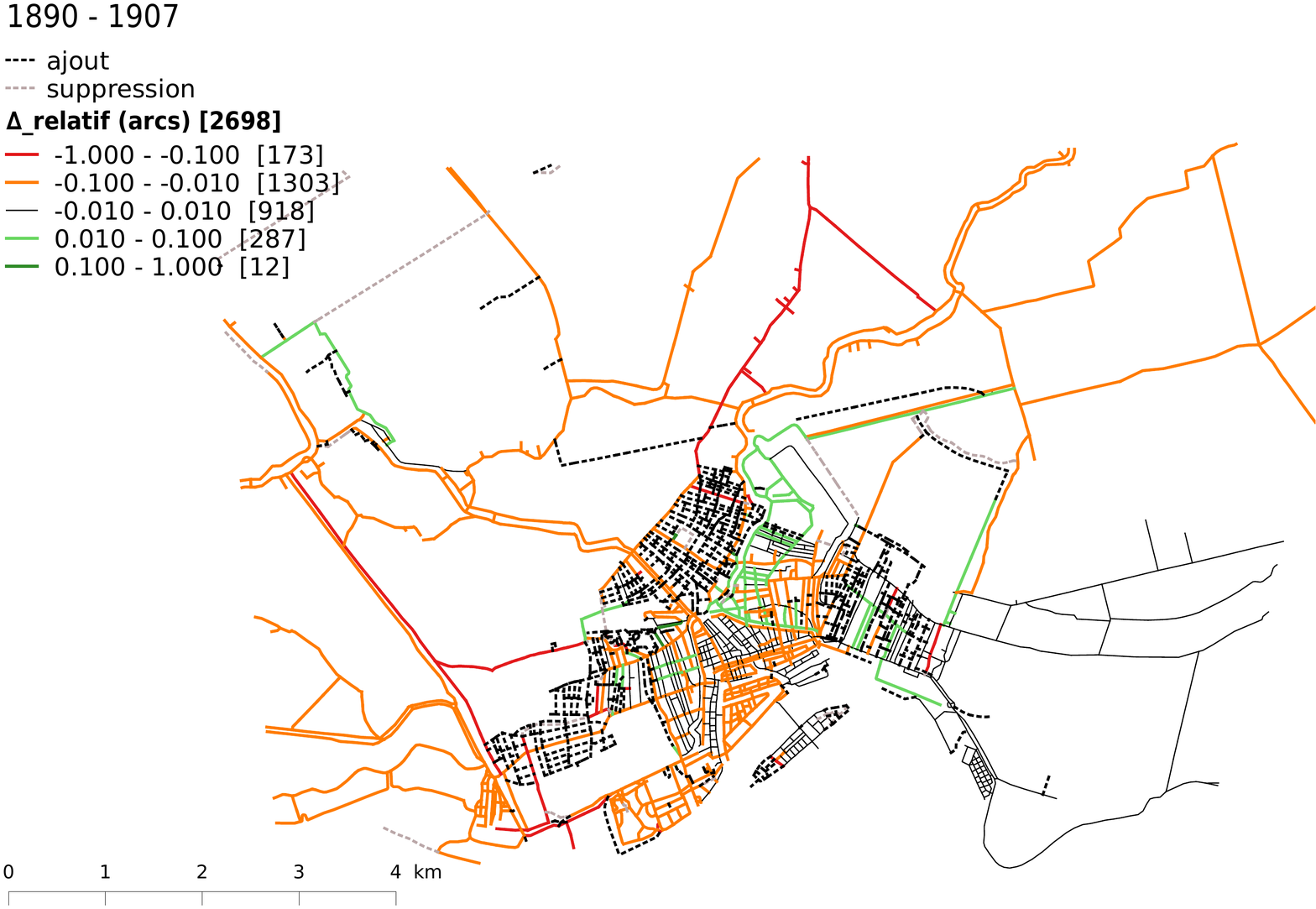}
        \caption{Étude cartographique de $\Delta_{relatif}$ sur la période 1890 - 1907.}
        \label{fig:diff_rd_1890}
    \end{figure}

\begin{figure}[c]
    \centering
        \includegraphics[width=0.8\textwidth]{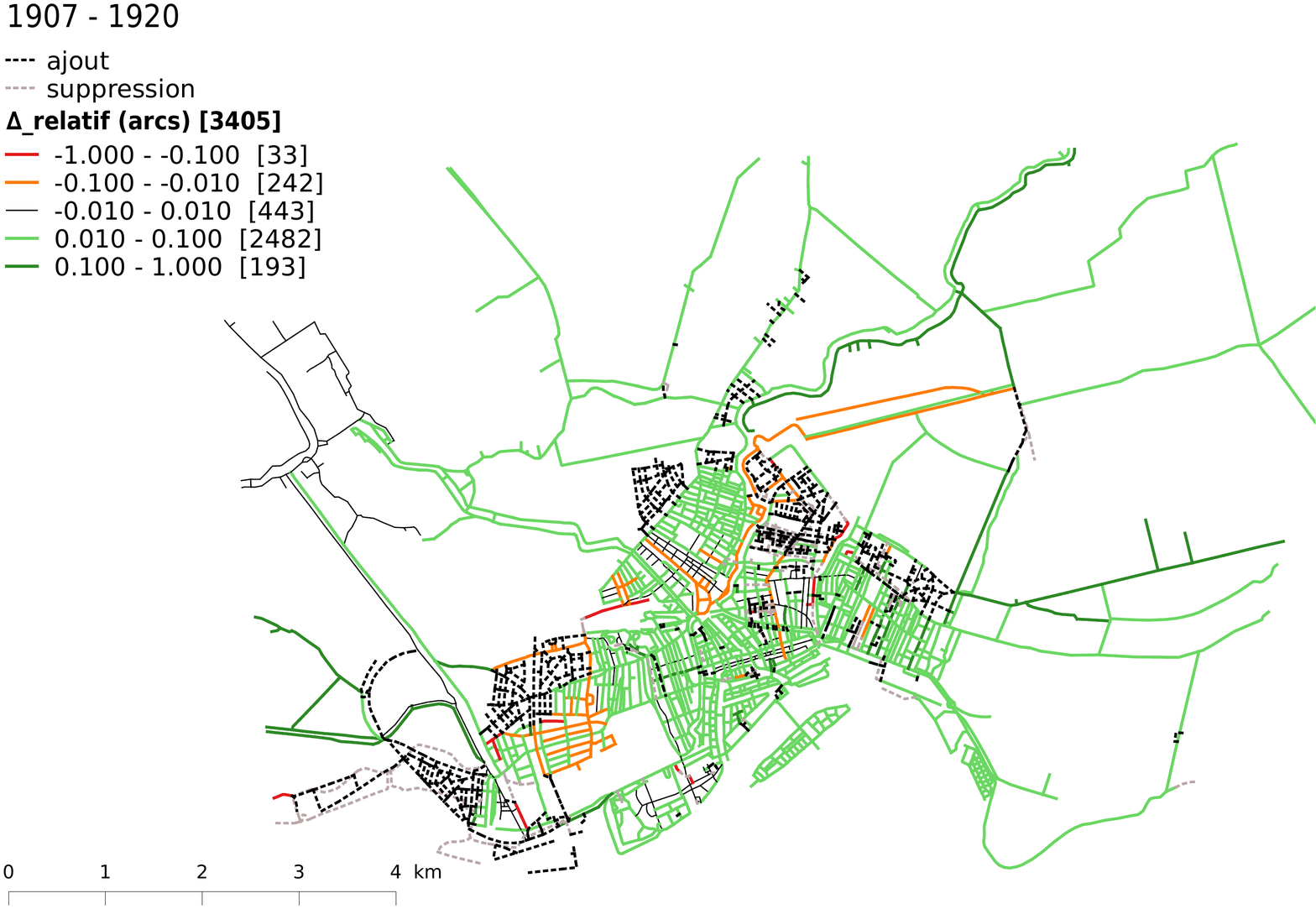}
        \caption{Étude cartographique de $\Delta_{relatif}$ sur la période 1907 - 1920.}
        \label{fig:diff_rd_1907}
    \end{figure}

\begin{figure}[c]
     \centering
        \includegraphics[width=0.8\textwidth]{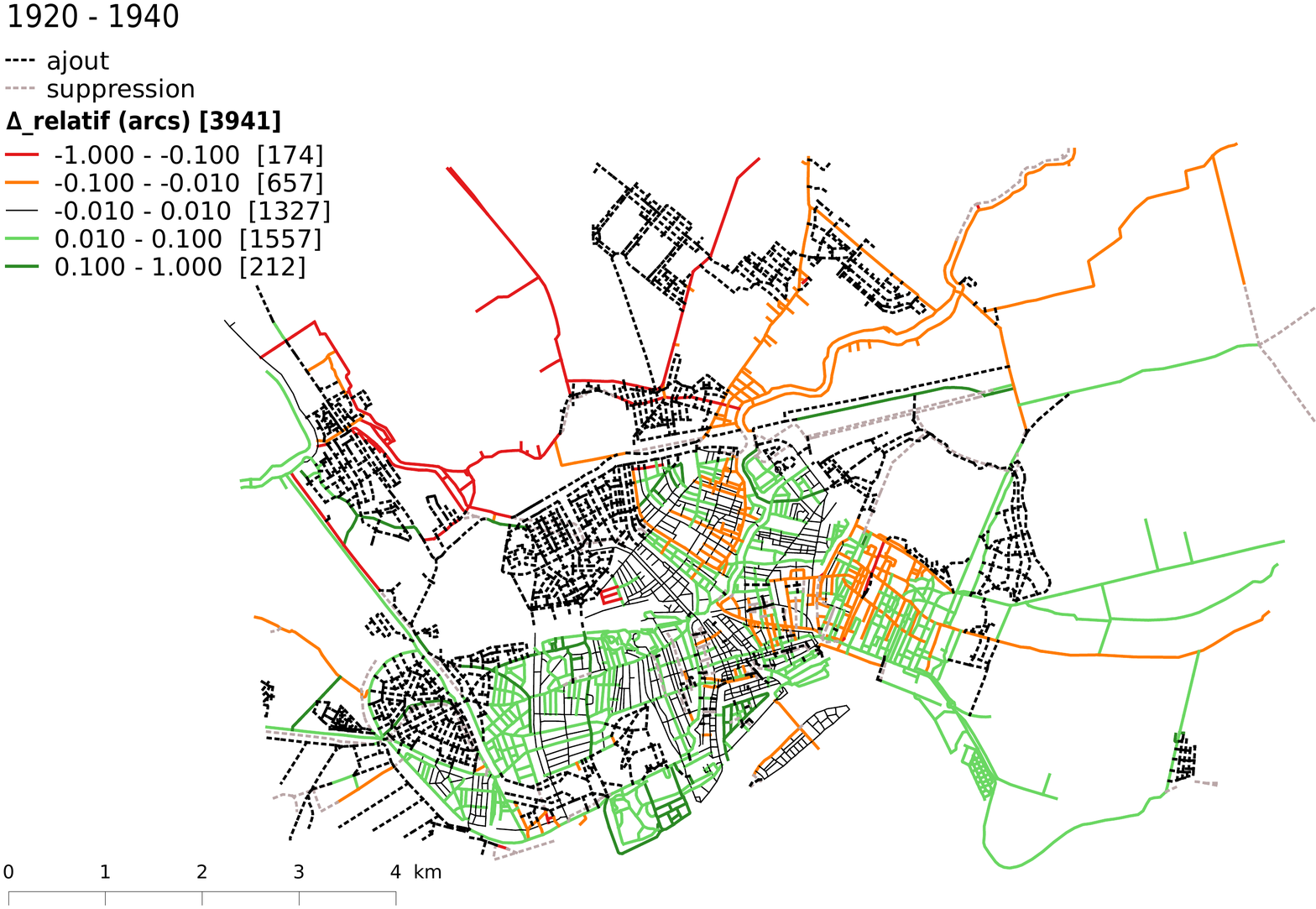}
        \caption{Étude cartographique de $\Delta_{relatif}$ sur la période 1920 - 1940.}
        \label{fig:diff_rd_1920}
    \end{figure}

Nous pouvons aussi remarquer que pendant certaines périodes (plus nettement observables sur Rotterdam redécoupé), les extensions urbaines sont globalement positives (figures \ref{fig:diff_rd_1374}, \ref{fig:diff_rd_1600}, \ref{fig:diff_rd_1907}, \ref{fig:diff_rd_1920} et \ref{fig:diff_rd_1940}), ou, au contraire, globalement négatives (figures \ref{fig:diff_rd_1570}, \ref{fig:diff_rd_1625} et \ref{fig:diff_rd_1890}). Ces inversions d'effets s'expliquent  \textit{a priori} par des différences structurelles ou topologiques dans les modifications apportées, dont il serait très intéressant d'approfondir l'étude. Nous pouvons déduire de nos observations sur ces graphes que lorsque la densification est ponctuelle et bien connectée avec l'ensemble, l'impact est globalement positif (figure \ref{fig:diff_rd_1600}). Lorsque la restructuration est massive, et que les nouvelles voies créées ne tiennent pas forcément compte de la cohérence pré-existante, l'impact est globalement négatif (figure \ref{fig:diff_rd_1625} : entre 1625 et 1890, l'inondation majeure entraîne une restructuration complète, incluant notamment la campagne avoisinante).

\begin{table}
\begin{center}
{ \small
\begin{tabular}{|c|c|r|r|r|r|}
\hline

période & durée & $N_{arcs(ajoutés)}$ & $L_{ajoutée}$ & $N_{arcs(retirés)}$ & $L_{retirée}$ \\ \hline

\multicolumn{6}{|c|}{Réseau entier}   \\ \hline

1374 - 1570 & 196 ans & 644 & 57 535 m & 7 & 263 m \\ \hline
1570 - 1600 & 30 ans & 308 & 15 660 m & 118 & 6 488 m \\ \hline
1600 - 1625 & 25 ans & 269 & 14 895 m & 18 & 1 554 m \\ \hline
1625 - 1890 & 265 ans & 1474 & 144 618 m & 335 & 50 353 m \\ \hline
1890 - 1907 & 17 ans & 1046 & 86 042 m & 94 & 15 462 m \\ \hline
1907 - 1920 & 13 ans & 1033 & 87 907 m & 315 & 29 411 m \\ \hline
1920 - 1940 & 20 ans & 2450 & 222 650 m & 441 & 57 559 m \\ \hline
1940 - 1955 & 15 ans & 2332 & 192 209 m & 1150 & 77 384 m \\ \hline

\multicolumn{6}{|c|}{Réseau découpé} \\ \hline

1374 - 1570 & 196 ans & 413 & 33 626 m & 4 & 217 m \\ \hline
1570 - 1600 & 30 ans & 191 & 9 008 m & 39 & 1616 m \\ \hline
1600 - 1625 & 25 ans & 269 & 14 895 m & 18 & 1554 m \\ \hline
1625 - 1890 & 265 ans & 1227 & 119 815 m & 222 & 37430 m \\ \hline
1890 - 1907 & 17 ans & 975 & 78 693 m & 84 & 12710 m \\ \hline
1907 - 1920 & 13 ans & 791 & 65 921 m & 268 & 23140 m \\ \hline
1920 - 1940 & 20 ans & 1 996 & 186 556 m & 255 & 34253 m \\ \hline
1940 - 1955 & 15 ans & 2 013 & 164 549 m & 1 092 & 71271 m \\ \hline

\end{tabular}
}
\end{center}
\caption{Détail du nombre de modifications du réseau pour chaque période d'étude du nord de Rotterdam (entier et redécoupé).}
\label{tab:rott_delta1}
\end{table}

\begin{table}
\begin{center}
{ \small
\begin{tabular}{|c|r|r|r|r|r|}
\hline

période & $N_{arcs(modifiés)}$ & $\overline{\Delta_{relatif}}$ & $\sigma(\Delta_{relatif})$  & $max \vert \Delta_{relatif}\vert$ & $\sum \vert \Delta_{relatif} \vert$ \\ \hline

\multicolumn{6}{|c|}{Réseau entier}   \\ \hline

1374 - 1570 & 651 & -0.0567 & 0.1212 & 0.7662 & 139.44 \\ \hline
1570 - 1600 & 426 & -0.0546 & 0.1241 & 0.9404 & 185.12 \\ \hline
1600 - 1625 & 287 & 0.0911 & 0.0941 & 0.3907 & 207.92  \\ \hline
1625 - 1890 & 1 809 & -0.0491 & 0.0617 & 0.2872 & 125.31\\ \hline
1890 - 1907 & 1 140 & -0.0433 & 0.0521 & 0.3144 & 159.27 \\ \hline
1907 - 1920 & 1 348 & 0.04467 & 0.0567 & 0.2971 & 215.63 \\ \hline
1920 - 1940 & 2 891 & 0.0543 & 0.0710 & 0.6344 & 315.22  \\ \hline
1940 - 1955 & 3 482 & -0.0498 & 0.0909 & 0.7473 & 476.04  \\ \hline

\multicolumn{6}{|c|}{Réseau découpé} \\ \hline

1374 - 1570 & 417 & 0.0192 & 0.1190 & 0.7506 & 78.58  \\ \hline
1570 - 1600 & 230 & -0.1005 & 0.1071 & 0.9503 & 163.86  \\ \hline
1600 - 1625 & 287  & 0.1158 & 0.1140 & 0.3927 & 187.34  \\ \hline
1625 - 1890 & 1449 & -0.0793 & 0.0616 & 0.4893 & 131.23  \\ \hline
1890 - 1907 & 1059 & -0.0271 & 0.0495 & 0.2761 & 93.98  \\ \hline
1907 - 1920 & 1059 & 0.0327 & 0.0508 & 0.3337 & 137.87 \\ \hline
1920 - 1940 & 2251 & 0.0131 & 0.0625 & 0.6588 & 155.12  \\ \hline
1940 - 1955 & 3105 & 0.0120 & 0.0998 & 0.7344 & 335.15  \\ \hline

\end{tabular}
}
\end{center}
\caption{Détail statistique des variations relatives $\Delta_{relatif}$ pour chaque période d'étude du nord de Rotterdam (entier et redécoupé).}
\label{tab:rott_delta2}
\end{table}

\clearpage
\subsubsection{Cartes Rotterdam Nord entier}

     \begin{figure}[h]
     \centering
        \includegraphics[width=0.55\textwidth]{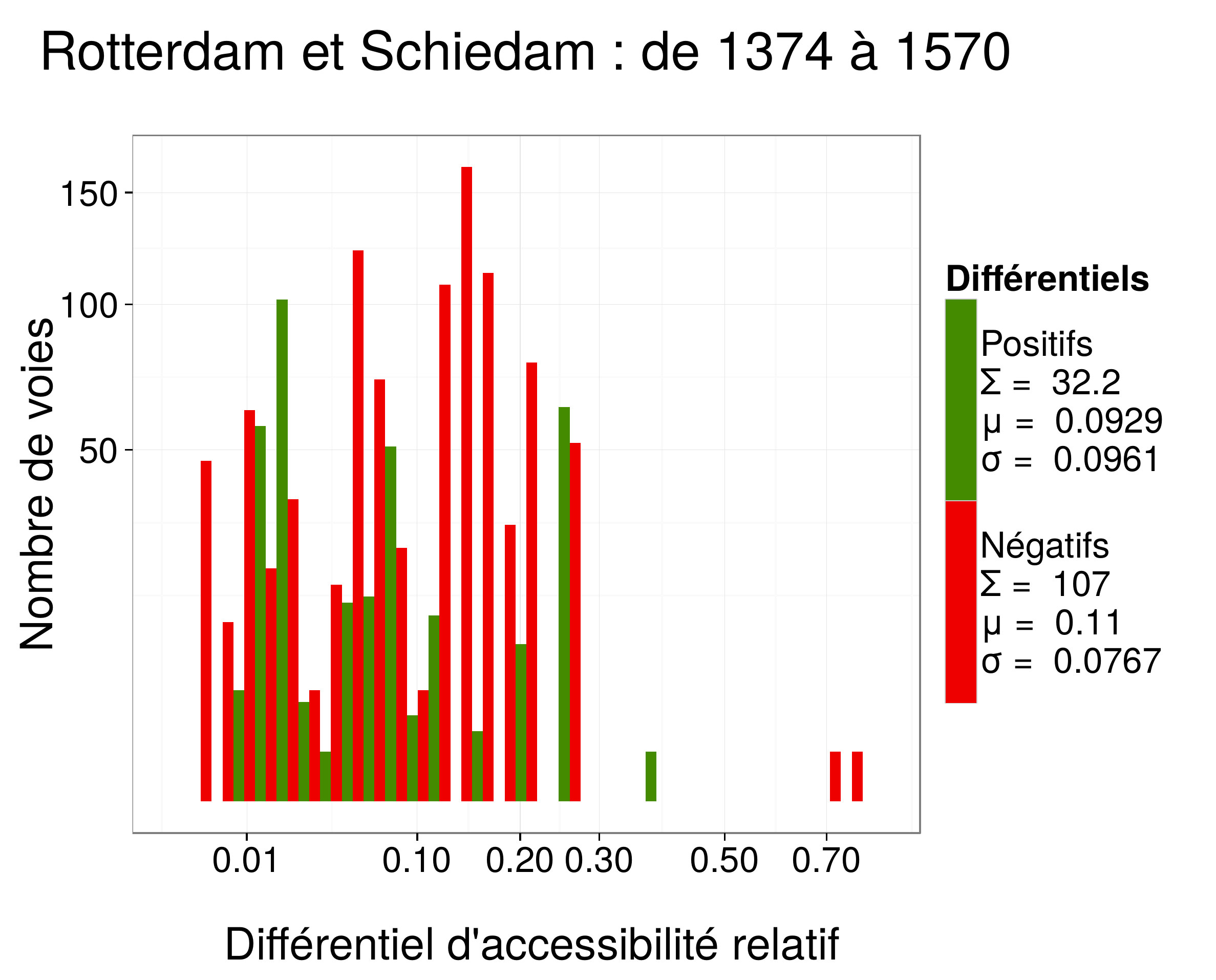}
        \caption{Étude statistique de $\Delta_{relatif}$ sur la période 1374 - 1570. \\ $\Sigma$ : somme ; $\mu$ : moyenne ; $\sigma$ : écart-type}
        \label{fig:diff_re2_1374_stat}
    \end{figure}

\clearpage 
   \begin{figure}[c]
   \centering
        \includegraphics[width=0.8\textwidth]{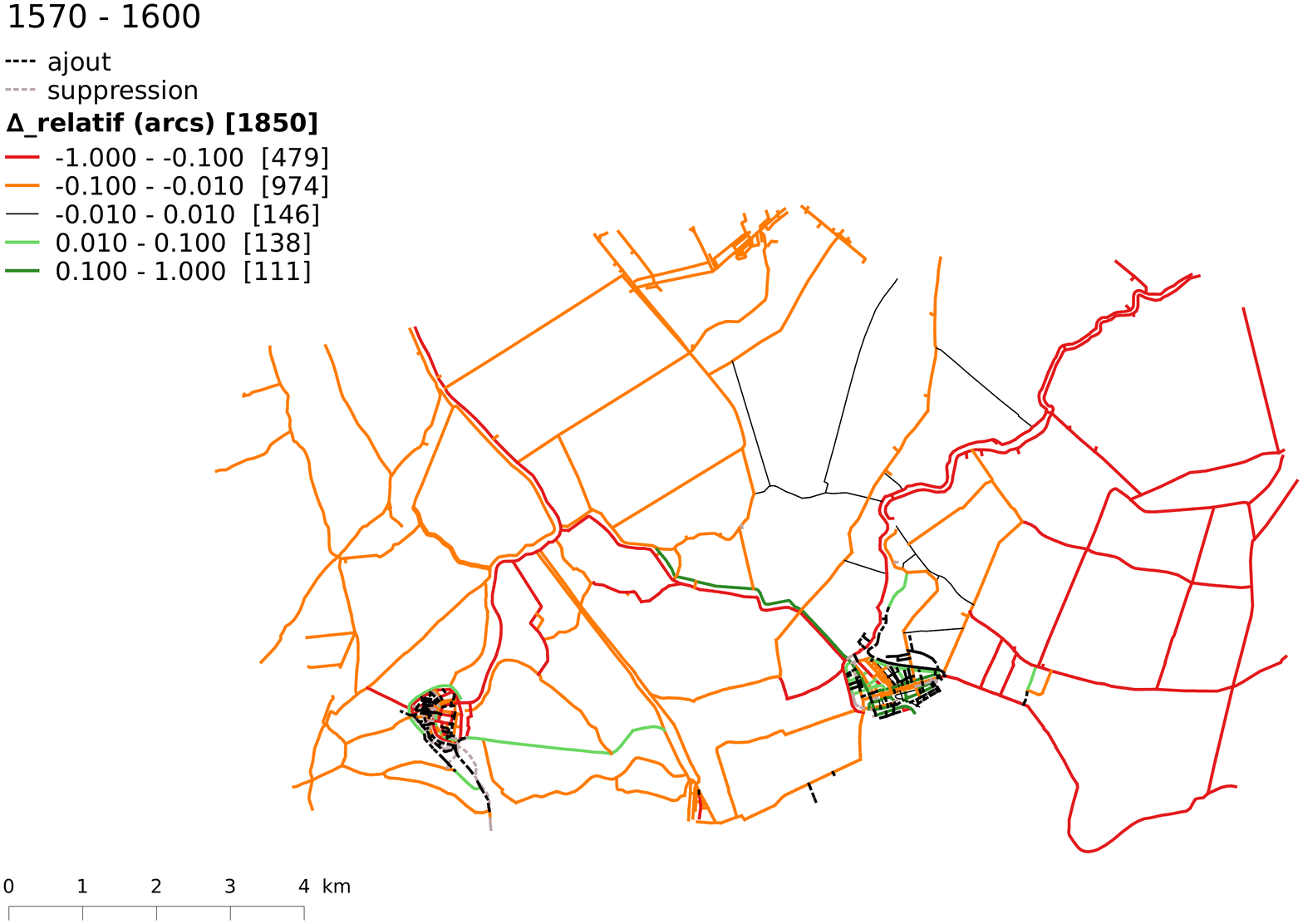}
        \caption{Étude cartographique de $\Delta_{relatif}$ sur la période 1570 - 1600.}
        \label{fig:diff_re2_1570}
    \end{figure}
    
     \begin{figure}[c]
     \centering
        \includegraphics[width=0.6\textwidth]{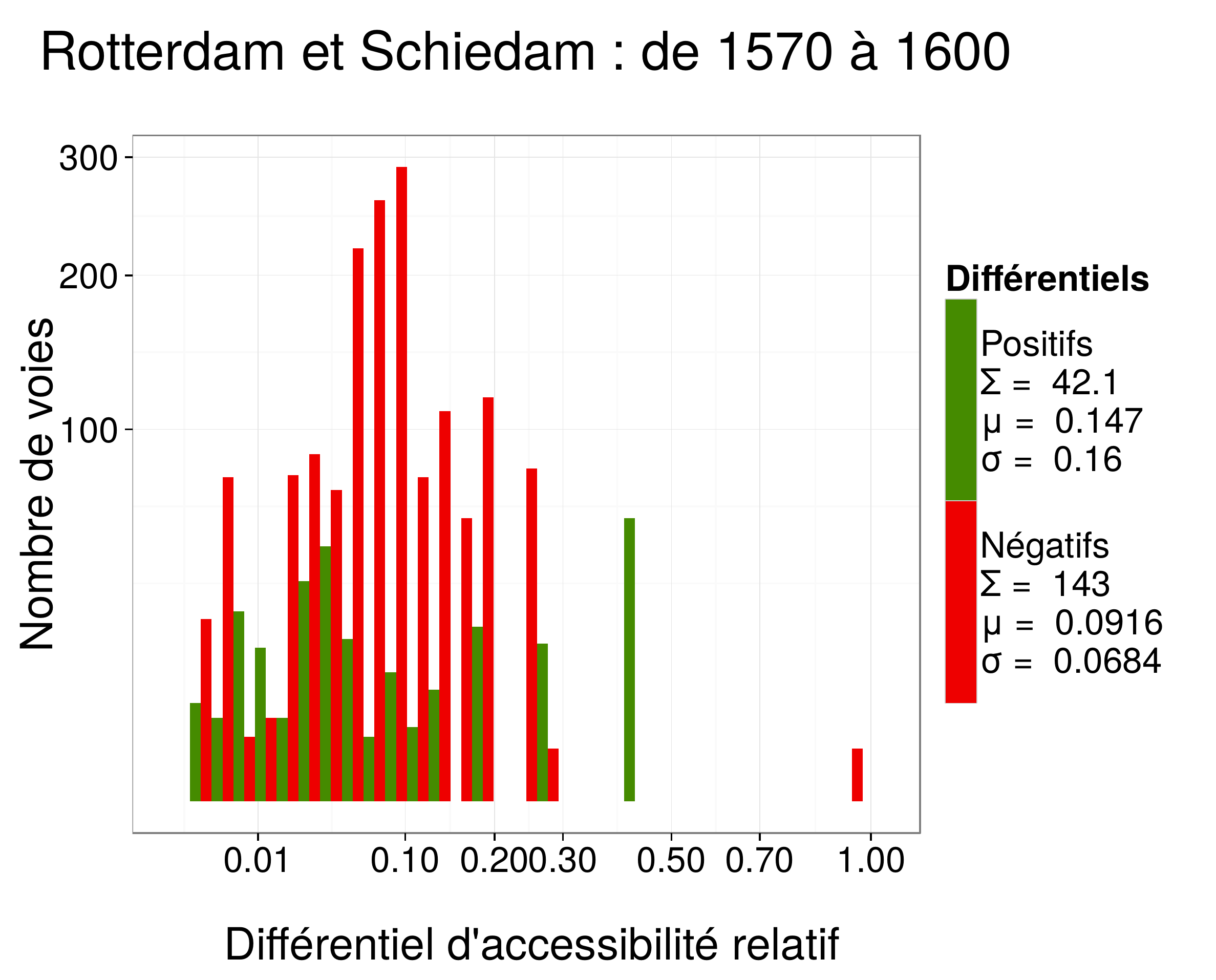}
        \caption{Étude statistique de $\Delta_{relatif}$ sur la période 1570 - 1600. \\ $\Sigma$ : somme ; $\mu$ : moyenne ; $\sigma$ : écart-type}
        \label{fig:diff_re2_1570_stat}
    \end{figure}

\clearpage 
   \begin{figure}[c]
   \centering
        \includegraphics[width=0.8\textwidth]{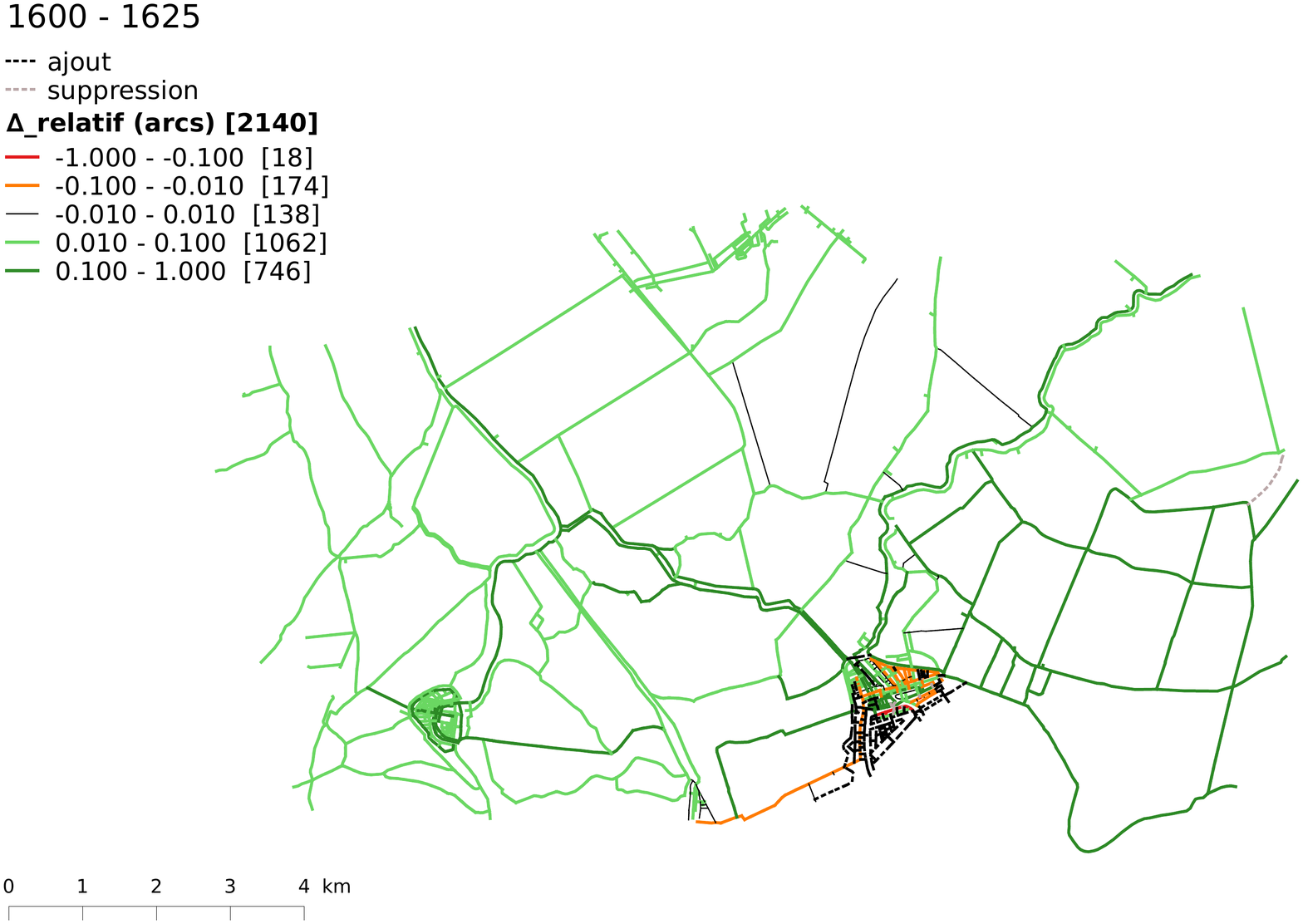}
        \caption{Étude cartographique de $\Delta_{relatif}$ sur la période 1600 - 1625.}
        \label{fig:diff_re2_1600}
    \end{figure}
    
     \begin{figure}[c]
     \centering
        \includegraphics[width=0.6\textwidth]{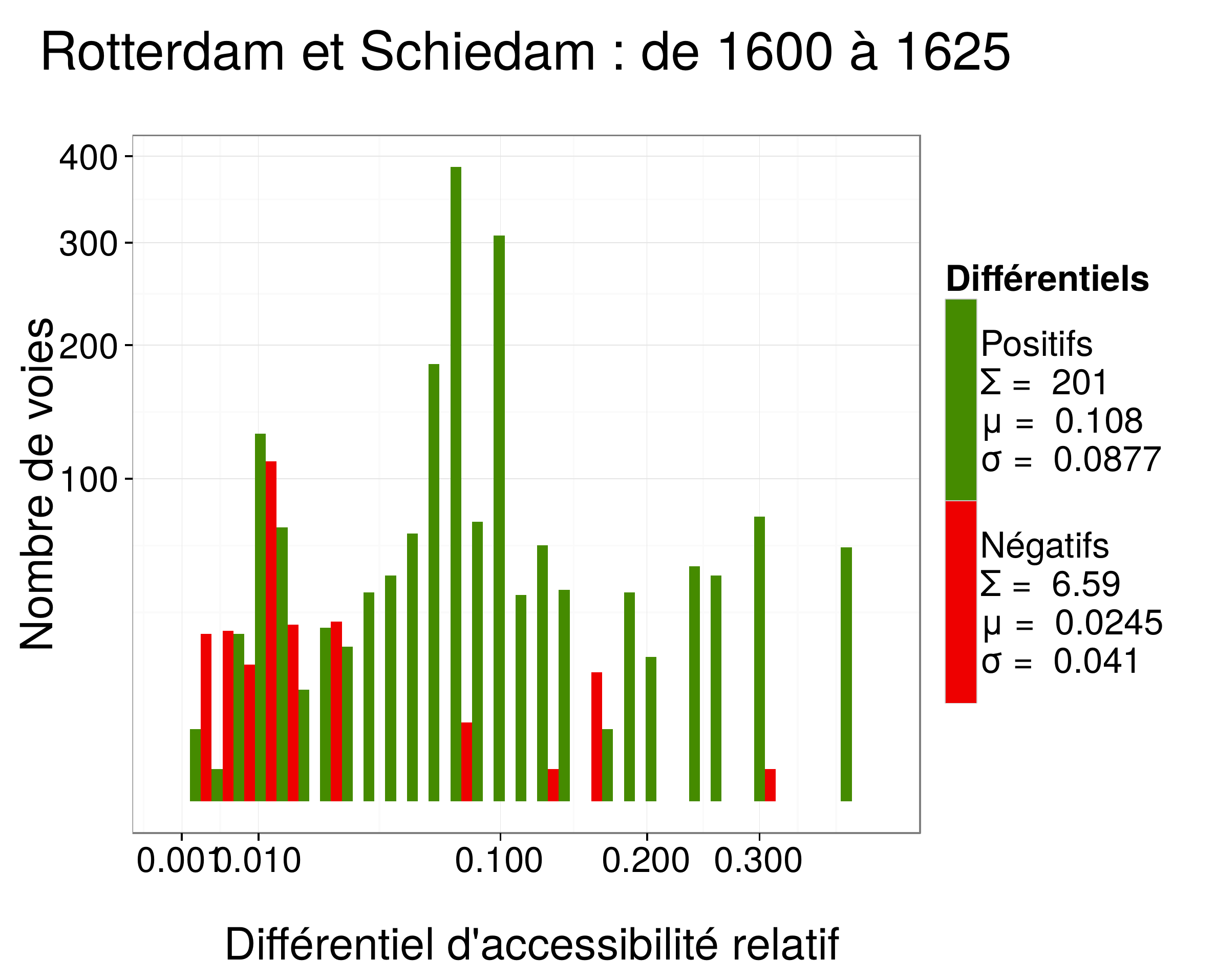}
        \caption{Étude statistique de $\Delta_{relatif}$ sur la période 1600 - 1625. \\ $\Sigma$ : somme ; $\mu$ : moyenne ; $\sigma$ : écart-type}
        \label{fig:diff_re2_1600_stat}
    \end{figure}

\clearpage 
    \begin{figure}[c]
    \centering
        \includegraphics[width=0.8\textwidth]{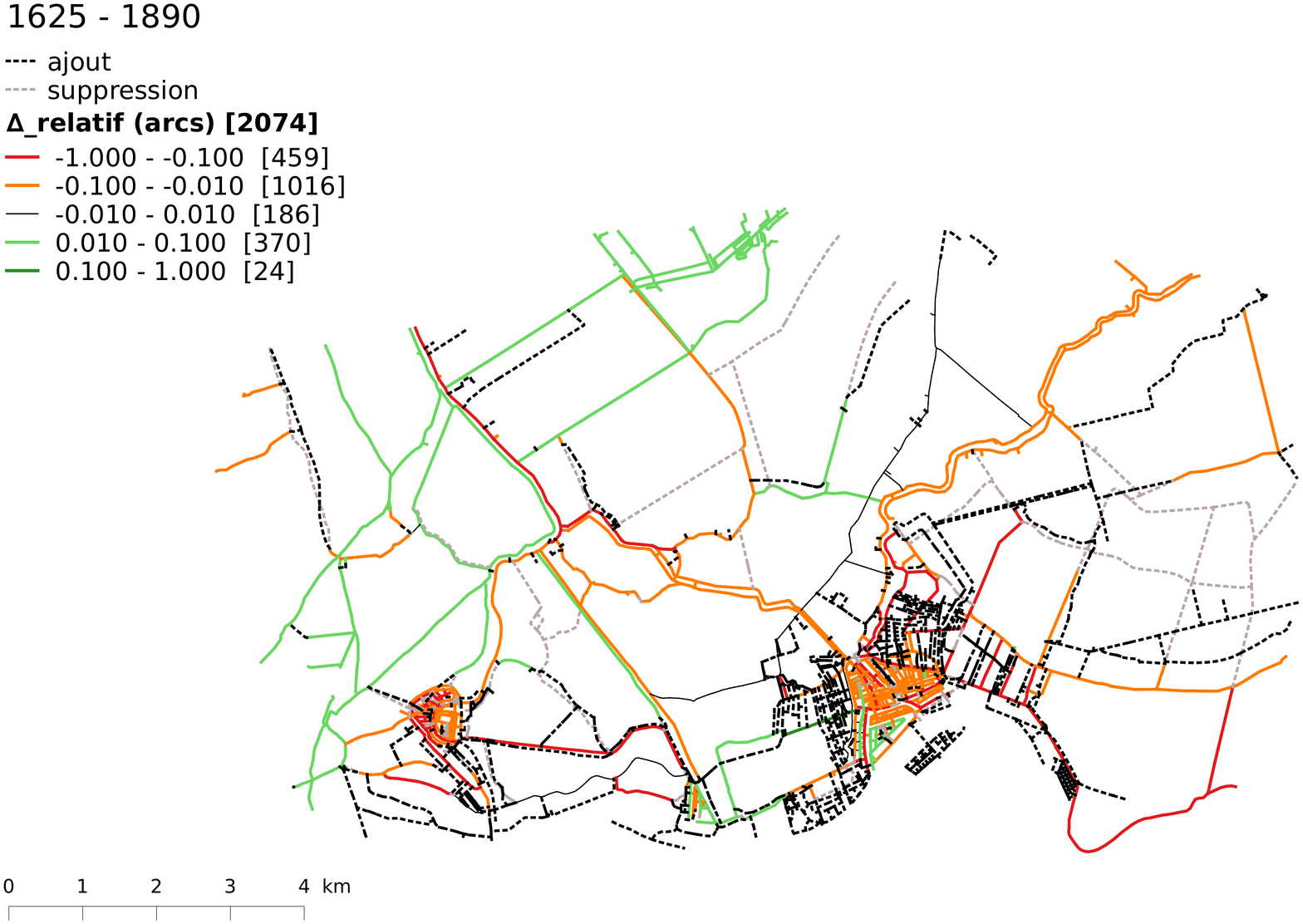}
        \caption{Étude cartographique de $\Delta_{relatif}$ sur la période 1625 - 1890.}
        \label{fig:diff_re2_1625}
    \end{figure}
    
     \begin{figure}[c]
     \centering
        \includegraphics[width=0.6\textwidth]{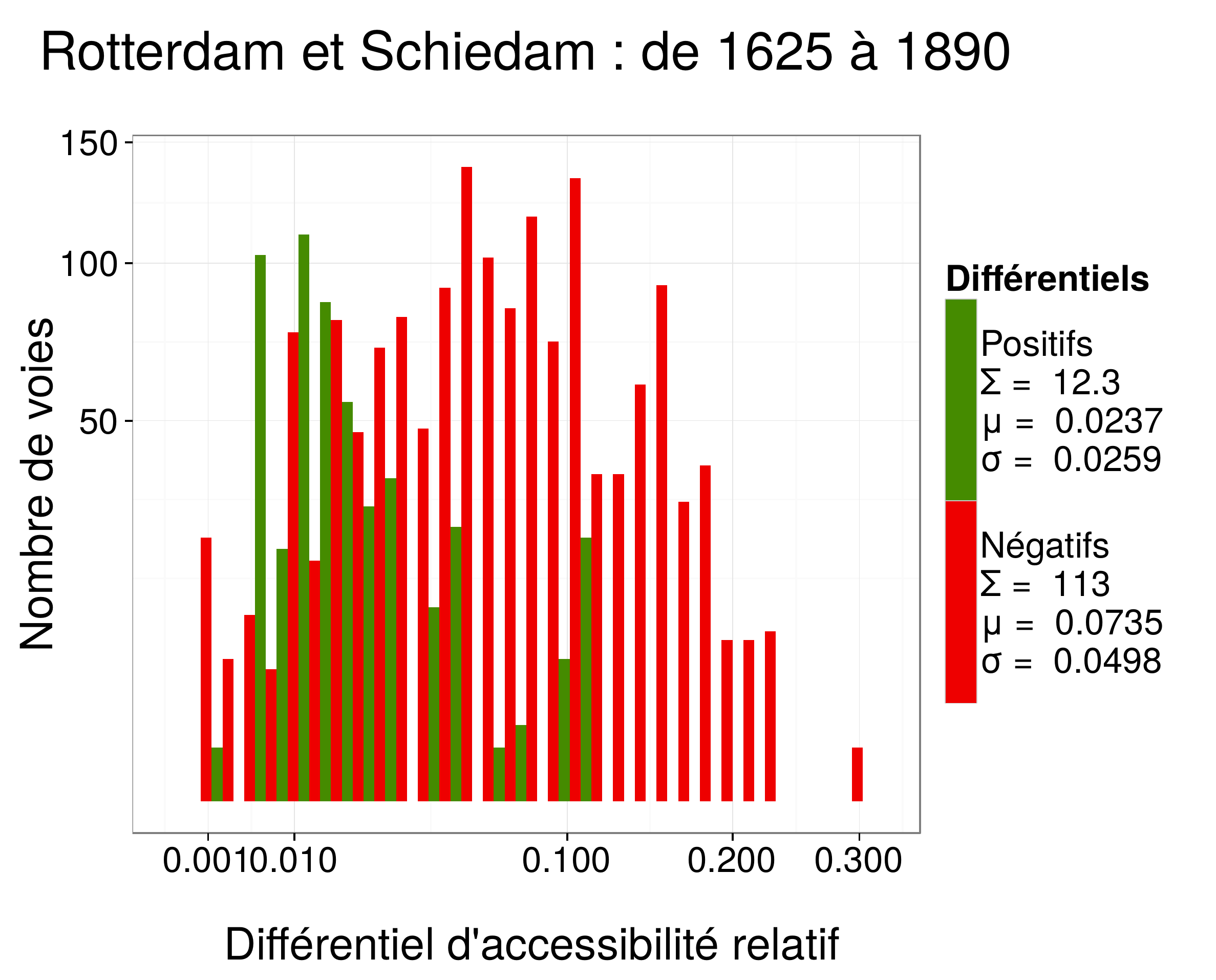}
        \caption{Étude statistique de $\Delta_{relatif}$ sur la période 1625 - 1890. \\ $\Sigma$ : somme ; $\mu$ : moyenne ; $\sigma$ : écart-type}
        \label{fig:diff_re2_1625_stat}
    \end{figure}

\clearpage 
     \begin{figure}[c]
     \centering
        \includegraphics[width=0.8\textwidth]{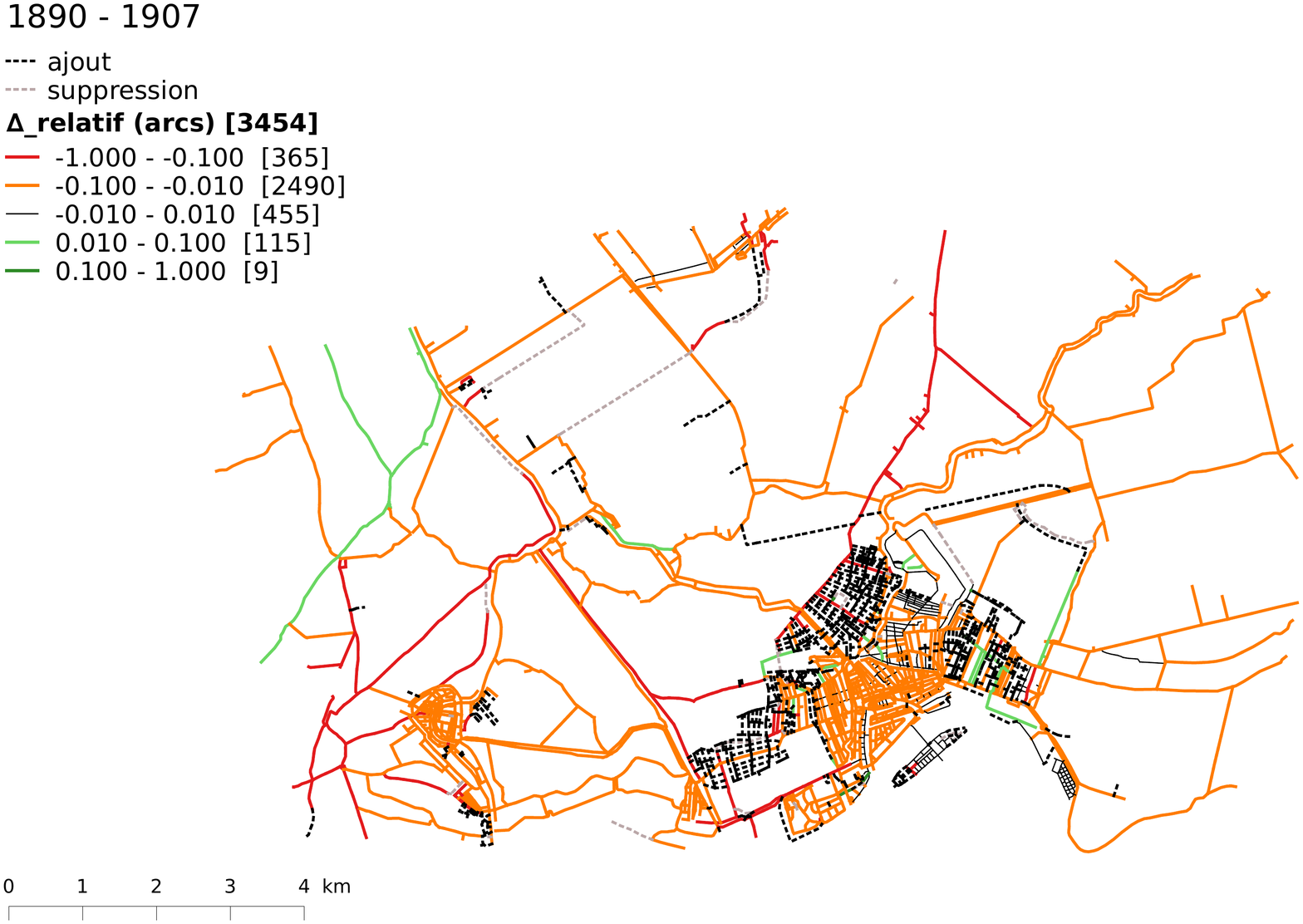}
        \caption{Étude cartographique de $\Delta_{relatif}$ sur la période 1890 - 1907.}
        \label{fig:diff_re2_1890}
    \end{figure}
    
     \begin{figure}[c]
     \centering
        \includegraphics[width=0.6\textwidth]{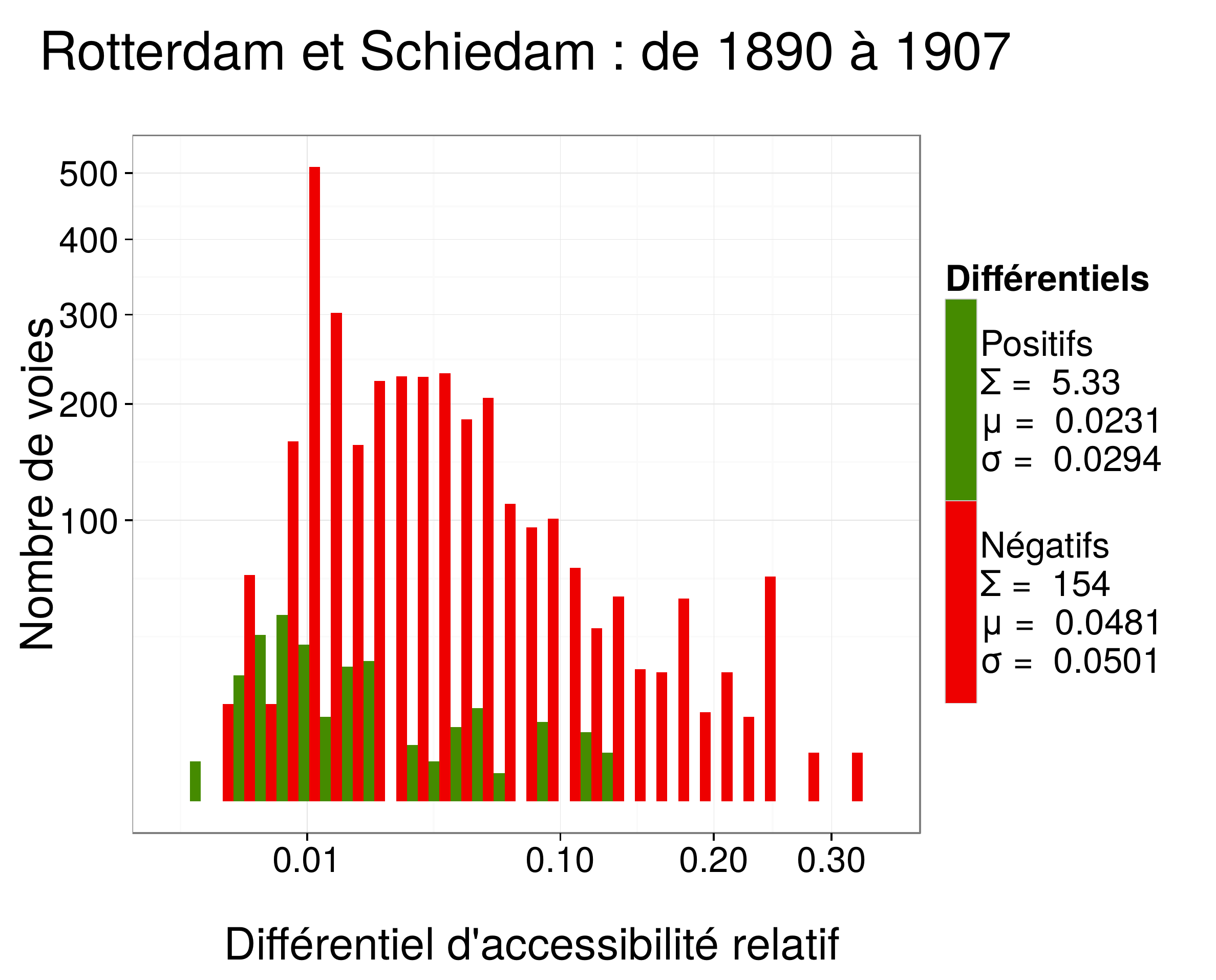}
        \caption{Étude statistique de $\Delta_{relatif}$ sur la période 1890 - 1907. \\ $\Sigma$ : somme ; $\mu$ : moyenne ; $\sigma$ : écart-type}
        \label{fig:diff_re2_1890_stat}
    \end{figure}

\clearpage 
    \begin{figure}[c]
    \centering
        \includegraphics[width=0.8\textwidth]{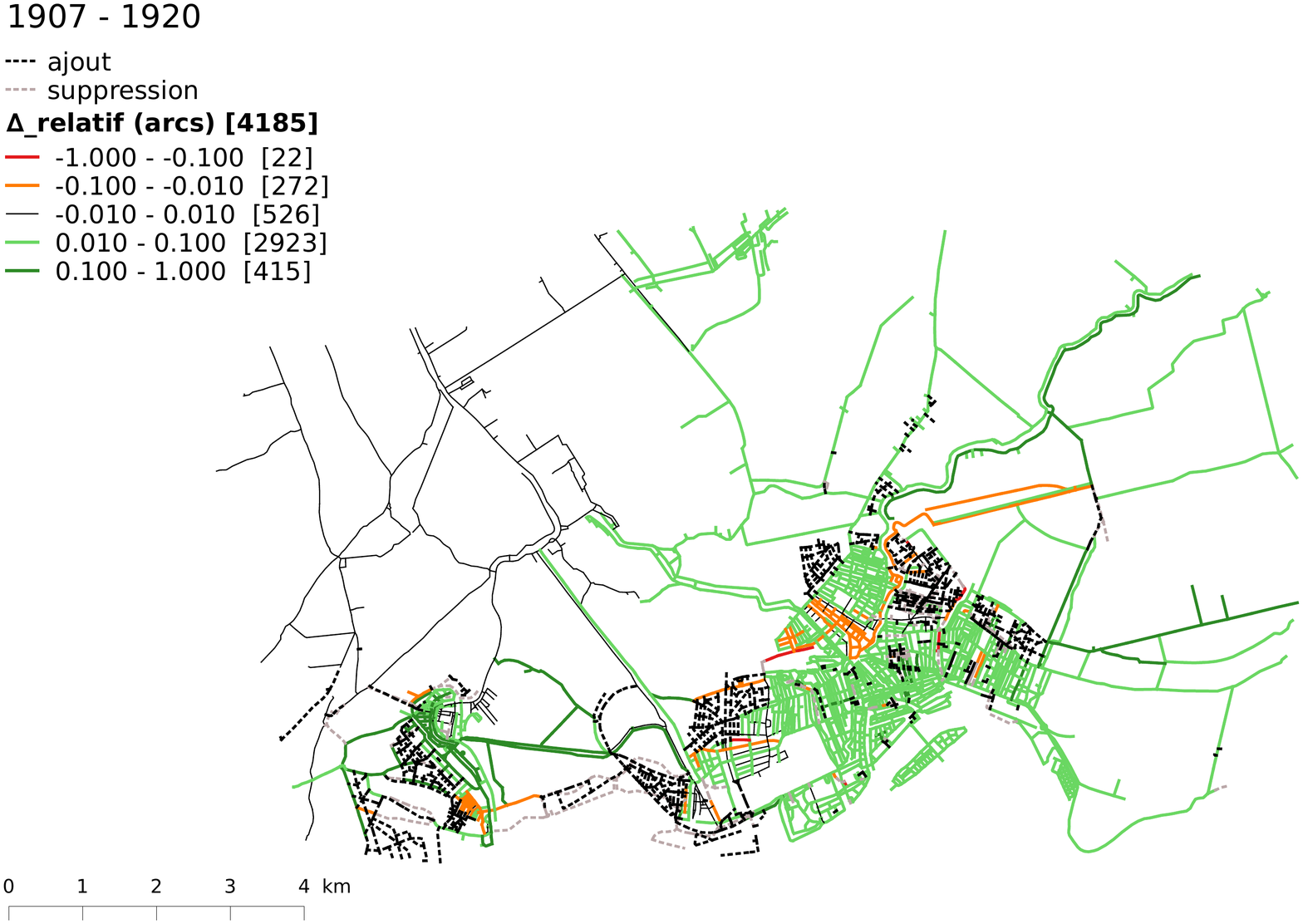}
        \caption{Étude cartographique de $\Delta_{relatif}$ sur la période 1907 - 1920.}
        \label{fig:diff_re2_1907}
    \end{figure}
    
     \begin{figure}[c]
     \centering
        \includegraphics[width=0.6\textwidth]{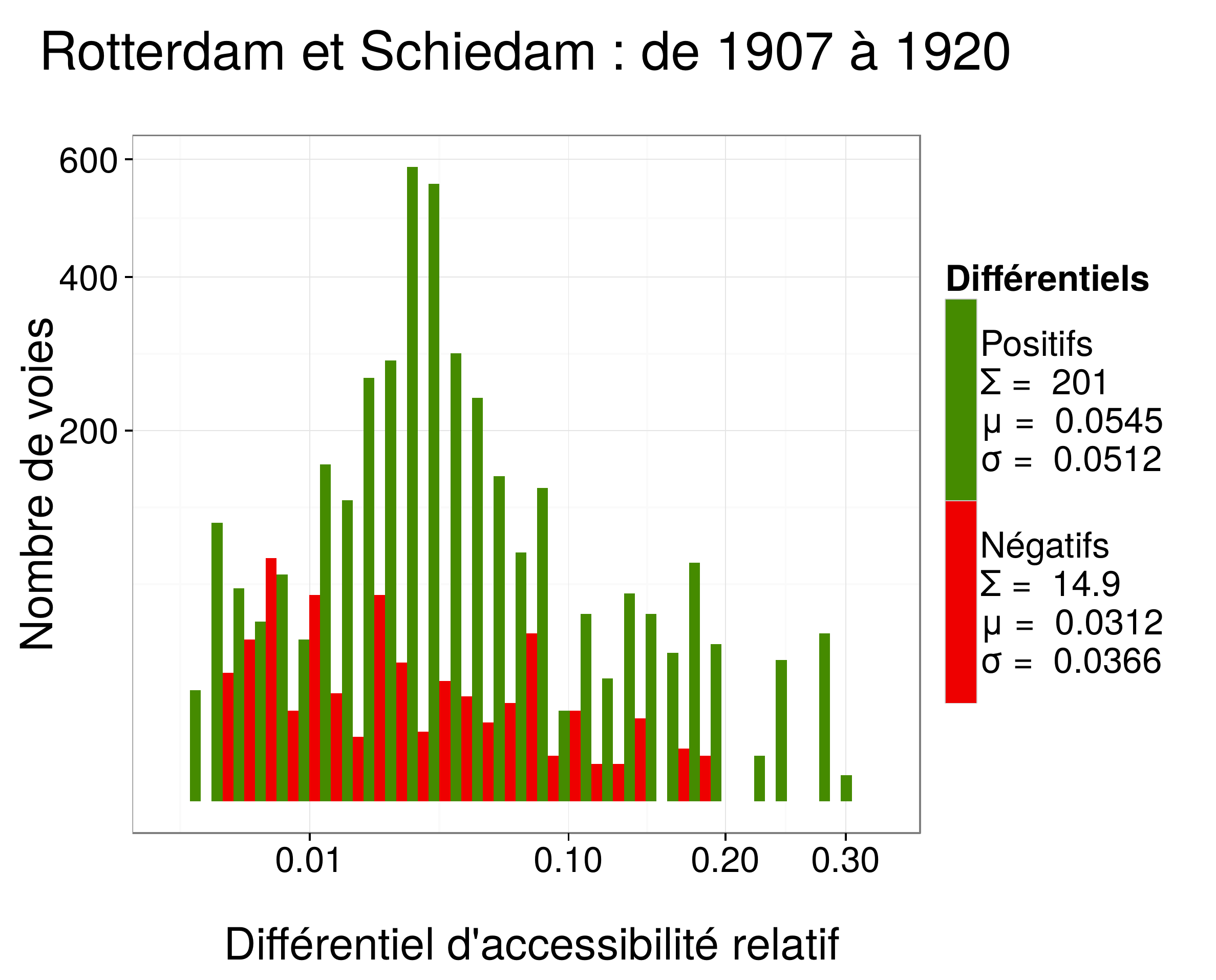}
        \caption{Étude statistique de $\Delta_{relatif}$ sur la période 1907 - 1920. \\ $\Sigma$ : somme ; $\mu$ : moyenne ; $\sigma$ : écart-type}
        \label{fig:diff_re2_1907_stat}
    \end{figure}

\clearpage 
     \begin{figure}[c]
     \centering
        \includegraphics[width=0.8\textwidth]{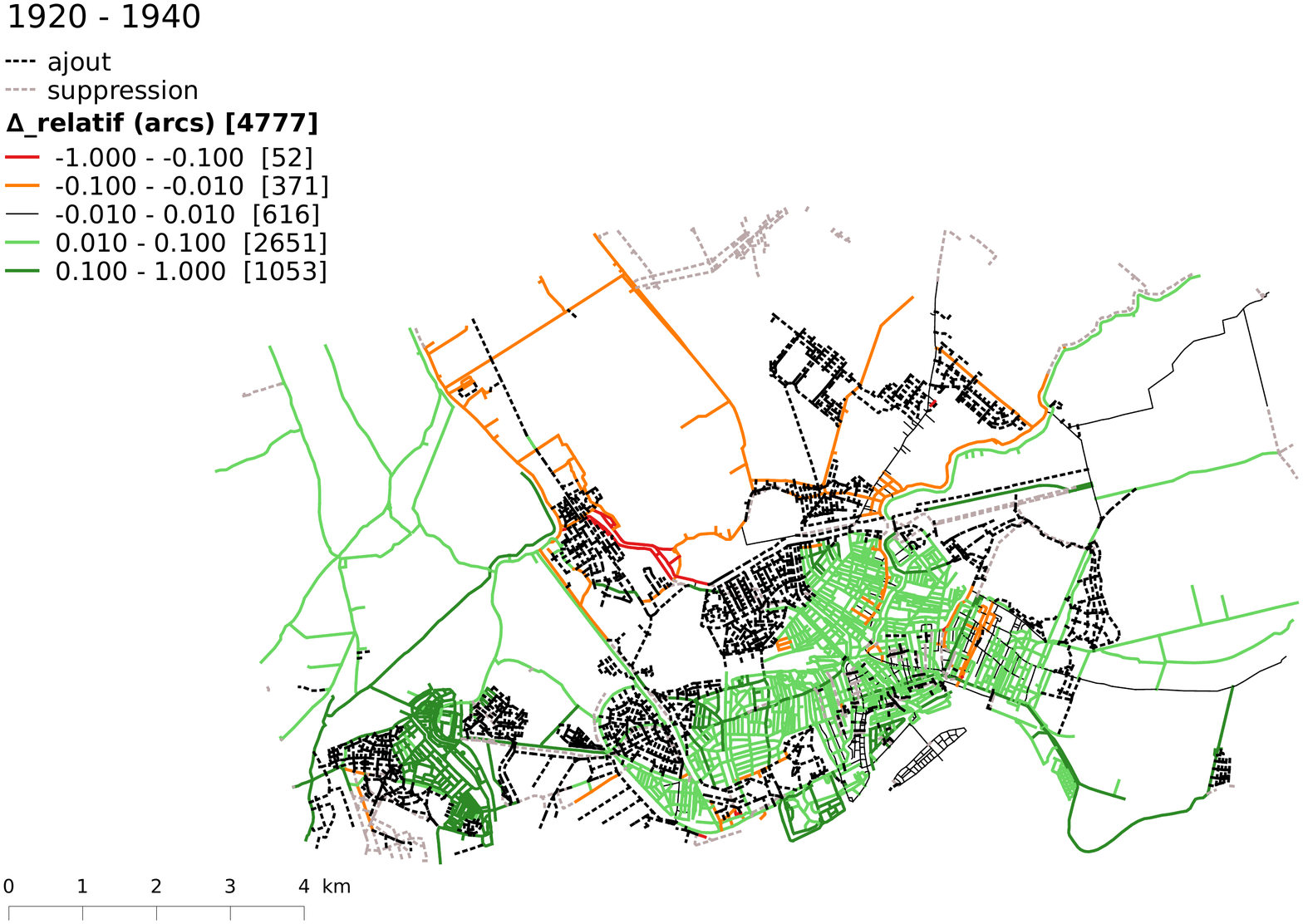}
        \caption{Étude cartographique de $\Delta_{relatif}$ sur la période 1920 - 1940.}
        \label{fig:diff_re2_1920}
    \end{figure}
    
     \begin{figure}[c]
     \centering
        \includegraphics[width=0.6\textwidth]{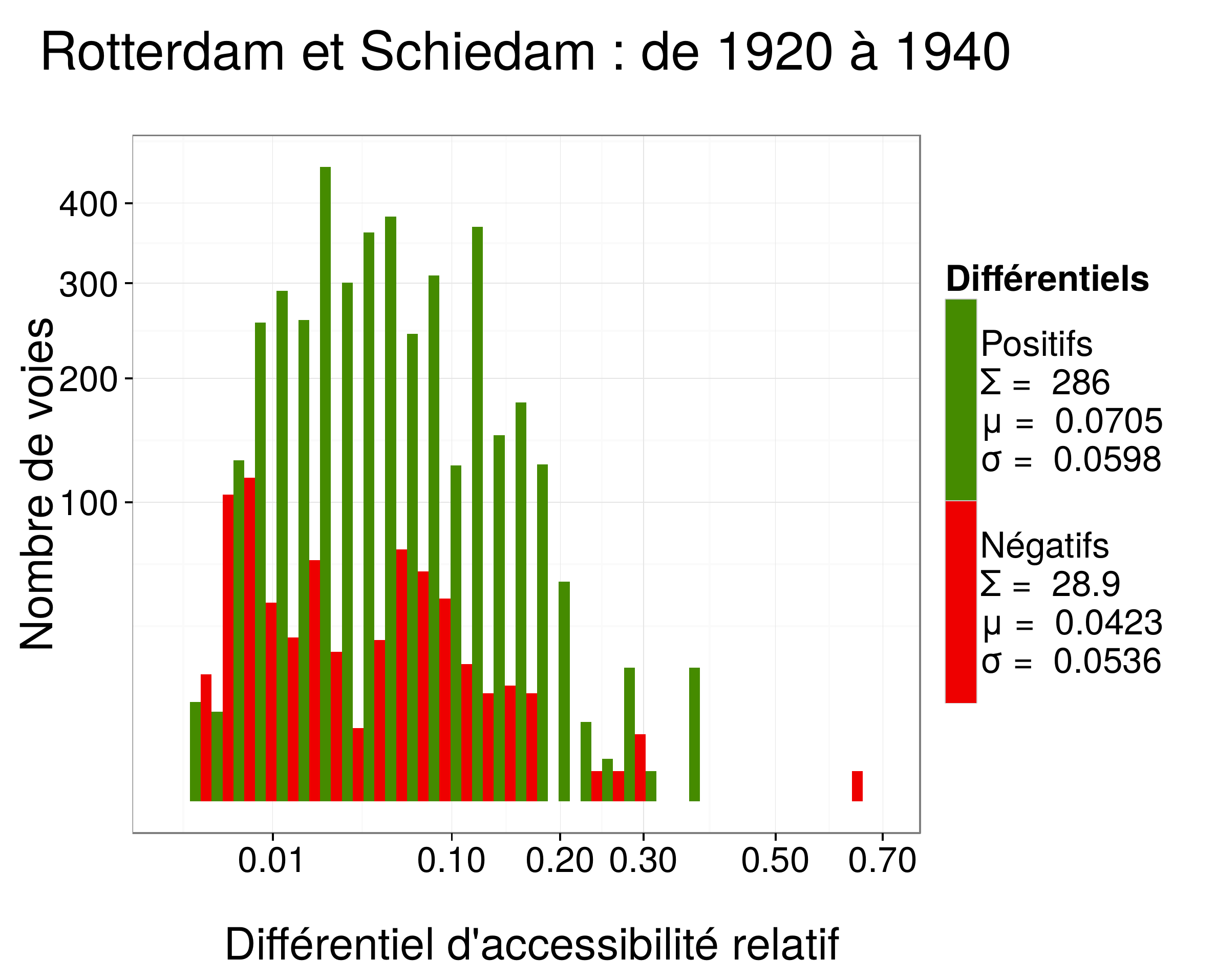}
        \caption{Étude statistique de $\Delta_{relatif}$ sur la période 1920 - 1940. \\ $\Sigma$ : somme ; $\mu$ : moyenne ; $\sigma$ : écart-type}
        \label{fig:diff_re2_1920_stat}
    \end{figure}

\clearpage 
    
     \begin{figure}[c]
     \centering
        \includegraphics[width=0.6\textwidth]{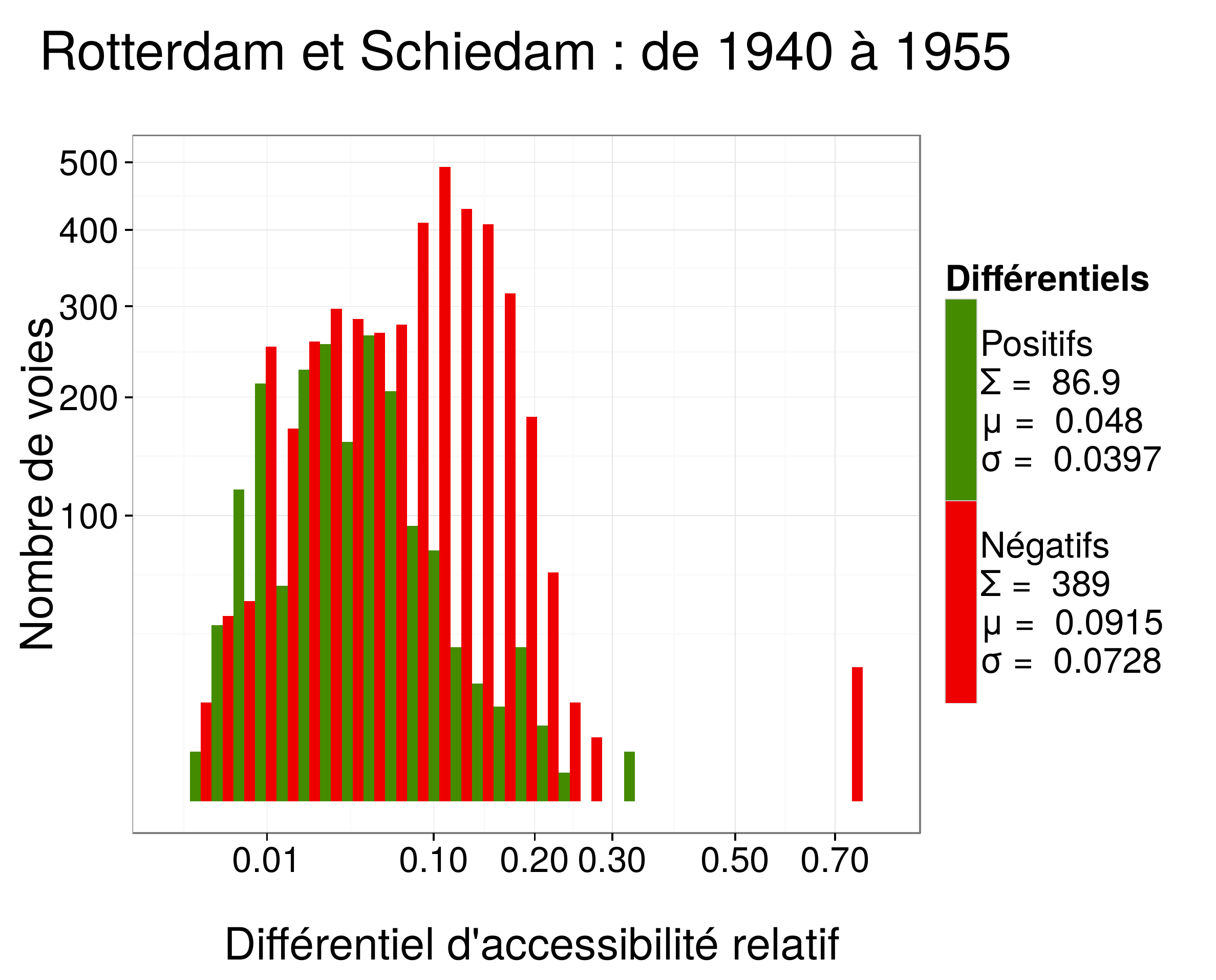}
        \caption{Étude statistique de $\Delta_{relatif}$ sur la période 1940 - 1955. \\ $\Sigma$ : somme ; $\mu$ : moyenne ; $\sigma$ : écart-type}
        \label{fig:diff_re2_1940_stat}
    \end{figure}

\clearpage
\subsubsection{Cartes Rotterdam Nord redécoupé}

	 \begin{figure}[h]
	 \centering
        \includegraphics[width=0.55\textwidth]{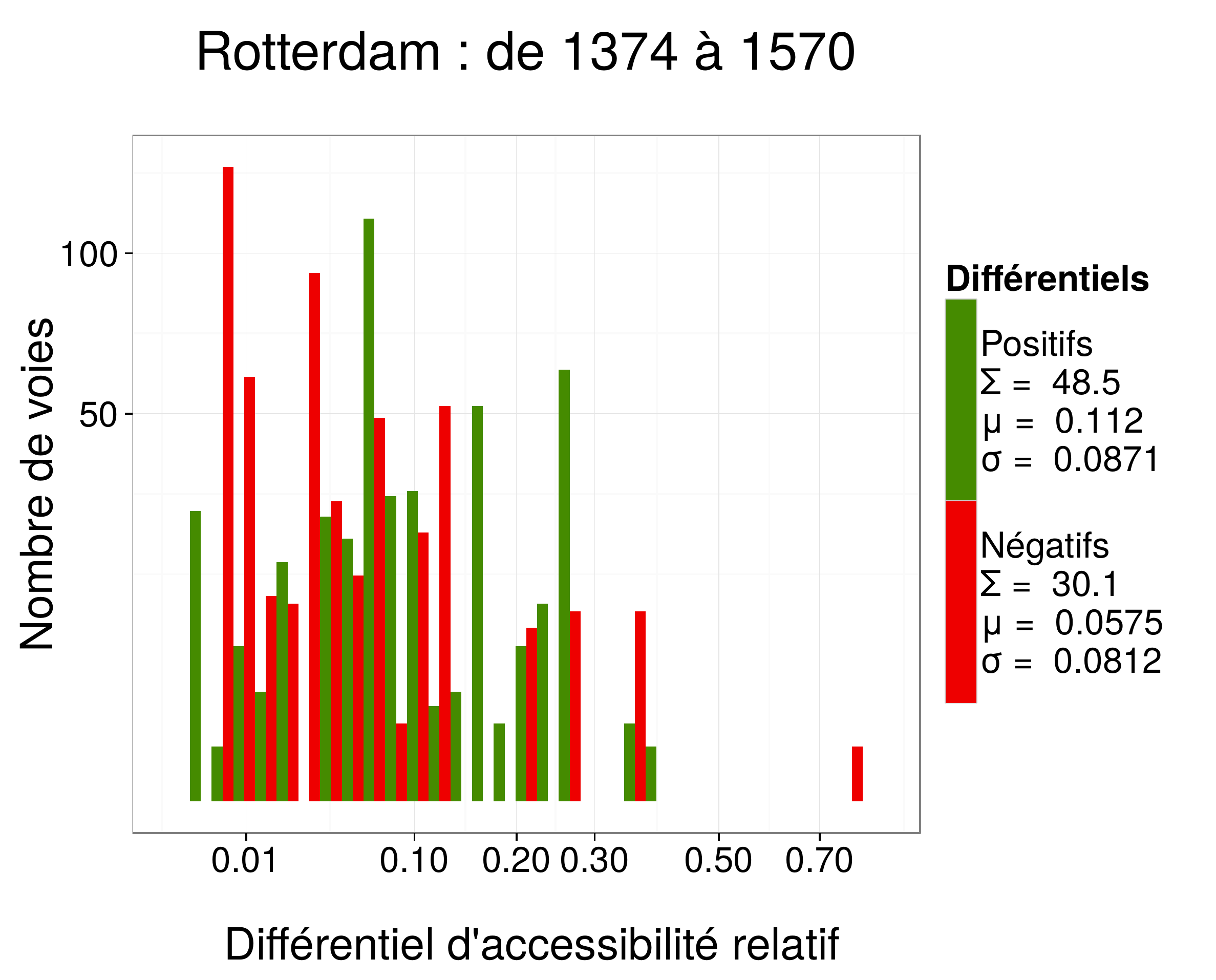}
        \caption{Étude statistique de $\Delta_{relatif}$ sur la période 1374 - 1570. \\ $\Sigma$ : somme ; $\mu$ : moyenne ; $\sigma$ : écart-type}
        \label{fig:diff_rd_1374_stat}
    \end{figure}

\clearpage 

	 \begin{figure}[c]
	 \centering
        \includegraphics[width=0.6\textwidth]{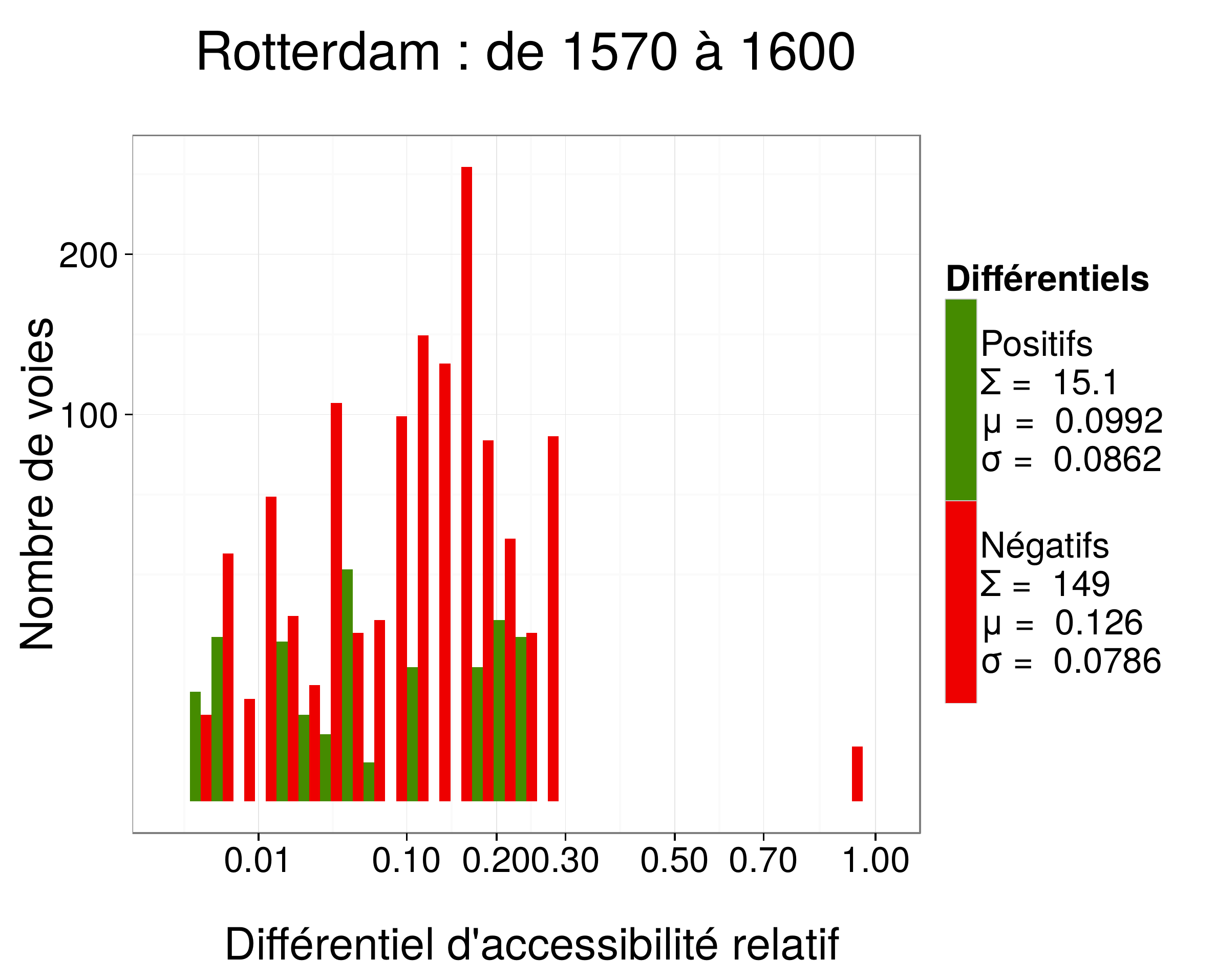}
        \caption{Étude statistique de $\Delta_{relatif}$ sur la période 1570 - 1600. \\ $\Sigma$ : somme ; $\mu$ : moyenne ; $\sigma$ : écart-type}
        \label{fig:diff_rd_1570_stat}
    \end{figure}

\clearpage 

	 \begin{figure}[c]
	 \centering
        \includegraphics[width=0.6\textwidth]{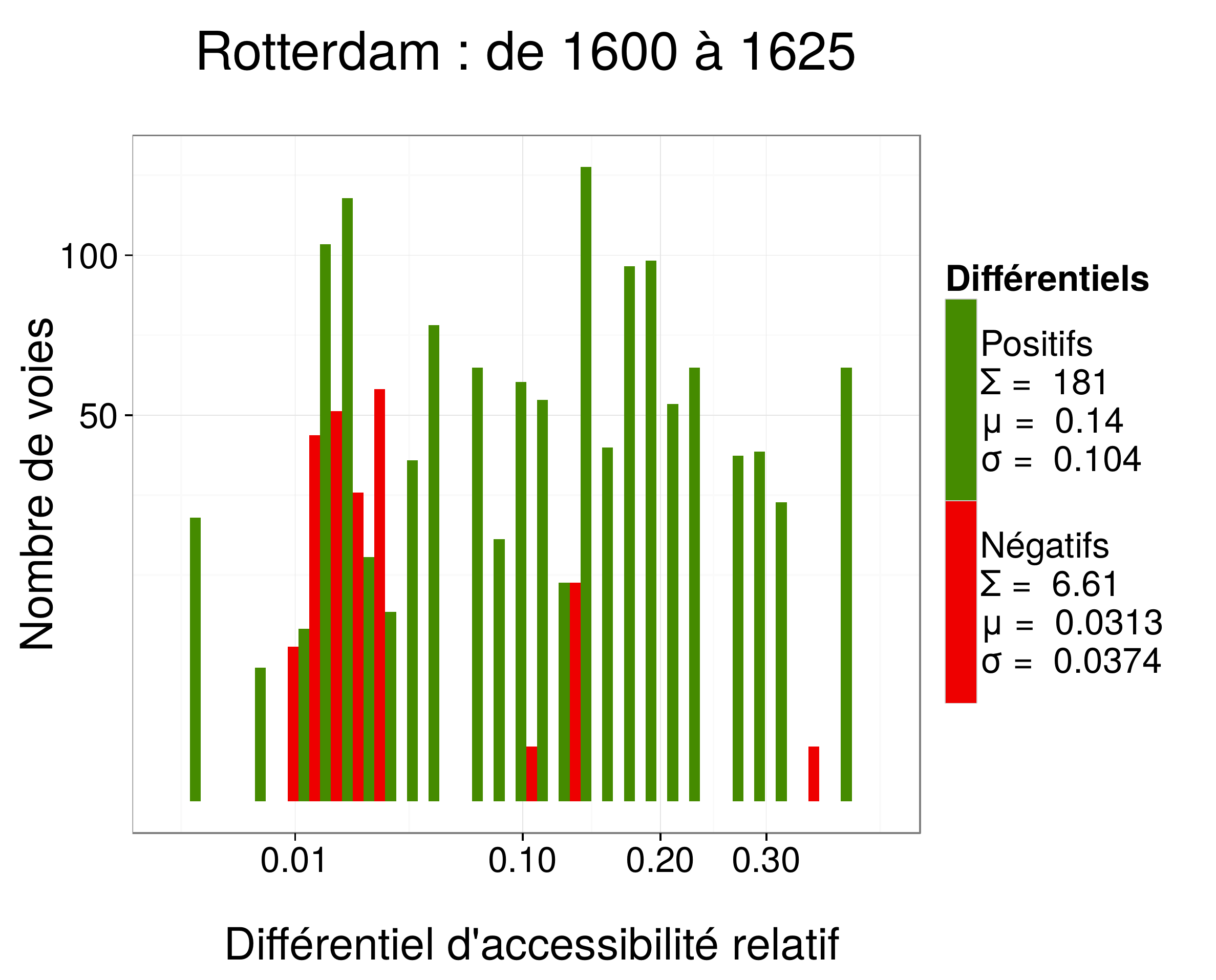}
        \caption{Étude statistique de $\Delta_{relatif}$ sur la période 1600 - 1625. \\ $\Sigma$ : somme ; $\mu$ : moyenne ; $\sigma$ : écart-type}
        \label{fig:diff_rd_1600_stat}
    \end{figure}

\clearpage 

	 \begin{figure}[c]
	 \centering
        \includegraphics[width=0.6\textwidth]{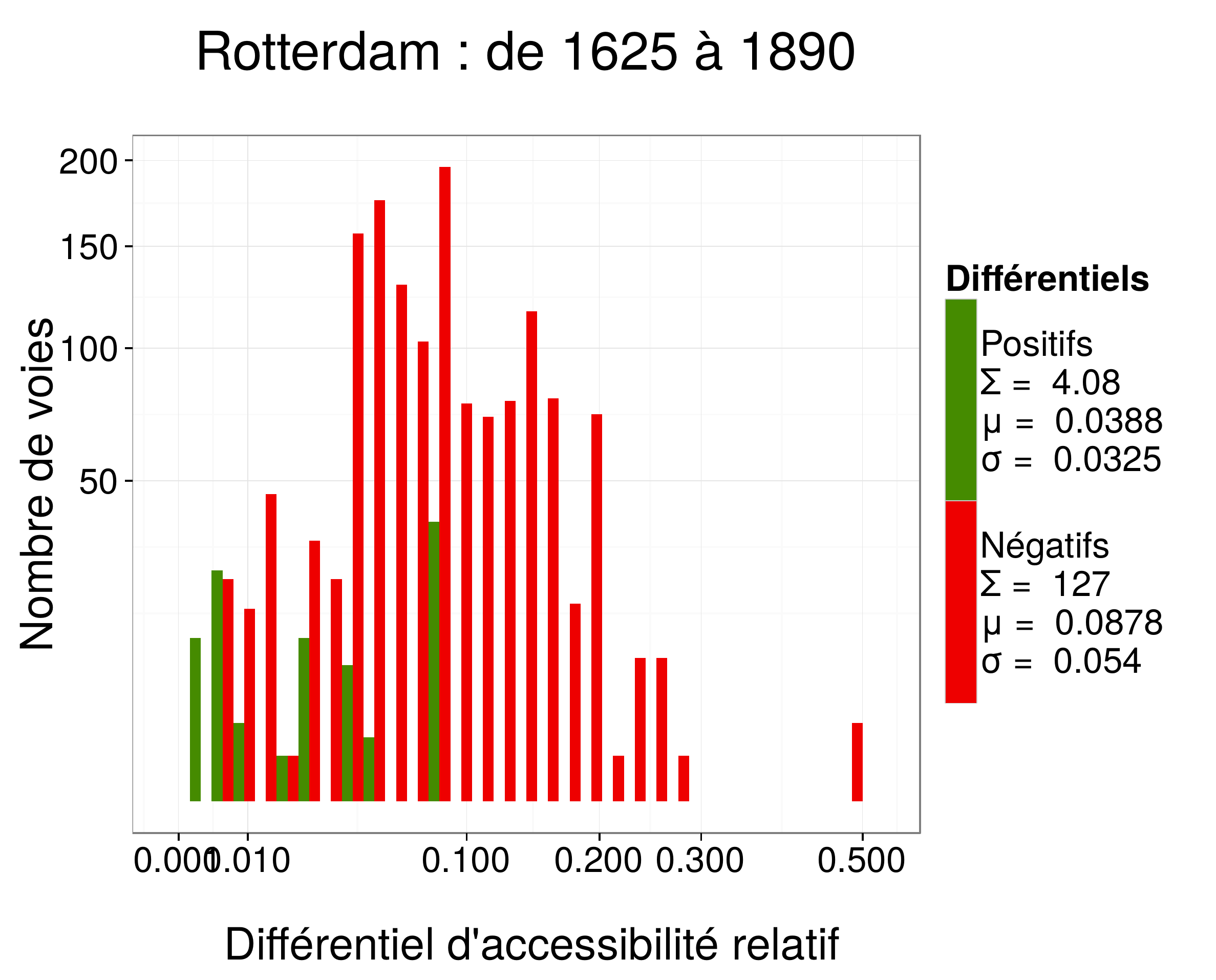}
        \caption{Étude statistique de $\Delta_{relatif}$ sur la période 1625 - 1890. \\ $\Sigma$ : somme ; $\mu$ : moyenne ; $\sigma$ : écart-type}
        \label{fig:diff_rd_1625_stat}
    \end{figure}

\clearpage 

	 \begin{figure}[c]
	 \centering
        \includegraphics[width=0.6\textwidth]{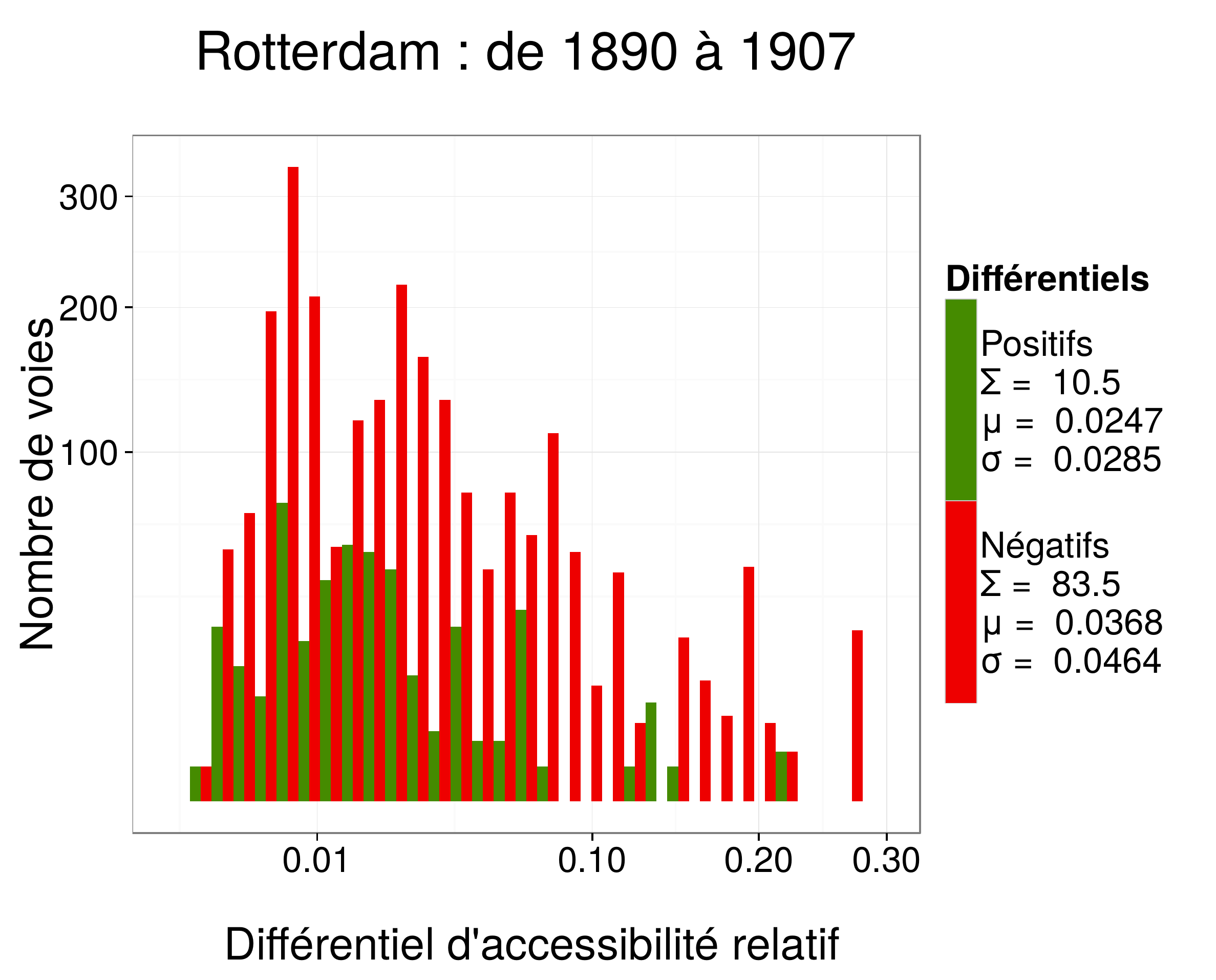}
        \caption{Étude statistique de $\Delta_{relatif}$ sur la période 1890 - 1907. \\ $\Sigma$ : somme ; $\mu$ : moyenne ; $\sigma$ : écart-type}
        \label{fig:diff_rd_1890_stat}
    \end{figure}

\clearpage 

	 \begin{figure}[c]
	 \centering
        \includegraphics[width=0.6\textwidth]{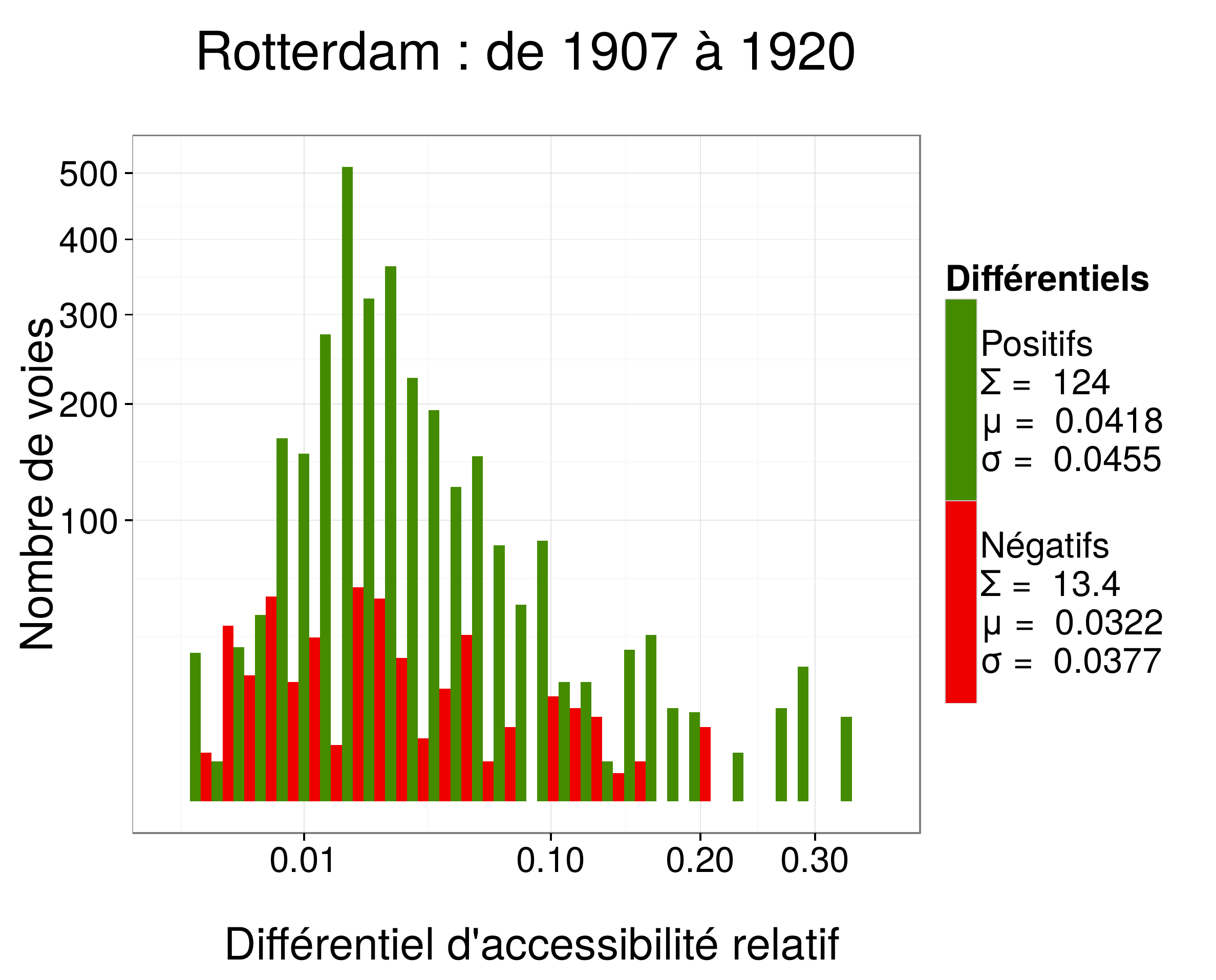}
        \caption{Étude statistique de $\Delta_{relatif}$ sur la période 1907 - 1920. \\ $\Sigma$ : somme ; $\mu$ : moyenne ; $\sigma$ : écart-type}
        \label{fig:diff_rd_1907_stat}
    \end{figure}

\clearpage 

	 \begin{figure}[c]
	 \centering
        \includegraphics[width=0.6\textwidth]{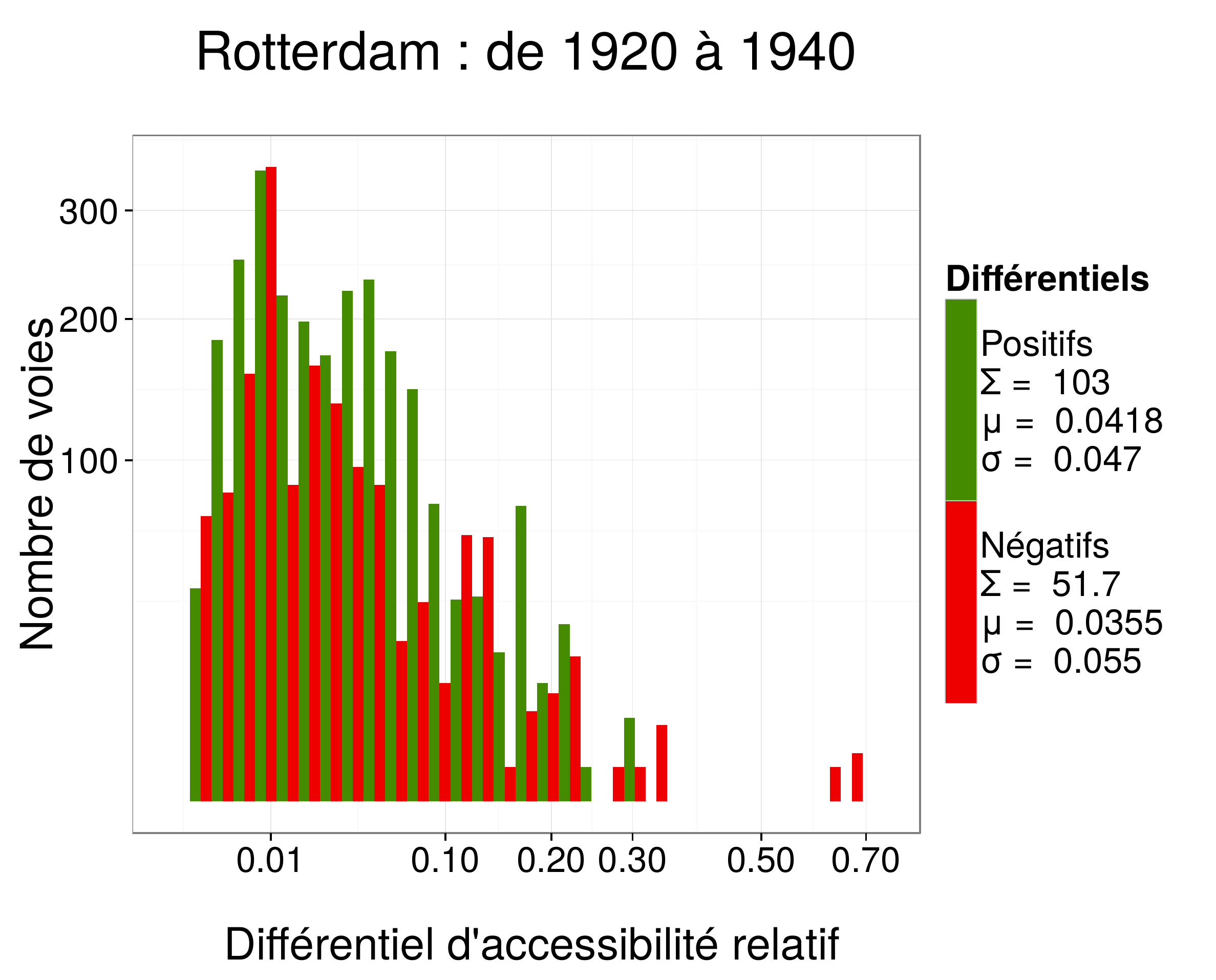}
        \caption{Étude statistique de $\Delta_{relatif}$ sur la période 1920 - 1940. \\ $\Sigma$ : somme ; $\mu$ : moyenne ; $\sigma$ : écart-type}
        \label{fig:diff_rd_1920_stat}
    \end{figure}

\clearpage 

\FloatBarrier
\subsection{Analyse de projets urbains}

L'analyse quantitative que nous avons faite sur différentes périodes peut également être faite dans une perspective de prospection afin d'établir la pertinence de projets urbains. Le service d'urbanisme d'Avignon nous en a proposé plusieurs afin que nous évaluions leurs impacts sur l'accessibilité de l'ensemble du territoire autour de la ville. Nous présentons deux de ces projets ici : \enquote{Leo2}, qui consiste à la création d'un pont sur la Durance, et \enquote{Raoul}, qui établit une connexion au Sud de la Ville. Nous cartographions l'impact de ces deux projets sur le territoire selon une échelle de couleurs légèrement différente : nous détaillons les variations les plus petites afin de pouvoir apprécier l'influence générale des projets. 

\begin{figure}[c]
     \centering
        \includegraphics[width=0.8\textwidth]{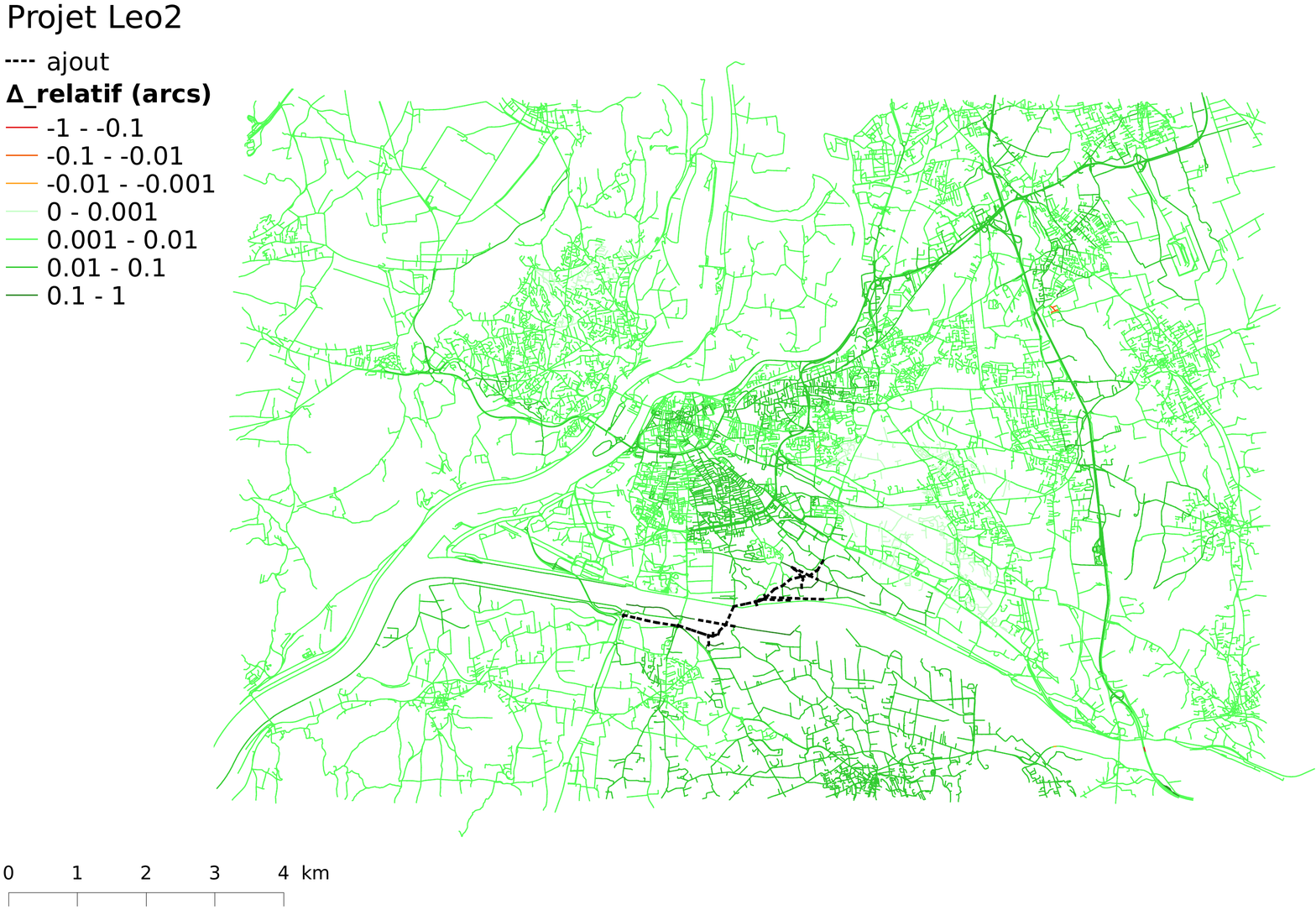}
        \caption{Carte de $\Delta_{relatif}$ sur le réseau du territoire d'Avignon avec et sans projet \enquote{Leo2}.}
        \label{fig:proj_avleo2_diff}
    \end{figure}

\begin{figure}[c]
     \centering
        \includegraphics[width=0.8\textwidth]{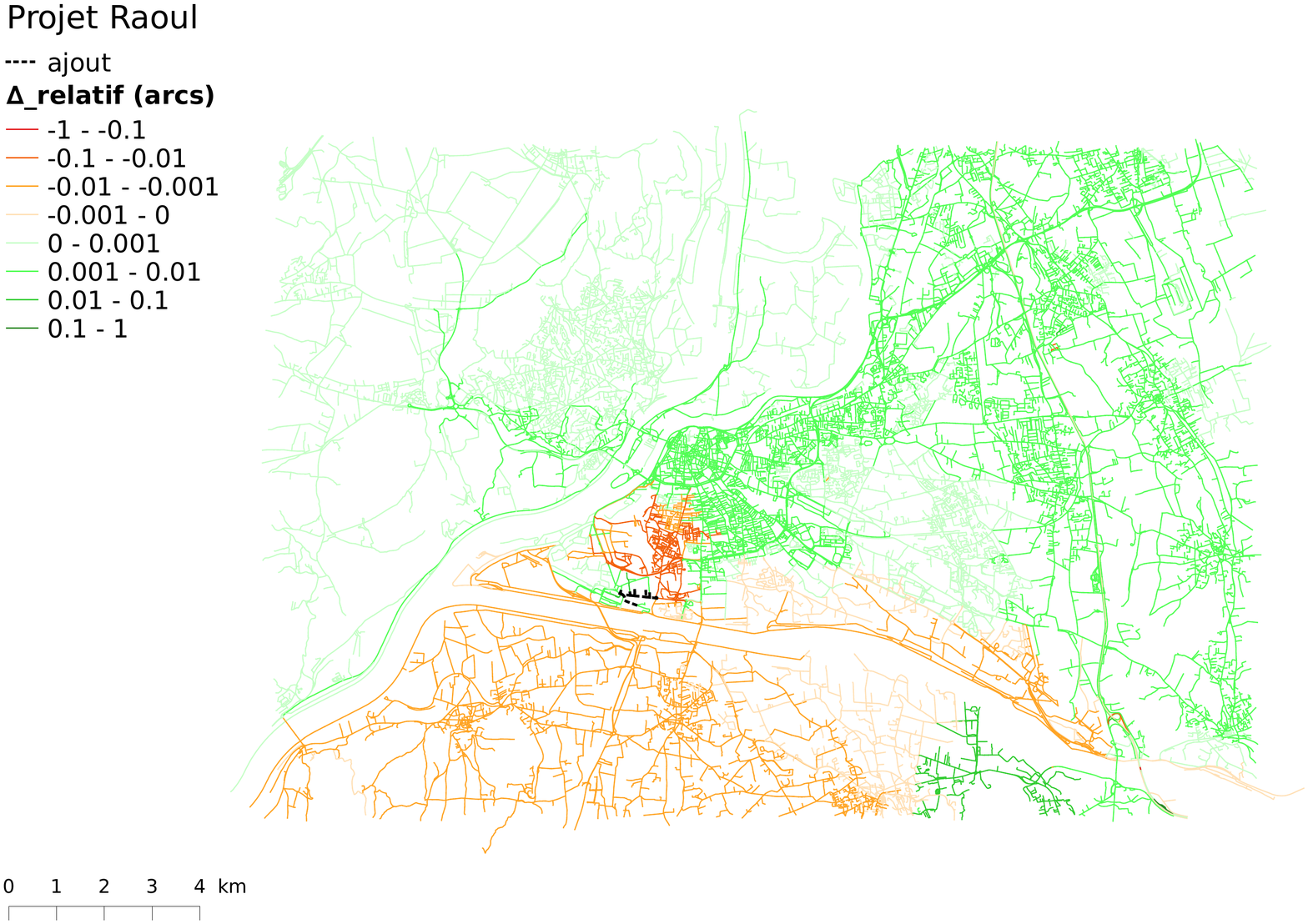}
        \caption{Carte de $\Delta_{relatif}$ sur le réseau du territoire d'Avignon avec et sans projet \enquote{Raoul}.}
        \label{fig:proj_avraoul_diff}
    \end{figure}

Dans notre modèle, le projet \enquote{Leo2} a un impact globalement positif sur l'accessibilité du territoire (figure \ref{fig:proj_avleo2_diff}). Il renforce les axes de desserte principaux de l'espace et améliore l'accessibilité topologique de l'Ouest du Rhône (vers Villeneuve-lès-Avignon) et du Nord-Est du territoire (vers Le Pontet). Le projet \enquote{Raoul}, quant à lui, n'a pas un impact aussi positif que le précédent (figure \ref{fig:proj_avraoul_diff}). Sur le modèle, il améliore globalement l'accès au Nord et au Sud-Est de la carte mais dessert celui du Sud-Ouest en allongeant les distances topologiques pour cette zone. Ceci s'explique par la modification de la géométrie d'une voie qui raccordait la partie au Nord du projet à la structure de contournement de la ville. Ce changement, ajoute un \textit{tournant} supplémentaire entre les deux parties du graphe, qui, sommé sur toutes les voies, engendre l'influence observée. Ces deux projets, à l'échelle d'un large territoire autour de la ville d'Avignon, n'ont donc pas le même apport topologique pour l'espace qu'ils concernent. 

Les cartes de closeness \enquote{brute} sont en annexe \ref{ann:sec_cartedia_projurb}. Dans le tableau réunissant les informations statistiques sur les influences des deux projets sur le territoire d'Avignon (tableau \ref{tab:proj_delta}) nous remarquons que si l'impact du projet \enquote{Raoul} est moins homogène que celui créé par \enquote{Leo2}, il est également beaucoup moins important. En sommant les variations relatives pour chaque voie, nous obtenons pour ce projet une variation totale de 68.45 contre 300.69 pour le projet \enquote{Leo2}. En moyenne, la variation de l'accessibilité des voies avec le projet \enquote{Raoul} ne représente que 0.7\% de leur valeur finale. L'impact négatif sur les accessibilités est donc à considérer relativement : il s'agit principalement de très faibles variations.

\begin{table}
\begin{center}
{ \small
\begin{tabular}{|c|r|r|r|r|r|r|}
\hline

projet & $N_{arcs(ajoutés)}$ & $L_{ajoutée}$ & $\overline{\Delta_{relatif}}$ & $\sigma(\Delta_{relatif})$  & $max \vert \Delta_{relatif}\vert$ & $\sum \vert \Delta_{relatif} \vert$ \\ \hline

Leo2 & 37 & 4 721 m & 0.0107 & 0.0161 & 0.5970 & 300.69 \\ \hline
Raoul & 13 & 1 146 m & 0.0007 & 0.0079 & 0.1216 & 68.45 \\ \hline

\end{tabular}
}
\end{center}
\caption{Détail statistique des variations relatives $\Delta_{relatif}$ pour les deux projets présentés.}
\label{tab:proj_delta}
\end{table}

        \begin{figure}[c]
         \centering
        \includegraphics[width=0.6\textwidth]{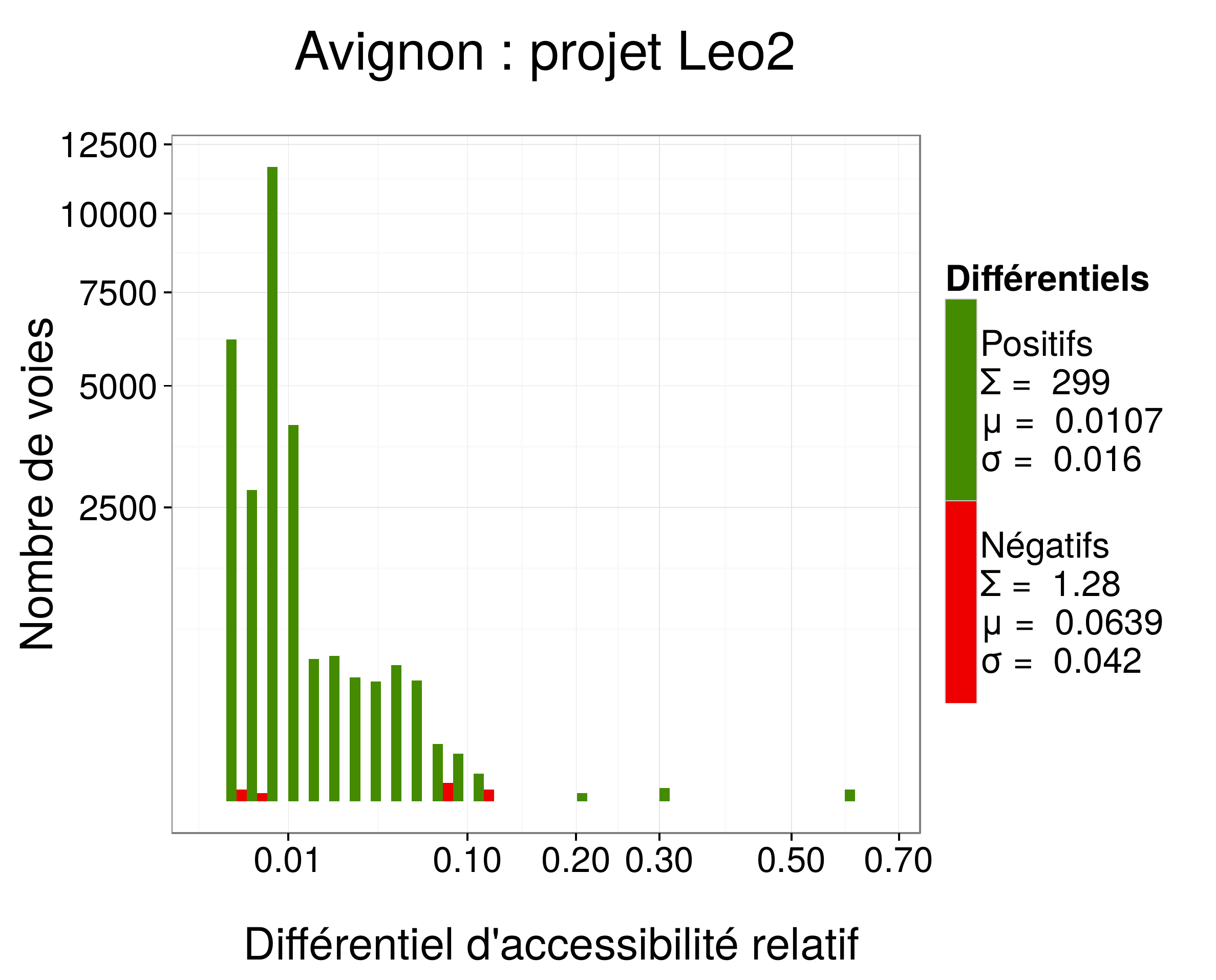}
        \caption{Détail statistique de $\Delta_{relatif}$ sur le réseau du territoire d'Avignon avec et sans projet \enquote{Leo2}.}
        \label{fig:proj_avleo2_diff_stat}
    \end{figure}  
    
    \begin{figure}[c]
     \centering
        \includegraphics[width=0.6\textwidth]{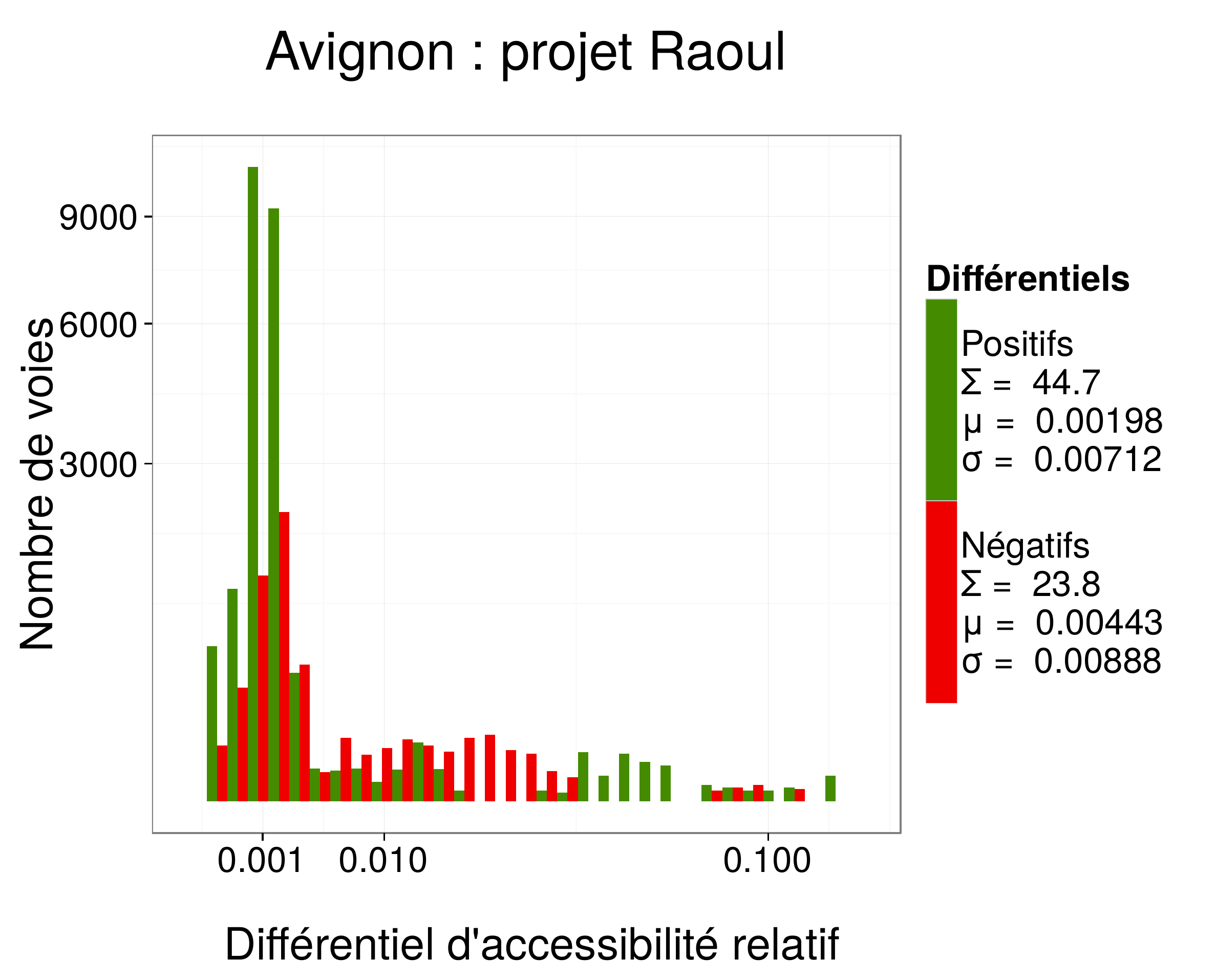}
        \caption{Détail statistique de $\Delta_{relatif}$ sur le réseau du territoire d'Avignon avec et sans projet \enquote{Raoul}.}
        \label{fig:proj_avraoul_diff_stat}
    \end{figure}
    
\FloatBarrier

Cette possibilité d'appuyer les décisions urbaines par des données quantitatives intéresse les collectivités territoriales (notamment celle d'Avignon, ville au cœur de notre étude). C'est une méthode d'appui capable de nous renseigner sur la proximité topologique entre les voies d'un territoire. Si elle est utile à une analyse prospective, elle l'est également pour évaluer l'impact de projets appartenant au passé.

\begin{figure}[h]
    \centering
        \includegraphics[width=0.8\textwidth]{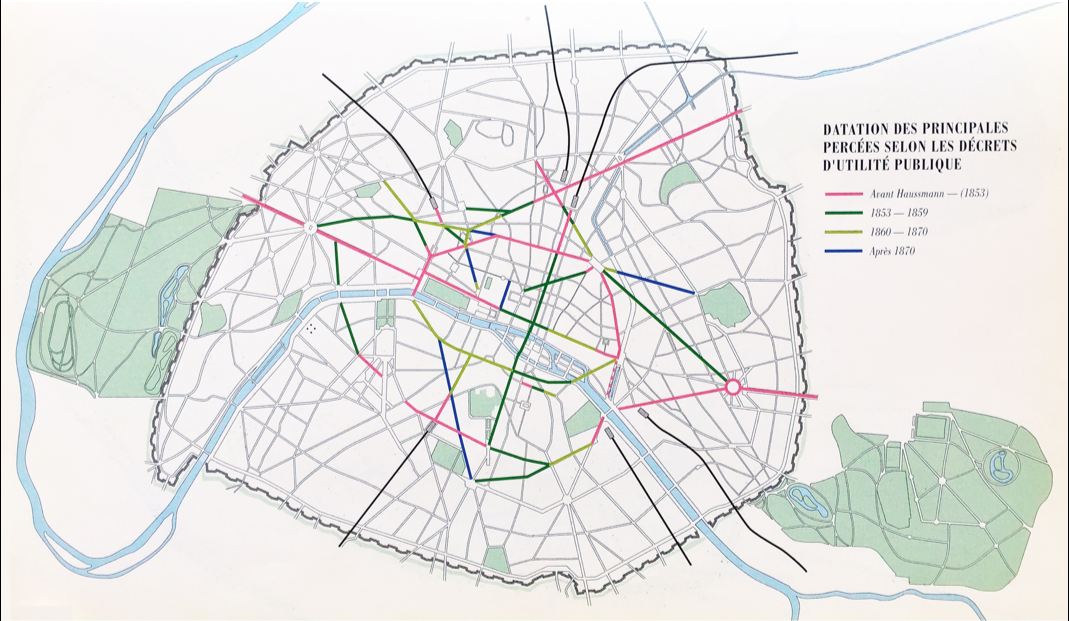}
        \caption{Détail des percements extraits. \\ source : \citep{pinon1991paris}}
        \label{fig:proj_paris_perc}
    \end{figure}

\begin{figure}[c]
    \centering
        \includegraphics[width=0.8\textwidth]{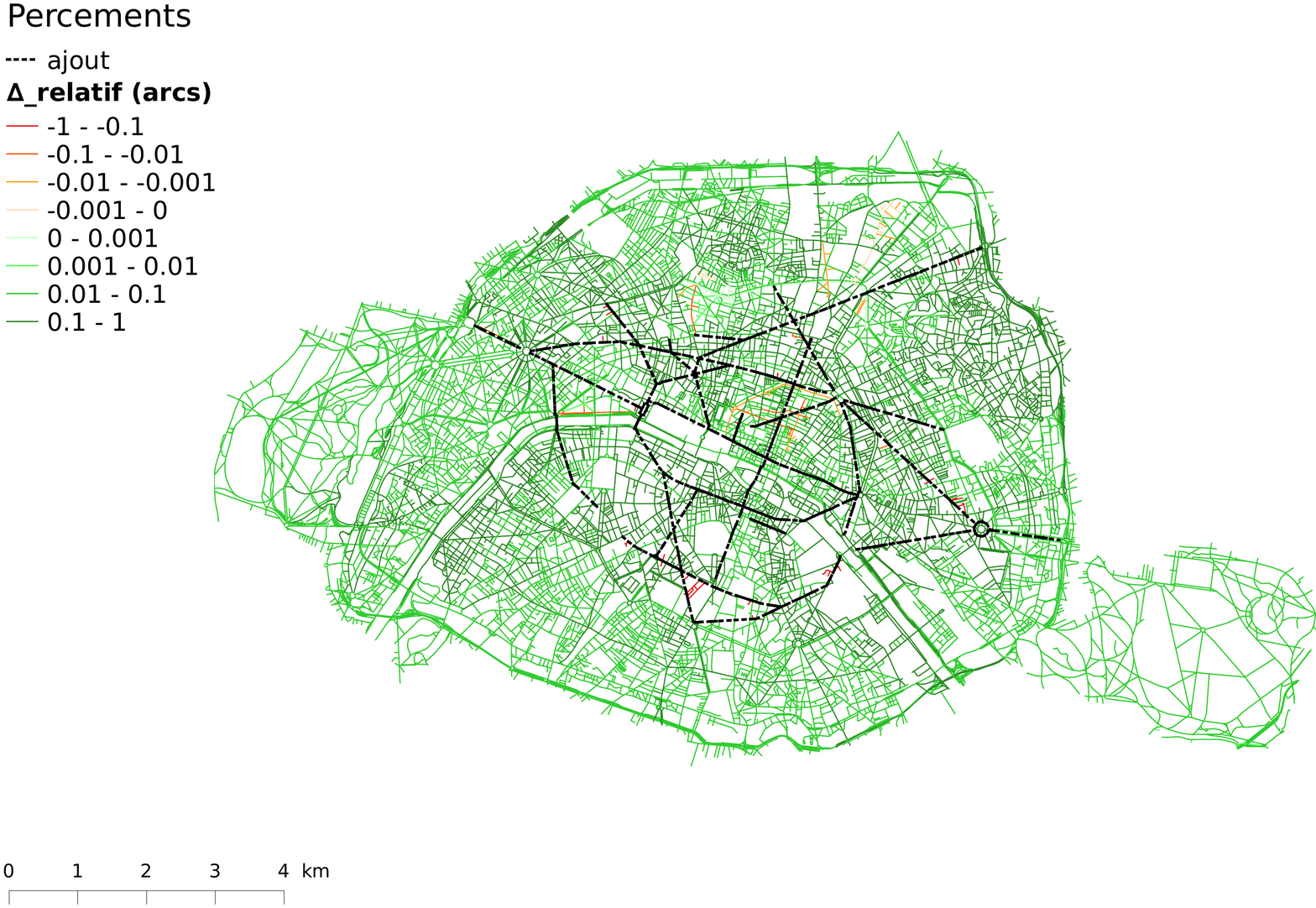}
        \caption{Étude cartographique de $\Delta_{relatif}$ sur sur le réseau de Paris avec et sans les percements.}
        \label{fig:proj_paris_diff}
    \end{figure}

Nous nous sommes ainsi penchés sur les percements dits \enquote{Haussmanniens} au sein de la ville de Paris (figure \ref{fig:proj_paris_perc}). Nous avons utilisé la carte d'Alphand Poubelle (1885) pour en extraire les percements (vectorisation faite par l'équipe de Maurizio Gribaudi, extraction faite par Babak Atashinbar). Le $\Delta_{relatif}$ résultant des calculs faits entre les réseaux avant et après percements se révèle être largement positif (figure \ref{fig:proj_paris_diff} et tableau \ref{tab:proj_paris}). Nous pouvons observer que toute la ville profite d'une amélioration d'accessibilité procurée par les percements, avec de grande valeurs. Nous notons, sur certains quartiers, une influence d'ensemble, indépendante de leur proximité aux percements (comme à Belleville, dans le prolongement de la rue de Turbigo, ou Montmartre). Nous pouvons donc visualiser quantitativement l'amélioration de l'accessibilité offerte par les grandes avenues, et attester rétrospectivement de leur action bénéfique sur l'accessibilité au sein de la capitale.

\begin{table}
\begin{center}
{ \small
\begin{tabular}{|c|r|r|r|r|r|r|}
\hline
projet & $N_{arcs(ajoutés)}$ & $L_{ajoutée}$ & $\overline{\Delta_{relatif}}$ & $\sigma(\Delta_{relatif})$  & $max \vert \Delta_{relatif}\vert$ & $\sum \vert \Delta_{relatif} \vert$ \\ \hline

Percements & 847 & 58 951 m & 0.0996 & 0.0902 & 0.9819 & 2306.43 \\ \hline

\end{tabular}
}
\end{center}
\caption{Détail statistique des variation relatives $\Delta_{relatif}$ pour les percements Haussmanniens.}
\label{tab:proj_paris}
\end{table}

\clearpage

\begin{figure}[c]
    \centering
        \includegraphics[width=0.6\textwidth]{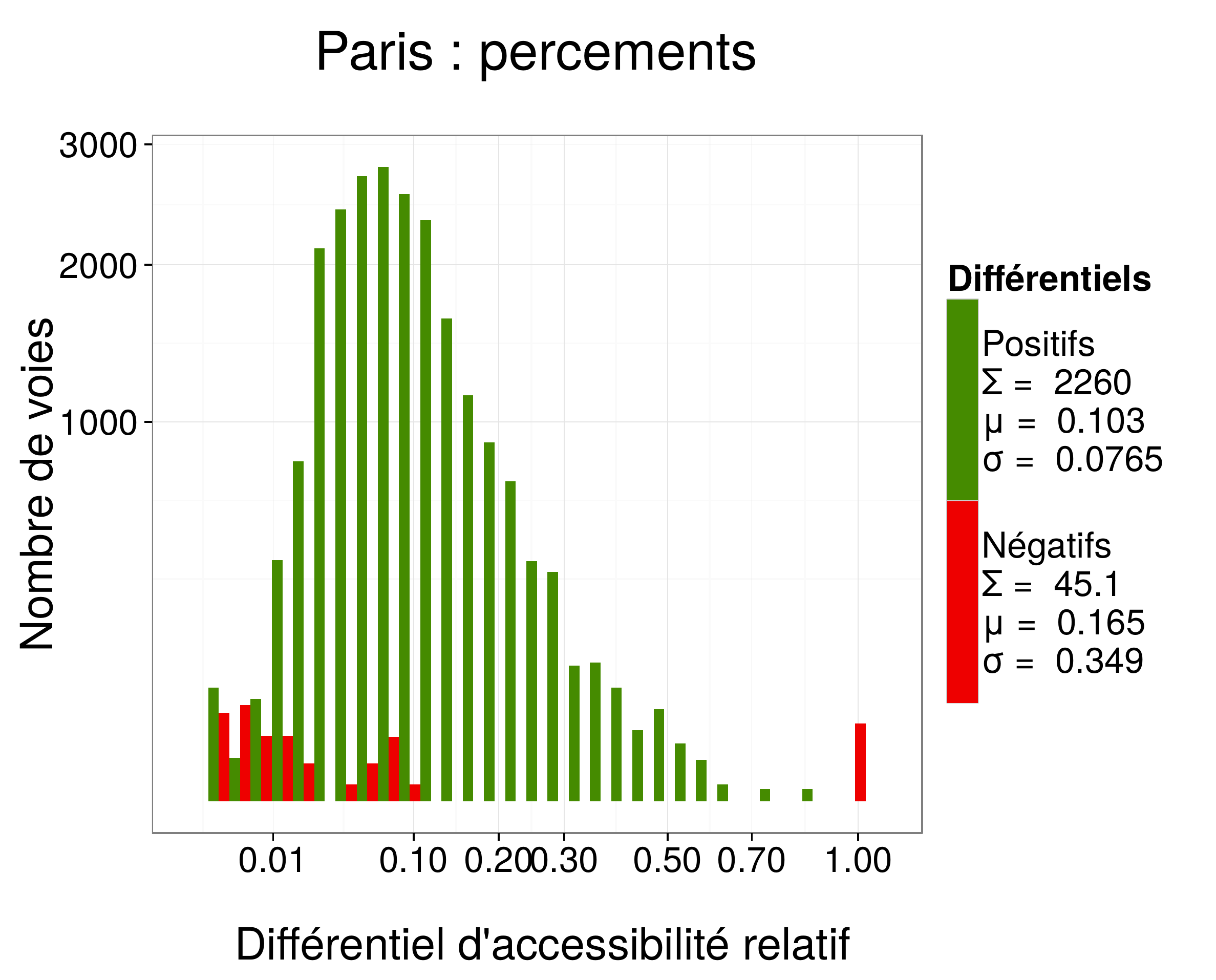}
        \caption{Étude statistique de $\Delta_{relatif}$ sur sur le réseau de Paris avec et sans les percements.}
        \label{fig:proj_paris_diffstat}
    \end{figure}

\clearpage{\pagestyle{empty}\cleardoublepage}
\chapter{Synthèse}
\minitoc
\markright{Synthèse de la deuxième partie}

\FloatBarrier
\section{Étude de robustesse du modèle utilisé}

La première partie de cette thèse a été consacrée à la construction d'une méthodologie de lecture. Avant d'en étudier l'application, un des objectifs de cette deuxième partie est d'en quantifier les limites. Nous avons ainsi identifié quatre points problématiques liés à la qualité et au choix des données :
\begin{itemize}
\item la précision de la numérisation des intersections
\item la finesse de vectorisation des arcs (et problèmes de généralisation liés)
\item les choix liés aux types des données
\item les choix liés aux délimitations spatiales des données
\end{itemize}

L'élaboration de la voie, à partir de règles locales, est fondée sur la géométrie du filaire numérisé aux intersections. Elle est donc sensible aux discontinuités. L'intégration dans le réseau d'un rond-point ou d'un décrochement aura donc pour effet de briser la continuité des voies et de les sectionner en plusieurs objets. Pour pallier ce problème, nous avons affiné notre méthode de construction pour permettre d'effacer les petites discontinuités qui viennent impacter nos analyses. Celle-ci est fondée sur la définition de \enquote{zones tampons}, que nous appelons également \textit{places}, autour des intersections. Suivant les paramètres choisis nous pouvons faire disparaître les décrochements ou ronds-points, joignant ainsi les axes entrecoupés. Il nous est également possible de prendre en compte les vraies places de la ville si nous en possédons la vectorisation.

Il est ainsi possible d'adapter la construction des voies à la problématique de recherche. La paramétrisation des zones tampons nécessite cependant de prendre en compte l'échelle du réseau sur lequel l'analyse est faite. Idéalement, nous pourrions projeter un ajustement automatique de la géométrie des zones tampons en fonction de l'indicateur d'espacement calculé sur le réseau. Les zones tampons auraient ainsi un diamètre faible dans les parties denses du graphe, qui pourra être plus important dans celles où le réseau est plus étiré. Il serait ainsi possible d'avoir des diamètres hétérogènes, en fonction de la partie du graphe sur laquelle le calcul est opéré.

D'autre part, les données que nous utilisons portent un attribut de \textit{type} spécifié dans les bases de données sources. Celui-ci est plus ou moins renseigné et rigoureux suivant l'utilisation de données professionnelles (comme celles issues de la \copyright BDTOPO de l'IGN pour les réseaux viaires) ou collaboratives (comme celles issues d'OpenStreetMap). Nous avons quantifié l'impact de la prise en compte de données secondaires, indépendantes des grandes structures. Nous avons ainsi pu conclure que le choix de leur inclusion, ou non, au graphe sur lequel est fait le calcul, a peu d'influence. Cela implique également que si un détail existe sur le réseau physique et que sa numérisation a été omise, cela n'aura pas d'impact sur les résultats observés.

Outre l'analyse qualitative de données, nous avons voulu tester l'incidence de leur découpage spatial. Dans les indicateurs primaires que nous avons définis comme pertinents pour l'analyse du graphe à travers les voies, deux sont calculés localement (le degré et l'orthogonalité). Nous pouvons donc considérer que ces deux indicateurs ne sont pas sensibles aux effets de bord car ils dépendent de la propre géométrie de l'objet et de celle de son entourage direct. Ne seront impactées par le découpage que les voies qui sont sectionnées par celui-ci.

L'indicateur primaire sur lequel nous portons notre attention est celui calculé en tenant compte de tout le réseau : la closeness. Nous travaillons ici sur sa valeur pondérée par la longueur (indicateur d'accessibilité). Nous quantifions les différences d'accessibilité d'un réseau en ajoutant autour de celui-ci un graphe plus ou moins large. Nous considérons quatre territoires différents, aux histoires et logiques de construction \textit{a priori} éloignées : Avignon, Paris, Barcelone et New-York. Nous délimitons plusieurs graphes au sein et/ou autour de ces villes dont les découpages sont très différents (circulaires, rectangulaires, selon les limites naturelles, etc). Tous les graphes étudiés révèlent une très faible variation d'accessibilité, quelle que soit l'étendue du réseau qui leur est ajoutée. Ce résultat contraste avec celui obtenu sur les arcs, où l'indicateur mettait en avant essentiellement le centre de l'emprise de découpage (cf Partie I, chapitres 3 et 4). Nous pouvons donc en conclure la stabilité de notre indicateur global, par son application à la voie, dans le cas d'un découpage \textit{rationnel} du réseau (sans briser les continuités évidentes).

Nous avons enfin étudié la sensibilité de notre modèle à la finesse de la vectorisation du réseau. Nous avons pour cela considéré un réseau très bruité : celui issu de la vectorisation automatique, à partir d'une photographie, de craquelures dans de l'argile. Le graphe obtenu fait apparaître un effet \enquote{bulle de savon} à chaque intersection, où les arcs décrochent leur dernier segment dans une direction perpendiculaire à ceux qui leur sont proches. Pour pallier ce biais numérique, nous appliquons la méthode complémentaire développée, en utilisant des zones tampons. Cependant, l'aspect bruité des arcs, comprenant de nombreux points annexes, peut les faire dévier au contact des zones calculées. Nous utilisons donc en complément une méthode de généralisation, visant à réduire les points annexes, pour lisser la géométrie des arcs. En effet, si la granularité de la numérisation entre le premier et le dernier segment des arcs n'a pas d'importance, l'orientation des segments aux extrémités est déterminante dans la construction de la voie. Le caractère multi-échelle de la voie assure la robustesse de l'analyse, il est donc nécessaire d'avoir des données dont le nombre de points annexes est réduit à l'essentiel pour pouvoir reconstituer de grandes continuités.

La voie est donc un objet robuste au choix des données sur lesquelles elle est construite. Elle possède deux qualités assurant sa stabilité spatiale : une construction locale et des géométries multi-échelles. Cela lui permet de corréler des indicateurs locaux et globaux ; et d'assurer la stabilité des indicateurs, dont le calcul tient compte de l'ensemble du graphe, au découpage du réseau.

\FloatBarrier
\section{Comparaison de graphes spatiaux}

Une fois la pertinence de notre méthodologie établie, nous avons choisi de l'appliquer à un ensemble de réseaux spatiaux. En effet, si notre travail a été bâti sur les réseaux viaires, il a été conçu comme générique pour être appliqué à n'importe quel graphe spatialisé. Nous construisons donc les voies sur un total de quarante réseaux spatiaux, avec les paramètres qui ont été déterminés dans la première partie, et une zone tampon paramétrée à 1 unité (mètre pour les réseaux viaires) autour des nœuds pour s'assurer que les légères imperfections de numérisation ne viennent pas perturber le calcul. Le rayon choisi est plutôt faible, par rapport à la longueur moyenne des arcs, pour avoir un diamètre égal sur tous les graphes sans introduire de biais dans les données.

En plus des indicateurs primaires définis dans la première partie (valeurs moyennes des indicateurs de degré, de closeness et d'orthogonalité), nous introduisons quatre coefficients pour caractériser l'ensemble du graphe. Nous utilisons ainsi le coefficient d'organicité défini par Thomas Courtat \citep{courtat2011mathematics} qui qualifie le degré des nœuds. Nous lui ajoutons trois coefficients construits en fonction des voies : le coefficient de maillance, qui quantifie la part du linéaire maillé du réseau ; le coefficient d'hétérogénéité, qui met en évidence l'hétérogénéité de répartition des intersections ; et le coefficient de réduction qui évalue de combien le nombre d'éléments du graphe a été réduit en passant des arcs aux voies.

Nous comparons ainsi les réseaux géographiques (viaires, ferrés et hydrographiques) sur des tailles de territoire variables, réseaux artificiels (générés via un programme), réseaux de craquelures dans de l'argile et réseaux biologiques (feuille, corail). Si la fonction et l'histoire de construction de ces réseaux sont très différentes, leur caractère spatial rend leur analyse topologique comparable. En effet, une caractéristique commune à tous ces graphes est d'avoir un degré moyen de sommet autour de 3. Cela traduit leur contrainte commune : s'inscrire dans un espace à deux (pour certains trois) dimensions dont l'échelle d'utilisation porte la contrainte d'emprise. Ainsi, pour qu'un réseau viaire soit praticable, les rues devront avoir une certaine largeur, ce qui conditionnera la taille des intersections. C'est un facteur structurel qui limite le nombre maximum d'arcs connectés à un même sommet. Celui-ci pourra varier en fonction du diamètre des zones tampons paramétré lors de la création des voies : plus ce diamètre est important, plus un grand nombre de voies pourront s'intersecter sur un même sommet. Il en est de même pour la connexion entre un fleuve et son affluent ou pour les intersections des veinures d'une feuille. Nous observons également que le degré moyen des nœuds est lié au coefficient de réduction : plus un réseau a de nœuds de degré important, plus il aura d'arcs par voies. En effet, plus les arcs connectés à un sommet sont nombreux, plus la contrainte spatiale leur impose d'être alignés. Il est donc ainsi plus facile de reconstituer des continuités sans dépasser l'angle seuil fixé.

Les différences entre ces réseaux sont plus portées par leur géométrie respective que par leur topologie. Les réseaux ferrés se distinguent ainsi par leur coefficient de maillance très faible. Les réseaux hydrographiques ont des arcs très longs et un faible coefficient de réduction.

La comparaison des réseaux viaires ne fait pas ressortir de tendance particulière liée à un continent. La diversité au sein des réseaux (même dans ceux les plus quadrillés) leur donne des caractéristiques selon lesquelles ils sont difficilement séparables en différents groupes. Quelques cas particuliers se détachent, comme le réseau de Manhattan, dont les voies très longues et traversantes lui donnent des caractéristiques particulières (très faible organicité, très faible hétérogénéité, forte closeness et fort degré moyen pour les voies).

Cette caractérisation, faite sur plusieurs réseaux spatiaux et dans différentes villes, peut également être pensée sur de plus petits territoires. Ainsi, au sein d'une ville, il est possible grâce à ces coefficients et indicateurs de caractériser certains quartiers. Nous pouvons ainsi déterminer leur \textit{efficacité} relative (en terme de proximité topologique), leur degré de maillance, d'hétérogénéité ou de réduction. Clément Bresch, stagiaire au sein de l'équipe MorphoCity, a entamé cette application sur les villes de Paris et Grenoble.

\FloatBarrier
\section{Quantification du changement}

Nous avons mis en place une quantification pour comparer les graphes d'un même territoire à plusieurs dates. La méthodologie sur laquelle nous nous appuyons a pour contrainte principale d'utiliser des données qui doivent être de même emprise d'une année sur l'autre \citep{bordin2006methode}. En effet, la quantification des différences structurelles nécessite en premier lieu la distinction, à travers la géométrie, des arcs qui ont été modifiés (ajoutés, supprimés ou déplacés) de ceux qui n'ont subi aucune transformation.

Le raisonnement mis en place, pour quantifier le changement, a pour but de supprimer le \enquote{bruit} dû aux ajouts ou suppressions de géométries afin de pouvoir identifier les modifications d'accessibilité réelles entre les objets du graphe (chemins les plus courts ajoutés ou retirés). Nous l'appliquons sur deux villes : Avignon et Rotterdam. La méthode de vectorisation (régressive) utilisée permet d'obtenir une base de données \textit{panchroniques} : l'ensemble des géométries est réuni dans une table \enquote{totale}. Chaque année qui a été traitée par la vectorisation d'une carte ancienne constitue un attribut dans cette table. Celui-ci, traduit sous forme d'un booléen, indique la présence ou l’absence des géométries pour chacune de ces dates. 

La comparaison des cartes deux à deux nous permet de visualiser l'évolution de l'accessibilité dans le temps. Il s'agit d'une représentation des différences entre deux états statiques mettant en avant une \textit{cinématique} plus ou moins bénéfique aux proximités topologiques entre objets du territoire. Cette perception du changement est un premier pas vers l'étude de la morphogenèse des réseaux. Elle nous permet de mieux visualiser au sein d'un territoire les dynamiques d'ajouts, de suppressions ou de modifications de géométries. Elle permet également de nous rendre compte que l'évolution d'un réseau ne se fait pas forcément vers une optimisation de son accessibilité globale.

Nous avons ainsi pu constater, à Avignon, l'apport majeur de la création de l'avenue de la République et du Cours Jean Jaurès à la proximité entre les voies. Les ruptures dans la structure extérieure circulaire lui étaient, quant à elles, très préjudiciables. Cette observation nous pousse à nous poser la question des limites du réseau dans ce cas d'étude également. Si le cours à l'extérieur des remparts avait été numérisé, la dépréciation de l'accessibilité topologique n'aurait sûrement pas été aussi marquée. Mais cette suggestion reste à l'état d'hypothèse car nous ne disposons pas d'une base de données panchroniques de l'extra-muros.

Pour Rotterdam, l'analyse diachronique montre, dès la première période, un renforcement de l'accessibilité vers le centre de la ville au détriment de Schiedam. Cela suggère que le développement du réseau entre 1374 et 1570 prédisposait déjà un des deux centres à se développer plus que l'autre et à englober finalement le second. Le rapprochement structurel entre différentes périodes est donc porteur de sens. La considération des propriétés topologiques et topographiques nous permet, dans une première approche, d'avoir l'intuition d'une forme de développement d'un territoire.

Cette méthodologie est également utile dans une analyse prospective. Elle permet d'évaluer la proposition de projets urbains en quantifiant leurs impacts sur l'accessibilité de tout un territoire. C'est de cette manière que nous avons pu répondre au service d'urbanisme d'Avignon sur l'\textit{efficacité théorique} des projets qu'ils nous ont soumis. Cette étude structurelle peut ainsi venir en appui à celles urbaines effectuées par les professionnels de la ville. L'analyse de projets peut également être faite dans le passé, pour étudier l'impact des transformations pensées par de célèbres urbanistes, architectes ou préfets. Nous avons pu de cette manière montrer l'impact grandement positif en terme de distances topologiques des aménagements du réseau viaire fait par Haussmann à Paris.

\begin{figure}[h]
    \centering
        \includegraphics[width=0.6\textwidth]{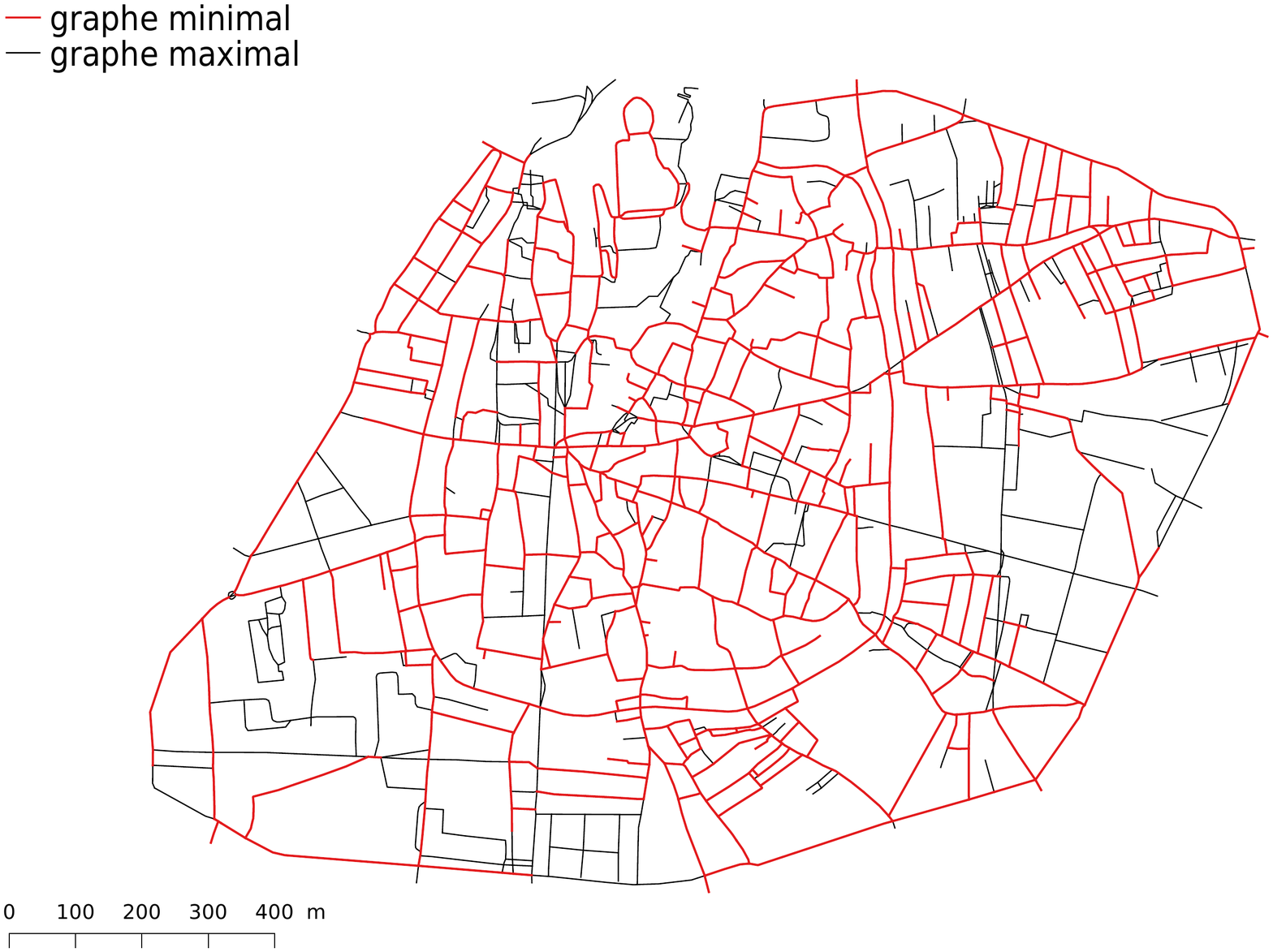}
    \caption{Carte des graphes minimal et maximal du réseau viaire intra-muros d'Avignon.}
    \label{fig:avi_minimaxi}
\end{figure}

Grâce à la création d'une base de données \textit{panchronique}, il nous est également possible de reconstruire des cartes \enquote{maximales} ou \enquote{minimales} du territoire. Une carte \enquote{maximale} réunira toutes les géométries qui ont été observées depuis la plus ancienne carte vectorisée à la plus récente. Une carte \enquote{minimale} réunira celles qui ont été appariées sur toutes les périodes et donc jamais supprimées (figure \ref{fig:avi_minimaxi}). Le réseau minimal nous indique une \textit{base géométrique pérenne} du territoire, alors que celui maximal nous donne un aperçu de toutes les parties du réseau qui ont été sujettes à transformation. Cette utilisation de notre modèle de données ouvre des perspectives de recherche sur la pérennité des structures viaires et la spécificité des géométries \textit{squelettes}, qui portent les transformations.

\clearpage{\pagestyle{empty}\cleardoublepage}
\part{Lire les réseaux viaires : Exploration qualitative des résultats obtenus}
\markboth{Lire les réseaux viaires}{Lire les réseaux viaires}

\thispagestyle{empty}
~\vfill
{\itshape Les formes sont un langage dont nous recherchons la grammaire. Elles expriment un inconscient géographique.}
~\vfill

\clearpage{\pagestyle{empty}\cleardoublepage}
\chapter{La ville : un système complexe aux lectures multiples}
\minitoc
\markright{La ville : un système complexe aux lectures multiples}

L'ensemble du travail que nous présentons ici a été réalisé au sein d'une équipe de recherche pluridisciplinaire (MorphoCity)\footnote{Cette équipe est soutenue par le CNRS - PIRVE (Programme Interdisciplinaire de Recherches Ville-Environnement) et par l'ANR-MN-2012 \emph{MoNuMoVi}. Elle est dirigée par Ph. Bonnin (CNRS-AUS-LAVUE) et S. Douady (CNRS-MSC), et se compose de P. Bordin (ESTP - Geospective), E. Degouys (doctorante AUS-LAVUE), C.-N. Douady (AUS-LAVUE), J.-P. Frey (CRH-LAVUE), C. Lagesse (doctorante MSC) et P. Vincent (Z-Studio). Ont également participé, B. Atashinbar (post-doctorant), A. Bonnin (ingénieur d'étude), C. Bresch (stagiaire), R. Brigand (post-doctorant), T. Courtat (doctorant), C. Lavigne (post-doctorant), R. Pousse (stagiaire), Wang X. (doctorante) et M. Watteaux (post-doctorante).}. Celle-ci réunit anthropologues, sociologues, urbanistes, architectes, archéo-géographes, physiciens, informaticiens et géomaticiens. Le développement de cette thèse s'est donc fait en collaboration constante avec des spécialistes de la ville, qui l'expérimentent dans leur travail en tant qu'objet anthropique.

Les échanges avec les thématiciens ont eu pour but de traduire en objets quantifiables, des modes d’analyse des structures urbaines que leurs disciplines ont mis au point savamment, durant des décennies de pratique. L'objectif est de rendre applicables les outils offerts par la théorie des graphes et la science des réseaux à un objet dont la complexité rend une analyse exhaustive impossible.

Dans cette partie, nous souhaitons retranscrire la richesse des échanges entre les différentes disciplines. Le développement de la méthodologie de caractérisation des réseaux spatiaux, fait pendant cette étude, s'est retrouvé au cœur de la problématique de recherche de l'équipe, et a suscité de nombreux débats. Nous voulons faire ici une synthèse de ces discussions afin de montrer les multiples questions de recherches, autour de la ville, auxquelles le travail présenté ici est lié.


\FloatBarrier
\section{Les différentes approches}
 
Dans le méandre d'une rivière, sur la côte d'une mer ou d'un océan, sur une surface plane de préférence, propice à l'agriculture, la vie humaine se développe. Afin d'apprivoiser son environnement, l'Homme s'y établit en y apposant sa culture. Il écrit sur le territoire les chemins qu'il aime y prendre, de manière plus ou moins pérenne. Il construit autour de ces chemins les bâtiments qui abriteront l'essence de la société qu'il crée : l'habitat, le commerce, les institutions politiques, culturelles ou religieuses. Ces bâtiments s'organisent selon un découpage précis de l'espace qu'ils occupent : les parcelles cadencent les constructions. De ces trois unités que l'on peut considérer comme élémentaires (les routes, les parcelles, le bâti) naissent des interactions complexes qui donnent forme à la ville.

Il est possible de faire plusieurs lectures de cet objet complexe. Certains proposent une lecture finaliste, en fonction de son utilisation. D'autres, une lecture pragmatique, en fonction de ses éléments en interaction. Les partisans de la lecture finaliste peuvent eux-mêmes être divisés entre deux traditions : celle fondée sur les projets, la fabrique de l'espace, et celle fondée sur le récit, suivie par les archéologues et archéo-géographes \citep{conzen1990making}. Pour parvenir à comprendre l'histoire des formes, les archéo-géographes ont mis en place une théorie des scénarios. Celle-ci considère chaque lieu avec son Histoire et sa temporalité propre. Les chercheurs de cette discipline ont pour philosophie de rester proches de leur objet, en considérant sa nature comme, sinon unique, du moins spécifique. Ils font donc appel à de multiples facteurs pour tenter de comprendre le processus d’émergence observé entre eux. Parmi les chercheurs adoptant une lecture pragmatique, nous trouvons les structuralistes, qui quantifient la ville à travers ses formes pour en lire les structures. Dans cette approche, la question de la répartition des fonctions reste en suspens.

Cependant, la frontière entre quantification mathématique et fonctionnalité empirique est ténue. La création d'un réseau viaire, par exemple, résulte d'un équilibre subtil entre desserte fine de l'espace et accès rapide aux points importants, stratégiques. Pour comprendre les dynamiques d'un tel développement, il est possible de reconstituer l'histoire en faisant l'hypothèse de scénarios. Dans une modélisation, la ville peut alors être considérée comme un ensemble d'individus dont les interactions, selon des contrainte spatiales et sociales, définissent un système \citep{franccois2002espace}. Il est également possible d'essayer de simplifier ce système, en n'en gardant qu'une information structurelle, afin de l'analyser quantitativement \citep{barthelemy2014discussion}. Ces deux démarches correspondent à deux postures prises devant un problème de recherche : celle voulant conserver un grand nombre de paramètres pour être fidèle à la réalité (descriptive) et celle ne conservant qu'un minimum de paramètres (simplificatrice). En modélisation informatique, elles sont appelée KIDS (pour \textit{Keep It Descriptive Stupid}) et KISS (pour \textit{Keep It Simple Stupid}) \citep{edmonds2005kiss, livet2014diversite}. A. Banos et L. Sanders ont ainsi travaillé sur les différentes modélisations des systèmes spatiaux en géographie \citep{banos2013modeliser}. Plus spécifiquement, M. Batty a examiné les simulations mathématiques permettant de décrire la croissance urbaine \citep{batty2009urban} ; de Dios Ortùzar et Willumsen ont étudiés celles des réseaux de transport \citep{ortuzar2001modelling}.

La pluralité des facteurs qui participent à la création et au développement de la ville ne rend pas son étude facile. Si l'on s'en tient à une méthodologie descriptive, l'ensemble des attributs qu'elle regroupe multiplie les approches que l'on peut en avoir et rend chaque lieu unique et indépendant de tous les autres. À l'inverse, des programmes de recherche, parmi lesquels nous nous situons, se sont penchés sur l'exploitation automatique de critères morphologiques pour analyser les trames (programmes ALPAGE, MODELSPACE et MONUMOVI). Il est alors question de confondre des espaces différents pour les analyser selon des critères comparables. La spécificité du lieu est perdue, mais les informations apportées n'en sont pas moins intéressantes. En nous reportant à la structure nous gommons, certes, l'ensemble complexe des attributs historiques et archéo-géographiques du territoire, mais nous parvenons à saisir des informations non négligeables, qui confirment ou complètent les connaissances acquises avec le temps par les spécialistes du lieu. 

L'approche structuraliste permet de retrouver des critères partagés par plusieurs lieux dont l'histoire est proche. Stephan Marshall, en 2004, distingue des caractéristiques propres à la ville moderne. Il décrit son réseau viaire comme un arbre hiérarchisé avec un centre-ville peu structurant. C'est en cela qu'elle se distingue de la ville ancienne, construite autour d'une place centrale, avec des rues qui sont organisées autour de celle-ci ou qui rejoignent la campagne environnante \citep{marshall2004streets}.

La diversité de la forme urbaine interroge les démarches de modélisation des pratiques sociales. Certains s'opposent à la réduction de la ville à sa forme, et considèrent les pratiques sociales et politiques comme essentielles à sa compréhension \citep{viala2005contre}. Ces critères peuvent donner lieu au développement d'indicateurs, prenant en compte certaines pratiques sociales. C. Grasland et G. Hamez s'interrogent ainsi sur les caractéristiques à retenir pour construire un indicateur de cohésion territoriale \citep{grasland2005vers}. Les chercheurs veulent permettre une évaluation objective et reproductible, et appuient l’importance de la dimension spatiale dans sa représentation.

Au sein de notre approche, les indicateurs quantitatifs que nous développons et que nous appliquons à l'objet urbain participent également à une pluralité méthodologique, qui confronte le quantitatif au qualitatif. Ainsi, le calcul de l'indicateur de \textit{betweenness} sur un réseau est souvent associé à une notion de flux. La démonstration de cette qualité reste empirique mais n'en est pas moins pertinente : un arc emprunté un grand nombre de fois pour parcourir un chemin entre deux points pris aléatoirement sur un graphe a de grandes chances de se retrouver au cœur de l'activité des déplacements urbains.

Nous avons choisi dans ce travail une approche méthodologique, pragmatique. La description exhaustive étant impossible, nous avons choisi de briser le lien avec la recherche d'une analogie complexe et de nous concentrer sur une unique information : la géométrie du graphe viaire. Nous faisons l'hypothèse que celle-ci est capable de porter de l'information pertinente pour lire la ville. La démarche que nous suivons est celle du croquis, qui s'oppose à la photo en ne faisant ressortir que l'essentiel d'un paysage (partie du territoire telle qu'elle est perçue et représentée par ses habitants). Notre volonté est donc d'extraire les lignes  qui \textit{portent} la ville.

\FloatBarrier
\section{Le choix de la voie}

Reconsidérons les trois unités géométriques élémentaires de la ville : la ligne (le réseau viaire), la surface (les parcelles), le volume (les bâtiments). Parmi celles-ci, sur le long terme (plusieurs siècles), seule la ligne se renforce. En effet, les bâtiments sont souvent détruits pour faire place à des constructions plus récentes, plus performantes en terme de nombre de logements fournis,  de qualité, ou de performances énergétiques. Les parcelles, plus robustes que le bâti, sont divisées, réunies ou redessinées pour satisfaire les transformations territoriales d'un champ en lotissement ou d'une banlieue en centre-ville. Entre ces éléments, le réseau viaire apparaît comme étant celui le plus pérenne : vecteur d'un accès indispensable, support du développement, son emprise est souvent renforcée. Ainsi un chemin de terre entre deux champs pourra devenir la route qui mène à la ville à proximité. Il subit bien évidemment, lui aussi, des transformations dans le temps, mais celles-ci sont plutôt de nature correctrice  (rues redressées), de changement de type (rues privatisées ou rendues piétonnes), ou d'expansion. La densification urbaine fait apparaître de nouvelles rues, rares sont celles complètement remodelées ou supprimées. Lorsque cela se produit à l'échelle globale, dans une ville, l'événement est suffisamment rare pour acquérir une renommée mondiale, comme celle d'Haussmann à Paris ou de Cerdà à Barcelone au XIX\textsuperscript{ème} siècle.

La ville se développe au sein de l'Histoire d'un lieu, selon une organisation socio-politique définie par le contexte historique et influencée par le contexte spatial. La croissance urbaine, du fait de l'évolution démographique, est corrélée avec la croissance des réseaux de déplacement, qu'ils soient viaires ou ferroviaires \citep{pumain1982chemin}. Les réseaux de transports sont ainsi porteurs du contexte social qu'ils abritent \citep{hanson1986geography}.

Les usagers circulent dans la ville selon une représentation mentale du territoire qu'ils occupent. Le squelette de cette représentation est constitué par le réseau qui façonne la forme urbaine : celui des rues. C'est une structure rassurante qui donne un référentiel spatial dans lequel se situer. Celui-ci peut être plus ou moins cartésien, des routes sinueuses des villes de l'ancienne Europe, jusqu'à celles rectilignes, participant à un réseau maillé, caractéristiques d'un développement planifié dans le Nouveau Monde ou ailleurs.

\begin{figure}[h]
    \centering
        \includegraphics[width=0.8\textwidth]{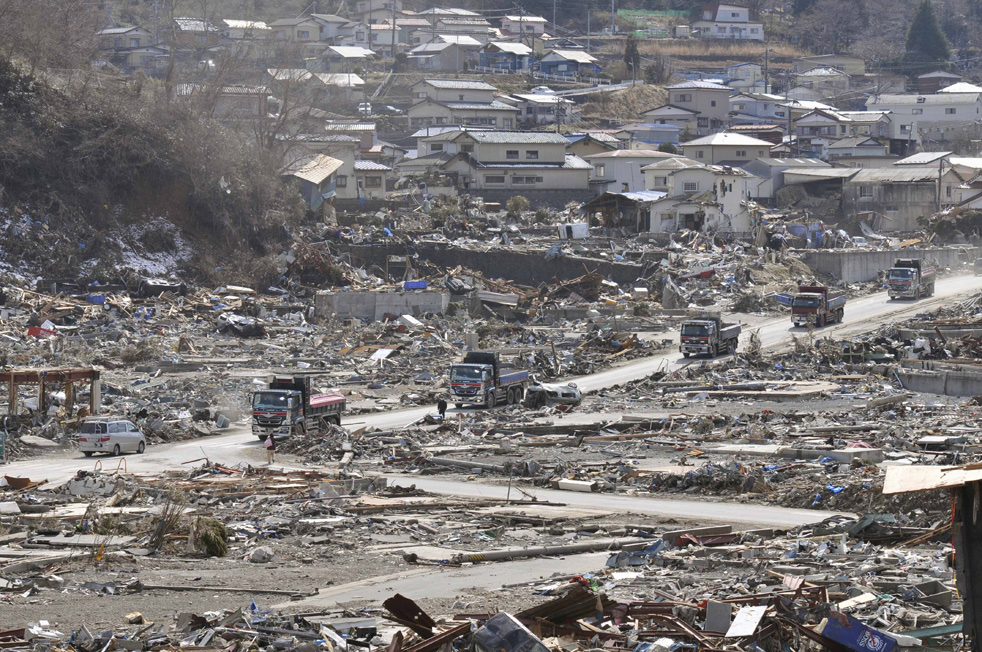}
    
    \caption{Camions transportant des vivres et équipements au milieu des décombres de la ville d'Onagawa, dans la préfecture de Miyagi, au Nord-Est du Japon. La photographie a été prise le 18 mars 2011, une semaine après le tremblement de terre et le tsunami lié. \\ crédit : AP / Kyodo News}
    \label{fig:jap_after1}
\end{figure}

\begin{figure}[h]
    \centering
        \includegraphics[width=0.8\textwidth]{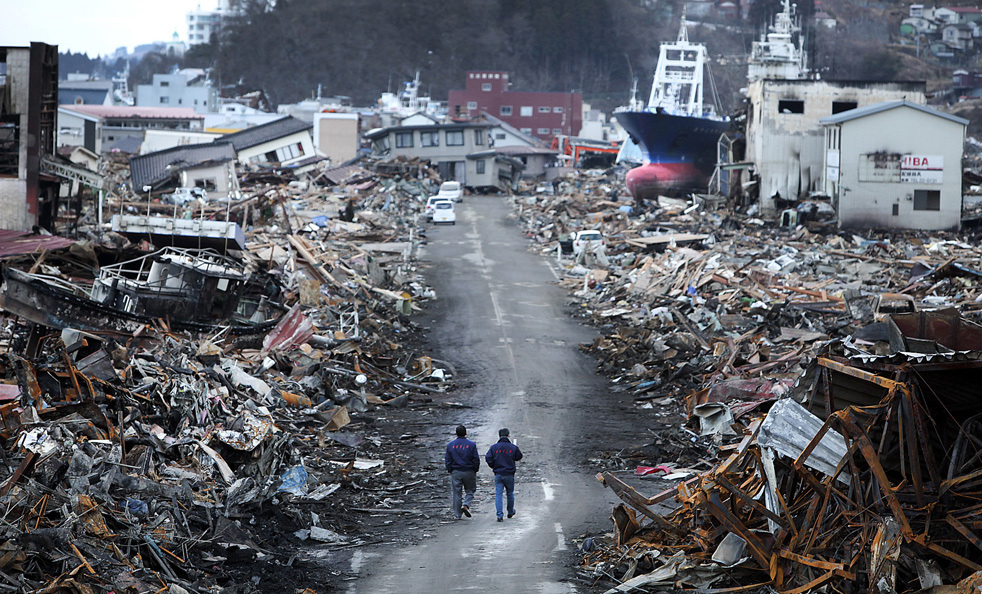}
    
    \caption{Le 17 mars 2011, une semaine après le tsunami, à Kesennuma, au Japon. \\ crédit : Los Angeles Times / Brian Van Der Brug}
    \label{fig:jap_after2}
\end{figure}

Dans les cas les plus extrêmes, comme par exemple suite à une catastrophe naturelle, le réseau viaire est celui indispensable à la reconstruction de tout un territoire (figure \ref{fig:jap_after1}). C'est donc le premier à être dégagé et remis en état et, pour poser un cadre connu sur un paysage dévasté, il suit les mêmes tracés que ceux précédant la catastrophe (figure \ref{fig:jap_after2}). Le filaire du réseau des rues est donc un des éléments les plus pérennes dans un contexte en constante transformation. C'est un élément géographique qui, de tous ceux composant nos villes, est susceptible de porter une information transcendant le temps d'observation. Seules des interventions majeures dissimulent l'information passée bien qu'elle transparaisse tout de même en certains endroits (lire le cas de Paris dans le chapitre suivant).

Dans ce travail, nous avons voulu nous libérer de la notion de \textit{rue}, imprécise et fluctuante, qui est attachée à la toponymie d'un lieu. La "rue" est un concept manipulé par beaucoup de disciplines avec parfois des  interprétations très différentes. Un sociologue l'associera à une ambiance, un géomaticien à une géométrie ou un morceau d'itinéraire, un physicien à un élément de connexion dont l'efficacité est quantifiable, un urbaniste à un objet de desserte, modulable, partie d'un patrimoine... L'un considérera l'objet en détail avec une largeur, une surface ; l'autre considérera uniquement son axe central et le troisième les points qu'il sert à relier. Nous avons voulu ici nous émanciper de tous les concepts, idées et raisonnements liés à la \textit{rue} en définissant un nouvel objet, que nous avons caractérisé selon des paramètres précis. Nous l’avons appelé \emph{voie}.

\begin{figure}[h]
    \centering
	\includegraphics[width=0.4\textwidth]{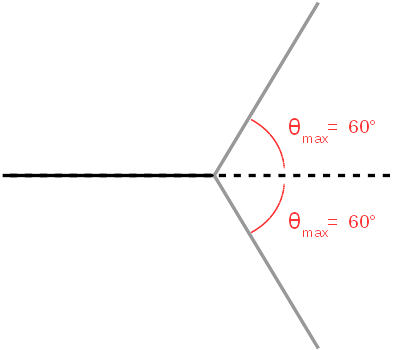}
	\caption{Schéma illustrant la déviation maximale pour conservation de la continuité au sein d'une voie.}
	\label{fig:deviation60}
\end{figure}

La voie est pensée comme une ligne continue. Elle est créée dans notre méthodologie pour objectiver un sentiment de perspective. Les lignes du réseau viaire sur le territoire se construisent au fil de leur histoire pour relier des lieux dont l'accessibilité est jugée importante. Nous choisissons les données représentant l'ensemble du filaire, carrossable ou simplement piéton, pour apposer le concept de voie sur un graphe dont la forme est indépendante des aménagements urbains, liés à une période temporelle restreinte. La voie est la perspective créée, à partir d'un utilisateur positionné sur le réseau, selon un angle seuil de \enquote{vision} paramétré avec une ouverture de 120\degre{} répartie équitablement autour de la ligne droite (dans le prolongement exact du tronçon du cheminement avant l'intersection, figure \ref{fig:deviation60}). Nous avons vu dans la première partie de ce travail que les arcs d'une ville sont majoritairement alignés ou perpendiculaires. La perspective offerte, sur la même voie, propose donc dans la plupart des cas un cheminement proche de la ligne droite.

Nous définissons le \textit{tournant} à travers la voie. \textit{Tourner} équivaut à changer de voie. Dans la paramétrisation choisie, l'angle seuil a été fixé à 60\degre{}. Un changement de voie correspond donc à faire un virage de plus de 60\degre{} à une intersection (figure \ref{fig:deviation60}). Cependant, si un arc, entre deux intersections, fait des lacets dont les virages dépassent ce seuil, l'utilisateur n'aura pas d'autres choix que de tourner et donc nous ne considérerons pas qu'il puisse prendre une décision quant à son itinéraire : ce n'est pas un tournant au sens où nous l'entendons dans cette étude.

Le réseau viaire peut être un élément révélateur sous sa forme filaire si nous disposons des bons outils pour en lire la structure. Pour conserver l'idée de persistance dans le temps notre méthode ne fait pas intervenir la largeur des voies. En effet, lors d'aménagements urbains, c'est souvent sur la largeur de la voirie que sont faites les interventions pour concrétiser l'importance dans le réseau de certaines rues. Nous cherchons au contraire à reconstruire \textit{a posteriori} cette information en identifiant les voies qui participent à la structure du graphe viaire. En suivant la même idée d'intemporalité, nous ne considérons pas l'aspect directionnel de la voie (le graphe sur lequel nous travaillons est non orienté) : une voie à double sens sera donc étudiée de la même manière qu'une voie à sens unique. Nous faisons ainsi abstraction dans nos recherches de tous les attributs non géométriques ou non topographiques du réseau viaire. Nous ne le caractérisons pas relativement à son contexte socio-politique mais essayons plutôt de retrouver celui-ci dans la lecture des propriétés géométriques. 

La voie est donc un objet morphologique, dépouillé de toute autre information. Elle se fond dans le paysage urbain, structurant autour d'elle le découpage parcellaire et avec lui toute l'occupation du sol. La topographie n'est pas étrangère à sa géométrie. Nous pouvons lire les dénivelés d'une ville en soulignant les  voies courtes et très connectées et en les opposant aux voies longues peu connectées (indicateur d'espacement). Dans les graphes viaires que nous utilisons, l'urbanisme souterrain est visualisable partiellement à travers la non-planarité de nos réseaux : un croisement entre deux arcs ne donnera pas lieu à la création d'un sommet si le point ne correspond pas à une intersection dans la réalité. Le développement de la ville se fait, en effet, à des niveaux différents mais interconnectés dont il est important de tenir compte. Les villes ont évolué d'un développement en étalement à une dynamique de densification : la verticalisation des organisations spatiales est donc une donnée importante.

Nous avons choisi dans ce travail de trouver un critère universel qui puisse nous aider à décrire un espace urbain, sans nécessairement en connaître exhaustivement la genèse. Nous travaillons sur les réseaux viaires car ils portent en eux une topologie et une topographie qui n'est pas étrangère à la culture et au lieu dans lesquels ils s'incluent. La schématisation de la ville à travers son réseau viaire nous oblige à faire abstraction de diverses informations, qui peuvent être considérées comme des paramètres fondamentaux par les sociologues et urbanistes. Notre problématique est de voir jusqu'à quel point la structure viaire peut être révélatrice de l'Histoire de la ville dans laquelle elle se situe. Ce ne peut être une démarche isolée, d'où la nécessité d'une collaboration constante avec les personnes travaillant sur le terrain. Elle nous est utile pour lire et connaître les limites des indicateurs calculés à partir de la topologie et de la topographie du réseau. Il est primordial de ne pas oublier l'objet sur lequel nous travaillons au profit de sa structure mathématique si nous voulons le comprendre. Les caractéristiques apportées par les indicateurs sont à mettre en perspective avec l'information mise de côté.

\FloatBarrier
\section{L'apport des outils de modélisation}

L'analyse morphologique d'un territoire peut être faite à travers une abstraction de celui-ci. La représentation sous forme de graphe donne accès à de nombreux outils mathématiques \citep{gibbons1985algorithmic}. Les formes modélisées sur différents espaces, à différentes époques, permettent d'établir des comparaisons, pour détecter d'éventuelles tendances. Grâce aux analyses quantitatives, il est possible d'identifier des propriétés communes, partagées par différents graphes viaires. Nous retrouvons plusieurs exemples, où la géométrie du réseau répond à des lois qui traversent les continents. Ainsi, les recherches de Serge Salat sur Paris montrent que la longueur des rues est inversement proportionnelle à leur largeur : ces deux attributs suivent une loi de Pareto \citep{adamic2000zipf, newman2005power}. Cette caractéristique se retrouve également dans d'autres villes \citep{gabaix1999zipf}. Elle montre qu'une information quantifiée peut dépasser l'Histoire d'un lieu, d'un tissu à l'autre \citep{salat2011villes, salat2014breaking}. De la même manière, nous avons montré dans la partie précédente que des propriétés topologiques et structurelles étaient partagées par plusieurs réseaux spatiaux, dont certaines au delà de leur nature. Ainsi, le degré moyen des nœuds d'un tissu urbain, qu'il soit planifié ou organique, se situe autour de 3. De plus, quel que soit le réseau spatial considéré, plus le degré moyen des sommets est élevé, plus les voies créées sur celui-ci regroupent un nombre important d'arcs.

Au delà de ces analyses topologiques et géométriques statistiques, des outils ont été développés spécifiquement pour l'étude de l'information spatialisée. Ainsi, les Systèmes d'Information Géographique (SIG) gèrent des données sous forme d'objets, ayant une géométrie et des attributs liés à celle-ci. Ils permettent d'intégrer des fichiers de formes (\textit{shapefiles}), chacun correspondant à une caractéristique partagée par tous ses éléments (réseau viaire, réseau hydrographique). Il est possible dans de tels logiciels de superposer les différents fichiers de formes sur un même territoire afin d'en étudier les ressemblances ou disparités. 

Ces logiciels apportent un potentiel d'analyse très important. Dans les services qu'ils proposent, interfaçables avec des systèmes de gestion de base de données, nous retrouvons des concepts au fondement de l'analyse topologique. Egenhoffer réunit ainsi la grammaire nécessaire à la lecture des interactions entre entités géographiques, délimitées par leur géométrie \citep{egenhofer1990categorizing}. Puis Clementini \textit{et al.} définit un cadre conceptuel pour modéliser les interactions spatiales \citep{clementini2008cadre} et regroupent neuf interactions élémentaires (\textit{Dimensionally Extended nine-Intersection Model (DE-9IM)}) entre deux géométries fondées sur l'intérieur, l'extérieur ou les frontières de leur intersection. La définition d'une topologie propre à la géométrie peut ainsi compléter celle de la théorie des graphes pour aider à la compréhension de problématiques spatiales \citep{linial1995geometry}.

\begin{figure}[h]
    \centering
    \includegraphics[width=0.8\textwidth]{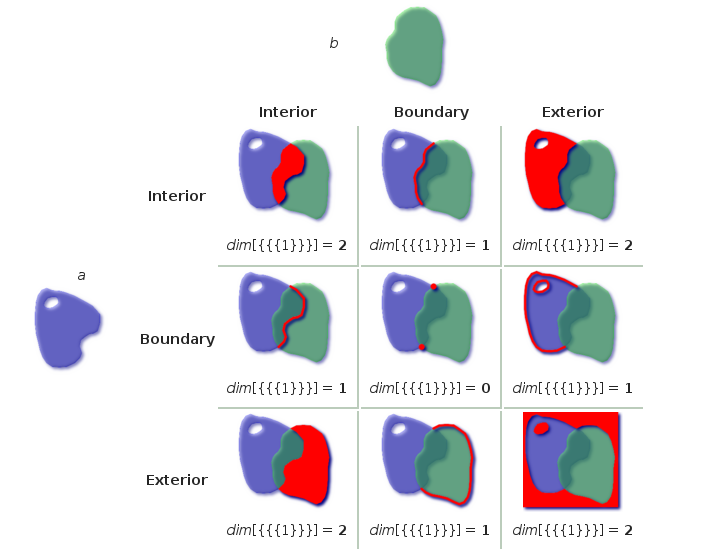}
    \caption{ Dimensionally Extended nine-Intersection Model (DE-9IM) \\ source : documentation PostGIS (Chapter 4. Using PostGIS: Data Management and Queries)}
    \label{fig:9IM}
\end{figure}

Ce modèle illustre le lien fort qui existe entre l'analyse de l'information géographique et les théories propres aux études topologiques. La théorie des graphes est renforcée par ces travaux, qui ajoutent aux recherches de la science des réseaux celles de la science des formes.

\FloatBarrier
\section{La représentation sous forme de graphe}

Quand Euler (1736) a résolu le problème des sept ponts de Konigsberg à l'aide d'un graphe, il n'imaginait peut être pas l'étendue du champ d'application de la théorie qu'il venait de créer. Il a retenu d'un environnement complexe, uniquement les données nécessaires à la résolution du problème et fait ainsi un saut fondamental dans l'abstraction. L'application de la méthode portait sur un environnement géographique bien qu'aucune information d'emprise ne transparaisse dans la modélisation. Plus tard, Dijkstra \citep{dijkstra1959note} a retenu les inscriptions spatiales de points stratégiques du réseau pour calculer le plus court chemin entre eux. Il s'est ainsi appuyé sur la notion de graphe pour construire une partition ponctuelle de l'espace, afin de pouvoir quantifier les distances entre différentes positions.

À l'origine, un graphe regroupe les relations (sous forme binaire : existence ou inexistence) des éléments d'un ensemble. Des extensions sont possibles : les relations peuvent être orientées, les arcs valués, les nœuds pondérés... Mais il n'y a pas d'information de forme ou de position \textit{a priori} contenue. Ainsi, on définit un graphe planaire comme un graphe où il existe une configuration telle qu'aucun arc n'en croise un autre. Cette configuration est potentielle, les sommets peuvent donc être disposés de façon quelconque, facilitant la visualisation. Une autre illustration de la possibilité de déplacement des sommets non spatialisés sont les \textit{core-diagram} qui ramènent tous les nœuds sur un cercle afin de mieux visualiser les arcs entre eux. Une discipline s'est développée autour de ce concept de visualisation d'un grand nombre de données sur un graphe : la \textit{data visualisation}. Un graphe spatialisé ne pourra pas bénéficier des mêmes techniques : ses sommets ont une position fixe les uns par rapport aux autres. Ainsi, ce type de graphe ne pourra être planaire que si les croisements à des altitudes différentes peuvent être réagencés, dans une conception topologique, sans intersection entre arcs. Dans un graphe routier à la connexité forte, il est peu probable qu'un réseau qui admet un pont ou un tunnel ait un graphe planaire. Si nous considérons la géométrie comme fixe, pour un graphe planaire, chaque intersection entre deux arcs donnera lieu à la création d'un sommet. De ce fait, les graphes spatiaux construits dans un espace à trois dimensions ne peuvent pas être considérés \textit{a priori} comme planaires.

D'un point de vue mathématique, un graphe est transposable sous forme de matrice. Sous cette forme, les informations géographiques ne sont pas transposables, seules peuvent être résumées l'information binaire de relation (symbolisée par un 0 ou un 1 entre deux arcs) ou les distances topologiques entre objets (la matrice regroupe alors toutes les distances topologiques entre arcs). Dans notre cas, l'association de plusieurs matrices est nécessaire si l'on ne veut pas perdre l'information de position des sommets et des points annexes. Il est de même nécessaire d'établir des liens entre ces matrices afin de comprendre comment sont positionnés et reliés les sommets et selon quelle géométrie.

Les problèmes usuels auxquels la théorie des graphes tente de répondre portent sur la catégorisation, le dénombrement, le calcul de chemins, de flots, l'optimisation des répartitions, la décomposition ou factorisation. Des problématiques connues portent sur la réduction de graphes à l'essentiel en ne conservant que les nœuds au dessus d'un certain degré (\textit{k-core}) ou bien se penchent sur la logique des colorations (problème des quatre couleurs) \citep{dorogovtsev2006k, errera1925contribution}. Lorsque l'on traite d'un graphe spatialisé, les questions qui se posent prennent en compte différents types de distances. Elles sont évoquées dans ce travail pour répondre à des problématiques portant sur l'accessibilité. Les logiques spatiales vont de pair avec celles de déplacement car elles incluent une information métrique, tangible, pour chaque arc. Selon les distances étudiées, la perception du réseau est modulée. Nous avons montré dans ce travail que si celles-ci sont calculées sur un objet multi-échelle, telle que la voie, la connectivité l'emporte sur la métrique. En effet, l'information géométrique porte en elle la topologie. Les indicateurs de théorie des graphes, initialement mis au point pour de l'information non spatialisée, lui sont donc applicables. Les graphes spatiaux ont donc, accessibles pour leur étude, tous les indicateurs calculés sur leur topologie, en plus de ceux faisant intervenir leurs propriétés géométriques. Le champ d'analyse est de ce fait élargi, et possible à de multiples niveaux de complexité.

\clearpage{\pagestyle{empty}\cleardoublepage}
\chapter{La lecture proposée par la voie}
\minitoc
\markright{La lecture proposée par la voie}

Nous proposons, dans ce chapitre, une lecture de la ville à travers la voie et les indicateurs que nous avons développés à partir de celle-ci. Cette approche nous permet de lire un territoire \enquote{vu du ciel}, sans \textit{a priori} en connaître l'Histoire. Nous utilisons des données issues de la \copyright BDTOPO de l'IGN pour Paris (2015) et Avignon (2014) et des données OpenStreetMap pour les autres villes (toutes extraites en 2015). Nous confrontons les résultats obtenus avec les informations et connaissances que les spécialistes des différents lieux ont partagées avec nous.

\FloatBarrier
\section{Les enjeux de la représentation cartographique}

Dans le regard que l'on porte sur une carte, il ne faut pas oublier que ce qui y est restitué résulte de choix de représentation. Dans toute lecture cartographique il est nécessaire de replacer l'objet dans son contexte de création pour interpréter l'information qui y est montrée. Ainsi, une carte élaborée pour un roi pourra indiquer son château de manière démesurée par rapport au village qui l'entoure. De la même manière, une carte créée par un groupe militant pour une cause (humanitaire, sociale ou commerciale) pourra arranger la représentation de manière à servir son propos.

Dans notre cas, la construction de notre objet de lecture, la voie, a été voulue objective. Plusieurs méthodes de construction ont été testées, et a été retenue celle qui donnait les meilleurs résultats (en terme de convergence et de nombre d'objets créés). La méthode est robuste au sens de lecture du réseau, et l'objet construit rend le calcul d'indicateurs robuste aux découpages de l'échantillon spatial.

Nous avons choisi, tout au long de ce travail, de présenter nos résultats selon une échelle de couleur en dix classes  regroupant des longueurs de réseau équivalentes. Nous avons opté pour cette solution afin de créer une hiérarchisation perceptible à l’œil nu. Cependant, cela atténue les nuances pouvant exister au sein d'un même quartier.

\begin{figure}[h]
    \centering
    \begin{subfigure}[t]{.48\linewidth}
        \includegraphics[width=\textwidth]{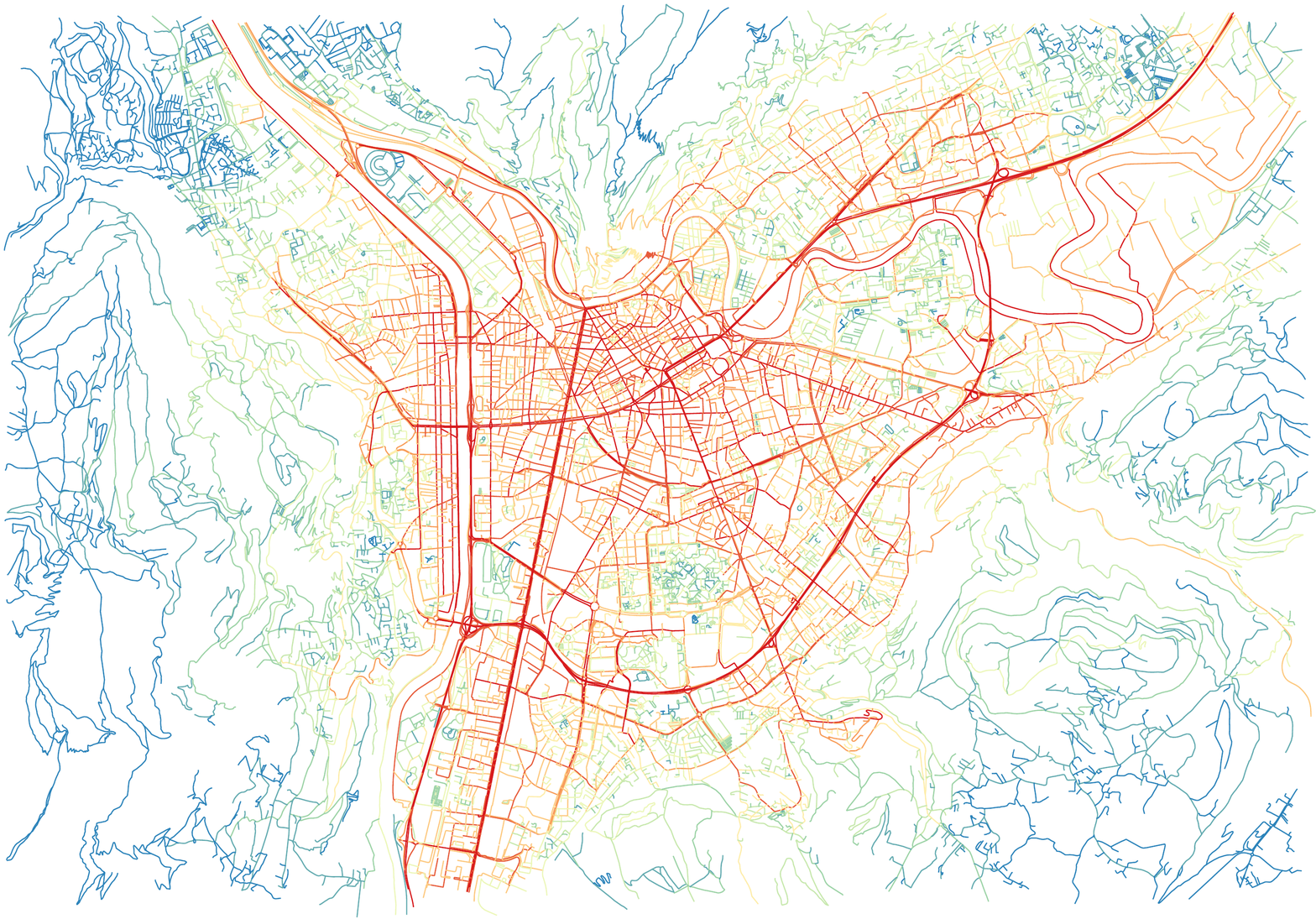}
        \caption{Représentation de l'accessibilité par classe de longueur équivalente.}
    \end{subfigure}
	~
    \begin{subfigure}[t]{.48\linewidth}
        \includegraphics[width=\textwidth]{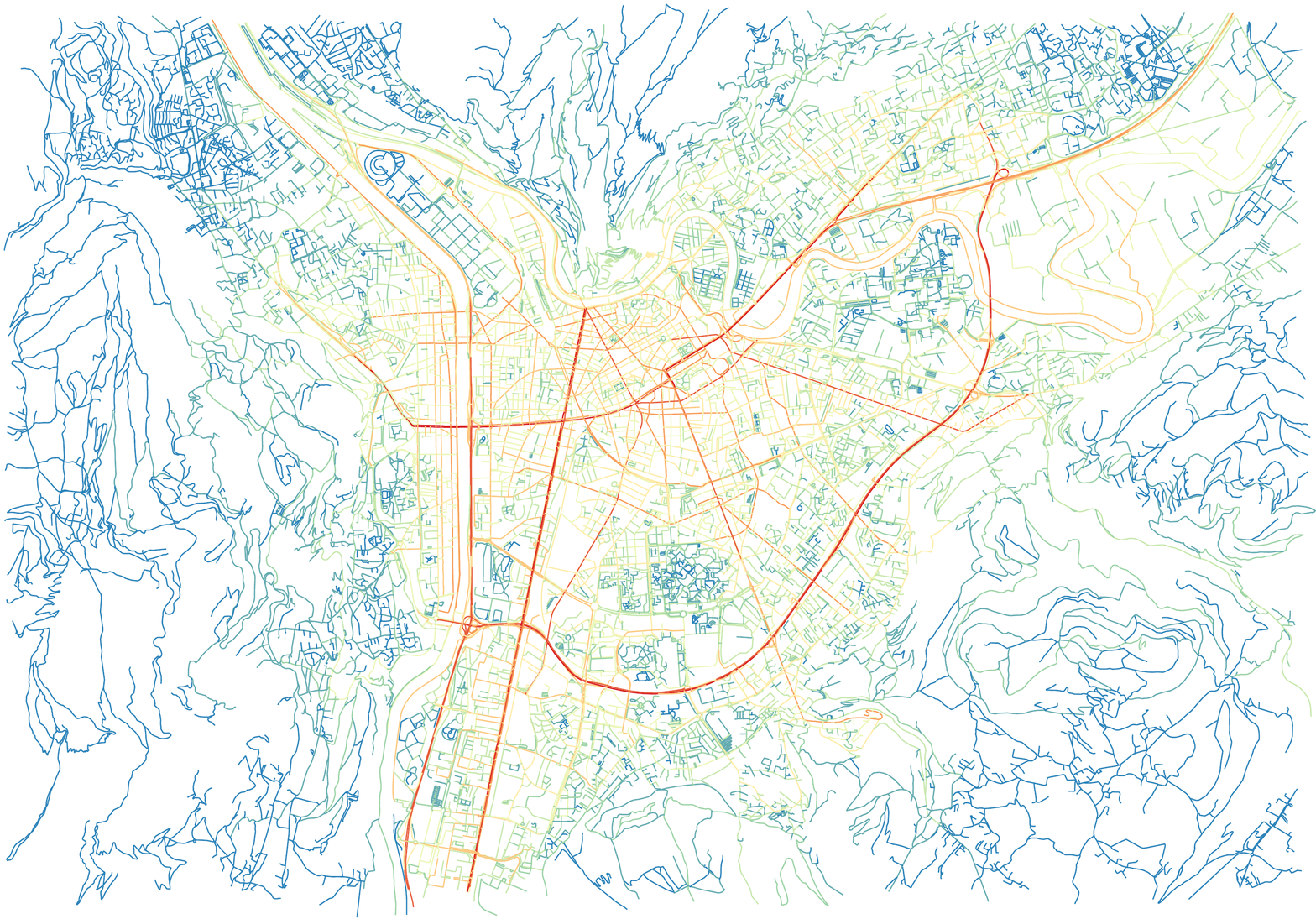}
        \caption{Représentation de l'accessibilité étalée par la fonction puissance 6 : $f(x) = x^6$ en échelle linéaire (intervalles égaux).}
    \end{subfigure}
    
    \caption{Différentes représentations du même indicateur (l'accessibilité) sur la ville de Grenoble. \\
    source : cartes réalisées par C. Bresch}
    \label{fig:grenobleClement}
\end{figure}

Clément Bresch, stagiaire dans l'équipe de recherche, a ainsi exploré les différentes méthodes de représentation qu'il était possible de choisir pour faire ressortir différents types d'informations. Dans l'exemple que nous donnons figure \ref{fig:grenobleClement}, nous observons que par la simple application à l'indicateur d'accessibilité d'une fonction de puissance, l'information que la carte fait ressortir est complètement différente. L'accent est mis sur les voies les plus accessibles, dans ce cas, et le reste de la ville est noyé dans les nuances de bleu.

\FloatBarrier
\section{Lecture de villes médiévales}

\subsection{Avignon}

Nous avons eu l'opportunité, durant l'ensemble de cette recherche, d'entretenir des liens étroits avec le service d'urbanisme d'Avignon. Nous avons ainsi pu comprendre l'histoire de cette ville et apprendre à donner du sens aux résultats que nous avons obtenus sur son graphe.

\begin{figure}[h]
    \centering
        \includegraphics[width=0.8\textwidth]{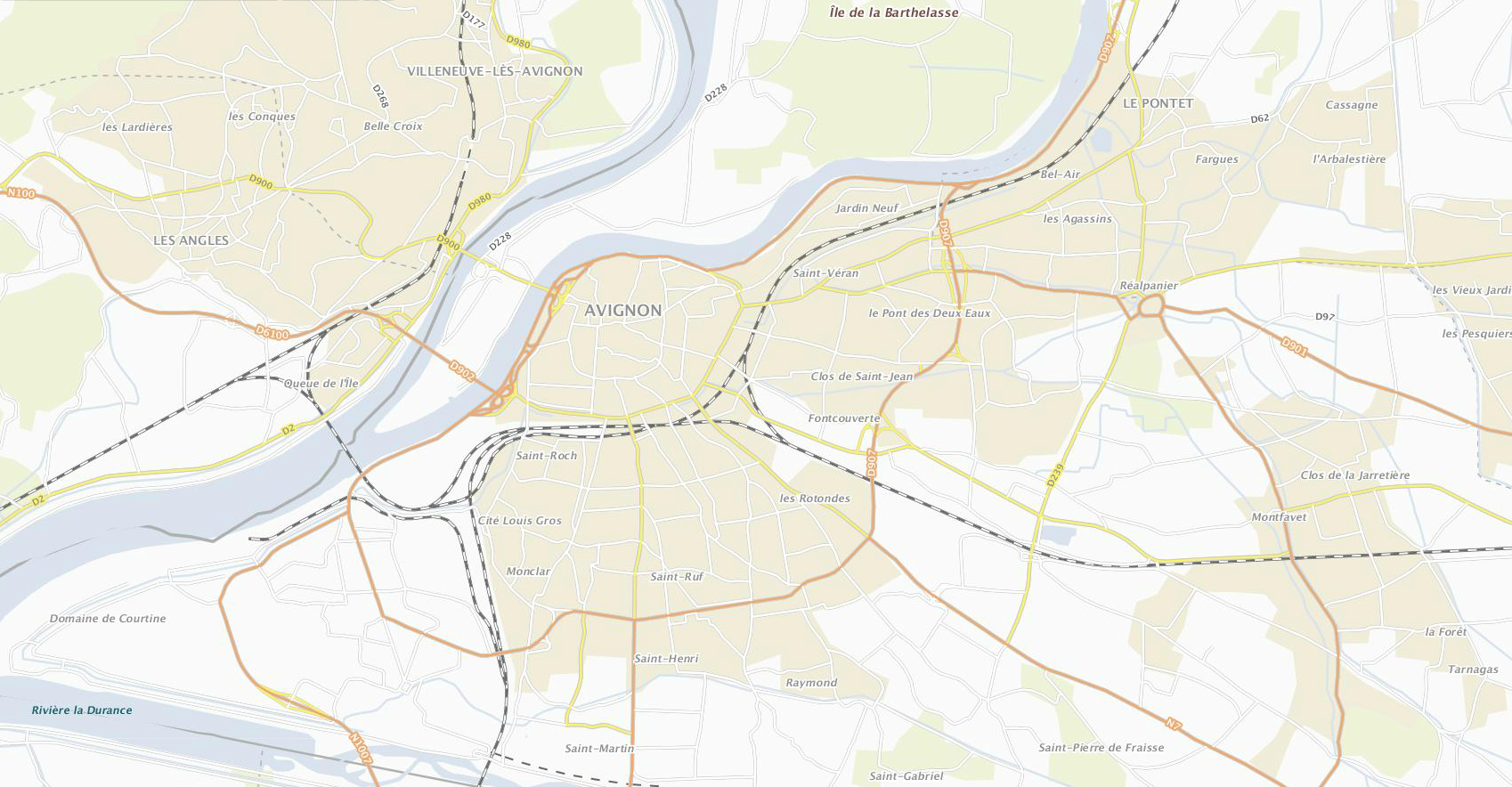}
    
    \caption{Carte du territoire d'Avignon (Plan IGN 2015).}
    \label{fig:av_plan_large}
\end{figure}

\begin{figure}[h]
    \centering
        \includegraphics[width=0.8\textwidth]{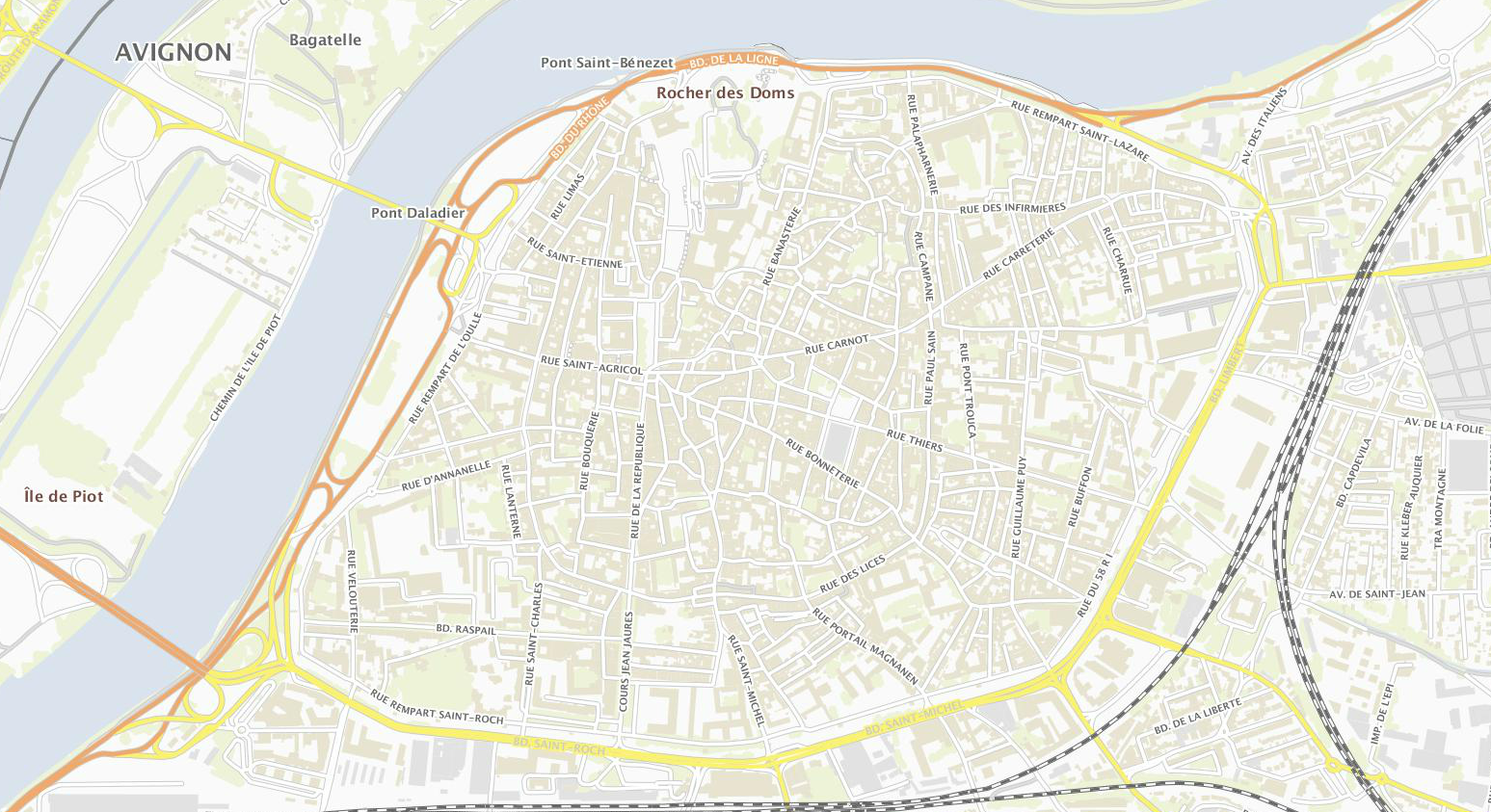}
    
    \caption{Carte de l'intra-murros d'Avignon (Plan IGN 2015).}
    \label{fig:av_plan_intra}
\end{figure}

L'intra-muros d'Avignon (figure \ref{fig:av_plan_intra}) est constitué d'un tissu médiéval partiellement remodelé, structuré par deux enceintes. La première date du XI\textsuperscript{ème} siècle et a été remplacée par une continuité de petites rues. La seconde date du XIV\textsuperscript{ème} siècle et a été conservée. Des boulevards ont été créés à l'intérieur ainsi qu'à l'extérieur pour faciliter la circulation. Cet espace a été très peu modifié depuis cette époque. La trame suivie pour l'extra-muros est radiale, orientée vers le Sud-Est. Certaines discontinuités de forme apparaissent, correspondant à des lotissements plus récents.

\begin{figure}[h]
    \centering
        \includegraphics[width=\textwidth]{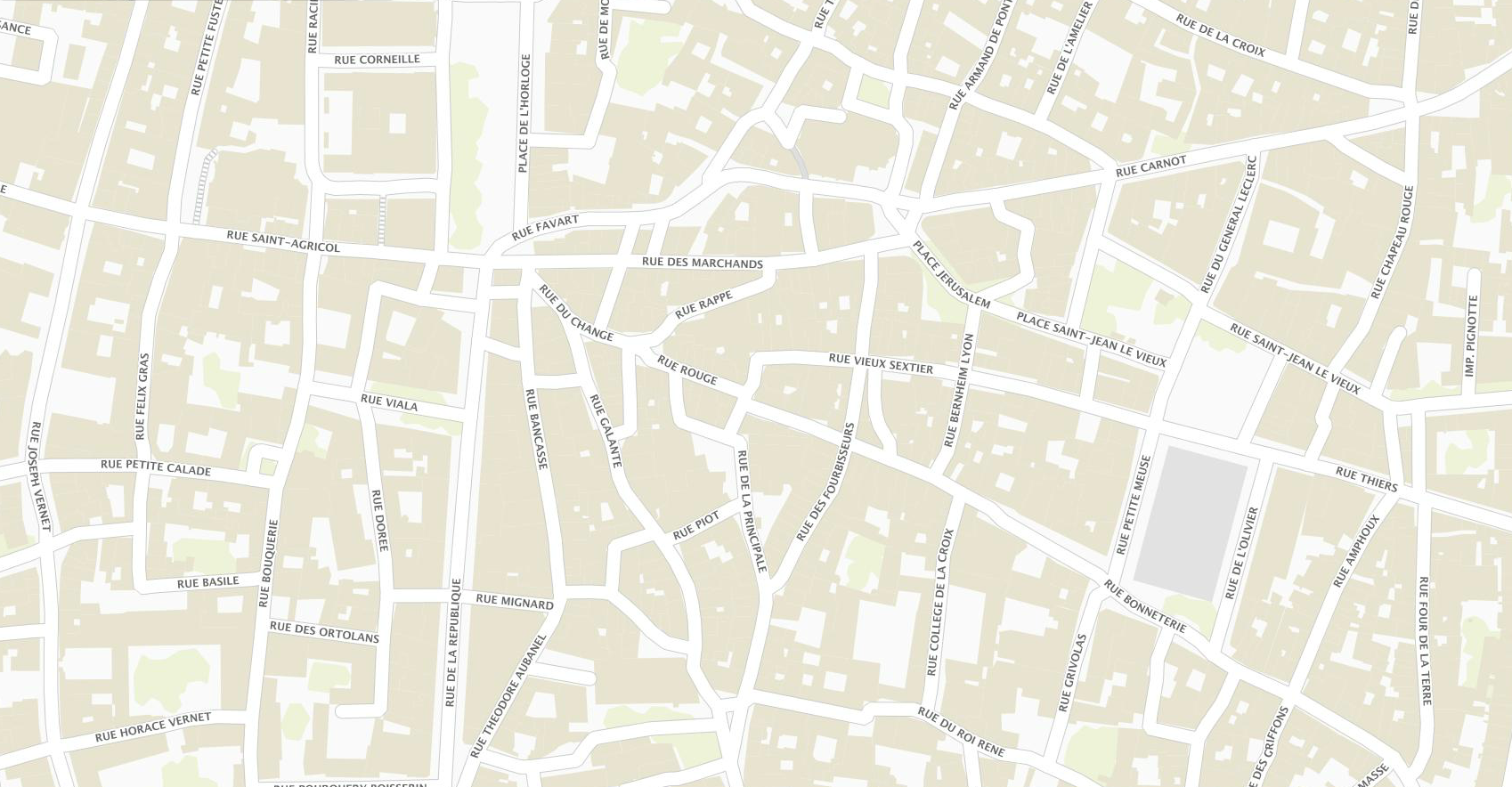}
    
    \caption{Détail du centre d'Avignon (Plan IGN 2015).}
    \label{fig:av_plan_zoom}
\end{figure}

Nous retrouvons dans le quartier intra-muros de la ville certains points d'articulation qui arbitrent la construction de la voie et ainsi la lecture que l'on peut faire du réseau. C'est le cas par exemple de la jonction entre l'avenue de la République (percée nouvelle, orientée Nord/Sud) et la rue Carnot (médiévale, orientée Est/Ouest) qui se fait sur la place de l'Horloge et avec la rue Favart en direction de la rue Carreterie (axe traditionnel)(figure \ref{fig:av_plan_zoom}).

\begin{figure}[h]
    \centering
   \begin{subfigure}[t]{0.6\linewidth}
        \includegraphics[width=\textwidth]{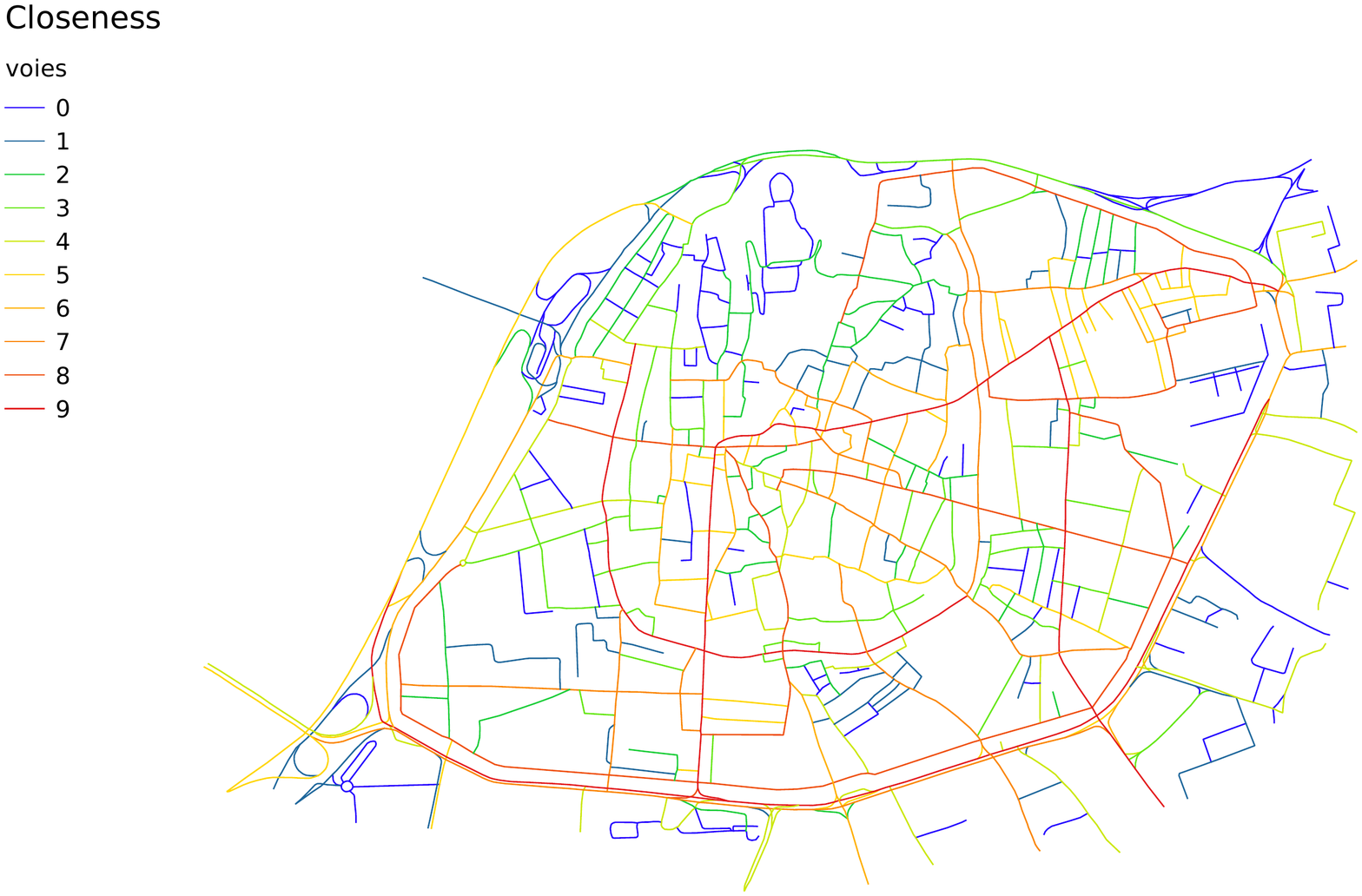}
        \caption{Sur le réseau brut, avec association d'arcs à chaque intersection.}
        \label{fig:avignon_rtopo1_2}
    \end{subfigure}
	
   \begin{subfigure}[t]{0.6\linewidth}
        \includegraphics[width=\textwidth]{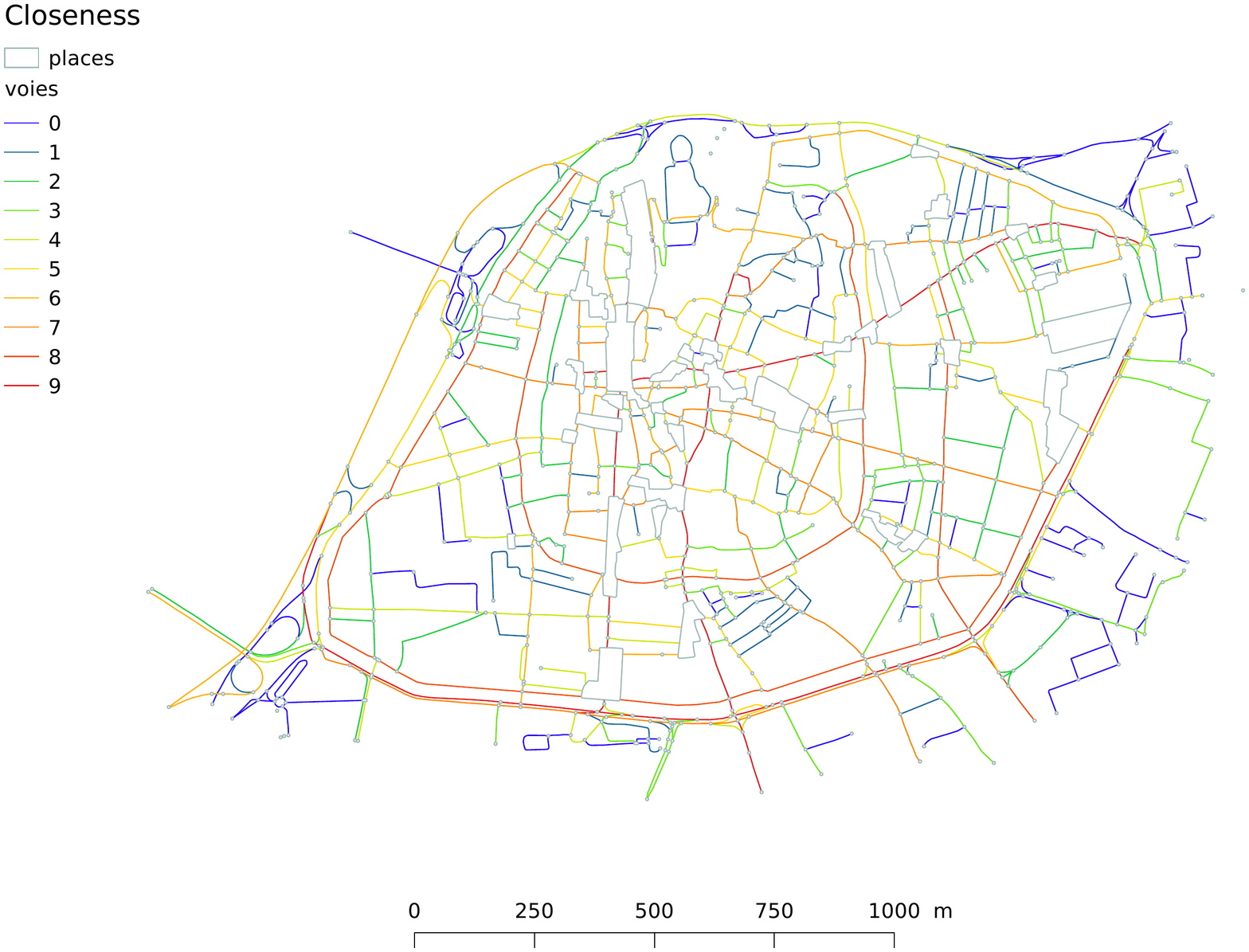}
        \caption{Sur le réseau \textit{amélioré} avec construction des voies avec les places.}
        \label{fig:avignon_rtopo2_2}
    \end{subfigure}
    
     \caption{Calcul de la closeness.}
    \label{fig:avignon_rtopo_2}
\end{figure}

Lorsque nous construisons les voies sur le graphe, ces rues sont réunies en un même objet. En observant plus précisément les données, nous remarquons que les arcs sont liés par une courbe douce. Celle-ci correspond, sur le terrain, à une \enquote{canalisation} du flux piétonnier et automobile sur la place, faite par des plots. Dans ce cas, les cartes construites avec le calcul des indicateurs mettent en évidence une structure continue très centrale (en terme d'accessibilité) dans le réseau. Cela correspond en effet au transit automobile dans la ville (orienté ainsi du Sud vers l'Est et réciproquement) et participe aux chemins les plus simples empruntés (lire ensuite, enquête menée par E. Degouys). Cependant, si l'on reprend les calculs en positionnant les places vectorisées, l'avenue de la République disparaît au profit de structures plus anciennes, sur lesquelles s'est effectivement construit le reste du réseau (figure \ref{fig:avignon_rtopo_2}).

\FloatBarrier

\begin{figure}[h]
    \centering
        \includegraphics[width=0.6\textwidth]{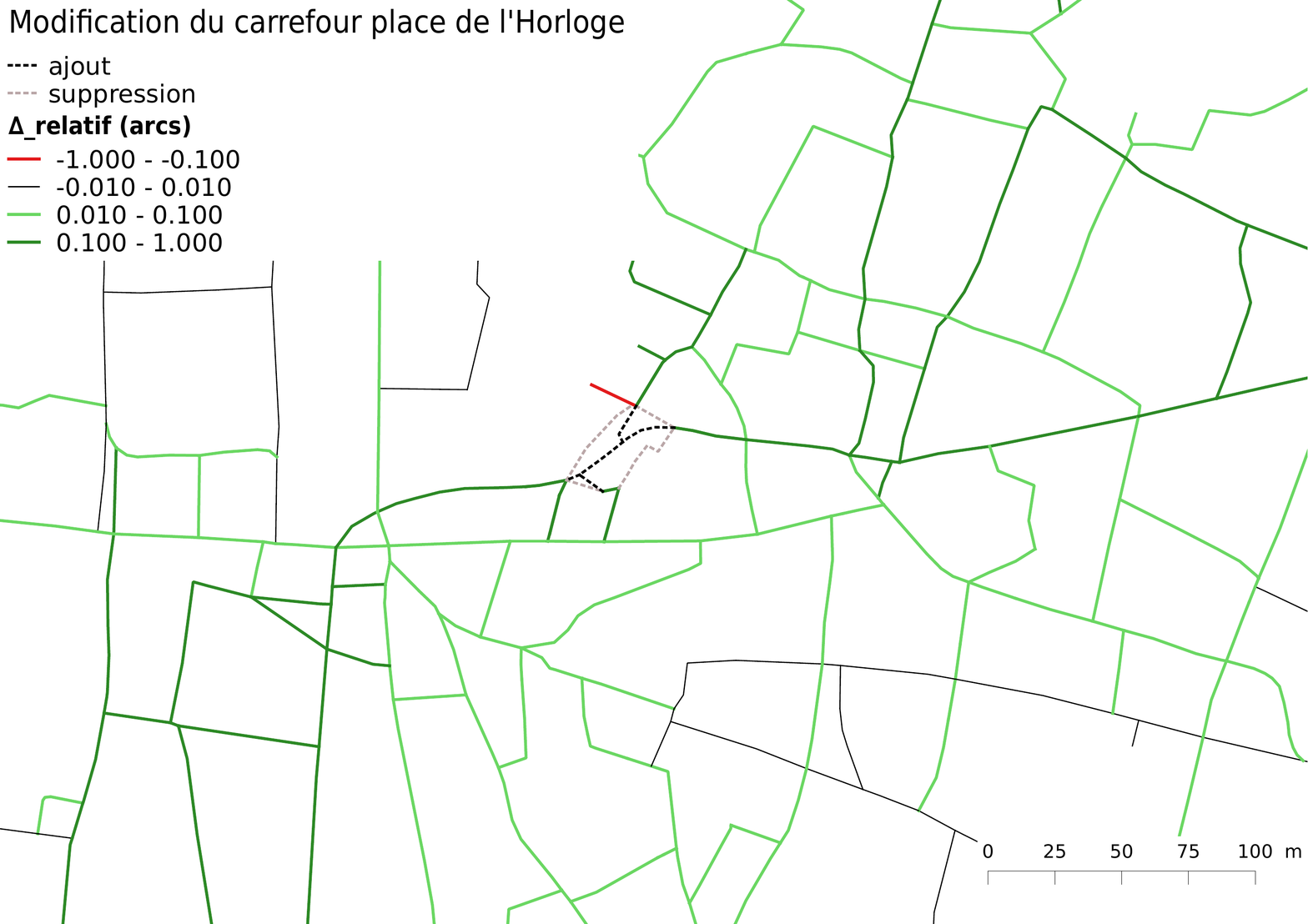}  
    
    \caption{Étude cartographique de $\Delta_{relatif}$ de la modification du centre du graphe d'Avignon intra-murros (extrait de la carte présentée dans le chapitre 4 de la deuxième partie).}
    \label{fig:proj_avcentre_detail}
\end{figure}

Ce point d'articulation fonctionne de la même manière que celui situé à proximité (vers l'Est), entre la rue Carnot et la rue Favart. Ce dernier résulte d'un remaniement. Un projet urbain a supprimé l'îlot présent à la jonction des deux rues pour organiser le trafic intra-muros de façon continue entre les deux voies (figure \ref{fig:proj_avcentre_detail}). Nous remarquons que ce \textit{pliage} d'un axe majeur d'Avignon correspond au coude du Rhône, dont la rive accueille un très faible développement urbain. Le territoire au Sud-Est vers lequel s'oriente \textit{le pliage} connaît, à l'inverse, l'essentiel du développement.

Selon ce que l'on veut observer, l'une ou l'autre méthodologie a donc sa pertinence. Les questions de recherche guident les choix méthodologiques qui arbitrent l'interprétation.

\FloatBarrier

\begin{figure}[h]
    \centering
    \begin{subfigure}[t]{.48\linewidth}
        \includegraphics[width=\textwidth]{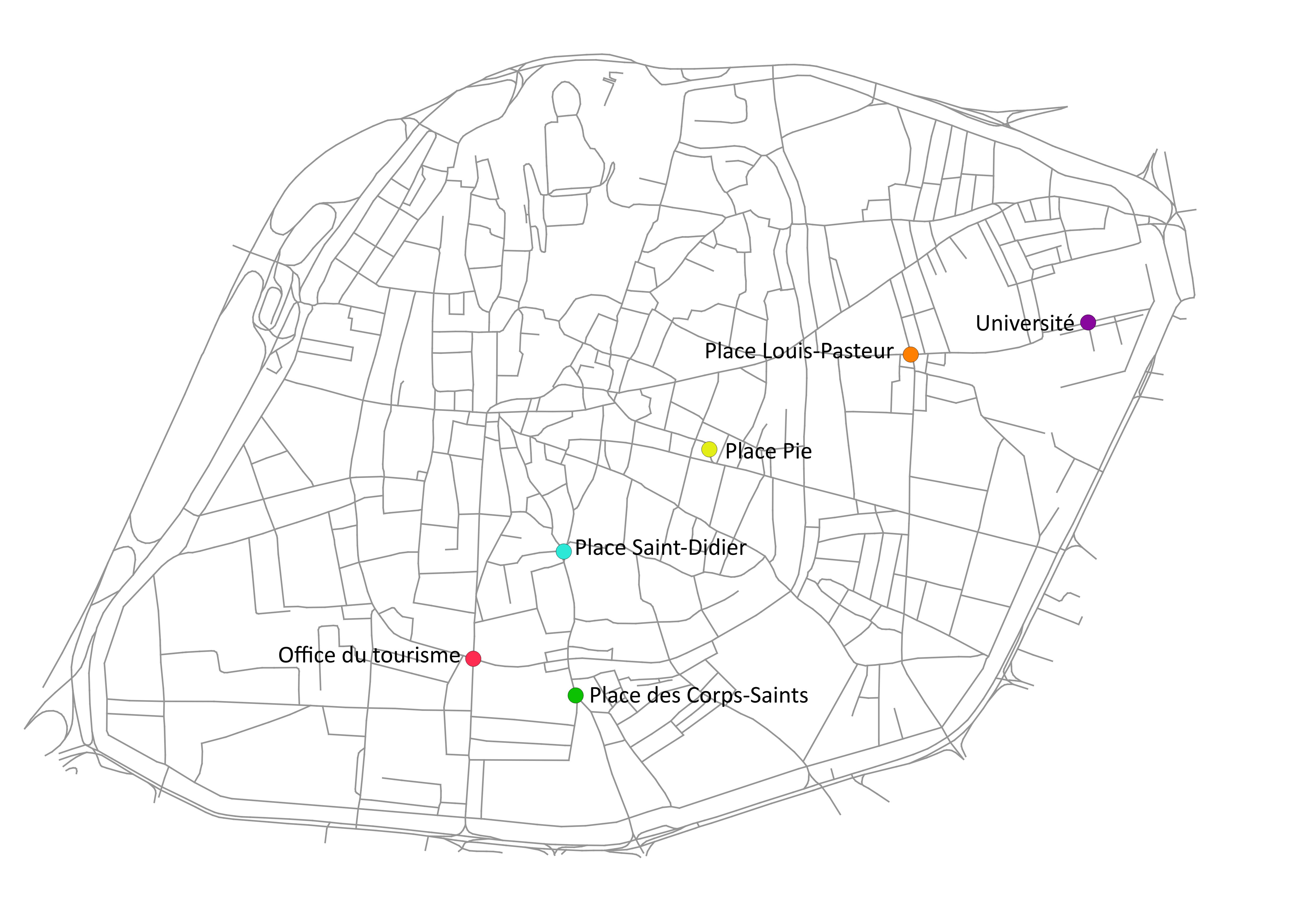}
        \caption{Points choisis comme origine ou destination sur le réseau intra-muros d'Avignon.}
    \end{subfigure}
    ~
    \begin{subfigure}[t]{.48\linewidth}
        \includegraphics[width=\textwidth]{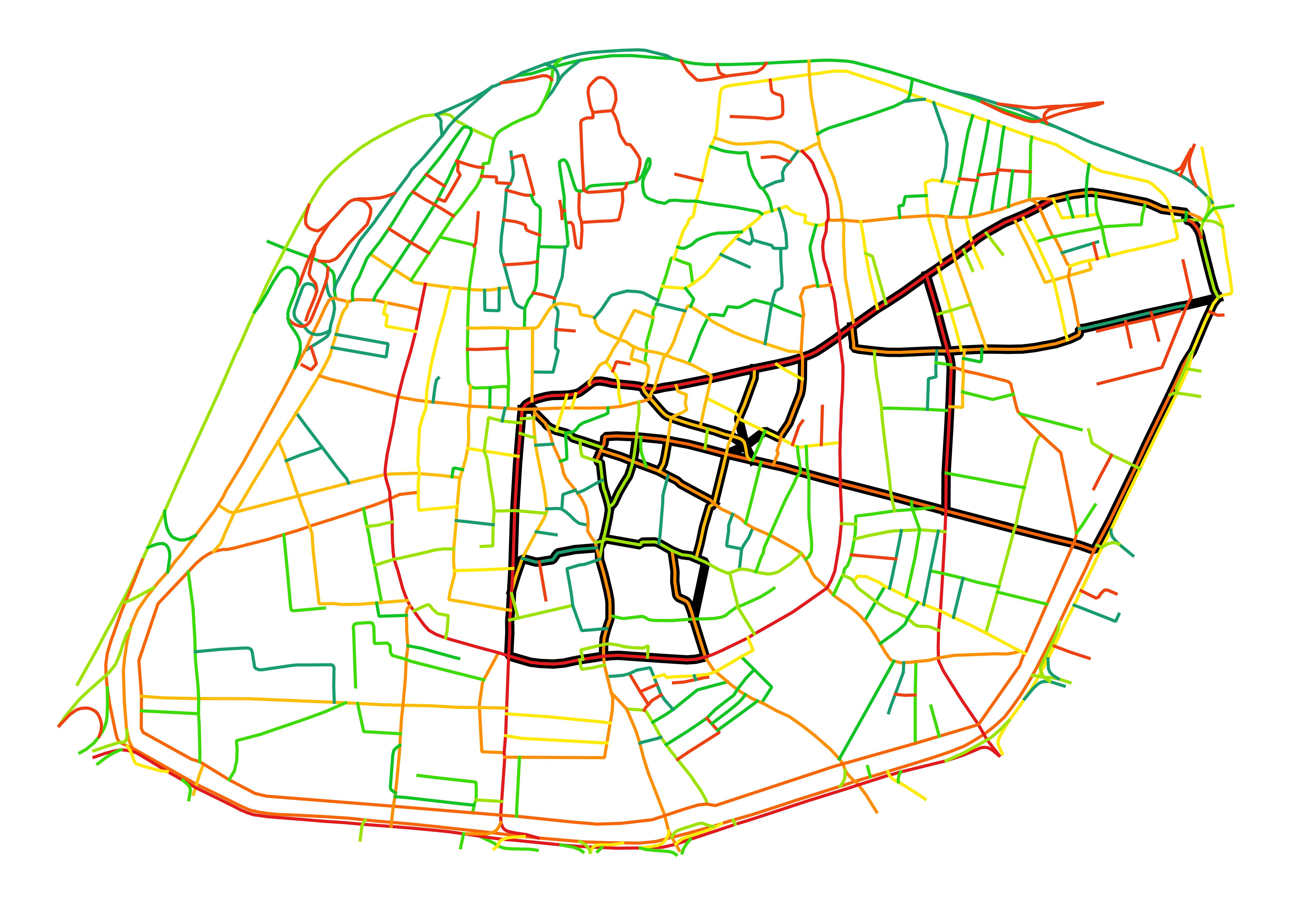}
        \caption{Réponses entendues superposées à la carte de closeness calculée sur le réseau intra-murros d'Avignon.}
    \end{subfigure}
    \caption{Illustration de l'enquête de terrain menée par E. Degouys sur l'intra-muros d'Avignon.\\
    source : cartes réalisées par E. Degouys}
    \label{fig:avignonEstelle1}
\end{figure}

Estelle Degouys, doctorante au sein de l'équipe MorphoCity, poursuit un travail approfondi sur les choix d'itinéraires des piétons dans l'intra-muros d'Avignon. Elle a choisi une série de points stratégiques et recueille les témoignages des usagers sur les chemins qu'ils prendraient eux-même ou indiqueraient à d'autres entre deux de ces points. Les trajets évoqués sont souvent les plus simples, corrélés aux voies révélées par l'indicateur de closeness (figure \ref{fig:avignonEstelle1}). Cependant, il est nécessaire de repositionner l'information recueillie dans son contexte et de comprendre si c'est la simplicité du déplacement, ou d'énonciation du déplacement, qui prime. En effet, la longueur de l'information transmise peut être un facteur important dans de telles enquêtes, la personne interrogée préférant minimiser le vocabulaire d'action employé si elle a peu de temps. Transmettre une information simple est également une stratégie efficace pour ne pas perdre son interlocuteur. Le chemin le plus simple réunit donc plusieurs avantages dans son énonciation.

\begin{figure}[h]
    \centering
    \begin{subfigure}[t]{.4\linewidth}
        \includegraphics[width=\textwidth]{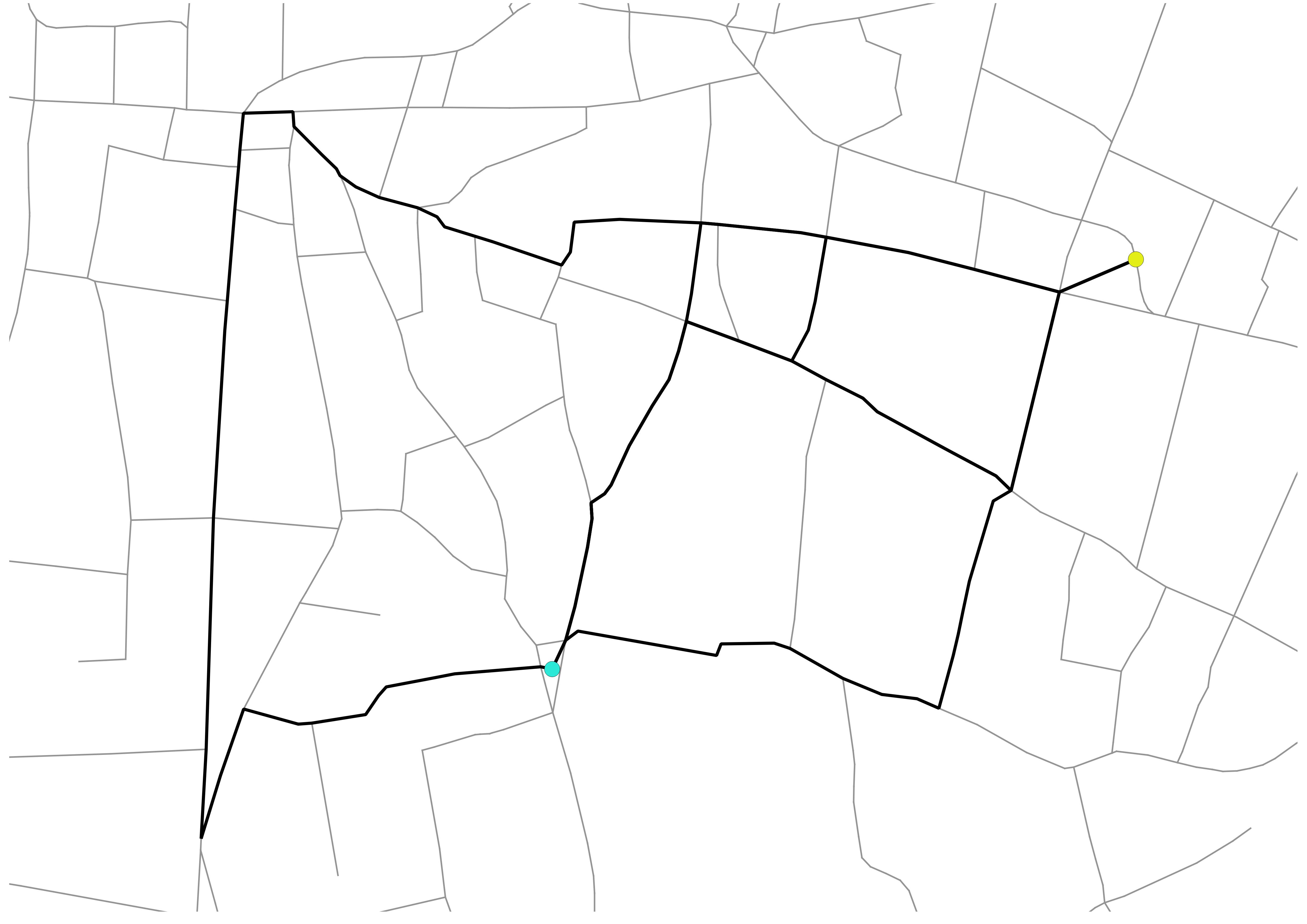}
        \caption{Itinéraires évoqués entre la Place Pie et la Place Saint-Didier.}
        \label{fig:avignonEstelle2_1}
    \end{subfigure}
	~
    \begin{subfigure}[t]{.4\linewidth}
        \includegraphics[width=\textwidth]{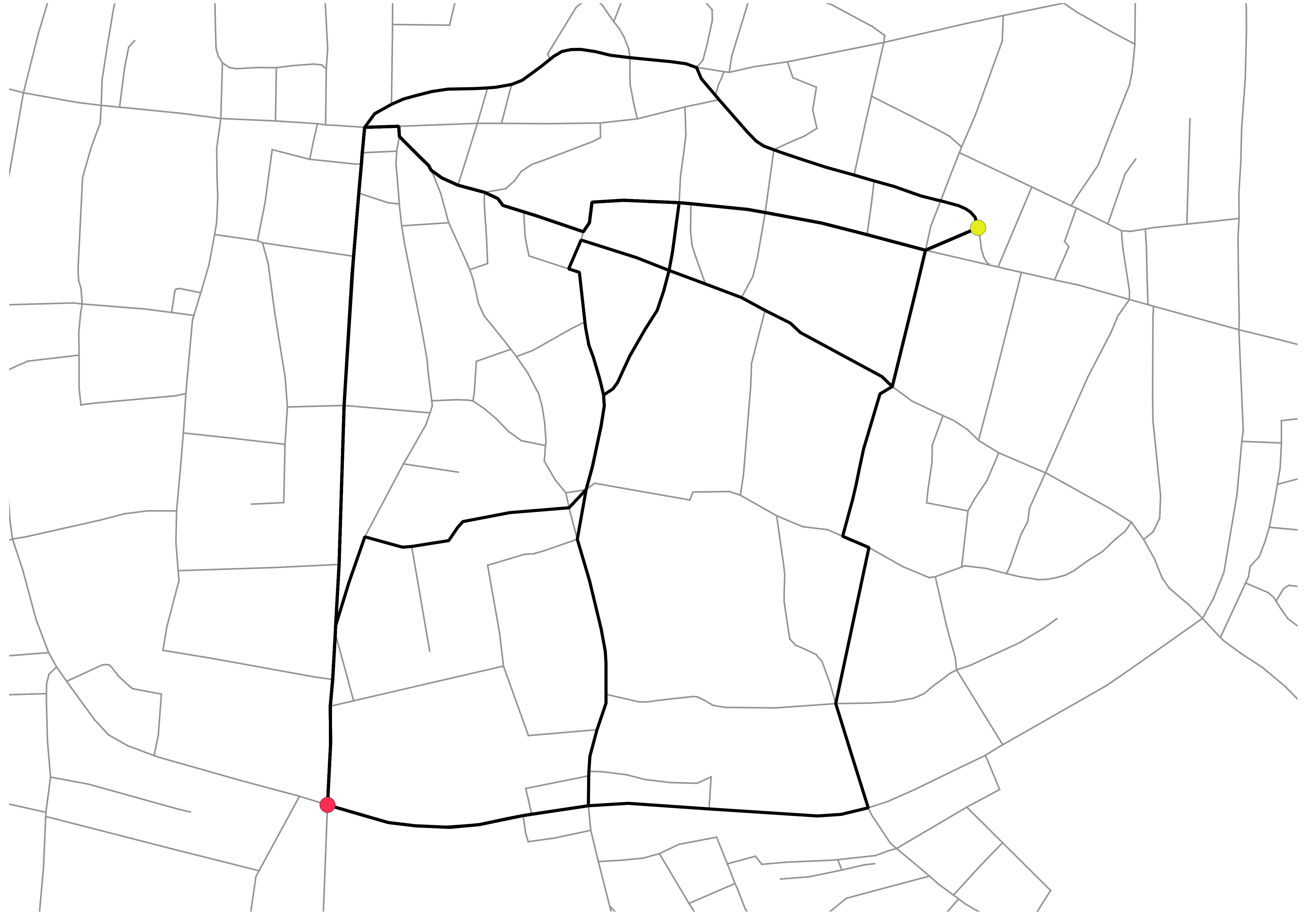}
        \caption{Itinéraires évoqués entre la Place Pie et l'Office du tourisme.}
        \label{fig:avignonEstelle2_2}
    \end{subfigure}
    \caption{Illustration des itinéraires évoqués entre deux points (origine-destination) sur un total de 13 personnes sondées. Enquête de terrain menée par E. Degouys sur l'intra-murros d'Avignon. \\
    source : cartes réalisées par E. Degouys}
    \label{fig:avignonEstelle2}
\end{figure}

Dans l'enquête menée, pour des points d'origine et de destination identiques, les itinéraires évoqués diffèrent (figure \ref{fig:avignonEstelle2}). Dans tous les cas, si l'on considère un couple de points (origine-destination), et que l'on sélectionne l'ensemble des rues utiles au déplacement selon les témoignages, nous retrouvons dans la plupart des cas les arêtes du rectangle dont la diagonale est formée par les deux points (figure \ref{fig:avignonEstelle2_1}), qui parfois s'élargit vers une structure de forte closeness (figure \ref{fig:avignonEstelle2_2}). Ces arêtes correspondent aux chemins les plus simples, qui représentent le moins de changements de voie. Ils reviennent de manière forte dans les discours \citep{degouysAPitineraire}.

\FloatBarrier

Si nous élargissons notre territoire d'étude jusqu'au boulevard hors les murs qui contourne la ville ancienne, nous pouvons faire ressortir plusieurs structures. Ce graphe de travail correspond à celui présenté dans l'étude de l'impact des effets de bord (échantillon 2).

\begin{figure}[h]
    \centering
   \begin{subfigure}[t]{\linewidth}
         \centering
  		 \includegraphics[width=\textwidth]{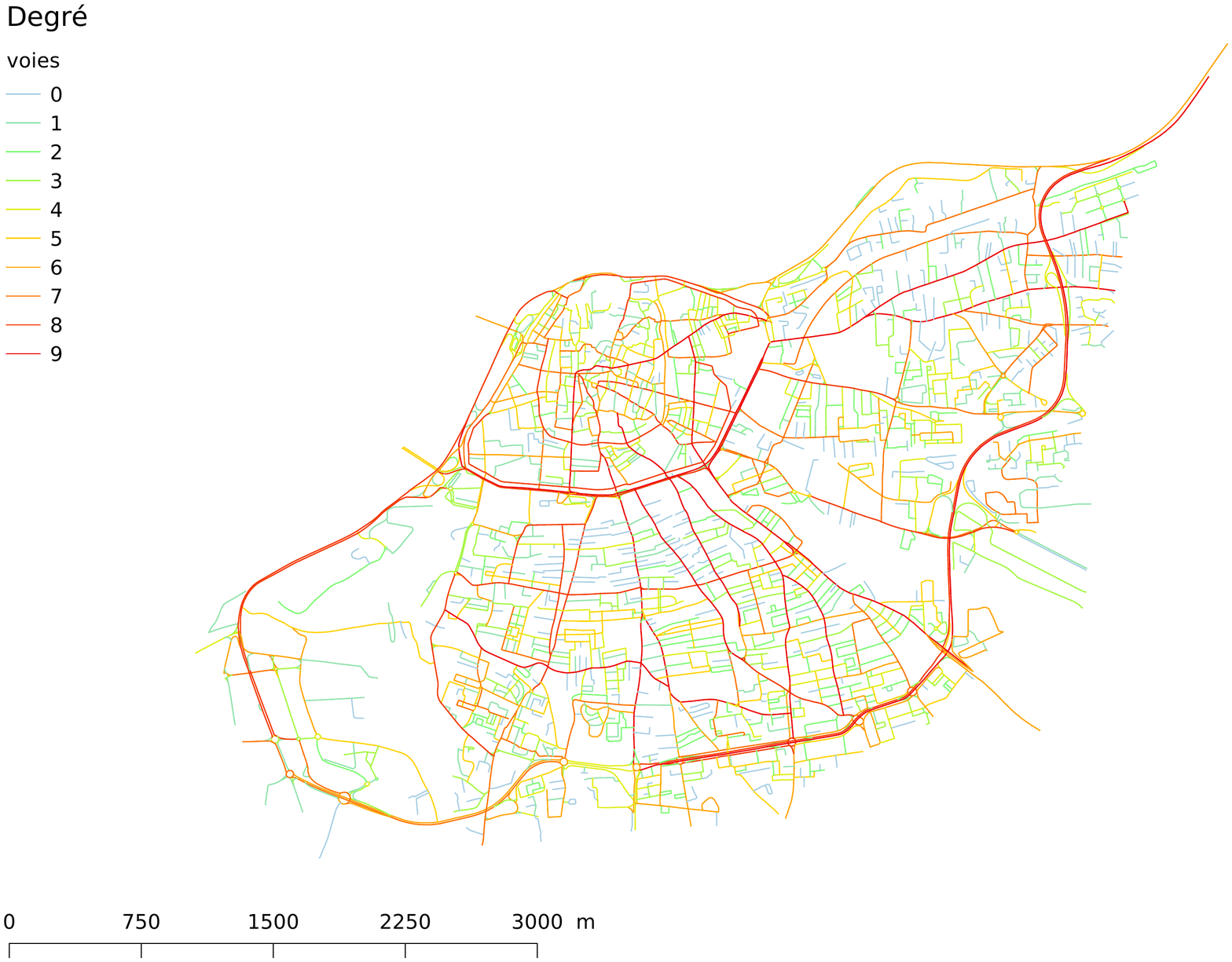}
   		 \caption{Cartographie de l'indicateur.}
   		 \label{fig:avignon_deg}
    \end{subfigure}
	
   \begin{subfigure}[t]{\linewidth}
       \centering
 	   \includegraphics[width=\textwidth]{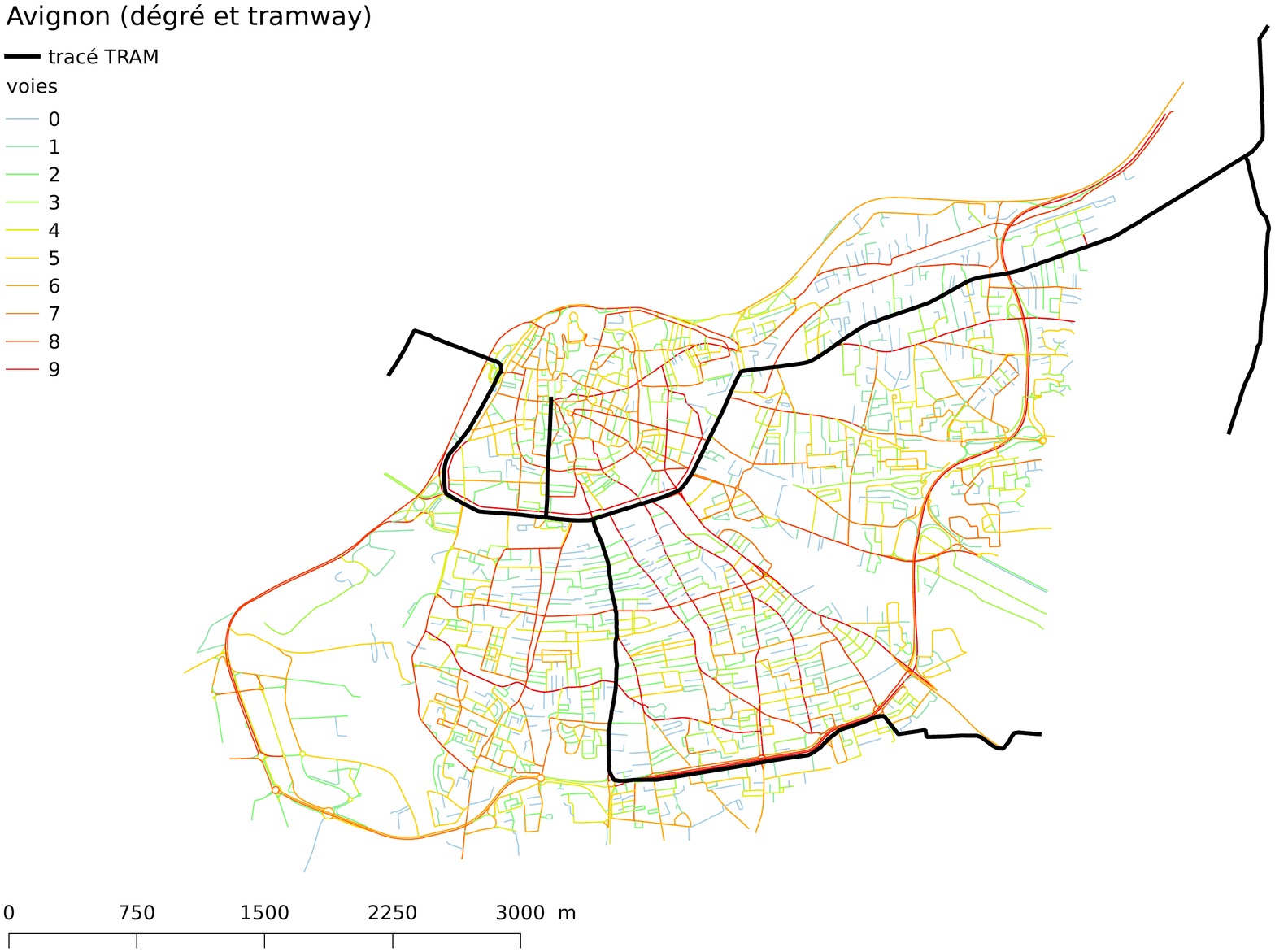}
 	   \caption{Superposition du tracé en projet pour la construction du tramway.}
    	\label{fig:avignon_deg_tram}
  	\end{subfigure}
    
     \caption{Calcul du degré sur les voies d'Avignon (échantillon 2).}
    \label{fig:avignon_deg_et_tram}
\end{figure}

Le degré des voies fait ressortir une structure principale maillée qui recouvre l'ensemble du territoire communal (figure \ref{fig:avignon_deg}). Les axes qui ressortent avec un degré fort sont ceux qui assurent une circulation rapide d'un bout à l'autre du territoire. Ce sont ces mêmes axes qui ont été choisis comme support au développement d'une ligne de tramway (en projet). En effet, les voies incluent l'alignement dans leur construction, et les plus connectées sont souvent celles qui optimisent la desserte du territoire. La superposition des lignes du projet avec le graphe met en évidence la nette corrélation entre nos résultats quantitatifs et les choix du service d'urbanisme de la commune (figure \ref{fig:avignon_deg_tram}).

\FloatBarrier

\begin{figure}[h]
    \centering
    \includegraphics[width=\textwidth]{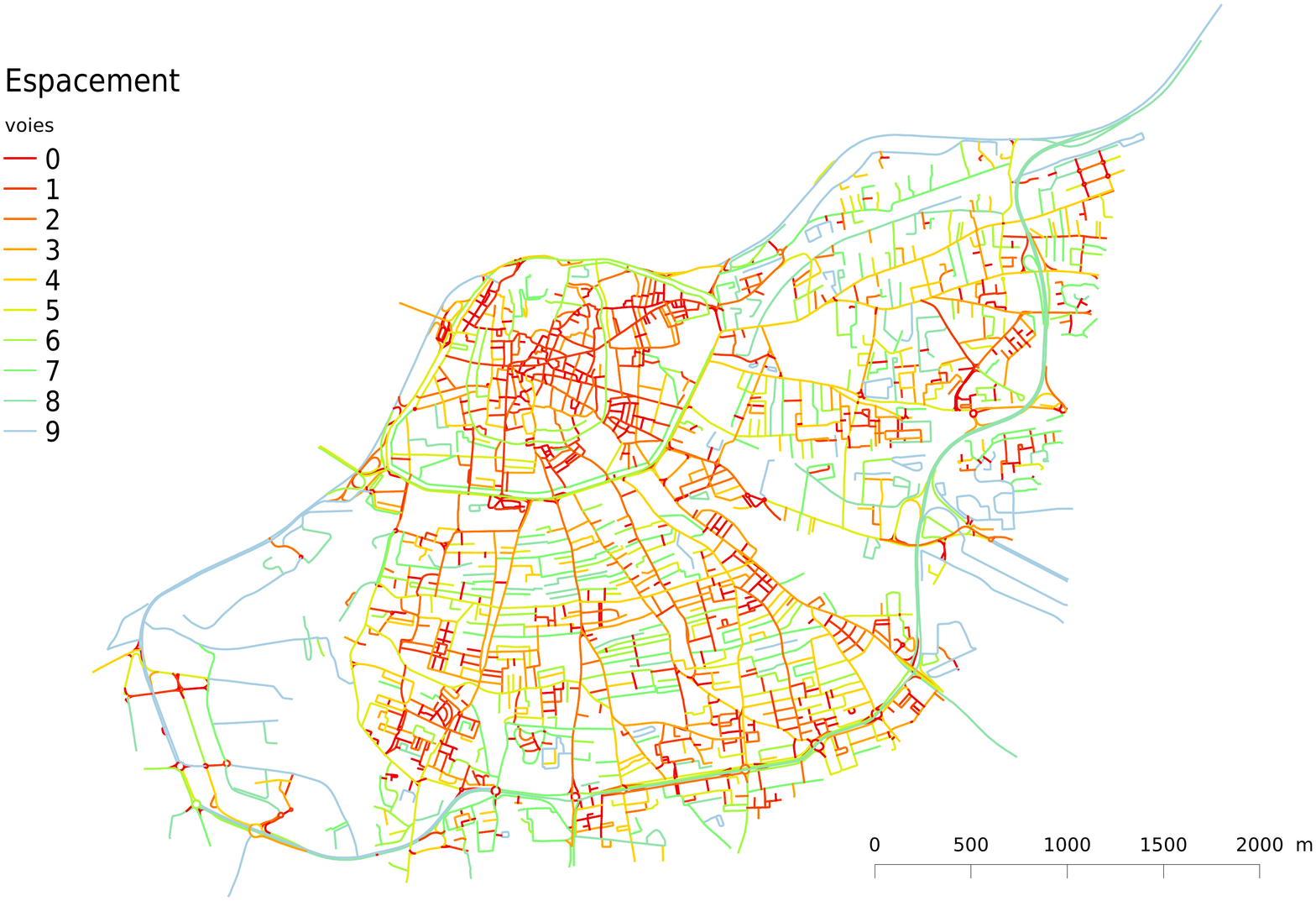}
    \caption{Calcul de l'espacement sur les voies d'Avignon (échantillon 2).}
    \label{fig:avignon_esp}
\end{figure}

L'indicateur d'espacement, inverse de la densité linéaire du réseau, fait ressortir les parties denses du tissu (figure \ref{fig:avignon_esp}). Sur le terrain, E. Degouys et Ph. Bonnin, ont relevé sa nette corrélation avec les centres-villes et les quartiers résidentiels. En zone péri urbaine, l'espacement fait ressortir des densités résidentielles au réseau local très serré (cités / lotissements). En effet, dans ces parties de la ville, le territoire nécessite une desserte locale fine. Cela se traduit donc par des arcs plus courts et plus connectés, ayant pour but un recouvrement optimal de l'espace. Nous avons vu précédemment que la lecture de cet indicateur sur les voies ou les arcs est équivalente. En effet, sa valeur correspond à la distance moyenne entre deux intersections, le calcul sur les voies n'apporte donc pas d'information supplémentaire.

\FloatBarrier

L'indicateur de closeness met en avant la proximité topologique entre les objets. Plus une voie aura un indicateur de closeness élevé, plus elle permettra d'accéder à l'ensemble du réseau avec un nombre minimum de changements de voie (qualifiés de \enquote{tournants}). Or, les structures qui permettent d'accéder rapidement à l'ensemble du réseau sont issues de deux catégories de construction bien distinctes : les plus anciennes (squelette historique) et les plus récentes (stratégies de traversée ou de contournement de la ville).

\begin{figure}[h]
    \centering
    \includegraphics[width=\textwidth]{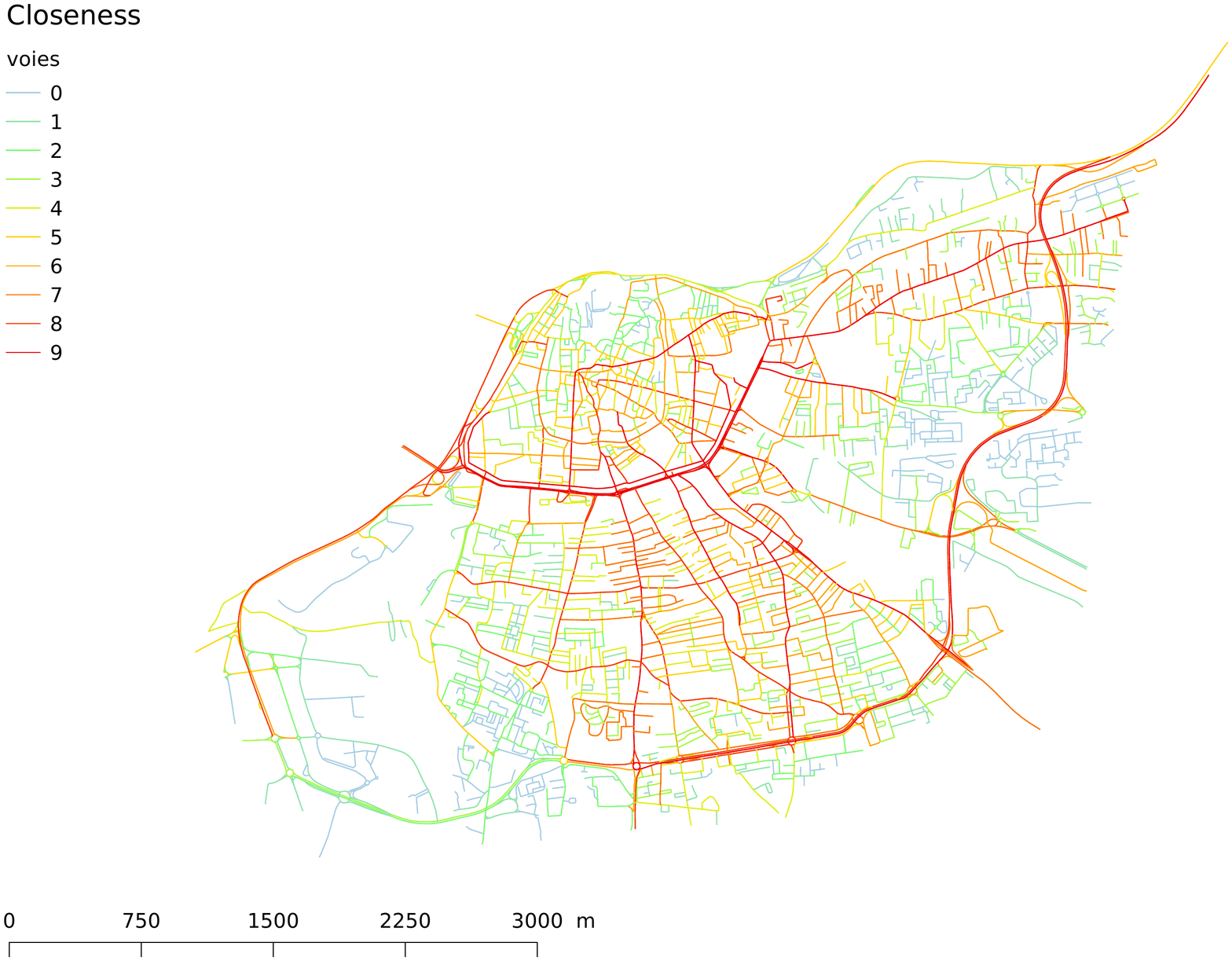}
    \caption{Calcul de la closeness sur les voies d'Avignon (échantillon 2).}
    \label{fig:avignon_clo}
\end{figure}

Ainsi, à Avignon (figure \ref{fig:avignon_clo}), nous observons des structures anciennes avec une forte closeness : les voies qui longent le Rhône ou la Durance ; la muraille extérieure au centre de la ville, structure ancienne contournée par des boulevards ; la rue Carreterie et la route de Lyon, qui ouvrent le centre de la ville vers l'Est ; ainsi que les avenues Saint-Ruf, Pierre Semard et la route de Montfavet, qui permettent une traversée du tissu vers le Sud. Dans la toponymie de ces rues, nous retrouvons leur caractère ancien de jonction entre les points d'intérêt du territoire, plus ou moins proches (\textit{route de Lyon, route de Montfavet}). À ces structures anciennes, viennent s'ajouter des voies qui correspondent à des créations plus récentes. Nous retrouvons ainsi le périphérique qui contourne le centre-ville à l'Est. Nous remarquons que toutes les structures que nous lisons avec un fort indicateur de closeness, sont également celles mises en valeur (trait plus épais) sur la carte de l'IGN (figure \ref{fig:av_plan_large}). Notre résultat mathématique vient donc rejoindre une cartographie établie à partir des propriétés qualitatives des routes d'un territoire.

Par ailleurs, sur la figure \ref{fig:avignon_clo}), nous distinguons trois parties du graphe qui ressortent avec un coefficient de closeness particulièrement faible. Il s'agit du palais des Papes (au Nord), d'une Zone Urbaine Sensible (au Sud-Ouest) et d'un quartier résidentiel aisé (à l'Est). Ces trois espaces caractérisés de la même manière par l'indicateur, ont des stratégies d'isolement du réseau très différentes. Le premier, le palais, est situé sur un rocher. Le dénivelé important rend son accès escarpé et minimise les chemins d'accès. Le second, la ZUS, subit son isolement. Cette partie du graphe ne suit pas les mêmes logiques géométriques que le reste du réseau et est de ce fait en rupture avec celui-ci. Le troisième, le quartier résidentiel aisé, a choisi son isolement. Sa géométrie est également en rupture avec celle de son environnement mais cela résulte d'une volonté particulière. L'objectif des personnes habitant ce quartier est de minimiser le passage traversant des voitures pour permettre, par exemple, à leurs enfants de jouer en sécurité près de la rue. La géométrie complexe du réseau à cet endroit dissuade les visiteurs étrangers de s'y aventurer au risque de s'y perdre. 

\begin{figure}[h]
    \centering
    \includegraphics[width=\textwidth]{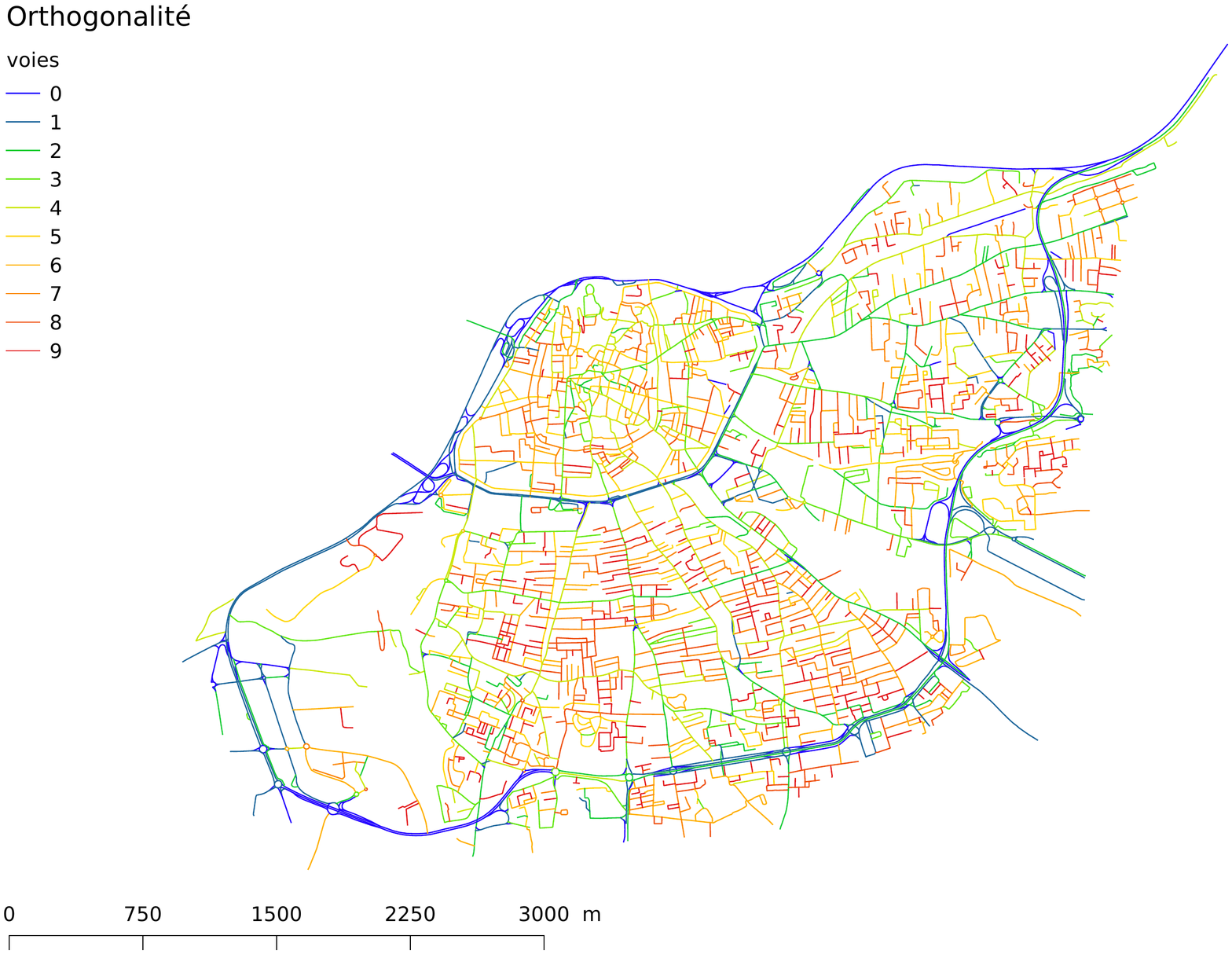}
    \caption{Calcul de l'orthogonalité sur les voies d'Avignon (échantillon 2).}
    \label{fig:avignon_ortho}
\end{figure}

L'indicateur d'orthogonalité permet quant à lui d'opposer les structures souples aux plus maillées (figure \ref{fig:avignon_ortho}). Il met ainsi en valeur les voies de contournement de la ville (valeurs les plus faibles) : leurs  connexions avec le reste du réseau se font avec des angles faibles qui permettent aux véhicules de s'introduire plus facilement sur des parcours où la vitesse de circulation est souvent plus élevée. Les structures traversantes du tissu apparaissent avec un indicateur d'orthogonalité intermédiaire (en vert). En effet, le plus souvent, ces voies sont longues et de degré important. Elles réunissent donc des angles de connexion variés ce qui leur donne une moyenne intermédiaire pour cet indicateur. Enfin, les valeurs les plus fortes de l'orthogonalité font ressortir les voies les plus locales, dont l'inclusion dans le réseau regroupe des angles proches de la perpendiculaire.

\begin{figure}[h]
    \centering
    \includegraphics[width=\textwidth]{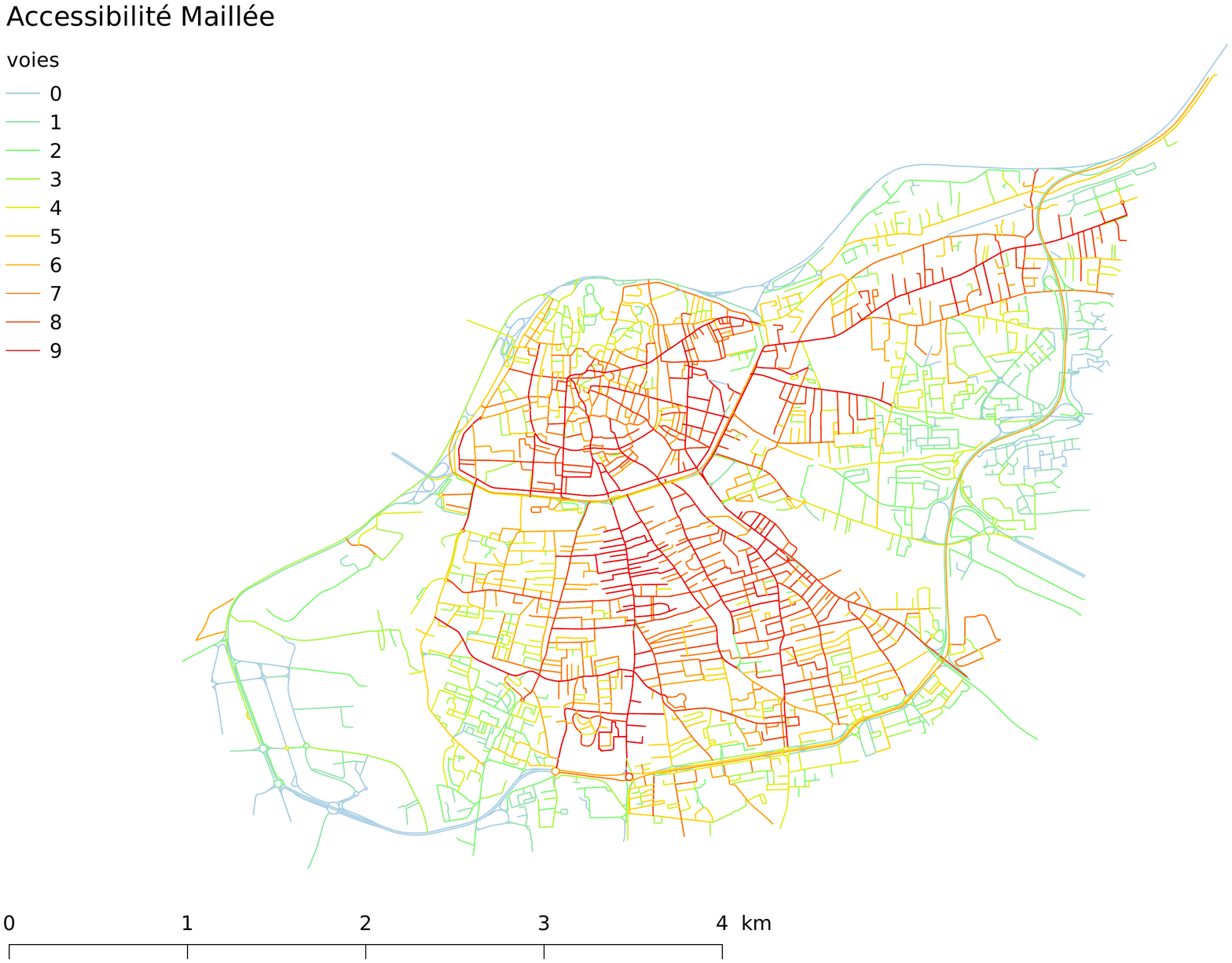}
    \caption{Calcul de l'accessibilité maillée sur les voies d'Avignon (échantillon 2).}
    \label{fig:avignon_roo}
\end{figure}

En combinant les informations de closeness et d'orthogonalité, nous obtenons l'indicateur que nous avons baptisé accessibilité maillée (figure \ref{fig:avignon_roo}). Celui-ci nous permet de retrouver la trame centrale historique de la ville. Il met en valeur les structures \enquote{en peigne} qui sont dans la plupart des cas les plus anciennes. Il permet d'isoler celles-ci des voies plus récentes (structures de contournement identifiées avec une valeur d'orthogonalité faible).

L'analyse topologique et géométrique du réseau nous permet ainsi de retracer une partie de son histoire. L'intérêt de notre travail est confirmé par les spécialistes de la ville qui nous aident à lire dans nos résultats des informations pertinentes. L'approche d'un territoire par ce biais mathématique est donc révélatrice de certaines de ses propriétés qualitatives. Nous verrons dans la suite de ce chapitre comment nos indicateurs nous permettent d'appréhender d'autres réseaux viaires de notre panel de recherche. Nous avons regroupé des informations d'utilisation et d'histoire afin de les confronter à nos résultats.

\FloatBarrier
\subsection{Brive-la-Gaillarde}

\begin{figure}[h]
    \centering
	\includegraphics[width=0.8\textwidth]{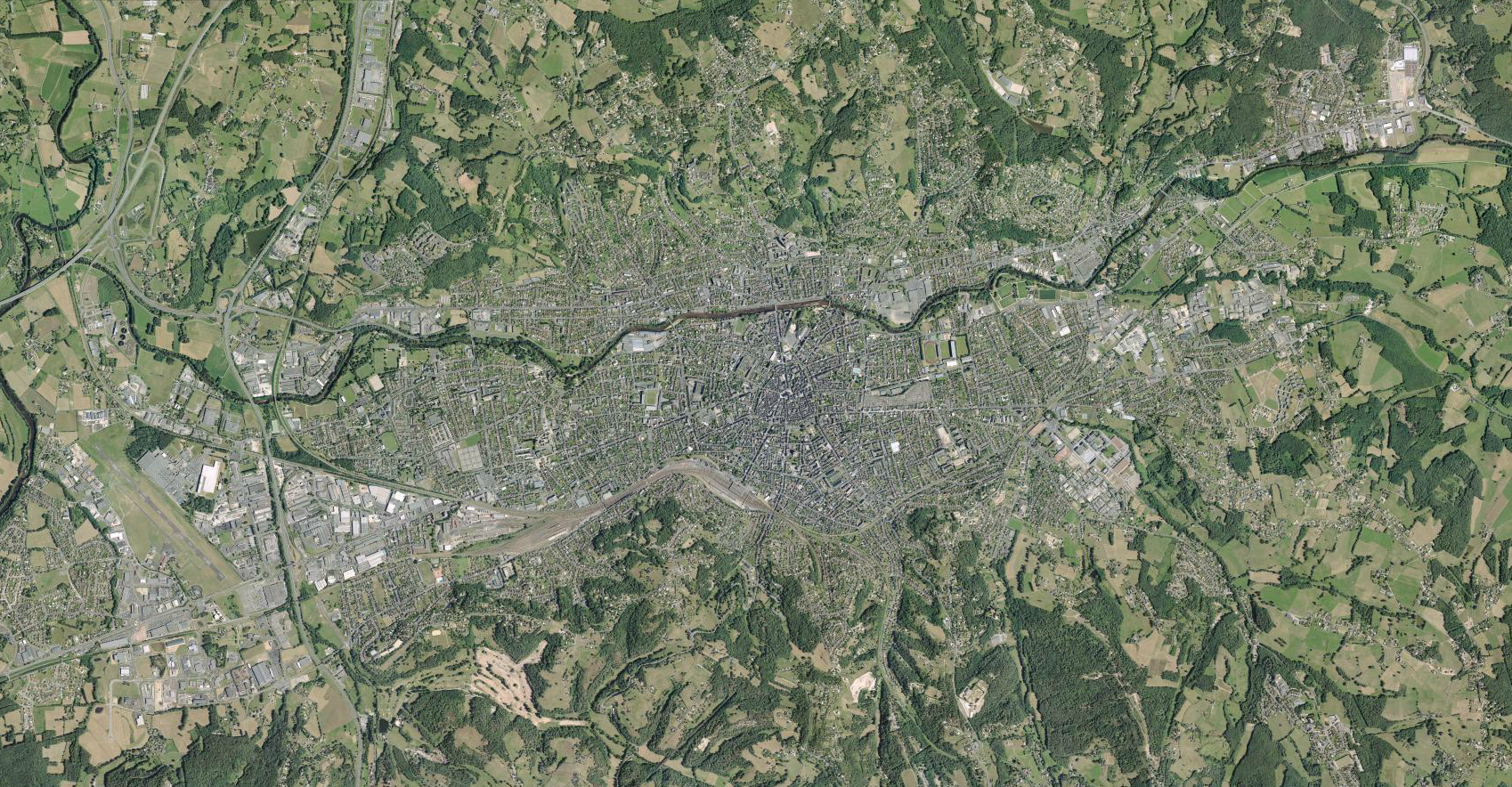}
	\caption{Vue aérienne du territoire de Brive-la-Gaillarde (Photographie aérienne IGN 2015).}
	\label{fig:brive_carte}
\end{figure}

\begin{figure}[h]
    \centering
	\includegraphics[width=\textwidth]{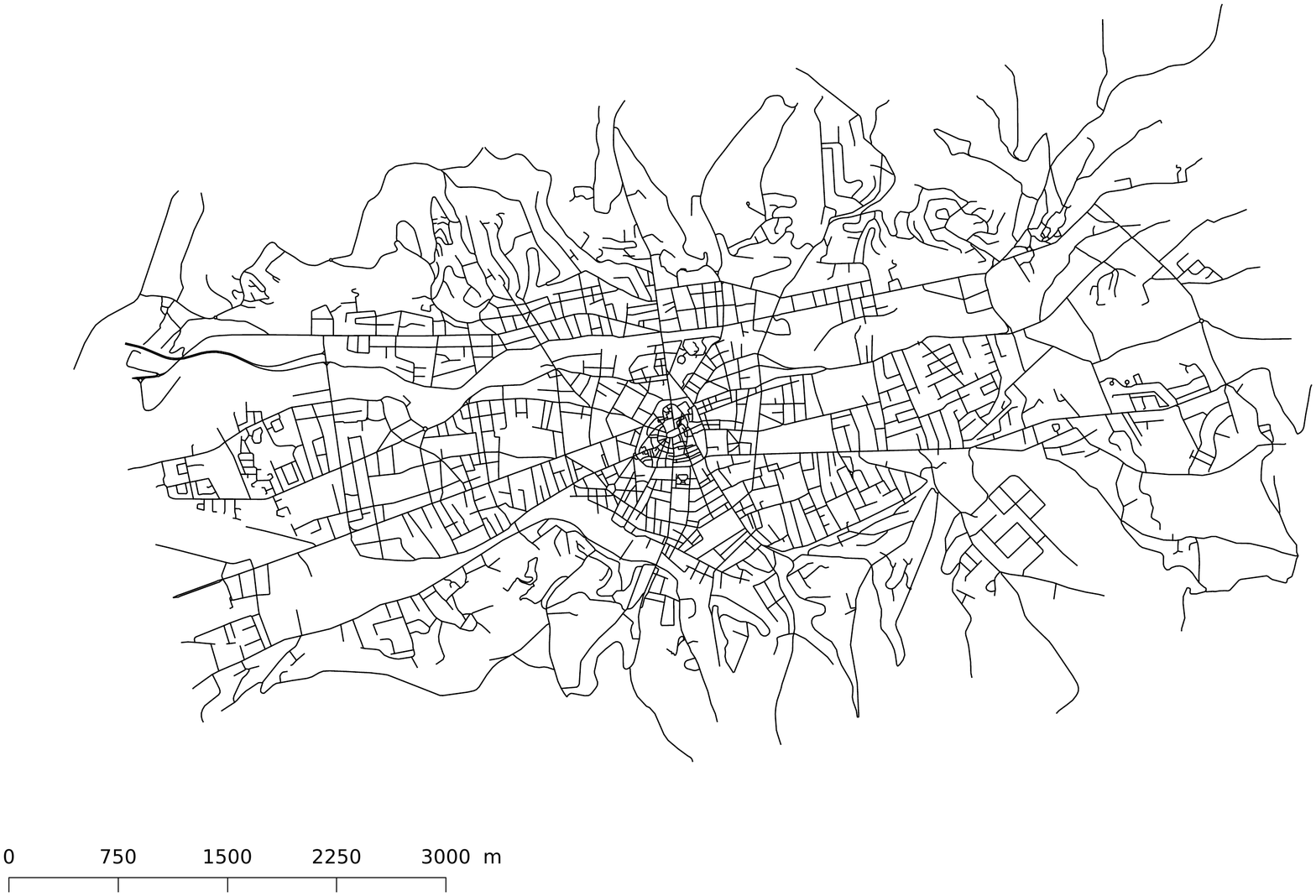}
	\caption{Réseau viaire de Brive-la-Gaillarde}
	\label{fig:brive_brut}
\end{figure}

\begin{figure}[h]
    \centering
    \includegraphics[width=\textwidth]{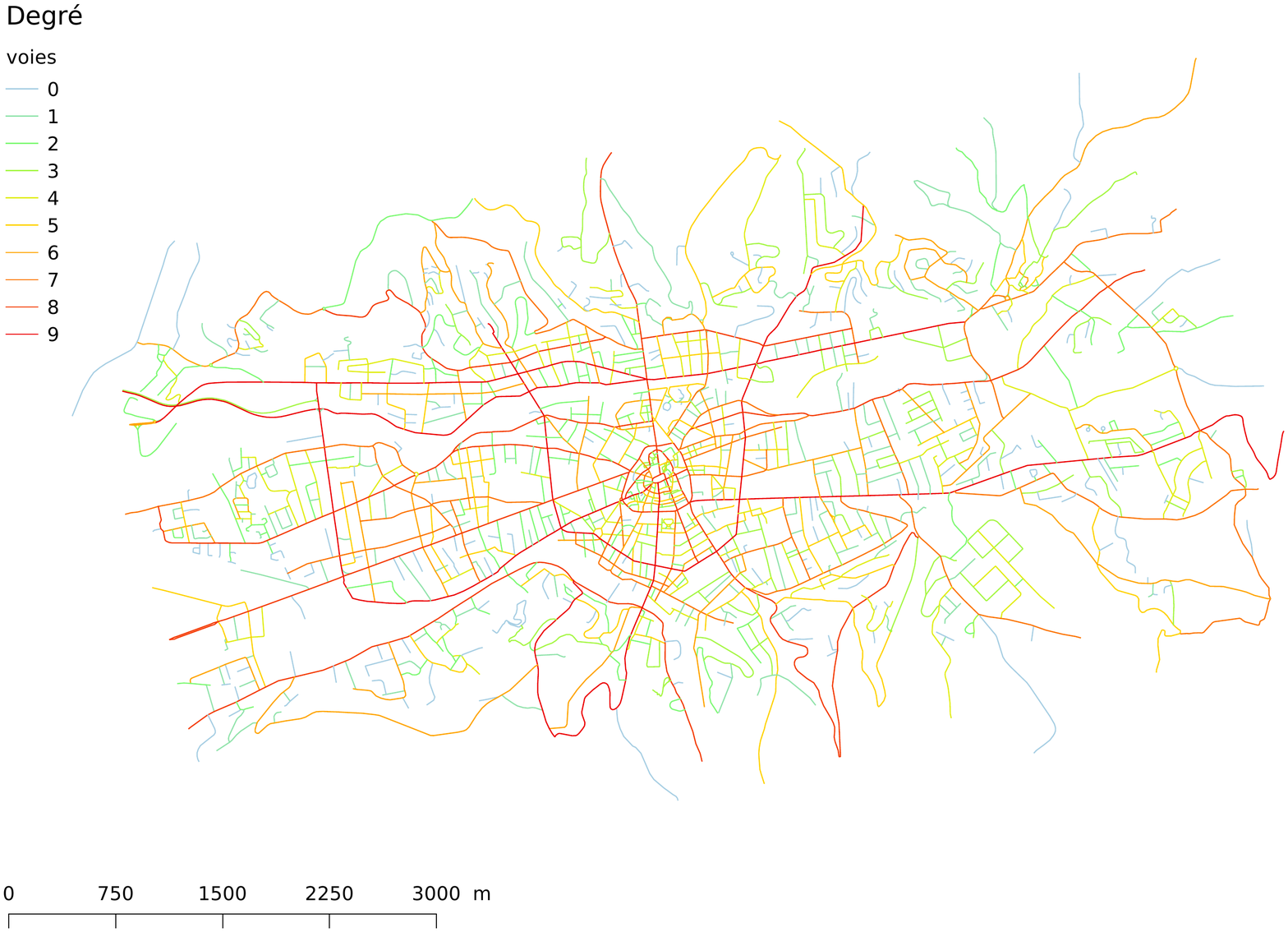}
    \caption{Calcul du degré sur les voies d'Avignon (échantillon 2).}
    \label{fig:brive_deg}
\end{figure}

La ville de Brive-la-Gaillarde est structurée par un contexte géographique fort : construite au fond d'une plaine, elle est traversée au Nord par une rivière, la Corrèze, et elle est limitée au Nord et au Sud par des collines qui lui imposent un étalement d'Est en Ouest (figure \ref{fig:brive_carte}). Le filaire brut du réseau viaire (figure \ref{fig:brive_brut}) permet de déceler l'hétérogénéité des structures mais non d'établir des liens entre elles. L'indicateur de degré, appliqué aux voies, les hiérarchise (figure \ref{fig:brive_deg}). Comme pour Avignon, il met en valeur les axes importants, depuis ceux quasi-circulaires du centre ancien à ceux radiaux qui ouvrent la ville vers le territoire environnant. La ville, construite autour d'un noyau fortifié, s'ouvre vers l'extérieur grâce à ces axes qui traversent la Corrèze par des ponts, créés au fil du temps. Cette forme géographique a fondé la toponymie du lieu : Brive-la-Gaillarde du latin \textit{briva} (pont, passerelle) et \textit{gaillard} (fort, solide). Le maillage très orthogonal du réseau au fond de la plaine avait été mis en évidence par un coefficient de maillance élevé (cf chapitre 3, deuxième partie).

C.-N. Douady analyse l'interaction de deux structures comme la signature de l'évolution radio-concentrique de la ville : construction historique d'un double réseau, concentrique puis radial \citep{douady2014trace}. Il distingue ainsi d'une part un maillage orienté Nord-Sud / Est-Ouest dans la plaine, selon les grandes directions du relief local ; d'autre part, au centre, une organisation radio-concentrique constituée en trois temps : couches concentriques du noyau historique, puis percées radiales du XIX\textsuperscript{ème} siècle, et enfin rocade sub-urbaine, la cohérence étant assurée par un centre commun (place de la Collégiale). La convergence classique des grandes voies historiques de liaison longue distance, habituellement en étoile, est ici ajustée au maillage de la plaine par les contraintes de la topographie, qu'illustre le contraste entre les formes géométriques en plaine et les tracés sinueux sur les coteaux Nord et Sud.

\begin{figure}[h]
    \centering
    \includegraphics[width=\textwidth]{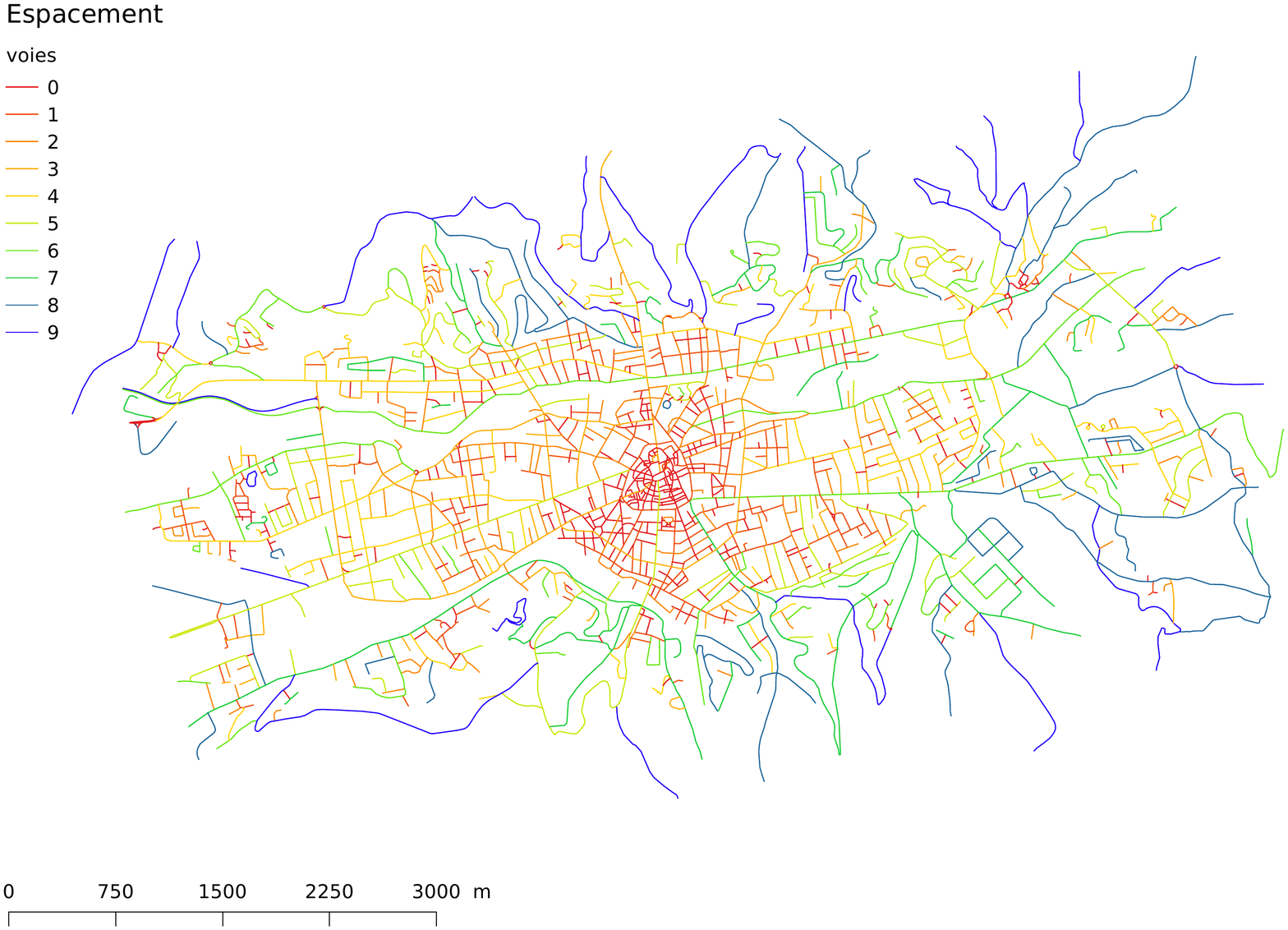}
    \caption{Calcul de l'espacement sur les voies d'Avignon (échantillon 2).}
    \label{fig:brive_esp}
\end{figure}

L'application de l'indicateur d'espacement permet de lire les contraintes topographiques. Ses valeurs sont très faibles au fond de la plaine, où le graphe est dense et très quadrillé ; elles sont intermédiaires pour les axes radiaux ; et elles sont élevées lorsque les voies deviennent sinueuses et peu connectées, là où le terrain devenant plus escarpé (figure \ref{fig:brive_esp}). 

La lecture de la ville à travers les indicateurs permet de retrouver la structure ancienne fortifiée et les radiales décrites plus haut. Elle nous propose également de lire la topographie du territoire à travers la forme du graphe viaire.

L'effacement de ronds-points ou de l'agencement des entrées sur le boulevard périphérique (ancienne muraille) a rendu possible l'observation historique. En effet, ces aménagements coupent certaines continuités historiques. Pour cela, nous avons utilisé le traitement par \enquote{zones tampons} décrit dans la deuxième partie. Nous avons ainsi pu retrouver certains grands axes historiques. La légitimité d'une telle manipulation reste une question dont la réponse dépend de notre problématique de recherche \citep{douady2015limites}. En effet, les axes de cheminement anciens ne pourront pas être identifiés de la même manière que ceux plus récents, où les structures sont adaptées à la circulation automobile (les giratoires altèrent délibérément la continuité de deux voies à leur intersection, privilégiant la sécurité sur la continuité historique).

\FloatBarrier
\section{Lecture de capitales} 

\FloatBarrier
\subsection{Téhéran (Centre)}

Nous avons eu la chance, au cours de nos recherches, d'être accueillis par le centre de recherche Nazar, à Téhéran, afin de présenter nos travaux appliqués à la ville. Cela nous a permis de faire une analyse structurelle complète de la capitale iranienne et de la confronter au terrain, que nous avons arpenté pendant trois jours \citep{douady2015peut, lagesse2015lire}.

\begin{figure}[h]
    \centering
        \includegraphics[width=0.8\textwidth]{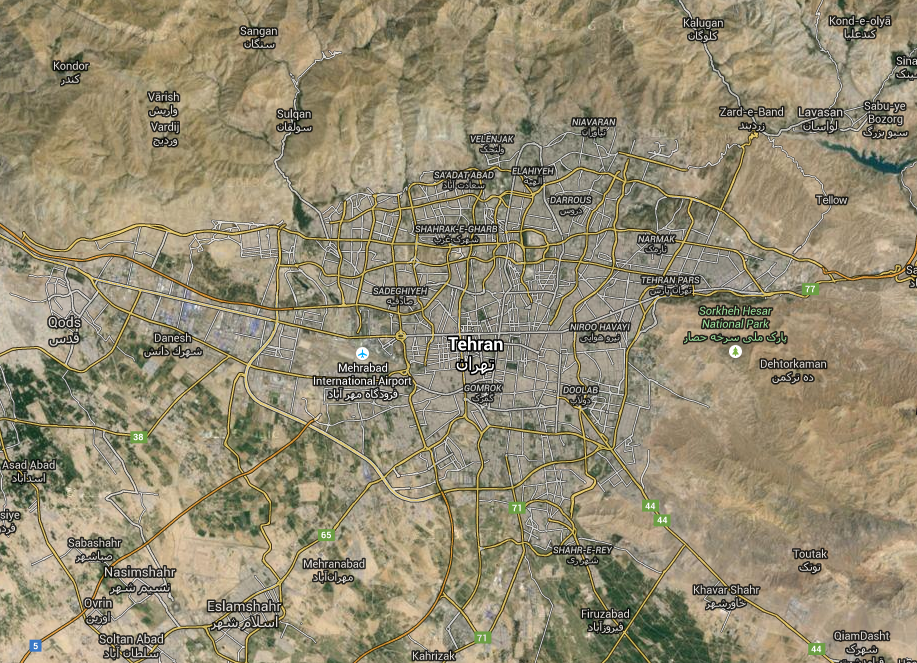}
    
    \caption{Vue aérienne de Téhéran. (\copyright GoogleMaps 2015).}
    \label{fig:teh_plan1}
\end{figure}

\begin{figure}[h]
    \centering
        \includegraphics[width=0.8\textwidth]{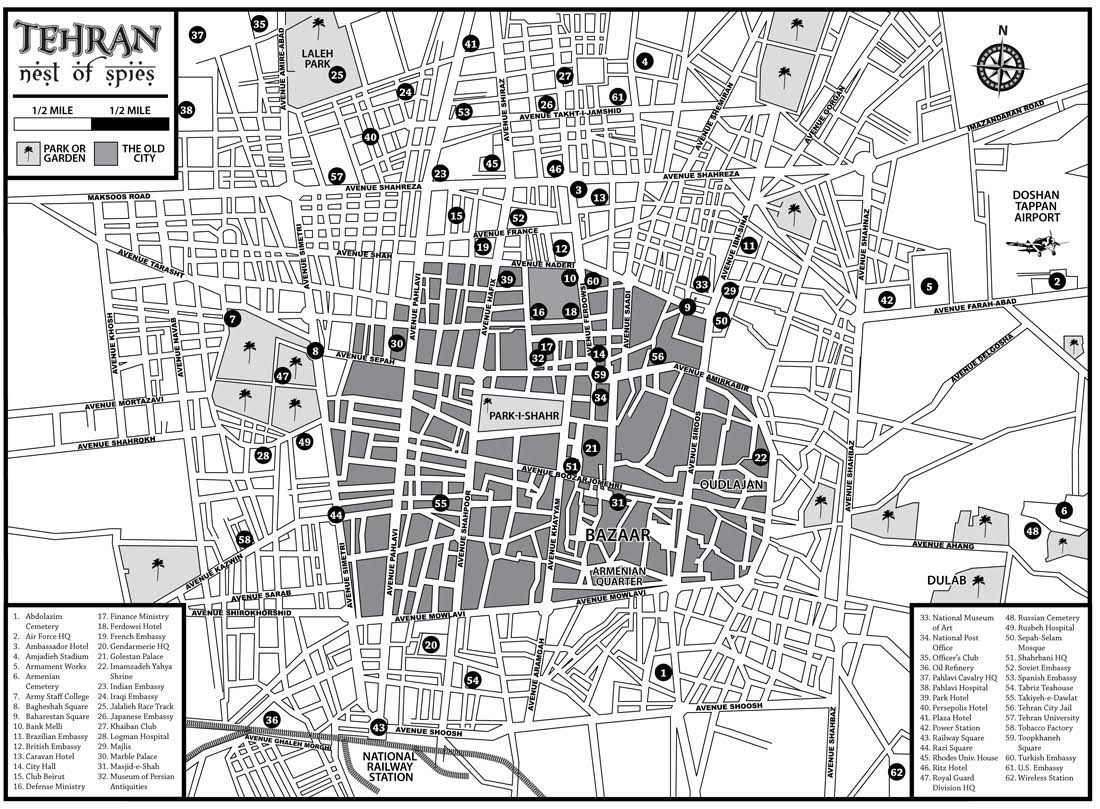}
    
    \caption{Plan du Centre de Téhéran avant 1979 (pré-révolution). \\ source : centre de recherche Nazar : institut iranien de recherche sur l'art, l'architecture et l'urbanisme}
    \label{fig:teh_plan2}
\end{figure}

Téhéran est une ville construite selon de fortes contraintes géographiques. Entre les montagnes, au Nord, et le désert, au Sud, la ville s'organise sur un terrain de faible déclivité, selon des axes très marqués (figure \ref{fig:teh_plan1}). En effet, la ville s'étend sur un territoire 686,3 $km^2$ que des autoroutes urbaines traversent de part en part. Dans la présentation que nous faisons ici, nous circonscrivons notre étude au centre de la ville (figure \ref{fig:teh_plan2}).

\begin{figure}[h]
    \centering
    \includegraphics[width=0.6\textwidth]{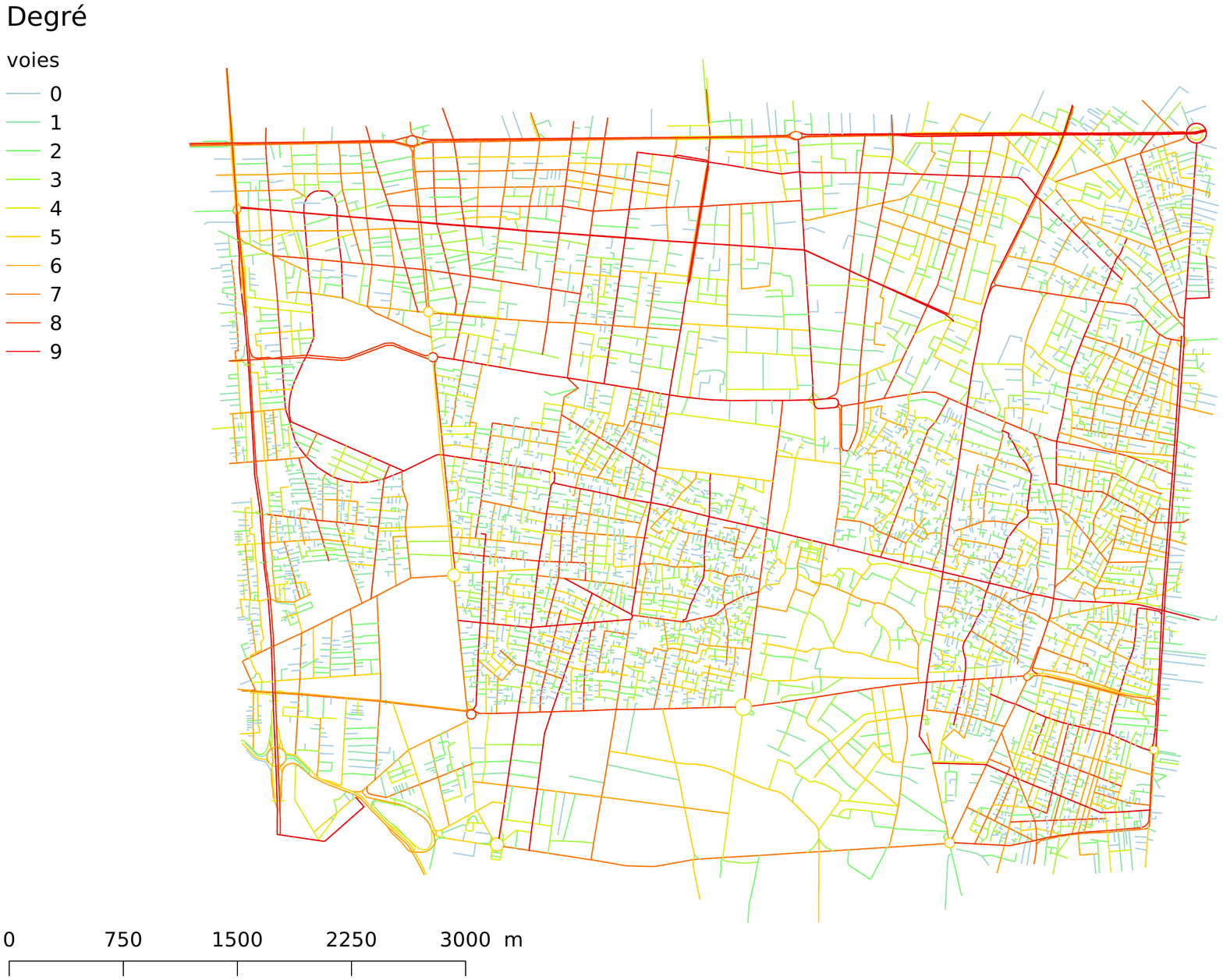}
    \caption{Calcul du degré sur les voies du centre de Téhéran.}
    \label{fig:teh_deg}
\end{figure}

\begin{figure}[h]
    \centering
    \includegraphics[width=0.6\textwidth]{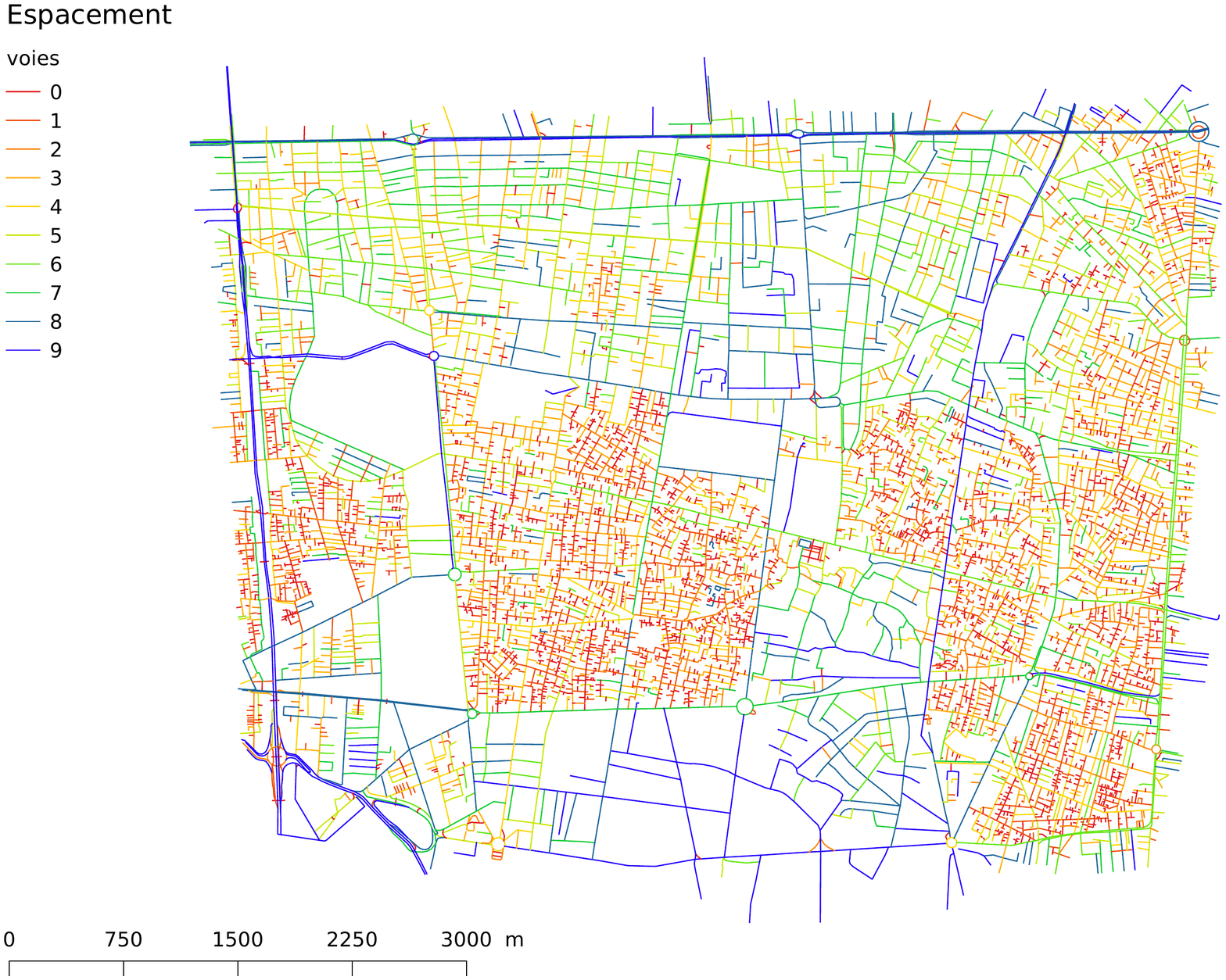}
    \caption{Calcul de l'espacement sur les voies du centre de Téhéran.}
    \label{fig:teh_esp}
\end{figure}

Ce centre s'organise autour d'axes routiers très empruntés, que l'indicateur de degré des voies fait ressortir  (figure \ref{fig:teh_deg}). À l'inverse, l'indicateur d'espacement classe les autoroutes aux connexions souples avec les voies contournant les parcs (au Sud) et met l'accent sur les centres denses, qui ressortent en rouge (figure \ref{fig:teh_esp}). Ces parties du graphe, avec un coefficient d'espacement faible, correspondent à des quartiers de vie (bazars) ou d'habitation. Les petites ruelles sont sinueuses et isolent du bruit des grands axes routiers. Elles imposent une fracture à la fois géométrique et culturelle entre le règne des véhicules et celui des piétons.

Pourtant les autoroutes ne sont jamais loin. Si l'on considère la portion de celle passant au Nord de notre échantillon (figure \ref{fig:teh_deg}), par exemple, et que l'on évalue les distances topologiques des autres voies par rapport à celle-ci, nous observons que la voie la plus éloignée en est à neuf \textit{tournants}. De plus, cela concerne uniquement trois voies : la majorité relient l'autoroute considérée en trois ou quatre \textit{tournants}. Les raccordements à la structure maillée que l'on observe avec la carte de degré permettent donc la traversée du graphe spatial avec de faibles distances topologiques, par rapport au nombre de voies du réseau.

L'évaluation de la closeness moyenne de la ville nous permet de quantifier cette efficacité à raccorder rapidement une structure \textit{maillante} (voir raisonnement établi par Pailhous dans le chapitre suivant). En effet, cet indicateur est l'inverse de la somme de l'ensemble des distances topologiques entre voies sur le graphe. Il permet donc d'identifier celles qui permettent d'accéder rapidement à l'ensemble du réseau. Pour l'ensemble de la ville de Téhéran, la valeur moyenne de la closeness est de 0.12. Pour cet échantillon particulier, elle est de 0.16. En effet, si l'on considère l'ensemble du graphe de la ville, les routes d'accès vers la montagne ou de circulation d'Est en Ouest font baisser la proximité topologique moyenne sur tout le réseau. Si l'on se réfère à l'étude comparative que nous avons faite dans le chapitre 3 de la deuxième partie, nous observons que ces moyennes ne sont pas parmi les meilleures. Cela signifie que les distances topologiques observées ne sont pas particulièrement faibles par rapport à celles calculées sur les villes de notre panel de recherche. Nos réseaux viaires sont ainsi construits : on peut les traverser de part en part en suivant un nombre minimum de voies.

\begin{figure}[h]
    \centering
    \includegraphics[width=0.6\textwidth]{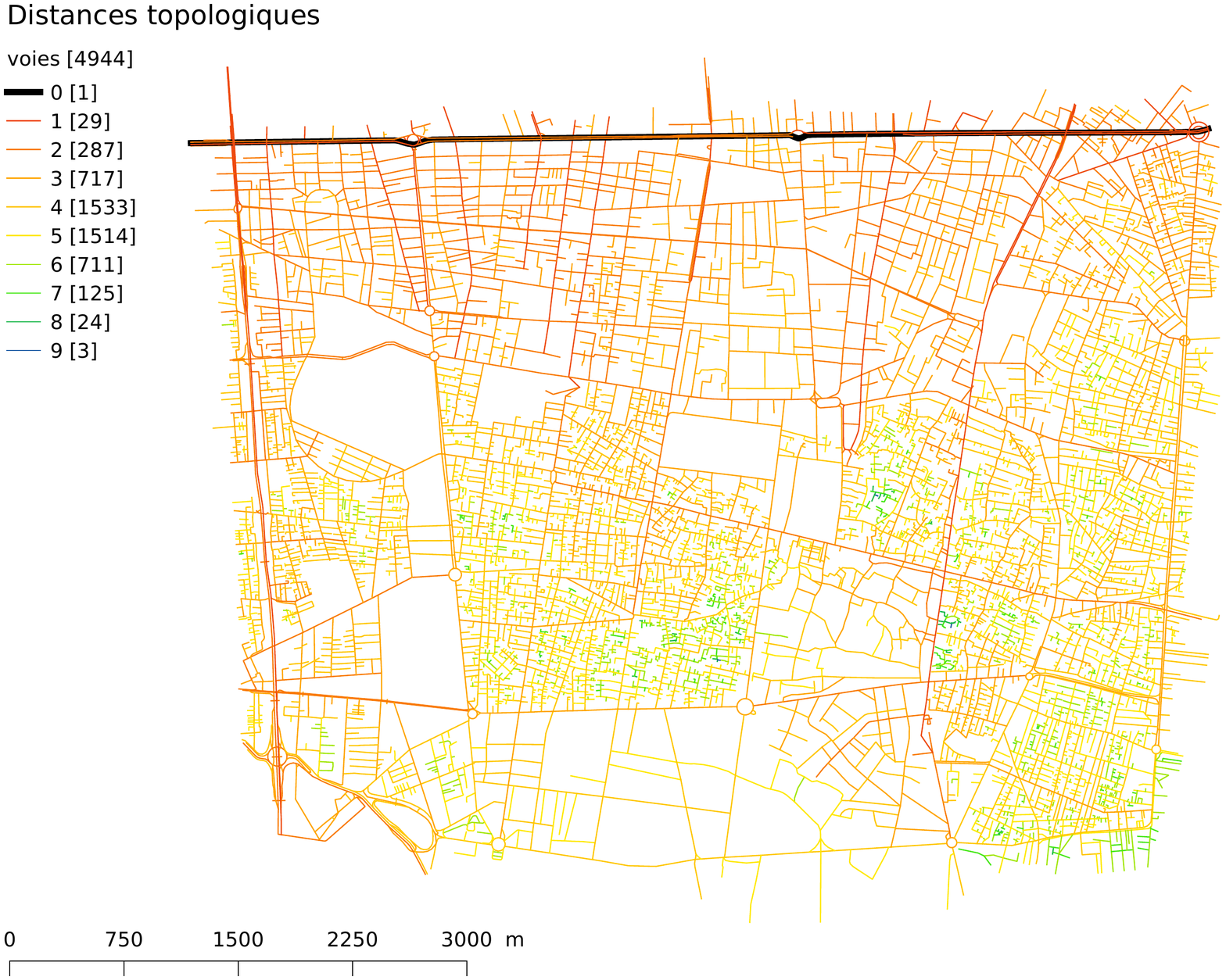}
    \caption{Calcul des distances topologique à partir d'une voie (représentée en noir) sur les voies du centre de Téhéran.}
    \label{fig:teh_dtopo}
\end{figure}

\begin{figure}[h]
    \centering
    \includegraphics[width=0.6\textwidth]{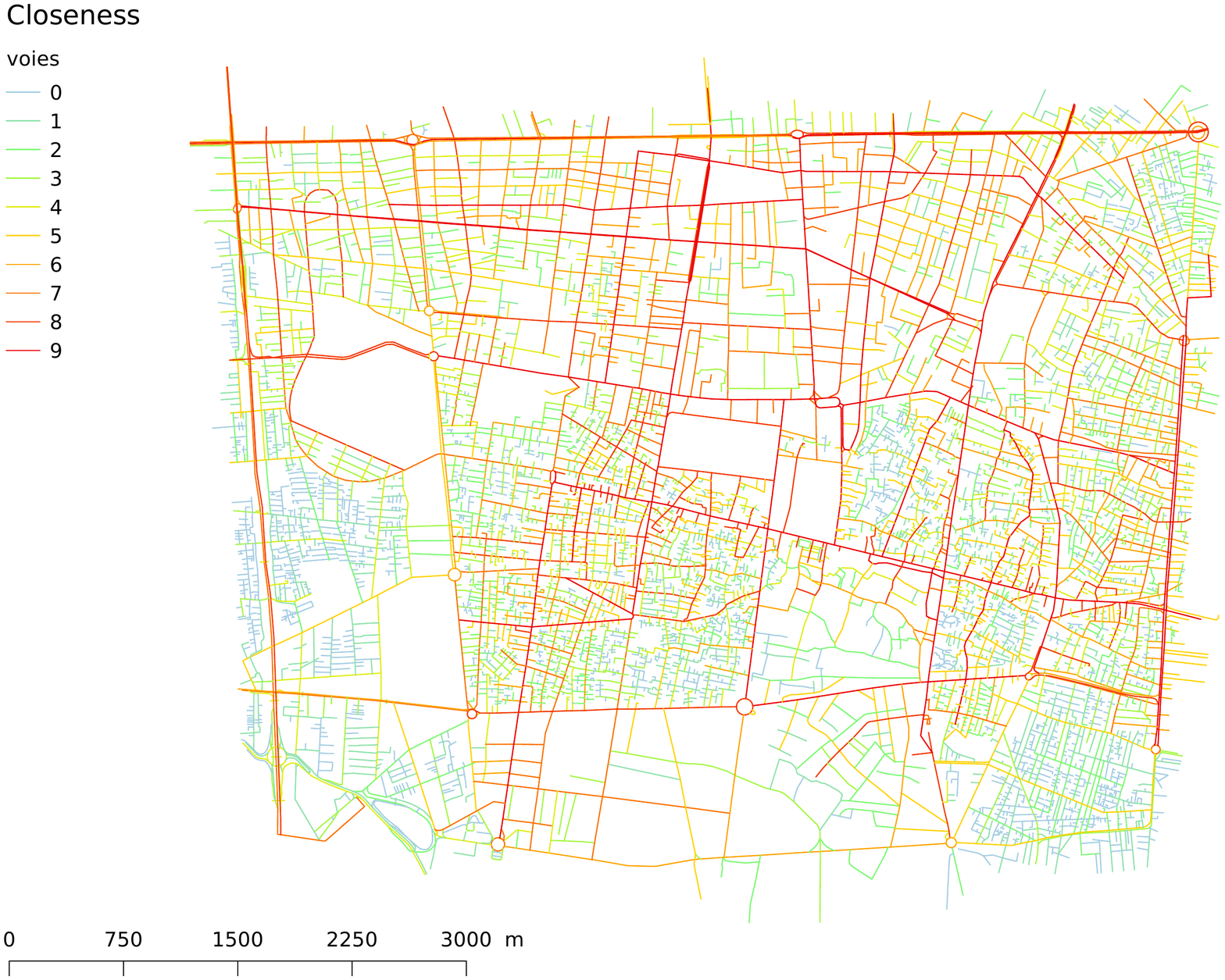}
    \caption{Calcul de la closeness sur les voies du centre de Téhéran.}
    \label{fig:teh_clo}
\end{figure}

Si l'on considère enfin la carte de closeness calculée sur le réseau du centre de Téhéran, nous retrouvons une structure proche de celle mise en valeur par l'étude du degré (figure \ref{fig:teh_clo}). Cette carte exprime l'accessibilité de chaque voie par rapport à l'ensemble du réseau. Nous voyons ainsi des quartiers particulièrement isolés (au Sud-Est et à l'Ouest) bien que des voies de degré important les traversent (figure \ref{fig:teh_deg}).

Sur le terrain, les voies de closeness forte (très accessibles car proches des grands axes) et de degré faible (car elles sont courtes et donc peu connectées) affichent un contraste fort avec celles de closeness et de degré forts (autoroutes urbaines). Nous nous rendons ainsi compte que même une proximité topologique réduite à 1 exprime un important décalage dans les ambiances viaires. Voilà pourquoi, sur un éventail de neuf degrés topologiques depuis une route principale, peuvent s'exprimer toutes les facettes de la diversité urbaine. En arpentant les rues de Téhéran, nous nous rendons également compte que l'indicateur de closeness est corrélé à une typologie d'habitations et de commerces. En effet, les voies ayant une forte proximité avec le reste du réseau regroupent un bâti élevé (immeubles) et de grands commerces, sûrs de remplir leurs surfaces de clients par leur position stratégique. Au contraire, les quartiers identifiés par une closeness faible, regroupent des habitations plus basses (maisons individuelles ou partagées en quelques appartements) et des commerces de proximité.

\FloatBarrier
\subsection{Paris}

\begin{figure}[h]
    \centering
        \includegraphics[width=0.6\textwidth]{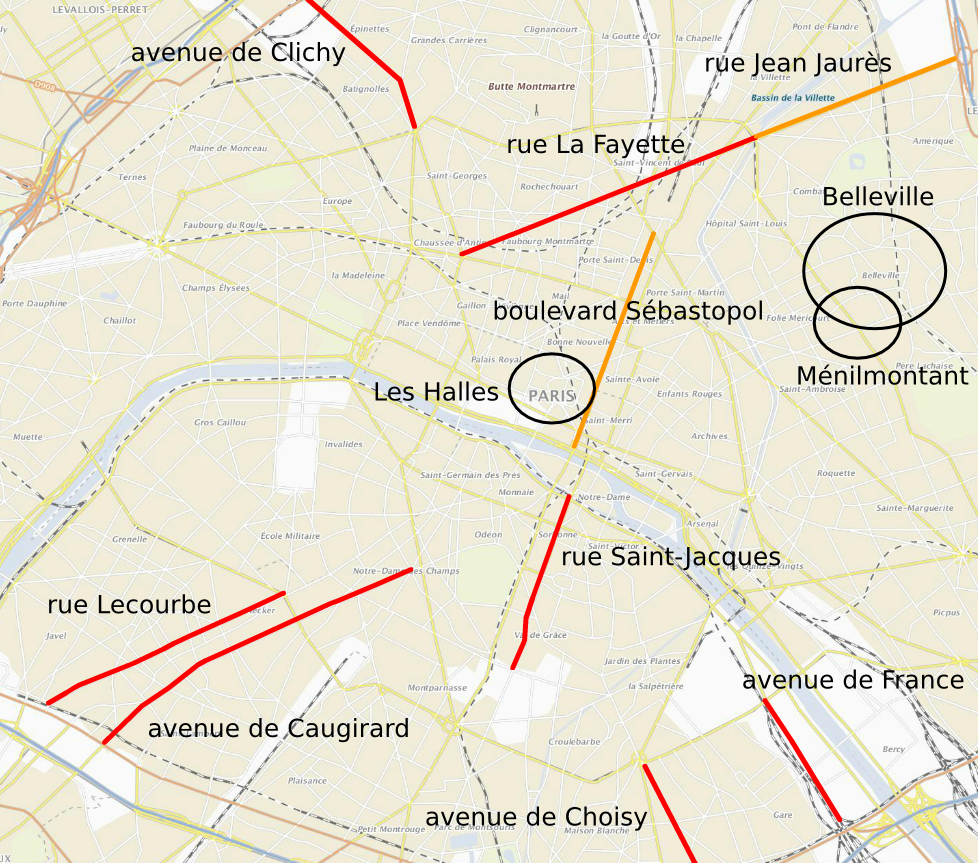}    
    \caption{Mise en situation des rues évoquées au sein de la ville de Paris. (fond : Plan IGN 2015).}
    \label{fig:paris_plan1}
\end{figure}

\begin{figure}[h]
    \centering
    \includegraphics[width=\textwidth]{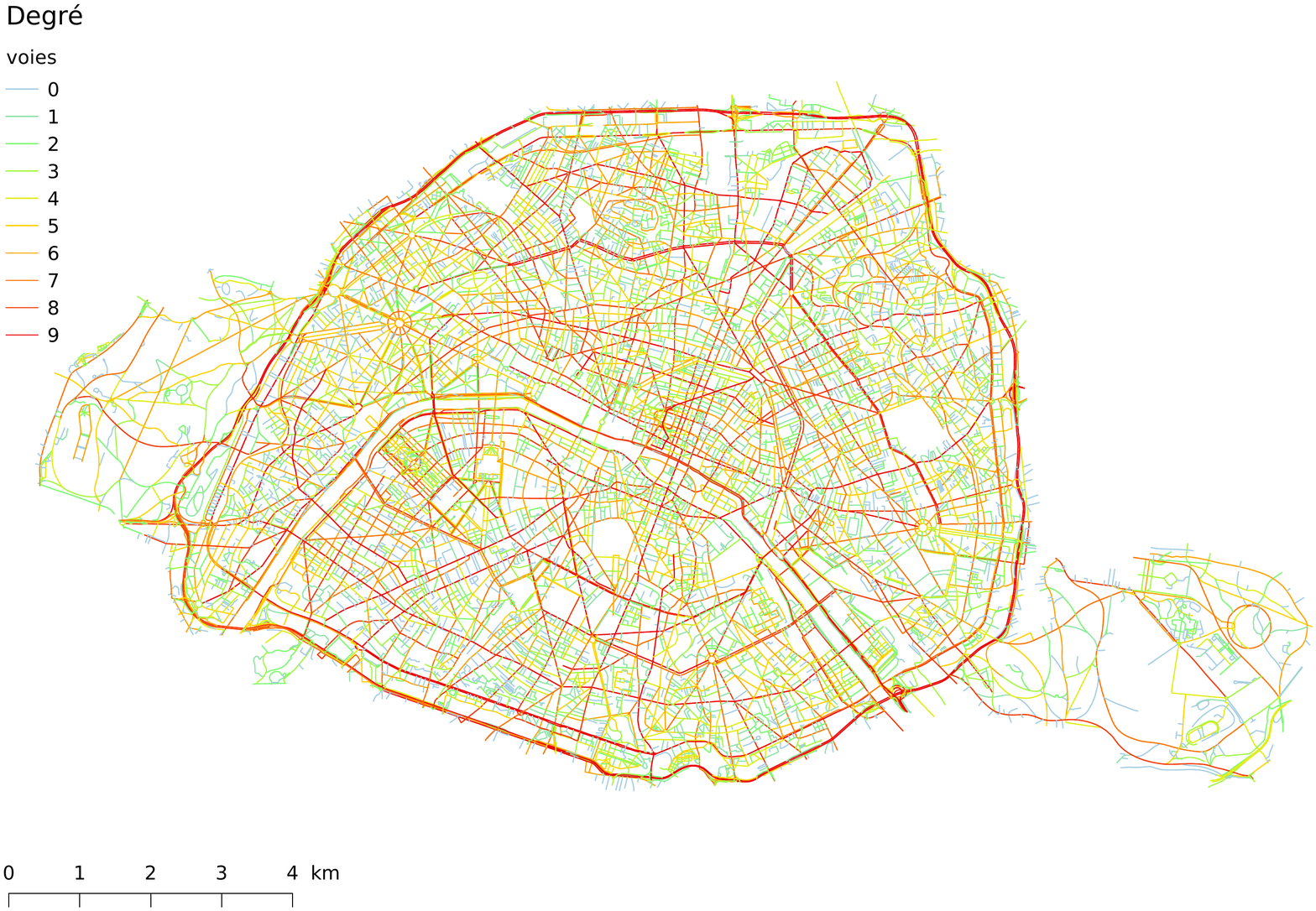}
    \caption{Calcul du degré sur les voies de Paris.}
    \label{fig:paris_deg}
\end{figure}

\begin{figure}[h]
    \centering
    \includegraphics[width=\textwidth]{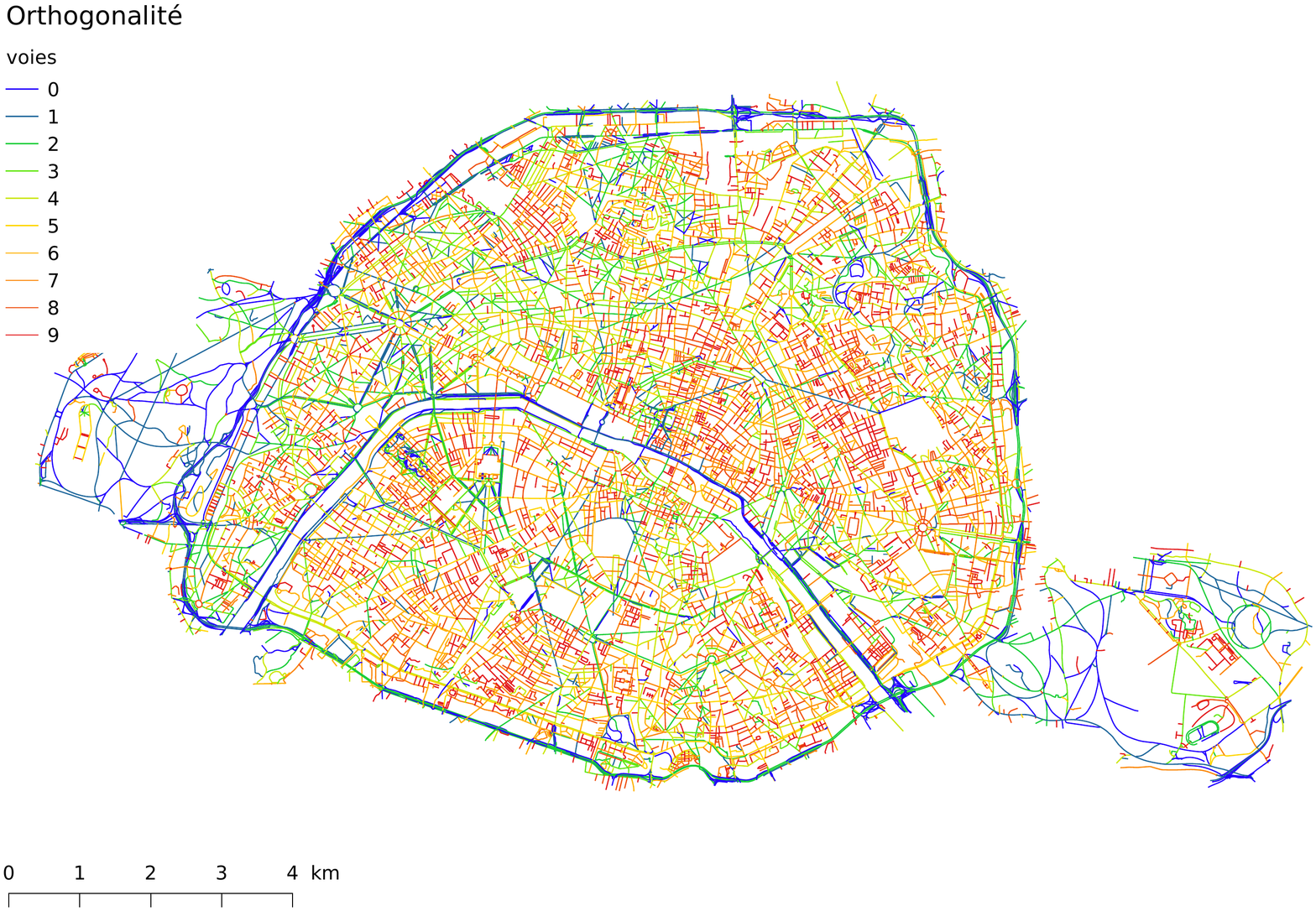}
    \caption{Calcul de l'orthogonalité sur les voies de Paris.}
    \label{fig:paris_ortho}
\end{figure}

\begin{figure}[h]
    \centering
    \includegraphics[width=\textwidth]{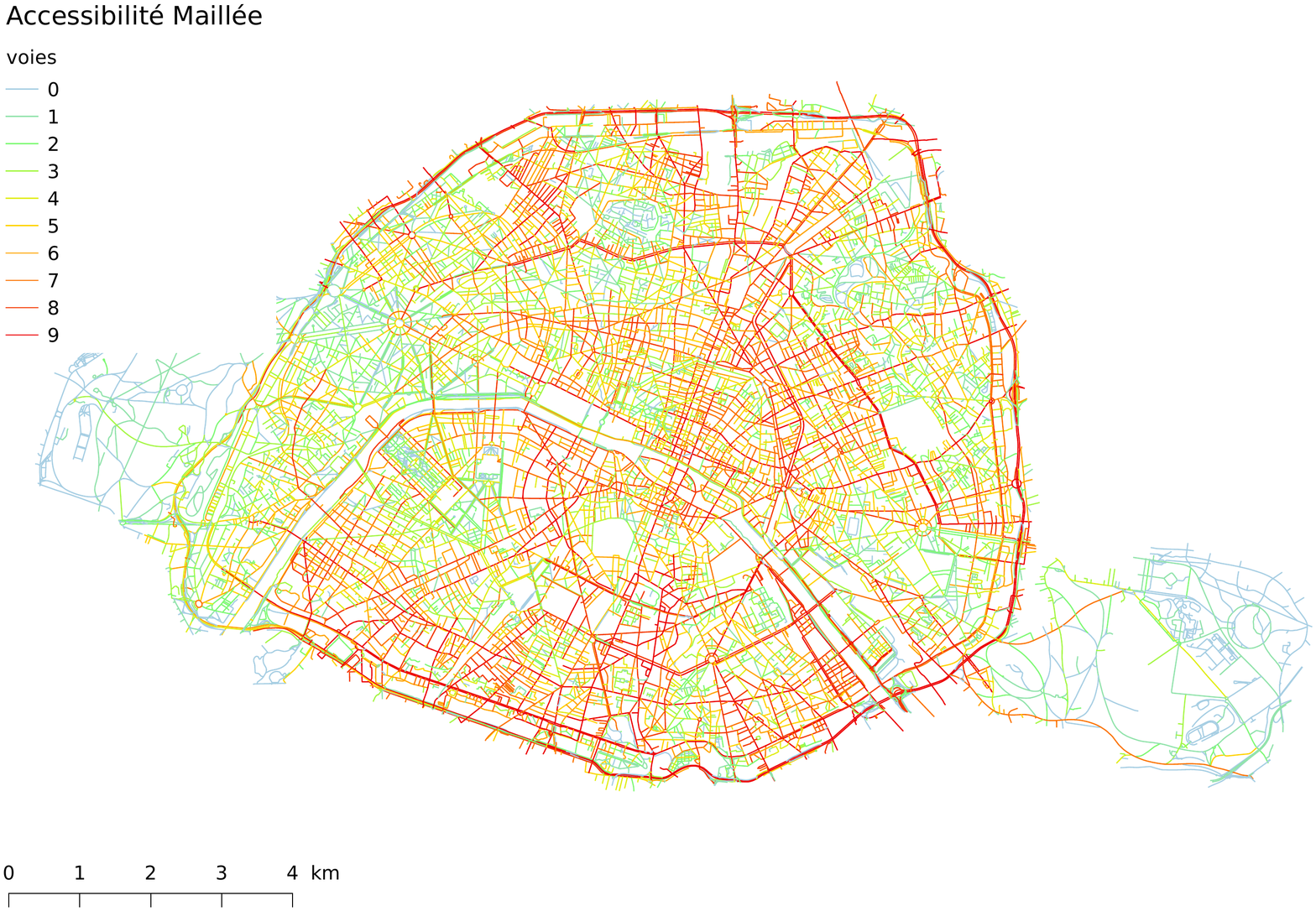}
    \caption{Calcul de l'accessibilité maillée sur les voies de Paris.}
    \label{fig:paris_roo}
\end{figure}

\begin{figure}[h]
    \centering
    \includegraphics[width=\textwidth]{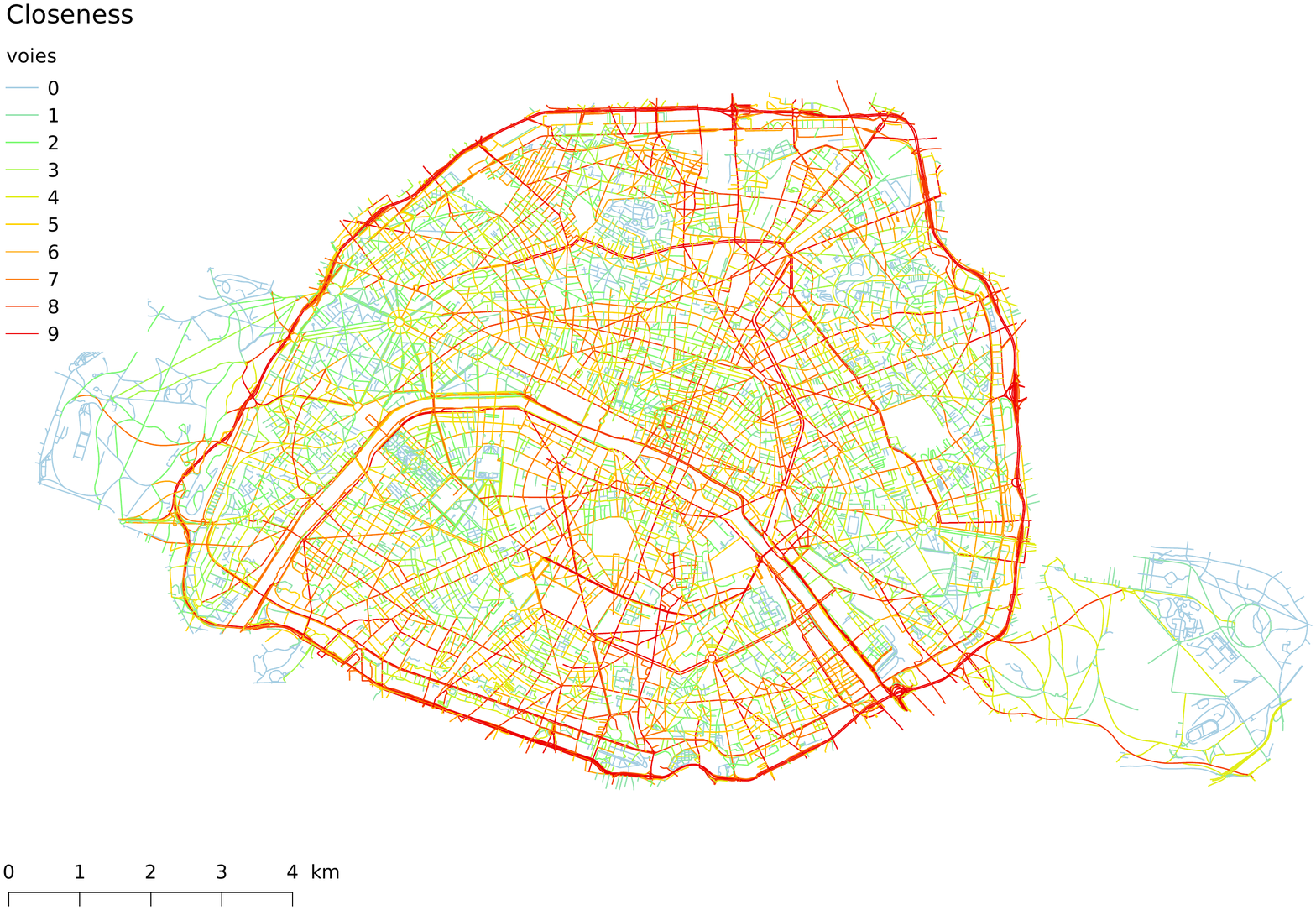}
    \caption{Calcul de la closeness sur les voies de Paris.}
    \label{fig:paris_clo}
\end{figure}

Les formes de la ville de Paris ont été étudiées dans de multiples ouvrages, se concentrant tour à tour sur ses rues, ses parcelles, son bâti, et souvent en analysant leurs liens avec l'Histoire \citep{rouleau1975trace, noizet2013paris}. La ville a également fait l'objet d'un projet réunissant sur une même plate-forme internet l'ensemble des données géographiques qui lui sont liées \citep{noizet2008alpage}.
Notre but est ici de proposer une visualisation de nos résultats sur la capitale. Nous en avons, par ailleurs, déjà montré quelques cartes, Paris ayant servi de support à la présentation de la construction de nos indicateurs (chapitre 3 de la première partie). Nous revenons ici sur les cartes de degré (figure \ref{fig:paris_deg}), d'orthogonalité (figure \ref{fig:paris_ortho}), de closeness (figure \ref{fig:paris_clo}) et d'accessibilité maillée (figure \ref{fig:paris_roo}). En effet, l'espacement a déjà fait l'objet d'une interprétation lors de l'explication de sa construction.

La carte du degré des voies de la capitale fait ressortir la structure maillée qui permet les déplacements rapides intra-muros (figure \ref{fig:paris_deg}). Cet indicateur oppose deux logiques de déplacement : ceux qui ont pour objectif la traversée de l'espace (sur les voies de plus fort degré, représentées en rouge) ; et ceux qui ont pour objectif une desserte fine du territoire (au second plan, de degré faible, en bleu clair). L'indicateur de degré oppose donc le global au local. Il fait ressortir l'essence même de la voie : son caractère multi-échelle. C'est grâce à la méthodologie de sa construction qu'un indicateur propre à l'objet peut être équivalent à des indicateurs plus complexes tels que la betweenness (où l'indicateur d'utilisation qui lui est équivalent). L'interprétation de l'utilisation de la structure maillée peut donc être équivalente à celle que nous pourrions faire à partir de la betweenness : ce sont les voies qui ont le plus fort potentiel d'utilisation lors des trajets d'un bout à l'autre de la capitale. Ce raisonnement est une approche quantitative de celui proposé par Pailhous dans son analyse des stratégies de déplacement \citep{pailhous1970representation}. Nous développerons cet aspect dans le chapitre suivant.

\begin{figure}[h]
    \centering
    \includegraphics[width=0.8\textwidth]{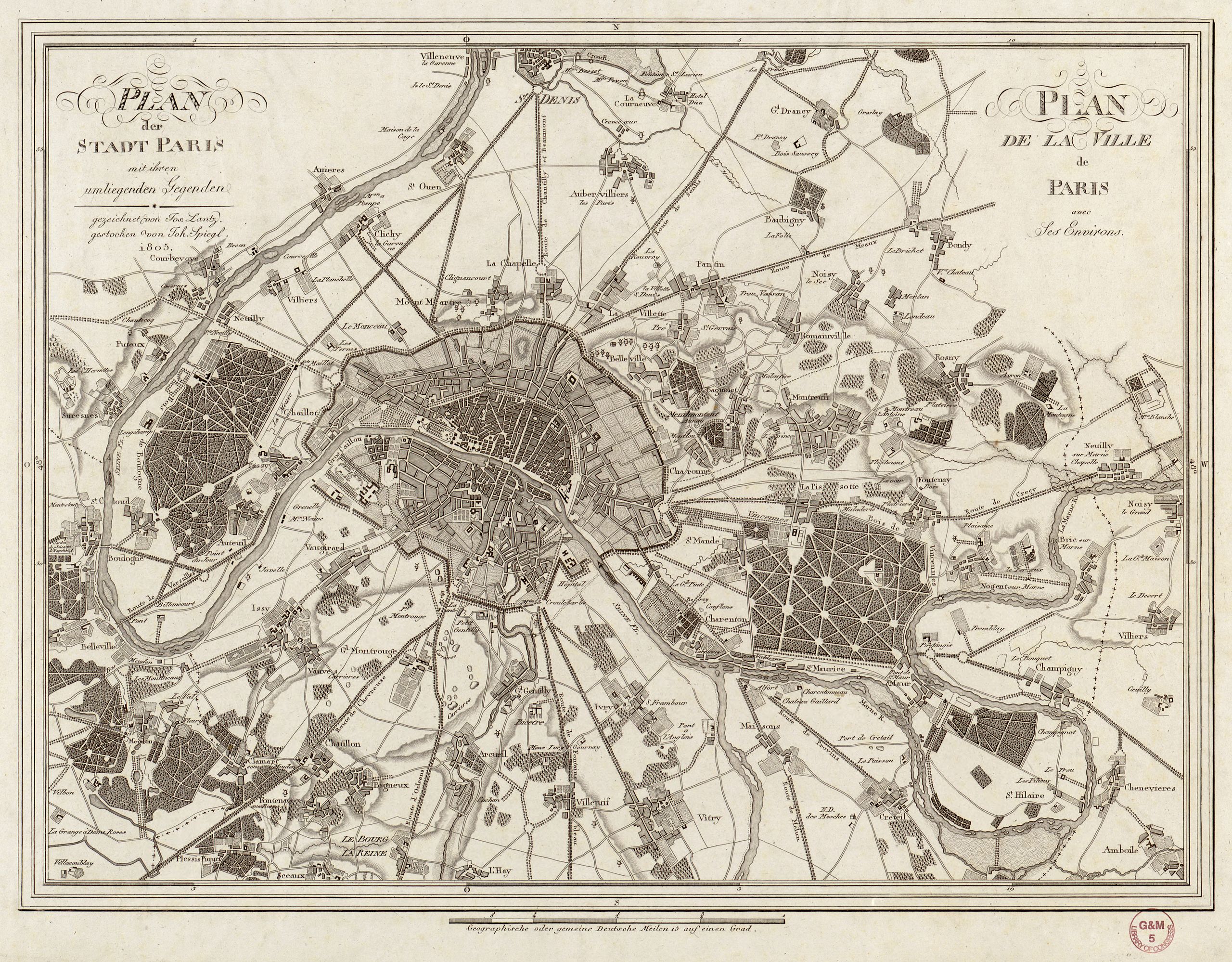}
    \caption{Plan de la ville de Paris et de ses environs en 1805 \\ source : J. Lantz}
    \label{fig:paris1805_plan}
\end{figure}

Le calcul de l'orthogonalité (figure \ref{fig:paris_ortho}) fait ressortir d'autres types de structures. Les axes de circulation rapide (boulevard périphérique) ou, à l'intérieur de la capitale, les chemins plus anciens (bords de Seine) ont des connexions à angle faible avec les voies qui les traversent. Ils apparaissent donc avec un faible coefficient d'orthogonalité (en bleu). Les allées dans les parcs, comme à Téhéran, sont classées de la même manière. Les parties de la ville denses et quadrillées ressortent au contraire avec une orthogonalité très forte. Nous retrouvons ainsi le quartier autour des Halles, centre historique, mais également les hameaux qui ont été happés par la capitale (Ménilmontant, Belleville, Clichy ; cf figure \ref{fig:paris1805_plan}). Les percées opérées par Haussmann sur la capitale sont, quant à elles, classées avec une orthogonalité moyenne (en vert). En effet, elles traversent abruptement le tissu en le recoupant avec des angles de connexion variés. Cela nous permet de les identifier grâce à cet indicateur.

L'indicateur d'accessibilité maillée appliqué à Paris fait ressortir toutes les structures \enquote{en peigne} de la ville (figure \ref{fig:paris_roo}). Nous voyons ainsi apparaître un axe Nord-Sud formé par la rue Saint-Jacques (parallèle au boulevard Saint-Michel mais plus ancienne que celui-ci) et le boulevard Sébastopol qui remonte jusqu'à la gare de l'Est. Nous retrouvons également les structures circulaires de contournement de la ville, dont la géométrie est maillée avec les radiales qui ouvrent le centre vers la périphérie (ce qui rappelle la structure radio-concentrique observée à Brive-la-Gaillarde). Enfin, des quartiers ressortent avec une accessibilité maillée forte liée à une structure en peigne : autour de l'avenue de Clichy (au Nord-Ouest) ou de celle de Choisy (au Sud-Est) ainsi que le quartier récent proche de l'avenue de France (également au Sud-Est).

\begin{figure}[h]
    \centering
    \includegraphics[width=0.6\textwidth]{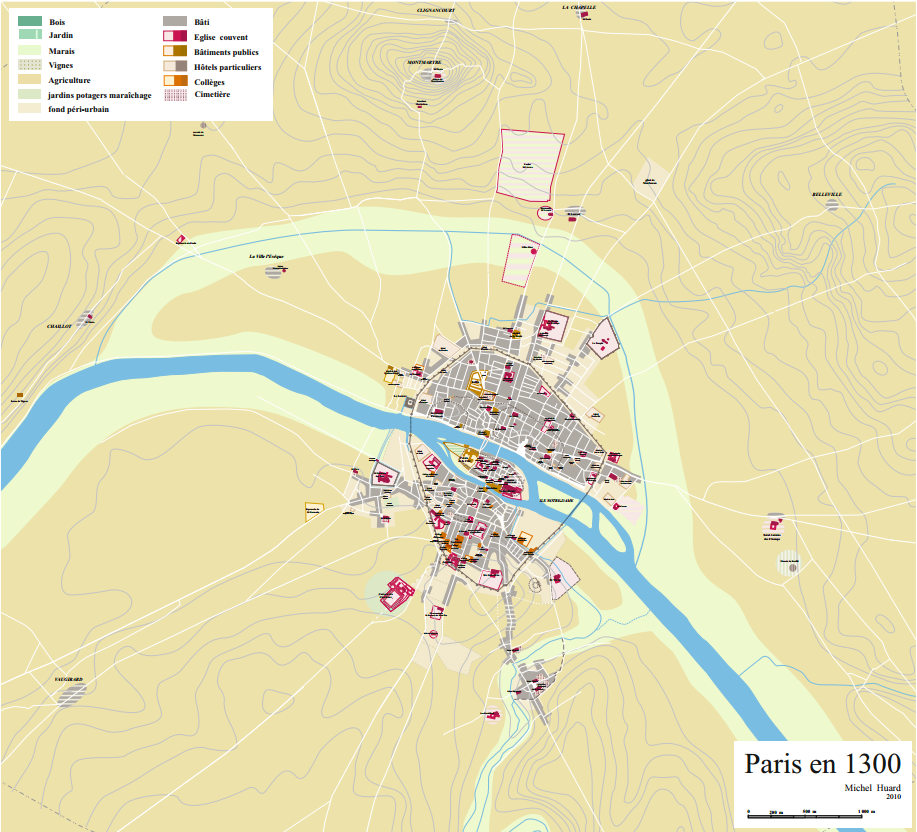}
    \caption{Plan de la ville de Paris et de ses environs en 1300 \\ source : Atlas historique de Paris - M. Huard}
    \label{fig:paris1300_plan}
\end{figure}

L'étude de la closeness hiérarchise les voies de manière plus contrastée (figure \ref{fig:paris_clo}). Elle met en valeur des structures  anciennes qui délimitaient la ville et sa relation avec l'hydrographie (bords de Seine, petite ceinture). Elle permet également de retrouver des rues historiques de liaison de la ville vers l'extérieur \citep{huard2013atlas}. Nous voyons ainsi se détacher au Sud-Ouest, la rue Lecourbe et la rue de Vaugirard qui existaient aux prémices de la création de la capitale (figure \ref{fig:paris1300_plan}). Au Nord-Est, la rue Lafayette est une voie récente qui prolonge des structures plus anciennes : la route d'Allemagne, puis la rue Jean-Jaurès (liaisons avec le Nord-Est). Nous retrouvons également de cette manière la rue Saint-Jacques, voie historique de liaison de la capitale vers le Sud. Son utilisation est aujourd'hui réduite au profit de celle du boulevard Saint-Michel, qui lui est parallèle, mais dont la continuité est entrecoupée de places et de ronds-points. Ceux-ci forcent la création de plusieurs voies, et créent donc des structures moins efficaces (en terme de proximité topologique). Comme nous l'avons vu, nous pourrions effacer ces discontinuités en traitant la construction des voies avec des \enquote{zones tampons} de diamètres adaptés.

Cependant, toutes les voies de closeness élevée ne peuvent pas être considérées comme des structures historiques. En effet, le périphérique est une infrastructure construite récemment qui suit une logique de proximité topologique minimale avec le reste du réseau (cas comparable à celui de la voie formée par la rue de la République et le cours Jean-Jaurès à Avignon).

\FloatBarrier
\section{Lecture de villes planifiées}

Il existe des villes planifiées de toutes époques et sur tous les continents. En Aquitaine les villefranches, petits villages aux allures vernaculaires, ont été dessinées sur une seule période. Après les deux épisodes de peste du début XIV\textsuperscript{ème} siècle, les seigneurs ont imaginé cette solution pour attirer des habitants sur leurs terres. Nous étudierons ici deux villes de notre panel de recherche : une dont la planification est partielle, Barcelone, et l'autre dont le quadrillage s'ancre dans la tradition du pays, Kyoto.

\FloatBarrier
\subsection{Barcelone}

\begin{figure}[h]
    \centering
        \includegraphics[width=0.8\textwidth]{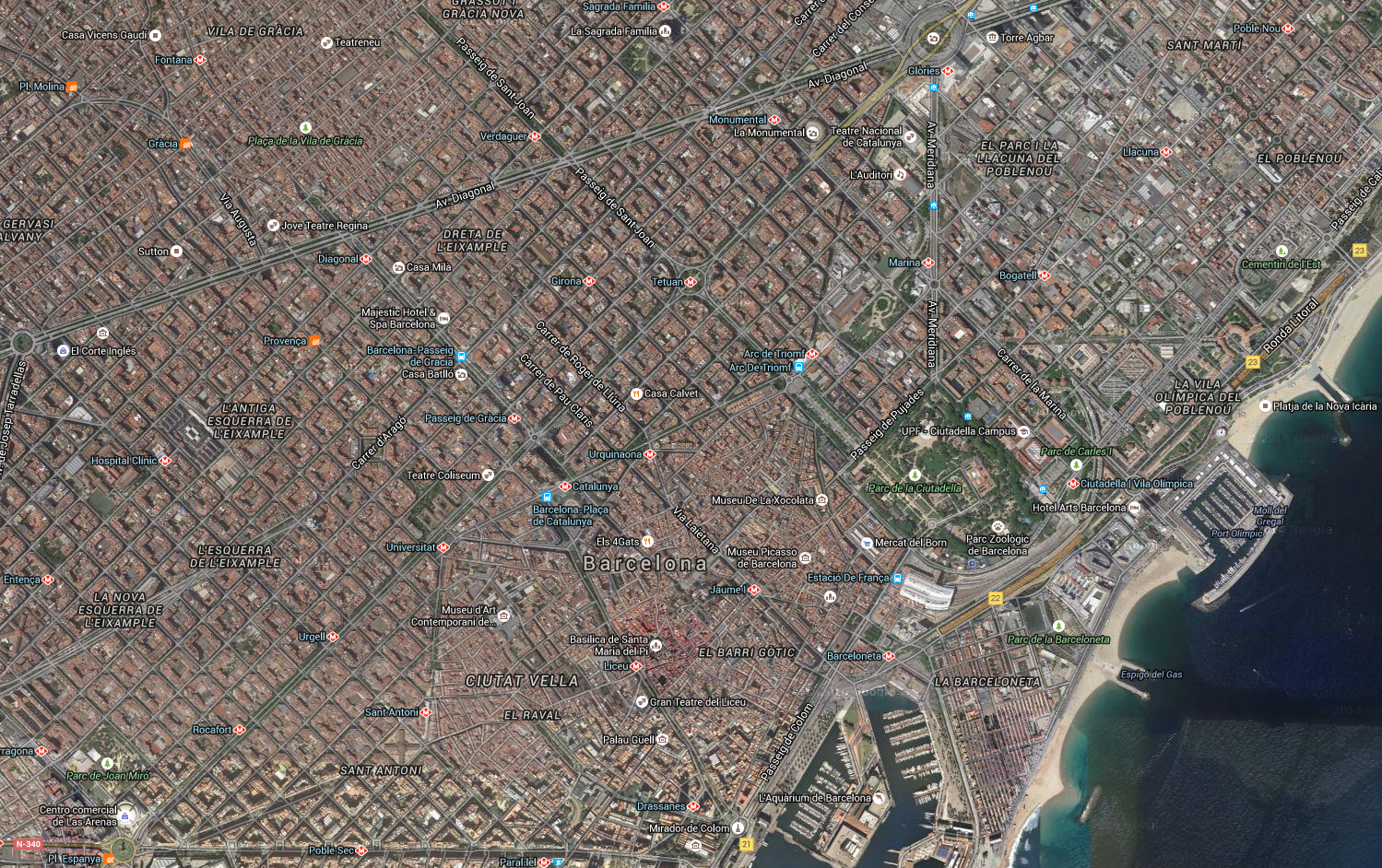}    
    \caption{Carte du centre de Barcelone. (\copyright GoogleMaps 2015).}
    \label{fig:barce_plan1}
\end{figure}

Barcelone a fait l'objet d'une extension de son centre-ville au XIX\textsuperscript{ème} siècle.  Ildefons Cerdà, en 1860, dessina le Nord-Est de la ville sous la forme d'un quadrillage régulier, entrecoupé d'une structure rayonnant autour d'une place centrale (La plaça de les Glories Catalenes) avec six voies. Il donna ainsi à la ville une signature particulière, autant par la géométrie de son système viaire, que par celle de ses îlots. Sa création vient directement s'appuyer sur la partie historique de la ville, dont les rues sinueuses contrastent avec la régularité du plan Cerdà (figure \ref{fig:barce_plan1}). L'extension est rattachée au plan ancien par trois voies qui s'y introduisent. Le reste du graphe s'y juxtapose sans chercher à s'y coordonner.

\begin{figure}[h]
    \centering
    \includegraphics[width=\textwidth]{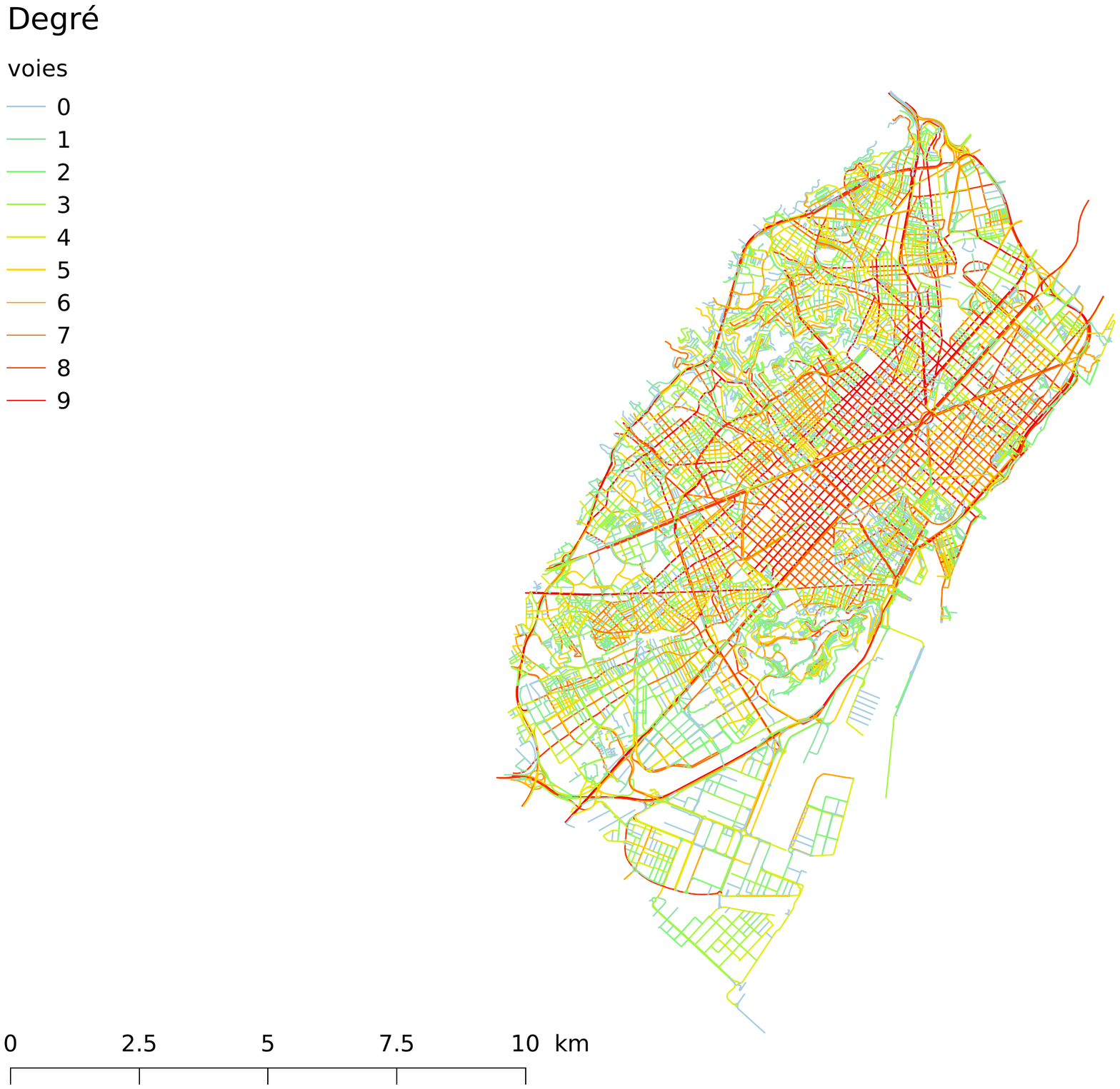}
    \caption{Calcul du degré sur les voies de Barcelone.}
    \label{fig:barcelone_deg}
\end{figure}

\begin{figure}[h]
    \centering
    \includegraphics[width=\textwidth]{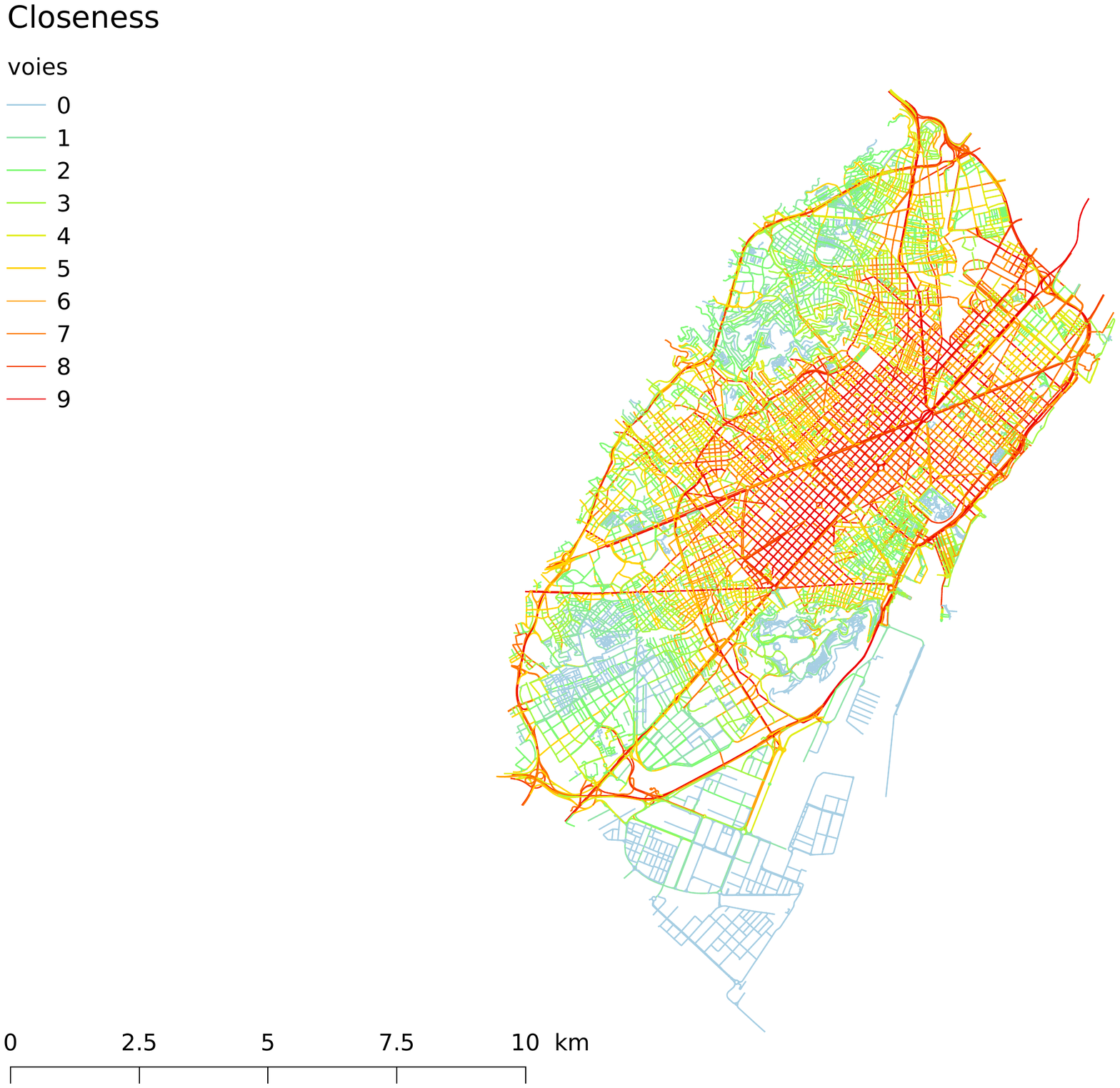}
    \caption{Calcul de la closeness sur les voies de Barcelone.}
    \label{fig:barcelone_clo}
\end{figure}

L'indicateur de closeness, appliqué au graphe de Barcelone, identifie la structure quadrillée comme étant la plus proche de l'ensemble du réseau, en nombre de changements de voie (figure \ref{fig:barcelone_clo}). C'est donc par celle-ci que se font la plupart des liens pour joindre les voies du graphe. C'est également cette structure qui est mise en valeur par l'indicateur de degré comme faisant partie du maillage principal de la ville, auquel viennent s'ajouter des voies de traversée ou de contournement mises en valeur de la même manière sur les deux cartes (figure \ref{fig:barcelone_deg}). Le centre historique, lui, est noyé dans les voies de faible degré et de proximité faible, qui sont aussi celles relevées sur le port ou sur les hauteurs de la ville (au Nord). Dans ce cas d'application, l'indicateur de closeness ne met donc pas en avant les voies anciennes mais celles redessinées pour créer un réseau pensé comme idéal, en terme d'efficacité dans les déplacements.

\begin{figure}[h]
    \centering
    \includegraphics[width=\textwidth]{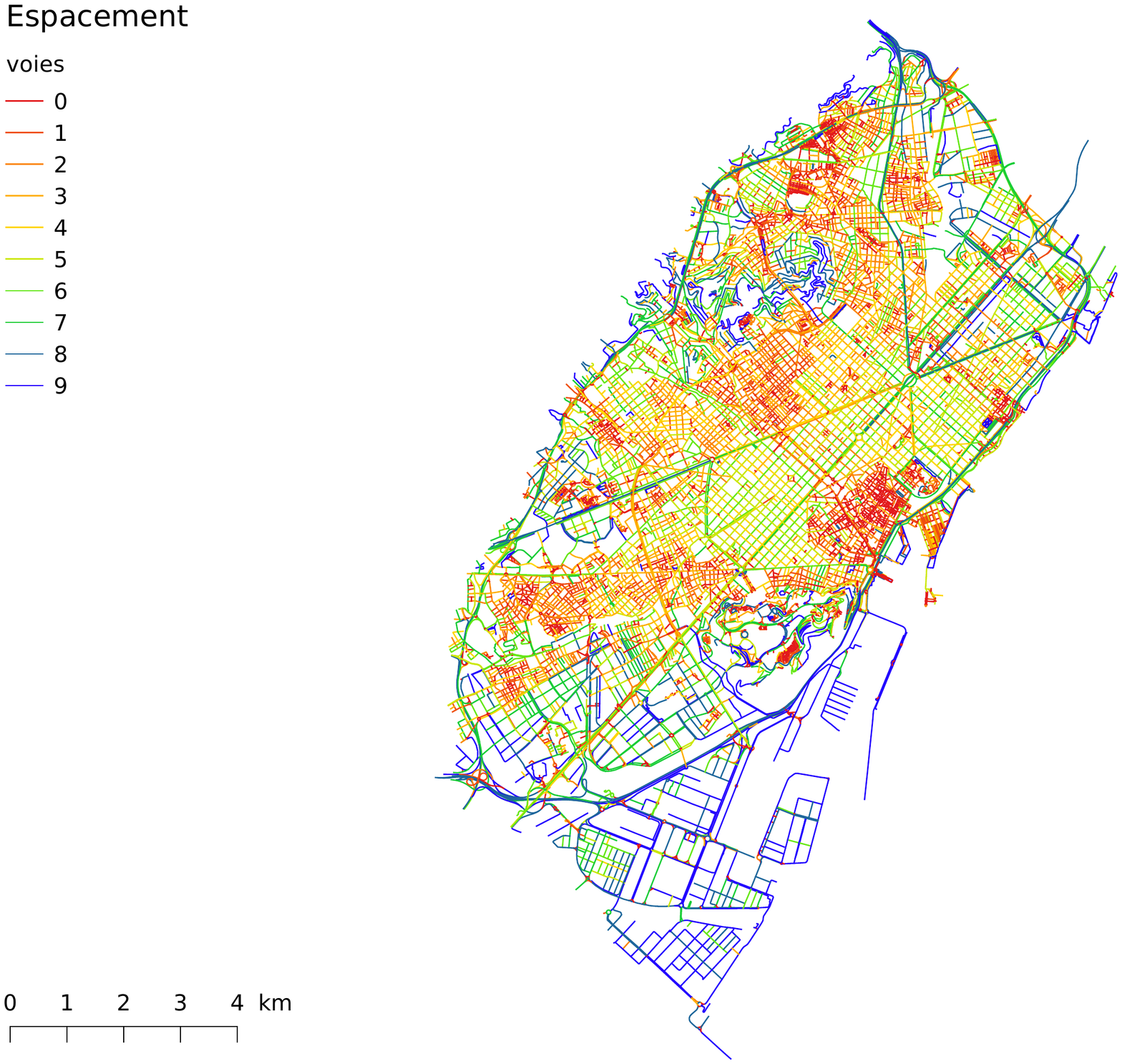}
    \caption{Calcul de l'espacement sur les voies de Barcelone.}
    \label{fig:barcelone_esp}
\end{figure}

En revanche, l'indicateur d'espacement s'oppose aux précédents pour mettre en valeur les structures anciennes : il relève avec des connexions très proches le centre ancien de la ville ainsi que les quartiers qui ont été moins touchés par la replanification. Les voies du port ressortent avec les coefficients d'espacement les plus forts, de même que les voies sinueuses qui abordent les zones escarpées des parcs Güell et Guinardo  (lecture du relief rappelant celle introduite pour Brive-la-Gaillarde). Le territoire couvert par le réseau régulier ressort dans les valeurs moyennes d'espacement : sa densité linéaire est intermédiaire entre celle très forte des centres anciens et celle faible de la zone portuaire et des chemins des parcs (figure \ref{fig:barcelone_esp}).

\FloatBarrier
\subsection{Kyoto}

\begin{figure}[h]
    \centering
        \includegraphics[width=0.8\textwidth]{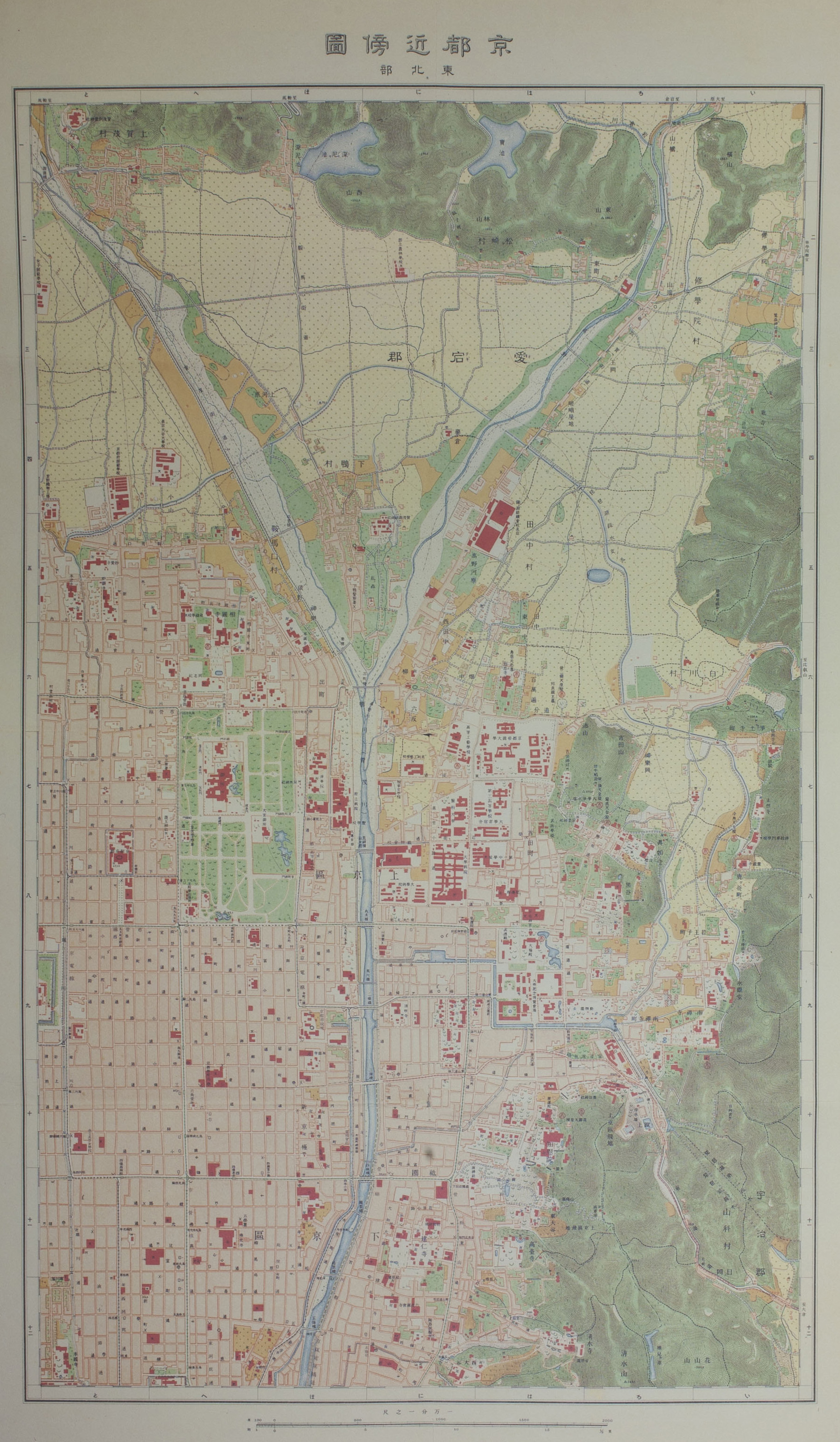}    
    \caption{Carte historique de Kyoto en 1915 (partie centrale et Nord-Est)}
    \label{fig:kyoto_plan1}
\end{figure}

Kyoto a été la troisième capitale japonaise, sous le nom d’Heian-Kyô, après Heijokyô et Fujiwarakyô. Fondée en 784 sur le modèle des capitales chinoises parfaitement \enquote{réglées}, elle a donc une structure planifiée. 

Cependant, la ville s’est déplacée de près d'un kilomètre vers l’est dans la même plaine enserrée de collines, le premier site étant trop marécageux lors des pluies diluviennes de la mousson. Le même quadrillage a subi de multiples périples depuis lors (scission en deux, redressement des avenues après divers abandons et guerres civiles, etc). Pourtant, la partie centrale de la ville, correspondant à la vieille ville, demeure « quadrillée » comme à l’origine \citep{bonnin1999aspect, bonnin2012vivre, bonnin2014etude}.

\begin{figure}[h]
    \centering
    \includegraphics[width=\textwidth]{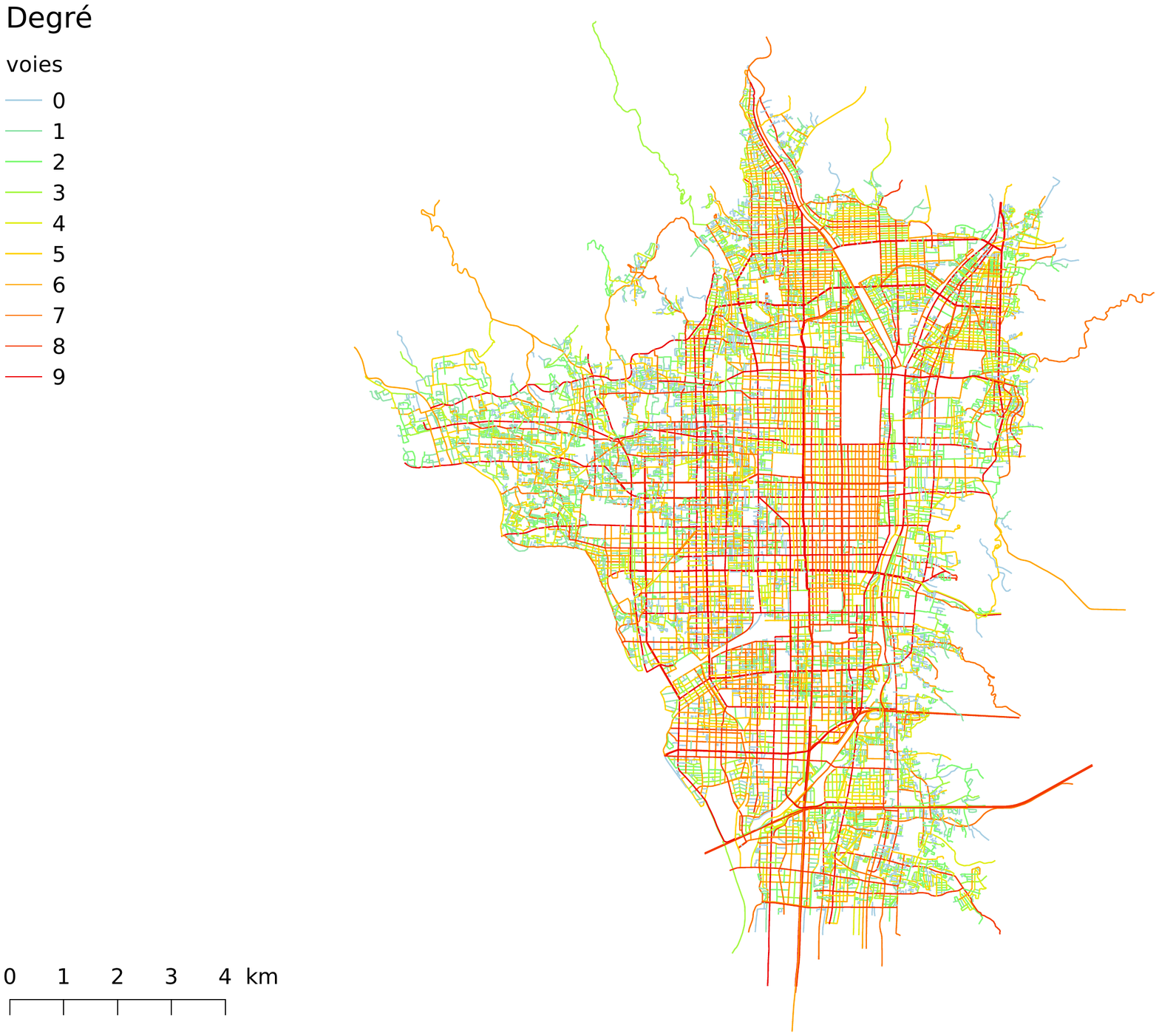}
    \caption{Calcul du degré sur les voies de Kyoto.}
    \label{fig:kyoto_deg}
\end{figure}

\begin{figure}[h]
    \centering
    \includegraphics[width=\textwidth]{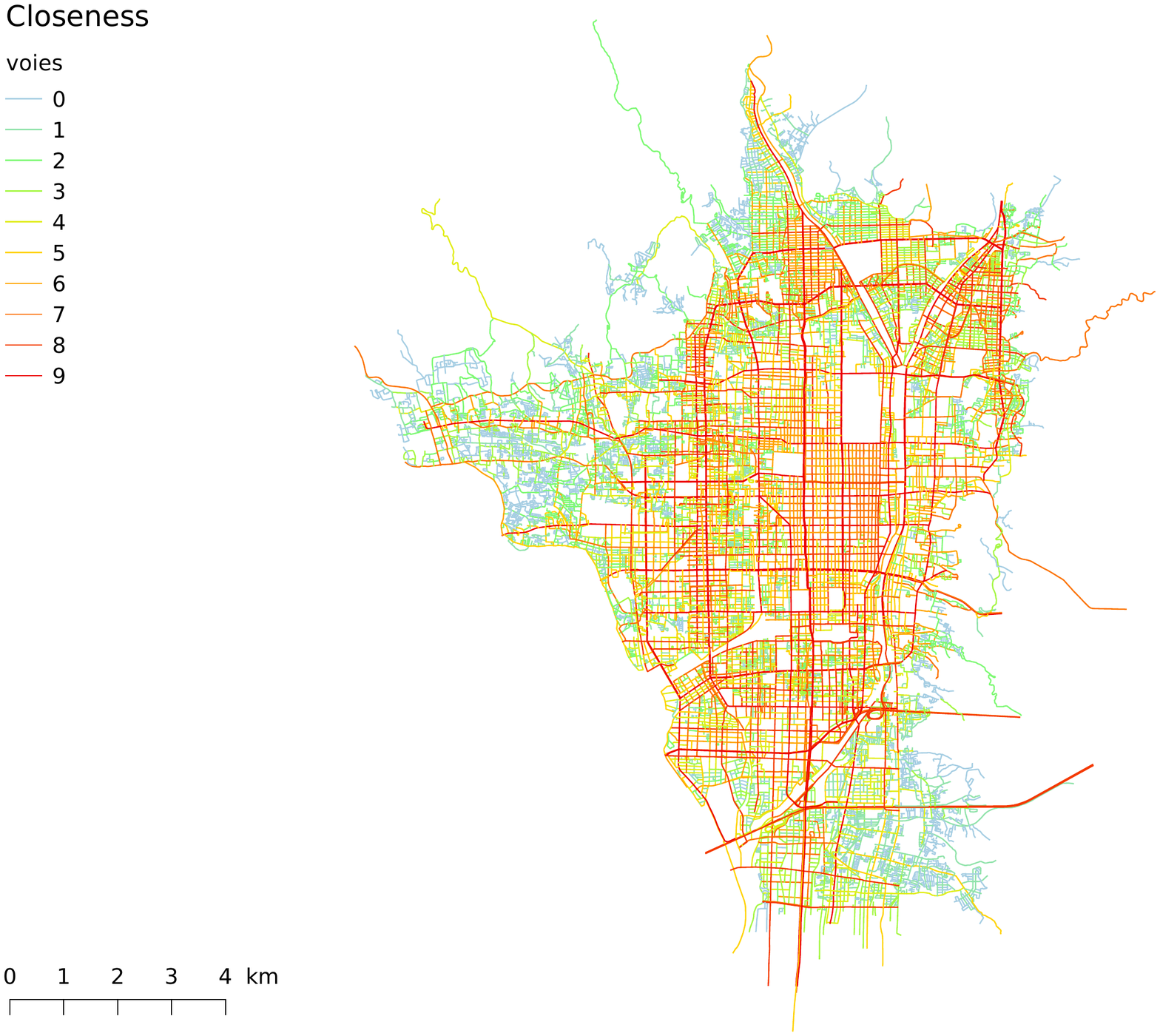}
    \caption{Calcul de la closeness sur les voies de Kyoto.}
    \label{fig:kyoto_clo}
\end{figure}

Comme pour le graphe de Barcelone, les résultats donnés par les indicateurs de degré et de closeness sont très proches (figures \ref{fig:kyoto_deg} et \ref{fig:kyoto_clo}). Nous avons pu remarquer ce résultat sur les villes à la structure fortement quadrillée avec les corrélations des indicateurs étudiées sur le réseau de Manhattan. Les deux indicateurs mettent ainsi en valeur la structure régulière de la ville. L'indicateur de degré met en avant les voies du maillage principal, qui s'étendent jusqu'aux bords de l'échantillon. L'indicateur de closeness identifie la plupart de ces mêmes voies comme celles assurant une transition topologique rapide d'un bout à l'autre du graphe. Il resserre cependant l'étendue des voies aux coefficients les plus importants autour du centre-ville et selon des axes Nord-Sud. Cet indicateur fait ainsi ressortir comme plus isolées les parties de la ville qui font la transition avec la montagne environnante à l'Ouest et au Sud-Est.

\begin{figure}[h]
    \centering
    \includegraphics[width=\textwidth]{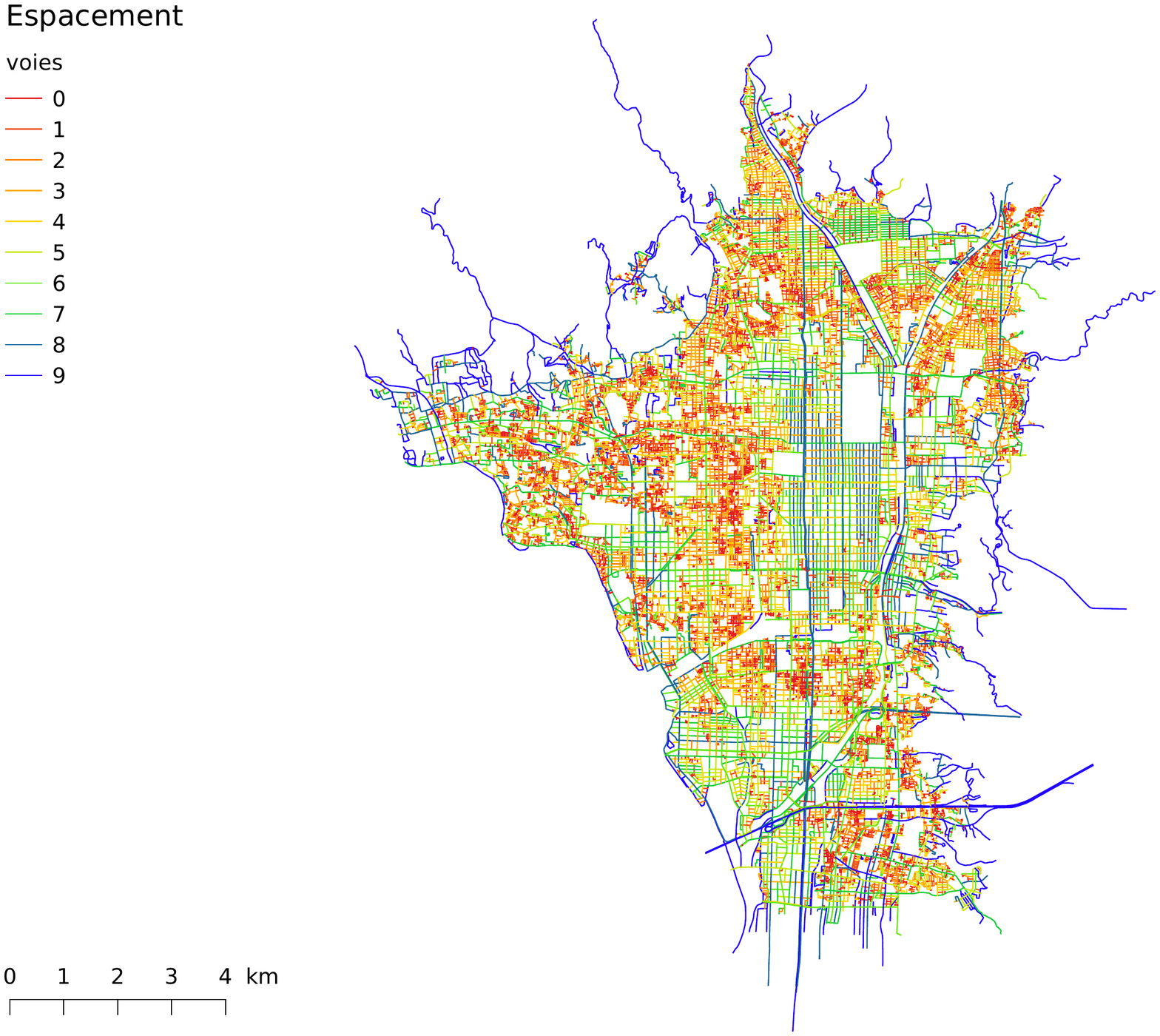}
    \caption{Calcul de l'espacement sur les voies de Kyoto.}
    \label{fig:kyoto_esp}
\end{figure}

Kyoto est un exemple de ville où le contexte hydrographique disparaît sous le réseau. La ville est en effet traversée par une rivière (\textit{Katsura River}, figure \ref{fig:kyoto_plan1}) dont il est difficile de distinguer les contours en n'observant que le graphe viaire. L'orographie transparaît à travers les voies en bord de réseau dont la closeness est faible mais la cartographie de l'indicateur de degré ne le met pas en évidence. L'application du coefficient d'espacement est, sur ce territoire, le plus révélateur de son contexte géographique : les voies qui grimpent sur le relief ont en effet un coefficient d'espacement fort (figure \ref{fig:kyoto_esp}). Cet indicateur fait également ressortir les quartiers très denses (en rouge) et les oppose à ceux quadrillés de manière plus espacée (en vert).

\begin{figure}[h]
    \centering
    \includegraphics[width=\textwidth]{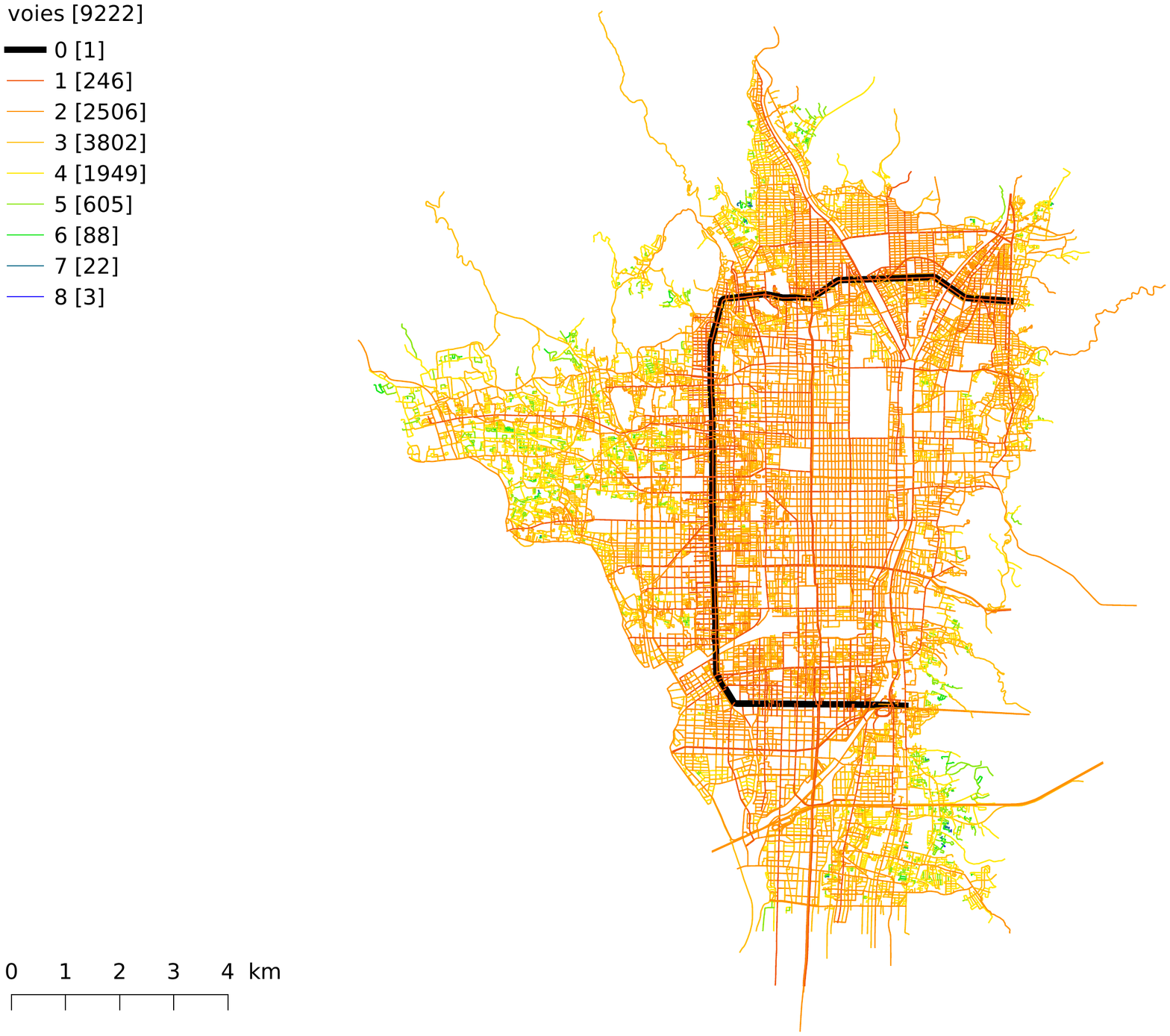}
    \caption{Calcul des distances topologiques à partir de la voie de plus fort degré (représentée en noir) sur les voies de Kyoto.}
    \label{fig:kyoto_dtopo}
\end{figure}

Enfin, nous avons également calculé sur cette ville les distances topologiques de toutes les voies vers celle au plus fort degré (retenue par notre programme). Nous représentons le résultat de ce calcul sur la figure \ref{fig:kyoto_dtopo}. Nous obtenons ainsi un nombre de changements de voie maximum de 8, qui ne concerne que trois voies, la majorité se situant à trois ou quatre tournants de la voie considérée (représentée en noir). L'efficacité en terme de proximité topologique de l'ensemble du réseau (calculée en faisant la moyenne de l'indicateur de closeness sur la ville) est de 0.19, ce qui est meilleur que Barcelone (0.15) ou Téhéran (0.12) mais moins bon que Manhattan (0.29).

\FloatBarrier
\section{Les limites de notre approche}

La perception que nous avons de la ville est difficilement saisissable par une vue aérienne. Afin de mieux comprendre les fonctionnalités potentielles des rues entremêlées nous extrayons le graphe viaire de ce paysage complexe pour en caractériser la structure. À partir de l'étude des propriétés géométriques de ce réseau, nous parvenons à hiérarchiser les liens entre eux pour mieux comprendre leurs relations et leurs implications dans le développement et l'utilisation de l'espace urbain. Cependant, ces résultats s'enrichissent lorsqu'ils sont mis en regard avec le lieu, sa culture et la temporalité dans laquelle il s'inscrit. Nos indicateurs, appliqués au réseau des rues, permettent d'en faire une première lecture.

Ces analyses sont dépendantes de l'échelle à laquelle elles sont appliquées. Celle-ci est définie par le détail du filaire numérisé sur lequel sont appliqués les calculs. Si nous l'augmentons exagérément, un parterre de fleurs sur un trottoir pourrait introduire une déviation. À l'inverse, si l'information considérée est trop globale, la ville elle-même deviendrait un nœud d'un réseau plus large, et le détail de la déviation à chaque intersection perdrait sa pertinence. Nous situons notre travail à l'échelle humaine, car c'est l'Homme qui construit et habite les villes. Il agit par petites itérations successives, projets urbains modelant son réseau ; ou bien par une action d'occupation de tout un espace, qui se traduit par un projet de planification globale. Cette échelle est celle où le rond-point peut poser question, de la même manière qu'un léger décrochement aux carrefours (problèmes évoqués en partie II). Le rayon des \textit{places}, zones tampons choisies, est notre moyen de répondre à cette question d'échelle, pour placer le \enquote{curseur} de l'étude un peu plus localement ou un peu plus globalement afin de saisir les axes qui nous intéressent. Les méthodes de fusion automatiques ont le grand avantage de permettre des calculs sur de larges espaces (dès lors que l'on en possède le réseau viaire vectorisé) mais peuvent aussi occasionner des simplifications un peu trop fortes. Ainsi, en voulant supprimer un large rond-point, le rayon de la \textit{place} paramétré pourra faire disparaître un décrochement dû à un pâté de maison, à un autre endroit. La paramétrisation de ce rayon dépend du réseau mais suppose l'homogénéité de celui-ci. Sans cela, il faudrait imaginer une paramétrisation automatique indicée sur l'indicateur d'espacement de chaque voie.

La seule méthode valable, pour pallier de manière sûre les erreurs de vectorisation et supprimer les aménagements non voulus, est la reprise manuelle du réseau suite à une enquête sur le terrain. Cela limite les terrains de recherche mais permet une analyse approfondie. Notons tout de même que seuls les grands points d'articulation seraient essentiels dans l'enquête car la méthode est robuste aux petites erreurs ou petits ajouts dans la vectorisation et les zones tampons permettent de supprimer les aménagements urbains à échelle réduite non désirés dans le calcul.

D'autre part, la méthode de construction des voies peut parfois se révéler inappropriée à la réalité du terrain. Ainsi, les routes de montagnes constituent une des limites de notre méthodologie. En effet, si elles font un lacet simple, cela ne posera pas de problème ; mais si d'autres routes viennent se connecter sur les lacets, elles entrecouperont la route principale par de multiples intersections. La voie créée par notre programme ne sera pas celle principale, historique, d’ascension de la montagne : plusieurs voies se créeront, coupées à chaque lacet avec carrefour.

La modélisation choisie peut également se retrouver inappropriée pour certaines villes qui, de manière historique, ne fonctionnent pas sans un autre réseau spatial sur lequel le réseau viaire viendrait s'appuyer. C'est le cas de Venise, par exemple, dont les îles bien connues portent le fonctionnement en symbiose des routes et des canaux qui permettent les déplacements. L'analyse du réseau viaire sur ce territoire n'a donc pas la même pertinence que pour les autres villes. La solution proposée consiste à étudier les deux réseaux (viaire et hydraulique) fusionnés \citep{perna2011characterization}.

Enfin, lire le temps à travers les voies centrales dans le réseau, pose des problèmes complexes. En effet, dans les villes, une forme ancienne, historique, sur laquelle le réseau est venu se construire (comme par exemple l'enceinte du XIV\textsuperscript{ème} siècle d'Avignon) aura certaines caractéristiques communes avec une voie récente qui redécoupe l'espace pour assurer son accessibilité (à l'image de l'avenue de la République et du cours Jean-Jaurès, toujours à Avignon). Nous pouvons cependant identifier des trames qui aident à lire la ville et sa probable évolution.

C'est pour ces raisons qu'il est nécessaire de ne pas se perdre dans l'abstraction et les statistiques de l'objet observé et de conserver un lien avec sa nature. Chaque ville est unique mais porte en elle, à travers son réseau viaire, une information quantifiable dont certaines propriétés traversent les continents. Ainsi, si le degré des voies permet d'avoir l'intuition de leur utilisation, et si leur closeness permet de retracer une partie de la croissance de la ville, nous avons vu que ces premières interprétations ne peuvent pas être faites sans maintenir un échange entre les sciences thématiques, expertes de villes particulières, et celles quantitatives, qui forcent l'abstraction pour étendre leurs études. 

\clearpage{\pagestyle{empty}\cleardoublepage}
\chapter{Les enjeux de la forme}
\minitoc
\markright{Les enjeux de la forme}

\FloatBarrier
\section{Les stratégies de morphologie}

À différentes époques, en différents lieux, les formes naissent dans des contextes particuliers. Lorsqu'il s'agit de formes viaires, cela peut faire intervenir des stratégies de repli, ou au contraire, d'ouverture. Ces différentes manières de penser la forme donnent prise au temps sur elles. Dès lors qu'elles sont créées dans un but particulier, elles sont liées au contexte historique et contiennent une empreinte de celui-ci.

La sinuosité des rues à l'intérieur des remparts de certaines villes médiévales pouvait être dissuasive. La même stratégie est utilisée de nos jours, dans les lotissements aisés, afin de dissuader les voitures de s'y introduire et ainsi de permettre aux enfants de jouer près de la rue en sécurité. Dans les villages japonais, la rue principale faisait une baïonnette étudiée pour briser la course des cavaliers.

Nous avons vu dans le chapitre précédent que cet isolement au réseau (volonté d'augmenter la distance topologique entre une partie du graphe et le reste de celui-ci), se traduit par un indicateur de closeness faible. Nous retrouvons également des cas où l'isolement est subi : c'est notamment le cas des Zones Urbaines Sensibles, où l'enclavement nuit au bien-être des habitants.

Dans un contexte de loisirs et de découvertes, la courbure des rues (sur un tronçon ou à une intersection) peut également inciter à aller voir plus loin et participer à une stratégie touristique. Les études de concavité et de convexité, d'élargissement, de piétonnisation d'un secteur relève de stratégies urbaines complexes mais ne restent pas étrangères à l'histoire des lieux. Nous développerons le lien entre passé et présent d'une trame viaire dans le quatrième chapitre de cette partie.

La forme peut donc avoir pour vocation de cacher, pour perdre les ennemis étrangers à la ville, en se rétrécissant et se courbant. Ou au contraire, dans des stratégies de protection plus récentes, de s'ouvrir pour exposer l'espace entre le bâti à la vue de tous et diminuer les zones d'obscurité de la ville. Dans le contexte actuel, l'élément rassurant de la ville n'est plus les rues tortueuses mais les grands axes. Ils structurent l'espace comme ils structurent la représentation de celui-ci pour celui qui s'y déplace. Ils apposent un cadre : élément fixe du mouvement de l'utilisateur qui lui permet de se situer, élément pérenne du développement de toute une ville. Ces grands axes sont ceux empruntés en voiture selon la stratégie dite \enquote{du chauffeur de taxi} que nous décrirons dans le paragraphe suivant. Comme ils sont connus et rassurants, l'utilisateur aura tendance à les prendre également à pied au détriment d'autre parcours dont l'ambiance aurait pu être qualifiée de plus agréable. Ce choix s’opère-t-il par l'effet de la connaissance préalable, rassurante, du parcours ou bien par souci de simplicité ? Quelle qu'en soit la raison, la corrélation observée entre les parcours est révélatrice d'un penchant pour une stratégie commune \citep{cristofol2013measuring}.

La forme pensée diffère souvent de la forme produite. Il est donc nécessaire de porter notre attention sur la différence de points de vue. La forme est souvent considérée comme parfaite au moment de sa création puis dégradée avec le temps. Par exemple, les grands projets urbains sont souvent associés à un créateur et à une époque (Haussmann à Paris, Cerdà à Barcelone). L'évolution dans le temps est considéré comme une altération de l’œuvre. On observe cependant dans les faits un cheminement inverse : la forme se perfectionne avec des transformations régulières. 

Cela dépend bien évidemment de l'Histoire du lieu étudié. Les villes planifiées (comme New-York, San-Francisco) ont une structure récente, construite d'une même main. Alors que dans la vieille Europe, les structures se peaufinent avec le temps (Avignon, Paris), lorsqu'on ne leur impose pas une ré-écriture globale (Barcelone). Les formes sont un langage, une expression du temps (plus ou moins long) sur un espace. Selon l’œil qui les étudie elles peuvent décrire une structure sociale, une griffe d'urbaniste, ou une morphogenèse complexe.

Nous avons tenté d'établir une grammaire de lecture des formes dans les chapitres précédents afin de saisir l'information qu'elles peuvent apporter. Néanmoins, cette méthodologie vient en complément de travaux spécifiques sur un lieu en proposant une quantification objective des structures.

\FloatBarrier
\section{Les stratégies d'itinéraires}

Le réseau viaire, avant d'être l'empreinte d'une évolution passée sur le présent, est un réseau de déplacements. Il coordonne points stratégiques (centres-villes), distribution locale (lotissements) et circulation rapide. Il peut être considéré à différentes échelles où les questions posées ne seront pas les mêmes. Nous avons choisi de considérer dans notre travail tous les chemins, qu'ils soient revêtus de terre, de goudron ou de pavés. Nous avons exclu les pistes cyclables, escaliers, sentiers, allées et circuits. Nous avons donc un extrait du réseau viaire que nous considérons comme pouvant être emprunté par tous. Les données dont nous disposons sont constituées par l'axe supposé numérisé au centre de la rue. Elles correspondraient donc mieux à la perspective depuis un véhicule, mais peuvent être appliquées aussi bien à celle de personnes à pied ou à vélo.

La recherche d'itinéraires par un utilisateur du réseau peut se faire selon différentes stratégies. B. Marchand fut un des premiers à explorer quantitativement le comportement des piétions selon leur perception de leur environnement \citep{marchand1974pedestrian}. Il compare ainsi des cartes mentales construites à partir de témoignages et des cartes topographiques. Le chercheur montre que l'utilisateur a une perception beaucoup plus symétrique de l'environnement dans lequel il se déplace qu'il ne l'est en réalité. Cette rationalisation de l'espace révèle la nécessité de simplification dans la conception des itinéraires.

Dans la recherche que nous présentons ici, nous appuyons l'importance de la continuité rectiligne dans l'analyse des structures urbaines. La portée de celle-ci dans les cheminements a fait l'objet de nombreux travaux \citep{hillier1993natural, dalton2001secret}. E. Degouys, doctorante au sein de notre équipe de recherche, approfondit cette problématique. Elle travaille sur les éventuelles corrélations entre les stratégies de déplacement et les résultats mis en avant par les indicateurs que nous avons présentés  \citep{degouysAPitineraire}.

Les stratégies de déplacement ont été étudiées précisément pour les piétons, entre autres par Foltête, Genre-Grandpierre et Piombini \citep{genre2003morphologie, foltete2007urban} et pour les chauffeurs de taxis par Pailhous \citep{pailhous1970representation}. Des travaux en physique ont également eu la volonté de se détacher de l'analyse purement structurelle pour modéliser les stratégies de cheminement \citep{lee2012exploring}.

\begin{figure}[h]
    \centering
    \includegraphics[width=\textwidth]{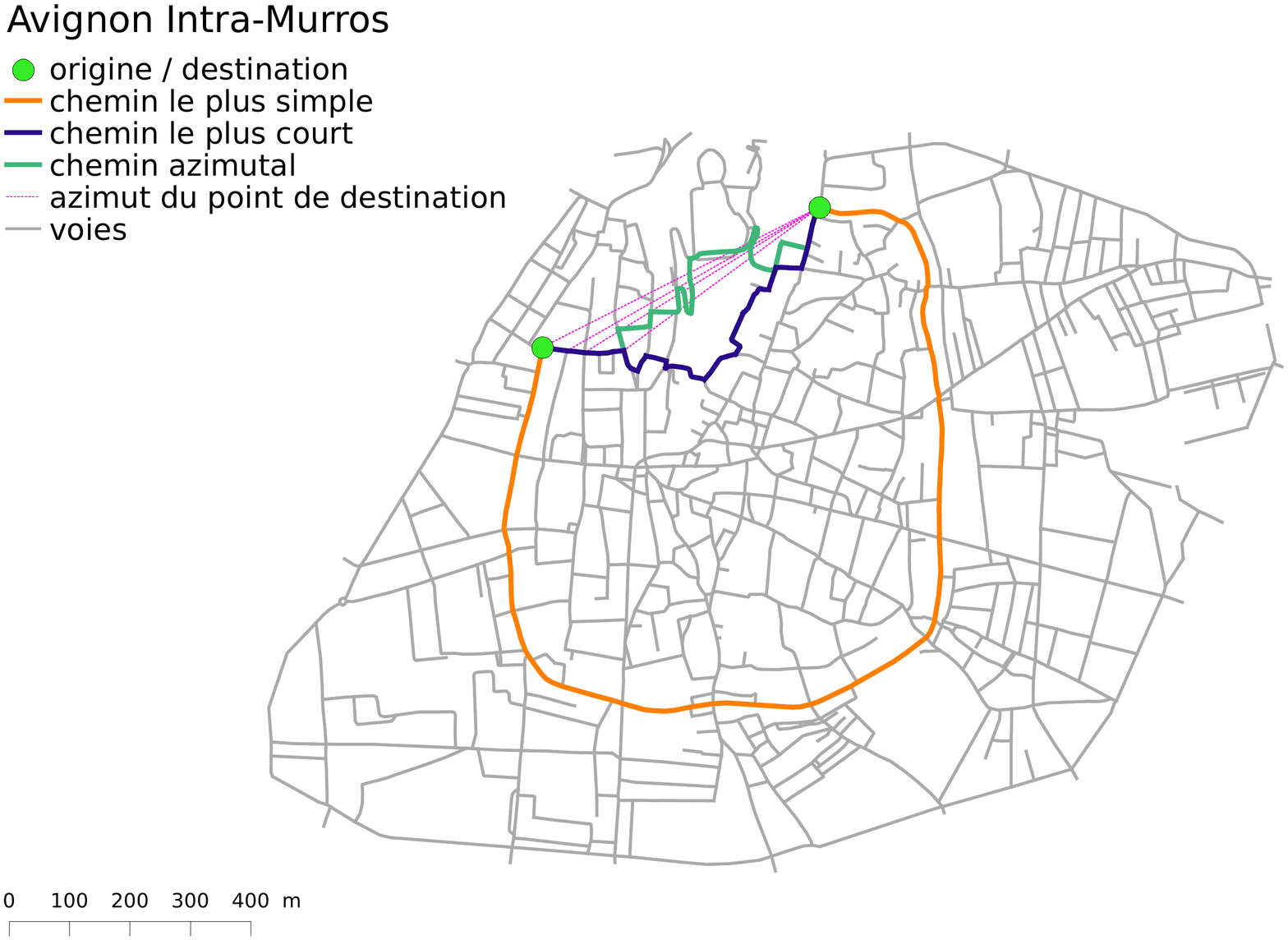}
    \caption{Illustration des trois stratégies d'itinéraire entre deux points de l'intra-murros de la ville d'Avignon. Pour le chemin azimutal, nous avons détaillé l'azimut du point de destination pour les quatre premières intersections}
    \label{fig:distances2}
\end{figure}

Nous faisons référence ici à trois types de parcours, tous liés à la géométrie du réseau (figure \ref{fig:distances2}) :

\begin{itemize}
\item le chemin le plus court, en linéaire le long du réseau
\item le chemin le plus simple, qui minimise le nombre de changements de voie
\item le chemin conservant l'azimut au point d'arrivée le plus proche à chaque intersection (qualifié de \textit{chemin azimutal})
\end{itemize}

Le choix de l'un de ces itinéraires dépend du but fonctionnel du trajet (résidence-travail, loisirs, etc.) et de la connaissance du réseau par l'usager. Ainsi, une personne qui ne connaît pas la ville aura tendance à choisir les chemins les plus simples alors que celle qui la connaît bien et veut minimiser son temps de parcours préférera le plus court. Bien évidemment, un grand nombre de paramètres intervient dans le choix d'un itinéraire, notamment par un piéton. Golledge en dresse une liste non exhaustive comprenant en plus des trois évoqués précédemment (minimisation de la longueur, du nombre de changements de direction et de détours), la minimisation du nombre d'obstacles, des externalités négatives (bruit, pollution), des efforts, du coût (réel ou perçu), de la dangerosité, du nombre de tronçons, la maximisation des externalités positives (esthétique, ambiance), etc  \citep{golledge1997spatial, golledge1999human}. Parmi ces critères, certains sont purement qualitatifs, comme l'ambiance urbaine, qui est difficilement saisie à travers des paramètres de modélisation. D'autres peuvent être quantifiés, comme le nombre de tronçons parcourus, la distance ou le nombre de changements de voie.

Dans les faits, les décisions prises par l'utilisateur du réseau ne peuvent pas tenir compte de l'ensemble des critères liés à son déplacement. Chaque individu hiérarchise ses propres critères selon son appréhension de la ville et du déplacement. Il établit son choix parmi au plus cinq d'entre eux, qu'il juge les plus importants. Si plus de paramètres interviennent, le choix n'est plus rationnel. Cet ensemble de critères de choix appuie la notion d'adhérence mise en avant par Amar \citep{amar1993pour} pour qualifier les différents modes de déplacement : le marcheur est en interaction forte avec son environnement. Nous pouvons dès lors considérer qu'un carrefour modifié inclut une pénibilité forte pour un utilisateur à ne pas négliger. La solution d'effacement de ces discontinuités ne serait donc pas pertinente dans une analyse des déplacements.

L'étude menée par Piombini et Foltête \citep{foltete2007urban} procède par corrélation des choix faits avec un ensemble de choix possibles. Ils considèrent tous les chemins éloignés au plus à 120\% du plus court chemin. Selon les micro-économistes, les individus sont capables de faire le choix optimum parmi cet ensemble de choix. Cette approche est controversée, notamment par les psychologues qui considèrent qu'un individu ne cherche pas à optimiser son parcours, mais à minimiser ses pertes sur celui-ci. La question du moment du choix de l'itinéraire reste en suspens : est-ce avant ou lors du déplacement ? Le chemin azimutal, qui consiste à choisir à chaque intersection l'option qui minimise l'angle fait avec la destination, laisse supposer une construction d'itinéraires avec le cheminement, \textit{recalculée} dès qu'une prise de décision doit être opérée. Il laisse ainsi supposer une recherche d'itinéraires plus efficace au cours du déplacement. 

C'est le fonctionnement suggéré par Pailhous. Il le compléta par l'idée de plusieurs \textit{niveaux de réseau}  \citep{pailhous1970representation}. Il considère les stratégies d'un chauffeur de taxi, qui part d'un tissu local pour se raccorder au plus vite aux grands axes avant de rejoindre à nouveau le tissu local afin d'arriver à destination. Il décrit ainsi trois types de réseau : celui principal, qui permet une circulation rapide ; le réseau secondaire qui permet une desserte fine de l'espace ; et enfin un réseau tertiaire très peu utilisé. Le chauffeur de taxi a ainsi une représentation topologique et topographique du réseau principal : il sait comment y accéder au plus vite et jusqu'où il peut être emprunté. Ces trois niveaux de réseau peuvent être illustrés par ceux que nous avons présentés dans le chapitre précédent, correspondant à des graphes \textit{kcore} exclusifs.

\begin{figure}[h]
    \centering
	\includegraphics[width=0.8\textwidth]{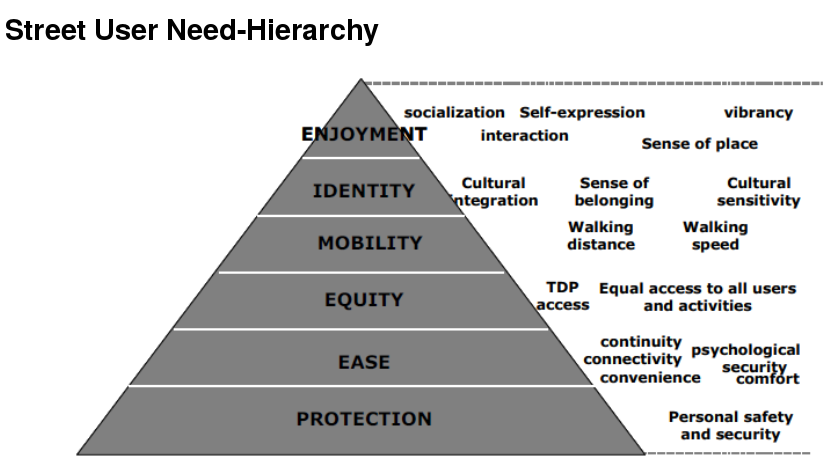}
	\caption{Pyramide de hiérarchisation des besoins de l'utilisateur du réseau viaire. \\ source : \citep{mateo2005street}}
	\label{fig:pyramideMateoBabiano}
\end{figure}

Les besoins du piéton dans ses déplacements au sein de la ville ont été étudiés par Mateo-Babiano \textit{et al.} \citep{mateo2005street, mateo2010sidewalk}. Les chercheurs dressent une pyramide de besoin sur le même modèle que celle construite à partir des travaux de Maslov sur les besoins d'un être humain \citep{maslow1943theory}. Ils définissent ainsi les besoins prioritaires de l'utilisateur constituant le socle de la pyramide (la protection, la mobilité) qu'ils raffinent jusqu'aux besoins plus superficiels (la réalisation, le plaisir) (cf figure \ref{fig:pyramideMateoBabiano}).

En pratique, les stratégies choisies par les utilisateurs du réseau recoupent en grand nombre les critères listés ci-dessus. Les méthodes d'analyse quantitative font appel à des enquêtes sur le terrain. Elles sont fondées sur deux types de protocoles : les préférences révélées (observation des actes) d'une part, les préférences déclarées d'autre part. Cependant, les déclarations peuvent inclure des reconstitutions  \textit{a posteriori} des choix opérés par la personne sondée. L'idéal est un protocole croisé, faisant intervenir les deux méthodes. C'est sur cette idée que A. Piombini est venu s'appuyer sur les travaux de syntaxe spatiale pour construire une analyse des itinéraires piétonniers. 

\begin{figure}[h]
    \centering
	\includegraphics[width=0.6\textwidth]{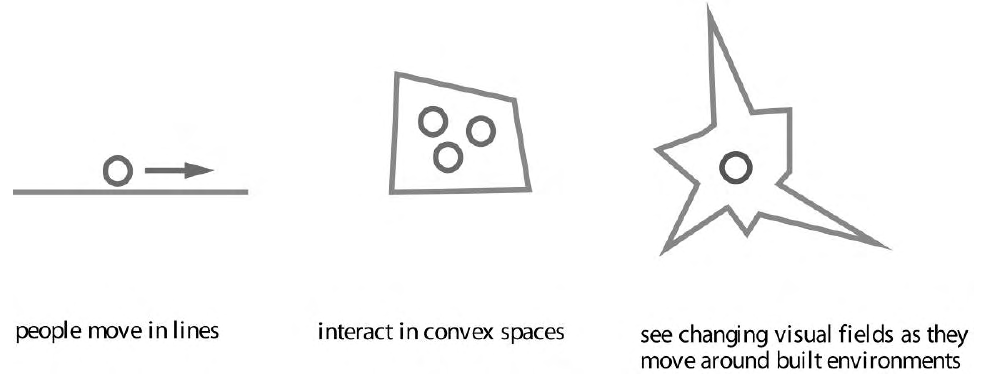}
	\caption{Différentes perceptions de l'espace par un piéton. \\ source : \citep{hillier2007city}}
	\label{fig:spaceHillier}
\end{figure}

Comme nous l'avons observé dans la partie précédente, l'étude des tronçons (arcs) du réseau viaire les uns indépendamment des autres n'est pas révélatrice de structures. Elle ne traduit pas de notion de continuité et donc de mouvement. Les travaux de syntaxe spatiale se placent parmi les premiers à reconstituer des continuités à travers des \textit{lignes de perspectives}. Ils positionnent ces éléments les uns par rapport aux autres afin de retracer un mouvement naturel fondé sur l'idée de la propension des piétons à se déplacer où leur regard porte. Ils ré-appréhendent donc le réseau depuis le point de vue de ceux-ci. L'\textit{effort} mis en œuvre est considéré être principalement celui du changement de direction, pour aller au delà de la perspective et changer ainsi d'espace. Il n'est plus dépendant du déplacement physique lié à la distance métrique mais aux changements de ligne de perspective. Les psychologues nomment ce processus \textit{reprogrammation motrice} \citep{hillier1976space, hillier1984social, hillier1993natural}. Les piétons se déplacent donc selon les visions du réseau viaire qui leur sont offertes. Celles-ci sont regroupées en trois catégories : la ligne de perspective rectiligne qui induit un mouvement en ligne droite, l'espace convexe (comme une place publique par exemple) qui est ouvert et offre une visibilité large, et enfin les différentes lignes d'ouvertures à 360\degres  lorsque, à une intersection, le piéton considère les espaces laissés libres par les bâtiments (figure \ref{fig:spaceHillier}, \citep{hillier2007city}). L'attraction qui induit le déplacement est donc liée aux endroits où le regard porte. Dans cette analyse, chaque itinéraire a un sens unique, l'étude n'est pas réversible car la perception dépend de l'orientation du cheminement.

Les plans sont donc transcrits sous forme de lignes axiales, où chaque segment est considéré indépendamment. Un arc peut donc être démultiplié s'il est sinueux selon plusieurs lignes différentes. Cette démultiplication apparaît également sur les ronds-points, qui posent problème dans l'analyse de certains plans. Le nombre de lignes axiales trop important contraint parfois à la suppression de certains aménagements urbains (lacets, giratoires) qui altèrent la compréhension de l'ensemble.

Le nombre de tournants pour accéder au reste du réseau est compté à partir de ces lignes axiales. Un indice d'intégration global est calculé en sommant les distances à partir d'un axe vers tous les autres. Ces distances sont comptées en nombre de changements (chemins les plus simples). Le calcul de cette intégration suit la même méthodologie que notre calcul de rayon topologique. Cependant, l'objet sur lequel il est appliqué diffère, les axes des segments ne sont pas multi-échelles comme la voie, ce qui implique une plus grande sensibilité aux effets de bord. Dans ce travail précurseur, une corrélation a été établie entre les axes qualifiés de centraux dans l'approche morphologique et ceux qui le sont dans l'observation fonctionnelle. La situation de la ligne de perspective au sein du réseau selon le facteur d'intégration est plus révélatrice qu'une analyse paysagère du même réseau. Les piétons sont plus sensibles à l'efficacité en terme de simplicité de leur itinéraire qu'à leur environnement \citep{piombini2006modelisation}. Dès lors, nous pouvons nous poser la question de la cause et de l'effet. Les commerces s'installent-ils sur les axes centraux ou la ville se développe-t-elle de manière à faciliter l'accès aux commerces ? Ce que nous pouvons confronter à cette problématique est une analyse statistique des éléments au fondement du réseau selon cette théorie : les tronçons et leurs lignes axiales.

Le lien entre morphologie et implantation pourrait être approfondi en étudiant la corrélation entre les prix fonciers et les axes détectés comme centraux. Les aménagements qualitatifs apportent en effet, en eux même, une quantification de la centralité des tronçons. Une voie élargie, sur laquelle sont développés des aménagements paysagés, pourra être considérée comme porteuse de flux, puisque les professionnels de la ville la traitent comme telle. Des données qualitatives pourraient ainsi venir compléter les analyses de flux établies sur les structures topologiques \citep{genre2001structure}.

Dans la lecture statique des réseaux viaires, il faut garder à l'esprit l'aspect polysémique de la ville pour ne pas nous enfermer dans un unique sens de lecture. En effet, la façon dont on pense la ville, dont on établit nos choix est plurielle. Elle dépend de notre moyen de déplacement (tourner le volant n'implique pas la même volonté que de tourner son corps pour changer de direction) et de nos stratégies d'itinéraires. Le cheminement est attaché au lieu dans lequel il s'inscrit, le choix d'un parcours dépend de notre connaissance du réseau et de nos choix précédents. L'influence du premier parcours sur ceux qui suivront n'est pas négligeable. L'exploration de nouveaux tronçons pouvant s'apparenter à un effort assimilable à celui du changement de direction.

Les nouvelles technologies offrent à l'étude des déplacements de nouvelles perspectives. Ainsi, il est désormais possible de recueillir des données sur la mobilité des usagers : la précision des GPS, en constante amélioration, permet de suivre leurs trajectoires. A. Noulas \textit{et al} ont ainsi révélé des comportements partagés, et identifié une loi universelle de déplacements urbains faisant intervenir un nouveau type de distance \citep{noulas2012tale}. En effet, les chercheurs ne considèrent plus la distance géographique entre deux points (origine-destination), mais celle comptée selon les lieux de densité importante. Ils arrivent ainsi à reproduire par une simulation numérique les distributions observées dans les données de mouvement collectées.

\FloatBarrier
\section{Les stratégies de modulation : retour des urbanistes face aux indicateurs développés}

Les itinéraires dépendant de la forme du réseau viaire des villes, il est logique que les personnes qui peuvent agir sur cette forme le fassent de manière à inciter les individus à prendre des tracés particuliers. Les interventions peuvent se faire de plusieurs manières : il peut s'agir d'imposer un sens de circulation ou de rendre un tronçon piéton (modifications d'itinéraires applicables uniquement aux véhicules), ou d'agir sur le réseau lui-même pour en modifier le tracé, par des projets urbains. Nous avons ainsi vu à Avignon que la modification du carrefour proche de la place de l'Horloge, en détruisant un îlot bâti, avait considérablement modifié l'accessibilité des voies intra-muros de la ville. En modifiant la proximité topologique des voies, et ainsi les chemins les plus simples, les stratégies de déplacement peuvent être impactées.

Les interactions entre l'espace urbain et les réseaux de déplacement sont donc nombreuses. En particulier, la conception du schéma viaire peut avoir une incidence directe sur les trajets de ses utilisateurs \citep{henson2003conception}. Ainsi, la sensibilité de notre modèle aux discontinuités locales peut se révéler riche de sens. En effet, elle traduit une stratégie utilisée par les aménageurs, lorsqu'ils modifient les deniers mètres d’une voie ou créent des rond-points, pour changer localement les connexions. Cette sensibilité dans la méthode numérique correspond donc \textit{a priori} à une sensibilité des usagers.

L'équipe de recherche MorphoCity maintient des relations étroites avec le service d'urbanisme de la commune d'Avignon. Celui-ci porte plusieurs projets, dont ceux évoqués dans la deuxième partie (modifications viaires) et celui de création d'une ligne de tramway. Ils ont lu nos cartes en appréciant l'information qu'elles leur apportaient. En étudiant l'impact des projets de constructions viaires sur l'accessibilité du territoire, nous étions en mesure de leur proposer une quantification objective de leur action sur le graphe routier. Cette nouvelle approche dans leur processus de prise de décision a été fortement appréciée et introduite comme une nouvelle donnée. La modulation qu'ils peuvent apporter au réseau viaire via la construction de quelques tronçons a un impact direct sur les chemins les plus simples du territoire. Les stratégies de déplacement des utilisateurs peuvent s'en trouver modifiées.

Les analyses que nous faisons sur le réseau peuvent également être mise en relation avec le trafic routier. Une voie de degré important, comme nous l'avons vu, sera plus susceptible d'être empruntée pour traverser le réseau. En ajoutant des voies au maillage principal, les services d'urbanisme sont susceptibles de décongestionner les axes les plus empruntés. Ces premières hypothèses ouvrent des champs de recherches à explorer, qui viennent compléter des travaux faits sur les transports et leurs liens avec les structures \citep{genre2010new, banos2011geographical, salat2011villes}.

Nous avons également initié et développé des relations avec un cabinet d'urbanisme : INGEROP. Le pôle urbain de celui-ci travaille notamment sur la ville de Cergy-Pontoise pour laquelle nous lui avons fourni les cartes de nos indicateurs. Ce cabinet nous a fait un retour très complet sur la possibilité d'intégration de nos résultats de recherche dans leur processus de prise de décision. Ils travaillent selon un Plan Local de Déplacement, élaboré pour cinq années (dans le cadre de celui de l'Île-de-France, fourni par le STIF).

\begin{figure}[h]
    \centering
	\includegraphics[width=\textwidth]{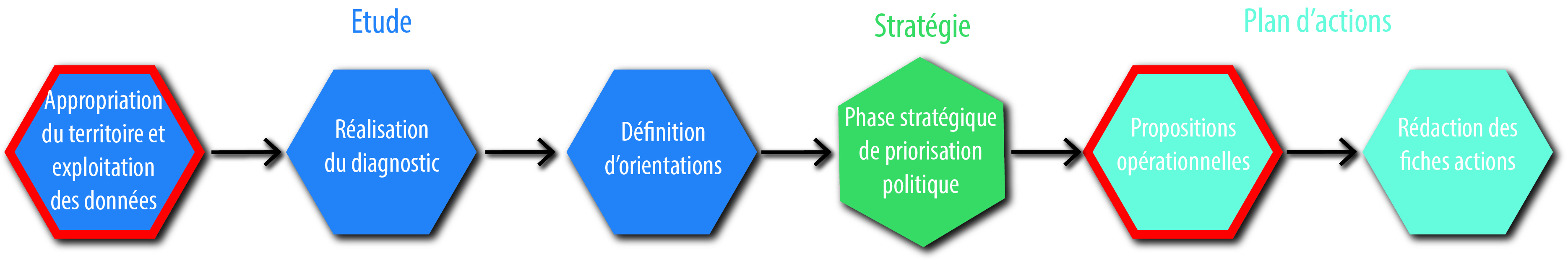}
	\caption{Schéma illustrant les différentes étapes du processus d'élaboration du diagnostique territorial. Celles où les résultats présentés dans cette thèse peuvent intervenir sont encadrées en rouge. \\ source : INGEROP (Majdoline Ouldelhkim, Laure Cochet)}
	\label{fig:processusINGEROP}
\end{figure}

Ils jugent l'intervention de nos résultats pertinente à deux étapes du processus d'élaboration du diagnostic territorial : celle de l'appropriation du territoire ; et celle des propositions opérationnelles (figure \ref{fig:processusINGEROP}). Nos travaux, dans un premier temps, peuvent les aider à avoir une vision objective, quantitative, du territoire. Ils peuvent s'appuyer sur celle-ci lors de la découverte de leur terrain d'étude pour en détecter les structures. Dans un second temps, ils sont utiles à la prise de recul, lors de la priorisation des propositions opérationnelles, avant les décisions d'action. Ils peuvent, de cette manière, appuyer leurs décisions sur des données objectives.

C'est dans ce cadre et cette dynamique de complémentarité que nous avons pensé et développé nos indicateurs appliqués à la ville. La réalisation d'études de mobilité nécessite la territorialisation de l'analyse et du plan d'actions et donc d'aller au delà de la géométrie de la ville. La lecture que nous faisons ici du réseau viaire est complémentaire de celle faite par les professionnels du terrain. Elle leur apporte une information objective qui a pour vocation d'enrichir leur regard sans prétendre s'y substituer. Ainsi, cette méthodologie pourra venir compléter celles mises en place par les scientifiques pour aider les décideurs dans la conception de politiques d'aménagements \citep{genre2000forme, baron2010villes}.

\FloatBarrier
\section{Les stratégies universelles} 

En extrayant la géométrie du réseau viaire des villes, nous permettons à d'autres systèmes, telles que des craquelures dans de l'argile, d'y devenir comparables. En effet, les craquelures sont elles aussi des empreintes laissées par le temps sur un territoire, à une échelle plus réduite. Selon des dynamiques, elles sont guidées par l'hétérogénéité du milieu, en composition ou en épaisseur.

Entre réseaux maillés, les analogies d'analyse de croissance sont tentantes. Ainsi, les réseaux de craquelures se structurent parfois par de grandes lignes traversantes qui apparaissent au début du processus, ne sont pas sans rappeler la naissance d'une ville. Entre deux intersections, une craquelure importante qui élargit et affirme sa ligne droite peut nous faire penser à certaines opérations d'urbanisme. De même, lorsqu'une tension apparaît entre une ancienne et profonde craquelure, et une autre nouvelle et fine. Celle-ci, pour s'y raccorder, détourne sa ligne droite afin de se connecter perpendiculairement à la craquelure précédente.

Réseaux sanguins, veinures de feuilles, graphes viaires, quel que soit l'organisme qu'ils soutiennent, ou la nature des échanges qu'ils rendent possibles, ces réseaux sont porteurs de vie. En effet, la ville grandit de ses échanges avec l'extérieur, elle utilise des ressources, en externalise d'autres, et fonctionne en cela comme un système vivant. Le réseau viaire porte ainsi les flux qui la nourrissent et permettent son développement.

À l'origine de ce développement, nous pouvons nous questionner sur les premières dynamiques d'exploration d'un territoire. Que se passerait-il sans réseaux pré-existants ? Si nous n'étions guidés que par les propriétés \enquote{navigables} de l'environnement dans lequel nous nous situons ? Nous suivrions probablement un itinéraire azimutal, prenant, à chaque point de décision, la direction globale de notre destination.

Nous agirions alors comme des fourmis avec la particularité d'avoir une connaissance globale de l'espace. Celles-ci se dispersent sur le territoire avant de hiérarchiser leurs déplacements selon un principe de renforcement qui leur est propre. L’efficacité du cheminement, pour les fourmis, est déterminée par le temps mis pour faire un aller-retour : les phéromones déposées ayant eu moins de temps pour s'évaporer entre deux passages. Le chemin le plus court en temps de parcours est donc le plus attracteur et est renforcé à chaque nouveau passage.

Il en est de même avec le physarum, champignon capable de créer un réseau optimal entre ses différents points d'alimentation. Il commence par explorer aléatoirement l'espace avant de concentrer ses flux entre les points stratégiques de son alimentation. Des chercheurs sont ainsi parvenus à recréer des réseaux de transport en disposant stratégiquement de la nourriture (céréales) sur une surface plane \citep{keller1970initiation, nakagaki2001smart, tero2007mathematical, takamatsu2009environment}. Ce champignon est sensible à la lumière, il est donc possible de jouer sur ses stratégies de parcours pour moduler le réseau qu'il crée.

Nous avons, dans cette dynamique d'analogies, comparé plusieurs réseaux spatiaux dont nous avons pu trouver ou créer les données : des réseaux de veinures de feuilles, des réseaux de craquelures dans de l'argile, des réseaux de cours d'eau et des réseaux de voies ferrées. Dans ces analogies, nous recherchions certaines propriétés partagées avec le réseau des rues. En effet, nous avons pensé et développé les indicateurs pour caractériser le squelette viaire, mais l'utilisation d'une information très dépouillée les rendait applicables à tous graphes spatialisés.

\begin{figure}[h]
    \centering
    \includegraphics[width=0.9\textwidth]{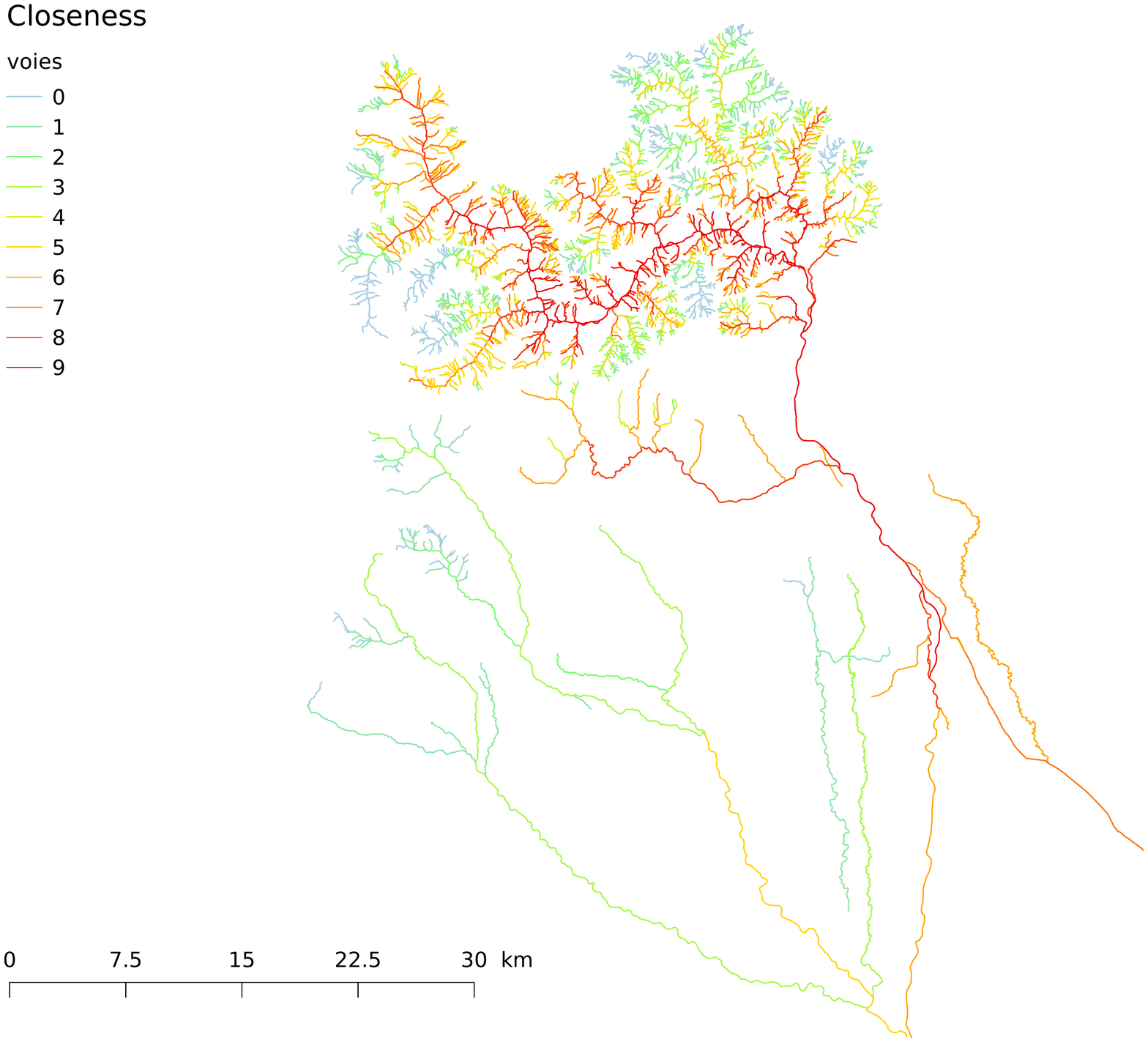}
    \caption{Calcul de l'indicateur de closeness sur le graphe extrait du réseau hydrographique présent au Nord de l'Italie. (Données OpenStreetMap)}
    \label{fig:norditalie_clo}
\end{figure}

Seuls certains indicateurs trouvent leur pertinence à travers leur application à ces réseaux de différentes natures. Les plus génériques (longueur, degré) s'appliquent aisément à tous. La closeness en revanche, n'est révélatrice que lorsqu'elle est appliquée sur des éléments qui structurent le réseau de part en part, à plusieurs échelles. Ainsi, un réseau de craquelures ne fera apparaître de hiérarchisation pertinente de ses voies que s'il admet un découpage à l'échelle globale, et que celui-ci est conservé lors de sa numérisation (suppression de l'effet \enquote{bulle de savon}). D'autre part, les voies créées sur des réseaux hydrauliques ou ferrés perdent le caractère structurant évoqué, car ces réseaux sont très peu maillés (le réseau hydrographique est en arbre). La partie du graphe ayant la meilleure proximité topologique avec le reste du graphe est donc celle formée des voies les plus longues (\enquote{tronc} de l'arbre) (figure \ref{fig:norditalie_clo}). Nous trouvons ici une limite d'application de notre méthodologie de caractérisation globale à un réseau spatial, autre qu'un réseau viaire. L'orthogonalité est un indicateur qui n'a pas de sens également, lorsqu'il est appliqué à des réseaux qui s'étendent sur plusieurs centaines de kilomètres : la précision de vectorisation des intersections à cette échelle est relative.

\FloatBarrier

Un des objets de ce travail est ainsi de synthétiser les critères inhérents aux réseaux spatiaux. Des attributs du graphe des rues que l'on retrouve d'une ville à l'autre, bien qu'elles soient construites dans des contextes très différents ; ou même de ceux qui se retrouvent entre réseaux spatiaux de natures dissemblables. Par exemple, nous avons montré dans la partie précédente que le degré moyen d'une intersection est d'environ 3, que ce soit sur un réseau viaire, à Manhattan, Nairobi ou Avignon, ou sur un réseau biologique, celui d'une feuille ou d'une gorgone. De même, l'histogramme du logarithme des longueurs des rues suit de manière répétée la forme d'une courbe gaussienne, ce qui traduit une dynamique de découpage par fractionnement successif. Dans ce mécanisme, chaque longueur est une fraction aléatoire de celle dont elle est issue, ce qui donne cette forme à la courbe de manière répétée. Ce découpage est donc indépendant du réseau viaire considéré.

Ces propriétés purement topologiques ou géométriques témoignent, pour le réseau viaire, d'une logique universelle d'appropriation d'un territoire. Elles le rendent porteur d'une structure culturelle complexe. Son filaire est en effet, quel que soit l'endroit observé, contraint par l'espace en trois dimensions dans lequel nous vivons et soumis à notre échelle d'utilisation : le nombre de tronçons raccordés à une intersection ne peut pas être trop important car c'est physiquement impossible. De même il est façonné suivant les logiques spatiales de l'esprit humain, qui se positionne volontiers dans un repère cardinal : Nord / Sud / Est / Ouest ou devant / derrière / droite / gauche. Ces contraintes physiques ou psychologiques traversent le temps et l'espace en donnant à nos villes des formes dont certaines propriétés sont persistantes, au delà des époques, des cultures et des continents.


\clearpage{\pagestyle{empty}\cleardoublepage}
\chapter{Les transmissions de la forme}
\minitoc
\markright{Les transmissions de la forme}

\FloatBarrier
\section{Un cadre de lecture} 

Pour décrire un lieu sur une durée importante, il faut construire un cadre, que l'on suppose fixe. Celui-ci peut s'articuler autour du réseau hydrographique par exemple. La création de formes par l'Homme est un acte social, d'affranchissement quant à la nature, mais qui est contraint par les limites physiques qu'elle impose (orographiques, telluriques...). La société est donc, dans sa production, soumise à la dualité de sa volonté de prendre possession de l'espace et à celle de se limiter aux lois physiques imposées par celui-ci. À travers les époques, elle a souvent été à la recherche de la forme parfaite. Et cette perfection était plus liée aux formes géométriques créées qu'à leur accord avec le lieu dans lequel elles s'inscrivent. Le seul élément naturel allié à l'idée de perfection était le soleil, dont les principaux axes de construction de certaines villes veulent retracer la course (notamment à Rome et en Chine). Le principe de régularité se retrouve dans l'antiquité à la fois dans les bâtiments mais également dans la structure viaire, imposant à la ville une structure quadrillée.

\begin{figure}[h]
    \centering
    \includegraphics[width=0.8\textwidth]{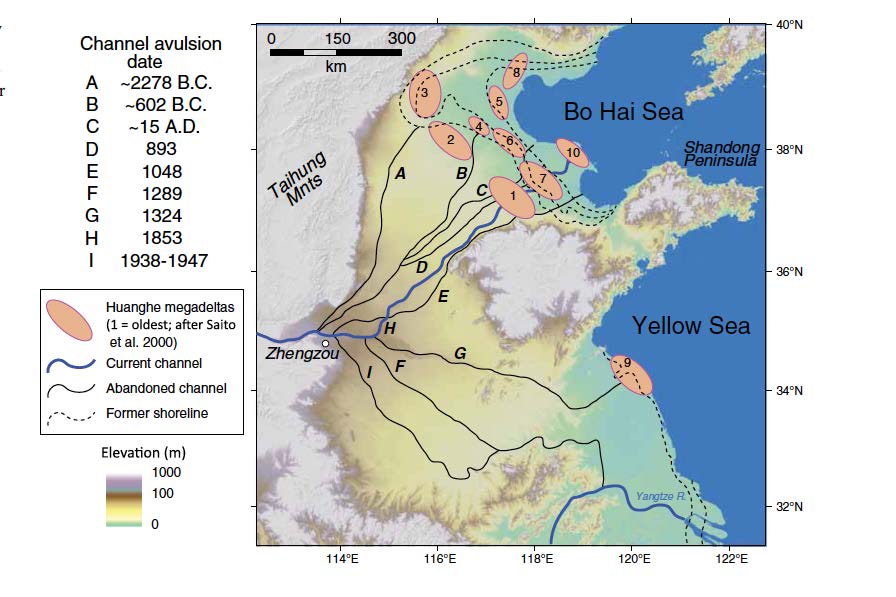}
    \caption{Différents tracés du Fleuve Jaune (Chine) au cours du temps. \\ source : http://openscienceworld.com}
    \label{fig:fleuvejaune}
\end{figure}

Observé donc, sur un long terme, le cadre que l'on considère pour la lecture d'un territoire peut varier et donc constituer un repère mouvant. Sur certains espaces, la notion du temps peut devenir cyclique. Ainsi, en Asie, le Fleuve Jaune, long de plusieurs milliers de kilomètres, s'émancipe régulièrement des digues qui contraignent son parcours pour s'étendre sur un espace grand comme la France. Les alluvions transportées redessinent entièrement le paysage (figure \ref{fig:fleuvejaune}). L'espace en est complètement transformé et reprend sa construction à partir des zones asséchées restantes. Il en est de même pour tous les paysages qui subissent une destruction du couvert végétal pour diverses raisons (période de rhexistasie). Un paysage de dunes est ainsi en constante mouvance, les barkanes (amas de sable) étant sans cesse déplacées par le vent. Cette situation s'oppose à celle de biostasie, où la végétation tient suffisamment la terre pour que le relief reste stable (à l'échelle de temps humain). Ces exemples illustrent l'impermanence des formes naturelles qu'il est nécessaire de prendre en compte pour comprendre celles créées par l'Homme. Les mécanismes de construction de l'espace prennent en leur sein le monde en mouvement. La culture fait forme, contrainte par la nature.

Les motifs parcellaires nous offrent également un tableau de lecture du temps. Des villes proches (comme Avignon et l'Ile-sur-Sorgue par exemple) ont des structures de parcelles très différentes, ce qui montre que la proximité géographique n'est pas forcément révélatrice d'une similarité morphologique. Les découpages s'organisent souvent en fonction de \textit{traces} passées sur lesquelles nous reviendrons dans cette partie. Les parcelles font l'objet de recherches très documentées, que nous ne pouvons citer exhaustivement ici \citep{juillard1953formes, noizet2013paris}. Elles structurent l'espace autour du réseau viaire. L'observation dans le temps montre qu'elles sont plus sujettes aux transformations (fusion ou division) que le réseau autour duquel elles s'organisent : moelle épinière de leur développement. Celui-ci distribue leur accès, condition première de leur existence.

Les formes constituent un héritage de dynamiques de très longue durée. Les mécanismes fonctionnent par type et par seuil. L'aboutissement de ces forces est de nature hybride, combiné des formes à l'instant $t-1$, légèrement transformées pour créer celles de l'instant $t$. Elles allient le passé et le présent, le physique et le social. Leur emprise est résiliente, donnant parfois une transformation d'un ru en rue, au travers d'une trame en perpétuelle évolution.

Cela pose la question de la distinction des formes planifiées par rapport aux formes émergentes (organiques). Elles sont souvent en conflit les unes avec les autres et ne répondent pas aux mêmes caractéristiques. Certains indicateurs permettent d'en deviner les contours. Nous avons ainsi vu dans les chapitres précédents que l'indicateur d'orthogonalité  distingue les formes maillées de celles traversantes ou encore celles de contournement. L'indicateur d'espacement nous permet de voir transparaître l'orographie entre les lignes. Autant de valeurs quantifiées qui mettent en valeur des sous-parties de graphe aux géométries particulières. Enfin, nous avons observé dans la partie précédente que plus la longueur des voies est corrélée à leur degré, plus on peut supposer une organisation globale planifiée (coefficient d’hétérogénéité). Cela permet, au travers d'un coefficient, d'avoir l'intuition de la forme globale du réseau.

\FloatBarrier
\section{La représentation du temps}

La relation entre les réseaux et le temps est complexe et s'applique à de nombreuses questions de recherches \citep{bahoken2013reseaux}. Nous nous intéressons ici aux réseaux spatialisés. Lorsque nous appliquons notre raisonnement aux villes, nous extrayons d'une structure complexe son réseau viaire. En plus d'imposer une limite dans l'espace à notre réseau, nous lui imposons également une position dans le temps : le graphe sur lequel nous travaillons, est une image à un instant $t$ précis d'un réseau en perpétuelle mutation. P. Hallot a ainsi travaillé sur les états spatio-temporels d'un objet, les relations entre ces états et leurs évolutions \citep{hallot2012identite}.

Le temps peut être considéré de différentes manières. Dans une structure productrice de données comme l'IGN, c'est un paramètre d'obsolescence. Il est nécessaire de faire disparaître le passé au profit du présent pour avoir des données les plus représentatives possible de l'existant. Cependant le moment présent appartenant continuellement au passé, à l'échelle de quelques mois pour les constructions d'ajustement viaires, c'est une mission difficile à mener.

Depuis leur naissance jusqu'au début du XXI\textsuperscript{ème} siècle, les Systèmes d'Information Géographique se sont peu préoccupés de la prise en compte de l'aspect temporel des données. Leur traitement statique prévalait pour apporter de l'information sur la situation la plus actuelle possible. Puis, les recherches menées sur l'évolution des informations géographiques dans le temps, par les sciences thématiques, ont poussé les sciences techniques à développer des outils prenant en compte cette quatrième dimension. À l'\emph{entre-deux}, les géomaticiens ont donc dû répondre aux questions de modélisation d'un phénomène continu sur des données discrètes, à l'aide de modèles plus ou moins complexes \citep{bordin2006methode}.

Les données, en plus d'être dépendantes du moment de leur saisie, le sont également de ce pourquoi elles ont été saisies. Ainsi, d'une année sur l'autre ($t$ à $t+1$), entre deux réseaux viaires d'une même emprise, les changements observés ne sont pas de manière certaine ceux qui ont été opérés sur le terrain entre les deux dates. Si les spécifications changent, et que, par exemple, les allées dans un cimetière doivent être intégrées aux données viaires, elles apparaîtront toutes à $t+1$ mais cela ne veut pas dire que certaines d'entre elles n'étaient pas présentent à $t$.

Pour suivre les changements dans les bases de données, il est nécessaire de ne pas considérer l'état instantané comme unique information mais également les liens entre les différents états. Cela implique l'existence d'un même identifiant d'un filaire à l'autre et/ou des géométries de même emprise. En effet, pour comparer le réseau viaire de différentes époques, il est nécessaire de vectoriser des cartes anciennes. Si chaque vectorisation est faite indépendamment des autres, bien que les cartes soient géoréférencées dans un premier temps, les géométries saisies seront différentes sur chaque carte et il sera très difficile de reconstituer les correspondances d'une carte à l'autre (problématique évoquée en début de chapitre 4, partie II). Ces problèmes d'appariement nécessitent une paramétrisation complexe. Ils constituent l'objet de thèses qui étudient précisément la gestion des imperfections de vectorisation \citep{dumenieu2013methode}.

Établir un lien entre des objets dans le temps implique la reconnaissance des objets d'une carte à l'autre. Dès lors, se pose la question de savoir ce qui établit l'identité d'un objet dans le temps. Les différents tronçons qui constituent le réseau des rues portent chacun un identifiant, une géométrie et plusieurs attributs. Le changement est-il défini par un changement de géométrie ou un simple changement attributaire ? Un tronçon prolongé est-il toujours le même tronçon ? Un chemin qui est goudronné est-il toujours le même objet ? La réponse à ces questions réside dans la définition de ce que l'on veut observer. Il y aurait donc une base spatio-temporelle à construire par question. Le travail de P. Bordin portant sur la conceptualisation de ces problématiques et l'élaboration d'une méthodologie pour y répondre nous apporte des éléments de réflexion. L'analyse faite ici rejoint la démarche d'une partition maintenue constante dans une approche rétrospective \citep{bordin2011vers}.

Dans notre travail, nous avons défini et construit deux bases de données \textit{panchroniques} (sur Avignon intra-muros et la partie Nord de Rotterdam). Comme nous l'avons vu (partie II, chapitre 4), notre étude porte sur la géométrie, c'est donc sur celle-ci que nous concentrons la description du changement. Nous codons de manière binaire l'existence ou non existence d'un arc sur une carte vectorisée. Afin de conserver toujours la même géométrie dans le temps nous avons opté pour une méthode de numérisation régressive. Partant du filaire actuel, les cartes anciennes sont étudiées pour ajouter (si disparition) ou supprimer (si création) un tronçon. Nous assurons ainsi l'appariement de nos données d'une année sur l'autre. Nous créons ainsi un unique fichier de formes \enquote{total}, avec l'ensemble du filaire qui a existé sur l'intervalle de temps étudié, et une base attributaire spécifiant les périodes de présence de chacun des tronçons.

Nous avons pris la décision de considérer l'emprise d'un arc comme constante d'une carte sur l'autre même si sa géométrie a été légèrement perturbée. Cependant, certains aménagements doivent être pris en compte. Ainsi, un arc prolongé est considéré comme supprimé à une date et ajouté avec une géométrie plus longue à une autre. Les identifiants dans la base totale sont donc dédoublés, il y en a un pour chaque géométrie. Ces spécifications sont liées à notre étude. Selon les problèmes de recherches, elles peuvent être adaptées différemment. Ainsi, si notre attention s'était portée sur la qualité du revêtement des routes, nous aurions considéré cet attribut comme venant s'ajouter à la géométrie et pouvant lui aussi créer des ruptures dans des arcs (goudronnés partiellement par exemple) dont les géométries seraient alors découpées. Il est évidemment possible d'augmenter comme souhaité le nombre d'attributs observés dans le temps, mais cela créera des géométries \enquote{minimales} de plus en plus réduites et augmentera le nombre d'éléments dans la base de données ainsi que le coût d'élaboration de cette base.

Cette méthodologie permet de saisir automatiquement les changements et de pouvoir les étudier en focalisant l'analyse sur l'attribut objet de la recherche. Dans la définition de ce protocole de recherche, la cohérence des données est fondamentale. Elle est assurée par la construction rétroactive de la base de données qui assure le partage de la même primitive géographique pour chaque objet, d'une année sur l'autre. Nous pouvons dès lors saisir non pas la dynamique des objets (qui fait intervenir des forces), mais leur cinématique : leur évolution \citep{bordin2006methode}. À l'identique du cinéma, nous procédons par différentes images (ou \textit{snapshot}) qui sont statiques dans l'absolu mais étroitement liées avec celles qui les précèdent et leur succèdent. L'apposition de ces \textit{snapshots} les uns à côté des autres en maintenant leur \textit{liant} (attributs sur lesquels est observé le changement) nous apporte l'information nécessaire à la description d'une évolution spatiale à travers le temps. Cette première quantification dessine les prémices d'une analyse plus approfondie de la morphogenèse viaire.

\begin{figure}[h]
    \centering
	\includegraphics[width=0.7\textwidth]{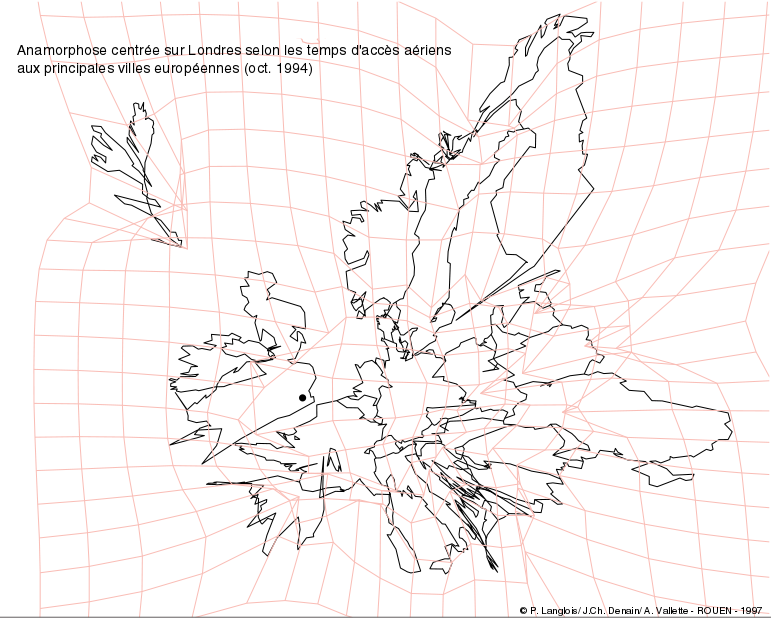}
	\caption{Carte en anamorphose centrée sur Londre, cartographiée selon les temps de parcours aériens vers les principales villes européennes. \\ source : \citep{langlois1996cartographie}}
	\label{fig:anamorphose}
\end{figure}

Il est également possible d'imaginer des graphes où la dimension spatiale disparaît au profit de la dimension temporelle. C'est le cas de la logique d'anamorphose par exemple (figure \ref{fig:anamorphose}). Le poids donné aux arcs n'est plus une distance métrique mais la durée de parcours associée à cette distance métrique (distance temporelle). La longueur des arcs peut être représentée suivant cette valuation qui déforme l'information géographique pour faire ressortir un autre type de caractérisation \citep{langlois1996cartographie}.

\FloatBarrier
\section{La vie d'un objet géographique}

P. Bordin pose dans son travail la question de l'\textit{objet suffisamment lui-même}. Cela peut s'interpréter comme une problématique portant sur la persistance du cycle de vie d'un objet géographique. Quand peut-on estimer qu'il apparaît ou qu'il disparaît ? Une ruine est-elle toujours le même bâtiment que celui décrit précédemment comme étant une maison ? Un objet géographique est défini par sa géométrie (information quantitative) et par ses attributs (informations qualitatives). Dans le cas de la ruine, l'emprise (et donc la géométrie) n'est pas modifiée mais l'occupation sera différente. Où se situe donc le seuil où le bâtiment ne sera plus considéré comme tel ? Pour répondre à ces questions il est nécessaire de fixer des limites plus ou moins arbitraires. Pour l'IGN, par exemple, si un bâtiment ne possède que deux murs ou moins il sera exclu de la base de données qui le référence (\copyright BDTOPO).

Lorsqu'il s'agit du réseau viaire, nous pouvons parfois retracer la naissance d'une rue à partir d'un chemin, qui, à force d'être emprunté, gagne de l'importance dans les itinéraires piétons (dans un premier temps). Son utilisation va petit à petit transformer son statut de chemin de terre à celui de route empierrée puis goudronnée afin de permettre d'autres types d'utilisation. Les limites de parcelles agricoles peuvent par exemple suivre cette évolution. Il est possible également de trouver des traces de centuriations romaines, héritage passé inscrit dans l'inconscient du paysage, que certaines de nos routes suivent comme une structure pré-établie servant de cadre plus ou moins avoué aux futurs développements. 

P. Bordin et M. Watteaux étudièrent la question en lui donnant le nom de \textit{trace} \citep{bordin2014observer}. La trace a en effet trois définitions précises données par le dictionnaire (Hachette 2006). C'est \enquote{une suite de marques, d'empreintes laissées par le passage d'un homme, d'un animal ou d'une chose} ; ce qui correspond dans le temps aux cheminements successifs sur un même parcours dans notre problématique. C'est également \enquote{une marque laissée par une action, un événement passé} ; la trace géographique est en effet emblématique d'une culture forte appliquée sur un territoire à une époque, à l'image de l'occupation romaine. C'est enfin en géométrie \enquote{le lieu d'intersection (d'une droite, d'un plan) avec un plan de projection}. Cette dernière définition s'applique moins bien à notre cas d'étude même si cela peut métaphoriquement se rapprocher de la projection d'un inconscient géométrique sur la surface de notre territoire. Cette notion de trace a été utilisée et expliquée dans le livre collectif de notre équipe de recherche, dont l'élaboration a été encadrée par Clément-Noël Douady : \textit{De la Trace à la Trame} \citep{douady2014trace}. Cet ouvrage n'a pas été construit sur le modèle d'un livre scientifique, sa vocation étant plutôt de donner, à un instant fixé, un aperçu des questions et points de vues au sein d'une équipe de recherche pluridisciplinaire.

Le concept d'une emprise géographique maintenue à travers le temps, qui peut être effacée puis ressurgir dans de nouvelles constructions, se retrouve dans plusieurs disciplines. Les archéo-géographes établissent une distinction entre flux (itinéraires), tracés et modelés. C'est dans la notion de  \textit{modelé} que l'on retrouve les différents aspects qu’une voie peut prendre avec le temps : chemin, limite de parcelle, limite de commune, etc. Une distinction est donc faite entre traces et formes, la trace étant, dans cette discipline, un aménagement matériel attesté lors d'une fouille archéologique, qui ne connaîtra pas forcément de transmission dans le temps. Elle est donc observée à un instant $t$ et est synonyme de vestige. Gérard Chouquer et Sandrine Robert ont travaillé sur ces questions afin d’étayer ces concepts dans leur discipline \citep{chouquer2000etude, robert2003comment, robert2006resilience}.

\begin{figure}[h]
    \centering
	\includegraphics[width=0.7\textwidth]{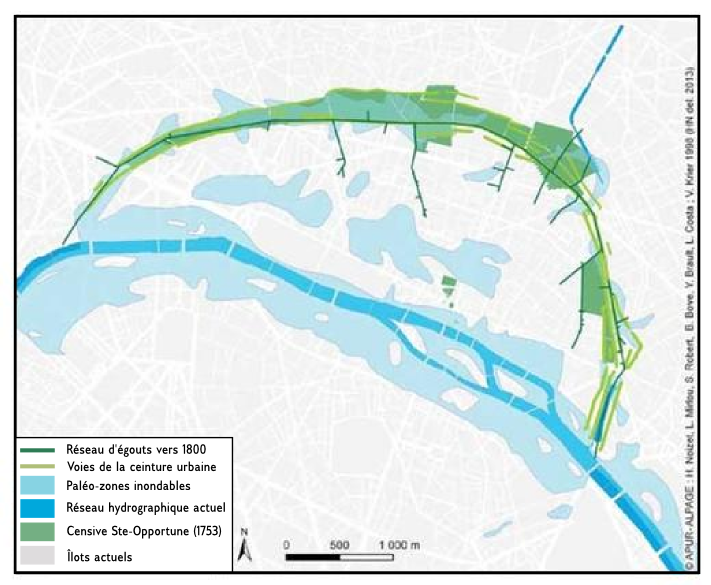}
	\caption{Illustration des différentes utilisations du paléo-chenal de la Seine à Paris. \\ source : \citep{noizet2013resilience}}
	\label{fig:paleochenal}
\end{figure}

En étudiant le territoire de la ville de Paris, par exemple, une structure très ancienne a été mise en évidence par le groupe travaillant sur le projet ALPAGE (AnaLyse diachronique de l'espace urbain PArisien : approche GEomatique) \citep{noizet2008alpage} : un paléo-chenal qui contournait le quartier du marais (nommé ainsi car c'en était un) sur la rive droite de la Seine (figure \ref{fig:paleochenal}). La forme physique de ce bras hydrographique a traversé les années \citep{noizet2013resilience}. Il est devenu une zone marécageuse au néolithique, puis fut drainé et mis en culture au Moyen-Âge (ceinture maraîchère). Il fut transformé en égout durant l'époque moderne (autour de 1740) pour devenir ensuite l'emprise des premières murailles de la ville. C'est aujourd'hui une succession de boulevards constituant la petite ceinture du Paris actuel, sur lesquels viennent s'aligner les parcelles (depuis le début du XIXème, plan Vasserot). L'inconscient géométrique d'un territoire transcende donc la nature de la forme observée. La structure d'un réseau viaire a donc pu être celle d'un réseau de déplacements comme celle d'un réseau hydrographique. De la même manière, avec des qualités plus proches, l'emprise d'un lotissement a pu être celle d'un château, d'une abbaye ou d'un cimetière. Le territoire est donc marqué par les époques qu'il traverse. Celles-ci le façonnent de la même manière qu'une personne est façonnée par les événements de sa vie. Par un procédé affirmé (conscient) ou plus subtil, les choix réalisés ne le sont pas sans lien avec l'histoire du lieu. En suivant ce raisonnement, nous pouvons nous demander s'il est possible de considérer qu'un objet géographique a une durée de vie limitée dans le temps, ou s'il s'agit plutôt d'une mise entre parenthèses d'un état entre deux points temporels, laissant toujours la porte ouverte à une réutilisation possible de l'emprise.

\FloatBarrier
\section{La dualité du plein et du vide} 

Dans la lecture des villes statiques, nous observons des éléments de structure forts, caractéristiques d'une époque et d'une culture. Ainsi, dans les villes médiévales (comme Brive-la-Gaillarde ou Avignon) nous relevons des lignes d'enclosure, qui définissent un \enquote{intérieur} et un \enquote{extérieur}. Ces lignes sont propices aux ruptures géométriques. Elles peuvent définir, par exemple, une différence de trame entre un centre ancien et un développement hors les murs plus récent. Ces différentes logiques géométriques se trouvent réunies par des lignes radiales permettant la circulation de l'une à l'autre puis vers un extérieur plus lointain \citep{watteaux2003plan}. Ce sont des structures schématiques héritées d'époques distinctes qui, comme nous l'avons vu, se trouvent souvent ré-empruntées.

\begin{figure}[h]
    \centering
    \begin{subfigure}[t]{.45\linewidth}
        \includegraphics[width=\textwidth]{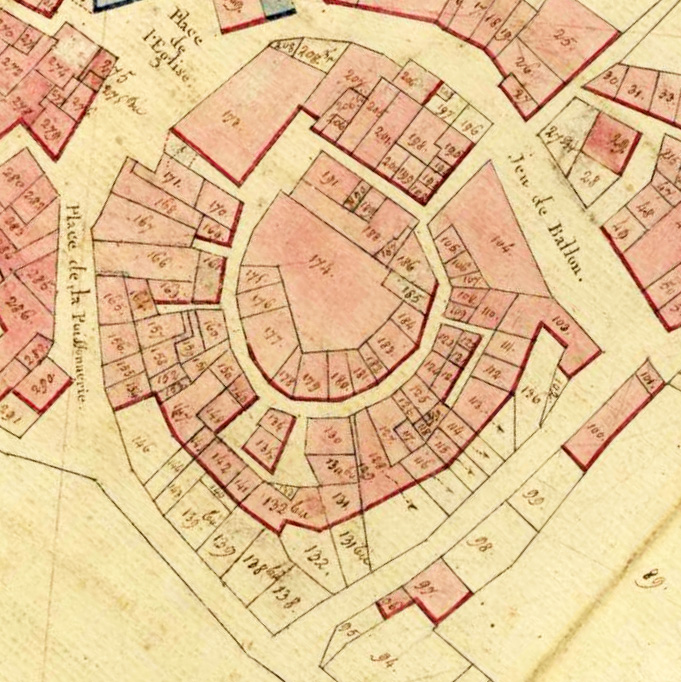}
        \caption{Cadastre actuel. source : \copyright{Geoportail}}
    \end{subfigure}
	~
    \begin{subfigure}[t]{.45\linewidth}
        \includegraphics[width=\textwidth]{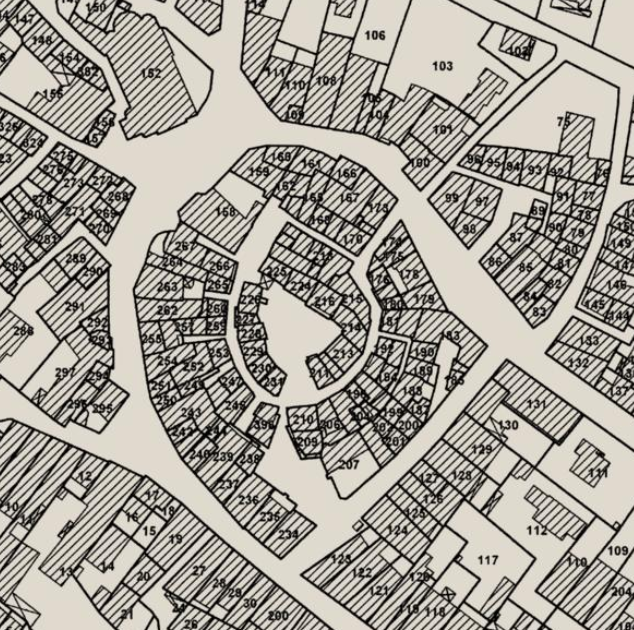}
        \caption{Cadastre Napoléonien (1815). source : Archives départementales de l'Hérault}
    \end{subfigure}
    \caption{Transmission de la forme au centre de la Commune de Fabrègues (Languedoc-Roussillon) et basculement du plein vers le vide. }
    \label{fig:fabregues}
\end{figure}

Il est commun d'observer des \textit{fermetures} qui se transforment en \textit{ouvertures} avec le temps \citep{cnd2014deconstruire}. Ainsi, les murailles des villes médiévales se transforment pour la plupart en boulevards (phénomène observé à Paris ou à Brive-la-Gaillarde par exemples). Si les remparts sont conservés, des boulevards sont créés à l'intérieur et à l'extérieur de ceux-ci (phénomène observé à Avignon). Dans une logique de coupure et de protection lors de leurs constructions, elles deviennent un support pour des structures facilitant la circulation. Pour les bâtiments, il est aussi possible d'observer d'anciennes emprises de châteaux qui se transforment en places publiques (comme par exemple à  Fabrègues au Sud-Ouest de Montpellier, figure \ref{fig:fabregues}), ou bien des amphithéâtres effacés au profit d'un espace vide et central (comme à Martina Franca dans les Pouilles, Italie du Sud). 

Les formes sont conservées et transmises, elles peuvent se retourner (du plein vers le vide), et permettre ainsi à différentes périodes de tirer partie des mêmes tracés. Cette question d'inversion est complexe et demanderait des recherches plus poussées. Elle illustre la capacité de la forme à perdurer. Ainsi, la transmission implique parfois un renversement de la fonctionnalité du lieu : de lieu de séparation, de protection (muraille, château), il peut être transformé en lieu d'union, de circulation (boulevard, place). La dualité entre le plein et le vide, à la fois opposés, complémentaires et sources l'un de l'autre est présente jusque dans la toponymie. Ainsi, le mot \textit{boulevard} vient du vieux néerlandais \textit{bolwerk} qui signifie \enquote{digue}, \enquote{bastion}, ou \enquote{rempart}.

\clearpage{\pagestyle{empty}\cleardoublepage}
\chapter{Ouverture}
\minitoc
\markright{Ouverture}

Les structures sont au cœur de notre travail, elles s'apposent au territoire qu'elles découpent selon différentes logiques. Certains indicateurs nous proposent un point de vue sur le réseau viaire, permettant de mettre en avant ses cohérences spatiales. Nous soumettons dans ce chapitre quatre pistes de lecture de la ville, à travers trois indicateurs et une distance : le degré, la closeness, l'espacement et la distance topologique. Ce sont des prémices de recherches à approfondir.

\FloatBarrier
\section{Indicateur de degré et structures maillées}

Nous avons vu dans l'ensemble des exemples précédents que l'indicateur de degré établit une structure maillée avec l'ensemble des voies de degré important. Cette \textit{grille} créée serait également obtenue en appliquant les indicateurs corrélés au degré sur les voies du graphe (longueur, betweenness ou utilisation). Cela signifie que ce maillage est également celui qui participe le plus à l'ensemble des chemins les plus courts calculés sur le graphe. Nous pouvons donc avoir l'intuition de son lien avec l'utilisation réelle du réseau.

En créant la voie par association d'arcs, nous construisons un hypergraphe qui nous donne une structure stable de lecture de l'espace. Nous pouvons poursuivre ce raisonnement en construisant, à partir de cet hypergraphe, un nouveau réseau dont les éléments seraient définis en associant les voies de degré proche entre elles. Cette nouvelle structure maillée pourrait elle-même servir aux calculs de nouvelles distances topologiques et autres indicateurs. Nous pourrions créer ainsi un emboîtement de sous-parties de l'hypergraphe des voies. L'enchaînement de ces sous-parties reconstituerait les générations hiérarchiques des différentes échelles de lecture de l'espace urbain.

En théorie des graphes, il existe la notion de \textit{k-core}. Établir le \textit{k-core} d'un graphe consiste à n'en retenir que les sommets dont le degré est supérieur à un degré \textit{k} fixé. Dans notre cas, appliqué au \textit{line graph} des voies, cela consiste à ne conserver que les voies au dessus d'un certain degré. Avec différentes valeurs de \textit{k} fixées, cela nous permet d'établir les différentes générations énoncées plus haut.

Pour choisir les seuils de degré à fixer pour établir les différents \textit{k-core} des graphes de nos villes, nous nous référons à la distribution du degré des voies dans chacun des graphes. Nous établissons notre choix sur l'écart type des courbes. Ainsi, par exemple, pour le centre de Téhéran, les seuils fixés sont à $k=3$ et $k=28$ et pour Kyoto $k=5$ et $k=50$. Dans la représentation cartographique que nous faisons de ces graphes nous choisissons de créer une partition de l'espace. Ainsi, une voie faisant partie du graphe n'admettant que des degrés au dessus de 50 pour Kyoto ne fera pas partie des autres graphes. Cela diffère de la notion de \textit{k-core} où les voies se recouvriraient : une voie appartenant à un \textit{k-core} appartiendrait aussi au \textit{(k+1)-core}. Sur les deux villes que nous prenons en exemple dans ce paragraphe, trois niveaux de structures se détachent, de la plus maillée, globale, à la structure locale dont la géométrie est moins régulière.

\begin{figure}[h]
    \centering
    \includegraphics[width=0.8\textwidth]{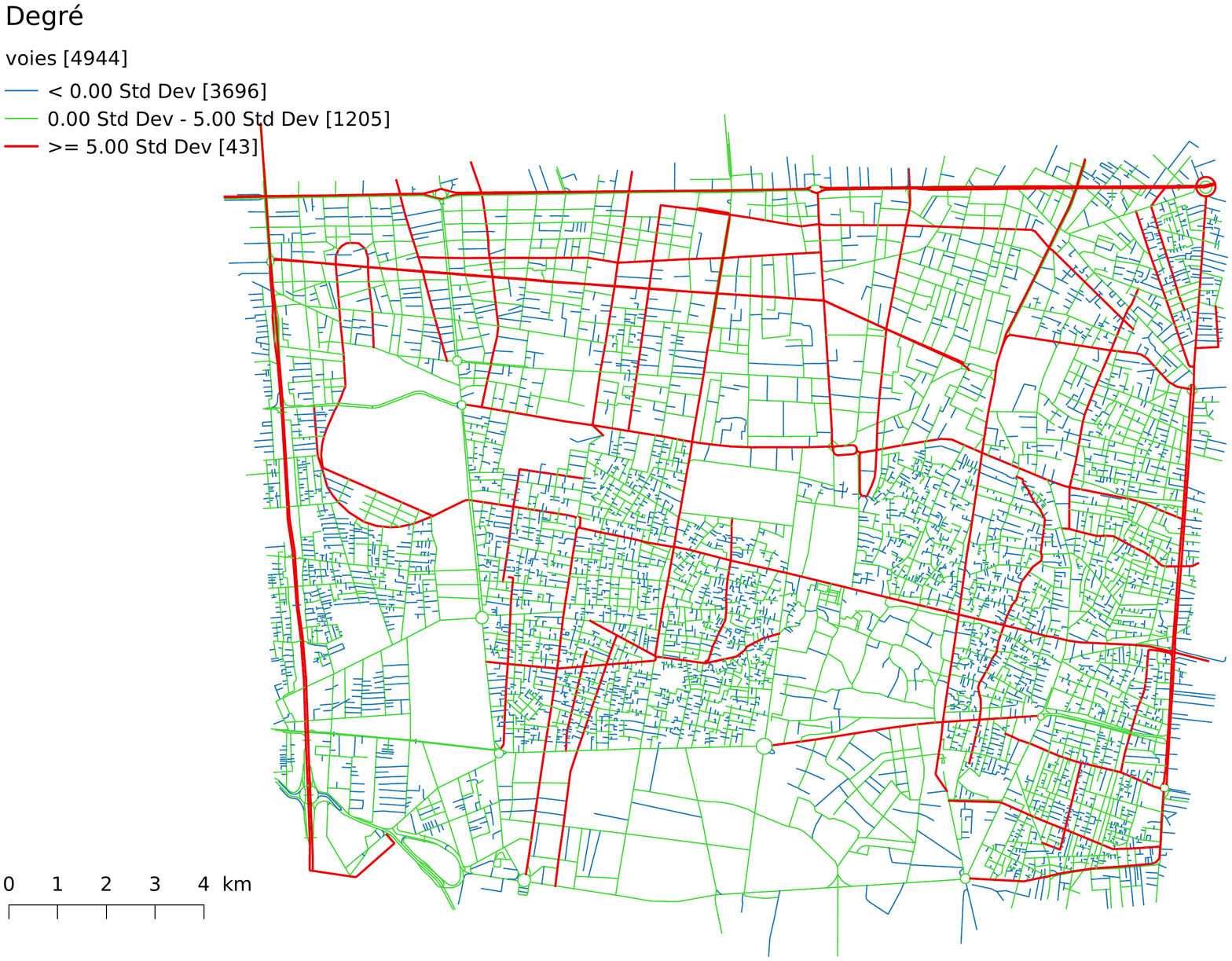}
    \caption{Représentation de trois générations de graphes sur le centre de Téhéran, établies selon les degrés des voies. La première classe regroupe des voies de degrés 1 et 2, le deuxième de degrés 3 à 28 et la troisième de 29 à 81.}
    \label{fig:teh_kcore}
\end{figure}

\begin{figure}[h]
    \centering
    \includegraphics[width=\textwidth]{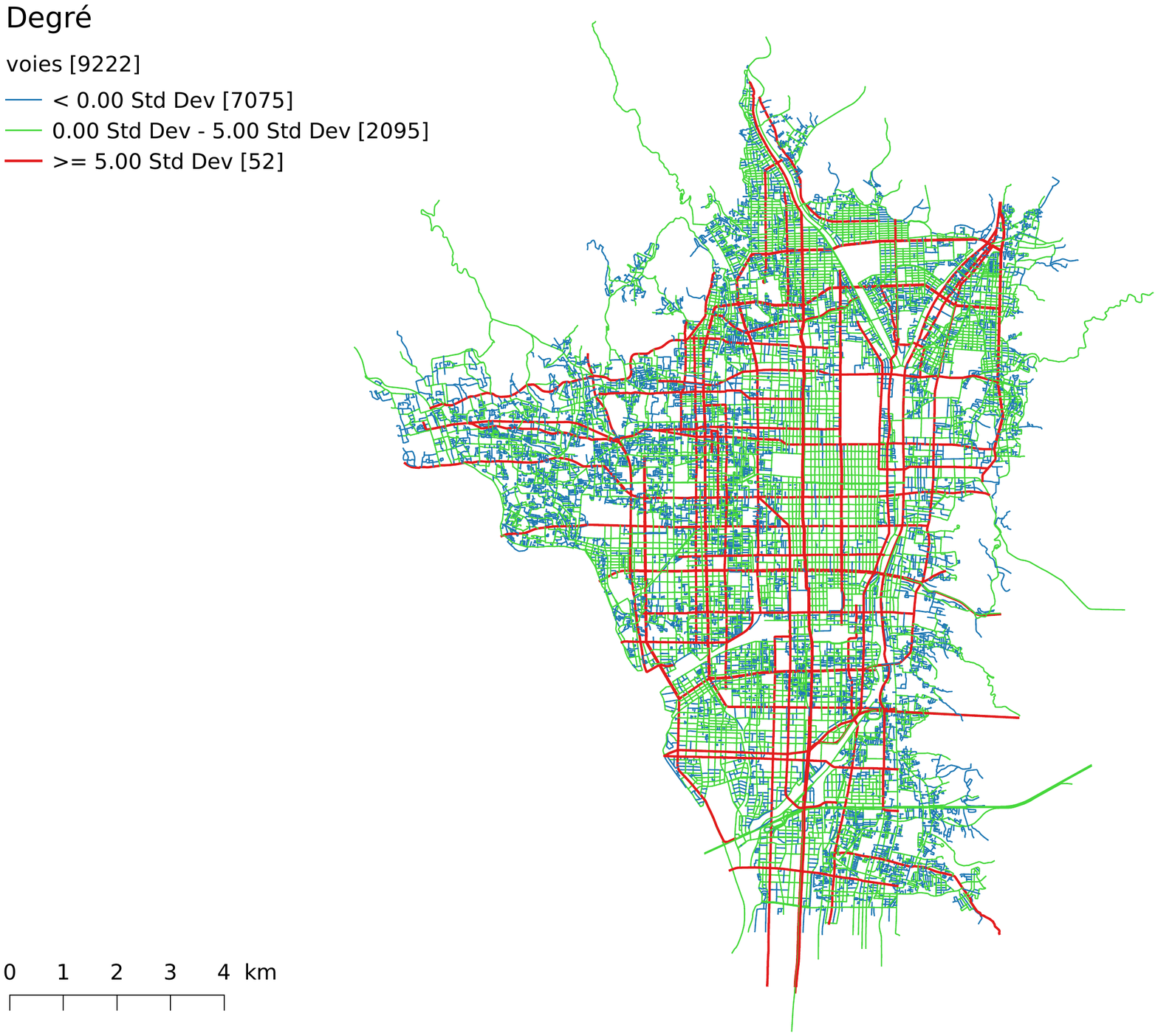}
    \caption{Représentation de trois générations de graphes sur le centre de Kyoto, établies selon les degrés des voies. La première classe regroupe des voies de degrés 1 à 4, le deuxième de degrés 5 à 49 et la troisième de 50 à 251.}
    \label{fig:kyoto_kcore}
\end{figure}

L'analyse fine de ces différents sous-graphes et de leurs liens avec l'utilisation du réseau nécessiterait d'étendre ce travail de recherche en circonscrivant nos terrains d'études et en approfondissant la collecte de données de flux et d'habitudes piétonnes. Cela fait l'objet d'une seconde thèse, en anthropologie urbaine, développée par E. Degouys. Celle-ci a entre autres terrains d'étude les villes d'Avignon et de Kyoto. Ces travaux viennent en complément de ceux déjà réalisés sur l'étude du lien existant entre structures urbaines et transports \citep{foltete2002structures, banos2011geographical}.

\FloatBarrier
\section{Indicateur de closeness et efficacité des quartiers}

Dans la deuxième partie, nous comparons les propriétés d'un panel de quarante réseaux spatiaux. Nous avons calculé différents indicateurs sur ceux-ci, et notamment la moyenne de l'indicateur de closeness sur les voies. Nous avons alors expliqué la capacité de cet indicateur à décrire l'\textit{efficacité} d'un réseau. À l'échelle à laquelle nous avons fait nos calculs, les réseaux mélangeaient plusieurs types de géométries et ne nous permettaient pas de qualifier précisément l'efficacité des différentes trames.

Nous pouvons reporter nos calculs sur des graphes plus restreints, au sein, par exemple, d'une même ville. Il est ainsi possible de qualifier les proximités topologiques de ces réseaux et de les comparer à celle d'un réseau \enquote{idéal} (moyenne d'indicateur de closeness égal à 1).

\begin{figure}[h]
    \centering
    \begin{subfigure}[t]{.3\linewidth}
        \includegraphics[width=\textwidth]{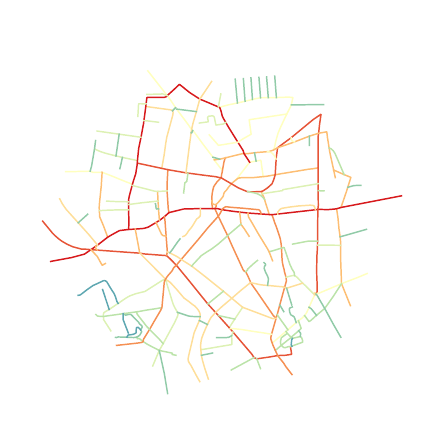}
        \caption{Quartier de Belleville. \\Efficacité : 0,3302.}
    \end{subfigure}
	~
    \begin{subfigure}[t]{.3\linewidth}
        \includegraphics[width=\textwidth]{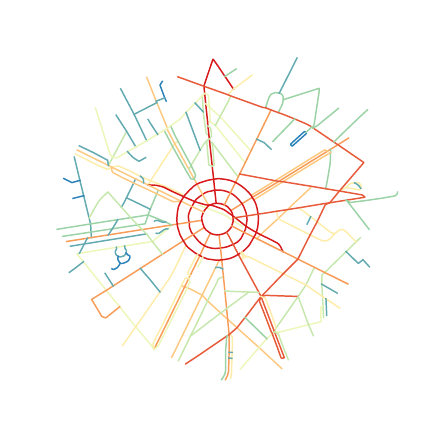}
        \caption{Quartier de l'Étoile. \\Efficacité : 0,3424.}
    \end{subfigure}
    ~
    \begin{subfigure}[t]{.3\linewidth}
        \includegraphics[width=\textwidth]{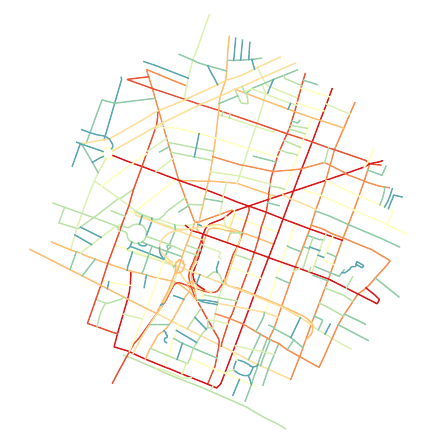}
        \caption{Quartier des Halles. \\Efficacité : 0,3476.}
    \end{subfigure}
    
    \begin{subfigure}[t]{.6\linewidth}
        \includegraphics[width=\textwidth]{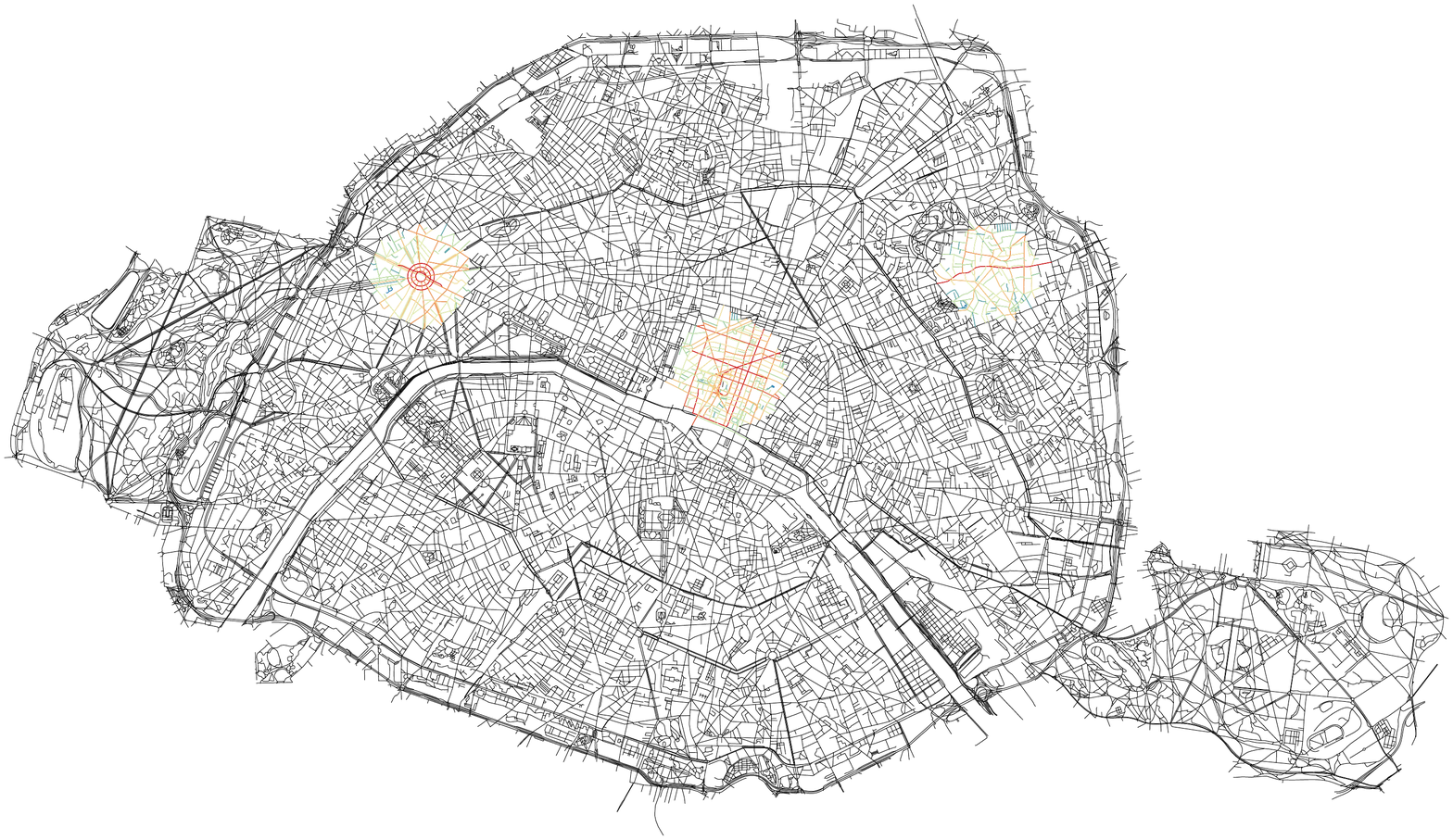}
    \caption{Positionnement des quartiers sur le graphe de Paris.}
    \end{subfigure}
    
    \caption{Calculs d'accessibilité indépendant sur les voies de 3 échantillons de la ville de Paris.
	 \\Échantillons de surface $\sim$ 0,42 $km^2$.    
     \\ source : cartes réalisées par C. Bresch}
    \label{fig:paris_eff_CB}
\end{figure}

\begin{figure}[h]
    \centering
    \begin{subfigure}[t]{.3\linewidth}
        \includegraphics[width=\textwidth]{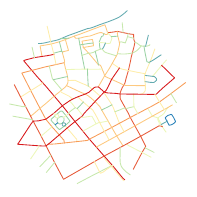}
        \caption{Quartier extrait du centre-ville. \\Efficacité : 0,3983.}
    \end{subfigure}
	~
    \begin{subfigure}[t]{.3\linewidth}
        \includegraphics[width=\textwidth]{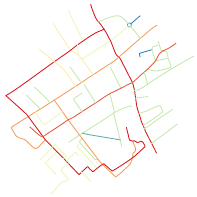}
        \caption{Quartier extrait du grand ensemble de Villeneuve (entre les communes de Grenoble et d'Échirolles). \\Efficacité : 0,2443.}
    \end{subfigure}
    ~
    \begin{subfigure}[t]{.3\linewidth}
        \includegraphics[width=\textwidth]{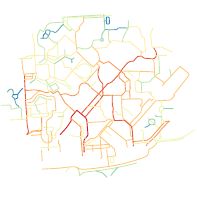}
        \caption{Quartier extrait de la commune de Meylan. \\Efficacité : 0,4319.}
    \end{subfigure}
    
    \begin{subfigure}[t]{.6\linewidth}
        \includegraphics[width=\textwidth]{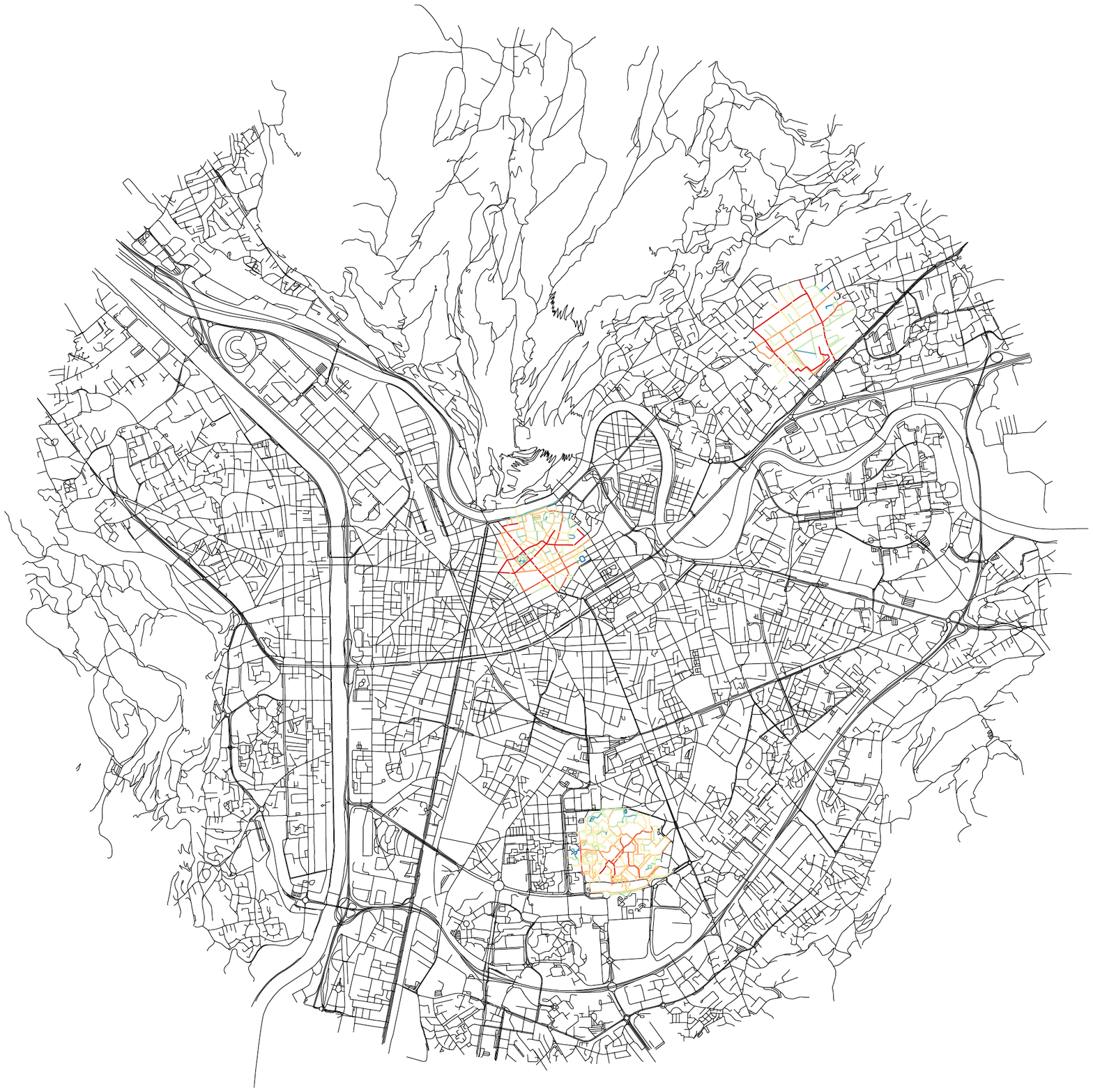}
    \caption{Positionnement des quartiers sur le graphe de Grenoble.}
    \end{subfigure}
    
    \caption{Calculs d'accessibilité indépendant sur les voies de 3 échantillons de la ville de Grenoble. 
	\\Échantillons de surface $\sim$ 0,64 $km^2$.      
    \\ source : cartes réalisées par C. Bresch}
    \label{fig:grenoble_eff_CB}
\end{figure}

C'est le raisonnement qu'a suivi Clément Bresch, stagiaire au sein de l'équipe MorphoCity. Il  a ainsi extrait trois quartiers différents des graphes viaires de Paris et de Grenoble (figures \ref{fig:paris_eff_CB} et \ref{fig:grenoble_eff_CB}). En quantifiant l'efficacité de ces quartiers, il a obtenu des résultats contrastés. Ainsi, à Paris, les quartiers de Belleville, de l'Étoile et des Halles ont tous les trois une moyenne de proximité topologique avec l'échantillon de leur graphe similaire (autour de 0,34). Nous pouvons donc en déduire une logique comparable de connexion entre voies dans ces trois quartiers. En revanche, à Grenoble, nous remarquons que le graphe extrait du réseau viaire du grand ensemble de Villeuneuve a une efficacité beaucoup moins importante que celle du centre de Grenoble ou de la commune de Meylan. Ces deux dernières sont comparables, il s'agit en effet dans les deux cas de centres urbains.

Nous pouvons présumer des possibilités de recherches offertes par ces caractérisations. Nous pourrions grâce à celles-ci tenter de comprendre les logiques géométriques plus ou moins favorables au développement d'un quartier. C'est une des pistes qui nous permettraient de mieux comprendre, par exemple, l'enclavement des Zones Urbaines Sensibles \citep{cristofol2013measuring}. Nous pourrions également tester des transformations du tissu pour améliorer son accessibilité, en utilisant la méthodologie décrite au chapitre 4 de la partie II.

\FloatBarrier
\section{Indicateur d'espacement et logiques de déplacement}

Grâce aux indicateurs que nous développons, nous déterminons une hiérarchisation des voies, selon des critères différents, faisant apparaître des structures globales mais également des effets de localité. L'indicateur d'espacement, en particulier, nous permet de faire ressortir différents types de réseaux, selon leur densité. Ainsi, sur le réseau élargi autour de Paris, selon les valeurs de l'indicateur sélectionnées, des trames très différentes sont mises en valeur.

\begin{figure}[h]
\centering
        \includegraphics[width=0.8\textwidth]{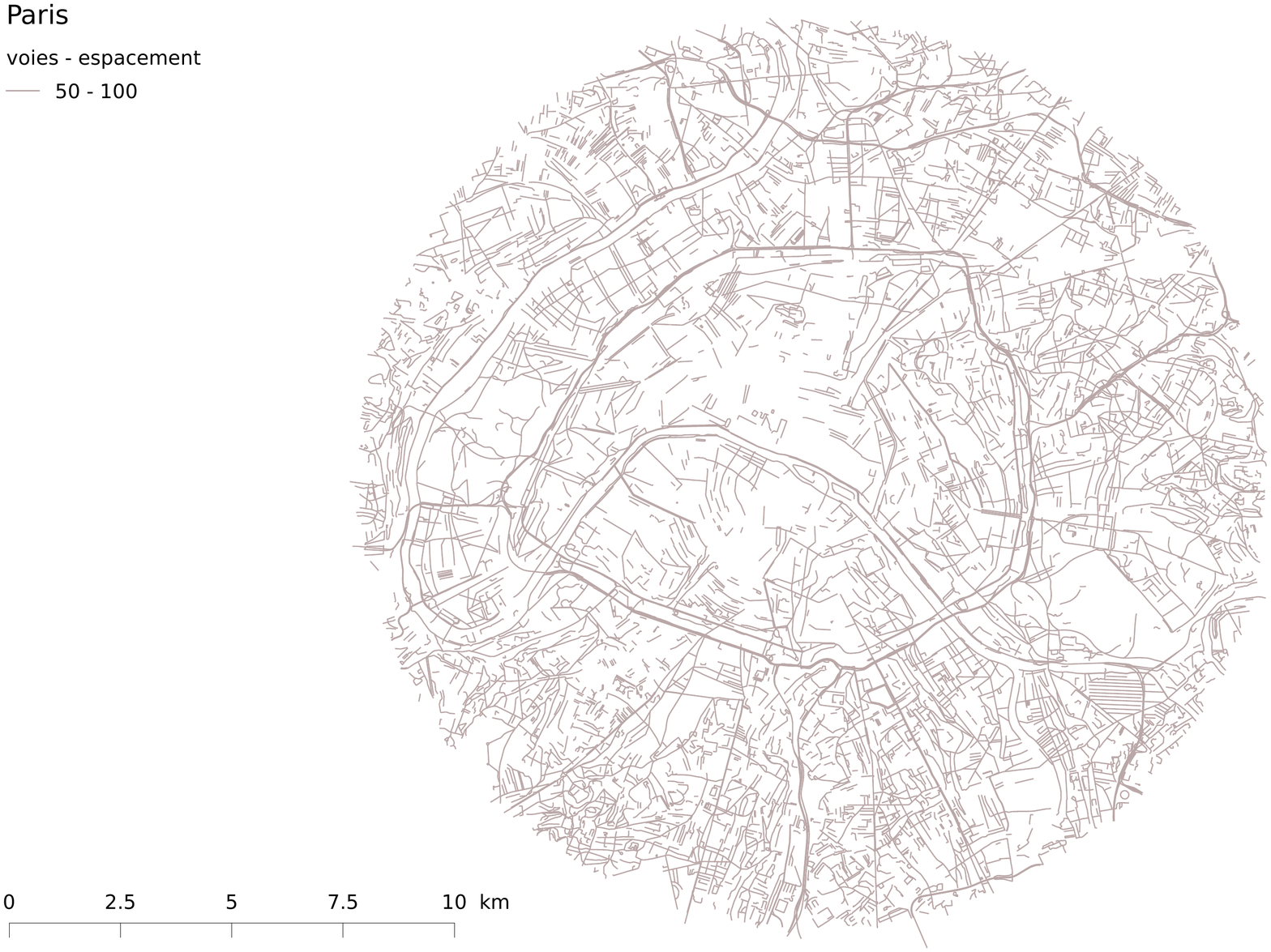}
        \caption{Voies - espacement entre 50 et 100 (mètres).}
        \label{fig:paris_esp_50_100}
\end{figure}

\begin{figure}[h]
\centering
        \includegraphics[width=0.8\textwidth]{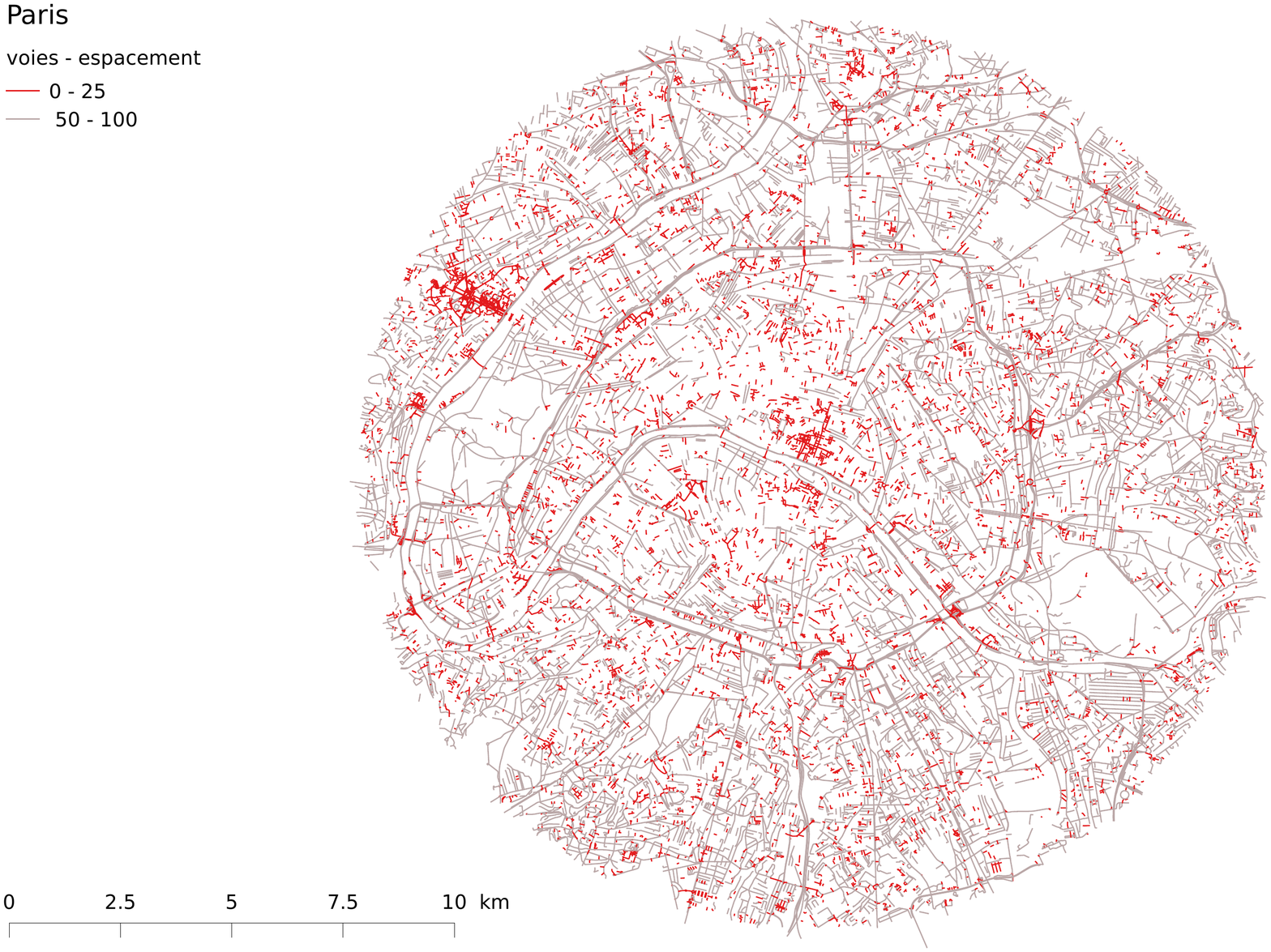}
        \caption{Voies - ajout à\ref{fig:paris_esp_50_100} de l'espacement entre 0 et 25 (mètres).}
        \label{fig:paris_esp_0_50}
\end{figure}

\begin{figure}[h]
\centering
        \includegraphics[width=0.8\textwidth]{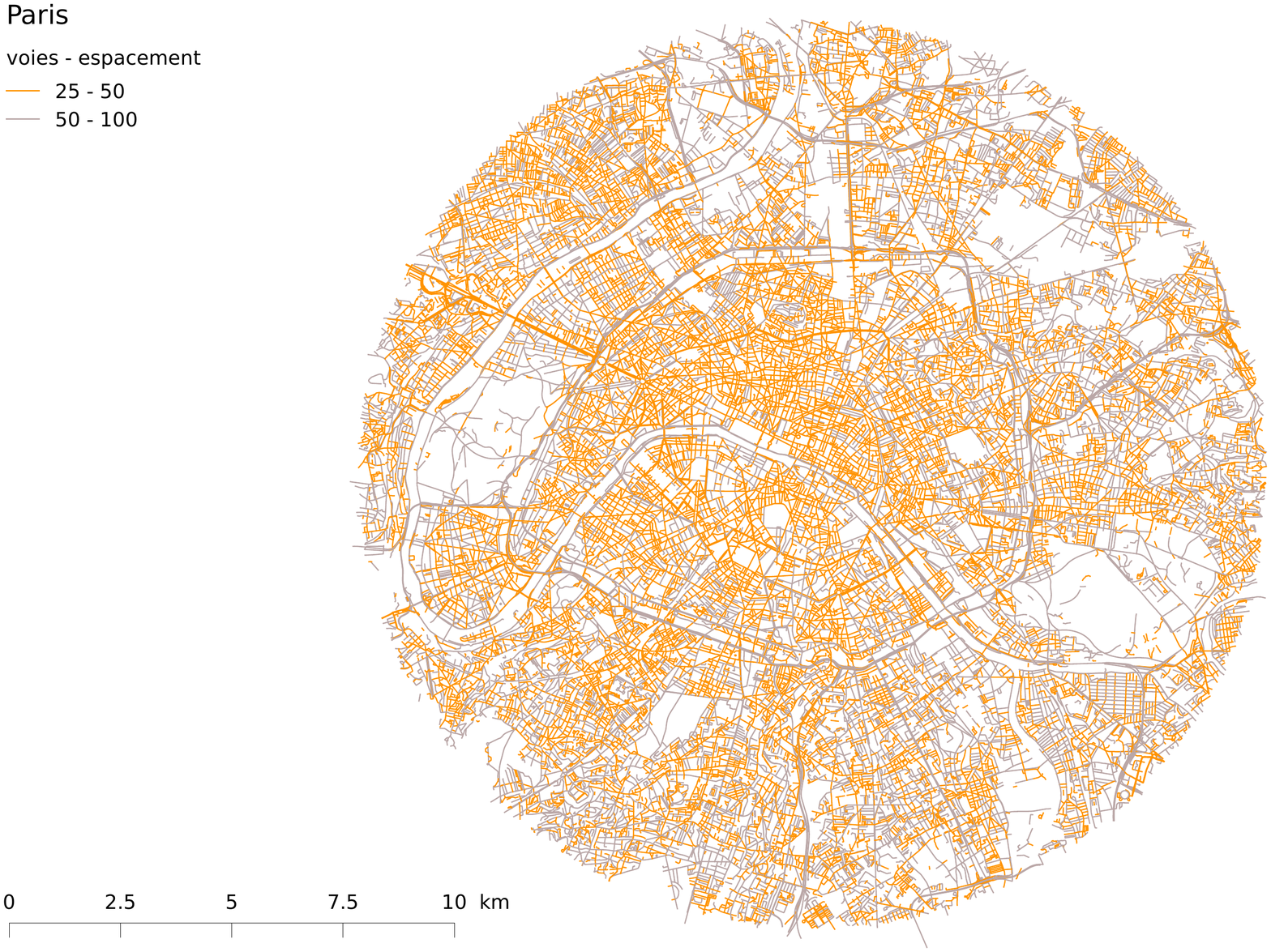}
        \caption{Voies - ajout à\ref{fig:paris_esp_50_100} de l'espacement entre 25 et 50 (mètres).}
        \label{fig:paris_esp_25_50}
\end{figure}

\begin{figure}[h]
\centering
        \includegraphics[width=0.8\textwidth]{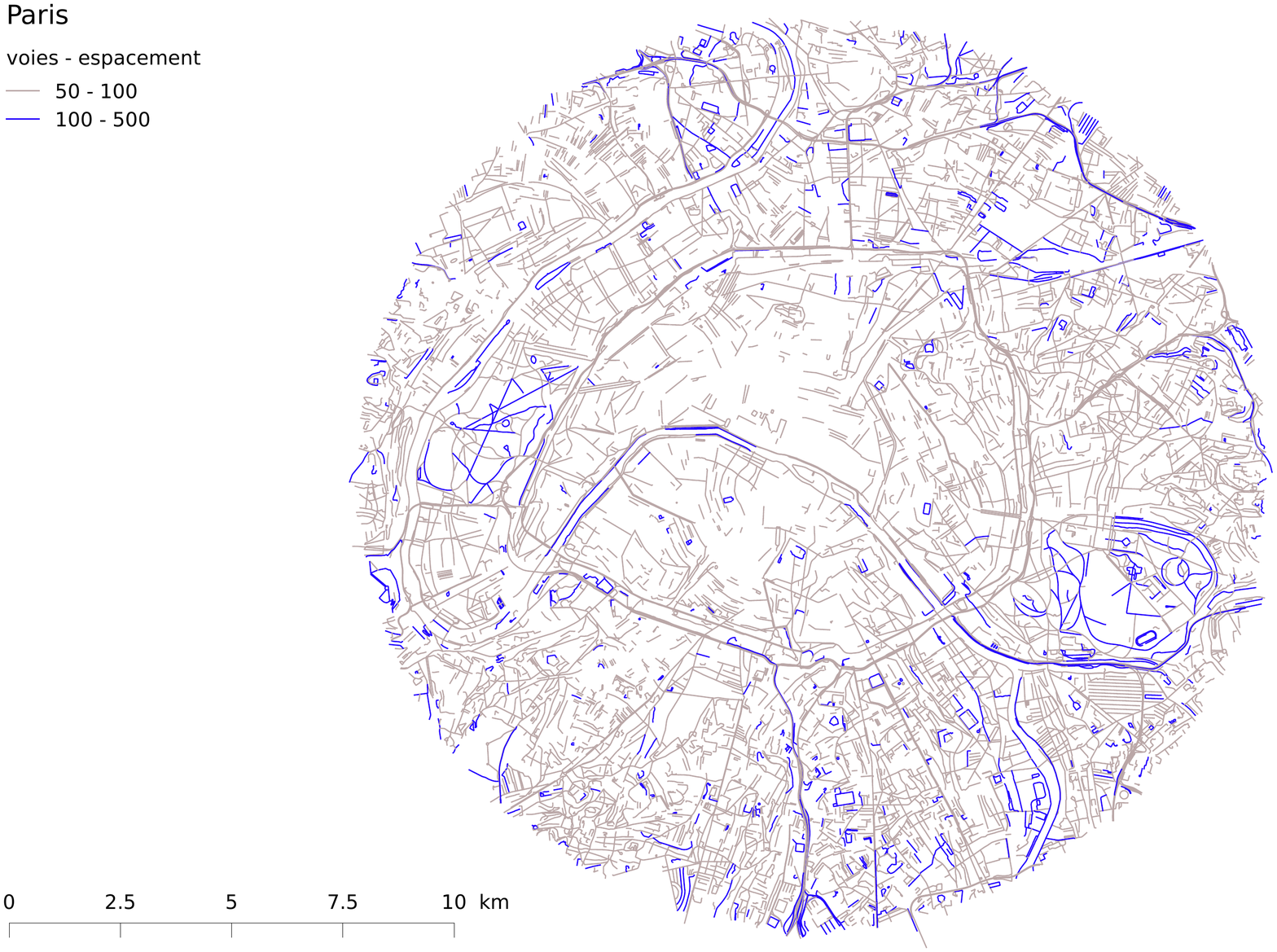}
        \caption{Voies - ajout à\ref{fig:paris_esp_50_100} de l'espacement entre 100 et 500 (mètres).}
        \label{fig:paris_esp_100_500} 
\end{figure}

L'espacement appliqué aux voies correspond à la distance moyenne entre les intersections de celles-ci. Sur Paris, nous étudions différents intervalles de valeurs. Nous ne retenons dans un premier temps que les voies dont les connexions sont espacées en moyenne de 50 à 100 mètres (figure \ref{fig:paris_esp_50_100}). Nous remarquons que celles-ci nous donnent les principales lignes du territoire. Elles dessinent le contexte hydrographique et la ceinture périphérique de la capitale. Lorsque nous ajoutons à ces lignes \textit{de repère} les voies espacées de 0 à 50 mètres, nous faisons ressortir les zones de très forte densité (figure \ref{fig:paris_esp_0_50}). Ainsi, ressort particulièrement le centre ancien de la capitale (les Halles) et le pôle de La Défense. Une toute autre trame est mise en valeur lorsque nous ajoutons les voies dont les connexions sont espacées de 25 à 50 mètres (figure \ref{fig:paris_esp_25_50}). Nous observons dans ce cas des géométries très maillées qui traversent le territoire de part en part. Enfin, en ajoutant les voies dont les connexions sont les plus espacées (entre 100 et 500 mètres), ce sont principalement les voies à l'intérieur des bois, des parcs ou au bord du fleuve qui ressortent (figure \ref{fig:paris_esp_100_500}).

Les quatre cartes nous donnent donc un aperçu de quatre facettes très différentes du territoire. Elles identifient différents modes de circulation au sein de celui-ci. Cette première analyse demande à être développée, notamment pour comparer les différents seuils de lecture selon les tissus étudiés. Ces recherches pourront être complétées par des études de flux pour établir des corrélations (ou anti-corrélations) avec les pistes de lecture apportées par ces cartes.

\FloatBarrier
\section{Distances topologiques : horizon d'une voie}

Les distances topologiques entre voies permettent de déterminer l'accessibilité topologique de chaque objet au sein du réseau. Si l'on se positionne à partir d'un objet, et que l'on observe la distribution des distances topologiques entre cet objet et l'ensemble des autres du graphe, nous obtenons son \textit{horizon}.

Nous avons tracé les \textit{horizons} de trois voies, choisies selon la valeur de leur indicateur de closeness (maximale, minimale ou moyenne) sur sept graphes viaires différents. Nous présentons les résultats sur trois d'entre eux ici, le reste est reporté en annexe \ref{ann:chap_distr_dtopo}.

\begin{figure}[c]
    \centering
    \includegraphics[width=0.7\textwidth]{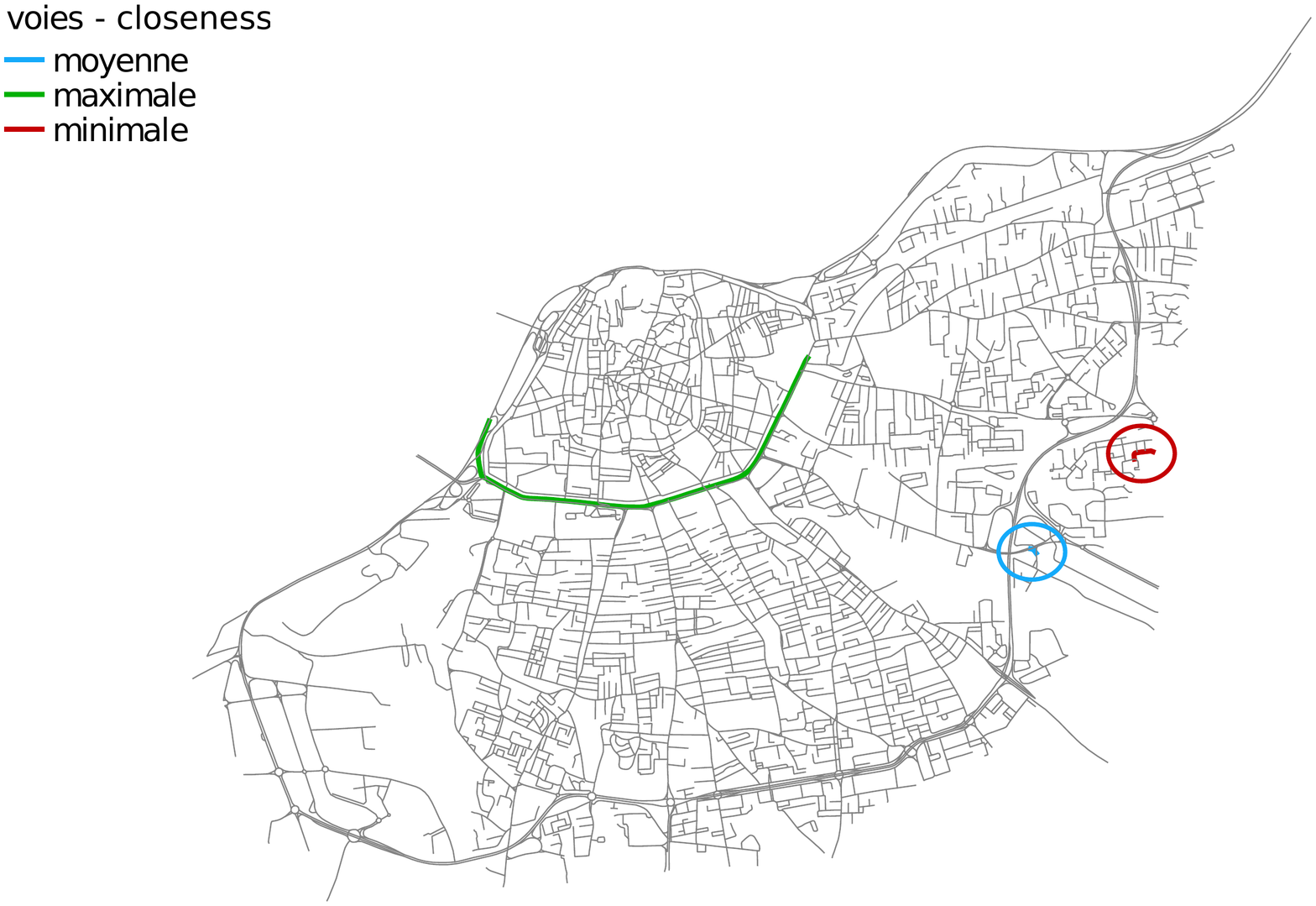}
    \caption{Situation des voies considérées sur le graphe d'Avignon}
    \label{fig:pos_avi}
\end{figure}

\begin{figure}[c]
    \centering
    \includegraphics[width=0.7\textwidth]{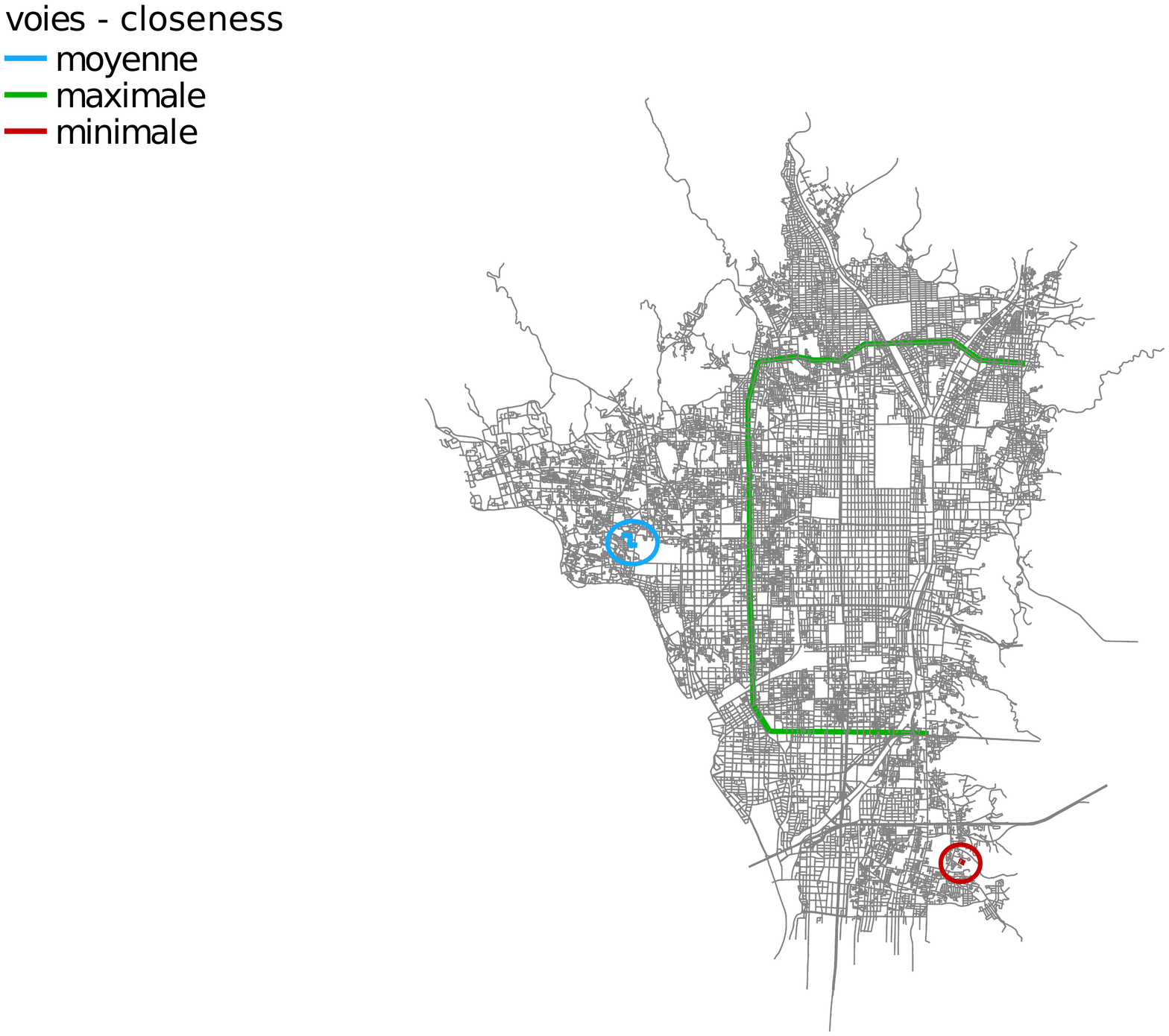}
    \caption{Situation des voies considérées sur le graphe de Kyoto}
    \label{fig:pos_kyo}
\end{figure}

\begin{figure}[c]
    \centering
    \includegraphics[width=0.7\textwidth]{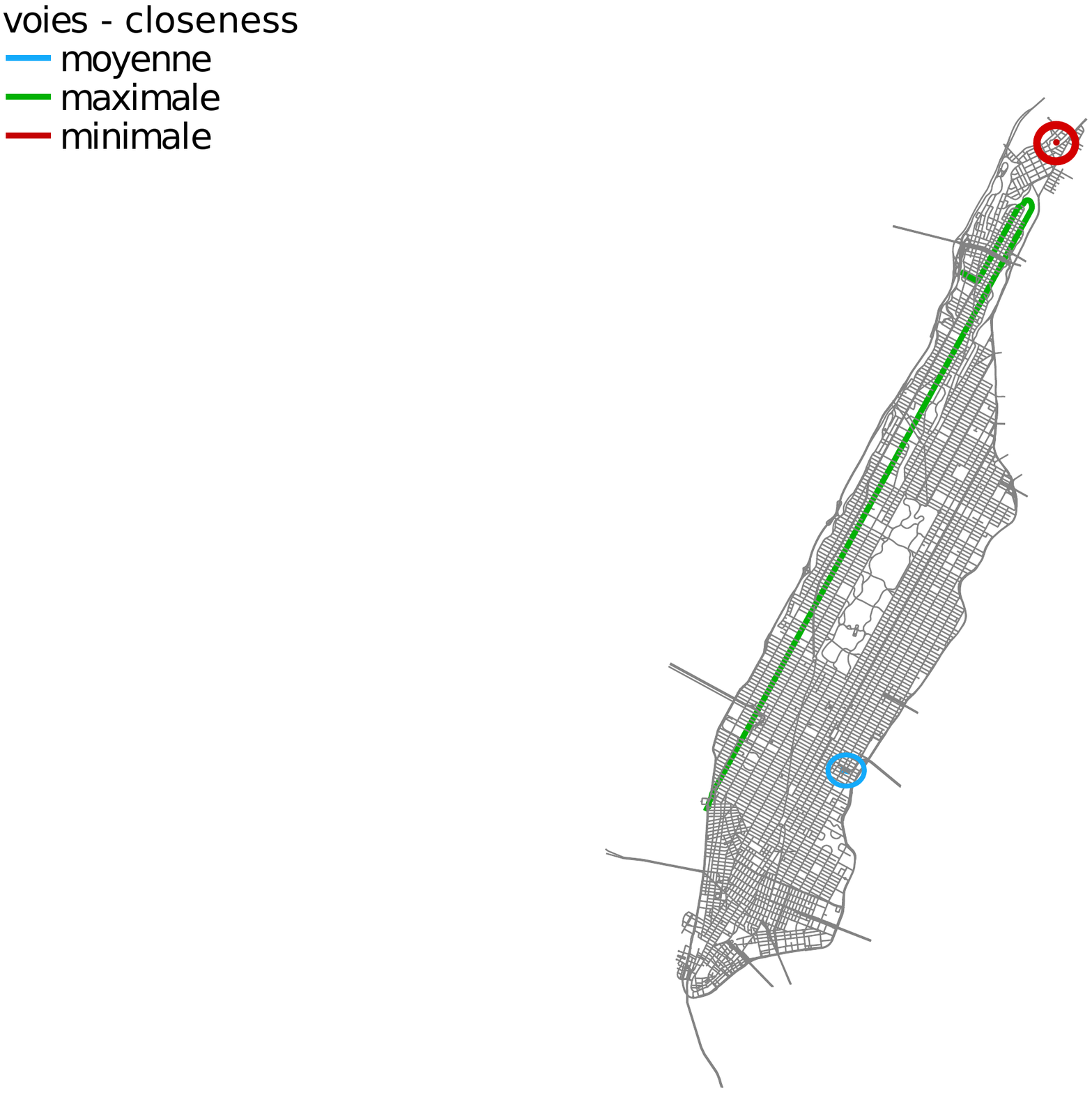}
    \caption{Situation des voies considérées sur le graphe de Manhattan}
    \label{fig:pos_man}
\end{figure}

Pour les trois graphes étudiés, Avignon, Kyoto et Manhattan, nous cartographions la position des trois voies choisies (figures \ref{fig:pos_avi}, \ref{fig:pos_kyo} et \ref{fig:pos_man}). Pour chacune de ces voies, nous traçons l'histogramme des distributions des distances topologiques des autres voies. Plus la distribution admet de valeurs faibles, plus cela sera révélateur de la \textit{centralité} de la voie étudiée. L'indicateur de closeness, en sommant toutes les distances topologiques pour une voie vers l'ensemble du graphe, correspond à l'inverse de l'intégrale de ces courbes.

Nous traçons, pour chacune de ces distributions, la courbe gaussienne de même moyenne et écart type (en noir). À  celle-ci nous ajoutons une seconde courbe, identique pour une ville sur les trois distributions. Cette courbe correspond à une valeur théorique. Elle est construite selon le raisonnement introduit par S. Douady et A. Perna dans \citep{perna2011characterization}.

Ainsi, nous calculons la moyenne et l'écart-type de cette courbe théorique selon les équations \ref{eq:moy} et \ref{eq:std} \citep{perna2011characterization} où $t$ correspond au logarithme en base 2 du nombre de faces ($N_{faces}$) du graphe étudié (équation \ref{eq:t}). Nous déterminons $N_{faces}$ avec la formule d'Euler (équation \ref{eq:eul}). Nous traçons, avec ces paramètres, les courbes gaussiennes théoriques, dont la distance avec les courbes obtenues à partir des données est significative. En effet, plus la médiane de la courbe des données réelles, tracée pour la voie de closeness maximale, se rapproche des valeurs faibles (et ainsi de celle de la courbe théorique), plus cela dénote de l'\textit{efficacité} du réseau.

\begin{equation}
\bar{r} = 2 + \lfloor  \frac{t-2}{6}  \rfloor
\label{eq:moy}
\end{equation}

\begin{equation}
\sigma^2 = 1,020 \times \frac{\sqrt{t+1}}{4}
\label{eq:std}
\end{equation}

\begin{equation}
t = \log_2(N_{faces})
\label{eq:t}
\end{equation}

\begin{equation}
N_{sommets}-N_{arcs}+N_{faces}=1
\label{eq:eul}
\end{equation}

\begin{figure}[c]
    \centering
    \begin{subfigure}[t]{0.45\textwidth}
        \centering
        \includegraphics[width=\textwidth]{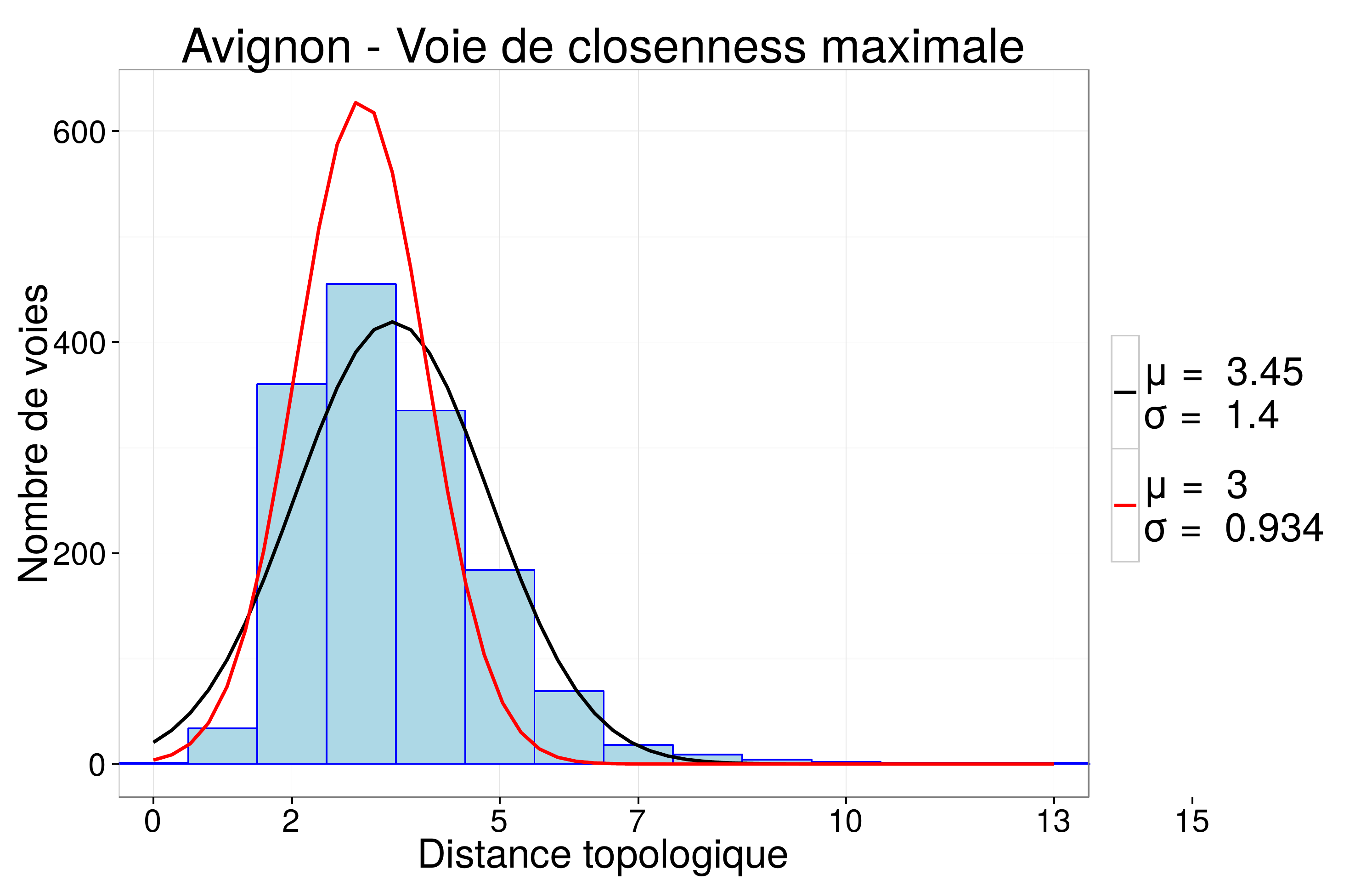}
        \caption{Pour la voie de closeness maximale}
        \label{fig:clo_avi_1}
    \end{subfigure}
    ~
    \begin{subfigure}[t]{0.45\textwidth}
        \centering
        \includegraphics[width=\textwidth]{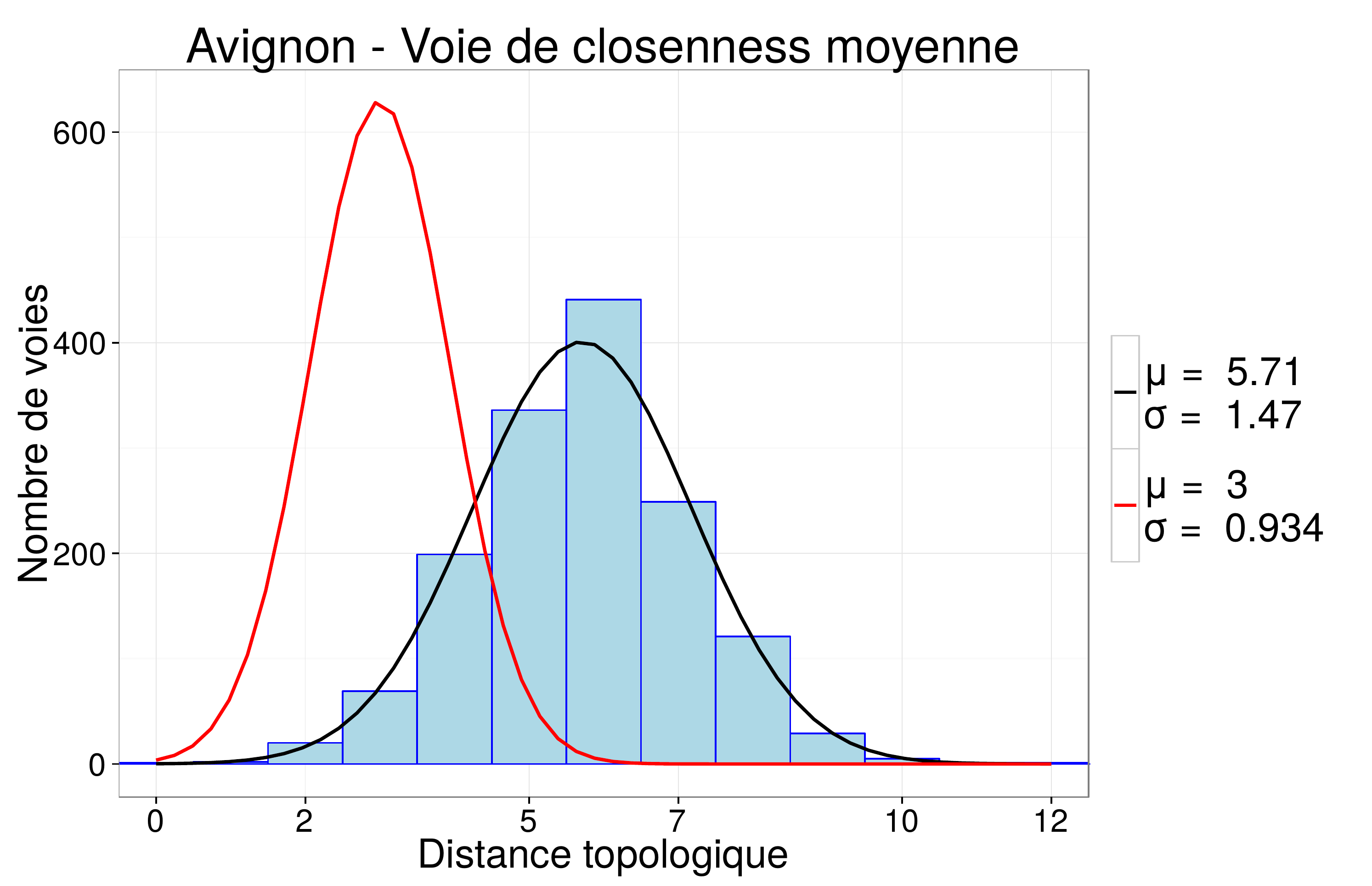}
        \caption{Pour une voie de closeness moyenne}
        \label{fig:clo_avi_2}
    \end{subfigure}
     ~
    \begin{subfigure}[t]{0.45\textwidth}
        \centering
        \includegraphics[width=\textwidth]{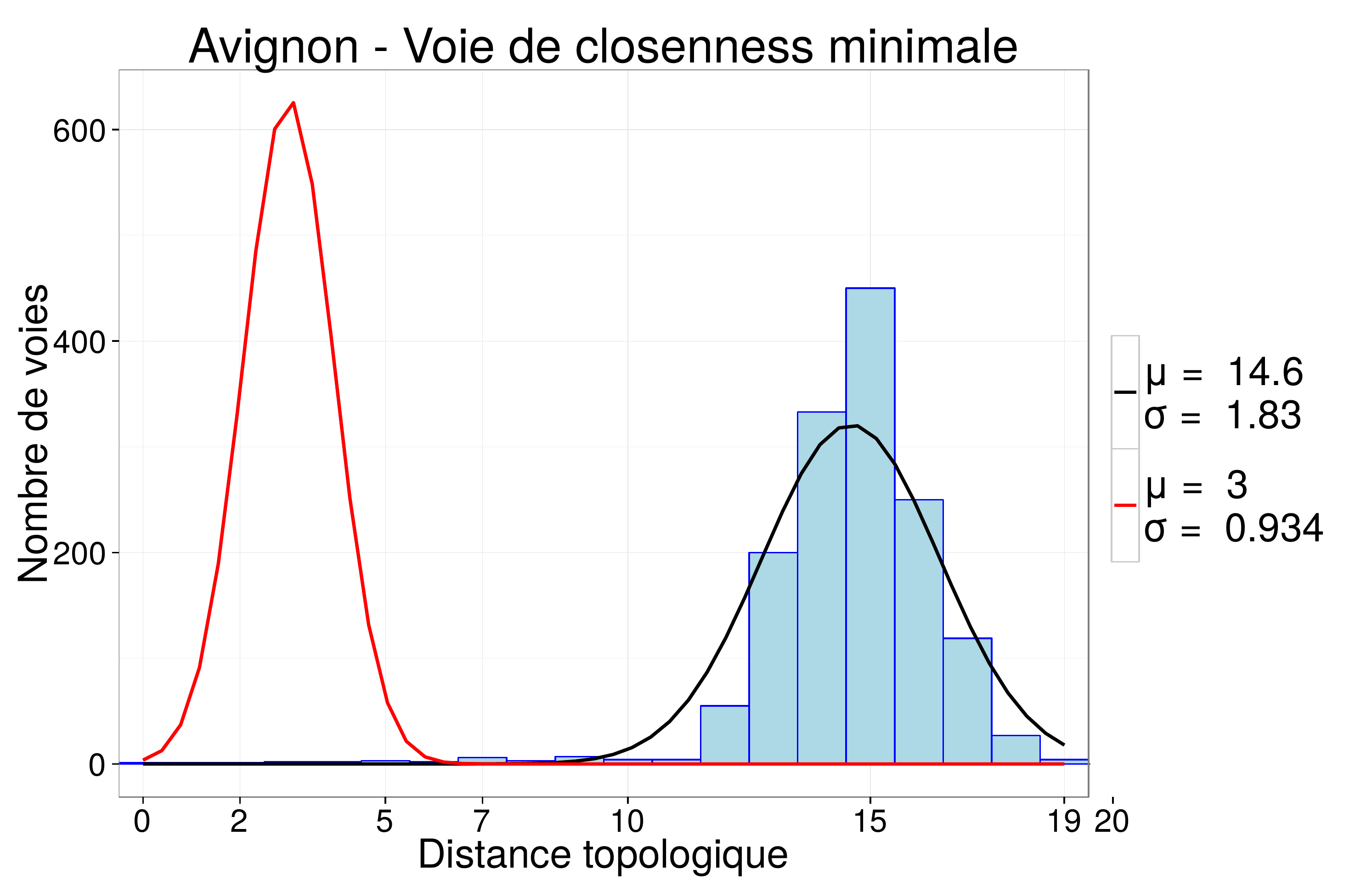}
        \caption{Pour la voie de closeness minimale}
        \label{fig:clo_avi_3}
    \end{subfigure}
    \caption{Répartition des distances topologiques. En rouge, gaussienne théorique}
\end{figure}

\begin{figure}[c]
    \centering
    \begin{subfigure}[t]{0.45\textwidth}
        \centering
        \includegraphics[width=\textwidth]{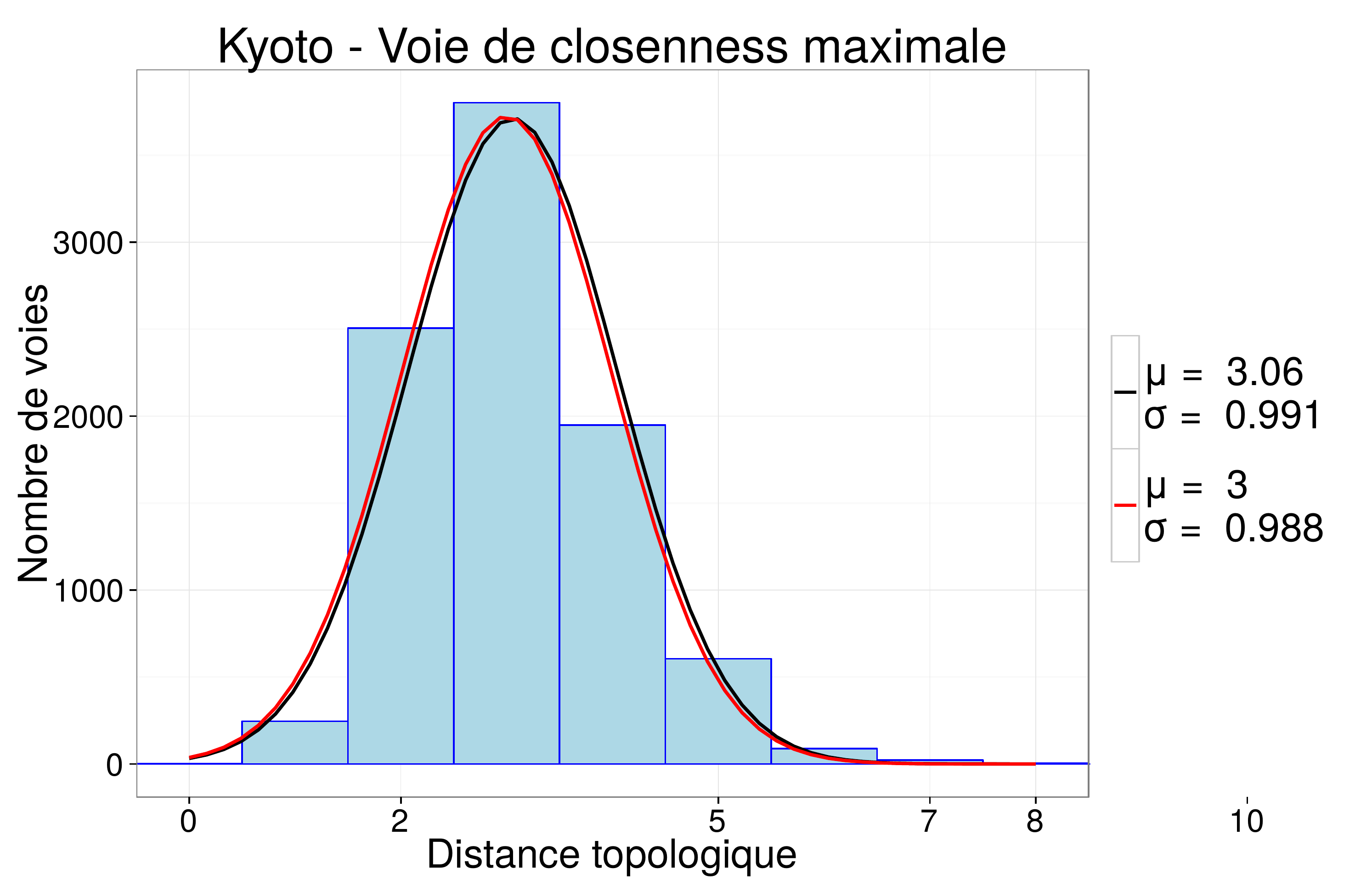}
        \caption{Pour la voie de closeness maximale}
        \label{fig:clo_kyo_1}
    \end{subfigure}
    ~
    \begin{subfigure}[t]{0.45\textwidth}
        \centering
        \includegraphics[width=\textwidth]{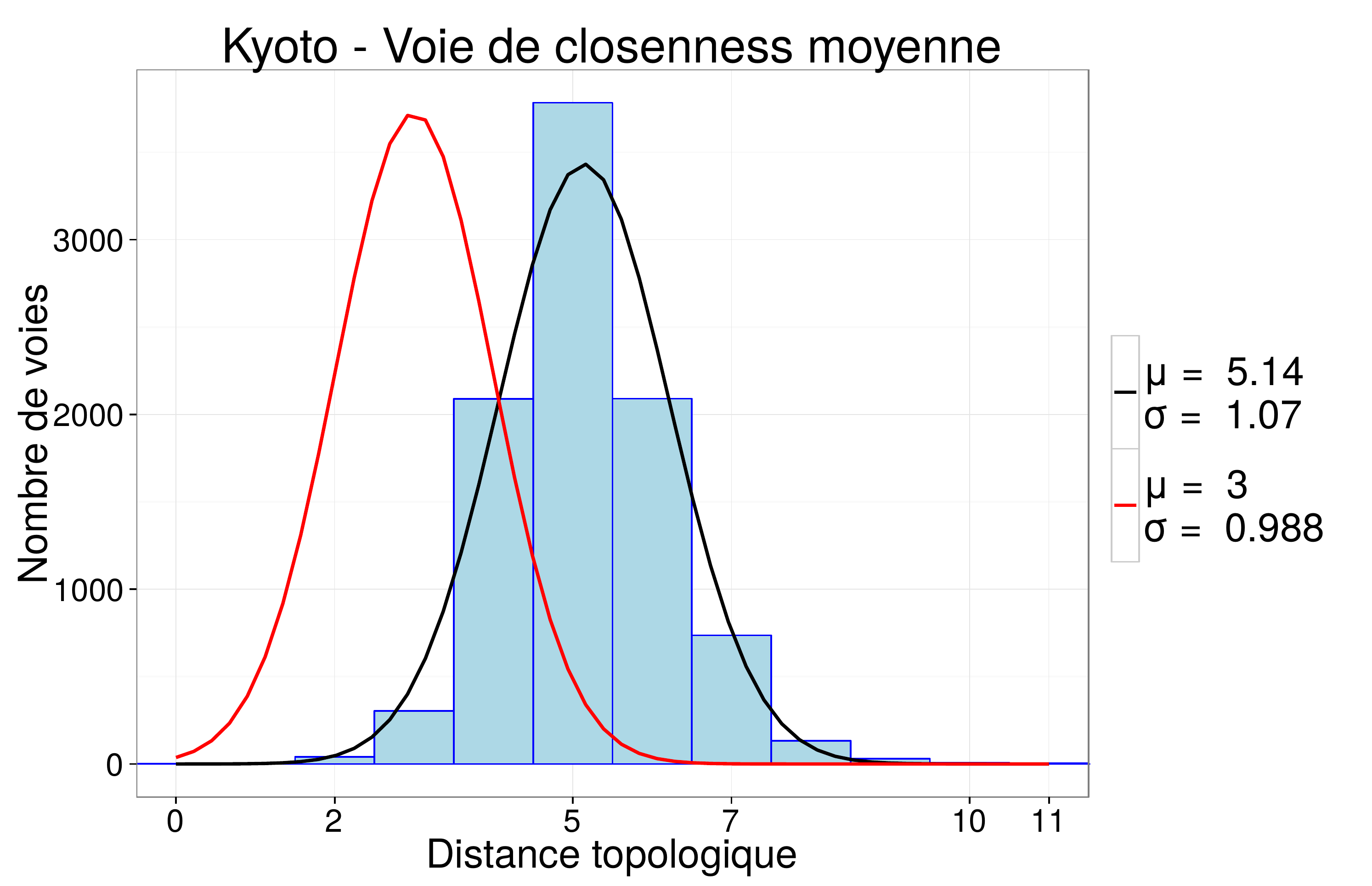}
        \caption{Pour une voie de closeness moyenne}
        \label{fig:clo_kyo_2}
    \end{subfigure}
    ~
    \begin{subfigure}[t]{0.45\textwidth}
        \centering
        \includegraphics[width=\textwidth]{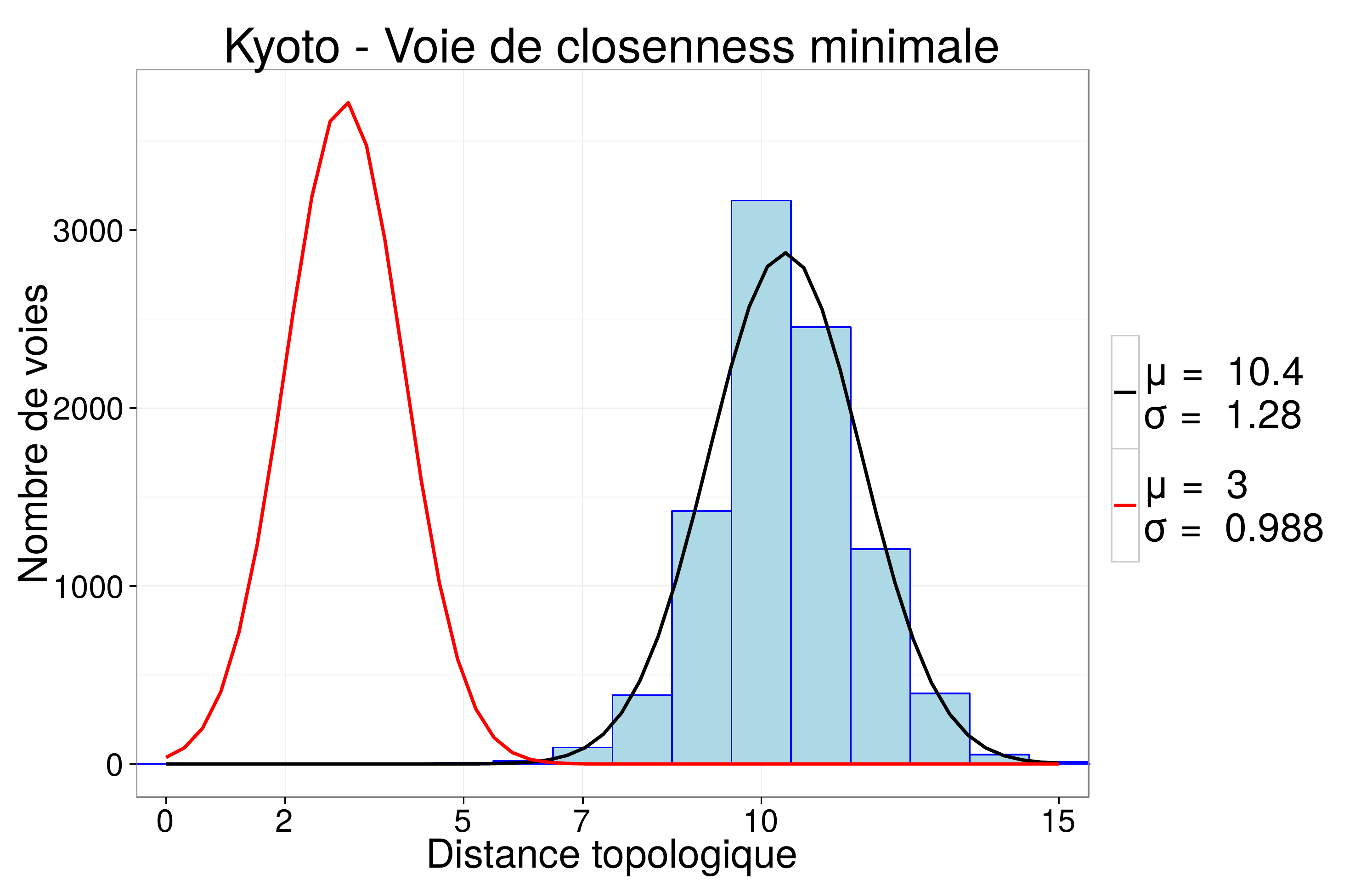}
        \caption{Pour la voie de closeness minimale}
        \label{fig:clo_kyo_3}
    \end{subfigure}
    \caption{Répartition des distances topologiques. En rouge, gaussienne théorique}
\end{figure}

\begin{figure}[c]
    \centering
    \begin{subfigure}[t]{0.45\textwidth}
        \centering
        \includegraphics[width=\textwidth]{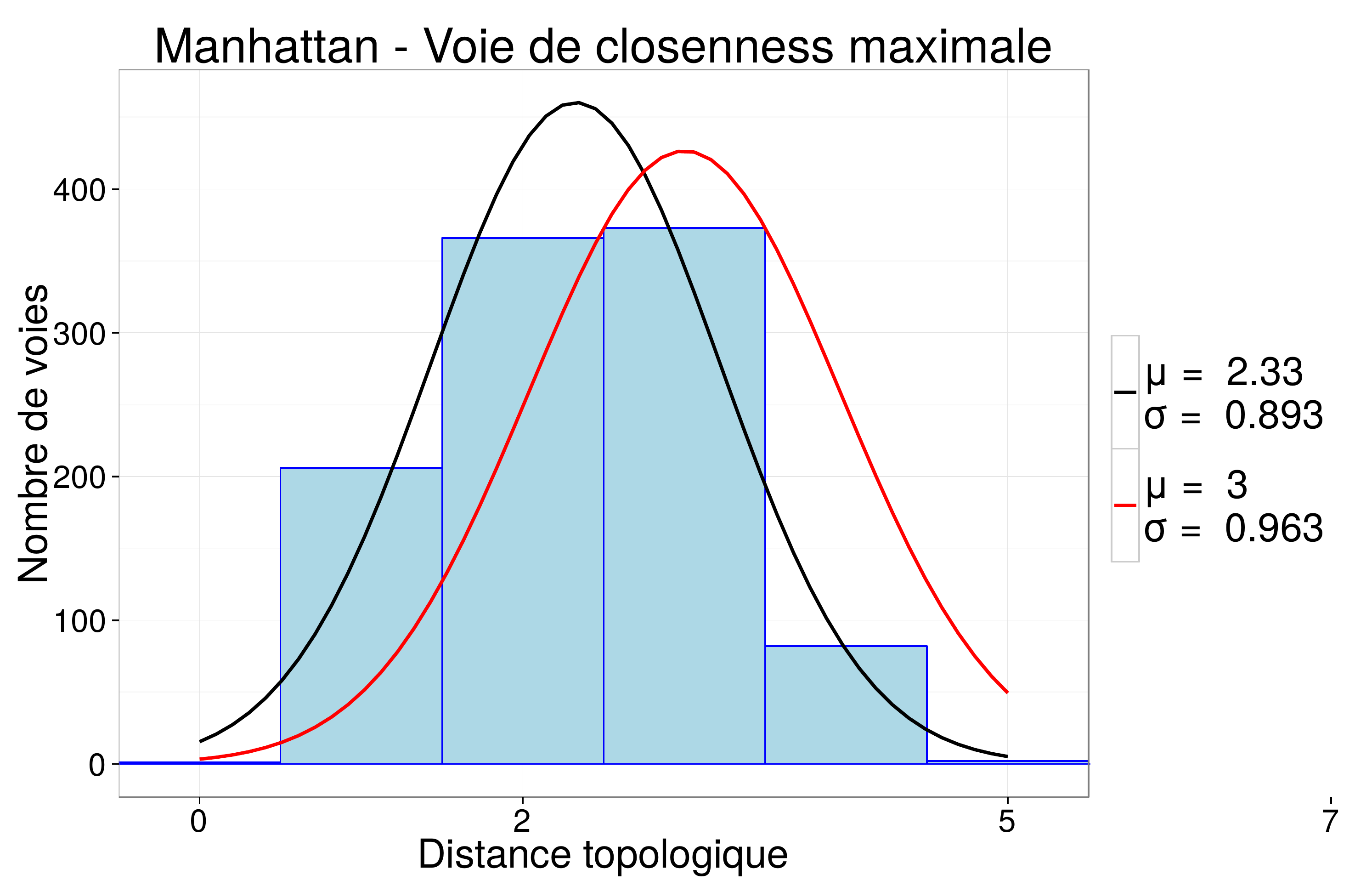}
        \caption{Pour la voie de closeness maximale}
        \label{fig:clo_man_1}
    \end{subfigure}
    ~
    \begin{subfigure}[t]{0.45\textwidth}
        \centering
        \includegraphics[width=\textwidth]{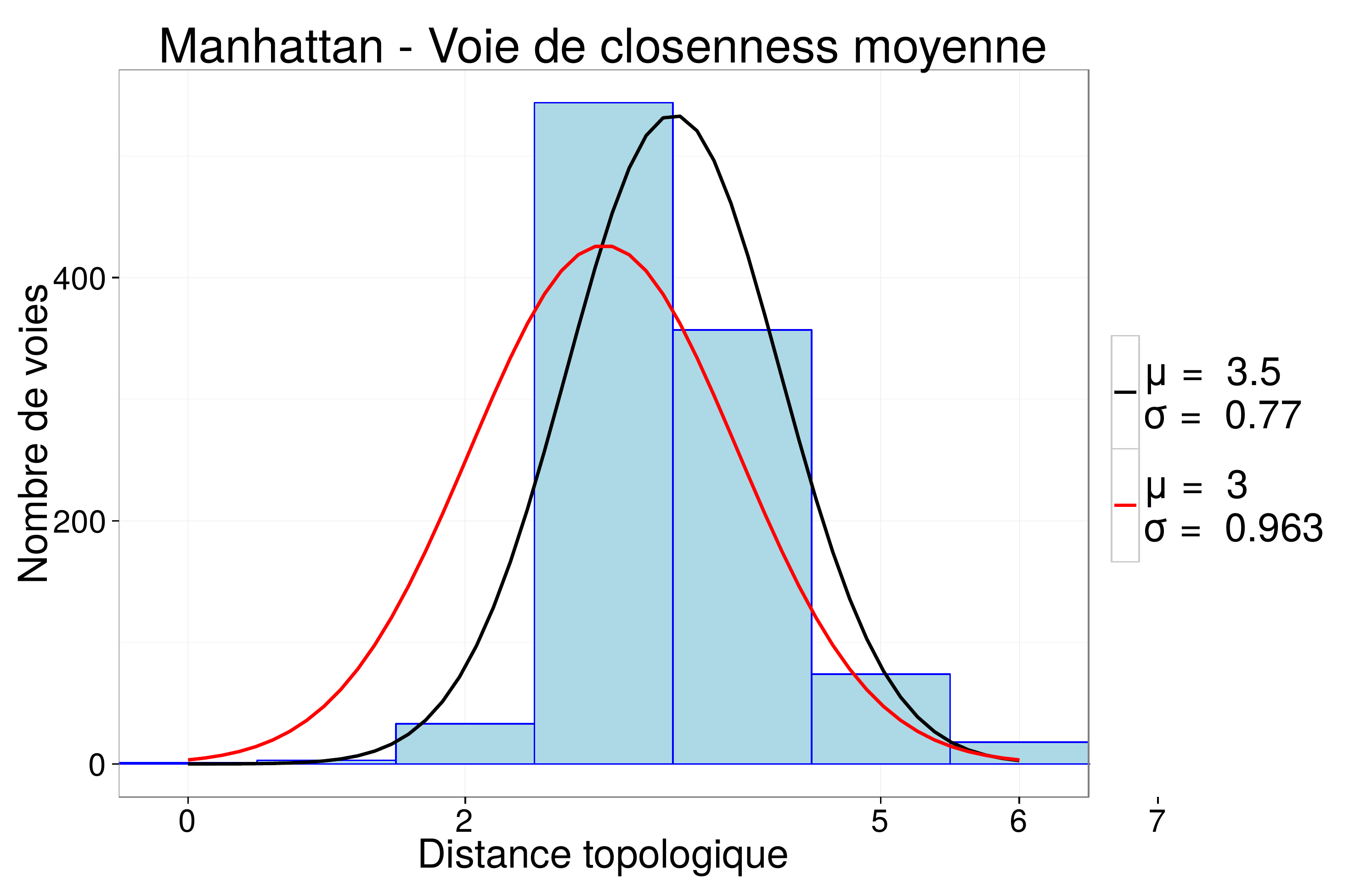}
        \caption{Pour une voie de closeness moyenne}
         \label{fig:clo_man_2}
    \end{subfigure}
    ~
    \begin{subfigure}[t]{0.45\textwidth}
        \centering
        \includegraphics[width=\textwidth]{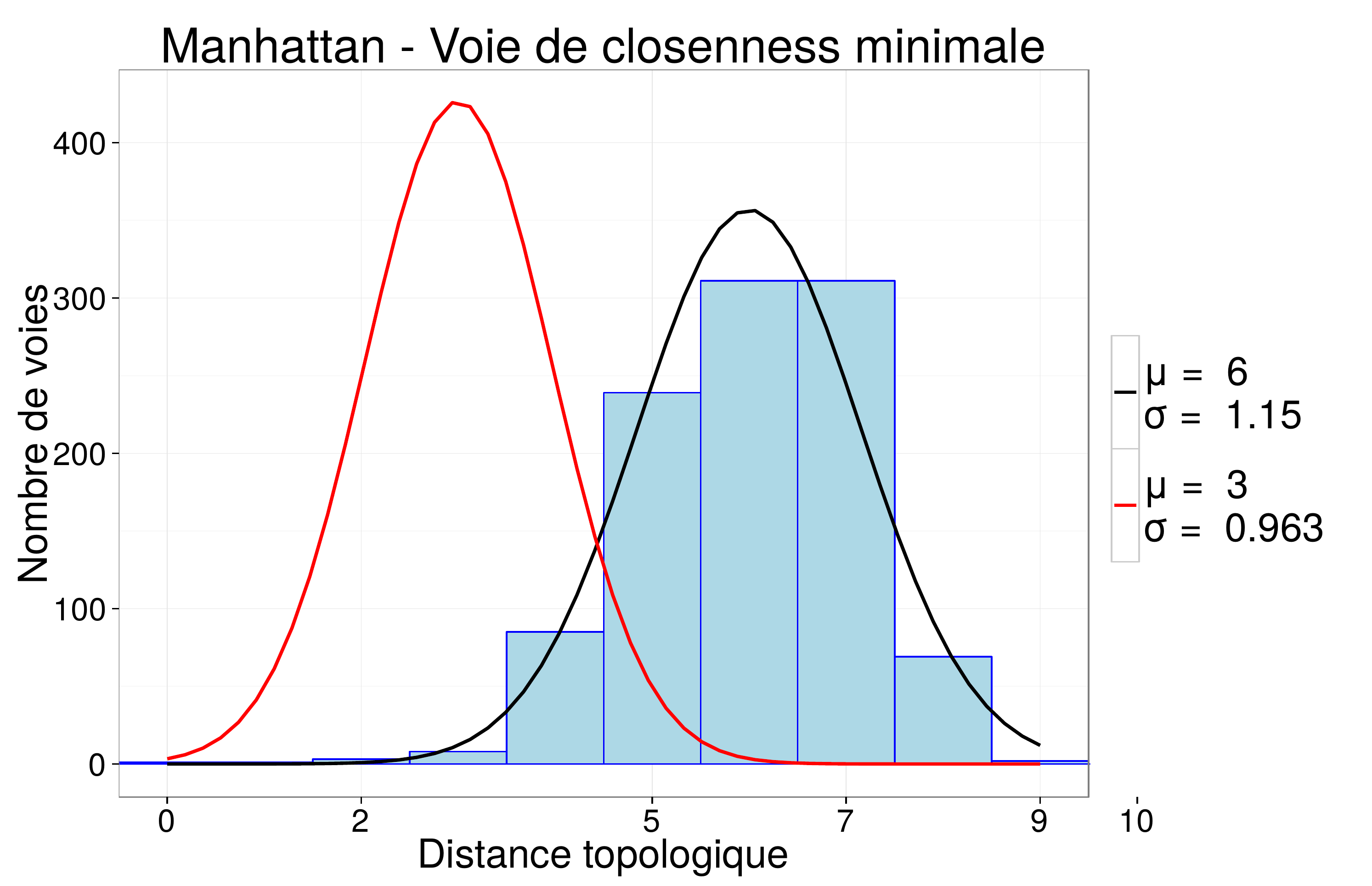}
        \caption{Pour la voie de closeness minimale}
         \label{fig:clo_man_3}
    \end{subfigure}
    \caption{Répartition des distances topologiques. En rouge, gaussienne théorique}
\end{figure}

Sur les trois villes étudiées, nous observons que plus la voie choisie a une valeur de closeness faible, plus la distribution des valeurs des distances topologiques est concentrée sur de fortes valeurs. Cette réaction est attendue puisque les deux informations sont liées. Ce qui est intéressant réside dans le comportement de la courbe correspondant à la voie de plus forte closeness. Celle-ci, pour Avignon est légèrement décalée sur des valeurs plus importantes, signe d'une voie à la proximité topologique un peu moins optimale que ce que la courbe aurait fait prédire (figure \ref{fig:clo_avi_1}). En revanche, sur Kyoto, les deux courbes sont confondues (figure \ref{fig:clo_kyo_1}). Cela indique que Kyoto suit un modèle de découpes successives en \enquote{Mondrian} proche de celui théorique. À l'extrême, la distribution de distances topologiques à partir de la voie de plus forte closeness sur le graphe de Manhattan est, elle, plus \textit{efficace} que le modèle théorique (figure \ref{fig:clo_man_1}).

Ces trois graphes, et les voies de plus forte closeness liées, sont donc représentatifs de trois logiques de découpage distinctes. Celles-ci conditionnent les distances topologiques d'une voie vers l'ensemble du réseau, et permettent une desserte de l'espace plus ou moins efficace. Il est intéressant de remarquer que la distribution des \textit{horizons} reste très proche d'une gaussienne. Il faudrait étudier comment la valeur moyenne et l'écart-type varient lorsque l'on parcourt le graphe vers les voies les moins accessibles.

Ces recherches, comme toutes celles présentées dans ce chapitre, sont une ouverture vers des méthodes de lecture des graphes viaires plus approfondies. Elles demandent à être poursuivies.

\FloatBarrier

\clearpage

\FloatBarrier

\clearpage

\FloatBarrier


\clearpage{\pagestyle{empty}\cleardoublepage}
\chapter*{Conclusion et Perspectives de recherche}
\markboth{Conclusion et Perspectives}{Conclusion et Perspectives}
\addcontentsline{toc}{part}{Conclusion et Perspectives de recherche}

\FloatBarrier
\section*{Synthèse de l'exploration de la spatialité}

L'information spatiale a fait son chemin à travers les siècles, pour venir compléter une théorie initialement imaginée pour symboliser les interactions relatives entre objets non spatialisés. Nous nous sommes ici emparés des outils de description des graphes pour les étoffer et les adapter au caractère spatial de notre problématique.

\subsection*{L'abstraction spatiale}

Nous commençons par extraire les \enquote{squelettes} des réseaux physiques que nous considérons. Nous conservons de la multitude d'informations qu'ils regroupent leurs axes, géométrie minimale symbolisant leurs tracés. Nous déconstruisons cette géométrie pour la retranscrire sous forme de graphe : chaque intersection devenant un sommet et le tronçon entre deux intersections, un arc. Dès lors, nous avons une abstraction d'une réalité physique sur laquelle nous pouvons appliquer des raisonnements propres à la topologie des réseaux (comment les objets sont connectés entre eux) et y ajouter l'information topographique (avec quelle \textit{forme} s'établit cette connexion).

Pour pouvoir approfondir notre étude de la spatialité, nous construisons un nouvel objet, appelé \textit{voie}, par association d'arcs à chaque sommet. De proche en proche, la voie peut, selon des critères géométriques, regrouper un nombre important d'arcs et ainsi traverser l'ensemble du graphe. Cette idée de \textit{superstructure axiale}, qui vient s'apposer au réseau, avait déjà été formalisée. Les axes, symbolisant des perspectives, ont ainsi été au fondement des travaux de syntaxe spatiale \citep{hillier1976space}. Cependant, nous cherchons ici à établir un objet dont la construction est locale, à chaque sommet, et indépendante du sens de lecture du graphe. De plus, nous faisons une étude approfondie de la paramétrisation de la construction de la voie, afin d'établir les seuils géométriques utiles à la création d'un objet robuste et significatif.

\subsection*{Les caractérisations spatiales}

Par la suite, la voie nous sert de support au développement d'un certain nombre d'indicateurs. Certains de ceux-ci sont issus de travaux en théorie des graphes. Nous les complétons en en créant de nouveaux, pour une caractérisation des réseaux tenant compte de leurs propriétés géométriques. En les appliquant aux arcs, puis aux voies, nous parvenons à mettre en avant les avantages de l'objet multi-échelle construit, dans la caractérisation de la spatialité. Ainsi, les indicateurs appliqués à l'hypergraphe formé par les voies, montrent des corrélations étonnantes, entre caractérisation locale (calculée à partir de la voie et de son voisinage direct) et globale (calculée en tenant compte de l'ensemble du réseau). La voie permet ainsi d'alléger de manière significative les temps de calcul. Non seulement elle regroupe les arcs, et permet ainsi de considérer un nombre moins important d'objets sur le graphe ; mais elle permet également de limiter une étude globale à une caractérisation locale.

Lorsque la caractérisation du graphe, faite sur son ensemble, apporte une information originale, qui ne peut être retrouvée par une analyse locale, la voie est garante de la stabilité de la lecture. Ainsi, l'étude des proximités topologiques, quand elle est faite sur les voies, n'est plus sensible au découpage de l'échantillon spatial (alors qu'elle l'était significativement lors de son application aux arcs). Des découpages de réseaux très différents révèlent de très faibles variations de l'indicateur. Cela signifie que la voie apporte une cohérence globale au graphe étudié, indépendamment de ses limites (sous réserve que les grands alignements \textit{a priori} continus ne soient pas interrompus). Ainsi, en élargissant le graphe autour de l'échantillon, nous ne modifions pas les \textit{chemins} à l'intérieur de celui-ci. Nous pouvons en conclure que les distances topologiques les plus simples, entre voies, passent rarement au delà du contour du graphe découpé. Cette grande stabilité au découpage du réseau l'est également à l'ajout d'arcs secondaires dans le graphe. Si les structures principales ne sont pas impactées, l'ajout ou le retrait d'arcs peu connectés ne modifiera pas de manière significative les proximités topologiques entre les voies du graphe.

\subsection*{Les sensibilités du modèle développé}

La sensibilité de notre méthodologie réside dans la vectorisation des intersections du graphe. Et, autour de celle-ci, dans l'orientation du premier et du dernier segment des arcs. En effet, la voie est construite à chaque intersection et est donc impactée par les éventuels décrochements à leur proximité. Pour limiter cet effet, nous avons développé une extension de la méthode de construction faisant intervenir des \enquote{zones tampons} qui permettent d'élargir les zones d'intersection et ainsi d'effacer éventuellement les légères discontinuités. Les diamètres de ces zones, paramétrables, doivent être évalués en fonction de la problématique de recherche et du tissu considéré. En effet, selon la question à laquelle nous souhaitons répondre, il sera utile ou non de conserver les aménagements du réseau (ronds-points, etc).

\subsection*{Les lois de la spatialité}

Une fois la méthodologie et ses sensibilités établies, nous évaluons ses potentiels de caractérisation sur un panel de quarante réseaux spatiaux. Nous montrons ainsi les propriétés communes auxquelles les lois physiques d’inscription dans un espace à deux ou trois dimensions contraignent tous les graphes spatiaux. Ainsi, la construction des voies sur chacun de ces graphes obéira à des critères communs. Par exemple, plus les sommets ont un degré important, plus les arcs qui leur sont connectés seront susceptibles d'être appariés sur une même voie : la contrainte spatiale force l'alignement. Plus un graphe admettra de nœuds de fort degré, plus ses voies pourront être longues et très connectées. Cette propriété est liée au caractère spatial du graphe et peut donc être retrouvée sur des réseaux de types très différents (géographiques, biologiques, etc). En revanche, au delà de la topologie, la géométrie de tels graphes leur est plus spécifique. Ainsi, les angles de connexion entre voies d'un réseau hydrographique ou ferré seront beaucoup plus faibles que ceux de réseaux viaires ou de craquelures.

\subsection*{Les cinématiques spatiales}

Au delà de la caractérisation statique des graphes, nous comparons des réseaux de différentes dates sur un même territoire. Afin d'en déceler les évolutions structurelles, nous établissons une méthodologie de quantification des changements, qui s'appuie sur celle de suivi temporel par emprise identique, formalisée dans \citep{bordin2006methode}. Notre but est de déceler les modifications de proximités topologiques (et ainsi de logiques d'accès) au sein des voies du graphe. Nous observons ainsi, par quantifications successives sur des couples de dates, une \textit{cinématique} d'évolution, naissant de la juxtaposition d'informations statiques. À la différence du résultat observé pour les effets de bord, les changements d'une année sur l'autre \textit{à l'intérieur du graphe} ont un réel impact sur la proximité entre les objets. Celui-ci ne se révèle pas toujours positif : la densification  de certaines zones (comme sur l'exemple du Nord de Rotterdam) ou les coupures de certaines structures globales (comme l'enceinte de l'intra-muros d'Avignon) peuvent diminuer la proximité des voies du graphe. Le remodelage permanent de ces graphes n'aboutit donc pas toujours à une amélioration de l'accessibilité globale. 
Les tendances de changements entre les couples d'années montrent une prédisposition topologique au développement de certains quartiers. Ainsi, entre 1374 et 1570, les modifications du réseau viaire joignant deux noyaux urbains voisins (Rotterdam et Schiedam) et de la campagne aux alentours, renforçaient l'accès au centre du village de Rotterdam, au détriment de celui de Schiedam.

Pour poursuivre ce travail de comparaison nous aurions besoin d'étendre la vectorisation de cartes anciennes. À Paris ou New-York, des cartes historiques ont été géo-référencés sur une période longue, ce qui facilite leur traduction sous forme vectorielle, ouvrant des perspectives de recherche intéressantes. Cependant, pour y appliquer la méthodologie de quantification mise en place dans ce travail, il est nécessaire que les données s'appuient sur l'emprise d'un graphe unique.

\subsection*{Les cohérences spatiales}

\begin{figure}[h]
    \centering
    \begin{subfigure}[t]{.6\linewidth}
        \includegraphics[width=\textwidth]{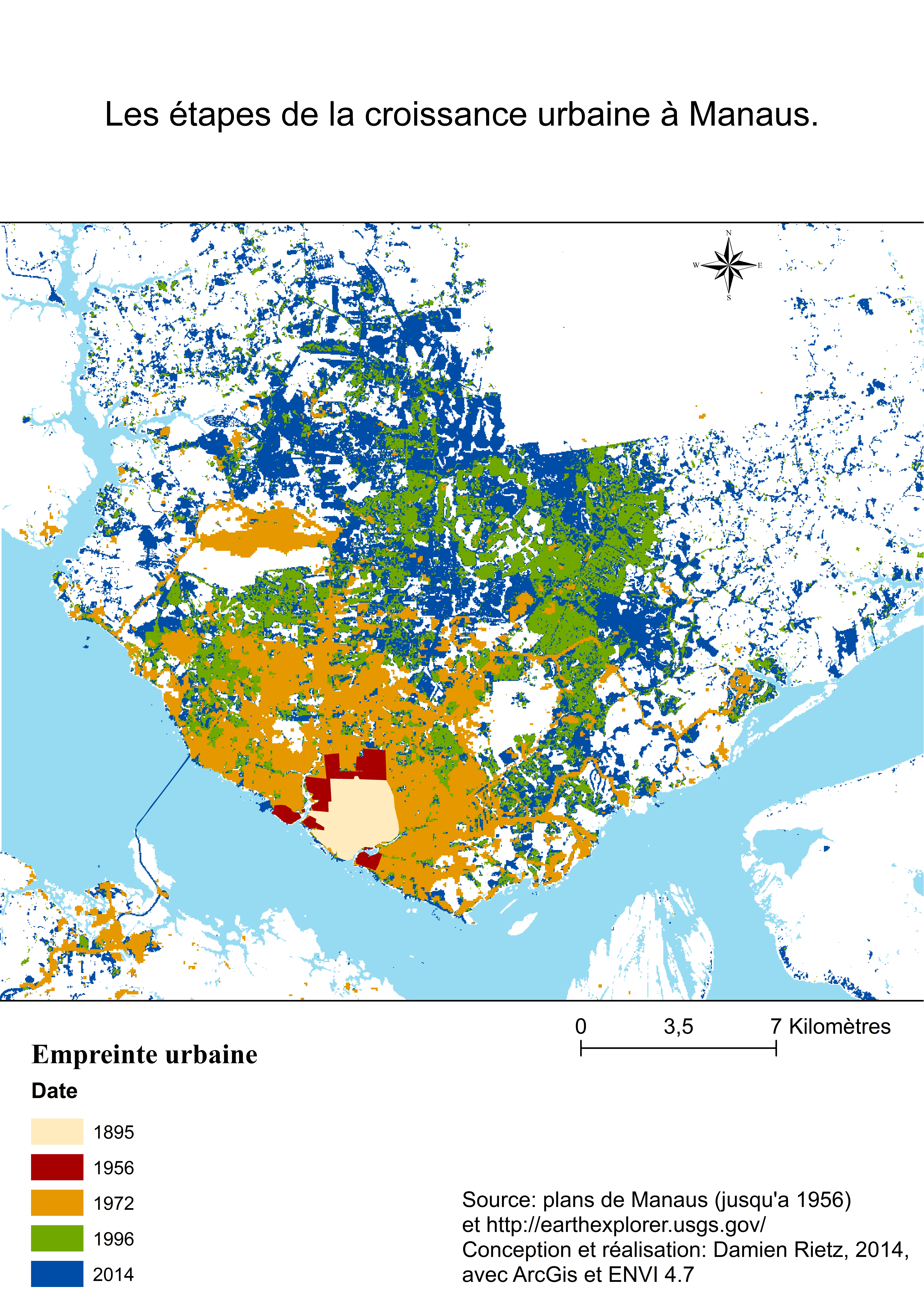}
        \caption{Évolution de Manaus entre 1895 et 2014. \\ source : D. Rietz}
    \end{subfigure}
    ~
    \begin{subfigure}[t]{.8\linewidth}
        \includegraphics[width=\textwidth]{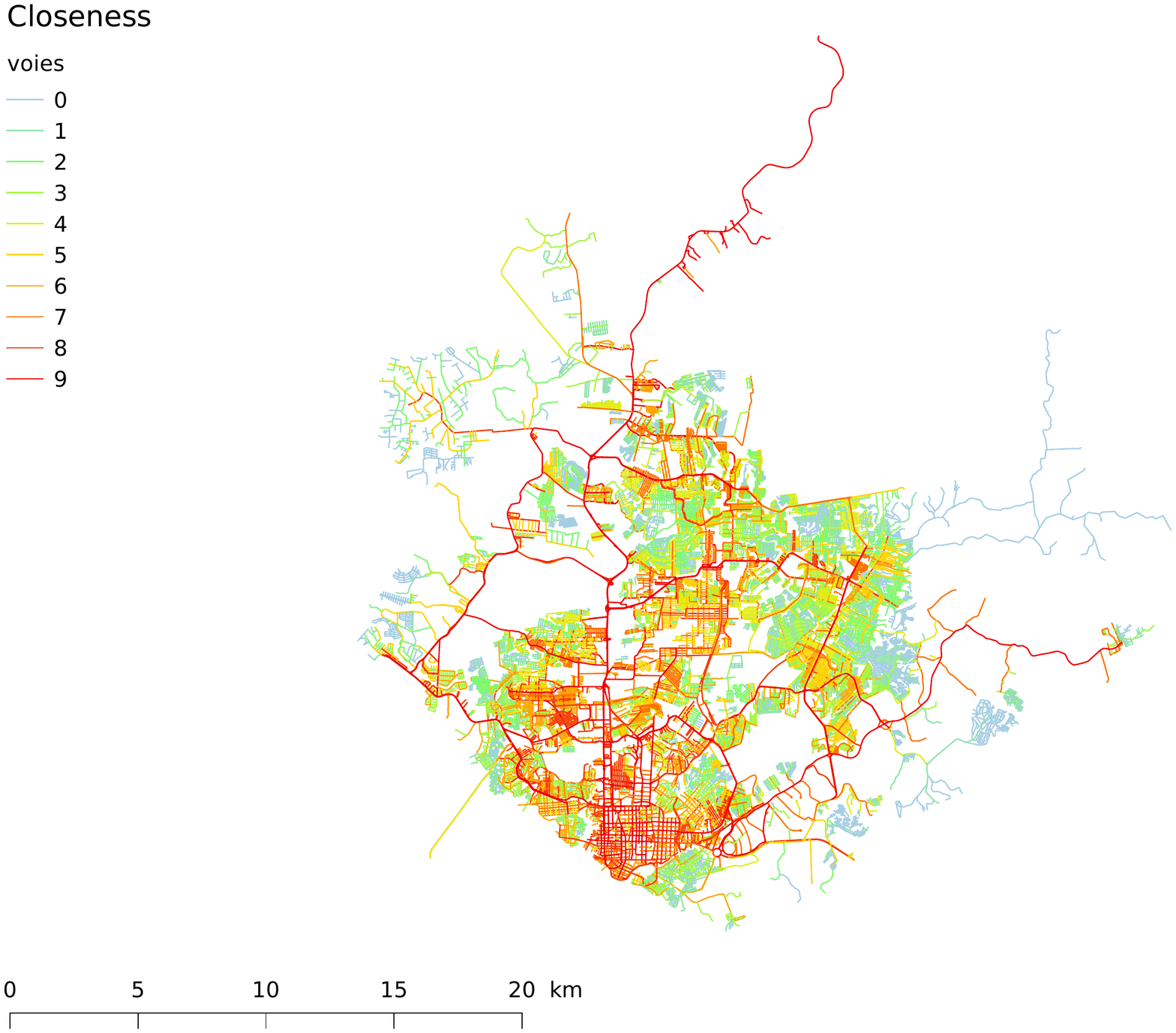}
        \caption{Indicateur de closeness calculé sur le graphe viaire actuel de la ville.}
    \end{subfigure}
    \caption{Parallèle entre la croissance de la ville de Manaus et l'indicateur de closeness.}
    \label{fig:manaus_croiss}
\end{figure}

L'ajout de nouveaux éléments dans un graphe viaire, au cours du temps, se fait souvent en cohérence avec l'existant. Par ce processus de développement, les nouvelles voies sont construites de façon à minimiser leurs distances topologiques avec de plus anciennes. Le squelette historique de la ville s'en trouve renforcé et ressort donc avec l'indicateur de closeness le plus fort. Le nouveau réseau s'appuie sur sa structure, et donc, plus une voie est ancienne, plus elle est susceptible de desservir l'ensemble du territoire de manière optimale. Ainsi, la lecture du réseau viaire proposée par l'indicateur de closeness offre un parallèle très intéressant avec l'observation de la croissance des villes. Sur le graphe viaire de Manaus, les données du développement de la ville entre 1895 et 2014 correspondent aux résultats de l'indicateur appliqué aux voies : celles dont la valeur de closeness est la plus importante correspondent aux quartiers les plus anciens (figure \ref{fig:manaus_croiss}).

\begin{figure}[h]
    \centering
    \begin{subfigure}[t]{.6\linewidth}
        \includegraphics[width=\textwidth]{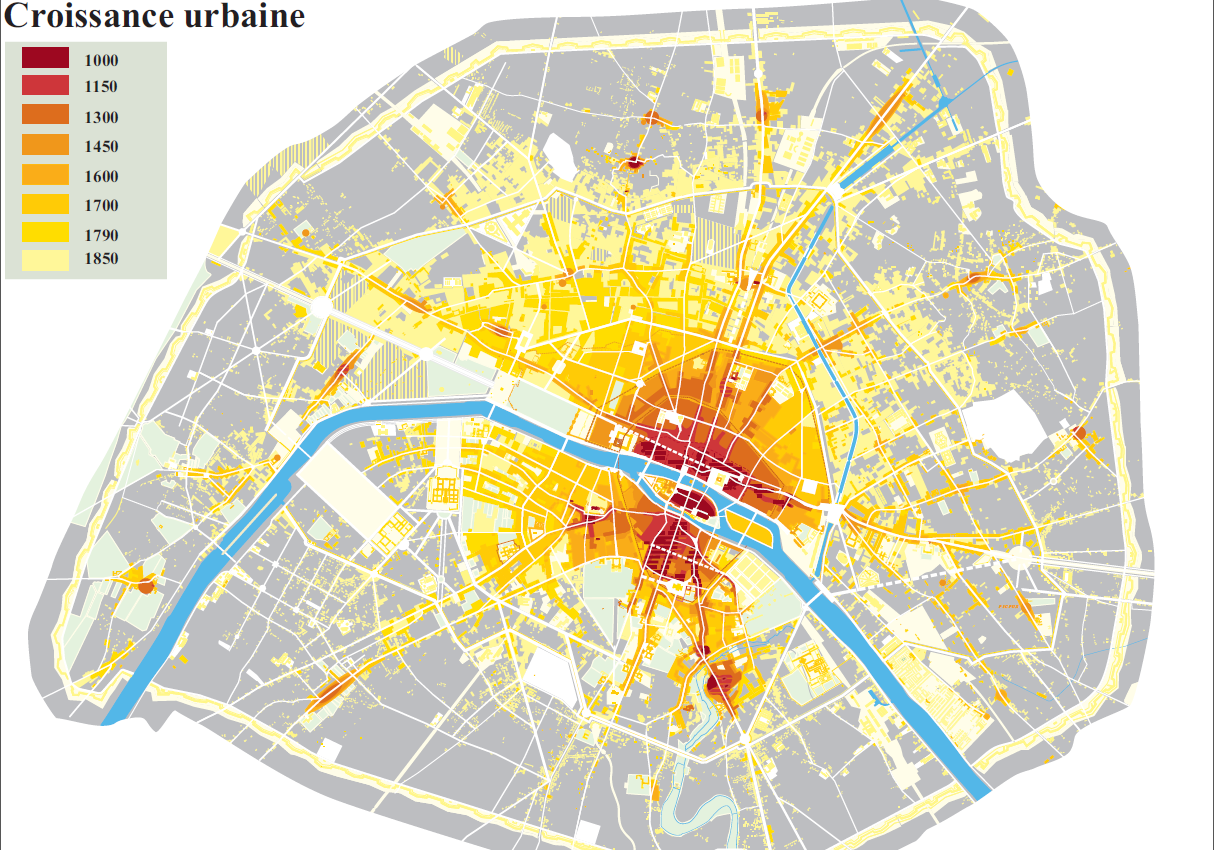}
        \caption{Évolution de Paris entre l'an 1000 et l'année 1850. \\ source : Atlas historique de Paris - M. Huard}
    \end{subfigure}
    ~
    \begin{subfigure}[t]{.8\linewidth}
        \includegraphics[width=\textwidth]{images/cities/paris_3_closeness.pdf}
        \caption{Indicateur de closeness calculé sur le graphe viaire actuel de la ville.}
    \end{subfigure}
    \caption{Parallèle entre la croissance de la ville de Paris et l'indicateur de closeness.}
    \label{fig:paris_croiss}
\end{figure}

Cette théorie a bien évidemment ses limites, notamment lors de la création de grandes structures planifiées, construites explicitement pour desservir efficacement le territoire. C'est ainsi le cas du périphérique à Paris, dont la fin de construction date des années 1970, qui a été conçu pour desservir efficacement l'ensemble de la capitale et donc assurer des distances topologiques faibles avec le graphe de l'intra-muros. En comparant la carte de l'indicateur de closeness de la ville à celle de sa croissance entre 1000 et 1850 (figure \ref{fig:paris_croiss}) nous retrouvons ainsi avec de fortes valeurs la structure englobante du boulevard qui vient s'ajouter à celles des grands axes de traversée de la ville (plus ou moins anciens).

L'estimation faite avec l'indicateur de closeness sera également plus éloignée de la dynamique de développement observée, lorsque la ville comprend des parties de son graphe entièrement planifiées. Celles-ci sont souvent créées par extension unitaire, indépendamment du centre ancien. C'est le cas à Barcelone, où la proximité topologique de la vieille ville est moins importante que celle de la structure qui lui a été ajoutée selon les plans de Cerdà. Cette partie du graphe a été pensée isolément de la structure ancienne, sans y accorder sa géométrie. Seules quelques voies s'introduisent dans ce territoire pré-existant dont la logique est en complète opposition : le centre de Barcelone est médiéval, et, par sa nature géométrique, peu accessible depuis l'extérieur. Cela peut s'interpréter comme une stratégie de protection. De la même manière, à Téhéran, les maisons du centre ancien comme les bazars sont éloignés des grands axes. Dans ces cas d'application, la sur-structure identifiée sur le graphe assure une desserte efficace du territoire indépendante des centres anciens.

Nous pouvons donc en conclure que la closeness fait ressortir la partie du graphe la plus cohérente. Usuellement, les arcs qui viennent s'ajouter au réseau avec le temps privilégient la cohérence avec les structures anciennes, et non entre eux (comme à Manaus, et la plupart du temps dans les villes médiévales). Ils renforcent ainsi la proximité topologique du centre ancien. Mais il est également possible qu'un ajout planifié d'un grand nombre d'arcs se détache des logiques géométriques pré-existantes pour imposer au graphe leur propre cohérence (comme à Barcelone, et dans les villes à la structure majoritairement planifiée).

Ces recherches ouvrent un premier champ de perspectives lié à l'analyse historique des villes. Cette étude de cohérence pourra être croisée avec celle de cinématique des changements afin de nourrir les travaux portant sur la croissance des villes. Il serait ainsi intéressant d'étudier les variations dans le temps de l'accessibilité du reste du graphe par rapport à une voie, \textit{centrale} dans le graphe le plus ancien. Nous pourrions ainsi observer le renforcement de cette centralité, ou au contraire, sa fluctuation. Cela nous permettrait de mieux comprendre les dynamiques temporelles et d'étoffer leur lien avec les cohérences spatiales. Les échanges pluridisciplinaires pourront, une fois encore, nourrir cette problématique.

\FloatBarrier
\section*{Perspectives}

Le travail que nous avons présenté ici est un noyau offrant de multiples perspectives de développement. Certaines ont été explorées au cours de cette recherche, et nécessiteraient d'être approfondies à sa suite. Nous en présentons quelques unes dans ce paragraphe final.

\FloatBarrier
\subsection*{Un pas vers la morphogenèse} 

Les dynamiques spatio-temporelles des réseaux viaires peuvent être de plusieurs types : accumulation, rupture, décalage... Il est impossible de les décrire exhaustivement, que ce soit individuellement ou avec leurs interactions croisées.  À l'aide de la quantification mathématique des structures viaires, M. Barthelemy a mené des travaux sur la modélisation de la morphogenèse urbaine. Il part d'un réseau minimal et choisit aléatoirement des points autour de celui-ci qu'il raccorde au réseau pré-existant. Il montre ainsi que, si les points sont choisis dans l'espace en fonction d'une courbe gaussienne, alors le graphe obtenu a une distribution de surfaces correspondant à ce qui est observé dans la réalité. Le raccordement est dans un premier temps fait avec le dernier sommet posé du réseau puis perpendiculairement au réseau \citep{barthelemy2008modeling, barthelemy2009co}. Cela aboutit à un graphe s'apparentant à un réseau viaire. Thomas Courtat a poursuivi cette théorie en ajoutant un potentiel lié au réseau lui même, influençant la position des nouveaux sommets placés au sein du graphe \citep{courtat2011mathematics}. 

À partir de ces travaux, au cours du développement de la recherche présentée ici, nous avons pu réfléchir aux différentes dynamiques qui régissent l’expansion d'un graphe viaire. Nous en avons retrouvé deux, déjà présentes dans les travaux de T. Courtat et de E. Strano \textit{et al.} \citep{strano2012elementary} : une dynamique d'expansion, où le graphe s'ouvre vers le territoire alentour, afin de l'explorer et de permettre un parcours rapide sur celui-ci ; et une dynamique de division, qui a pour but une desserte fine de l'espace. Ces deux dynamiques peuvent être décelées à travers les grilles de lecture que nous avons définies à partir des \textit{k-core} dans la troisième partie. La poursuite de leur caractérisation est un des débouchés de ce travail.

\begin{figure}[h]
    \centering
    \begin{subfigure}[t]{.22\linewidth}
        \includegraphics[width=\textwidth]{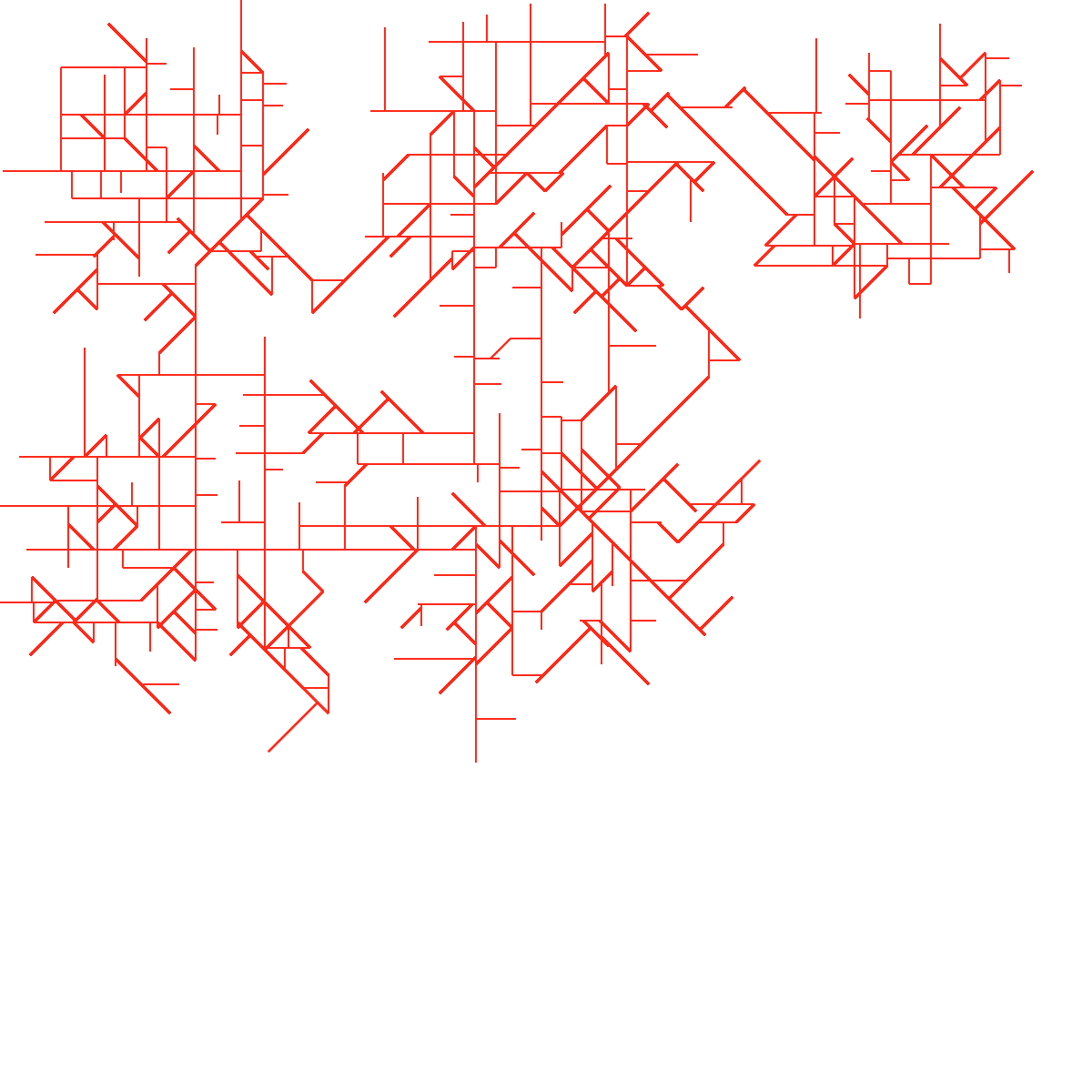}
        \caption{400 itérations.}
    \end{subfigure}
    ~
    \begin{subfigure}[t]{.22\linewidth}
        \includegraphics[width=\textwidth]{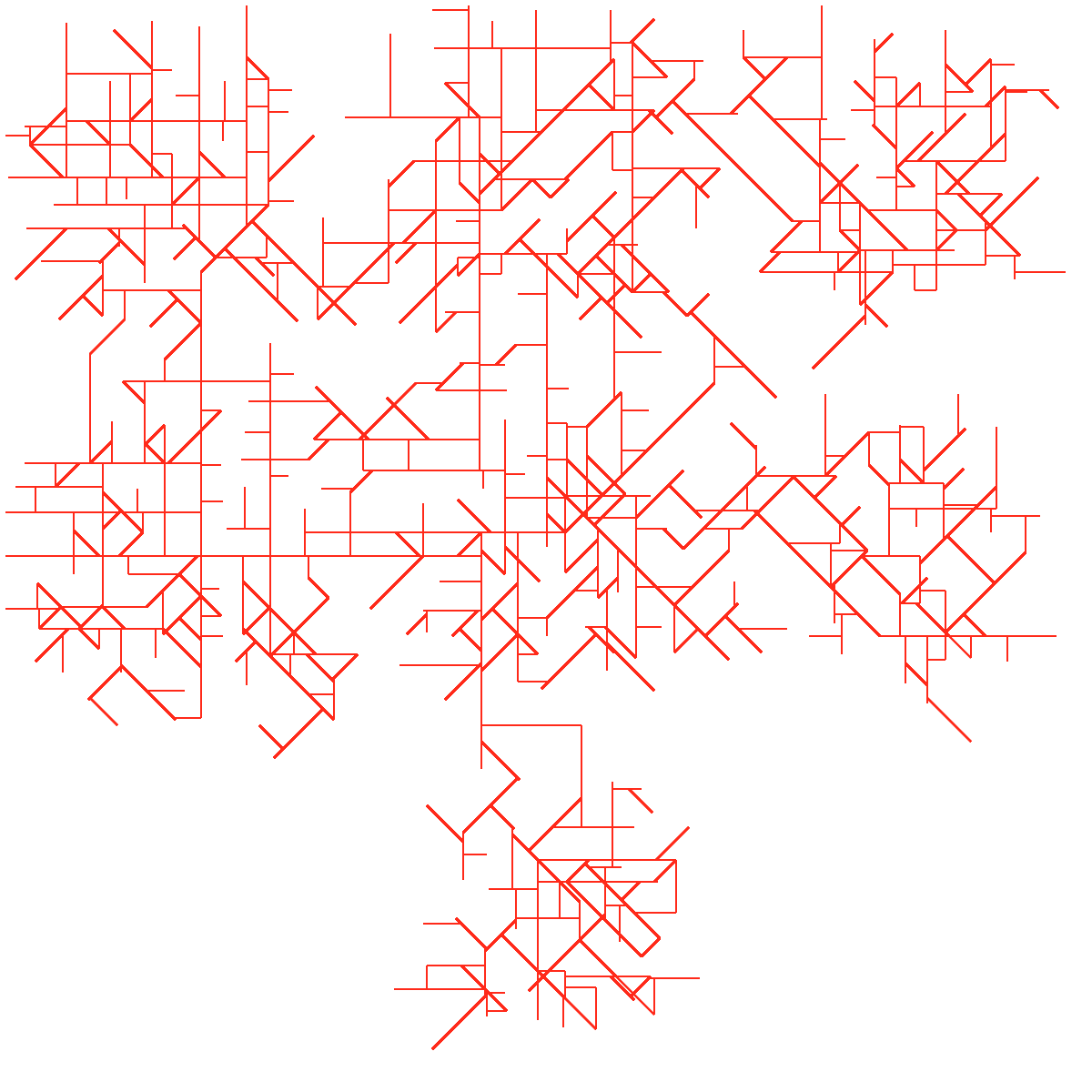}
        \caption{600 itérations.}
    \end{subfigure}
    ~
    \begin{subfigure}[t]{.22\linewidth}
        \includegraphics[width=\textwidth]{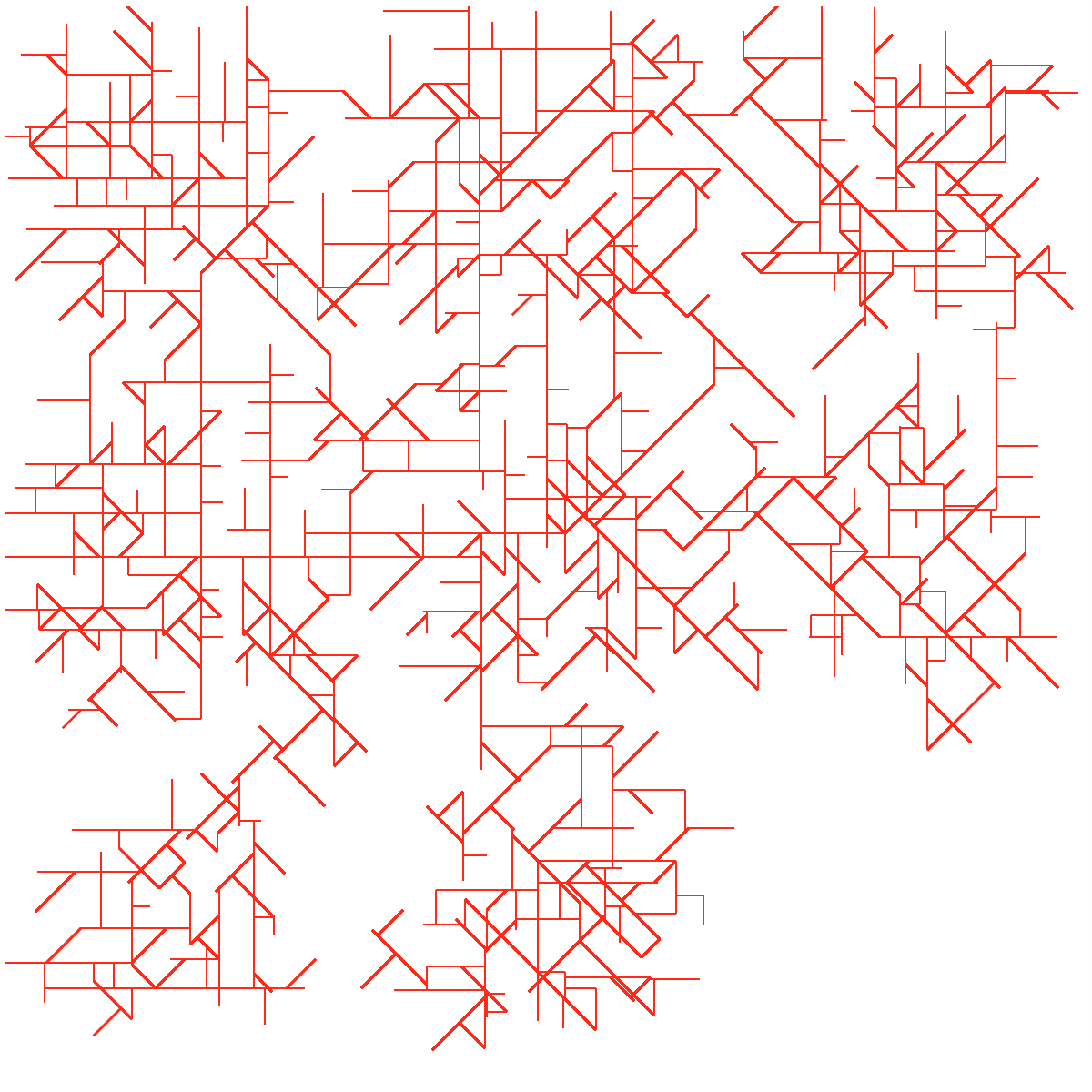}
        \caption{800 itérations.}
    \end{subfigure}
    ~
    \begin{subfigure}[t]{.22\linewidth}
        \includegraphics[width=\textwidth]{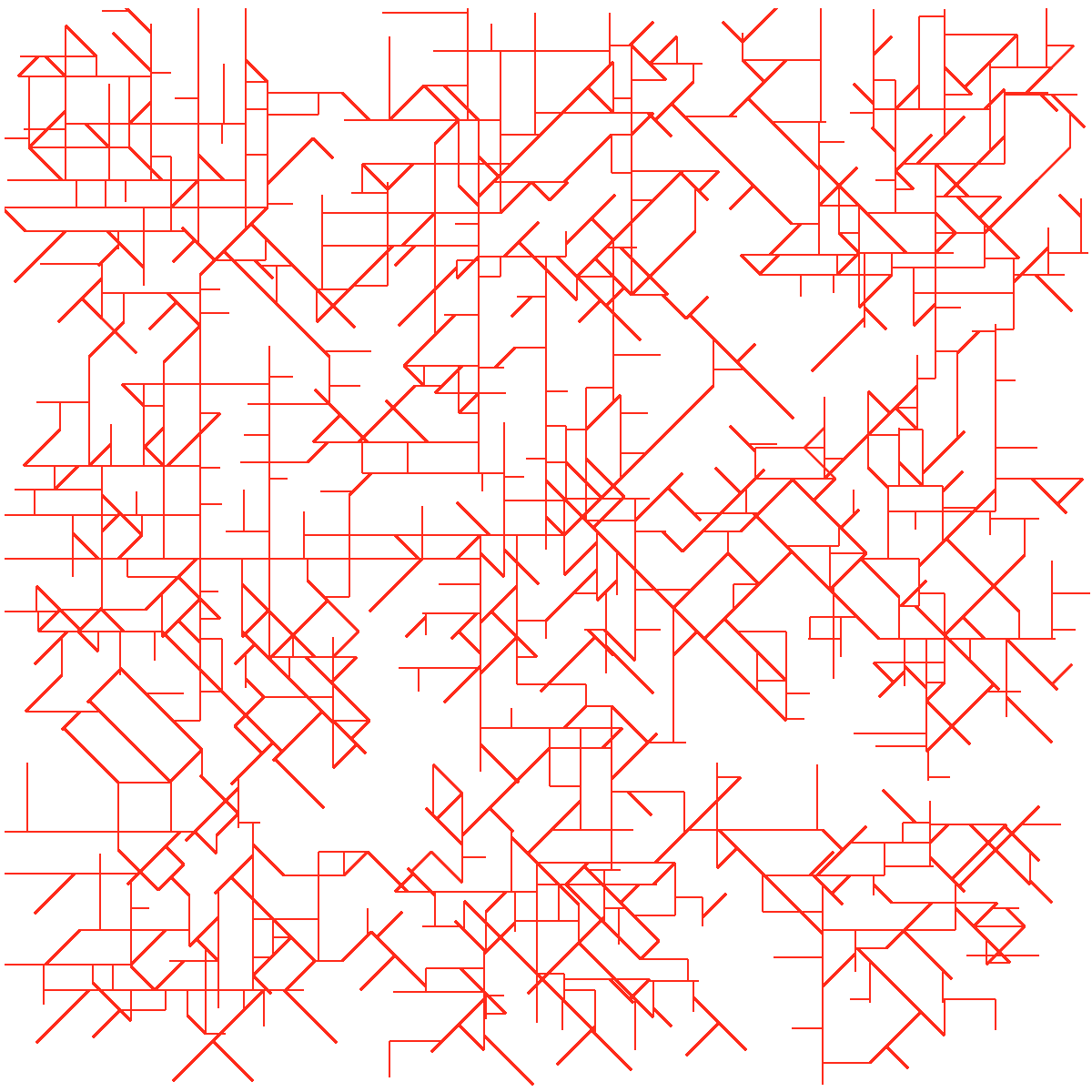}
        \caption{1000 itérations.}
    \end{subfigure}

    \caption{Premier modèle de morphogenèse. Aucune contrainte géographique ; Définition de 9 zones \textit{prolifiques} (Centre, Nord, Est, Ouest, Sud, Nord-Est, Nord-Ouest, Sud-Est, Sud-Ouest). Visualisation de la croissance du réseau sur 1000 itérations.}
    \label{fig:morpho_SF}
\end{figure}

Afin de formaliser ce processus d'exploration / division, nous avons imaginé un modèle. Ses prémices ont été posés lors d'une école thématique à l'Institut des Systèmes Complexes de Santa-Fe avec un groupe pluridisciplinaire de doctorants et post-doctorants \citep{lagesse2014can}. Nous avons participé à ce projet en construisant un modèle spécifique aux développements des villes, avec Alberto Antonioni. À partir de la géométrie d'un réseau pré-existant (initié par un unique sommet), nous avons défini une probabilité d'apparition d'un nouveau nœud et une stratégie de raccordement en fonction de sa position. Nous avons également fixé des contraintes de développement, en positionnant des zones où l'accès était moins probable ou interdit (intégration de contraintes géographiques). À celles-ci nous avons opposé des zones où le développement était plus propice (pouvant correspondre à des zones favorables à l'agriculture). Ces travaux ont abouti à un premier modèle de croissance, prémices de futurs développements, pour décrire la morphogenèse viaire (figure \ref{fig:morpho_SF}). Ce modèle de croissance devra être complexifié. La perspective souhaitée est de pouvoir reconstruire un modèle numérique de terrain avec des contraintes géographiques auxquelles répondrait la croissance du réseau.

Au sein du groupe avec lequel ce travail a été réalisé, certains (au profil plus biologiste) ont étudié plus particulièrement un modèle de croissance d'un réseau d'exploration créé par des fourmis. Nous avons ainsi pu comparer les structures de deux graphes provenant de modèles très différents (l'un multi-agent, l'autre fondé uniquement sur des probabilités liées à une géométrie). 

\FloatBarrier
\subsection*{Étude de la robustesse structurelle des réseaux viaires} 

\begin{figure}[h]
    \centering
    \begin{subfigure}[t]{.45\linewidth}
        \includegraphics[width=\textwidth]{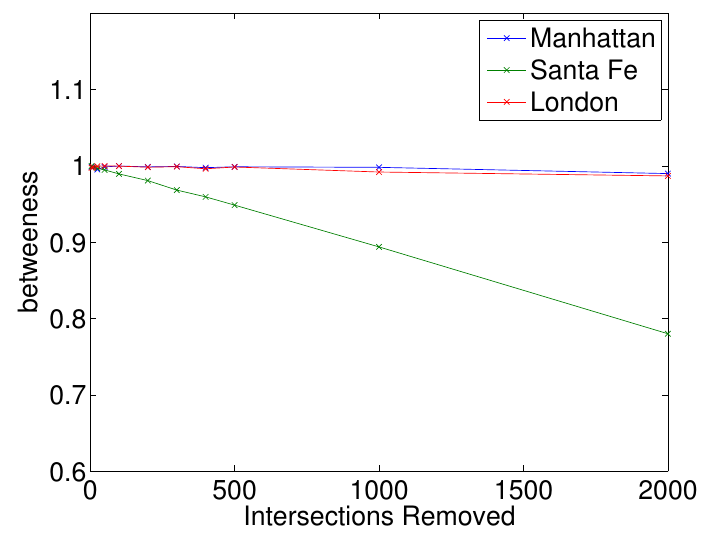}
        \caption{Retrait d'intersections.}
        \label{fig:robust_betw_1}
    \end{subfigure}    
    ~
    \begin{subfigure}[t]{.45\linewidth}
        \includegraphics[width=\textwidth]{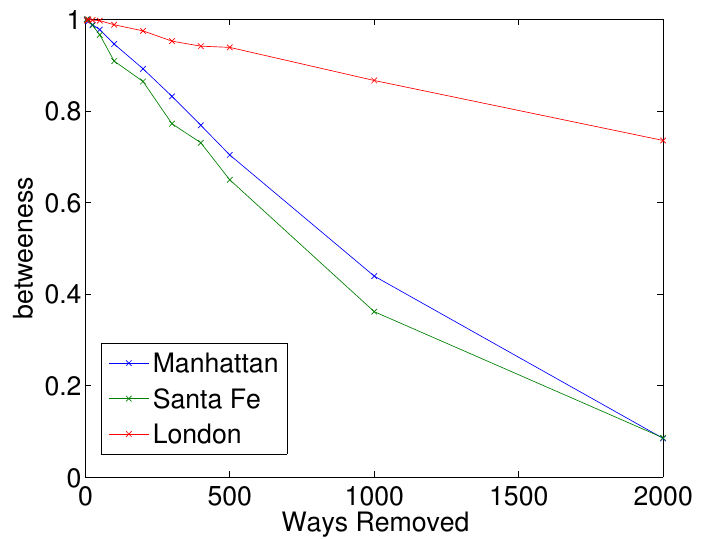}
        \caption{Retrait de voies.}
        \label{fig:robust_betw_2}
    \end{subfigure}

    \caption{Comparaison de la robustesse de la moyenne de betweenness sur les graphes viaires de Manhattan, Santa Fe et Londres au retrait aléatoire d'intersections ou de voies. \\ source : \citep{lagesse2014structural}}
    \label{fig:robust_betw}
\end{figure}

\begin{figure}[h]
    \centering
    \begin{subfigure}[t]{.45\linewidth}
        \includegraphics[width=\textwidth]{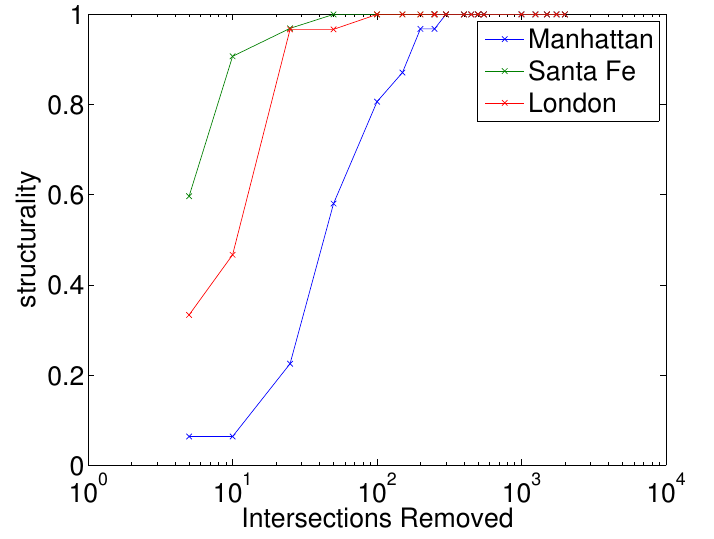}
        \caption{Retrait d'intersections.}
        \label{fig:robust_struct_1}
    \end{subfigure}
    ~
    \begin{subfigure}[t]{.45\linewidth}
        \includegraphics[width=\textwidth]{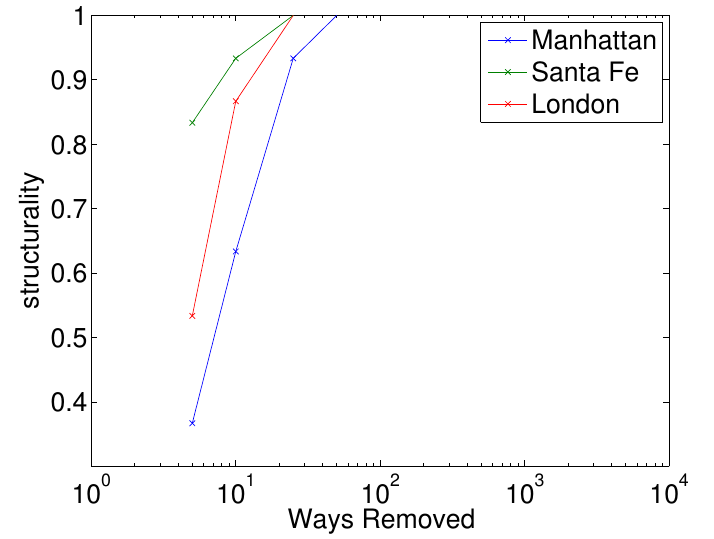}
        \caption{Retrait de voies.}
        \label{fig:robust_struct_2}
    \end{subfigure}

    \caption{Comparaison de la robustesse de la moyenne d'accessibilité sur les graphes viaires de Manhattan, Santa Fe et Londres au retrait aléatoire d'intersections ou de voies. \\ source : \citep{lagesse2014structural}}
    \label{fig:robust_struct}
\end{figure}

L'école thématique suivie à l'Institut des Systèmes Complexes de Santa-Fe a également été propice au développement d'un deuxième travail, mené avec Alireza Goudarzi \citep{lagesse2014structural}. Nous avons, dans celui-ci, étudié la robustesse de plusieurs réseaux viaires au retrait de voies ou d'intersections. Nous avons ainsi analysé, sur les villes de Manhattan, Santa Fe et Londres, les variations des moyennes des indicateurs de betweenness et d'accessibilité (appelée alors \textit{structuralité}) au cours de retraits successifs d'éléments du graphe. Les échantillons spatiaux que nous avons utilisés sont sensiblement identiques à ceux définis pour ces trois villes dans le panel de recherche du chapitre 3 de la partie II. Nous avons ainsi pu montrer la grande stabilité de la moyenne de betweenness au retrait d'intersections pour les réseaux de Manhattan et de Santa Fe, due à leur géométrie maillée (figure \ref{fig:robust_betw_1}). En revanche, le réseau de Santa Fe, moins maillé, a une moyenne de betweenness beaucoup moins robuste au retrait d'intersections. Lorsque la robustesse de cet indicateur est étudiée pour le retrait des voies, Londres se détache comme étant la ville restant la plus connectée (figure \ref{fig:robust_betw_2}). En menant cette étude sur l'indicateur d'accessibilité (dont l'augmentation de la valeur signifie une baisse de proximité entre les objets du graphe), nous observons des variations au comportement relativement similaire lors du retrait d'intersections ou de voies (figure \ref{fig:robust_struct}).

Ces travaux n'en sont qu'à leurs premiers pas. Ils seront poursuivis en utilisant une partie du travail présenté dans cette thèse, notamment les coefficients calculés sur les graphes (présentés dans le chapitre 3 de la partie II). Nous pourrons ainsi travailler sur les corrélations éventuelles entre coefficients (maillance, hétérogénéité, organicité ou réduction) et la robustesse des graphes au retrait de voies ou d'intersections. Ces recherches s'inscrivent dans le cadre de celles menées sur la vulnérabilité des réseaux \citep{albert2000error, holme2002attack, guillaume2005comparison}, et notamment des réseaux spatiaux \citep{gleyze2005vulnerabilite}.

\FloatBarrier
\subsection*{Détermination automatique des limites d'un réseau}

Tout réseau spatial s'inscrit, par définition, dans un espace. Lorsqu'il s'agit d'une feuille, les limites sont facilement identifiables et il est facile de savoir où l’échantillon s'arrête. Lorsque l'on travaille sur un réseau de craquelures sur une plaque d'argile, nous créons un effet de bord en limitant notre expérience aux bordures de la plaque, qui déterminent son \textit{territoire}. Le problème devient plus ardu lorsque nous considérons le réseau viaire. Les limites de celui-ci sont les mers et les océans. Il ne nous est pas possible de le considérer dans son intégralité, le volume de données étant alors trop important. Il est de ce fait nécessaire de fixer nous même les limites de notre échantillon de recherche. Celles-ci peuvent avoir un impact non négligeable sur les calculs des indicateurs et l'interprétation que nous pouvons en faire. 


Nous avons répondu à ce problème avec la voie, objet géographique multi-échelle, qui permet d'assurer la stabilité du calcul des indicateurs qui lui sont appliqués sur un échantillon spatial. Nous avons proposé notre analyse de stabilité, en précisant que la conservation des structures \textit{a priori} continues était garante de la robustesse de l'étude.

Le découpage d'un territoire toujours plus restreint amène un \enquote{décrochage} de la stabilité. Cette transition est intéressante et ouvre un champ de recherche non négligeable, qui pourra faire suite à ces travaux. Pour assurer la pertinence d'un découpage spatial, il serait intéressant d'apposer des limites rétroactivement. Pour cela il faudrait imaginer un processus qui calcule les voies, puis un indicateur global sur celles-ci (la closeness semble être la plus pertinente), et tester itérativement ses variations par suppression alternative d'un ou plusieurs arcs. Il serait ainsi possible de définir quantitativement le sous-échantillon le plus stable à partir d'un échantillon initial volontairement découpé plus largement. Dans cette première description, le temps de calcul nécessaire à la réalisation du protocole est problématique. Il faudra donc imaginer des algorithmes permettant de faire une analyse efficace, ou organiser une parallélisation des calculs.

Nous pouvons ainsi élargir ce travail vers une détection automatique des limites \textit{idéales} des réseaux spatiaux.

\FloatBarrier
\subsection*{Analyses approfondies d'autres types de réseaux spatiaux}

Nous pouvons ouvrir ce travail à l'analyse de différents réseaux spatiaux pour lesquels il pourrait offrir une caractérisation utile. 

Une première application porte sur les réseaux d'exploitation. Les réseaux enterrés en sont un exemple. P. Bordin travaille notamment sur la caractérisation de ces réseaux et sur la qualification des données qui les représentent \citep{bordin2013vers}. Un réseau de canalisation peut être pertinemment décrit par la nature de ses connexions. Comme c'est un réseau qui fait intervenir du matériel, nécessitant un entretien, nous pouvons imaginer une modélisation sous forme de graphe multiplexe. Les deux niveaux du graphe total représenteraient l'un, le graphe spatial, l'autre, le graphe des unités de matériel de canalisation. Ce second graphe serait uniquement topologique : les sommets représentant les unités de matériel (avec leurs attributs, dont la vétusté peut être un exemple) et les arcs symbolisant la jonction entre deux unités, sur une durée déterminée (qui peut être formalisée par une valuation de l'arc). Les connexions entre les deux niveaux de graphes peuvent symboliser le positionnement d'une unité de matériel à un endroit donné du réseau, sur une durée qui peut être ici également formalisée par une pondération. Daniella Malnar, doctorante à l'Université Paris 1 et à l'IRC (Institut de Recherche en Constructibilité), travaille ainsi sur le réseau de canalisations des fontaines de Versailles, dont les tronçons sont parfois déplacés. Cette modélisation pourrait lui être utile dans le suivi temporel de l'exploitation du réseau.

D'autre part, nous pourrons approfondir l'étude de réseaux biologiques. Le physarum, par exemple, forme des réseaux aux trajectoires optimisées. Il serait intéressant de comparer la structure de tels réseaux à celles de cheminements d'animaux, optimisant par différentes stratégies leurs trajectoires (les fourmis par exemple). En positionnant stratégiquement des portions de nourriture, des chercheurs sont parvenus à faire croître un physarum de manière identique à des réseaux de transport \citep{takamatsu2009environment}. Il serait ainsi intéressant de comparer les dynamiques de croissance de ces deux réseaux spatiaux. Adrien Hallou, doctorant à l'Université Paris Diderot, a ainsi cherché à reconstituer le réseau autoroutier français avec un physarum, établissant son parcours entre les villes (points sur lesquels étaient déposées des céréales).  Sa répulsion à un certain type de lumière permet de conditionner sa trajectoire. Nous avons ainsi réalisé une carte orographique de la France en échelle de gris, laissant donc plus ou moins passer de lumière en fonction du relief. Nous voulons ainsi observer si cela permet de retrouver un réseau plus proche de celui existant. Nous projetons de travailler sur la modélisation de la croissance de l'organisme et de la comparer à celle d'un réseau viaire.

Par la suite, les comparaisons des différentes structures des réseaux spatiaux pourront aboutir à une caractérisation automatique de ceux-ci. En effet, nous avons vu dans ce travail que s'ils observent certains comportements topologiques similaires, leurs géométries diffèrent. Nous pouvons ainsi penser à la définition d'une \textit{typologie} des réseaux spatiaux. Dans un premier temps, au sein des réseaux viaires, nous pourrions attribuer un certain type de caractéristiques géométriques à chaque tissu (quartier planifié, centre ville médiéval, lotissement, zone industrielle, etc). Ainsi, une des perspectives de ce travail serait de retrouver automatiquement dans un tissu viaire les différents types de quartier. Raphaël Charbey, ingénieur d'étude au LIAFA (Laboratoire d'Informatique Algorithmique: Fondements et Applications), a ainsi travaillé à la détection de motifs topologiques simples dans les réseaux. Nous avons pu discuter de l'application possible de ses résultats aux graphes spatiaux.

En élargissant ce raisonnement, nous pouvons imaginer une caractérisation automatique de la nature d'un réseau spatial. L'analyse d'un graphe pourrait ainsi aboutir à la définition de probabilité de type (biologique, viaire, hydrographique, etc) et de sous-types (feuille...). Nous pourrions ainsi retrouver des réseaux viaires qui ont plus ou moins de corrélations avec d'autres types de réseaux spatiaux. L'analyse de critères géométriques permettrait alors une différentiation qui est plus complexe à réaliser lorsque seule l'information topologique est étudiée \citep{beguin1997morphologie}.

Enfin, nous pourrons compléter ces recherches par l'ajout aux réseaux spatiaux de réseaux aléatoires, ayant une inscription spatiale. C'est une partie du travail débuté avec Romain Pousse, pour parvenir à déceler les mécanismes qui rendent la géométrie d'un réseau artificiel comparable à celle d'un tissu viaire. Cette comparaison demande à être approfondie et pourra être ouverte à tout type de réseaux spatiaux.

\FloatBarrier
\subsection*{Ouverture à une étude anthropologique et urbaine}

Les travaux que nous présentons ici proposent une lecture de la ville à travers ses caractéristiques topologiques et géométriques. La pertinence anthropologique de cette lecture mérite d'être exploré à la fois par des enquêtes sur des terrains spécifiques, mais également par des discussions approfondies avec les personnes spécialisées dans la structure et l'histoire de villes spécifiques. C'est notamment un des objectifs de la thèse menée par Estelle Degouys.

Nous avons ainsi, durant ces travaux, noué des liens avec le service d'urbanisme de la ville d'Avignon. Nous avons également eu d'autres contacts pour approfondir la lecture historique de villes sur différents continents, comme Cuzco (Manuel Guerra), Manaus (Damien Rietz), ou Téhéran (centre de recherche Nazar). Ces contacts sont des atouts précieux pour comprendre la lecture structurelle proposée par nos indicateurs sur ces villes. Tous nous offrent des perspectives d'analyses urbaines passionnantes.

Une recherche approfondie menée sur des tissus spécifiques nous permettrait également de travailler sur la notion d'échelle. En effet, dans ces travaux nous avons considéré principalement des villes dans leur ensemble. Il serait intéressant d'étudier plus précisément les proximités locales. Par exemple, les horizons topologiques d'un point, d'un arc ou d'une voie sur le réseau peuvent révéler des propriétés locales pertinentes (idée recueillie auprès de Philippe Bonnin). La confrontation entre cet horizon local et celui à l'échelle de toute une ville peut également apporter de nouvelles perspectives de recherche.

D'autre part, l'étude des réseaux viaires à l'échelle d'un quartier permettrait de nous pencher sur des points d'articulation précis de la ville. Comme nous l'avons vu, notamment sur le réseau viaire d'Avignon, certains carrefours portent dans leur géométrie beaucoup d'informations. Il serait ainsi intéressant d'approfondir leurs analyses (travail entamé avec Clément-Noël Douady). Cela ouvre des questions de recherche qui peuvent remettre en question la construction de l'hypergraphe des voies. Nous pourrions, par exemple, imaginer une modélisation où un arc peut appartenir à plusieurs voies.

La confrontation des études menées sur le territoire permet d'en explorer les spécificités géométriques et de les comparer avec des particularités anthropologiques. Les Zones Urbaines Sensibles sont ainsi souvent des quartiers marqués par des caractéristiques morphologiques fortes, liées à la notion d'enclavement, étudiée par Anna Cristofol (encadrée par Patricia Bordin). Nous avons réfléchi ensemble à l'apport d'une caractérisation spatiale pour la compréhension de ce phénomène. Ce travaille demande à être poursuivi.

Enfin, l'étude présentée ici, qui s'inscrit dans une démarche structurelle intemporelle, peut être adaptée à l'étude des flux. Nous avons voulu nous écarter de cette notion dans un premier temps, afin de rendre l'analyse des réseaux indépendante de leur temporalité. Cependant, cela reste une perspective d'application très intéressante.

\FloatBarrier
\subsection*{Vers le partage de la caractérisation spatiale}

Au cours de la recherche que nous avons présentée ici, nous avons développé un programme informatique (en langage C++) qui nous a permis de décomposer le réseau en objets élémentaires, y construire les voies et calculer nos indicateurs. Nous présentons en annexe \ref{ann:chap_tuto_1} une documentation simplifiée de ce programme, où nous trouvons notamment son modèle de données et la visualisation de l'interface développée. Ce déploiement est très dense, spécifique à cette recherche, et complexe à manipuler.

Pour permettre une utilisation plus large de nos indicateurs, nous avons également développé, en collaboration avec la société Oslandia, un plugin qui est directement intégrable dans un Système d'Information Géographique libre (QGIS). Celui-ci permet de nettoyer et d'analyser des fichiers de formes (\textit{shapefiles}). Le plugin construit les voies et calcule les indicateurs principaux sur celles-ci, ainsi que sur les arcs du réseau. Un tutoriel présentant ses fonctionnalités est disponible en annexe \ref{ann:chap_tuto_2}.

Enfin, nous voulons également mener ces travaux de recherche vers la réalisation d'un \enquote{jeu intelligent} (\textit{smart game}). Les prémices de celui-ci ont été posé avec Ells Campbell (qui a réalisé le \textit{smart game} : \textit{Vax - A game about epidemics}). Le jeu a pour but de proposer à ses utilisateurs la création d'une ville, à travers les axes de son réseau viaire. Il permet de gagner des points en fonction des ressources acquises : plus la ville grandit, plus cela attire de population, plus cela procure de ressources mais plus cela nécessite de développement. Pour étendre le réseau, il est nécessaire d'utiliser des ressources. Si l'expansion ne suit pas l'affluence de population, alors la ville créée sera de moins en moins attractive. L'idée est de garder un modèle très simple, où la création de nouveaux tronçons du réseau viaire répond à quelques contraintes géométriques imposées, mais où l'utilisateur est libre d'exploiter l'espace comme il l'entend, aux limites près du contexte géographique (aléatoire) dans lequel il est inclus. Plus le réseau sera efficace (en terme de proximités topologiques), plus la ville pourra accueillir de population. En collectant des données à partir de ce jeu, il serait ainsi possible à la fois d'observer les logiques que donnent les utilisateurs aux réseaux viaires qu'ils créent ; mais également de déterminer les caractéristiques topologiques des réseaux les plus efficaces. Sur un grand nombre de réseaux créés, nous pourrions ainsi disposer de statistiques importantes sur des réseaux théoriques. L'élaboration de ce \enquote{jeu intelligent} pourrait ainsi participer à la compréhension des structures viaires de nos villes.


\clearpage{\pagestyle{empty}\cleardoublepage} 

\thispagestyle{empty}
~\vfill
{\itshape Il faudrait pouvoir filmer les paysages de dos.}
\begin{flushright}
  - Jean-Luc \textsc{Godard}\label{cit:godard}
\end{flushright}
~\vfill

\clearpage{\pagestyle{empty}\cleardoublepage}
\addcontentsline{toc}{chapter}{Bibliographie}
\bibliographystyle{apalike}
\markboth{Bibliographie}{Bibliographie}
\bibliography{biblio}

\begin{thebibliography}{}

\bibitem[Adamic, 2000]{adamic2000zipf}
Adamic, L.~A. (2000).
\newblock Zipf, power-laws, and pareto-a ranking tutorial.
\newblock {\em Xerox Palo Alto Research Center, Palo Alto, CA, http://ginger.
  hpl. hp. com/shl/papers/ranking/ranking. html}.

\bibitem[Albert and Barab{\'a}si, 2002]{albert2002statistical}
Albert, R. and Barab{\'a}si, A.-L. (2002).
\newblock Statistical mechanics of complex networks.
\newblock {\em Reviews of modern physics}, 74(1):47.

\bibitem[Albert et~al., 1999]{albert1999internet}
Albert, R., Jeong, H., and Barab{\'a}si, A.-L. (1999).
\newblock Internet: Diameter of the world-wide web.
\newblock {\em Nature}, 401(6749):130--131.

\bibitem[Albert et~al., 2000]{albert2000error}
Albert, R., Jeong, H., and Barab{\'a}si, A.-L. (2000).
\newblock Error and attack tolerance of complex networks.
\newblock {\em nature}, 406(6794):378--382.

\bibitem[Amar, 1993]{amar1993pour}
Amar, G. (1993).
\newblock Pour une {\'e}cologie urbaine des transports.
\newblock In {\em Les annales de la recherche urbaine}, number 59-60, page~15.

\bibitem[Amaral et~al., 2000]{amaral2000classes}
Amaral, L. A.~N., Scala, A., Barthelemy, M., and Stanley, H.~E. (2000).
\newblock Classes of small-world networks.
\newblock {\em Proceedings of the national academy of sciences},
  97(21):11149--11152.

\bibitem[Antonioni et~al., 2014]{lagesse2014can}
Antonioni, A., Brummer, A., Edwards, M., Furtado, B.~A., Holdener, J.,
  Kalyuzhny, M., Lagesse, C., LaScalaGruenewald, D., Liu, Y., and Mehta, R.
  (2014).
\newblock Can simple models reproduce complex transportation networks: {H}uman
  cities and ant colonies.
\newblock {\em Santa Fe Institute 2014 Complex Systems Summer School
  Proceedings}.

\bibitem[Arthur, 1994]{arthur1994increasing}
Arthur, W.~B. (1994).
\newblock {\em Increasing returns and path dependence in the economy}.
\newblock University of Michigan Press.

\bibitem[Bahoken and Drevelle, 2013]{bahoken2013reseaux}
Bahoken, F. and Drevelle, M. (2013).
\newblock Les r{\'e}seaux dans le temps et dans l’espace. synth{\`e}se de la
  seconde journ{\'e}e du groupe fmr.
\newblock {\em Netcom. R{\'e}seaux, communication et territoires},
  (27-3/4):411--426.

\bibitem[Banos and Sanders, 2013]{banos2013modeliser}
Banos, A. and Sanders, L. (2013).
\newblock Mod{\'e}liser et simuler les syst{\`e}mes spatiaux en g{\'e}ographie.
\newblock {\em Mod{\'e}liser et Simuler--Epist{\'e}mologies et Pratiques des
  Mod{\`e}les et des Simulations, Ebook des {\'e}ditions Mat{\'e}riologiques},
  1:833--863.

\bibitem[Banos and Th{\'e}venin, 2011]{banos2011geographical}
Banos, A. and Th{\'e}venin, T. (2011).
\newblock {\em Geographical Information and Urban Transport Systems}.
\newblock Wiley Online Library.

\bibitem[Barab{\'a}si and Albert, 1999]{barabasi1999emergence}
Barab{\'a}si, A.-L. and Albert, R. (1999).
\newblock Emergence of scaling in random networks.
\newblock {\em science}, 286(5439):509--512.

\bibitem[Barab{\'a}si and Frangos, 2002]{barabasi2002linked}
Barab{\'a}si, A.-L. and Frangos, J. (2002).
\newblock {\em Linked: the new science of networks science of networks}.
\newblock Basic Books.

\bibitem[Barab{\'a}si and Bonabeau, 2003]{barabasi2003scale}
Barab{\'a}si, B. A.-L. and Bonabeau, E. (2003).
\newblock Scale-free.
\newblock {\em Scientific American}.

\bibitem[Baron et~al., 2010]{baron2010villes}
Baron, M., Cunningham-Sabot, E., Grasland, C., Rivi{\`e}re, D., and Van~Hamme,
  G. (2010).
\newblock Villes et r{\'e}gions europ{\'e}ennes en d{\'e}croissance.

\bibitem[Barrat et~al., 2004]{barrat2004architecture}
Barrat, A., Barthelemy, M., Pastor-Satorras, R., and Vespignani, A. (2004).
\newblock The architecture of complex weighted networks.
\newblock {\em Proceedings of the National Academy of Sciences of the United
  States of America}, 101(11):3747--3752.

\bibitem[Barrat et~al., 2008]{barrat2008dynamical}
Barrat, A., Barthelemy, M., and Vespignani, A. (2008).
\newblock {\em Dynamical processes on complex networks}.
\newblock Cambridge University Press.

\bibitem[Barth{\'e}lemy, 2011]{barthelemy2011spatial}
Barth{\'e}lemy, M. (2011).
\newblock Spatial networks.
\newblock {\em Physics Reports}, 499(1):1--101.

\bibitem[Barthelemy, 2014]{barthelemy2014discussion}
Barthelemy, M. (2014).
\newblock Discussion: Social and spatial networks.
\newblock {\em Nouvelles de l'arch{\'e}ologie}, (135):51--61.

\bibitem[Barthelemy et~al., 2013]{barthelemy2013self}
Barthelemy, M., Bordin, P., Berestycki, H., and Gribaudi, M. (2013).
\newblock Self-organization versus top-down planning in the evolution of a
  city.
\newblock {\em Scientific reports}, 3.

\bibitem[Barth{\'e}lemy and Flammini, 2006]{barthelemy2006optimal}
Barth{\'e}lemy, M. and Flammini, A. (2006).
\newblock Optimal traffic networks.
\newblock {\em Journal of Statistical Mechanics: Theory and Experiment},
  2006(07):L07002.

\bibitem[Barth{\'e}lemy and Flammini, 2008]{barthelemy2008modeling}
Barth{\'e}lemy, M. and Flammini, A. (2008).
\newblock Modeling urban street patterns.
\newblock {\em Physical review letters}, 100(13):138702.

\bibitem[Barth{\'e}lemy and Flammini, 2009]{barthelemy2009co}
Barth{\'e}lemy, M. and Flammini, A. (2009).
\newblock Co-evolution of density and topology in a simple model of city
  formation.
\newblock {\em Networks and spatial economics}, 9(3):401--425.

\bibitem[Batty, 2007]{batty2007cities}
Batty, M. (2007).
\newblock {\em Cities and complexity: understanding cities with cellular
  automata, agent-based models, and fractals}.
\newblock The MIT press.

\bibitem[Batty, 2009]{batty2009urban}
Batty, M. (2009).
\newblock Urban modeling.
\newblock {\em International Encyclopedia of Human Geography. Oxford, UK:
  Elsevier}.

\bibitem[Bavelas, 1948]{bavelas1948mathematical}
Bavelas, A. (1948).
\newblock A mathematical model for group structures.
\newblock {\em Human organization}, 7(3):16--30.

\bibitem[Bavelas, 1950]{bavelas1950communication}
Bavelas, A. (1950).
\newblock Communication patterns in task-oriented groups.
\newblock {\em The Journal of the Acoustical Society of America}, pages
  725--730.

\bibitem[Bavelas, 1951]{bavelas1951reseaux}
Bavelas, A. (1951).
\newblock {\em R{\'e}seaux de communications au sein de groupes plac{\'e}s dans
  des conditions exp{\'e}rimentales de travail}.

\bibitem[Beauchamp, 1965]{beauchamp1965improved}
Beauchamp, M.~A. (1965).
\newblock An improved index of centrality.
\newblock {\em Behavioral Science}, 10(2):161--163.

\bibitem[Beguin and Thomas, 1997]{beguin1997morphologie}
Beguin, H. and Thomas, I. (1997).
\newblock Morphologie du r{\'e}seau de communication et localisations optimales
  d'activit{\'e}s. quelle mesure pour exprimer la forme d'un r{\'e}seau?
\newblock {\em Cybergeo: European Journal of Geography}.

\bibitem[Ben-Naim et~al., 2004]{ben2004complex}
Ben-Naim, E., Frauenfelder, H., and Toroczkai, Z. (2004).
\newblock {\em Complex networks}, volume 650.
\newblock Springer Science \& Business Media.

\bibitem[Berge, 1973]{berge1973graphes}
Berge, C. (1973).
\newblock Graphes et hypergraphes.

\bibitem[Bertolini, 2007]{bertolini2007evolutionary}
Bertolini, L. (2007).
\newblock Evolutionary urban transportation planning? an exploration.
\newblock {\em Applied Evolutionary Economics and Economic Geography}, page
  279.

\bibitem[Black, 1971]{black1971iterative}
Black, W.~R. (1971).
\newblock An iterative model for generating transportation networks*.
\newblock {\em Geographical Analysis}, 3(3):283--288.

\bibitem[Blondel et~al., 2008]{blondel2008fast}
Blondel, V.~D., Guillaume, J.-L., Lambiotte, R., and Lefebvre, E. (2008).
\newblock Fast unfolding of communities in large networks.
\newblock {\em Journal of Statistical Mechanics: Theory and Experiment},
  2008(10):P10008.

\bibitem[Boccaletti et~al., 2006]{boccaletti2006complex}
Boccaletti, S., Latora, V., Moreno, Y., Chavez, M., and Hwang, D.-U. (2006).
\newblock Complex networks: Structure and dynamics.
\newblock {\em Physics reports}, 424(4):175--308.

\bibitem[Bohn et~al., 2002]{bohn2002constitutive}
Bohn, S., Andreotti, B., Douady, S., Munzinger, J., and Couder, Y. (2002).
\newblock Constitutive property of the local organization of leaf venation
  networks.
\newblock {\em Physical Review E}, 65(6):061914.

\bibitem[Bohn et~al., 2005a]{bohn2005four}
Bohn, S., Douady, S., and Couder, Y. (2005a).
\newblock Four sided domains in hierarchical space dividing patterns.
\newblock {\em Physical review letters}, 94(5):054503.

\bibitem[Bohn et~al., 2005b]{bohn2005hierarchical}
Bohn, S., Pauchard, L., and Couder, Y. (2005b).
\newblock Hierarchical crack pattern as formed by successive domain divisions.
\newblock {\em Physical Review E}, 71(4):046214.

\bibitem[Bon, 1979]{bon1979allometry}
Bon, R. (1979).
\newblock Allometry in topologic structure of transportation networks.
\newblock {\em Quality \& Quantity}, 13(4):307--326.

\bibitem[Bonnin, 2014]{bonnin2014etude}
Bonnin, Philippe et~Degouys, E. (2014).
\newblock {\em Étude de cas / KYOTO : de quelles voies parle-t-on ?}, pages
  251--257.
\newblock L'Harmattan.

\bibitem[Bonnin, 2012]{bonnin2012vivre}
Bonnin, P. (2012).
\newblock Vivre en lisi{\`e}re.
\newblock {\em M{\'e}sologiques, Etudes des milieux urbains}, pages 15--p.

\bibitem[Bonnin and Adachi, 1999]{bonnin1999aspect}
Bonnin, P. and Adachi, F. (1999).
\newblock Un aspect de la pens{\'e}e japonaise sur le paysage urbain.
\newblock {\em G{\'e}ographie et cultures}, (29):25--38.

\bibitem[Bonnin and Douady, 2013]{bonnin2013distance}
Bonnin, P. and Douady, S. (2013).
\newblock Distance et temps dans la morphogen{\`e}se du r{\'e}seau des rues.
\newblock {\em Les r{\'e}seaux dans le temps et dans l'espace, Actes de la
  deuxi{\`e}me journ{\'e}e d'{\'e}tude du groupe fmr}, pages 47--73.

\bibitem[Bordin, 2006]{bordin2006methode}
Bordin, P. (2006).
\newblock {\em M{\'e}thode d’observation multi-niveaux pour le suivi de
  ph{\'e}nom{\`e}nes g{\'e}ographiques avec un SIG}.
\newblock PhD thesis, Th{\`e}se de doctorat: Marne la Vall{\'e}e,
  Universit{\'e} de Marne la Vall{\'e}e, IGN.

\bibitem[Bordin, 2010]{bordin2010integration}
Bordin, P. (2010).
\newblock L’intégration du temps dans les {S}{I}{G}.
\newblock {\em XYZ}, (122):51--58.

\bibitem[Bordin, 2011]{bordin2011vers}
Bordin, P. (2011).
\newblock Vers une base de données d'occupation des sols à grande échelle.
\newblock {\em XYZ}, (128):17--24.

\bibitem[Bordin, 2013]{bordin2013vers}
Bordin, P. (2013).
\newblock Vers une base de connaissance sur les réseaux enterr{'e}s.
\newblock In {\em Conférence Internationale Géomatique}.

\bibitem[Bordin, 2014]{bordin2014observer}
Bordin, P. (2014).
\newblock {\em Observer des évolutions. {M}odéliser ou non les changements},
  pages 169--172.
\newblock L'Harmattan.

\bibitem[Bouttier et~al., 2003]{bouttier2003geodesic}
Bouttier, J., Di~Francesco, P., and Guitter, E. (2003).
\newblock Geodesic distance in planar graphs.
\newblock {\em Nuclear Physics B}, 663(3):535--567.

\bibitem[Brandes, 2001]{brandes2001faster}
Brandes, U. (2001).
\newblock A faster algorithm for betweenness centrality*.
\newblock {\em Journal of Mathematical Sociology}, 25(2):163--177.

\bibitem[Brandes, 2008]{brandes2008variants}
Brandes, U. (2008).
\newblock On variants of shortest-path betweenness centrality and their generic
  computation.
\newblock {\em Social Networks}, 30(2):136--145.

\bibitem[Bryan and O'Kelly, 1999]{bryan1999hub}
Bryan, D.~L. and O'Kelly, M.~E. (1999).
\newblock Hub-and-spoke networks in air transportation: an analytical review.
\newblock {\em Journal of regional science}, 39(2):275--295.

\bibitem[Buhl et~al., 2006]{buhl2006topological}
Buhl, J., Gautrais, J., Reeves, N., Sol{\'e}, R., Valverde, S., Kuntz, P., and
  Theraulaz, G. (2006).
\newblock Topological patterns in street networks of self-organized urban
  settlements.
\newblock {\em The European Physical Journal B-Condensed Matter and Complex
  Systems}, 49(4):513--522.

\bibitem[Camacho et~al., 2002]{camacho2002robust}
Camacho, J., Guimer{\`a}, R., and Amaral, L. A.~N. (2002).
\newblock Robust patterns in food web structure.
\newblock {\em Physical Review Letters}, 88(22):228102.

\bibitem[Cardillo et~al., 2006]{cardillo2006structural}
Cardillo, A., Scellato, S., Latora, V., and Porta, S. (2006).
\newblock Structural properties of planar graphs of urban street patterns.
\newblock {\em Physical Review E}, 73(6):066107.

\bibitem[Chartrand et~al., 1998]{chartrand1998graph}
Chartrand, G., Kubicki, G., and Schultz, M. (1998).
\newblock Graph similarity and distance in graphs.
\newblock {\em Aequationes Mathematicae}, 55(1-2):129--145.

\bibitem[Chouquer, 2000]{chouquer2000etude}
Chouquer, G. (2000).
\newblock {\em L'{\'e}tude des paysages: essais sur leurs formes et leur
  histoire}.
\newblock Editions Errance.

\bibitem[Clementini and Laurini, 2008]{clementini2008cadre}
Clementini, E. and Laurini, R. (2008).
\newblock Un cadre conceptuel pour mod{\'e}liser les relations spatiales.
\newblock {\em Revue des Nouvelles Technologies de l\&}, 8217:1--17.

\bibitem[Cohn and Marriott, 1958]{cohn1958networks}
Cohn, B.~S. and Marriott, M. (1958).
\newblock Networks and centres of integration in indian civilization.
\newblock {\em Journal of social Research}, 1(1):1--9.

\bibitem[Conzen, 1990]{conzen1990making}
Conzen, M.~P. (1990).
\newblock {\em The making of the American landscape}.
\newblock Routledge.

\bibitem[Courtat, 2012]{courtat2012walk}
Courtat, T. (2012).
\newblock {\em Walk on City Maps-Mathematical and Physical phenomenology of the
  City, a Geometrical approach}.
\newblock PhD thesis, Universit{\'e} Paris-Diderot-Paris VII.

\bibitem[Courtat et~al., 2011]{courtat2011mathematics}
Courtat, T., Gloaguen, C., and Douady, S. (2011).
\newblock Mathematics and morphogenesis of cities: A geometrical approach.
\newblock {\em Physical Review E}, 83(3):036106.

\bibitem[Cristofol and Bordin, 2013]{cristofol2013measuring}
Cristofol, A. and Bordin, P. (2013).
\newblock Measuring the geographical isolation of urban areas : definition and
  application of an indicator of confiment by surfaces ratio ({C}{S}{R}).
\newblock In {\em European Colloquium of Theoretical and Quantitative
  Geography}.

\bibitem[Crucitti et~al., 2006a]{crucitti2006centralityin}
Crucitti, P., Latora, V., and Porta, S. (2006a).
\newblock Centrality in networks of urban streets.
\newblock {\em Chaos: an interdisciplinary journal of nonlinear science},
  16(1):015113.

\bibitem[Crucitti et~al., 2006b]{crucitti2006centralitymeasures}
Crucitti, P., Latora, V., and Porta, S. (2006b).
\newblock Centrality measures in spatial networks of urban streets.
\newblock {\em Physical Review E}, 73(3):036125.

\bibitem[Cs{\'a}nyi and Szendr{\H{o}}i, 2004]{csanyi2004fractal}
Cs{\'a}nyi, G. and Szendr{\H{o}}i, B. (2004).
\newblock Fractal--small-world dichotomy in real-world networks.
\newblock {\em Physical Review E}, 70(1):016122.

\bibitem[Curie et~al., 2010]{curie2010simulation}
Curie, F., Perret, J., and Ruas, A. (2010).
\newblock Simulation of urban blocks densification.
\newblock In {\em 13th AGILE International Conference on Geographic Information
  Science, Guimaraes, Portugal}.

\bibitem[Dalton, 2001]{dalton2001secret}
Dalton, R. (2001).
\newblock The secret is to follow your nose: route path selection and
  angularity.

\bibitem[de~Dios~Ort{\'u}zar and Willumsen, 2001]{ortuzar2001modelling}
de~Dios~Ort{\'u}zar, J. and Willumsen, L.~G. (2001).
\newblock {\em Modelling transport}.
\newblock John Wiley \& Sons.

\bibitem[de~Solla~Price, 1965]{yu1965networks}
de~Solla~Price, D.~J. (1965).
\newblock Networks of scientific papers.
\newblock {\em Science}, 149:510--515.

\bibitem[Degouys, 2013]{estelledegouys2013}
Degouys, E. (2013).
\newblock Figures de réseaux viaires.
\newblock {\em Rapport de {M}aster {E}spaces urbains et démarches de projet,
  {P}arcours et {E}spaces publics, {I}nstitut d'{U}rbanisme de {P}aris,
  {U}{P}{E}{C}}.

\bibitem[Degouys, 2015]{degouysAPitineraire}
Degouys, E. (2015).
\newblock Itin{\'e}raires pi{\'e}tons et forme des villes. {S}trat{\'e}gies
  d’usage et caract{\'e}risation du r{\'e}seau viaire en milieu urbain.
\newblock {\em GUEZ, A., et al., Repr{\'e}senter la transformation. Ou comment
  saisir les espaces-temps habit{\'e}s.}

\bibitem[Des~Cars, 1991]{pinon1991paris}
Des~Cars, Jean et~Pinon, P. (1991).
\newblock Paris-haussmann : {\guillemotright} le pari d ‘haussmann
  {\guillemotleft}.
\newblock {\em Paris: Picard}, page 349.

\bibitem[Dijkstra, 1959]{dijkstra1959note}
Dijkstra, E.~W. (1959).
\newblock A note on two problems in connexion with graphs.
\newblock {\em Numerische mathematik}, 1(1):269--271.

\bibitem[Dill, 2004]{dill2004measuring}
Dill, J. (2004).
\newblock Measuring network connectivity for bicycling and walking.
\newblock In {\em 83rd Annual Meeting of the Transportation Research Board,
  Washington, DC}.

\bibitem[Dorogovtsev et~al., 2006]{dorogovtsev2006k}
Dorogovtsev, S.~N., Goltsev, A.~V., and Mendes, J. F.~F. (2006).
\newblock K-core organization of complex networks.
\newblock {\em Physical review letters}, 96(4):040601.

\bibitem[Dorogovtsev and Mendes, 2002]{dorogovtsev2002evolution}
Dorogovtsev, S.~N. and Mendes, J.~F. (2002).
\newblock Evolution of networks.
\newblock {\em Advances in physics}, 51(4):1079--1187.

\bibitem[Douady, 2015a]{douady2015limites}
Douady, C.-N. et~Lagesse, C. (2015a).
\newblock Limites et surprises de l’approche mathématique ; À la recherche
  de la continuité graphique.
\newblock {\em LAVUE}.

\bibitem[Douady, 2014]{cnd2014deconstruire}
Douady, C.-N. (2014).
\newblock {\em Déconstruire la mosaïque urbaine}, pages 9--142.
\newblock L'Harmattan.

\bibitem[Douady and Morphocity, 2014]{douady2014trace}
Douady, C.-N. and Morphocity (2014).
\newblock {\em De la Trace à la Trame. La voie, lecture du développement
  urbain.}
\newblock L'Harmattan.

\bibitem[Douady, 2015b]{douady2015peut}
Douady, S. (2015b).
\newblock Peut-on lire une ville à travers ses lignes ?
\newblock {\em Manzar}, 6(30).

\bibitem[Ducruet, 2010]{ducruet2010mesures}
Ducruet, C. (2010).
\newblock Les mesures locales d'un r{\'e}seau.
\newblock {\em Groupe f.m.r.}

\bibitem[Dumenieu et~al., 2013]{dumenieu2013methode}
Dumenieu, B., Perret, J., and Ruas, A. (2013).
\newblock Une m{\'e}thode de construction de donn{\'e}es spatio-temporelles
  pour l$\backslash$'{\'e}tude de l$\backslash $'espace urbain ancien.
\newblock In {\em Colloque International de Géomatique et d'Analyse Spatiale
  SAGEO'13 (23-26 September 2013) Key: citeulike:12701132}.

\bibitem[Dunne et~al., 2002a]{dunne2002food}
Dunne, J.~A., Williams, R.~J., and Martinez, N.~D. (2002a).
\newblock Food-web structure and network theory: the role of connectance and
  size.
\newblock {\em Proceedings of the National Academy of Sciences},
  99(20):12917--12922.

\bibitem[Dunne et~al., 2002b]{dunne2002network}
Dunne, J.~A., Williams, R.~J., and Martinez, N.~D. (2002b).
\newblock Network structure and biodiversity loss in food webs: robustness
  increases with connectance.
\newblock {\em Ecology letters}, 5(4):558--567.

\bibitem[Edmonds and Moss, 2005]{edmonds2005kiss}
Edmonds, B. and Moss, S. (2005).
\newblock {\em From KISS to KIDS--an ‘anti-simplistic’modelling approach}.
\newblock Springer.

\bibitem[Egenhofer and Herring, 1990]{egenhofer1990categorizing}
Egenhofer, M.~J. and Herring, J. (1990).
\newblock Categorizing binary topological relations between regions, lines, and
  points in geographic databases.
\newblock {\em The}, 9:94--1.

\bibitem[Errera, 1925]{errera1925contribution}
Errera, A. (1925).
\newblock Une contribution au probl{\`e}me des quatre couleurs.
\newblock {\em Bulletin de la Soci{\'e}t{\'e} Math{\'e}matique de France},
  53:42--55.

\bibitem[Foltete and Piombini, 2007]{foltete2007urban}
Foltete, J.-C. and Piombini, A. (2007).
\newblock Urban layout, landscape features and pedestrian usage.
\newblock {\em Landscape and Urban Planning}, 81(3):225--234.

\bibitem[Foltête et~al., 2002]{foltete2002structures}
Foltête, J.-C., Genre-{G}randpierre, C., Houot, H., and Flitti, M. (2002).
\newblock Structures urbaines, offres de transport et comportement de
  mobilit{\'e}.
\newblock In {\em Rapport de recherche ACI Ville–99V358, Minist{\`e}re de
  l’Enseignement de la Recherche et de la Technologie}.

\bibitem[Fran{\c{c}}ois et~al., 2002]{franccois2002espace}
Fran{\c{c}}ois, J.-C., Grasland, C., and Le~Goix, R. (2002).
\newblock L'espace compte.
\newblock {\em L’Espace g{\'e}ographique}, (4):355--356.

\bibitem[Freeman, 1977]{freeman1977set}
Freeman, L.~C. (1977).
\newblock A set of measures of centrality based on betweenness.
\newblock {\em Sociometry}, pages 35--41.

\bibitem[Fullerton, 1975]{fullerton1975development}
Fullerton, B. (1975).
\newblock {\em The development of British transport networks}.
\newblock Oxford University Press.

\bibitem[Gabaix, 1999]{gabaix1999zipf}
Gabaix, X. (1999).
\newblock Zipf's law and the growth of cities.
\newblock {\em American Economic Review}, pages 129--132.

\bibitem[Garrison, 1960]{garrison1960connectivity}
Garrison, W.~L. (1960).
\newblock Connectivity of the interstate highway system.
\newblock {\em Papers in Regional Science}, 6(1):121--137.

\bibitem[Garrison and Levinson, 2005]{garrison2005transportation}
Garrison, W.~L. and Levinson, D.~M. (2005).
\newblock {\em The transportation experience: policy, planning, and
  deployment}.
\newblock Oxford university press.

\bibitem[Garrison and Marble, 1962]{garrison1962structure}
Garrison, W.~L. and Marble, D.~F. (1962).
\newblock The structure of transportation networks.
\newblock Technical report, DTIC Document.

\bibitem[Gastner and Newman, 2006]{gastner2006spatial}
Gastner, M.~T. and Newman, M.~E. (2006).
\newblock The spatial structure of networks.
\newblock {\em The European Physical Journal B-Condensed Matter and Complex
  Systems}, 49(2):247--252.

\bibitem[Gauthier, 1966]{gauthier1966highway}
Gauthier, H.~L. (1966).
\newblock Highway development and urban growth in sao paulo, brazil: a network
  analysis.

\bibitem[Genre-Grandpierre, 2000]{genre2000forme}
Genre-Grandpierre, C. (2000).
\newblock {\em Forme et fonctionnement des r{\'e}seaux de transport: approche
  fractale et r{\'e}flexions sur l'am{\'e}nagement des villes}.
\newblock PhD thesis, Besan{\c{c}}on.

\bibitem[Genre-{G}randpierre, 2001]{genre2001structure}
Genre-{G}randpierre, C. (2001).
\newblock La structure topologique et fonctionnelle des réseaux routiers
  urbains comme déterminant de la géographie des flux de déplacements.
\newblock In {\em Actes du colloque Géopoint 2000, Avignon}, pages 61--67.

\bibitem[Genre-{G}randpierre and Banos, 2010]{genre2010new}
Genre-{G}randpierre, C. and Banos, A. (2010).
\newblock {\em New types of metrics for urban road networks explored with {S}3:
  an agent-based simulation platform}, volume 325, pages 267--286.
\newblock BAI QUAN, FUKUTA NAOKI.

\bibitem[Genre-Grandpierre and Folt{\^e}te, 2003]{genre2003morphologie}
Genre-Grandpierre, C. and Folt{\^e}te, J.-C. (2003).
\newblock Morphologie urbaine et mobilit{\'e} en marche {\`a} pied.
\newblock {\em Cybergeo: European Journal of Geography}.

\bibitem[Gibbons, 1985]{gibbons1985algorithmic}
Gibbons, A. (1985).
\newblock {\em Algorithmic graph theory}.
\newblock Cambridge University Press.

\bibitem[Gleyze, 2005]{gleyze2005vulnerabilite}
Gleyze, J.-F. (2005).
\newblock {\em La vuln{\'e}rabilit{\'e} structurelle des r{\'e}seaux de
  transport dans un contexte de risques}.
\newblock PhD thesis, Universit{\'e} Paris-Diderot-Paris VII.

\bibitem[Golledge, 1997]{golledge1997spatial}
Golledge, R.~G. (1997).
\newblock {\em Spatial behavior: A geographic perspective}.
\newblock Guilford Press.

\bibitem[Golledge, 1999]{golledge1999human}
Golledge, R.~G. (1999).
\newblock Human wayfinding and cognitive maps.
\newblock {\em Wayfinding behavior: Cognitive mapping and other spatial
  processes}, pages 5--45.

\bibitem[Grasland and Hamez, 2005]{grasland2005vers}
Grasland, C. and Hamez, G. (2005).
\newblock Vers la construction d'un indicateur de coh{\'e}sion territoriale
  europ{\'e}en?

\bibitem[Guillaume et~al., 2005]{guillaume2005comparison}
Guillaume, J.-L., Latapy, M., and Magnien, C. (2005).
\newblock Comparison of failures and attacks on random and scale-free networks.
\newblock In {\em Principles of Distributed Systems}, pages 186--196. Springer.

\bibitem[Guimera et~al., 2005]{guimera2005worldwide}
Guimera, R., Mossa, S., Turtschi, A., and Amaral, L.~N. (2005).
\newblock The worldwide air transportation network: Anomalous centrality,
  community structure, and cities' global roles.
\newblock {\em Proceedings of the National Academy of Sciences},
  102(22):7794--7799.

\bibitem[Hage and Harary, 1995]{hage1995eccentricity}
Hage, P. and Harary, F. (1995).
\newblock Eccentricity and centrality in networks.
\newblock {\em Social networks}, 17(1):57--63.

\bibitem[Haggett and Chorley, 1969]{haggett1969network}
Haggett, P. and Chorley, R.~J. (1969).
\newblock {\em Network analysis in geography}, volume~67.
\newblock Edward Arnold London.

\bibitem[Hallot, 2012]{hallot2012identite}
Hallot, P. (2012).
\newblock {\em L'identit{\'e} à travers l'espace et le temps. Vers une
  définition de l'identité et des relations spatio-temporelles entre objets
  géographiques.}
\newblock PhD thesis, Universit{\'e} de Li{\`e}ge.

\bibitem[Hamaina et~al., 2012]{hamaina2012caracterisation}
Hamaina, R., Leduc, T., and Moreau, G. (2012).
\newblock Caract{\'e}risation des tissus urbains {\`a} partir de l’analyse
  structurelle des r{\'e}seaux viaires.
\newblock {\em Cybergeo: European Journal of Geography}.

\bibitem[Hanson and Giuliano, 1986]{hanson1986geography}
Hanson, S. and Giuliano, G. (1986).
\newblock {\em The geography of urban transportation}.
\newblock Guilford Press.

\bibitem[Harary and Norman, 1960]{harary1960some}
Harary, F. and Norman, R.~Z. (1960).
\newblock Some properties of line digraphs.
\newblock {\em Rendiconti del Circolo Matematico di Palermo}, 9(2):161--168.

\bibitem[Henson and Essex, 2003]{henson2003conception}
Henson, R. and Essex, S. (2003).
\newblock Conception, organisation et {\'e}valuation de r{\'e}seaux de
  transport locaux durables.
\newblock {\em Revue internationale des sciences sociales}, 176(2):243--260.

\bibitem[Hillier, 1996]{hillier1996space}
Hillier, B. (1996).
\newblock Space is the machine: a configurational theory of architecture.
\newblock {\em Cambridge, Cambridge University Press}.

\bibitem[Hillier, 1999]{hillier1999centrality}
Hillier, B. (1999).
\newblock Centrality as a process: accounting for attraction inequalities in
  deformed grids.
\newblock {\em Urban Design International}, 4(3-4):107--127.

\bibitem[Hillier, 2006]{hillier2006studying}
Hillier, B. (2006).
\newblock Studying cities to learn about minds how geometric intuitions shape
  urban space and make it work.

\bibitem[Hillier, 2009]{hillier2009spatial}
Hillier, B. (2009).
\newblock Spatial sustainability in cities: organic patterns and sustainable
  forms.

\bibitem[Hillier and Hanson, 1984]{hillier1984social}
Hillier, B. and Hanson, J. (1984).
\newblock The social logic of space.
\newblock {\em Cambridge, Cambridge University Press}.

\bibitem[Hillier and Iida, 2005]{hillier2005network}
Hillier, B. and Iida, S. (2005).
\newblock Network and psychological effects in urban movement.
\newblock In {\em Spatial information theory}, pages 475--490. Springer.

\bibitem[Hillier et~al., 1976]{hillier1976space}
Hillier, B., Leaman, A., Stansall, P., and Bedford, M. (1976).
\newblock Space syntax.
\newblock {\em Environment and Planning B: Planning and Design}, 3(2):147--185.

\bibitem[Hillier et~al., 1993]{hillier1993natural}
Hillier, B., Penn, A., Hanson, J., Grajewski, T., and Xu, J. (1993).
\newblock Natural movement-or, configuration and attraction in urban pedestrian
  movement.
\newblock {\em Environ Plann B}, 20(1):29--66.

\bibitem[Hillier and Vaughan, 2007]{hillier2007city}
Hillier, B. and Vaughan, L. (2007).
\newblock The city as one thing.
\newblock {\em Progress in Planning}, 67(3):205--230.

\bibitem[Holme et~al., 2002]{holme2002attack}
Holme, P., Kim, B.~J., Yoon, C.~N., and Han, S.~K. (2002).
\newblock Attack vulnerability of complex networks.
\newblock {\em Physical Review E}, 65(5):056109.

\bibitem[Holme and Saram{\"a}ki, 2012]{holme2012temporal}
Holme, P. and Saram{\"a}ki, J. (2012).
\newblock Temporal networks.
\newblock {\em Physics reports}, 519(3):97--125.

\bibitem[Hu and Zhu, 2009]{hu2009empirical}
Hu, Y. and Zhu, D. (2009).
\newblock Empirical analysis of the worldwide maritime transportation network.
\newblock {\em Physica A: Statistical Mechanics and its Applications},
  388(10):2061--2071.

\bibitem[Huard, 2013]{huard2013atlas}
Huard, M. (2013).
\newblock Atlas {H}istorique de {P}aris.
\newblock http://paris-atlas-historique.fr.

\bibitem[Ito et~al., 2001]{ito2001comprehensive}
Ito, T., Chiba, T., Ozawa, R., Yoshida, M., Hattori, M., and Sakaki, Y. (2001).
\newblock A comprehensive two-hybrid analysis to explore the yeast protein
  interactome.
\newblock {\em Proceedings of the National Academy of Sciences},
  98(8):4569--4574.

\bibitem[Janelle, 1969]{janelle1969spatial}
Janelle, D.~G. (1969).
\newblock Spatial reorganization: a model and concept.
\newblock {\em Annals of the Association of American Geographers},
  59(2):348--364.

\bibitem[Jeong et~al., 2001]{jeong2001lethality}
Jeong, H., Mason, S.~P., Barab{\'a}si, A.-L., and Oltvai, Z.~N. (2001).
\newblock Lethality and centrality in protein networks.
\newblock {\em Nature}, 411(6833):41--42.

\bibitem[Jeong et~al., 2000]{jeong2000large}
Jeong, H., Tombor, B., Albert, R., Oltvai, Z.~N., and Barab{\'a}si, A.-L.
  (2000).
\newblock The large-scale organization of metabolic networks.
\newblock {\em Nature}, 407(6804):651--654.

\bibitem[Jiang and Claramunt, 2002]{jiang2002integration}
Jiang, B. and Claramunt, C. (2002).
\newblock Integration of space syntax into {G}{I}{S}: new perspectives for
  urban morphology.
\newblock {\em Transactions in GIS}, 6(3):295--309.

\bibitem[Jiang and Claramunt, 2004]{jiang2004topological}
Jiang, B. and Claramunt, C. (2004).
\newblock Topological analysis of urban street networks.
\newblock {\em Environment and Planning B}, 31(1):151--162.

\bibitem[John, 2000]{john2000social}
John, S. (2000).
\newblock Social network analysis: A handbook.

\bibitem[Juillard, 1953]{juillard1953formes}
Juillard, E. (1953).
\newblock Formes de structure parcellaire dans la plaine d'alsace.
\newblock {\em Bulletin de l'Association de g{\'e}ographes fran{\c{c}}ais},
  30(232-233):72--77.

\bibitem[Kalapala et~al., 2003]{kalapala2003structure}
Kalapala, V., Sanwalani, V., and Moore, C. (2003).
\newblock The structure of the united states road network.
\newblock {\em Preprint, University of New Mexico}.

\bibitem[Kansky and Danscoine, 1989]{kansky1989measures}
Kansky, K. and Danscoine, P. (1989).
\newblock Measures of network structure.
\newblock {\em Flux}, 5(1):89--121.

\bibitem[Kansky, 1963]{kansky1963structure}
Kansky, K.~J. (1963).
\newblock {\em Structure of transportation networks: relationships between
  network geometry and regional characteristics}.
\newblock PhD thesis, University of Chicago.

\bibitem[Keller and Segel, 1970]{keller1970initiation}
Keller, E.~F. and Segel, L.~A. (1970).
\newblock Initiation of slime mold aggregation viewed as an instability.
\newblock {\em Journal of Theoretical Biology}, 26(3):399--415.

\bibitem[Krausz, 1943]{krausz1943demonstration}
Krausz, J. (1943).
\newblock Démonstration nouvelle d'un théorème de whitney sur les réseaux.
\newblock {\em Mat. Fiz. Lapok}, 50:75--89.

\bibitem[Lachene, 1965]{lachene1965networks}
Lachene, R. (1965).
\newblock Networks and the location of economic activities.
\newblock In {\em Papers of the Regional Science Association}, volume~14, pages
  183--196. Springer.

\bibitem[Lagesse, 2015]{lagesse2015lire}
Lagesse, C. (2015).
\newblock Lire la ville à travers ses rues. méthodologie de construction
  d’un objet complexe et d’indicateurs liés. application à un quartier de
  la ville de téhéran.
\newblock {\em Manzar}, 6(30).

\bibitem[Lagesse et~al., 2015]{lagesse2015spatial}
Lagesse, C., Bordin, P., and Douady, S. (2015).
\newblock A spatial multi-scale object to analyze road networks.
\newblock {\em Network Science}, 3(01):156--181.

\bibitem[Lagesse and Goudarzi, 2014]{lagesse2014structural}
Lagesse, C. and Goudarzi, A. (2014).
\newblock Structural robustness in road networks.
\newblock {\em Santa Fe Institute 2014 Complex Systems Summer School
  Proceedings}.

\bibitem[L{\"a}mmer et~al., 2006]{lammer2006scaling}
L{\"a}mmer, S., Gehlsen, B., and Helbing, D. (2006).
\newblock Scaling laws in the spatial structure of urban road networks.
\newblock {\em Physica A: Statistical Mechanics and its Applications},
  363(1):89--95.

\bibitem[Langlois and Denain, 1996]{langlois1996cartographie}
Langlois, P. and Denain, J.-C. (1996).
\newblock Cartographie en anamorphose.
\newblock {\em Cybergeo: European Journal of Geography}.

\bibitem[Latora and Marchiori, 2002]{latora2002boston}
Latora, V. and Marchiori, M. (2002).
\newblock Is the boston subway a small-world network?
\newblock {\em Physica A: Statistical Mechanics and its Applications},
  314(1):109--113.

\bibitem[Leavitt, 1951]{leavitt1951some}
Leavitt, H.~J. (1951).
\newblock Some effects of certain communication patterns on group performance.
\newblock {\em The Journal of Abnormal and Social Psychology}, 46(1):38.

\bibitem[Lee and Lee, 1998]{lee1998new}
Lee, K. and Lee, H.-Y. (1998).
\newblock A new algorithm for graph-theoretic nodal accessibility measurement.
\newblock {\em Geographical Analysis}, 30(1):1--14.

\bibitem[Lee and Holme, 2012]{lee2012exploring}
Lee, S.~H. and Holme, P. (2012).
\newblock Exploring maps with greedy navigators.
\newblock {\em Physical review letters}, 108(12):128701.

\bibitem[Lee~Rodgers and Nicewander, 1988]{lee1988thirteen}
Lee~Rodgers, J. and Nicewander, W.~A. (1988).
\newblock Thirteen ways to look at the correlation coefficient.
\newblock {\em The American Statistician}, 42(1):59--66.

\bibitem[Levinson, 2005]{levinson2005evolution}
Levinson, D. (2005).
\newblock The evolution of transport networks.
\newblock {\em Handbook of Transport Strategy, Policy and Institutions},
  6:175--188.

\bibitem[Levinson and Yerra, 2006]{levinson2006self}
Levinson, D. and Yerra, B. (2006).
\newblock Self-organization of surface transportation networks.
\newblock {\em Transportation Science}, 40(2):179--188.

\bibitem[Liebowitz and Margolis, 1995]{liebowitz1995path}
Liebowitz, S.~J. and Margolis, S.~E. (1995).
\newblock Path dependence, lock-in, and history.
\newblock {\em Journal of Law, Economics, and Organization}, 11(1):205--226.

\bibitem[Linial et~al., 1995]{linial1995geometry}
Linial, N., London, E., and Rabinovich, Y. (1995).
\newblock The geometry of graphs and some of its algorithmic applications.
\newblock {\em Combinatorica}, 15(2):215--245.

\bibitem[Livet et~al., 2014]{livet2014diversite}
Livet, P., Phan, D., and Sanders, L. (2014).
\newblock Diversit{\'e} et compl{\'e}mentarit{\'e} des mod{\`e}les multi-agents
  en sciences sociales.
\newblock {\em Revue fran{\c{c}}aise de sociologie}, 55(4):689--729.

\bibitem[Louf and Barthelemy, 2013]{louf2013modeling}
Louf, R. and Barthelemy, M. (2013).
\newblock Modeling the polycentric transition of cities.
\newblock {\em Physical review letters}, 111(19):198702.

\bibitem[Lowe and Moryadas, 1975]{lowe1975geography}
Lowe, J.~C. and Moryadas, S. (1975).
\newblock The geography of movement.

\bibitem[Mapping{H}istory, 2014]{mappinghistory}
Mapping{H}istory (2014).
\newblock Mapping {H}istory (formerly {M}apsplusmotion).
\newblock Accessed: 2014.

\bibitem[Marchand, 1974]{marchand1974pedestrian}
Marchand, B. (1974).
\newblock Pedestrian traffic planning and the perception of the urban
  environment: a french example.
\newblock {\em Environment and Planning A}, 6(5):491--507.

\bibitem[Mariolis, 1975]{mariolis1975interlocking}
Mariolis, P. (1975).
\newblock Interlocking directorates and control of corporations: The theory of
  bank control.
\newblock {\em Social Science Quarterly}, pages 425--439.

\bibitem[Marshall, 2004]{marshall2004streets}
Marshall, S. (2004).
\newblock {\em Streets and patterns}.
\newblock Routledge.

\bibitem[Maslov and Sneppen, 2002]{maslov2002specificity}
Maslov, S. and Sneppen, K. (2002).
\newblock Specificity and stability in topology of protein networks.
\newblock {\em Science}, 296(5569):910--913.

\bibitem[Maslow, 1943]{maslow1943theory}
Maslow, A.~H. (1943).
\newblock A theory of human motivation.
\newblock {\em Psychological review}, 50(4):370.

\bibitem[Mateo-Babiano and Ieda, 2005]{mateo2005street}
Mateo-Babiano, I. and Ieda, H. (2005).
\newblock Street space renaissance: A spatio-historical survey of two asian
  cities.
\newblock {\em Journal of the Eastern Asia Society for Transportation Studies},
  6:4317--4332.

\bibitem[Mateo-Babiano et~al., 2010]{mateo2010sidewalk}
Mateo-Babiano, I.~B., Hitoshi, I., and Eng, D. (2010).
\newblock Sidewalk sustainability through need-assessment of street users in
  asian cities.
\newblock In {\em 12th World Conference on Transport Research (WCTR 2010)},
  pages 1--19. World Conference on Transport Research Society.

\bibitem[Mermet, 2011]{mermet2011aide}
Mermet, {\'E}. (2011).
\newblock {\em Aide {\`a} l'exploration des propri{\'e}t{\'e}s structurelles
  d'un r{\'e}seau de transport. Conception d'un modèle pour l'analyse, la
  visualisation et l'exploration d'un r{\'e}seau de transport.}
\newblock PhD thesis, Universit{\'e} Paris-Est.

\bibitem[Milo et~al., 2002]{milo2002network}
Milo, R., Shen-Orr, S., Itzkovitz, S., Kashtan, N., Chklovskii, D., and Alon,
  U. (2002).
\newblock Network motifs: simple building blocks of complex networks.
\newblock {\em Science}, 298(5594):824--827.

\bibitem[Mizruchi, 1982]{mizruchi1982american}
Mizruchi, M.~S. (1982).
\newblock {\em The American corporate network, 1904-1974}, volume 138.
\newblock Sage Publications, Inc.

\bibitem[Mohring, 1961]{mohring1961land}
Mohring, H. (1961).
\newblock Land values and the measurement of highway benefits.
\newblock {\em The Journal of Political Economy}, pages 236--249.

\bibitem[Montoya and Sol{\'e}, 2002]{montoya2002small}
Montoya, J.~M. and Sol{\'e}, R.~V. (2002).
\newblock Small world patterns in food webs.
\newblock {\em Journal of theoretical biology}, 214(3):405--412.

\bibitem[Moreno et~al., 1934]{moreno1934shall}
Moreno, J.~L., Jennings, H.~H., et~al. (1934).
\newblock Who shall survive?

\bibitem[Morrill, 1965]{morrill1965migration}
Morrill, R.~L. (1965).
\newblock {\em Migration and the spread and growth of urban settlement},
  volume~26.
\newblock Royal University of Lund, Sweden, Dept. of Geography;[for sale at]
  CWK Gleerup.

\bibitem[Nakagaki, 2001]{nakagaki2001smart}
Nakagaki, T. (2001).
\newblock Smart behavior of true slime mold in a labyrinth.
\newblock {\em Research in Microbiology}, 152(9):767--770.

\bibitem[Newell, 1980]{newell1980traffic}
Newell, G.~F. (1980).
\newblock {\em Traffic flow on transportation networks}.
\newblock Number Monograph.

\bibitem[Newman, 2010]{newman2010networks}
Newman, M. (2010).
\newblock {\em Networks: an introduction}.
\newblock Oxford University Press.

\bibitem[Newman et~al., 2006]{newman2006structure}
Newman, M., Barabasi, A.-L., and Watts, D.~J. (2006).
\newblock {\em The structure and dynamics of networks}.
\newblock Princeton University Press.

\bibitem[Newman, 2001]{newman2001scientific}
Newman, M.~E. (2001).
\newblock Scientific collaboration networks.
\newblock {\em Physical review E}, 64(1):016131.

\bibitem[Newman, 2003]{newman2003structure}
Newman, M.~E. (2003).
\newblock The structure and function of complex networks.
\newblock {\em SIAM review}, 45(2):167--256.

\bibitem[Newman, 2005]{newman2005power}
Newman, M.~E. (2005).
\newblock Power laws, pareto distributions and zipf's law.
\newblock {\em Contemporary physics}, 46(5):323--351.

\bibitem[Noizet et~al., 2013a]{noizet2013paris}
Noizet, H., Bove, B., and Costa, L. (2013a).
\newblock Paris de parcelles en pixels.

\bibitem[Noizet et~al., 2008]{noizet2008alpage}
Noizet, H., Dallo, A., Blary, G.-X., Costa, L., and Pouget, F. (2008).
\newblock Alpage: towards the setting-up of a collaborative work tool.
\newblock {\em Archeologia e Calcolatori}, 19:86--95.

\bibitem[Noizet et~al., 2013b]{noizet2013resilience}
Noizet, H., Mirlou, L., and Robert, S. (2013b).
\newblock La r{\'e}silience des formes.
\newblock {\em Etudes rurales}, (1):191--219.

\bibitem[Noulas et~al., 2012]{noulas2012tale}
Noulas, A., Scellato, S., Lambiotte, R., Pontil, M., and Mascolo, C. (2012).
\newblock A tale of many cities: universal patterns in human urban mobility.
\newblock {\em PloS one}, 7(5):e37027.

\bibitem[P., 2002]{bordin2002sig}
P., B. (2002).
\newblock {\em {S}{I}{G}, outil concepts et données}.
\newblock Hermès Sciences – Lavoisier.

\bibitem[Padgett and Ansell, 1993]{padgett1993robust}
Padgett, J.~F. and Ansell, C.~K. (1993).
\newblock Robust action and the rise of the medici, 1400-1434.
\newblock {\em American journal of sociology}, pages 1259--1319.

\bibitem[Pailhous, 1970]{pailhous1970representation}
Pailhous, J. (1970).
\newblock {\em Repr{\'e}sentation de l'espace urbain et cheminements :
  l'exemple du chauffeur de taxi}.
\newblock Presses Universitaires de France.

\bibitem[Parlebas, 1972]{parlebas1972centralite}
Parlebas, P. (1972).
\newblock Centralit{\'e} et compacit{\'e} d'un graphe.
\newblock {\em Mathematiques et sciences humaines}, 39:5--26.

\bibitem[Penn, 2003]{penn2003space}
Penn, A. (2003).
\newblock Space syntax and spatial cognition or why the axial line?
\newblock {\em Environment and behavior}, 35(1):30--65.

\bibitem[Perna et~al., 2011]{perna2011characterization}
Perna, A., Kuntz, P., and Douady, S. (2011).
\newblock Characterization of spatial networklike patterns from junction
  geometry.
\newblock {\em Physical Review E}, 83(6):066106.

\bibitem[Perret et~al., 2010]{perret2010systeme}
Perret, J., Curie, F., Gaffuri, J., and Ruas, A. (2010).
\newblock Un syst{\`e}me multi-agent pour la simulation des dynamiques
  urbaines.
\newblock In {\em JFSMA}, pages 205--213.

\bibitem[Piombini, 2006]{piombini2006modelisation}
Piombini, A. (2006).
\newblock {\em Mod{\'e}lisation des choix d'itin{\'e}raires p{\'e}destres en
  milieu urbain. Approche g{\'e}ographique et paysag{\`e}re}.
\newblock PhD thesis, Universit{\'e} de Franche-Comt{\'e}.

\bibitem[Podani et~al., 2001]{podani2001comparable}
Podani, J., Oltvai, Z.~N., Jeong, H., Tombor, B., Barab{\'a}si, A.-L., and
  Szathmary, E. (2001).
\newblock Comparable system-level organization of archaea and eukaryotes.
\newblock {\em Nature genetics}, 29(1):54--56.

\bibitem[{P}olytechnique (France), 1856]{ecole1856journal}
{P}olytechnique (France), E. (1856).
\newblock {\em Journal}.
\newblock Number vol.~21. Ecole polytechnique.

\bibitem[Porta et~al., 2006]{porta2006network}
Porta, S., Crucitti, P., and Latora, V. (2006).
\newblock The network analysis of urban streets: a dual approach.
\newblock {\em Physica A: Statistical Mechanics and its Applications},
  369(2):853--866.

\bibitem[Pousse, 2015]{romainpousse2015}
Pousse, R. (2015).
\newblock Développement théorique et numérique de processus de
  multiplication du réseau viaire \enquote{rigide} reproduisant les
  statistiques observées sur des villes.
\newblock {\em Rapport de Master2}.

\bibitem[Pred, 1966]{pred1966spatial}
Pred, A.~R. (1966).
\newblock {\em The spatial dynamics of US urban-industrial growth, 1800-1914:
  interpretive and theoretical essays}, volume~6.
\newblock Cambridge, Mass.: MIT Press.

\bibitem[Pumain, 1982]{pumain1982chemin}
Pumain, D. (1982).
\newblock Chemin de fer et croissance urbaine en france au xix e si{\`e}cle.
\newblock In {\em Annales de g{\'e}ographie}, pages 529--550. JSTOR.

\bibitem[Rapoport and Horvath, 1961]{rapoport1961study}
Rapoport, A. and Horvath, W.~J. (1961).
\newblock A study of a large sociogram.
\newblock {\em Behavioral Science}, 6(4):279--291.

\bibitem[Rietveld, 1995]{rietveld1995some}
Rietveld, P. (1995).
\newblock Some notes on interconnectivity in transport networks.
\newblock In {\em Overcoming isolation}, pages 18--28. Springer.

\bibitem[Rietveld, 1997]{rietveld1997policy}
Rietveld, P. (1997).
\newblock Policy aspects of interconnectivity in networks.
\newblock {\em Networks in transport and communication: a policy approach},
  pages 177--92.

\bibitem[Rimmer, 1967]{rimmer1967changing}
Rimmer, P.~J. (1967).
\newblock The changing status of new zealand seaports, 1853--1960.
\newblock {\em Annals of the Association of American Geographers},
  57(1):88--100.

\bibitem[Robert, 2003]{robert2003comment}
Robert, S. (2003).
\newblock Comment les formes du pass{\'e} se transmettent-elles?

\bibitem[Robert, 2006]{robert2006resilience}
Robert, S. (2006).
\newblock R{\'e}silience des r{\'e}seaux routiers: l'exemple du val-d'oise.
\newblock {\em Bulletin AGER}, (15, ann{\'e}e 2005):8--14.

\bibitem[Rodrigue et~al., 2004]{rodrigue2004transport}
Rodrigue, J.-P., Comtois, C., and Slack, B. (2004).
\newblock Transport geography on the web.
\newblock {\em Dept. of Economics \& Geography, Hofstra}.

\bibitem[Rouleau, 1975]{rouleau1975trace}
Rouleau, B. (1975).
\newblock Le trac{\'e} des rues de paris. formation, typologie, fonctions.

\bibitem[Sabidussi, 1966]{sabidussi1966centrality}
Sabidussi, G. (1966).
\newblock The centrality index of a graph.
\newblock {\em Psychometrika}, 31(4):581--603.

\bibitem[Salat et~al., 2014]{salat2014breaking}
Salat, S., Bourdic, L., and Labbe, F. (2014).
\newblock Breaking symmetries and emerging scaling urban structures: A
  morphological tale of 3 cities: Paris, new york and barcelona.
\newblock {\em International Journal of Architectural Research: ArchNet-IJAR},
  8(2):77--93.

\bibitem[Salat et~al., 2011]{salat2011villes}
Salat, S., des morphologies~urbaines (France), L., and Labb{\'e}, F. (2011).
\newblock {\em Les villes et les formes: sur l'urbanisme durable}.
\newblock Hermann.

\bibitem[Scellato et~al., 2006]{scellato2006backbone}
Scellato, S., Cardillo, A., Latora, V., and Porta, S. (2006).
\newblock The backbone of a city.
\newblock {\em The European Physical Journal B-Condensed Matter and Complex
  Systems}, 50(1):221--225.

\bibitem[Schweitzer et~al., 1997]{schweitzer1997optimization}
Schweitzer, F., Ebeling, W., Rose, H., and Weiss, O. (1997).
\newblock Optimization of road networks using evolutionary strategies.
\newblock {\em Evolutionary Computation}, 5(4):419--438.

\bibitem[Sen et~al., 2003]{sen2003small}
Sen, P., Dasgupta, S., Chatterjee, A., Sreeram, P., Mukherjee, G., and Manna,
  S. (2003).
\newblock Small-world properties of the indian railway network.
\newblock {\em Physical Review E}, 67(3):036106.

\bibitem[Shimbel, 1953]{shimbel1953structural}
Shimbel, A. (1953).
\newblock Structural parameters of communication networks.
\newblock {\em The bulletin of mathematical biophysics}, 15(4):501--507.

\bibitem[Sole and Montoya, 2001]{sole2001complexity}
Sole, R.~V. and Montoya, M. (2001).
\newblock Complexity and fragility in ecological networks.
\newblock {\em Proceedings of the Royal Society of London B: Biological
  Sciences}, 268(1480):2039--2045.

\bibitem[Sol{\'e} and Pastor-Satorras, 2006]{sole20067}
Sol{\'e}, R.~V. and Pastor-Satorras, R. (2006).
\newblock 7 complex networks in genomics and proteomics.
\newblock {\em Handbook of graphs and networks: From the genome to the
  internet}.

\bibitem[Sporns, 2002]{sporns2002network}
Sporns, O. (2002).
\newblock Network analysis, complexity, and brain function.
\newblock {\em Complexity}, 8(1):56--60.

\bibitem[Sporns et~al., 2000]{sporns2000theoretical}
Sporns, O., Tononi, G., and Edelman, G.~M. (2000).
\newblock Theoretical neuroanatomy: relating anatomical and functional
  connectivity in graphs and cortical connection matrices.
\newblock {\em Cerebral Cortex}, 10(2):127--141.

\bibitem[Stelling et~al., 2002]{stelling2002metabolic}
Stelling, J., Klamt, S., Bettenbrock, K., Schuster, S., and Gilles, E.~D.
  (2002).
\newblock Metabolic network structure determines key aspects of functionality
  and regulation.
\newblock {\em Nature}, 420(6912):190--193.

\bibitem[Strano et~al., 2012]{strano2012elementary}
Strano, E., Nicosia, V., Latora, V., Porta, S., and Barth{\'e}lemy, M. (2012).
\newblock Elementary processes governing the evolution of road networks.
\newblock {\em Scientific reports}, 2.

\bibitem[Taaffe, 1996]{taaffe1996geography}
Taaffe, E.~J. (1996).
\newblock {\em Geography of transportation}.
\newblock Morton O'kelly.

\bibitem[Taaffe et~al., 1963]{taaffe1963transport}
Taaffe, E.~J., Morrill, R.~L., and Gould, P.~R. (1963).
\newblock Transport expansion in underdeveloped countries: a comparative
  analysis.
\newblock {\em Geographical review}, pages 503--529.

\bibitem[Takamatsu et~al., 2009]{takamatsu2009environment}
Takamatsu, A., Takaba, E., and Takizawa, G. (2009).
\newblock Environment-dependent morphology in plasmodium of true slime mold
  physarum polycephalum and a network growth model.
\newblock {\em Journal of theoretical biology}, 256(1):29--44.

\bibitem[Taylor et~al., 1995]{taylor1995hub}
Taylor, G.~D., Harit, S., and English, J.~R. (1995).
\newblock {\em Hub and spoke networks in truckload trucking: Configuration and
  operational concerns}.
\newblock University of Arkansas, Mack-Blackwell National Rural Transportation
  Study Center.

\bibitem[Tero et~al., 2007]{tero2007mathematical}
Tero, A., Kobayashi, R., and Nakagaki, T. (2007).
\newblock A mathematical model for adaptive transport network in path finding
  by true slime mold.
\newblock {\em Journal of theoretical biology}, 244(4):553--564.

\bibitem[Touya, 2010]{touya2010road}
Touya, G. (2010).
\newblock A road network selection process based on data enrichment and
  structure detection.
\newblock {\em Transactions in GIS}, 14(5):595--614.

\bibitem[Turner, 2001]{turner2001angular}
Turner, A. (2001).
\newblock Angular analysis.
\newblock In {\em Proceedings of the 3rd international symposium on space
  syntax}, pages 30--1.

\bibitem[Uetz et~al., 2000]{uetz2000comprehensive}
Uetz, P., Giot, L., Cagney, G., Mansfield, T.~A., Judson, R.~S., Knight, J.~R.,
  Lockshon, D., Narayan, V., Srinivasan, M., Pochart, P., et~al. (2000).
\newblock A comprehensive analysis of protein--protein interactions in
  saccharomyces cerevisiae.
\newblock {\em Nature}, 403(6770):623--627.

\bibitem[Vaughan, 1987]{vaughan1987urban}
Vaughan, R. (1987).
\newblock {\em Urban spatial traffic patterns}.

\bibitem[Viala, 2005]{viala2005contre}
Viala, L. (2005).
\newblock Contre le d{\'e}terminisme de la forme urbaine: une approche totale
  de la forme de la ville.
\newblock {\em Espaces et soci{\'e}t{\'e}s}, (122):99--114.

\bibitem[Viana et~al., 2013]{viana2013simplicity}
Viana, M.~P., Strano, E., Bordin, P., and Barthelemy, M. (2013).
\newblock The simplicity of planar networks.
\newblock {\em Scientific reports}, 3.

\bibitem[Vorono{\"\i}, 1908]{voronoi1908nouvelles}
Vorono{\"\i}, G. (1908).
\newblock Nouvelles applications des param{\`e}tres continus {\`a} la
  th{\'e}orie des formes quadratiques. deuxi{\`e}me m{\'e}moire. recherches sur
  les parall{\'e}llo{\`e}dres primitifs.

\bibitem[Wang et~al., 2009]{wang2009spatiotemporal}
Wang, J., Jin, F., Mo, H., and Wang, F. (2009).
\newblock Spatiotemporal evolution of china’s railway network in the 20th
  century: An accessibility approach.
\newblock {\em Transportation Research Part A: Policy and Practice},
  43(8):765--778.

\bibitem[Wasserman, 1994]{wasserman1994social}
Wasserman, S. (1994).
\newblock {\em Social network analysis: Methods and applications}, volume~8.
\newblock Cambridge university press.

\bibitem[Watteaux, 2003]{watteaux2003plan}
Watteaux, M. (2003).
\newblock Le plan radio-quadrill{\'e} des terroirs non planifi{\'e}s.
\newblock {\em Etudes rurales}, 167(3):187--214.

\bibitem[Watts and Strogatz, 1998]{watts1998collective}
Watts, D.~J. and Strogatz, S.~H. (1998).
\newblock Collective dynamics of ‘small-world’networks.
\newblock {\em nature}, 393(6684):440--442.

\bibitem[Weisstein, 2008]{weisstein2008floyd}
Weisstein, E.~W. (2008).
\newblock Floyd-warshall algorithm.

\bibitem[White et~al., 2004]{white2004does}
White, H.~D., Wellman, B., and Nazer, N. (2004).
\newblock Does citation reflect social structure?: Longitudinal evidence from
  the “globenet” interdisciplinary research group.
\newblock {\em Journal of the American Society for information Science and
  Technology}, 55(2):111--126.

\bibitem[White et~al., 1986]{white1986structure}
White, J., Southgate, E., Thomson, J., and Brenner, S. (1986).
\newblock The structure of the nervous system of the nematode caenorhabditis
  elegans: the mind of a worm.
\newblock {\em Phil. Trans. R. Soc. Lond}, 314:1--340.

\bibitem[Whitney, 1932]{whitney1932congruent}
Whitney, H. (1932).
\newblock Congruent graphs and the connectivity of graphs.
\newblock {\em American Journal of Mathematics}, pages 150--168.

\bibitem[Williams et~al., 2002]{williams2002two}
Williams, R.~J., Berlow, E.~L., Dunne, J.~A., Barab{\'a}si, A.-L., and
  Martinez, N.~D. (2002).
\newblock Two degrees of separation in complex food webs.
\newblock {\em Proceedings of the National Academy of Sciences},
  99(20):12913--12916.

\bibitem[Xie and Levinson, 2007]{xie2007measuring}
Xie, F. and Levinson, D. (2007).
\newblock Measuring the structure of road networks.
\newblock {\em Geographical analysis}, 39(3):336--356.

\bibitem[Xie and Levinson, 2009]{xie2009modeling}
Xie, F. and Levinson, D. (2009).
\newblock Modeling the growth of transportation networks: a comprehensive
  review.
\newblock {\em Networks and Spatial Economics}, 9(3):291--307.

\bibitem[Yamins et~al., 2003]{yamins2003growing}
Yamins, D., Rasmussen, S., and Fogel, D. (2003).
\newblock Growing urban roads.
\newblock {\em Networks and Spatial Economics}, 3(1):69--85.

\bibitem[Yerra and Levinson, 2005]{yerra2005emergence}
Yerra, B.~M. and Levinson, D.~M. (2005).
\newblock The emergence of hierarchy in transportation networks.
\newblock {\em The Annals of Regional Science}, 39(3):541--553.

\bibitem[Zhang, 2006]{zhang2006search}
Zhang, L. (2006).
\newblock {\em Search, Information, Learning, and Knowledge in Travel
  Decision-making: A Positive Approach for Travel Behavior and Demand
  Analysis}.
\newblock ProQuest.

\bibitem[Zhang and Levinson, 2005]{zhang2005road}
Zhang, L. and Levinson, D. (2005).
\newblock Road pricing with autonomous links.
\newblock {\em Transportation Research Record: Journal of the Transportation
  Research Board}, (1932):147--155.

\bibitem[Zhang and Levinson, 2007]{zhang2007economics}
Zhang, L. and Levinson, D. (2007).
\newblock The economics of transportation network growth.
\newblock In {\em Essays on transport economics}, pages 317--339. Springer.

\bibitem[Zhang et~al., 2004]{zhang2004model}
Zhang, L., Levinson, D., et~al. (2004).
\newblock A model of the rise and fall of roads.
\newblock In {\em 50th North American Regional Science Council Annual Meeting,
  Philadelphia, Pa}.

\end{thebibliography}

\clearpage{\pagestyle{empty}\cleardoublepage}
\part*{Annexes}
\markboth{Annexes}{Annexes}
\addcontentsline{toc}{part}{Annexes}

\appendix

\setlength{\parskip}{0pt}
\newgeometry{left=2.5cm,right=2.5cm,top=2cm,bottom=2cm}

\FloatBarrier 
\chapter{Indicateurs}\label{ann:chap_indicateurs}

\section{Indicateurs locaux}\label{ann:sec_indloc}


 
\begin{figure}[c]
    \centering
    \includegraphics[width=\textwidth]{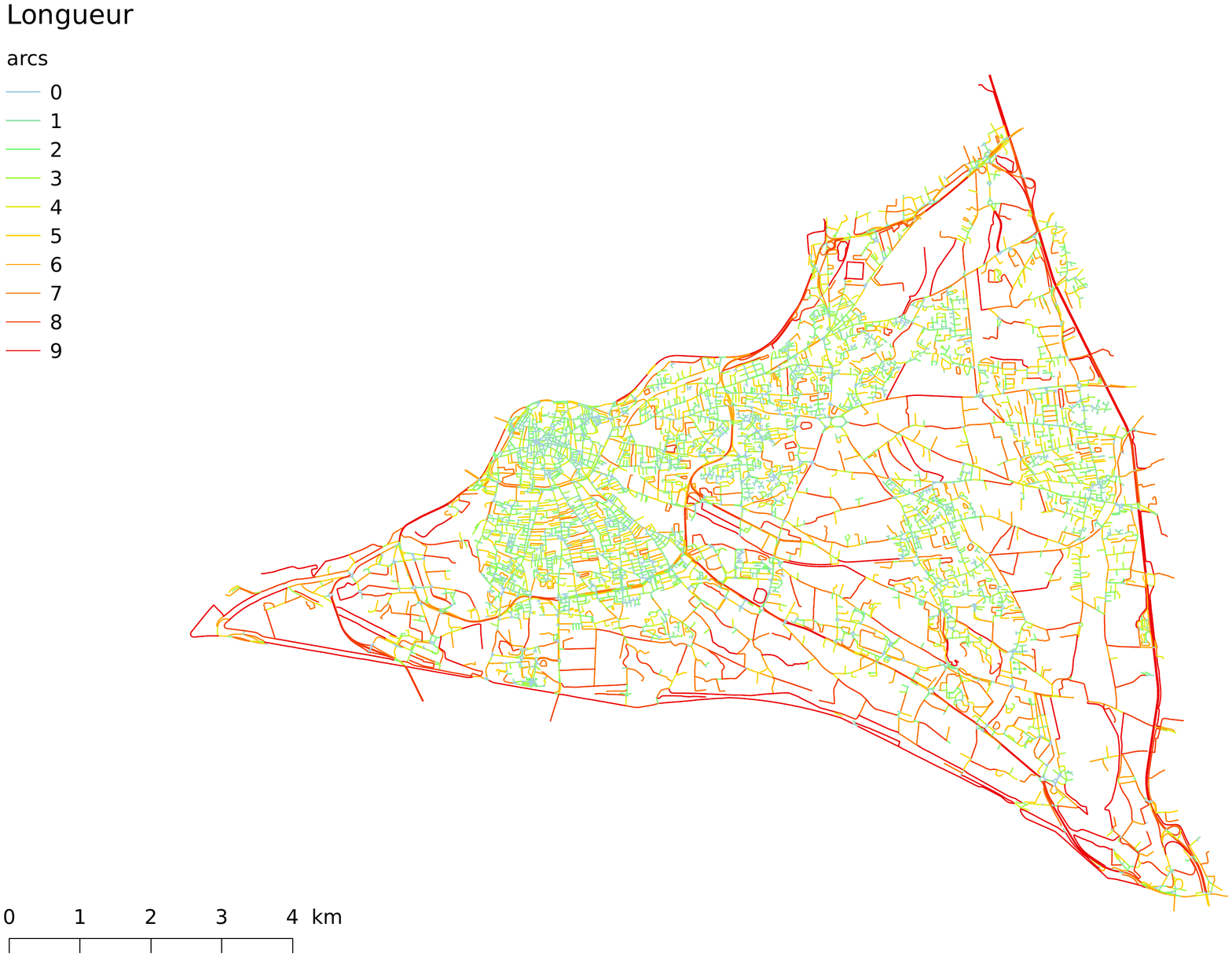}
    \caption{Indicateur de {\large \textbf{longueur}} calculé sur les {\large \textbf{arcs}} du graphe viaire d'{\large \textbf{Avignon}}.}
    \label{fig:arcs_longueur_av}
\end{figure}

\begin{figure}[c]
    \centering
    \includegraphics[width=\textwidth]{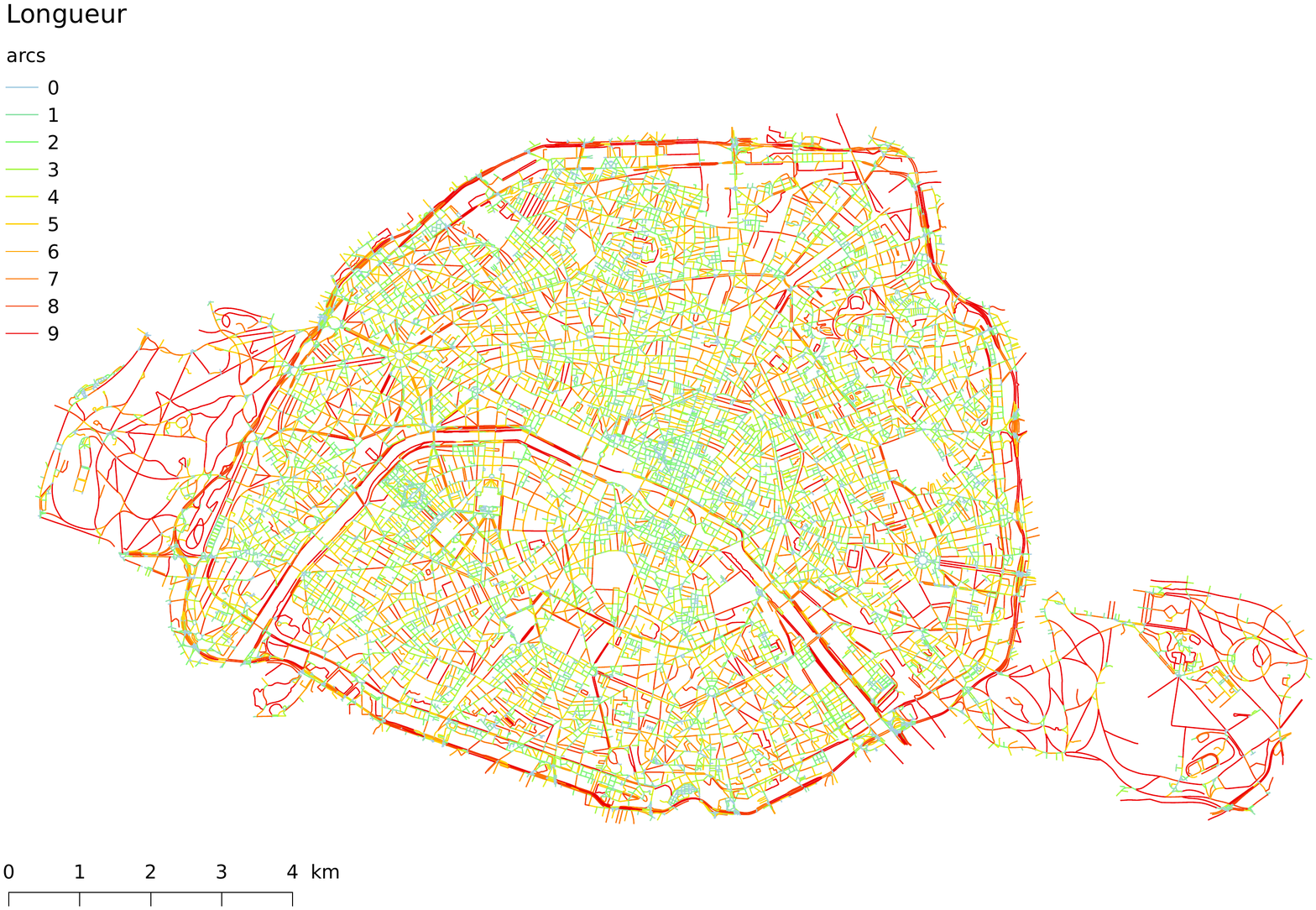}
    \caption{Indicateur de {\large \textbf{longueur}} calculé sur les {\large \textbf{arcs}} du graphe viaire de {\large \textbf{Paris}}.}
    \label{fig:arcs_longueur_par}
\end{figure}

\FloatBarrier

\begin{figure}[c]
    \centering
    \includegraphics[width=\textwidth]{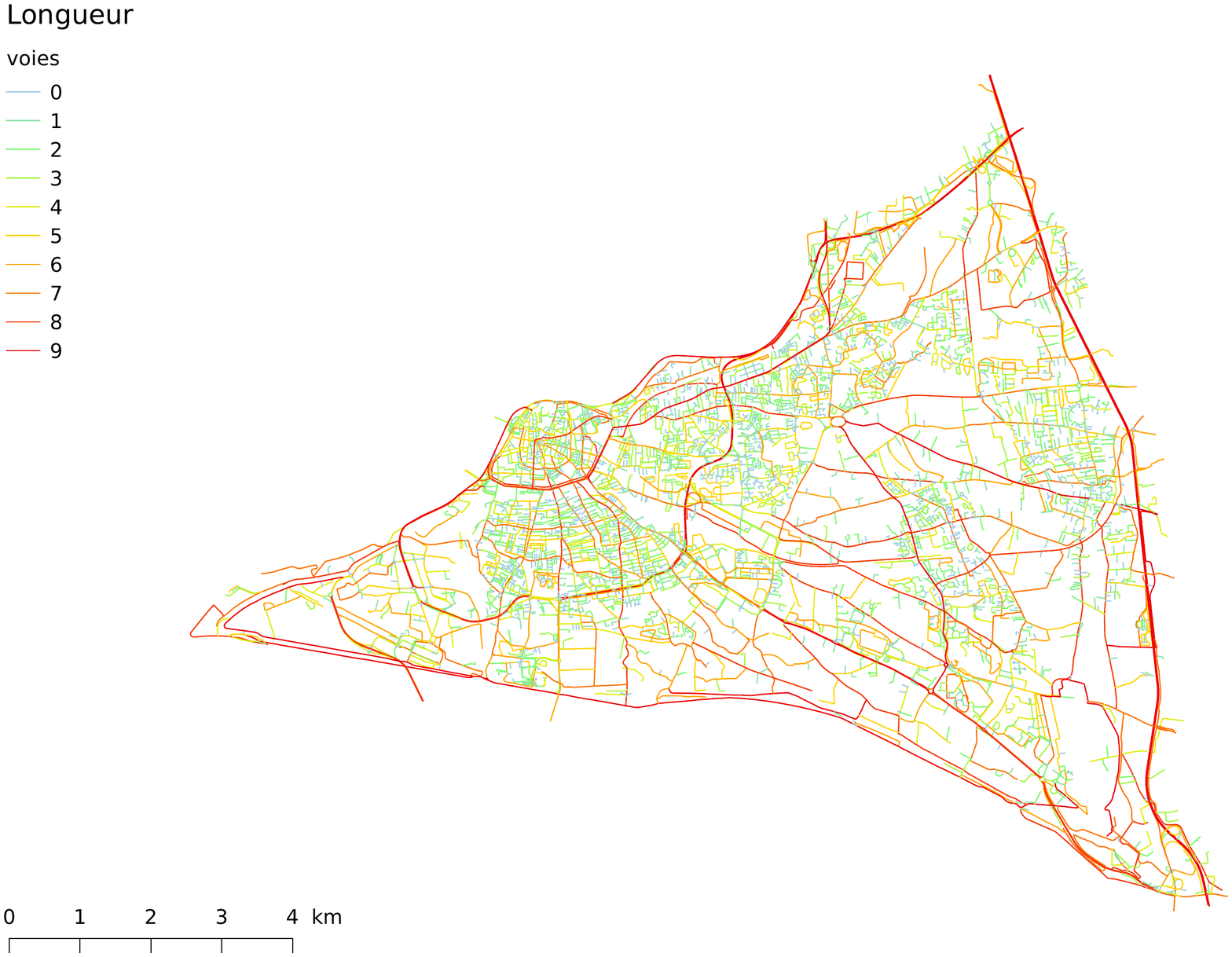}
    \caption{Indicateur de {\large \textbf{longueur}} calculé sur les {\large \textbf{voies}} du graphe viaire d'{\large \textbf{Avignon}}.}
    \label{fig:voies_longueur_av}
\end{figure}

\begin{figure}[c]
    \centering
    \includegraphics[width=\textwidth]{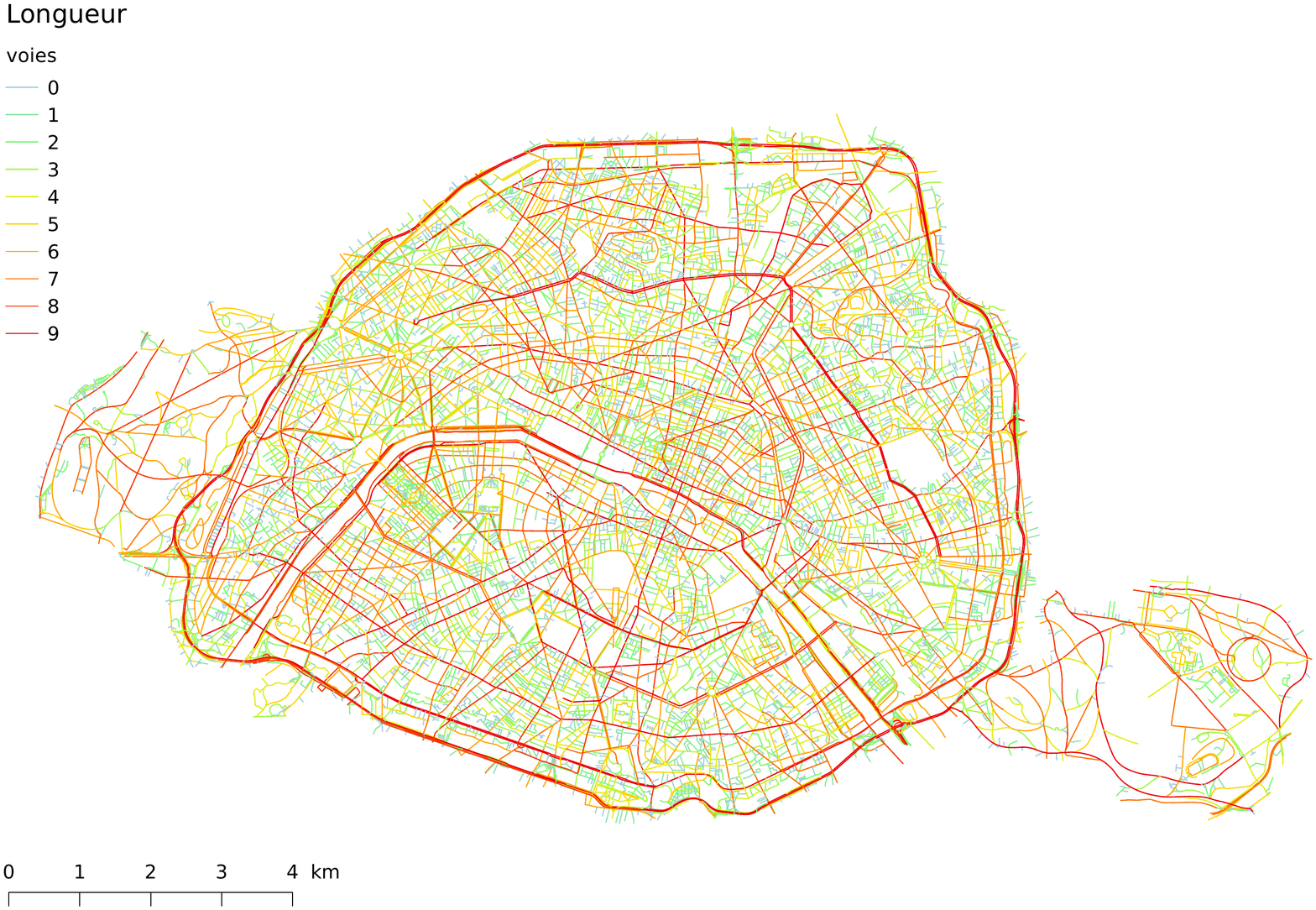}
    \caption{Indicateur de {\large \textbf{longueur}} calculé sur les {\large \textbf{voies}} du graphe viaire de {\large \textbf{Paris}}.}
    \label{fig:voies_longueur_par}
\end{figure}

\FloatBarrier

\FloatBarrier

\begin{figure}[c]
    \centering
    \includegraphics[width=\textwidth]{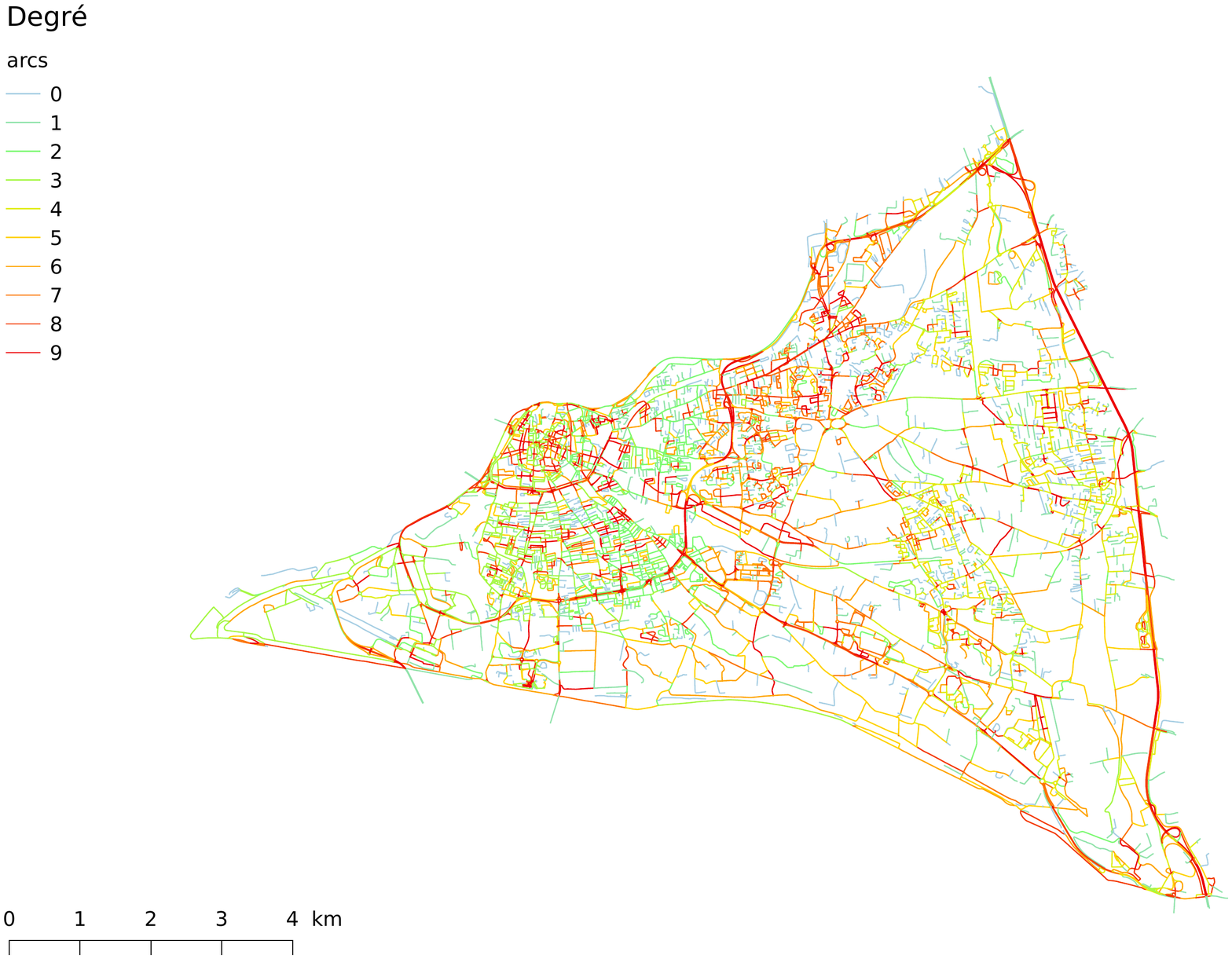}
    \caption{Indicateur de {\large \textbf{degré}} calculé sur les {\large \textbf{arcs}} du graphe viaire d'{\large \textbf{Avignon}}.}
    \label{fig:arcs_degre_av}
\end{figure}

\begin{figure}[c]
    \centering
    \includegraphics[width=\textwidth]{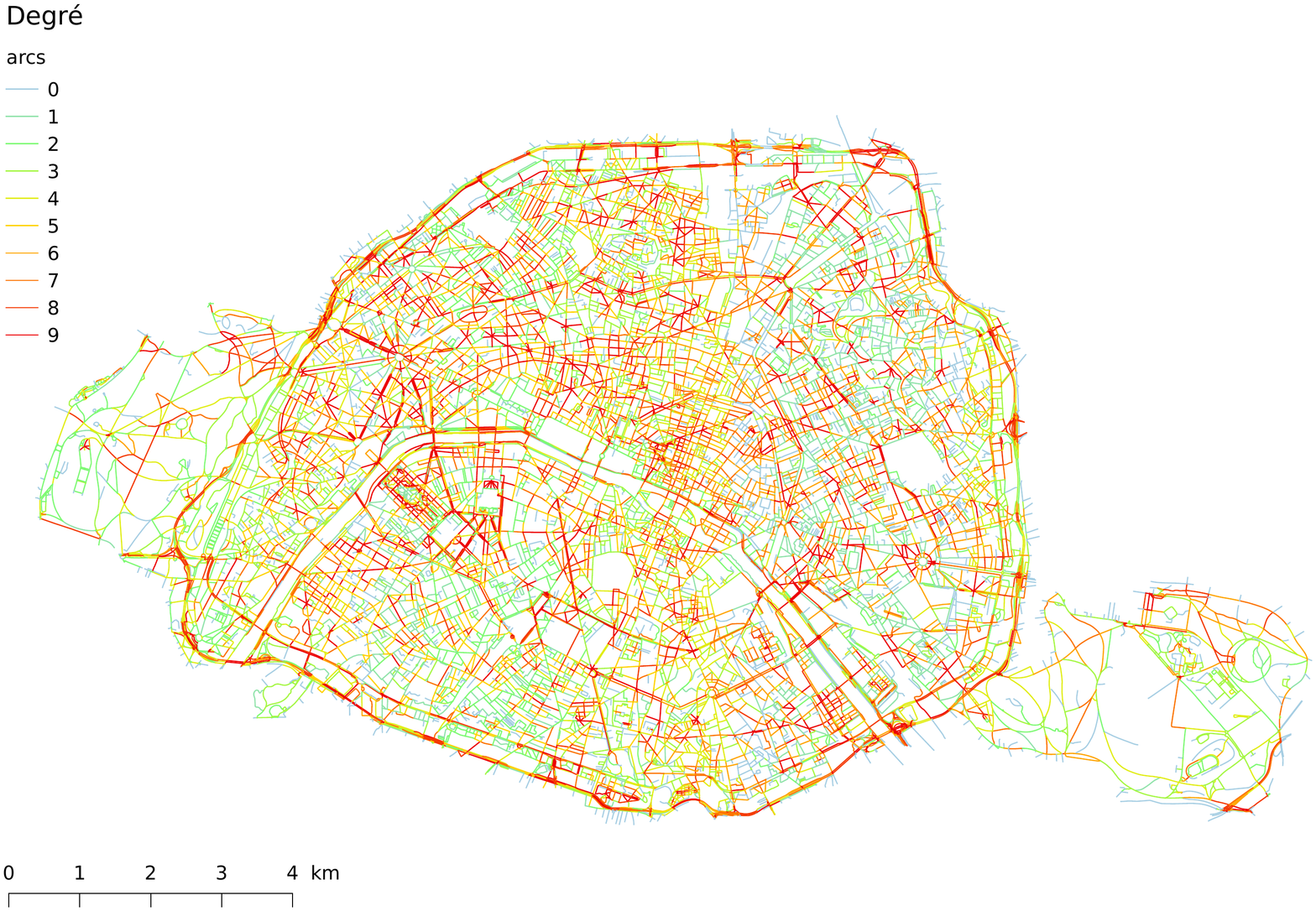}
    \caption{Indicateur de {\large \textbf{degré}} calculé sur les {\large \textbf{arcs}} du graphe viaire de {\large \textbf{Paris}}.}
    \label{fig:arcs_degre_par}
\end{figure}

\FloatBarrier

\begin{figure}[c]
    \centering
    \includegraphics[width=\textwidth]{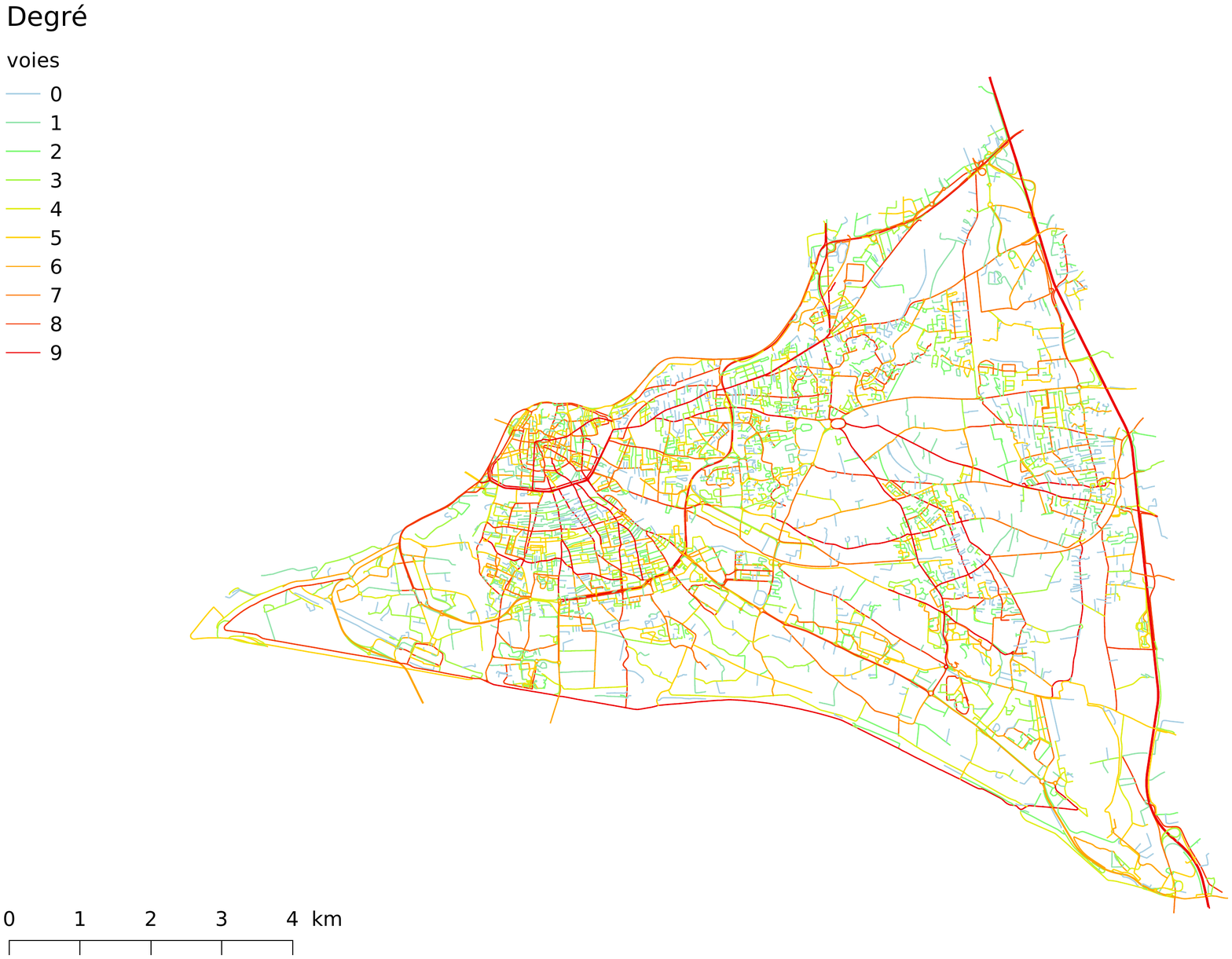}
    \caption{Indicateur de {\large \textbf{degré}} calculé sur les {\large \textbf{voies}} du graphe viaire d'{\large \textbf{Avignon}}.}
    \label{fig:voies_degre_av}
\end{figure}

\begin{figure}[c]
    \centering
    \includegraphics[width=\textwidth]{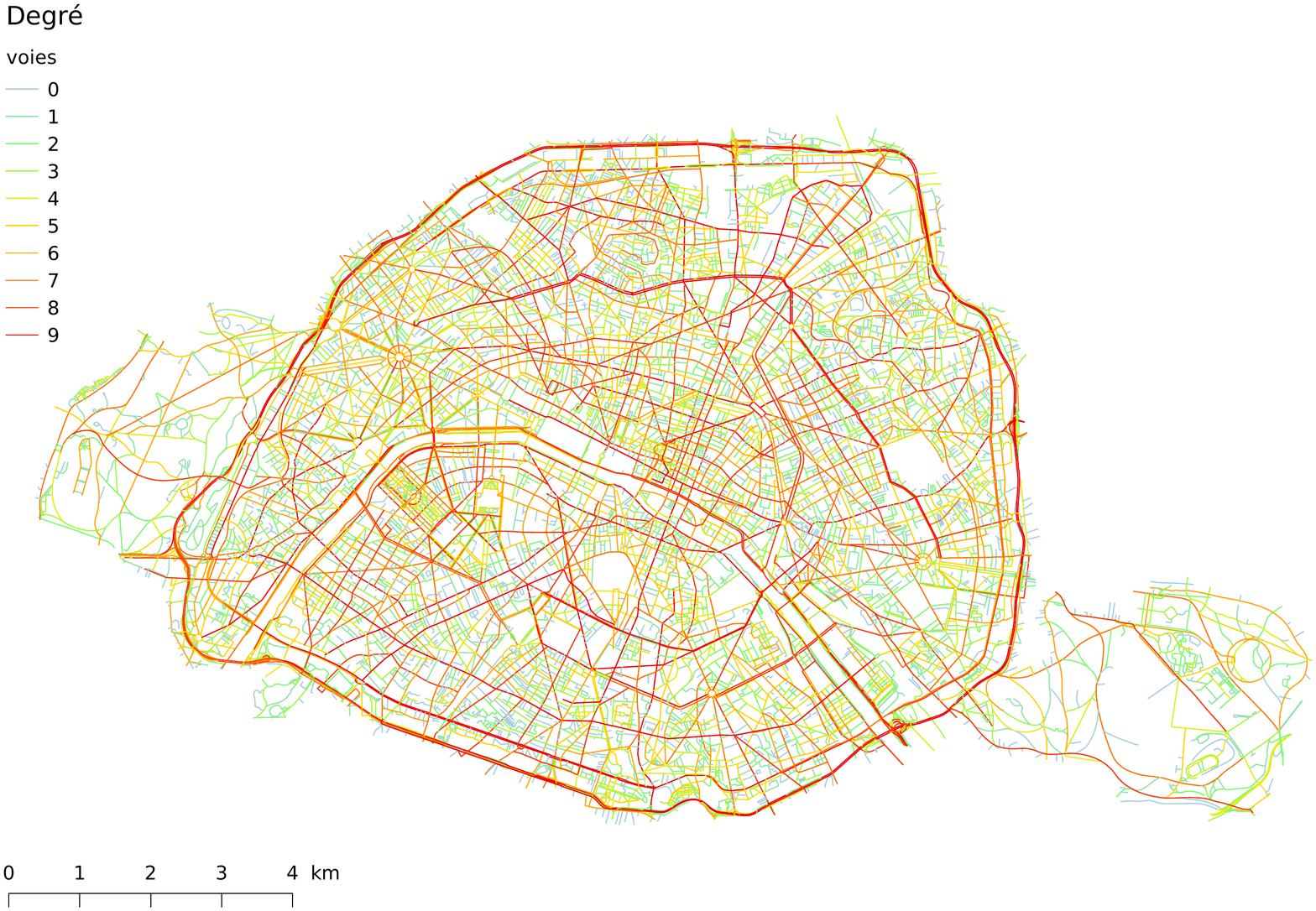}
    \caption{Indicateur de {\large \textbf{degré}} calculé sur les {\large \textbf{voies}} du graphe viaire de {\large \textbf{Paris}}.}
    \label{fig:voies_degre_par}
\end{figure}

\FloatBarrier

\begin{figure}[c]
    \centering
    \includegraphics[width=\textwidth]{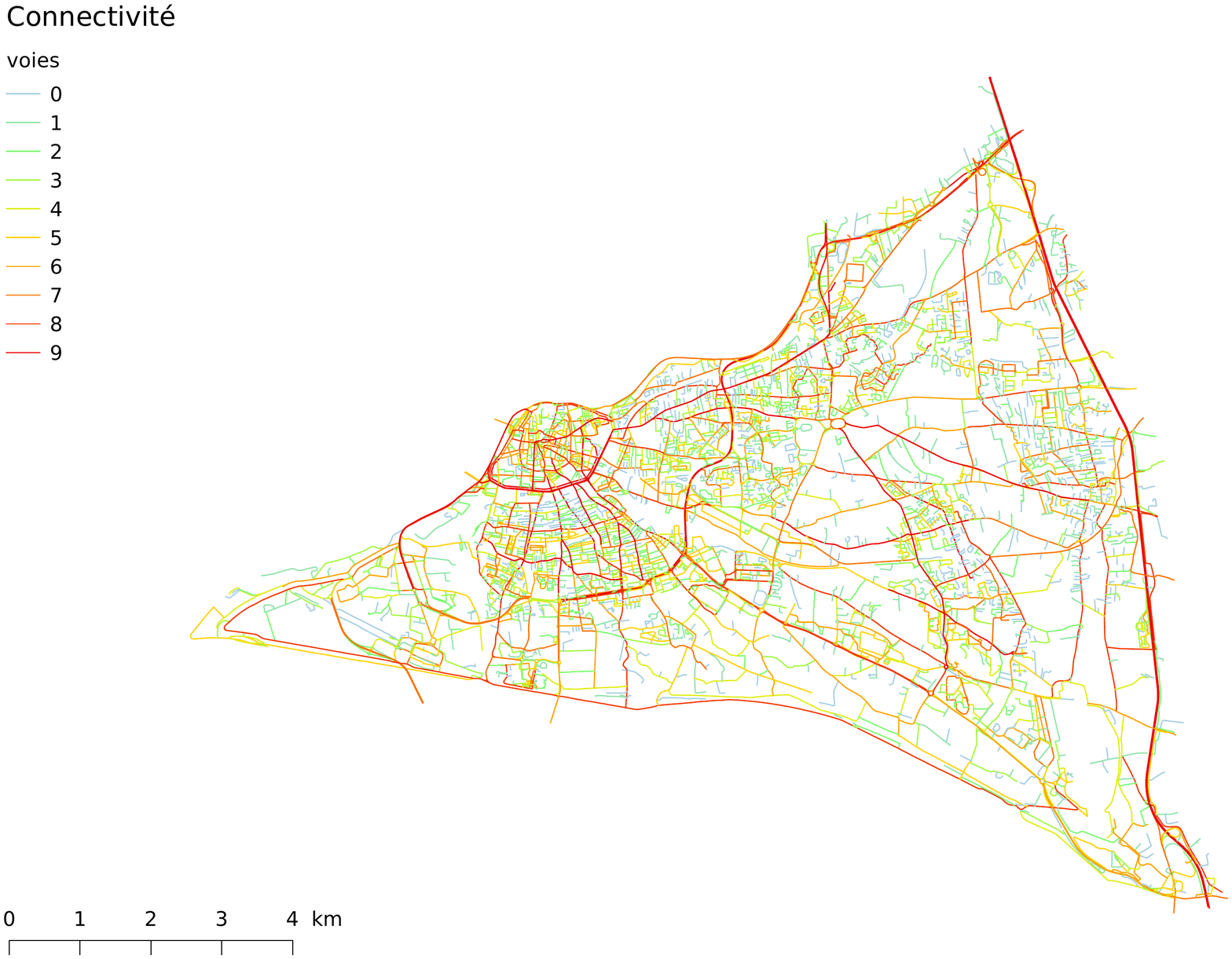}
    \caption{Indicateur de {\large \textbf{connectivité}} calculé sur les {\large \textbf{voies}} du graphe viaire d'{\large \textbf{Avignon}}.}
    \label{fig:voies_connectivite_av}
\end{figure}

\begin{figure}[c]
    \centering
    \includegraphics[width=\textwidth]{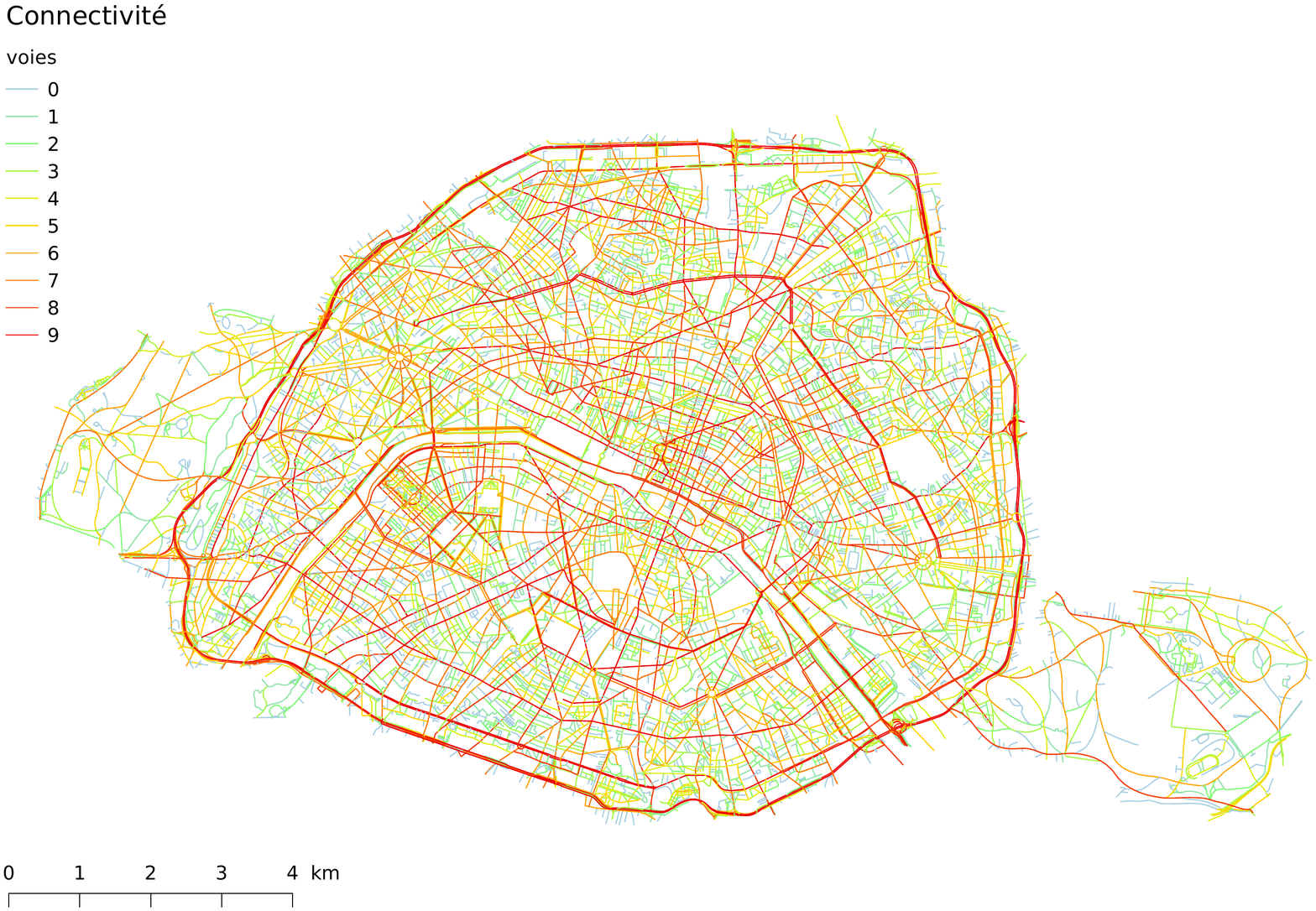}
    \caption{Indicateur de {\large \textbf{connectivité}} calculé sur les {\large \textbf{voies}} du graphe viaire de {\large \textbf{Paris}}.}
    \label{fig:voies_connectivite_par}
\end{figure}

\FloatBarrier

\begin{figure}[c]
    \centering
    \includegraphics[width=\textwidth]{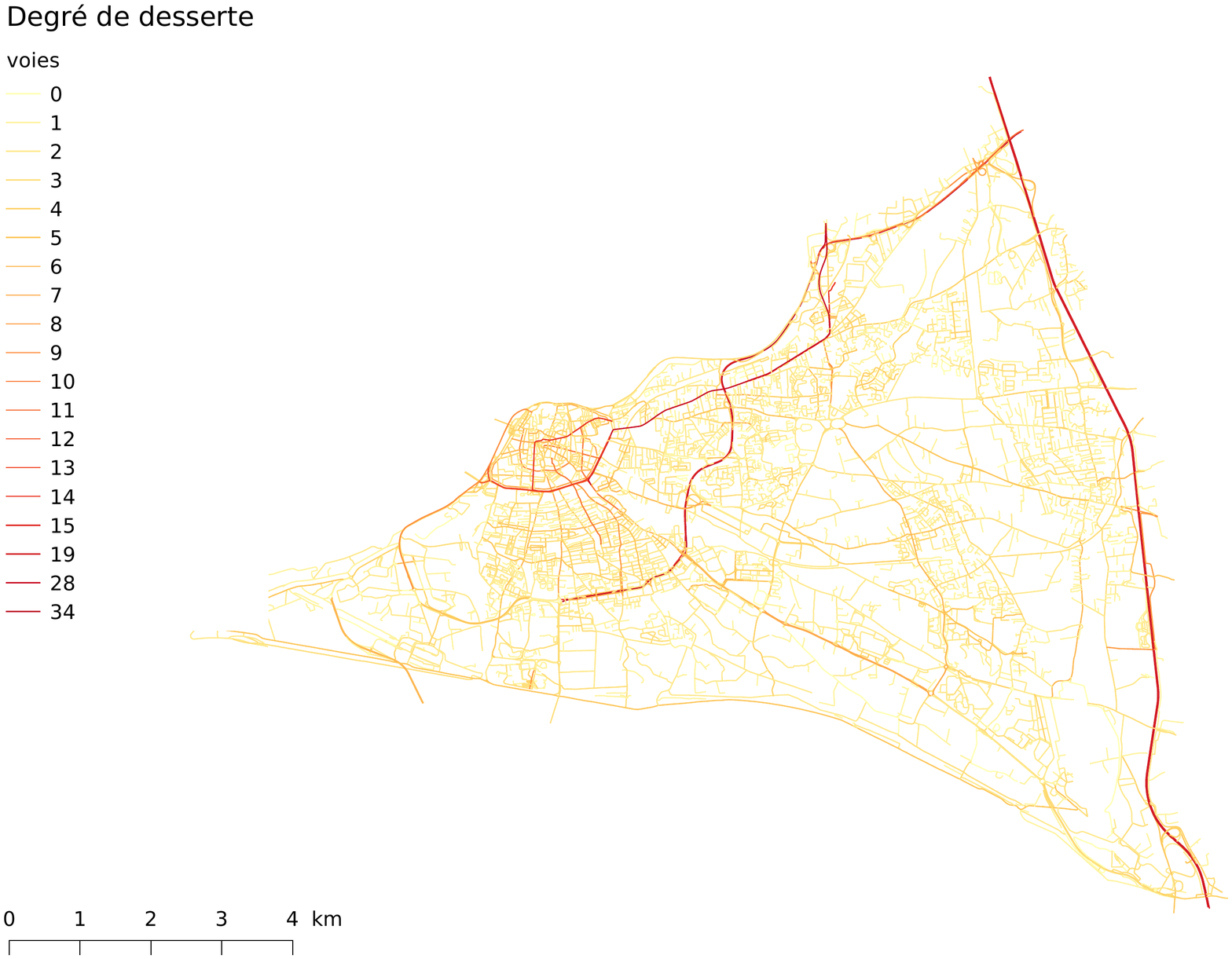}
    \caption{Indicateur de {\large \textbf{degré de desserte}} calculé sur les {\large \textbf{voies}} du graphe viaire d'{\large \textbf{Avignon}}.}
    \label{fig:voies_degdesserte_av}
\end{figure}

\begin{figure}[c]
    \centering
    \includegraphics[width=\textwidth]{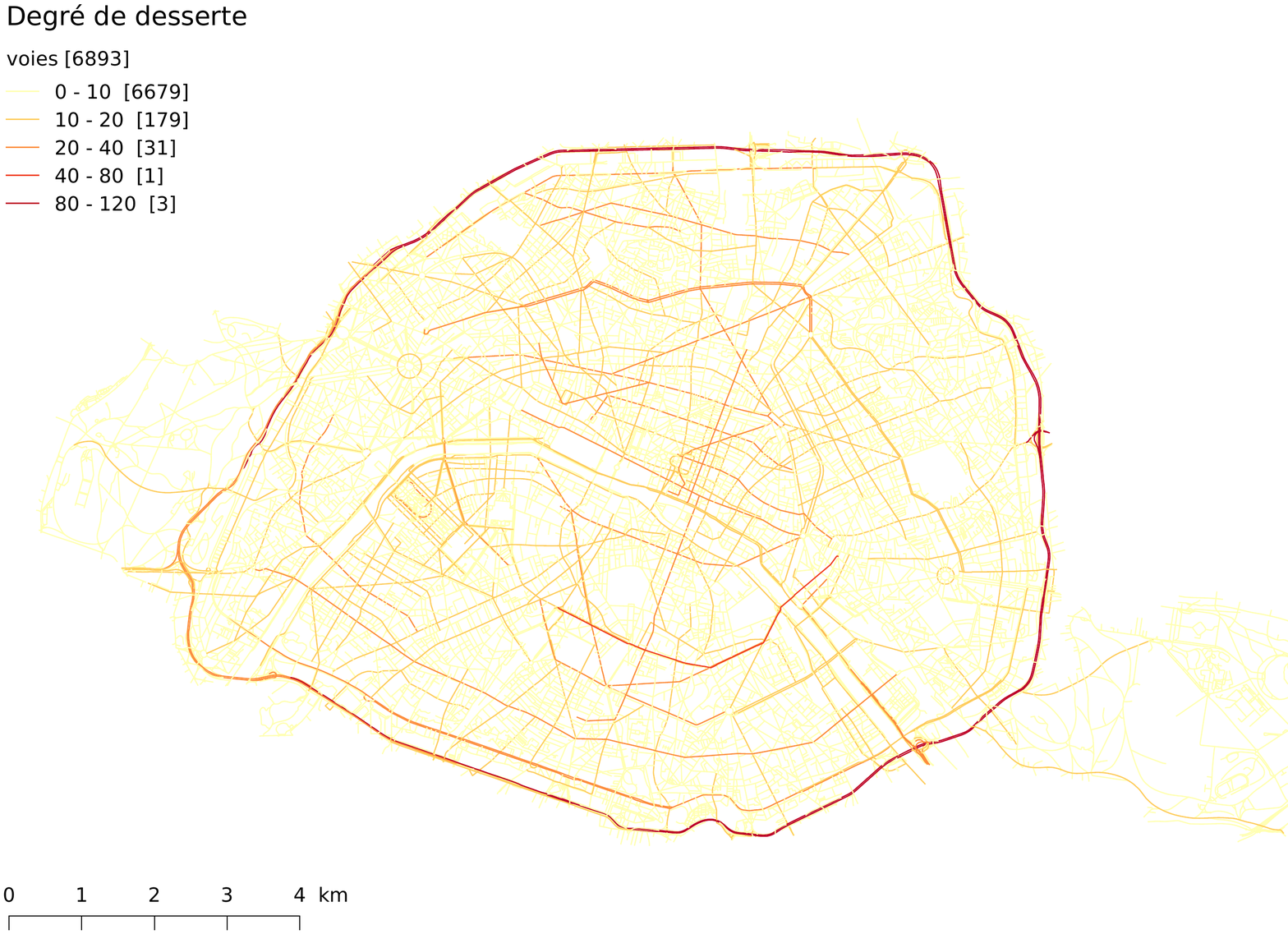}
    \caption{Indicateur de {\large \textbf{degré de desserte}} calculé sur les {\large \textbf{voies}} du graphe viaire de {\large \textbf{Paris}}.}
    \label{fig:voies_degdesserte_par}
\end{figure}

\FloatBarrier

\FloatBarrier

\begin{figure}[c]
    \centering
    \includegraphics[width=\textwidth]{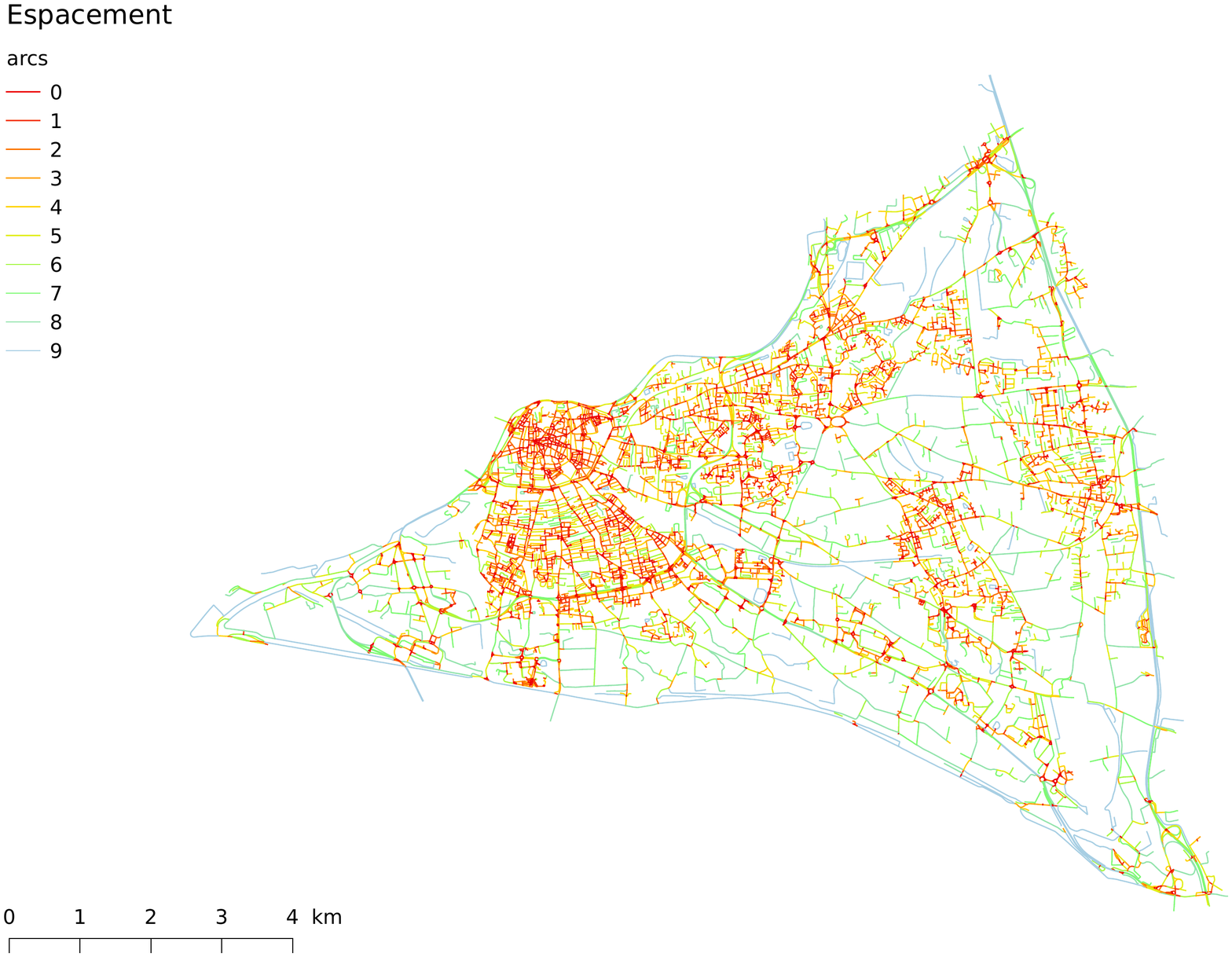}
    \caption{Indicateur d'{\large \textbf{espacement}} calculé sur les {\large \textbf{arcs}} du graphe viaire d'{\large \textbf{Avignon}}.}
    \label{fig:arcs_espacement_av}
\end{figure}

\begin{figure}[c]
    \centering
    \includegraphics[width=\textwidth]{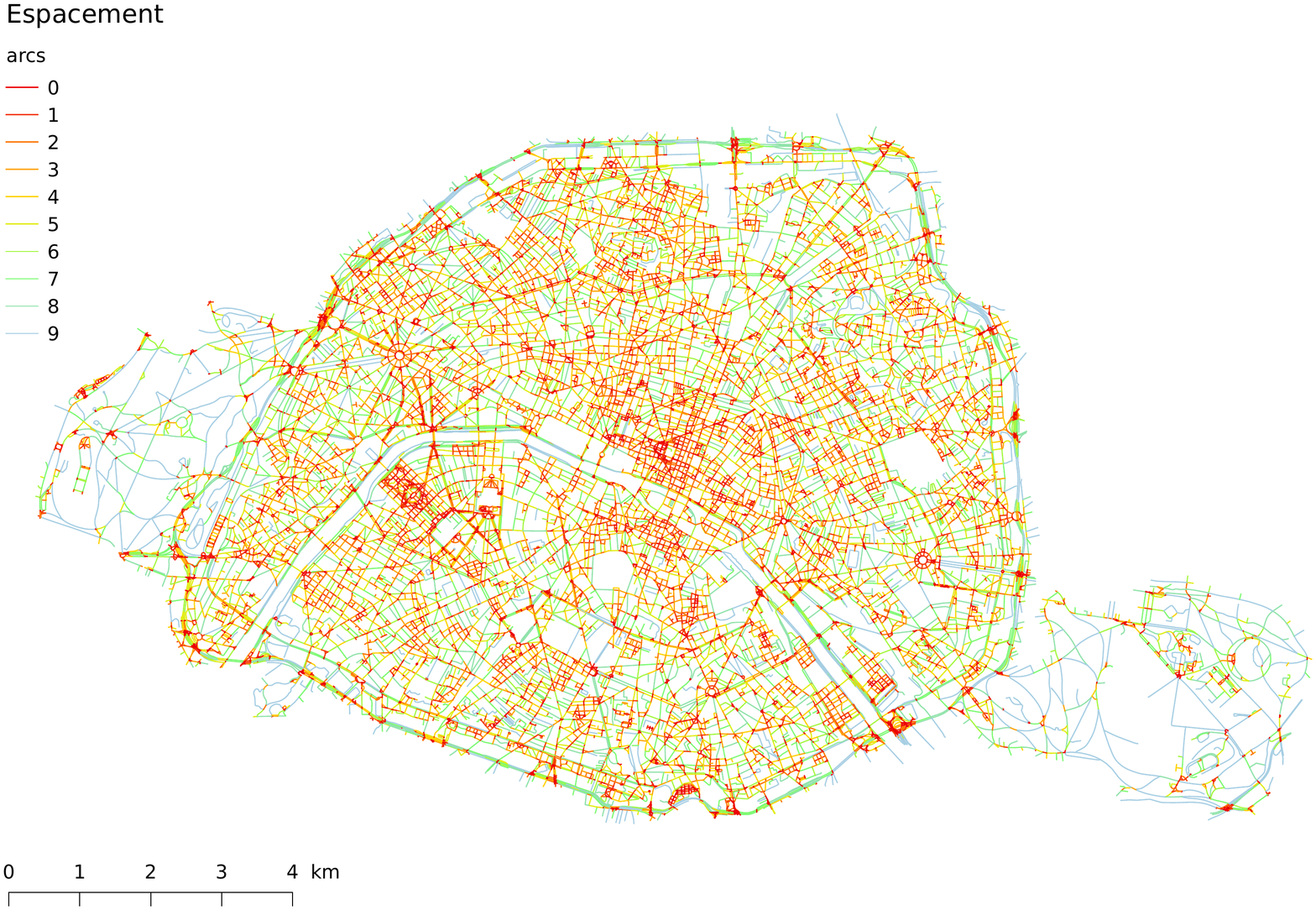}
    \caption{Indicateur d'{\large \textbf{espacement}} calculé sur les {\large \textbf{arcs}} du graphe viaire de {\large \textbf{Paris}}.}
    \label{fig:arcs_espacement_par}
\end{figure}

\FloatBarrier

\begin{figure}[c]
    \centering
    \includegraphics[width=\textwidth]{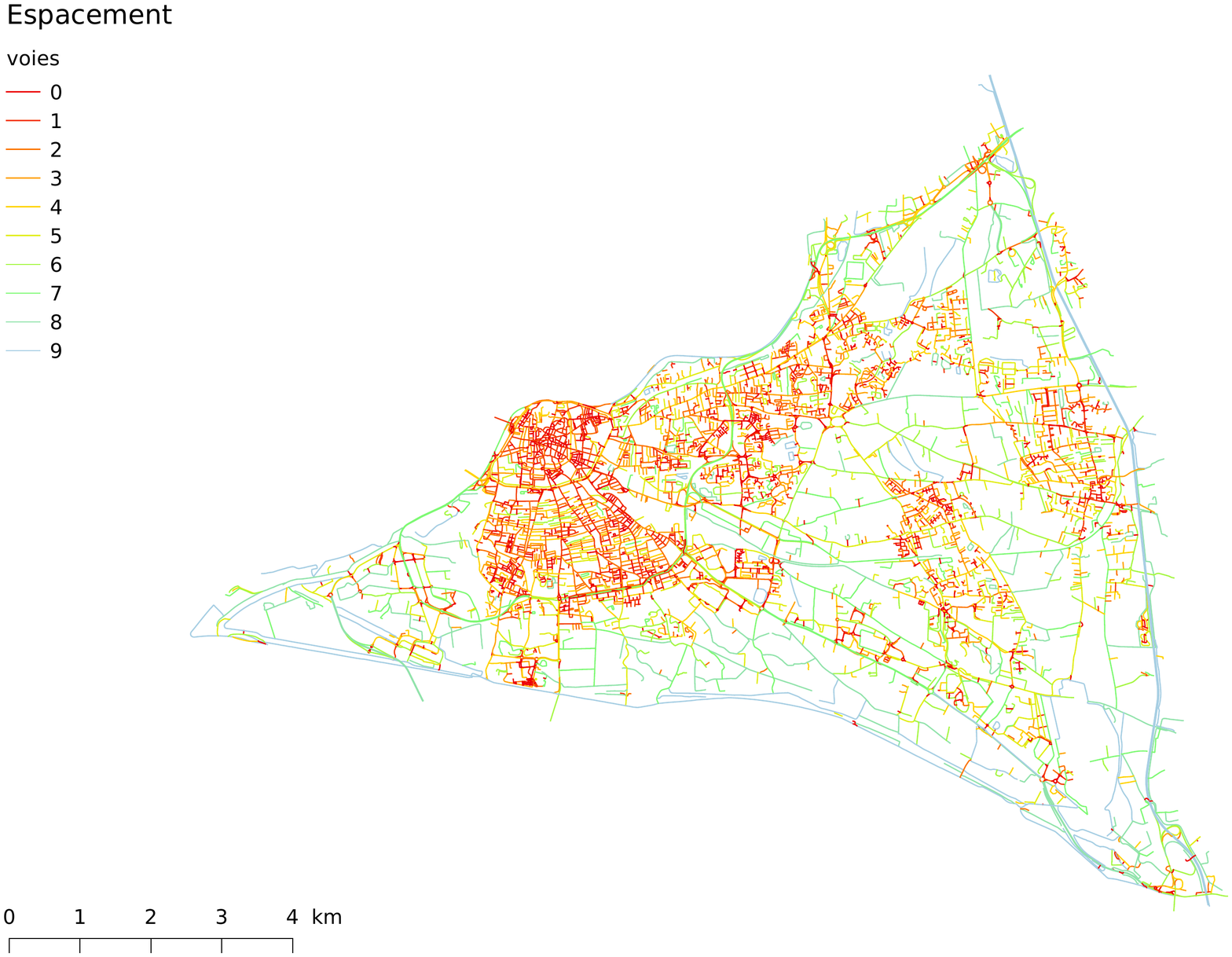}
    \caption{Indicateur d'{\large \textbf{espacement}} calculé sur les {\large \textbf{voies}} du graphe viaire d'{\large \textbf{Avignon}}.}
    \label{fig:voies_espacement_av}
\end{figure}

\begin{figure}[c]
    \centering
    \includegraphics[width=\textwidth]{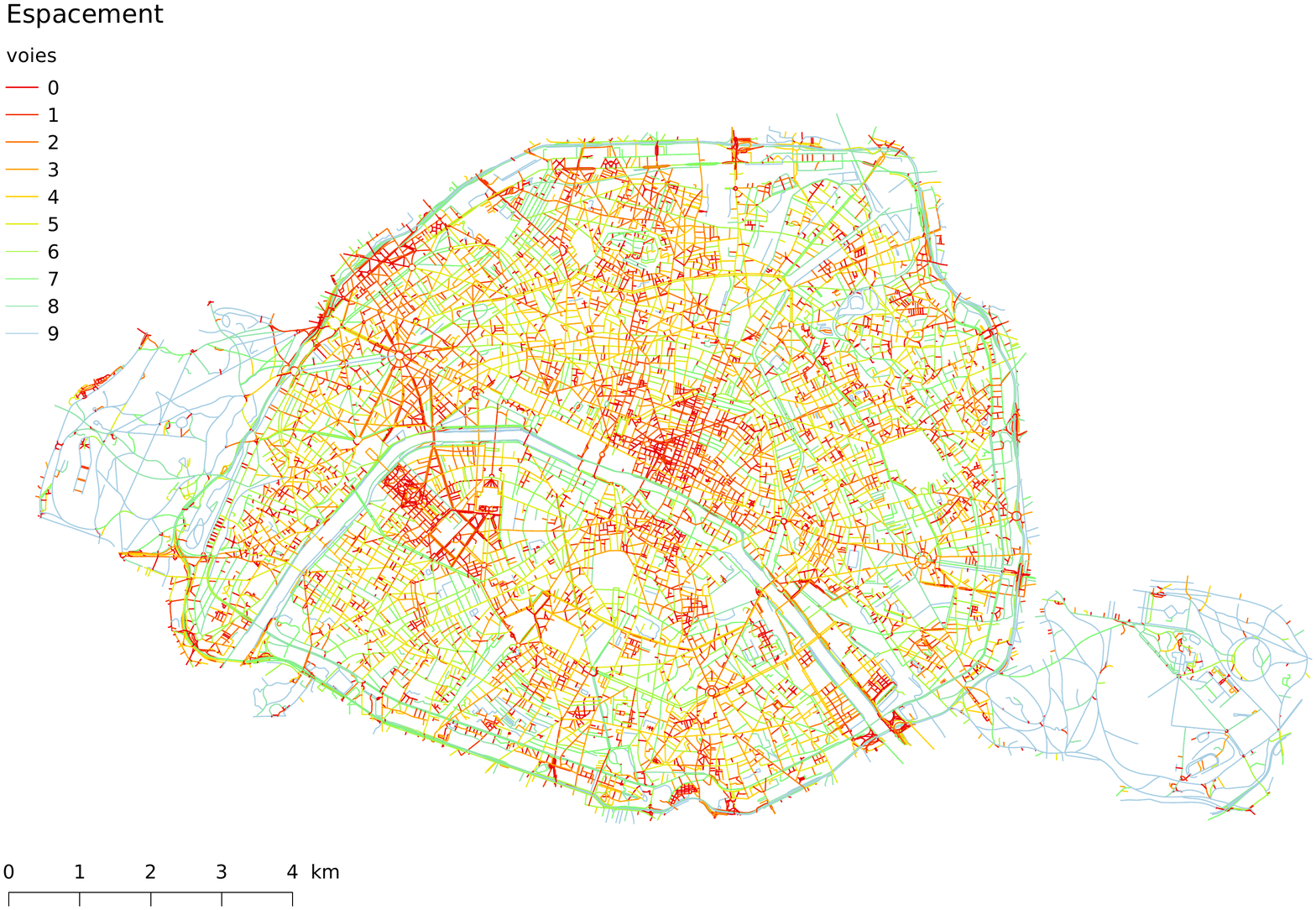}
    \caption{Indicateur d'{\large \textbf{espacement}} calculé sur les {\large \textbf{voies}} du graphe viaire de {\large \textbf{Paris}}.}
    \label{fig:voies_espacement_par}
\end{figure}

\FloatBarrier

\FloatBarrier

\begin{figure}[c]
    \centering
    \includegraphics[width=\textwidth]{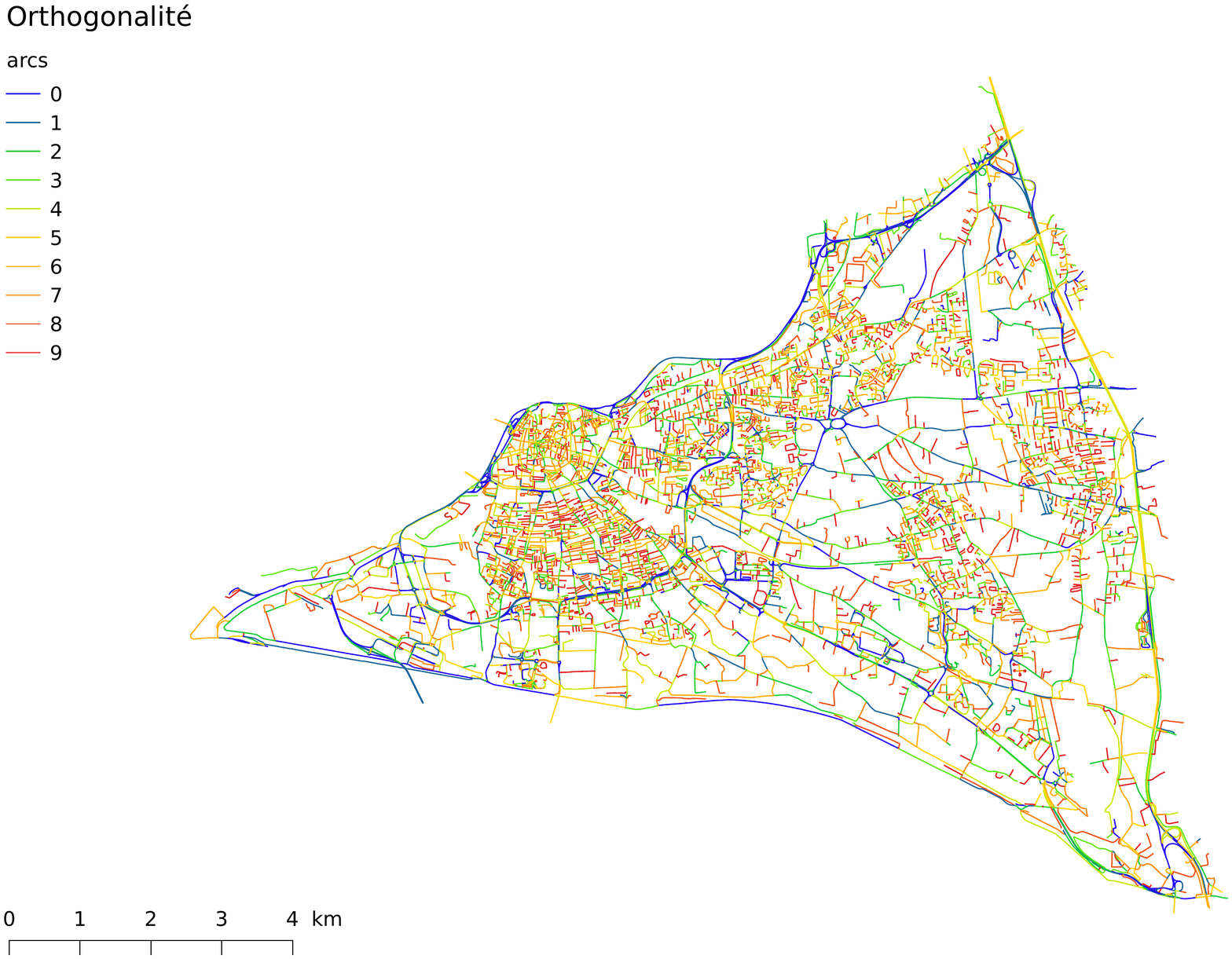}
    \caption{Indicateur d'{\large \textbf{orthogonalité}} calculé sur les {\large \textbf{arcs}} du graphe viaire d'{\large \textbf{Avignon}}.}
    \label{fig:arcs_ortho_av}
\end{figure}

\begin{figure}[c]
    \centering
    \includegraphics[width=\textwidth]{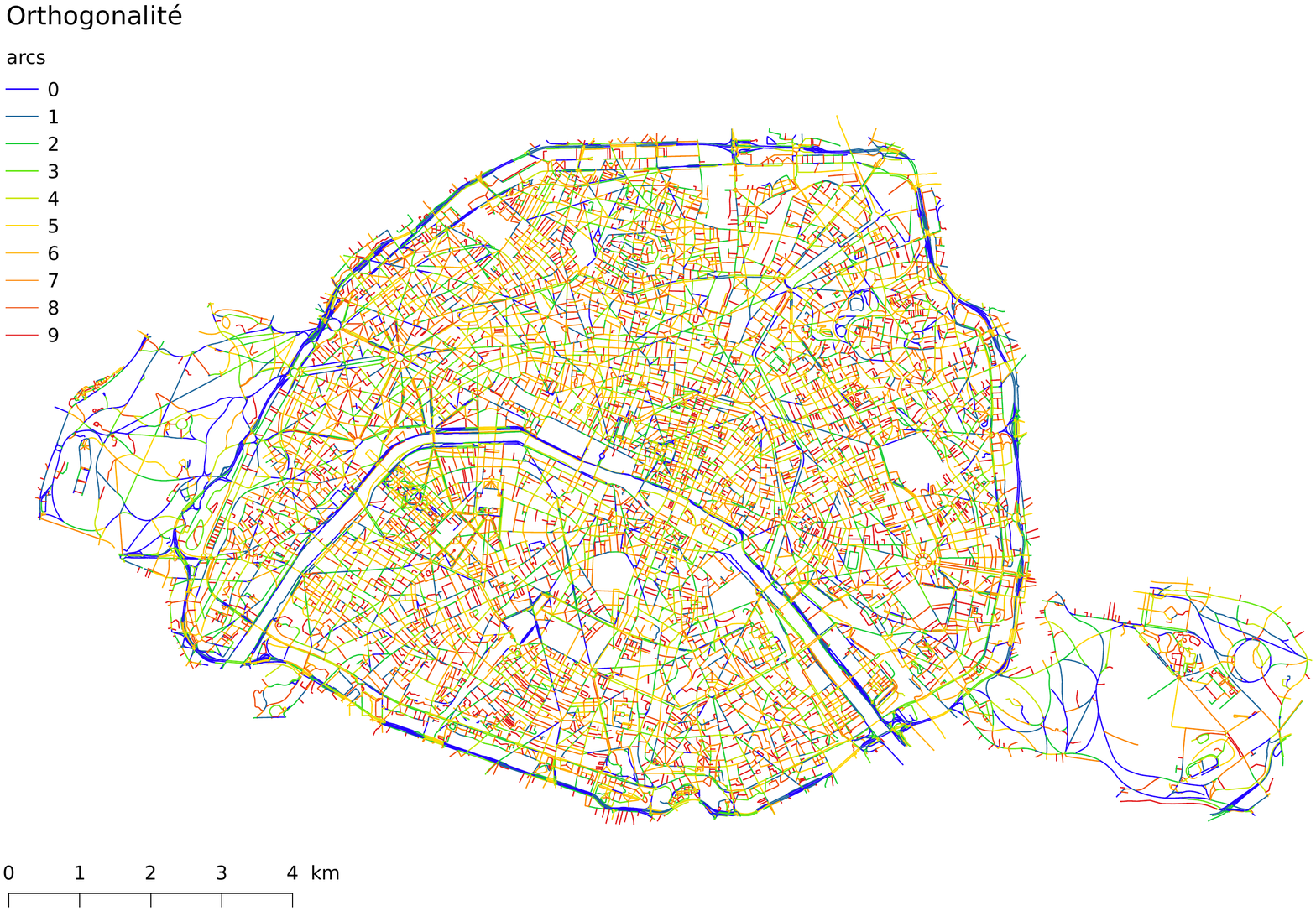}
    \caption{Indicateur d'{\large \textbf{orthogonalité}} calculé sur les {\large \textbf{arcs}} du graphe viaire de {\large \textbf{Paris}}.}
    \label{fig:arcs_ortho_par}
\end{figure}

\FloatBarrier

\begin{figure}[c]
    \centering
    \includegraphics[width=\textwidth]{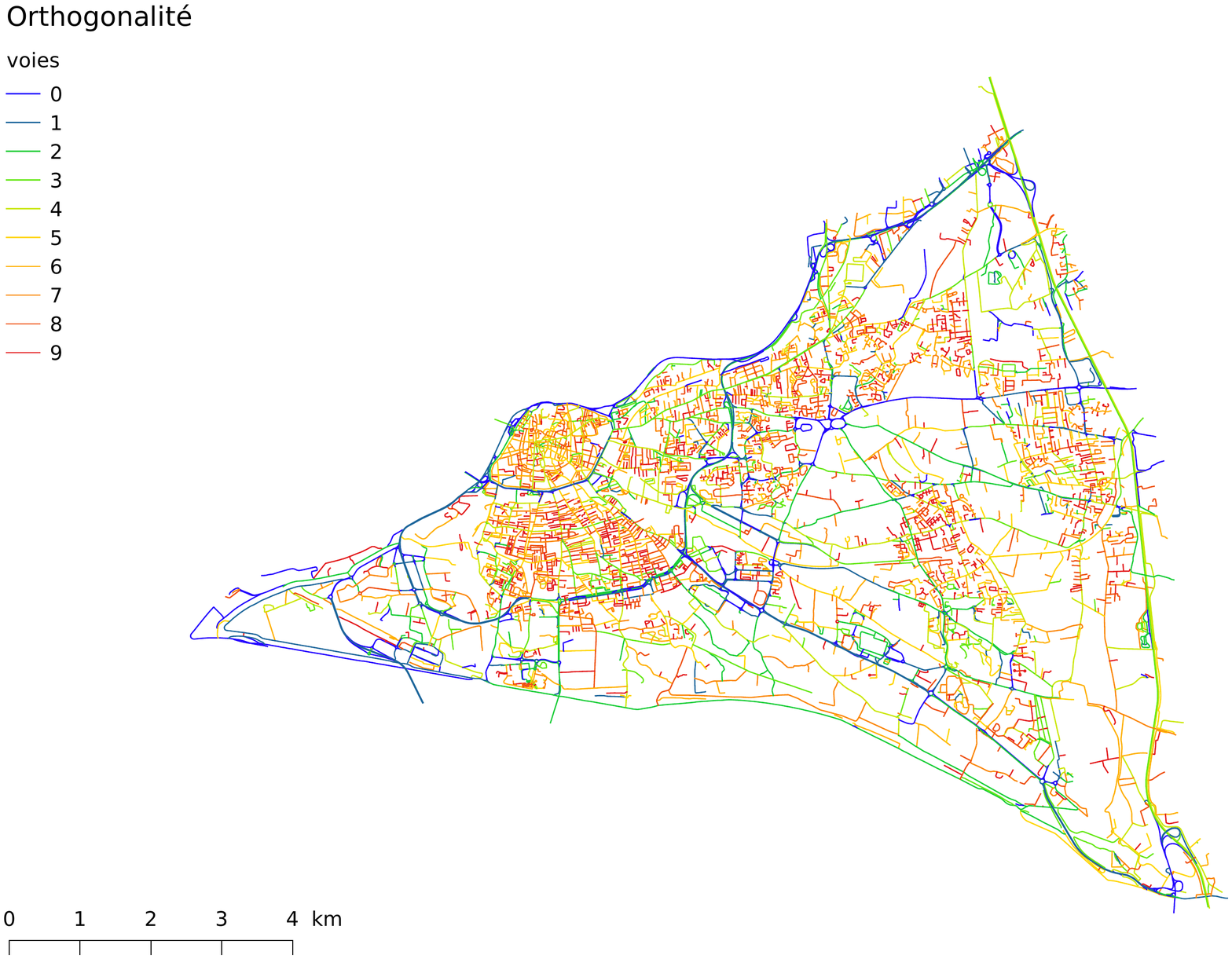}
    \caption{Indicateur d'{\large \textbf{orthogonalité}} calculé sur les {\large \textbf{voies}} du graphe viaire d'{\large \textbf{Avignon}}.}
    \label{fig:voies_ortho_av}
\end{figure}

\begin{figure}[c]
    \centering
    \includegraphics[width=\textwidth]{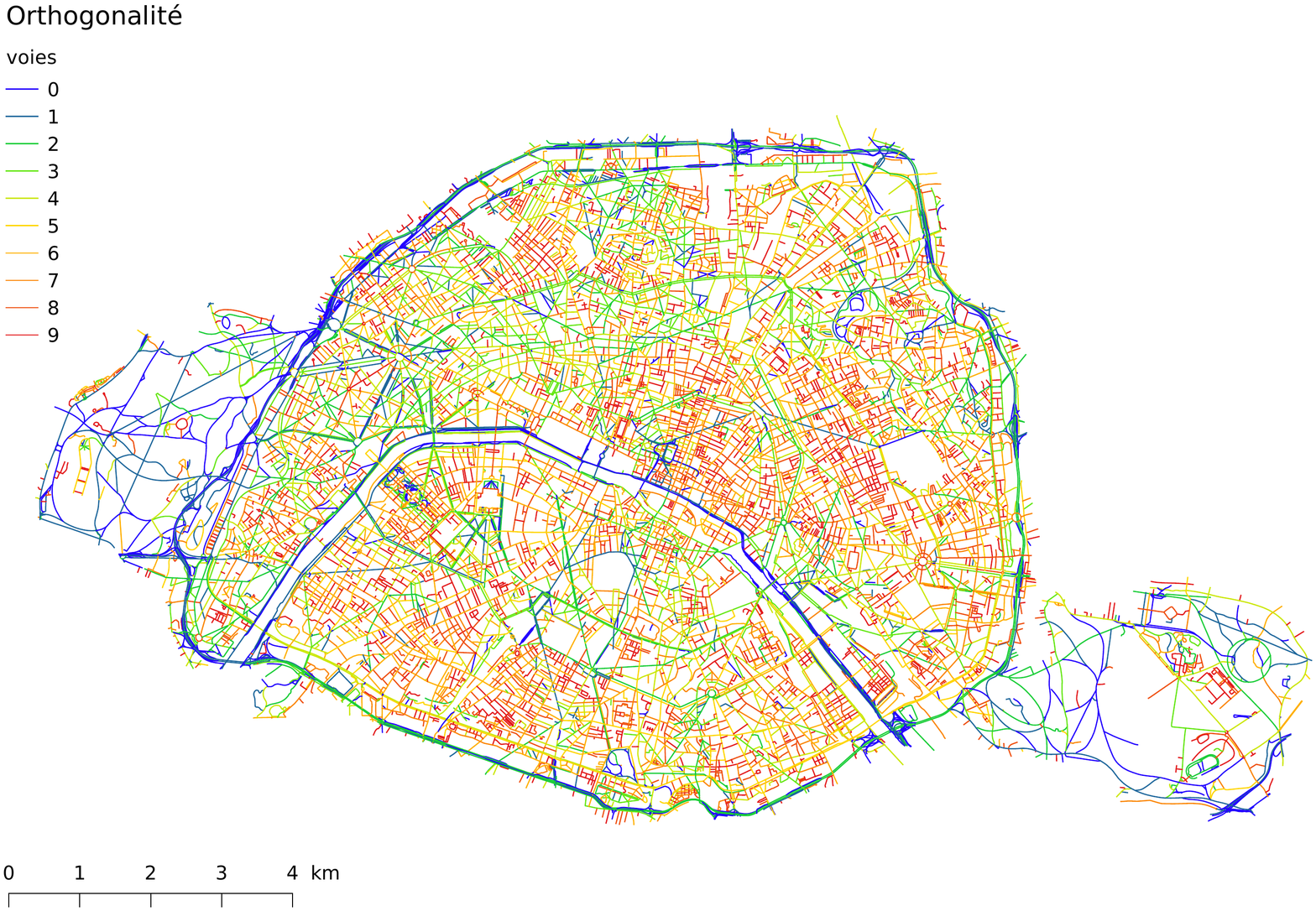}
    \caption{Indicateur d'{\large \textbf{orthogonalité}} calculé sur les {\large \textbf{voies}} du graphe viaire de {\large \textbf{Paris}}.}
    \label{fig:voies_ortho_par}
\end{figure}

\FloatBarrier 
\section{Indicateurs globaux}\label{ann:sec_indglo}

\FloatBarrier

\FloatBarrier

\begin{figure}[c]
    \centering
    \includegraphics[width=\textwidth]{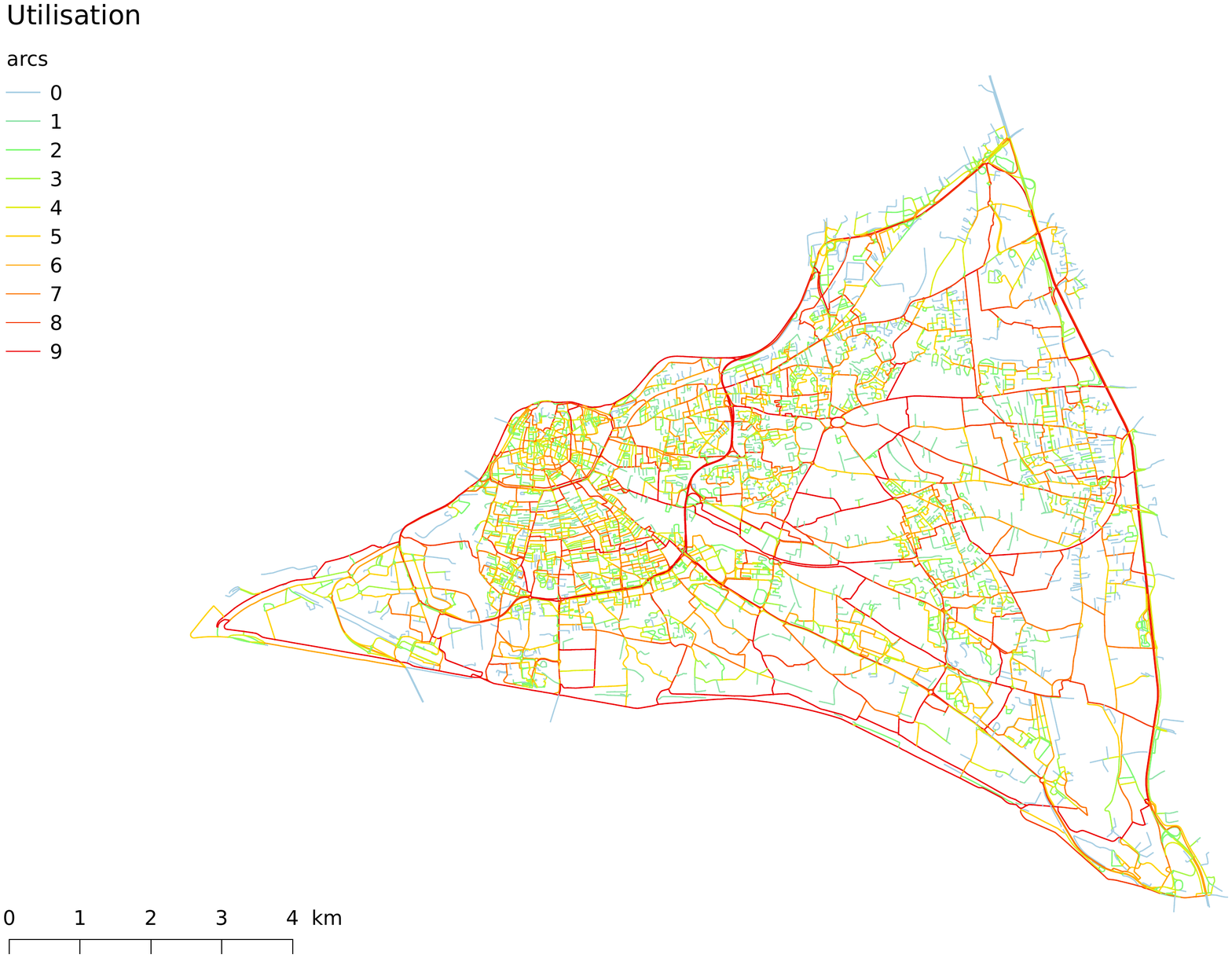}
    \caption{Indicateur d'{\large \textbf{utilisation}} calculé sur les {\large \textbf{arcs}} du graphe viaire d'{\large \textbf{Avignon}}.}
    \label{fig:arcs_use_av}
\end{figure}

\begin{figure}[c]
    \centering
    \includegraphics[width=\textwidth]{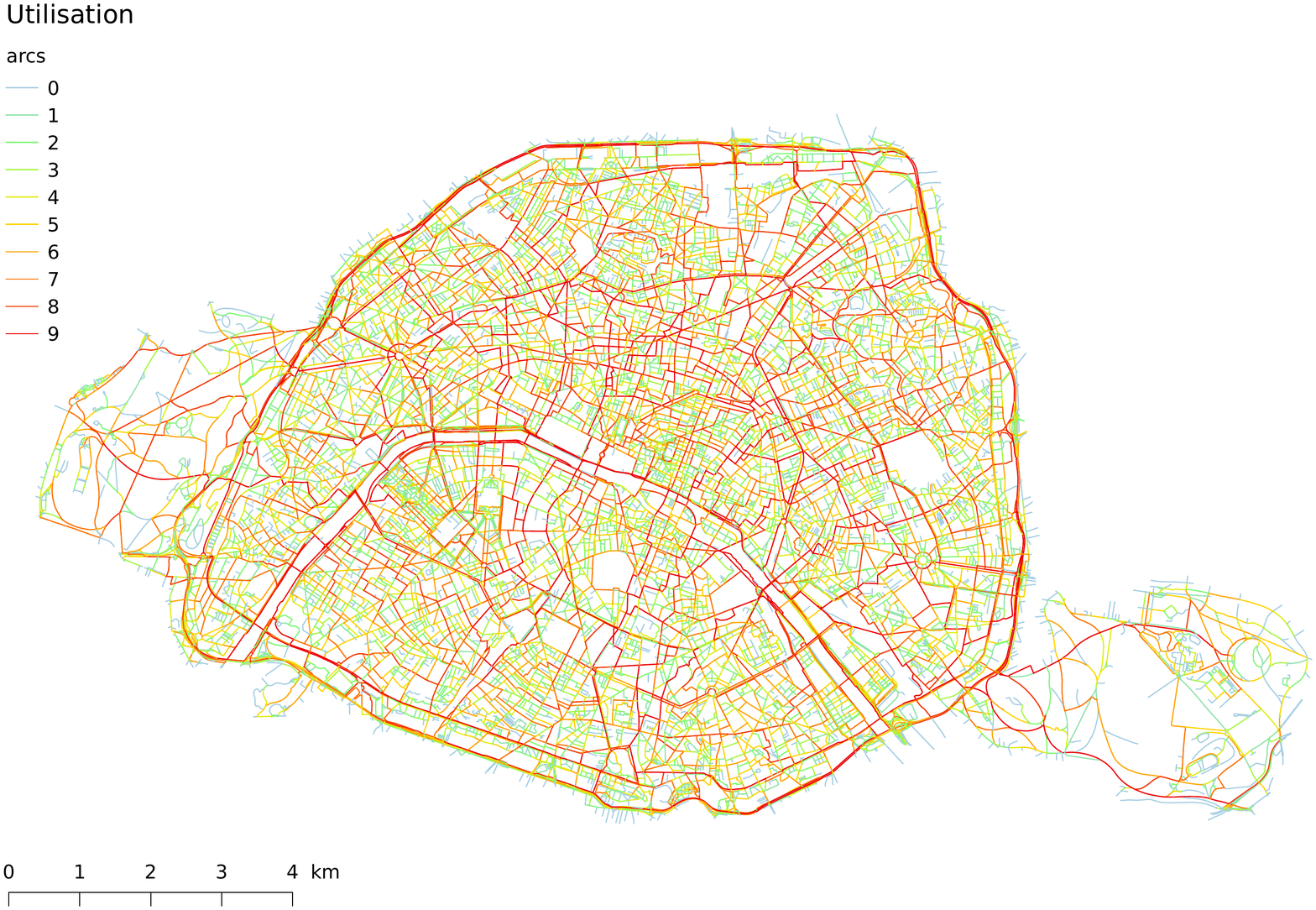}
    \caption{Indicateur d'{\large \textbf{utilisation}} calculé sur les {\large \textbf{arcs}} du graphe viaire de {\large \textbf{Paris}}.}
    \label{fig:arcs_use_par}
\end{figure}

\FloatBarrier

\begin{figure}[c]
    \centering
    \includegraphics[width=\textwidth]{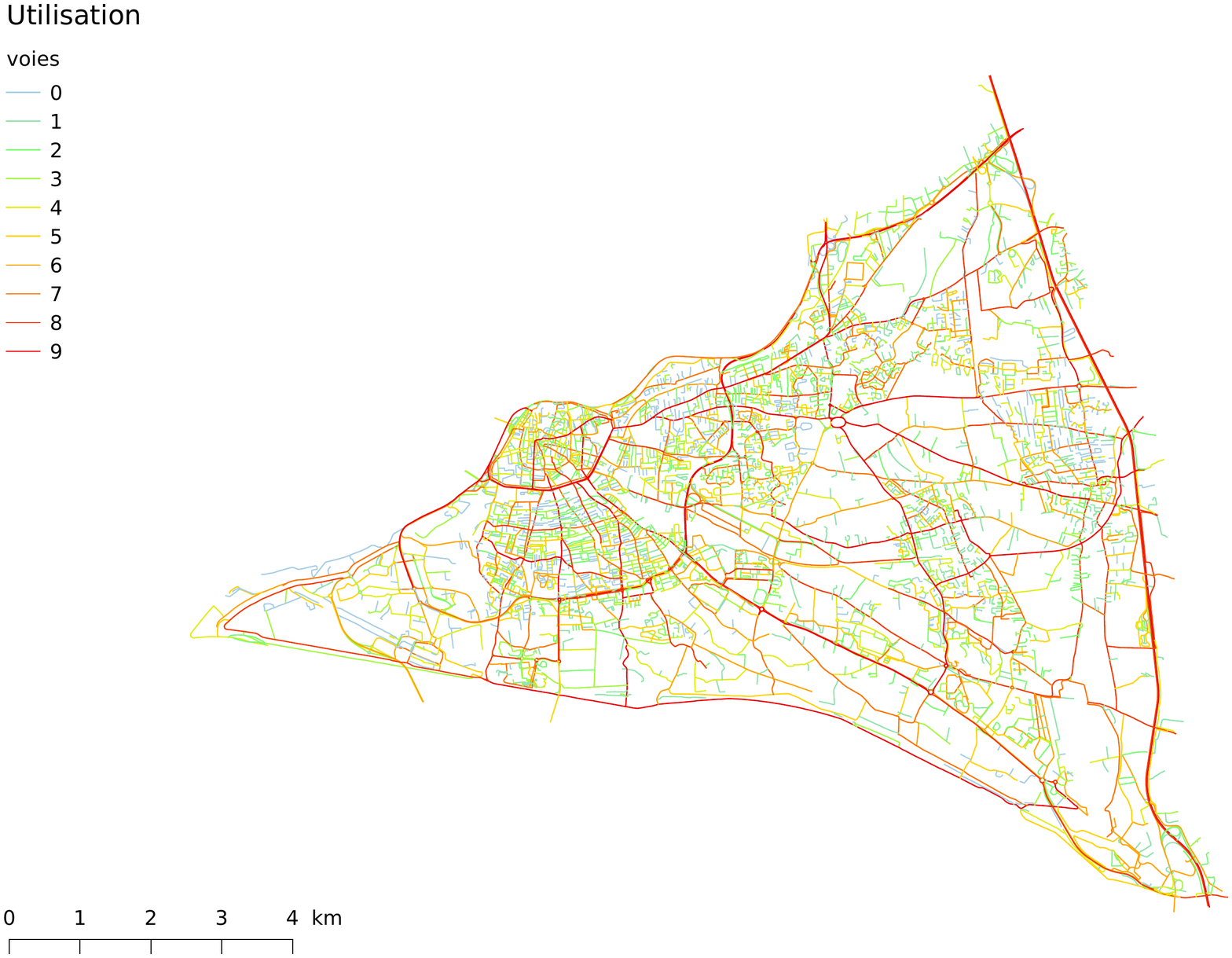}
    \caption{Indicateur d'{\large \textbf{utilisation}} calculé sur les {\large \textbf{voies}} du graphe viaire d'{\large \textbf{Avignon}}.}
    \label{fig:voies_use_av}
\end{figure}

\begin{figure}[c]
    \centering
    \includegraphics[width=\textwidth]{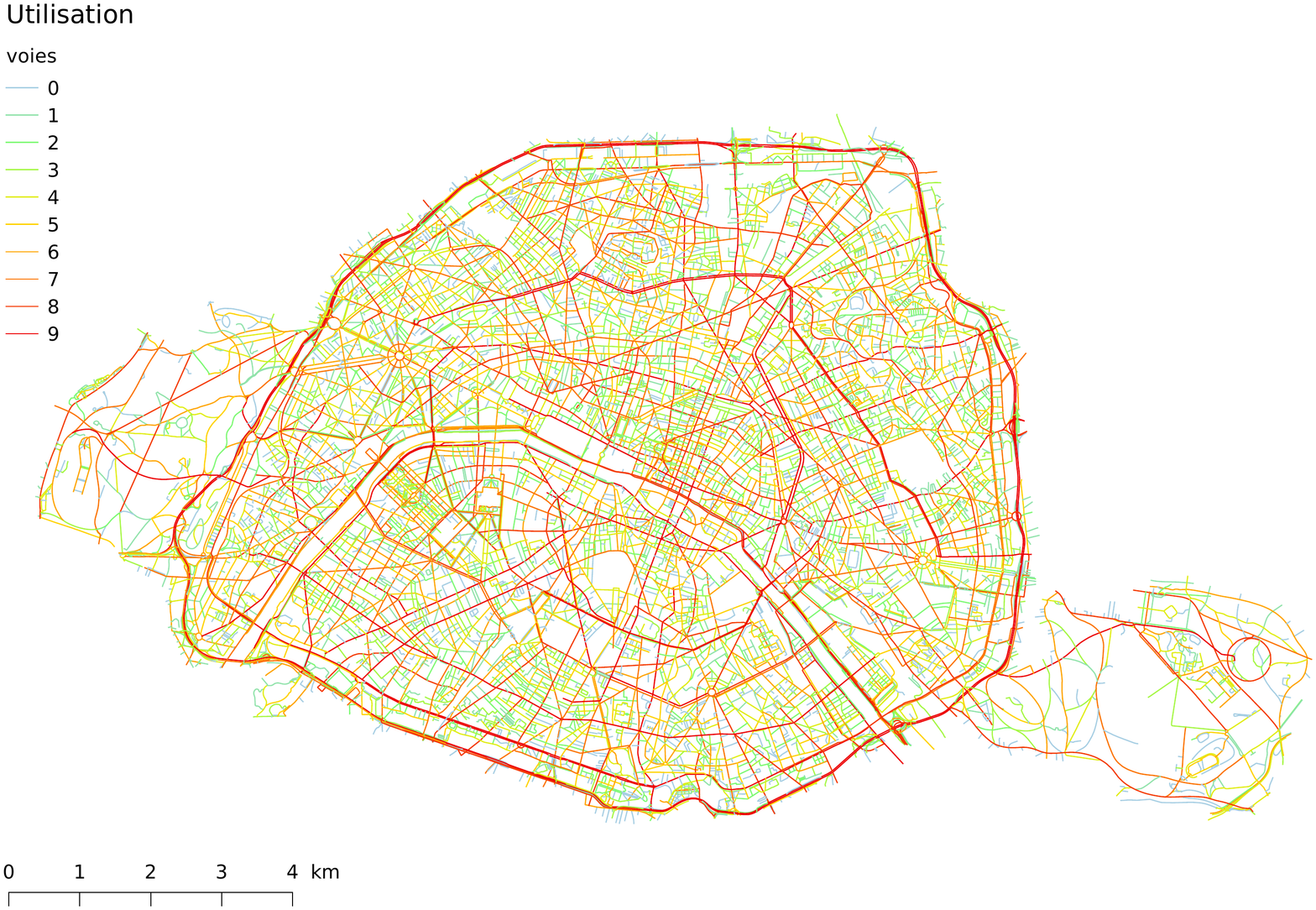}
    \caption{Indicateur d'{\large \textbf{utilisation}} calculé sur les {\large \textbf{voies}} du graphe viaire de {\large \textbf{Paris}}.}
    \label{fig:voies_use_par}
\end{figure}

\FloatBarrier

\FloatBarrier

\begin{figure}[c]
    \centering
    \includegraphics[width=\textwidth]{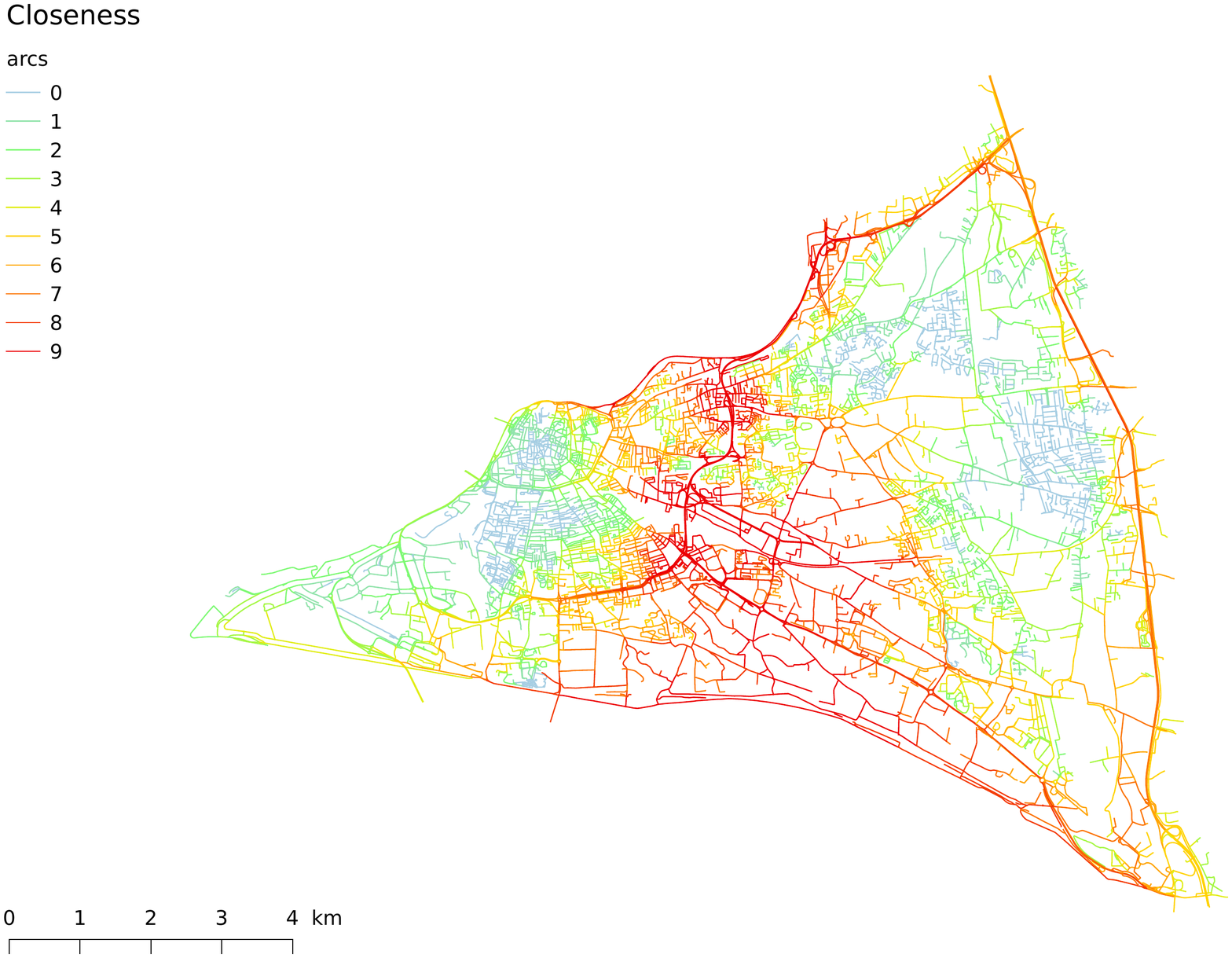}
    \caption{Indicateur de {\large \textbf{closeness}} calculé sur les {\large \textbf{arcs}} du graphe viaire d'{\large \textbf{Avignon}}.}
    \label{fig:arcs_closeness_av}
\end{figure}

\begin{figure}[c]
    \centering
    \includegraphics[width=\textwidth]{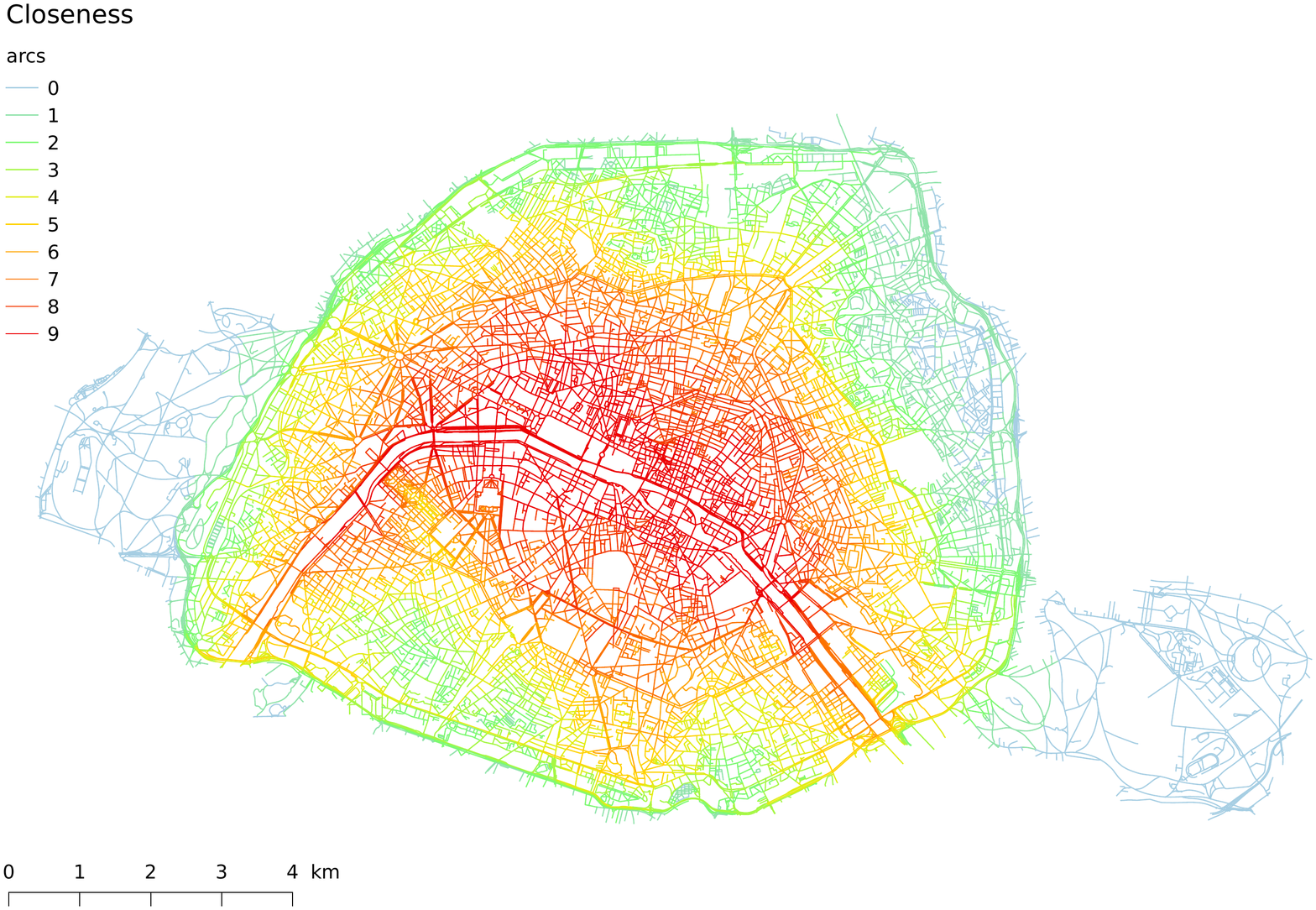}
    \caption{Indicateur de {\large \textbf{closeness}} calculé sur les {\large \textbf{arcs}} du graphe viaire de {\large \textbf{Paris}}.}
    \label{fig:arcs_closeness_par}
\end{figure}

\FloatBarrier

\begin{figure}[c]
    \centering
    \includegraphics[width=\textwidth]{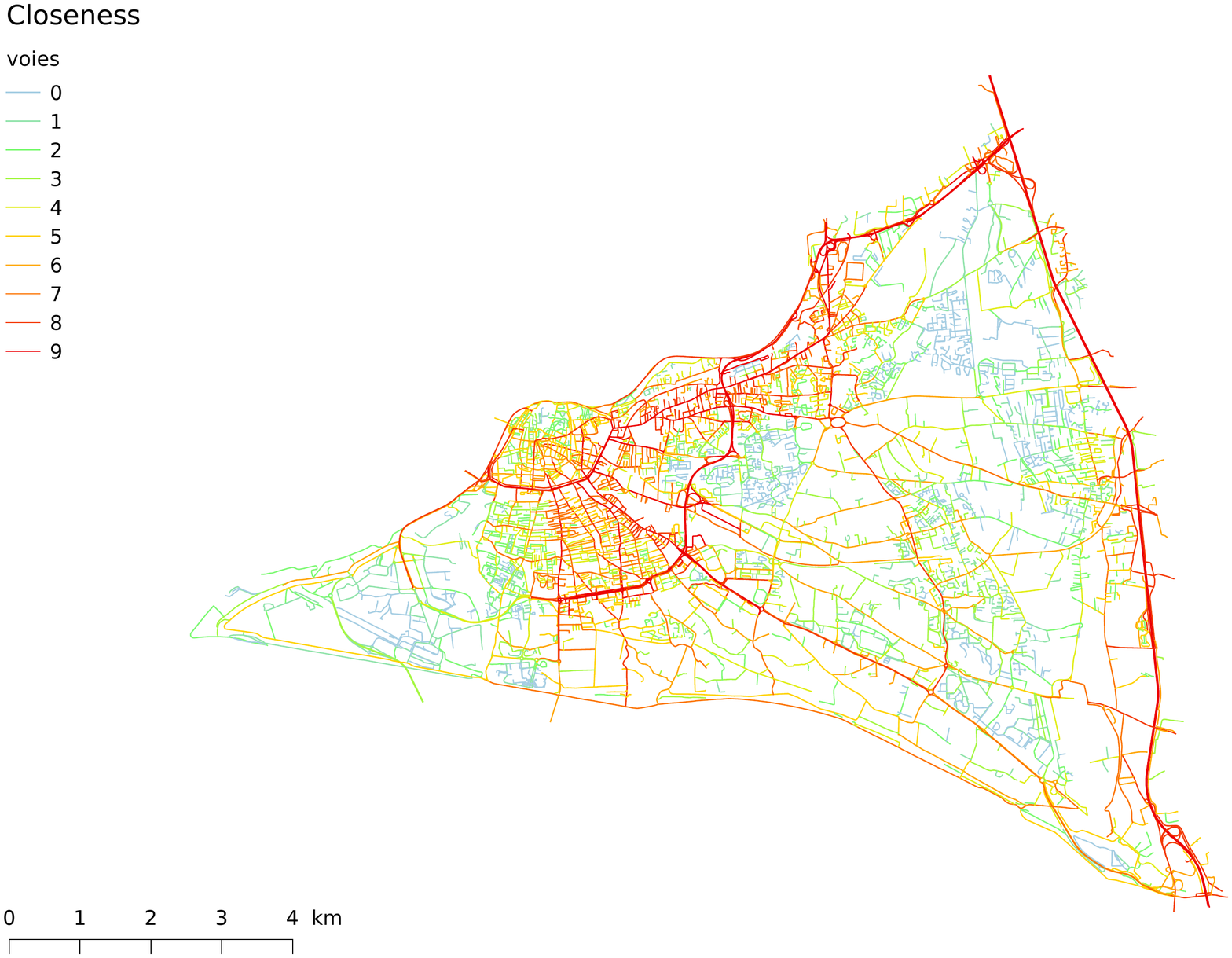}
    \caption{Indicateur de {\large \textbf{closeness}} calculé sur les {\large \textbf{voies}} du graphe viaire d'{\large \textbf{Avignon}}.}
    \label{fig:voies_closeness_av}
\end{figure}

\begin{figure}[c]
    \centering
    \includegraphics[width=\textwidth]{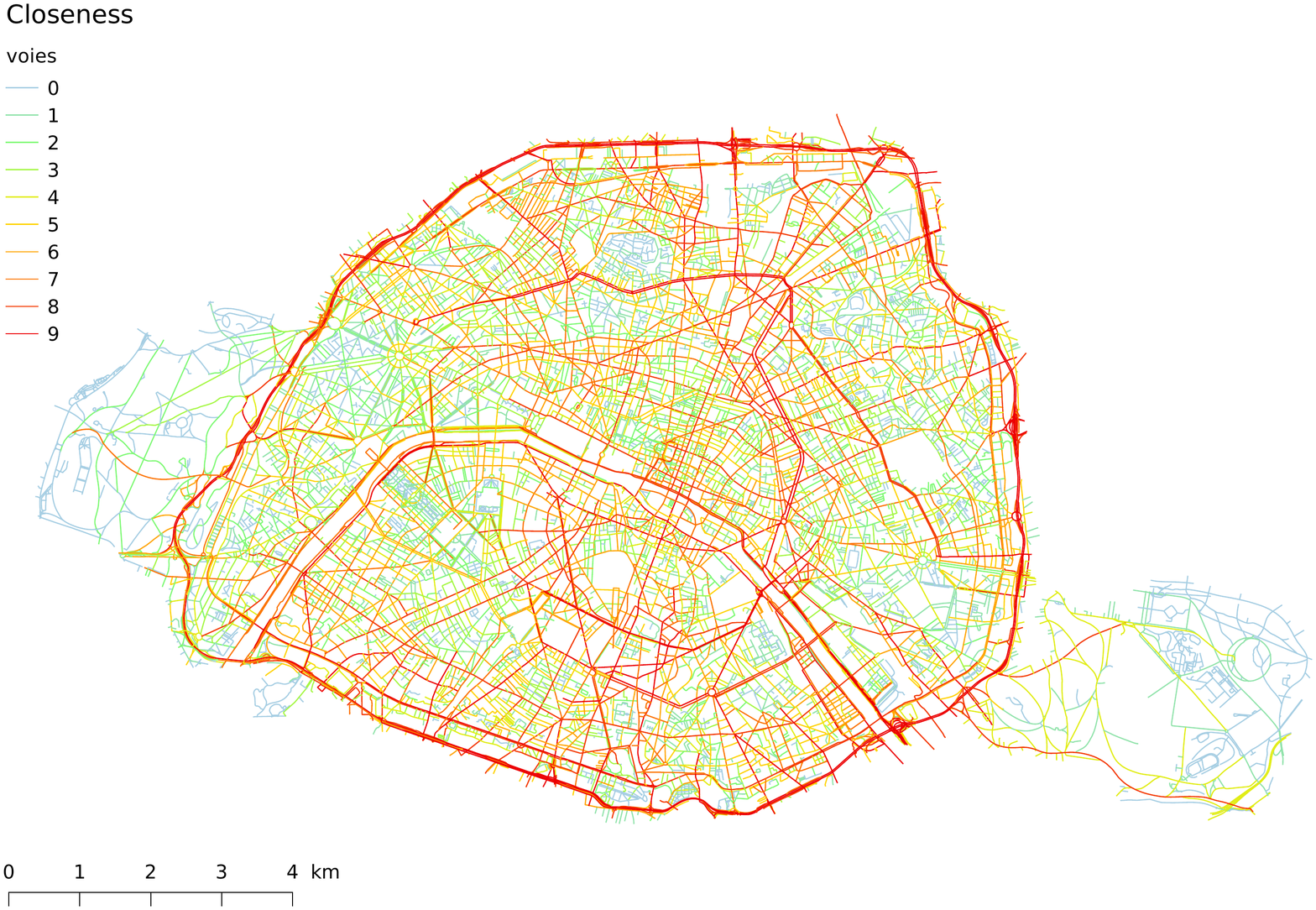}
    \caption{Indicateur de {\large \textbf{closeness}} calculé sur les {\large \textbf{voies}} du graphe viaire de {\large \textbf{Paris}}.}
    \label{fig:voies_closeness_par}
\end{figure}

\FloatBarrier

\FloatBarrier

\begin{figure}[c]
    \centering
    \includegraphics[width=\textwidth]{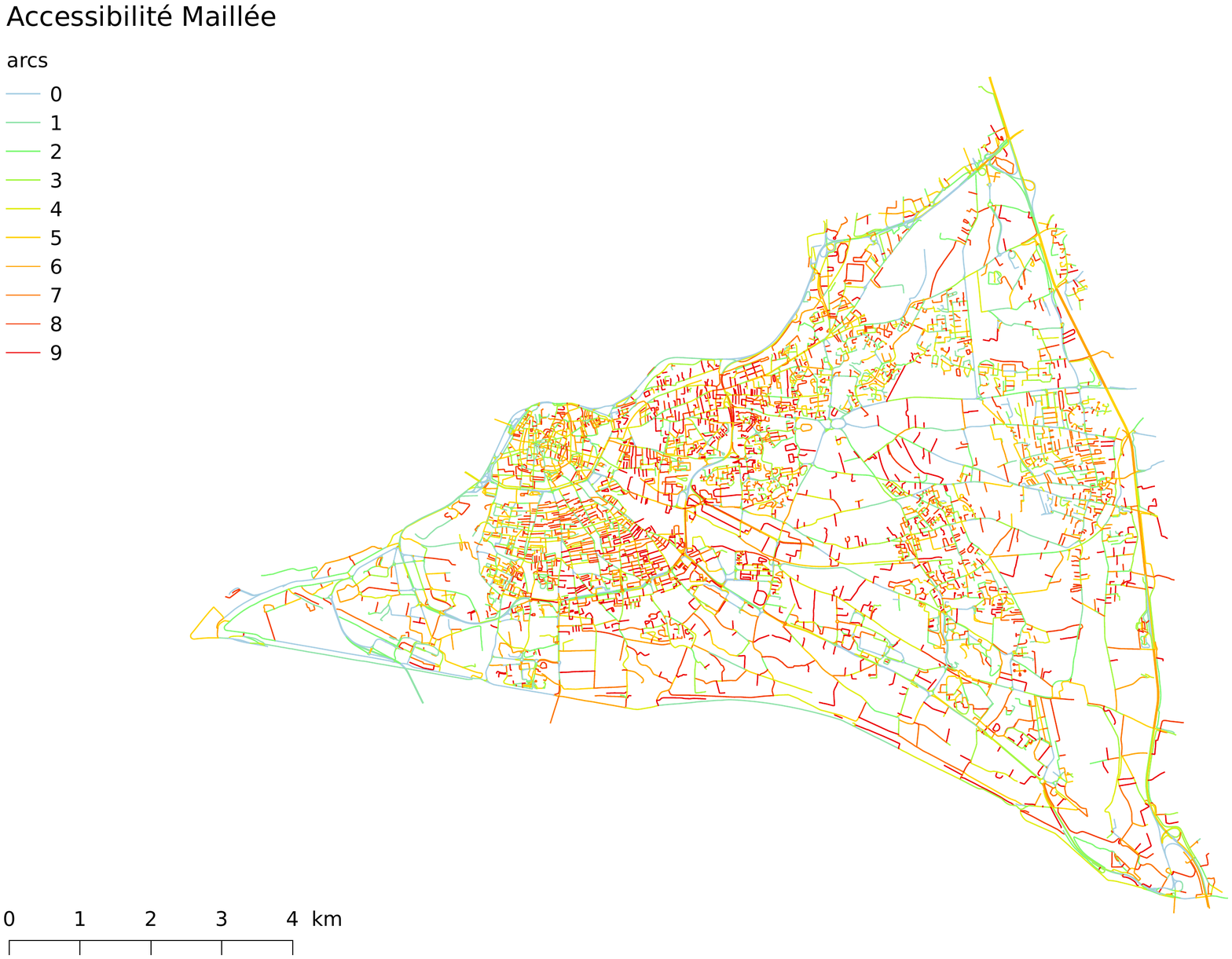}
    \caption{Indicateur d'{\large \textbf{accessibilité maillée}} calculé sur les {\large \textbf{arcs}} du graphe viaire d'{\large \textbf{Avignon}}.}
    \label{fig:arcs_roo_av}
\end{figure}

\begin{figure}[c]
    \centering
    \includegraphics[width=\textwidth]{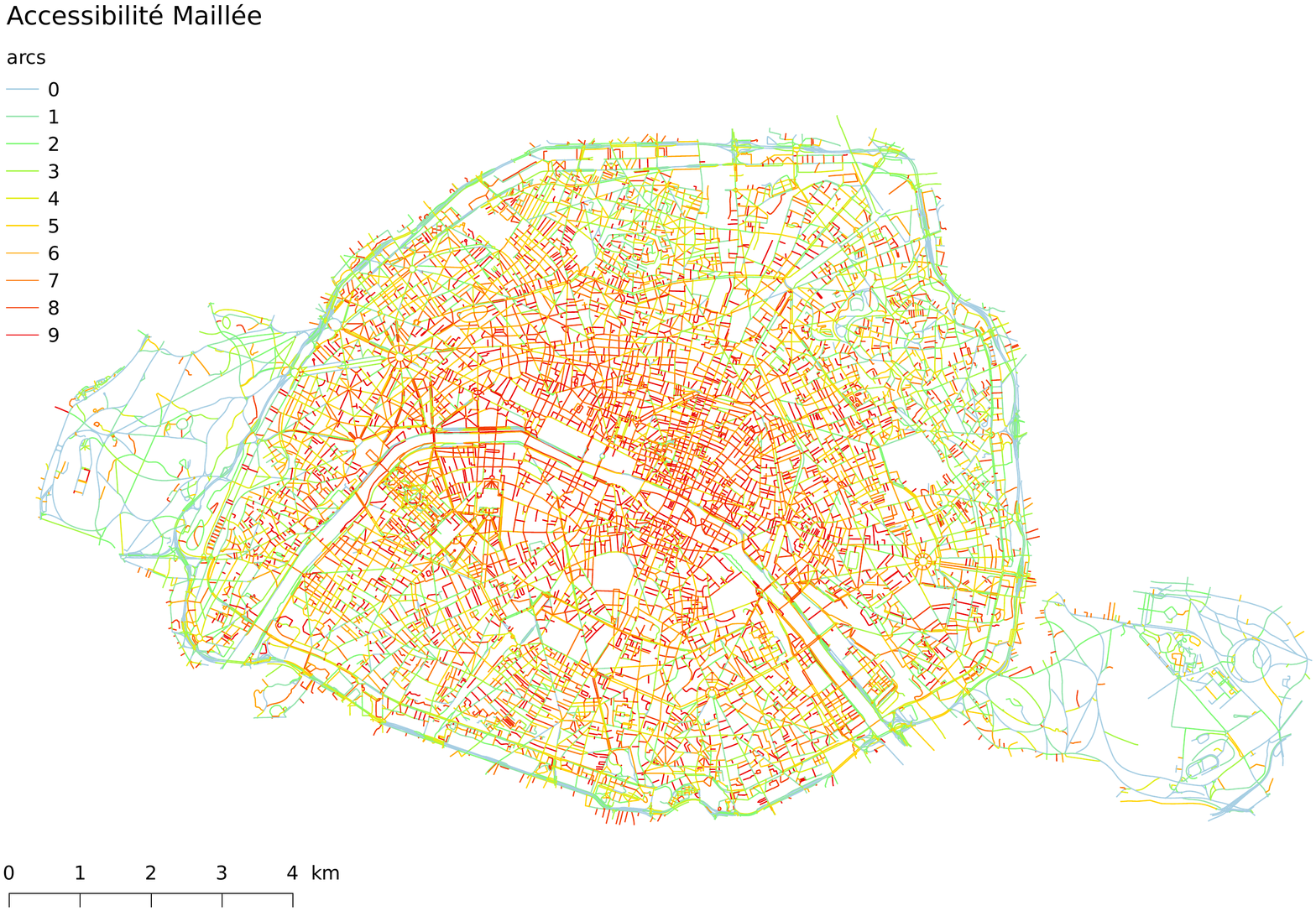}
    \caption{Indicateur d'{\large \textbf{accessibilité maillée}} calculé sur les {\large \textbf{arcs}} du graphe viaire de {\large \textbf{Paris}}.}
    \label{fig:arcs_roo_par}
\end{figure}

\FloatBarrier

\begin{figure}[c]
    \centering
    \includegraphics[width=\textwidth]{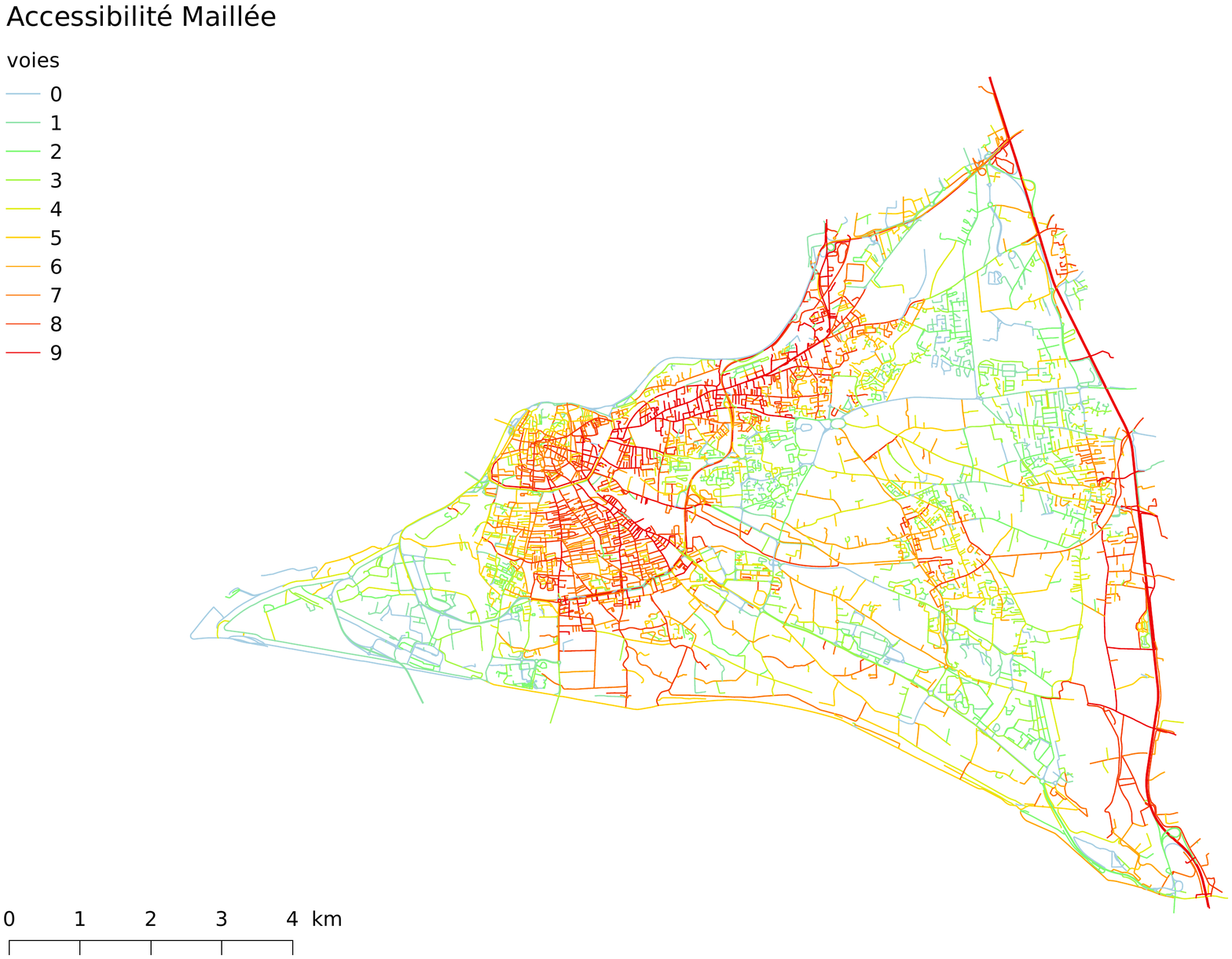}
    \caption{Indicateur d'{\large \textbf{accessibilité maillée}} calculé sur les {\large \textbf{voies}} du graphe viaire d'{\large \textbf{Avignon}}.}
    \label{fig:voies_roo_av}
\end{figure}

\begin{figure}[c]
    \centering
    \includegraphics[width=\textwidth]{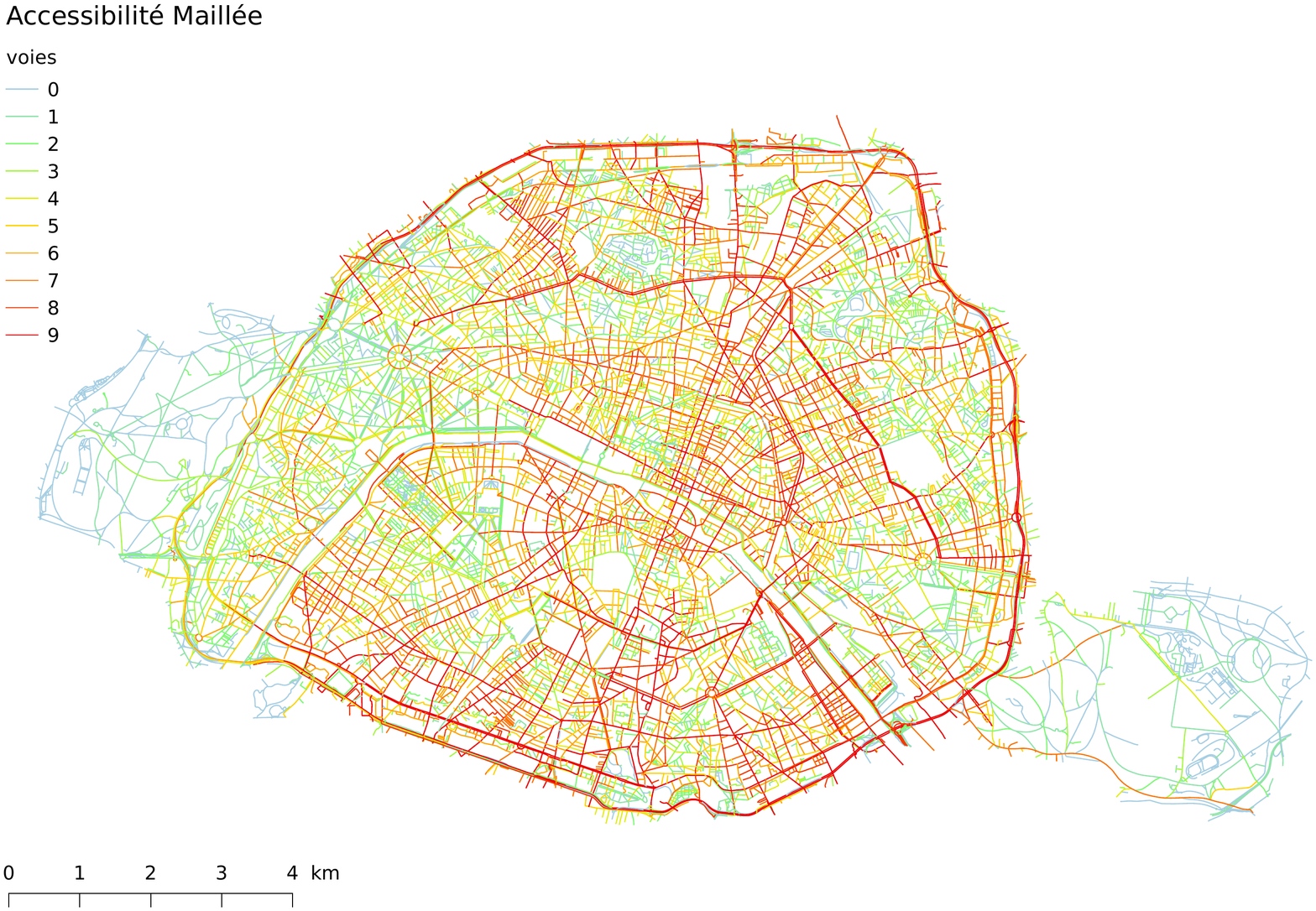}
    \caption{Indicateur d'{\large \textbf{accessibilité maillée}} calculé sur les {\large \textbf{voies}} du graphe viaire de {\large \textbf{Paris}}.}
    \label{fig:voies_roo_par}
\end{figure}

\FloatBarrier

\FloatBarrier

\begin{figure}[c]
    \centering
    \includegraphics[width=\textwidth]{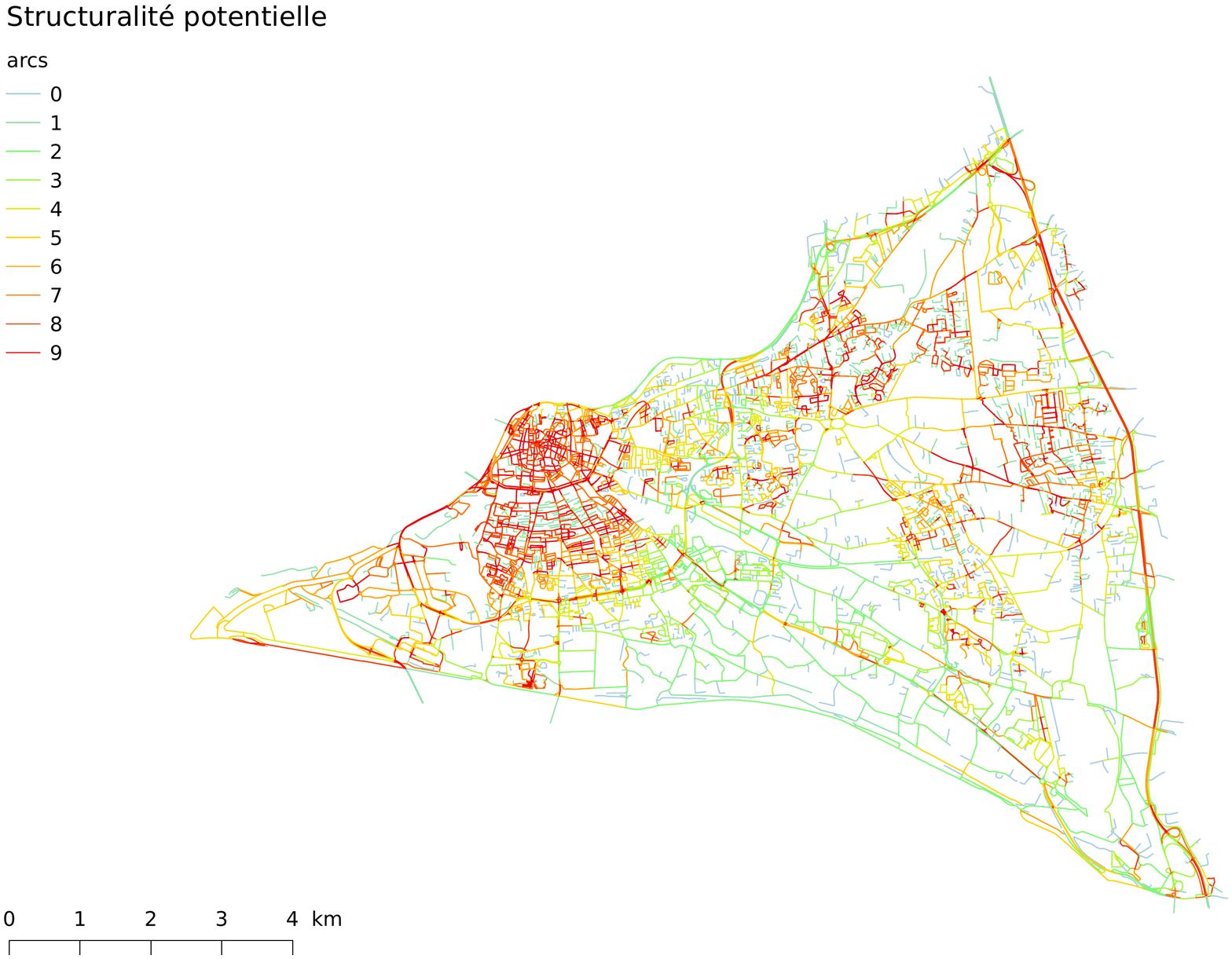}
    \caption{Indicateur de {\large \textbf{structuralité potentielle}} calculé sur les {\large \textbf{arcs}} du graphe viaire d'{\large \textbf{Avignon}}.}
    \label{fig:arcs_structpot_av}
\end{figure}

\begin{figure}[c]
    \centering
    \includegraphics[width=\textwidth]{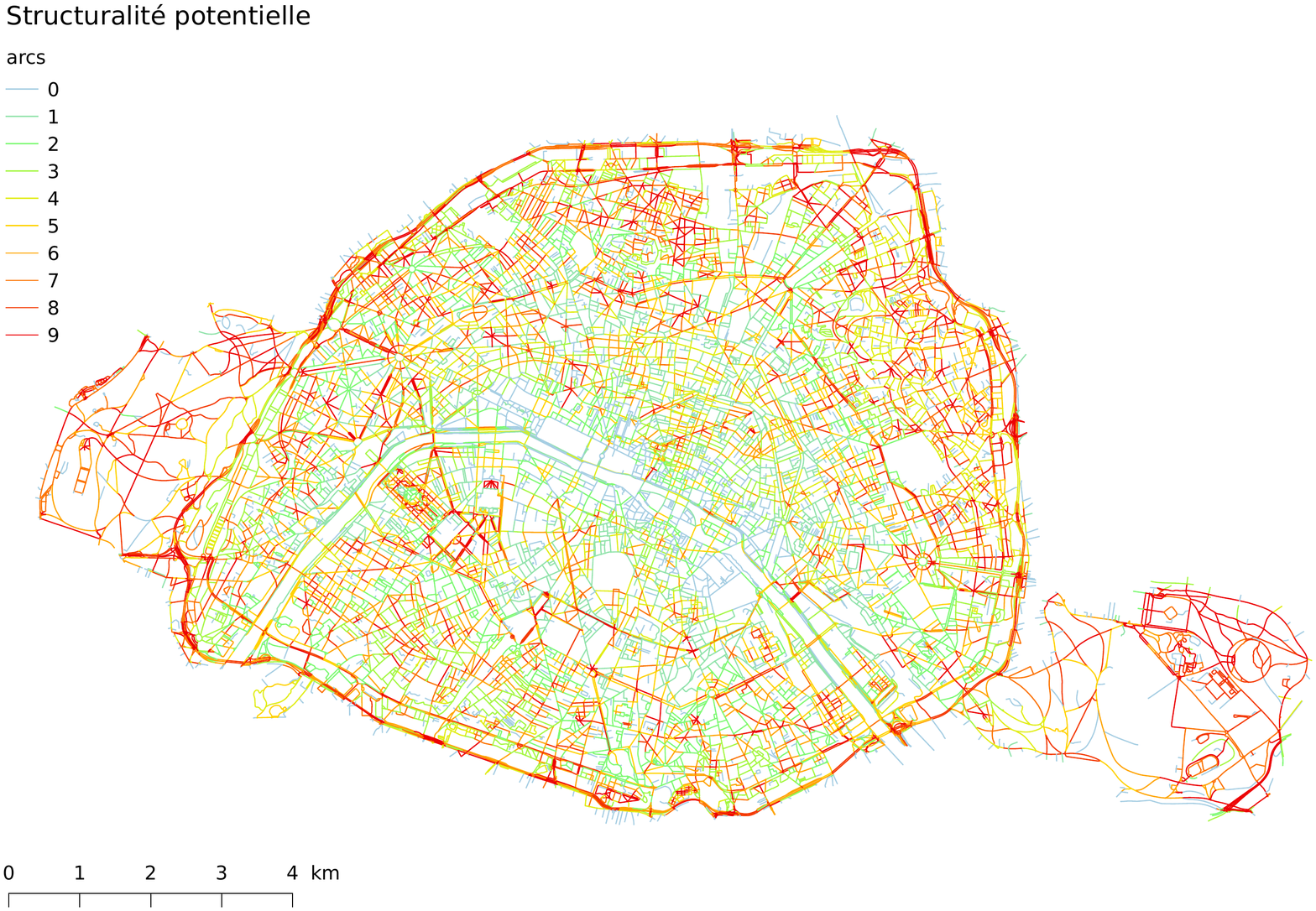}
    \caption{Indicateur de {\large \textbf{structuralité potentielle}} calculé sur les {\large \textbf{arcs}} du graphe viaire de {\large \textbf{Paris}}.}
    \label{fig:arcs_structpot_par}
\end{figure}

\FloatBarrier

\begin{figure}[c]
    \centering
    \includegraphics[width=\textwidth]{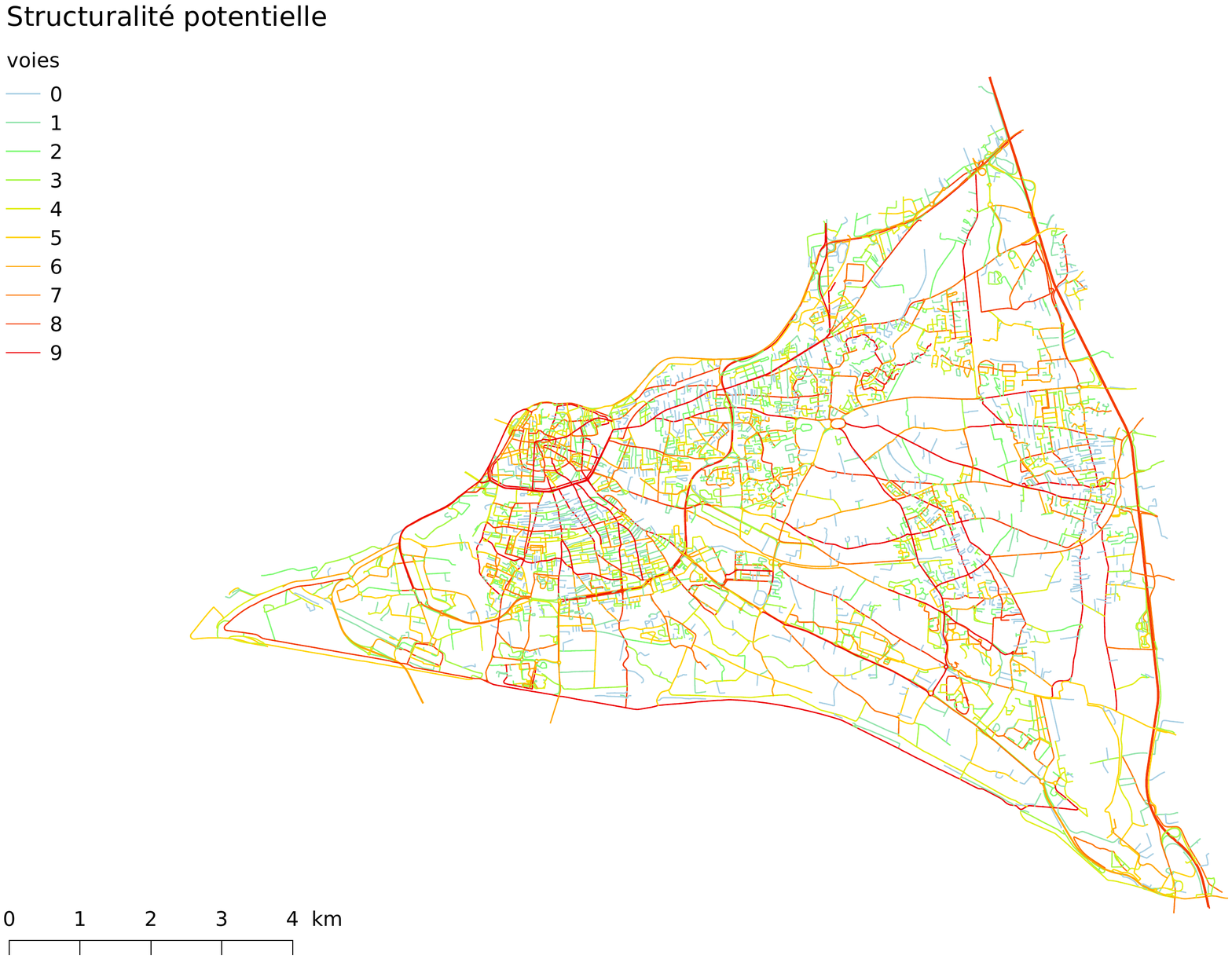}
    \caption{Indicateur de {\large \textbf{structuralité potentielle}} calculé sur les {\large \textbf{voies}} du graphe viaire d'{\large \textbf{Avignon}}.}
    \label{fig:voies_structpot_av}
\end{figure}

\begin{figure}[c]
    \centering
    \includegraphics[width=\textwidth]{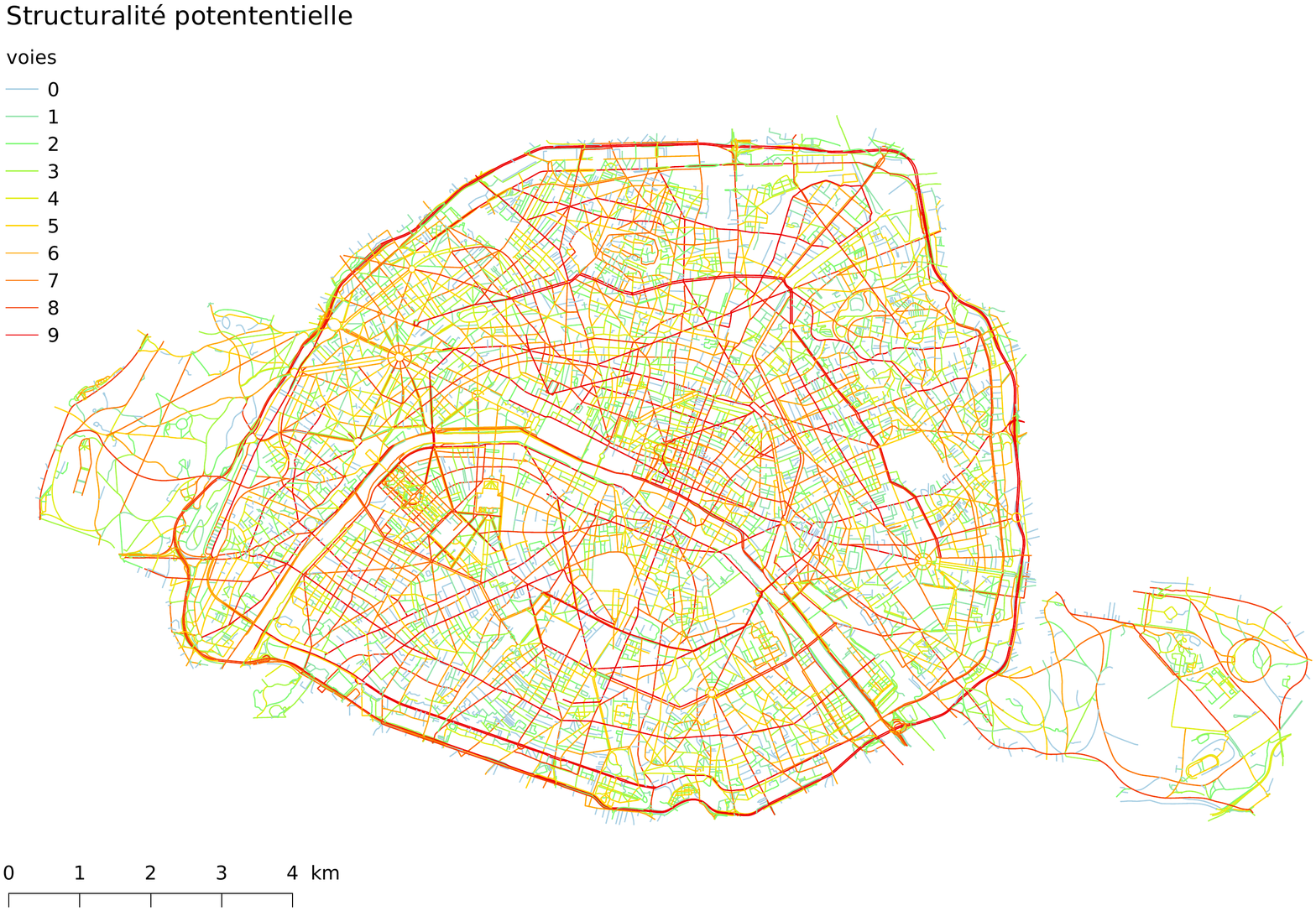}
    \caption{Indicateur de {\large \textbf{structuralité potentielle}} calculé sur les {\large \textbf{voies}} du graphe viaire de {\large \textbf{Paris}}.}
    \label{fig:voies_structpot_par}
\end{figure}

\FloatBarrier 

\chapter{Comparaison des indicateurs}

\section{Comparaison sur les arcs} 


\begin{table}
\begin{center}
\begin{tabular}{|r|c|c|c|c|c|c|c|c|}
\hline
 & \rotatebox{-90}{longueur} & \rotatebox{-90}{degré} & \rotatebox{-90}{structuralité potentielle    } & \rotatebox{-90}{betweenness} & \rotatebox{-90}{utilisation} & \rotatebox{-90}{accessibilité} & \rotatebox{-90}{closeness} & \rotatebox{-90}{orthogonalité} \\ \hline
longueur & 1,00 & -0,18 & -0,20 & 0,07 & 0,05 & -0,17 & 0,10 & 0,13 \\ \hline
degré & -0,18 & 1,00 & 0,75 & 0,46 & 0,45 & -0,06 & 0,08 & -0,19 \\ \hline
structuralité potentielle & -0,20 & 0,75 & 1,00 & 0,37 & 0,37 & 0,43 & -0,36 & -0,09 \\ \hline
betweenness & 0,07 & 0,46 & 0,37 & 1,00 & 0,96 & -0,24 & 0,28 & -0,34 \\ \hline
utilisation & 0,05 & 0,45 & 0,37 & 0,96 & 1,00 & -0,23 & 0,27 & -0,34 \\ \hline
accessibilité & -0,17 & -0,06 & 0,43 & -0,24 & -0,23 & 1,00 & -0,91 & 0,19 \\ \hline
closeness & 0,10 & 0,08 & -0,36 & 0,28 & 0,27 & -0,91 & 1,00 & -0,17 \\ \hline
orthogonalité & 0,13 & -0,19 & -0,09 & -0,34 & -0,34 & 0,19 & -0,17 & 1,00 \\ \hline
\end{tabular}
\end{center}
\caption{Calcul de la corrélation de Pearson pour les  {\large \textbf{indicateurs primaires}} calculés sur les {\large \textbf{arcs}} du graphe viaire d'{\large \textbf{Avignon}}.}
\label{tab:corr_avignon_arc}
\end{table}

\begin{table}
\begin{center}
\begin{tabular}{|r|c|c|c|c|c|c|c|c|}
\hline
 & \rotatebox{-90}{longueur} & \rotatebox{-90}{degré} & \rotatebox{-90}{structuralité potentielle    } & \rotatebox{-90}{betweenness} & \rotatebox{-90}{utilisation} & \rotatebox{-90}{accessibilité} & \rotatebox{-90}{closeness} & \rotatebox{-90}{orthogonalité} \\ \hline
longueur & 1,00 & -0,01 & -0,14 & 0,25 & 0,24 & -0,18 & 0,18 & -0,01 \\ \hline
degré & -0,01 & 1,00 & 0,53 & 0,22 & 0,19 & -0,10 & 0,10 & 0,00 \\ \hline
structuralité potentielle & -0,14 & 0,53 & 1,00 & 0,02 & 0,00 & 0,69 & -0,69 & 0,19 \\ \hline
betweenness & 0,25 & 0,22 & 0,02 & 1,00 & 0,89 & -0,22 & 0,23 & -0,03 \\ \hline
utilisation & 0,24 & 0,19 & 0,00 & 0,89 & 1,00 & -0,21 & 0,22 & -0,06 \\ \hline
accessibilité & -0,18 & -0,10 & 0,69 & -0,22 & -0,21 & 1,00 & -0,99 & 0,22 \\ \hline
closeness & 0,18 & 0,10 & -0,69 & 0,23 & 0,22 & -0,99 & 1,00 & -0,21 \\ \hline
orthogonalité & -0,01 & 0,00 & 0,19 & -0,03 & -0,06 & 0,22 & -0,21 & 1,00 \\ \hline
\end{tabular}
\end{center}
\caption{Calcul de la corrélation de Pearson pour les {\large \textbf{indicateurs primaires}} calculés sur les {\large \textbf{arcs}} du graphe viaire de {\large \textbf{Manhattan}}.}
\label{tab:corr_manhattan_arc}
\end{table}

\FloatBarrier

\begin{table}
\begin{center}
\small
\begin{tabular}{|r|c|c|c|c|c|}
\hline
 & $\frac{longueur}{utilisation}$ & $\frac{orthogonalite}{utilisation}$ & $\frac{structuralite}{utilisation}$ & $\frac{utilisation}{closeness}$ & $\frac{longueur}{degre}$ \\ \hline
utilisation & -0,85 & -0,96 & -0,96 & 0,99 & -0,15 \\ \hline
longueur & 0,42 & -0,04 & -0,11 & 0,05 & 0,94 \\ \hline
orthogonalité & 0,39 & 0,52 & 0,32 & -0,34 & 0,23 \\ \hline
degré & -0,50 & -0,42 & -0,32 & 0,45 & -0,41 \\ \hline
structuralité potentielle & -0,42 & -0,33 & -0,20 & 0,40 & -0,43 \\ \hline
closeness & -0,19 & -0,27 & -0,33 & 0,18 & 0,05 \\ \hline
\hline
& $\frac{longueur}{structuralite}$  & $\frac{longueur}{closeness}$ & $\frac{orthogonalite}{longueur}$ & $\frac{orthogonalite}{degre}$ & $\frac{orthogonalite}{structuralite}$ \\ \hline 
utilisation & -0,12 & 0,03 & -0,16 & -0,52 & -0,48 \\ \hline
longueur & 0,94 & 0,99 & -0,91 & 0,15 & 0,21 \\ \hline
orthogonalité & 0,20 & 0,15 & 0,21 & 0,84 & 0,82 \\ \hline
degré & -0,40 & -0,19 & 0,12 & -0,55 & -0,55 \\ \hline
structuralité potentielle & -0,46 & -0,17 & 0,17 & -0,45 & -0,57 \\ \hline
closeness & 0,13 & 0,01 & -0,15 & -0,21 & 0,02 \\ \hline
 \hline
 & $\frac{orthogonalite}{closeness}$ & $\frac{structuralite}{closeness}$ & $\frac{degre}{closeness}$ & $\frac{degre}{structuralite}$ & $\frac{degre}{utilisation}$ \\ \hline
utilisation & -0,39 & 0,30 & 0,35 & 0,19 & 0,00 \\ \hline
longueur & 0,09 & -0,21 & -0,17 & 0,18 & 0,03 \\ \hline
orthogonalité & 0,95 & -0,09 & -0,10 & -0,17 & 0,01 \\ \hline
degré & -0,20 & 0,64 & 0,75 & 0,03 & 0,08 \\ \hline
structuralité potentielle & 0,01 & 0,96 & 0,97 & -0,46 & -0,03 \\ \hline
closeness & -0,43 & -0,53 & -0,40 & 0,91 & -0,03 \\ \hline

\end{tabular}

\end{center}
\caption{Calcul de la corrélation de Pearson pour les {\large \textbf{indicateurs composés}} calculés sur les {\large \textbf{arcs}} du graphe viaire d'{\large \textbf{Avignon}}.}
\label{tab:corr_avignon_comb}
\end{table}

\begin{table}
\begin{center}
\small
\begin{tabular}{|r|c|c|c|c|c|c|}
\hline
 & $\frac{longueur}{utilisation}$ & $\frac{orthogonalite}{utilisation}$ & $\frac{structuralite}{utilisation}$ & $\frac{utilisation}{closeness}$ & $\frac{longueur}{degre}$ \\ \hline
utilisation & -0,77 & -0,60 & -0,99 & 0,99 & 0,18 \\ \hline
longueur & 0,29 & -0,21 & -0,26 & 0,23 & 0,96 \\ \hline
orthogonalité & 0,04 & 0,63 & 0,09 & -0,03 & -0,01 \\ \hline
degré & -0,19 & -0,22 & -0,10 & 0,17 & -0,21 \\ \hline
structuralité potentielle & -0,09 & 0,09 & 0,15 & 0,08 & -0,27 \\ \hline
closeness & -0,11 & -0,33 & -0,33 & 0,12 & 0,14 \\ \hline
\hline
& $\frac{longueur}{structuralite}$  & $\frac{longueur}{closeness}$ & $\frac{orthogonalite}{longueur}$ & $\frac{orthogonalite}{degre}$ & $\frac{orthogonalite}{structuralite}$ \\ \hline 
utilisation & 0,21 & 0,20 & -0,16 & -0,14 & -0,04 \\ \hline
longueur & 0,94 & 0,96 & -0,57 & -0,03 & 0,02 \\ \hline
orthogonalité & -0,06 & 0,05 & 0,64 & 0,81 & 0,76 \\ \hline
degré & -0,18 & -0,04 & -0,09 & -0,38 & -0,30 \\ \hline
structuralité potentielle & -0,40 & 0,02 & 0,18 & -0,09 & -0,33 \\ \hline
closeness & 0,34 & -0,06 & -0,29 & -0,23 & 0,14 \\ \hline
 \hline
 & $\frac{orthogonalite}{closeness}$ & $\frac{structuralite}{closeness}$ & $\frac{degre}{closeness}$ & $\frac{degre}{structuralite}$ & $\frac{degre}{utilisation}$ \\ \hline
utilisation & -0,13 & 0,30 & -0,01 & 0,19 & -0,01 \\ \hline
longueur & -0,09 & -0,21 & -0,13 & 0,18 & 0,05 \\ \hline
orthogonalité & 0,86 & -0,09 & 0,17 & -0,22 & 0,07 \\ \hline
degré & -0,06 & 0,64 & 0,56 & 0,09 & -0,02 \\ \hline
structuralité potentielle & 0,43 & 0,96 & 0,99 & -0,70 & 0,00 \\ \hline
closeness & -0,56 & -0,53 & -0,66 & 0,99 & -0,03 \\ \hline
\end{tabular}
\end{center}
\caption{Calcul de la corrélation de Pearson pour les {\large \textbf{indicateurs composés}} calculés sur les {\large \textbf{arcs}} du graphe viaire de {\large \textbf{Manhattan}}.}
\label{tab:corr_manhattan_comb}
\end{table}

\FloatBarrier 
\section{Comparaison sur les voies} 

\begin{landscape}
\begin{table}
\begin{center}

\begin{tabular}{|r|c|c|c|c|c|c|c|c|c|c|}
\hline
 & \rotatebox{-90}{longueur} & \rotatebox{-90}{degré} & \rotatebox{-90}{nombre d'arcs} & \rotatebox{-90}{connectivité} & \rotatebox{-90}{structuralité potentielle    } & \rotatebox{-90}{betweenness} & \rotatebox{-90}{utilisation} & \rotatebox{-90}{accessibilité} & \rotatebox{-90}{closeness} & \rotatebox{-90}{orthogonalité} \\ \hline
 
longueur & 1,00 & 0,62 & 0,74 & 0,59 & 0,60 & 0,61 & 0,60 & -0,25 & 0,24 & -0,13 \\ \hline
degré & 0,62 & 1,00 & 0,84 & 0,95 & 0,96 & 0,83 & 0,81 & -0,34 & 0,34 & -0,35 \\ \hline
nombre d'arcs & 0,74 & 0,84 & 1,00 & 0,77 & 0,83 & 0,79 & 0,78 & -0,27 & 0,27 & -0,18 \\ \hline
connectivité & 0,59 & 0,95 & 0,77 & 1,00 & 0,92 & 0,77 & 0,75 & -0,39 & 0,39 & -0,38 \\ \hline
structuralité potentielle & 0,60 & 0,96 & 0,83 & 0,92 & 1,00 & 0,83 & 0,81 & -0,14 & 0,14 & -0,31 \\ \hline
betweenness & 0,61 & 0,83 & 0,79 & 0,77 & 0,83 & 1,00 & 0,93 & -0,31 & 0,30 & -0,20 \\ \hline
utilisation & 0,60 & 0,81 & 0,78 & 0,75 & 0,81 & 0,93 & 1,00 & -0,30 & 0,29 & -0,21 \\ \hline
accessibilité & -0,25 & -0,34 & -0,27 & -0,39 & -0,14 & -0,31 & -0,30 & 1,00 & -0,99 & 0,24 \\ \hline
closeness & 0,24 & 0,34 & 0,27 & 0,39 & 0,14 & 0,30 & 0,29 & -0,99 & 1,00 & -0,22 \\ \hline
orthogonalité & -0,13 & -0,35 & -0,18 & -0,38 & -0,31 & -0,20 & -0,21 & 0,24 & -0,22 & 1,00 \\ \hline

\end{tabular}
\end{center}
\caption{Calcul de la corrélation de Pearson pour les {\large \textbf{indicateurs primaires}} calculés sur les {\large \textbf{voies}} du graphe viaire d'{\large \textbf{Avignon}}.}
\label{tab:corr_avignon_voie}
\end{table}
\end{landscape}
\begin{landscape}

\begin{table}
\begin{center}

\begin{tabular}{|r|c|c|c|c|c|c|c|c|c|c|}
\hline
 & \rotatebox{-90}{longueur} & \rotatebox{-90}{degré} & \rotatebox{-90}{nombre d'arcs} & \rotatebox{-90}{connectivité} & \rotatebox{-90}{structuralité potentielle    } & \rotatebox{-90}{betweenness} & \rotatebox{-90}{utilisation} & \rotatebox{-90}{accessibilité} & \rotatebox{-90}{closeness} & \rotatebox{-90}{orthogonalité} \\ \hline
 
longueur & 1,00 & 0,92 & 0,92 & 0,92 & 0,88 & 0,81 & 0,83 & -0,77 & 0,75 & 0,03 \\ \hline
degré & 0,92 & 1,00 & 0,99 & 0,99 & 0,99 & 0,87 & 0,88 & -0,69 & 0,68 & -0,08 \\ \hline
nombre d'arcs & 0,92 & 0,99 & 1,00 & 0,98 & 0,97 & 0,87 & 0,88 & -0,70 & 0,68 & -0,08 \\ \hline
connectivité & 0,92 & 0,99 & 0,98 & 1,00 & 0,97 & 0,85 & 0,86 & -0,72 & 0,70 & -0,05 \\ \hline
structuralité potentielle & 0,88 & 0,99 & 0,97 & 0,97 & 1,00 & 0,89 & 0,89 & -0,60 & 0,59 & -0,11 \\ \hline
betweenness & 0,81 & 0,87 & 0,87 & 0,85 & 0,89 & 1,00 & 0,96 & -0,64 & 0,66 & -0,08 \\ \hline
utilisation & 0,83 & 0,88 & 0,88 & 0,86 & 0,89 & 0,96 & 1,00 & -0,66 & 0,66 & -0,04 \\ \hline
accessibilité & -0,77 & -0,69 & -0,70 & -0,72 & -0,60 & -0,64 & -0,66 & 1,00 & -0,98 & -0,07 \\ \hline
closeness & 0,75 & 0,68 & 0,68 & 0,70 & 0,59 & 0,66 & 0,66 & -0,98 & 1,00 & 0,05 \\ \hline
orthogonalité & 0,03 & -0,08 & -0,08 & -0,05 & -0,11 & -0,08 & -0,04 & -0,07 & 0,05 & 1,00 \\ \hline

\end{tabular}
\end{center}
\caption{Calcul de la corrélation de Pearson pour les {\large \textbf{indicateurs primaires}} calculés sur les {\large \textbf{voies}} du graphe viaire de {\large \textbf{Manhattan}}.}
\label{tab:corr_manhattan_voie}
\end{table}
\end{landscape}

\begin{landscape}
\begin{table}
\begin{center}

\begin{tabular}{|r|c|c|c|c|c|c|c|c|c|c|}
\hline
 & \rotatebox{-90}{longueur} & \rotatebox{-90}{degré} & \rotatebox{-90}{nombre d'arcs} & \rotatebox{-90}{connectivité} & \rotatebox{-90}{structuralité potentielle    } & \rotatebox{-90}{betweenness} & \rotatebox{-90}{utilisation} & \rotatebox{-90}{accessibilité} & \rotatebox{-90}{closeness} & \rotatebox{-90}{orthogonalité} \\ \hline
 
longueur & 1,00 & 0,84 & 0,87 & 0,84 & 0,83 & 0,75 & 0,75 & -0,49 & 0,49 & -0,26 \\ \hline
degré & 0,84 & 1,00 & 0,91 & 0,98 & 0,98 & 0,86 & 0,85 & -0,54 & 0,55 & -0,33 \\ \hline
nombre d'arcs & 0,87 & 0,91 & 1,00 & 0,89 & 0,90 & 0,82 & 0,82 & -0,50 & 0,50 & -0,27 \\ \hline
connectivité & 0,84 & 0,98 & 0,89 & 1,00 & 0,96 & 0,83 & 0,81 & -0,56 & 0,57 & -0,34 \\ \hline
structuralité potentielle & 0,83 & 0,98 & 0,90 & 0,96 & 1,00 & 0,87 & 0,86 & -0,41 & 0,42 & -0,30 \\ \hline
betweenness & 0,75 & 0,86 & 0,82 & 0,83 & 0,87 & 1,00 & 0,97 & -0,51 & 0,51 & -0,19 \\ \hline
utilisation & 0,75 & 0,85 & 0,82 & 0,81 & 0,86 & 0,97 & 1,00 & -0,50 & 0,51 & -0,17 \\ \hline
accessibilité & -0,49 & -0,54 & -0,50 & -0,56 & -0,41 & -0,51 & -0,50 & 1,00 & -1,00 & 0,27 \\ \hline
closeness & 0,49 & 0,55 & 0,50 & 0,57 & 0,42 & 0,51 & 0,51 & -1,00 & 1,00 & -0,26 \\ \hline
orthogonalité & -0,26 & -0,33 & -0,27 & -0,34 & -0,30 & -0,19 & -0,17 & 0,27 & -0,26 & 1,00 \\ \hline

\end{tabular}
\end{center}
\caption{Calcul de la corrélation de Pearson pour les {\large \textbf{indicateurs primaires}} calculés sur les {\large \textbf{voies}} du graphe viaire de {\large \textbf{Paris}}.}
\label{tab:corr_paris_voie}
\end{table}
\end{landscape}

\begin{landscape}
\begin{table}
\begin{center}

\begin{tabular}{|r|c|c|c|c|c|c|c|c|c|c|}
\hline
 & \rotatebox{-90}{longueur} & \rotatebox{-90}{degré} & \rotatebox{-90}{nombre d'arcs} & \rotatebox{-90}{connectivité} & \rotatebox{-90}{structuralité potentielle    } & \rotatebox{-90}{betweenness} & \rotatebox{-90}{utilisation} & \rotatebox{-90}{accessibilité} & \rotatebox{-90}{closeness} & \rotatebox{-90}{orthogonalité} \\ \hline
 
longueur & 1,00 & 0,81 & 0,82 & 0,81 & 0,78 & 0,67 & 0,67 & -0,43 & 0,43 & -0,15 \\ \hline
degré & 0,81 & 1,00 & 0,94 & 0,97 & 0,96 & 0,81 & 0,81 & -0,47 & 0,48 & -0,15 \\ \hline
nombre d'arcs & 0,82 & 0,94 & 1,00 & 0,91 & 0,90 & 0,78 & 0,79 & -0,46 & 0,47 & -0,09 \\ \hline
connectivité & 0,81 & 0,97 & 0,91 & 1,00 & 0,92 & 0,75 & 0,76 & -0,53 & 0,53 & -0,13 \\ \hline
structuralité potentielle & 0,78 & 0,96 & 0,90 & 0,92 & 1,00 & 0,82 & 0,81 & -0,27 & 0,28 & -0,16 \\ \hline
betweenness & 0,67 & 0,81 & 0,78 & 0,75 & 0,82 & 1,00 & 0,97 & -0,36 & 0,37 & -0,10 \\ \hline
utilisation & 0,67 & 0,81 & 0,79 & 0,76 & 0,81 & 0,97 & 1,00 & -0,40 & 0,41 & -0,09 \\ \hline
accessibilité & -0,43 & -0,47 & -0,46 & -0,53 & -0,27 & -0,36 & -0,40 & 1,00 & -1,00 & 0,05 \\ \hline
closeness & 0,43 & 0,48 & 0,47 & 0,53 & 0,28 & 0,37 & 0,41 & -1,00 & 1,00 & -0,05 \\ \hline
orthogonalité & -0,15 & -0,15 & -0,09 & -0,13 & -0,16 & -0,10 & -0,09 & 0,05 & -0,05 & 1,00 \\ \hline

\end{tabular}
\end{center}
\caption{Calcul de la corrélation de Pearson pour les {\large \textbf{indicateurs primaires}} calculés sur les {\large \textbf{voies}} du graphe viaire de {\large \textbf{Barcelone}}.}
\label{tab:corr_barcelone_voie}
\end{table}
\end{landscape}

\FloatBarrier

\begin{table}
\begin{center}
\small
\begin{tabular}{|r|c|c|c|c|c|c|}
\hline
 & $\frac{longueur}{degre}$ & $\frac{longueur}{closeness}$ & $\frac{degre}{closeness}$ & $\frac{orthogonalite}{longueur}$ & $\frac{orthogonalite}{degre}$ & $\frac{orthogonalite}{closeness}$ \\ \hline
longueur & 0,59 & 0,99 & 0,60 & -0,97 & -0,54 & -0,15 \\ \hline
degré & -0,19 & 0,58 & 0,96 & -0,63 & -0,92 & -0,38 \\ \hline
structuralité potentielle & -0,21 & 0,59 & 1,00 & -0,60 & -0,89 & -0,23 \\ \hline
utilisation & -0,05 & 0,57 & 0,80 & -0,60 & -0,73 & -0,25 \\ \hline
closeness & -0,04 & 0,11 & 0,14 & -0,27 & -0,38 & -0,71 \\ \hline
orthogonalité & 0,18 & -0,11 & -0,32 & 0,29 & 0,63 & 0,77 \\ \hline
$\frac{orthogonalite}{closeness}$ & 0,18 & -0,06 & -0,24 & 0,30 & 0,61 & 1,00 \\ \hline
$\frac{longueur}{degre}$ & 1,00 & 0,62 & -0,21 & -0,54 & 0,24 & 0,18 \\ \hline
\end{tabular}
\end{center}
\caption{Calcul de la corrélation de Pearson pour les {\large \textbf{indicateurs composés}} calculés sur les {\large \textbf{voies}} du graphe viaire d'{\large \textbf{Avignon}}.}
\label{tab:corr_voies_avignon_comb}
\end{table}

\begin{table}
\begin{center}
\small
\begin{tabular}{|r|c|c|c|c|c|c|}
\hline
 & $\frac{longueur}{degre}$ & $\frac{longueur}{closeness}$ & $\frac{degre}{closeness}$ & $\frac{orthogonalite}{longueur}$ & $\frac{orthogonalite}{degre}$ & $\frac{orthogonalite}{closeness}$ \\ \hline
longueur & 0,60 & 0,99 & 0,89 & -0,98 & -0,85 & -0,25 \\ \hline
degré & 0,32 & 0,91 & 0,99 & -0,90 & -0,95 & -0,27 \\ \hline
structuralité potentielle & 0,27 & 0,88 & 0,99 & -0,87 & -0,94 & -0,24 \\ \hline
utilisation & 0,31 & 0,82 & 0,87 & -0,81 & -0,83 & -0,23 \\ \hline
closeness & 0,50 & 0,69 & 0,59 & -0,72 & -0,61 & -0,37 \\ \hline
orthogonalité & 0,26 & 0,02 & -0,10 & 0,09 & 0,28 & 0,76 \\ \hline
$\frac{orthogonalite}{closeness}$ & -0,01 & -0,22 & -0,23 & 0,35 & 0,46 & 1,00 \\ \hline
$\frac{longueur}{degre}$ & 1,00 & 0,59 & 0,28 & -0,57 & -0,24 & -0,01 \\ \hline
\end{tabular}
\end{center}
\caption{Calcul de la corrélation de Pearson pour les {\large \textbf{indicateurs composés}} calculés sur les {\large \textbf{voies}} du graphe viaire de {\large \textbf{Manhattan}}.}
\label{tab:corr_voies_manhattan_comb}
\end{table}

\begin{table}
\begin{center}
\small
\begin{tabular}{|r|c|c|c|c|c|c|}
\hline
 & $\frac{longueur}{degre}$ & $\frac{longueur}{closeness}$ & $\frac{degre}{closeness}$ & $\frac{orthogonalite}{longueur}$ & $\frac{orthogonalite}{degre}$ & $\frac{orthogonalite}{closeness}$ \\ \hline
longueur & 0,49 & 0,99 & 0,83 & -0,98 & -0,82 & -0,39 \\ \hline
degré & 0,02 & 0,82 & 0,98 & -0,83 & -0,96 & -0,47 \\ \hline
structuralité potentielle & 0,00 & 0,82 & 1,00 & -0,81 & -0,95 & -0,37 \\ \hline
utilisation & 0,05 & 0,72 & 0,83 & -0,72 & -0,80 & -0,35 \\ \hline
closeness & 0,09 & 0,41 & 0,41 & -0,50 & -0,56 & -0,72 \\ \hline
orthogonalité & 0,01 & -0,25 & -0,31 & 0,37 & 0,50 & 0,78 \\ \hline
$\frac{orthogonalite}{closeness}$ & -0,02 & -0,33 & -0,37 & 0,48 & 0,60 & 1,00 \\ \hline
$\frac{longueur}{degre}$ & 1,00 & 0,51 & 0,01 & -0,48 & -0,01 & -0,02 \\ \hline
\end{tabular}
\end{center}
\caption{Calcul de la corrélation de Pearson pour les {\large \textbf{indicateurs composés}} calculés sur les {\large \textbf{voies}} du graphe viaire de {\large \textbf{Paris}}.}
\label{tab:corr_voies_paris_comb}
\end{table}

\begin{table}
\begin{center}
\small
\begin{tabular}{|r|c|c|c|c|c|c|}
\hline
 & $\frac{longueur}{degre}$ & $\frac{longueur}{closeness}$ & $\frac{degre}{closeness}$ & $\frac{orthogonalite}{longueur}$ & $\frac{orthogonalite}{degre}$ & $\frac{orthogonalite}{closeness}$ \\ \hline
longueur & 0,62 & 0,99 & 0,79 & -0,97 & -0,78 & -0,30 \\ \hline
degré & 0,13 & 0,77 & 0,97 & -0,77 & -0,93 & -0,32 \\ \hline
structuralité potentielle & 0,11 & 0,78 & 1,00 & -0,73 & -0,90 & -0,18 \\ \hline
utilisation & 0,12 & 0,64 & 0,79 & -0,63 & -0,74 & -0,24 \\ \hline
closeness & 0,12 & 0,31 & 0,30 & -0,43 & -0,48 & -0,66 \\ \hline
orthogonalité & -0,07 & -0,14 & -0,16 & 0,29 & 0,39 & 0,68 \\ \hline
$\frac{orthogonalite}{closeness}$ & -0,11 & -0,22 & -0,20 & 0,43 & 0,53 & 1,00 \\ \hline
$\frac{longueur}{degre}$ & 1,00 & 0,64 & 0,12 & -0,63 & -0,15 & -0,11 \\ \hline
\end{tabular}
\end{center}
\caption{Calcul de la corrélation de Pearson pour les {\large \textbf{indicateurs composés}} calculés sur les {\large \textbf{voies}} du graphe viaire de {\large \textbf{Barcelone}}.}
\label{tab:corr_voies_barcelone_comb}
\end{table}

\FloatBarrier 
\chapter{Cartes de corrélation}\label{ann:chap_cartes_corr}

\section{Cartes de corrélation sur les arcs}\label{ann:sec_corr_arcs} 

\subsection{Indicateurs primaires}\label{ann:ssec_arcs_indprim}

\begin{figure}[h]\centering
	\begin{subfigure}[t]{0.45\textwidth}
		\includegraphics[width=\linewidth]{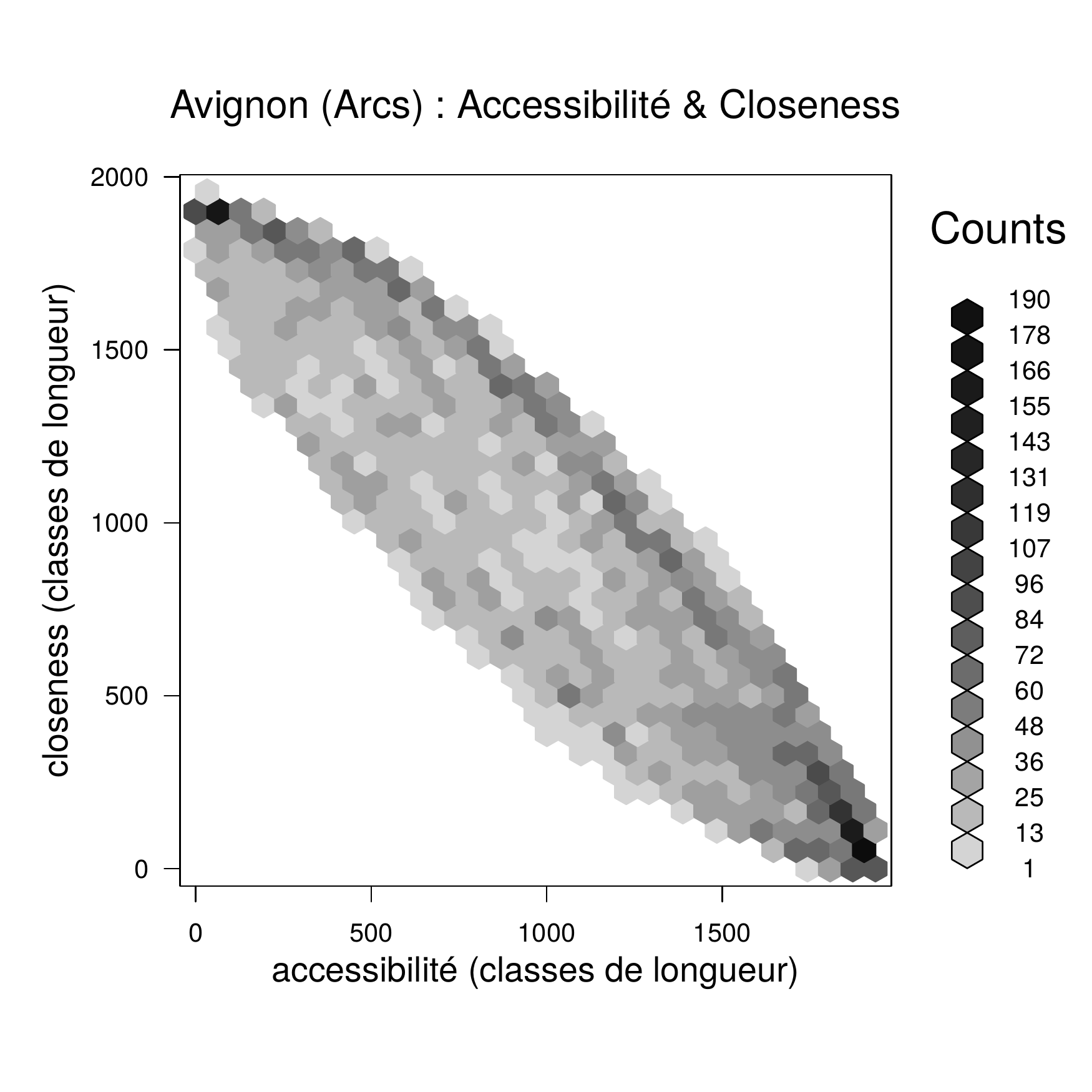}
	\end{subfigure}
	~
	\begin{subfigure}[t]{0.45\textwidth}
		\includegraphics[width=\linewidth]{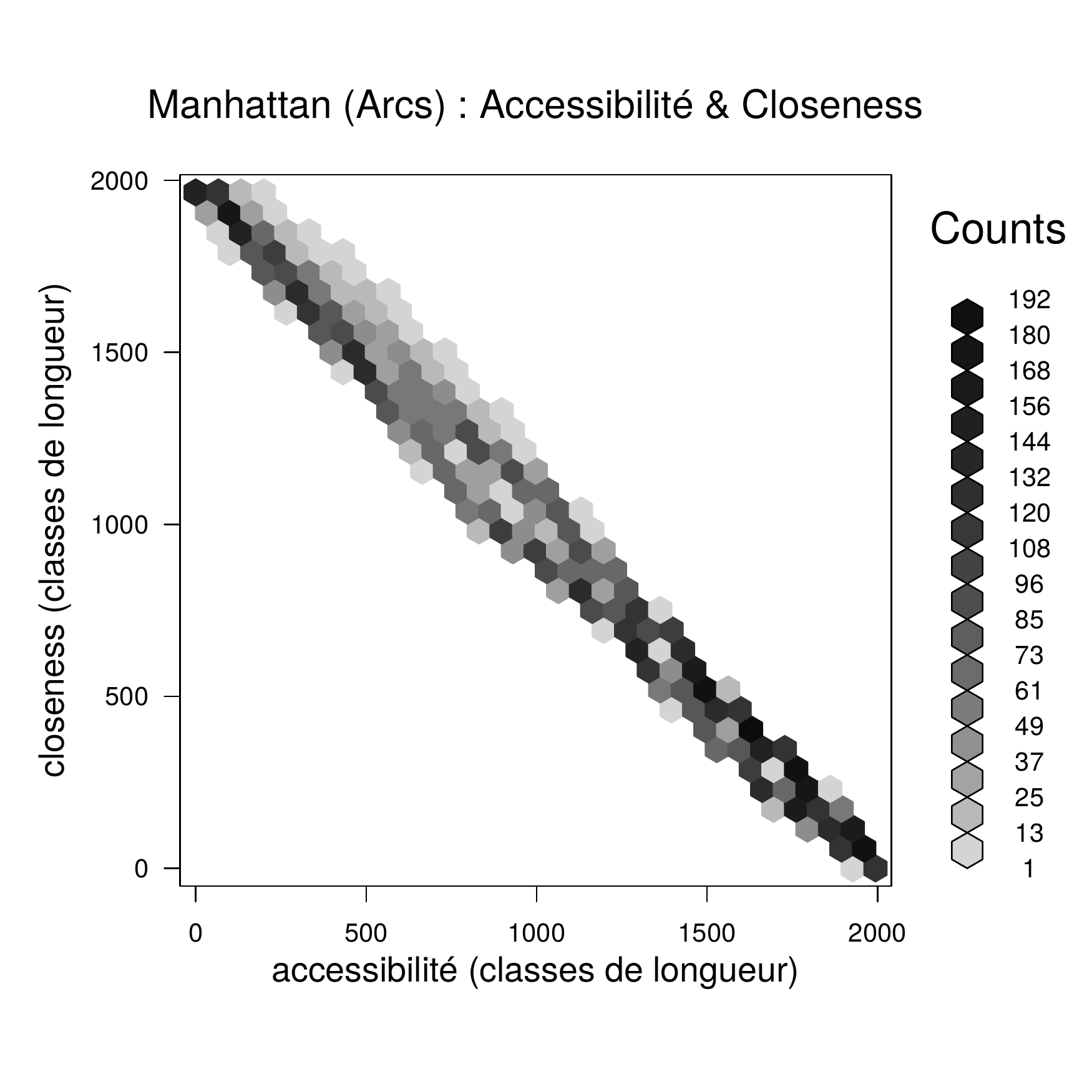}
	\end{subfigure}
	\caption{Accessibilité et closeness}
\end{figure}

\begin{figure}[h]\centering
	\begin{subfigure}[t]{0.45\textwidth}
		\includegraphics[width=\linewidth]{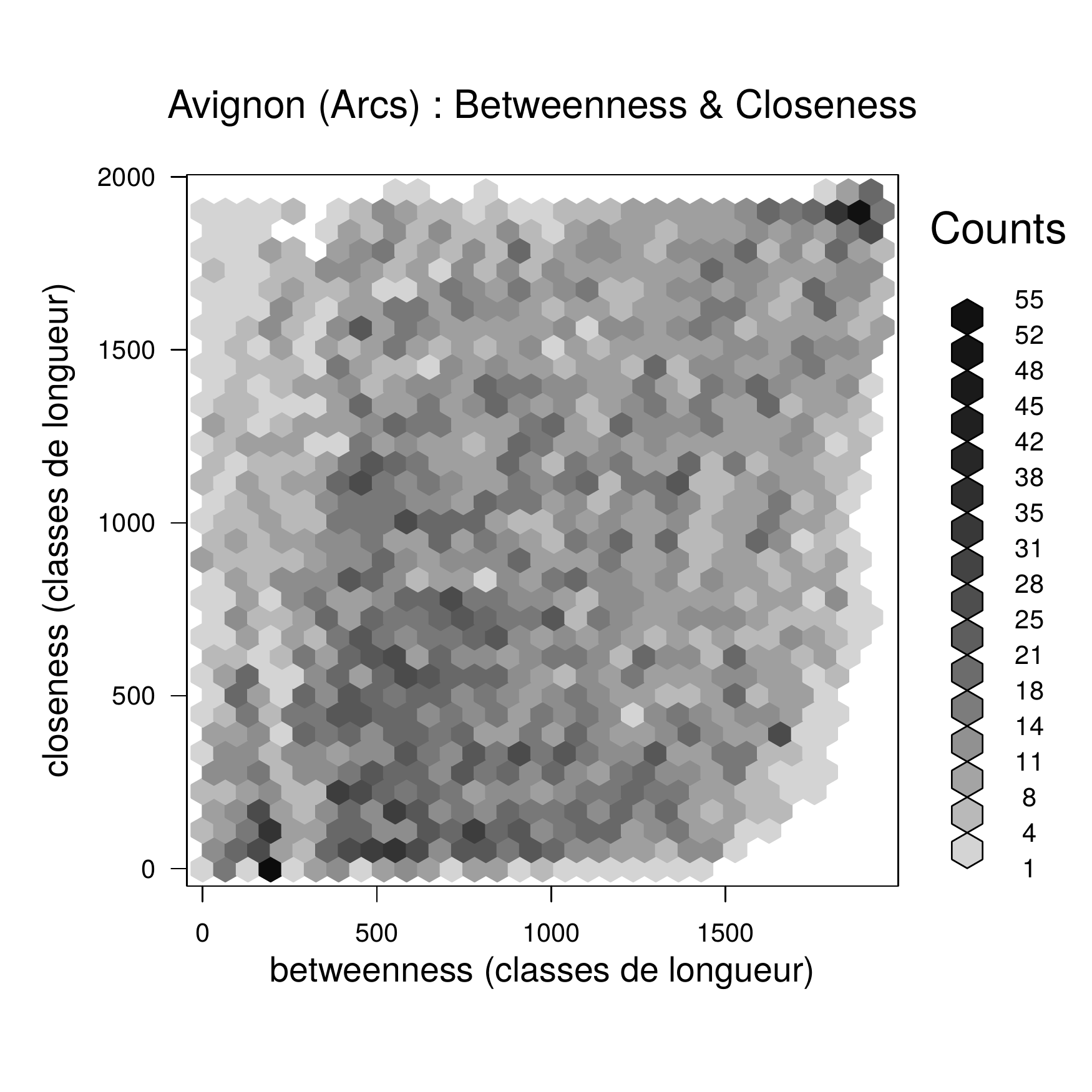}
	\end{subfigure}
	~
	\begin{subfigure}[t]{0.45\textwidth}
		\includegraphics[width=\linewidth]{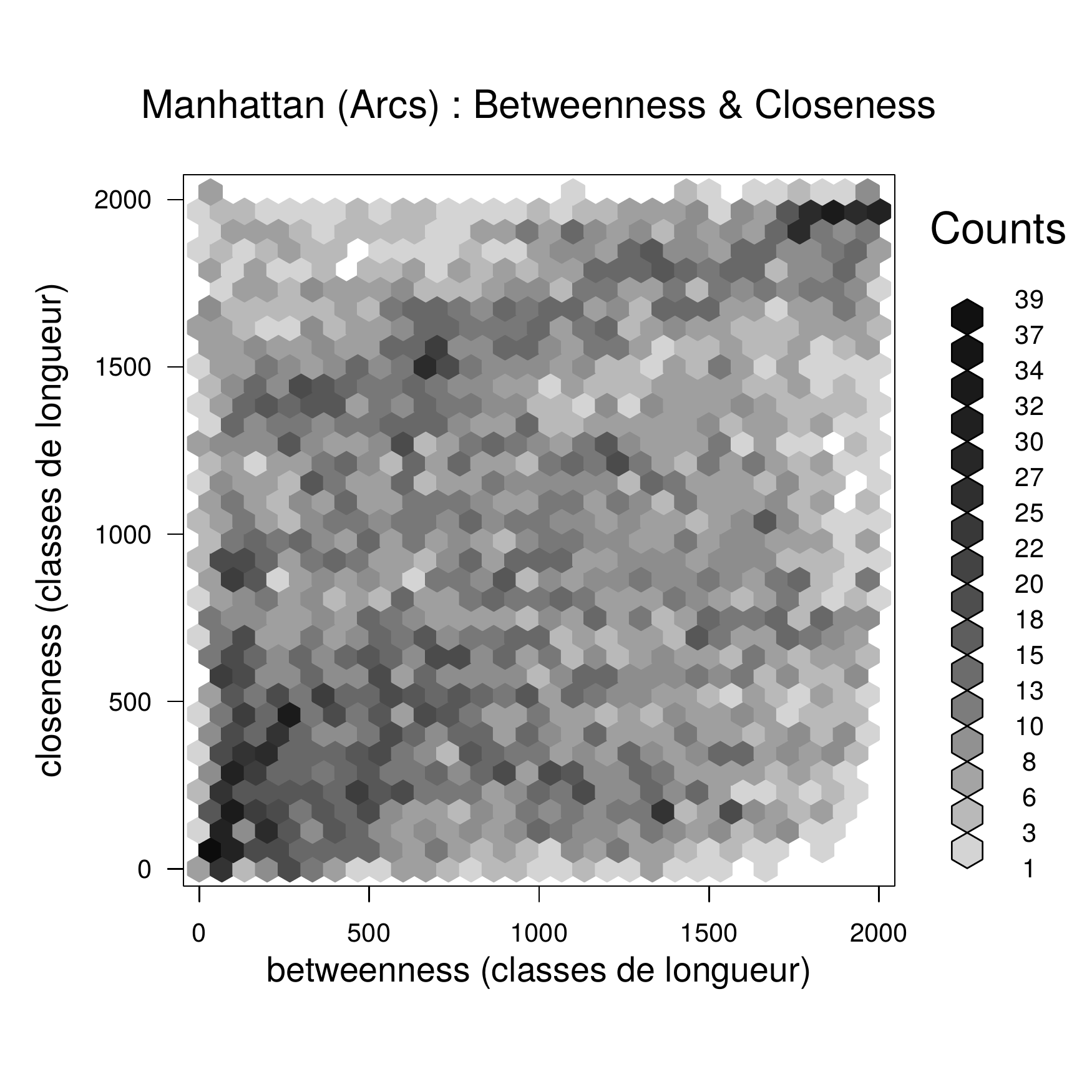}
	\end{subfigure}
	\caption{Betweenness et closeness}
\end{figure}

\begin{figure}[h]\centering
	\begin{subfigure}[t]{0.45\textwidth}
		\includegraphics[width=\linewidth]{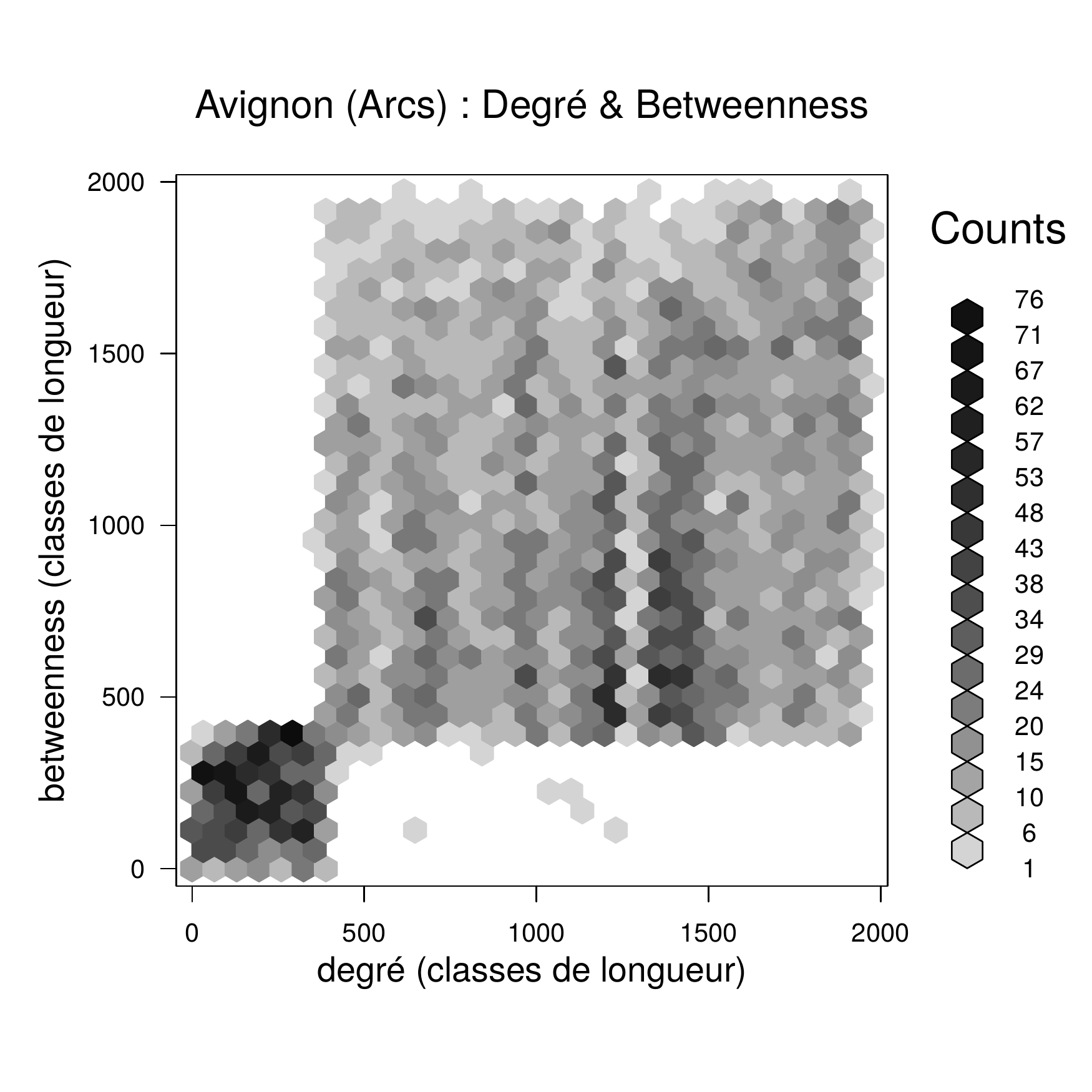}
	\end{subfigure}
	~
	\begin{subfigure}[t]{0.45\textwidth}
		\includegraphics[width=\linewidth]{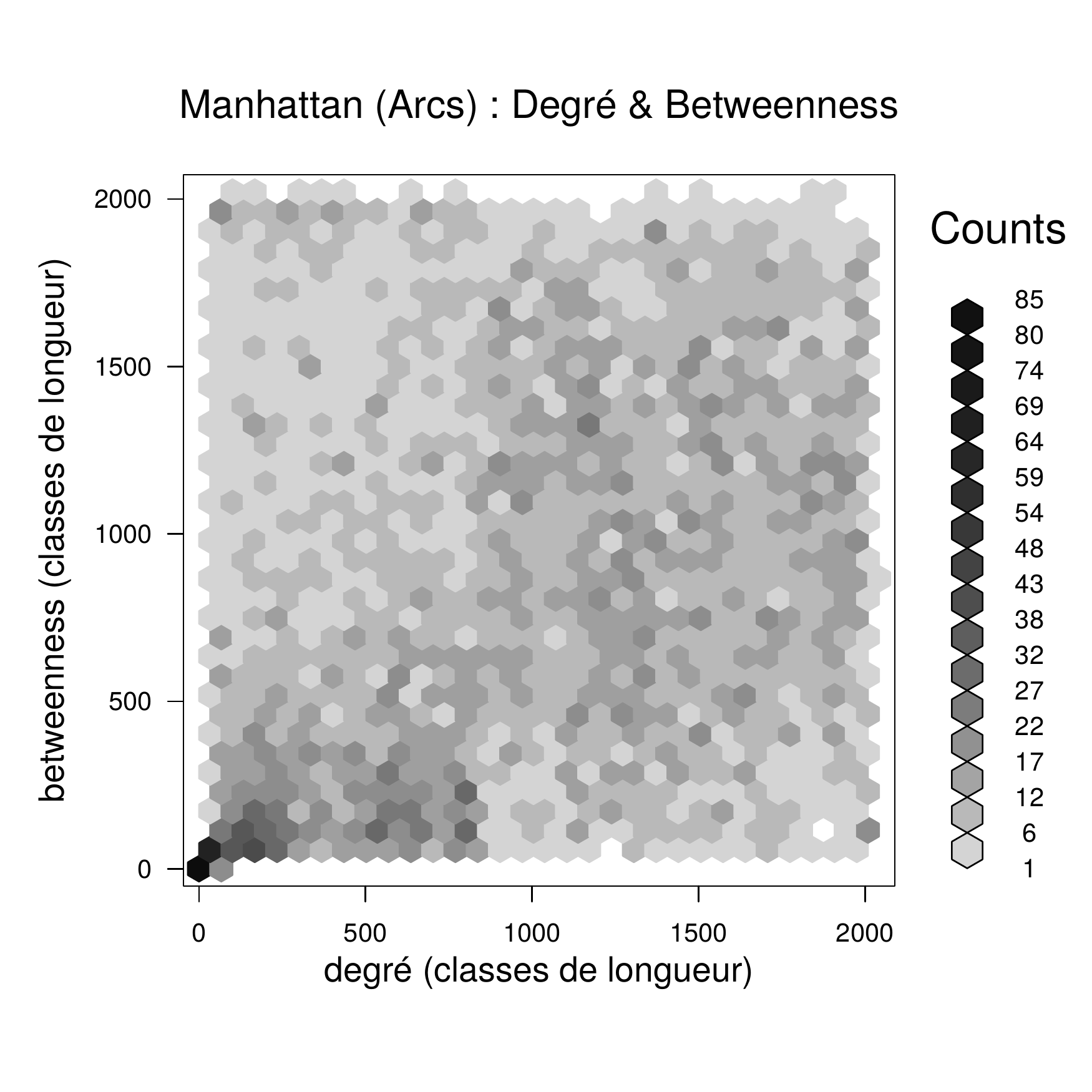}
	\end{subfigure}
	\caption{Degré et betweenness}
\end{figure}

\begin{figure}[h]\centering
	\begin{subfigure}[t]{0.45\textwidth}
		\includegraphics[width=\linewidth]{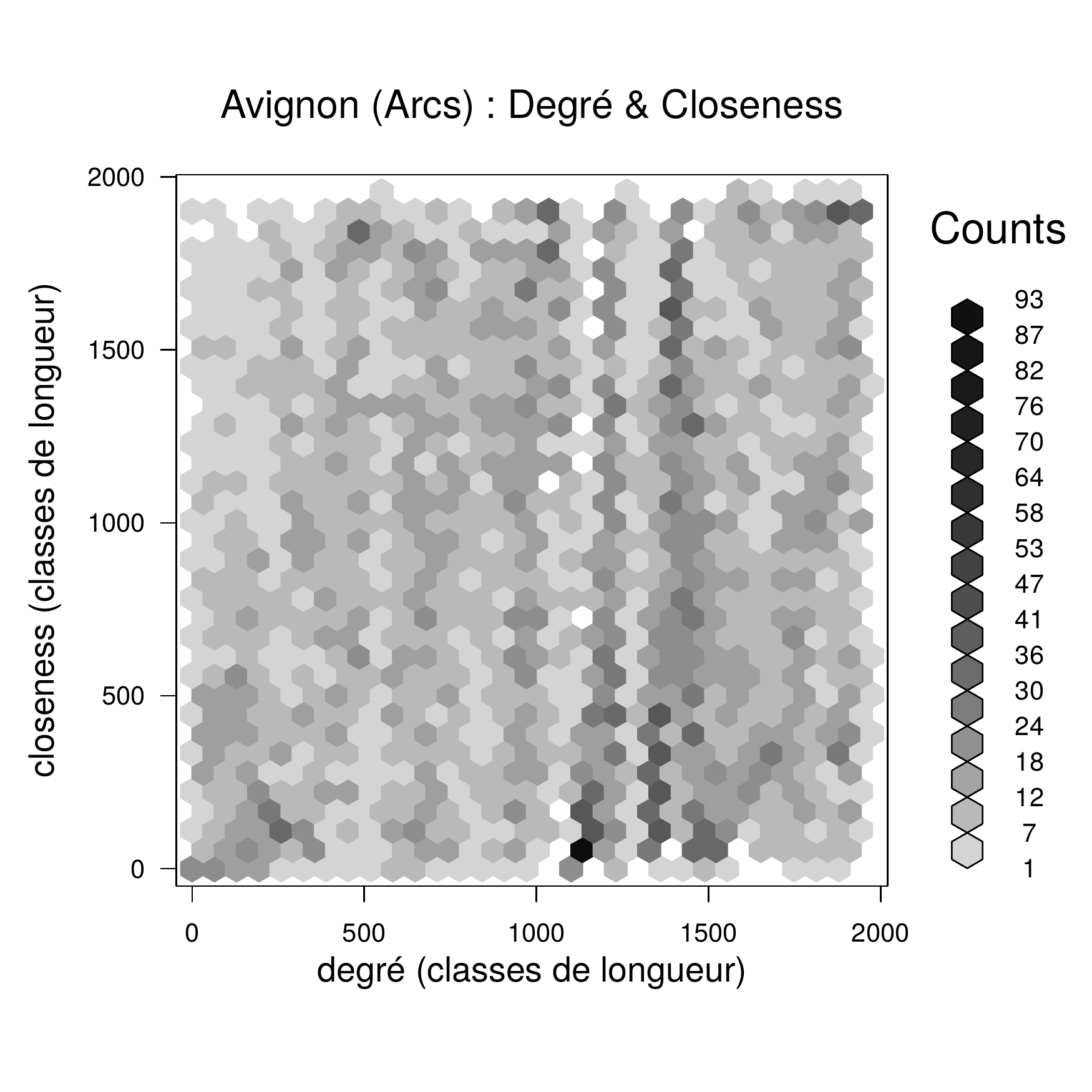}
	\end{subfigure}
	~
	\begin{subfigure}[t]{0.45\textwidth}
		\includegraphics[width=\linewidth]{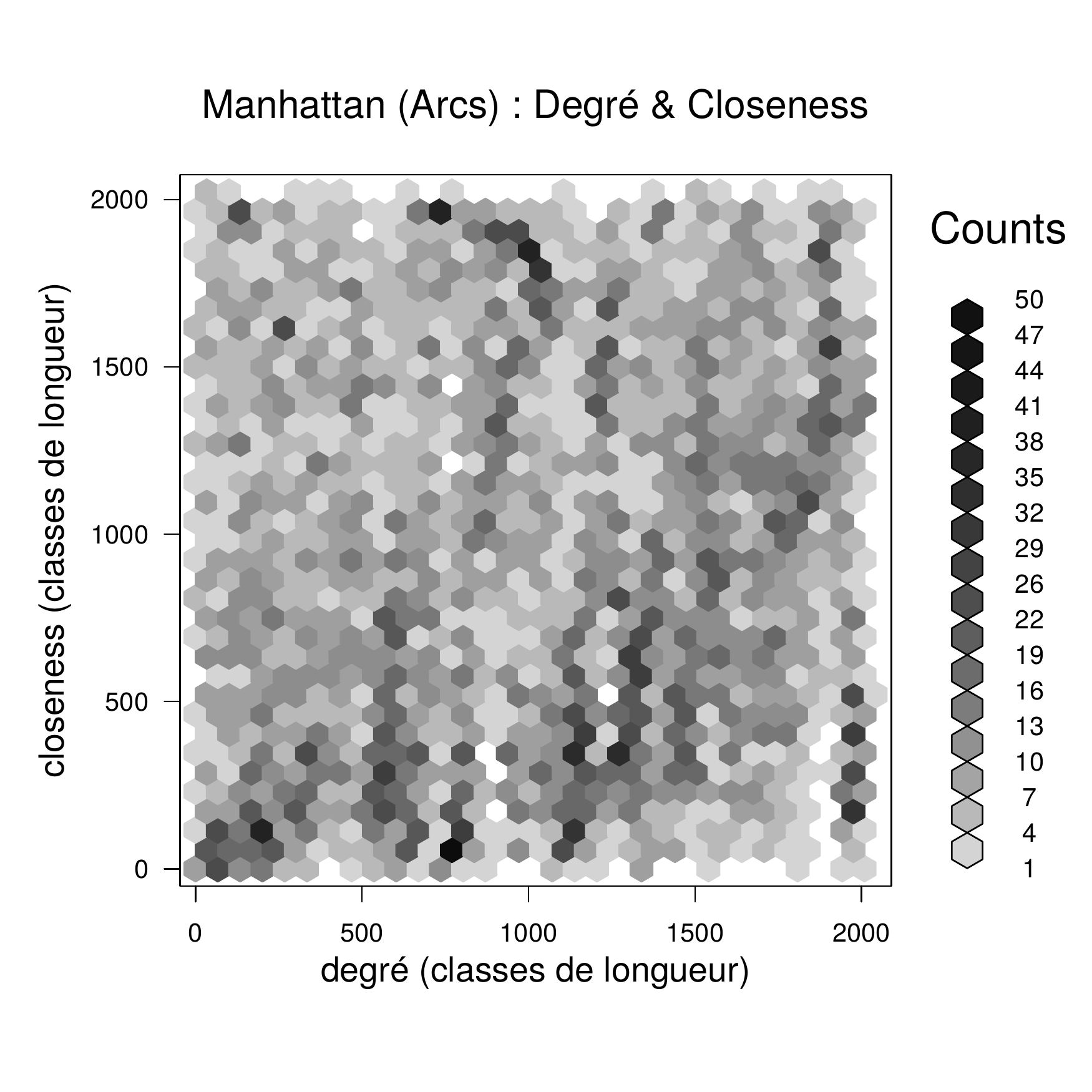}
	\end{subfigure}
	\caption{Degré et closeness}
\end{figure}

\begin{figure}[h]\centering
	\begin{subfigure}[t]{0.45\textwidth}
		\includegraphics[width=\linewidth]{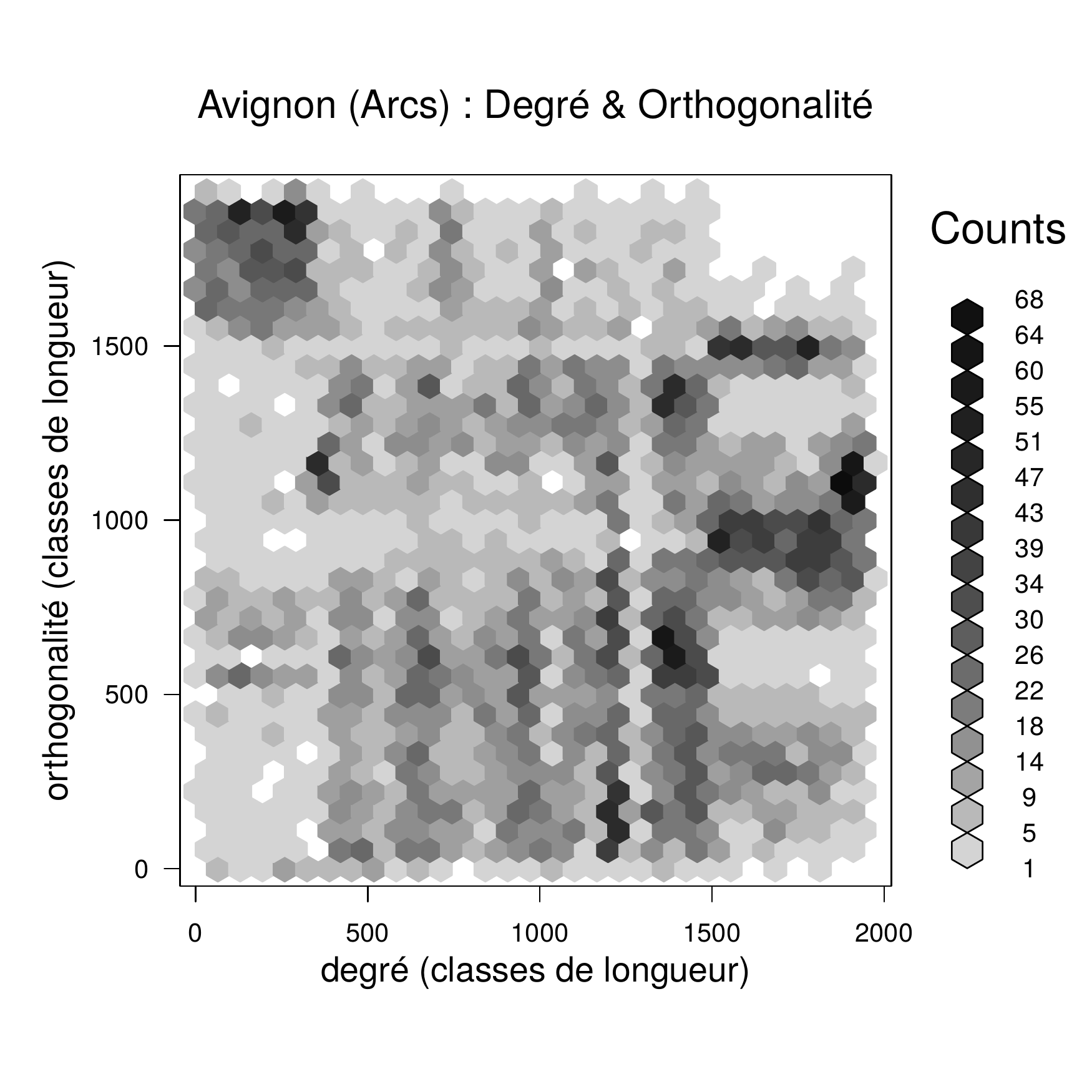}
	\end{subfigure}
	~
	\begin{subfigure}[t]{0.45\textwidth}
		\includegraphics[width=\linewidth]{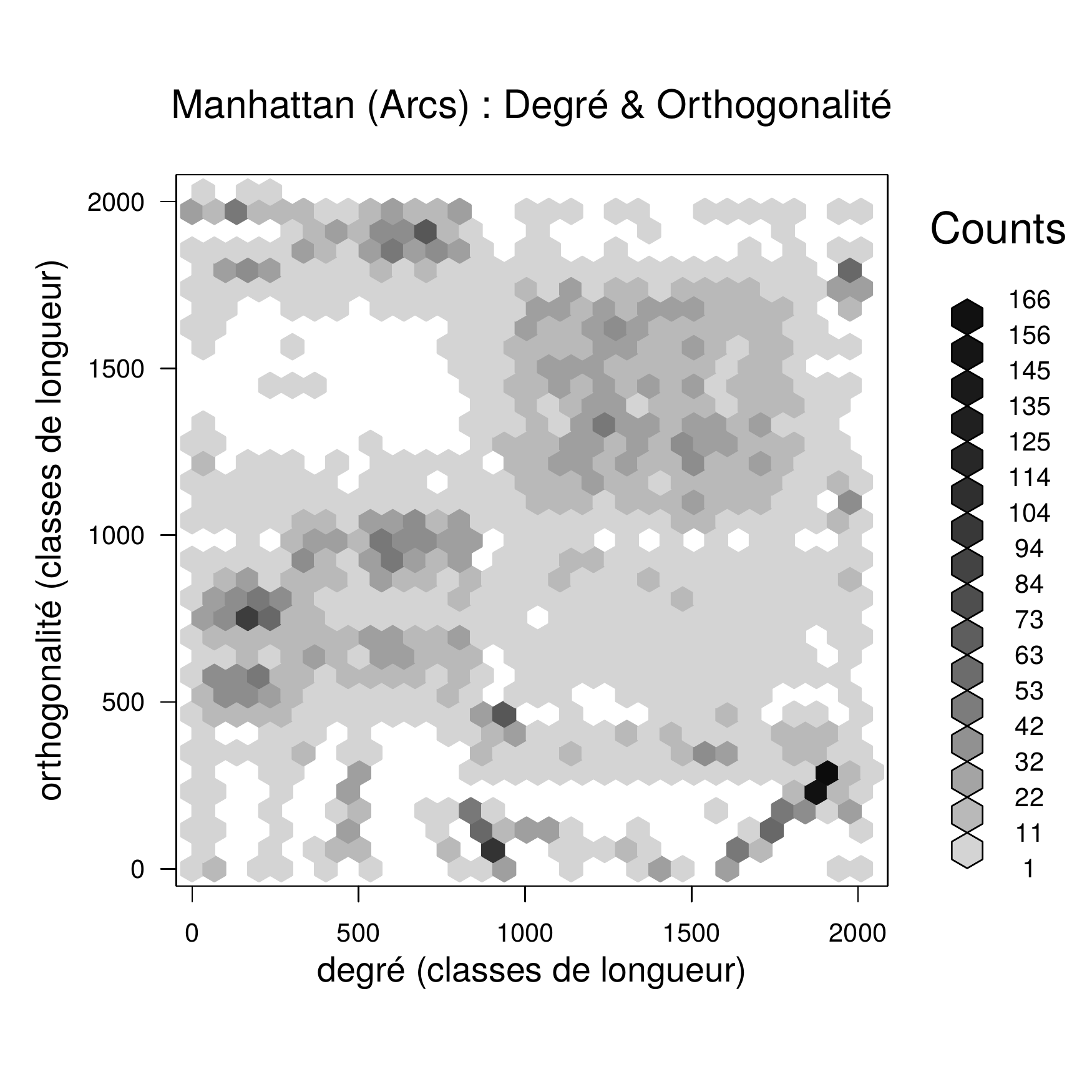}
	\end{subfigure}
	\caption{Degré et orthogonalité}
\end{figure}

\begin{figure}[h]\centering
	\begin{subfigure}[t]{0.45\textwidth}
		\includegraphics[width=\linewidth]{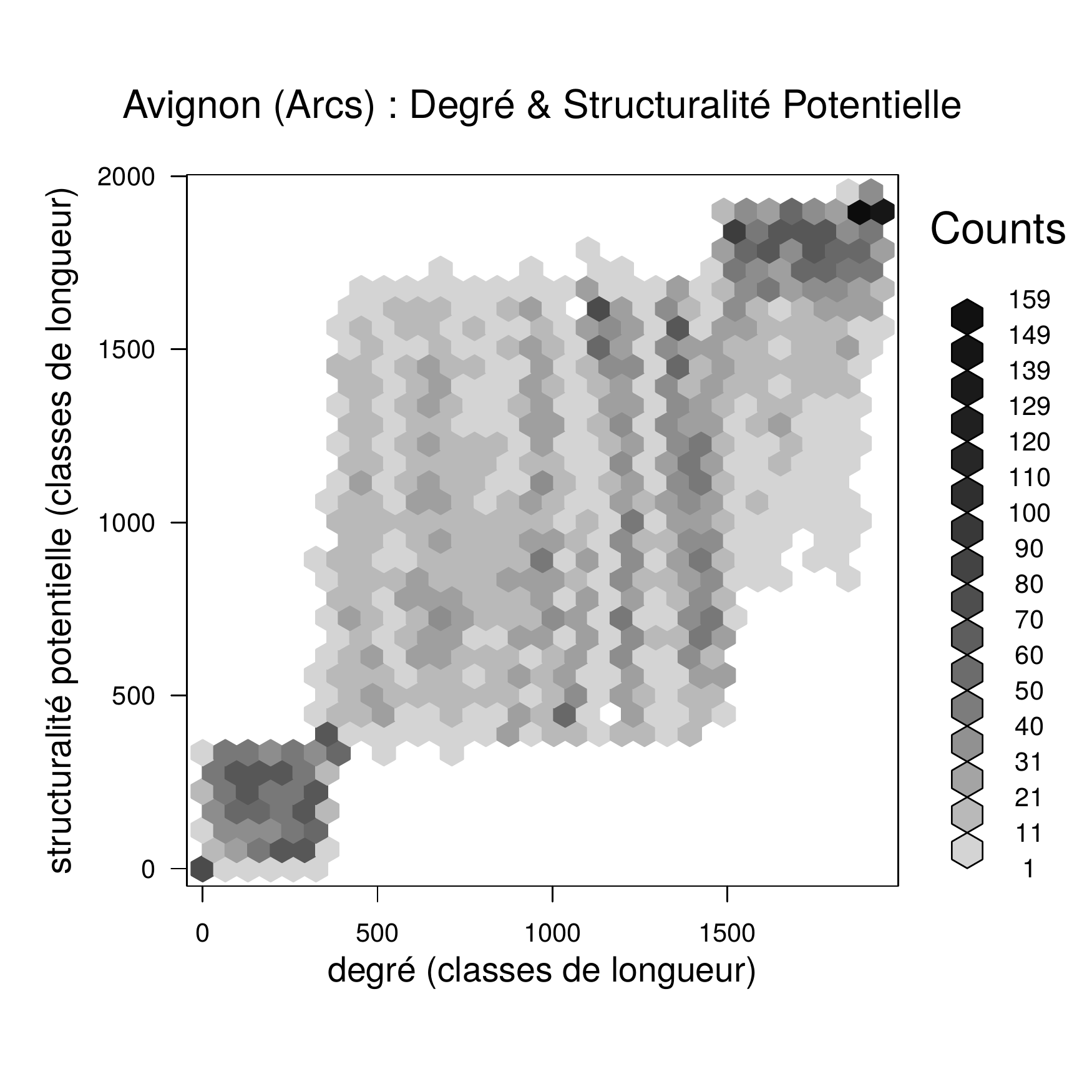}
	\end{subfigure}
	~
	\begin{subfigure}[t]{0.45\textwidth}
		\includegraphics[width=\linewidth]{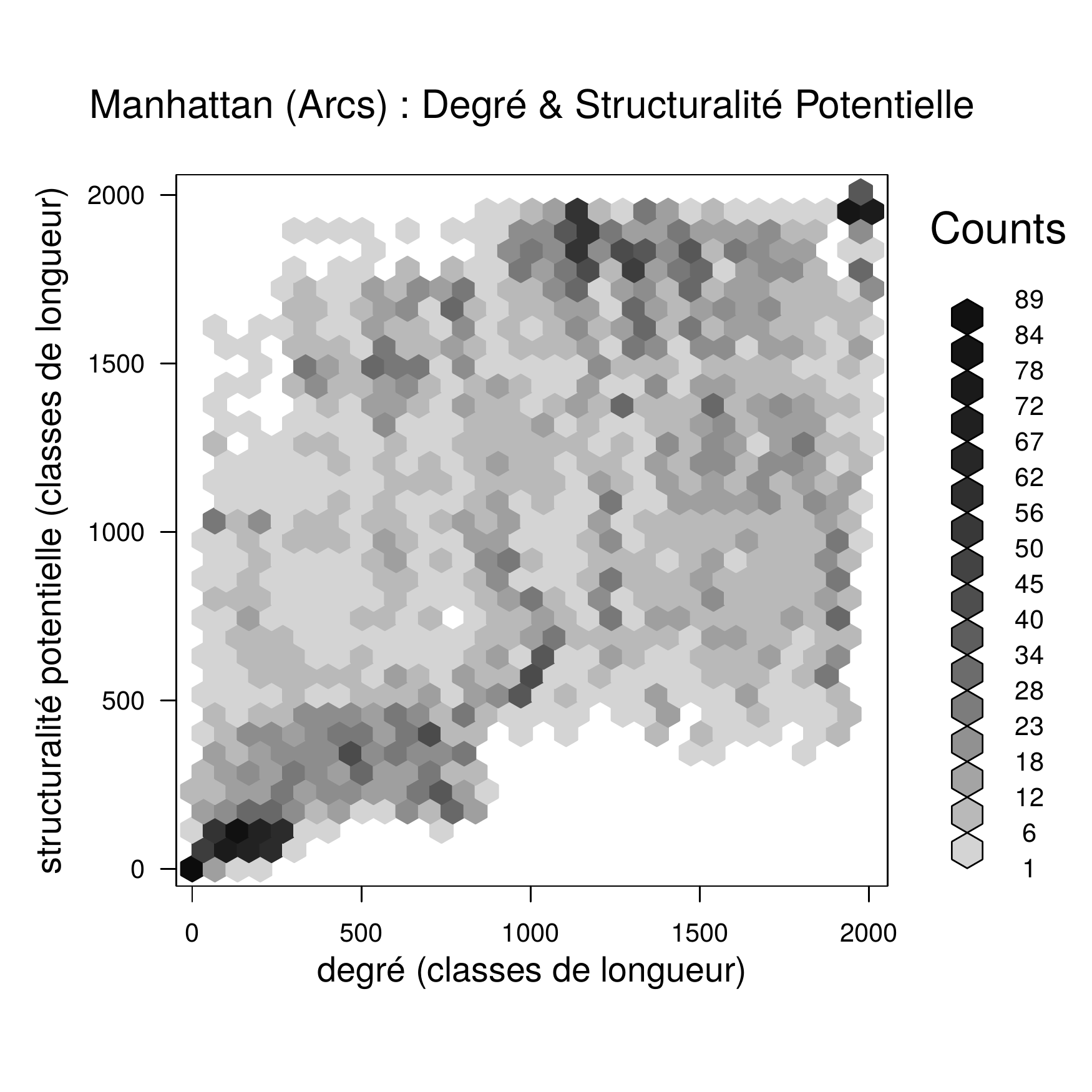}
	\end{subfigure}
	\caption{Degré et structuralité potentielle}
\end{figure}

\begin{figure}[h]\centering
	\begin{subfigure}[t]{0.45\textwidth}
		\includegraphics[width=\linewidth]{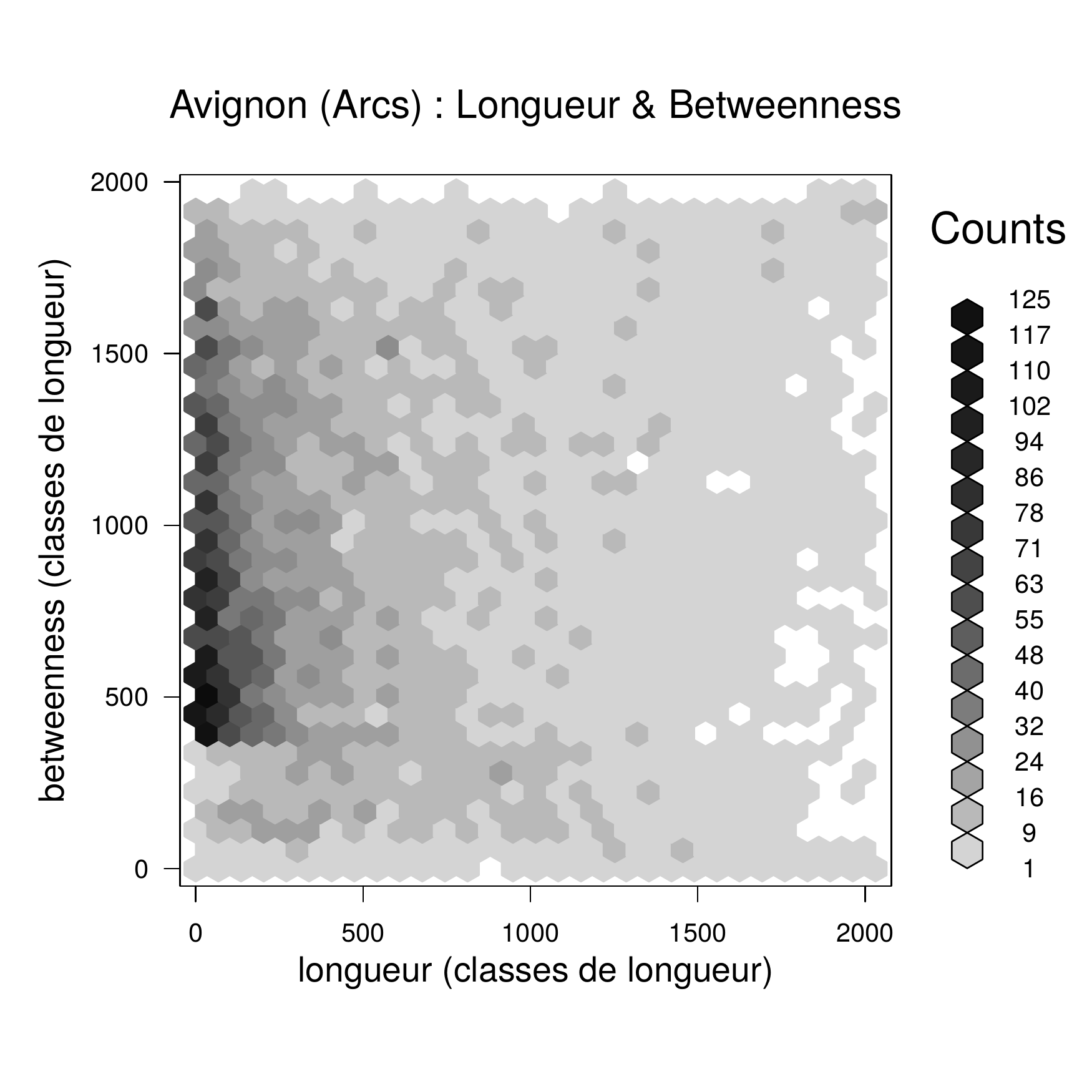}
	\end{subfigure}
	~
	\begin{subfigure}[t]{0.45\textwidth}
		\includegraphics[width=\linewidth]{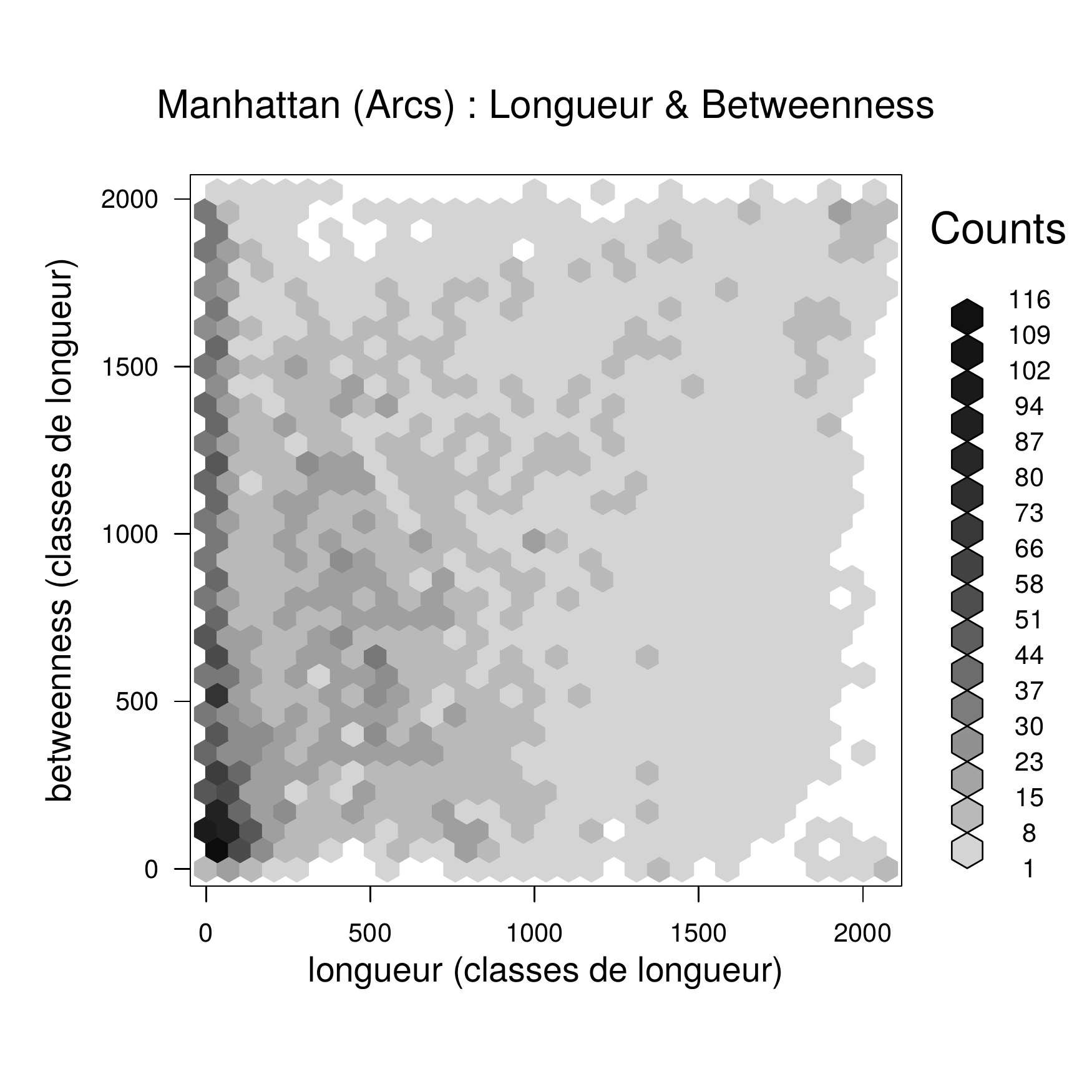}
	\end{subfigure}
	\caption{Longueur et betweenness}
\end{figure}

\begin{figure}[h]\centering
	\begin{subfigure}[t]{0.45\textwidth}
		\includegraphics[width=\linewidth]{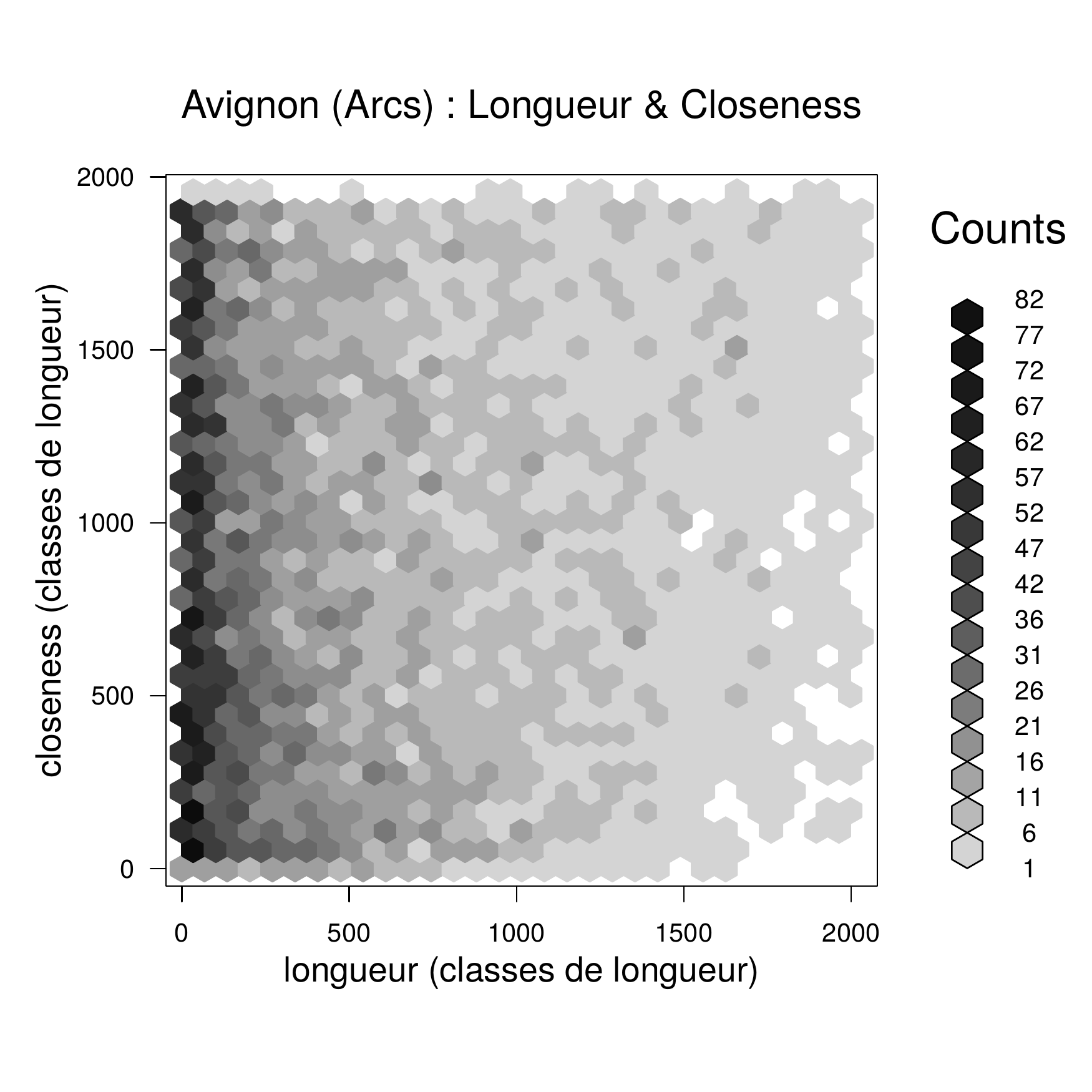}
	\end{subfigure}
	~
	\begin{subfigure}[t]{0.45\textwidth}
		\includegraphics[width=\linewidth]{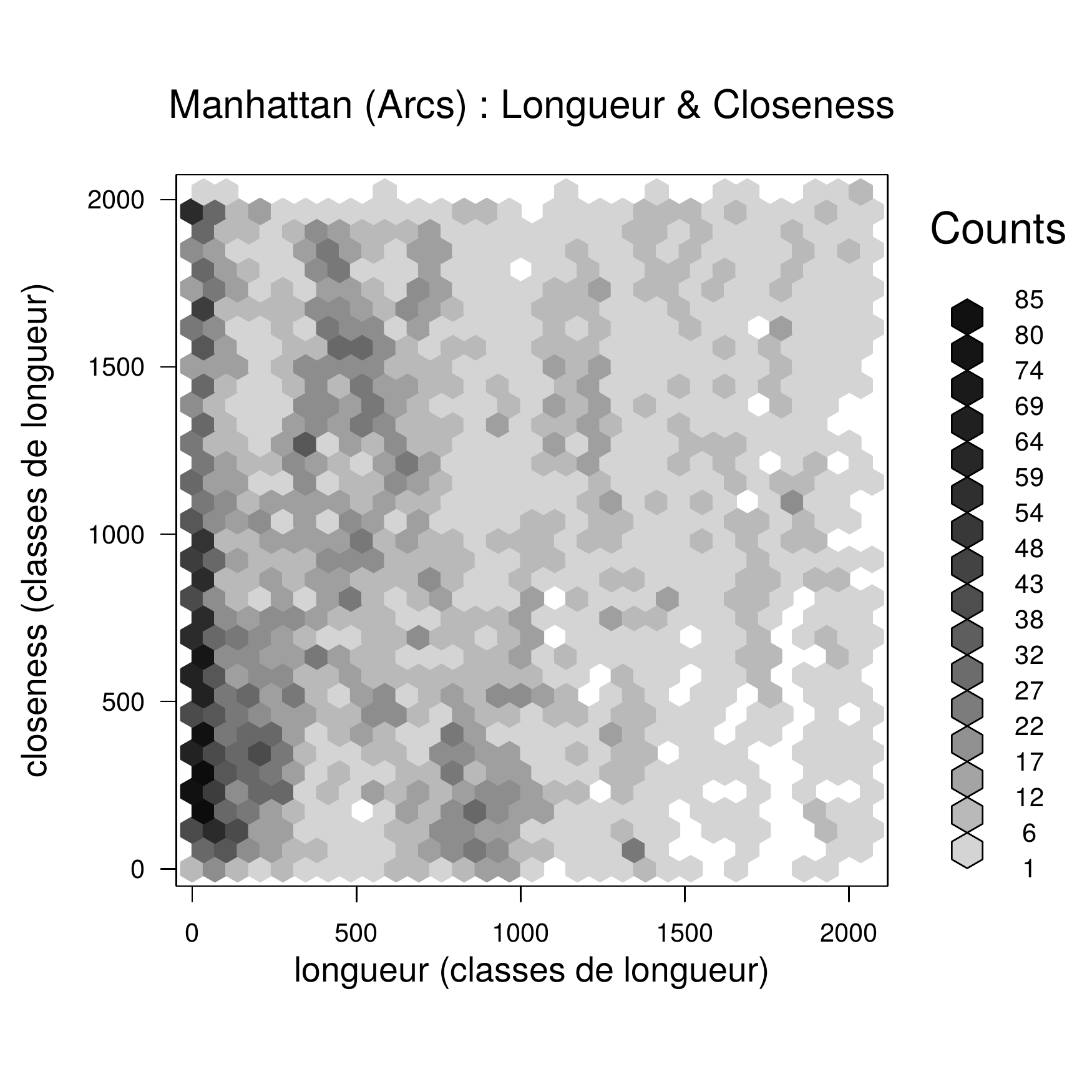}
	\end{subfigure}
	\caption{Longueur et closeness}
\end{figure}

\begin{figure}[h]\centering
	\begin{subfigure}[t]{0.45\textwidth}
		\includegraphics[width=\linewidth]{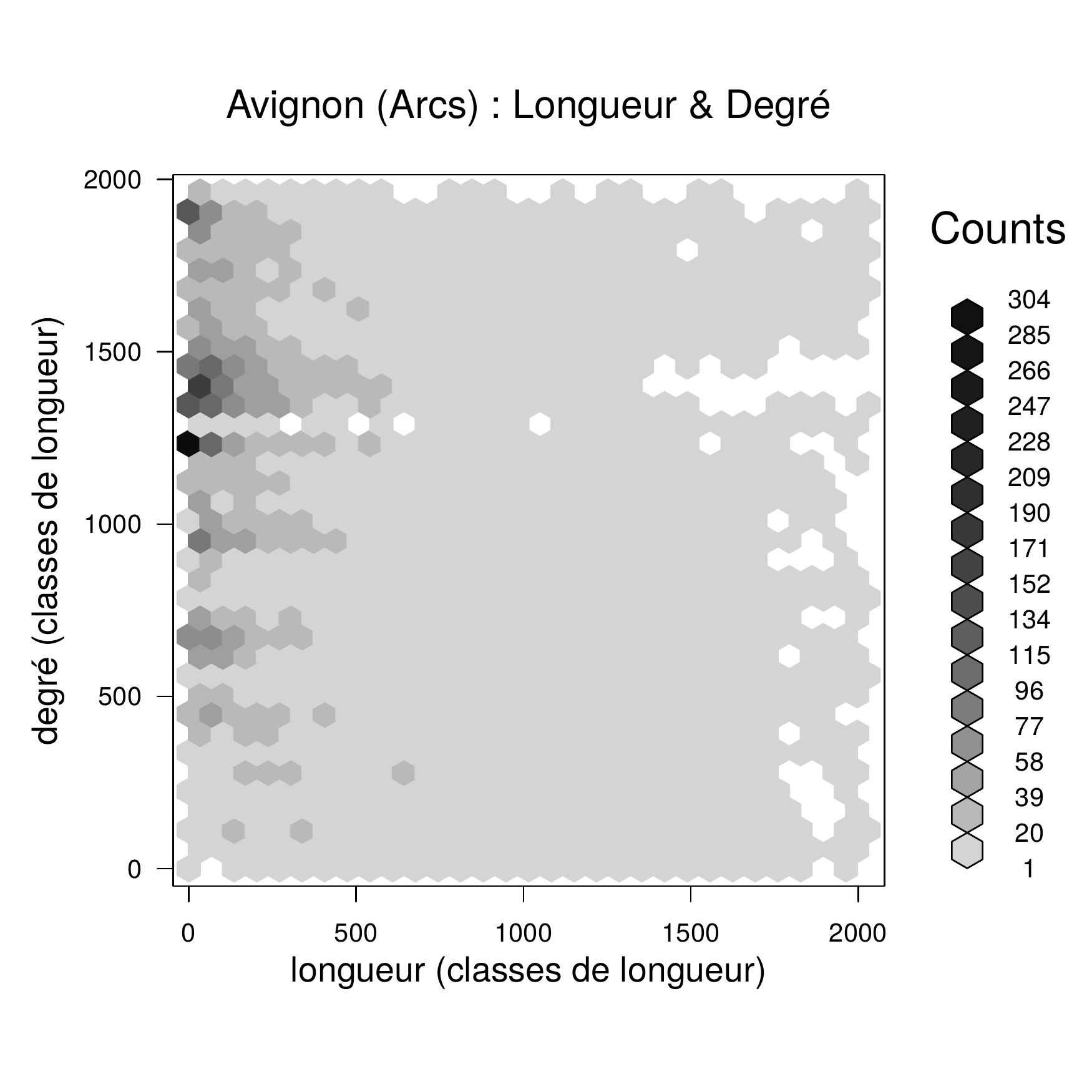}
	\end{subfigure}
	~
	\begin{subfigure}[t]{0.45\textwidth}
		\includegraphics[width=\linewidth]{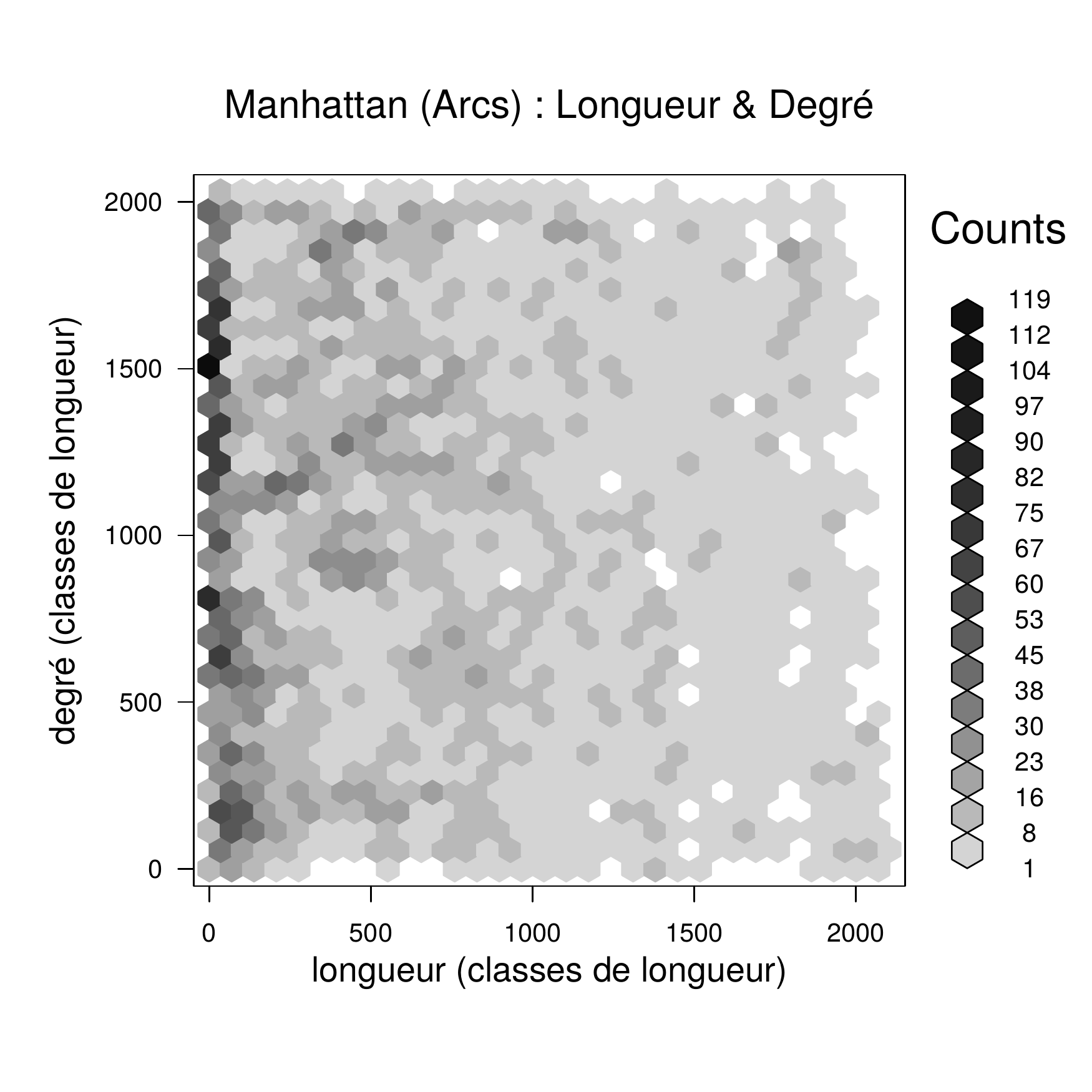}
	\end{subfigure}
	\caption{Longueur et degré}
\end{figure}

\begin{figure}[h]\centering
	\begin{subfigure}[t]{0.45\textwidth}
		\includegraphics[width=\linewidth]{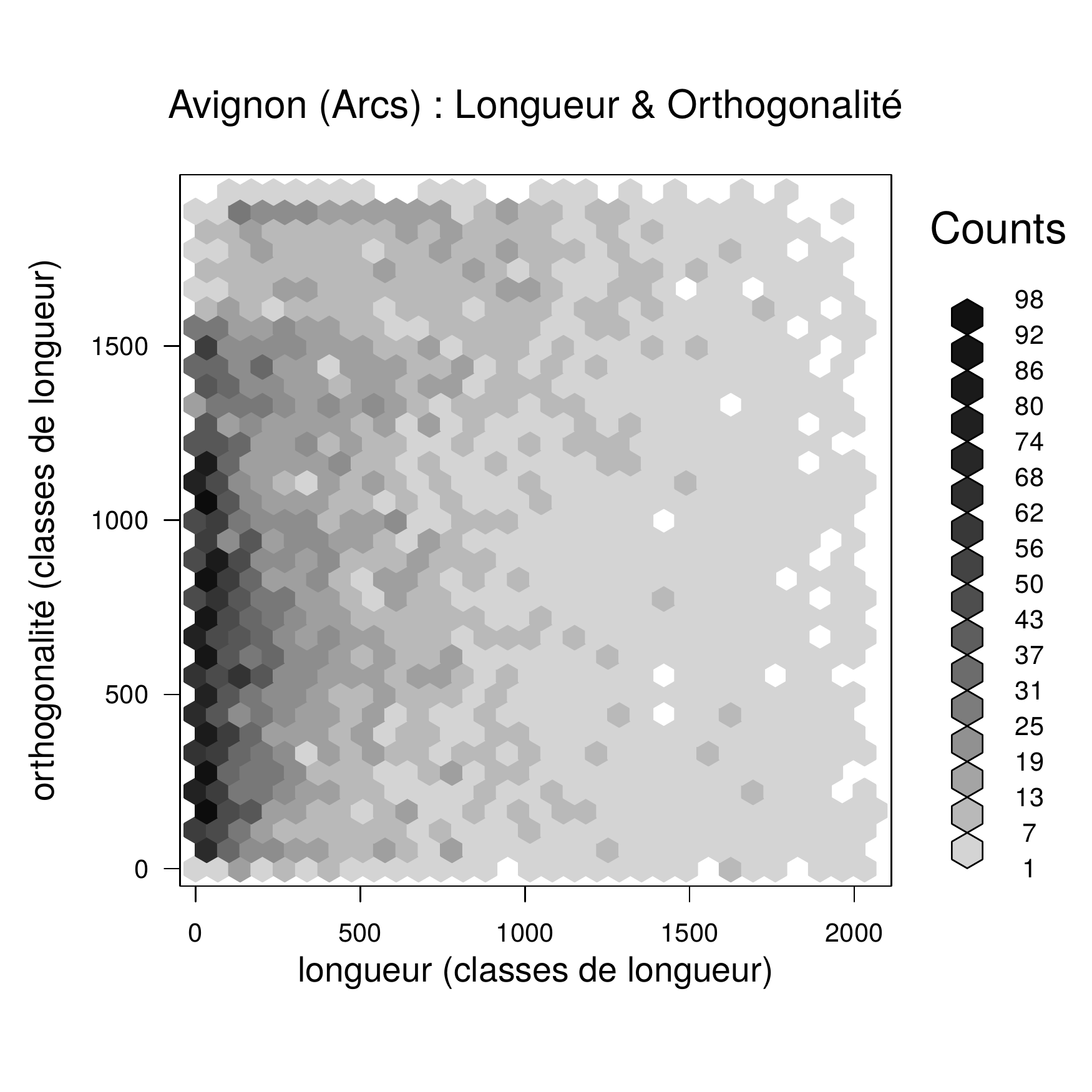}
	\end{subfigure}
	~
	\begin{subfigure}[t]{0.45\textwidth}
		\includegraphics[width=\linewidth]{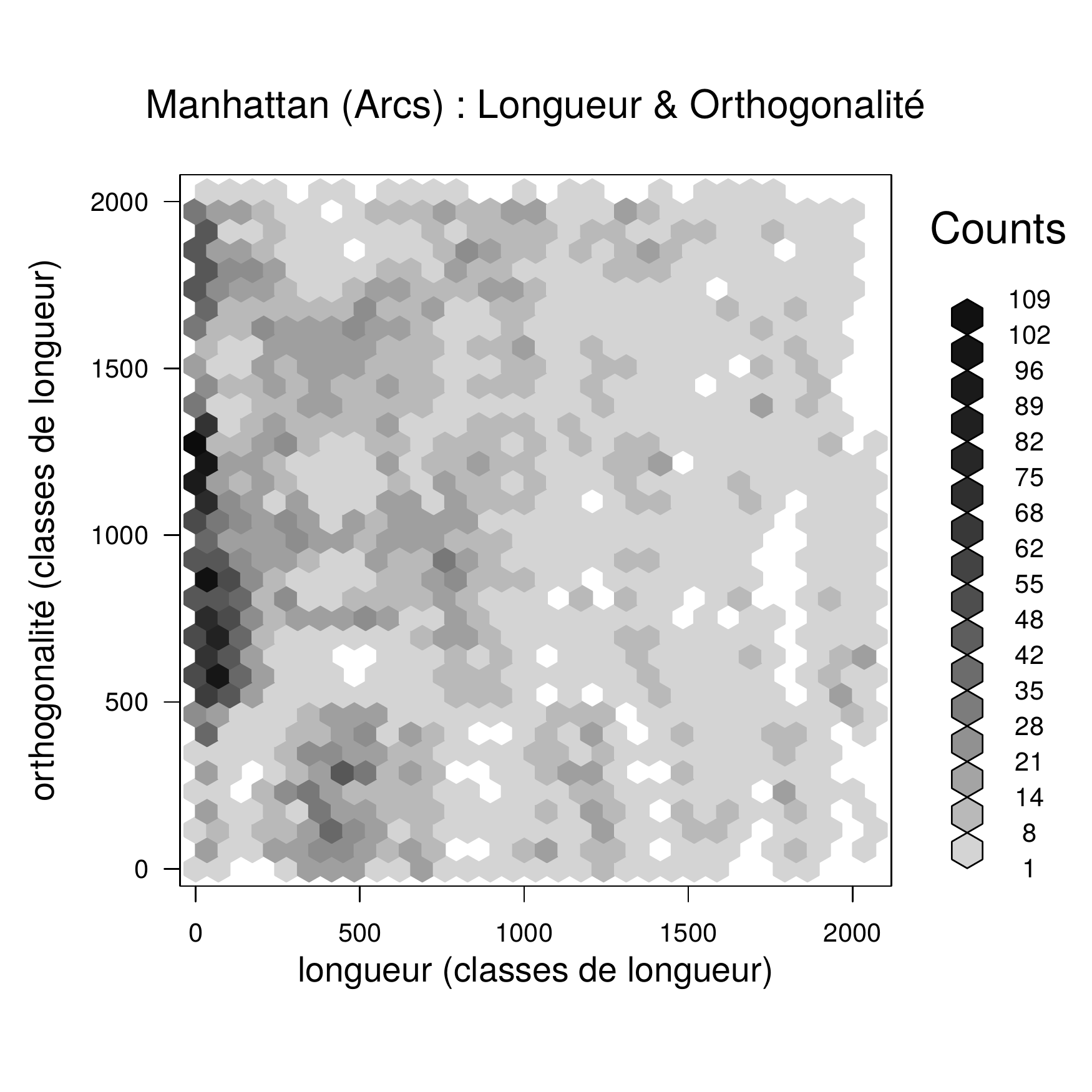}
	\end{subfigure}
	\caption{Longueur et orthogonalité}
\end{figure}

\begin{figure}[h]\centering
	\begin{subfigure}[t]{0.45\textwidth}
		\includegraphics[width=\linewidth]{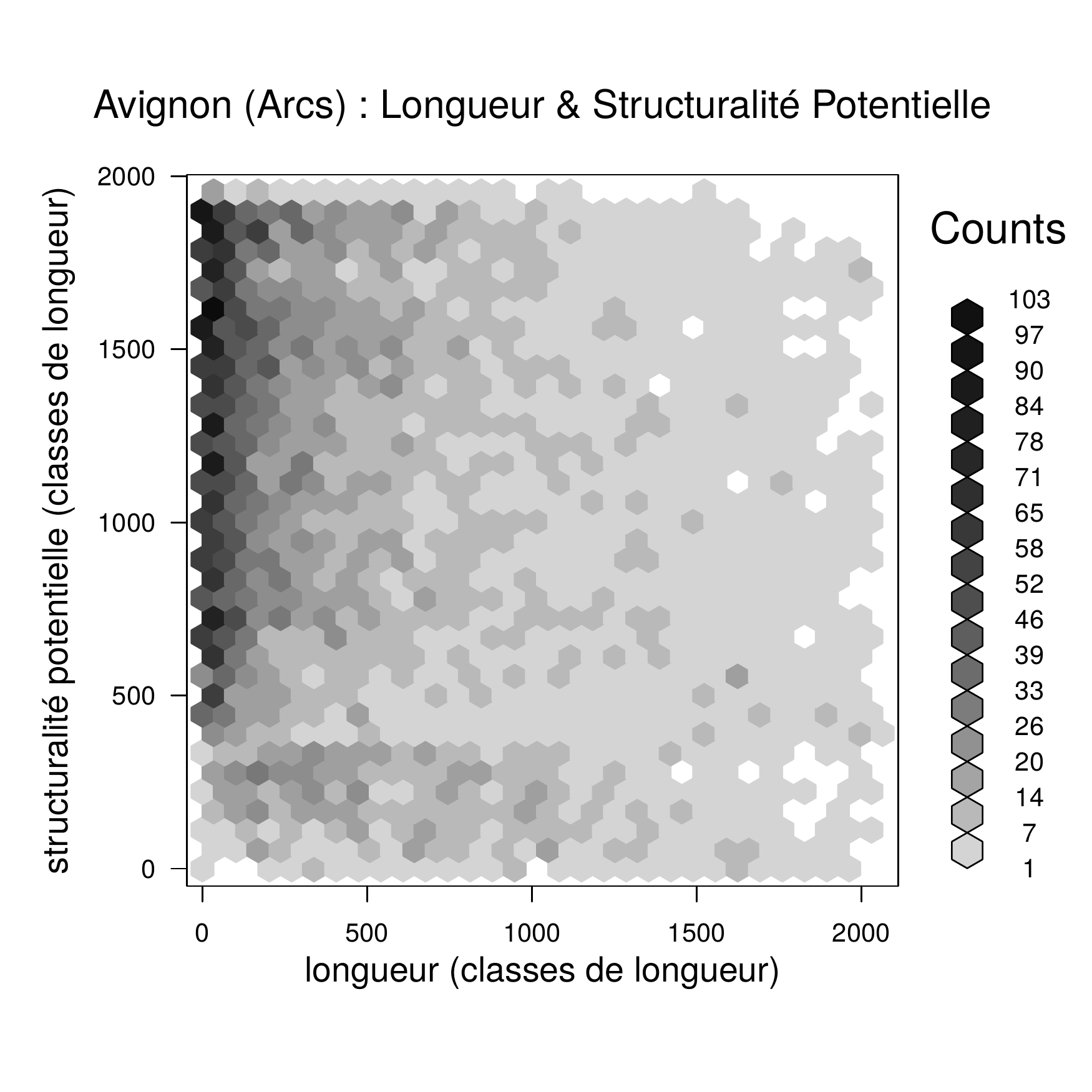}
	\end{subfigure}
	~
	\begin{subfigure}[t]{0.45\textwidth}
		\includegraphics[width=\linewidth]{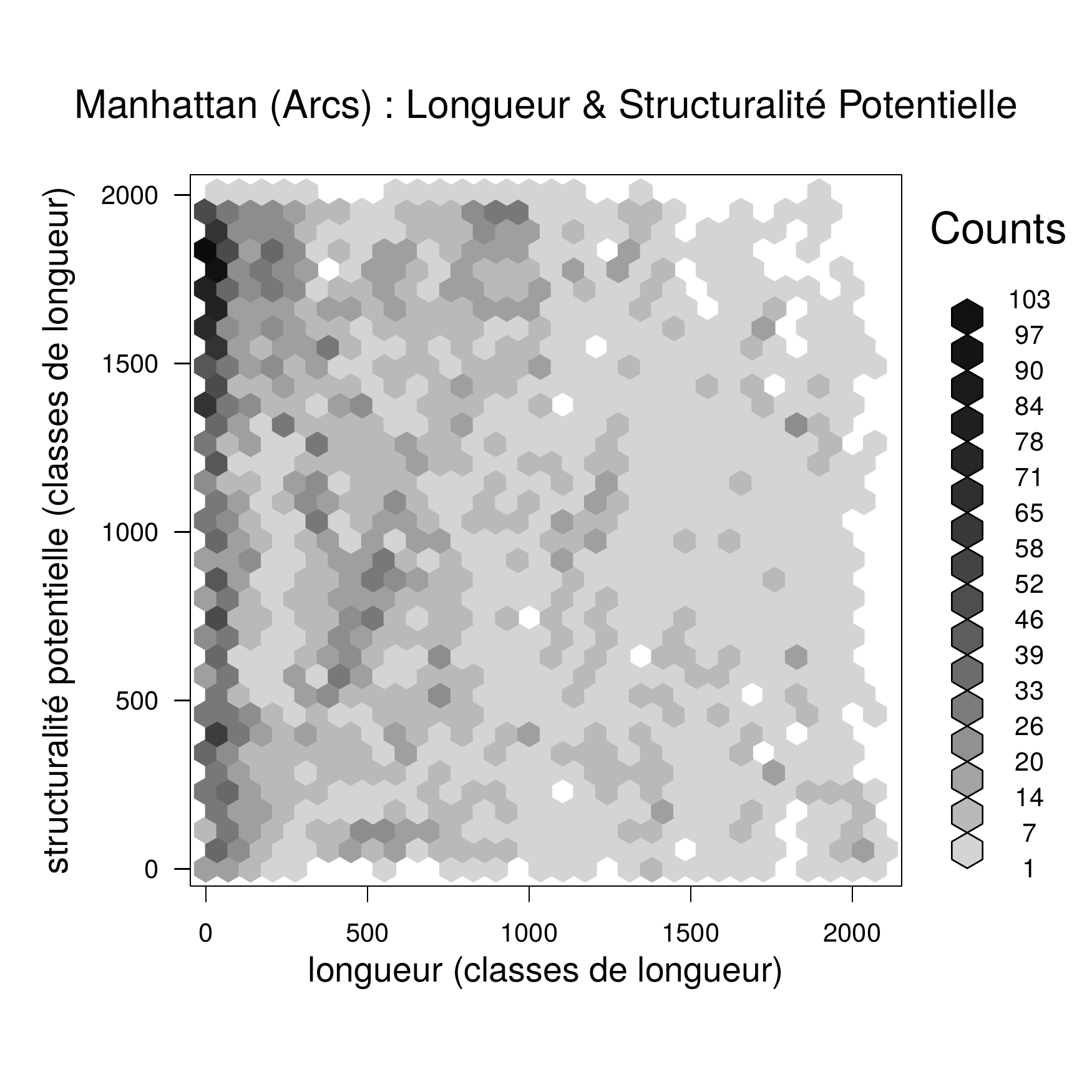}
	\end{subfigure}
	\caption{Longueur et structuralité potentielle}
\end{figure}

\begin{figure}[h]\centering
	\begin{subfigure}[t]{0.45\textwidth}
		\includegraphics[width=\linewidth]{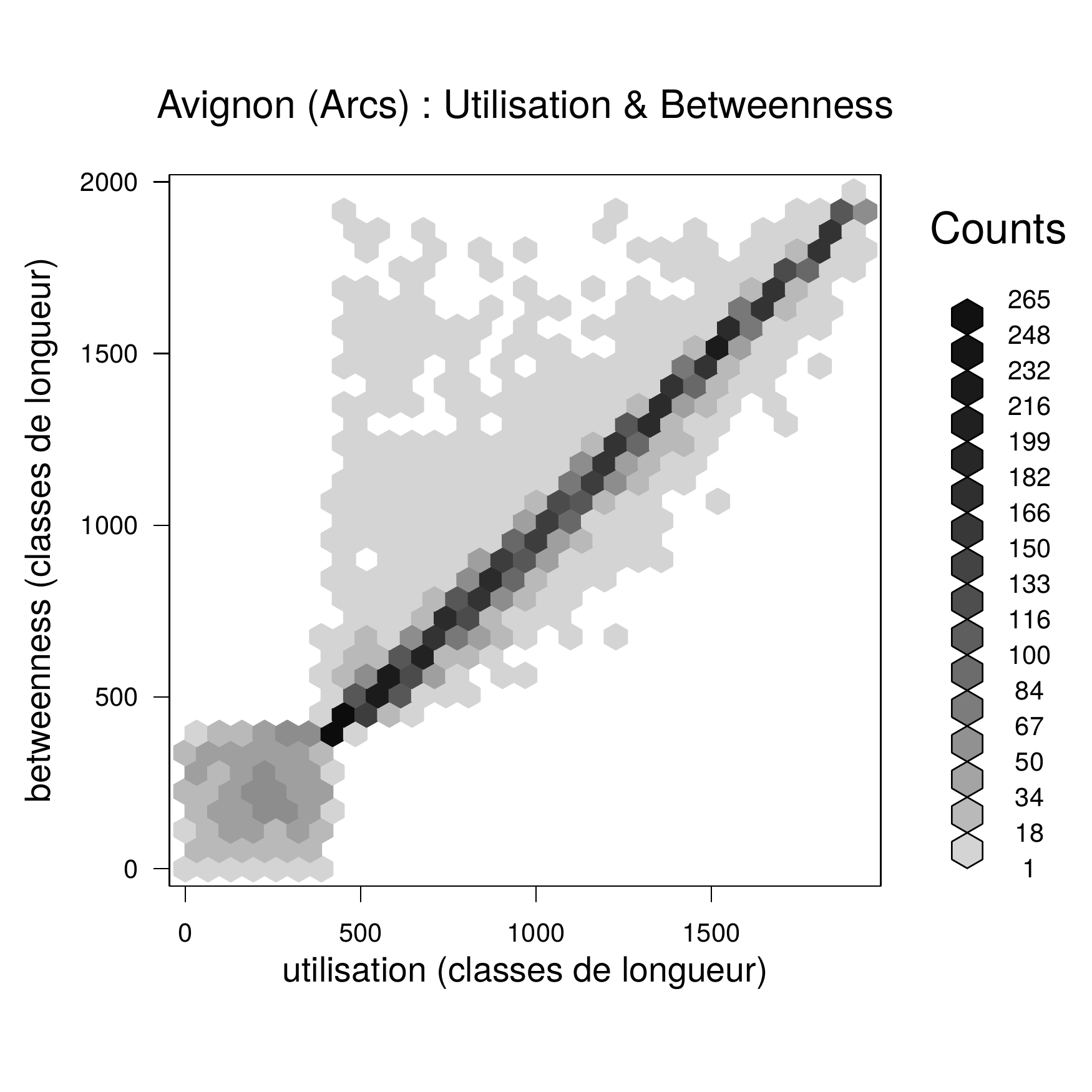}
	\end{subfigure}
	~
	\begin{subfigure}[t]{0.45\textwidth}
		\includegraphics[width=\linewidth]{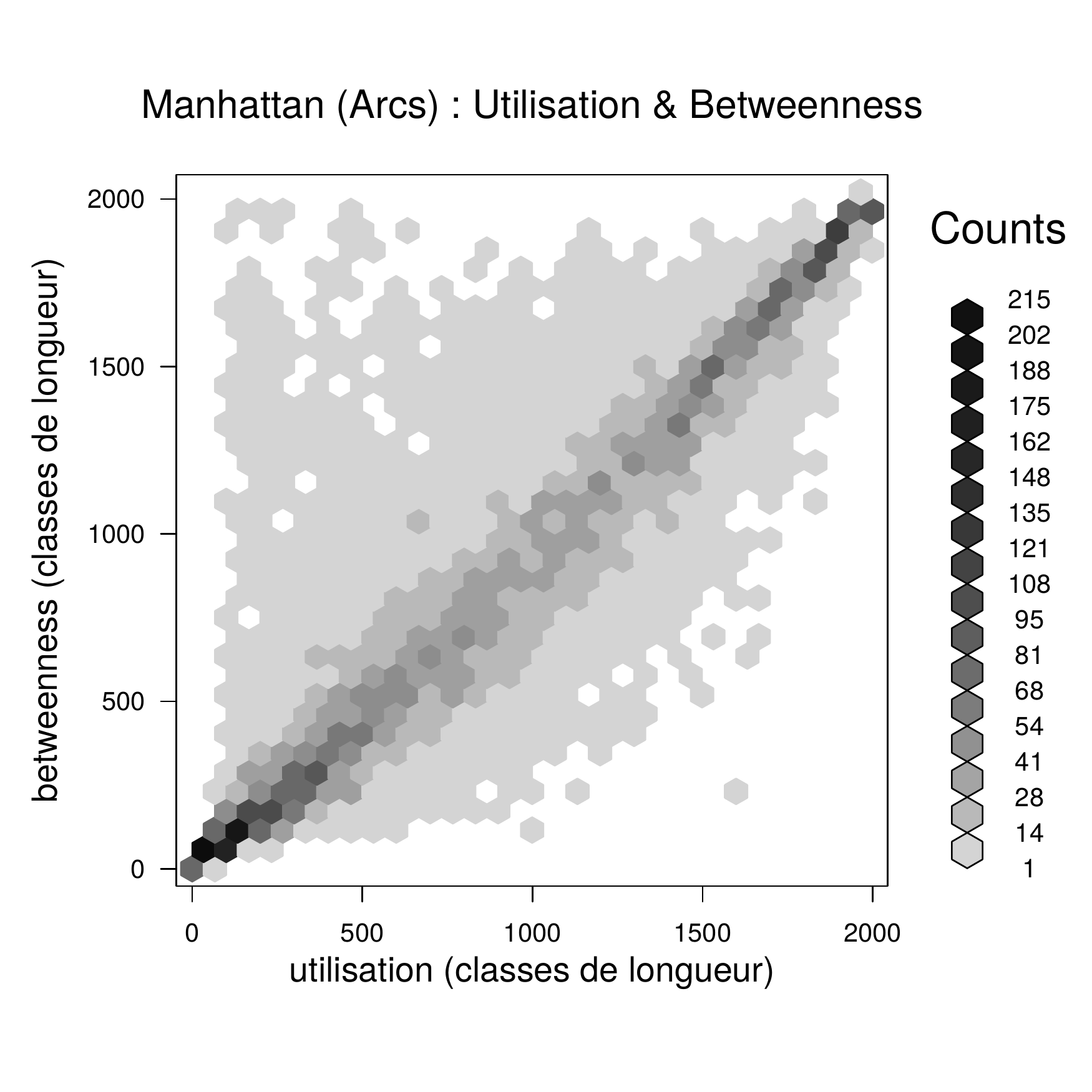}
	\end{subfigure}
	\caption{Utilisation et betweenness}
\end{figure}

\begin{figure}[h]\centering
	\begin{subfigure}[t]{0.45\textwidth}
		\includegraphics[width=\linewidth]{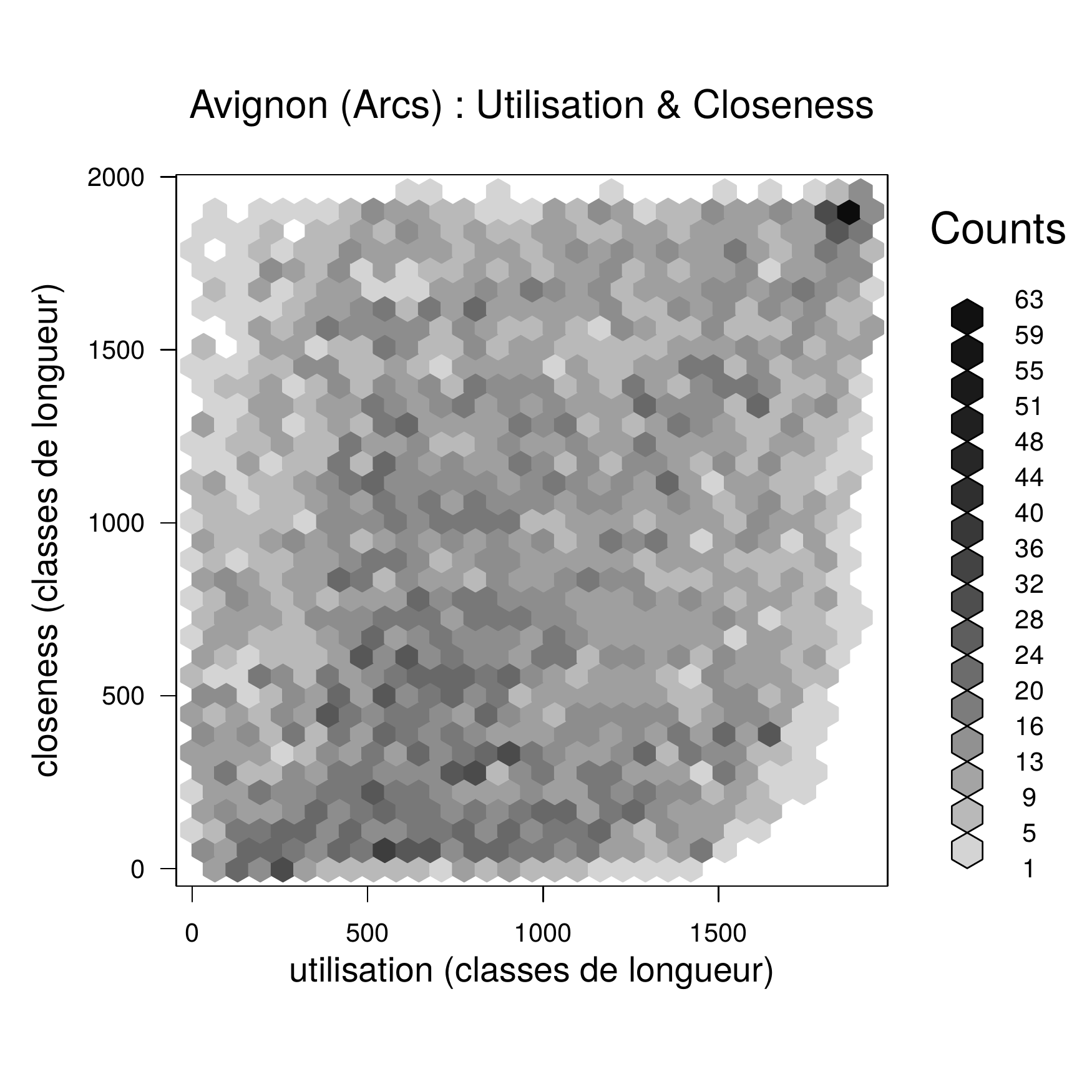}
	\end{subfigure}
	~
	\begin{subfigure}[t]{0.45\textwidth}
		\includegraphics[width=\linewidth]{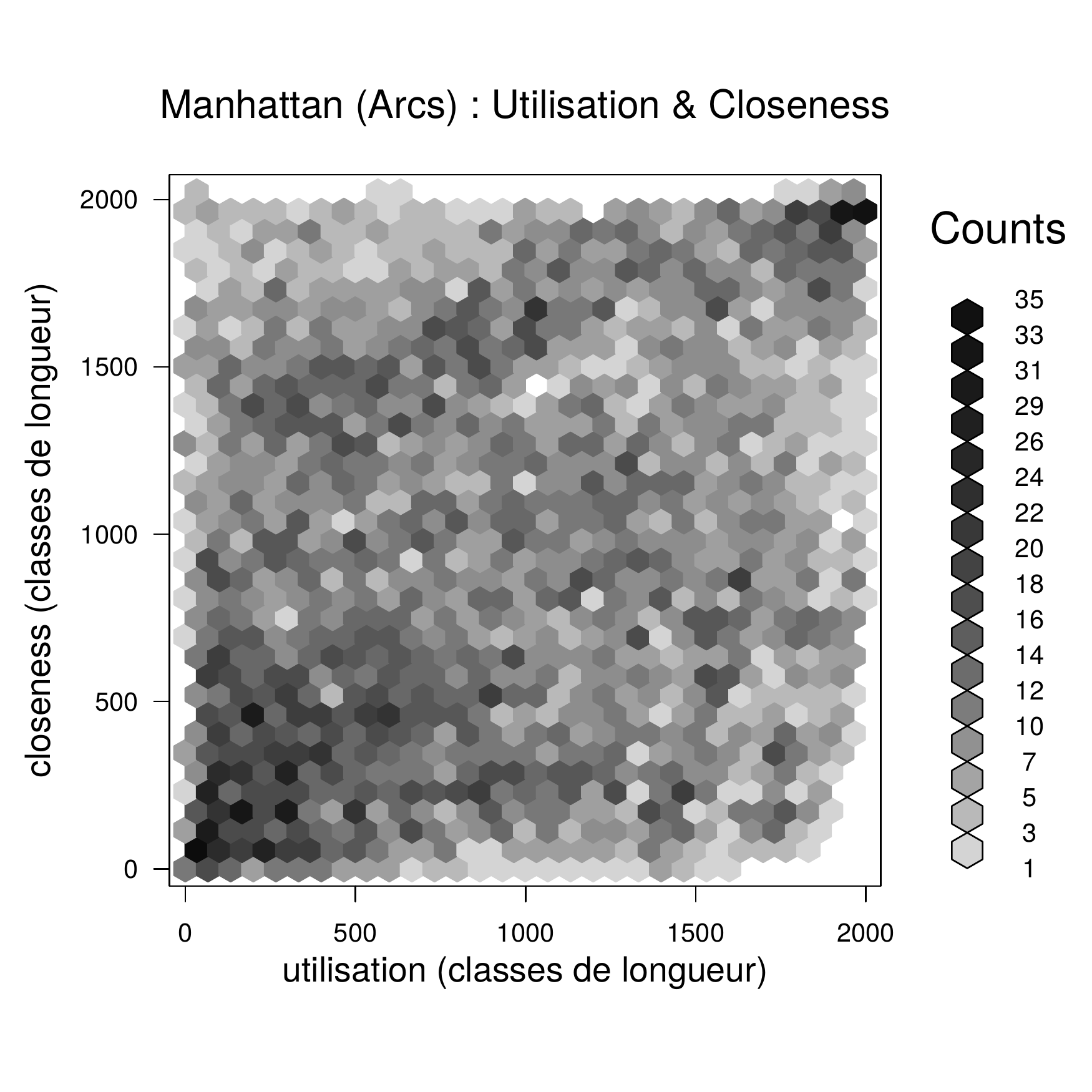}
	\end{subfigure}
	\caption{Utilisation et closeness}
\end{figure}

\begin{figure}[h]\centering
	\begin{subfigure}[t]{0.45\textwidth}
		\includegraphics[width=\linewidth]{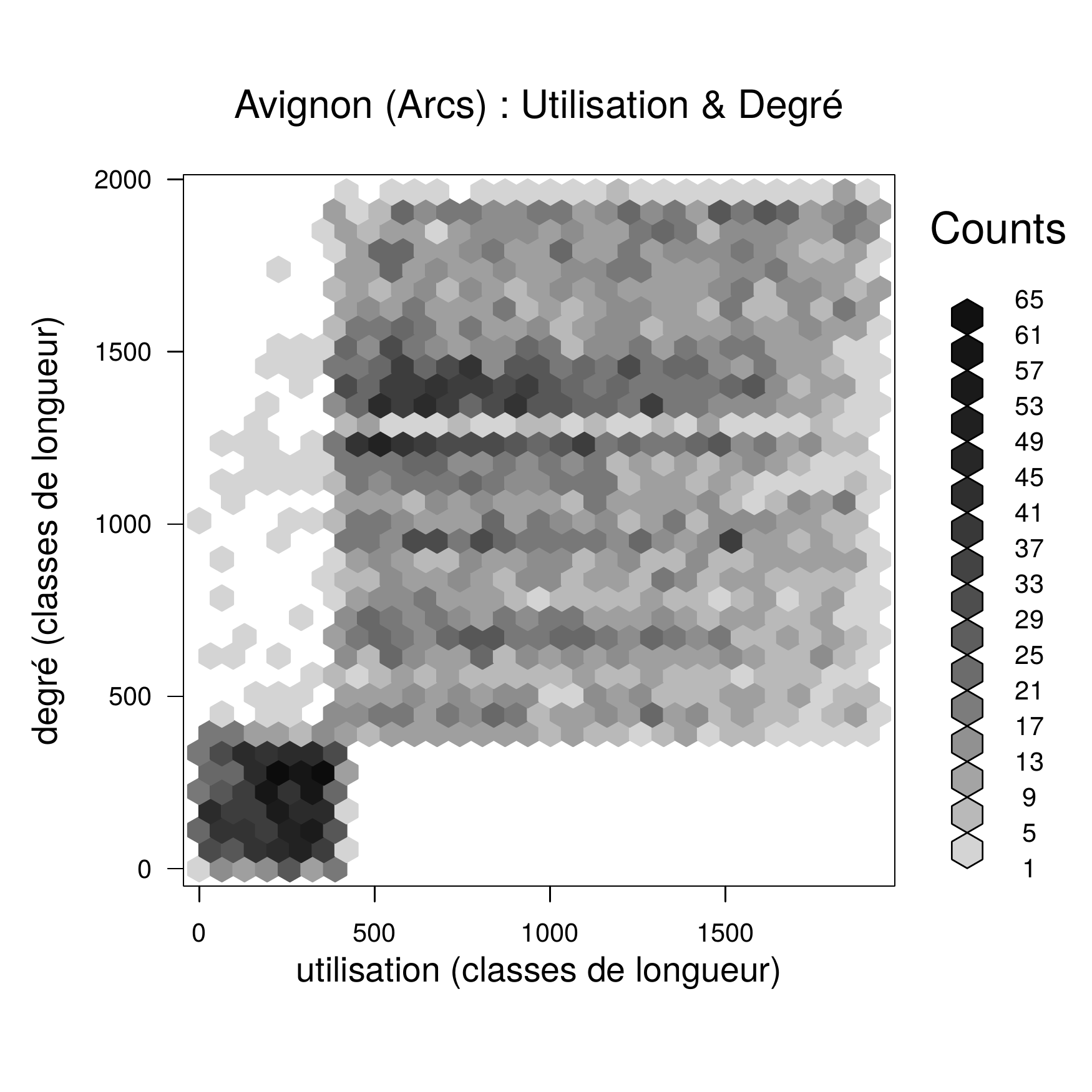}
	\end{subfigure}
	~
	\begin{subfigure}[t]{0.45\textwidth}
		\includegraphics[width=\linewidth]{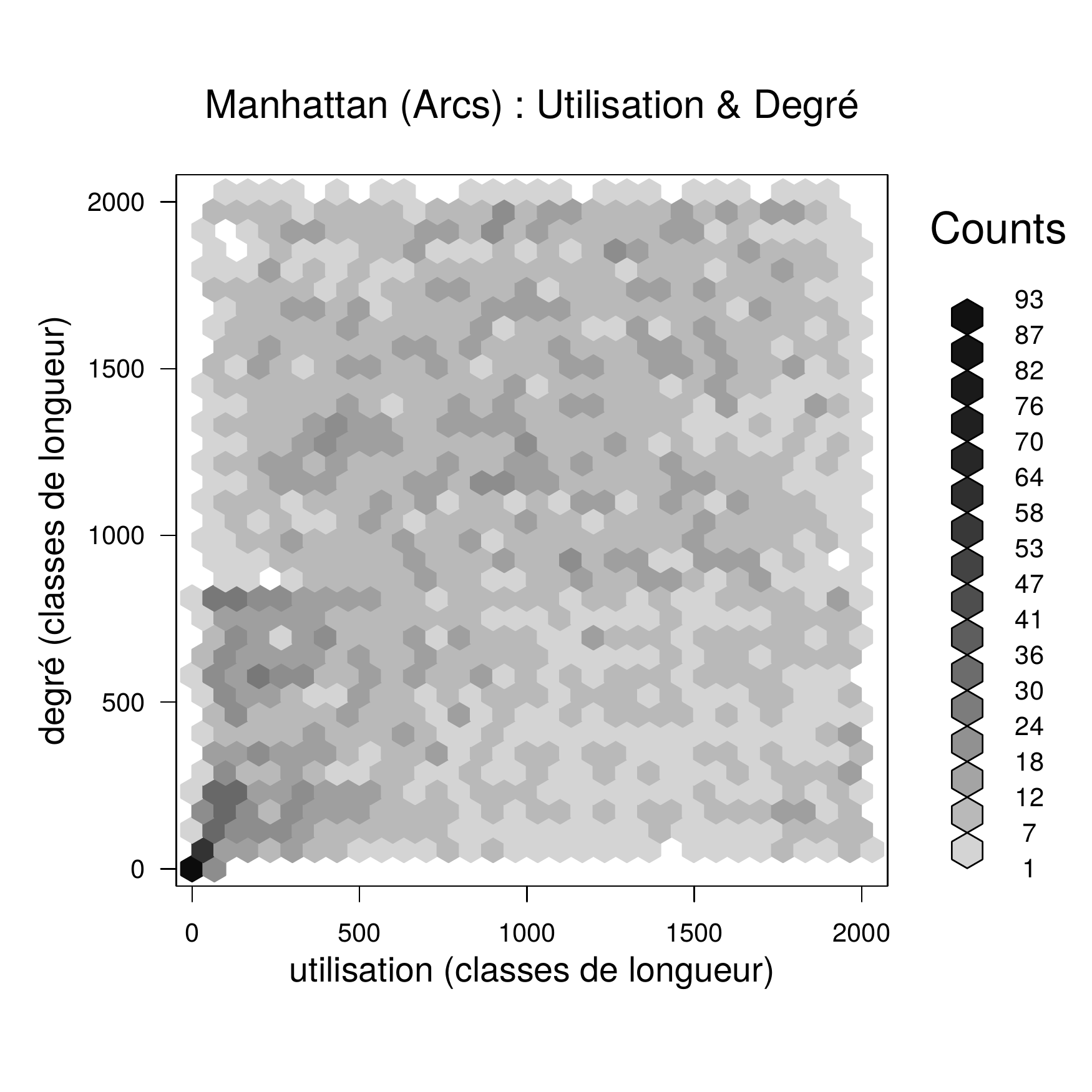}
	\end{subfigure}
	\caption{utilisation et degré}
\end{figure}

\begin{figure}[h]\centering
	\begin{subfigure}[t]{0.45\textwidth}
		\includegraphics[width=\linewidth]{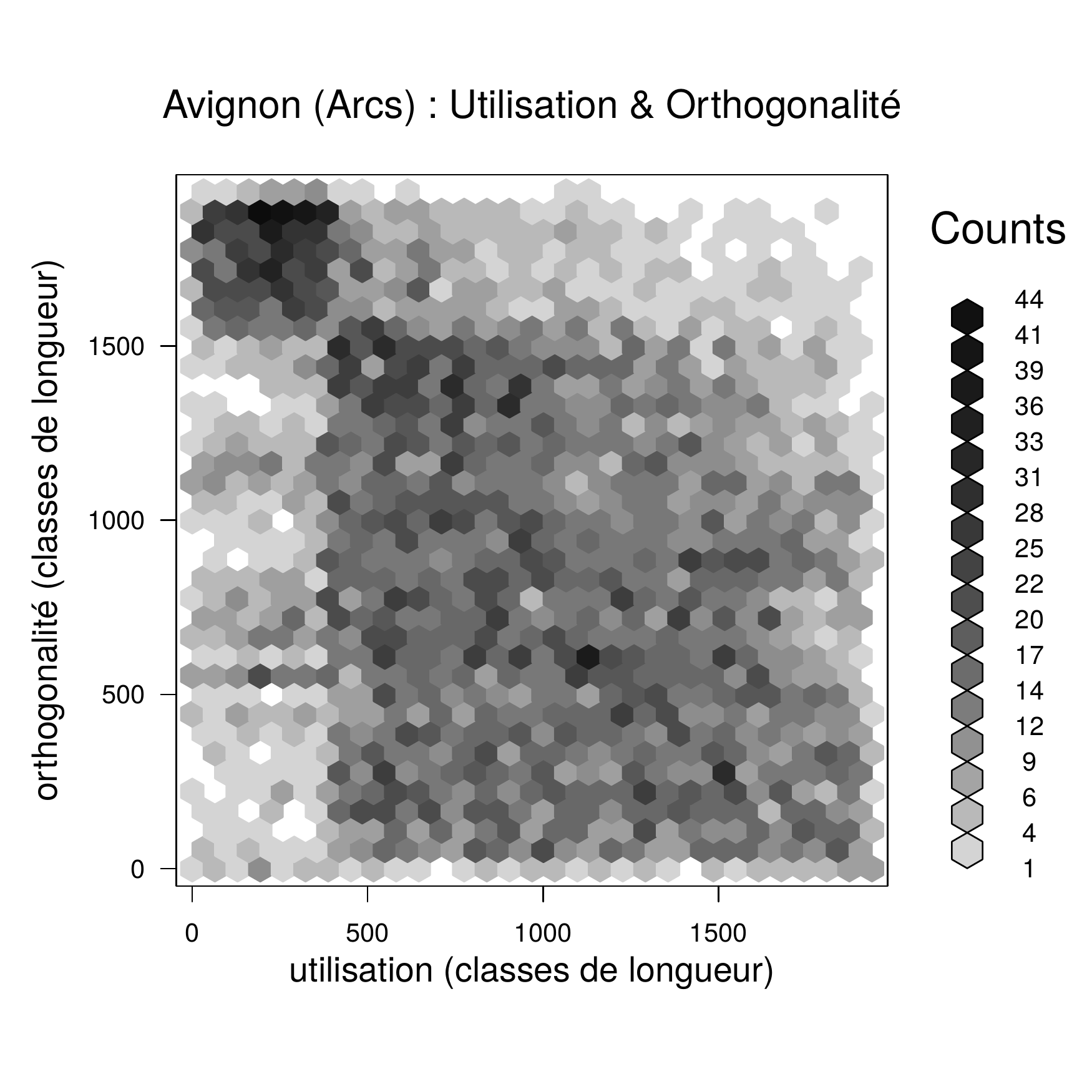}
	\end{subfigure}
	~
	\begin{subfigure}[t]{0.45\textwidth}
		\includegraphics[width=\linewidth]{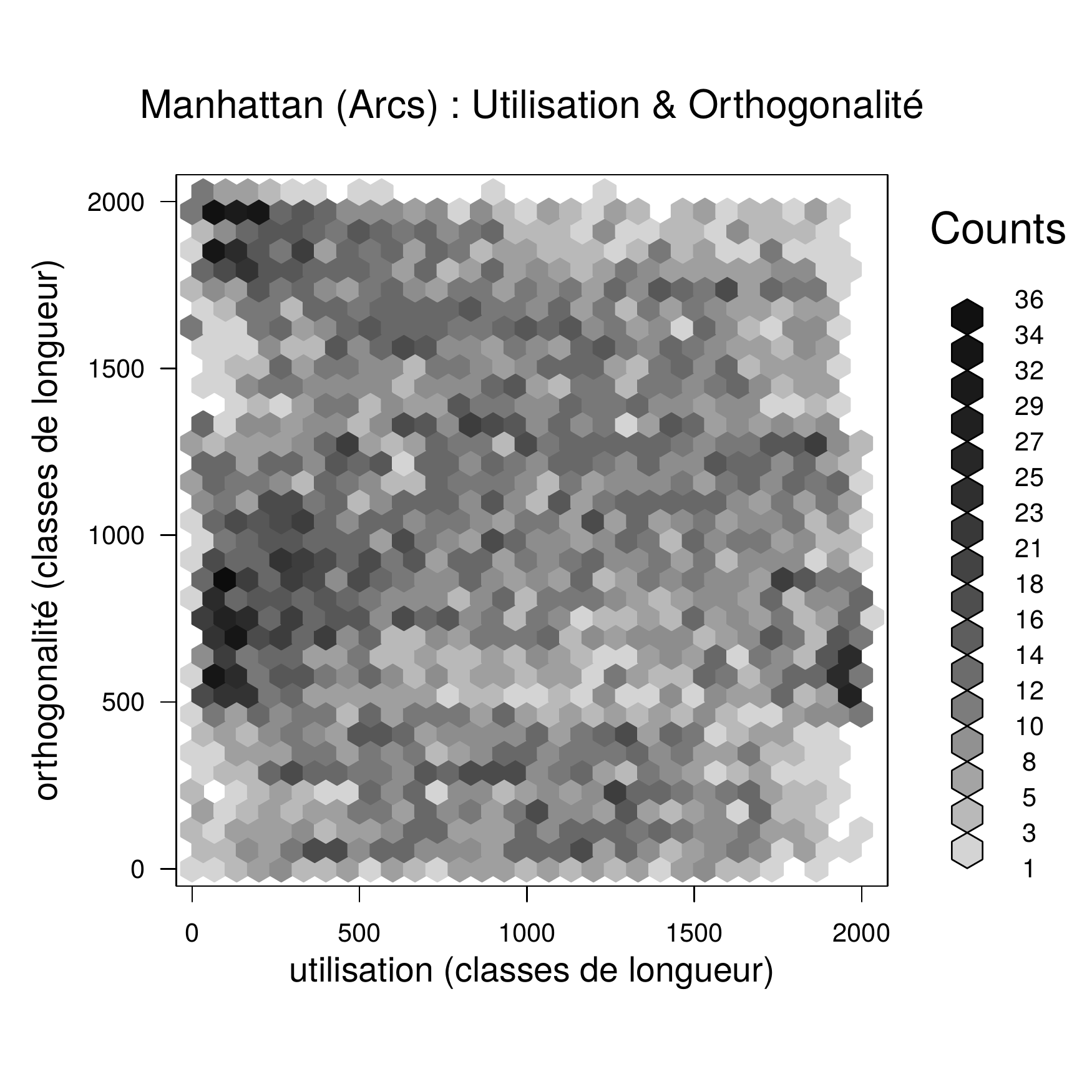}
	\end{subfigure}
	\caption{Utilisation et orthogonalité}
\end{figure}

\FloatBarrier 
\subsection{Indicateurs composés}\label{ann:ssec_arcs_indcomp}

\begin{figure}[h]\centering
    \begin{subfigure}[t]{0.45\textwidth}
        \includegraphics[width=\linewidth]{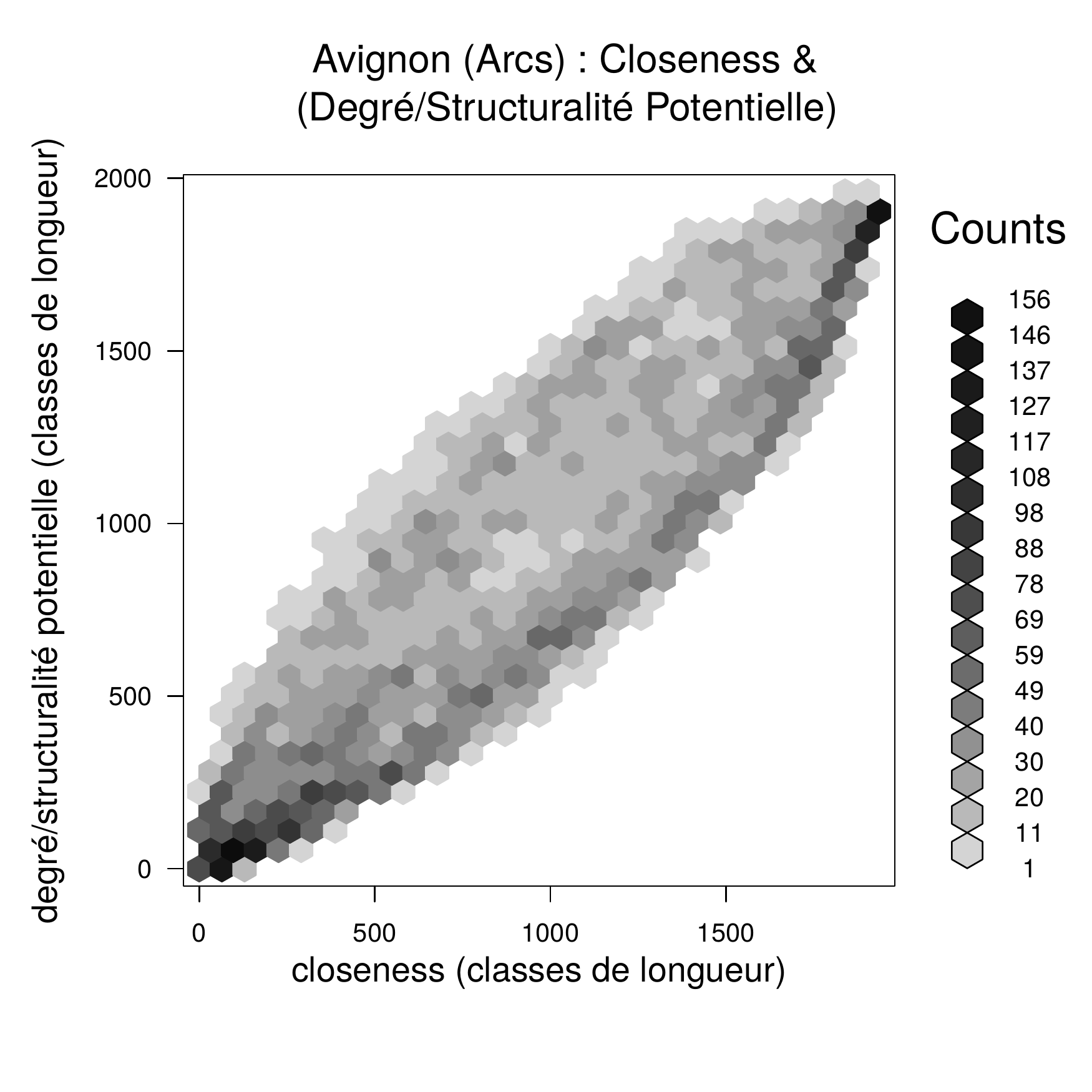}
    \end{subfigure}
    ~
    \begin{subfigure}[t]{0.45\textwidth}
        \includegraphics[width=\linewidth]{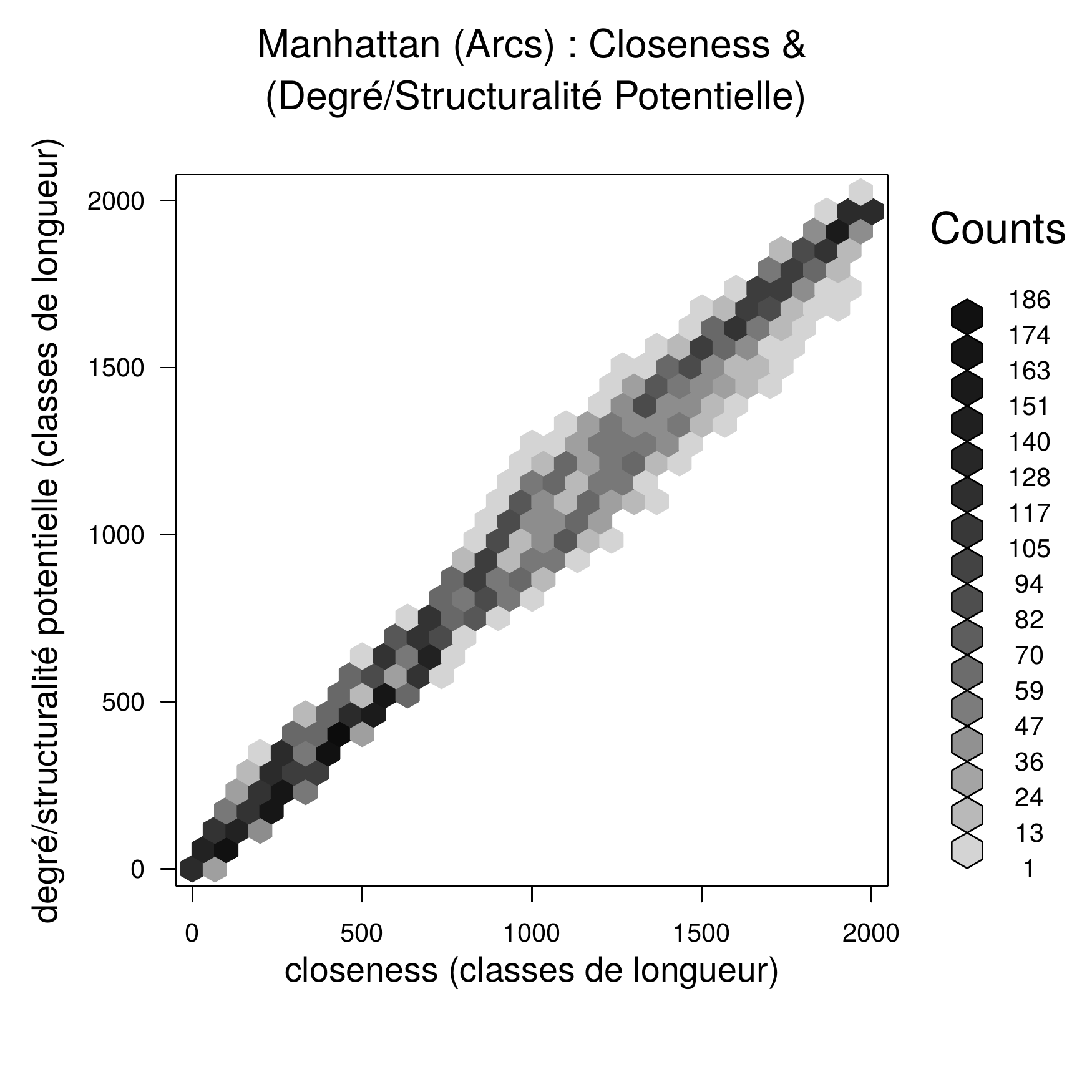}
    \end{subfigure}
    \caption{Closeness et degré sur structuralité}
\end{figure}

\begin{figure}[h]\centering
    \begin{subfigure}[t]{0.45\textwidth}
        \includegraphics[width=\linewidth]{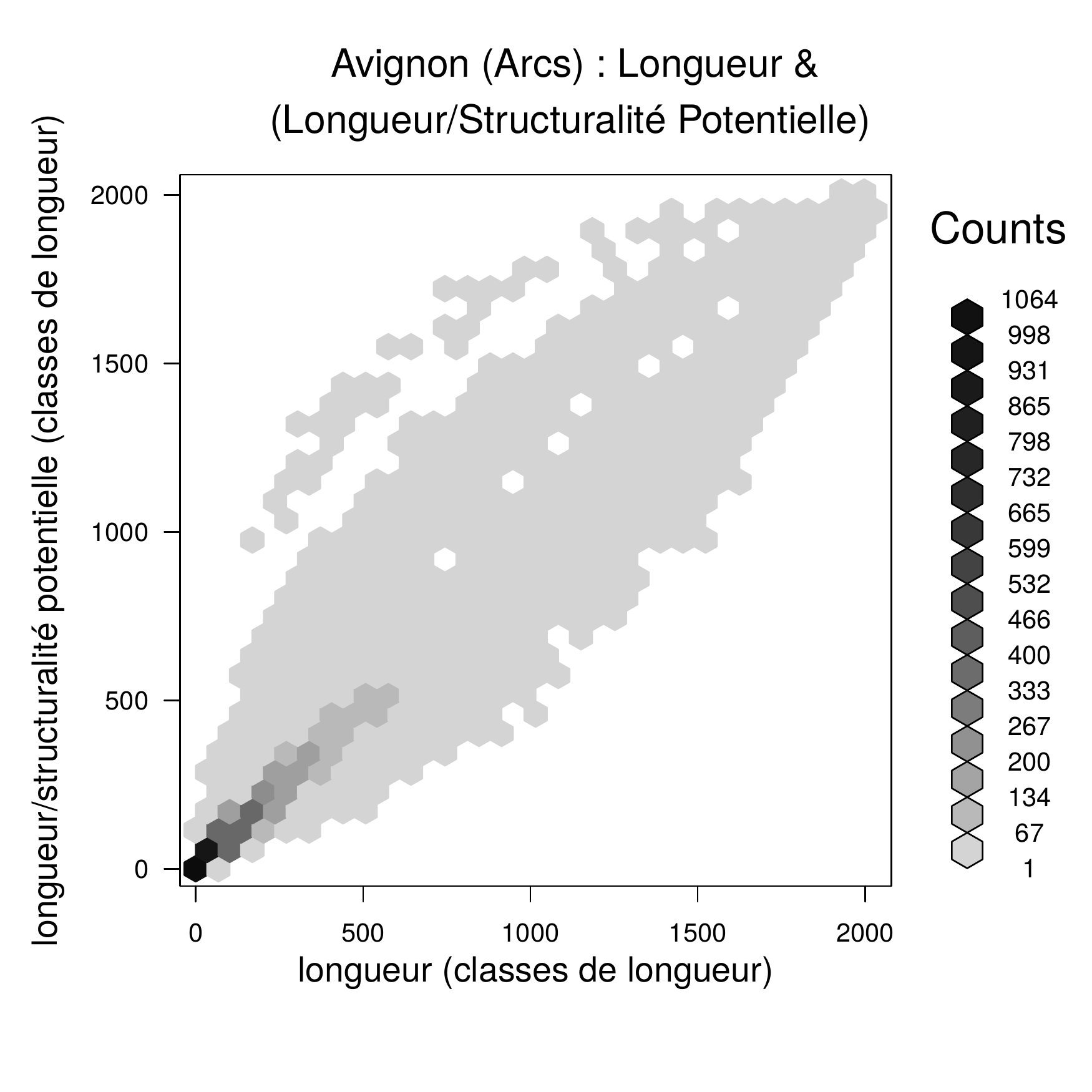}
    \end{subfigure}
    ~
    \begin{subfigure}[t]{0.45\textwidth}
        \includegraphics[width=\linewidth]{images/cartes_hexbin/ind_comp/arcs_Manhattan_length_los.pdf}
    \end{subfigure}
    \caption{Longueur et longueur sur structuralité}
\end{figure}

\begin{figure}[h]\centering
    \begin{subfigure}[t]{0.45\textwidth}
        \includegraphics[width=\linewidth]{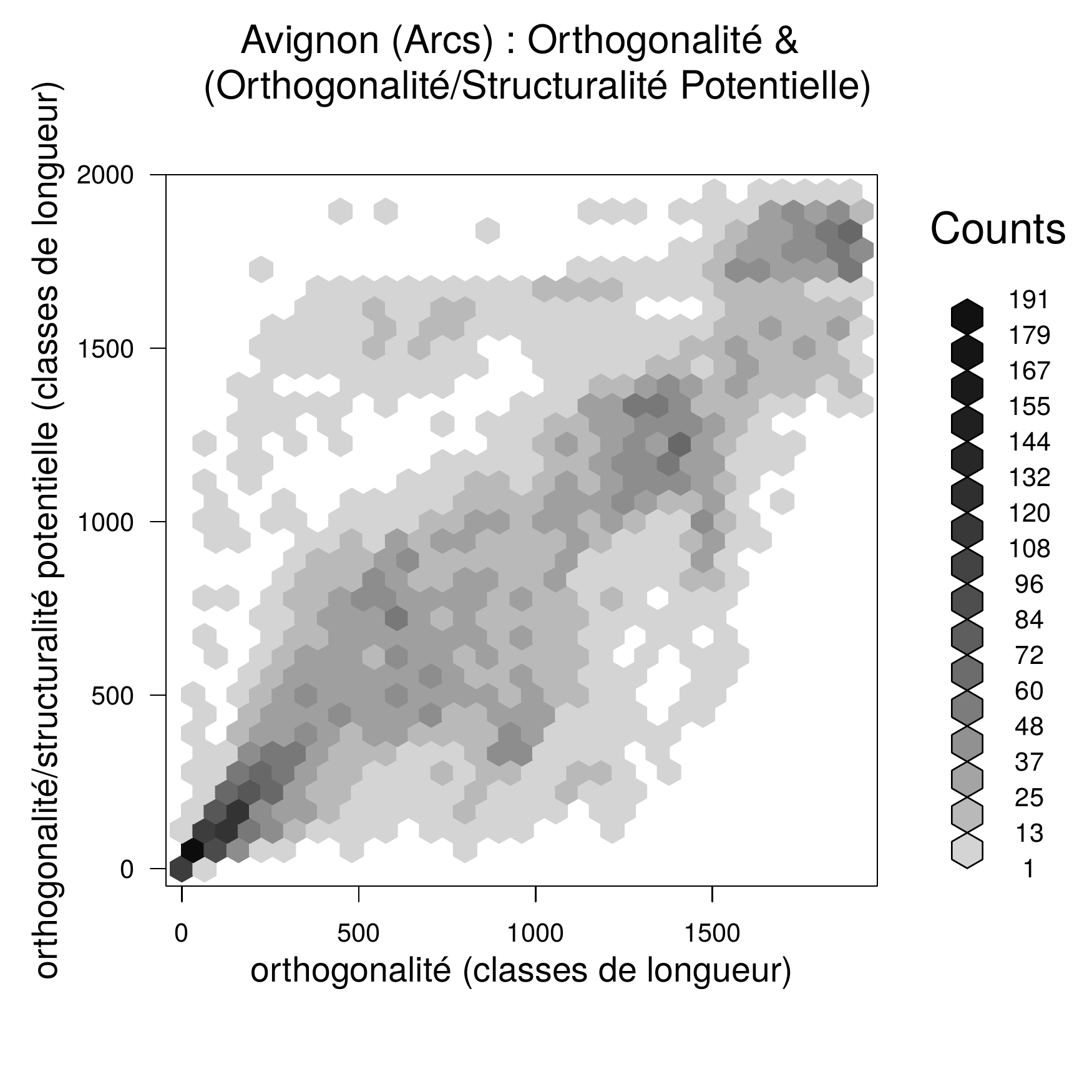}
    \end{subfigure}
    ~
    \begin{subfigure}[t]{0.45\textwidth}
        \includegraphics[width=\linewidth]{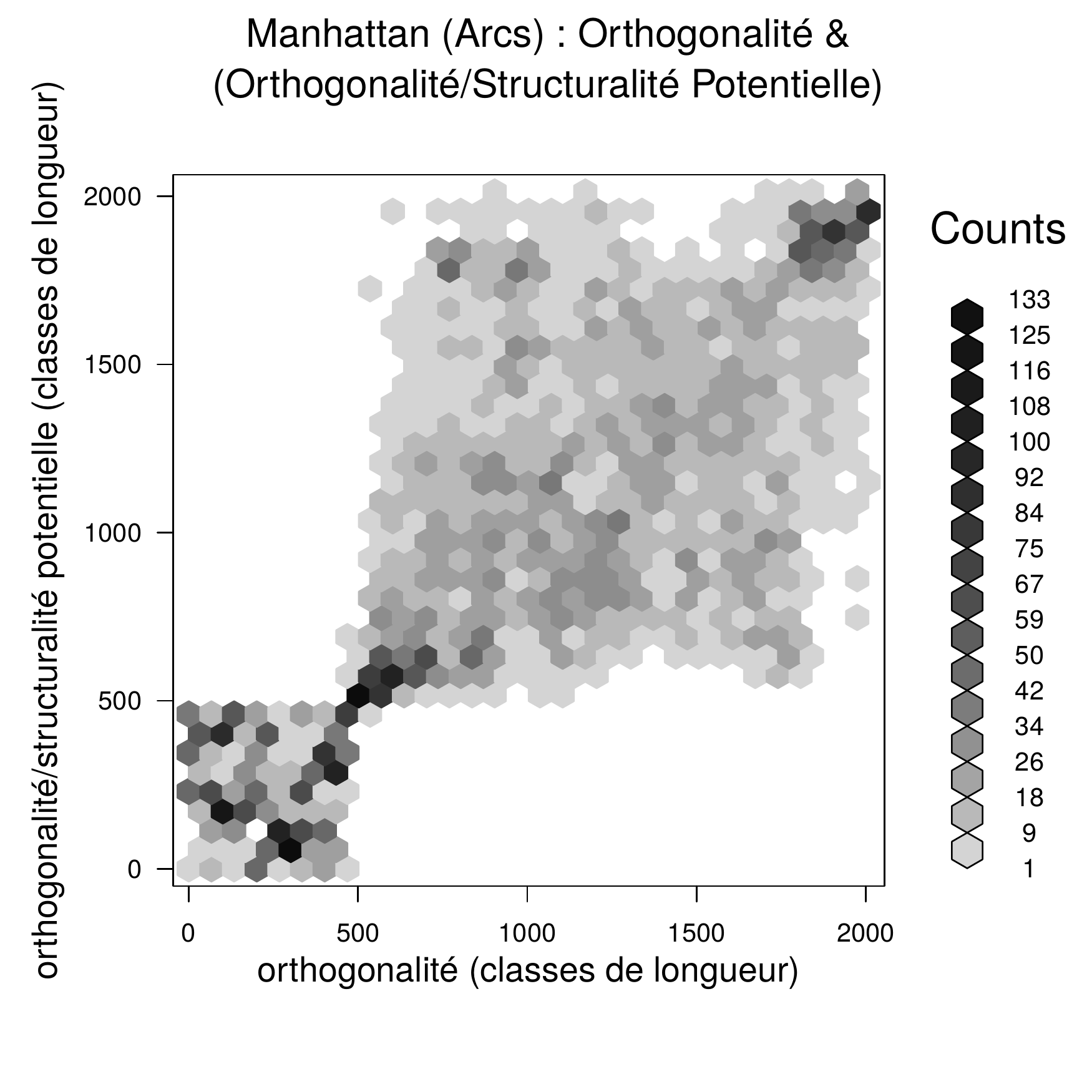}
    \end{subfigure}
    \caption{Orthogonalité et orthogonalité sur structuralité}
\end{figure}

\begin{figure}[h]\centering
    \begin{subfigure}[t]{0.45\textwidth}
        \includegraphics[width=\linewidth]{images/cartes_hexbin/ind_comp/arcs_Avignon_use_lou.pdf}
    \end{subfigure}
    ~
    \begin{subfigure}[t]{0.45\textwidth}
        \includegraphics[width=\linewidth]{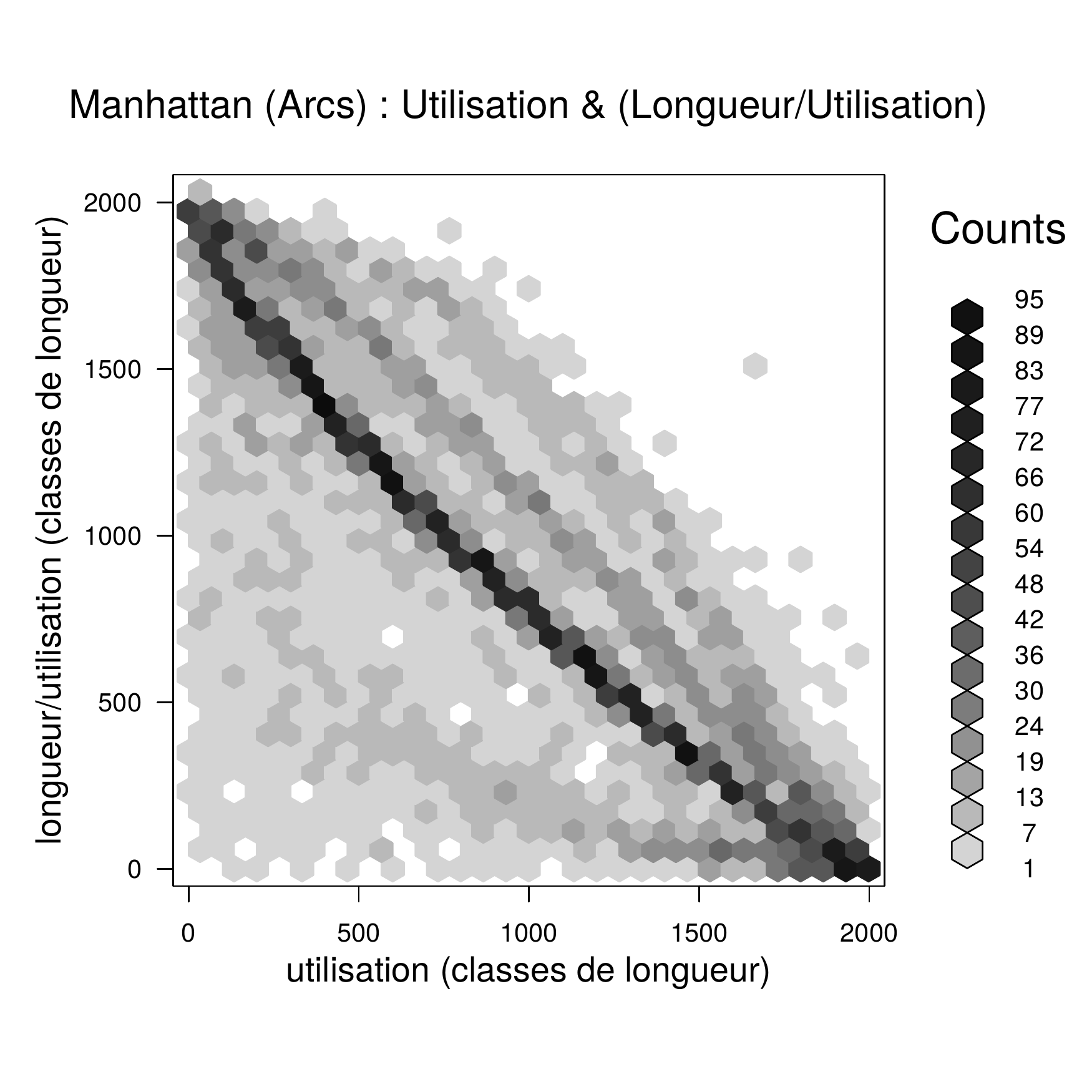}
    \end{subfigure}
    \caption{Utilisation et longueur sur utilisation}
\end{figure}

\begin{figure}[h]\centering
    \begin{subfigure}[t]{0.45\textwidth}
        \includegraphics[width=\linewidth]{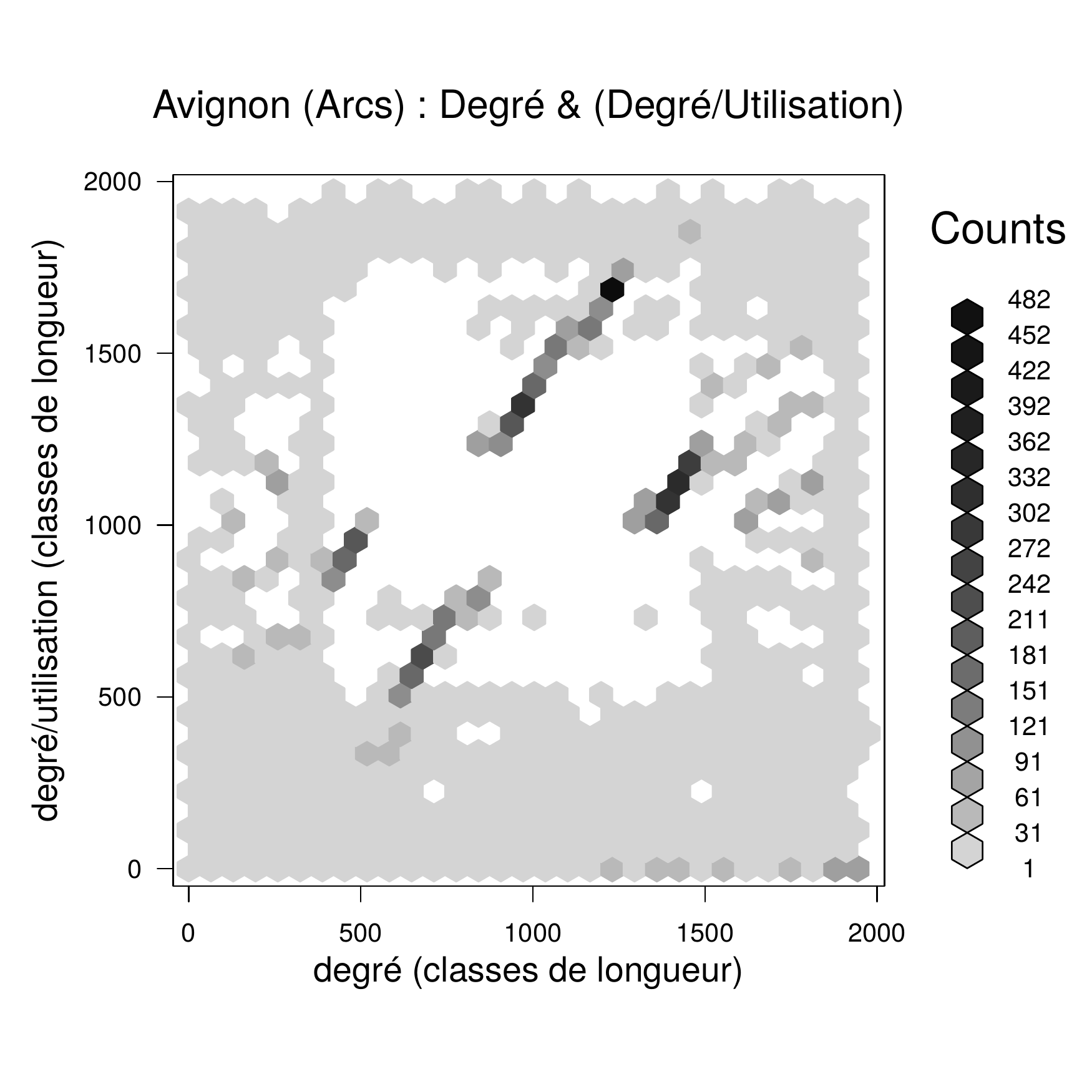}
    \end{subfigure}
    ~
    \begin{subfigure}[t]{0.45\textwidth}
        \includegraphics[width=\linewidth]{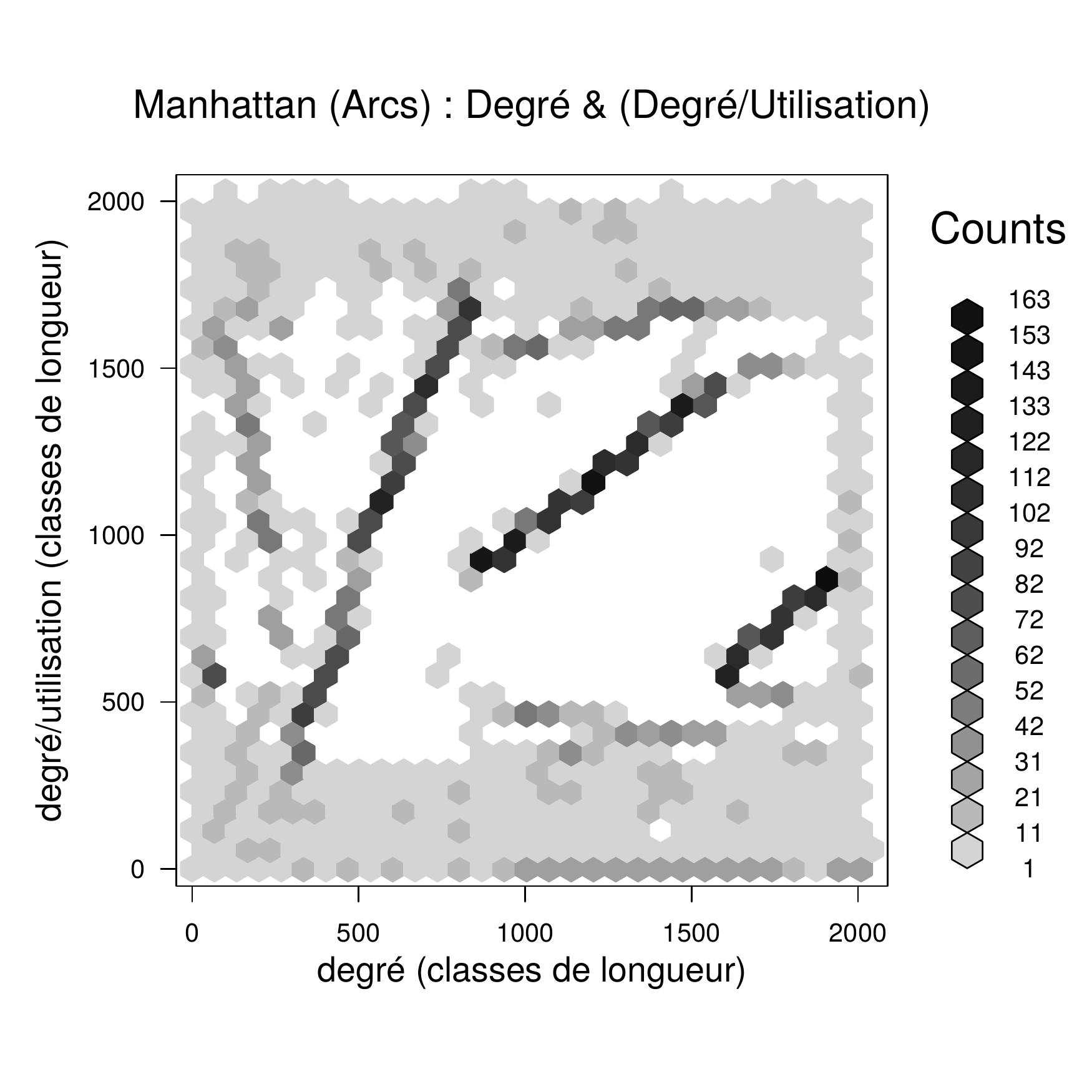}
    \end{subfigure}
    \caption{Degré et degré sur utilisation}
\end{figure}

\begin{figure}[h]\centering
    \begin{subfigure}[t]{0.45\textwidth}
        \includegraphics[width=\linewidth]{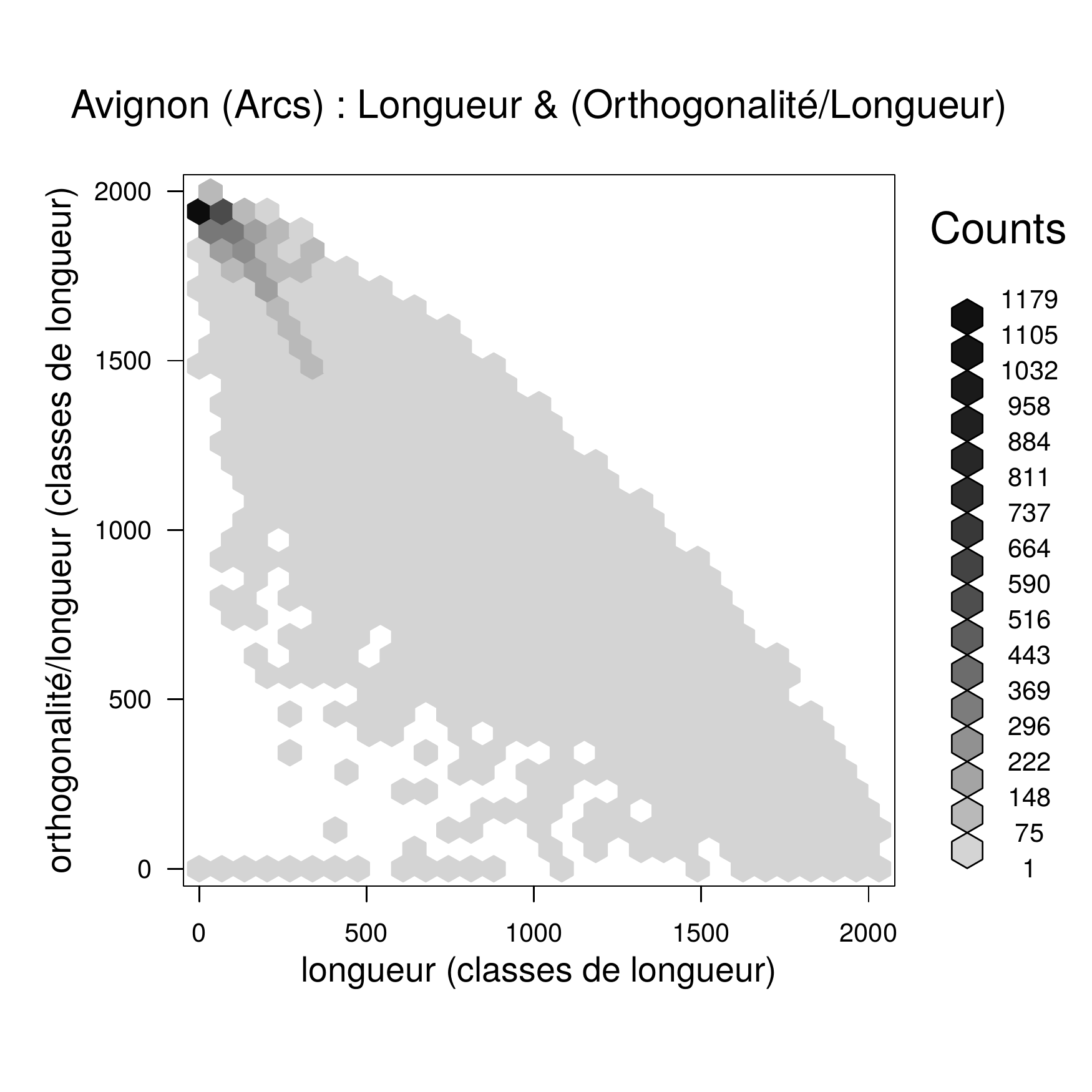}
    \end{subfigure}
    ~
    \begin{subfigure}[t]{0.45\textwidth}
        \includegraphics[width=\linewidth]{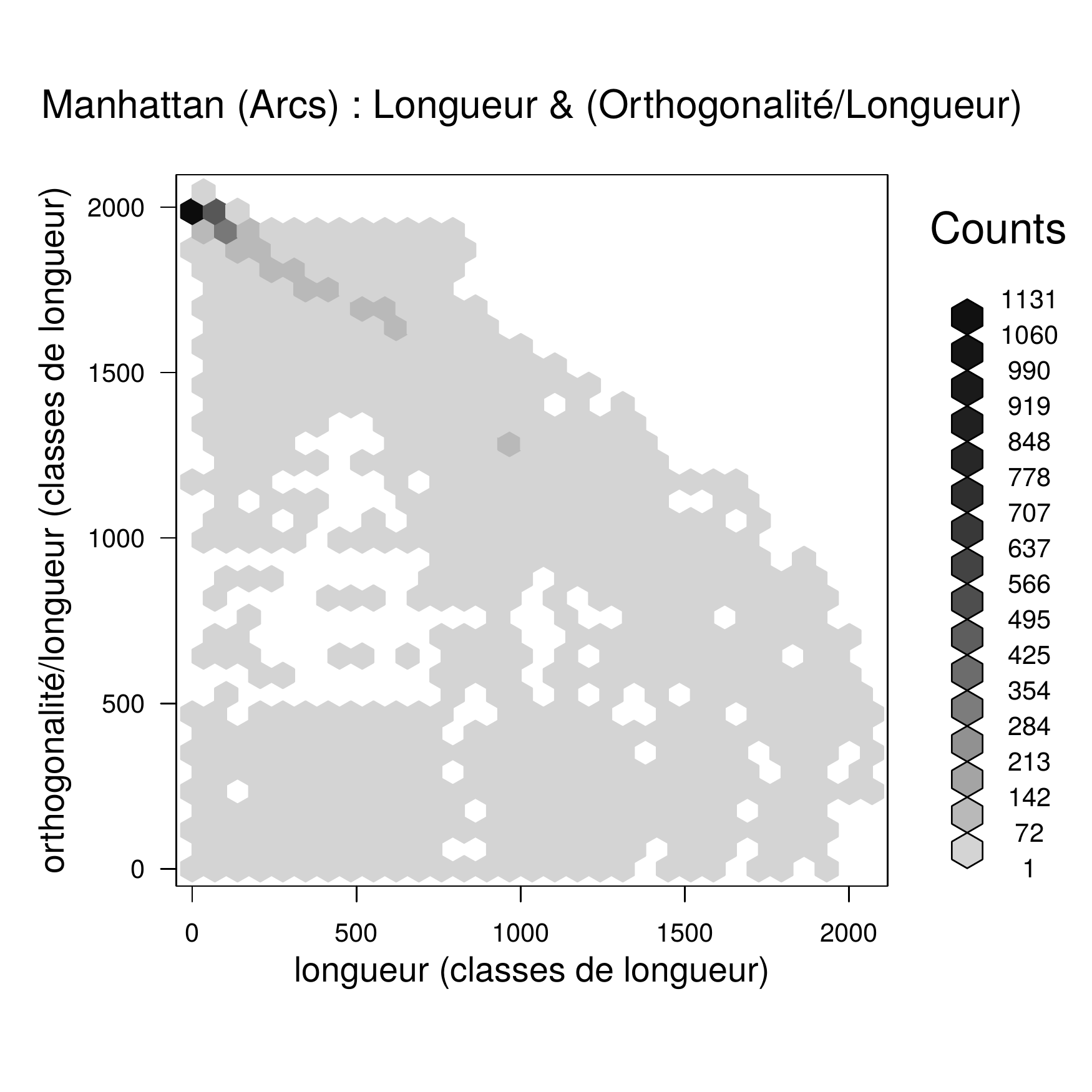}
    \end{subfigure}
    \caption{Longueur et orthogonalité sur longueur}
\end{figure}

\begin{figure}[h]\centering
    \begin{subfigure}[t]{0.45\textwidth}
        \includegraphics[width=\linewidth]{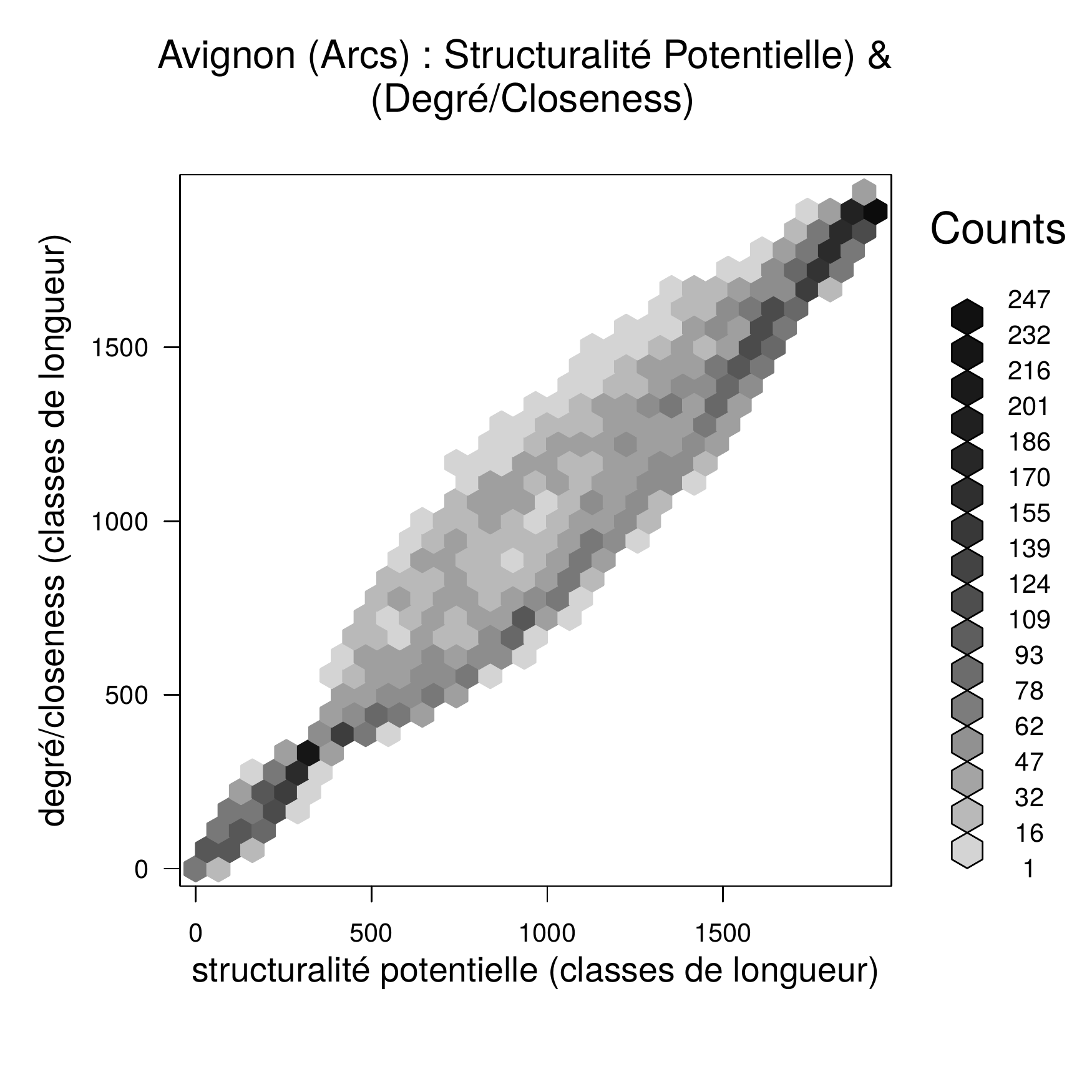}
    \end{subfigure}
    ~
    \begin{subfigure}[t]{0.45\textwidth}
        \includegraphics[width=\linewidth]{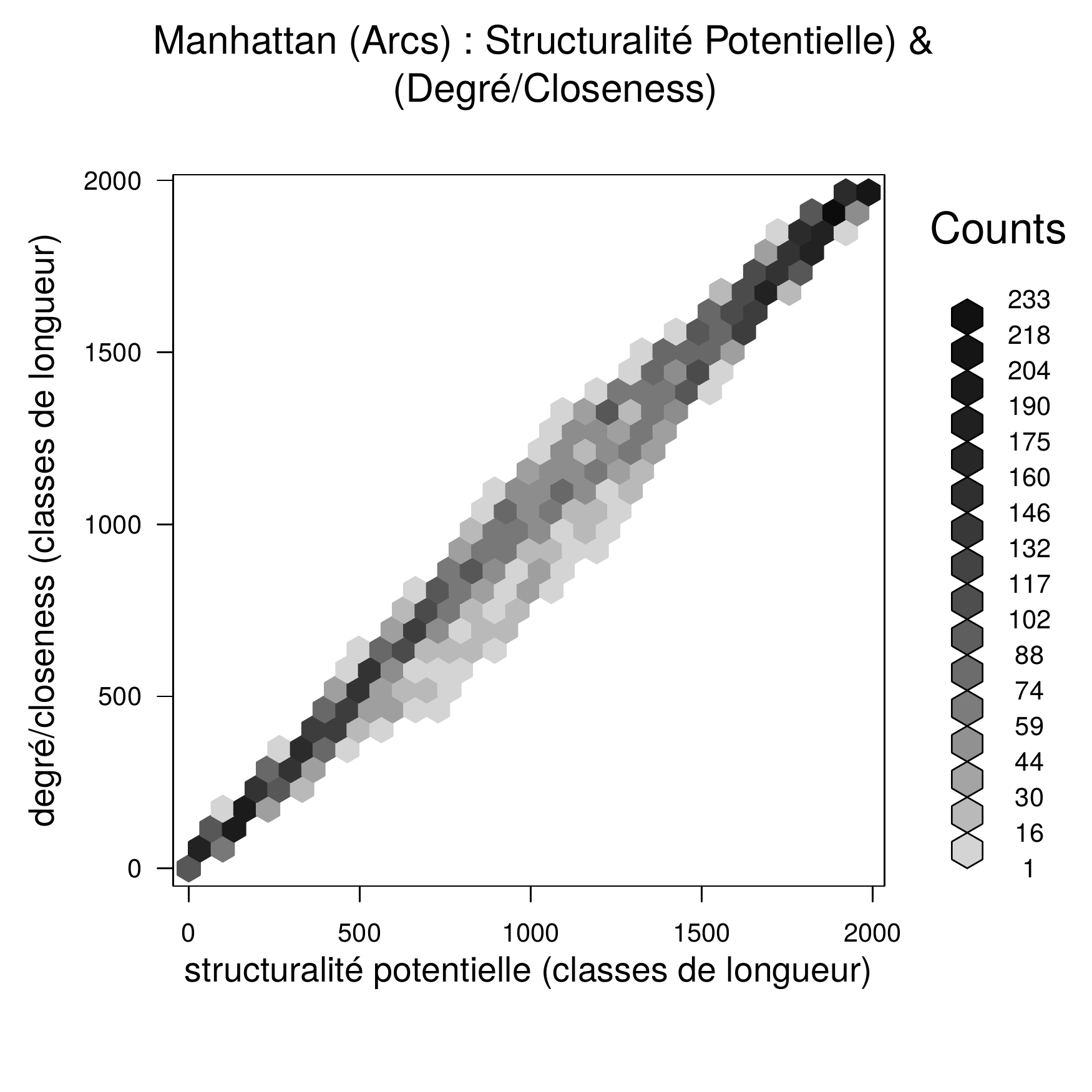}
    \end{subfigure}
    \caption{Structuralité potentielle et doc.pdf}
\end{figure}

\begin{figure}[h]\centering
    \begin{subfigure}[t]{0.45\textwidth}
        \includegraphics[width=\linewidth]{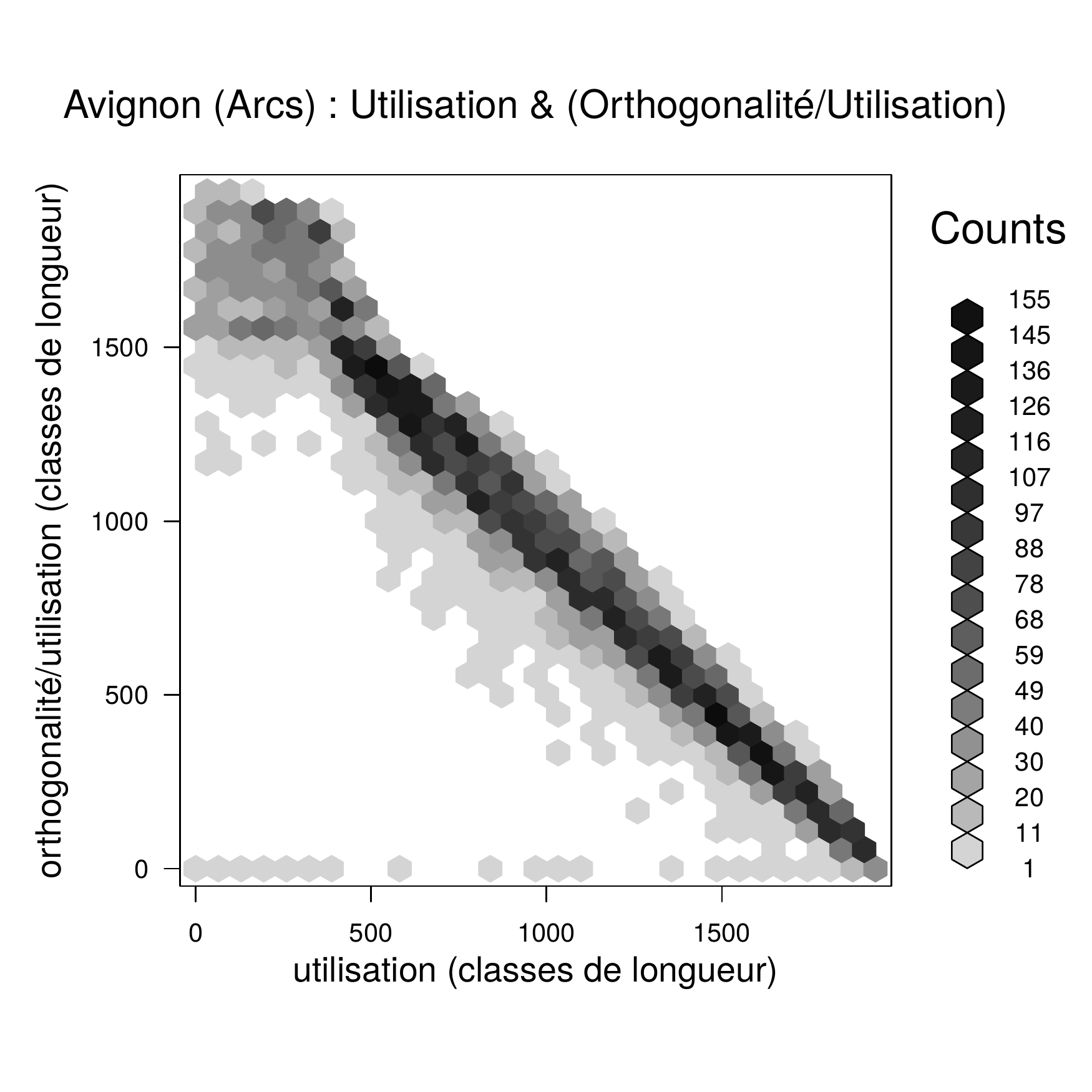}
    \end{subfigure}
    ~
    \begin{subfigure}[t]{0.45\textwidth}
        \includegraphics[width=\linewidth]{images/cartes_hexbin/ind_comp/arcs_Manhattan_use_oou.pdf}
    \end{subfigure}
    \caption{Utilisation et orthogonalité sur utilisation}
\end{figure}

\begin{figure}[h]\centering
    \begin{subfigure}[t]{0.45\textwidth}
        \includegraphics[width=\linewidth]{images/cartes_hexbin/ind_comp/arcs_Avignon_length_loc.pdf}
    \end{subfigure}
    ~
    \begin{subfigure}[t]{0.45\textwidth}
        \includegraphics[width=\linewidth]{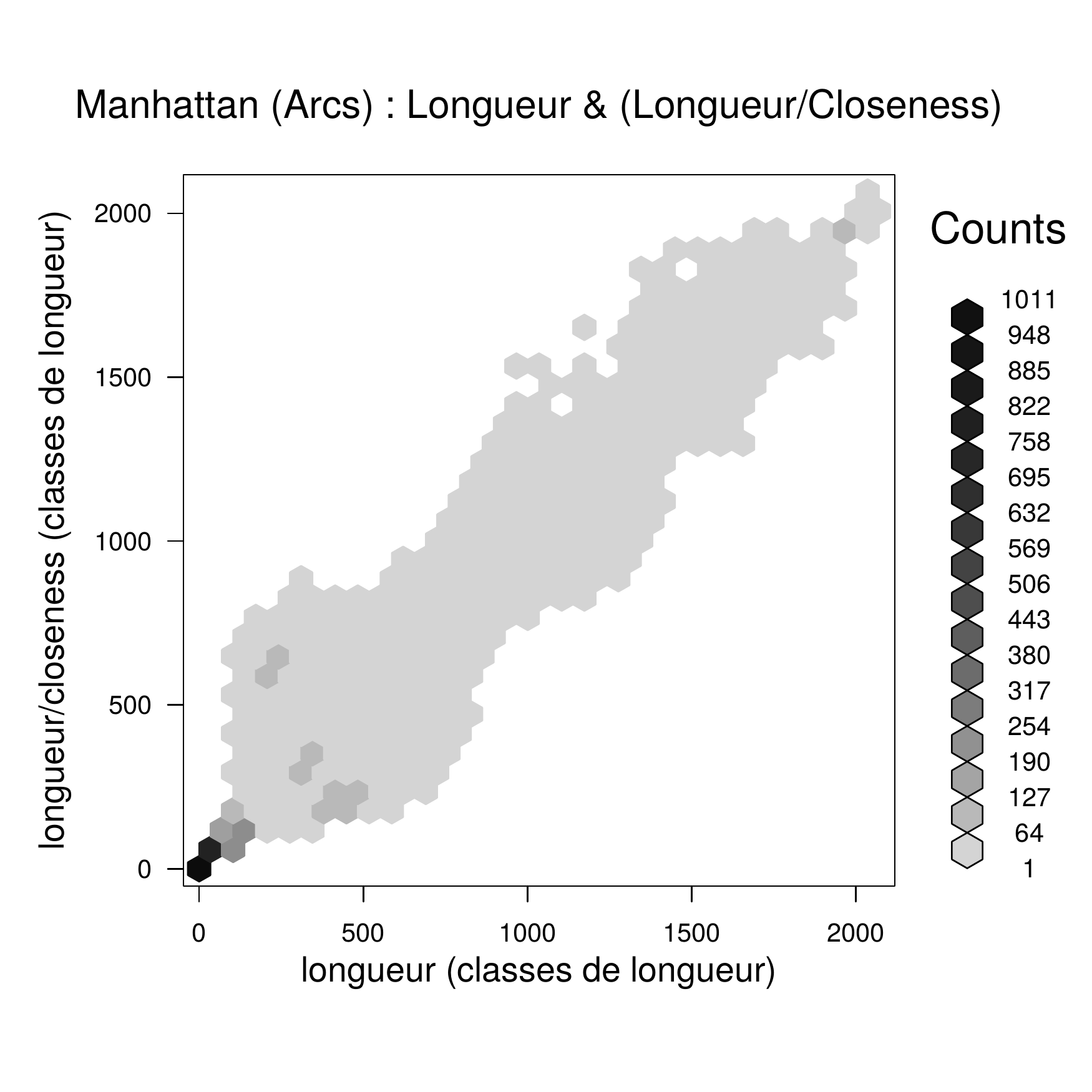}
    \end{subfigure}
    \caption{Longueur et longueur sur closeness}
\end{figure}

\begin{figure}[h]\centering
    \begin{subfigure}[t]{0.45\textwidth}
        \includegraphics[width=\linewidth]{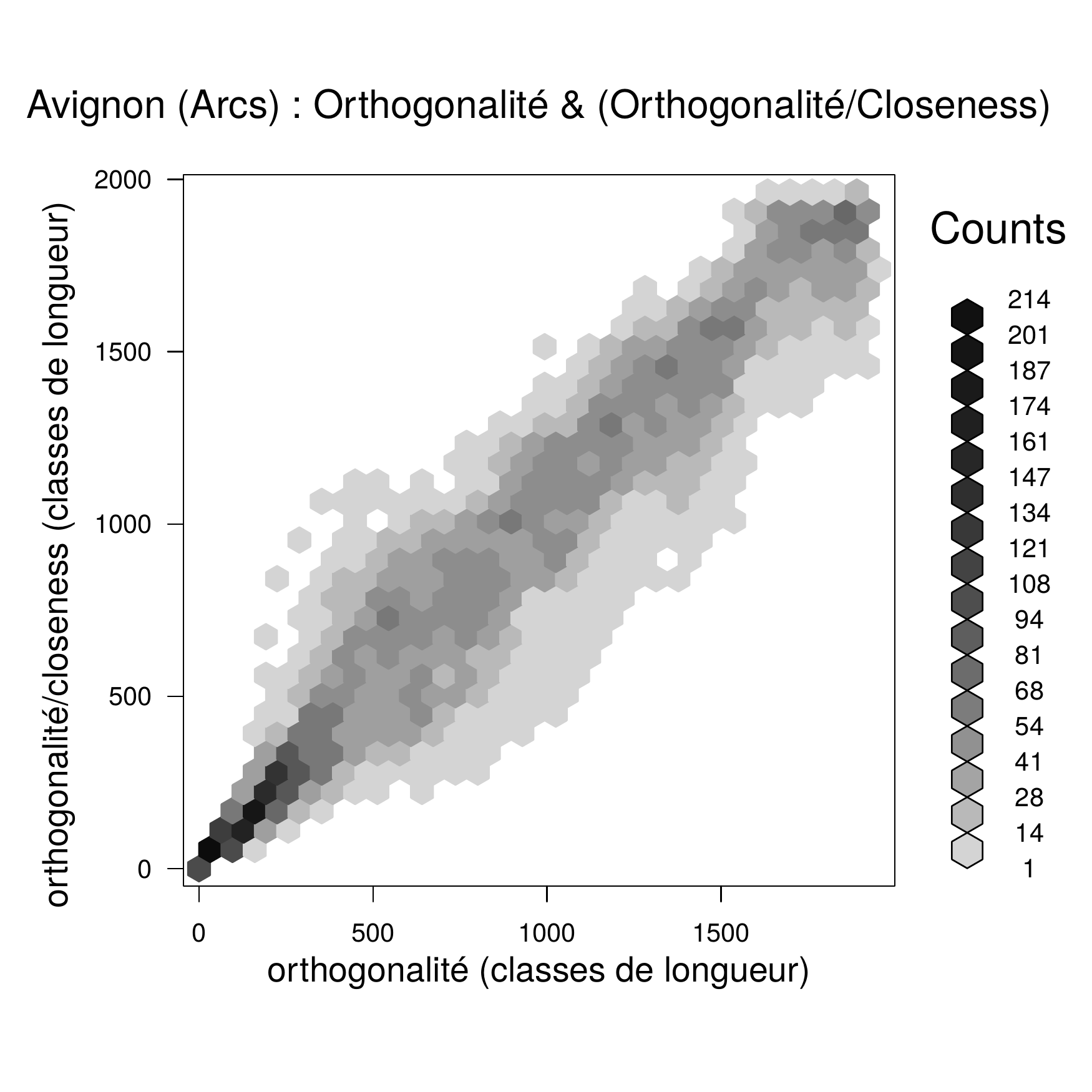}
    \end{subfigure}
    ~
    \begin{subfigure}[t]{0.45\textwidth}
        \includegraphics[width=\linewidth]{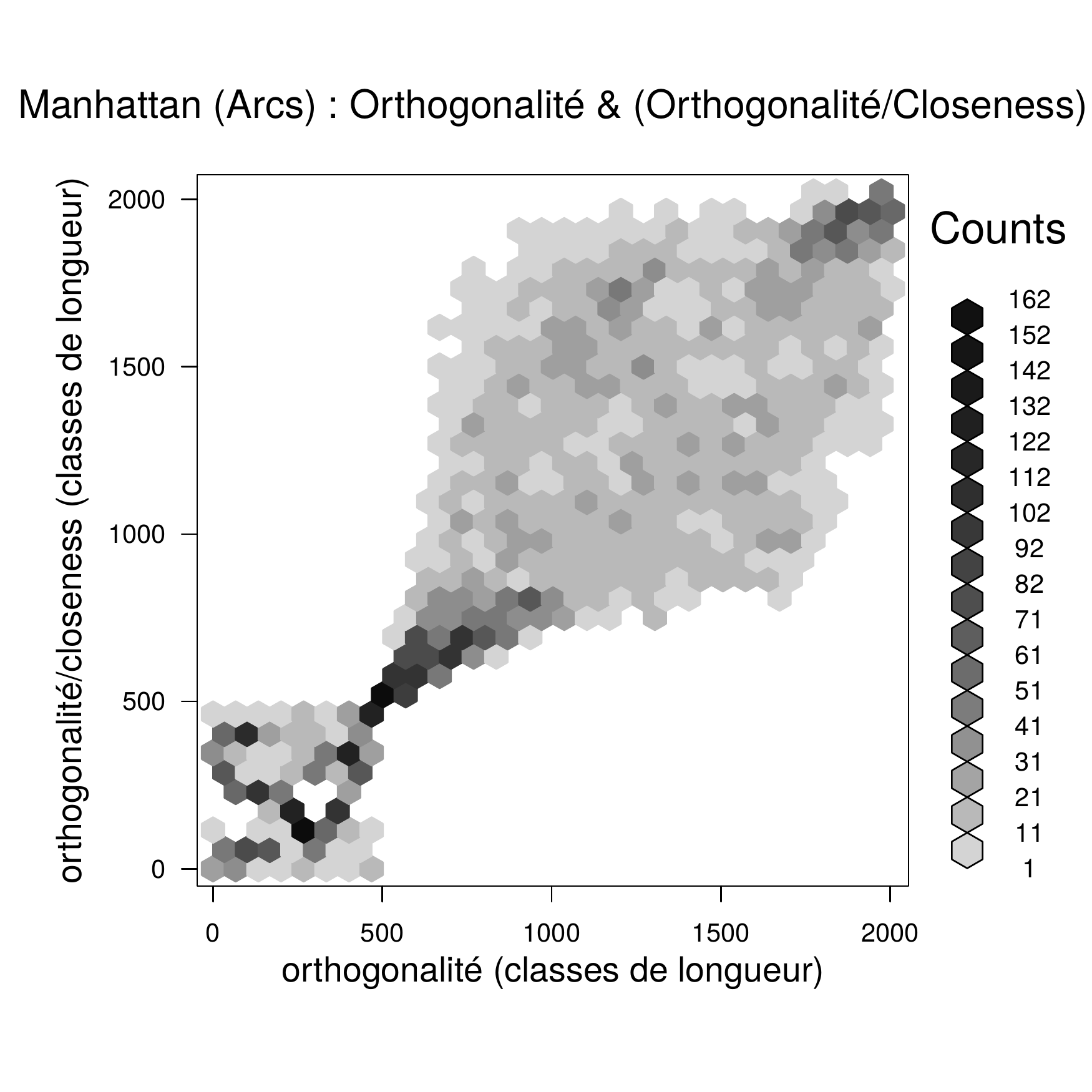}
    \end{subfigure}
    \caption{Orthogonalité et orthogonalité sur closeness}
\end{figure}

\begin{figure}[h]\centering
    \begin{subfigure}[t]{0.45\textwidth}
        \includegraphics[width=\linewidth]{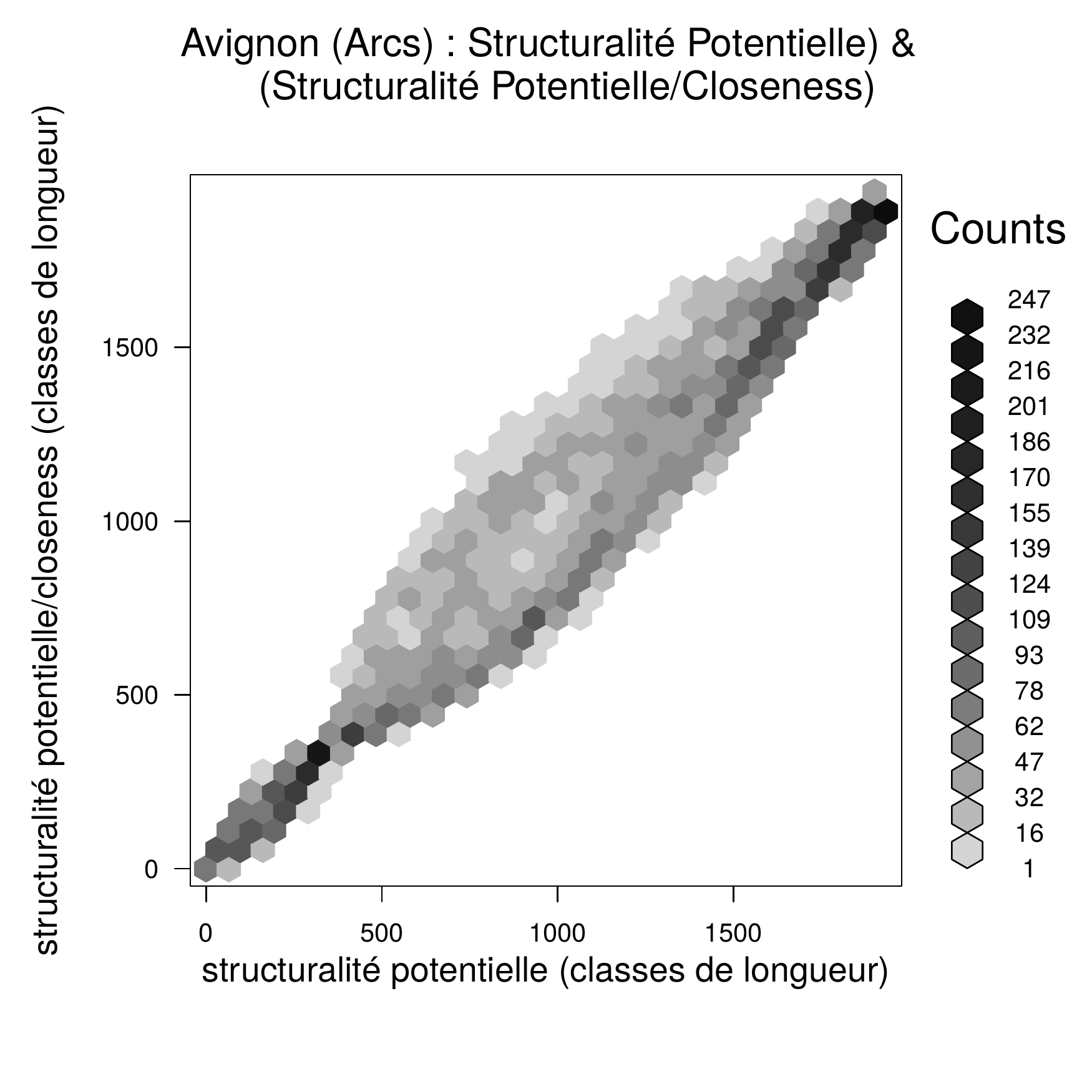}
    \end{subfigure}
    ~
    \begin{subfigure}[t]{0.45\textwidth}
        \includegraphics[width=\linewidth]{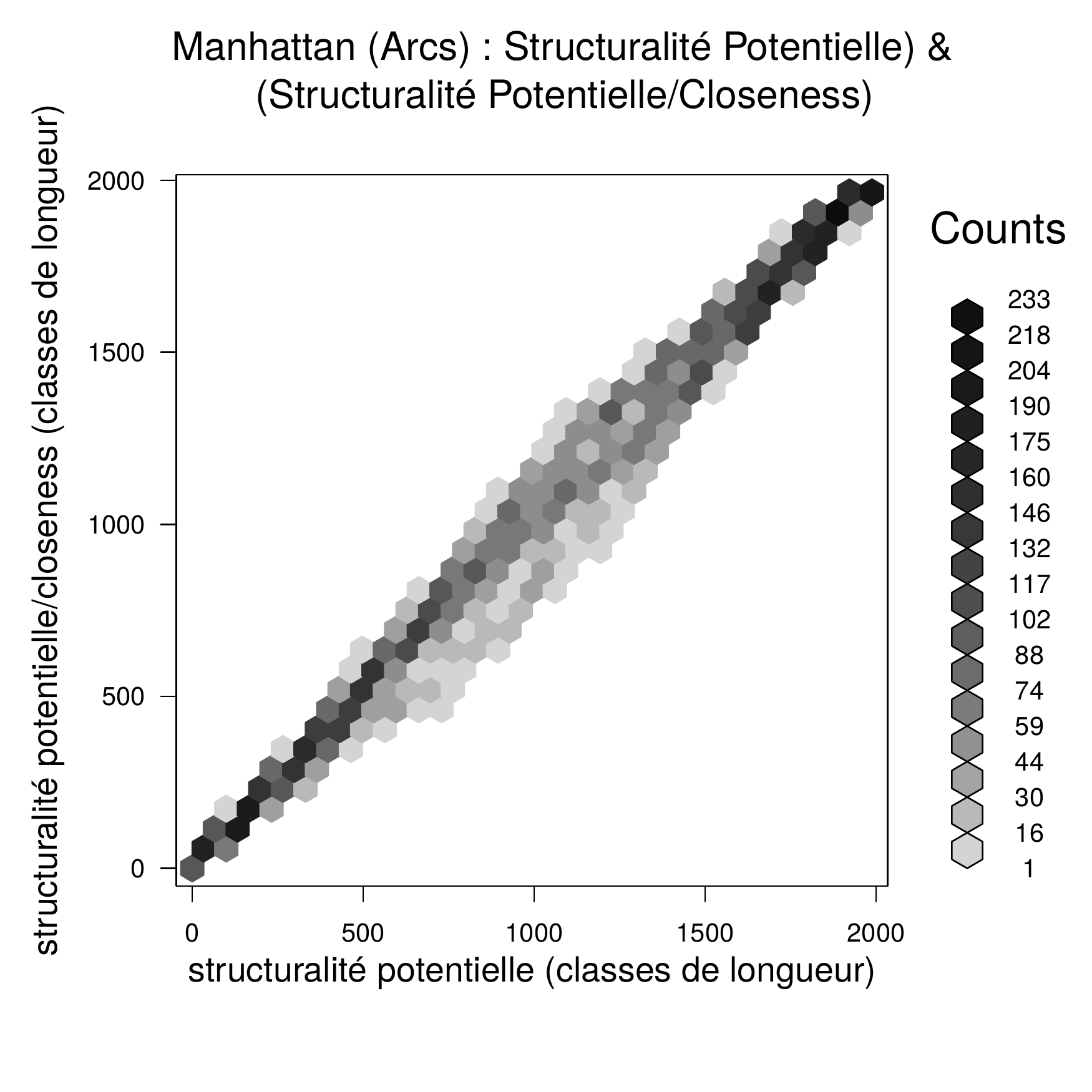}
    \end{subfigure}
    \caption{Structuralité potentielle et soc.pdf}
\end{figure}

\begin{figure}[h]\centering
    \begin{subfigure}[t]{0.45\textwidth}
        \includegraphics[width=\linewidth]{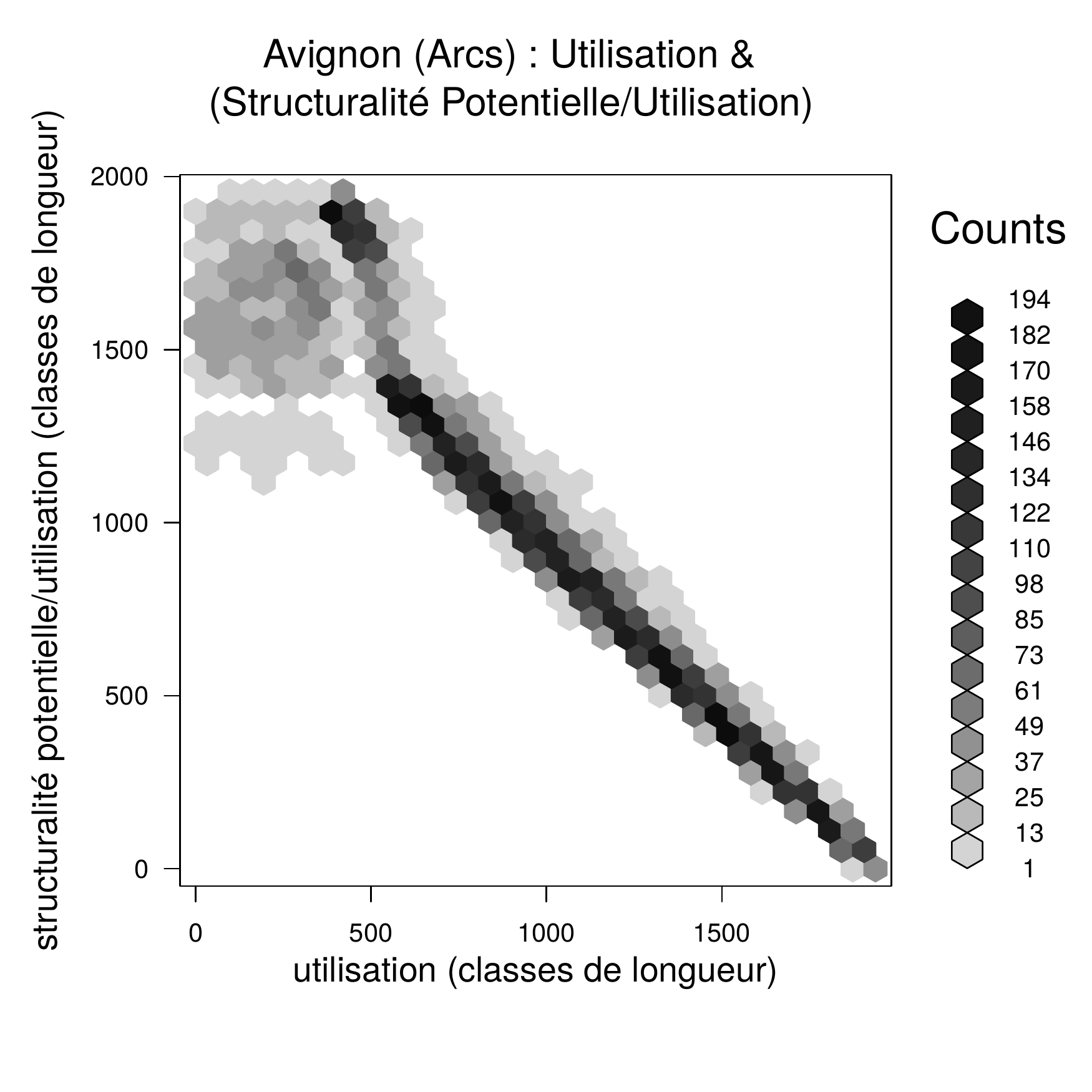}
    \end{subfigure}
    ~
    \begin{subfigure}[t]{0.45\textwidth}
        \includegraphics[width=\linewidth]{images/cartes_hexbin/ind_comp/arcs_Manhattan_use_sou.pdf}
    \end{subfigure}
    \caption{Utilisation et structuralité sur utilisation}
\end{figure}

\begin{figure}[h]\centering
    \begin{subfigure}[t]{0.45\textwidth}
        \includegraphics[width=\linewidth]{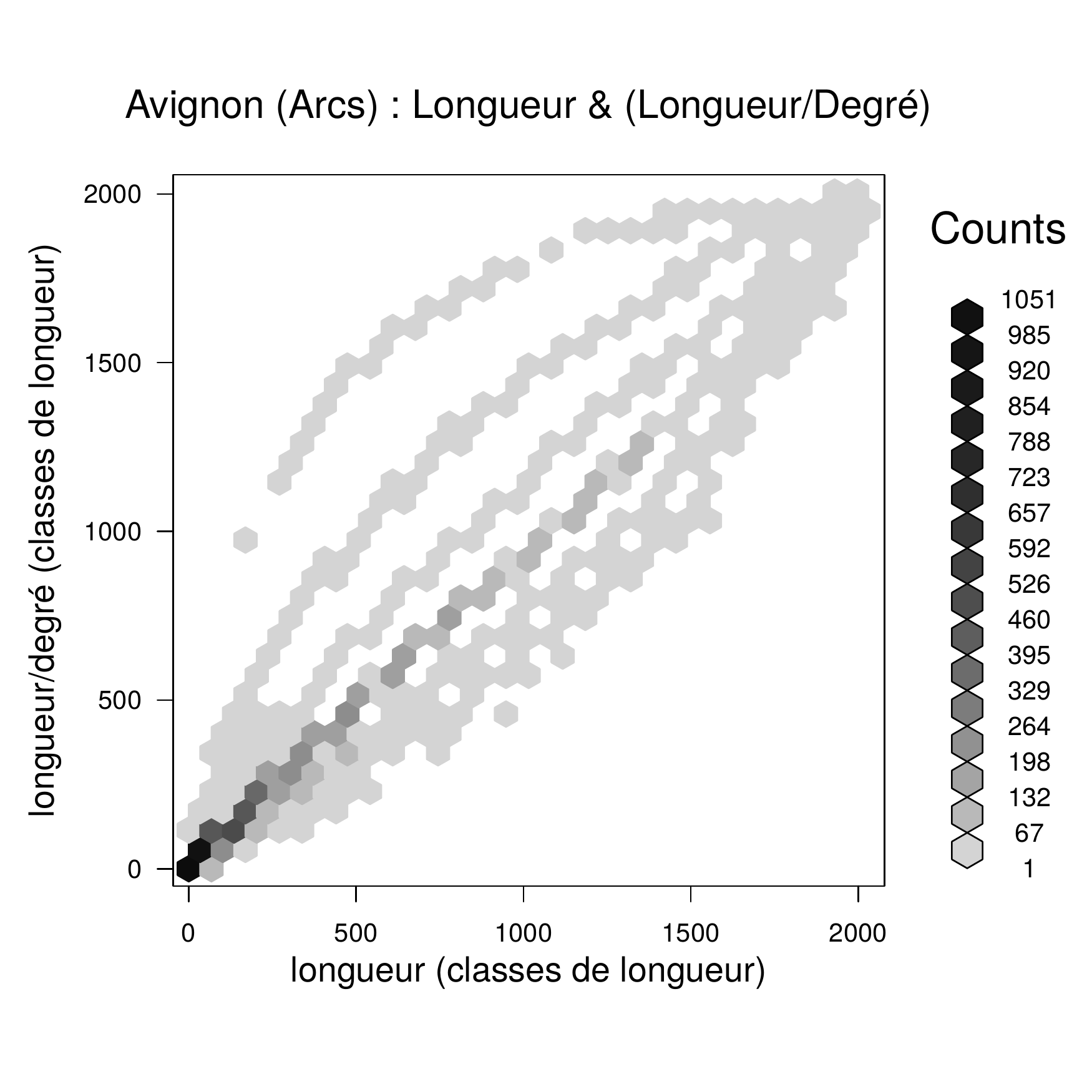}
    \end{subfigure}
    ~
    \begin{subfigure}[t]{0.45\textwidth}
        \includegraphics[width=\linewidth]{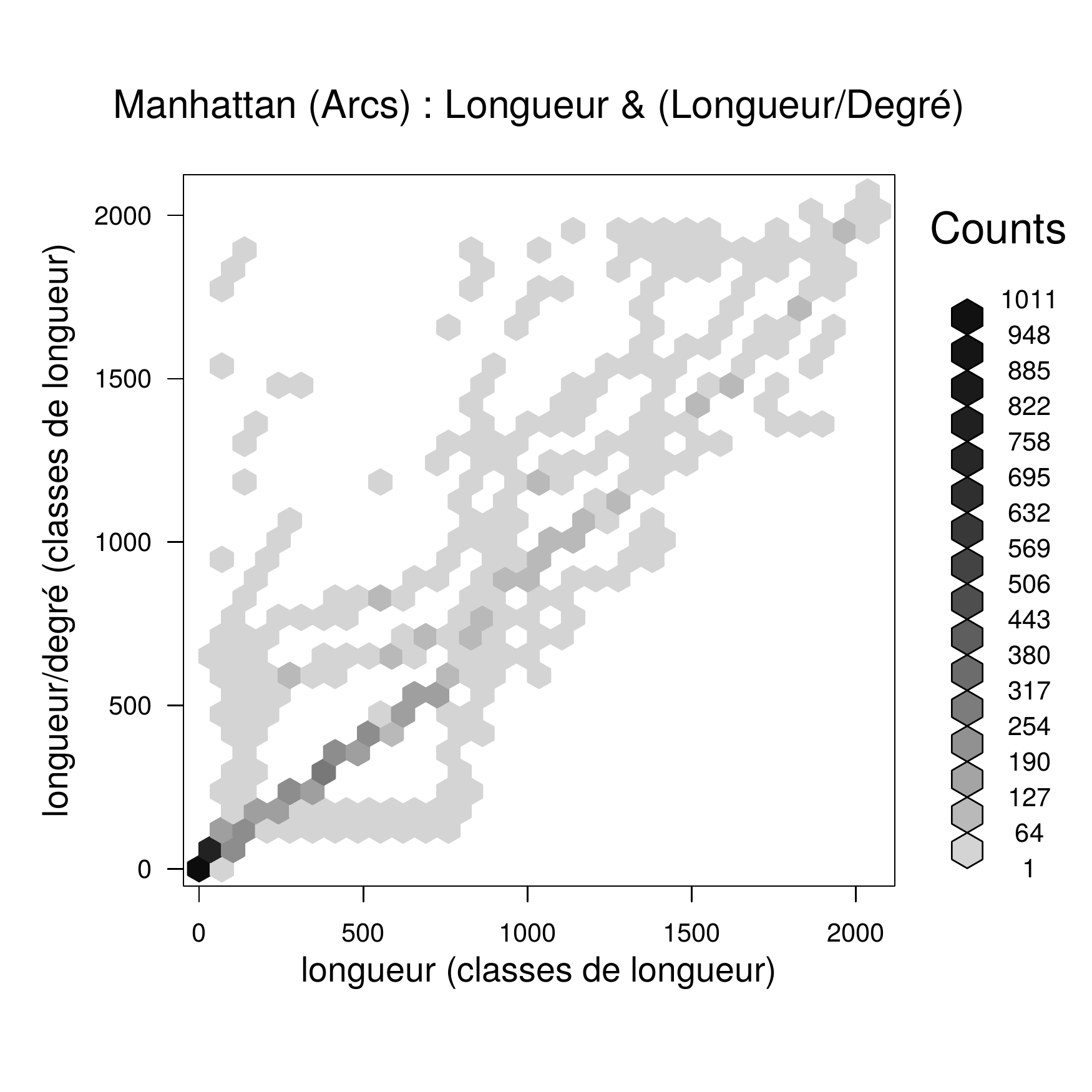}
    \end{subfigure}
    \caption{Longueur et longueur sur degré}
\end{figure}

\begin{figure}[h]\centering
    \begin{subfigure}[t]{0.45\textwidth}
        \includegraphics[width=\linewidth]{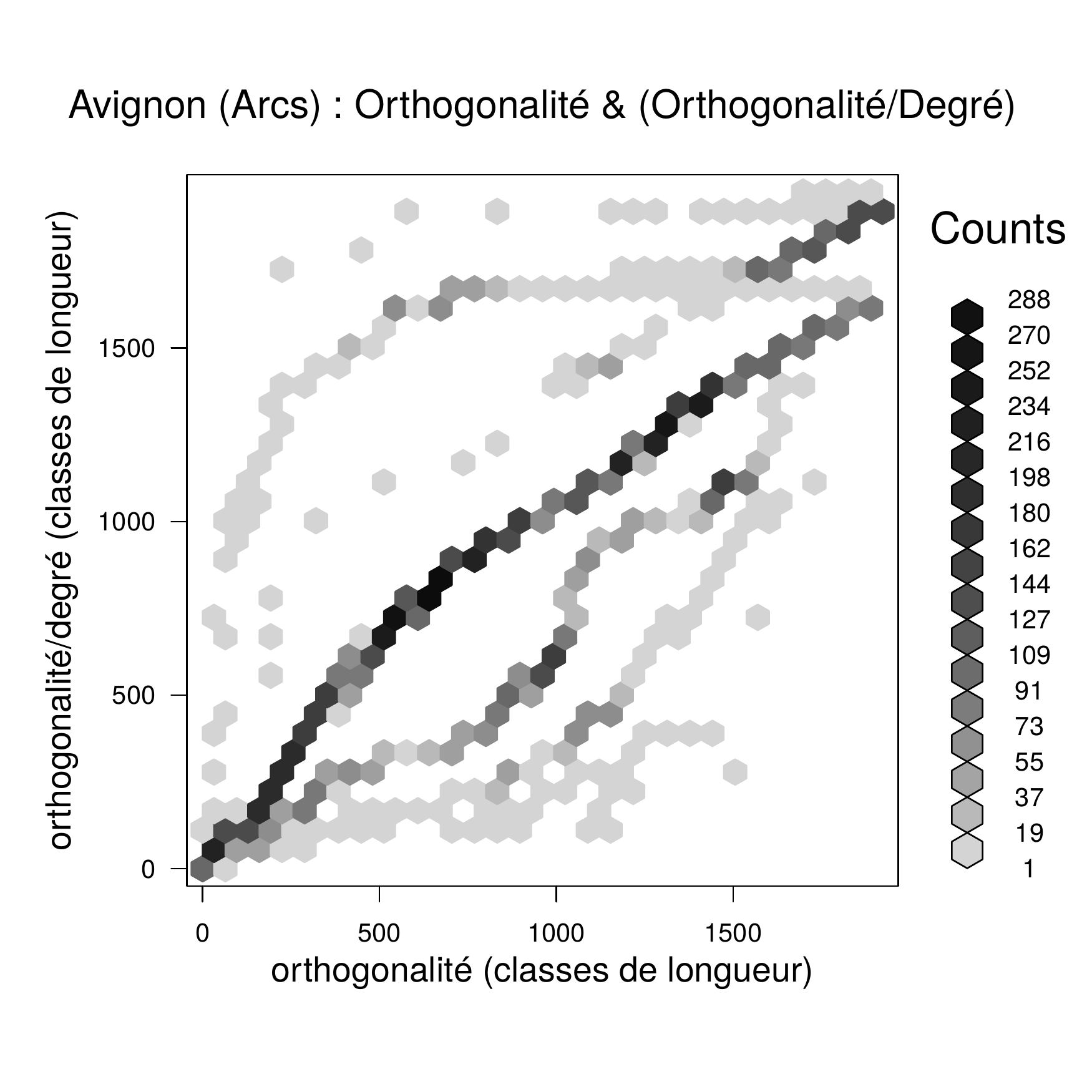}
    \end{subfigure}
    ~
    \begin{subfigure}[t]{0.45\textwidth}
        \includegraphics[width=\linewidth]{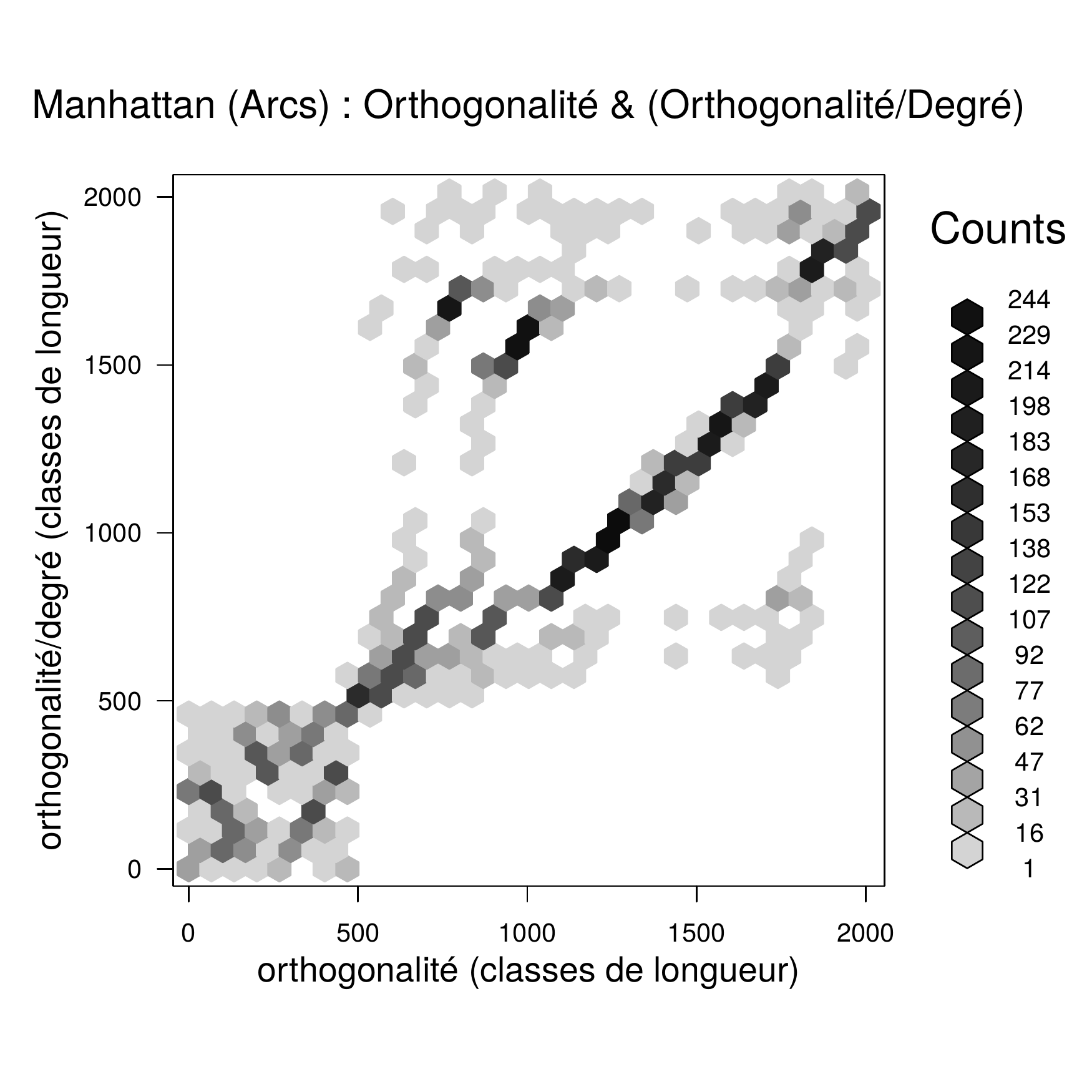}
    \end{subfigure}
    \caption{Orthogonalité et orthogonalité sur degré}
\end{figure}

\begin{figure}[h]\centering
    \begin{subfigure}[t]{0.45\textwidth}
        \includegraphics[width=\linewidth]{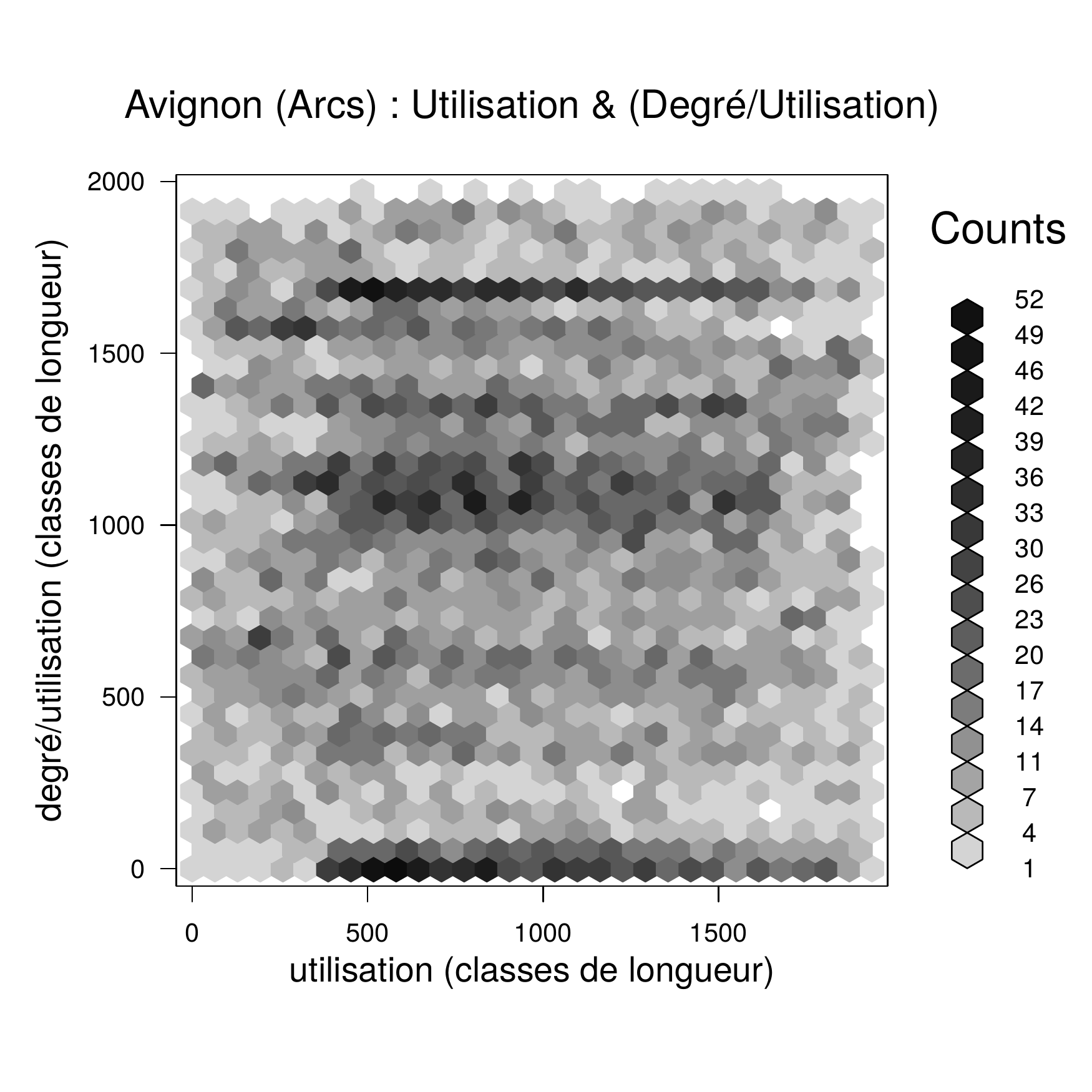}
    \end{subfigure}
    ~
    \begin{subfigure}[t]{0.45\textwidth}
        \includegraphics[width=\linewidth]{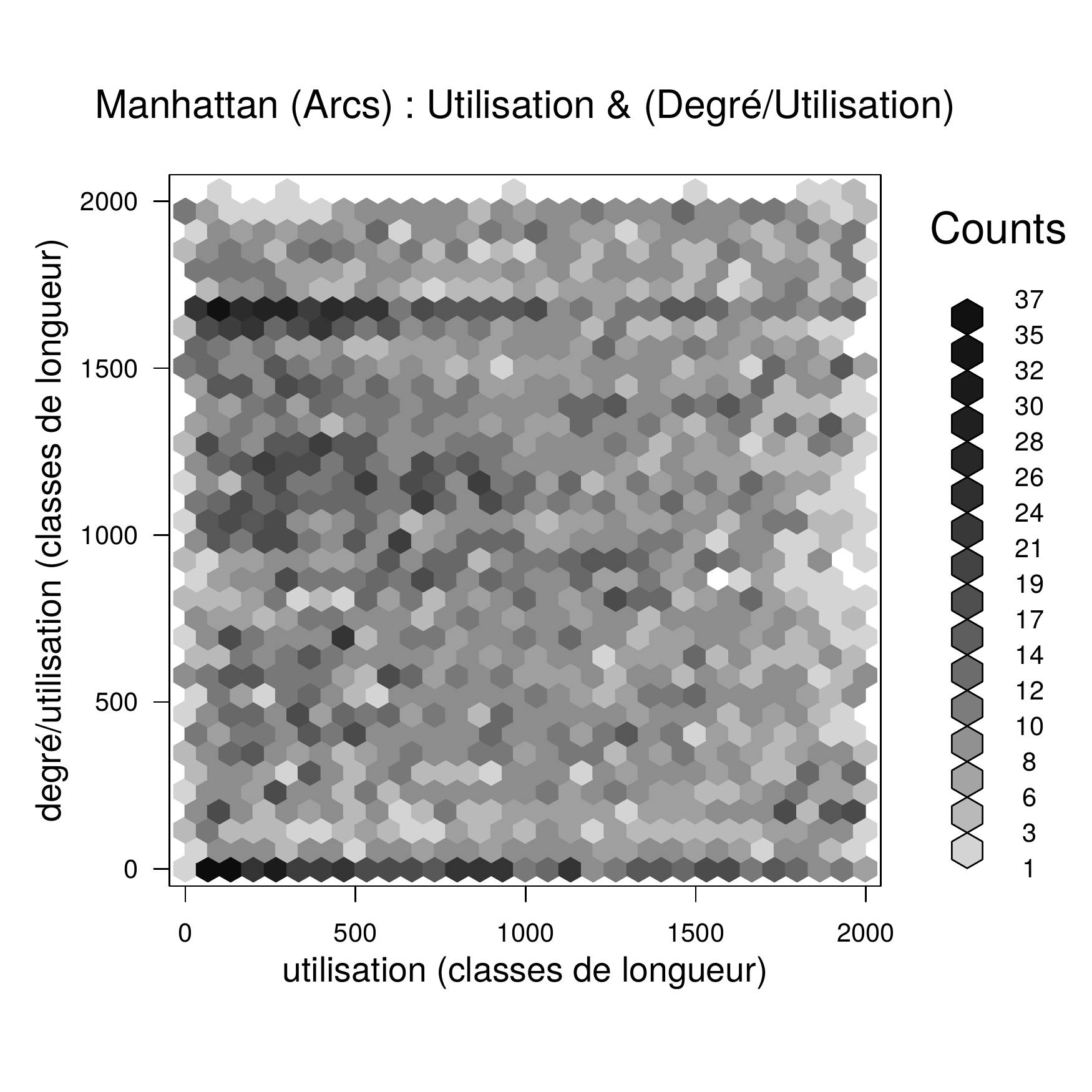}
    \end{subfigure}
    \caption{Utilisation et degré sur utilisation}
\end{figure}

\begin{figure}[h]\centering
    \begin{subfigure}[t]{0.45\textwidth}
        \includegraphics[width=\linewidth]{images/cartes_hexbin/ind_comp/arcs_Avignon_use_uoc.pdf}
    \end{subfigure}
    ~
    \begin{subfigure}[t]{0.45\textwidth}
        \includegraphics[width=\linewidth]{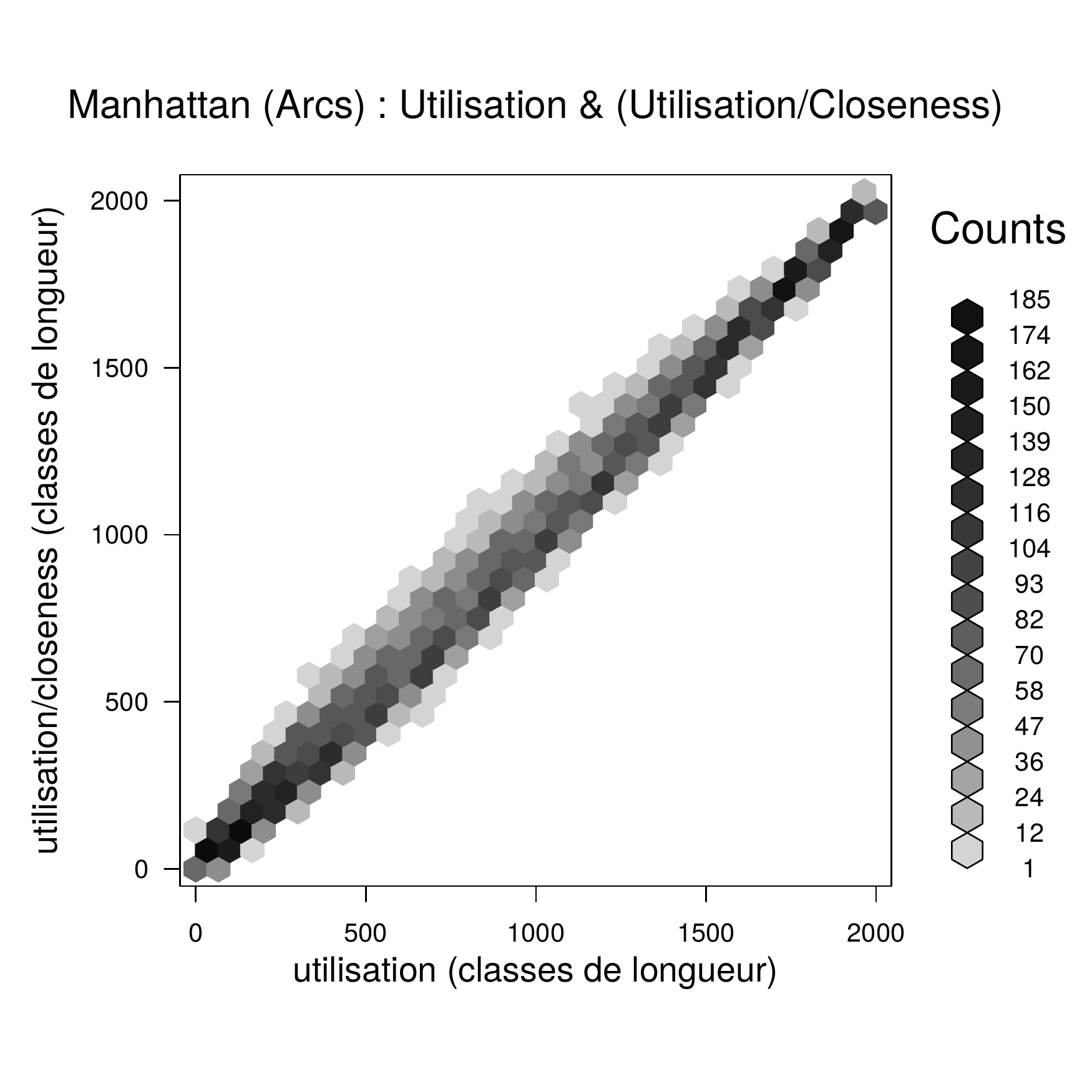}
    \end{subfigure}
    \caption{Utilisation et $\frac{utilisation}{closeness}$}
\end{figure}

\FloatBarrier 
\section{Cartes de corrélation sur les voies}\label{A14_hexb_voies}\label{ann:sec_corr_voies} 

\subsection{Indicateurs primaires}\label{ann:ssec_voies_indprim}

\begin{figure}[h]\centering
    \begin{subfigure}[t]{0.45\textwidth}
        \includegraphics[width=\linewidth]{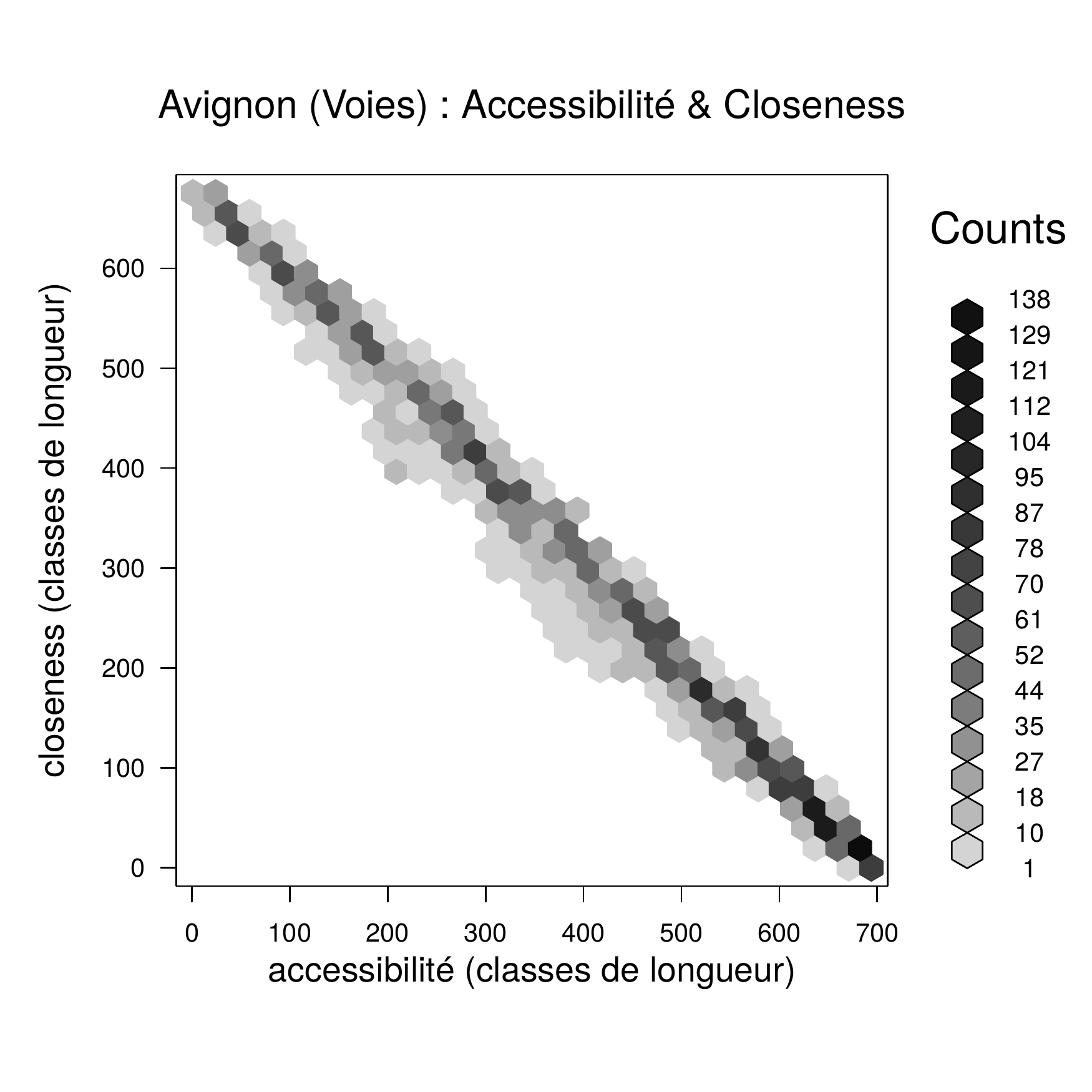}
    \end{subfigure}
    ~
    \begin{subfigure}[t]{0.45\textwidth}
        \includegraphics[width=\linewidth]{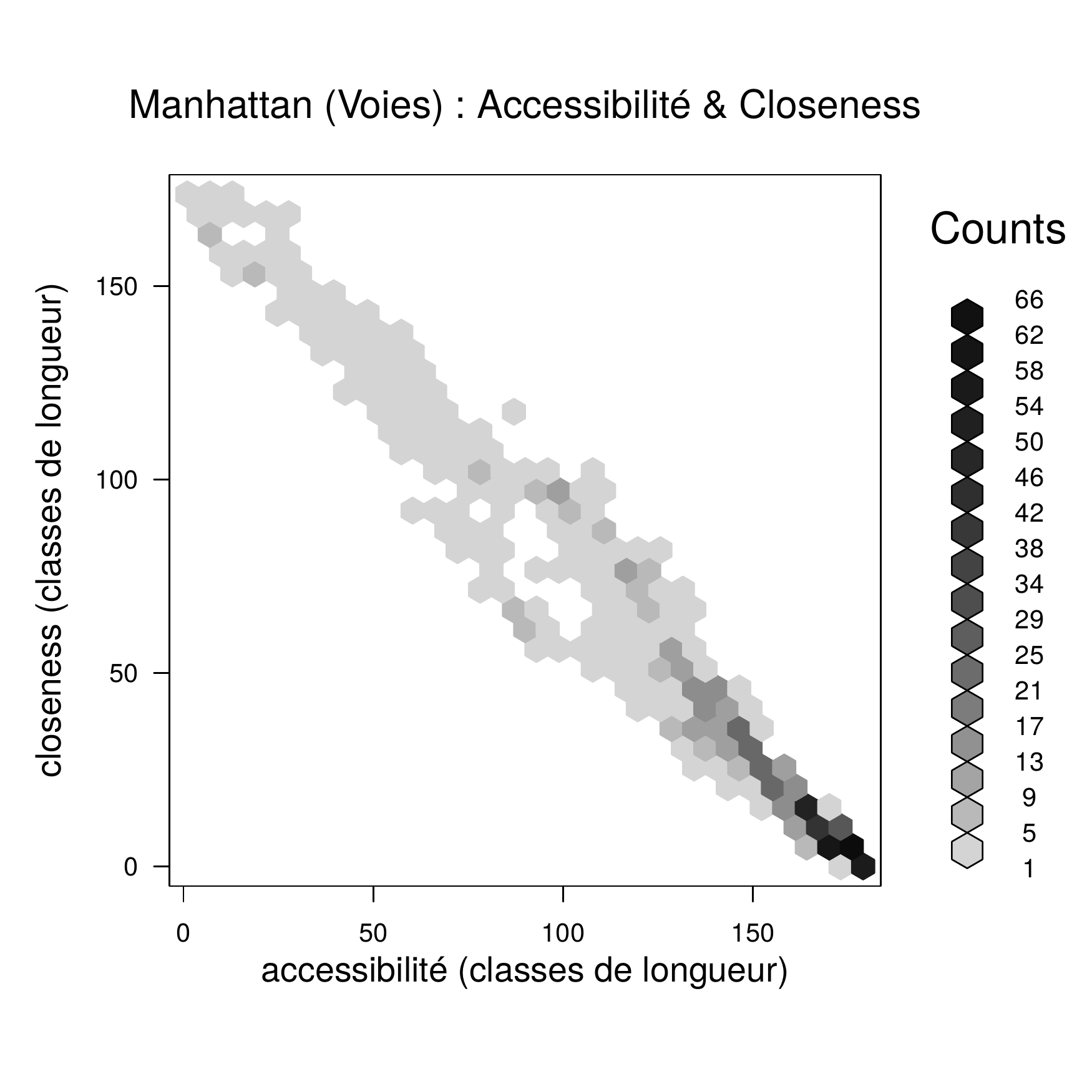}
    \end{subfigure}

    \begin{subfigure}[t]{0.45\textwidth}
        \includegraphics[width=\linewidth]{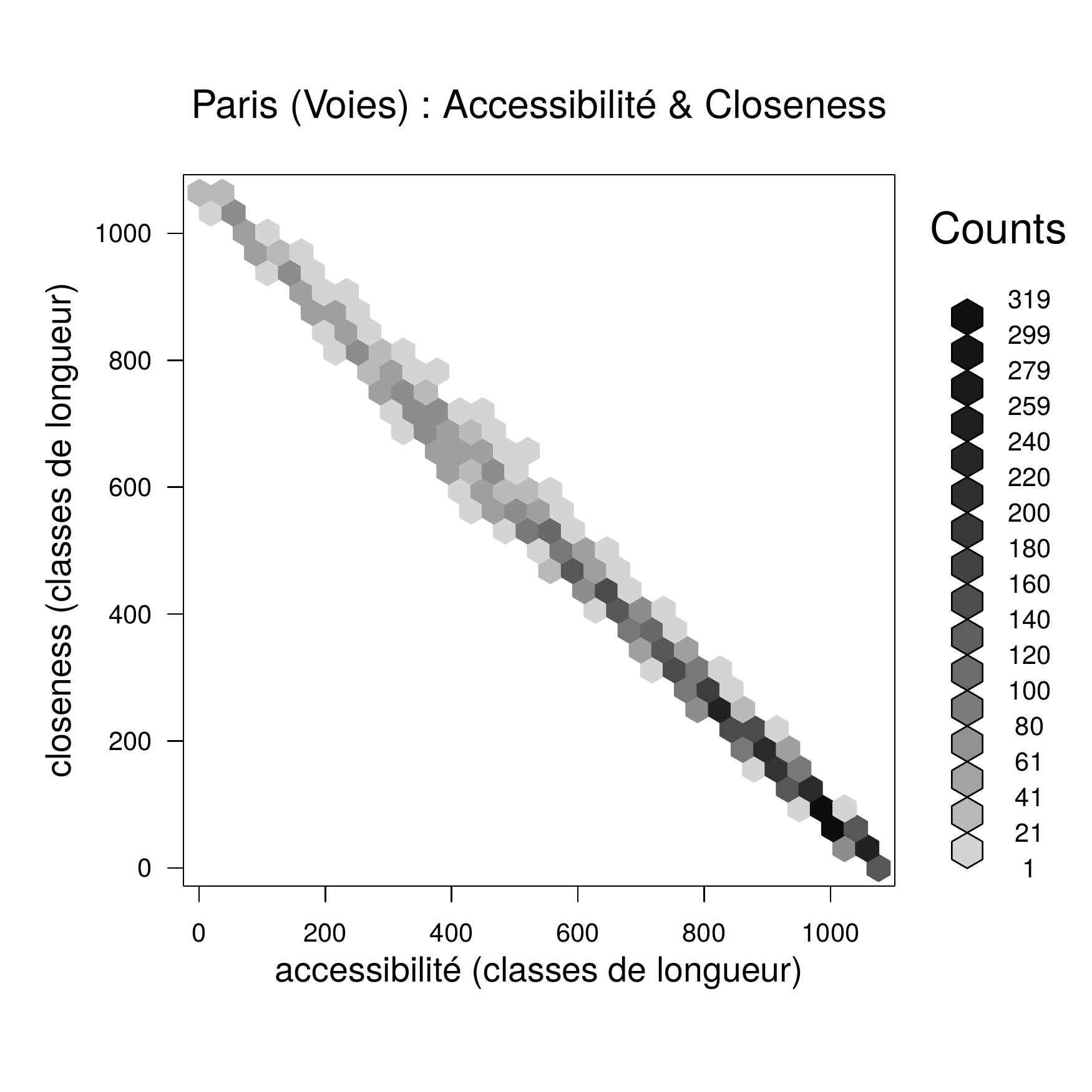}
    \end{subfigure}
    ~
    \begin{subfigure}[t]{0.45\textwidth}
        \includegraphics[width=\linewidth]{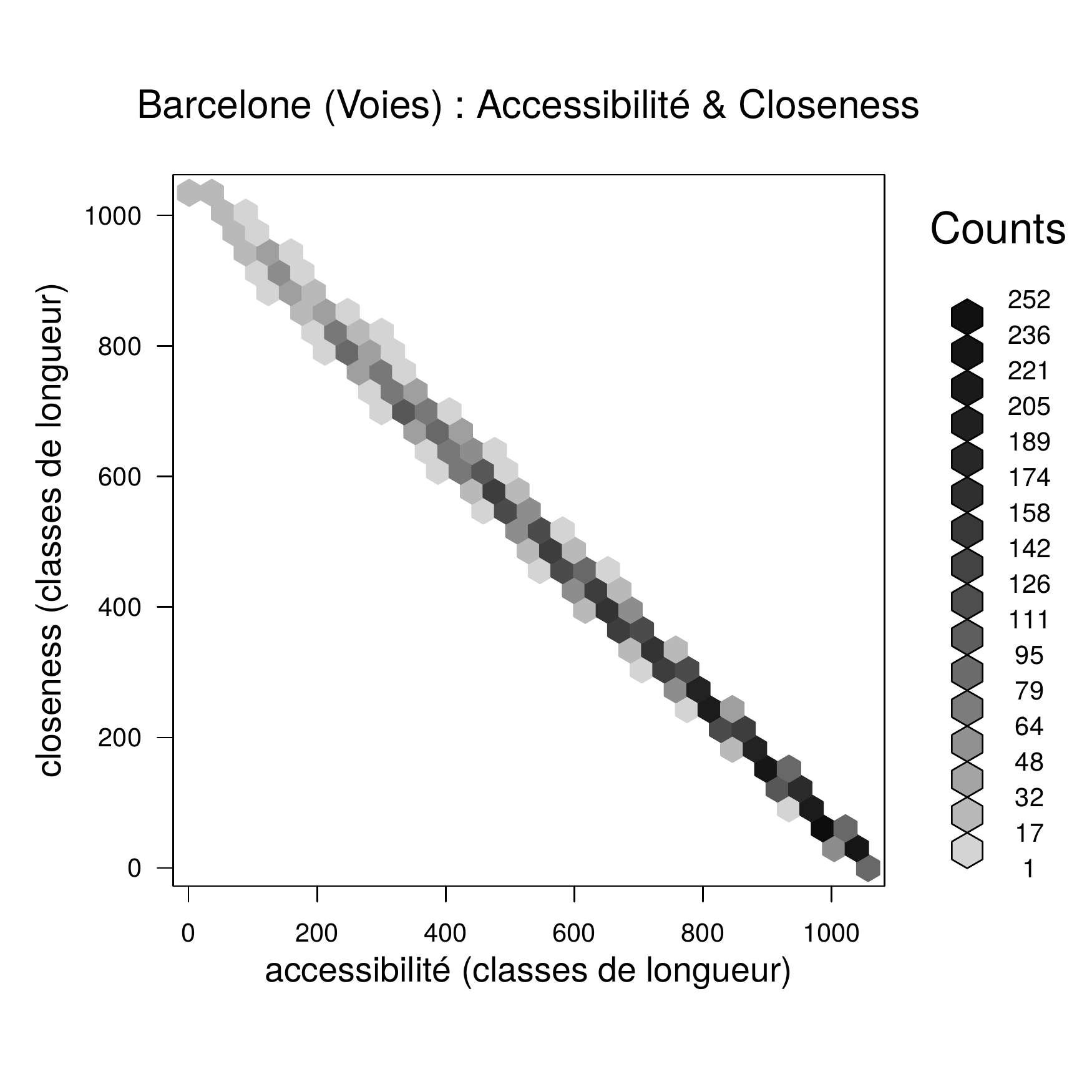}
    \end{subfigure}
    \caption{Accessibilité et closeness}
\end{figure}

\begin{figure}[h]\centering
    \begin{subfigure}[t]{0.45\textwidth}
        \includegraphics[width=\linewidth]{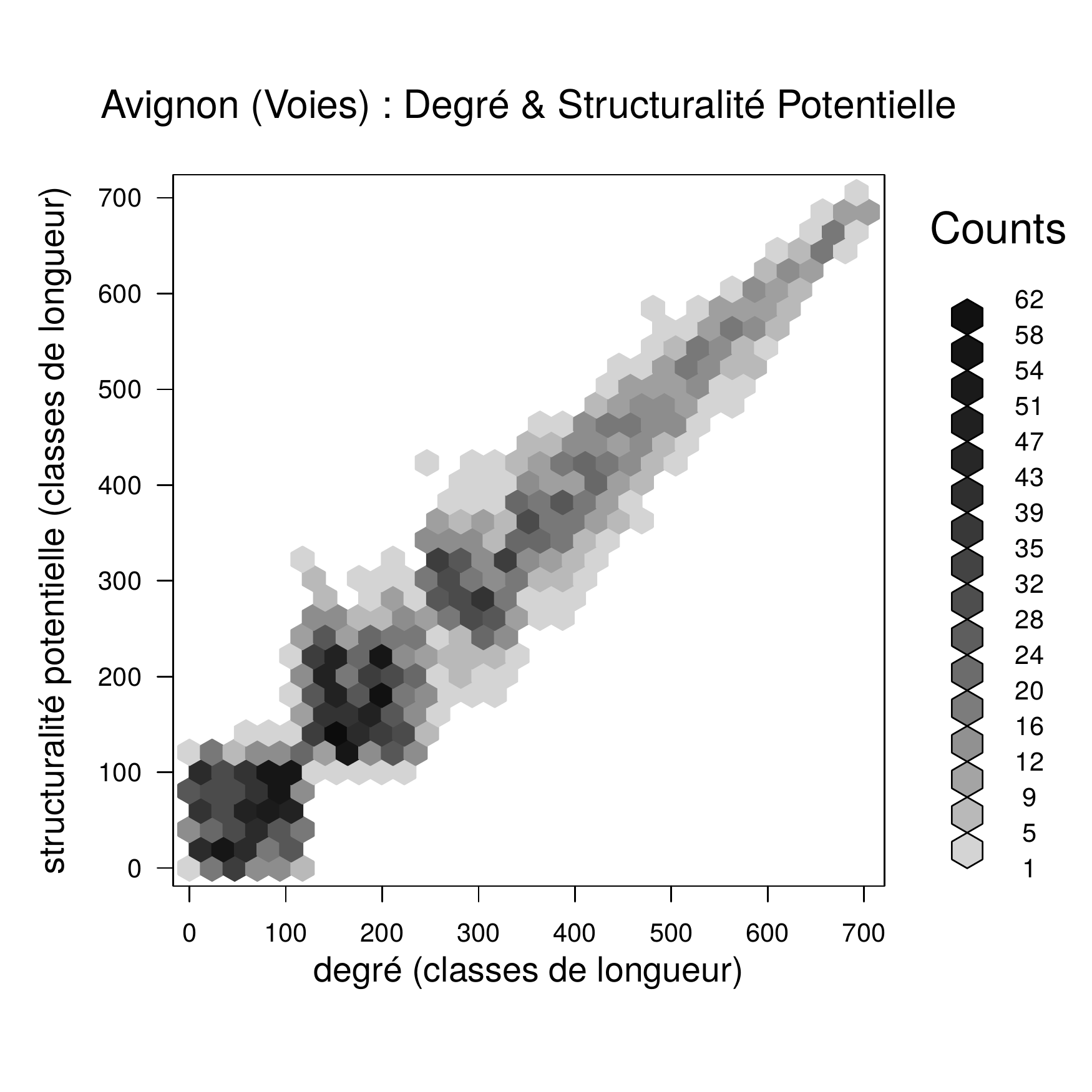}
    \end{subfigure}
    ~
    \begin{subfigure}[t]{0.45\textwidth}
        \includegraphics[width=\linewidth]{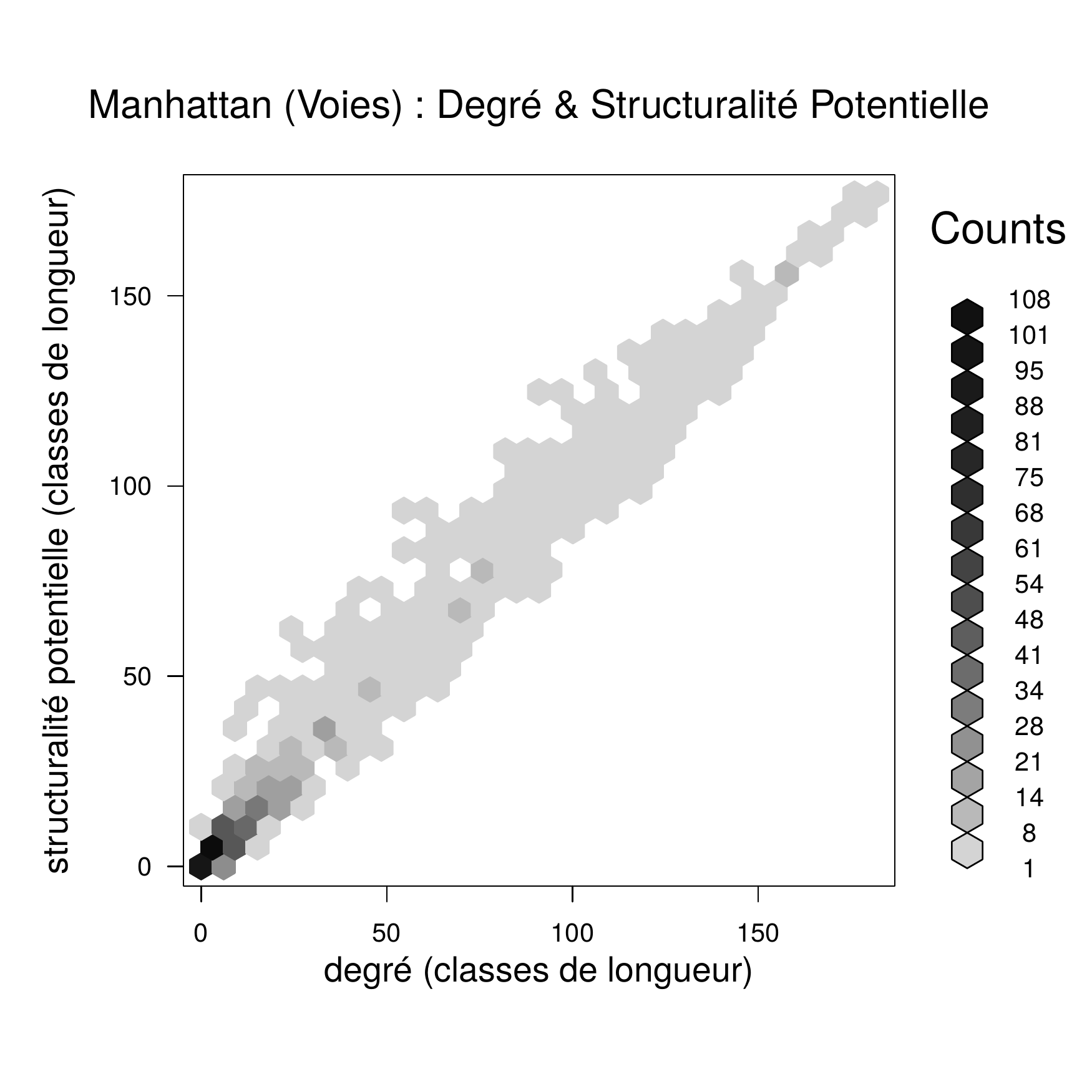}
    \end{subfigure}

    \begin{subfigure}[t]{0.45\textwidth}
        \includegraphics[width=\linewidth]{images/cartes_hexbin/paris_degree_structpot.pdf}
    \end{subfigure}
    ~
    \begin{subfigure}[t]{0.45\textwidth}
        \includegraphics[width=\linewidth]{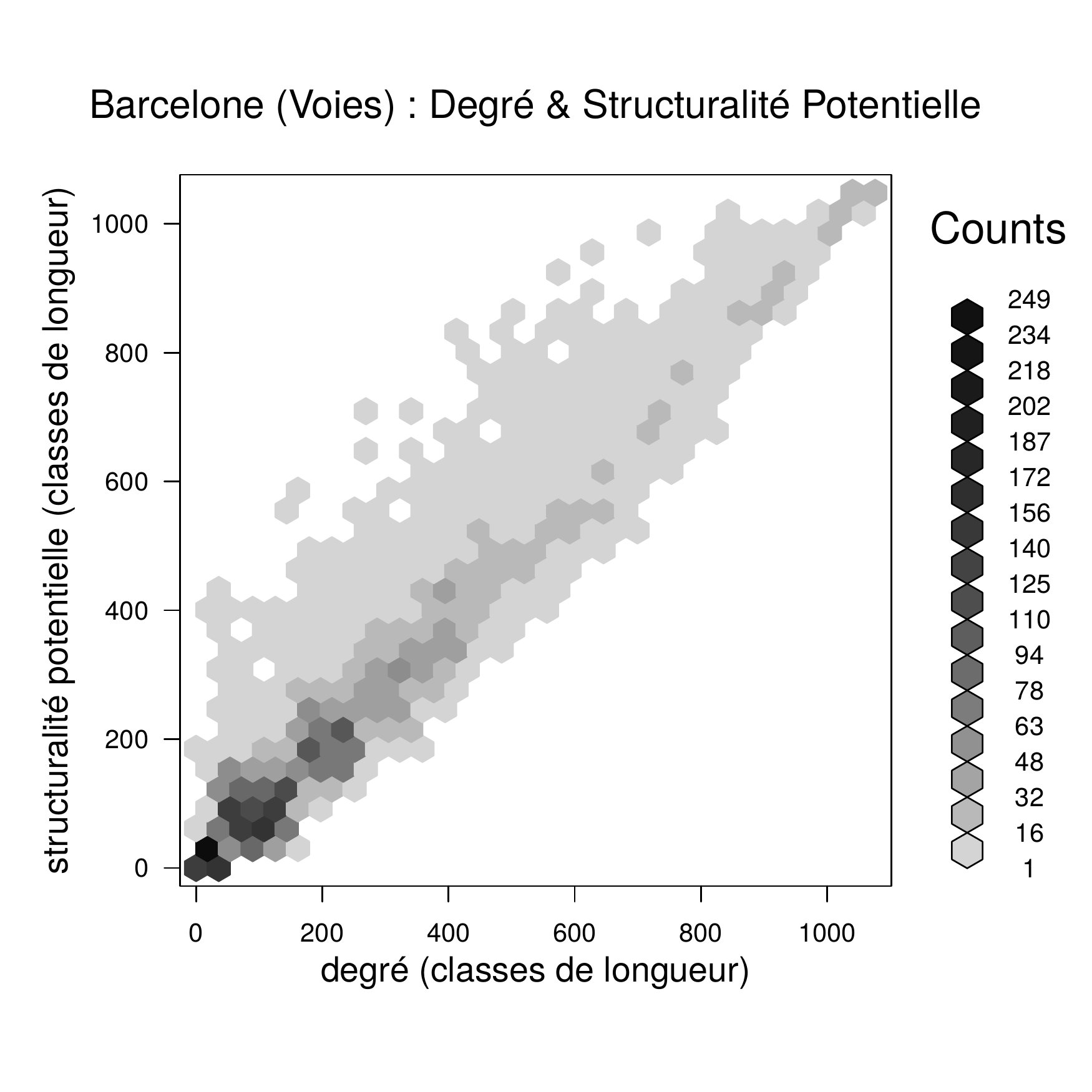}
    \end{subfigure}
    \caption{Degré et structuralité potentielle}
\end{figure}

\begin{figure}[h]\centering
    \begin{subfigure}[t]{0.45\textwidth}
        \includegraphics[width=\linewidth]{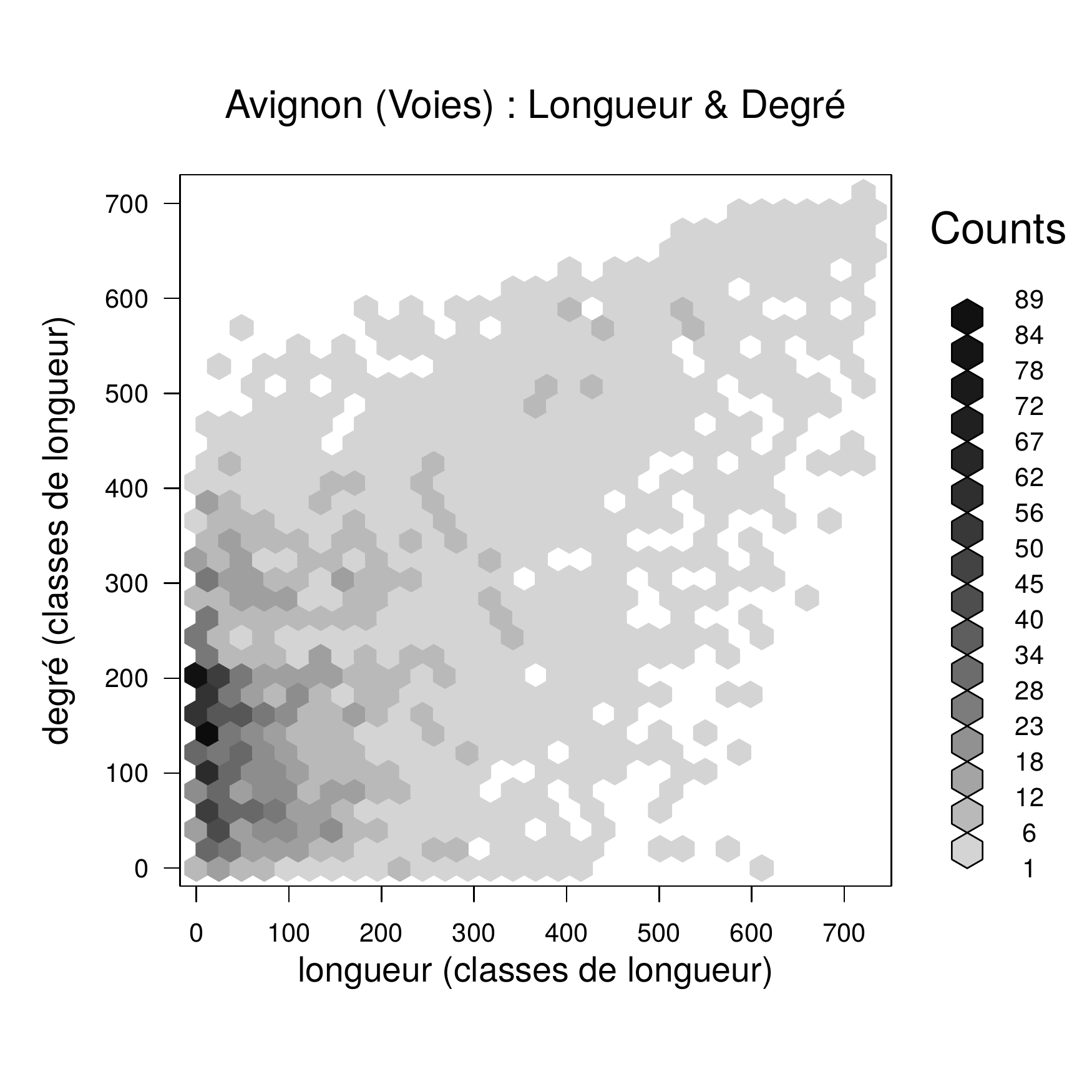}
    \end{subfigure}
    ~
    \begin{subfigure}[t]{0.45\textwidth}
        \includegraphics[width=\linewidth]{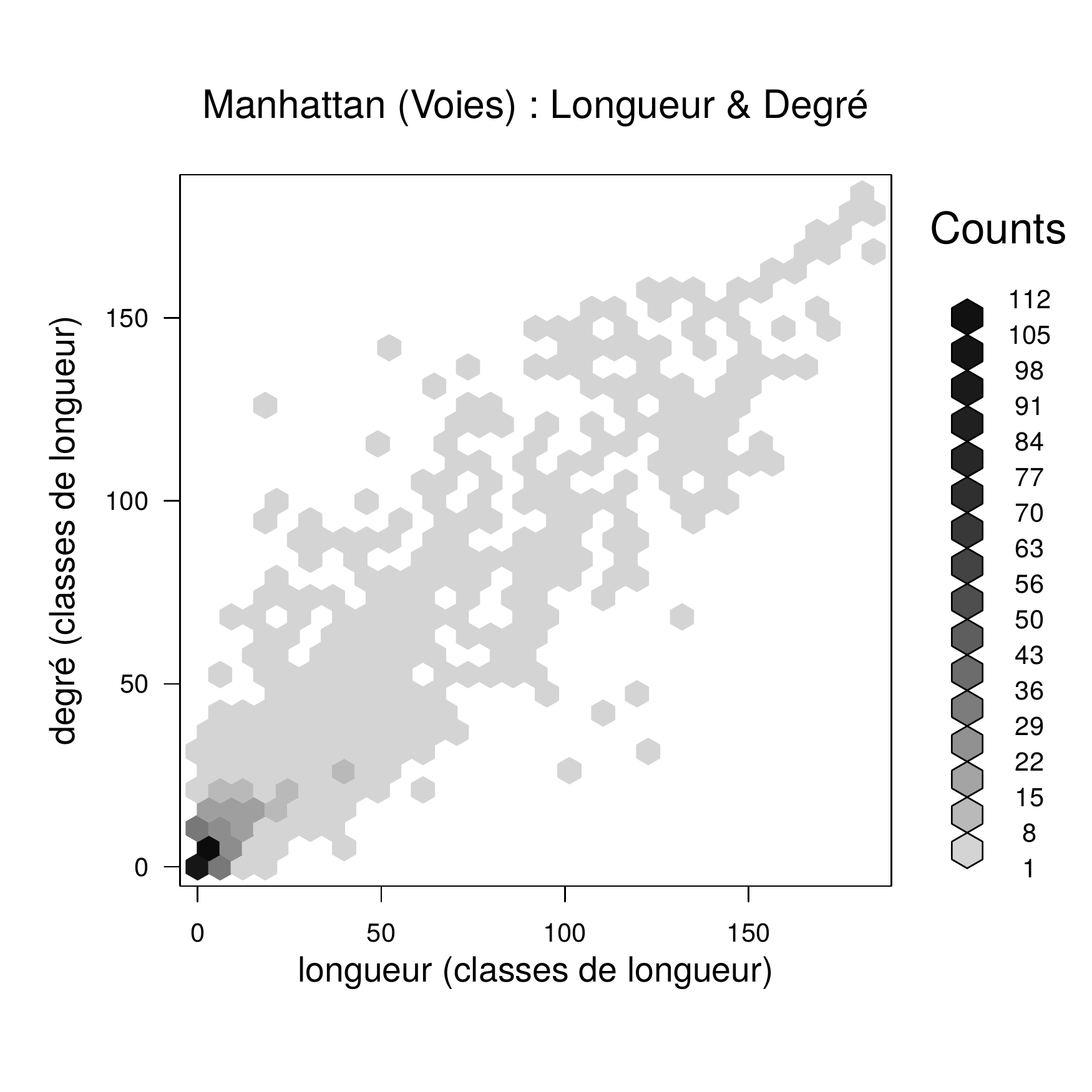}
    \end{subfigure}

    \begin{subfigure}[t]{0.45\textwidth}
        \includegraphics[width=\linewidth]{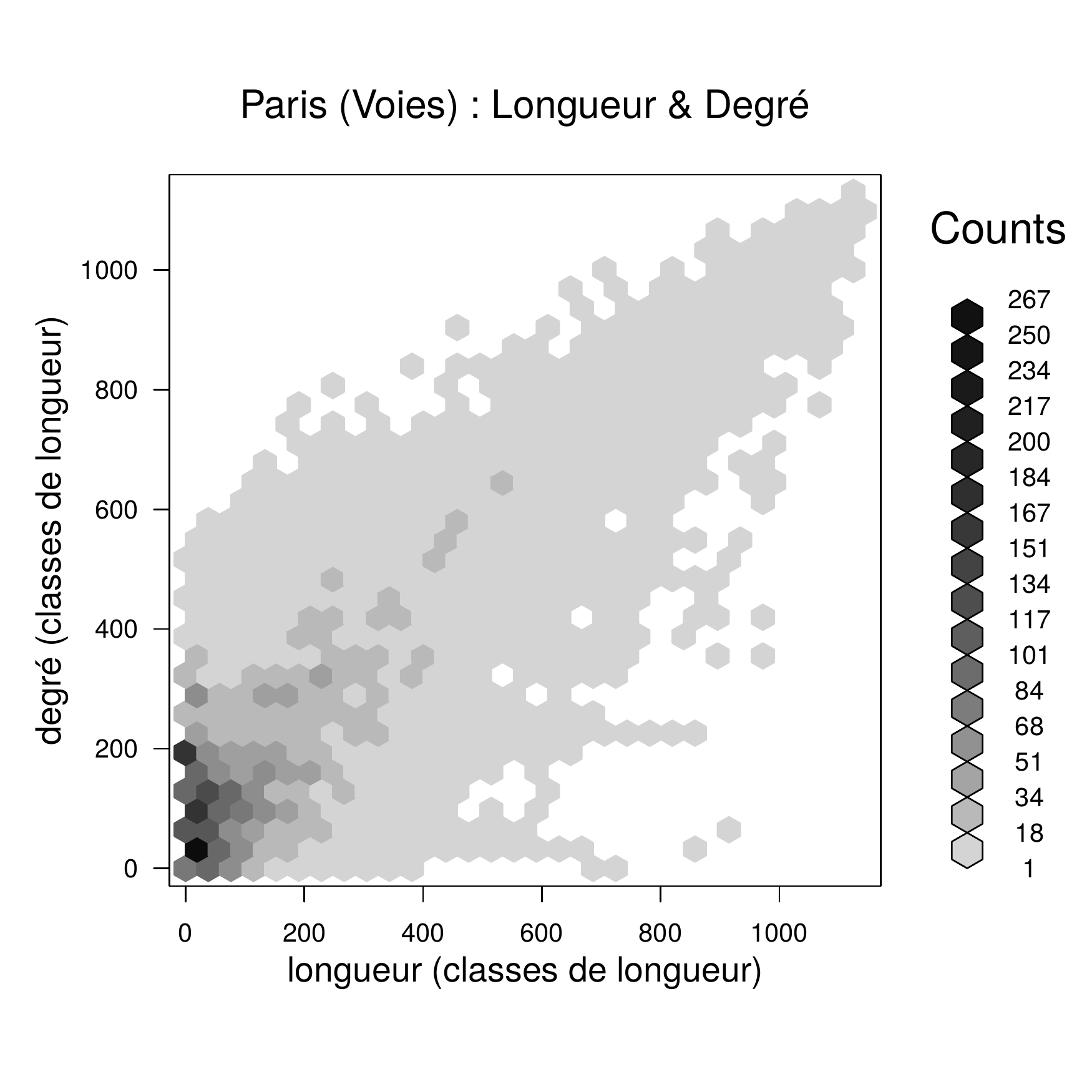}
    \end{subfigure}
    ~
    \begin{subfigure}[t]{0.45\textwidth}
        \includegraphics[width=\linewidth]{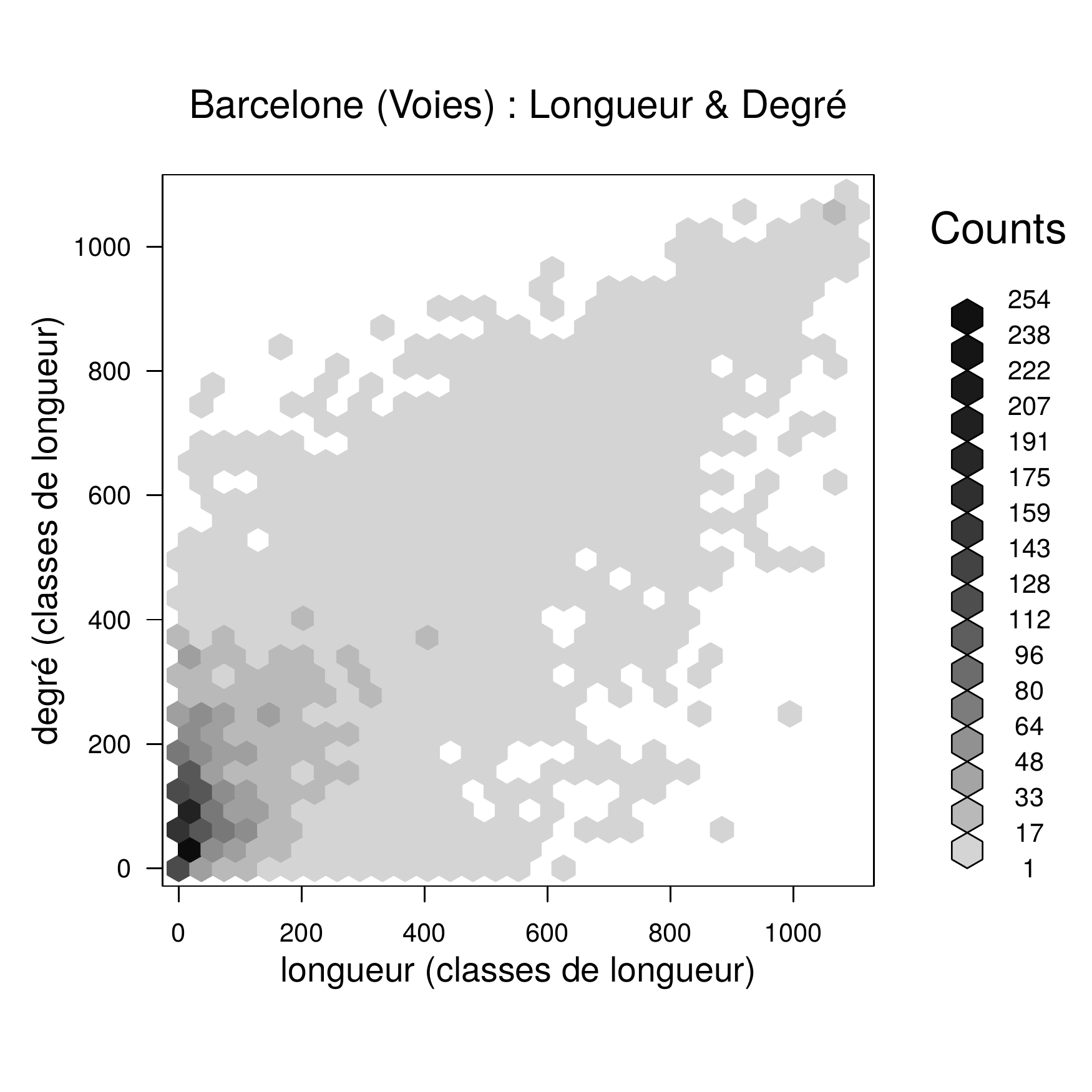}
    \end{subfigure}
    \caption{Longueur et degré}
\end{figure}

\begin{figure}[h]\centering
    \begin{subfigure}[t]{0.45\textwidth}
        \includegraphics[width=\linewidth]{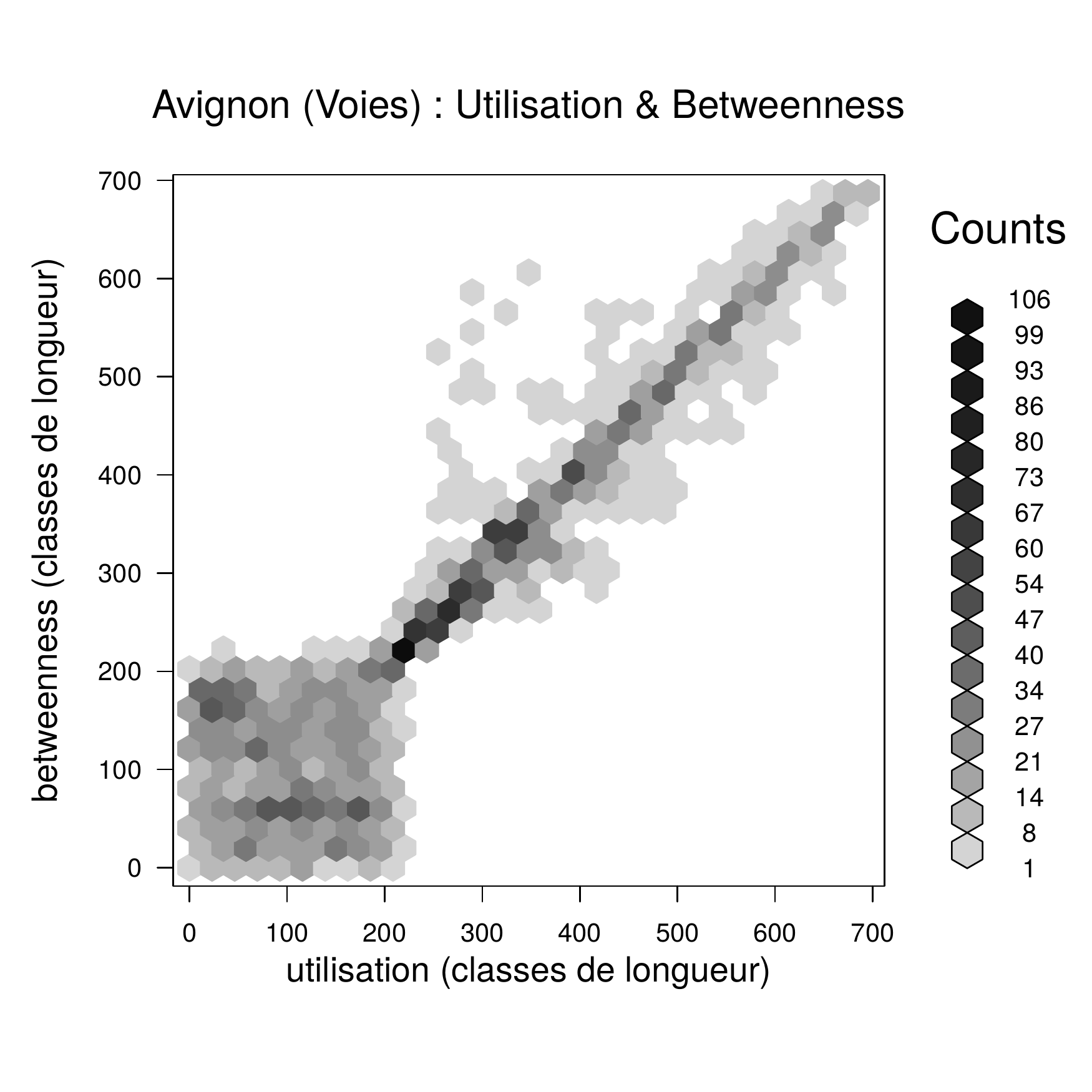}
    \end{subfigure}
    ~
    \begin{subfigure}[t]{0.45\textwidth}
        \includegraphics[width=\linewidth]{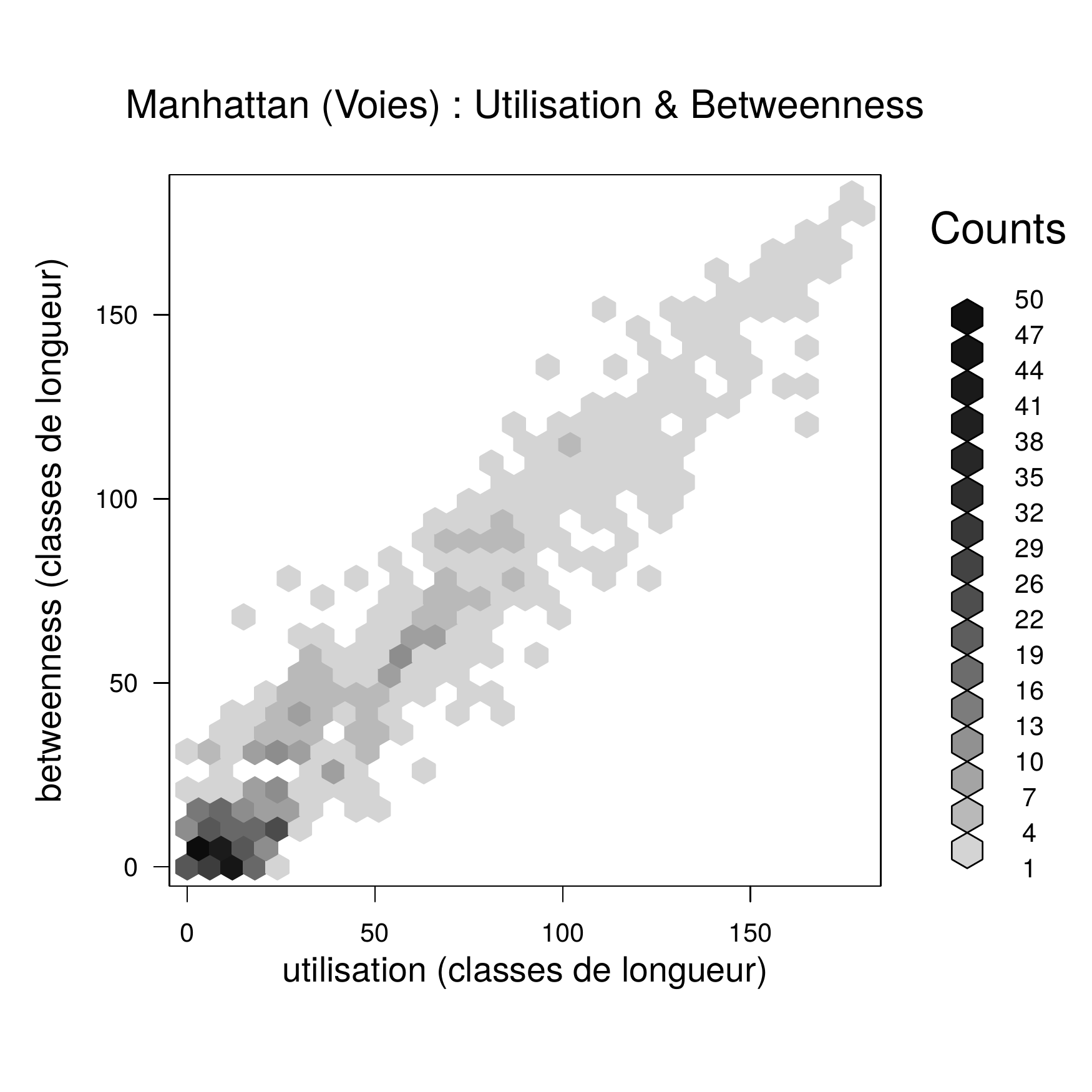}
    \end{subfigure}

    \begin{subfigure}[t]{0.45\textwidth}
        \includegraphics[width=\linewidth]{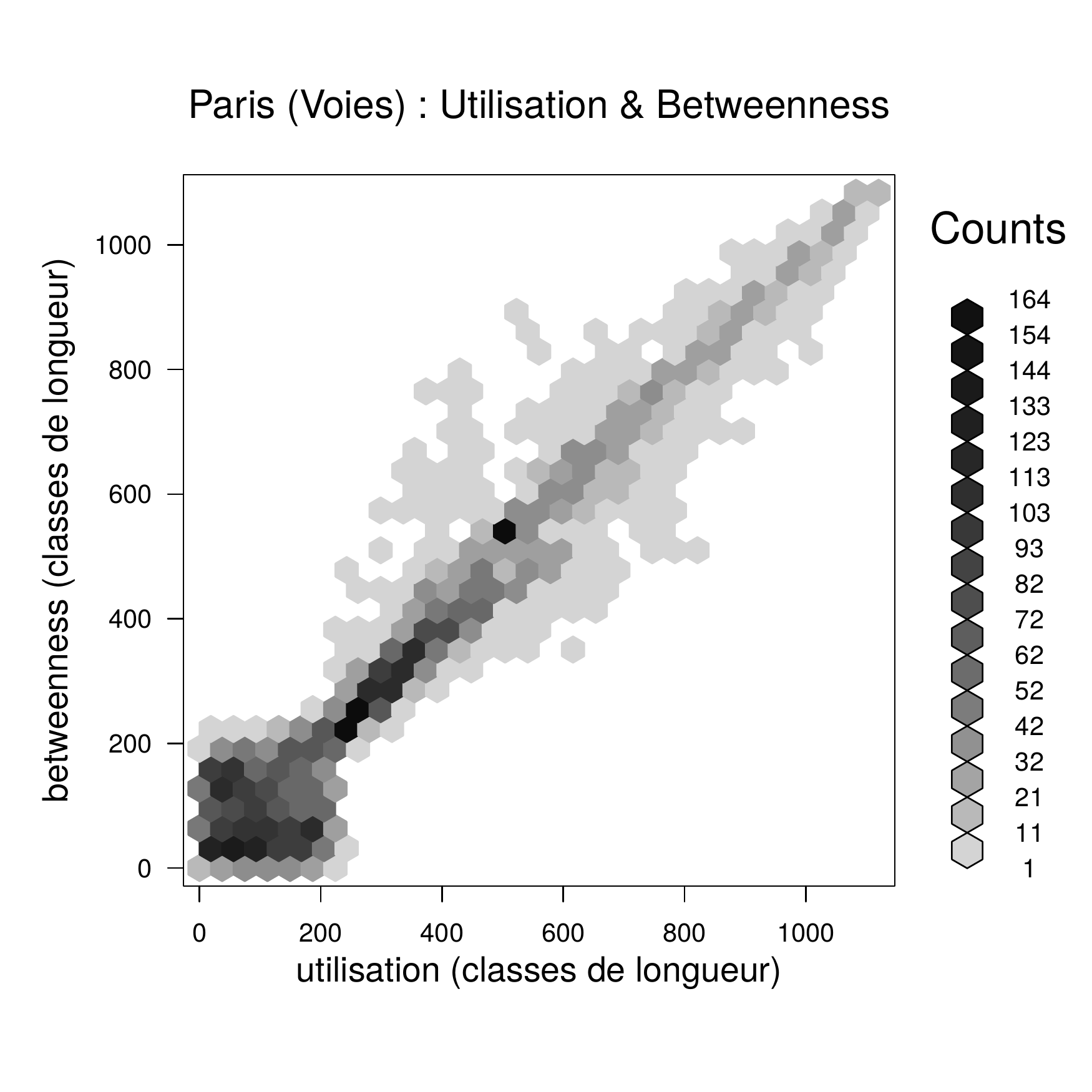}
    \end{subfigure}
    ~
    \begin{subfigure}[t]{0.45\textwidth}
        \includegraphics[width=\linewidth]{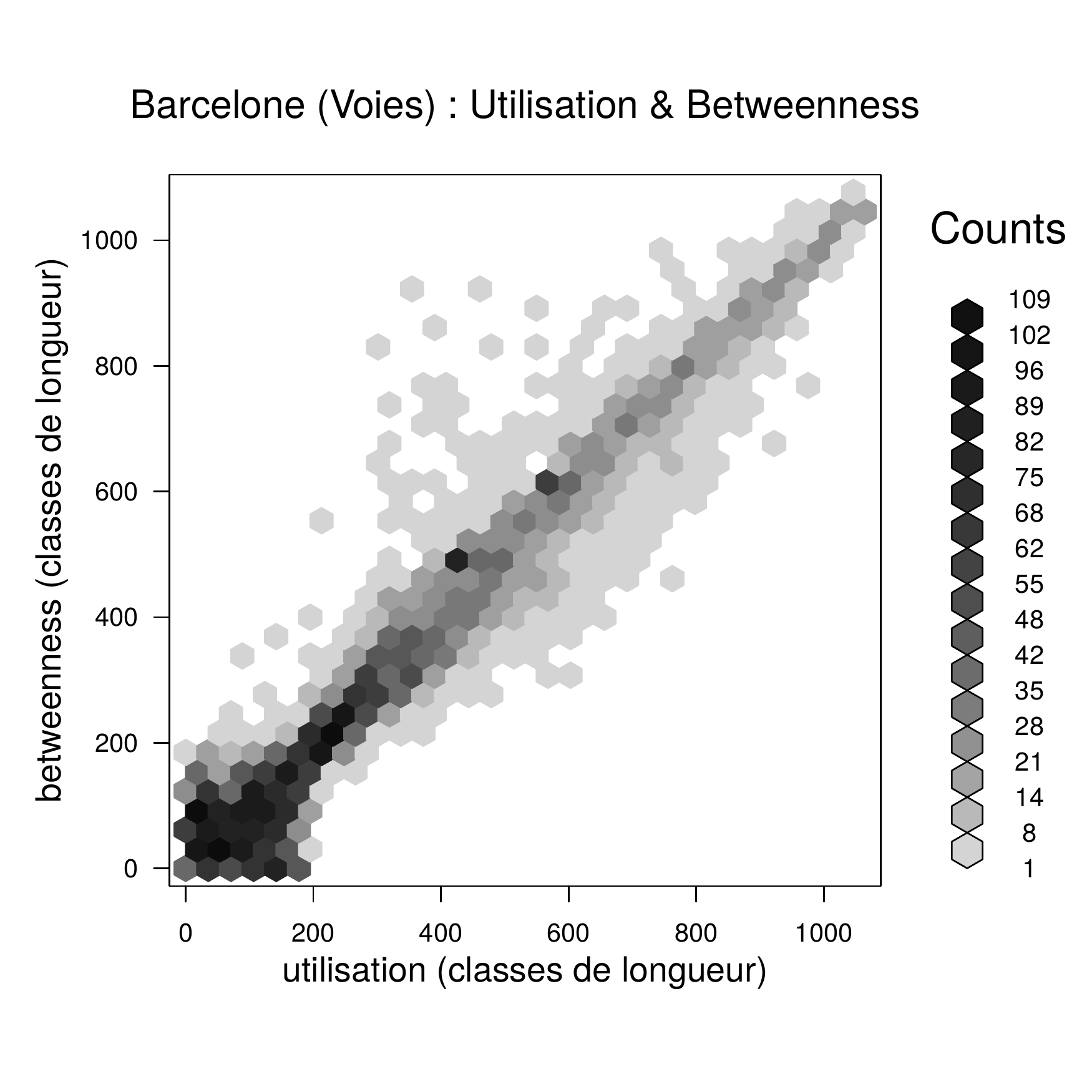}
    \end{subfigure}
    \caption{Utilisation et betweenness}
\end{figure}

\begin{figure}[h]\centering
    \begin{subfigure}[t]{0.45\textwidth}
        \includegraphics[width=\linewidth]{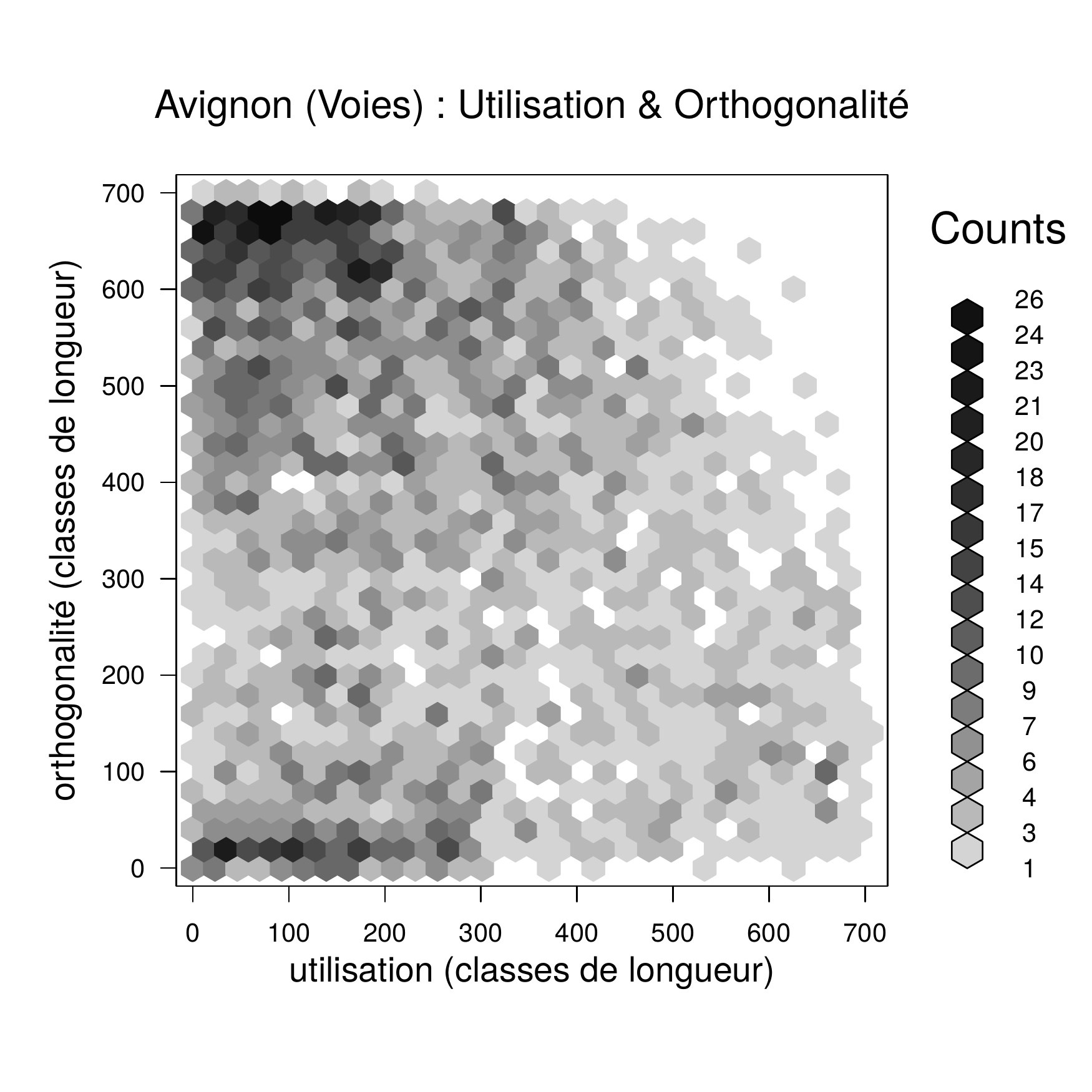}
    \end{subfigure}
    ~
    \begin{subfigure}[t]{0.45\textwidth}
        \includegraphics[width=\linewidth]{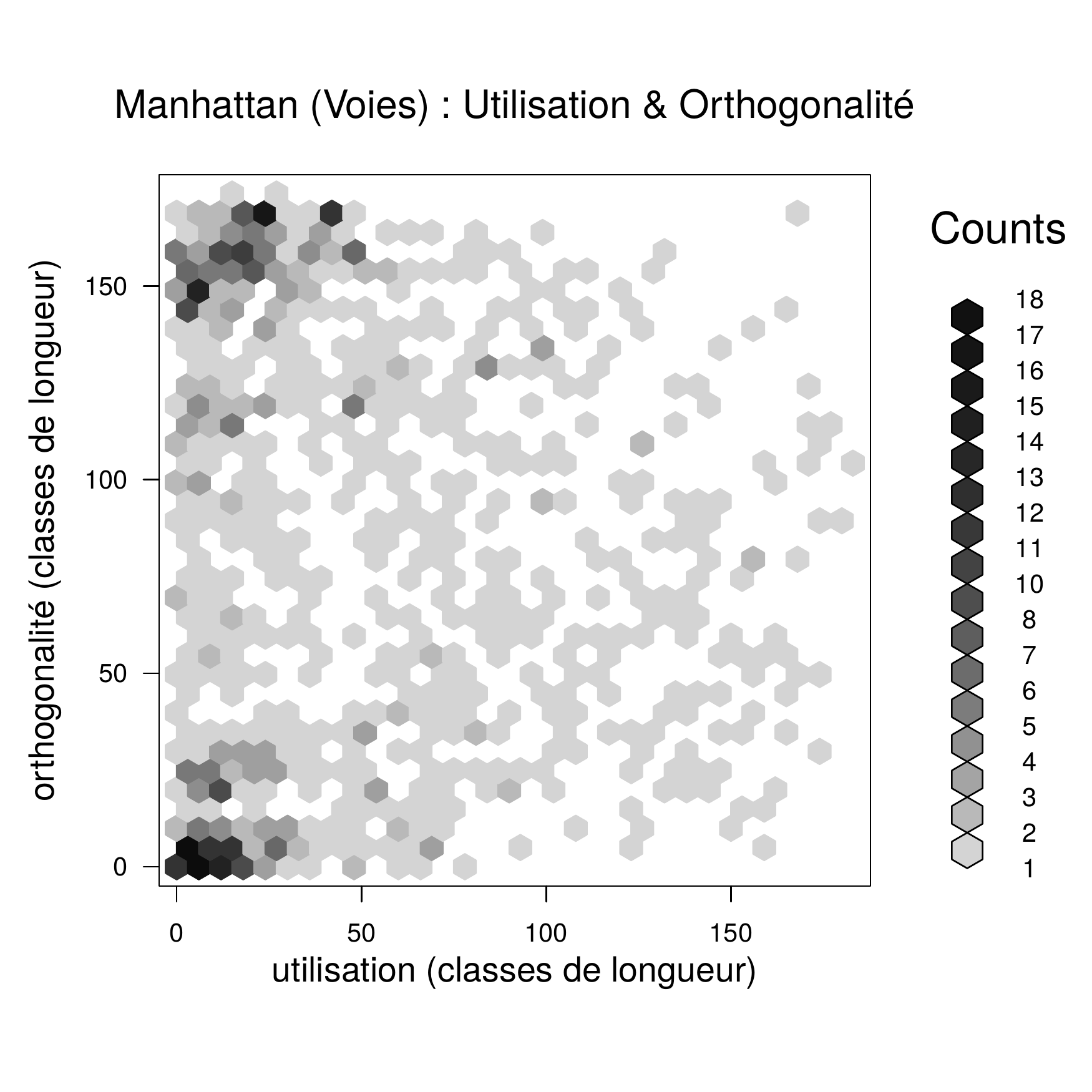}
    \end{subfigure}

    \begin{subfigure}[t]{0.45\textwidth}
        \includegraphics[width=\linewidth]{images/cartes_hexbin/paris_use_ortho.pdf}
    \end{subfigure}
    ~
    \begin{subfigure}[t]{0.45\textwidth}
        \includegraphics[width=\linewidth]{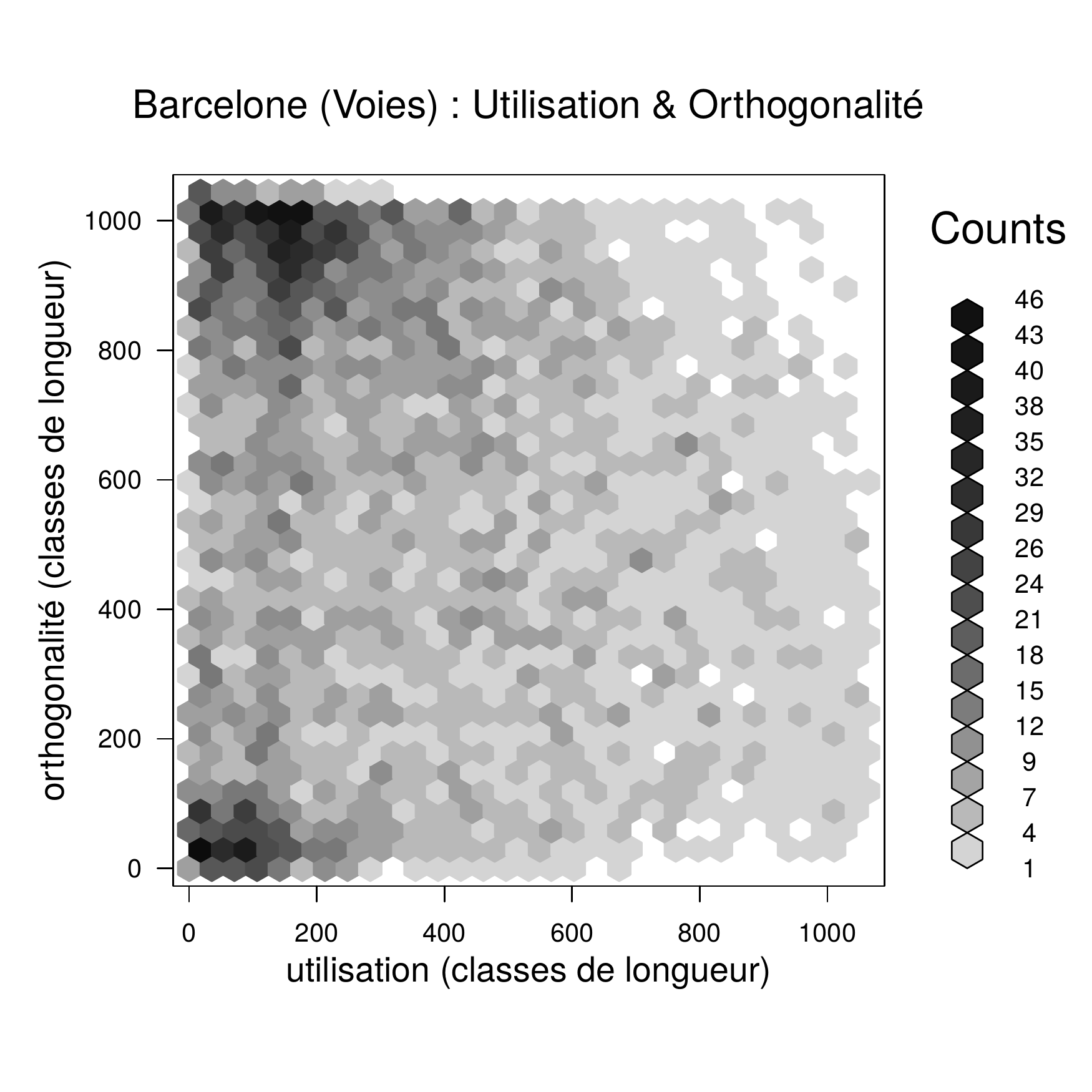}
    \end{subfigure}
    \caption{Utilisation et orthogonalité}
\end{figure}

\begin{figure}[h]\centering
    \begin{subfigure}[t]{0.45\textwidth}
        \includegraphics[width=\linewidth]{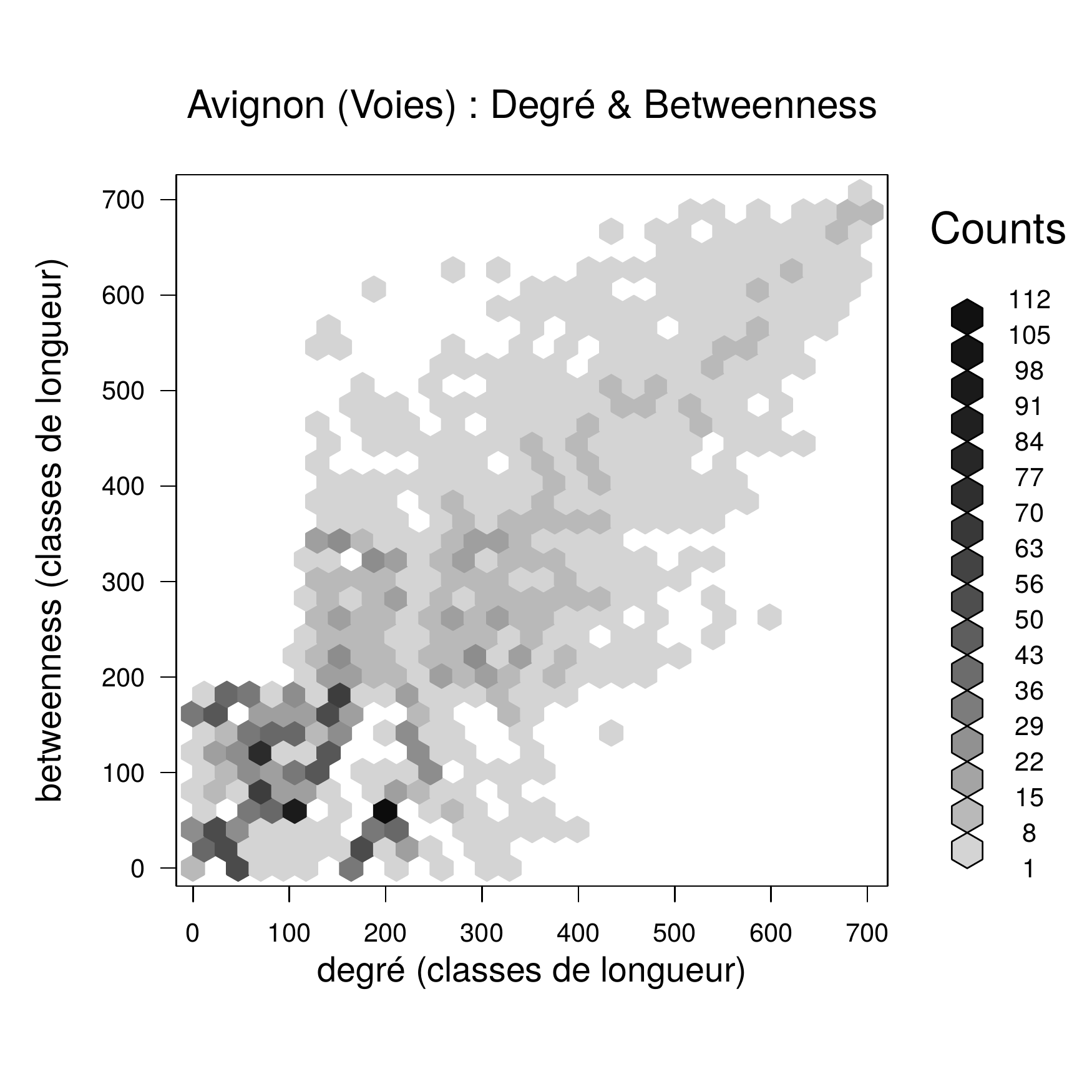}
    \end{subfigure}
    ~
    \begin{subfigure}[t]{0.45\textwidth}
        \includegraphics[width=\linewidth]{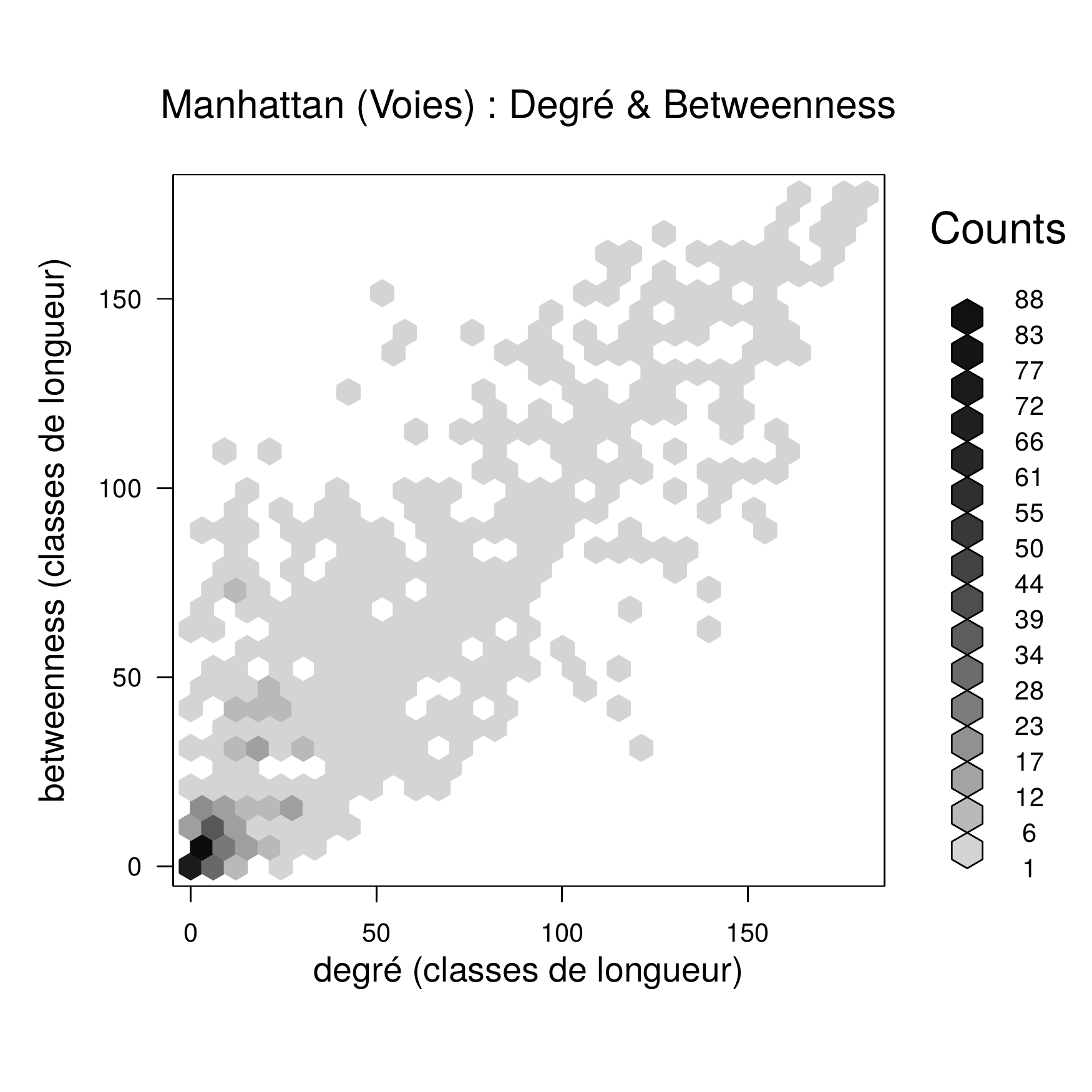}
    \end{subfigure}

    \begin{subfigure}[t]{0.45\textwidth}
        \includegraphics[width=\linewidth]{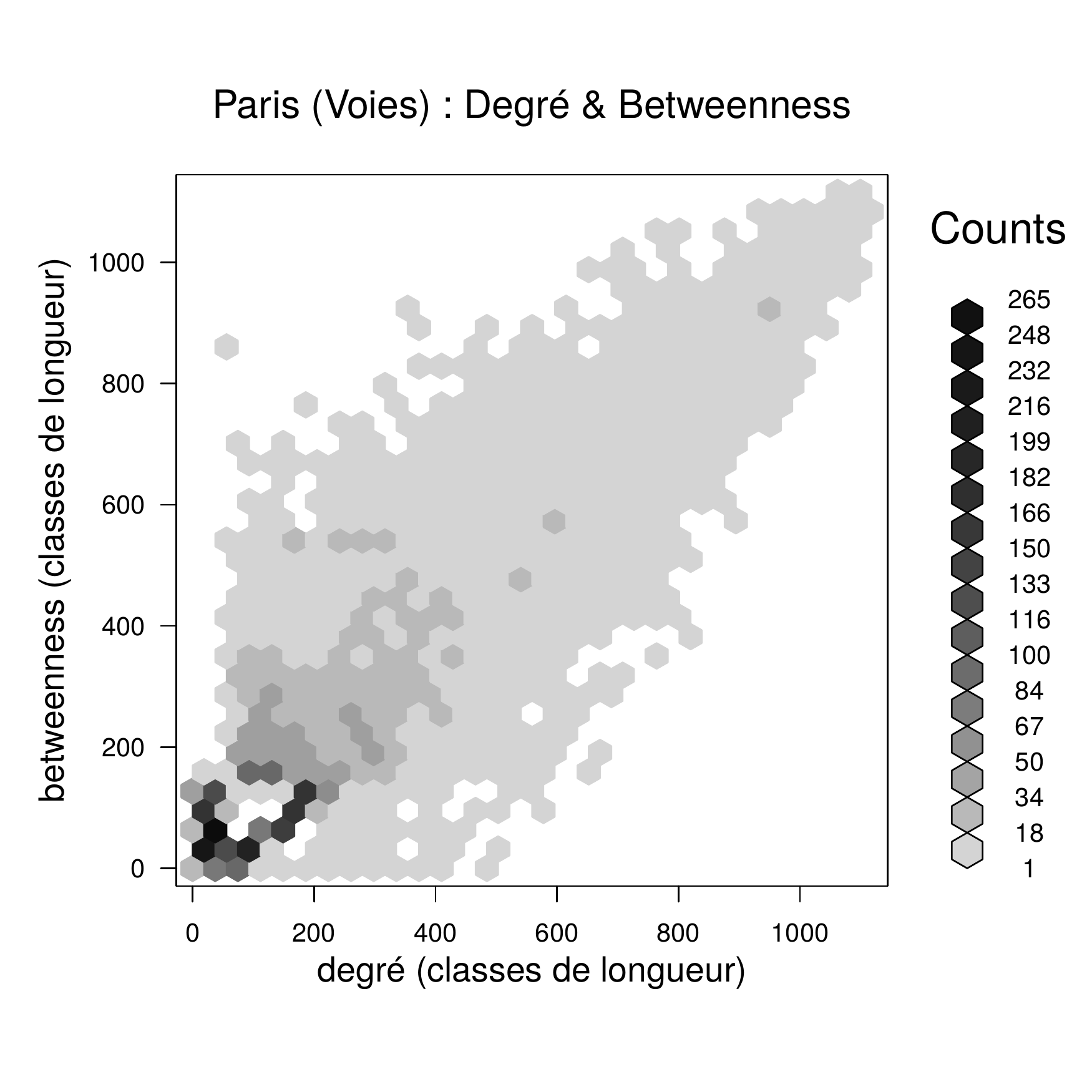}
    \end{subfigure}
    ~
    \begin{subfigure}[t]{0.45\textwidth}
        \includegraphics[width=\linewidth]{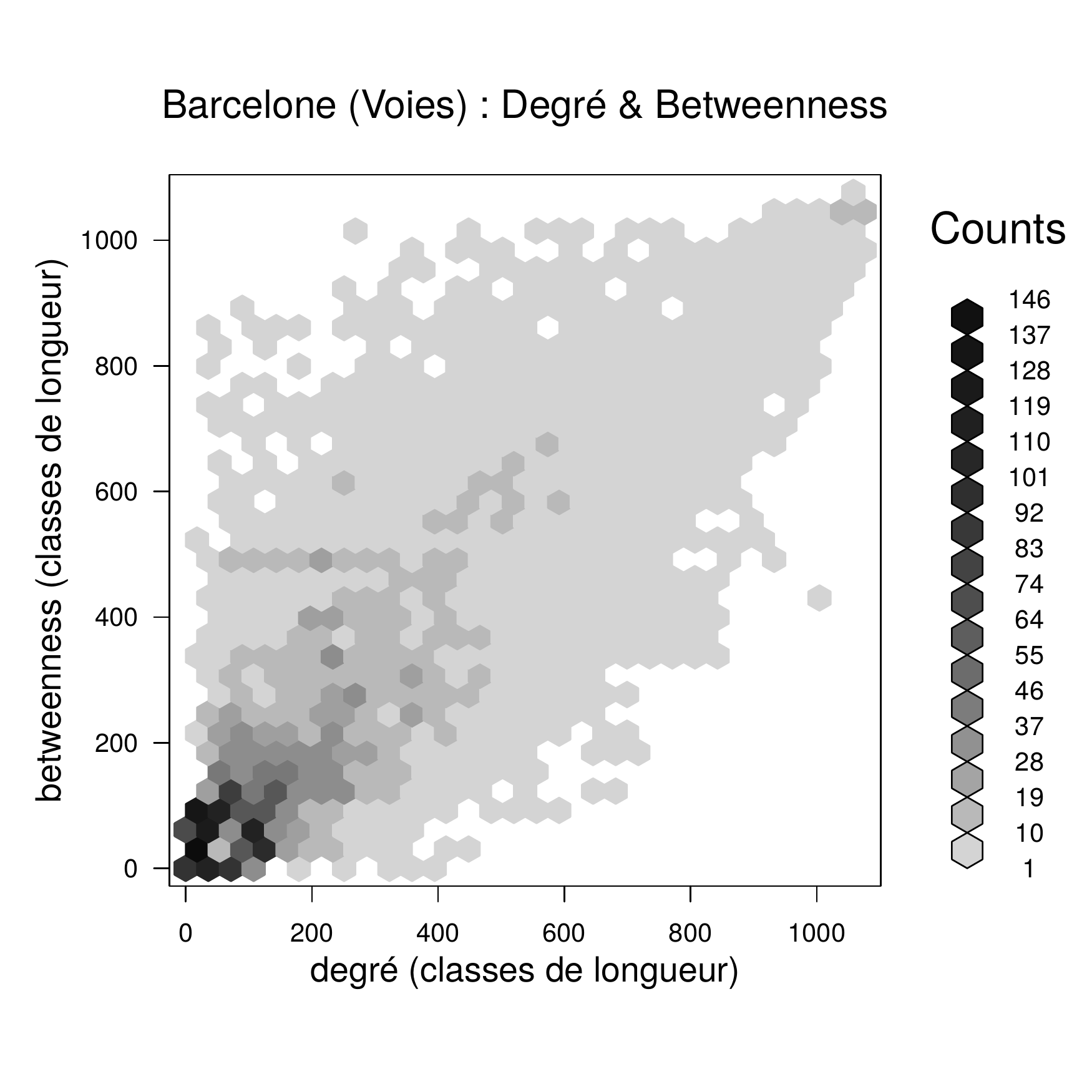}
    \end{subfigure}
    \caption{Degré et betweenness}
\end{figure}

\begin{figure}[h]\centering
    \begin{subfigure}[t]{0.45\textwidth}
        \includegraphics[width=\linewidth]{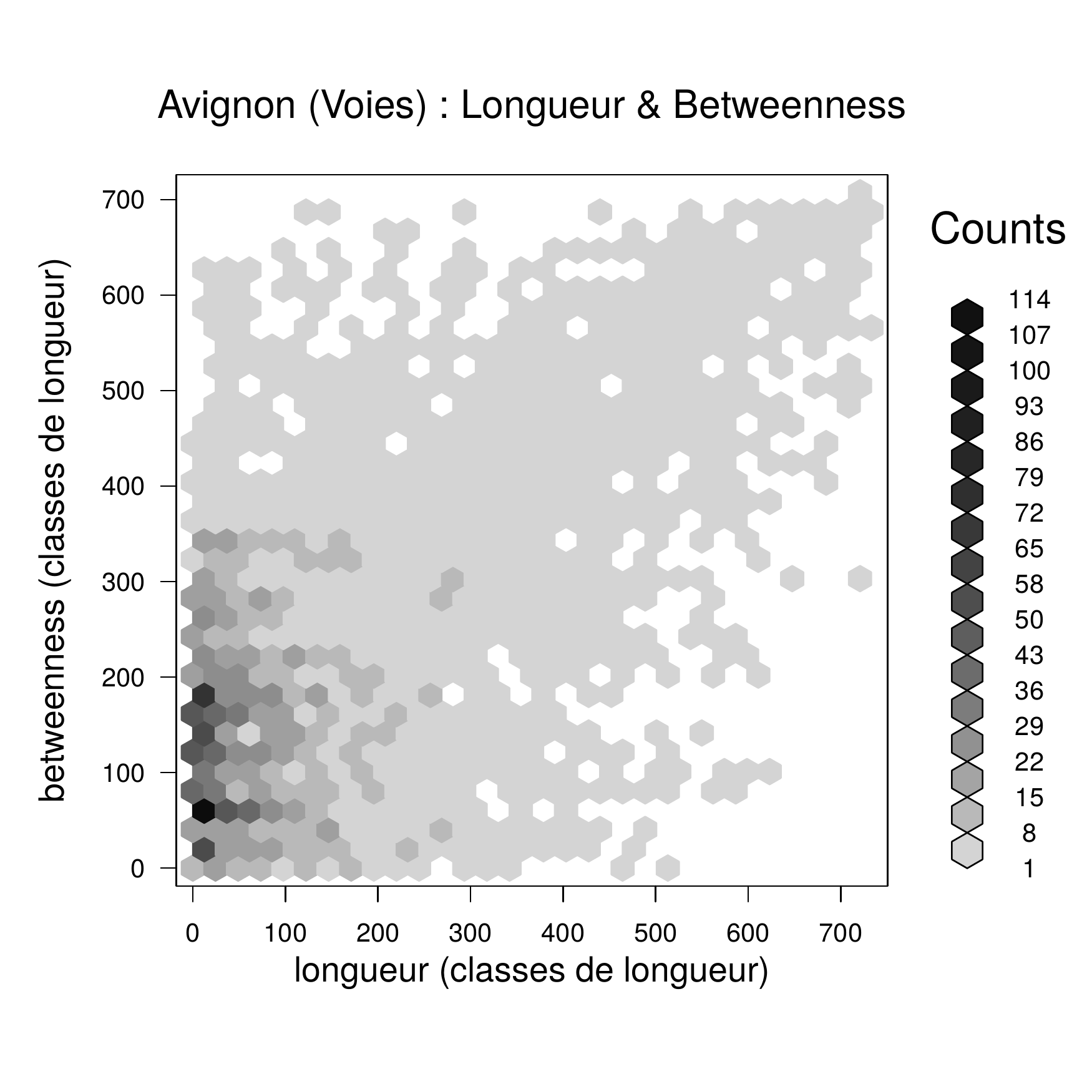}
    \end{subfigure}
    ~
    \begin{subfigure}[t]{0.45\textwidth}
        \includegraphics[width=\linewidth]{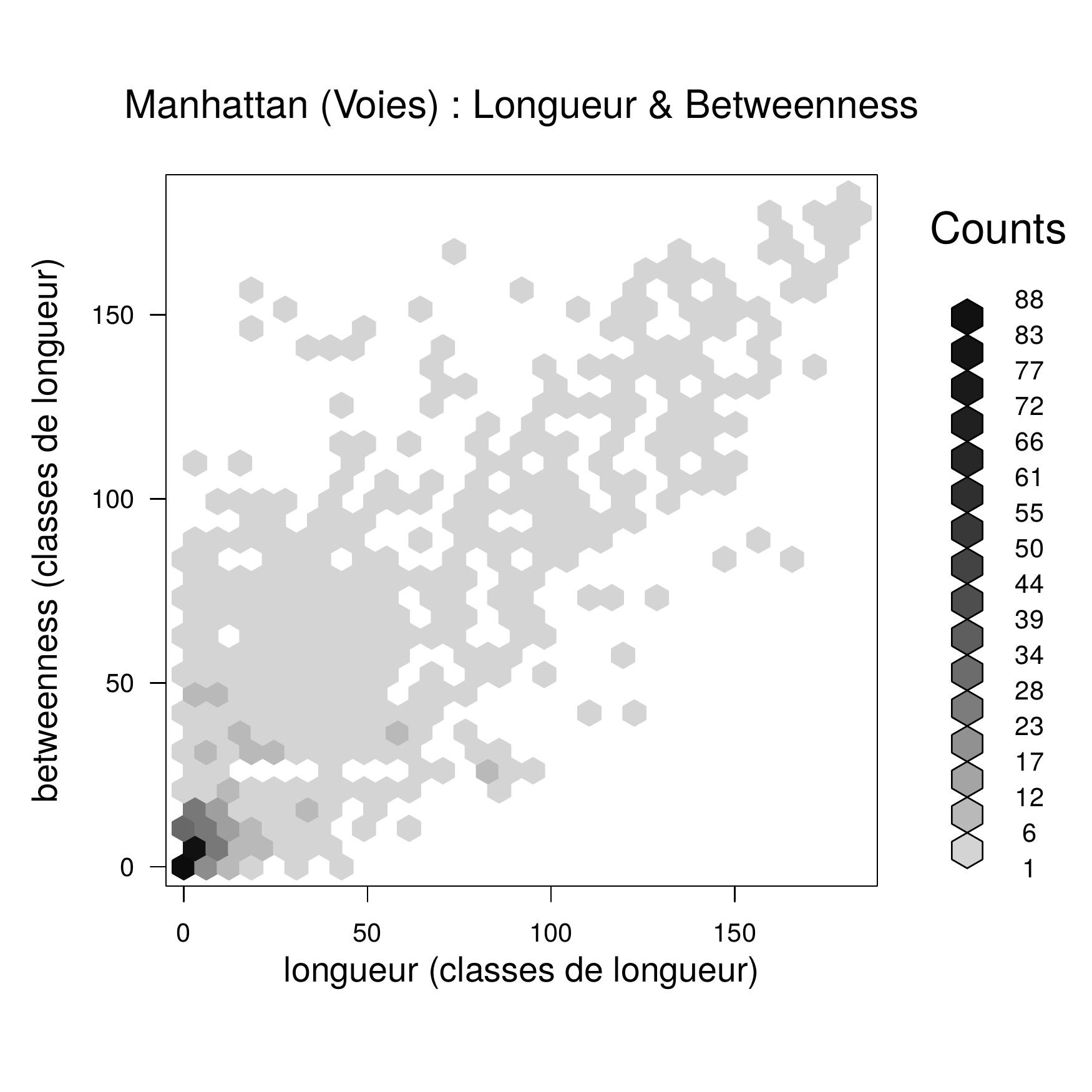}
    \end{subfigure}

    \begin{subfigure}[t]{0.45\textwidth}
        \includegraphics[width=\linewidth]{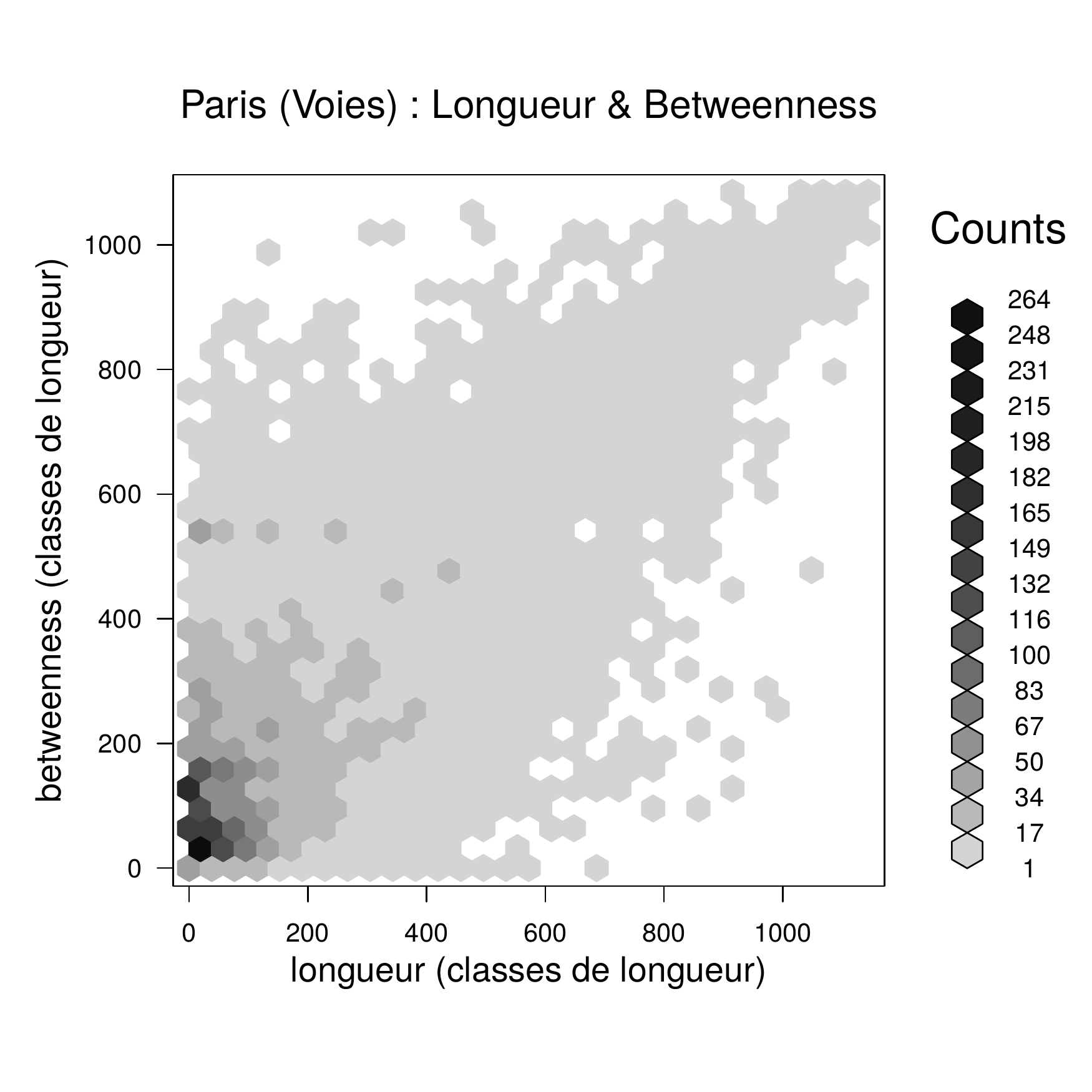}
    \end{subfigure}
    ~
    \begin{subfigure}[t]{0.45\textwidth}
        \includegraphics[width=\linewidth]{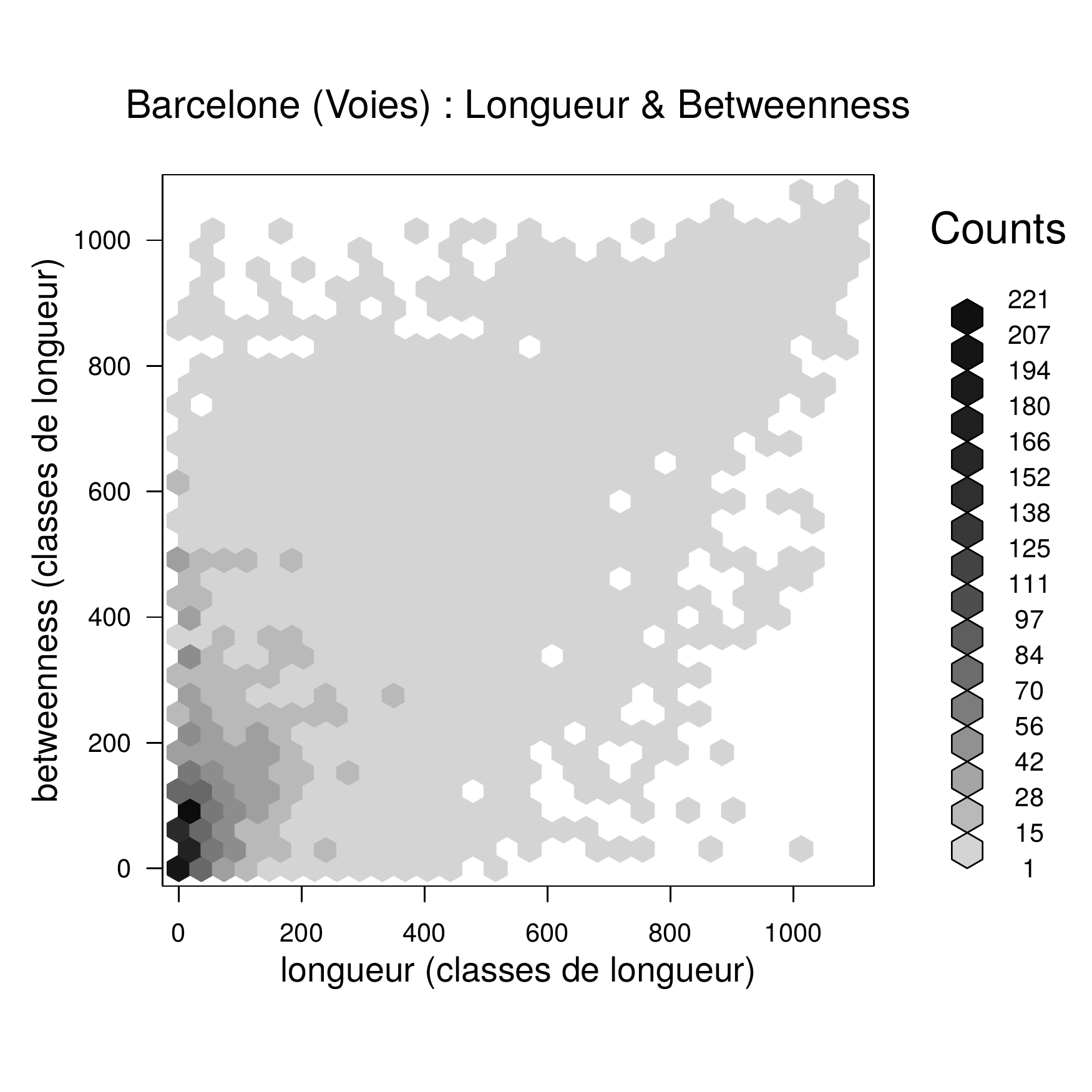}
    \end{subfigure}
    \caption{Longueur et betweenness}
\end{figure}

\begin{figure}[h]\centering
    \begin{subfigure}[t]{0.45\textwidth}
        \includegraphics[width=\linewidth]{images/cartes_hexbin/avignon_length_ortho.pdf}
    \end{subfigure}
    ~
    \begin{subfigure}[t]{0.45\textwidth}
        \includegraphics[width=\linewidth]{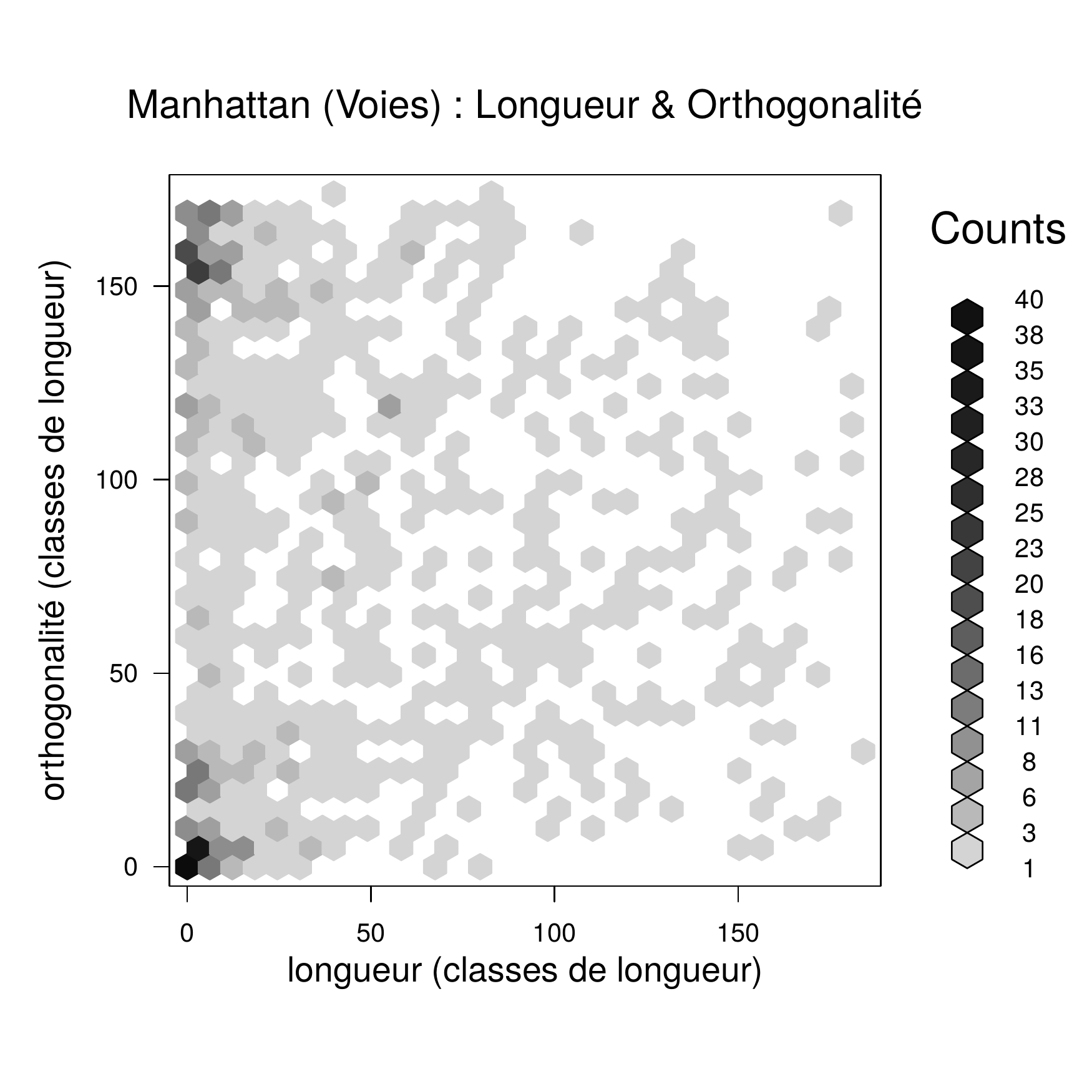}
    \end{subfigure}

    \begin{subfigure}[t]{0.45\textwidth}
        \includegraphics[width=\linewidth]{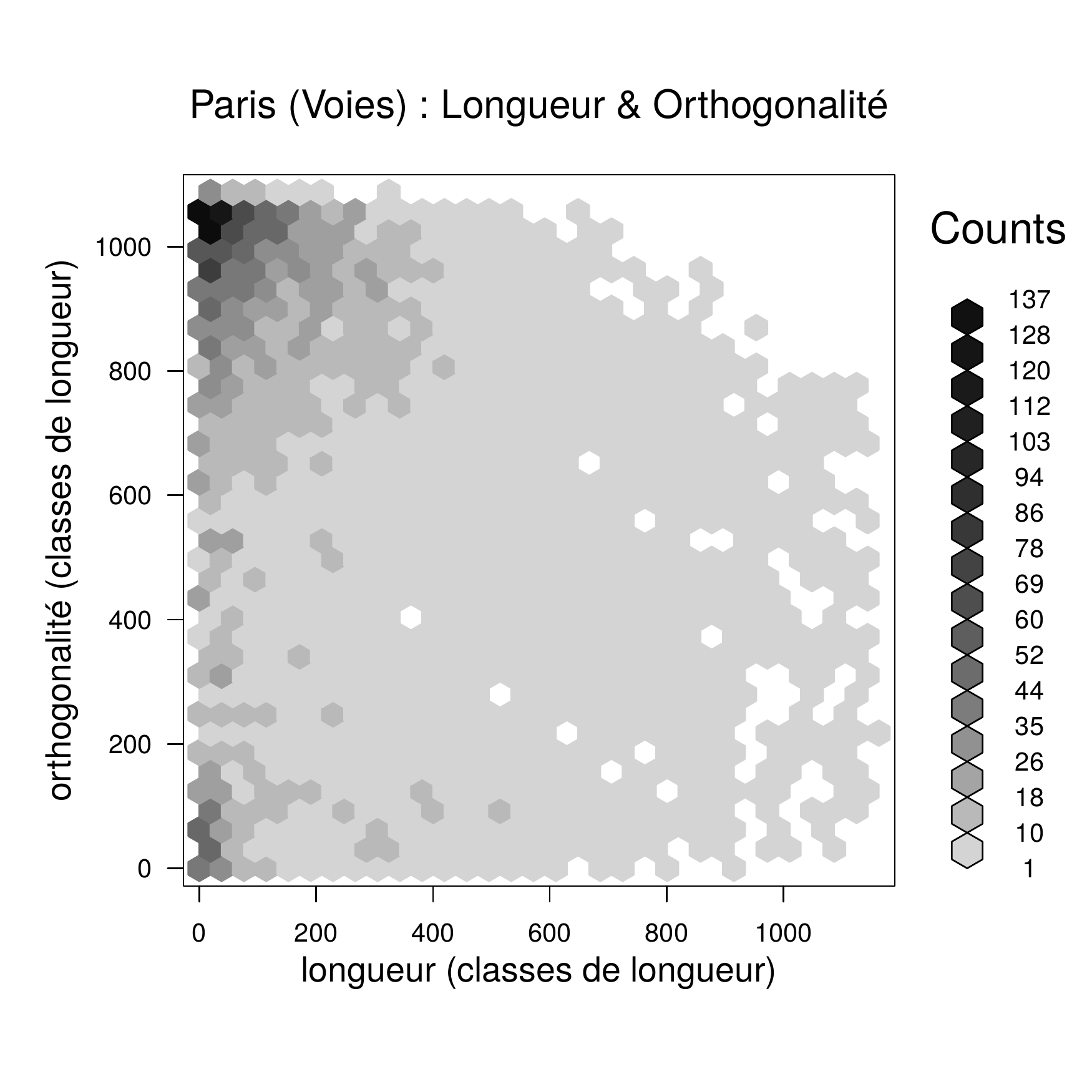}
    \end{subfigure}
    ~
    \begin{subfigure}[t]{0.45\textwidth}
        \includegraphics[width=\linewidth]{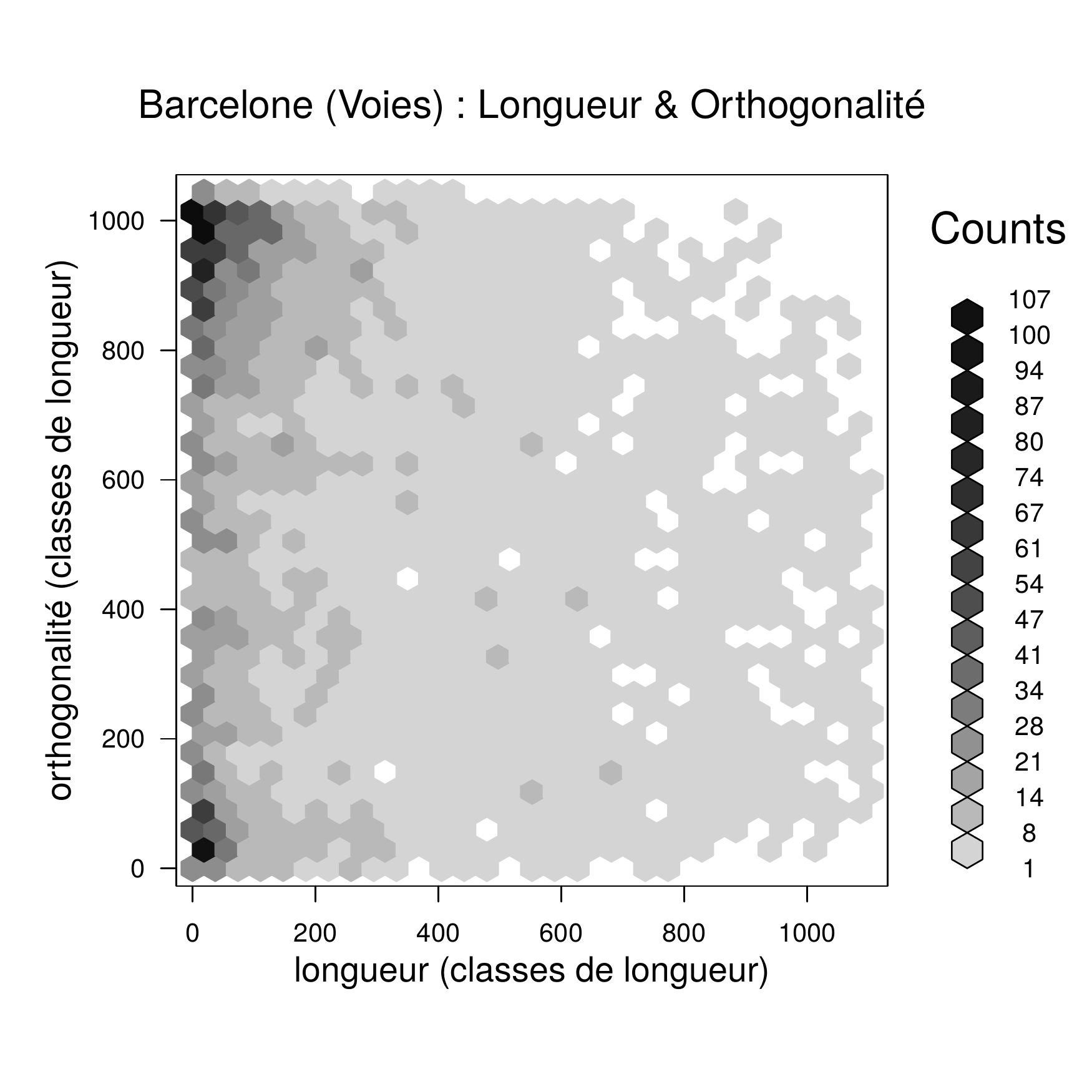}
    \end{subfigure}
    \caption{Longueur et orthogonalité}
\end{figure}

\begin{figure}[h]\centering
    \begin{subfigure}[t]{0.45\textwidth}
        \includegraphics[width=\linewidth]{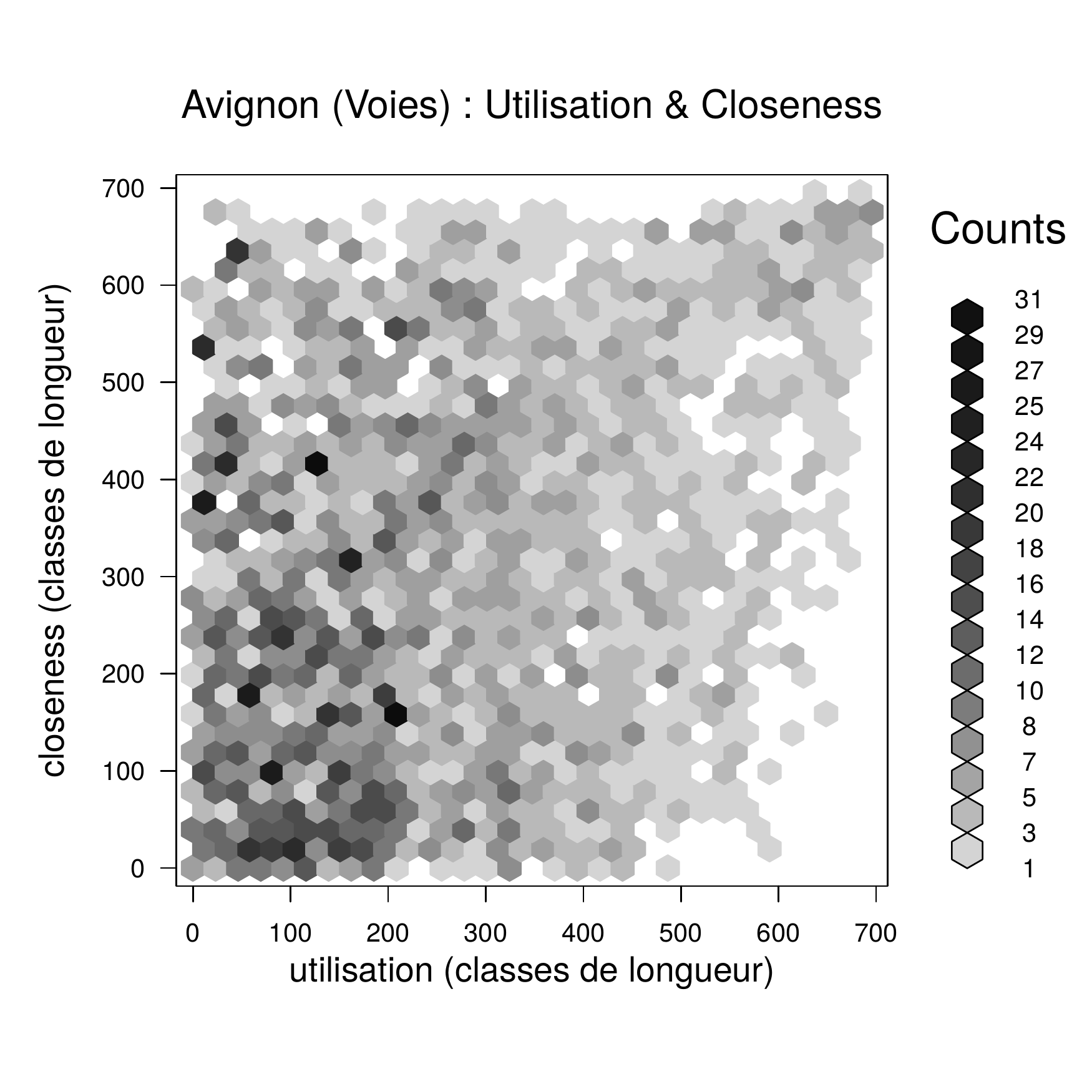}
    \end{subfigure}
    ~
    \begin{subfigure}[t]{0.45\textwidth}
        \includegraphics[width=\linewidth]{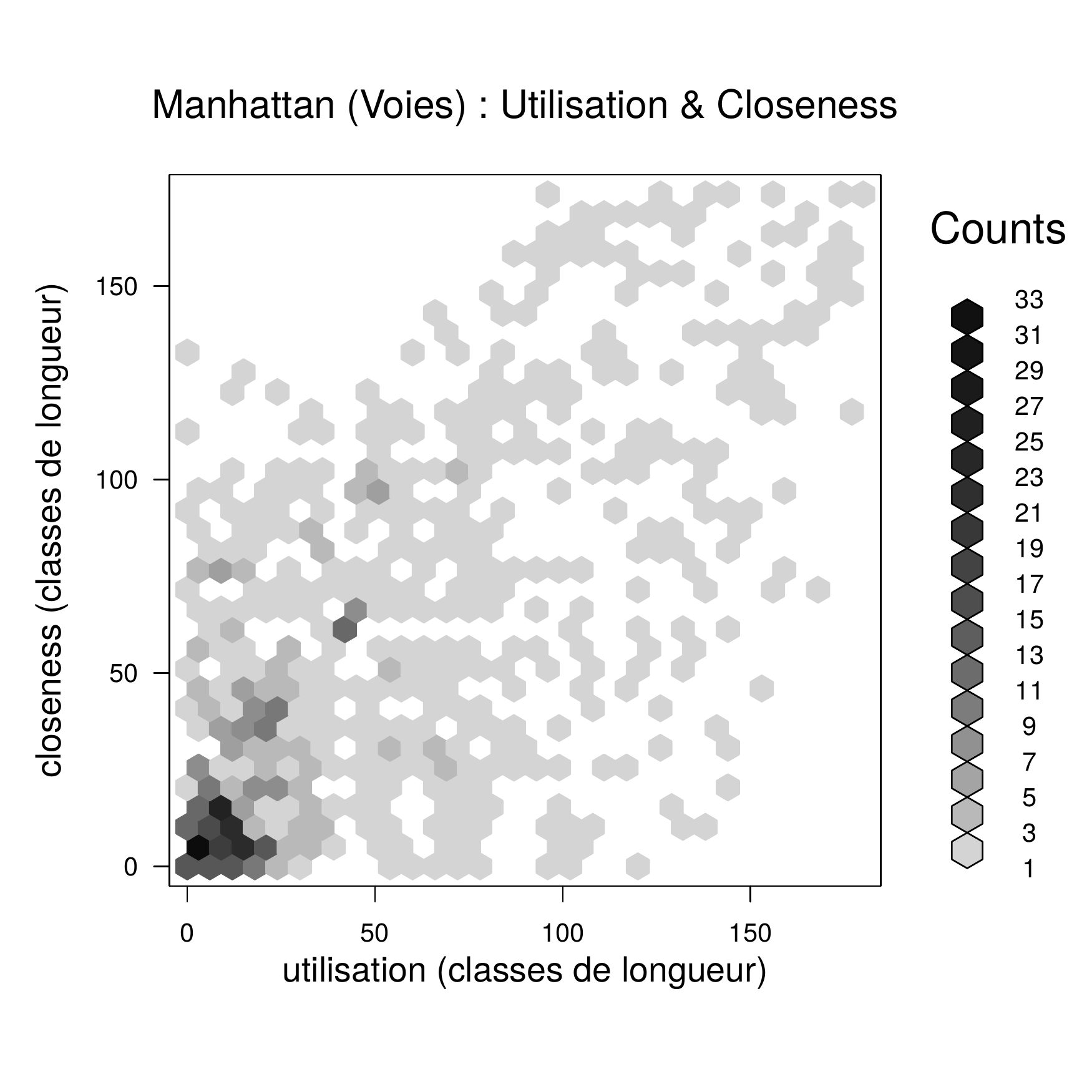}
    \end{subfigure}

    \begin{subfigure}[t]{0.45\textwidth}
        \includegraphics[width=\linewidth]{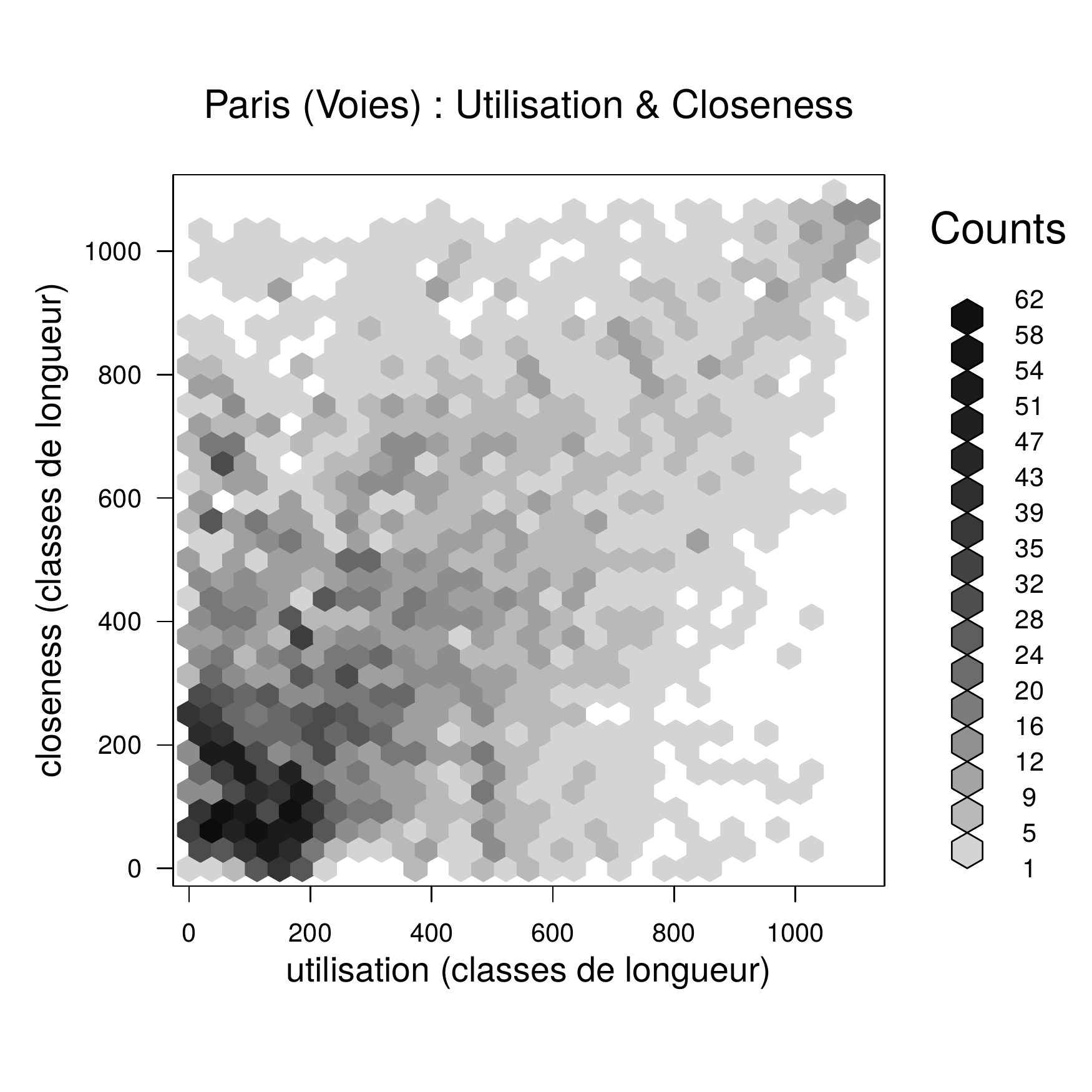}
    \end{subfigure}
    ~
    \begin{subfigure}[t]{0.45\textwidth}
        \includegraphics[width=\linewidth]{images/cartes_hexbin/barcelone_use_clo.pdf}
    \end{subfigure}
    \caption{Utilisation et closeness}
\end{figure}

\begin{figure}[h]\centering
    \begin{subfigure}[t]{0.45\textwidth}
        \includegraphics[width=\linewidth]{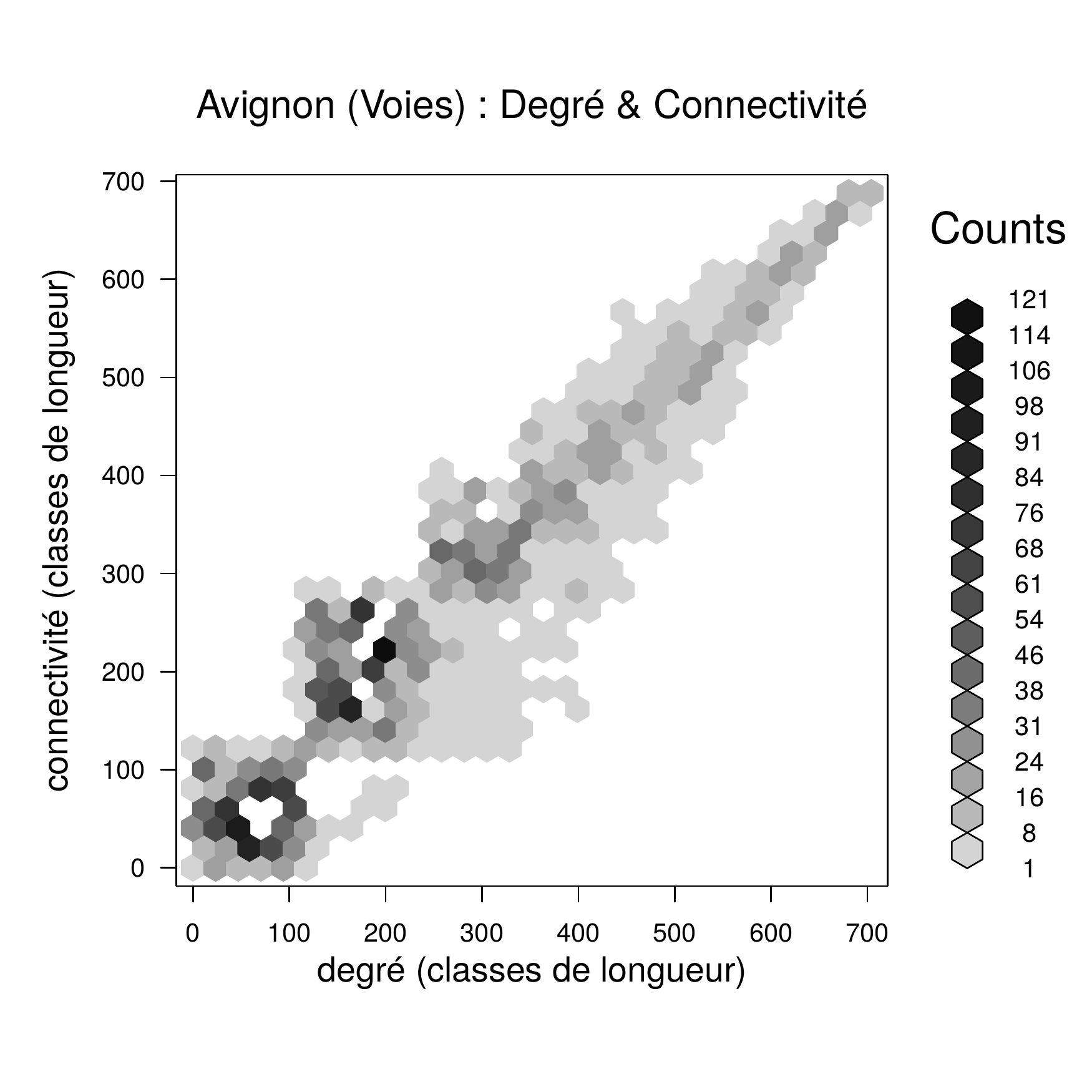}
    \end{subfigure}
    ~
    \begin{subfigure}[t]{0.45\textwidth}
        \includegraphics[width=\linewidth]{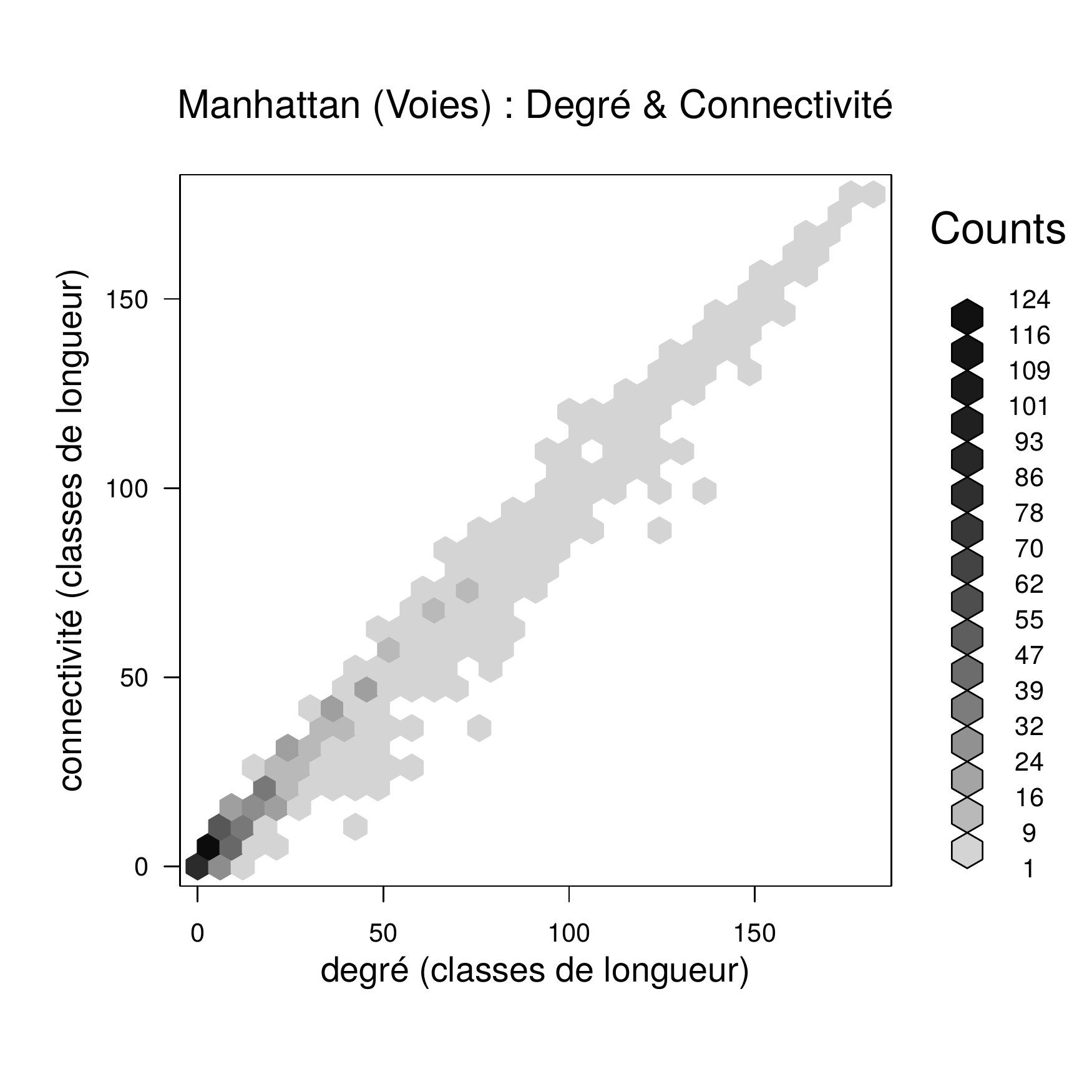}
    \end{subfigure}

    \begin{subfigure}[t]{0.45\textwidth}
        \includegraphics[width=\linewidth]{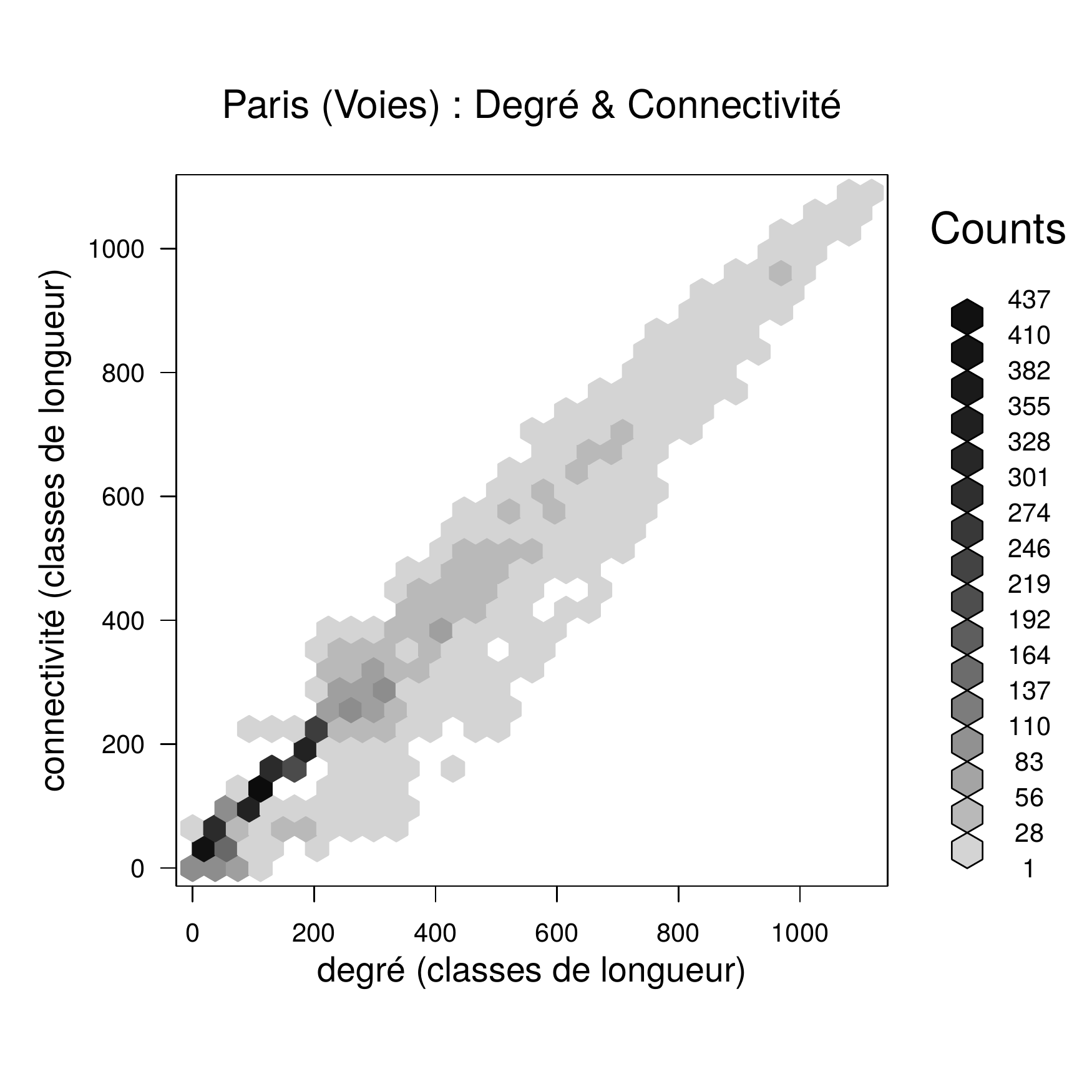}
    \end{subfigure}
    ~
    \begin{subfigure}[t]{0.45\textwidth}
        \includegraphics[width=\linewidth]{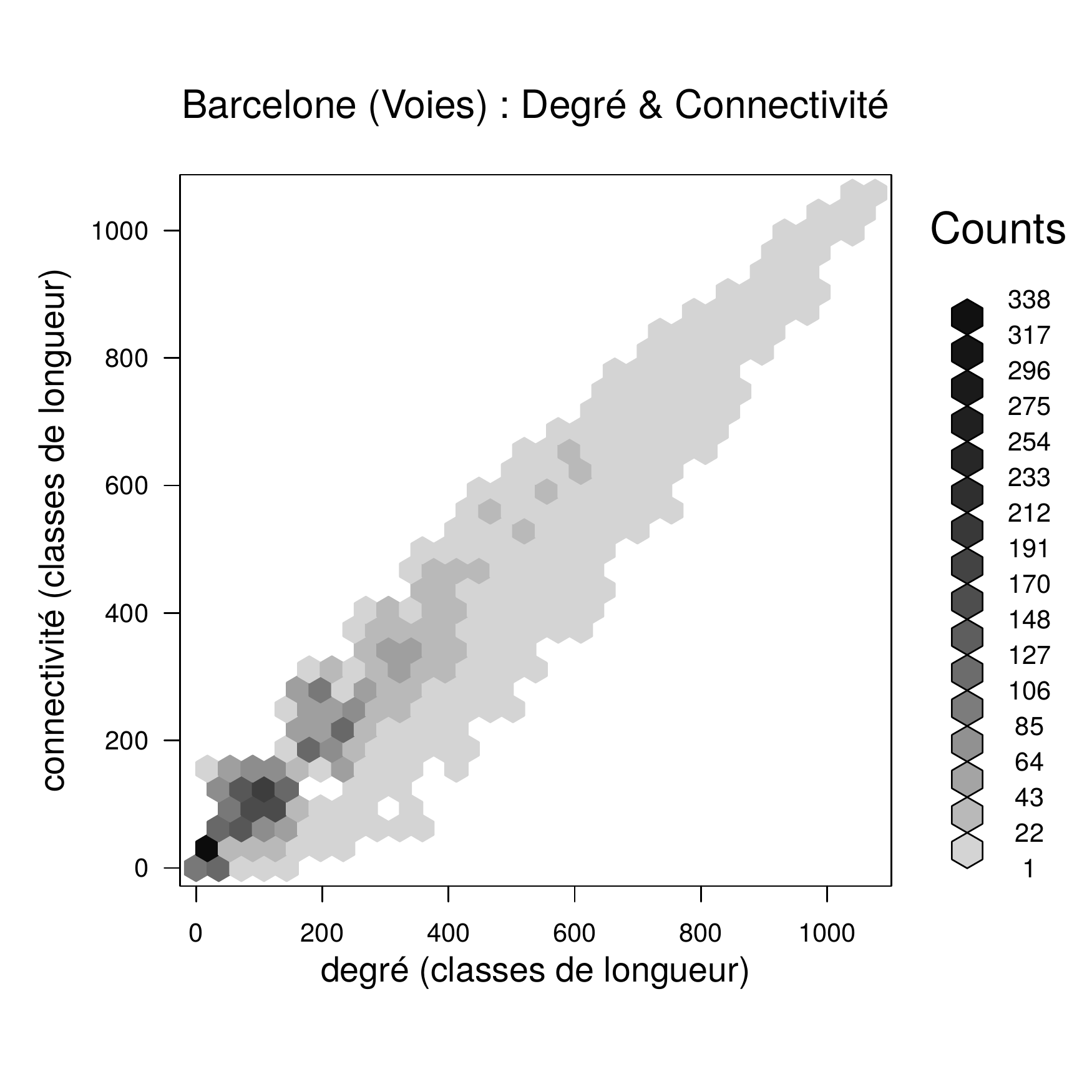}
    \end{subfigure}
    \caption{Degré et connectivité}
\end{figure}

\begin{figure}[h]\centering
    \begin{subfigure}[t]{0.45\textwidth}
        \includegraphics[width=\linewidth]{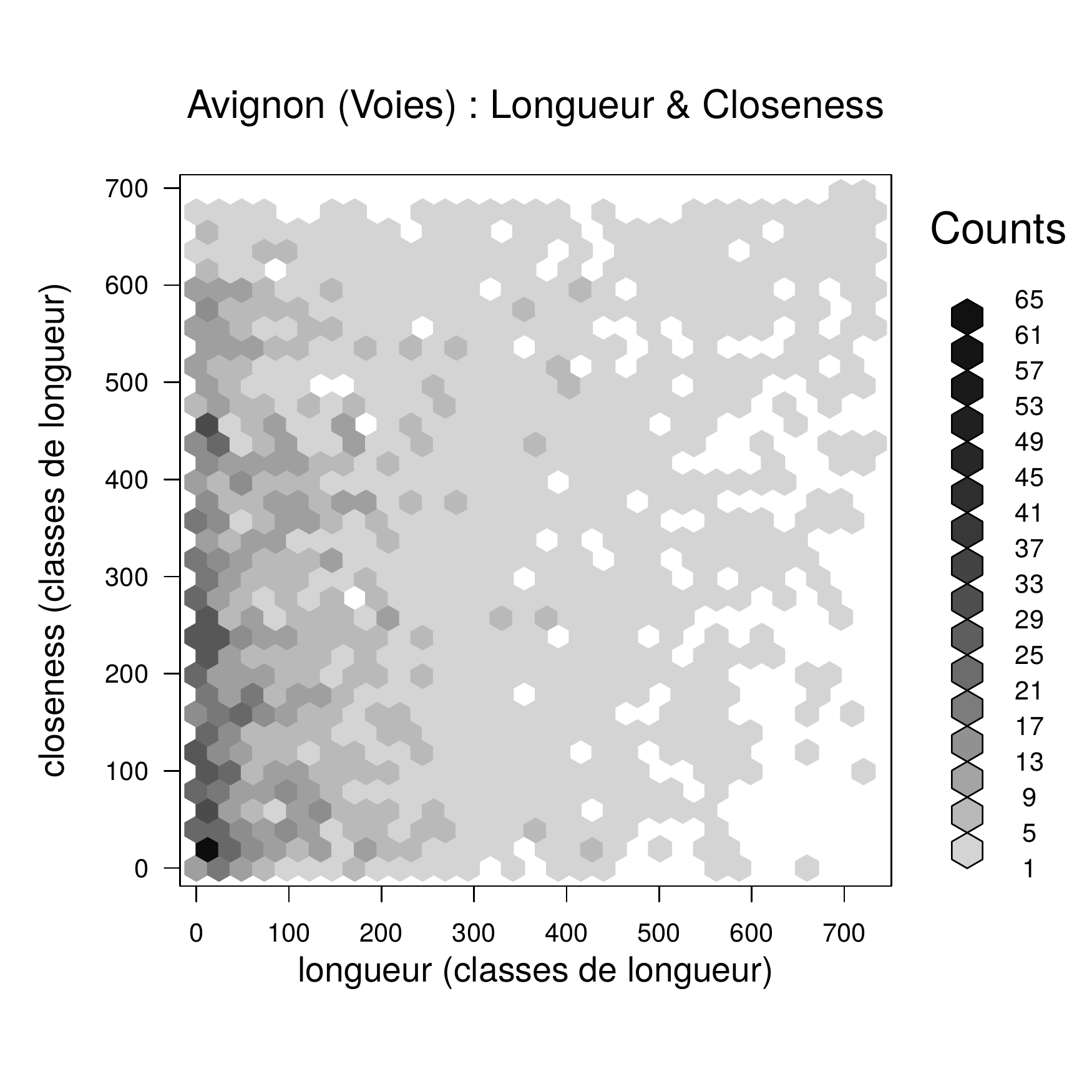}
    \end{subfigure}
    ~
    \begin{subfigure}[t]{0.45\textwidth}
        \includegraphics[width=\linewidth]{images/cartes_hexbin/manhattan_length_clo.pdf}
    \end{subfigure}

    \begin{subfigure}[t]{0.45\textwidth}
        \includegraphics[width=\linewidth]{images/cartes_hexbin/paris_length_clo.pdf}
    \end{subfigure}
    ~
    \begin{subfigure}[t]{0.45\textwidth}
        \includegraphics[width=\linewidth]{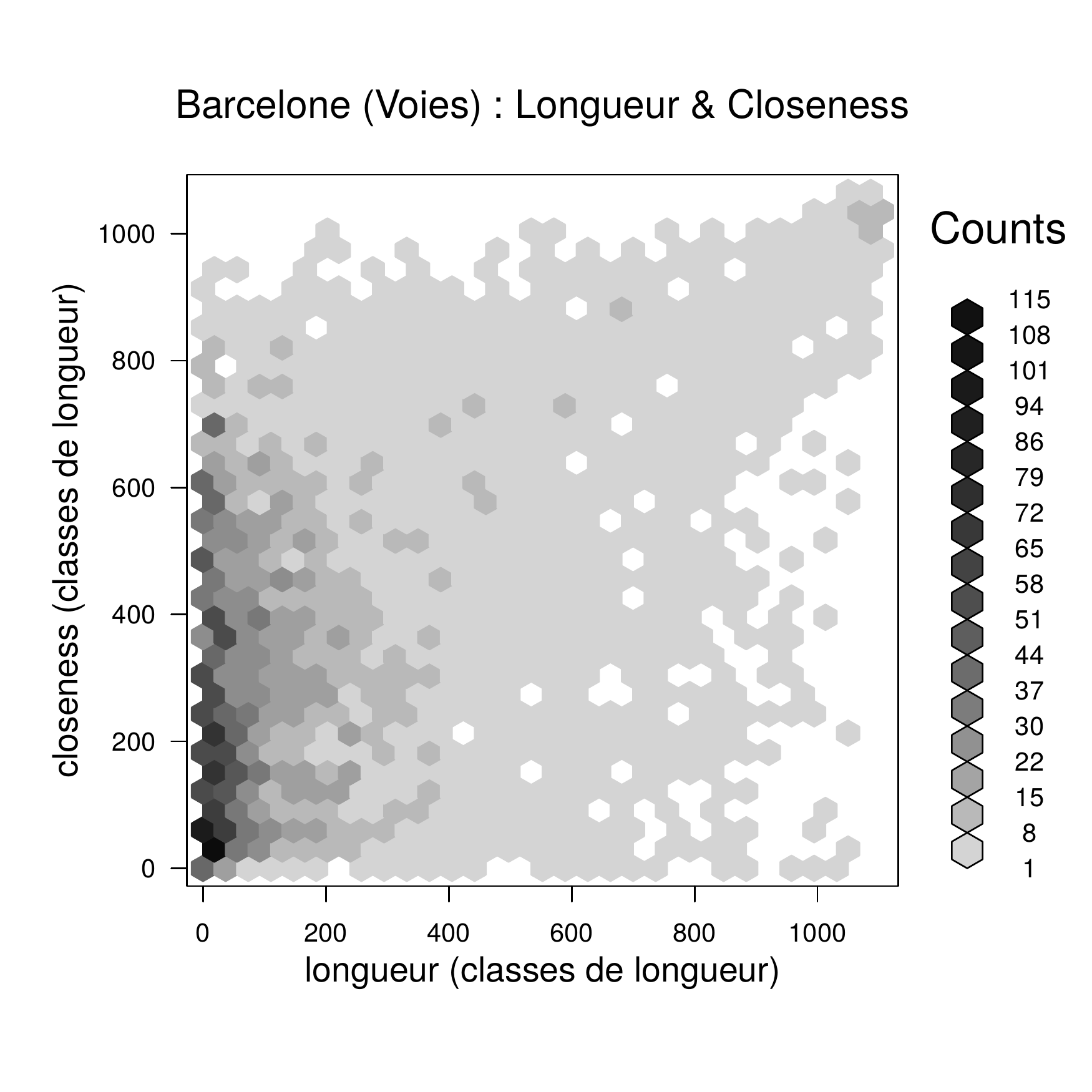}
    \end{subfigure}
    \caption{Longueur et closeness}
\end{figure}

\begin{figure}[h]\centering
    \begin{subfigure}[t]{0.45\textwidth}
        \includegraphics[width=\linewidth]{images/cartes_hexbin/avignon_ortho_clo.pdf}
    \end{subfigure}
    ~
    \begin{subfigure}[t]{0.45\textwidth}
        \includegraphics[width=\linewidth]{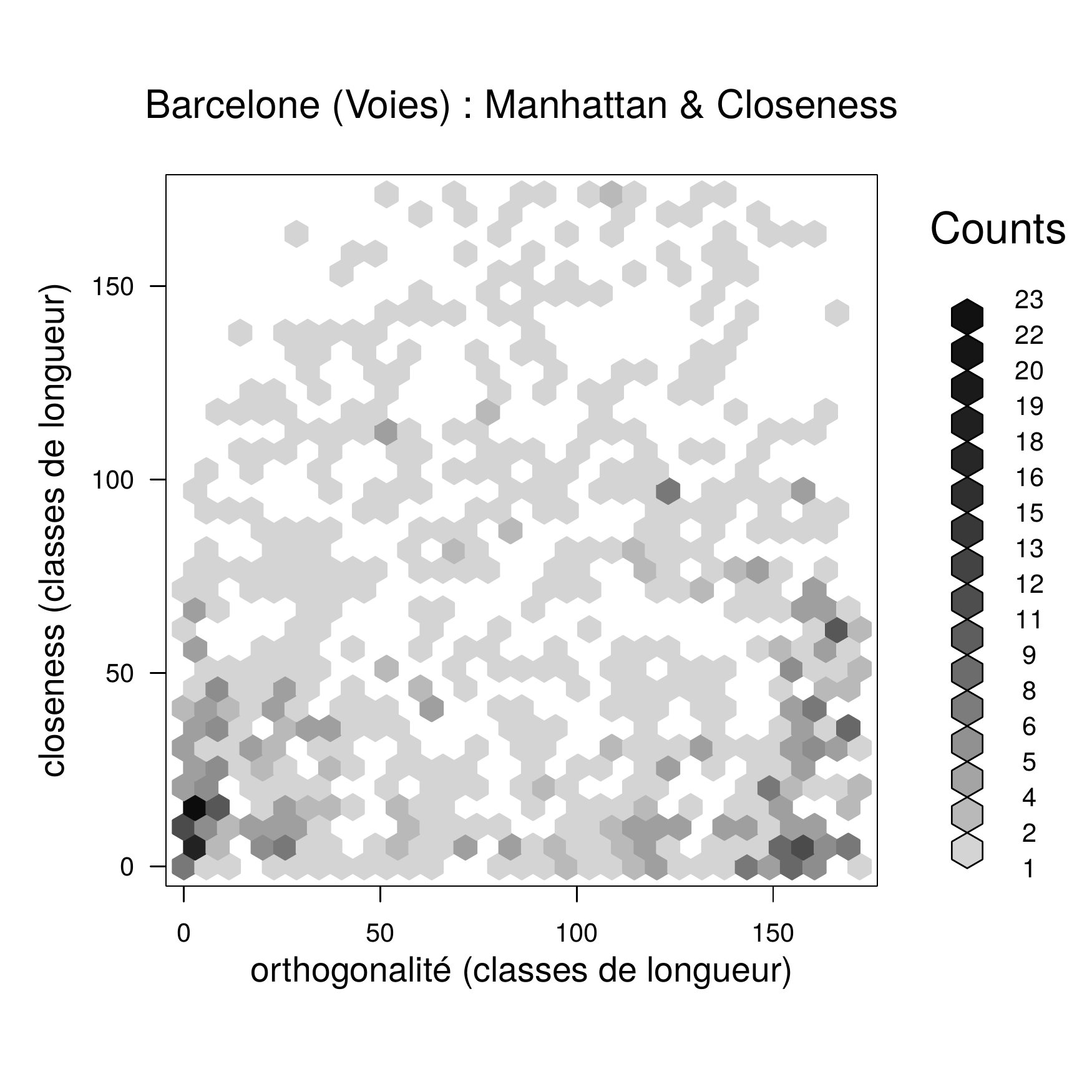}
    \end{subfigure}

    \begin{subfigure}[t]{0.45\textwidth}
        \includegraphics[width=\linewidth]{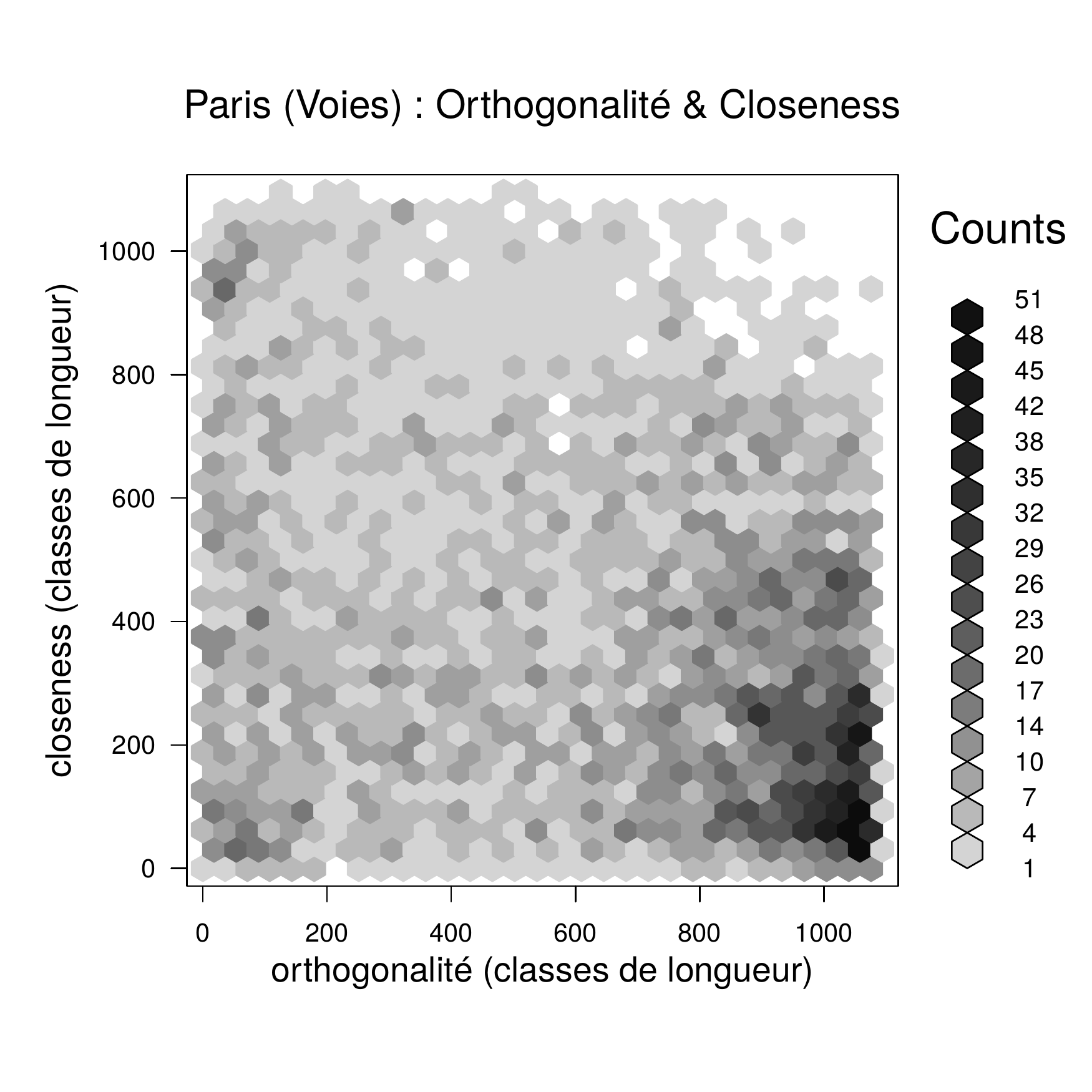}
    \end{subfigure}
    ~
    \begin{subfigure}[t]{0.45\textwidth}
        \includegraphics[width=\linewidth]{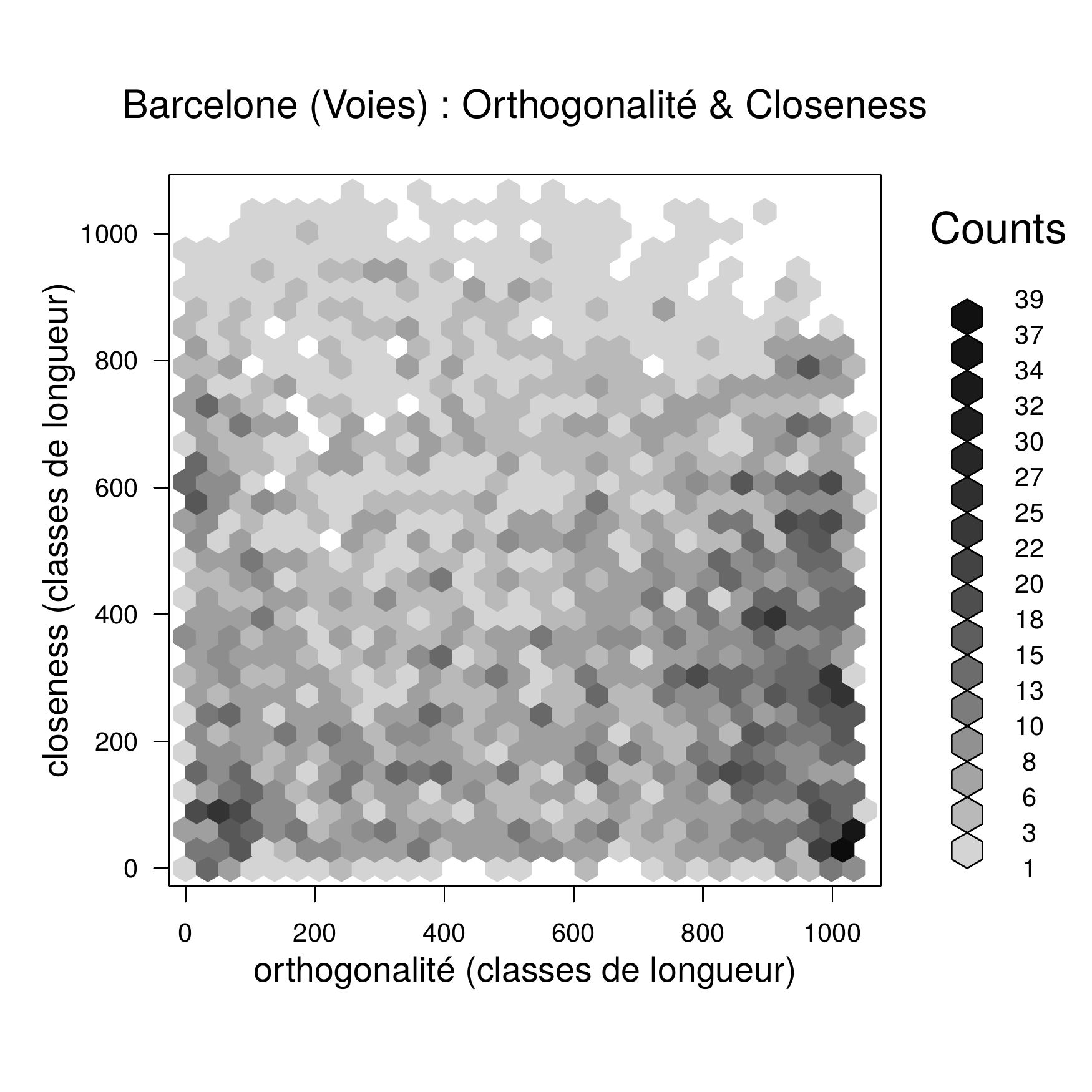}
    \end{subfigure}
    \caption{Orthogonalité et closeness}
\end{figure}

\begin{figure}[h]\centering
    \begin{subfigure}[t]{0.45\textwidth}
        \includegraphics[width=\linewidth]{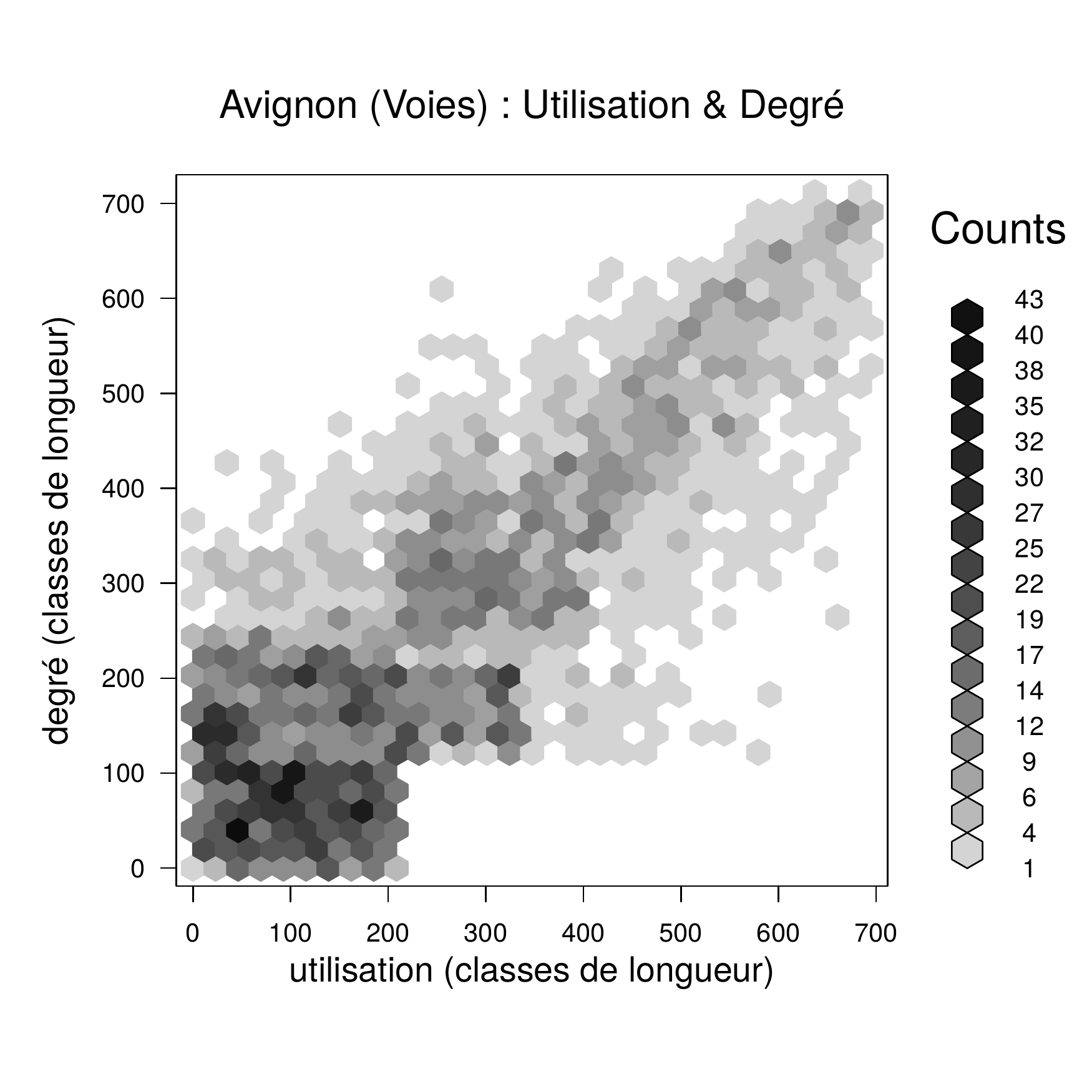}
    \end{subfigure}
    ~
    \begin{subfigure}[t]{0.45\textwidth}
        \includegraphics[width=\linewidth]{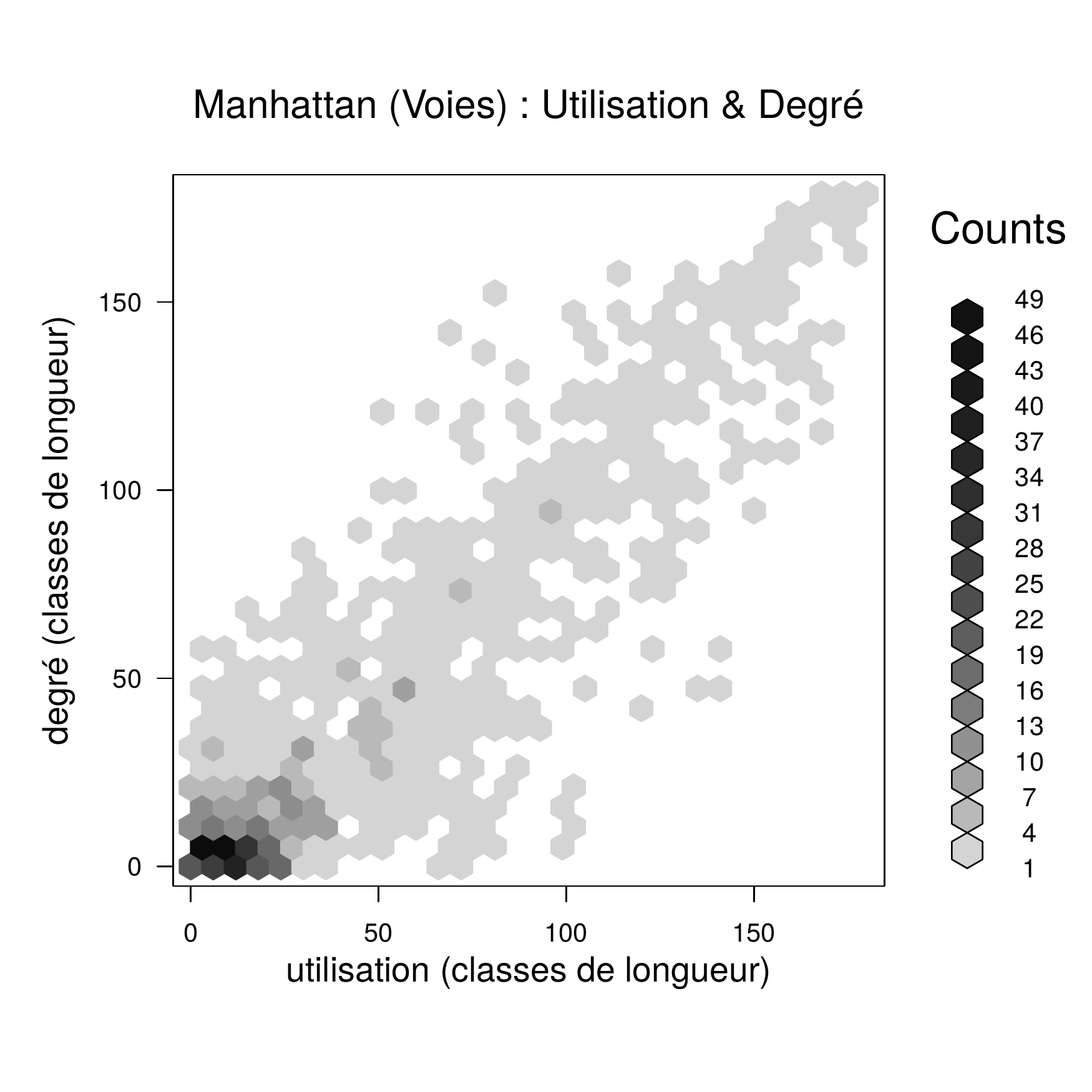}
    \end{subfigure}

    \begin{subfigure}[t]{0.45\textwidth}
        \includegraphics[width=\linewidth]{images/cartes_hexbin/paris_use_degree.pdf}
    \end{subfigure}
    ~
    \begin{subfigure}[t]{0.45\textwidth}
        \includegraphics[width=\linewidth]{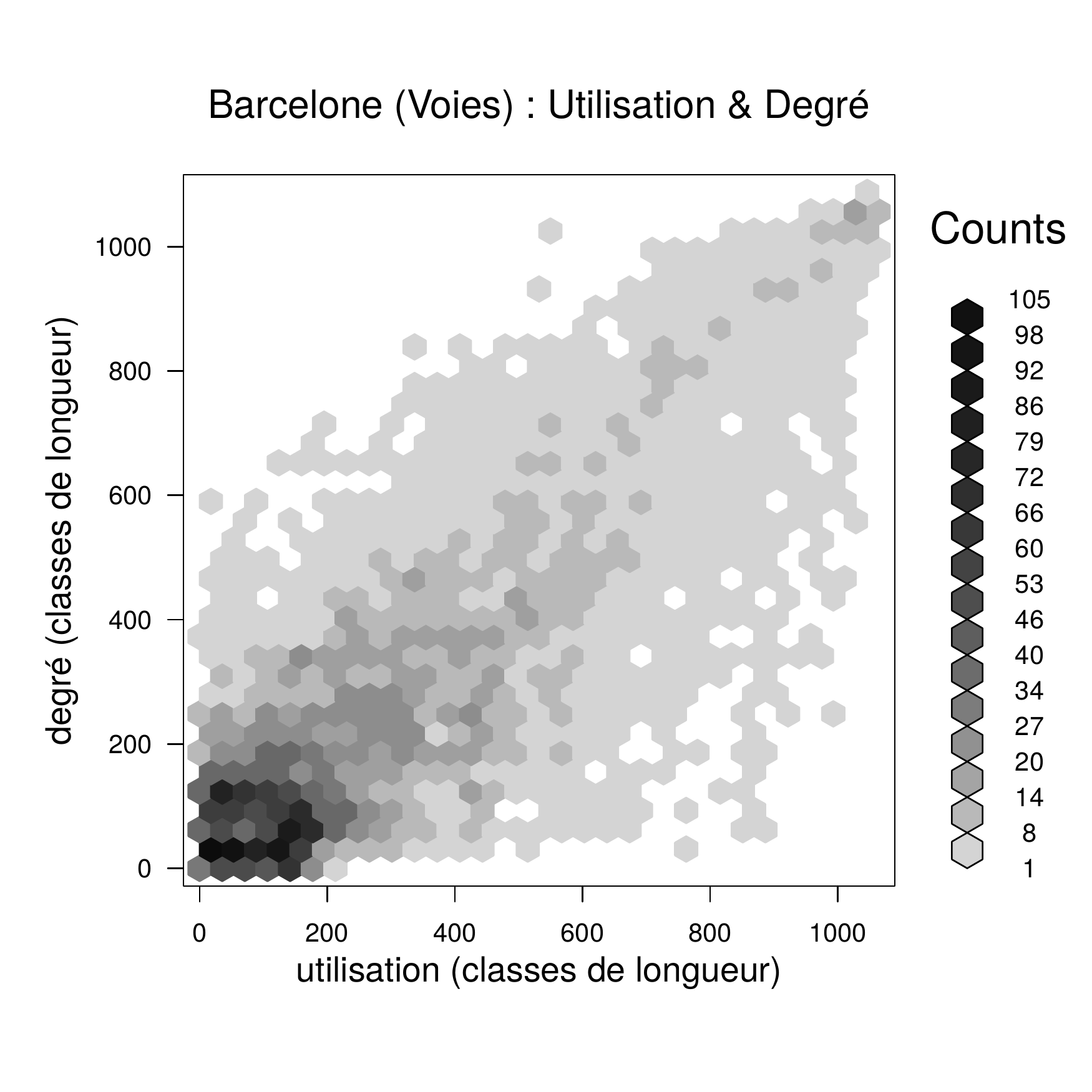}
    \end{subfigure}
    \caption{Utilisation et degré}
\end{figure}

\FloatBarrier 
\subsection{Indicateurs composés}\label{ann:ssec_voies_indcomp}

\begin{figure}[h]\centering
    \begin{subfigure}[t]{0.45\textwidth}
        \includegraphics[width=\linewidth]{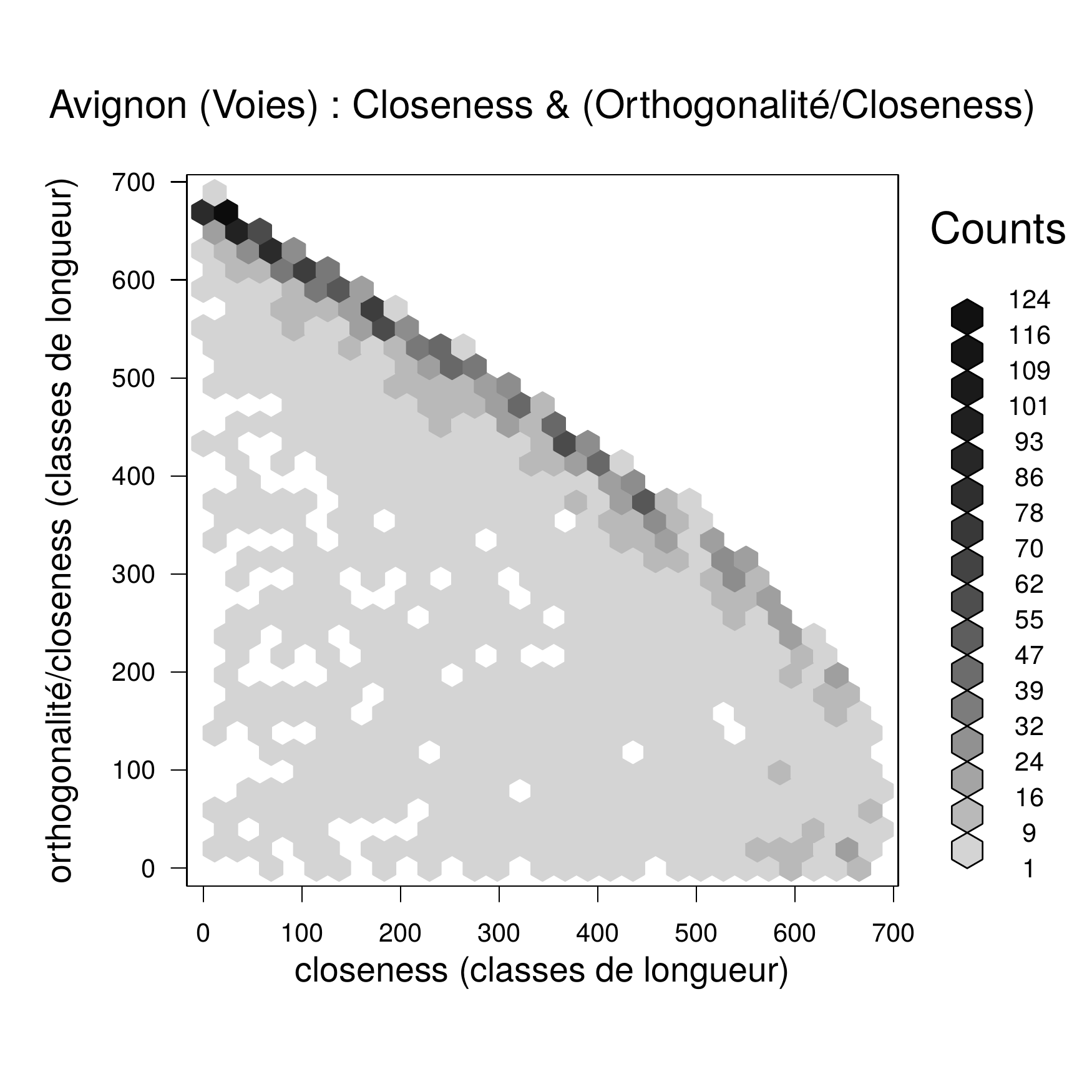}
    \end{subfigure}
    ~
    \begin{subfigure}[t]{0.45\textwidth}
        \includegraphics[width=\linewidth]{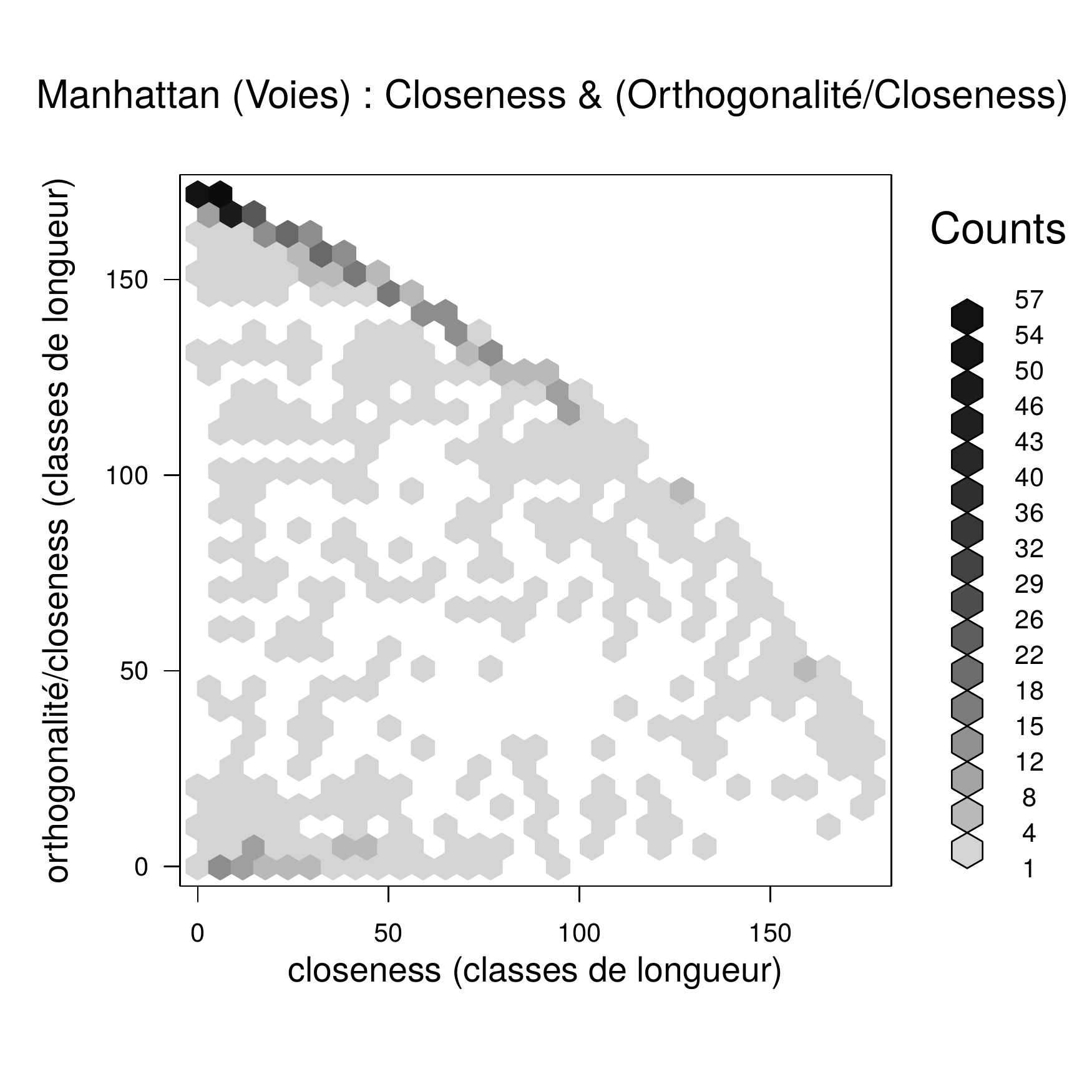}
    \end{subfigure}

    \begin{subfigure}[t]{0.45\textwidth}
        \includegraphics[width=\linewidth]{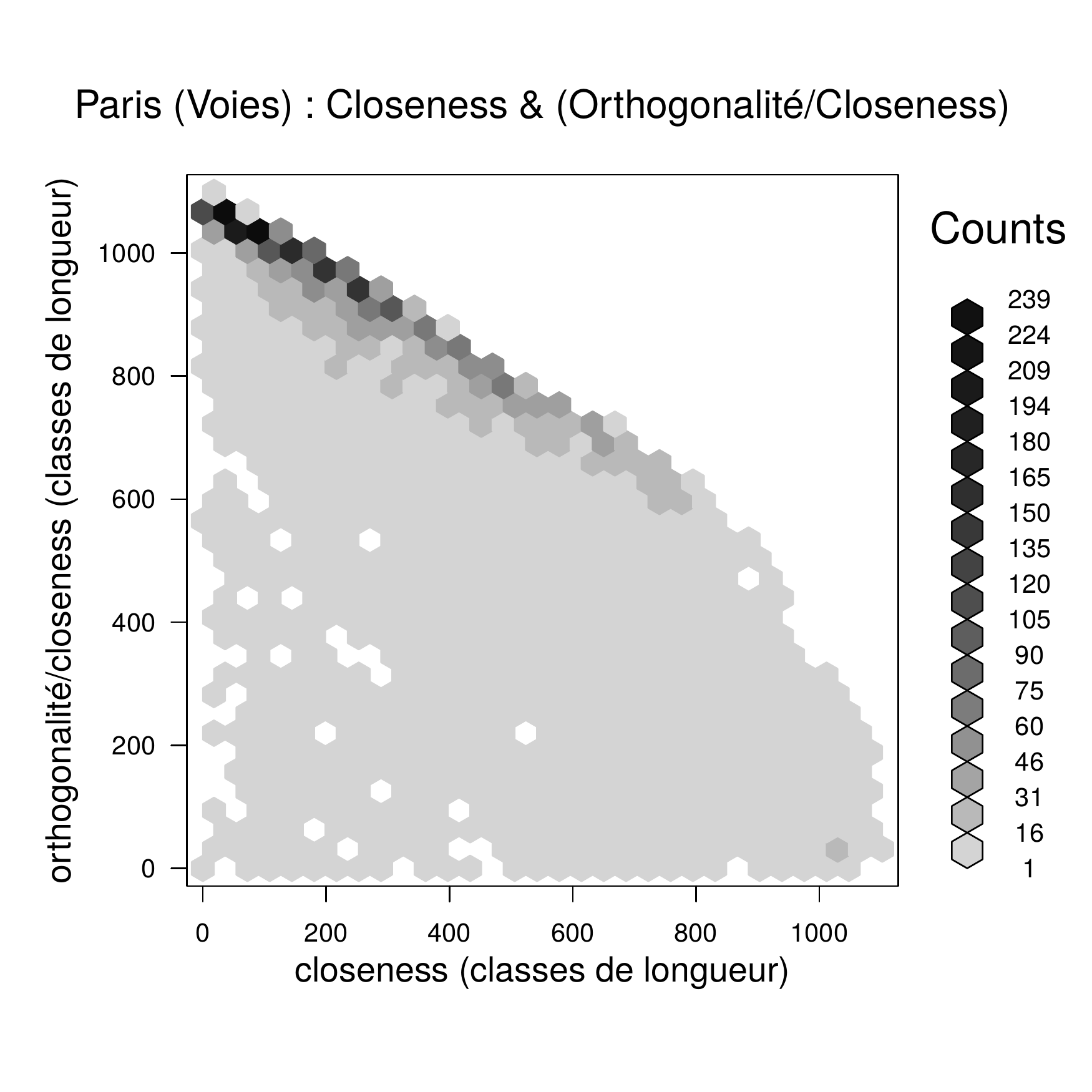}
    \end{subfigure}
    ~
    \begin{subfigure}[t]{0.45\textwidth}
        \includegraphics[width=\linewidth]{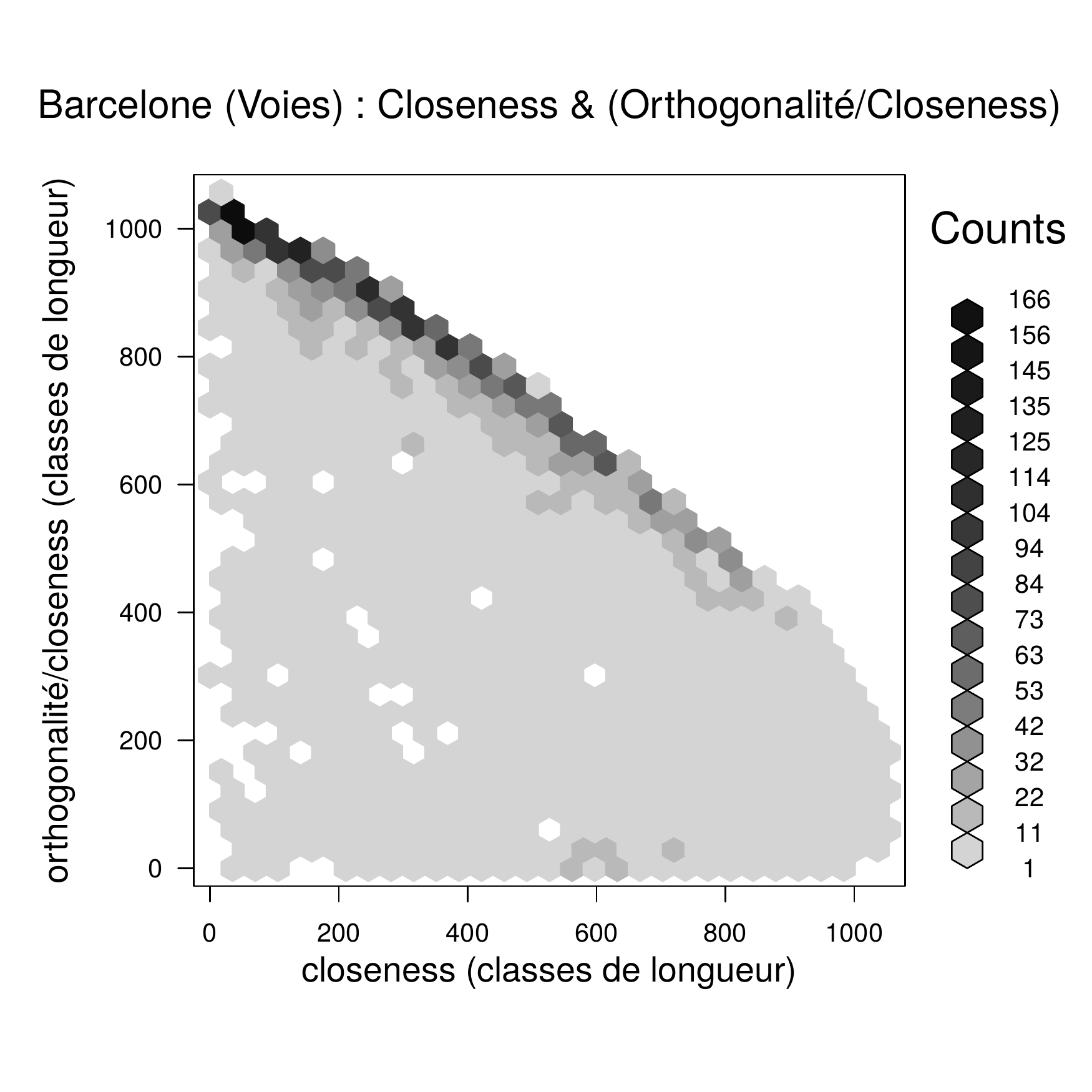}
    \end{subfigure}
    \caption{Closeness et orthogonalité sur closeness}
\end{figure}

\begin{figure}[h]\centering
    \begin{subfigure}[t]{0.45\textwidth}
        \includegraphics[width=\linewidth]{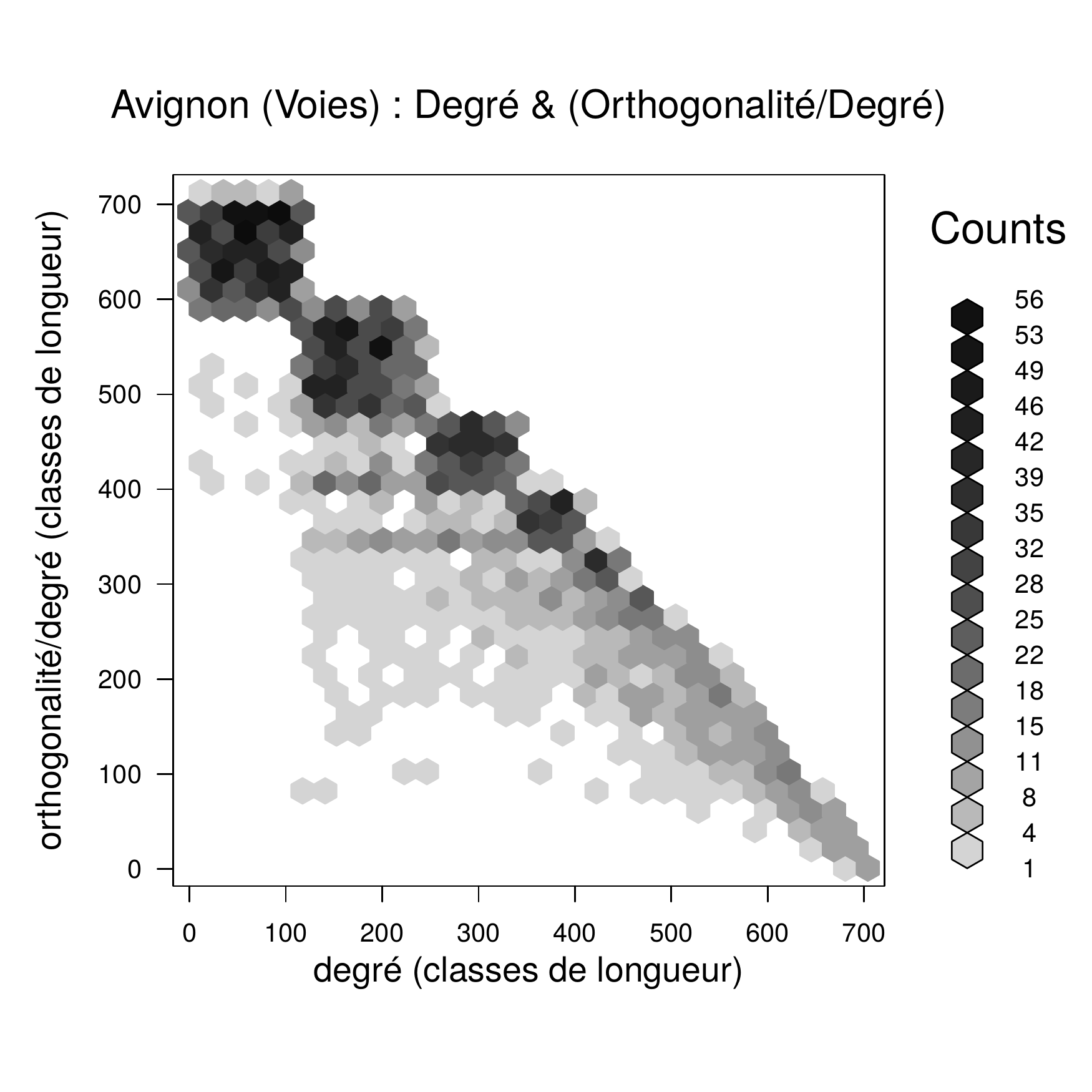}
    \end{subfigure}
    ~
    \begin{subfigure}[t]{0.45\textwidth}
        \includegraphics[width=\linewidth]{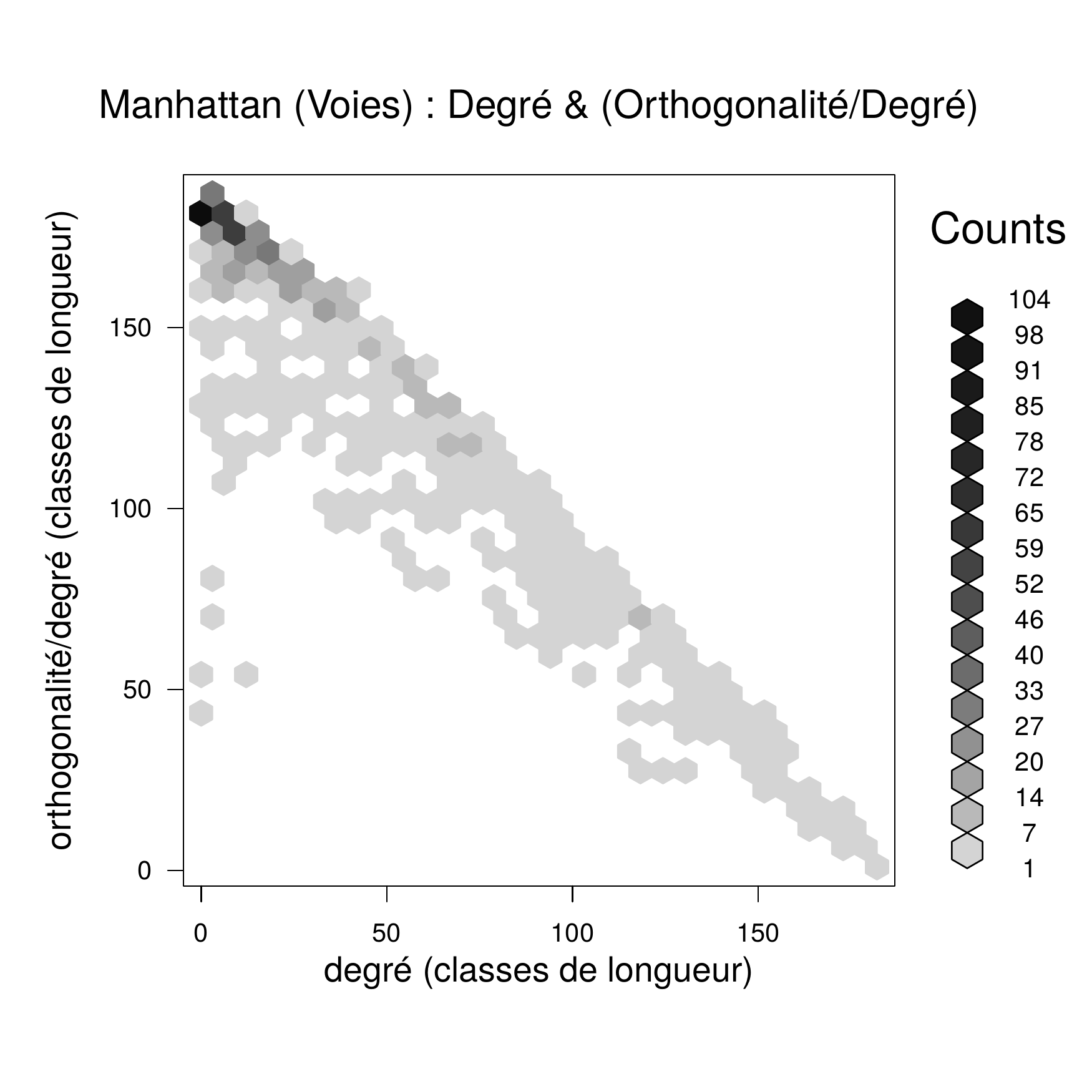}
    \end{subfigure}

    \begin{subfigure}[t]{0.45\textwidth}
        \includegraphics[width=\linewidth]{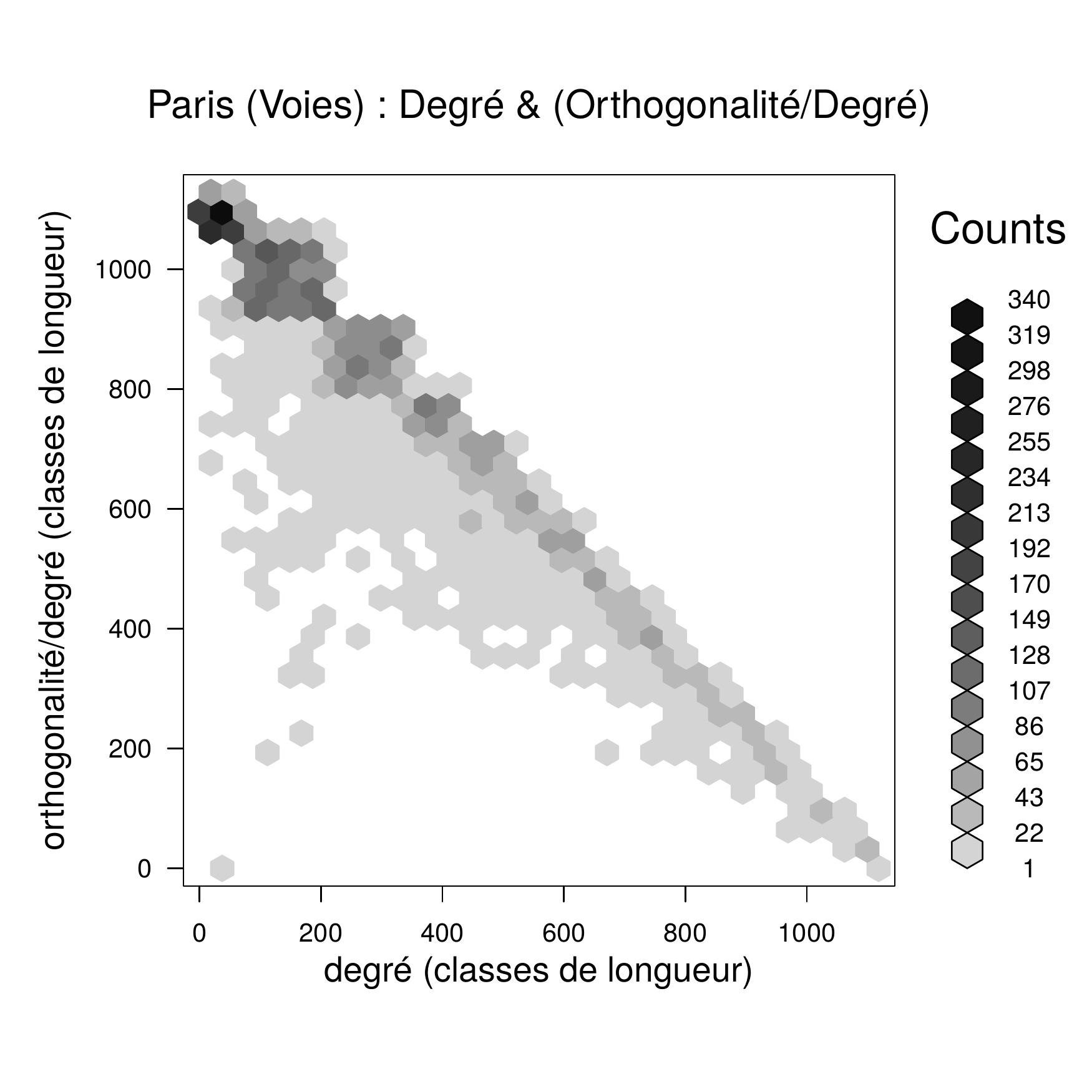}
    \end{subfigure}
    ~
    \begin{subfigure}[t]{0.45\textwidth}
        \includegraphics[width=\linewidth]{images/cartes_hexbin/ind_comp/voies_Barcelone_degree_ood.pdf}
    \end{subfigure}
    \caption{Degré et orthogonalité sur degré}
\end{figure}

\begin{figure}[h]\centering
    \begin{subfigure}[t]{0.45\textwidth}
        \includegraphics[width=\linewidth]{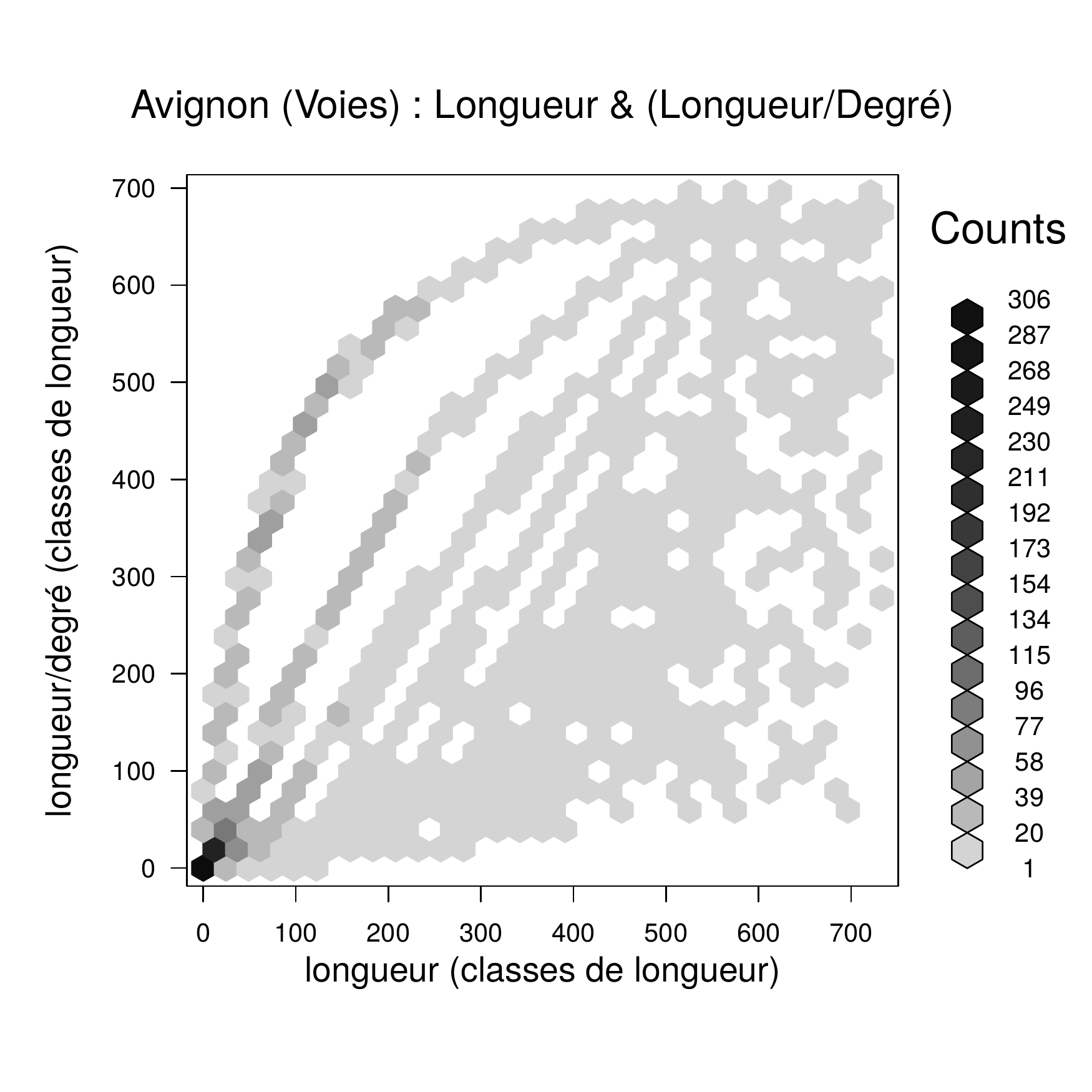}
    \end{subfigure}
    ~
    \begin{subfigure}[t]{0.45\textwidth}
        \includegraphics[width=\linewidth]{images/cartes_hexbin/ind_comp/voies_Manhattan_length_lod.pdf}
    \end{subfigure}

    \begin{subfigure}[t]{0.45\textwidth}
        \includegraphics[width=\linewidth]{images/cartes_hexbin/ind_comp/voies_Paris_length_lod.pdf}
    \end{subfigure}
    ~
    \begin{subfigure}[t]{0.45\textwidth}
        \includegraphics[width=\linewidth]{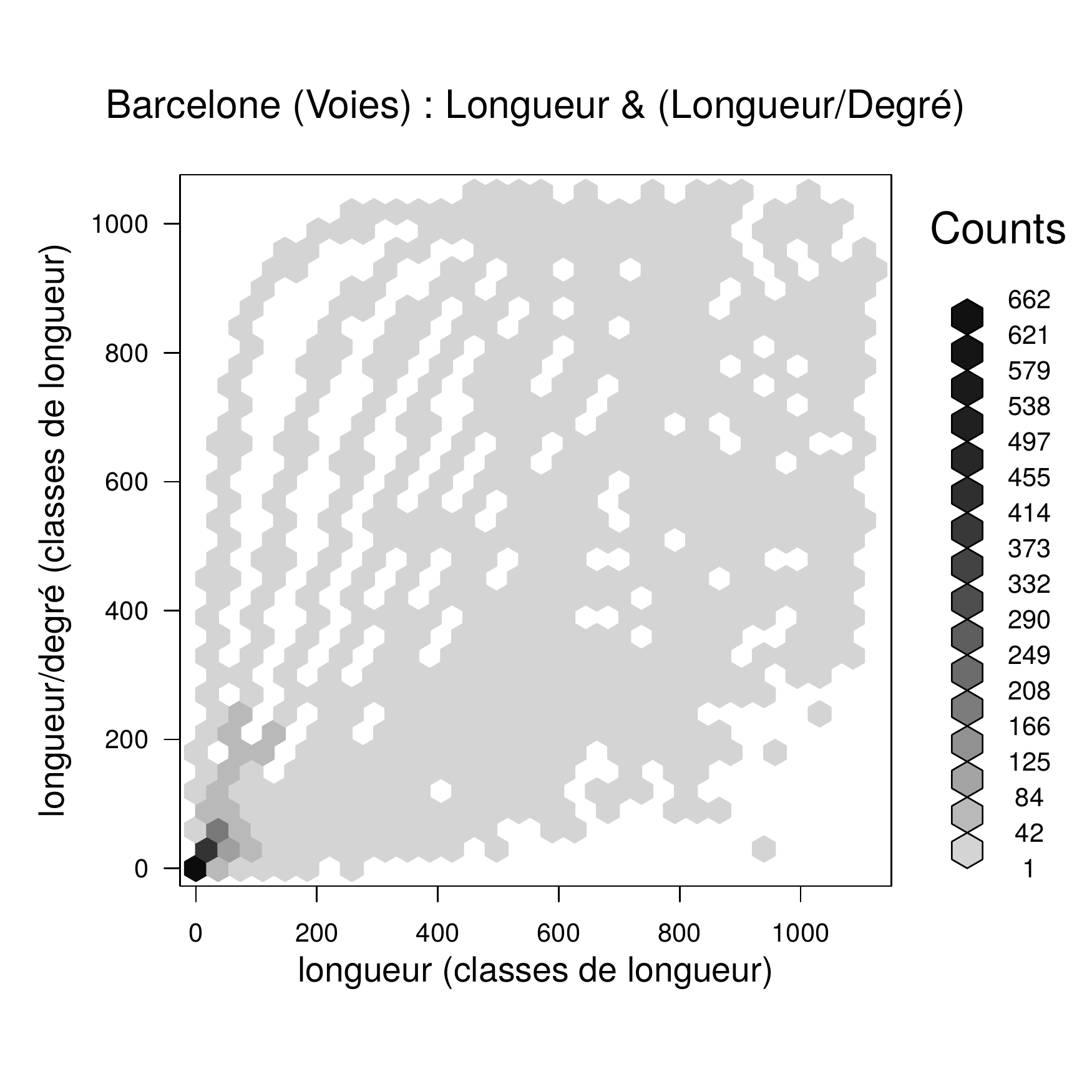}
    \end{subfigure}
    \caption{Longueur et longueur sur degré}
\end{figure}

\begin{figure}[h]\centering
    \begin{subfigure}[t]{0.45\textwidth}
        \includegraphics[width=\linewidth]{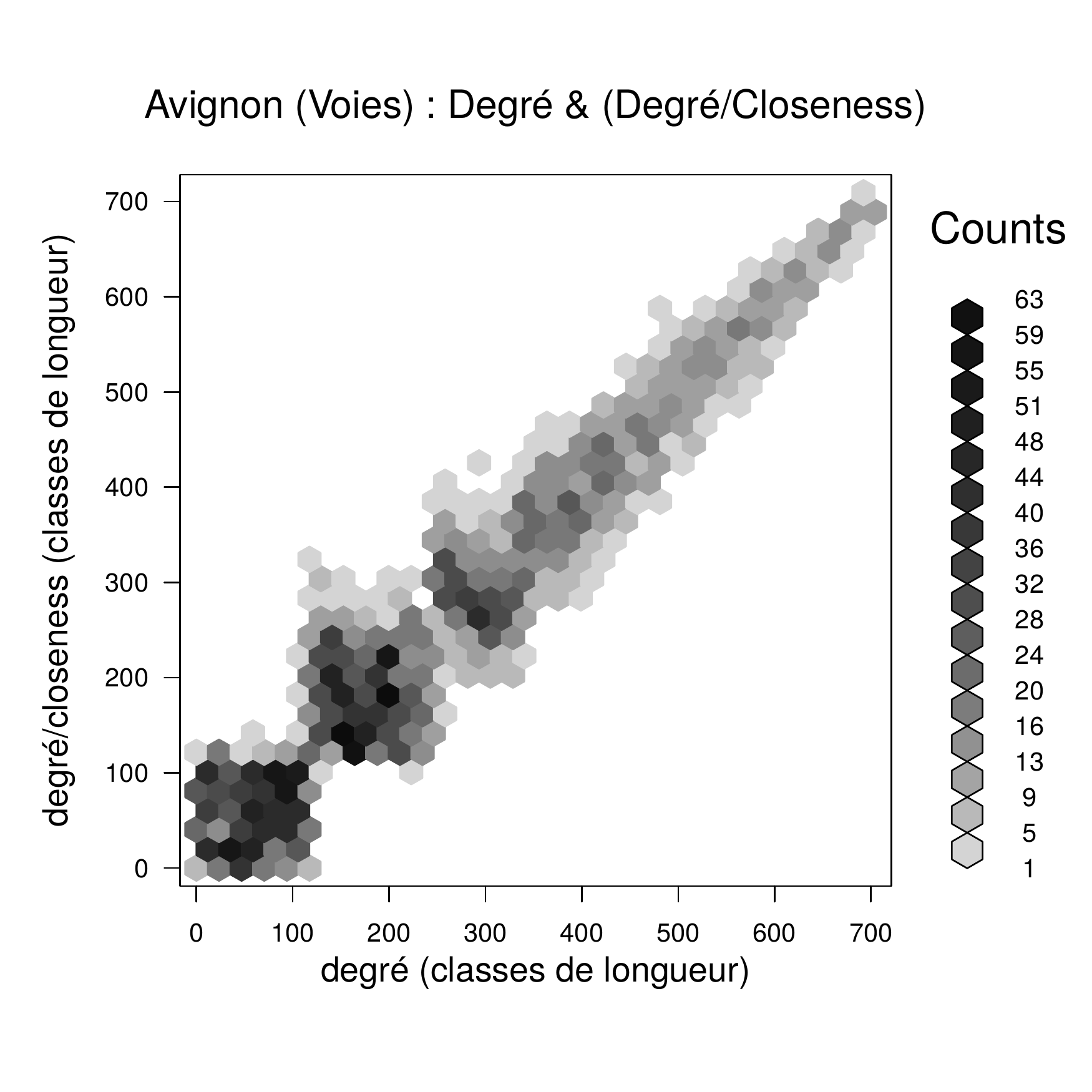}
    \end{subfigure}
    ~
    \begin{subfigure}[t]{0.45\textwidth}
        \includegraphics[width=\linewidth]{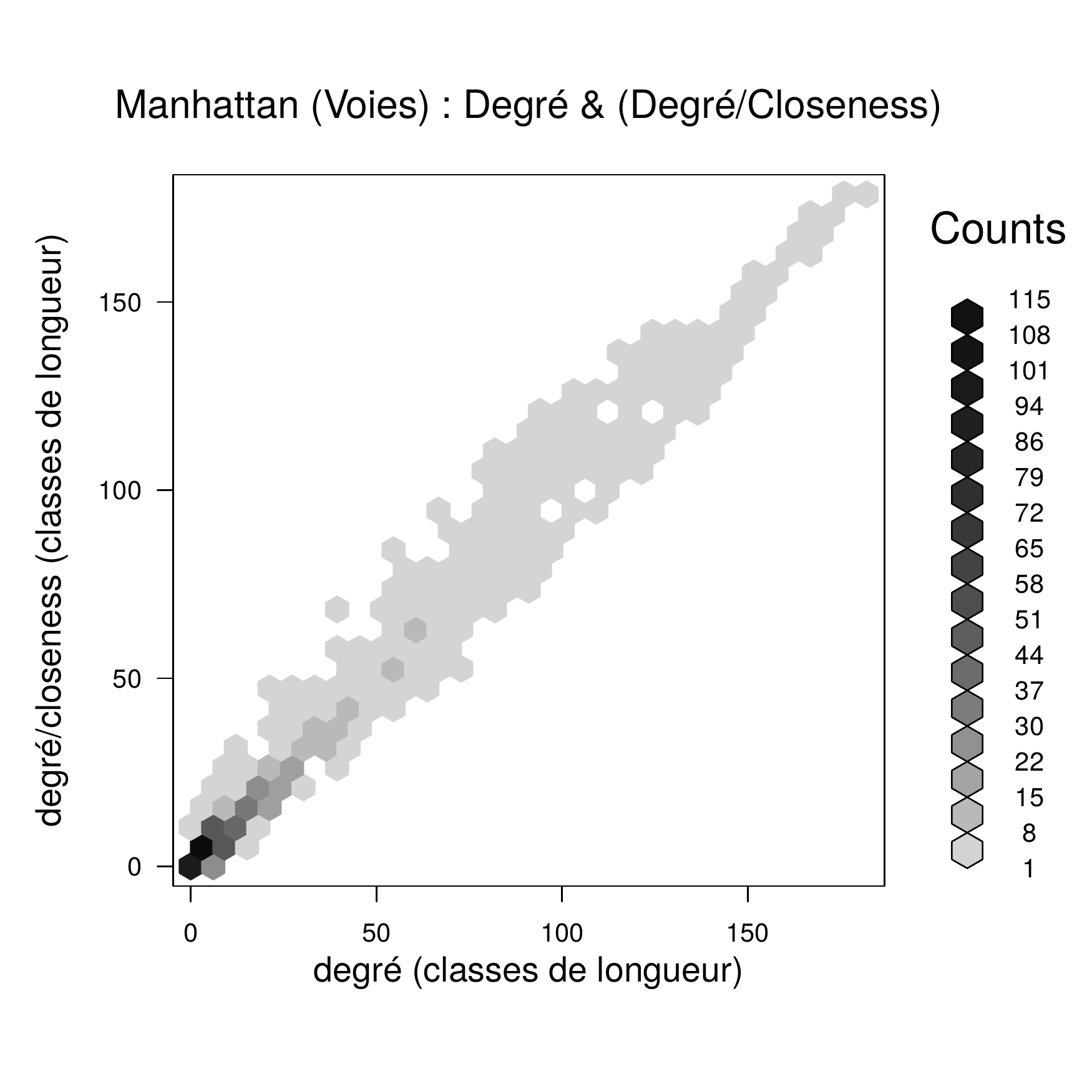}
    \end{subfigure}

    \begin{subfigure}[t]{0.45\textwidth}
        \includegraphics[width=\linewidth]{images/cartes_hexbin/ind_comp/voies_Paris_degree_doc.pdf}
    \end{subfigure}
    ~
    \begin{subfigure}[t]{0.45\textwidth}
        \includegraphics[width=\linewidth]{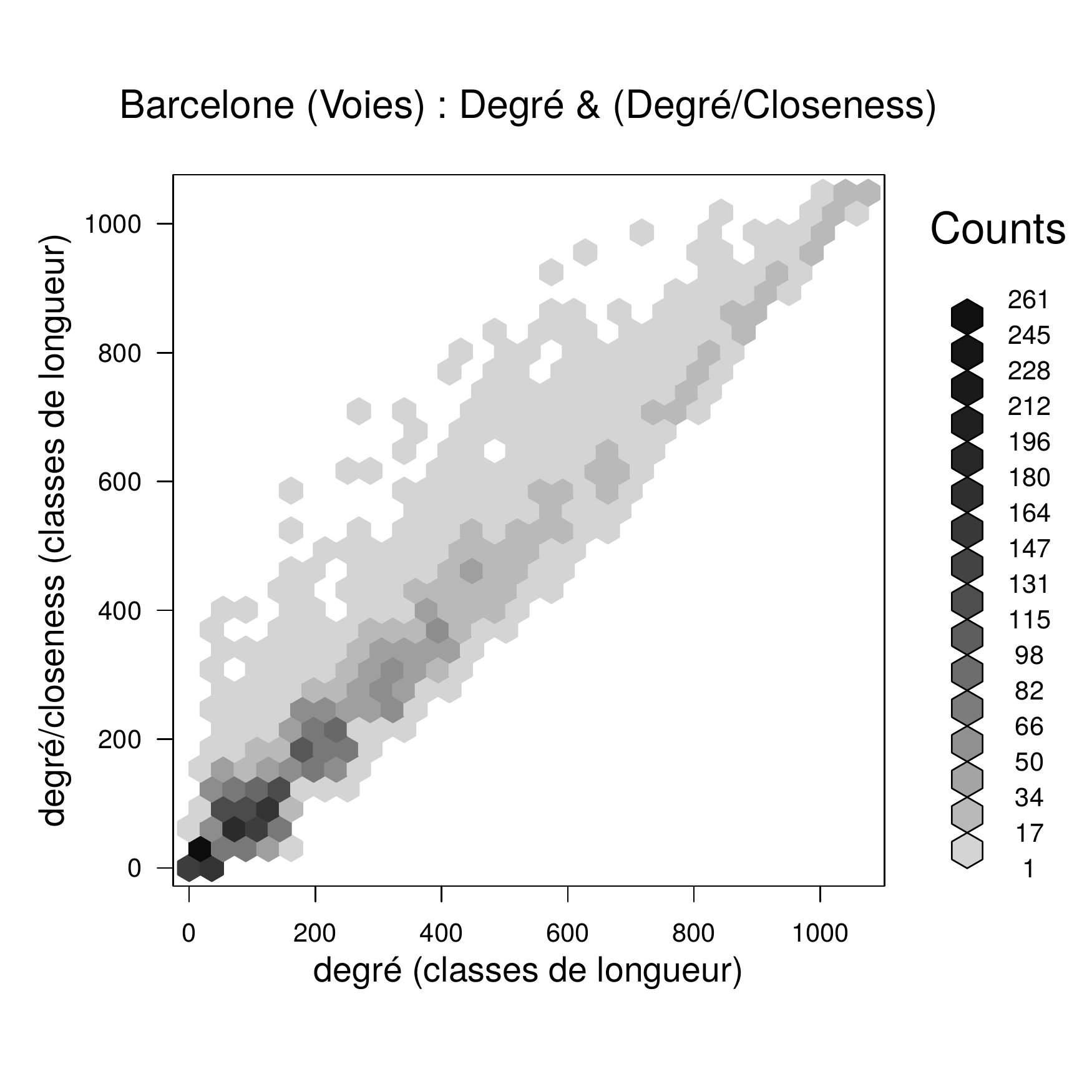}
    \end{subfigure}
    \caption{Degré et degré sur closeness}
\end{figure}

\begin{figure}[h]\centering
    \begin{subfigure}[t]{0.45\textwidth}
        \includegraphics[width=\linewidth]{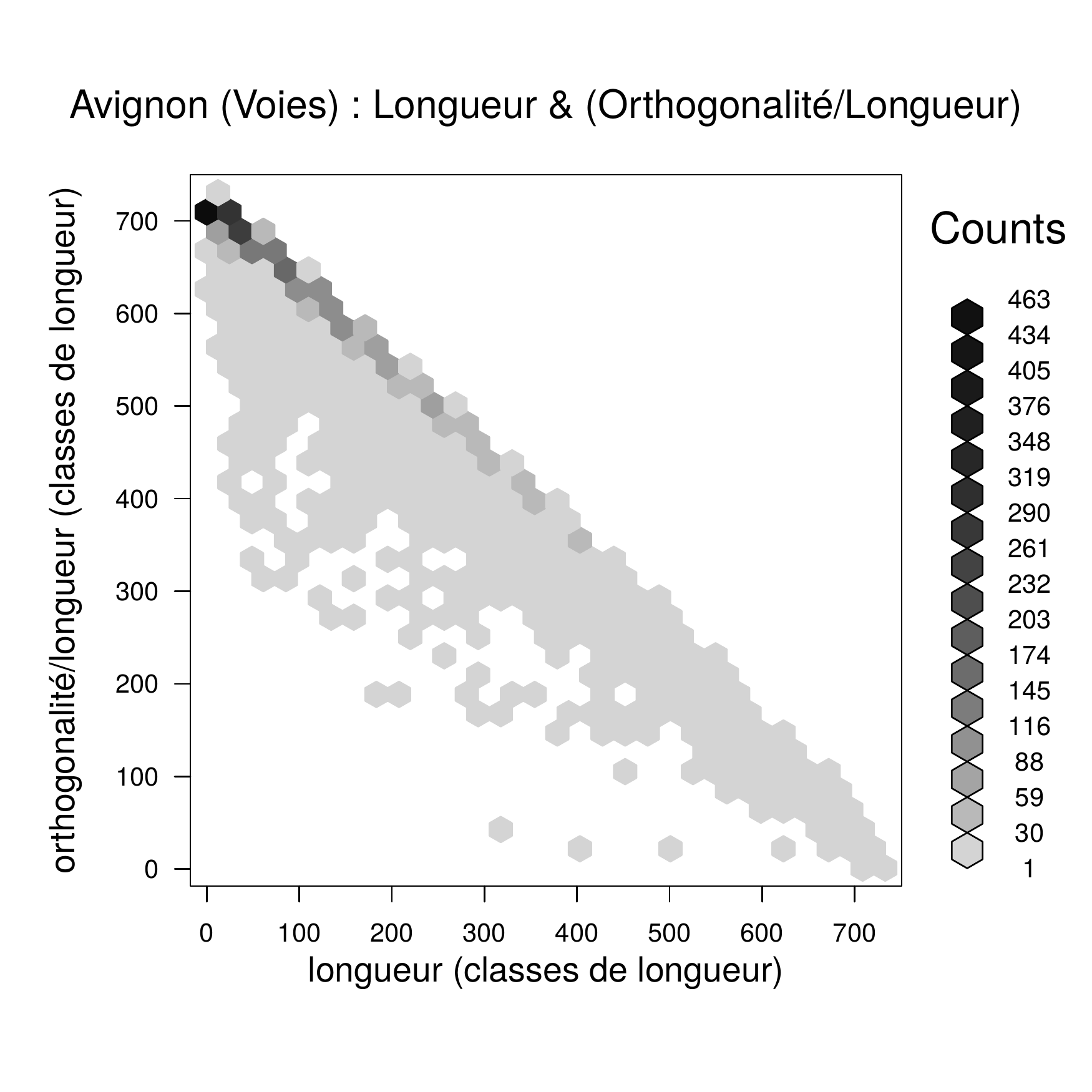}
    \end{subfigure}
    ~
    \begin{subfigure}[t]{0.45\textwidth}
        \includegraphics[width=\linewidth]{images/cartes_hexbin/ind_comp/voies_Manhattan_length_ool.pdf}
    \end{subfigure}

    \begin{subfigure}[t]{0.45\textwidth}
        \includegraphics[width=\linewidth]{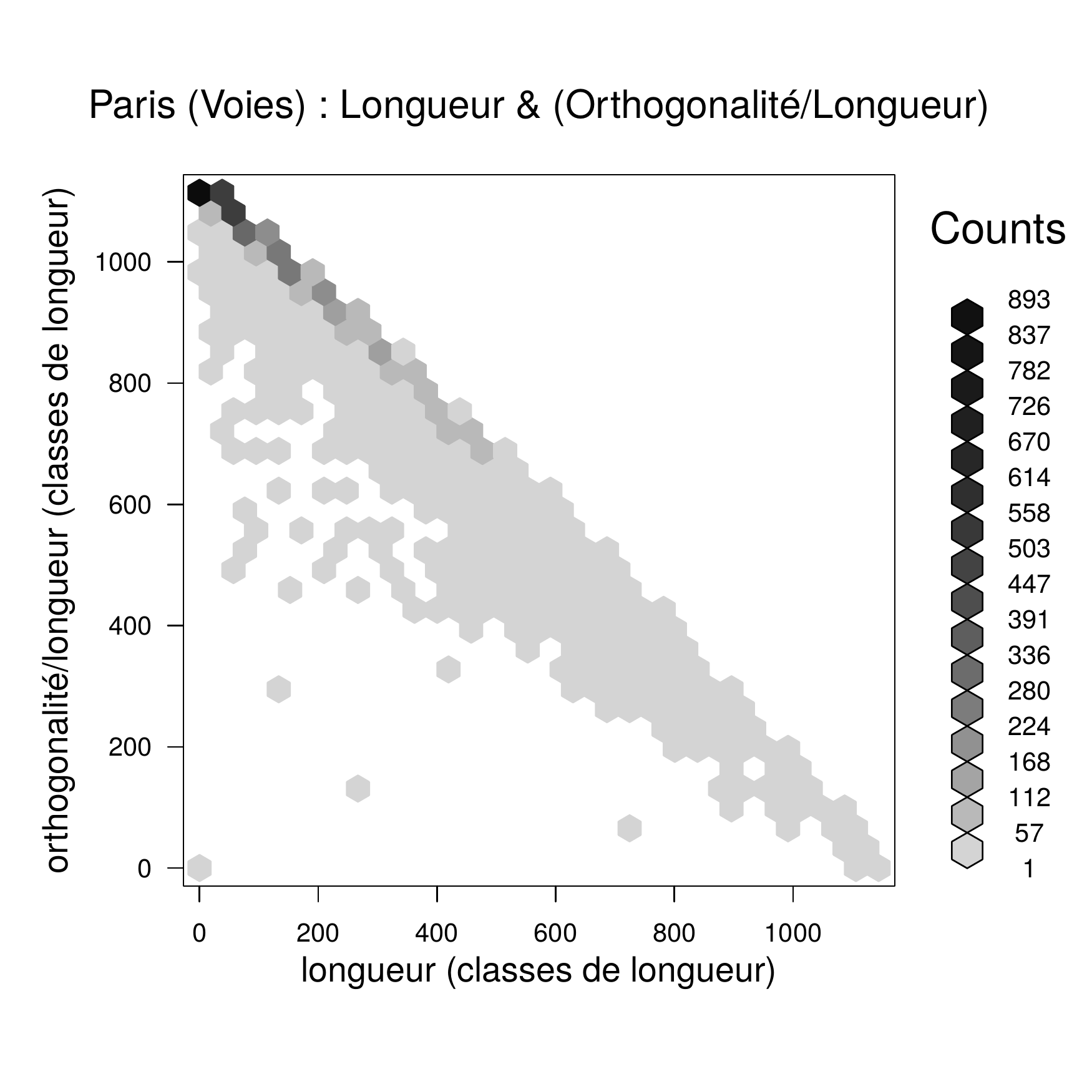}
    \end{subfigure}
    ~
    \begin{subfigure}[t]{0.45\textwidth}
        \includegraphics[width=\linewidth]{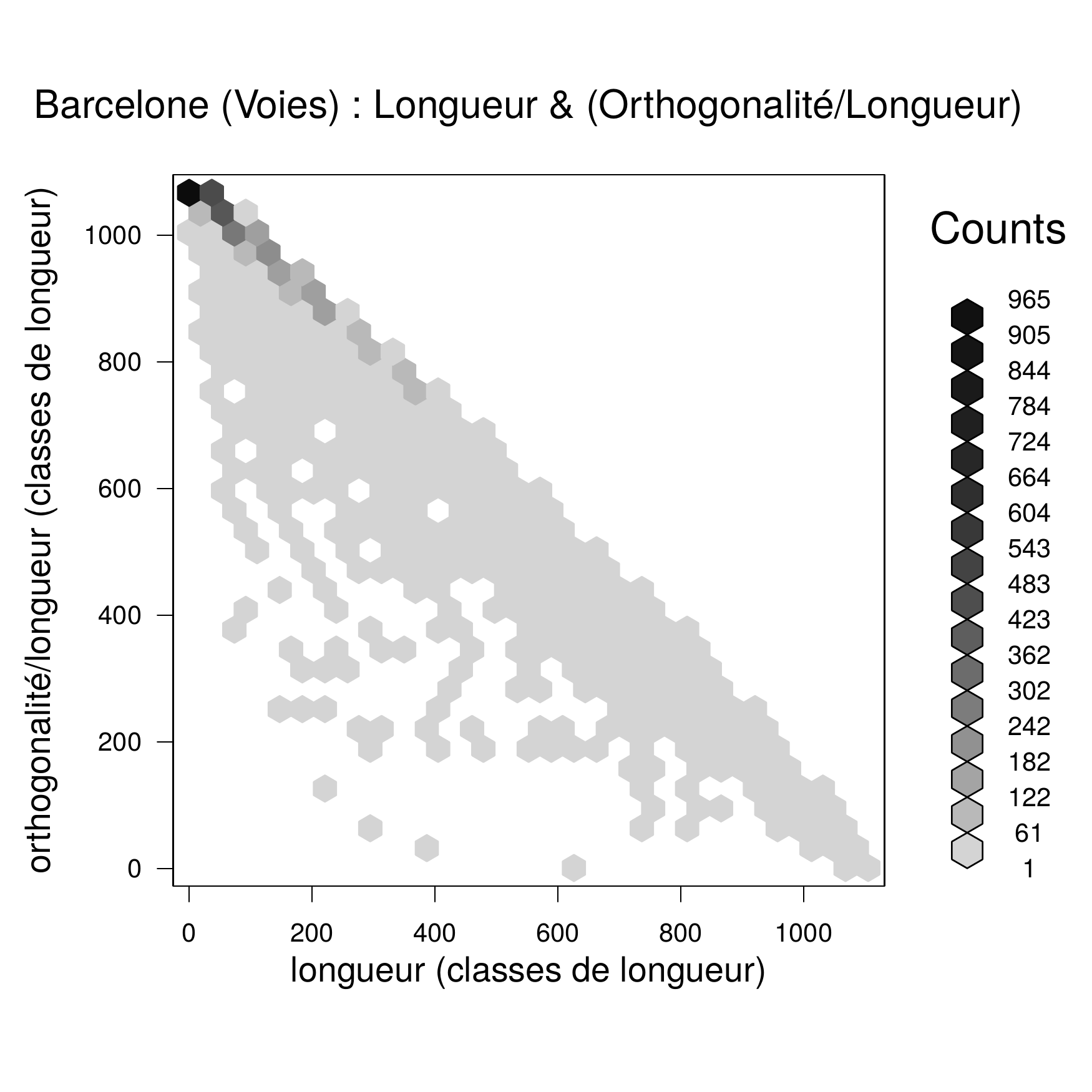}
    \end{subfigure}
    \caption{Longueur et orthogonalité sur longueur}
\end{figure}

\begin{figure}[h]\centering
    \begin{subfigure}[t]{0.45\textwidth}
        \includegraphics[width=\linewidth]{images/cartes_hexbin/ind_comp/voies_Avignon_degree_lod.pdf}
    \end{subfigure}
    ~
    \begin{subfigure}[t]{0.45\textwidth}
        \includegraphics[width=\linewidth]{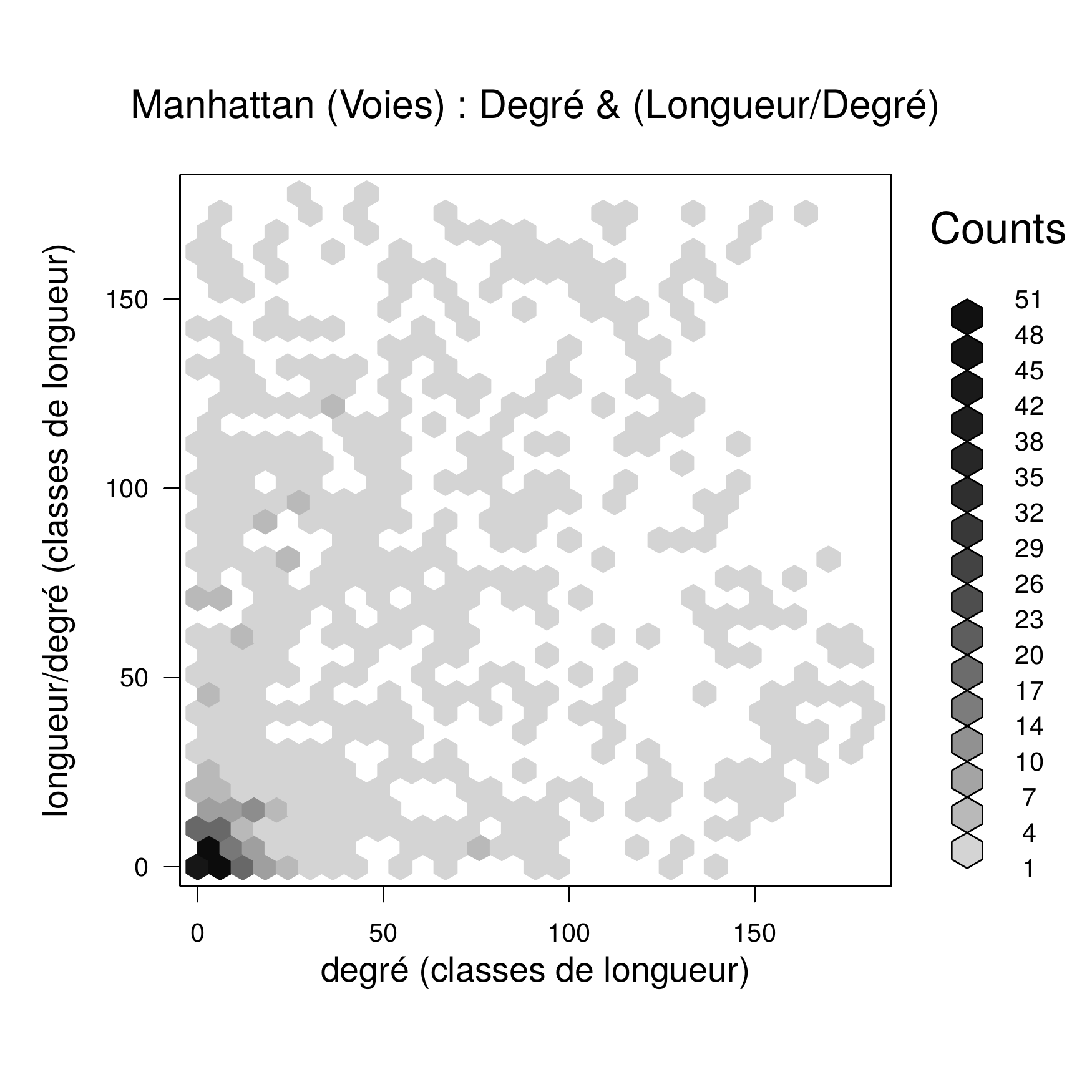}
    \end{subfigure}

    \begin{subfigure}[t]{0.45\textwidth}
        \includegraphics[width=\linewidth]{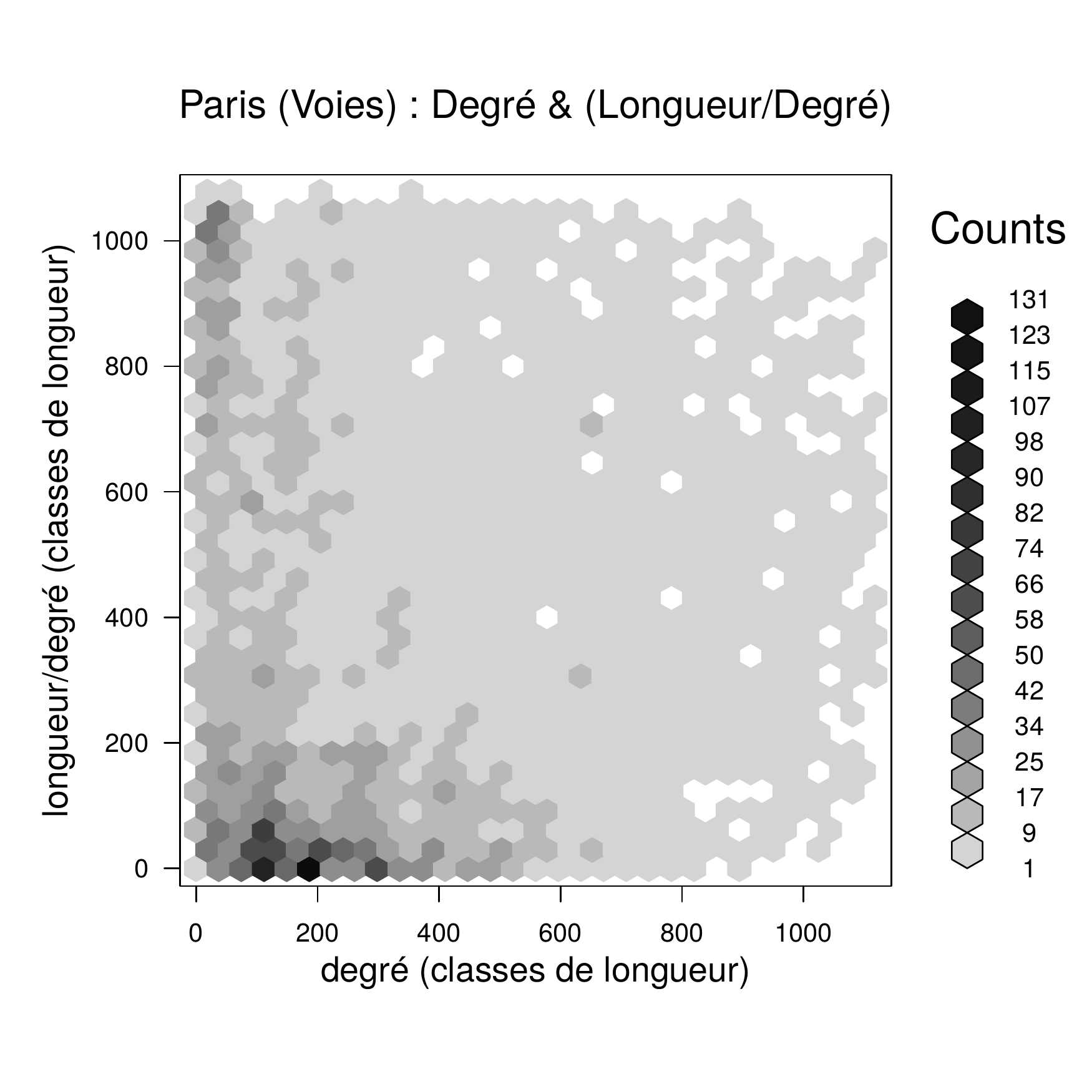}
    \end{subfigure}
    ~
    \begin{subfigure}[t]{0.45\textwidth}
        \includegraphics[width=\linewidth]{images/cartes_hexbin/ind_comp/voies_Barcelone_degree_lod.pdf}
    \end{subfigure}
    \caption{Degré et longueur sur degré}
\end{figure}

\begin{figure}[h]\centering
    \begin{subfigure}[t]{0.45\textwidth}
        \includegraphics[width=\linewidth]{images/cartes_hexbin/ind_comp/voies_Avignon_length_loc.pdf}
    \end{subfigure}
    ~
    \begin{subfigure}[t]{0.45\textwidth}
        \includegraphics[width=\linewidth]{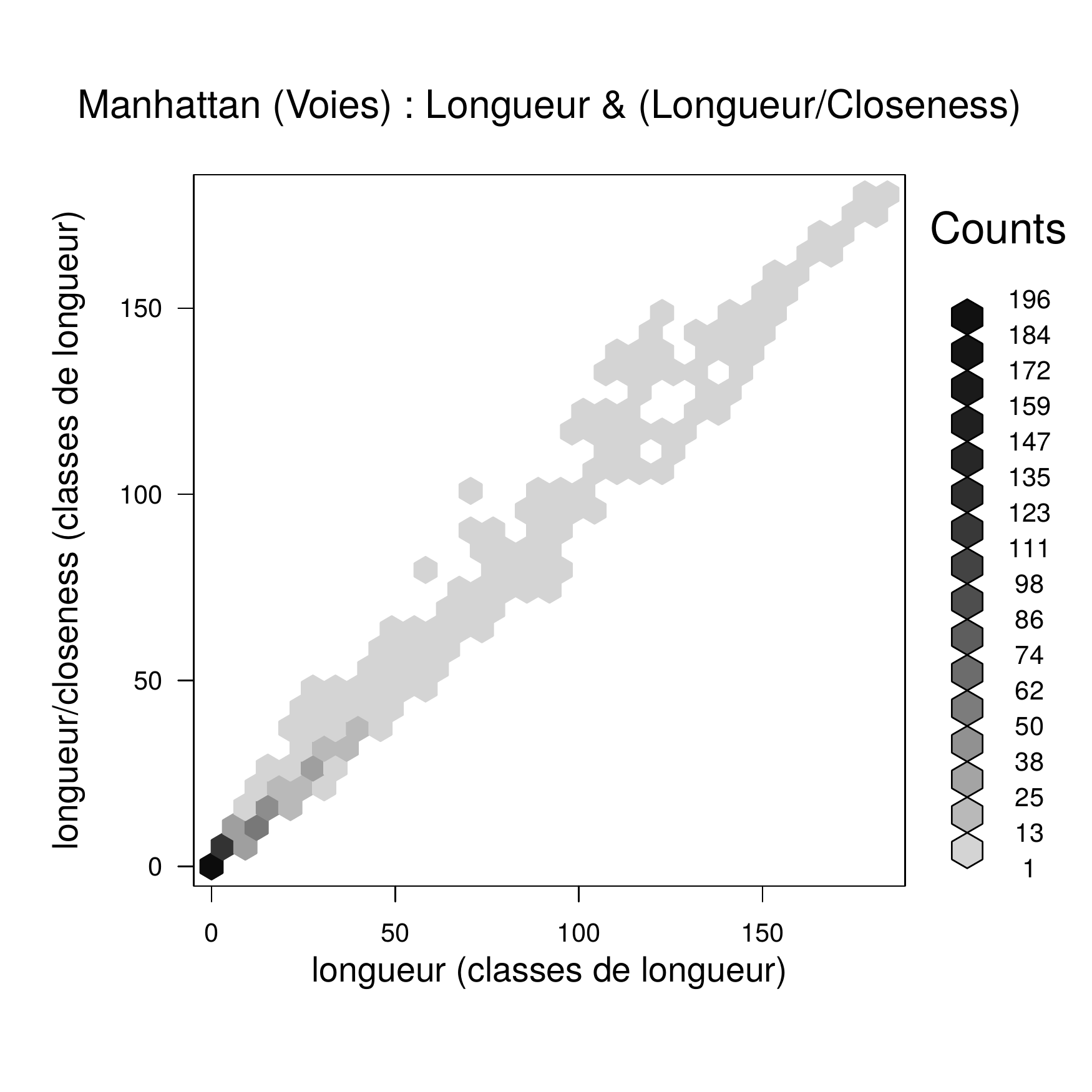}
    \end{subfigure}

    \begin{subfigure}[t]{0.45\textwidth}
        \includegraphics[width=\linewidth]{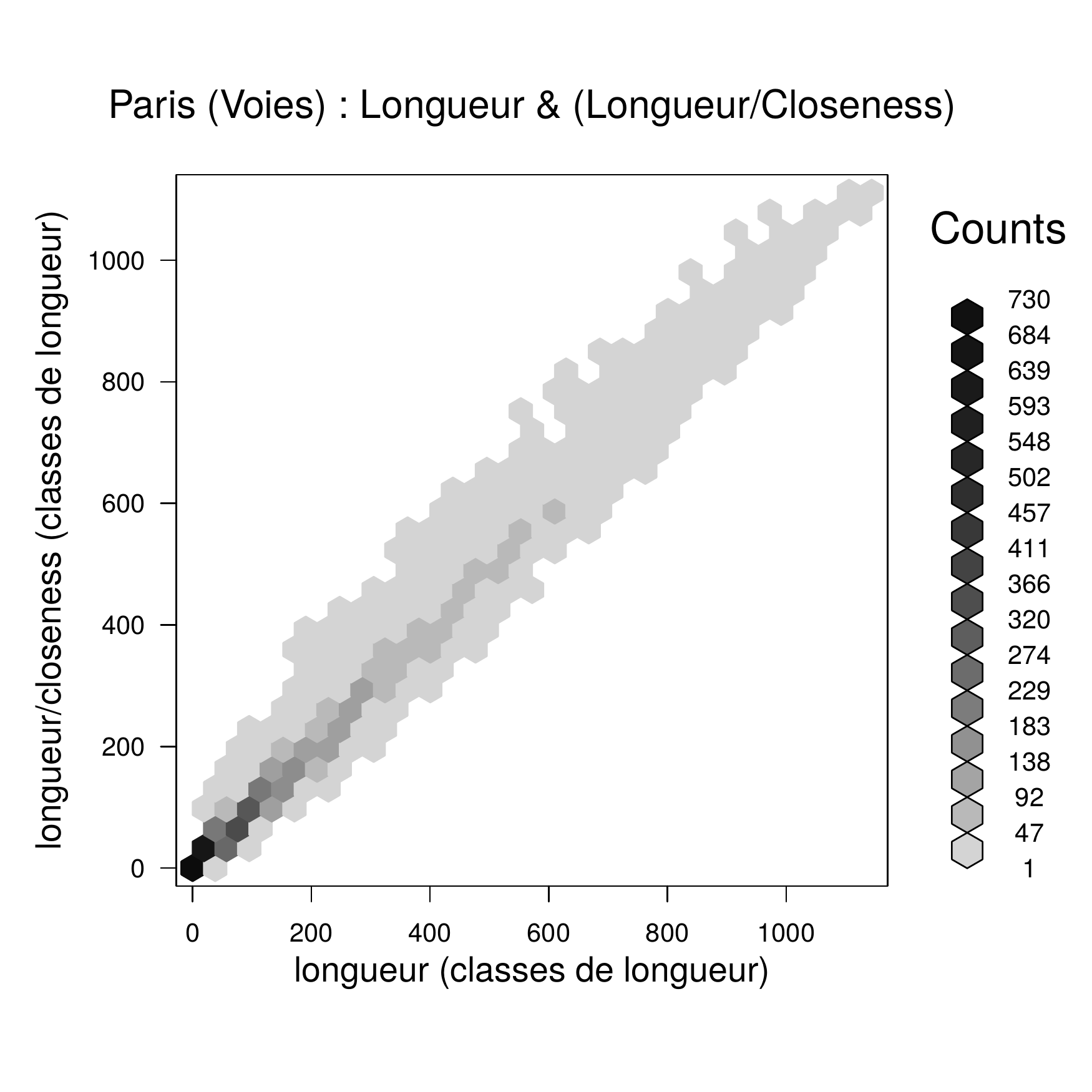}
    \end{subfigure}
    ~
    \begin{subfigure}[t]{0.45\textwidth}
        \includegraphics[width=\linewidth]{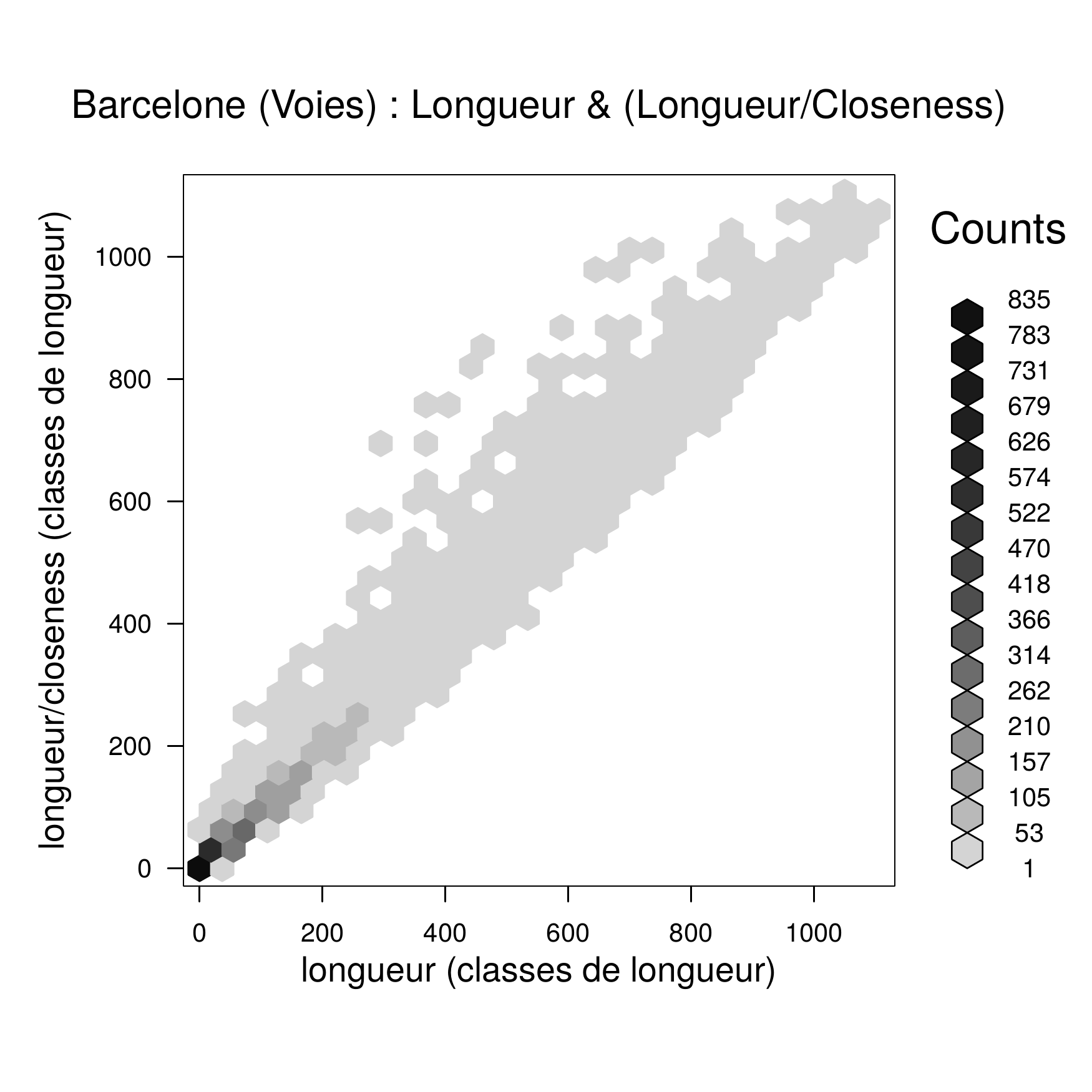}
    \end{subfigure}
    \caption{Longueur et longueur sur closeness}
\end{figure}

\begin{figure}[h]\centering
    \begin{subfigure}[t]{0.45\textwidth}
        \includegraphics[width=\linewidth]{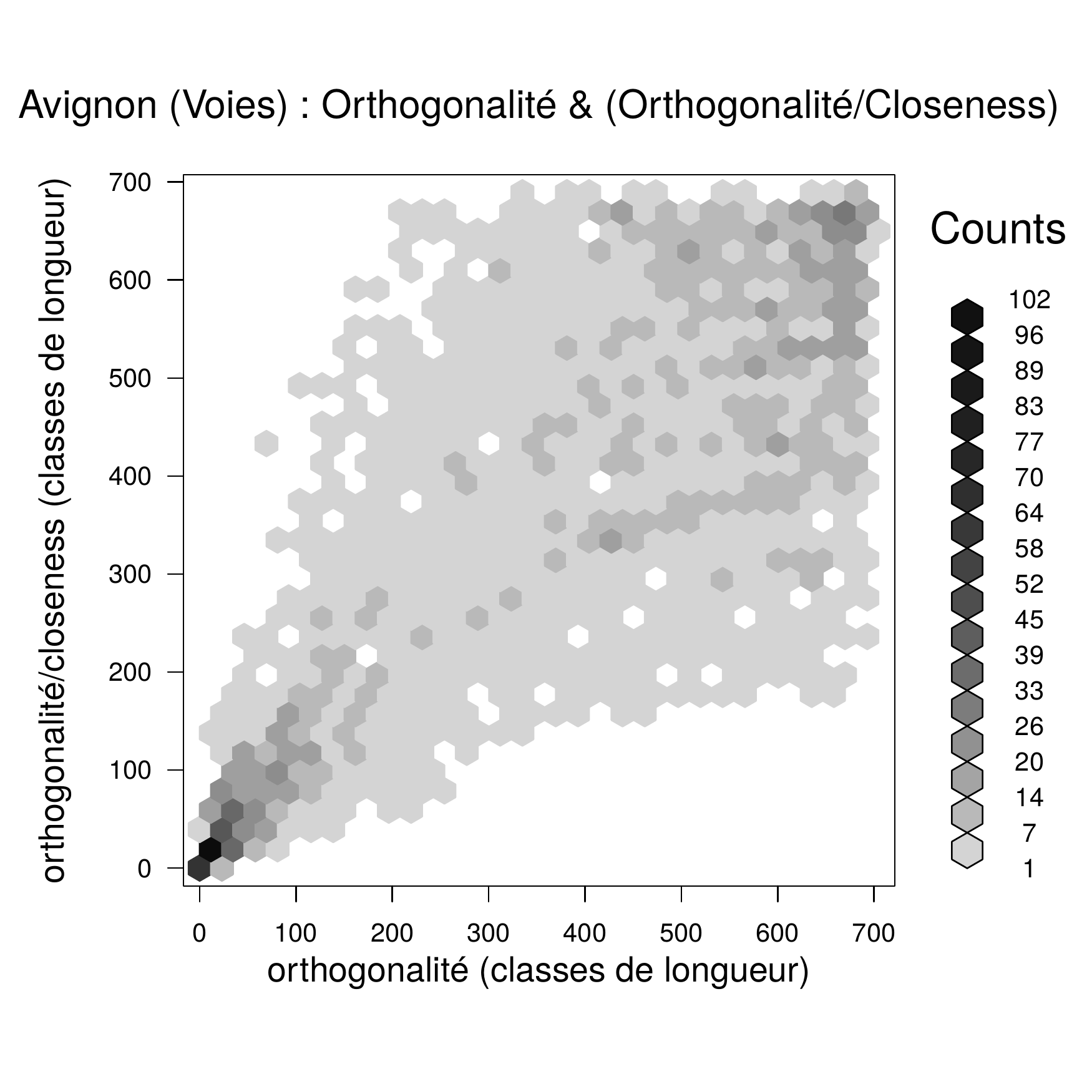}
    \end{subfigure}
    ~
    \begin{subfigure}[t]{0.45\textwidth}
        \includegraphics[width=\linewidth]{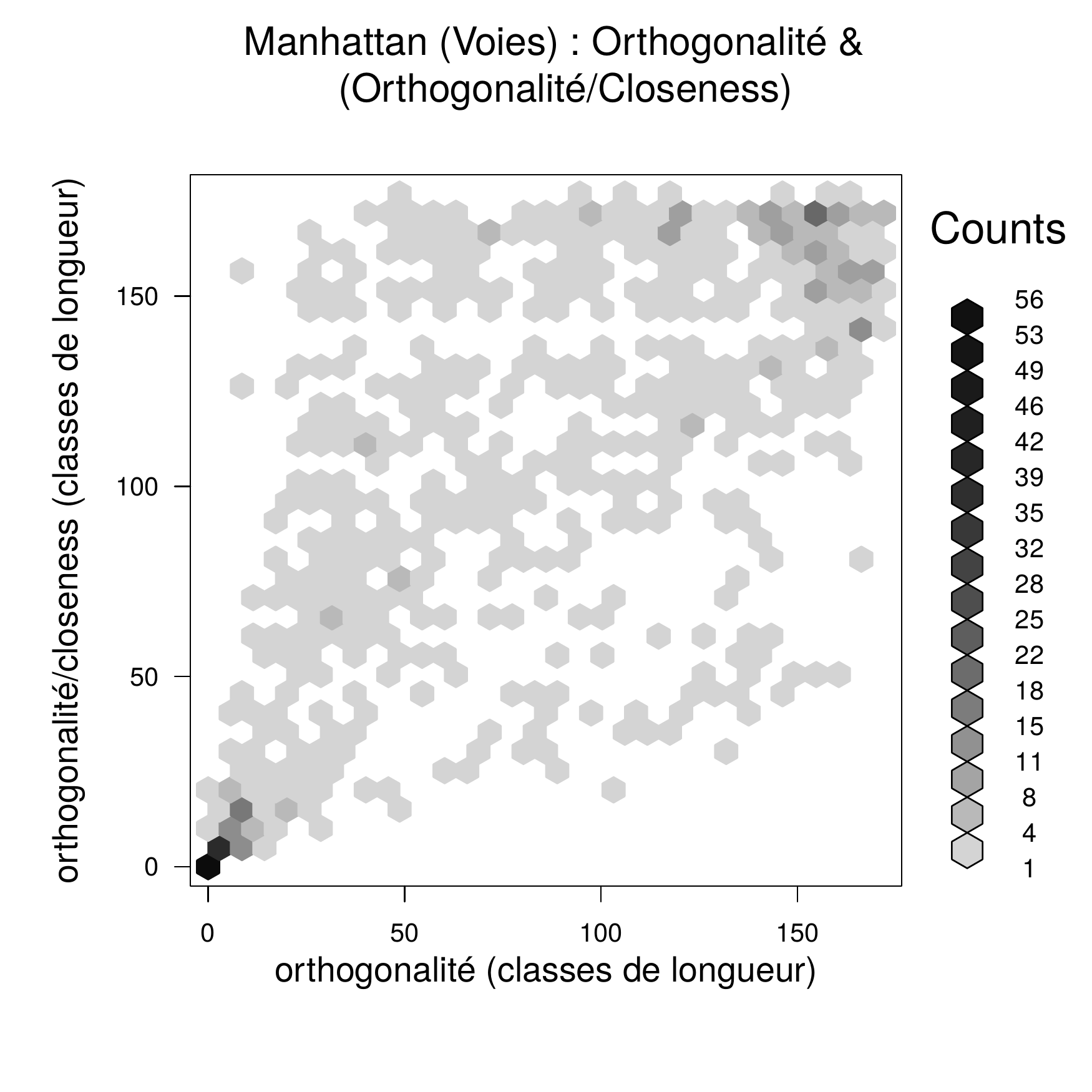}
    \end{subfigure}

    \begin{subfigure}[t]{0.45\textwidth}
        \includegraphics[width=\linewidth]{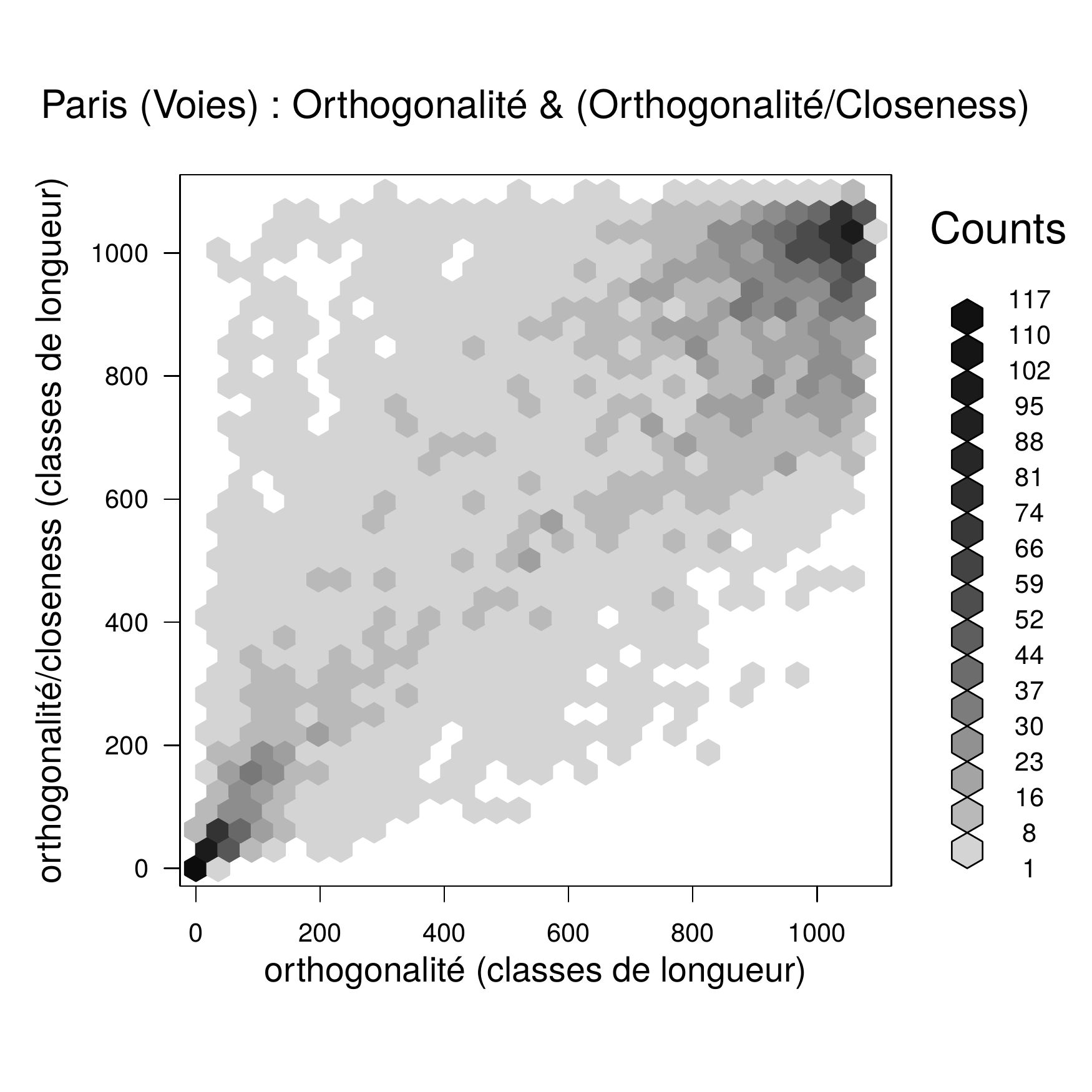}
    \end{subfigure}
    ~
    \begin{subfigure}[t]{0.45\textwidth}
        \includegraphics[width=\linewidth]{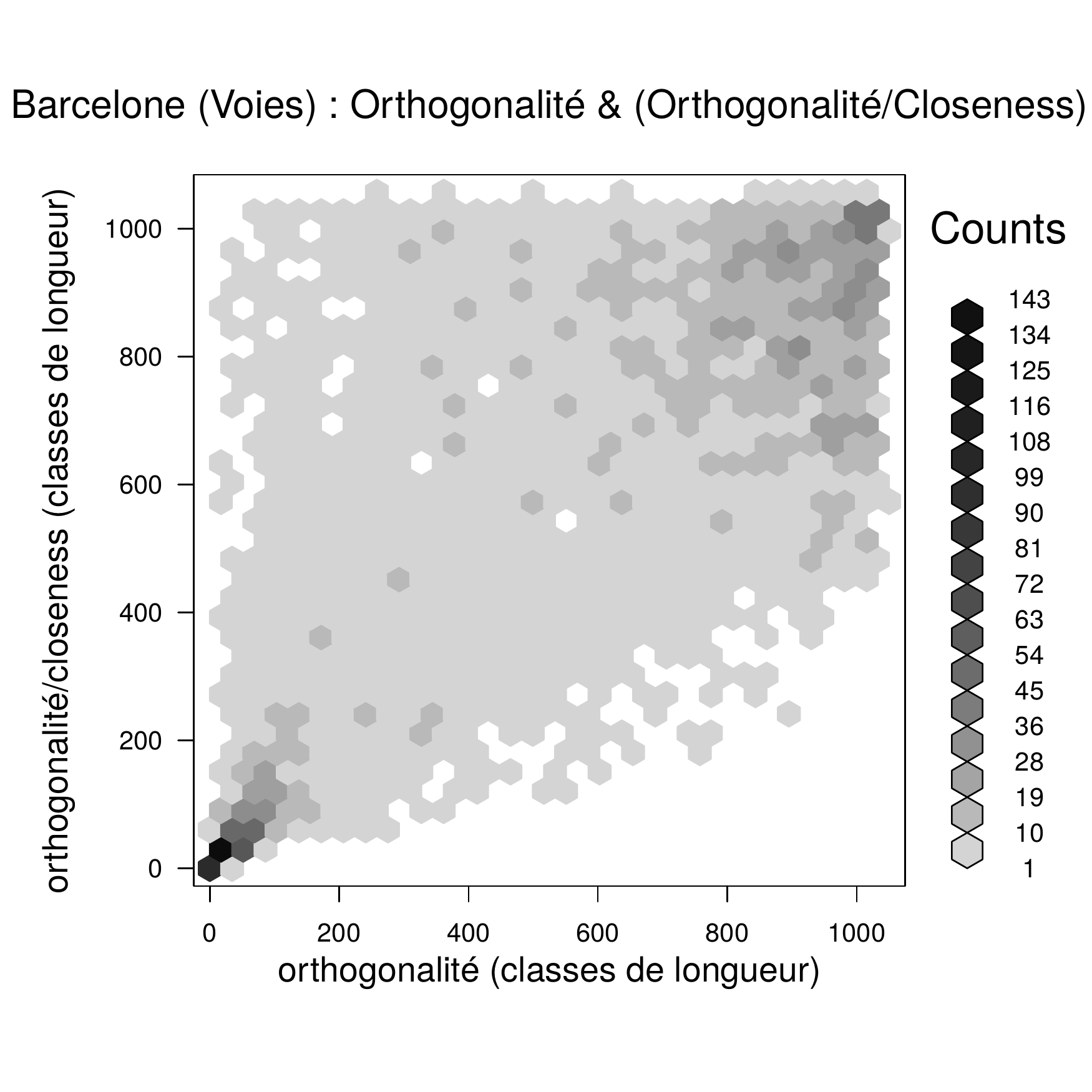}
    \end{subfigure}
    \caption{Orthogonalité et orthogonalité sur closeness}
\end{figure}

\begin{figure}[h]\centering
    \begin{subfigure}[t]{0.45\textwidth}
        \includegraphics[width=\linewidth]{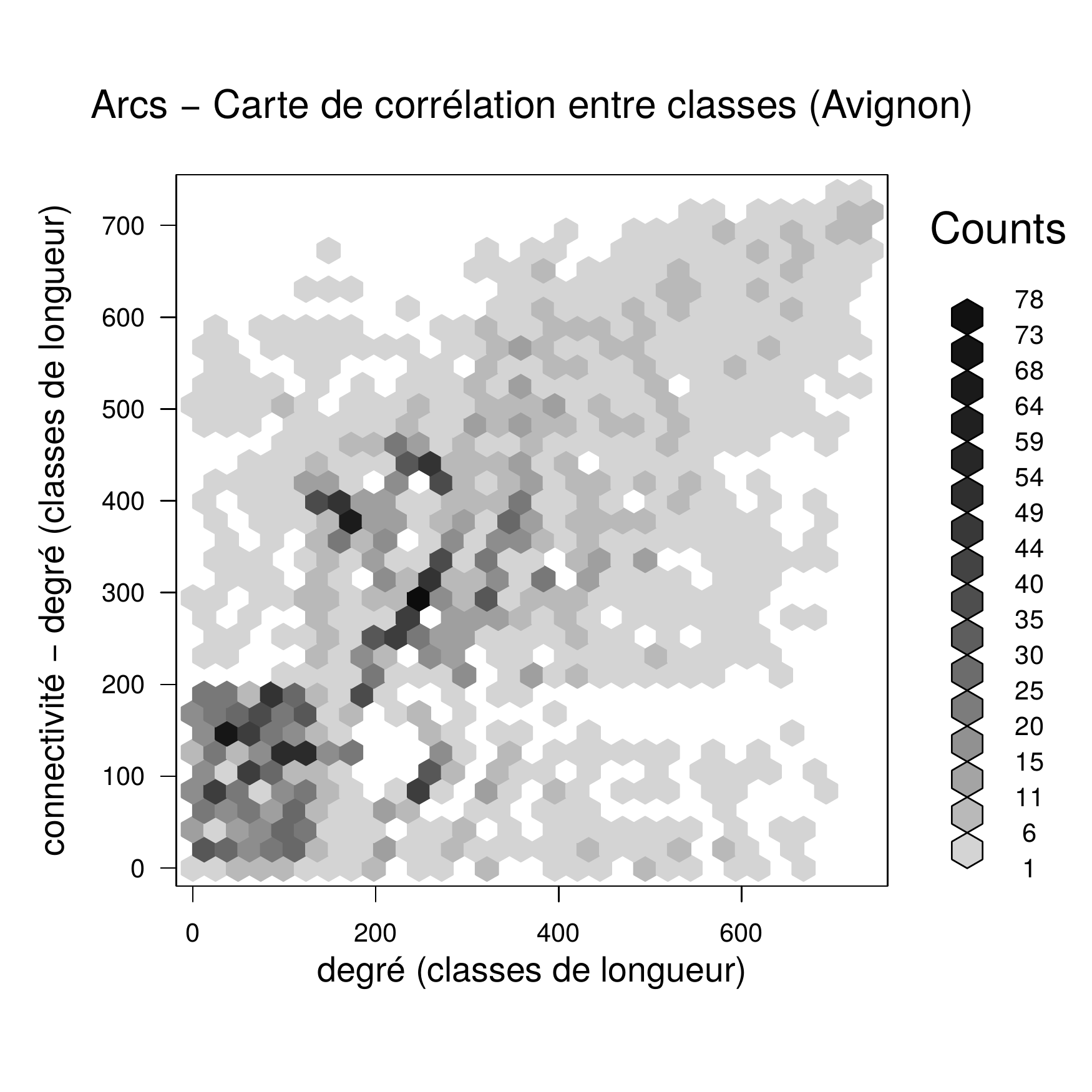}
    \end{subfigure}
    \caption{Degré et degré de desserte}
\end{figure}

\FloatBarrier

\FloatBarrier 
\chapter{Cartes d'effets de bord}\label{ann:chap_ebords}


\FloatBarrier
\section{Avignon}\label{ann:sec_eb_avignon}

\begin{figure}[t]
    \centering
    \includegraphics[width=0.8\textwidth]{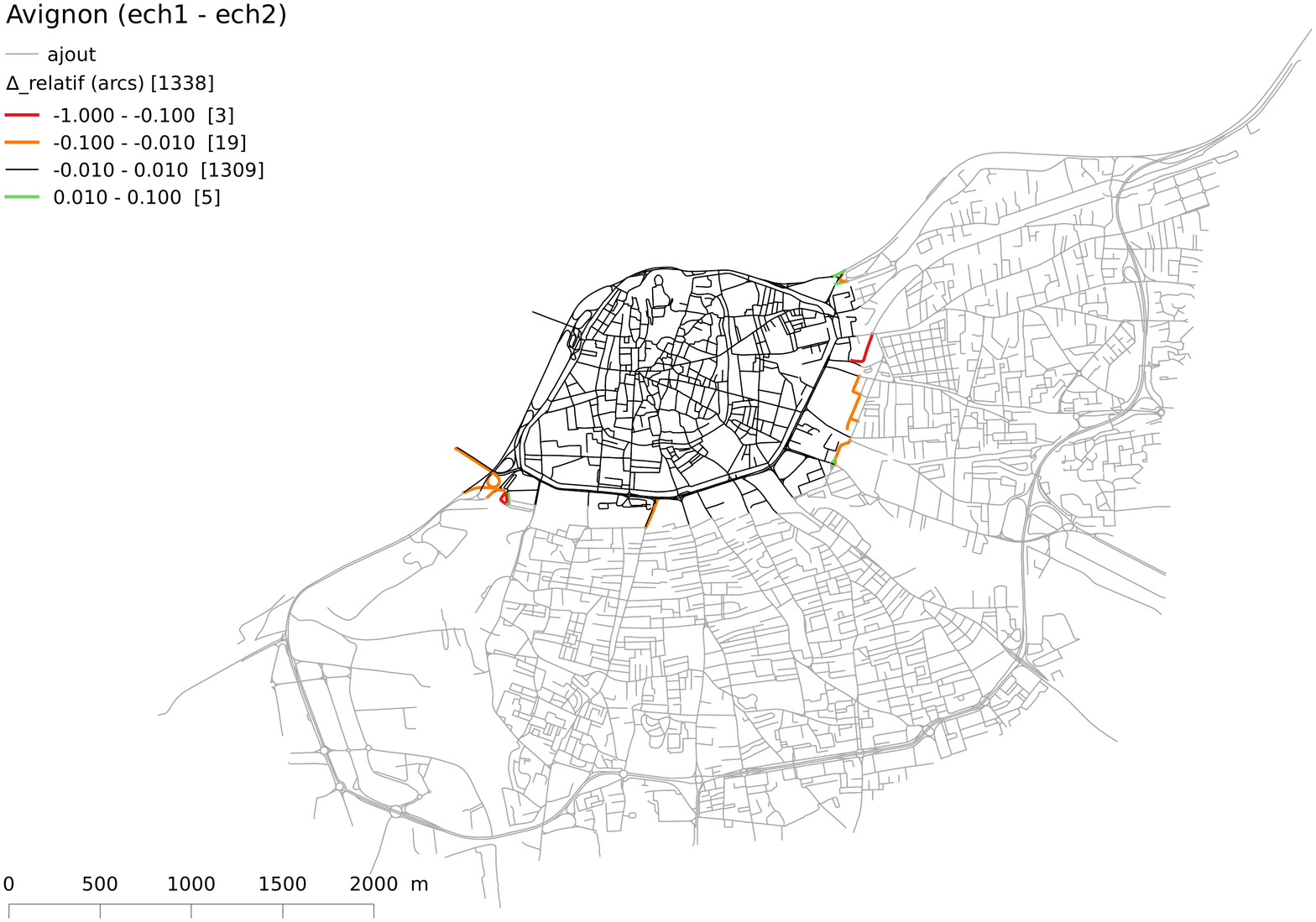}
    \caption{Carte de $\Delta_{relatif}$ calculé entre les échantillons 1 et 2 d'Avignon.}
    \label{fig_ann:diff_avignon_sm}
\end{figure}

\begin{figure}[b]
    \centering
    \includegraphics[width=0.8\textwidth]{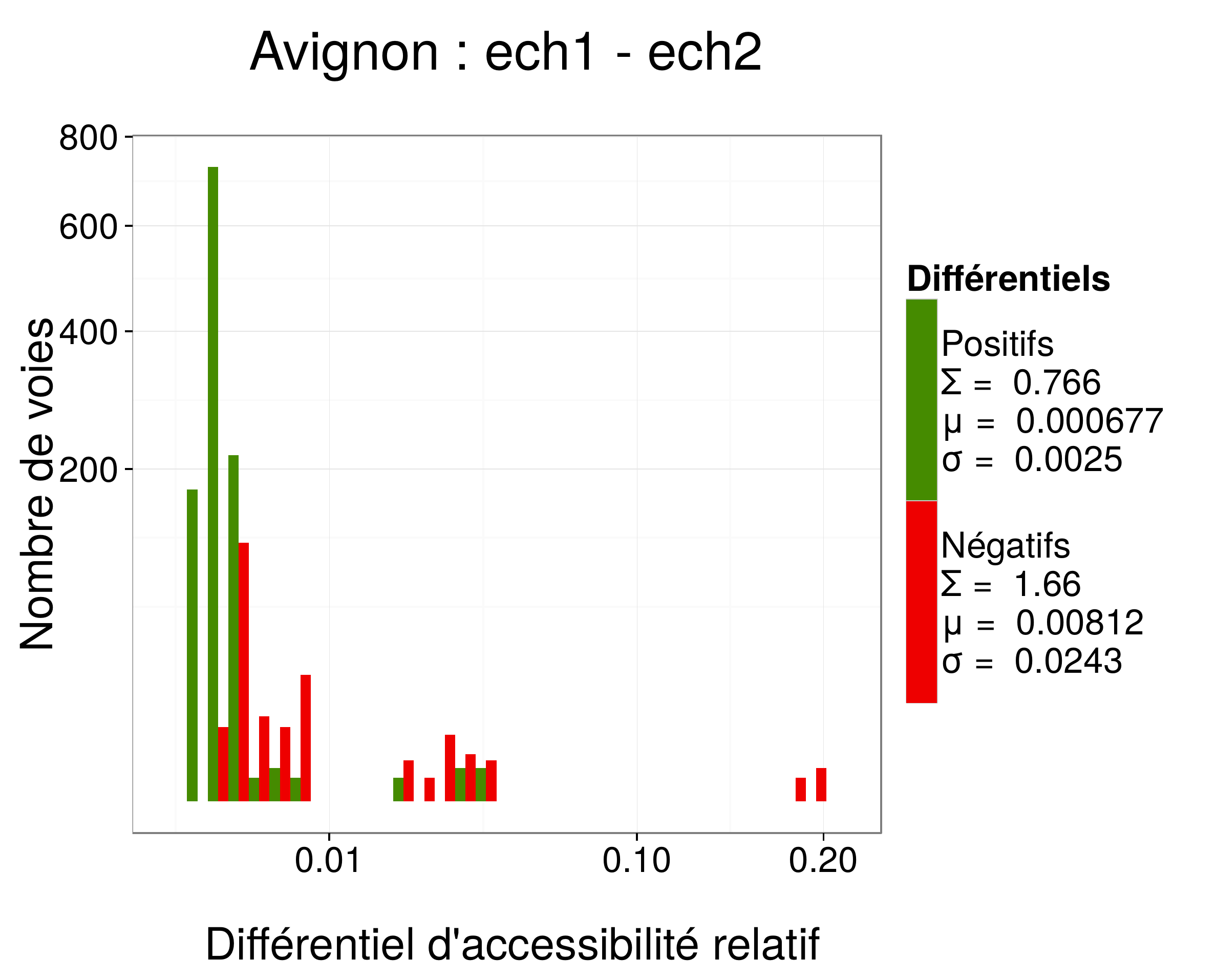}
    \caption{Répartition de $\Delta_{relatif}$ calculé entre les échantillons 1 et 2 d'Avignon.}
    \label{fig_ann:diff_hist_avignon_sm}
\end{figure}

\begin{figure}[t]
    \centering
    \includegraphics[width=0.8\textwidth]{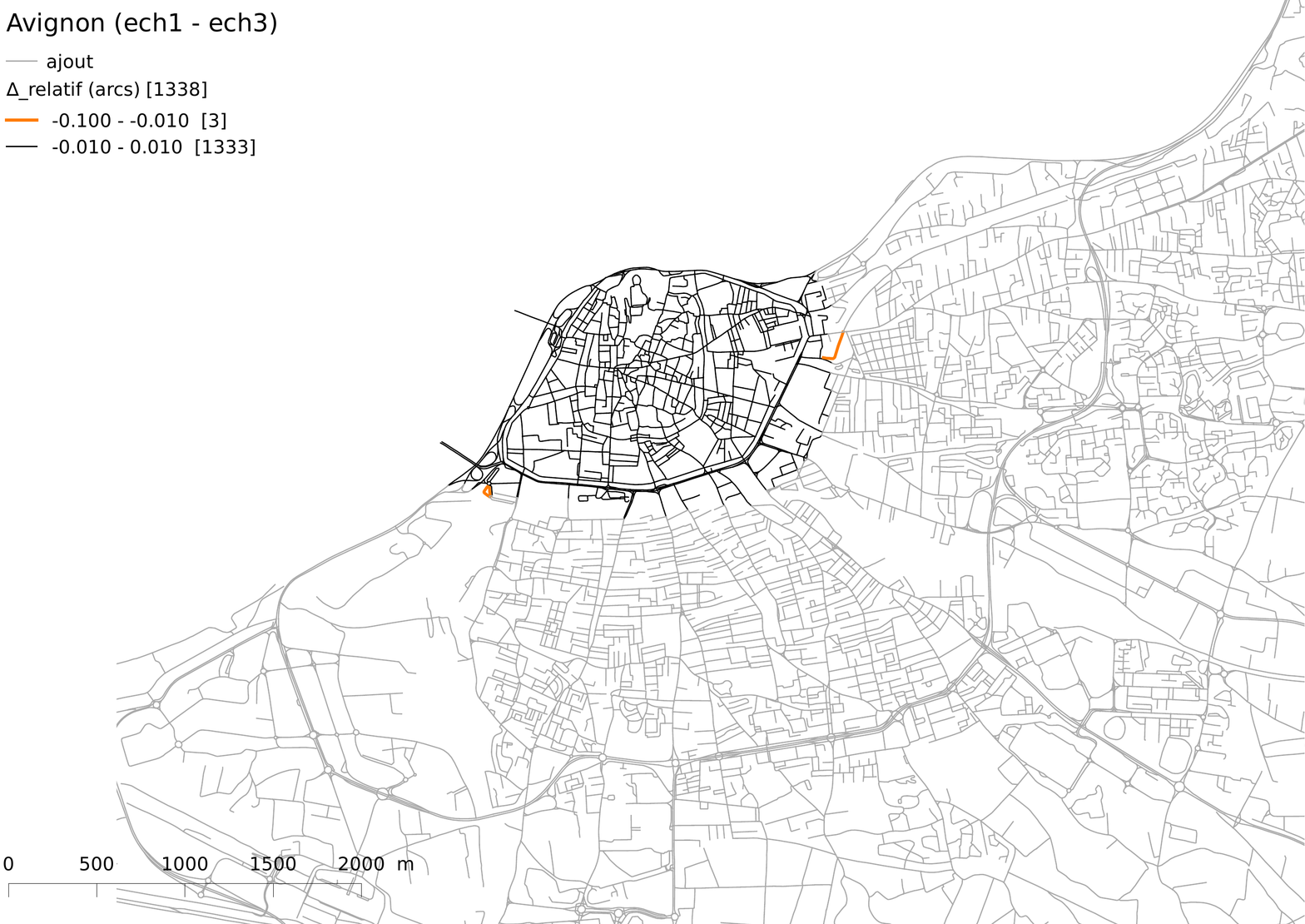}
    \caption{Carte de $\Delta_{relatif}$ calculé entre les échantillons 1 et 3 d'Avignon.}
    \label{fig_ann:diff_avignon_sl}
\end{figure}

\begin{figure}[b]
    \centering
    \includegraphics[width=0.8\textwidth]{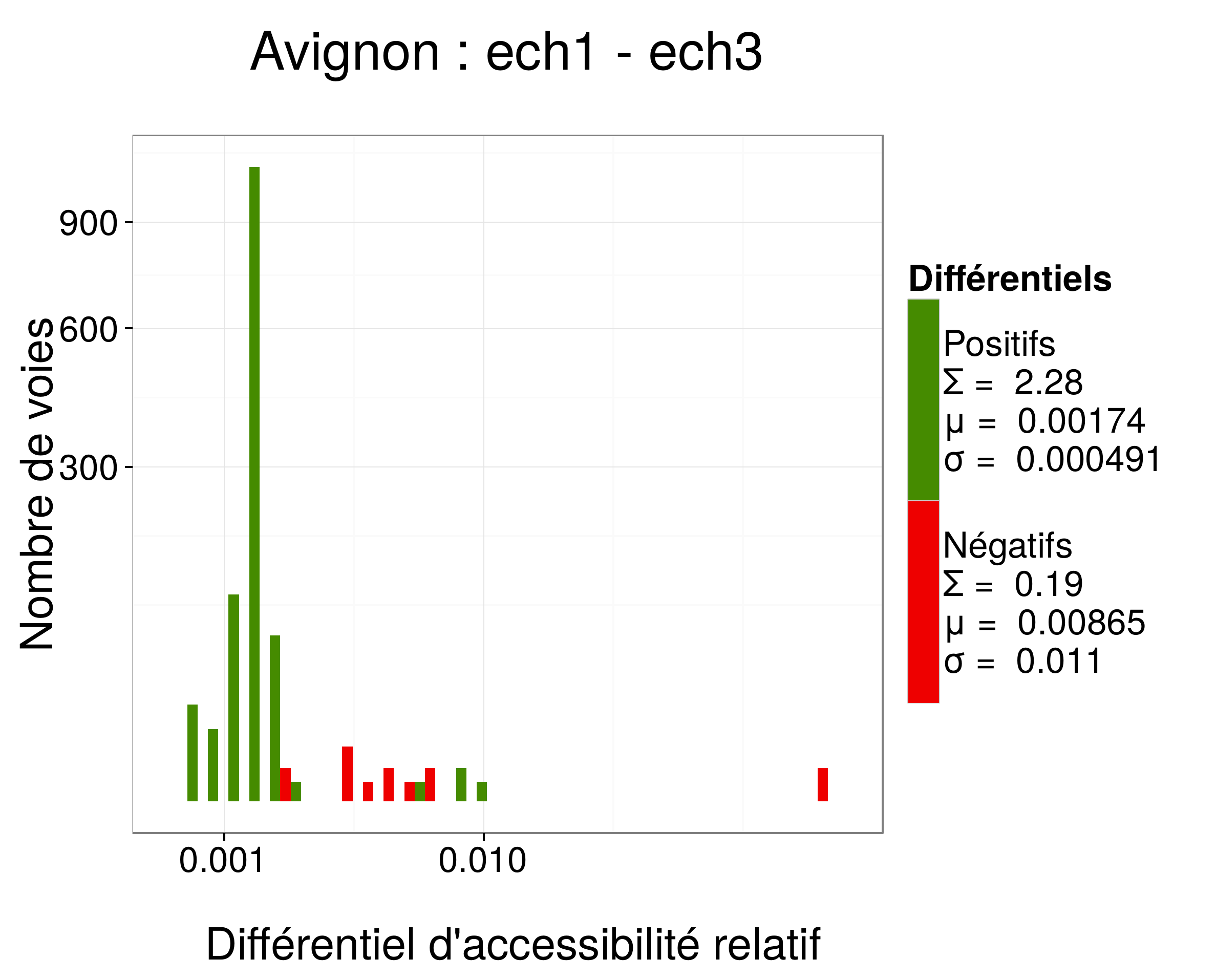}
    \caption{Répartition de $\Delta_{relatif}$ calculé entre les échantillons 1 et 3 d'Avignon.}
    \label{fig_ann:diff_hist_avignon_sl}
\end{figure}

\begin{figure}[t]
    \centering
    \includegraphics[width=0.8\textwidth]{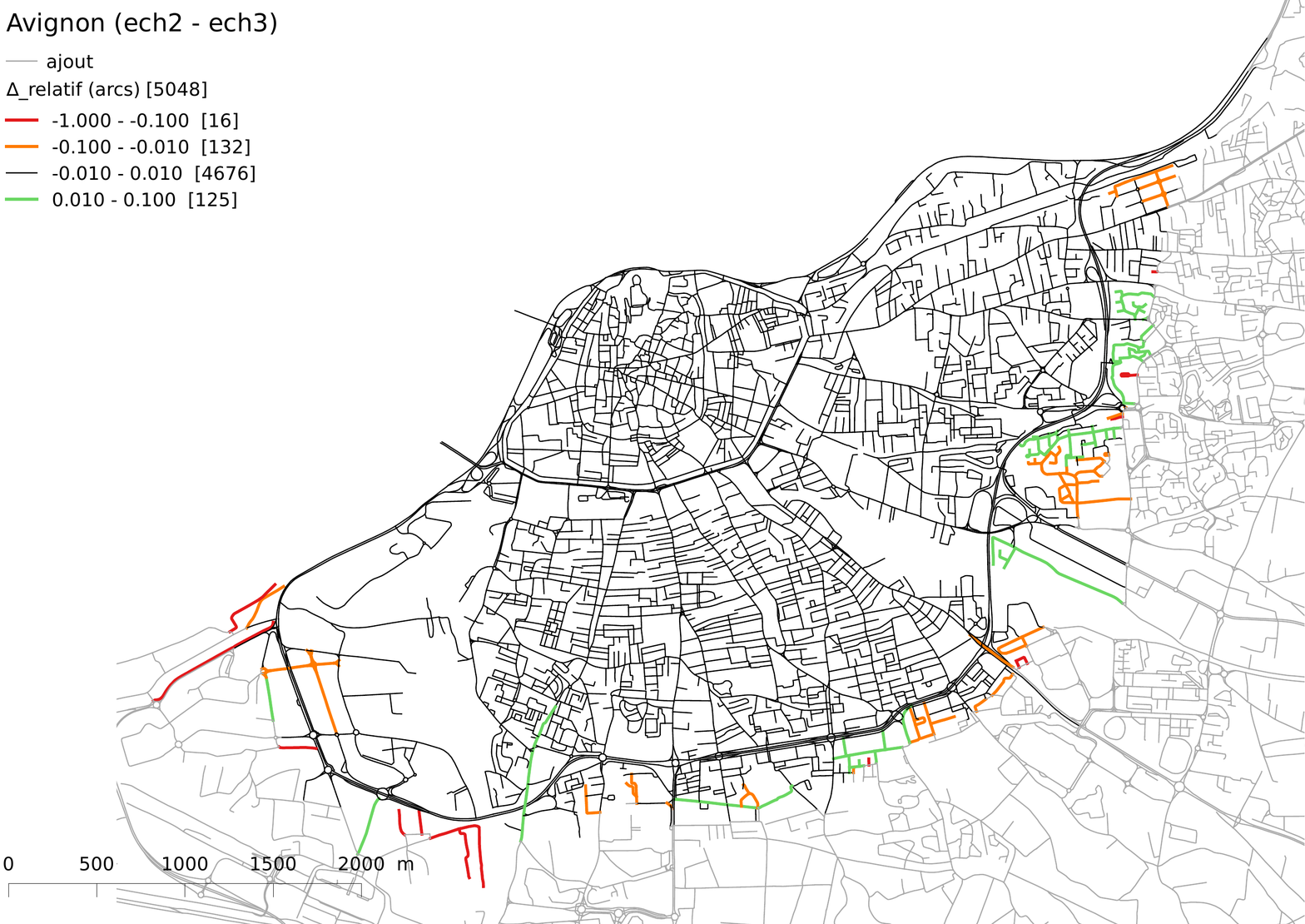}
    \caption{Carte de $\Delta_{relatif}$ calculé entre les échantillons 2 et 3 d'Avignon.}
    \label{fig_ann:diff_avignon_ml}
\end{figure}

\begin{figure}[b]
    \centering
    \includegraphics[width=0.8\textwidth]{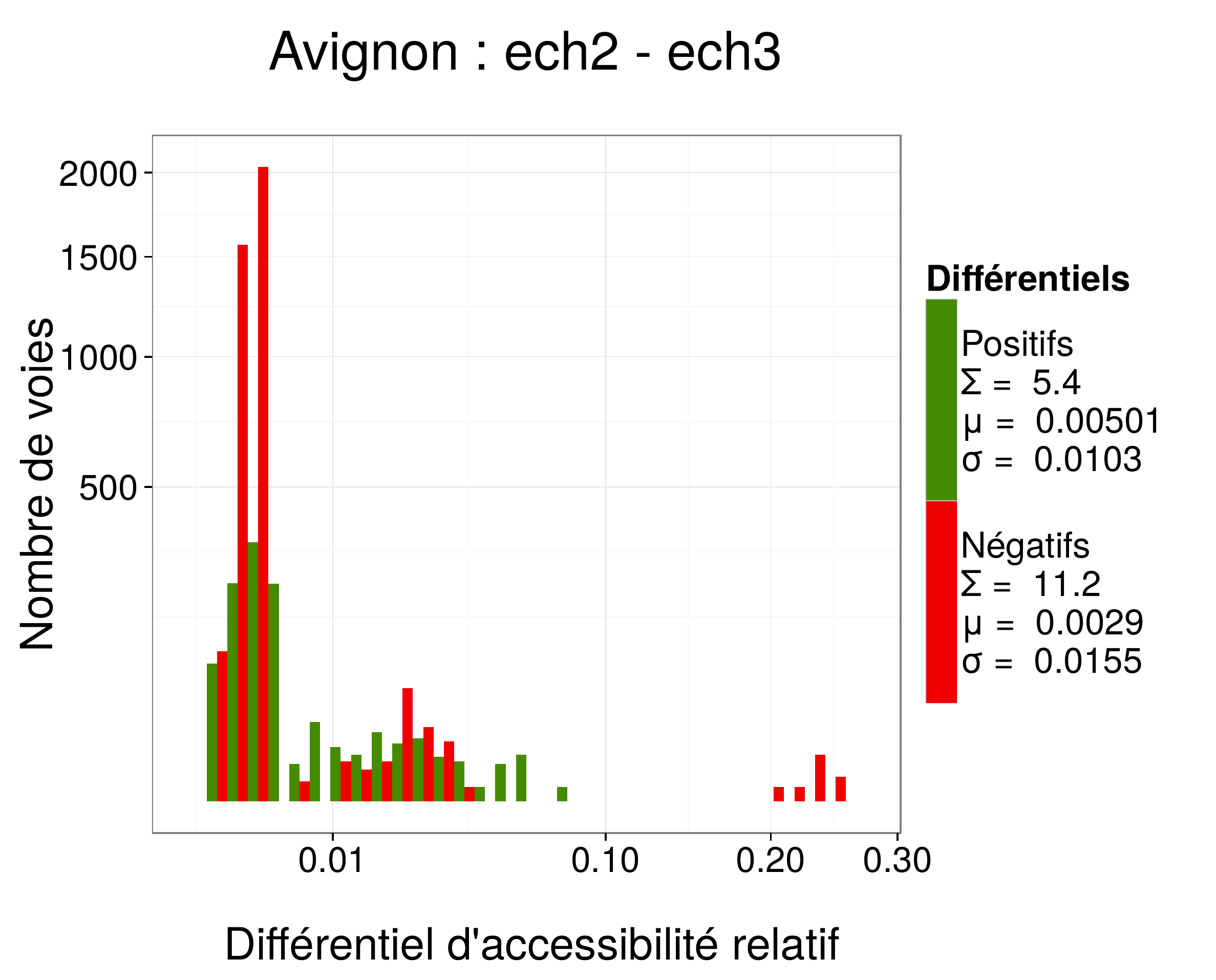}
    \caption{Répartition de $\Delta_{relatif}$ calculé entre les échantillons 2 et 3 d'Avignon.}
    \label{fig_ann:diff_hist_avignon_ml}
\end{figure}


\FloatBarrier
\section{Paris}\label{ann:sec_eb_paris}

\begin{figure}[t]
    \centering
    \includegraphics[width=0.8\textwidth]{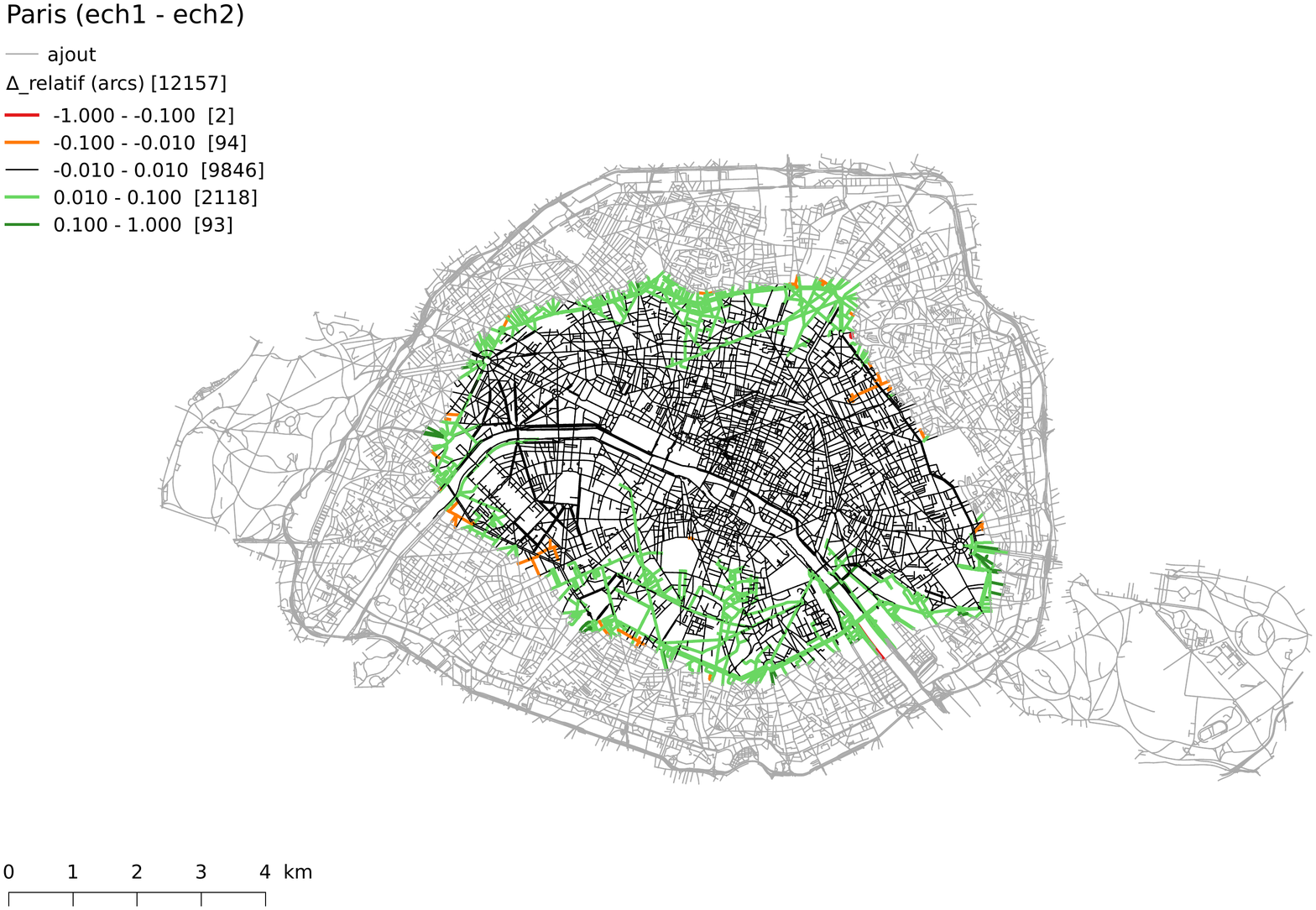}
    \caption{Carte de $\Delta_{relatif}$ calculé entre les échantillons 1 et 2 de Paris.}
    \label{fig_ann:diff_paris_sm}
\end{figure}

\begin{figure}[b]
    \centering
    \includegraphics[width=0.8\textwidth]{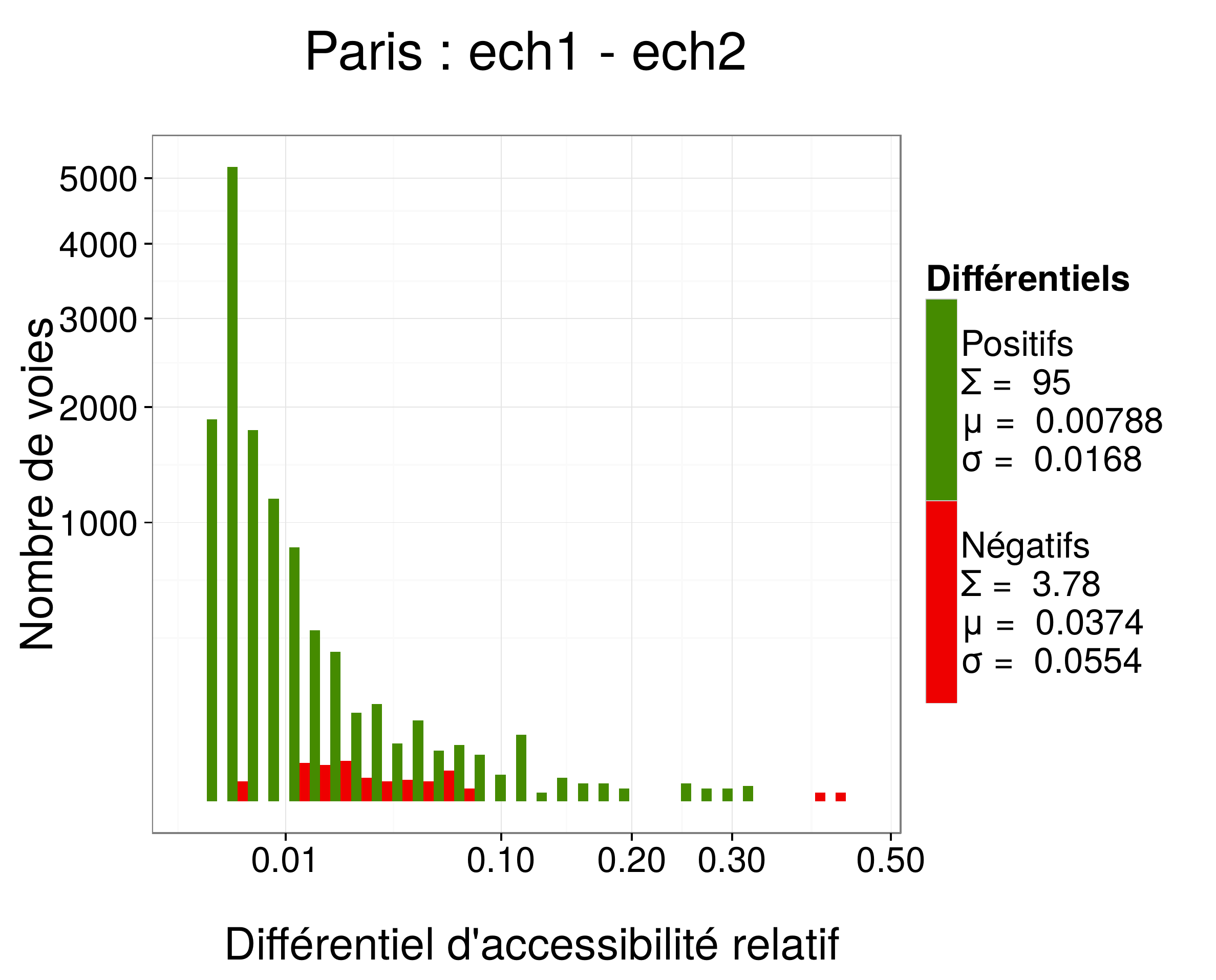}
    \caption{Répartition de $\Delta_{relatif}$ calculé entre les échantillons 1 et 2 de Paris.}
    \label{fig_ann:diff_hist_paris_ms}
\end{figure}

\begin{figure}[t]
    \centering
    \includegraphics[width=0.8\textwidth]{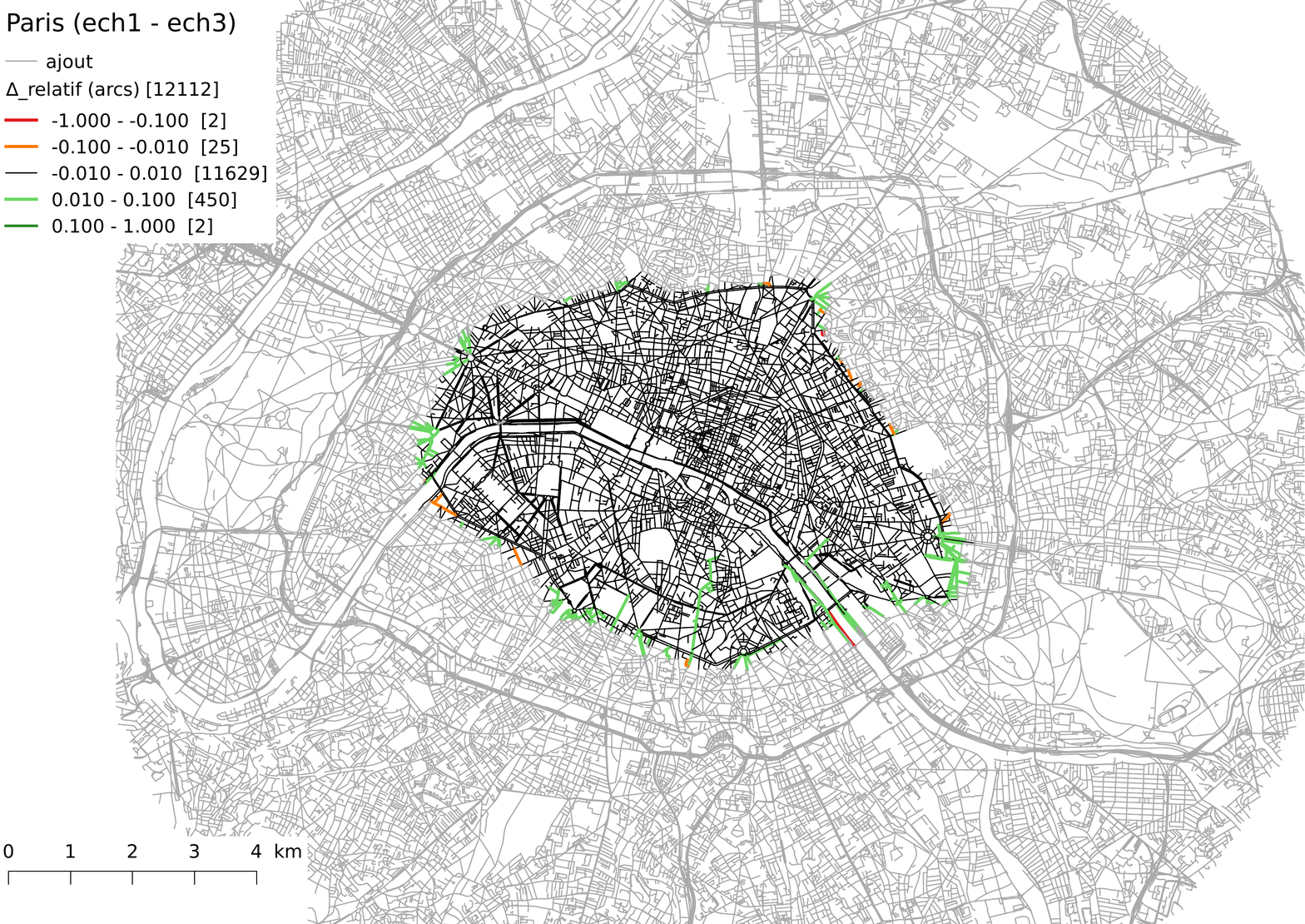}
    \caption{Carte de $\Delta_{relatif}$ calculé entre les échantillons 1 et 3 de Paris.}
    \label{fig_ann:diff_paris_sl}
\end{figure}

\begin{figure}[b]
    \centering
    \includegraphics[width=0.8\textwidth]{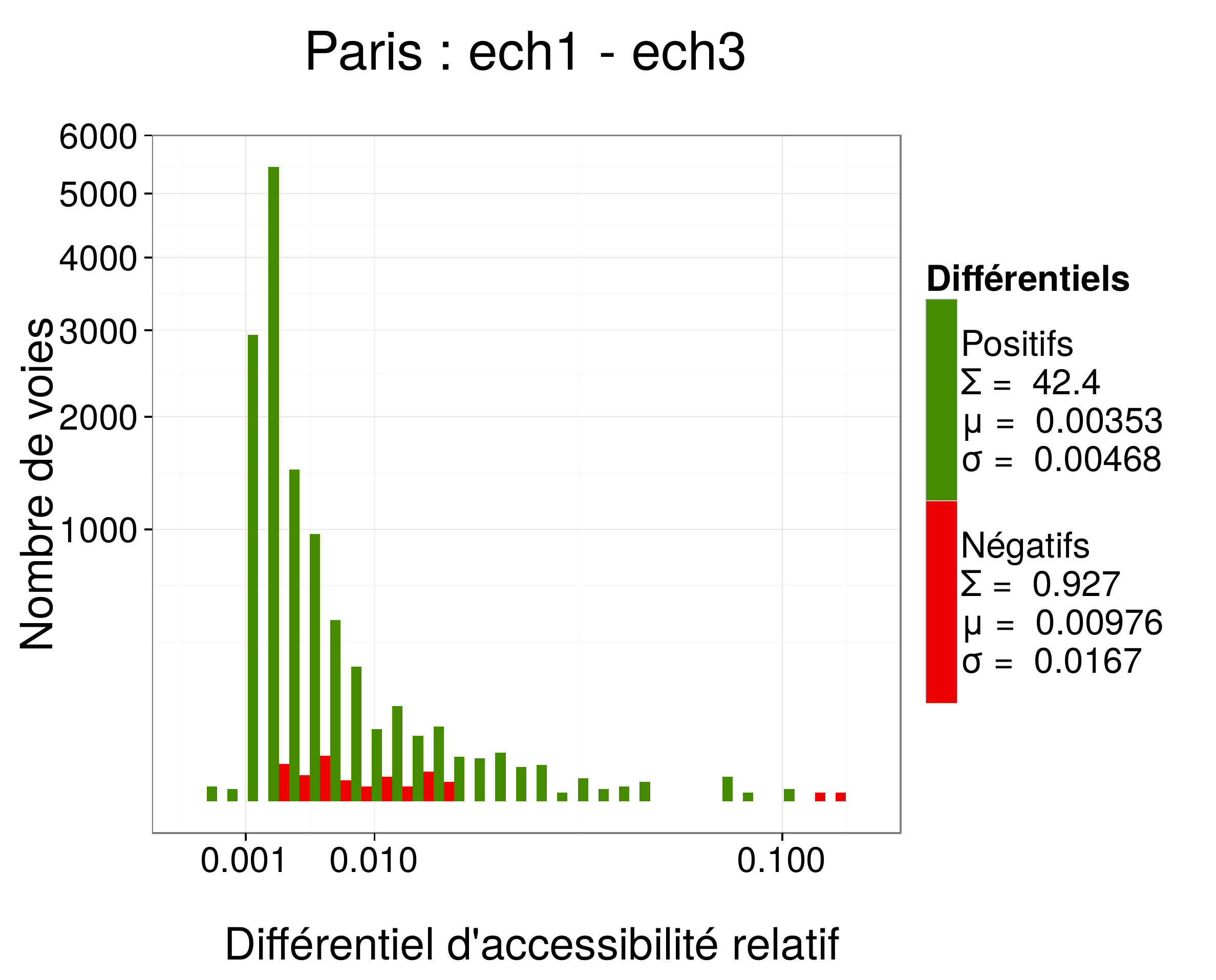}
    \caption{Répartition de $\Delta_{relatif}$ calculé entre les échantillons 1 et 3 de Paris.}
    \label{fig_ann:diff_hist_paris_sl}
\end{figure}

\begin{figure}[t]
    \centering
    \includegraphics[width=0.8\textwidth]{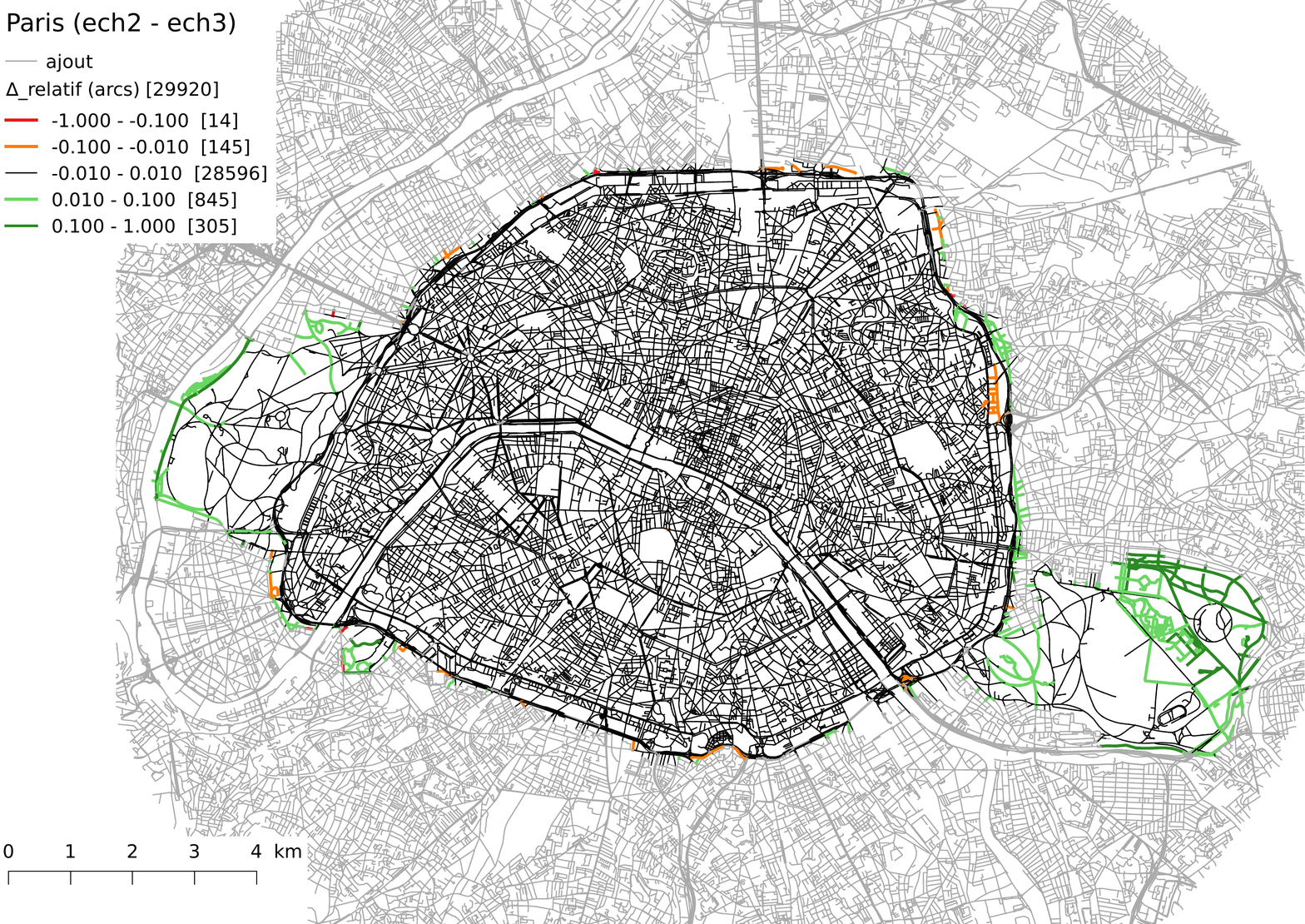}
    \caption{Carte de $\Delta_{relatif}$ calculé entre les échantillons 2 et 3 de Paris.}
    \label{fig_ann:diff_paris_ml}
\end{figure}

\begin{figure}[b]
    \centering
    \includegraphics[width=0.8\textwidth]{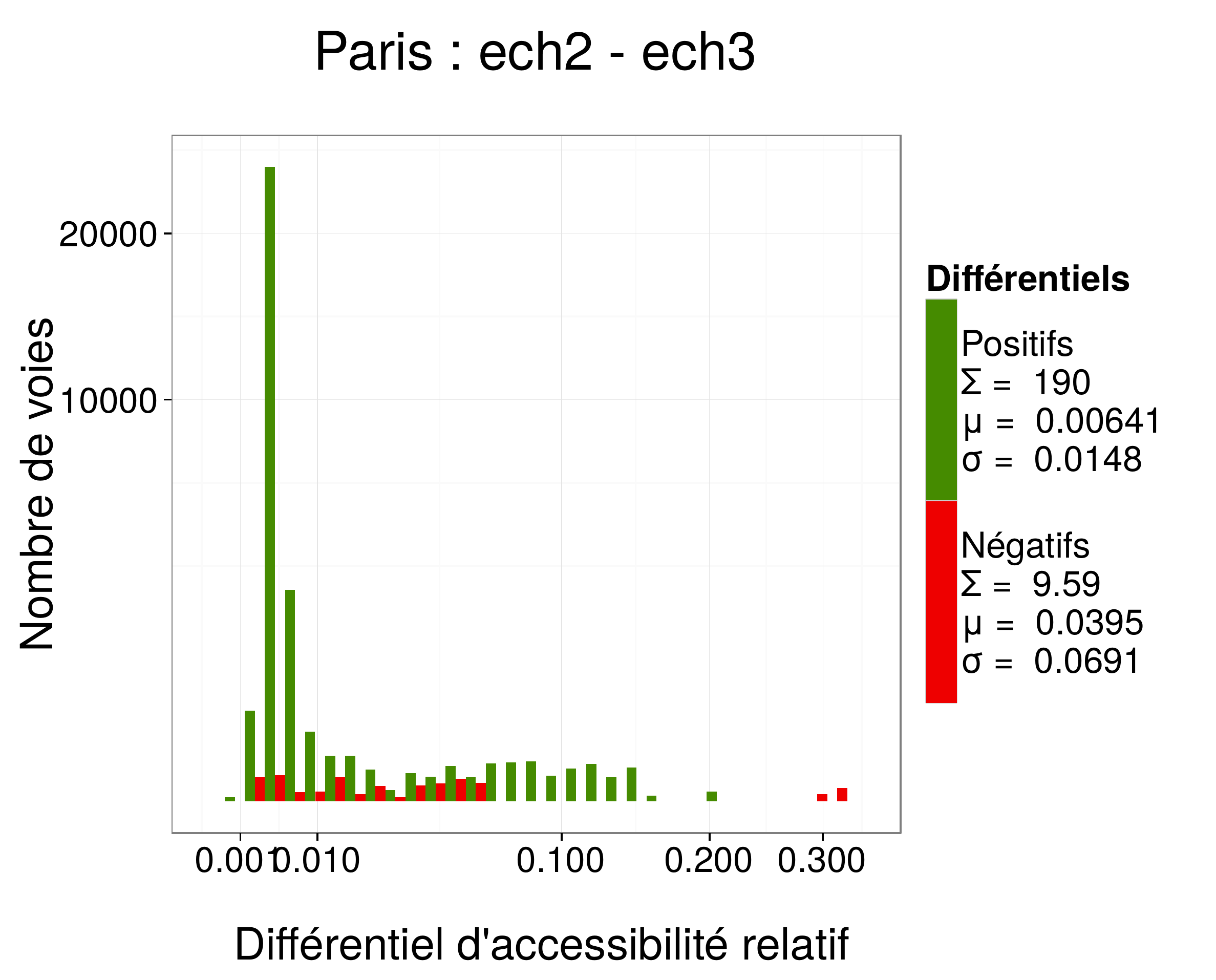}
    \caption{Répartition de $\Delta_{relatif}$ calculé entre les échantillons 2 et 3 de Paris.}
    \label{fig_ann:diff_hist_paris_ml}
\end{figure}


\FloatBarrier
\section{Barcelone}\label{ann:sec_eb_barcelone}

\begin{figure}[t]
    \centering
    \includegraphics[width=0.8\textwidth]{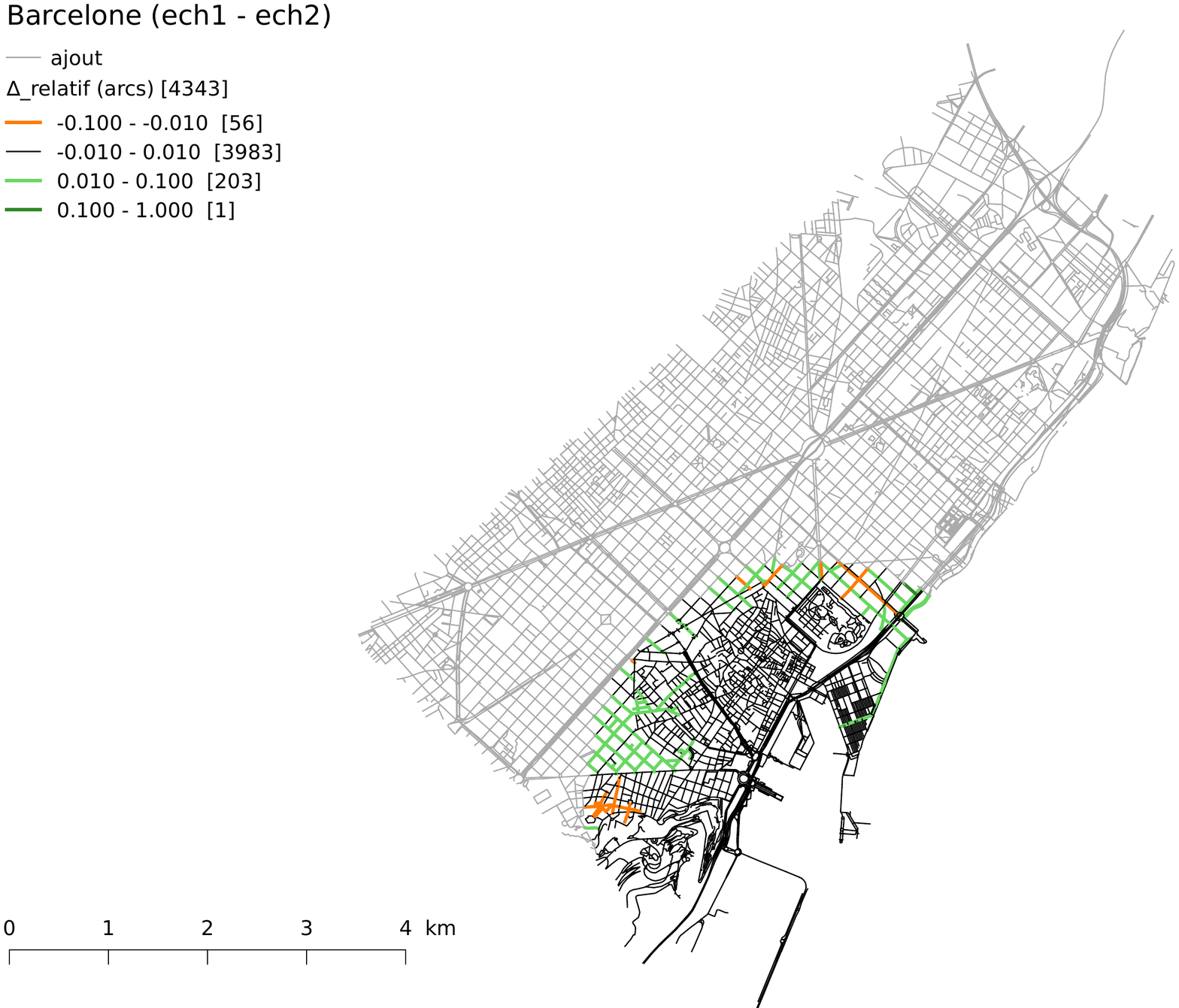}
    \caption{Carte de $\Delta_{relatif}$ calculé entre les échantillons 1 et 2 de Barcelone.}
    \label{fig_ann:diff_barcelone_sm}
\end{figure}

\begin{figure}[b]
    \centering
    \includegraphics[width=0.8\textwidth]{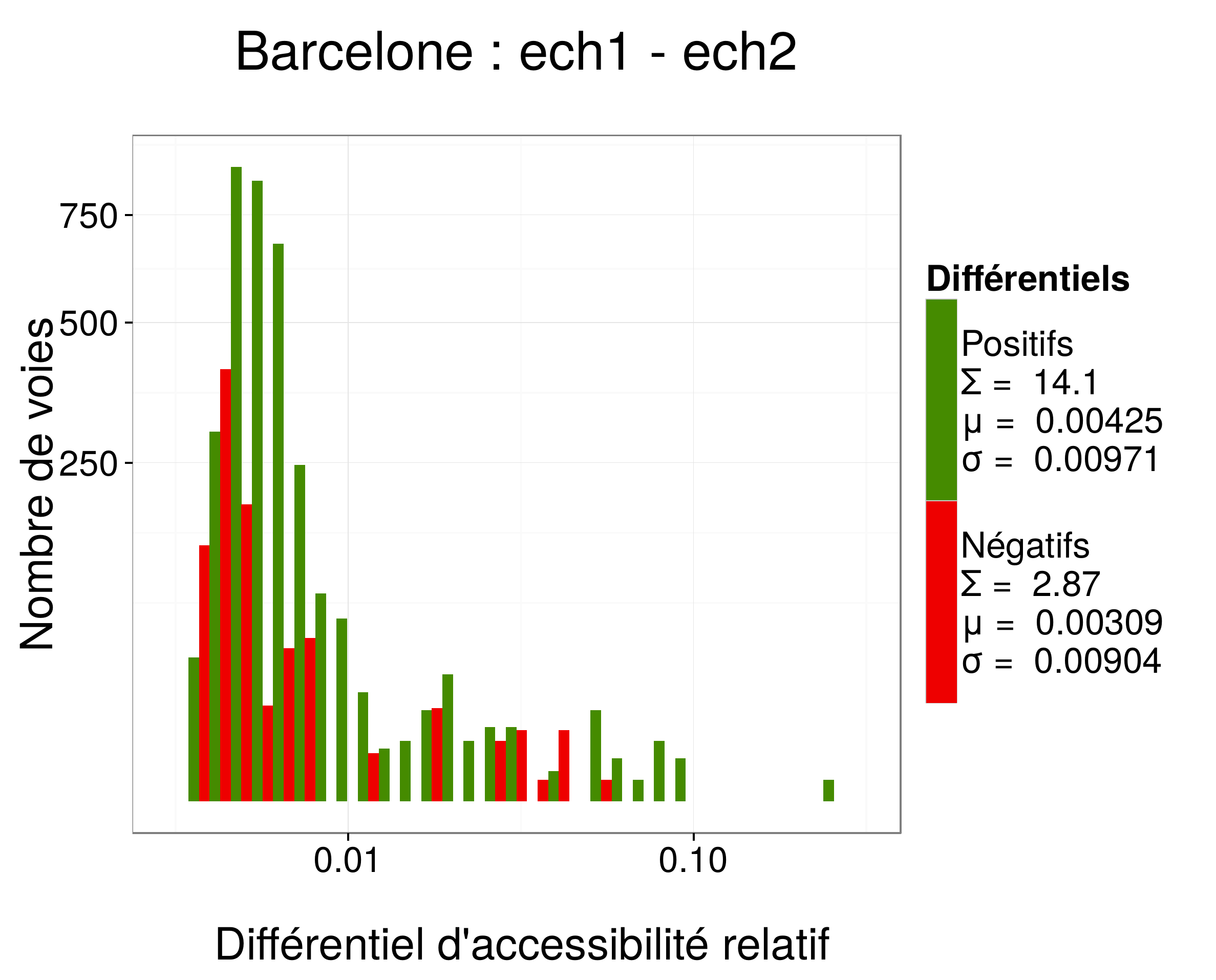}
    \caption{Répartition de $\Delta_{relatif}$ calculé entre les échantillons 1 et 2 de Barcelone.}
    \label{fig_ann:diff_hist_barcelone_sm}
\end{figure}

\begin{figure}[t]
    \centering
    \includegraphics[width=0.8\textwidth]{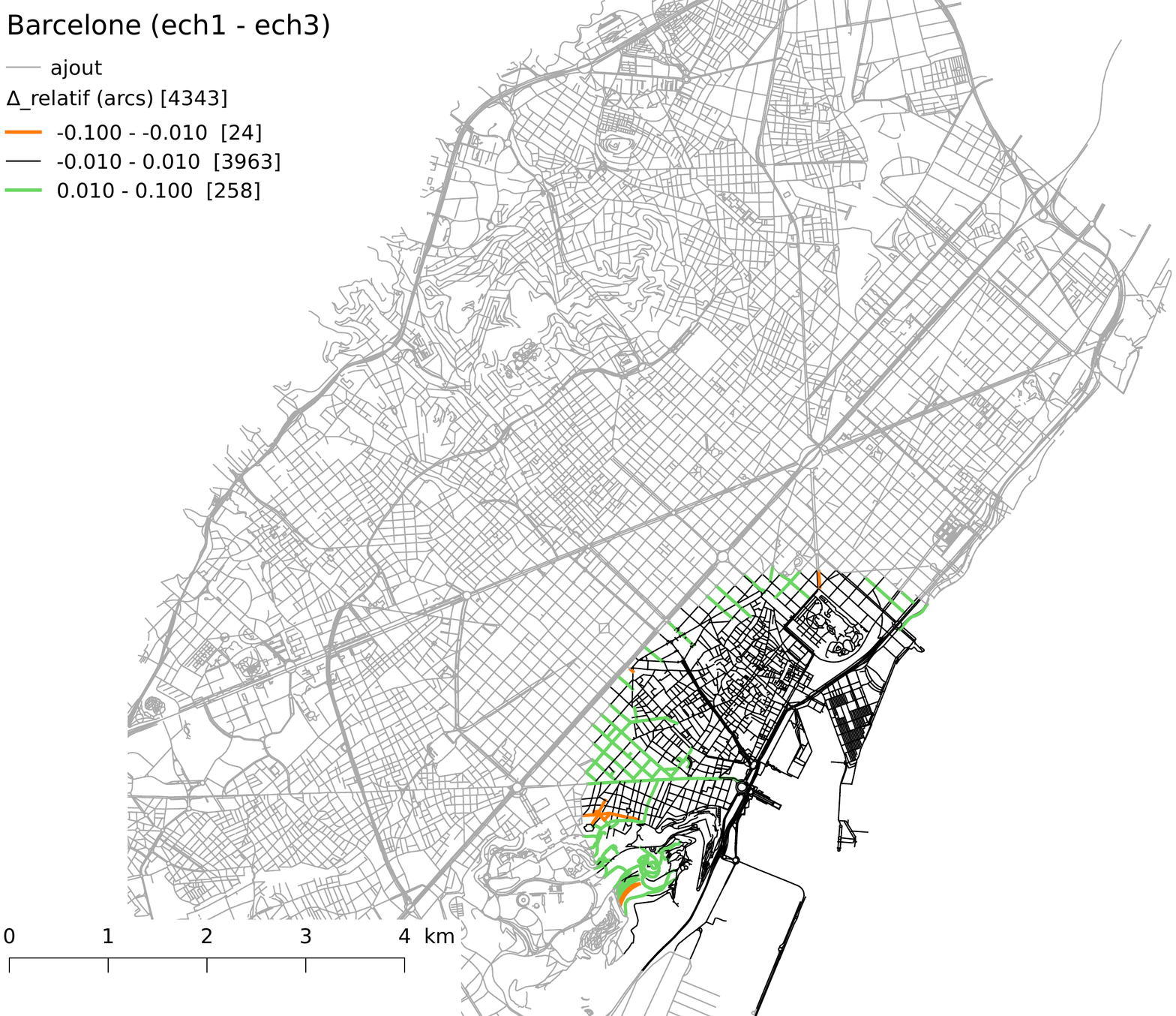}
    \caption{Carte de $\Delta_{relatif}$ calculé entre les échantillons 1 et 3 de Barcelone.}
    \label{fig_ann:diff_barcelone_sl}
\end{figure}

\begin{figure}[b]
    \centering
    \includegraphics[width=0.8\textwidth]{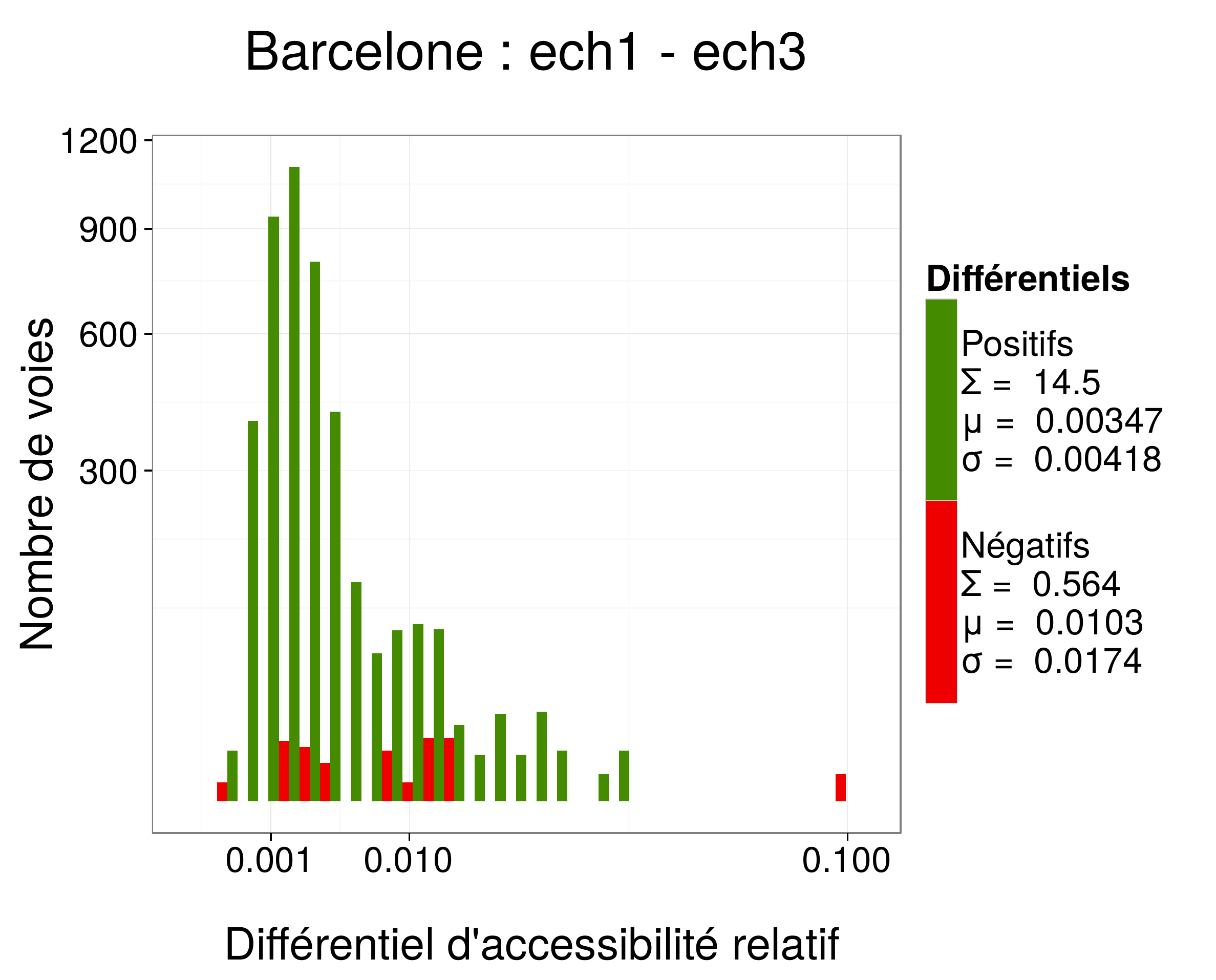}
    \caption{Répartition de $\Delta_{relatif}$ calculé entre les échantillons 1 et 3 de Barcelone.}
    \label{fig_ann:diff_hist_barcelone_sl}
\end{figure}

\begin{figure}[t]
    \centering
    \includegraphics[width=0.8\textwidth]{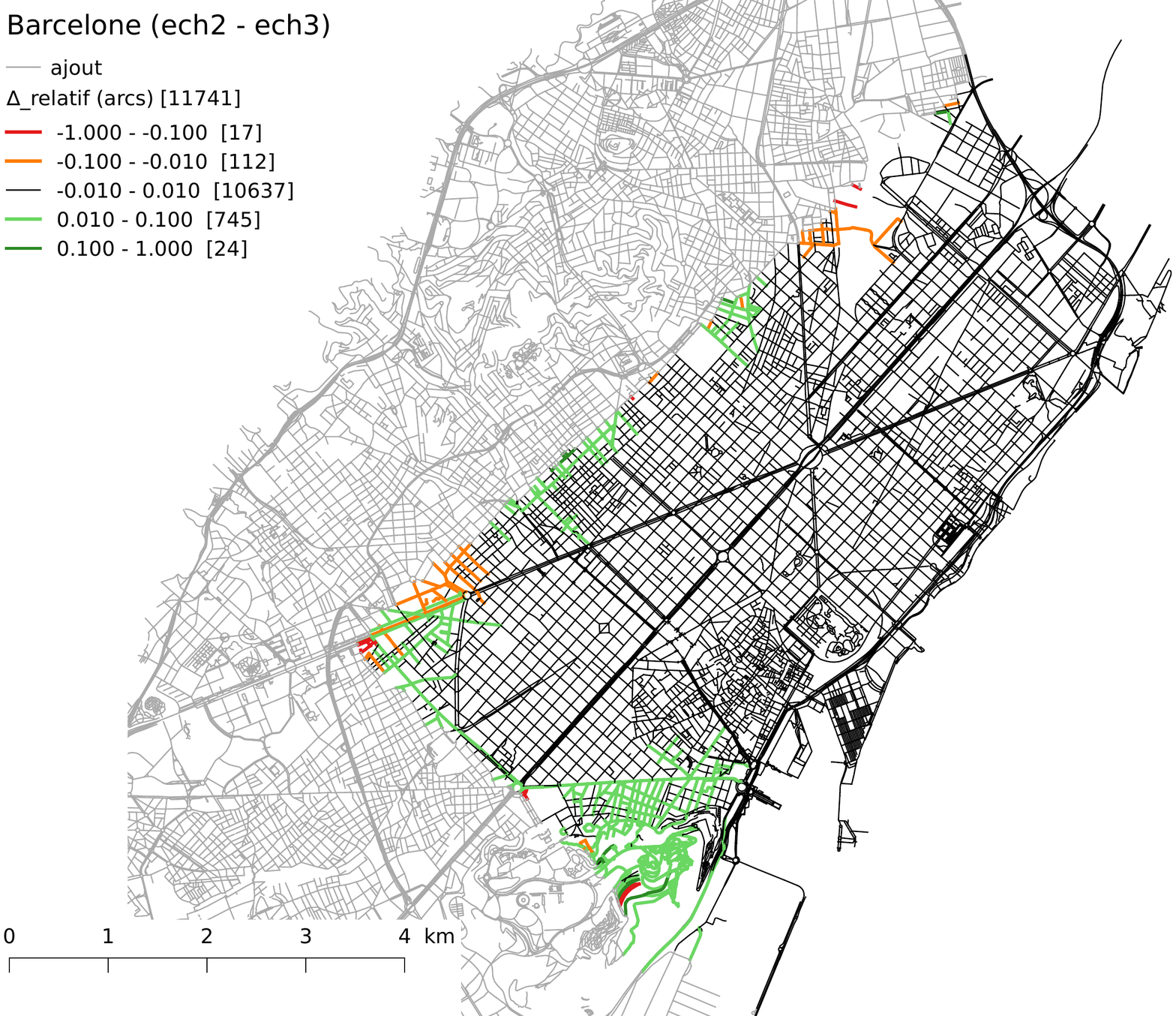}
    \caption{Carte de $\Delta_{relatif}$ calculé entre les échantillons 2 et 3 de Barcelone.}
    \label{fig_ann:diff_barcelone_ml}
\end{figure}

\begin{figure}[b]
    \centering
    \includegraphics[width=0.8\textwidth]{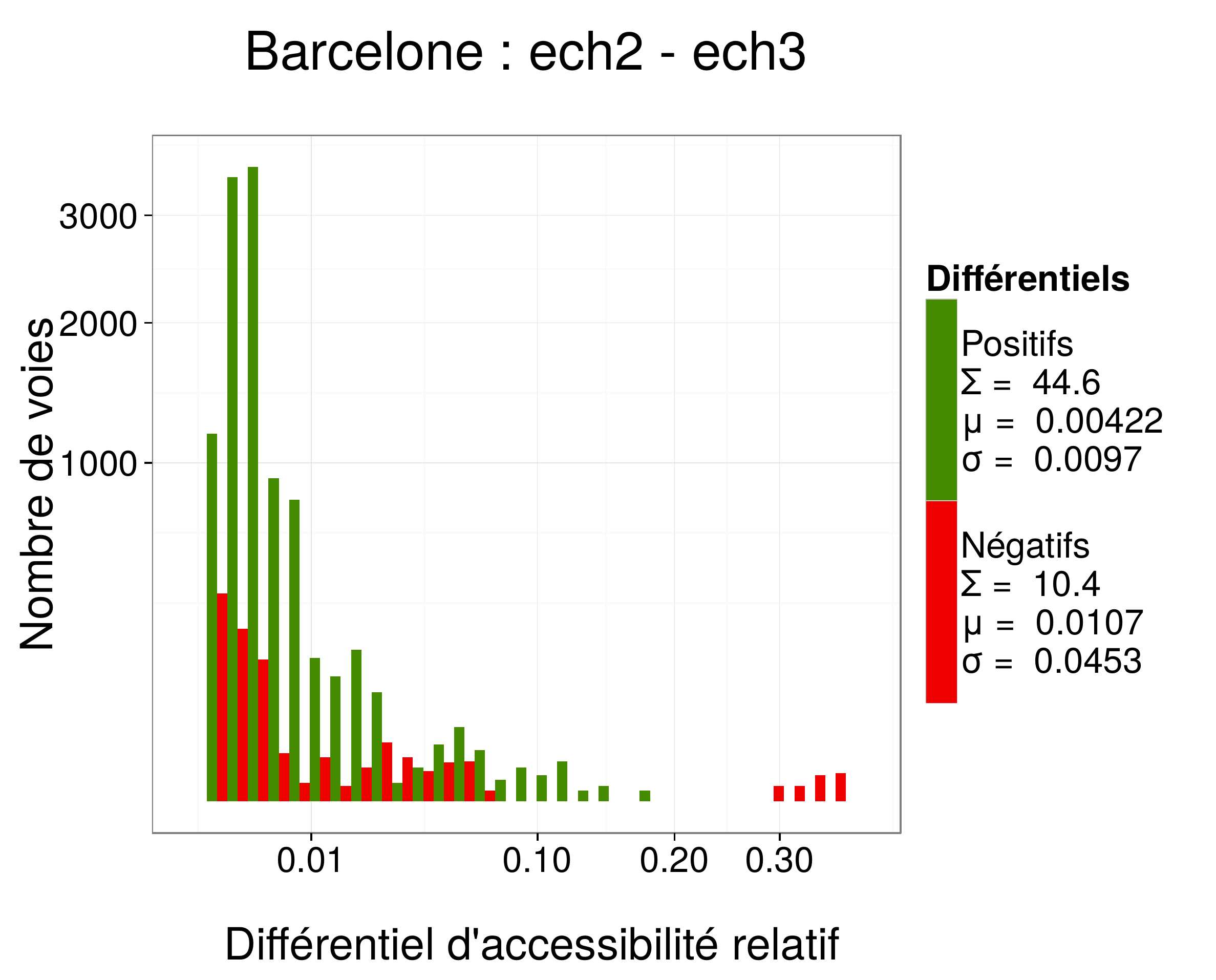}
    \caption{Répartition de $\Delta_{relatif}$ calculé entre les échantillons 2 et 3 de Barcelone.}
    \label{fig_ann:diff_hist_barcelone_ml}
\end{figure}


\FloatBarrier
\section{New-York}\label{ann:sec_eb_newyork}

\begin{figure}[t]
    \centering
    \begin{subfigure}[t]{0.8\textwidth}
    \centering
    \includegraphics[width=\textwidth]{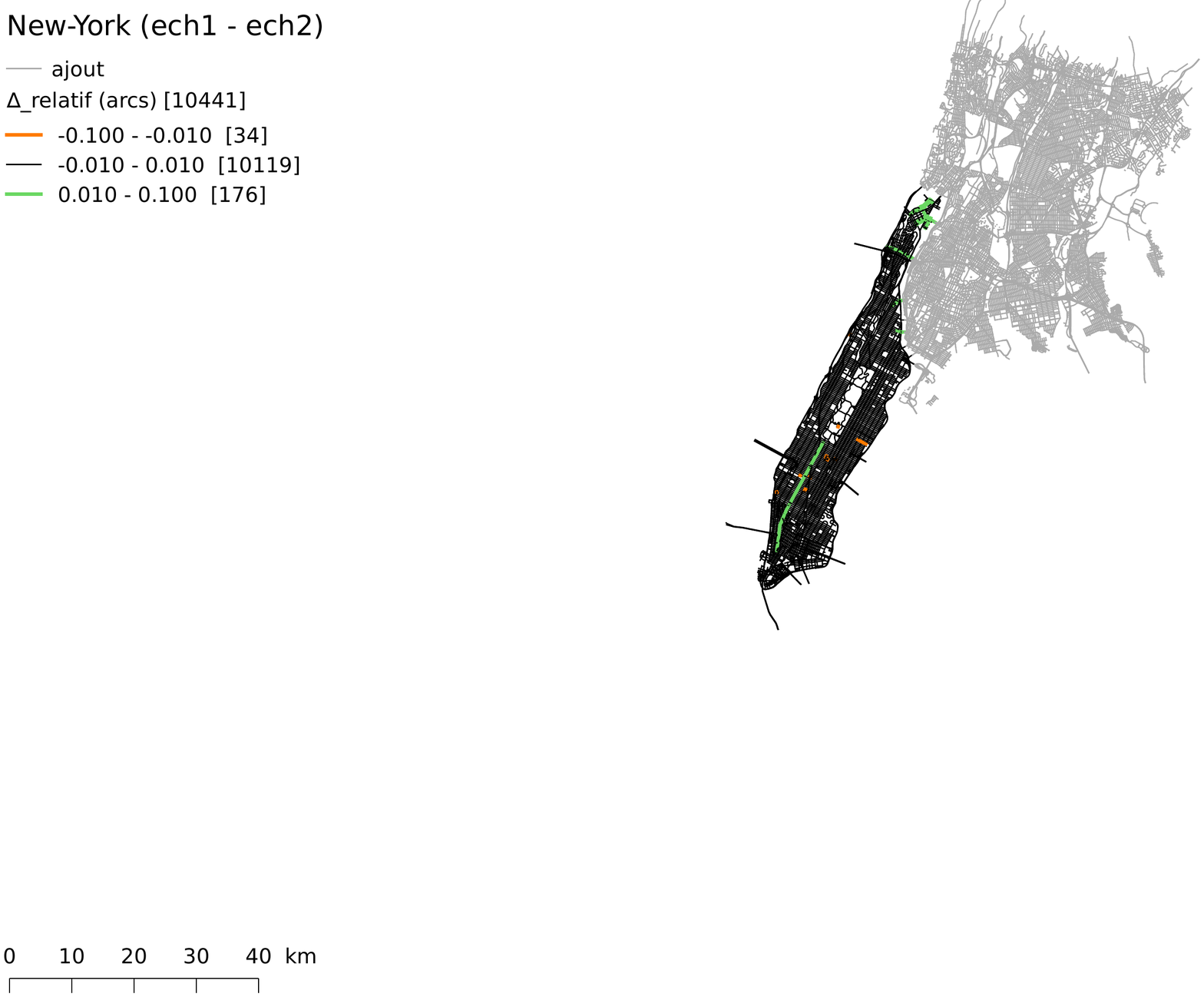}
    \caption{Carte de $\Delta_{relatif}$ calculé entre les échantillons 1 et 2 de New-York.}
    \label{fig_ann:diff_man_is}
	\end{subfigure}

    \begin{subfigure}[t]{0.8\textwidth}
    \centering
    \includegraphics[width=\textwidth]{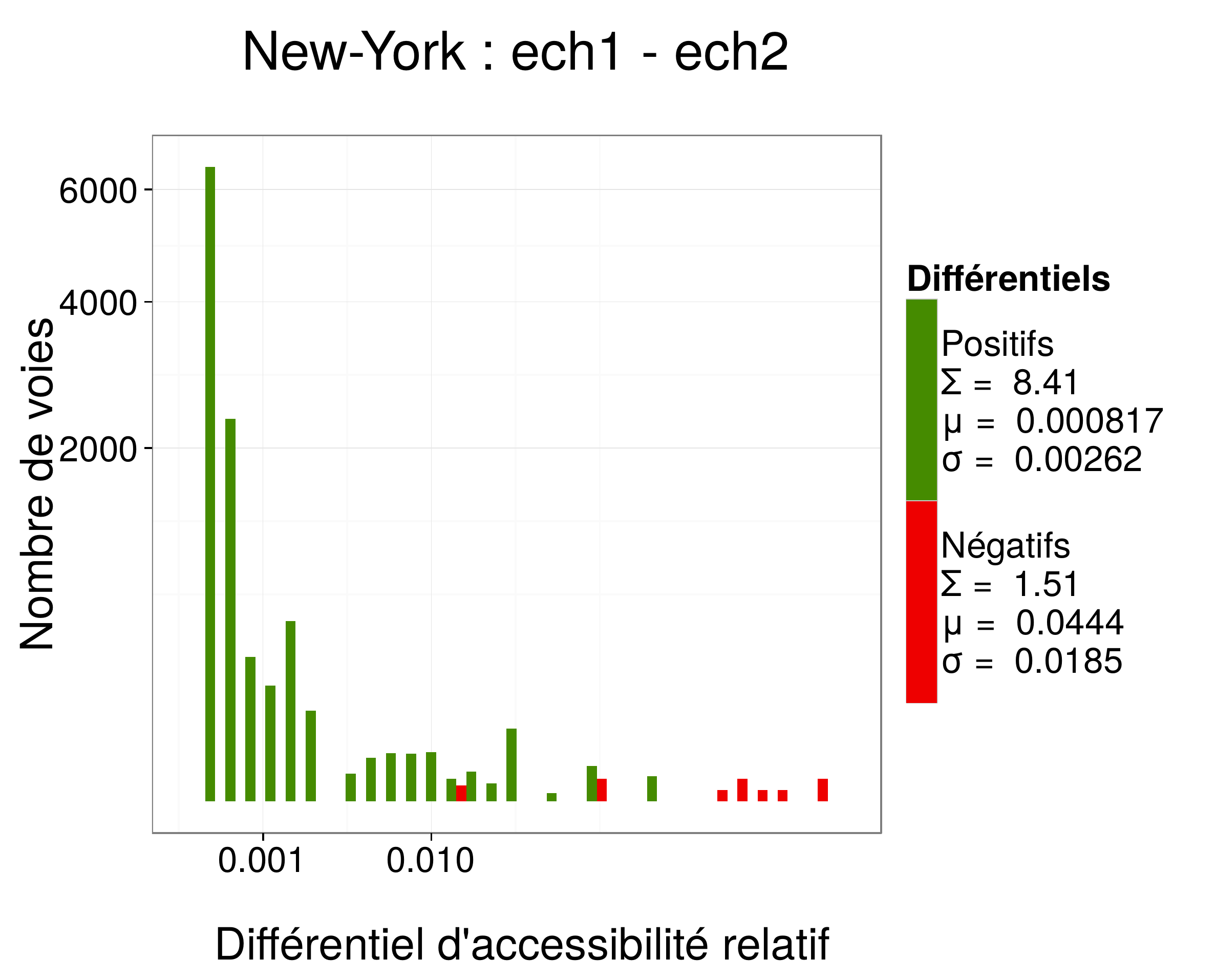}
    \caption{Répartition de $\Delta_{relatif}$ calculé entre les échantillons 1 et 2 de New-York.}
    \label{fig_ann:diff_hist_man_is}
    \end{subfigure}
\end{figure}

\begin{figure}[t]
    \centering
    \includegraphics[width=0.8\textwidth]{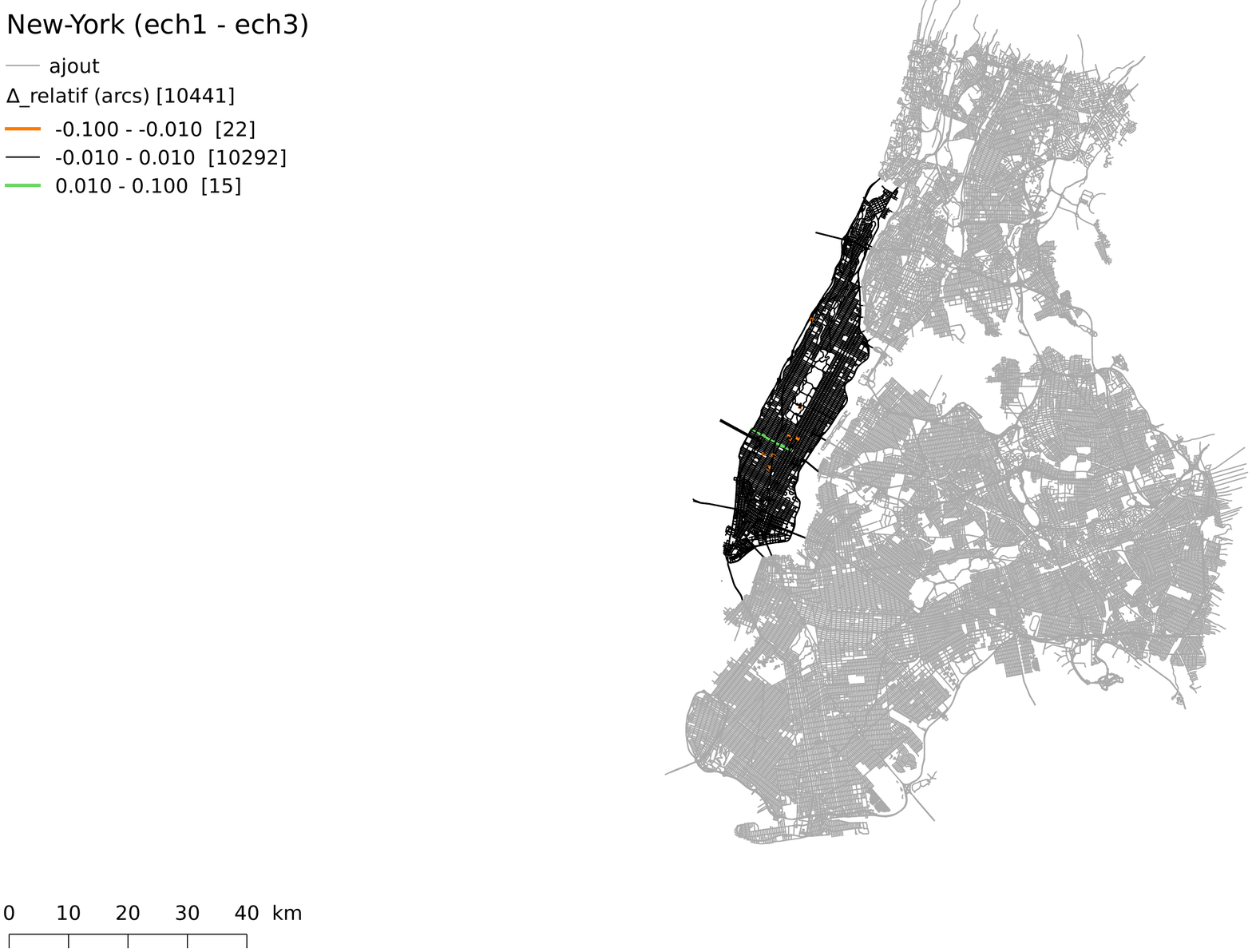}
    \caption{Carte de $\Delta_{relatif}$ calculé entre les échantillons 1 et 3 de New-York.}
    \label{fig_ann:diff_man_im}
\end{figure}

\begin{figure}[b]
    \centering
    \includegraphics[width=0.8\textwidth]{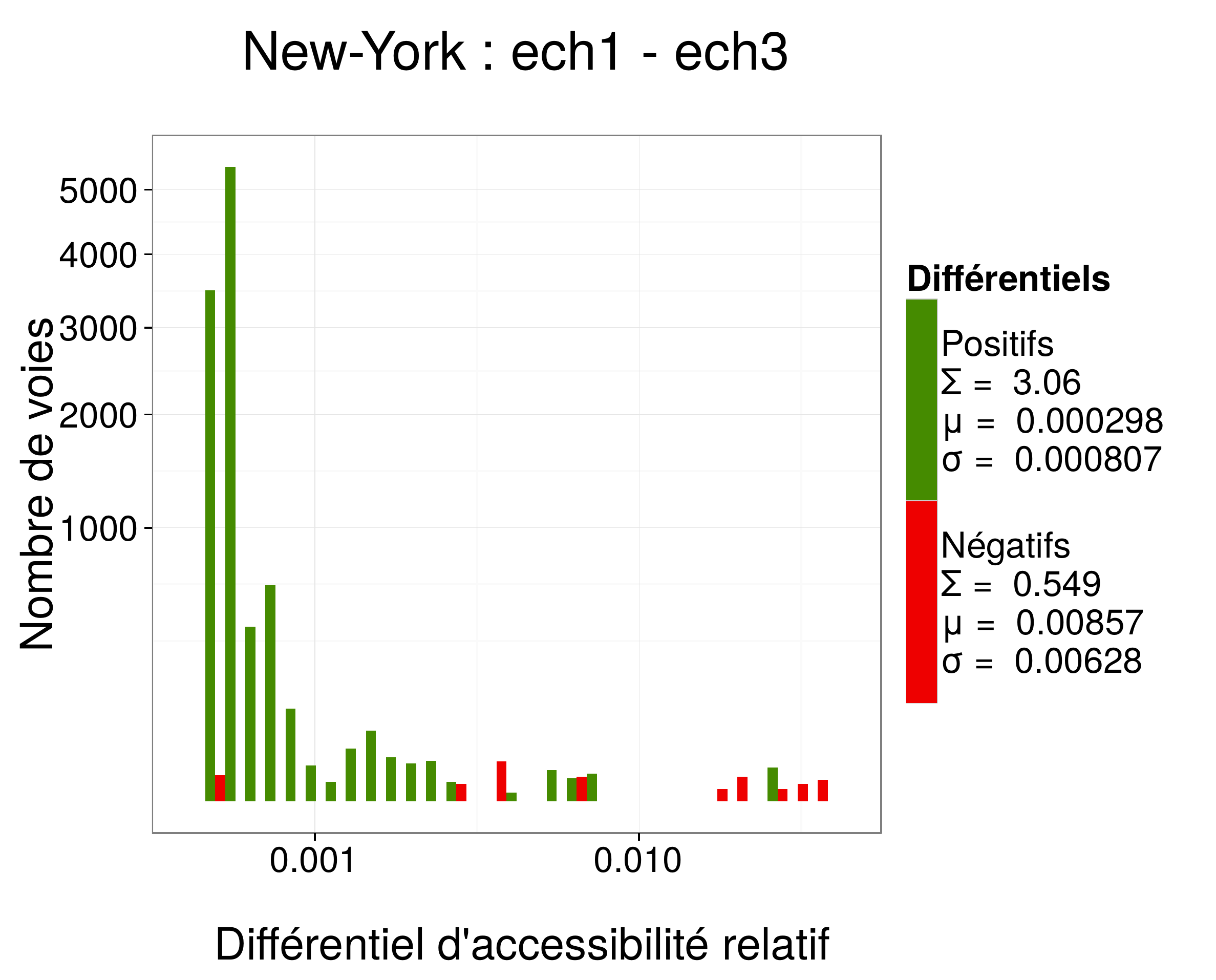}
    \caption{Répartition de $\Delta_{relatif}$ calculé entre les échantillons 1 et 3 de New-York.}
    \label{fig_ann:diff_hist_man_im}
\end{figure}

\begin{figure}[t]
    \centering
    \includegraphics[width=0.8\textwidth]{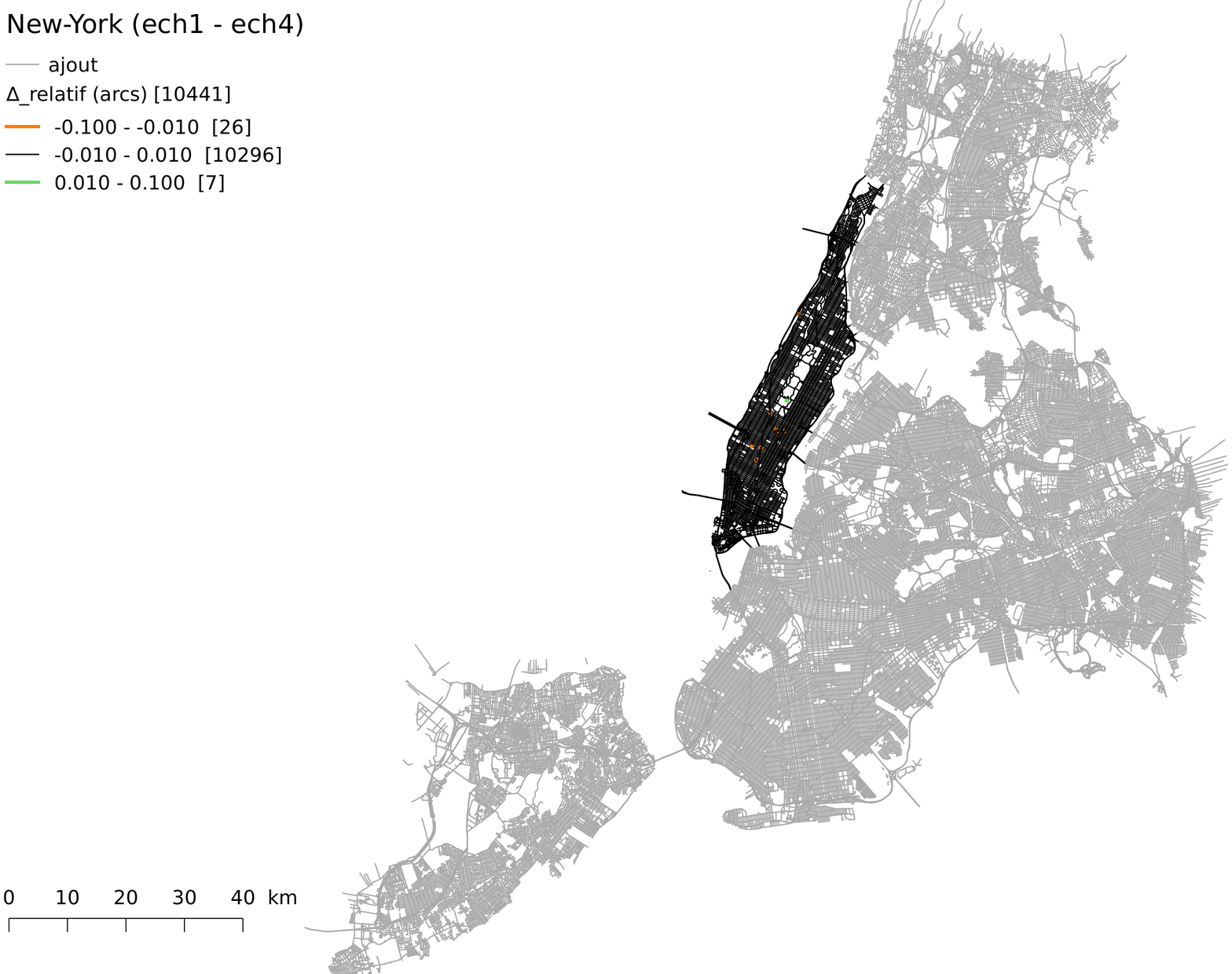}
    \caption{Carte de $\Delta_{relatif}$ calculé entre les échantillons 1 et 4 de New-York.}
    \label{fig_ann:diff_man_il}
\end{figure}

\begin{figure}[b]
    \centering
    \includegraphics[width=0.8\textwidth]{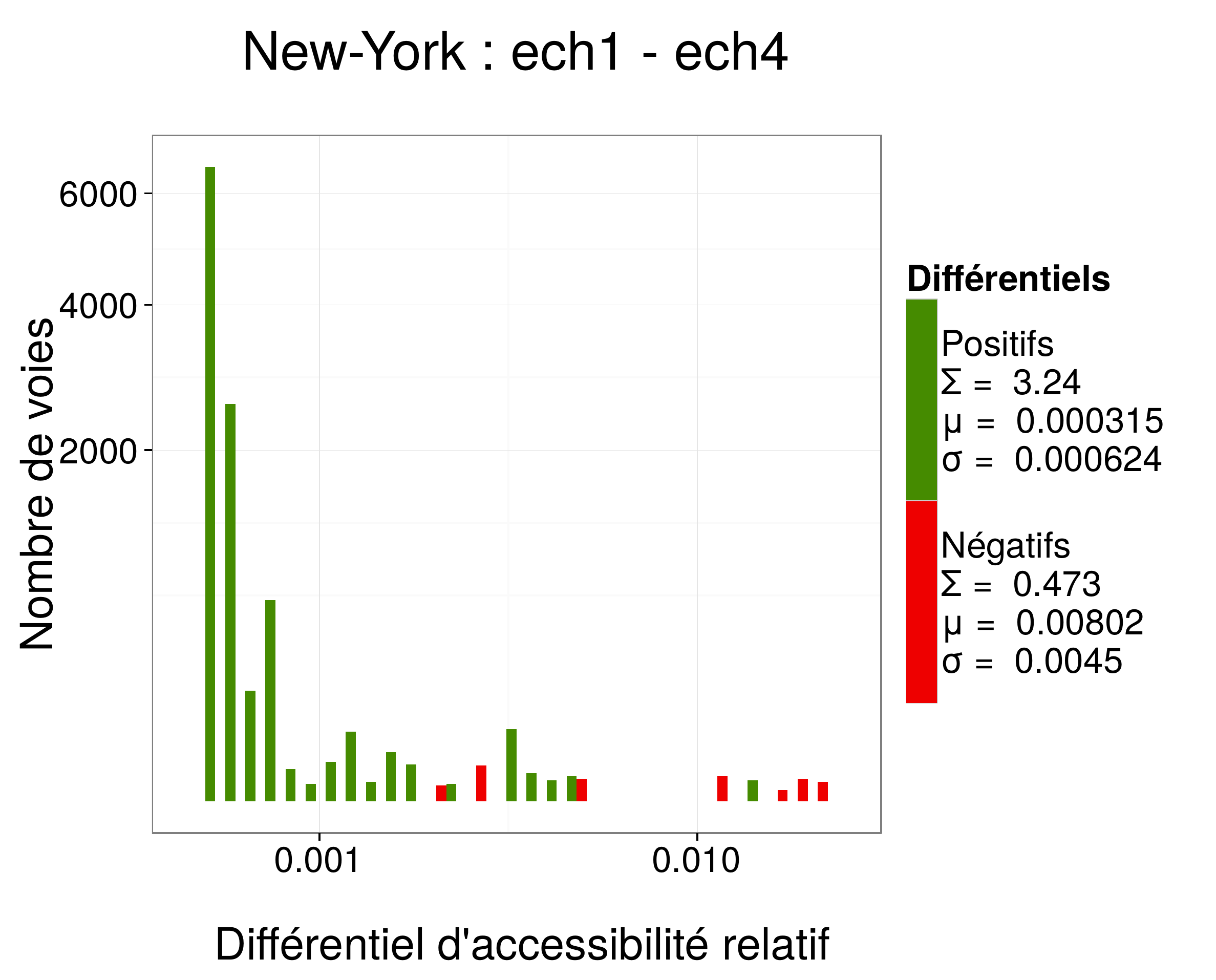}
    \caption{Répartition de $\Delta_{relatif}$ calculé entre les échantillons 1 et 4 de New-York.}
    \label{fig_ann:diff_hist_man_il}
\end{figure}

\begin{figure}[t]
    \centering
    \includegraphics[width=0.8\textwidth]{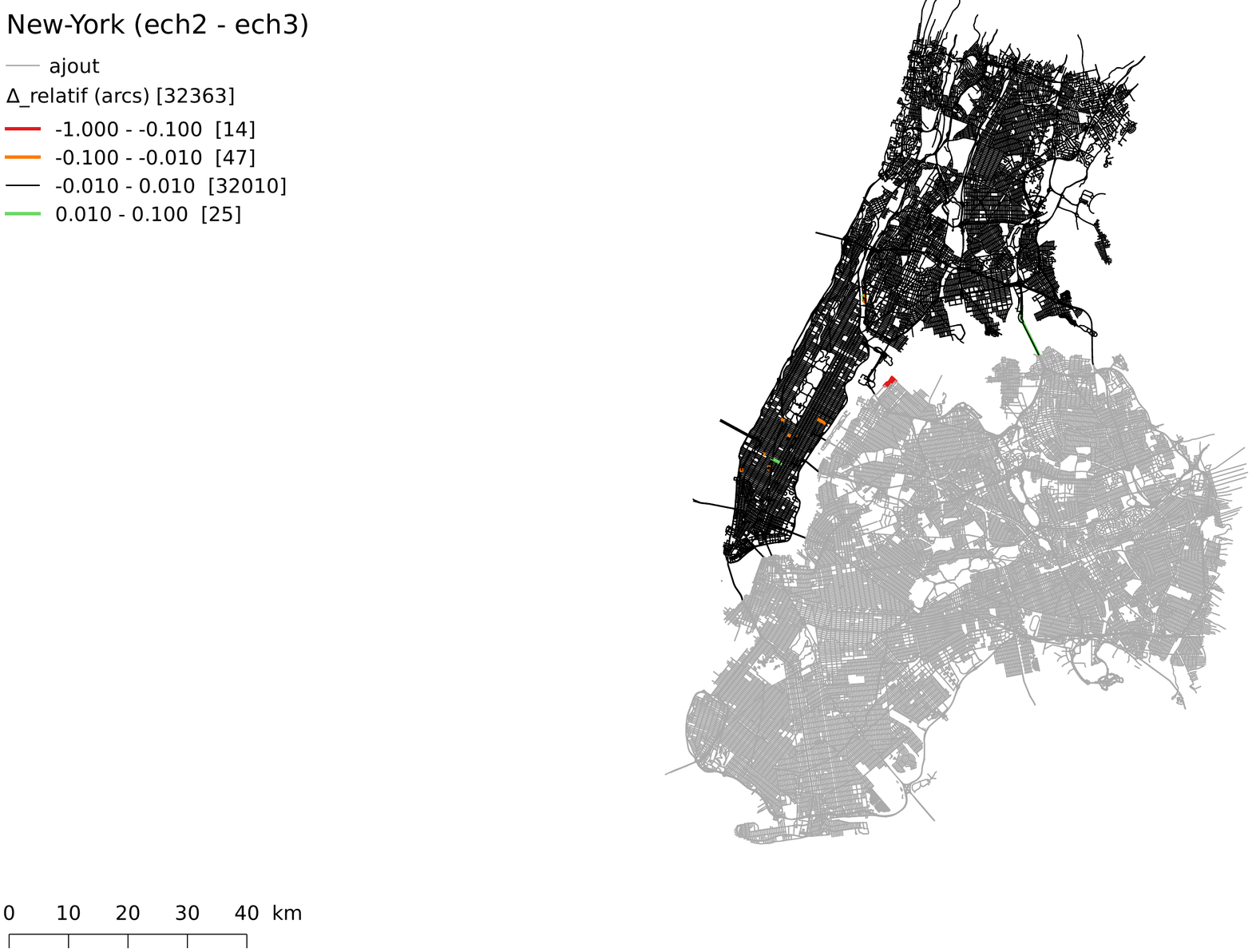}
    \caption{Carte de $\Delta_{relatif}$ calculé entre les échantillons 2 et 3 de New-York.}
    \label{fig_ann:diff_man_sm}
\end{figure}

\begin{figure}[b]
    \centering
    \includegraphics[width=0.8\textwidth]{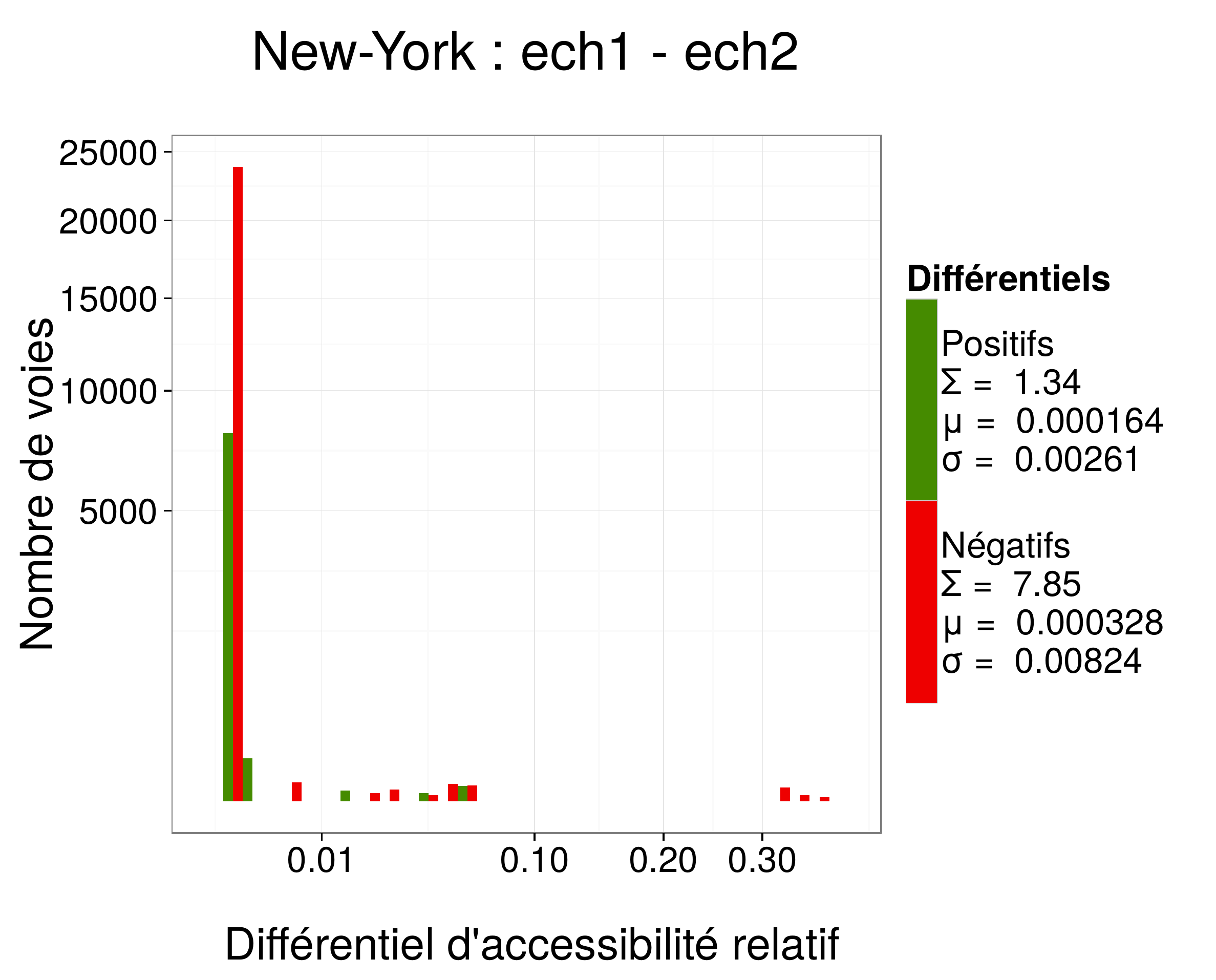}
    \caption{Répartition de $\Delta_{relatif}$ calculé entre les échantillons 2 et 3 de New-York.}
    \label{fig_ann:diff_hist_man_sm}
\end{figure}

\begin{figure}[t]
    \centering
    \includegraphics[width=0.8\textwidth]{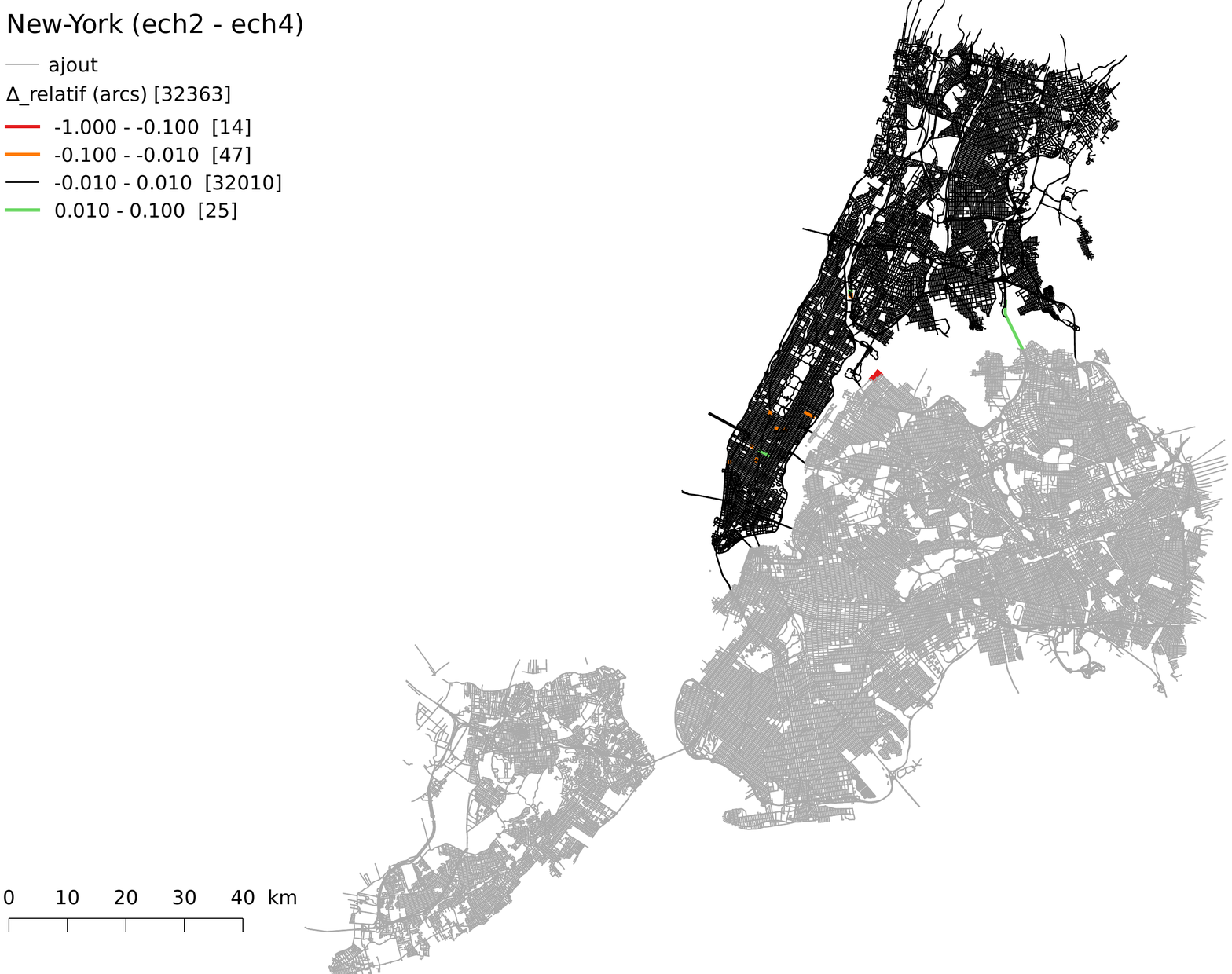}
    \caption{Carte de $\Delta_{relatif}$ calculé entre les échantillons 2 et 4 de New-York.}
    \label{fig_ann:diff_man_sl}
\end{figure}

\begin{figure}[b]
    \centering
    \includegraphics[width=0.8\textwidth]{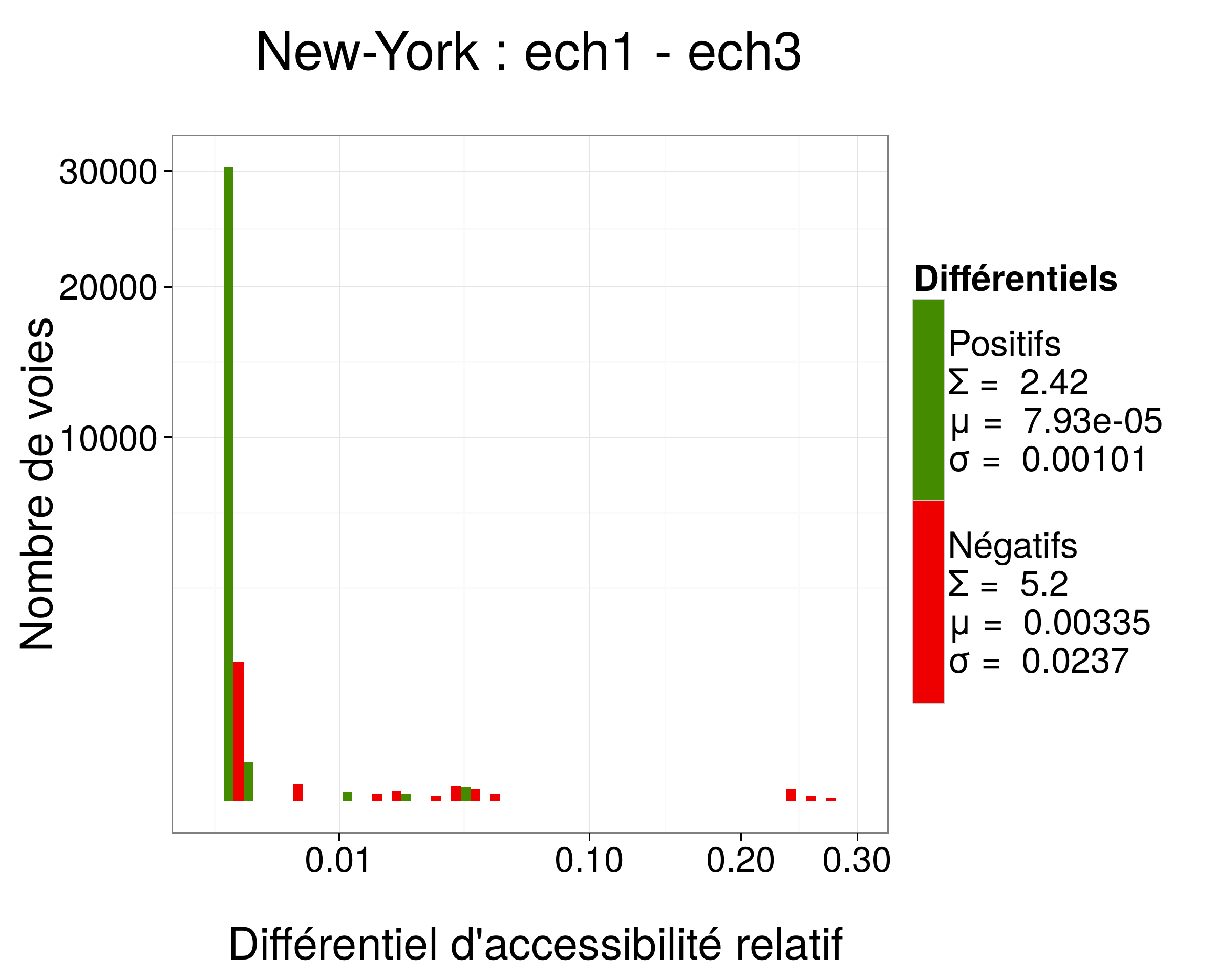}
    \caption{Répartition de $\Delta_{relatif}$ calculé entre les échantillons 2 et 4 de New-York.}
    \label{fig_ann:diff_hist_man_sl}
\end{figure}

\begin{figure}[t]
    \centering
    \includegraphics[width=0.8\textwidth]{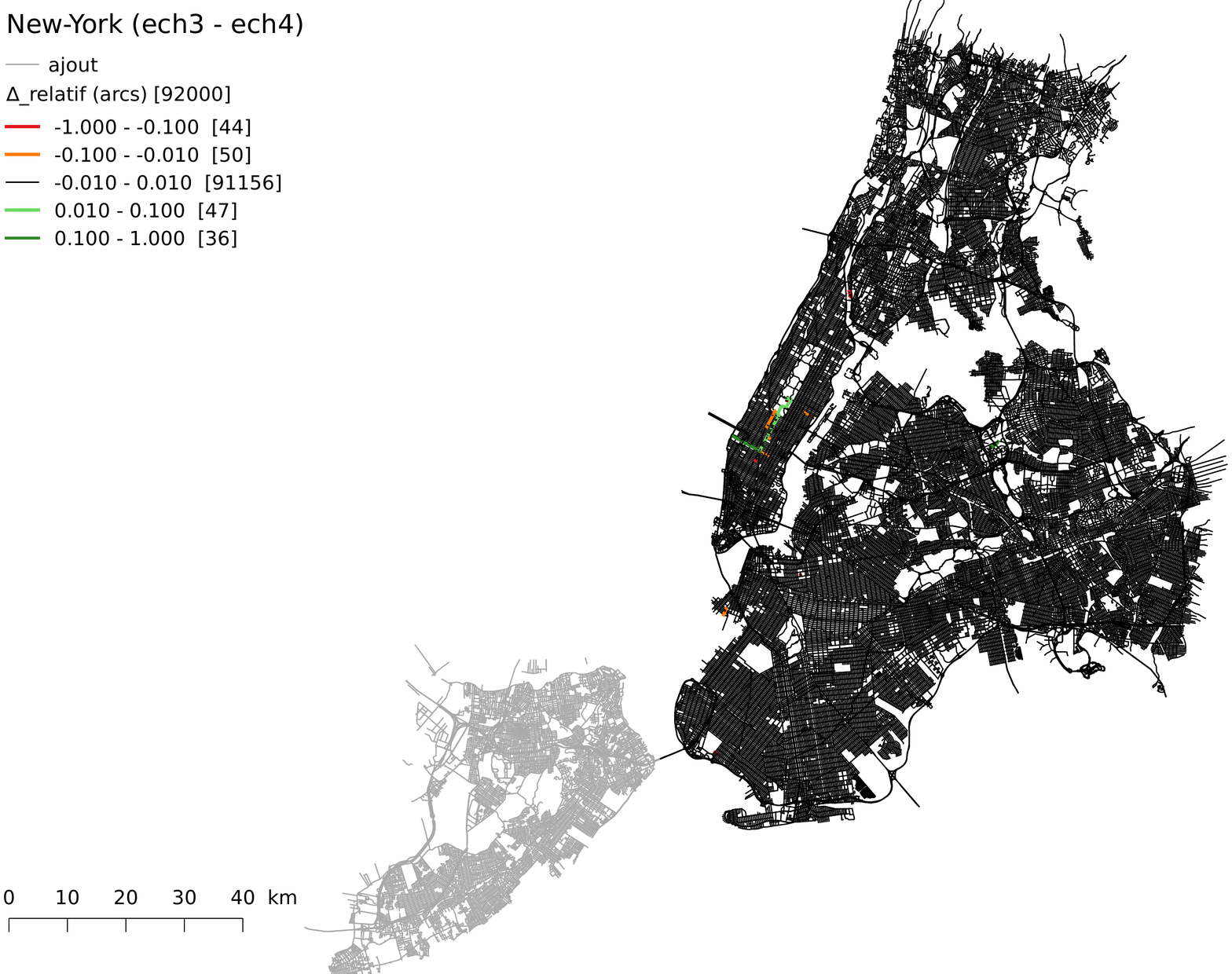}
    \caption{Carte de $\Delta_{relatif}$ calculé entre les échantillons 3 et 4 de New-York.}
    \label{fig_ann:diff_man_ml}
\end{figure}

\begin{figure}[b]
    \centering
    \includegraphics[width=0.8\textwidth]{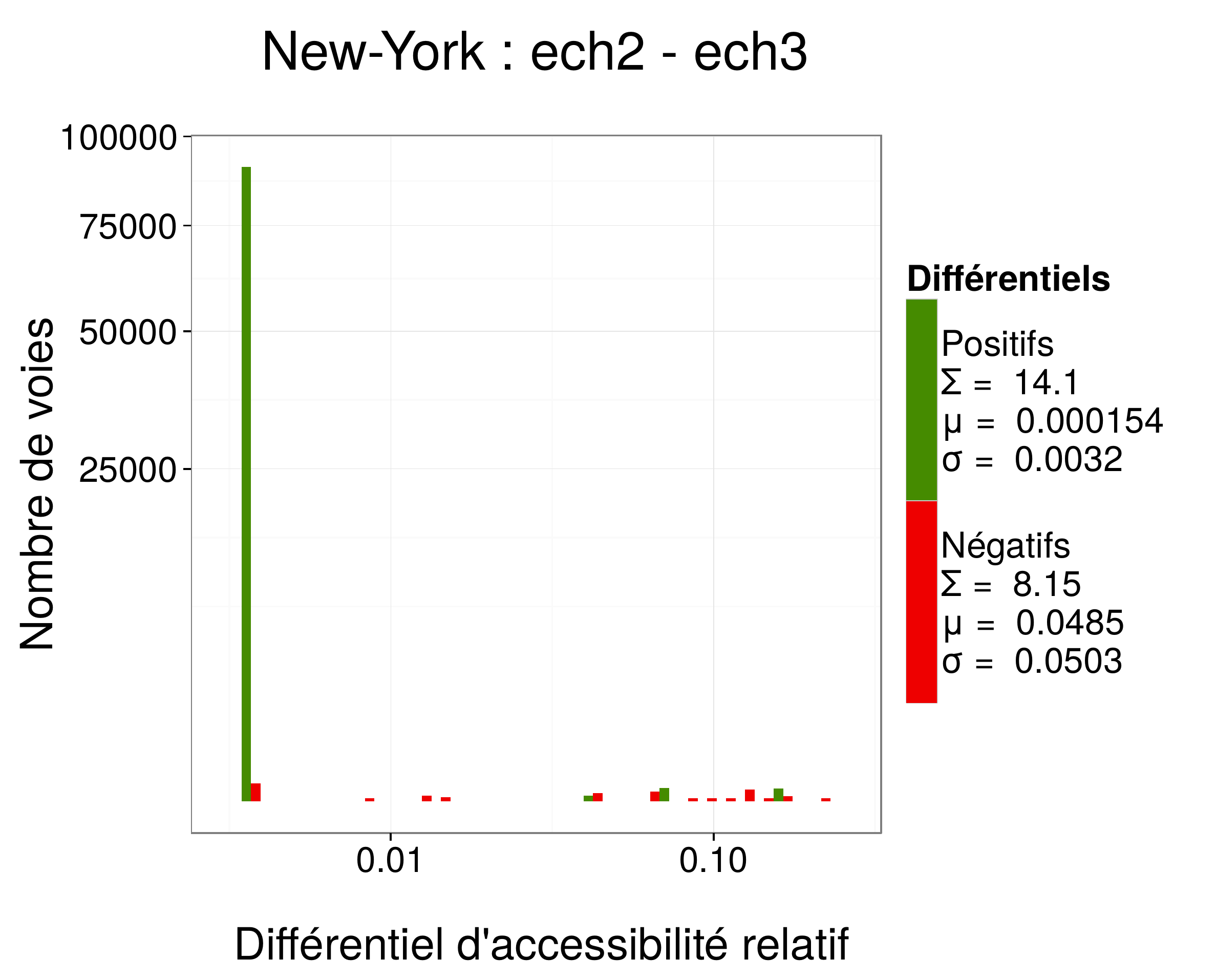}
    \caption{Répartition de $\Delta_{relatif}$ calculé entre les échantillons 3 et 4 de New-York.}
    \label{fig_ann:diff_hist_man_ml}
\end{figure}

\FloatBarrier 
\chapter{Graphes des villes du panel de recherche}\label{ann:chap_panel_graphe}


\begin{figure}[h]
	    \centering
	    \includegraphics[width=\textwidth]{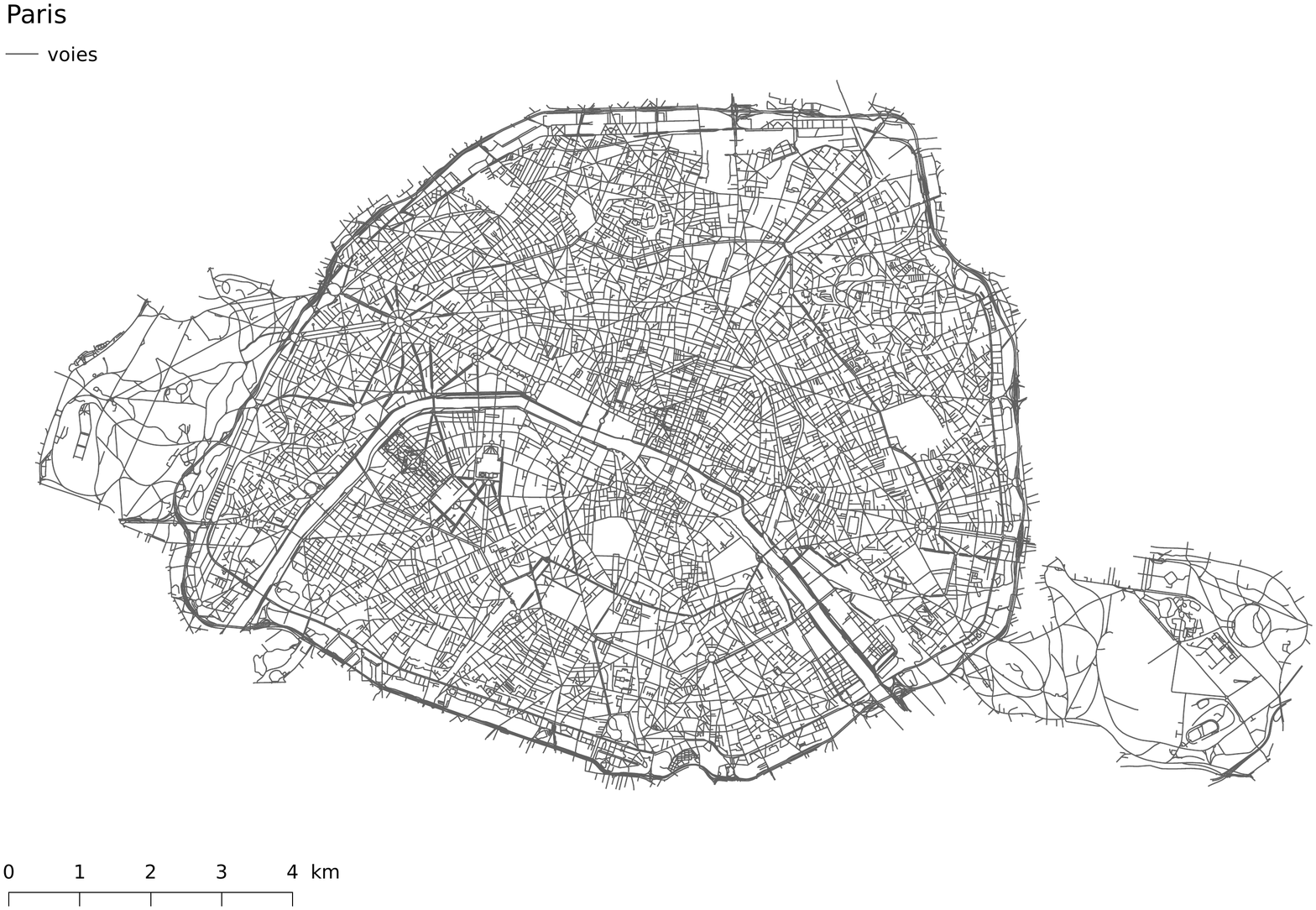}
	    \caption*{Graphe brut}
\end{figure}


\begin{figure}[h]
	    \centering
	    \includegraphics[width=\textwidth]{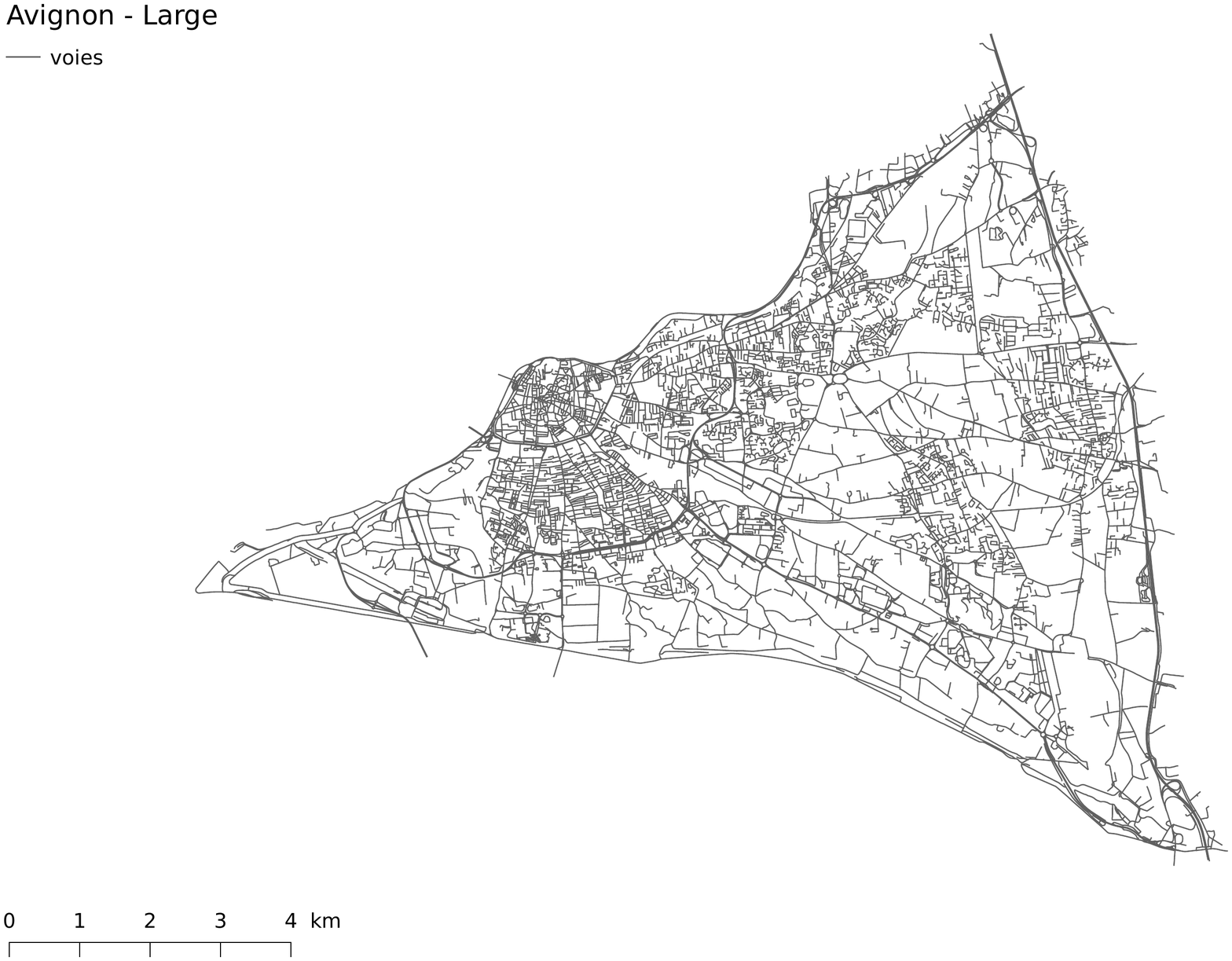}
	    \caption*{Graphe brut}
\end{figure}


\begin{figure}[h]
	    \centering
	    \includegraphics[width=\textwidth]{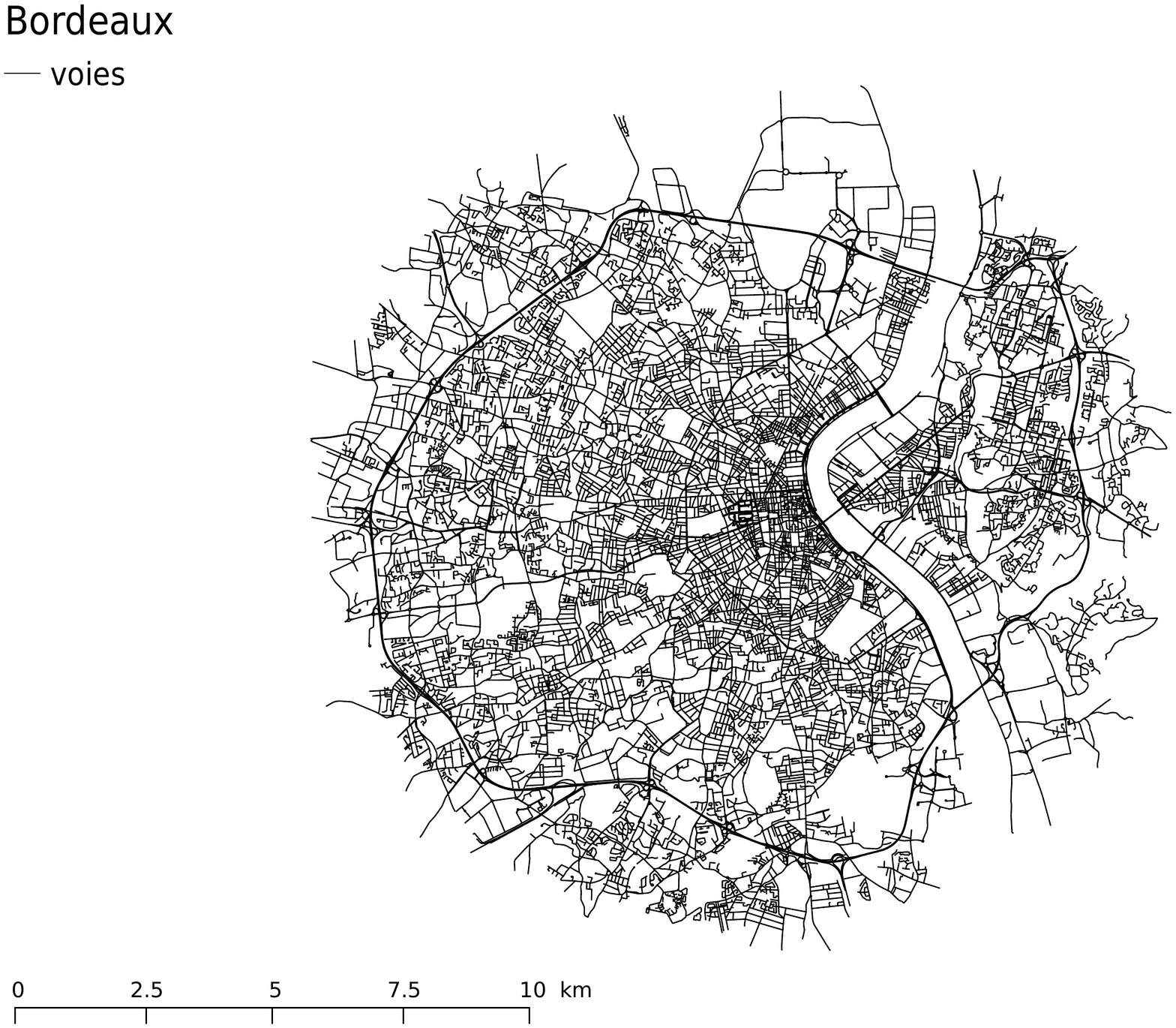}
	    \caption*{Graphe brut}
\end{figure}


\begin{figure}[h]
	    \centering
	    \includegraphics[width=\textwidth]{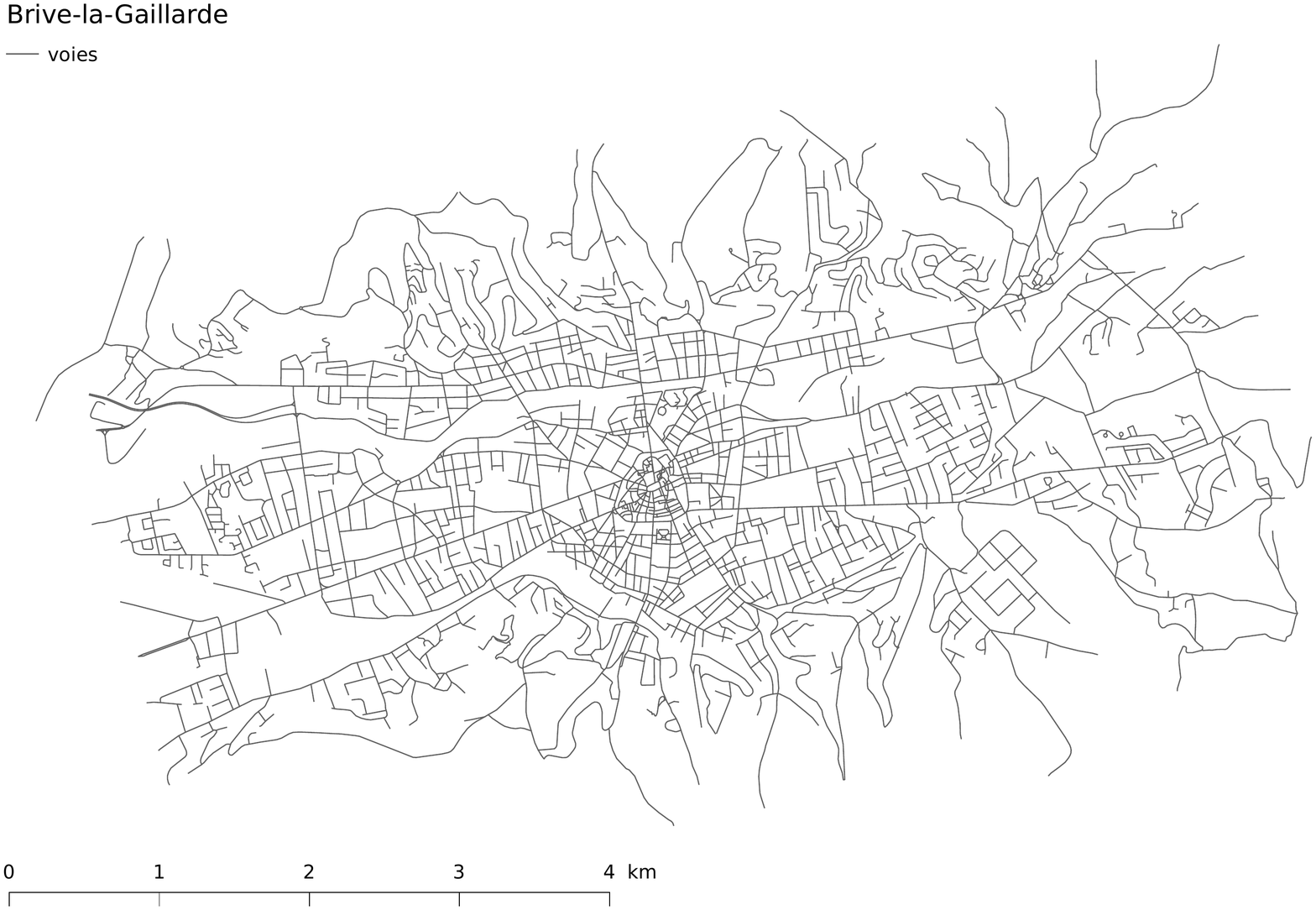}
	    \caption*{Graphe brut}
\end{figure}


\begin{figure}[h]
	    \centering
	    \includegraphics[width=\textwidth]{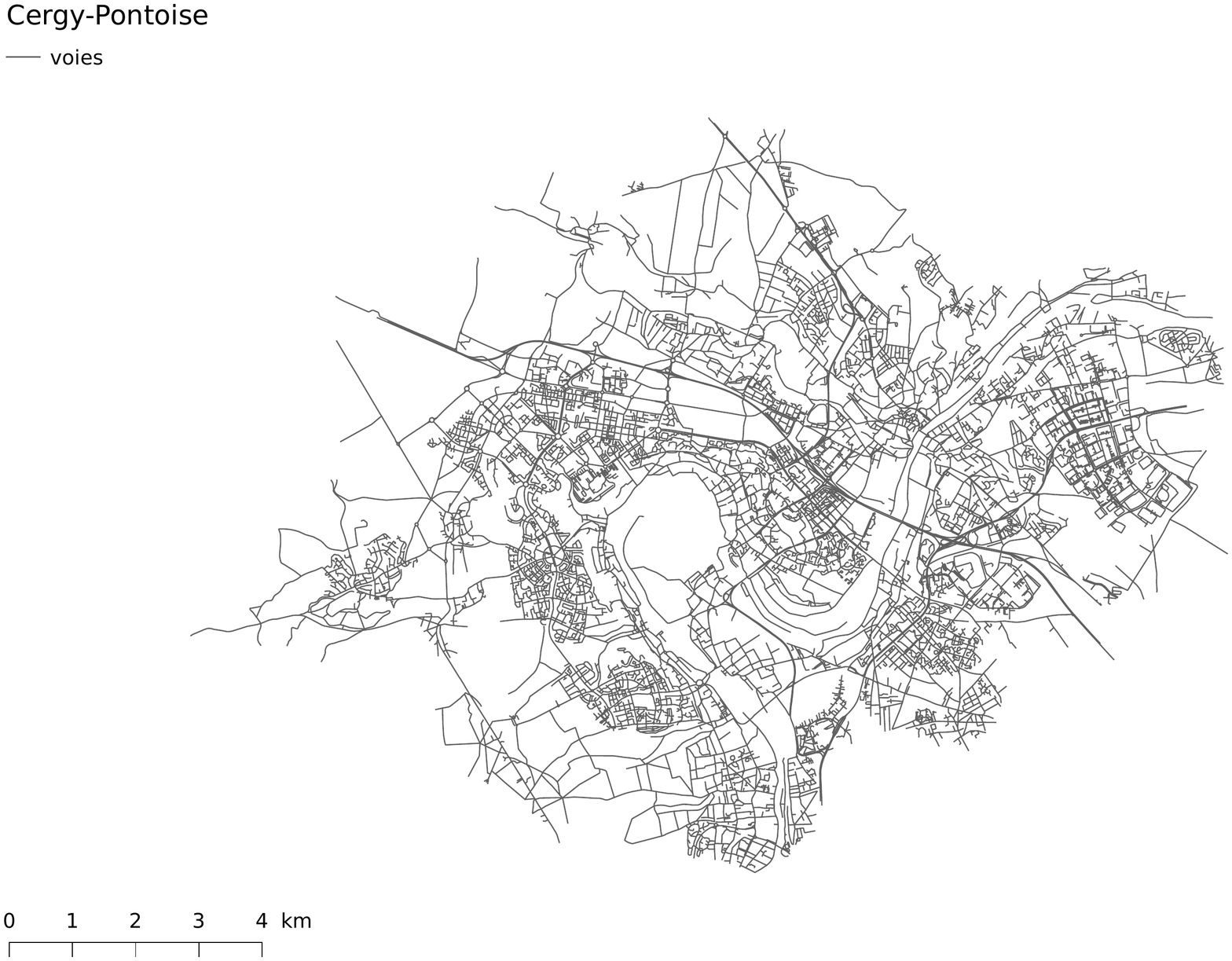}
	    \caption*{Graphe brut}
\end{figure}


\begin{figure}[h]
	    \centering
	    \includegraphics[width=\textwidth]{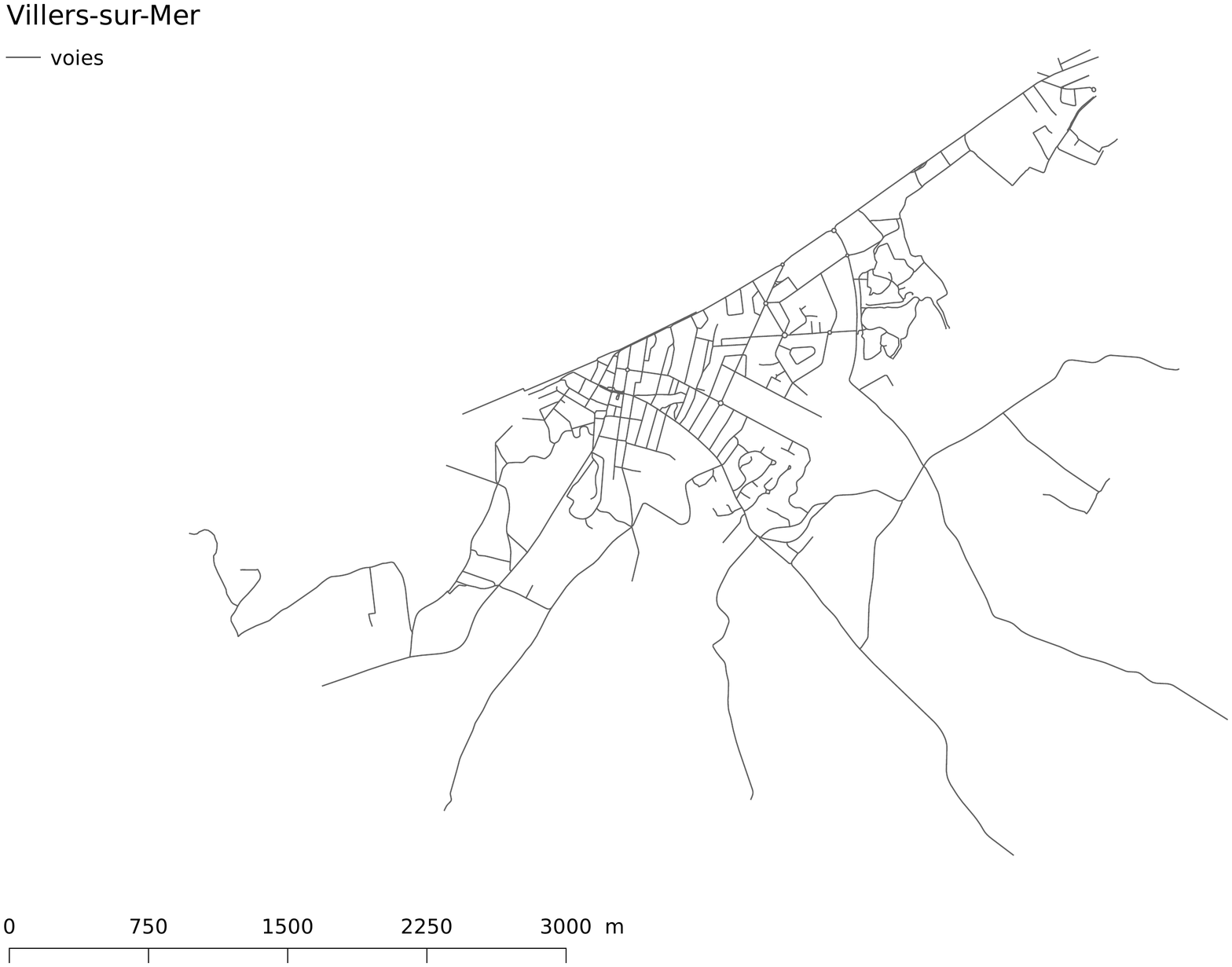}
	    \caption*{Graphe brut}
\end{figure}


\begin{figure}[h]
	    \centering
	    \includegraphics[width=\textwidth]{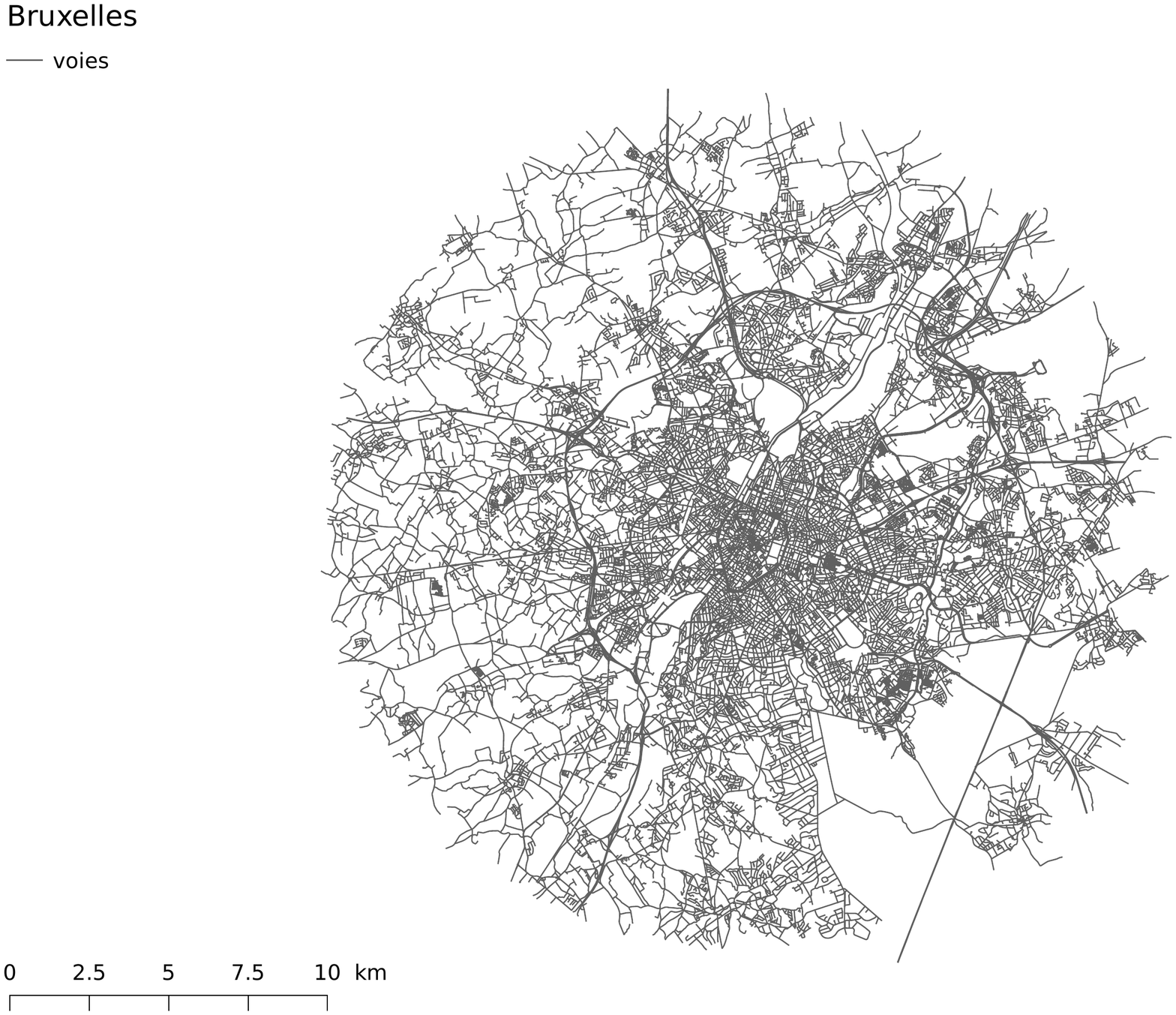}
	    \caption*{Graphe brut}
\end{figure}


\begin{figure}[h]
	    \centering
	    \includegraphics[width=\textwidth]{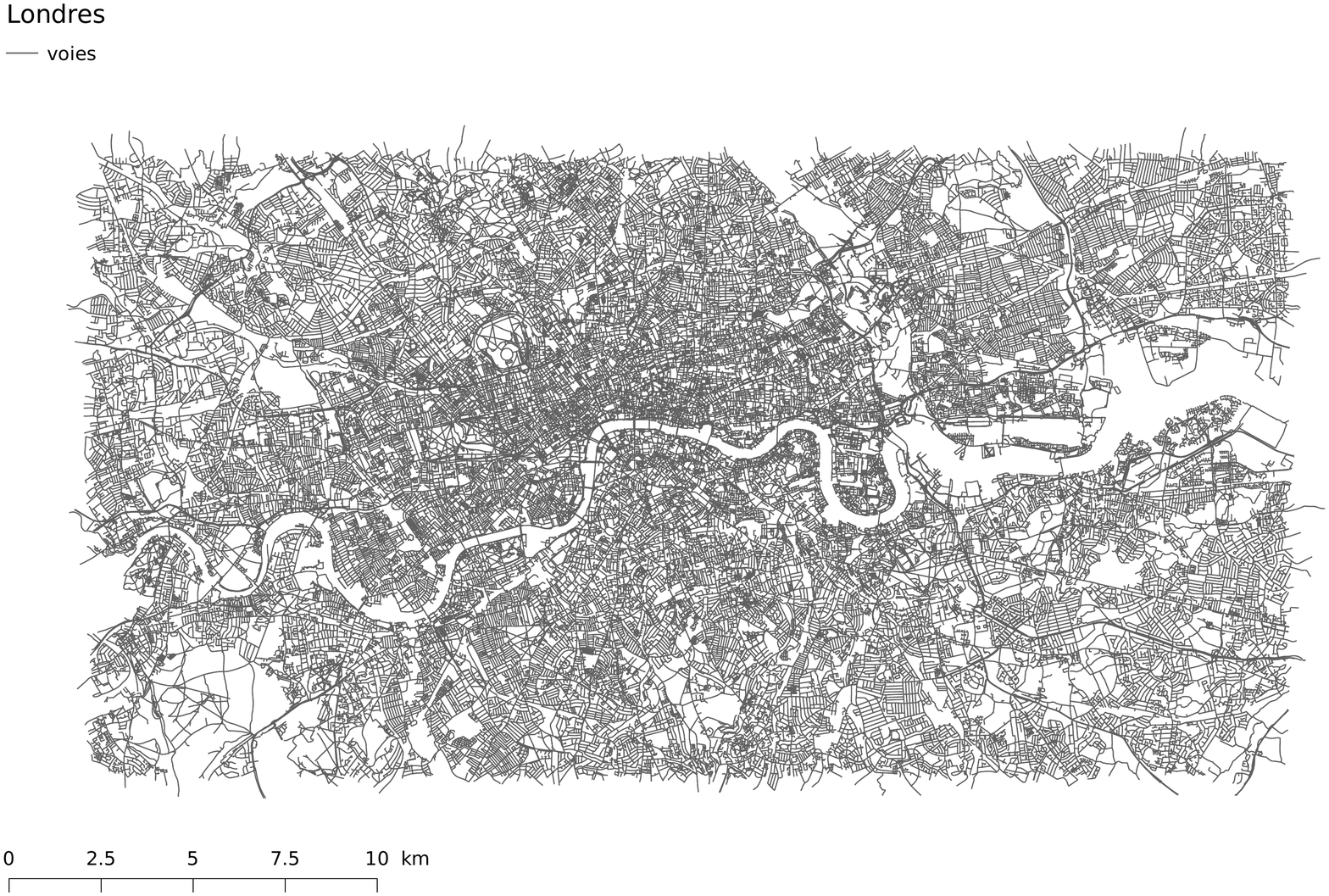}
	    \caption*{Graphe brut}
\end{figure}


\begin{figure}[h]
	    \centering
	    \includegraphics[width=\textwidth]{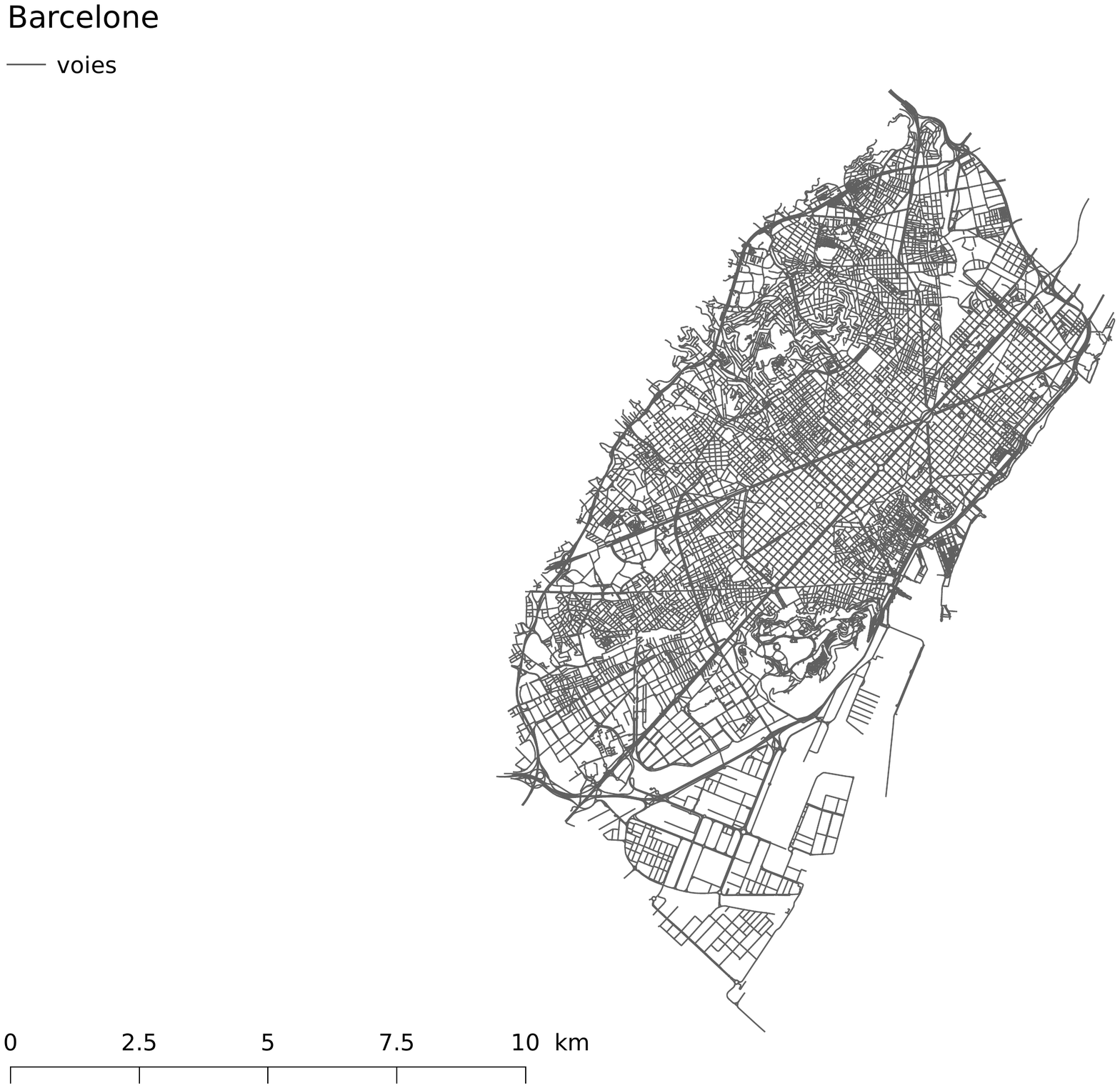}
	    \caption*{Graphe brut}
\end{figure}


\begin{figure}[h]
	    \centering
	    \includegraphics[width=\textwidth]{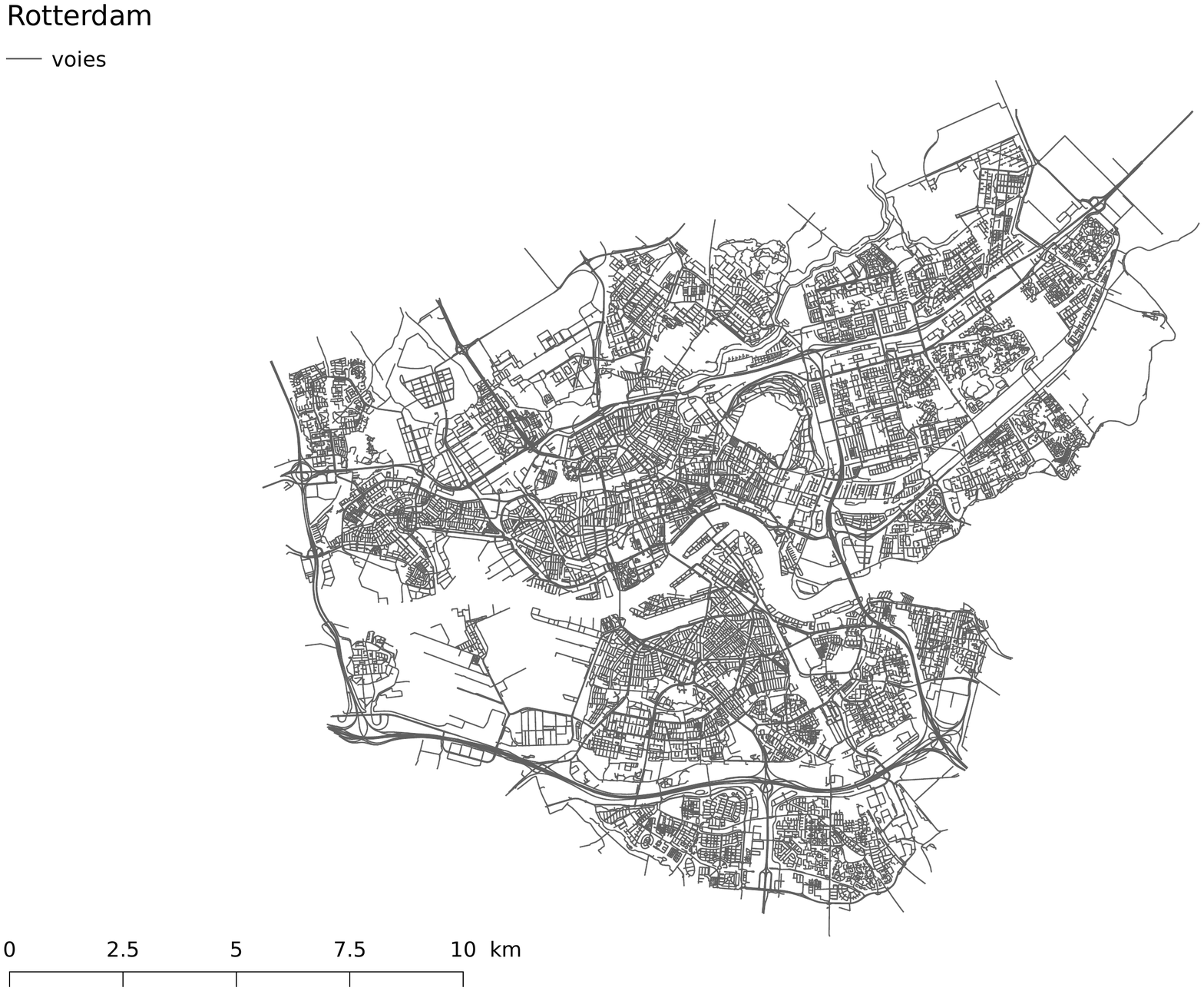}
	    \caption*{Graphe brut}
\end{figure}

\clearpage


\begin{figure}[h]
	    \centering
	    \includegraphics[width=\textwidth]{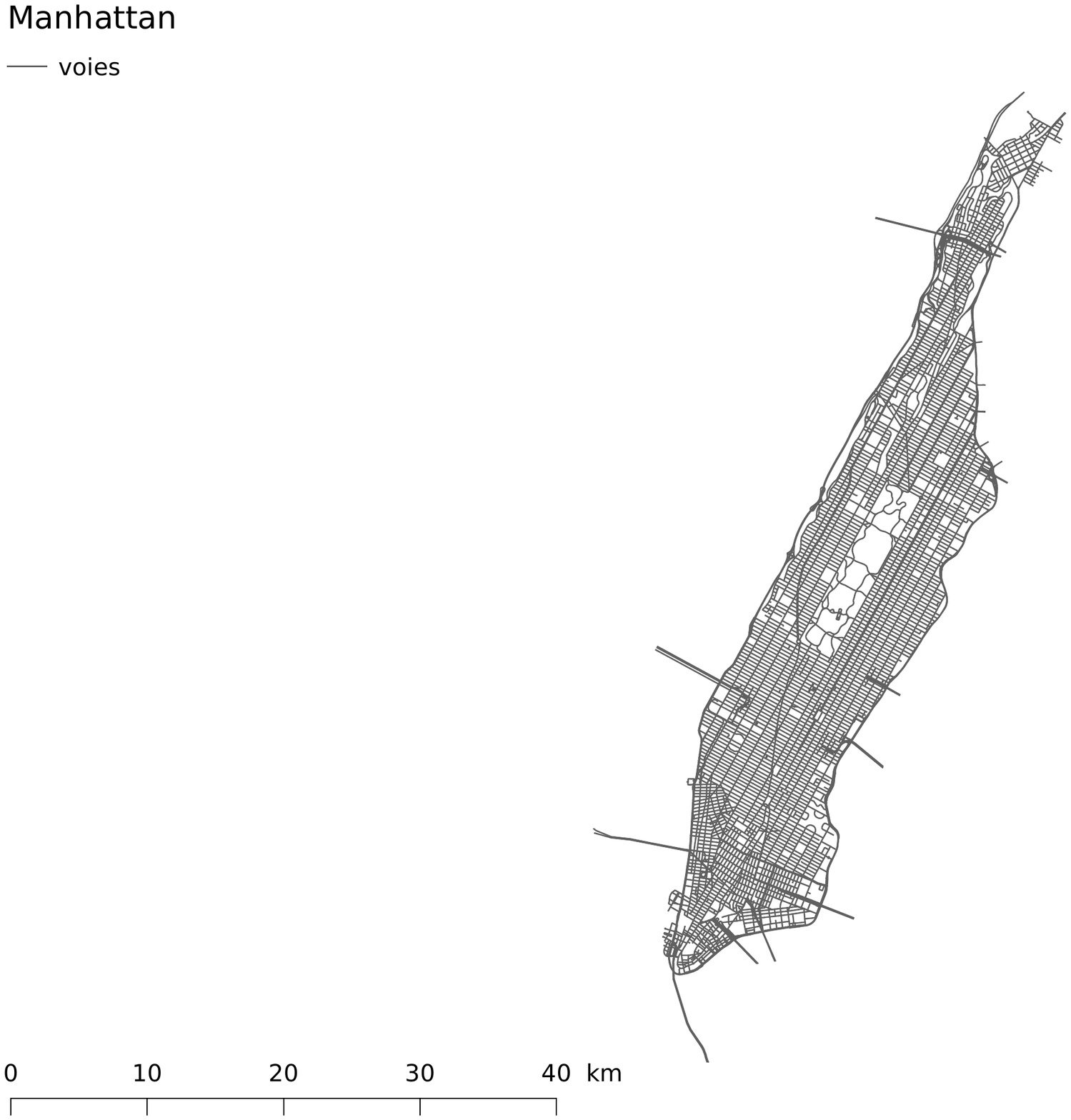}
	    \caption*{Graphe brut}
\end{figure}


\begin{figure}[h]
	    \centering
	    \includegraphics[width=\textwidth]{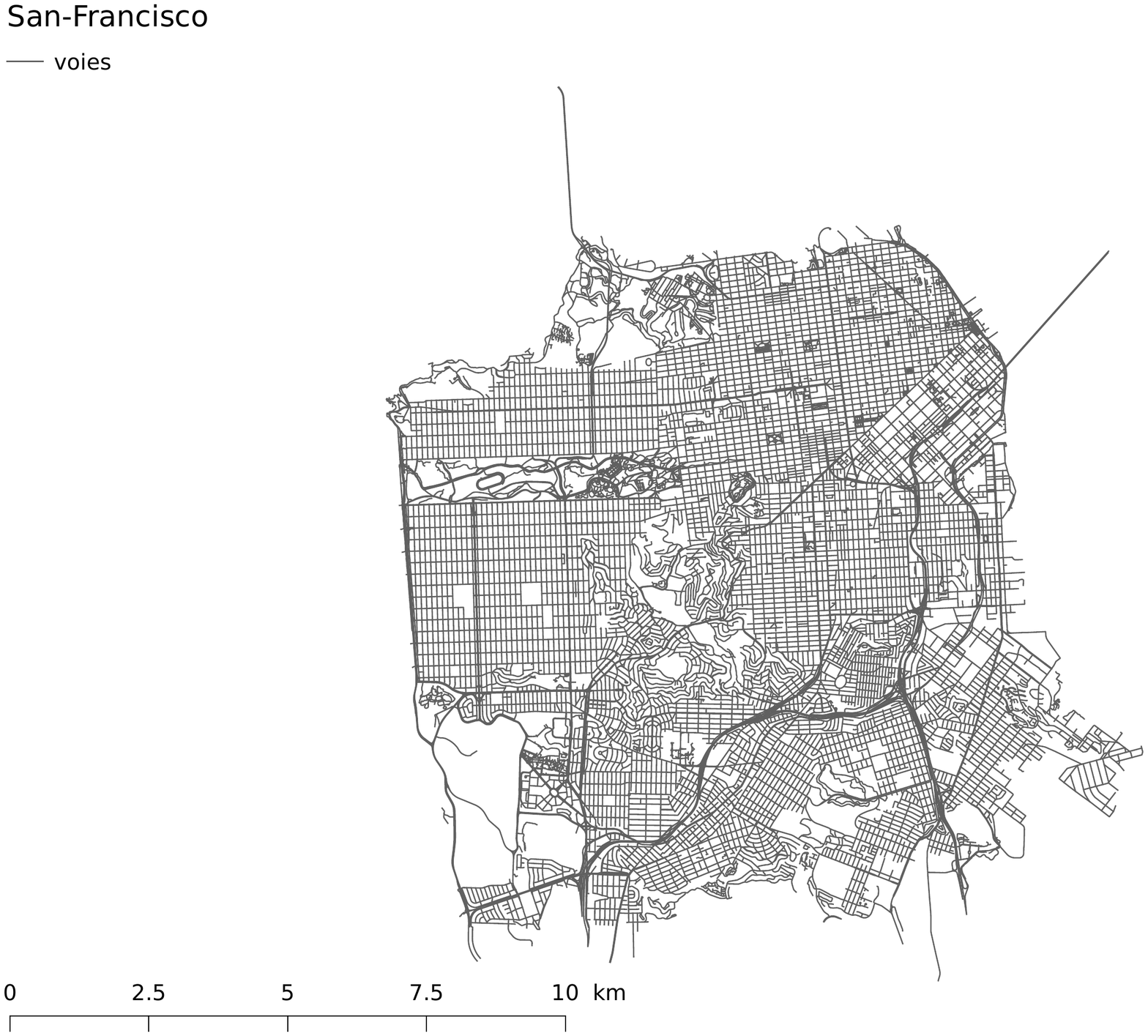}
	    \caption*{Graphe brut}
\end{figure}


\begin{figure}[h]
	    \centering
	    \includegraphics[width=\textwidth]{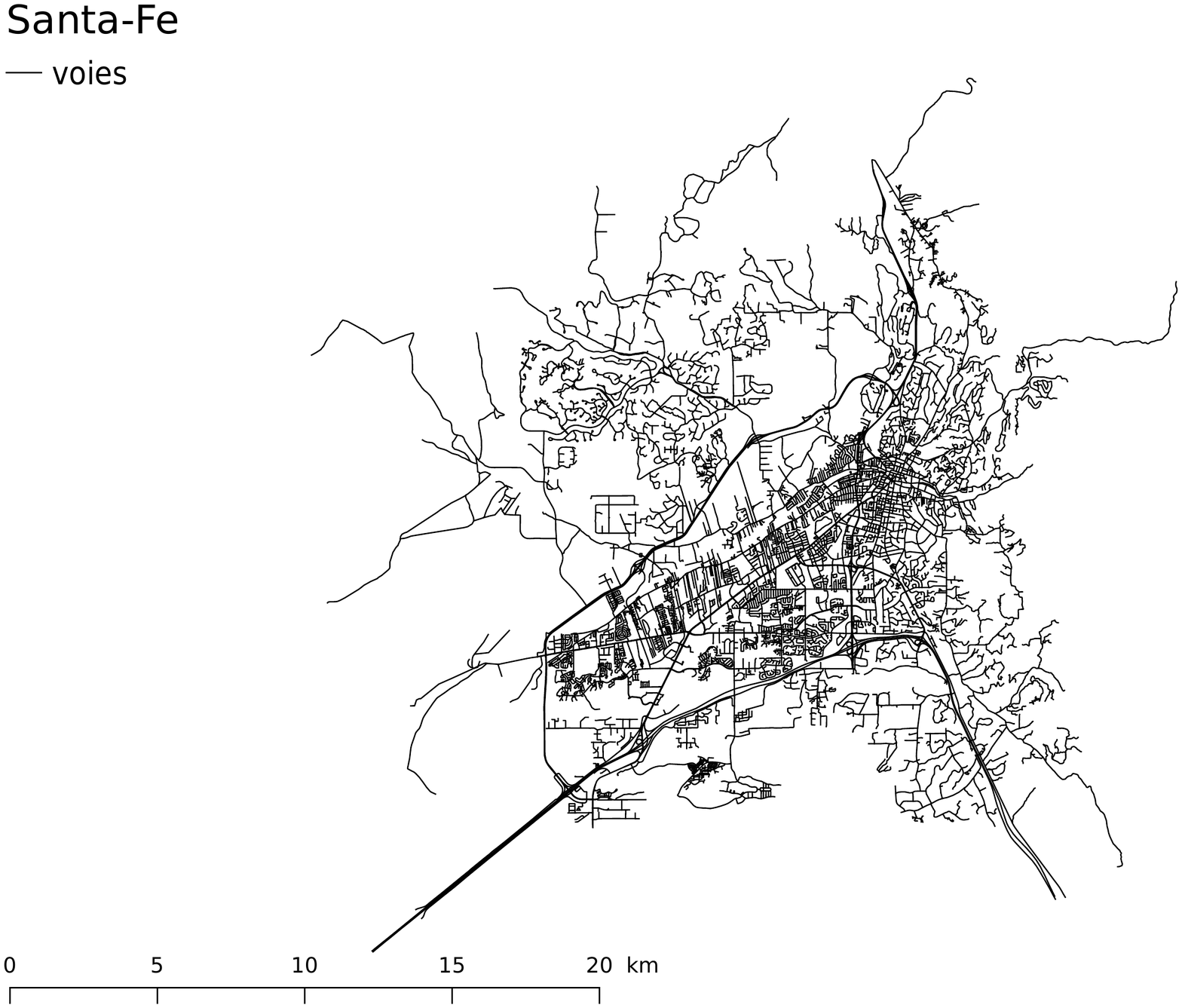}
	    \caption*{Graphe brut}
\end{figure}


\begin{figure}[h]
	    \centering
	    \includegraphics[width=\textwidth]{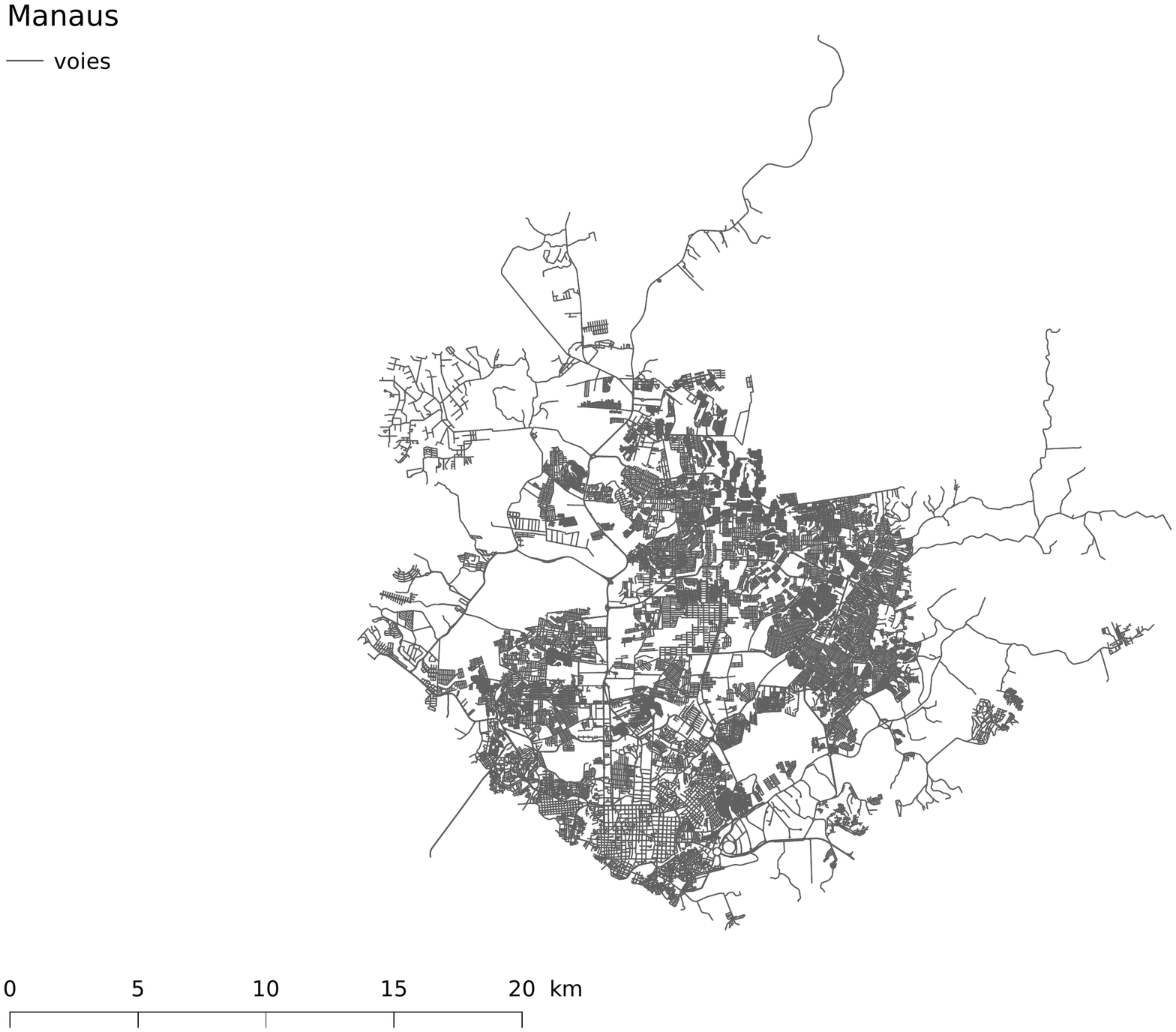}
	    \caption*{Graphe brut}
\end{figure}


\begin{figure}[h]
	    \centering
	    \includegraphics[width=\textwidth]{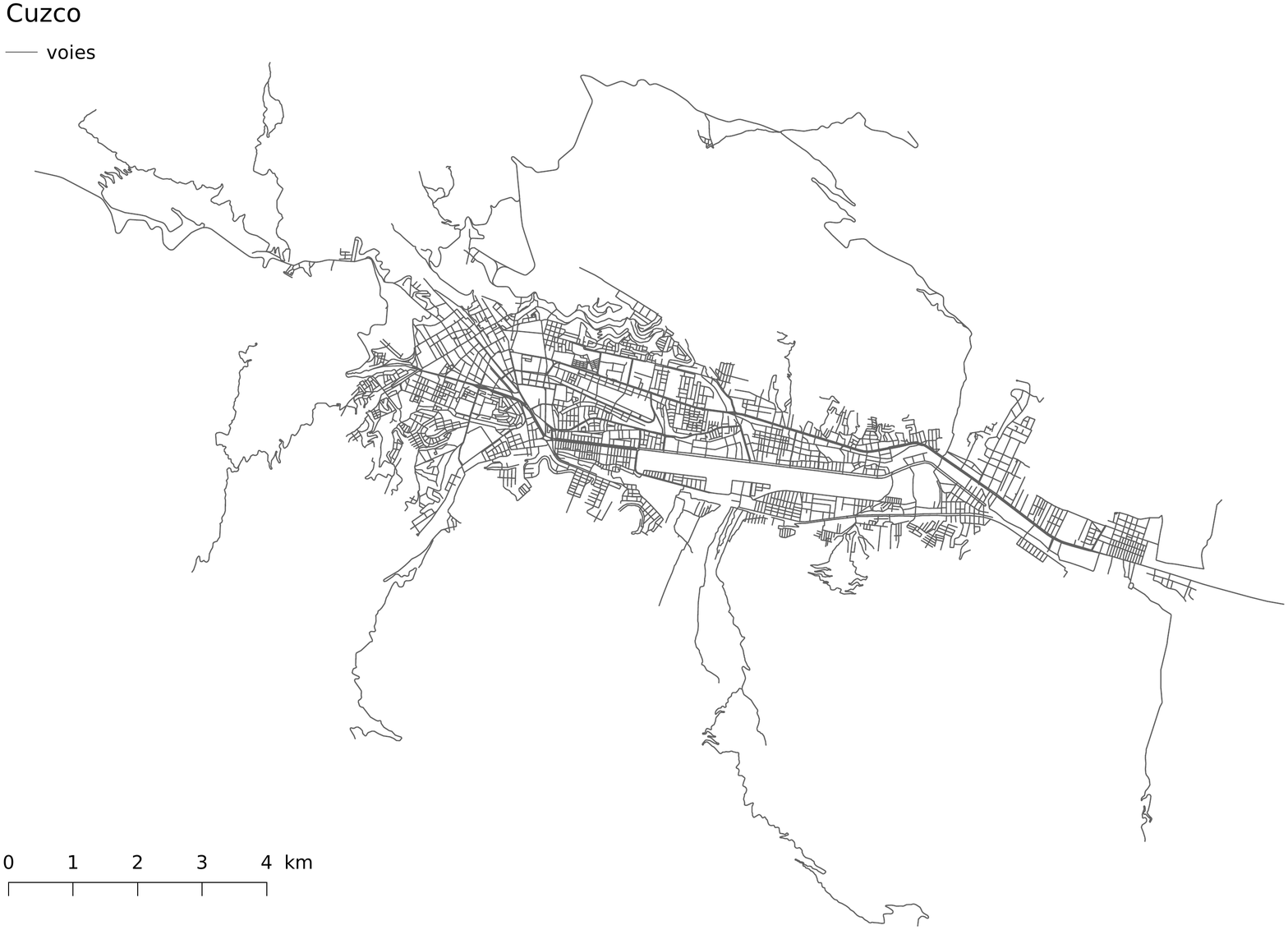}
	    \caption*{Graphe brut}
\end{figure}


\begin{figure}[h]
	    \centering
	    \includegraphics[width=\textwidth]{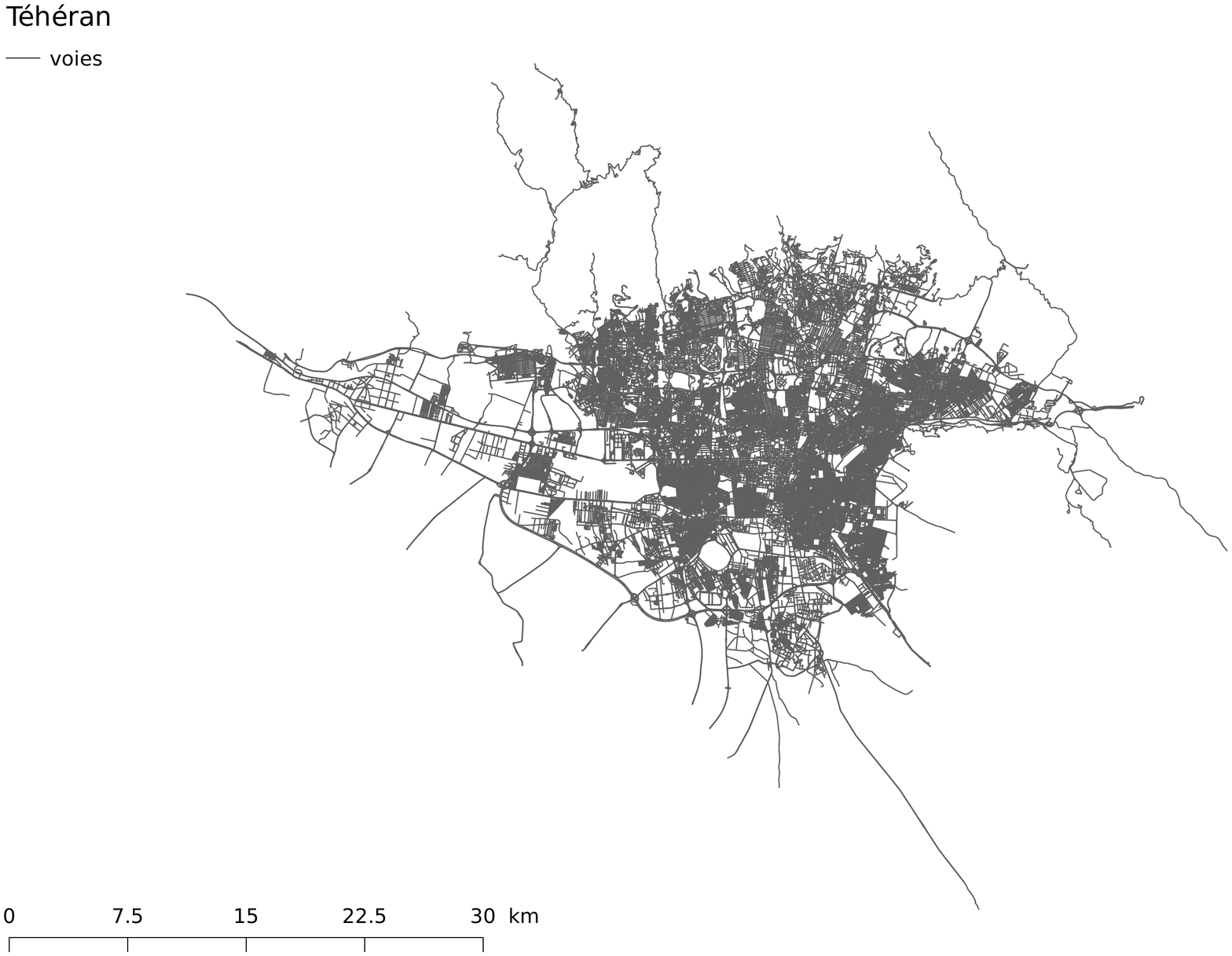}
	    \caption*{Graphe brut}
\end{figure}


\begin{figure}[h]
	    \centering
	    \includegraphics[width=\textwidth]{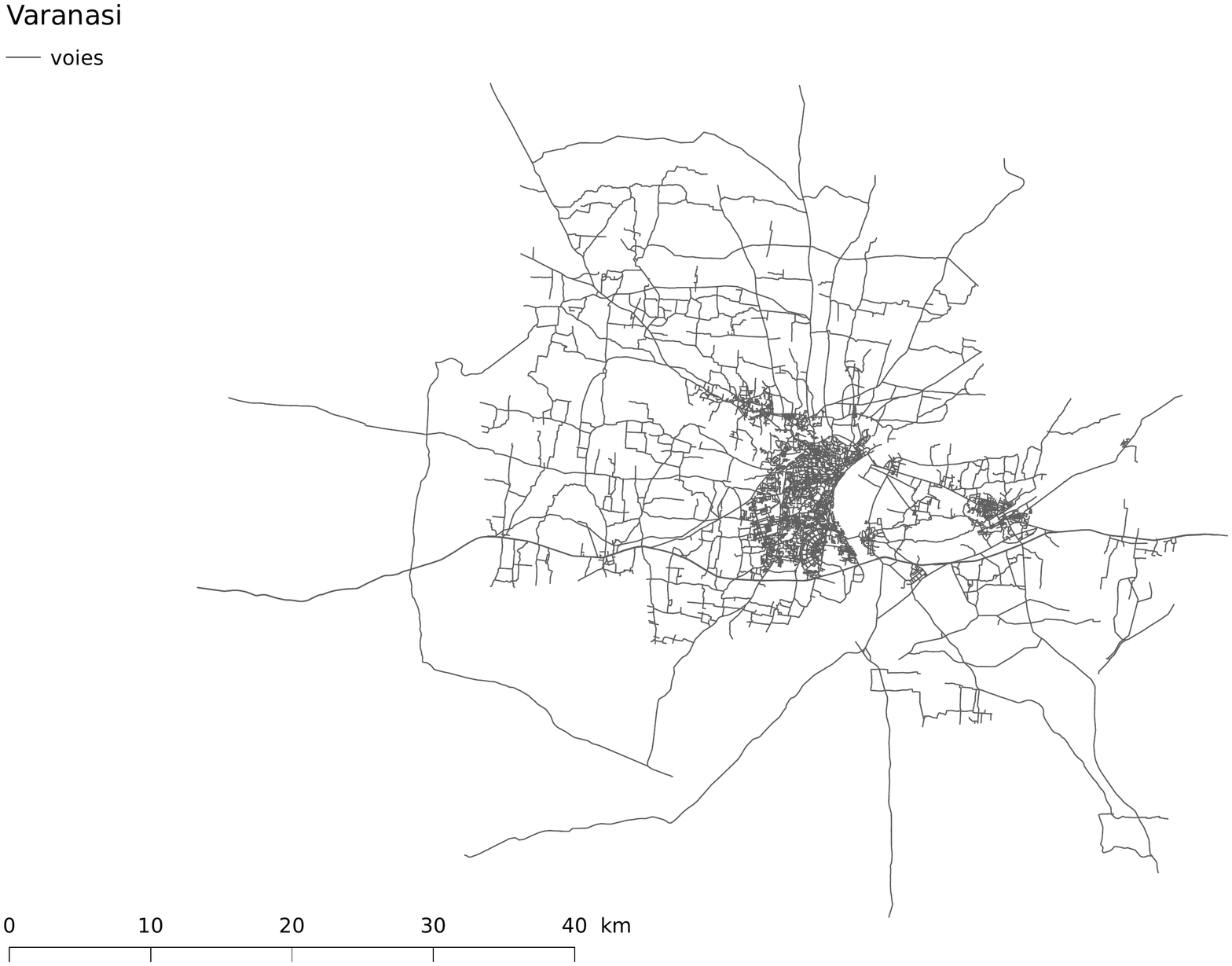}
	    \caption*{Graphe brut}
\end{figure}


\begin{figure}[h]
	    \centering
	    \includegraphics[width=\textwidth]{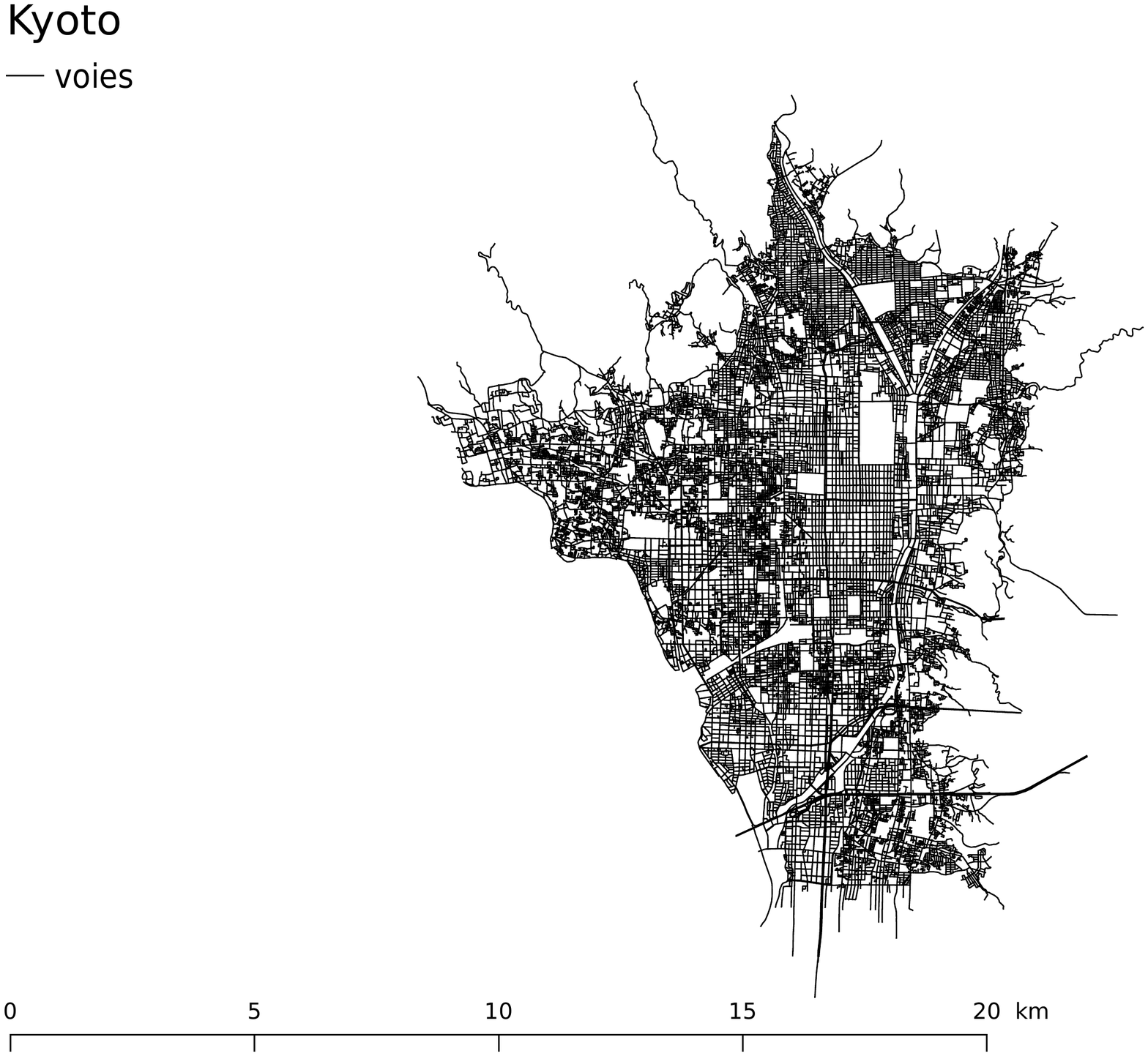}
	    \caption*{Graphe brut}
\end{figure}


\begin{figure}[h]
	    \centering
	    \includegraphics[width=\textwidth]{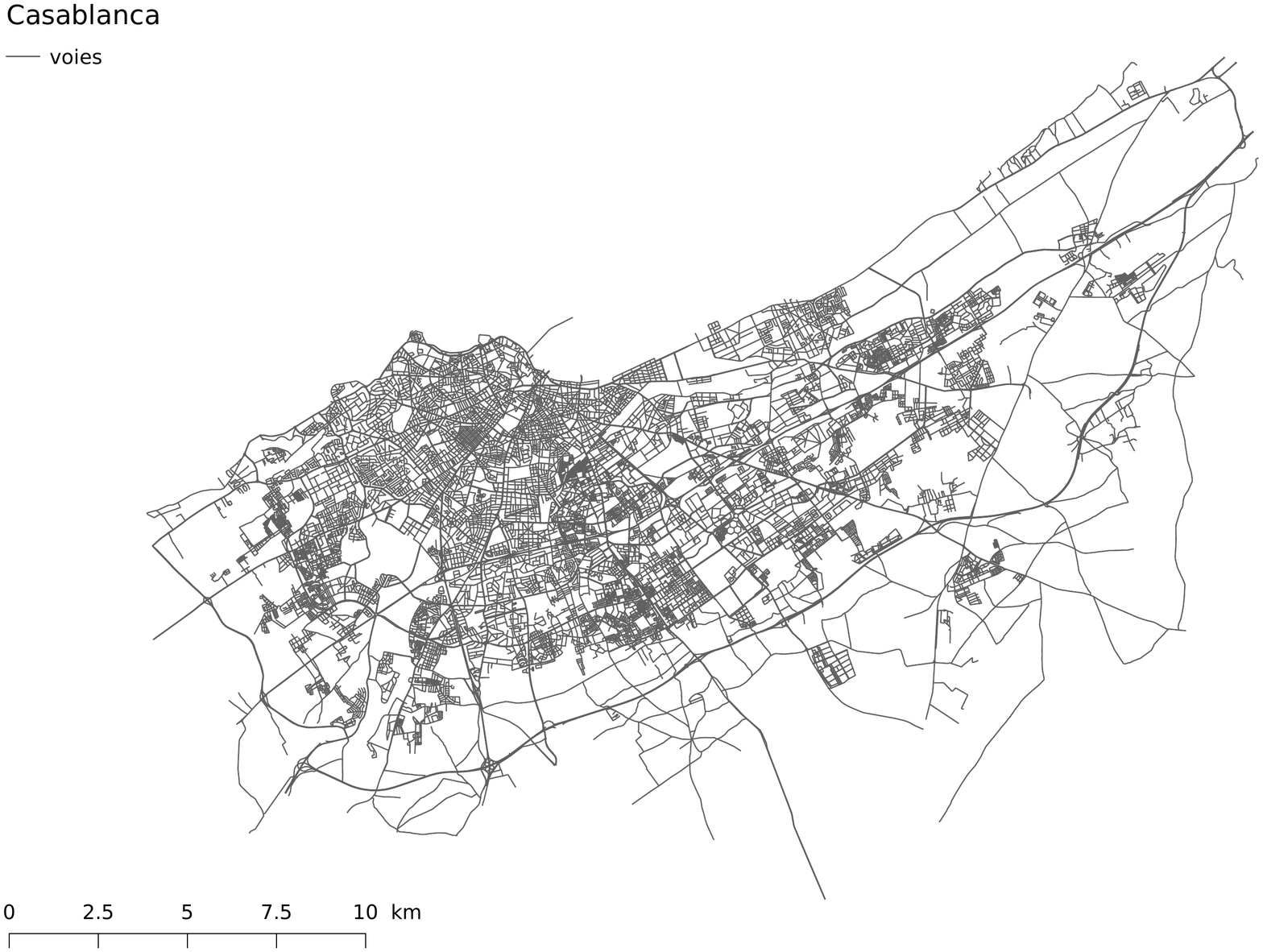}
	    \caption*{Graphe brut}
\end{figure}


\begin{figure}[h]
	    \centering
	    \includegraphics[width=\textwidth]{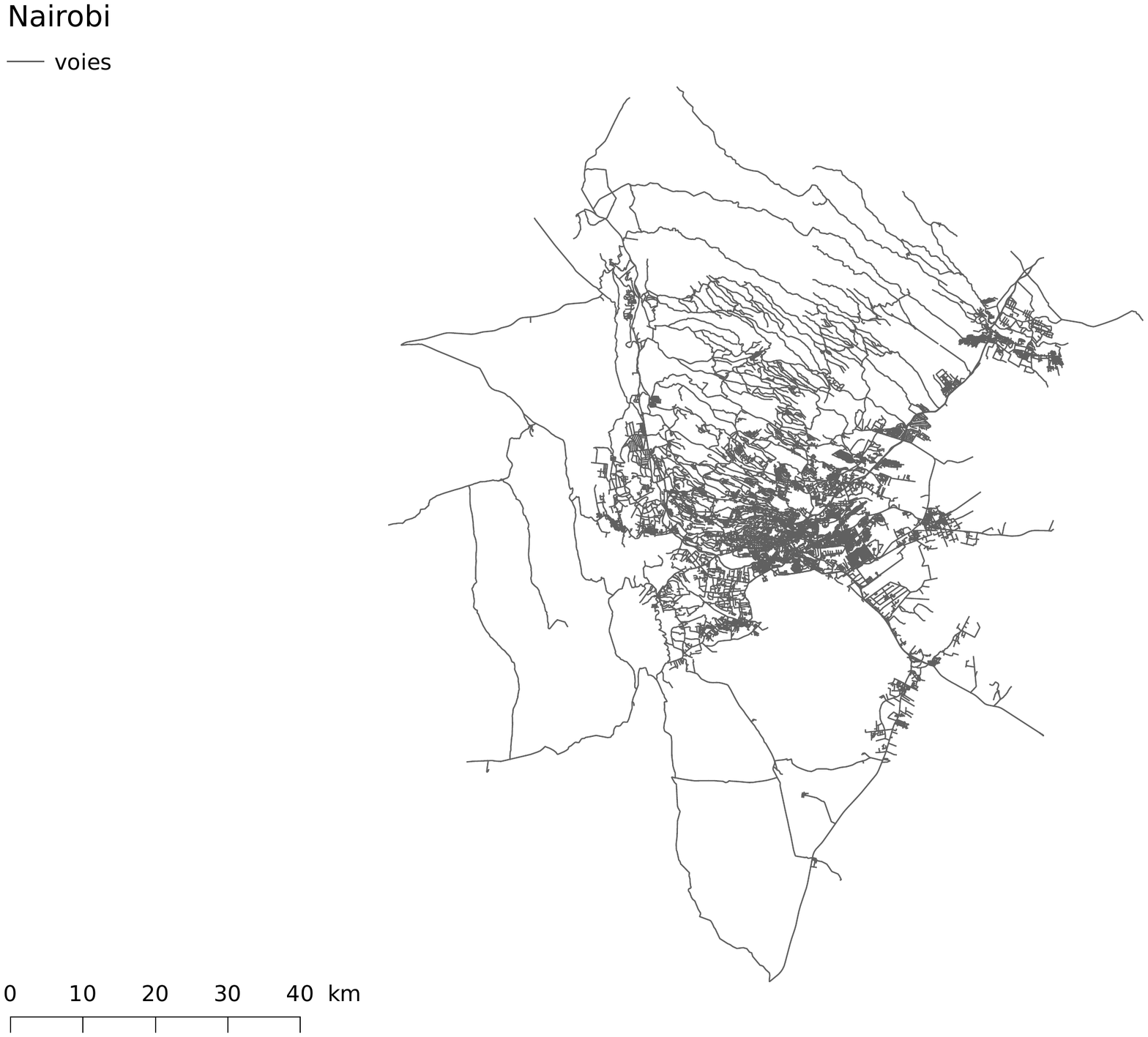}
	    \caption*{Graphe brut}
\end{figure}


\FloatBarrier 
\chapter{Cartes diachroniques de l'indicateur de closeness}\label{ann:chap_cartes-diachr}

\FloatBarrier 
\section{Avignon (1760 - 2014)}\label{ann:sec_cartedia_avignon}

\begin{figure}[h]
    \centering
    
     \begin{subfigure}[t]{.6\linewidth}
        \includegraphics[width=\textwidth]{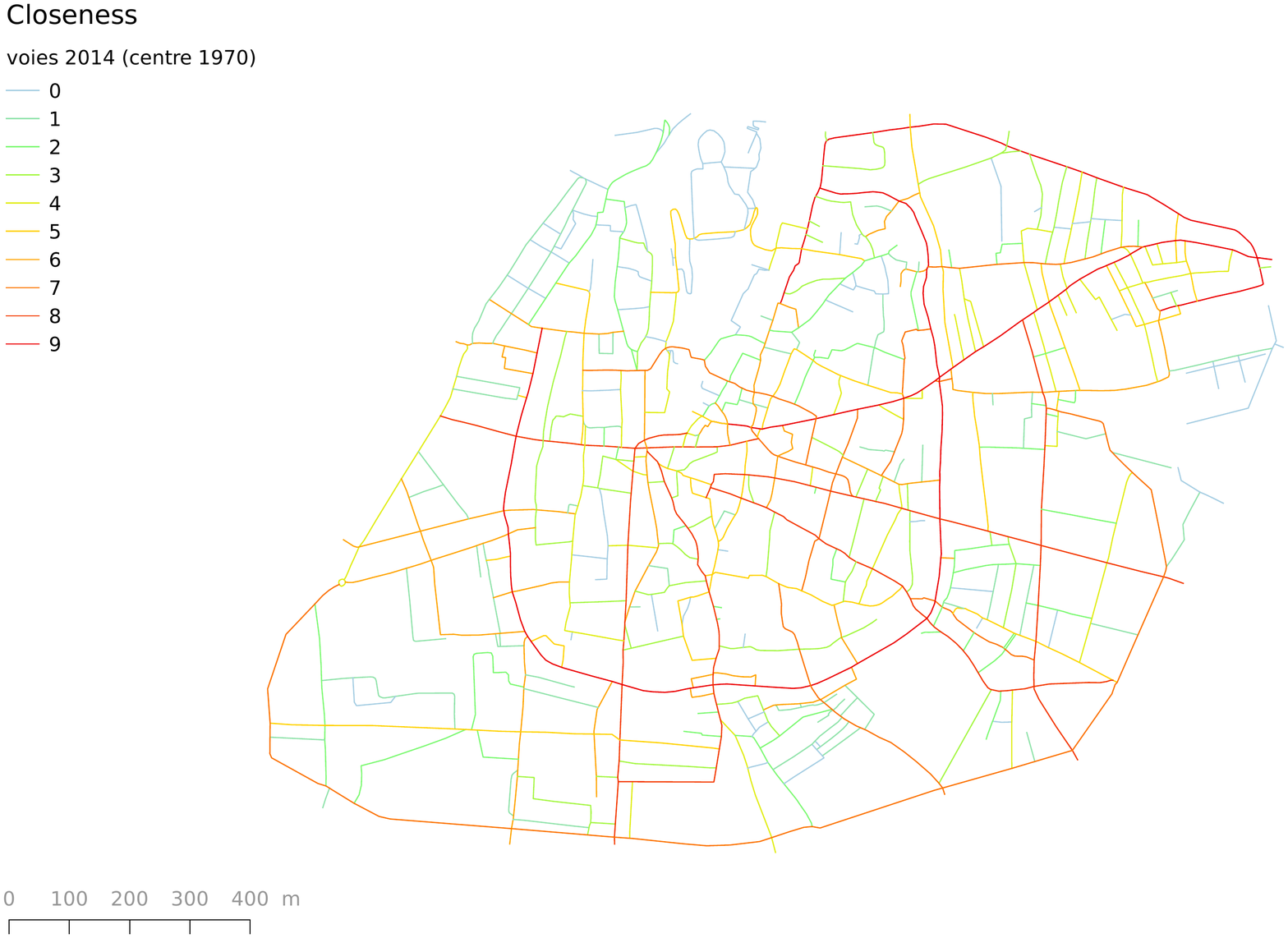}
        \caption{Closeness sur le graphe de 2014 avec intersection modifiée.}
        \label{fig:proj_avcentre_1}
    \end{subfigure}
     
    \begin{subfigure}[t]{.6\linewidth}
        \includegraphics[width=\textwidth]{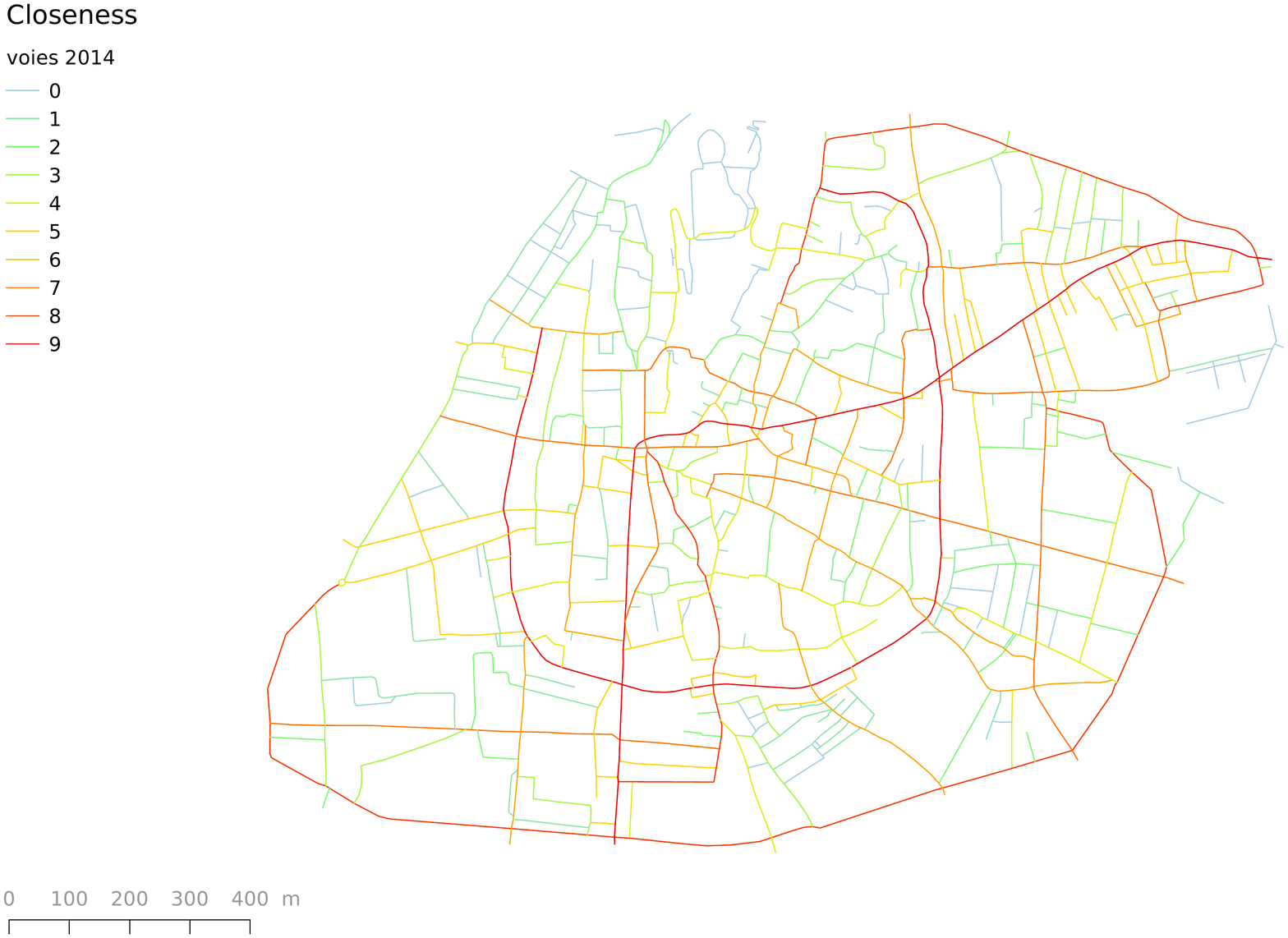}
        \caption{Closeness sur le graphe de 2014 sans intersection modifiée.}
        \label{fig:proj_avcentre_2}
    \end{subfigure}

    \caption{Closeness calculée sur le graphe d'Avignon intra-murros {\large \textbf{avant et après modification de l'intersection centrale}}.}
    \label{fig:clo_proj_avcentre}
\end{figure}

    \begin{figure}[h]
     \centering
        \includegraphics[width=\textwidth]{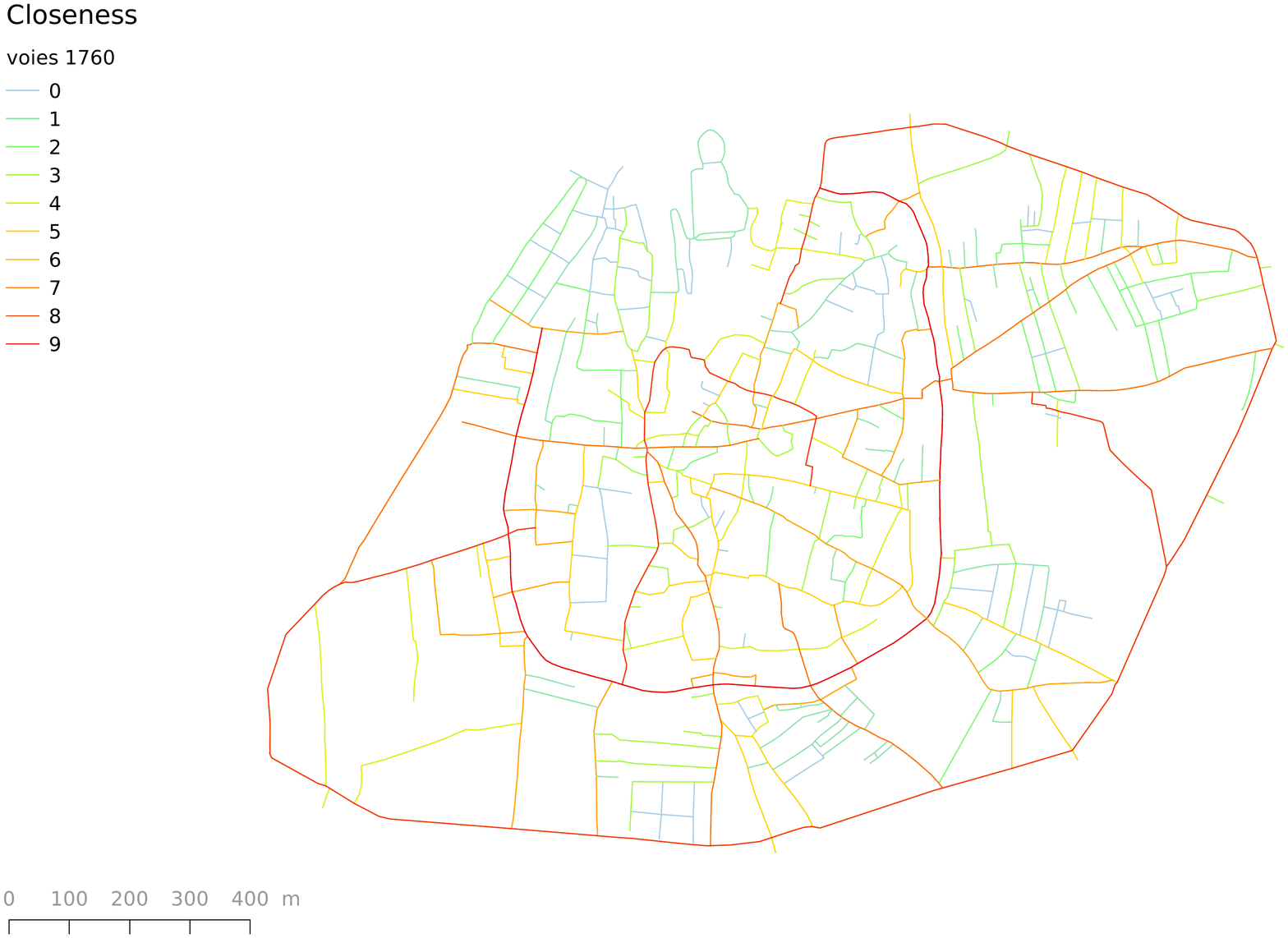}
        \caption{Closeness calculée sur le réseau viaire d'Avignon en {\large \textbf{1760}}.}
        \label{fig:clo_av_1760}
    \end{figure}

   \begin{figure}[h]
    \centering
        \includegraphics[width=\textwidth]{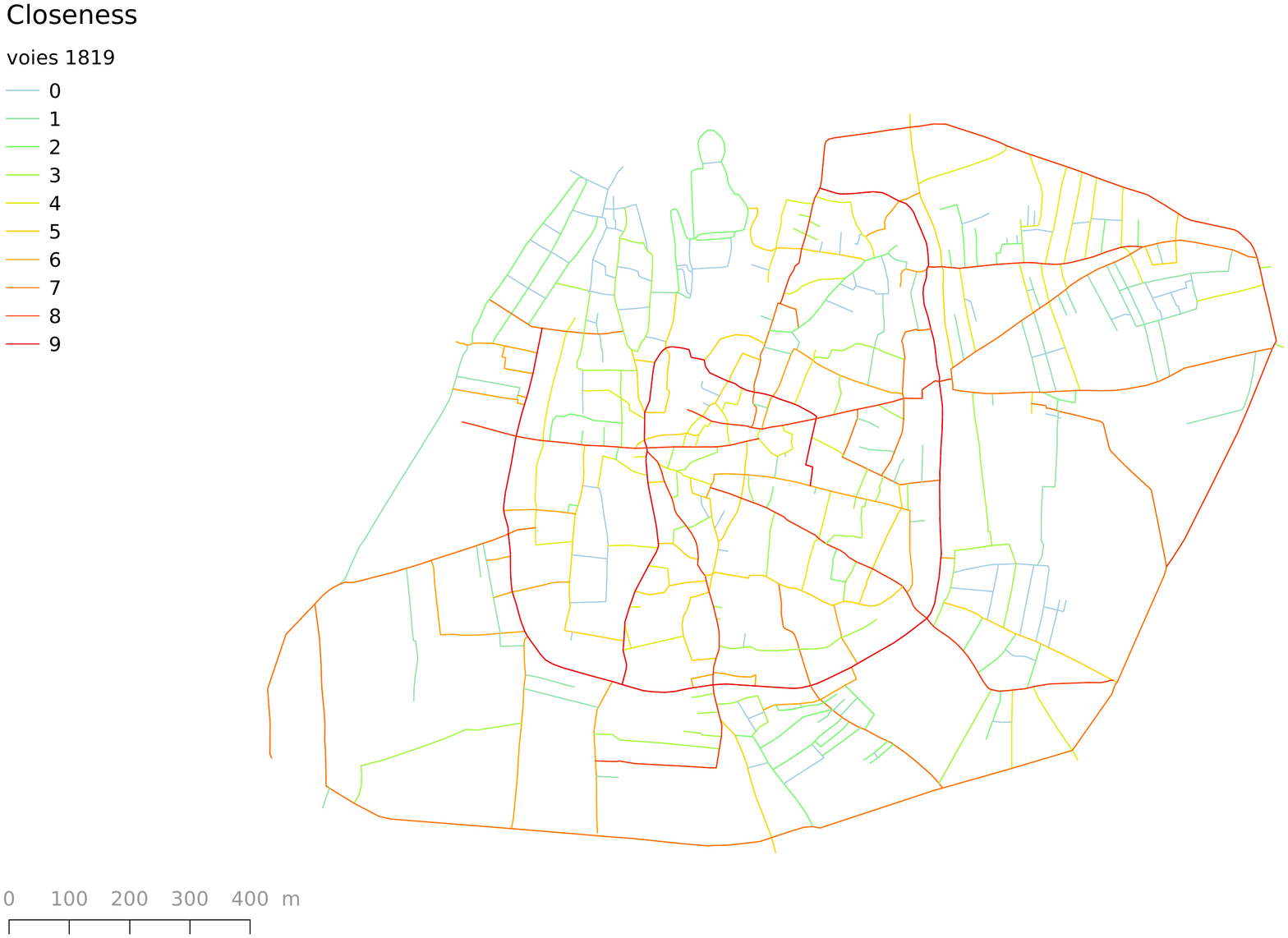}
        \caption{Closeness calculée sur les réseaux viaires d'Avignon en {\large \textbf{1819}}.}
        \label{fig:clo_av_1819}
    \end{figure}
    
   \begin{figure}[h]
    \centering
        \includegraphics[width=\textwidth]{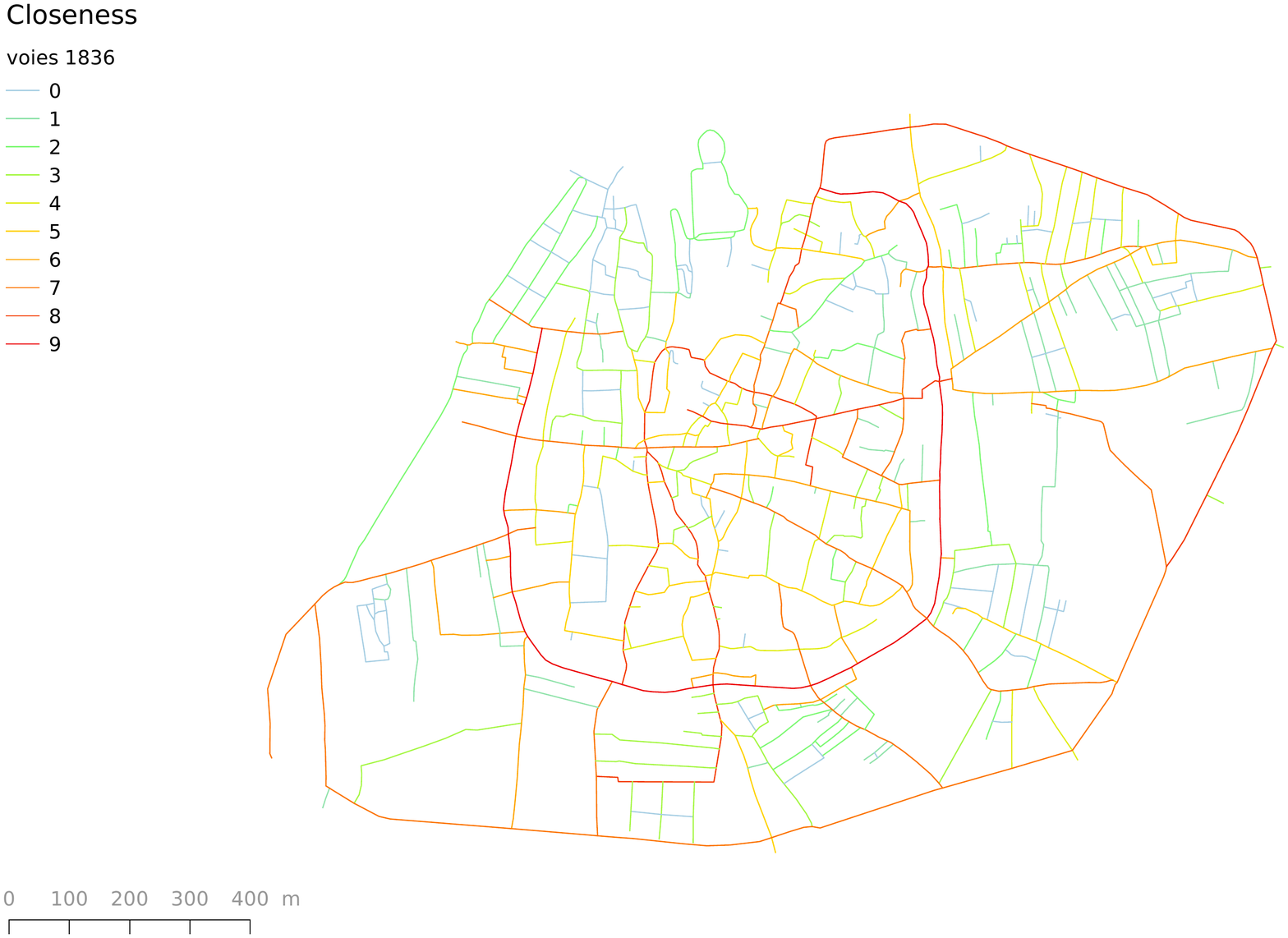}
        \caption{Closeness calculée sur le réseau viaire d'Avignon en {\large \textbf{1836}}.}
        \label{fig:clo_av_1836}
    \end{figure}

    \begin{figure}[h]
     \centering
        \includegraphics[width=\textwidth]{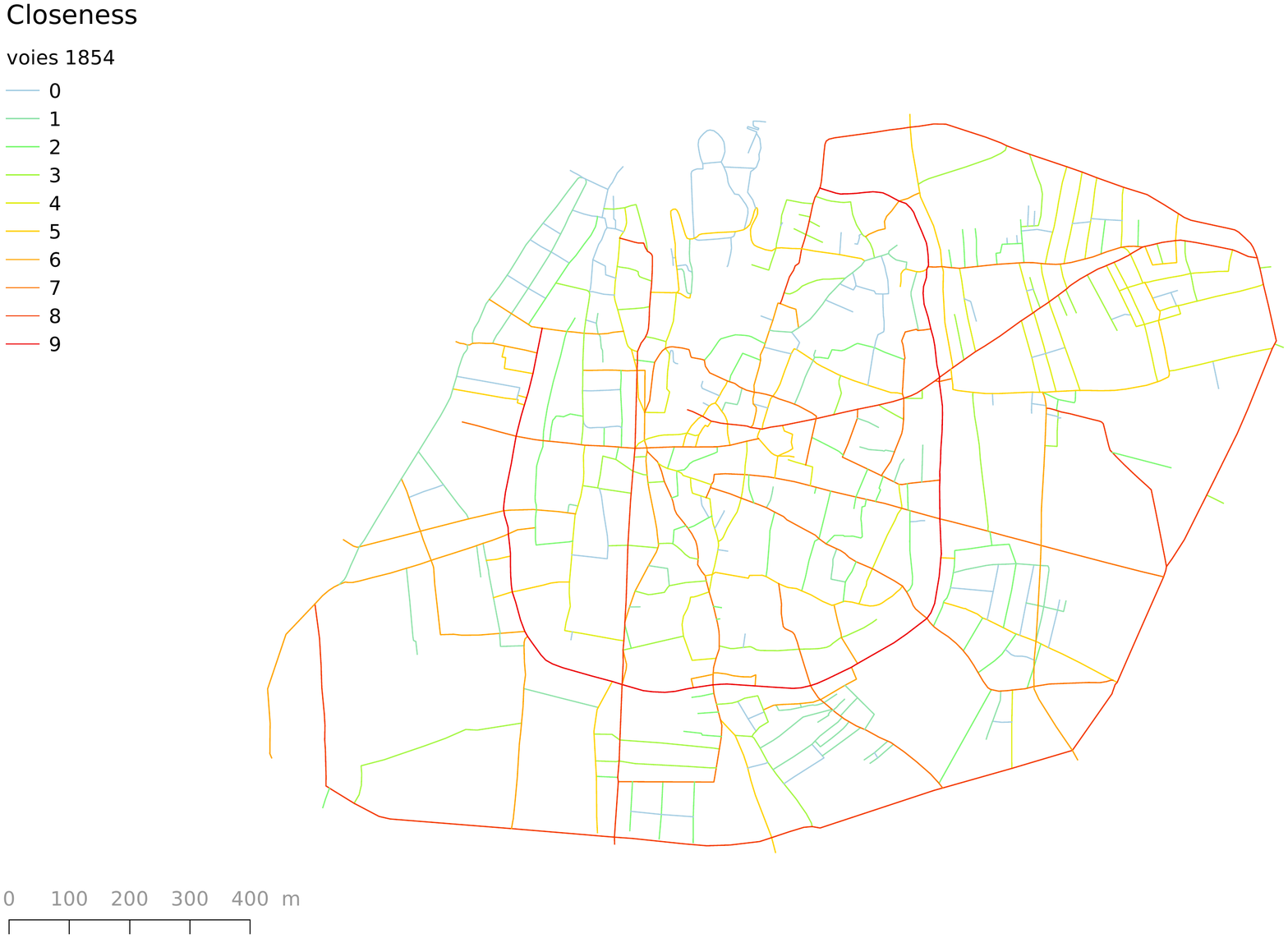}
        \caption{Closeness calculée sur le réseau viaire d'Avignon en {\large \textbf{1854}}.}
        \label{fig:clo_av_1854}
    \end{figure}    

	 \begin{figure}[h]
     \centering
        \includegraphics[width=\textwidth]{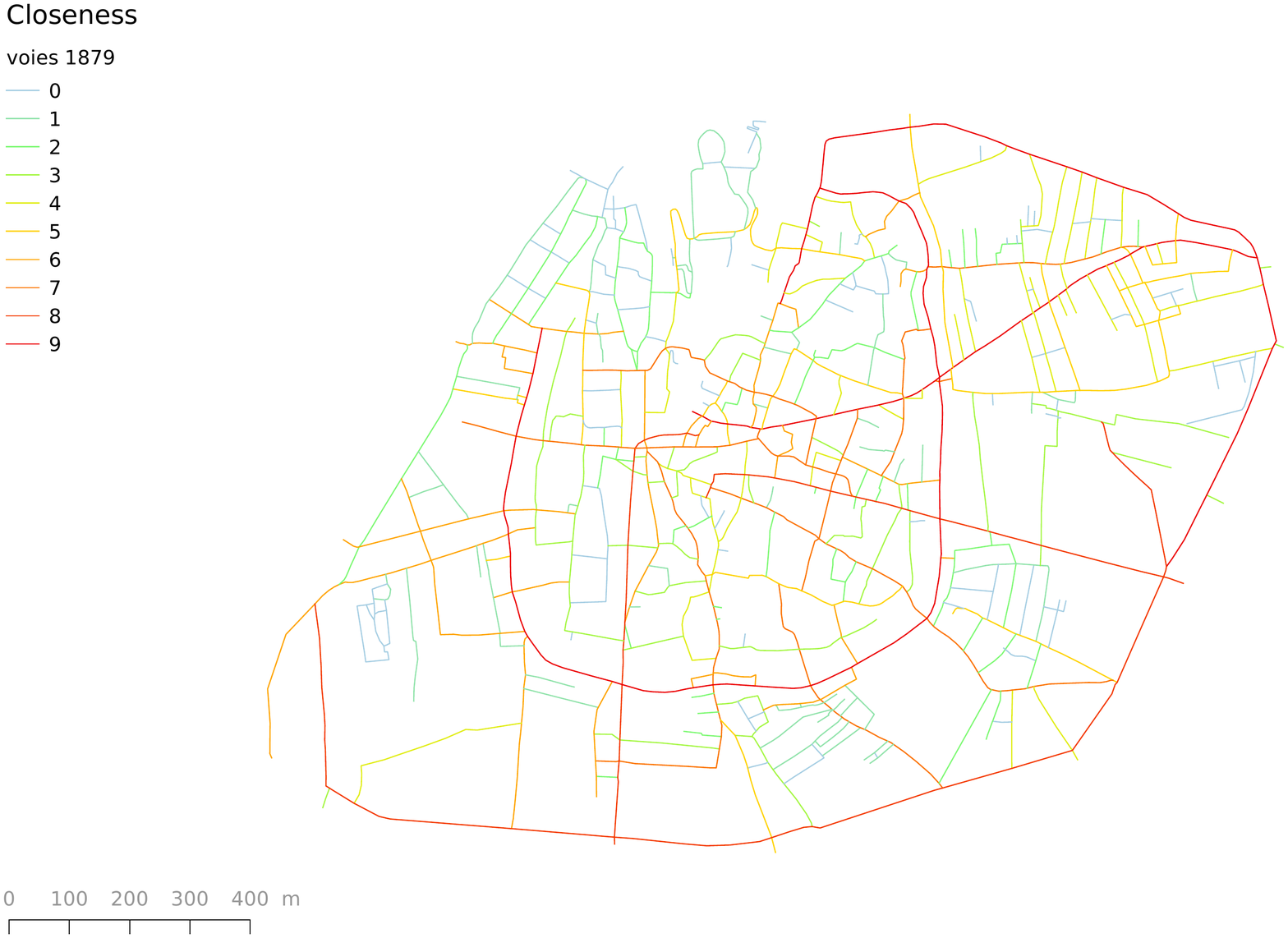}
        \caption{Closeness calculée sur le réseau viaire d'Avignon en {\large \textbf{1879}}.}
        \label{fig:clo_av_1879}
    \end{figure}

    \begin{figure}[h]
     \centering
        \includegraphics[width=\textwidth]{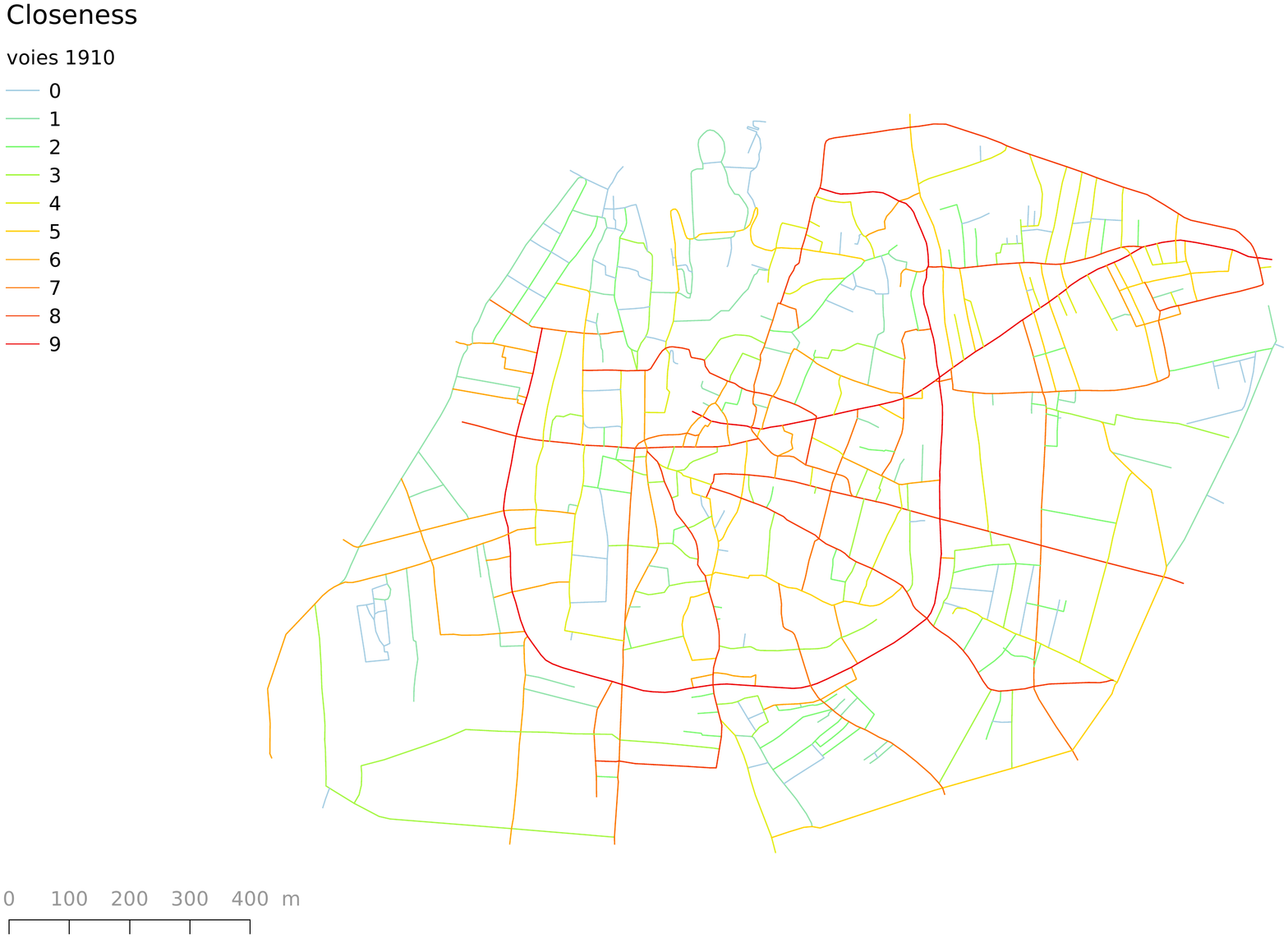}
        \caption{Closeness calculée sur le réseau viaire d'Avignon en {\large \textbf{1910}}.}
        \label{fig:clo_av_1910}
    \end{figure}
    
     \begin{figure}[h]
      \centering
        \includegraphics[width=\textwidth]{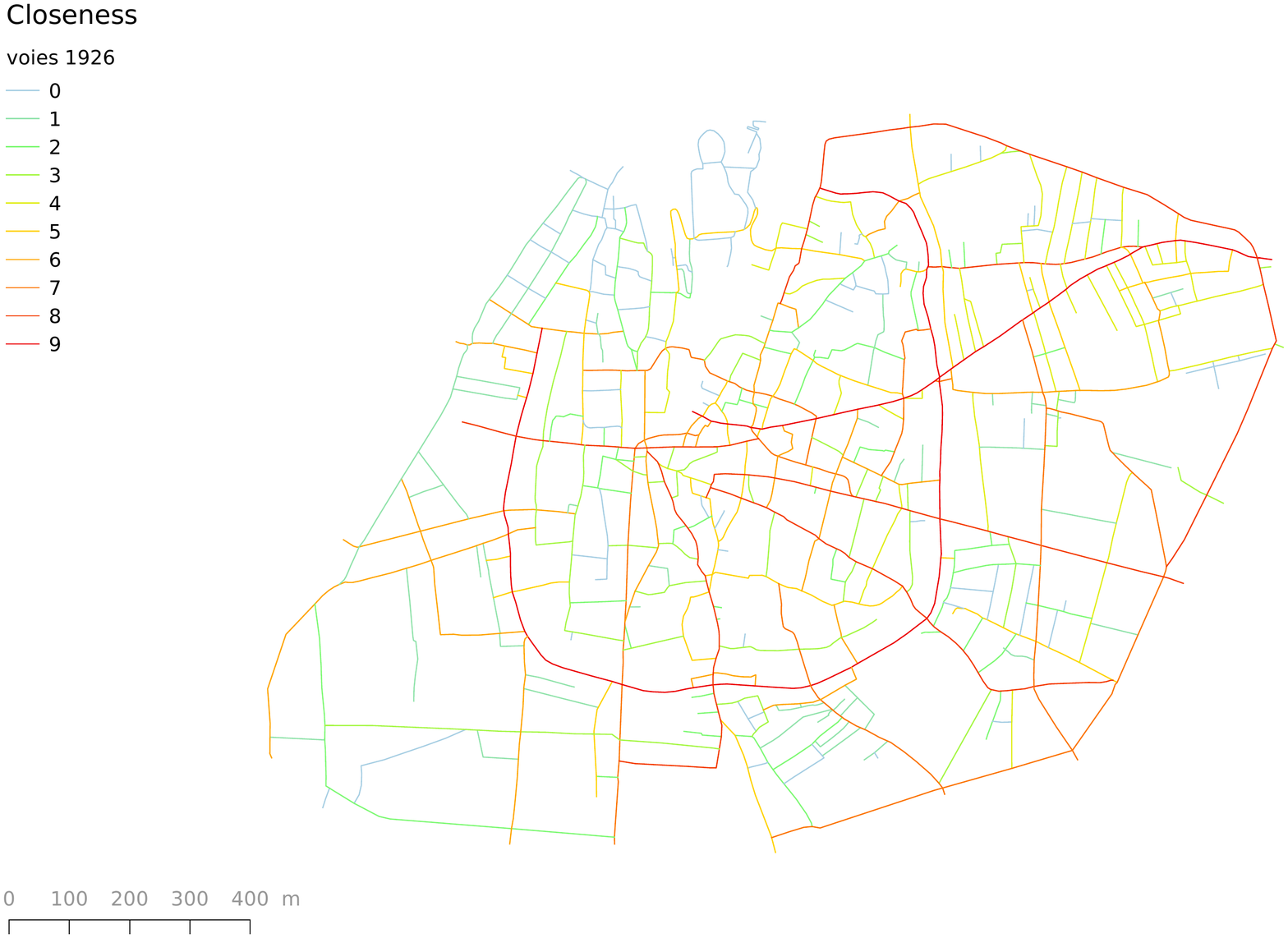}
        \caption{Closeness calculée sur le réseau viaire d'Avignon en {\large \textbf{1926}}.}
        \label{fig:clo_av_1926}
    \end{figure}

    \begin{figure}[h]
     \centering
        \includegraphics[width=\textwidth]{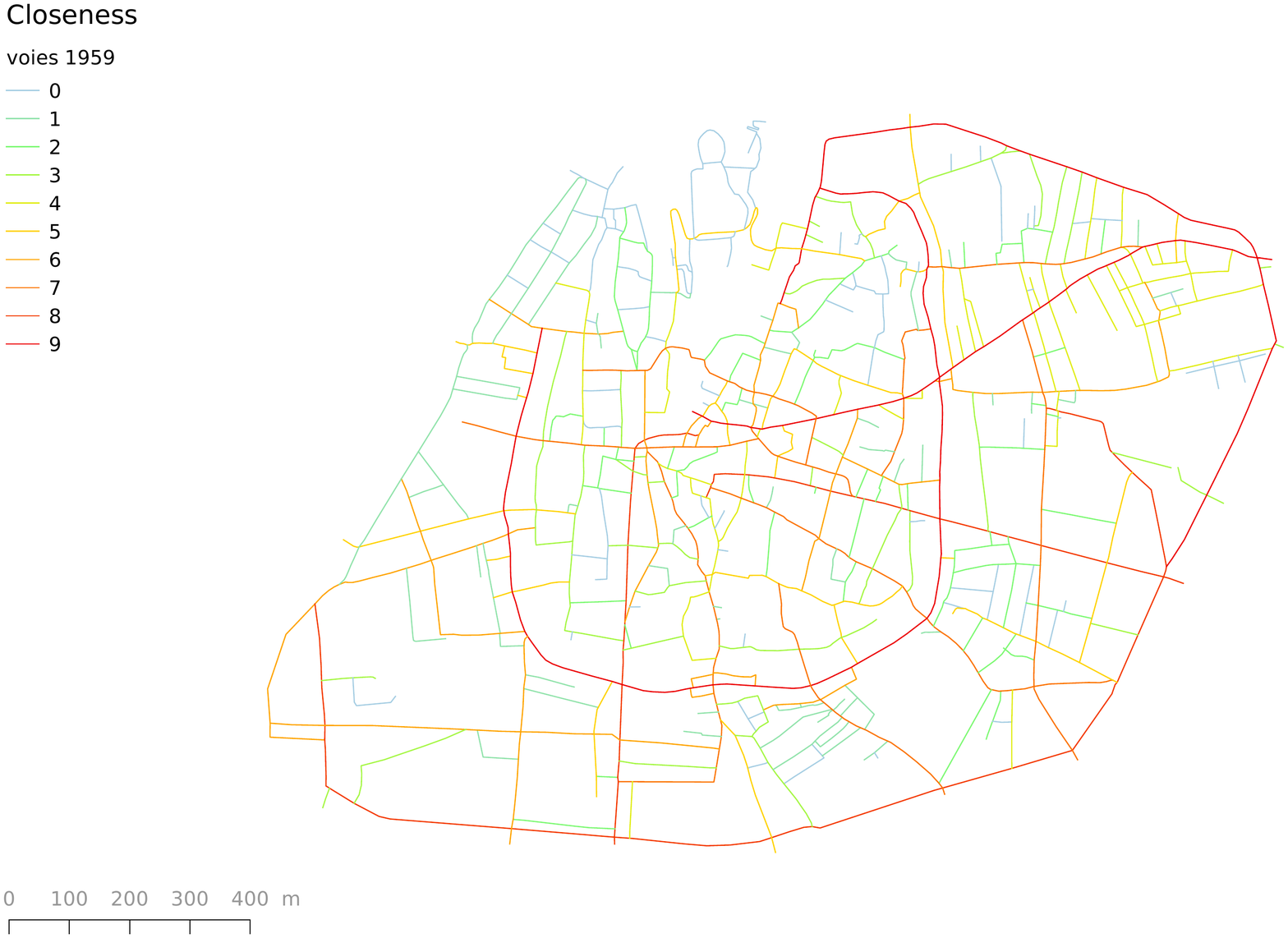}
        \caption{Closeness calculée sur le réseau viaire d'Avignon en {\large \textbf{1959}}.}
        \label{fig:clo_av_1959}
    \end{figure}
    
     \begin{figure}[h]
      \centering
        \includegraphics[width=\textwidth]{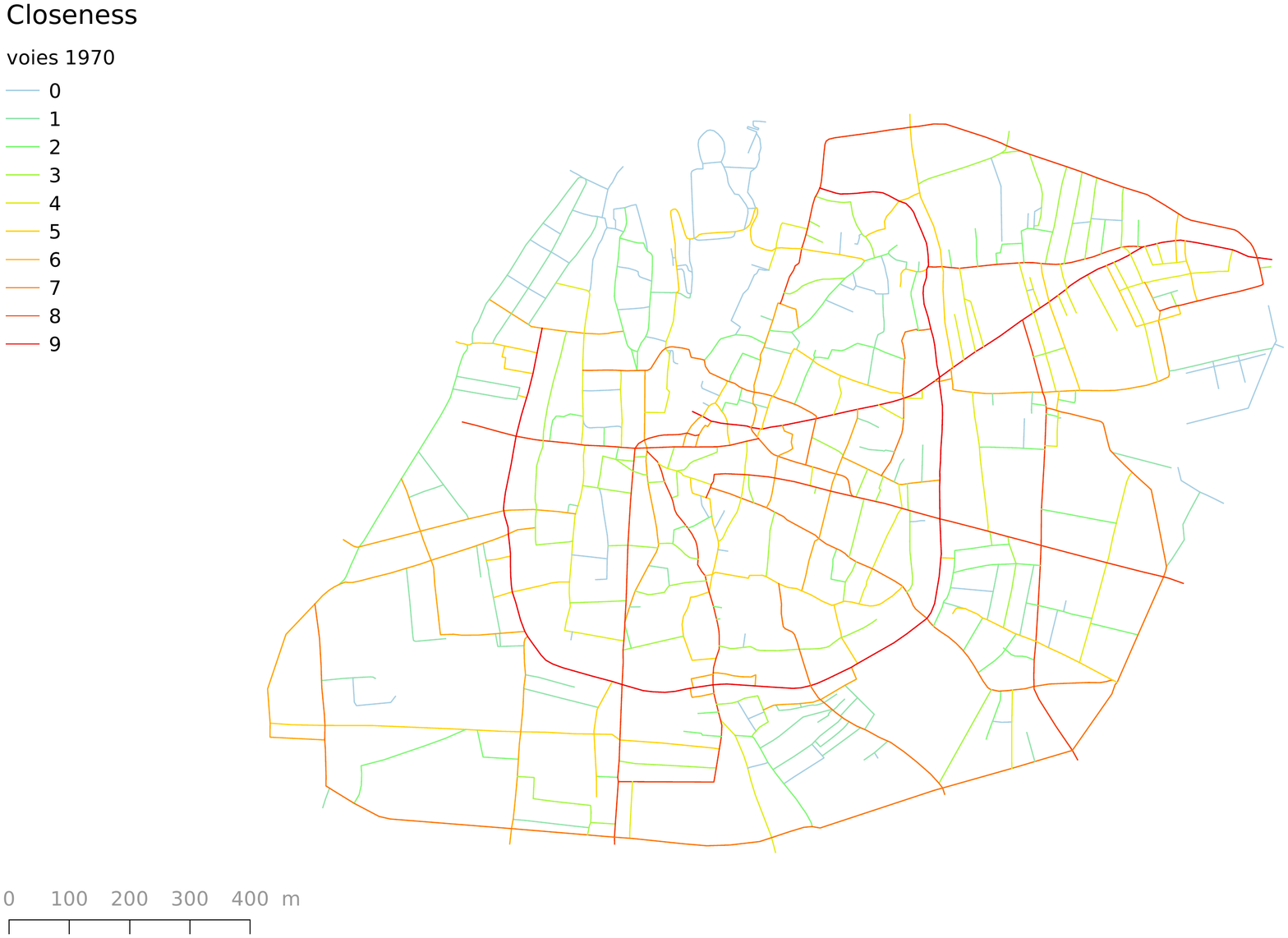}
        \caption{Closeness calculée sur le réseau viaire d'Avignon en {\large \textbf{1970}}.}
        \label{fig:clo_av_1970}
    \end{figure}
  
    \begin{figure}[h]
     \centering
        \includegraphics[width=\textwidth]{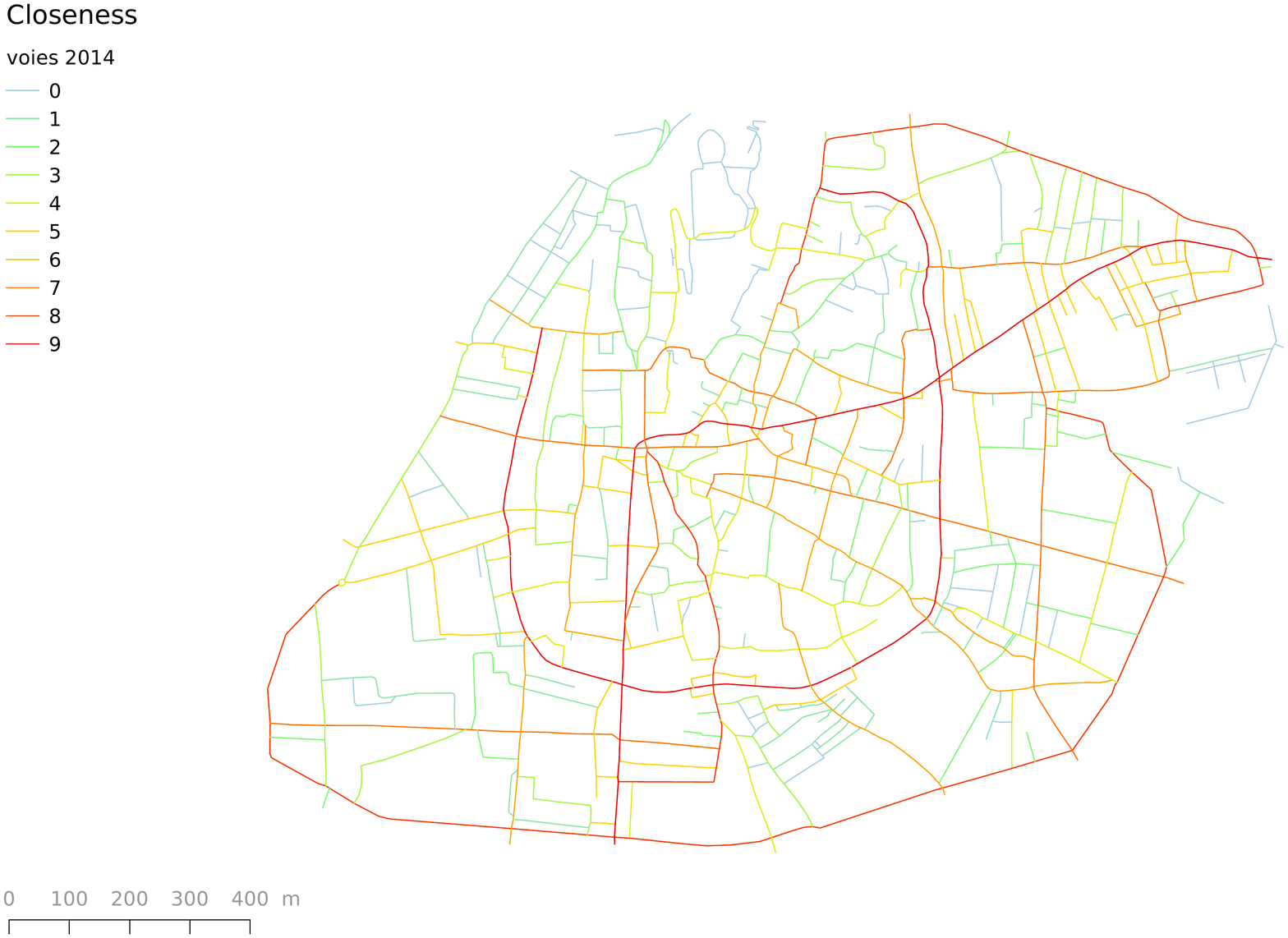}
        \caption{Closeness calculée sur le réseau viaire d'Avignon en {\large \textbf{2014}}.}
        \label{fig:clo_av_2014}
    \end{figure}

\FloatBarrier
\section{Rotterdam \& Schiedam (1374 - 1955)}\label{ann:sec_cartedia_rotterdamsch}

    \begin{figure}[h]
        \includegraphics[width=\textwidth]{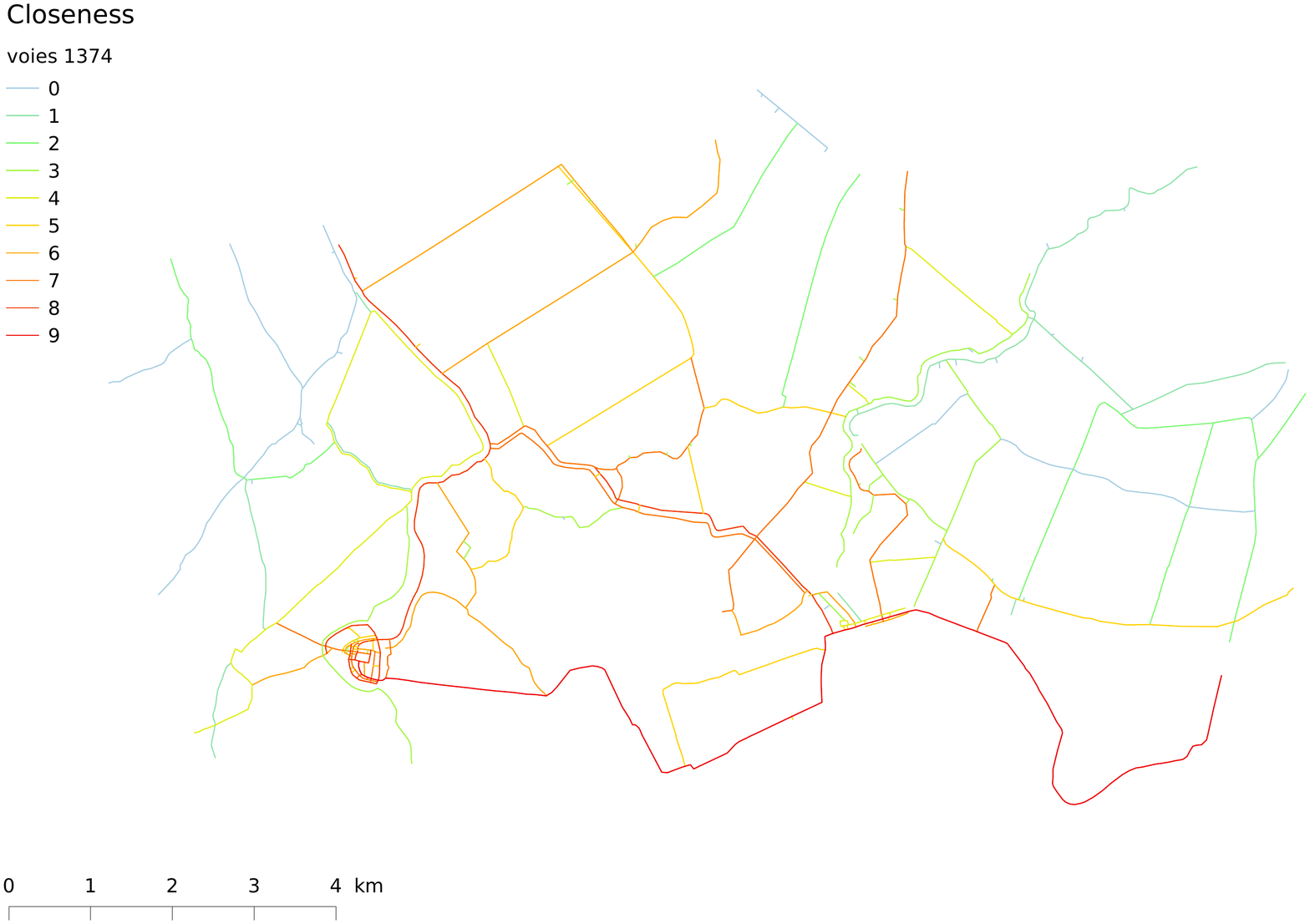}
        \caption{Closeness calculée sur le réseau viaire Nord de Rotterdam (entier) en {\large \textbf{1374}}.}
        \label{fig:clo_re2_1374}
    \end{figure}
    
   \begin{figure}[h]
        \includegraphics[width=\textwidth]{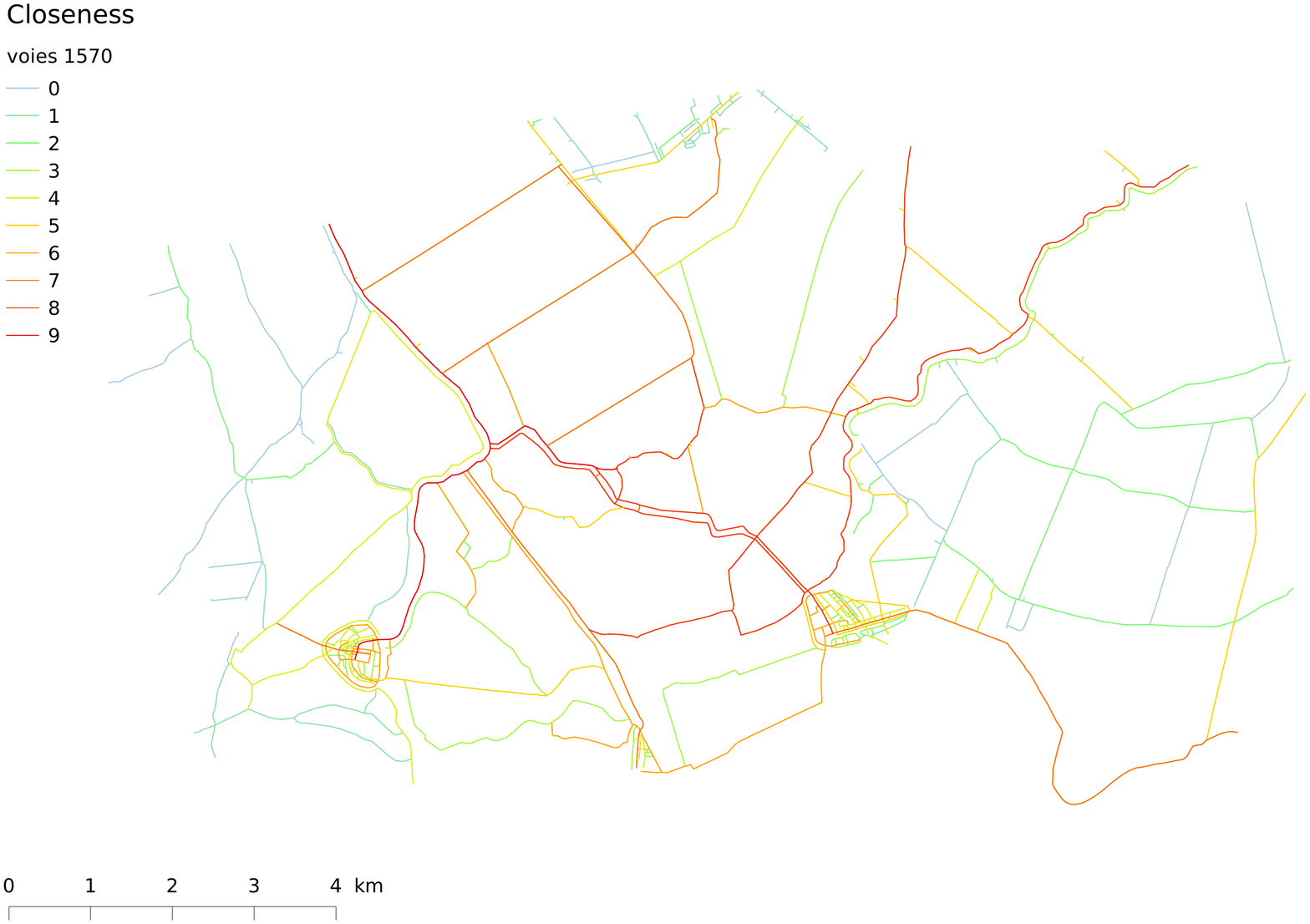}
        \caption{Closeness calculée sur le réseau viaire Nord de Rotterdam (entier) en {\large \textbf{1570}}.}
        \label{fig:clo_re2_1570}
    \end{figure}
    
   \begin{figure}[h]
        \includegraphics[width=\textwidth]{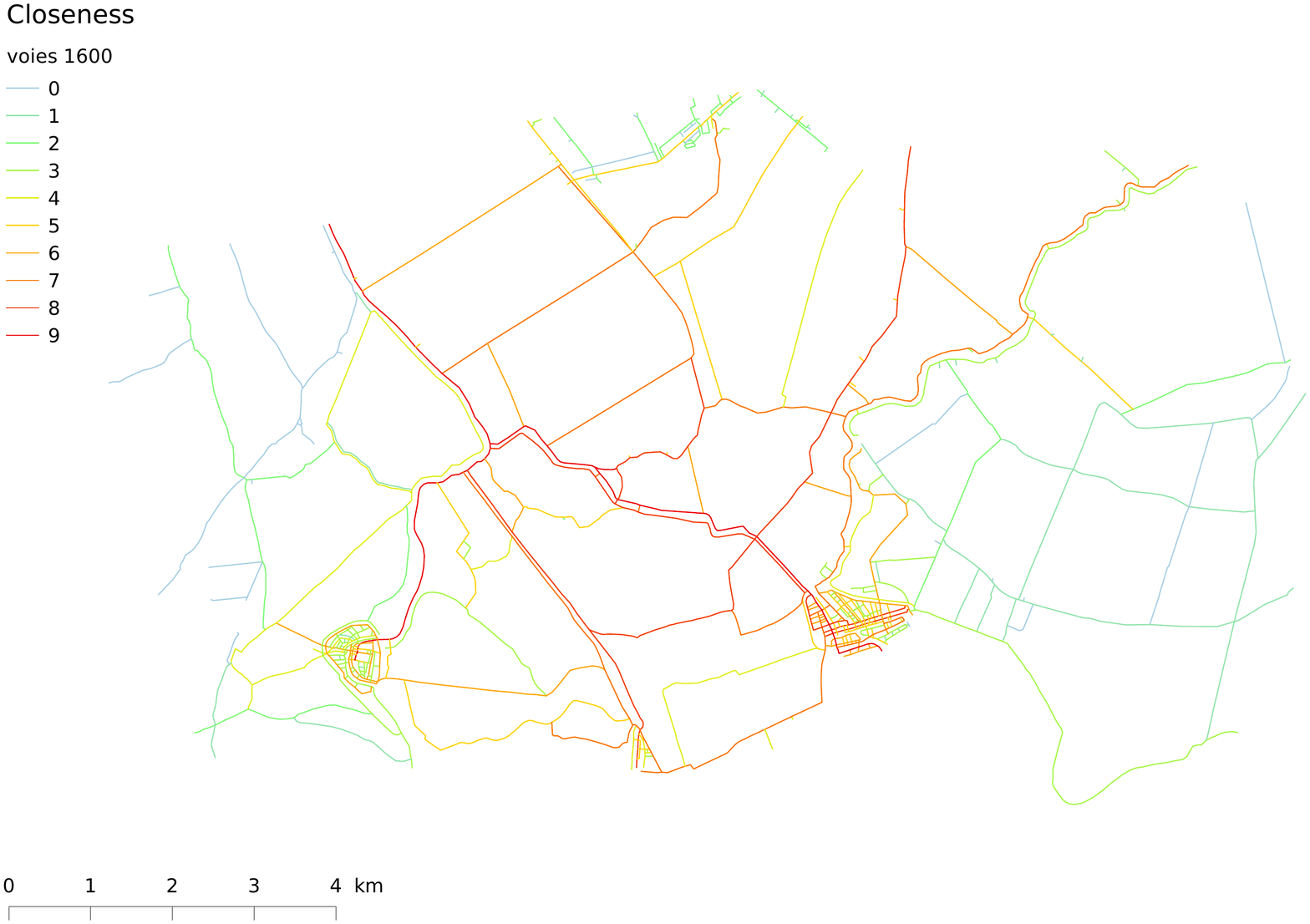}
        \caption{Closeness calculée sur le réseau viaire Nord de Rotterdam (entier) en {\large \textbf{1600}}.}
        \label{fig:clo_re2_1600}
    \end{figure}
        
    \begin{figure}[h]
        \includegraphics[width=\textwidth]{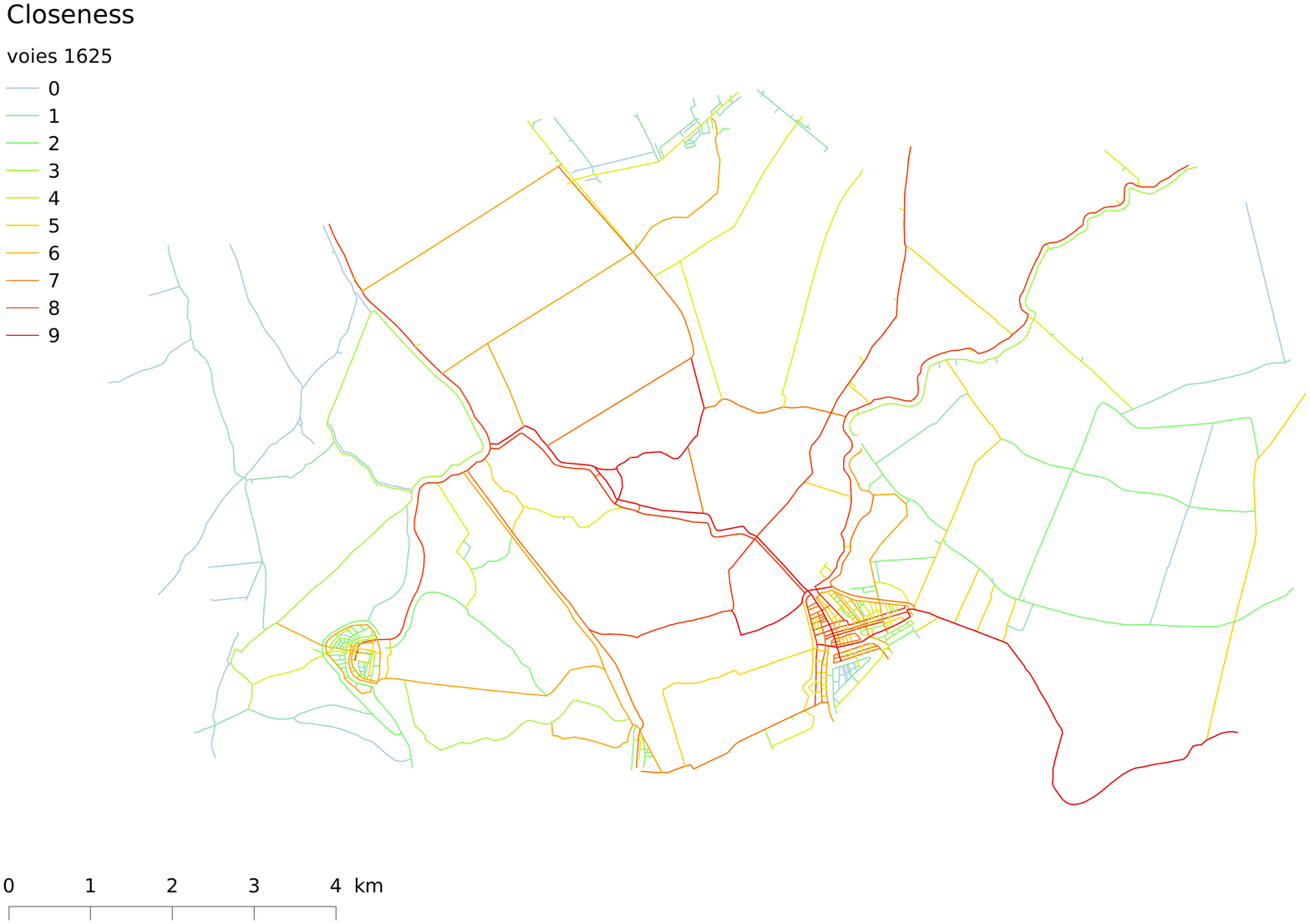}
        \caption{Closeness calculée sur le réseau viaire Nord de Rotterdam (entier) en {\large \textbf{1625}}.}
        \label{fig:clo_re2_1625}
    \end{figure}
    
     \begin{figure}[h]
        \includegraphics[width=\textwidth]{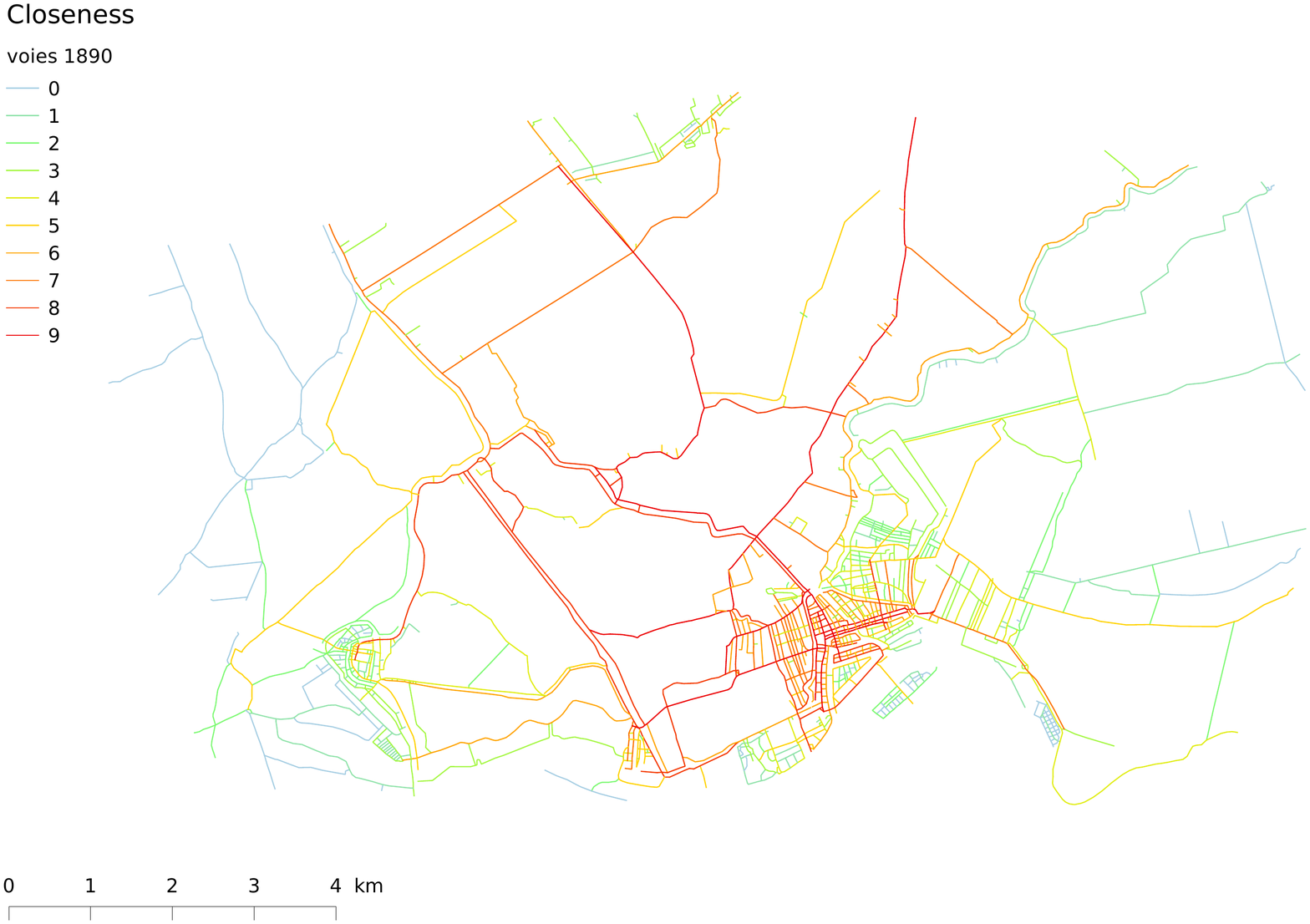}
        \caption{Closeness calculée sur le réseau viaire Nord de Rotterdam (entier) en {\large \textbf{1890}}.}
        \label{fig:clo_re2_1890}
    \end{figure}
        
    \begin{figure}[h]
        \includegraphics[width=\textwidth]{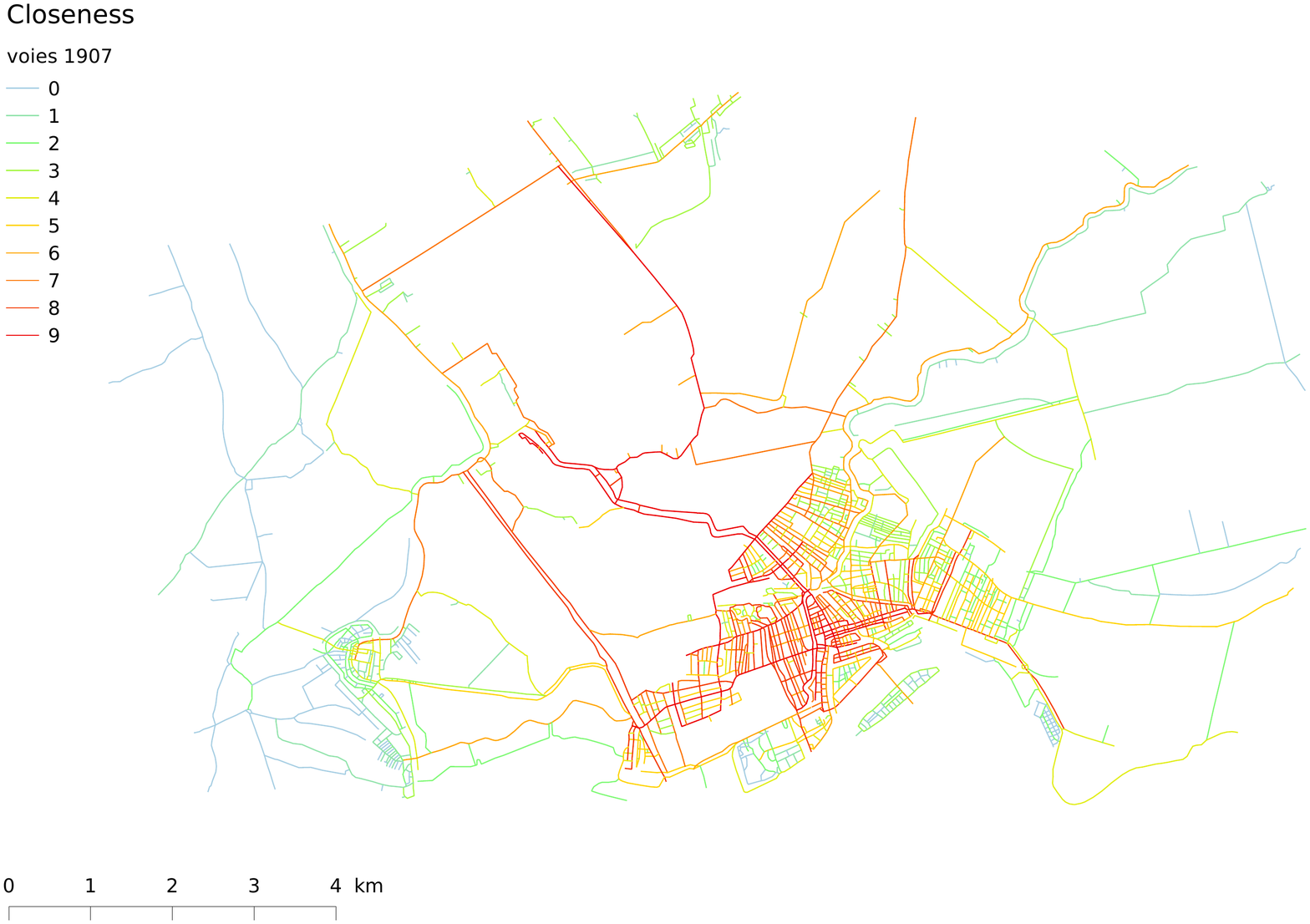}
        \caption{Closeness calculée sur le réseau viaire Nord de Rotterdam (entier) en {\large \textbf{1907}}.}
        \label{fig:clo_re2_1907}
    \end{figure}
    
     \begin{figure}[h]
        \includegraphics[width=\textwidth]{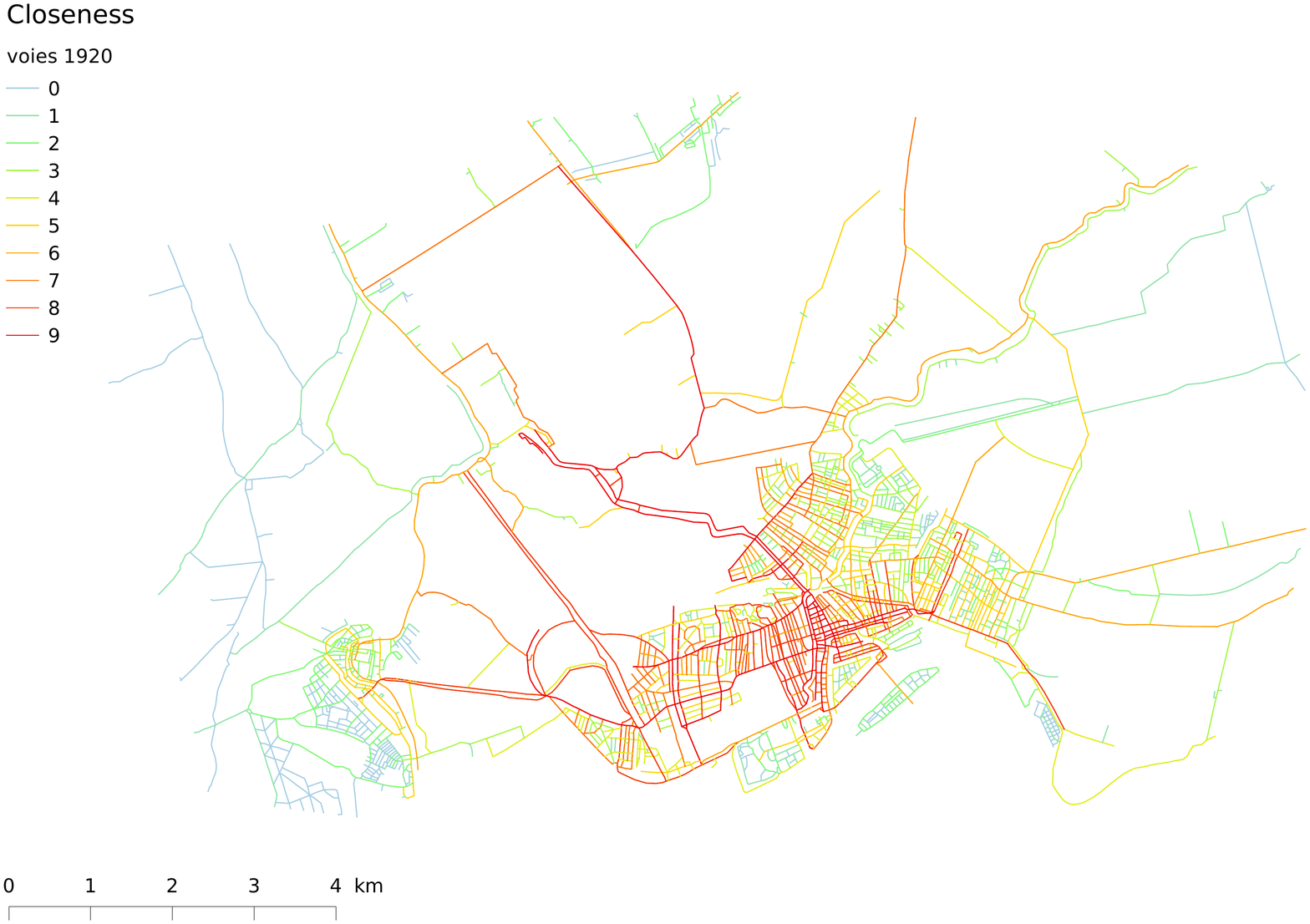}
        \caption{Closeness calculée sur le réseau viaire Nord de Rotterdam (entier) en {\large \textbf{1920}}.}
        \label{fig:clo_re2_1920}
    \end{figure}
        
    \begin{figure}[h]
        \includegraphics[width=\textwidth]{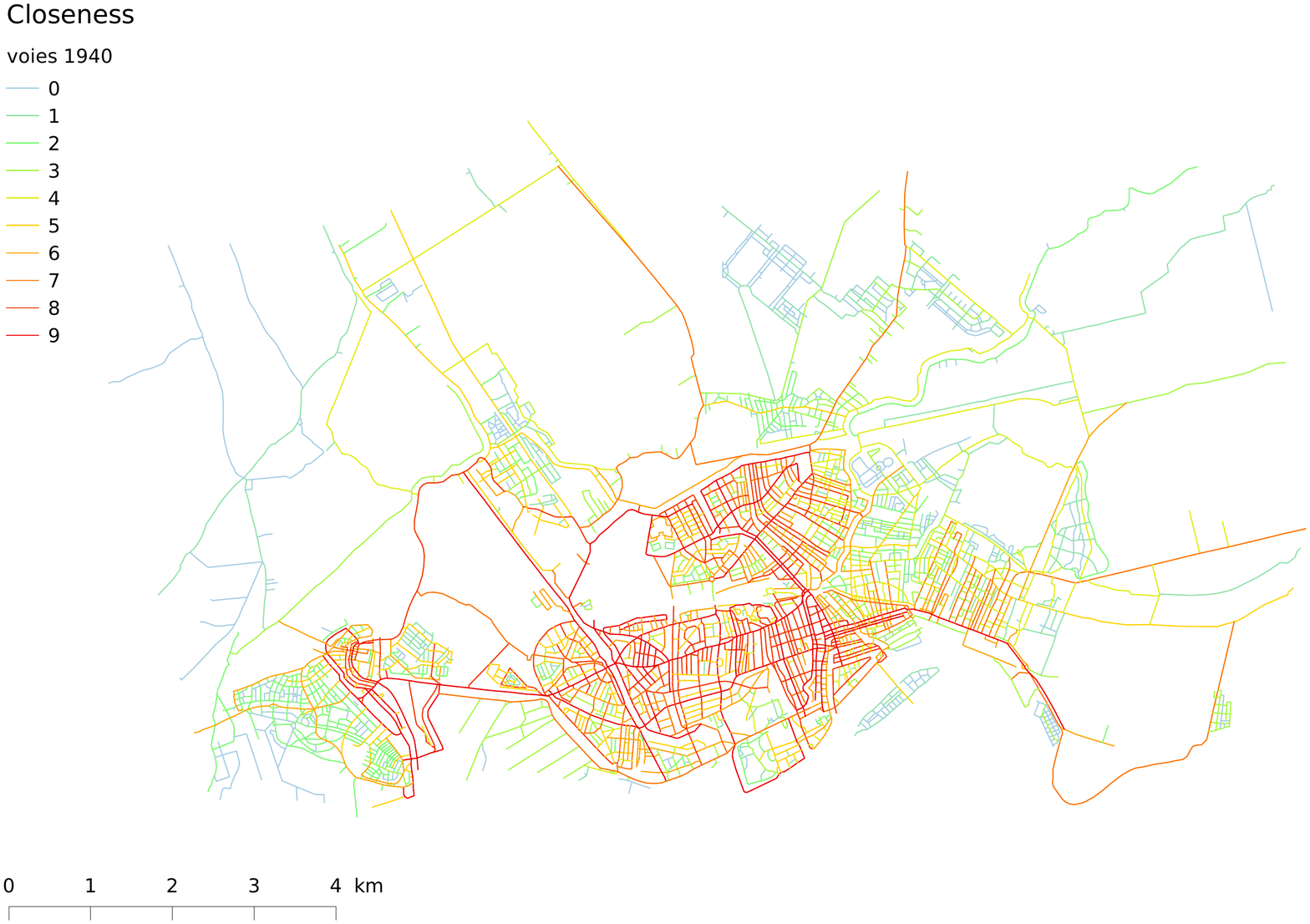}
        \caption{Closeness calculée sur le réseau viaire Nord de Rotterdam (entier) en {\large \textbf{1940}}.}
        \label{fig:clo_re2_1940}
    \end{figure}
    
     \begin{figure}[h]
        \includegraphics[width=\textwidth]{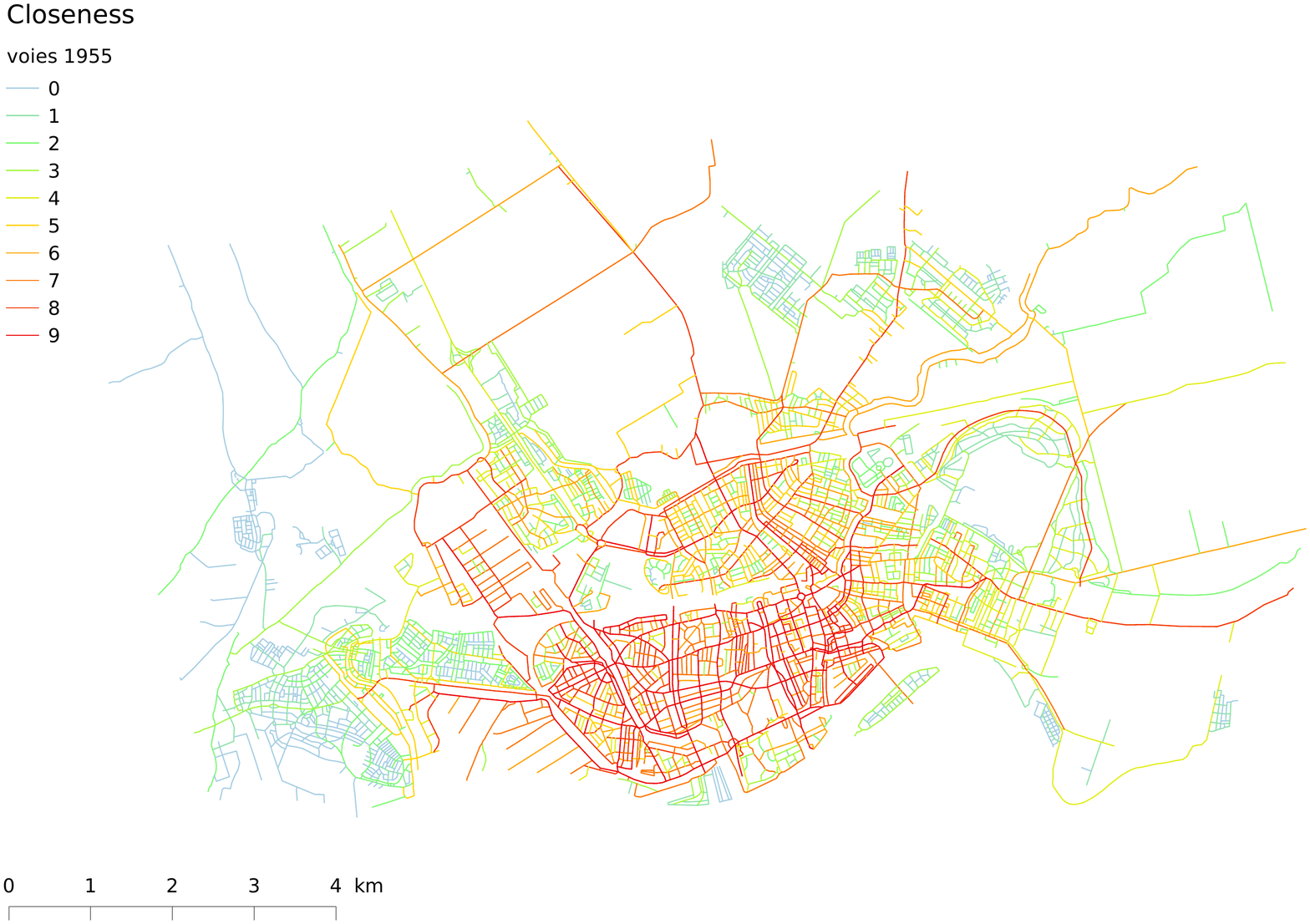}
        \caption{Closeness calculée sur le réseau viaire Nord de Rotterdam (entier) en {\large \textbf{1955}}.}
        \label{fig:clo_re2_1955}
    \end{figure}

\FloatBarrier
\section{Rotterdam (1374 - 1955)}\label{ann:ssec_cartedia_rotterdam}

    \begin{figure}[h]
        \includegraphics[width=\textwidth]{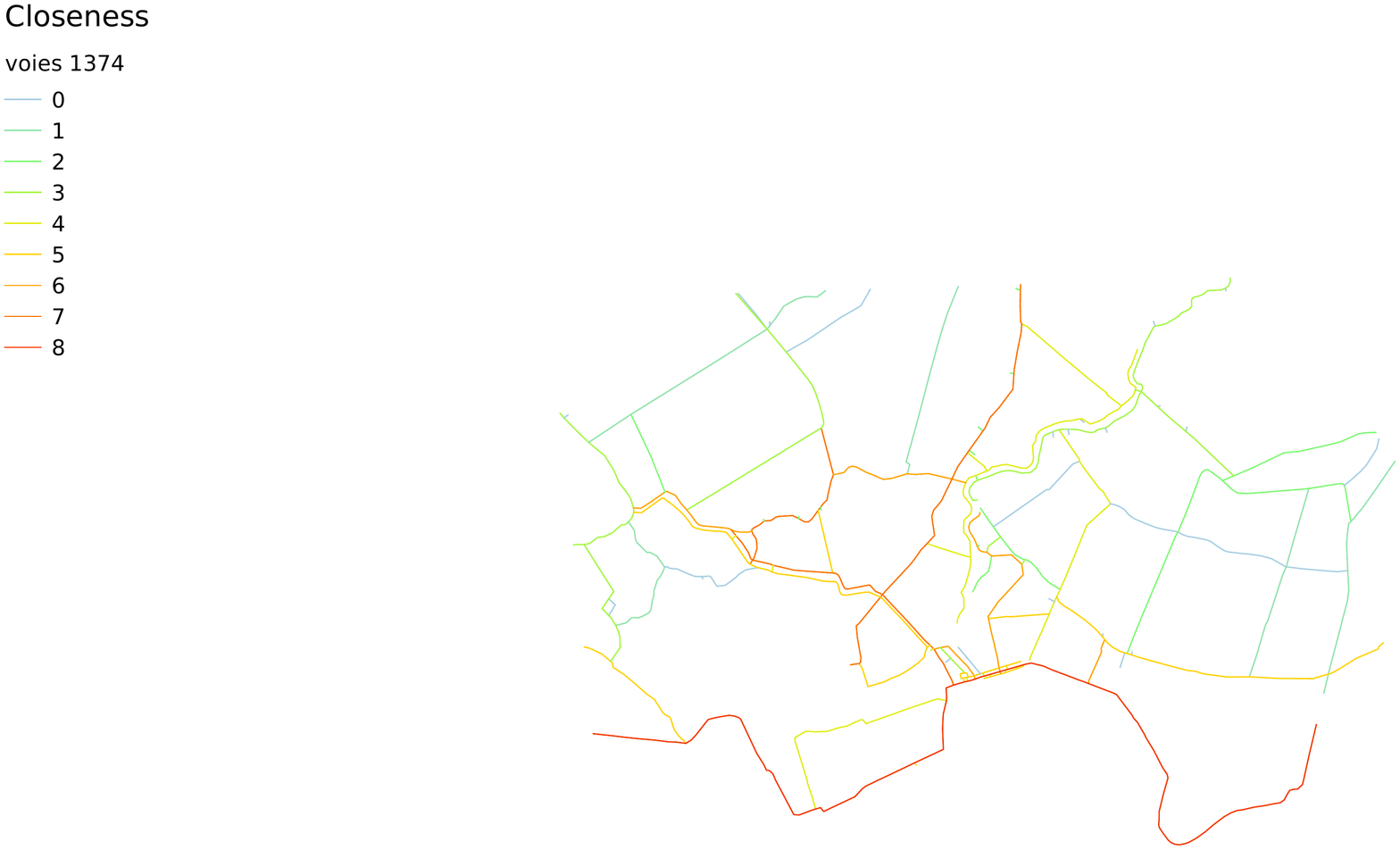}
        \caption{Closeness calculée sur le réseau viaire Nord de Rotterdam (découpé) en {\large \textbf{1374}}.}
        \label{fig:clo_rd_1374}
    \end{figure}
    
   \begin{figure}[h]
        \includegraphics[width=\textwidth]{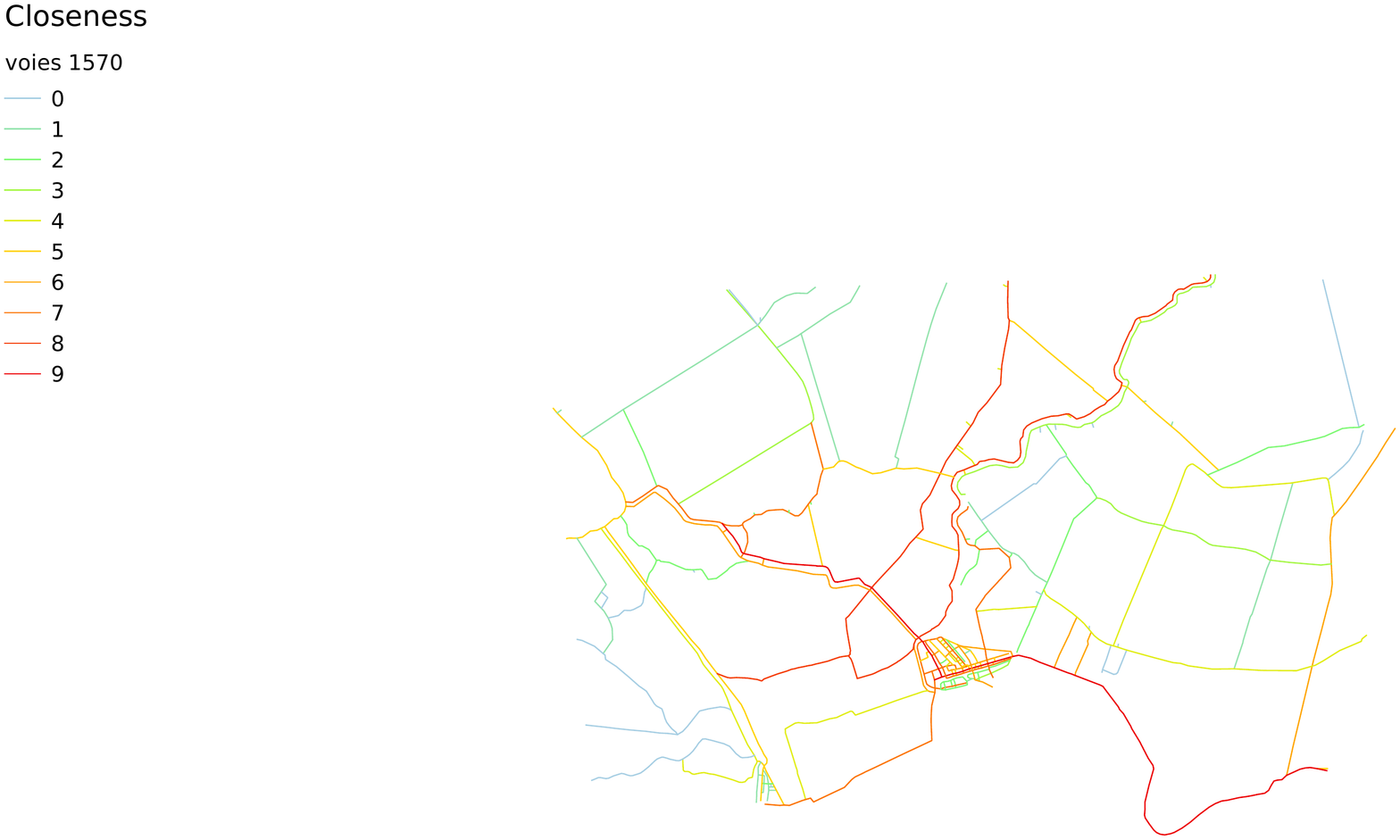}
        \caption{Closeness calculée sur le réseau viaire Nord de Rotterdam (découpé) en {\large \textbf{1570}}.}
        \label{fig:clo_rd_1570}
    \end{figure}
    
   \begin{figure}[h]
        \includegraphics[width=\textwidth]{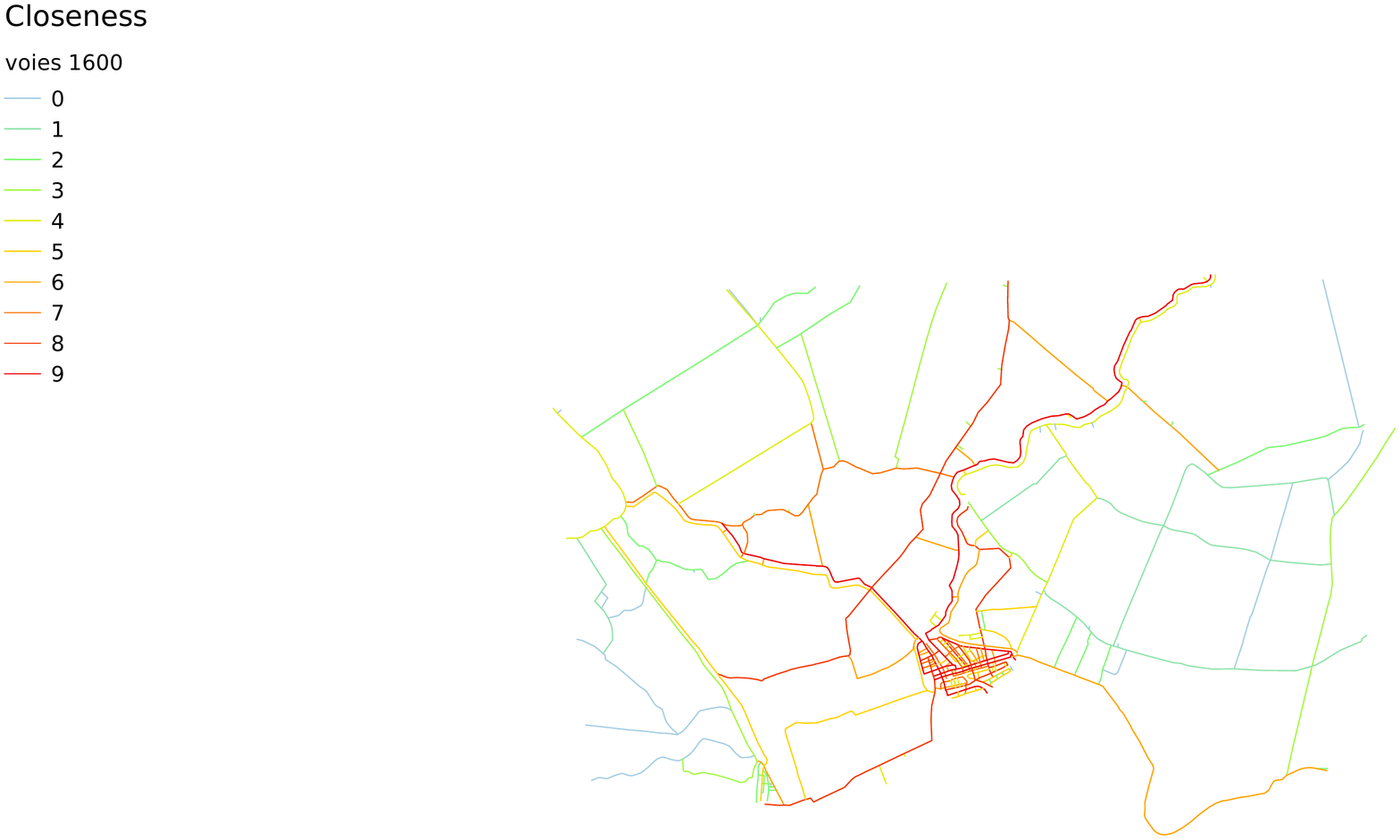}
        \caption{Closeness calculée sur le réseau viaire Nord de Rotterdam (découpé) en {\large \textbf{1600}}.}
        \label{fig:clo_rd_1600}
    \end{figure}
        
    \begin{figure}[h]
        \includegraphics[width=\textwidth]{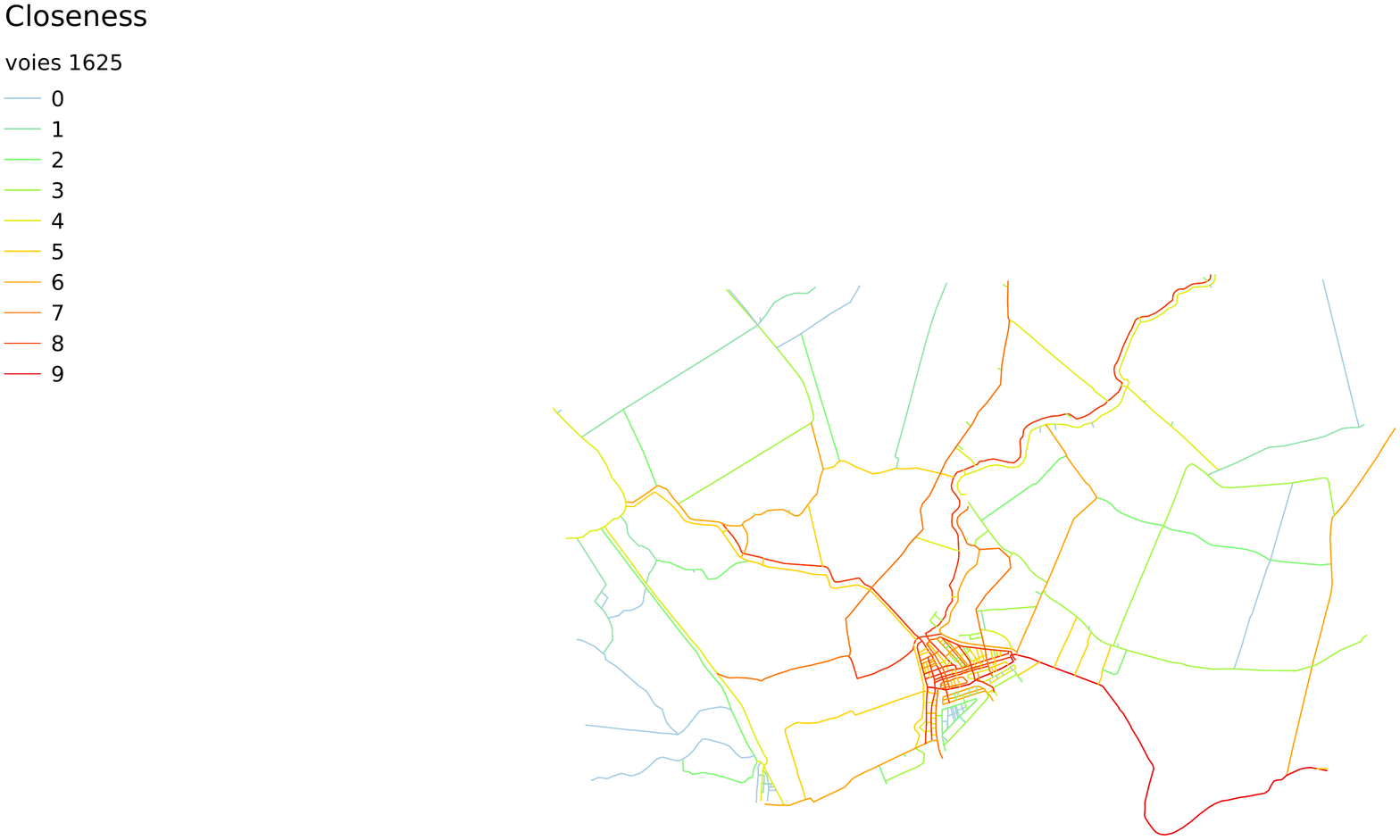}
        \caption{Closeness calculée sur le réseau viaire Nord de Rotterdam (découpé) en {\large \textbf{1625}}.}
        \label{fig:clo_rd_1625}
    \end{figure}
    
     \begin{figure}[h]
        \includegraphics[width=\textwidth]{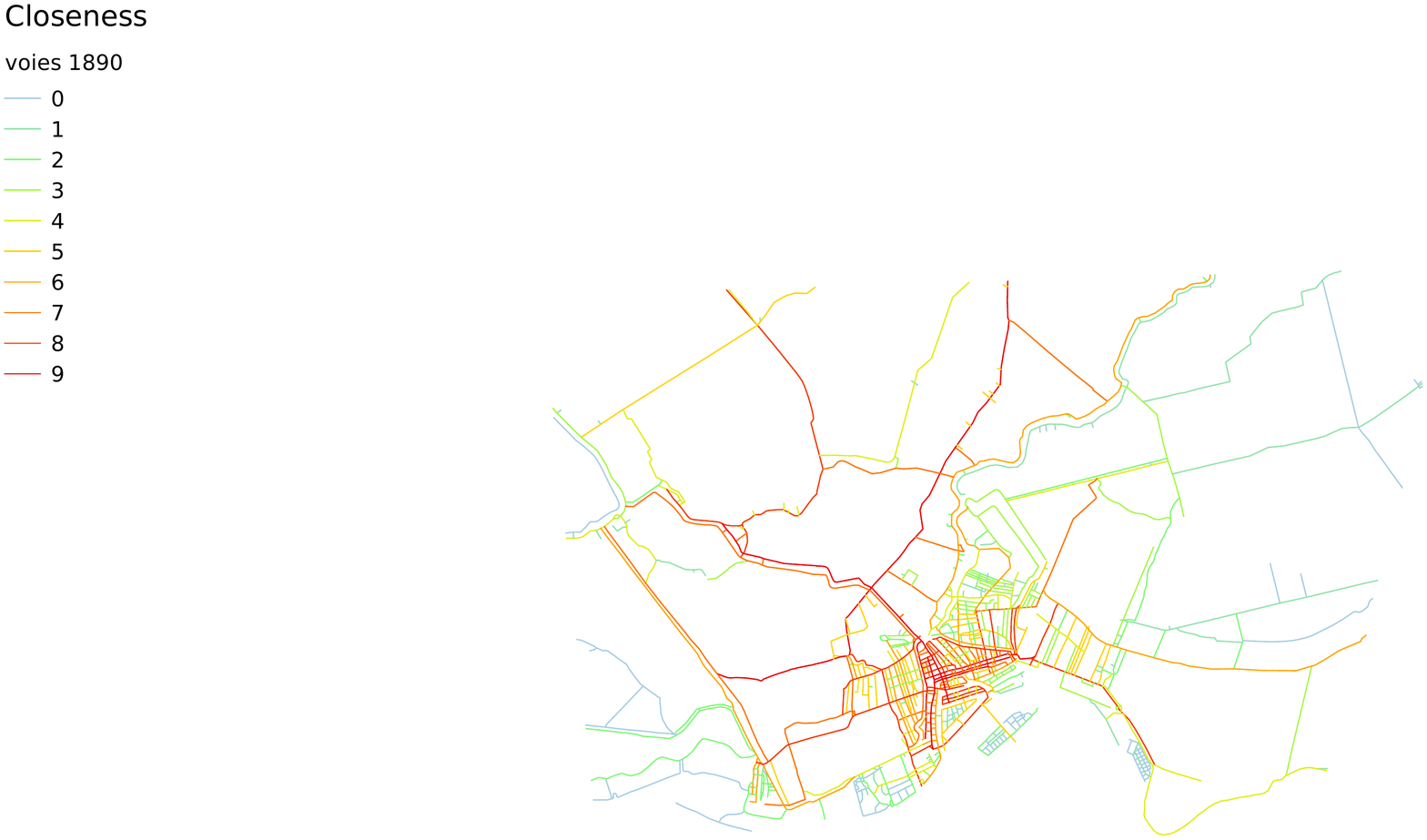}
        \caption{Closeness calculée sur le réseau viaire Nord de Rotterdam (découpé) en {\large \textbf{1890}}.}
        \label{fig:clo_rd_1890}
    \end{figure}
   
    \begin{figure}[h]
        \includegraphics[width=\textwidth]{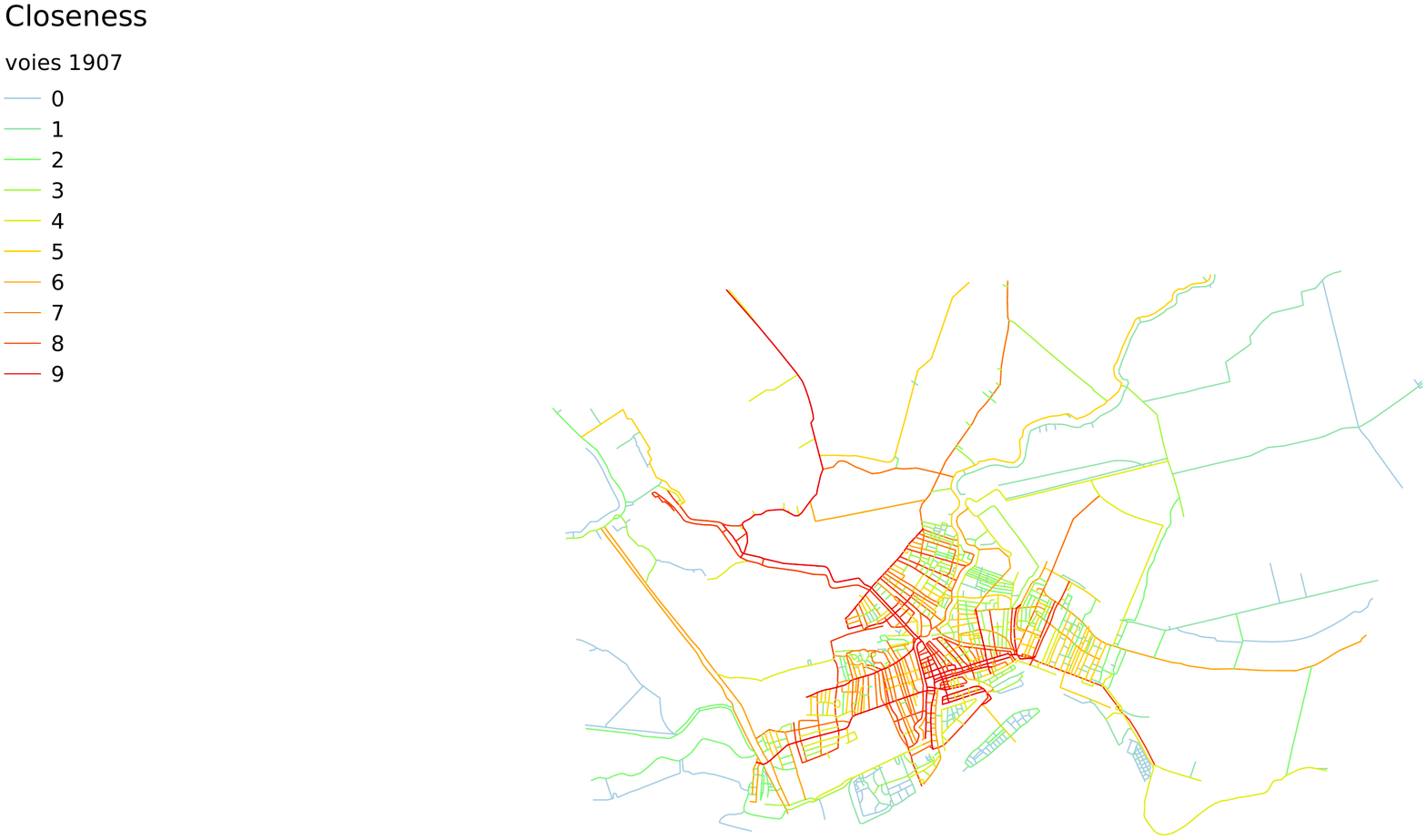}
        \caption{Closeness calculée sur le réseau viaire Nord de Rotterdam (découpé) en {\large \textbf{1907}}.}
        \label{fig:clo_rd_1907}
    \end{figure}
        
     \begin{figure}[h]
        \includegraphics[width=\textwidth]{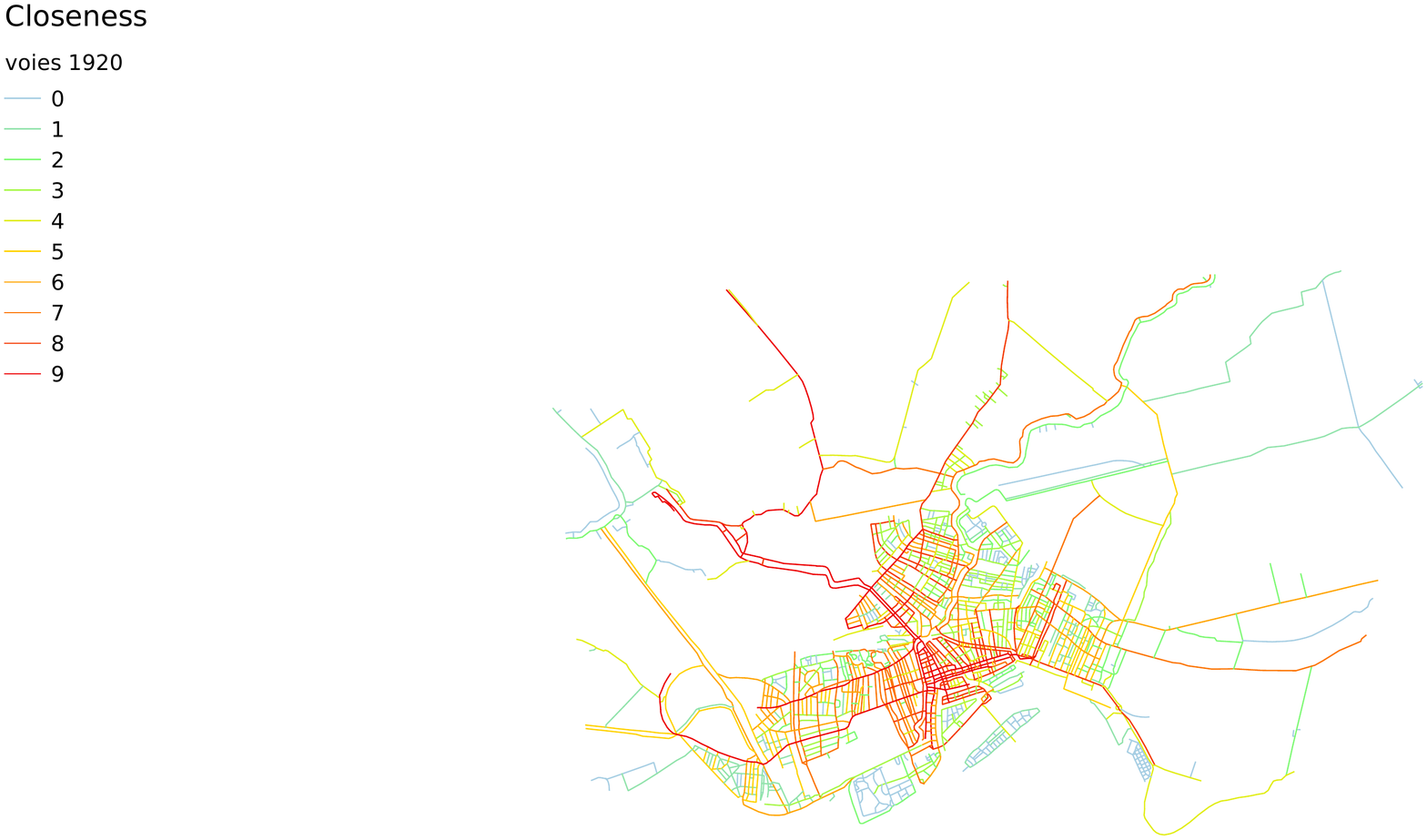}
        \caption{Closeness calculée sur le réseau viaire Nord de Rotterdam (découpé) en {\large \textbf{1920}}.}
        \label{fig:clo_rd_1920}
    \end{figure}
 
    \begin{figure}[h]
        \includegraphics[width=\textwidth]{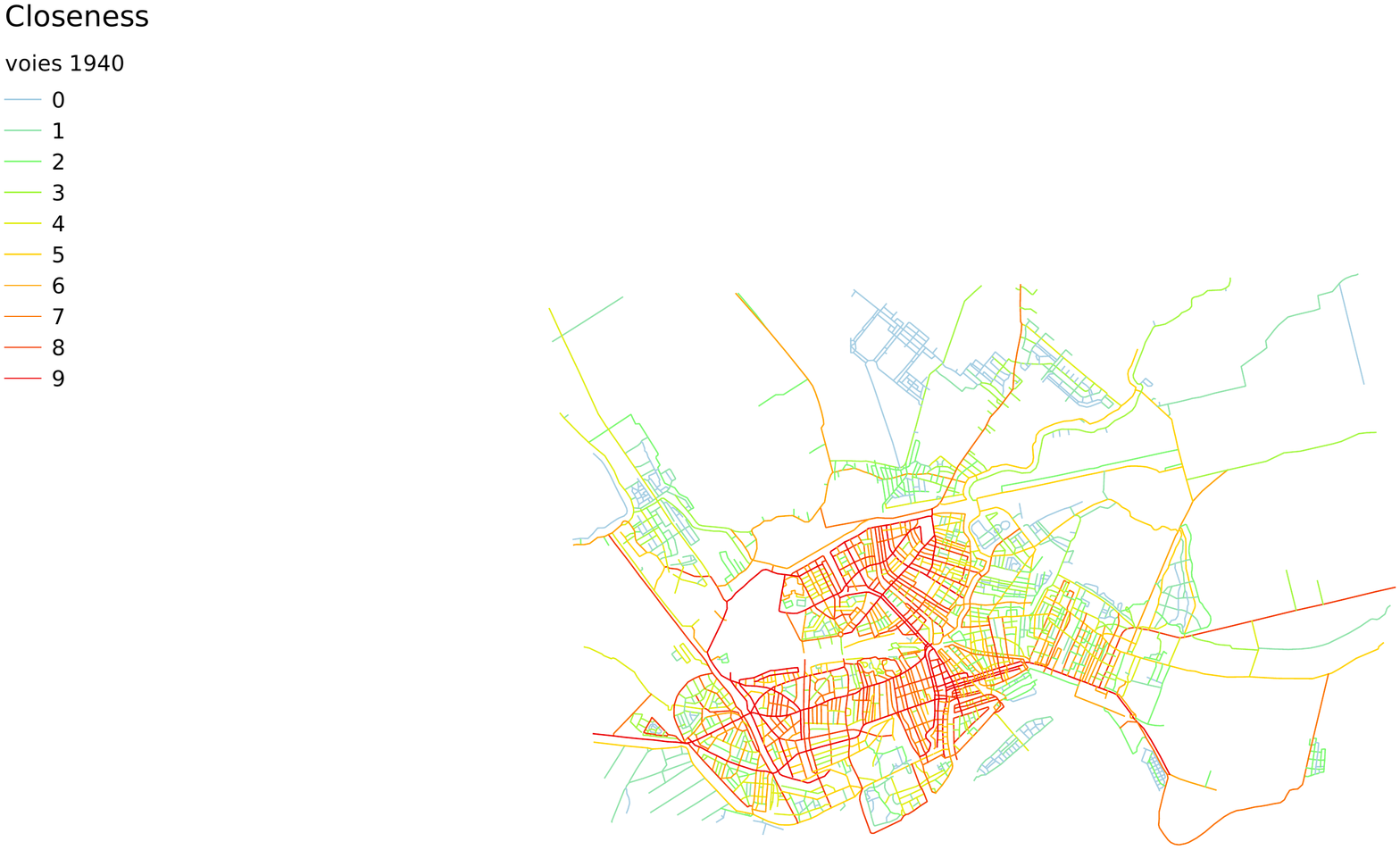}
        \caption{Closeness calculée sur le réseau viaire Nord de Rotterdam (découpé) en {\large \textbf{1940}}.}
        \label{fig:clo_rd_1940}
    \end{figure}
    
     \begin{figure}[h]
        \includegraphics[width=\textwidth]{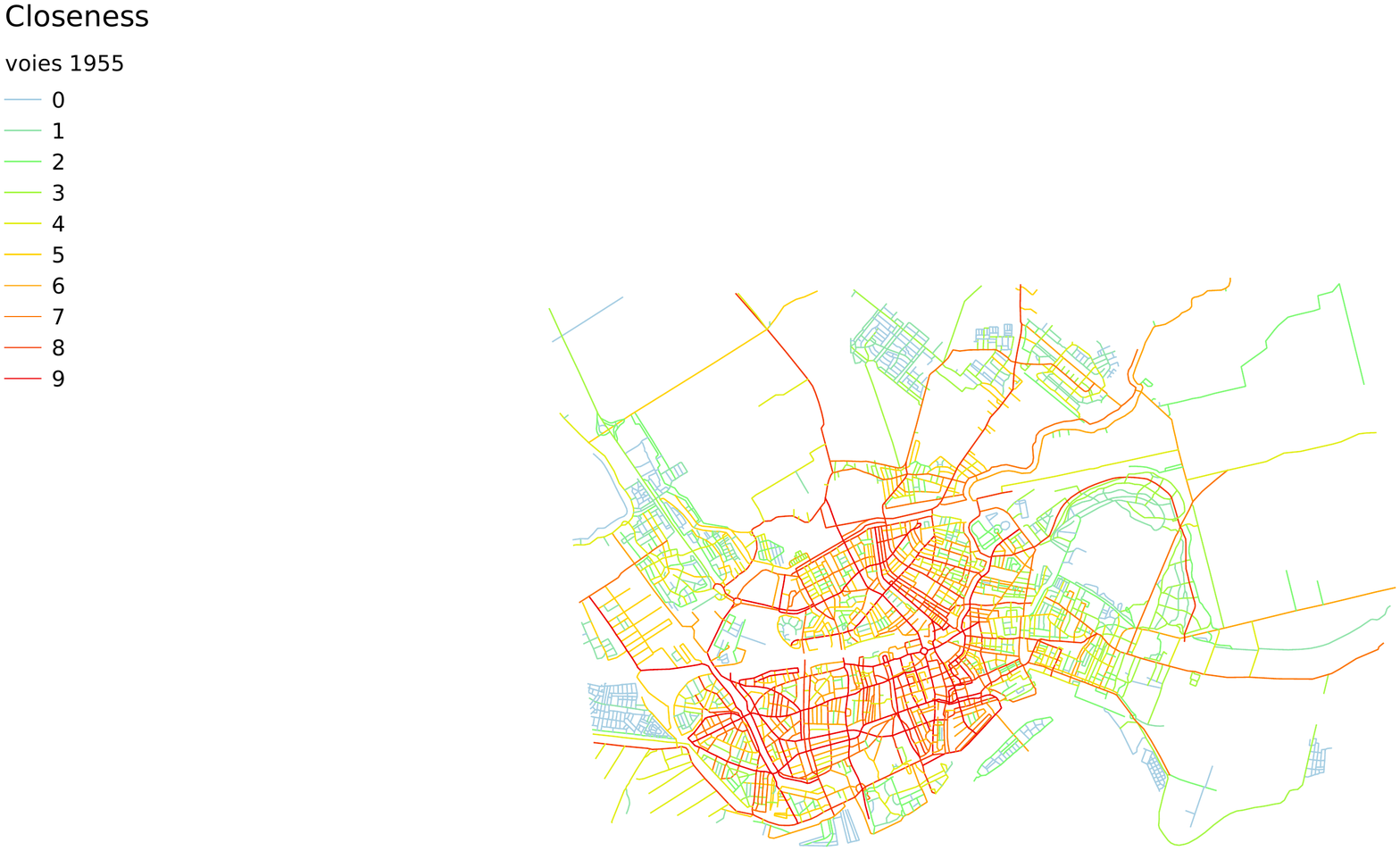}
        \caption{Closeness calculée sur le réseau viaire Nord de Rotterdam (découpé) en {\large \textbf{1955}}.}
        \label{fig:clo_rd_1955}
    \end{figure}

 \FloatBarrier
 \section{Projets urbains}\label{ann:sec_cartedia_projurb}
 
    \begin{figure}[h]
        \includegraphics[width=\textwidth]{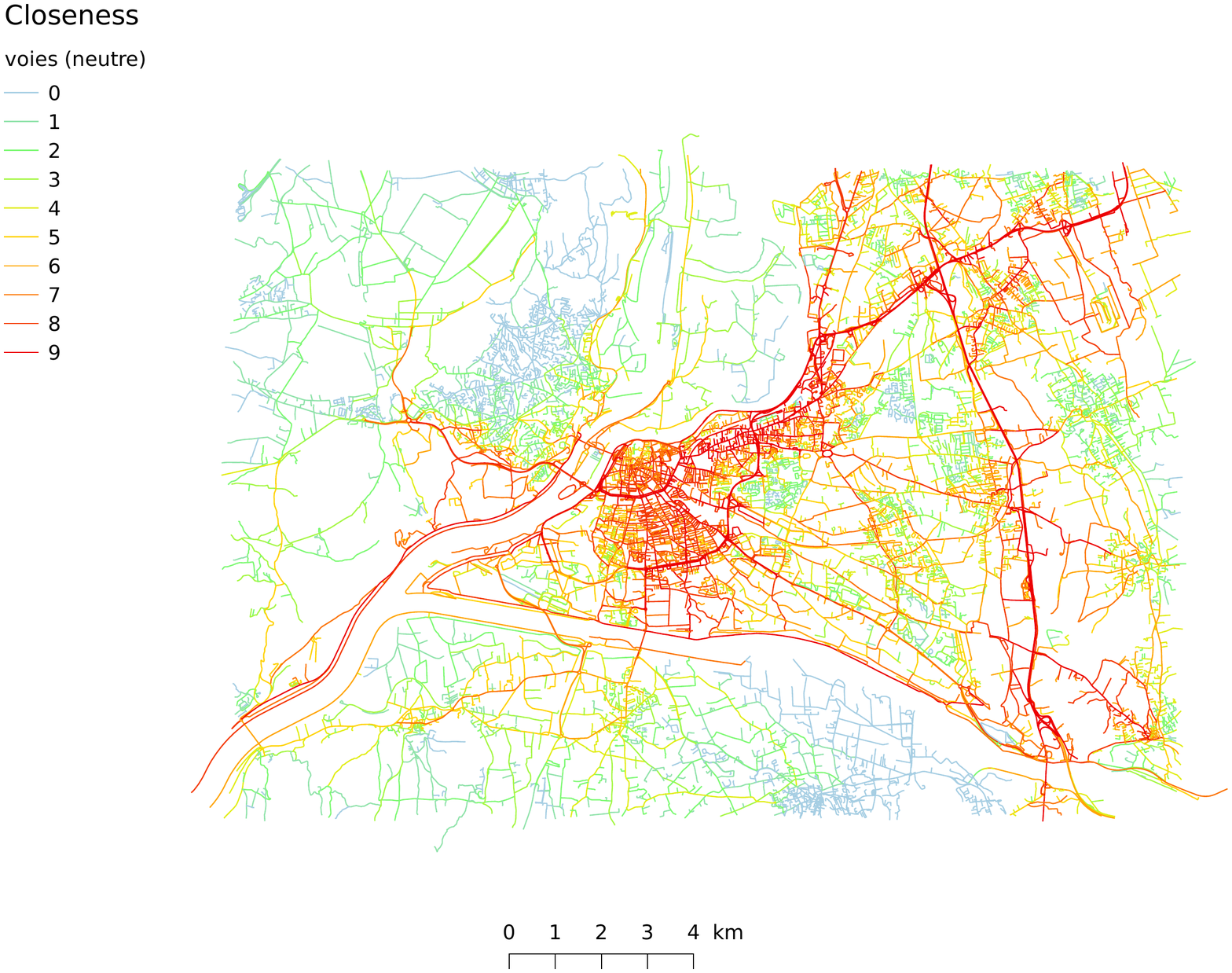}
        \caption{Closeness sur le graphe d'{\large \textbf{Avignon}} {\large \textbf{\enquote{neutre}}}.}
        \label{fig:proj_avleo2_1}
    \end{figure}
      
    \begin{figure}[h]
        \includegraphics[width=\textwidth]{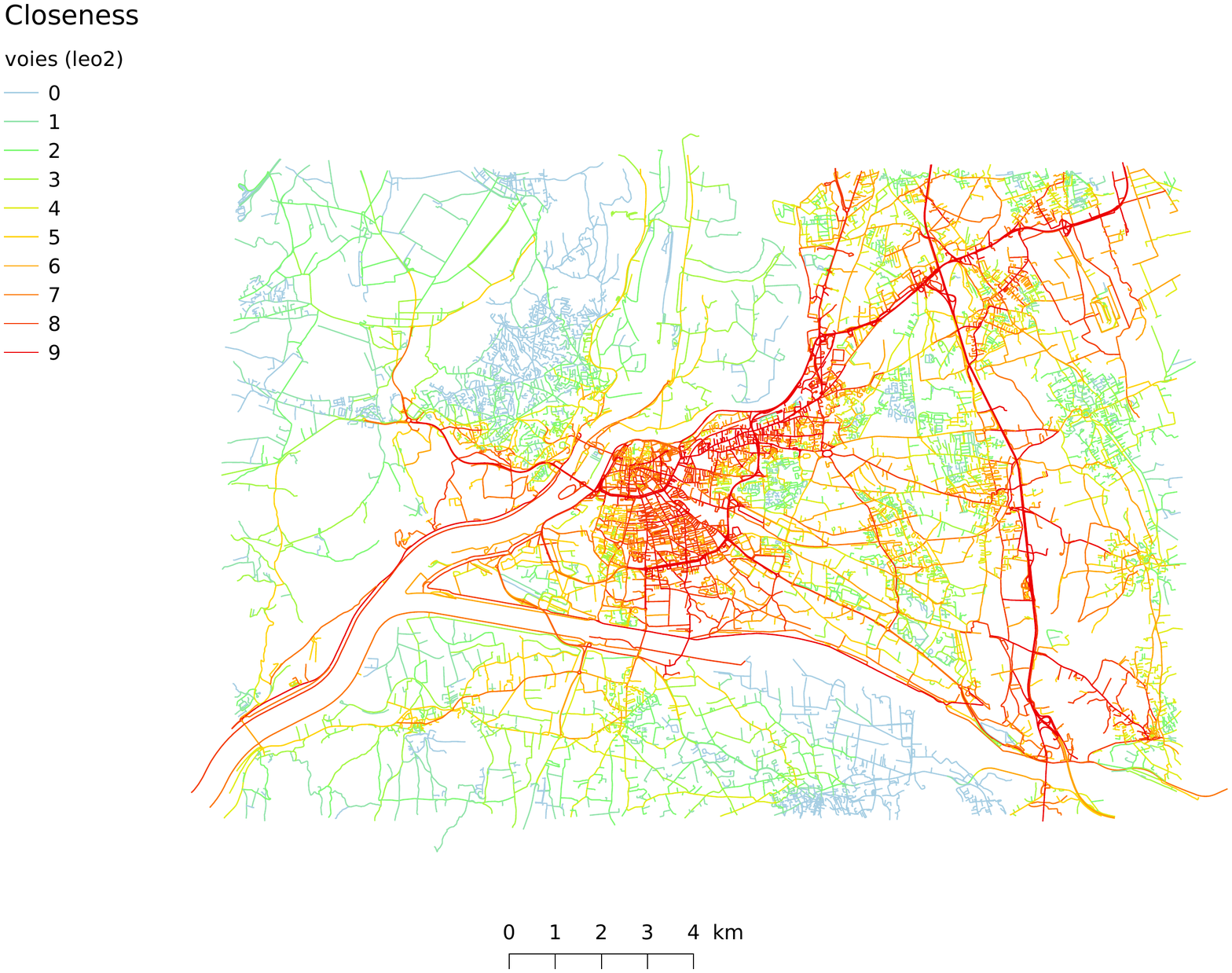}
        \caption{Closeness sur le graphe d'{\large \textbf{Avignon}} avec la réalisation du {\large \textbf{projet \enquote{Leo2}}}.}
        \label{fig:proj_avleo2_2}
    \end{figure}

    \begin{figure}[h]
        \includegraphics[width=\textwidth]{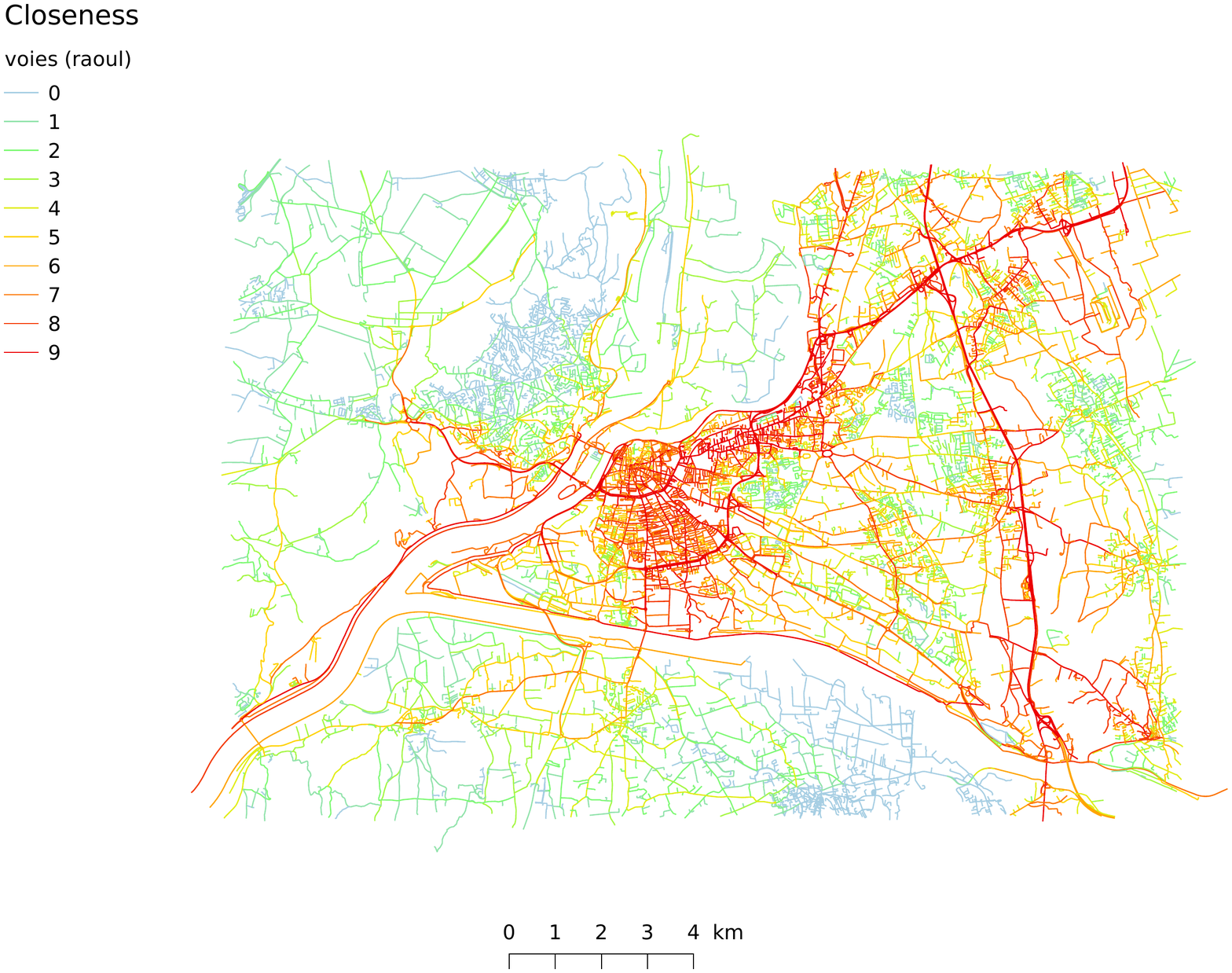}
        \caption{Closeness sur le graphe d'{\large \textbf{Avignon}} avec la réalisation du {\large \textbf{projet \enquote{Raoul}}}.}
        \label{fig:proj_avraoul_2}
    \end{figure} 
    
    \begin{figure}[h]
        \includegraphics[width=\textwidth]{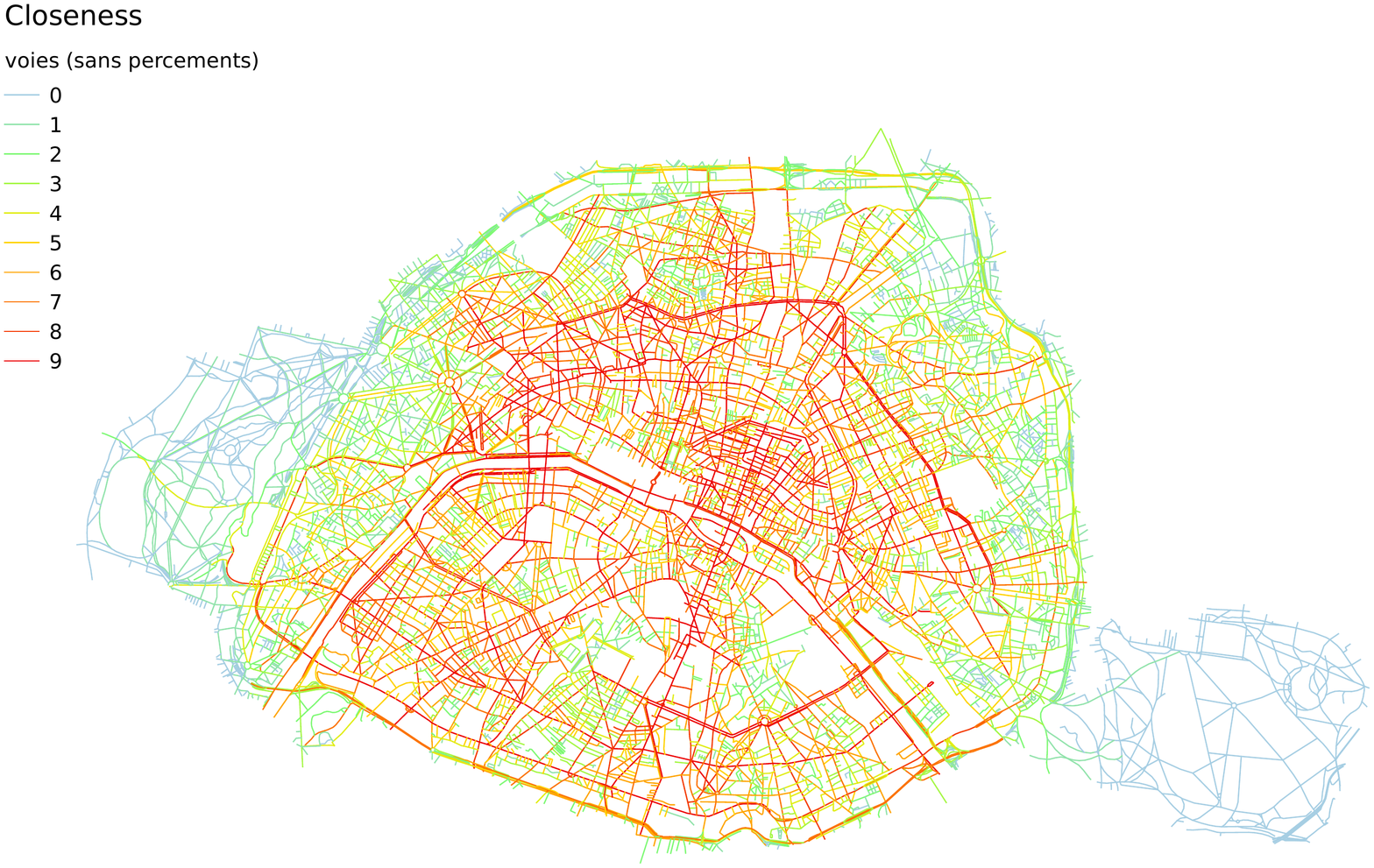}
        \caption{Closeness sur le graphe de {\large \textbf{Paris sans percements}}.}
        \label{fig:proj_paris_1}
    \end{figure}
   
    \begin{figure}[h]
        \includegraphics[width=\textwidth]{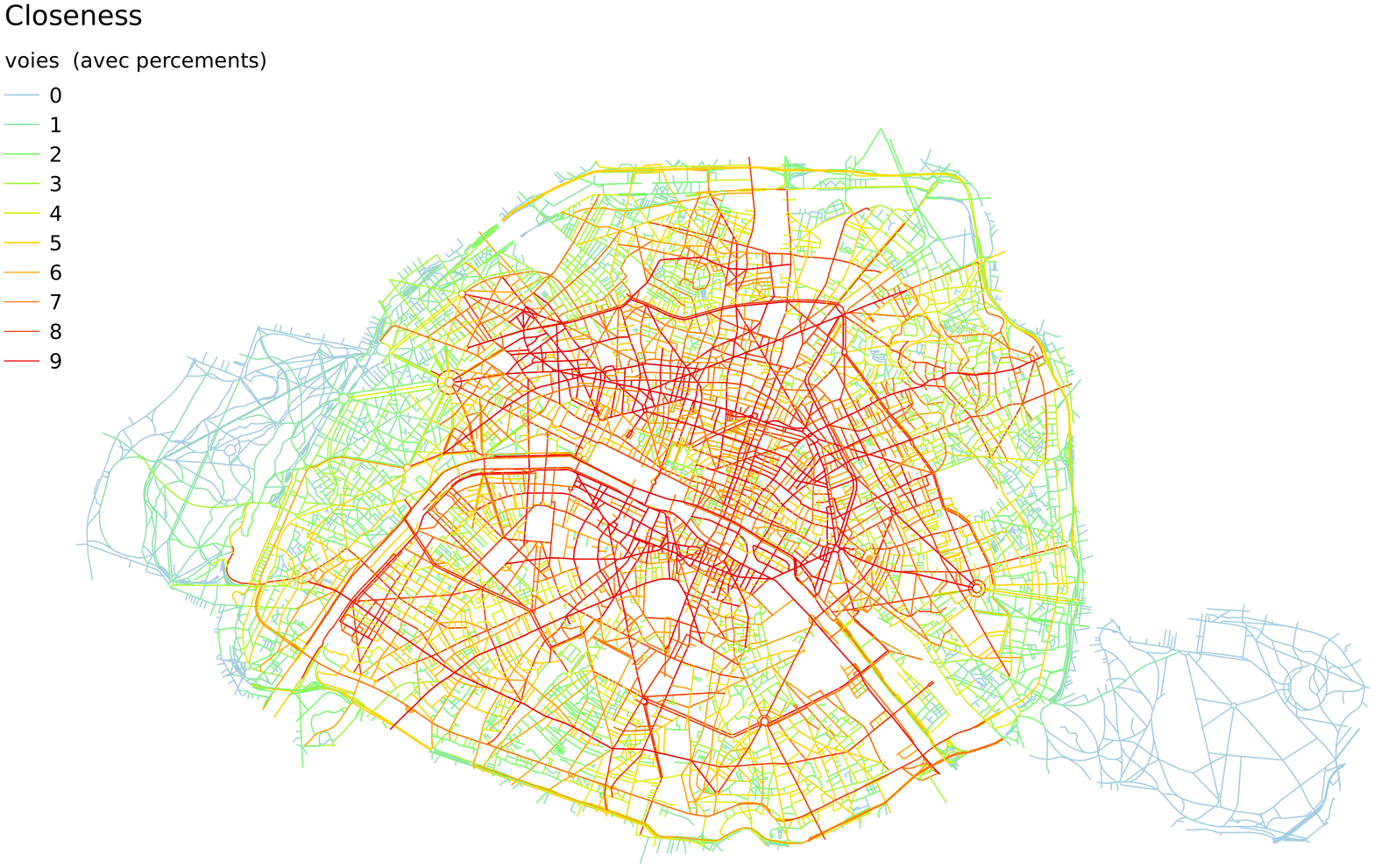}
        \caption{Closeness sur le graphe de {\large \textbf{Paris avec percements}}.}
        \label{fig:proj_paris_2}
    \end{figure}

\FloatBarrier 
\chapter{Distribution des distances topologiques}\label{ann:chap_distr_dtopo}

\clearpage

\section{Avignon}\label{ann:sec_dtopo_avignon}

\begin{figure}[h]
    \centering
    \includegraphics[width=\textwidth]{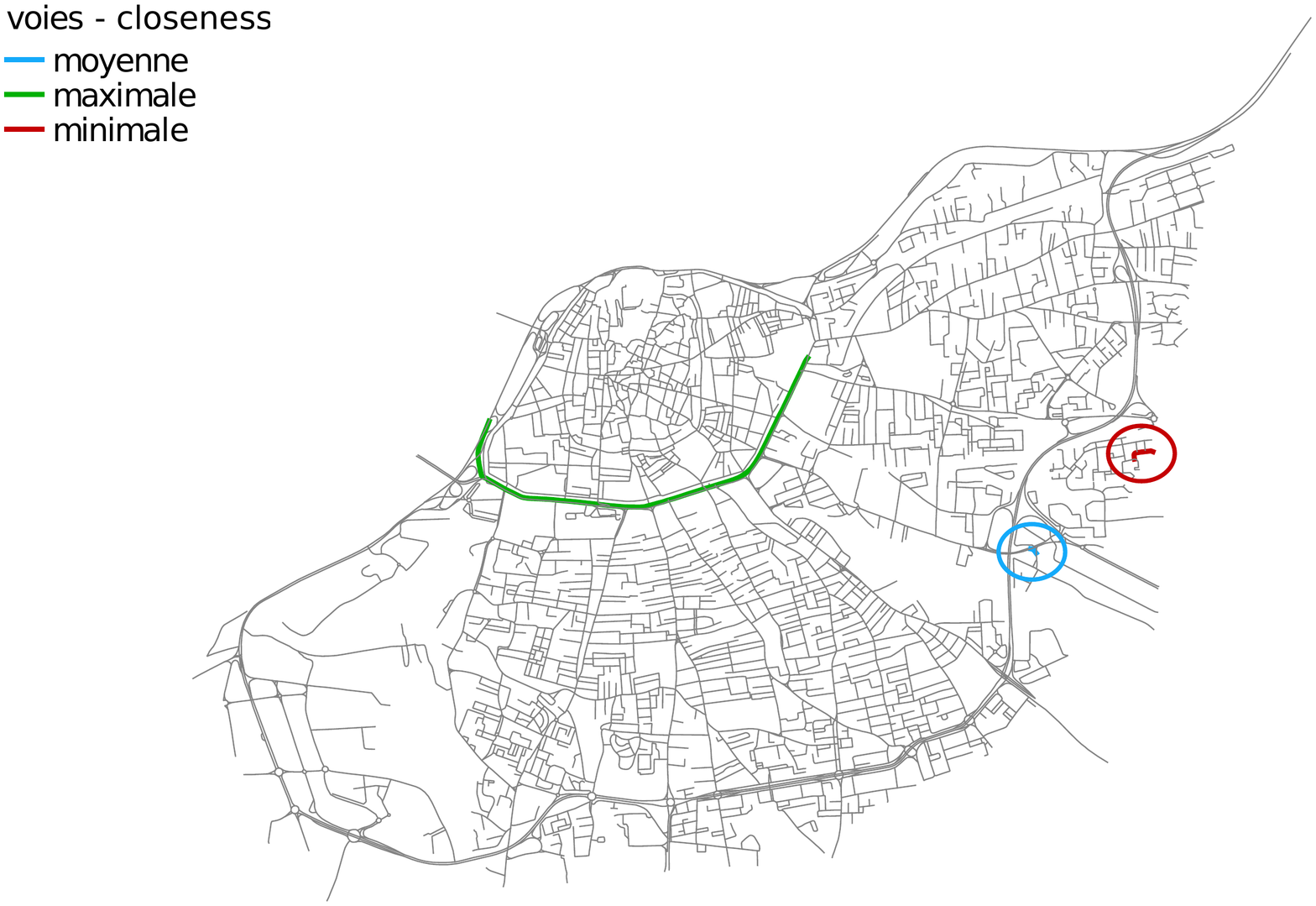}
    \caption{\large{\textbf{Situation des voies considérées}}}
\end{figure}

\begin{figure}[h]
    \centering
    \begin{subfigure}[t]{0.45\textwidth}
        \centering
        \includegraphics[width=\textwidth]{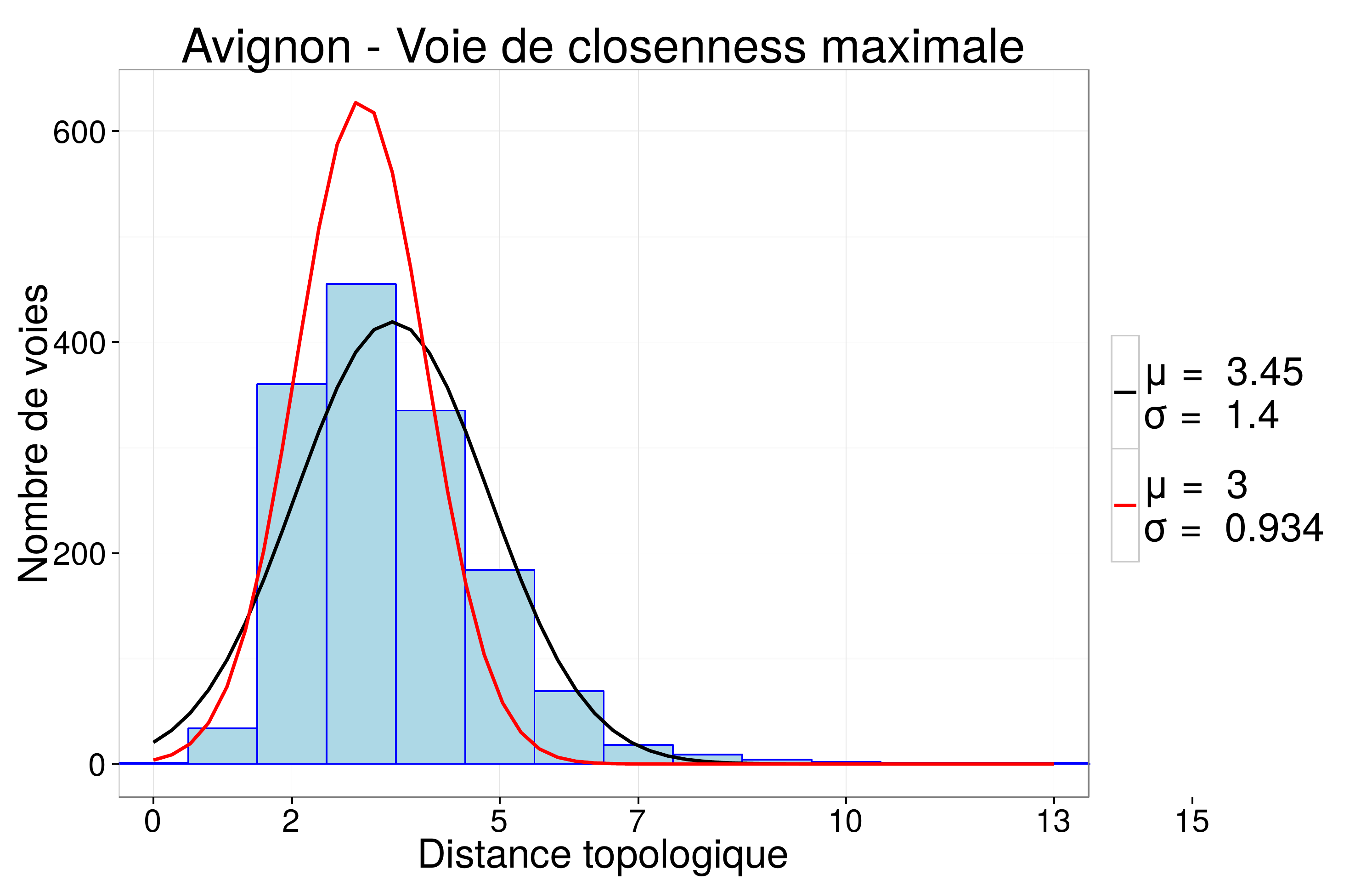}
        \caption*{Pour la voie de closeness maximale}
    \end{subfigure}
    ~
    \begin{subfigure}[t]{0.45\textwidth}
        \centering
        \includegraphics[width=\textwidth]{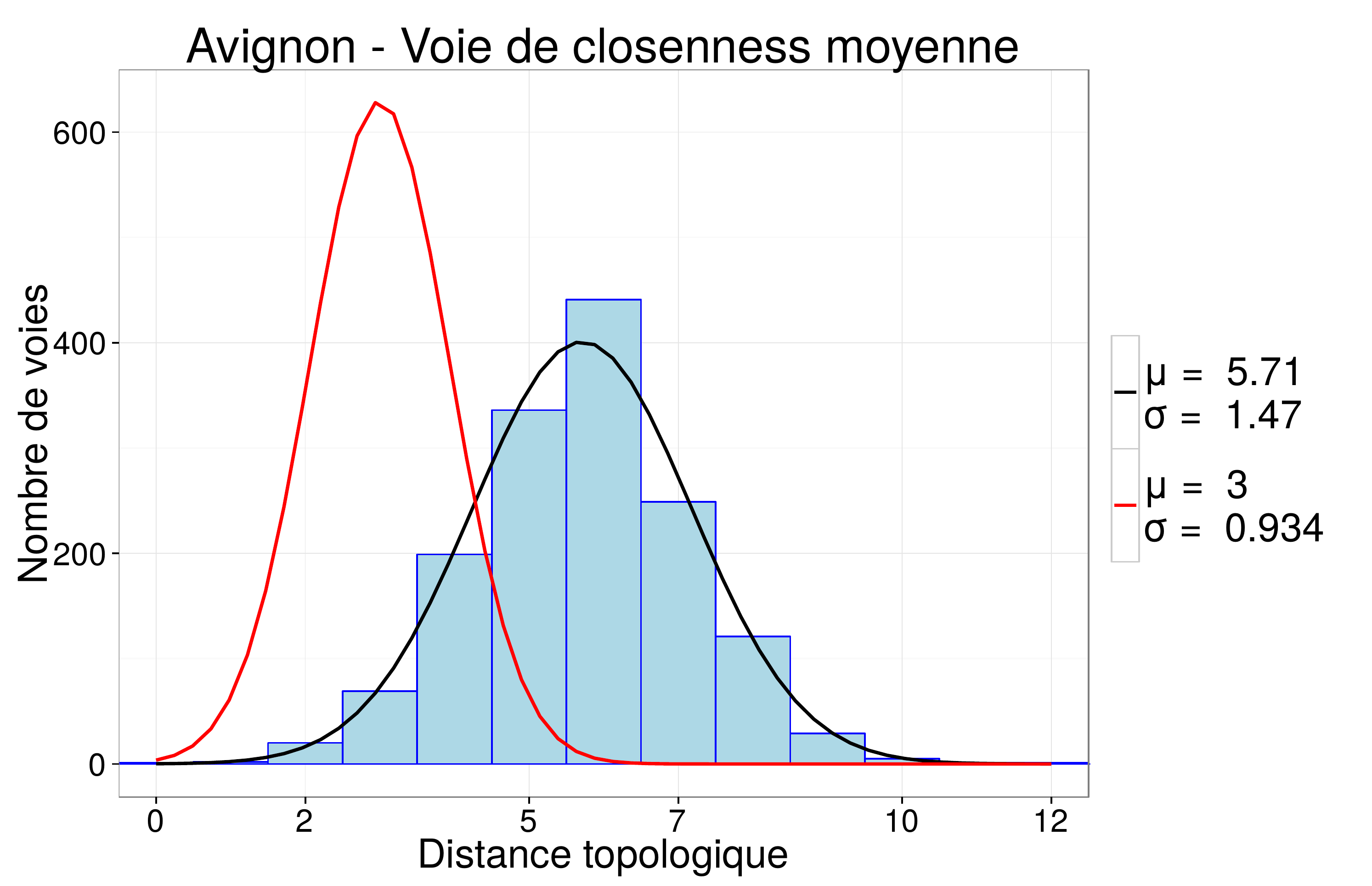}
        \caption*{Pour une voie de closeness moyenne}
    \end{subfigure}
    ~
    \begin{subfigure}[t]{0.45\textwidth}
        \centering
        \includegraphics[width=\textwidth]{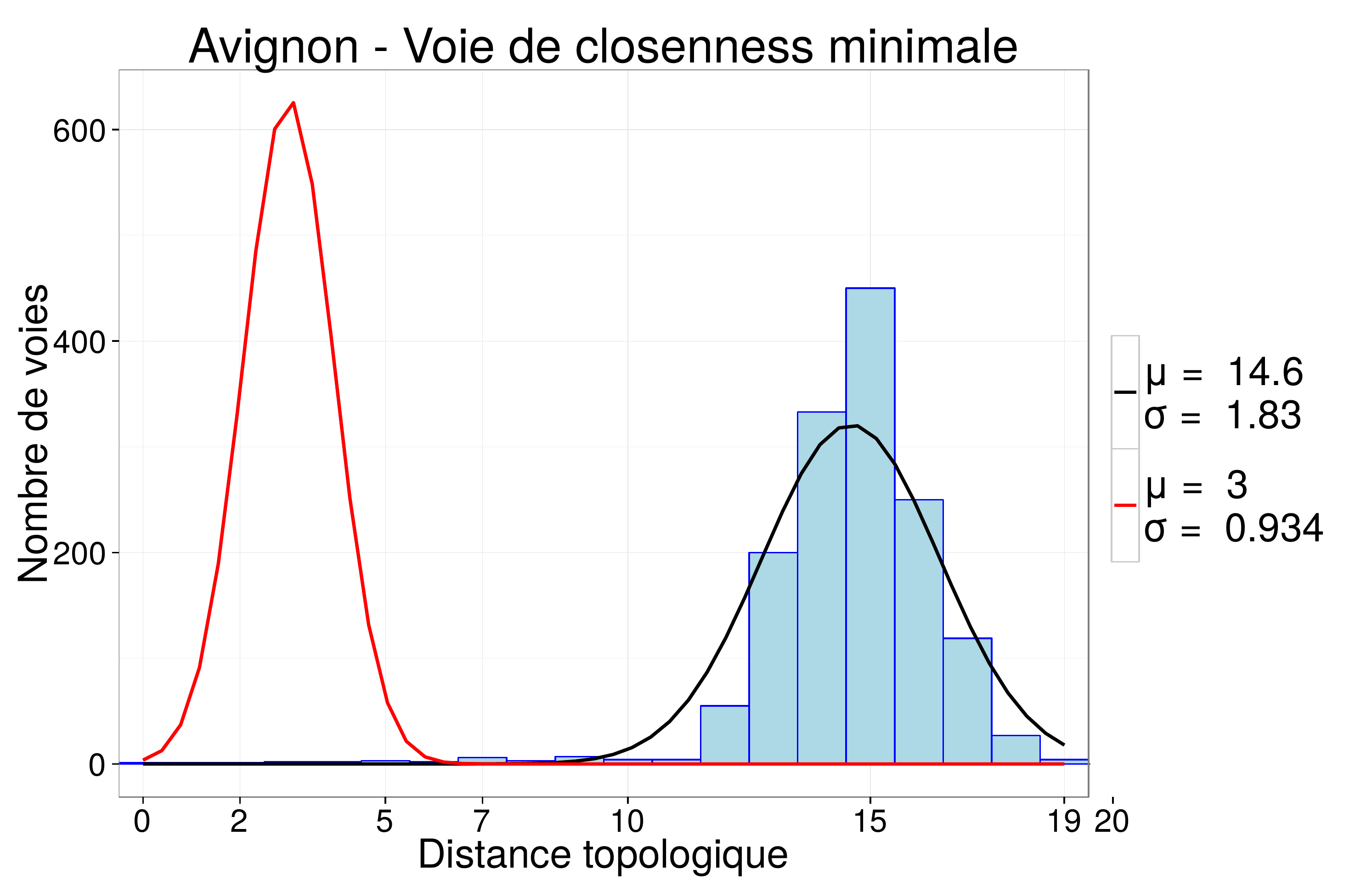}
        \caption*{Pour la voie de closeness minimale}
    \end{subfigure}
    \caption{Répartition des \large{\textbf{distances topologiques}}. En rouge, gaussienne théorique}
\end{figure}

\FloatBarrier

\clearpage

\section{Barcelone}\label{ann:sec_dtopo_barcelone}

\begin{figure}[h]
    \centering
    \includegraphics[width=\textwidth]{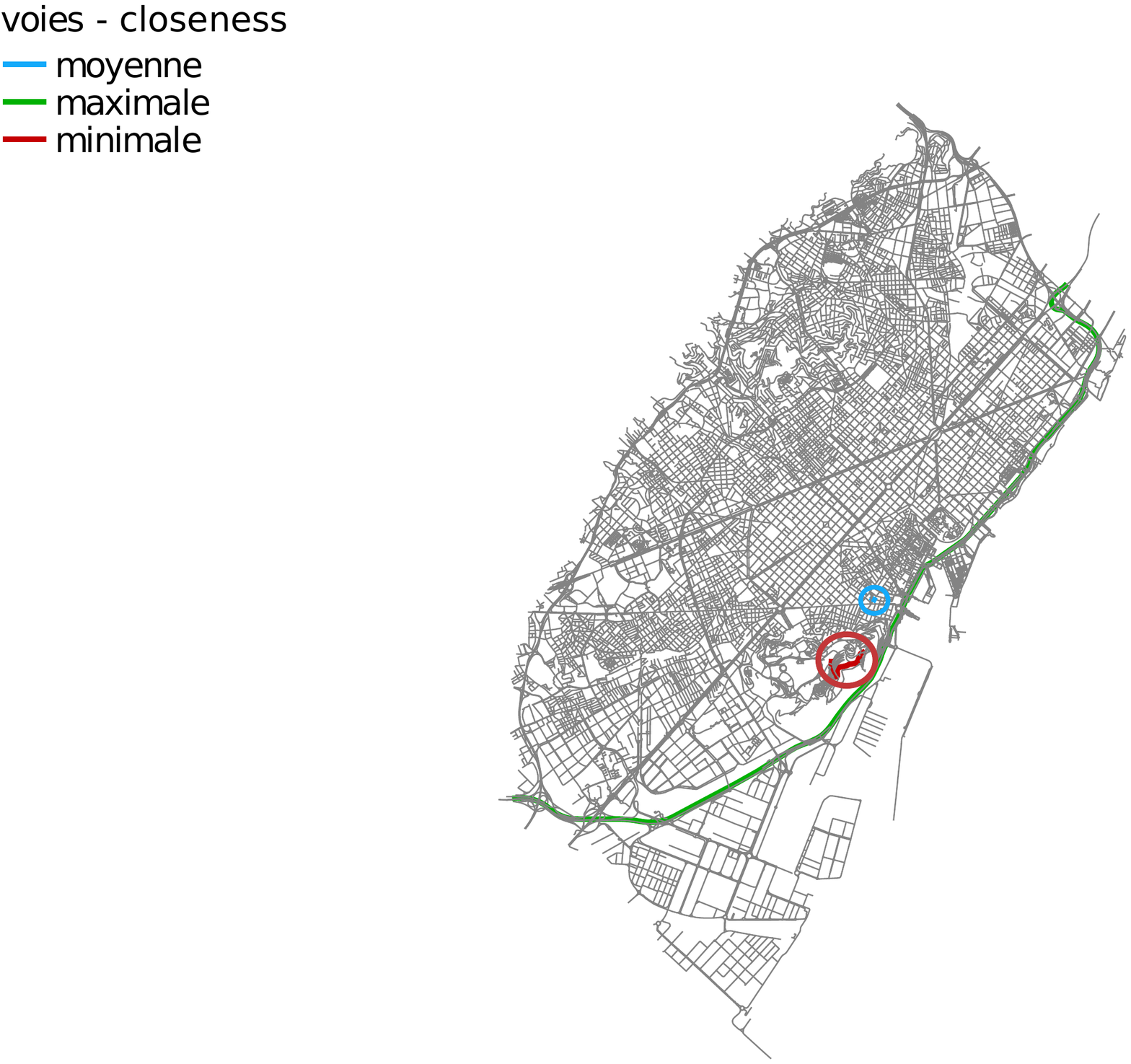}
    \caption{\large{\textbf{Situation des voies considérées}}}
\end{figure}

\begin{figure}[h]
    \centering
    \begin{subfigure}[t]{0.45\textwidth}
        \centering
        \includegraphics[width=\textwidth]{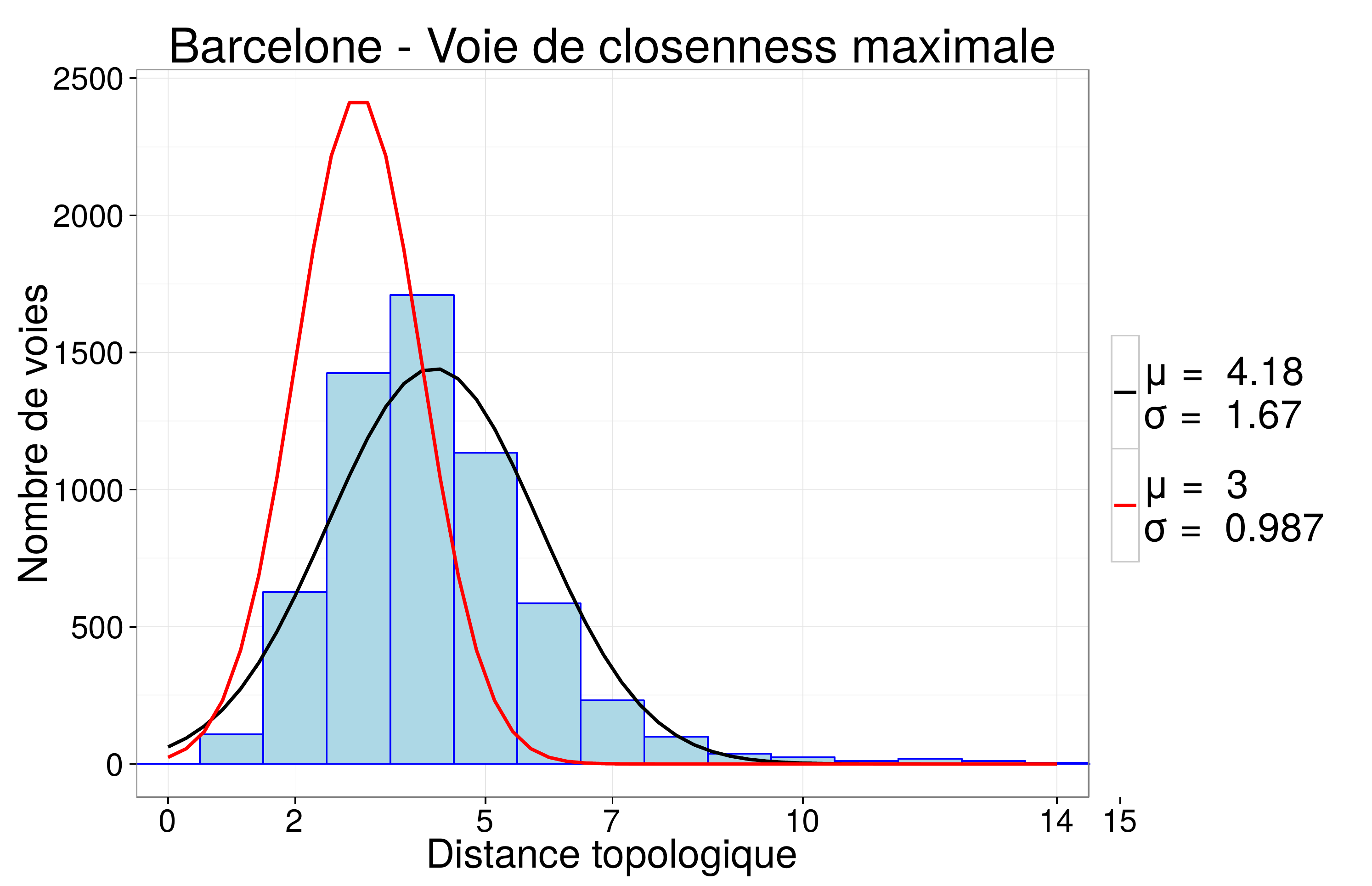}
        \caption{Pour la voie de closeness maximale}
    \end{subfigure}
    ~
    \begin{subfigure}[t]{0.45\textwidth}
        \centering
        \includegraphics[width=\textwidth]{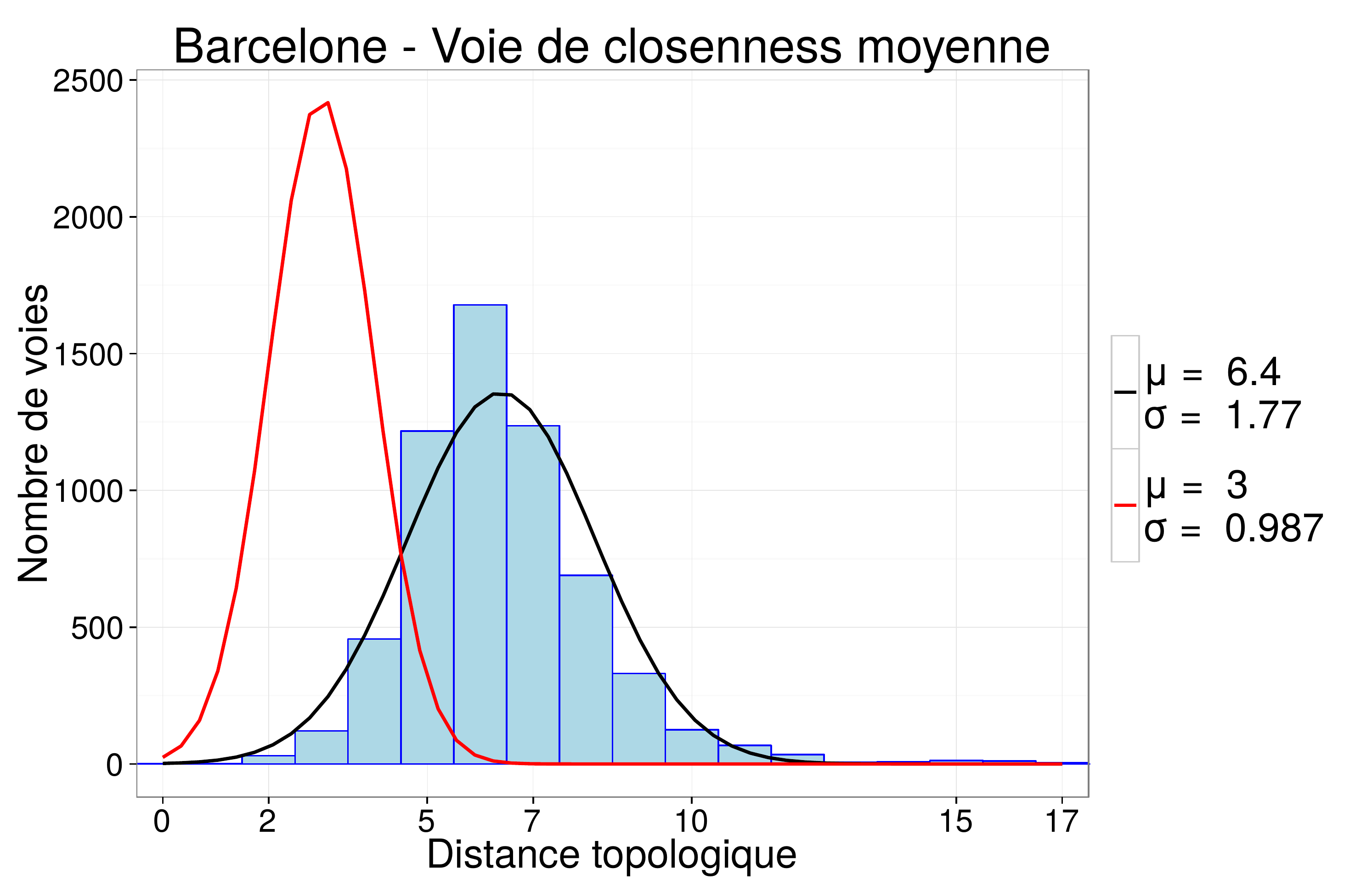}
        \caption{Pour une voie de closeness moyenne}
    \end{subfigure}
    ~
    \begin{subfigure}[t]{0.45\textwidth}
        \centering
        \includegraphics[width=\textwidth]{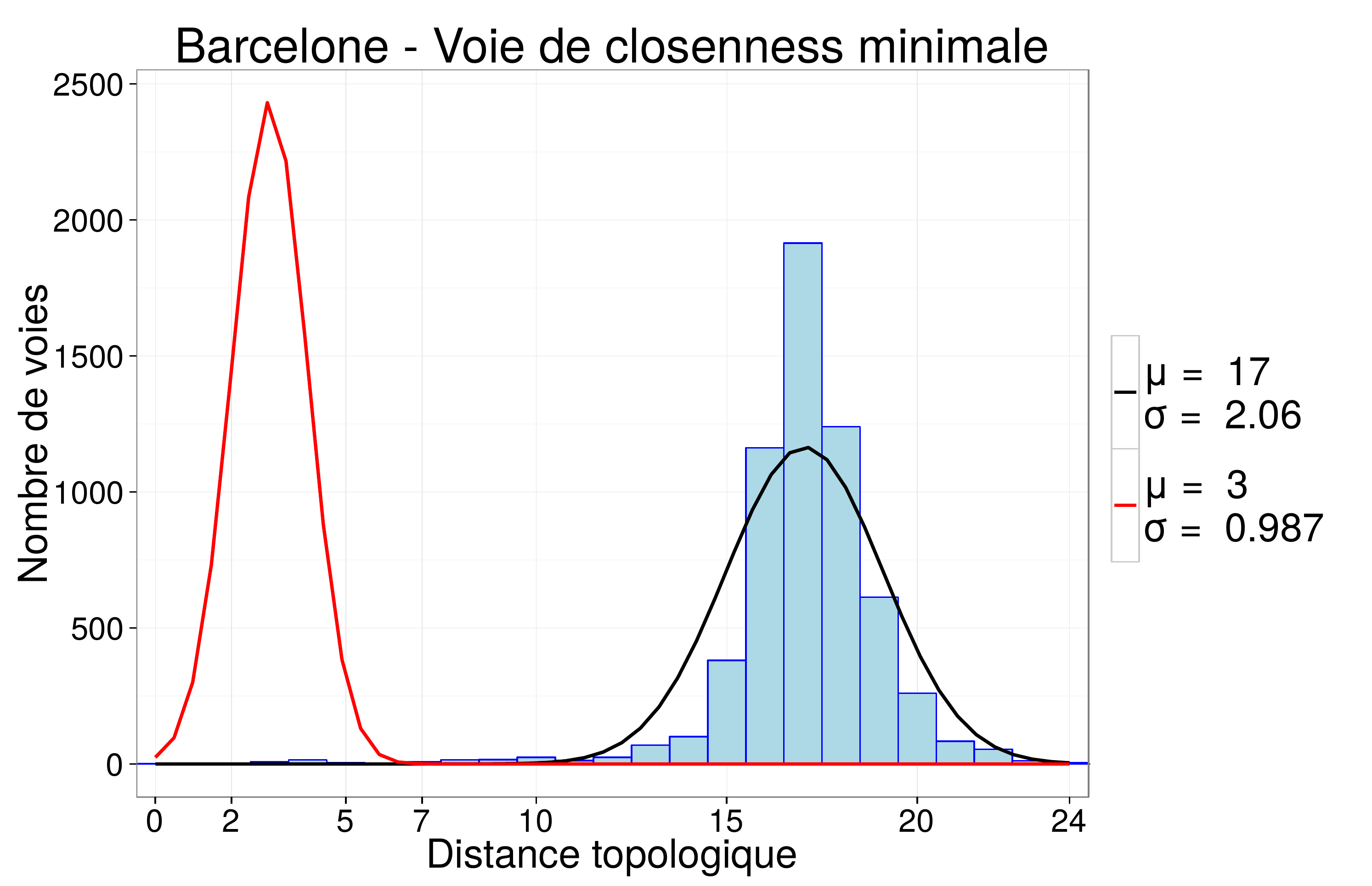}
        \caption{Pour la voie de closeness minimale}
    \end{subfigure}
    \caption{Répartition des \large{\textbf{distances topologiques}}. En rouge, gaussienne théorique}
\end{figure}

\FloatBarrier

\clearpage

\section{Kyoto}\label{ann:sec_dtopo_kyoto}

\begin{figure}[h]
    \centering
    \includegraphics[width=\textwidth]{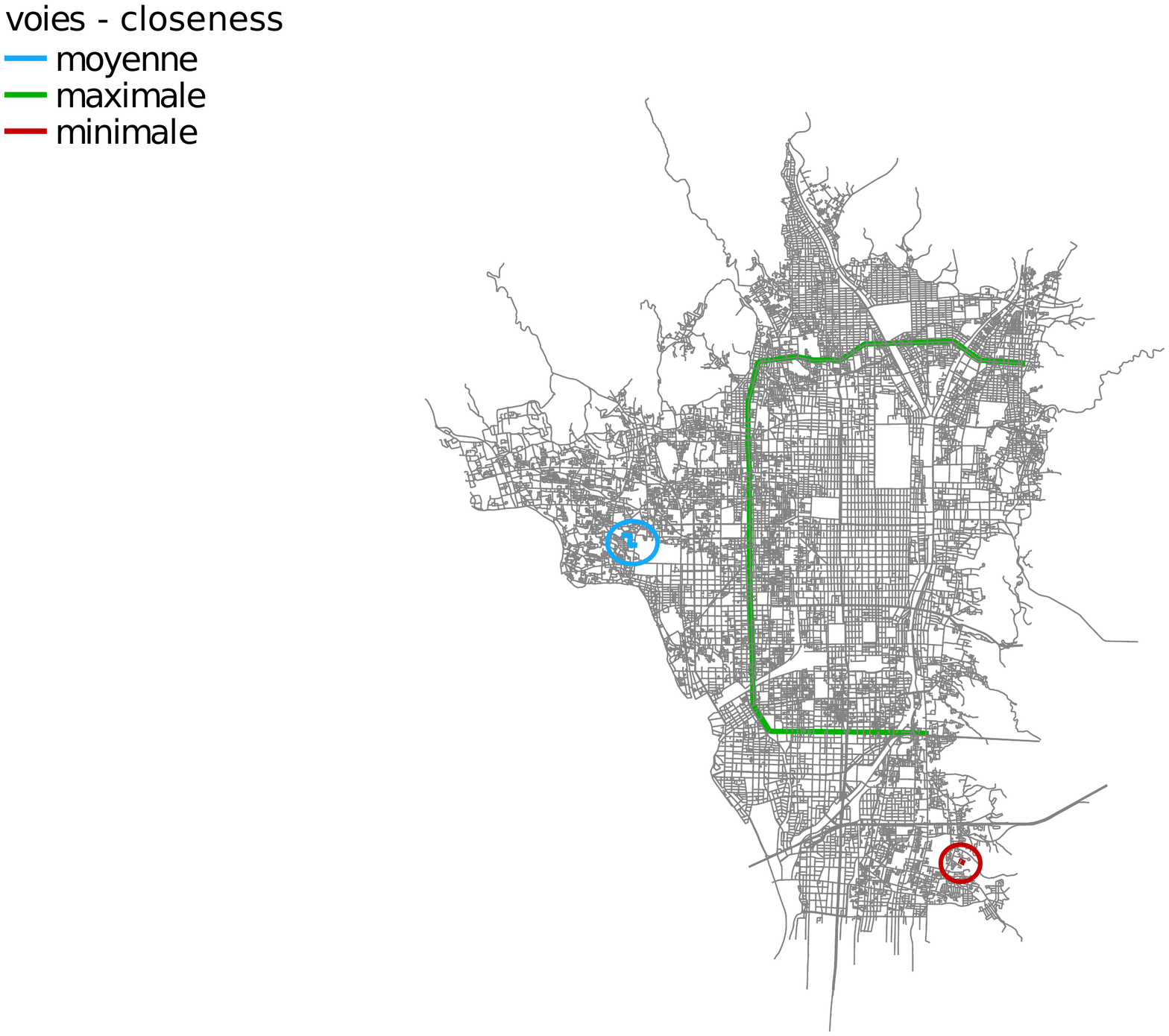}
    \caption{\large{\textbf{Situation des voies considérées}}}
\end{figure}

\begin{figure}[h]
    \centering
    \begin{subfigure}[t]{0.45\textwidth}
        \centering
        \includegraphics[width=\textwidth]{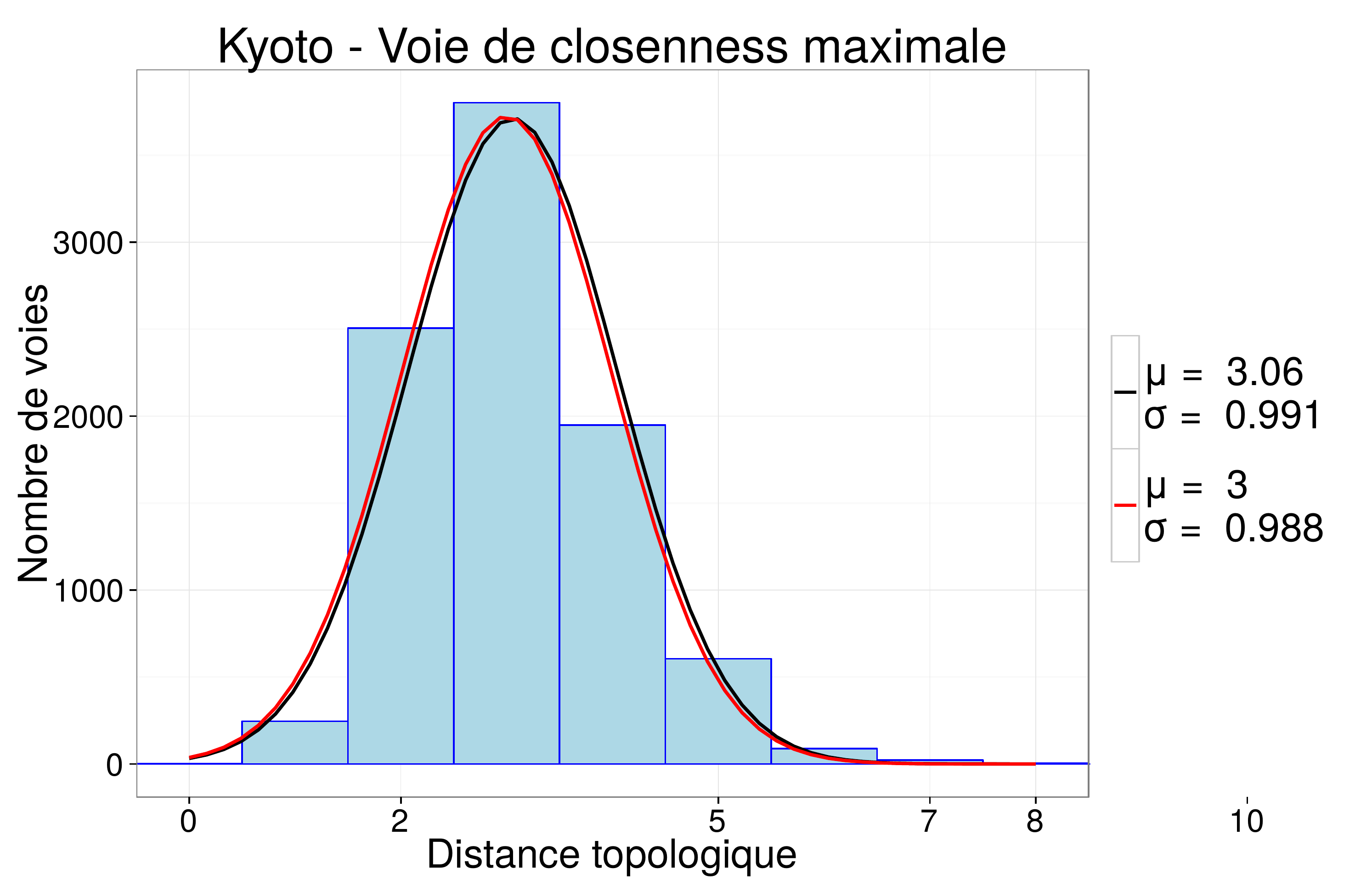}
        \caption{Pour la voie de closeness maximale}
    \end{subfigure}
    ~
    \begin{subfigure}[t]{0.45\textwidth}
        \centering
        \includegraphics[width=\textwidth]{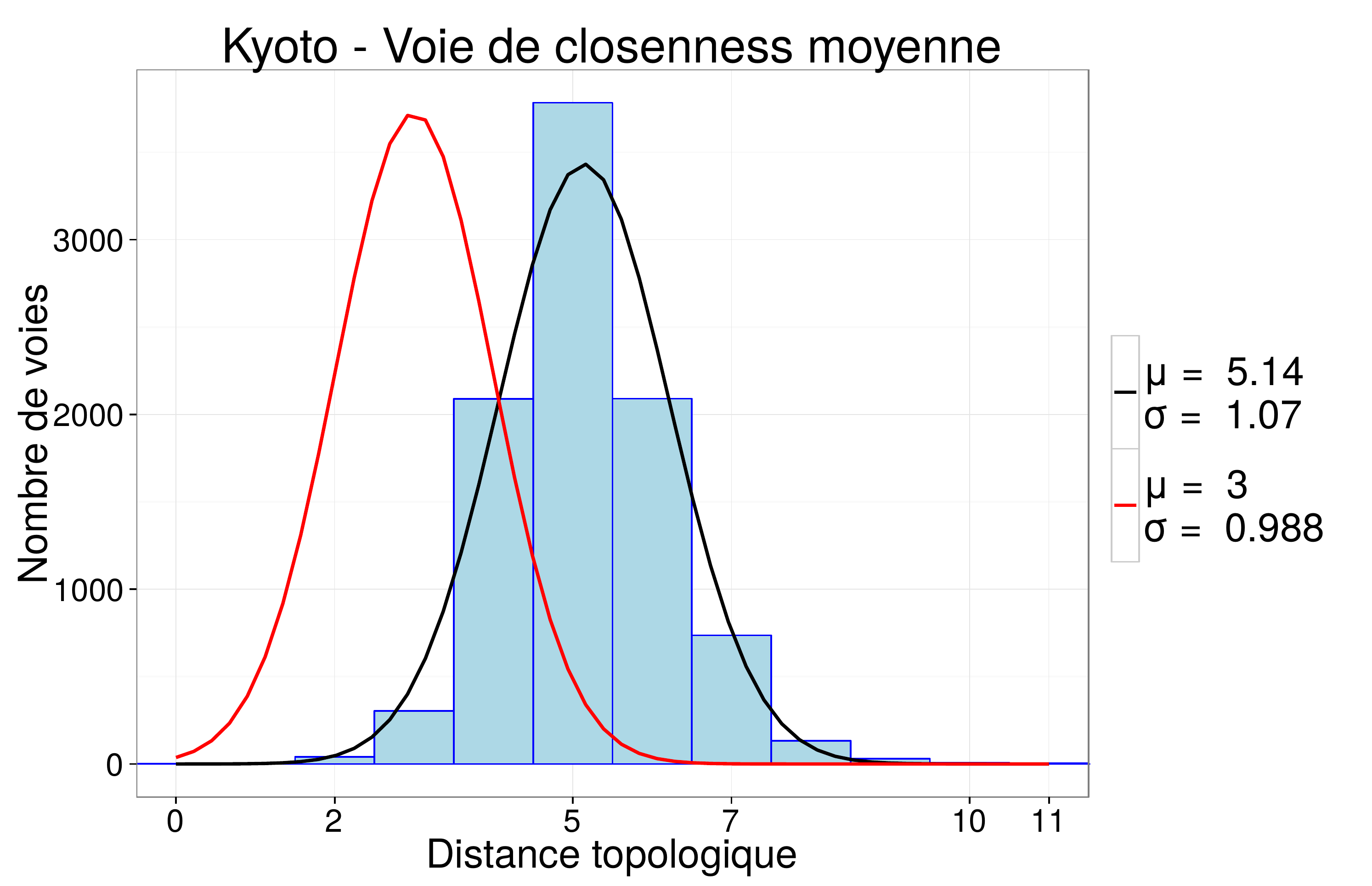}
        \caption{Pour une voie de closeness moyenne}
    \end{subfigure}
    ~
    \begin{subfigure}[t]{0.45\textwidth}
        \centering
        \includegraphics[width=\textwidth]{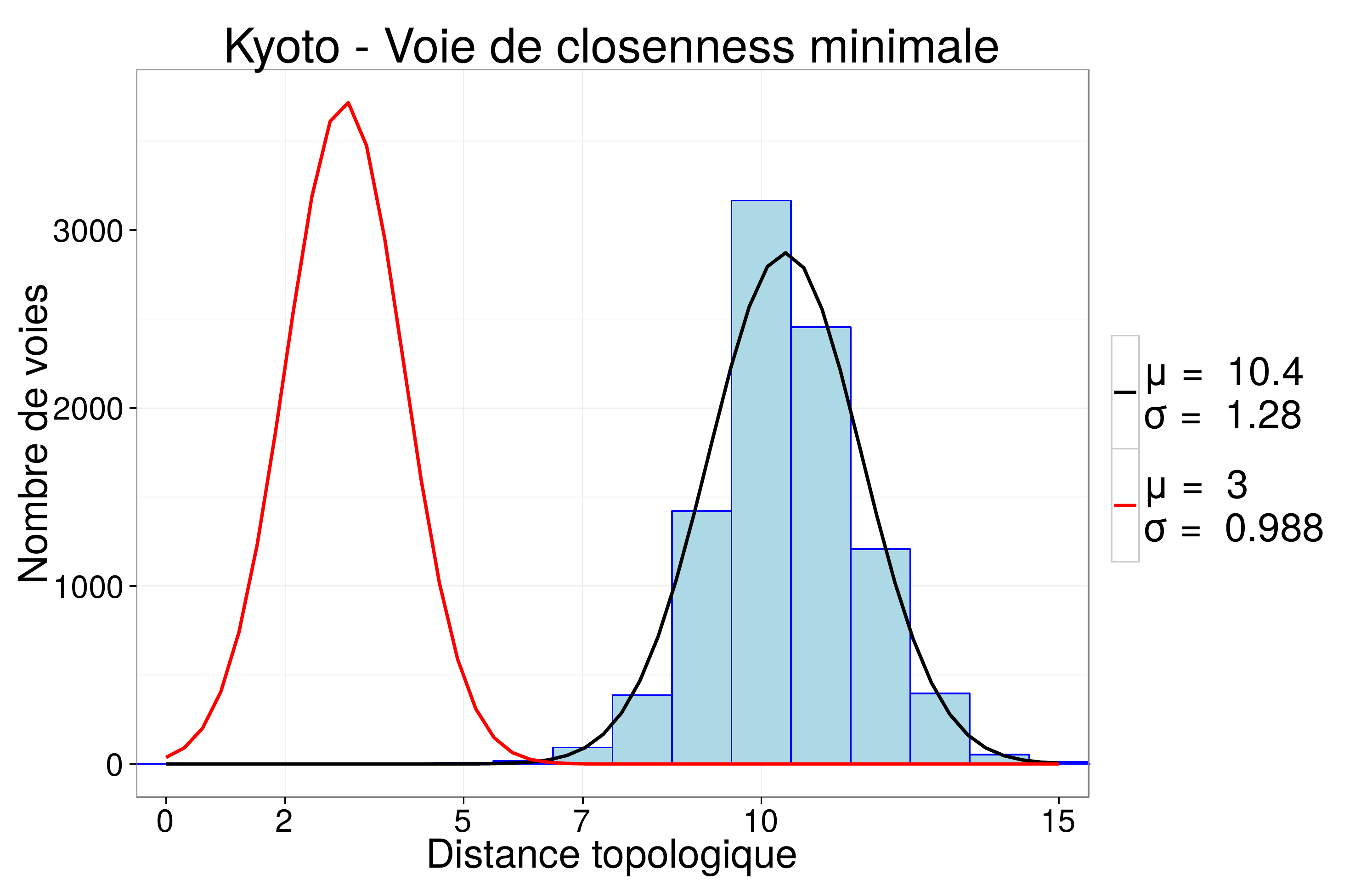}
        \caption{Pour la voie de closeness minimale}
    \end{subfigure}
    \caption{Répartition des \large{\textbf{distances topologiques}}. En rouge, gaussienne théorique}
\end{figure}

\FloatBarrier

\clearpage

\section{Manhattan}\label{ann:sec_dtopo_manhattan}

\begin{figure}[h]
    \centering
    \includegraphics[width=\textwidth]{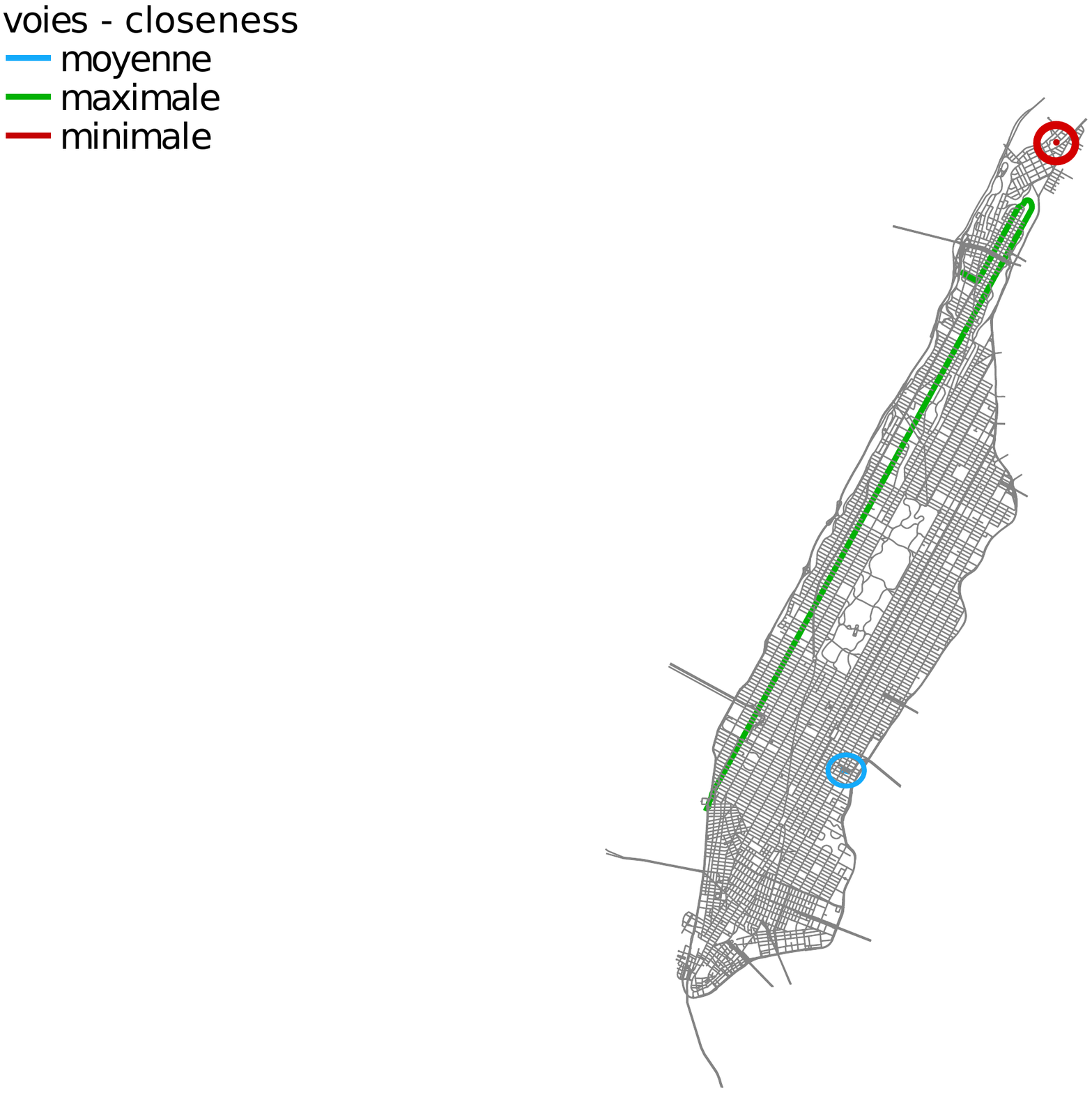}
    \caption{\large{\textbf{Situation des voies considérées}}}
\end{figure}

\begin{figure}[h]
    \centering
    \begin{subfigure}[t]{0.45\textwidth}
        \centering
        \includegraphics[width=\textwidth]{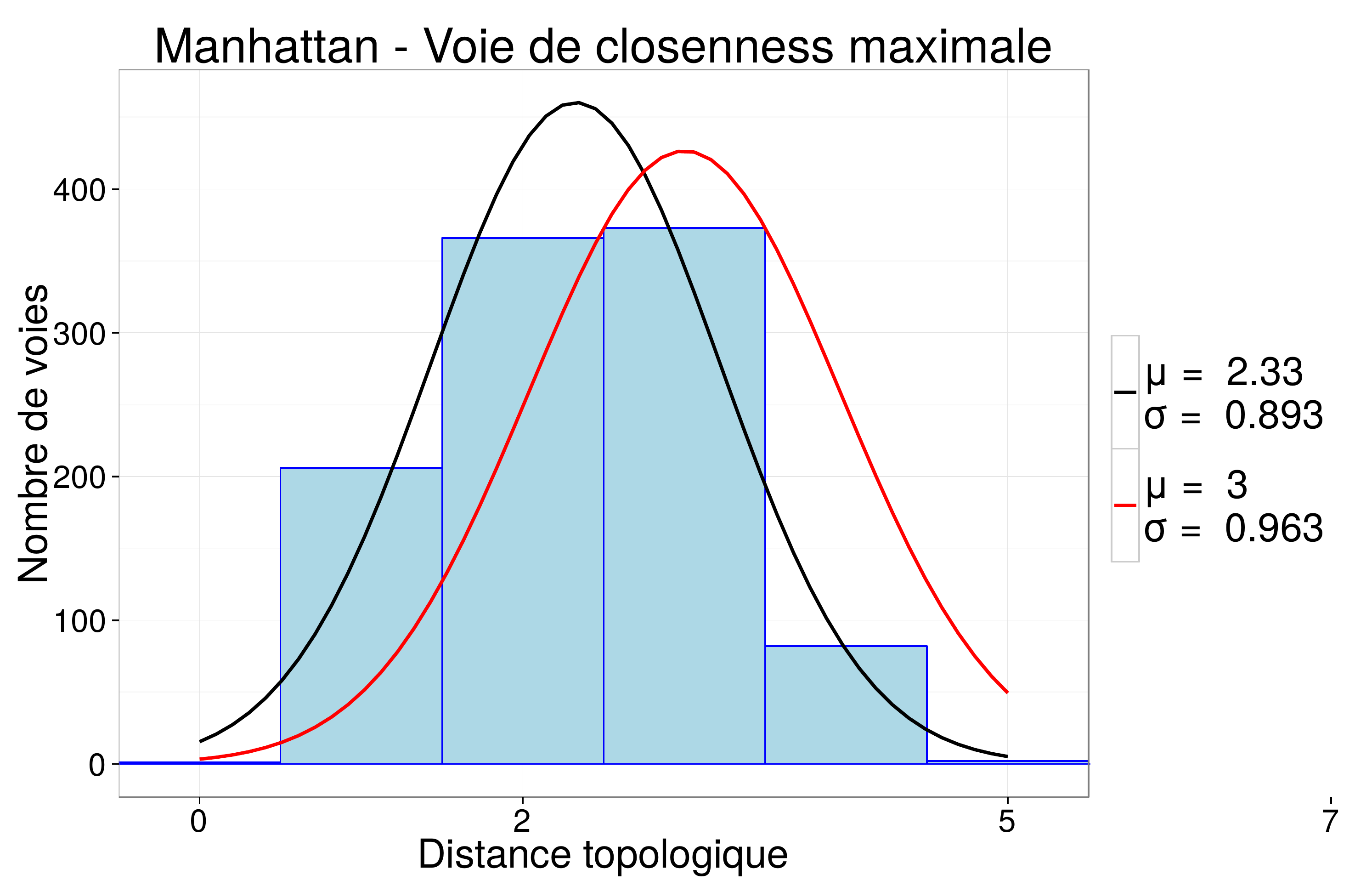}
        \caption{Pour la voie de closeness maximale}
    \end{subfigure}
    ~
    \begin{subfigure}[t]{0.45\textwidth}
        \centering
        \includegraphics[width=\textwidth]{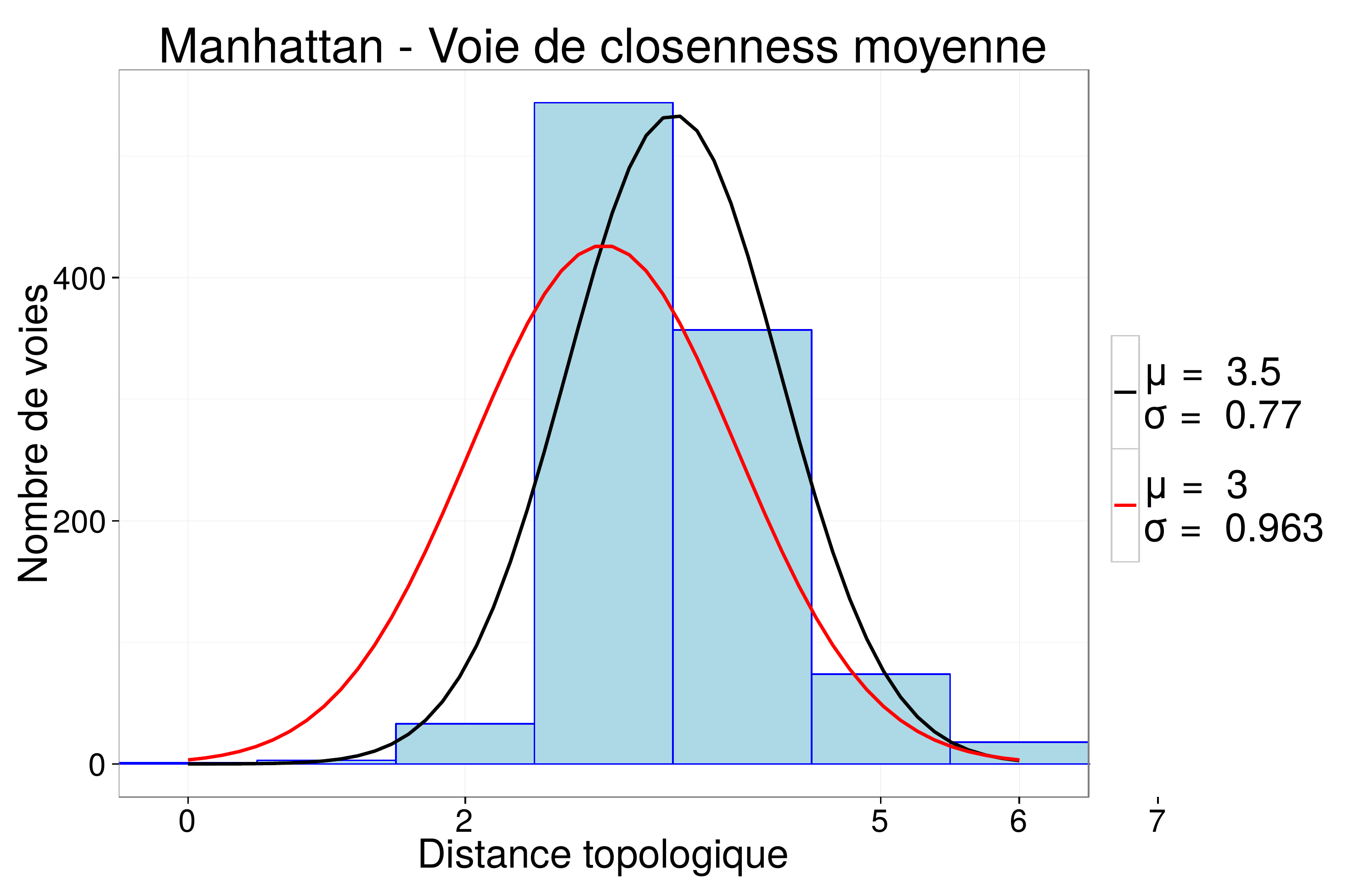}
        \caption{Pour une voie de closeness moyenne}
    \end{subfigure}
    ~
    \begin{subfigure}[t]{0.45\textwidth}
        \centering
        \includegraphics[width=\textwidth]{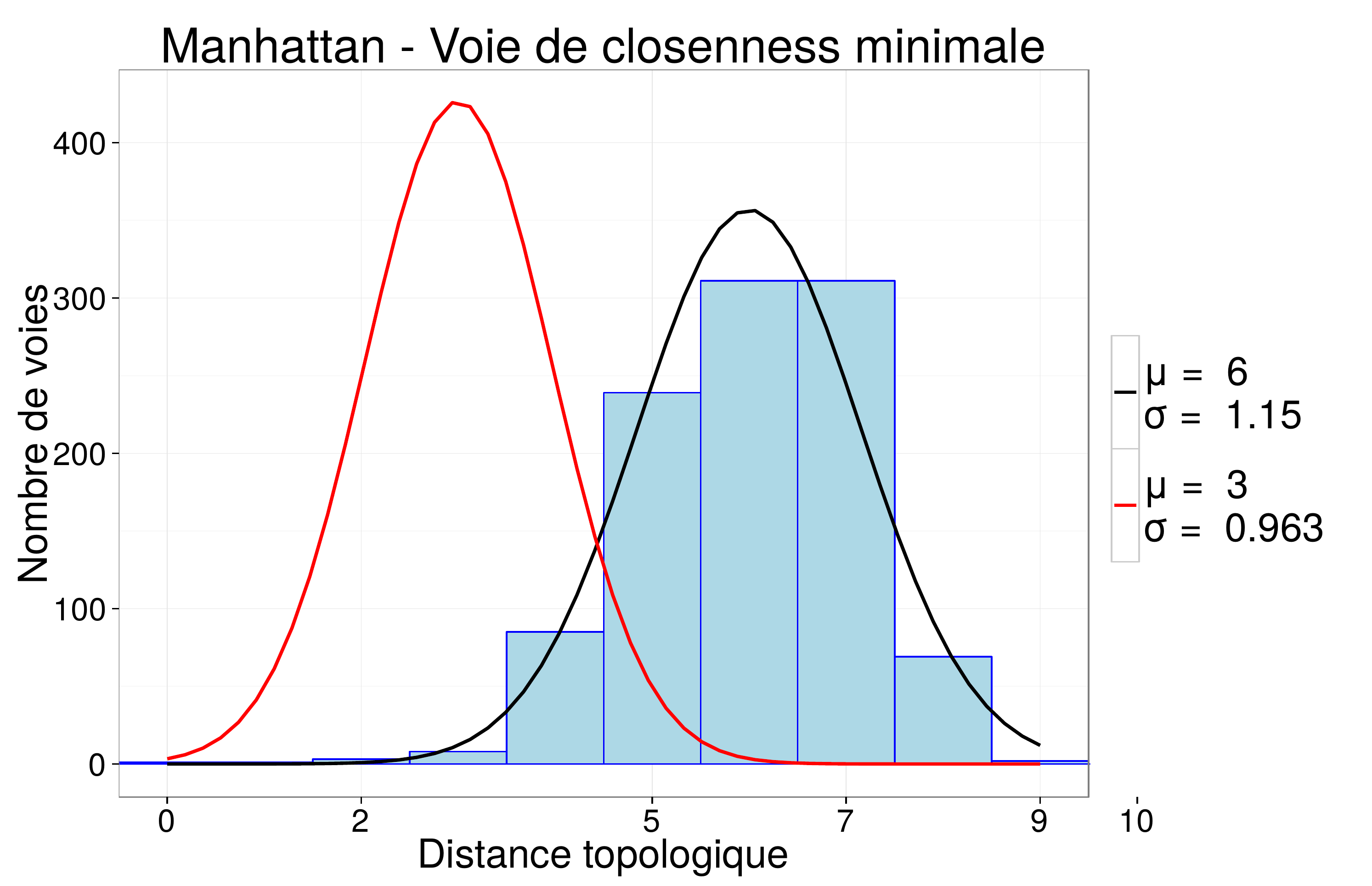}
        \caption{Pour la voie de closeness minimale}
    \end{subfigure}
    \caption{Répartition des \large{\textbf{distances topologiques}}. En rouge, gaussienne théorique}
\end{figure}

\FloatBarrier

\clearpage

\section{Paris}\label{ann:sec_dtopo_paris}

\begin{figure}[h]
    \centering
    \includegraphics[width=\textwidth]{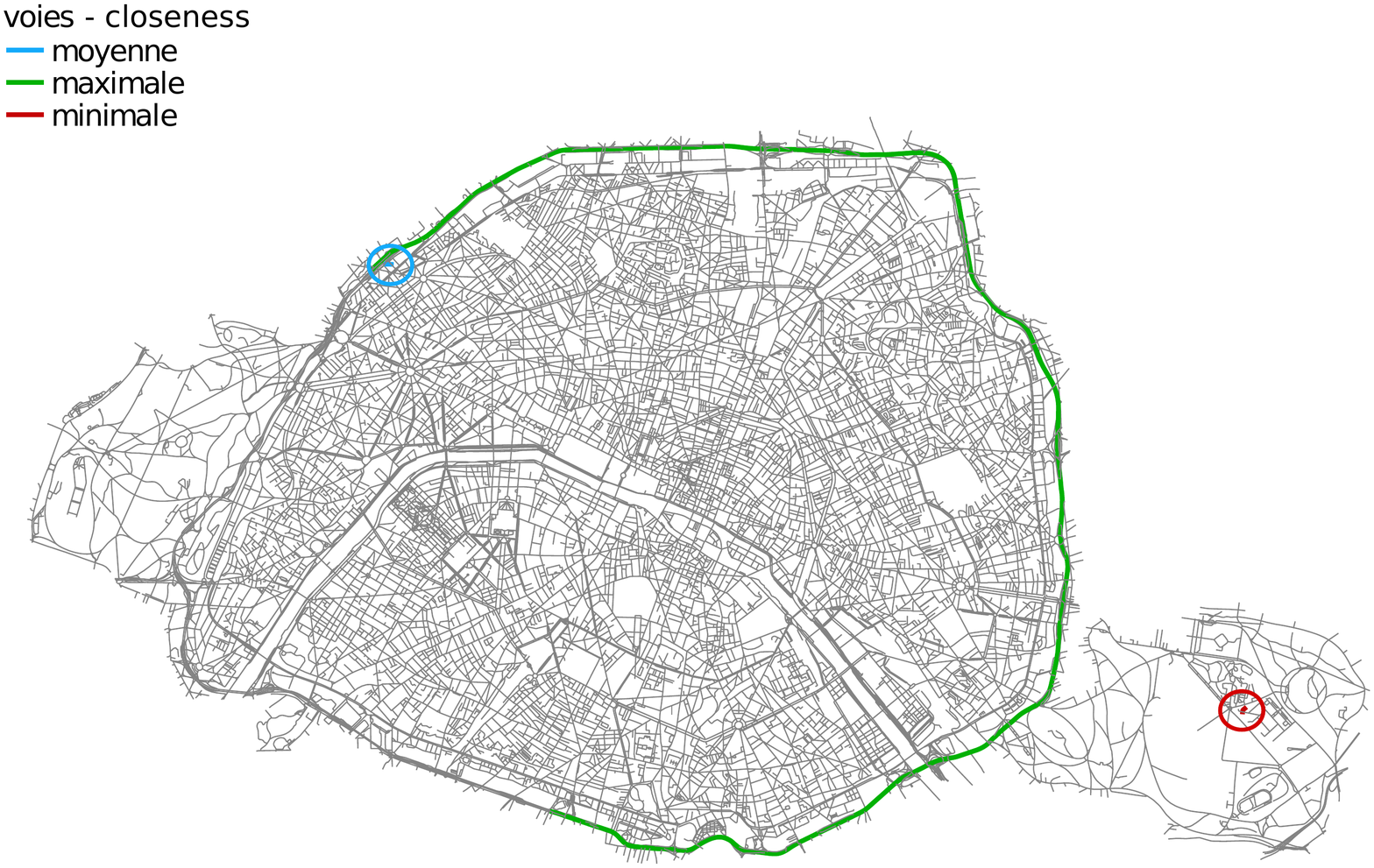}
    \caption{\large{\textbf{Situation des voies considérées}}}
\end{figure}

\begin{figure}[h]
    \centering
    \begin{subfigure}[t]{0.45\textwidth}
        \centering
        \includegraphics[width=\textwidth]{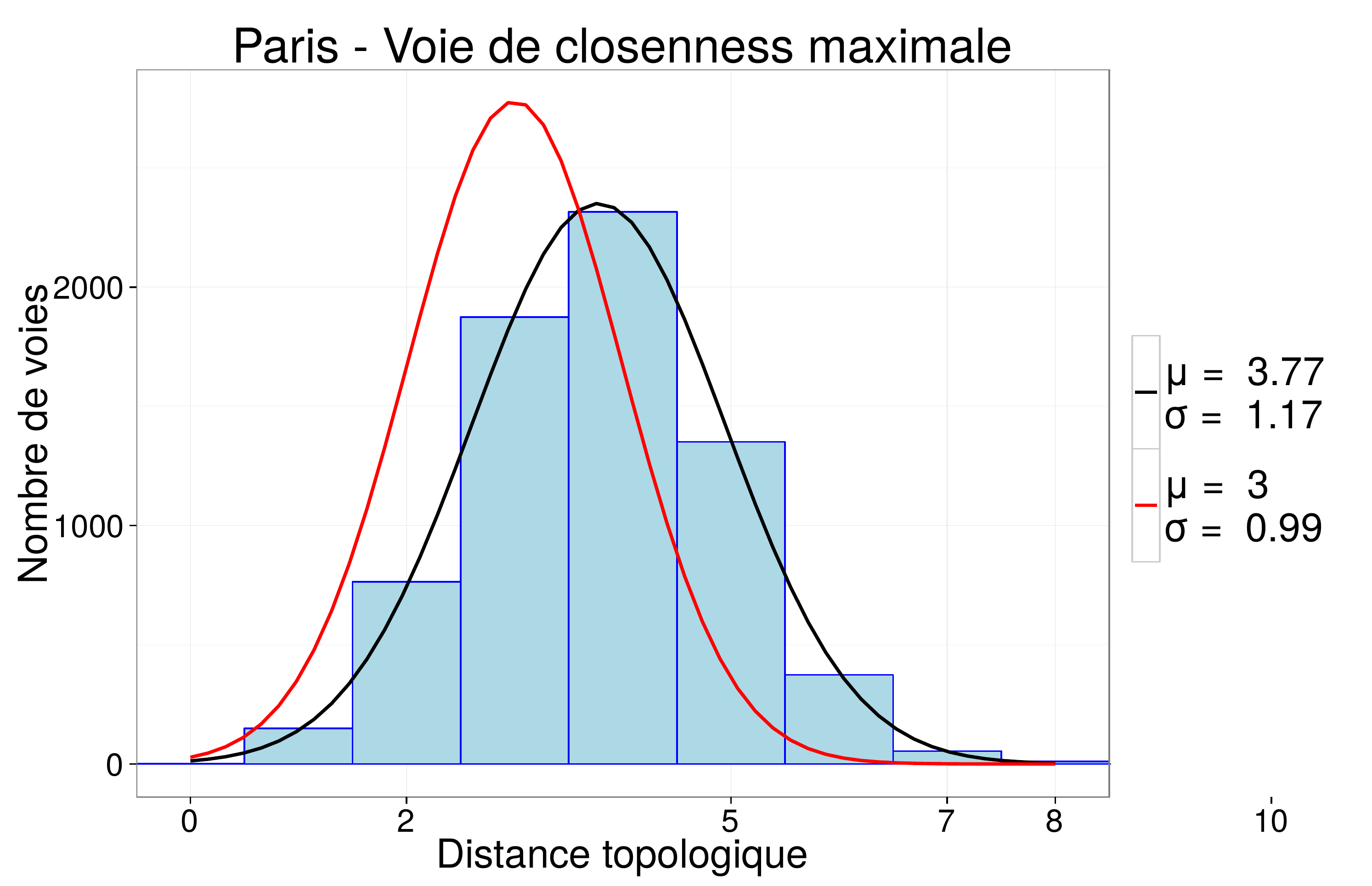}
        \caption{Pour la voie de closeness maximale}
    \end{subfigure}
    ~
    \begin{subfigure}[t]{0.45\textwidth}
        \centering
        \includegraphics[width=\textwidth]{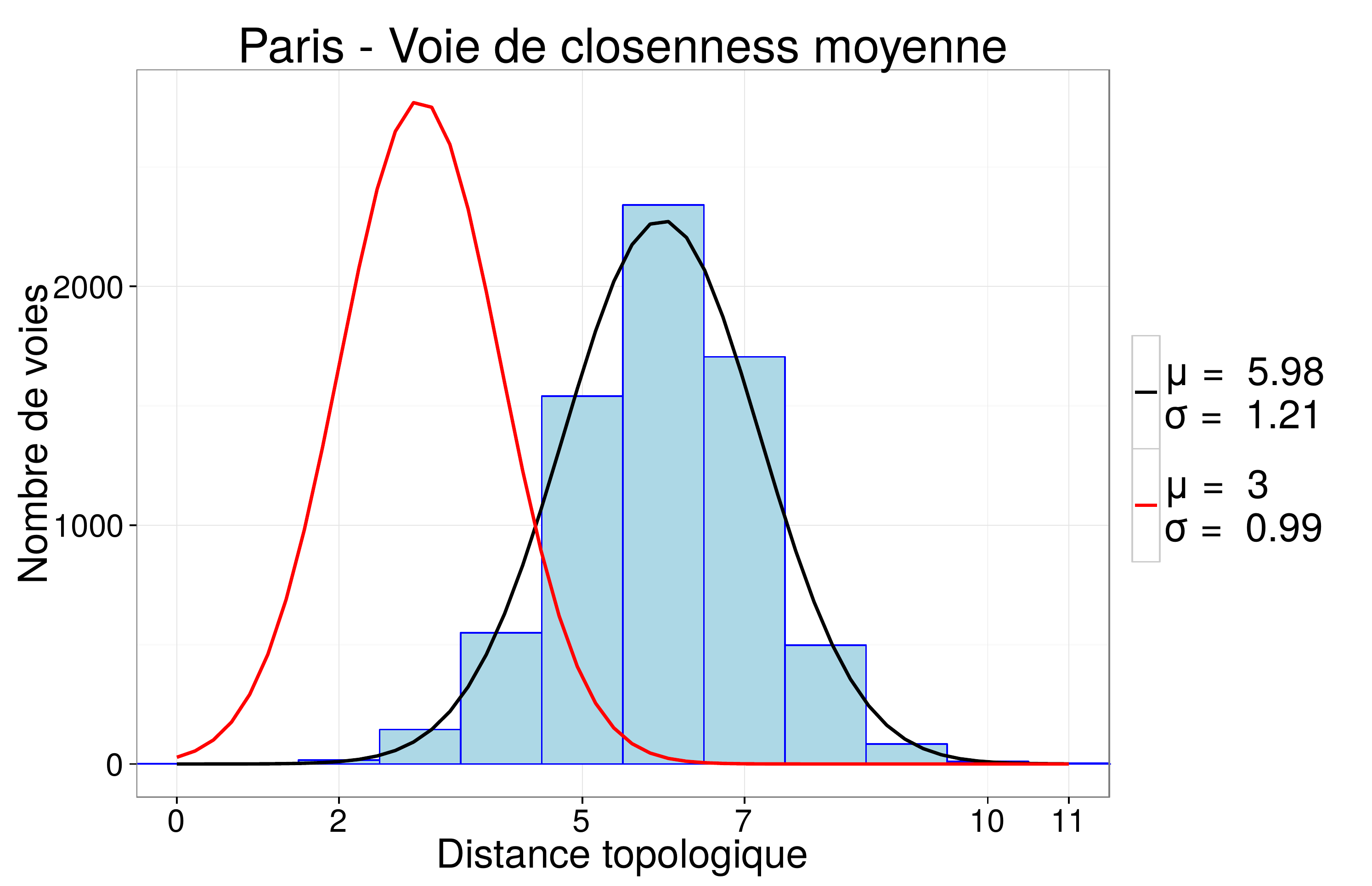}
        \caption{Pour une voie de closeness moyenne}
    \end{subfigure}
    ~
    \begin{subfigure}[t]{0.45\textwidth}
        \centering
        \includegraphics[width=\textwidth]{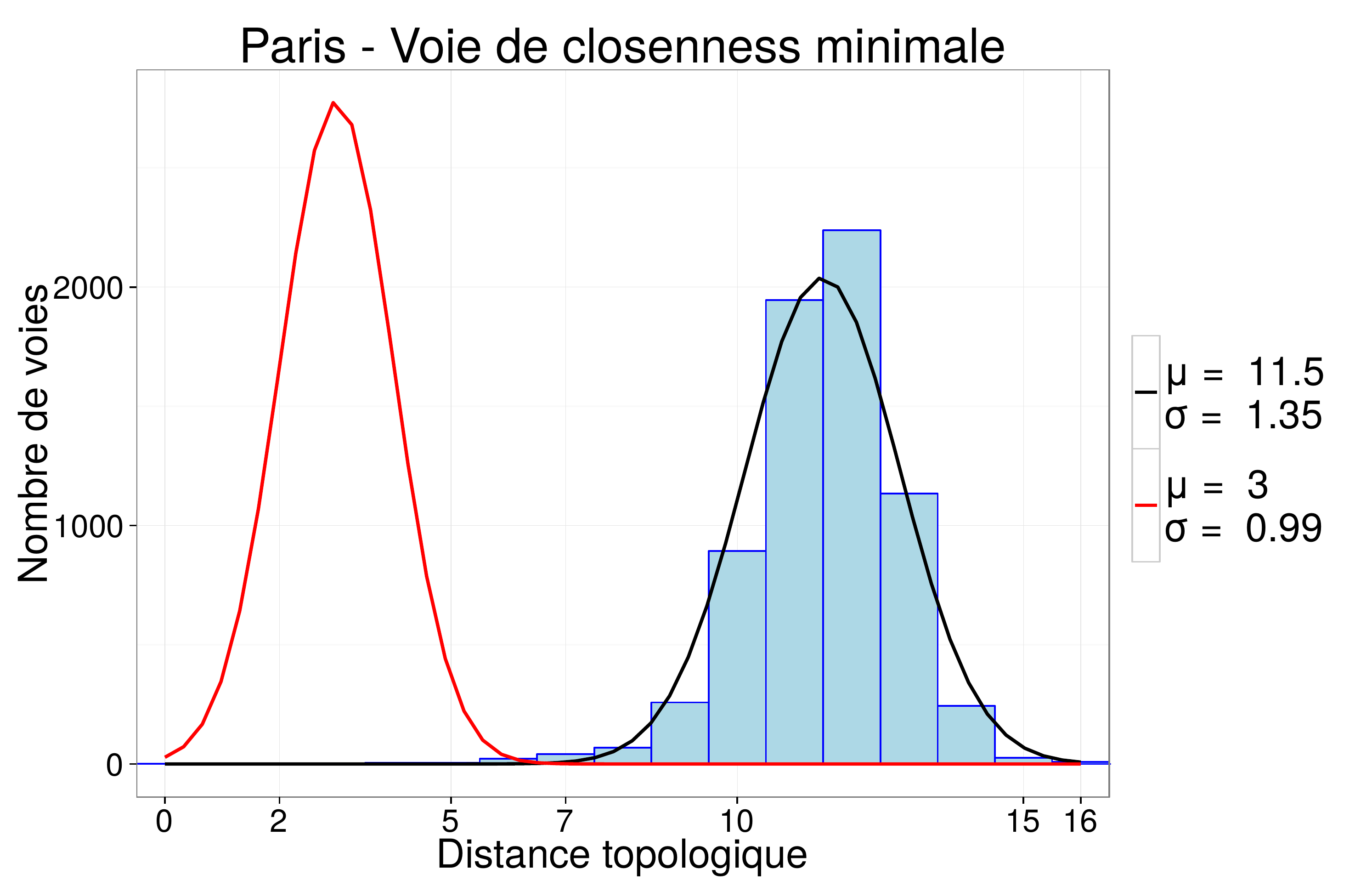}
        \caption{Pour la voie de closeness minimale}
    \end{subfigure}
    \caption{Répartition des \large{\textbf{distances topologiques}}. En rouge, gaussienne théorique}
\end{figure}

\FloatBarrier

\clearpage

\section{Téhéran}\label{ann:sec_dtopo_teheran}

\begin{figure}[h]
    \centering
    \includegraphics[width=\textwidth]{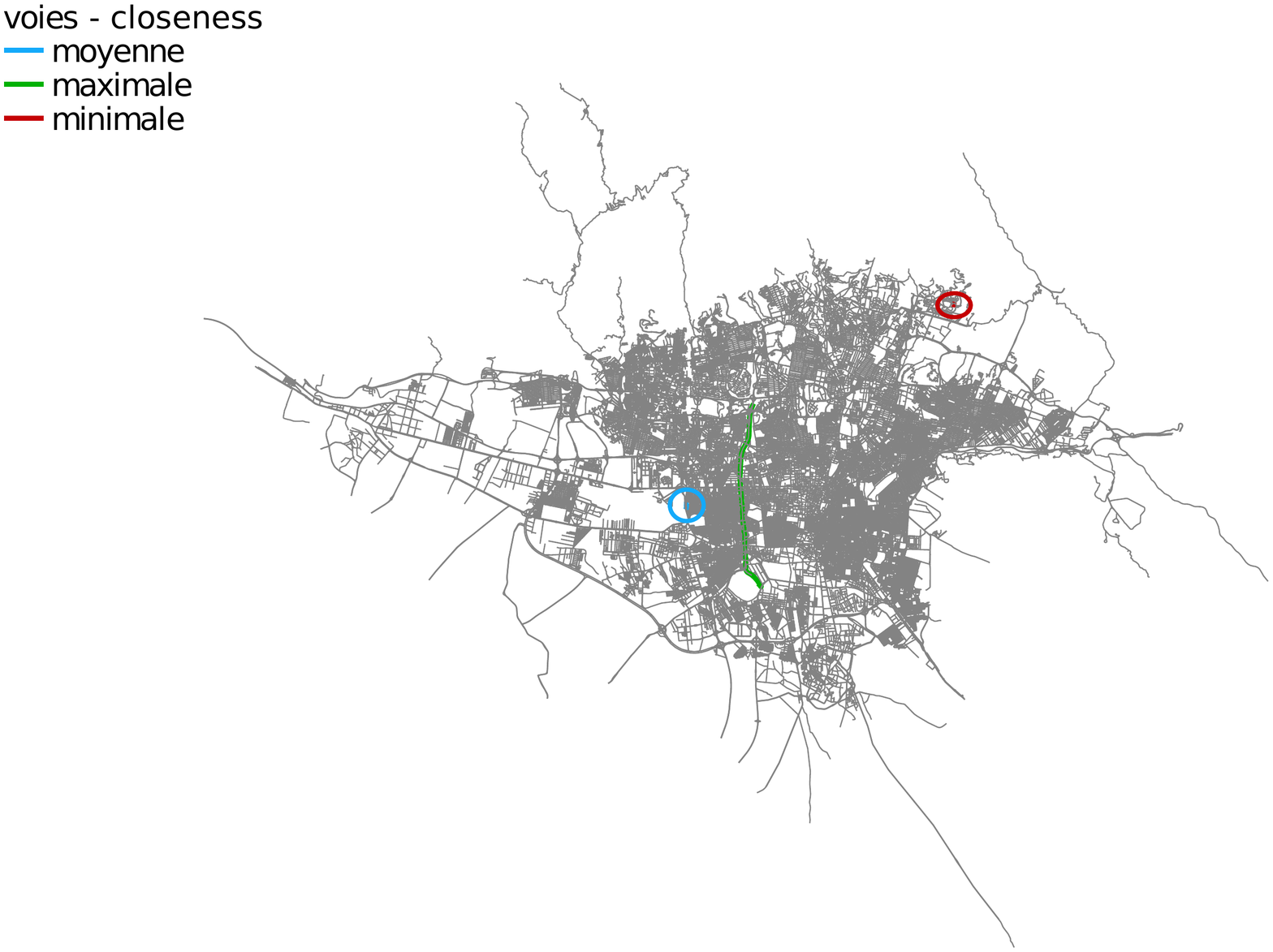}
    \caption{\large{\textbf{Situation des voies considérées}}}
\end{figure}

\begin{figure}[h]
    \centering
    \begin{subfigure}[t]{0.45\textwidth}
        \centering
        \includegraphics[width=\textwidth]{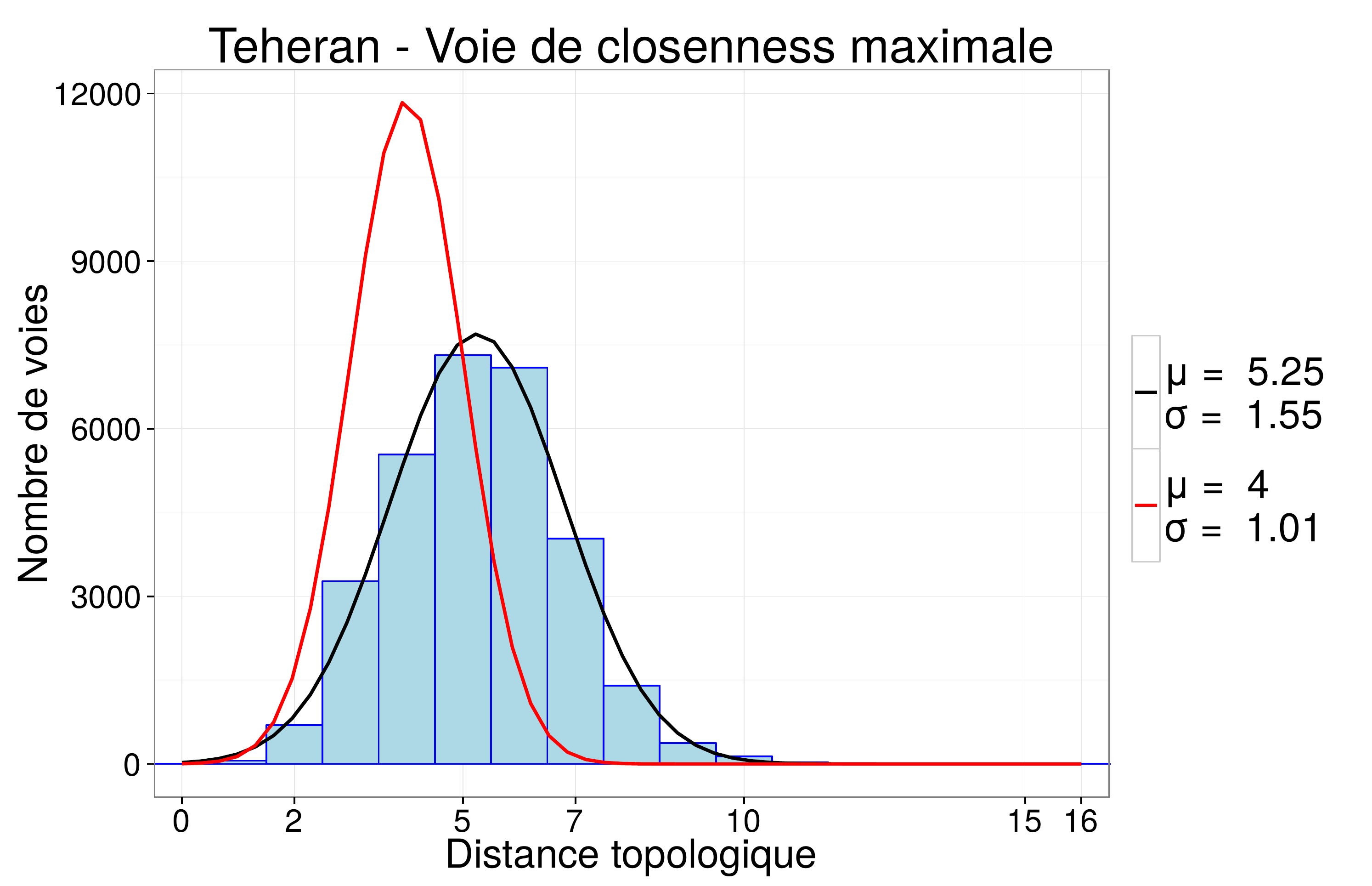}
        \caption{Pour la voie de closeness maximale}
    \end{subfigure}
    ~
    \begin{subfigure}[t]{0.45\textwidth}
        \centering
        \includegraphics[width=\textwidth]{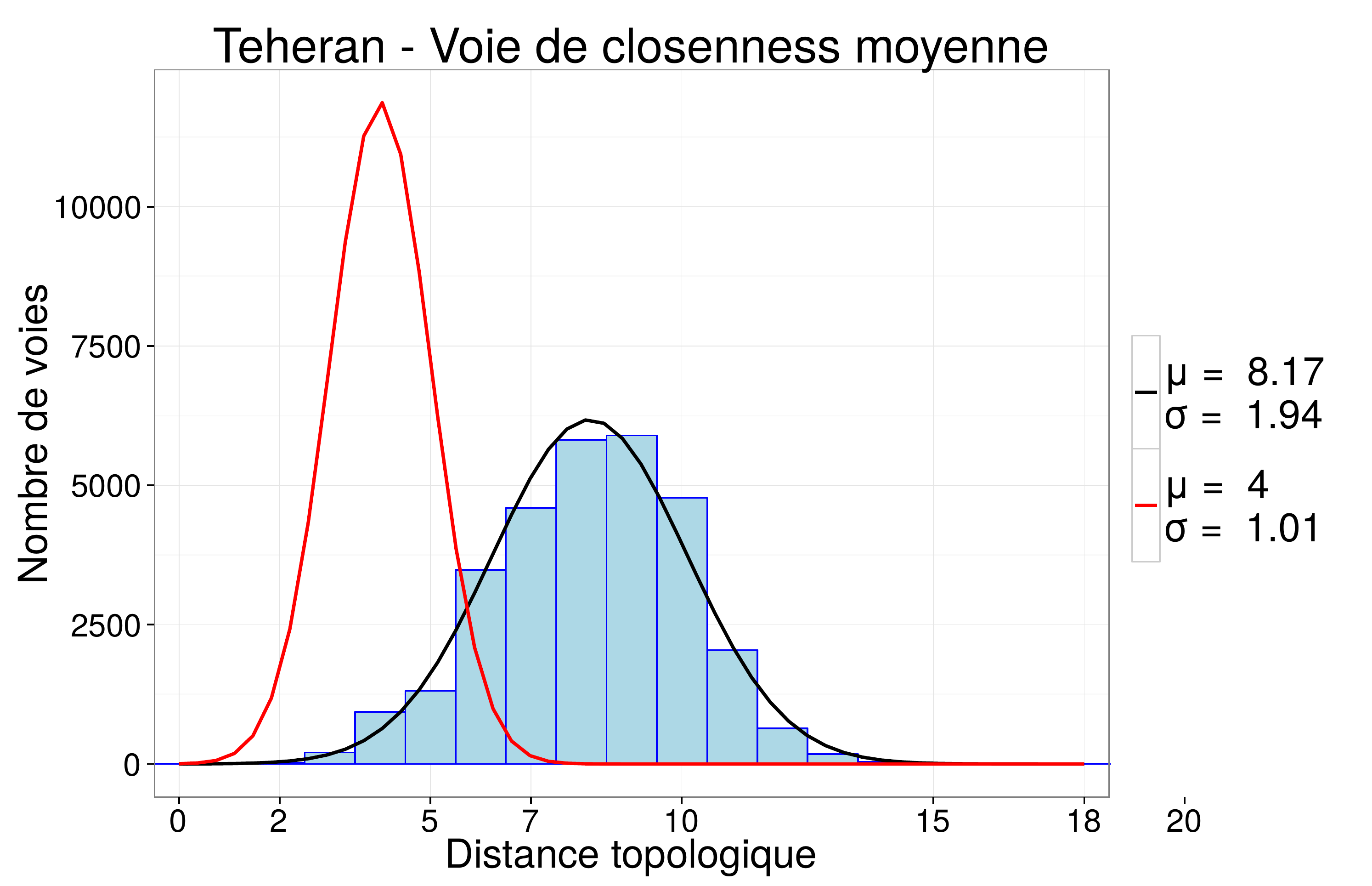}
        \caption{Pour une voie de closeness moyenne}
    \end{subfigure}
    ~
    \begin{subfigure}[t]{0.45\textwidth}
        \centering
        \includegraphics[width=\textwidth]{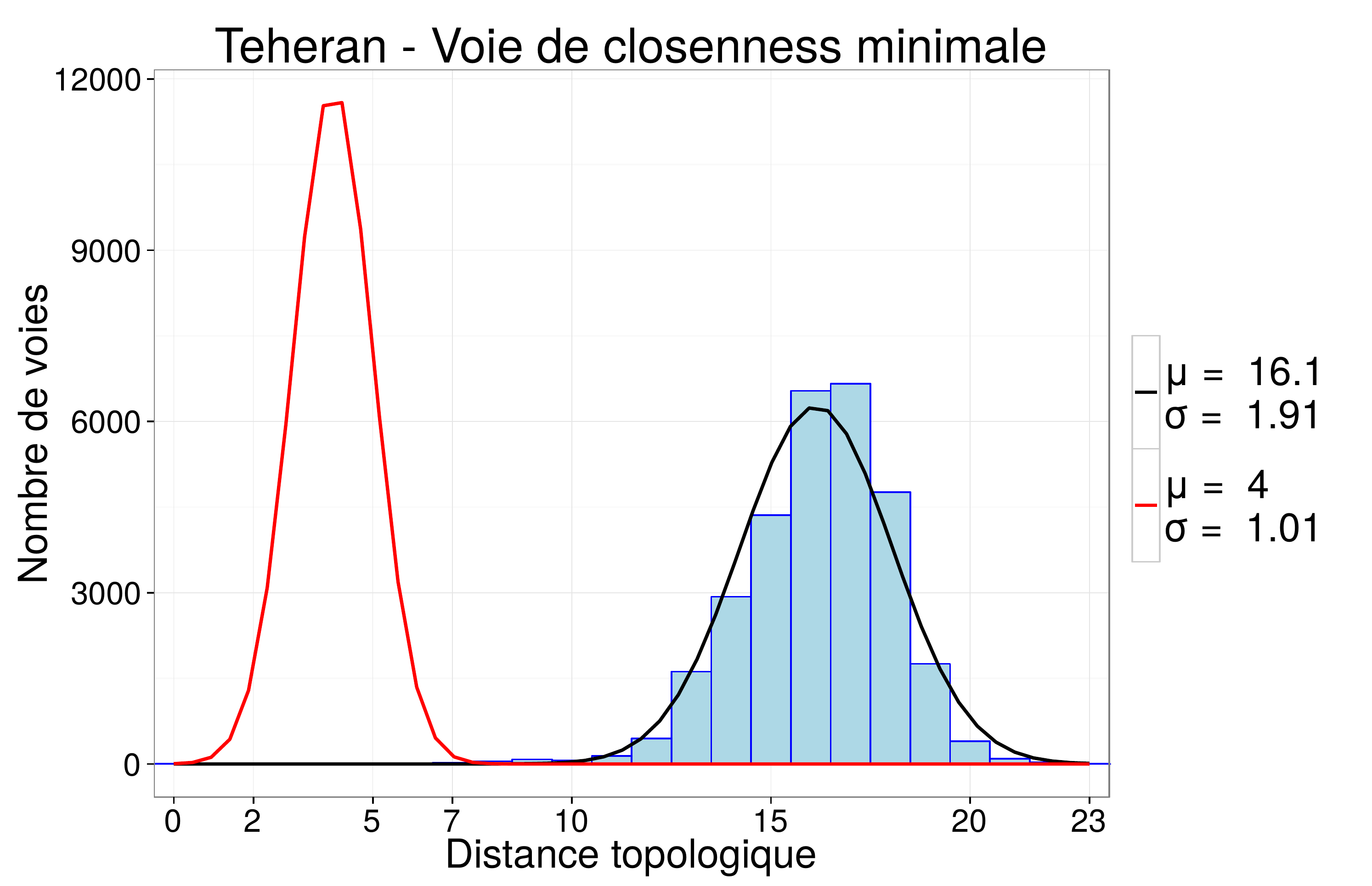}
        \caption{Pour la voie de closeness minimale}
    \end{subfigure}
    \caption{Répartition des \large{\textbf{distances topologiques}}. En rouge, gaussienne théorique}
\end{figure}

\FloatBarrier

\clearpage

\section{Téhéran-centre}\label{ann:sec_dtopo_teherancentre}

\begin{figure}[h]
    \centering
    \includegraphics[width=\textwidth]{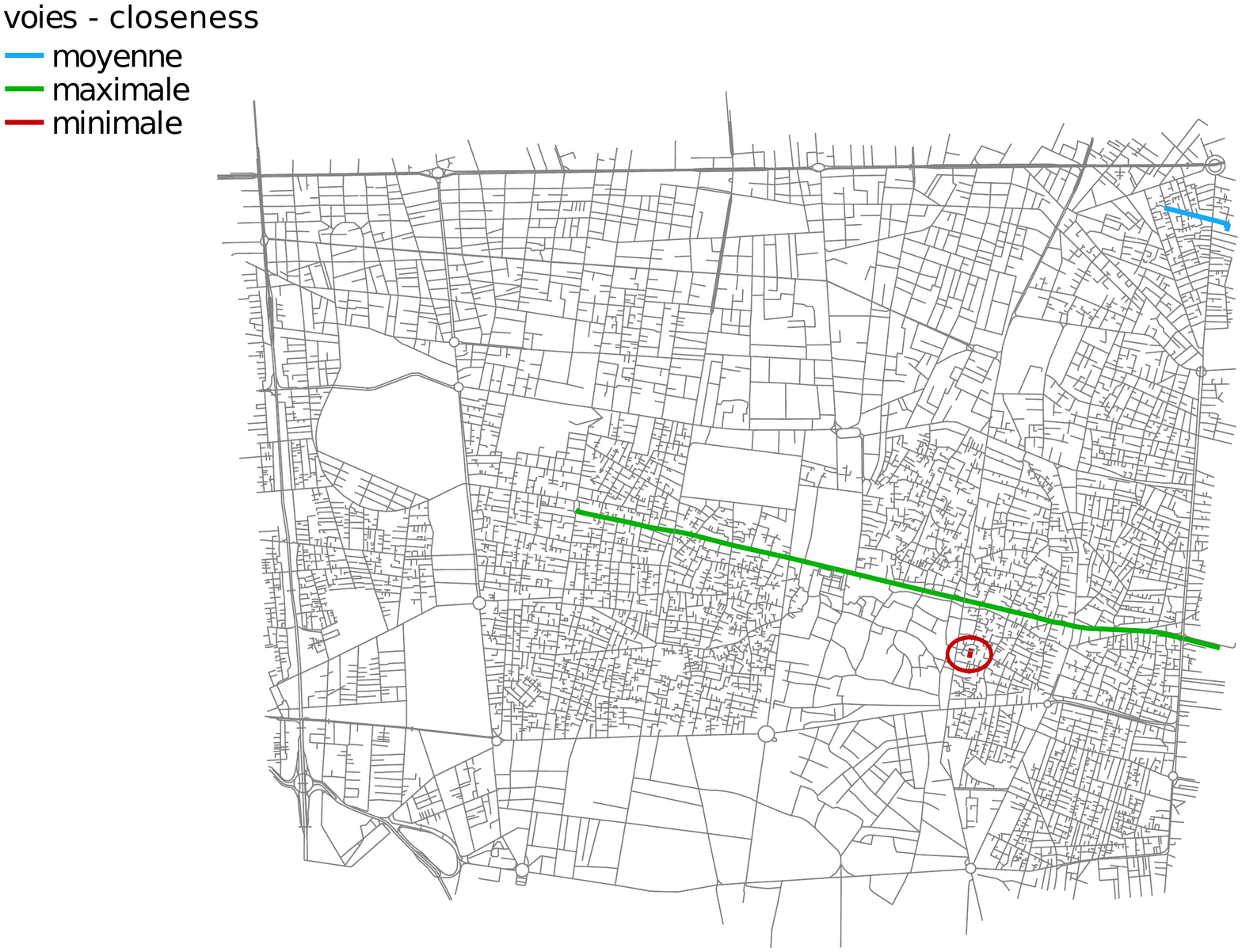}
    \caption{\large{\textbf{Situation des voies considérées}}}
\end{figure}

\begin{figure}[h]
    \centering
    \begin{subfigure}[t]{0.45\textwidth}
        \centering
        \includegraphics[width=\textwidth]{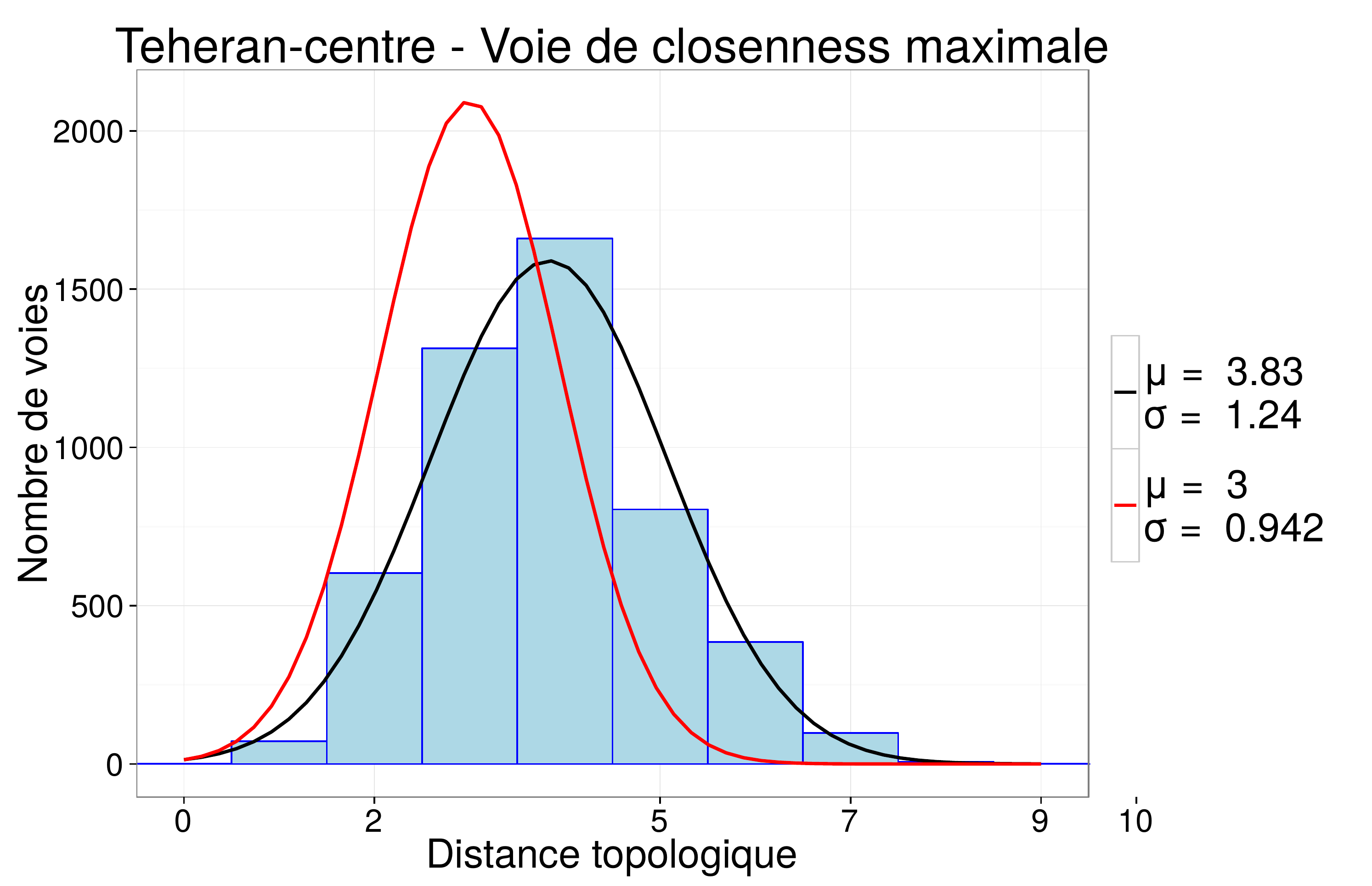}
        \caption{Pour la voie de closeness maximale}
    \end{subfigure}
    ~
    \begin{subfigure}[t]{0.45\textwidth}
        \centering
        \includegraphics[width=\textwidth]{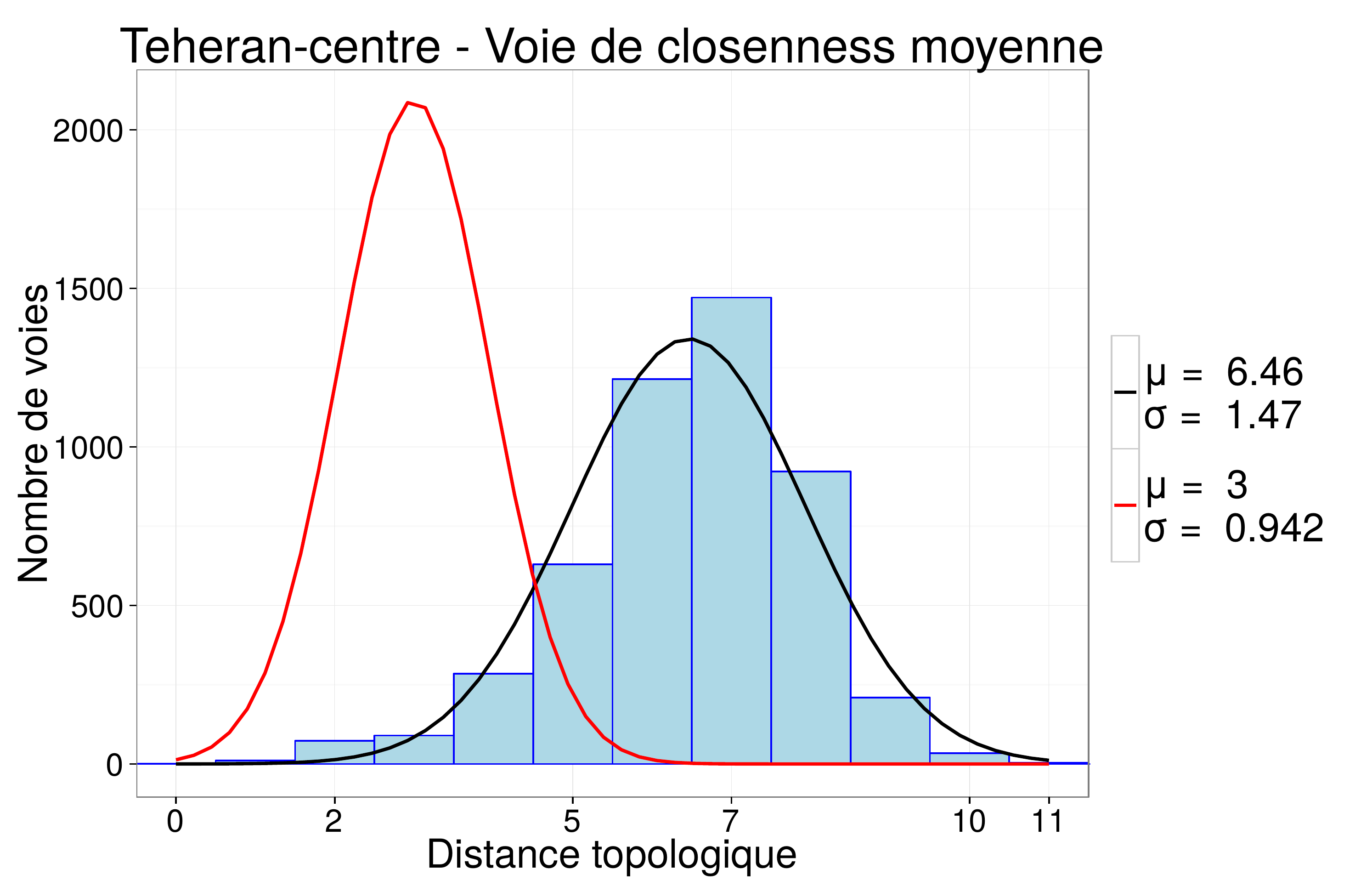}
        \caption{Pour une voie de closeness moyenne}
    \end{subfigure}
    ~
    \begin{subfigure}[t]{0.45\textwidth}
        \centering
        \includegraphics[width=\textwidth]{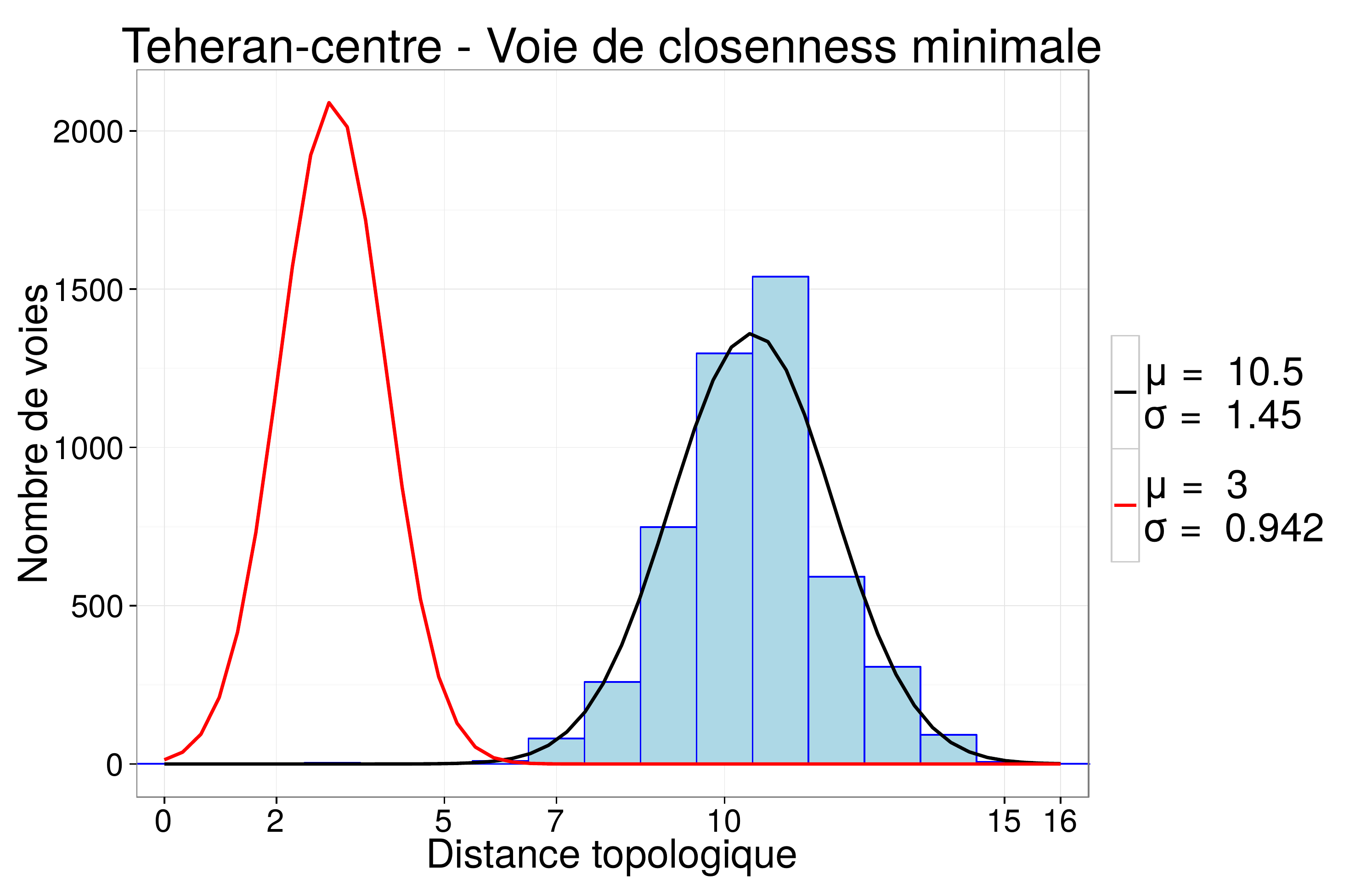}
        \caption{Pour la voie de closeness minimale}
    \end{subfigure}
    \caption{Répartition des \large{\textbf{distances topologiques}}. En rouge, gaussienne théorique}
\end{figure}

\FloatBarrier

\setlength{\parskip}{10pt}

\chapter{Présentation du logiciel de calcul des indicateurs}\label{ann:chap_tuto_1}

\section{Fonctionnement général}

Le programme a été écrit en C++, avec une interface graphique, permettant de :
\begin{itemize}
\item calculer les indicateurs à partir d'une donnée vecteur intégrée dans une base PostgreSQL. Cette fonctionnalité sera détaillée par la suite.
\item modifier une table déjà existante dans une base PostgreSQL : ajout d'une classification sur un des attributs, ajout d'un nouvel attribut composé de deux existants...
\end{itemize}

\begin{figure}[h]
\centering
\includegraphics[width=0.8\textwidth]{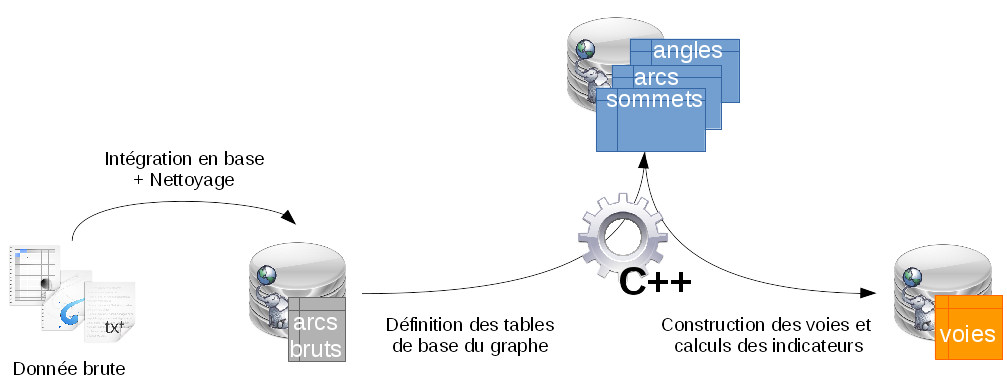}
\caption{Interactions entre le programme C++ et les données dans PostgreSQL}
\end{figure}

Le calcul des indicateurs se fait en deux étapes :

\begin{enumerate}
\item construction des tables de représentation (topologique et topographique) du graphe : les tables d'arcs (SIF) et de sommets (SXYZ) sont remplies à partir de la table brute des arcs. Les angles entre arcs à un sommet sont calculés pour préparer la construction des voies. Si ces tables existent déjà, il est possible de les réutiliser pour ne faire que l'étape liées de construction des voies.
\item construction des voies (en accord avec la méthode choisie) et calculs des indicateurs.
\end{enumerate}

Les indicateurs sont calculés en cascade, selon les choix de l’utilisateur.

\begin{figure}[h]
\centering
\includegraphics[width=0.8\textwidth]{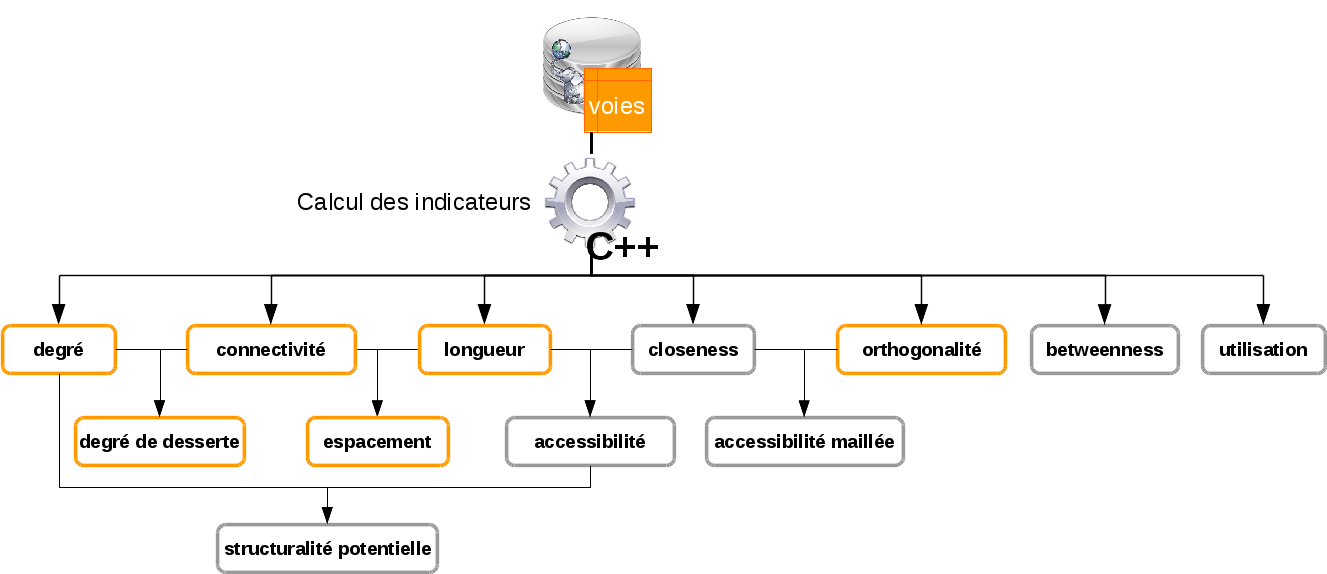}
\caption{Arborescence des indicateurs}
\end{figure}

\begin{figure}[h]
\centering
\includegraphics[width=0.8\textwidth]{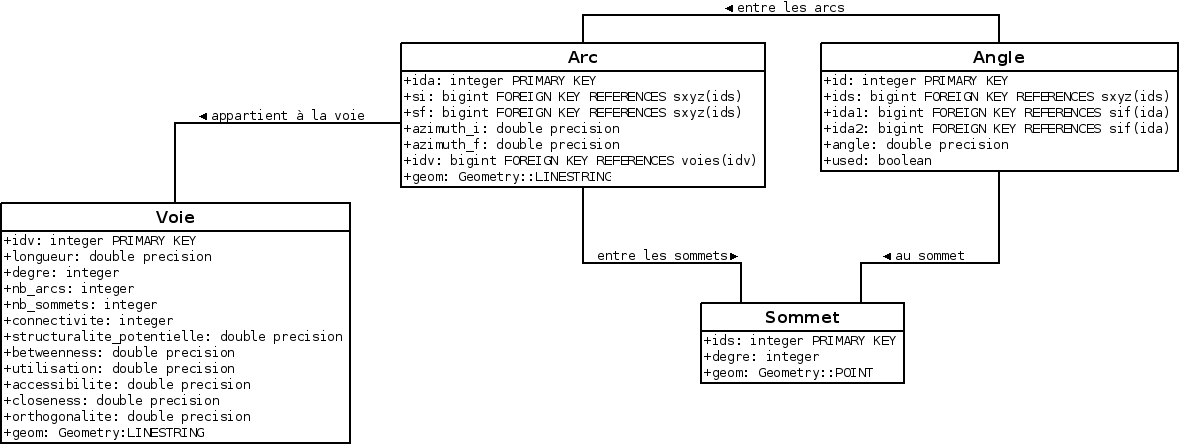}
\caption{Structure des tables créées lors du calcul d'indicateurs}
\label{fig:struct_tables}
\end{figure}

À la fin du traitement, nous avons donc les tables suivantes (figure \ref{fig:struct_tables}). C'est la table des voies qui contient les valeurs des indicateurs.

\section{Utilisation via son interface}

L'interface est composée de trois onglets. Le premier permet de configurer la base de données vecteur sur laquelle le calcul doit se faire, et de préciser les emplacements où écrire les résultats. Le deuxième onglet permet de configurer plus précisément les indicateurs à calculer, ainsi que leurs paramètres. Enfin, un dernier onglet permet de classifier des données selon un des attributs.

\subsection{Onglet de configuration de la source de données}

Que l'on veuille calculer les indicateurs ou ajouter une classification, il est indispensable de configurer la source de données.

Nous devons préciser (cadre 1) :

\begin{itemize}
\item L'hôte hébergeant la base de données PostgreSQL. Laisser 'localhost' dans le cas où la donnée est sur la même machine que celle où s'exécutera le programme.
\item Le nom de la base de données
\item Le nom d'utilisateur et le mot de passe pour se connecter (les droits de lecture et d'écriture doivent être accordés à cet utilisateur).
\item Le nom du schéma et de la table sur laquelle le calcul doit se faire.
\begin{itemize}
\item Dans le cas du calcul d'indicateurs, ce doit être la table des arcs bruts du réseau
\item Dans le cas d'une modification, ce doit être la table à laquelle on veut ajouter un attribut
\end{itemize}
\end{itemize}

Pour le calcul des indicateurs, des nouvelles données sont écrites, à la fois dans la base de données et sous forme de fichiers. Les nouvelles tables seront écrites dans la même base de données, mais peuvent être créées dans un schéma différent. Il est possible de préciser ce schéma (cadre 2). Les fichiers seront quant à eux écrits dans le dossier renseigné (cadre 2).

Dans le cas d'une modification de table, il n'y a pas de création de fichier ou de nouvelle table.

\begin{figure}[h]
\centering
\includegraphics[width=0.8\textwidth]{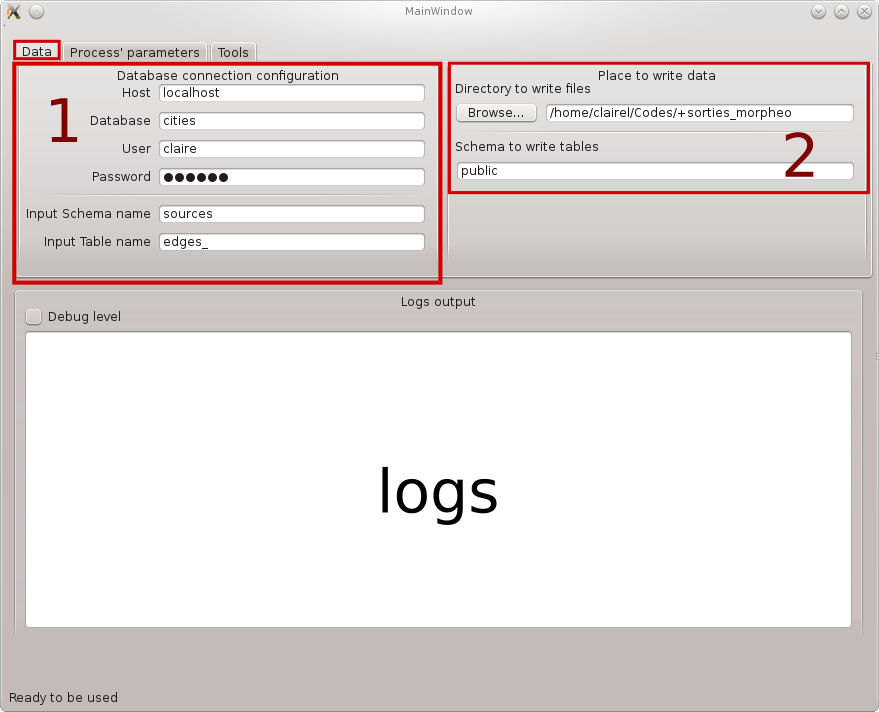}
\caption{Onglet de configuration de la source de données.}
\end{figure}

\FloatBarrier

\subsection{Onglet de configuration du calcul des indicateurs}

Lorsque l'on souhaite calculer les indicateurs sur un réseau, il est nécessaire de préciser les paramètres de ce calcul (les numéros correspondent à ceux des cadres) :

\begin{enumerate}
\item Comment construit-on les voies à partir des arcs : selon quelle méthode, avec quel angle seuil, quelle tolérance pour l'identification des sommets initiaux et finaux des arcs ?
\item Doit-on supprimer les tables à créer si elles sont déjà présentes dans la base ? Sous certaines conditions, il est possible de reprendre le travail à partir des tables déjà existantes.
\item Doit-on construire des tables supplémentaires ? Les rues sont reconstituées à partir des arcs en se basant sur la toponymie.
\item Quels indicateurs supplémentaires doit-on calculer ?
\end{enumerate}

\begin{figure}[h]
\centering
\includegraphics[width=0.8\textwidth]{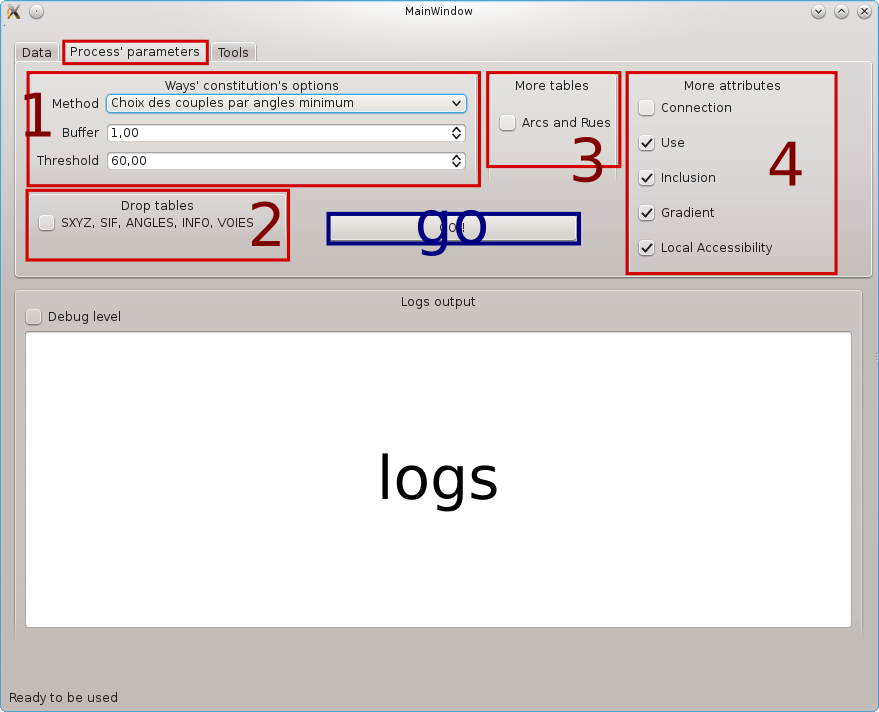}
\caption{Onglet de configuration du calcul des indicateurs}
\end{figure}

Des fichiers sont écrits, contenant les distances topologiques et les adjacences entre voies sous forme de matrices triangulaires.

\FloatBarrier

\subsection{Onglet de configuration du calcul de classification}

Il est possible d'ajouter de nouveaux attributs à une table déjà existante. Le choix du mode de calcul du nouvel attribut se fait dans le cadre 1.

\begin{itemize}
\item Combinaison de deux attributs : par addition, soustraction, multiplication, division ou différence absolue. Dans ce cas, on précise les deux attributs sources et le nom du nouveau (cadre 2) .
\item Classification selon un attribut. On doit alors préciser l'attribut source, le nom du nouveau, le nombre de classes et le sens de la classification (croissante ou décroissante).  La table doit contenir un attribut de longueur. En effet, la classification est faite de manière à ce que la somme des longueurs des objets dans une classe soit toujours la même.
\end{itemize}

\begin{figure}[h]
\centering
\includegraphics[width=0.8\textwidth]{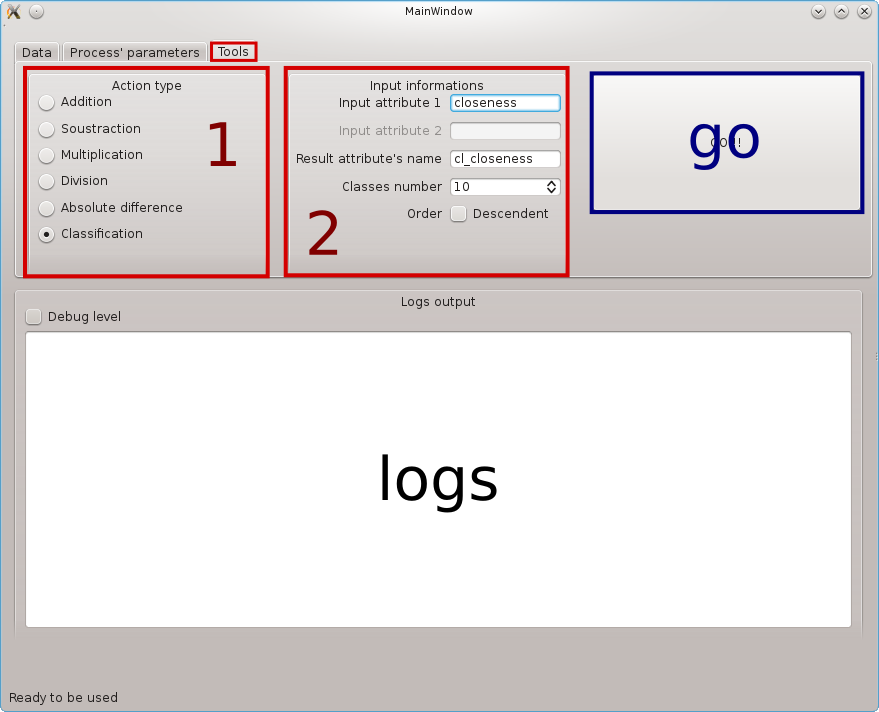}
\caption{Onglet de configuration du calcul de classification}
\end{figure}

\FloatBarrier
\setlist{nosep,nosep}
\setlength{\parskip}{10pt}

\chapter{Présentation du plugin QGIS \protect \footnote{Plugin développé en collaboration avec la société Oslandia.}}\label{ann:chap_tuto_2}

\section{Installation du plugin sous QGIS}

Après avoir installé QGIS sur votre ordinateur, et récupéré l'archive de Morpheo, décompressez-la et déposez les fichiers dans le dossier :

\begin{itemize}
\item Sous Linux : \enquote{\textasciitilde{}/.qgis2/python/plugins}
\item Sous Windows :\enquote{C:\textbackslash{}Users\textbackslash{}\{username\}\textbackslash{}.qgis2\textbackslash{}python\textbackslash{}plugins}
\item Sous Mac : dans les paquets QGIS, \enquote{contents / Resources / Python / Plugins}
\end{itemize}

\section{Utilisation du plugin sous QGIS}

Une fois QGIS ouvert, commencez par charger les données sources, à partir desquelles seront faits la construction des voies et le calcul les différents indicateurs.

\begin{figure}[H]
	\centering
	\includegraphics[width=0.7\textwidth]{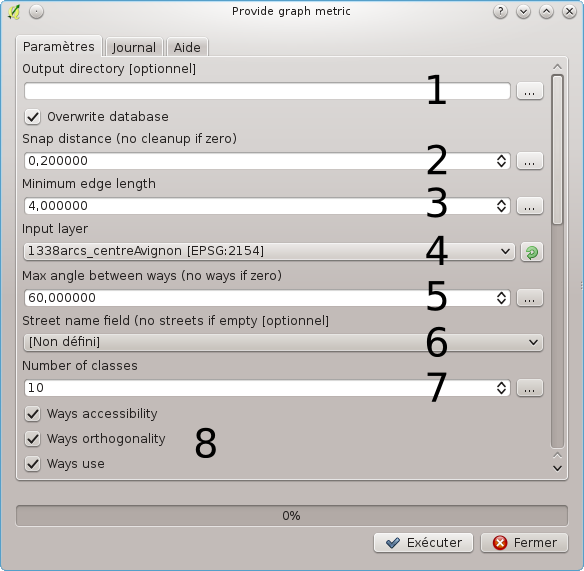}
	\caption{Interface de configuration du calcul des indicateurs.}
	\label{ann:tuto2_interface}
\end{figure}

Ensuite, configurez le traitement de ces données, via l'interface du plugin (figure \ref{ann:tuto2_interface}). Dans l'ordre :

\begin{enumerate}
\item Pour sauvegarder les données géographiques au format SQLite, renseigner le dossier dans lequel écrire (voire remplacer) ces fichiers.
\item L'option \enquote{Snap Distance} permet de corriger les légères erreurs de vectorisation et de connecter les géométries dont l'écart est moins important que le seuil fixé. Cette distance est dans l'unité du système de projection des données sources.
\item L'option \enquote{Minimum edge length} permet de supprimer des données les arcs trop petits. Si un arc a une longueur inférieur à cette valeur, il est supprimé et les arcs voisins se raccordent en son centroïde. La longueur est dans l'unité du système de projection des données sources.
\item Choix de la couche des données sources.
\item Choix de l'angle seuil utilisé pour la construction des voies.
\item Pour reconstruire les rues, préciser l'attribut contenant la toponymie.
\item Tous les indicateurs qui sont calculés donnent lieu à une classification : le nombre de classes à réaliser doit être précisé dans ce champ.
\item Choix des indicateurs à calculer pour les voies:
\begin{itemize}
\item l'accessibilité
\item l'orthogonalité
\item l’utilisation
\end{itemize}
\end{enumerate}

Sont calculés par défaut :
\begin{itemize}
\item la longueur
\item le degré
\item la connectivité
\item l'espacement
\end{itemize}

Il est également possible de calculer ces trois indicateurs pour les arcs, ainsi que pour les rues, en cochant les checkboxes appropriées.

\begin{figure}[H]
    \centering    
    \begin{subfigure}[t]{.2\linewidth}
        \includegraphics[width=\textwidth]{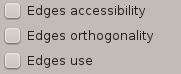}
    \end{subfigure}
     ~
    \begin{subfigure}[t]{.2\linewidth}
        \includegraphics[width=\textwidth]{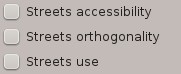}
    \end{subfigure}   
    
    \caption{Interface de choix d'indicateurs à calculer sur les arcs et les rues.}
\end{figure}

Les valeurs des indicateurs sont exportables au format CSV (sans les géométries des objets auxquels ils sont appliqués). Pour cela, il suffit de préciser les chemins vers les fichiers où écrire les données:

\begin{itemize}
\item sur les sommets
\item sur les arcs
\item sur les rues
\item sur les voies
\item sur les angles (pour chaque sommet, angle entre les arcs qui s'y intersectent ; le sommet et les arcs concernés sont précisés dans la table attributaire)
\end{itemize}

\begin{figure}[H]
	\centering
	\includegraphics[width=0.7\textwidth]{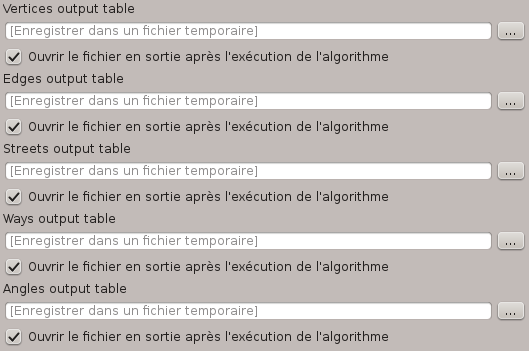}
	\caption{Interface de configuration des chemins vers les fichiers CSV en sortie.}
\end{figure}

Lorsque tout est paramétré, lancer le calcul en cliquant sur le bouton \enquote{Exécuter}. Des messages sont affichés dans l'onglet \enquote{Journal} afin de connaître l'avancement du traitement.

\begin{figure}[H]
	\centering
	\includegraphics[width=0.6\textwidth]{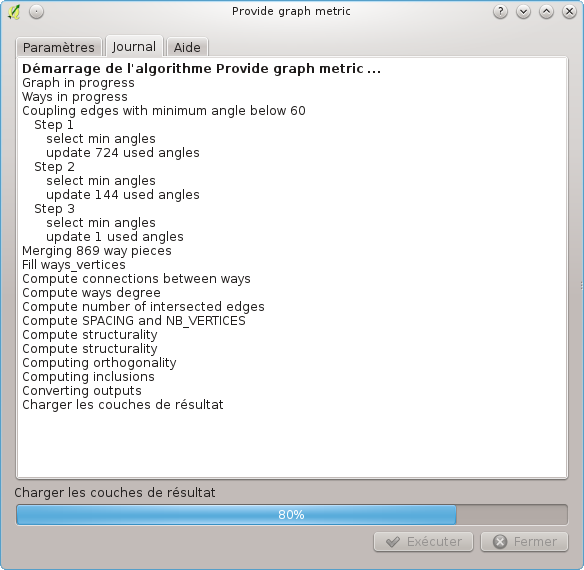}
	\caption{Affichage des logs pendant le traitement.}
\end{figure}

Lorsque le traitement est correctement terminé, plusieurs couches de données ont été ajoutées au projet QGIS :

\begin{itemize}
\item les données géométriques (potentiellement sauvegardées au format SQLite), avec leurs tables attributaires respectives contenant les indicateurs calculés.
\begin{itemize}
\item les sommets
\item les arcs
\item les voies
\end{itemize}
\item les données sous forme de table (potentiellement sauvegardées au format CSV)
\end{itemize}

\begin{figure}[H]
	\centering
	\includegraphics[width=0.4\textwidth]{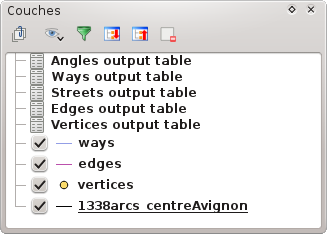}
	\caption{Couches générées lors du traitement.}
\end{figure}

\begin{figure}[H]
	\begin{subfigure}{0.45\textwidth}
		\centering
		\includegraphics[width=\textwidth]{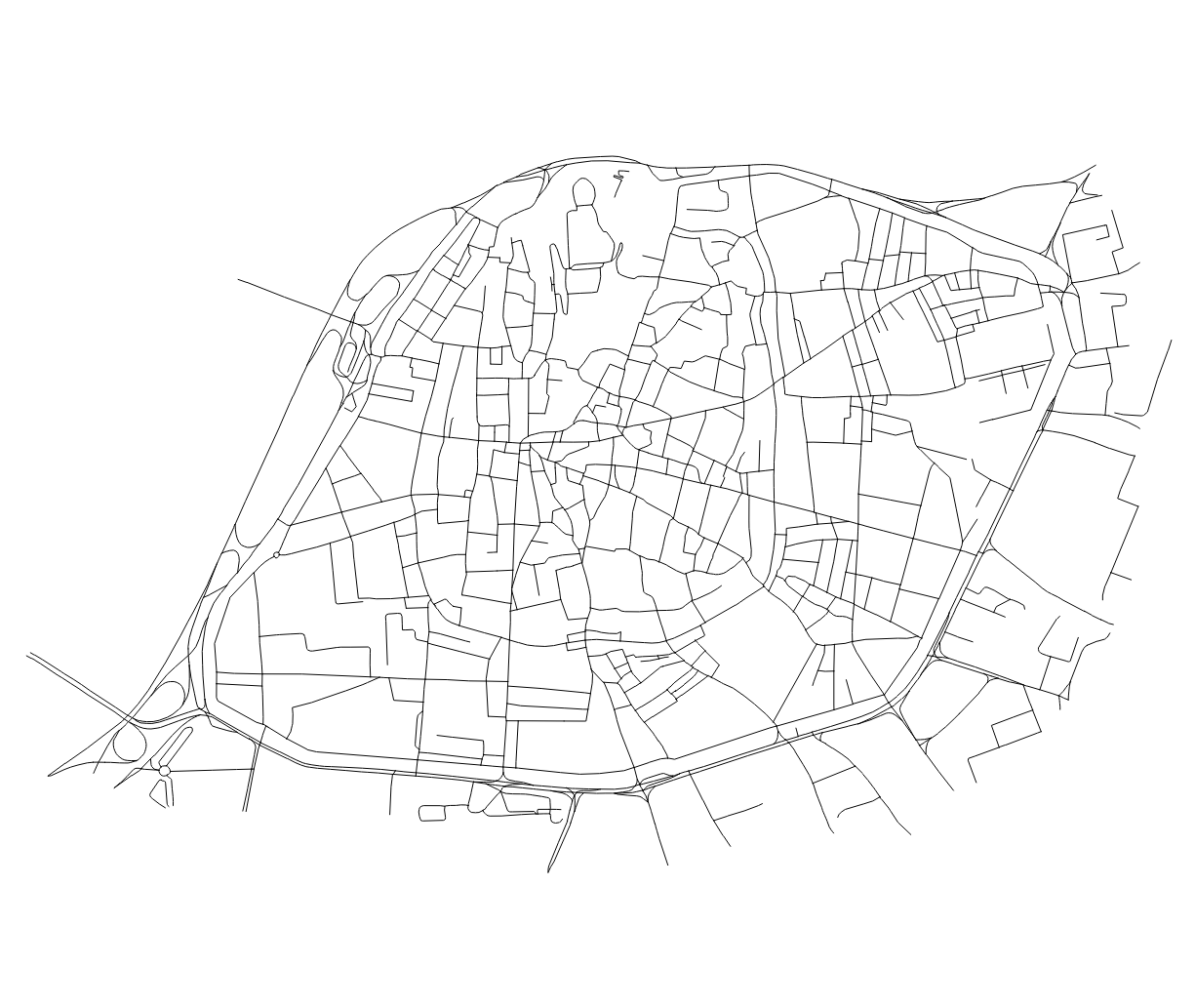}
		\caption{Cartographie avant traitement.}
	\end{subfigure}
	~
	\begin{subfigure}{0.45\textwidth}
		\centering
		\includegraphics[width=\textwidth]{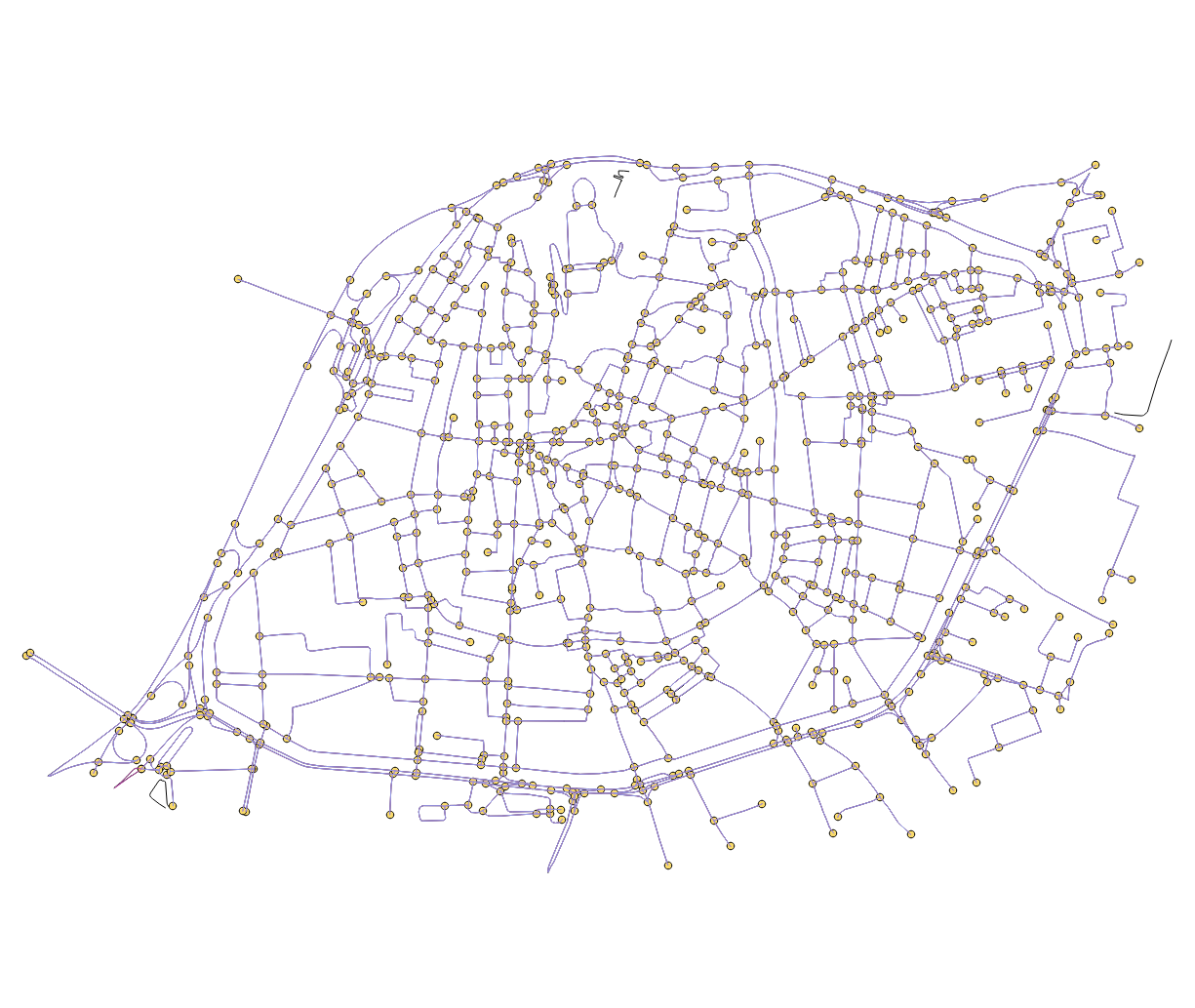}
		\caption{Cartographie après traitement.}
	\end{subfigure}
\end{figure}

\restoregeometry

\end{document}